\documentclass[12pt]{article}
\usepackage[linesnumbered,ruled,vlined]{algorithm2e}
\usepackage[english]{babel}
\usepackage[justification=centering]{caption}
\usepackage{wrapfig, color}
\usepackage[margin=1in]{geometry}
\usepackage{diagbox,threeparttable} 
\usepackage[sort, round]{natbib}
\usepackage{amsmath,amsthm, amssymb,enumitem,esvect,bbm, cases}
\usepackage{graphicx, wrapfig}
\usepackage{comment, setspace, empheq, url}
\usepackage[T1]{fontenc}
\usepackage{hyperref}
\hypersetup{colorlinks = true, citecolor=blue}
\usepackage{color}
\usepackage{textcomp, multirow}
\usepackage[linesnumbered,ruled,vlined]{algorithm2e}
\usepackage{algorithmic, pdflscape}
\usepackage{mathtools, nccmath}
\usepackage{standalone}
\usepackage[compact]{titlesec}  
\titlespacing*{\section}{0pt}{0pt}{0pt}
\titlespacing*{\subsection}{0pt}{0pt}{0pt}
\DeclarePairedDelimiter{\nint}\lfloor\rceil

\allowdisplaybreaks
\def\thm@space@setup{%
  \thm@preskip=0pt plus 1pt minus 1pt
  \thm@postskip=\thm@preskip 
}

\newtheorem{defn}{Definition}
\newtheorem{theorem}{Theorem}
\newtheorem{corollary}[theorem]{Corollary}
\newtheorem{lemma}[theorem]{Lemma}
\newtheorem{con}[theorem]{Condition}

\newcommand{\M}{\mathcal{M}}
\def\BibTeX{{\mathcal{M}m B\kerP-.05em{\sc i\kerP-.025em b}\kerP-.08em
    T\kerP-.1667em\lower.7ex\hbox{E}\kerP-.125emX}}

\begin{document}
\title{\LARGE{Privacy-preserving Inference of Group Mean Difference in Zero-inflated Right Skewed Data with Partitioning and Censoring }}
\author{Fang Liu$^{1}$\footnote{$\;$Liu and Zhou are co-first authors. Correspondence author: fang.liu.131@nd.edu}, Ruyu Zhou$^{1*}$, Yiming Paul Li$^2$, James Honaker$^2$, Milan Shen$^2$\\
$^1$\small Applied and Computational Mathematics \& Statistics, \\ 
\small University of Notre Dame, IN, USA, 45630\\
$^2$\small Meta Platforms, Inc. } 
\date{}
\maketitle\vspace{-25pt}

\begin{abstract}
We examine privacy-preserving  inferences of group mean differences in zero-inflated right-skewed (zirs) data. Zero inflation and right skewness are typical characteristics of ads clicks and purchases data collected from e-commerce and social media platforms, where we also want to preserve user privacy to ensure that individual data is protected. 
In this work, we develop likelihood-based and model-free approaches to analyzing zirs data with formal privacy guarantees.  We first apply partitioning and censoring  (PAC) to ``regularize'' zirs data to get the PAC data.  We expect inferences based on PAC to have better inferential properties and more robust privacy considerations compared to analyzing the raw data directly. We conduct theoretical analysis to establish the MSE consistency of the privacy-preserving estimators from the proposed approaches based on the PAC data and examine the rate of convergence in the number of partitions $P$ and privacy loss parameters.  The theoretical results  also suggest that it is the sampling error of PAC data rather than the sanitization error that is the limiting factor in the convergence rate. We conduct extensive simulation studies to compare the inferential utility of the proposed approach for different types of zirs data,  sample size and partition size combinations, censoring scenarios, mean differences,  privacy budgets, and privacy loss composition schemes. 
We also apply the methods to obtain privacy-preserving inference for the group mean difference in a real digital ads click-through data set. Based on the theoretical and empirical results, we make recommendations regarding the usage of these methods in practice. 

\vspace{6pt}
\noindent \textbf{keywords}: partitioning and censoring (PAC),  differential privacy,  group mean difference,  zero inflation right skewness (zirs), Bayesian, MSE consistency, privacy-preserving inference
\end{abstract} 

\maketitle

\setlength{\abovedisplayskip}{6pt}
\setlength{\belowdisplayskip}{6pt}  

\newpage
\section{Introduction}\label{sec:introduction}
\vspace{-3pt}
\subsection{Background}\label{sec:background}
When collecting and analyzing information collected from internet users such as social media, we want to preserve privacy for sensitive data while still providing a high-quality personalized experience for users. A state-of-the-art mathematical concept in privacy that can be used to achieve this goal is differential privacy (DP) \citet{dwork2006calibrating, dwork2006our}. Based on DP, privacy-preserving techniques can be developed to collect, share, and analyze individual data with formal privacy guarantees.

In this study, we focus on  privacy-preserving comparisons of the population means of two independent groups of zero-inflated right-skewed (zirs) data. This type of data often occurs in e-commerce or digital advertising when it relates to purchases or revenue. Since most user sessions usually do not lead to any engagement with products or ads on e-commerce platforms, or web and mobile applications such as Facebook, Youtube, Twitter, and Tiktok, data distributions for how many times each user has engaged often contain a large number of 0's. Among those who engage, the majority would not lead to actual sales of the underlying products, also resulting in a large amount of 0's in revenue \citep{ruleranalytics,wordstream}. In both cases (engagement counts or purchase values), collected data are zero-inflated. Putting aside the zero-inflation component, the rest of the data are often right-skewed, which can be modeled by Poisson or negative binomial distributions for click count data and by log-normal distributions for sales data, or other suitable distributions. 

In recent decades, large-scale online A/B testing has become the common industry practice for technology companies in testing any new features of the website, and similarly in the scenario of digital marketing \citep{kohavi_tang_xu_2020, Movahedi}. In those online experiments, users are randomly split into control and treatment groups each given slightly different experiences, such as incrementally changed back-end algorithms or marketing strategies. Data depicting user engagement is then collected during the experiment, transformed into relevant metrics of interest, and then analyzed and compared between these two groups. The goal is to determine whether the change implemented for the treatment group is resulting in a better user experience, represented by improved value on average for the key metric. In other words, we are interested in testing a null hypothesis of  there is no difference between the population means of the key metric for control and treatment users.

\subsection{Our Contributions}\label{sec:contribution}

We propose and compare several approaches to privacy-preserving inference (point and interval estimation) on the mean difference in an outcome between two groups of independent zirs data, coupled with data partitioning and censoring (PAC). The approaches we examine include likelihood-based and non-parametric approaches. In all approaches, we partition the raw individual-level data into non-overlapping subsets, obtain the partition-level means that can be reasonably assumed normally distributed, and apply double censoring (on both left and right tails) to the partition-level data (comprising partition means). This data processing step generates partitioned and censored (PAC) data, on which privacy-preserving approaches are  based on.   Our main contributions  can be summarized as follows.

\emph{First}, to our knowledge, this is the first work that obtains privacy-preserving inference from censored likelihood from aggregated partitioned data, effectively exploiting the normality assumption for partition-level statistics. We examine four ways for sanitizing the censored likelihood and compare their inferential utility. 
\emph{Second}, we formally prove the MSE-consistency of the estimators based on privacy-preserving censored likelihoods and examine the rate of convergence in  the number of partitions $P$ and privacy loss parameters, as well as the relation between $P$ and  the total sample size $n$. 
\emph{Third}, besides the likelihood-based approaches, we also obtain model-free privacy-preserving winsorized and trimmed mean group differences  when the censoring percentage is the same on both tails of the sampling distribution of mean difference, and compare the rates of convergence to consistency between the likelihood-based and model-free estimators, along with the privacy-preserving estimator without censoring. In all approaches, we use the variance combination rule in \citet{liu2016model} to obtain valid inferences for the privacy-preserving estimates.
\emph{Lastly}, we  run extensive simulation studies (16,320 simulation scenarios)  to examine the impacts of $n$, $P$, and their ratio on  privacy-preserving inference in the context of right-skewed distributions with high zero-inflation proportions for both continuous and discrete data and two types of privacy loss composition (Laplace mechanism with basic composition \citep{mcsherry2007mechanism} with $\epsilon$-DP guarantees, and Gaussian mechanism with tighter privacy loss composition based on $\rho$-zero concentrated DP \citep{bun2016concentrated}). We also examine the empirical Type-I error rates of the privacy-preserving inferences from each approach  when the truth is there is no difference in the means between two groups and power when there is a difference. \emph{Finally},  we apply the methods to obtain privacy-preserving inference for the group mean difference in a real digital ads click-through dataset. The theoretical analysis, the extensive simulation studies, and the case study shed light on the usage of these approaches in various data settings.

\subsection{Related Work} \label{sec:Relatedwork}

\citet{nissim2007smooth} propose a subsample-and-aggregate (SA) framework  to evaluate smooth sensitivity to calibrate the noise for DP guarantees, aiming for improved utility compared to using global sensitivity.   
\citet{smith2011privacy} employs the SA framework, uses averaging to aggregate the partition-level statistics, proves that there exists a differentially private estimator with the same asymptotic distribution as  the original estimator, and proposes the widened winsorized mean (WWM) as such a private estimator. 
\citet{d2015differential} apply SA-WWM  to obtain privacy-preserving standard errors for the mean difference between two groups. 
\citet{heifetz2017shade} utilizes SA-WWM to incorporate  privacy guarantees in  Apache Spark, a cluster computing framework. 
\citet{evans2020statistically} apply SA with censored partition-level statistics at user-prespecified cut-offs and bias correction that leverages the normality assumption of partition-level statistics to obtain approximately unbiased estimators for parameters. 
\citet{covington2021unbiased} use the bag of little bootstraps 
(BLB) technique \citep{kleiner2014scalable} and the differentially private CoinPress procedure \citep{biswas2020coinpress} to generate  unbiased estimators and pointwise confidence intervals with high probability.  
\cite{neel2019use} use random partitioning to develop theory and algorithms for using heuristics to solve computationally hard problems in DP. \citet{su2020utility} derive bounds for the mean squared errors of estimated linear regression coefficients based on the horizontally merged differentially private data synthesized in each partition.

Trimming/truncation and censoring/winsorization are common techniques used in statistical analysis to generate robust estimates in the presence of outliers and extreme-valued observations. In the DP setting, these ``bounding'' or ``clipping'' techniques may further help to decrease the sensitivity of statistics so that less noise would be needed for privacy guarantees. 
\citet{liu2019statistical} examines the impacts of truncation and winsorization on the statistical accuracy and validity of sanitized statistics via the Laplace mechanism.  \citet{alabi2020differentially} apply winsorization to reduce the noise calibrated to local sensitivity in differentially private linear regressions.  \citet{evans2020statistically} take into account censoring/winsorization when constructing estimates to correct biased estimates.

\citet{smith2011privacy,evans2020statistically} and \citet{covington2021unbiased} use partitioning or subsampling and generate partition/subsample-level statistics in the first step. However, the aggregation steps where the final private point and interval estimates in these works are formulated differently from ours. \citet{smith2011privacy}  uses WWM  to aggregate partition-level statistics and output mean estimates and examines symmetric winsorization  in that the same amount of data is censored on both tails of the distribution of the aggregated statistics whereas our likelihood-based privacy-preserving estimators are not subject to this constraint.  In addition, the widening procedure in WWM  seems somewhat ad-hoc from a statistical inferential perspective though justification is provided in the paper on its ``optimality'' in terms of ensuring just enough variance around the winsorized mean without over-sanitizing it. Finally, no procedure is provided to obtain interval estimation though one  may presumably construct an interval based on the established asymptotic normality of the WWM. The approach in \citet{evans2020statistically} also considers symmetric winsorization and leverages the normality assumption to correct the bias of the private estimator based on trimmed sample data. A simulation-based approach is proposed to estimate the uncertainty around the private  estimator, but no formal proof or empirical evidence is provided on whether the interval estimation approach would lead to correct coverage. \citet{covington2021unbiased} claim unbiasedness of the private estimator constructed via their approach. A key assumption underlying the claim is that the estimator by the BLB technique is unbiased, which is incorrect as BLB estimates are consistent but unbiased. In addition, the  proposed interval construction approach  only guarantees correct coverage with a certain probability; though it is a ``high probability'', as long as it is not 1, there will always be under-coverage and inflation in type-I error rates.

The rest of the paper is organized as follows. Section \ref{sec:prelim} provides an overview of basic concepts in DP. Section \ref{sec:method} proposes several approaches for sanitizing censored likelihood functions constructed from PAC data, and  privacy-preserving  trimmed mean and winsorized mean. Section \ref{sec:theory} conducts theoretical analysis that establishes MSE consistency  of privacy-preserving estimators from the proposed approaches based on PAC data and examines the rate of convergence in  the number of partitions, raw zirs data size $n$, and privacy loss parameters. Section \ref{sec:simulation} conducts simulation studies to examine the inferential utility of different approaches, bench-marked against the likelihood-based SA approach without censoring in a wide range of simulation settings.  Section \ref{sec:real} applies the procedures to a real digital ads dataset. Section \ref{sec:discussion}  provides some final remarks and discusses future directions.

\section{Preliminaries}\label{sec:prelim}
We provide a brief overview of the concepts of DP and some common randomized mechanisms to achieve DP guarantees. The overview is not comprehensive and we focus on the concepts and mechanisms  used or mentioned in this work. We also introduce an approach to obtaining valid inferences based on sanitized information. 

\subsection{Differential privacy}
DP is a mathematical concept for robust privacy guarantees. In layman's terms, DP implies the chance that an individual in a dataset can be re-identified or his or her personal information can be learned based on released sanitized information is low as the released information is about the same with or without that individual in the data.
\begin{defn}[$(\epsilon,\delta)$-DP\citep{dwork2006calibrating,dwork2006our}] \label{defn:dp}
A randomized algorithm $\M$ is of $(\epsilon,\delta)$-DP if for  all  pairs of neighboring datasets $(X,X')$ differing by one record and for all subsets $\mathcal{S}\subseteq$ image$(\M)$,
\begin{equation}\label{eqn:DP}
\Pr(\mathcal{M}(X)\in \mathcal{S}) \leq e^{\epsilon} \Pr(\mathcal{M}(X')\in \mathcal{S})+\delta.
\end{equation}
\end{defn}
$X$ and $X'$ differing by one record (denoted by $d(X,X')=1$) may refer to  that they are of the same size but differ in at least one attribute value in exactly one record, or that $X'$ has one record less than $X$ or vice versa. 
 
$\epsilon>0$ and $\delta\ge0$ are privacy budget or privacy loss parameters.  When $\delta=0$, $(\epsilon,\delta)$-DP reduces to pure $\epsilon$-DP. The smaller $\epsilon$ is, the more privacy protection there is on the individuals in the data as the outputs based on $X$ and $X'$ are more similar given their probability density/mass function ratio of the output is bounded with $(e^{-\epsilon}, e^{\epsilon})$. 
$\epsilon$ typically ranges from $10^{-3}$ to $10$ in empirical studies in the DP literature, depending on the type of information released, social perception of privacy, and expected accuracy of released data, among others. Real-life applications of DP often employ larger $\epsilon$ (higher privacy loss) for better utility (e.g., US Census uses $\epsilon$ of 19.61, and Apple Inc. sets $\epsilon$ at 2, 4, or 8  for different Apps).  When $\delta\ne0$, it is often set at a small value (e.g $o(n^{-1})$, where $n$ is the data sample size) and can be interpreted as the probability that pure $\epsilon$-DP is violated.

Every time a dataset is queried, there is a privacy cost (loss) on the individuals in the dataset. Data curators need to track the privacy cost of repeatedly releasing query results to ensure the overall privacy spending does not exceed a certain level. The basic privacy loss composition principle \citep{mcsherry2007mechanism} states that if mechanism $\mathcal{M}_j$ of $(\epsilon_j,\delta_j)$-DP for $j=1,\ldots,k$ is applied to the same data, then the overall privacy guarantee is $(\sum_j\epsilon_j,\sum_j\delta_j)$-DP. The bound on the privacy loss composited  under the basic composition principle with $(\epsilon,\delta)$-DP is not tight. Different extensions and variants of the DP definition have been developed, aiming at achieving tighter bounds for composited privacy loss, such as  $(\epsilon,\tau)$-concentrated DP (CDP) \citep{cPD}, $\rho$-zero-concentrated DP \citep{bun2016concentrated} (zCDP), R\'{e}nyi DP (RDP) \citep{mironov2017renyi}, and Gaussian DP (GDP) \citep{dong2019gaussian}.
Our approaches for privately comparing  means of two groups of zirs data are compatible with all the  mentioned DP concepts. WLOG, we illustrate the approaches using $\epsilon$-DP and $\rho$-zCDP in this work. 
\begin{defn}[$\rho$-zero concentrated DP \citep{bun2016concentrated}] \label{defn:zcdp}
A randomized algorithm $\M$ is of $\rho$-zCDP if for  all dataset pairs of neighboring data sets $(X,X')$ differing by one record, all $\alpha\in[1,\infty)$, and for all subsets $\mathcal{S}\subseteq$ image$(\M)$,
\begin{equation}\label{eqn:KL}
D_{\alpha}(\mathcal{M}(X)\|\mathcal{M}(X'))\leq \rho\alpha,
\end{equation}
where $D_{\alpha}(\mathcal{M}(X)\|\mathcal{M}(X'))$ is the R\'enyi divergence of order $\alpha$ of distribution $\mathcal{M}(X)$ from distribution $\mathcal{M}(X')$.\vspace{-6pt}
\end{defn}
$\rho$-zCDP can be converted to $(\epsilon,\delta)$-DP. Specifically, if $\mathcal{M}$ satisfies $\rho$-zCDP, then it also satisfies  $(\epsilon,\delta)$-DP with 
\begin{equation}\label{eqn:convert}
\epsilon=\rho+2\sqrt{\rho\log(\delta^{-1})} \mbox{ for any $\delta>0$}.
\end{equation}

Many mechanisms and procedures can be applied to achieve DP guarantees. In this work, we use  the Laplace mechanism for $\epsilon$-DP guarantees and the Gaussian mechanism for $\rho$-zCDP  guarantees. Let $\mathbf{s}=(s_1,\ldots,s_r)$ be statistics calculated from a dataset.  The \emph{Laplace mechanism} of  $\epsilon$-DP sanitizes $\mathbf{s}$ as in $\mathbf{s}^*=\mathbf{s}+\mathbf{e}$, where $\mathbf{e}$ contains $r$ independent samples from Laplace$\left(0,\Delta\epsilon^{-1}\right)$ and $\Delta_1=\mbox{max}_{\scriptstyle{X,X', d(X,X')=1}} \|\mathbf{s}(X)-\mathbf{s}(X')\|_1$ is the $\ell_1$ \textit{global sensitivity} of $\mathbf{s}$ (in general, one can define  $\ell_p$ sensitivity for $p\ge0$ \citep{liu2018generalized}) and represents the maximum change  in $\mathbf{s}$ between two neighboring data sets measured by $\ell_1$ norm. The larger the sensitivity, the more impact a single individual has on the value of  $\mathbf{s}$, and more noise would be needed to achieve pre-set privacy guarantees.

The \emph{Gaussian mechanism} can be used to achieve $(\epsilon,\delta)$-DP and $\rho$-zCDP guarantees. Specifically, sanitized $s^*_j\!=\!s_j\!+\!e_j$ for $j\!=\!1,\ldots,r$, where $e_j\!\sim\! \mathcal{N}(0,\sigma^2)$ and
\begin{numcases}{\sigma\geq }
c\Delta_2/ \epsilon, \mbox{ where $\epsilon<1$ and  $c^2>2\log(1.25/\delta)$, for $(\epsilon,\delta)$-DP} \label{eqn:Gaussian.DP}\\
\Delta_2/\sqrt{2\rho} \mbox{ for $\rho$-zCDP}. \label{eqn:Gaussian.zCDP}
\end{numcases}

Since the composition of privacy loss under $\rho$-zCDP is tighter than $(\epsilon,\delta)$-DP, we employ the Gaussian mechanism of $\rho$-zCDP in Eqn \eqref{eqn:Gaussian.zCDP}. One can always apply Eqn \eqref{eqn:convert} to convert the final composed privacy loss to $(\epsilon,\delta)$-DP. Specifically,  if mechanism $\mathcal{M}_j$ of $\rho_j$-zCDP for $j=1,\ldots,r$ is applied to the same data $X$, then the overall privacy guarantee is $(\sum_j\rho_j)$-zCDP, which implies ($\sum_j\rho_j+2\sqrt{\sum_j\rho_j\log(\delta^{-1})}, \delta)$-DP for any $\delta>0$.

The Laplace mechanism and Gaussian mechanism are common mechanisms for sanitizing continuous values. The exponential mechanism is a more general mechanism and can be used to sanitize  both categorical and numerical outputs.  
\begin{defn}[Exponential Mechanism \citep{mcsherry2007mechanism}] \label{def:exp}
Exponential mechanism of $\epsilon$-DP releases $s^{\ast}$ with probability $\frac{\exp\left(u(s^{\ast}|X)\frac{\epsilon}{2\Delta_u}\right)}{\int \exp\left(u(s^{\ast}|X)\frac{\epsilon}{2\Delta_u}\right) ds^{\ast}}$, where $u$ is a utility function  that assigns a score to each possible output $s^{\ast}$ from data $X$ and   $\Delta_u$ is the global sensitivity of $u$ (if $s^*$ is discrete, the integral in the denominator is replaced with summation). \vspace{-12pt}
\end{defn} 
\citet{cesar2021bounding}  show that the exponential mechanism of $\epsilon$-DP also satisfies $\epsilon^2/8$-zCDP. This suggests that one can apply the exponential mechanism in Definition \ref{def:exp} to achieve $\rho$-zCDP by setting $\epsilon=2\sqrt{2\rho}$.

\subsection{Accounting for Randomness from DP Sanitization in Inference}\label{sec:inference}
DP sanitization of statistics introduces an additional layer of variability to the statistics on top of the sampling variability. To obtain valid inference for a parameter given sanitized information, this additional source of variability needs to be accounted for. Otherwise, the uncertainty would be under-estimated, leading to invalid inference including inflated type I error rates in hypothesis testing and under-coverage in interval estimation.  

There are a few approaches to account for randomness from DP sanitization in inference. One may use the multiple sanitization (MS) procedure and the variance combination rule in \citet{liu2016model}.  Suppose the parameter of interest is $\theta$. In the MS procedure, one obtains $m>1$ sets of privacy-preserving  estimates $\hat{\theta}^{*(h)}$ of $\theta$ for $h=1,\ldots,m$. The final inference of $\theta$ is given by
\begin{align}
\bar{\theta}^*&= \textstyle m^{-1}\sum_{h=1}^m \hat{\theta}^{*(h)}\label{eqn:mean}\\
\bar{\theta}^*& \sim t_{\nu}(\bar{\beta}^*, m^{-1}b+w), 
\mbox{ where the degree of freedom $\nu =(m-1)(1+mw/b)^2$},\label{eqn:var}\\
&\qquad\qquad\qquad\qquad\qquad
w= \textstyle m^{-1}\sum_{h=1}^m\hat{v}^{*(h)}, \mbox{ and } 
b=\textstyle (m-1)^{-1} \sum_{h=1}^m (\hat{\theta}^{*(h)}-\bar{\theta}^*)^2.\notag
\end{align}
$w$ is the average of $\hat{v}^{*(h)}$ for for $h=1,\ldots,m$, the within-set variance of $\hat{\theta}^{*(h)}$, and $b$ is the between-set variance of $\hat{\theta}^{*(h)}$. Eqns \eqref{eqn:mean} and \eqref{eqn:var} are obtained by the law of total expectation and the law of total variance, respectively, and is the Monte Carlo (MC) approximations of $\mathbb{E}(\theta|\hat{\theta}^*)$ and $\mathbb{V}(\theta|\hat{\theta}^*)$, and $b/m$ is the correction with a  finite $m$ (accounting for the MC error). \citet{liu2016model} suggests $m\in[3,5]$ is a good choice, balancing utility and privacy considerations (too large a $m$ would spread the total  privacy budget too thin over $m$ sanitizations, whereas too small $m$ would yield unstable estimates for $w$ and $\bar{\theta}^*$).

Another approach to accounting for sanitization uncertainty for inferential purposes is to explicitly quantify the extra variability (either analytically or numerically) and add that onto the variance around $\theta^*$ from a single sanitization so there is no need to split the total privacy budget as in the MS procedure. \citet{miao2022privacy} propose an MC approach to estimate the extra variance without incurring additional privacy loss when it can be calculated analytically. Specifically, one first obtains a sanitized $\hat{\theta}^*$ estimate via a DP mechanism and the associated variance $v_1$; then applies the same mechanism $m$ times to $\hat{\theta}^*$  to obtain doubly-sanitized $\hat{\theta}^{**(h)}$ for $h=1,\ldots, m$, the sample variance  of which over the $m$ sets estimates the extra variance $v_2$. Since generating doubly-sanitized statistics costs no privacy, $m$ can be as large as  computationally permissible.  The total variance of $\hat{\theta}^*$ is $v_1+v_2$. When the DP mechanism itself is computationally inefficient, this approach can be computationally costly  as  the mechanism will be applied $m+1$ times.  A possible solution, in this case, is to use a small $m$ and then apply the finite-$m$ correction, in a similar manner as in the MS procedure, that is, the total variance for  $\hat{\theta}^*$ is thus $v_1+v_2+v_2/m$.

One may also directly model the randomized mechanism used for achieving DP during an inferential procedure. This  can be analytically or computationally difficult and also case-dependent. See \citet{karwa2017sharing,karwa2015private, charest2011can} for examples.

Considering the pros and cons of each of the above three approaches, we employ the MS procedure  in this work given its straightforwardness and generalizability. 

\section{Methods}\label{sec:method}
Zirs data can be modeled directly, for example, using zero-inflated negative binomial distribution for  engagement count data and  zero-inflated log-normal distribution for purchase value data. For data collected from the web or mobile applications, the data can be of large scale, and analyzing them directly can be computationally costly. In addition,  data need to be bounded to achieve DP guarantees for the privacy-preserving analysis. Given the right-skewness of the data with a long right tail, determining what bounds to use can be challenging -- bounds  too large would lead to too much noise injected   and possibly useless sanitized results, and bounds too small may introduce bias into sanitized results.

We address the analysis of large-scale zirs data and privacy-utility trade-off using a partition-and-aggregation framework with some modifications.  In a nutshell, we partition the raw data into non-overlapping subsets and then calculate  statistics  that are relevant for the subsequent inferential problem from each subset. The partitioning step offers several benefits if done properly, including approximate normality of the partition-level statistics and reduced computational cost from working with smaller datasets, among others.  We then explore different ways of incorporating formal privacy guarantees during the aggregation of the partition-level statistics with censoring. Censoring helps overcome the difficulty to have to choose bounds for sensitivity calculation when sanitizing some statistics. In what follows, we illustrate the problem and different ways of  incorporating privacy guarantees when obtaining privacy-preserving mean differences between two groups of zirs data.

Regarding data partitioning, it may occur naturally, for example, when data are collected and stored on different servers. Each server may send only aggregate statistics to a central server for subsequent analysis and inferences. Besides natural partitions, data may also be partitioned manually as a way to decrease computational costs from processing and analyzing large datasets, and, in privacy-preserving analysis, also helps lower privacy costs or improve the utility of  privacy-preserving outputs. 

\subsection{Problem Setting}\label{sec:setting}
Denote the variable of interest in group 0 and group 1 by $Y_0$ and $Y_1$, respectively; the number of partitions by $P$; observed data in group 0 in partition $j$ by $y_{ij0}$ for $i=1,\ldots,n_{j0}$ and $j=1,\ldots,P$ and that in group 1  by $y_{ij1}$ for $i=1,\ldots,n_{j1}$. Assume $y_{ij0}\sim f(\mu_0,\sigma_0^2)$ for $j=1,\ldots,P$ and $i=1,\ldots,n_{j0}$ and  $y_{ij1}\sim f(\mu_1,\sigma_1^2)$ for $j=1,\ldots,P$ and $i=1,\ldots,n_{j1}$, where $\mu_0$ and $\mu_1$ are the means of $Y_0$ and $Y_1$ in the two groups and $\sigma^2_0$ and $\sigma^2_1$ are their variances, and $f$ is some probability distribution. We are interested in the inference of $\mu_1-\mu_0$.

The sample means of $Y_0$ and $Y_1$ in partition  $j$ are $\bar{y}_{j0}= n^{-1}_{j0}\sum_{j=1}^{n_{j0}} y_{ij0}$ and $\bar{y}_{j1}= n^{-1}_{j1}\sum_{j=1}^{n_{j1}} y_{ij1}$, respectively. When $n_{j0}$ and $n_{j1}$ are large, by the central limit theorem (CLT), 
\begin{equation}\label{eqn:CLT}
\begin{aligned}
&\bar{y}_{j0} \sim \mathcal{N}(\mu_0, n^{-1}_{j0}\sigma_0^2),\quad 
\bar{y}_{j1}\sim \mathcal{N}(\mu_1, n^{-1}_{j1}\sigma_1^2), \mbox{ and}\\
&z_j\triangleq\bar{y}_{j1}-\bar{y}_{j0} \sim \mathcal{N}(\mu_1-\mu_0,\sigma_j^2), \mbox{ where } \sigma_j^2=n^{-1}_{j1}\sigma_1^2+n^{-1}_{j0}\sigma_0^2,
\end{aligned}
\end{equation}
approximately. $\sigma^2_j$ would be different across partitions if $n_{j0}$ or $n_{j1}$ are different across $j$ especially when the partitions are natural.  We focus on the case of $\sigma_j^2\equiv\sigma^2$ for $i=1,\ldots, P$ in this study, which is common in practice, and discuss extensions to the case of heterogeneous $\sigma^2_j$ in Section \ref{sec:hetero}. Based on the normality assumption for the partition-level statistic $\bar{y}_{j1}-\bar{y}_{j0}$, we can formulate the likelihood function of $\theta=\mu_1-\mu_0$,
\begin{align}
l(\theta,\sigma^2; z_1,\ldots,z_P) &= \textstyle \prod_{j=1}^P (2\pi\sigma^2)^{-1/2}\exp\left\{-(z_j-\theta)^2/(2\sigma^2)\right\},\notag\\
ll(\theta,\sigma^2; z_1,\ldots,z_P)&= \textstyle  C-\frac{1}{2}\sum_{j=1}^P\left\{\log(\sigma^2)+(z_j-\theta)^2/\sigma^2)\right\}.\label{eqn:ll}
\end{align}

\subsection{Privacy-preserving inference based on partitioned data} \label{sec:baseline}
Inference on $\theta$ can be obtained  from the log-likelihood in Eqn \eqref{eqn:ll} based on the partitioned data. To differentiate from the  partitioned and censored data used in the latter sections (\ref{sec:pac} to \ref{sec:wt}), we denote the partitioned data $z_j$ for $j=1,\ldots,P$ as \emph{Partitioning NO Censoring} data, \emph{PnoC} data  for short.  Privacy-preserving inference from PnoC data can be obtained after sanitizing the likelihood in Eqn \eqref{eqn:ll}.

Specifically, the log-likelihood  involves two statistics $s_1=\sum_{j=1}^Pz_j$ and $s_2=\sum_{j=1}^Pz^2_j$. As long as the two statistics are sanitized, so is the log-likelihood. We denote this approach by \emph{2S} (2-Statistic sanitization) and use it as a baseline method to benchmark the new privacy-preserving inferential methods introduced in Sections \ref{sec:pac} to \ref{sec:wt}. The sanitized likelihood can be maximized to obtain the privacy-preserving MLE for $\theta$ or be combined with some prior to obtain posterior inference on $\theta$. Denote the sanitized statistics of $s_1$ and $s_2$ by $s^*_1$ and $s^*_2$, respectively; the sanitized version of the log-likelihood in Eqn \eqref{eqn:ll} is
\begin{equation}\label{eqn:lls}
ll^*(\theta,\sigma^2; s_1^*,s_2^*)= C-\frac{P}{2}\log(\sigma^2)-\frac{P\theta^2}{2\sigma^2}+\frac{s_1^*\theta}{\sigma^2}- \frac{s_2^*}{2\sigma^2}.
\end{equation}
The privacy-preserving MLE  of $\theta$ and its asymptotic distribution are respectively 
\begin{align}\label{eqn:PP.MLE}
\hat{\theta}^* = s_1^*/P\mbox{ and }
\left(\frac{s_2^*-(s^*_1)^2/P}{P(P-1)}\right)^{-1/2}(\hat{\theta}^*-\theta) \sim t_{P-1} \mbox{ given $s_1^*$ and $s_2^*$}.
\end{align}
If the Bayesian framework is used, the posterior distribution of $\theta$ given $s_1^*$ and $s_2^*$ with prior $f(\theta,\sigma^2)\propto \sigma^{-2}$ is
\begin{align}\label{eqn:PP.bayes}
f(\theta|s_1^*, s_2^*) \sim t_{P-1}\left(\hat{\theta}^*, \frac{s_2^*-(s^*_1)^2/P}{P(P-1)}\right).
\end{align}

The MLE and posterior mean of $\theta$ and in Eqns \eqref{eqn:PP.MLE} and \eqref{eqn:PP.bayes} are conditional on $s_1^*$ and $s_2^*$ that involve not only the sampling error of the data but also the  randomness from sanitizing $s_1$ and $s_2$. The sampling distribution in Eqn \eqref{eqn:PP.MLE} and the posterior distribution  in Eqn \eqref{eqn:PP.bayes} are conditional of $s_1^*$ and $s_2^*$ and thus only involve the sampling error. If the inference of $\theta$ (hypothesis testing and  interval estimation) is based on Eqn \eqref{eqn:PP.MLE} and \eqref{eqn:PP.bayes} directly, it would only consider sampling variability but randomness from the statistics sanitization is ignored. To take into account both the sampling error and the sanitization variability,  we employ the MS approach and  generate $m$ sets of $(s_1^*(h), s_2^*(h))$ for $h=1,\ldots,m$, each at a privacy budget of $\epsilon/m$ in the case of $\epsilon$-DP or $\rho/m$ in the case of $\rho$-zCDP. In each sanitization, $\hat{\theta}^{*(h)}=s^{*(h)}_1/P$ and its estimated variance $w^{*(h)}=(s_2^{*(h)}-(s^{*(h)}_1)^2/P)/(P(P-1))$ are calculated, and the inferential rule in Eqns \eqref{eqn:mean} and \eqref{eqn:var} is  applied to obtain the final inference on $\theta$.

Regarding the sanitization of $s_1$ and $s_2$,  the Laplace mechanism can be used for $\epsilon$-DP guarantees and  the Gaussian mechanism can be employed for $\rho$-zCDP guarantees. In either case, the sanitization requires the $\ell_1$ or $\ell_2$ sensitivities of each statistic. Denote the global bounds on data $\mathbf{z}=(z_1,\ldots,z_P)$ by $(L,U)$\footnote{Theoretically, the support of a Gaussian distribution is unbounded. In practice, real-life data are almost always bounded. Bounding is needed for DP guarantees. The global bounds $L$ and $U$ can be chosen based on domain knowledge and prior information, where $\Pr(z\notin(L,U))$ is small and ignoble.}, then
\begin{equation}\label{eqn:s1s2}
\begin{cases}
\Delta_1(s_1)=\Delta_2(s_1)=U-L; \\
\Delta_2(s_2)=\Delta_2(s_2)=\max\{L^2,U^2\}.
\end{cases}
\end{equation}
$U$ and $L$ are the bounds on the partition-level statistic $Z$, the average difference between $Y_1$ and $Y_0$ in a partition. Suppose the global bound on $Y_1$ and $Y_0$ is $(L_Y, U_Y)$, then $L=L_Y-U_Y$ and $U=U_Y-L_Y$. If both $L_Y$ and $U_Y$ are nonnegative integers, which is the case for user engagement counts or purchase value data, then the global bounds ($L,U)$ would be narrower than $(L_Y, U_Y)$. Further averaging $Y_1-Y_0$ over the observations in each partition to get $Z$ would not further shrink the global bounds, but it would make taking tighter bounds than $(L, U)$ more ``reasonable'' and less harmful from an inferential perspective in the sense that the probability mass outside the tighter bounds is more ignorable with the average as its sampling distribution of  is much less dispersed than before taking the average. Though $P$ is smaller than the sample sizes of the raw data $n_0$ and $n_1$, the possibility that the bounds could be tighter and the global sensitivities would be smaller can outweigh the decrease in the number of data points.

\subsection{Privacy-preserving Inference based on Partitioned and Censored (PAC) Data} \label{sec:pac}
If the bounds $(L, U)$ in Eqn \eqref{eqn:s1s2} are hard to determine (insufficient prior information or domain knowledge), the sensitivities of $s_1$ and $s_2$ cannot be calculated. If the bounds are wide,  the sensitivities would be high and the amount of noise needed to obtain privacy-preserving inferences of $\theta$ based on Eqns \eqref{eqn:PP.MLE} or \eqref{eqn:PP.bayes} would be large, leading to imprecise inferences.  One approach to deal with this problem is to manually clip the data. 

Clipping could stand for censoring, winsorization, or trimming/truncation, which are all common techniques for protecting sensitive information, even before the DP era (winsorization and censoring might go by different names such as top and bottom coding). For example, suppose the global bounds of a sensitive attribute $Y$ are $[0, 100,000,000]$. If an individual $Y$ value is $>200,000$, it is re-coded as $>200,000$ with censoring and set at $200,000$ with winsorization when data is released; for $Y<10,000$, it is re-coded as $<10,000$ with censoring and set at $10,000$ with winsorization when data is released. For trimming and truncation, the values beyond $[10,000, 200,000]$ would not be released.

We focus on data censoring in this work. Censoring does not suppress data as trimming/truncation does, nor force out-of-bounds values to take  values at the bounds as winsorization does. Though censoring still leads to some information loss without publishing the observed individual-level beyond the bounds, the  information  about these individuals  is still accurate though it is coarsened.  Coarsening produces some privacy protection for extreme values. In addition, since the bounds on $Y$ for the uncensored data points are now public and local, statistics calculated from this uncensored subset of data are based on narrower bounds compared to the global bounds $(L,U)$, resulting in less perturbation on the statistics to achieve DP guarantees, partially offsetting the information loss due to censoring. 

Denote the cutoff bounds where the censoring occurs by $(l,u)$, where $L<l<u<U$. The censored data can be expressed with the pair $(z'_j,c_j)$, where
\begin{equation*}
\begin{cases}
c_j=-1\mbox{ and } z'_j=l & \mbox{if }z_j\le l\\
c_j=0 \mbox{ and } z'_j=z_j& \mbox{if }z_j\in(l,u)\\
c_j=1 \mbox{ and } z'_j=u & \mbox{if }z_j\ge u\\
\end{cases}.
\end{equation*}
Regarding the specification of $(l,u)$, one may leverage prior or domain knowledge to set $l$ and $u$ values independent of the collected data. This way of setting $(l,u)$ does not incur privacy costs; on the other hand, due to the sampling error of the collected data, the choice may lead to too much or too little censoring  on one or both sides of the data distribution. As an alternative, if the data curator has preferred left and right censoring percentages $(\alpha,\beta)$, she may calculate the sample quantiles at $(\alpha,1-\beta)$ from the observed data $\mathbf{z}=(z_1,\ldots,z_n)$ and use them as $l$ and $u$. Since those are sample statistics and used in calculating the sensitivities for $s_1$ and $s_2$ in Eqn \eqref{eqn:s1s2}, they need to be sanitized. We focus on the second scenario of setting $(l,u)$ in this work. 

We can formulate the censored likelihood of  $\theta$ based on $(z'_j,c_j)$, leveraging the approximate normality assumption on $z_j$.  
\begin{align}
l(\theta,\sigma^2) &=\!\!\!\!
\prod_{\{j:c_j=-1\}}\!\!\!\Phi\left(\frac{l-\theta}{\sigma}\right)\!\!
\prod_{\{j:c_j=1\}}\!\!\!\left(1-\Phi\left(\frac{u-\theta}{\sigma}\right)\right)\!\!
\prod_{\{j:c_j=0\}}\!\!\left\{ (2\pi\sigma^2)^{-1/2}\exp\left\{-\frac{(z'_j-\theta)^2}{2\sigma^2}\right\}\right\},\notag\\
ll(\theta)&= C+
P_l\log\!\bigg(\!\Phi\!\left(\!\frac{l-\theta}{\sigma}\right)\!\!\!\bigg)+
P_u\log\!\bigg(1-\!\Phi\!\left(\!\frac{u-\theta}{\sigma}\right)\!\!\!\bigg)
-\frac{P-P_l-P_u}{2}\left(\!\log(\sigma^2)+\frac{\theta^2}{\sigma^2}\right)\notag\\ 
& \textstyle\quad-(2\sigma^2)^{-1}\sum_{\{j:c_j=0\}}z'^2_j+\sigma^{-2}\theta\sum_{\{j:c_j=0\}} z'_j,\label{eqn:llc}
\end{align}
 where  $\Phi$ is the CDF of the standard normal distribution, $P_l=\#\{c_j=-1\}$, and $ P_u=\#\{c_j=1\}$. Let
\begin{equation}
\textstyle s'_1=\sum_{\{j:c_j=0\}} z'_j;\quad
s'_2= \sum_{\{j:c_j=0\}}z'^2_j.
\end{equation}
The sanitization of the likelihood function in Eqn \eqref{eqn:llc} would involve sanitization of either 6 statistics $\mathbf{s}'_6=(l, u, s'_1,s'_2,P_l, P_u)$ or 4 statistics $\mathbf{s}'_4=(l, u, s'_1,s'_2)$, with $P_l, P_u$ replaced by $\nint{P\alpha}$ and $\nint{P(1-\beta)}$, respectively. We refer to the former as the \emph{6S} approach and the latter  as the \emph{4S} approach. 6S is more aware of the intra-consistency with the statistics themselves than 4S (as it is likely $P_l\ne\nint{P\alpha}$ and $P_u\ne\nint{P(1-\beta)}$ given that $l,u$ are sample quantiles and $P$ is finite). But 4S leverages public information on the censoring proportions and sanitizes two fewer statistics, offering more utility on the sanitized statistics than 6S at the same privacy budget. One can even be more strict when maintaining the intra-consistency among the 6 statistics when sanitizing $\mathbf{s}'_6$, which we refer to as the ``doubling-down'' version of 6S, or \emph{6SDD} for short.  similarly, there is also a ``doubling-down'' version of 4S, refereed as the \emph{4SDD} approach. The details on 6SDD and 4SDD are provided in Sections \ref{sec:6} and \ref{sec:4}.

\begin{figure}[!htb]
\centering\vspace{-12pt}
\includegraphics[width=1\textwidth, trim={0 0 0 0},clip]{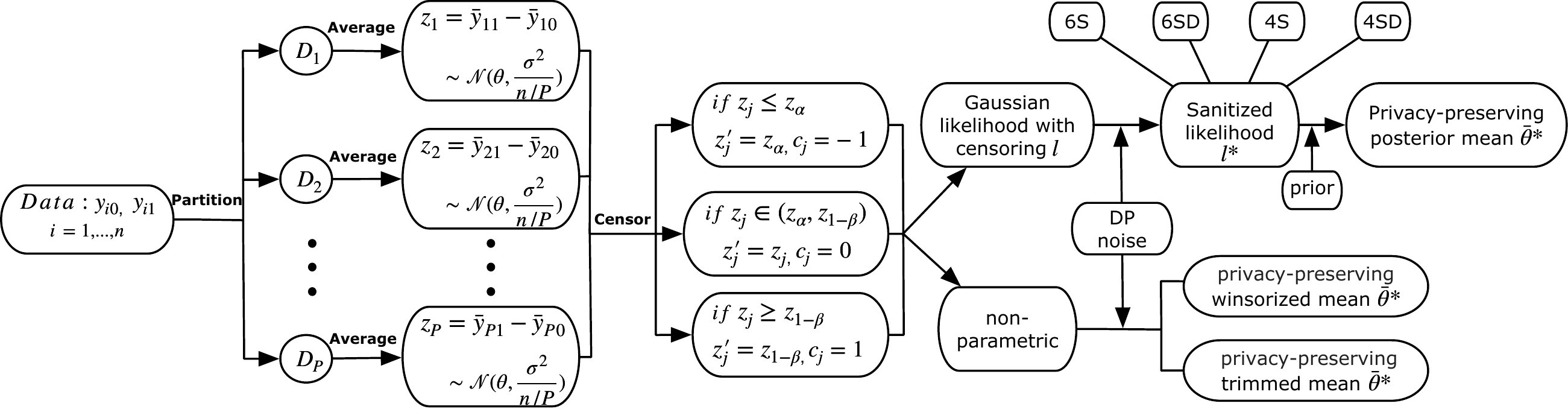} 
\caption{Approaches of privacy-preserving inference on group mean difference based on PAC data}\label{fig:procedurePAC}\vspace{-20pt}
\end{figure}

Altogether, there are 4 different approaches to sanitizing the statistics in the censored likelihood. Besides the likelihood-based privacy-preserving inference on $\theta$, we also develop the \emph{winsorized mean} and \emph{trimmed mean} approaches.  Figure \ref{fig:procedurePAC} presents the procedural steps in all our proposed approaches based on the PAC data.

\subsection{6S and 6SDD:  6-statistic sanitization of censored likelihood} \label{sec:6}
Both 6S and 6SDD  sanitize all  6 statistics that appear in the censored likelihood in Eqn \eqref{eqn:llc} to obtain a sanitized likelihood. To  sanitize $l$ and $u$, we apply the PrivateQuantile procedure in \citet{smith2011privacy} with some minor modifications, as shown in Algorithm \ref{alg:pq}. The algorithm can also be used to achieve $\rho$-zCDP by replacing $\epsilon$ with $\sqrt{8\rho}$ \citep{cesar2021bounding}. 

\begin{algorithm}[H]
\caption{PrivateQuantile of $\epsilon$-DP}\label{alg:pq}
\SetAlgoLined
\SetKwInOut{Input}{input}
\SetKwInOut{Output}{output}
\Input{data  $\mathbf{z}$, proportion $q\!\in\![0,1]$,  privacy loss $\epsilon$, global bounds {\small $(L,U)$} on $Z$.}
\Output{privacy preserving quantile $q^*_{q}$.}
sort $\mathbf{z}$ in ascending order $z_{(1)},\ldots,z_{(P)}$; let $z_0=L$, $z_{(P+1)}=U$\;
define $y_{(j)}\triangleq(z_{(j+1)}-z_{(j)})\exp(-\epsilon|j-q P|)$ for $j=0,\ldots,P$\;
sample $j^*\in\{1,\ldots,P-1\}$ with probability $\propto y_{(j)}$\;
draw $q^*_{q}$ from Unif($z_{(j^*)},z_{(j^*+1)}$)
\end{algorithm}

The sanitization of $s'_1,s'_2, P_l, P_u$ can be achieved through the Laplace or Gaussian mechanisms, depending on the DP guarantee type. The $\ell_1$ and $\ell_2$ sensitivities of  $s'_1,s'_2,P_l, P_u$ are
\begin{align}\label{eqn:sens0}
\begin{cases}
\Delta_1(s'_1)=\Delta_2(s'_2)=u-l\\ 
\Delta_2(s'_2)=\Delta_2(s'_2)=\max\{u^{2},l^{2}\}\\
\Delta_1(P_l)=\Delta_2(P_l)=1 \\
\Delta_2(P_u)=\Delta_2(P_u)=1
\end{cases},
\end{align}
respectively. Since the  $l$ and $u$ are statistics, being used directly in  the sensitivities of $s'_1$ and $s'_2$ would lead to privacy loss. Therefore, we replace  $l$ and $u$ in Eqn \eqref{eqn:sens0} with $l^*$ and $u^*$, the sanitized quantiles at $\alpha$ and $1-\beta$, respectively , from the PrivateQuantile algorithm.
\begin{align}\label{eqn:sens}
\begin{cases}
\Delta_1(s'_1)=\Delta_2(s'_2)=u^*-l^*\\ 
\Delta_2(s'_2)=\Delta_2(s'_2)=\max\{u^{*2},l^{2*}\}
\end{cases}.
\end{align}
The sensitivities, along with the allocated privacy budgets to each  statistic, can then be plugged into the Laplace or the Gaussian mechanisms to obtain sanitized $s'^*_1,s'^*_2, P_l$ and $P_u$.

A careful examination of the 6 sanitized statistics in $\mathbf{s}'^*_6$ suggests some internal inconsistency among them given how $s'_1,s'_2, P_l$ and $P_u$ are calculated in the first place. The values of all 4 statistics depend on either $l$ or $u$ and they are calculated using the original $l$ and $u$ instead of the sanitized $l^*$ and $u^*$. One may argue  
that $s'_1,s'_2, P_l$ and $P_u$ should be calculated based on sanitized $l^*$ and $u^*$ to maintain the intra-consistency among the statistics; that is, 
\begin{equation}\label{eq:dd6}
\textstyle s'_1=\sum_{\{j:z_j<l^*\}} z'_j;\quad
s'_2= \sum_{\{j:z_j>u^*\}}z'^2_j; \quad
P_l=\#{\{z_j<l^*\}}; \quad
P_u=\#{\{z_j>u^*\}}.
\end{equation}
However, the statistics  in Eqn \eqref{eq:dd6} are not automatically privacy-preserving despite the usage of $l^*$ and $u^*$ in their calculations as they still access the raw data $z_j$, and thus would still need to be sanitized. Therefore, though 6SDD, compared to 6S, makes more of an effort to honor the relationships among the statistics, the sanitized $P^*_l, P^*_u, s'^*_1,s'^*_2$ in the former are ``doubly'' perturbed and noisier than those from 6S.

Plugging in the 6 sanitized statistics $\mathbf{s}'^*_6=(P^*_l,P^*_u,l^*,u^*,s'^*_1,s'^*_2)$ obtained via either 6S or 6SDD in the likelihood function in Eqn \eqref{eqn:llc}, we obtain a sanitized log-likelihood. 
\begin{align}
ll^*(\theta,\sigma^2; \mathbf{s}'^*_6)&= C\!+\!
P^*_l\log\!\bigg(\!\Phi\!\left(\!\frac{l^*\!-\theta}{\sigma}\right)\!\!\!\bigg)\!+\!
P^*_u\log\!\bigg(\!1\!-\!\Phi\!\left(\!\frac{u^*\!-\theta}{\sigma}\right)\!\!\!\bigg)
\!-\!\frac{P\!-\!P^*_l\!-\!P^*_u}{2}\left(\!\log(\sigma^2)\!+\!\frac{\theta^2}{\sigma^2}\right)\notag\\ 
& \quad -(2\sigma^2)^{-1}s'^*_2+\sigma^{-2}\theta s'^*_1. \label{eqn:llcs}
\end{align}
To obtain privacy-preserving inferences on $\theta$, one may directly maximize  Eqn \eqref{eqn:llcs} to get the MLE and its variance. Repeating the sanitization/MLE process $m$ times, each at a privacy budget of $\epsilon/m$ in the case of $\epsilon$-DP or $\rho/m$ in the case of $\rho$-zCDP, we can apply the  inferential rule in Eqns \eqref{eqn:var} to release the final privacy-preserving inference on $\theta$. Since closed-form solutions on neither $\hat{\theta}_{\text{mle}}^{(h)*}$ nor its estimated variance $w^{(h)*}$ exist for $h=1,\ldots,m$, numerical solutions can be obtained instead. An alternative is to obtain Bayesian inference on $\theta$.  If we impose the Jeffereys' prior $f(\mu,\sigma^2)\propto\sigma^{-2}$, the privacy-preserving posterior distribution is
\begin{align}
f^*(\theta,\sigma^2|\mathbf{s}'^*_6)\propto&\;\sigma^{P_l+P_u-2-P}
\prod_{j=1}^{P^*_l}\Phi\!\left(\frac{l^*-\theta}{\sigma}\right) \prod_{j=1}^{P^*_u}\left(1-\Phi\!\left(\frac{u^*-\theta}{\sigma}\right)\right)
\exp\left(\frac{-(P\!-\!P^*_l\!-\!P^*_u)\theta^2}{2\sigma^2}\right)
\notag\\
&\;\times\exp\!\left(-(2\sigma^2)^{-1}s'^*_2+\sigma^{-2}\theta s'^*_1\right). \label{eqn:posterior6}
\end{align}
We can apply MCMC sampling  to draw samples of $(\theta,\sigma^2)$ from $f^*(\theta,\sigma^2|\mathbf{s}'^*_6)$. Since the posterior distribution satisfies DP, so do the samples from it per the immunity to post-processing property.  One can obtain the posterior mean and variance $(\hat{\theta}^*, w^*)$ of $\theta$ based on the MCMC samples.  Similar to the privacy-preserving inference using MLE, we repeat the sanitization/posterior sampling procedure $m$ times,  each at a privacy budget of $\epsilon/m$ in the case of $\epsilon$-DP or $\rho/m$ in the case of $\rho$-zCDP, leading to $m$ set of $(\hat{\theta}^{*(h)}, w^{*(h)})$ for $h=1,\ldots,m$, and then apply the inferential combination rule to obtain the final inference on $\theta$.

\subsection{4S and 4SDD:  4-statistic sanitization of censored likelihood} \label{sec:4}
4S and 4SDD sanitize 4 statistics $\mathbf{s}'_4=(l, u, s'_1,s'_2)$ that appear in the censored likelihood in Eqn \eqref{eqn:llc}, replacing $P_l, P_u$ with $\nint{P\alpha}$ and  $\nint{P\beta}$, respectively, to obtain a sanitized likelihood. 4SDD relative to 4S, is similar to 6SDD relative to 6S and calculates $s'_1,s'_2$ based on the sanitized $l^*$ and $u^*$.
After plugging in the 4 sanitized statistics $\mathbf{s}'^*_4=(l^*,u^*,s'^*_1,s'^*_2)$ in the likelihood function, we obtain a sanitized log-likelihood
\begin{align}
f^*(\theta,\sigma^2|\mathbf{s}'^*_4)\propto&
\;\sigma^{-(\nint{P(1-\alpha-\beta}+2)}
\prod_{j=1}^{\nint{P\alpha}}\!\!\Phi\!\left(\frac{l^*-\theta}{\sigma}\right)\! \prod_{j=1}^{\nint{P\beta}}\left(\!1\!-\!\Phi\!\left(\frac{u^*-\theta}{\sigma}\right)\!\!\right)
\exp\!\left(\!\frac{\nint{P(1-\alpha-\beta)}\theta^2}{2\sigma^2}\right)
\notag\\
&\;\times\exp\!\left(-(2\sigma^2)^{-1}s'^*_2+\sigma^{-2}\theta s'^*_1\right). \label{eqn:posterior4}
\end{align}
Similar to  6S and 6SDD, one can use either MLE or Bayesian inference to estimate $\theta$  and the MS procedure  to propagate sanitization variability in the final inference of $\theta$ in 4S and 4SDD.

Compared to 6S and 6SDD, 4S and 4SDD leverage the population-level and public information on $P$ and $(\alpha,\beta)$ to decrease the number of statistics to be sanitized,  potentially improving the inferential utility of $\theta$ at the same privacy budget. On the other hand, since $l$ and $u$ are sample quantiles and $P$ is finite, $(\nint{P\alpha},\nint{P\beta})$ are unlikely equal to $(P_l, P_u)$, leading to some inconsistency in the likelihood. In the 6S and 6SDD approaches, $(P_l, P_u)$ are sanitized whereas 4S and 4SDD use $(\nint{P\alpha},\nint{P\beta}$ $P_l)$ though $l$ and $u$ are sanitized, leading to further intra-inconsistency among the quantities in the likelihood. However, the loyalty of 4S to 4SDD to the population-level information can be rewarding for the utility of inference from the sanitized likelihood. In addition, the intra-inconsistency would not lead to biased inferences about population parameters, as $(\nint{P\alpha},\nint{P\beta}$ $P_l)$  are what the sanitized $(P^*_l, P^*_u)$ converge to at the population level (i.e., $P\rightarrow\infty$ and $\epsilon\rightarrow\infty$). When $P$ or $\epsilon$ is large, inferences about $\theta$ from 4S, 4SDD, 6S, and 6SDD  would be similar.

\subsection{Winsorized mean and trimmed mean when \texorpdfstring{$\alpha=\beta$}{S}}\label{sec:wt}
When $\alpha=\beta$,\footnote{Though theoretically asymmetric winsorization or trimming ($\alpha\ne\beta$) can be used, inferential properties of the associated mean estimators are not well studied and is seldom used even in the non-DP setting.} in addition to obtaining likelihood-based inference of $\theta$, we can use the winsorized mean and the trimmed mean approaches to estimate $\theta$ in a model-free manner.  The winsorized mean and its variance when $\alpha=\beta$ given data $\mathbf{z}'$ are 
\begin{align}
\hat{\theta}_{\text{w}}&=\!\textstyle\left(\sum_{j=1}^{P}z'_j\right)\!/P\!=\! \left(P_ll+P_uu+\sum_{j:c_j=0}z'_j\right)\!/P\!=\!
\alpha l+\beta u+s'_1/P,\label{eqn:mean.w}\\
\hat{\sigma}^2_{\text{w}} &= \frac{P-1}{P(P\!-\!P_l\!-\!P_u-1)^2}\left(\!P_l(l-\hat{\theta}_{\text{w}})^2\!+\! P_u(u-\hat{\theta}_{\text{w}})^2\!+\!\!\!\sum_{j:c_j=0}\!\!(z'_j-\hat{\theta}_{\text{w}})^2\!\right)\!\!\label{eqn:var.w}\\
&= \frac{P-1}{(P\!-\!P_l\!-\!P_u-1)^2}\left(\!\alpha(l-\hat{\theta}_{\text{w}})^2\!+\! \beta(u-\hat{\theta}_{\text{w}})^2\!+s'_2/P-(1-\alpha-\beta)\hat{\theta}_{\text{w}}^2\!\right).\notag
\end{align}
The trimmed mean and its variance when $\alpha=\beta$ given data  $z'_j$, where $j\in\{j:c_j=0\}$, are 
\begin{align}
\hat{\theta}_{\text{t}}&=\textstyle(P-P_l-P_u)^{-1} \sum_{j:c_j=0}z'_j= s'_1/(P-P_l-P_u),\label{eqn:mean.t}\\
\hat{\sigma}^2_{\text{t}}&= \frac{P(P-P_l-P_u-1)}{(P-P_l-P_u)(P-1)}\hat{\sigma}^2_{\text{w}}.\label{eqn:var.t}
\end{align}     
The privacy-preserving inferences based on the winsorized and trimmed means are straightforward as the means and variance estimates have closed form and the involved statistics are straightforward to sanitize. After sanitized $(l^*,u^*,s'^*_1, s'^*_2)$ are obtained, they can be plugged in Eqns \eqref{eqn:mean.w} to \eqref{eqn:var.t} to obtain privacy-preserving mean and variance estimates. 
\begin{align}
\!\!\!&\begin{cases}
\hat{\theta}^*_{\text{w}}=\alpha l^*+\beta u^*+s'^*_1/P \\
\hat{\sigma}^{2*}_{\text{w}} = \frac{P-1}{(P\!-\!P\alpha\!-\!P\beta-1)^2}\big(
\alpha(l^*-\hat{\theta}^*_{\text{w}})^2\!+\! \beta(u^*-\hat{\theta}^*_{\text{w}})^2\!+\! s'^*_2/P\!-\!(1-\alpha-\beta)(\hat{\theta}^*_{\text{w}})^2\big)
\end{cases}\!\!\label{eqn:w*}\\
\!\!\!&\begin{cases}
\hat{\theta}^*_{\text{t}}=s'^*_1/(P-P\alpha-P\beta)\\
\hat{\sigma}^{2*}_{\text{t}} \!=\!\frac{1}{(1-\alpha-\beta)(P-P\alpha-P\beta-1)} \big(\alpha(l^*\!-\!\hat{\theta}^*_{\text{t}})^2\!+\! \beta(u^*\!-\!\hat{\theta}^*_{\text{t}})^2\!+\! s'^*_2/P\!\!-\!(1\!-\!\alpha\!-\!\beta)(\hat{\theta}^*_{\text{t}})^2\big)\!\!\!\!
\end{cases}\!\!\!\!\!\!\label{eqn:t*}
\end{align}
To propagate sanitation randomness, we can apply the MS procedure and generate the final inferences using Eqns \eqref{eqn:mean} and \eqref{eqn:var}. 

Eqns \eqref{eqn:w*} to \eqref{eqn:t*} both involve 4 statistics. One can also formulate the 6-statistic version of the trimmed and winsorized mean approaches, substituting  $P^*_l$ and $P^*_u$ for $\nint{P\alpha}$ and $\nint{P\beta}$. Similar to the doubling-down versions 4SDD and 6SDD, if the calculations of $(P_l, P_u,s'_1,s'_2)$ use sanitized $(l^*,u^*)$, we also have doubling-down versions for the trimmed and winsorized mean approaches.

\subsection{Algorithmic steps of PAC-based approaches}
We present a step-by-step algorithm (Algorithm \ref{alg:pac}) for implementing the 6 PAC-based approaches introduced in this section. 

\small 
\begin{algorithm}[H]
\caption{Privacy-preserving inference of group mean difference on PAC data}\label{alg:pac}
\SetAlgoLined
\SetKwInOut{Input}{input}
\SetKwInOut{Output}{output}
\Input{data $\{\mathbf{y}_1,\mathbf{y}_0\}$, partition number $P$,  censoring proportions $\alpha,\beta$, confidence coefficient $\gamma$,  privacy loss $\epsilon$ (or $\rho)$, number of sanitizations $m$, method.}
\Output{privacy preserving point and interval estimates of group mean difference.}
randomly partition $\mathbf{y}_1$ to $P$ equal-sized portions and $\mathbf{y}_0$ to $P$ equal-sized portions\;
calculate averages $\bar{y}_{j1},\bar{y}_{j0}$ and 
define $z_j\triangleq\bar{y}_{j1}-\bar{y}_{j0}$ for each partition $j=1,\ldots,P$\;
\For{$h=1,\ldots,m$}{
\If{method = 6S or 6SDD or 4S or 4SDD}{
  construct the likelihood based on censored data (Eqn \eqref{eqn:llc})\;
  \If{method = 6S or 6SDD }{
    apply PrivateQuantile to  $\mathbf{z}$ to obtain the privacy-preserving quantiles $(l^*,u^*)$ at $\alpha$ and $1-\beta$, each with $1/(6m)$ of the total privacy budget \;
     \If{method = 6SDD}{
        recalculate $(P_l,P_u, s'_1,s'_2)$ based on the sanitized $(l^*,u^*)$\;}
    sanitize  $(P_l,P_u, s'_1,s'_2)$, each with $1/(6m)$ of the total privacy budget\;
    }
\If{method = 4S or 4SDD}{
   apply PrivateQuantile to  $\mathbf{z}$ to obtain privacy-preserving quantiles $(l^*,u^*)$ at $\alpha$ and $1-\beta$, each with $1/(4m)$ of the total privacy budget\;
     \If{method = 4SDD}{
        recalculate $(s'_1,s'_2)$ based on the sanitized $(l^*,u^*)$\;
    }
    sanitize $(s'_1,s'_2)$, each with $1/(4m)$ of the total privacy budget\;
}
  plug  the sanitized statistics in Eqn \eqref{eqn:llcs} to obtain a sanitized likelihood (Eqn \eqref{eqn:llcs})\;
  obtain estimate $\hat{\theta}^{*(h)}$ and estimated variance $w^{*(h)}$ from the sanitized likelihood via either MLE or Bayesian inferences.
  }
\If{method = winsorized mean or trimmed mean}{
  apply PrivateQuantile to  $\mathbf{z}$ to obtain the privacy-preserving quantiles $(l^*,u^*)$ at $\alpha$ and $1-\beta$, each with $1/(4m)$ of the total privacy budget\;
  sanitize $(s'_1,s'_2)$, each with $1/(4m)$ of the total privacy budget\;
  calculate estimate $\hat{\theta}^{*(h)}$ and estimated variance $w^{*(h)}$ via Eqn \eqref{eqn:w*} for winsorized mean and Eqn \eqref{eqn:t*} for trimmed mean.
}
}
use the inferential combination rule in Eqns \eqref{eqn:mean} and \eqref{eqn:var} to generate privacy-preserving inference on mean difference.
\end{algorithm}

\normalsize
\subsection{Heterogeneity variance across partitions}\label{sec:hetero}
The 6 PAC-based approaches presented above assume constant variance of the partition-level data. If partitions are created manually, one should aim at equal partitions if at all possible because it leads to clean and straightforward inferences in the setting of DP. As mentioned in Section \ref{sec:setting},  different $\sigma_j$  across partitions may occur when the partition sizes are unequal. In other words, each of the $P$ observations in $\mathbf{z}$ has a different variance $\mathbb{V}(z_j)=\sigma^2_j=n_{j0}^{-1}\sigma_0^2+n_{j1}^{-1}\sigma_1^2$, where $n_{j0}$ and $n_{j1}$ are a known constants. As a result, $s_1$ and $s_2$, the sum of $\mathbf{z}$ and the sum of squared $\mathbf{z}$ when $\sigma^2_j$ is constant, are replaced by  weighted sum $s_{w,1}$ and weighted sum of squares $s_{w,2}$, respectively, in the likelihood function of $(\theta,\sigma_0^2,\sigma_1^2)$ given $\mathbf{z}$, where the weight is proportional to $\sigma^{-2}_j$ for $j=1,\ldots,P$. The inference of $\theta$, without DP, is straightforward and its MLE is a weighted average of the $P$ observations in $\mathbf{z}$. Since $\sigma^{-2}_j$ involves unknown parameters $\sigma_0^2,\sigma_1^1$, ML estimates of $\sigma_0^2,\sigma_1^1$ can be plugged in to obtain the MLE of $\theta$. If the Bayesian framework is used, one can obtain posterior samples $(\theta,\sigma_0^2,\sigma_1^2)$  first, from which posterior samples on $\theta$ can be calculated.

Privacy-preserving inference on $\theta$  with heterogeneous $\sigma^2_j$  across partitions is not as straightforward though it can still be obtained. One first derives the sensitivities for $s_{w,1}$ and $s_{w,2}$ in the likelihood of $(\theta,\sigma_0^2,\sigma_1^2)$ to  obtain a privacy-preserving likelihood. Since  $s_{w,1}$ and $s_{w,2}$  involve both data and unknown parameters $\sigma_0^2$ and $\sigma_1^2$, one needs to bound not only the data but also $\sigma_0^2$ and $\sigma_1^2$.Thus the calculation of sensitivities for $s_{w,1}$ and $s_{w,2}$ may not be analytically straightforward and could be overly conservative, especially  when there is little prior knowledge of  $\sigma_0^2$ and $\sigma_1^2$ and one tends to choose conservatively wide bounds.   One then sanitizes $s_{w,1}$ and $s_{w,2}$ via either the Laplace or Gaussian mechanism at allocated privacy budgets. The other 4 statistics $(l, u, P_L, P_U)$ can be sanitized as in the case of constant $\sigma^2_j$: $(l, u)$ can be sanitized through PrivateQuantile;  $(P_L, P_U)$ are still counts with a sensitivity of 1 and can be sanitized through either the Laplace or Gaussian mechanisms. Once the privacy-preserving likelihood is obtained, one can obtain privacy-preserving inference on $\theta$ using similar approaches to 4S, 4SDD, 6S, and 6SDD in the constant $\sigma^2_j$ case.   

The two model-free approaches -- winsorized mean and trimmed mean -- can no longer be employed as estimators for $\theta$ when $\sigma^2_j$ is heterogeneous even in the original data without DP. To our knowledge, inferential properties of the two estimators when the sample data do not come from the population distribution are not established yet.

\section{Theoretical Analysis}\label{sec:theory}
We conduct a theoretical analysis of the approaches in Section \ref{sec:method}. We focus on establishing MSE consistency of the privacy-preserving estimate for the population parameter in each approach and examine the convergence rate in $P$ and privacy loss parameter $\rho$ or $\epsilon$. Though the theoretical analysis focuses on the asymptotics as  $P\rightarrow\infty$ and $\rho\rightarrow\infty$ or$\epsilon\rightarrow\infty$,  the sample size of the raw zirs data in each group ($n/P$) is also critical for the establishment of the theoretical properties, especially for the four likelihood-based approaches that utilize the asymptotic normality of the partition-level averages per the CLT (Eqn \eqref{eqn:CLT} in Section \ref{sec:setting}). In order for the CLT to hold, the size of  each partition $n/P$  should be large, especially considering the zero inflation and high skewness in zirs data. A necessary condition is $P=o(n)$, that is, $P$ increases at a smaller rate than $n$. The upper bound $o(n)$ could be tightened. \citet{smith2011privacy} assumes $P=o(\sqrt{n})$ when establishing optimal convergence of the asymptotic distribution as $n\rightarrow\infty$ of a privacy-preserving estimator that is constructed differently from our approaches.\footnote{The estimator in  \citet{smith2011privacy} is constructed by first obtaining a generic estimator for the parameter of interest, such as MLE, in each partition, then taking the averages of the estimates over the $P$ partitions.} Whether the bound $P=o(\sqrt{n})$ applies in our case needs further investigation. For the current analysis, we examine the case of $P\rightarrow\infty$, implying $b\rightarrow\infty$, and assume $P=o(n)$.

\normalsize

We first examine in Section \ref{sec:noC} the MSE consistency of the privacy-preserving estimator based on the partition-level data without censoring, the baseline method presented in Section \ref{sec:baseline}. We then examine the theoretical properties of outputs from PrivateQuantile, a key step in the construction of the PAC-based privacy-preserving estimators, the MSE consistency of which is established in Section \ref{sec:pacT}.

\subsection{MSE consistency of privacy-preserving estimators based on partitioned data}\label{sec:noC}
\begin{theorem}[MSE Consistency of likelihood-based methods based on partitioned data without censoring (the baseline method)]\label{thm:noC} Denote the global bounds on the raw data $\mathbf{y}$ by $(L,U)$, where $L$ and $U$ are both finite. The estimator based on the sanitized likelihood via the Laplace mechanism of $\epsilon$-DP in Eqn \eqref{eqn:PP.MLE} satisfies 
\begin{align}\label{eqn:noC}
&\mathbb{E}_{\mathcal{M},\mathbf{z}}(\hat{\theta}^* - \theta)^2 
=O\left(n^{-1}\right)+O\left(P^{-2}(U-L)^2\epsilon^{-2}\right)+O(\left(U-L)n^{-1/2}P^{-1}\epsilon^{-1}\right) \notag\\
=&\;O\left(n^{-1}\right)+O\left(P^{-2}\epsilon^{-2}\right)+O\left(n^{-1/2}P^{-1}\epsilon^{-1}\right) \notag\\
=&\;O\left(n^{-1}\right)+O\left(P^{-2}\right)+O\left(n^{-1/2}P^{-1}\right)\mbox{ 
if $\epsilon$ is regarded as constant} \notag\\
=&\begin{cases}
O(n^{-1}) & \mbox{ if } P\in (\Omega(\sqrt{n}), o(n))\\
O(P^{-2})=o(n^{-1}) & \mbox{ if } P = o(\sqrt{n})\\
\end{cases};
\end{align}
If the Gaussian mechanism of $\rho$-zCDP is used, then 
\begin{align}\label{eqn:noCzCDP}
&\mathbb{E}_{\mathcal{M},\mathbf{z}}(\hat{\theta}^* - \theta)^2  =O\left(n^{-1}\right)+O\left(P^{-2}(U-L)^2\rho^{-1}\right)+O(\left(U-L)n^{-1/2}P^{-1}\rho^{-1/2}\right)\notag\\
=&O\left(n^{-1}\right)+O\left(P^{-2}\rho^{-1}\right)+O\left(n^{-1/2}P^{-1}\rho^{-1/2}\right)\notag\\
=& O\left(n^{-1}\right)+O\left(P^{-2}\right)+O\left(n^{-1/2}P^{-1}\right)\mbox{ 
 if $\rho$ is regarded as constant}.\notag
\end{align}
leading to the same convergence rate in $P$ as $\epsilon$-DP in the Laplace mechanism in Eqn \eqref{eqn:noC}. 
\end{theorem}
The proof is provided in the supplementary materials.  Theorem \ref{thm:noC} offers several interesting insights. \emph{First}, it suggests that the convergence rate of MSE-consistent $\hat{\theta}^*$ depends on $(n,P)$ and how $P$ increases relative to $n$. Compared to the rate  $O(n^{-1})$ when directly sanitizing the sample mean of the raw data without partitioning\footnote{The rate is easy to derive in a similar fashion as in the proof for Theorem \ref{thm:noC} and is $O\left(n^{-1}+n^{-2}(U-L)^2\epsilon^{-2}+(U-L)n^{-3/2}\epsilon^{-1}\right)=O(n^{-1}+n^{-3/2}\epsilon^{-1})=O(n^{-1})$, which is the same as the rate in Eqn \eqref{eqn:noC} if $P\in (\Omega(\sqrt{n}), o(n))$ and also the same as the rate of the non-private estimator $\theta$.}, partitioning does seem to offer a faster convergence rate; however, as stated in Section \ref{sec:method}, partitioning and averaging at the partition level allows us to use a much narrower bound than $(L,U)$ and ``hide'' the extreme values in the raw data, without losing much information on the inference of $\theta$. \emph{Second}, given $n$ and $P$, the MSE of $\hat{\theta}^*$ goes to 0 in rate $O(\epsilon^{-1})$ if the Laplace mechanism of $\epsilon$-DP is used and in rate $O(\rho^{-1/2})$ if the Gaussian mechanism of $\rho$-zCDP is used.  \emph{Third}, the rate of the sanitized estimator  $\hat{\theta}^*$ converging to the population parameter $\theta$ in Eqn \eqref{eqn:noC} is at most at fast as $O(P^{-2})$, which happens to be the rate $\hat{\theta}^*$ converging to the original non-private estimator $\hat{\theta}$. Taken together with the rate of $\hat{\theta}$  converging to $\theta$, which is $O(n^{-1})$, this implies the limiting factor in the convergence of $\hat{\theta}^*$ to  $\theta$ is the rate of $\hat{\theta}$ converging to $\theta$ rather than due to the sanitization.  This is a recurrent theme across the theoretical results in Sections \ref{sec:PQ} and \ref{sec:pacT}.

\subsection{MSE Consistency of  PrivateQuantile}\label{sec:PQ}
The PrivateQuantile procedure is a critical step in all 6 approaches based on PAC data presented in  Section \ref{sec:method}. We first show that a sanitized sample quantile from PrivateQuantile is MSE consistent for the population quantile at the rate of $O(P^{-1})$, before proving the MSE consistency of the estimators from the approaches based on PAC data.
\begin{con}\label{con}
Let $x_{(1)} \leq x_{(2)} \leq \ldots \leq x_{(P)}$ be the order statistics of a random sample $x_1, \ldots, x_P$ from a continuous distribution $f_X$, and  $F_X^{-1}(q)\!=\!\inf \{x: F_X(x) \geq q\}$ be the quantiles at $q$, where $0\!<\!q\!<\!1$ and $F_X$ is the CDF. Assume $f_X$ is positive, finite, and continuous at $F_X^{-1}(q)$. 
\end{con}

\begin{lemma}[Asymptotic distribution of the spacing between two order statistics]\label{lemma:asym_gap}
Let $\mathbf{x}=(x_1,\ldots, x_P)$ be a sample from a continuous distribution $f_X$, and $x_{(\nint{qP})}$ be the sample quantile at $q$ and $x_{(\nint{qP}+1)}$ be the value immediately succeeding  $x_{(\nint{qP})}$. Given the regularity conditions in Condition \ref{con}, 
\begin{equation}\label{eqn:asym_gap}
 P\cdot(x_{(\nint{qP}+1)}-x_{(\nint{qP})})\cdot f_X(F_X^{-1}(q))\stackrel{d}{\longrightarrow}\mbox{exp}(1) \mbox{ as } P\rightarrow \infty
 \end{equation}
\end{lemma}
The proof of Lemma \ref{lemma:asym_gap} is provided in the supplementary materials. Based on  the results in Lemma \ref{lemma:asym_gap}, we can establish the MSE consistency of PrivateQuantile outputs to population quantiles, the proof of which is also provided in the supplementary materials.

\begin{theorem}[MSE consistency of PrivateQuantile]  \label{convergence_quantileDP}
Denote the sample data of size $P$ by $\mathbf{x}$ and let $x_{(\nint{qP})}^*$ be the sanitized $q^{th}$ sample quantile of $\mathbf{x}$  from the privateQuantile procedure $\mathcal{M}$ of $\epsilon$-DP. Under the regularity conditions in Condition \ref{con}, then 
\begin{equation}\label{Eqn:quantileDP_consistency}
\mathbb{E}_{\mathcal{M},\mathbf{x}}\left(x_{(\nint{qP})}^*-F_X^{-1}(q)\right)^2= O(P^{-1})+O(e^{-P\epsilon}P^{-3/2}).
\end{equation}
If privateQuantile procedure $\mathcal{M}$ of $\rho$-DP is used, then \begin{equation}\label{Eqn:quantileDP_consistency_zCDP}
\mathbb{E}_{\mathcal{M},\mathbf{x}}\left(x_{(\nint{qP})}^*-F_X^{-1}(q)\right)^2= O(P^{-1})+O(e^{-P\sqrt{\rho}}P^{-3/2}).
\end{equation}
\end{theorem}
The convergence rate of the sanitized quantile to the population quantile is $O(P^{-1})$ whereas the convergence rate of the sanitized quantile to the sample quantile is $O(e^{-P\epsilon}P^{-3/2})$ or $O(e^{-P\sqrt{\rho}}P^{-3/2})$, depending on the DP type. Similar to Theorem \ref{thm:noC}, the convergence rate of the sanitized quantile to  the sample quantile is faster in $P$ than the rate in Eqn \eqref{Eqn:quantileDP_consistency} and the limiting factor of the convergence of the sanitized quantile to the population quantile $\theta$ is the convergence rate of the sample quantile to the population quantile rather than due to the sanitization. 

The result in Theorem \ref{convergence_quantileDP} is a theoretical property of the PrivateQuantile procedure and also applied to the partition-level data $\mathbf{z}$, as long as the distribution of $\mathbf{z}$ satisfies Condition \ref{con}. 

The proof of Theorem \ref{convergence_quantileDP} also leads to several conclusions on which the Theorems in Section \ref{sec:pacT} are based. We present the conclusions in as Corollary \ref{coro}. 
\begin{corollary}\label{coro}
Let $l$ and $u$ be the sample quantiles at proportion $\alpha$ and $1-\beta$,   respectively from the observed data $\mathbf{x} = (x_1,x_2,\ldots,x_P)$. Their sanitized versions $l^*$ and $u^*$ via the PrivateQuantile procedure $\mathcal{M}$ of $\epsilon$-DP satisfy, as $P\rightarrow\infty$, 
\begin{flalign*}
\mbox{(1)}&\;\mathbb{E}_{\mathbf{x}} \mathbb{E}_{\mathcal{M}|\mathbf{x}}(l^*-l) \rightarrow \frac{1}{2e^{P\epsilon}Pf_X(F_X^{-1}(\alpha))};\; \mathbb{E}_{\mathbf{x}} \mathbb{E}_{\mathcal{M}|\mathbf{x}}(u^*-u)\rightarrow \frac{1}{2e^{P\epsilon}Pf_X(F_X^{-1}(1-\beta))};&&\\
\mbox{(2)}&\; \mathbb{E}_{\mathbf{x}} \mathbb{E}_{\mathcal{M}|\mathbf{x}}(l^*-l)^2 \rightarrow \frac{2}{3(e^{P\epsilon}Pf_X(F_X^{-1}(\alpha)))^2};\; \mathbb{E}_{\mathbf{x}} \mathbb{E}_{\mathcal{M}|\mathbf{x}}(u^*-u)^2\rightarrow \frac{2}{3(e^{P\epsilon}Pf_X(F_X^{-1}(1-\beta)))^2};\\
\mbox{(3)}&\; \mathbb{E}_{\mathbf{x}} \mathbb{E}_{\mathcal{M}|\mathbf{x}}(u^*-l^*)\rightarrow F_X^{-1}(1-\beta)-F_X^{-1}(\alpha) + O(P^{-1/2})+O(P^{-1}e^{-P\epsilon});\\
\mbox{(4)}&\; \mathbb{E}_{\mathbf{x}} \mathbb{E}_{\mathcal{M}|\mathbf{x}}(u^*-l^*)^2\rightarrow (F_X^{-1}(1-\beta)-F_X^{-1}(\alpha))^2 + O(P^{-1/2})++O(P^{-1}e^{-P\epsilon}).
\end{flalign*}
If the PrivateQuantile procedure $\mathcal{M}$ of $\rho$-zCDP is used, the expectations in (1) to (4) still hold, after replacing $\epsilon$ with $\sqrt{\rho}$.
\end{corollary}

\subsection{MSE-consistency of privacy-preserving  estimators based on PAC data}\label{sec:pacT}

We now establish  the MSE consistency of the likelihood-based estimators based on PAC data in Theorem \ref{thm:ll} and the trimmed mean and winsorized mean approaches in Theorem \ref{thm:w.t}. The proofs of both theorems  are available in the supplementary materials. 
\begin{theorem}[partition-level MSE consistency of likelihood-based estimators based on PAC data] \label{thm:ll}
Let $\mathbf{z}=(z_1, z_2, \ldots, z_P)$ denote the partition-level samples and $z_j\sim \mathcal{N}(\theta, \sigma^2(n/P)^{-1})$ for $j=1,\ldots,P$ per the CLT for large $n/P$.  Let $\hat{\theta}^*=g(\mathbf{s'}^*)$ denote the estimate of $\theta$ obtained via one of the four likelihood-based approaches (6S, 6SDD, 4S, 4SDD) given the sanitized censored Gaussian likelihood  in Eqn \eqref{eqn:llcs}, where $\mathbf{s'}^*$ is the sanitized statistic in the likelihood. 
Assume the censoring percentages on the left and right tails satisfy $5\alpha/4 + \beta <1$ and $g$ is a continuous function, under the regularity conditions in Condition \ref{con}, then
$\mathbb{E}_{\mathbf{z},\mathcal{M}} (\hat{\theta}^*-\theta)^2 
=O(P^{-1}+P^{-3/2}\epsilon^{-1})$ if the Laplace mechanism  of $\epsilon$-DP is used and $O(P^{-1}+P^{-3/2}\rho^{-1/2})$ if the Gaussian mechanism  of $\rho$-zCDP is used. The rates are simplified to $O(P^{-1})$ if $\epsilon$ and $\rho$ are treated as constant.
\end{theorem}

\begin{theorem}[partition-level MSE consistency of  winsorized and trimmed means  based on PAC data]\label{thm:w.t}
Under the regularity conditions in Condition \ref{con},  when the censoring percentages on the left and right tails are the same, i.e., $\alpha=\beta$, the privacy-preserving trimmed and winsorized means in Eqns \eqref{eqn:mean.w} and \eqref{eqn:mean.t} are MSE consistency in that $\mathbb{E}_{\mathbf{z},\mathcal{M}}(\hat{\theta}^*_t-\theta)^2 =O(P^{-1}+P^{-3/2}\epsilon^{-1})=O(P^{-1})$ if the Laplace mechanism  of $\epsilon$-DP is used and $O(P^{-1}+P^{-3/2}\rho^{-1/2})$ if the Gaussian mechanism  of $\rho$-zCDP is used;   and $\mathbb{E}_{\mathbf{z},\mathcal{M}}(\hat{\theta}^*_w - \theta)^2 =O(P^{-1}+P^{-3/2}\epsilon^{-1})=O(P^{-1})$ if the Laplace mechanism  of $\epsilon$-DP is used and $O(P^{-1}+P^{-3/2}\rho^{-1/2})$ if the Gaussian mechanism  of $\rho$-zCDP is used. The rates are simplified to $O(P^{-1})$ if $\epsilon$ and $\rho$ are treated as constant.
\end{theorem}

In both Theorems \ref{thm:ll} and \ref{thm:w.t}, the convergence rates of the sanitized estimators of $\hat{\theta}^*$,  $\hat{\theta}_w^*$ and  $\hat{\theta}_t^*$ to the nonprivate original estimators $\hat{\theta}, \hat{\theta}_w$ and  $\hat{\theta}_t$ t are $O(P^{-2}\epsilon^{-2})$ if the Laplace mechanism of $\epsilon$-DP is used and $O(P^{-2}\rho^{-1})$ if the Gaussian mechanism of $\rho$-zCDP is used.   Similar to Theorem \ref{thm:noC}, the convergence of the sanitized estimators to the original estimators are faster in terms of $P$ than the rates in Theorems \ref{thm:ll} and \ref{thm:w.t}, implying that the limiting factor of the convergence of the sanitized estimator to the population parameter is rather due to the sampling error than due to the sanitization.

\section{Simulation Studies}\label{sec:simulation}
We run extensive simulation studies to compare different methods -- 6S, 6SDD, 4S, 4SDD, trimmed mean,  and winsorized mean based on PAC data in the utility of privacy-preserving inference for $\theta$. We benchmark the results against the 2S method (Section \ref{sec:baseline}), the private-preserving inference based on partitioned data without censoring. We also include a na\"{i}ve method that ignores the censoring of PAC data and assumes normality on truncated data $z'_j$ for $j=1,\ldots,P-P_l-P_u$ as a negative control.  The na\"{i}ve method sanitizes $s'_1$ and $s'_2$, which are then plugged into the Gaussian likelihood function to formulate a sanitized likelihood. Since the sanitization of $(s'_1,s'_2)$ needs the bounds $(l,u)$, $l,u$ would be sanitized via the PrivateQuantile procedure before the sanitization of $(s'_1,s'_2)$.  Regarding the sample size $P_c=P-P_l-P_u$, we can either sanitize it or use the public information $\nint{P(1-\alpha-\beta)}$ without costing privacy\footnote{There are 5 statistics that need to be sanitized in  the former and 4 in the latter case.}, the corresponding sanitized likelihood functions are respectively
\begin{align}
&ll^*(\theta,\sigma^2; s'^*_1,s'^*_*,P^*_c)= C-\frac{P^*_c}{2}\log(\sigma^2)- \frac{\nint{P^*_c(1-\alpha-\beta)}\theta^2}{2\sigma^2}+ \frac{s'^*_2}{2\sigma^2}-\frac{s'^*_1\theta}{\sigma^2},\label{eqn:naive3}\\
&ll^*(\theta,\sigma^2; s'^*_1,s'^*_2)= C-\frac{\nint{P(1-\alpha-\beta)}}{2}\log(\sigma^2)- \frac{\nint{P(1-\alpha-\beta)}\theta^2}{2\sigma^2}+ \frac{s'^*_2}{2\sigma^2}-\frac{s'^*_1\theta}{\sigma^2}.\label{eqn:naive2}
\end{align}
Similar to the doubling-down cases of 6SDD and 4SDD to 6S and 4S, depending on whether the calculation of $s'_1,s'_2$ uses the observed $l,u$ or the sanitized $l^*,u^*$, the na\"{i}ve method also has its doubling-down version\footnote{This would not matter much as the likelihood is misspecified in the first place.}. Combining the sanitized likelihood in Eqns \eqref{eqn:naive2} or \eqref{eqn:naive3} with the Jeffreys' prior, we  obtain the closed-form posterior mean of $\theta$ and its variance, 
\begin{equation}
\theta^*= s'^*_1/P'_c;\; w^*=(P'_c-1)^{-1}(s'^*_2-(s'^*_1)^2/P'_c), 
\mbox{ where $P'_c =\nint{P(1-\alpha-\beta)}$ or $P^*_c$}.
\end{equation}
We also apply MS to account for the sanitization randomness and obtain the final privacy-preserving inference of $\theta$ in the  na\"{i}ve method.

\subsection{Simulation Setting}
We examine a total of 2040 simulation settings, a combination of 3 raw data  types of $Y$ (Gaussian, zero-inflated log-normal/ZILN, zero-inflated negative binomial/ZINB), 2 true values of $\theta$ (zero vs. non-zero), 2 censoring scenarios ($\alpha=\beta$ vs $\alpha\ne\beta$ ), 2 types of DP guarantees ($\epsilon$-DP, $\rho$-zCDP), 5 privacy loss values, and 17 settings of $n/P$. 

In each simulation setting, we run 12 methods when $\alpha=\beta$, which are \vspace{-12pt}
\begin{itemize}
\setlength\itemsep{-3pt}
\setlength\itemindent{-24pt}
\item[]1. original: non-private inference based on partitioned data without censoring
\item[]2. 2S: the baseline; privacy-preserving inference based on partitioned data without censoring
\item[]3 $\sim$ 6.  6S, 6SDD, 4S, 4SDD
\item[]7. trimmed mean
\item[]8. winsorized mean
\item[]9 $\sim$ 12. 4 na\"{i}ve approaches as presented at the beginning of Section \ref{sec:simulation}
\end{itemize}\vspace{-12pt}
and  10 methods when $\alpha\ne\beta$ (without  trimmed mean and winsorized mean). For all the likelihood-based methods, we used the Metropolis Hasting algorithm to obtain Bayesian inference for $\theta$. We run 500 repeats in each method and each simulation scenario and summarize the inferential results by bias, root mean squared error (RMSE), coverage probability of  95\% posterior intervals, widths of 95\% confidence intervals, Type-I error rate when $\theta=0$ ($H_0$ is true), and power when  $\theta\ne 0$  ($H_1$ is true). 

For ZILN, data $\mathbf{y}$ were simulated as follows,
\begin{align*}
b_{i0} \sim \text{Bern}(p_0); 
y_{i0}=0 \text{ if } b_{i0}=0 \text{ and } y_{i0}\sim\text{logN}(\mu_0,\sigma^2) \mbox{ if } b_{i0}=1, \\
b_{i1} \sim \text{Bern}(p_1); 
y_{i1}=0 \text{ if } b_{i1}=0 \text{ and } y_{i1}\sim\text{logN}(\mu_1,\sigma^2) \mbox{ if } b_{i1}=1.
\end{align*}
When $\theta=0$, we set $p_0=p_1=0.02$, $\mu_0=\mu_1=4.6$, and $\sigma=1$; when $\theta\ne0$,  we set $p_0=0.02,p_1=0.03, \mu_0=\mu_1=4.6$, and $\sigma=1$, leading to $\theta=p_1\exp(\mu_1+\sigma^2/2)-p_0\exp(\mu_0+\sigma^2/2)=1.64$ (the parameter values are chosen to mimic real-life data).  

For ZINB,  data $\mathbf{y}$ were simulated as follows, 
\begin{align*}
b_{i0} \sim \text{Bern}(p_0); 
y_{i0}=0 \text{ if } b_{i0}=0 \text{ and } y_{i0}\sim\text{NegBin}(\mu_0,\tau_0) \mbox{ if } b_{i0}=1, \\
b_{i1} \sim \text{Bern}(p_1); 
y_{i1}=0 \text{ if } b_{i1}=0 \text{ and } y_{i1}\sim\text{NegBin}(\mu_1,\tau_1) \mbox{ if } b_{i1}=1.
\end{align*}
When $\theta=0$, we set $p_0=p_1=0.02$, $\mu_0=\mu_1=3$, and $\tau_0=\tau_1=2$ (the variance of  NB distribution is $\mu+\mu^2/\tau$); when $\theta\ne0$, we set $p_0=0.02,p_1=0.03, \mu_0=\mu_1=3$, and $\tau_0=\tau_1=2$, leading to $\theta=p_1\mu_1-p_0\mu_0=0.03$. 

In addition to the two zirs data cases (ZILN and ZINB), we also include the Gaussian case to examine how the methods perform in a ``perfect'' scenario; that is, the Gaussian assumption on the partitioned data always hold, regardless of $n$ and $P$. The Gaussian data were simulated from $y_{i0}\sim\mathcal{N}(\mu_0,\sigma^2)$ and $y_{i0}\sim\mathcal{N}(\mu_1,\sigma^2)$. When $\theta=0$, we set $\mu_0=\mu_1=3.32$ and $\sigma=6$;  when $\theta\ne0$, we set $\mu_0=3.32,\mu_1=4.95$ and $\sigma=6$.

The examined  $(n, P)$ cases are listed in Table \ref{tab:nP}. The cases that are the closest to real-life setting is when $n=1,000,000$ and  $P\!\in\![100,1000]$. We include smaller $n$ and other $P$ values to examine how $n,P$, and the $n/P$ ratio affects the Gaussian assumption imposed on the partitioned data and the privacy-preserving inference, and to recommend a good combination of the two parameters for practical applications. Note the MSE consistency of the privacy-preserving estimators established section would require $p=o(n)$. 

\begin{table}[!htb]
\caption{The  $(n,P)$ scenarios examined in the simulation studies}\label{tab:nP}\vspace{-9pt}
\centering
\resizebox{\textwidth}{!}{
\begin{tabular}{@{\hspace{2pt}}c@{\hspace{2pt}}|
@{\hspace{2pt}}c@{\hspace{2pt}}|
@{\hspace{2pt}}c@{\hspace{4pt}}c@{\hspace{4pt}}c| c@{\hspace{4pt}}c@{\hspace{4pt}}c@{\hspace{4pt}}c@{\hspace{2pt}}c| c@{\hspace{2pt}}c@{\hspace{4pt}}c@{\hspace{4pt}}c@{\hspace{4pt}}c@{\hspace{4pt}} c@{\hspace{4pt}}c@{\hspace{4pt}}c@{\hspace{2pt}}}
\hline
$n$ & 1,000& \multicolumn{3}{c|}{10,000} & \multicolumn{5}{c|}{\emph{100,000}} & \multicolumn{8}{c}{\emph{1,000,000}} \\
\hline
$P$& 100 & 100 & 500 & 1,000 & \emph{100} & 500 & 1,000 & 5,000 & 10,000 & \emph{100} & \emph{300} & \emph{700} & \emph{1,000} & 5,000 & 10,000 & 50,000 & 100,000\\
\hline
$n/P$ & 10 & 100 & 20 & 10 & \emph{1,000} & 200 & 100 & 20 & 10 &
\emph{10,000} & \emph{3,333} & \emph{1,667} & \emph{1,000} & 200 & 100 & 20 & 10\\
\hline
\end{tabular}}
\resizebox{\textwidth}{!}{\begin{tabular}{l}
The results on the \emph{italic $(n,P$) scenarios} are presented in the main text; the rest are in the supplementary materials.\\
\hline
\end{tabular}}
\vspace{-12pt}
\end{table}

We examine a symmetric censoring case $\alpha=\beta=0.1$ and an asymmetric censoring case $\alpha=0.05,\beta=0.15$. For privacy guarantees, we set $\epsilon=0.5, 1, 2, 5, 50$ in the case of $\epsilon$-DP\footnote{$\epsilon=50$ is used to examine whether the privacy-preserving inferences in the proposed procedures converge to the original inference as privacy loss approaches $\infty$.} and $\rho= 0.005, 0.02, 0.08, 0.32, 1.28$ in the case of $\rho$-zCDP.  The $\rho$ values are chosen to yield a similar range on $\epsilon$ to the pure $\epsilon$-DP case.  Table \ref{tab:rhoeps} lists the corresponding $(\epsilon,\delta)$-DP guarantees for $\rho$-zCDP, and the associated scale parameters in the Laplace mechanism of $\epsilon$-DP and the Gaussian mechanism  of $\rho$-zCDP, respectively. Since we divide the total privacy budget by $k$, the number of statistics to be sanitized in each approach, the $\ell_1$ and $\ell_2$ global sensitivities  $\Delta_1$ and $\Delta_2$ are the same  for each scalar statistic. Table \ref{tab:rhoeps} suggests the variance of the Laplace mechanism seems to be notably smaller than that of the Gaussian mechanism in the examined simulation scenarios. 

\begin{table}[!htb]
\caption{Privacy loss parameters in $\rho$-zCDP and $\epsilon$-DP and the scale parameters of the associated  randomized mechanisms } \label{tab:rhoeps}\vspace{-6pt}
\centering
\resizebox{1\textwidth}{!}{
\begin{tabular}{@{\hspace{2pt}}c| c@{\hspace{3pt}}c@{\hspace{3pt}}c@{\hspace{3pt}} c@{\hspace{3pt}}c| c@{\hspace{3pt}}c@{\hspace{3pt}}c@{\hspace{3pt}} c@{\hspace{3pt}}c| c@{\hspace{3pt}}c@{\hspace{3pt}}c@{\hspace{3pt}} c@{\hspace{3pt}}c@{\hspace{3pt}}}
\hline
\multicolumn{15}{c}{Gaussian mechanism of $\rho$-zCDP} \\
\hline
n & \multicolumn{5}{c|}{1,000} & \multicolumn{5}{c|}{100,000} & \multicolumn{5}{c}{1,000,000} \\
\hline
$\rho$ & 0.005 & 0.02 & 0.08 & 0.32 & 1.28& 0.005 & 0.02 & 0.08 & 0.32 & 1.28& 0.005 & 0.02 & 0.08 & 0.32 & 1.28\\
$k\sigma/\Delta_1 ^{\dagger}$&$10$ &$5$  &$2.5$ &$1.25$&$0.625$ &$10$ &$5$  &$2.5$ &$1.25$&$0.625$ &$10$ &$5$  &$2.5$ &$1.25$&$0.625$\\
\hline
\multicolumn{15}{c}{corresponding $\epsilon$ value in $(\epsilon, \delta)$ for $\rho$-zCDP (setting $\delta = n^{-1}$)} \\
\hline
$\epsilon$ & 0.377 & 0.763 & 1.567 & 3.294 & 7.227 & 0.485 & 0.980 & 1.999 & 4.159 & 8.958 & 0.531 & 1.071 & 2.183 & 4.525 & 9.690\\
$\delta$ & \multicolumn{5}{c|}{$10^{-3}$} & \multicolumn{5}{c|}{$10^{-5}$} & \multicolumn{5}{c}{$10^{-6}$} \\
\hline
\multicolumn{15}{c}{Laplace mechanism of $\epsilon$-DP} \\
\hline
$\epsilon$ & 0.5 & 1 & 2 & 5 & 10 & 0.5 & 1 & 2 & 5 & 10 & 0.5 & 1 & 2 & 5 & 10\\
$k\sigma/\Delta_1 ^{\dagger}$ & 2.828 & 1.414 & 0.707 & 0.283 & 0.141 &  2.828 & 1.414 & 0.707 & 0.283 & 0.141 &  2.828 & 1.414 & 0.707 & 0.283 & 0.141 \\
\hline
\end{tabular}}
\resizebox{1\textwidth}{!}{
\begin{tabular}{l}
$ ^{\dagger}\; k$ is the number of statistics to be sanitized, $\Delta_1$ is the global sensitivity of a scalar statistic, and $\sigma$ is the SD \\ of the  Gaussian mechanism of $\rho$-zCDP and the SD of the Laplace mechanism of $\epsilon$-DP, respectively.\\
\hline
\end{tabular}}
\vspace{-6pt}
\end{table}

\subsection{Results}
Due to space limitations and to highlight the findings from the simulation scenarios that are closer to practical scenarios, we present a subset of the results from the thousands of simulation scenarios; the rest can be found in the supplementary materials. Specifically, among the 4 na\"{i}ve methods, we include the one presented in Eqn \eqref{eqn:naive2} with sanitization of 4 statistics ($s'_1, s'_2, l,u$) without doubling down (the performances of the  other three na\"{i}ve alternatives are similar). Among the 17 $(n,P)$ scenarios, we presented 5 pairs $(n,P) = (100k, 100), (1 \text{ million}, 100), (1 \text{ million}, 300), (1 \text{ million}, 700), (1 \text{ million}, 1k)$.

In summary, the performance of each method varies by the $(n,P)$ scenario, the raw data type,  the privacy budget, whether the underlying truth is $H_0$ or $H_1$, and whether the censoring  is symmetric. Overall speaking,  4S is the best performer, all simulation scenarios considered. Compared to the positive control 2S and PAC approaches 6S, 6SDD, and 4SDD, 4S is non-inferior in bias and CP and superior with smaller RMSE and narrower CIs, implying more precise inferences. When $\alpha=\beta$, winsorized and trimmed mean yield smaller bias, slightly narrower CIs, and higher power under $H_1$ than 4S for $(n=100k, P=100)$ in the case of ZINB and ZILN; in all the other scenarios, 4S  outperforms trimmed mean and outperforms or is similar to winsorized mean. The worst performer is the na\"{i}ve method, as expected, as it fails to account for the censoring when obtaining inference.  The Gaussian mechanism of $\rho$-zCDP does not seem to offer better statistical utility compared to the Laplace mechanism of  $\epsilon$-DP, likely due to the small number of sanitized statistics and relatively large $P$, where the advantage of $\rho$-zCDP is not obvious compared to the $\epsilon$-DP. This finding is consistent with Table \ref{tab:rhoeps}, which suggests the variance of the Laplace mechanism is notably smaller than that of the Gaussian mechanism in the examined simulation scenarios.

In what follows, we present the detailed results when the raw data $Y$ is Gaussian (Sections \ref{sec:0sN} to \ref{sec:1asN}), ZILN (Sections \ref{sec:0sZILN} to \ref{sec:1asZILN}), and ZINB (Sections \ref{sec:0sZINB} to \ref{sec:1asZINB}) when $H_0$ or $H_1$ is true and when the censoring is symmetric ($\alpha=\beta$) or asymmetric ($\alpha\ne\beta$).

\subsubsection{\texorpdfstring{$Y\sim$ Gaussian}{}, \texorpdfstring{$\theta=0$ ($H_0$)}{}, and \texorpdfstring{$\alpha=\beta$}{}}\label{sec:0sN}
\begin{figure}[!htb]
\hspace{0.45in}$\epsilon=0.5$\hspace{0.75in}$\epsilon=1$\hspace{0.8in}$\epsilon=2$
\hspace{0.8in}$\epsilon=5$\hspace{0.8in}$\epsilon=50$\\
\centering
\includegraphics[width=0.175\textwidth, trim={2.6in 0 2.6in 0},clip] {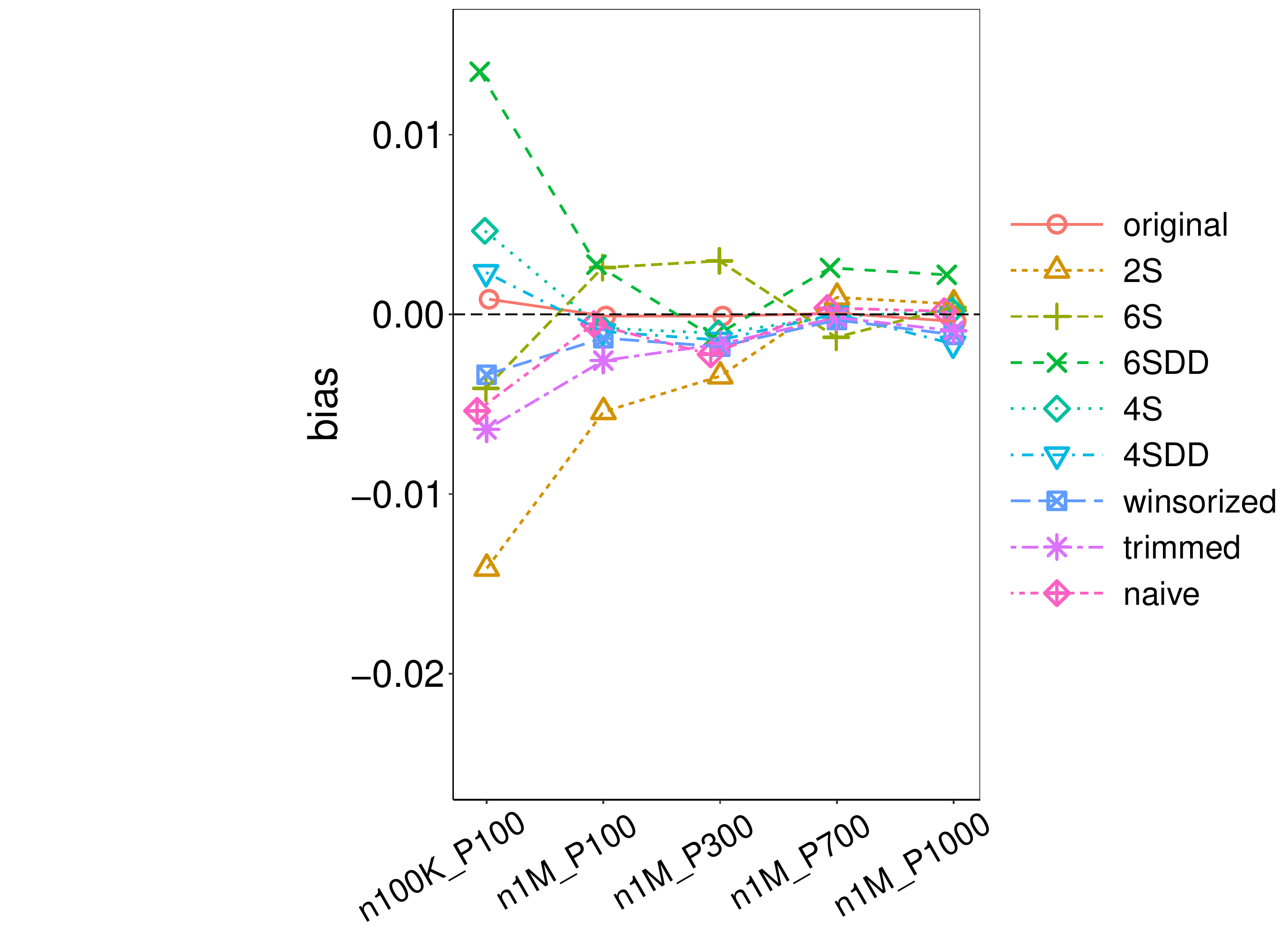}
\includegraphics[width=0.175\textwidth, trim={2.6in 0 2.6in 0},clip] {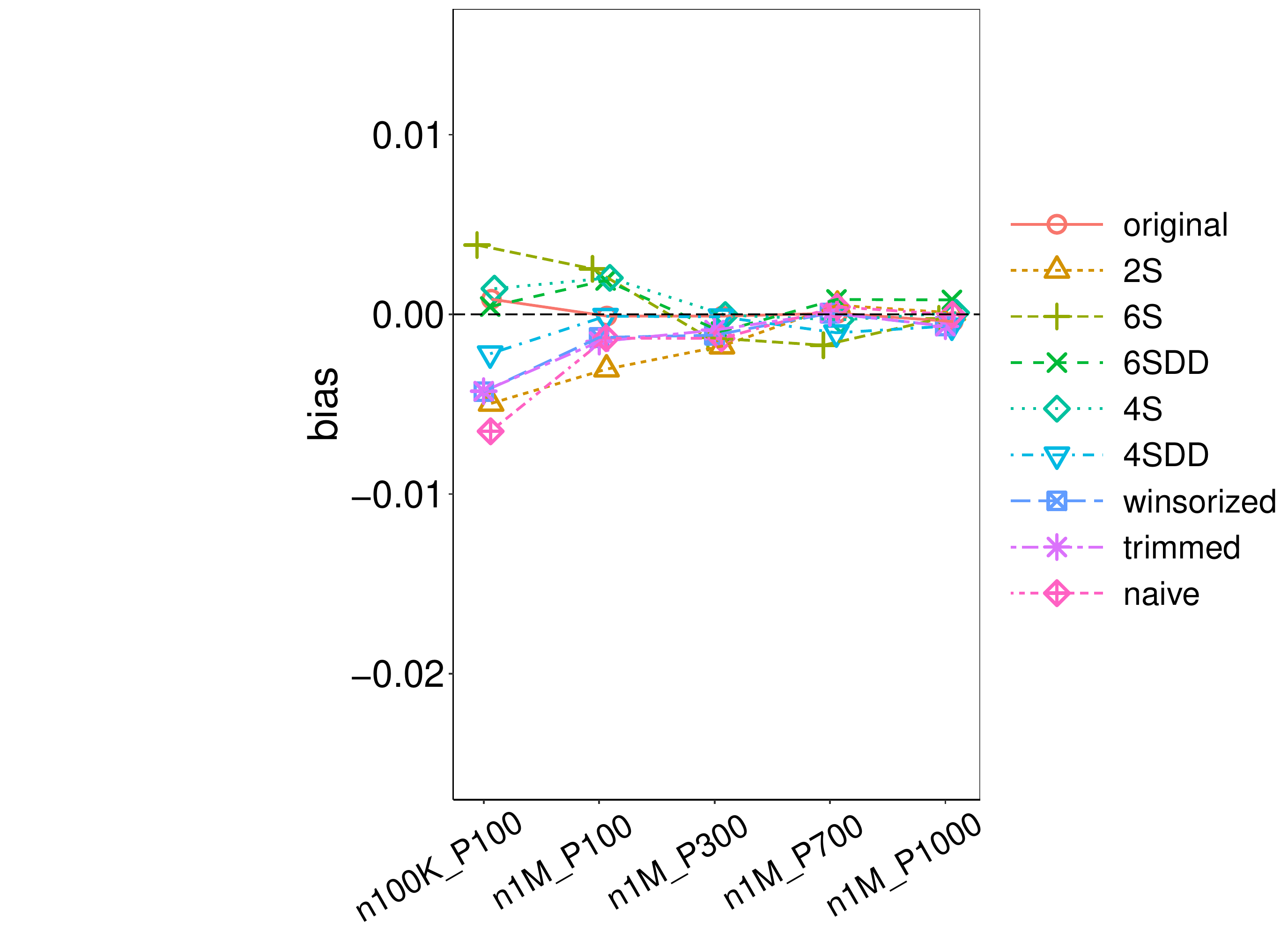}
\includegraphics[width=0.175\textwidth, trim={2.6in 0 2.6in 0},clip] {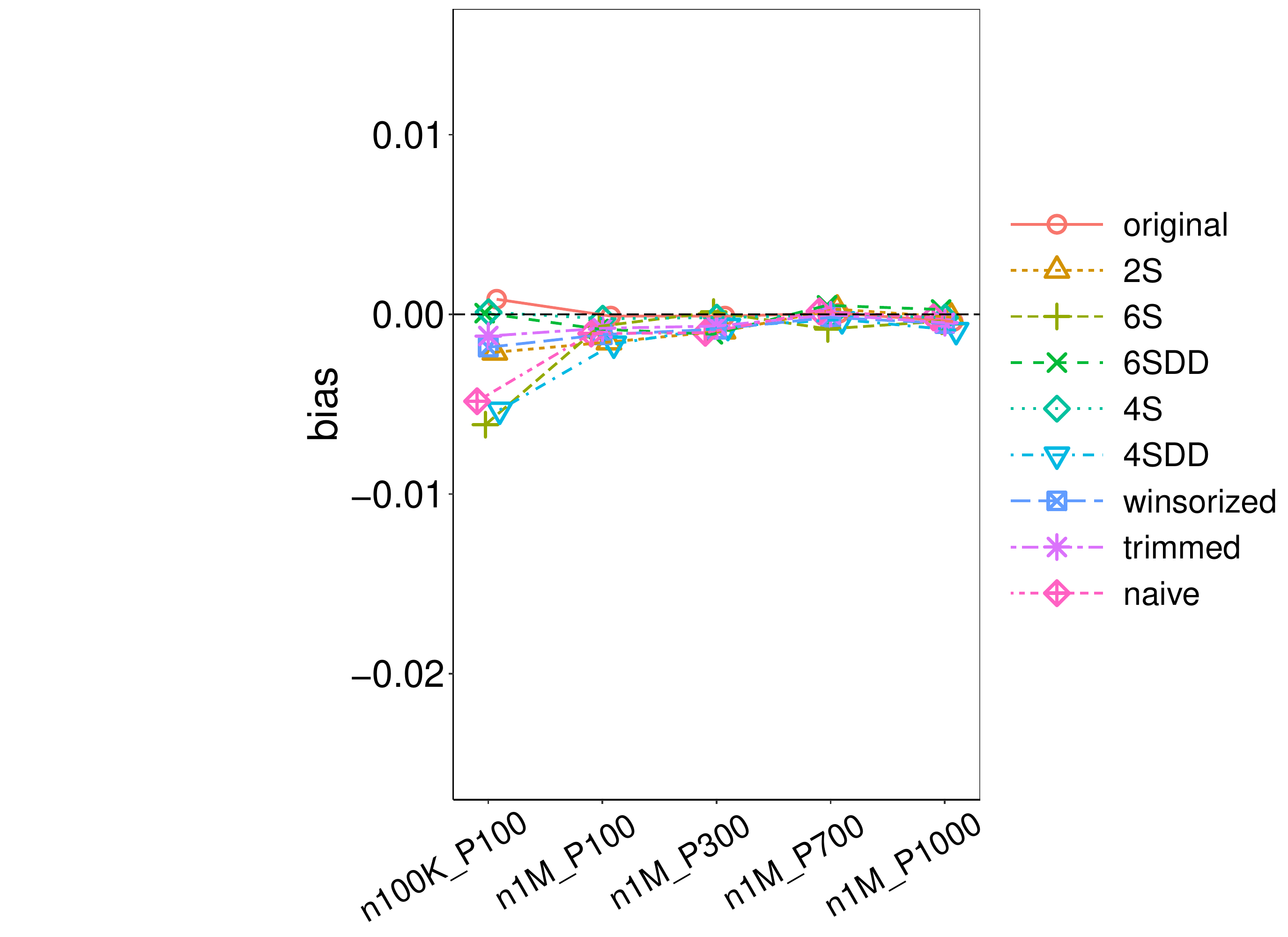}
\includegraphics[width=0.175\textwidth, trim={2.6in 0 2.6in 0},clip] {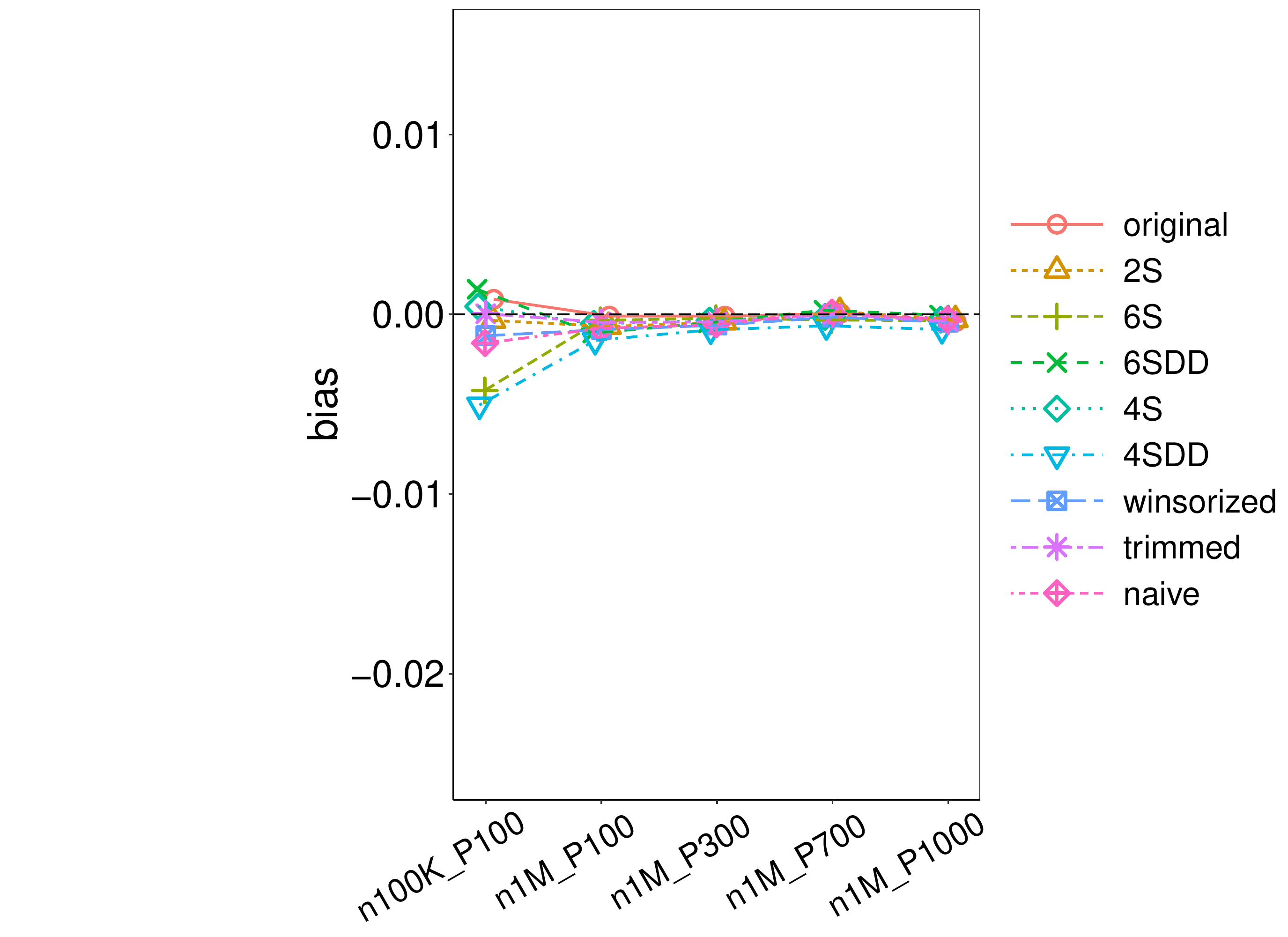}
\includegraphics[width=0.175\textwidth, trim={2.6in 0 2.6in 0},clip] {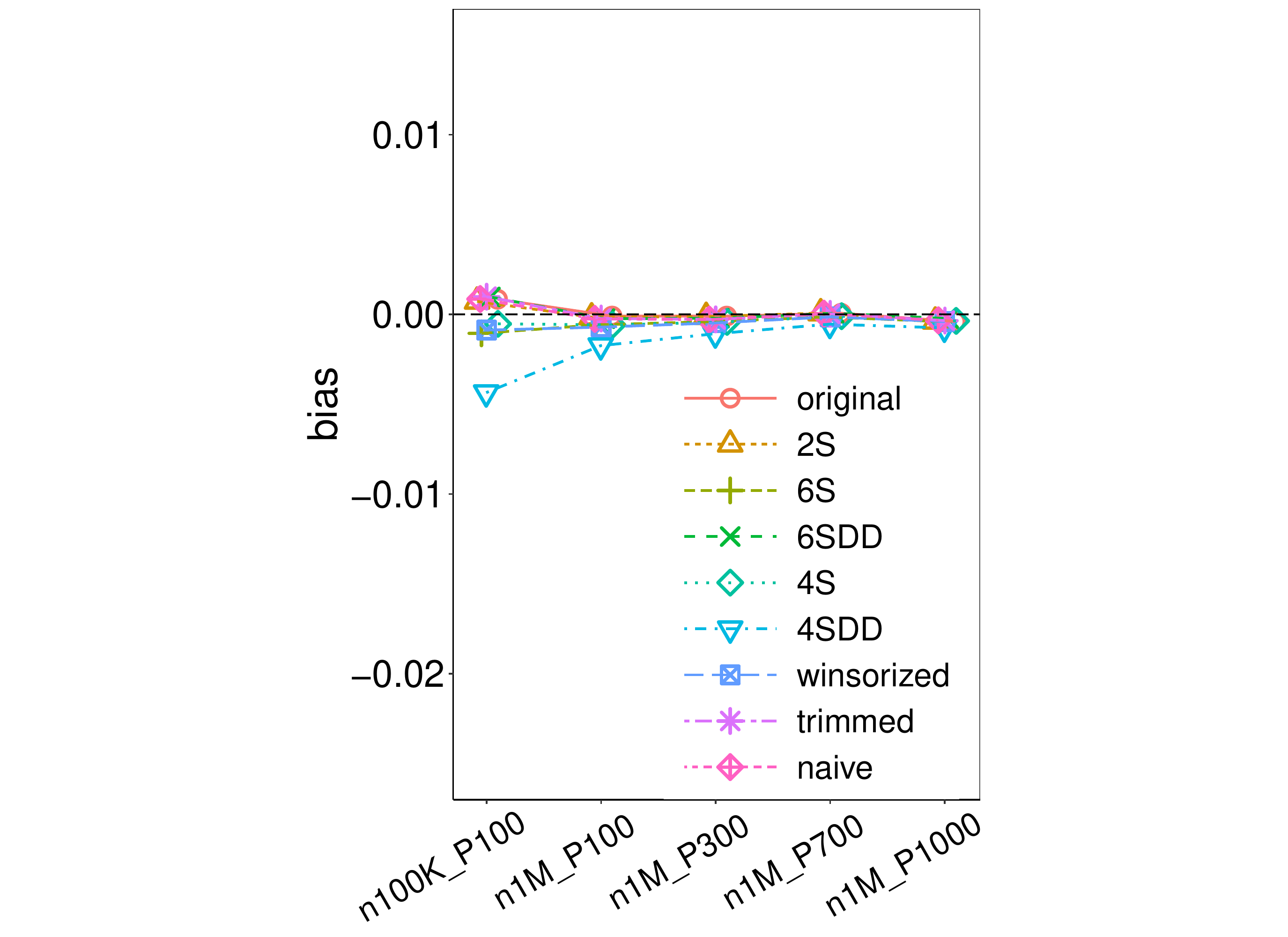}\\
\includegraphics[width=0.185\textwidth, trim={2.5in 0 2.6in 0},clip] {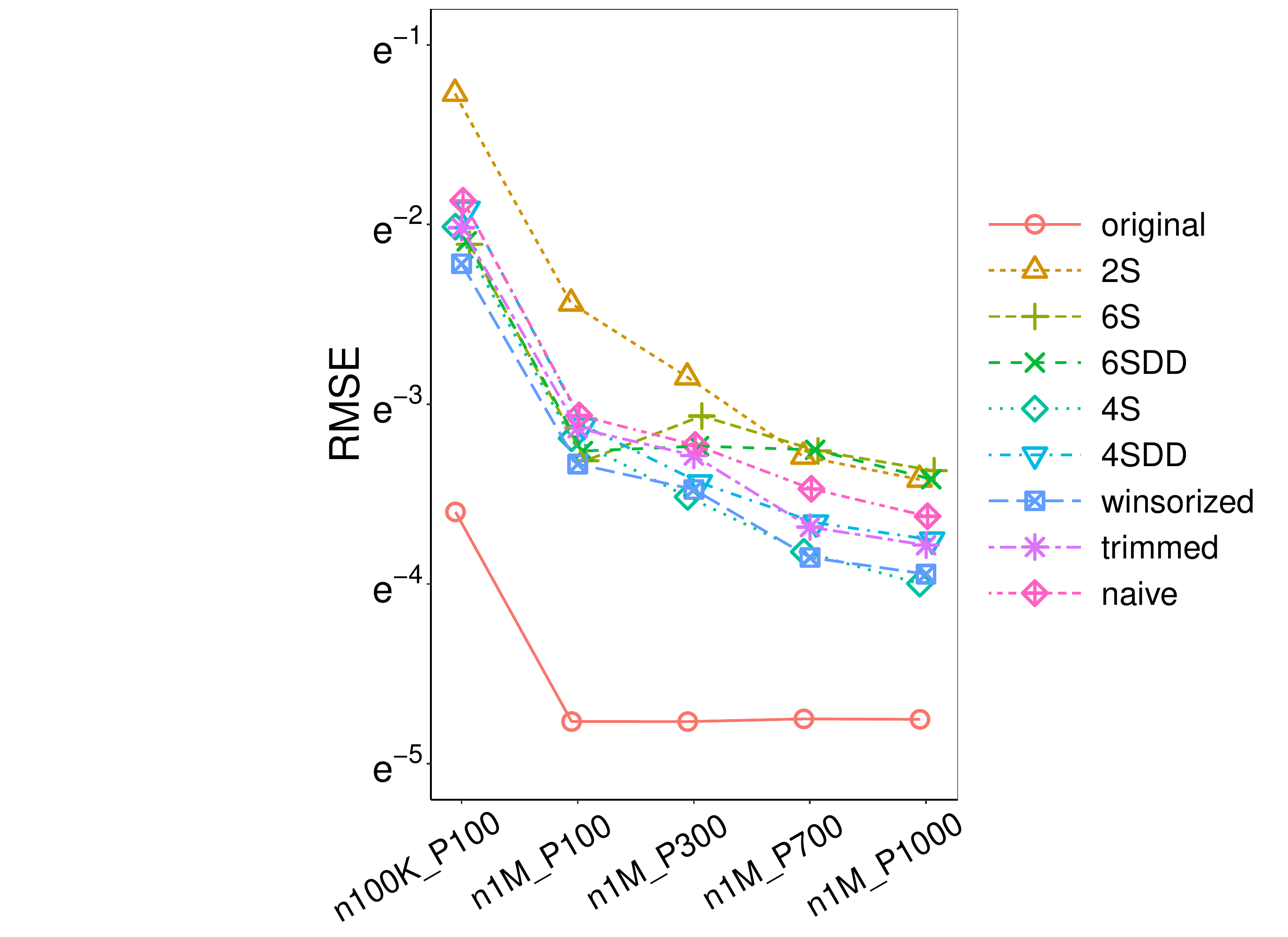}
\includegraphics[width=0.175\textwidth, trim={2.8in 0 2.6in 0},clip] {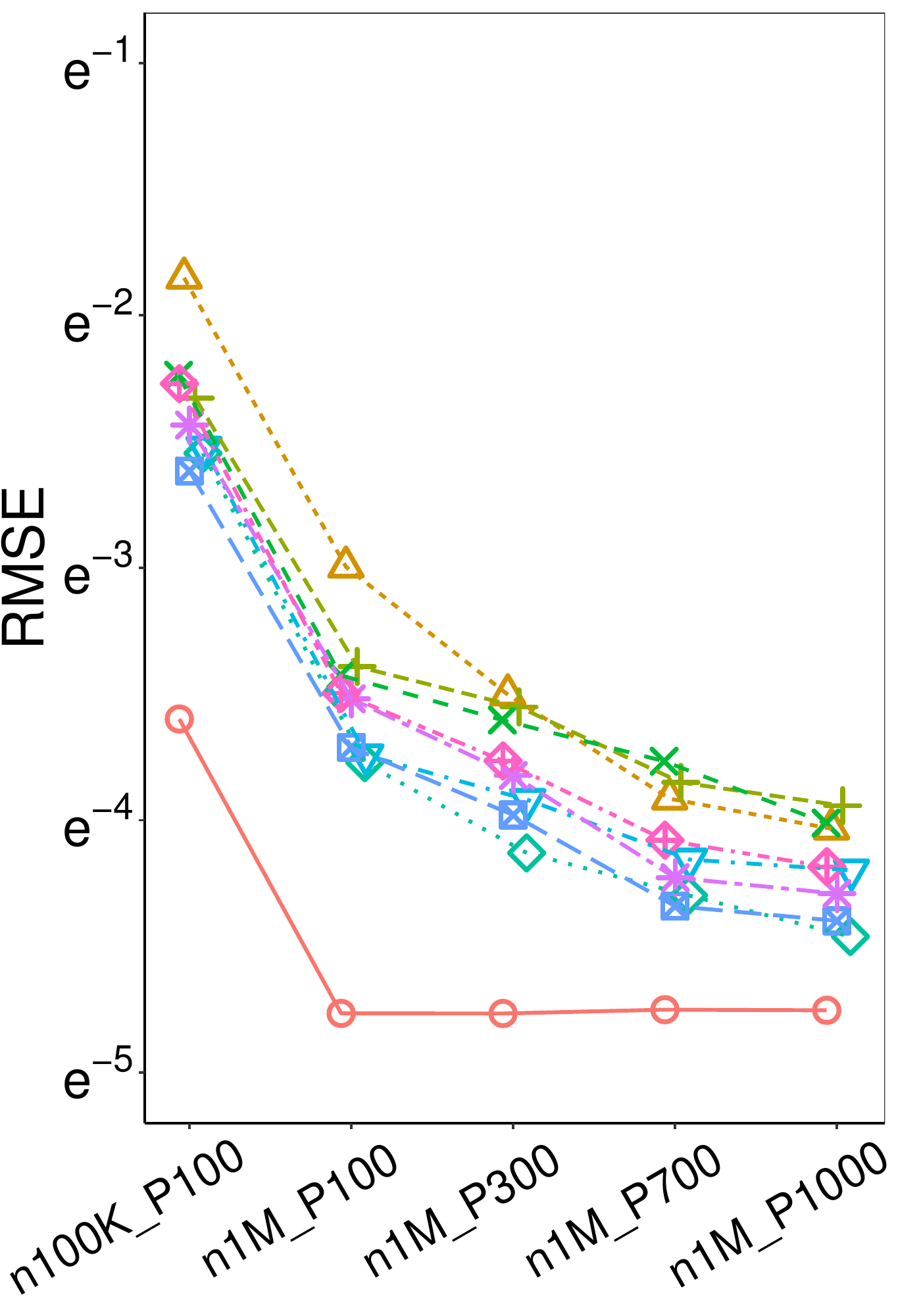}
\includegraphics[width=0.175\textwidth, trim={2.8in 0 2.6in 0},clip] {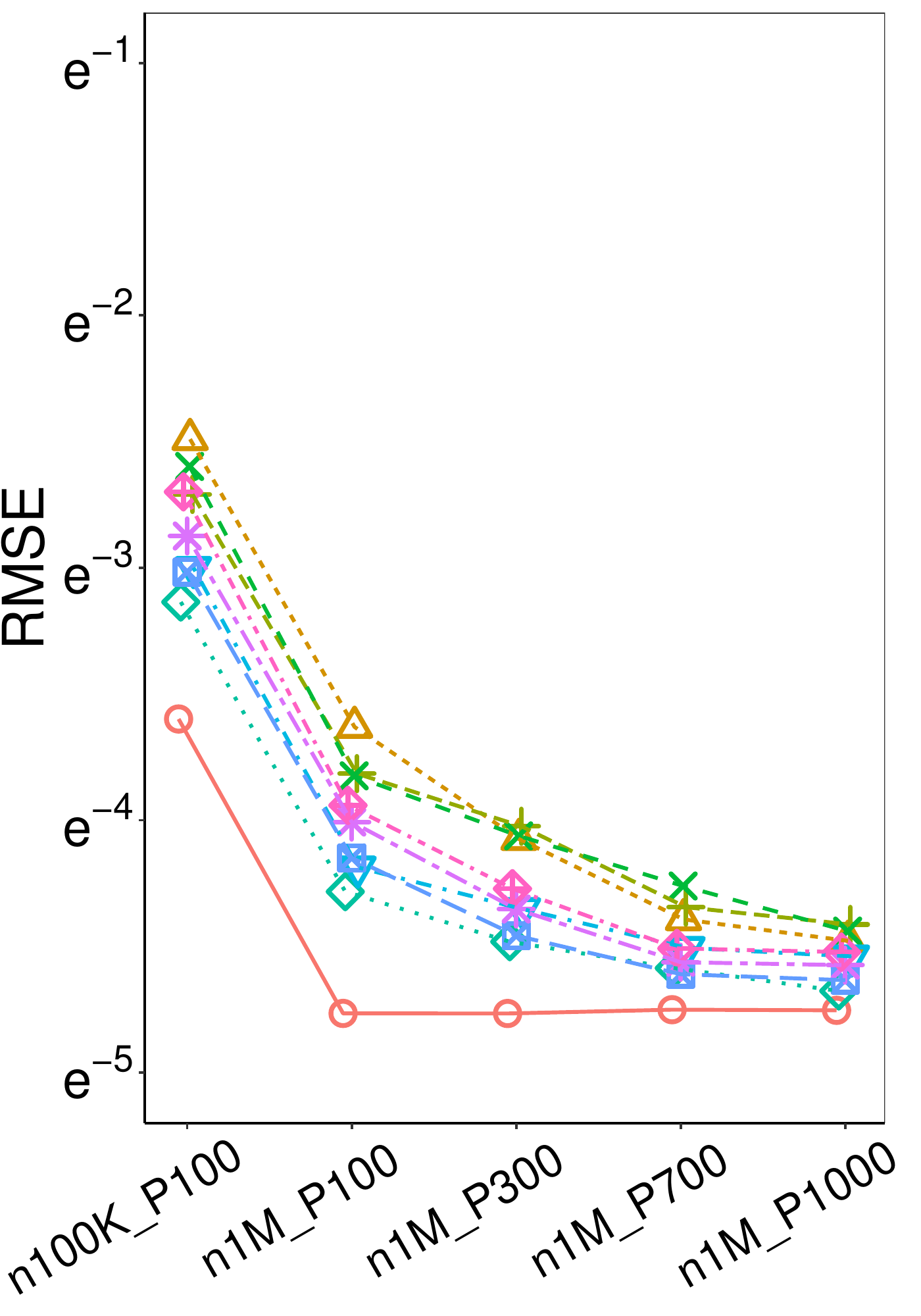}
\includegraphics[width=0.175\textwidth, trim={2.8in 0 2.6in 0},clip] {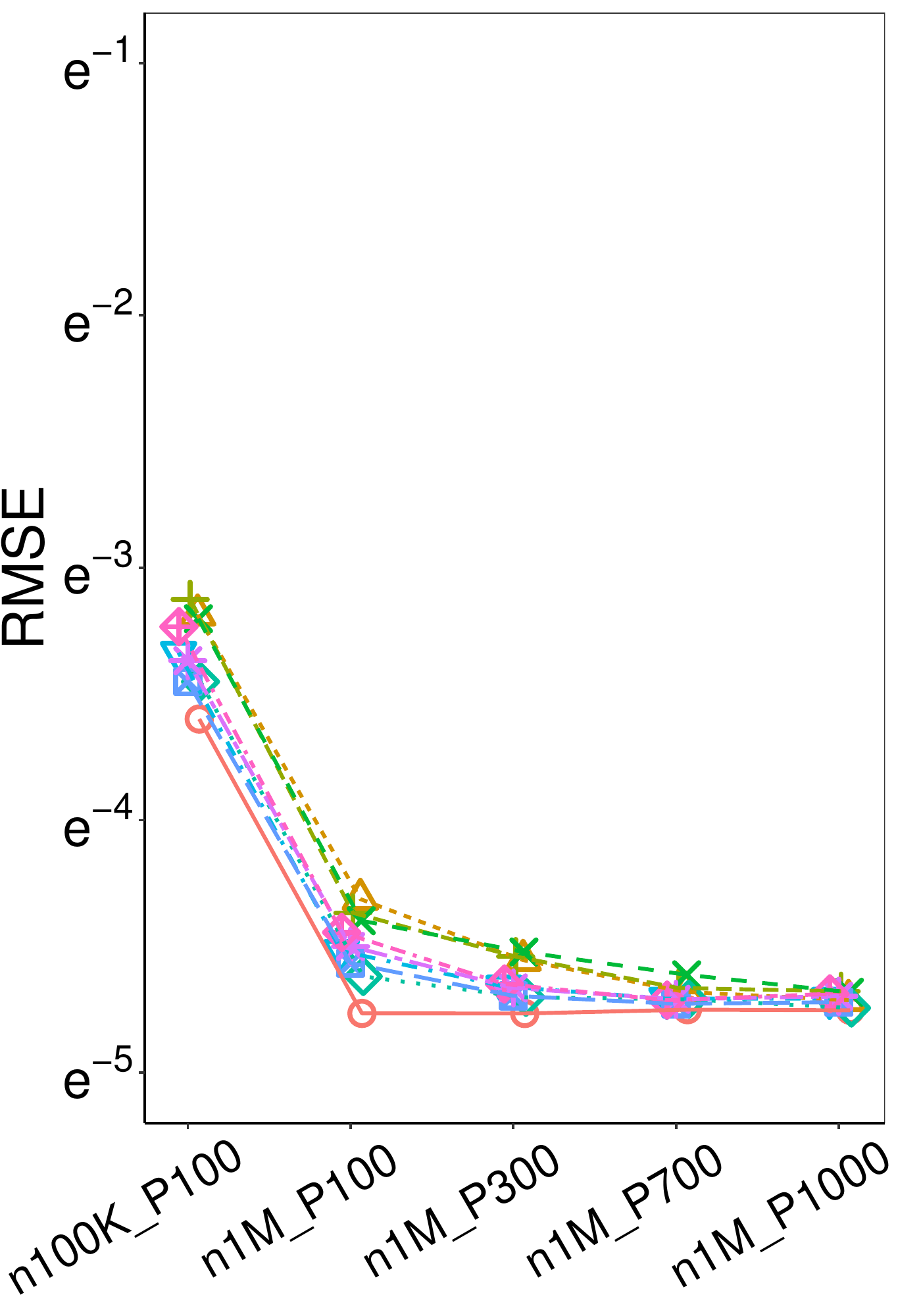}
\includegraphics[width=0.175\textwidth, trim={2.8in 0 2.6in 0},clip] {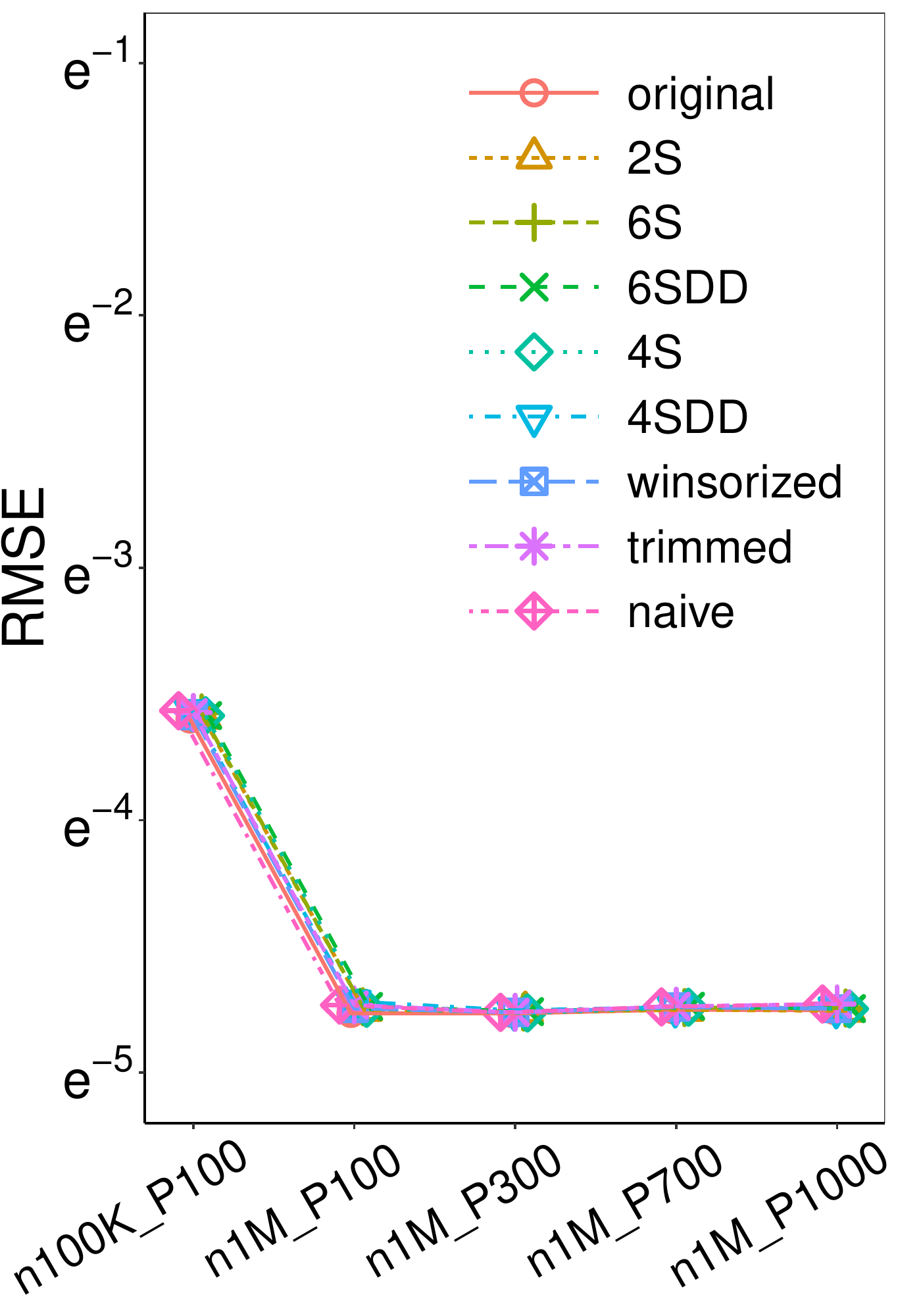}\\
\includegraphics[width=0.19\textwidth, trim={2.5in 0 2.6in 0},clip] {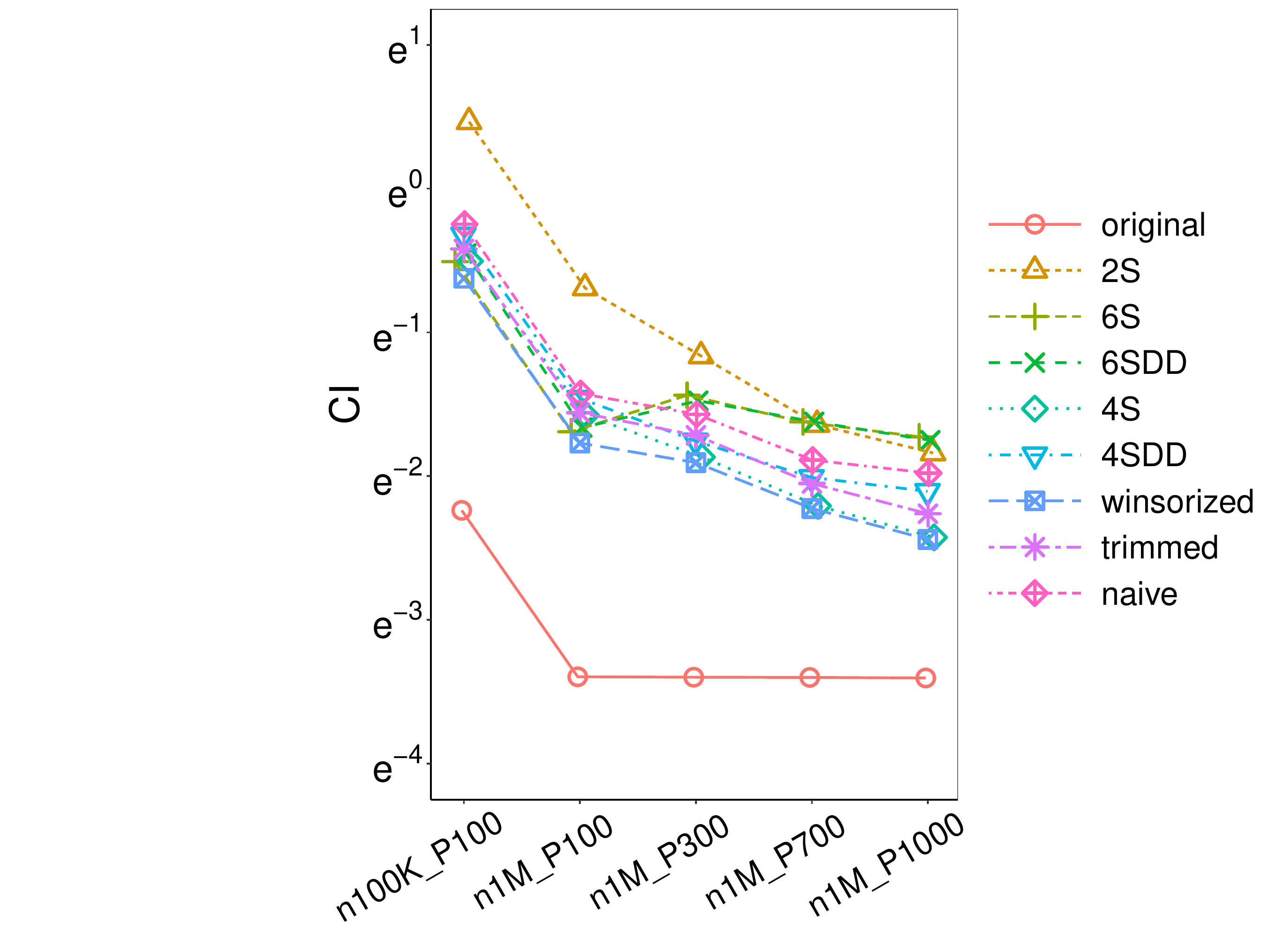}
\includegraphics[width=0.175\textwidth, trim={2.8in 0 2.6in 0},clip] {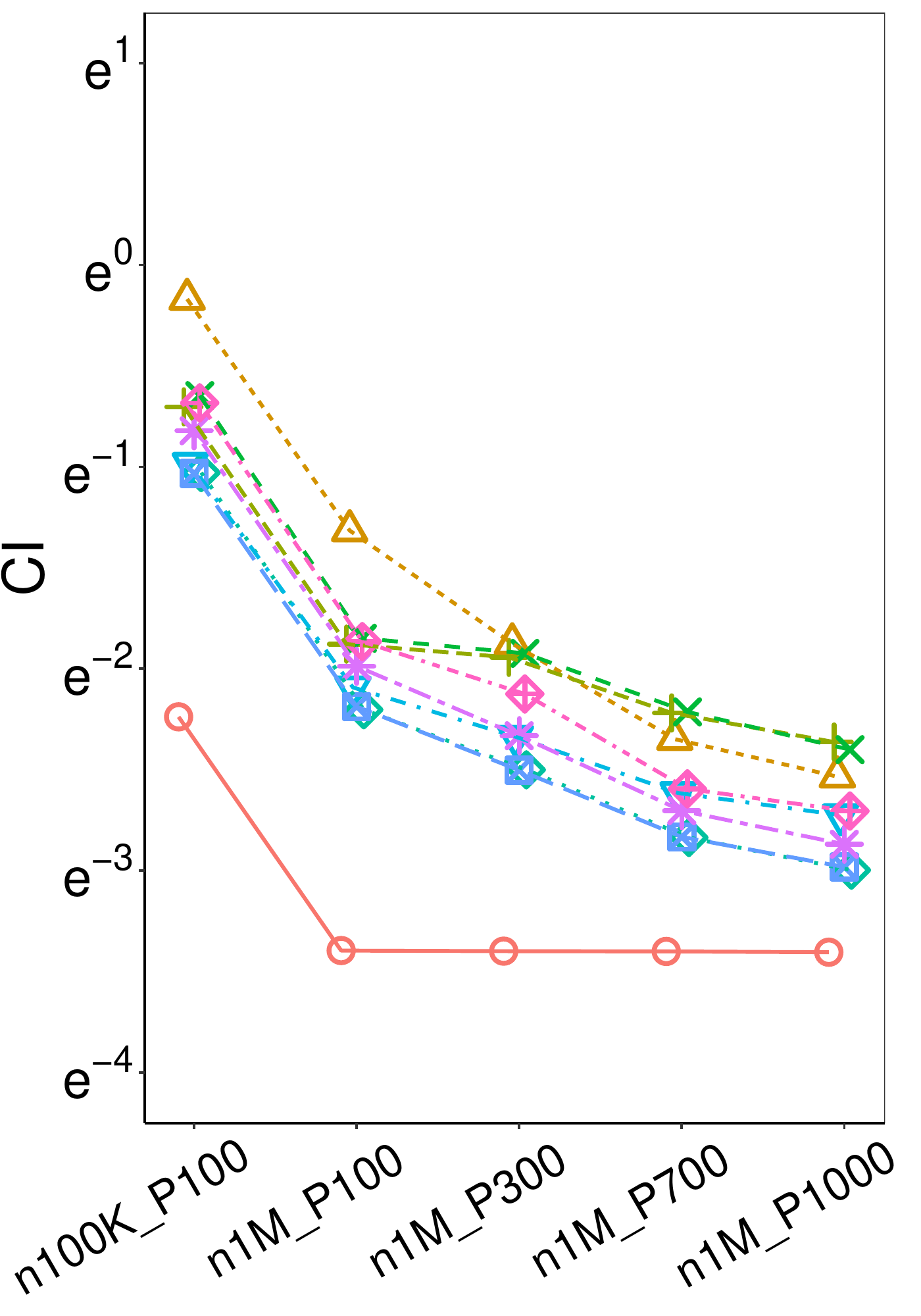}
\includegraphics[width=0.175\textwidth, trim={2.8in 0 2.6in 0},clip] {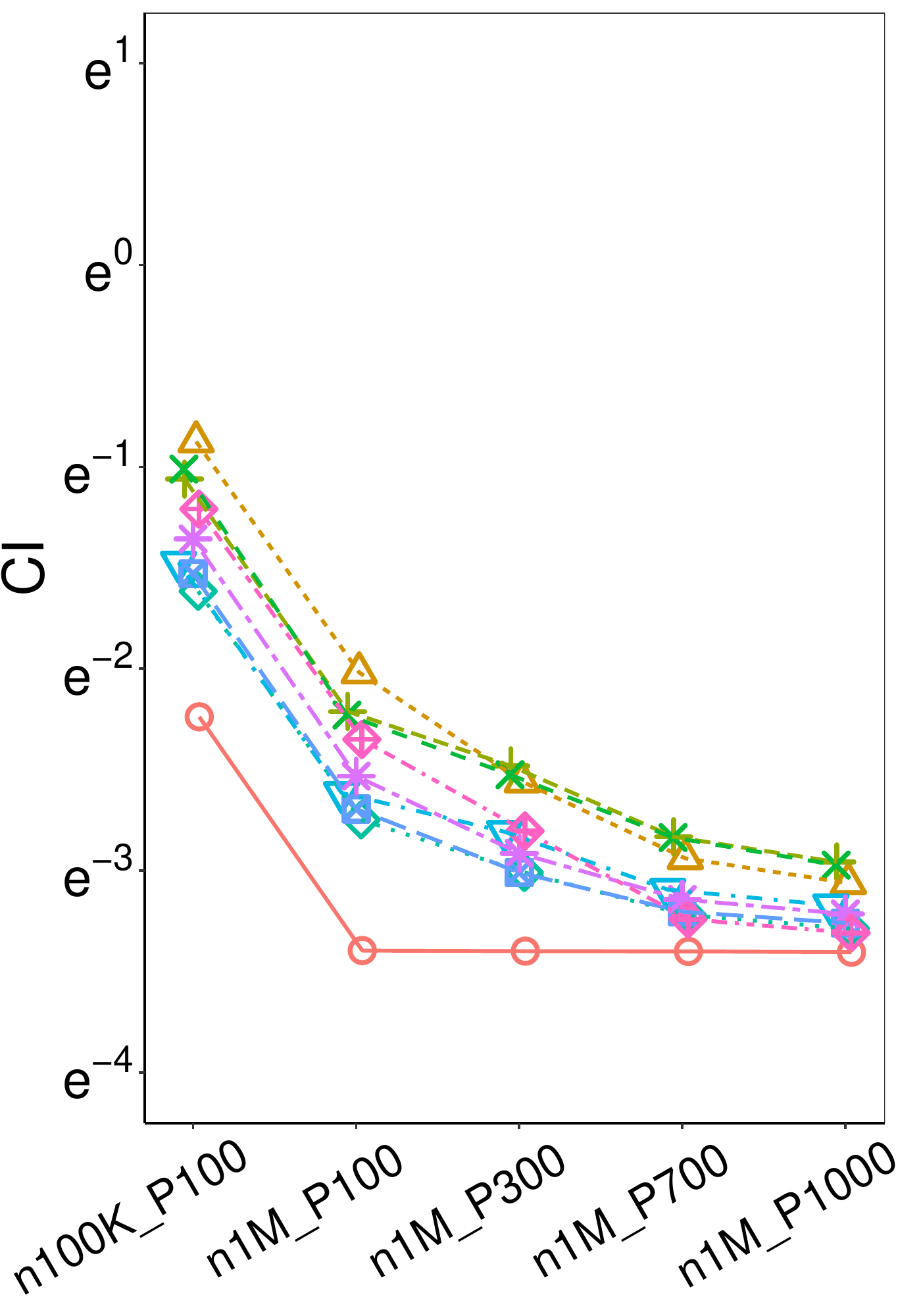}
\includegraphics[width=0.175\textwidth, trim={2.8in 0 2.6in 0},clip] {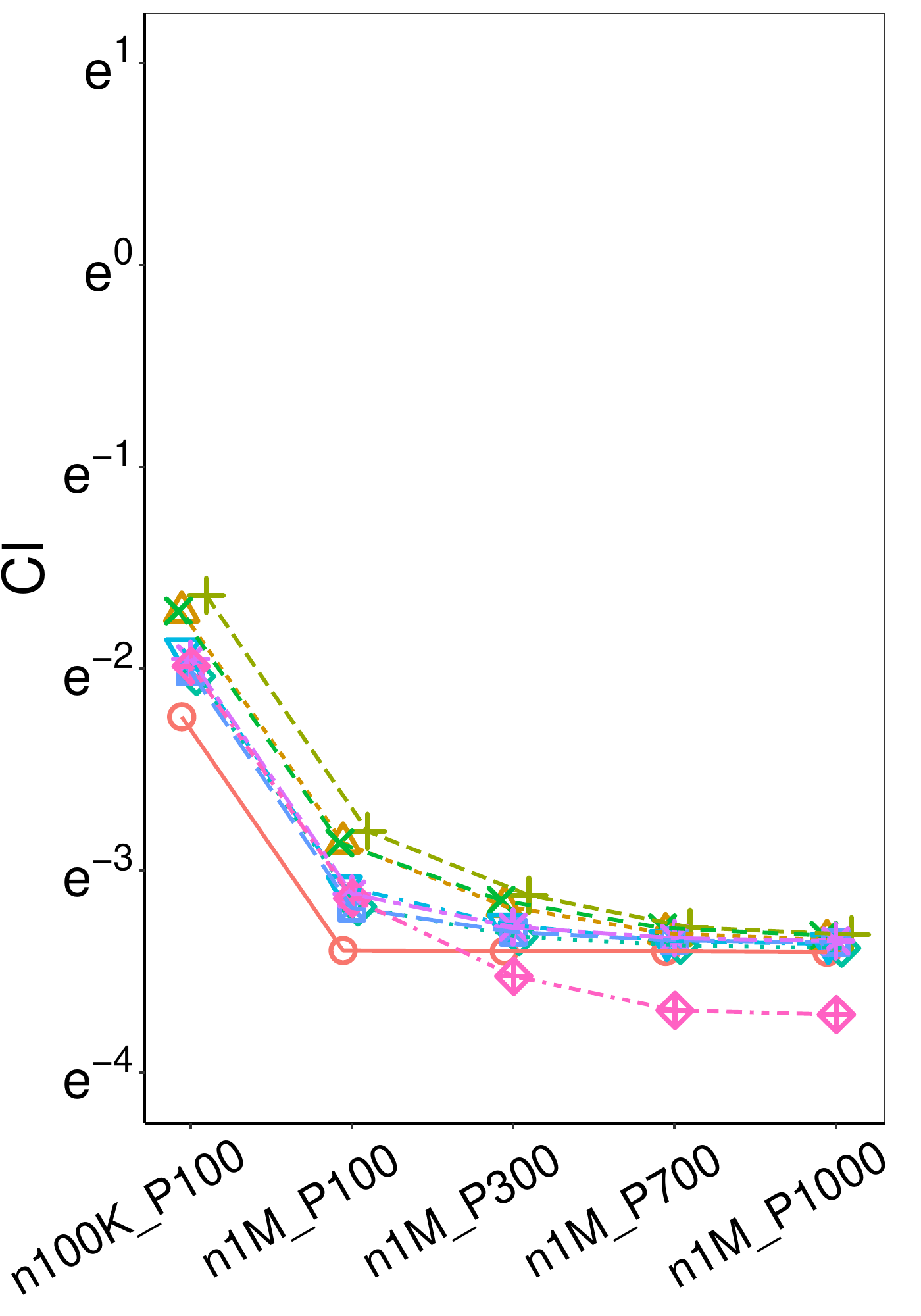}
\includegraphics[width=0.175\textwidth, trim={2.8in 0 2.6in 0},clip] {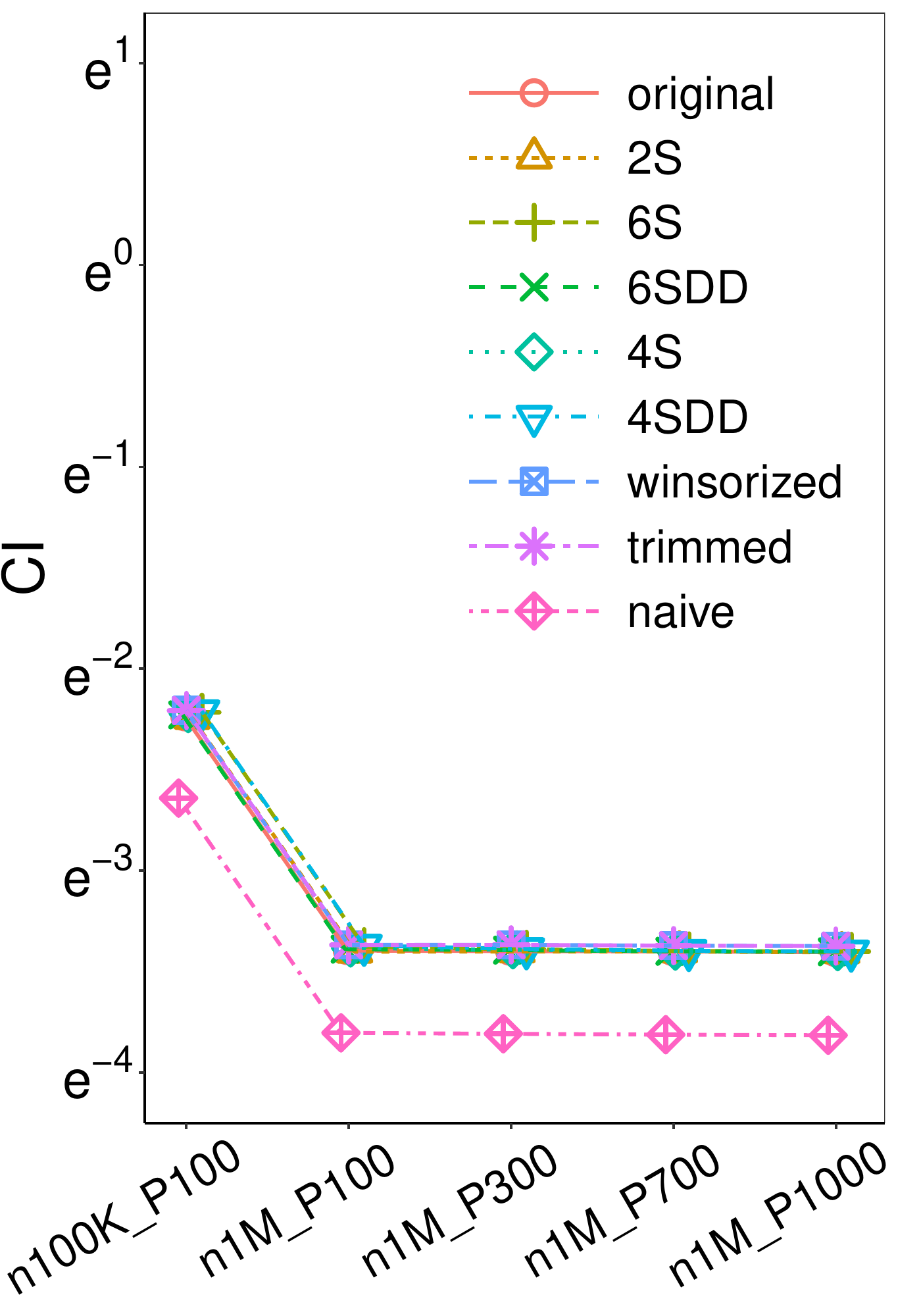}\\
\includegraphics[width=0.19\textwidth, trim={2.5in 0 2.6in 0},clip] {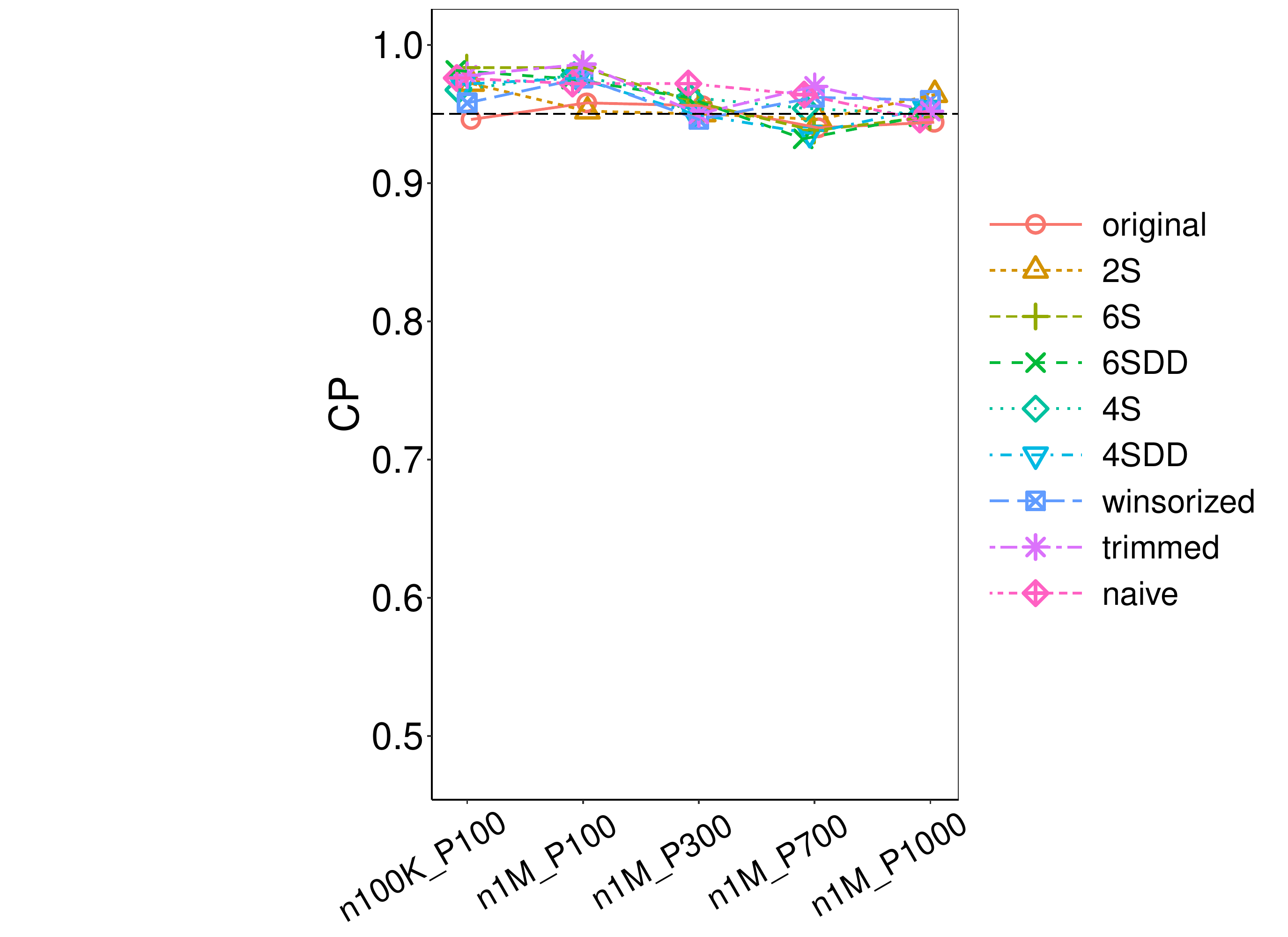}
\includegraphics[width=0.175\textwidth, trim={2.8in 0 2.6in 0},clip] {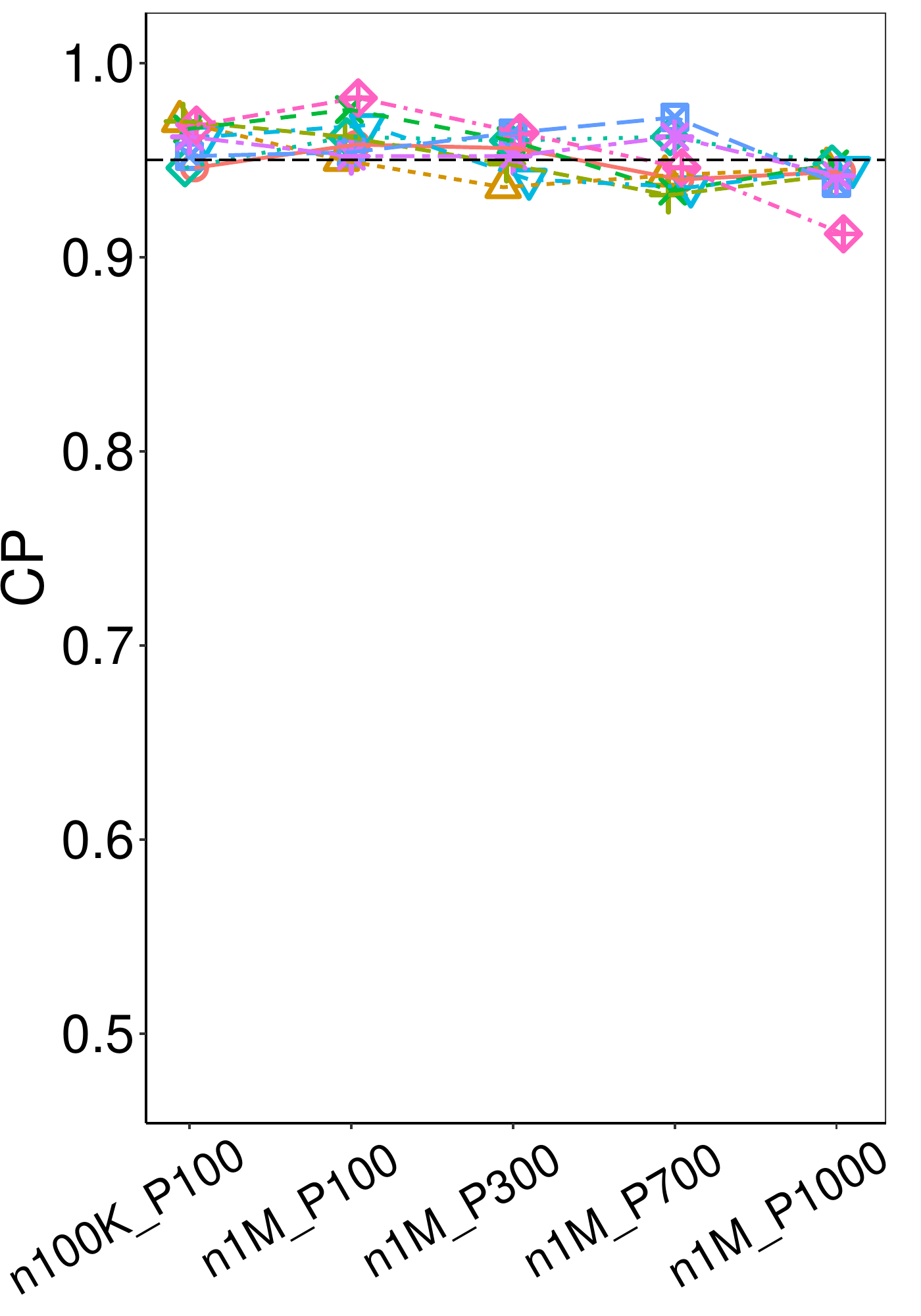}
\includegraphics[width=0.175\textwidth, trim={2.8in 0 2.6in 0},clip] {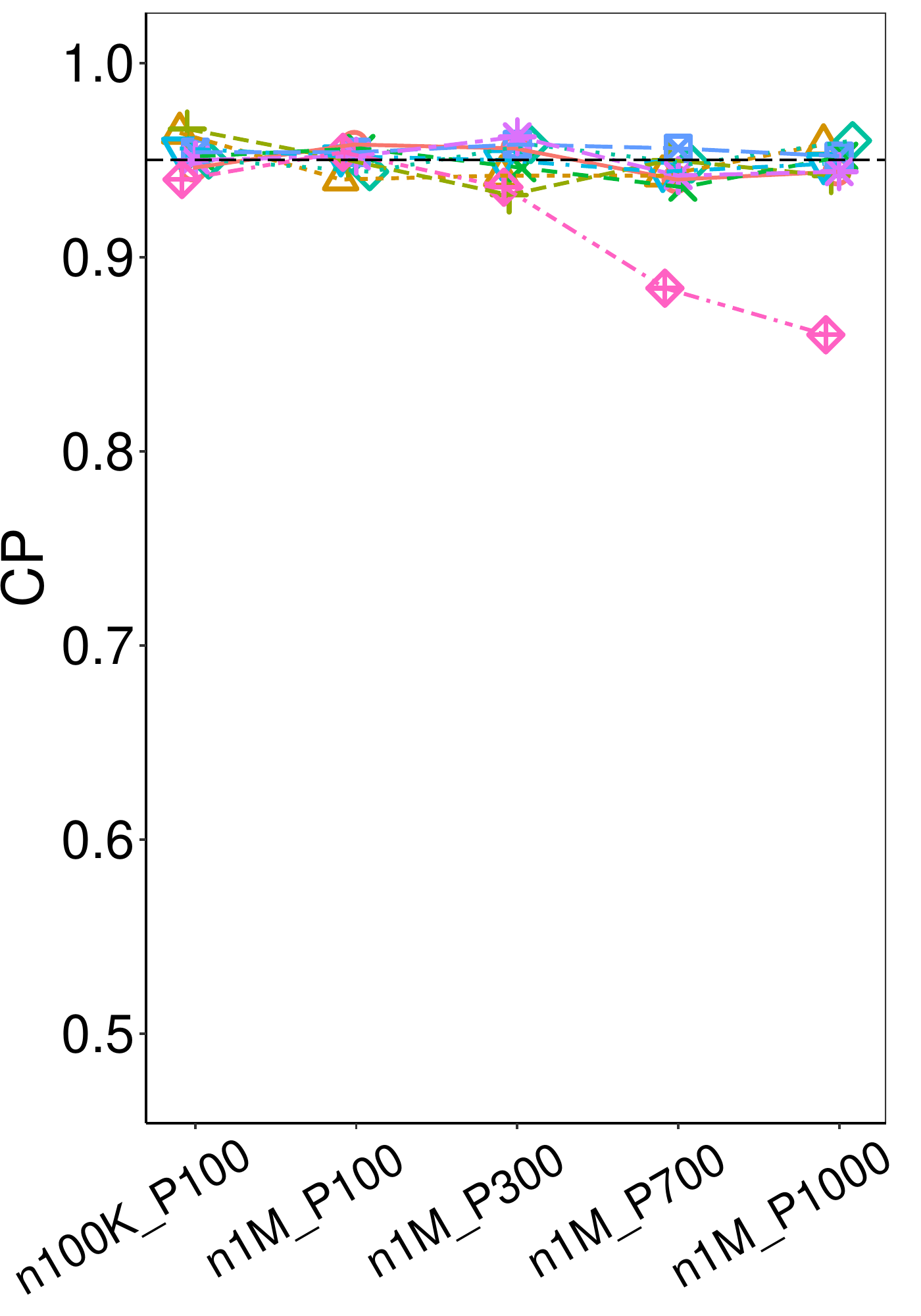}
\includegraphics[width=0.175\textwidth, trim={2.8in 0 2.6in 0},clip] {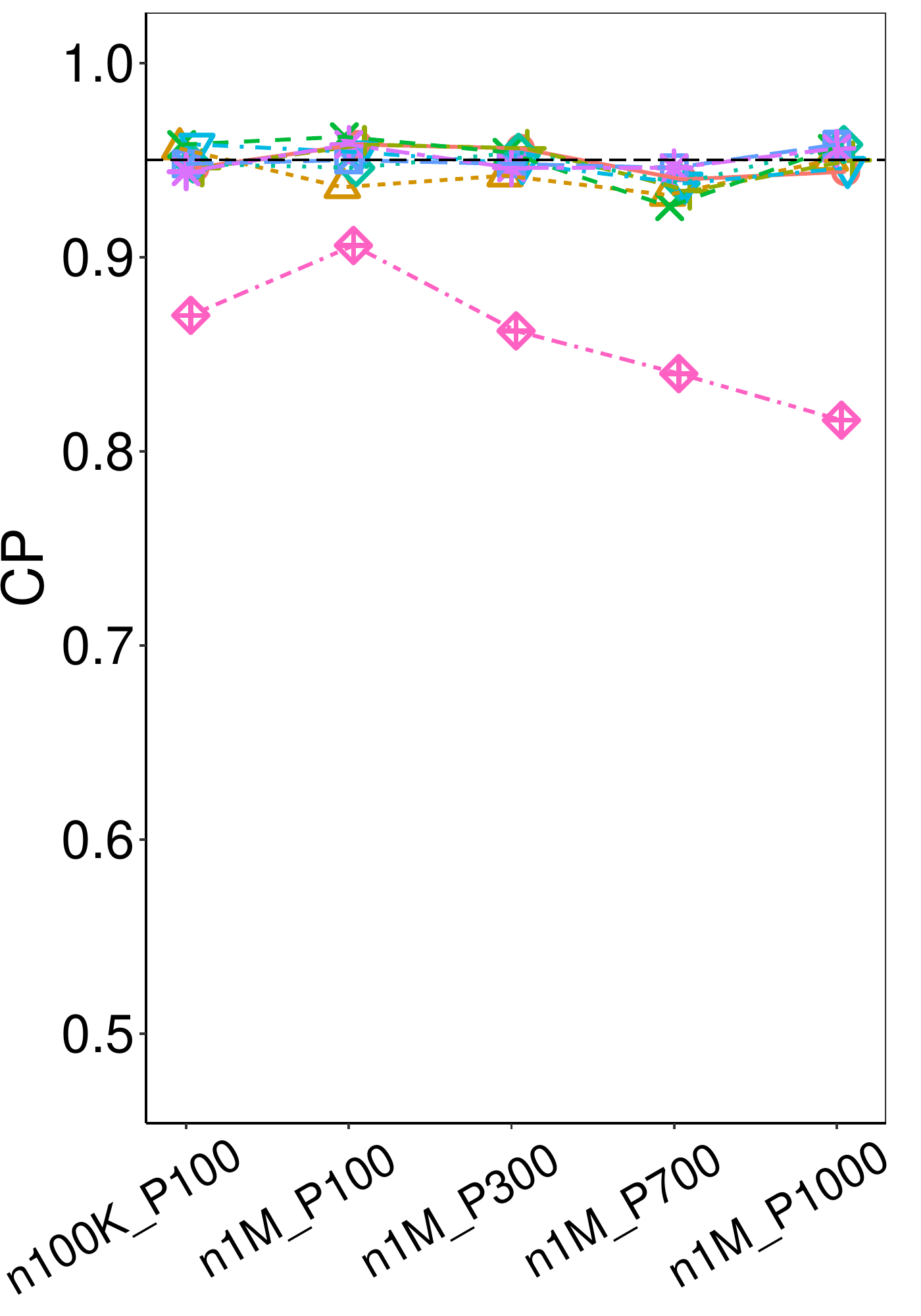}
\includegraphics[width=0.175\textwidth, trim={2.8in 0 2.6in 0},clip] {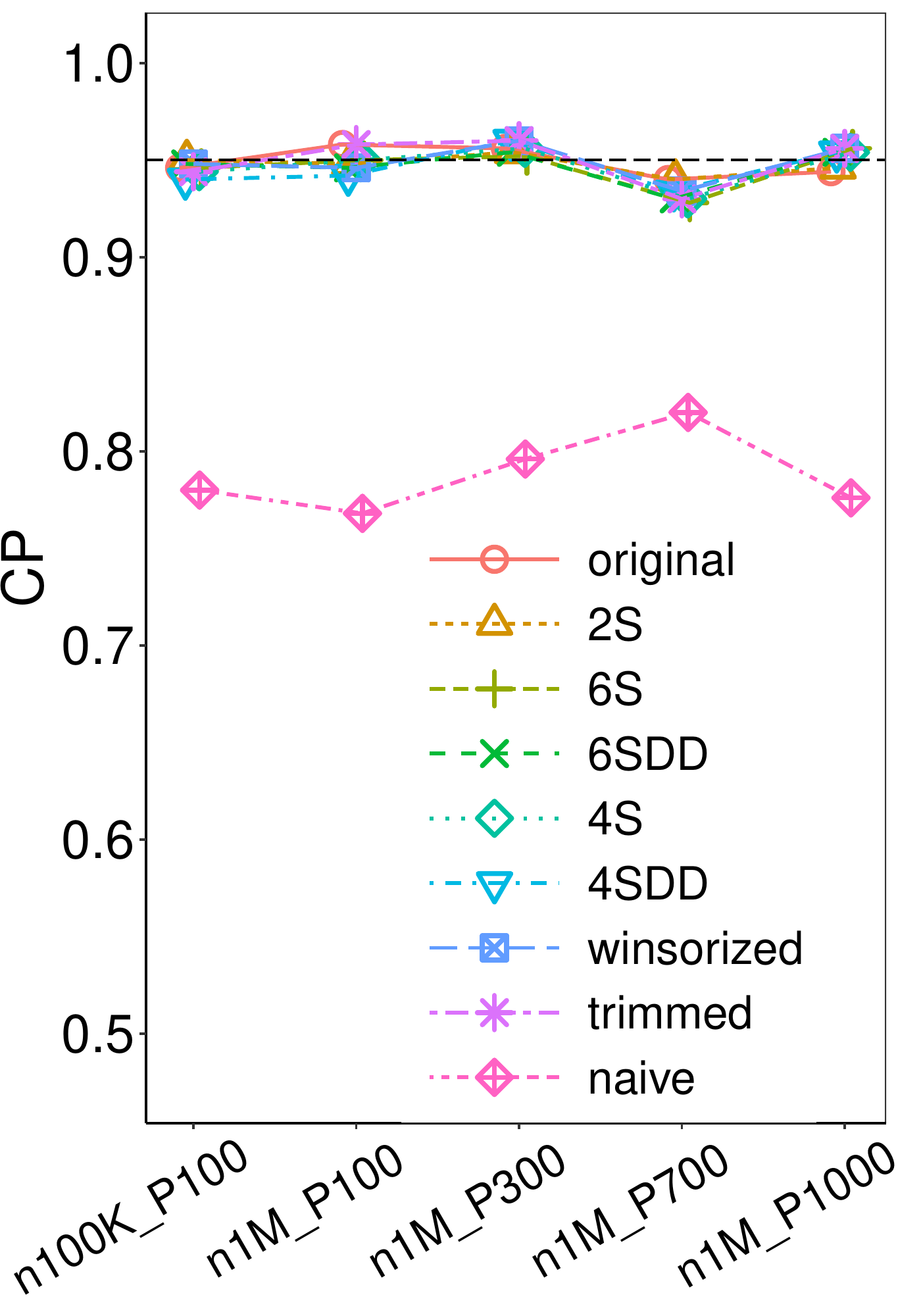}
\vspace{-9pt}
\caption{Simulation results with $\epsilon$-DP for Gaussian data  with $\alpha=\beta$ when and $\theta=0$ } \label{fig:0sDPN}

\end{figure}

\begin{figure}[!htb]
\hspace{0.4in}$\rho=0.005$\hspace{0.6in}$\rho=0.02$\hspace{0.8in}$\rho=0.08$
\hspace{0.65in}$\rho=0.32$\hspace{0.7in}$\rho=1.28$\\
\includegraphics[width=0.19\textwidth, trim={2.5in 0 2.6in 0},clip] {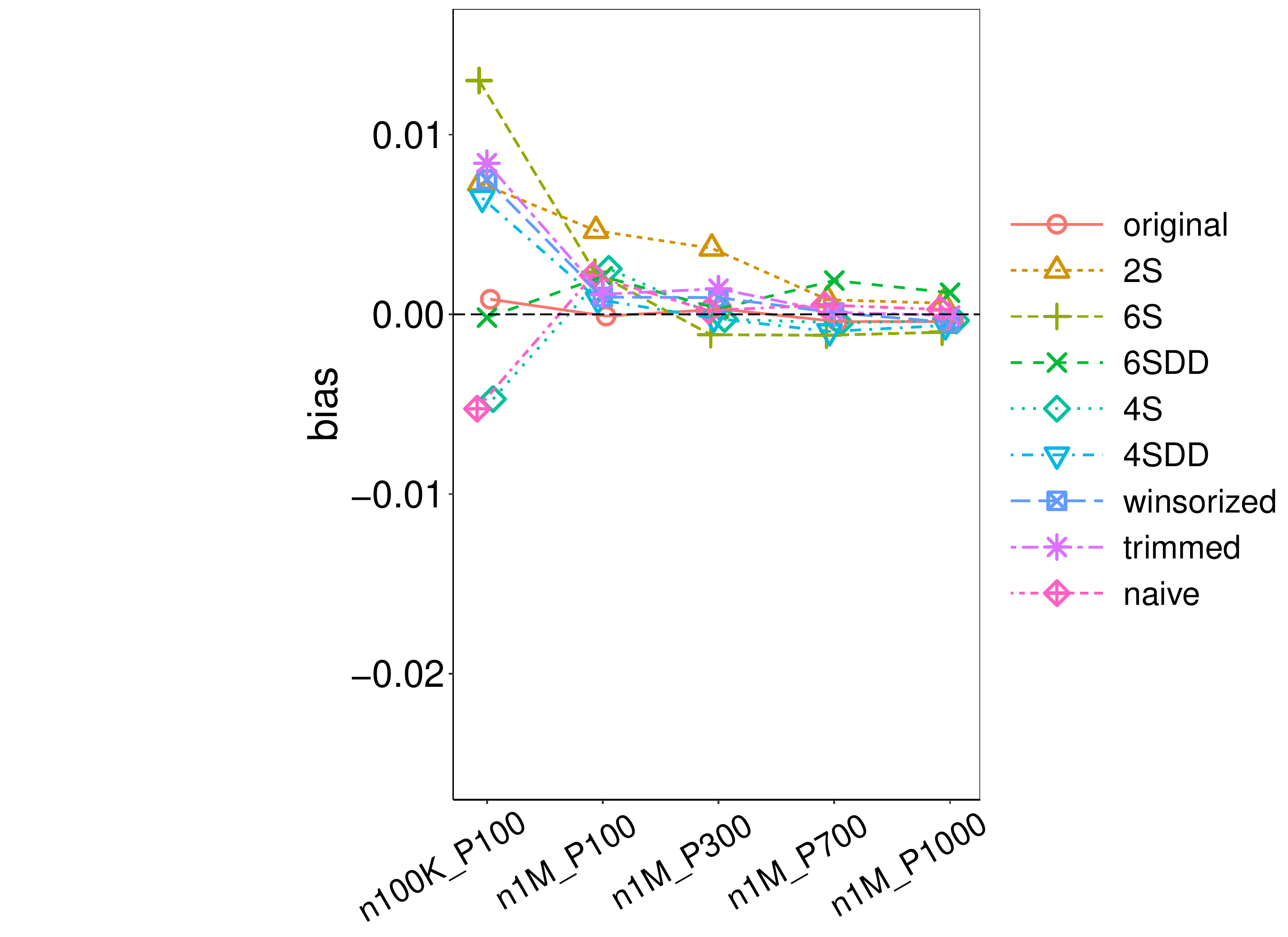}
\includegraphics[width=0.19\textwidth, trim={2.5in 0 2.6in 0},clip] {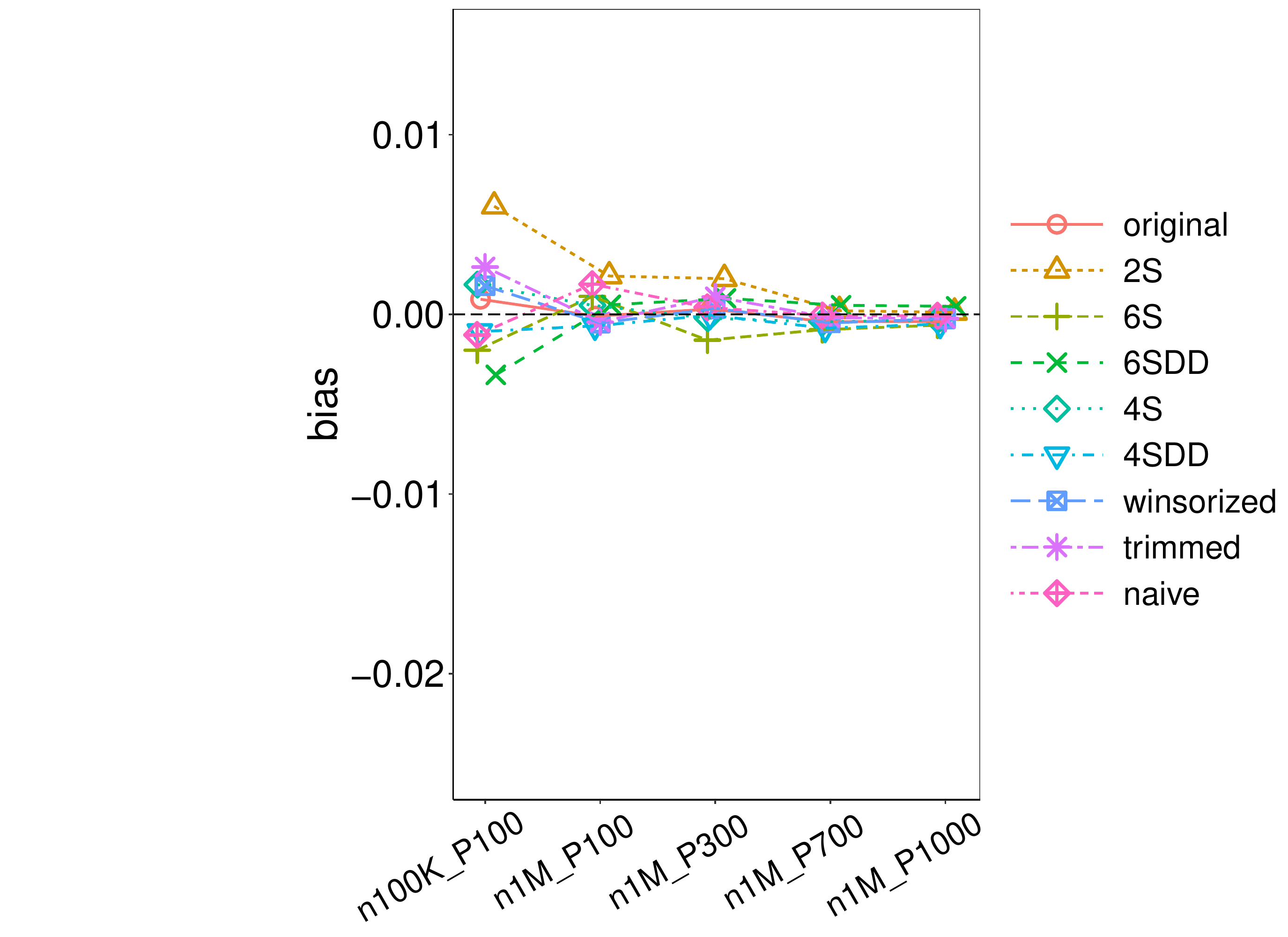}
\includegraphics[width=0.19\textwidth, trim={2.5in 0 2.6in 0},clip] {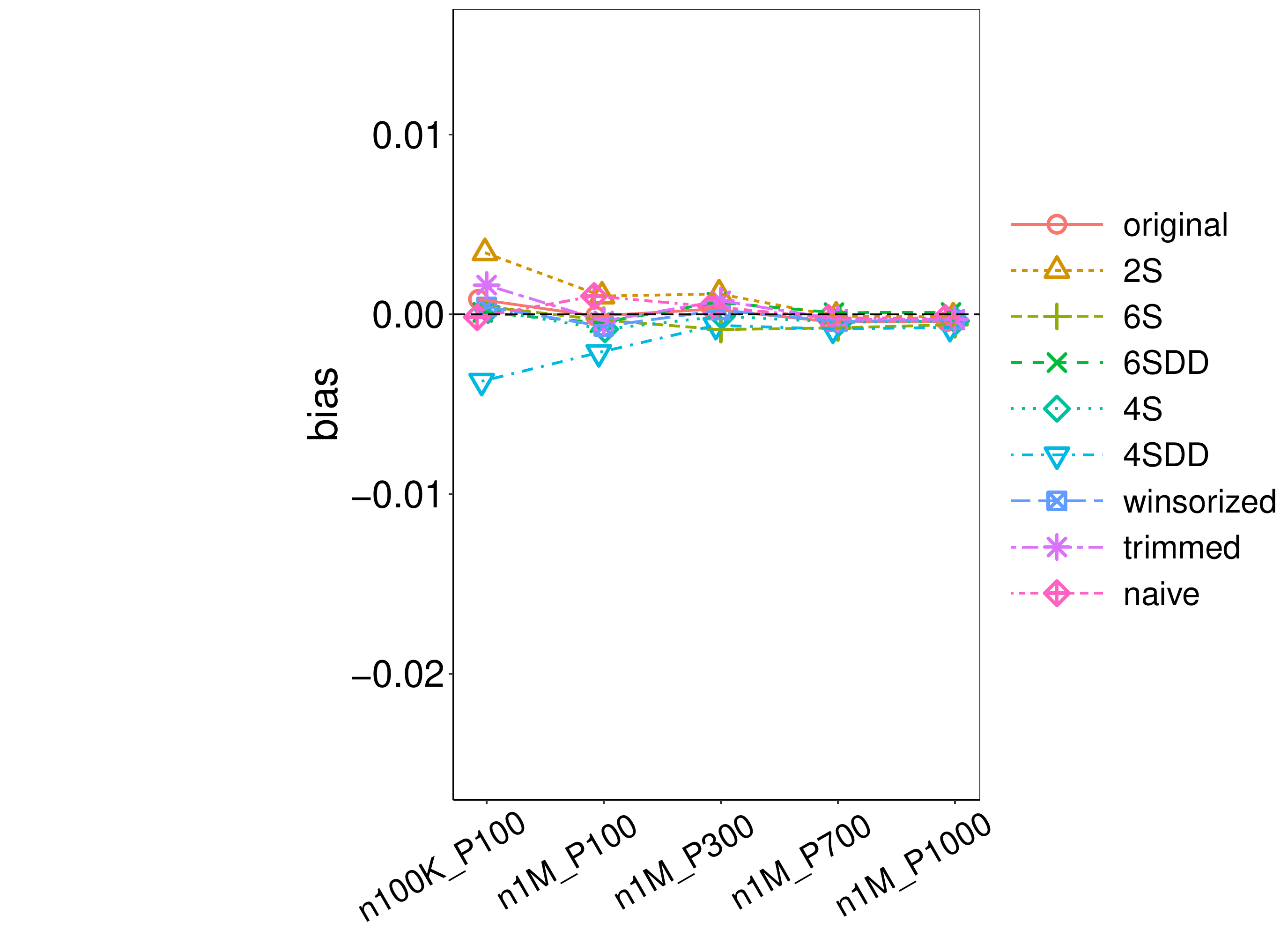}
\includegraphics[width=0.19\textwidth, trim={2.5in 0 2.6in 0},clip] {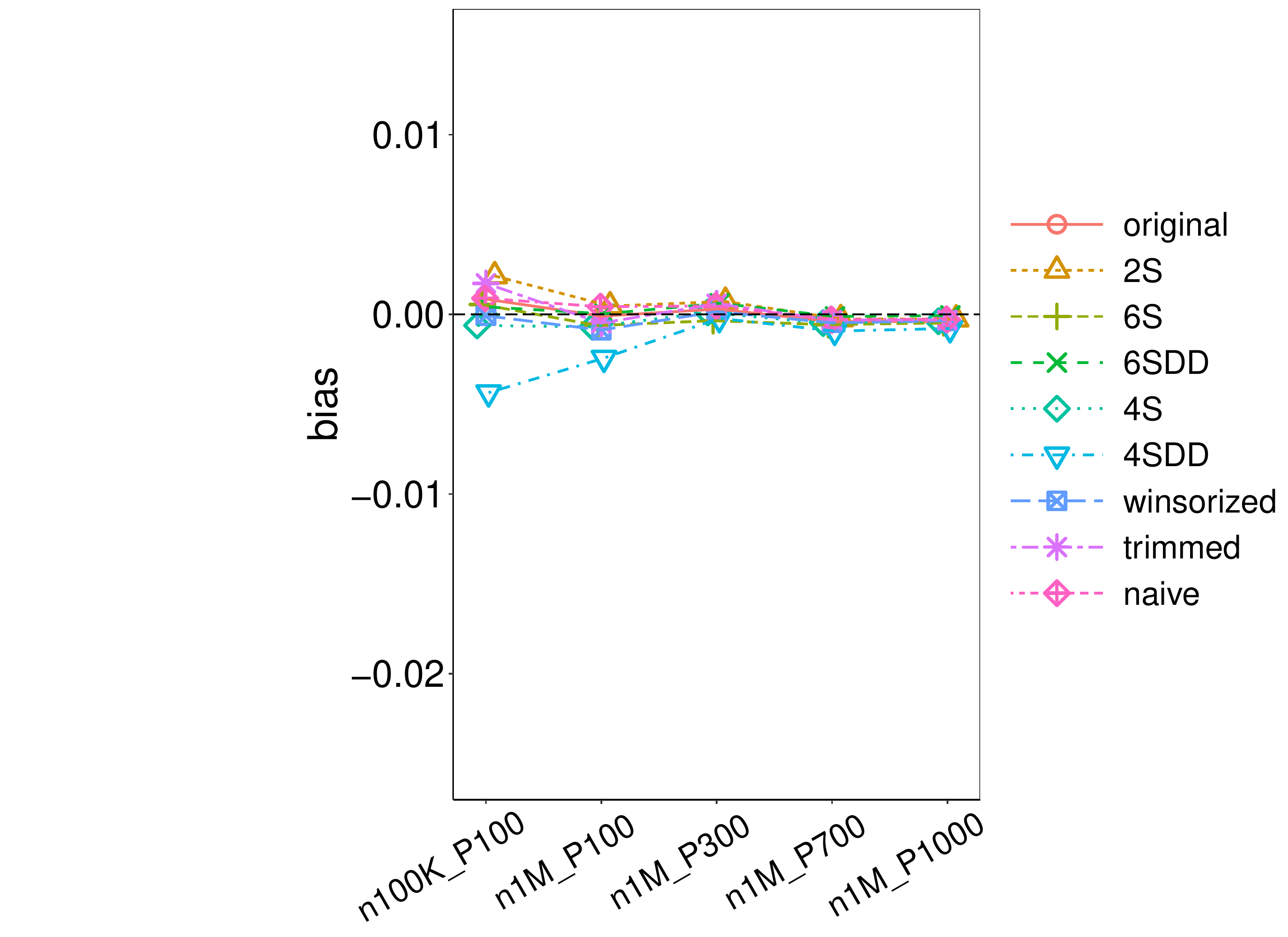}
\includegraphics[width=0.19\textwidth, trim={2.5in 0 2.6in 0},clip] {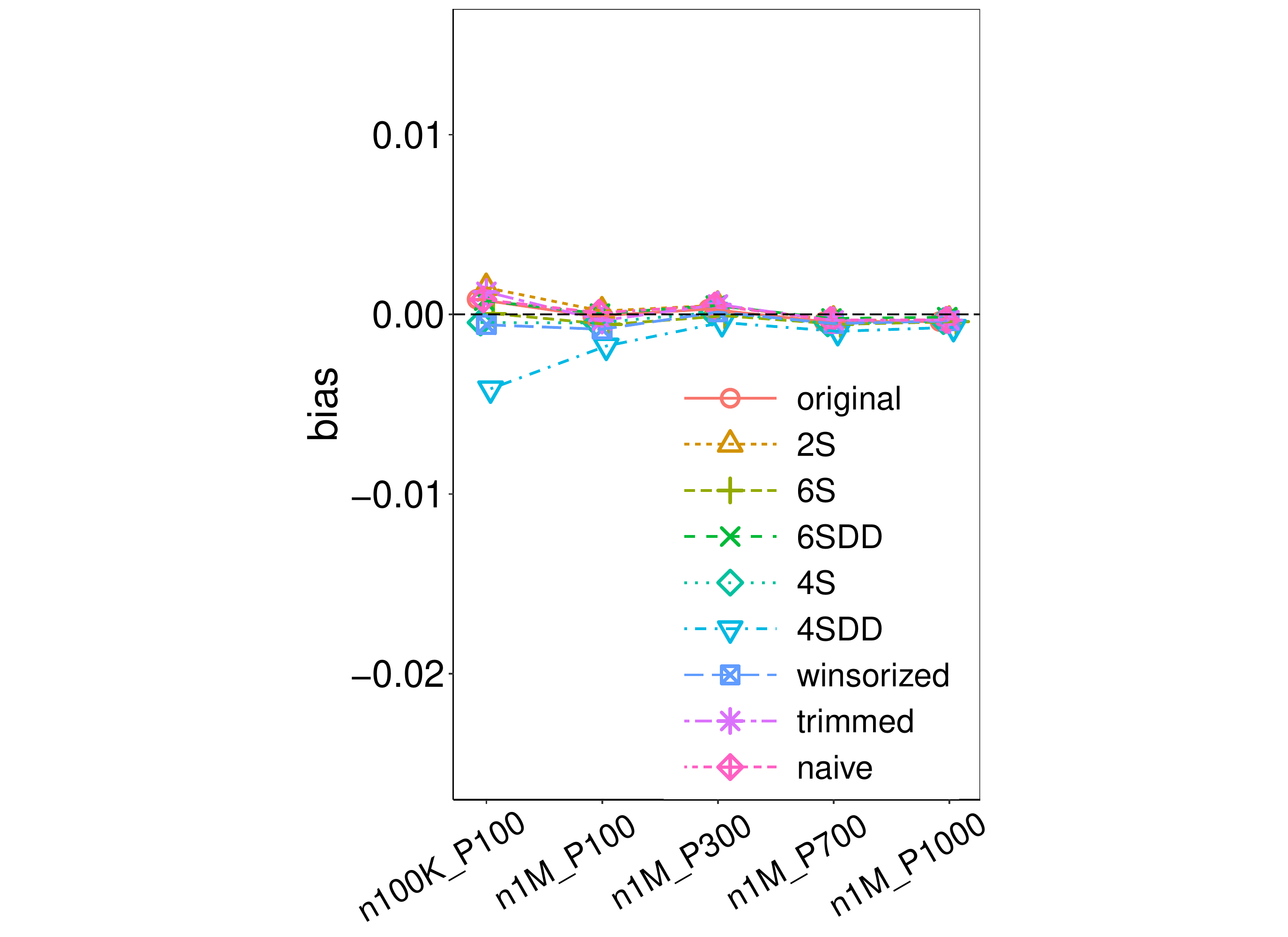}

\includegraphics[width=0.19\textwidth, trim={2.5in 0 2.6in 0},clip] {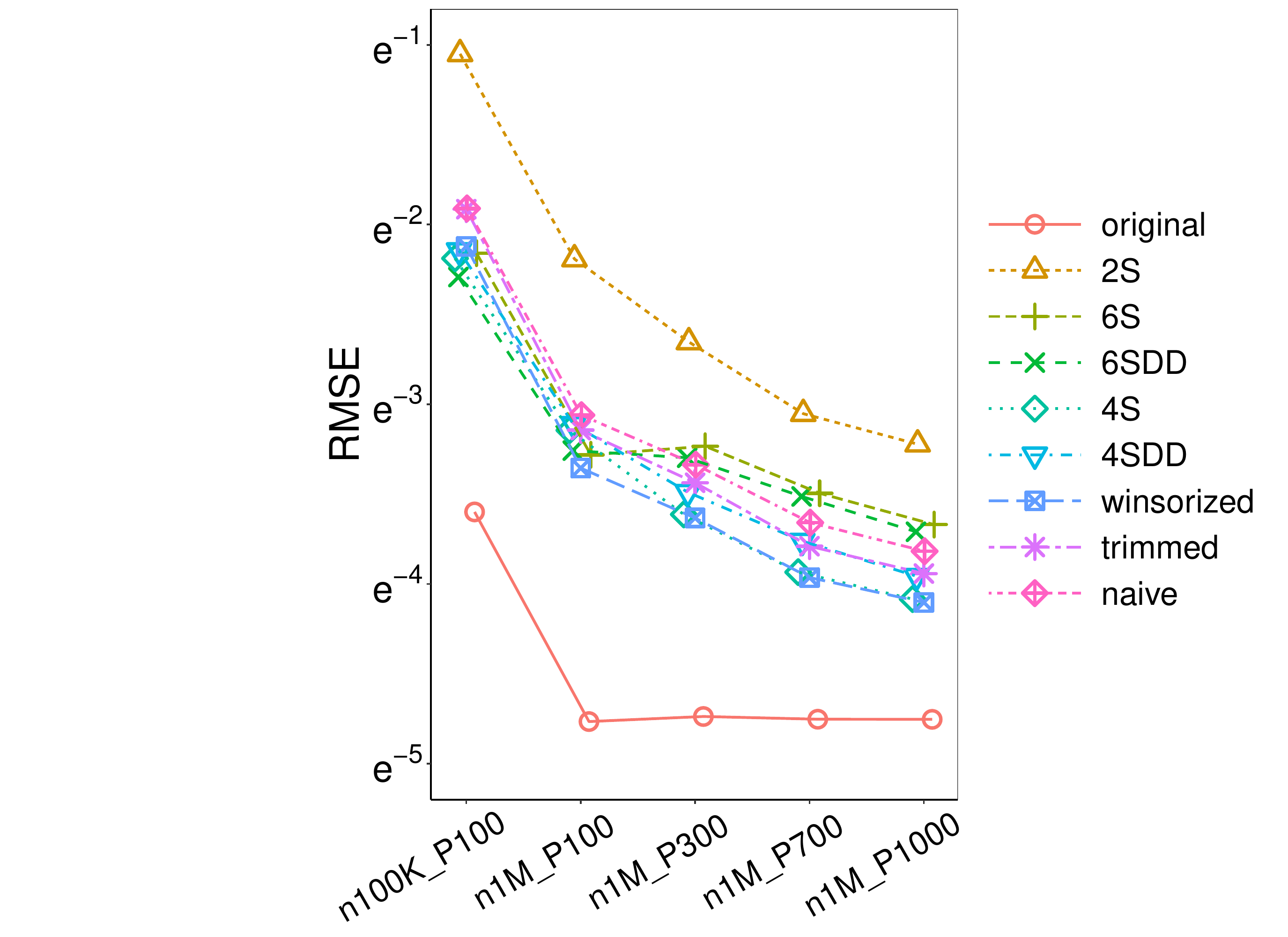}
\includegraphics[width=0.19\textwidth, trim={2.5in 0 2.6in 0},clip] {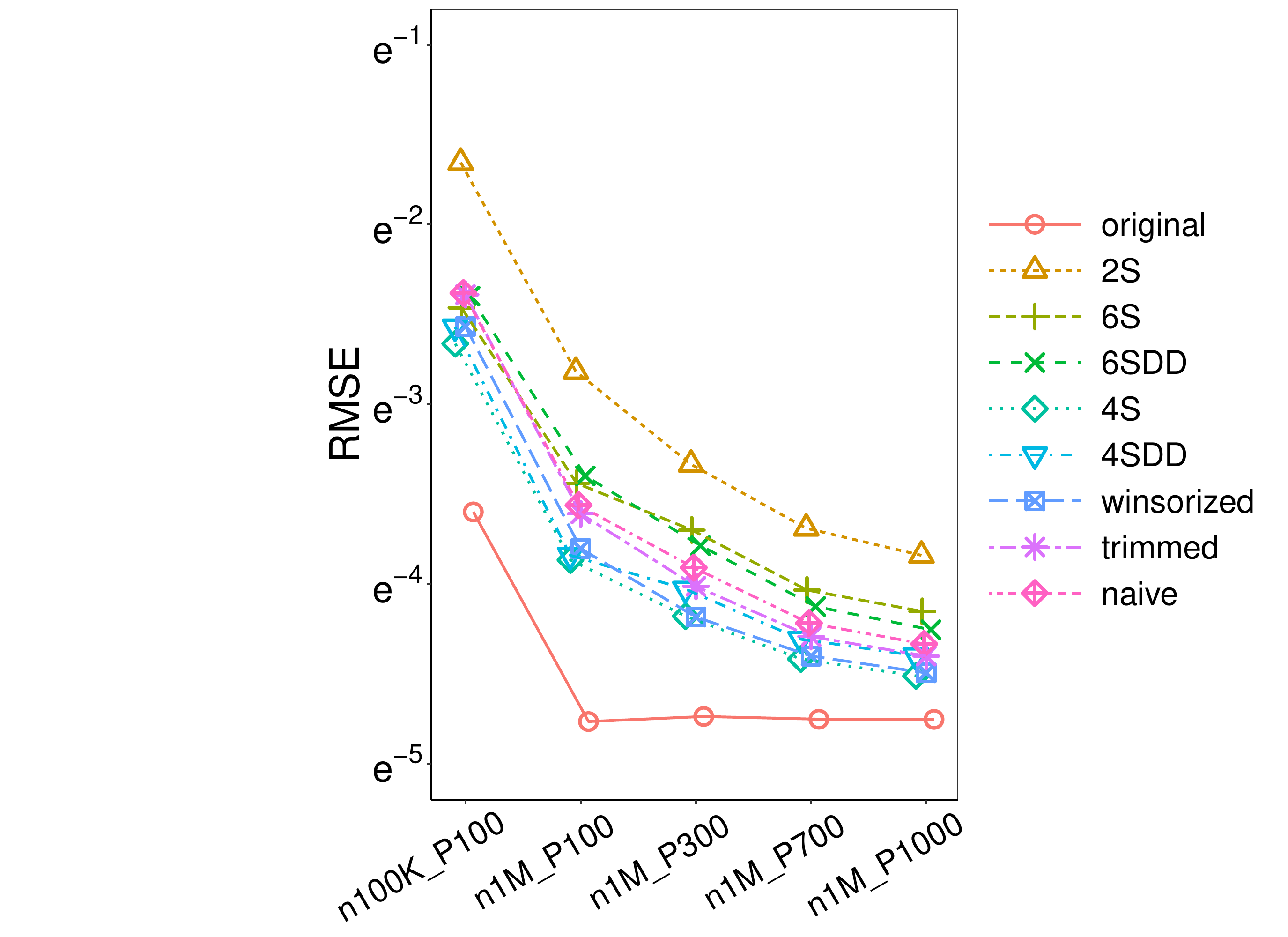}
\includegraphics[width=0.19\textwidth, trim={2.5in 0 2.6in 0},clip] {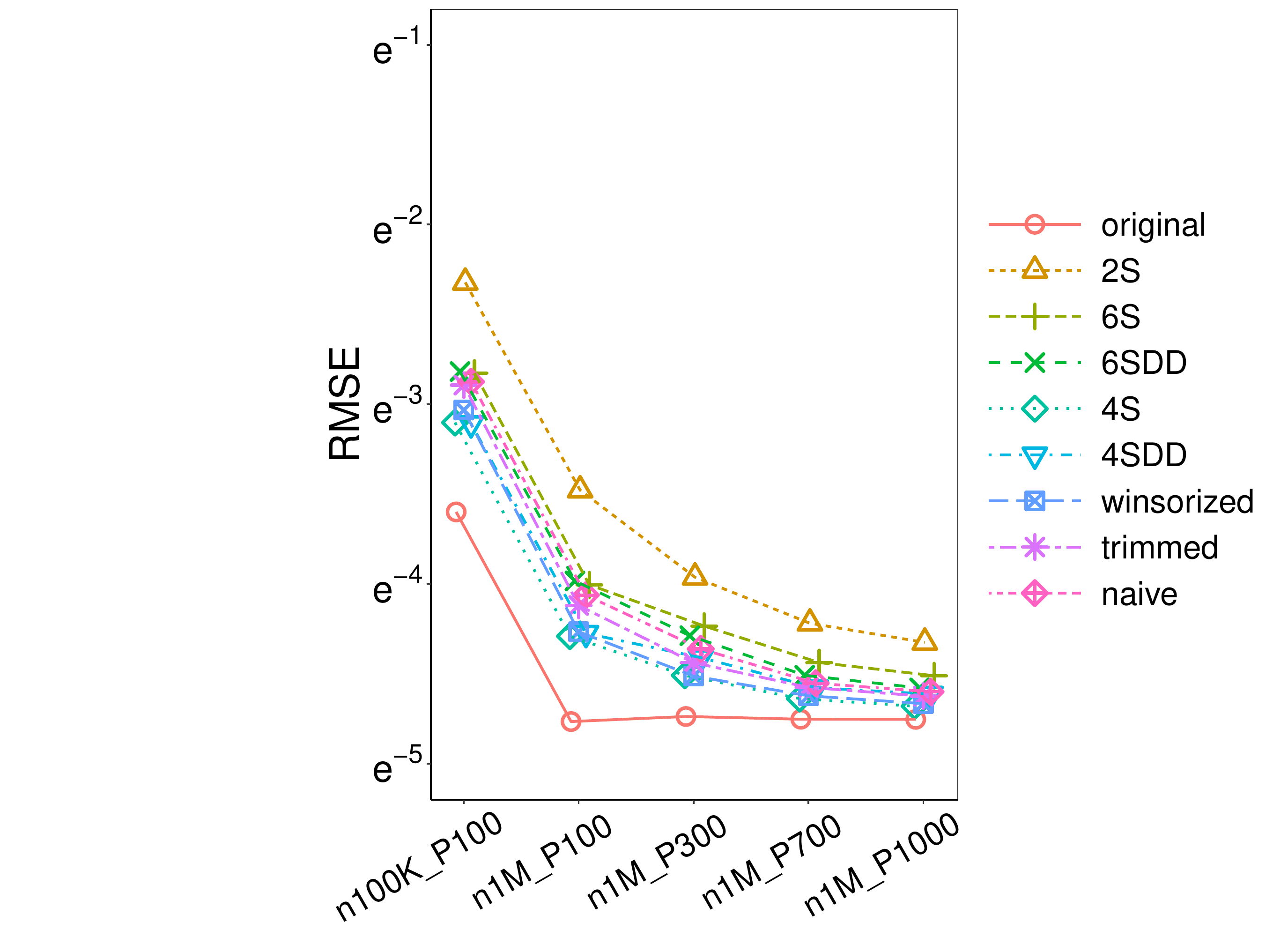}
\includegraphics[width=0.19\textwidth, trim={2.5in 0 2.6in 0},clip] {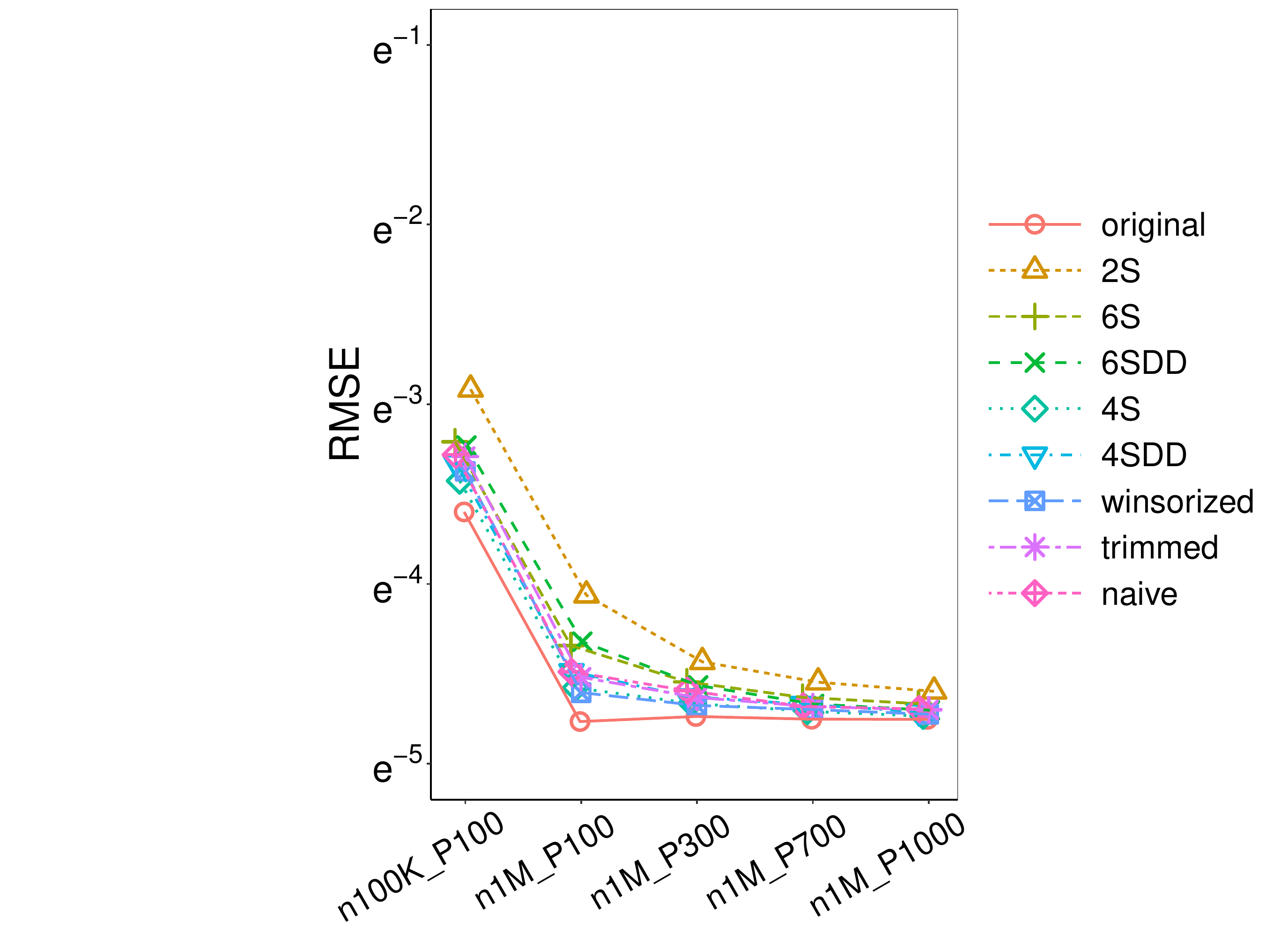}
\includegraphics[width=0.19\textwidth, trim={2.5in 0 2.6in 0},clip] {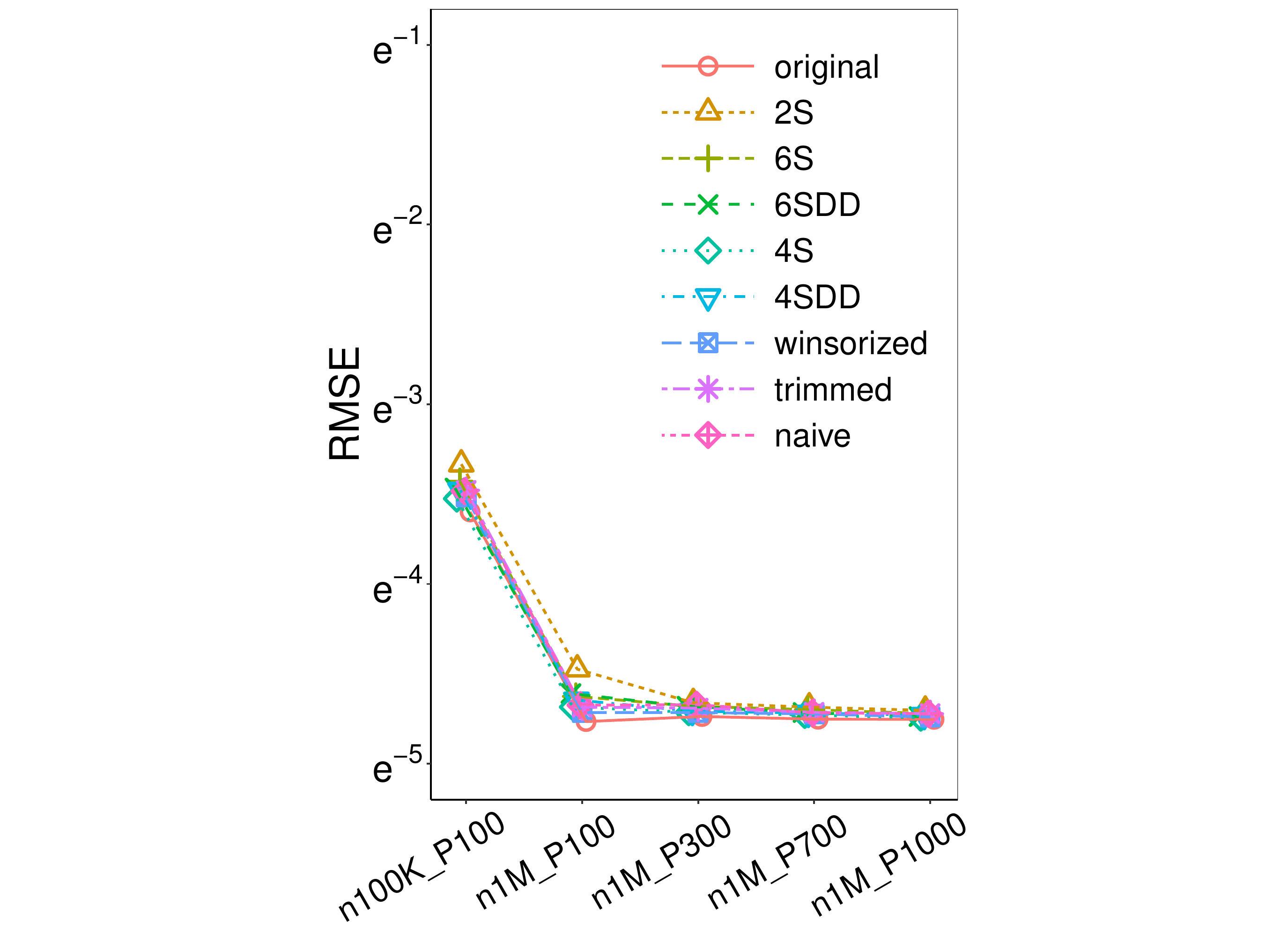}

\includegraphics[width=0.19\textwidth, trim={2.5in 0 2.6in 0},clip] {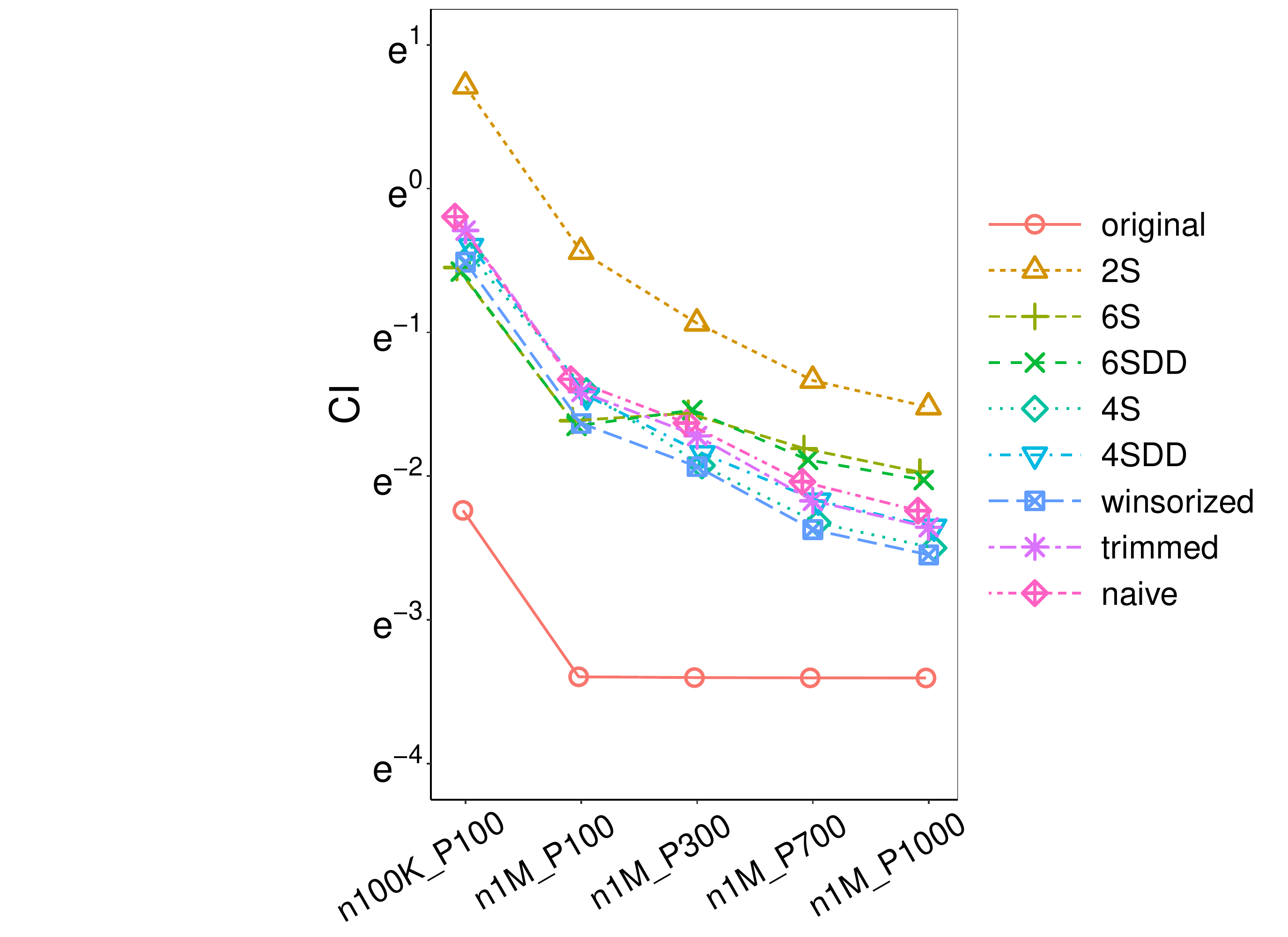}
\includegraphics[width=0.19\textwidth, trim={2.5in 0 2.6in 0},clip] {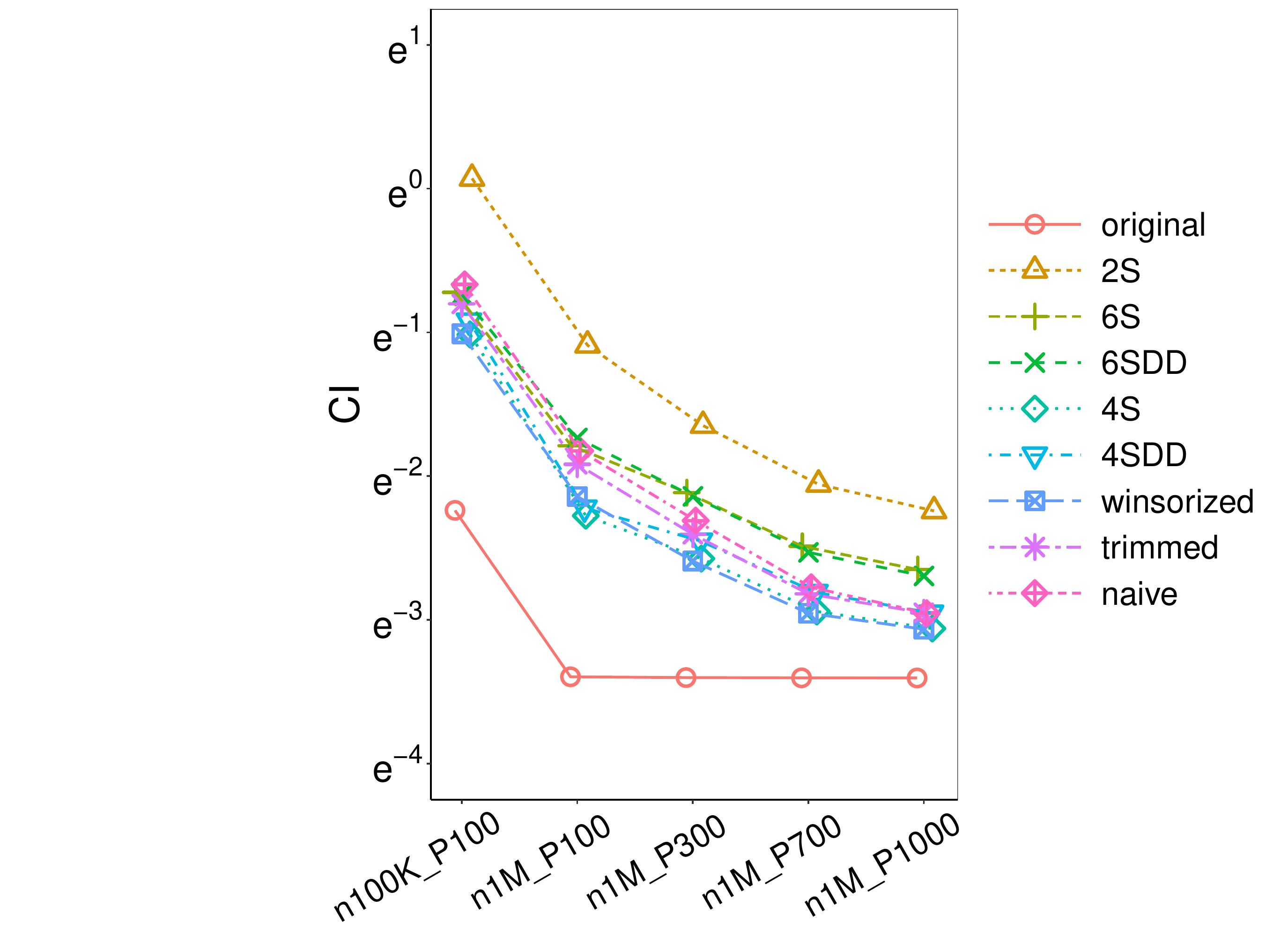}
\includegraphics[width=0.19\textwidth, trim={2.5in 0 2.6in 0},clip] {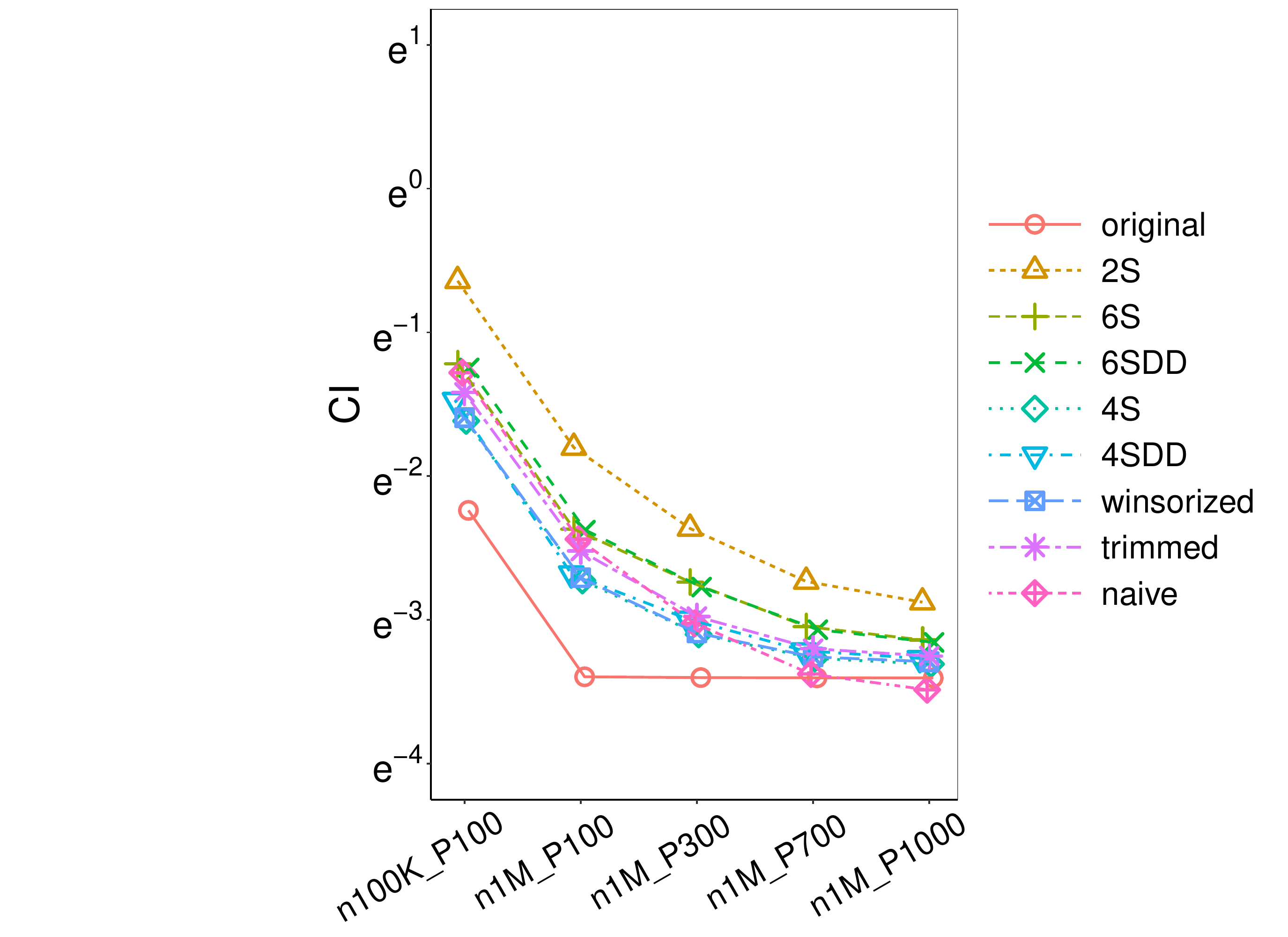}
\includegraphics[width=0.19\textwidth, trim={2.5in 0 2.6in 0},clip] {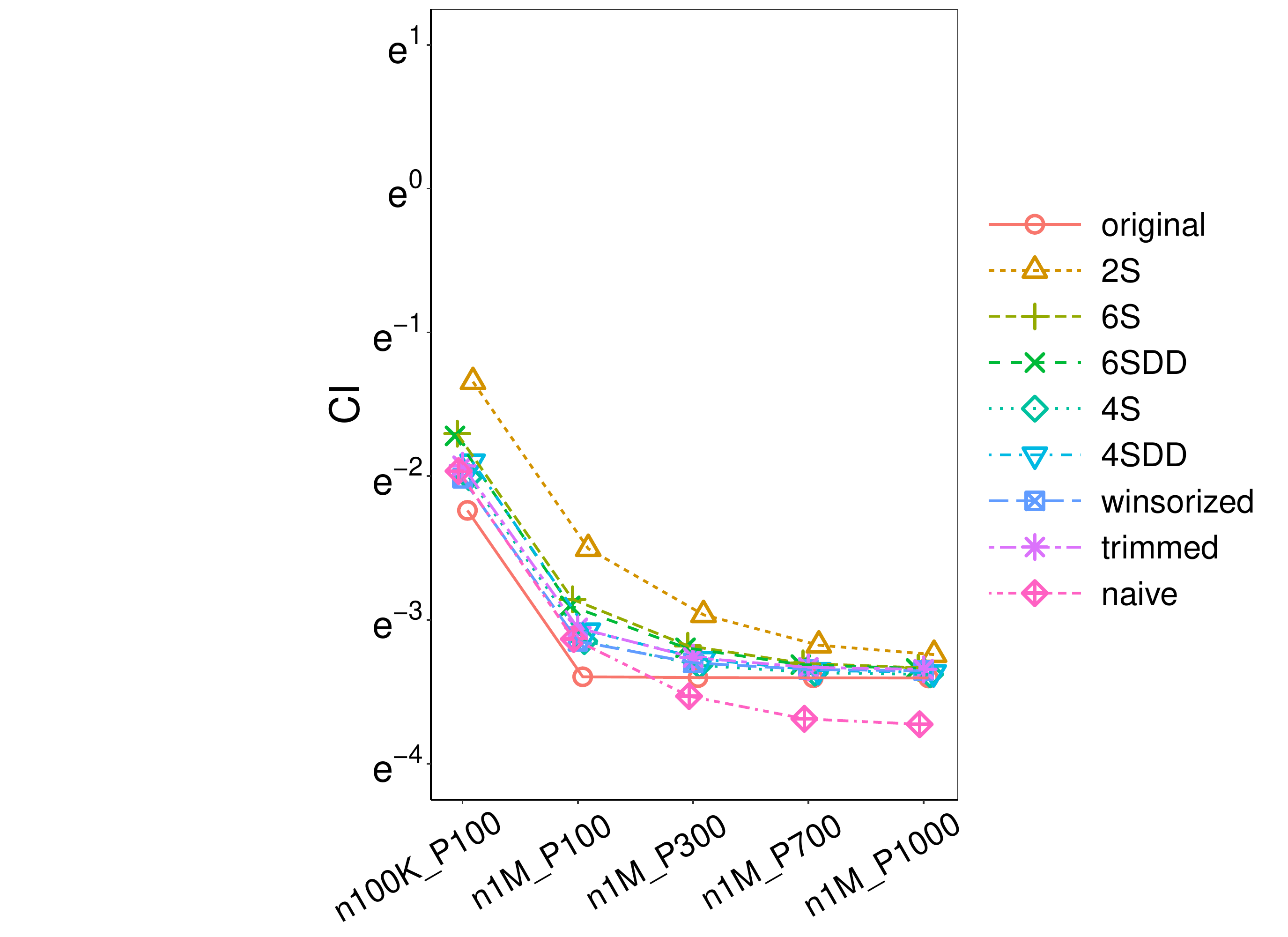}
\includegraphics[width=0.19\textwidth, trim={2.5in 0 2.6in 0},clip] {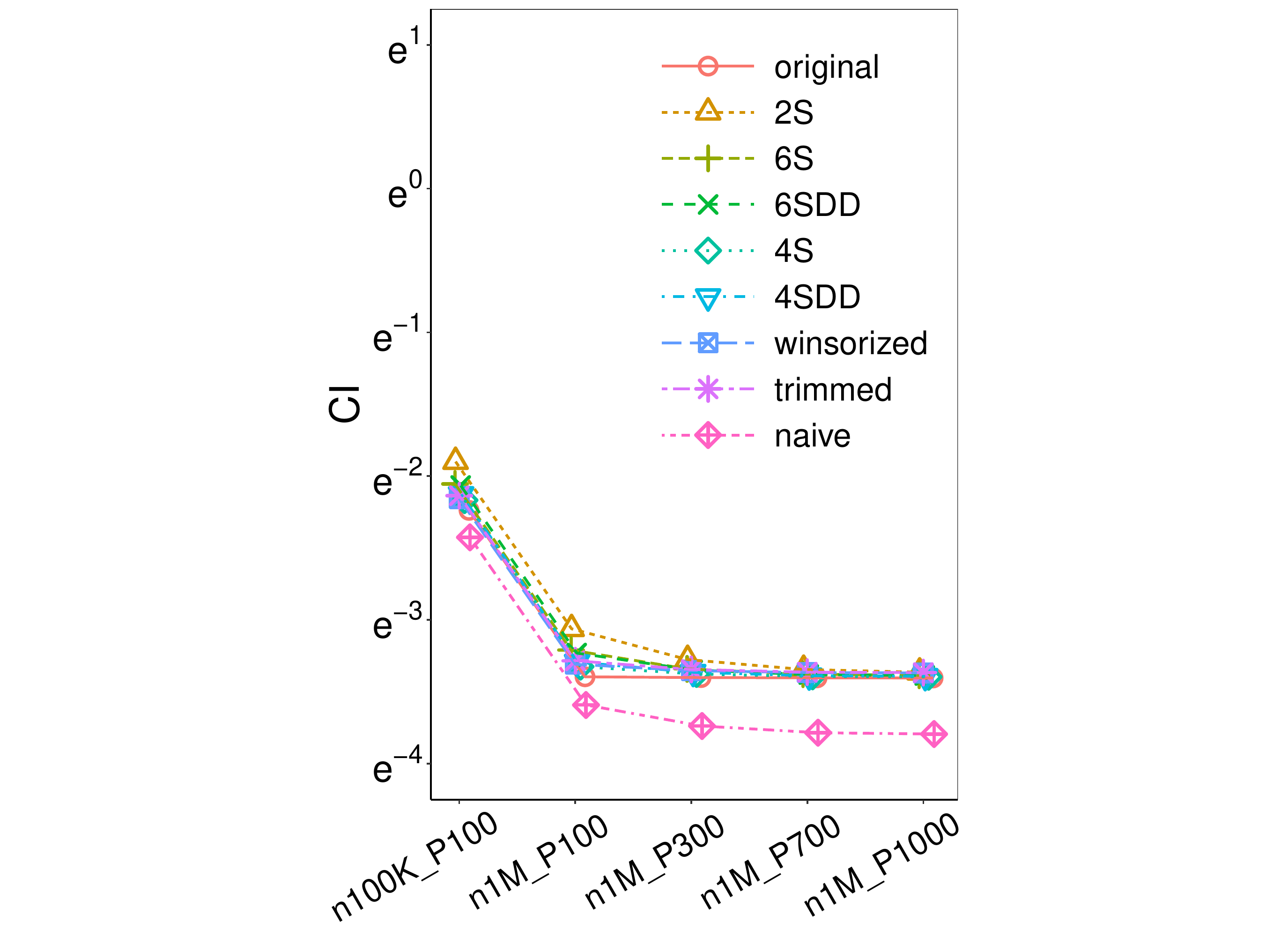}

\includegraphics[width=0.19\textwidth, trim={2.5in 0 2.6in 0},clip] {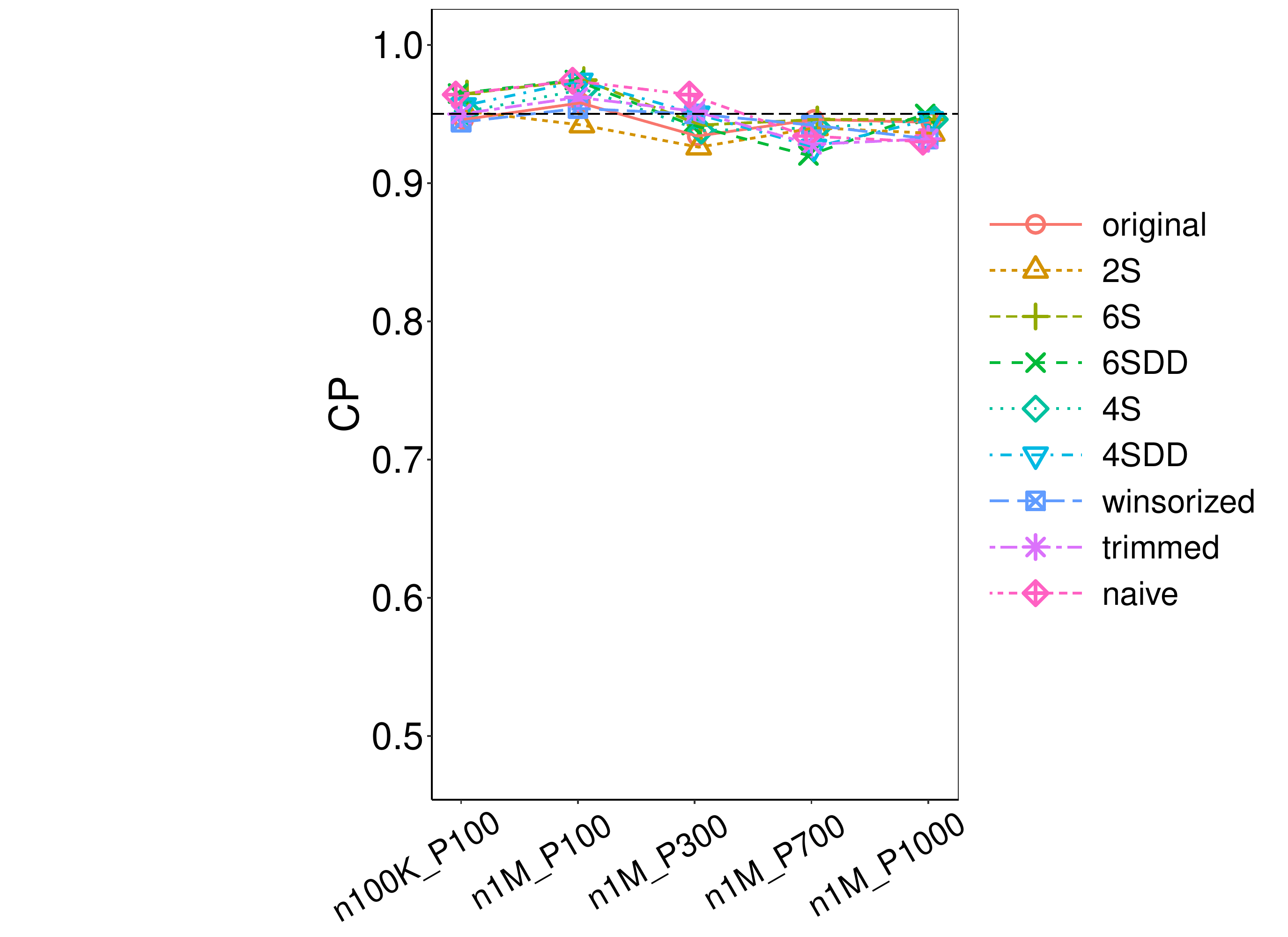}
\includegraphics[width=0.19\textwidth, trim={2.5in 0 2.6in 0},clip] {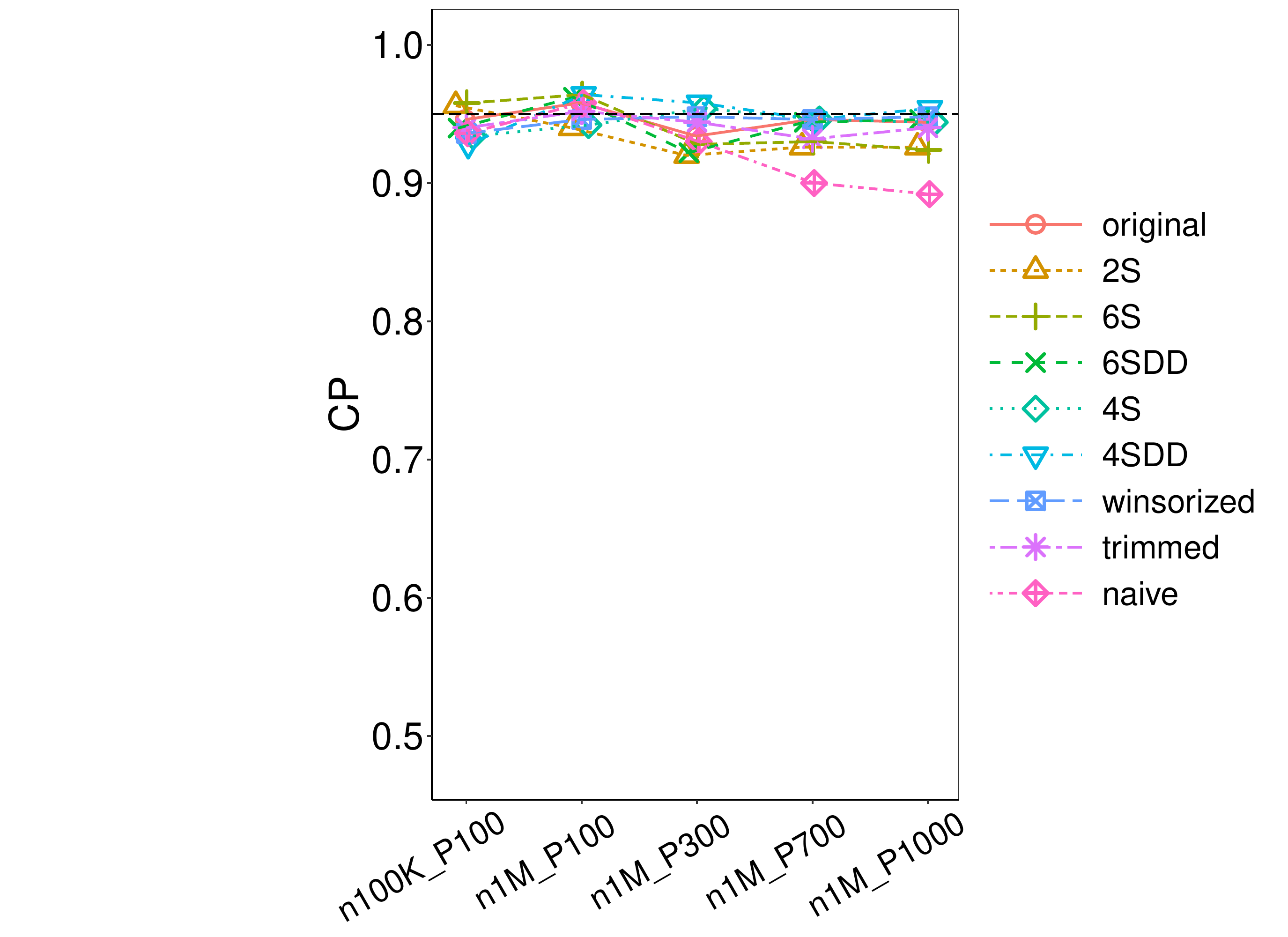}
\includegraphics[width=0.19\textwidth, trim={2.5in 0 2.6in 0},clip] {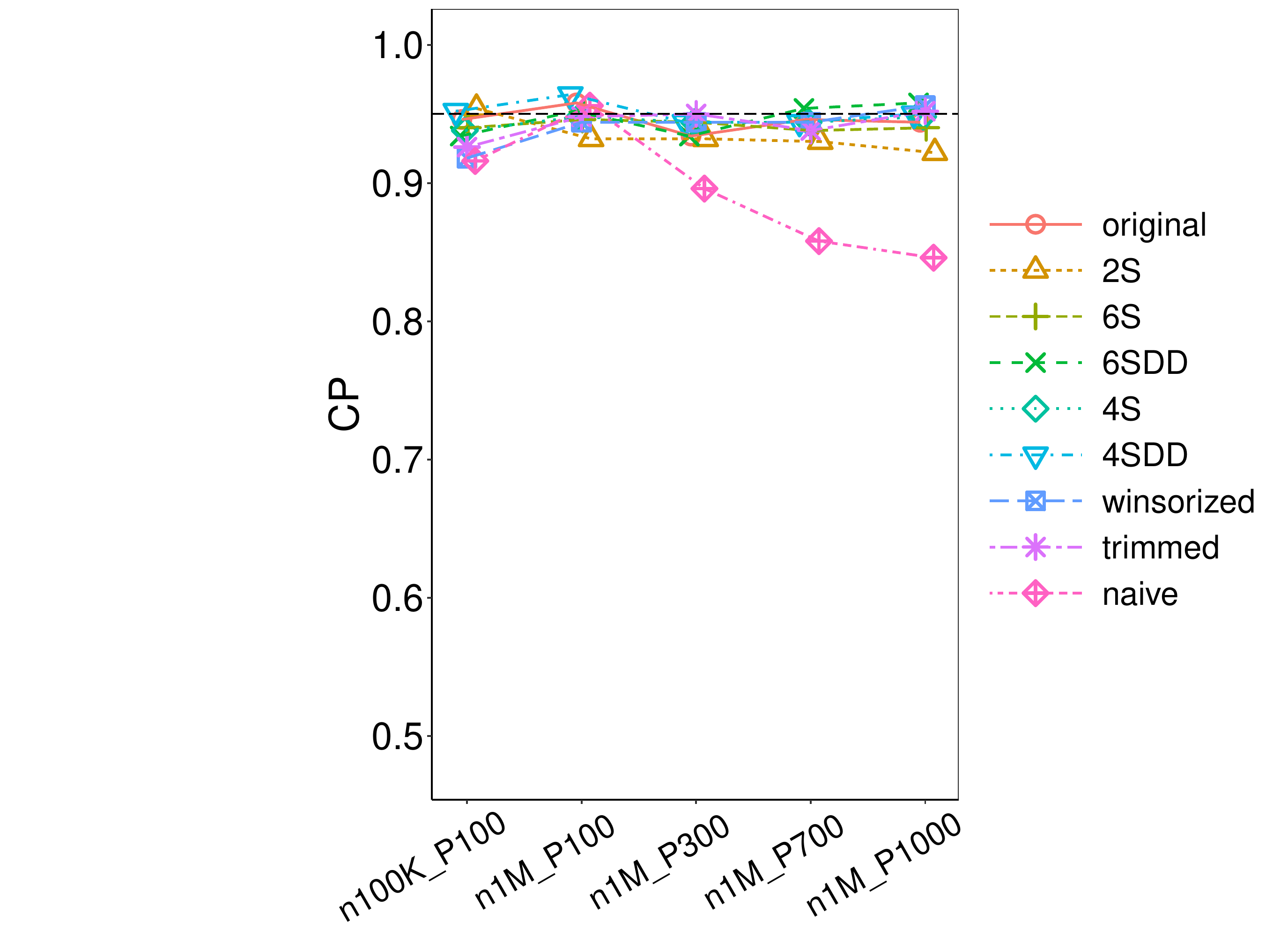}
\includegraphics[width=0.19\textwidth, trim={2.5in 0 2.6in 0},clip] {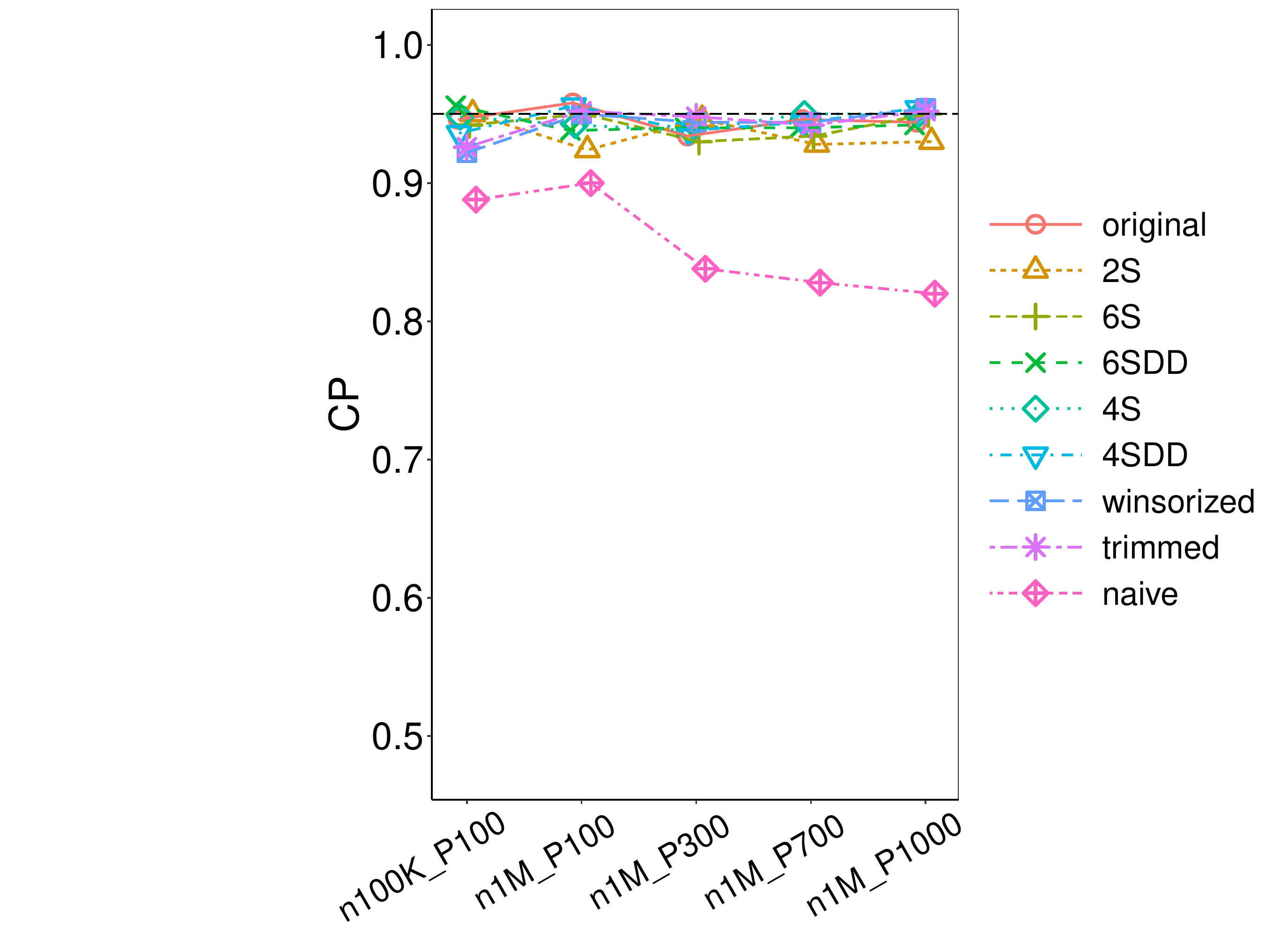}
\includegraphics[width=0.19\textwidth, trim={2.5in 0 2.6in 0},clip] {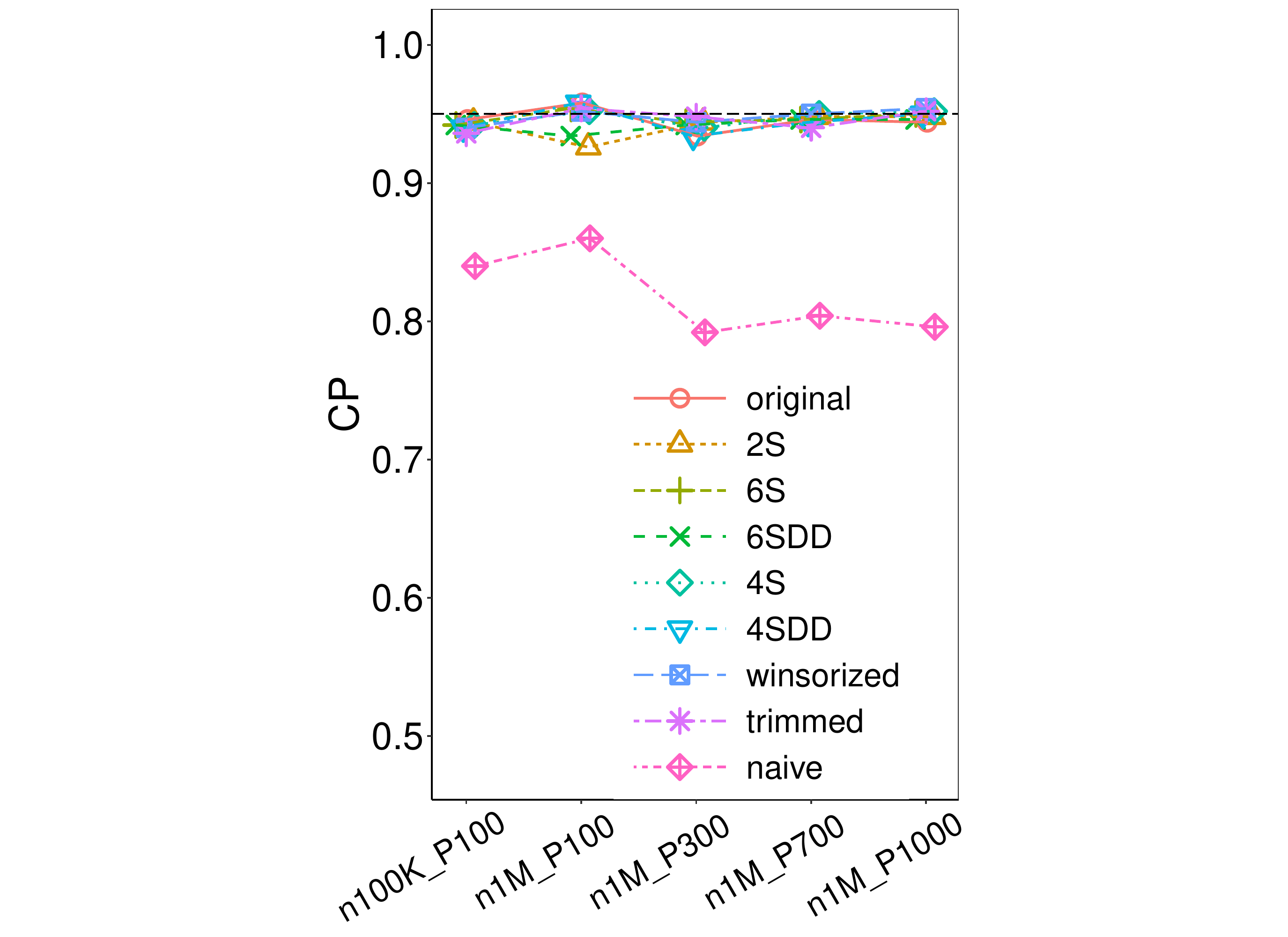}
\caption{Simulation results with $\rho$-zCDP for Gaussian data with $\alpha=\beta$ when $\theta=0$}\label{fig:0szCDPN}
\end{figure}

The results are presented in Figures \ref{fig:0sDPN} and \ref{fig:0szCDPN} and are summarized  as follows.  4S is the best performer, all metrics considered, and the na\"{i}ve method is the worst. 
4SDD, winsorized  mean, and trimmed means are similar to 4S in bias and are better than 6S and 6SDD. For RMSE and CI width,  winsorized mean is similar to 4S (the best group), followed by trimmed mean and 4SDD (the second best group), and then 6SDD and 6SDD; 2S is the worst, especially at $n=100k/p=100$, $n=1m/P=100$ and $\epsilon\le1$. All methods, except for the  na\"{i}ve method, provide nominal-level coverage (slight over-coverage at $n=100k/P=100$ and $n=1m/P=100$ for $\epsilon\le1$).  Finally, the Gaussian mechanism of $\rho$-zCDP does not seem to be advantageous in statistical utility compared to the Laplace mechanism of $\epsilon$-DP.  

\subsubsection{\texorpdfstring{$Y\sim$ Gaussian}{}, \texorpdfstring{$\theta\ne0$ ($H_1$)}{H1}, and \texorpdfstring{$\alpha=\beta$}{S}}\label{sec:1sN}
\begin{figure}[!htb]
\hspace{0.5in}$\epsilon=0.5$\hspace{0.8in}$\epsilon=1$\hspace{0.9in}$\epsilon=2$
\hspace{0.95in}$\epsilon=5$\hspace{0.9in}$\epsilon=50$

\includegraphics[width=0.19\textwidth, trim={2.5in 0 2.5in 0},clip] {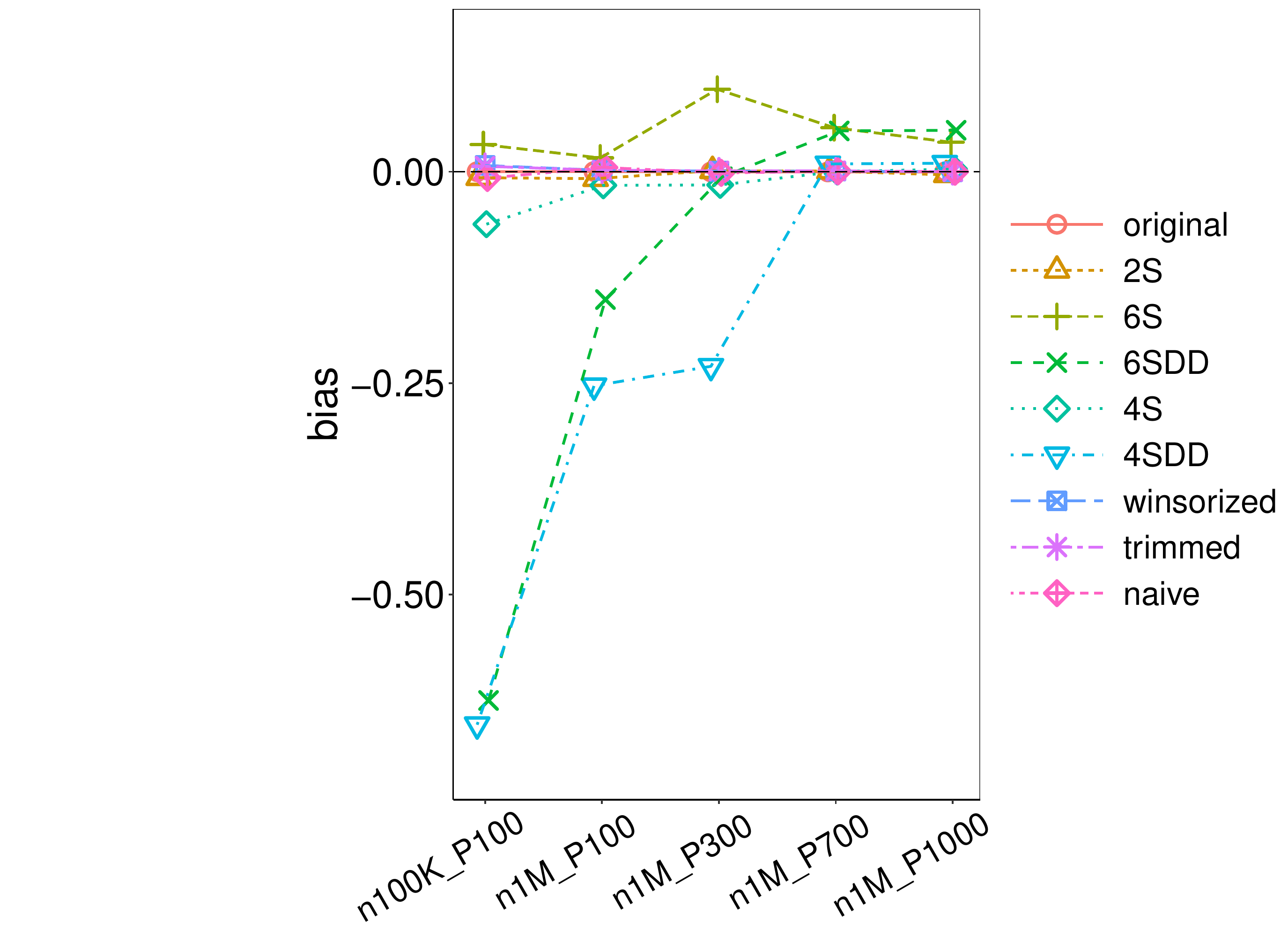}
\includegraphics[width=0.19\textwidth, trim={2.5in 0 2.5in 0},clip] {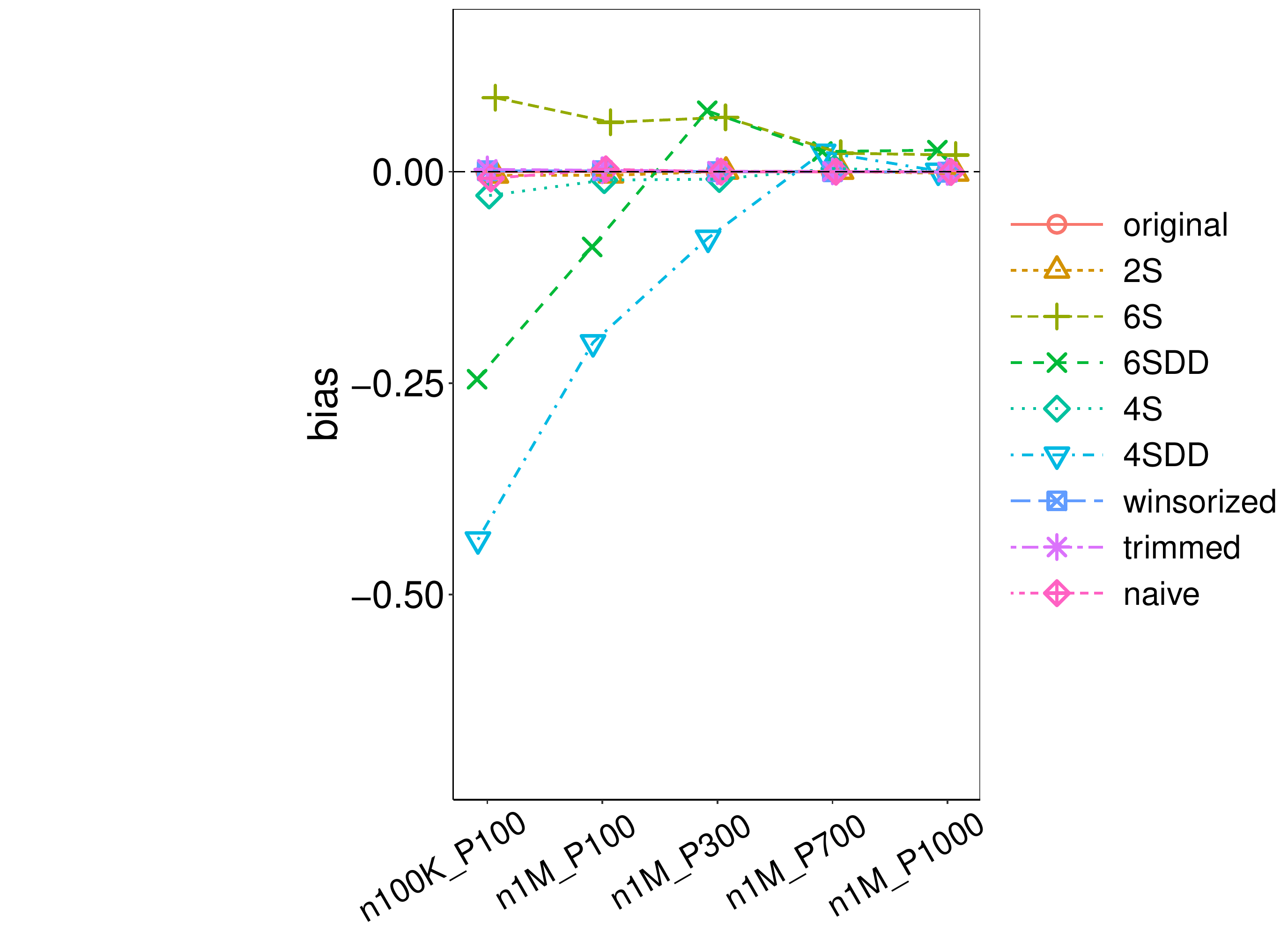}
\includegraphics[width=0.19\textwidth, trim={2.5in 0 2.5in 0},clip] {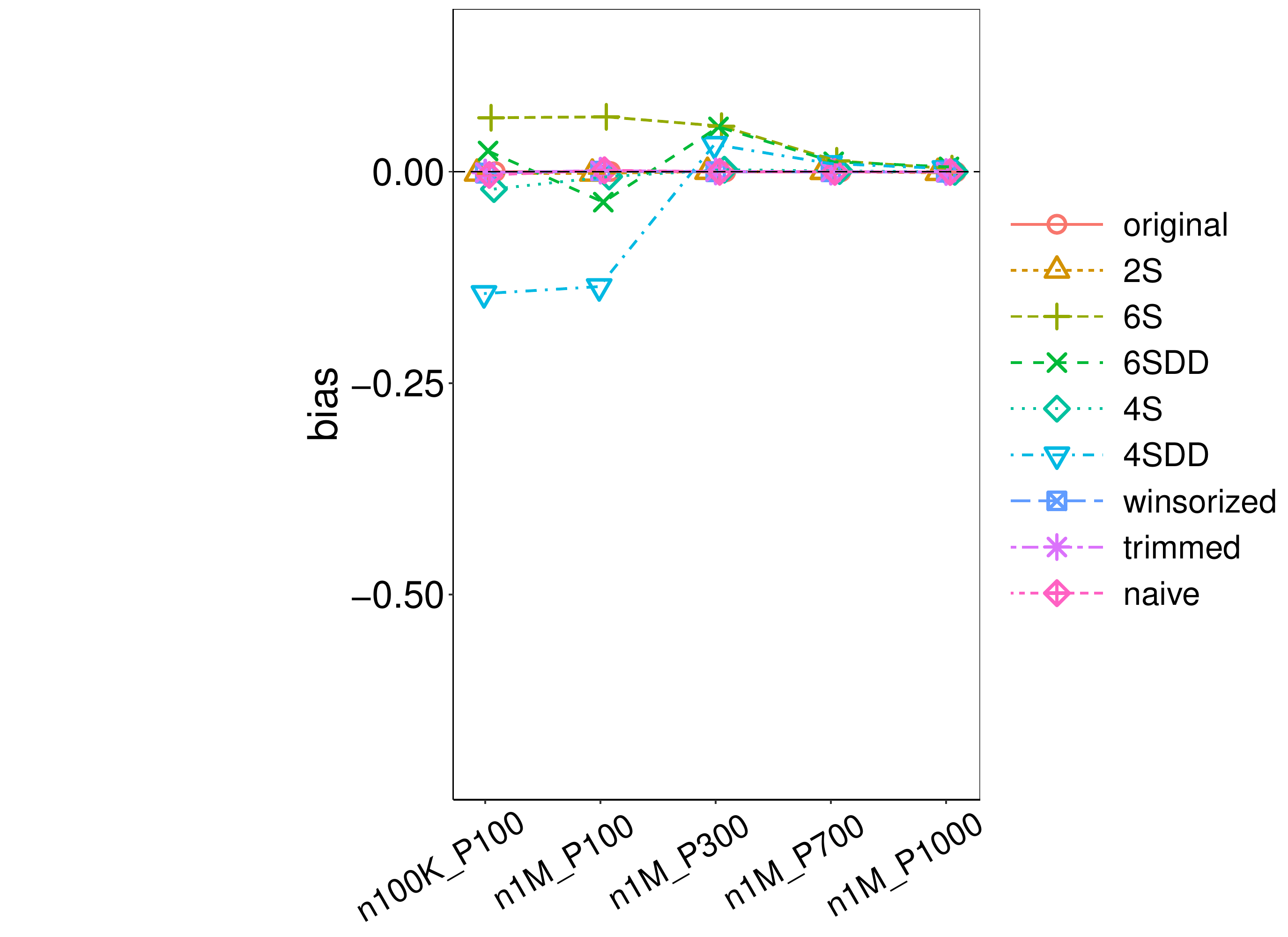}
\includegraphics[width=0.19\textwidth, trim={2.5in 0 2.5in 0},clip] {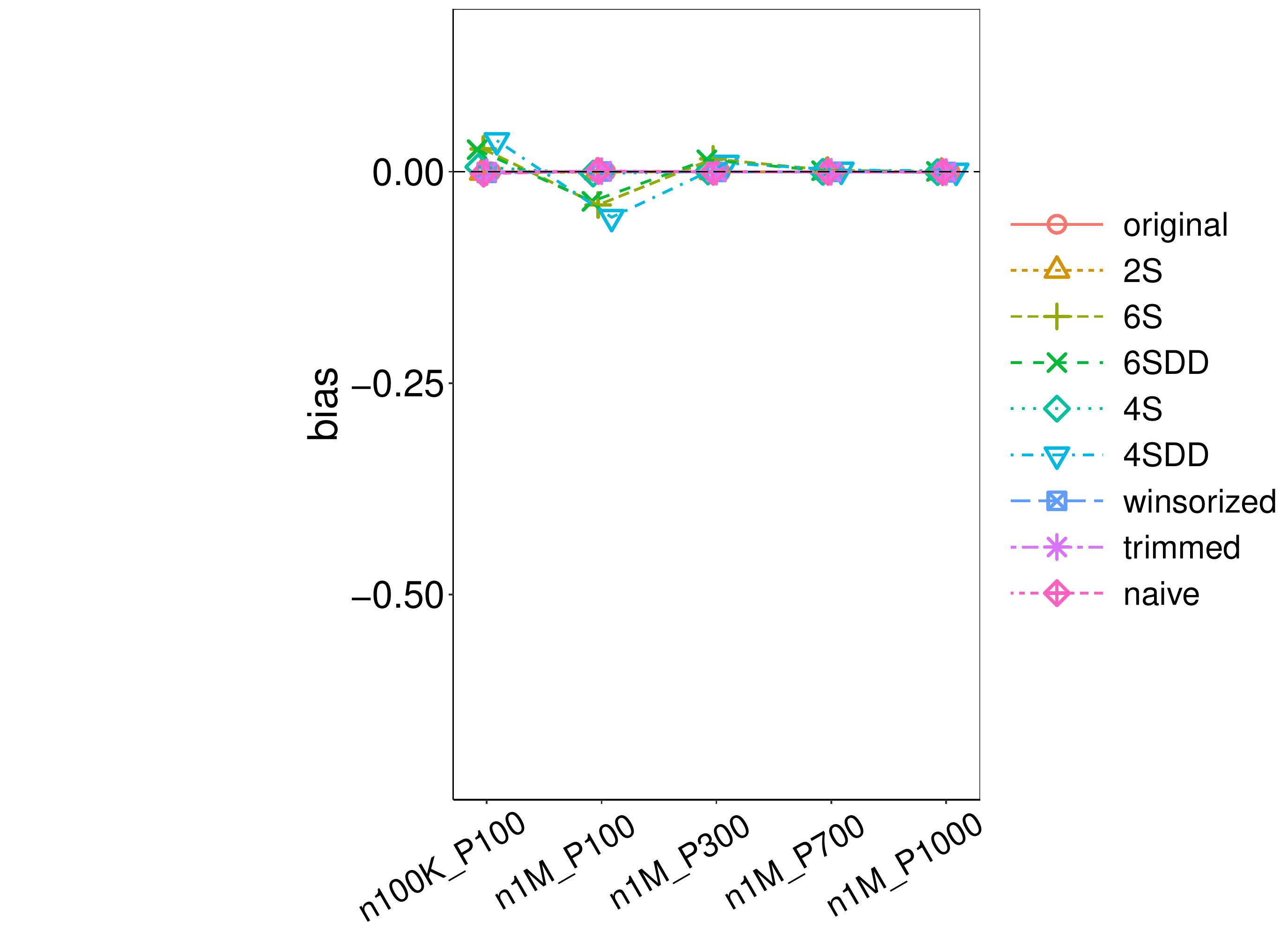}
\includegraphics[width=0.19\textwidth, trim={2.5in 0 2.5in 0},clip] {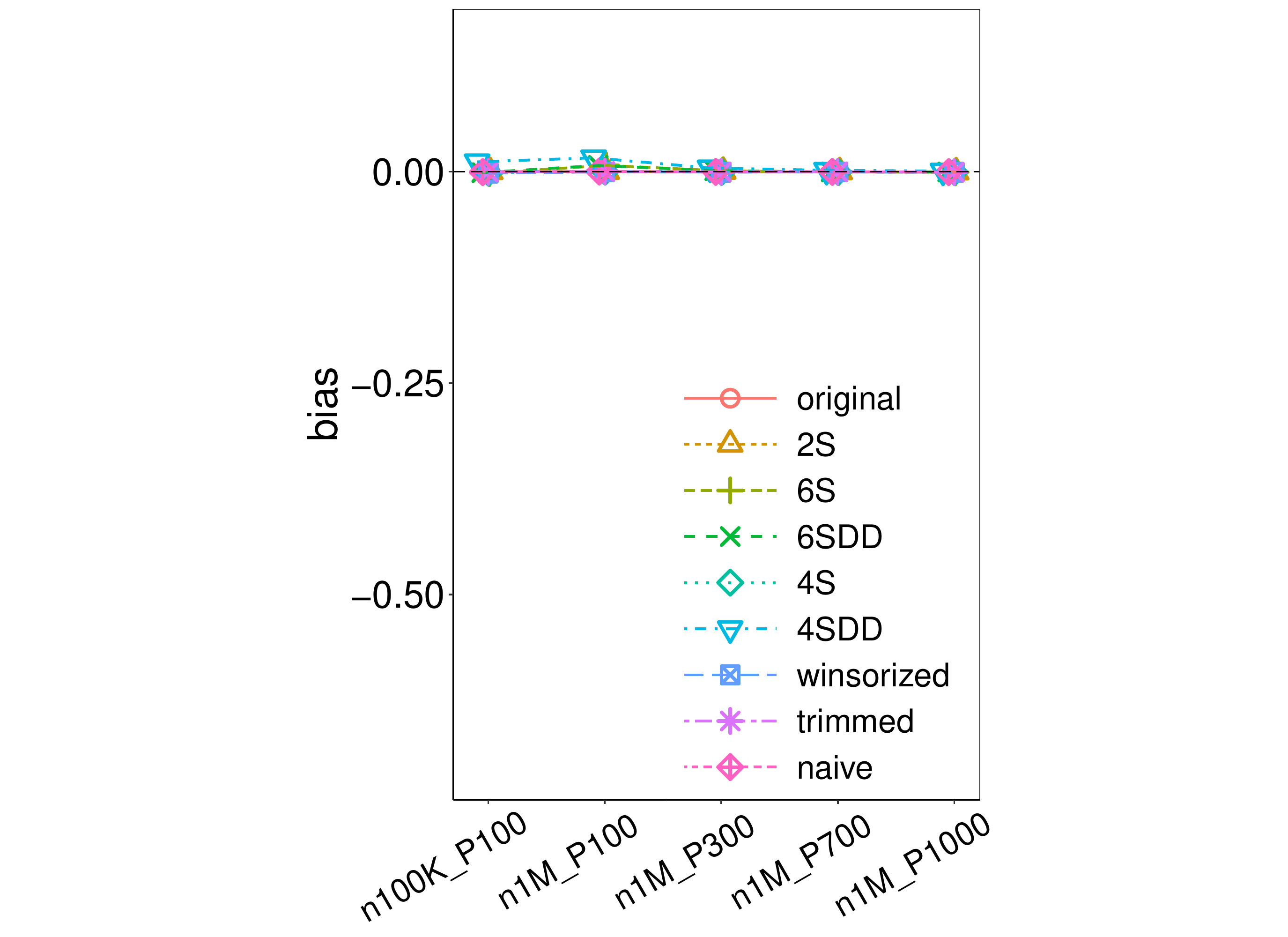}

\includegraphics[width=0.19\textwidth, trim={2.5in 0 2.6in 0},clip] {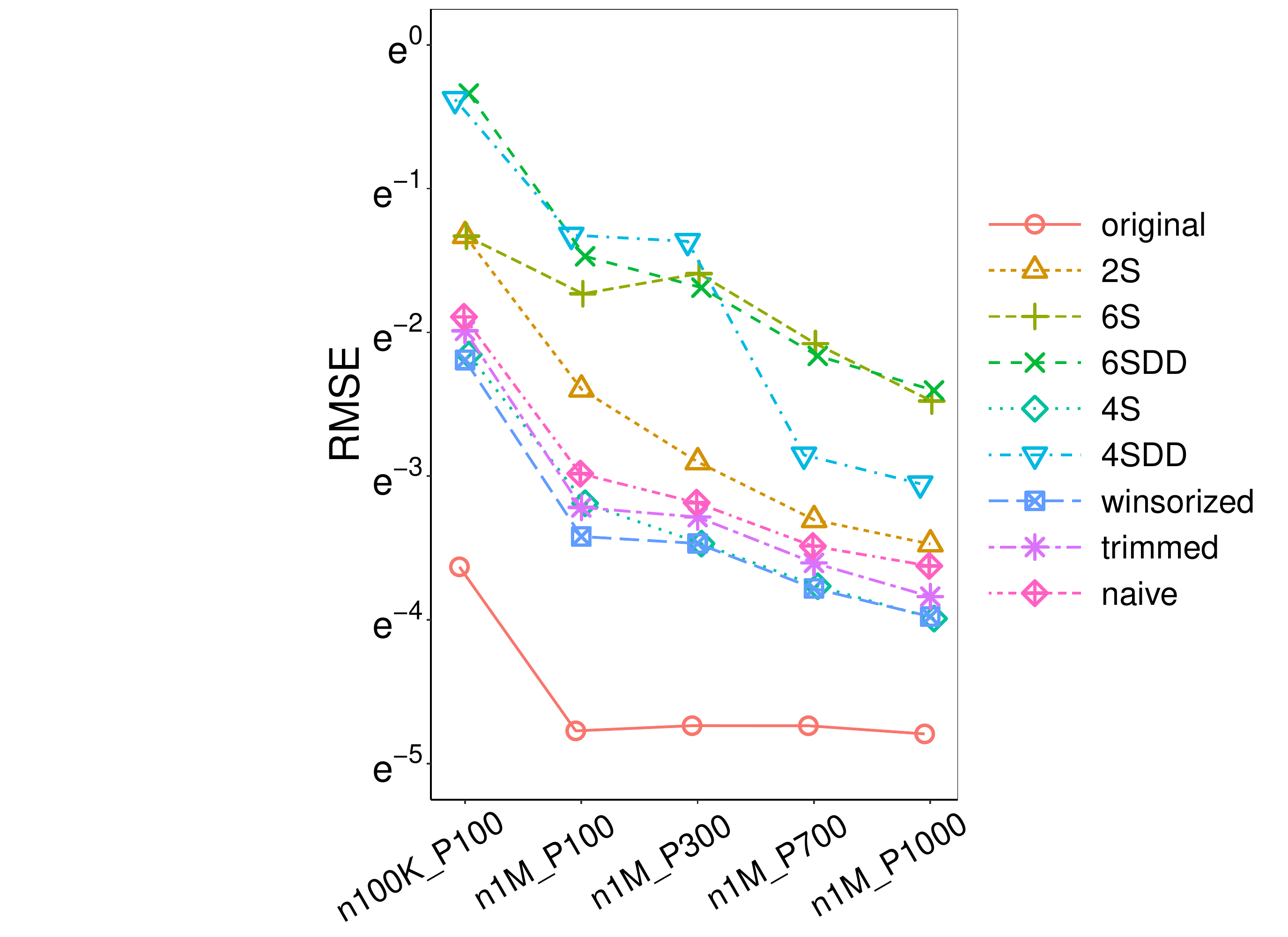}
\includegraphics[width=0.19\textwidth, trim={2.5in 0 2.6in 0},clip] {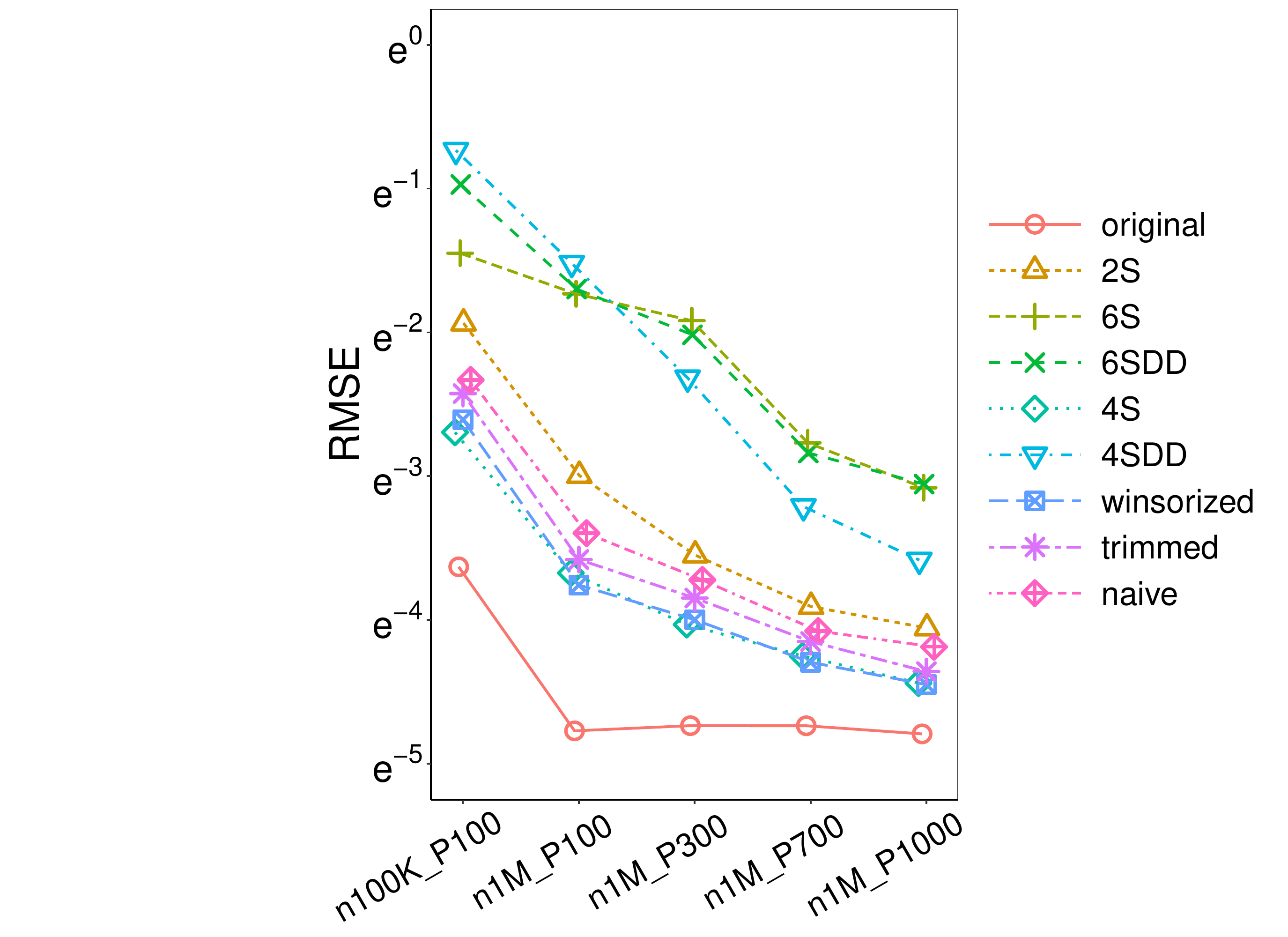}
\includegraphics[width=0.19\textwidth, trim={2.5in 0 2.6in 0},clip] {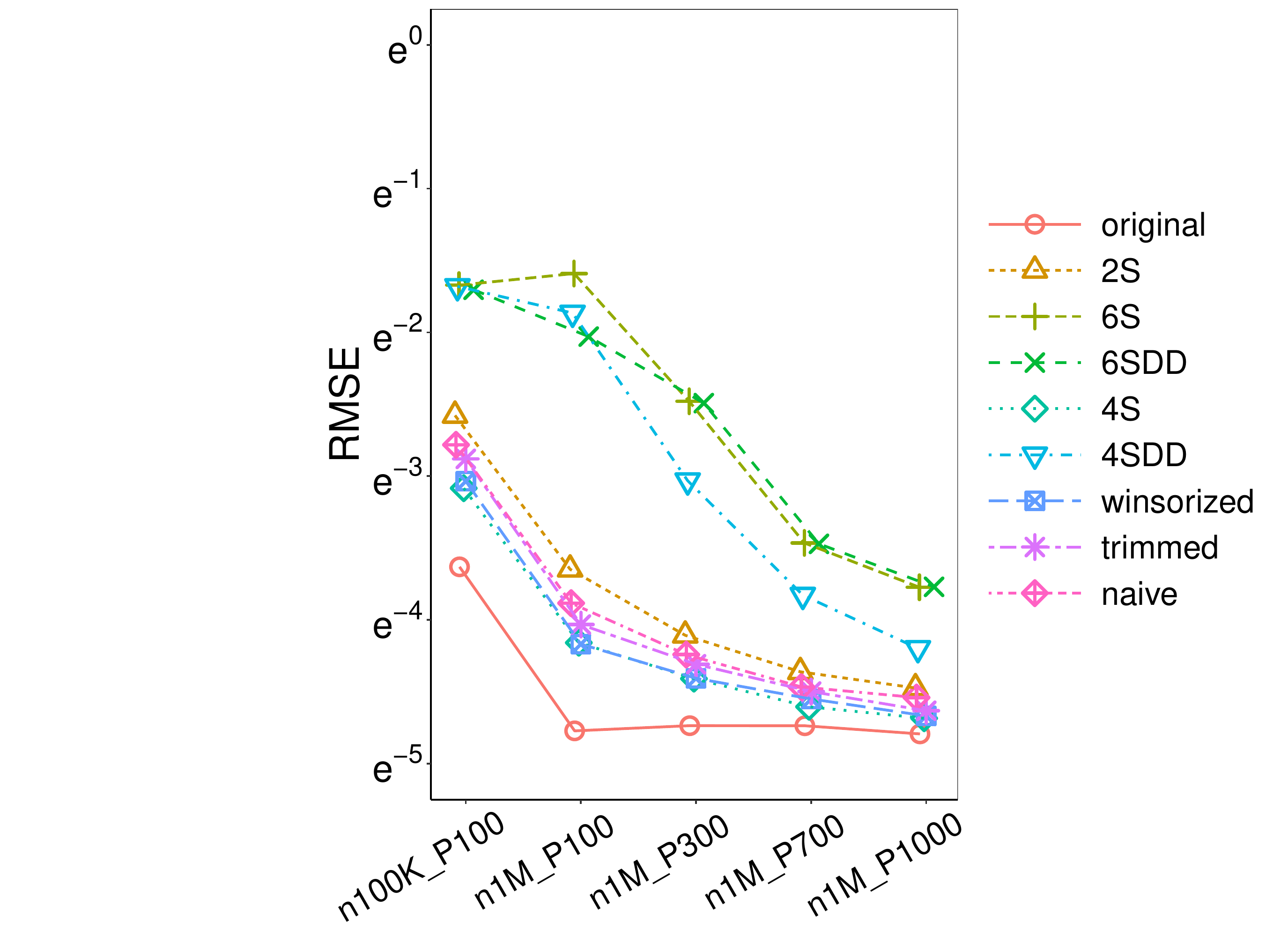}
\includegraphics[width=0.19\textwidth, trim={2.5in 0 2.6in 0},clip] {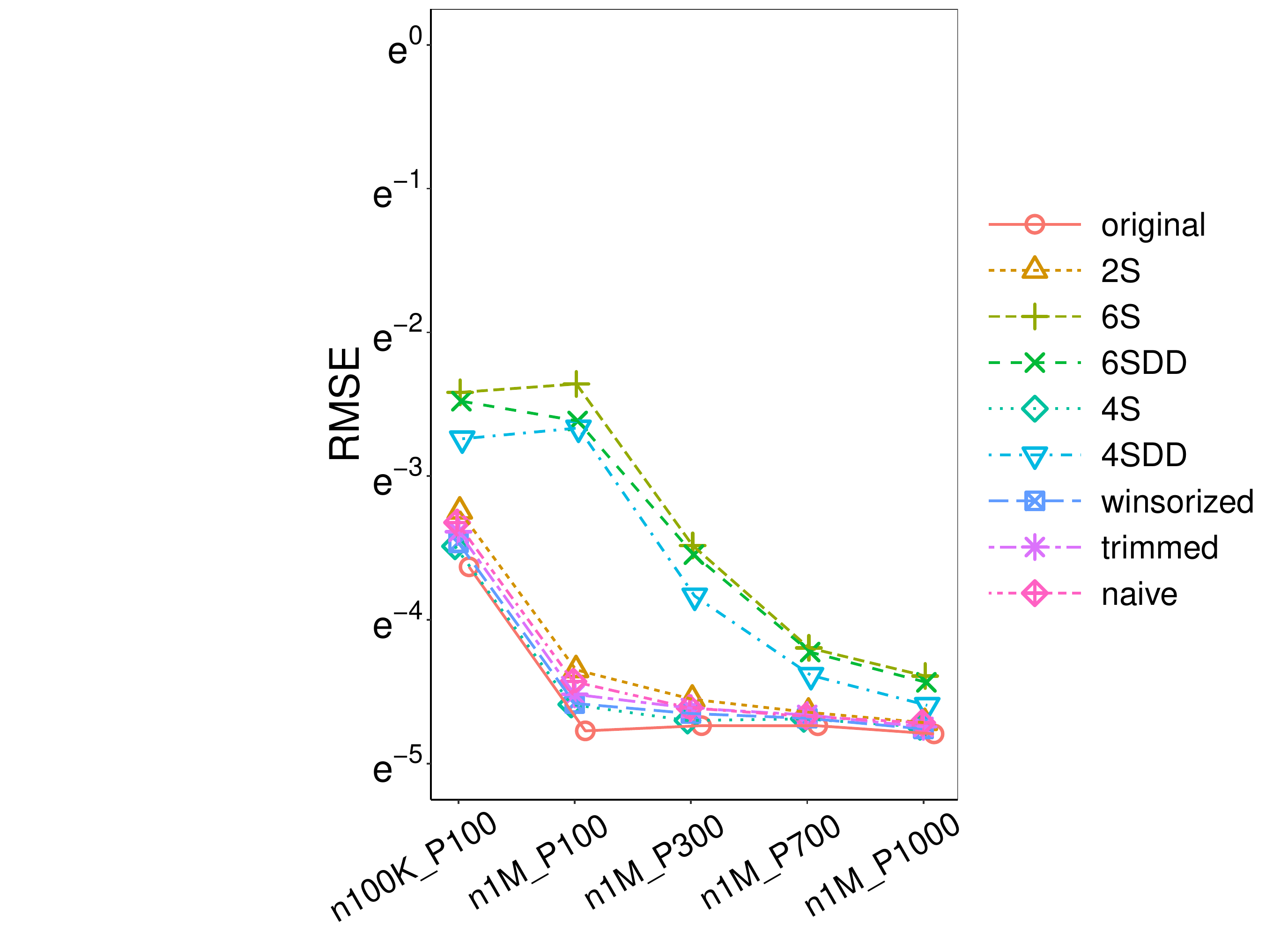}
\includegraphics[width=0.19\textwidth, trim={2.5in 0 2.6in 0},clip] {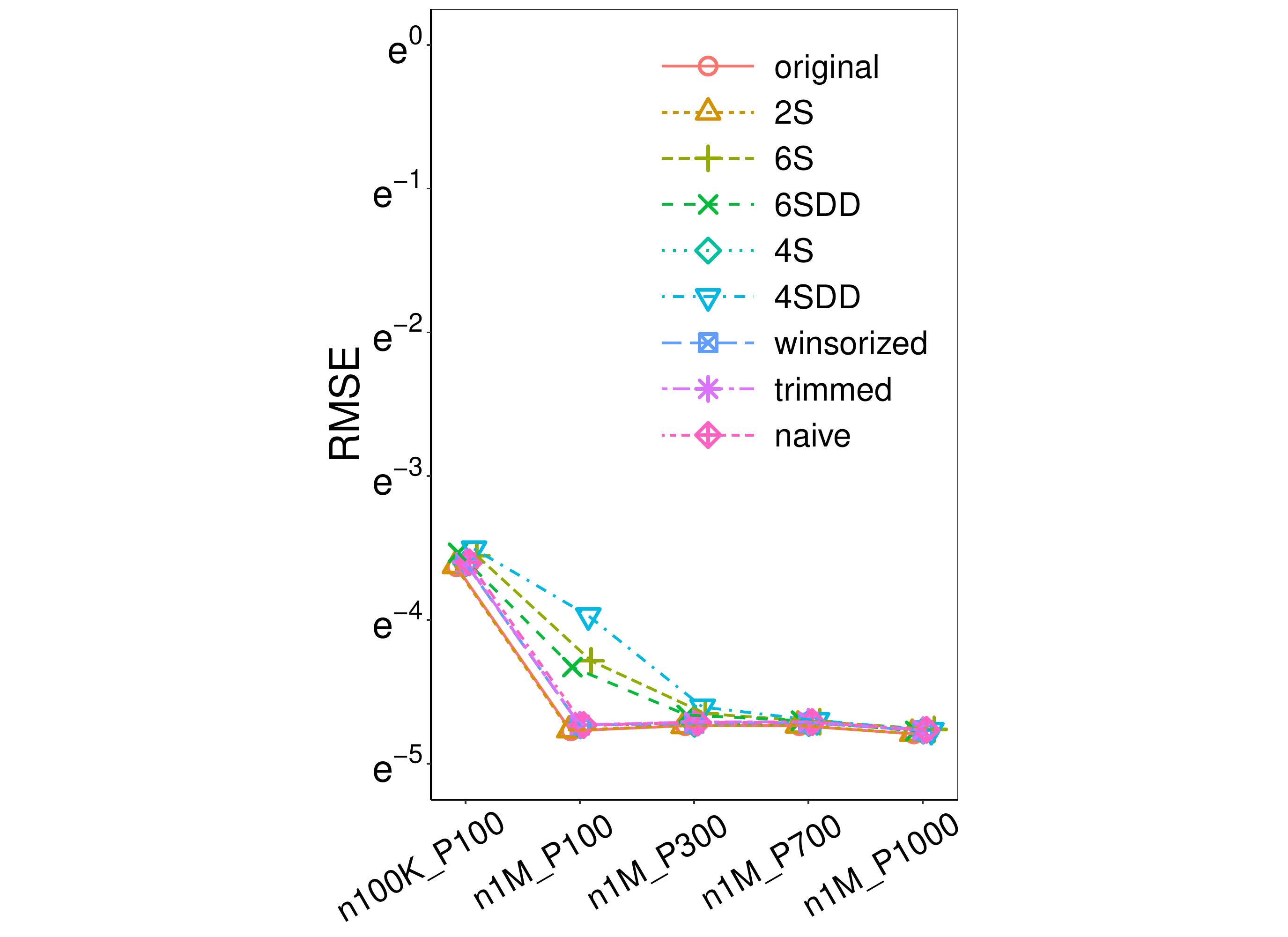}

\includegraphics[width=0.19\textwidth, trim={2.5in 0 2.6in 0},clip] {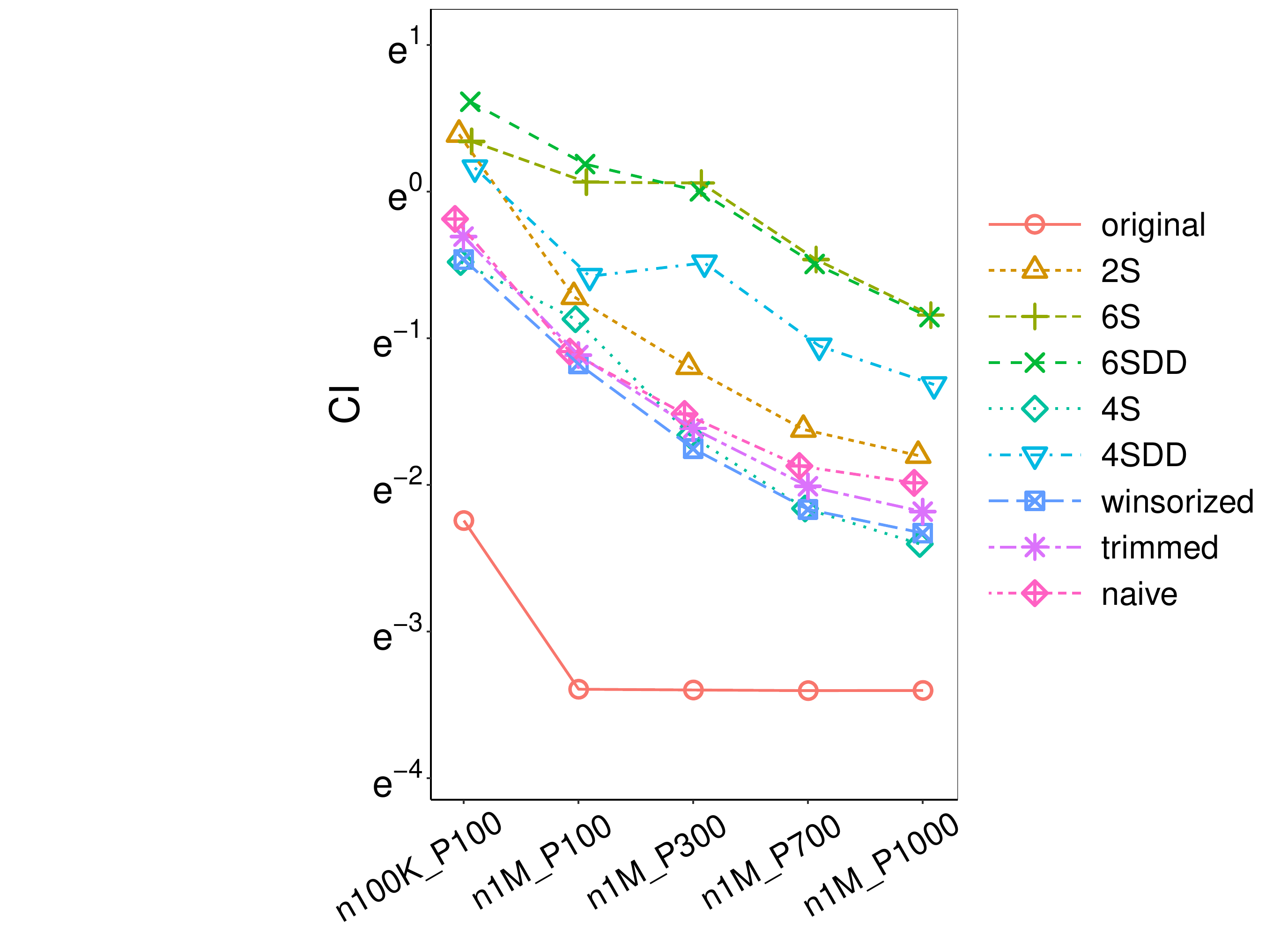}
\includegraphics[width=0.19\textwidth, trim={2.5in 0 2.6in 0},clip] {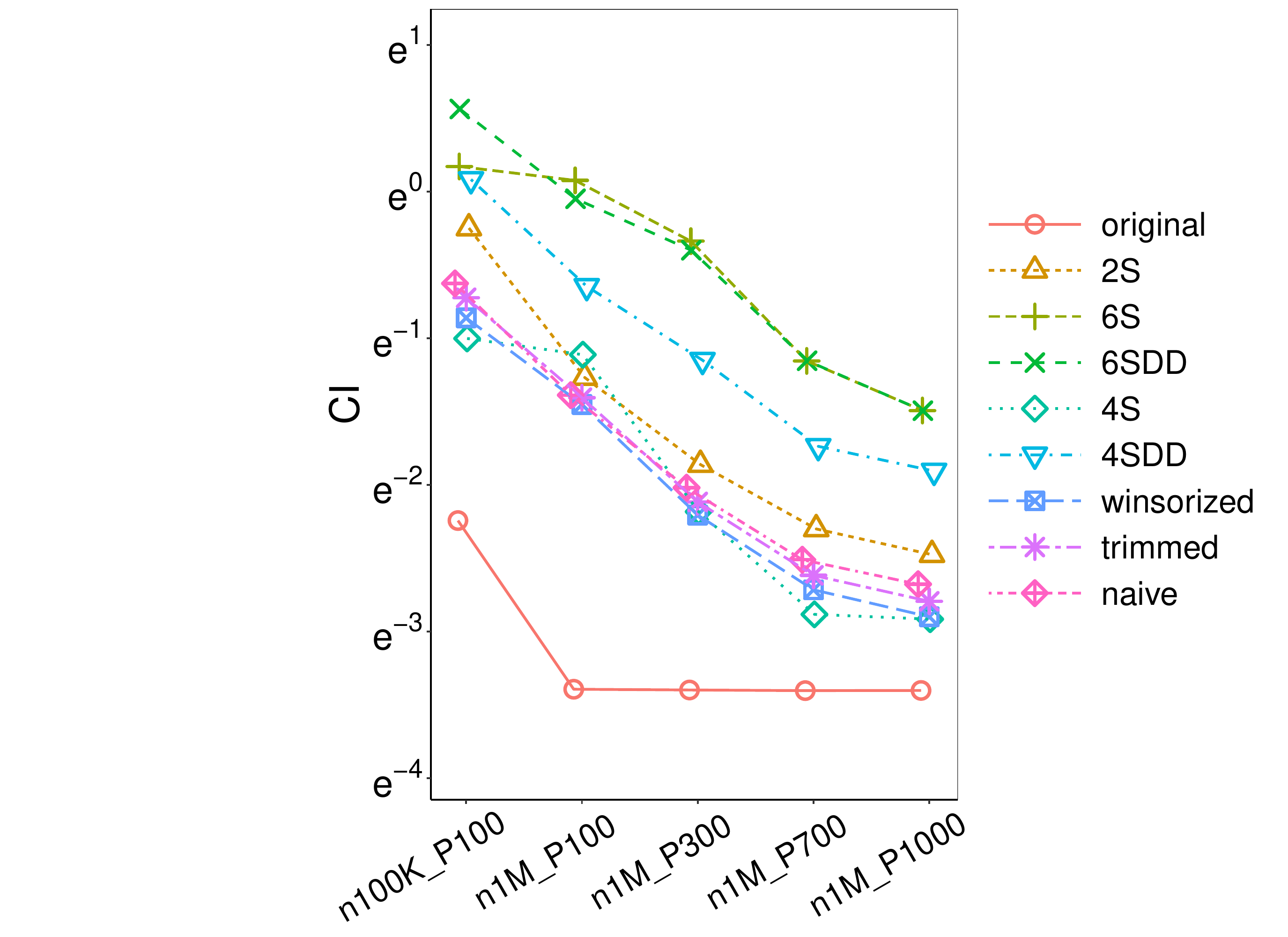}
\includegraphics[width=0.19\textwidth, trim={2.5in 0 2.6in 0},clip] {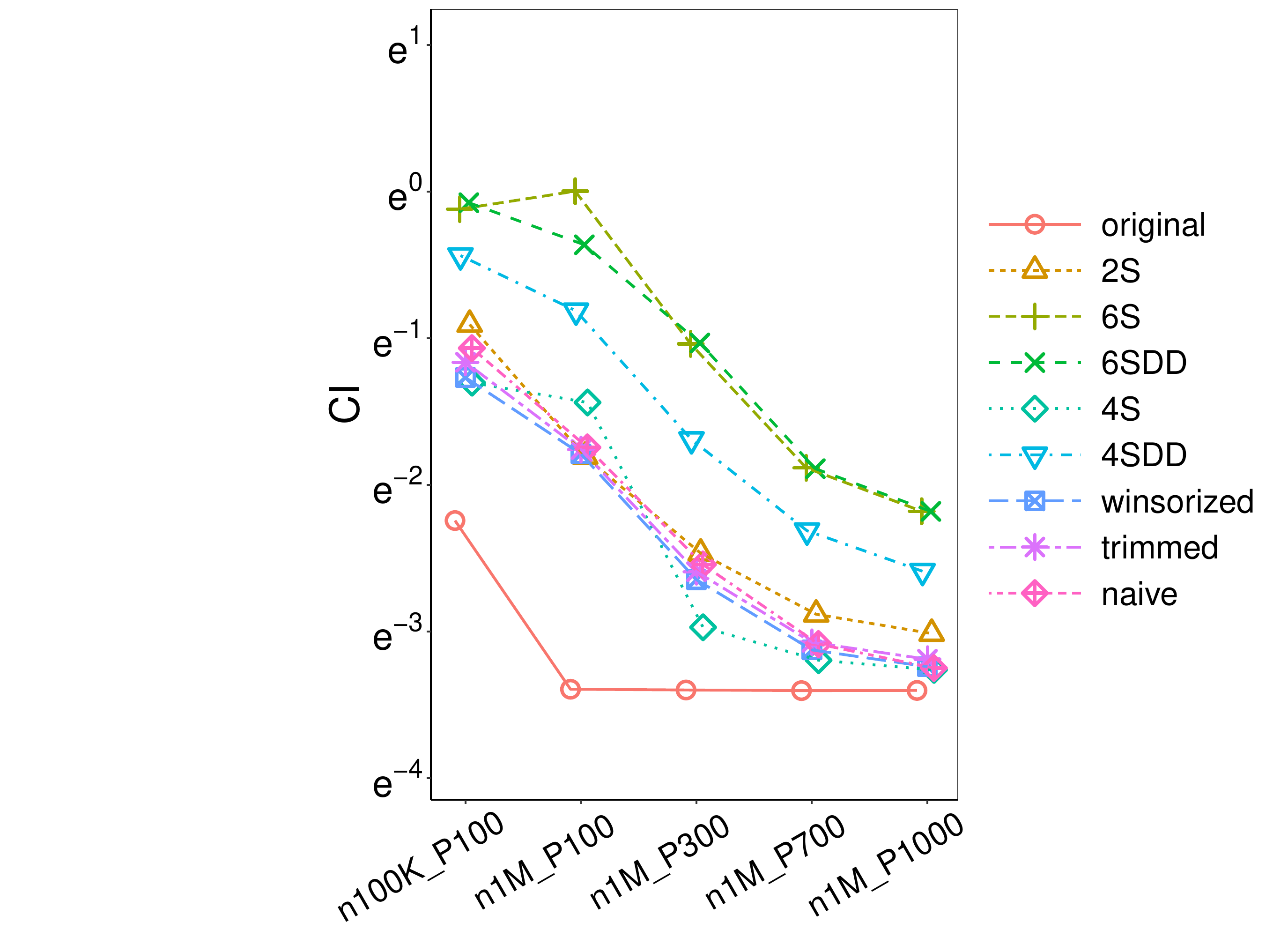}
\includegraphics[width=0.19\textwidth, trim={2.5in 0 2.6in 0},clip] {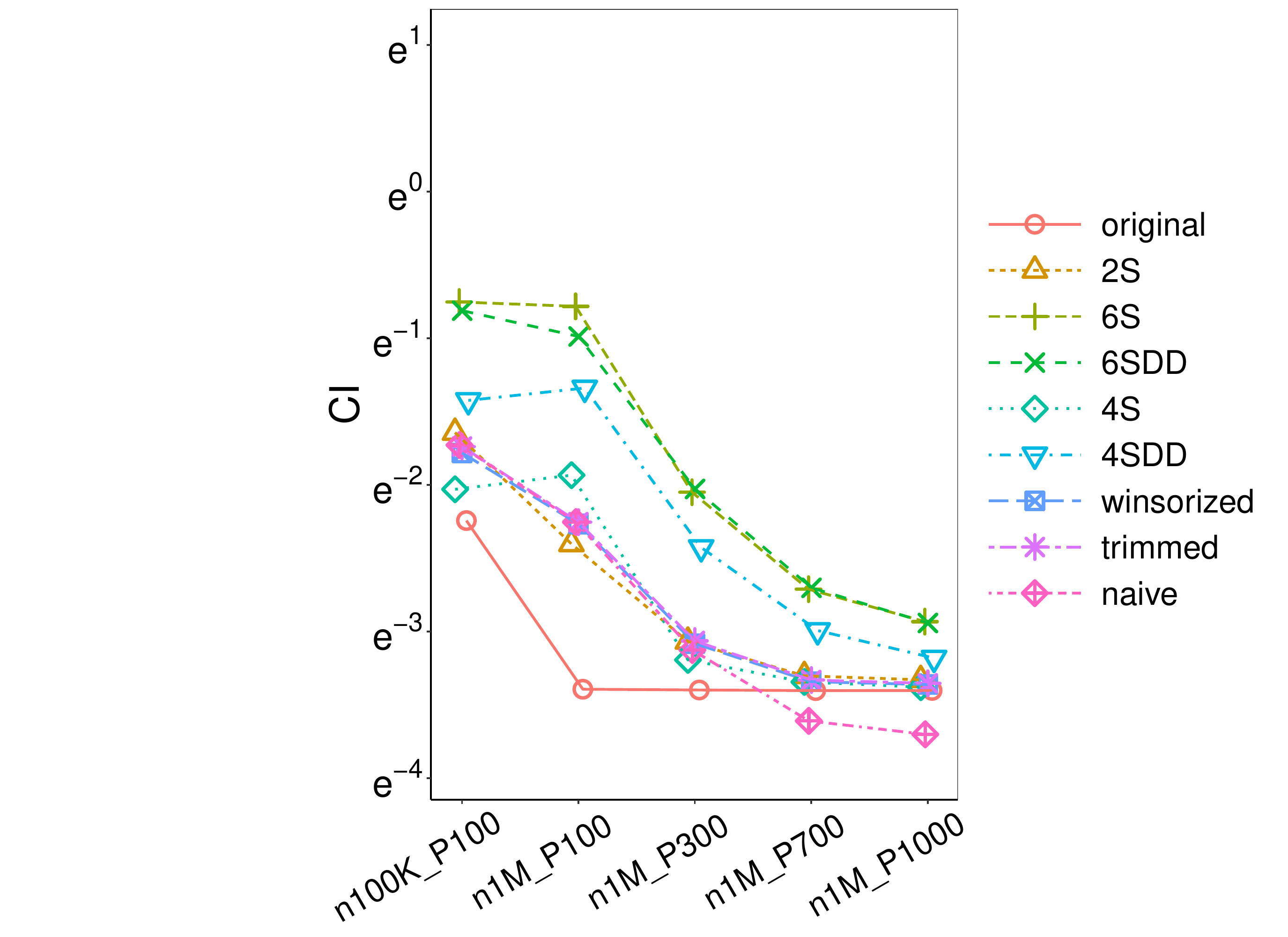}
\includegraphics[width=0.19\textwidth, trim={2.5in 0 2.6in 0},clip] {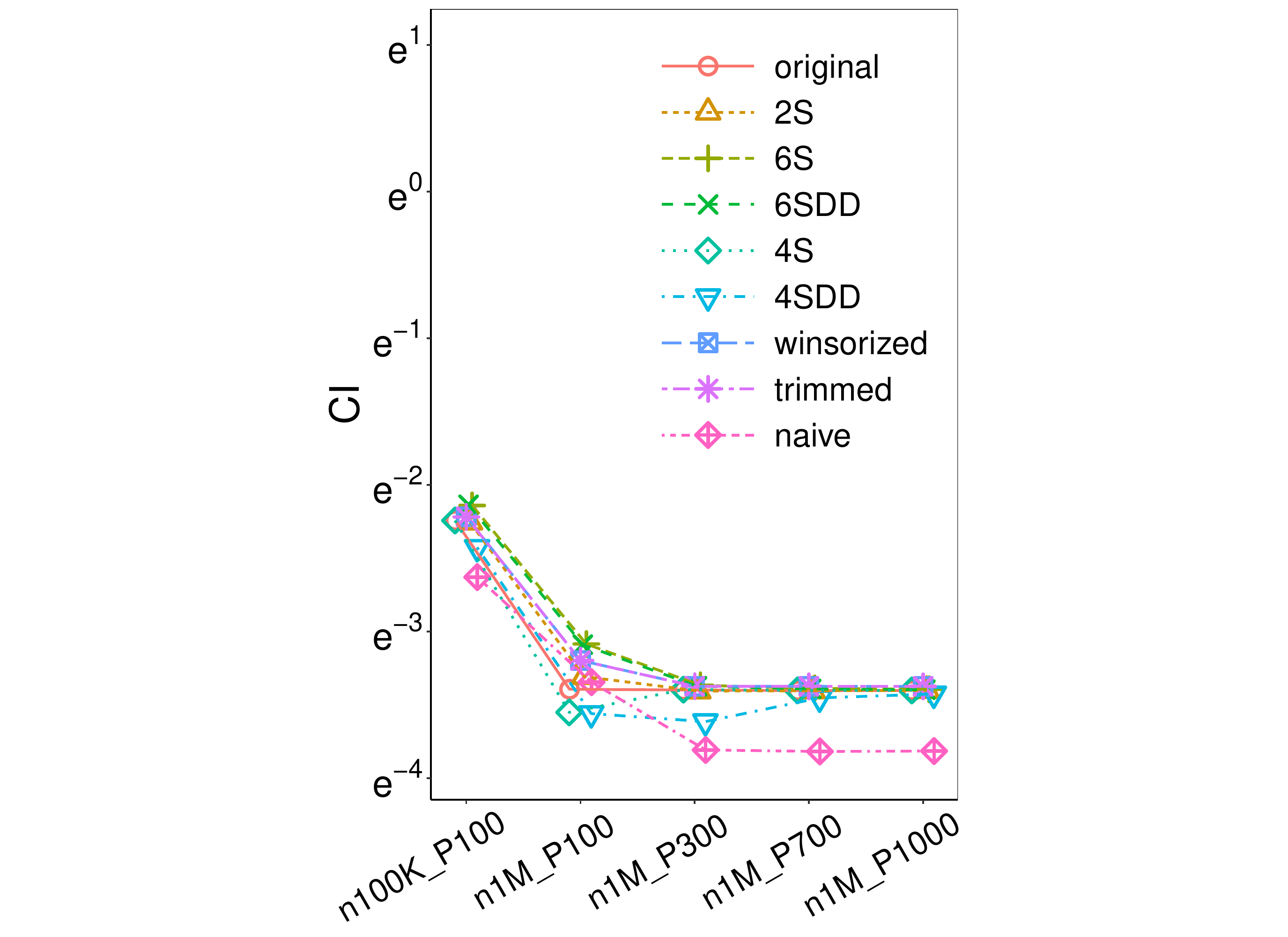}

\includegraphics[width=0.19\textwidth, trim={2.5in 0 2.6in 0},clip] {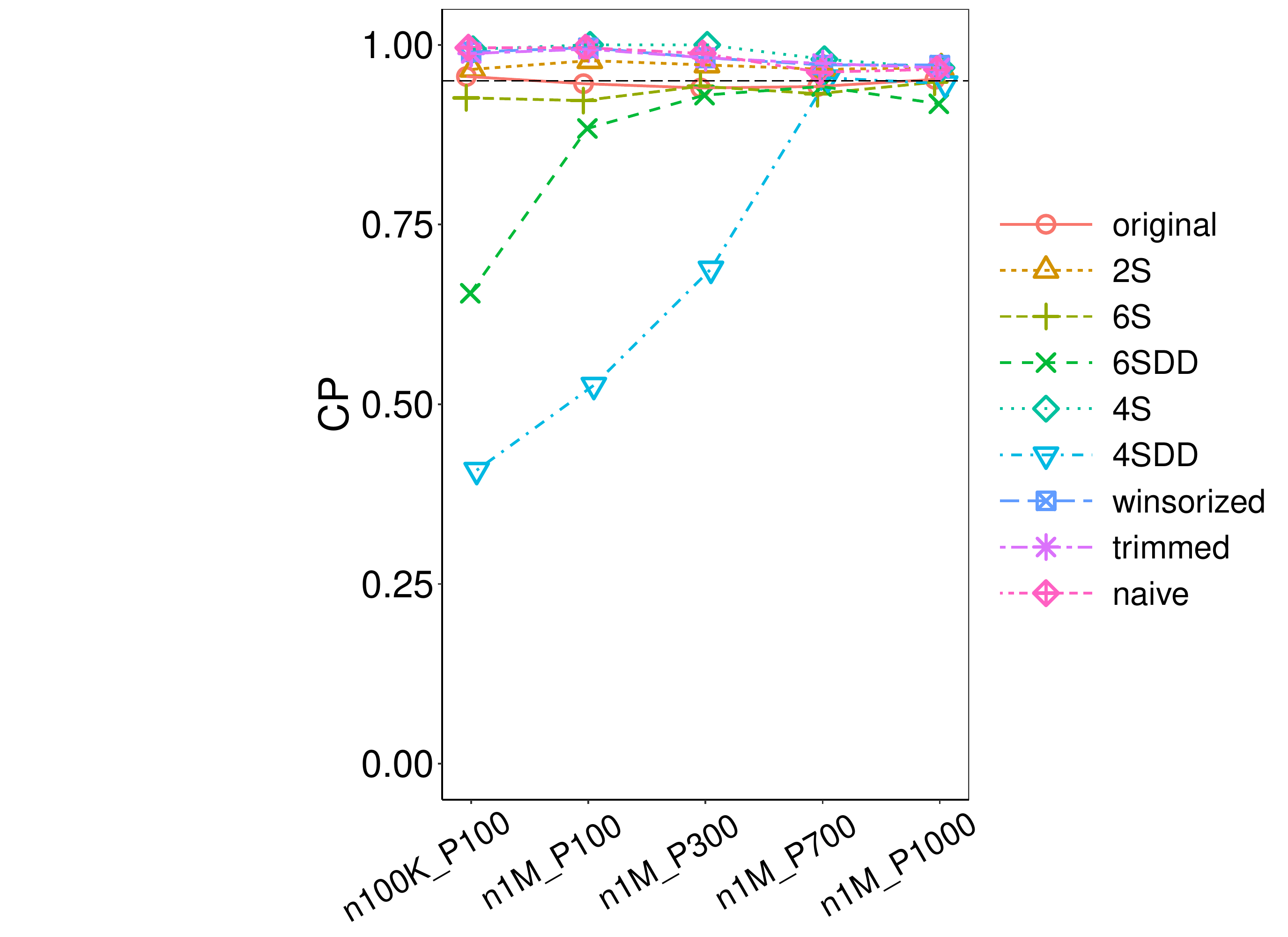}
\includegraphics[width=0.19\textwidth, trim={2.5in 0 2.6in 0},clip] {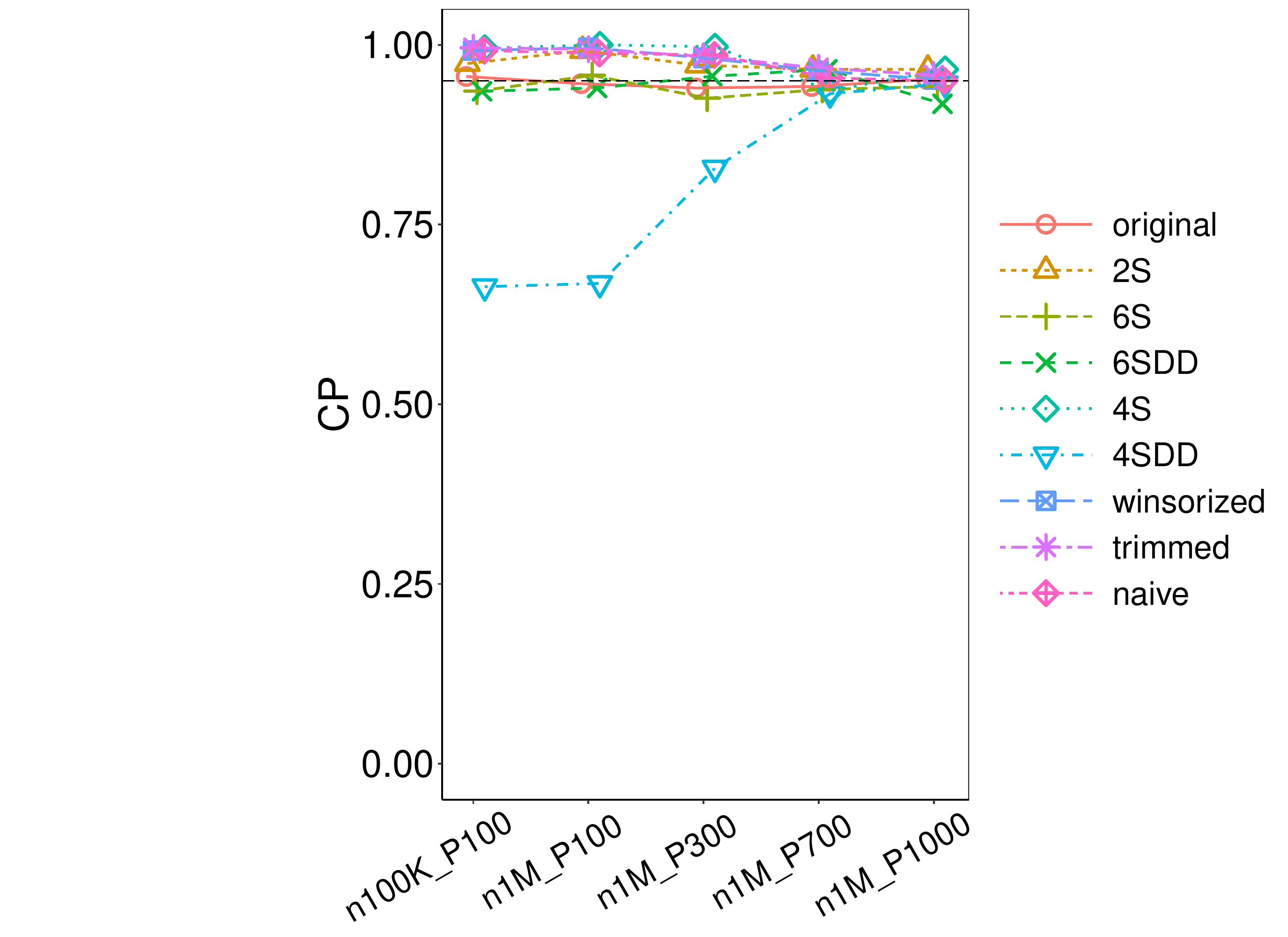}
\includegraphics[width=0.19\textwidth, trim={2.5in 0 2.6in 0},clip] {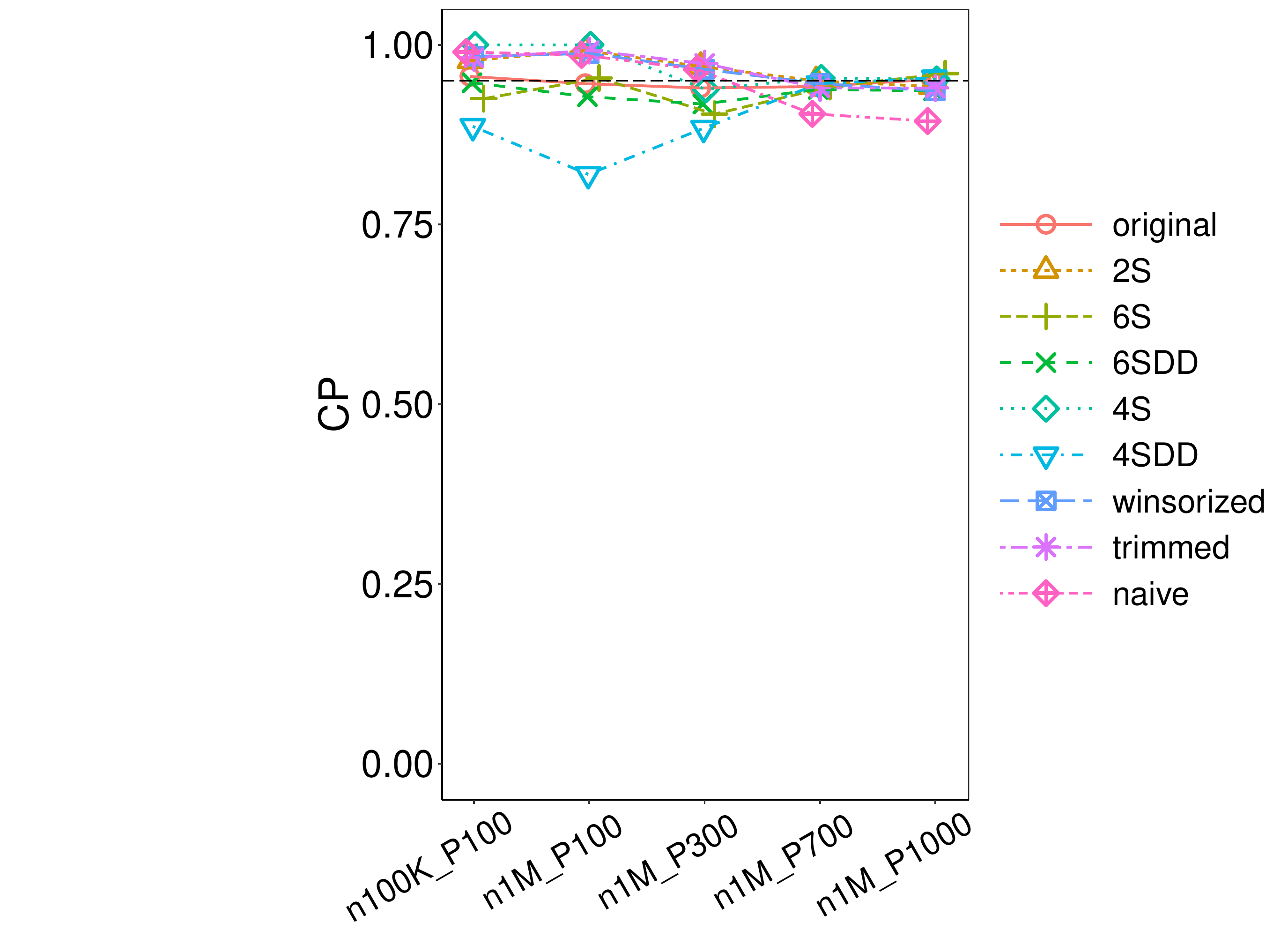}
\includegraphics[width=0.19\textwidth, trim={2.5in 0 2.6in 0},clip] {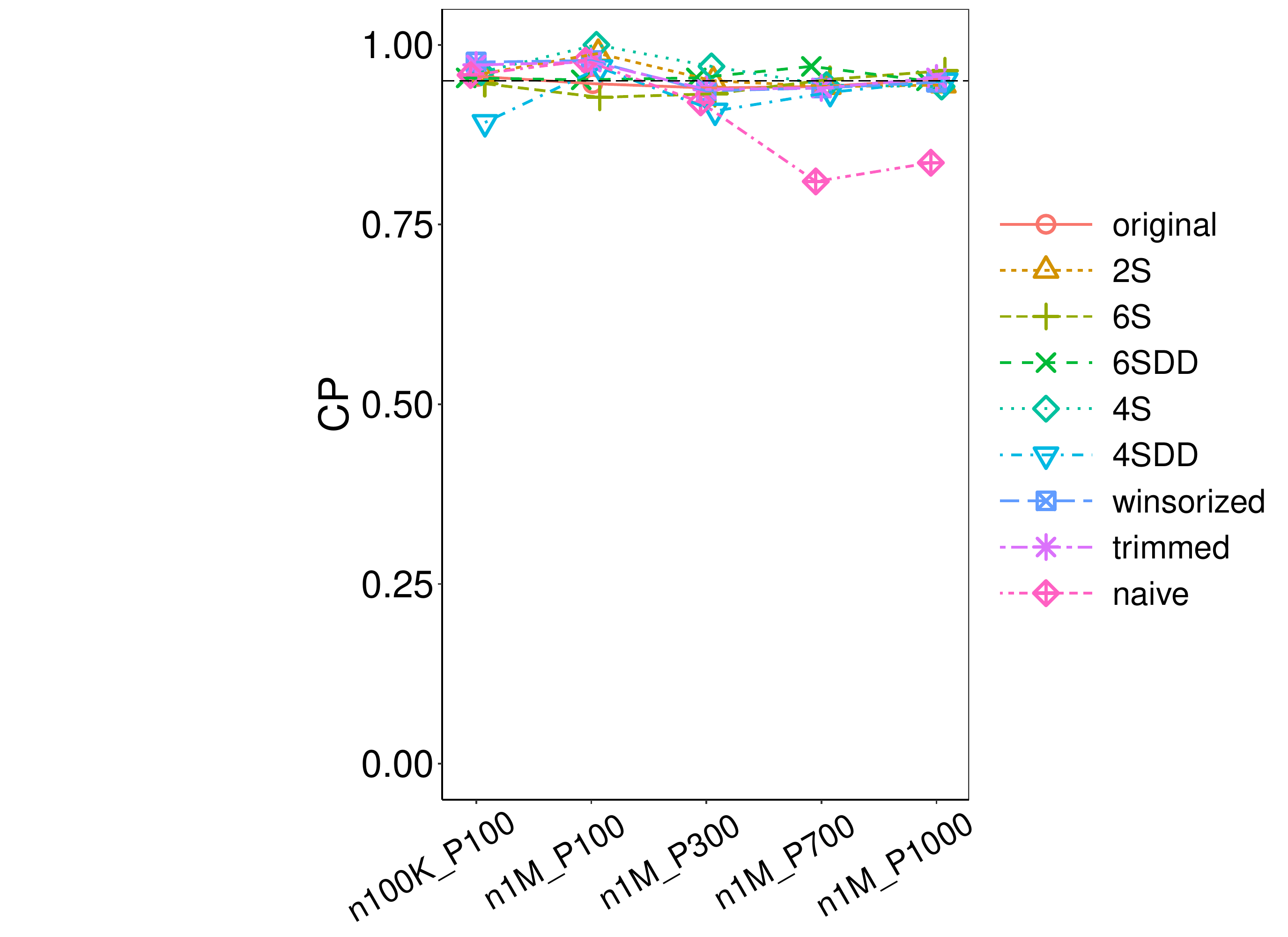}
\includegraphics[width=0.19\textwidth, trim={2.5in 0 2.6in 0},clip] {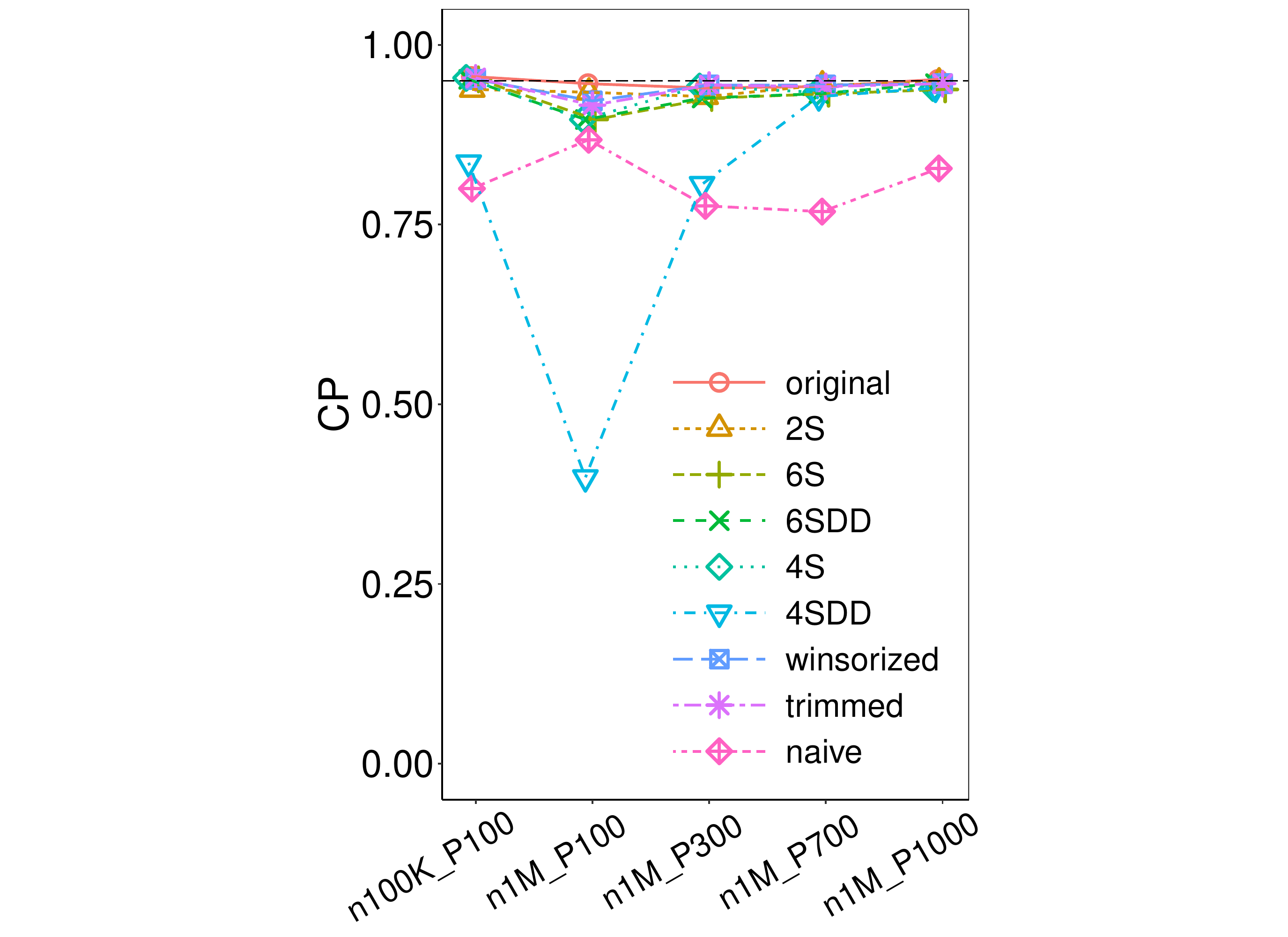}

\includegraphics[width=0.19\textwidth, trim={2.5in 0 2.6in 0},clip] {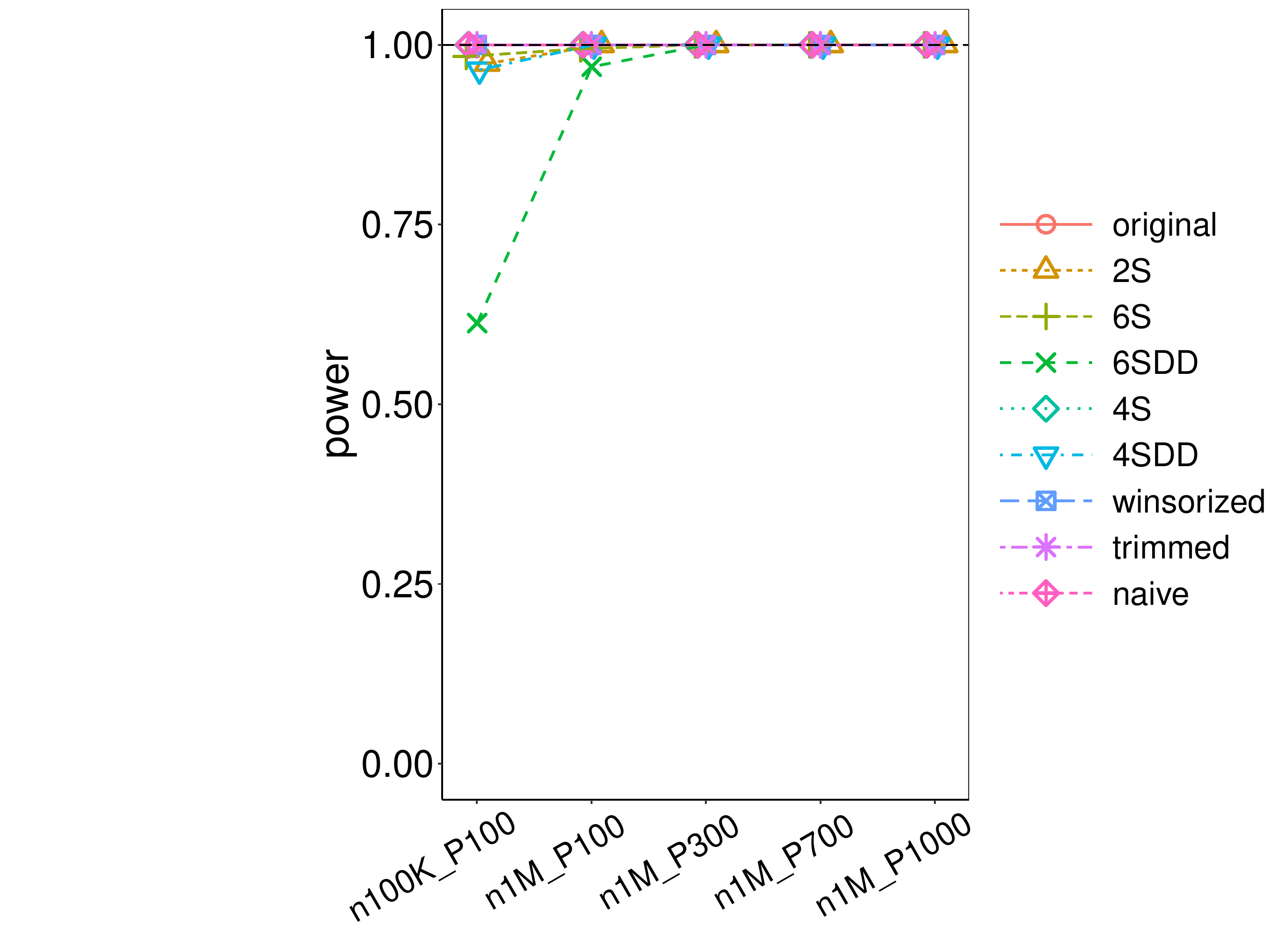}
\includegraphics[width=0.19\textwidth, trim={2.5in 0 2.6in 0},clip] {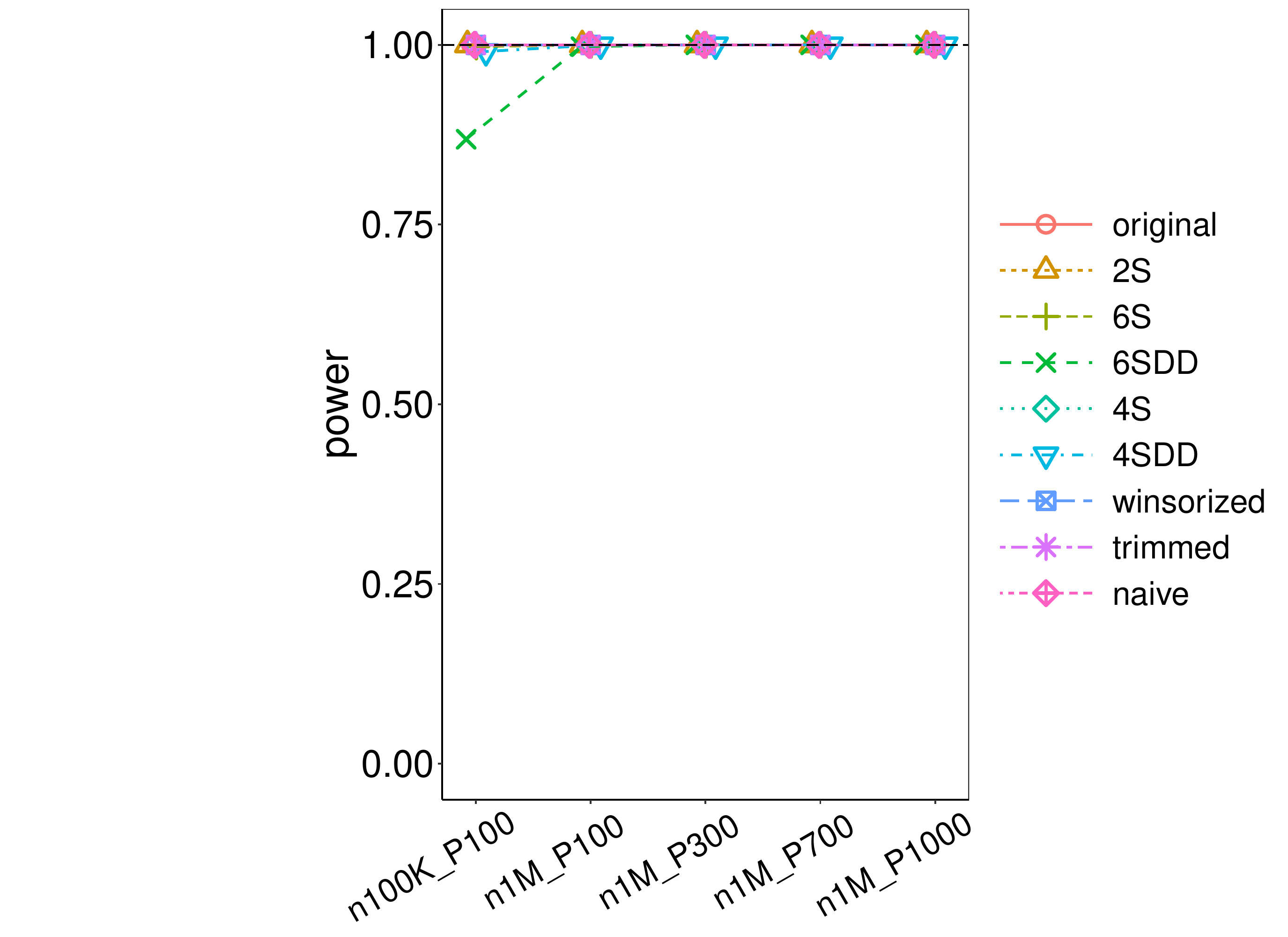}
\includegraphics[width=0.19\textwidth, trim={2.5in 0 2.6in 0},clip] {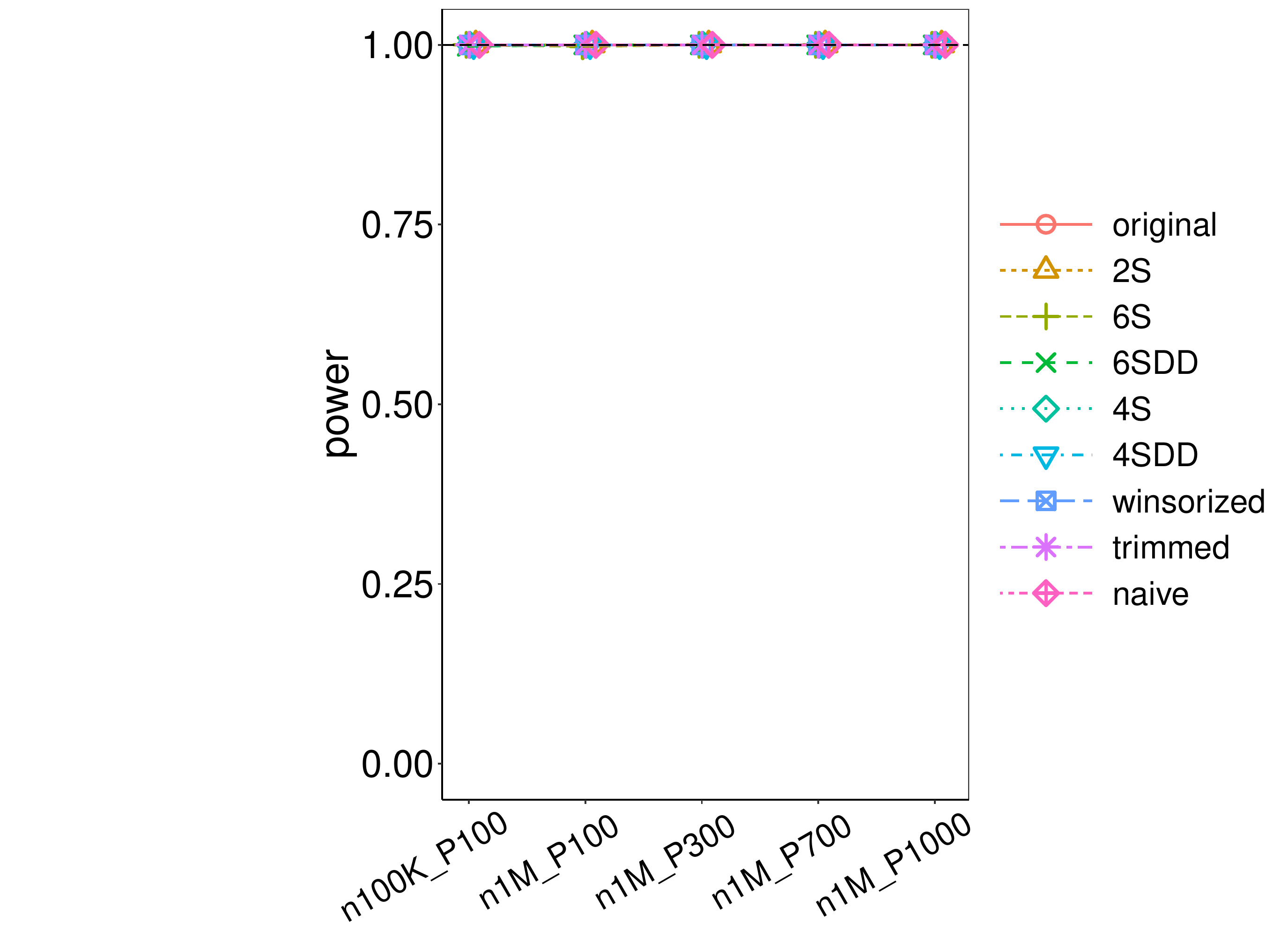}
\includegraphics[width=0.19\textwidth, trim={2.5in 0 2.6in 0},clip] {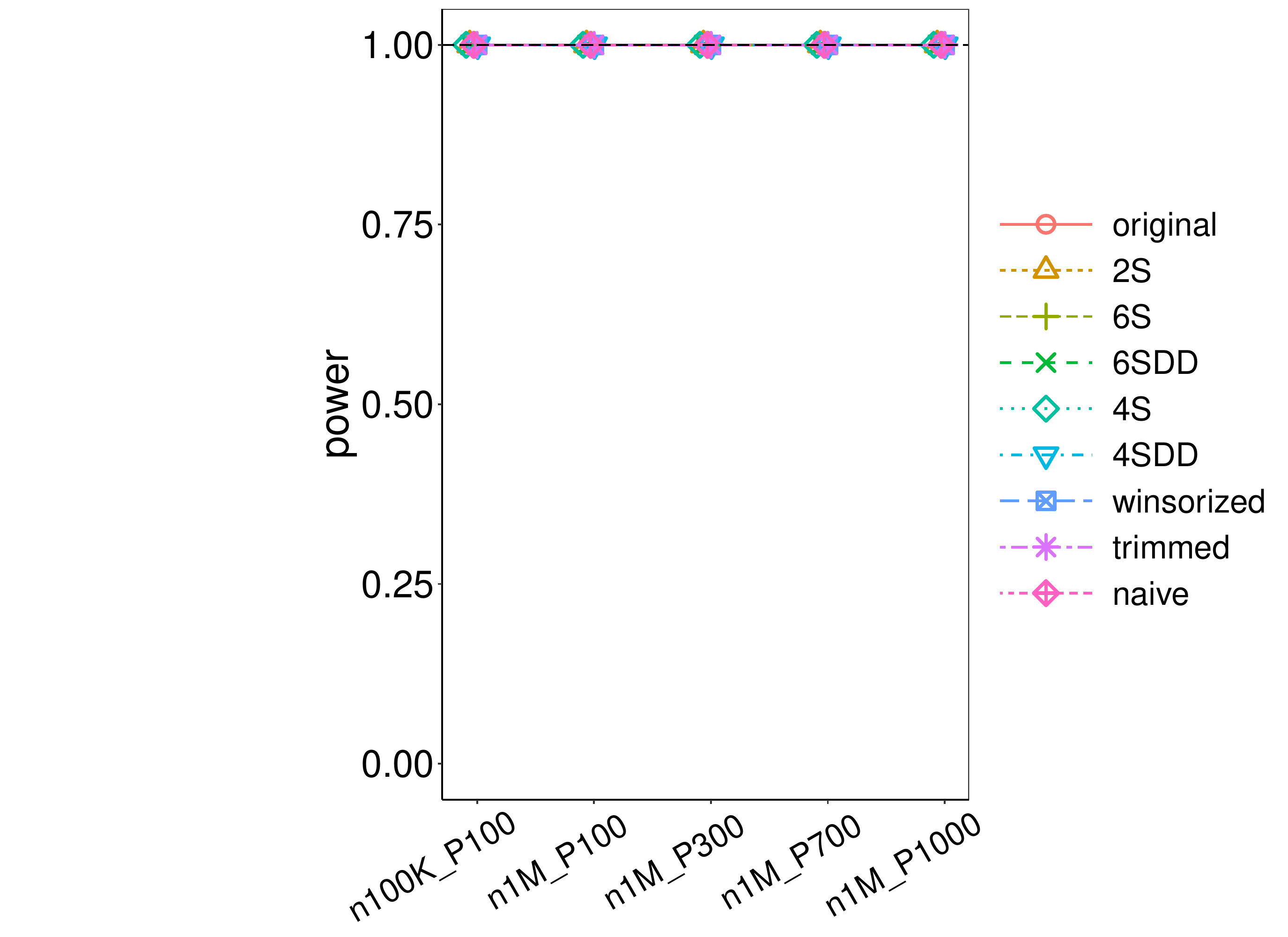}
\includegraphics[width=0.19\textwidth, trim={2.5in 0 2.6in 0},clip] {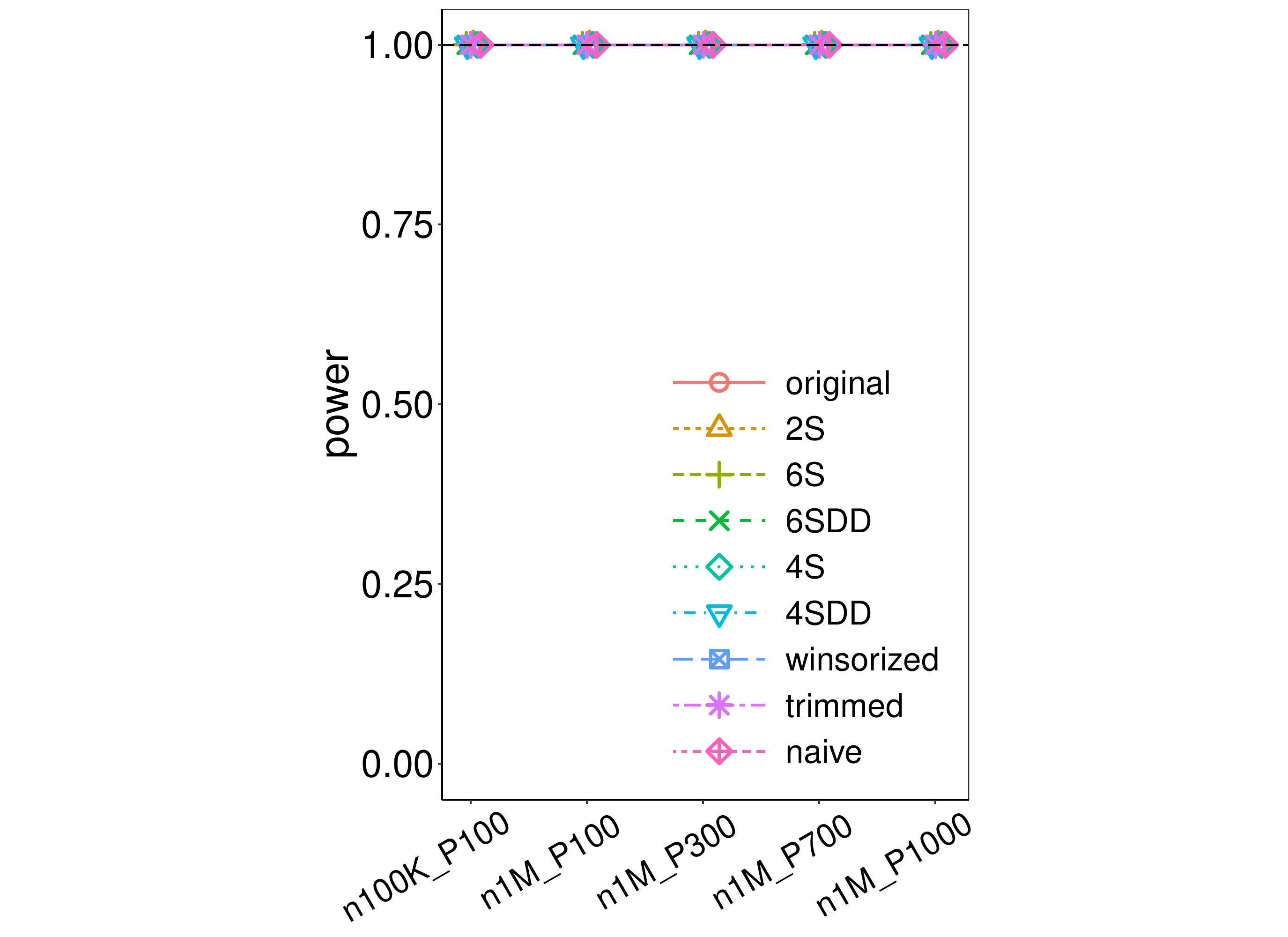}

\caption{Simulation results with $\epsilon$-DP for Gaussian data with  $\alpha=\beta$ when $\theta\ne0$} \label{fig:1sDPN}
\end{figure}

\begin{figure}[!htb]
\hspace{0.45in}$\rho=0.005$\hspace{0.65in}$\rho=0.02$\hspace{0.65in}$\rho=0.08$
\hspace{0.65in}$\rho=0.32$\hspace{0.65in}$\rho=1.28$

\includegraphics[width=0.19\textwidth, trim={2.5in 0 2.6in 0},clip] {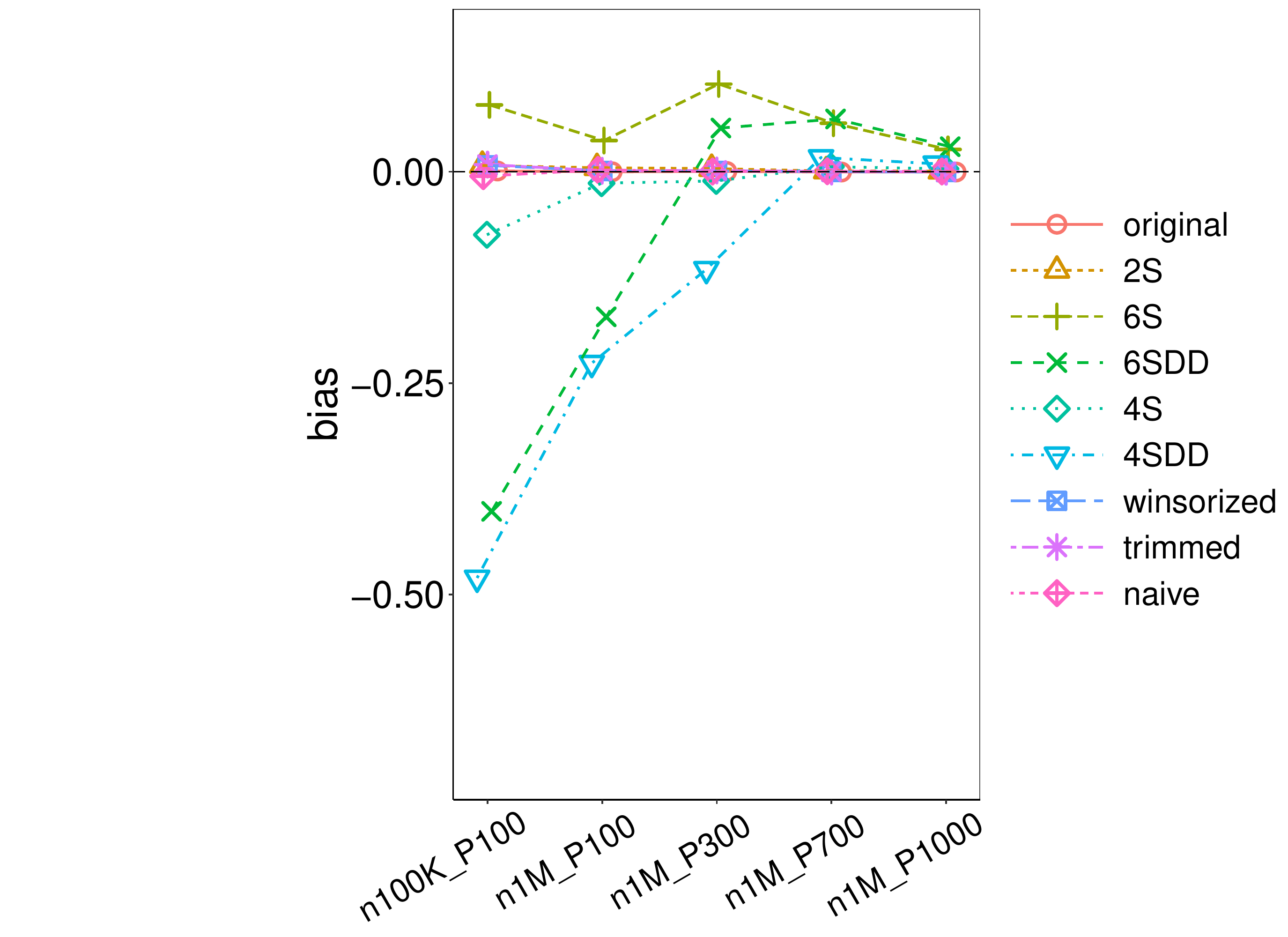}
\includegraphics[width=0.19\textwidth, trim={2.5in 0 2.6in 0},clip] {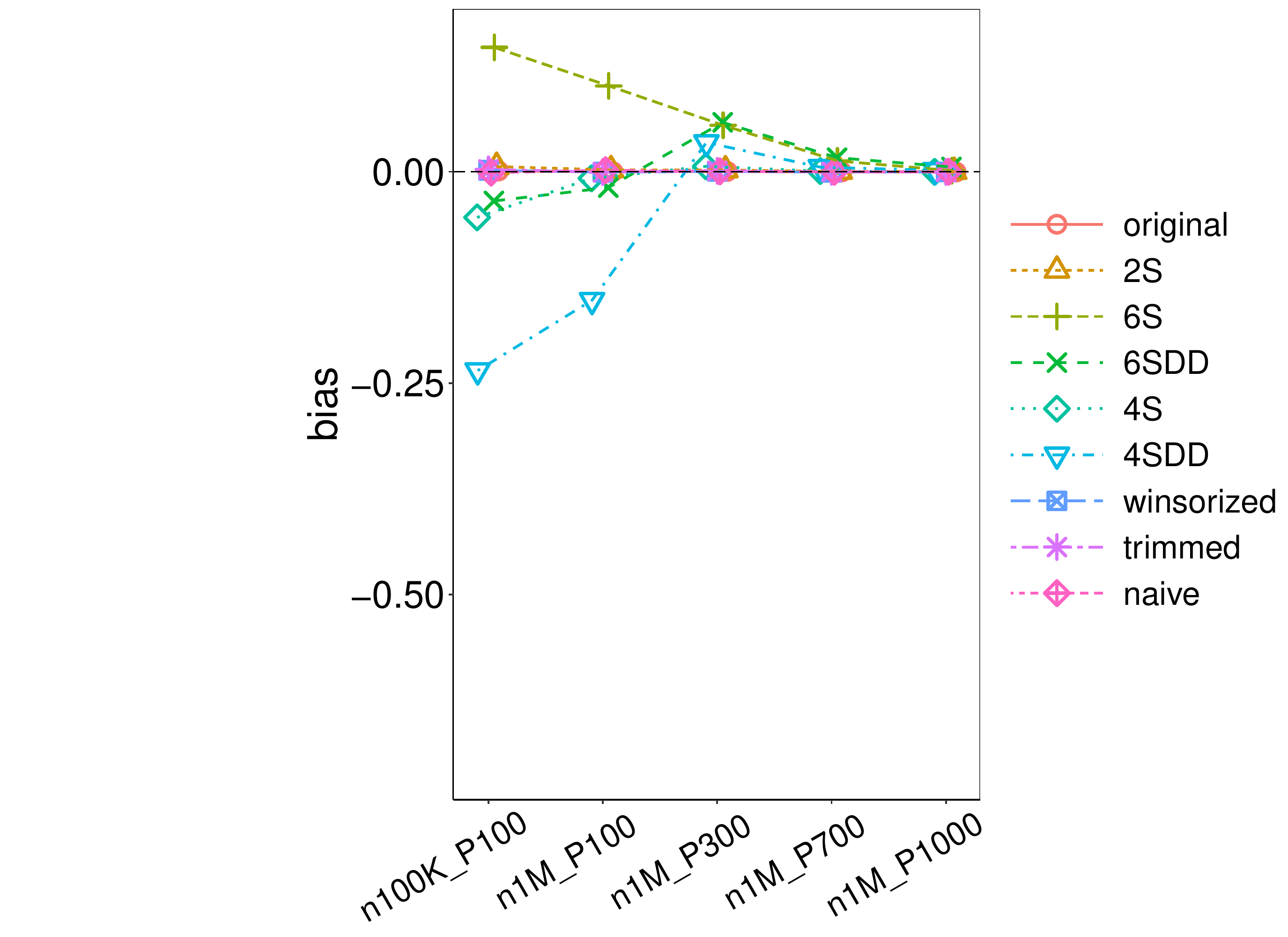}
\includegraphics[width=0.19\textwidth, trim={2.5in 0 2.6in 0},clip] {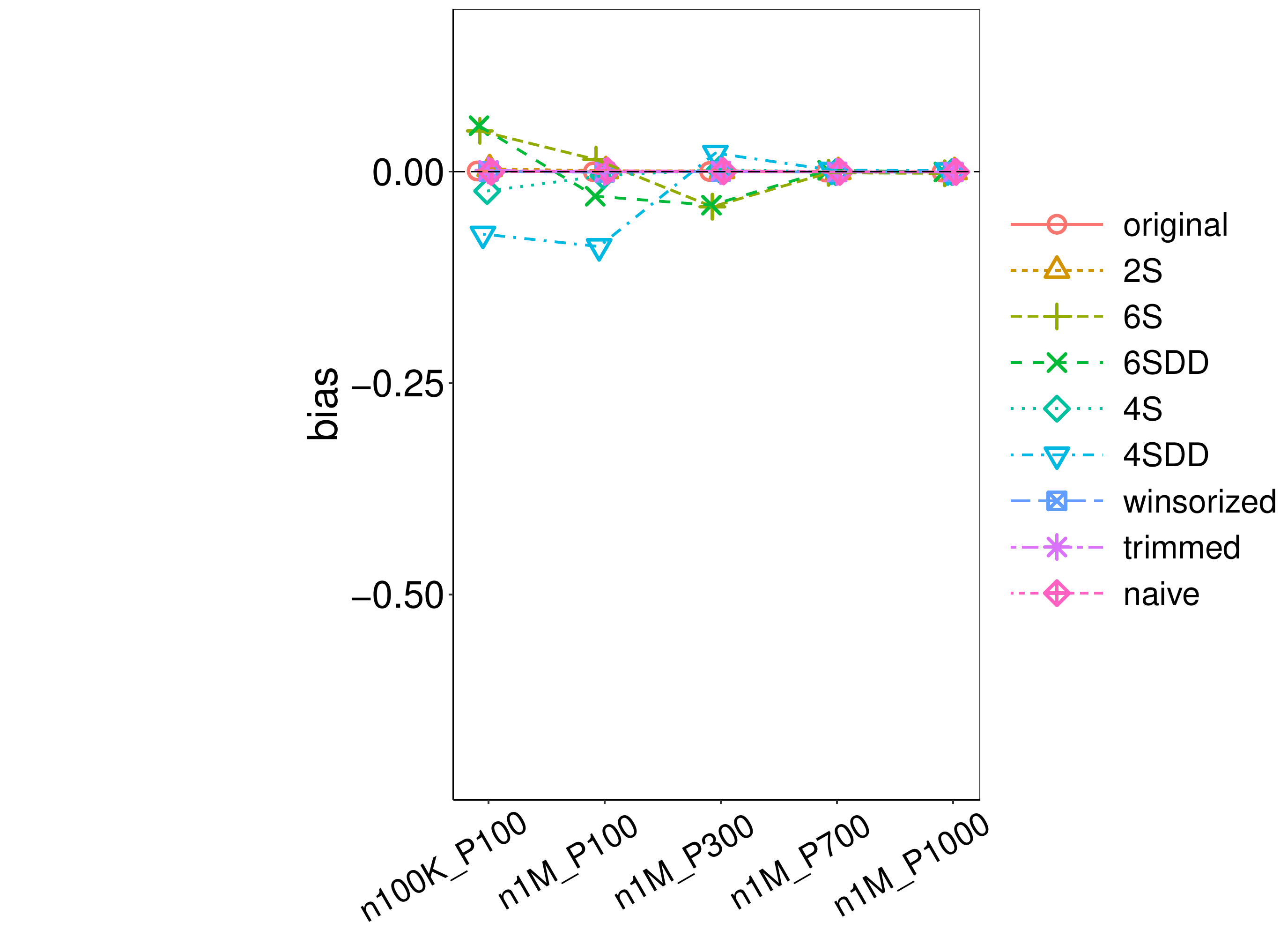}
\includegraphics[width=0.19\textwidth, trim={2.5in 0 2.6in 0},clip] {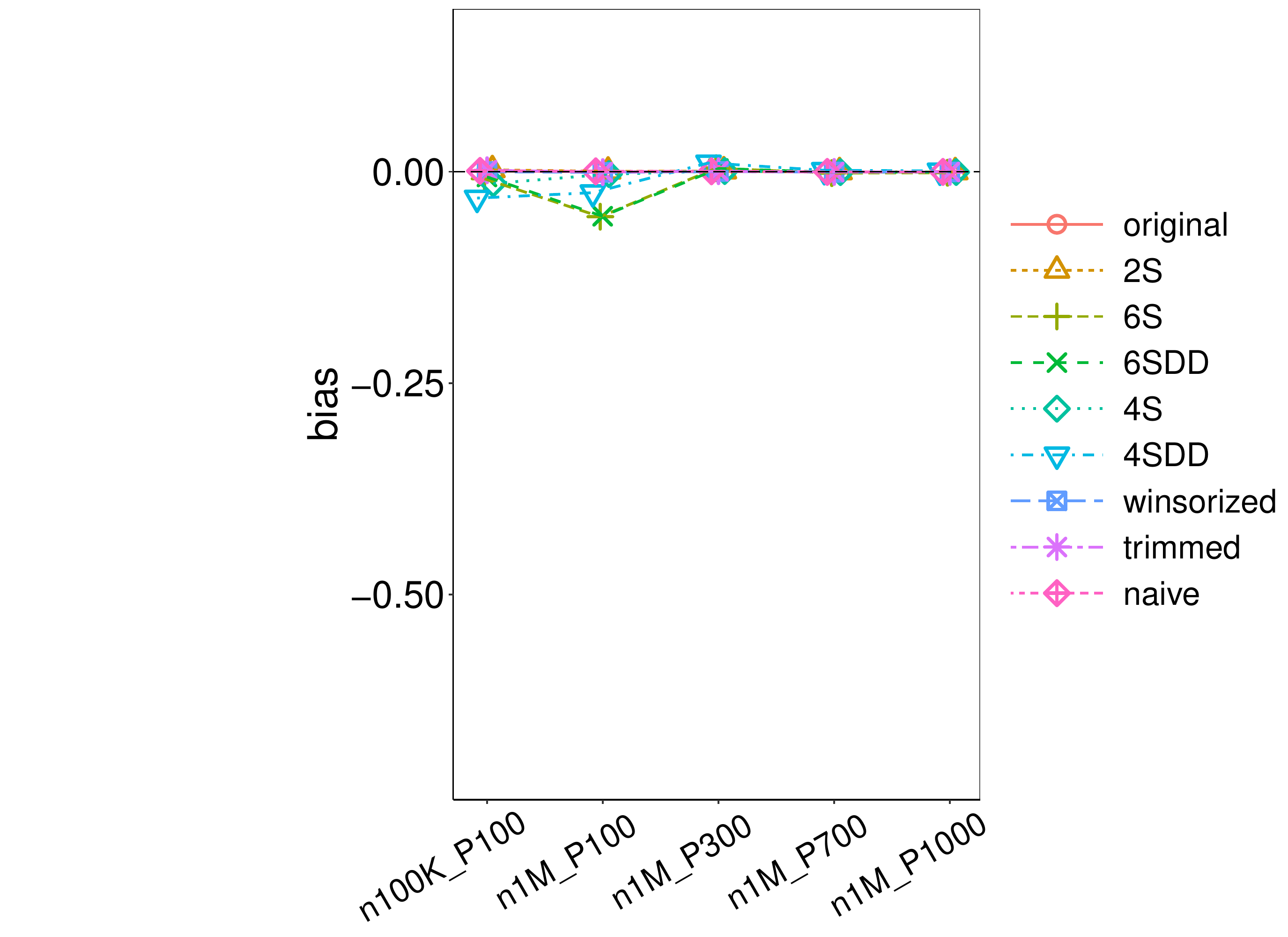}
\includegraphics[width=0.19\textwidth, trim={2.5in 0 2.6in 0},clip] {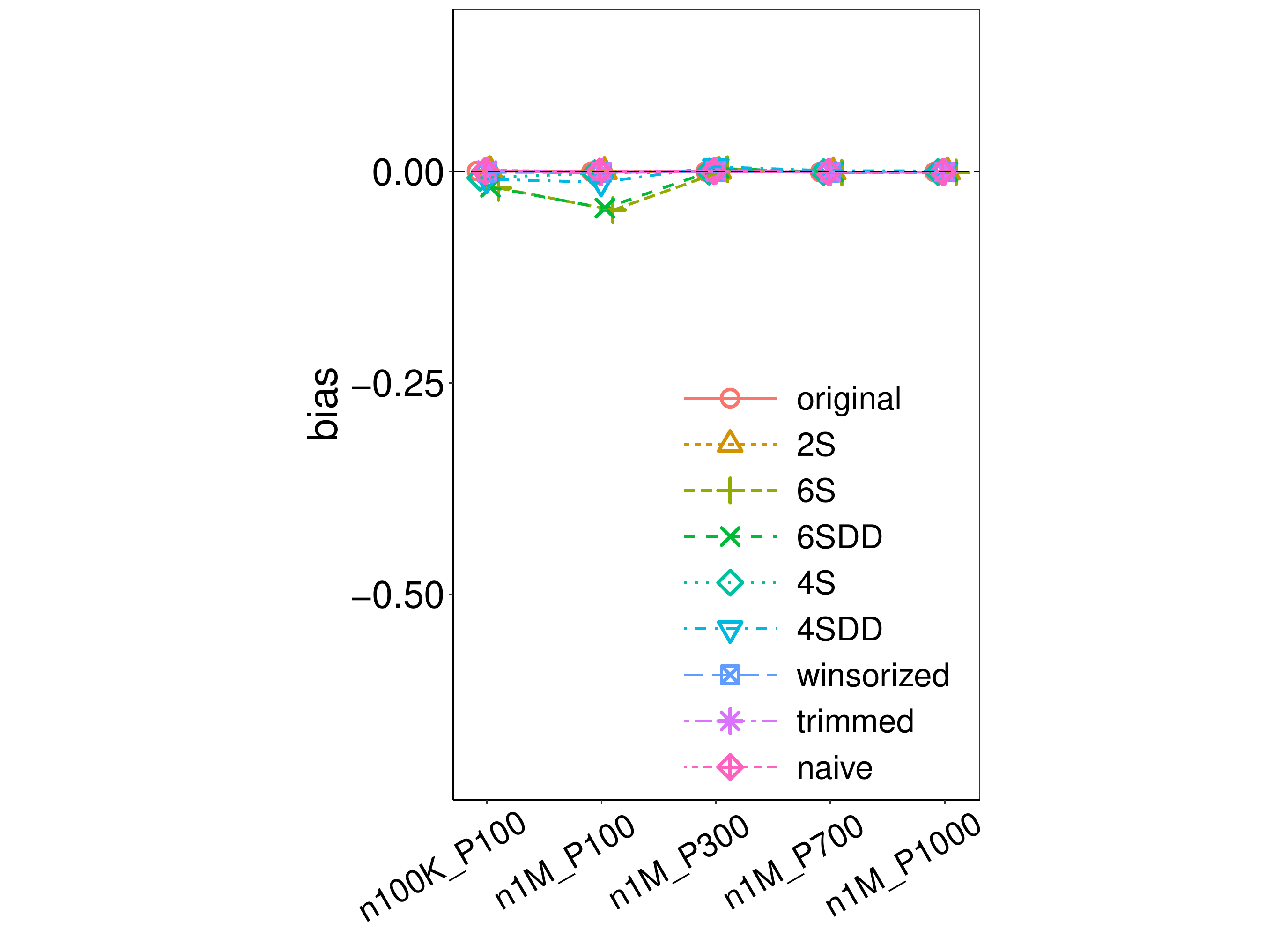}

\includegraphics[width=0.19\textwidth, trim={2.5in 0 2.6in 0},clip] {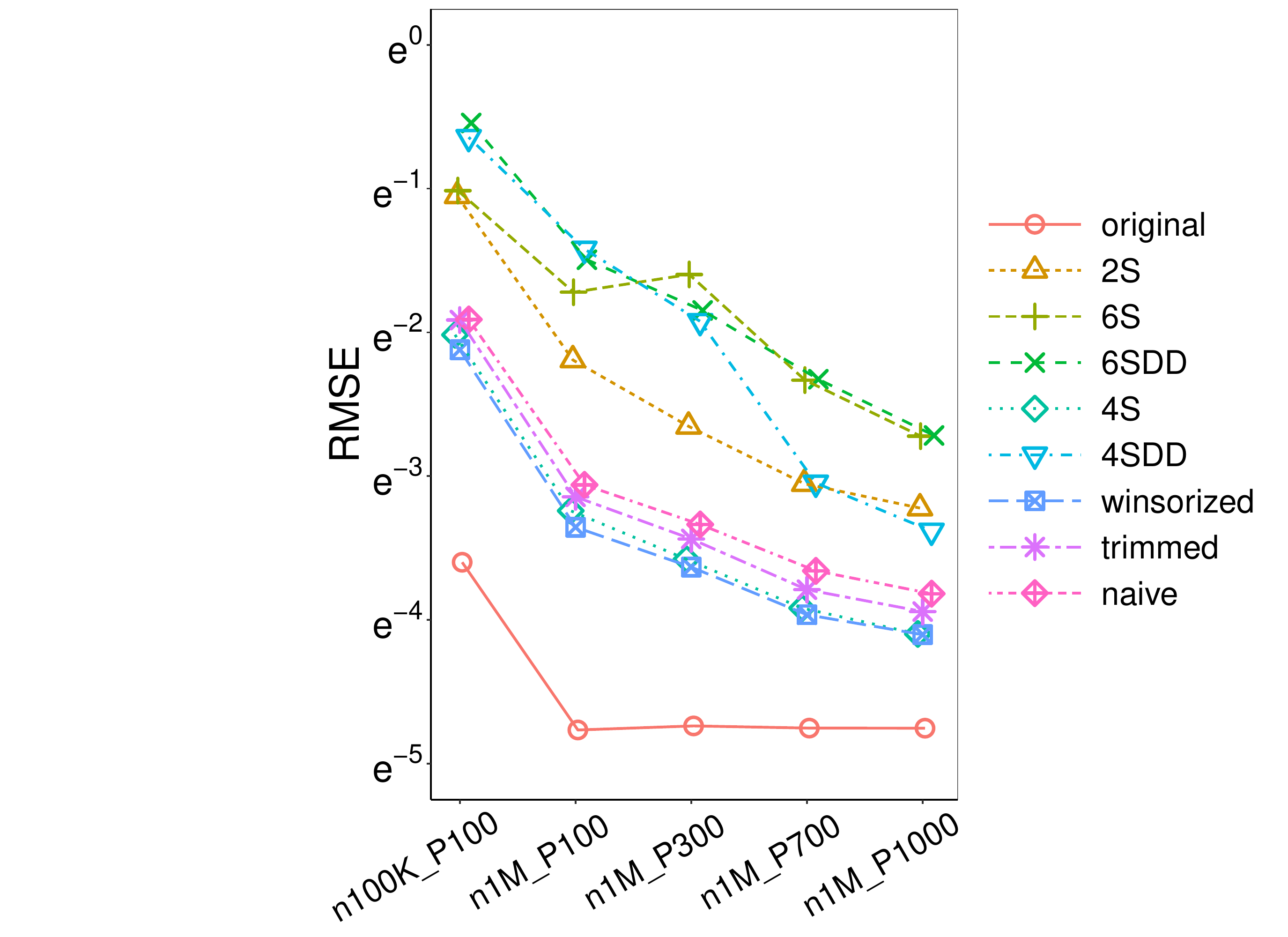}
\includegraphics[width=0.19\textwidth, trim={2.5in 0 2.6in 0},clip] {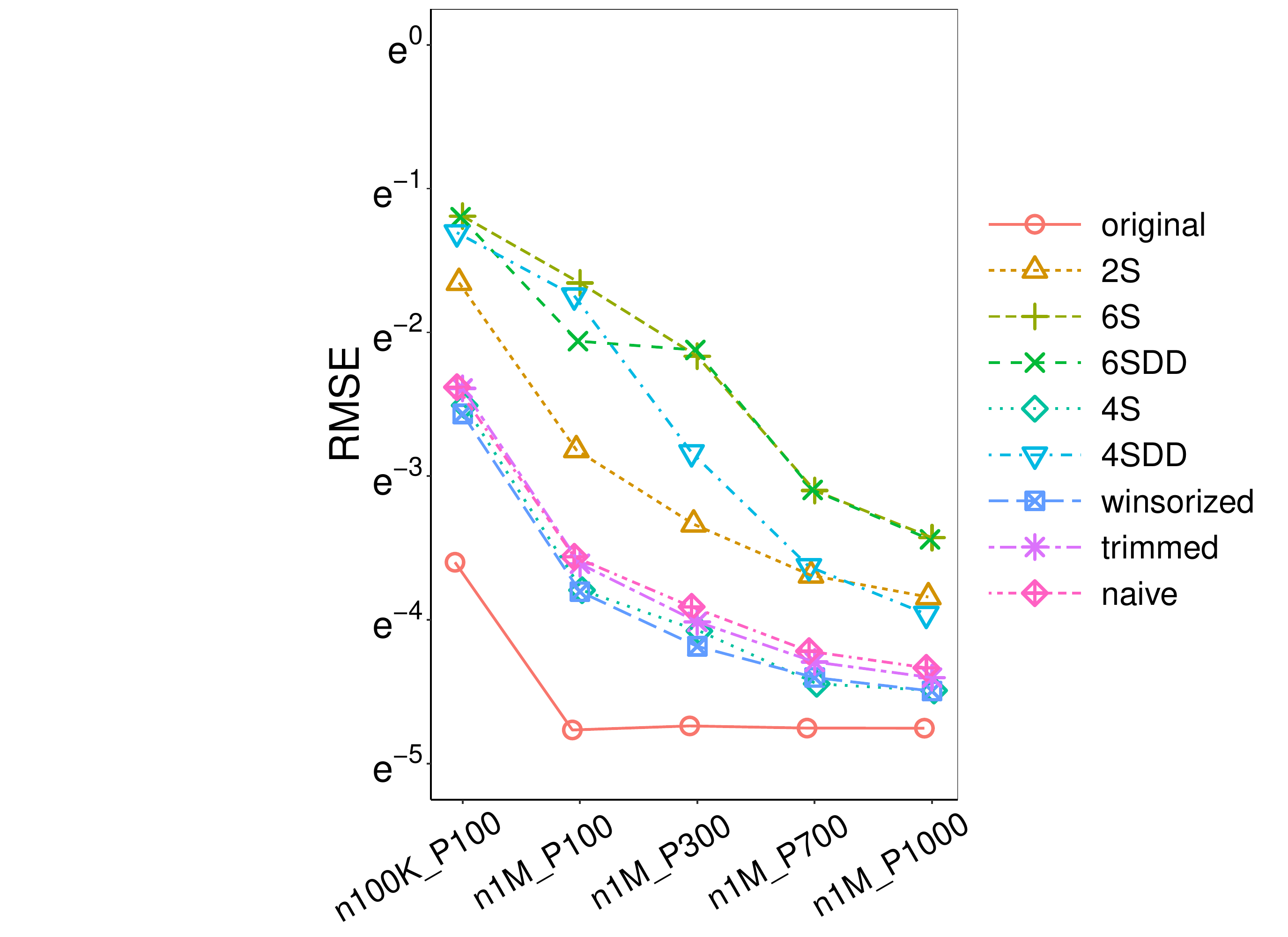}
\includegraphics[width=0.19\textwidth, trim={2.5in 0 2.6in 0},clip] {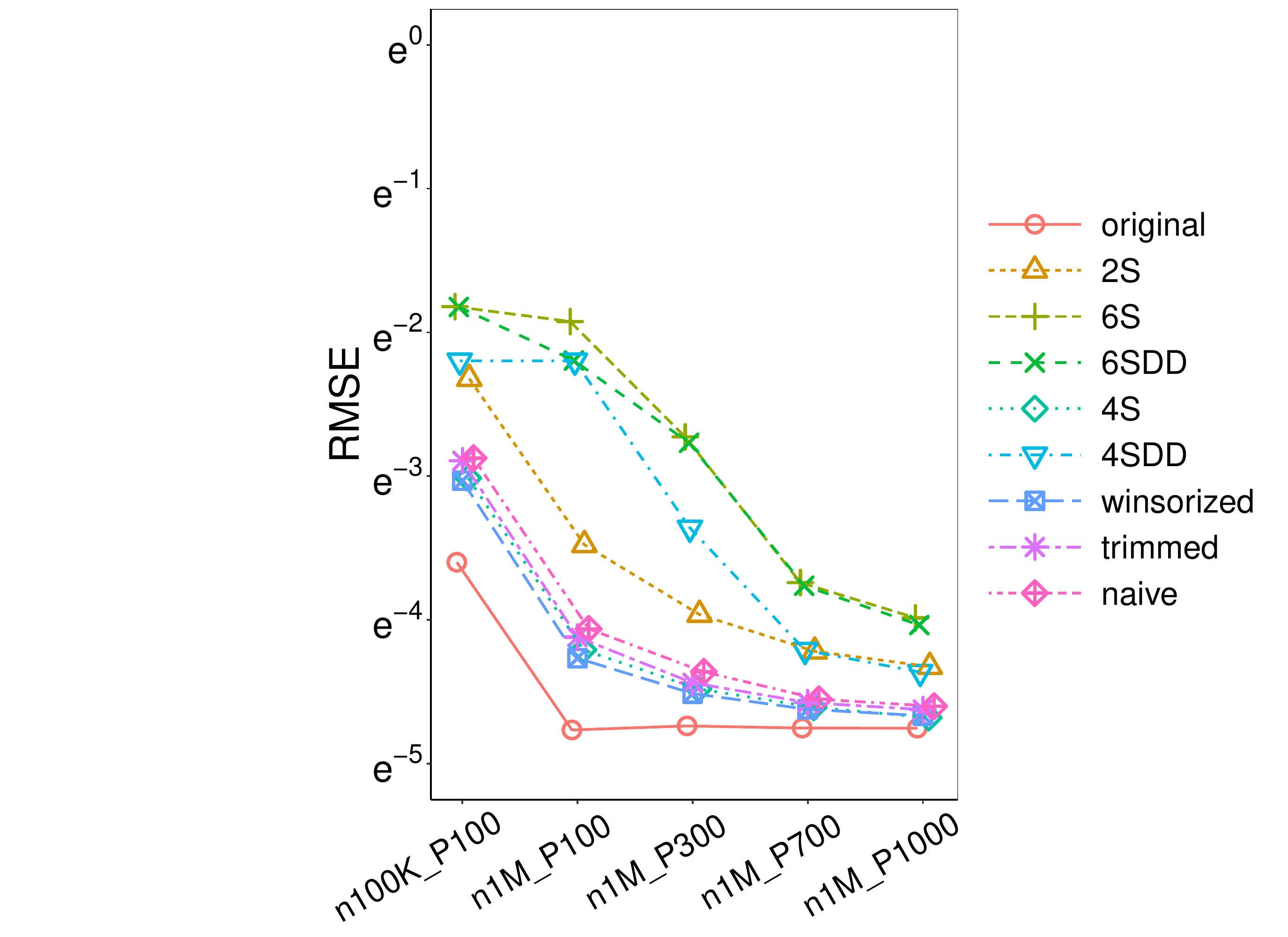}
\includegraphics[width=0.19\textwidth, trim={2.5in 0 2.6in 0},clip] {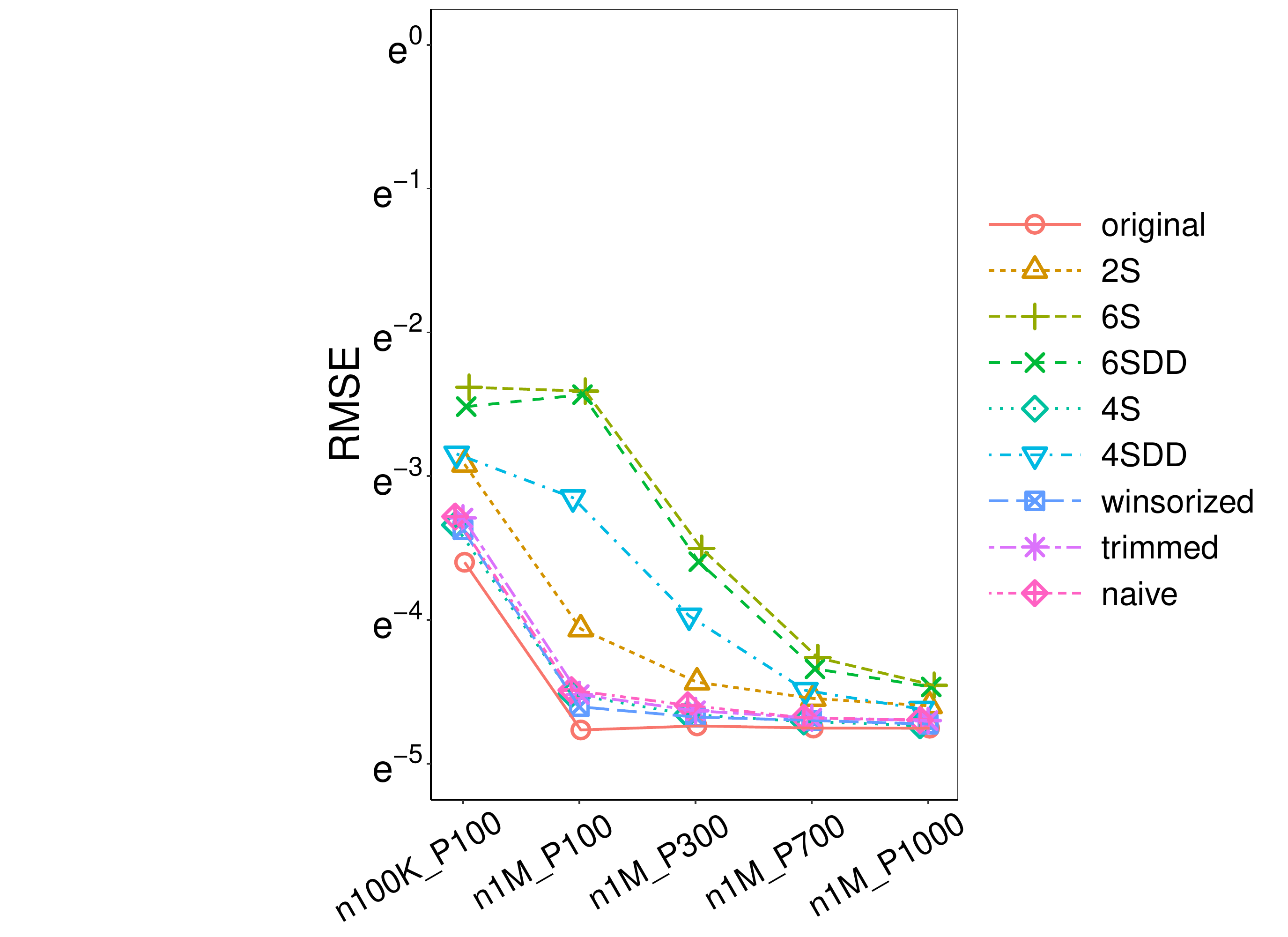}
\includegraphics[width=0.19\textwidth, trim={2.5in 0 2.6in 0},clip] {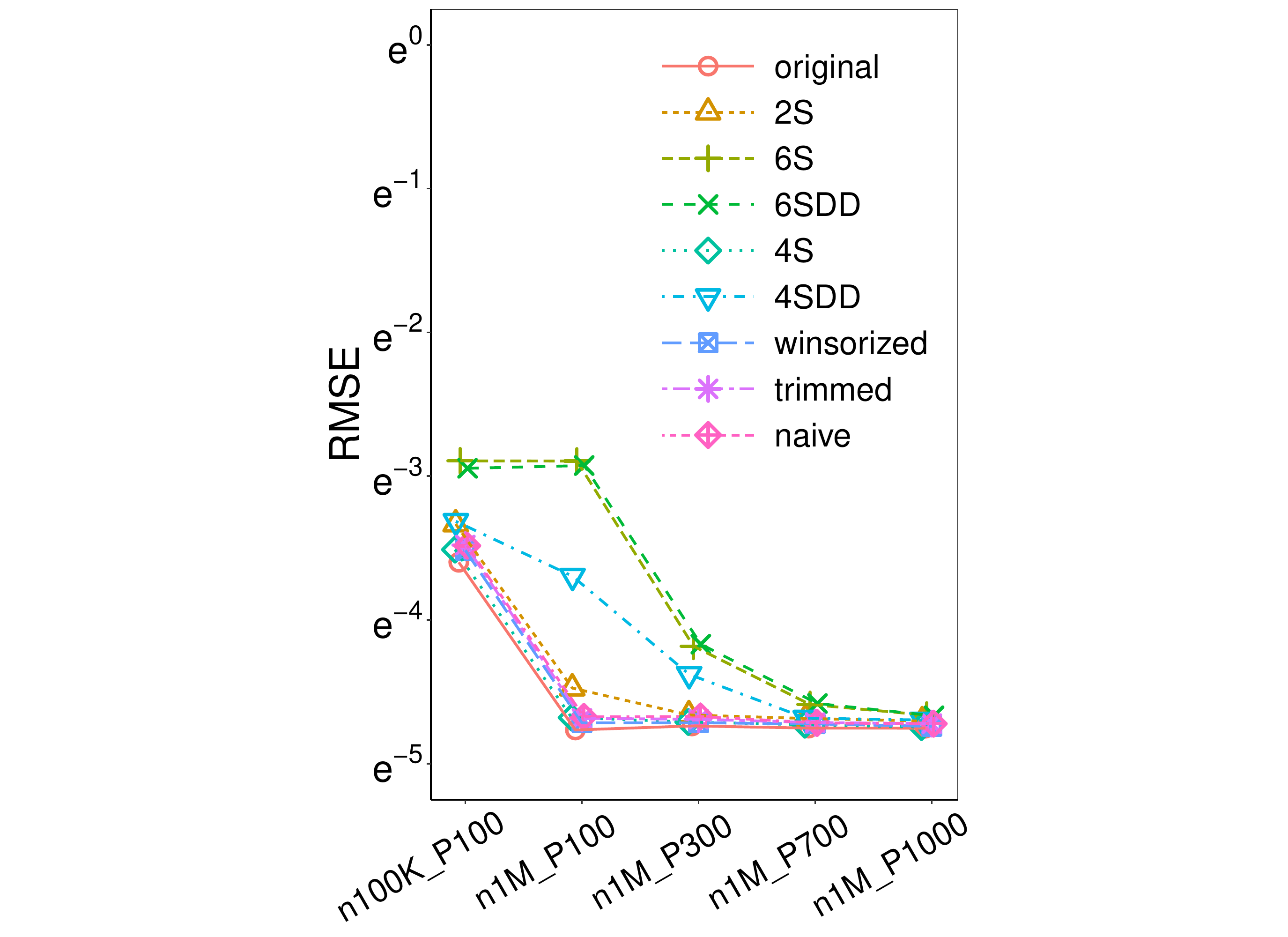}

\includegraphics[width=0.19\textwidth, trim={2.5in 0 2.6in 0},clip] {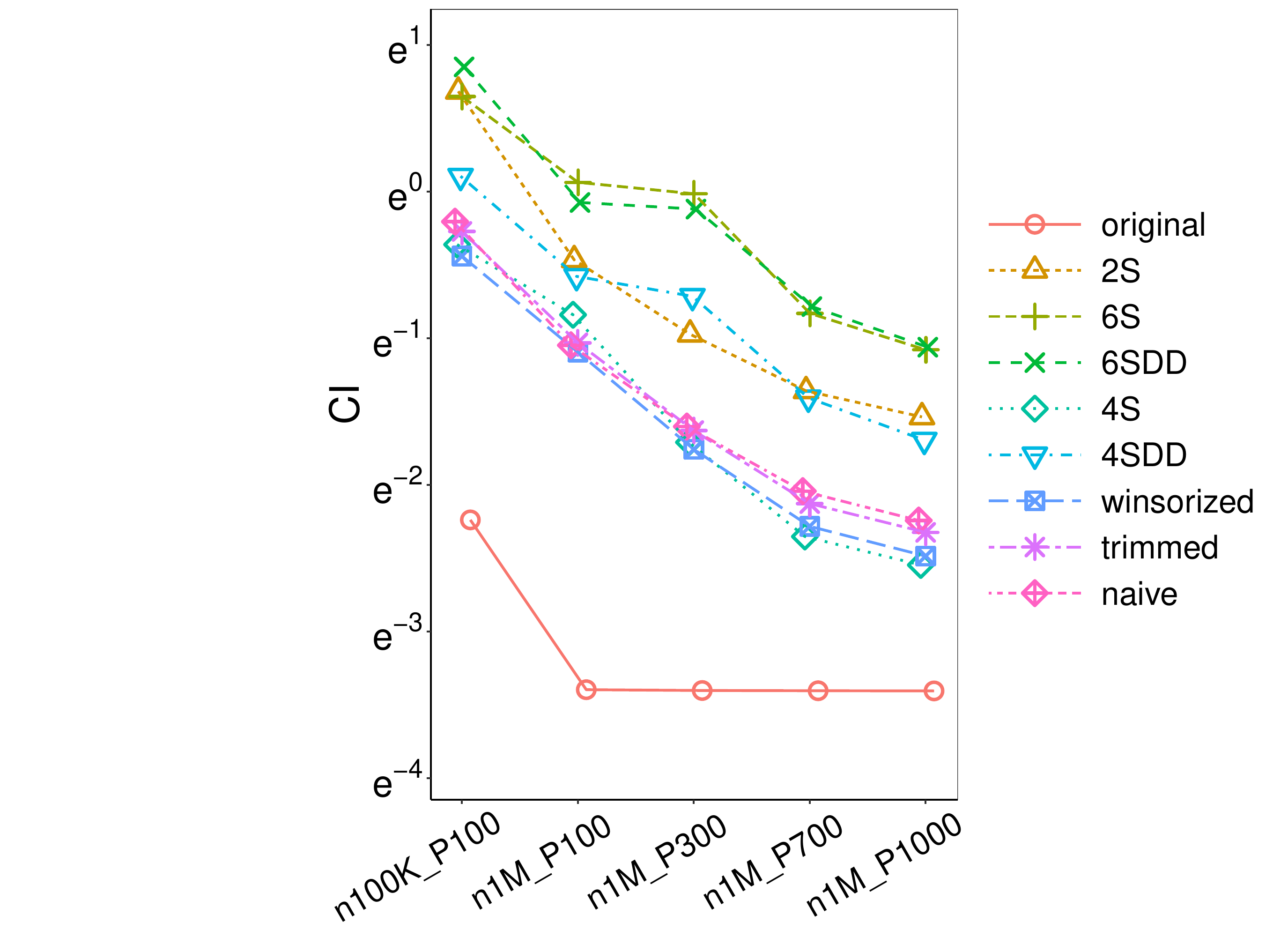}
\includegraphics[width=0.19\textwidth, trim={2.5in 0 2.6in 0},clip] {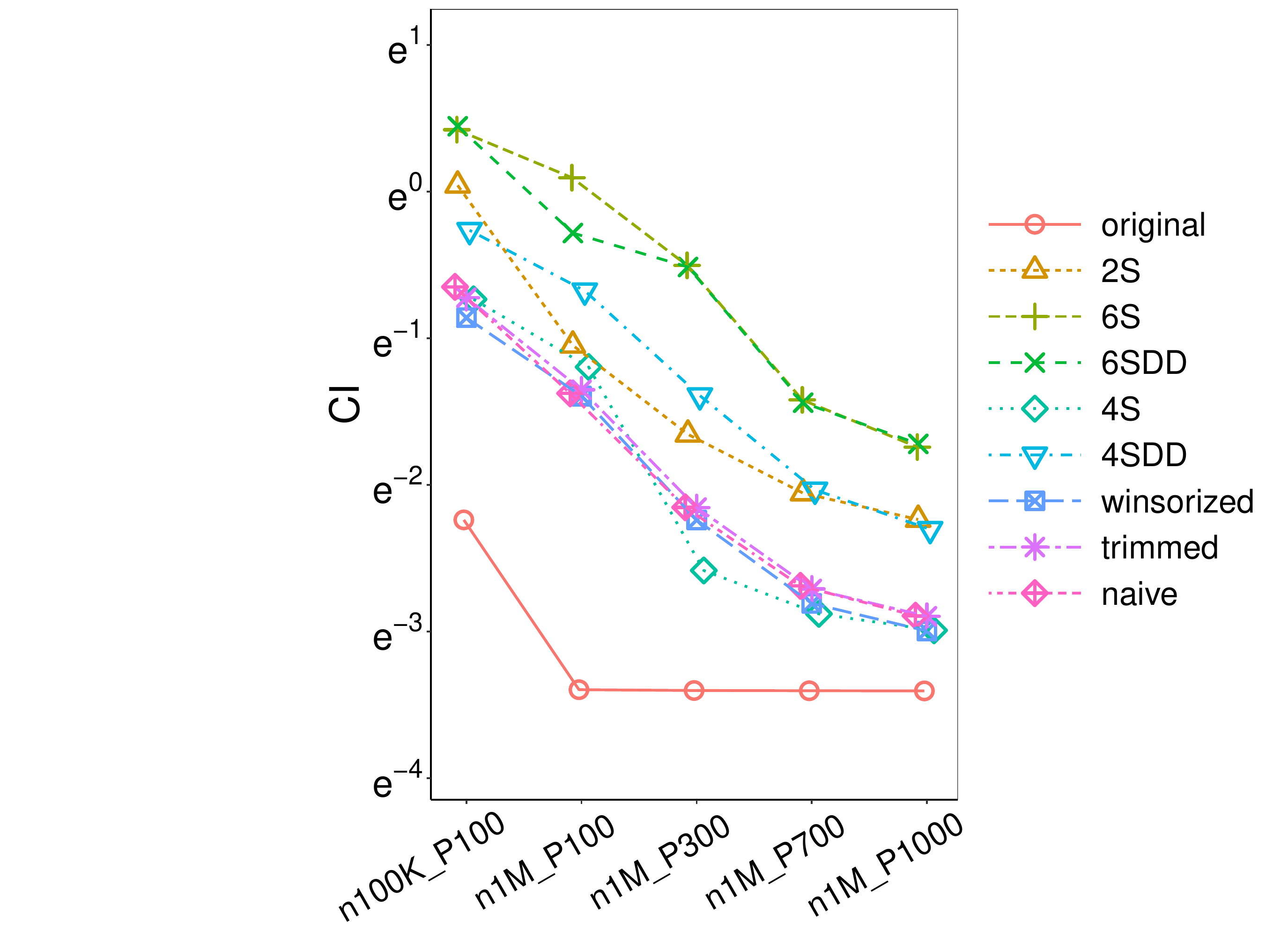}
\includegraphics[width=0.19\textwidth, trim={2.5in 0 2.6in 0},clip] {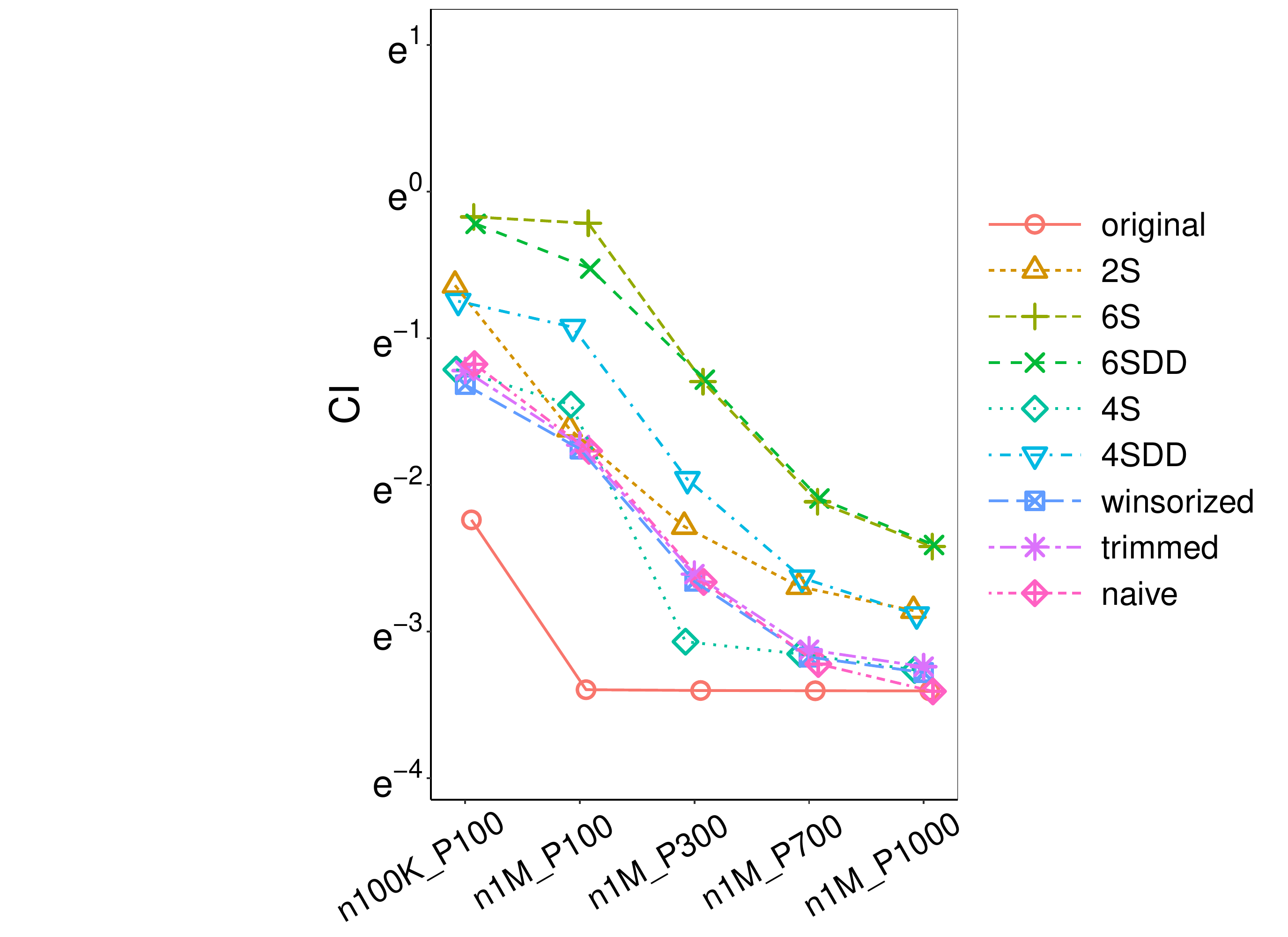}
\includegraphics[width=0.19\textwidth, trim={2.5in 0 2.6in 0},clip] {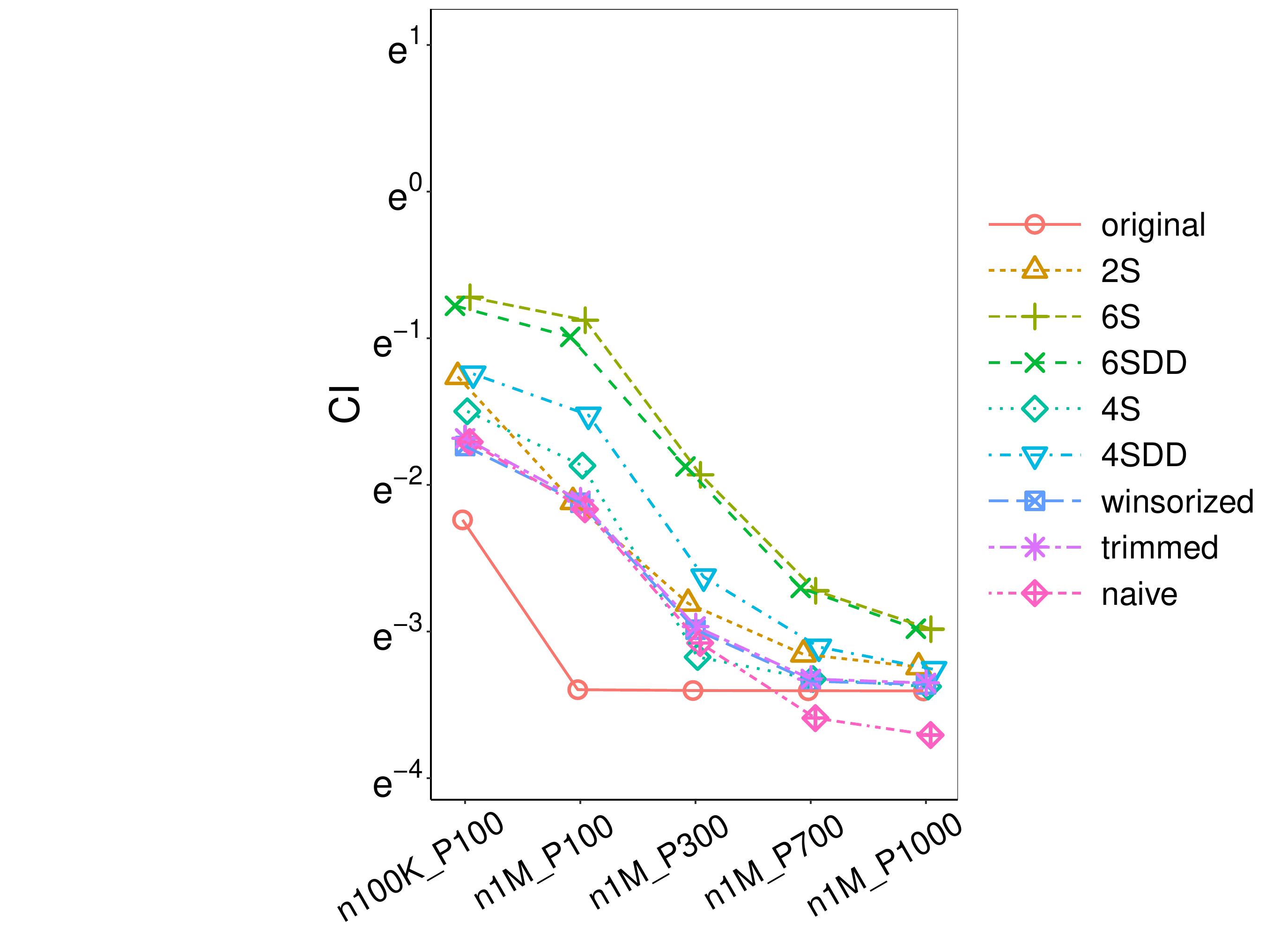}
\includegraphics[width=0.19\textwidth, trim={2.5in 0 2.6in 0},clip] {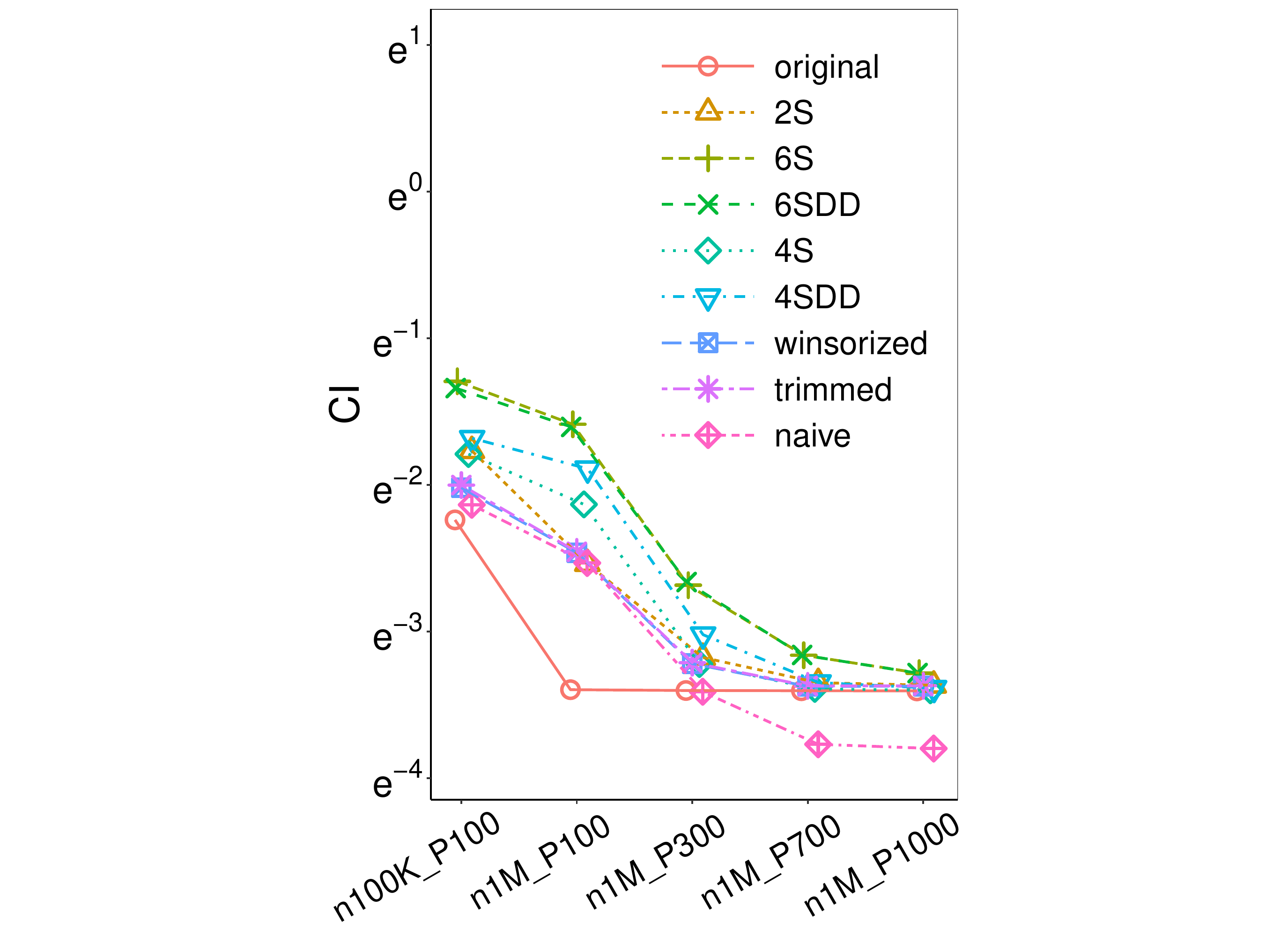}

\includegraphics[width=0.19\textwidth, trim={2.5in 0 2.6in 0},clip] {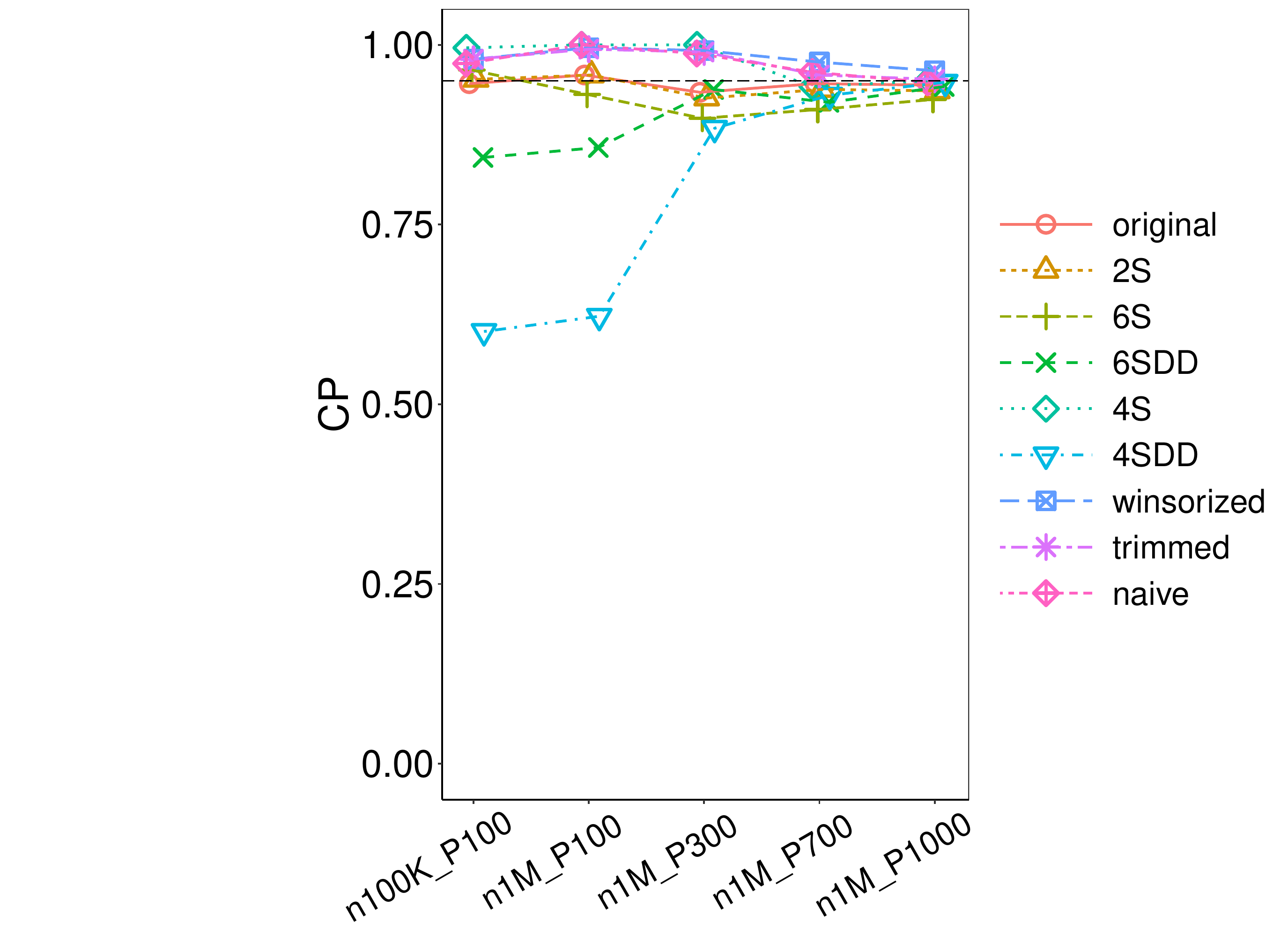}
\includegraphics[width=0.19\textwidth, trim={2.5in 0 2.6in 0},clip] {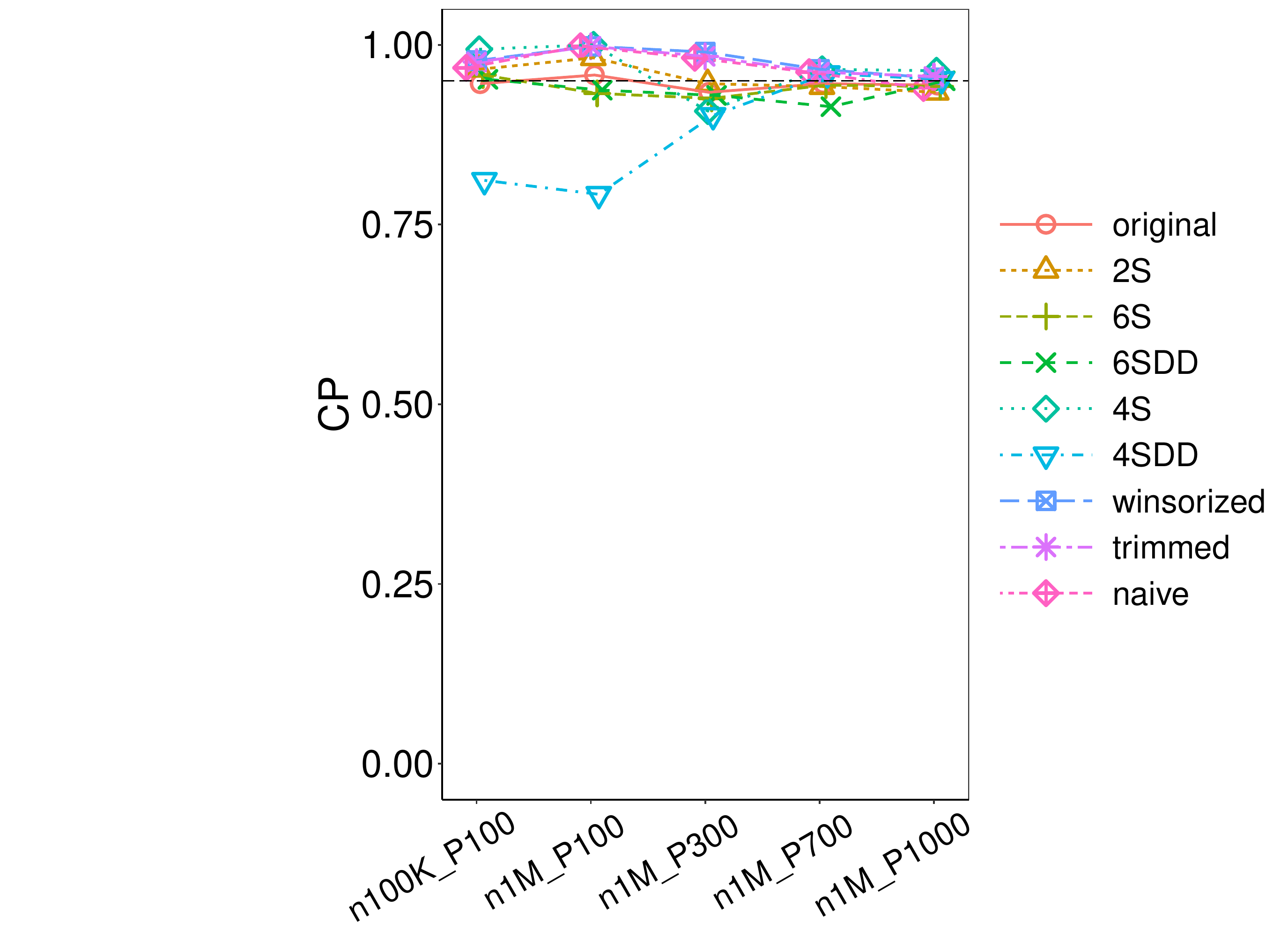}
\includegraphics[width=0.19\textwidth, trim={2.5in 0 2.6in 0},clip] {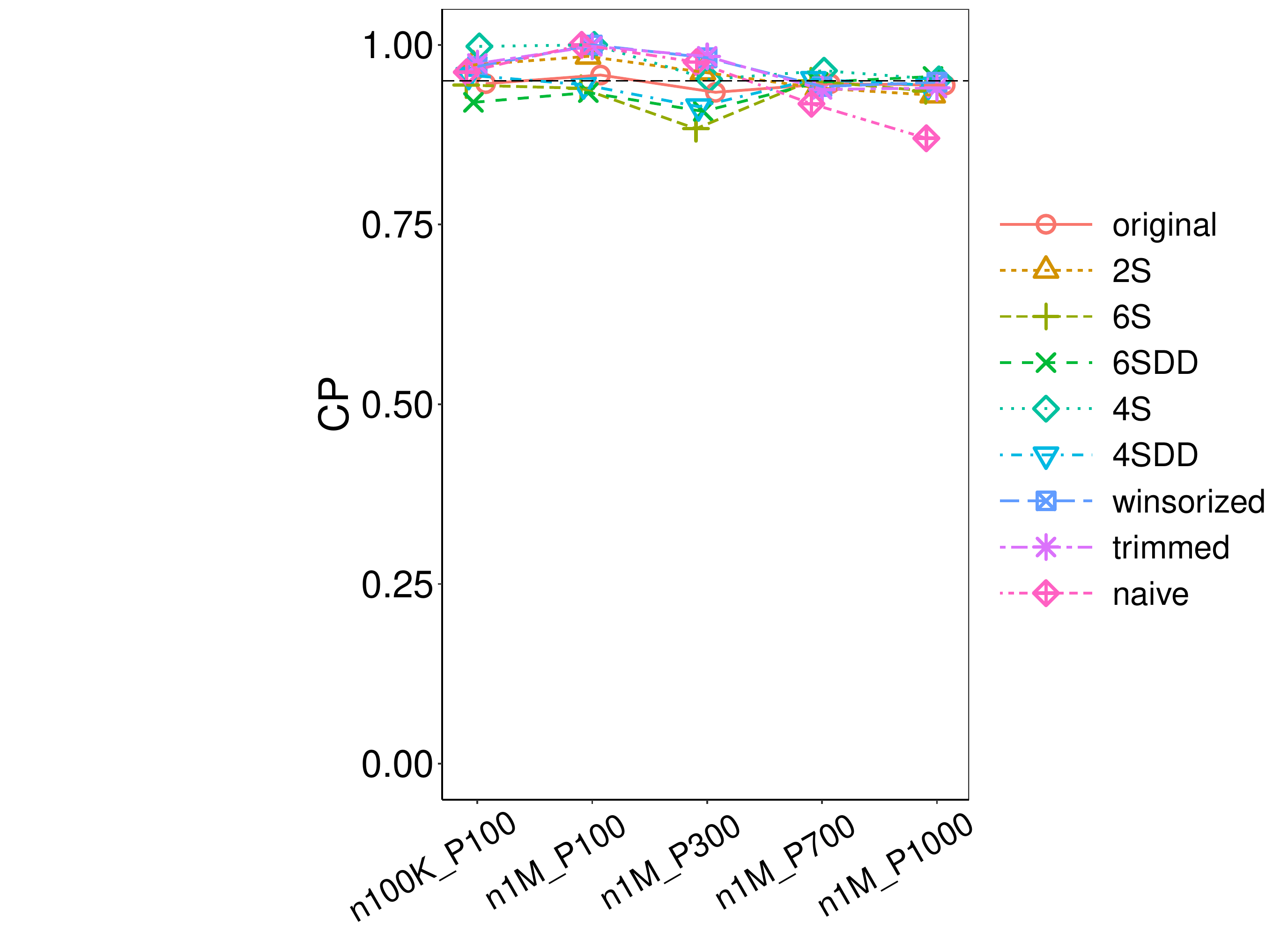}
\includegraphics[width=0.19\textwidth, trim={2.5in 0 2.6in 0},clip] {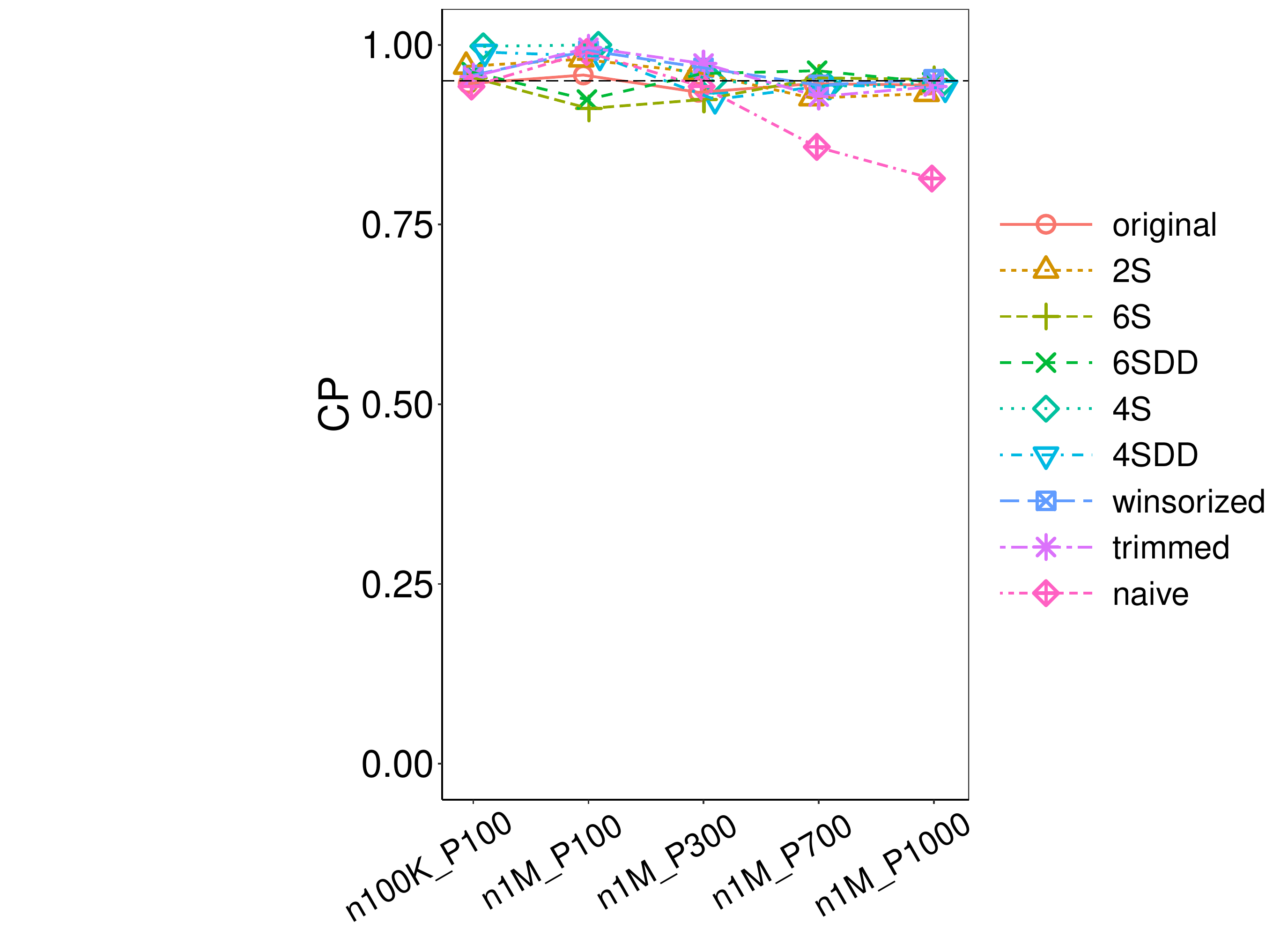}
\includegraphics[width=0.19\textwidth, trim={2.5in 0 2.6in 0},clip] {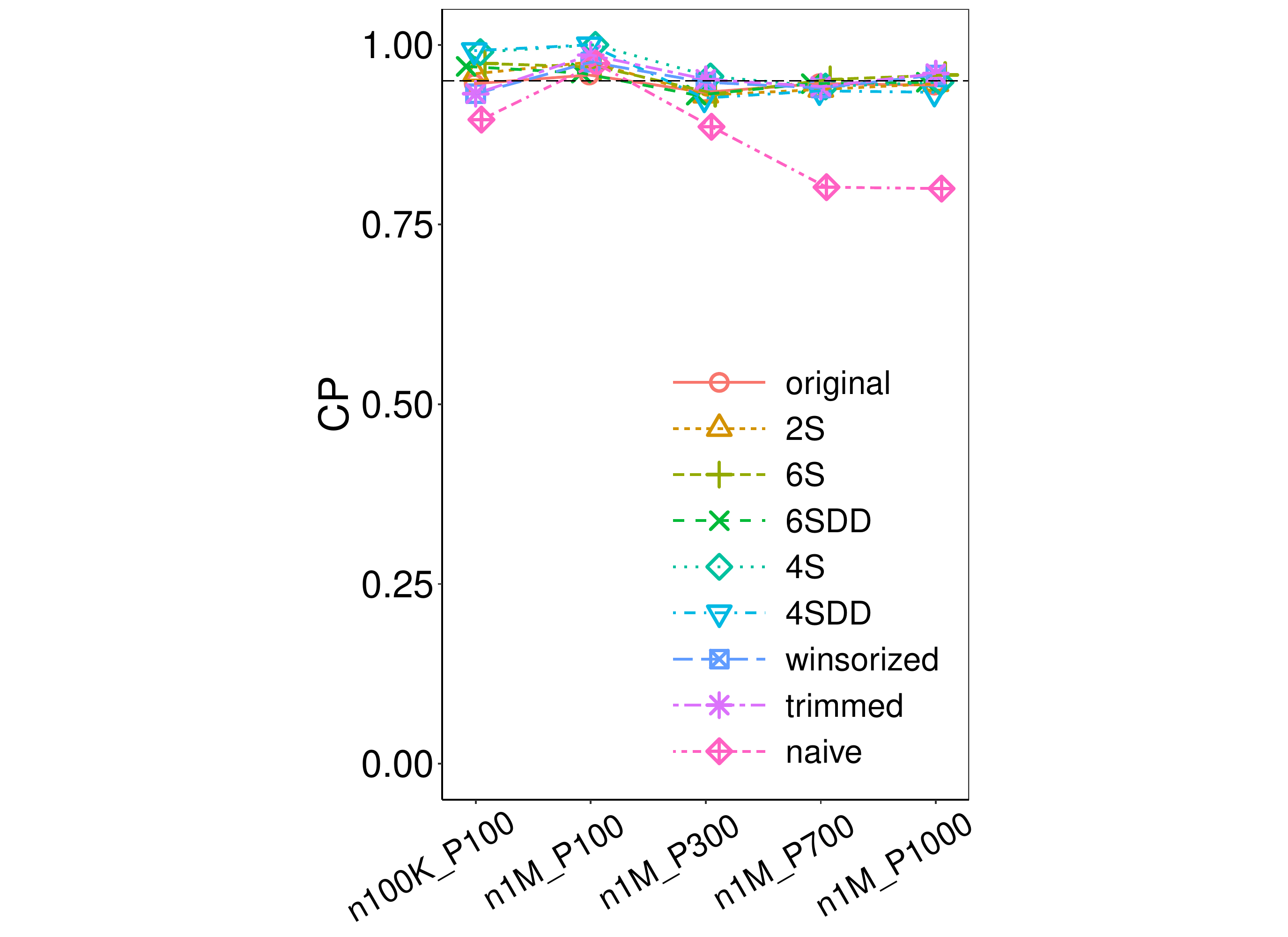}

\includegraphics[width=0.19\textwidth, trim={2.5in 0 2.6in 0},clip] {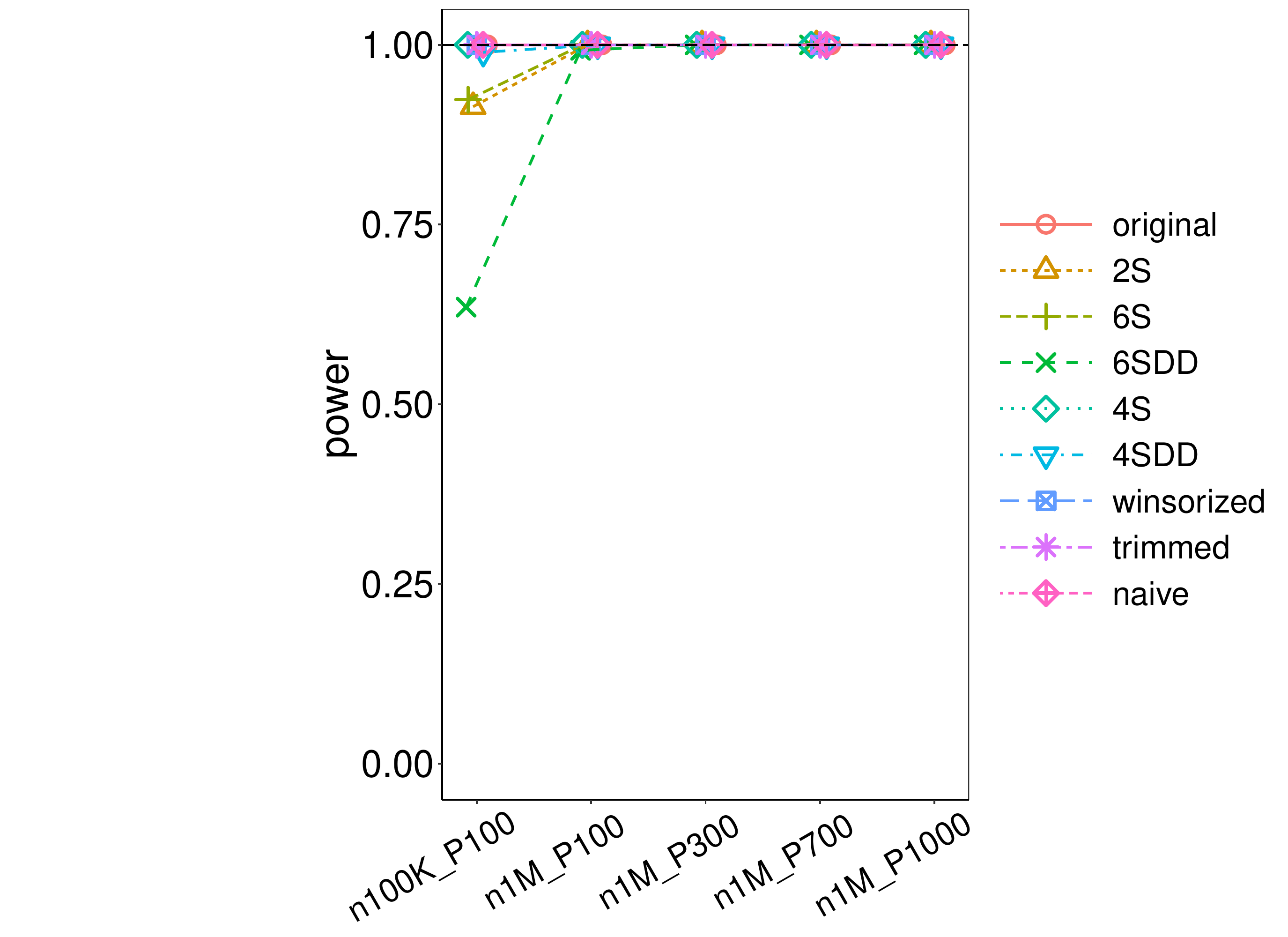}
\includegraphics[width=0.19\textwidth, trim={2.5in 0 2.6in 0},clip] {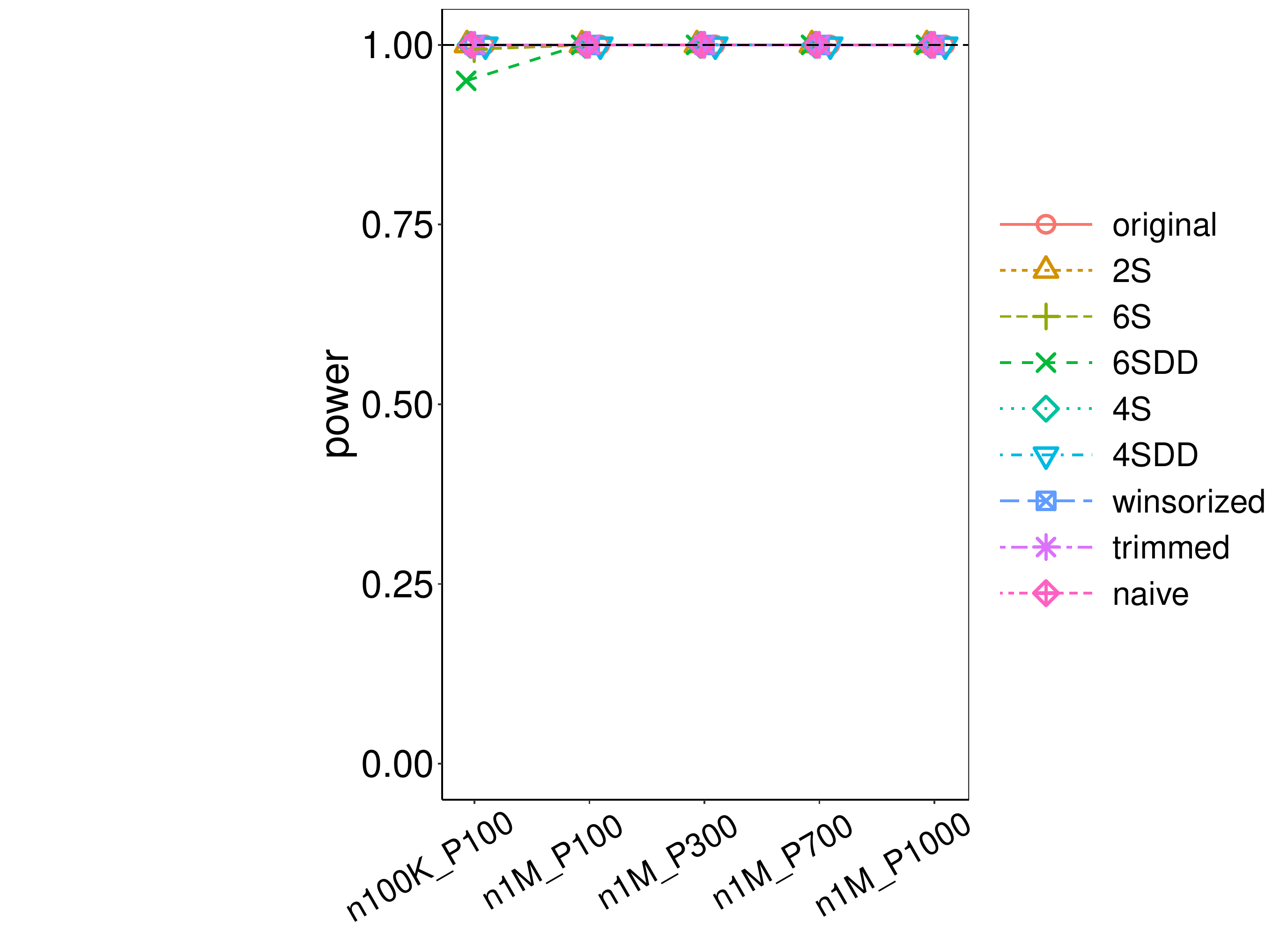}
\includegraphics[width=0.19\textwidth, trim={2.5in 0 2.6in 0},clip] {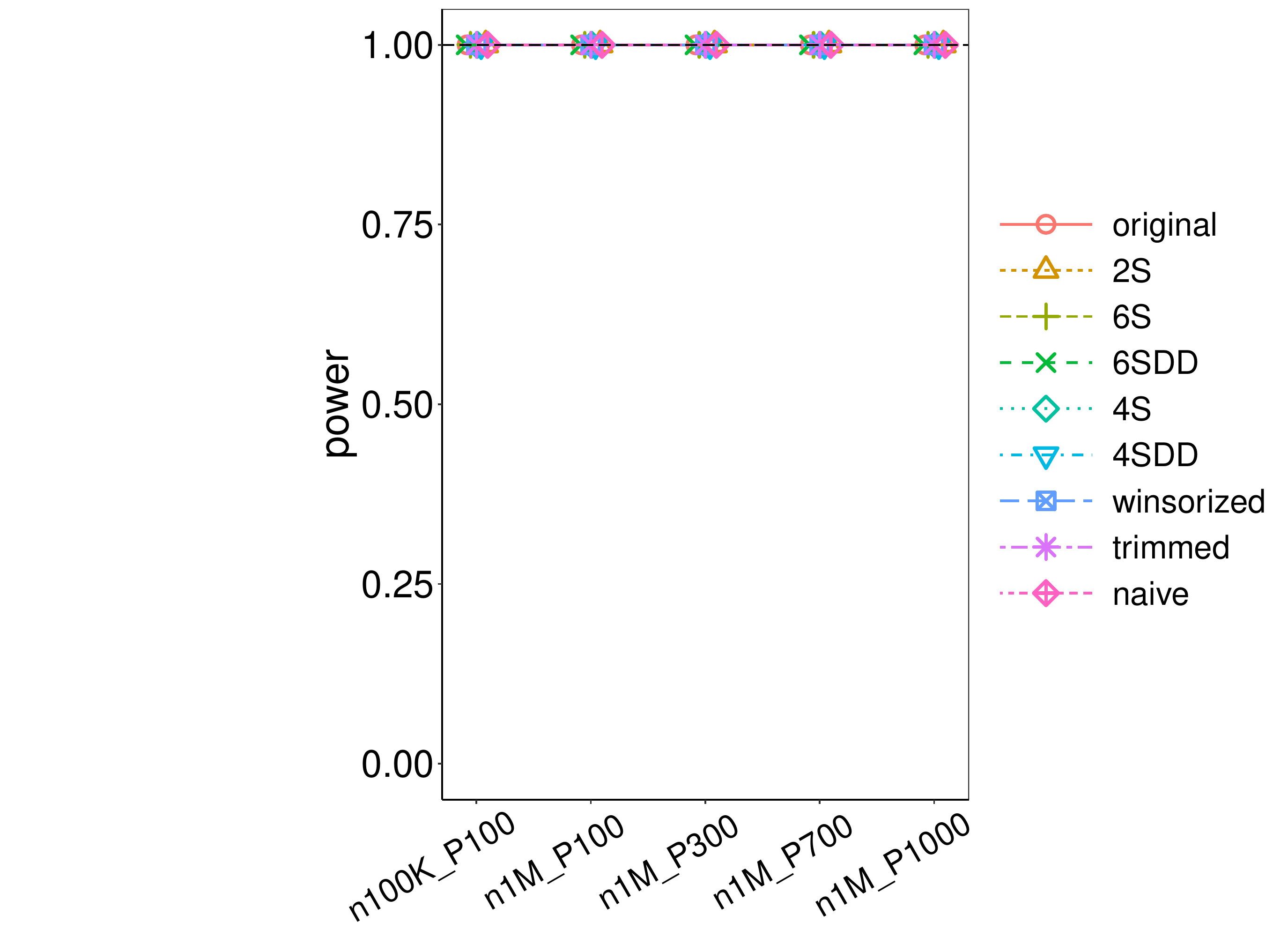}
\includegraphics[width=0.19\textwidth, trim={2.5in 0 2.6in 0},clip] {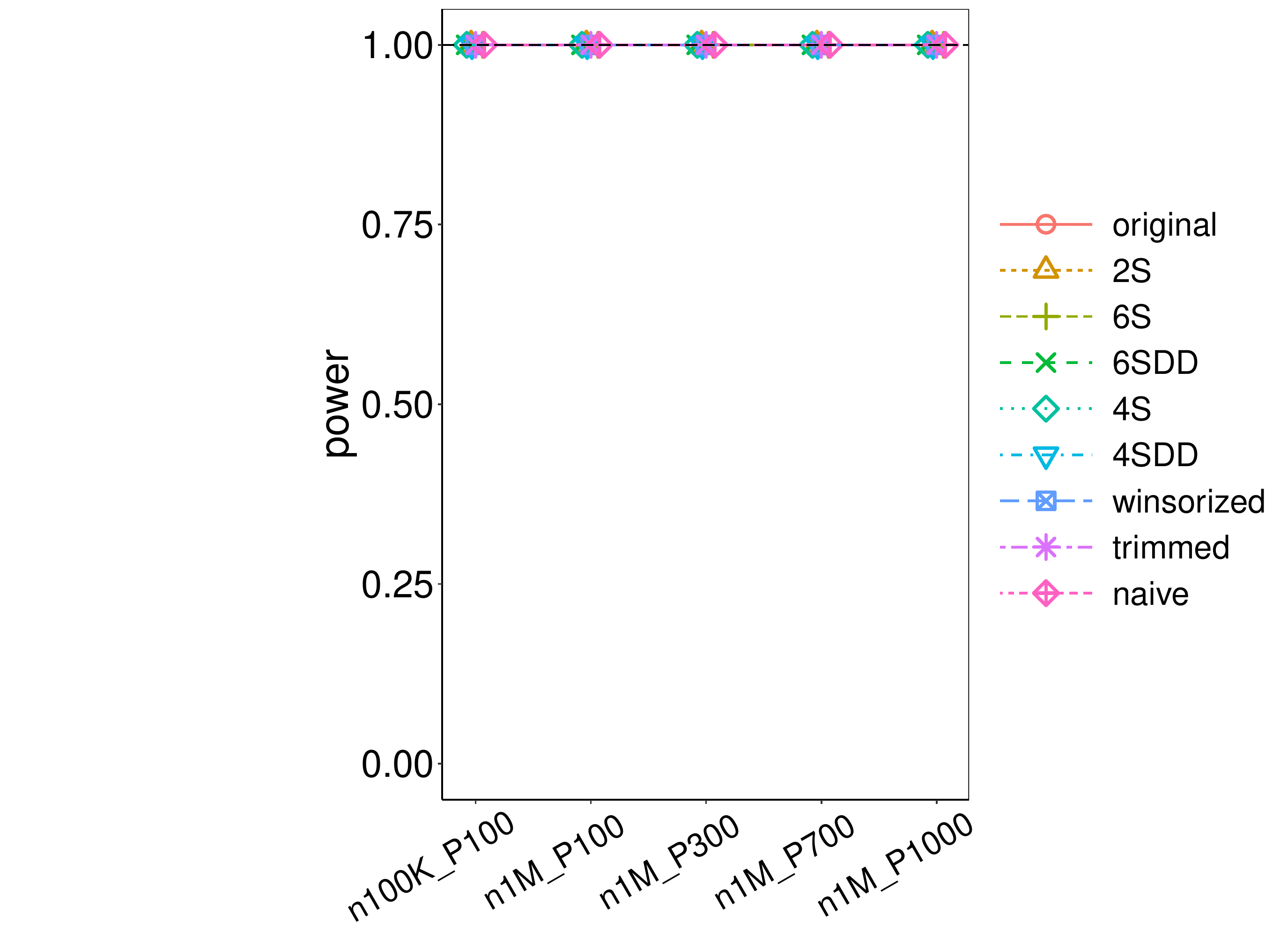}
\includegraphics[width=0.19\textwidth, trim={2.5in 0 2.6in 0},clip] {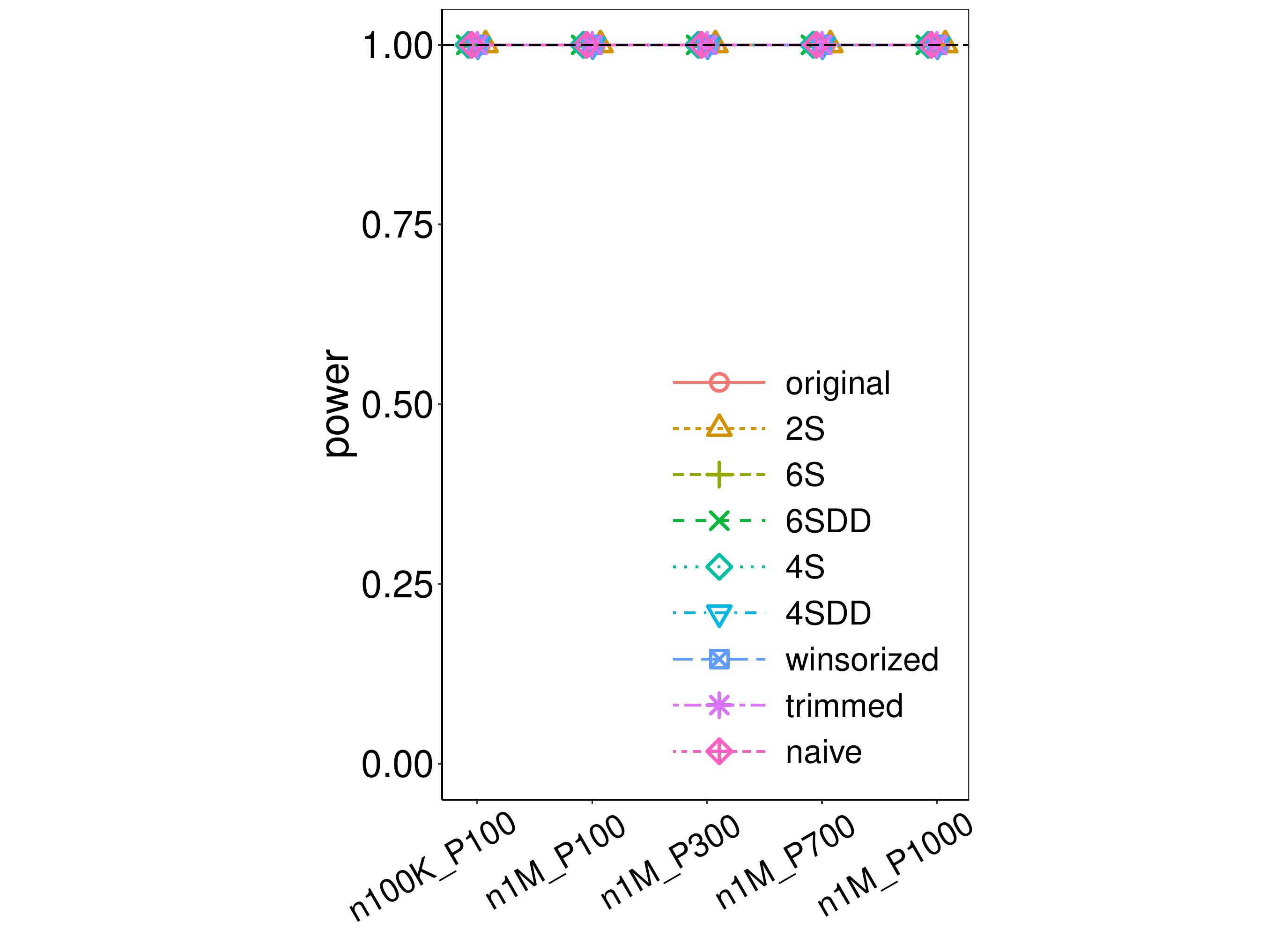}

\caption{Simulation results with $\rho$-zCDP for Gaussian data with  $\alpha=\beta$ when $\theta\ne0$}
\label{fig:1szCDPN}
\end{figure}

The results are presented in Figures \ref{fig:1sDPN} and \ref{fig:1szCDPN} and are summarized  as follows. The  winsorized mean is the best performer, all metrics considered. The relative performance of the other approaches depends on the evaluation metrics. For example, in terms of bias, trimmed mean, 2S, and the na\"{i}ve method are as close to 0 as winsorized mean, with 4S also performing well at $n=1m$; in terms of RMSE and CI width, 4S is similar to winsorized mean (the best group), followed by trimmed mean; regarding the CP, winsorized mean, 4S, trimmed mean and 2S provide nominal-level CP  all cases (slight over-coverage at $n=100k/P=100$, $n=1m/P=100$ and $n=1m/P=300$ for $\epsilon\le1$), but na\"{i}ve method can exhibit under-coverage at $n=1m/P=700$ and $n=1m/P=1000$ for $\epsilon\ge5$, and 4SDD produces under-coverage for almost all cases; all methods except for 6SDD at $n=100K/P=100$ can provide power close to 1. The Gaussian mechanism of $\rho$-zCDP does not seem to be advantageous in statistical utility compared to the Laplace mechanism of $\epsilon$-DP.

\subsubsection{\texorpdfstring{$Y\sim$ Gaussian}{}, \texorpdfstring{$\theta=0$ ($H_0$)}{H0 }, and \texorpdfstring{$\alpha\ne\beta$}{AS}}\label{sec:0asN}
\begin{figure}[!htb]
\hspace{0.5in}$\epsilon=0.5$\hspace{0.8in}$\epsilon=1$\hspace{0.9in}$\epsilon=2$
\hspace{0.95in}$\epsilon=5$\hspace{0.9in}$\epsilon=50$

\includegraphics[width=0.19\textwidth, trim={2.5in 0 2.6in 0},clip] {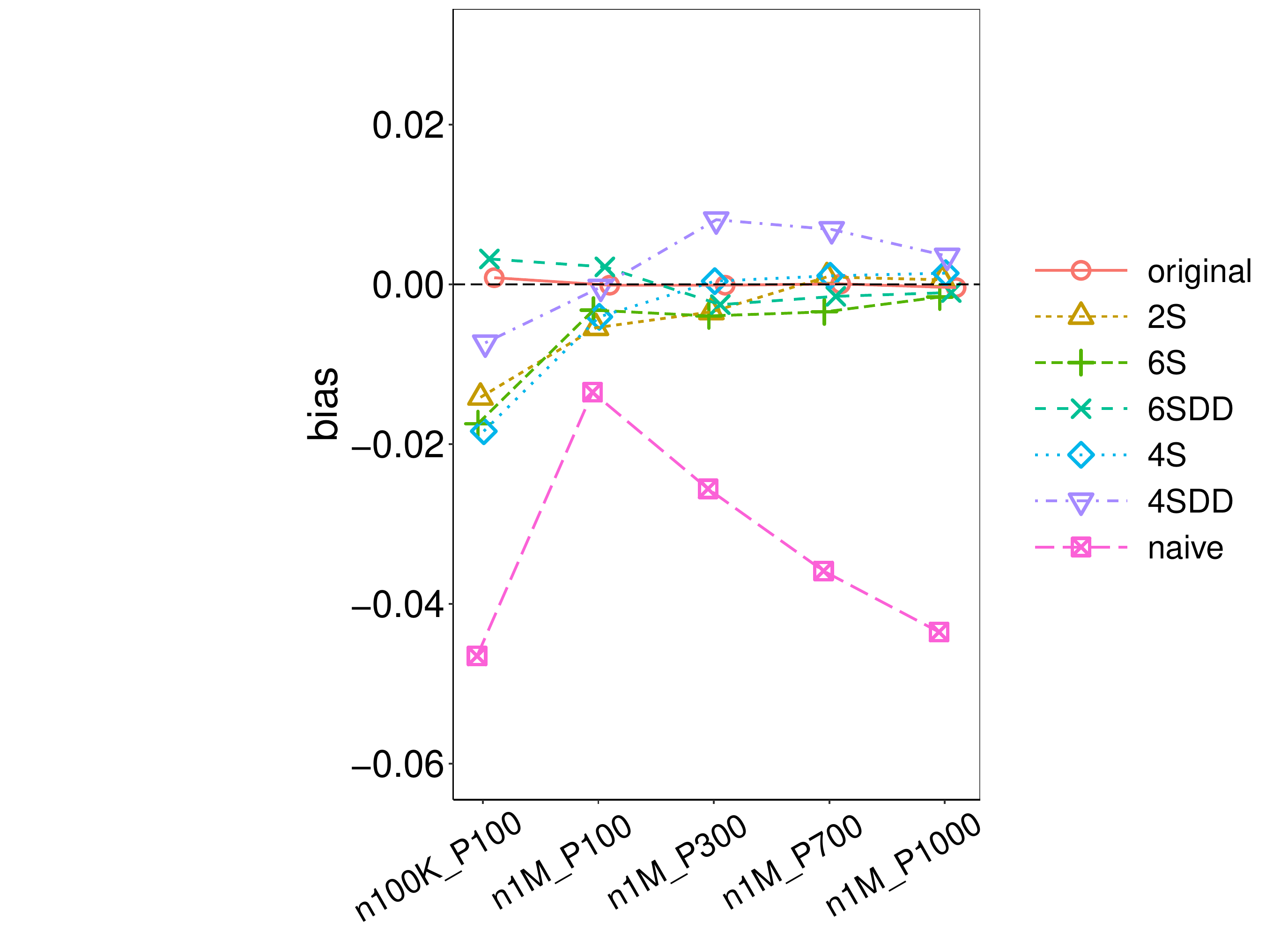}
\includegraphics[width=0.19\textwidth, trim={2.5in 0 2.6in 0},clip] {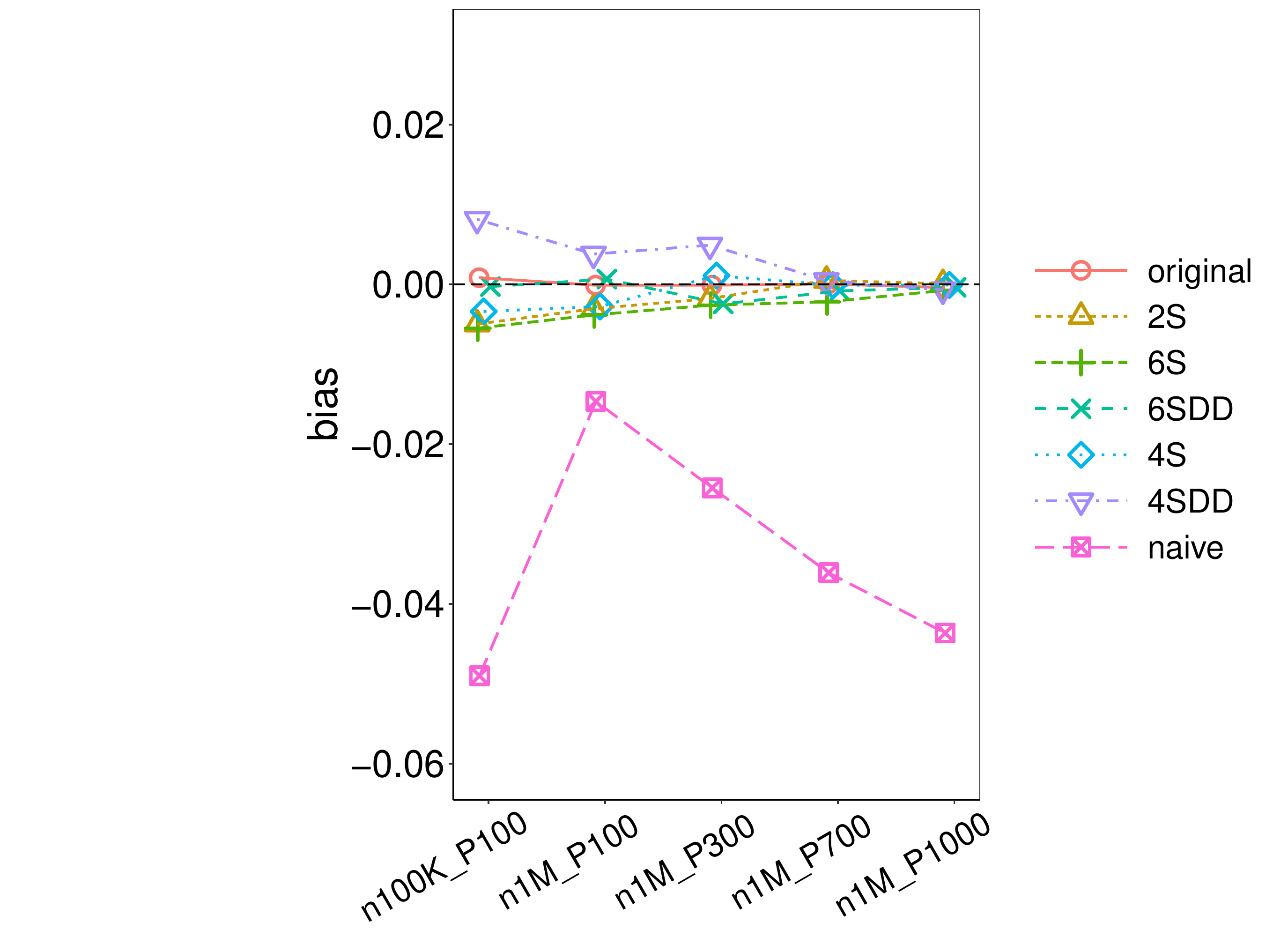}
\includegraphics[width=0.19\textwidth, trim={2.5in 0 2.6in 0},clip] {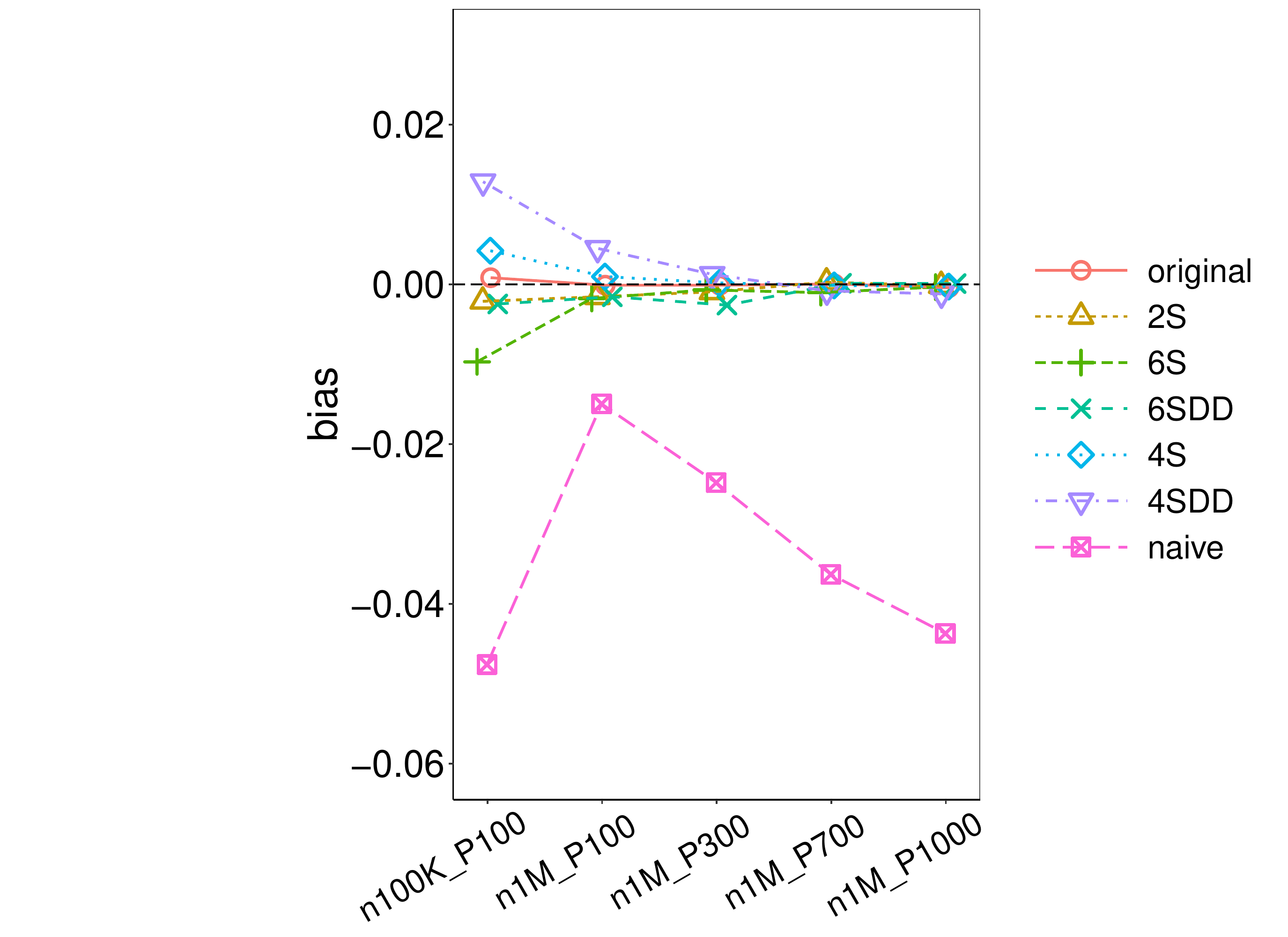}
\includegraphics[width=0.19\textwidth, trim={2.5in 0 2.6in 0},clip] {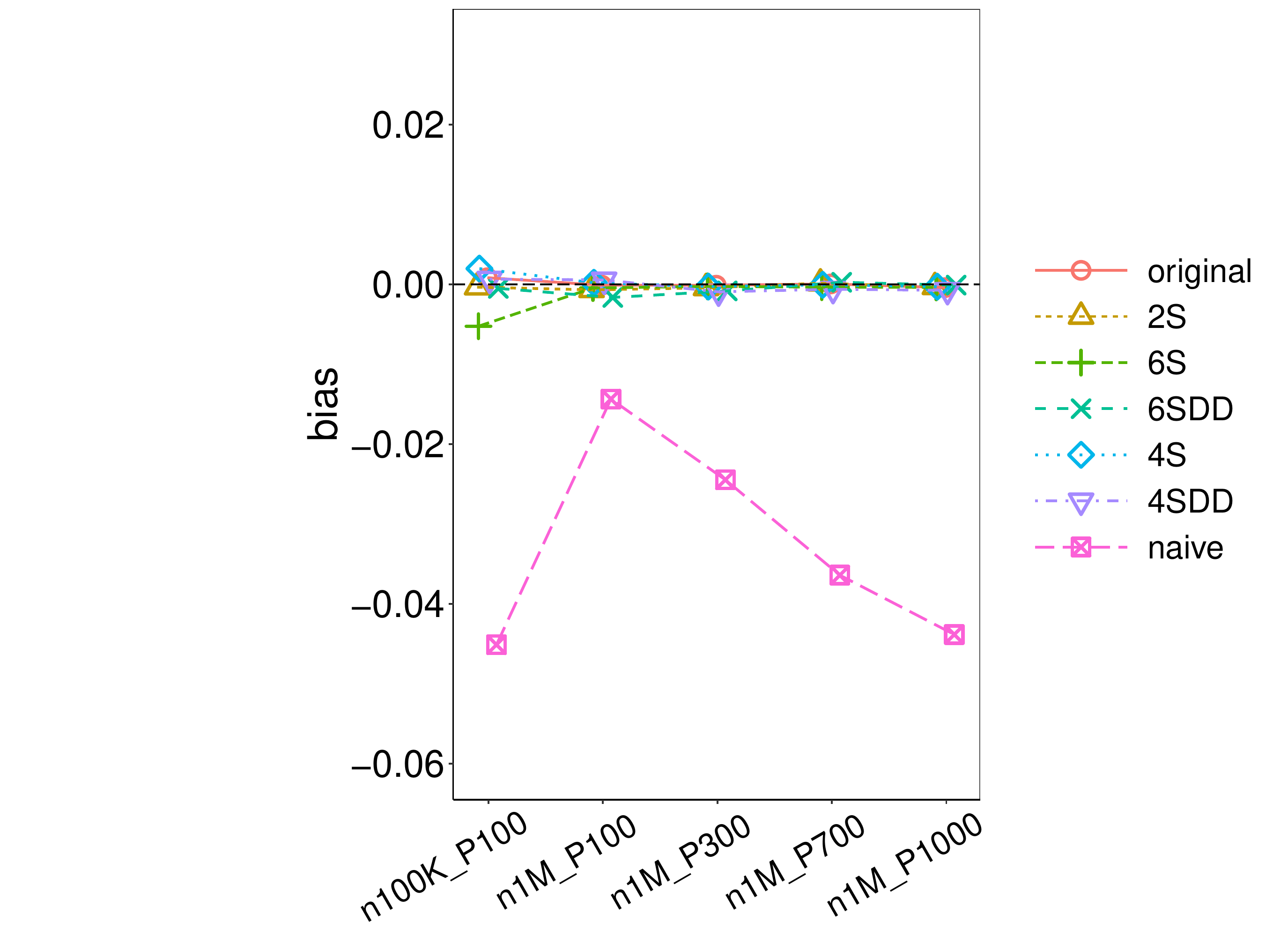}
\includegraphics[width=0.19\textwidth, trim={2.5in 0 2.6in 0},clip] {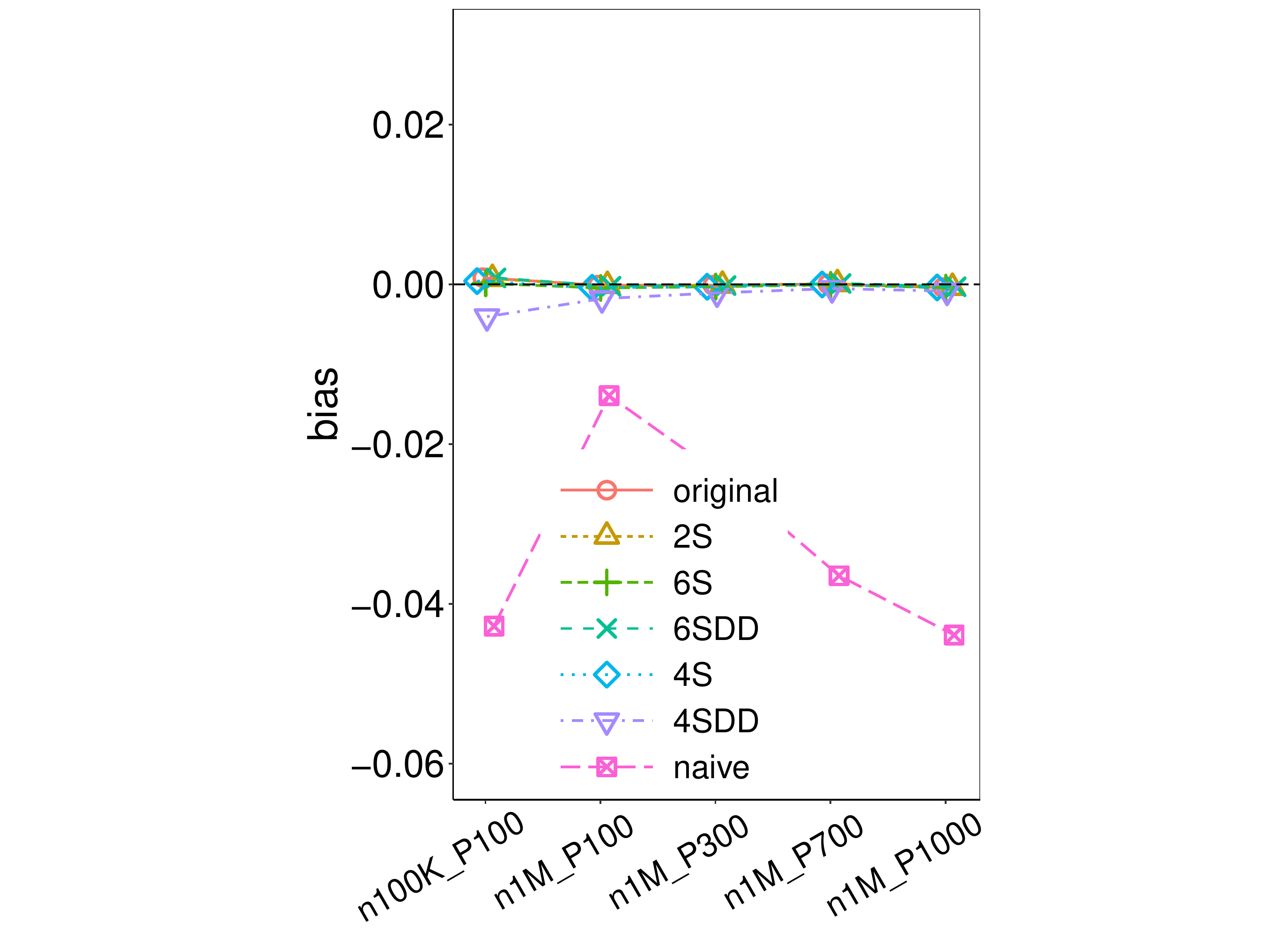}

\includegraphics[width=0.19\textwidth, trim={2.5in 0 2.6in 0},clip] {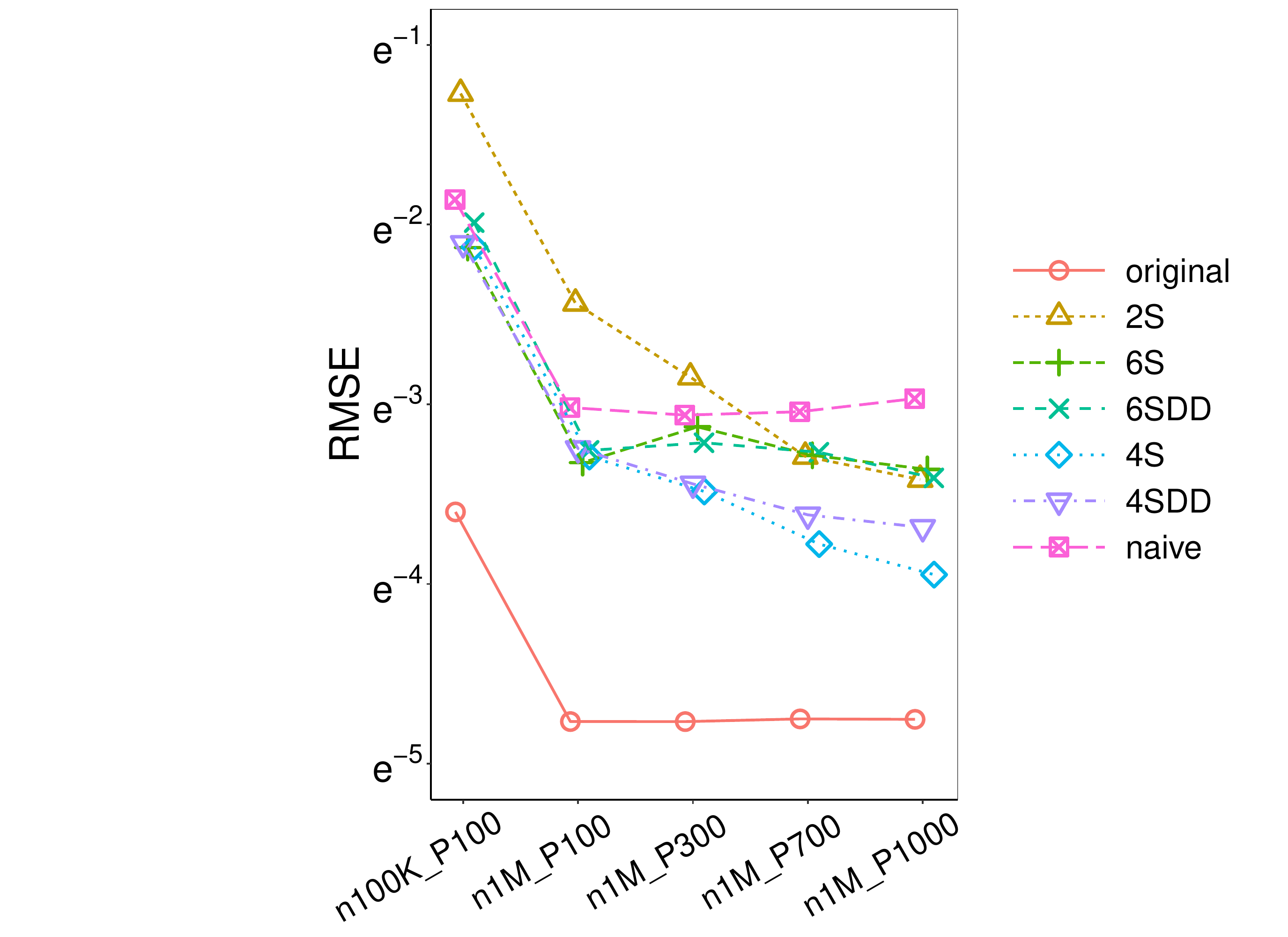}
\includegraphics[width=0.19\textwidth, trim={2.5in 0 2.6in 0},clip] {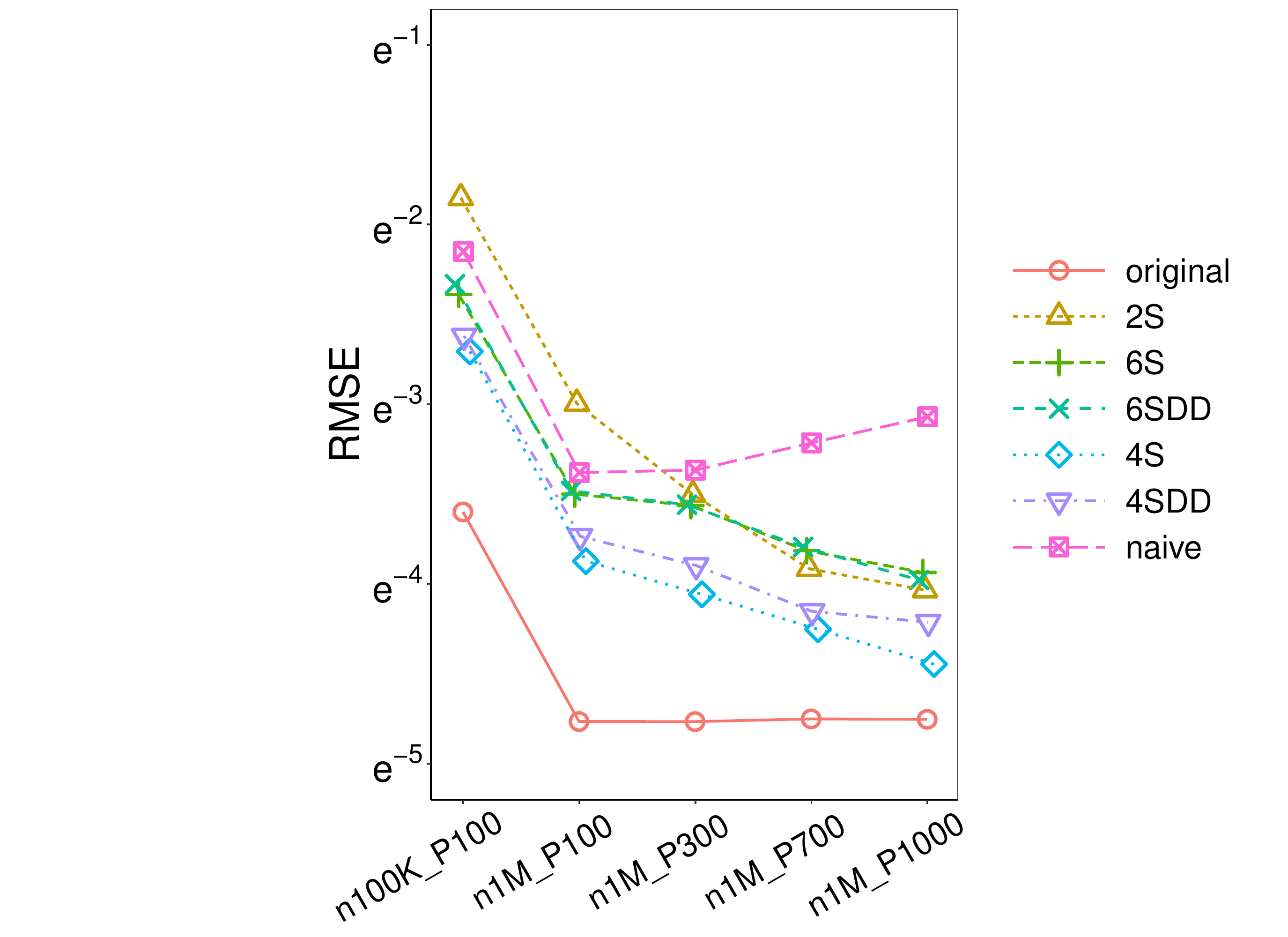}
\includegraphics[width=0.19\textwidth, trim={2.5in 0 2.6in 0},clip] {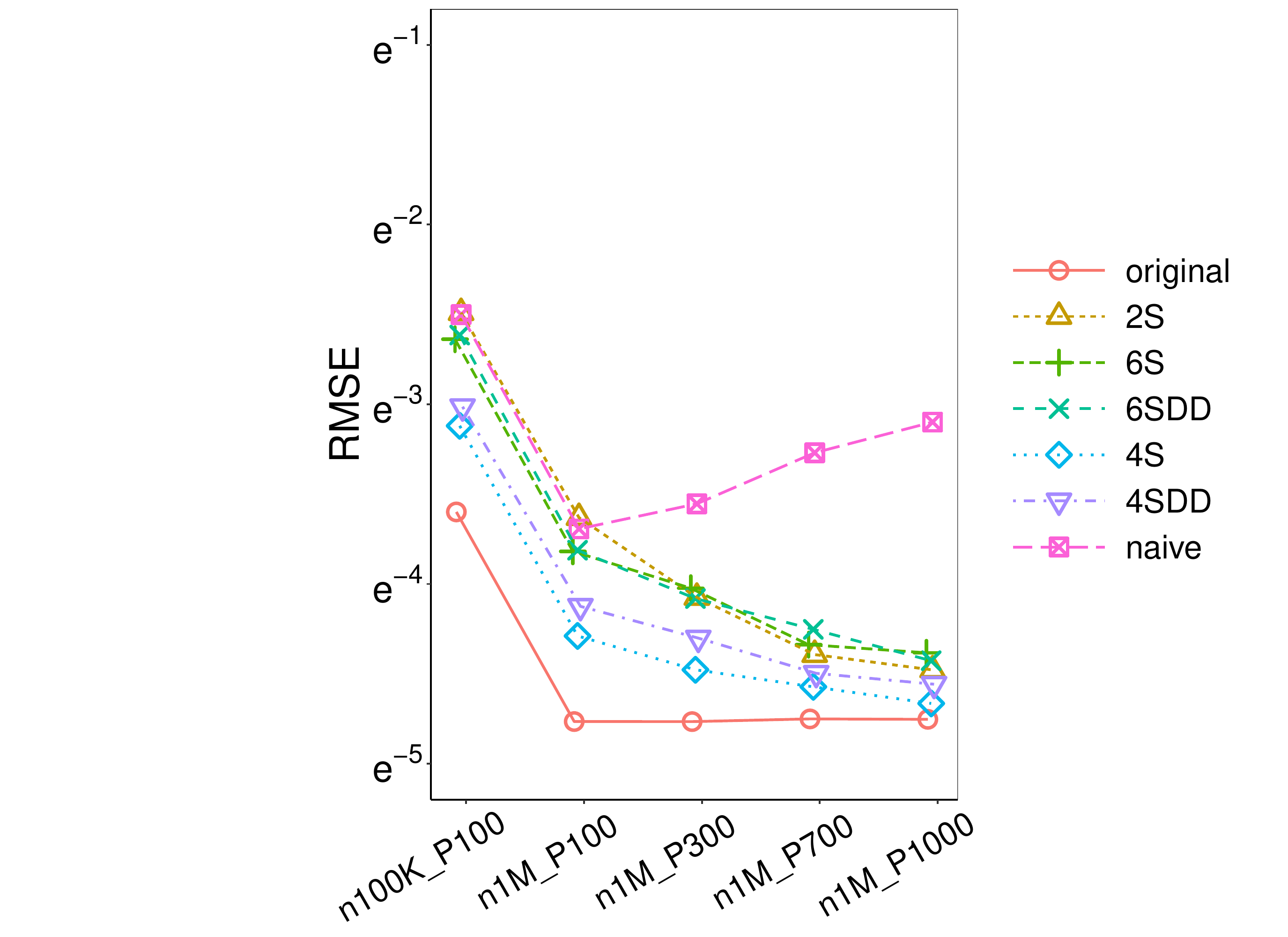}
\includegraphics[width=0.19\textwidth, trim={2.5in 0 2.6in 0},clip] {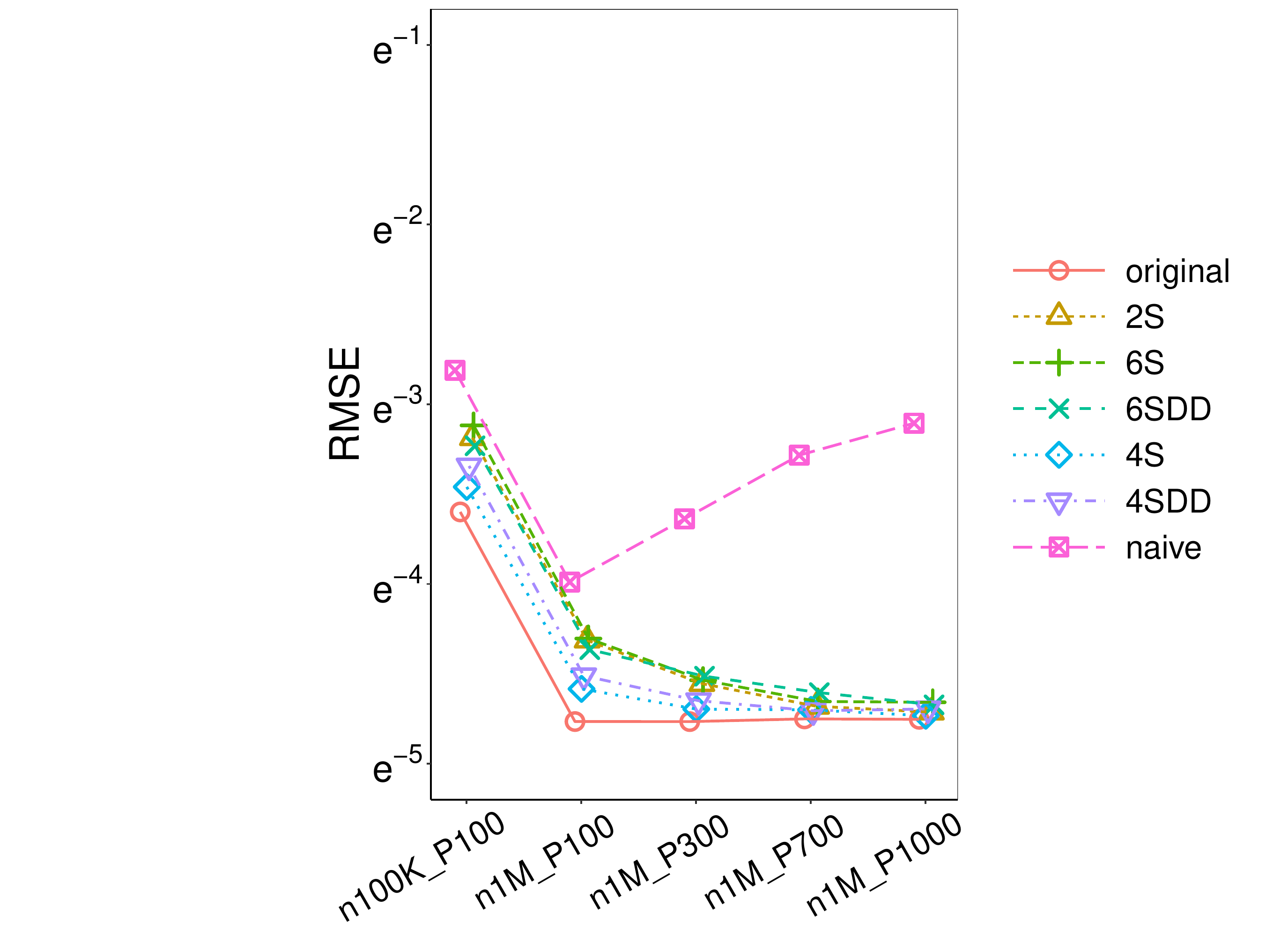}
\includegraphics[width=0.19\textwidth, trim={2.5in 0 2.6in 0},clip] {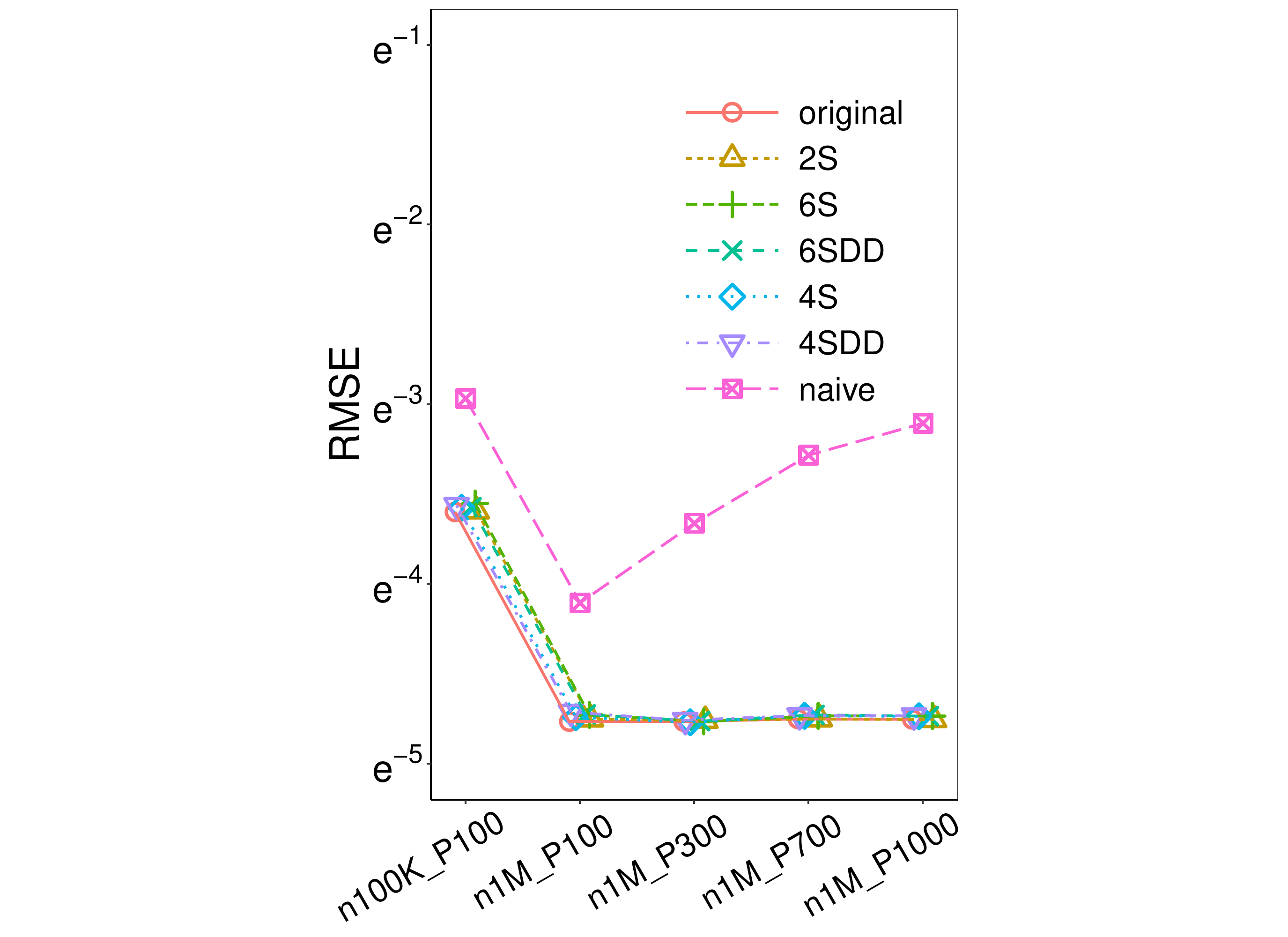}

\includegraphics[width=0.19\textwidth, trim={2.5in 0 2.6in 0},clip] {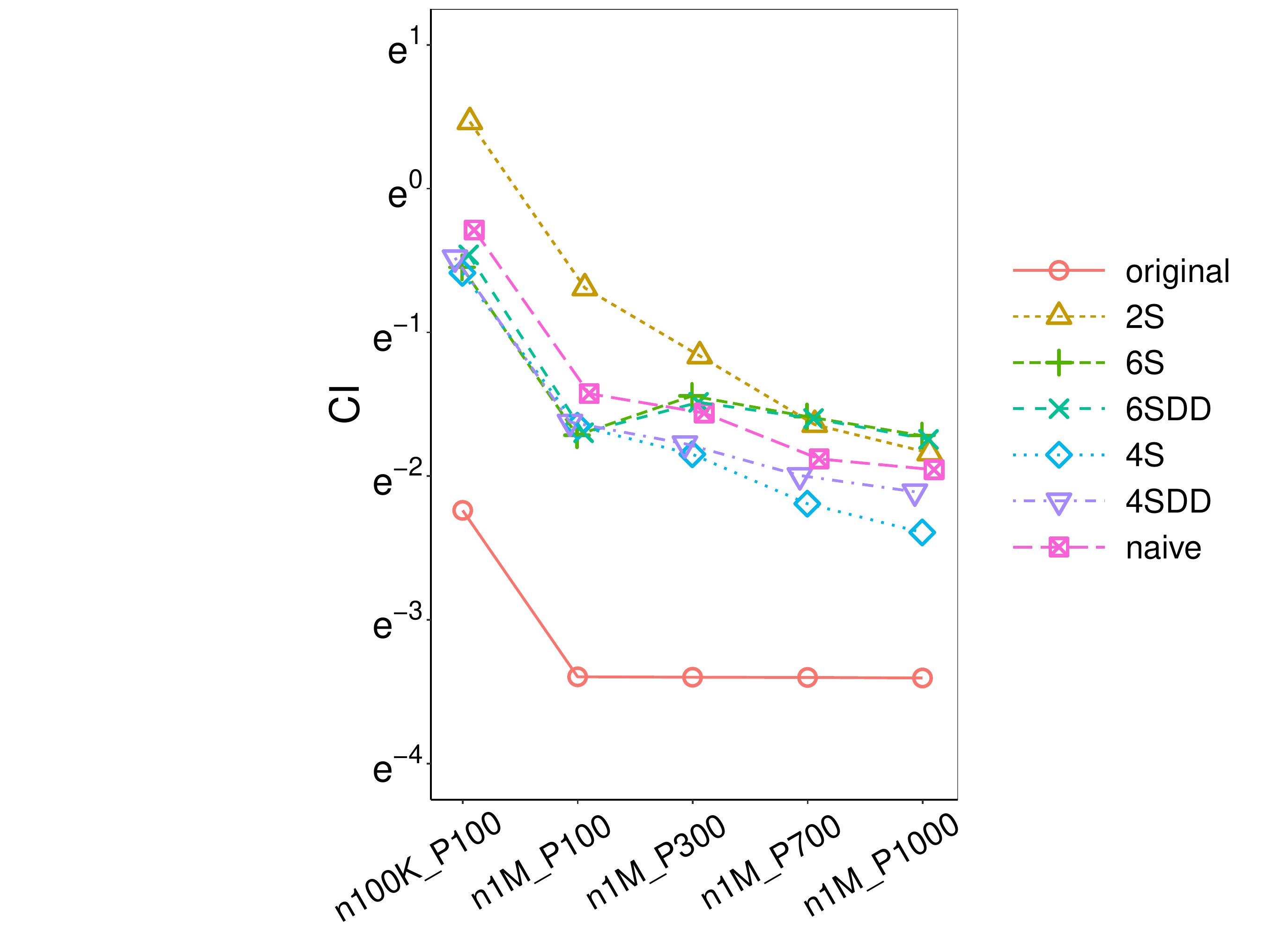}
\includegraphics[width=0.19\textwidth, trim={2.5in 0 2.6in 0},clip] {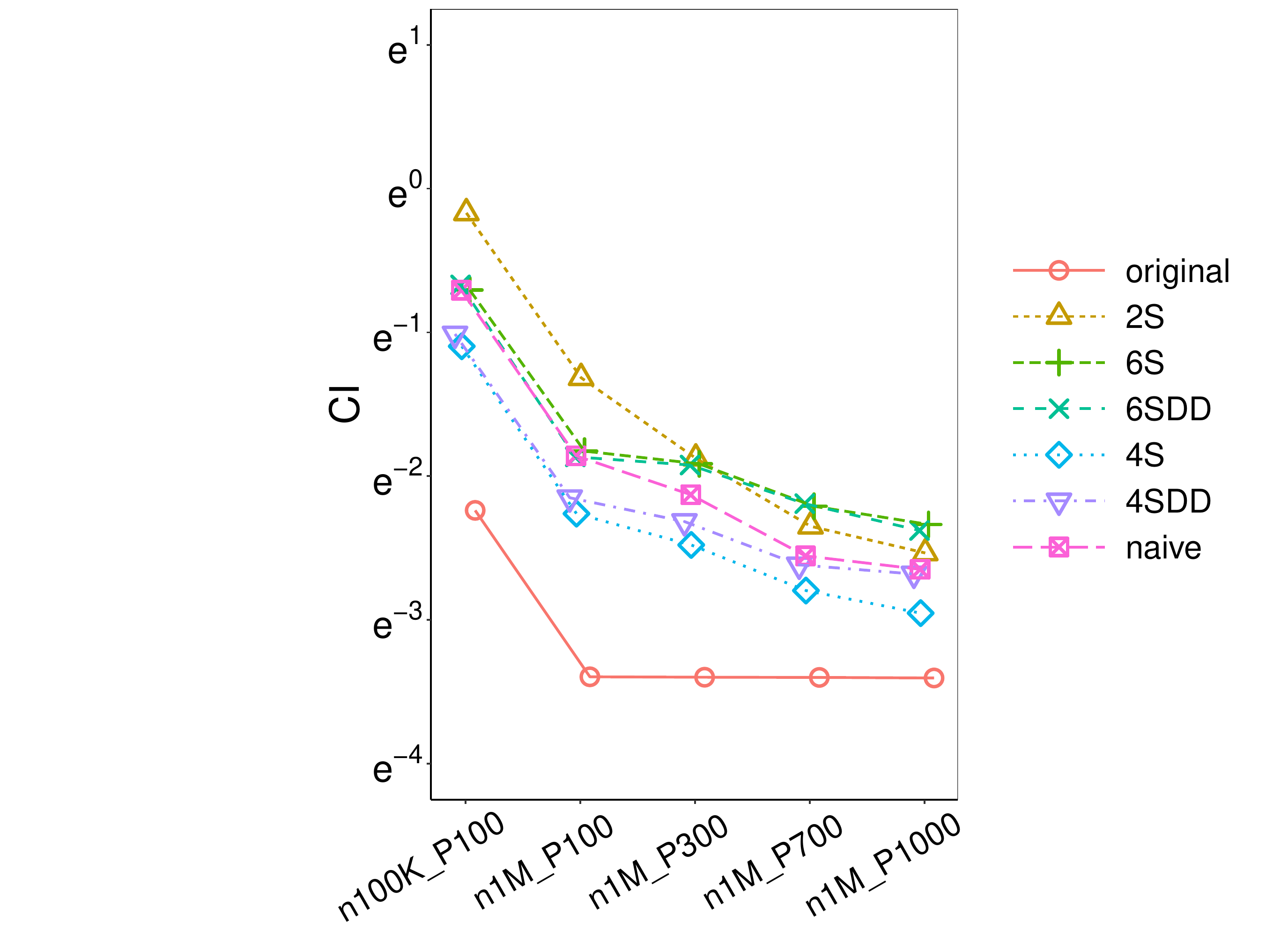}
\includegraphics[width=0.19\textwidth, trim={2.5in 0 2.6in 0},clip] {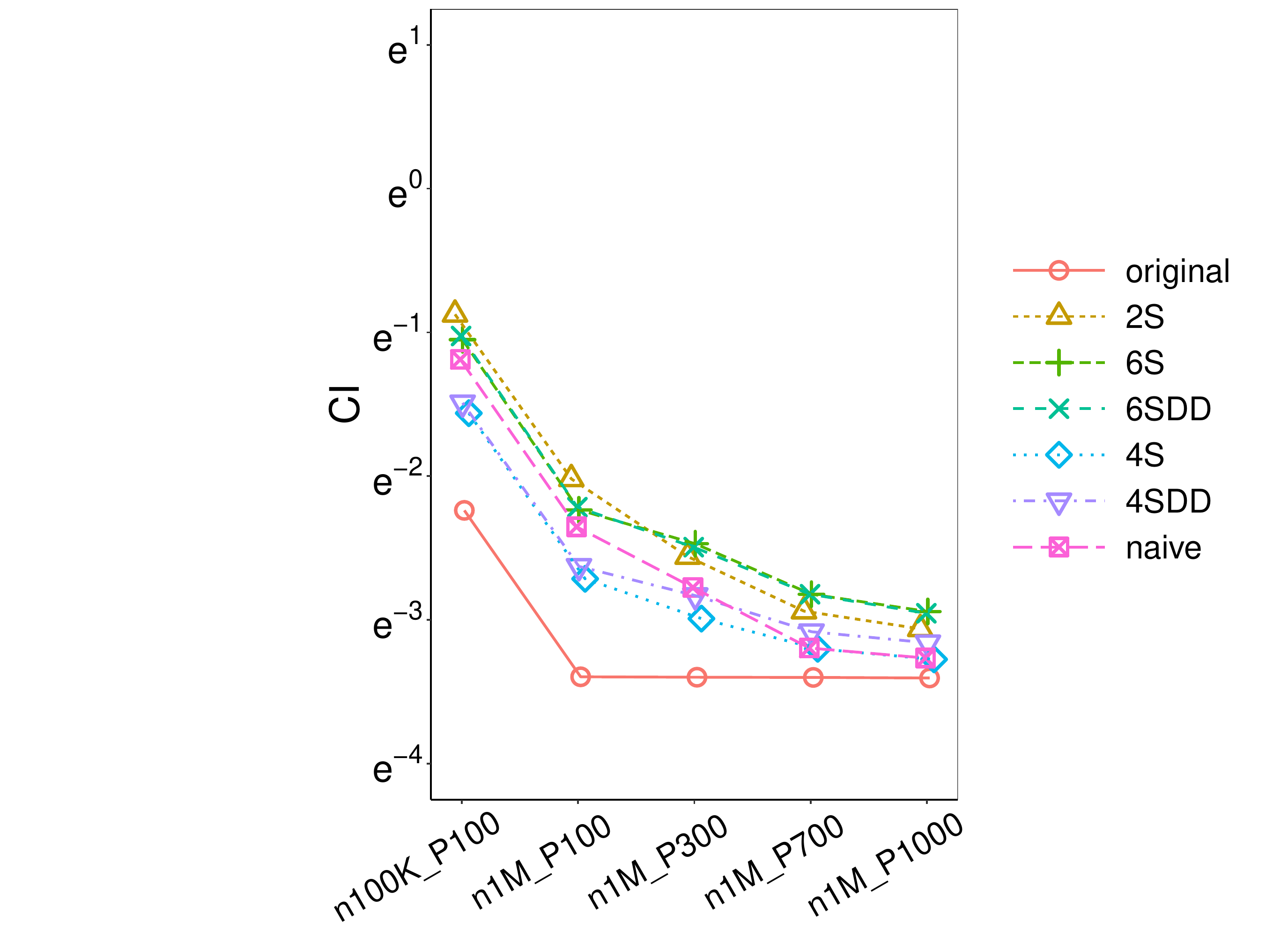}
\includegraphics[width=0.19\textwidth, trim={2.5in 0 2.6in 0},clip] {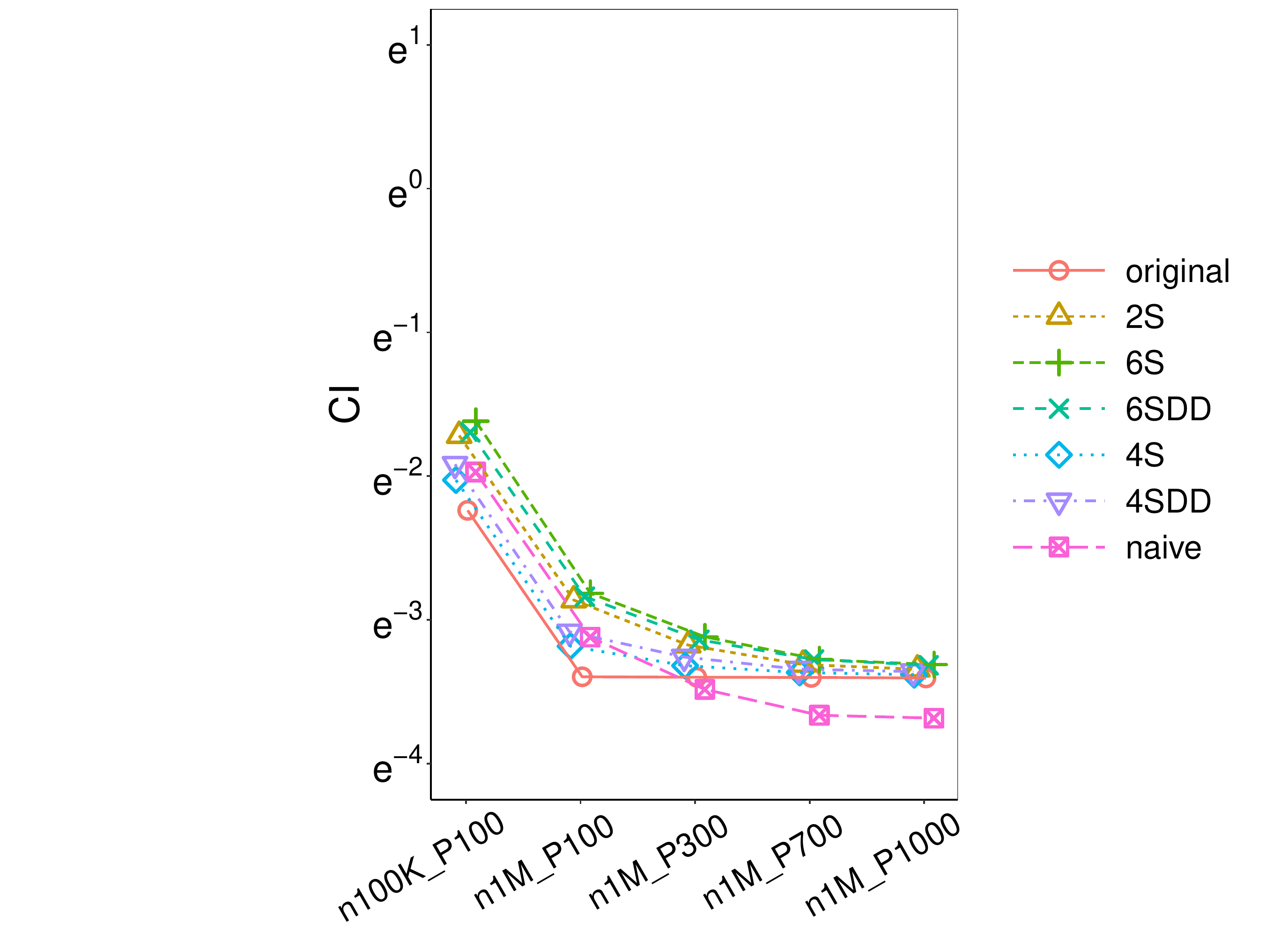}
\includegraphics[width=0.19\textwidth, trim={2.5in 0 2.6in 0},clip] {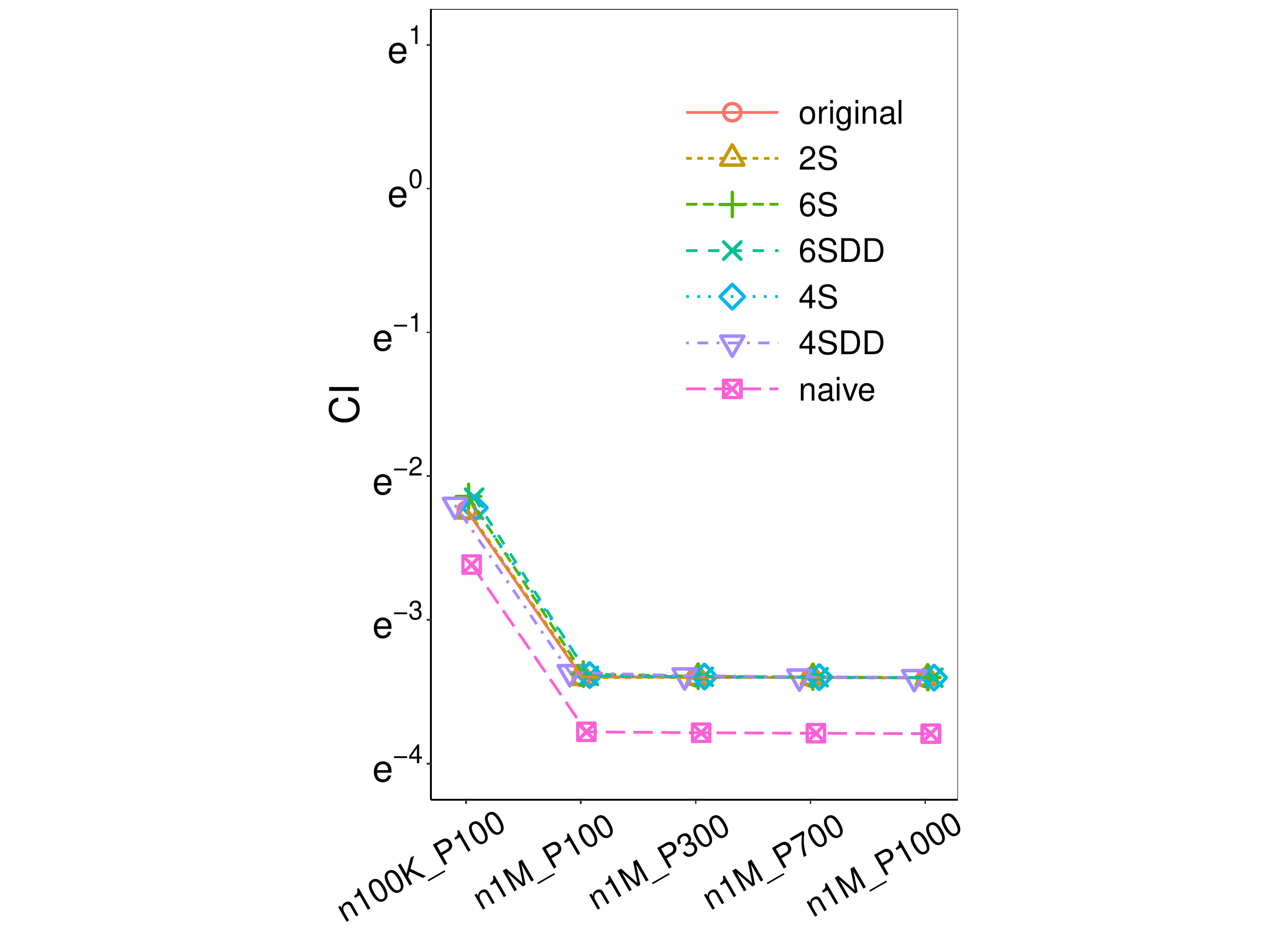}

\includegraphics[width=0.19\textwidth, trim={2.5in 0 2.6in 0},clip] {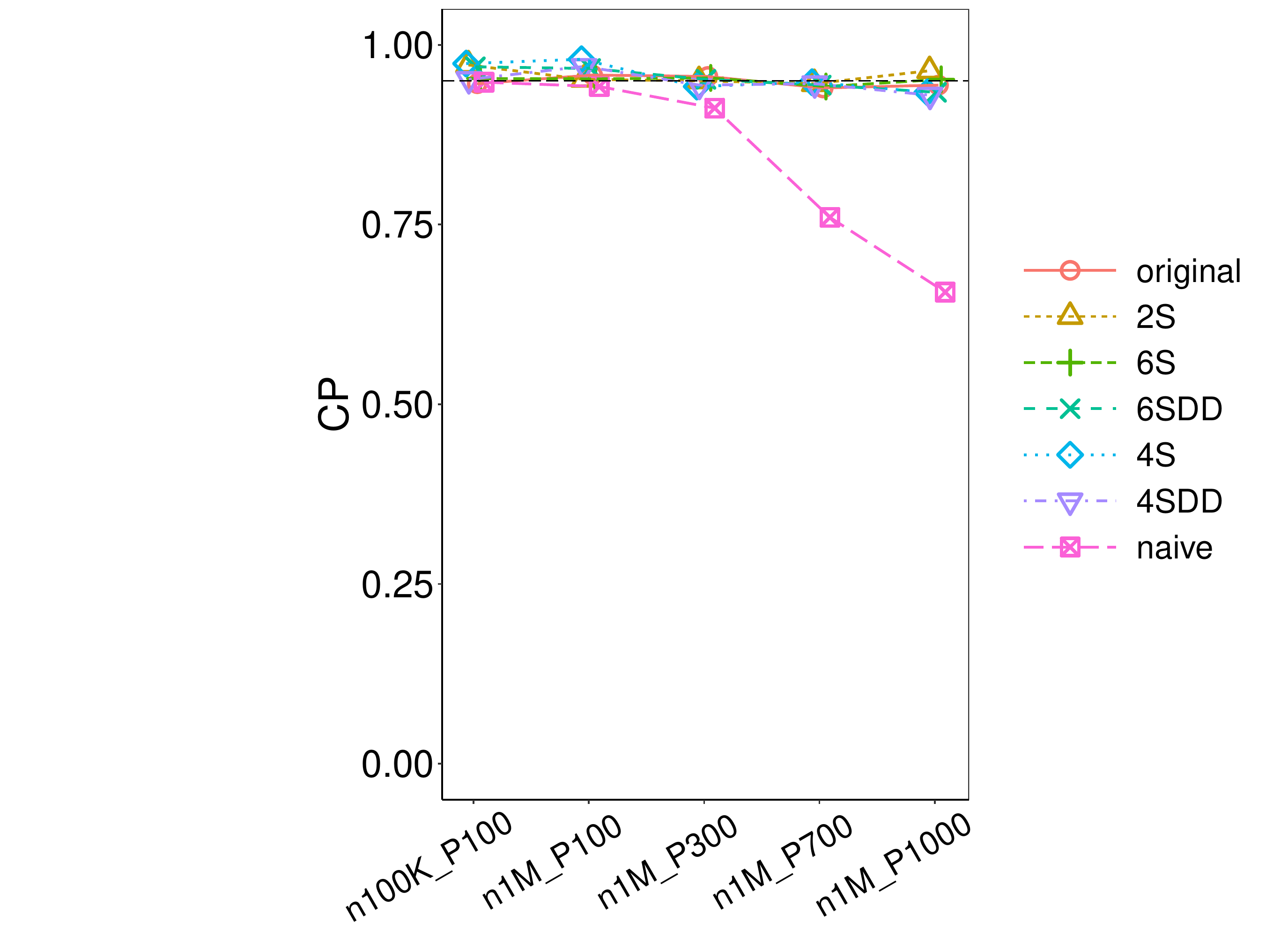}
\includegraphics[width=0.19\textwidth, trim={2.5in 0 2.6in 0},clip] {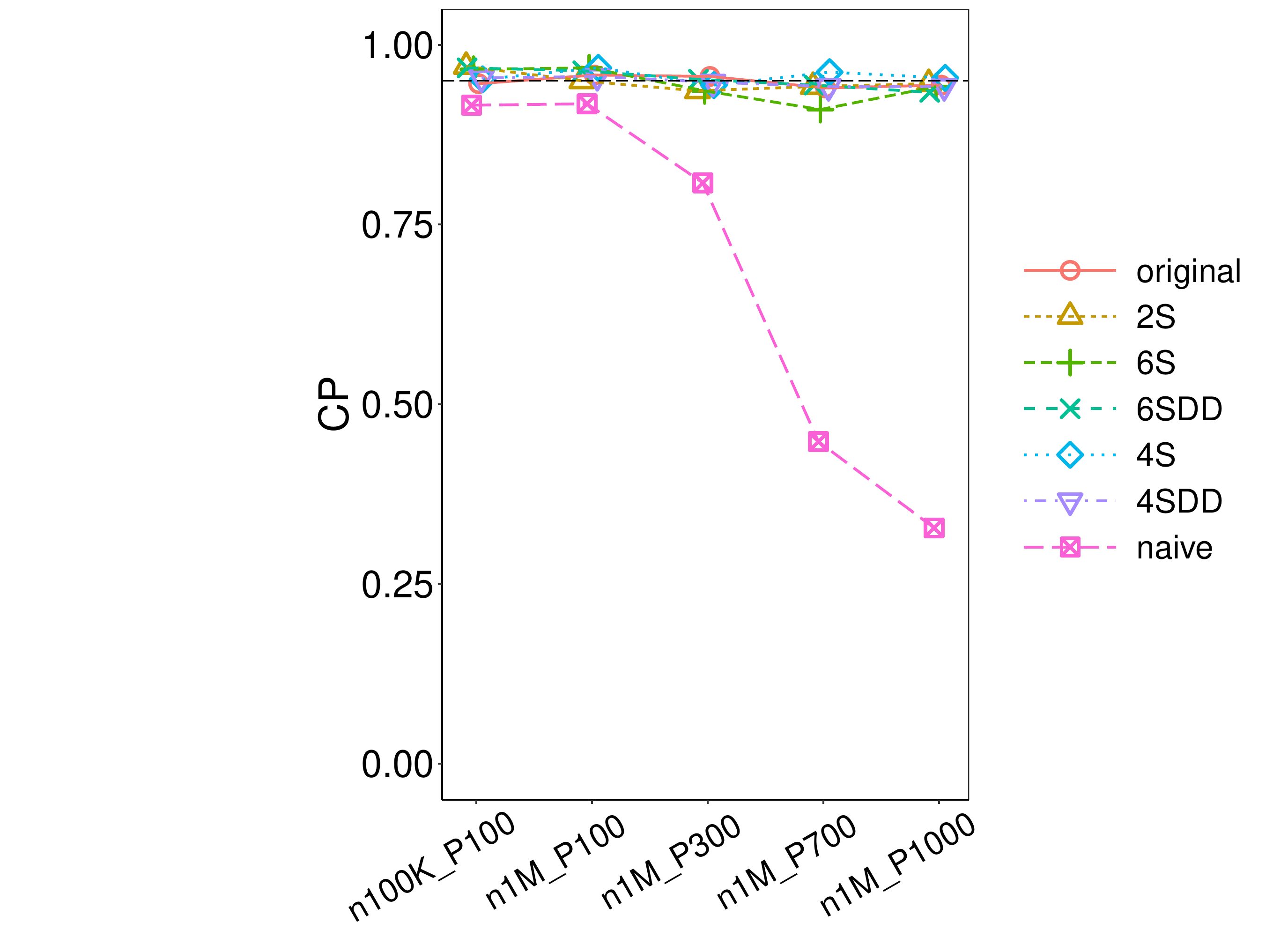}
\includegraphics[width=0.19\textwidth, trim={2.5in 0 2.6in 0},clip] {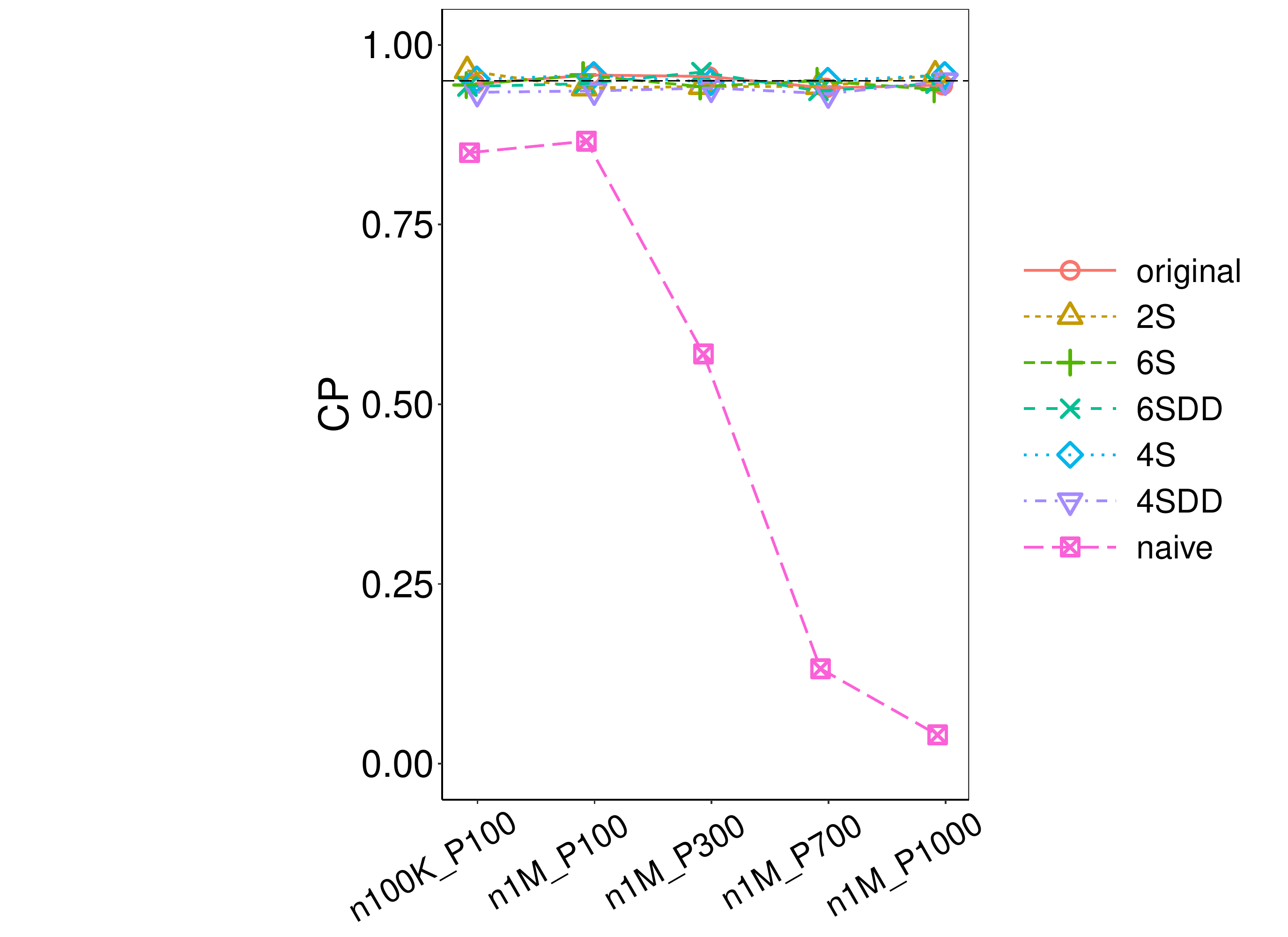}
\includegraphics[width=0.19\textwidth, trim={2.5in 0 2.6in 0},clip] {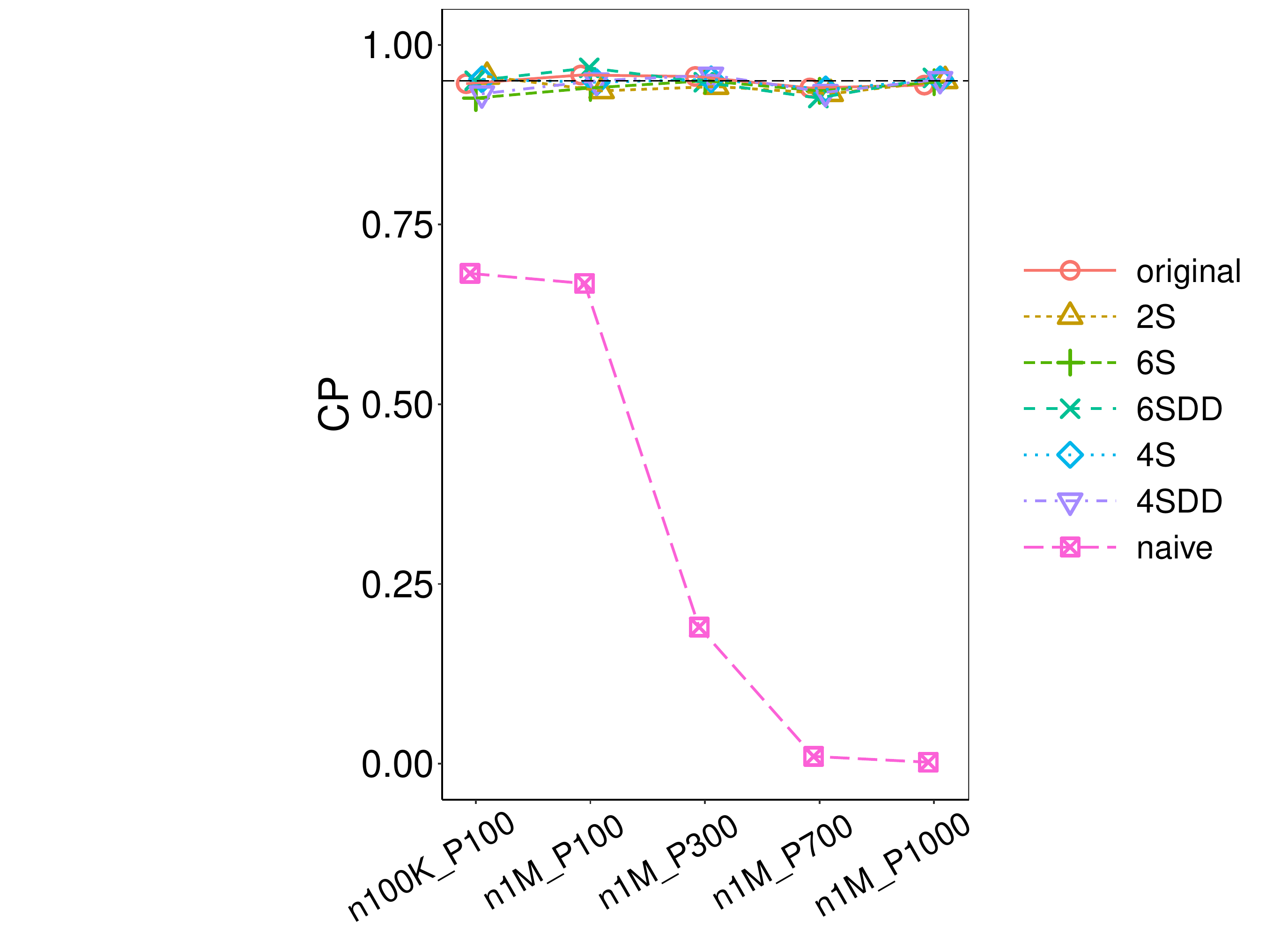}
\includegraphics[width=0.19\textwidth, trim={2.5in 0 2.6in 0},clip] {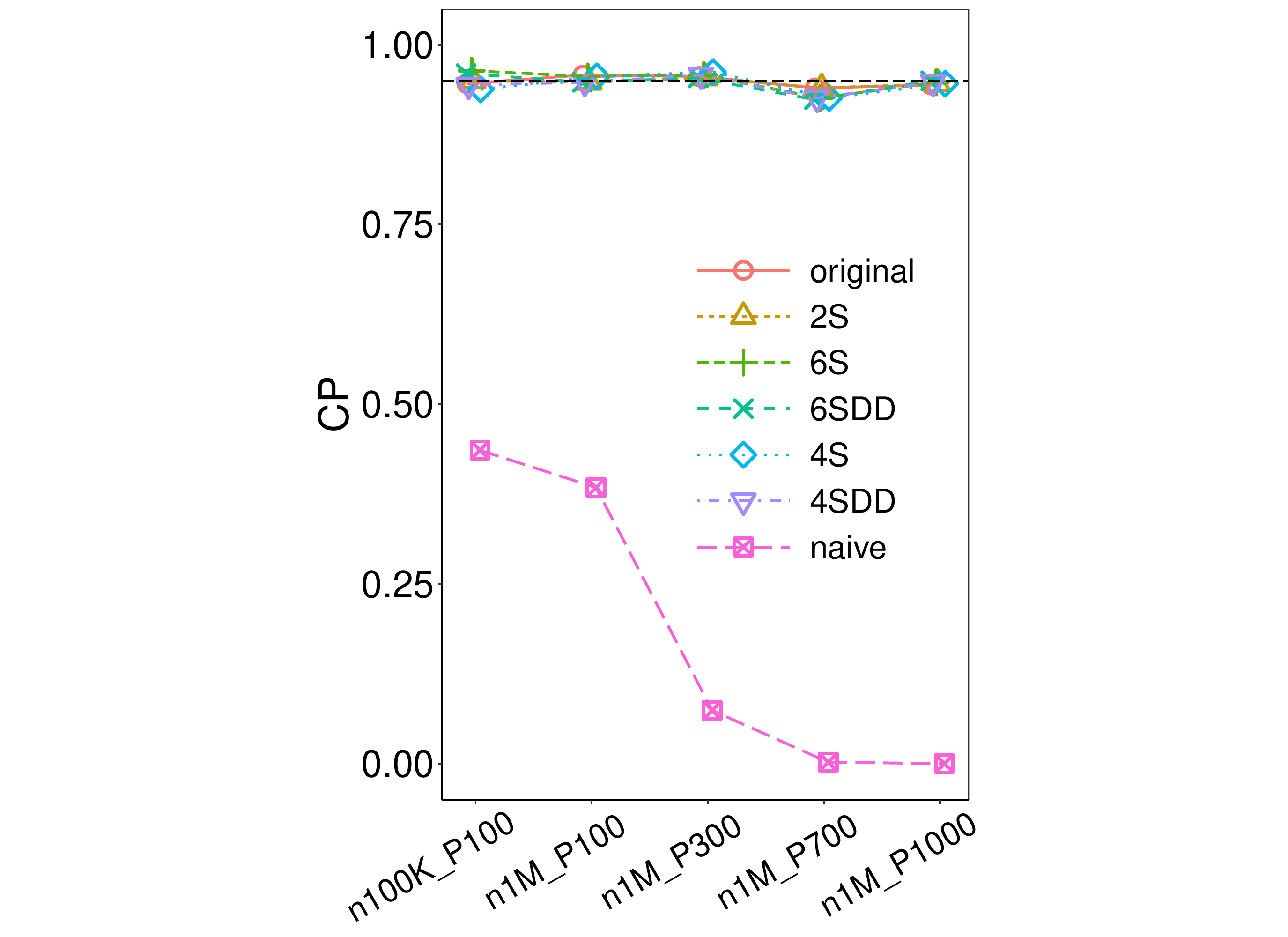}

\caption{Simulation results with $\epsilon$-DP for Gaussian data with  $\alpha\ne\beta$ when $\theta=0$} \label{fig:0asDPN}
\end{figure}

\begin{figure}[!htb]
\hspace{0.45in}$\rho=0.005$\hspace{0.65in}$\rho=0.02$\hspace{0.65in}$\rho=0.08$
\hspace{0.65in}$\rho=0.32$\hspace{0.65in}$\rho=1.28$

\includegraphics[width=0.19\textwidth, trim={2.5in 0 2.6in 0},clip] {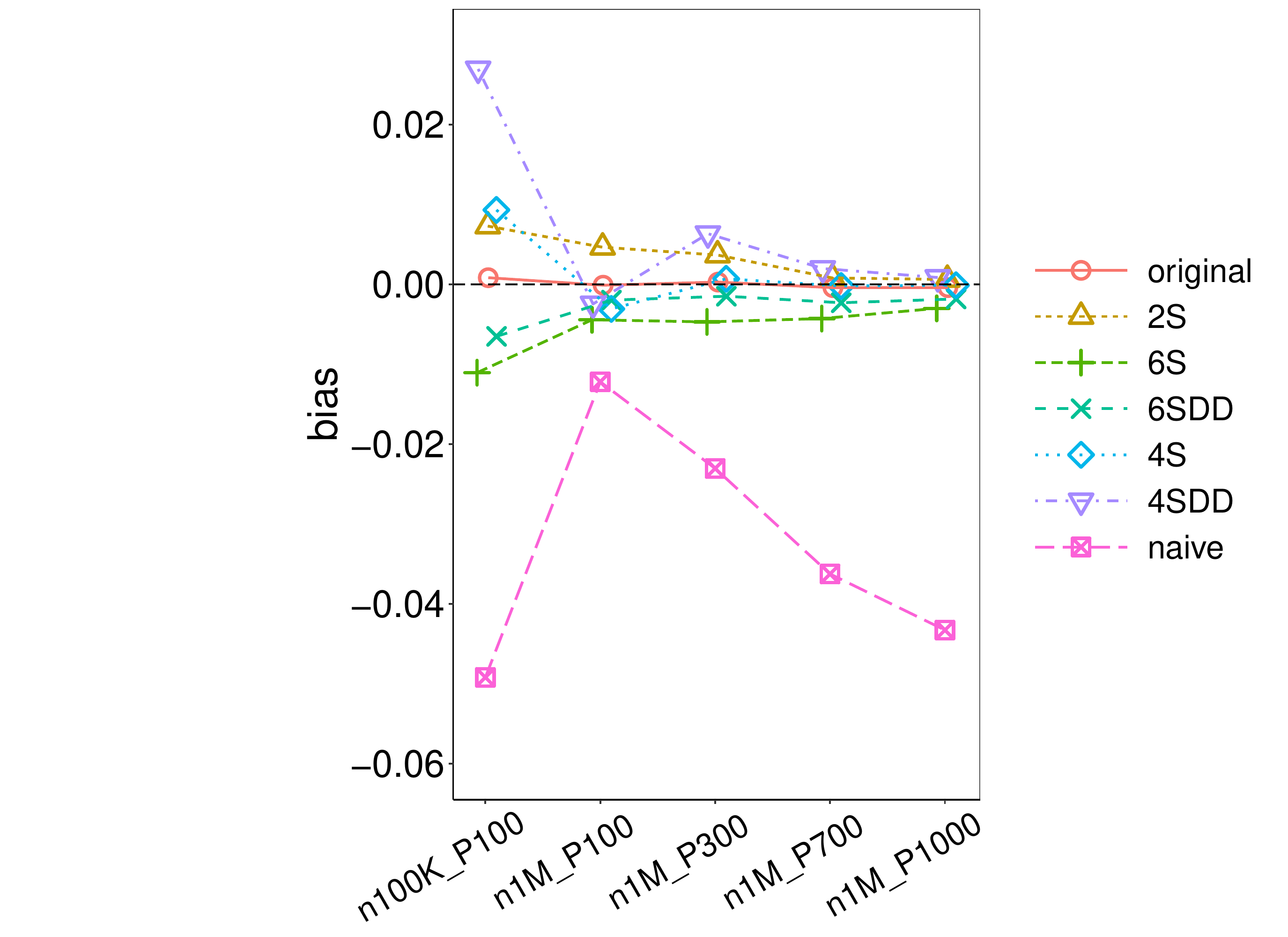}
\includegraphics[width=0.19\textwidth, trim={2.5in 0 2.6in 0},clip] {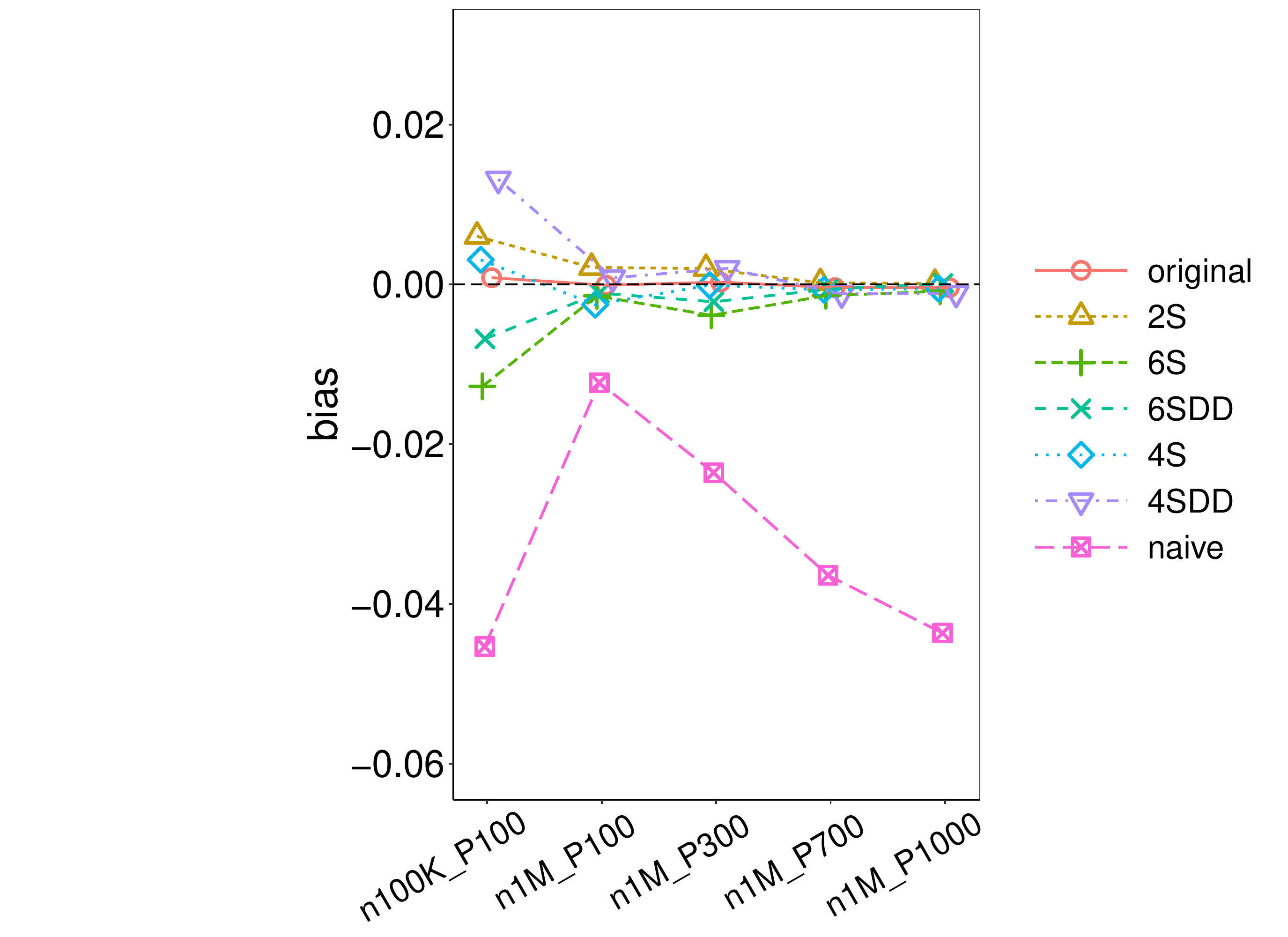}
\includegraphics[width=0.19\textwidth, trim={2.5in 0 2.6in 0},clip] {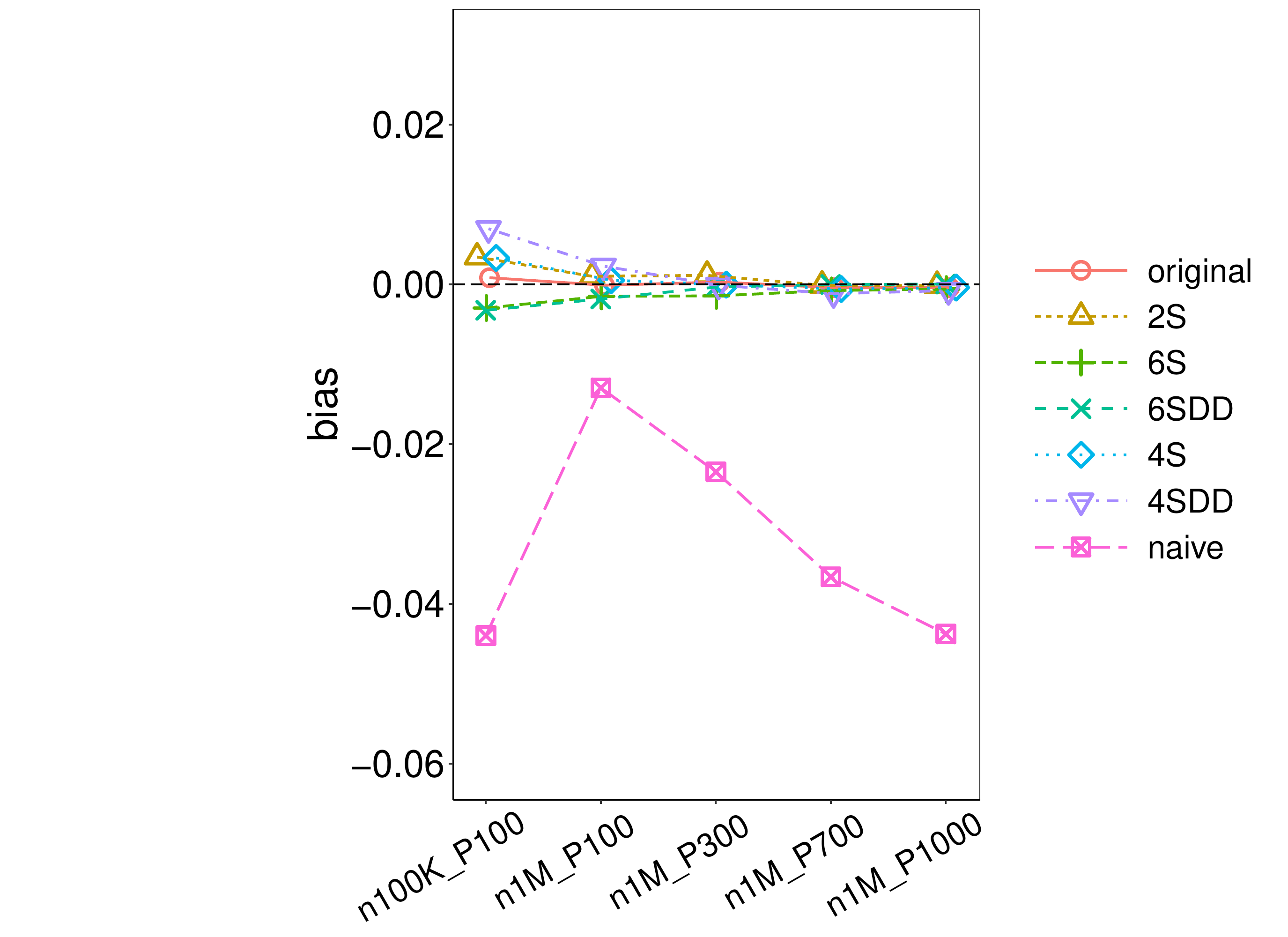}
\includegraphics[width=0.19\textwidth, trim={2.5in 0 2.6in 0},clip] {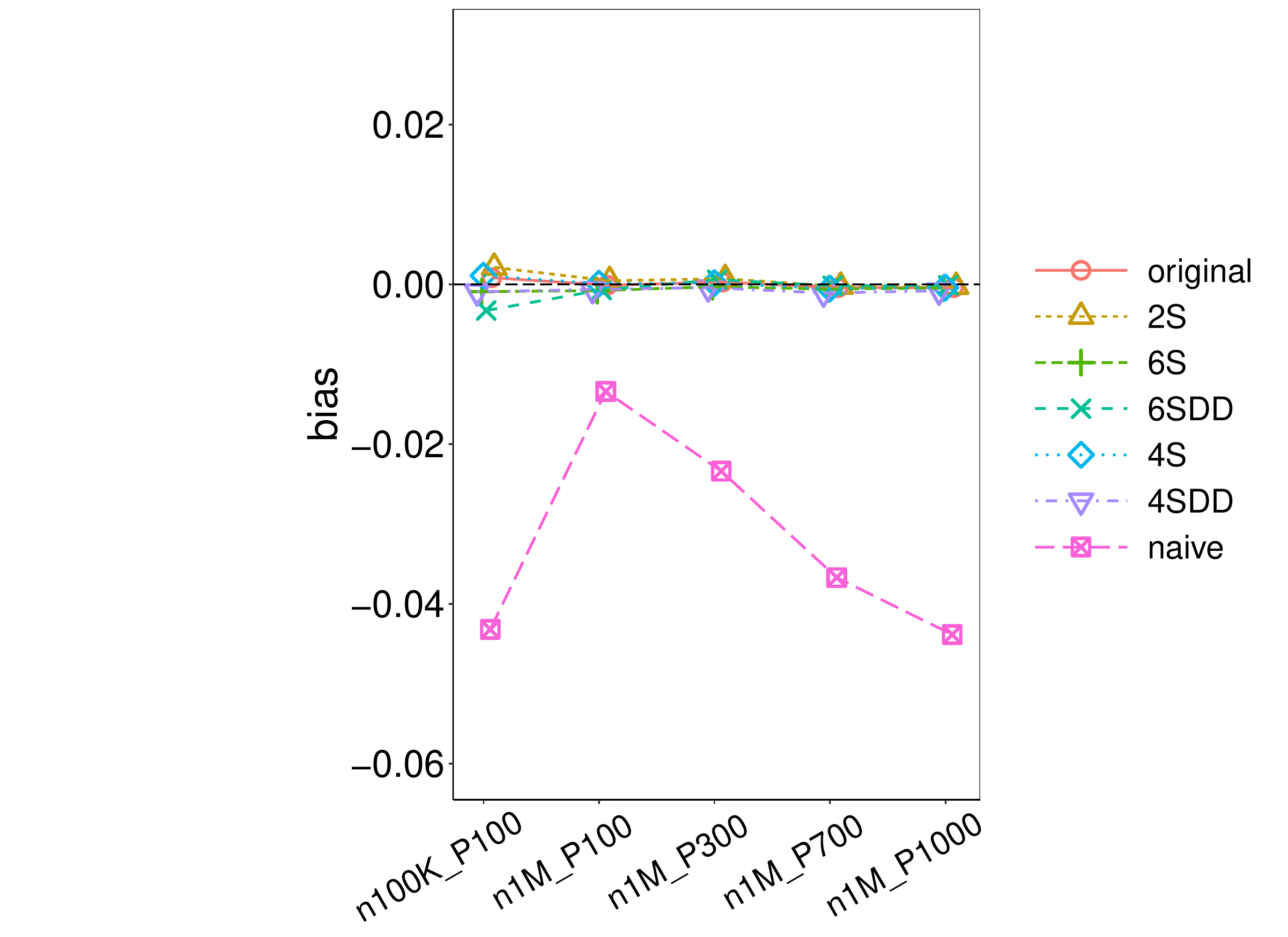}
\includegraphics[width=0.19\textwidth, trim={2.5in 0 2.6in 0},clip] {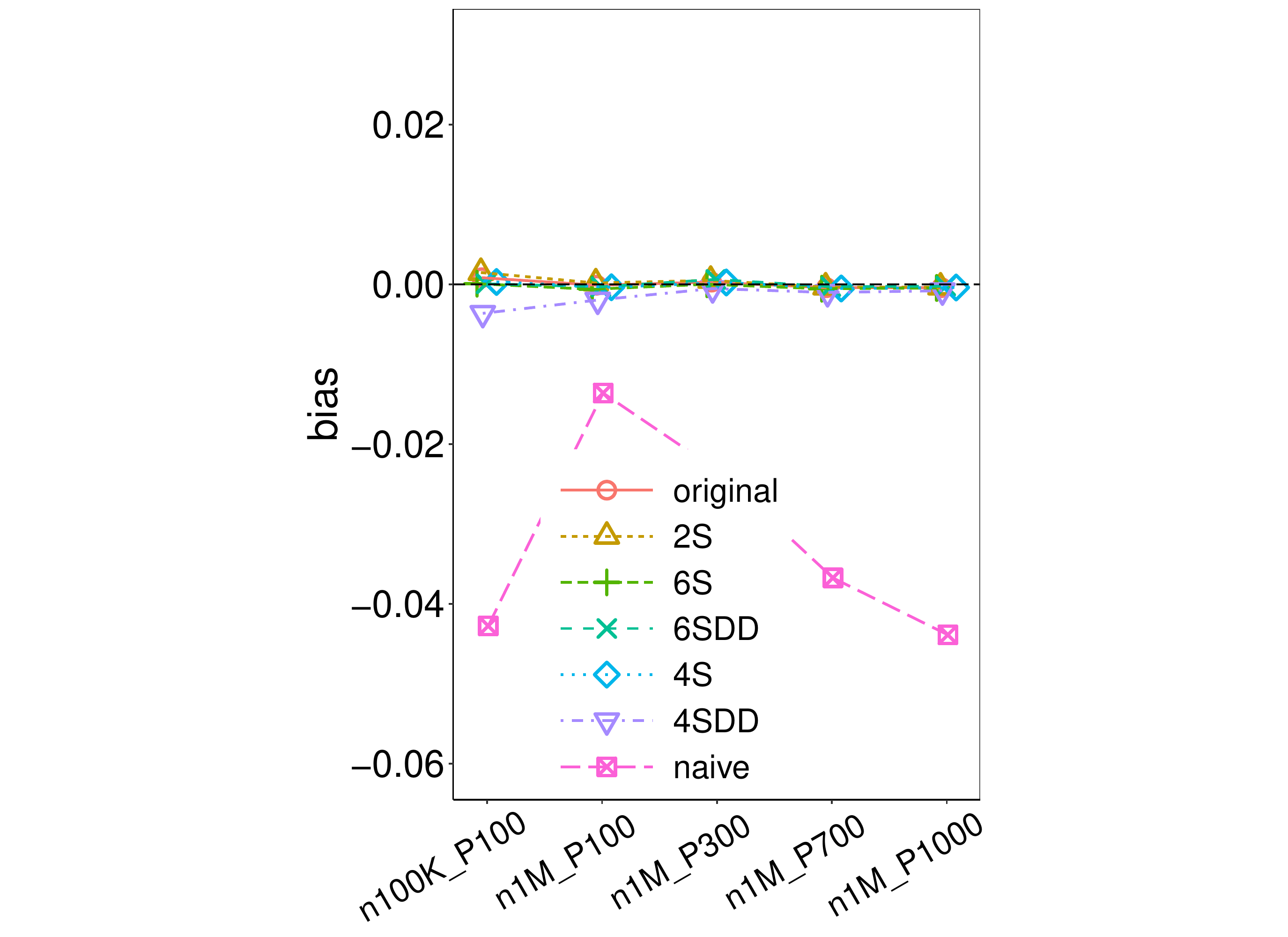}

\includegraphics[width=0.19\textwidth, trim={2.5in 0 2.6in 0},clip] {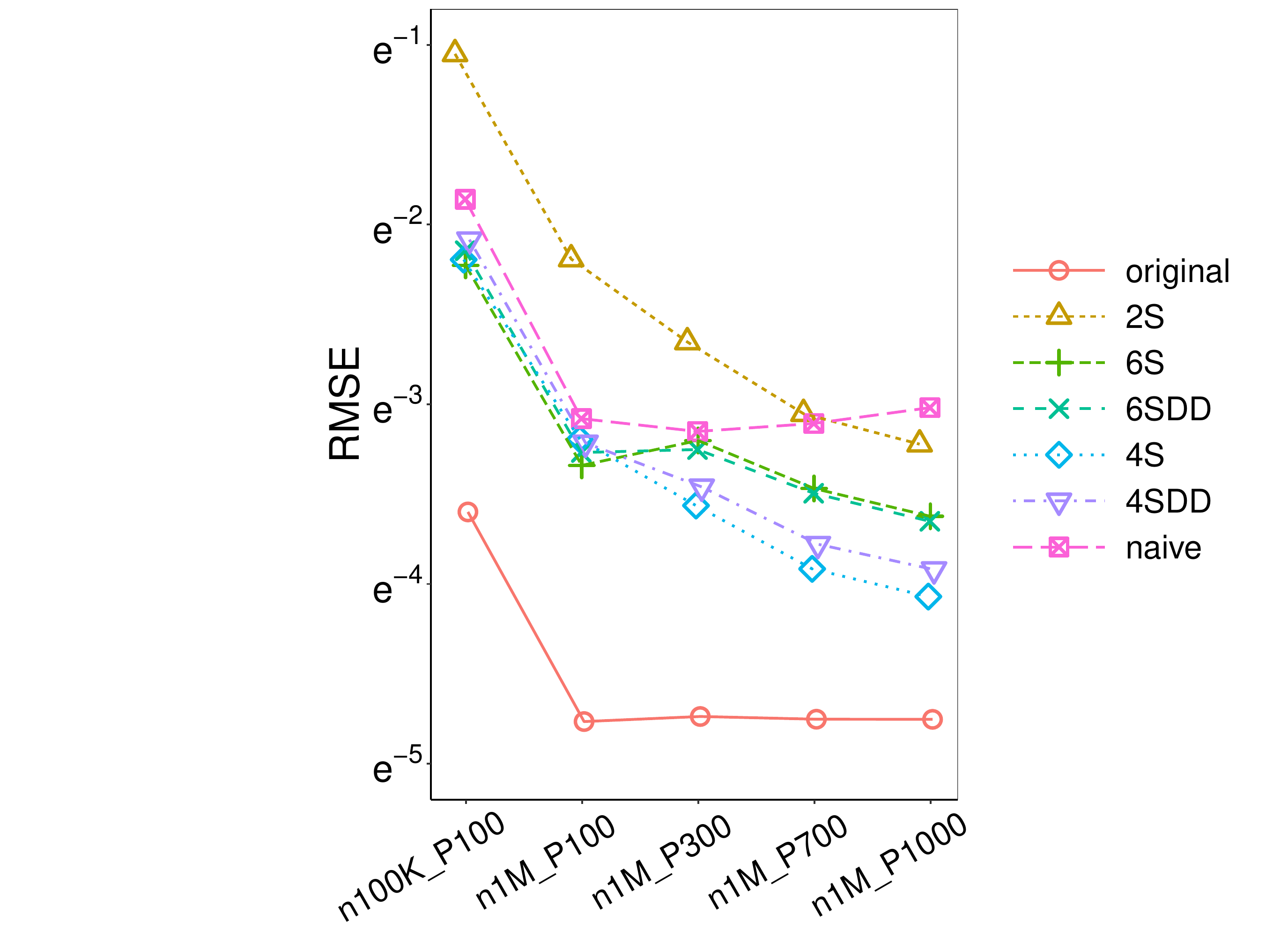}
\includegraphics[width=0.19\textwidth, trim={2.5in 0 2.6in 0},clip] {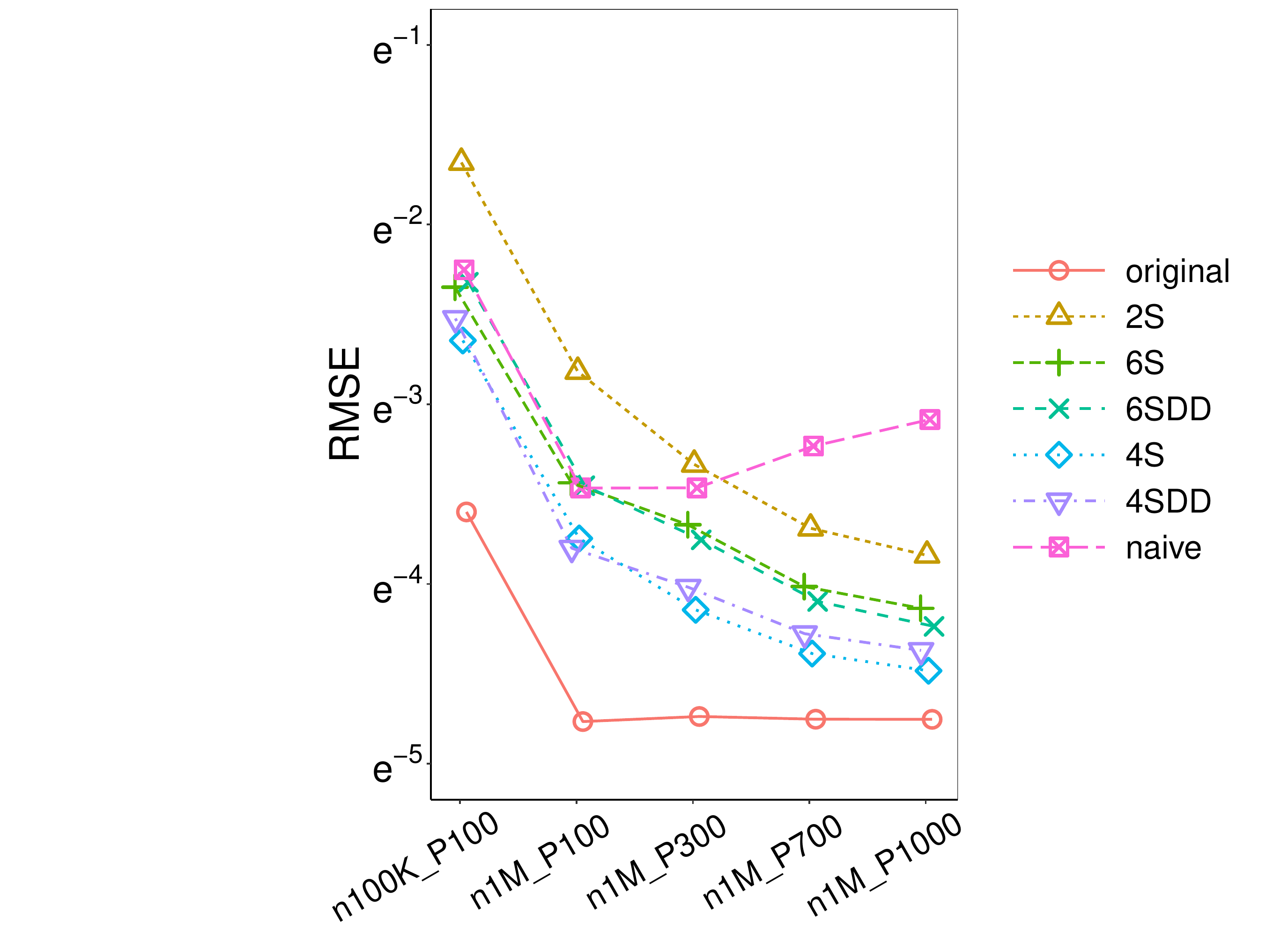}
\includegraphics[width=0.19\textwidth, trim={2.5in 0 2.6in 0},clip] {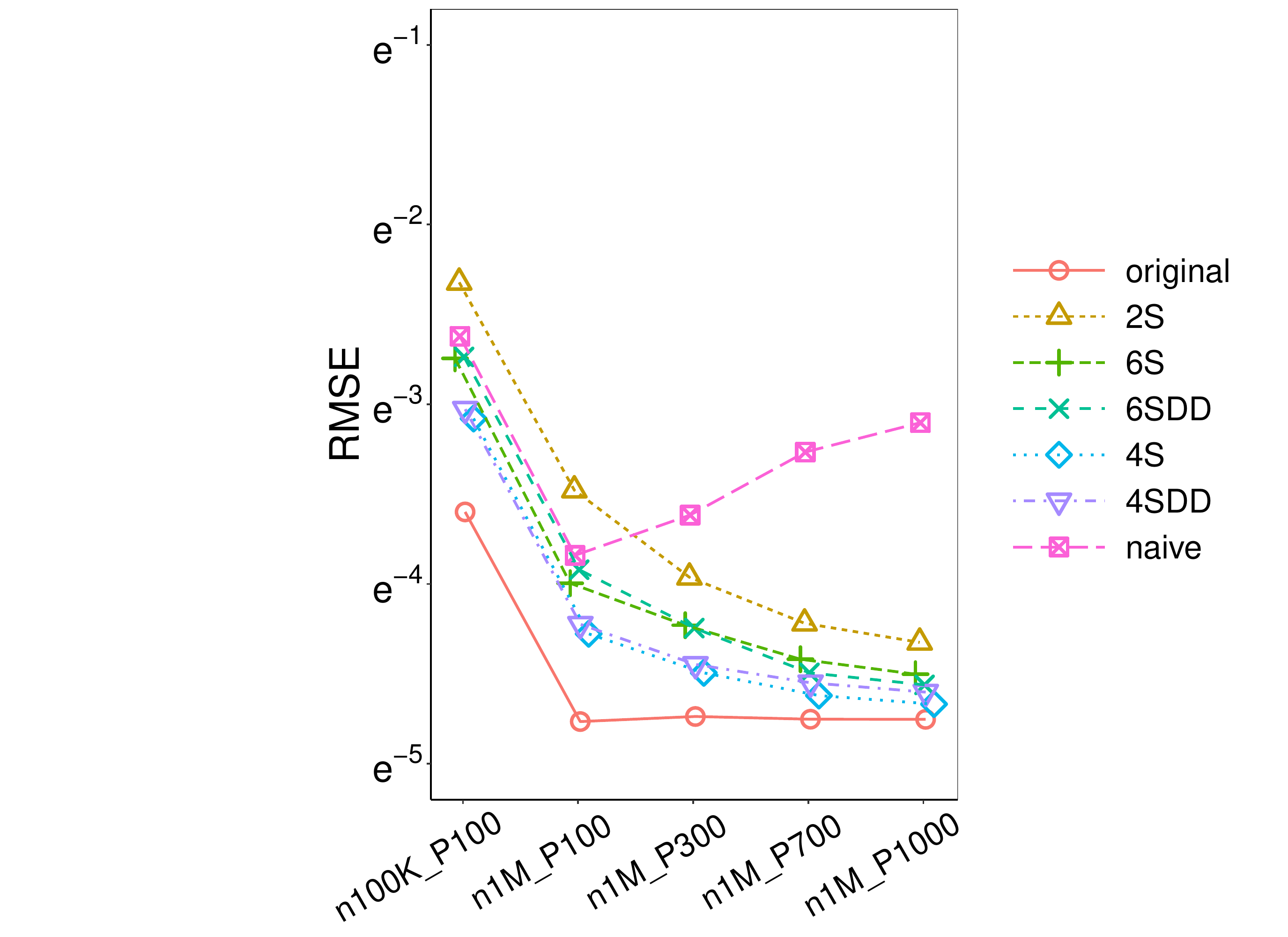}
\includegraphics[width=0.19\textwidth, trim={2.5in 0 2.6in 0},clip] {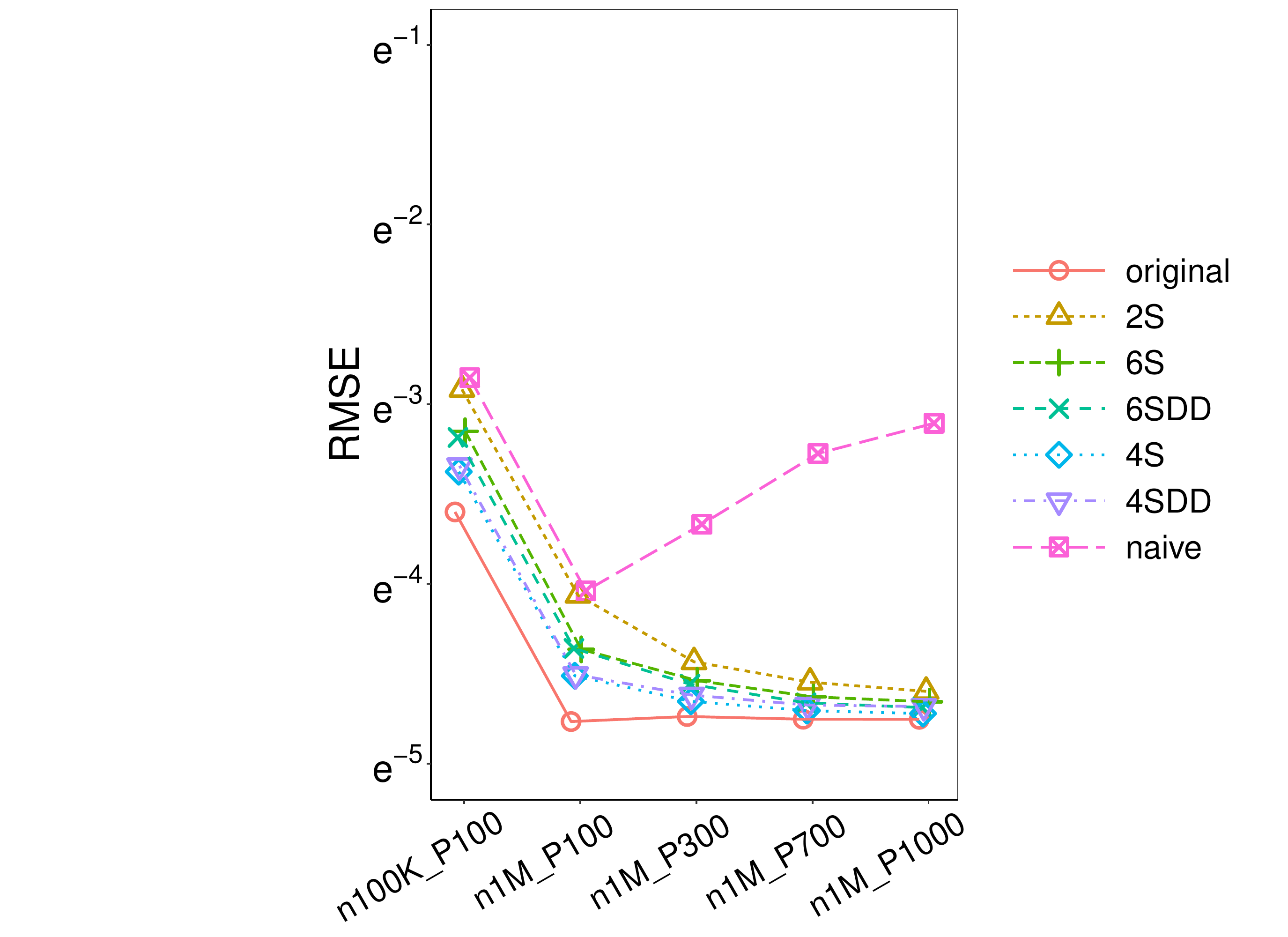}
\includegraphics[width=0.19\textwidth, trim={2.5in 0 2.6in 0},clip] {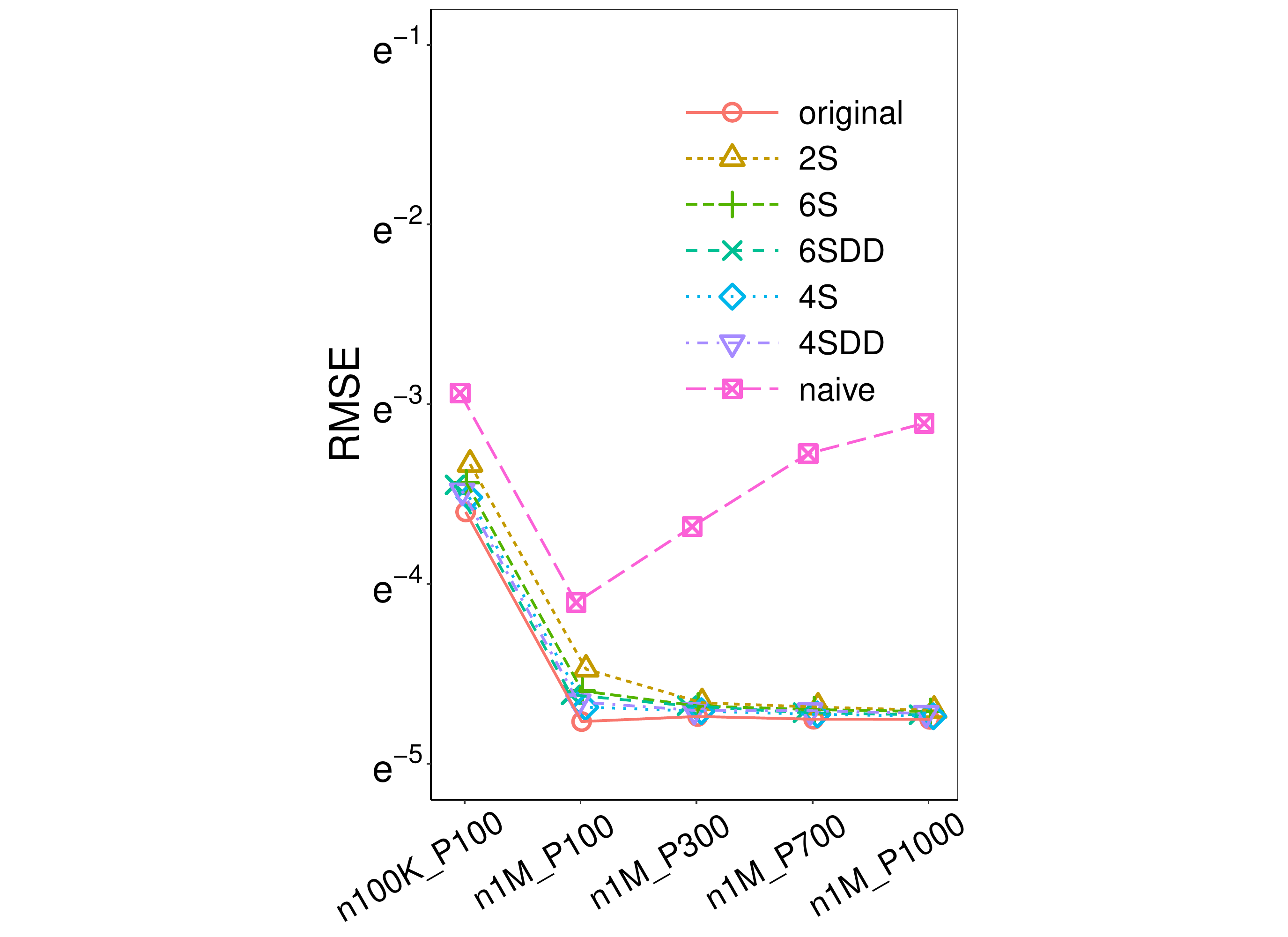}

\includegraphics[width=0.19\textwidth, trim={2.5in 0 2.6in 0},clip] {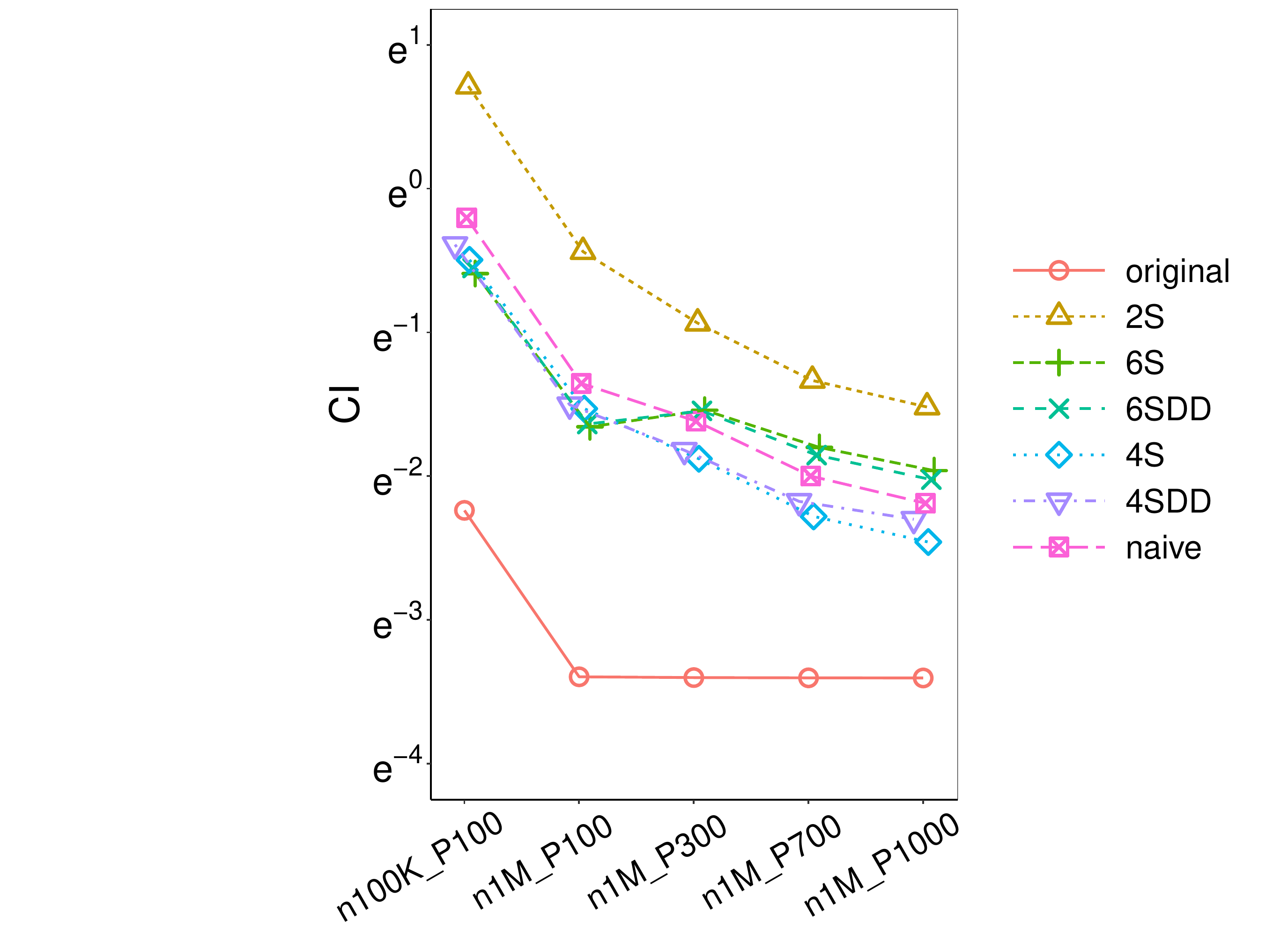}
\includegraphics[width=0.19\textwidth, trim={2.5in 0 2.6in 0},clip] {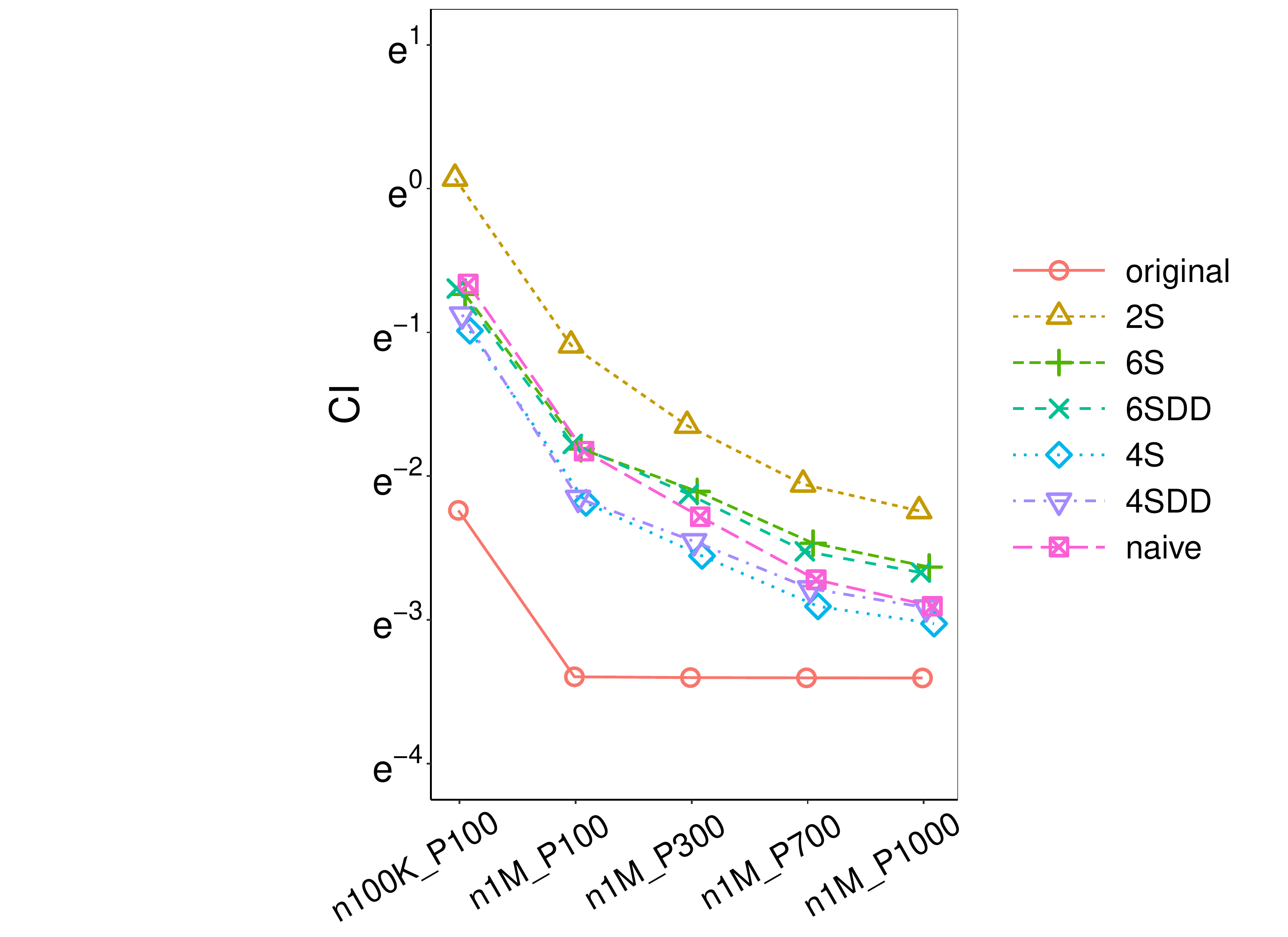}
\includegraphics[width=0.19\textwidth, trim={2.5in 0 2.6in 0},clip] {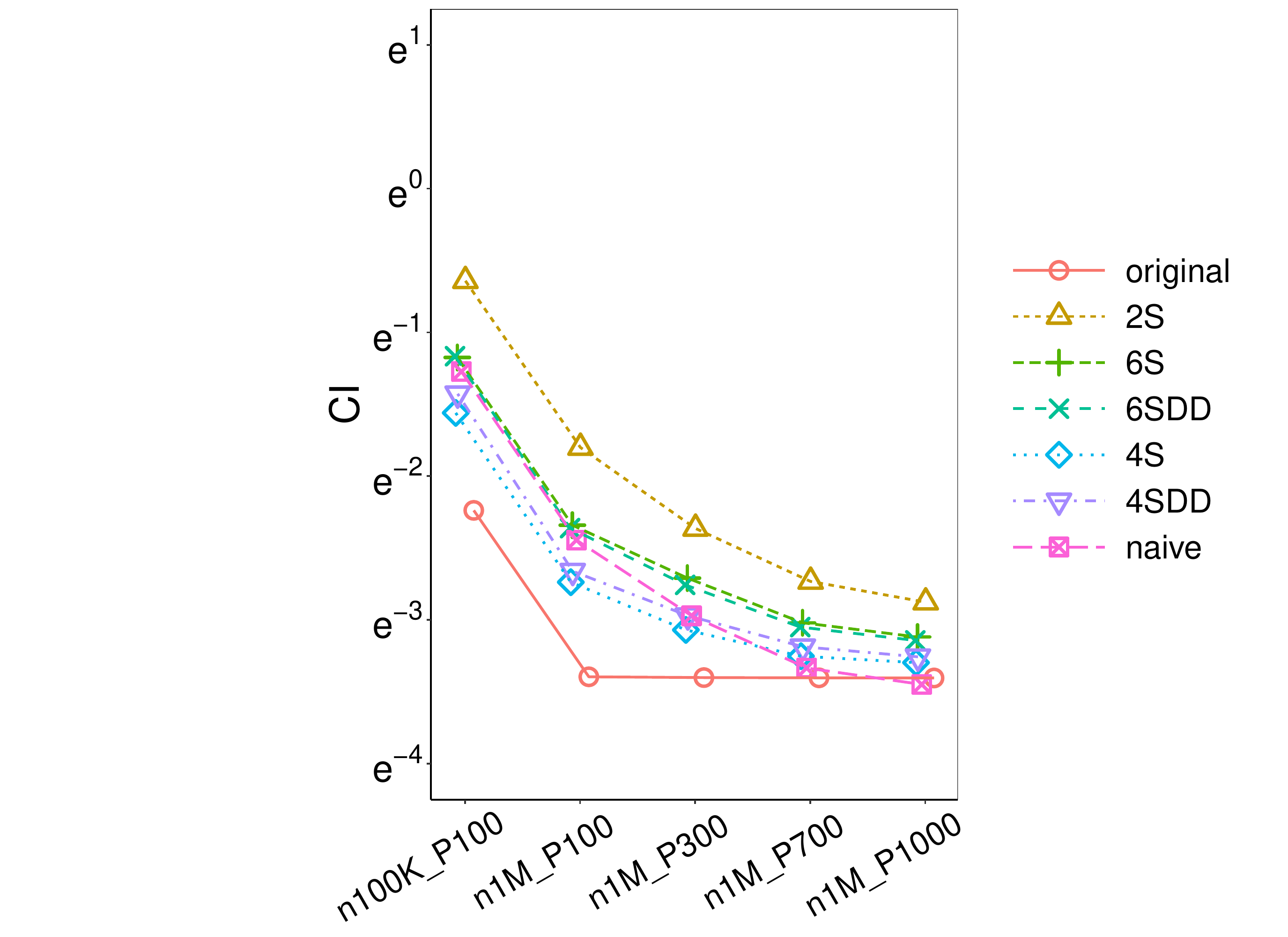}
\includegraphics[width=0.19\textwidth, trim={2.5in 0 2.6in 0},clip] {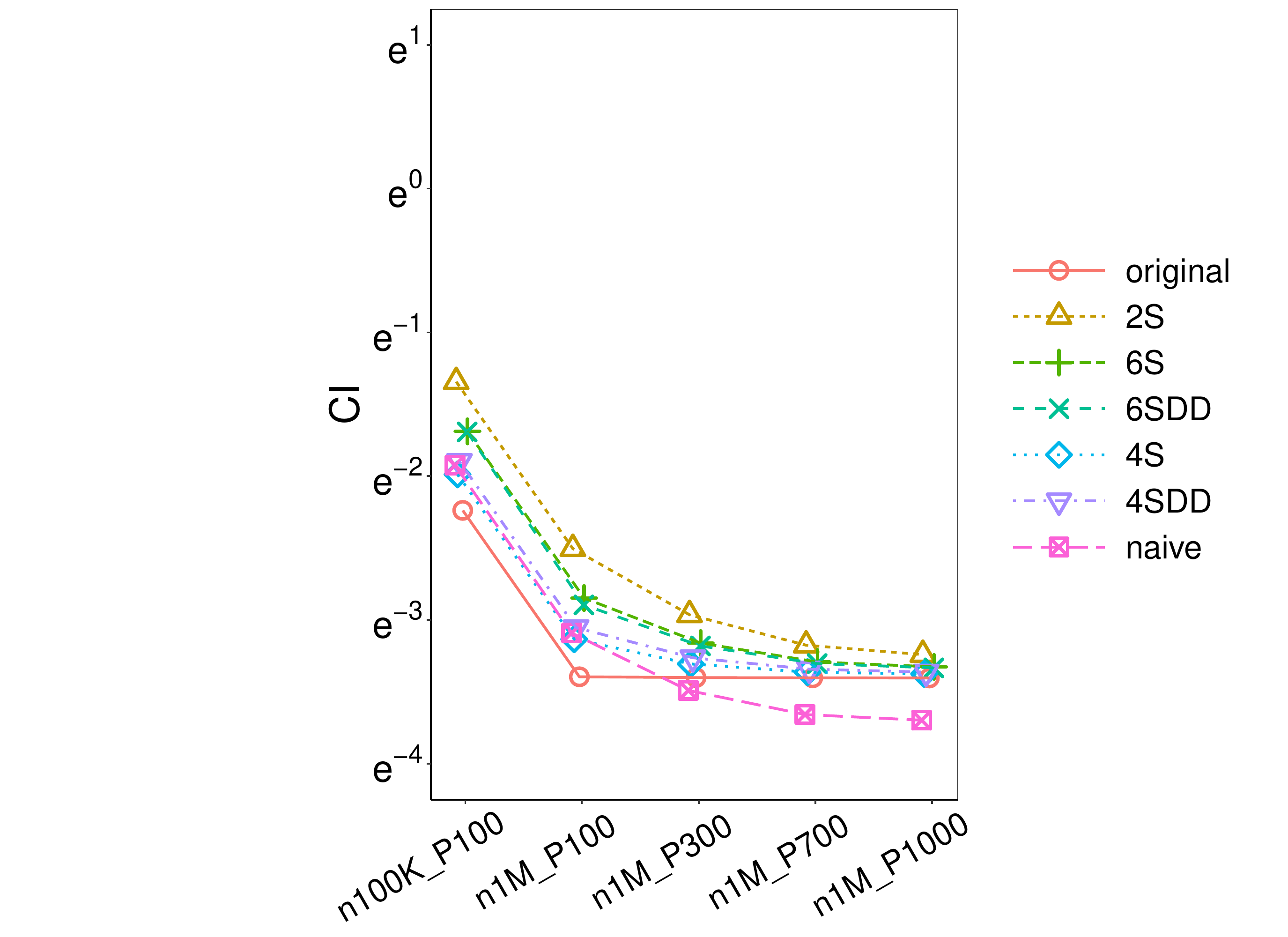}
\includegraphics[width=0.19\textwidth, trim={2.5in 0 2.6in 0},clip] {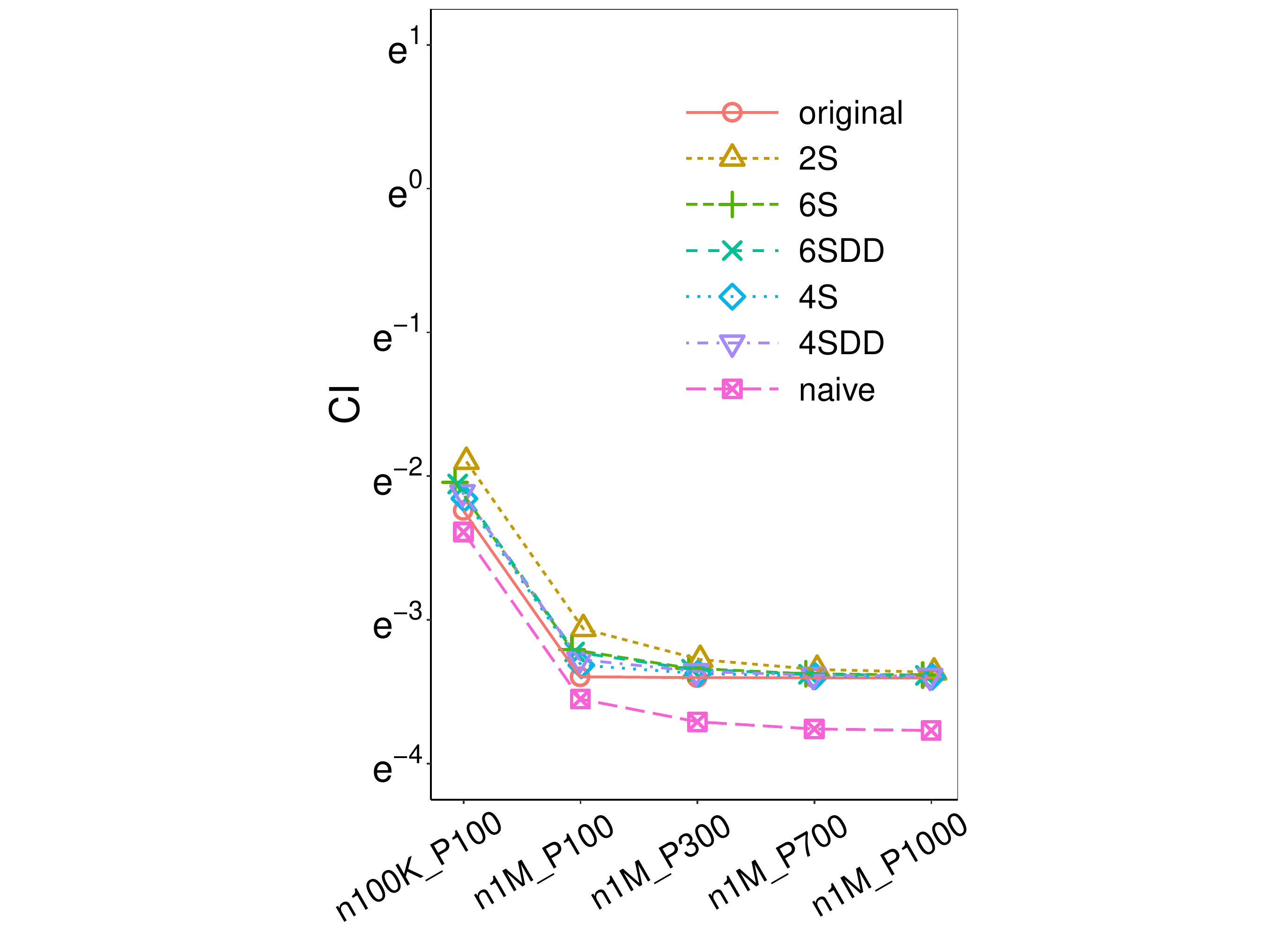}

\includegraphics[width=0.19\textwidth, trim={2.5in 0 2.6in 0},clip] {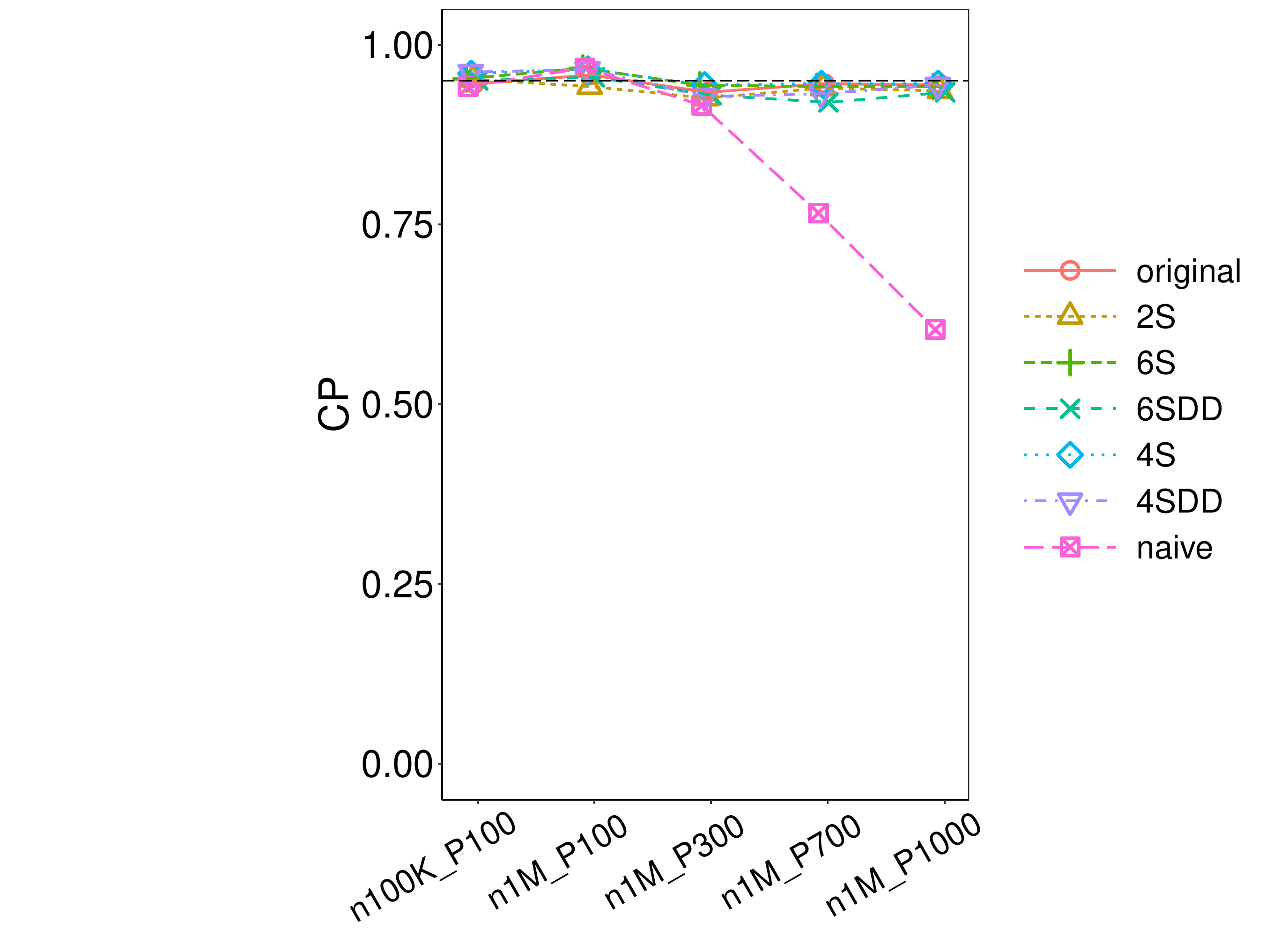}
\includegraphics[width=0.19\textwidth, trim={2.5in 0 2.6in 0},clip] {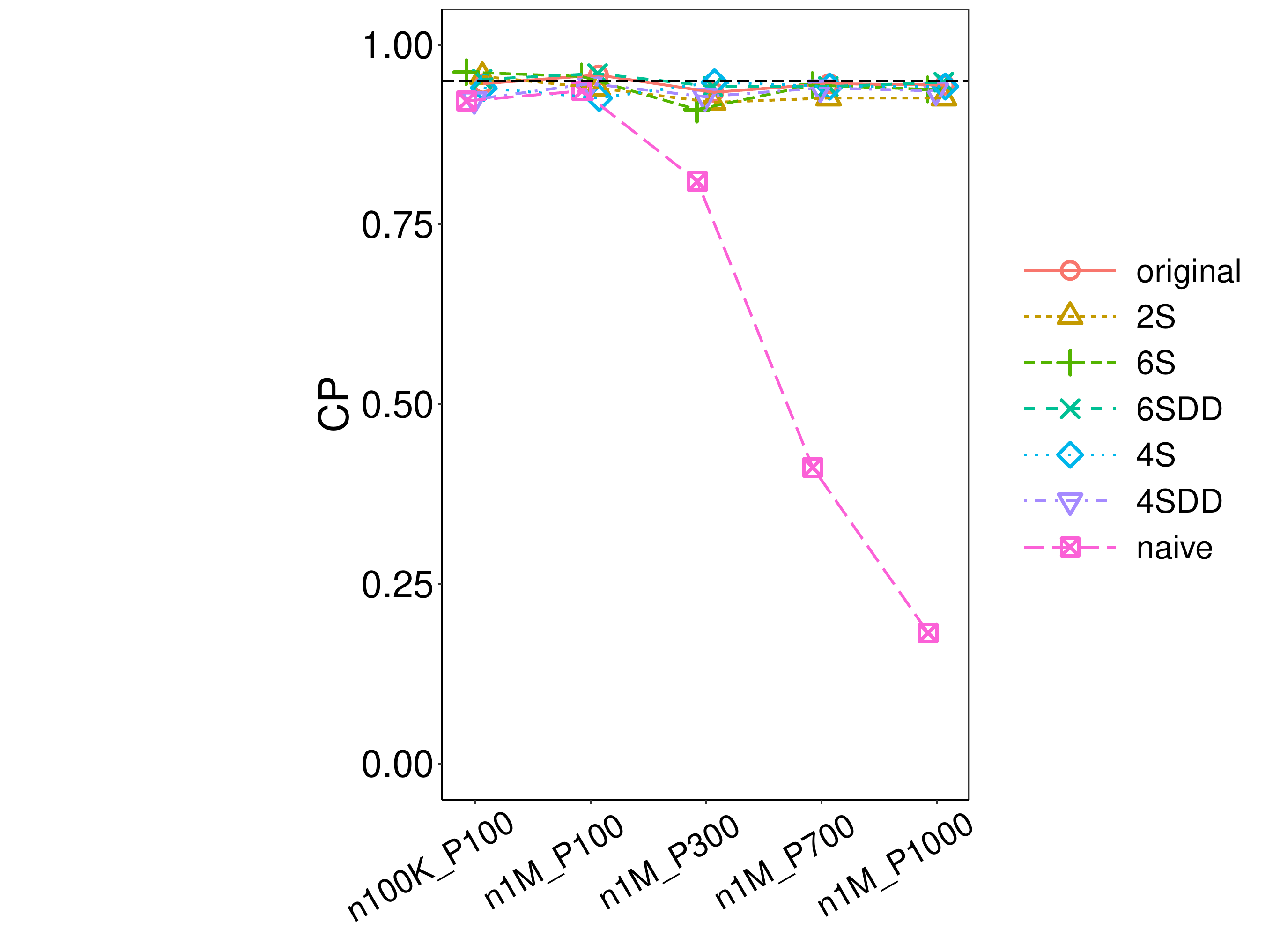}
\includegraphics[width=0.19\textwidth, trim={2.5in 0 2.6in 0},clip] {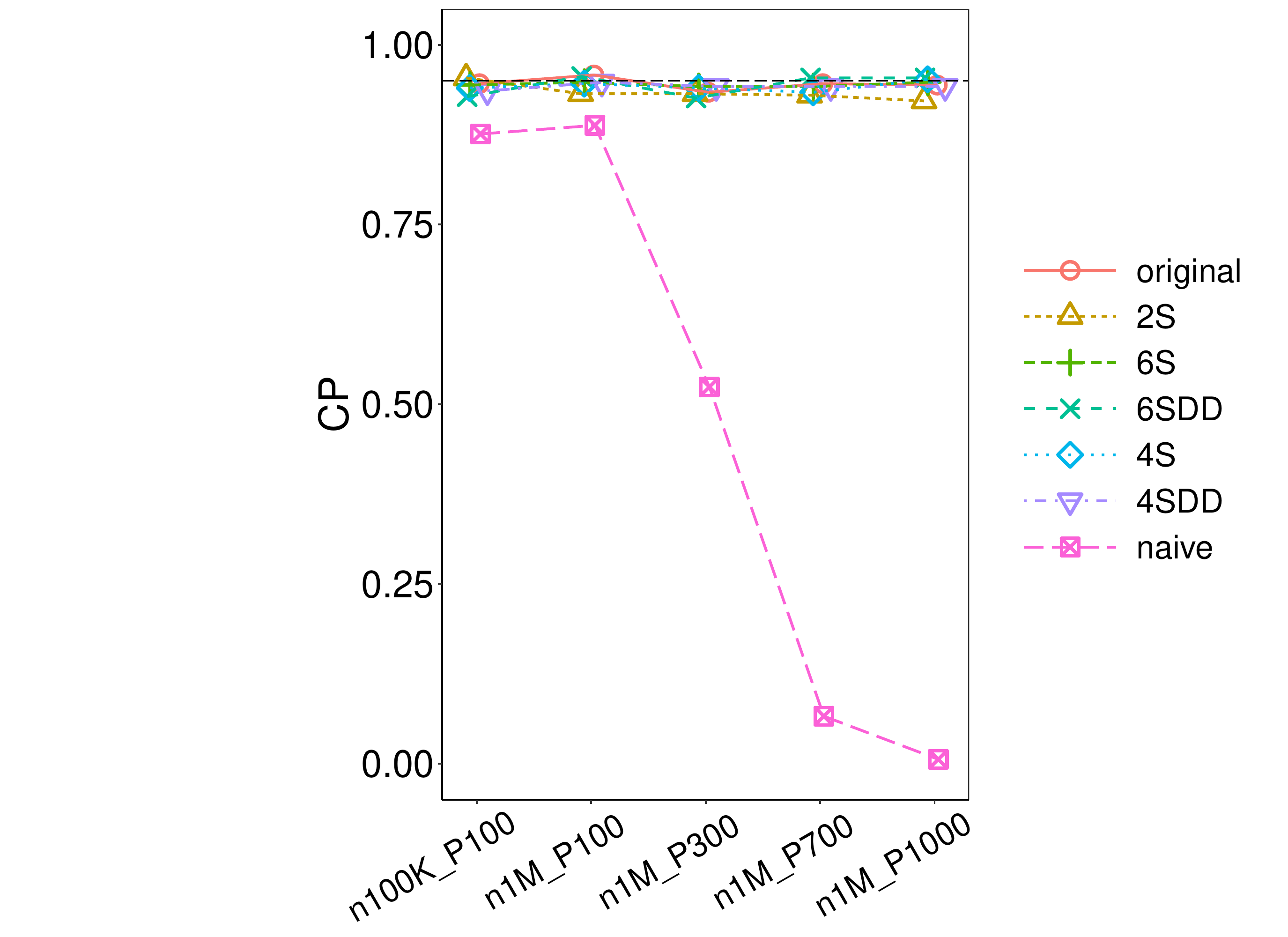}
\includegraphics[width=0.19\textwidth, trim={2.5in 0 2.6in 0},clip] {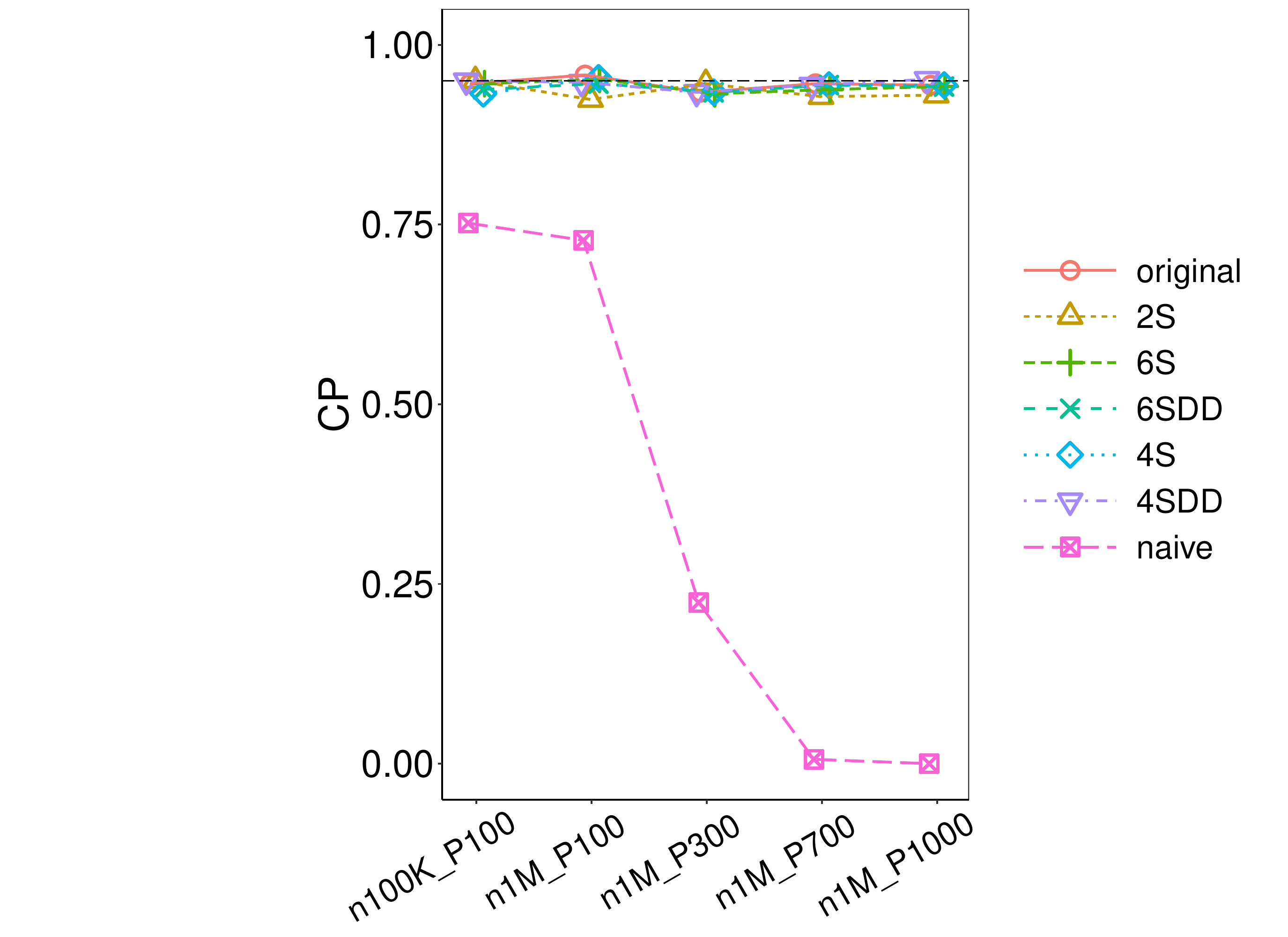}
\includegraphics[width=0.19\textwidth, trim={2.5in 0 2.6in 0},clip] {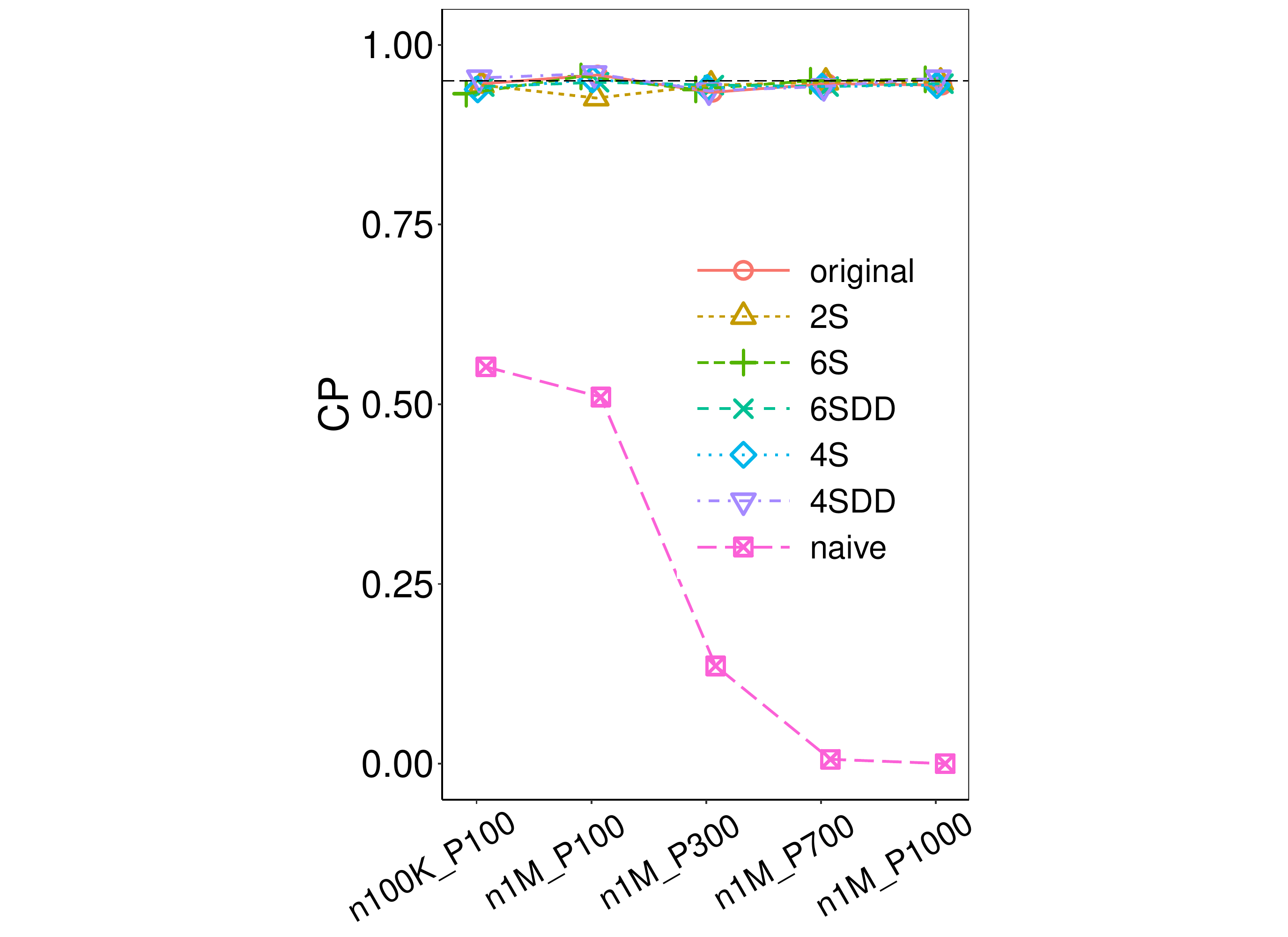}

\caption{Simulation results with $\rho$-zCDP for Gaussian data with  $\alpha\ne\beta$ when $\theta=0$}
\label{fig:0aszCDPN}
\end{figure}

The results are presented in Figures \ref{fig:0asDPN} and \ref{fig:0aszCDPN} and are summarized  as follows. 4S is the best performer by all metrics and the na\"{i}ve method is the worst. In between the two, some approaches are better than others, depending on the evaluation metrics. For example, in terms of bias, 6SDD approach 0 faster as $n$ increase, $P$ increases when $n=1m$, or $\epsilon$ increases, than 6S, 4S, 4SDD, and 2S; in terms of RMSE and CI width, 4S is the best performer, followed closely by 4SDD, and then 6S and 6SDD, all of which are superior or not inferior to 2S in all cases, and the na\"{i}ve method is the worst for the large RMSE and too narrow CIs when $\epsilon \ge 5$; all methods, except for the  na\"{i}ve method, provides nominal level CP. The Gaussian mechanism of $\rho$-zCDP does not seem to be advantageous in statistical utility compared to the Laplace mechanism of $\epsilon$-DP.

\subsubsection{\texorpdfstring{$Y\sim$ Gaussian}{ }, \texorpdfstring{$\theta\ne0$ ($H_1$)}{H1 }, and \texorpdfstring{$\alpha\ne\beta$}{AS}}\label{sec:1asN}
\begin{figure}[!htb]
\hspace{0.5in}$\epsilon=0.5$\hspace{0.8in}$\epsilon=1$\hspace{0.9in}$\epsilon=2$
\hspace{0.95in}$\epsilon=5$\hspace{0.9in}$\epsilon=50$

\includegraphics[width=0.19\textwidth, trim={2.5in 0 2.6in 0},clip] {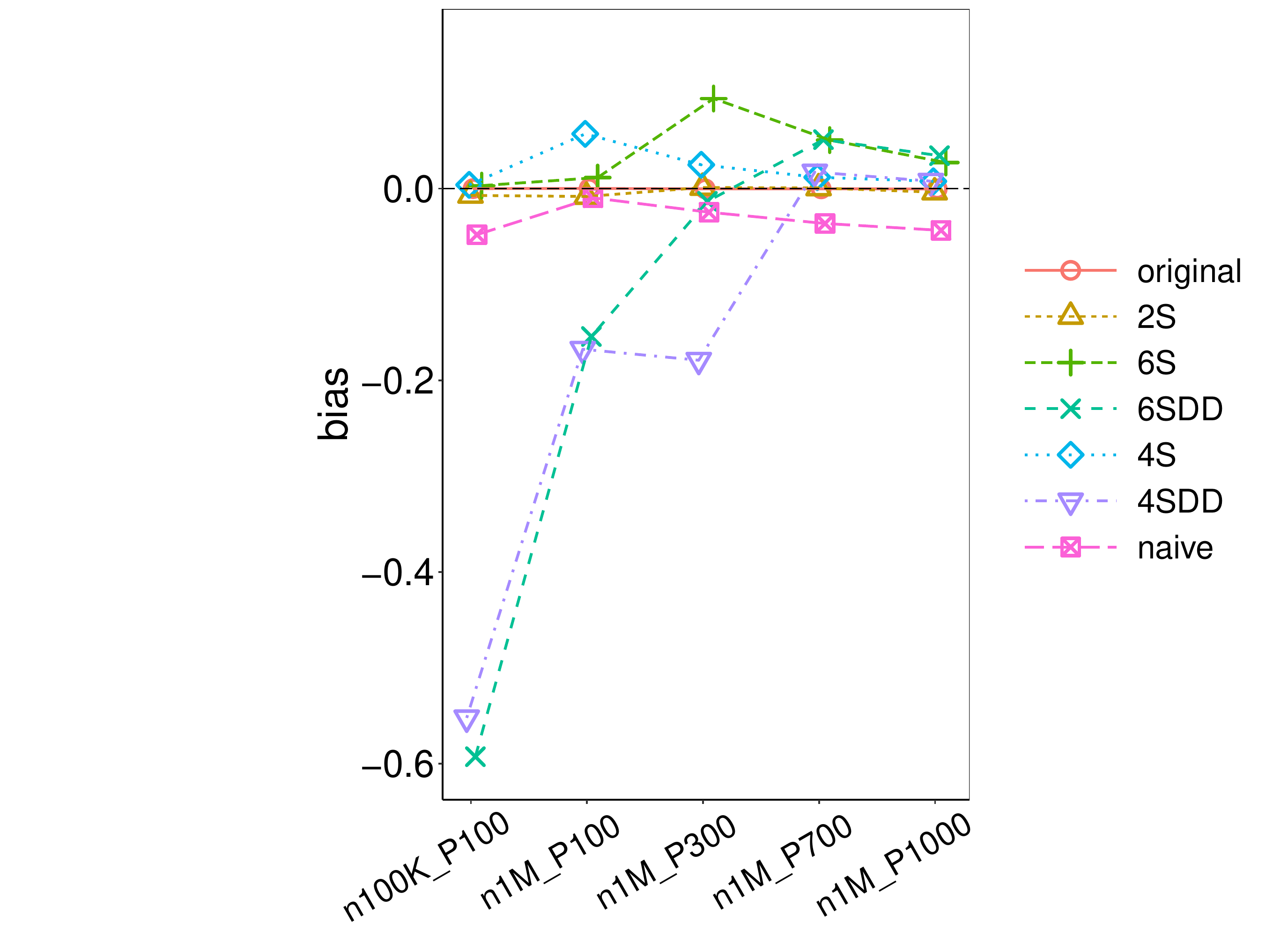}
\includegraphics[width=0.19\textwidth, trim={2.5in 0 2.6in 0},clip] {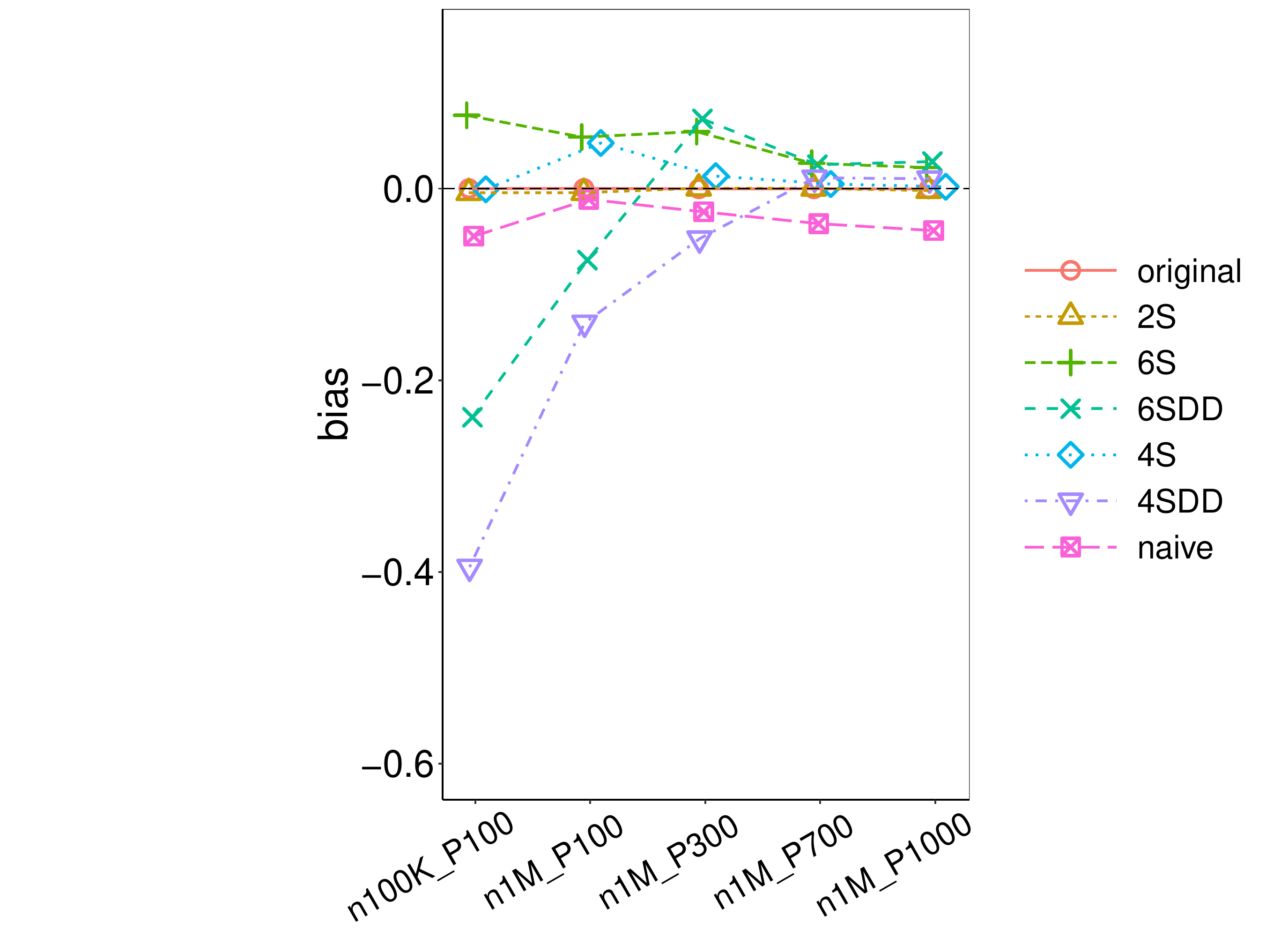}
\includegraphics[width=0.19\textwidth, trim={2.5in 0 2.6in 0},clip] {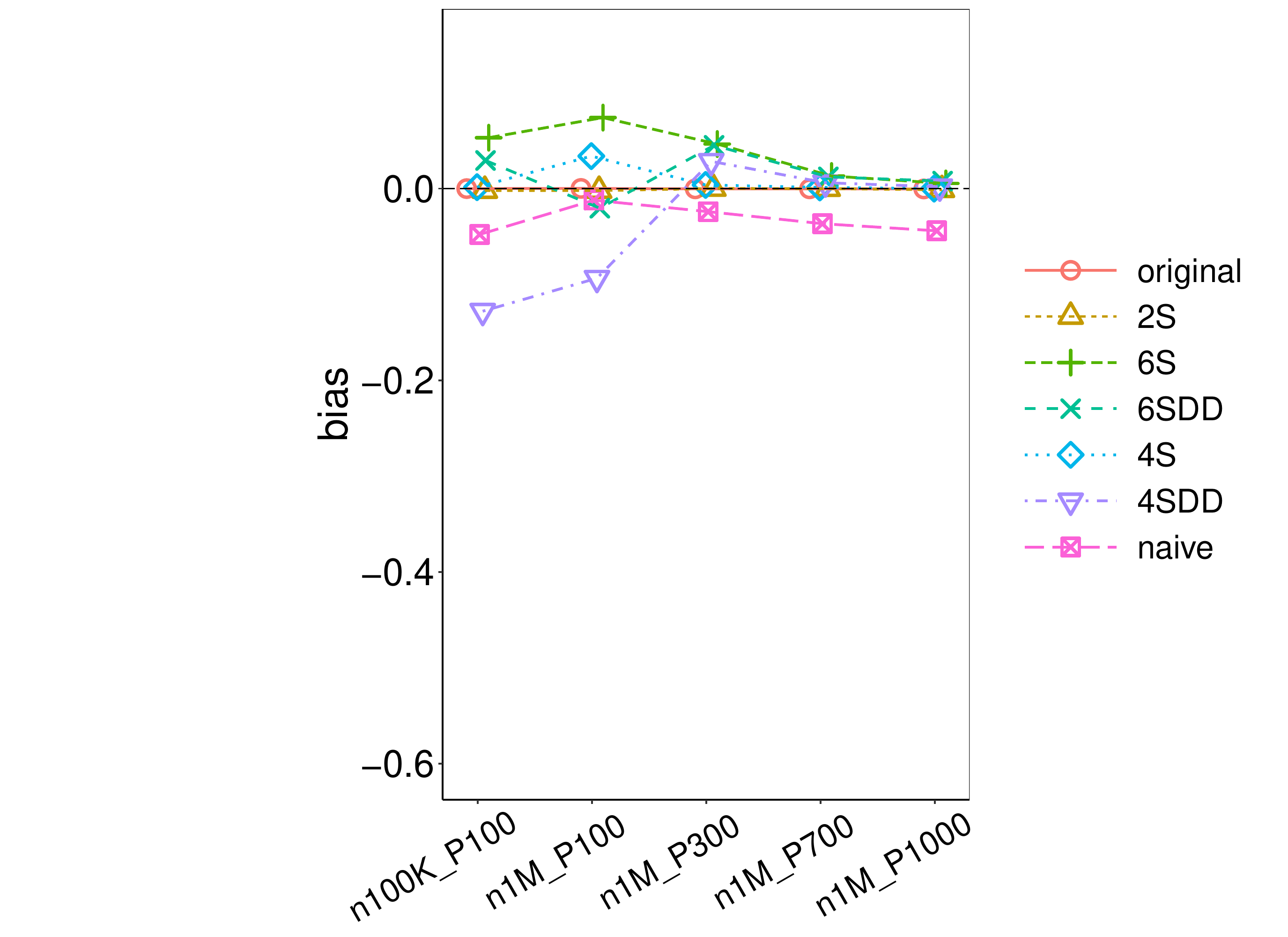}
\includegraphics[width=0.19\textwidth, trim={2.5in 0 2.6in 0},clip] {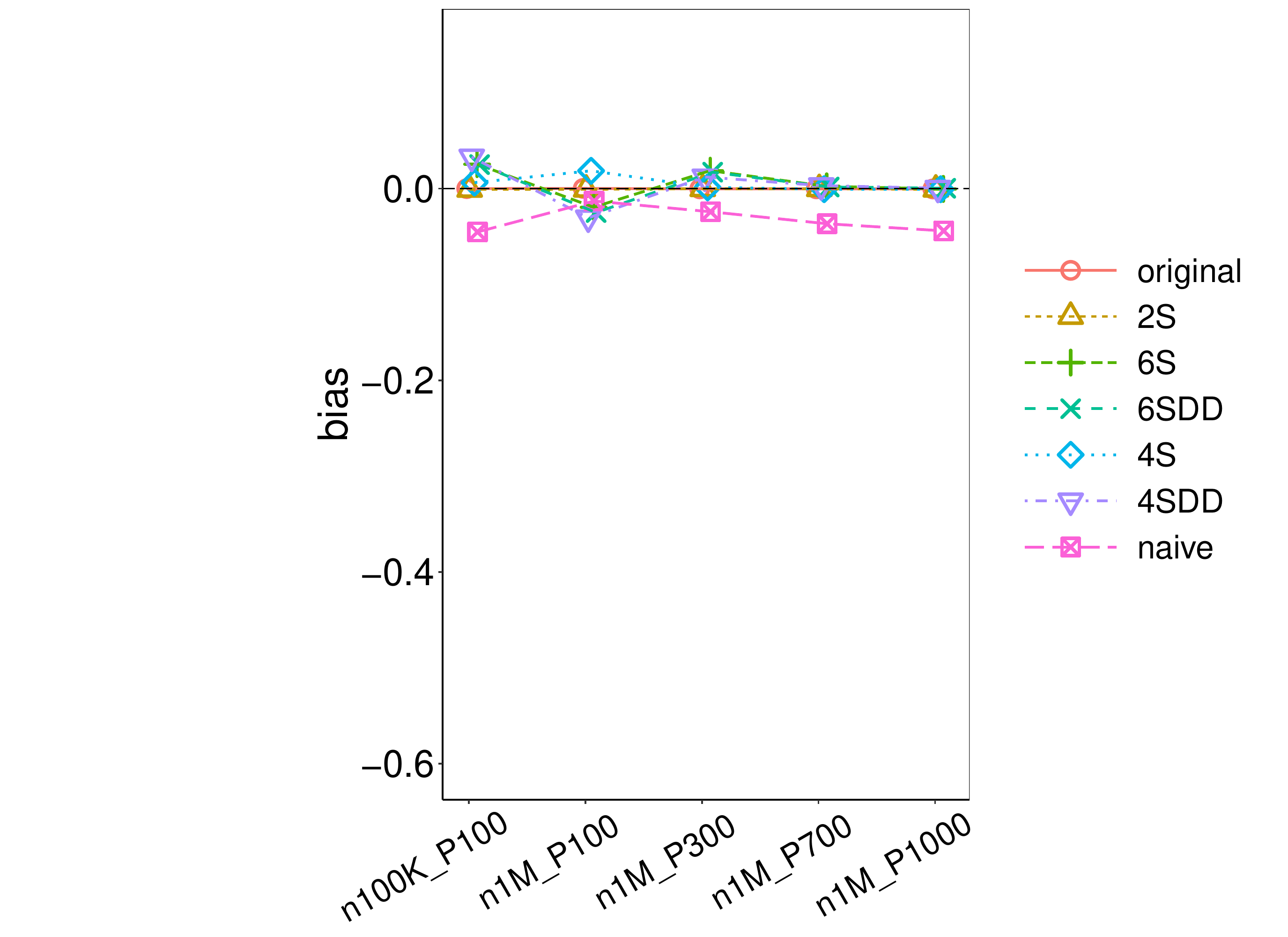}
\includegraphics[width=0.19\textwidth, trim={2.5in 0 2.6in 0},clip] {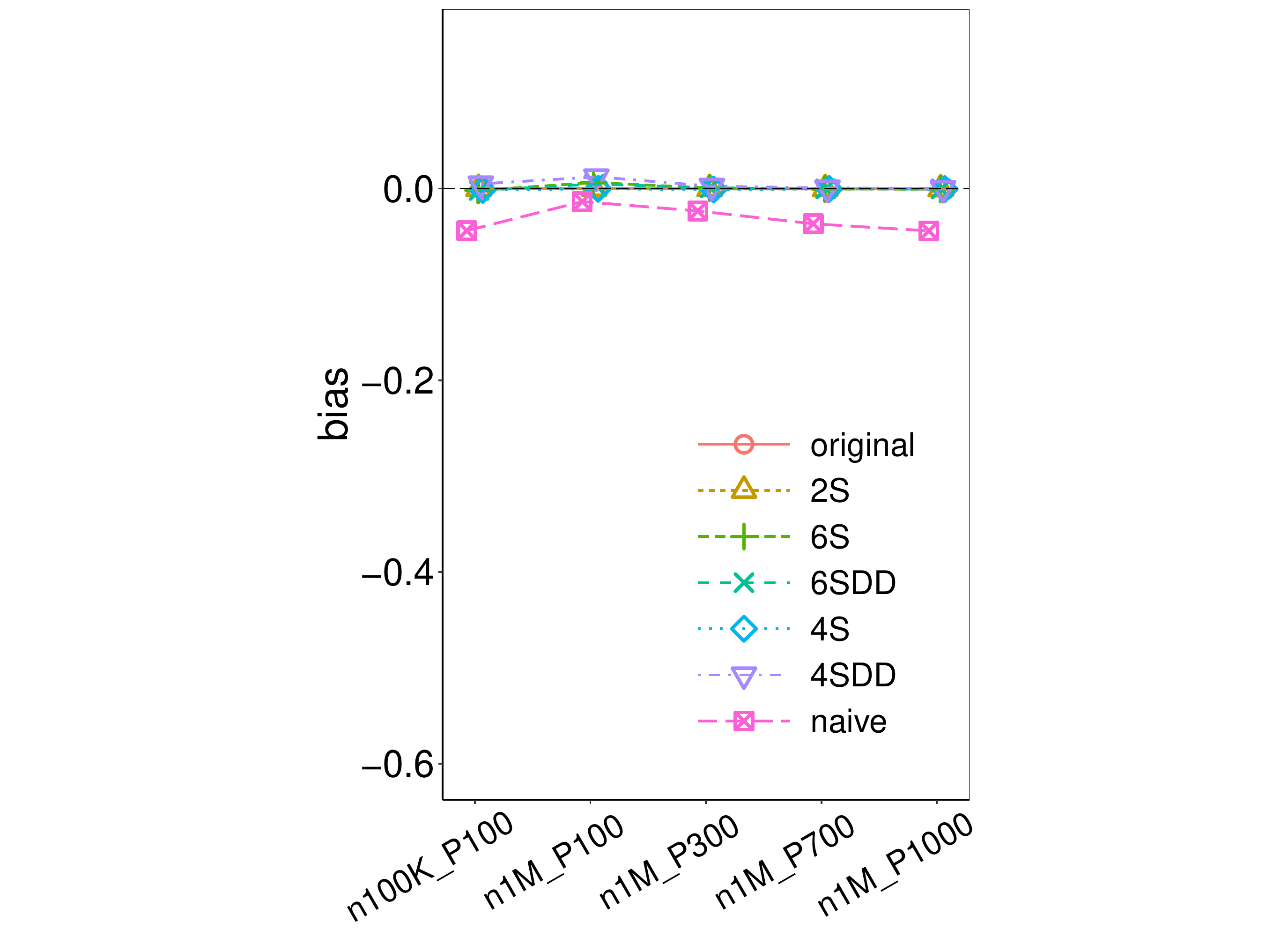}

\includegraphics[width=0.19\textwidth, trim={2.5in 0 2.6in 0},clip] {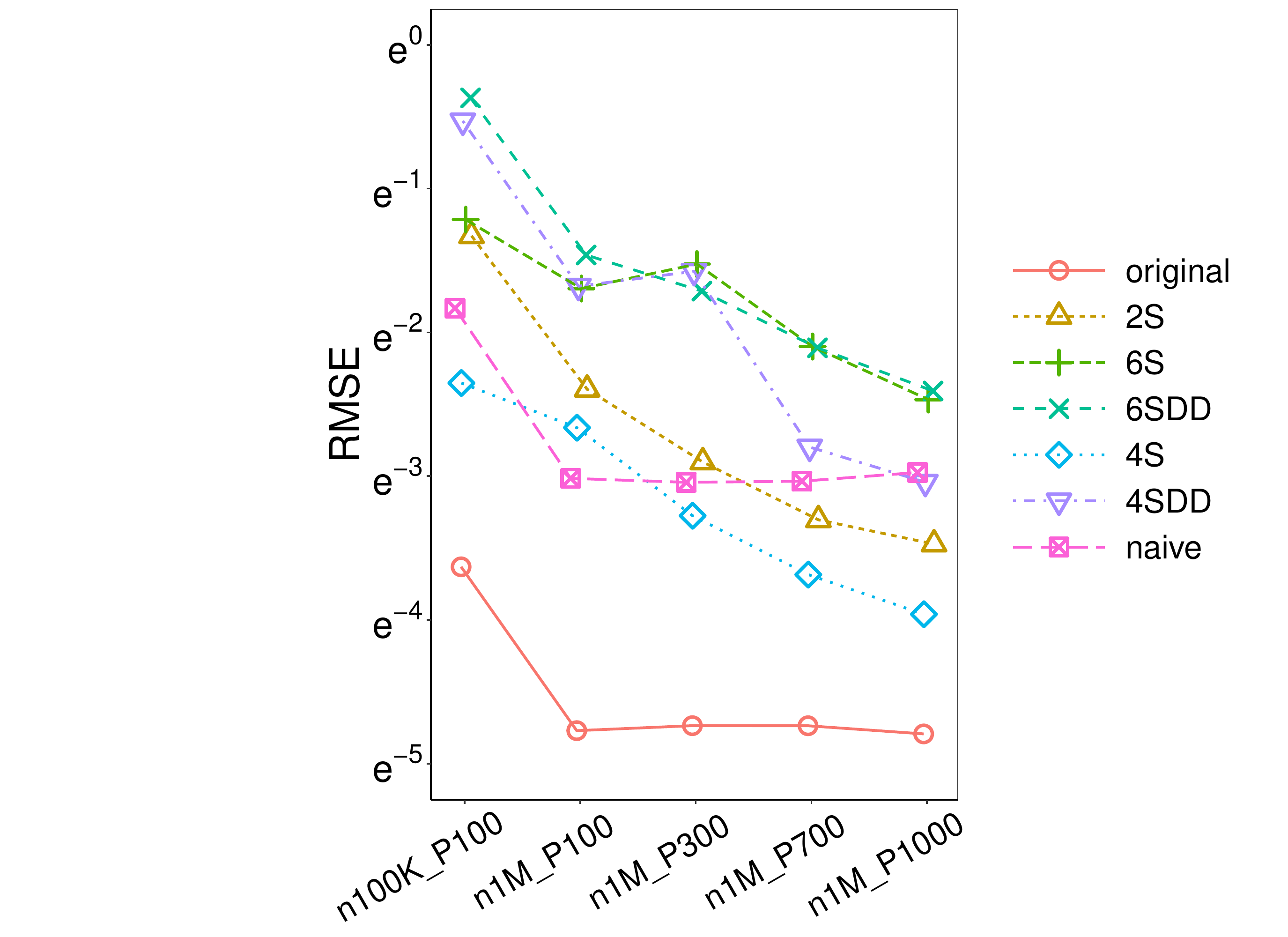}
\includegraphics[width=0.19\textwidth, trim={2.5in 0 2.6in 0},clip] {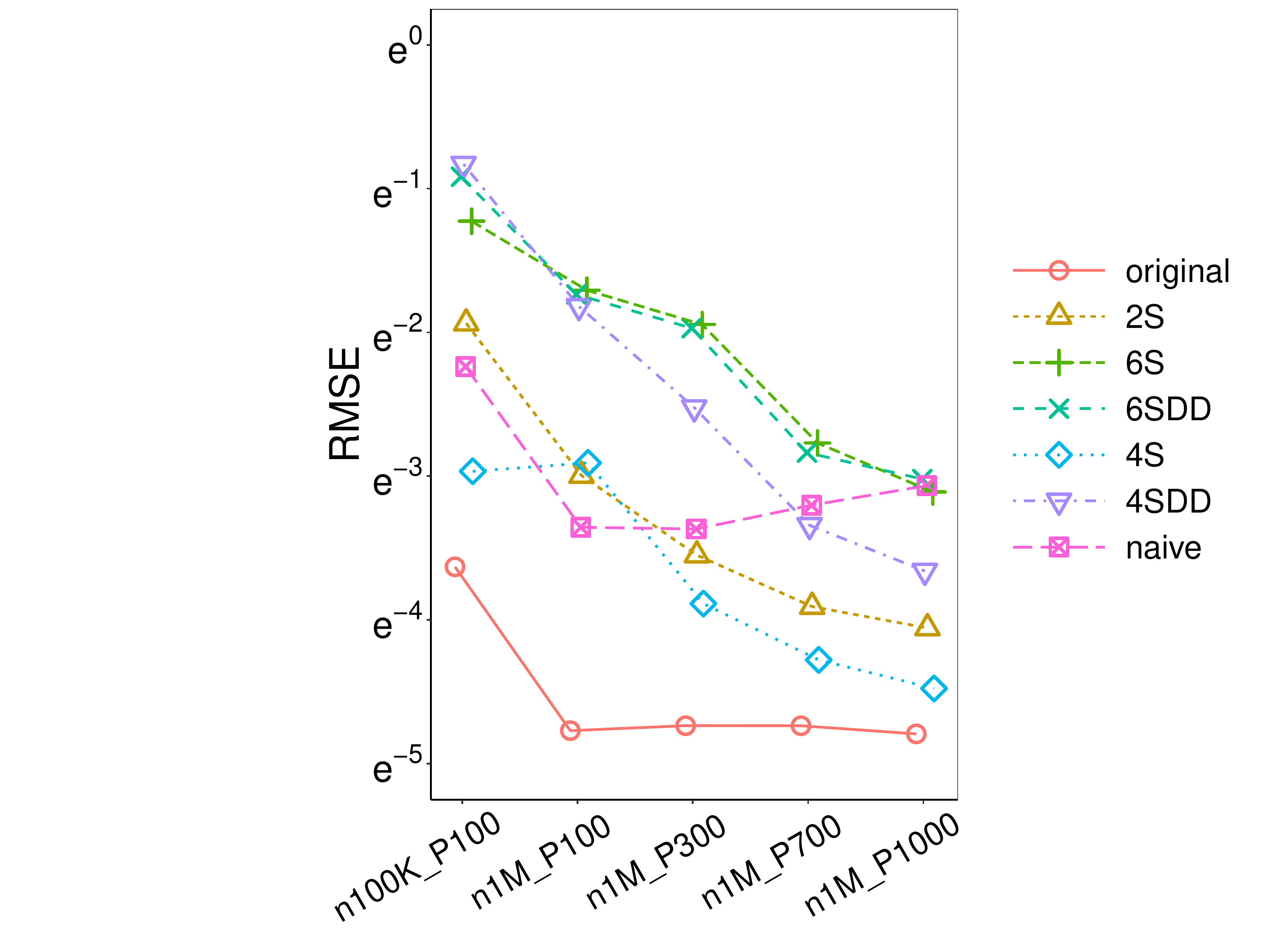}
\includegraphics[width=0.19\textwidth, trim={2.5in 0 2.6in 0},clip] {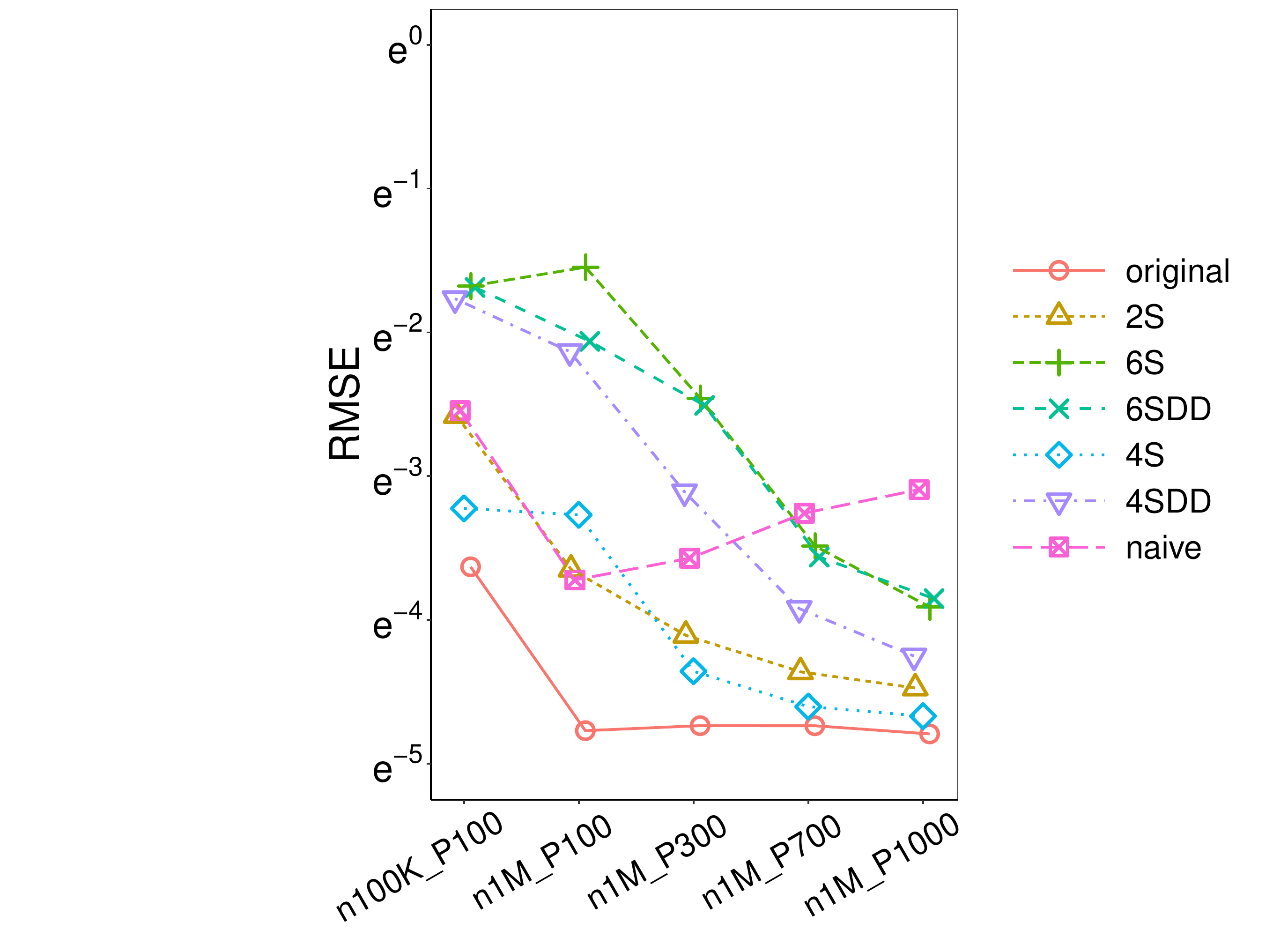}
\includegraphics[width=0.19\textwidth, trim={2.5in 0 2.6in 0},clip] {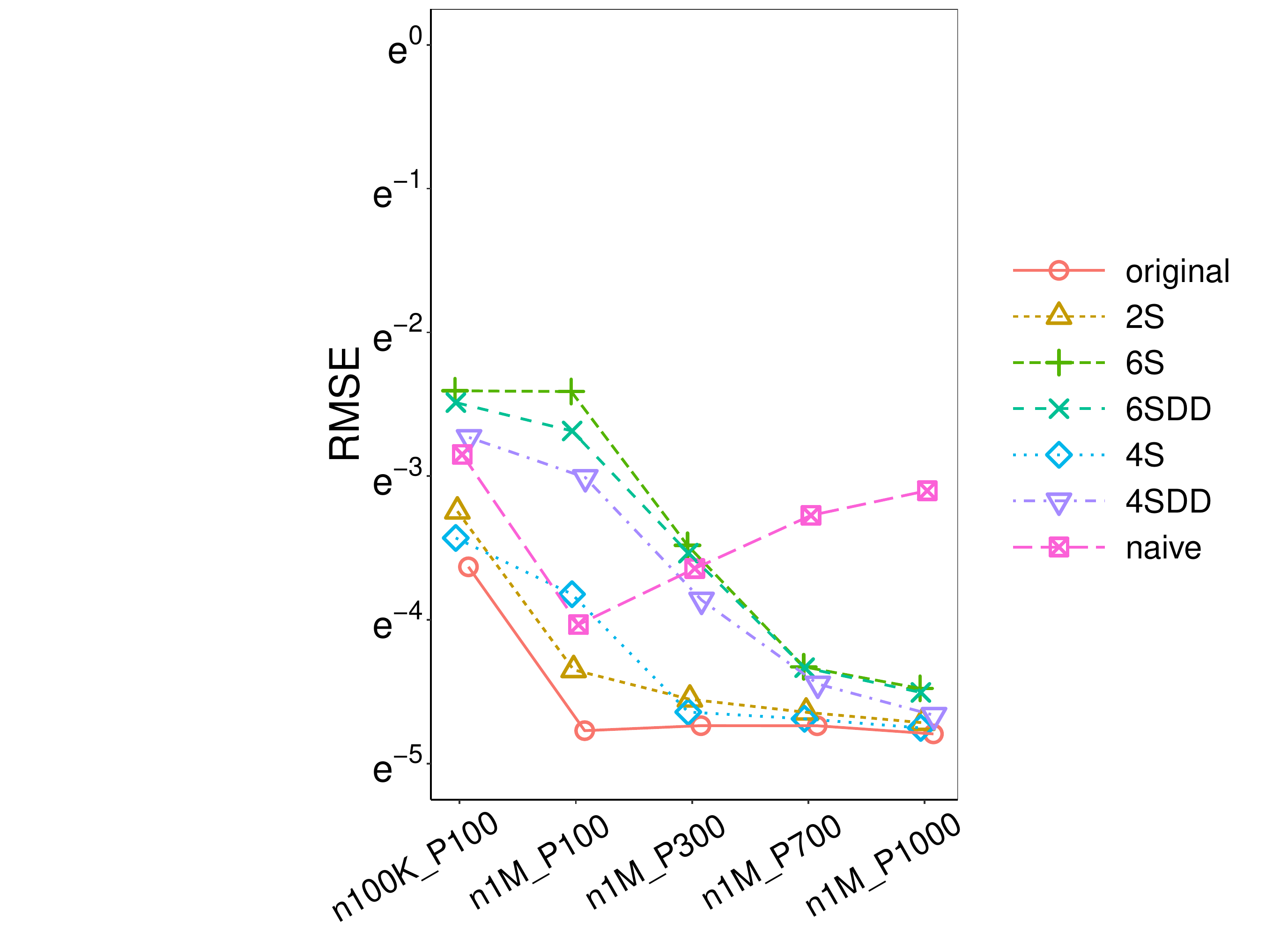}
\includegraphics[width=0.19\textwidth, trim={2.5in 0 2.6in 0},clip] {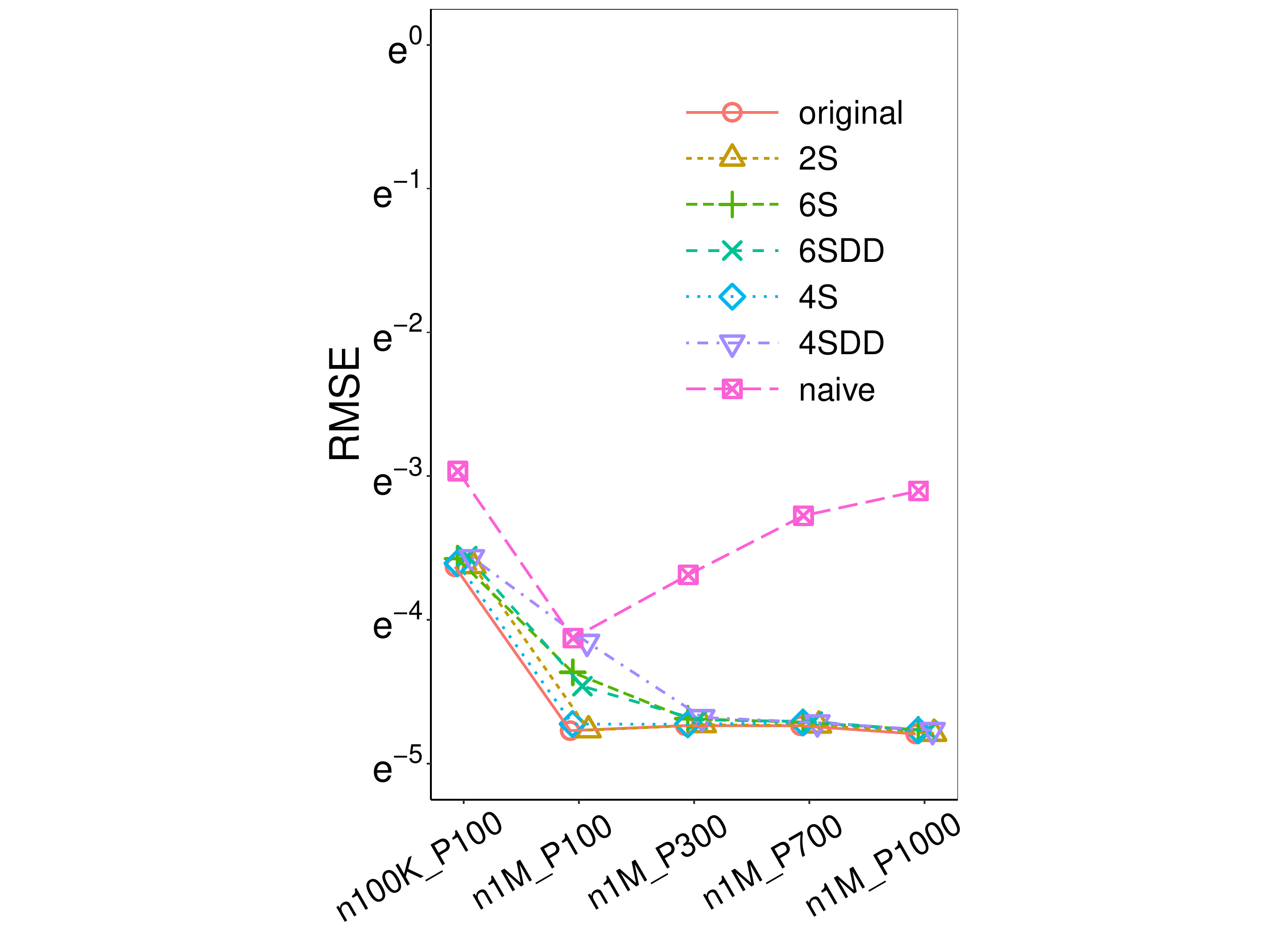}

\includegraphics[width=0.19\textwidth, trim={2.5in 0 2.6in 0},clip] {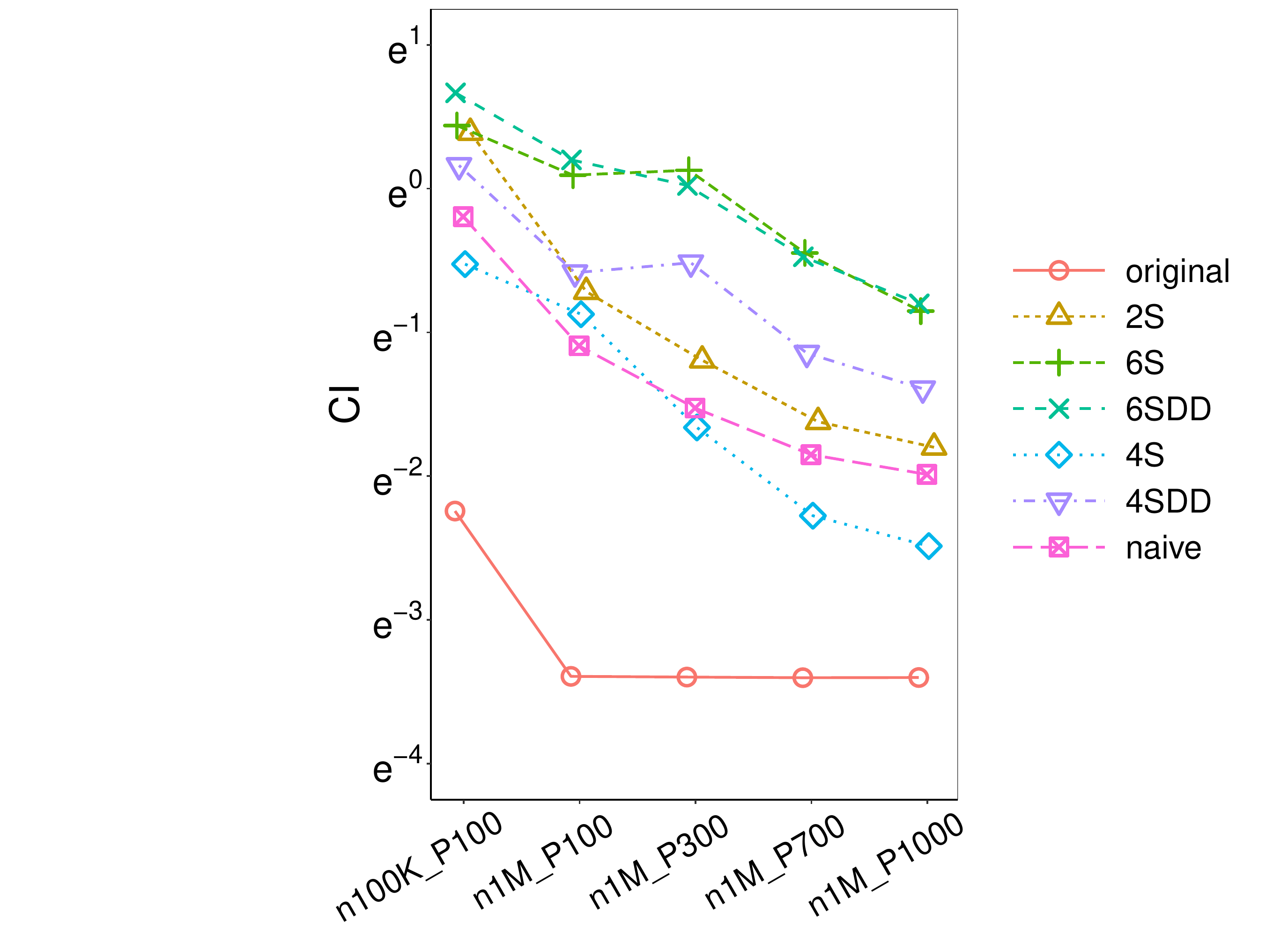}
\includegraphics[width=0.19\textwidth, trim={2.5in 0 2.6in 0},clip] {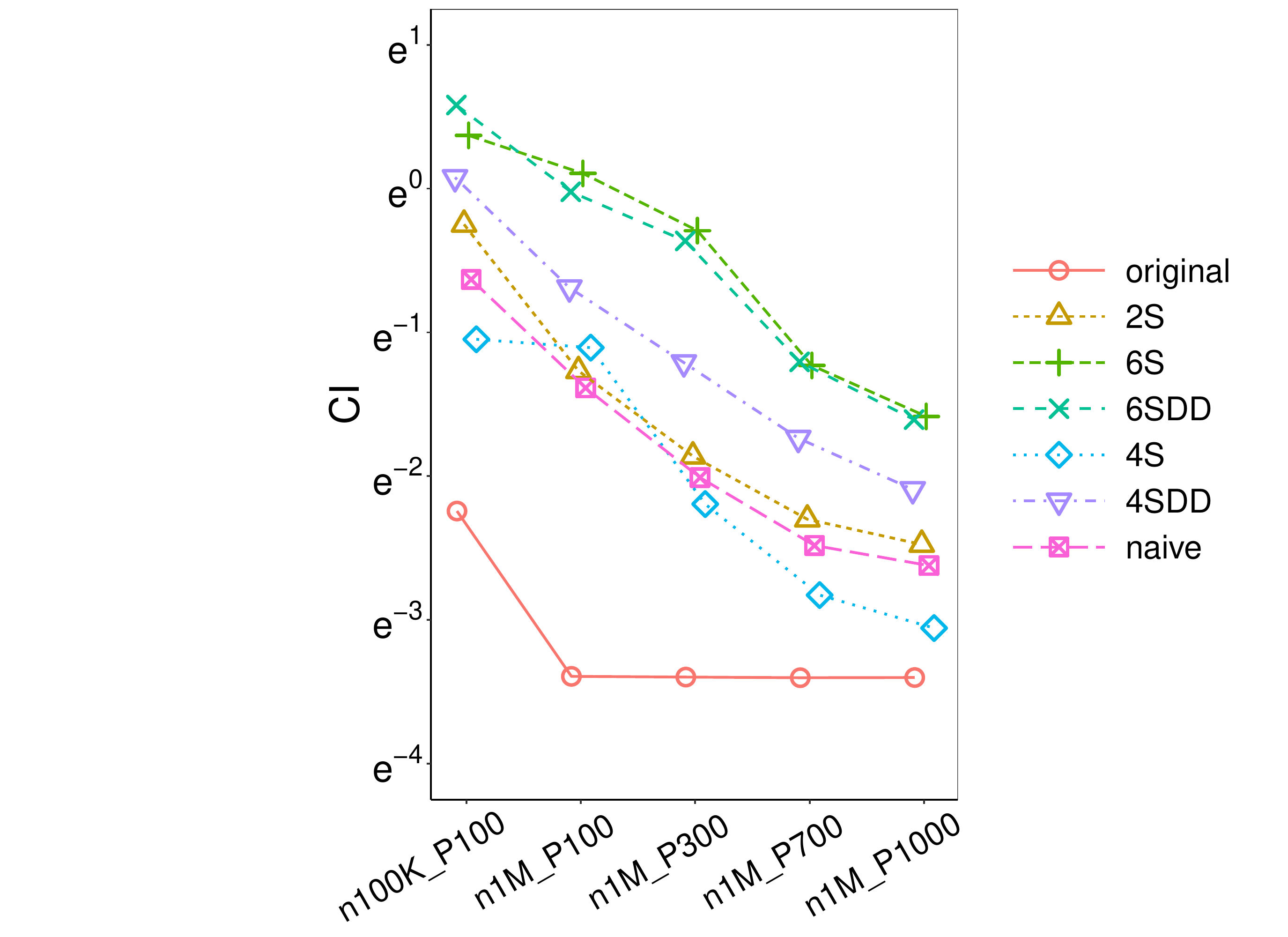}
\includegraphics[width=0.19\textwidth, trim={2.5in 0 2.6in 0},clip] {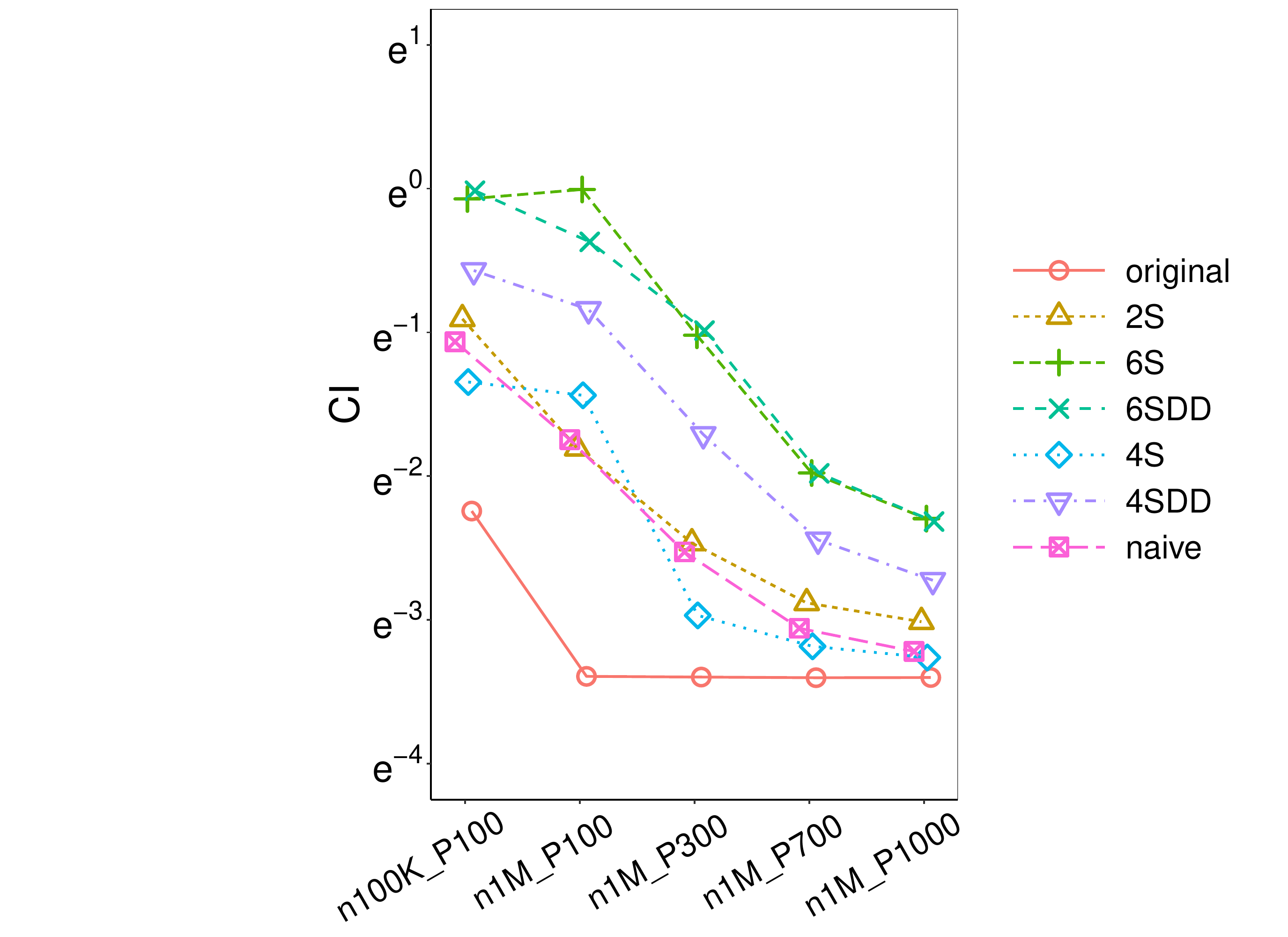}
\includegraphics[width=0.19\textwidth, trim={2.5in 0 2.6in 0},clip] {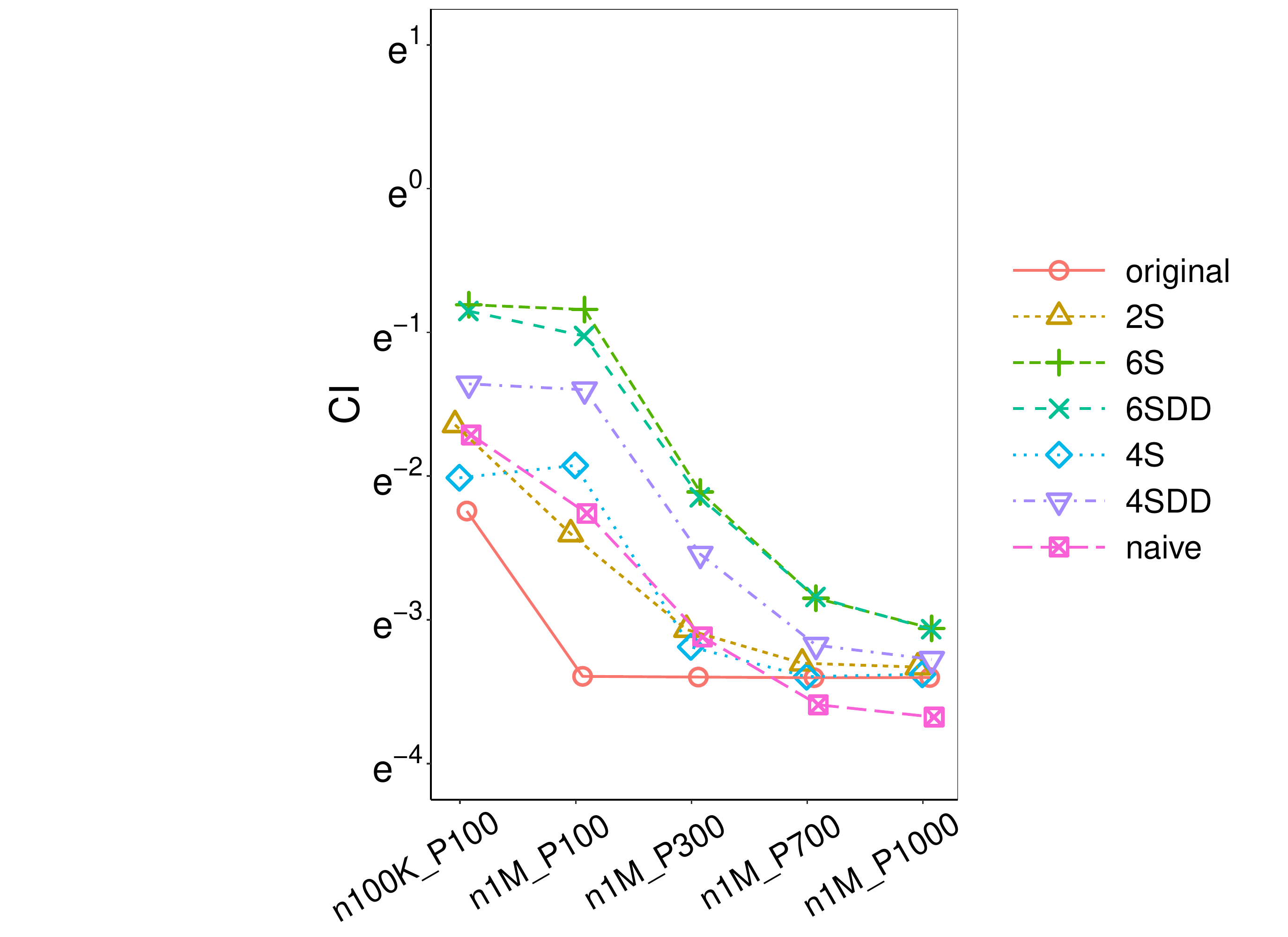}
\includegraphics[width=0.19\textwidth, trim={2.5in 0 2.6in 0},clip] {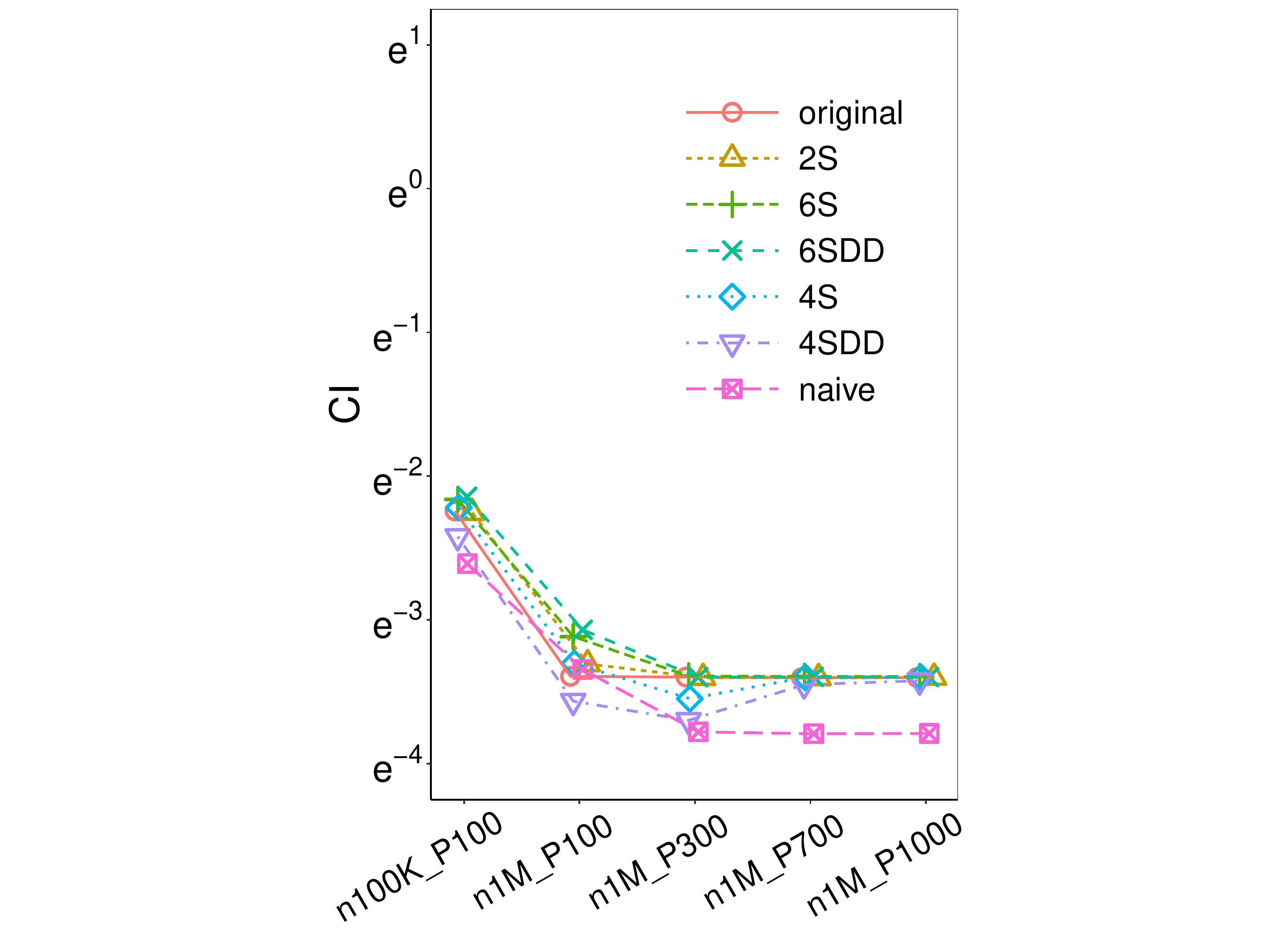}

\includegraphics[width=0.19\textwidth, trim={2.5in 0 2.6in 0},clip] {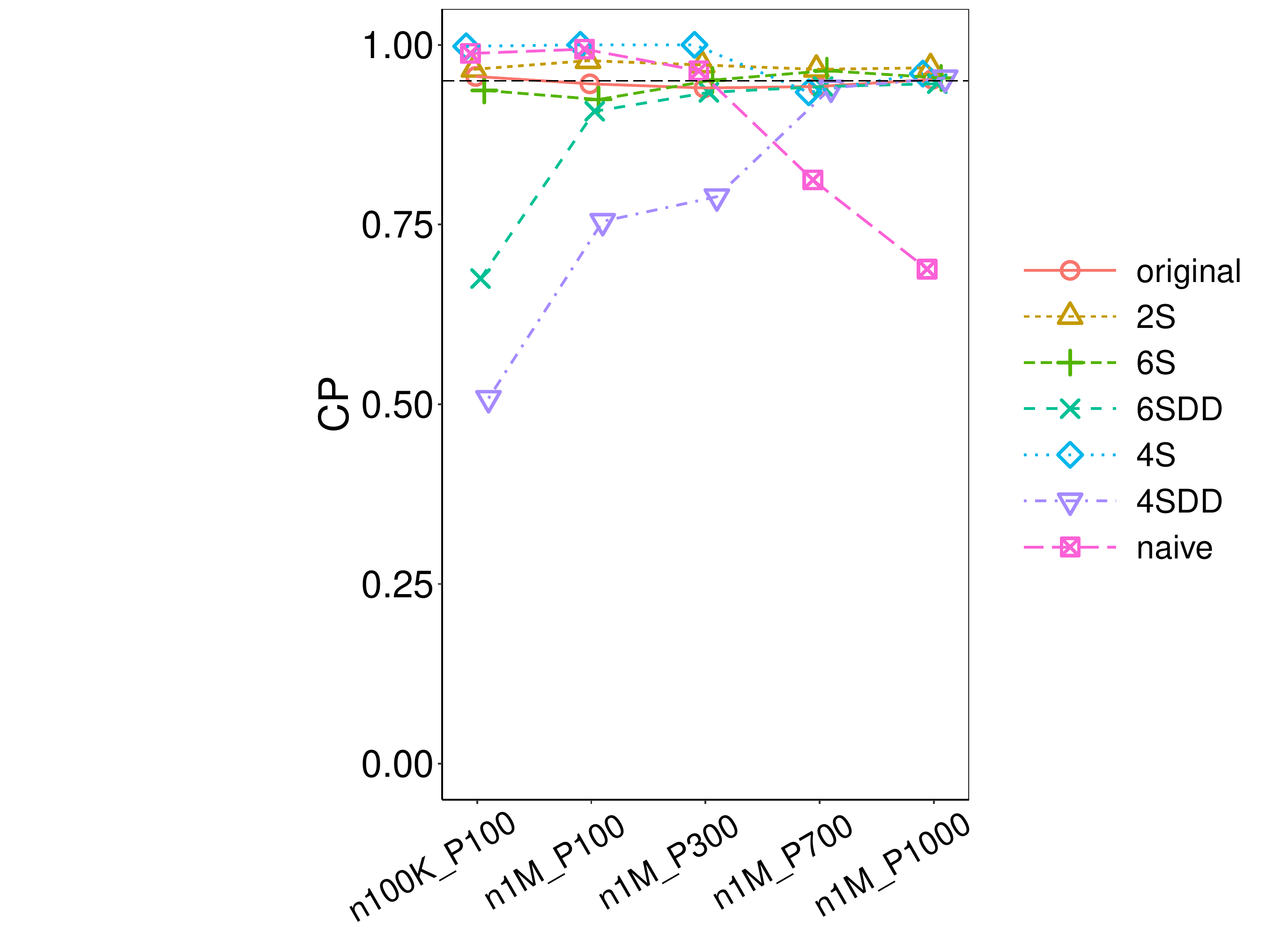}
\includegraphics[width=0.19\textwidth, trim={2.5in 0 2.6in 0},clip] {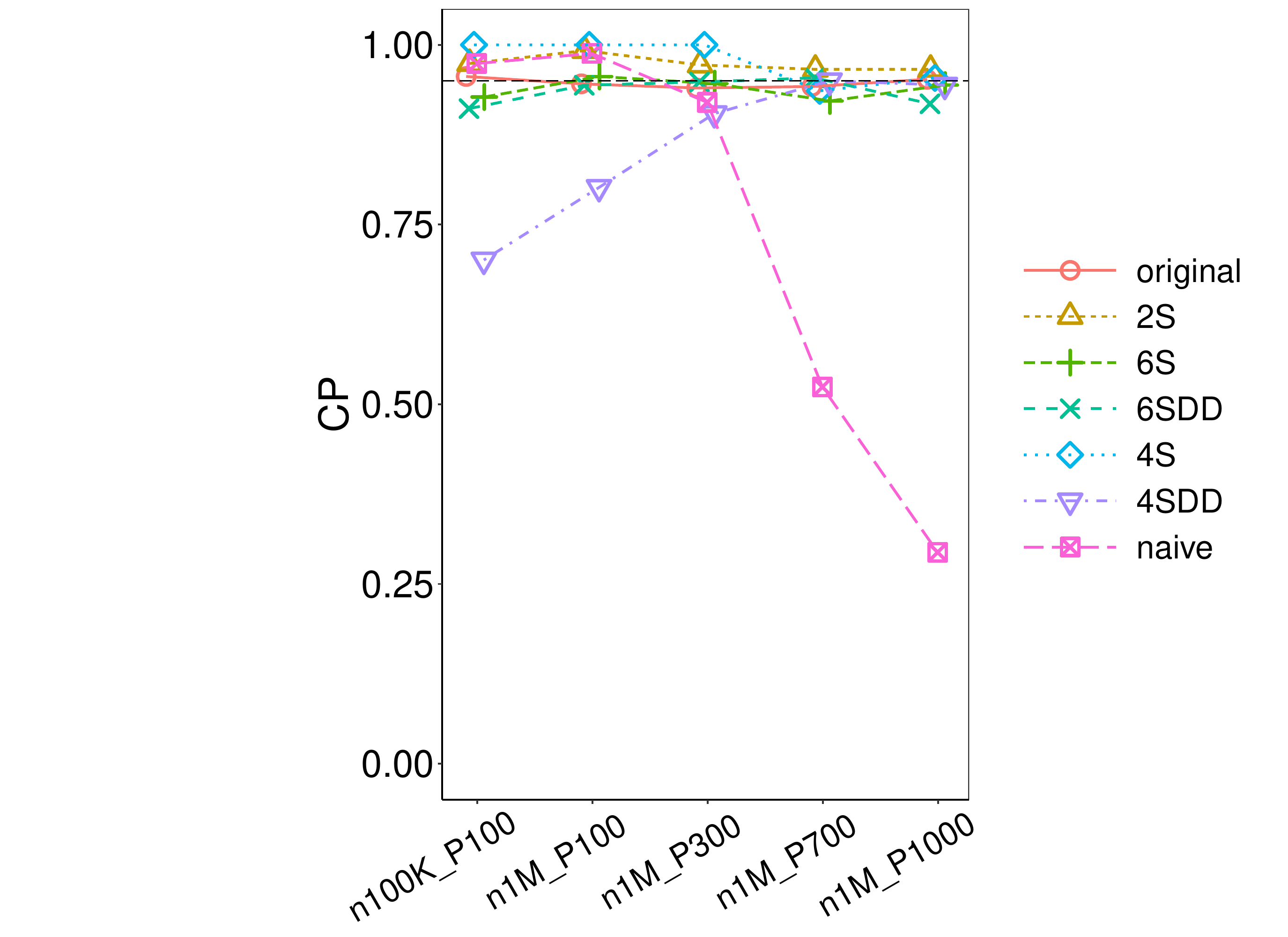}
\includegraphics[width=0.19\textwidth, trim={2.5in 0 2.6in 0},clip] {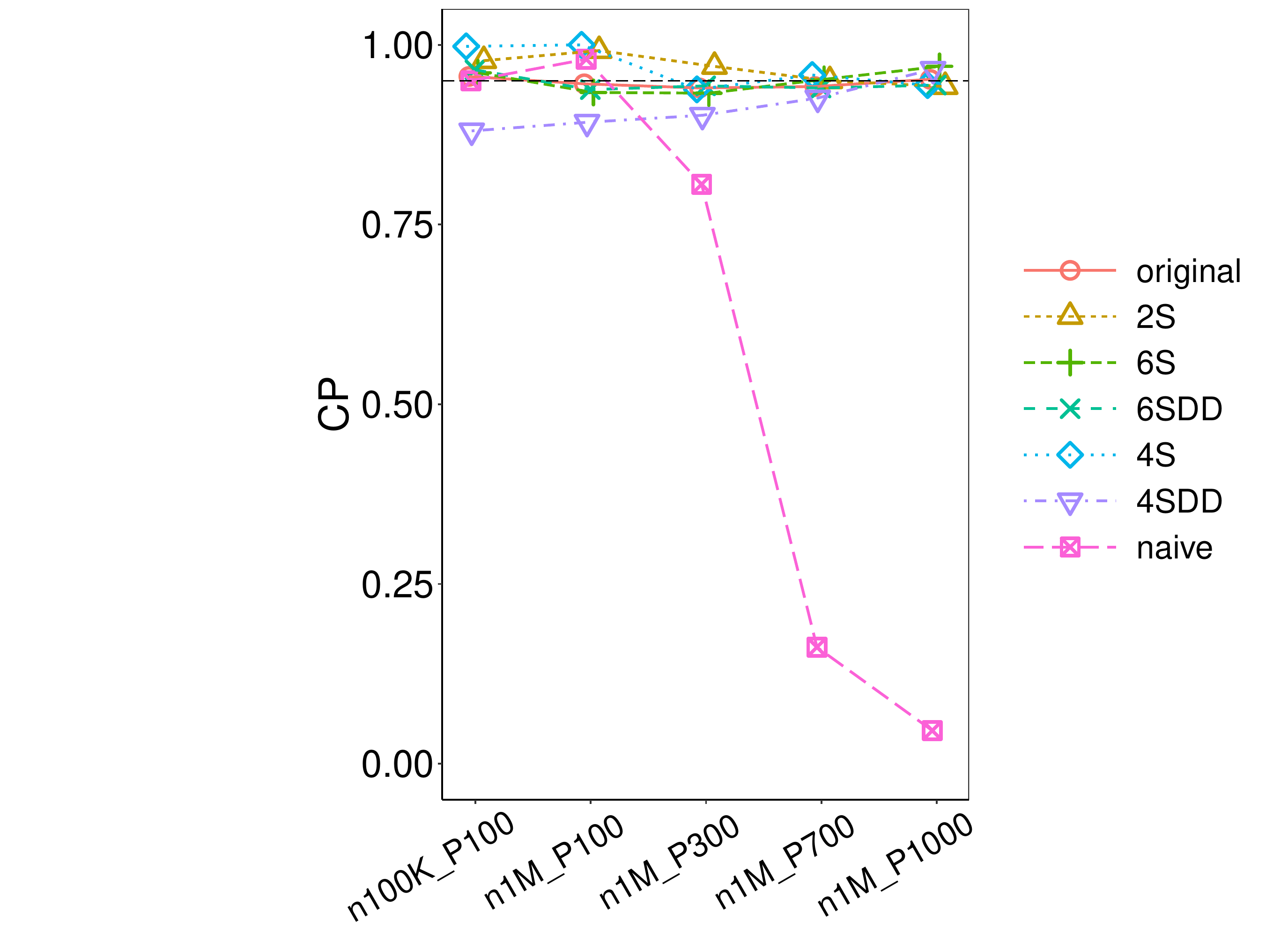}
\includegraphics[width=0.19\textwidth, trim={2.5in 0 2.6in 0},clip] {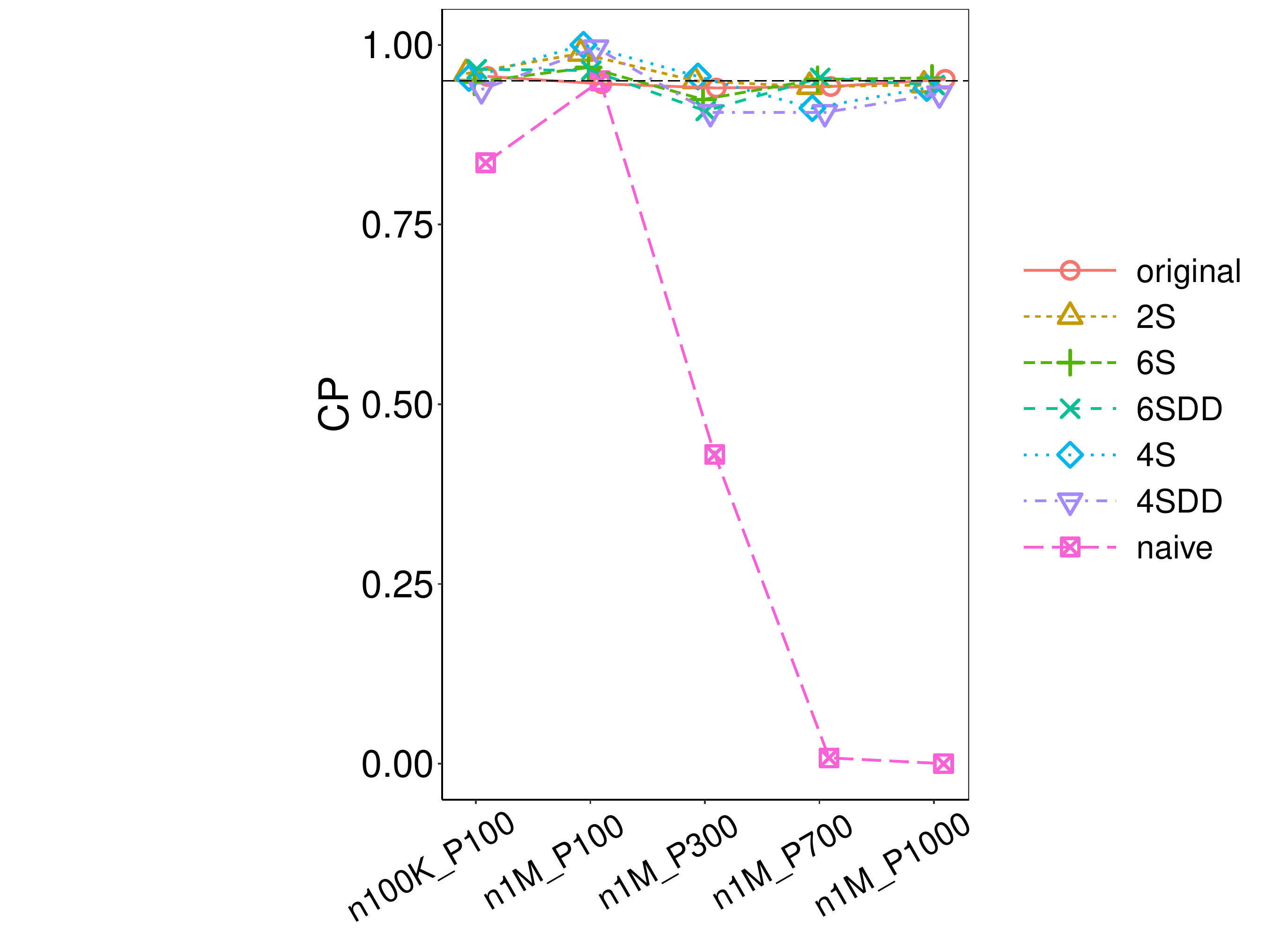}
\includegraphics[width=0.19\textwidth, trim={2.5in 0 2.6in 0},clip] {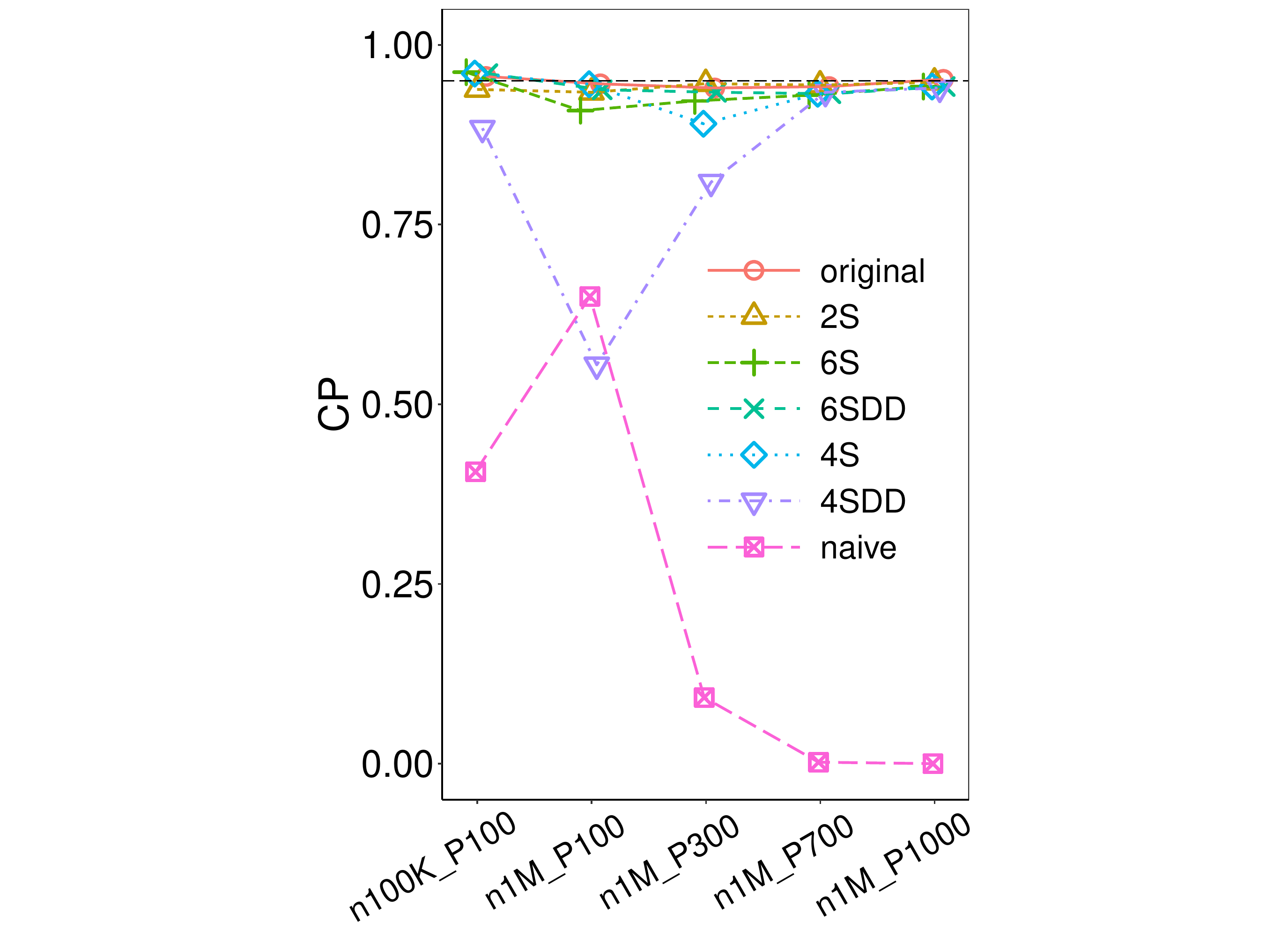}

\includegraphics[width=0.19\textwidth, trim={2.5in 0 2.6in 0},clip] {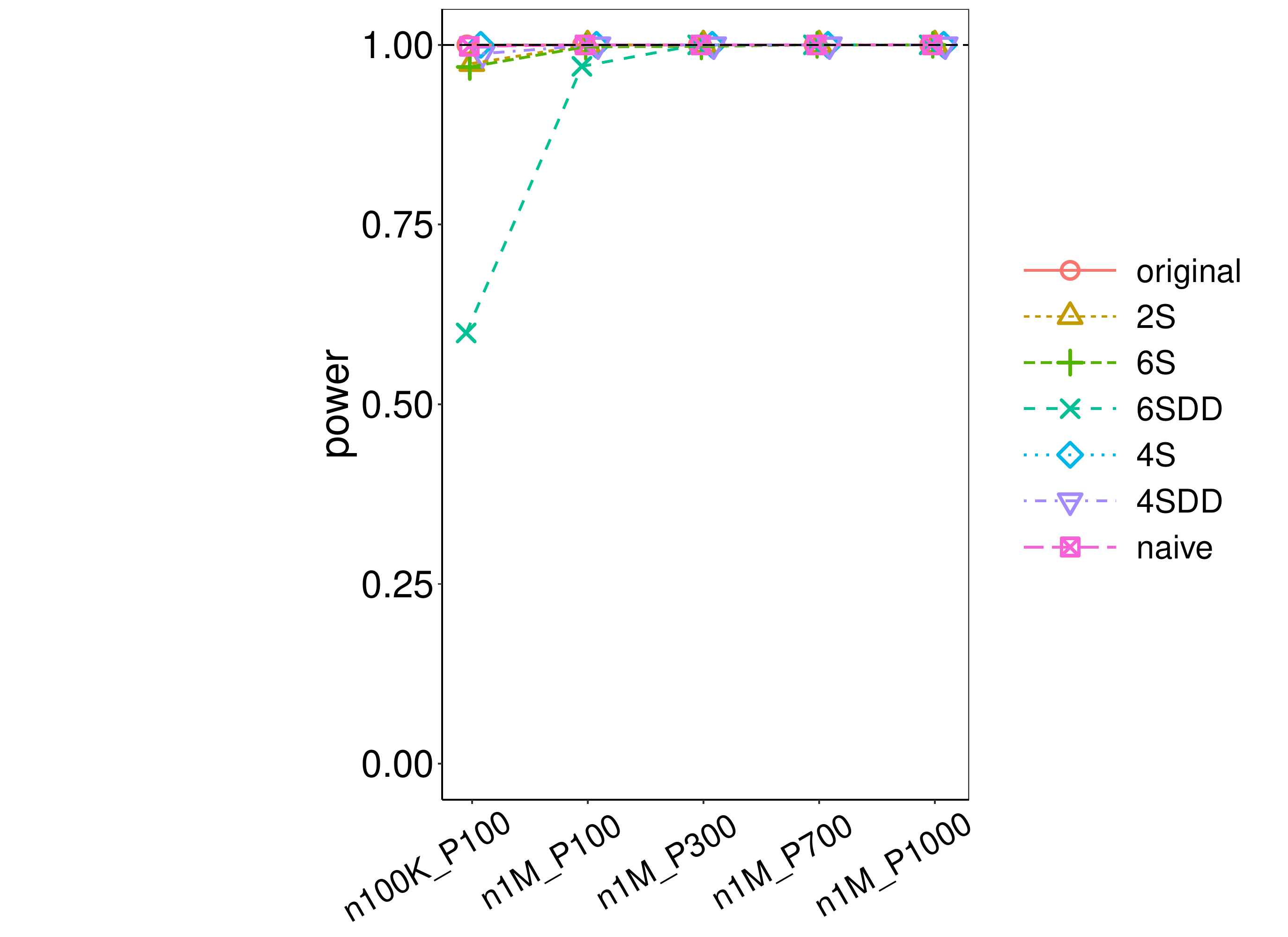}
\includegraphics[width=0.19\textwidth, trim={2.5in 0 2.6in 0},clip] {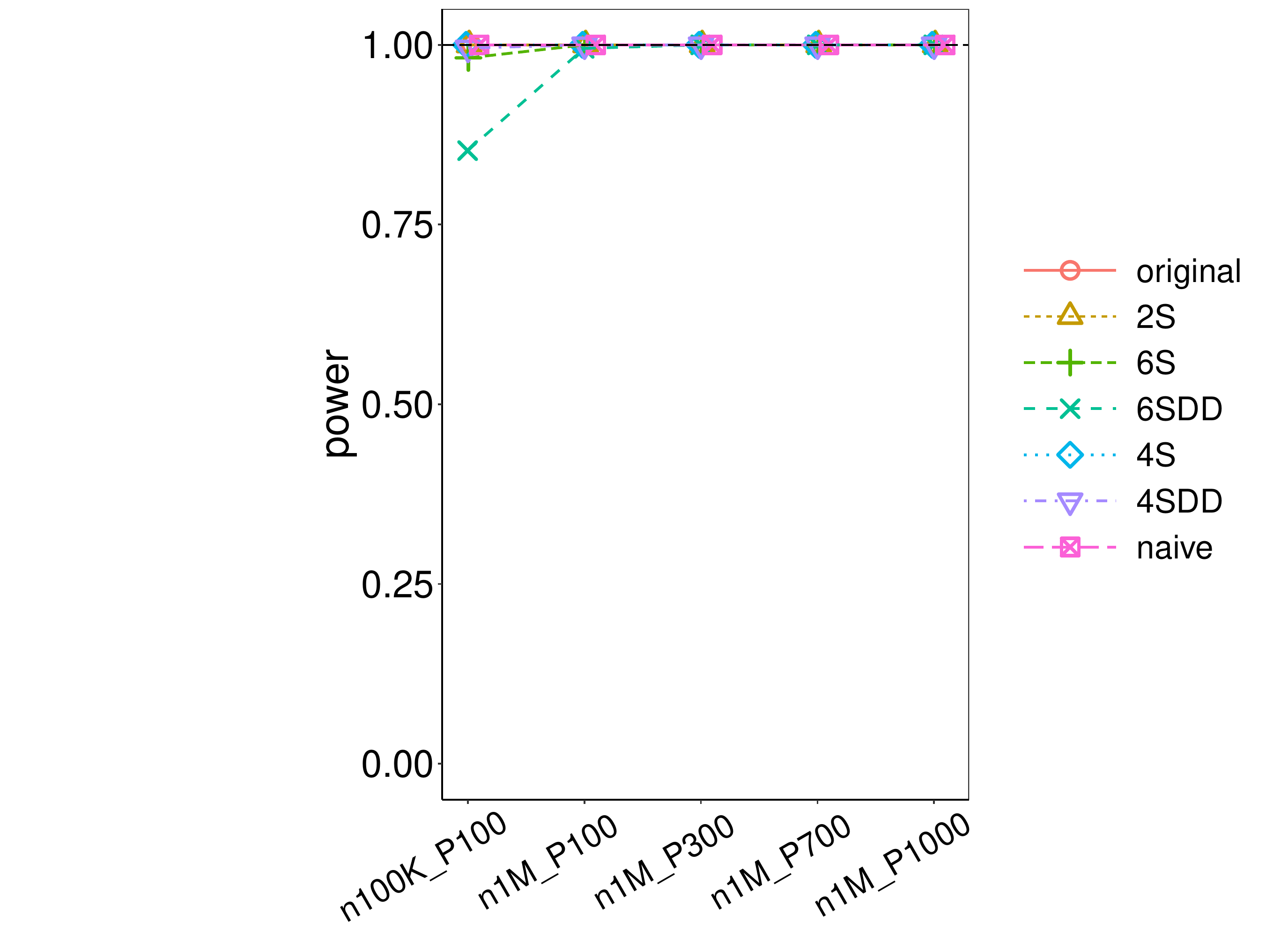}
\includegraphics[width=0.19\textwidth, trim={2.5in 0 2.6in 0},clip] {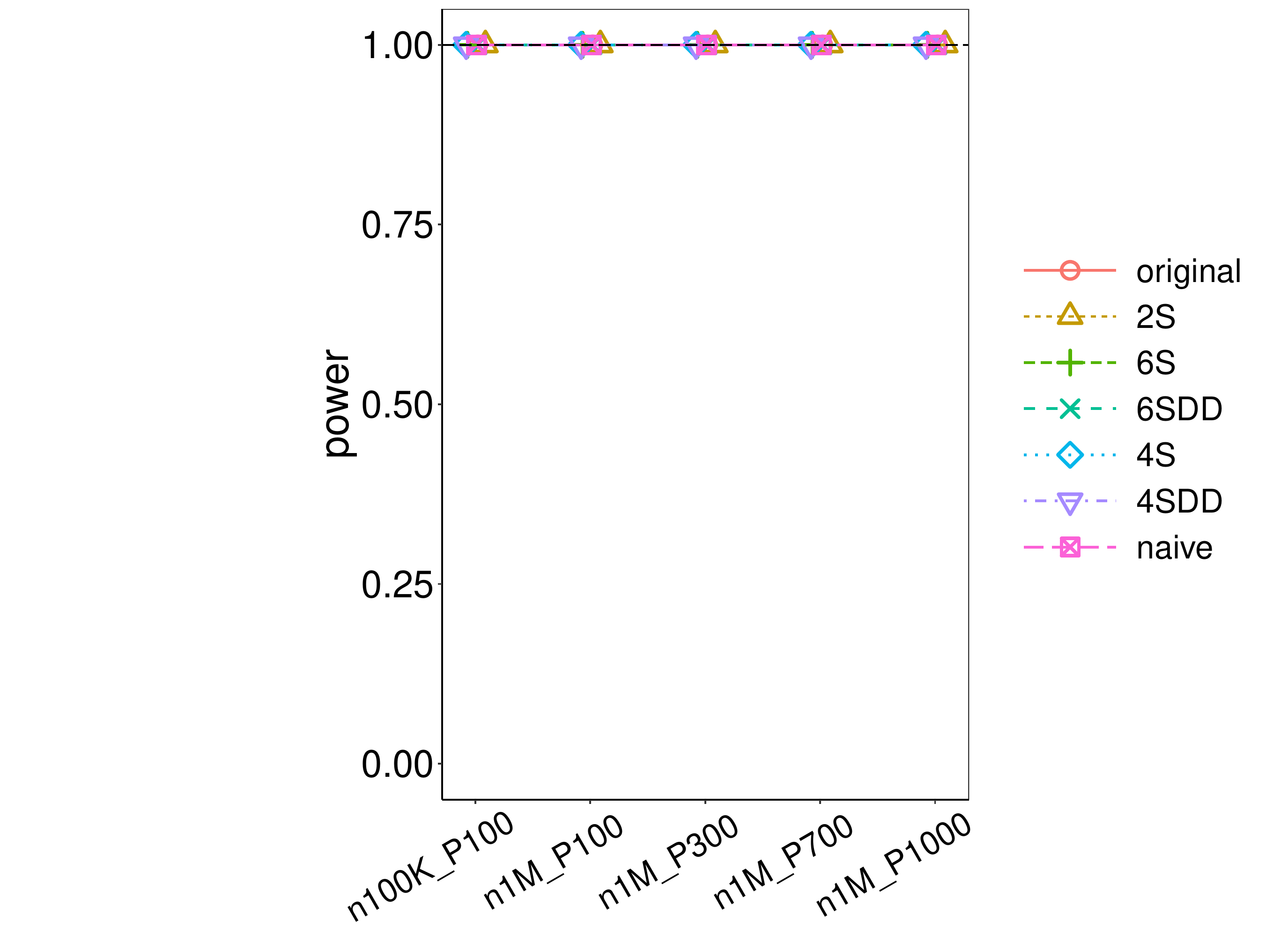}
\includegraphics[width=0.19\textwidth, trim={2.5in 0 2.6in 0},clip] {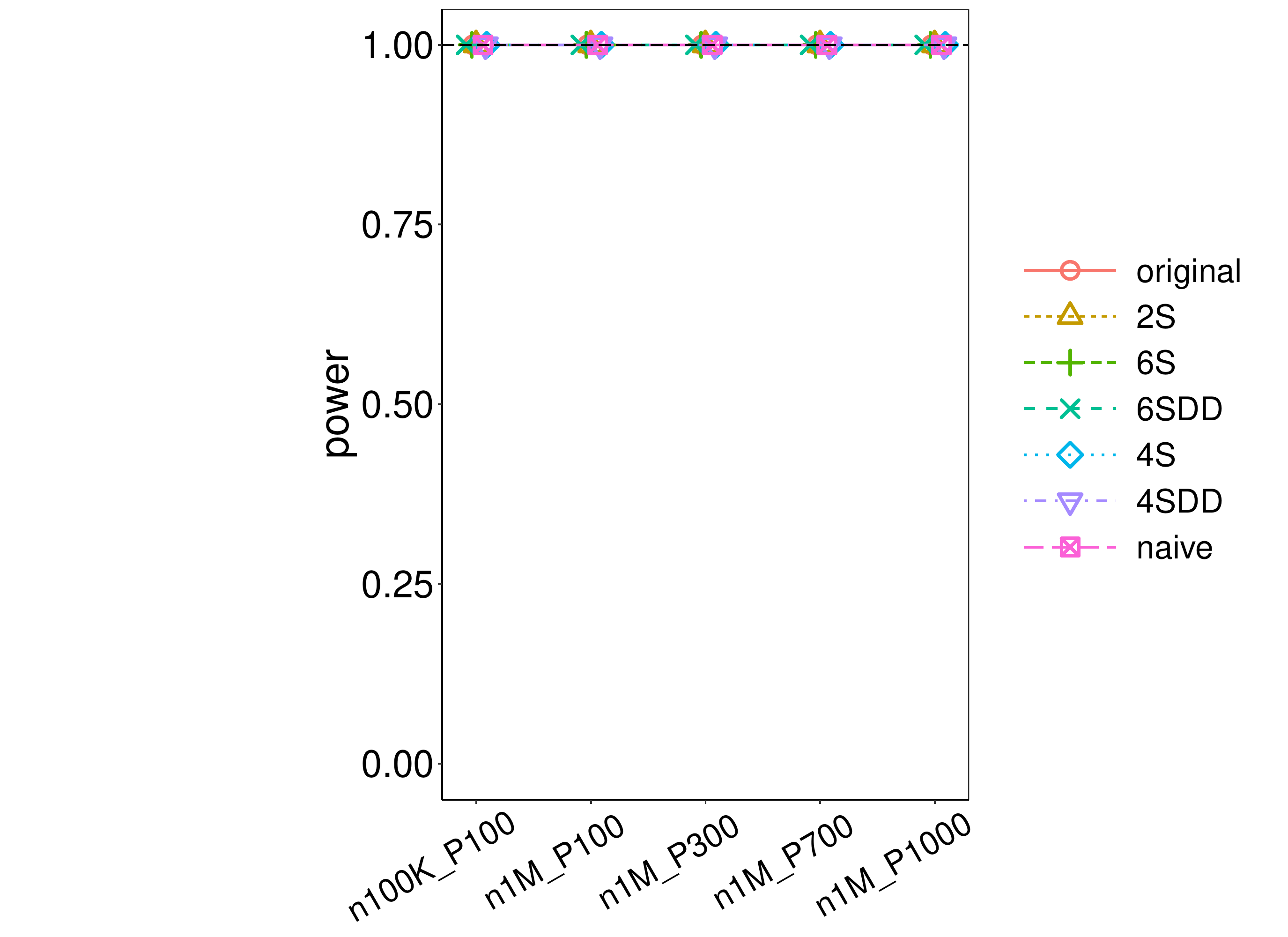}
\includegraphics[width=0.19\textwidth, trim={2.5in 0 2.6in 0},clip] {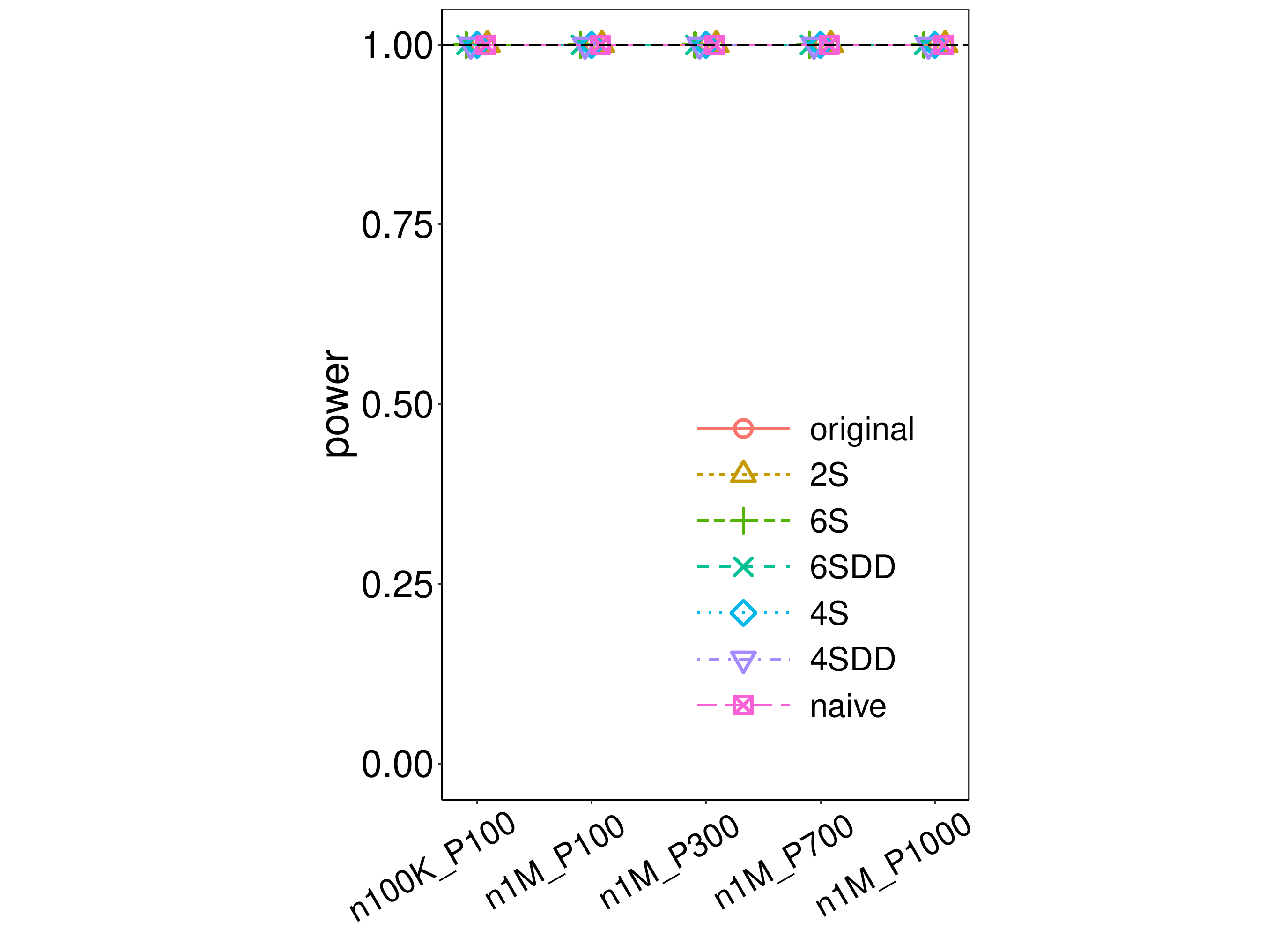}

\caption{Simulation results with $\rho$-DP for Gaussian data with  $\alpha\ne\beta$ when $\theta\ne0$} \label{fig:1asDPN}

\end{figure}

\begin{figure}[!htb]
\hspace{0.45in}$\rho=0.005$\hspace{0.65in}$\rho=0.02$\hspace{0.65in}$\rho=0.08$
\hspace{0.65in}$\rho=0.32$\hspace{0.65in}$\rho=1.28$

\includegraphics[width=0.19\textwidth, trim={2.5in 0 2.6in 0},clip] {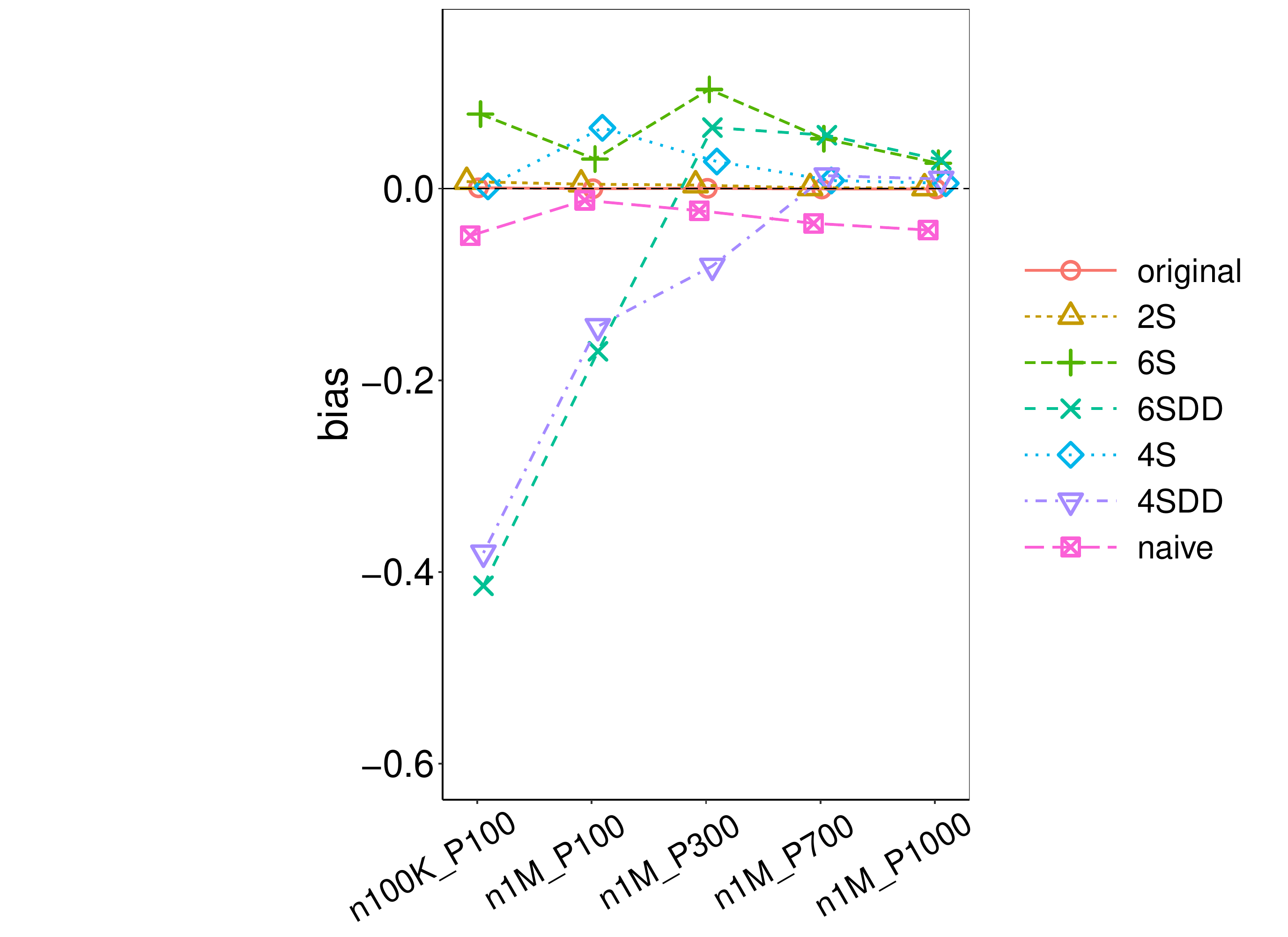}
\includegraphics[width=0.19\textwidth, trim={2.5in 0 2.6in 0},clip] {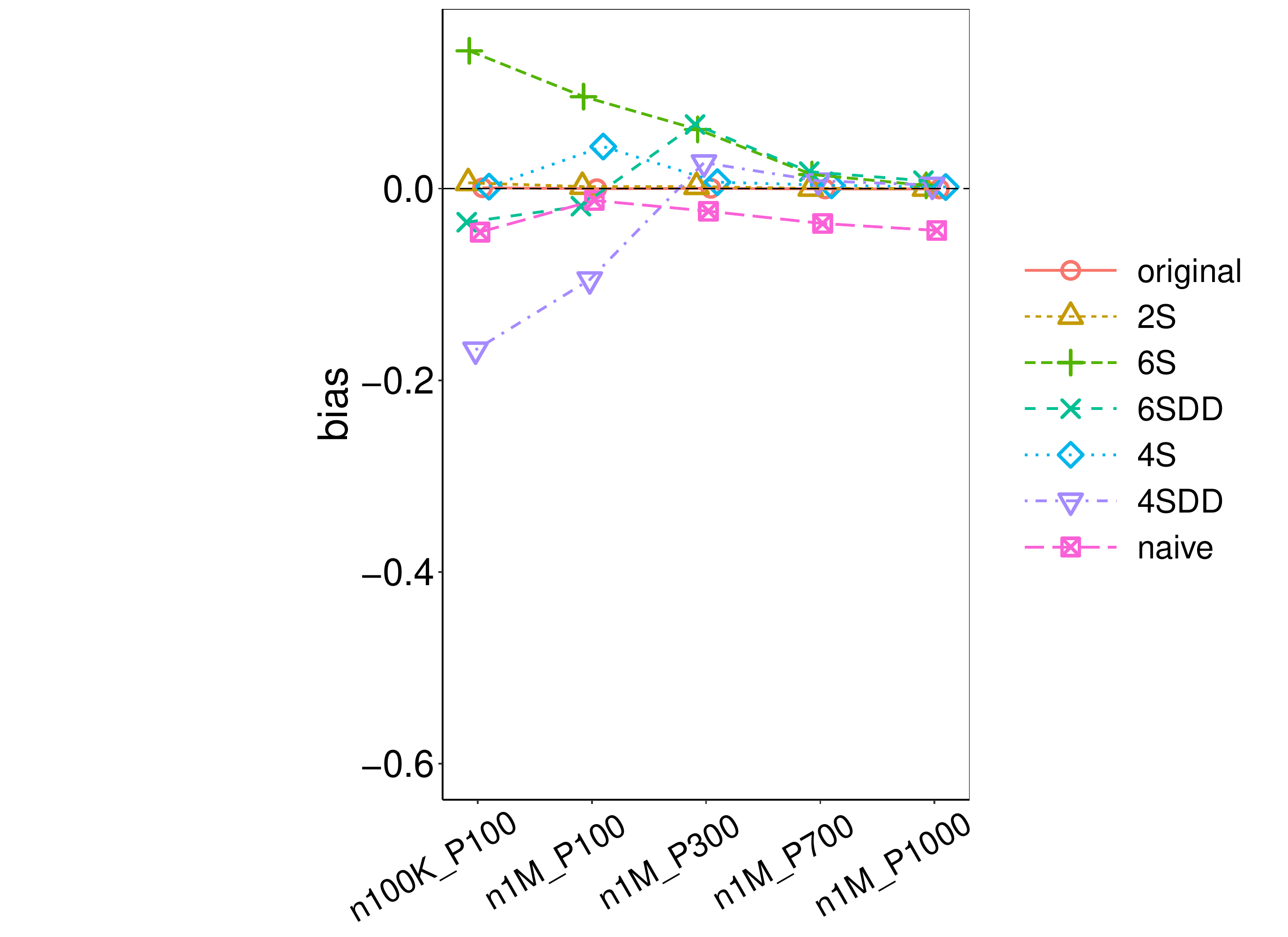}
\includegraphics[width=0.19\textwidth, trim={2.5in 0 2.6in 0},clip] {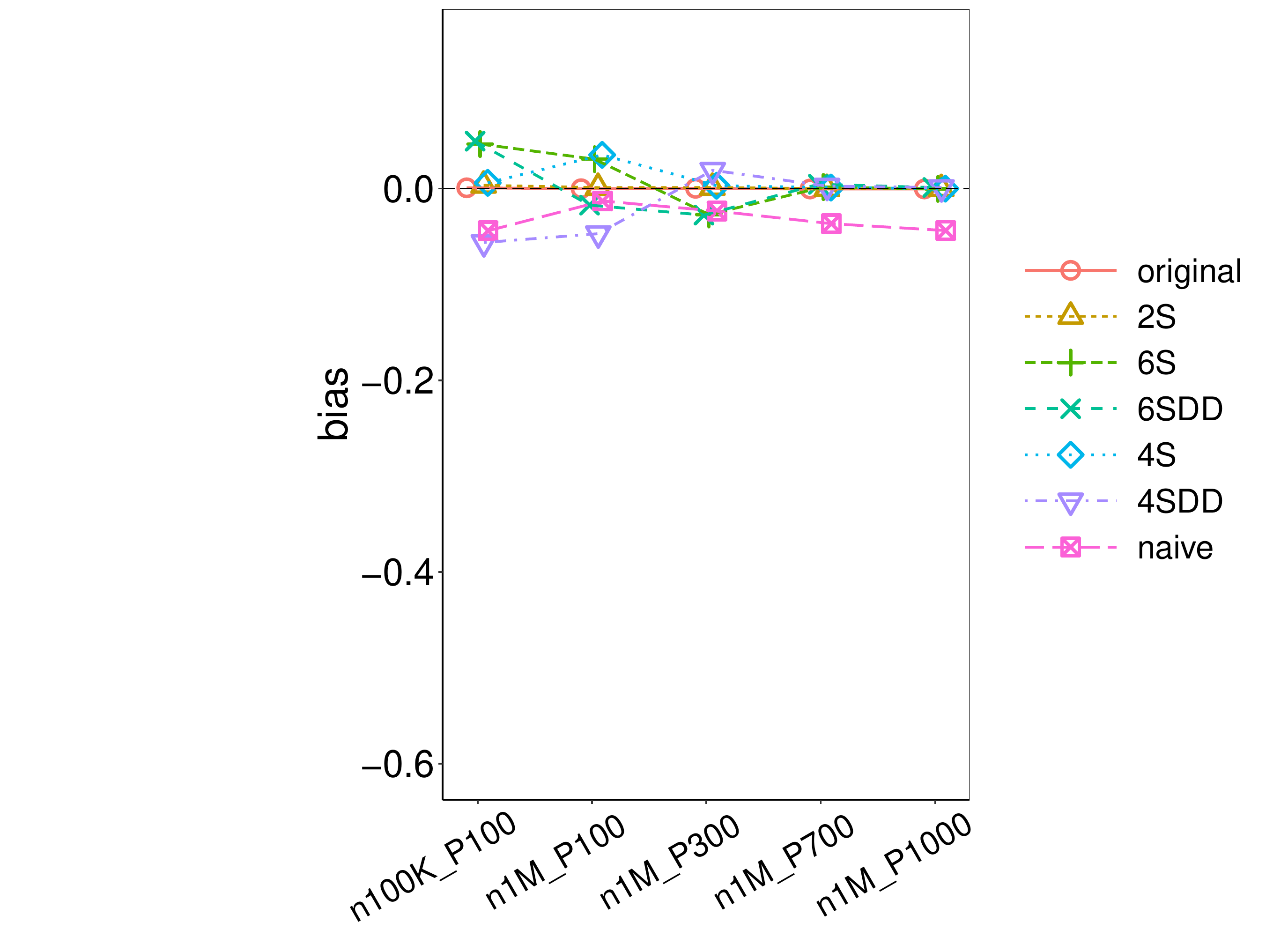}
\includegraphics[width=0.19\textwidth, trim={2.5in 0 2.6in 0},clip] {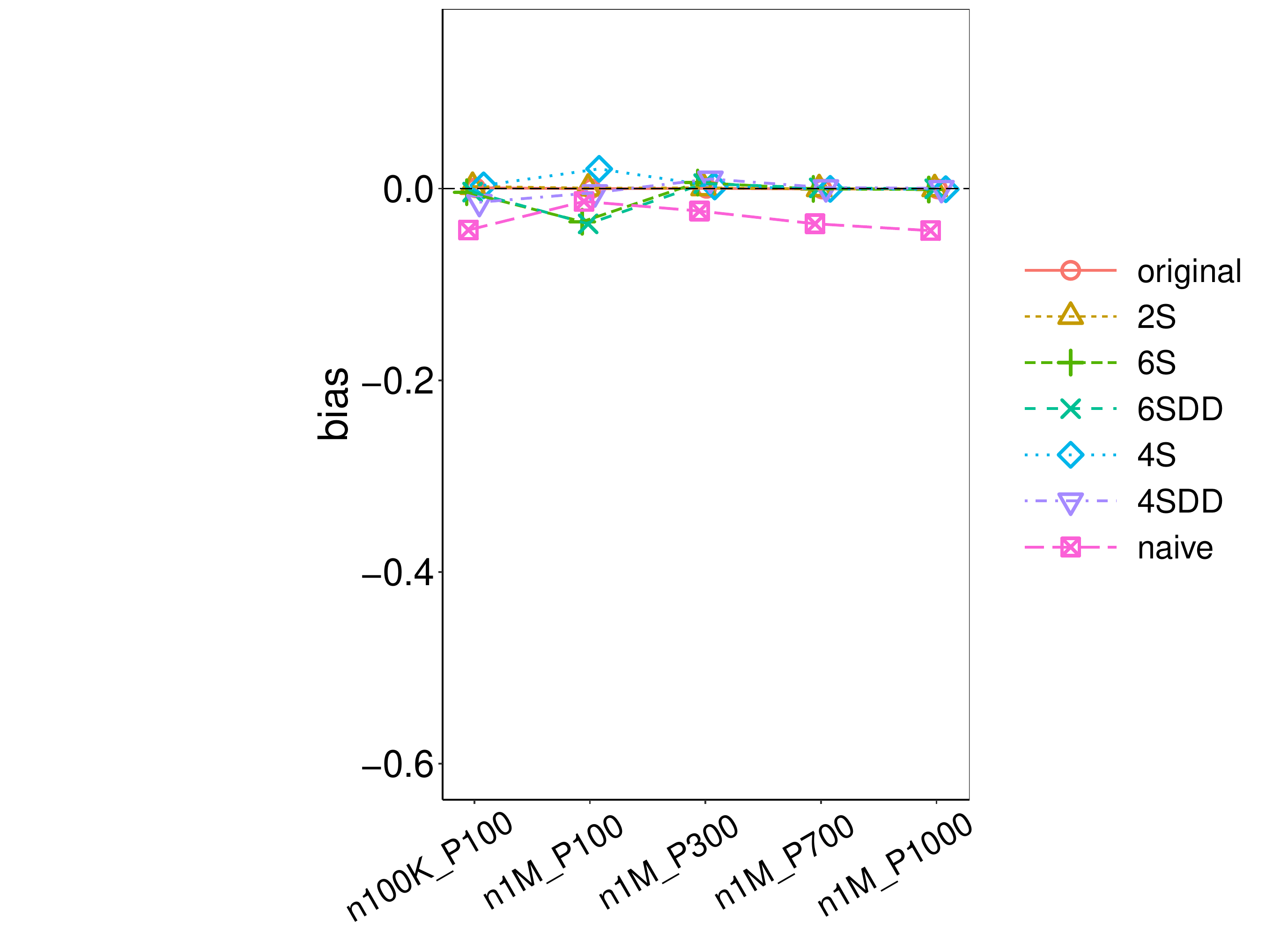}
\includegraphics[width=0.19\textwidth, trim={2.5in 0 2.6in 0},clip] {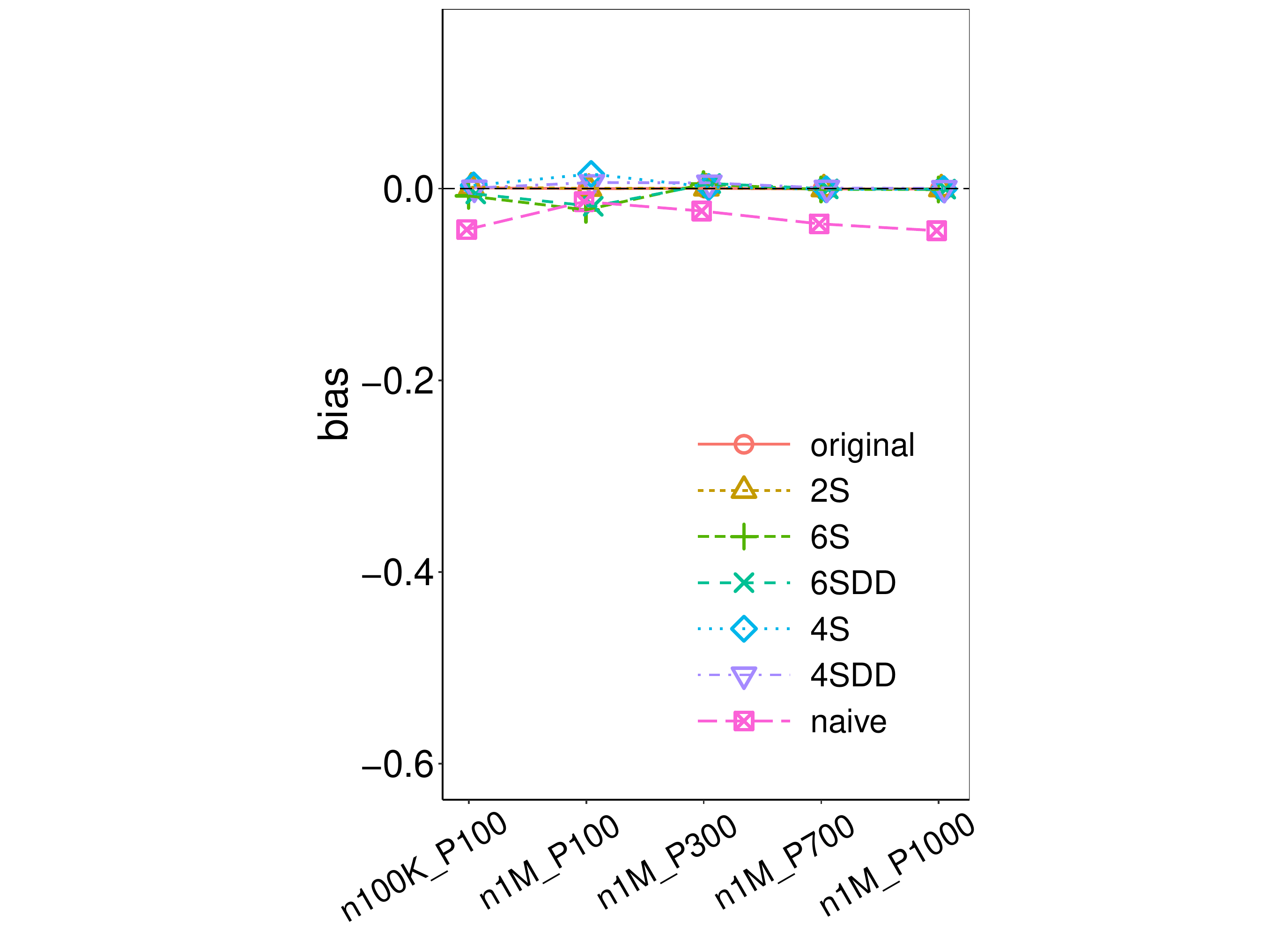}

\includegraphics[width=0.19\textwidth, trim={2.5in 0 2.6in 0},clip] {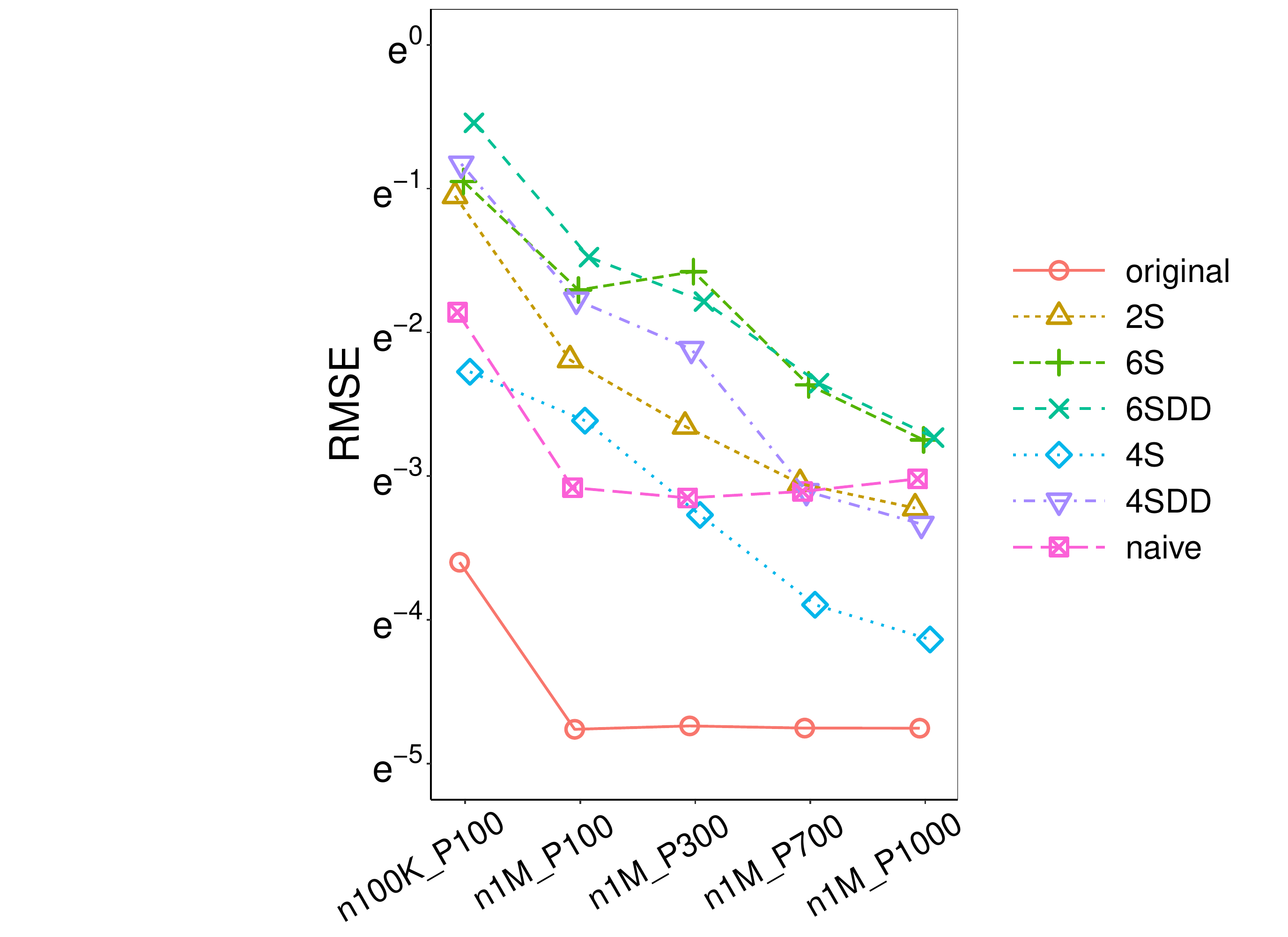}
\includegraphics[width=0.19\textwidth, trim={2.5in 0 2.6in 0},clip] {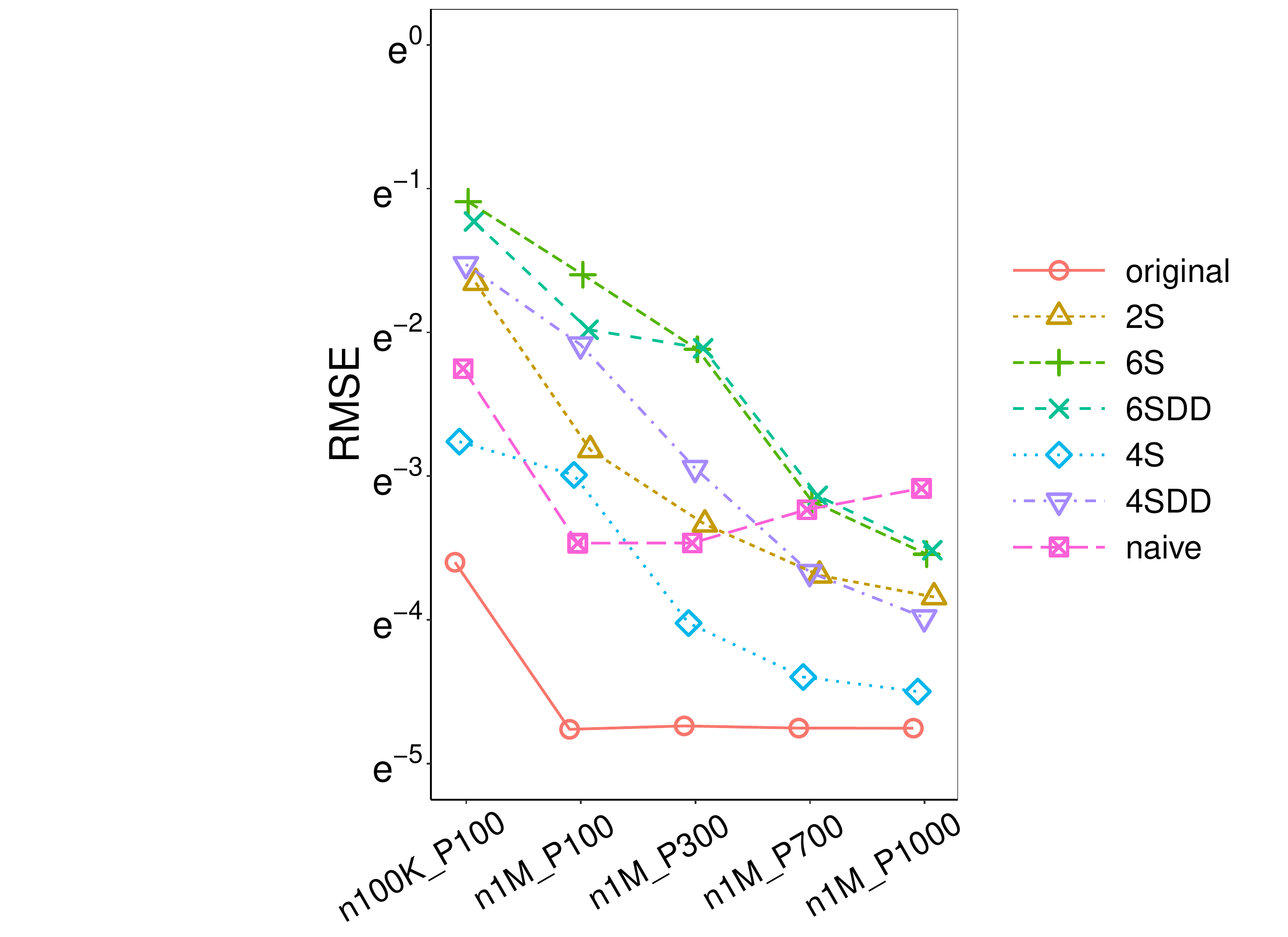}
\includegraphics[width=0.19\textwidth, trim={2.5in 0 2.6in 0},clip] {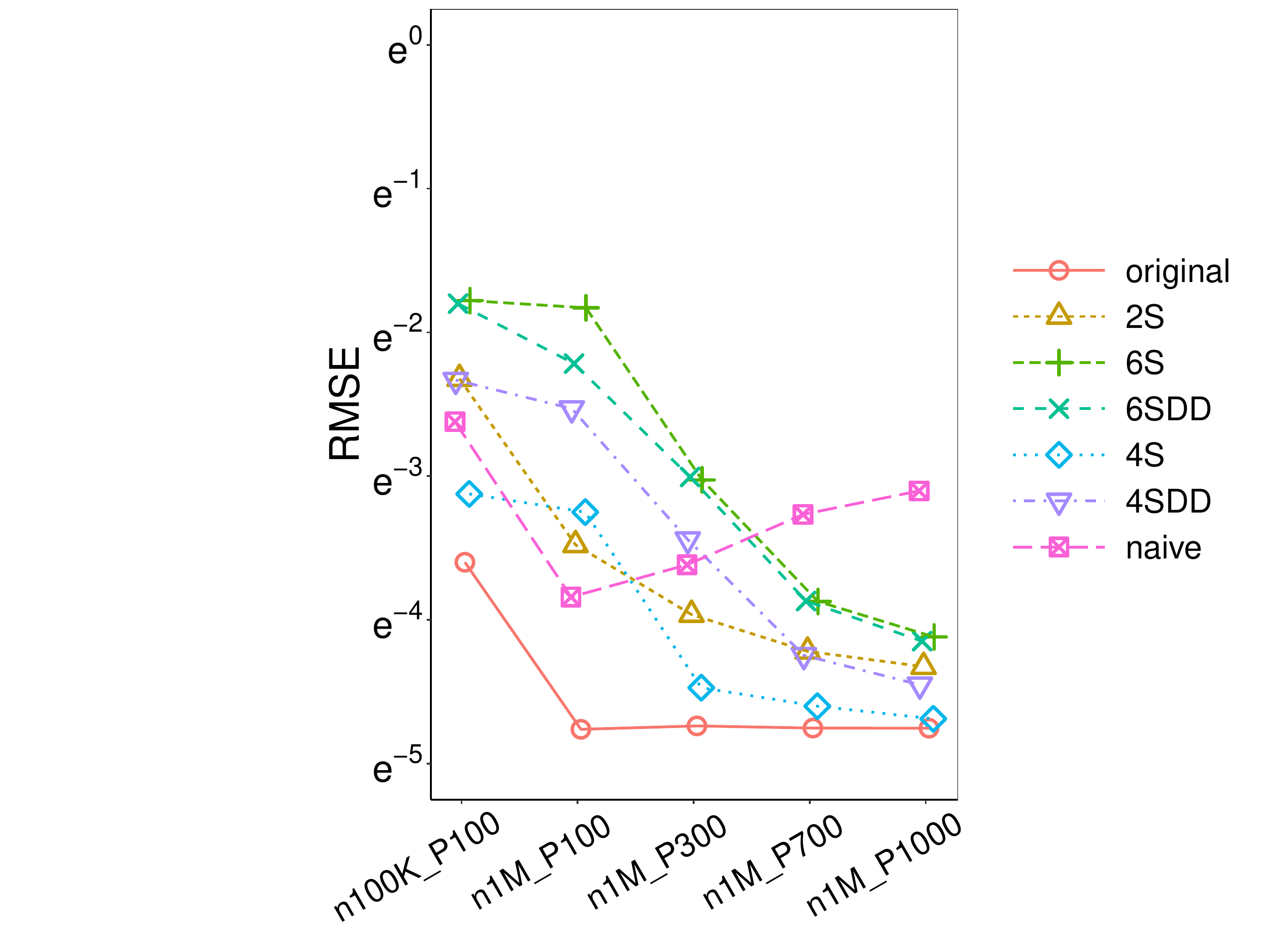}
\includegraphics[width=0.19\textwidth, trim={2.5in 0 2.6in 0},clip] {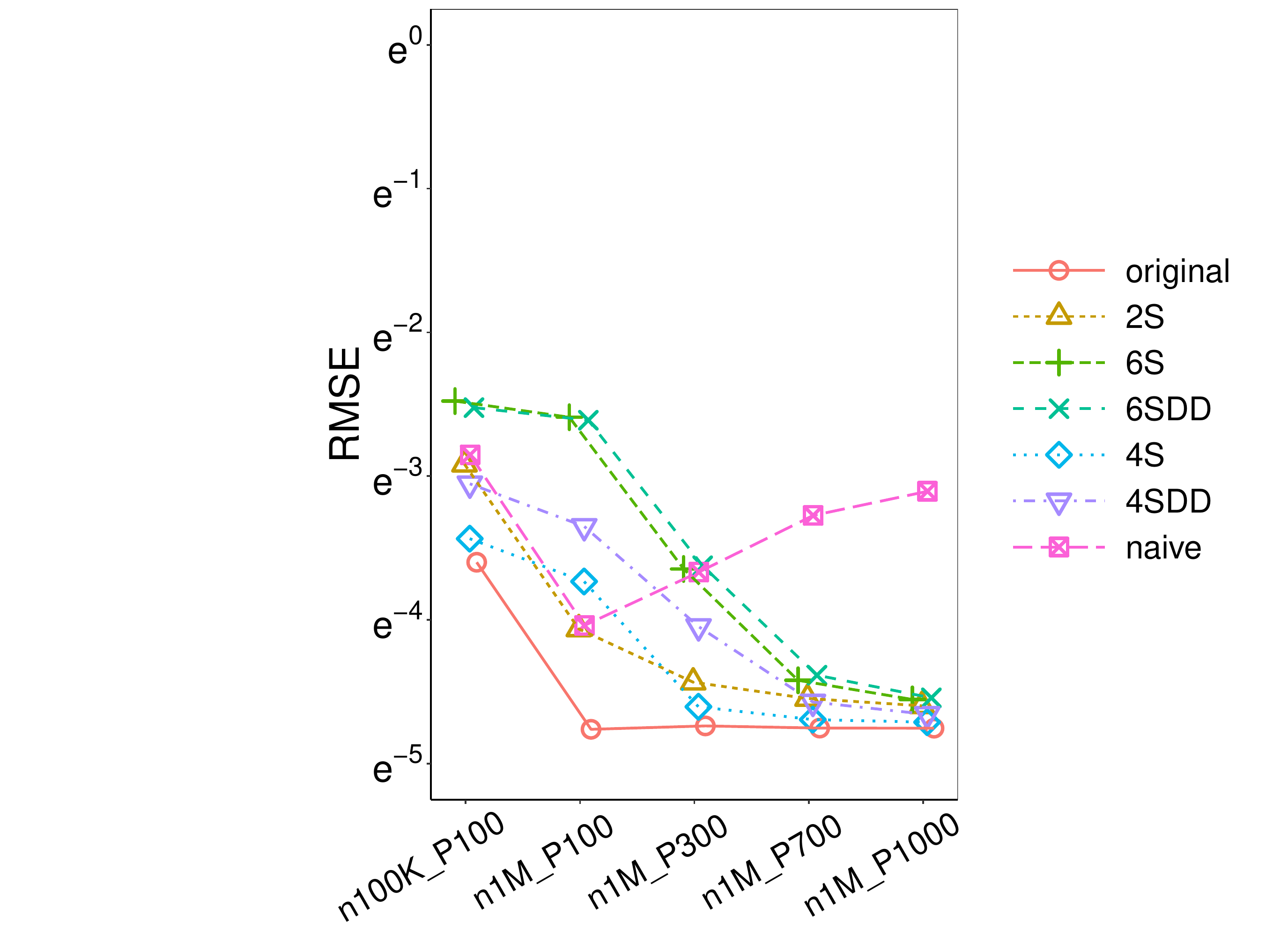}
\includegraphics[width=0.19\textwidth, trim={2.5in 0 2.6in 0},clip] {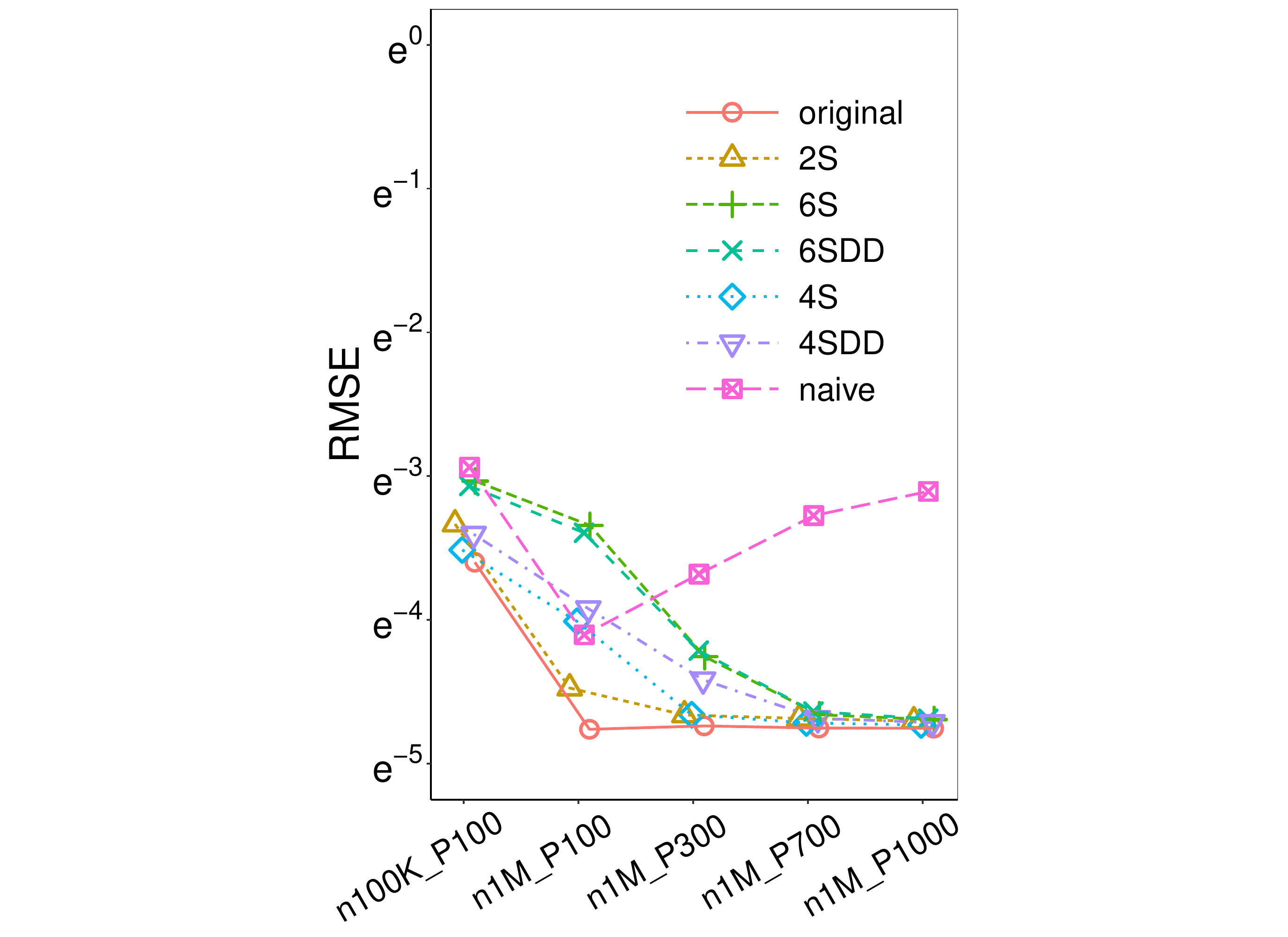}

\includegraphics[width=0.19\textwidth, trim={2.5in 0 2.6in 0},clip] {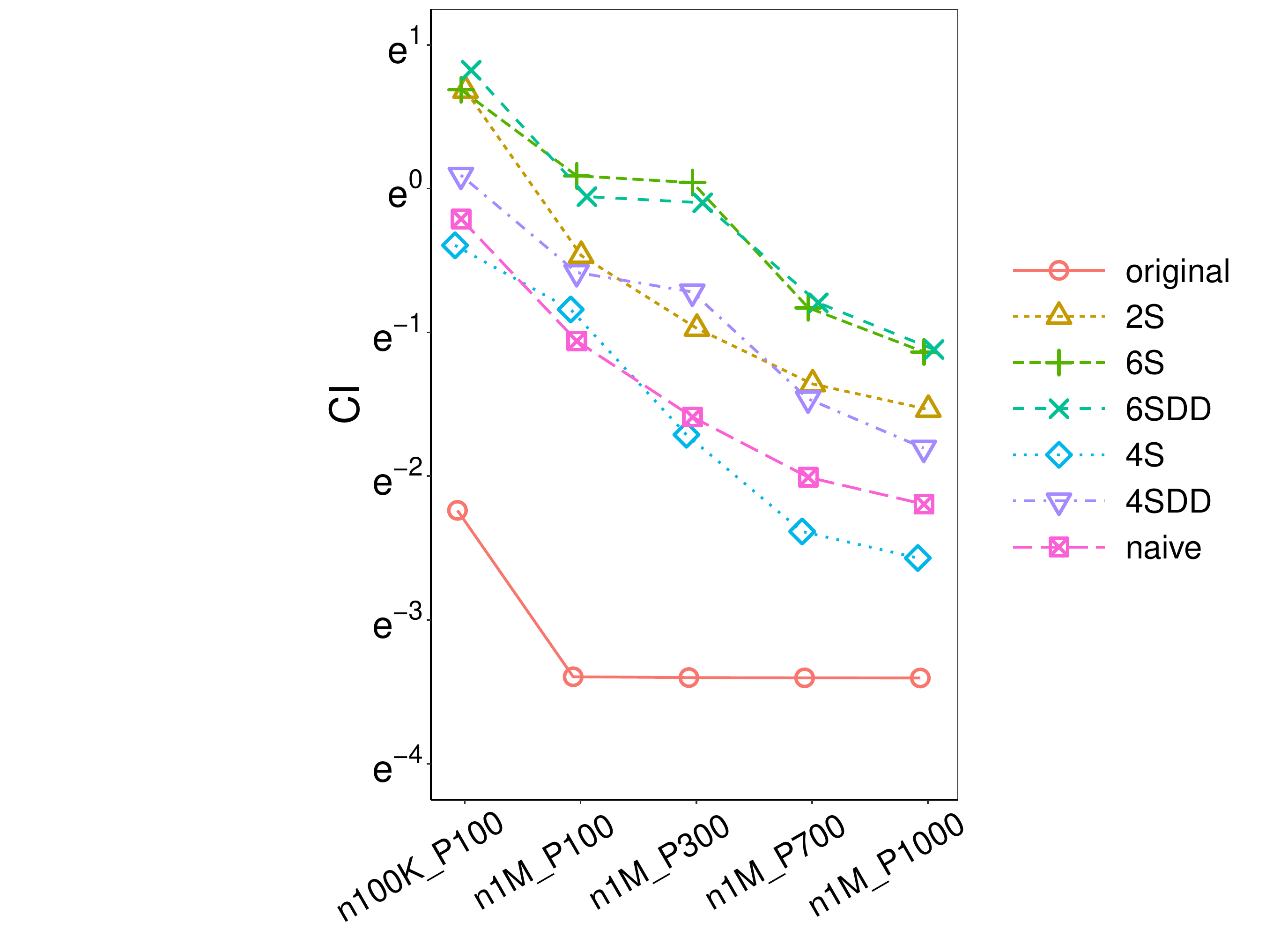}
\includegraphics[width=0.19\textwidth, trim={2.5in 0 2.6in 0},clip] {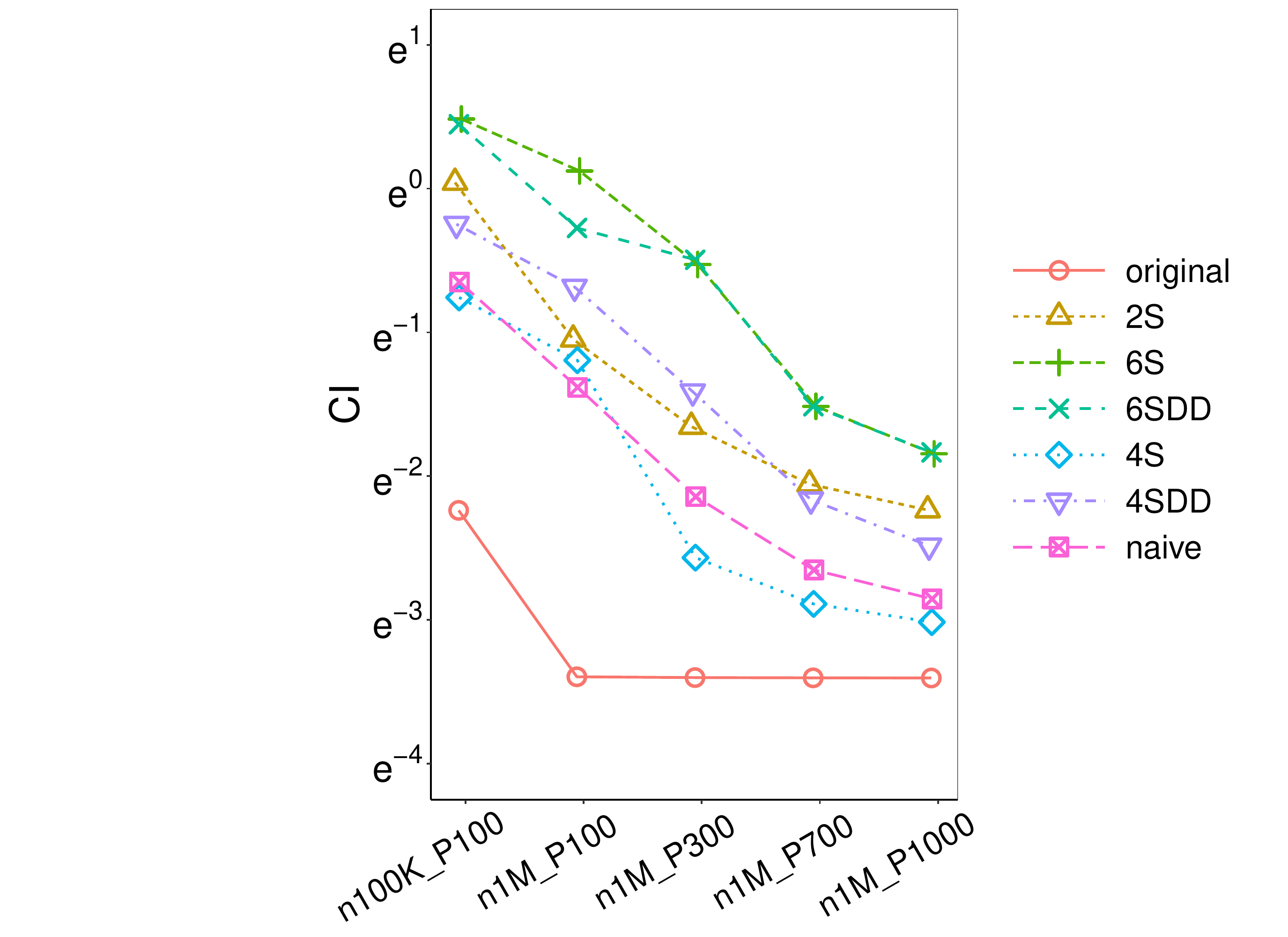}
\includegraphics[width=0.19\textwidth, trim={2.5in 0 2.6in 0},clip] {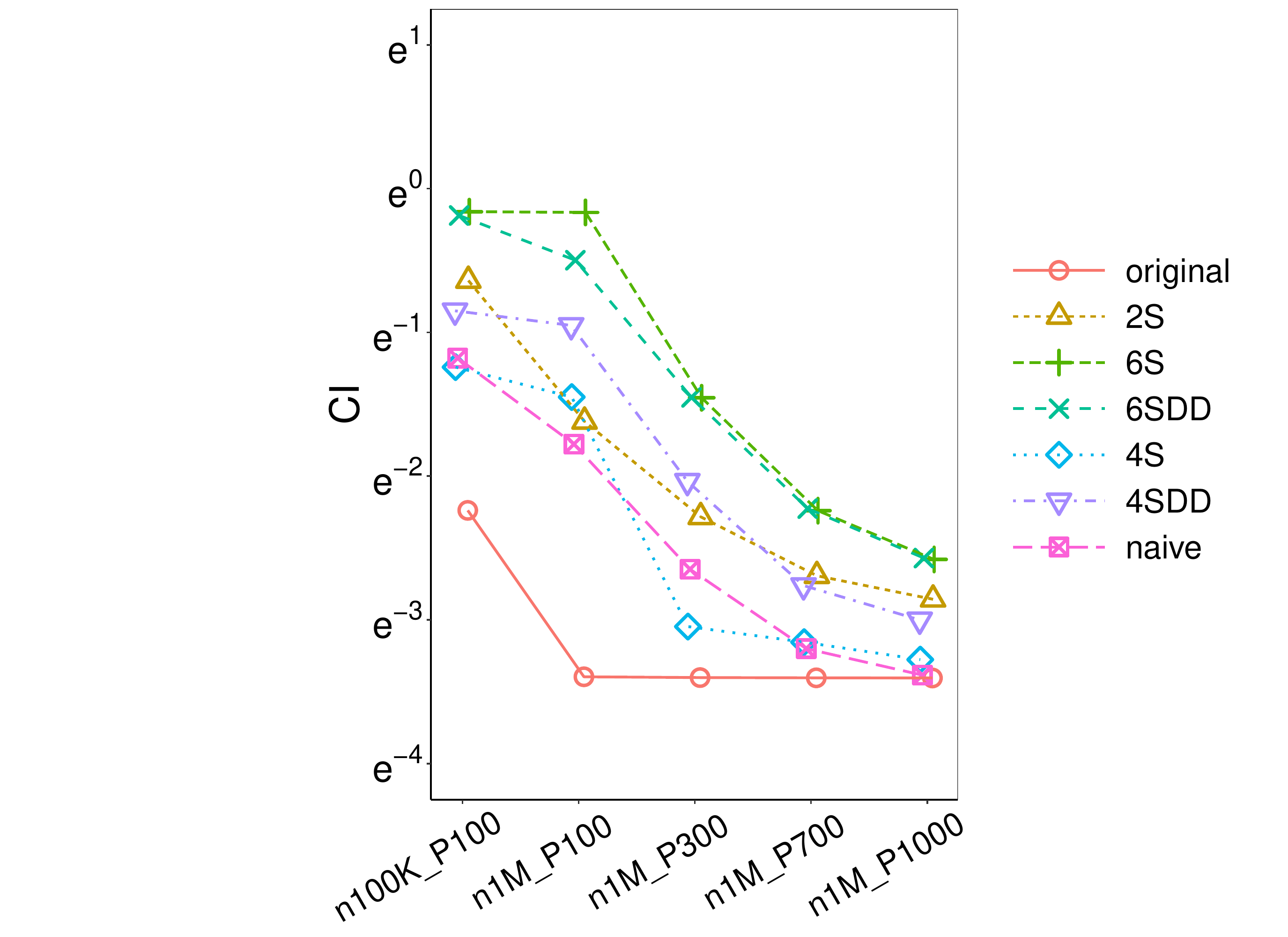}
\includegraphics[width=0.19\textwidth, trim={2.5in 0 2.6in 0},clip] {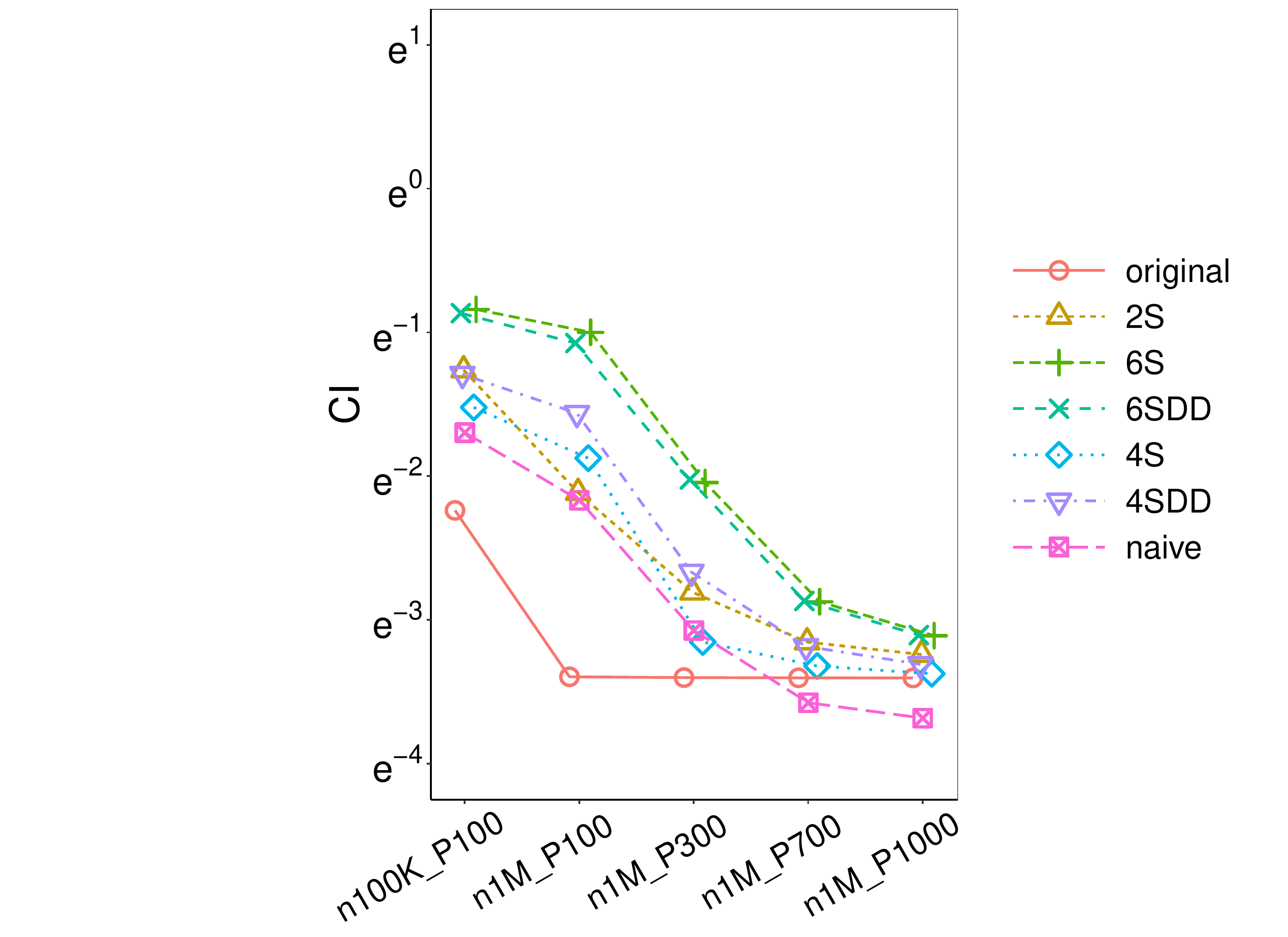}
\includegraphics[width=0.19\textwidth, trim={2.5in 0 2.6in 0},clip] {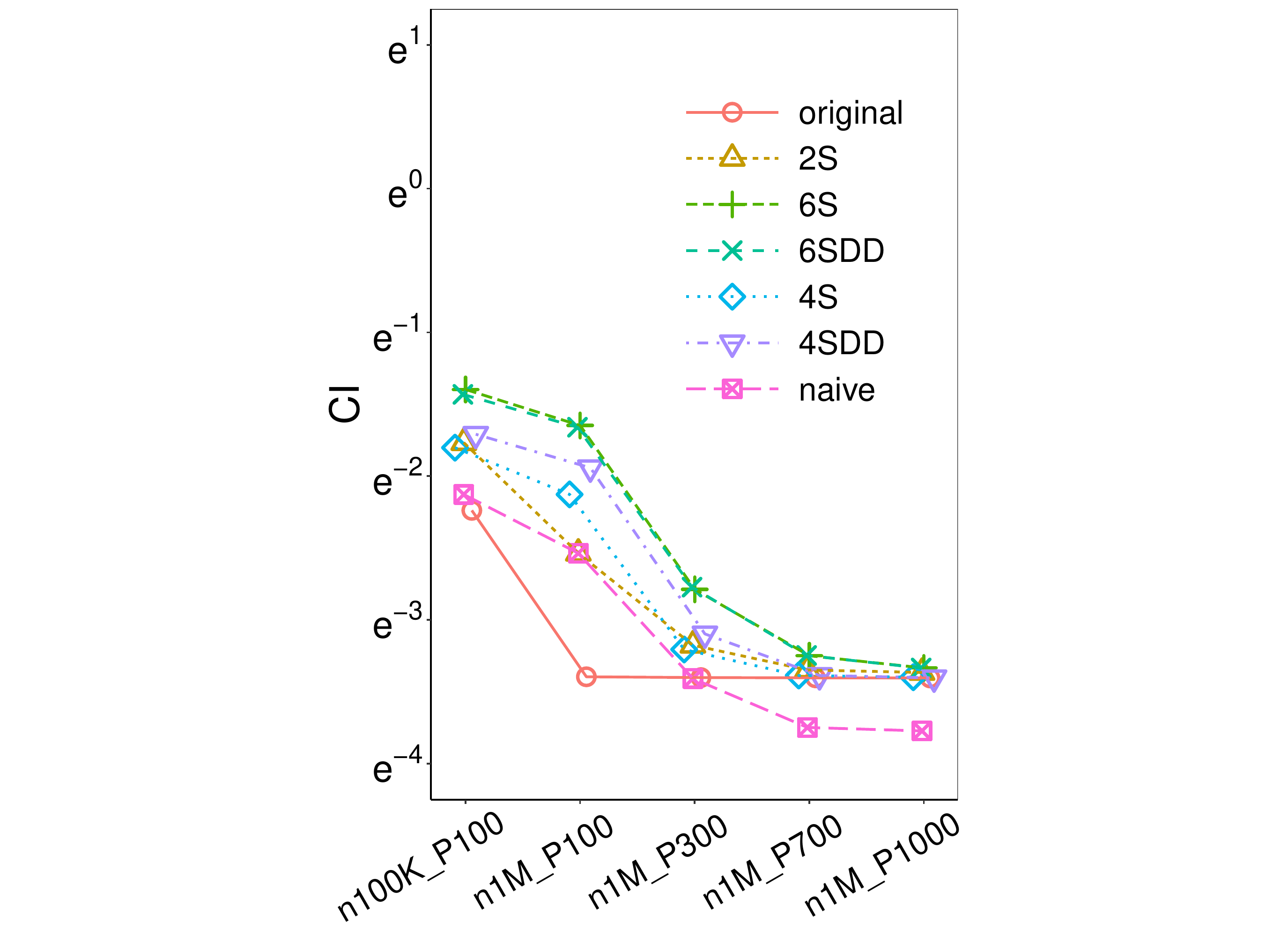}

\includegraphics[width=0.19\textwidth, trim={2.5in 0 2.6in 0},clip] {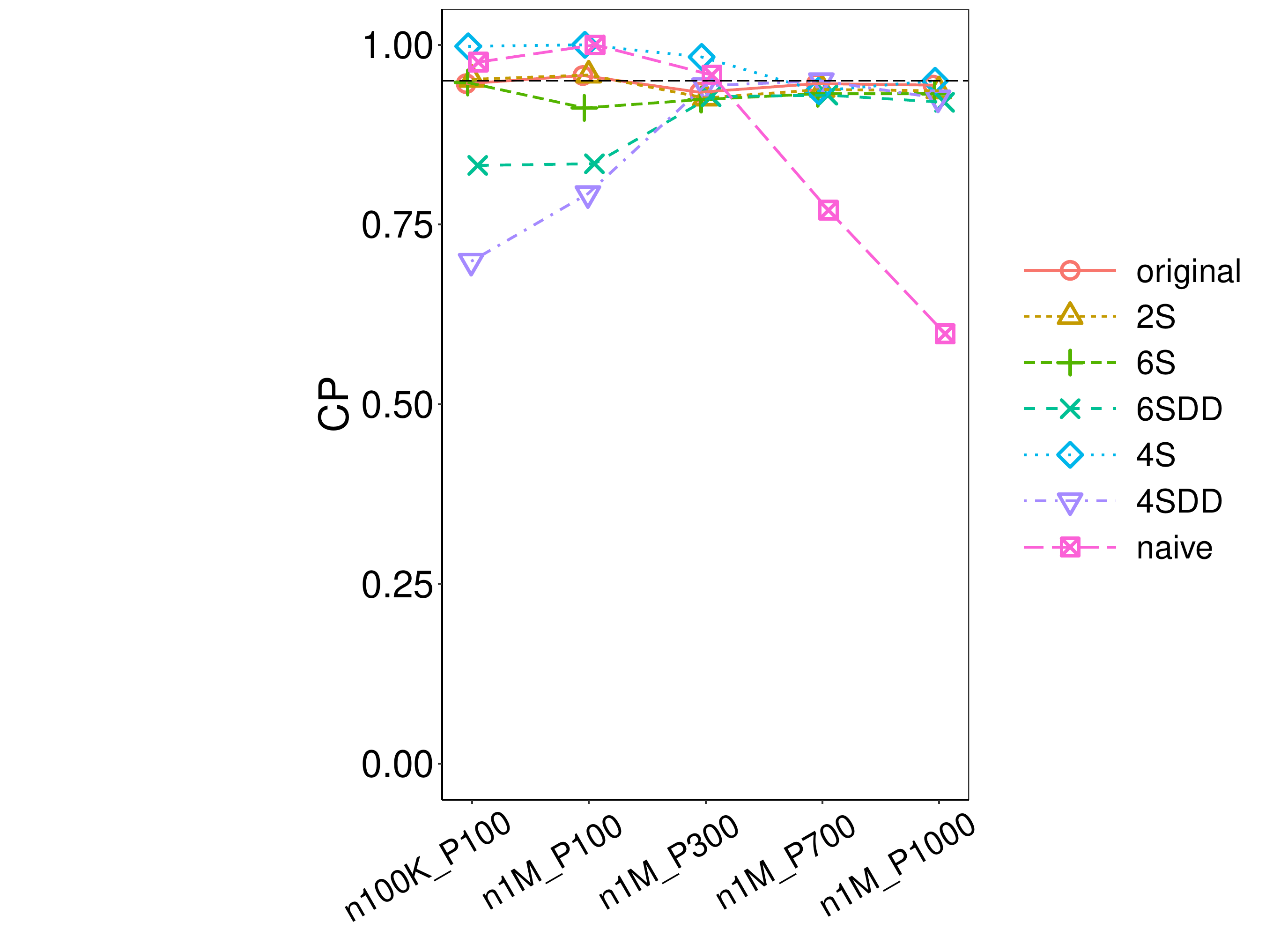}
\includegraphics[width=0.19\textwidth, trim={2.5in 0 2.6in 0},clip] {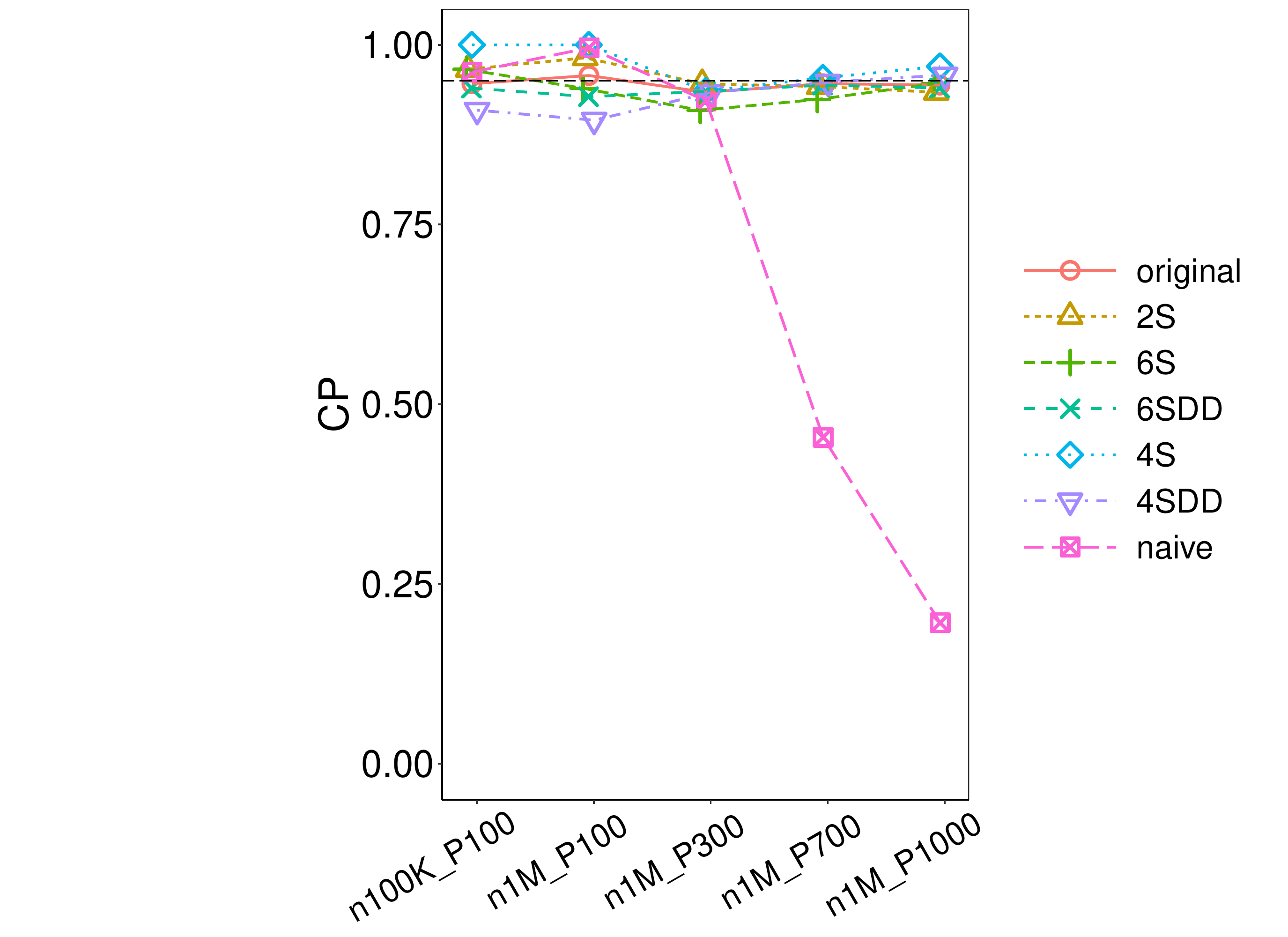}
\includegraphics[width=0.19\textwidth, trim={2.5in 0 2.6in 0},clip] {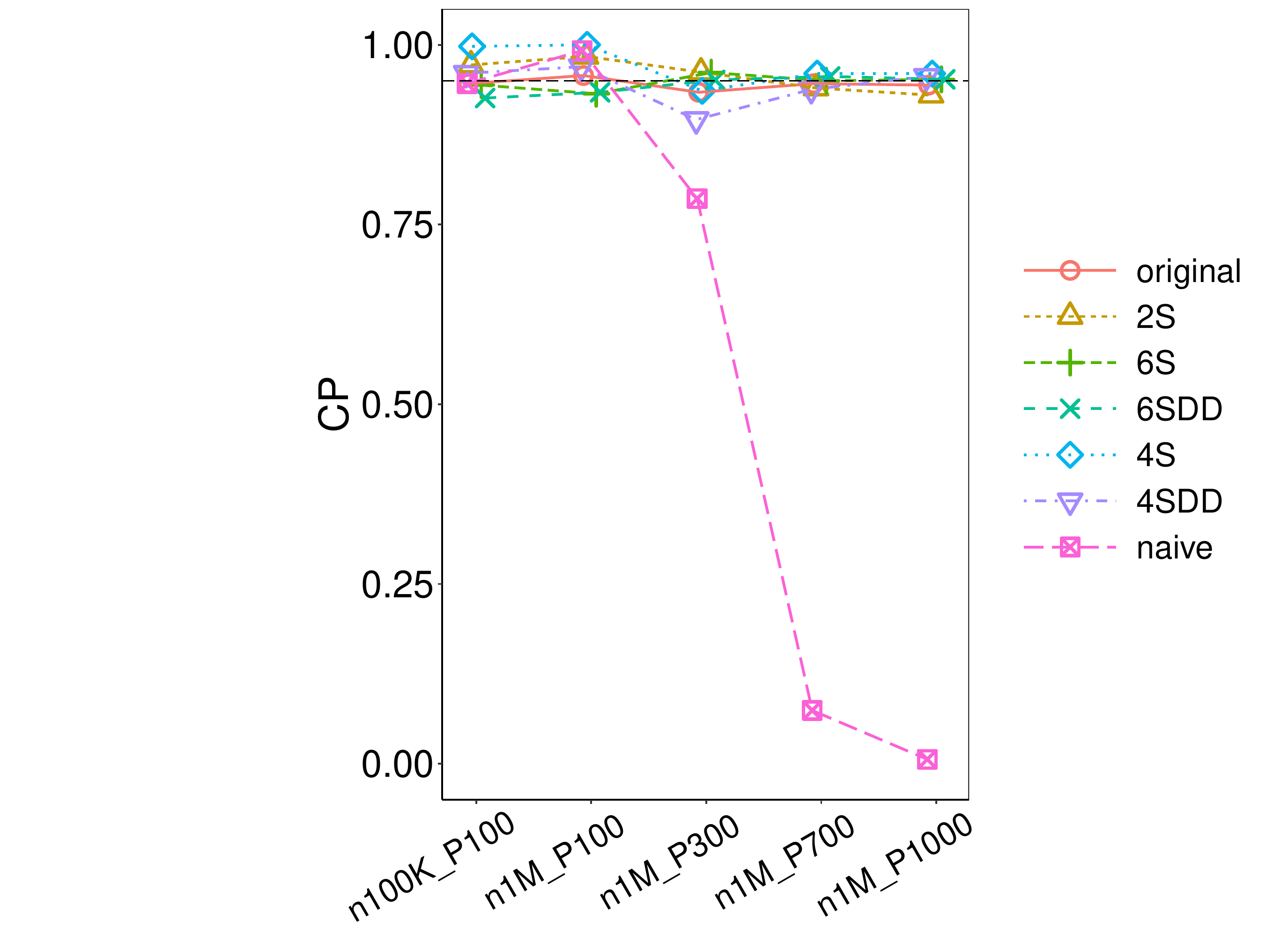}
\includegraphics[width=0.19\textwidth, trim={2.5in 0 2.6in 0},clip] {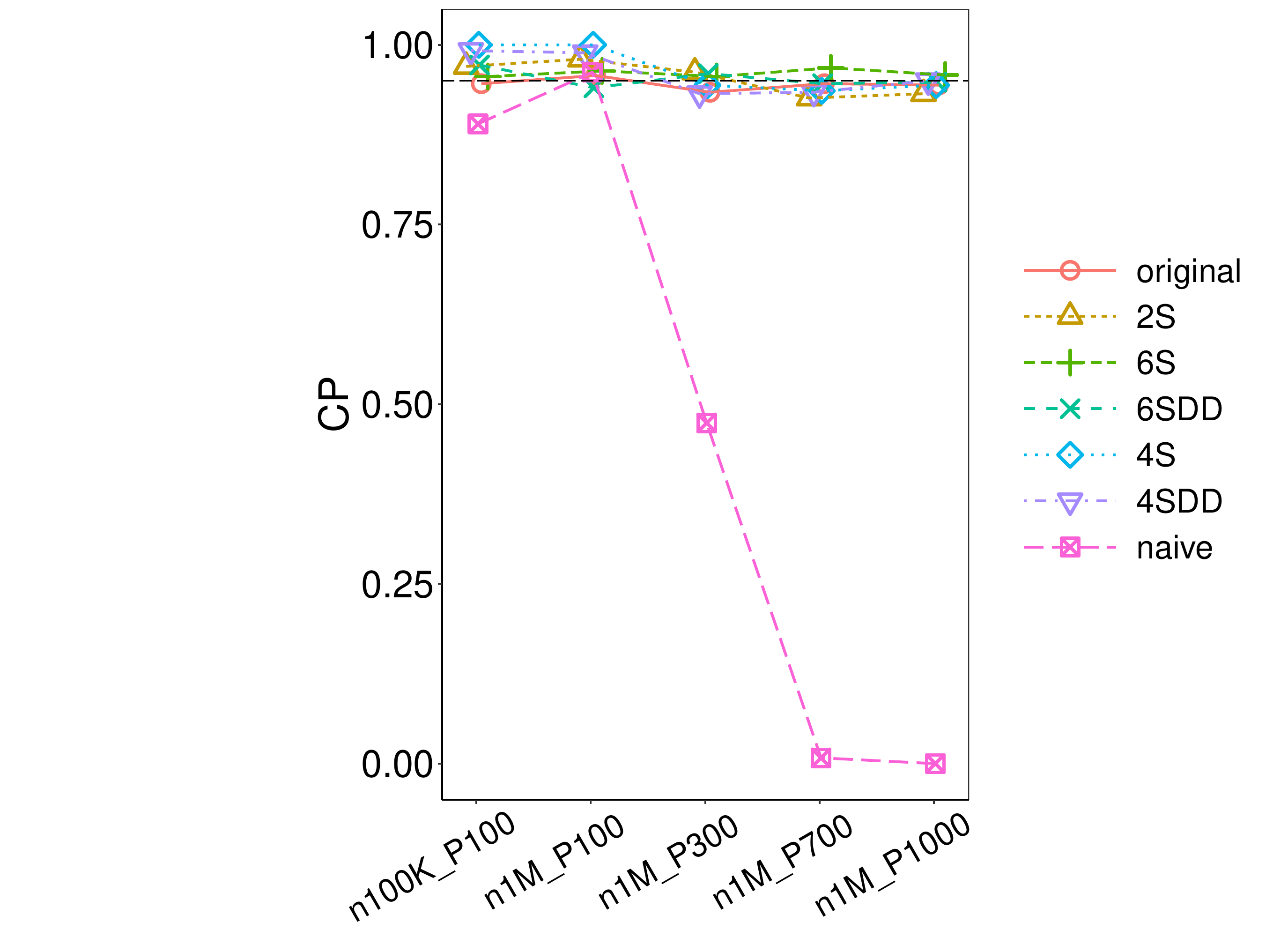}
\includegraphics[width=0.19\textwidth, trim={2.5in 0 2.6in 0},clip] {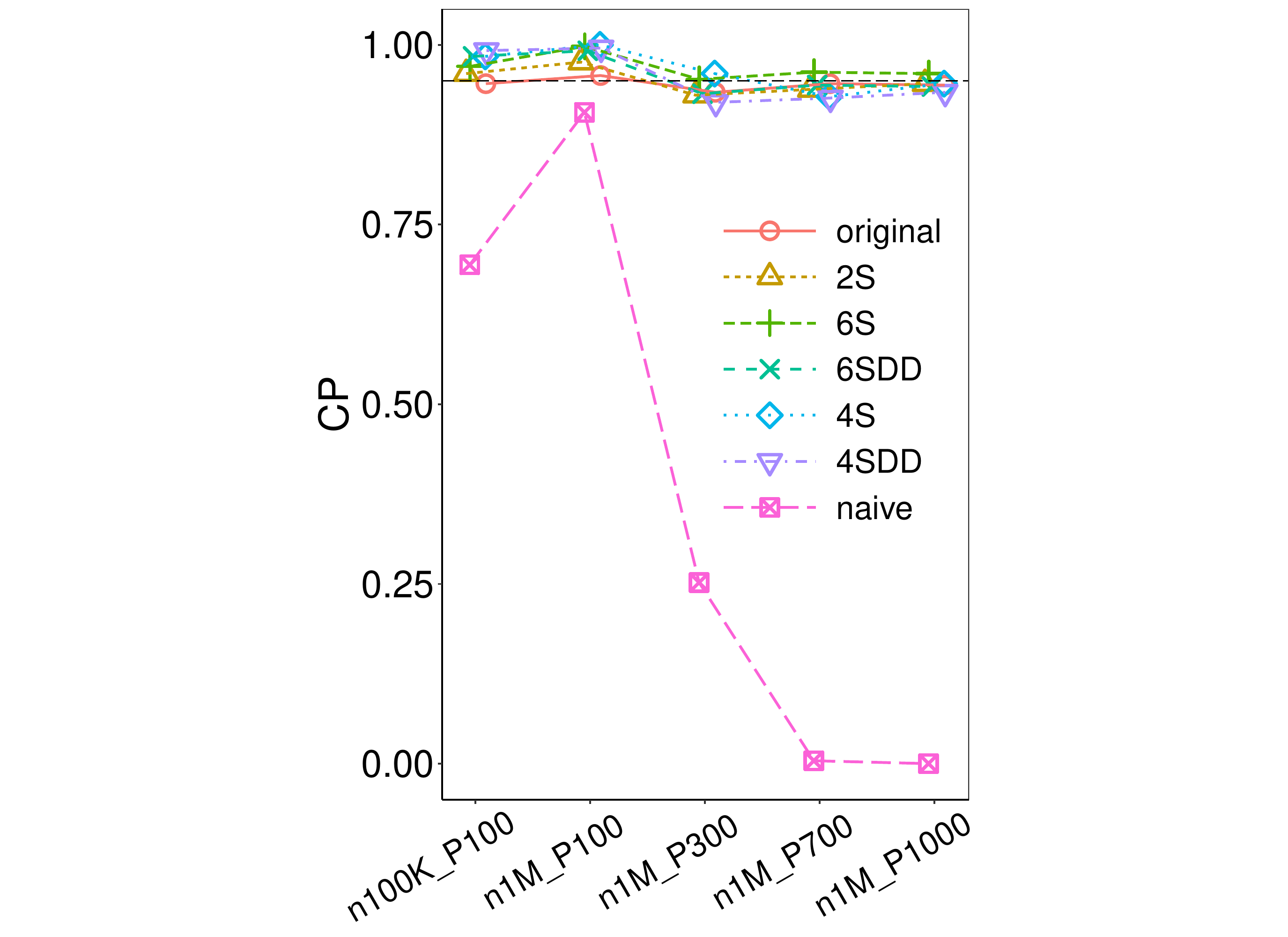}

\includegraphics[width=0.19\textwidth, trim={2.5in 0 2.6in 0},clip] {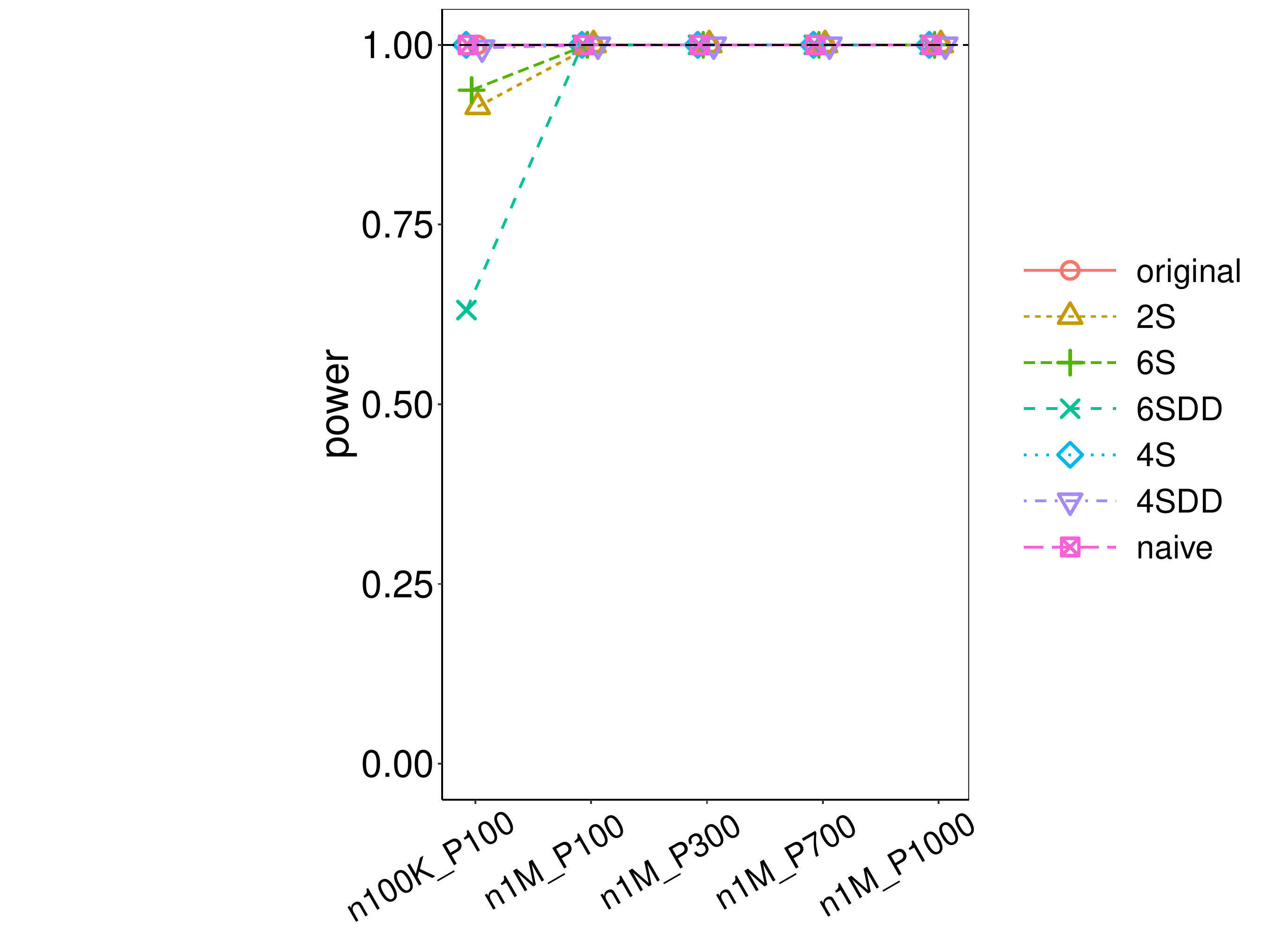}
\includegraphics[width=0.19\textwidth, trim={2.5in 0 2.6in 0},clip] {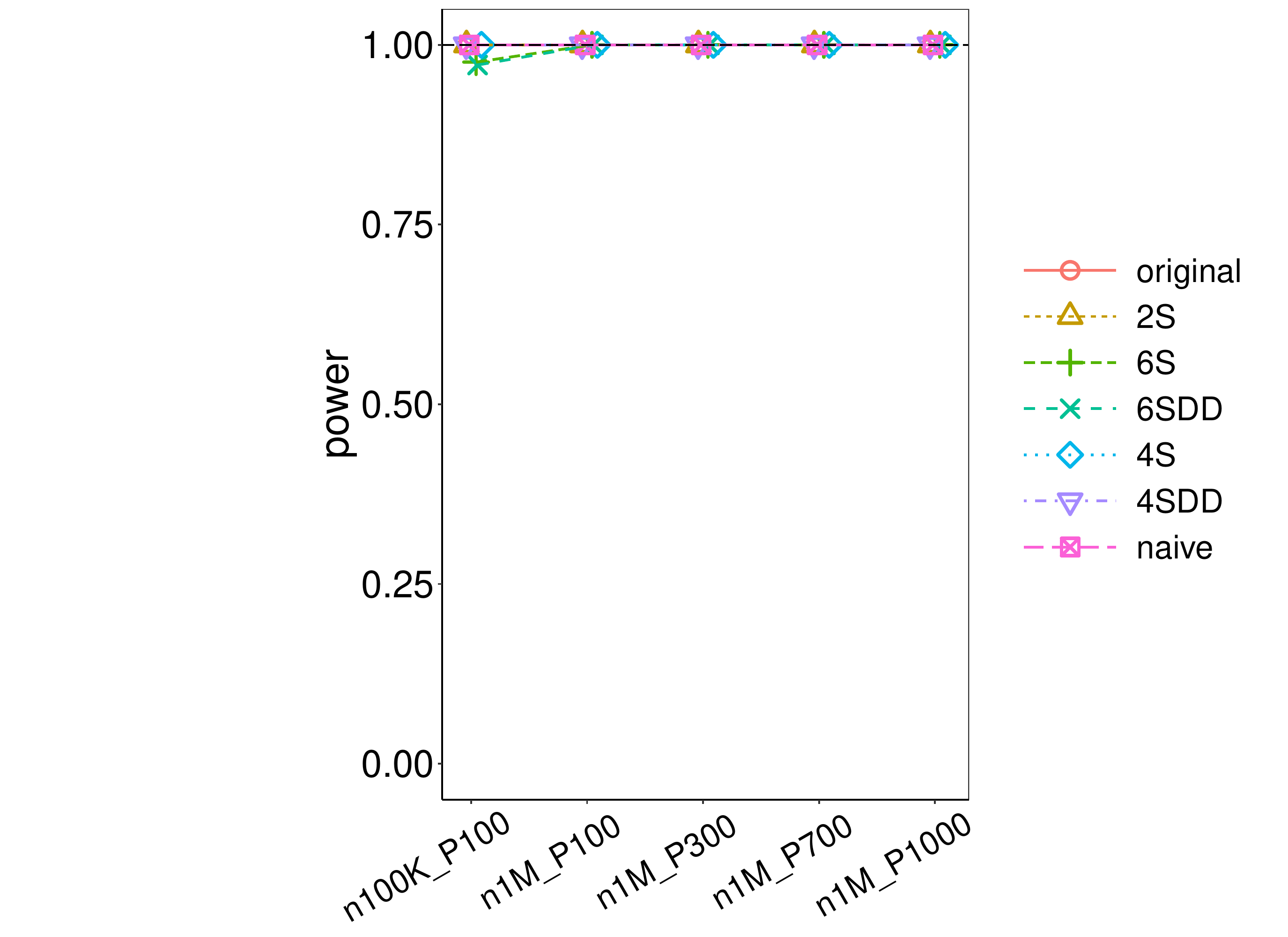}
\includegraphics[width=0.19\textwidth, trim={2.5in 0 2.6in 0},clip] {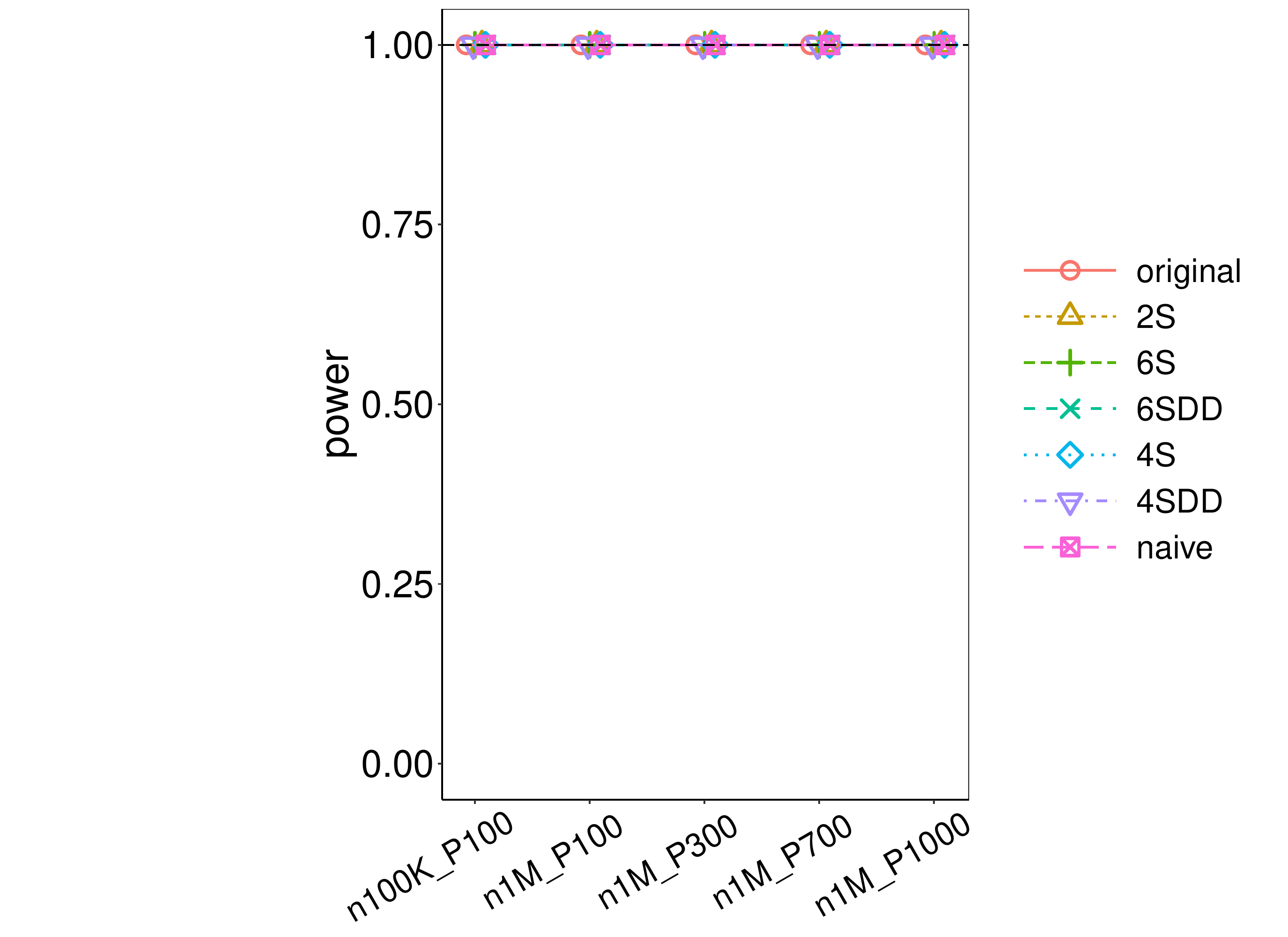}
\includegraphics[width=0.19\textwidth, trim={2.5in 0 2.6in 0},clip] {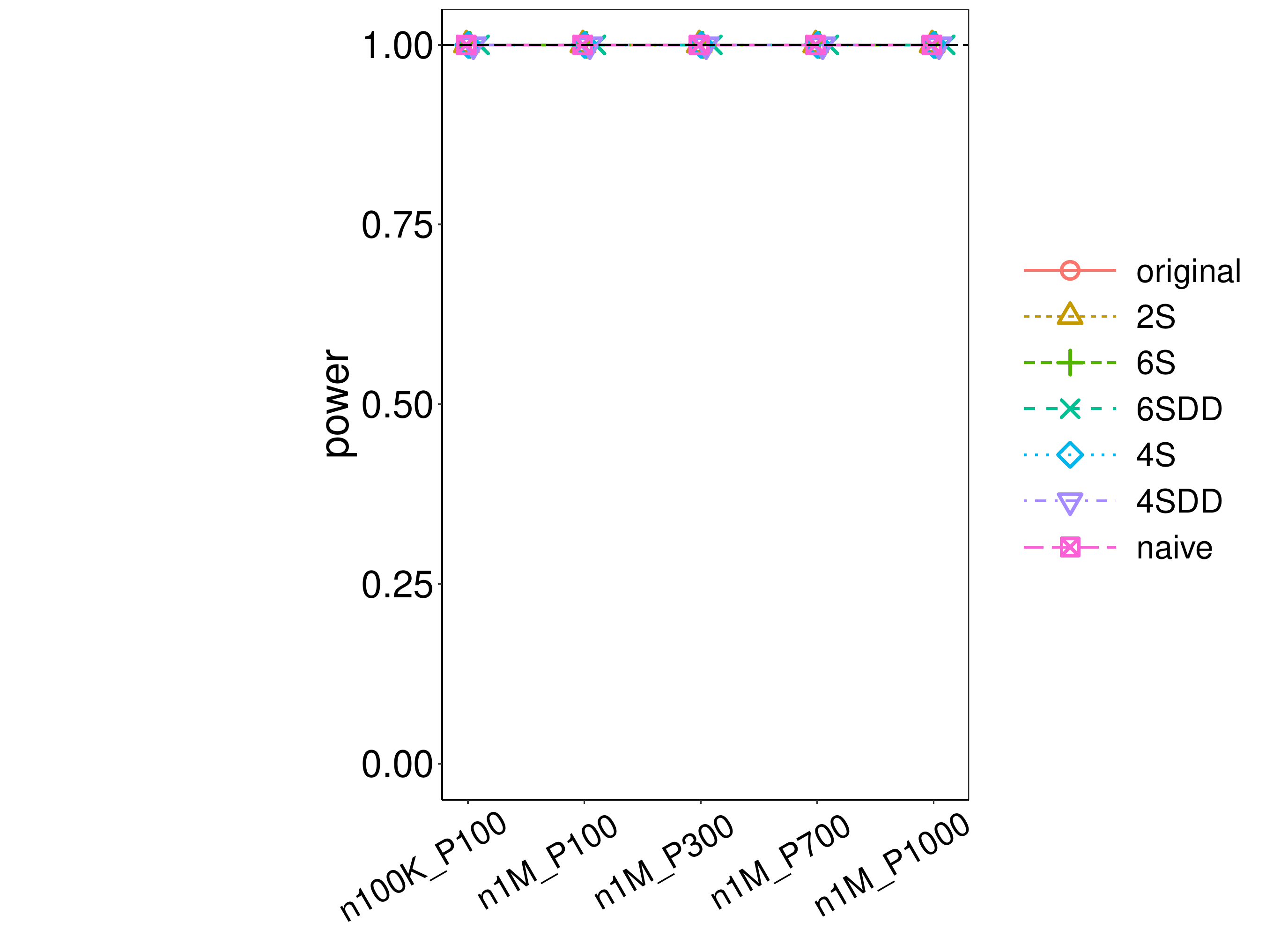}
\includegraphics[width=0.19\textwidth, trim={2.5in 0 2.6in 0},clip] {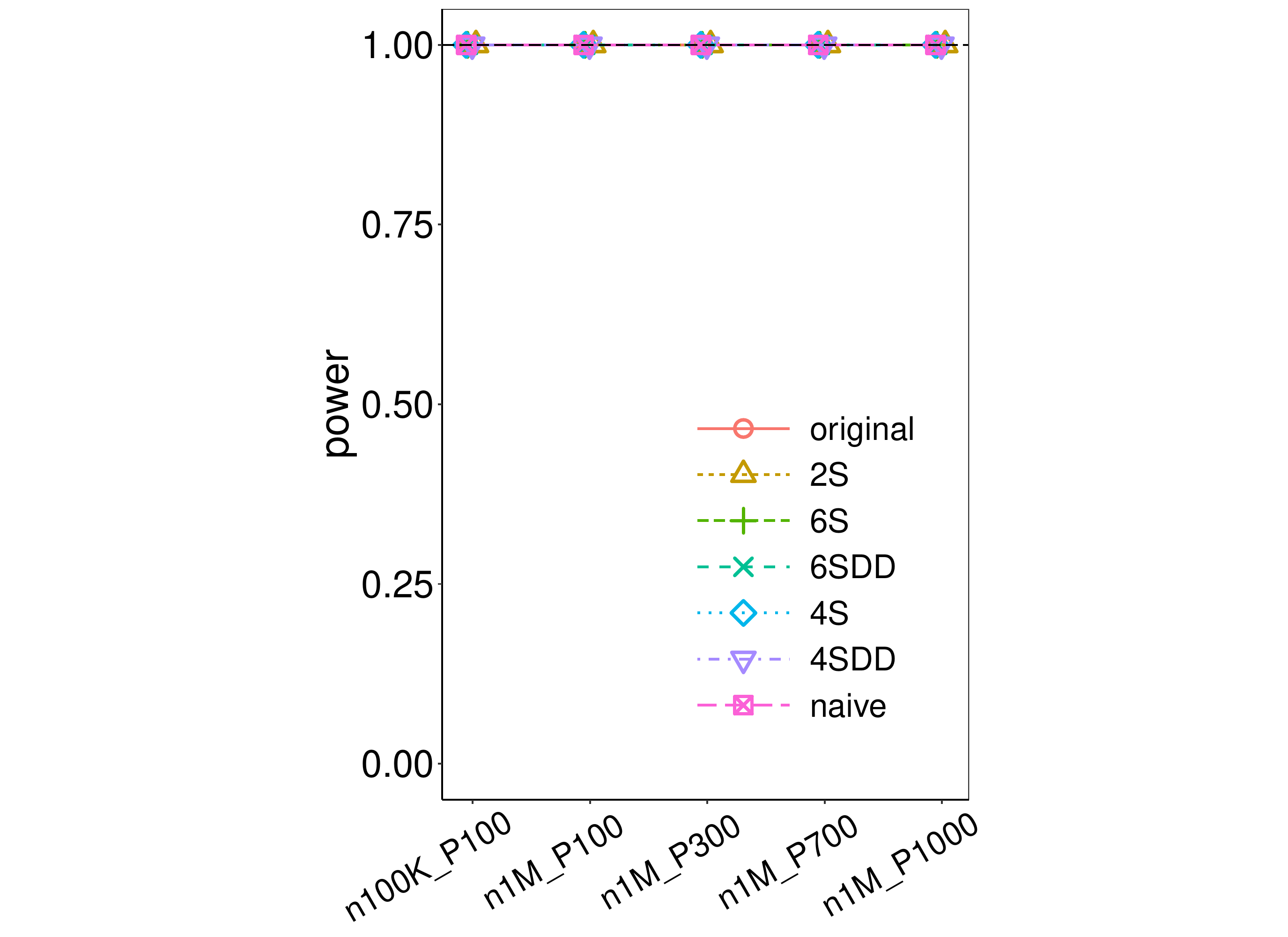}

\caption{Simulation results with $\rho$-zCDP for Gaussian data with  $\alpha\ne\beta$ when $\theta\ne0$}
\label{fig:1aszCDPN}
\end{figure}

The results are presented in Figures \ref{fig:1asDPN} and \ref{fig:1aszCDPN} and are summarized  as follows. 4S is the best performer by all metrics in almost all cases. At $n=100k/P=100$ and $n=1m/P=100$, the na\"{i}ve method is similar to 4S from all metrics, performs better than the remaining methods, but its RMSE becomes the largest, CI becomes too narrow and the CP approaches 0 at $n=1m$ as $P$ increases. As for other approaches, the performance depends on the evaluation metrics. For example, in terms of bias, 2S is closer to 0 than others; in terms of RMSE and CI width, 2S provides narrower CIs than 4SDD, 6S, and 6SDD, but wider than 4S; regarding CP, 4S, 6S, and 2S provide nominal level CP (4S shows slight over-coverage at $n=100k/P=100$, $n=1m/P=100$ and $n=1m/P=300$ for $\epsilon\le2$), while 4SDD and 6SDD exhibit significantly under-coverage at $n=100k/P=100$ and $n=1m/P=100$; all methods except for 6SDD at $n = 100K/P = 100$ provide power close to 1.  The Gaussian mechanism of $\rho$-zCDP does not seem to be advantageous in statistical utility compared to the Laplace mechanism of $\epsilon$-DP.

\subsubsection{\texorpdfstring{$Y\sim$ ZINB}{}, \texorpdfstring{$\theta=0$ ($H_0$)}{H0 }, and \texorpdfstring{$\alpha=\beta$}{S}}\label{sec:0sZINB}
\begin{figure}[!htb]
\hspace{0.5in}$\epsilon=0.5$\hspace{0.8in}$\epsilon=1$\hspace{0.9in}$\epsilon=2$
\hspace{0.95in}$\epsilon=5$\hspace{0.9in}$\epsilon=50$\\
\includegraphics[width=0.19\textwidth, trim={2.5in 0 2.5in 0},clip] {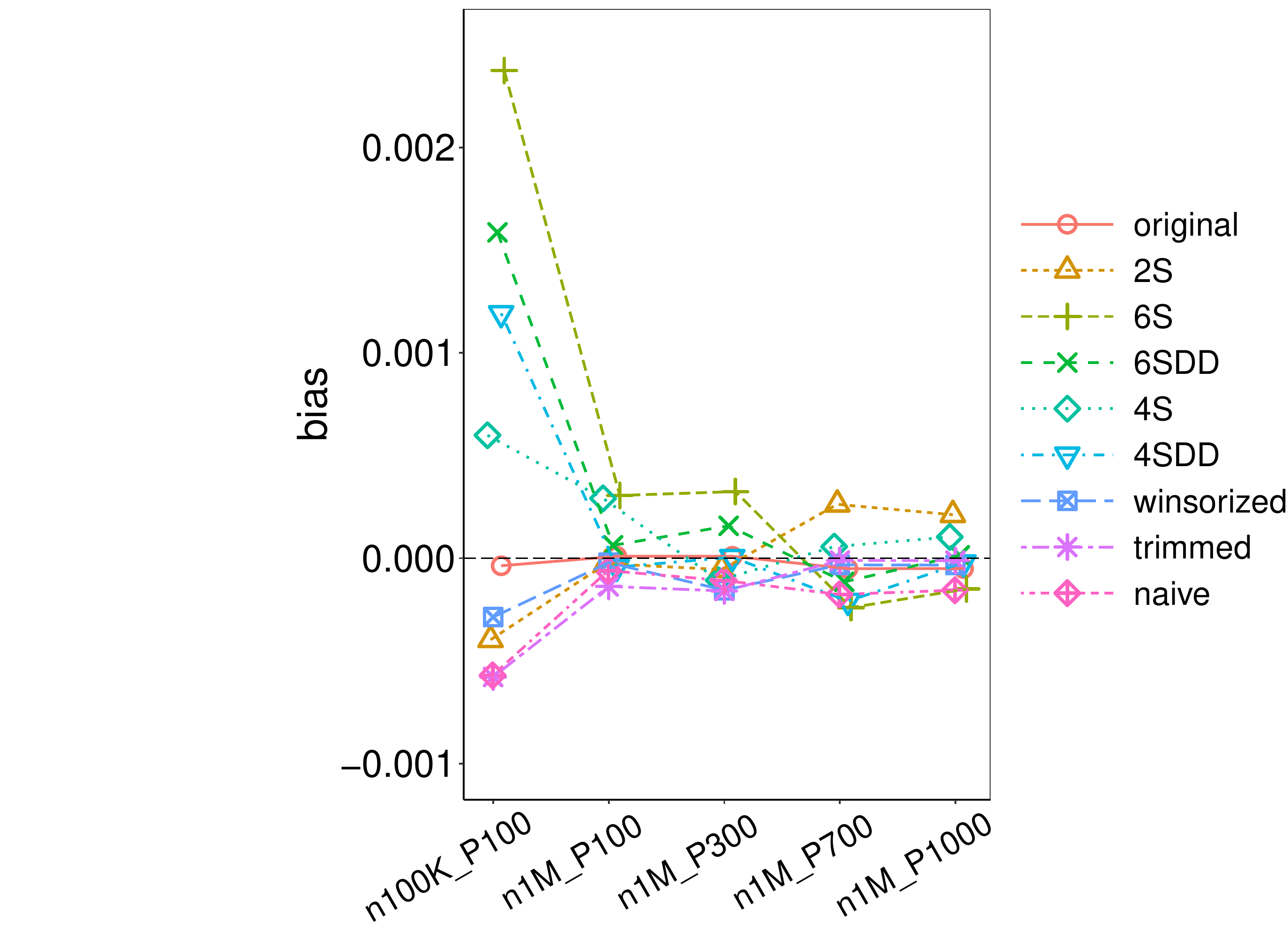}
\includegraphics[width=0.19\textwidth, trim={2.5in 0 2.5in 0},clip] {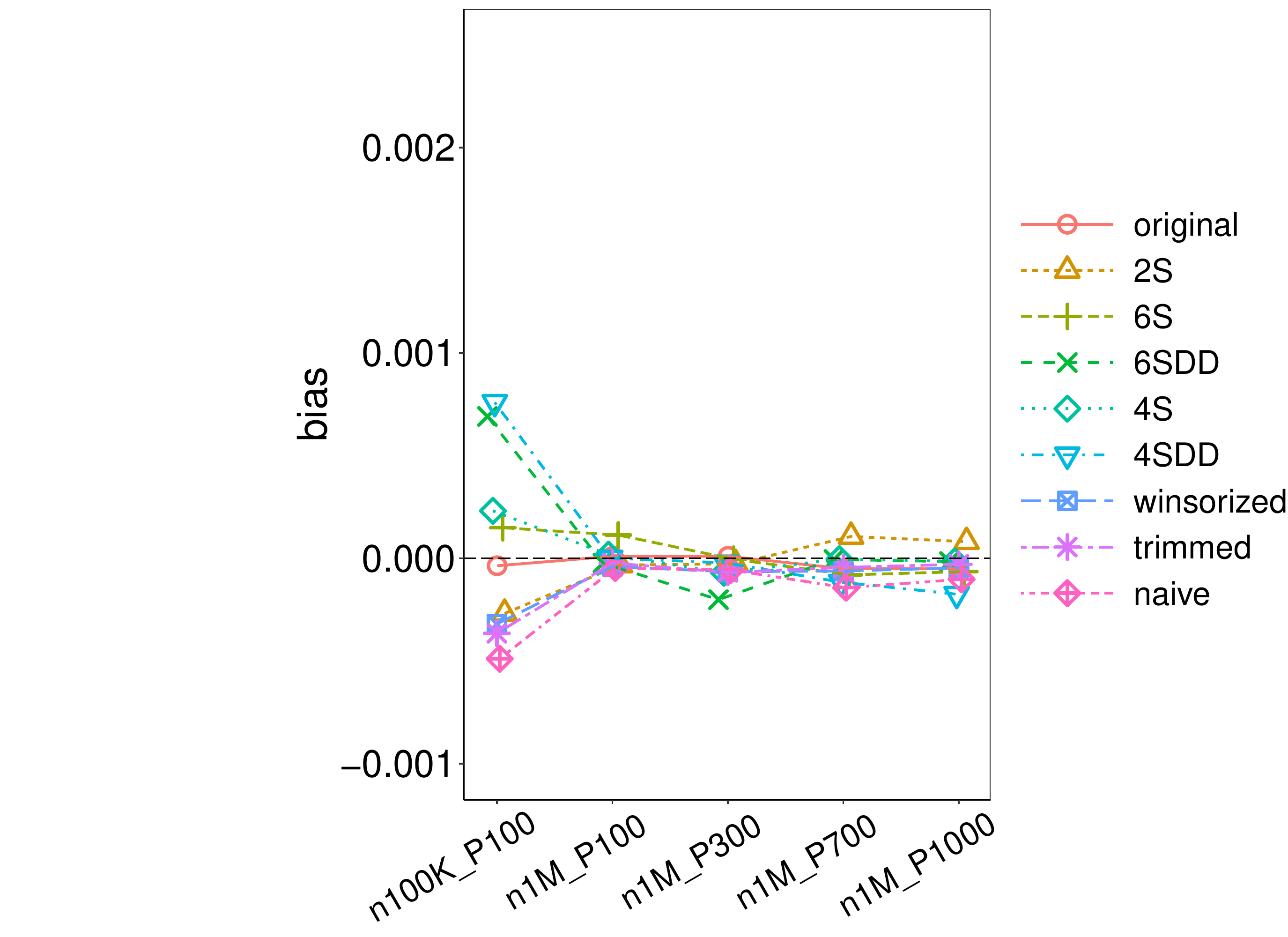}
\includegraphics[width=0.19\textwidth, trim={2.5in 0 2.5in 0},clip] {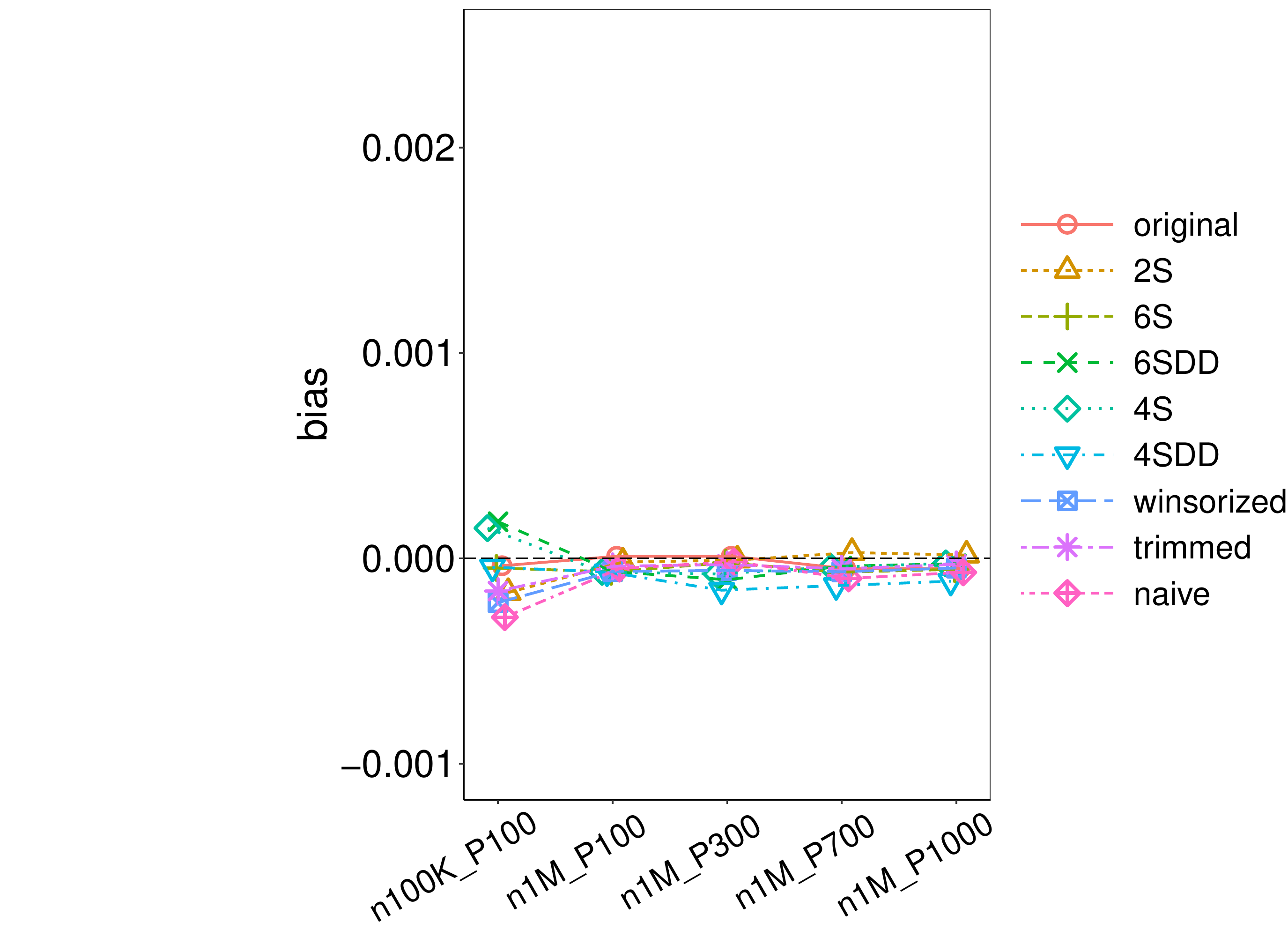}
\includegraphics[width=0.19\textwidth, trim={2.5in 0 2.5in 0},clip] {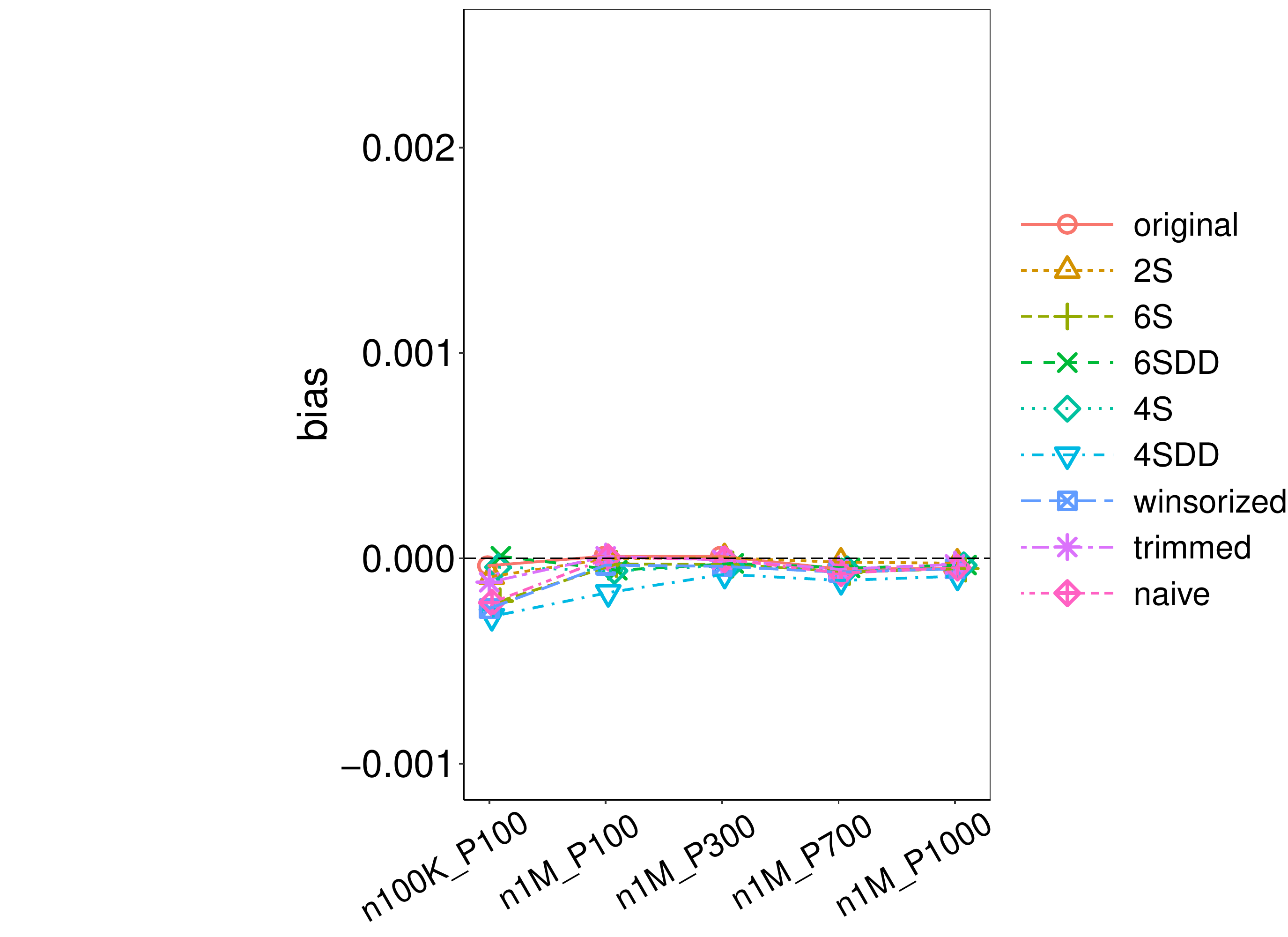}
\includegraphics[width=0.19\textwidth, trim={2.5in 0 2.5in 0},clip] {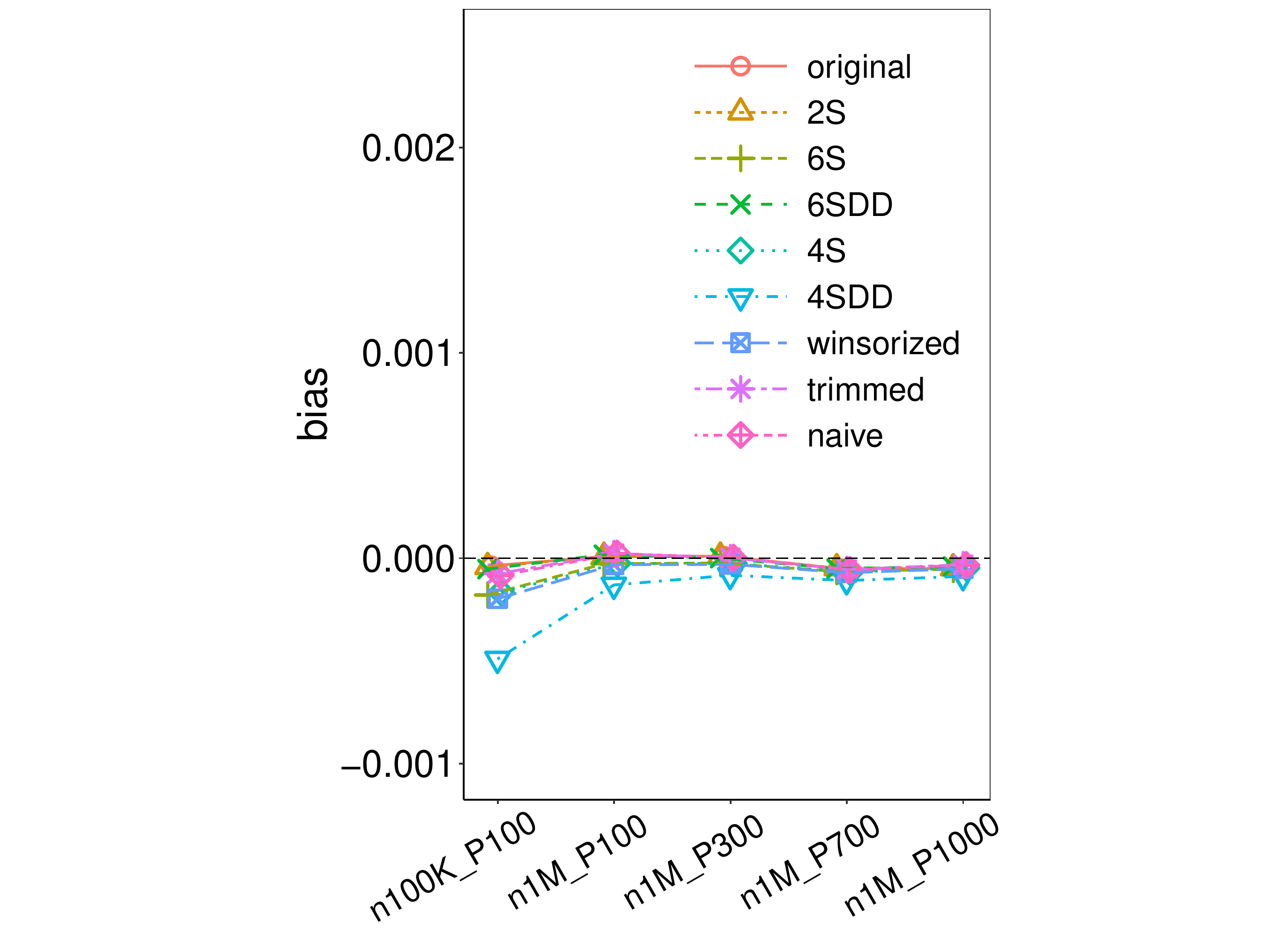}\\
\includegraphics[width=0.19\textwidth, trim={2.5in 0 2.6in 0},clip] {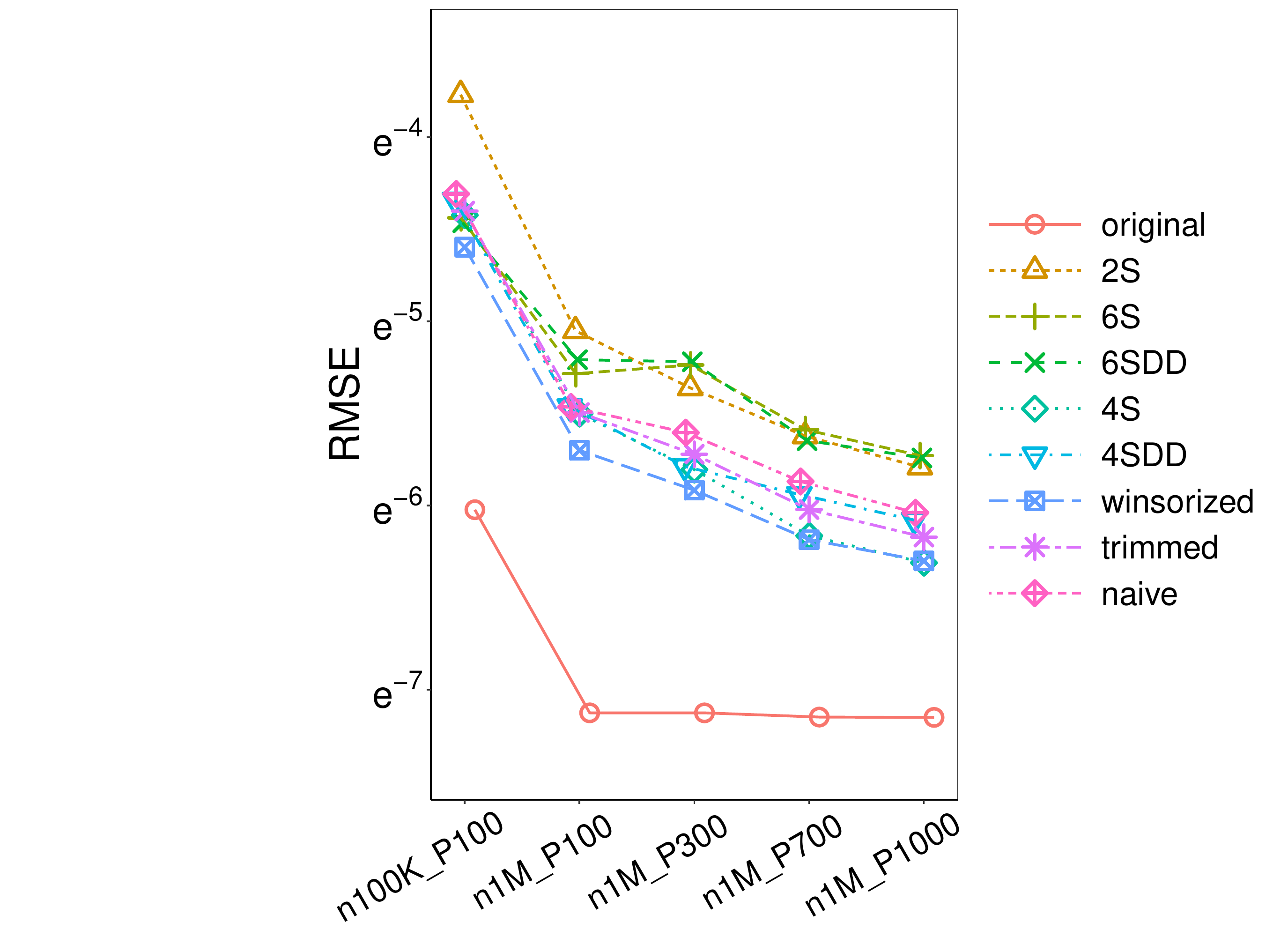}
\includegraphics[width=0.19\textwidth, trim={2.5in 0 2.6in 0},clip] {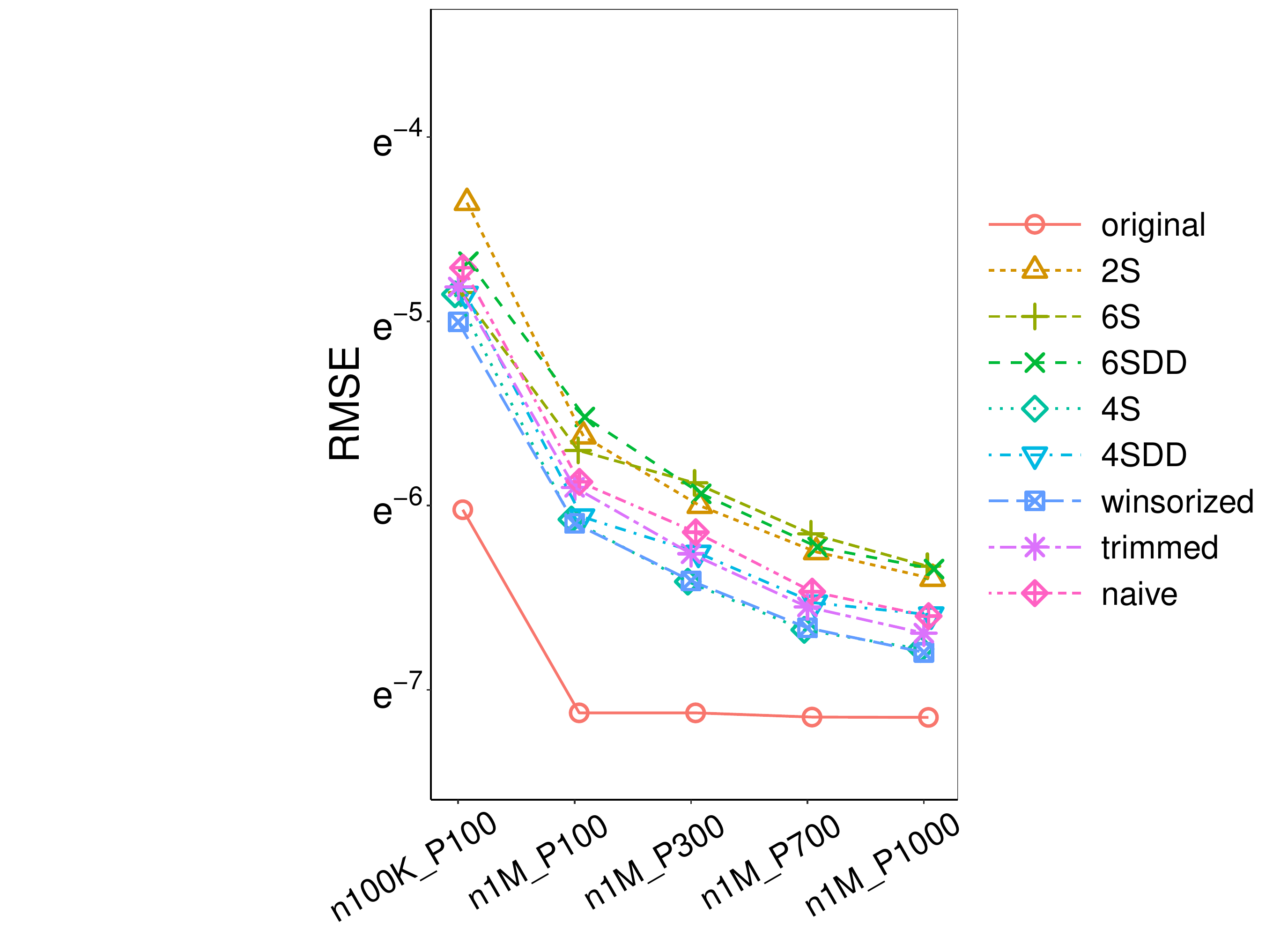}
\includegraphics[width=0.19\textwidth, trim={2.5in 0 2.6in 0},clip] {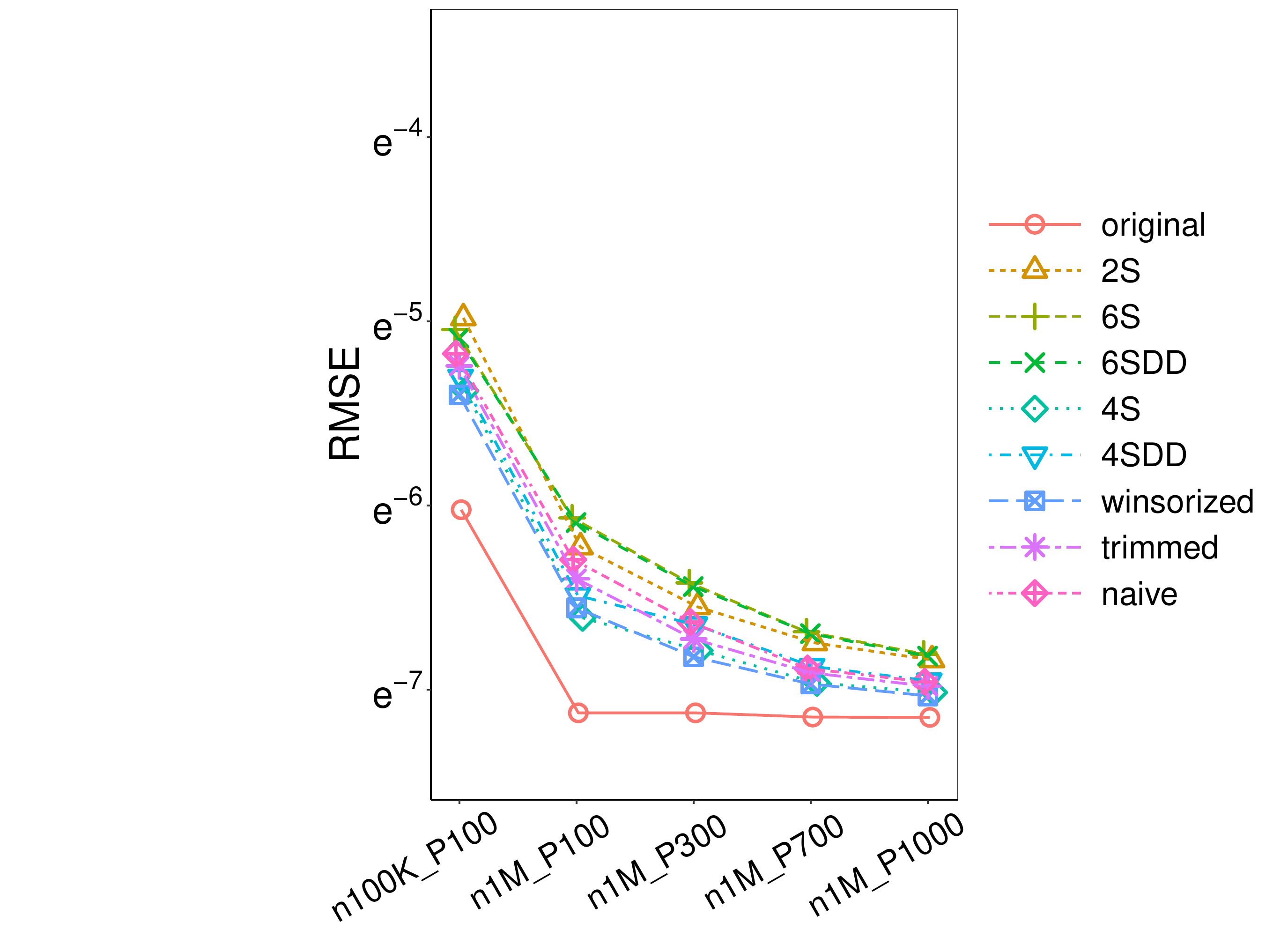}
\includegraphics[width=0.19\textwidth, trim={2.5in 0 2.6in 0},clip] {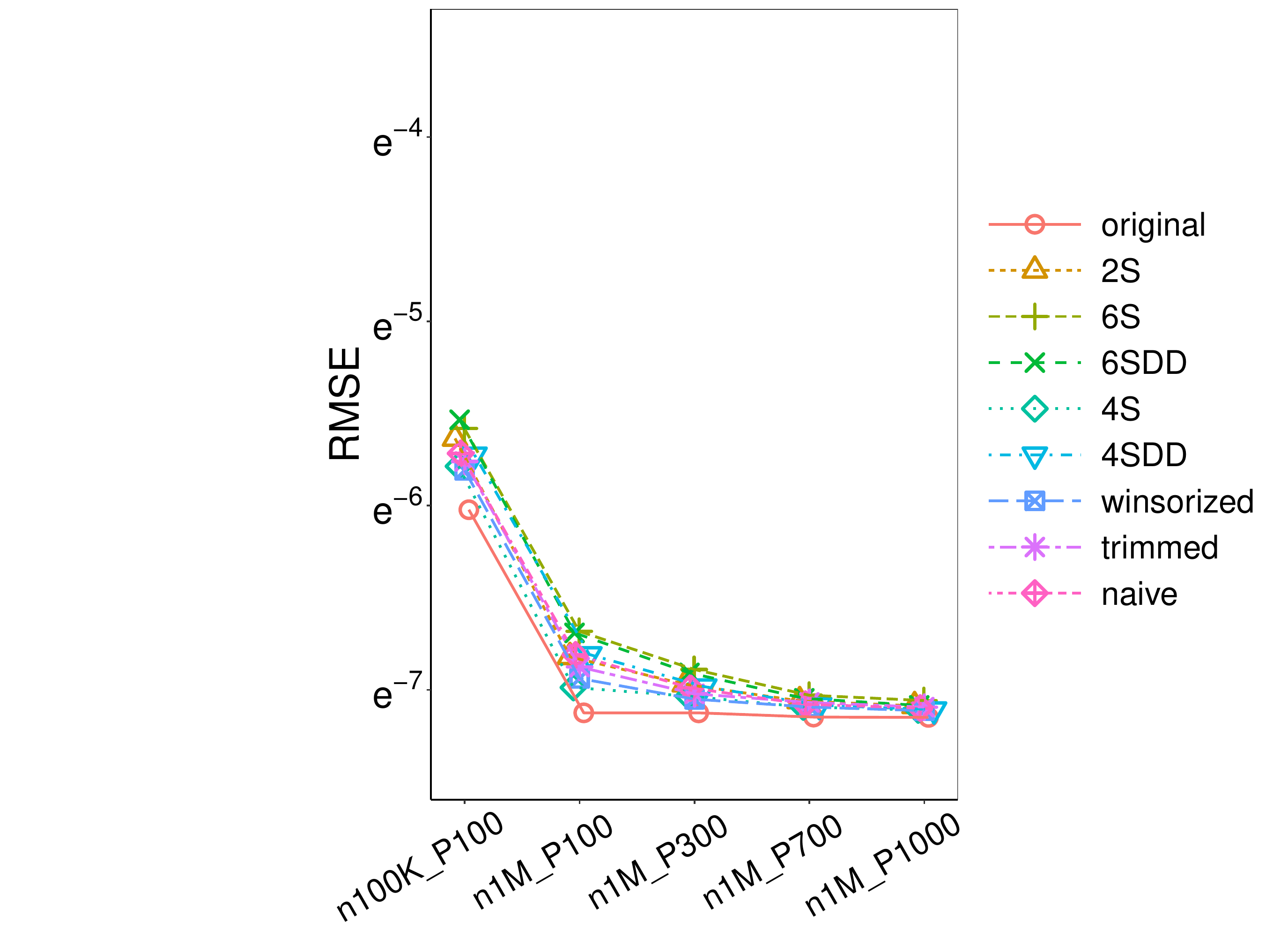}
\includegraphics[width=0.19\textwidth, trim={2.5in 0 2.6in 0},clip] {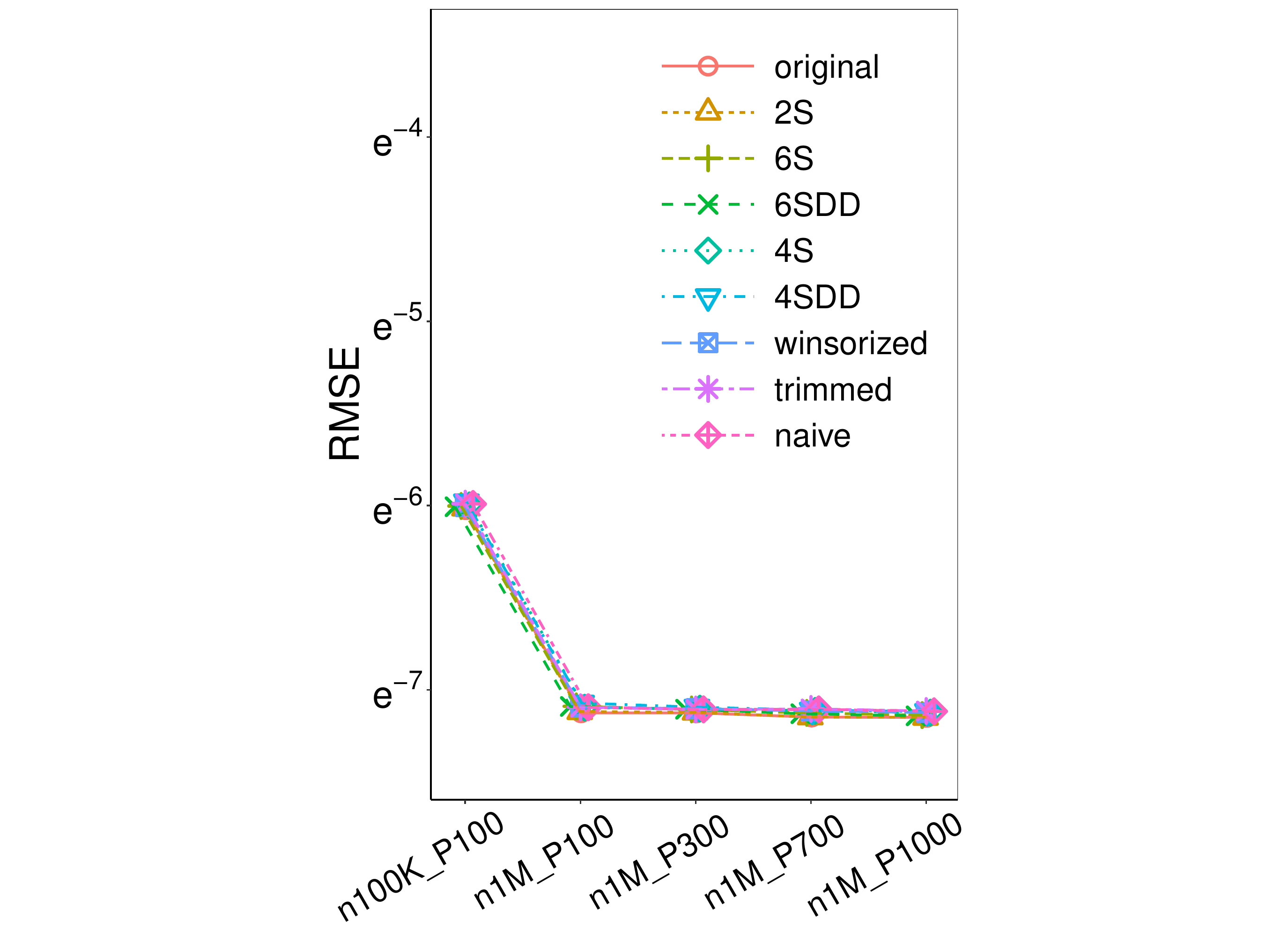}\\
\includegraphics[width=0.19\textwidth, trim={2.5in 0 2.6in 0},clip] {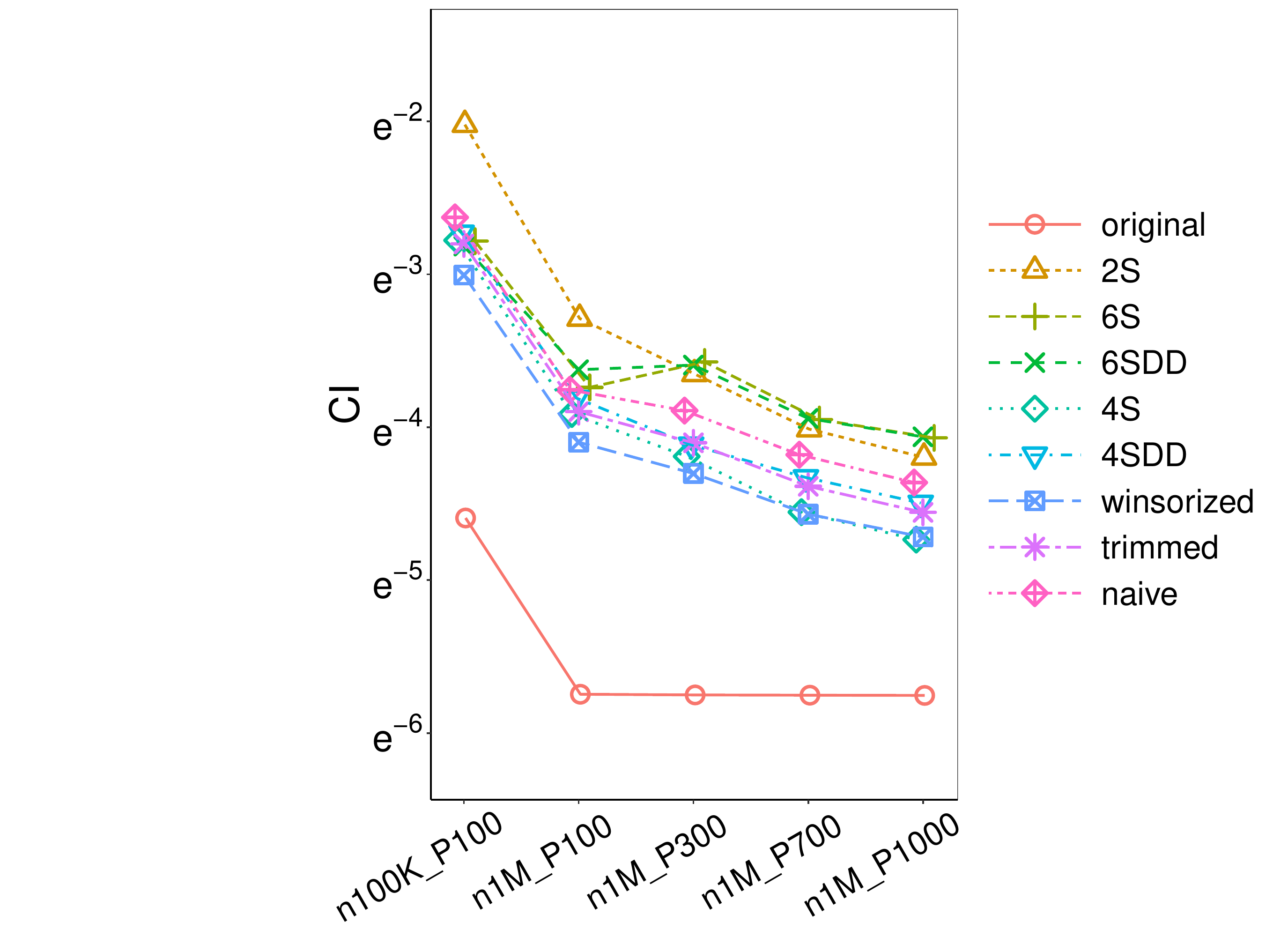}
\includegraphics[width=0.19\textwidth, trim={2.5in 0 2.6in 0},clip] {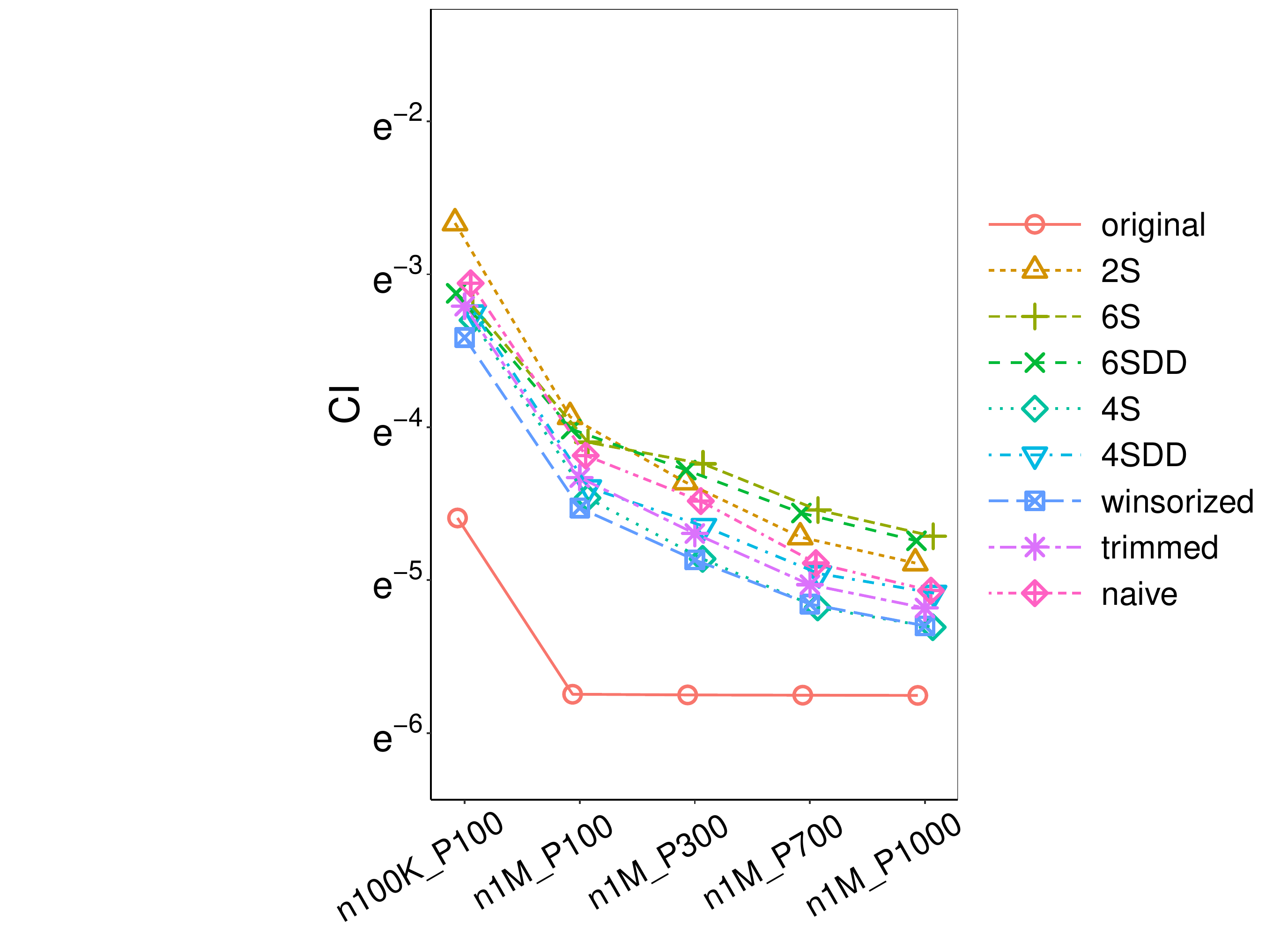}
\includegraphics[width=0.19\textwidth, trim={2.5in 0 2.6in 0},clip] {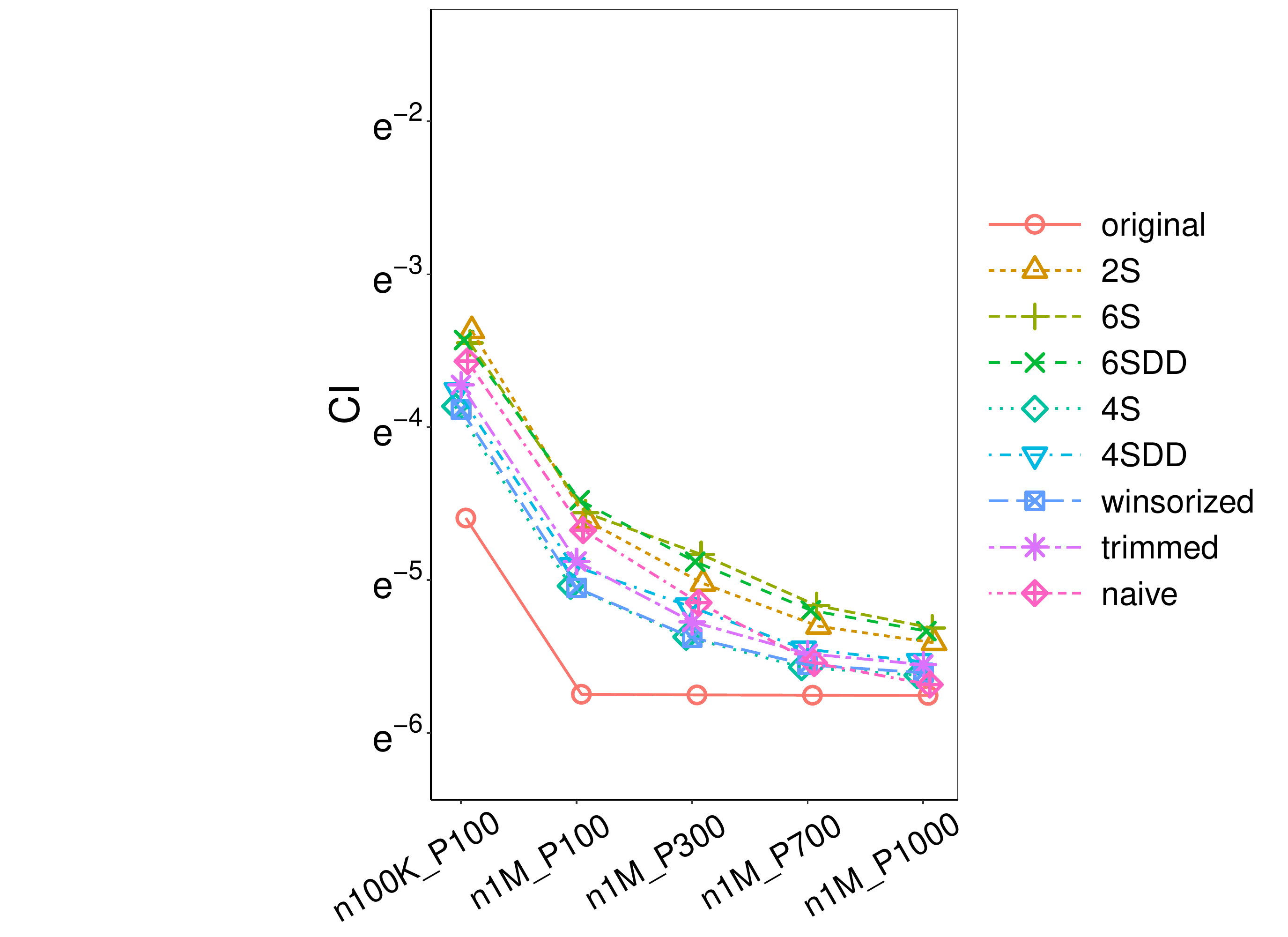}
\includegraphics[width=0.19\textwidth, trim={2.5in 0 2.6in 0},clip] {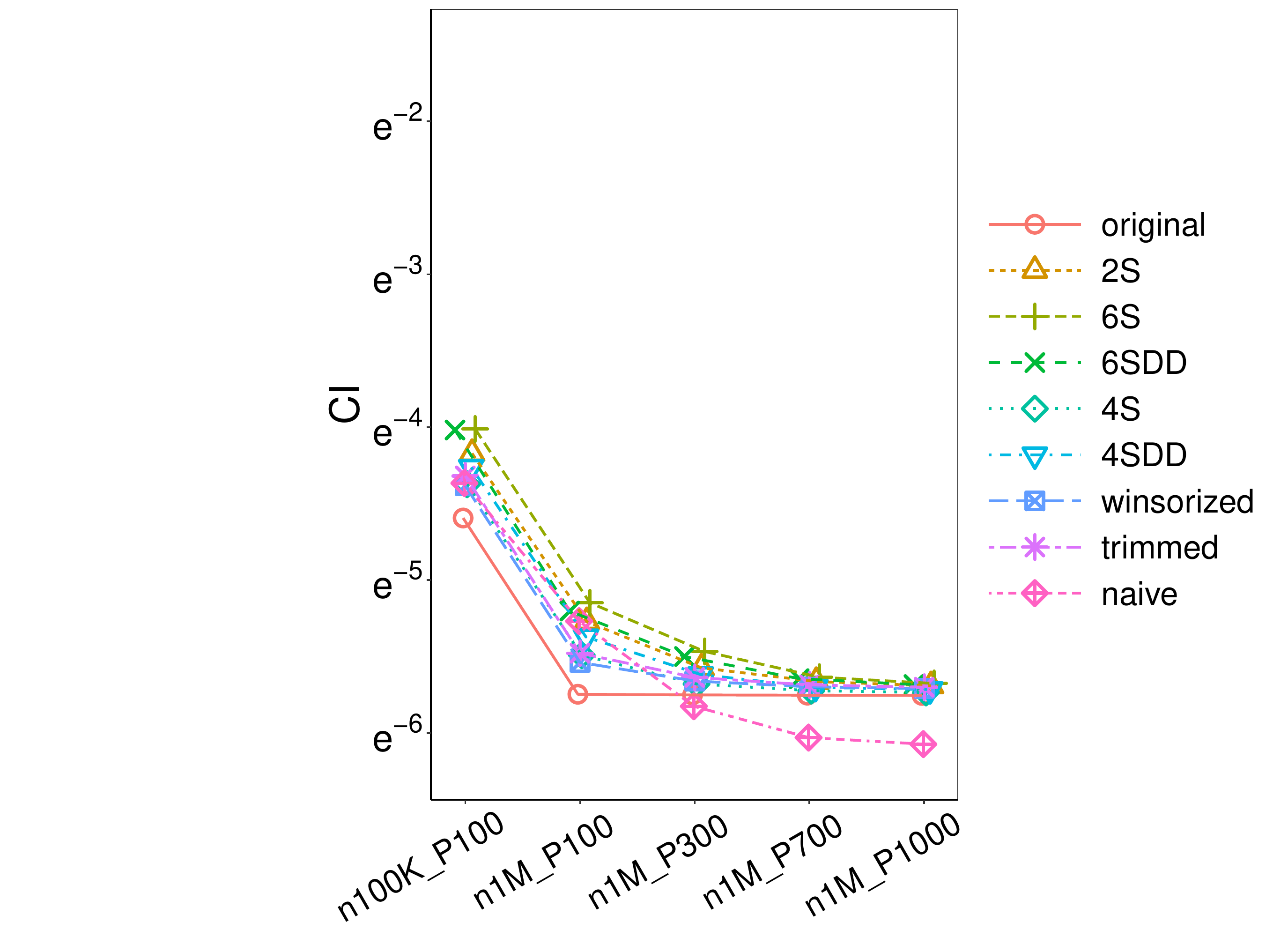}
\includegraphics[width=0.19\textwidth, trim={2.5in 0 2.6in 0},clip] {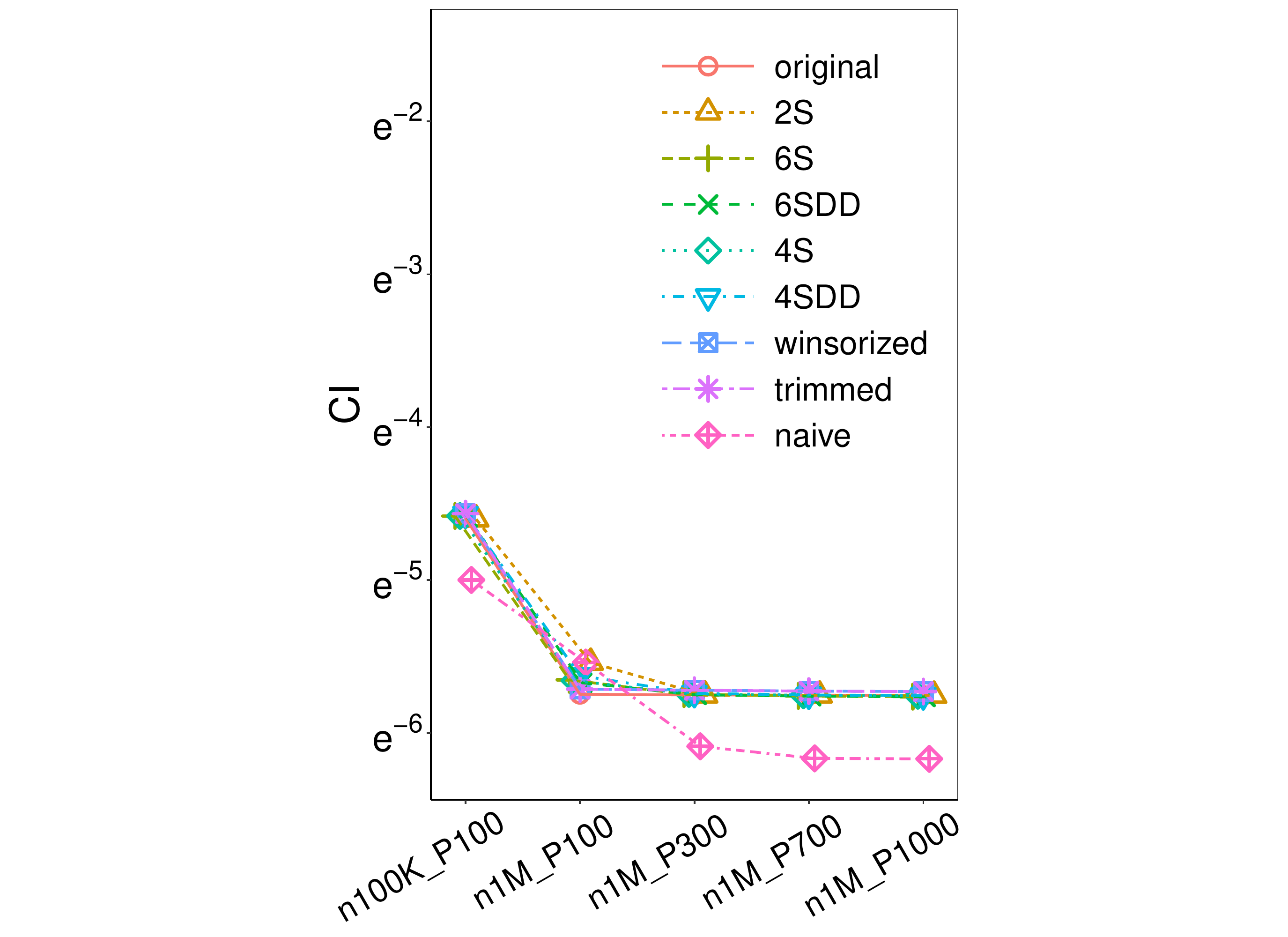}\\
\includegraphics[width=0.19\textwidth, trim={2.5in 0 2.6in 0},clip] {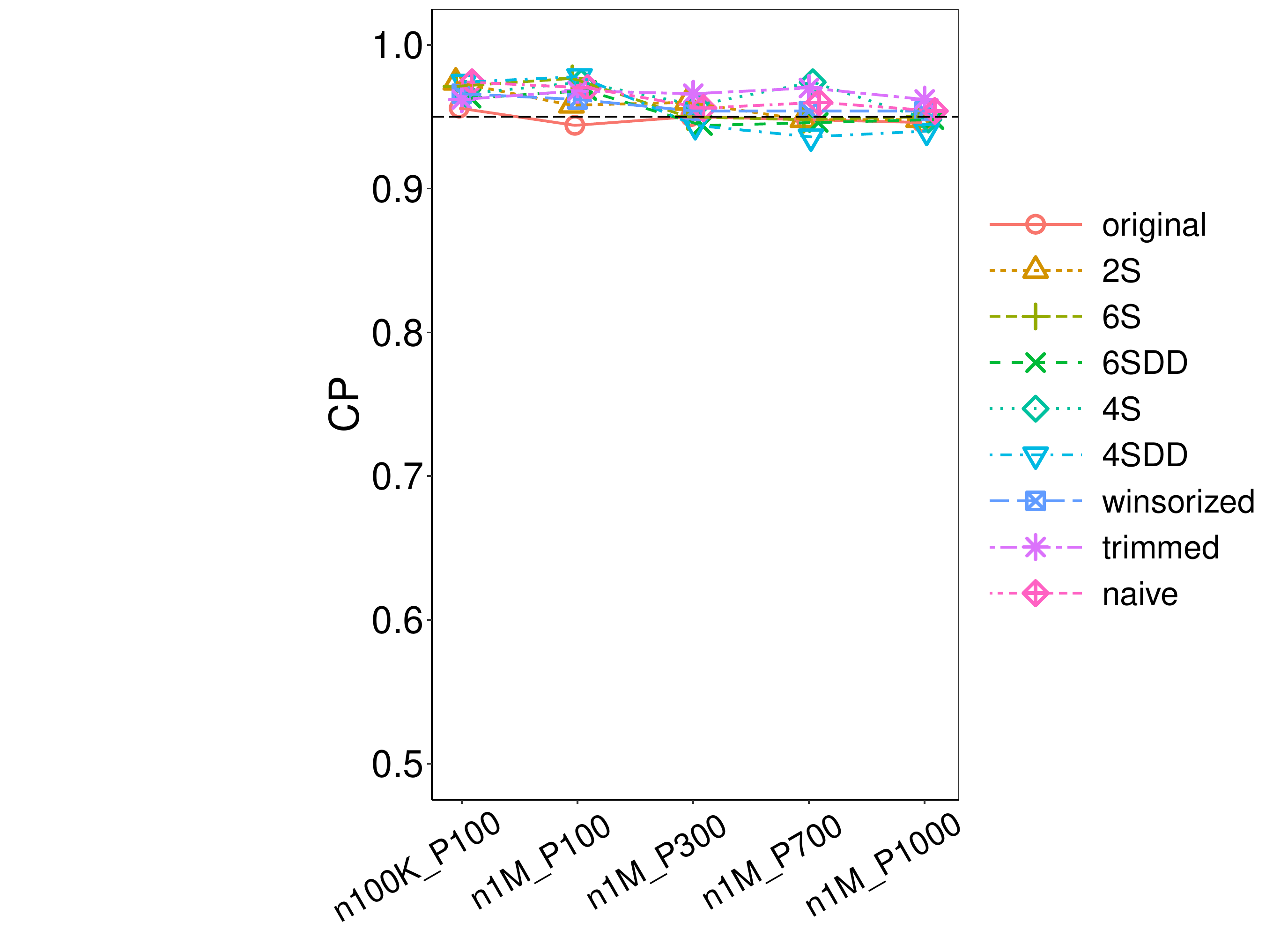}
\includegraphics[width=0.19\textwidth, trim={2.5in 0 2.6in 0},clip] {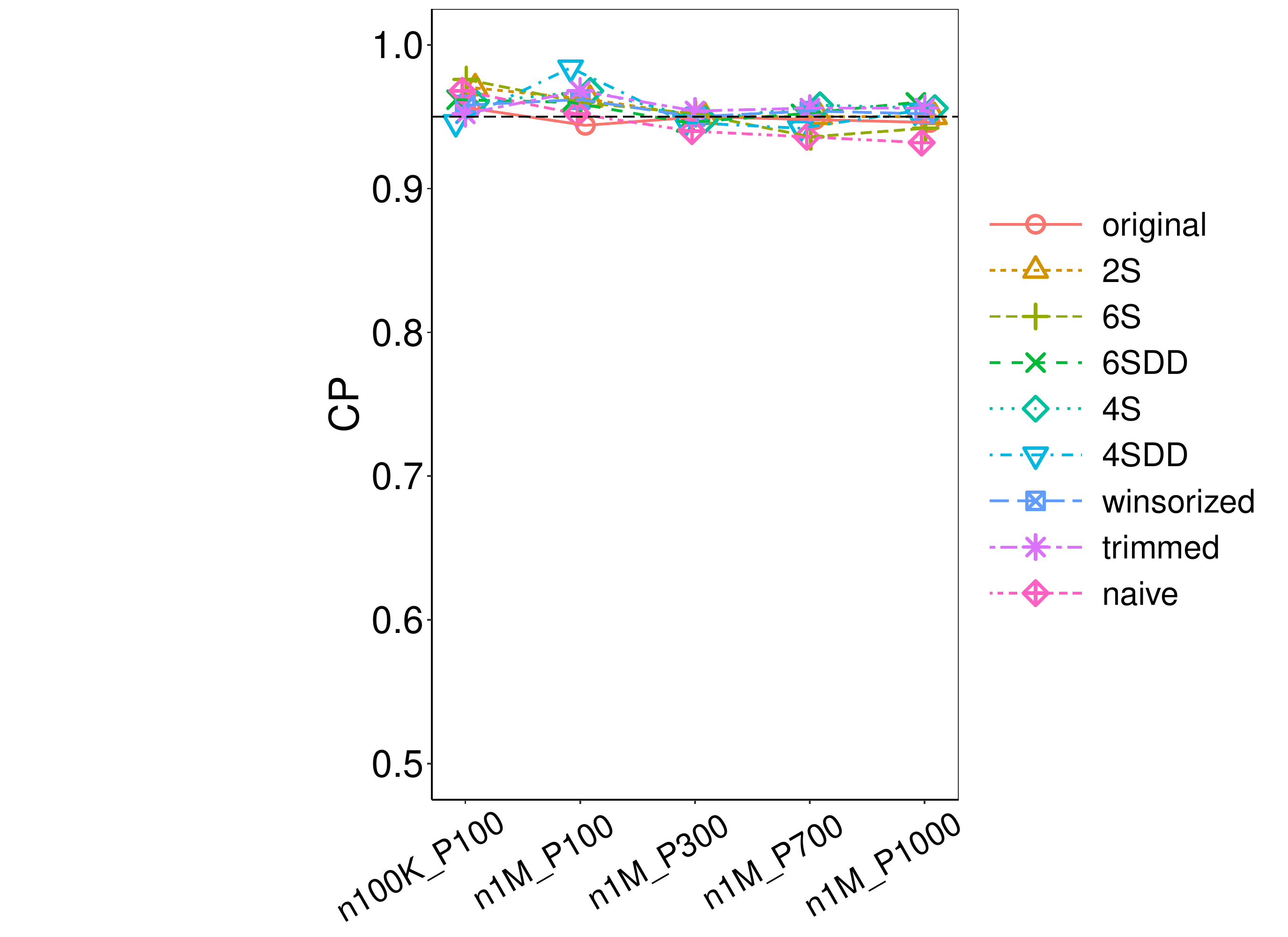}
\includegraphics[width=0.19\textwidth, trim={2.5in 0 2.6in 0},clip] {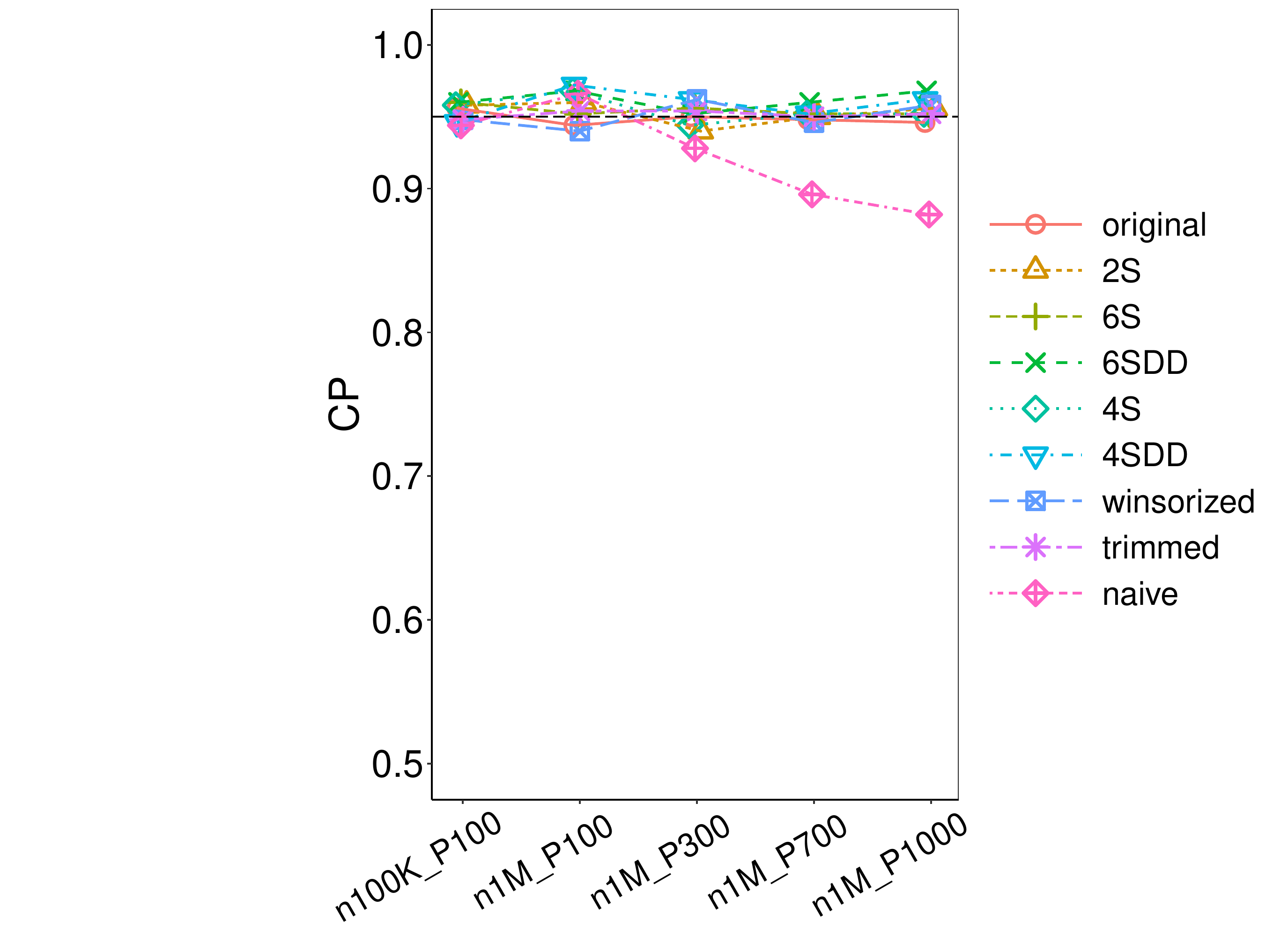}
\includegraphics[width=0.19\textwidth, trim={2.5in 0 2.6in 0},clip] {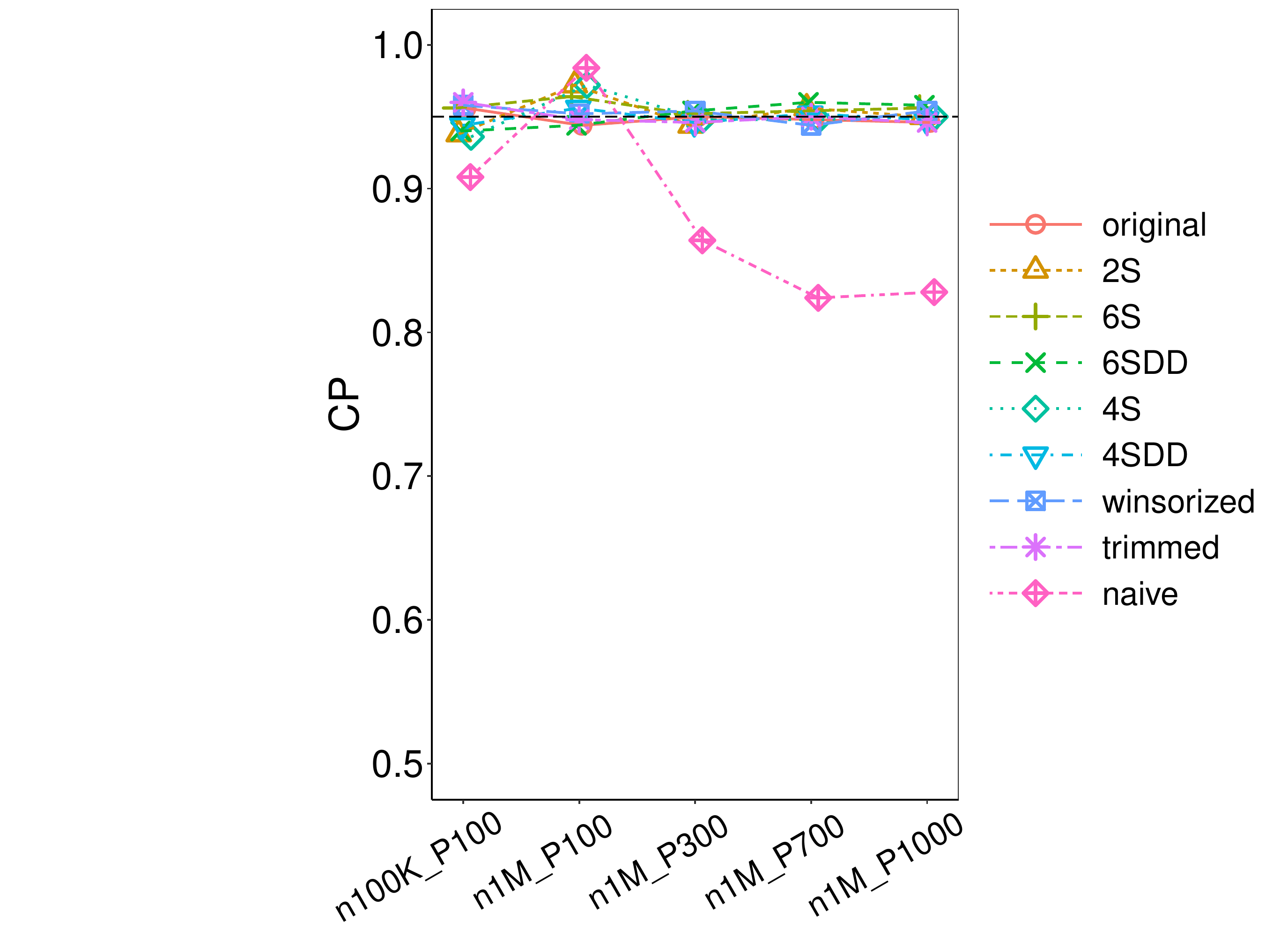}
\includegraphics[width=0.19\textwidth, trim={2.5in 0 2.6in 0},clip] {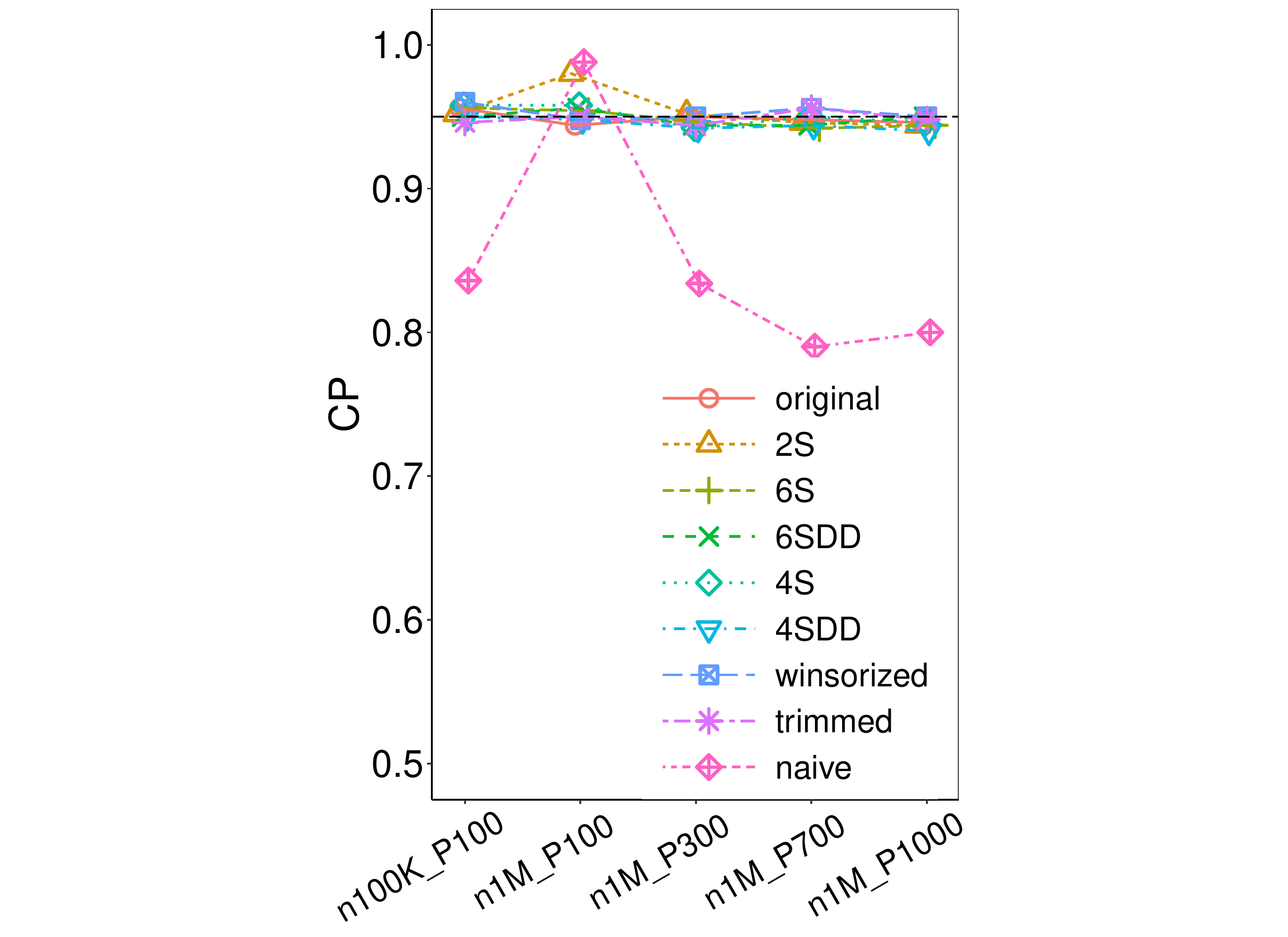}\\
\caption{Simulation results with $\epsilon$-DP for ZINB data with  $\alpha=\beta$ when $\theta=0$} \label{fig:0sDPZINB}

\end{figure}

\begin{figure}[!htb]
\hspace{0.45in}$\rho=0.005$\hspace{0.65in}$\rho=0.02$\hspace{0.65in}$\rho=0.08$
\hspace{0.65in}$\rho=0.32$\hspace{0.65in}$\rho=1.28$\\
\includegraphics[width=0.19\textwidth, trim={2.5in 0 2.5in 0},clip] {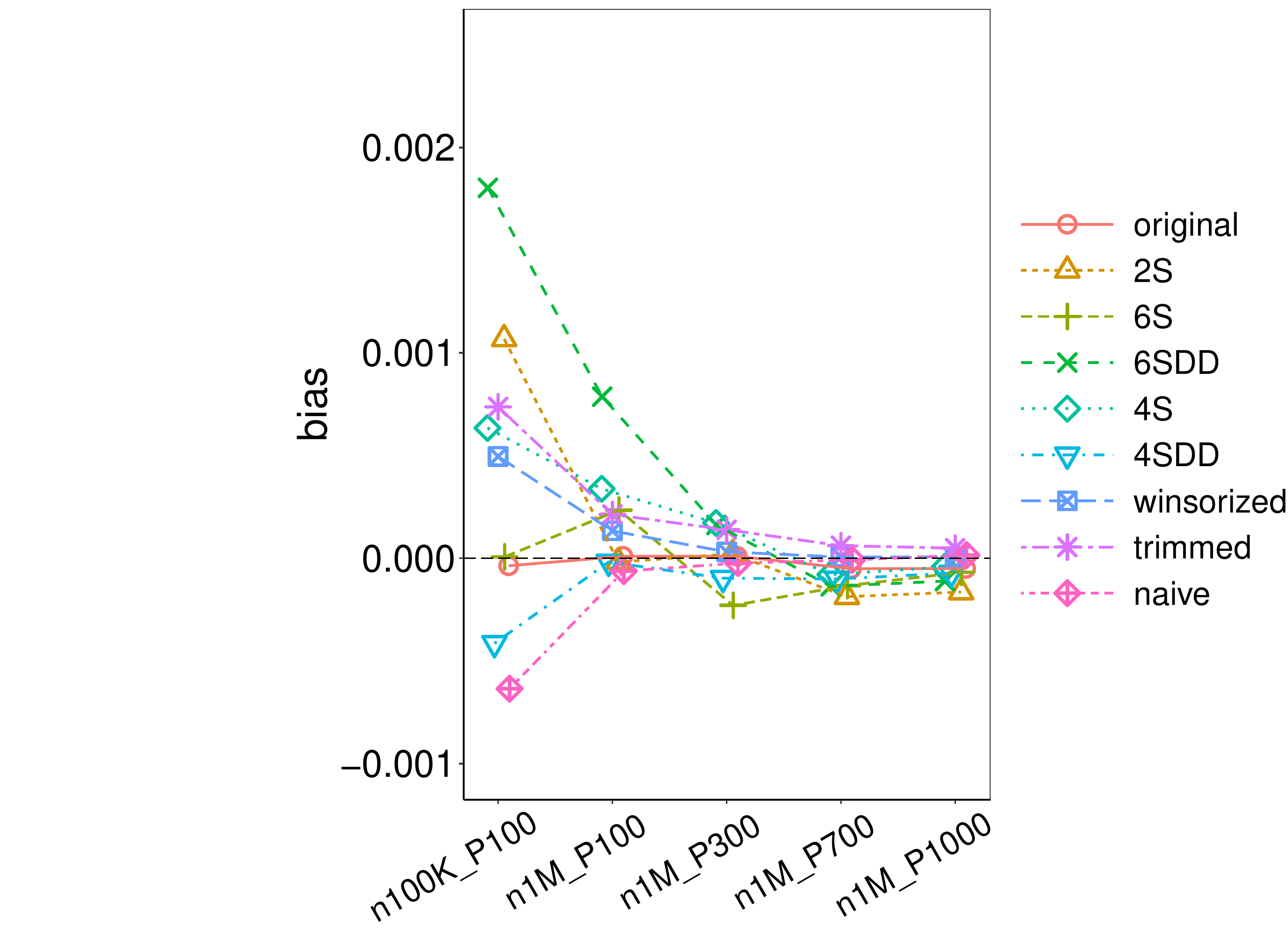}
\includegraphics[width=0.19\textwidth, trim={2.5in 0 2.5in 0},clip] {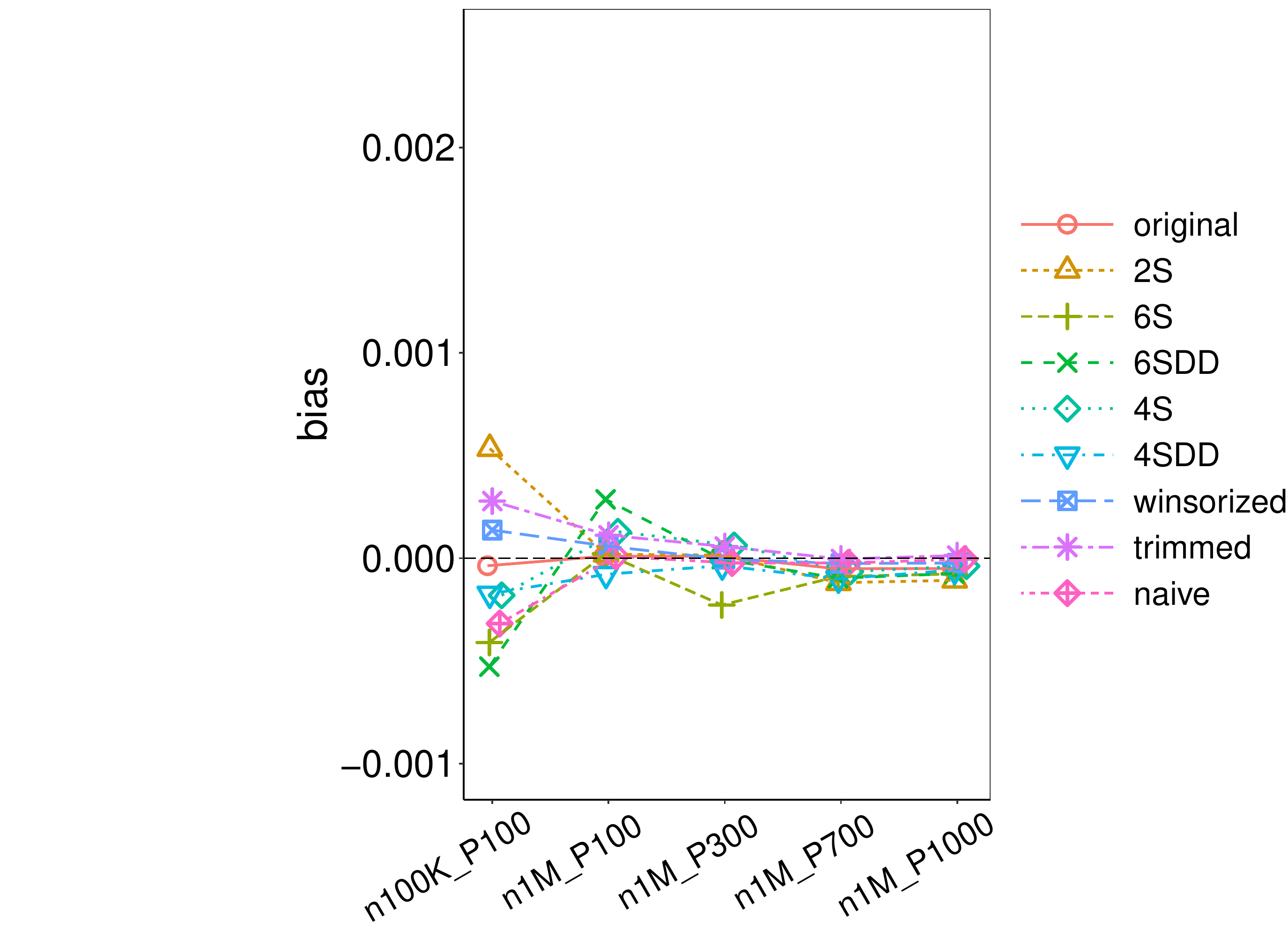}
\includegraphics[width=0.19\textwidth, trim={2.5in 0 2.5in 0},clip] {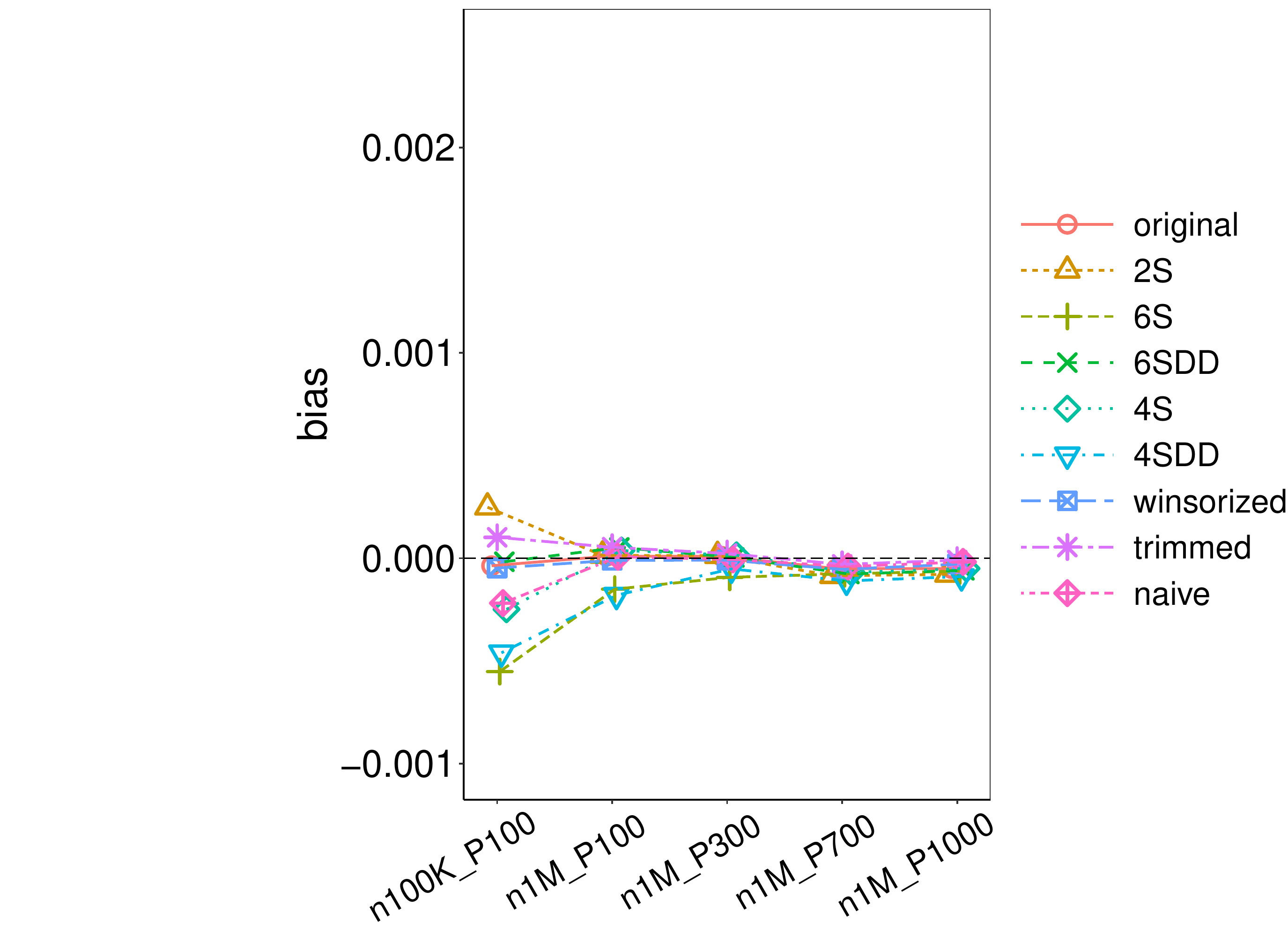}
\includegraphics[width=0.19\textwidth, trim={2.5in 0 2.5in 0},clip] {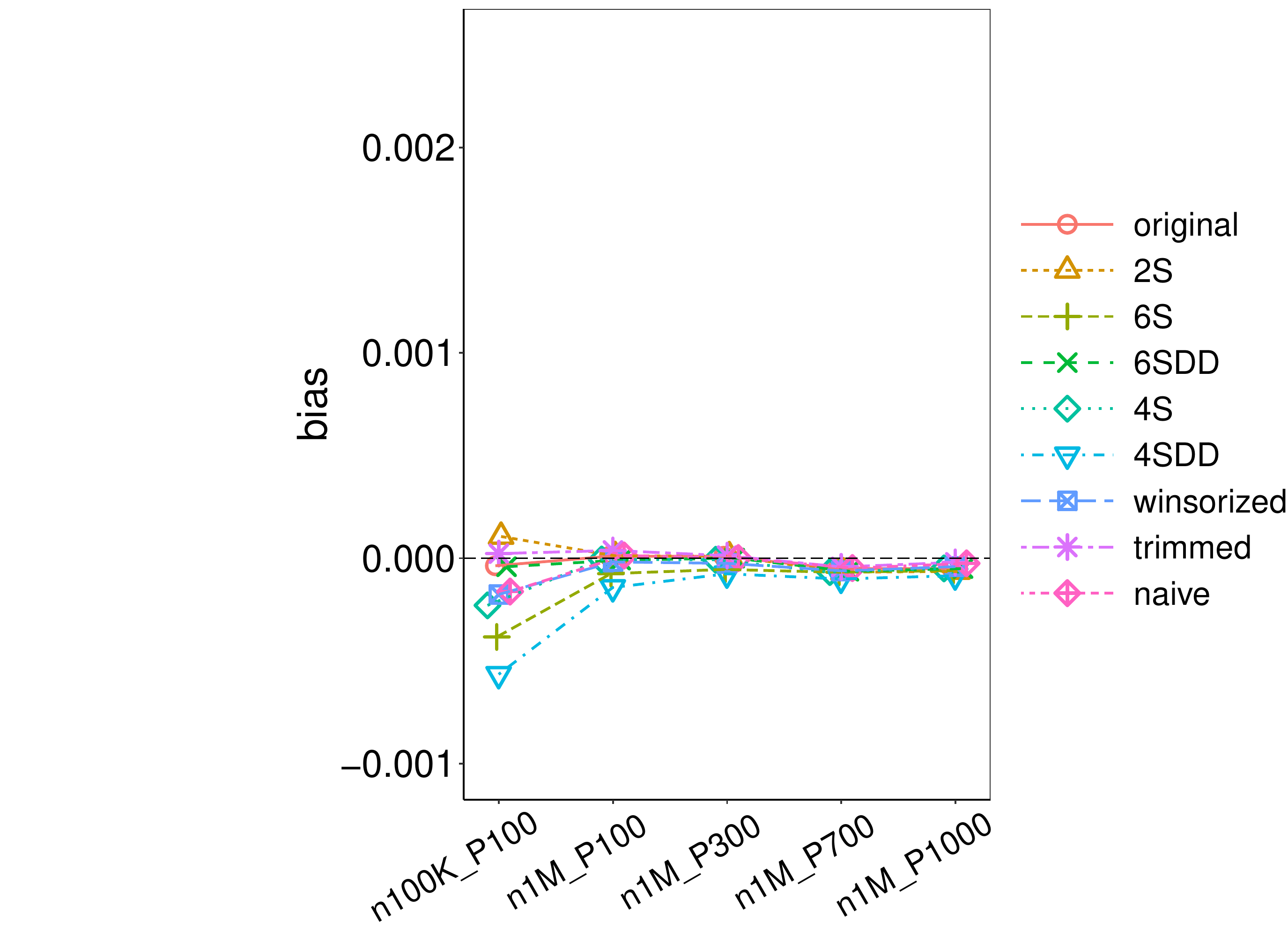}
\includegraphics[width=0.19\textwidth, trim={2.5in 0 2.5in 0},clip] {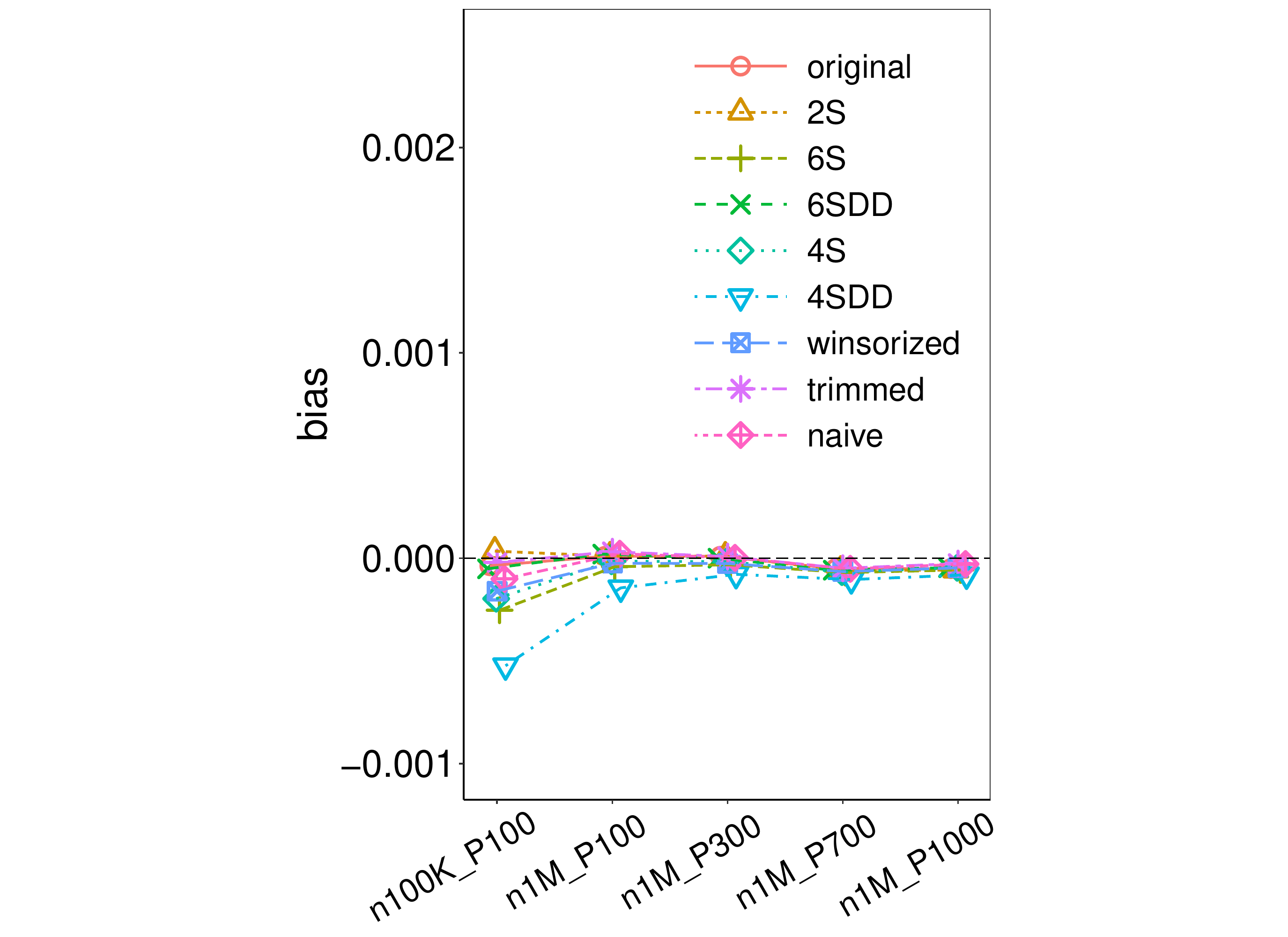}\\
\includegraphics[width=0.19\textwidth, trim={2.5in 0 2.6in 0},clip] {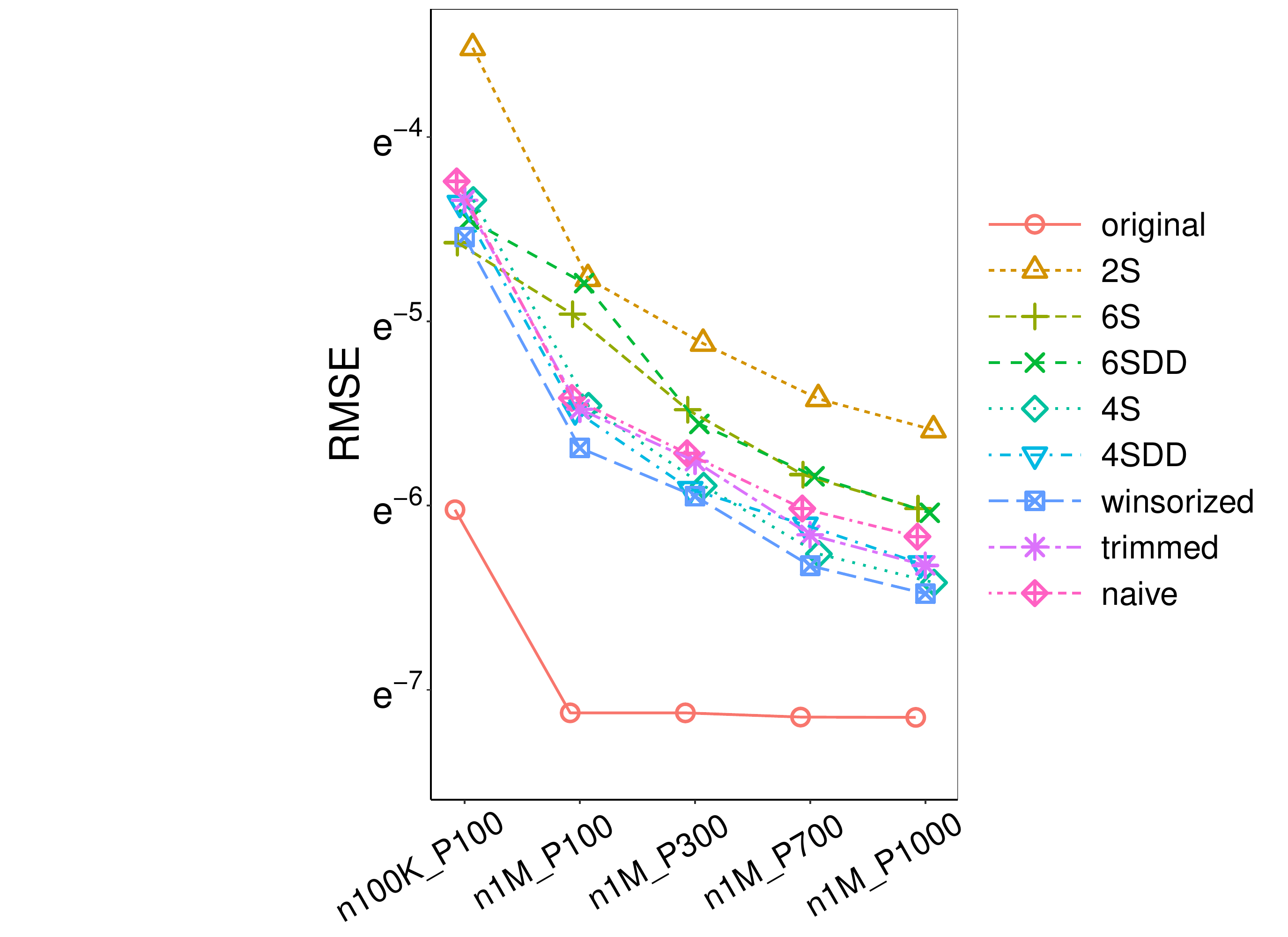}
\includegraphics[width=0.19\textwidth, trim={2.5in 0 2.6in 0},clip] {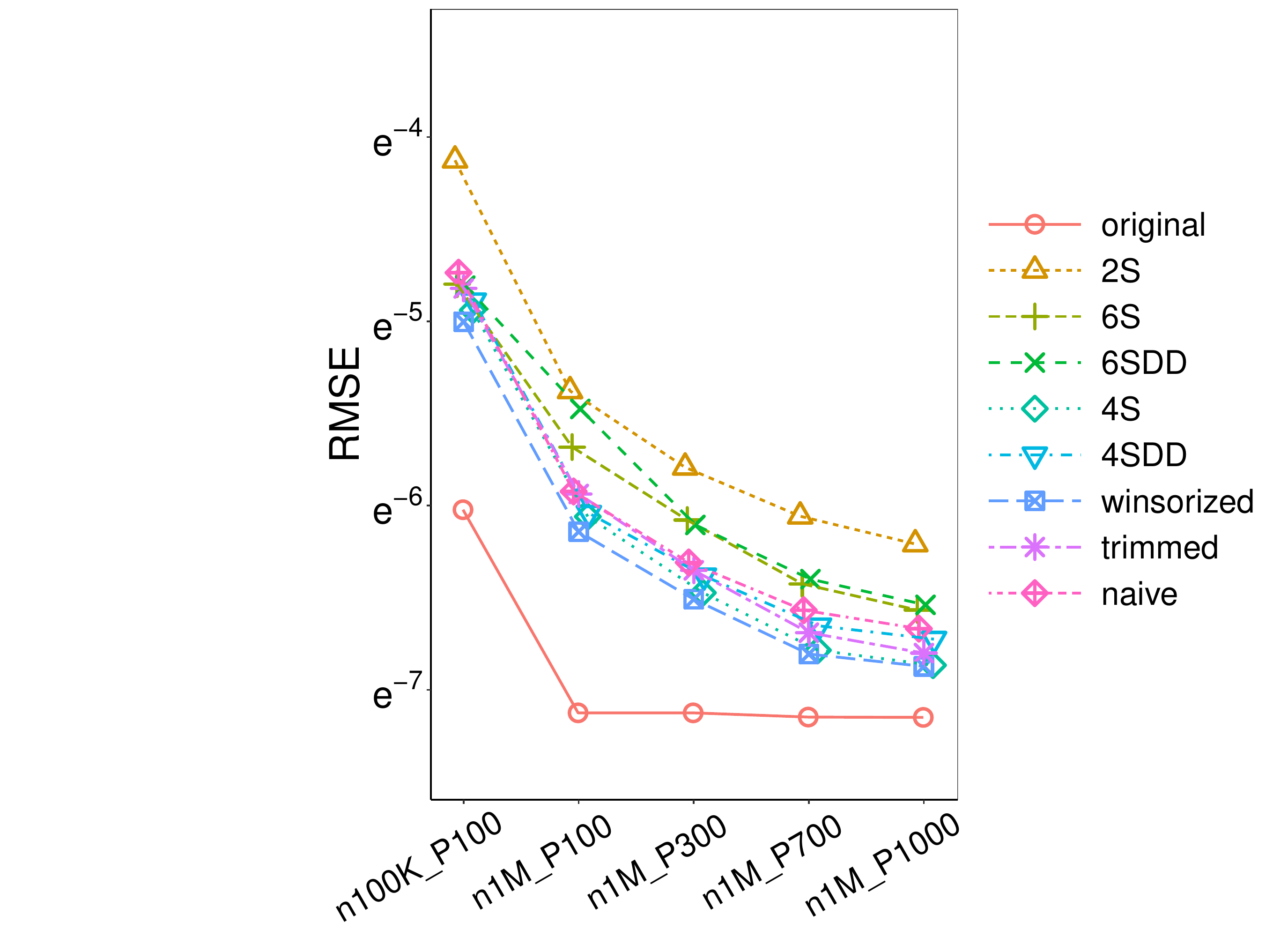}
\includegraphics[width=0.19\textwidth, trim={2.5in 0 2.6in 0},clip] {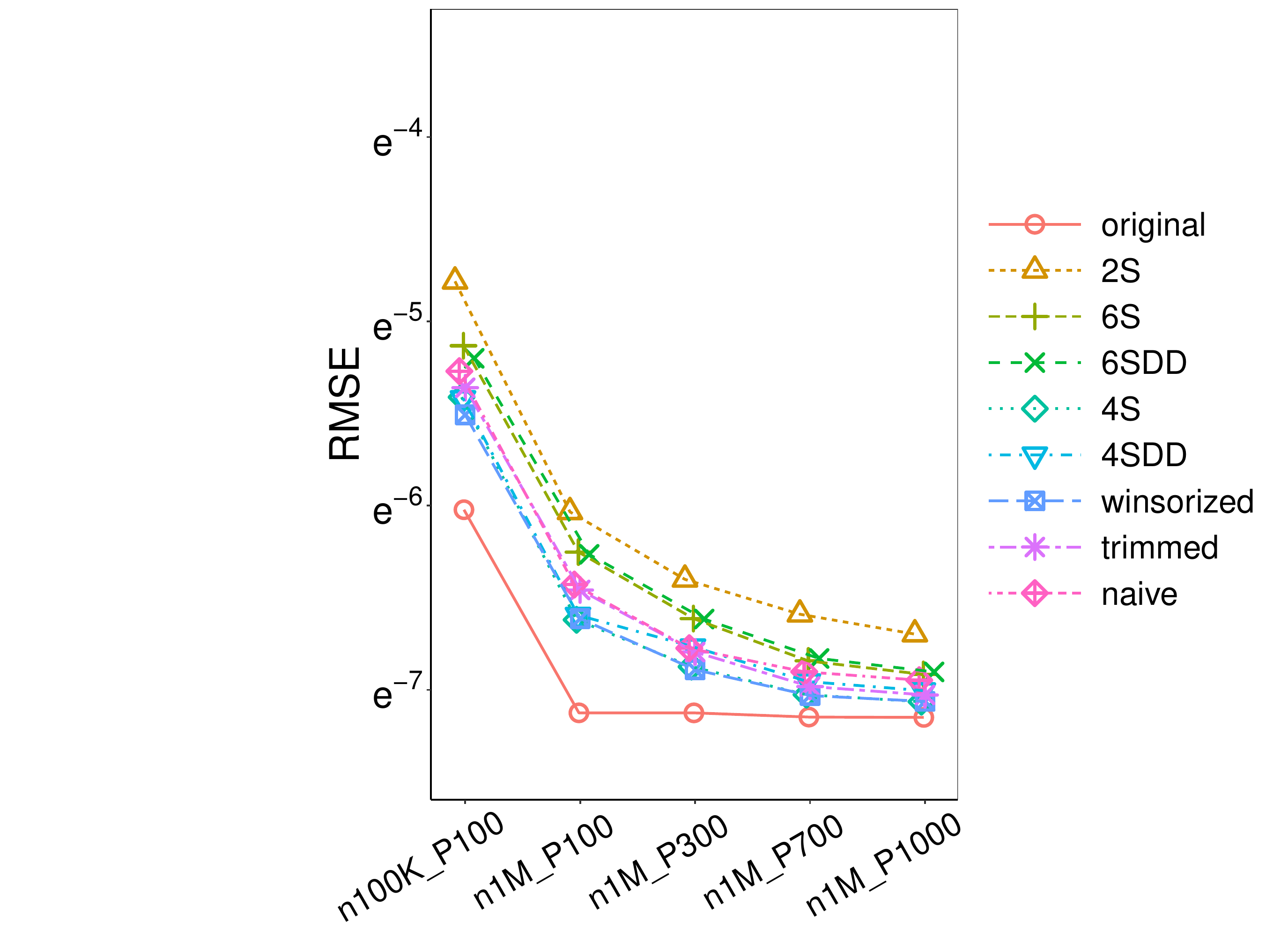}
\includegraphics[width=0.19\textwidth, trim={2.5in 0 2.6in 0},clip] {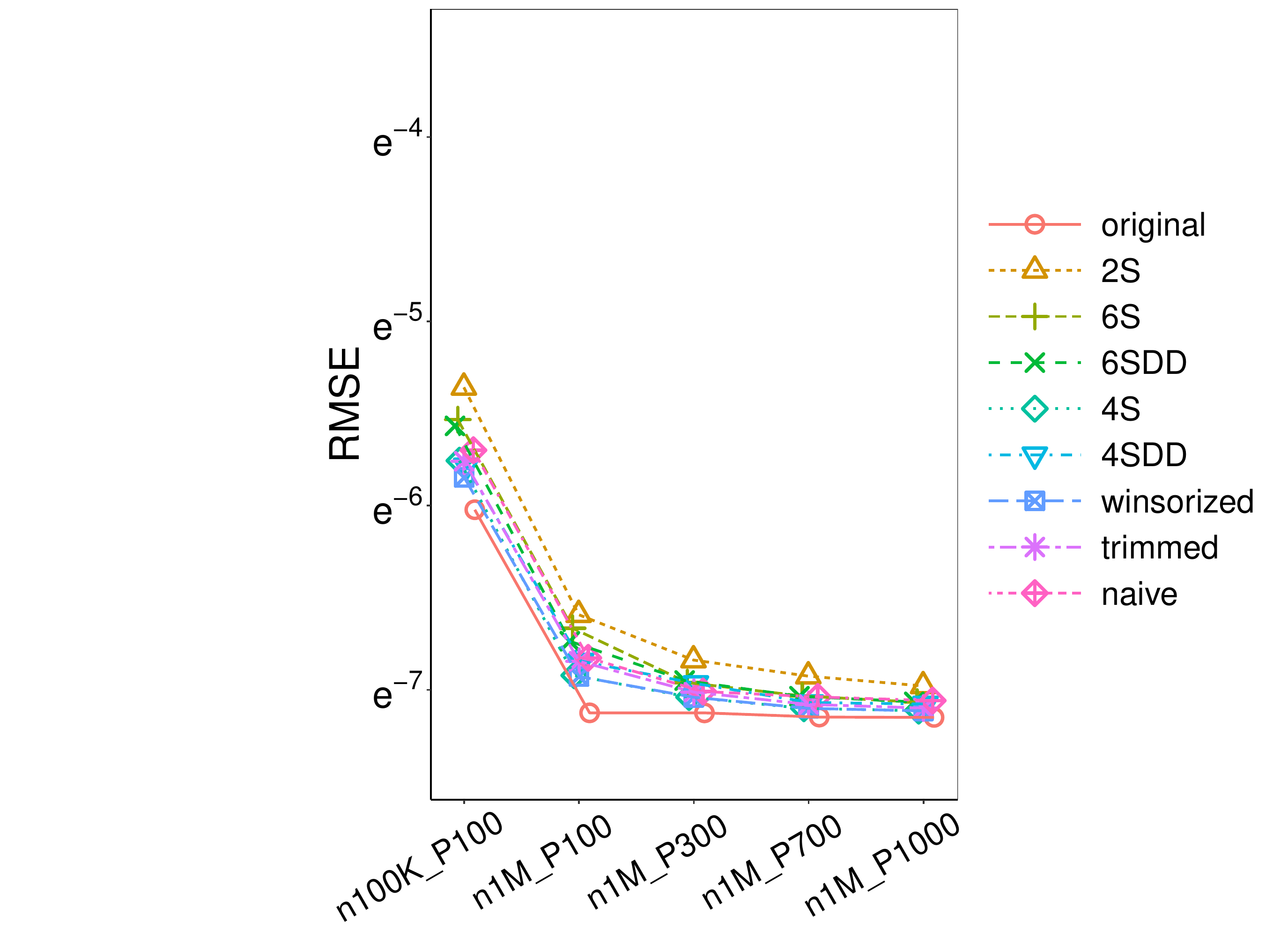}
\includegraphics[width=0.19\textwidth, trim={2.5in 0 2.6in 0},clip] {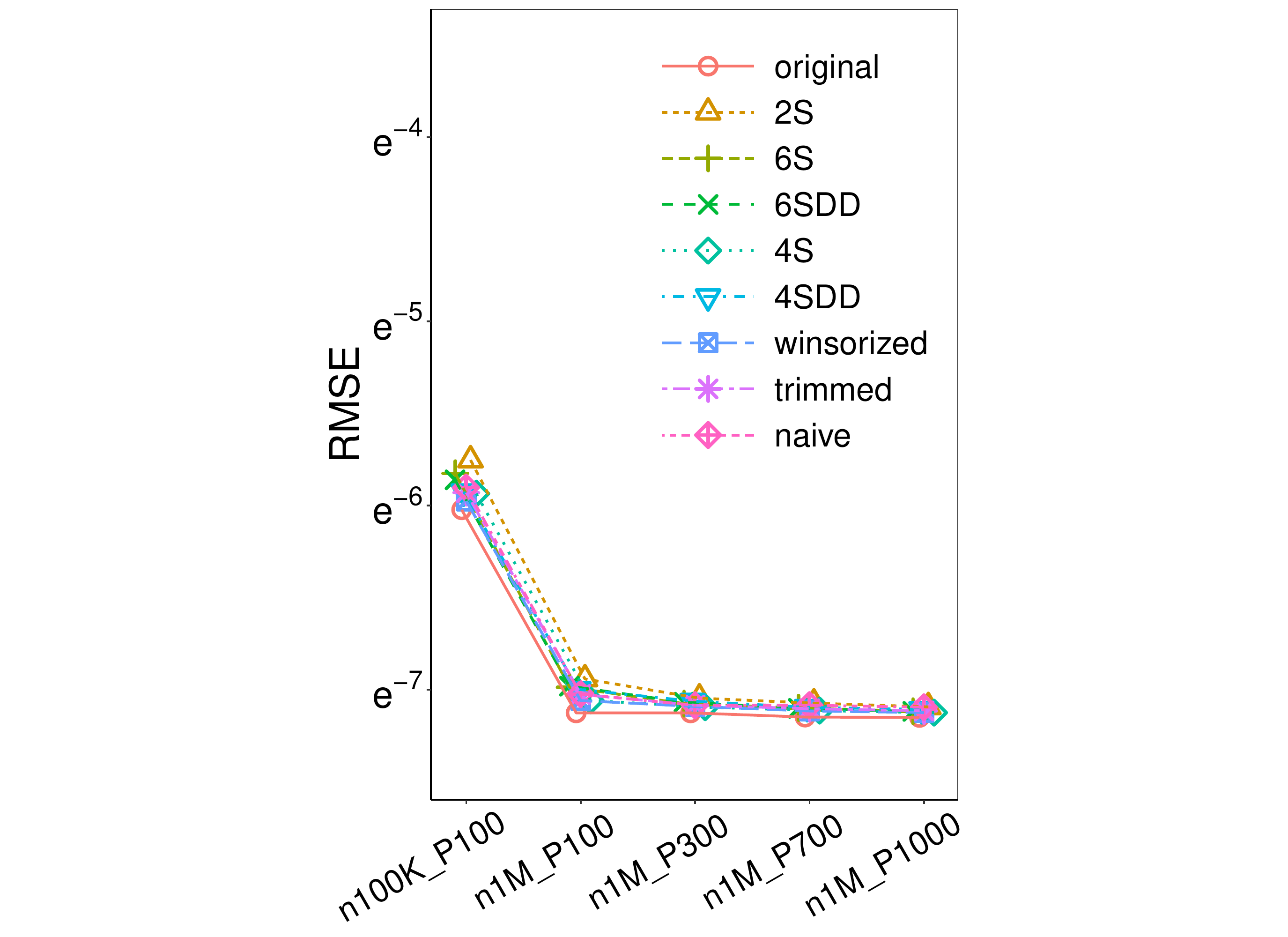}\\
\includegraphics[width=0.19\textwidth, trim={2.5in 0 2.6in 0},clip] {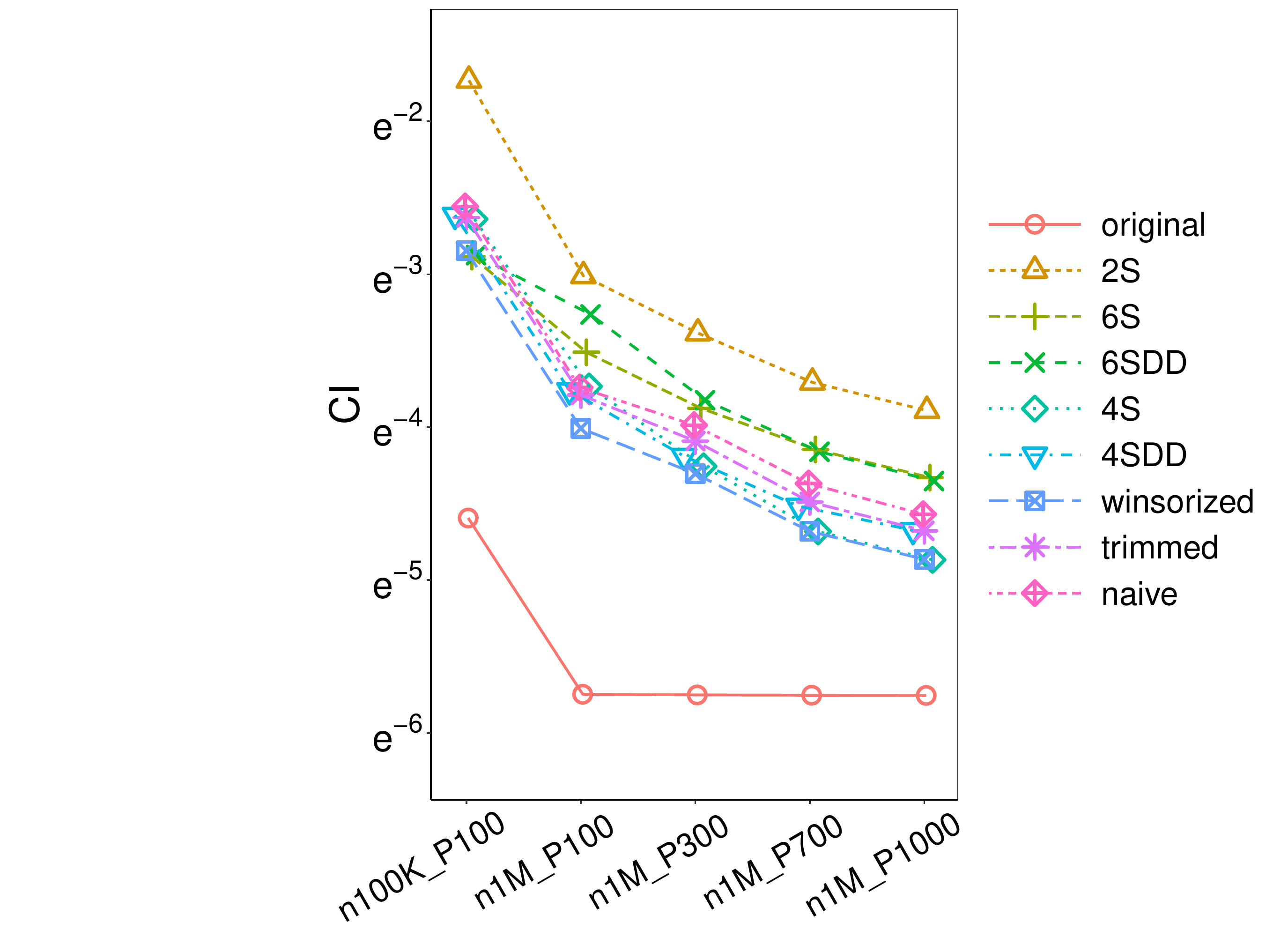}
\includegraphics[width=0.19\textwidth, trim={2.5in 0 2.6in 0},clip] {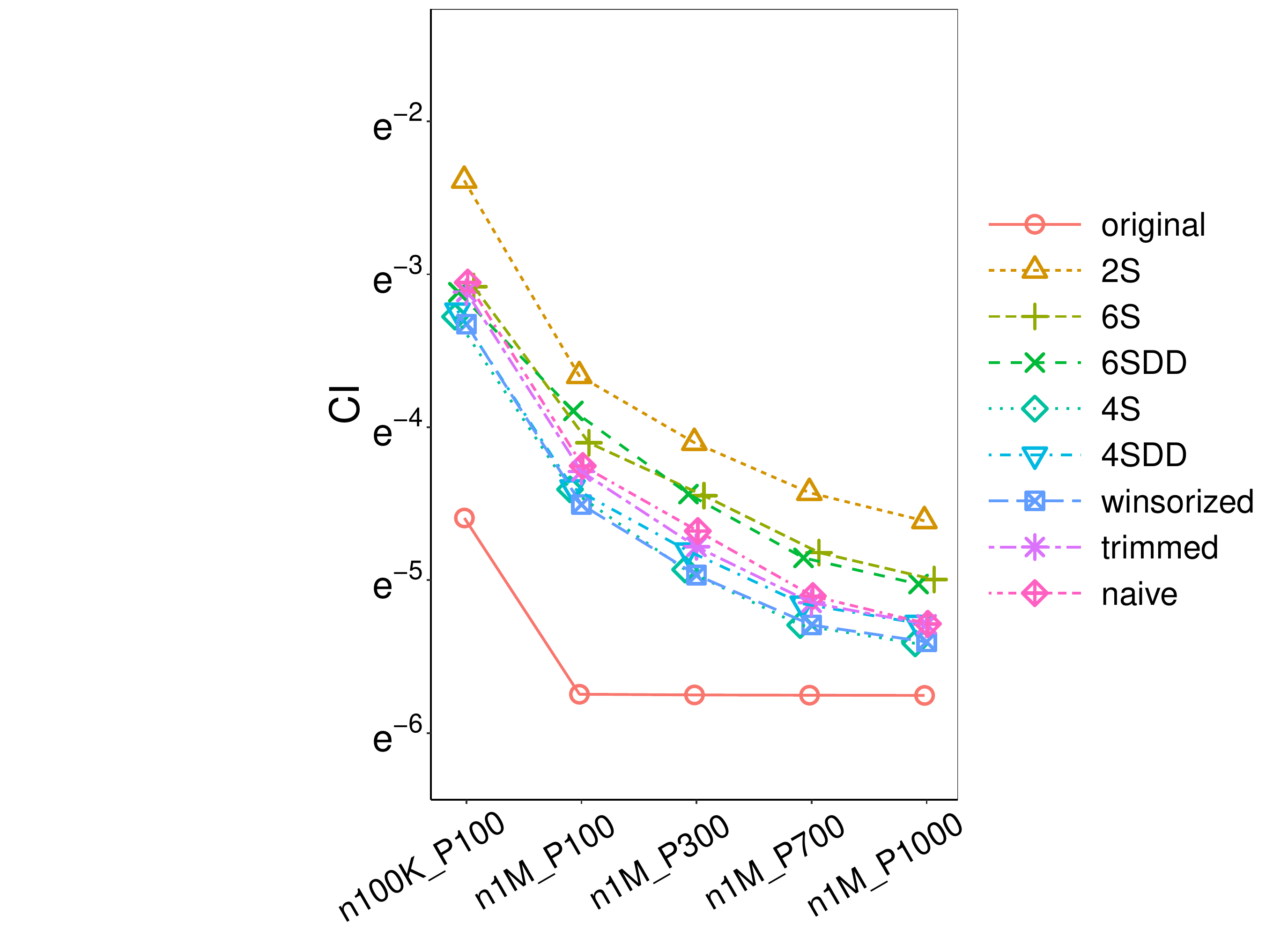}
\includegraphics[width=0.19\textwidth, trim={2.5in 0 2.6in 0},clip] {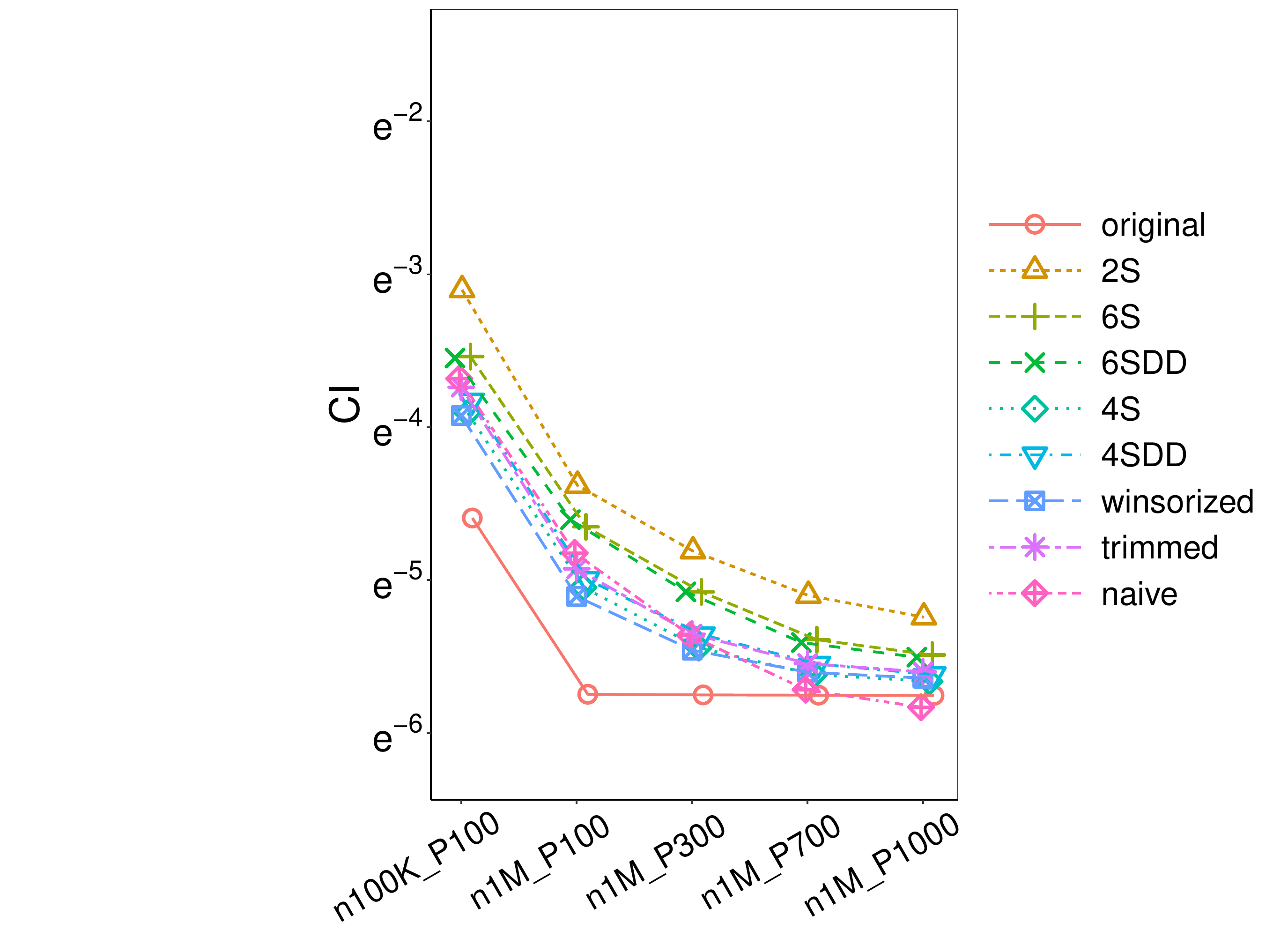}
\includegraphics[width=0.19\textwidth, trim={2.5in 0 2.6in 0},clip] {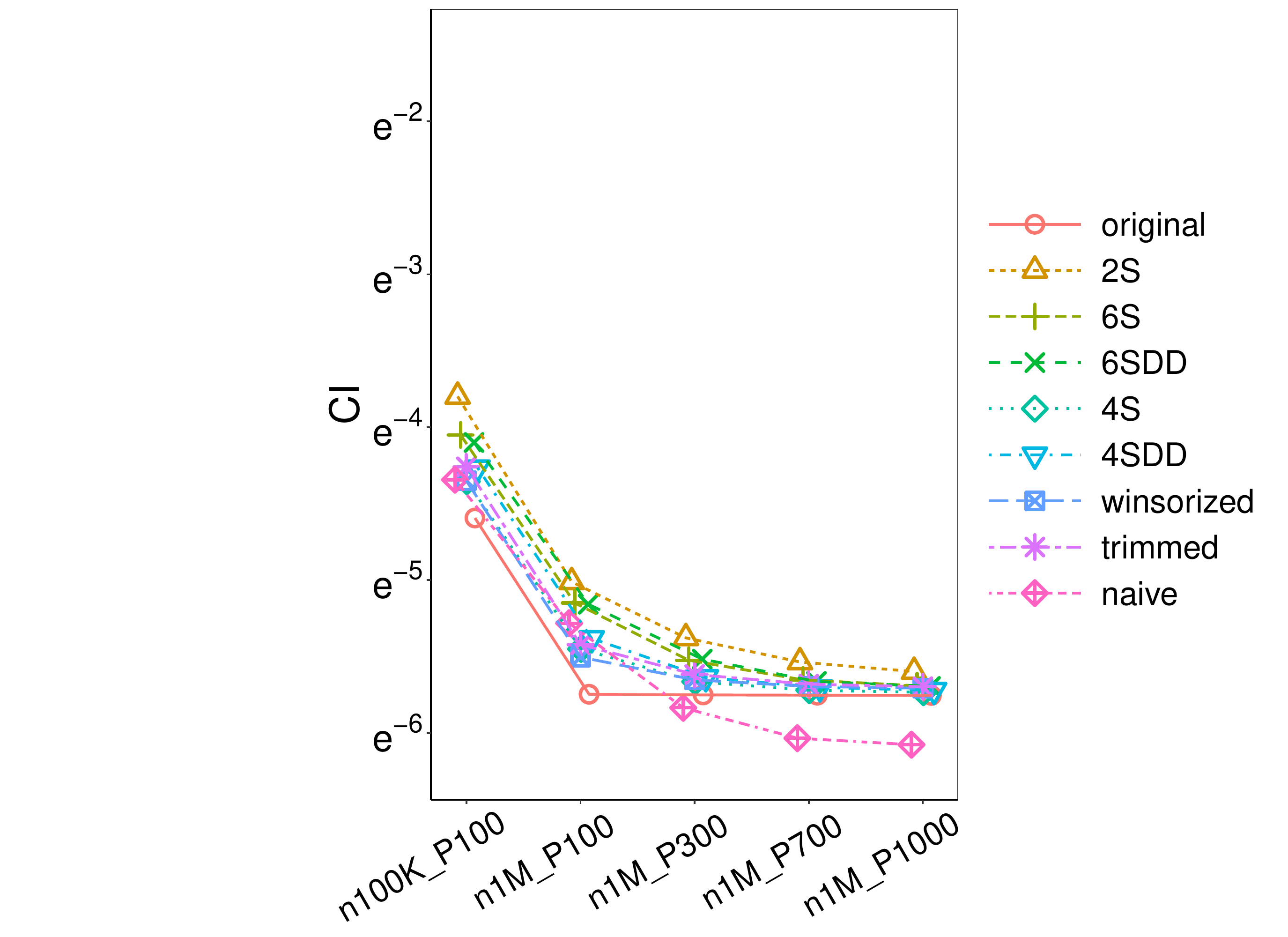}
\includegraphics[width=0.19\textwidth, trim={2.5in 0 2.6in 0},clip] {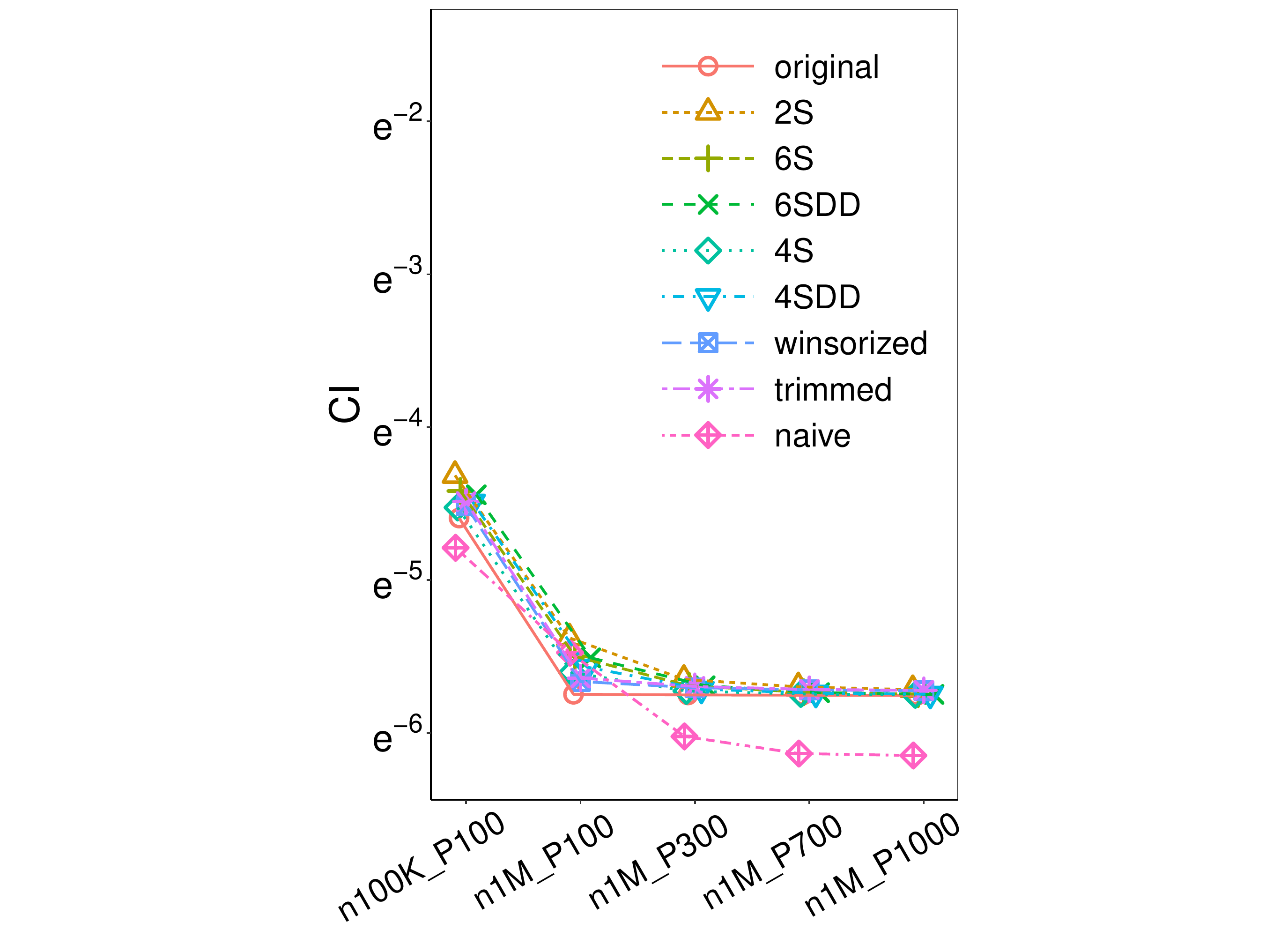}\\
\includegraphics[width=0.19\textwidth, trim={2.5in 0 2.6in 0},clip] {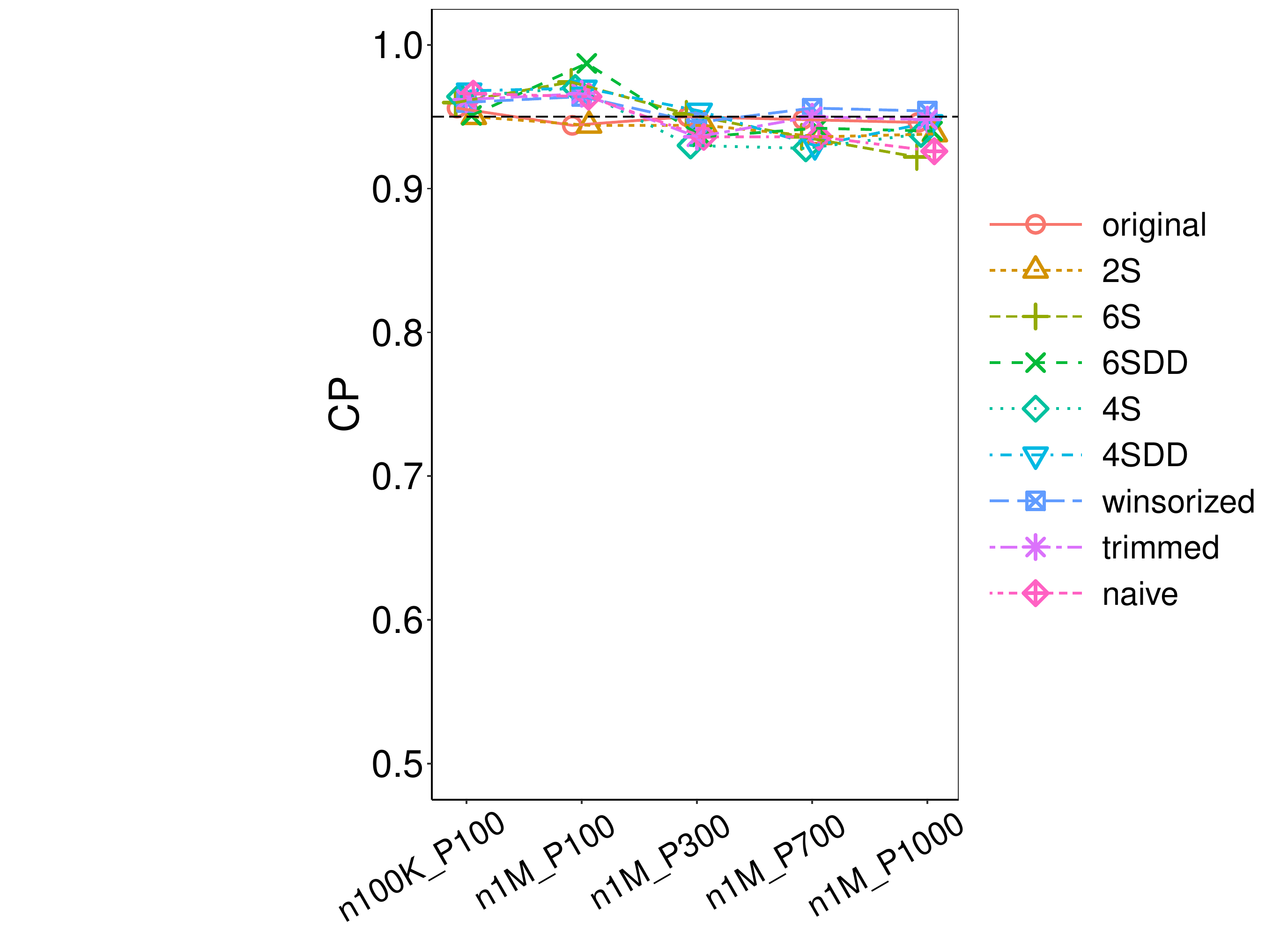}
\includegraphics[width=0.19\textwidth, trim={2.5in 0 2.6in 0},clip] {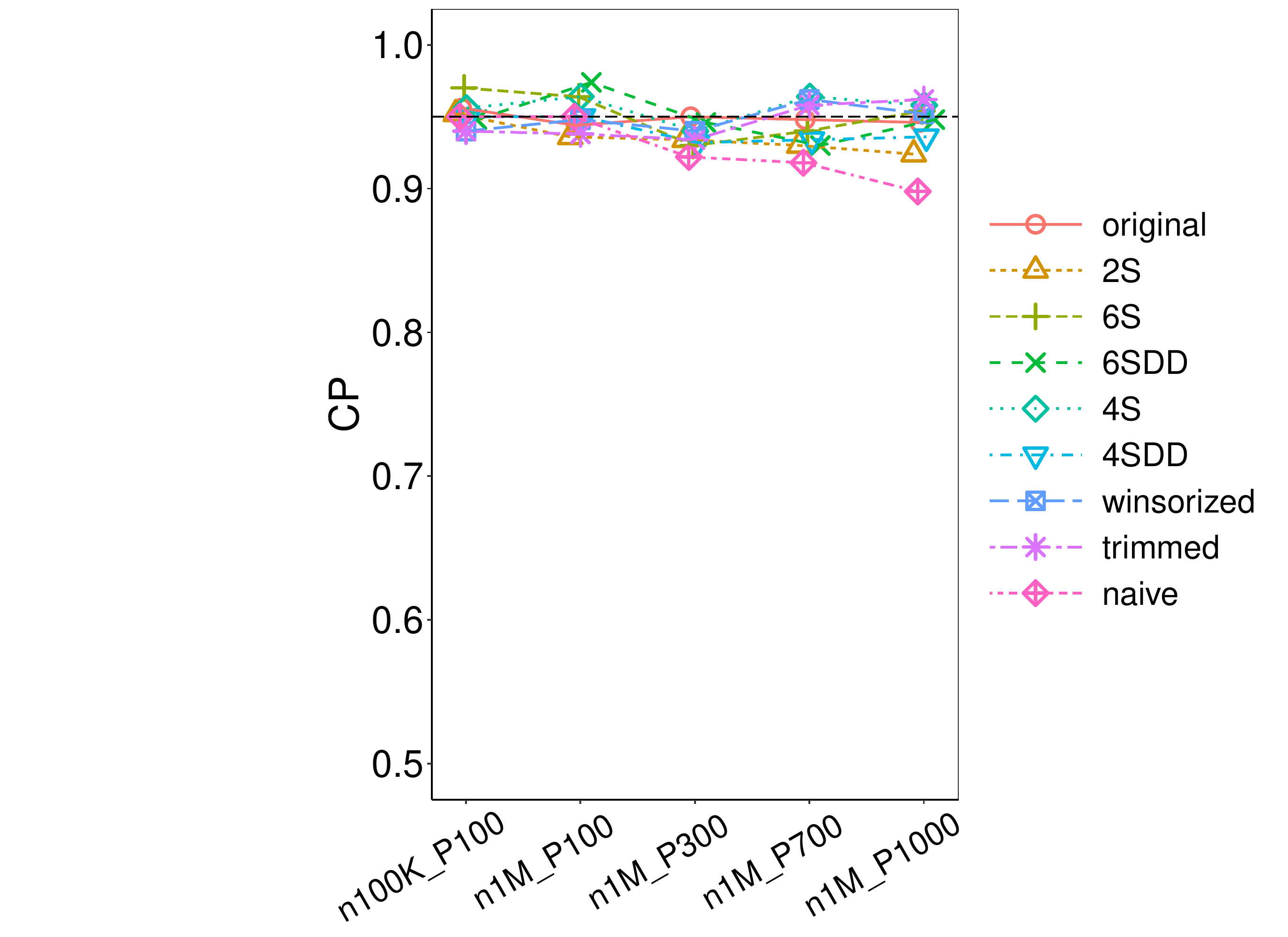}
\includegraphics[width=0.19\textwidth, trim={2.5in 0 2.6in 0},clip] {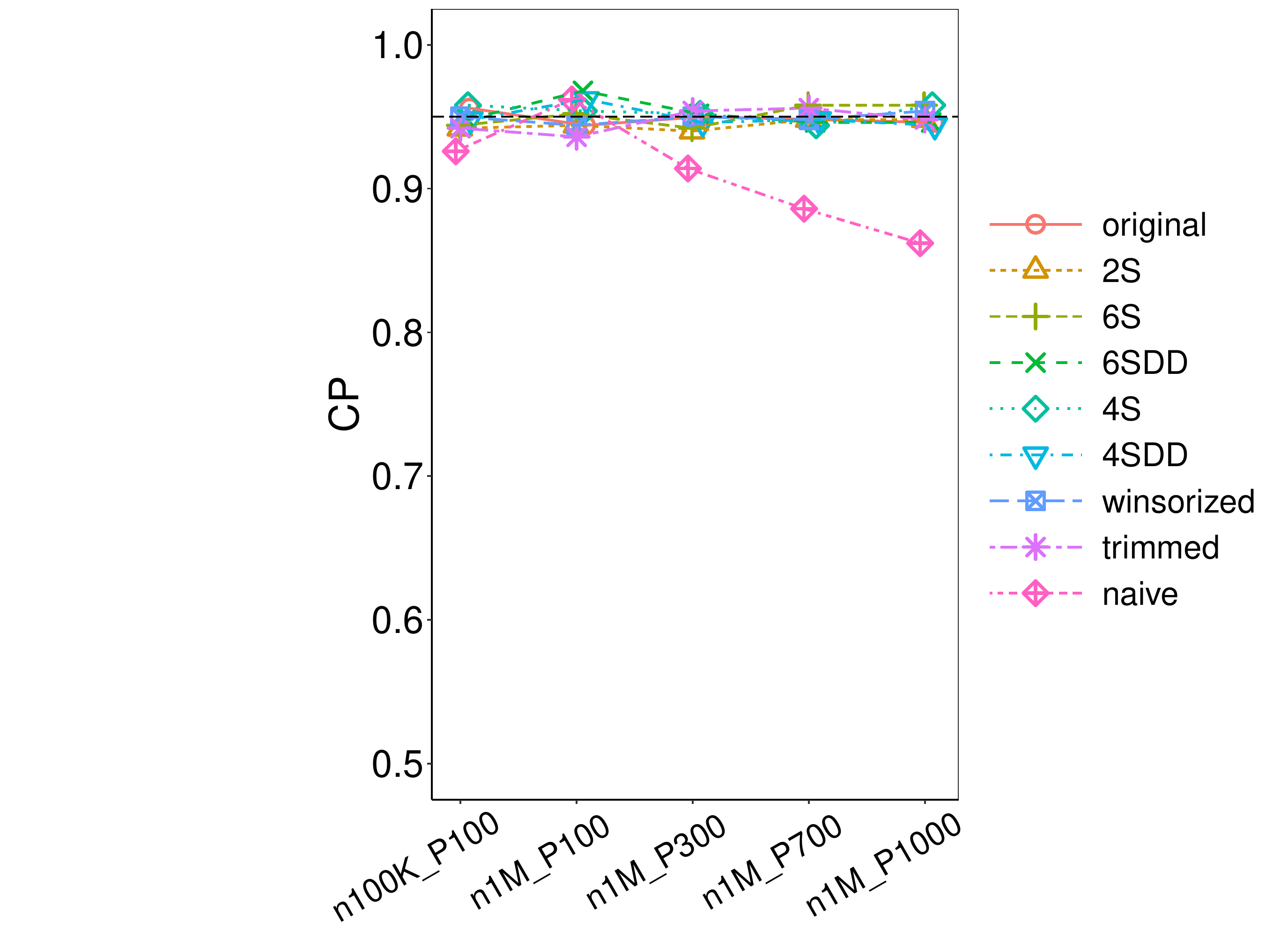}
\includegraphics[width=0.19\textwidth, trim={2.5in 0 2.6in 0},clip] {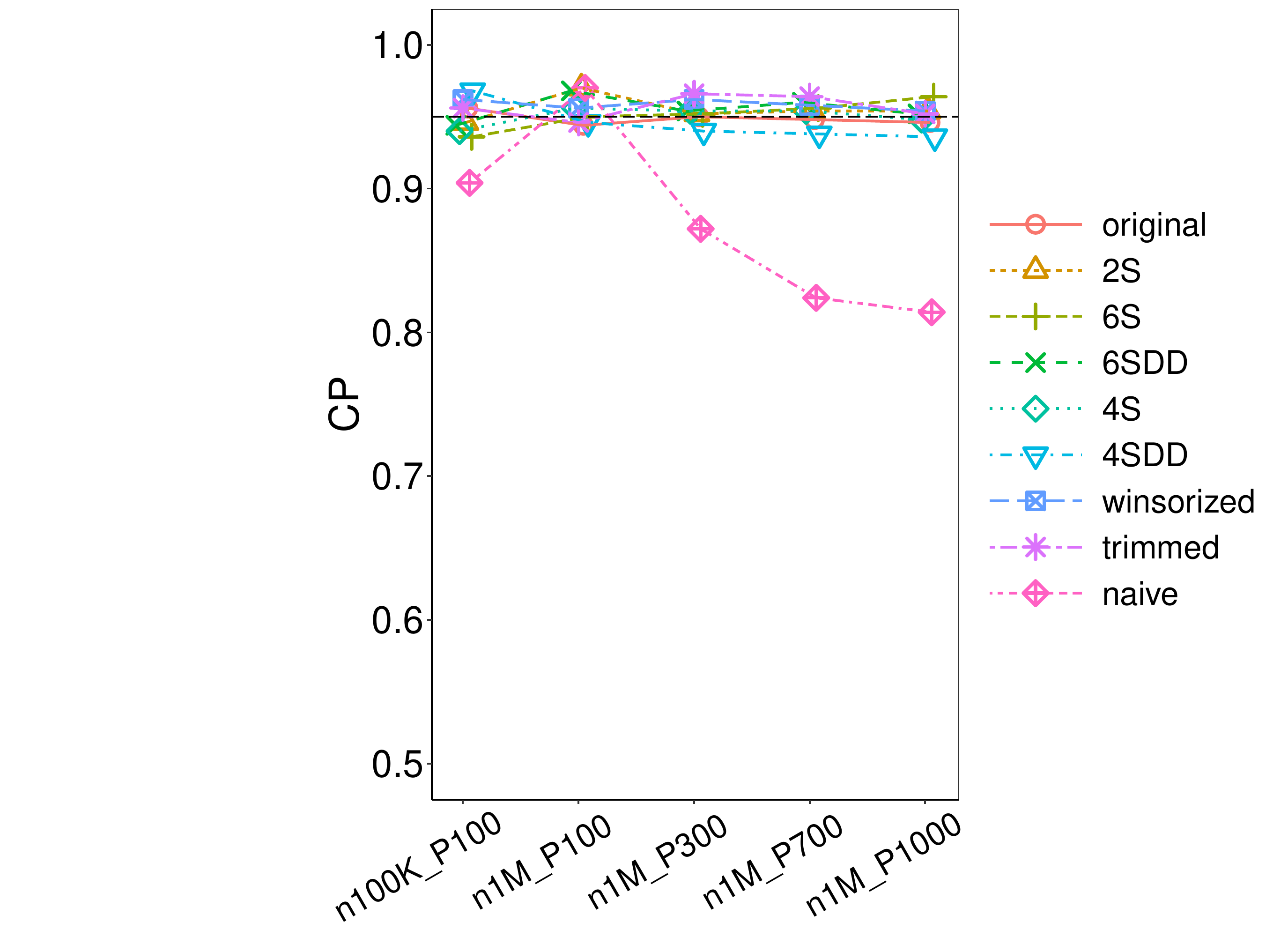}
\includegraphics[width=0.19\textwidth, trim={2.5in 0 2.6in 0},clip] {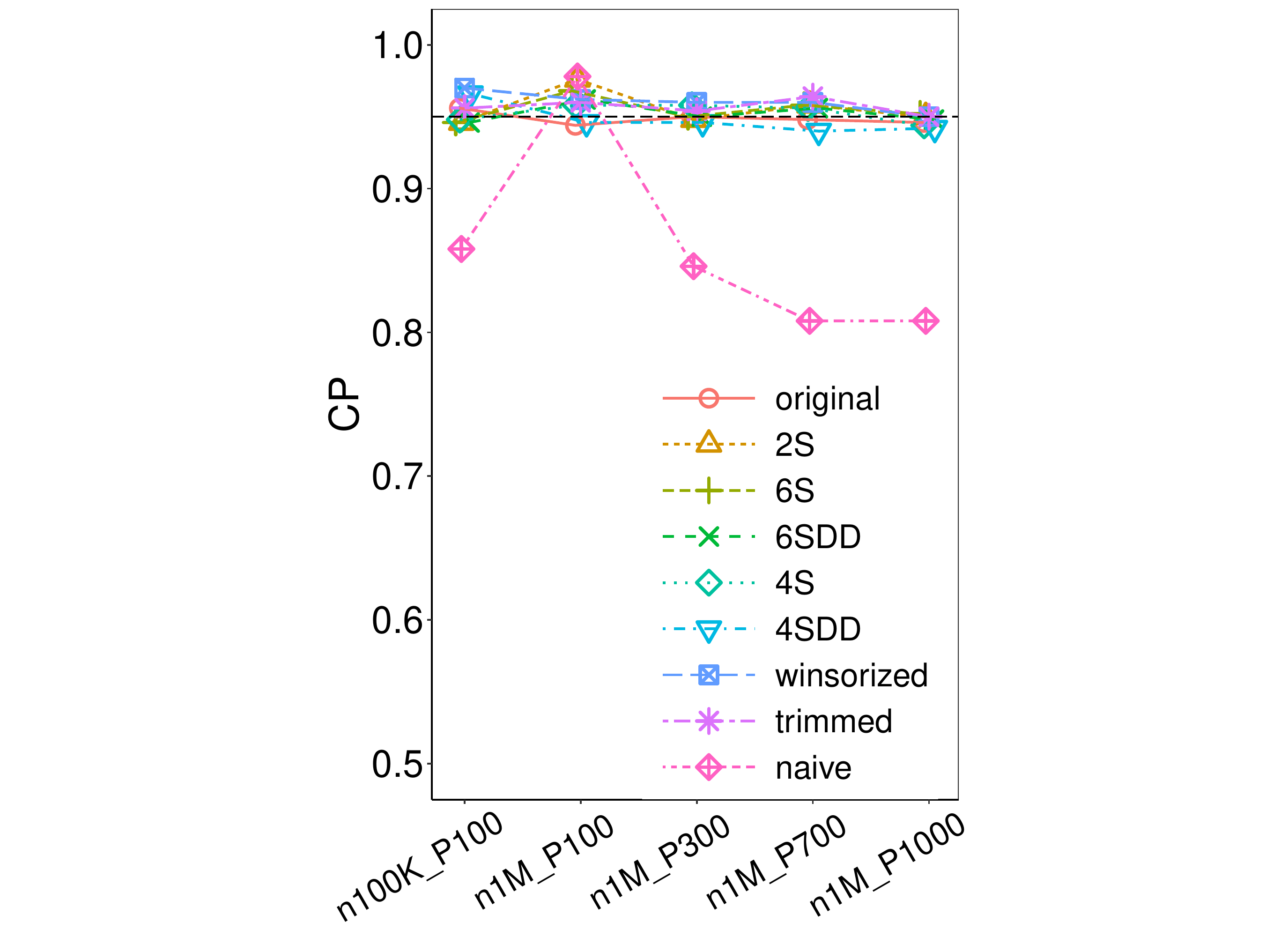}\\

\caption{Simulation results with $\rho$-zCDP for ZINB data with  $\alpha=\beta$ when $\theta=0$}
\label{fig:0szCDPZINB}
\end{figure}

The results are presented in Figures \ref{fig:0sDPZINB} and \ref{fig:0szCDPZINB} and are summarized  as follows.  4S is the best performer by all metrics and the na\"{i}ve method is the worst. In between the two, some approaches are better than others, depending on the evaluation metrics. For bias, winsorized mean and trimmed mean approach 0  than 4S, 6SDD, 6S, 4SDD, and 2S faster as $n, P$ or $\epsilon$ increases. As for  RMSE and CI width,  winsorized mean is similar to 4S (the best group), followed by trimmed mean and 4SDD (the second best group), and 2S is the worst especially for at $n=100k/p=100$, $n=1m/P=100$ and $\epsilon\le1$ but can be better than 6S and 6SDD for other cases. In terms of CP, all methods, except for the  na\"{i}ve method, provide nominal-level coverage or slight over-coverage.  The trimmed mean is slightly inferior to winsorized mean,  but  generally speaking, they are quite similar for almost all cases. The Gaussian mechanism of $\rho$-zCDP does not seem to be advantageous in statistical utility compared to the Laplace mechanism of $\epsilon$-DP. 

\subsubsection{\texorpdfstring{$Y\sim$ ZINB}{}, \texorpdfstring{$\theta\ne0$ ($H_1$)}{H1 }, and \texorpdfstring{$\alpha=\beta$}{S}}\label{sec:1sZINB}
\begin{figure}[!htb]
\hspace{0.5in}$\epsilon=0.5$\hspace{0.8in}$\epsilon=1$\hspace{0.9in}$\epsilon=2$
\hspace{0.95in}$\epsilon=5$\hspace{0.9in}$\epsilon=50$

\includegraphics[width=0.19\textwidth, trim={2.5in 0 2.5in 0},clip] {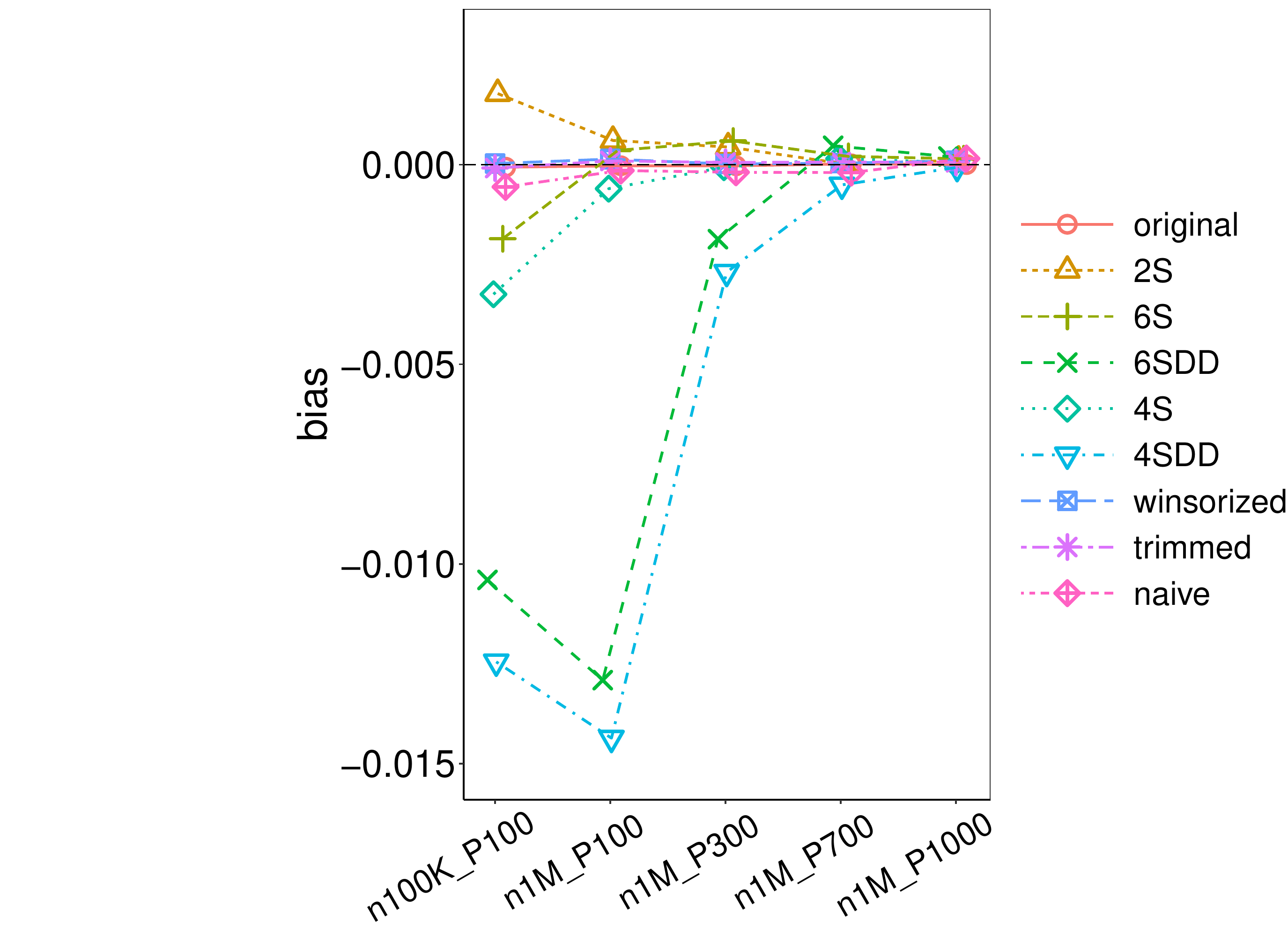}
\includegraphics[width=0.19\textwidth, trim={2.5in 0 2.5in 0},clip] {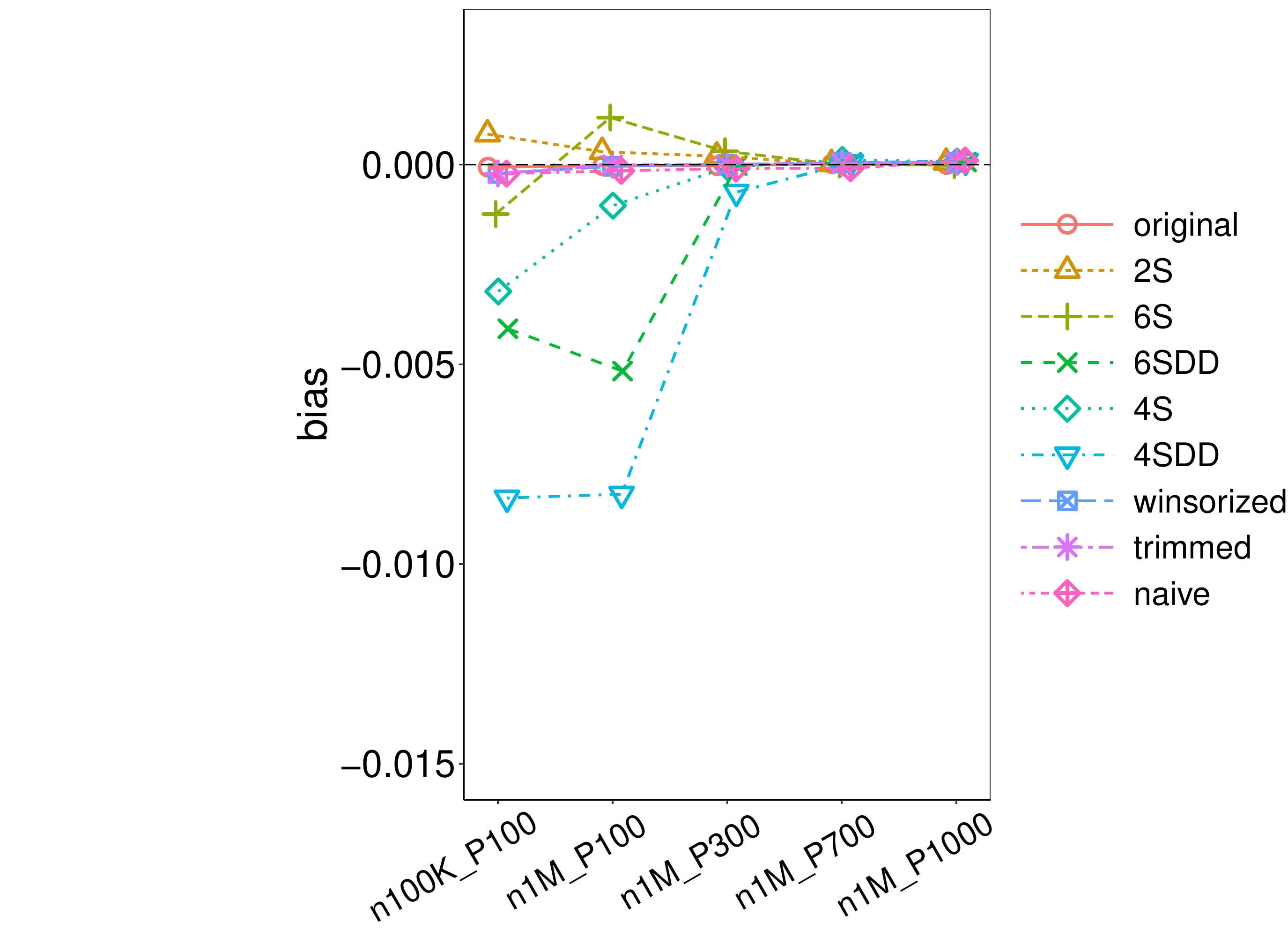}
\includegraphics[width=0.19\textwidth, trim={2.5in 0 2.5in 0},clip] {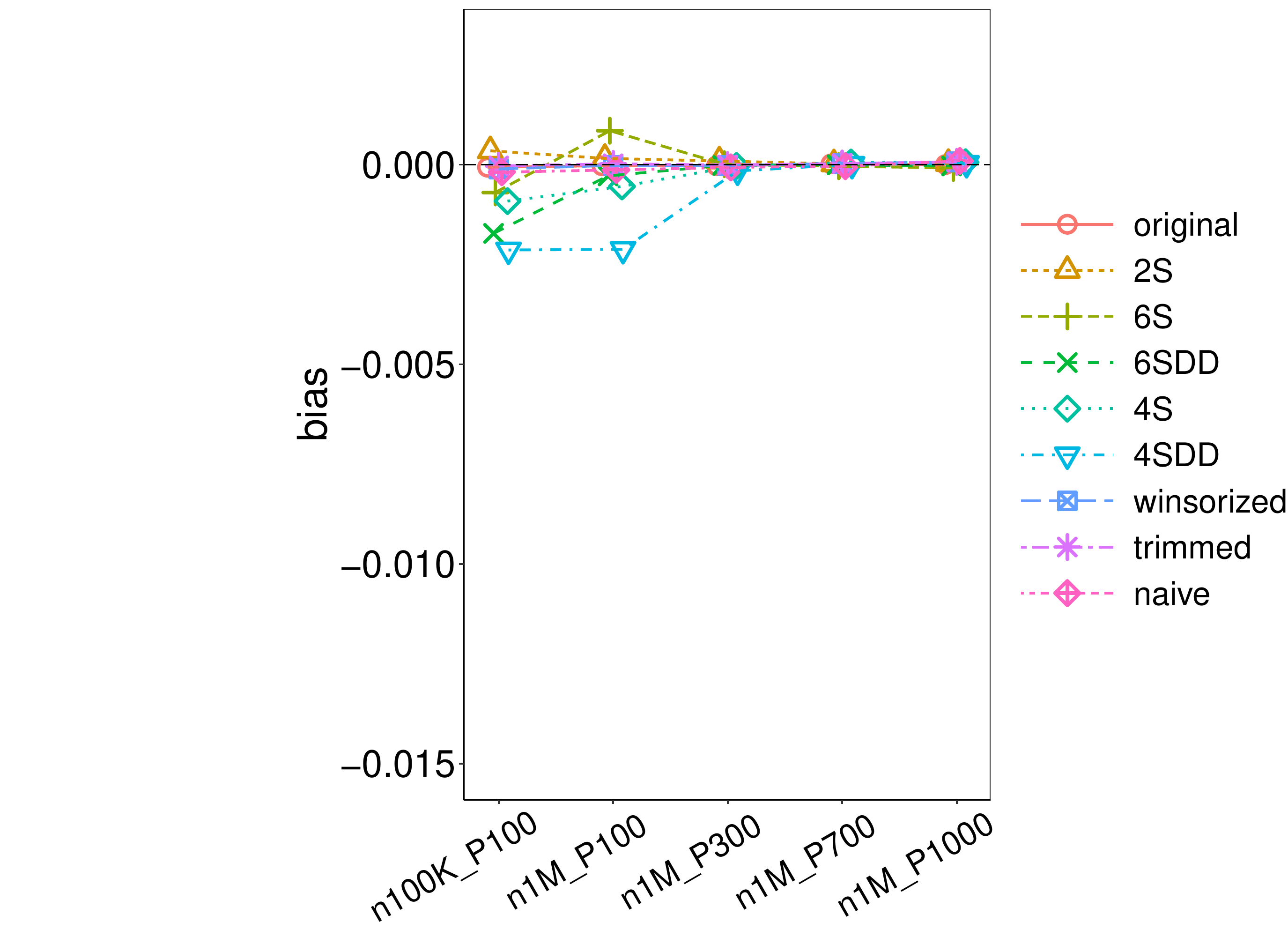}
\includegraphics[width=0.19\textwidth, trim={2.5in 0 2.5in 0},clip] {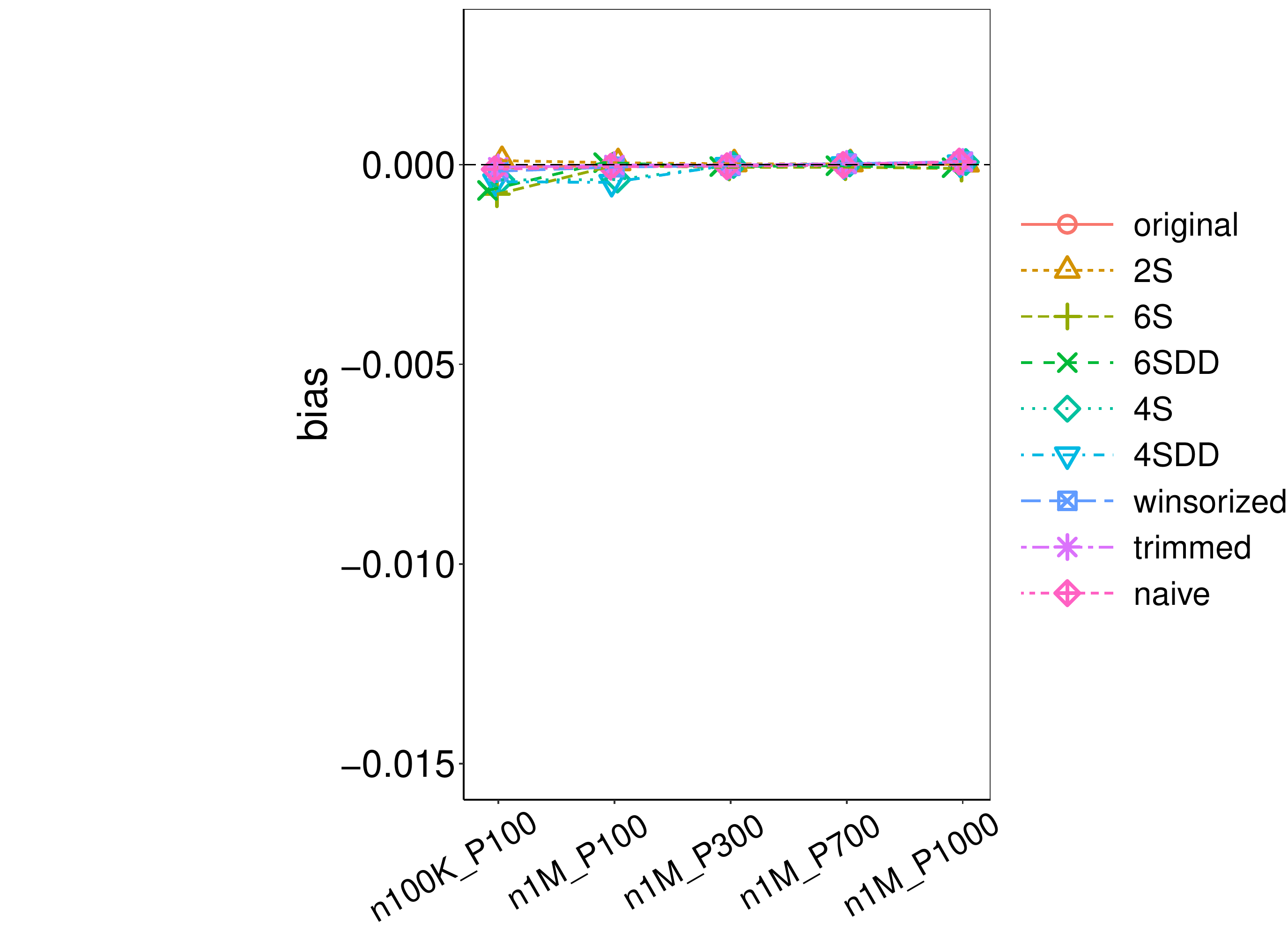}
\includegraphics[width=0.19\textwidth, trim={2.5in 0 2.5in 0},clip] {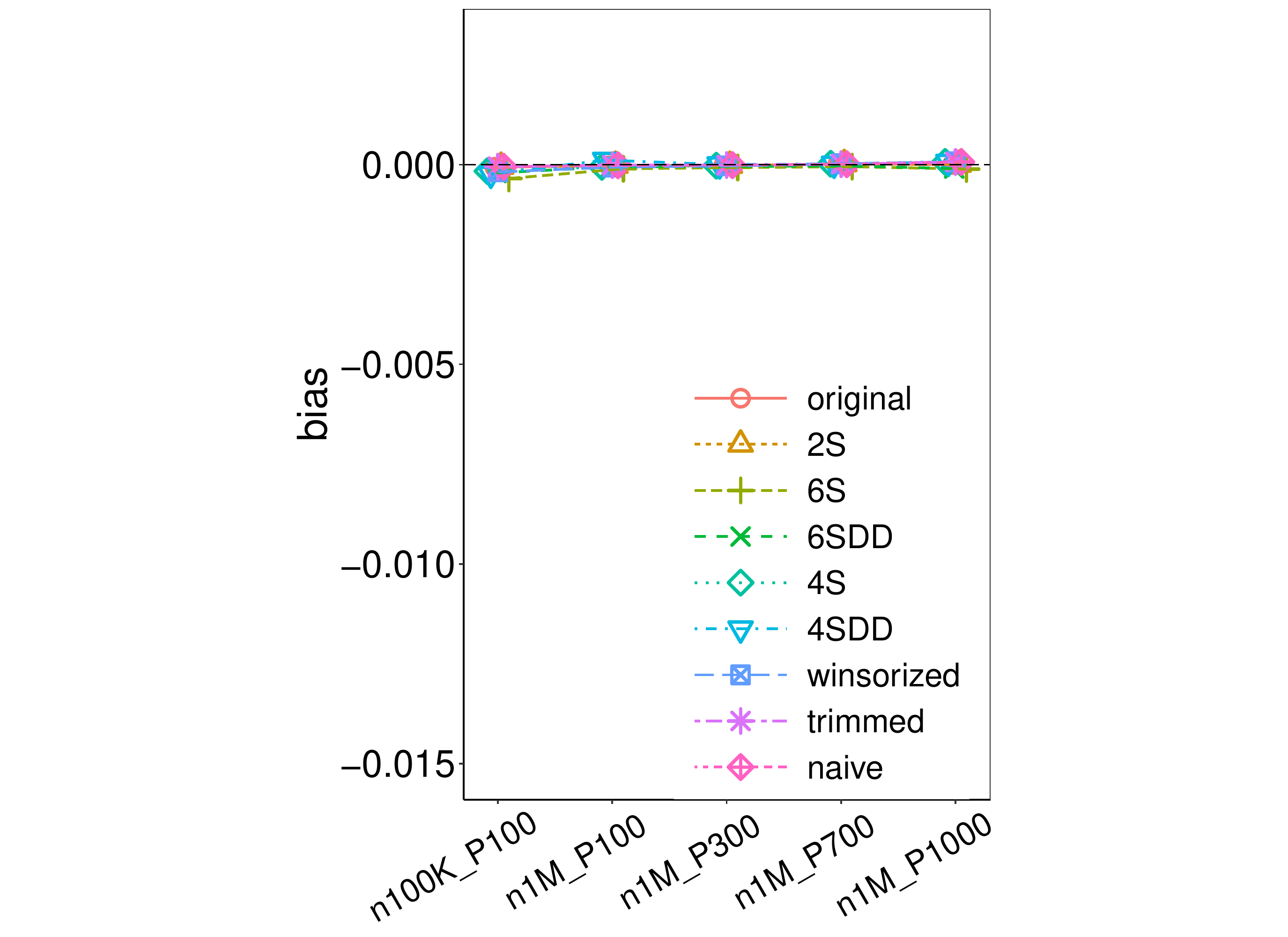}

\includegraphics[width=0.19\textwidth, trim={2.5in 0 2.6in 0},clip] {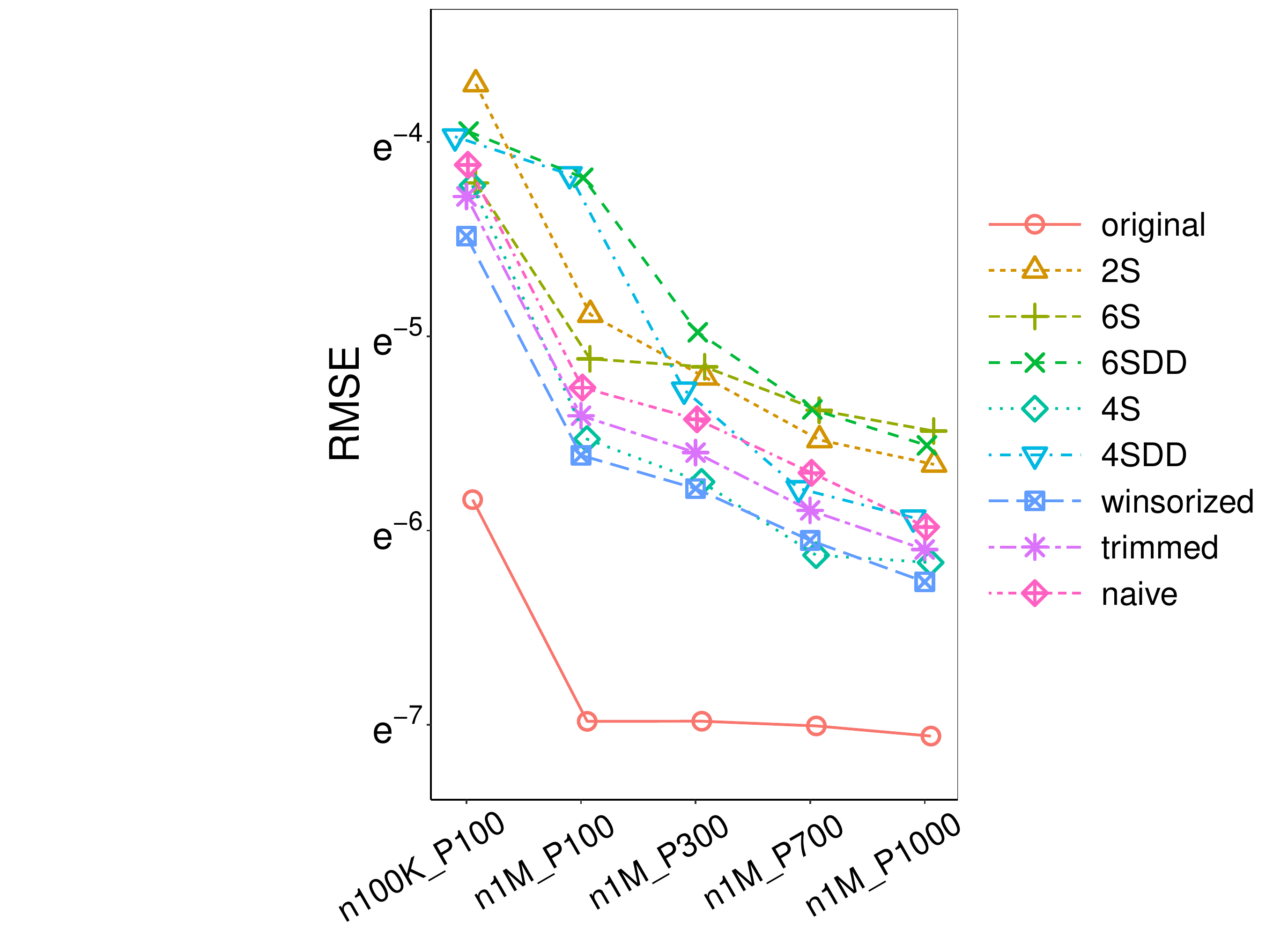}
\includegraphics[width=0.19\textwidth, trim={2.5in 0 2.6in 0},clip] {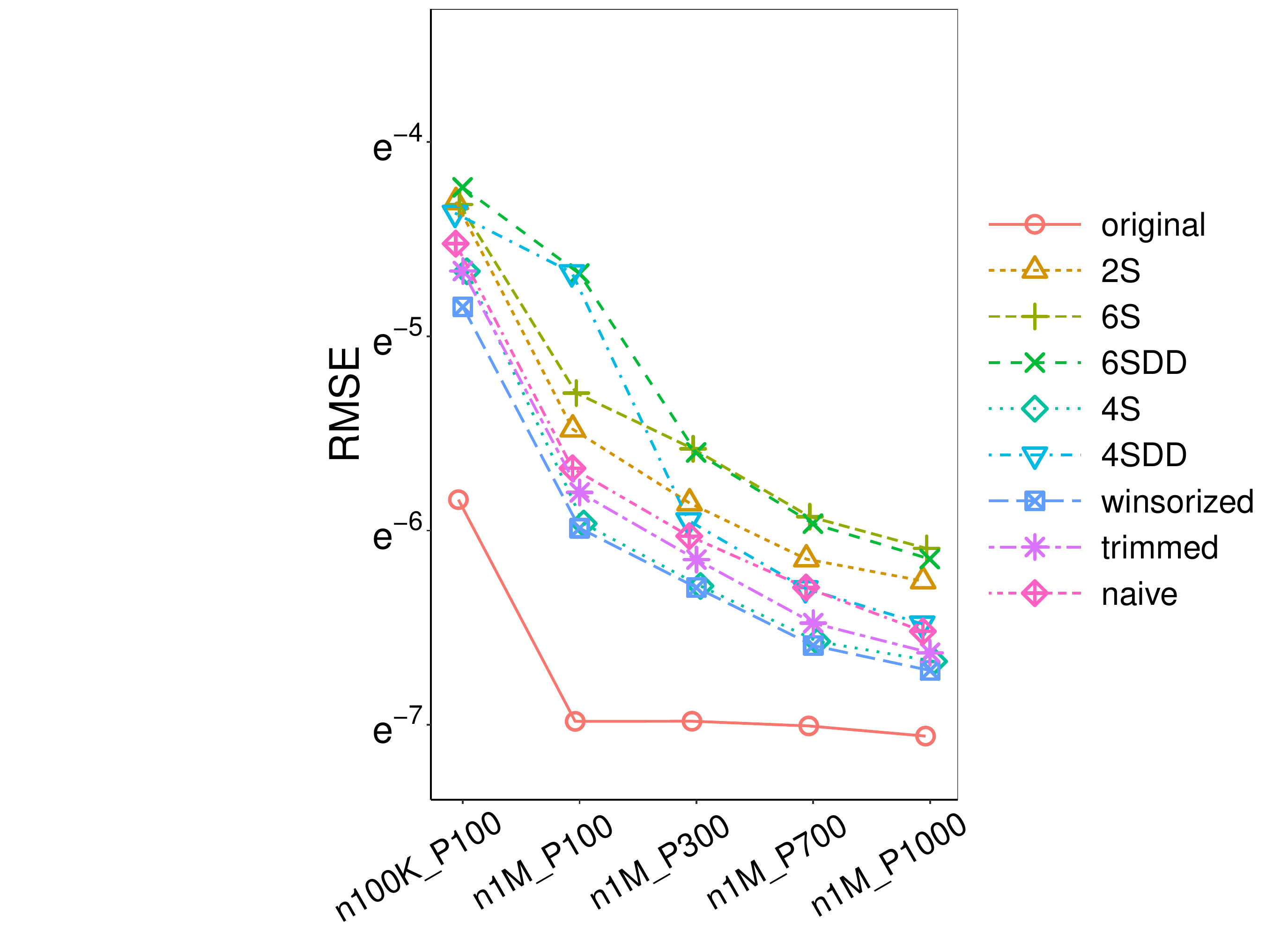}
\includegraphics[width=0.19\textwidth, trim={2.5in 0 2.6in 0},clip] {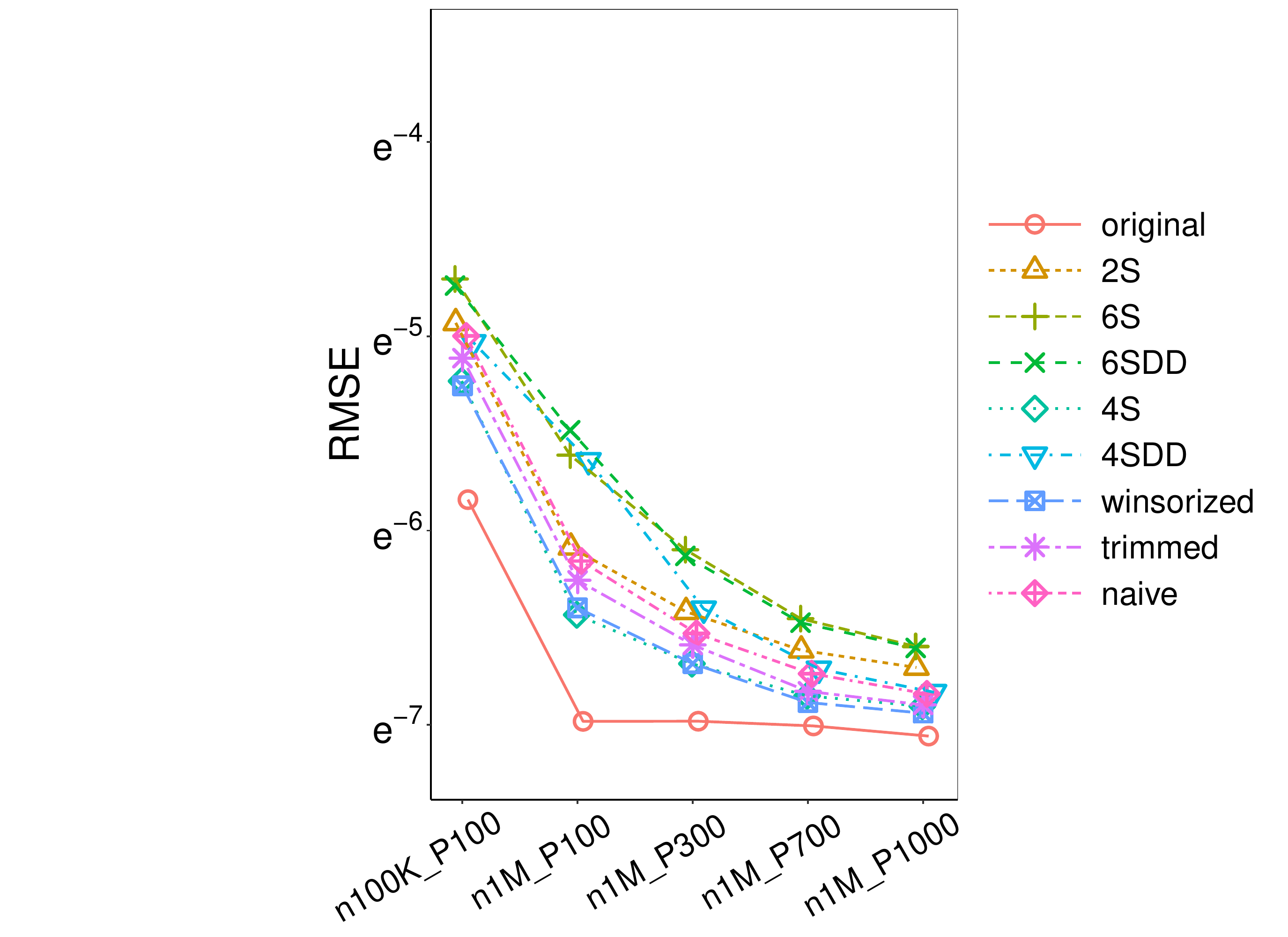}
\includegraphics[width=0.19\textwidth, trim={2.5in 0 2.6in 0},clip] {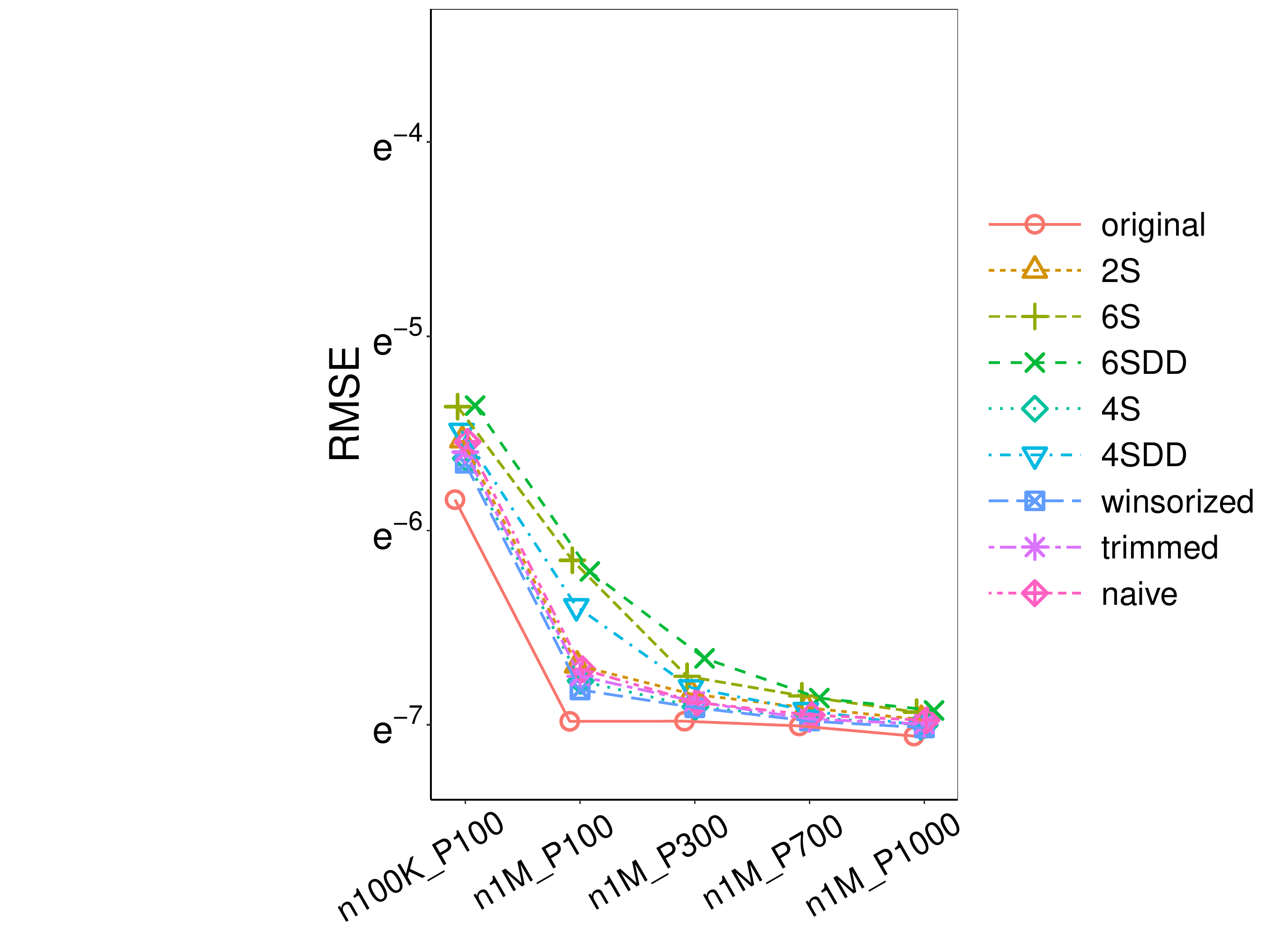}
\includegraphics[width=0.19\textwidth, trim={2.5in 0 2.6in 0},clip] {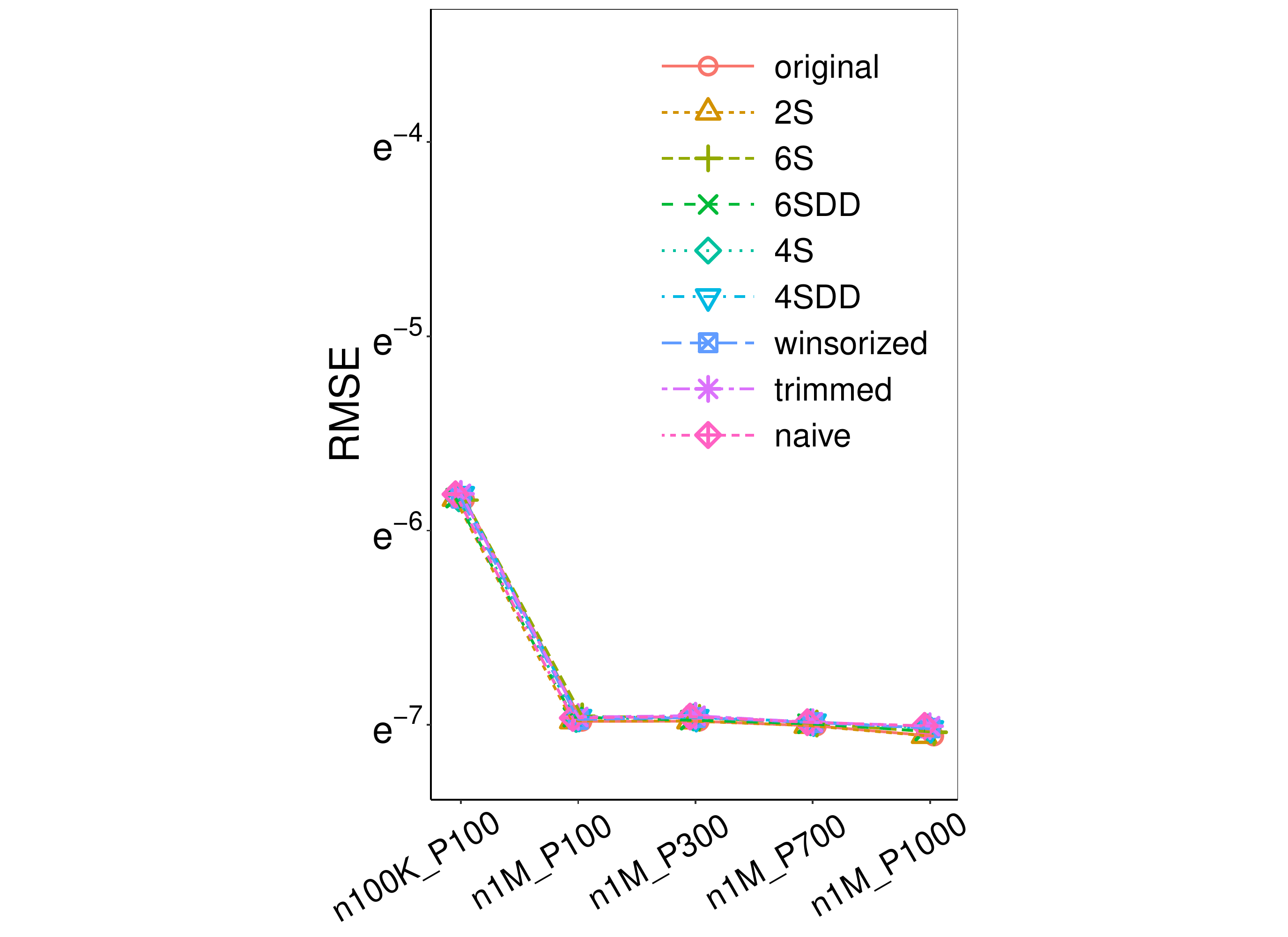}

\includegraphics[width=0.19\textwidth, trim={2.5in 0 2.6in 0},clip] {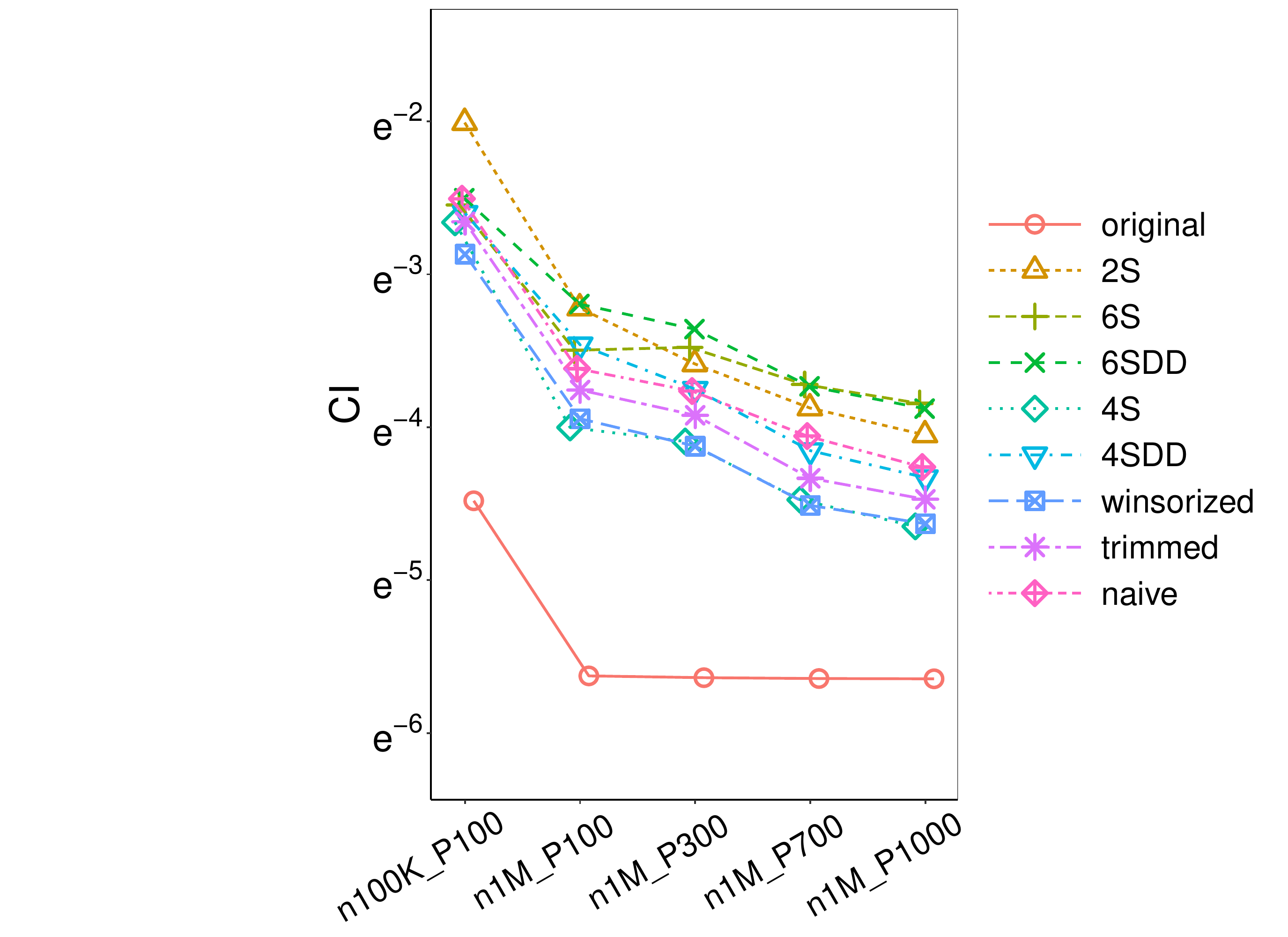}
\includegraphics[width=0.19\textwidth, trim={2.5in 0 2.6in 0},clip] {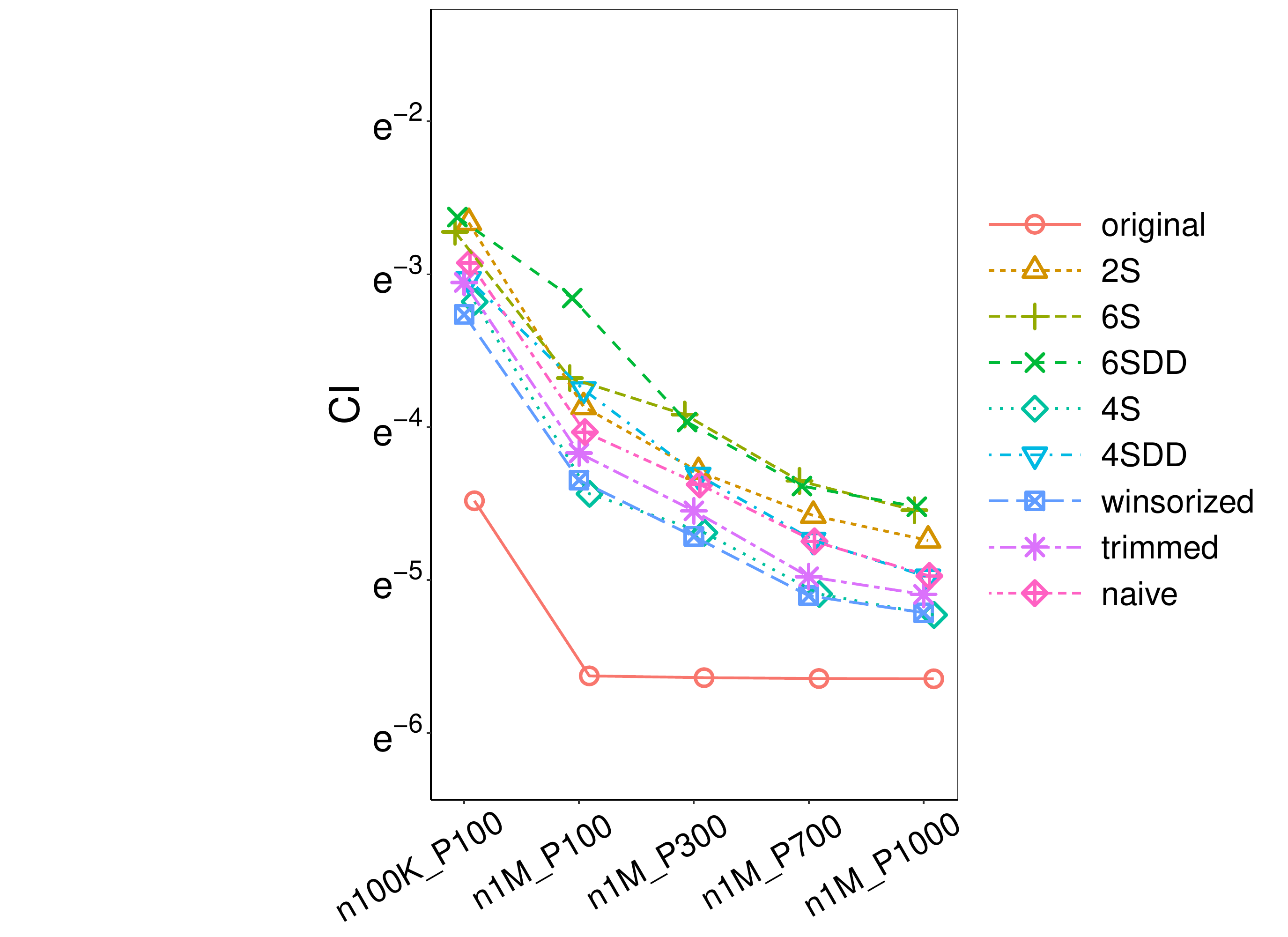}
\includegraphics[width=0.19\textwidth, trim={2.5in 0 2.6in 0},clip] {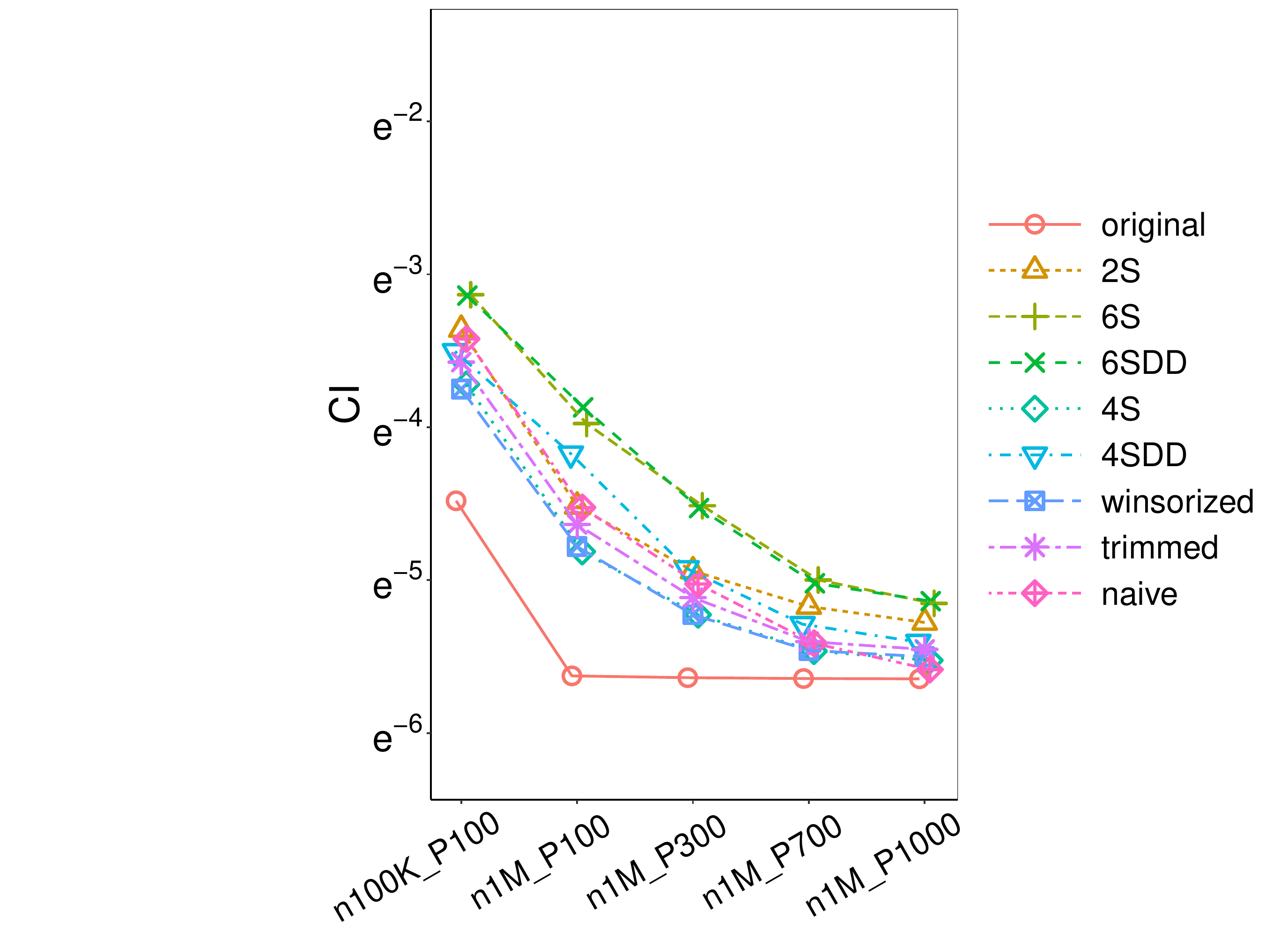}
\includegraphics[width=0.19\textwidth, trim={2.5in 0 2.6in 0},clip] {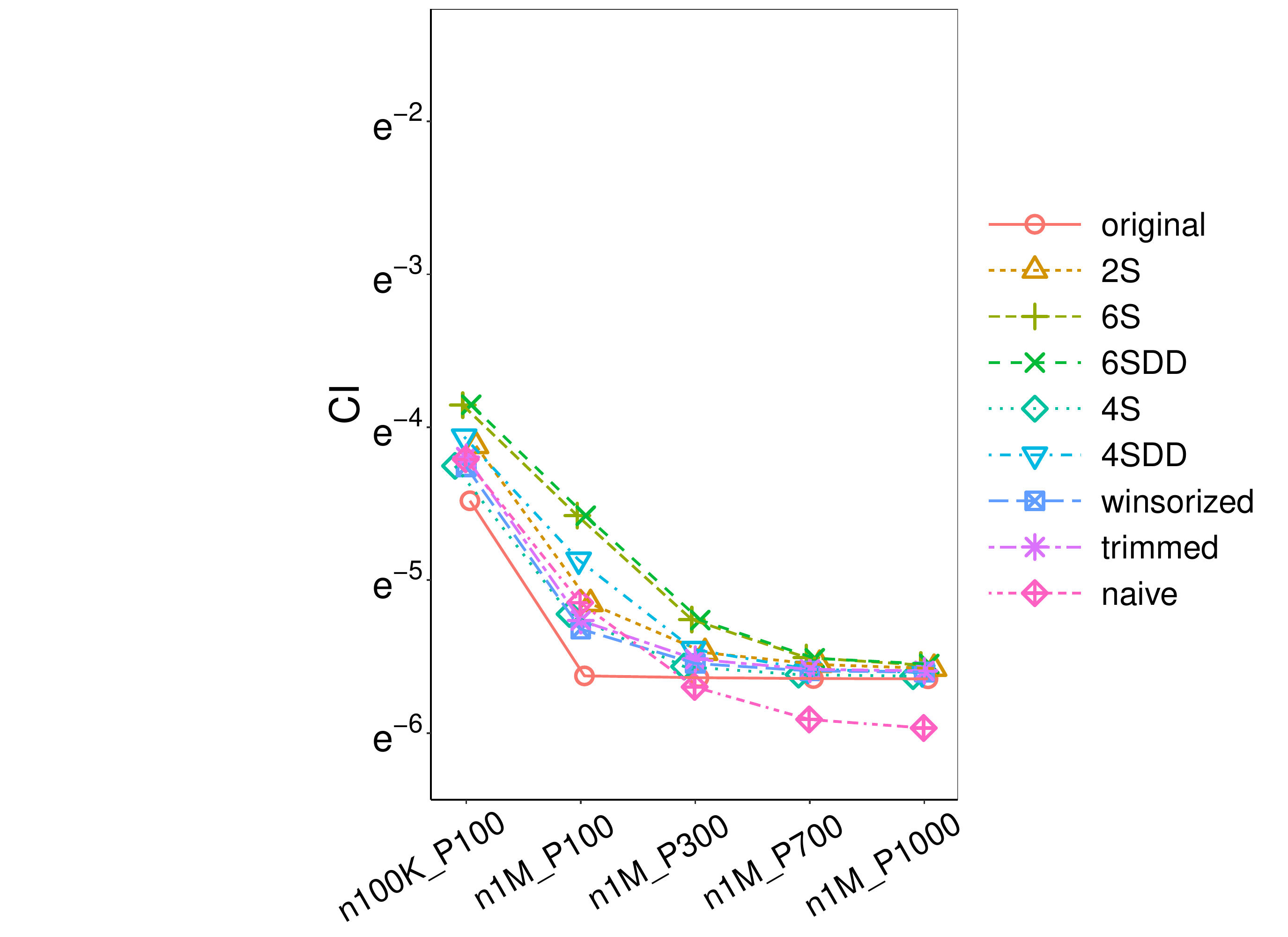}
\includegraphics[width=0.19\textwidth, trim={2.5in 0 2.6in 0},clip] {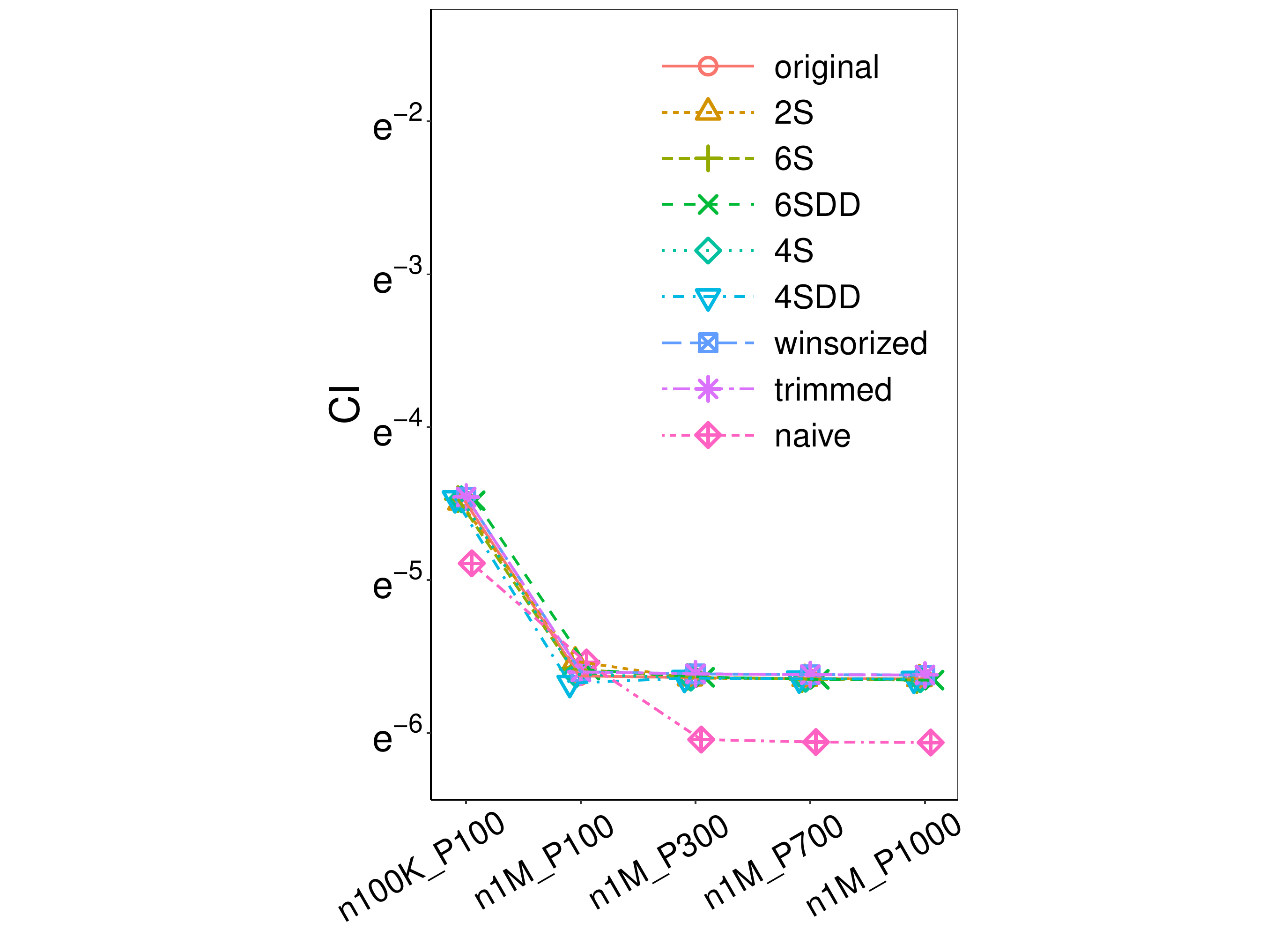}

\includegraphics[width=0.19\textwidth, trim={2.5in 0 2.6in 0},clip] {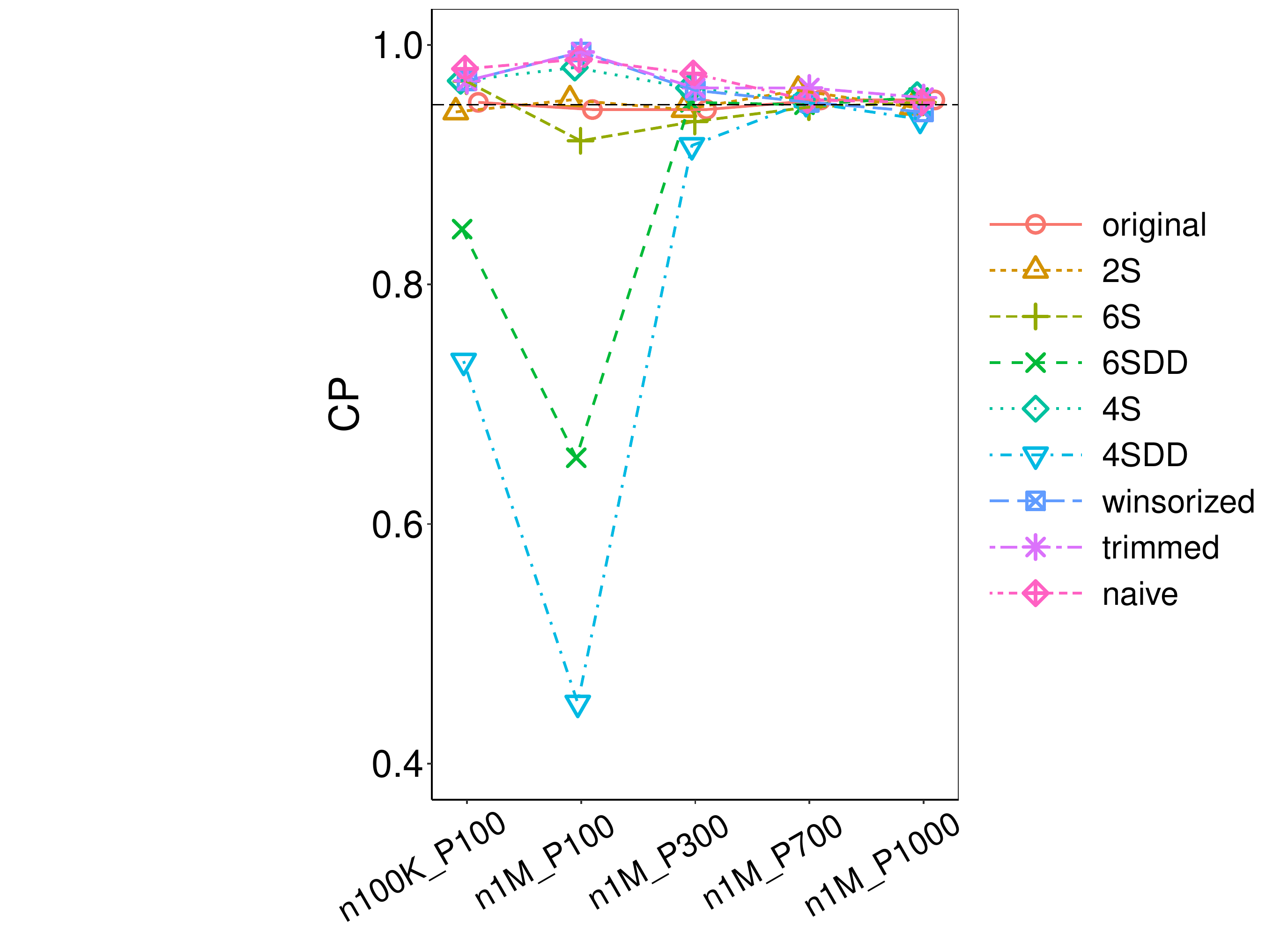}
\includegraphics[width=0.19\textwidth, trim={2.5in 0 2.6in 0},clip] {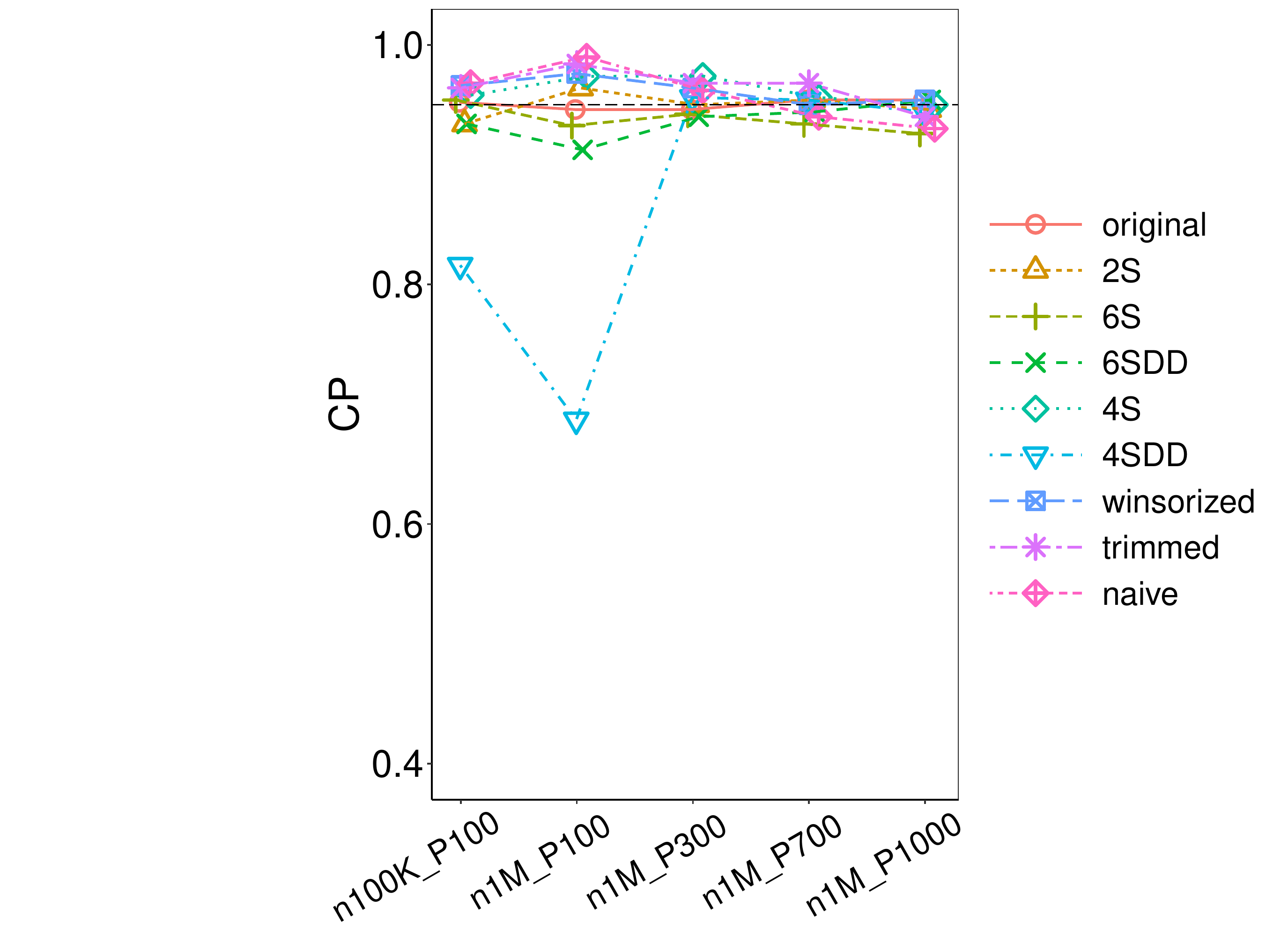}
\includegraphics[width=0.19\textwidth, trim={2.5in 0 2.6in 0},clip] {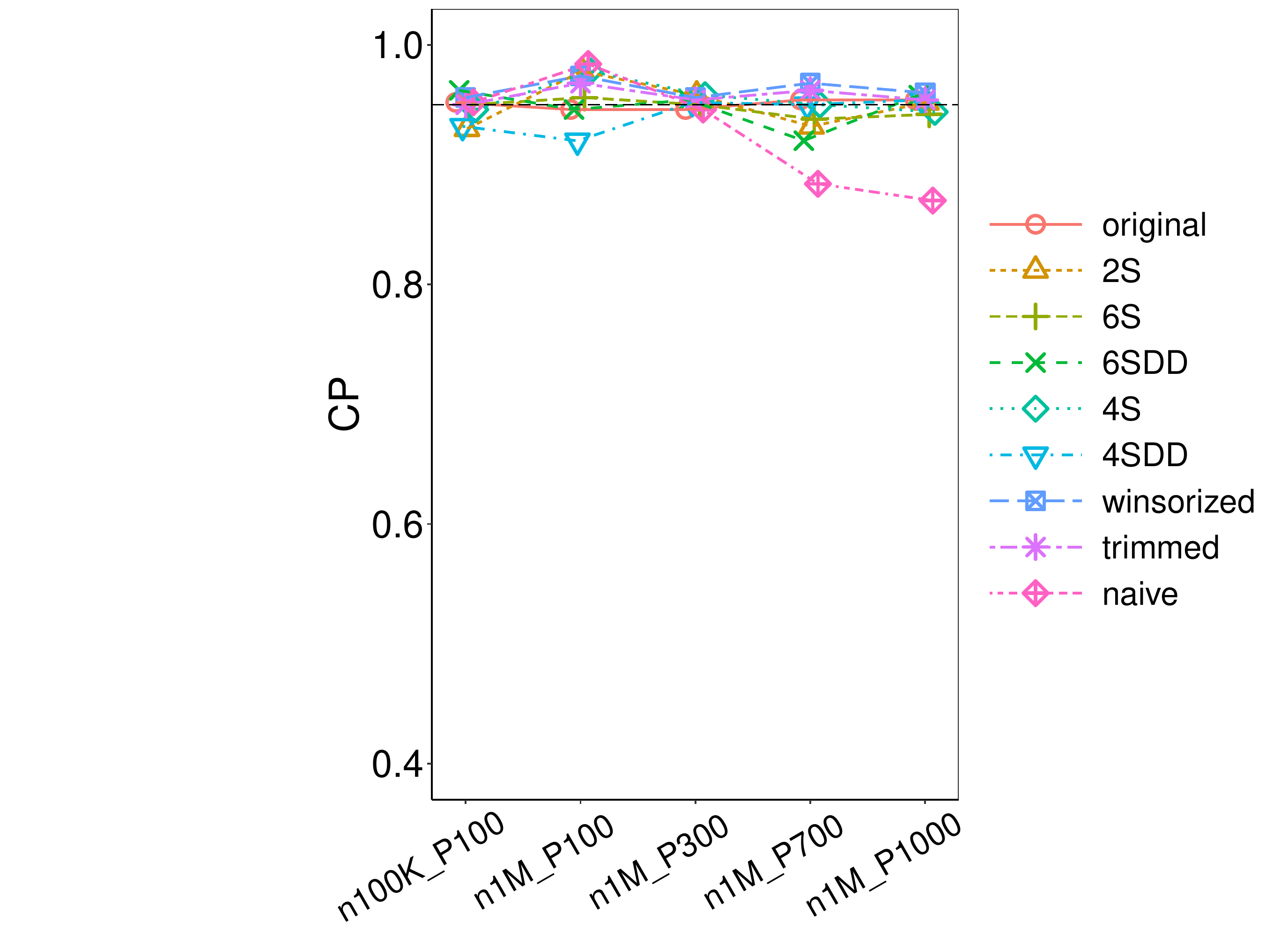}
\includegraphics[width=0.19\textwidth, trim={2.5in 0 2.6in 0},clip] {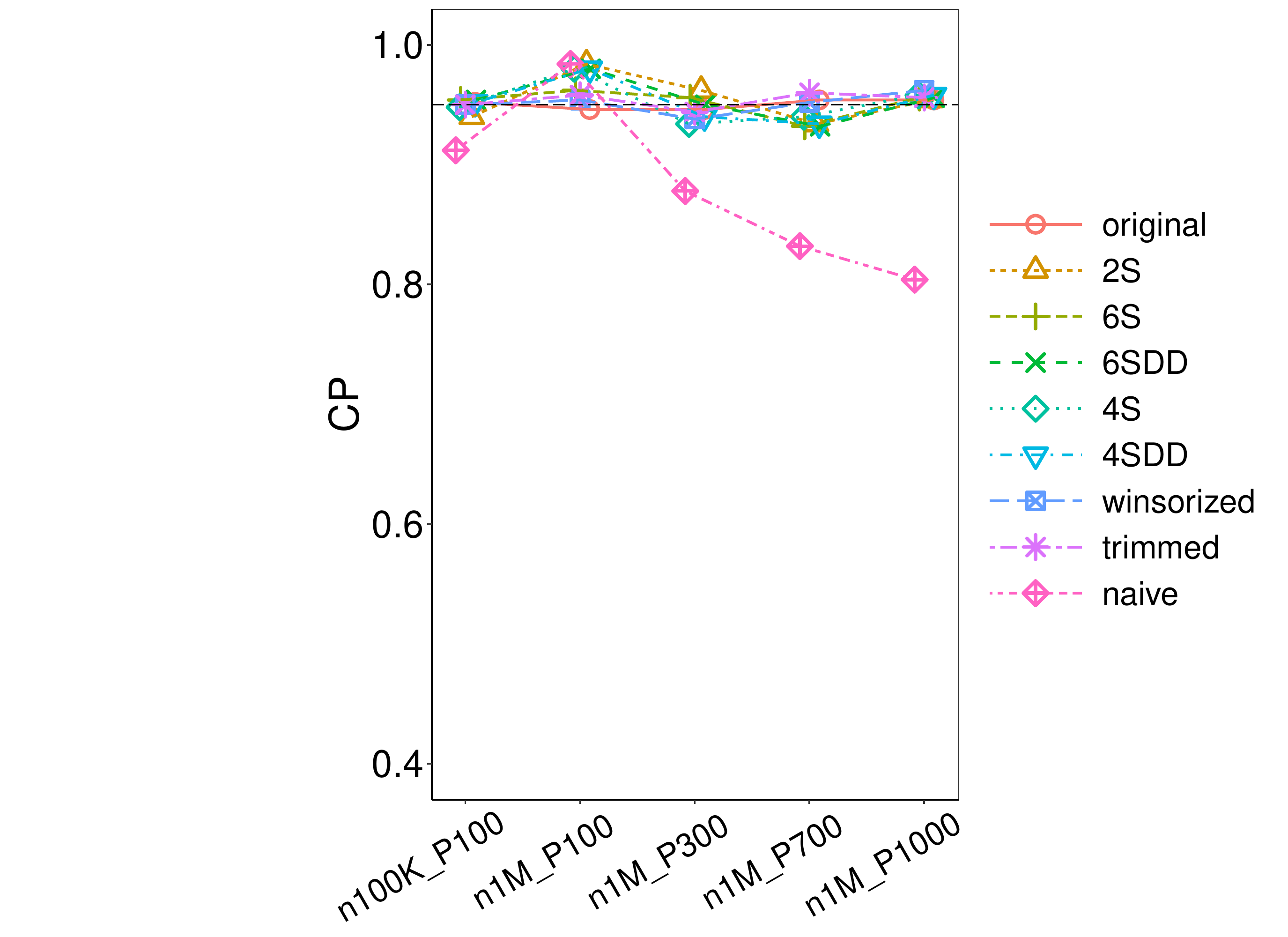}
\includegraphics[width=0.19\textwidth, trim={2.5in 0 2.6in 0},clip] {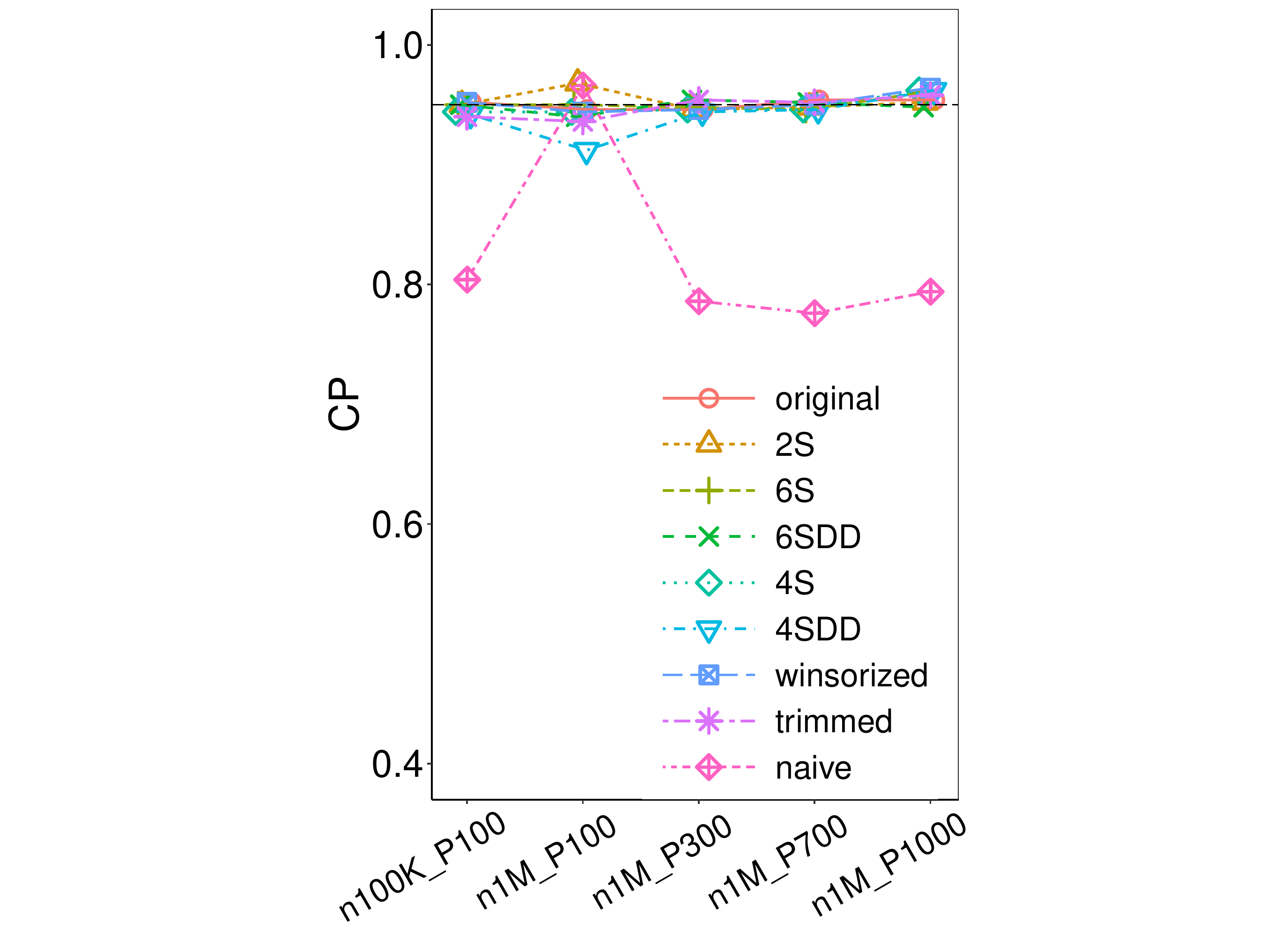}

\includegraphics[width=0.19\textwidth, trim={2.5in 0 2.6in 0},clip] {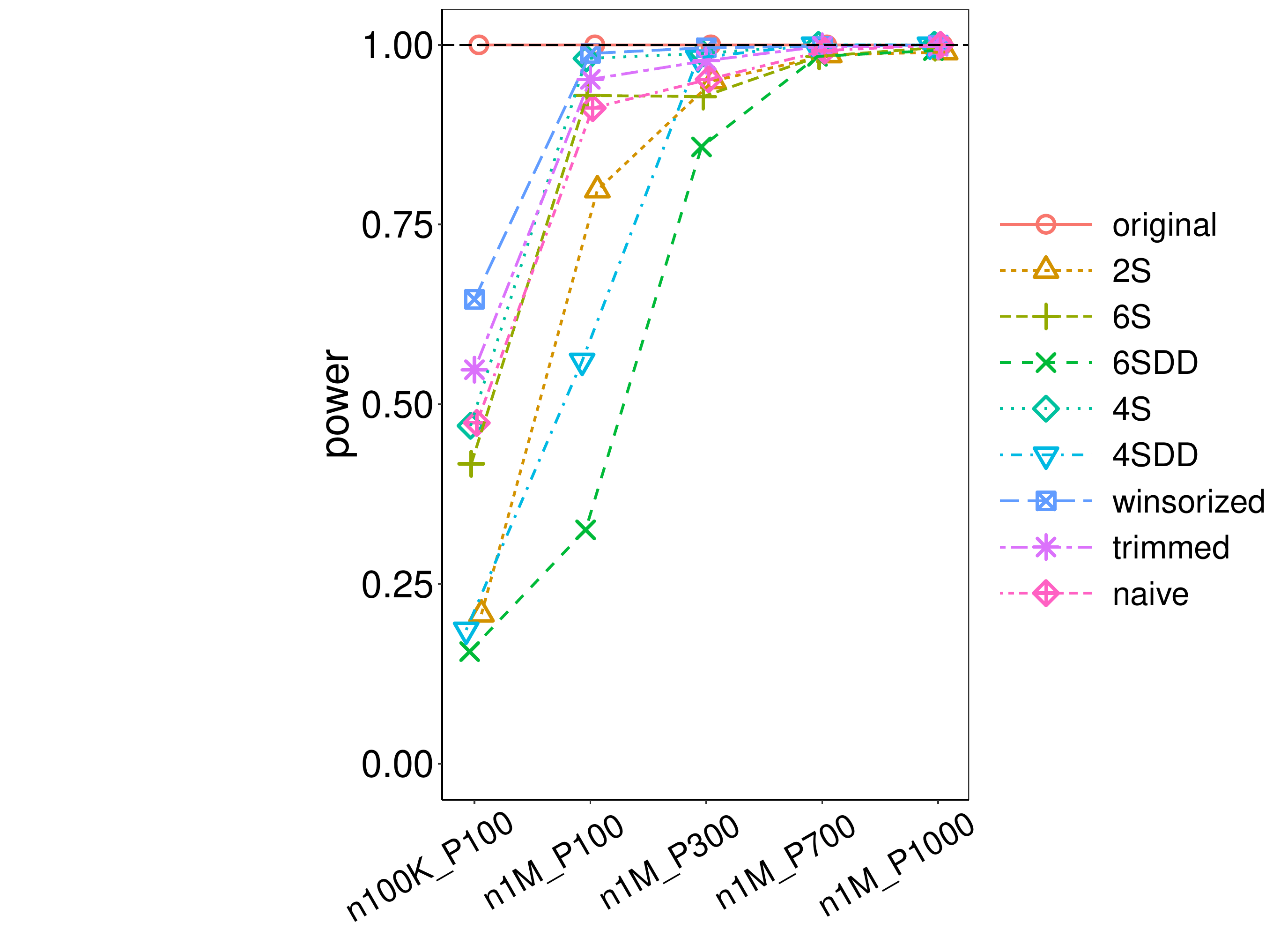}
\includegraphics[width=0.19\textwidth, trim={2.5in 0 2.6in 0},clip] {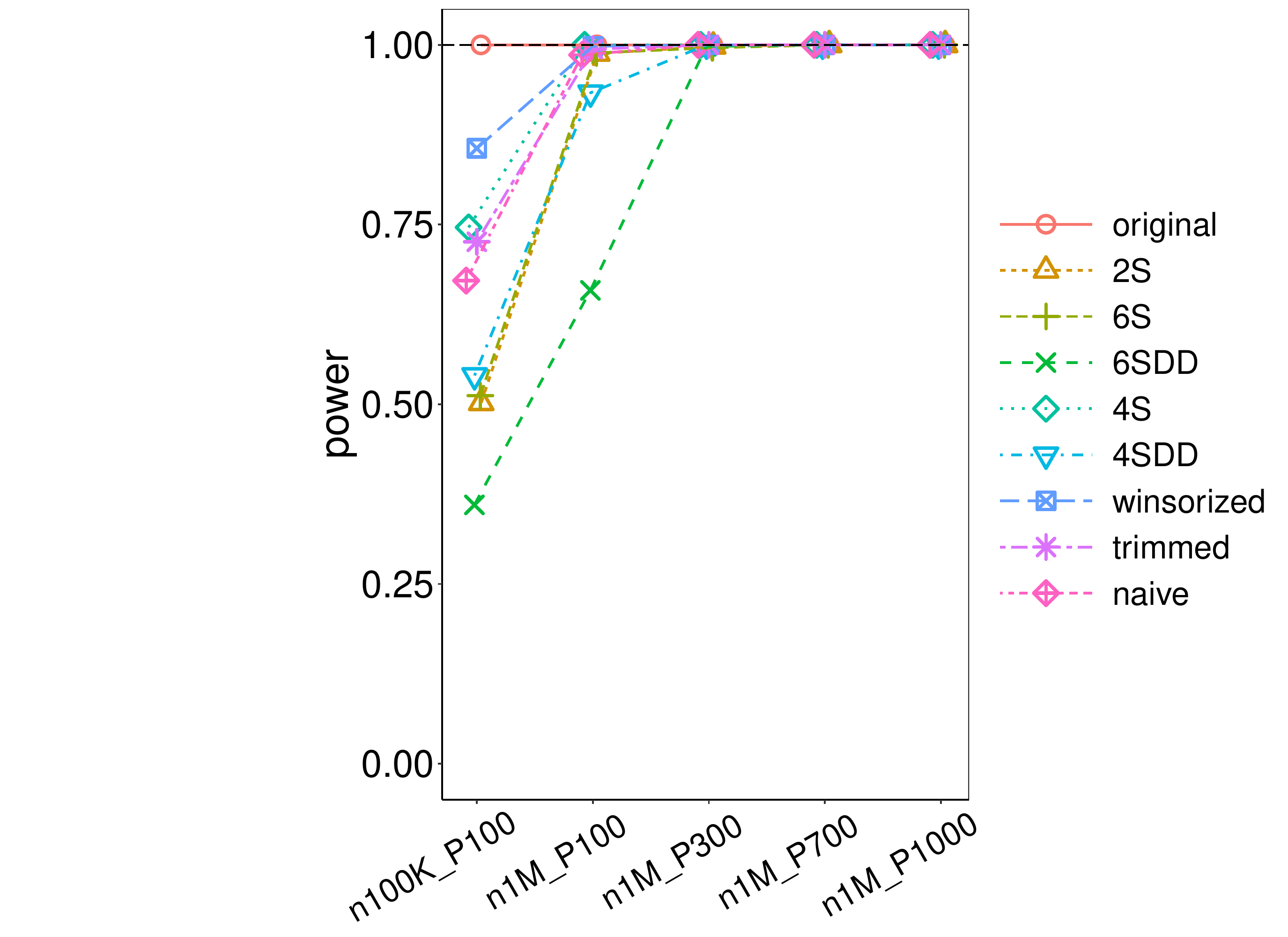}
\includegraphics[width=0.19\textwidth, trim={2.5in 0 2.6in 0},clip] {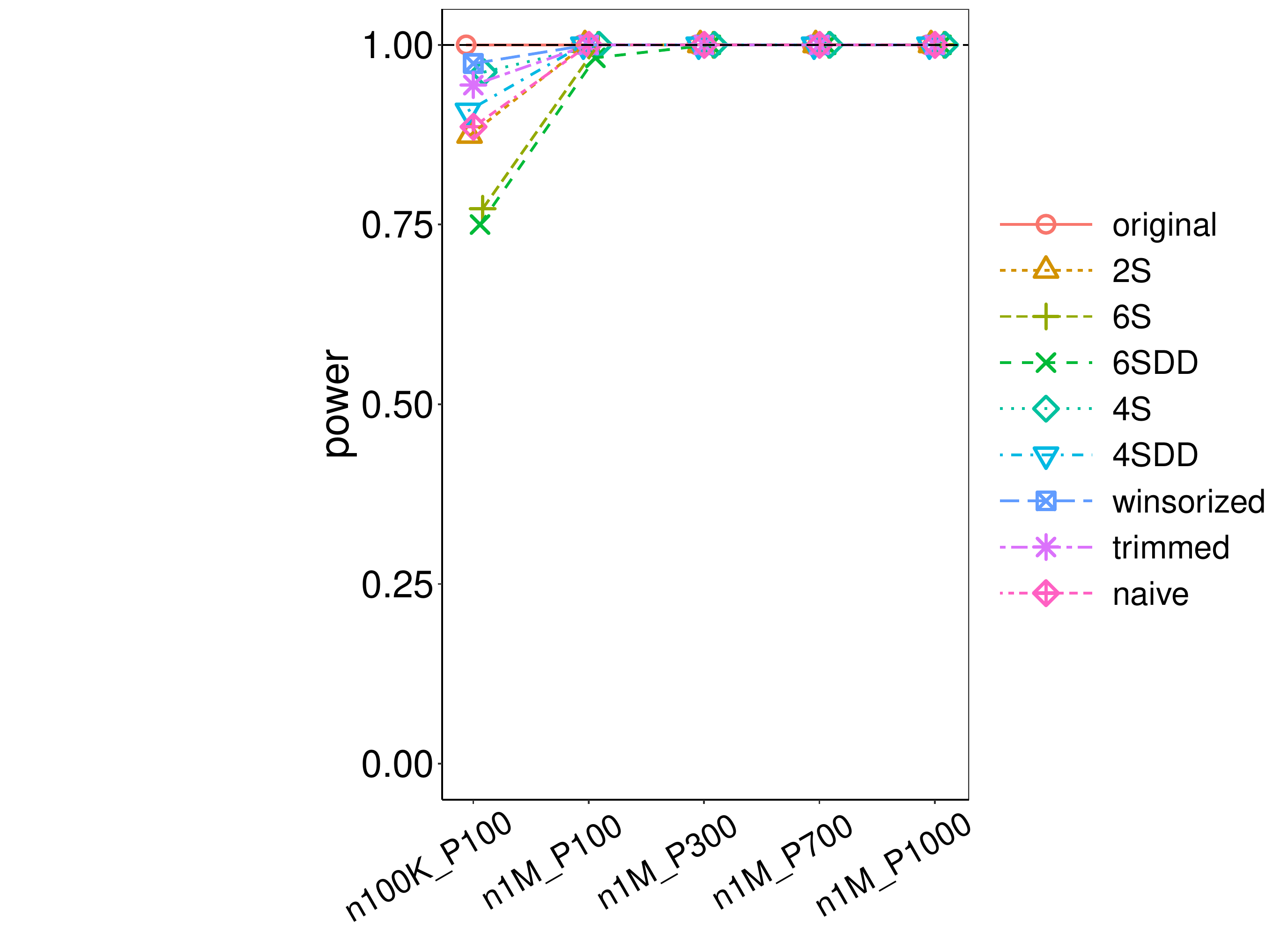}
\includegraphics[width=0.19\textwidth, trim={2.5in 0 2.6in 0},clip] {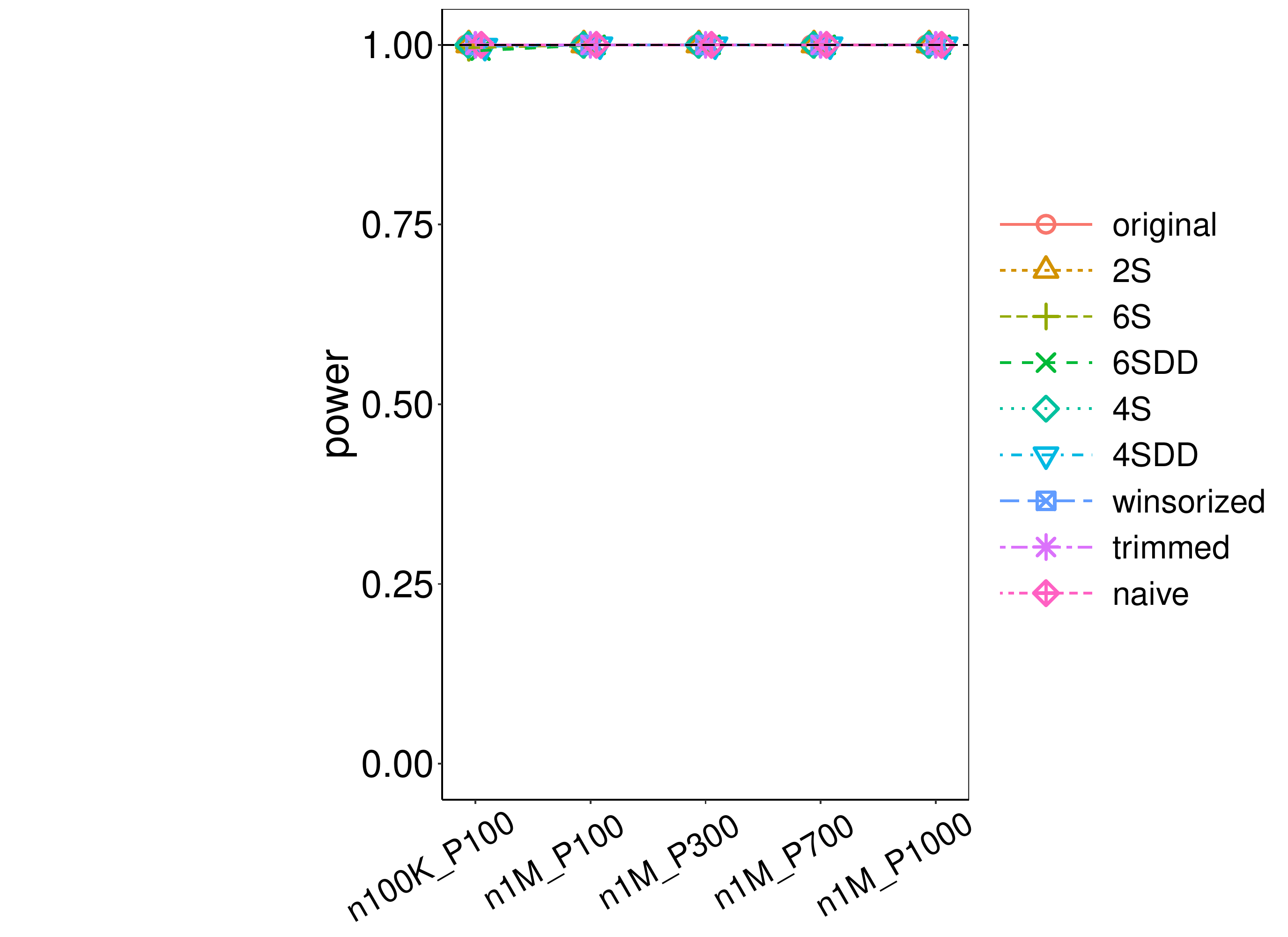}
\includegraphics[width=0.19\textwidth, trim={2.5in 0 2.6in 0},clip] {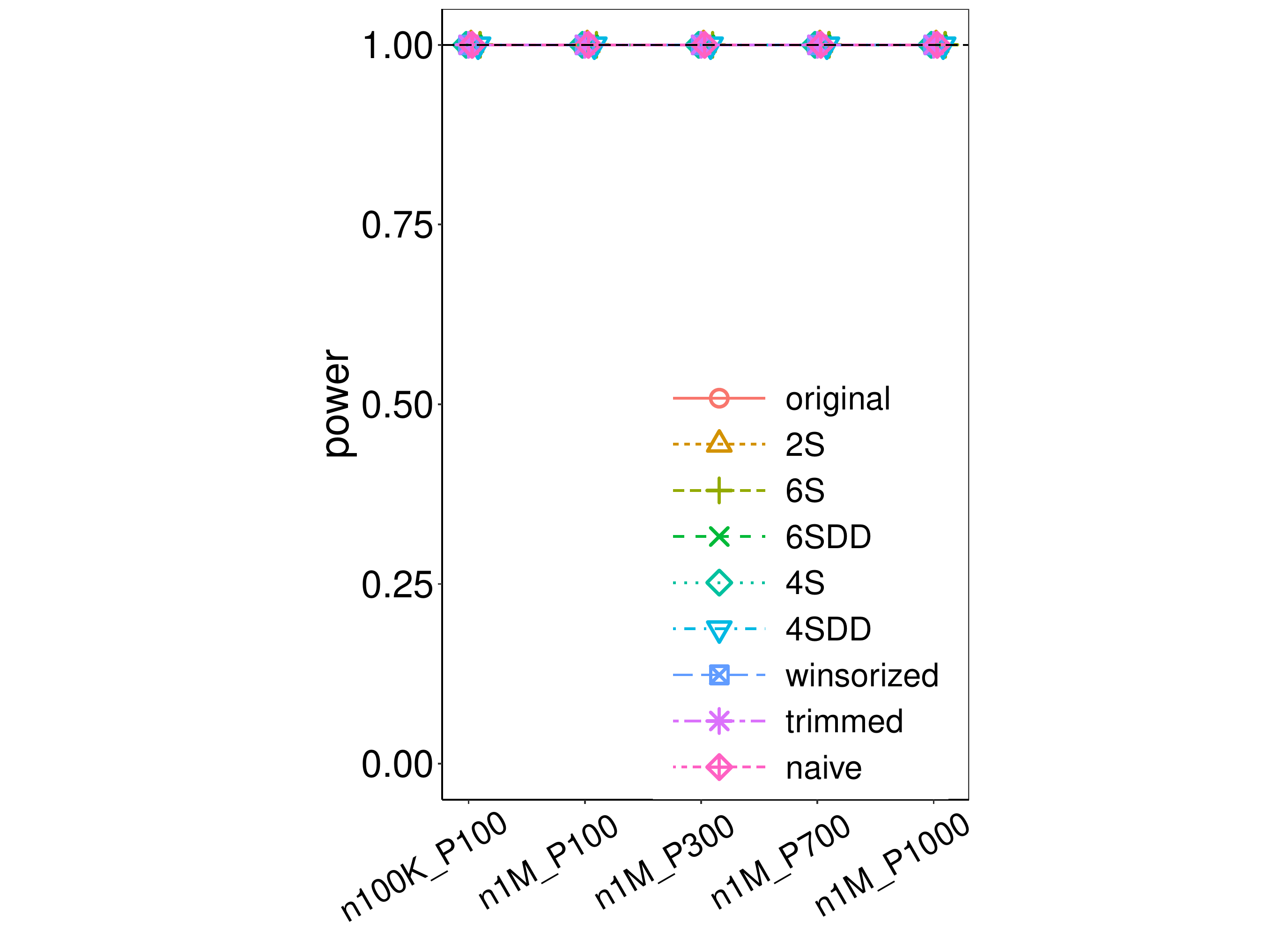}
\caption{Simulation results with $\epsilon$-DP for ZINB data with  $\alpha=\beta$ when $\theta\ne0$} \label{fig:1sDPZINB}
\end{figure}

\begin{figure}[!htb]
\hspace{0.45in}$\rho=0.005$\hspace{0.65in}$\rho=0.02$\hspace{0.65in}$\rho=0.08$
\hspace{0.65in}$\rho=0.32$\hspace{0.65in}$\rho=1.28$

\includegraphics[width=0.19\textwidth, trim={2.5in 0 2.5in 0},clip] {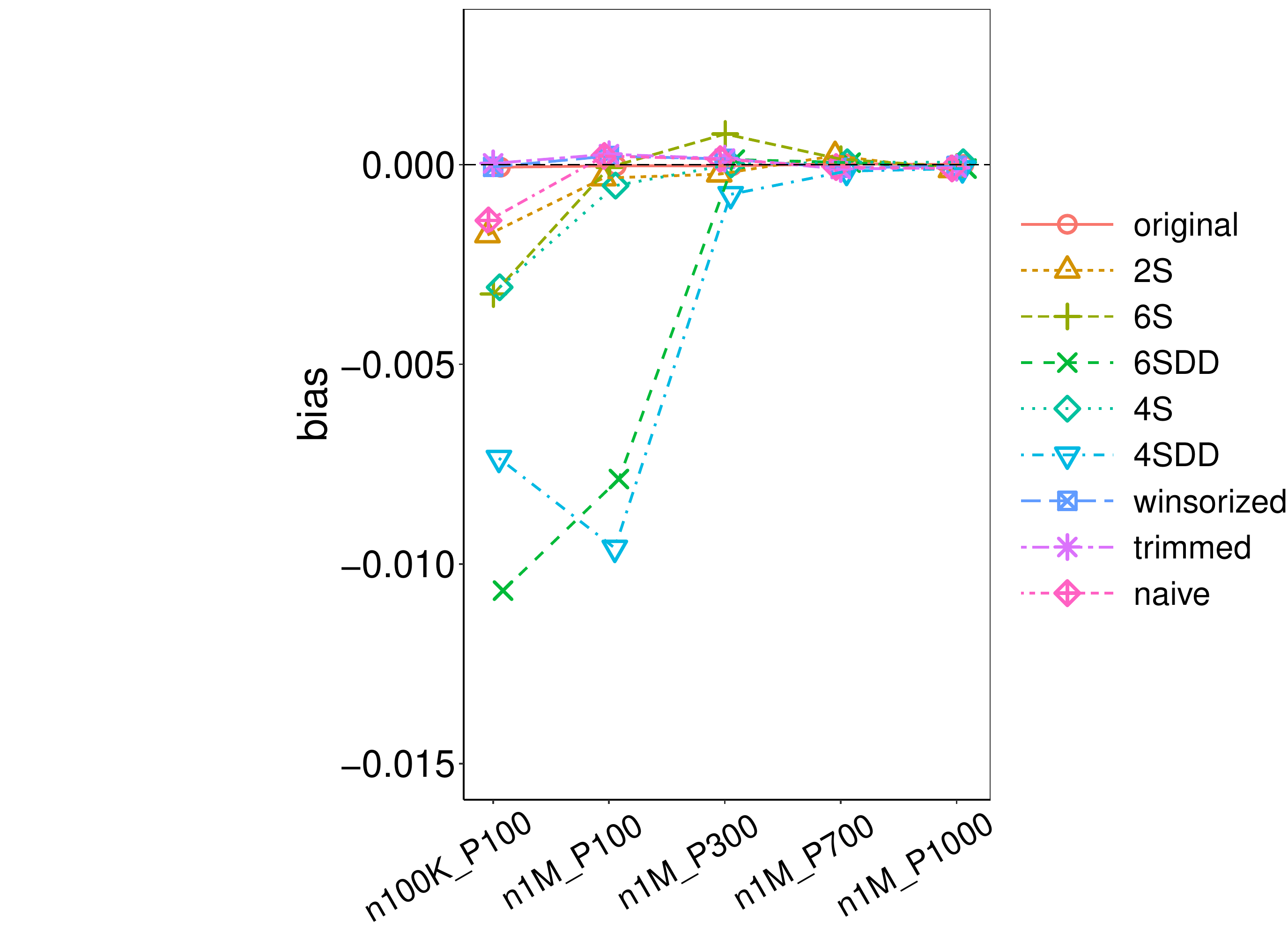}
\includegraphics[width=0.19\textwidth, trim={2.5in 0 2.5in 0},clip] {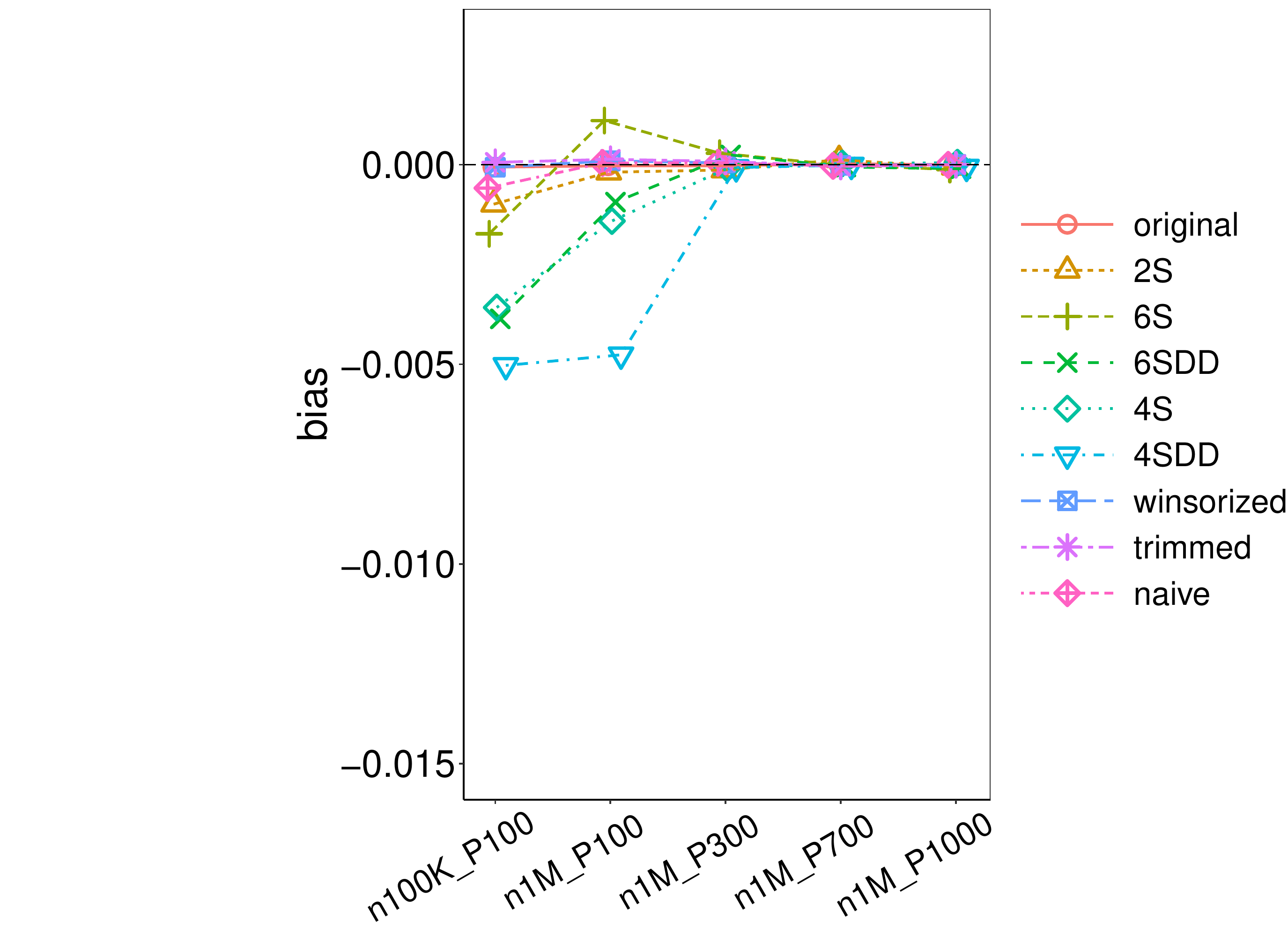}
\includegraphics[width=0.19\textwidth, trim={2.5in 0 2.5in 0},clip] {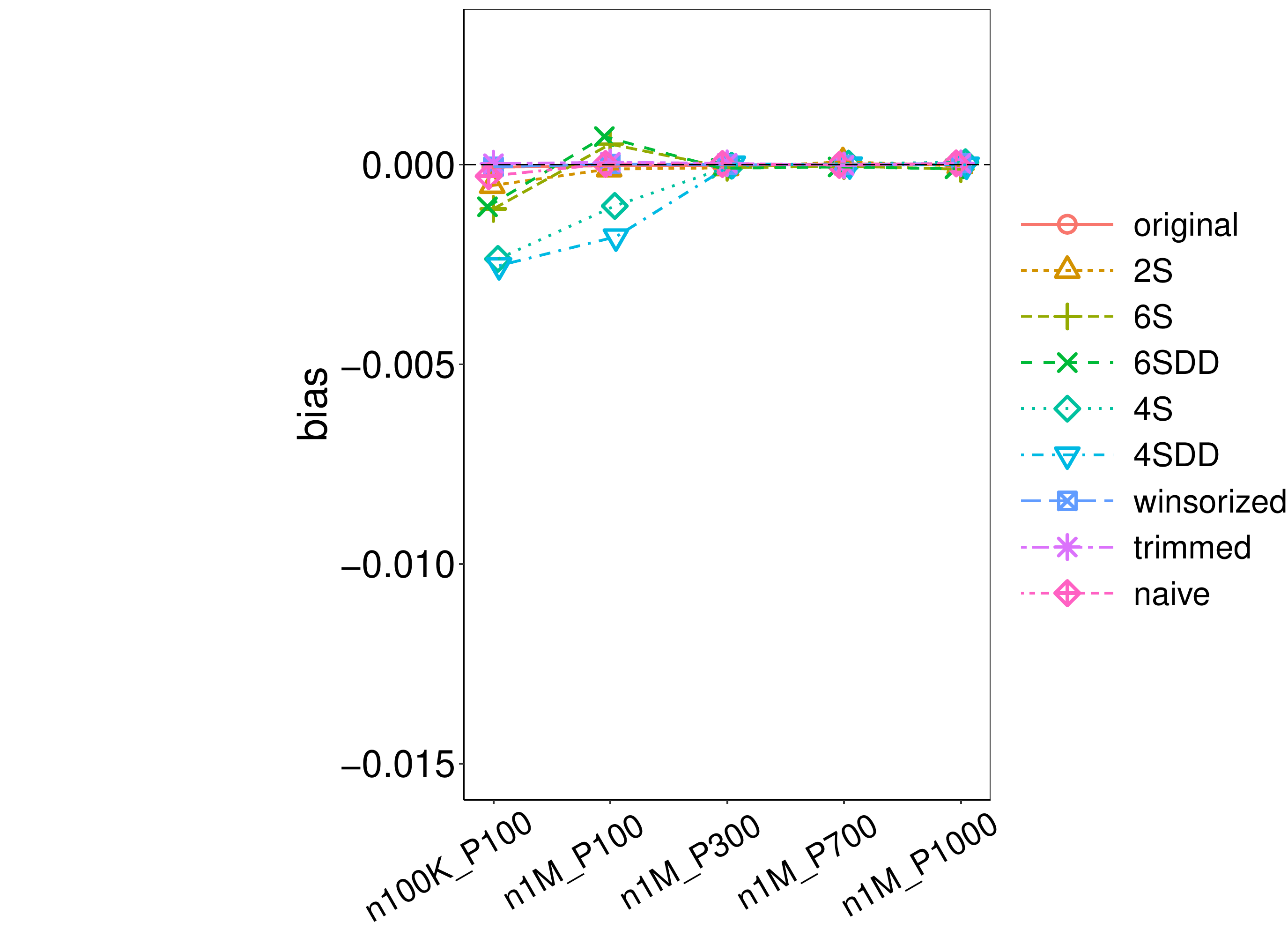}
\includegraphics[width=0.19\textwidth, trim={2.5in 0 2.5in 0},clip] {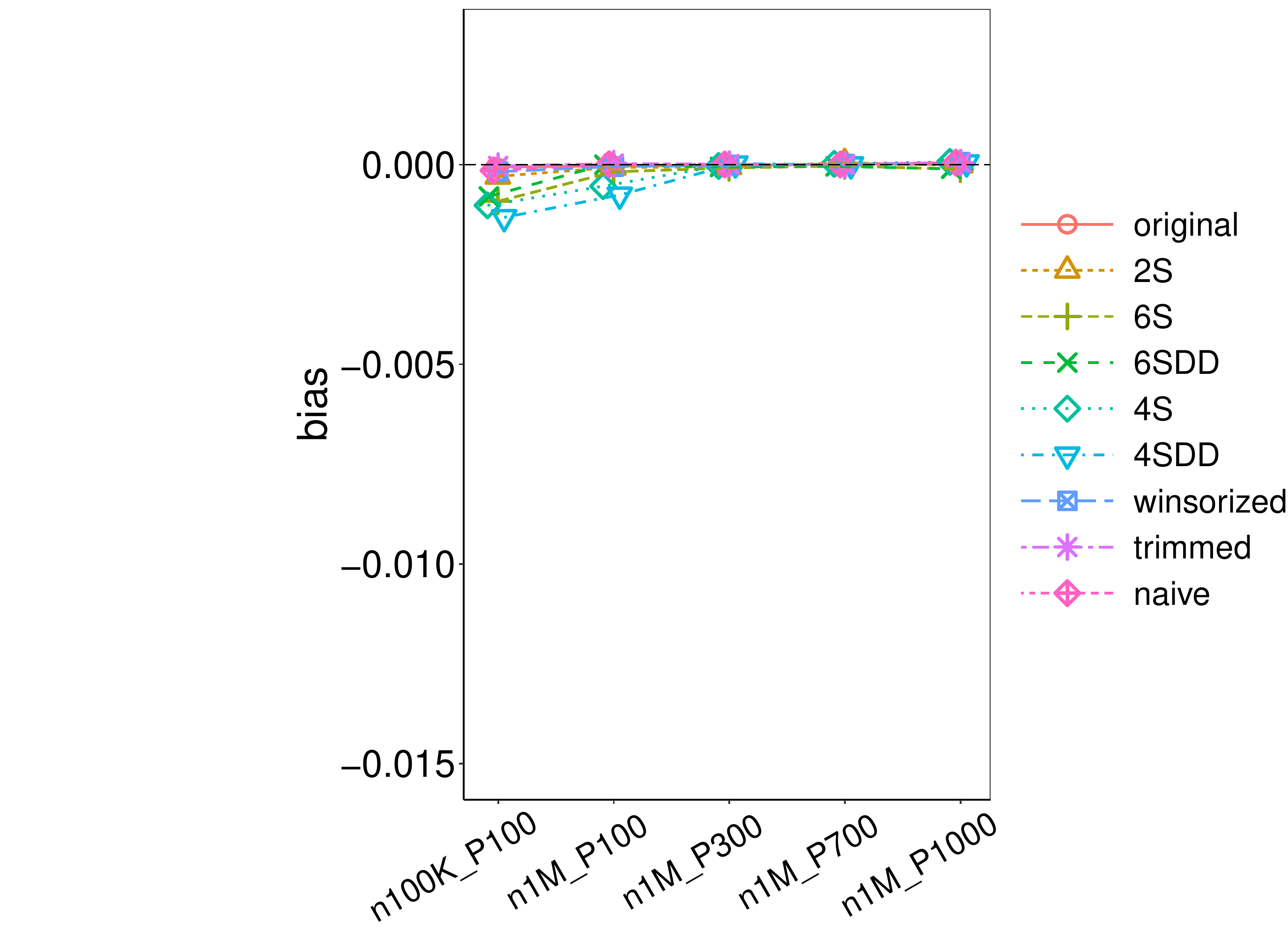}
\includegraphics[width=0.19\textwidth, trim={2.5in 0 2.5in 0},clip] {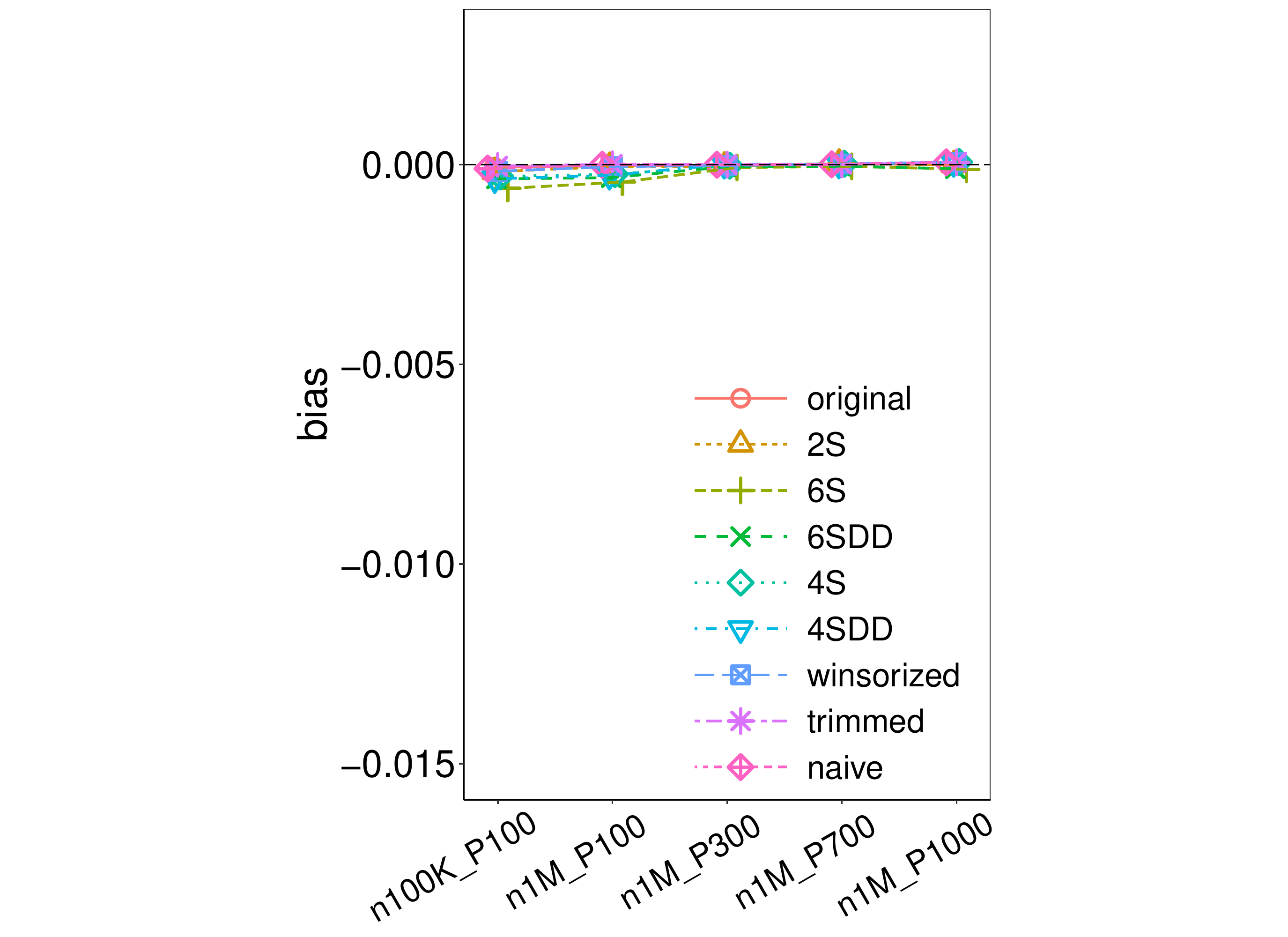}

\includegraphics[width=0.19\textwidth, trim={2.5in 0 2.6in 0},clip] {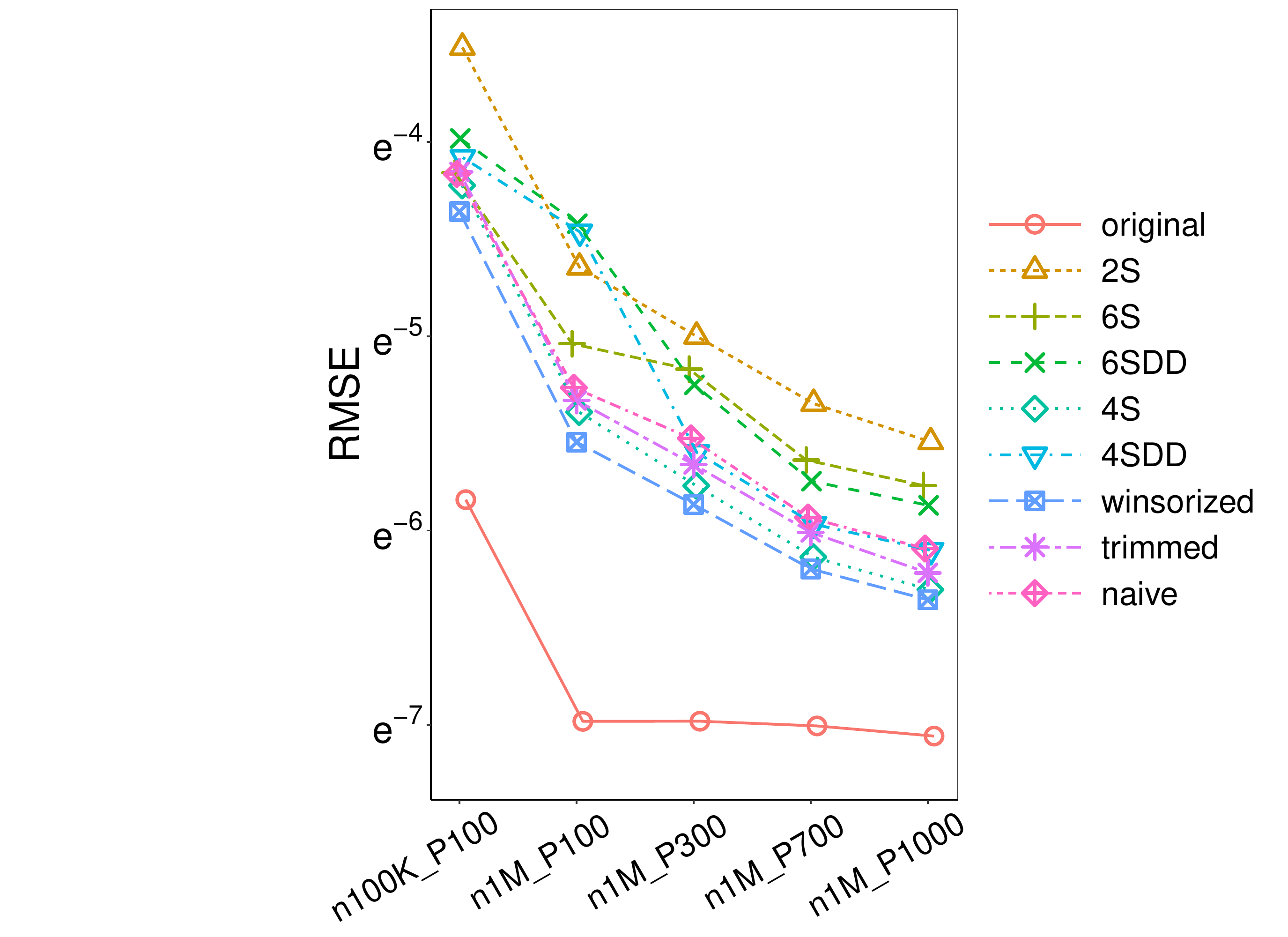}
\includegraphics[width=0.19\textwidth, trim={2.5in 0 2.6in 0},clip] {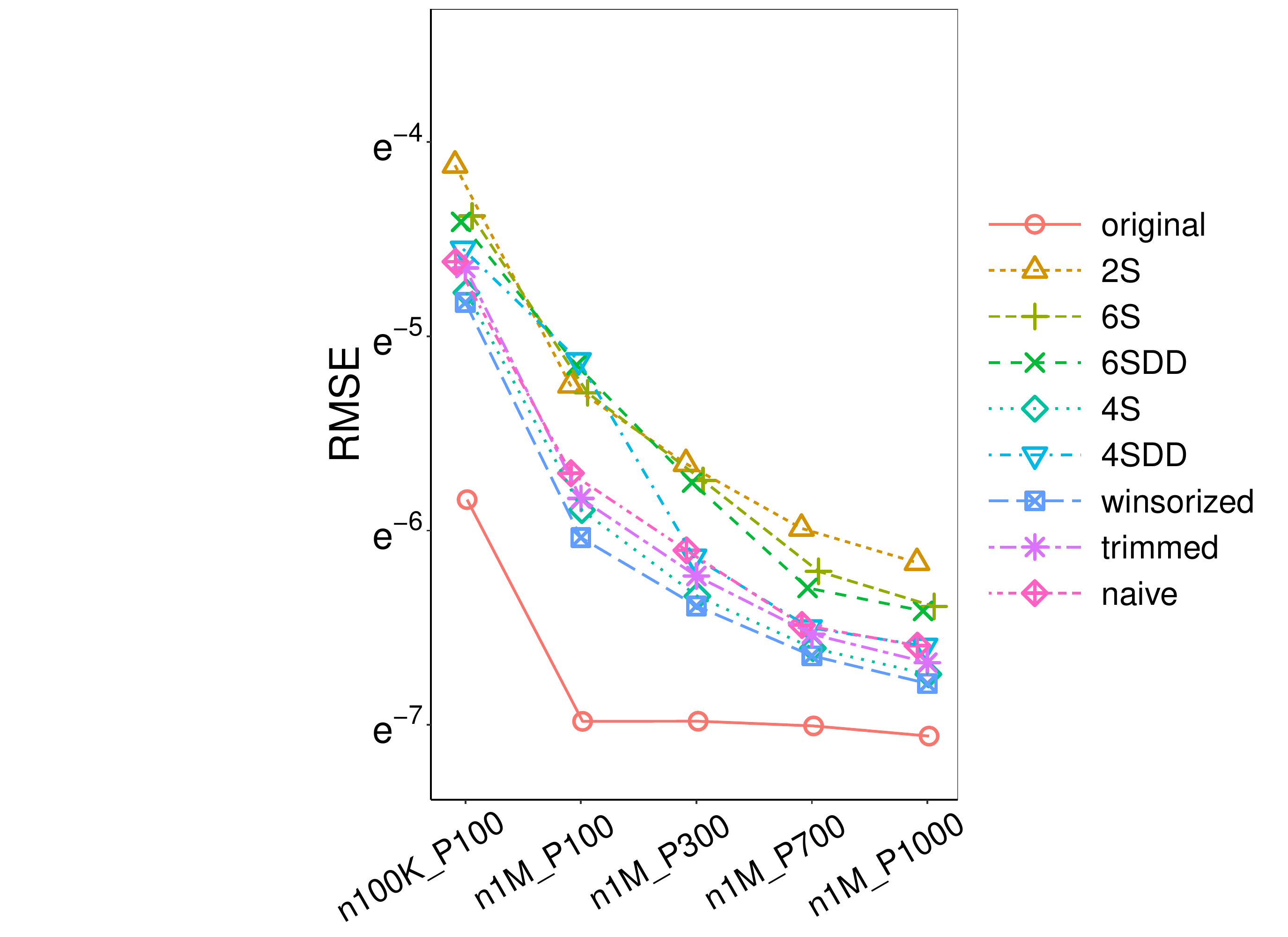}
\includegraphics[width=0.19\textwidth, trim={2.5in 0 2.6in 0},clip] {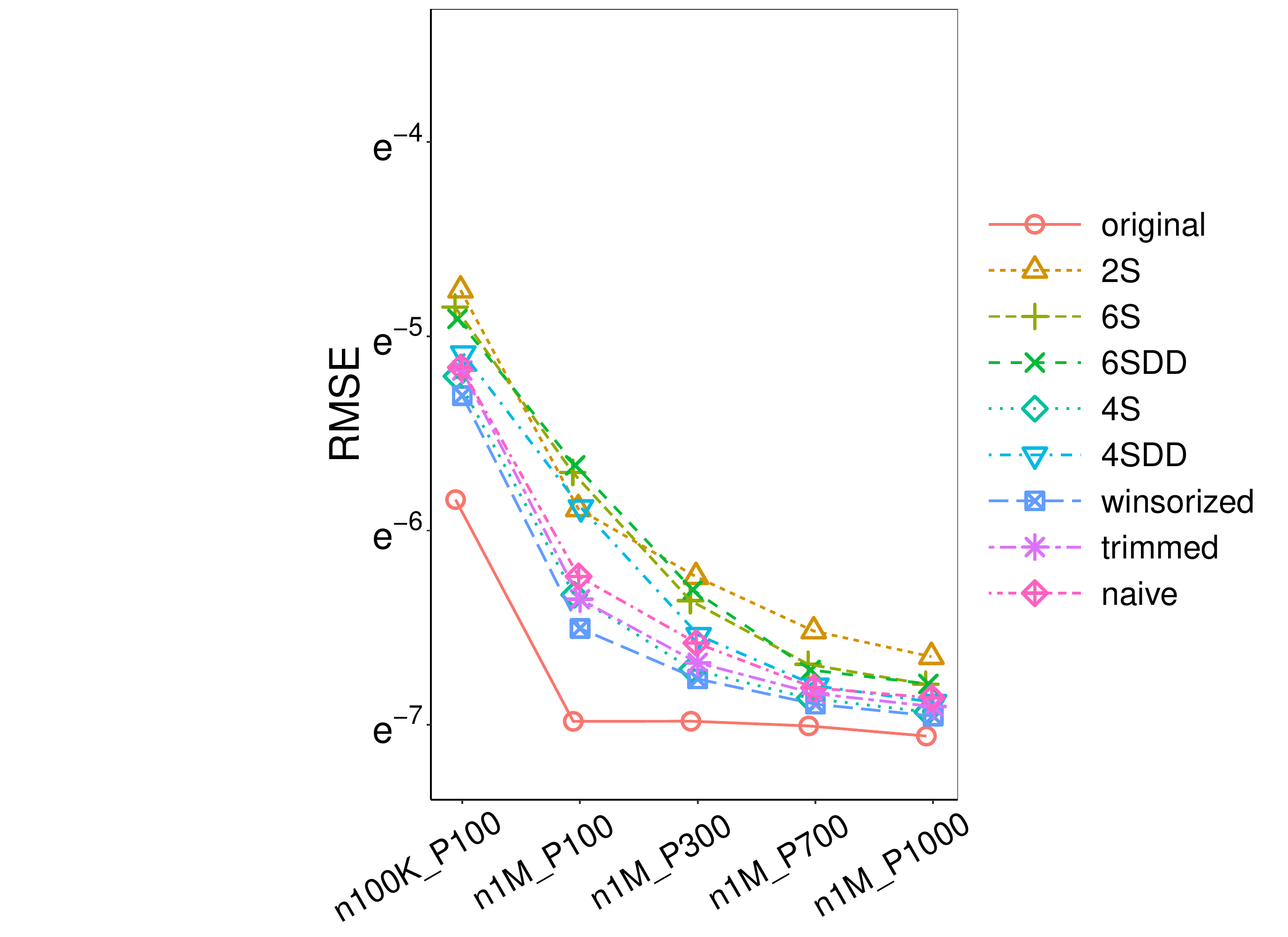}
\includegraphics[width=0.19\textwidth, trim={2.5in 0 2.6in 0},clip] {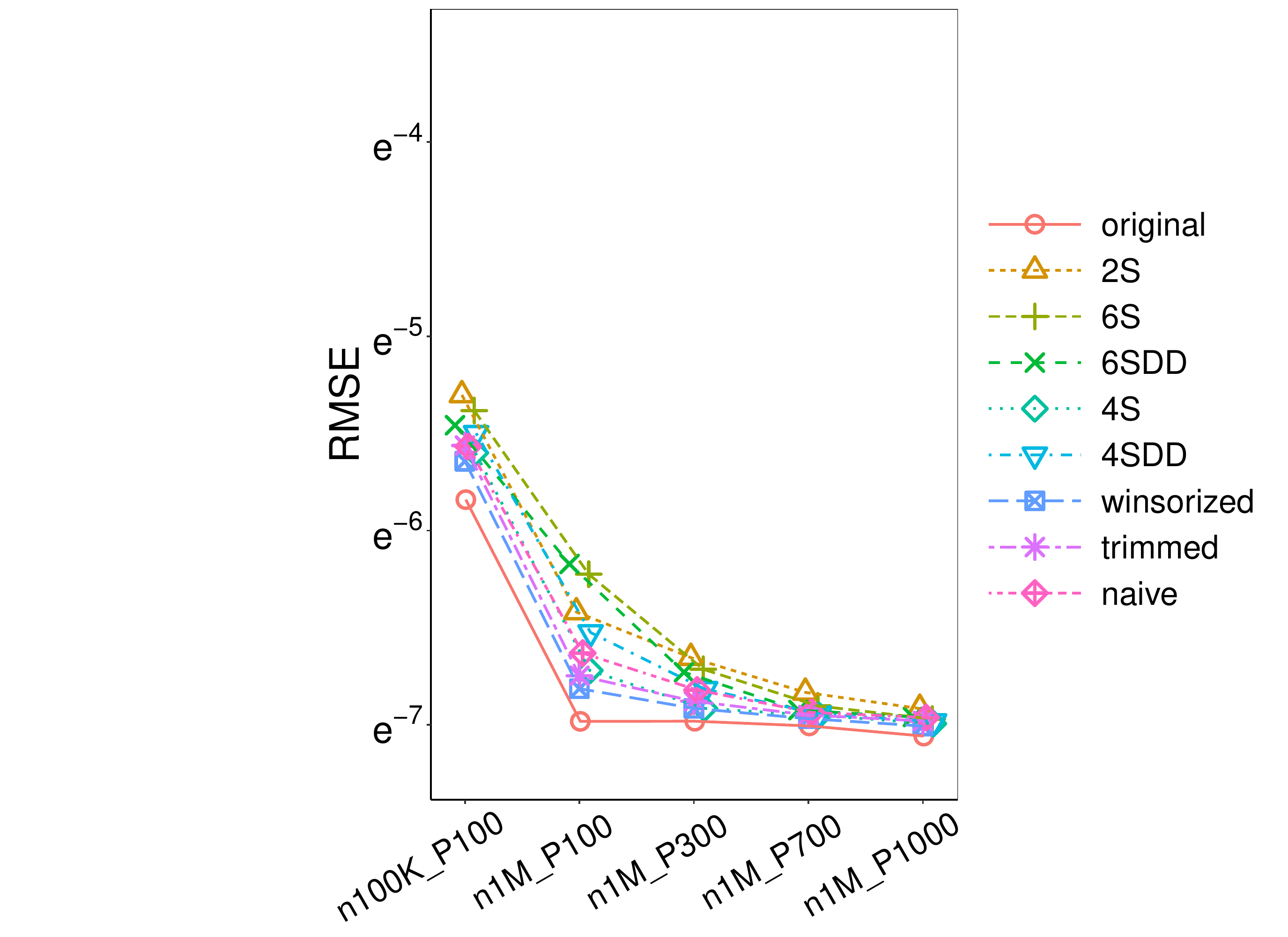}
\includegraphics[width=0.19\textwidth, trim={2.5in 0 2.6in 0},clip] {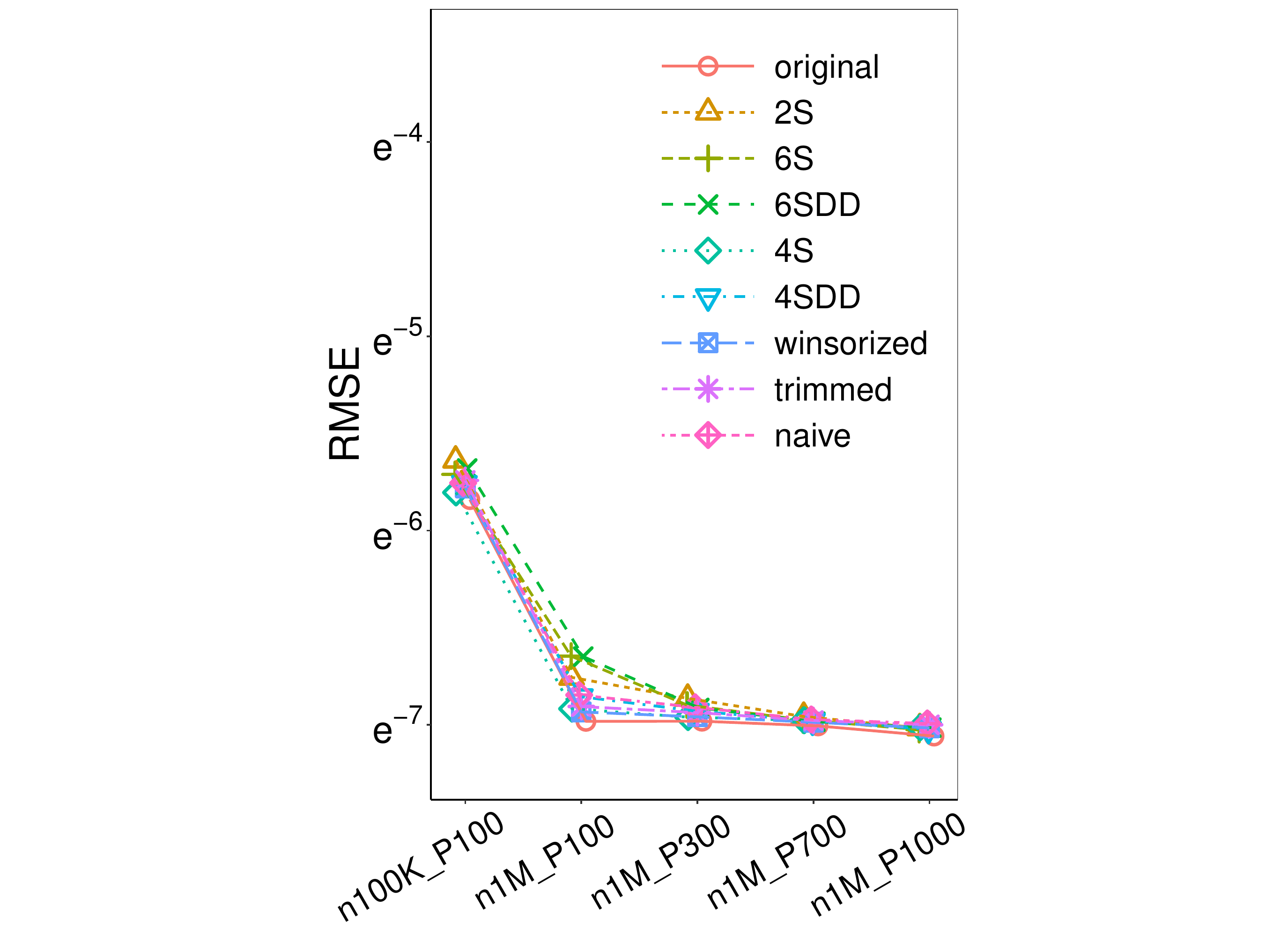}

\includegraphics[width=0.19\textwidth, trim={2.5in 0 2.6in 0},clip] {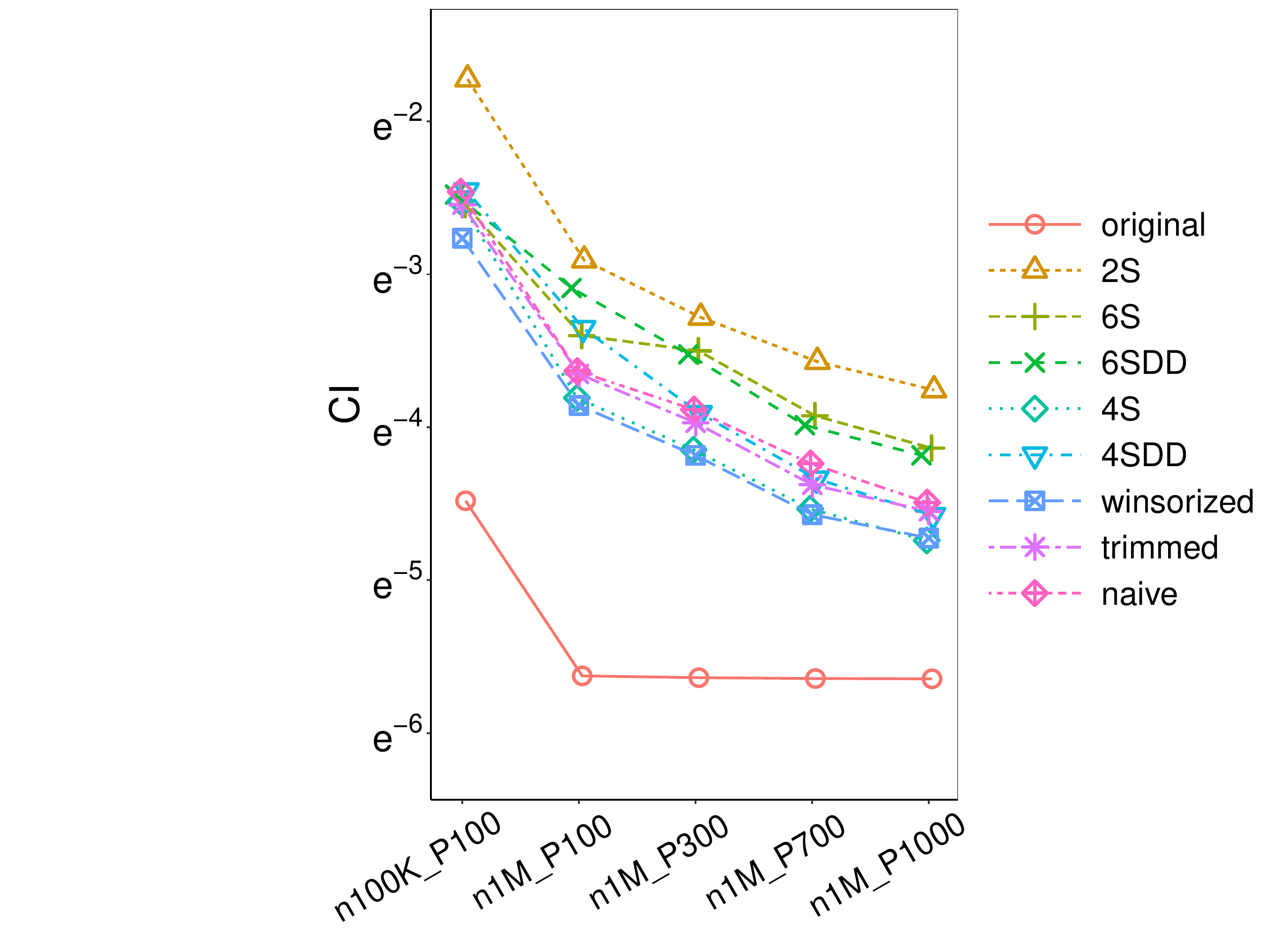}
\includegraphics[width=0.19\textwidth, trim={2.5in 0 2.6in 0},clip] {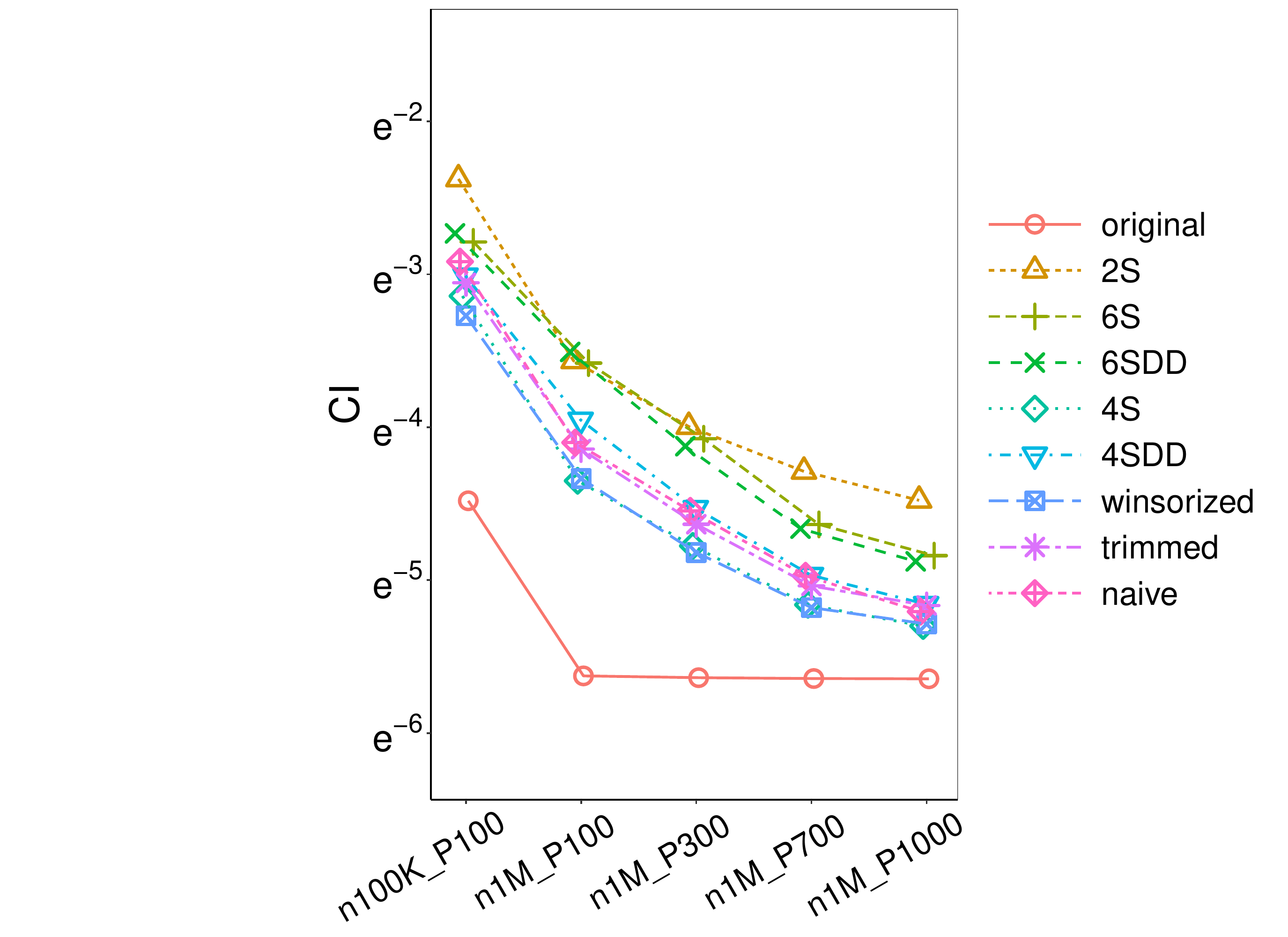}
\includegraphics[width=0.19\textwidth, trim={2.5in 0 2.6in 0},clip] {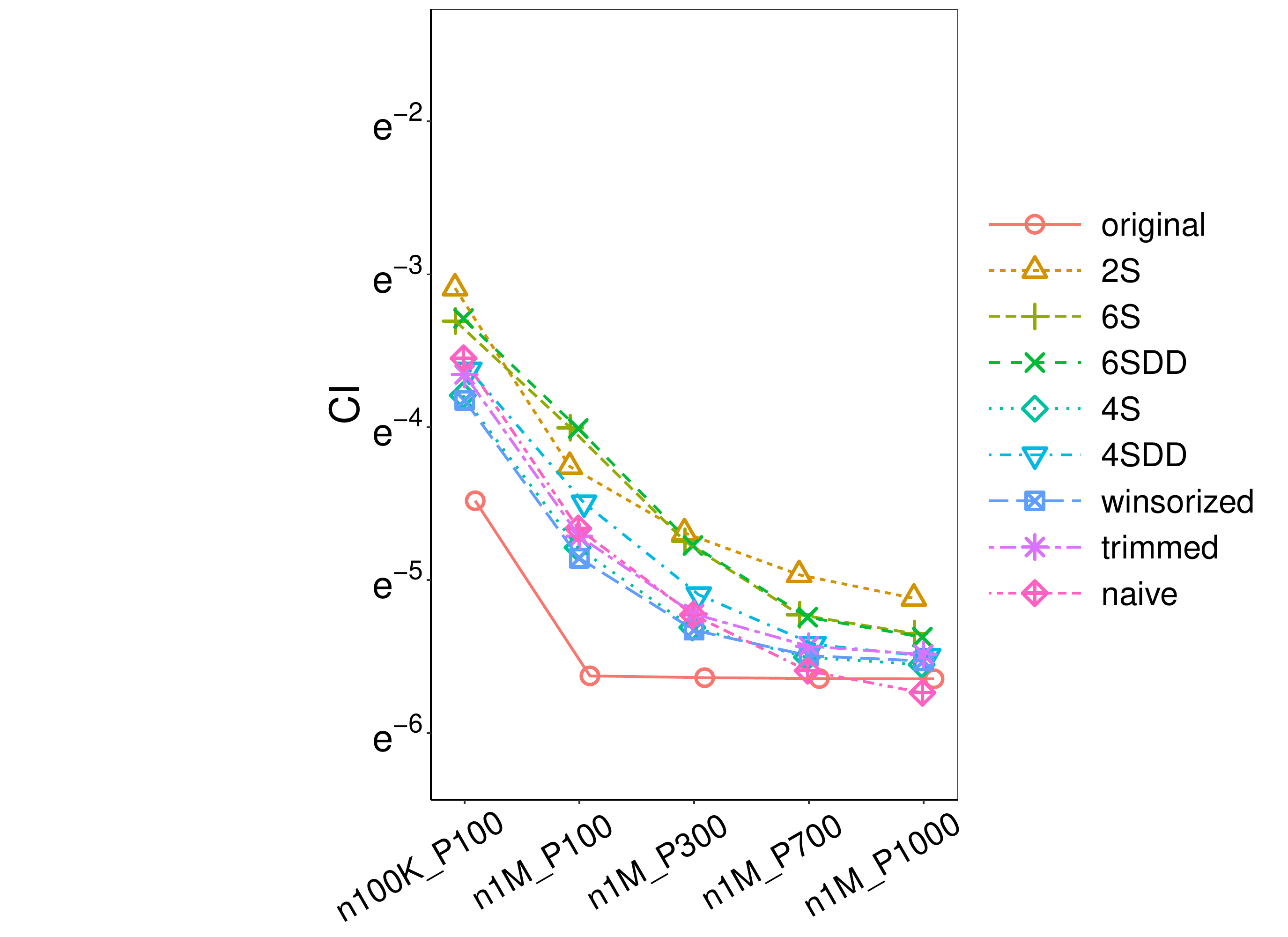}
\includegraphics[width=0.19\textwidth, trim={2.5in 0 2.6in 0},clip] {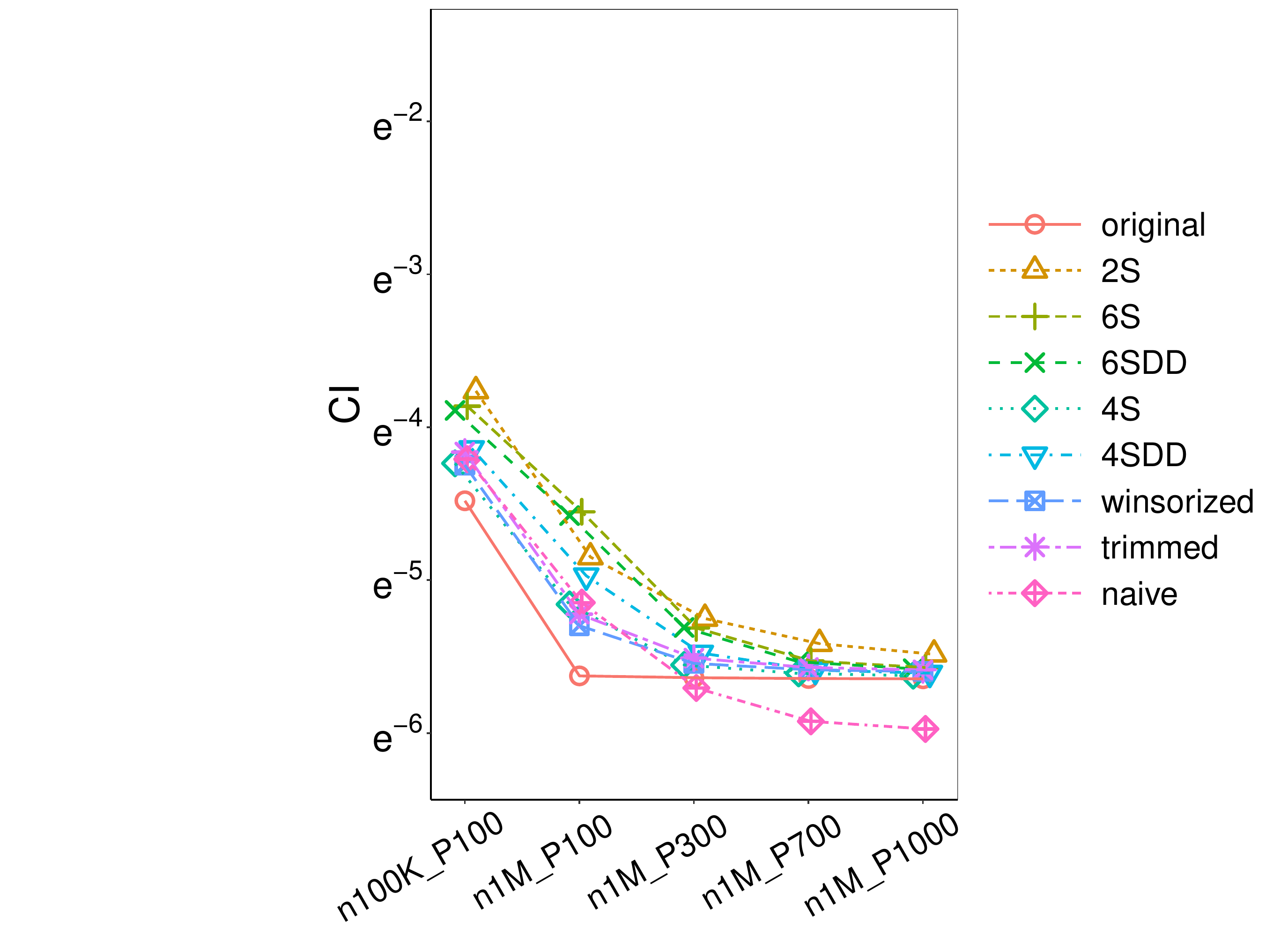}
\includegraphics[width=0.19\textwidth, trim={2.5in 0 2.6in 0},clip] {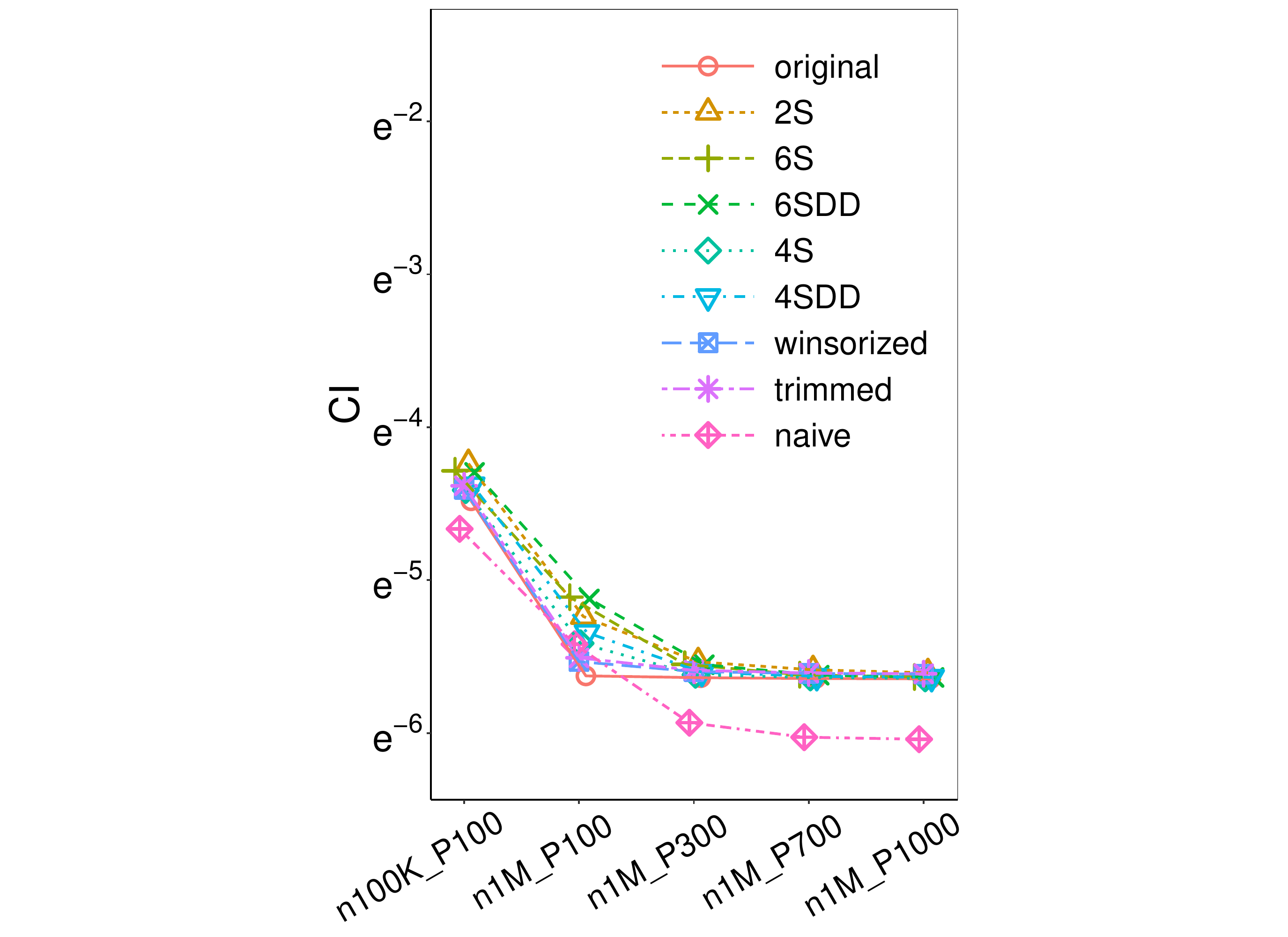}

\includegraphics[width=0.19\textwidth, trim={2.5in 0 2.6in 0},clip] {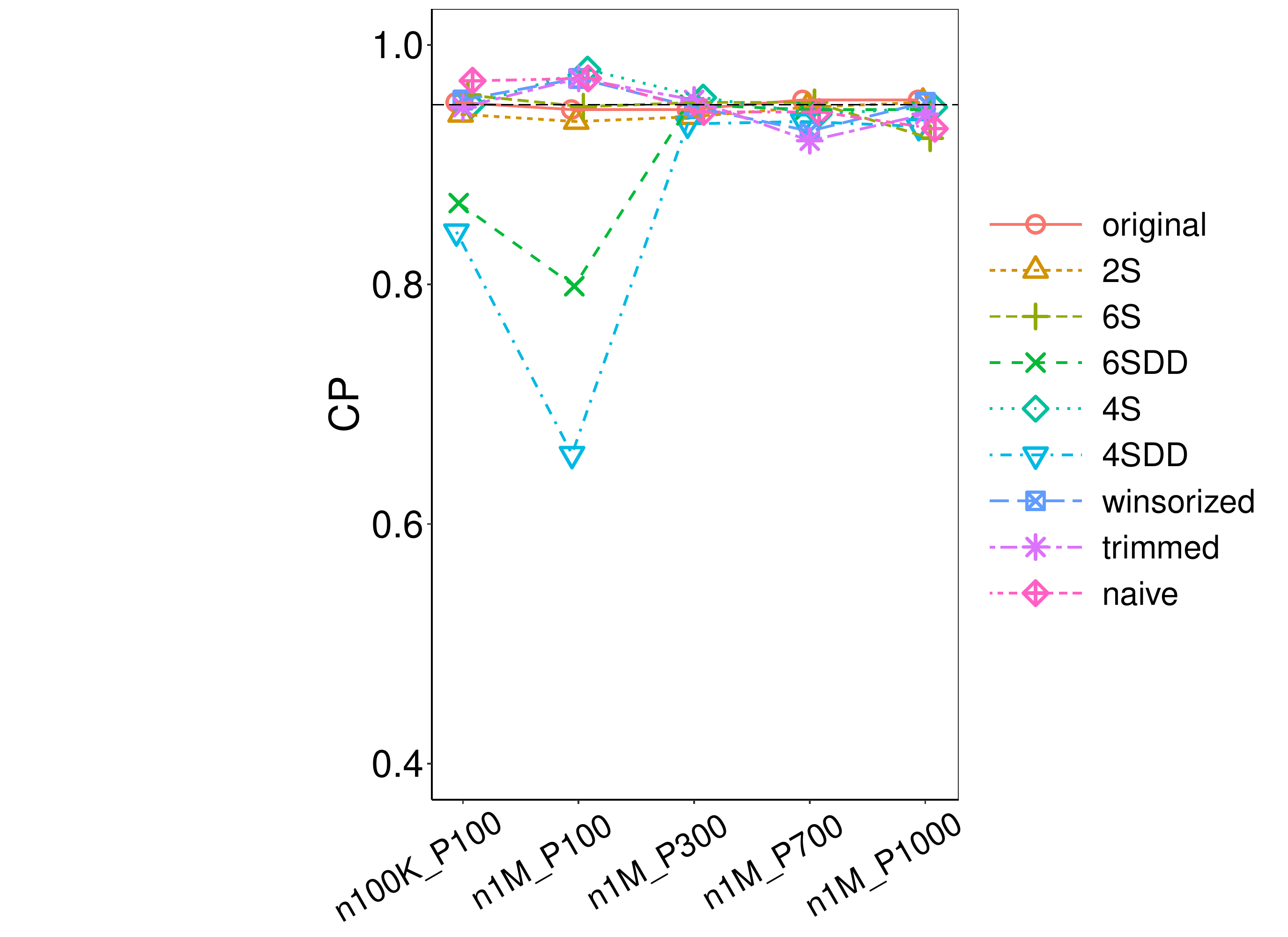}
\includegraphics[width=0.19\textwidth, trim={2.5in 0 2.6in 0},clip] {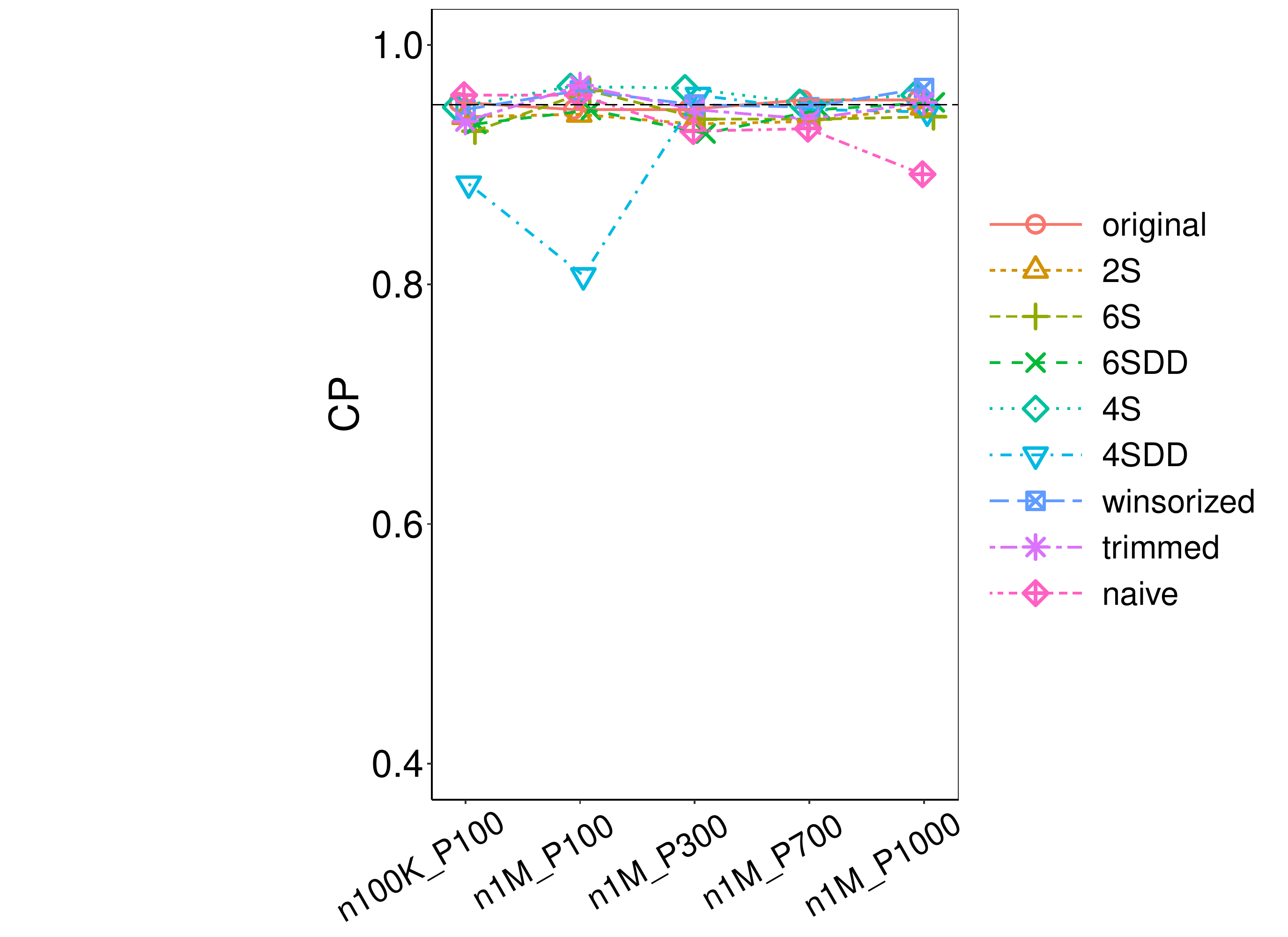}
\includegraphics[width=0.19\textwidth, trim={2.5in 0 2.6in 0},clip] {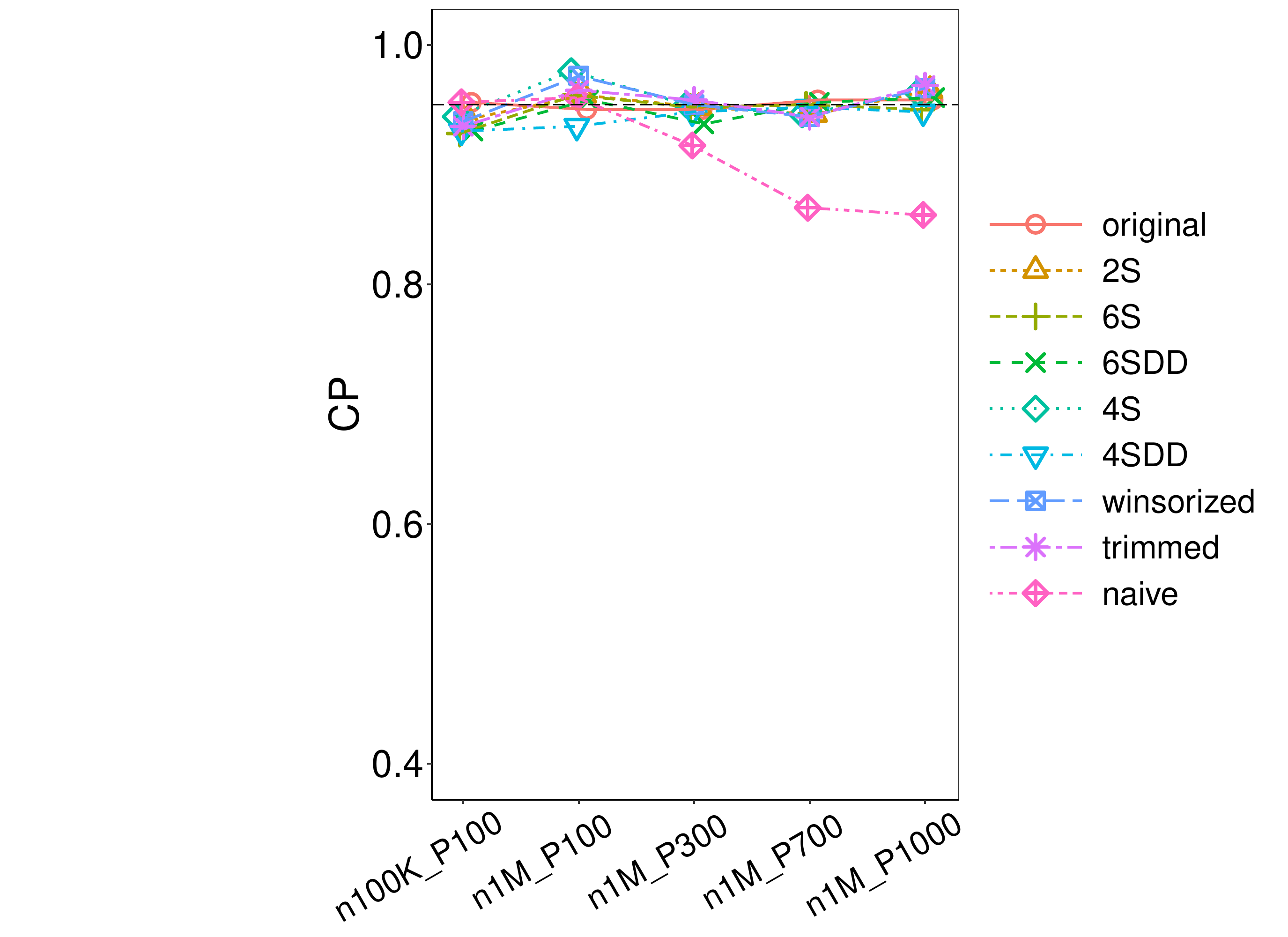}
\includegraphics[width=0.19\textwidth, trim={2.5in 0 2.6in 0},clip] {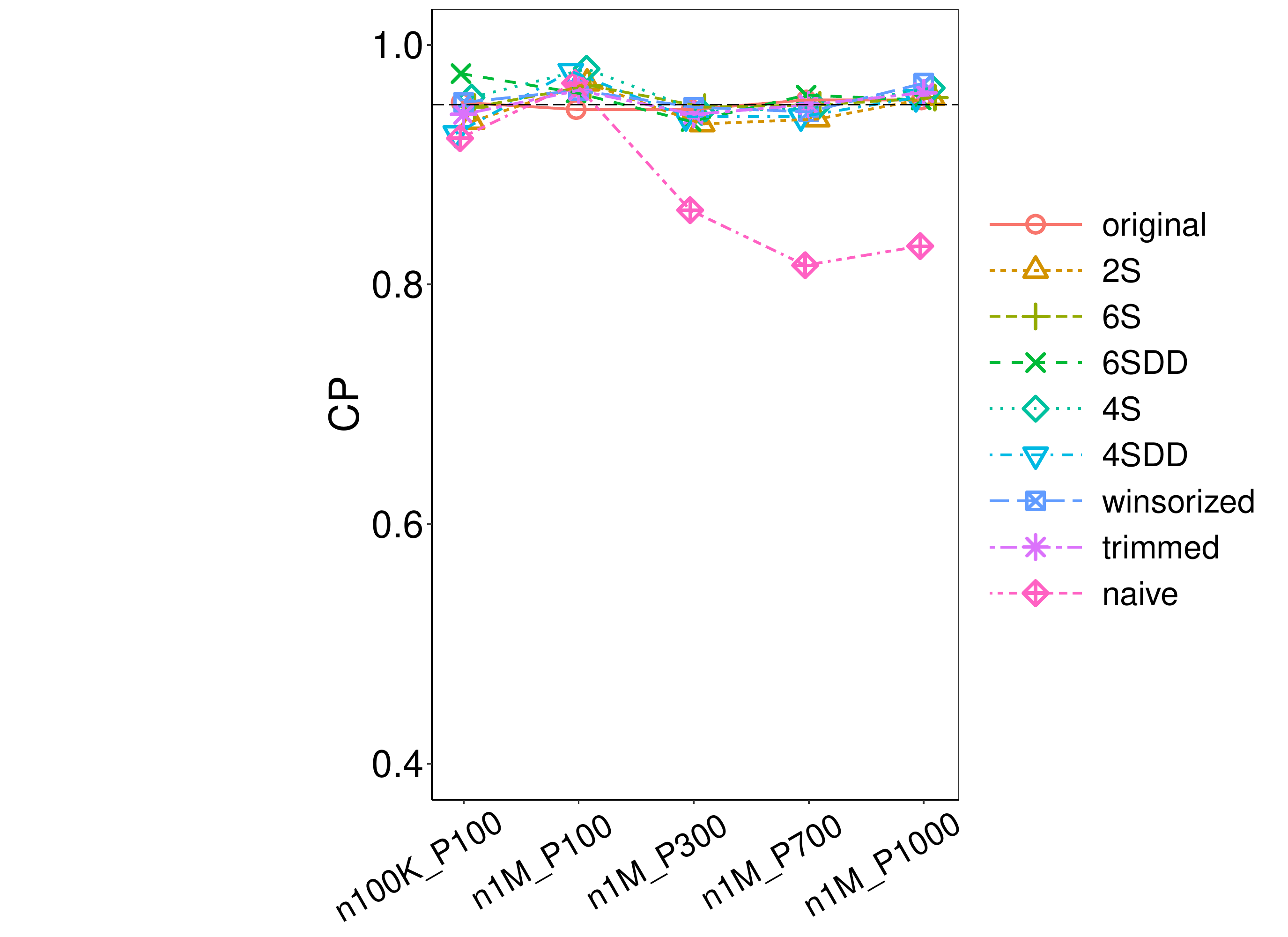}
\includegraphics[width=0.19\textwidth, trim={2.5in 0 2.6in 0},clip] {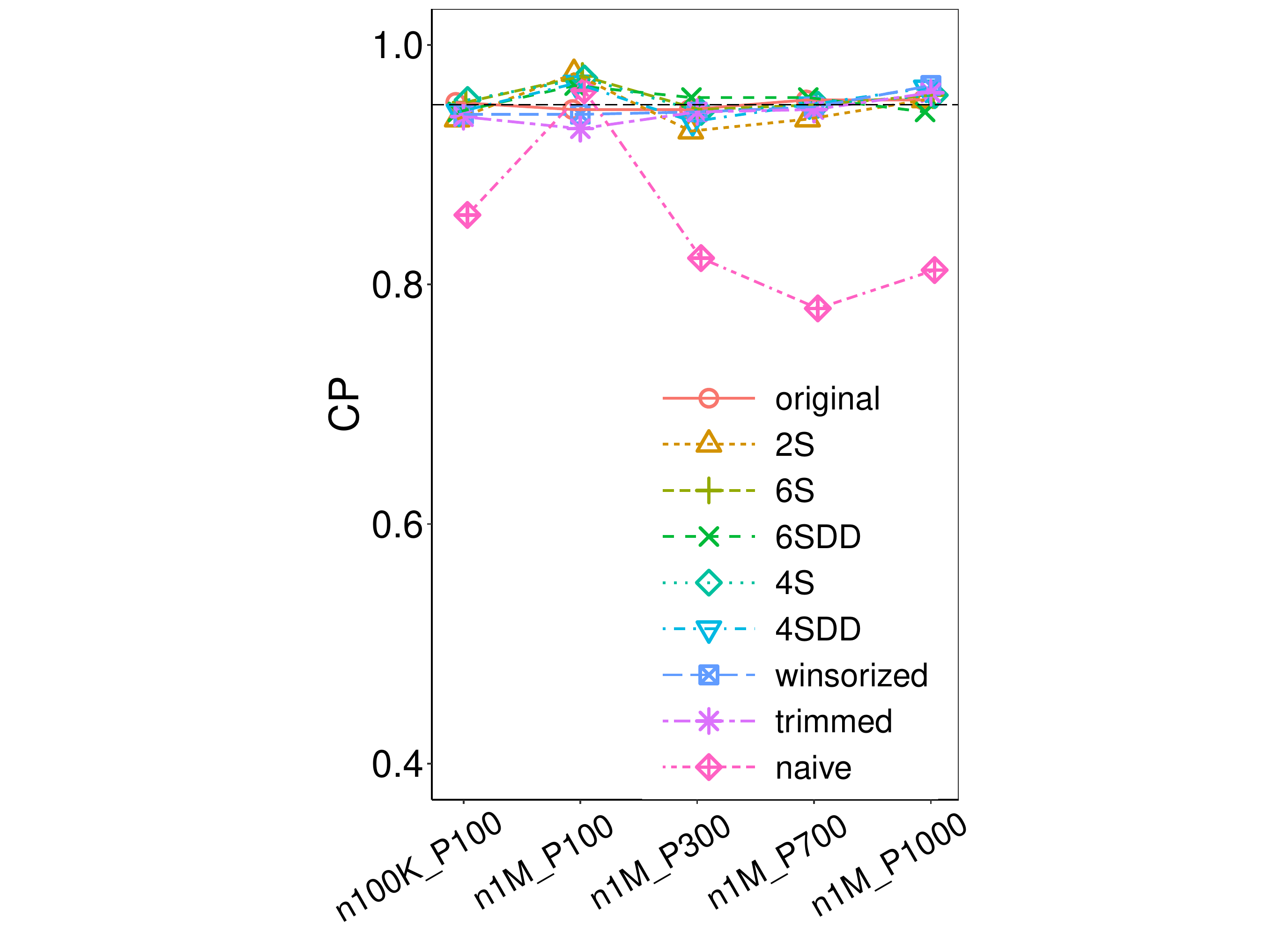}

\includegraphics[width=0.19\textwidth, trim={2.5in 0 2.6in 0},clip] {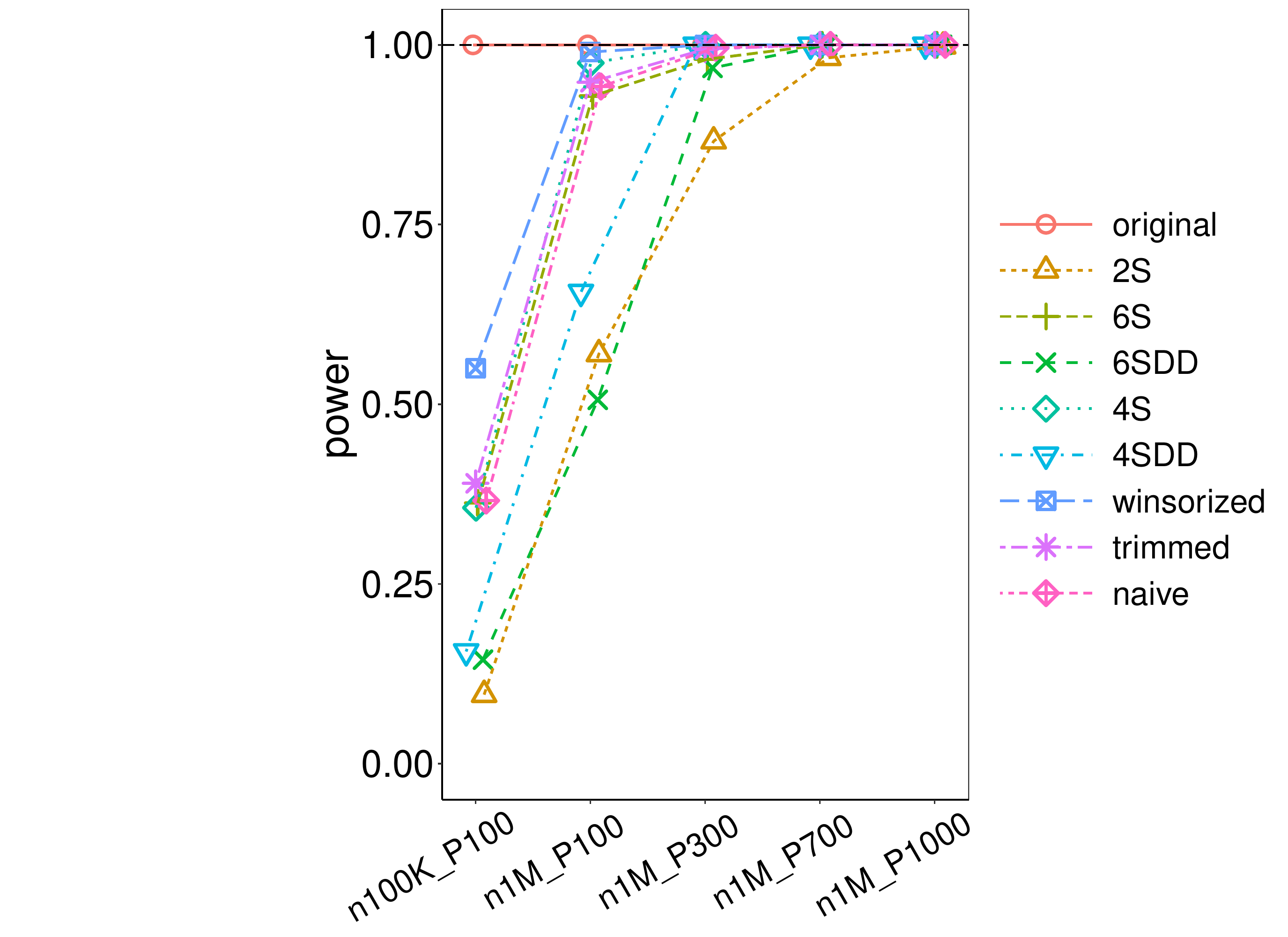}
\includegraphics[width=0.19\textwidth, trim={2.5in 0 2.6in 0},clip] {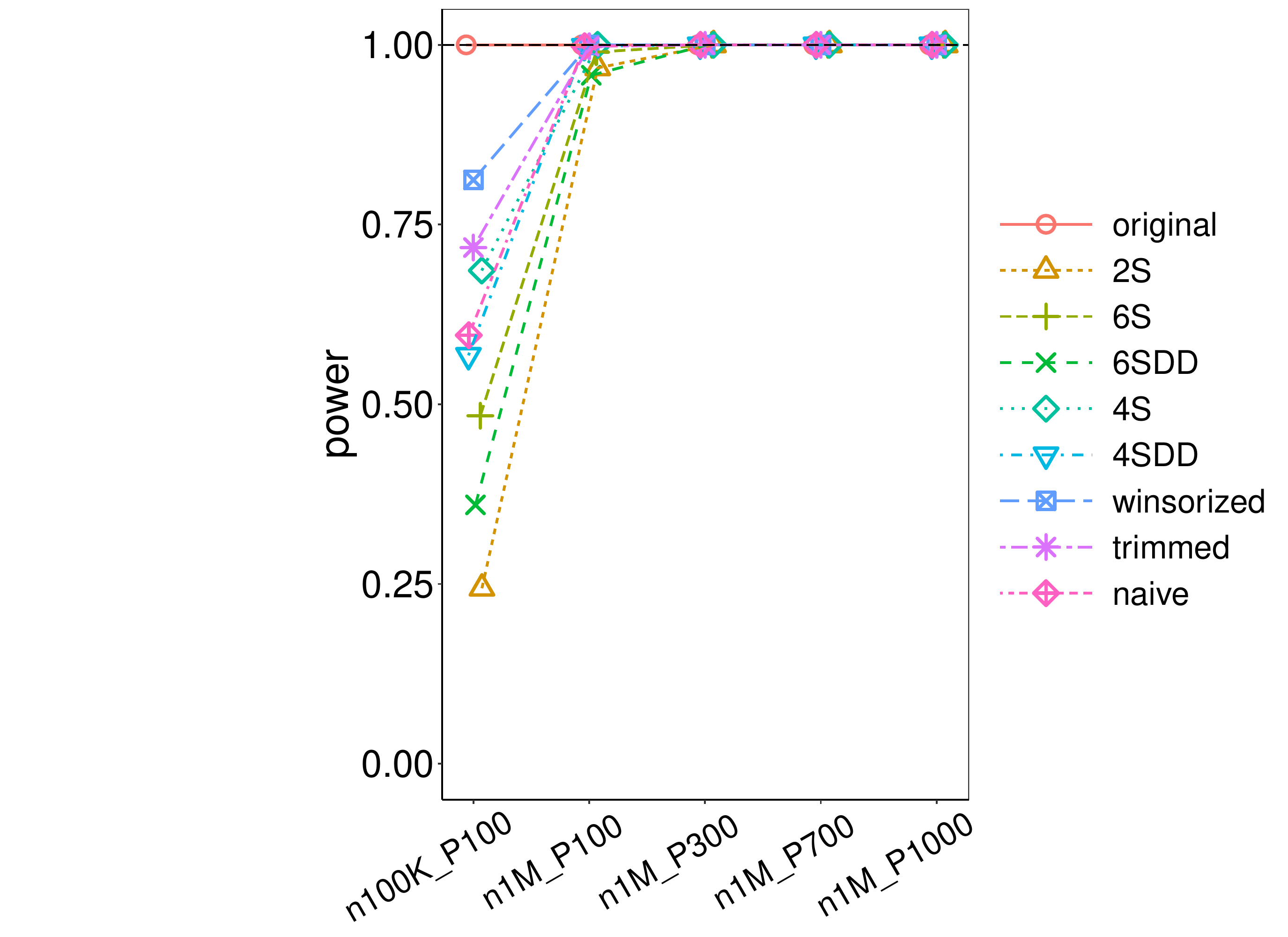}
\includegraphics[width=0.19\textwidth, trim={2.5in 0 2.6in 0},clip] {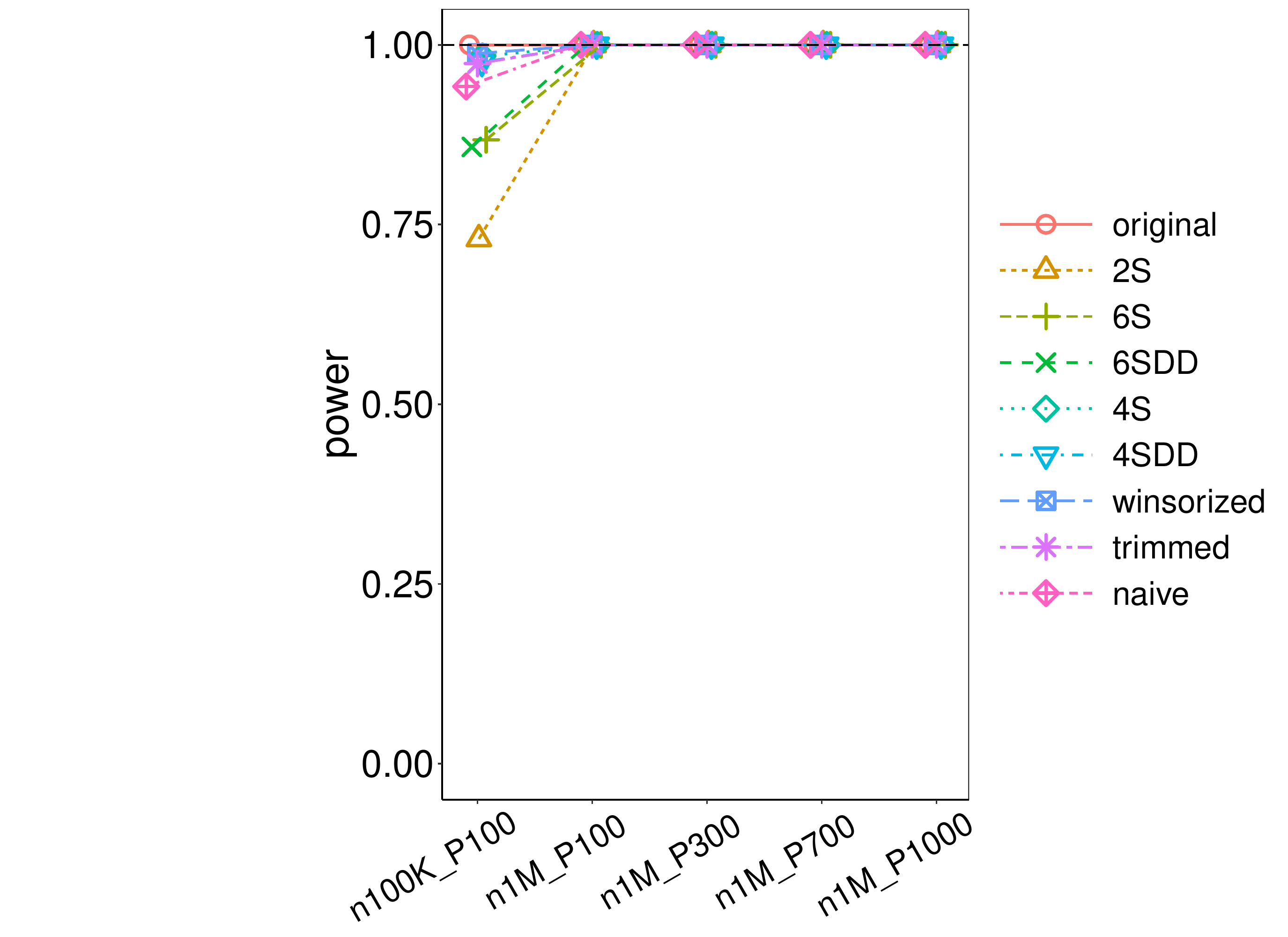}
\includegraphics[width=0.19\textwidth, trim={2.5in 0 2.6in 0},clip] {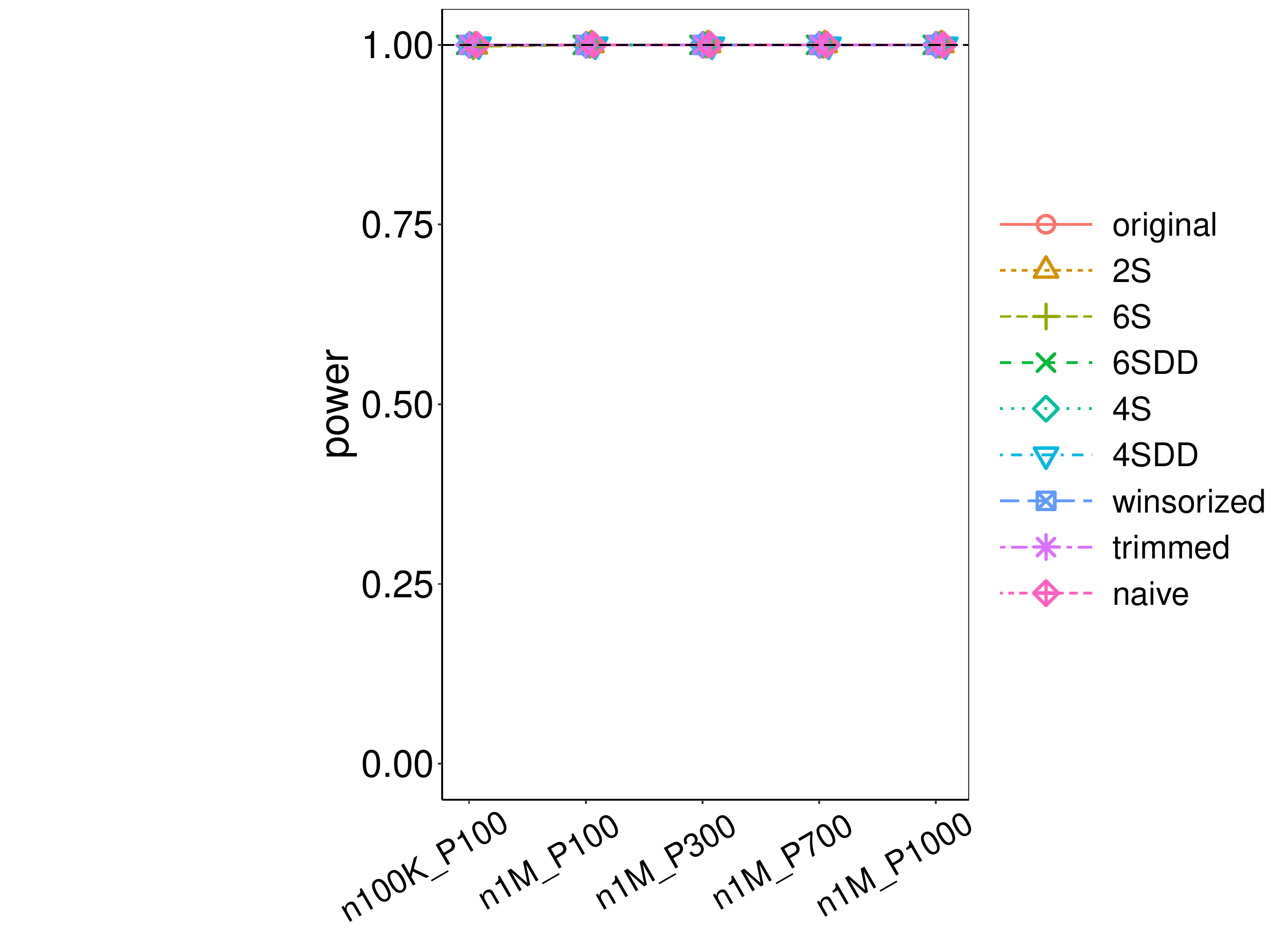}
\includegraphics[width=0.19\textwidth, trim={2.5in 0 2.6in 0},clip] {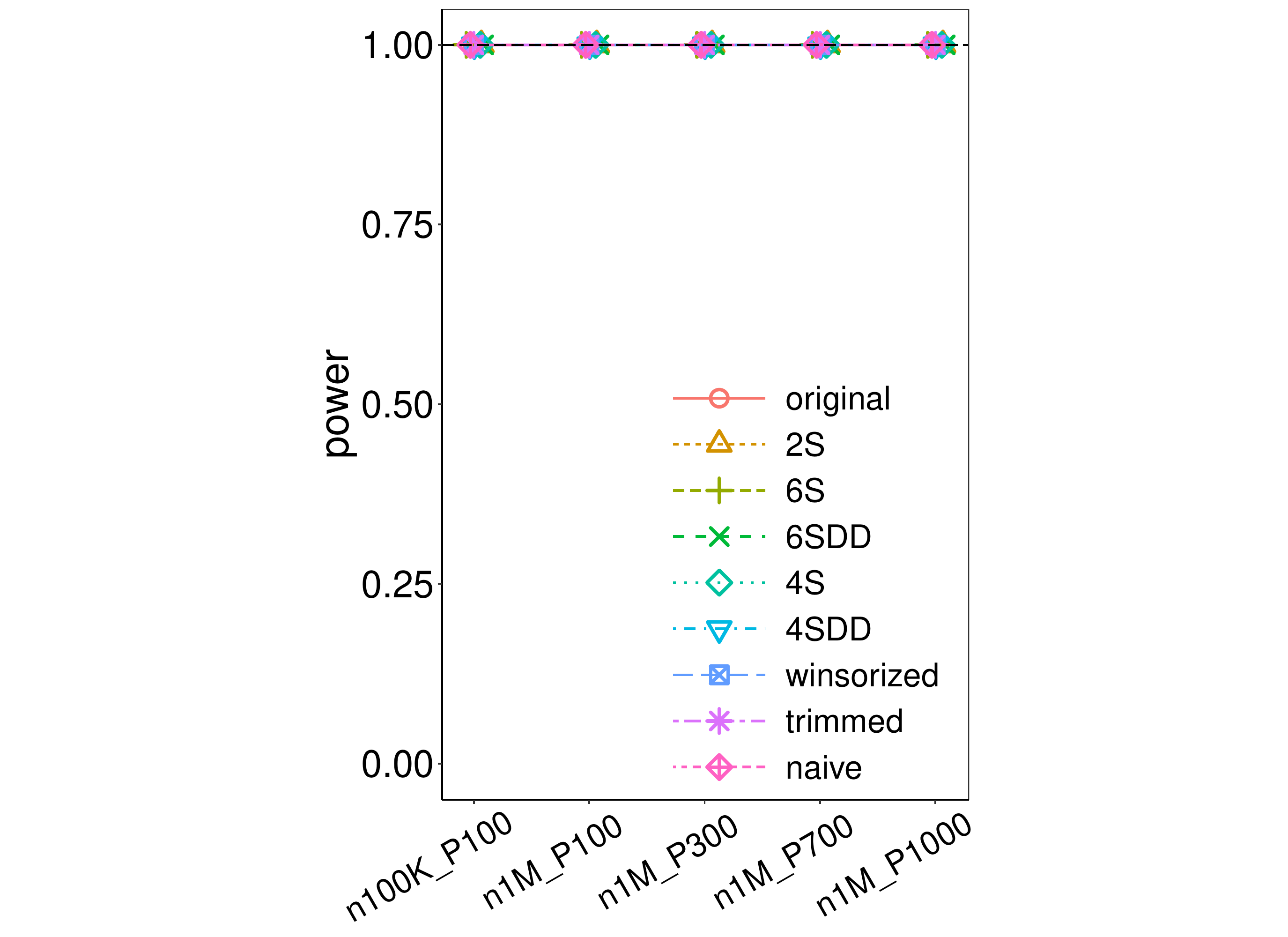}

\caption{Simulation results with $\rho$-zCDP for ZINB data with  $\alpha=\beta$ when $\theta\ne0$}
\label{fig:1szCDPZINB}
\end{figure}

The results are presented in Figures \ref{fig:1sDPZINB} and \ref{fig:1szCDPZINB} and are summarized  as follows. Winsorized mean is the best performer by all metrics. As for the other approaches, the performance  depends on the evaluation metrics. For example, in terms of bias, winsorized mean and trimmed mean  are the closest to 0, followed by 4S and 2S, 4SDD and 6SDD are the worst; in terms of RMSE and CI width, 4S and winsorized mean are the best group, followed by trimmed mean and the na\"{i}ve method (the second best group), and  2S, 6SDD, and 6SDD are the worst, especially at $n=1m/P=100$ and $\epsilon\le2$; regarding CP,  all methods except for 4SDD and 6SDD provide nominal-level coverage; in terms of power, when $n=100k/P=100$ and $n=1m/P=100$ if $\epsilon \le 2$, winsorized mean, 4S and trimmed mean are the most powerful and all methods can provide power close to 1 for the remaining cases. The Gaussian mechanism of $\rho$-zCDP does not seem to be advantageous in statistical utility compared to the Laplace mechanism of $\epsilon$ pure DP. 

\subsubsection{\texorpdfstring{$Y\sim$ ZINB}{}, \texorpdfstring{$\theta=0$ ($H_0$)}{H0 }, and \texorpdfstring{$\alpha\ne\beta$}{AS}}\label{sec:0asZINB}
\begin{figure}[!htb]
\hspace{0.5in}$\epsilon=0.5$\hspace{0.8in}$\epsilon=1$\hspace{0.9in}$\epsilon=2$
\hspace{0.95in}$\epsilon=5$\hspace{0.9in}$\epsilon=50$

\includegraphics[width=0.19\textwidth, trim={2.45in 0 2.45in 0},clip] {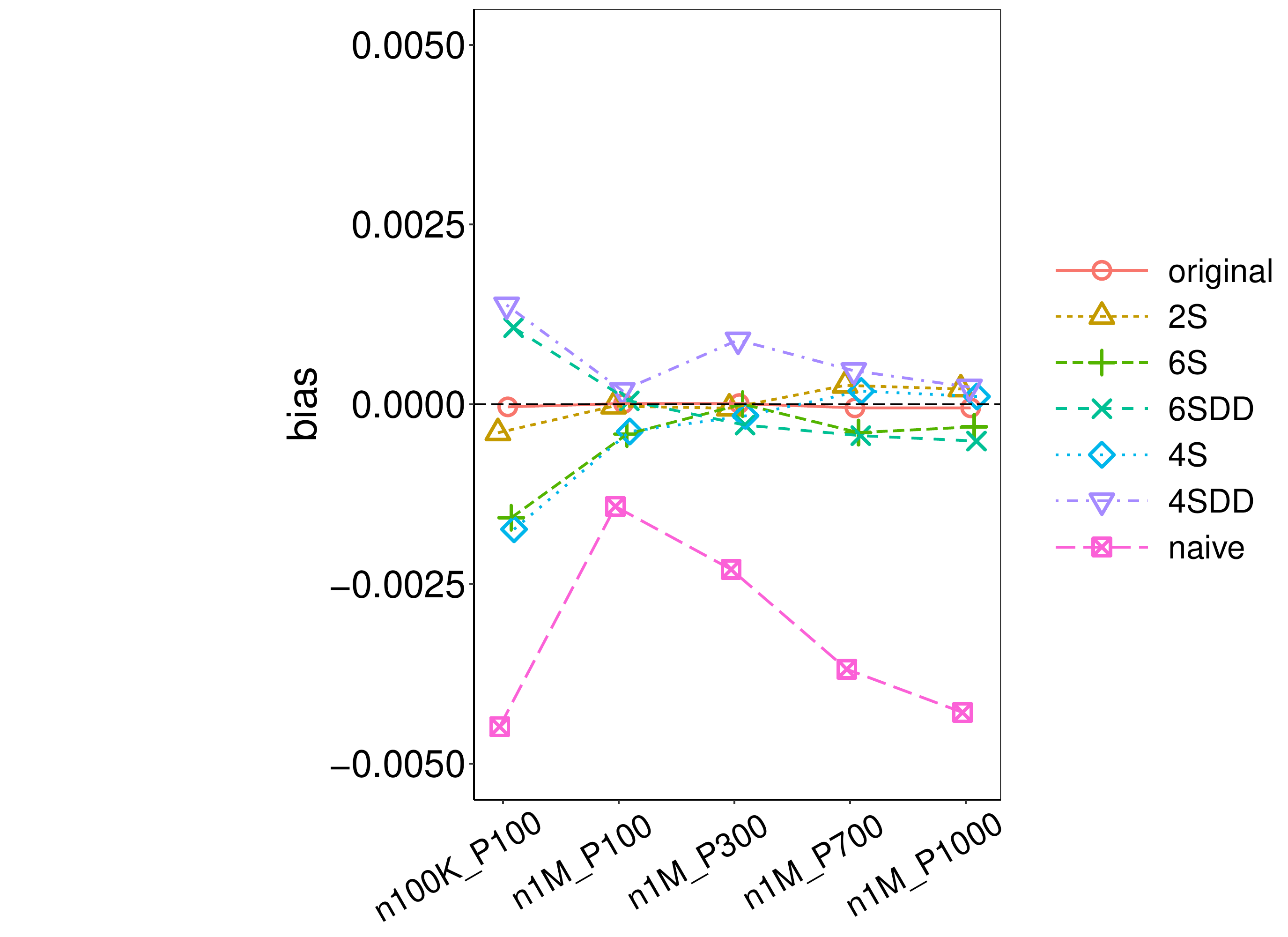}
\includegraphics[width=0.19\textwidth, trim={2.45in 0 2.45in 0},clip] {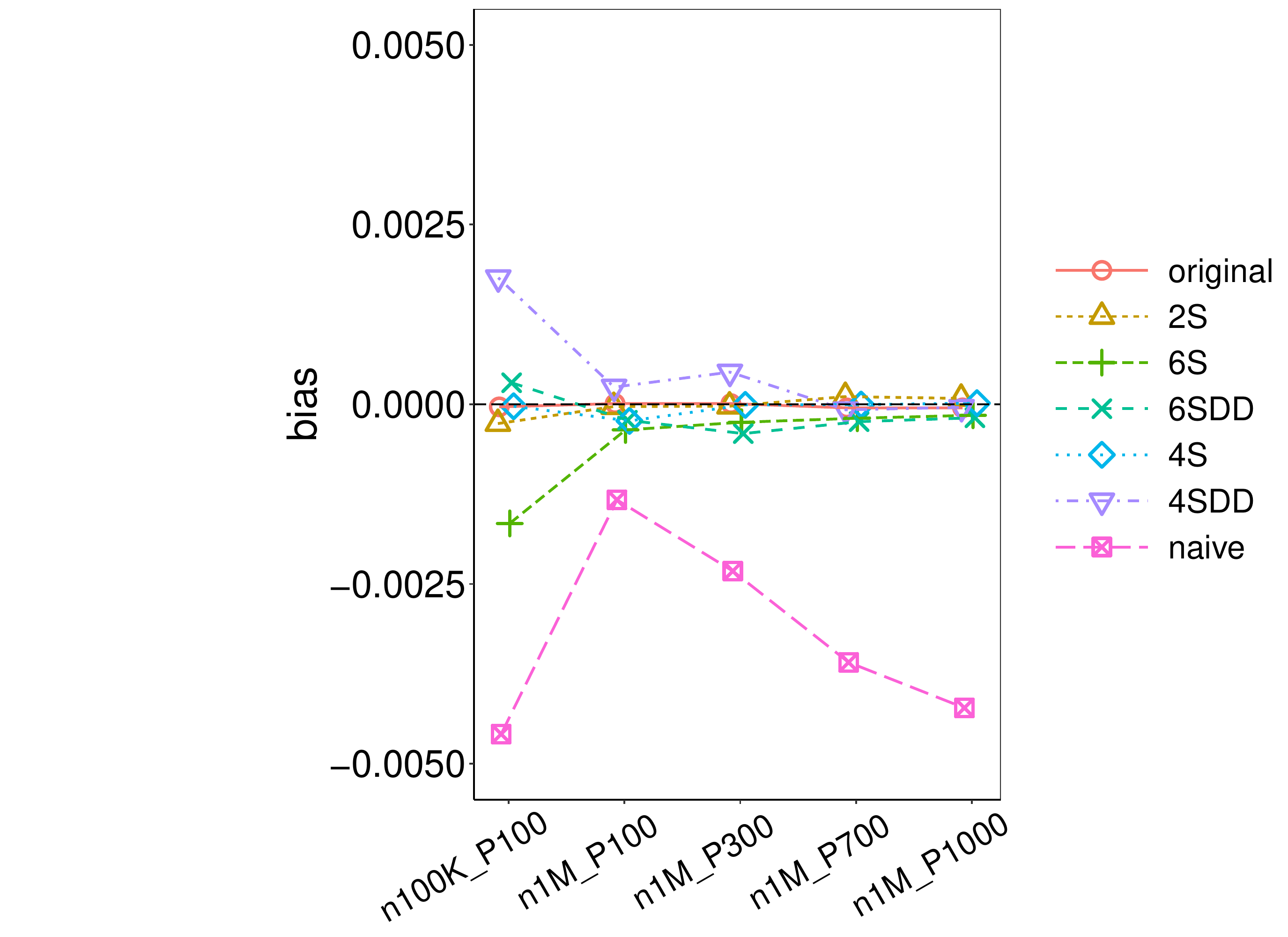}
\includegraphics[width=0.19\textwidth, trim={2.45in 0 2.45in 0},clip] {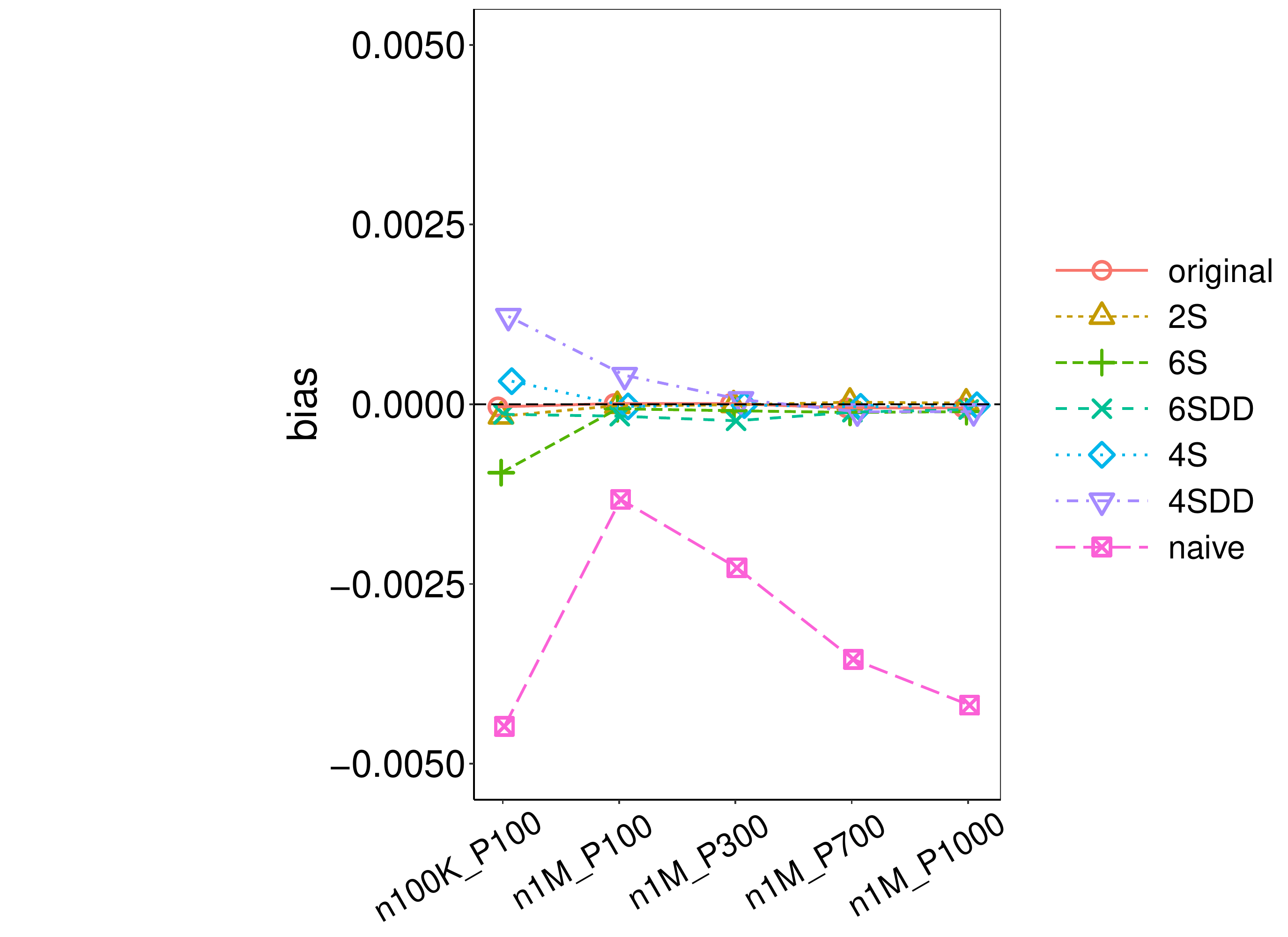}
\includegraphics[width=0.19\textwidth, trim={2.45in 0 2.45in 0},clip] {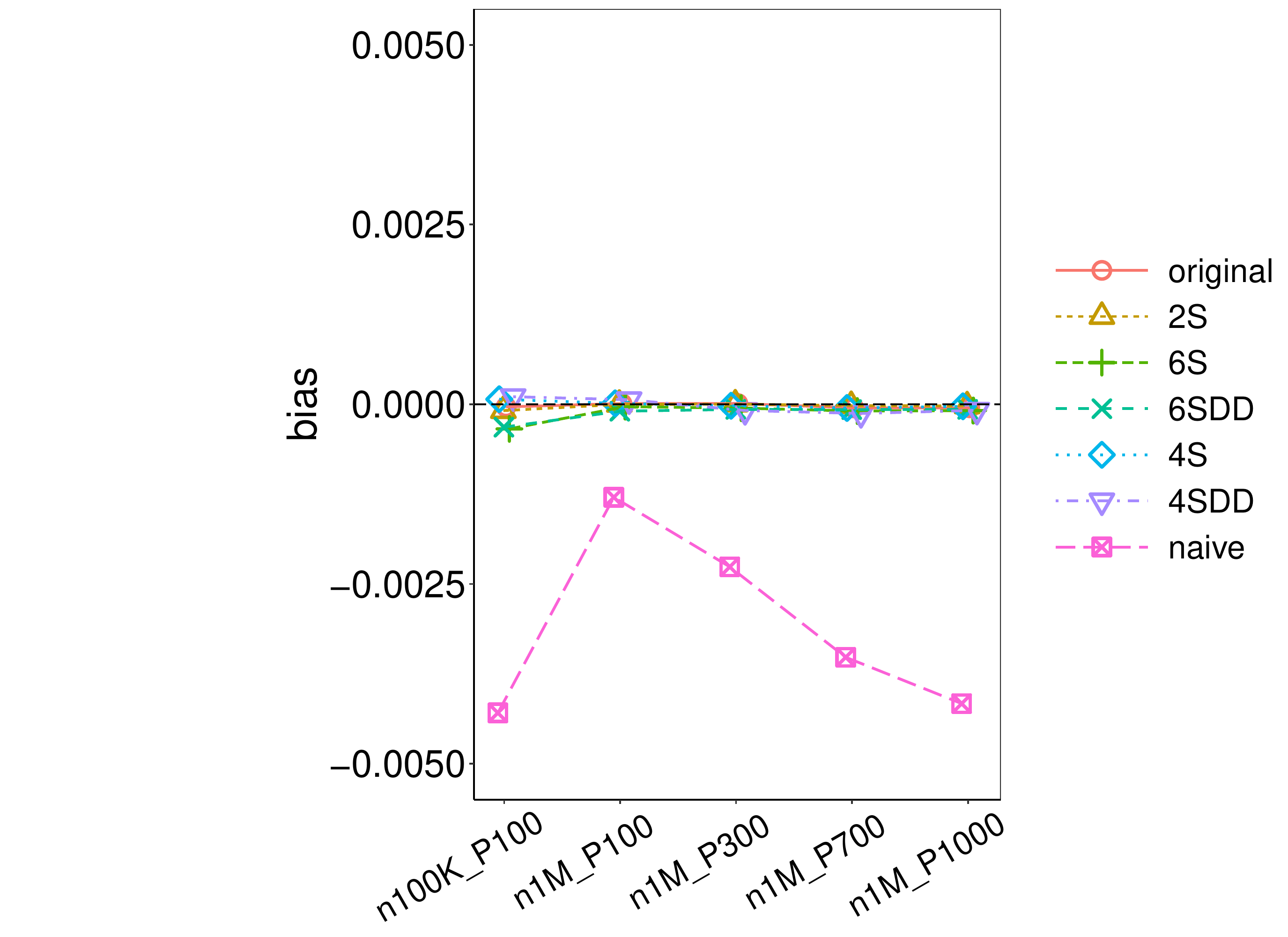}
\includegraphics[width=0.19\textwidth, trim={2.45in 0 2.45in 0},clip] {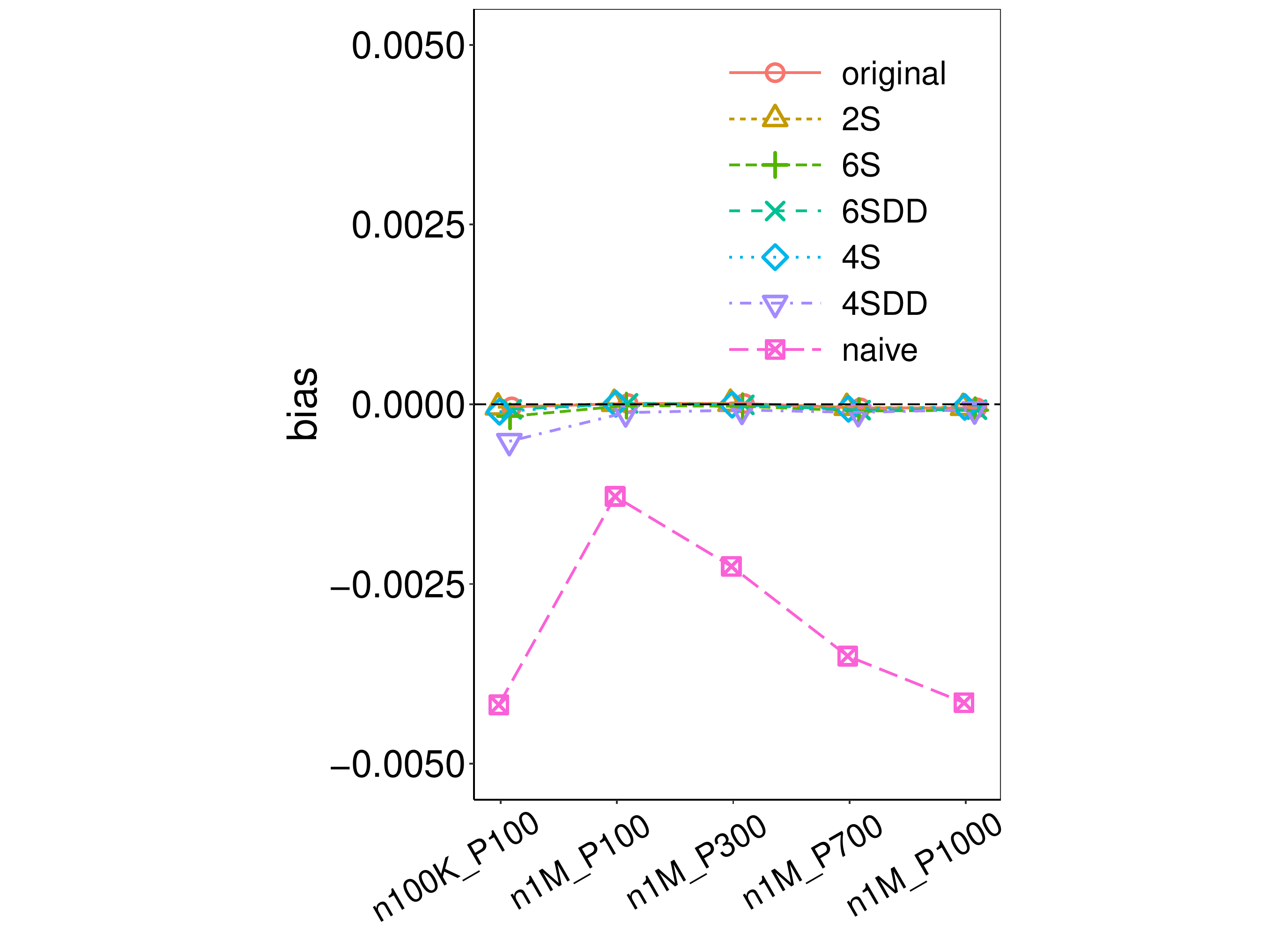}

\includegraphics[width=0.19\textwidth, trim={2.5in 0 2.6in 0},clip] {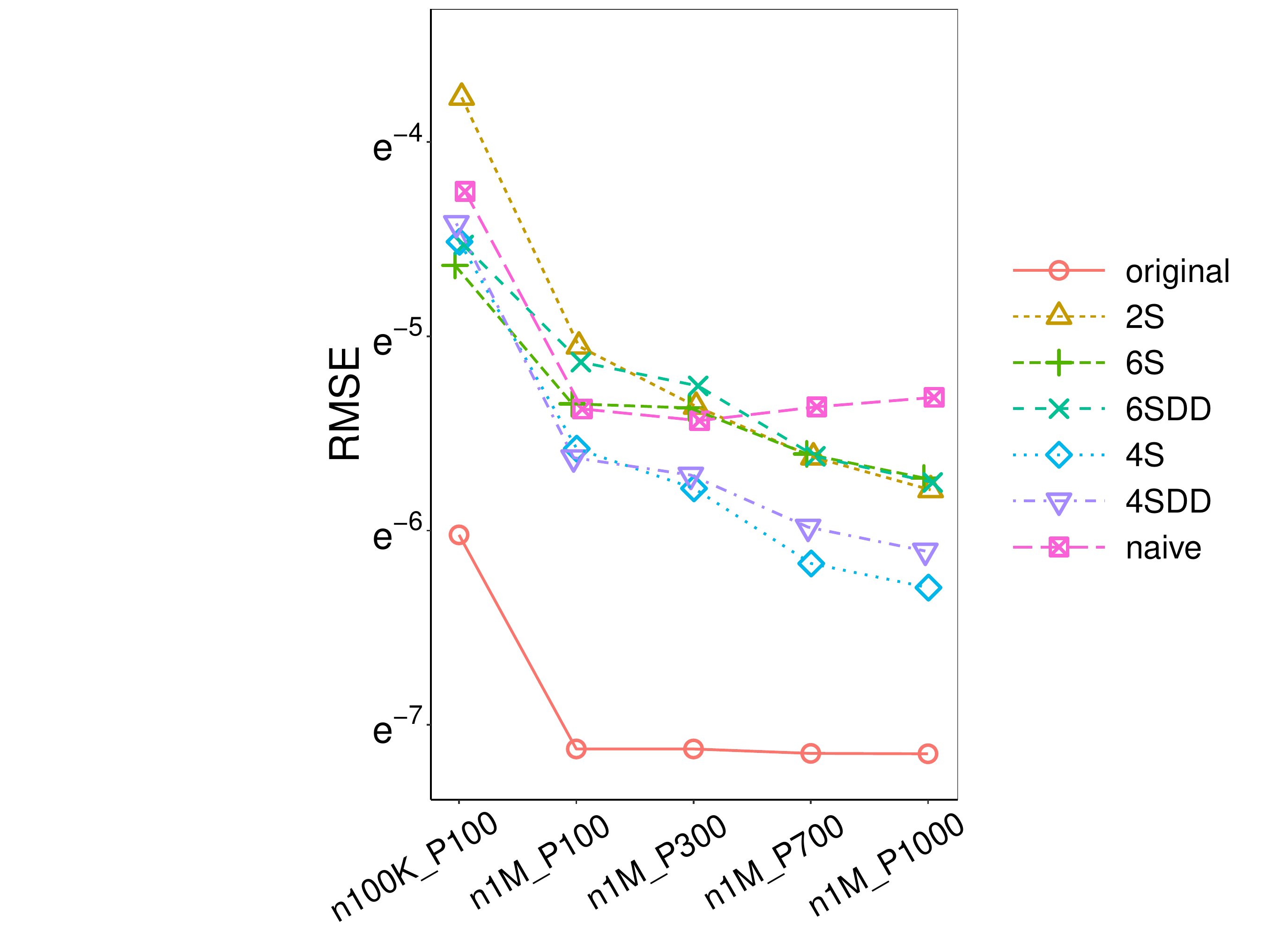}
\includegraphics[width=0.19\textwidth, trim={2.5in 0 2.6in 0},clip] {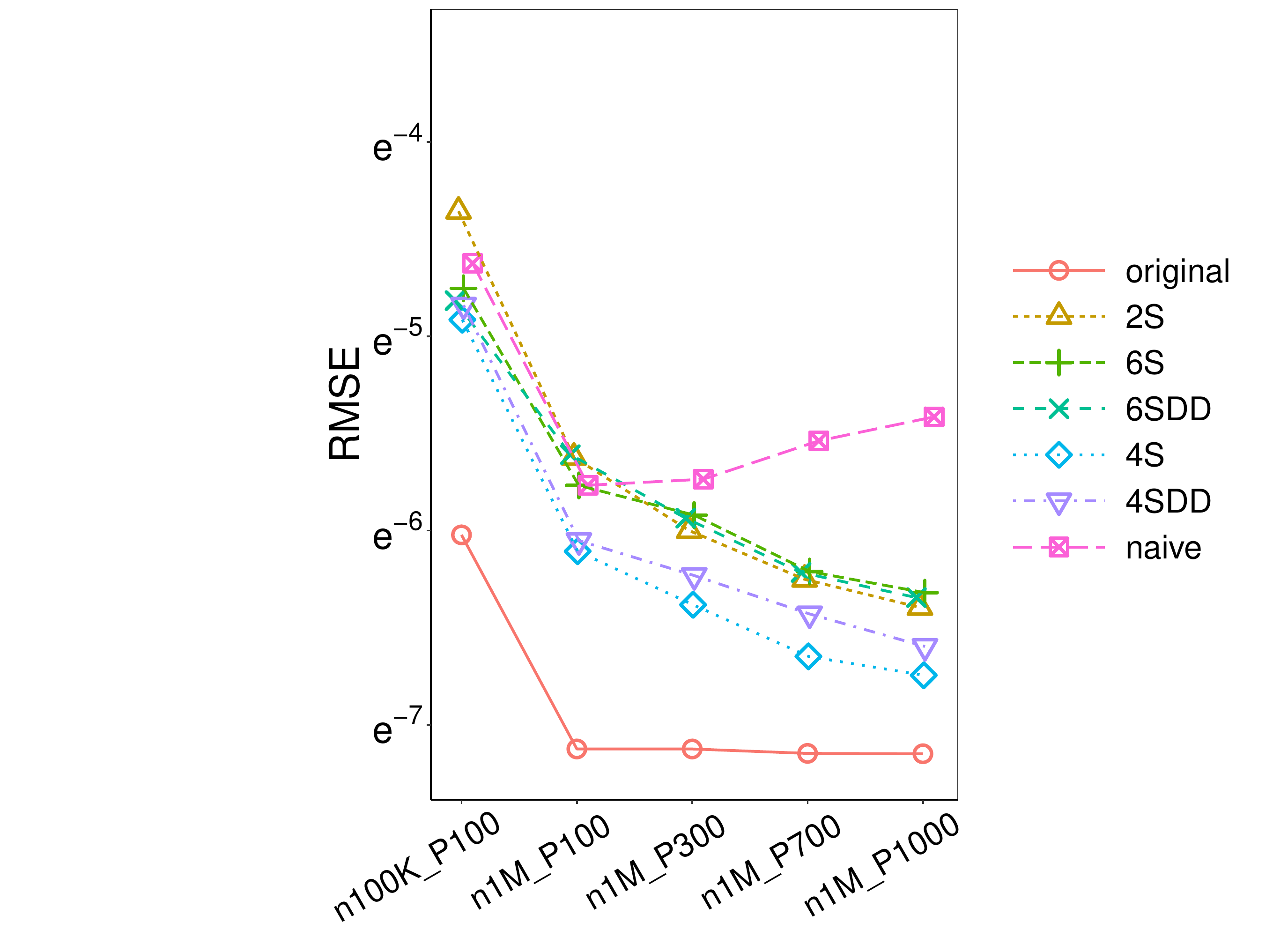}
\includegraphics[width=0.19\textwidth, trim={2.5in 0 2.6in 0},clip] {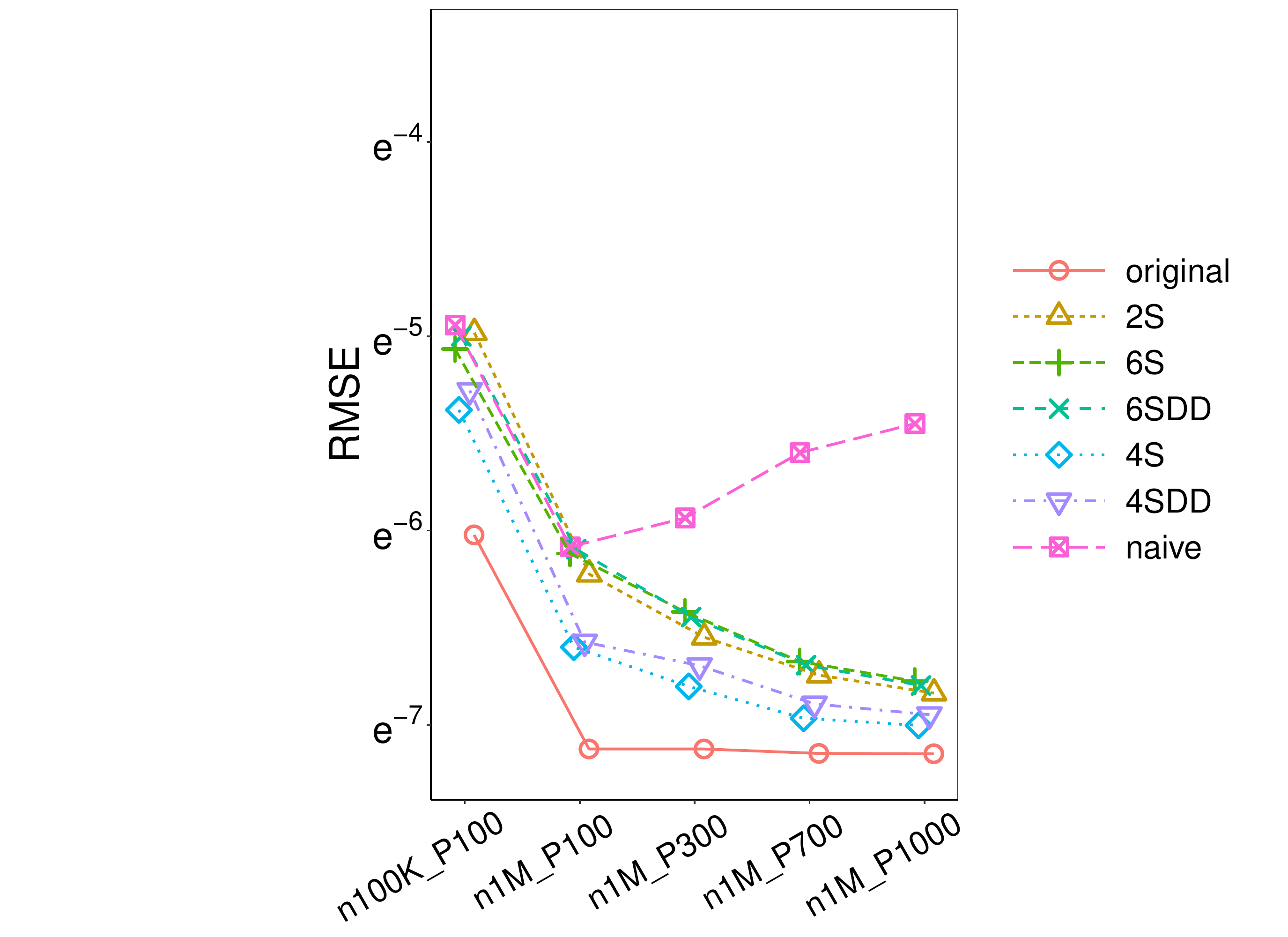}
\includegraphics[width=0.19\textwidth, trim={2.5in 0 2.6in 0},clip] {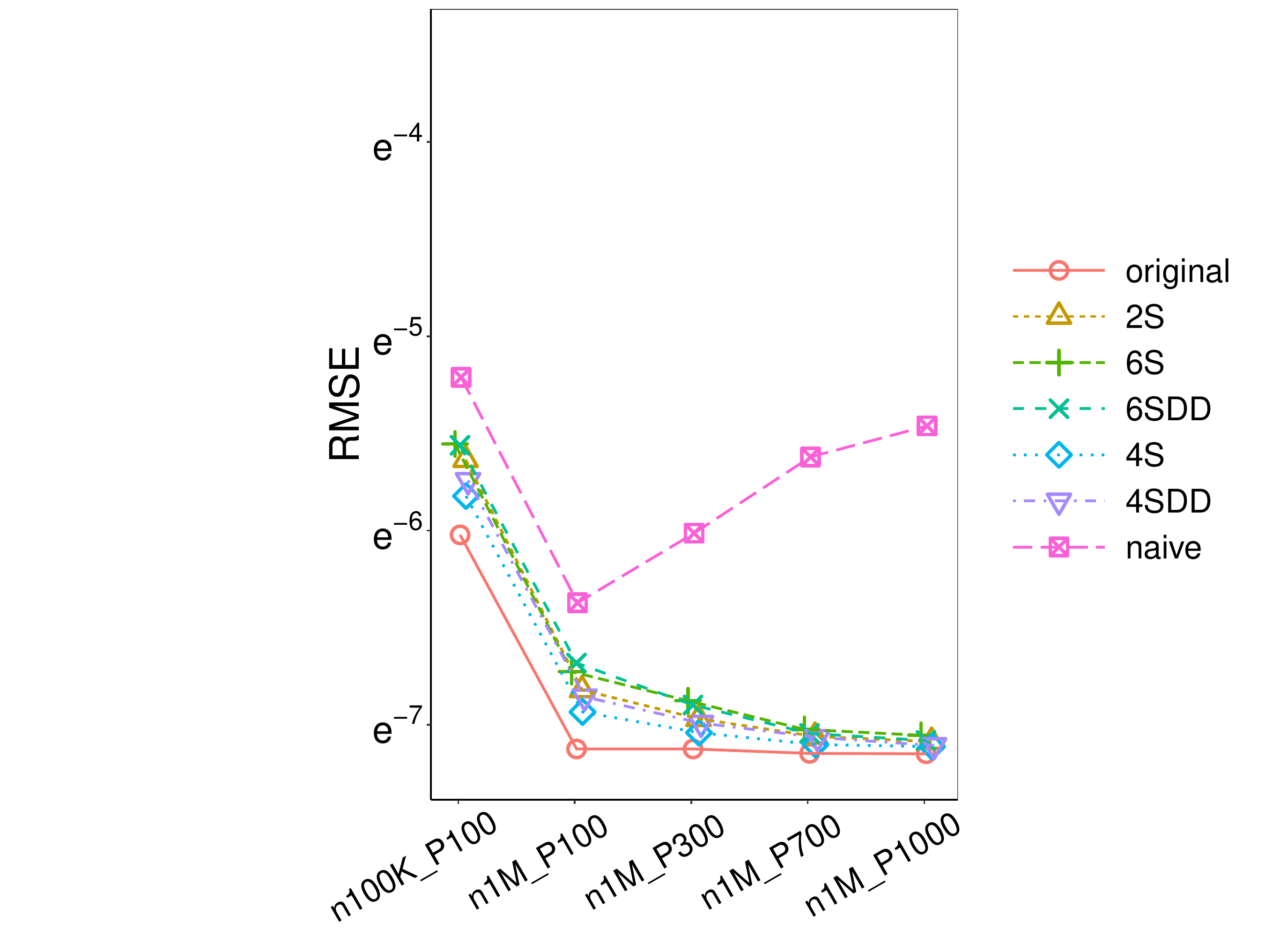}
\includegraphics[width=0.19\textwidth, trim={2.5in 0 2.6in 0},clip] {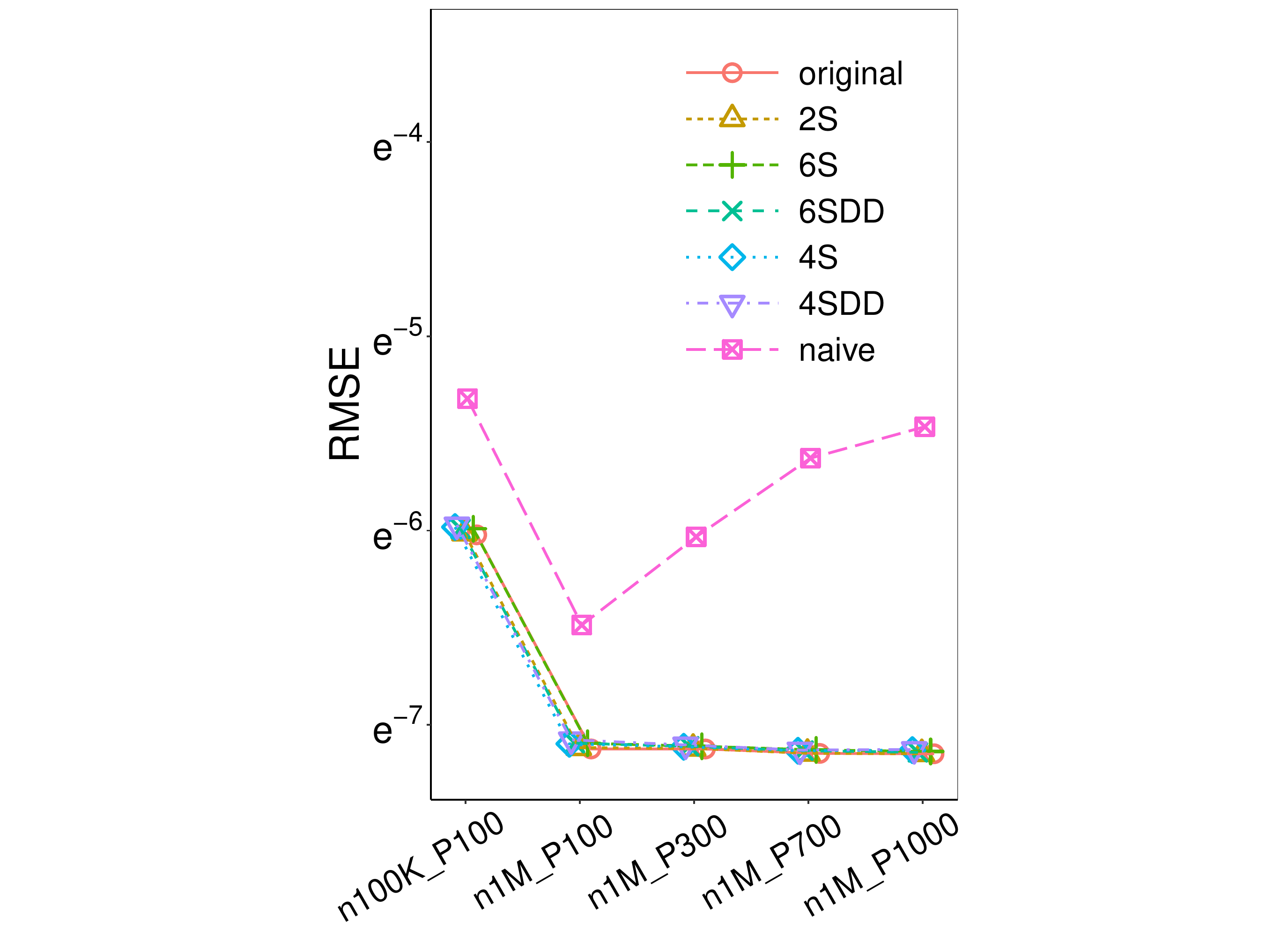}

\includegraphics[width=0.19\textwidth, trim={2.5in 0 2.6in 0},clip] {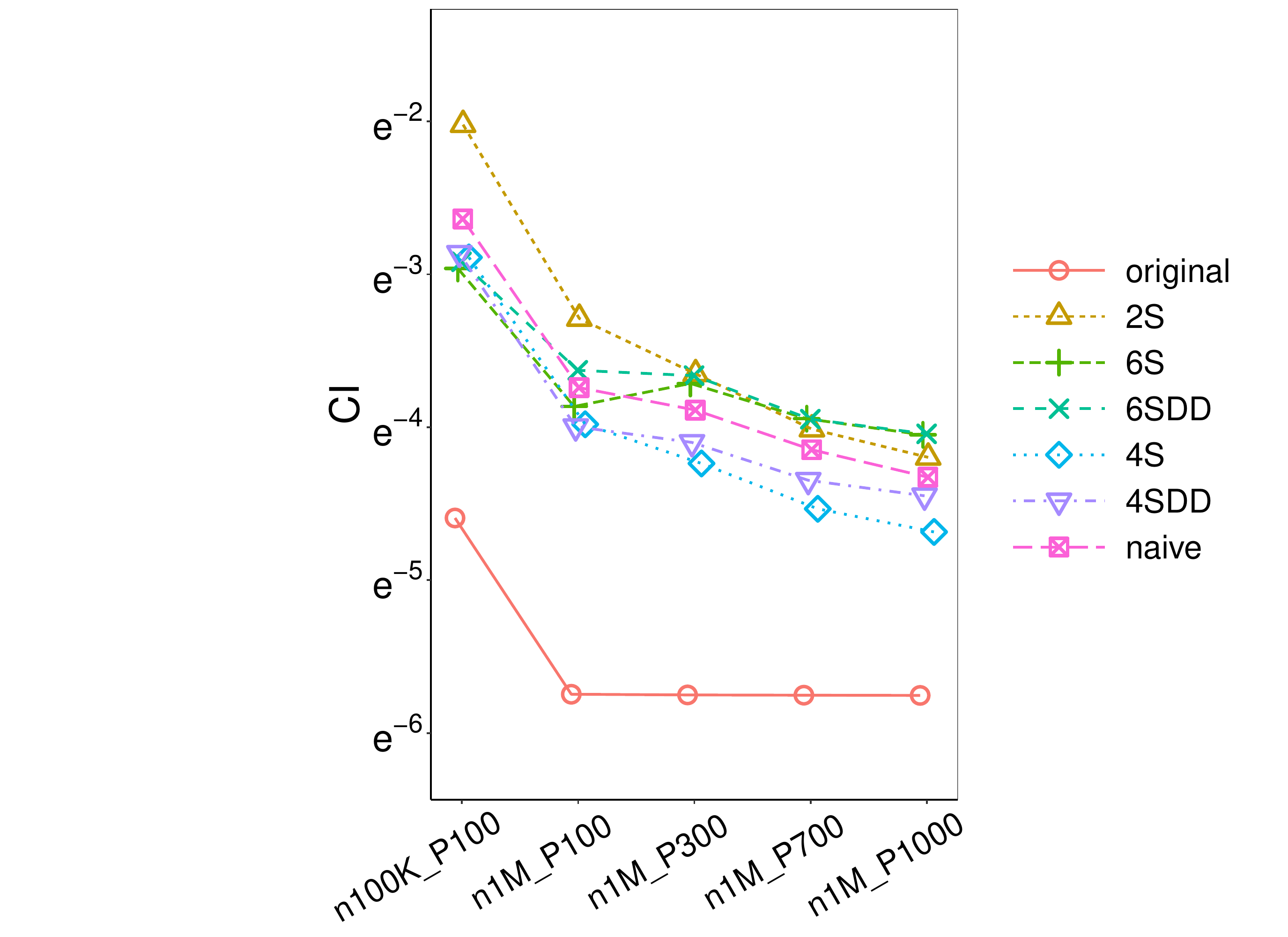}
\includegraphics[width=0.19\textwidth, trim={2.5in 0 2.6in 0},clip] {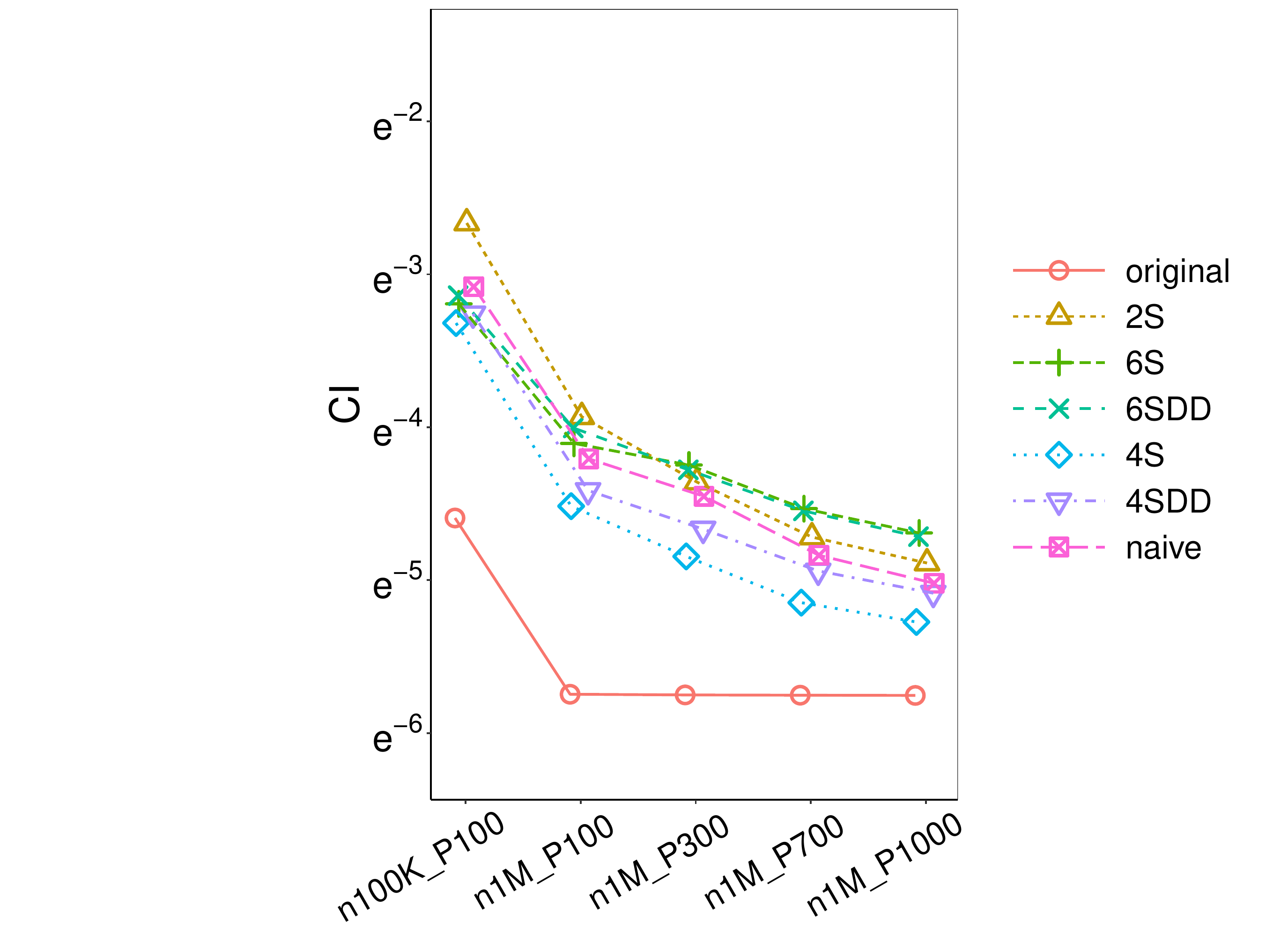}
\includegraphics[width=0.19\textwidth, trim={2.5in 0 2.6in 0},clip] {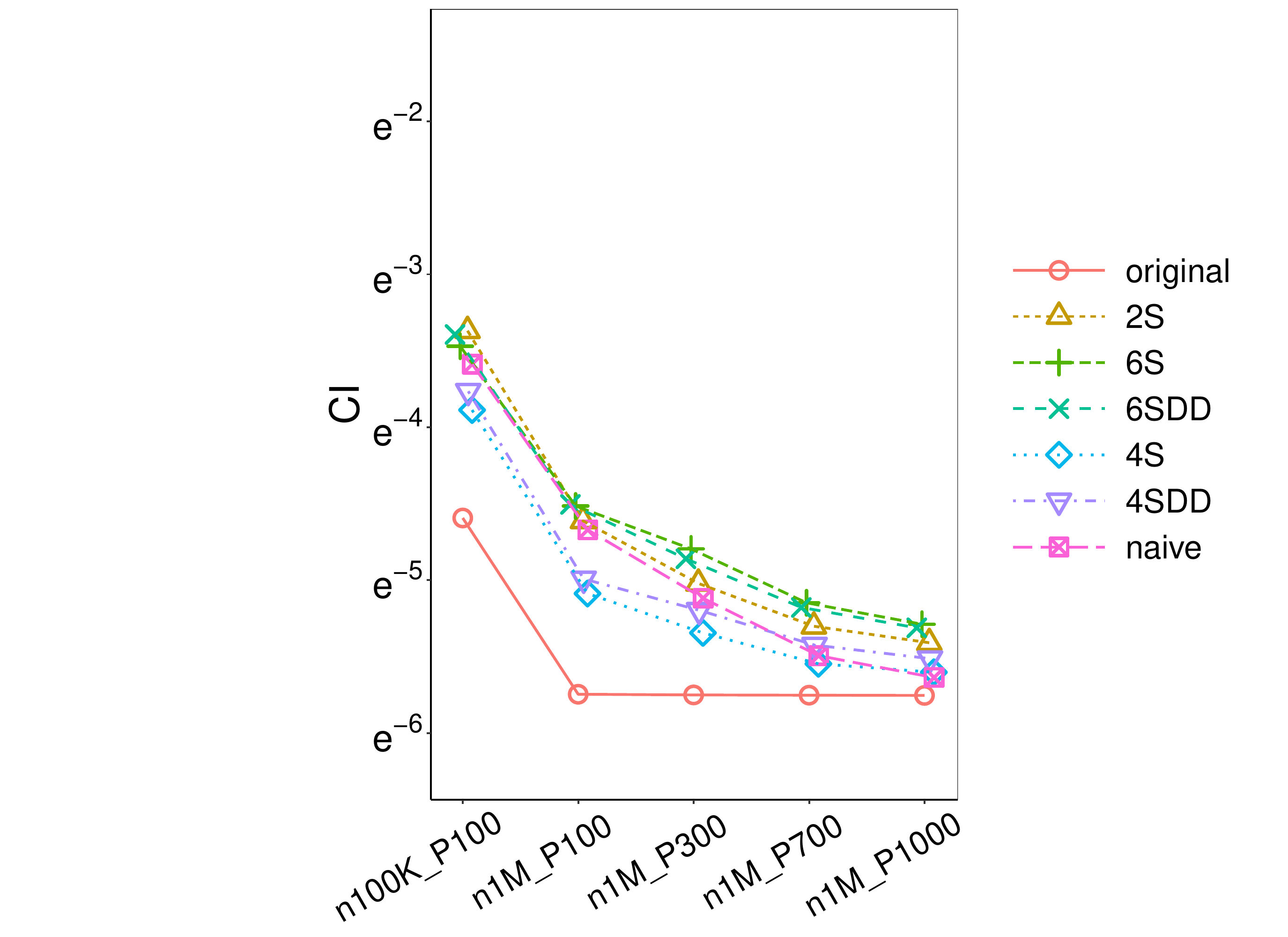}
\includegraphics[width=0.19\textwidth, trim={2.5in 0 2.6in 0},clip] {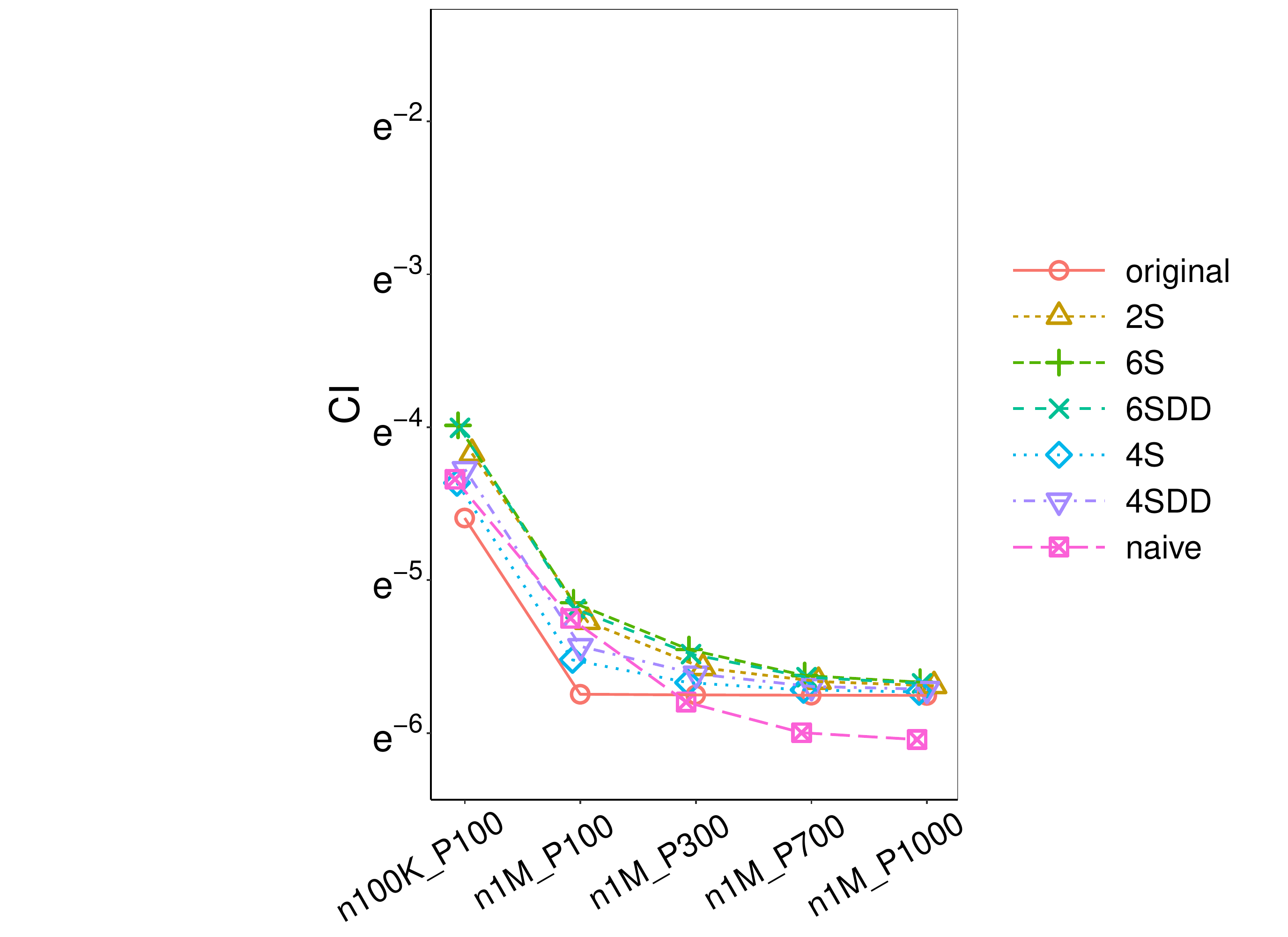}
\includegraphics[width=0.19\textwidth, trim={2.5in 0 2.6in 0},clip] {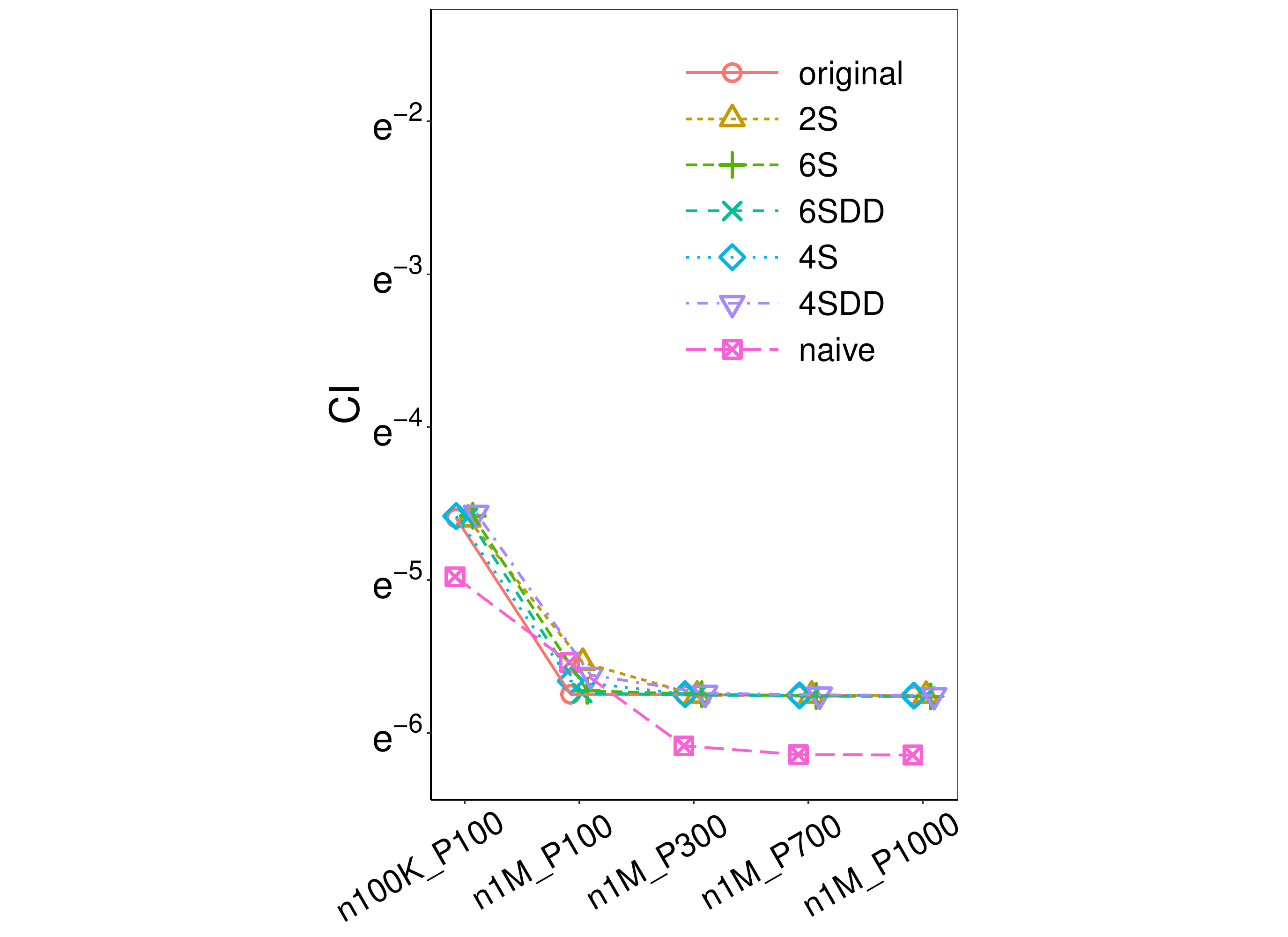}

\includegraphics[width=0.19\textwidth, trim={2.5in 0 2.6in 0},clip] {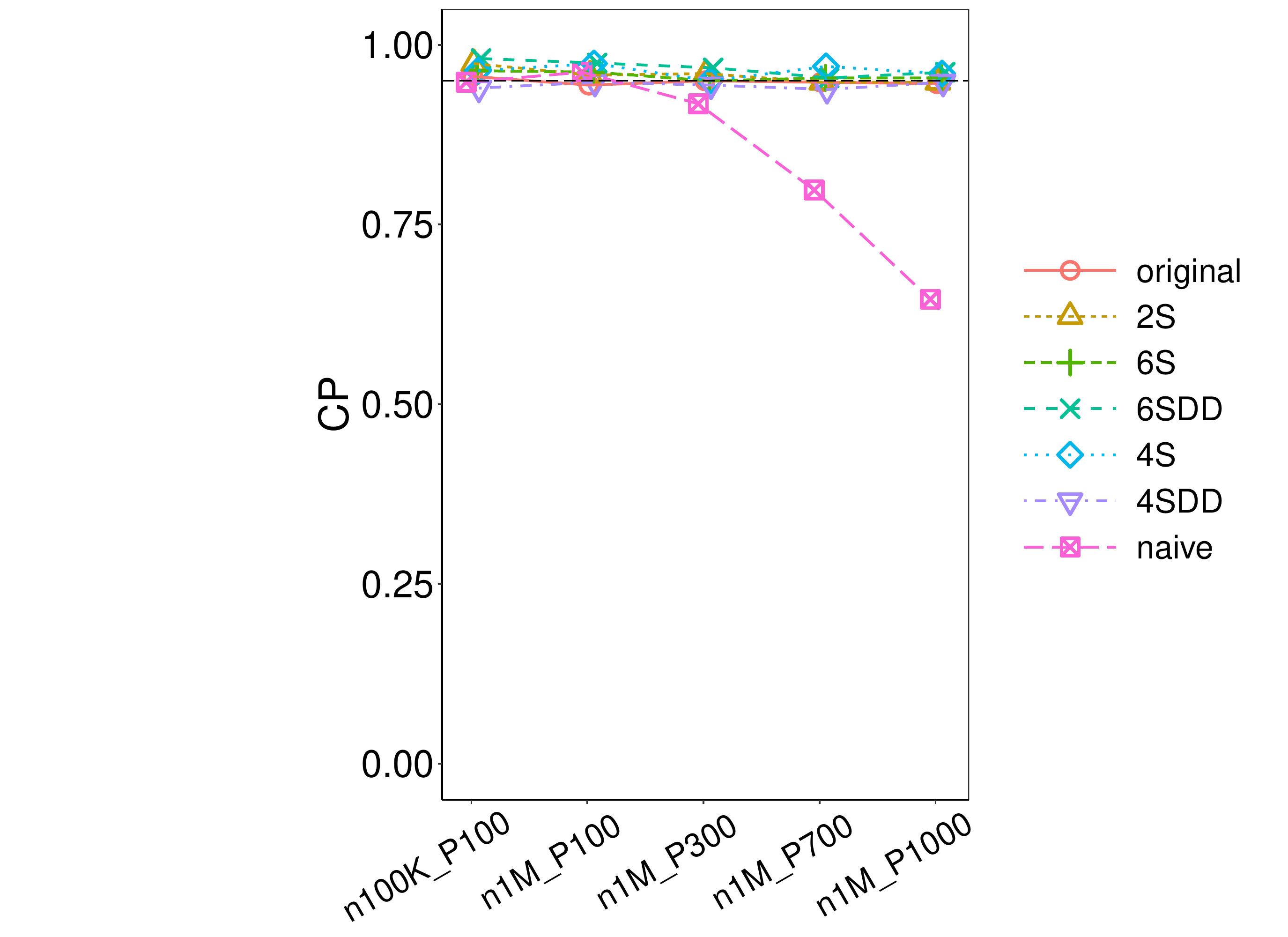}
\includegraphics[width=0.19\textwidth, trim={2.5in 0 2.6in 0},clip] {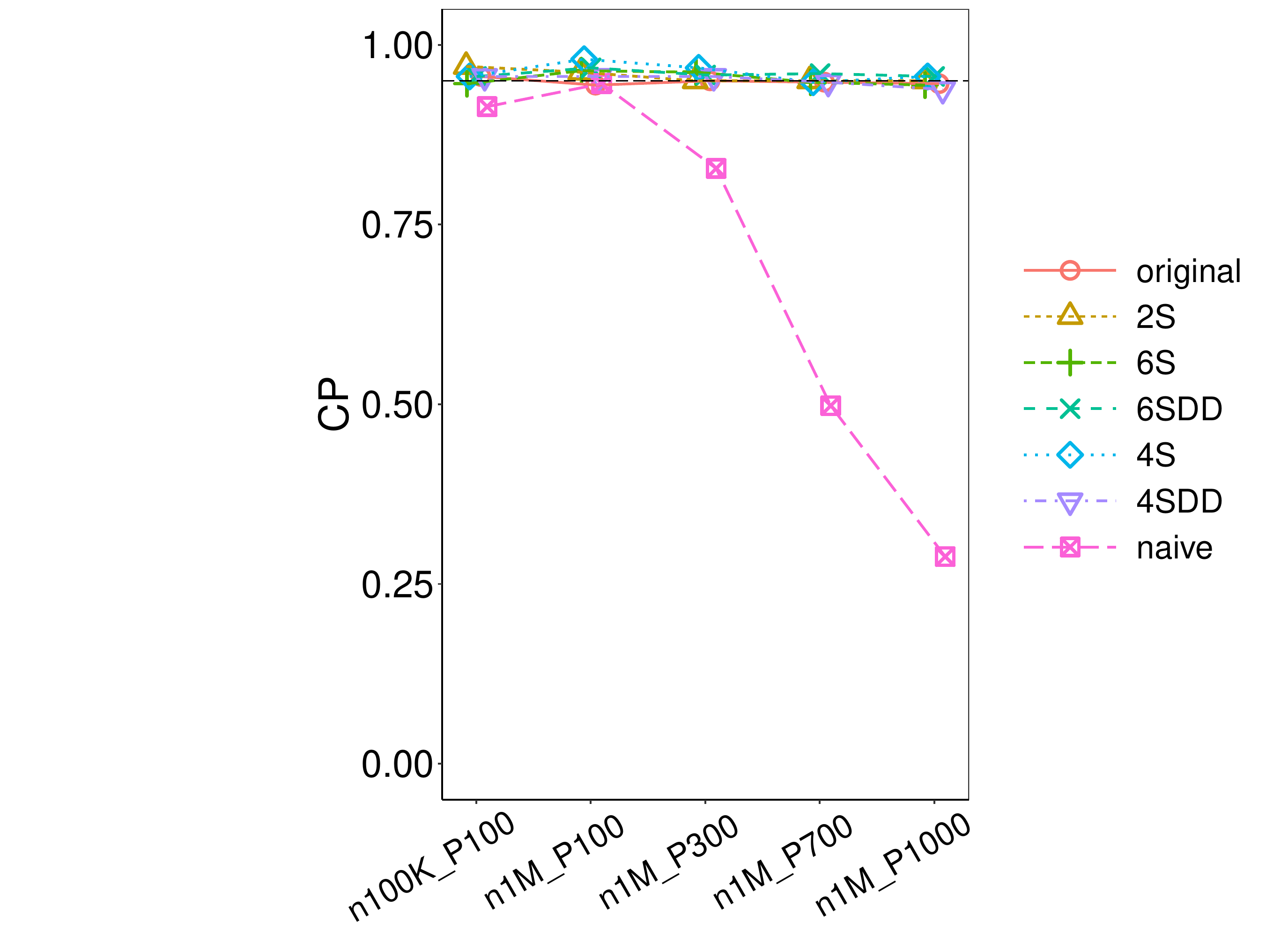}
\includegraphics[width=0.19\textwidth, trim={2.5in 0 2.6in 0},clip] {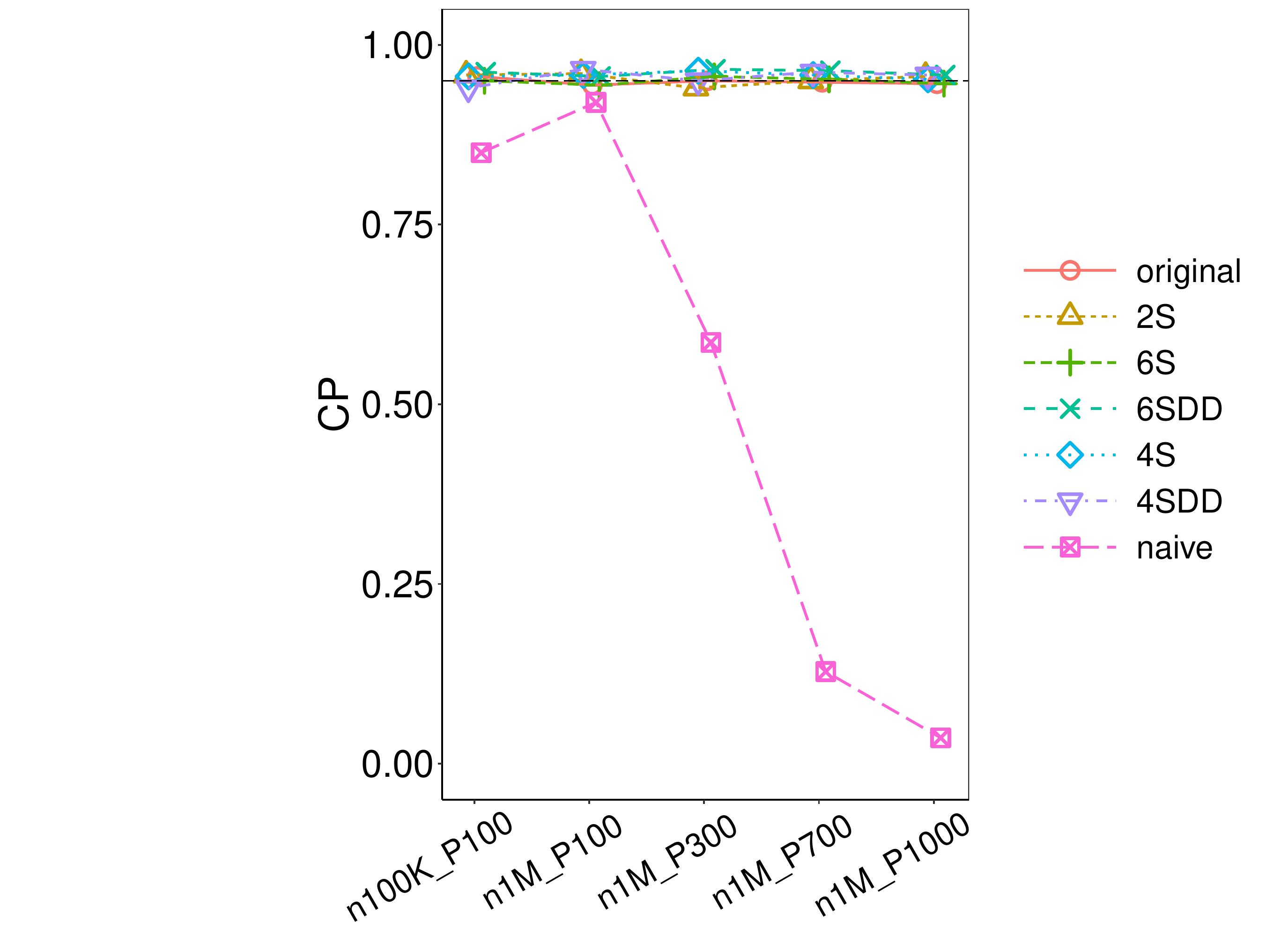}
\includegraphics[width=0.19\textwidth, trim={2.5in 0 2.6in 0},clip] {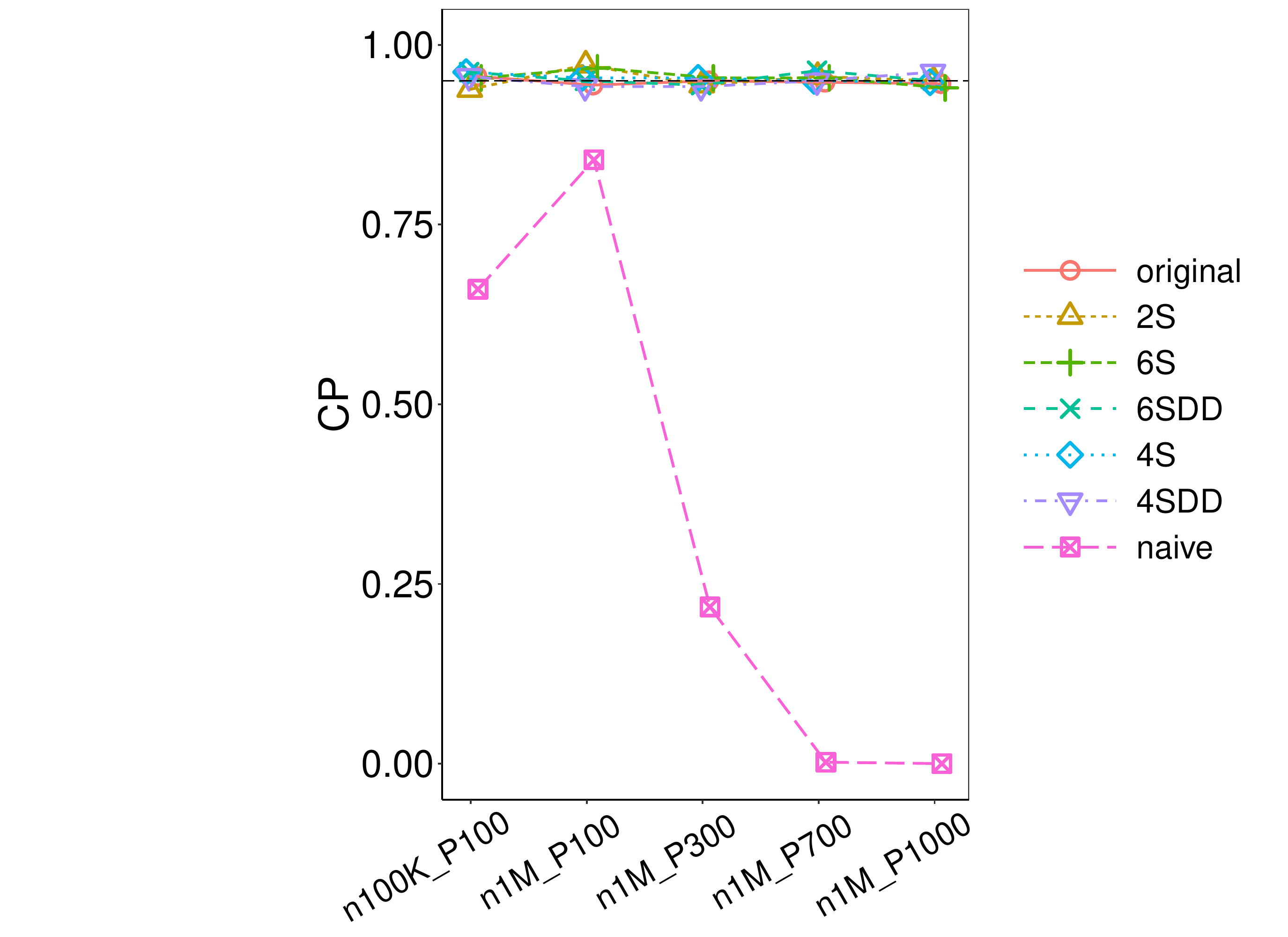}
\includegraphics[width=0.19\textwidth, trim={2.5in 0 2.6in 0},clip] {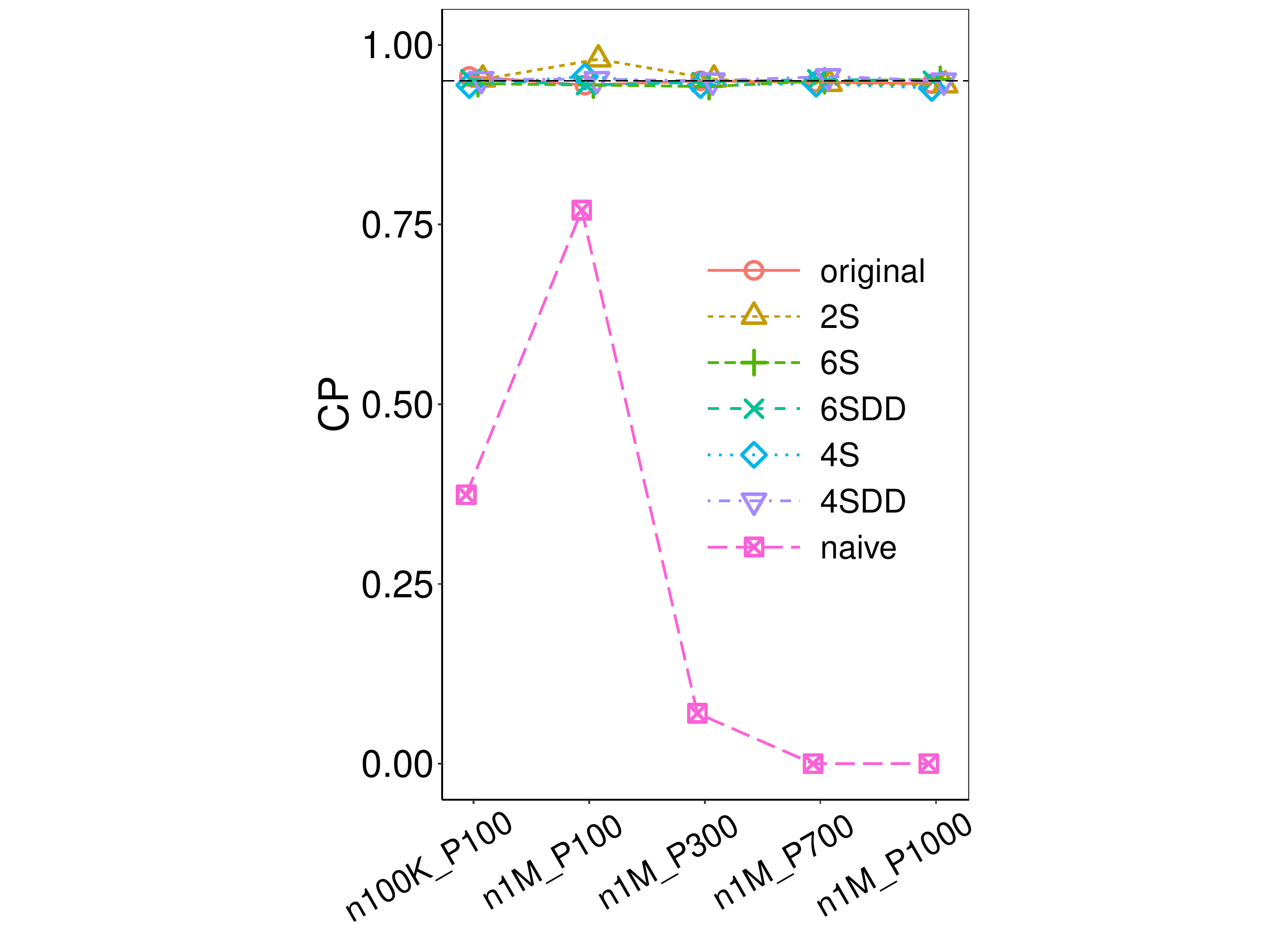}

\caption{Simulation results with $\epsilon$-DP for ZINB data with  $\alpha\ne\beta$ when $\theta=0$} \label{fig:0asDPZINB}
\end{figure}

\begin{figure}[!htb]
\hspace{0.45in}$\rho=0.005$\hspace{0.65in}$\rho=0.02$\hspace{0.65in}$\rho=0.08$
\hspace{0.65in}$\rho=0.32$\hspace{0.65in}$\rho=1.28$

\includegraphics[width=0.19\textwidth, trim={2.45in 0 2.45in 0},clip] {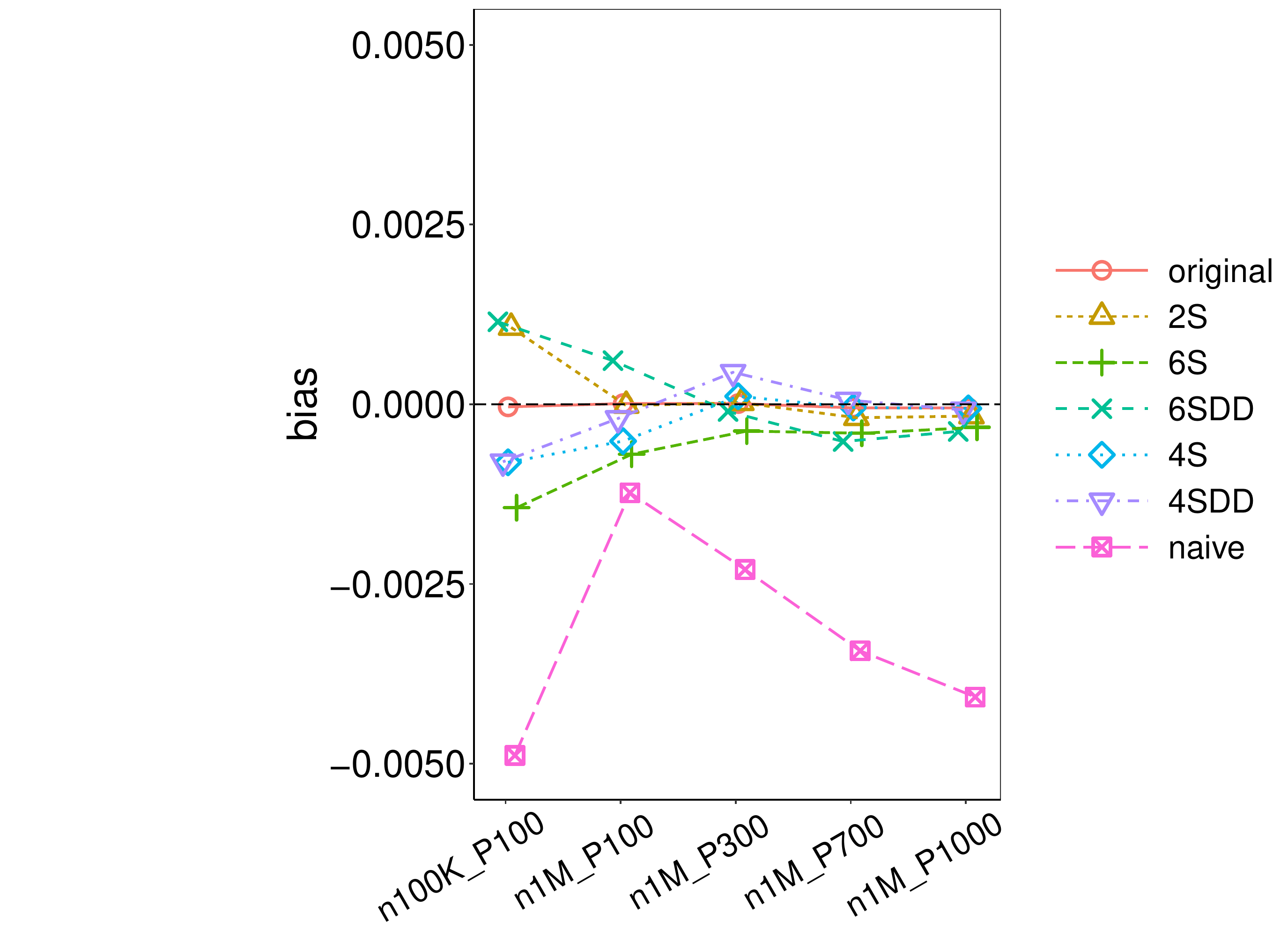}
\includegraphics[width=0.19\textwidth, trim={2.45in 0 2.45in 0},clip] {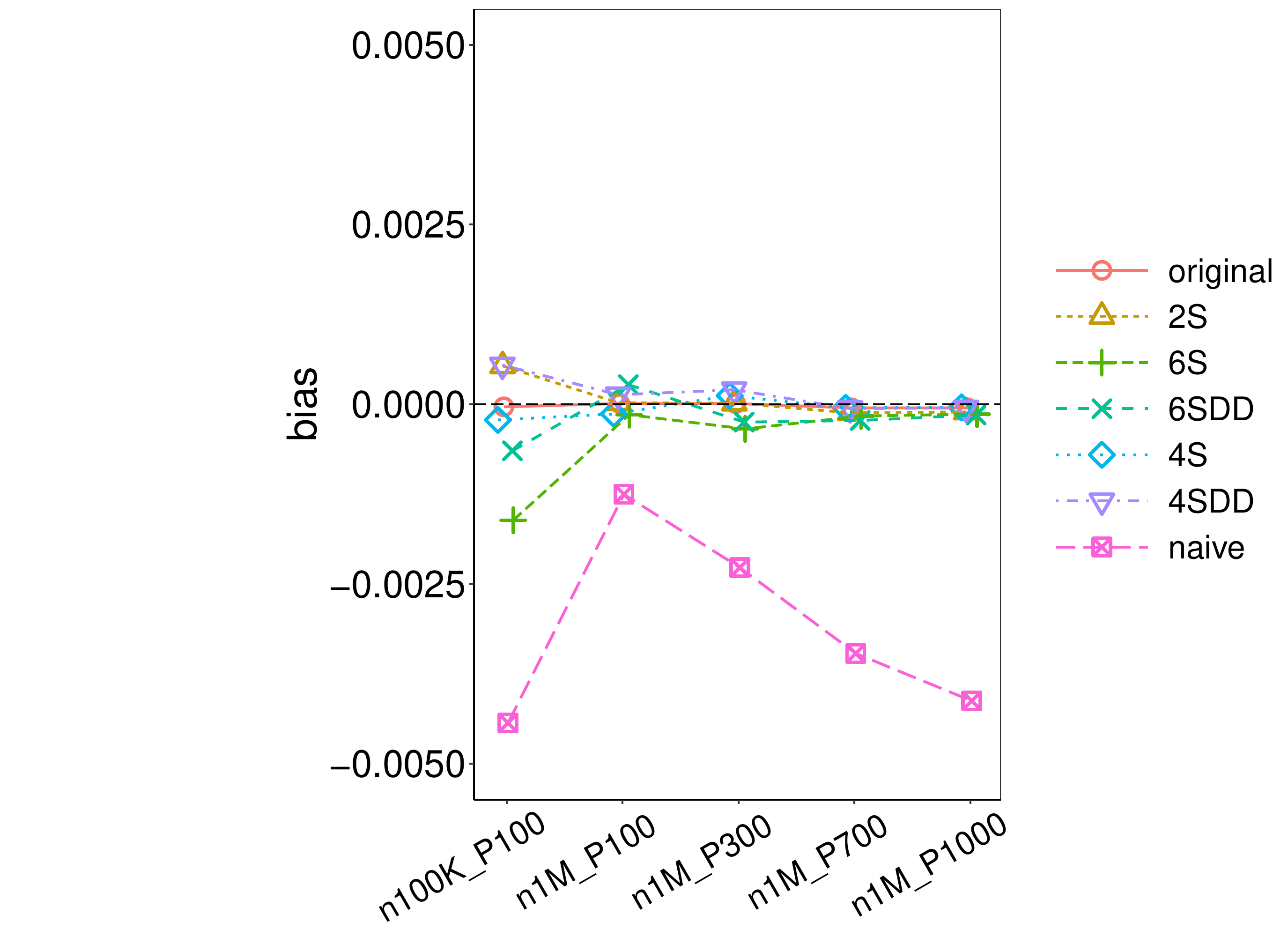}
\includegraphics[width=0.19\textwidth, trim={2.45in 0 2.45in 0},clip] {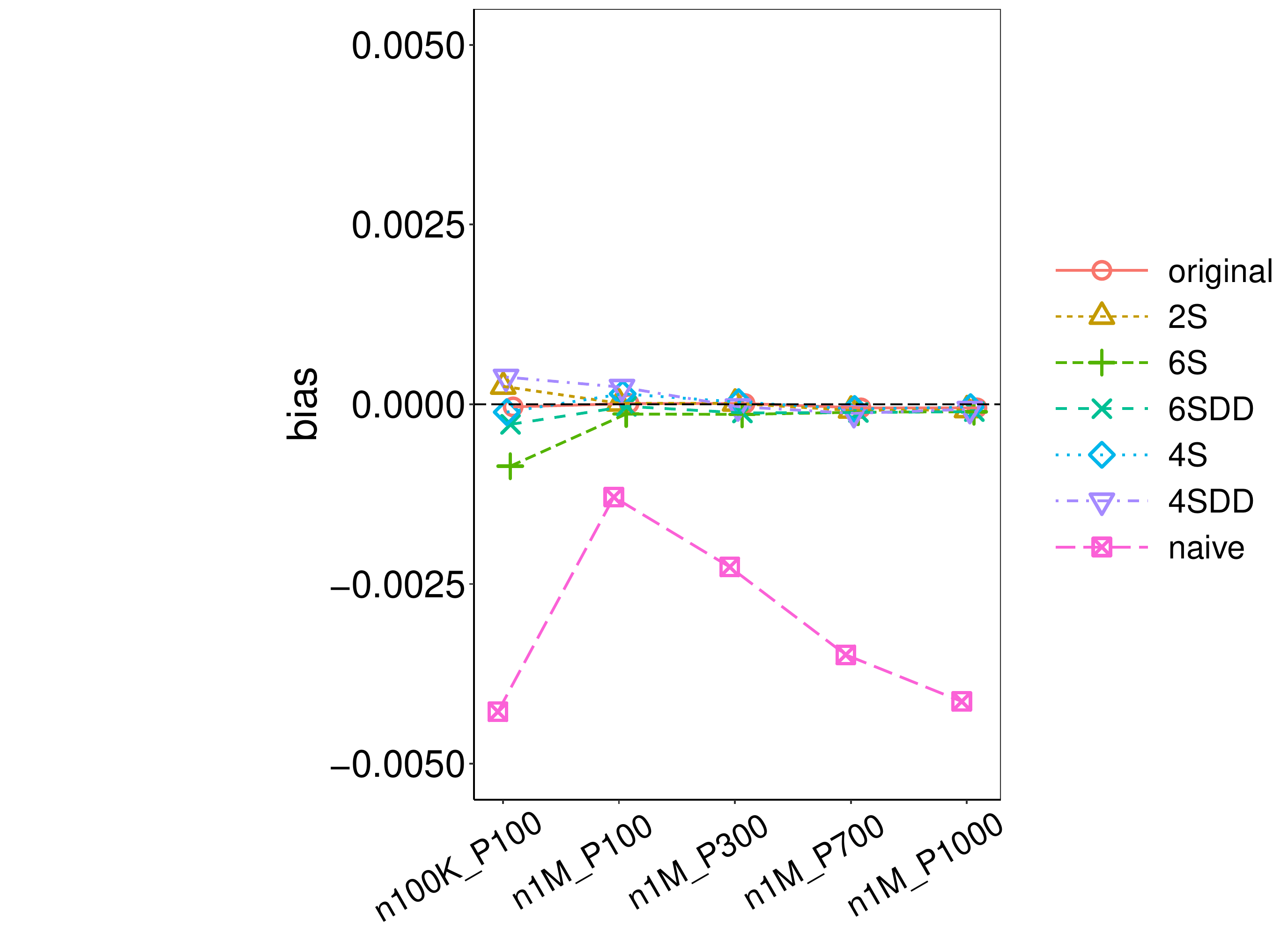}
\includegraphics[width=0.19\textwidth, trim={2.45in 0 2.45in 0},clip] {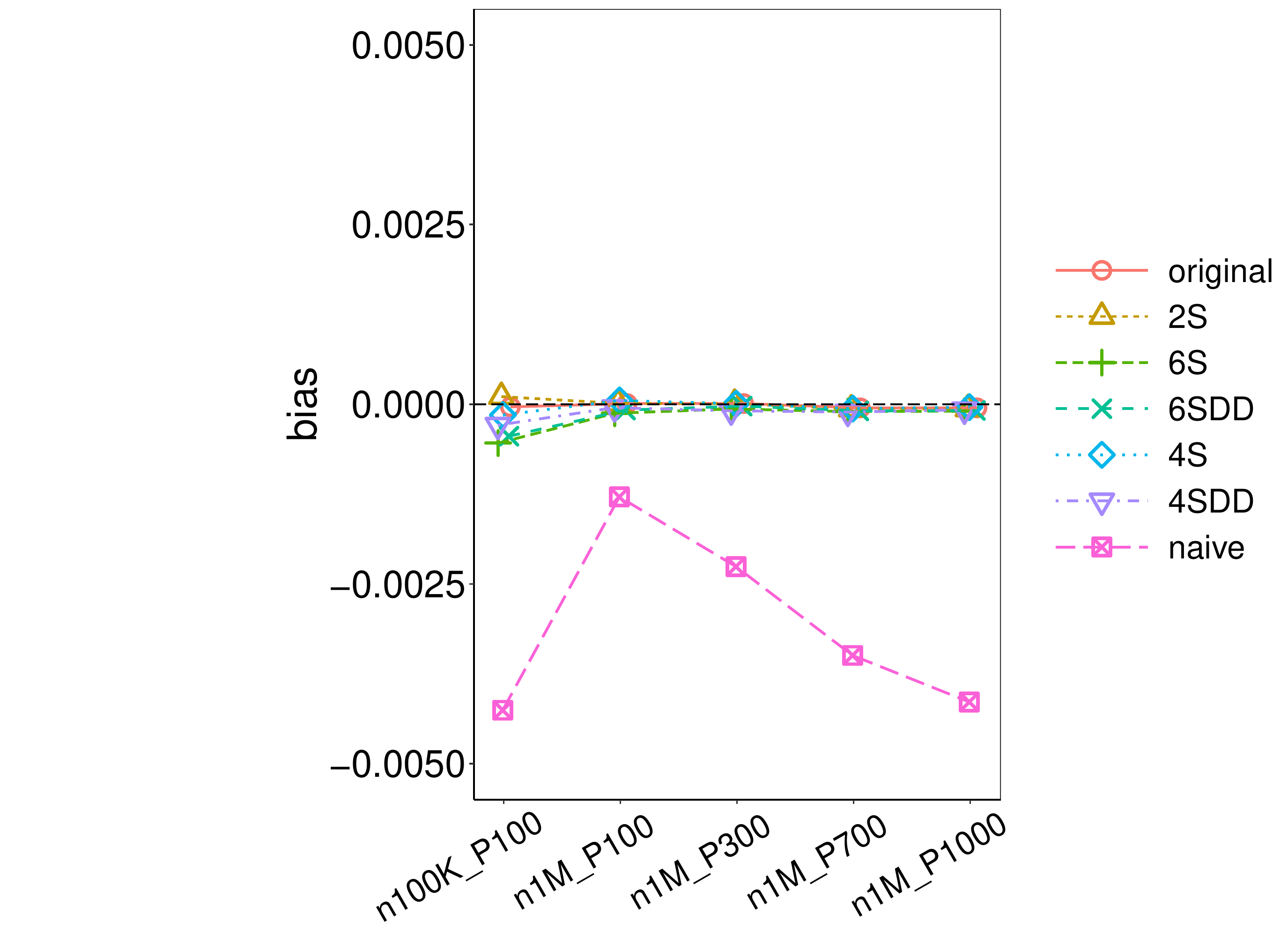}
\includegraphics[width=0.19\textwidth, trim={2.45in 0 2.45in 0},clip] {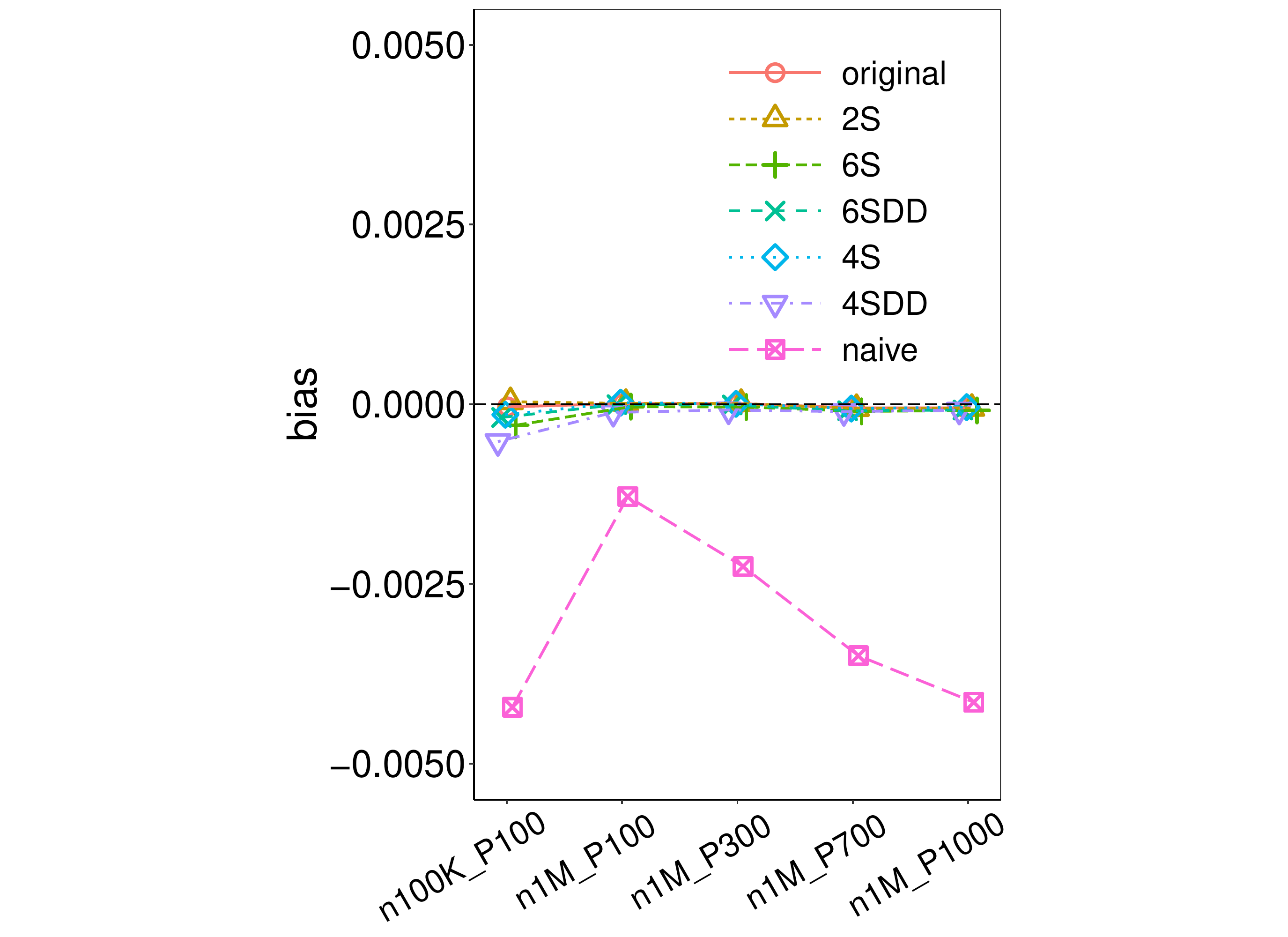}

\includegraphics[width=0.19\textwidth, trim={2.5in 0 2.6in 0},clip] {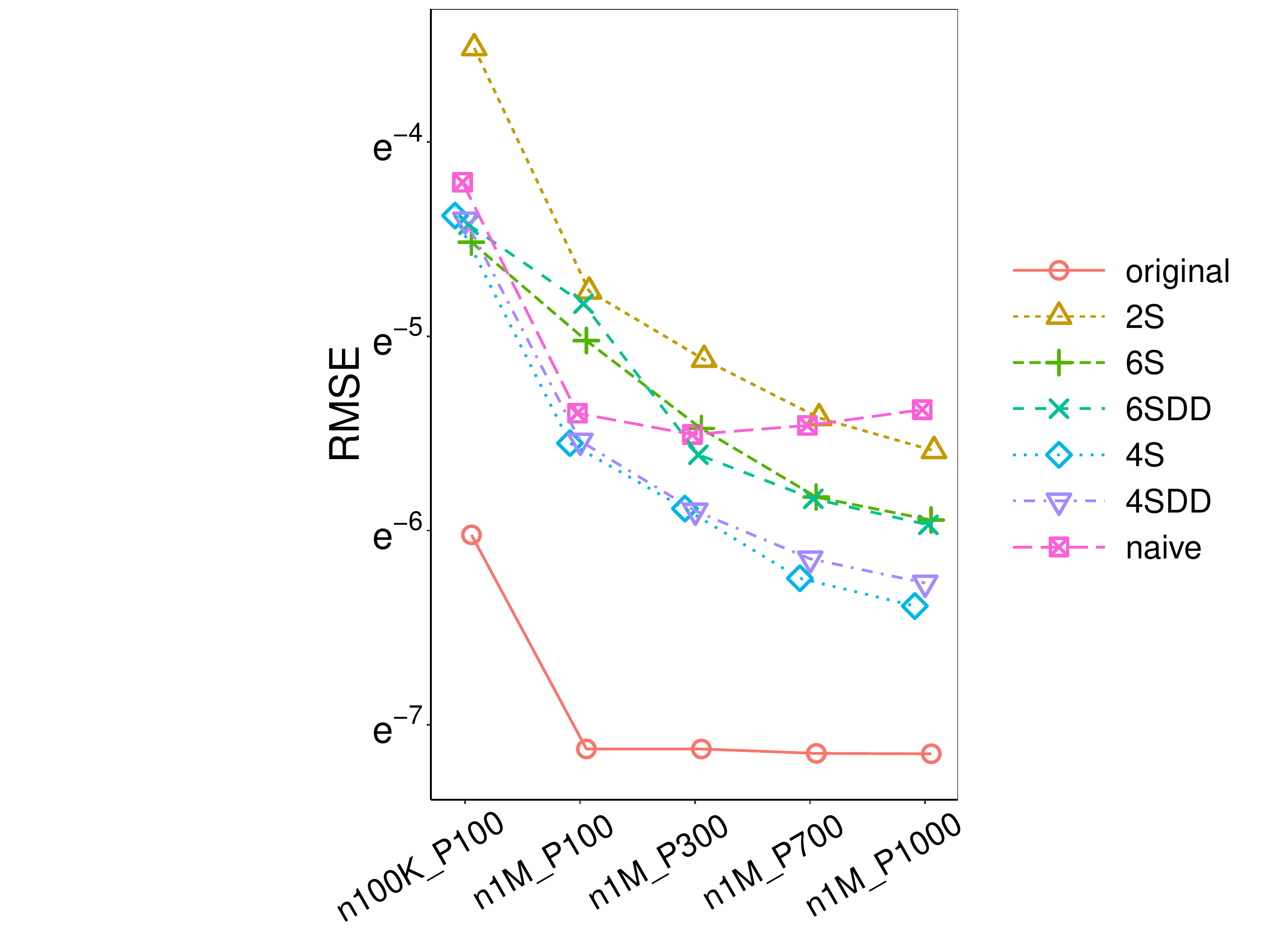}
\includegraphics[width=0.19\textwidth, trim={2.5in 0 2.6in 0},clip] {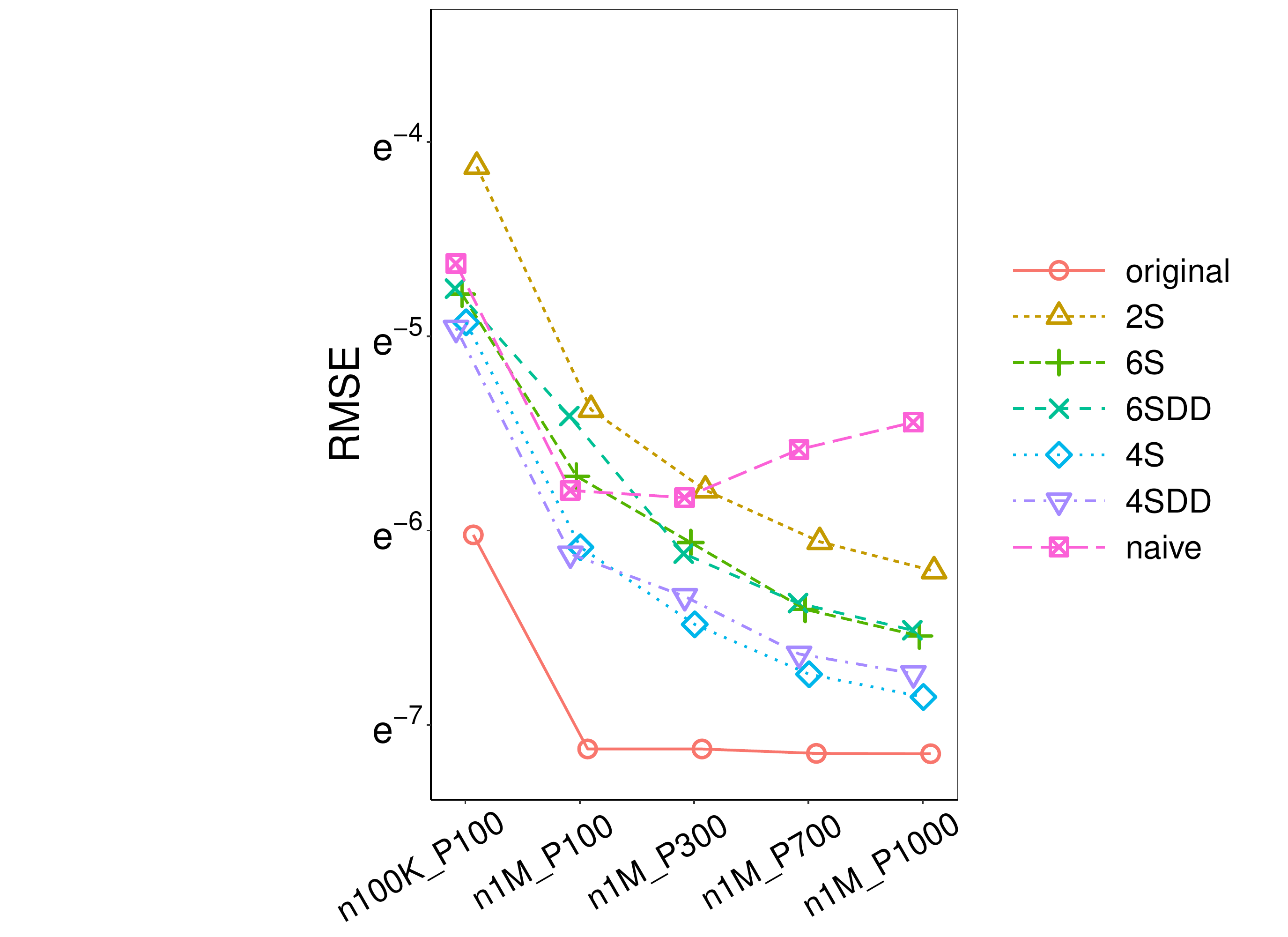}
\includegraphics[width=0.19\textwidth, trim={2.5in 0 2.6in 0},clip] {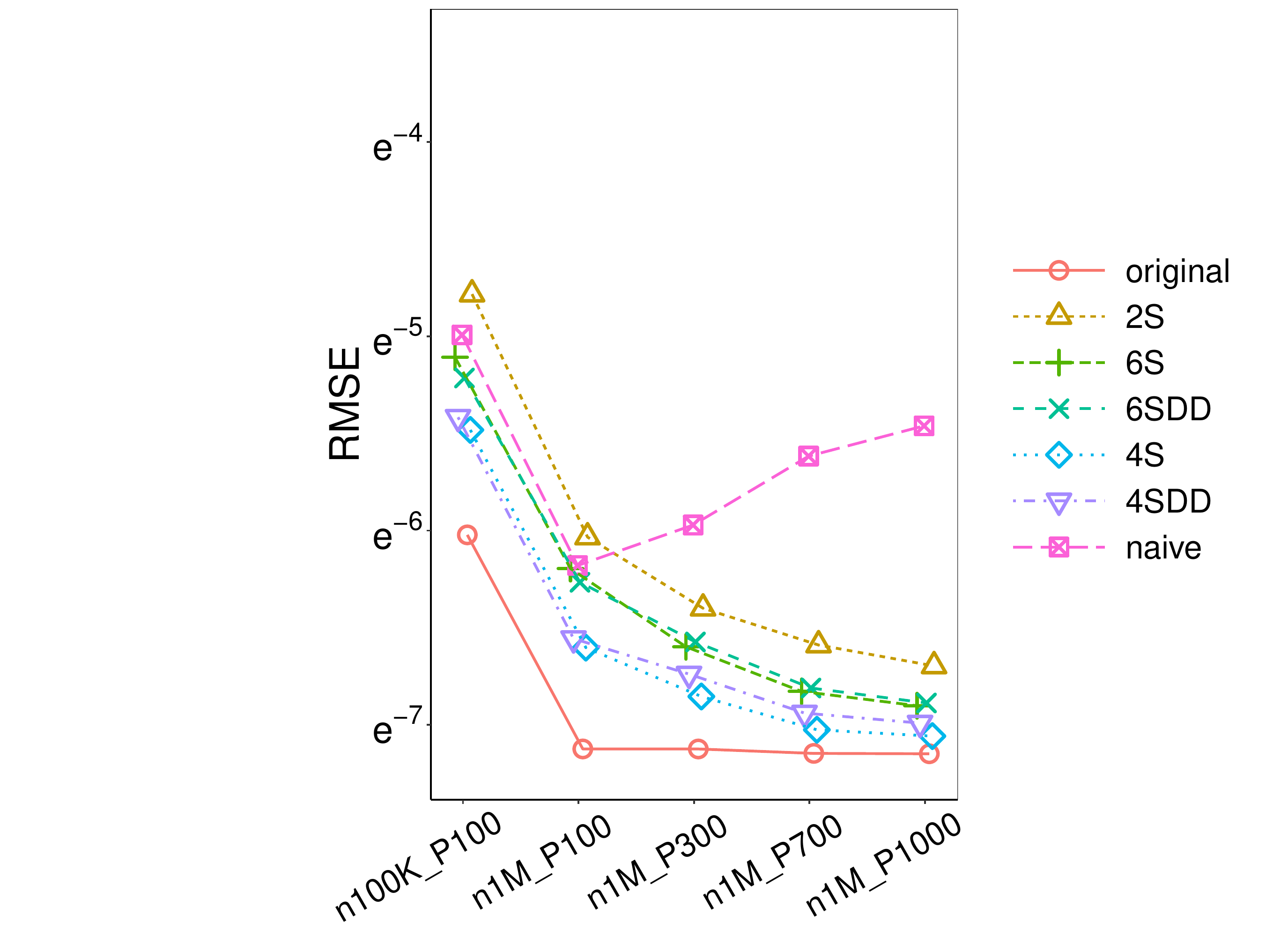}
\includegraphics[width=0.19\textwidth, trim={2.5in 0 2.6in 0},clip] {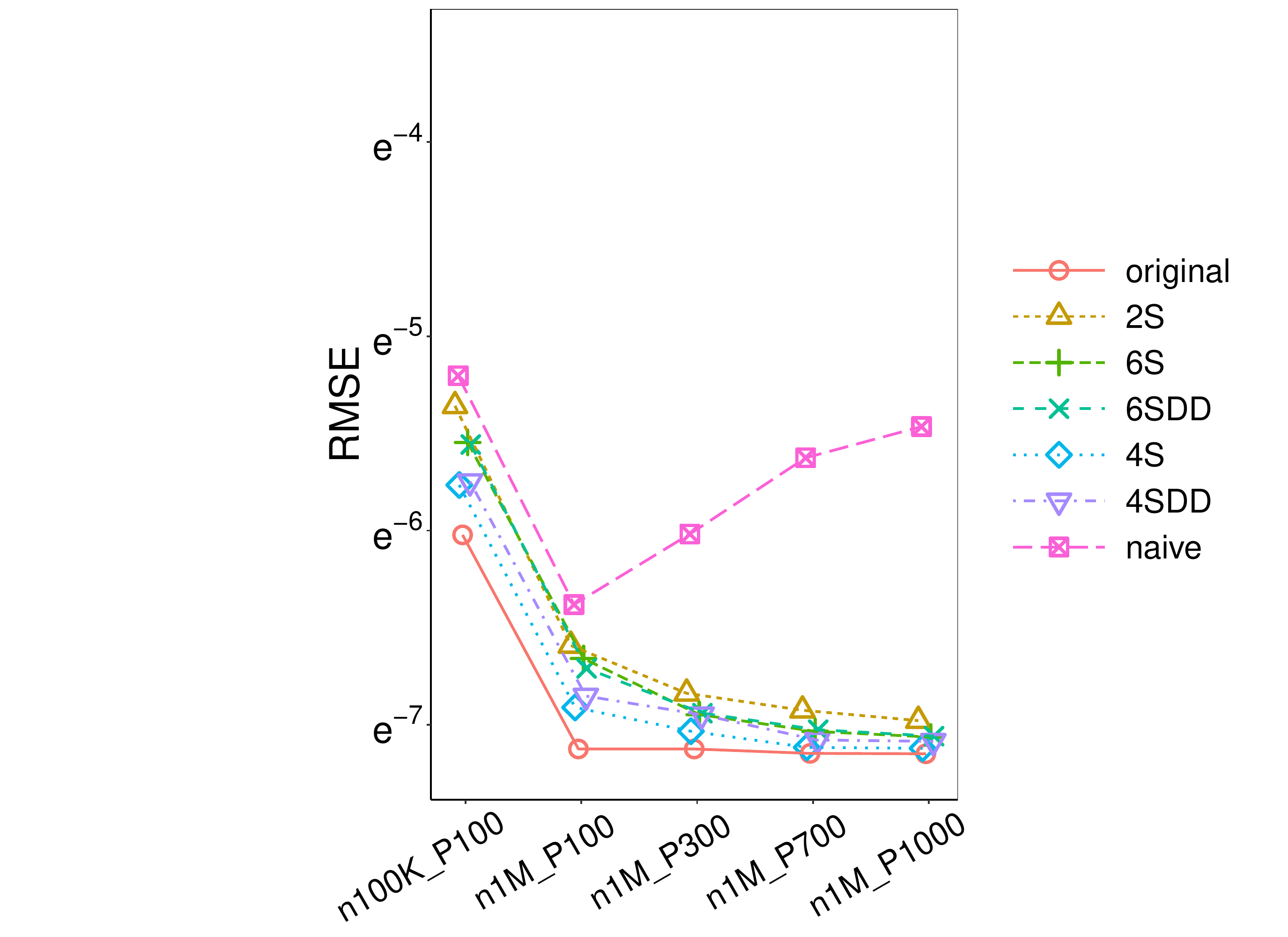}
\includegraphics[width=0.19\textwidth, trim={2.5in 0 2.6in 0},clip] {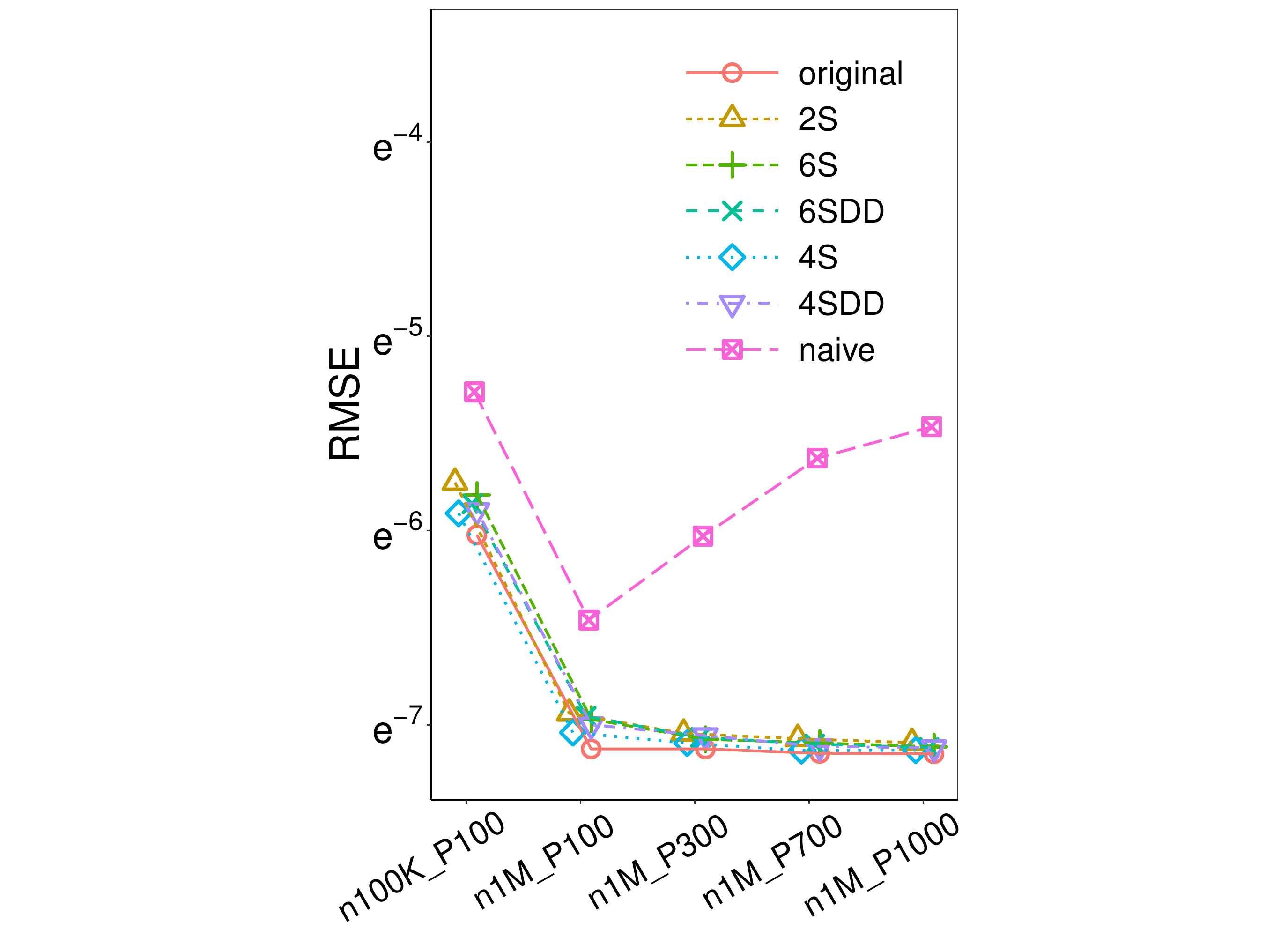}

\includegraphics[width=0.19\textwidth, trim={2.5in 0 2.6in 0},clip] {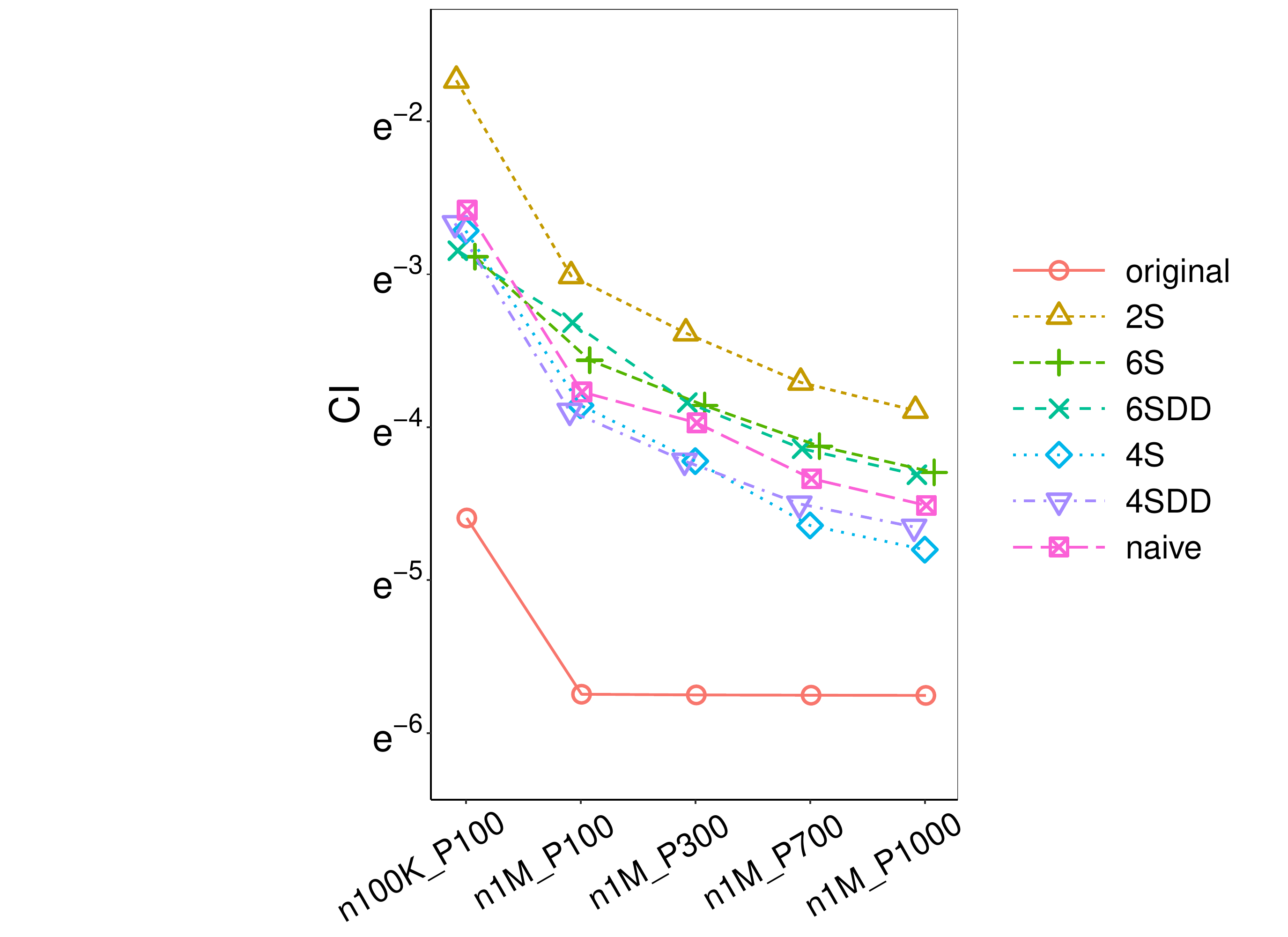}
\includegraphics[width=0.19\textwidth, trim={2.5in 0 2.6in 0},clip] {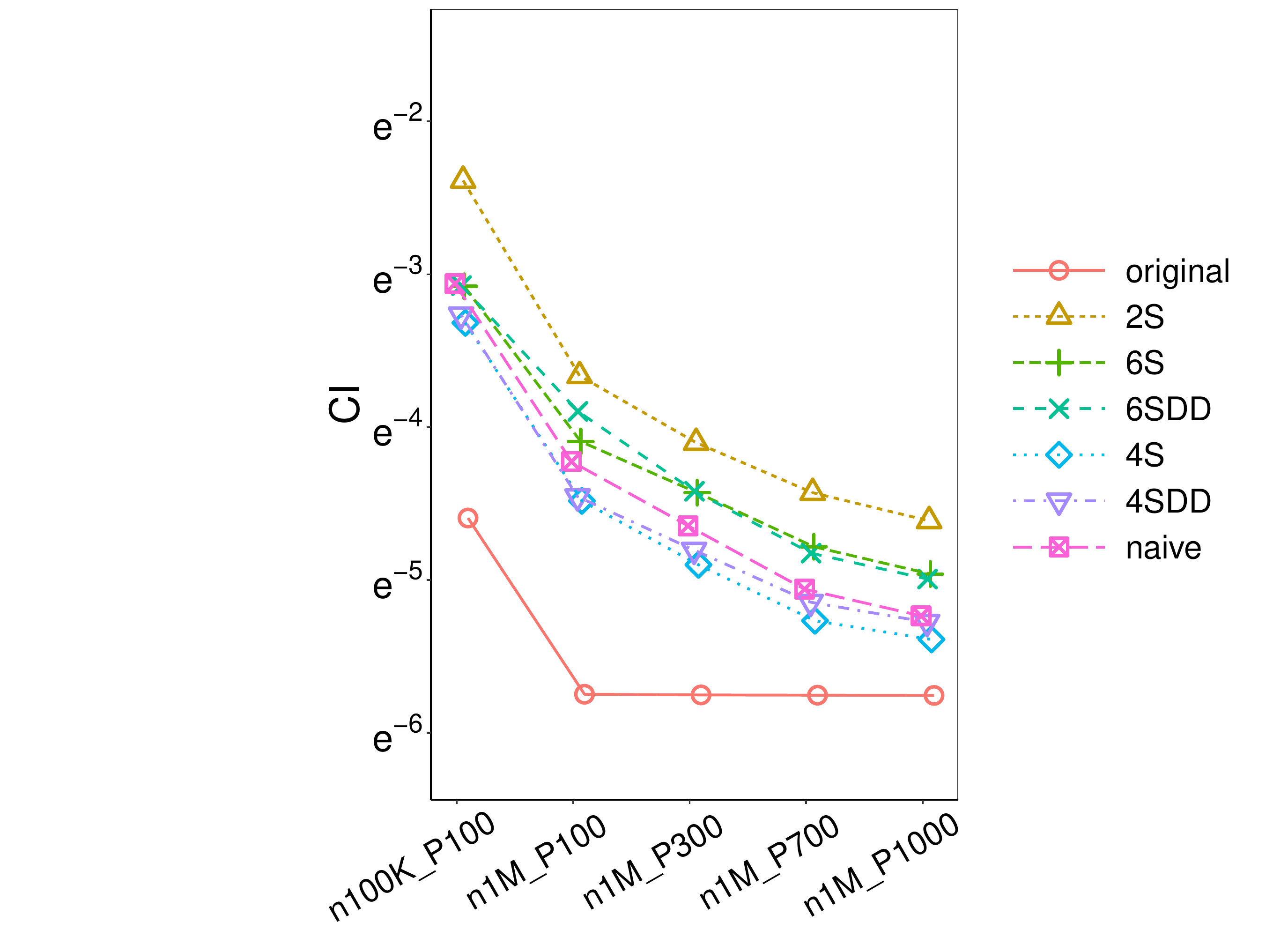}
\includegraphics[width=0.19\textwidth, trim={2.5in 0 2.6in 0},clip] {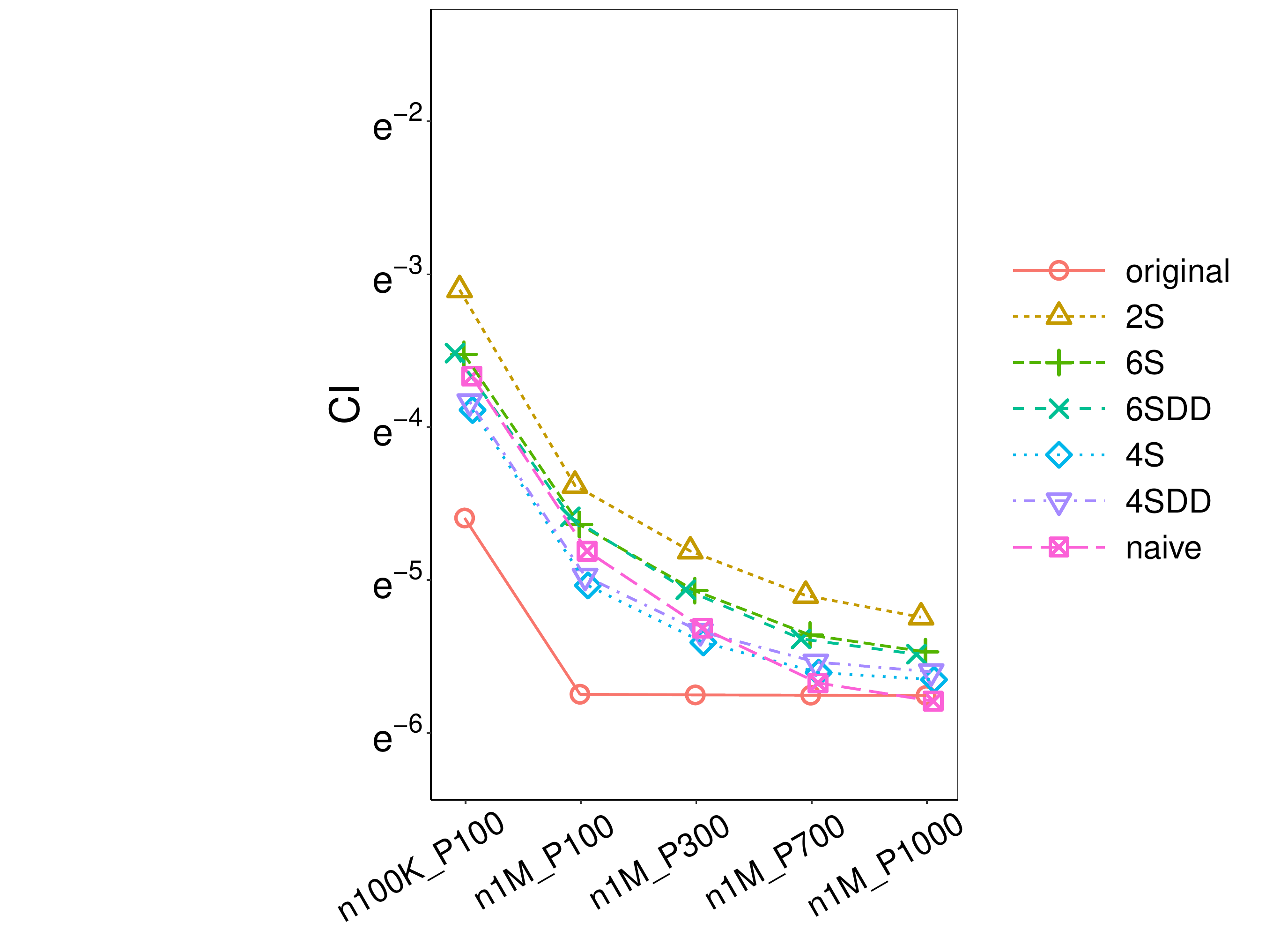}
\includegraphics[width=0.19\textwidth, trim={2.5in 0 2.6in 0},clip] {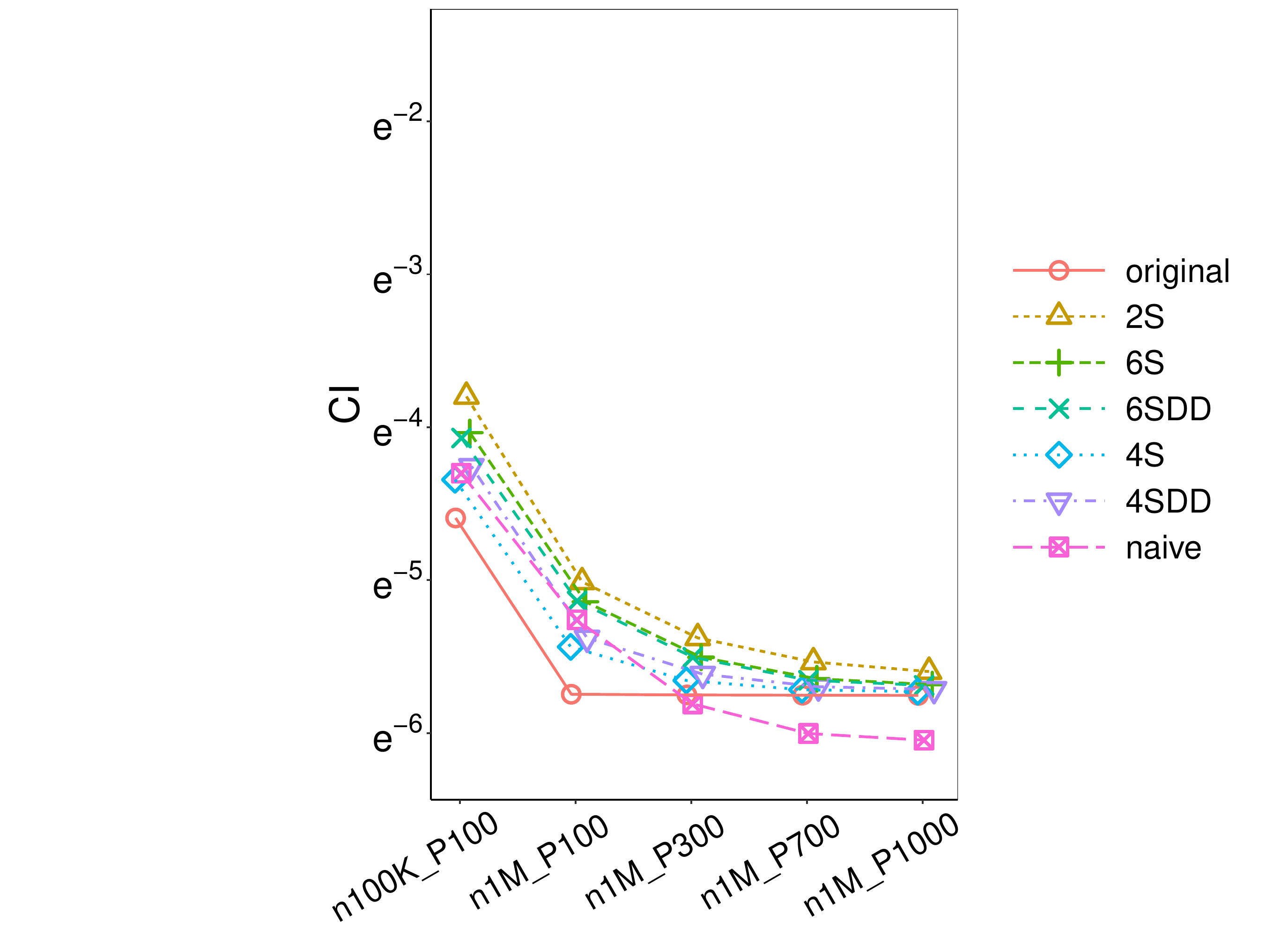}
\includegraphics[width=0.19\textwidth, trim={2.5in 0 2.6in 0},clip] {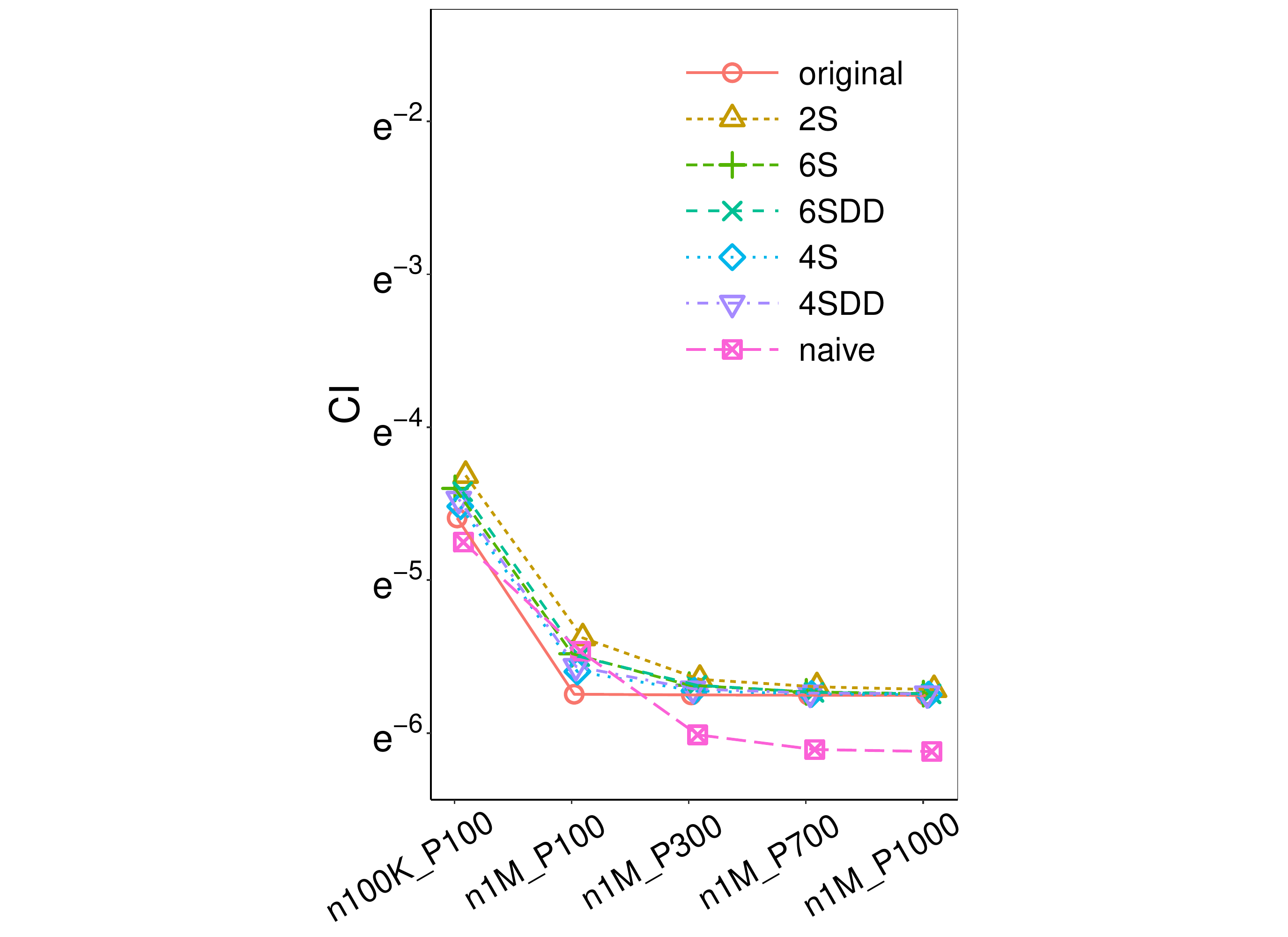}

\includegraphics[width=0.19\textwidth, trim={2.5in 0 2.6in 0},clip] {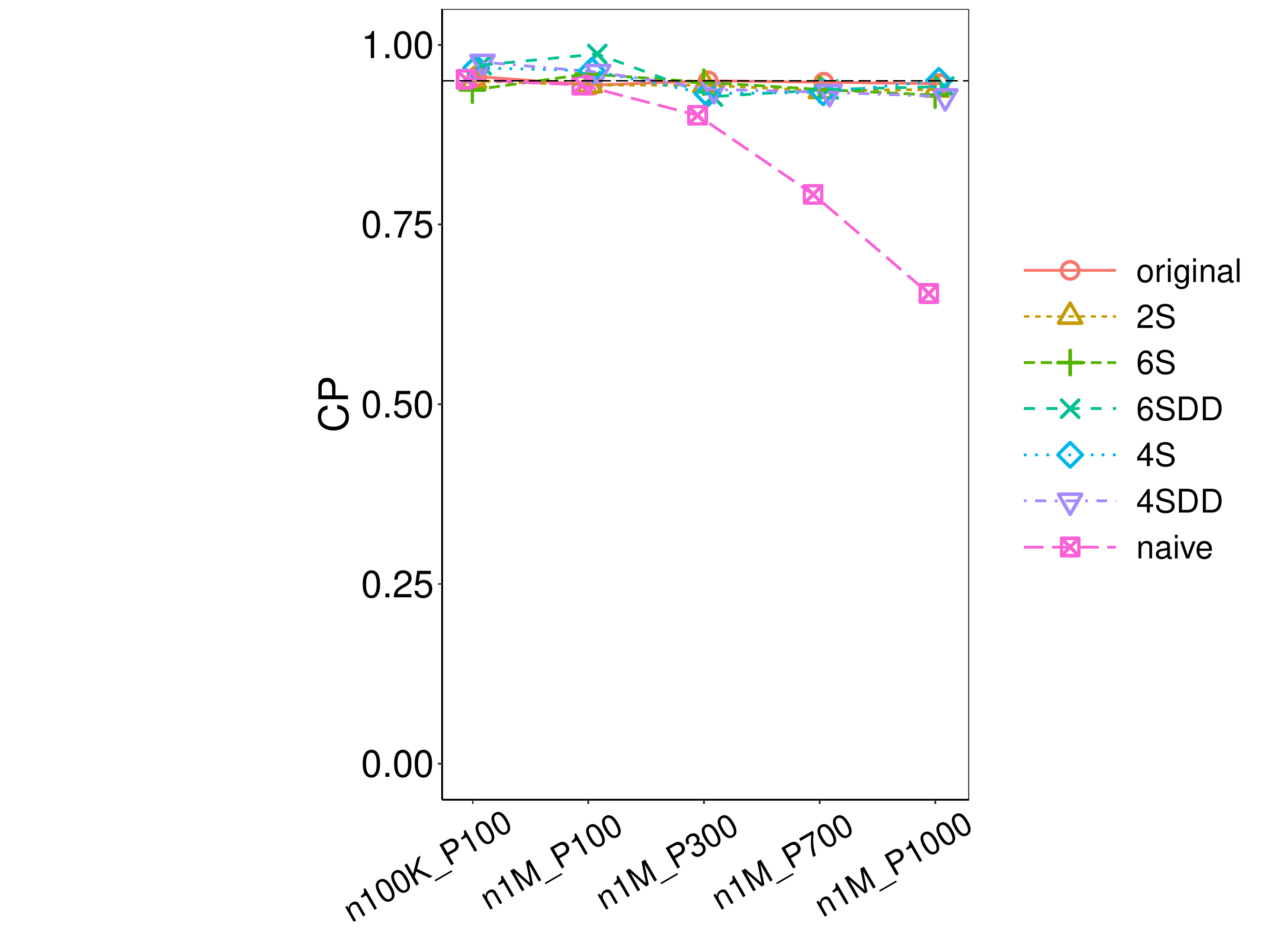}
\includegraphics[width=0.19\textwidth, trim={2.5in 0 2.6in 0},clip] {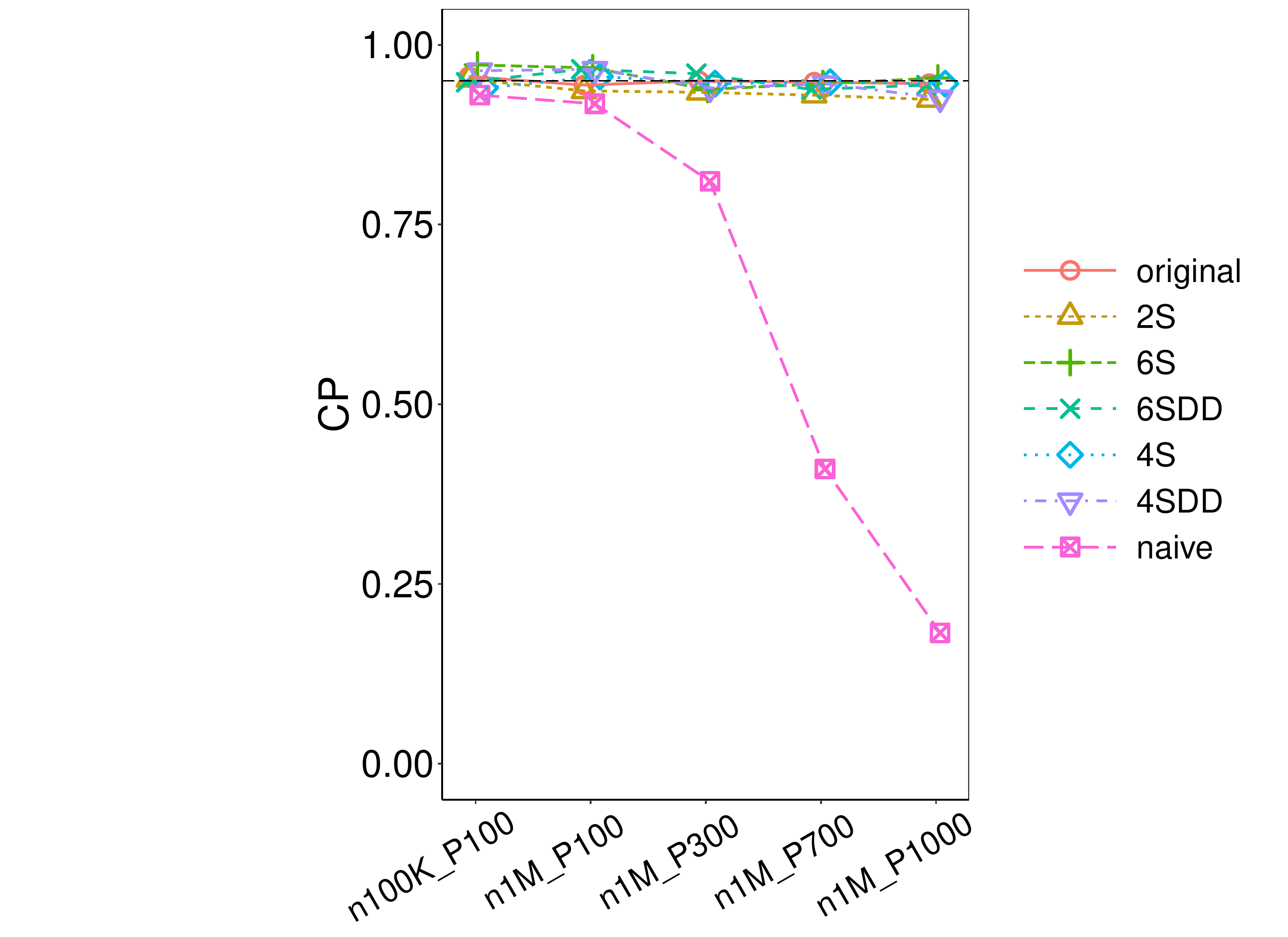}
\includegraphics[width=0.19\textwidth, trim={2.5in 0 2.6in 0},clip] {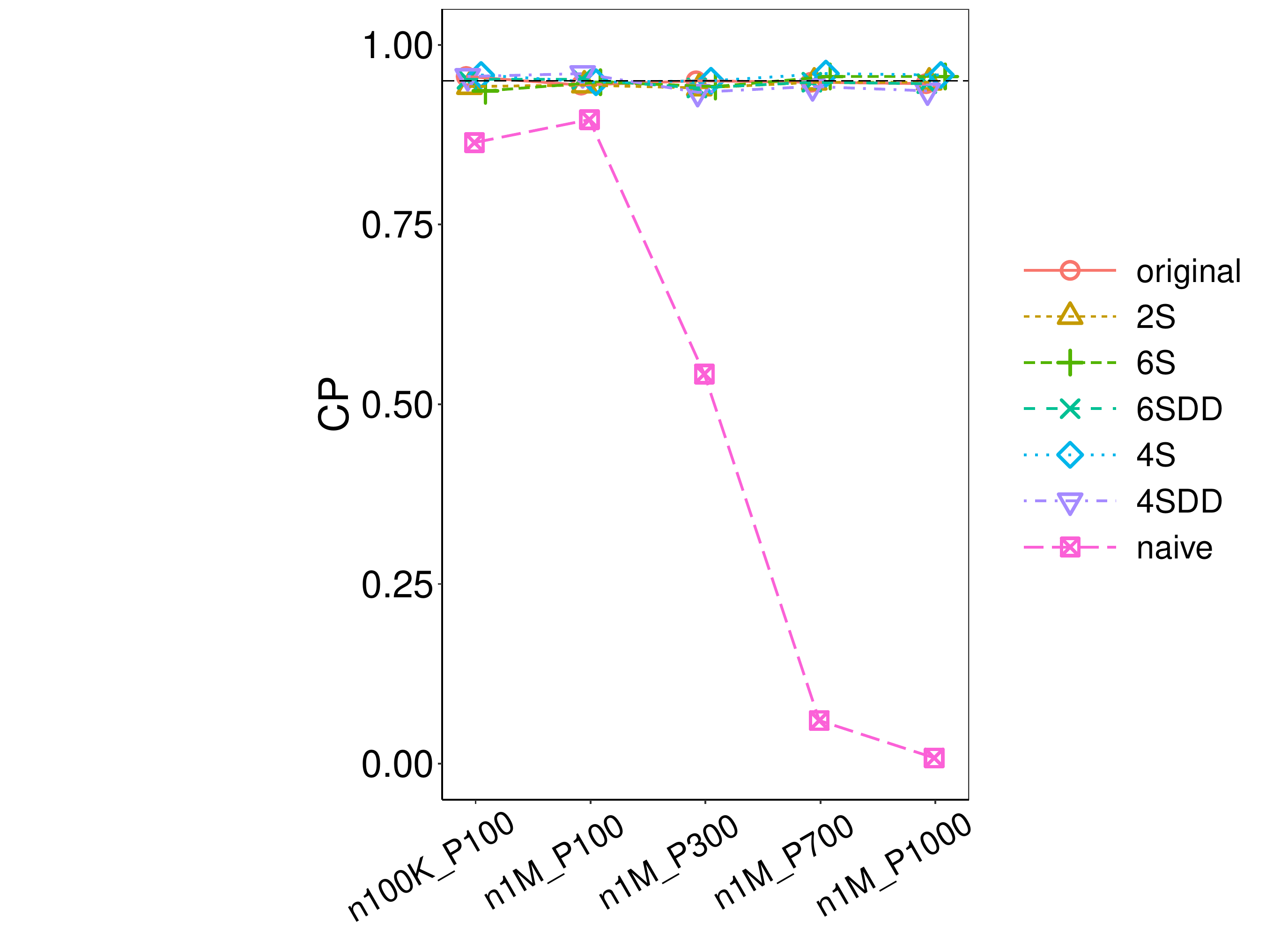}
\includegraphics[width=0.19\textwidth, trim={2.5in 0 2.6in 0},clip] {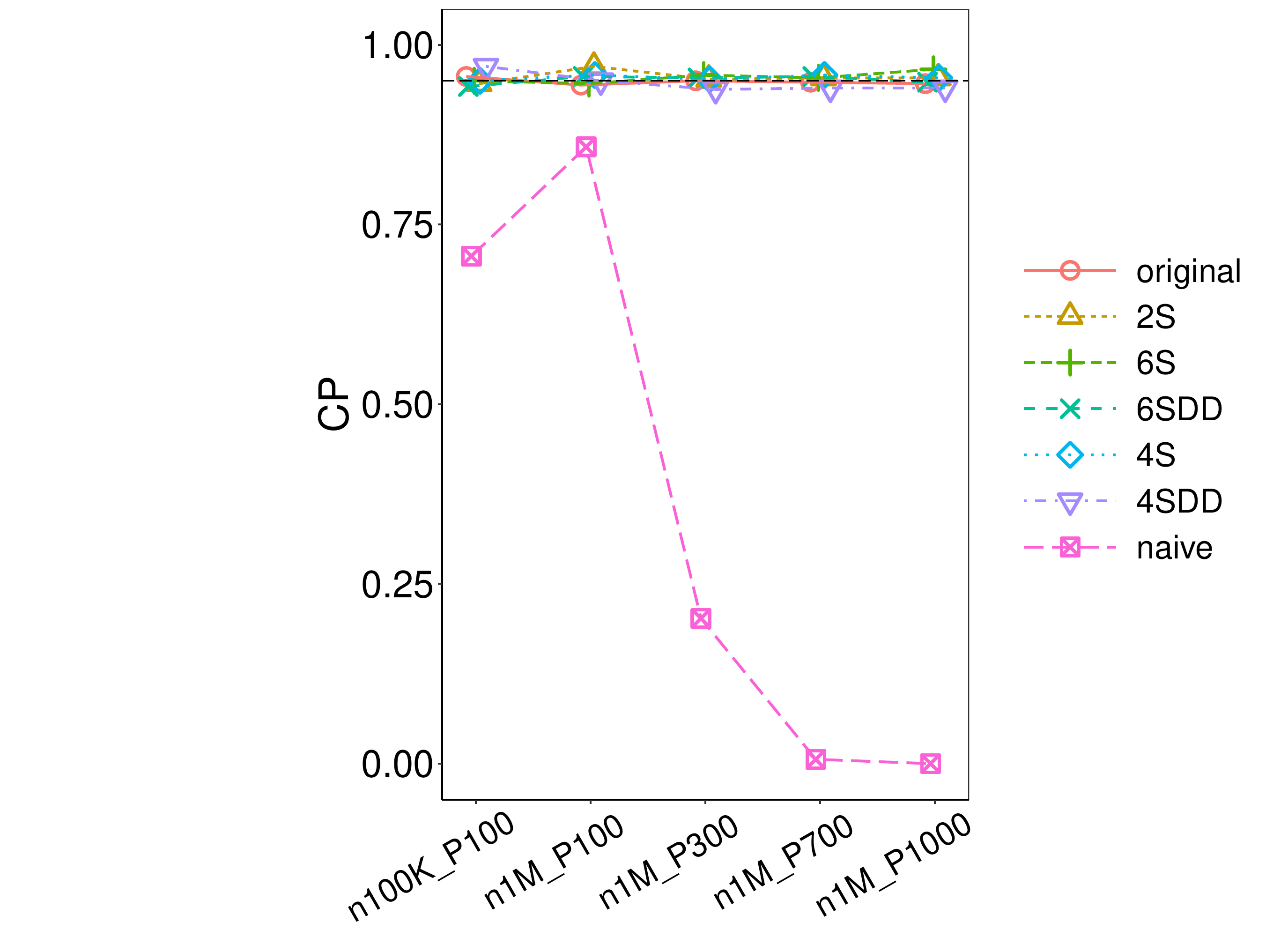}
\includegraphics[width=0.19\textwidth, trim={2.5in 0 2.6in 0},clip] {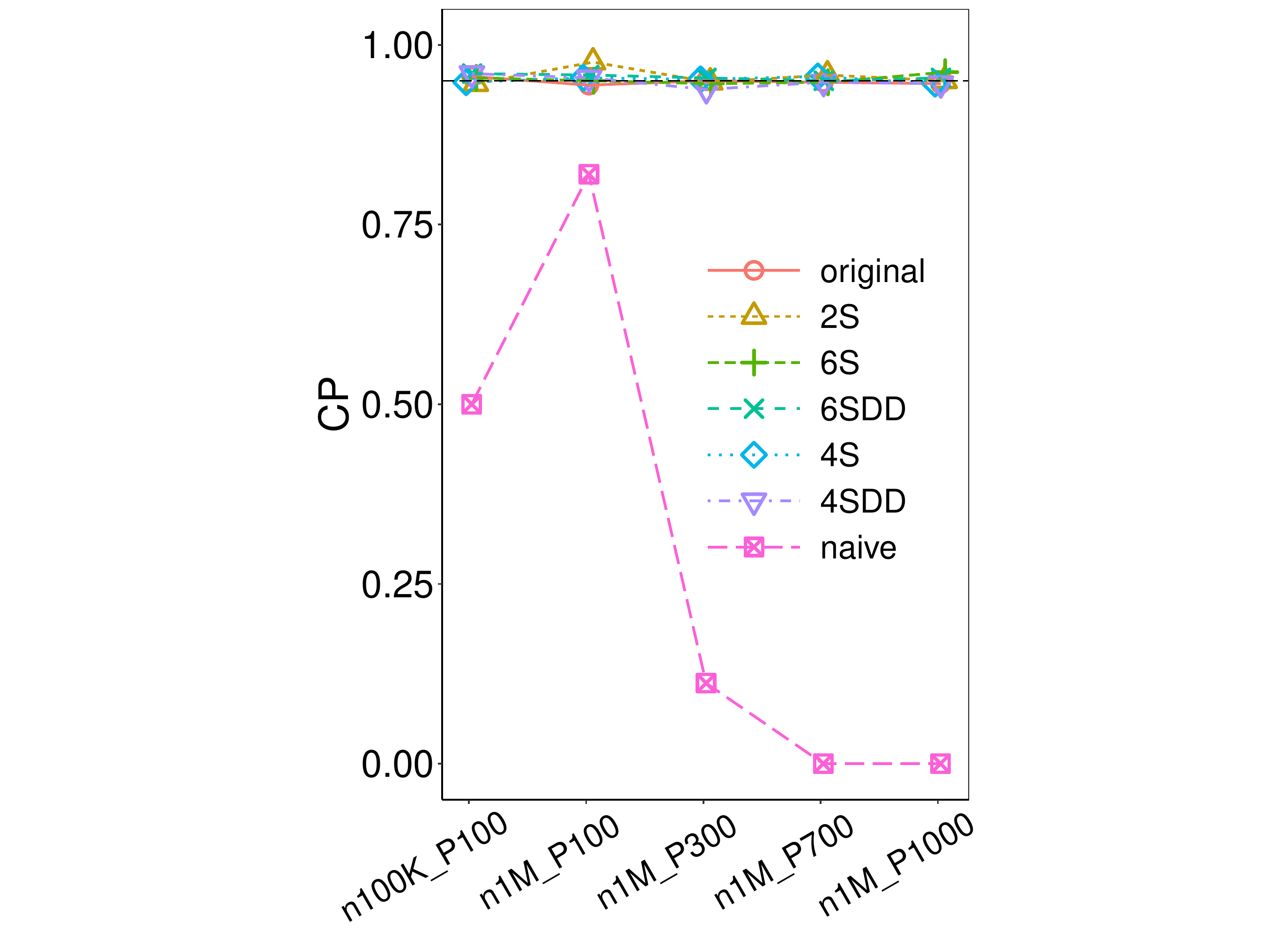}
\caption{Simulation results with $\rho$-zCDP for ZINB data with  $\alpha\ne\beta$ when $\theta=0$}
\label{fig:0aszCDPZINB}
\end{figure}

The results are presented in Figures \ref{fig:0asDPZINB} and \ref{fig:0aszCDPZINB} and are summarized  as follows. 4S is the best performer by most metrics and the na\"{i}ve method is the worst. In between the two, some approaches are better than others, depending on the evaluation metrics. For example, in terms of bias, 2S approaches 0 faster as $n$, $P$, or $\epsilon$ increases, than 4S, 6SDD, 6S, and 4SDD; in terms of RMSE and CI width, 4S performs the best and is followed closely by 4SDD, and 2S is the worst especially for at $n=100k/p=100$, $n=1m/P=100$ and $\epsilon\le1$ and the na\"{i}ve method is the worst for other cases; all methods, except for the  na\"{i}ve method, provides nominal level CP. The Gaussian mechanism of $\rho$-zCDP does not seem to be advantageous in statistical utility compared to the Laplace mechanism of $\epsilon$-DP.

\subsubsection{\texorpdfstring{$Y\sim$ ZINB}{}, \texorpdfstring{$\theta\ne0$ ($H_1$)}{H1}, and \texorpdfstring{$\alpha\ne\beta$}{AS}}\label{sec:1asZINB}
\begin{figure}[!htb]
\hspace{0.5in}$\epsilon=0.5$\hspace{0.8in}$\epsilon=1$\hspace{0.9in}$\epsilon=2$
\hspace{0.95in}$\epsilon=5$\hspace{0.9in}$\epsilon=50$

\includegraphics[width=0.19\textwidth, trim={2.45in 0 2.45in 0},clip] {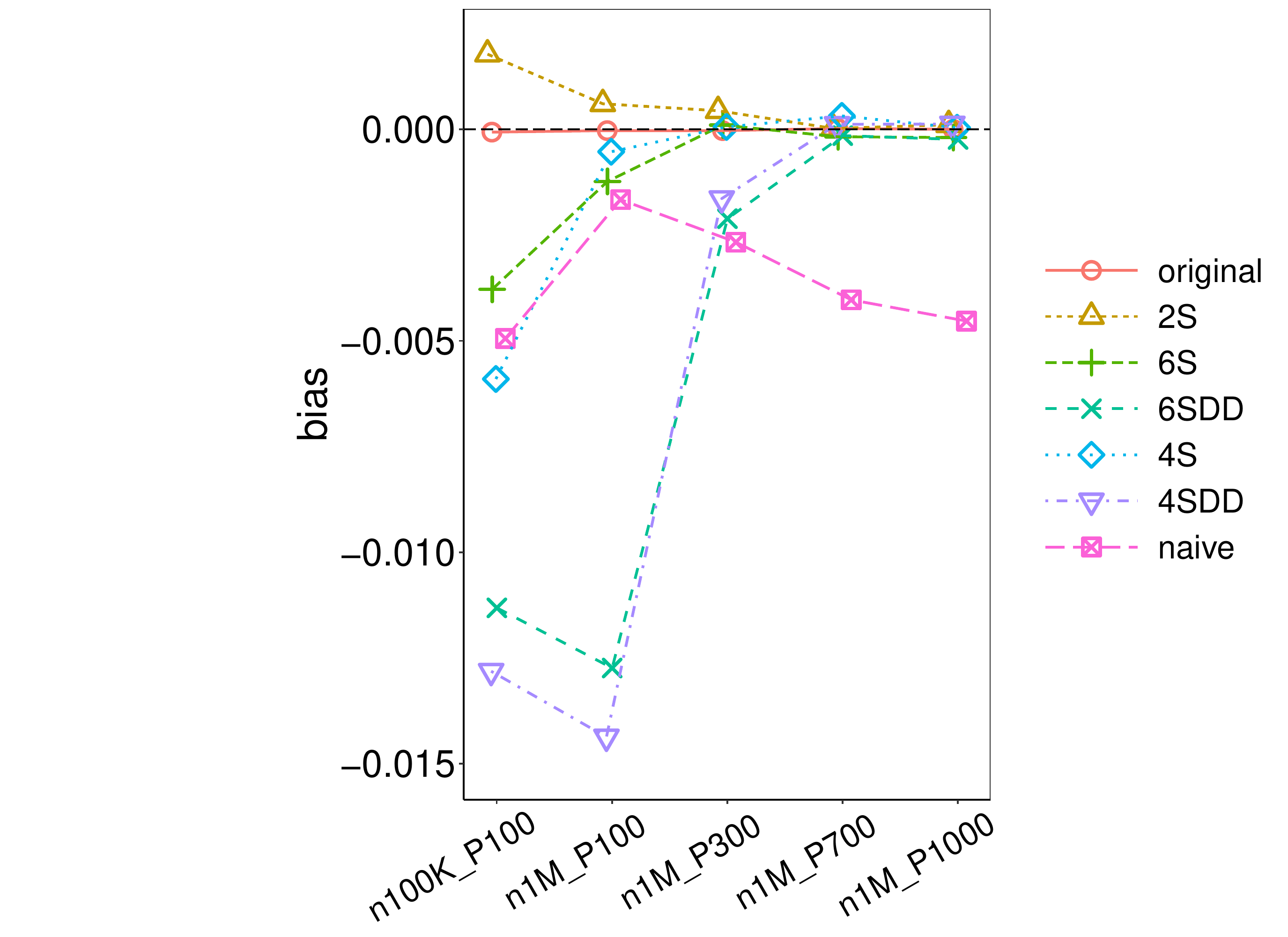}
\includegraphics[width=0.19\textwidth, trim={2.45in 0 2.45in 0},clip] {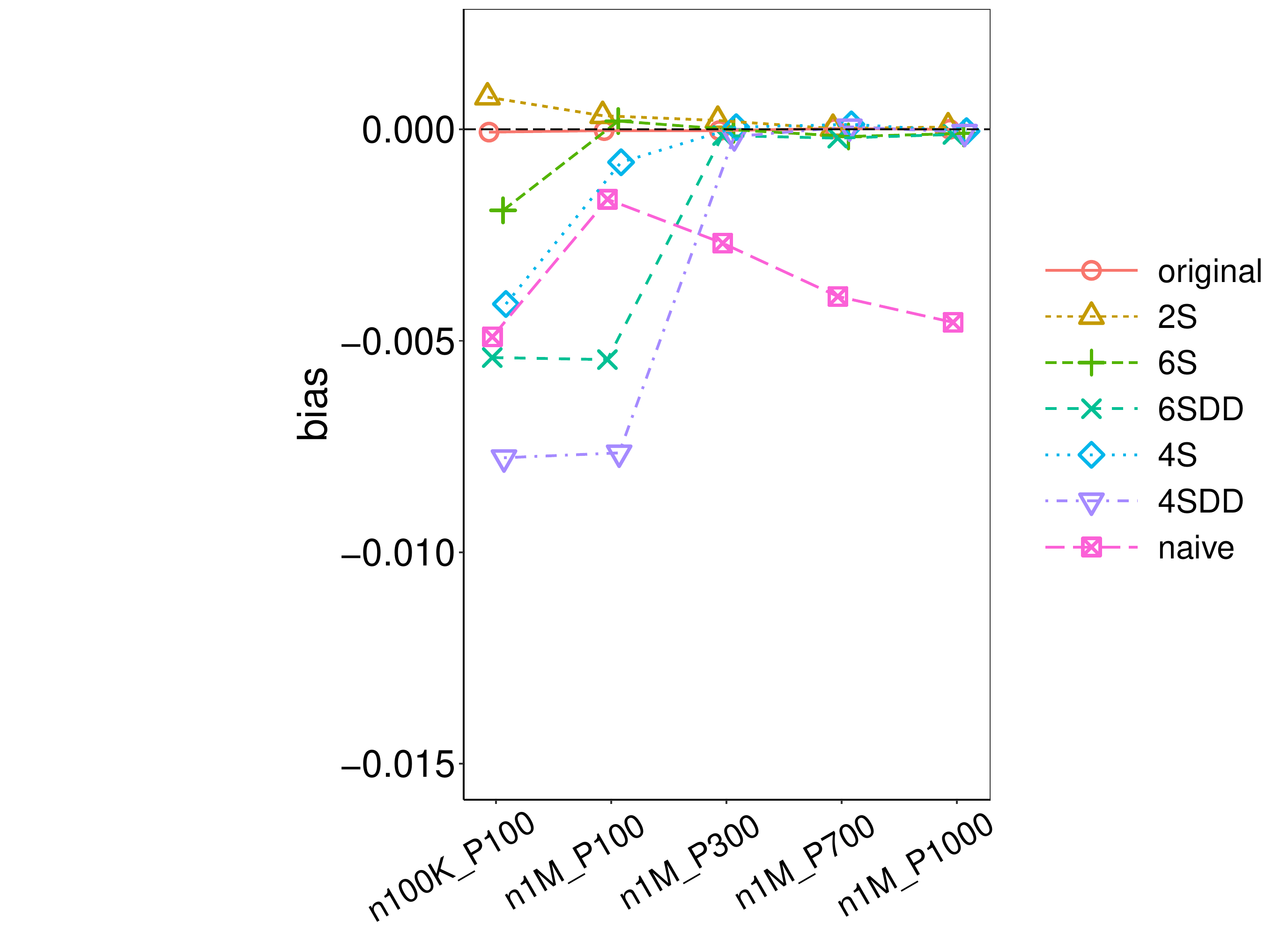}
\includegraphics[width=0.19\textwidth, trim={2.45in 0 2.45in 0},clip] {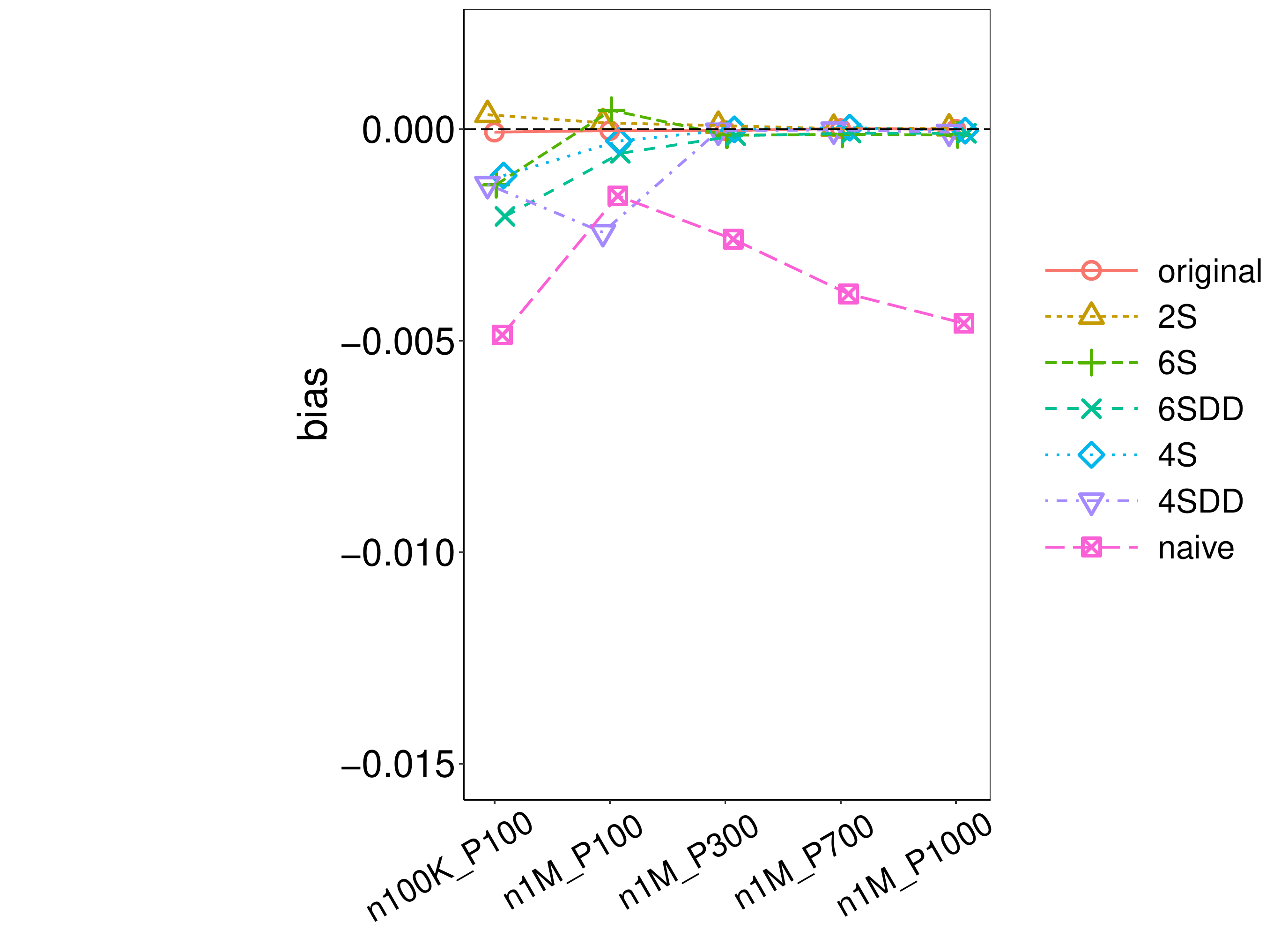}
\includegraphics[width=0.19\textwidth, trim={2.45in 0 2.45in 0},clip] {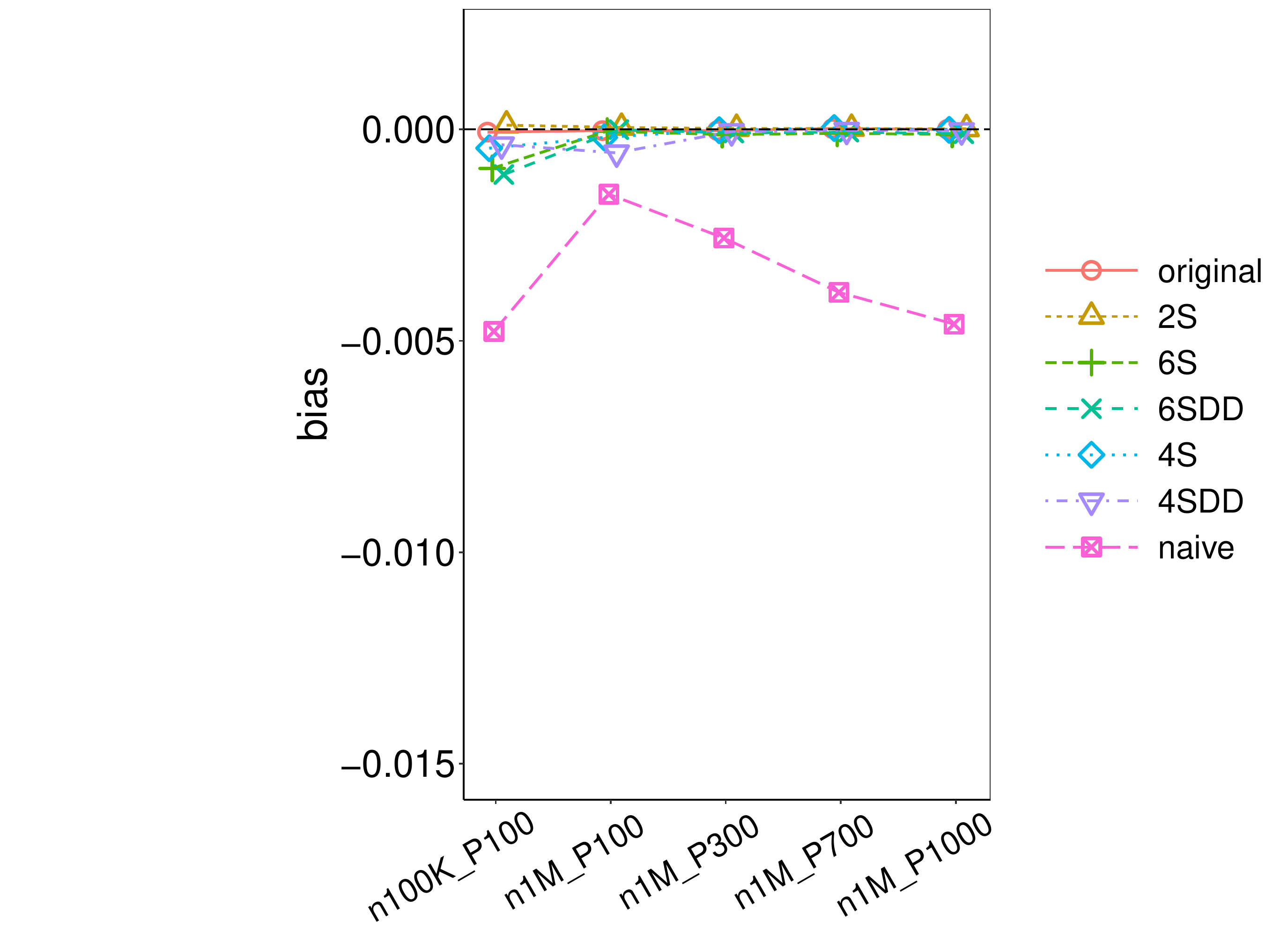}
\includegraphics[width=0.19\textwidth, trim={2.45in 0 2.45in 0},clip] {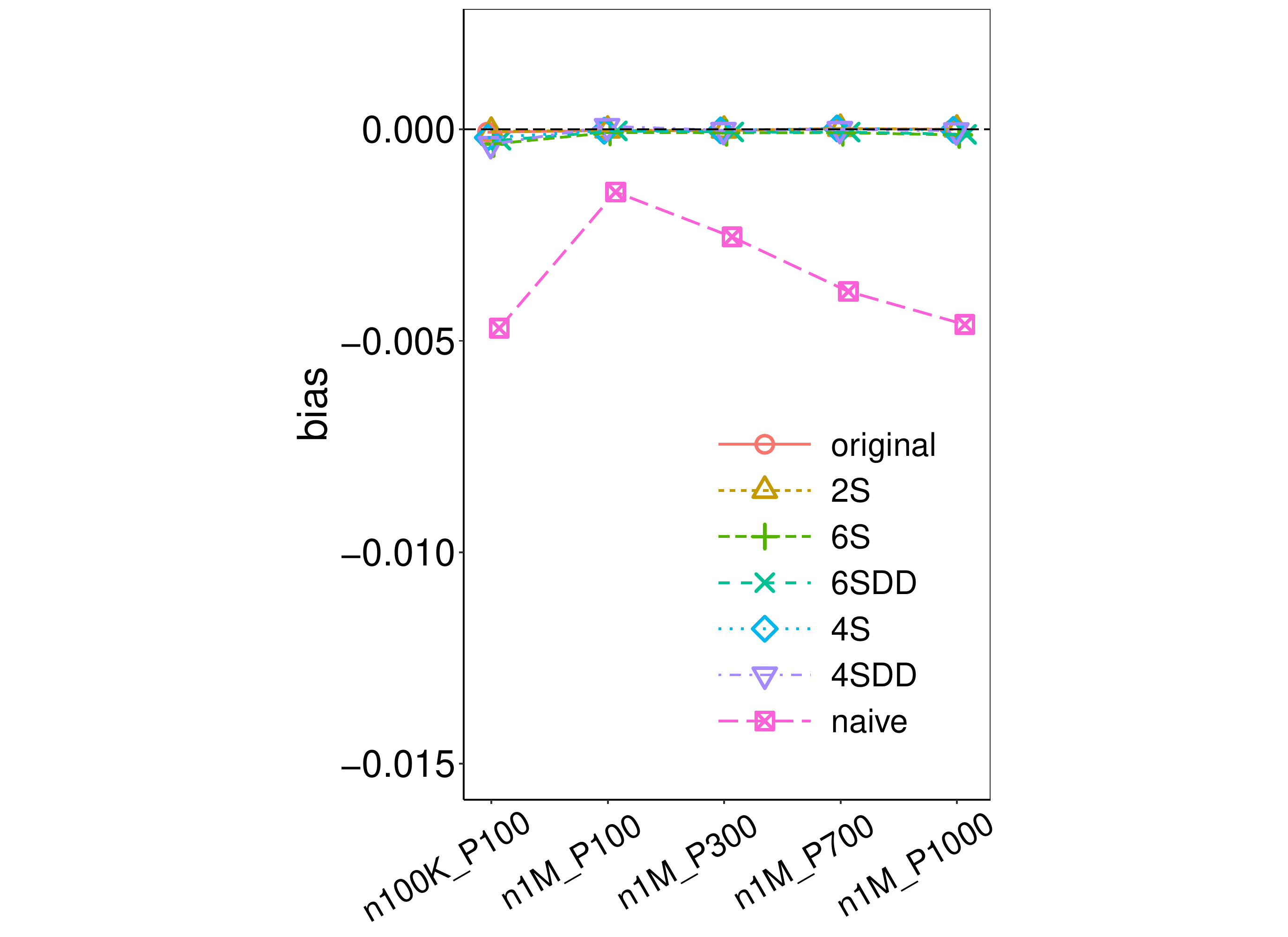}

\includegraphics[width=0.19\textwidth, trim={2.5in 0 2.6in 0},clip] {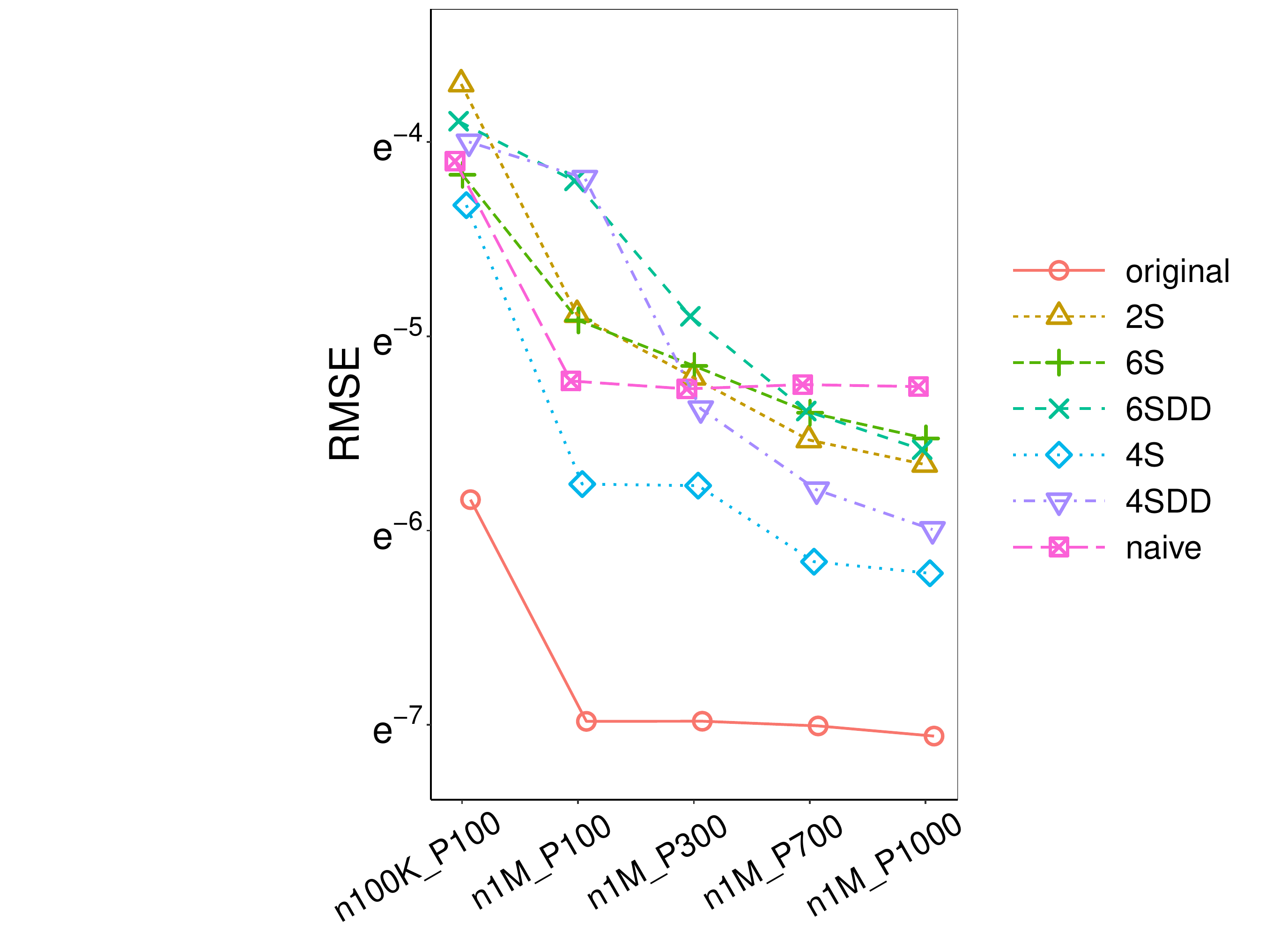}
\includegraphics[width=0.19\textwidth, trim={2.5in 0 2.6in 0},clip] {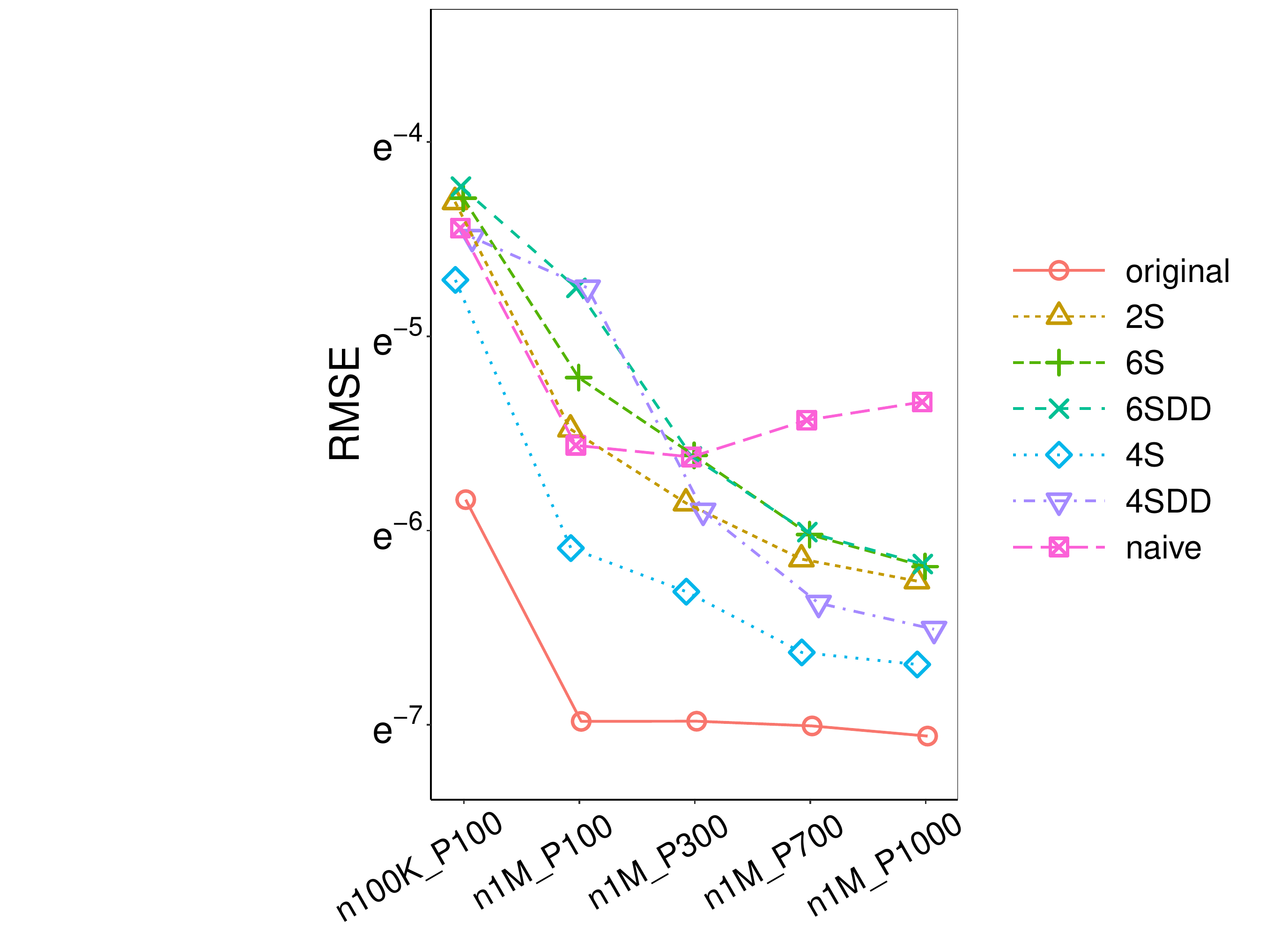}
\includegraphics[width=0.19\textwidth, trim={2.5in 0 2.6in 0},clip] {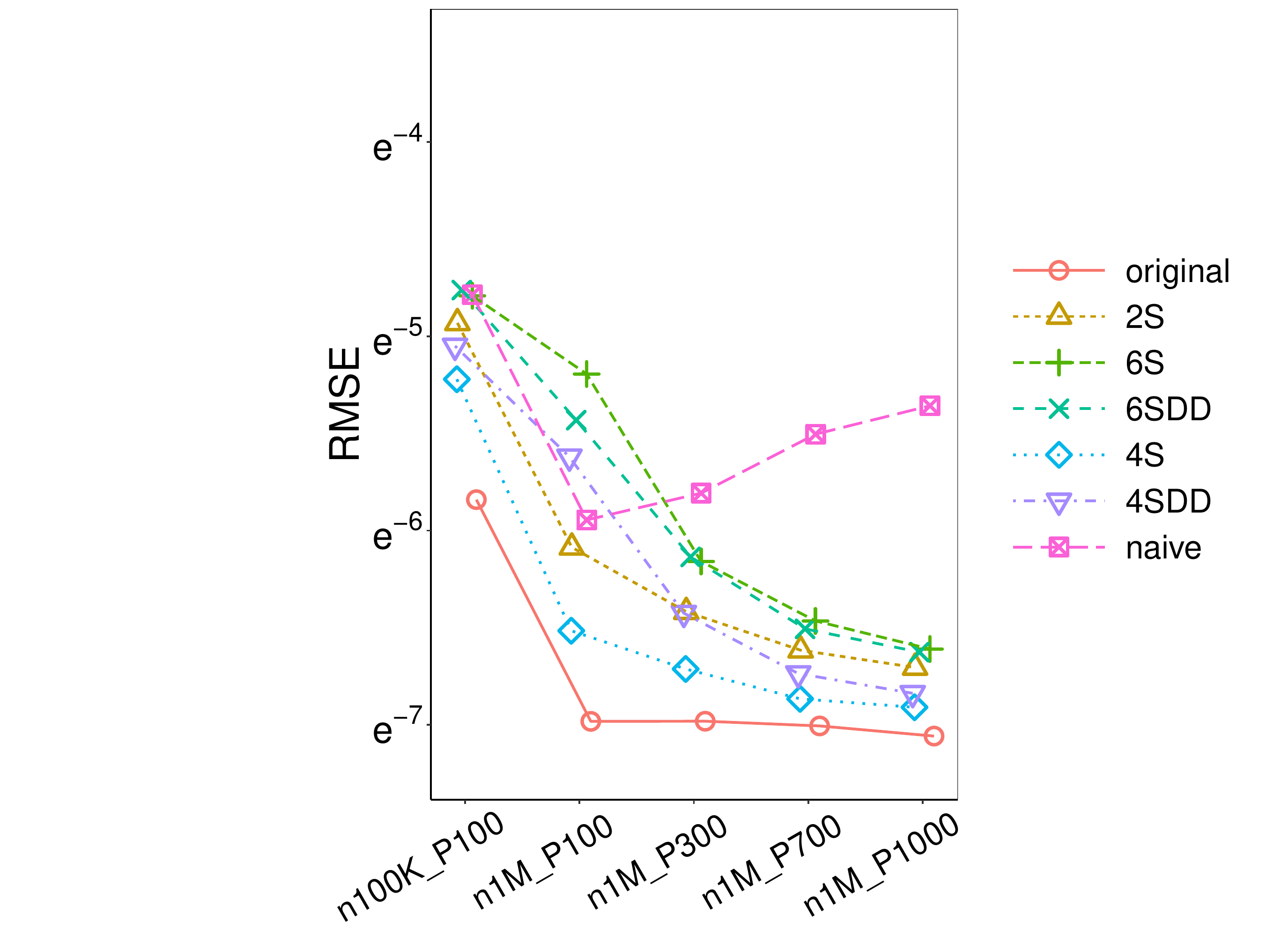}
\includegraphics[width=0.19\textwidth, trim={2.5in 0 2.6in 0},clip] {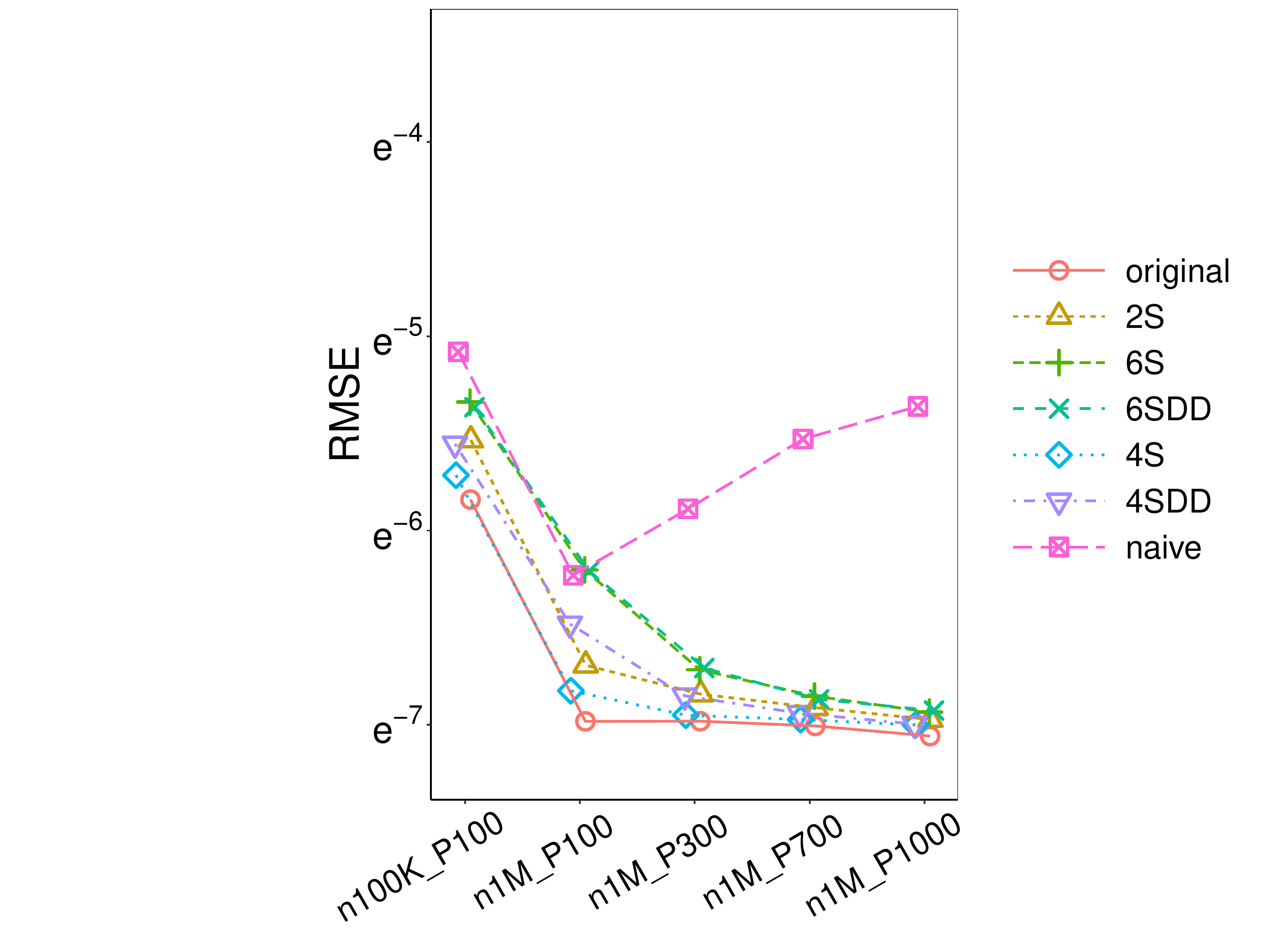}
\includegraphics[width=0.19\textwidth, trim={2.5in 0 2.6in 0},clip] {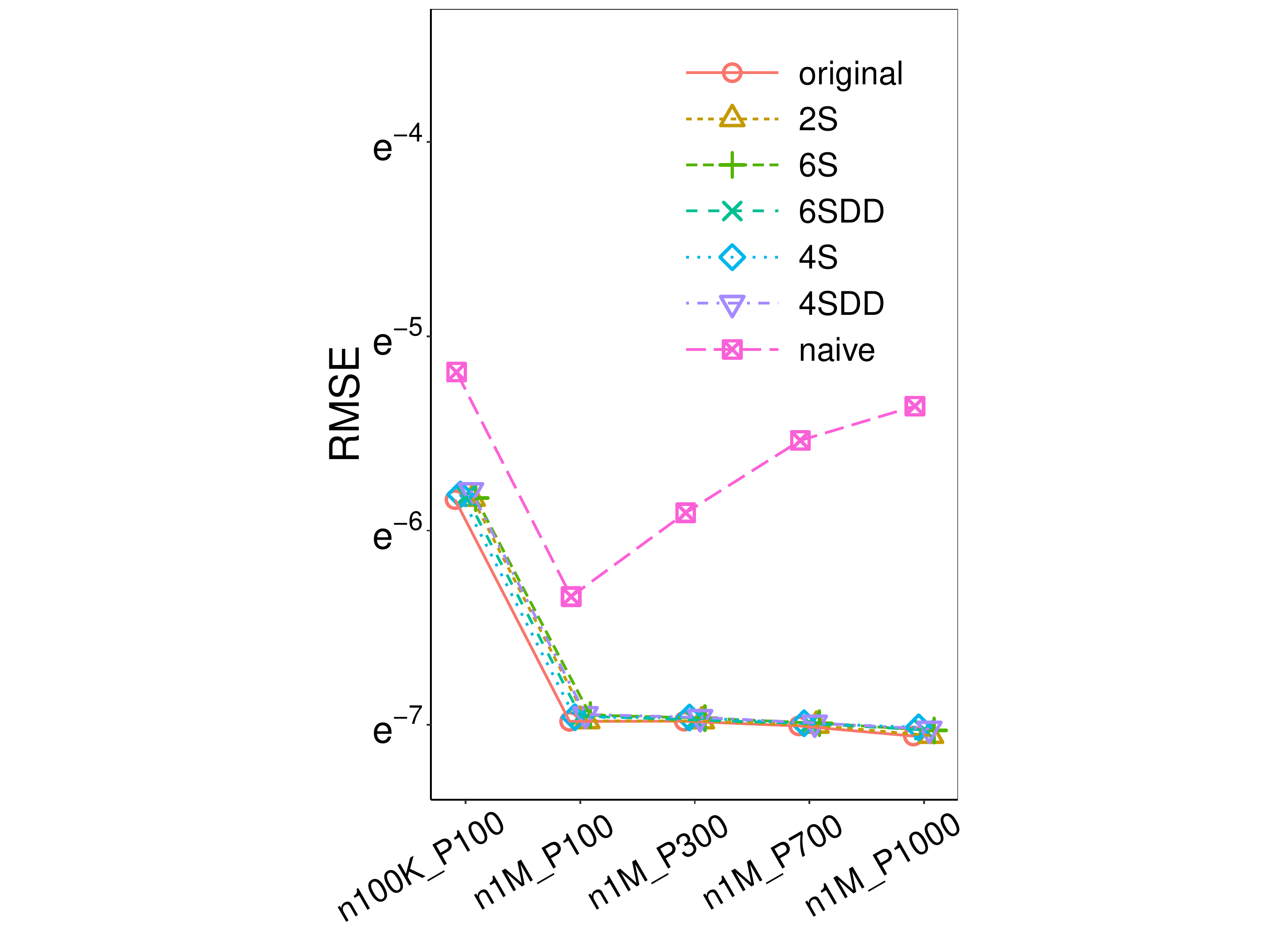}

\includegraphics[width=0.19\textwidth, trim={2.5in 0 2.6in 0},clip] {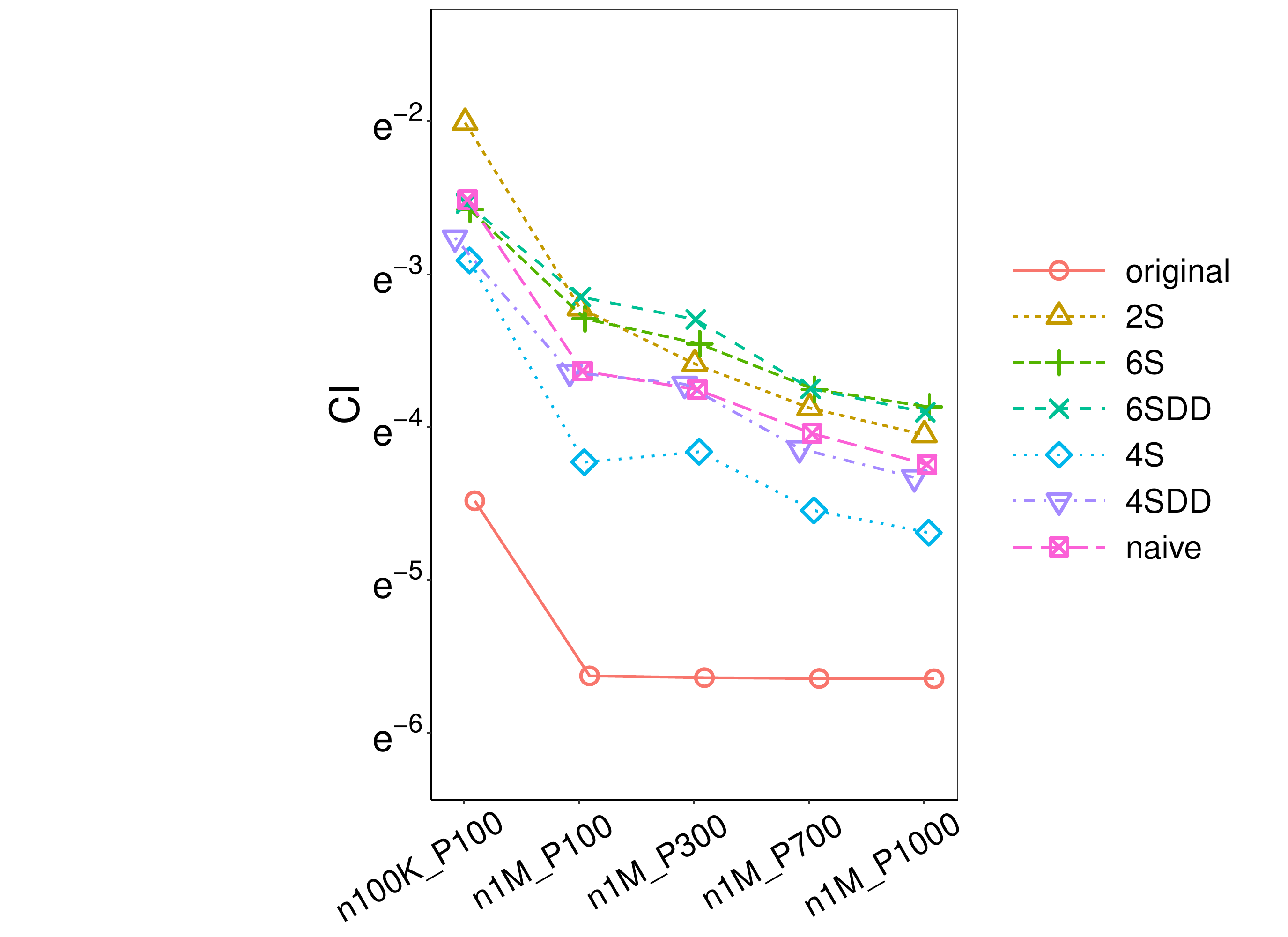}
\includegraphics[width=0.19\textwidth, trim={2.5in 0 2.6in 0},clip] {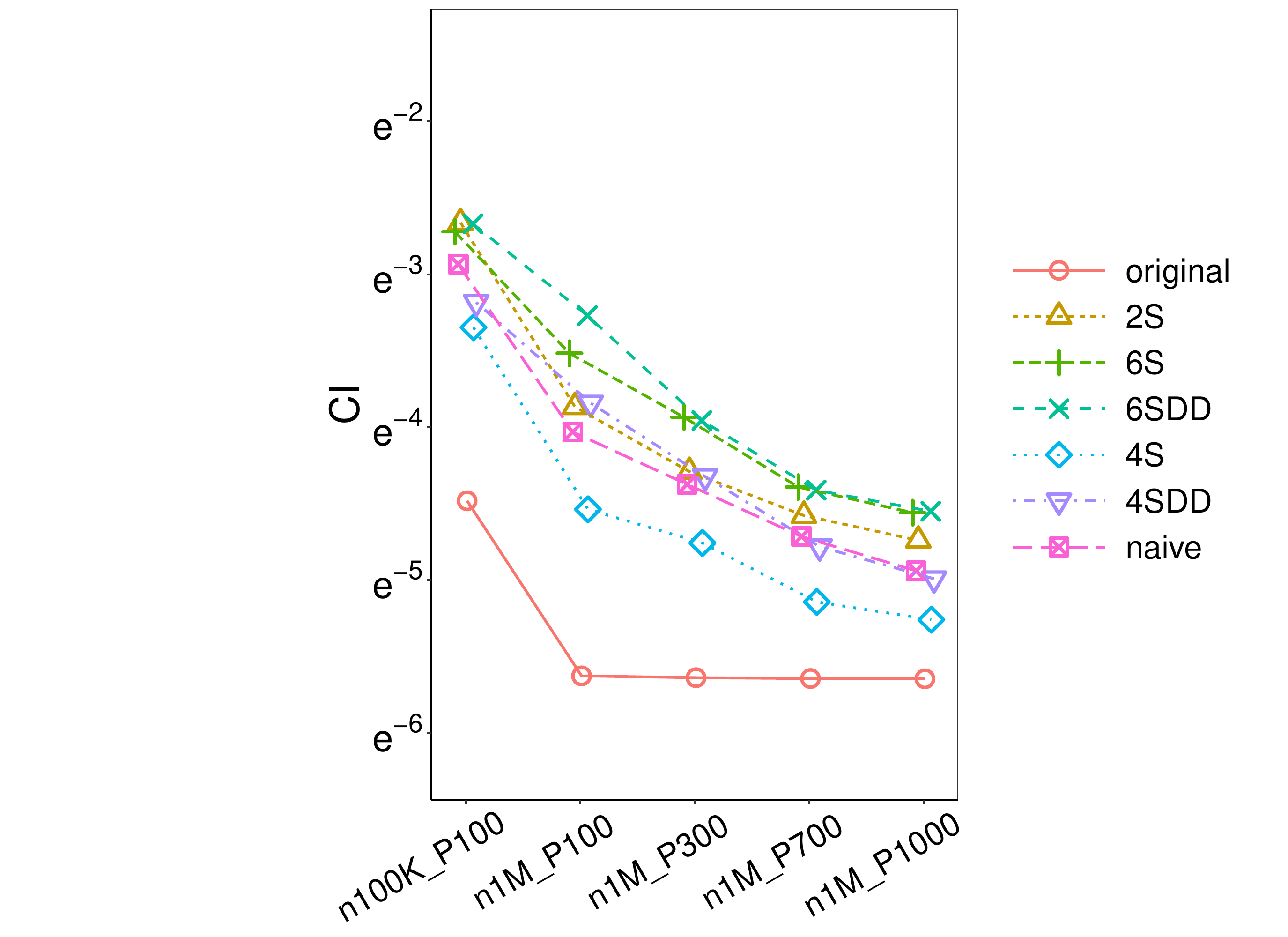}
\includegraphics[width=0.19\textwidth, trim={2.5in 0 2.6in 0},clip] {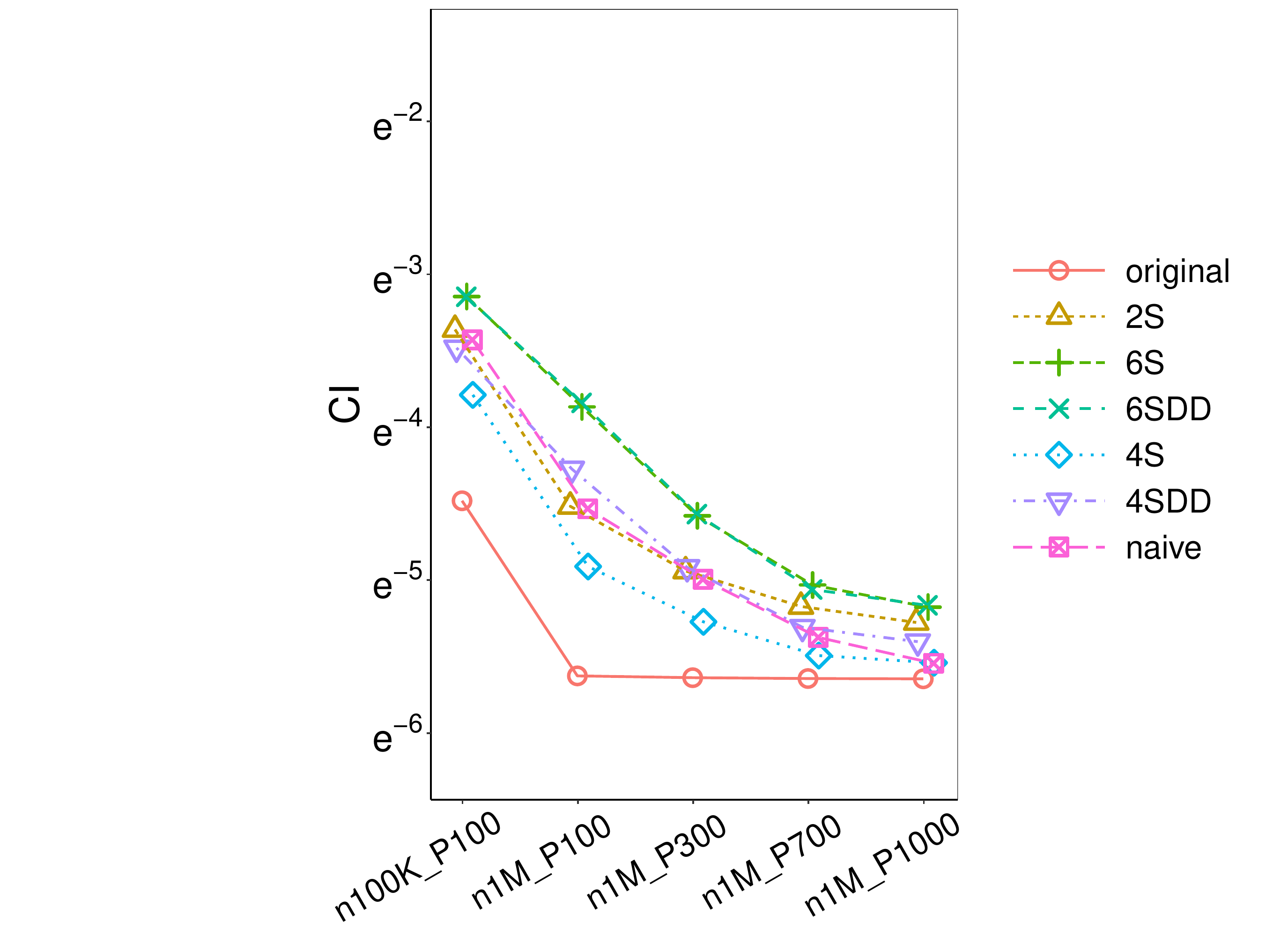}
\includegraphics[width=0.19\textwidth, trim={2.5in 0 2.6in 0},clip] {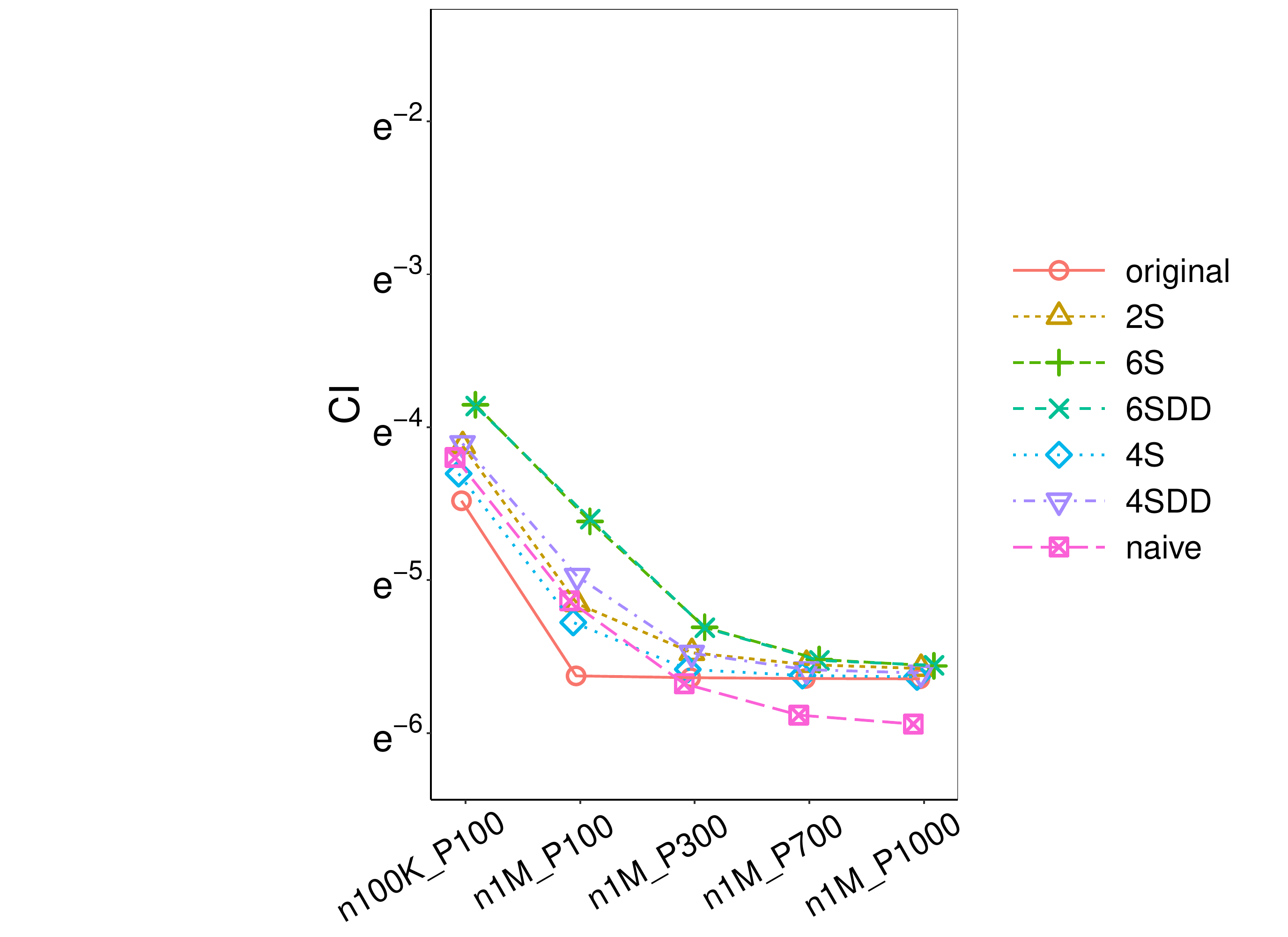}
\includegraphics[width=0.19\textwidth, trim={2.5in 0 2.6in 0},clip] {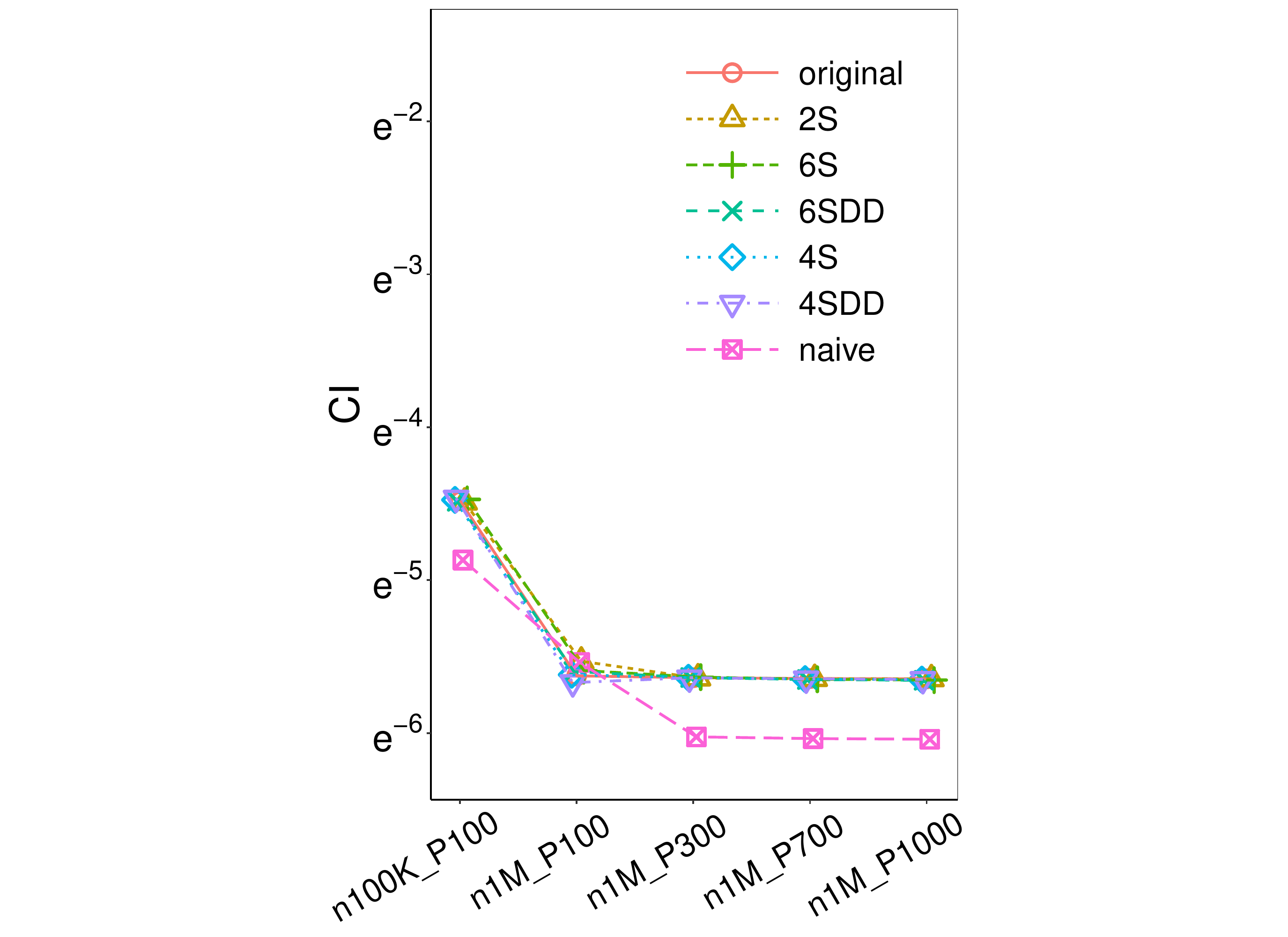}

\includegraphics[width=0.19\textwidth, trim={2.5in 0 2.6in 0},clip] {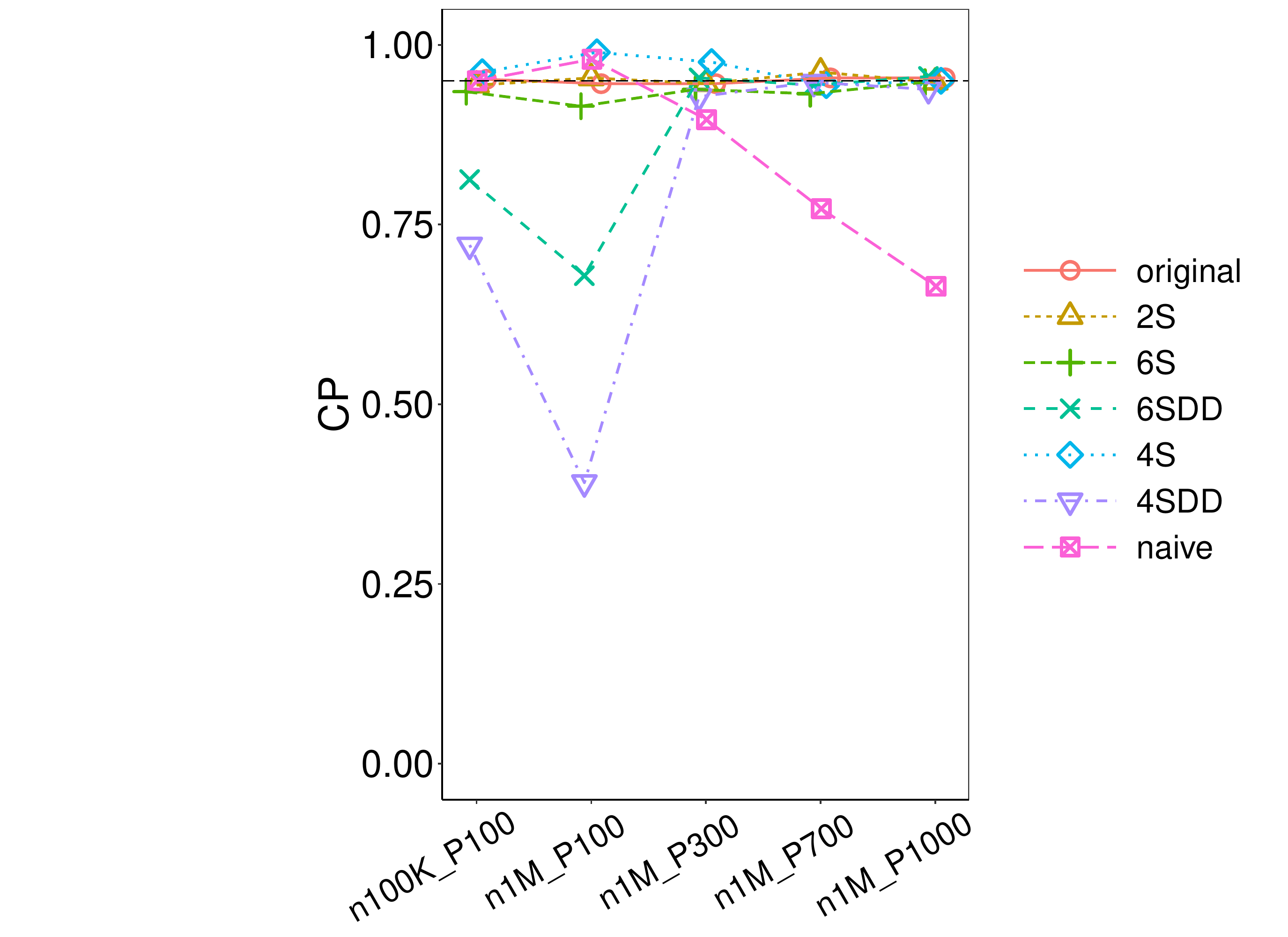}
\includegraphics[width=0.19\textwidth, trim={2.5in 0 2.6in 0},clip] {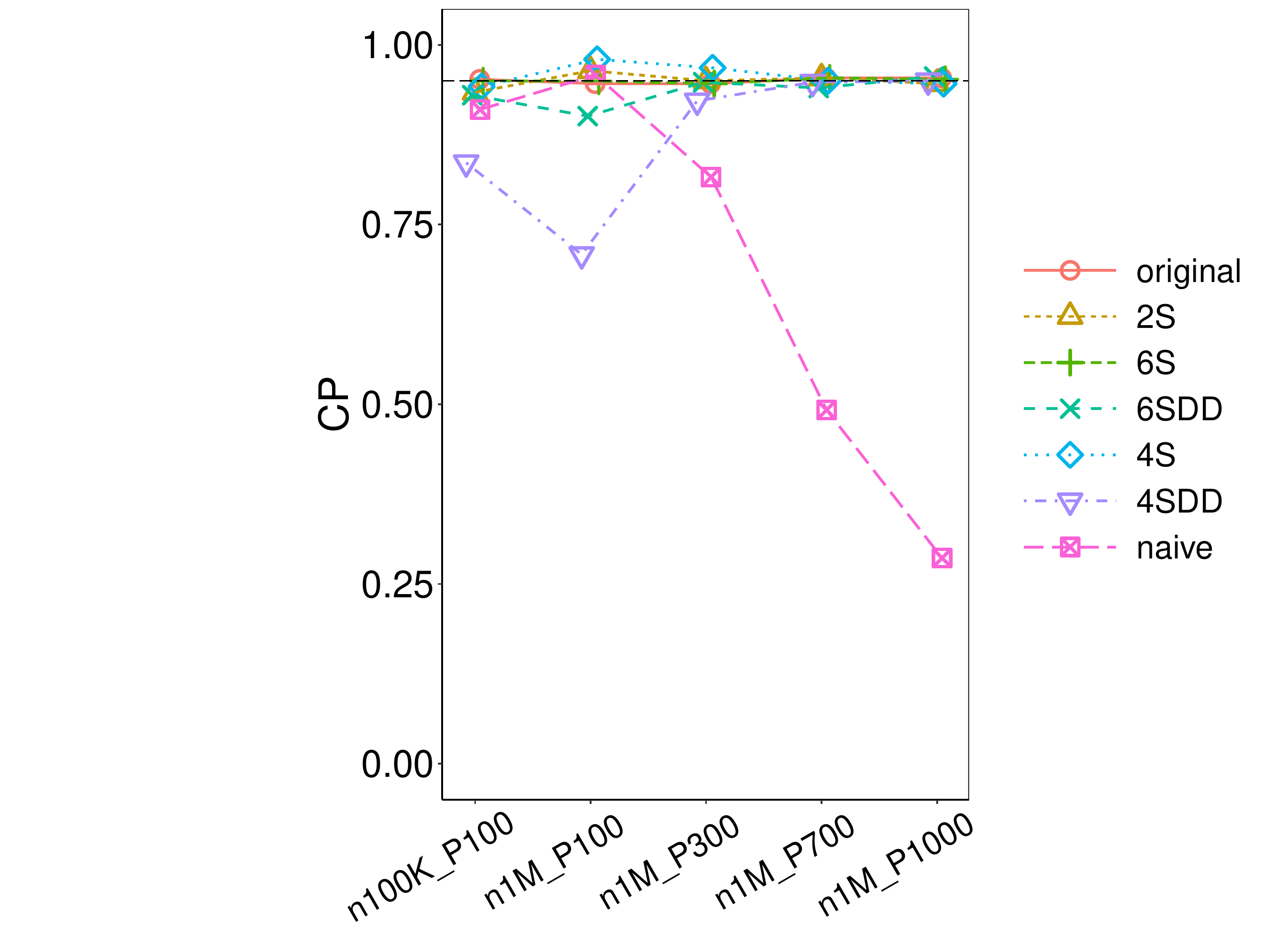}
\includegraphics[width=0.19\textwidth, trim={2.5in 0 2.6in 0},clip] {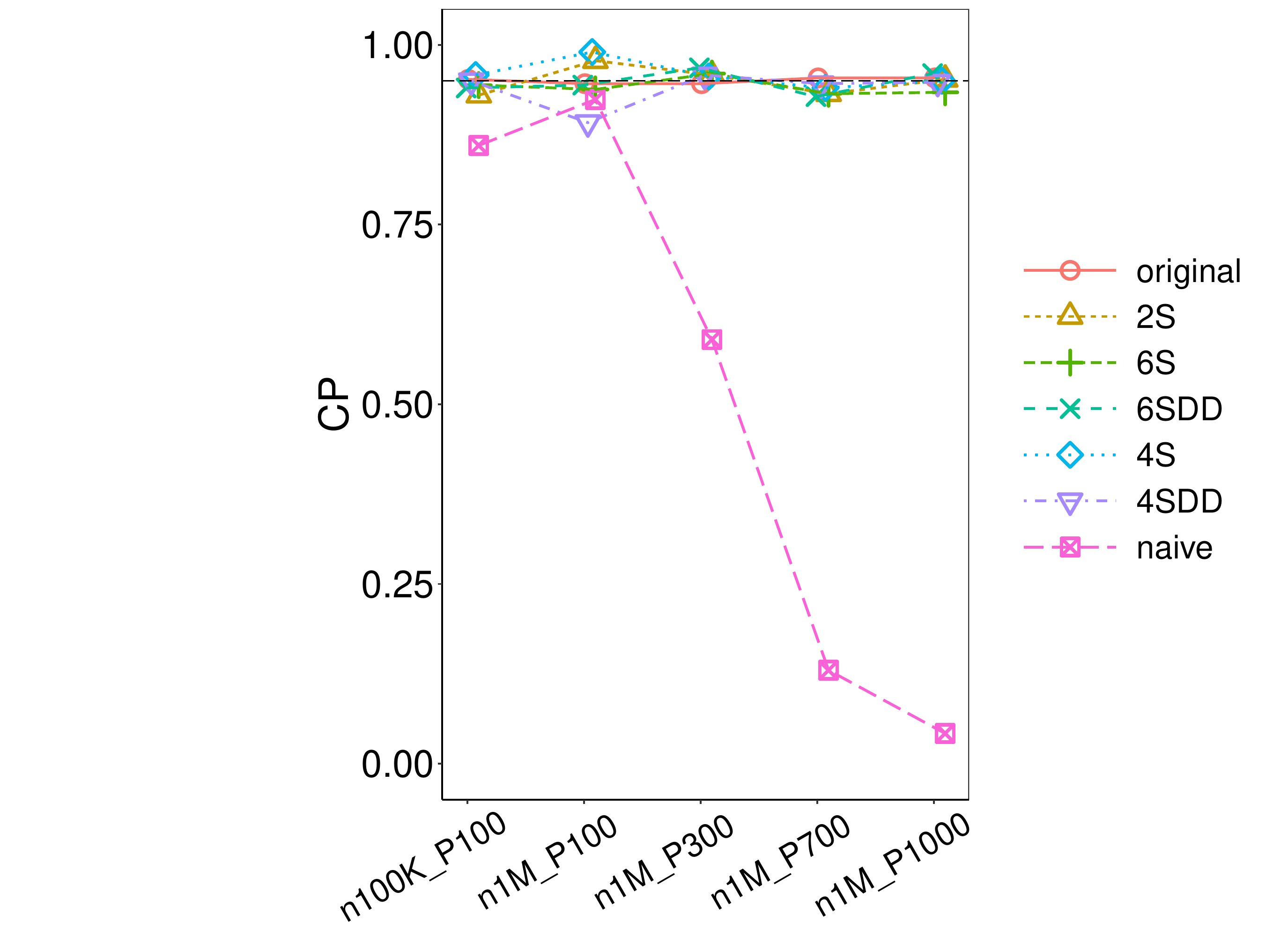}
\includegraphics[width=0.19\textwidth, trim={2.5in 0 2.6in 0},clip] {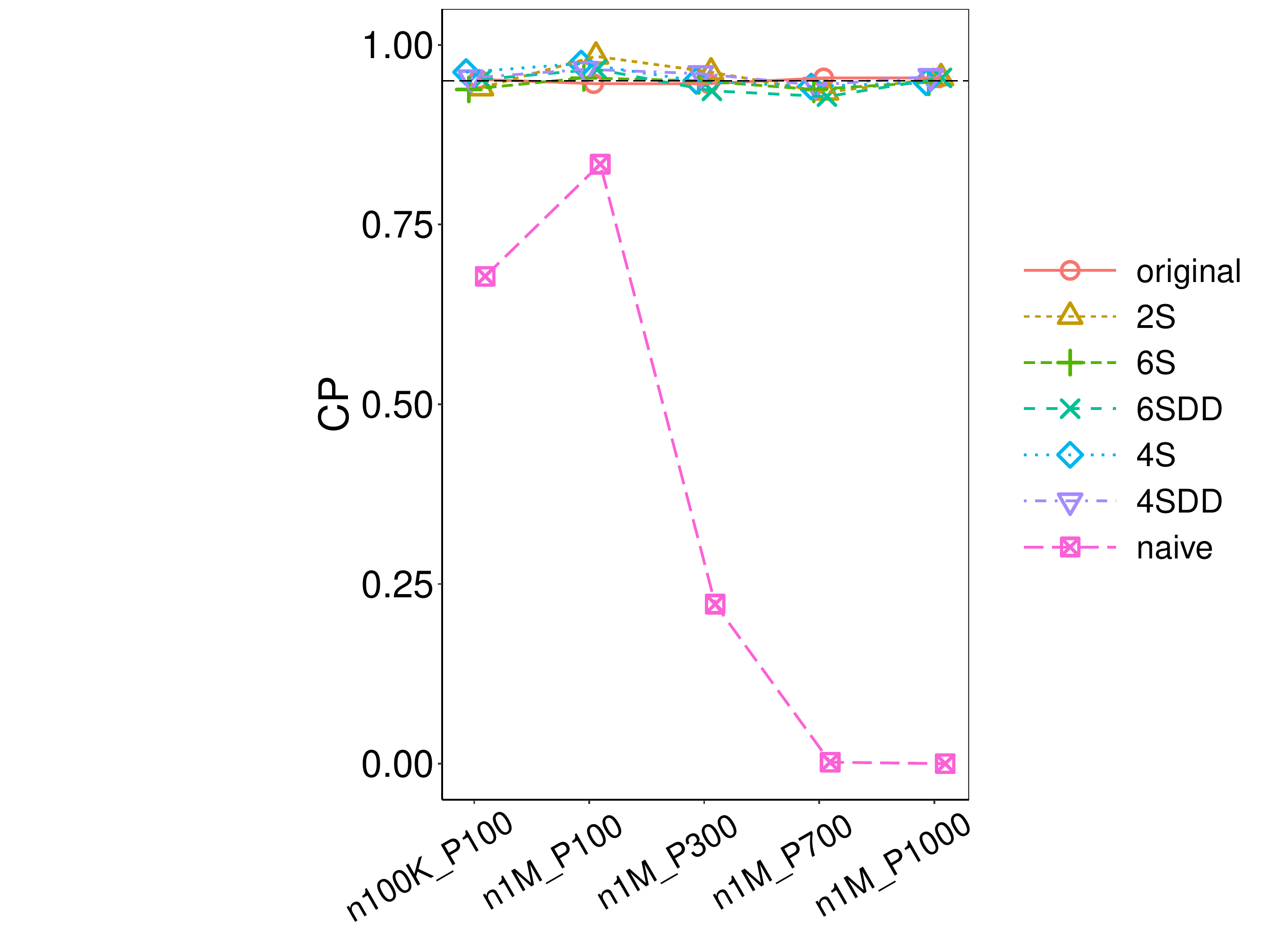}
\includegraphics[width=0.19\textwidth, trim={2.5in 0 2.6in 0},clip] {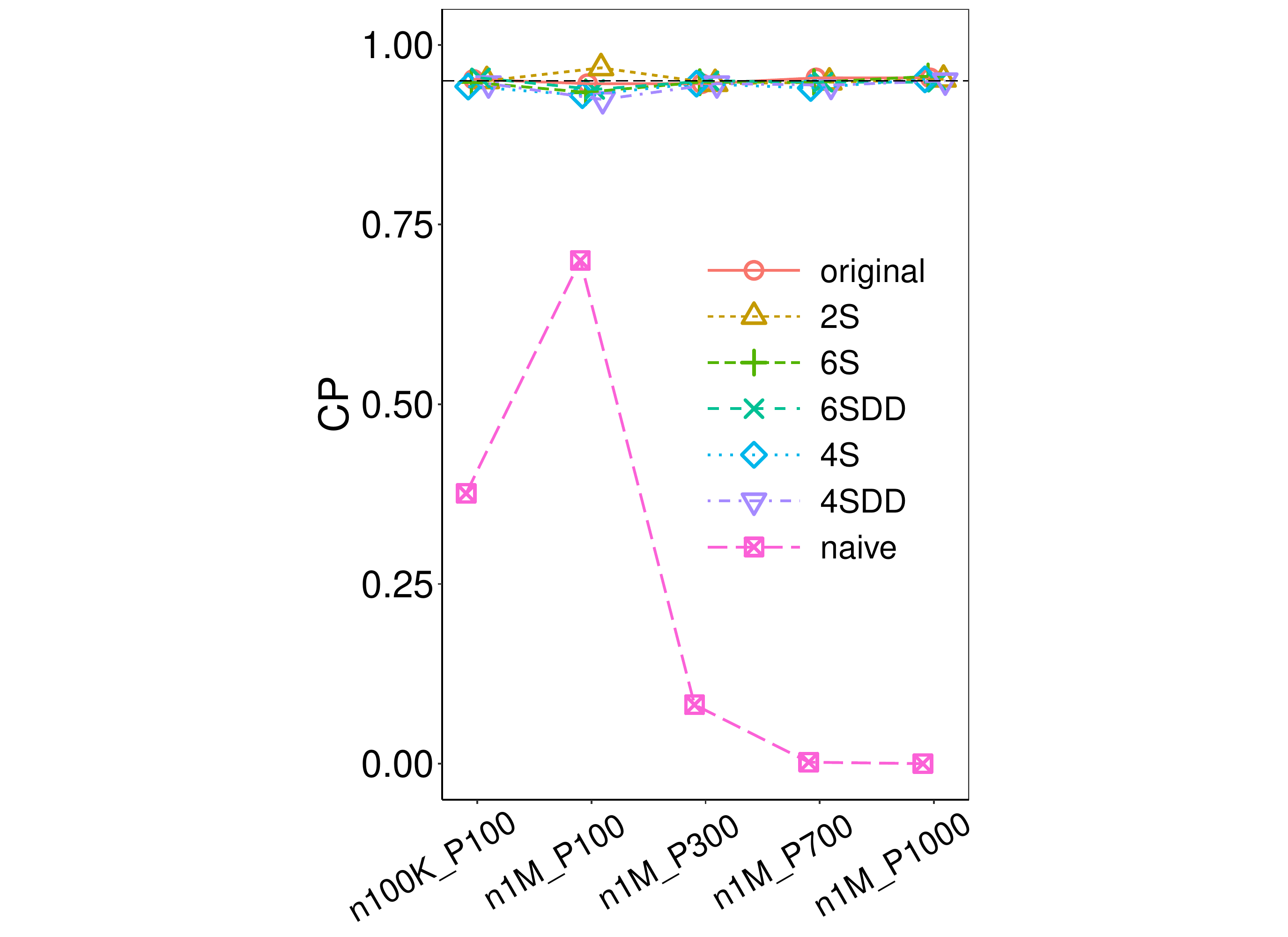}

\includegraphics[width=0.19\textwidth, trim={2.5in 0 2.6in 0},clip] {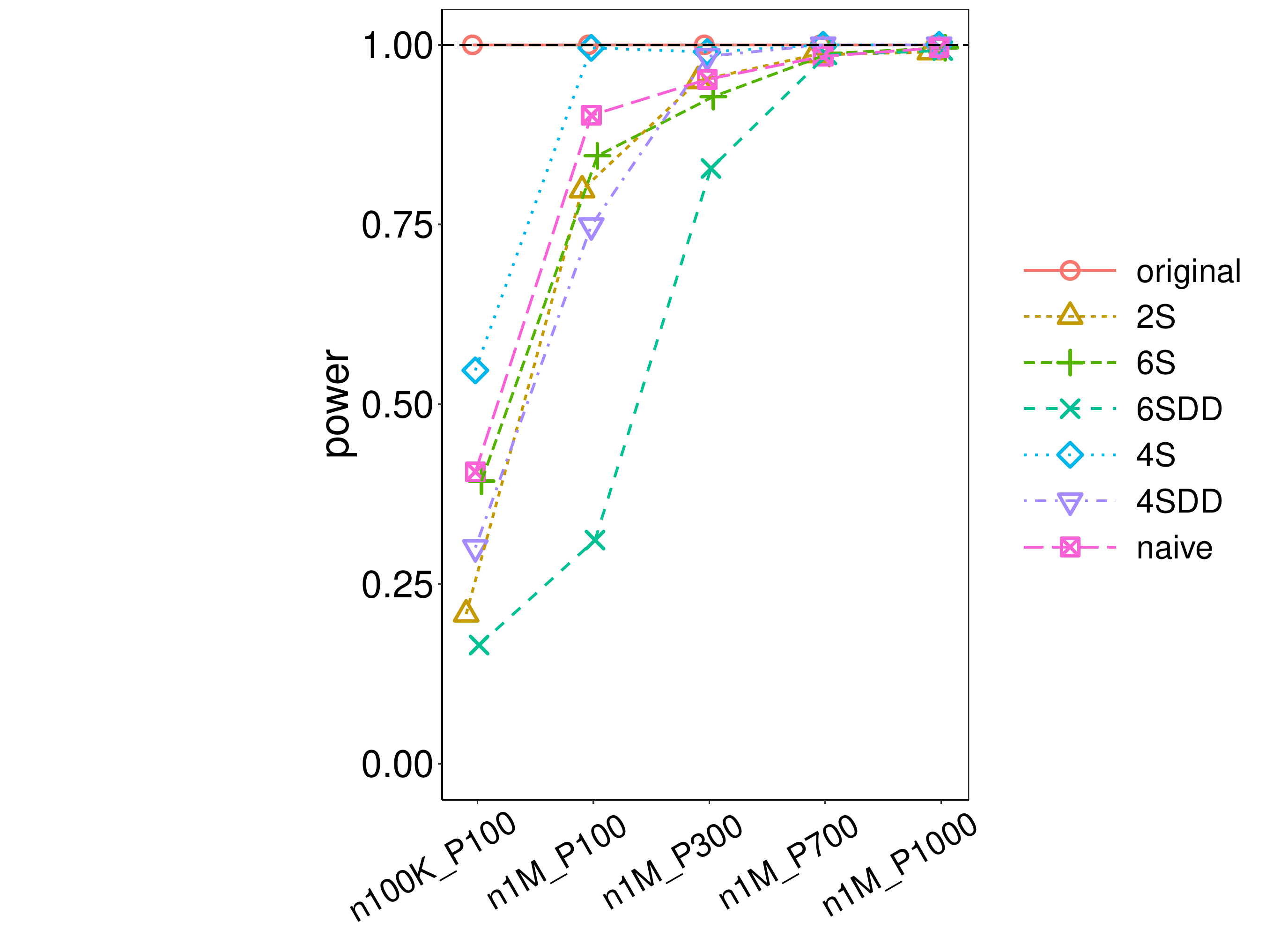}
\includegraphics[width=0.19\textwidth, trim={2.5in 0 2.6in 0},clip] {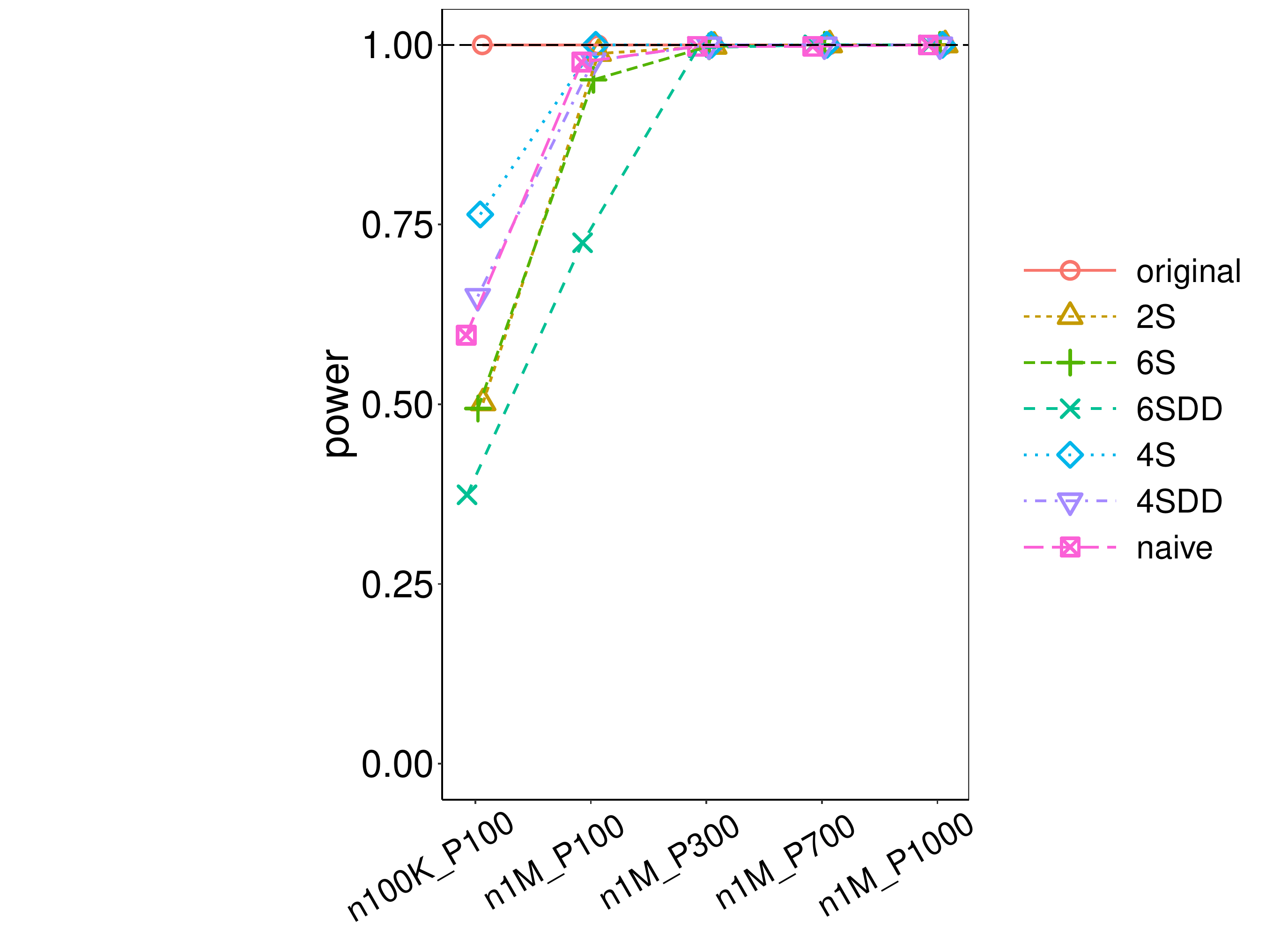}
\includegraphics[width=0.19\textwidth, trim={2.5in 0 2.6in 0},clip] {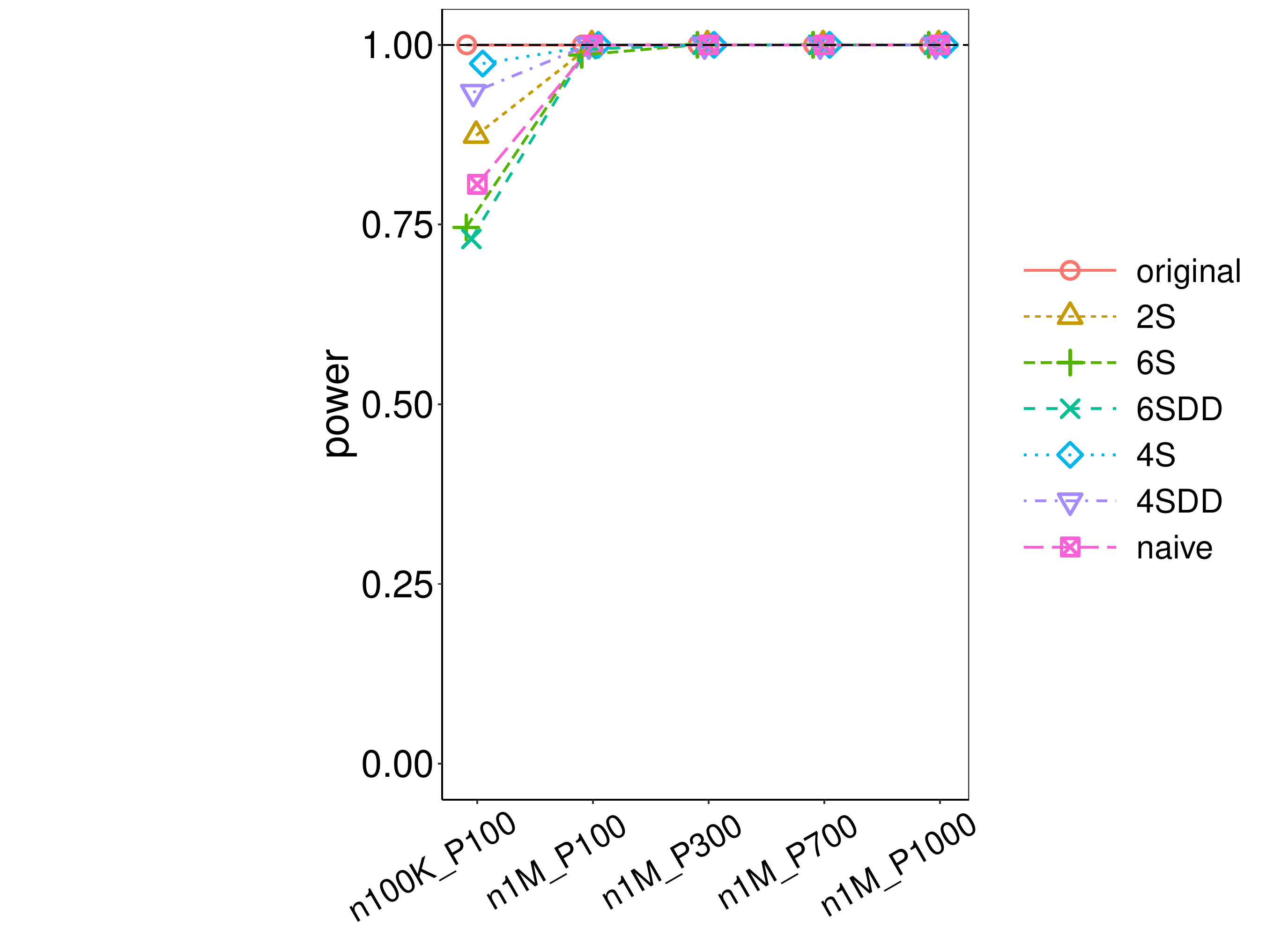}
\includegraphics[width=0.19\textwidth, trim={2.5in 0 2.6in 0},clip] {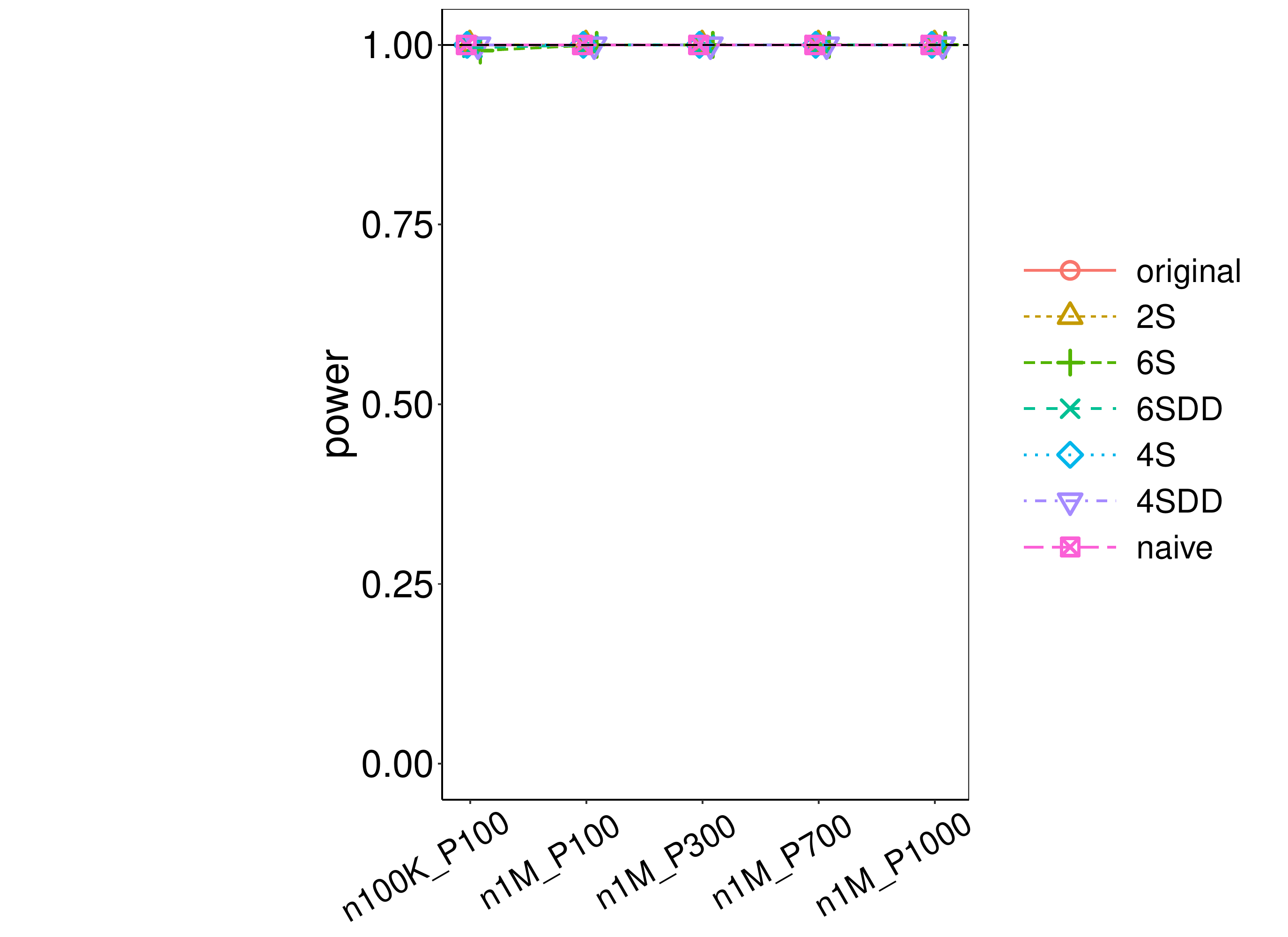}
\includegraphics[width=0.19\textwidth, trim={2.5in 0 2.6in 0},clip] {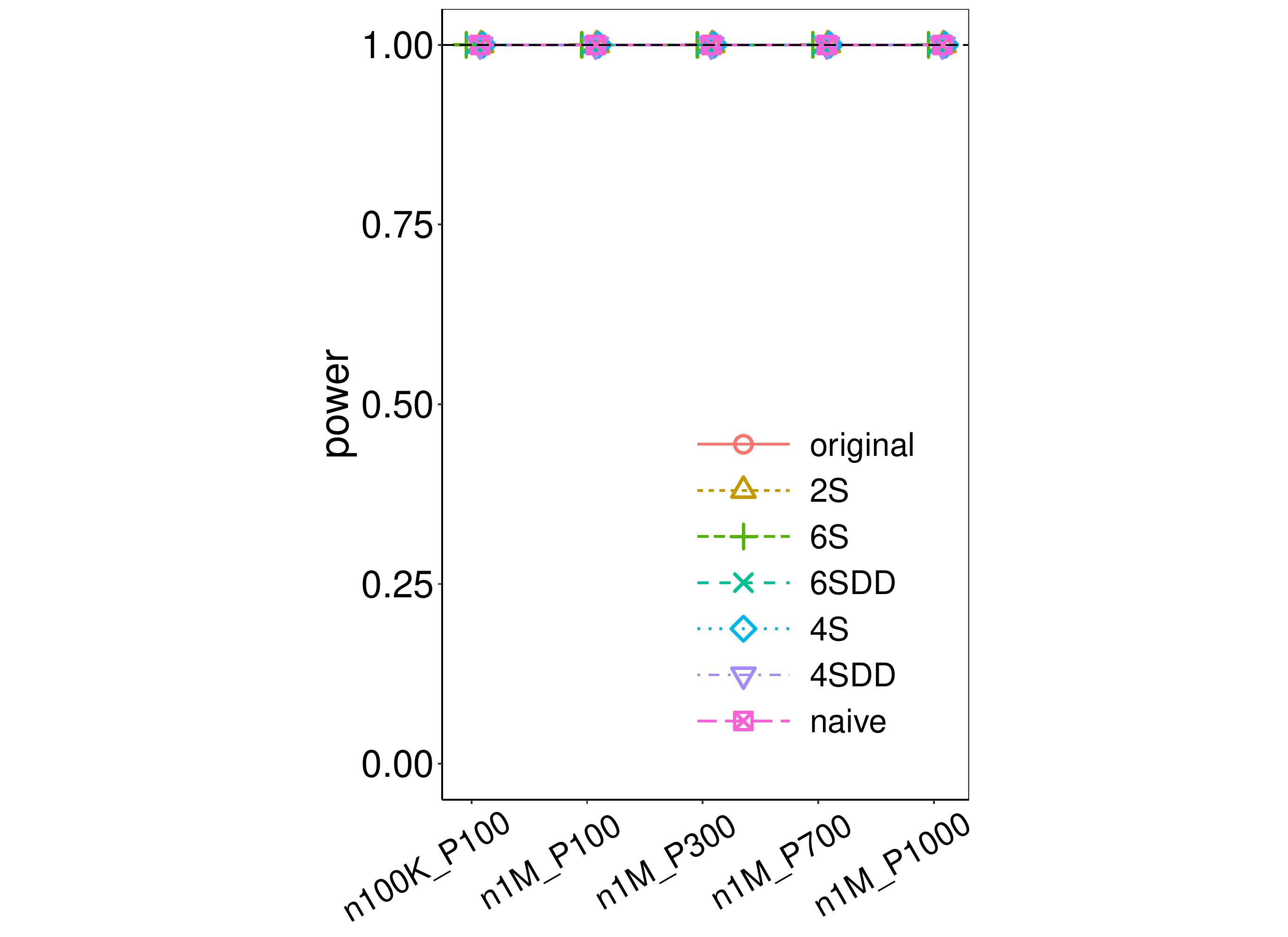}
\caption{Simulation results with $\epsilon$-DP for ZINB data with  $\alpha\ne\beta$ when $\theta\ne0$} \label{fig:1asDPZINB}

\end{figure}

\begin{figure}[!htb]
\hspace{0.45in}$\rho=0.005$\hspace{0.65in}$\rho=0.02$\hspace{0.65in}$\rho=0.08$
\hspace{0.65in}$\rho=0.32$\hspace{0.65in}$\rho=1.28$

\includegraphics[width=0.19\textwidth, trim={2.45in 0 2.45in 0},clip] {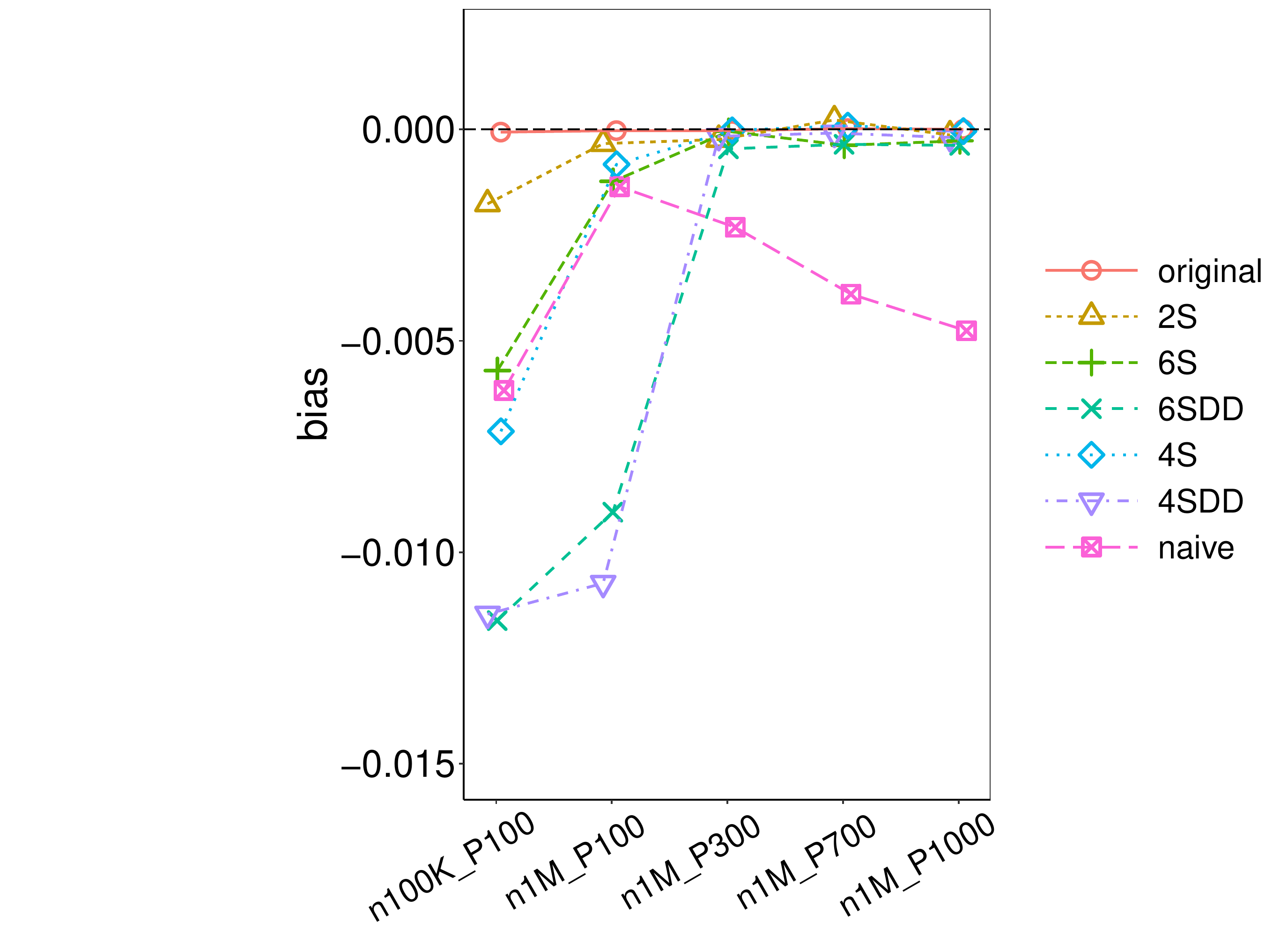}
\includegraphics[width=0.19\textwidth, trim={2.45in 0 2.45in 0},clip] {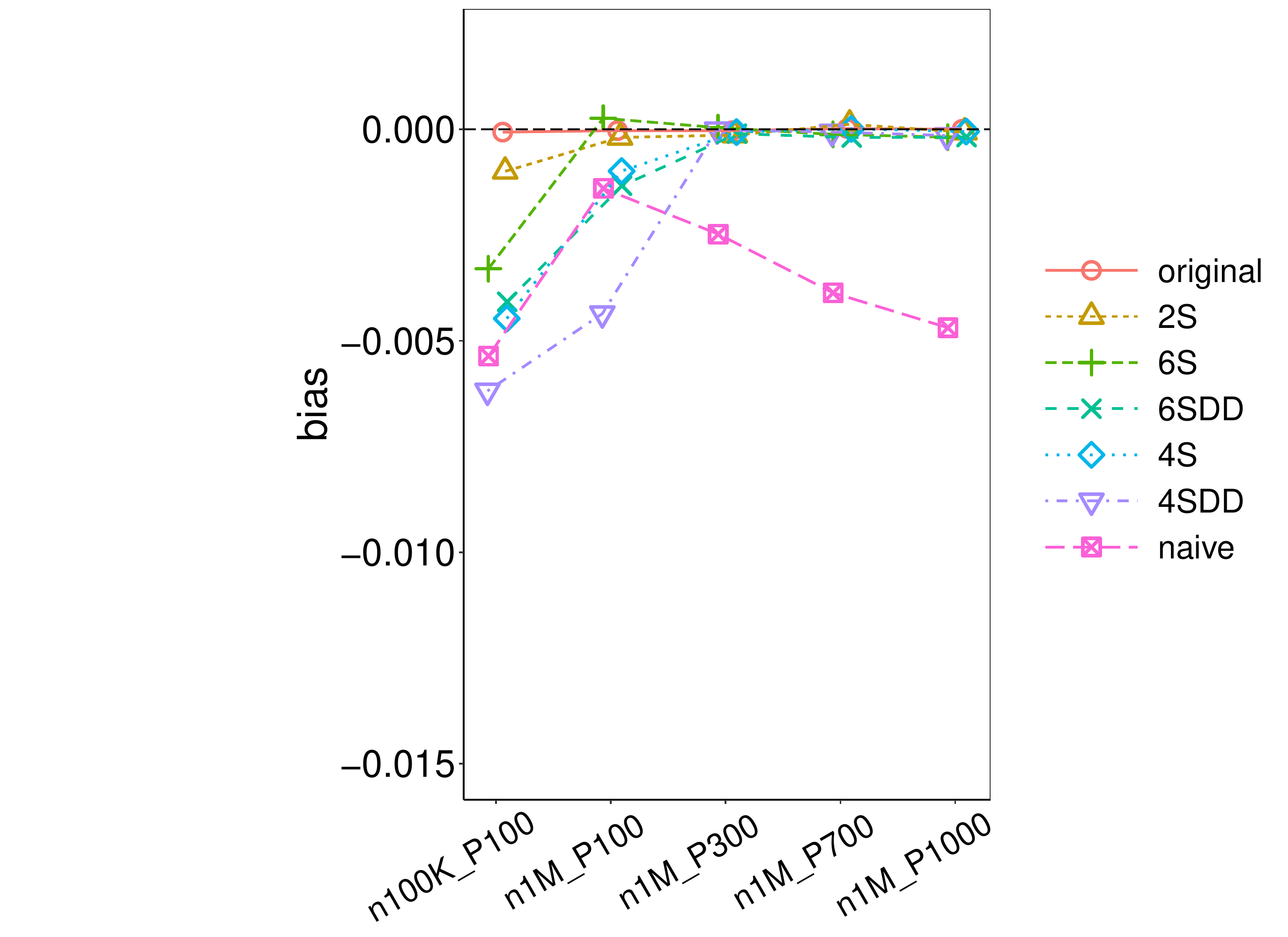}
\includegraphics[width=0.19\textwidth, trim={2.45in 0 2.45in 0},clip] {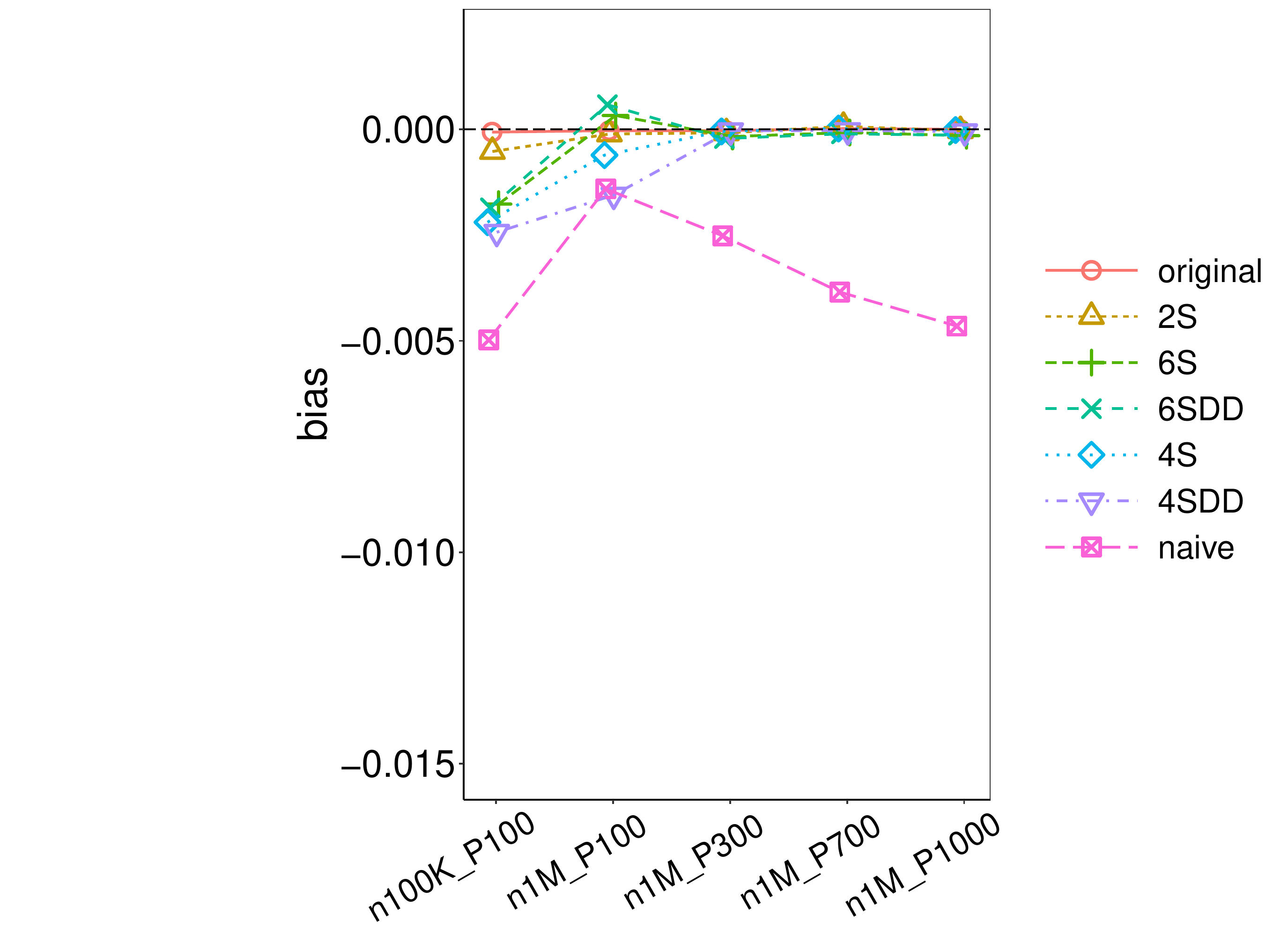}
\includegraphics[width=0.19\textwidth, trim={2.45in 0 2.45in 0},clip] {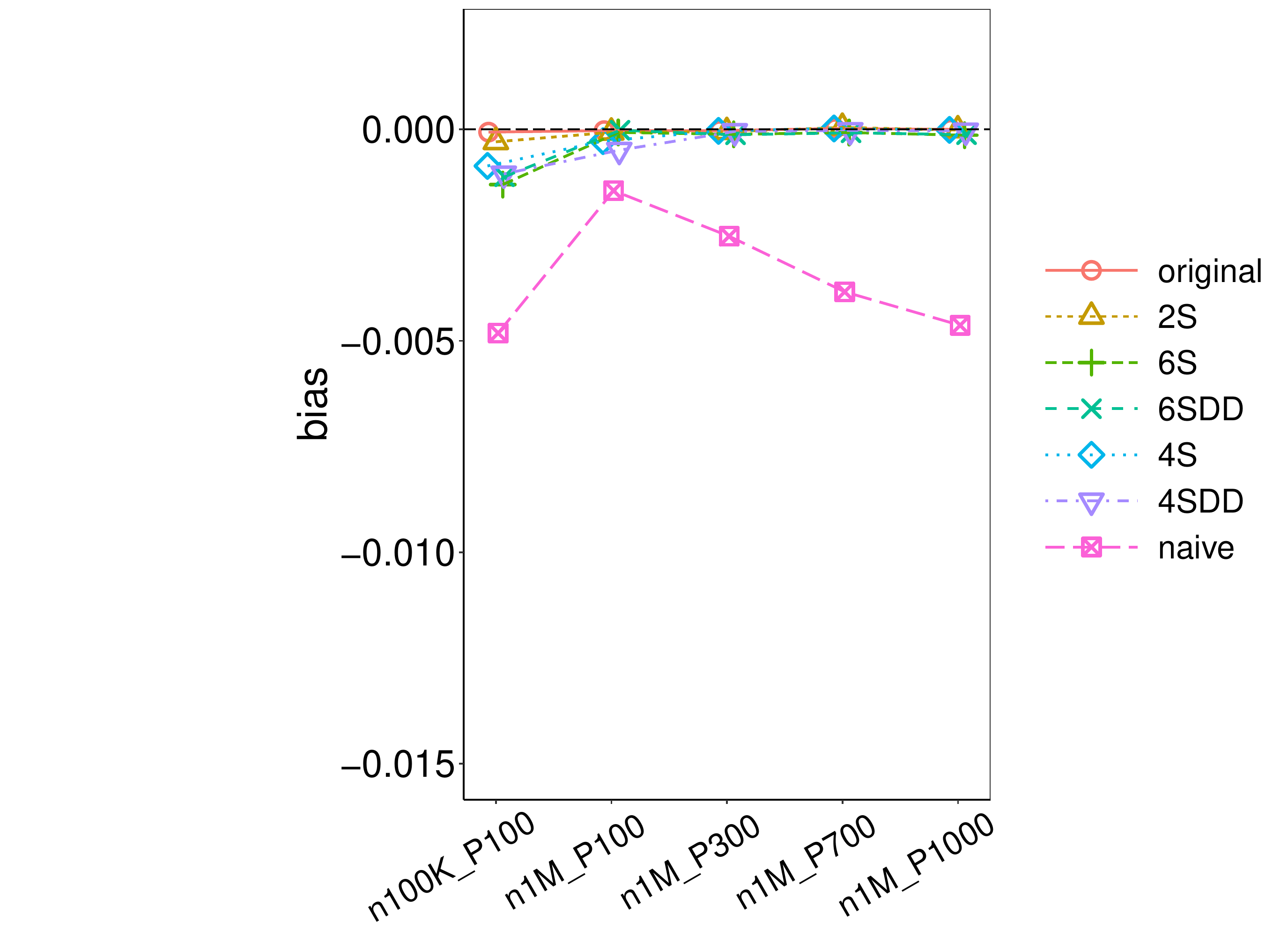}
\includegraphics[width=0.19\textwidth, trim={2.45in 0 2.45in 0},clip] {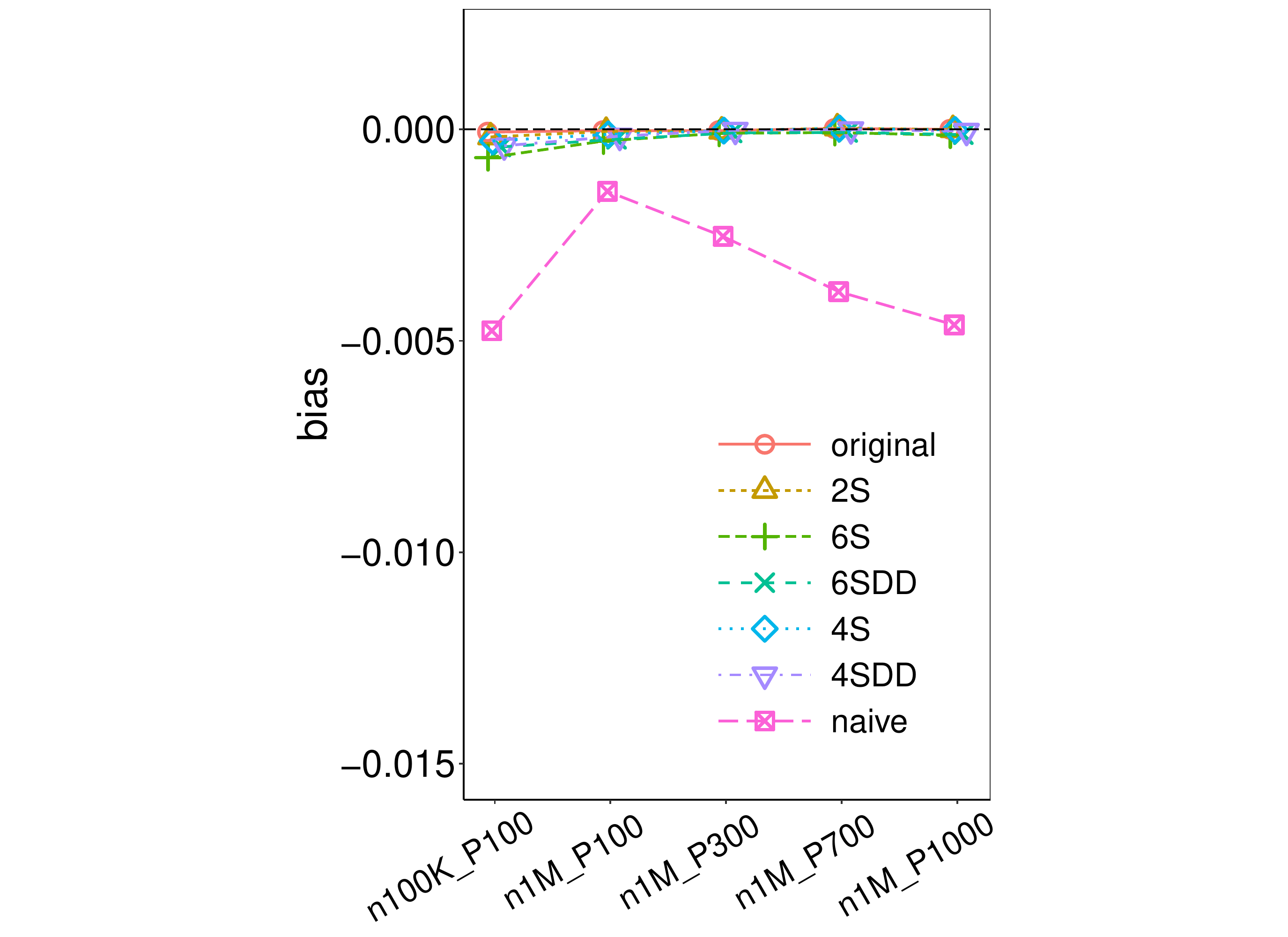}

\includegraphics[width=0.19\textwidth, trim={2.5in 0 2.6in 0},clip] {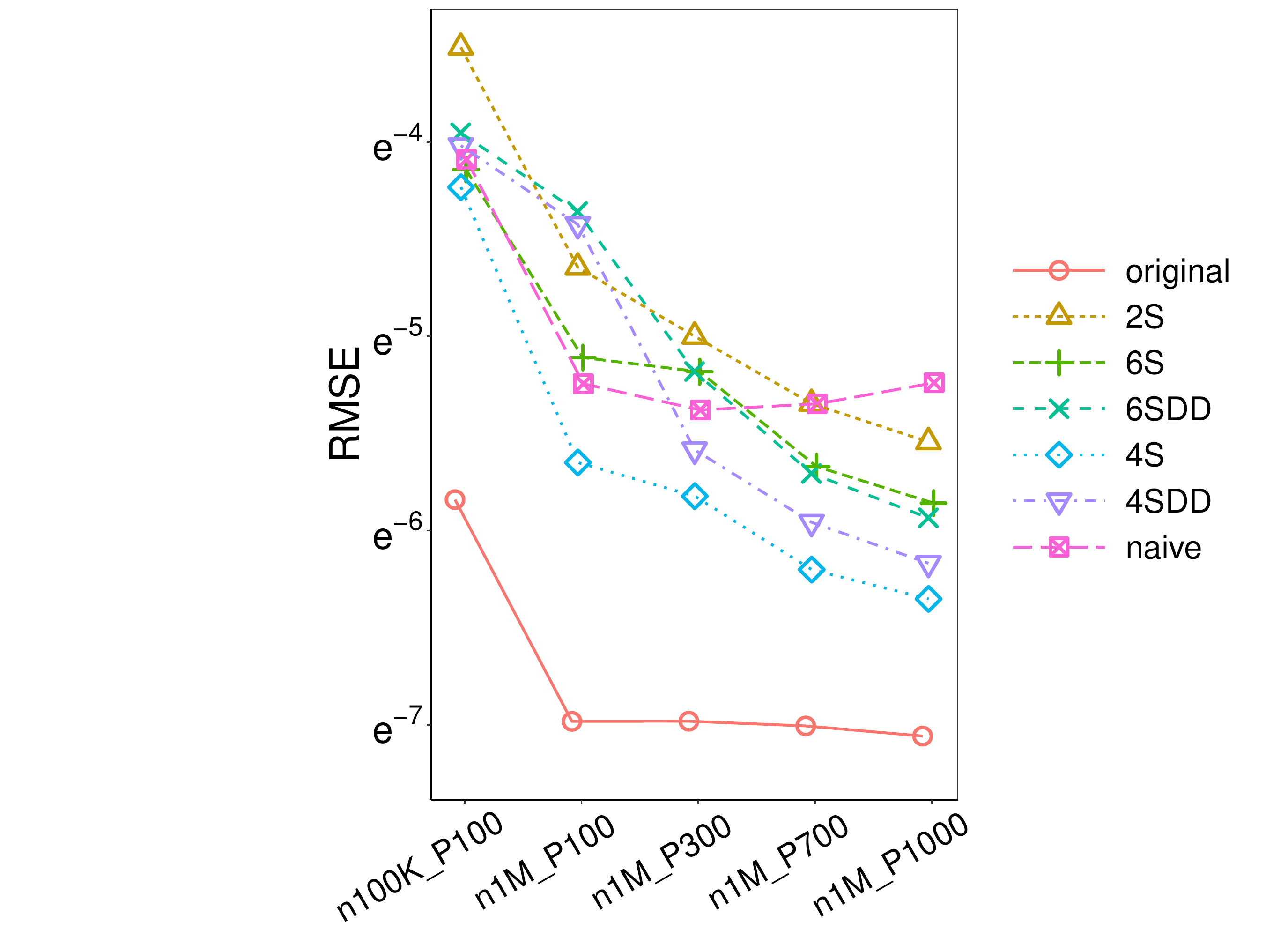}
\includegraphics[width=0.19\textwidth, trim={2.5in 0 2.6in 0},clip] {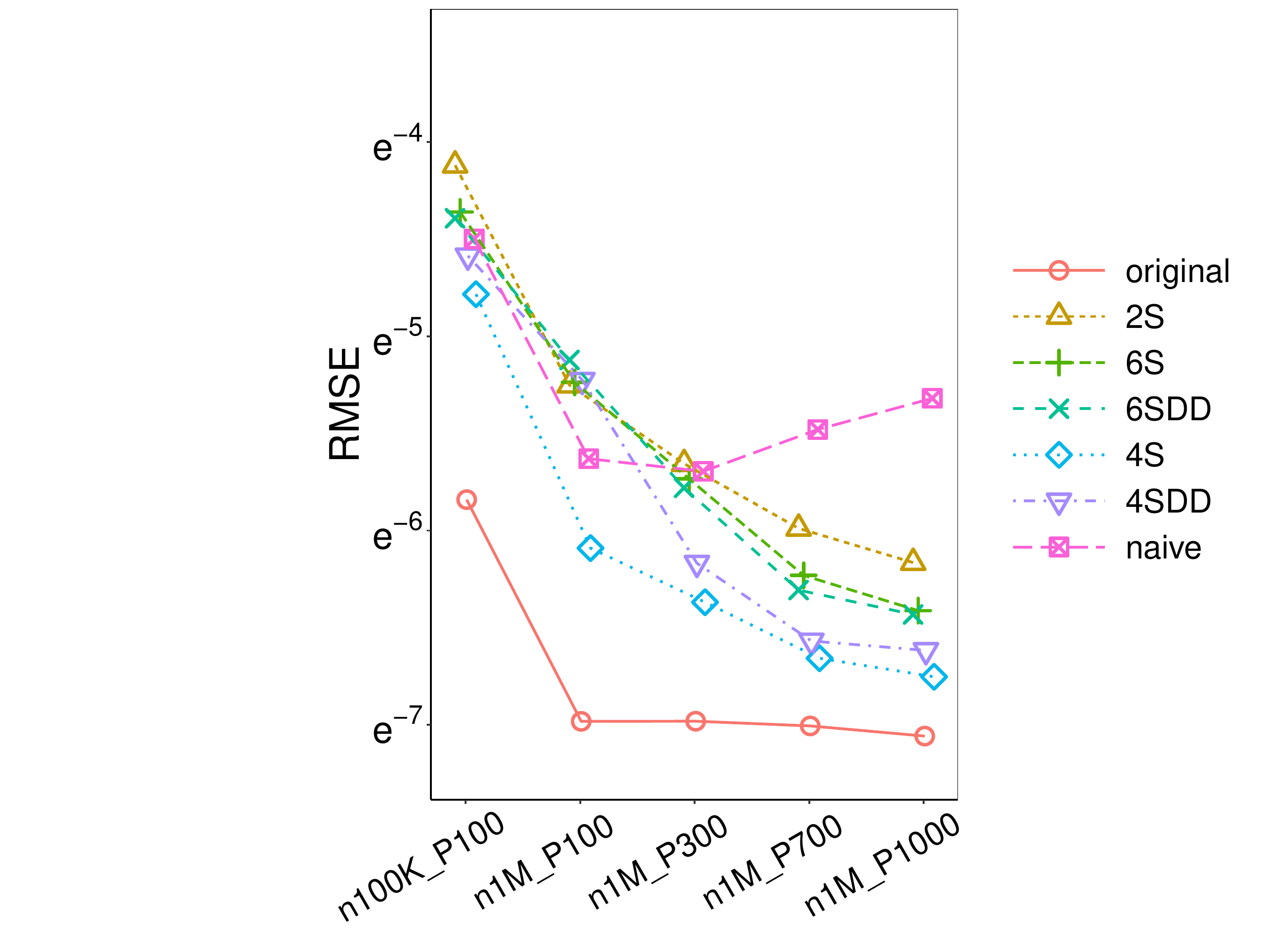}
\includegraphics[width=0.19\textwidth, trim={2.5in 0 2.6in 0},clip] {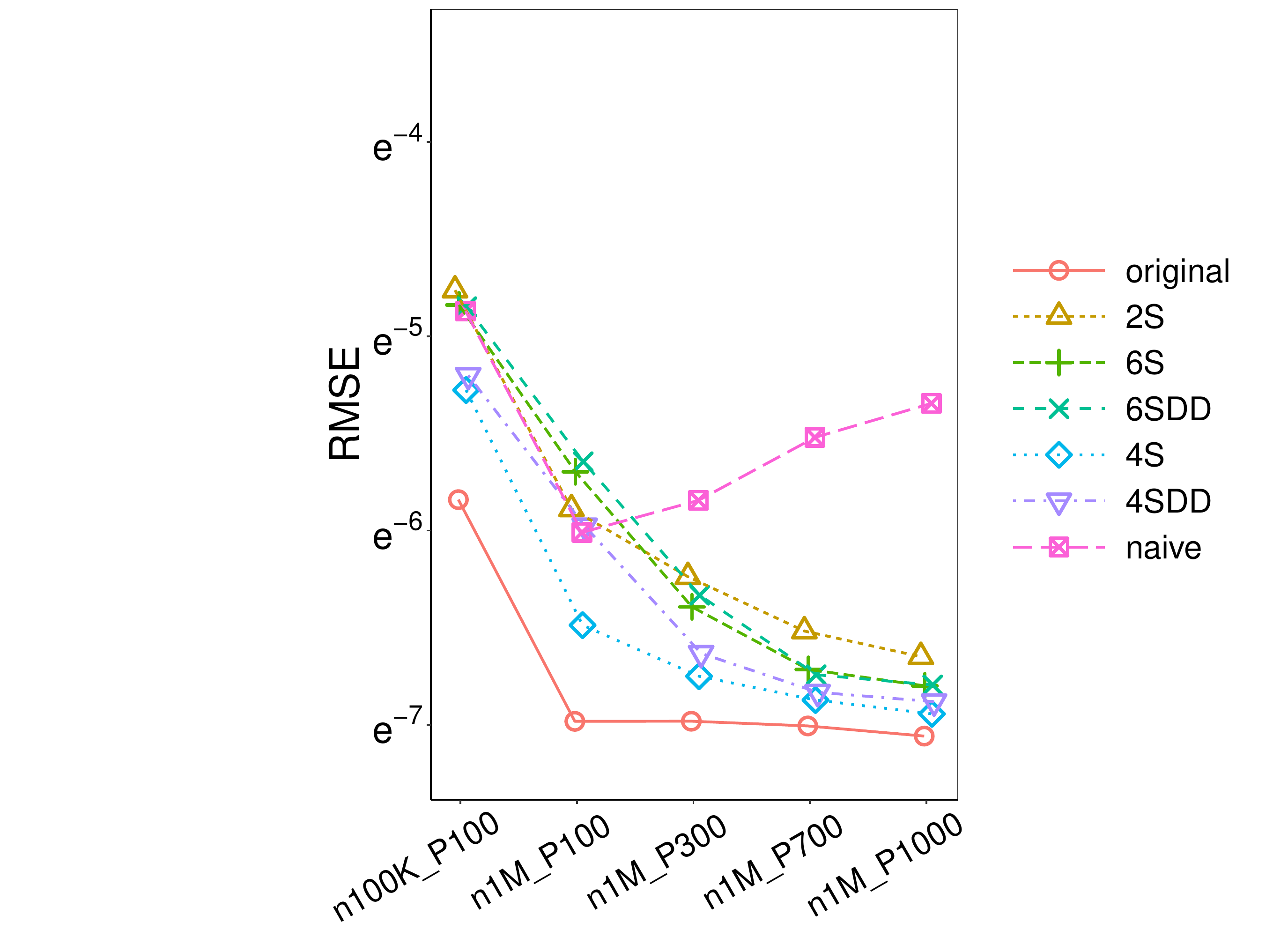}
\includegraphics[width=0.19\textwidth, trim={2.5in 0 2.6in 0},clip] {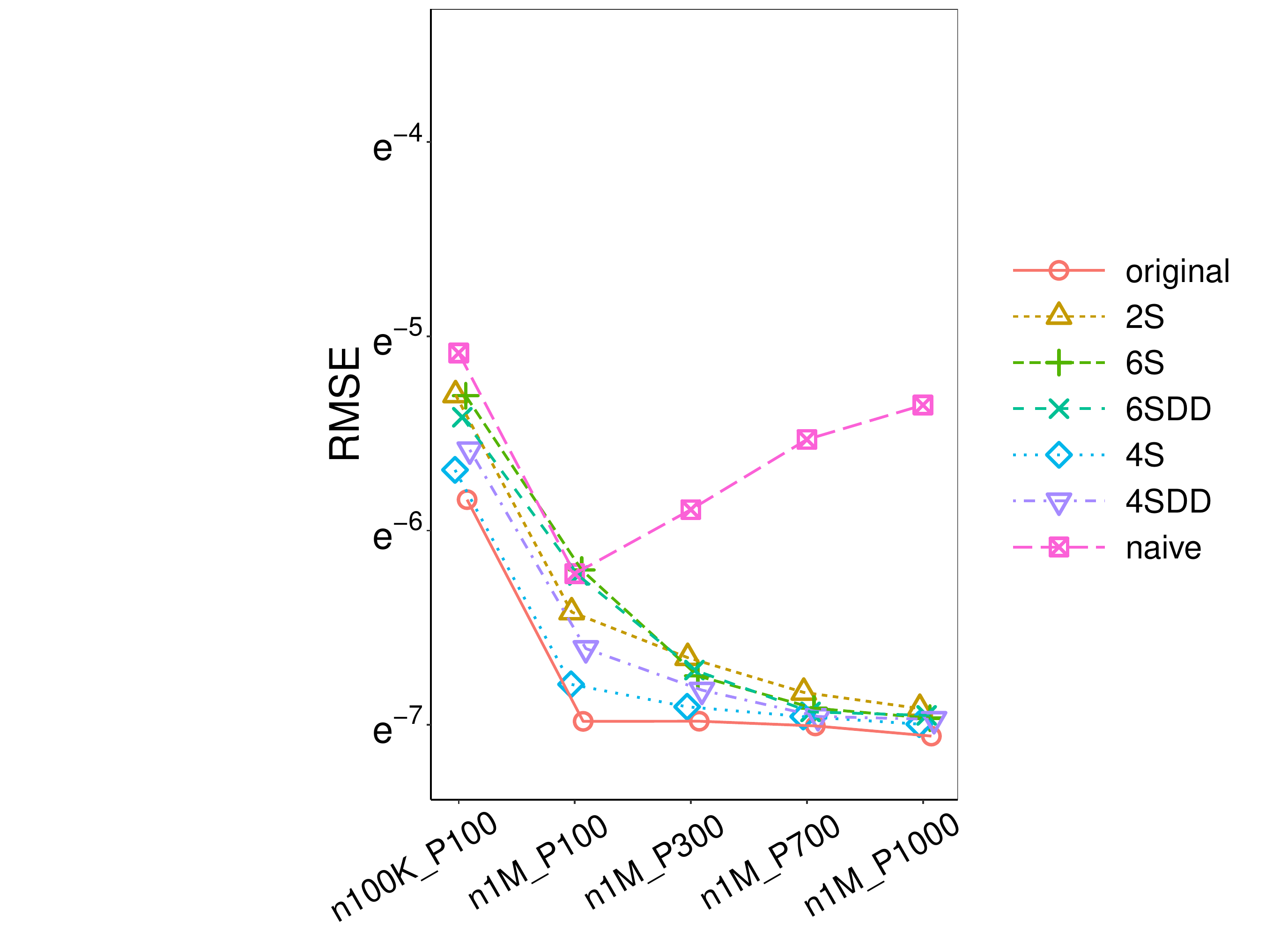}
\includegraphics[width=0.19\textwidth, trim={2.5in 0 2.6in 0},clip] {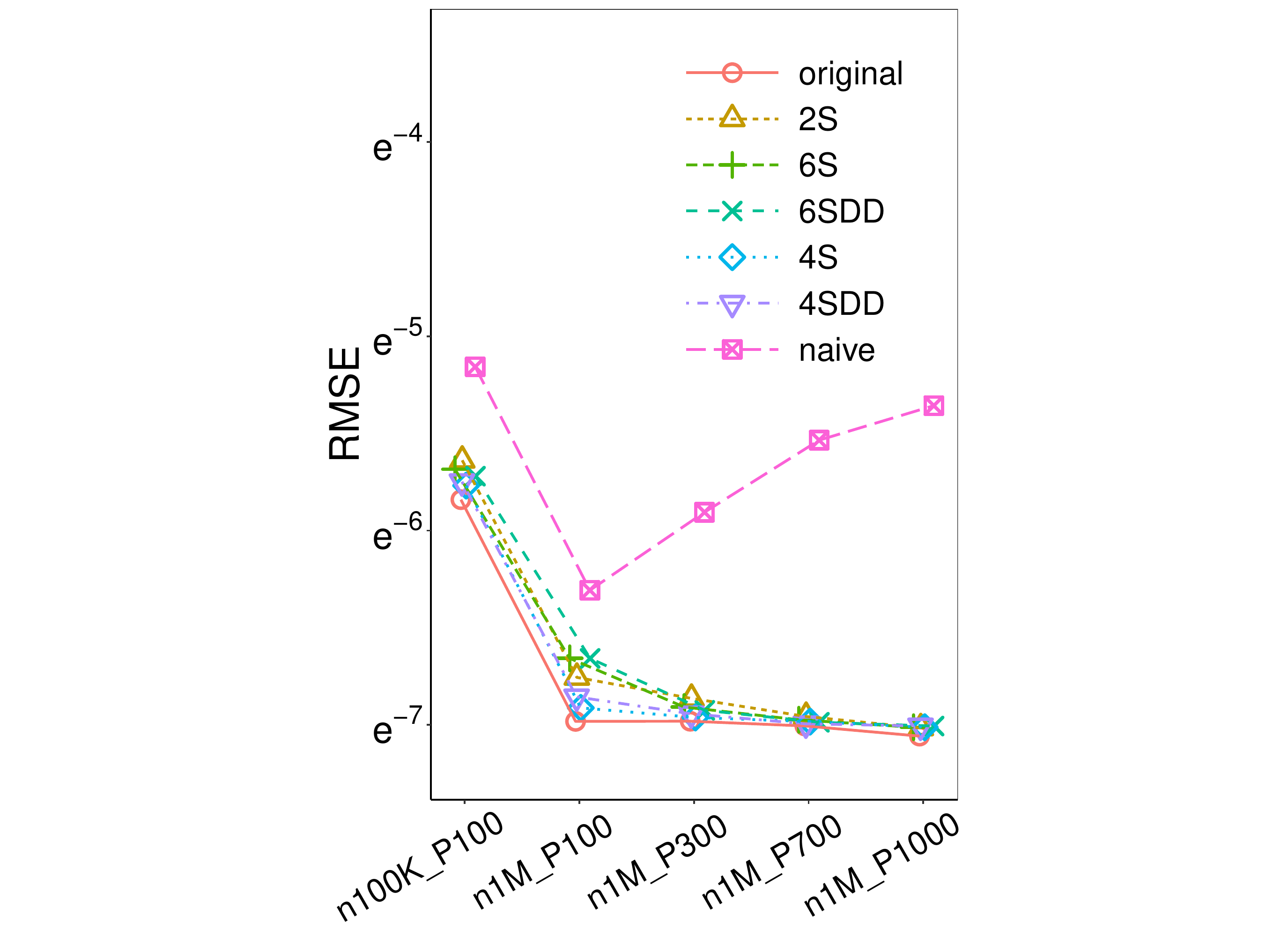}

\includegraphics[width=0.19\textwidth, trim={2.5in 0 2.6in 0},clip] {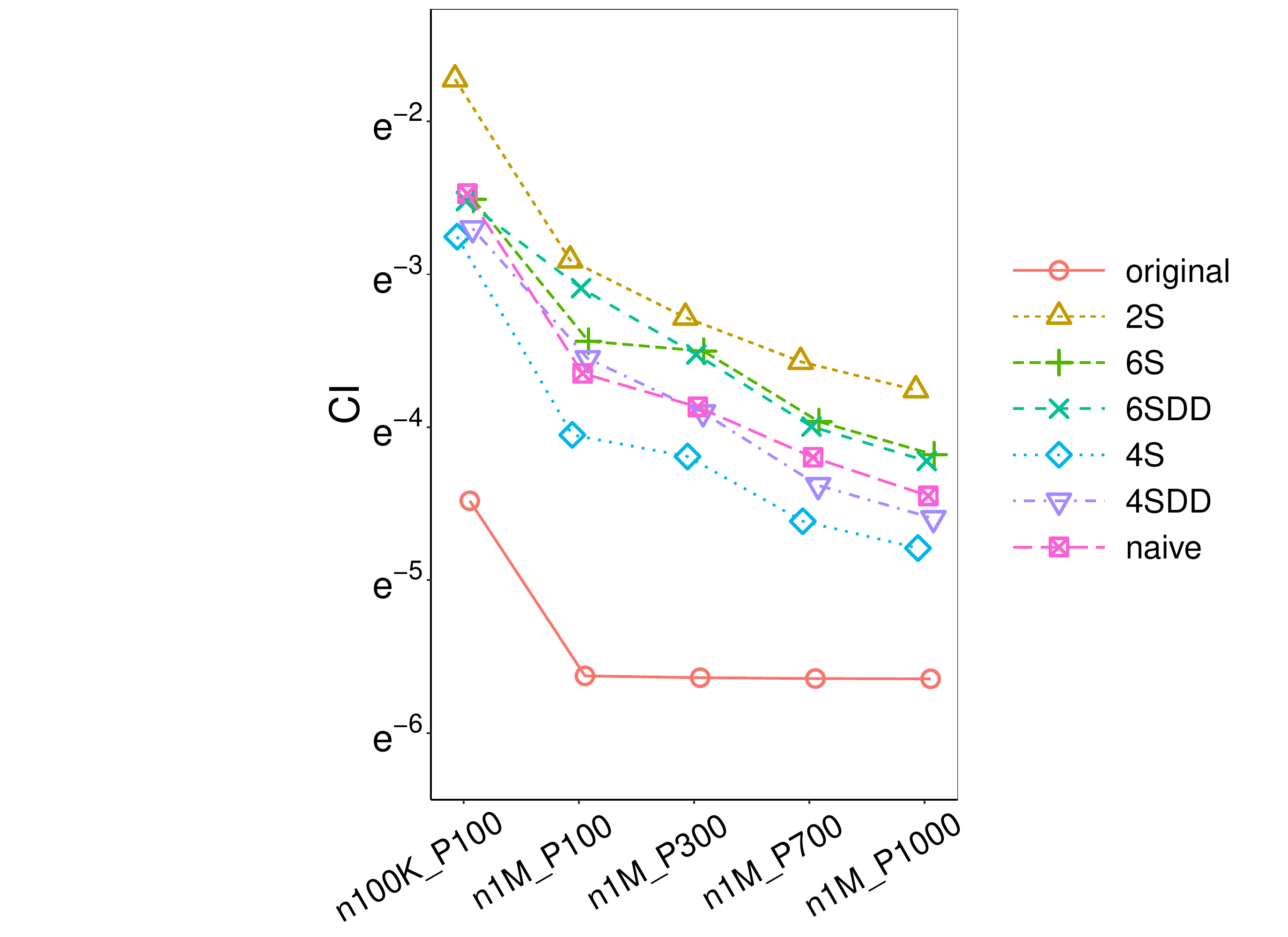}
\includegraphics[width=0.19\textwidth, trim={2.5in 0 2.6in 0},clip] {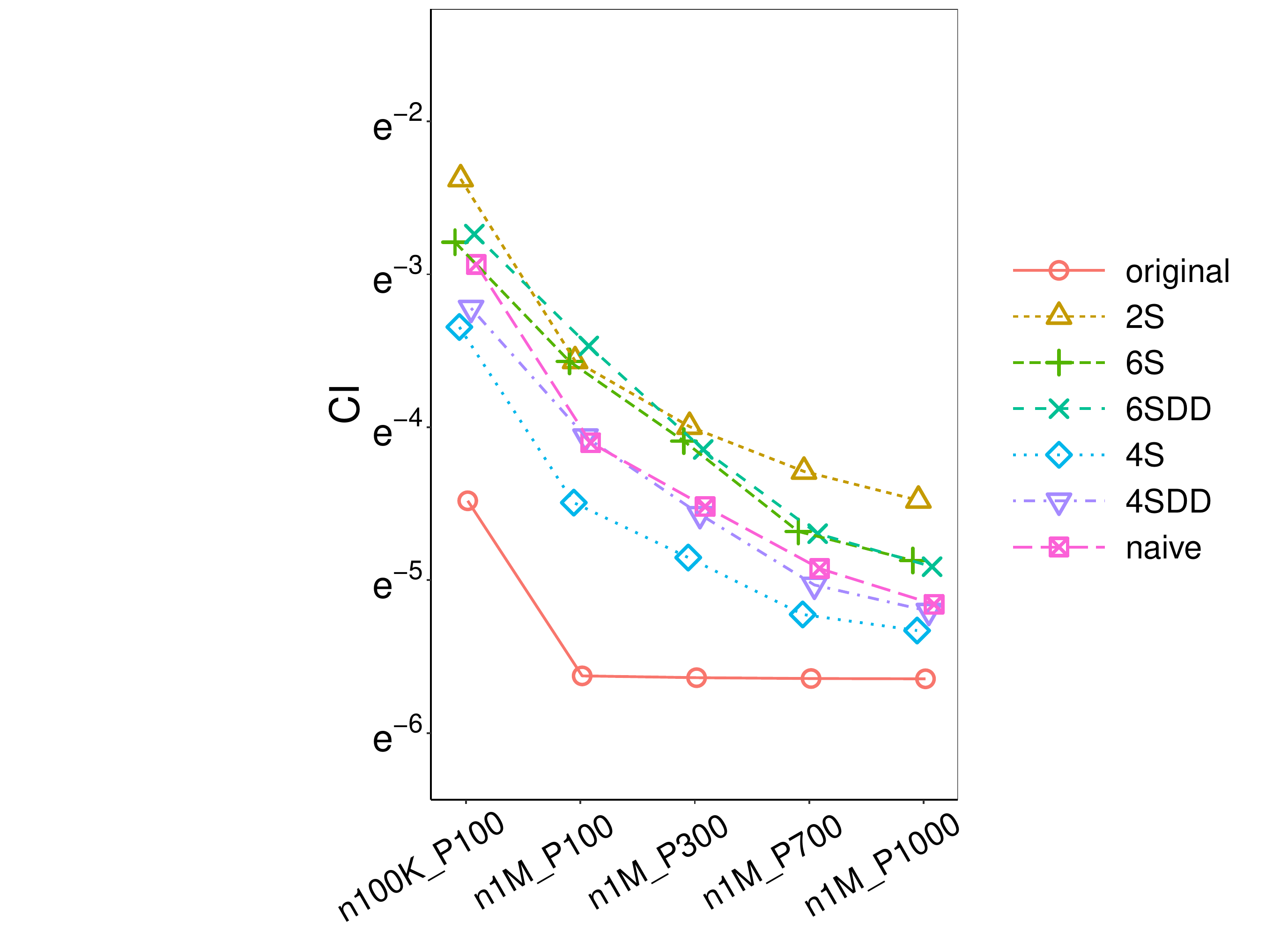}
\includegraphics[width=0.19\textwidth, trim={2.5in 0 2.6in 0},clip] {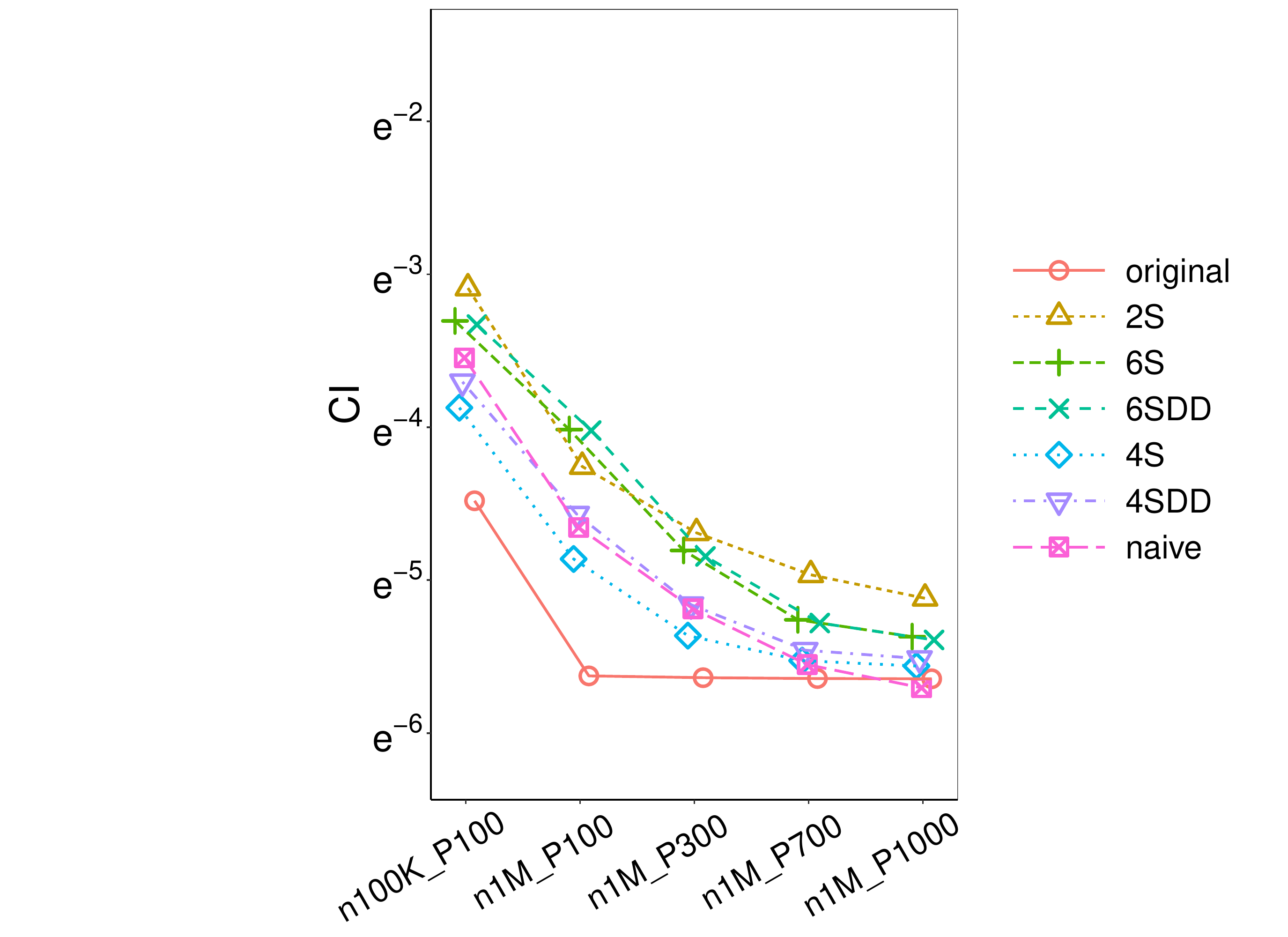}
\includegraphics[width=0.19\textwidth, trim={2.5in 0 2.6in 0},clip] {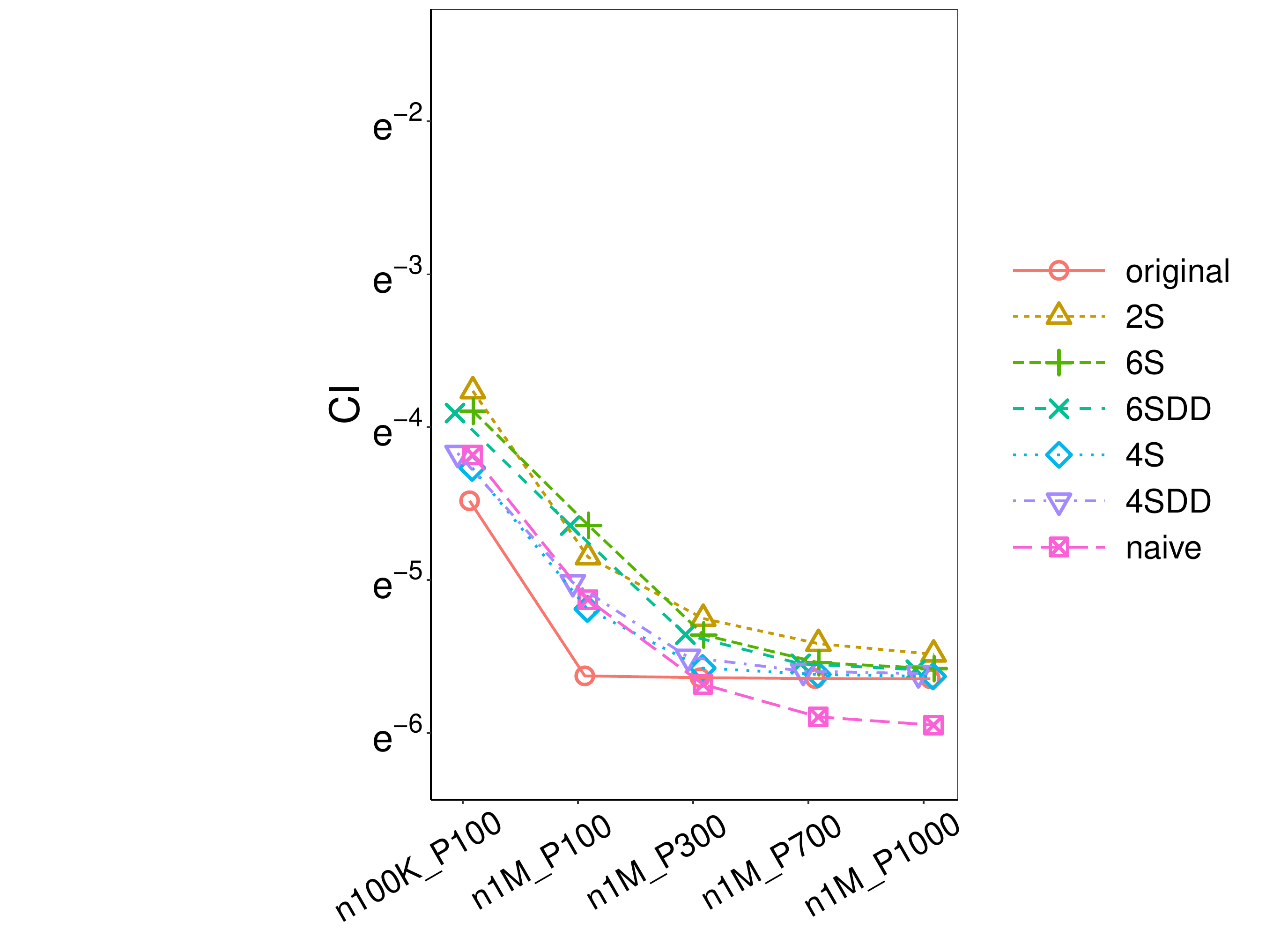}
\includegraphics[width=0.19\textwidth, trim={2.5in 0 2.6in 0},clip] {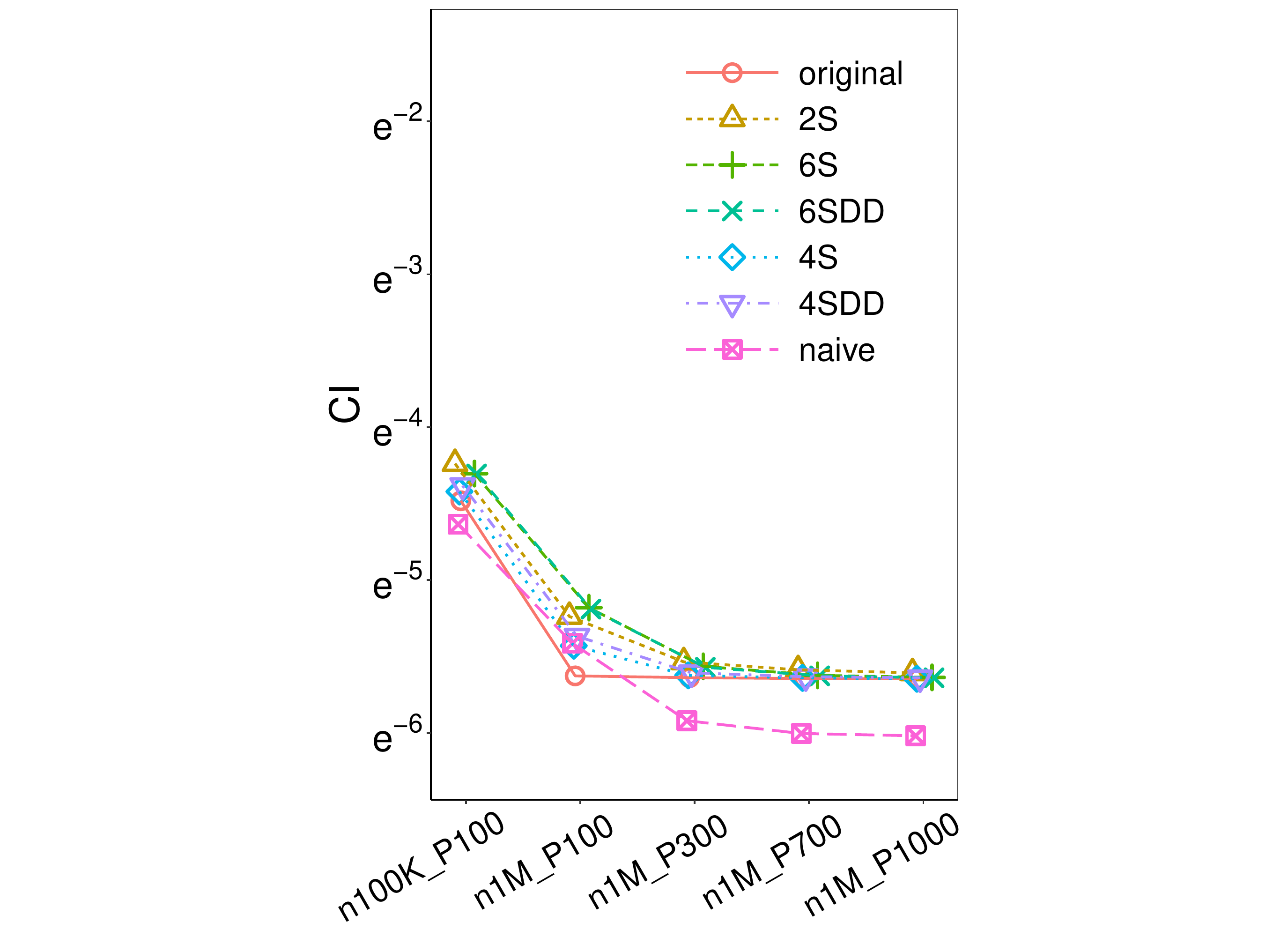}

\includegraphics[width=0.19\textwidth, trim={2.5in 0 2.6in 0},clip] {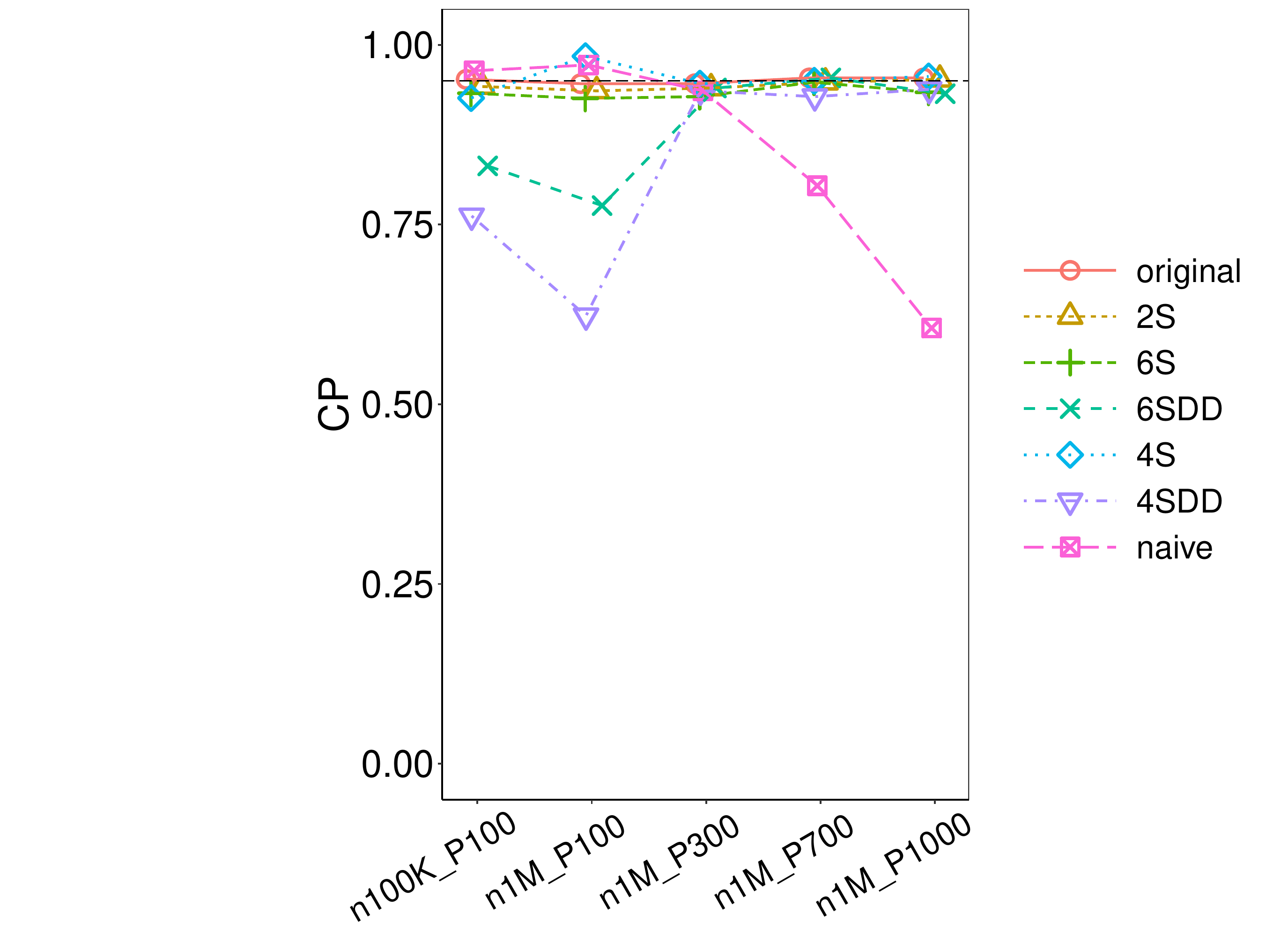}
\includegraphics[width=0.19\textwidth, trim={2.5in 0 2.6in 0},clip] {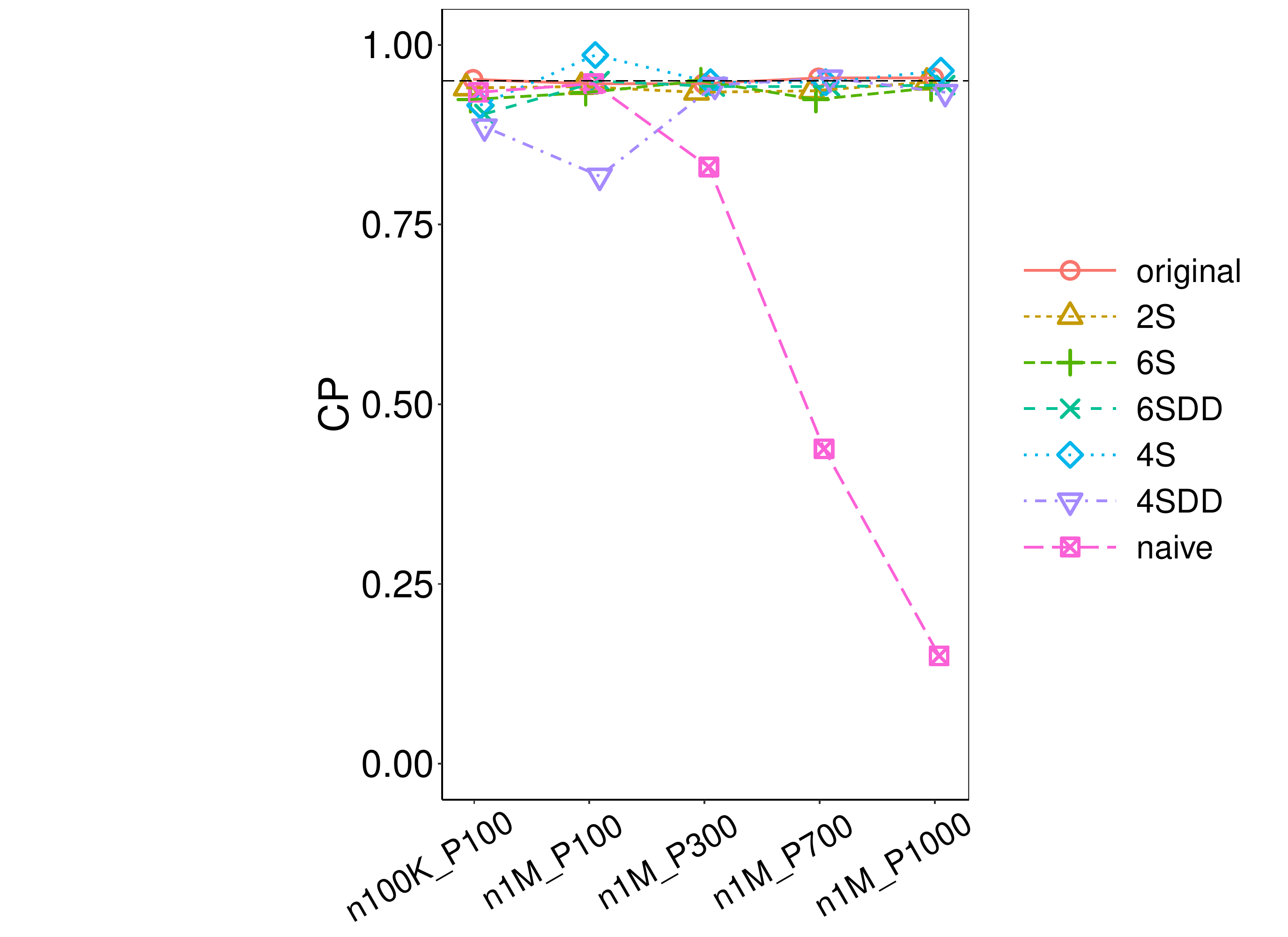}
\includegraphics[width=0.19\textwidth, trim={2.5in 0 2.6in 0},clip] {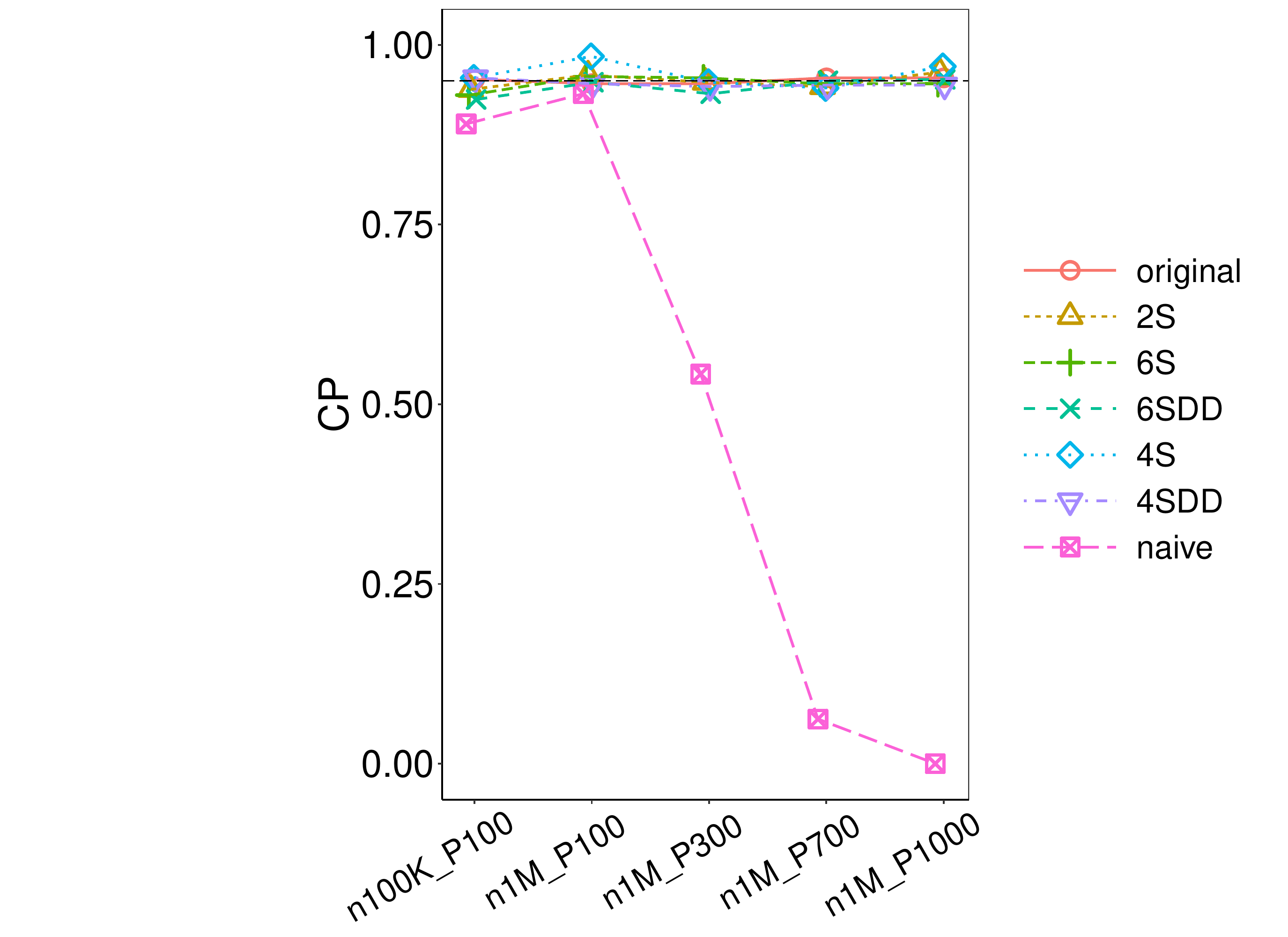}
\includegraphics[width=0.19\textwidth, trim={2.5in 0 2.6in 0},clip] {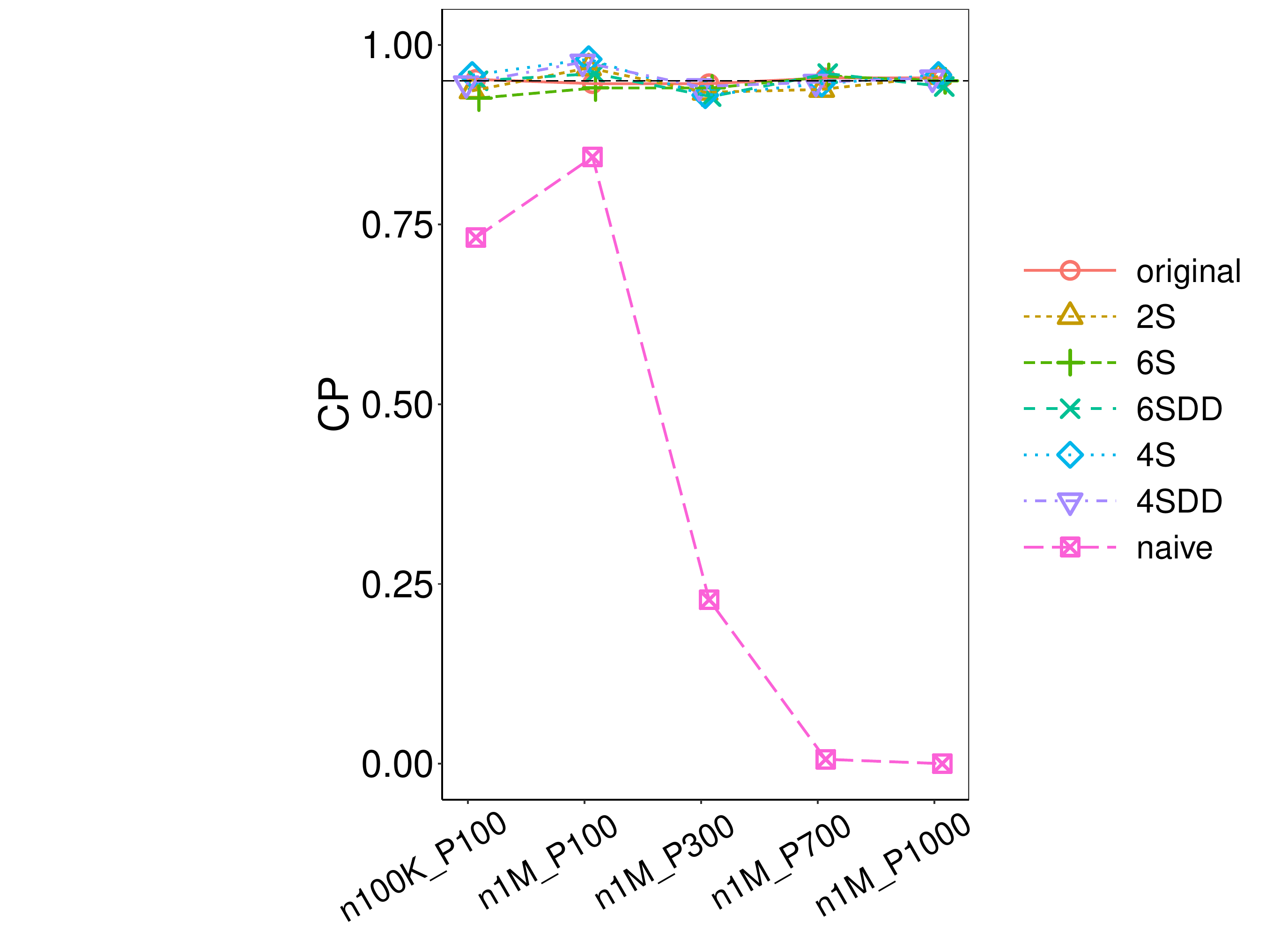}
\includegraphics[width=0.19\textwidth, trim={2.5in 0 2.6in 0},clip] {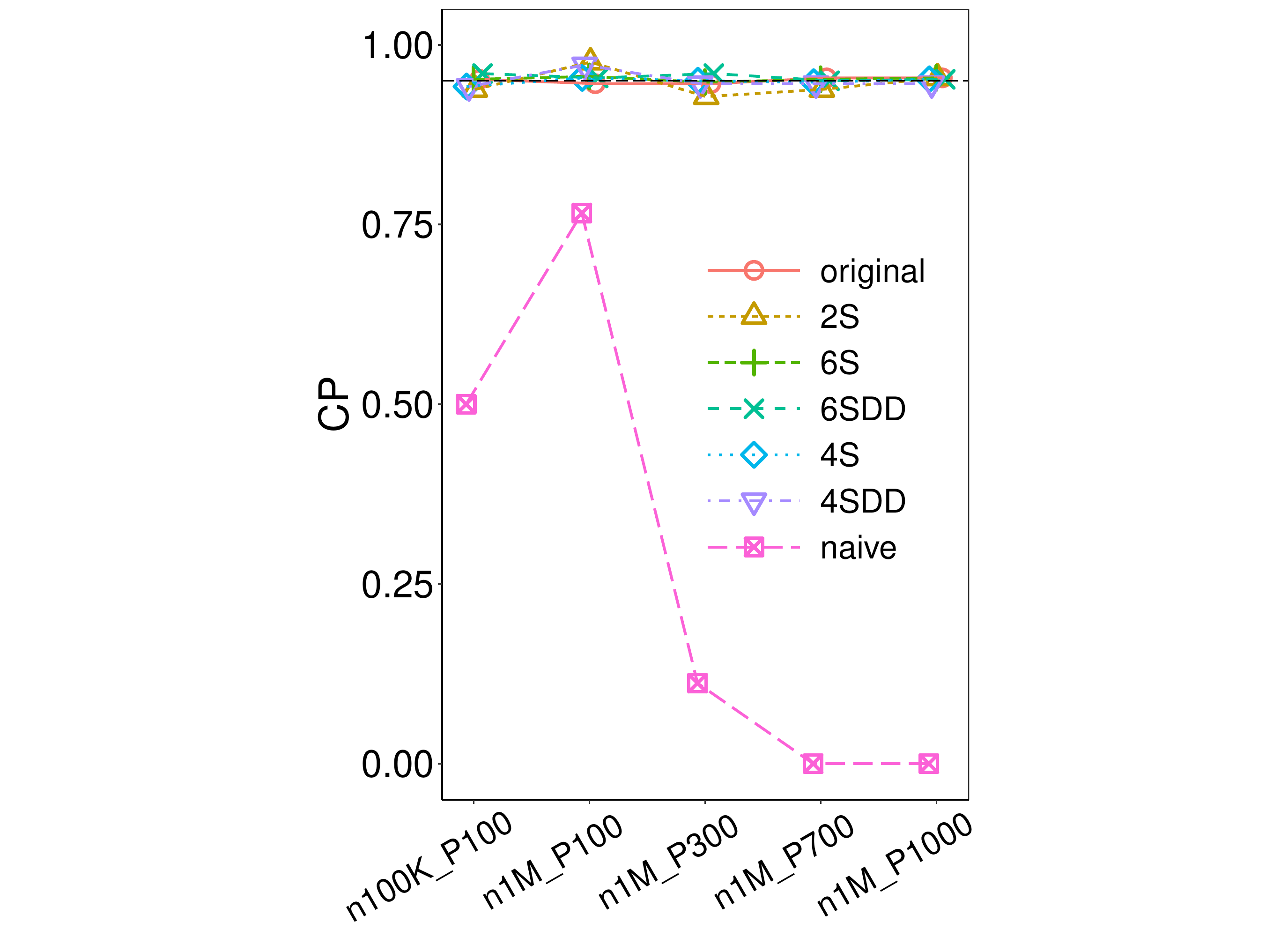}

\includegraphics[width=0.19\textwidth, trim={2.5in 0 2.6in 0},clip] {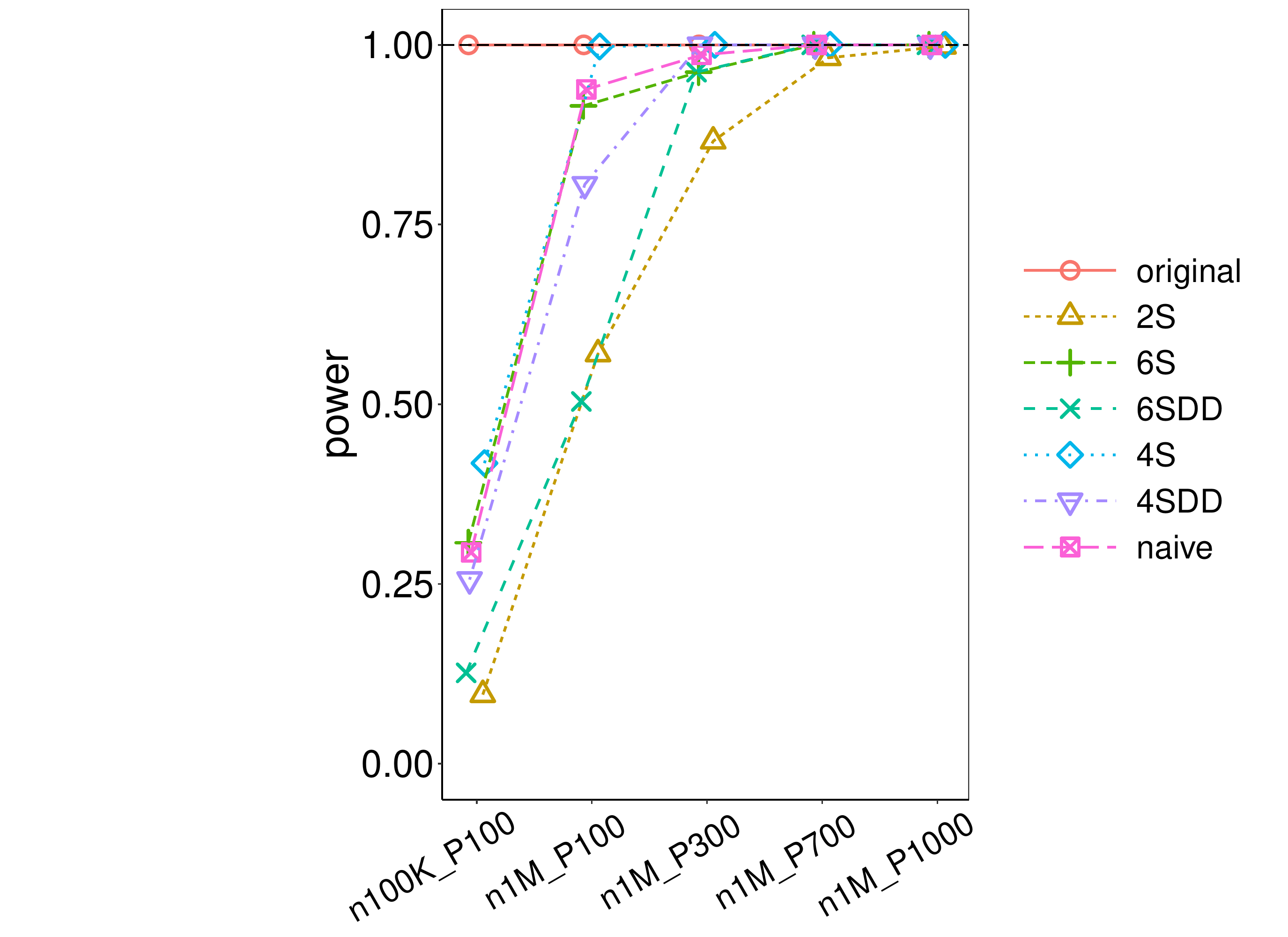}
\includegraphics[width=0.19\textwidth, trim={2.5in 0 2.6in 0},clip] {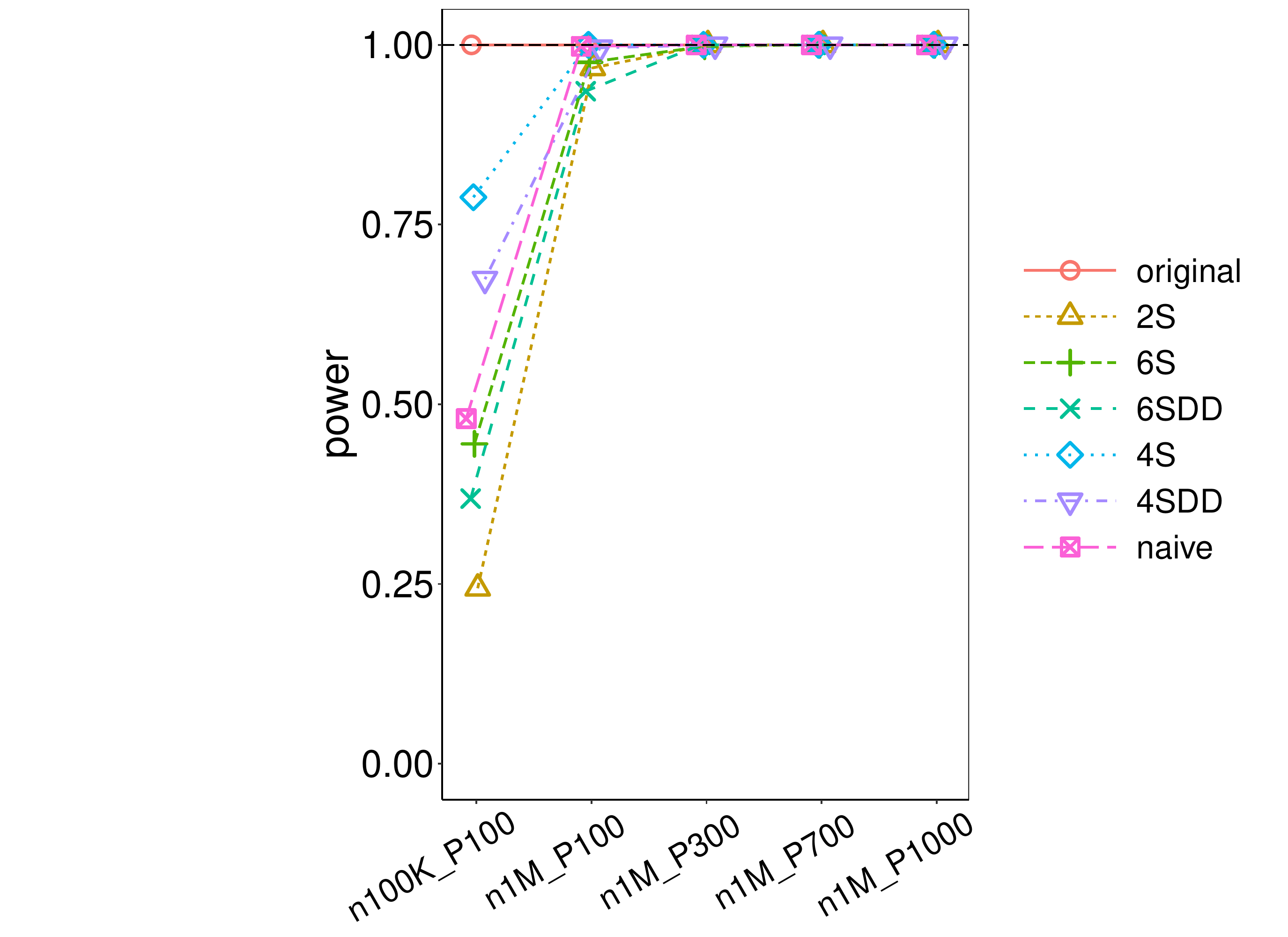}
\includegraphics[width=0.19\textwidth, trim={2.5in 0 2.6in 0},clip] {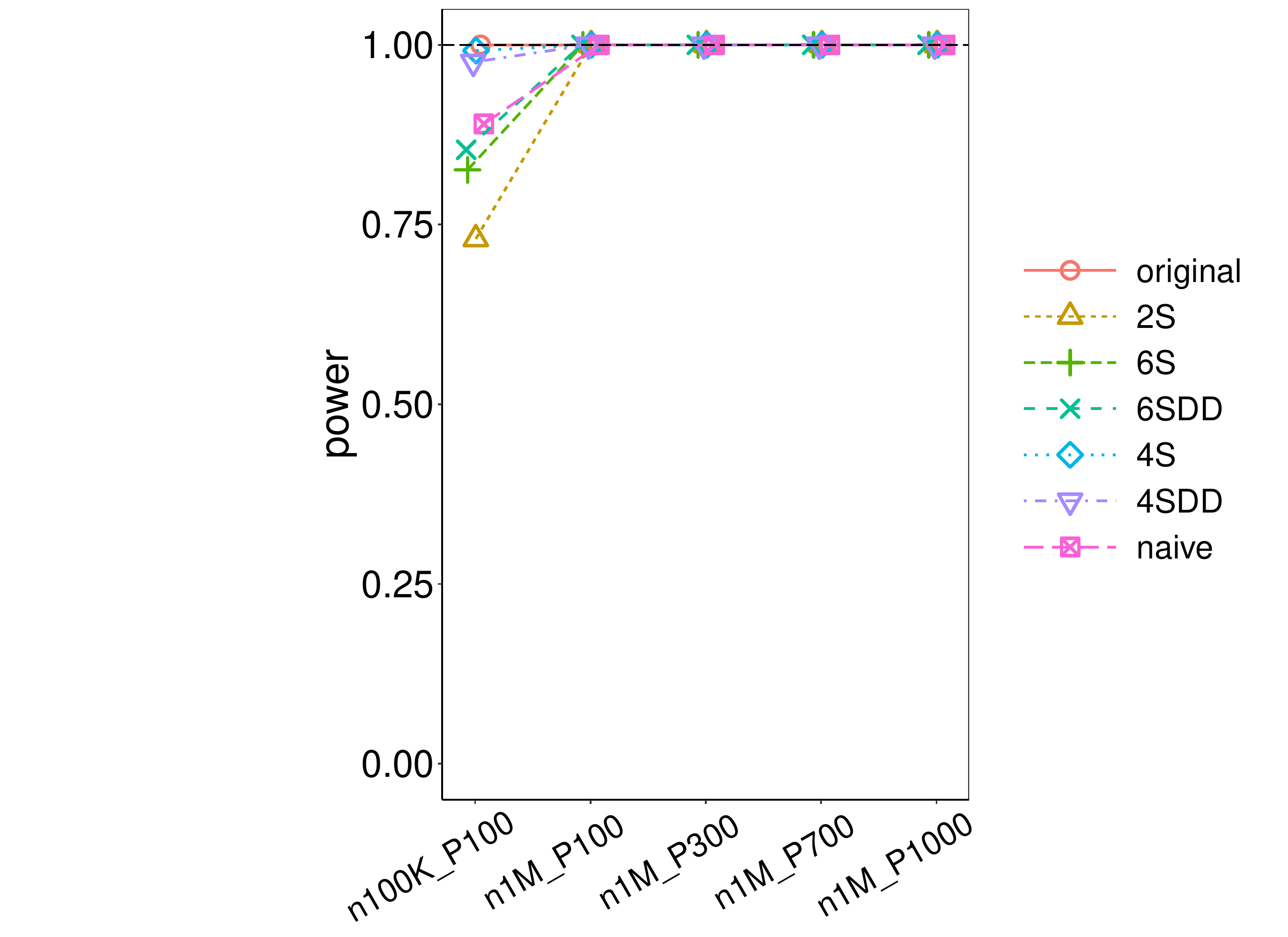}
\includegraphics[width=0.19\textwidth, trim={2.5in 0 2.6in 0},clip] {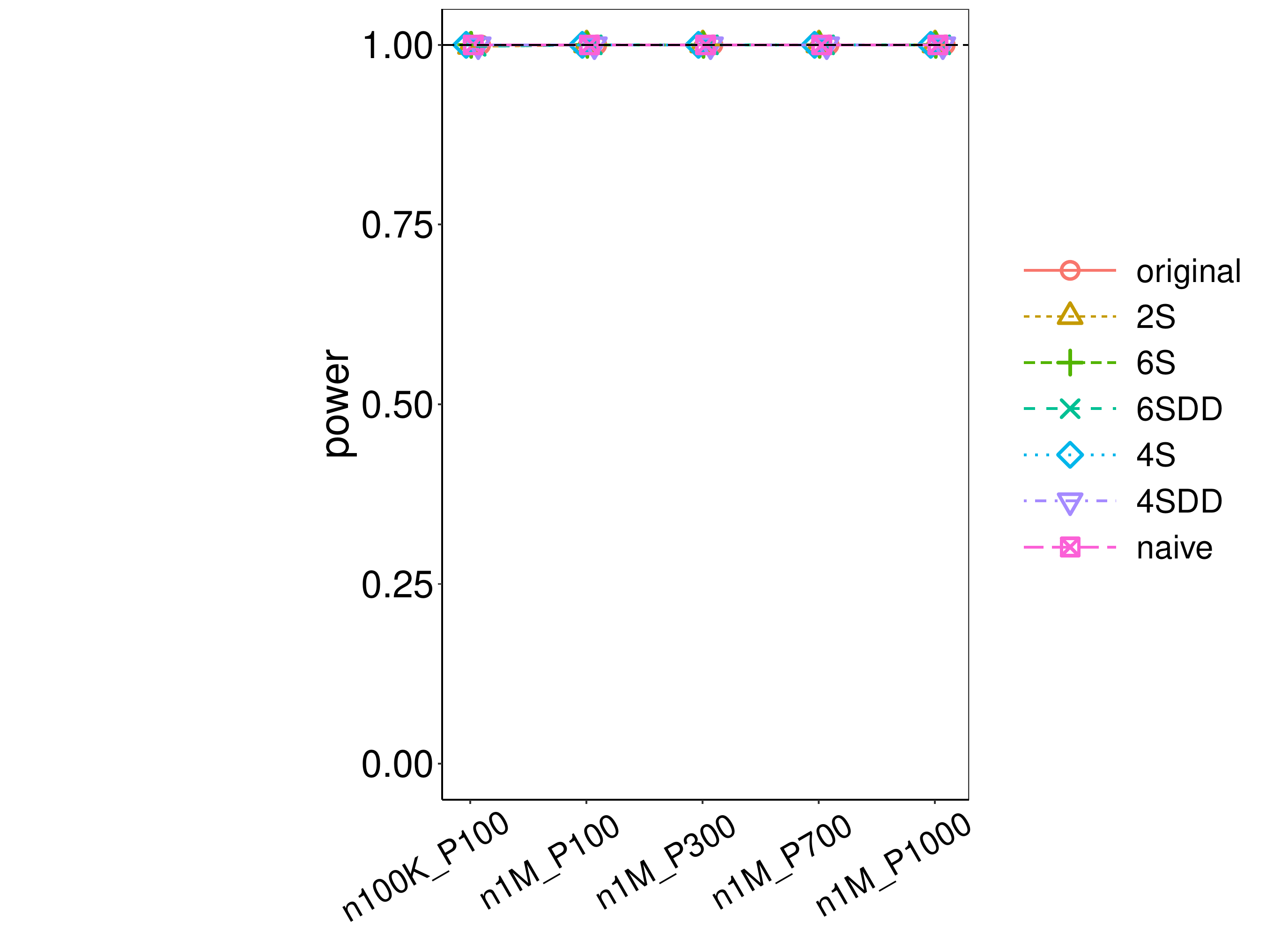}
\includegraphics[width=0.19\textwidth, trim={2.5in 0 2.6in 0},clip] {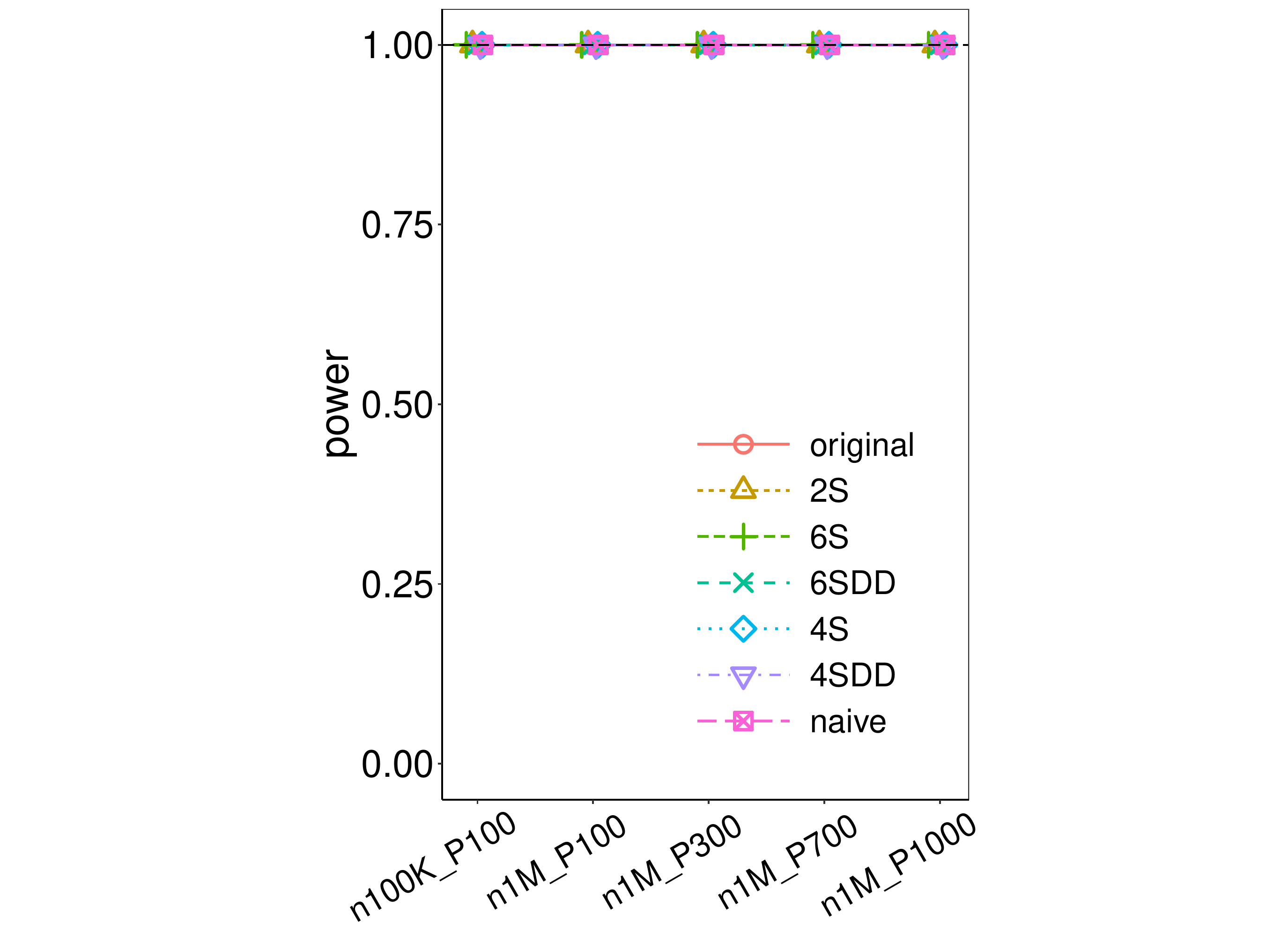}

\caption{Simulation results with $\rho$-zCDP for ZINB data with  $\alpha\ne\beta$ when $\theta\ne0$}
\label{fig:1aszCDPZINB}
\end{figure}

The results are presented in Figures \ref{fig:1asDPZINB} and \ref{fig:1aszCDPZINB} and are summarized  as follows. 4S is the best performer by all metrics and the na\"{i}ve method is the worst. In between the two, some approaches are better than others, depending on the evaluation metrics. In terms of bias, 6S approaches 0 faster than 2S, 6SDD, and 4SDD  as $n$ or $P$ increases when $n=1m$ or $\epsilon$ increases. In terms of RMSE and CI width,
4S significantly outperforms the others, with 6S and 6SDD being the worst for $\epsilon \le 5$. As for CP, 4S, 6S, and 2S provide nominal-level coverage (4S exhibits slight over-coverage at $n=100k/P=100$ and $n=1m/P=100$ for $\epsilon\le1$) while 4SDD and 6SDD shows significant under-coverage. In terms of power, 4S is the most powerful, all simulation settings considered. The Gaussian mechanism of $\rho$-zCDP does not seem to be advantageous in statistical utility compared to the Laplace mechanism of $\epsilon$-DP. 

\subsubsection{\texorpdfstring{$Y\sim$ ZILN}{}, \texorpdfstring{$\theta=0$ ($H_0$)}{H0 }, and \texorpdfstring{$\alpha=\beta$}{S}}\label{sec:0sZILN}
\begin{figure}[!htb]
\hspace{0.5in}$\epsilon=0.5$\hspace{0.8in}$\epsilon=1$\hspace{0.9in}$\epsilon=2$
\hspace{0.95in}$\epsilon=5$\hspace{0.9in}$\epsilon=50$

\includegraphics[width=0.19\textwidth, trim={2.5in 0 2.6in 0},clip] {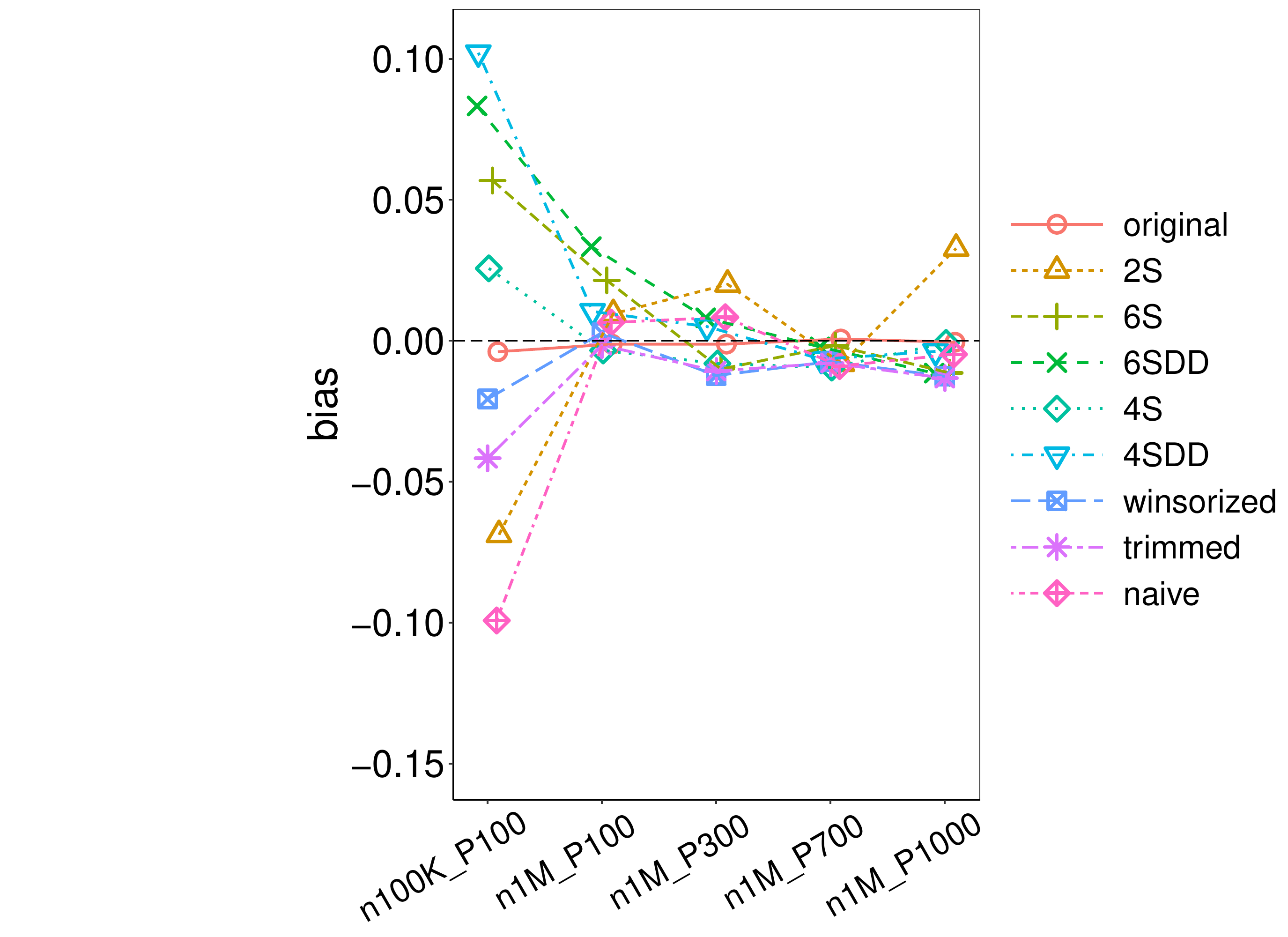}
\includegraphics[width=0.19\textwidth, trim={2.5in 0 2.6in 0},clip] {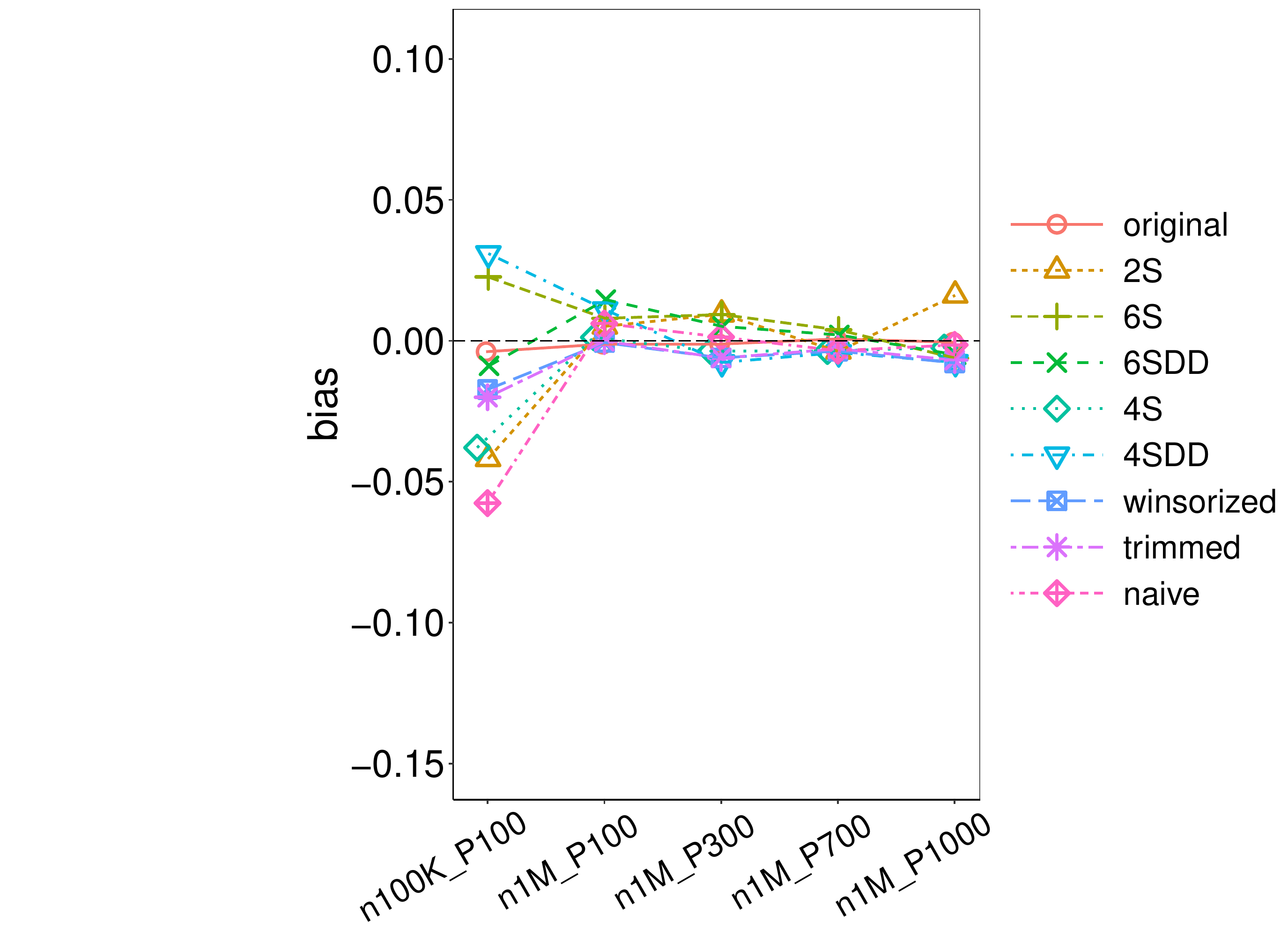}
\includegraphics[width=0.19\textwidth, trim={2.5in 0 2.6in 0},clip] {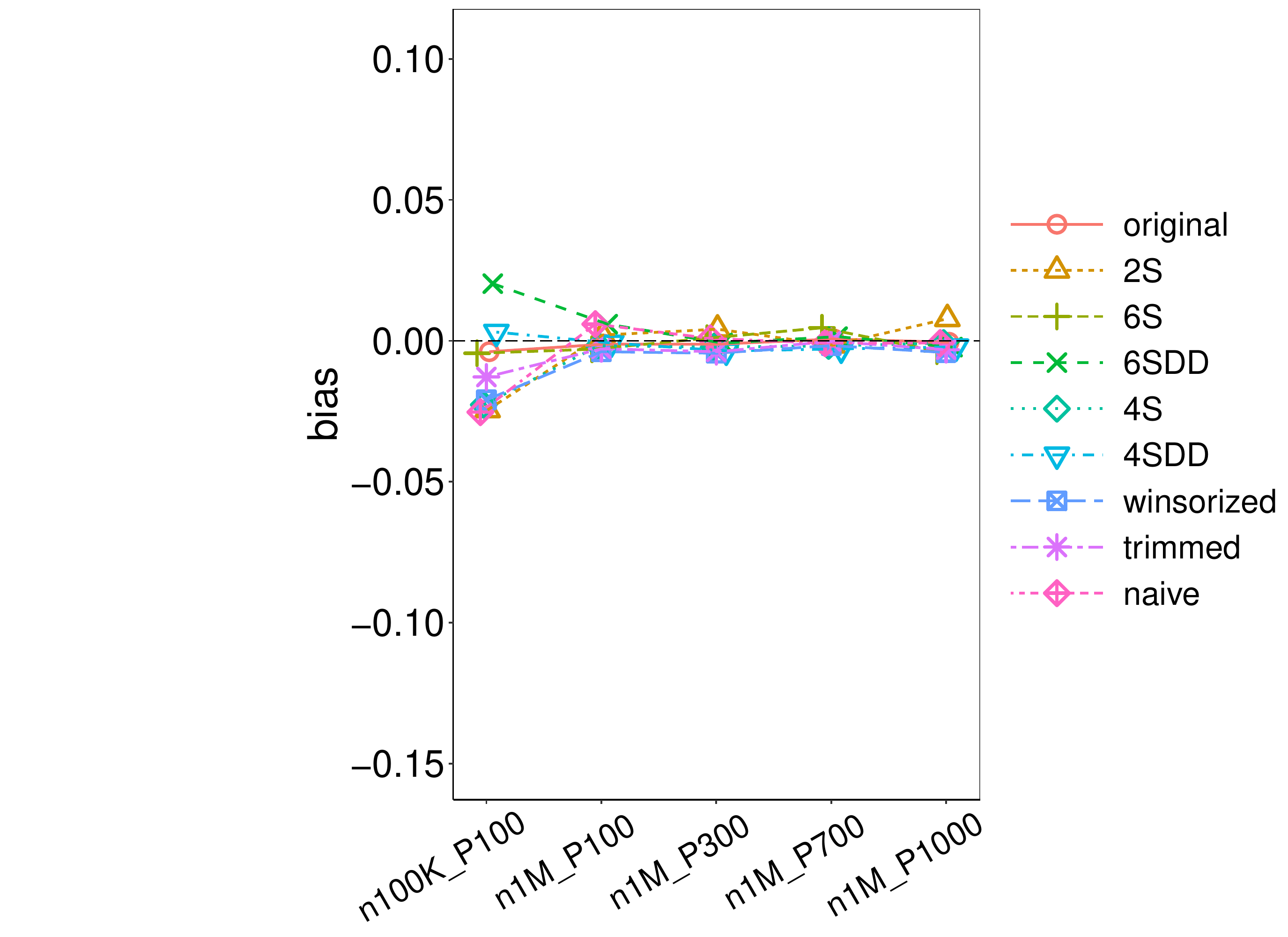}
\includegraphics[width=0.19\textwidth, trim={2.5in 0 2.6in 0},clip] {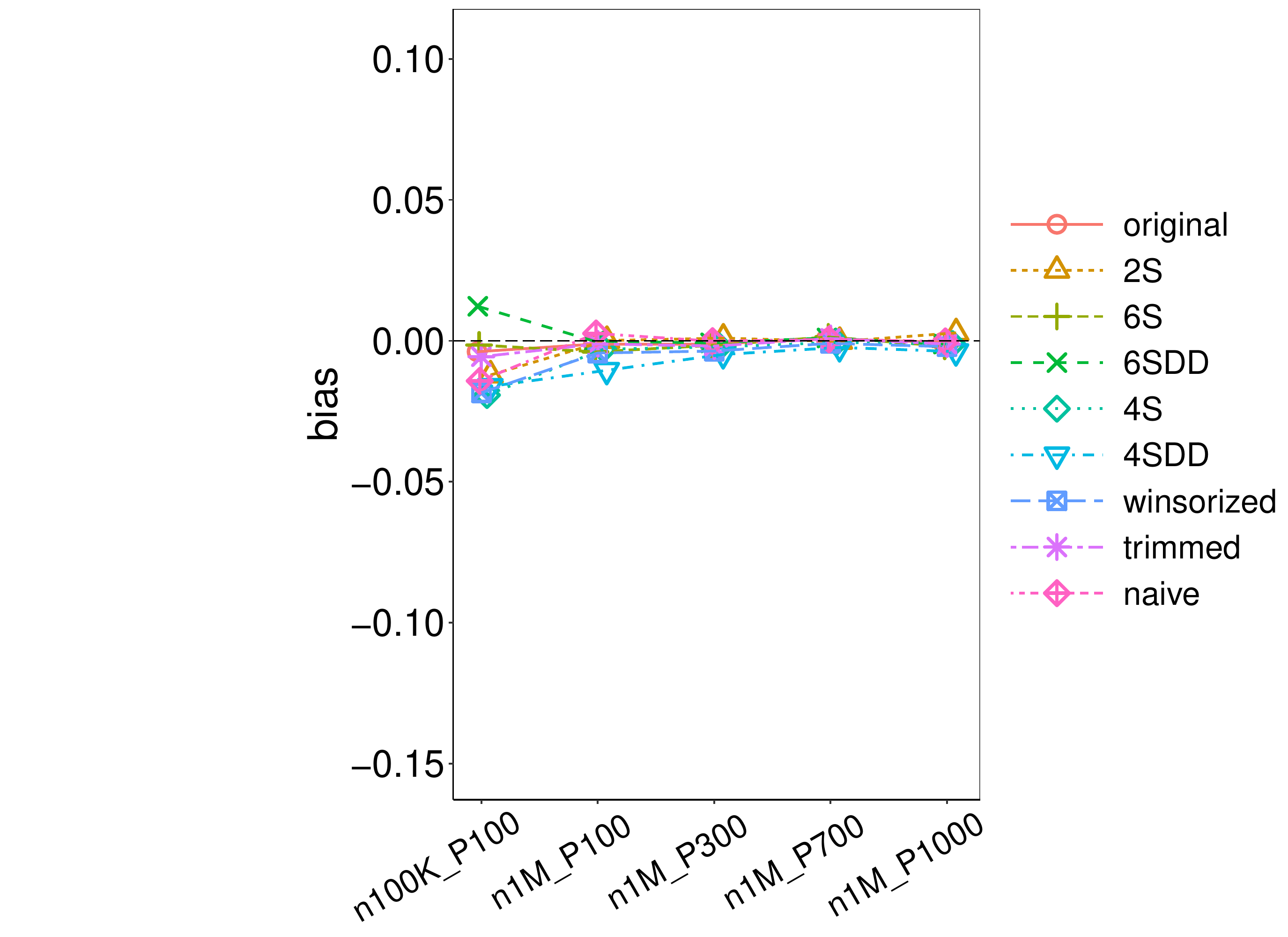}
\includegraphics[width=0.19\textwidth, trim={2.5in 0 2.6in 0},clip] {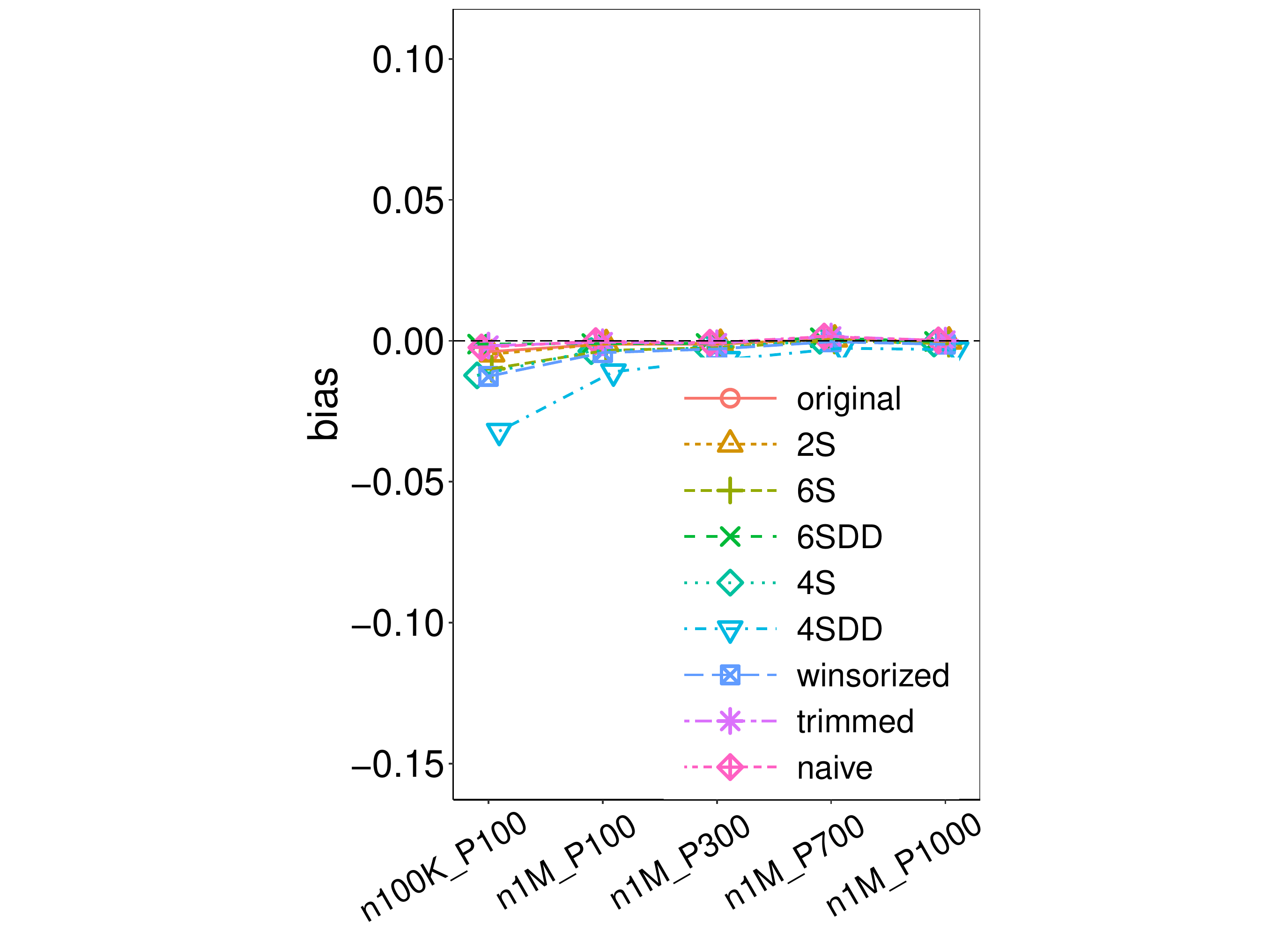}

\includegraphics[width=0.19\textwidth, trim={2.5in 0 2.6in 0},clip] {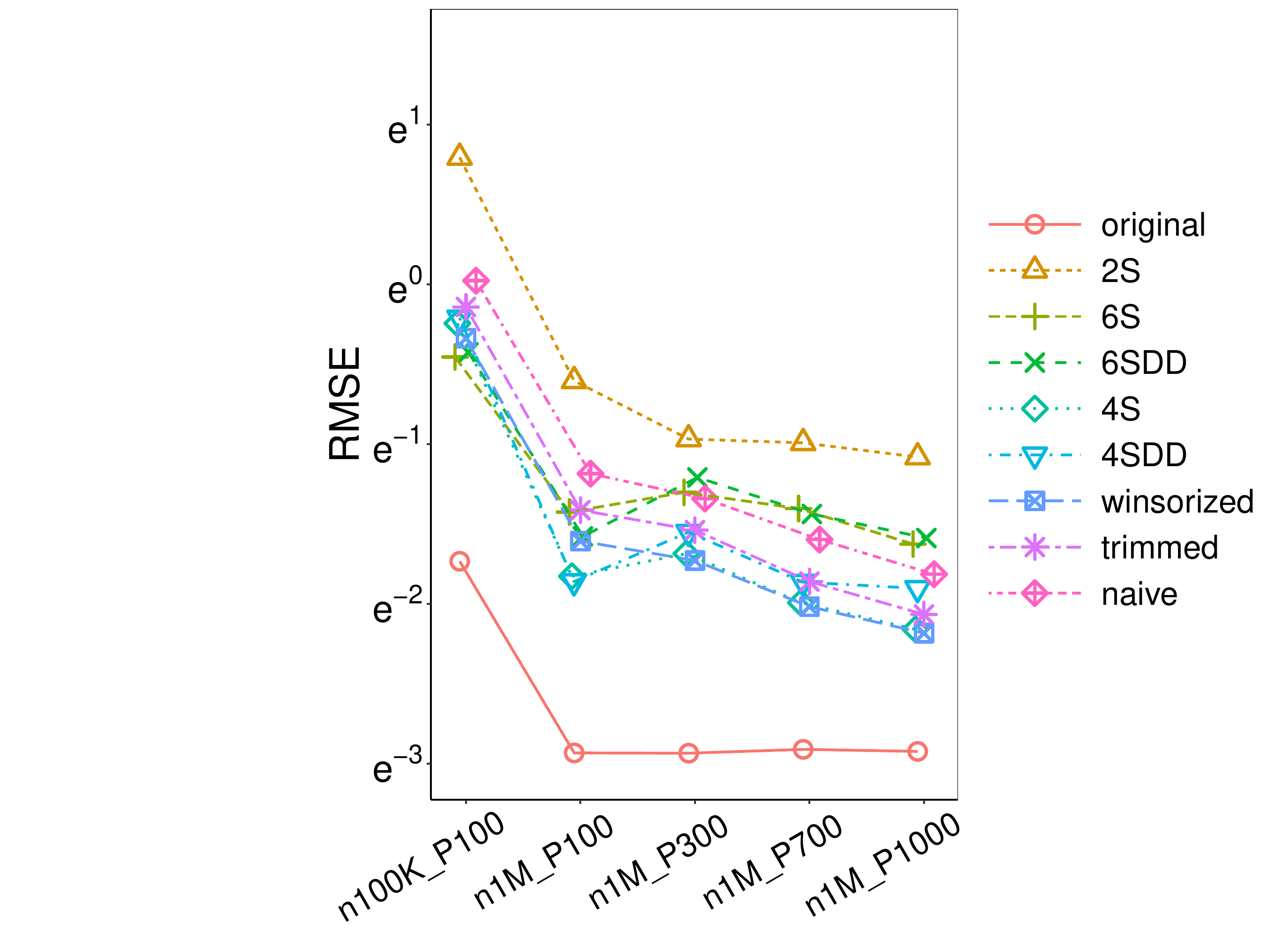}
\includegraphics[width=0.19\textwidth, trim={2.5in 0 2.6in 0},clip] {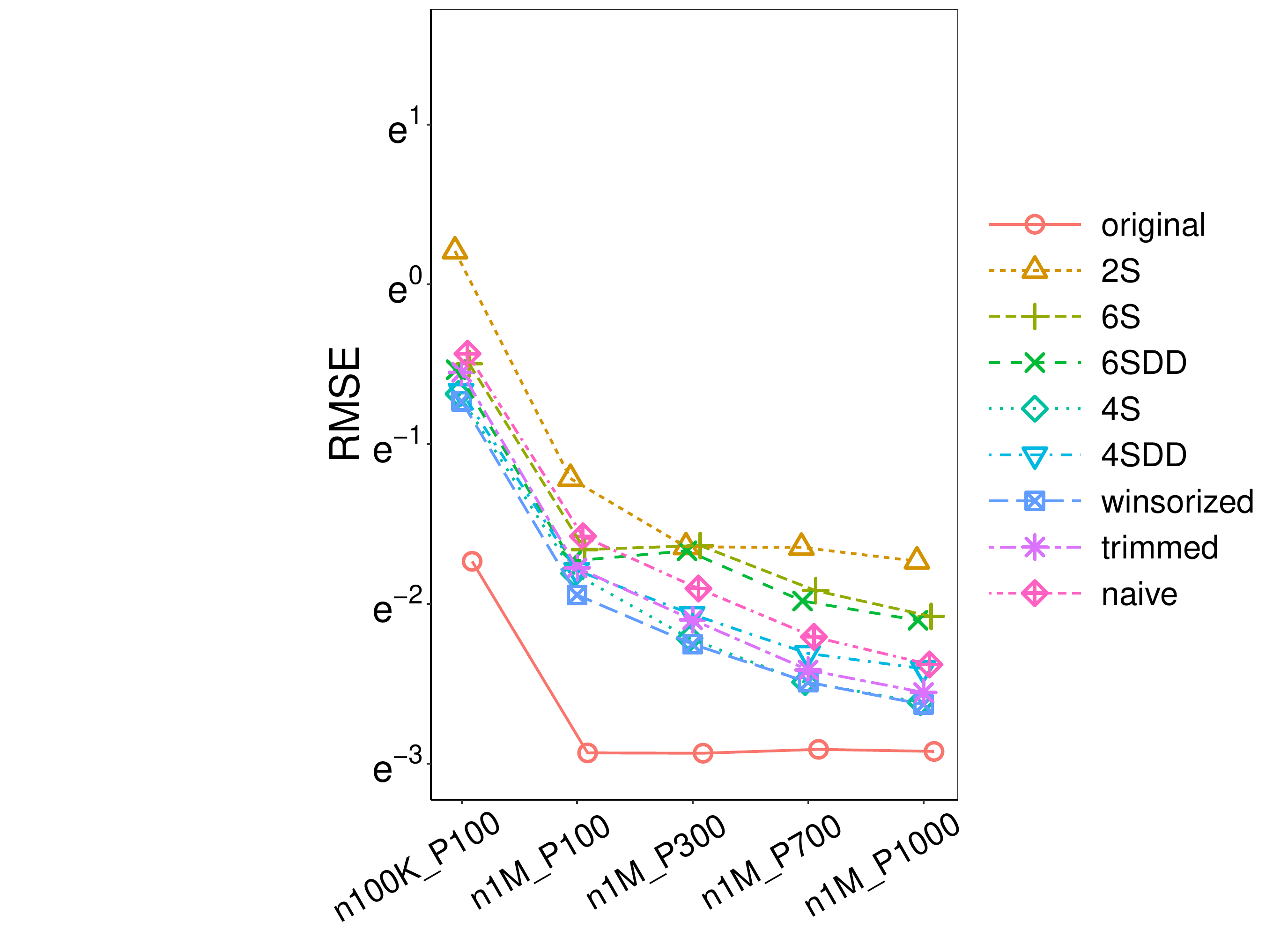}
\includegraphics[width=0.19\textwidth, trim={2.5in 0 2.6in 0},clip] {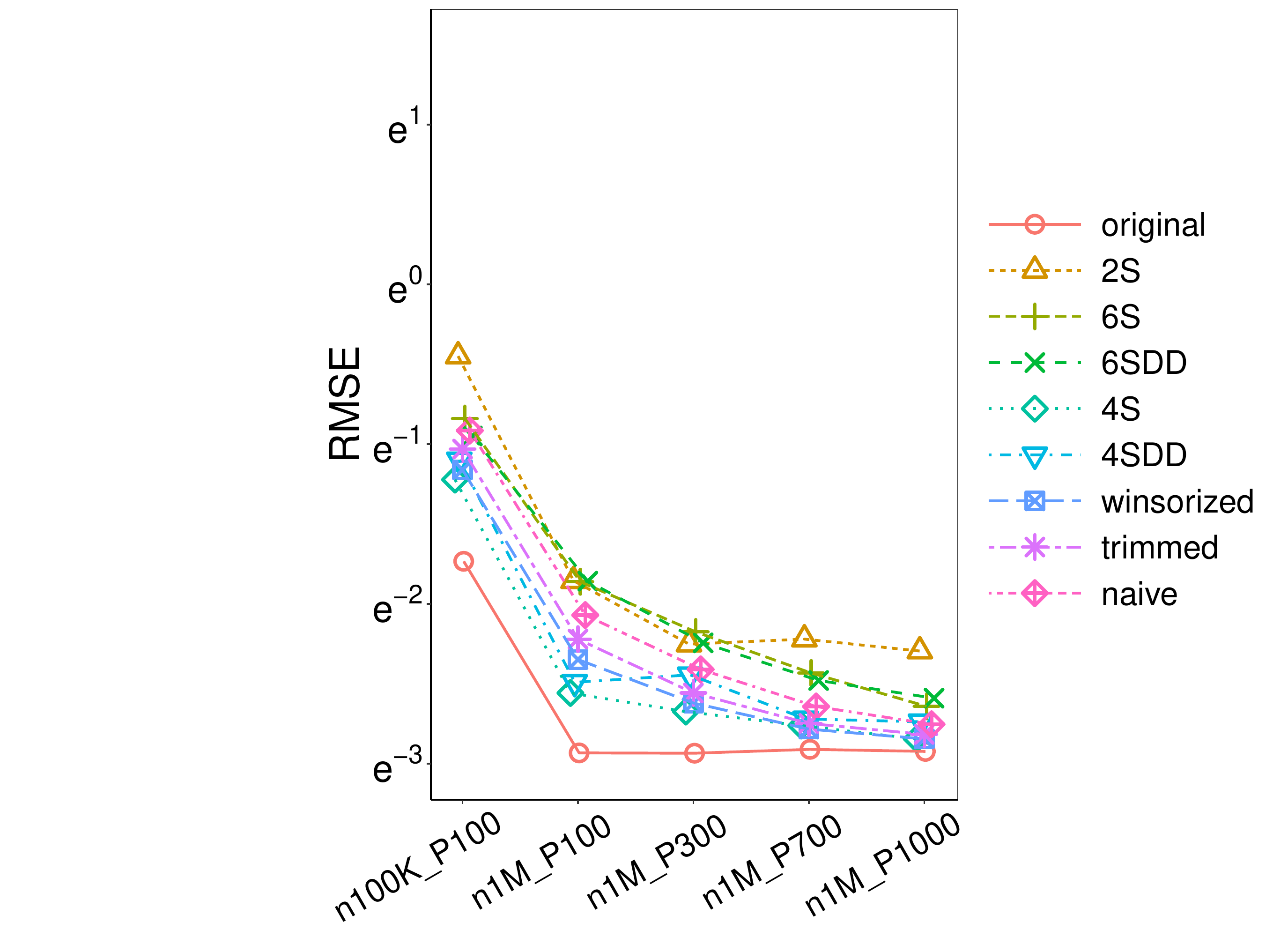}
\includegraphics[width=0.19\textwidth, trim={2.5in 0 2.6in 0},clip] {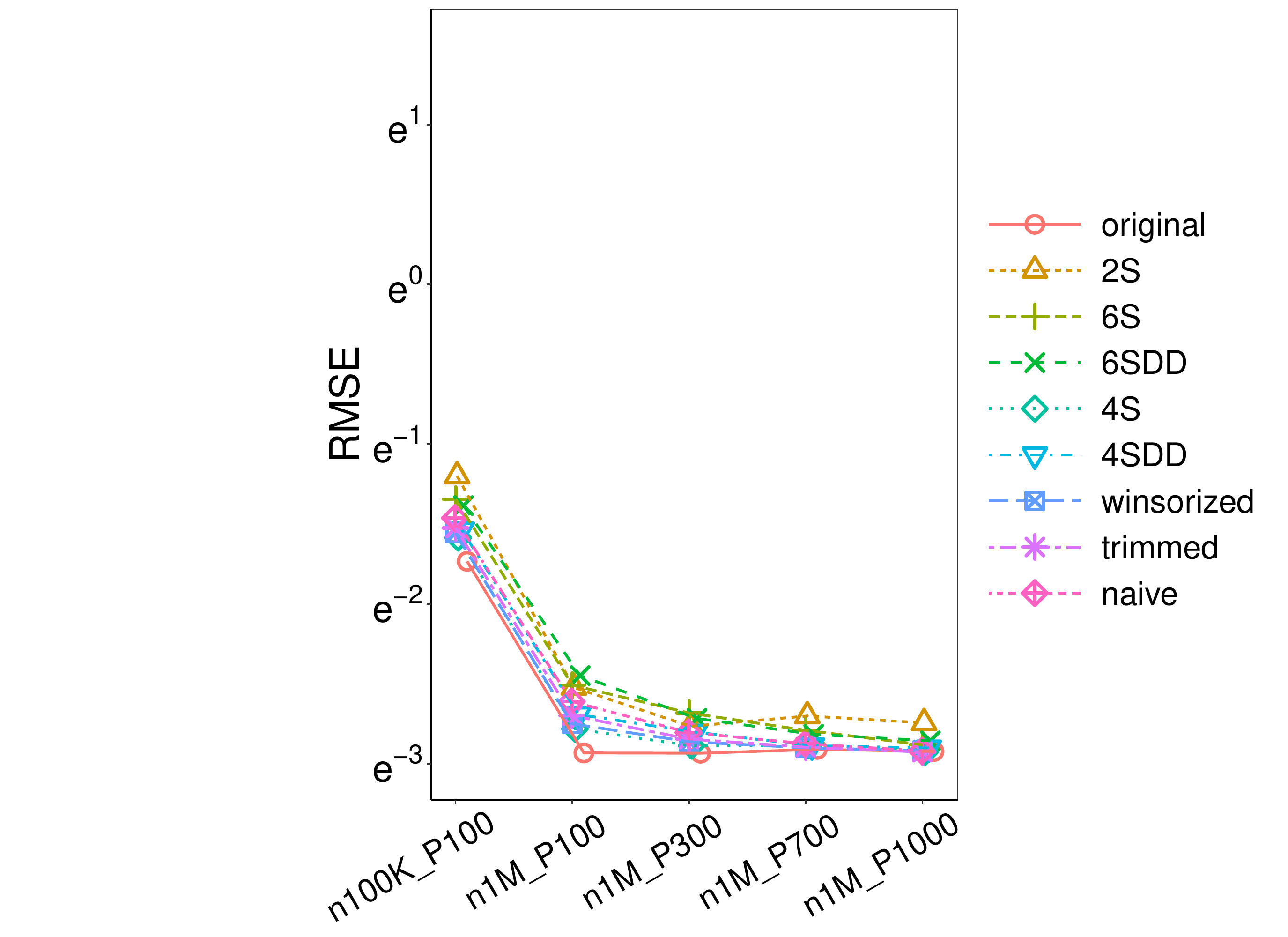}
\includegraphics[width=0.19\textwidth, trim={2.5in 0 2.6in 0},clip] {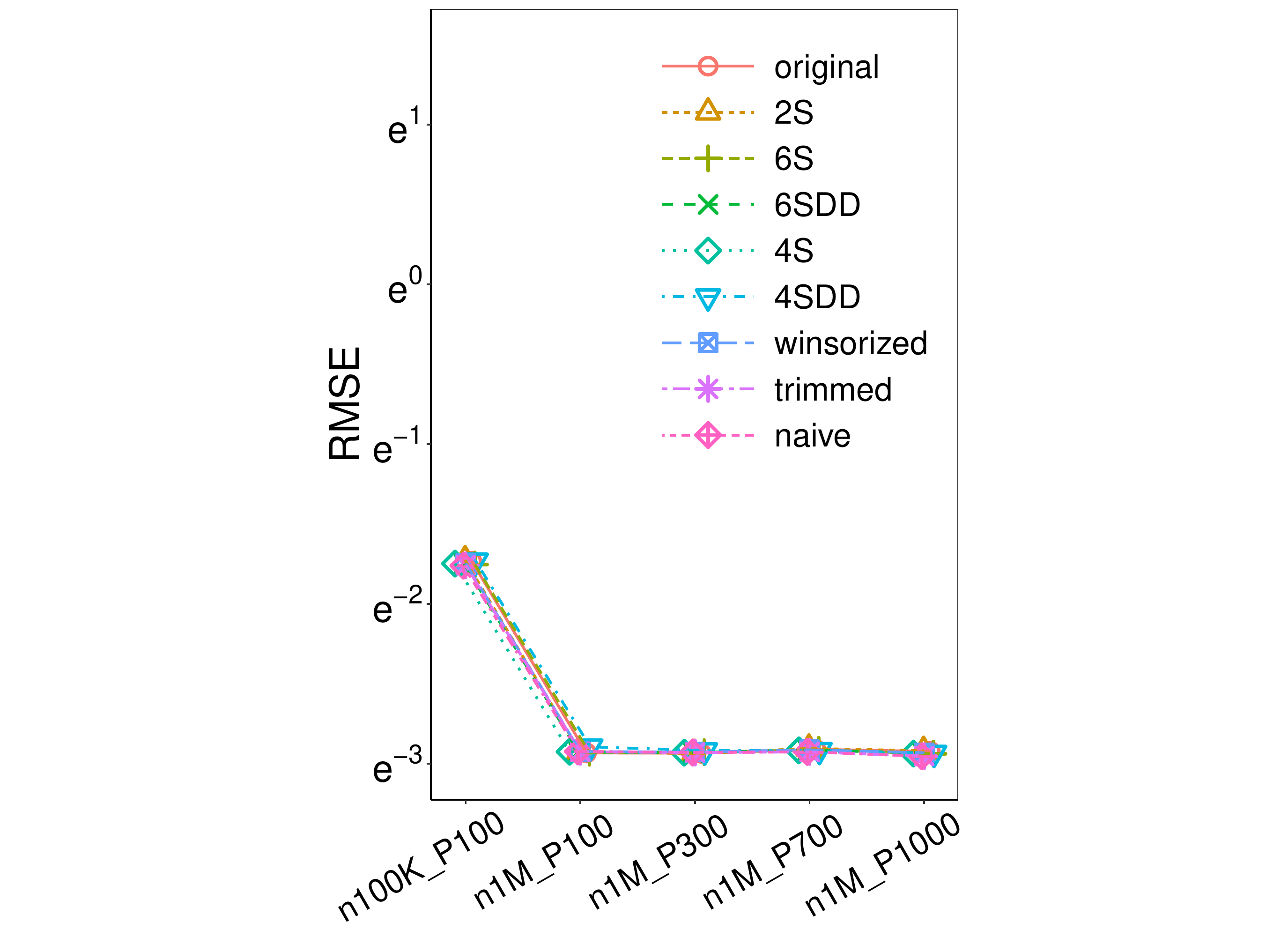}

\includegraphics[width=0.19\textwidth, trim={2.5in 0 2.6in 0},clip] {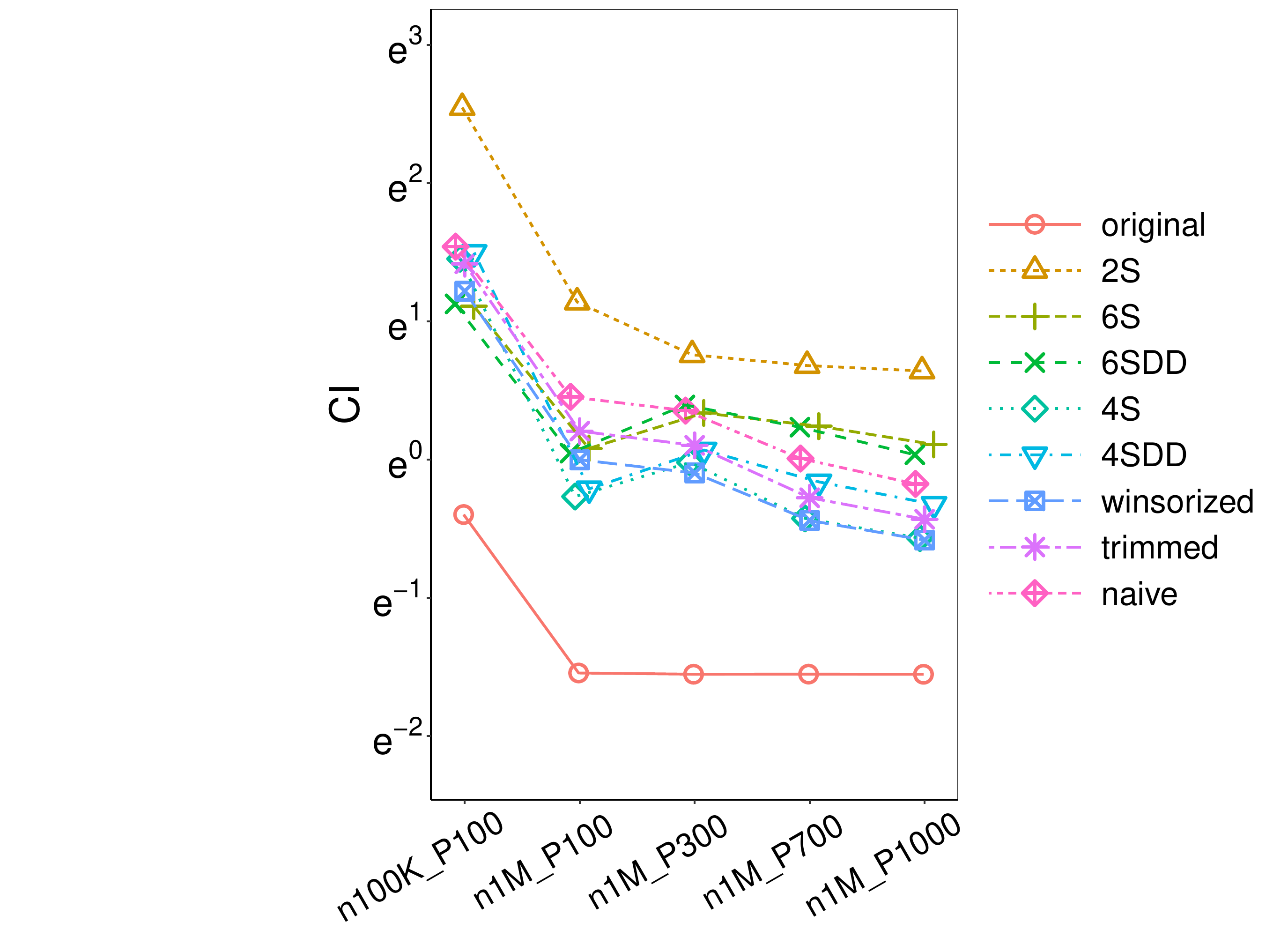}
\includegraphics[width=0.19\textwidth, trim={2.5in 0 2.6in 0},clip] {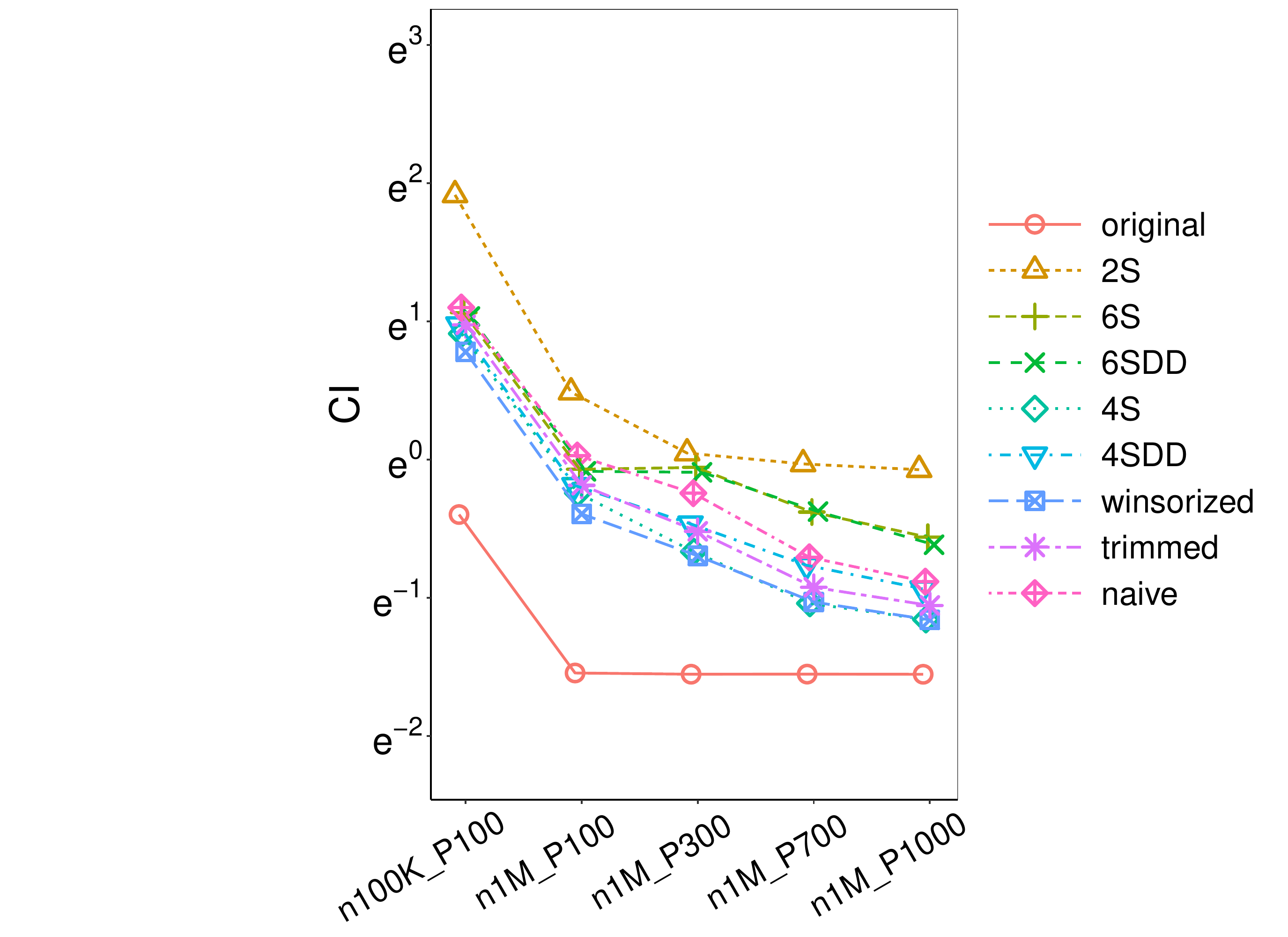}
\includegraphics[width=0.19\textwidth, trim={2.5in 0 2.6in 0},clip] {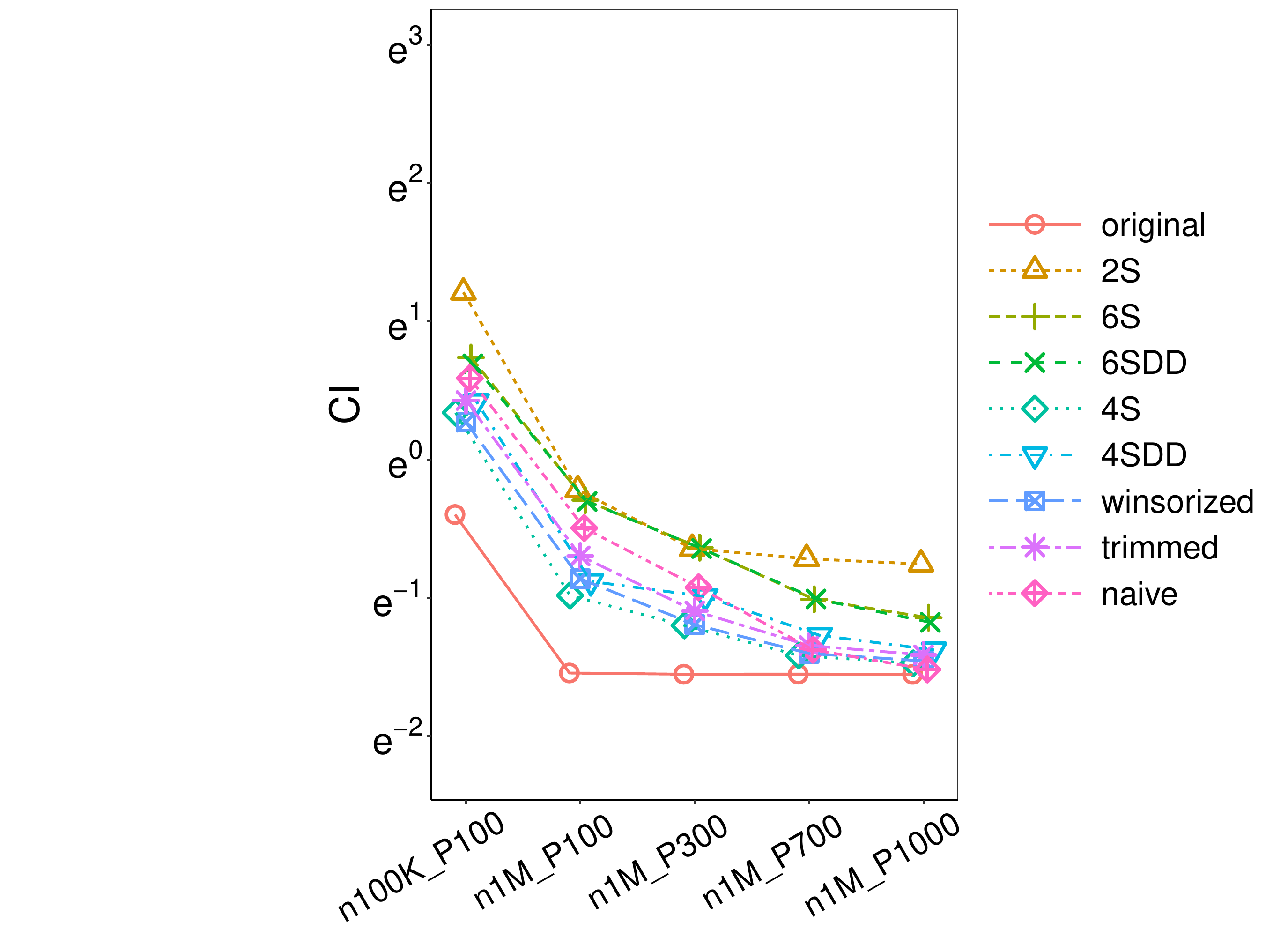}
\includegraphics[width=0.19\textwidth, trim={2.5in 0 2.6in 0},clip] {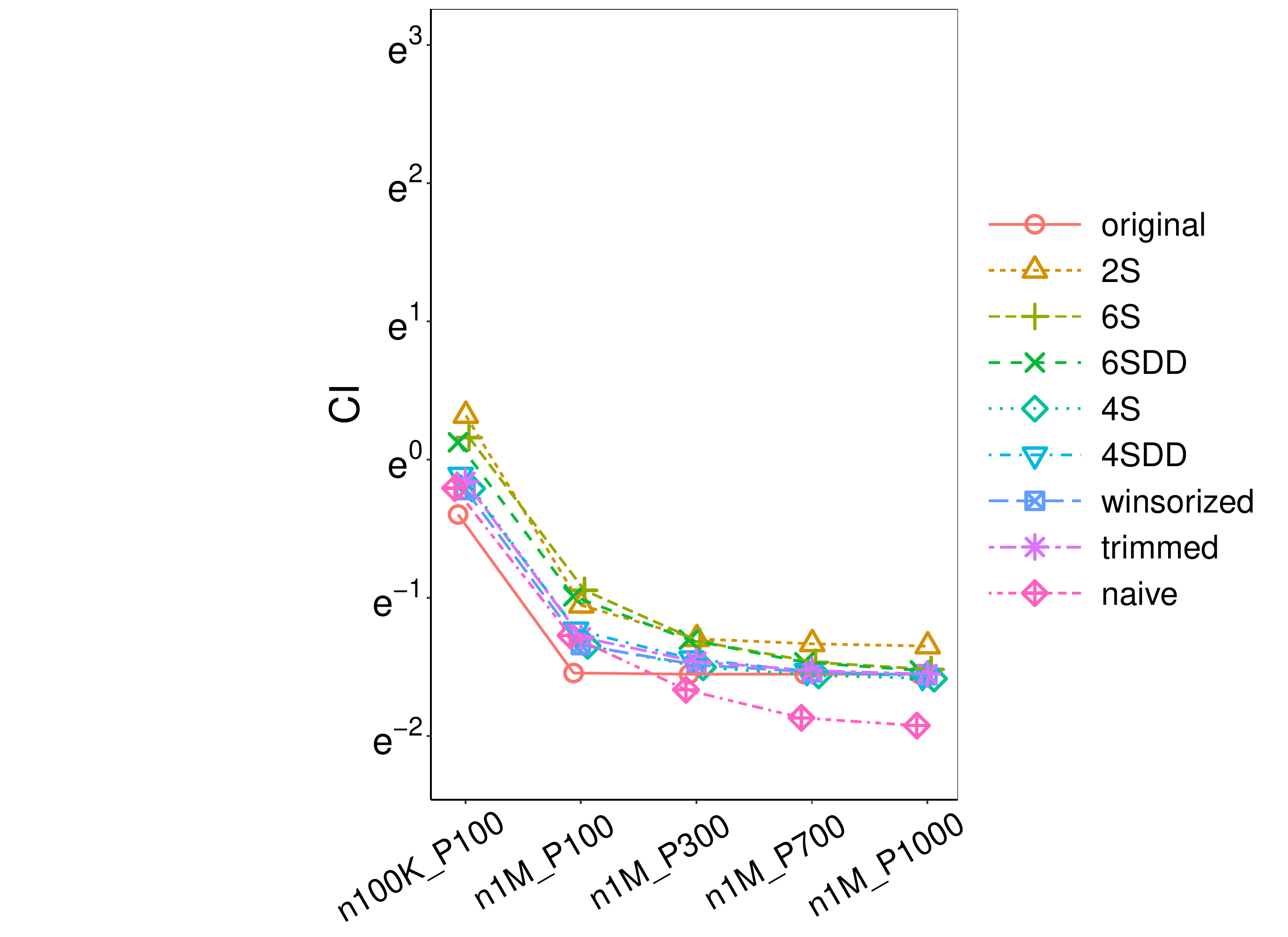}
\includegraphics[width=0.19\textwidth, trim={2.5in 0 2.6in 0},clip] {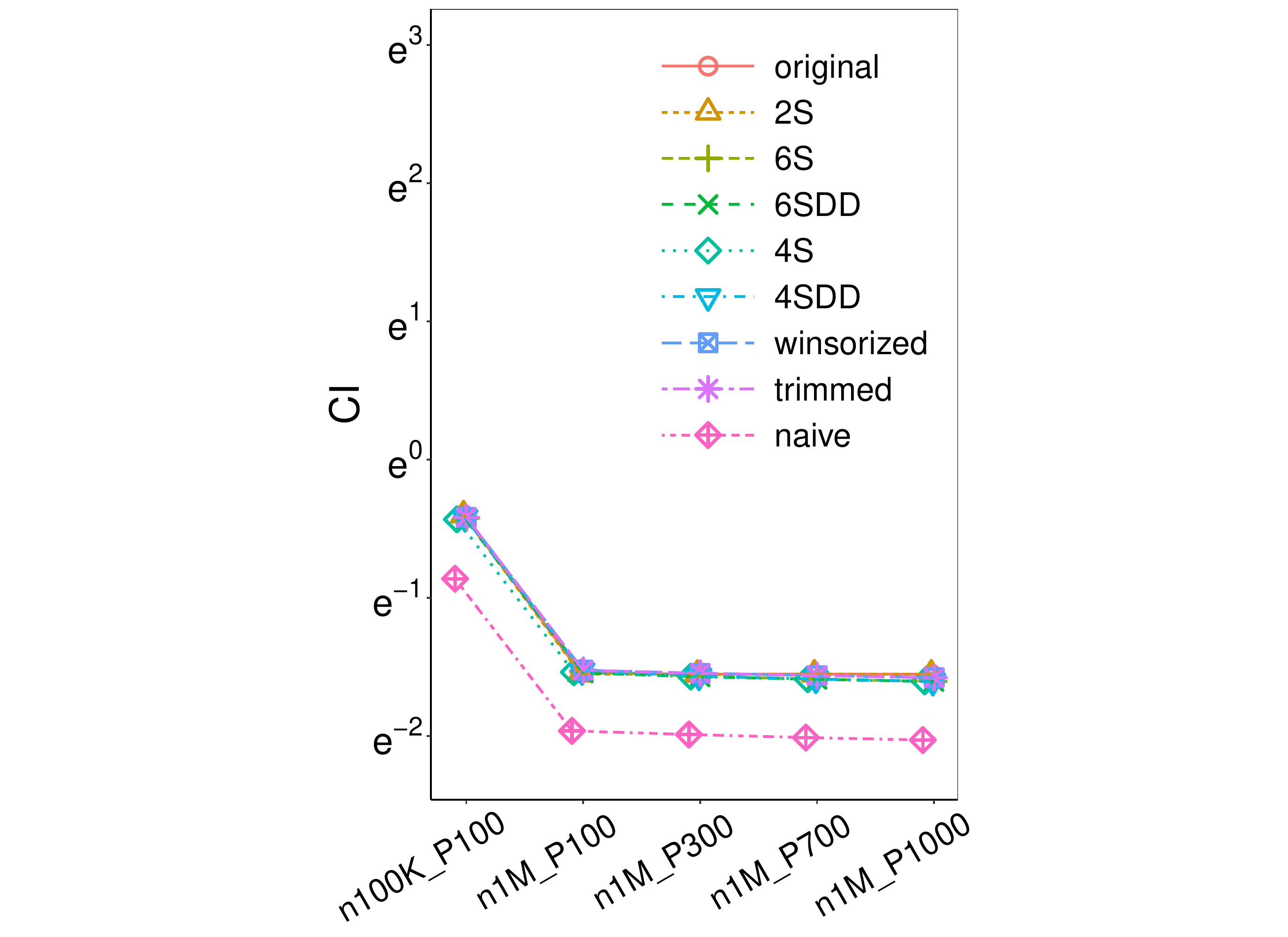}

\includegraphics[width=0.19\textwidth, trim={2.5in 0 2.6in 0},clip] {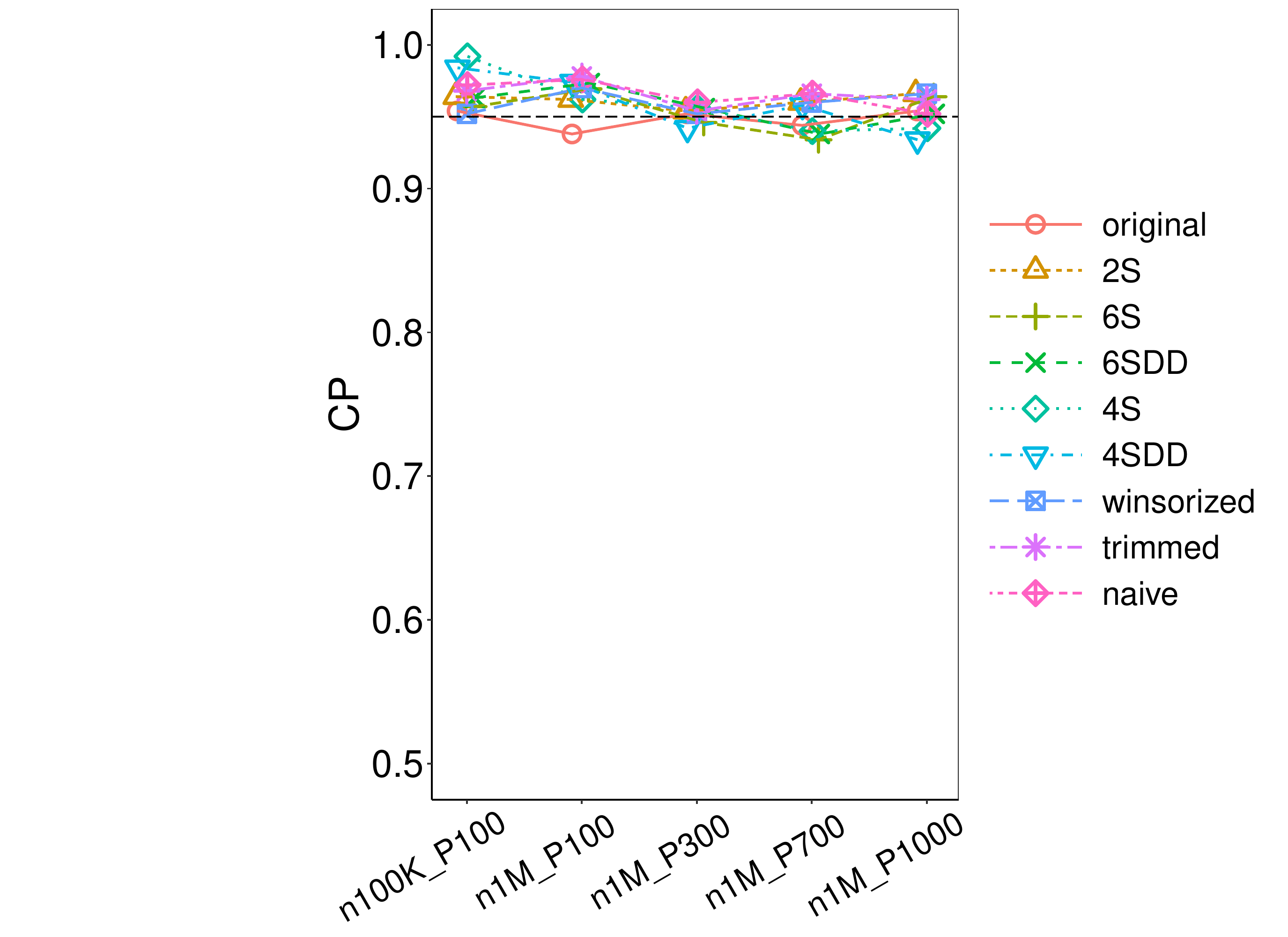}
\includegraphics[width=0.19\textwidth, trim={2.5in 0 2.6in 0},clip] {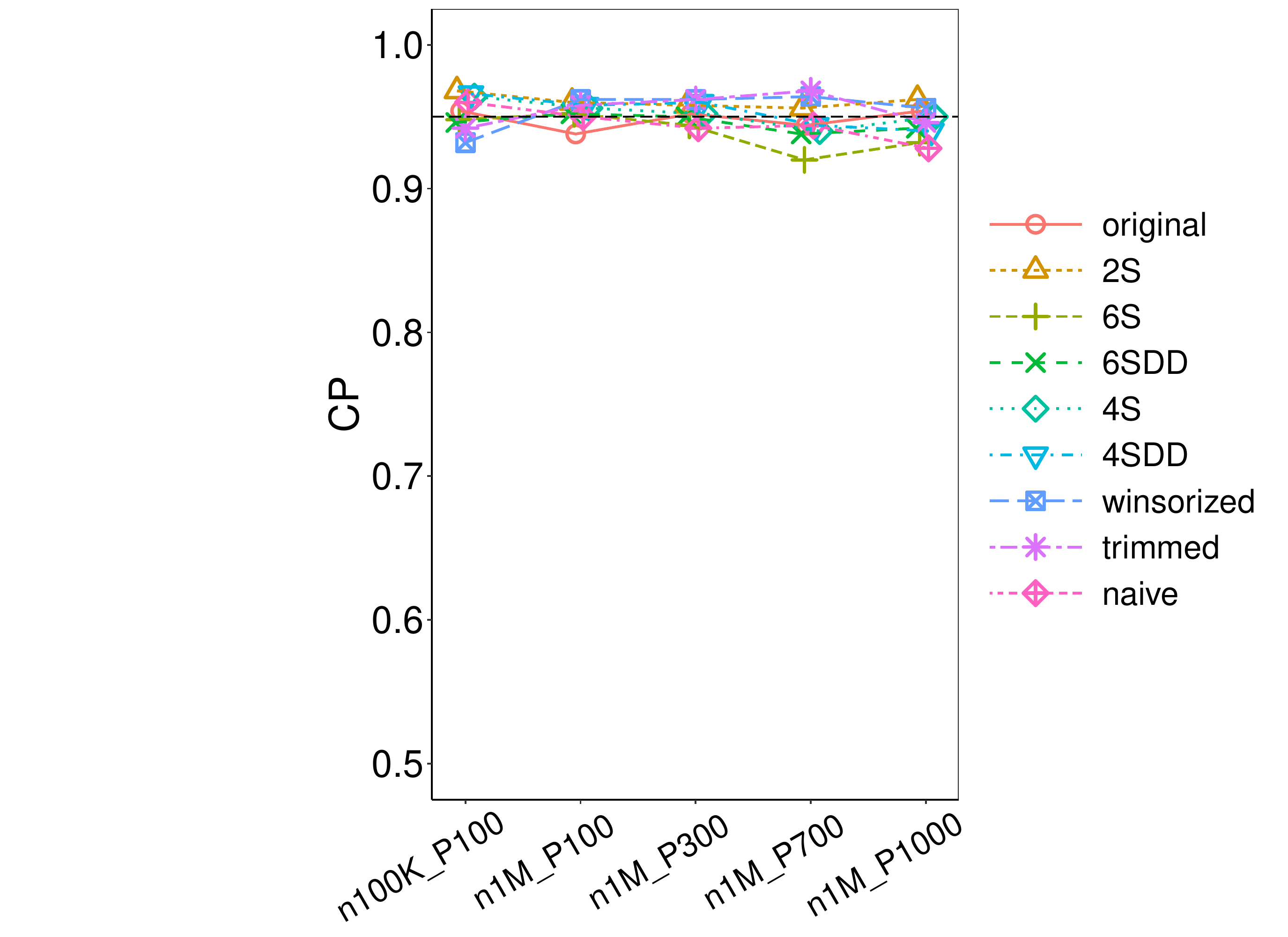}
\includegraphics[width=0.19\textwidth, trim={2.5in 0 2.6in 0},clip] {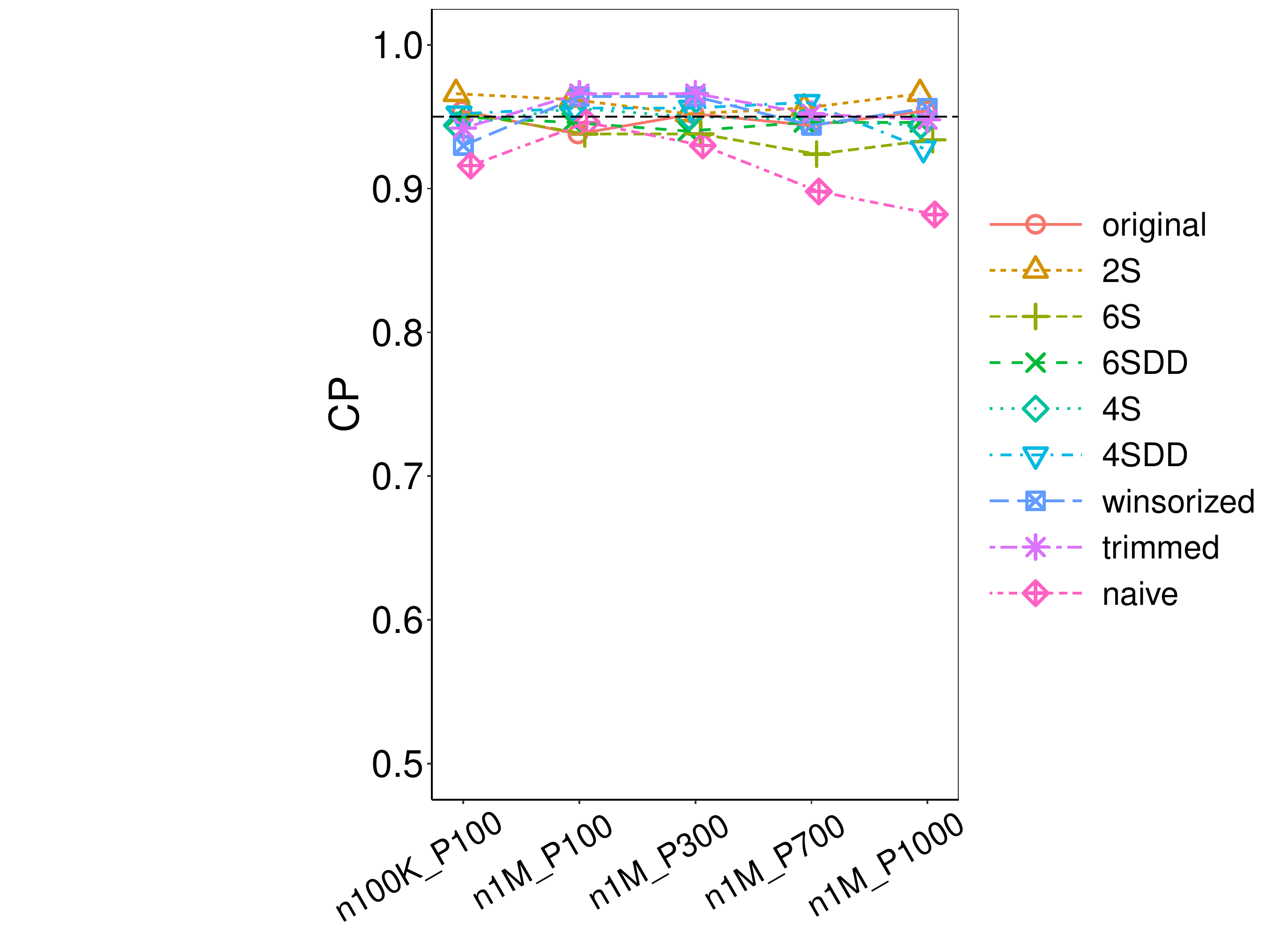}
\includegraphics[width=0.19\textwidth, trim={2.5in 0 2.6in 0},clip] {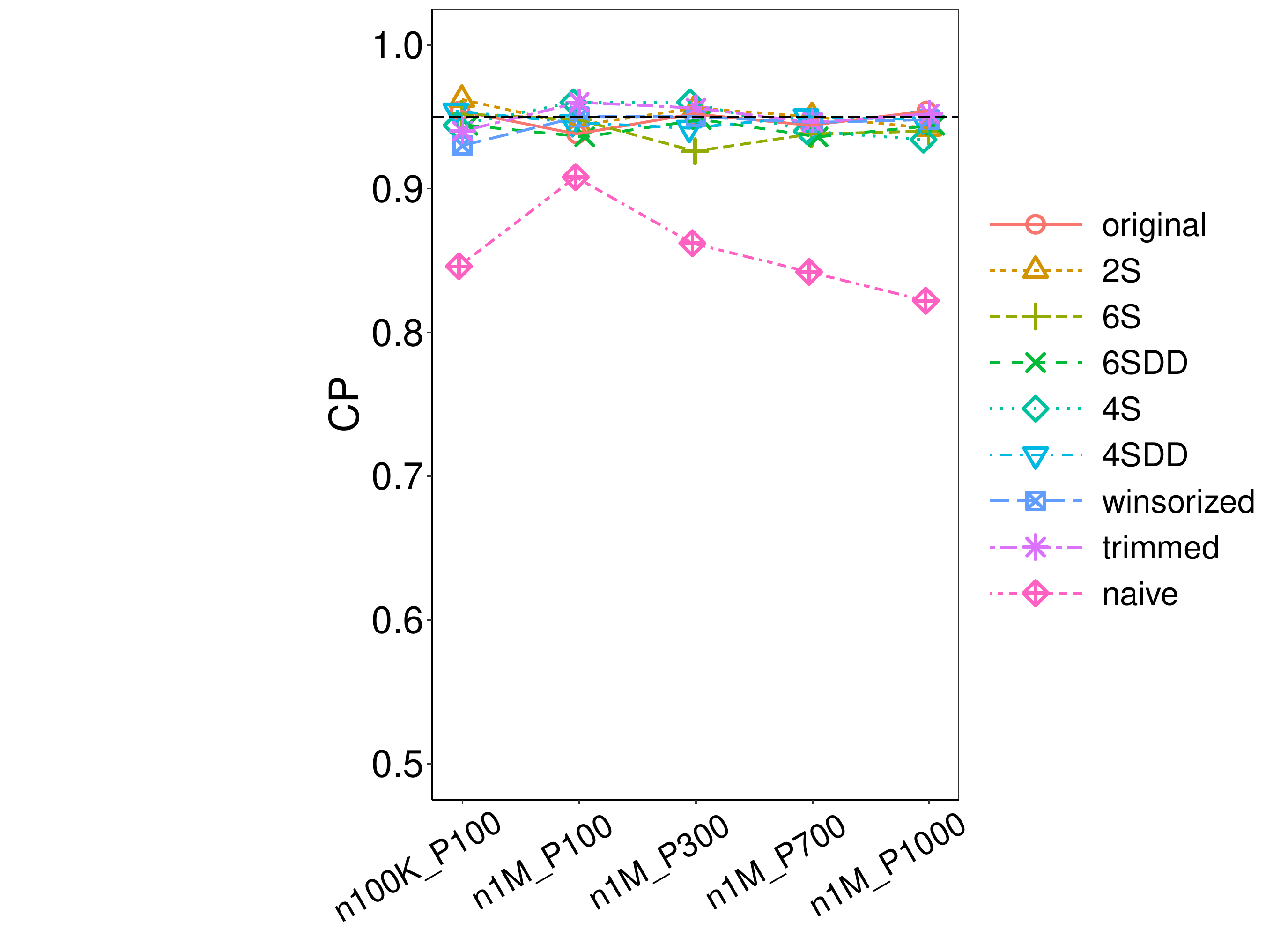}
\includegraphics[width=0.19\textwidth, trim={2.5in 0 2.6in 0},clip] {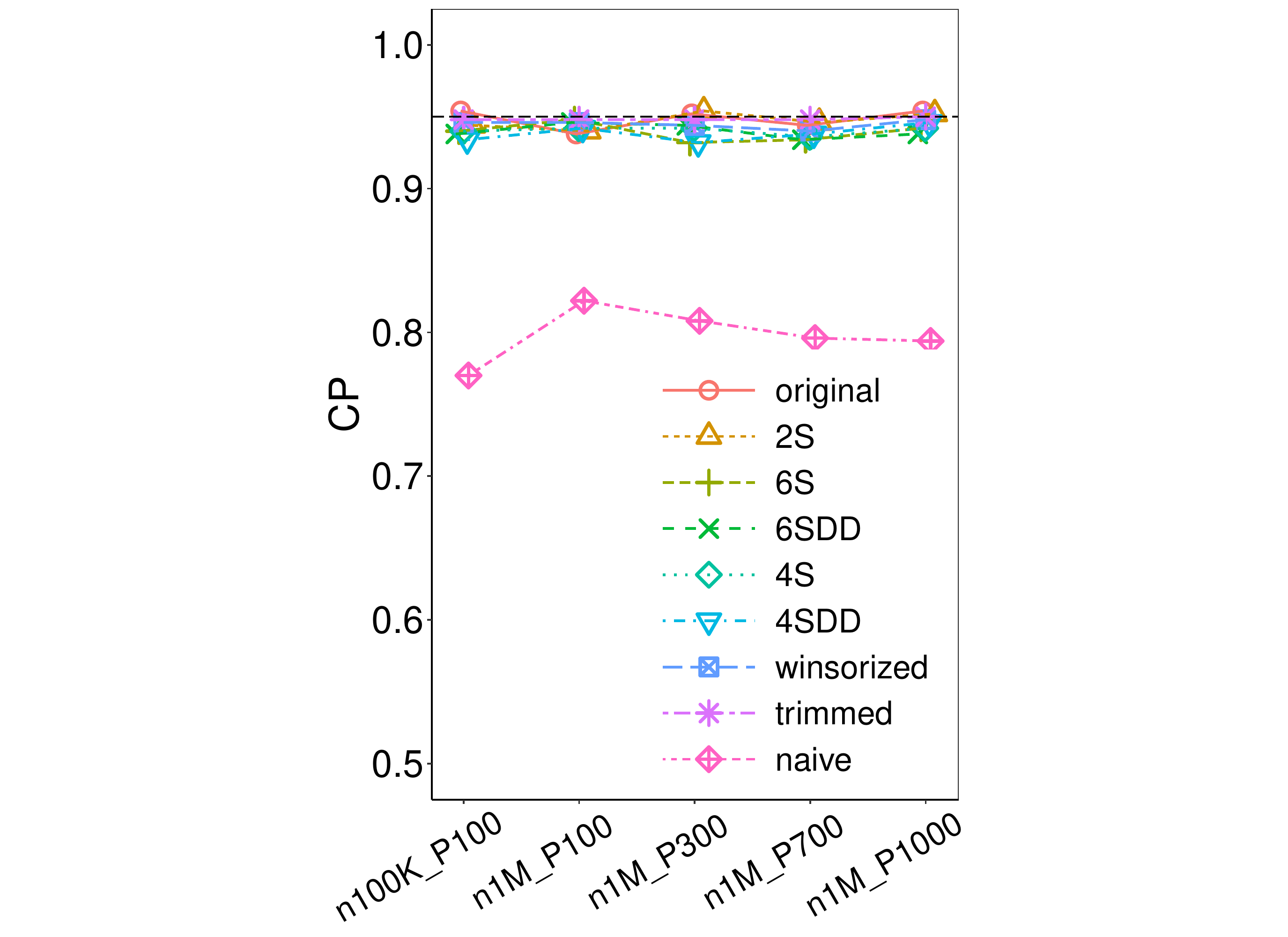}
\caption{Simulation results with $\epsilon$-DP for ZILN data with  $\alpha=\beta$ when $\theta=0$} \label{fig:0sDPZILN}
\end{figure}

\begin{figure}[!htb]
\hspace{0.45in}$\rho=0.005$\hspace{0.65in}$\rho=0.02$\hspace{0.65in}$\rho=0.08$
\hspace{0.65in}$\rho=0.32$\hspace{0.65in}$\rho=1.28$

\includegraphics[width=0.19\textwidth, trim={2.5in 0 2.6in 0},clip] {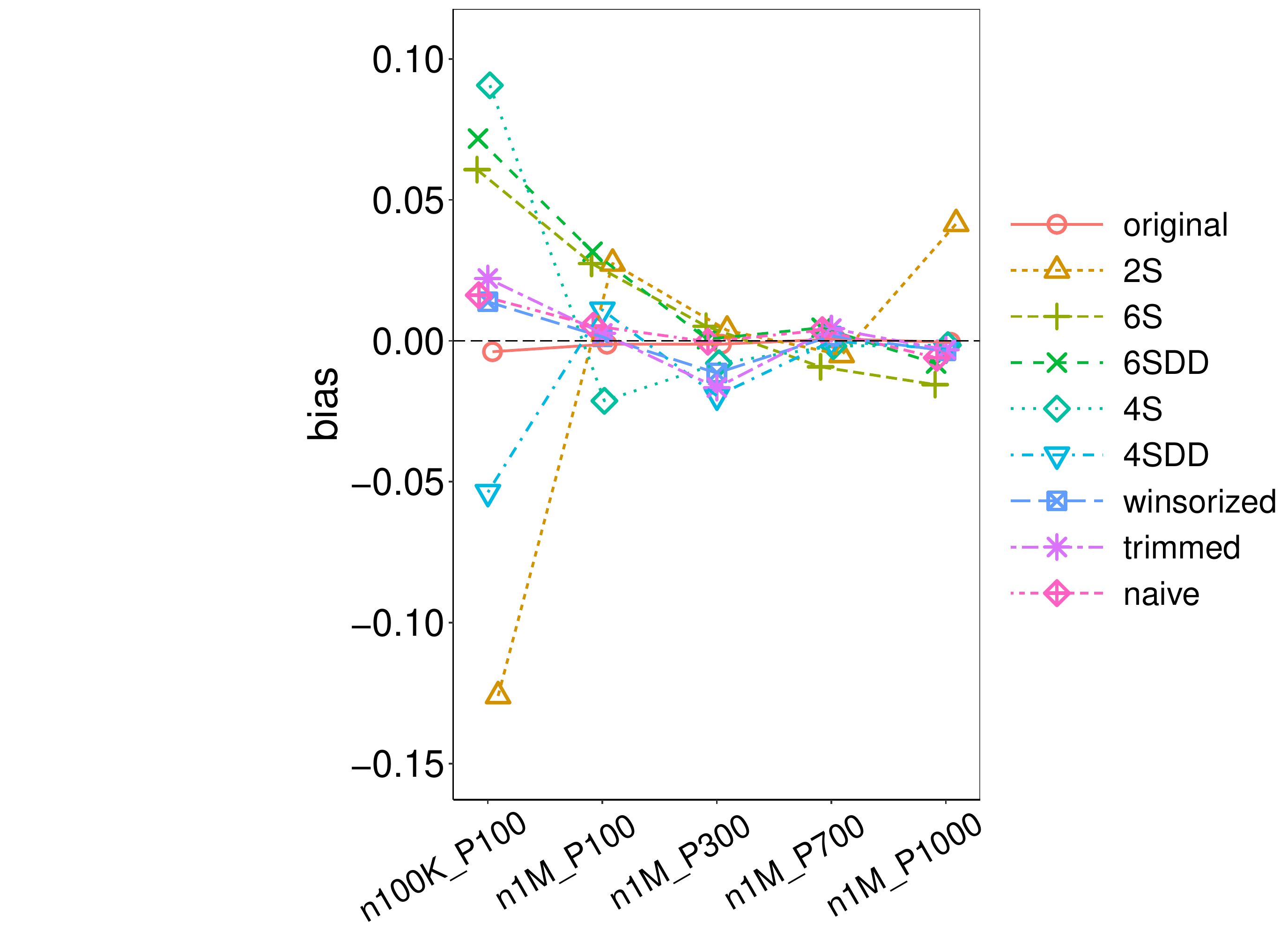}
\includegraphics[width=0.19\textwidth, trim={2.5in 0 2.6in 0},clip] {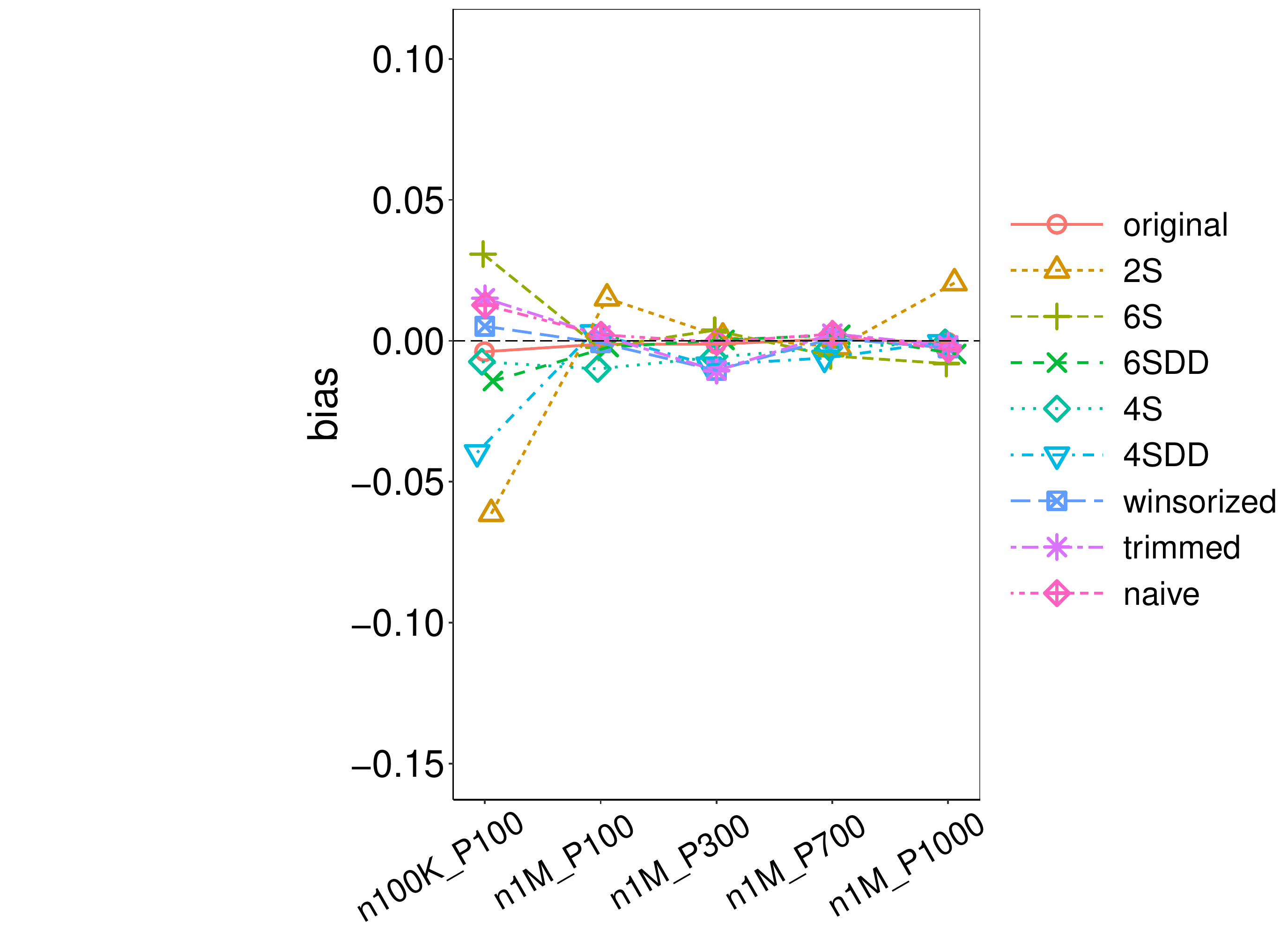}
\includegraphics[width=0.19\textwidth, trim={2.5in 0 2.6in 0},clip] {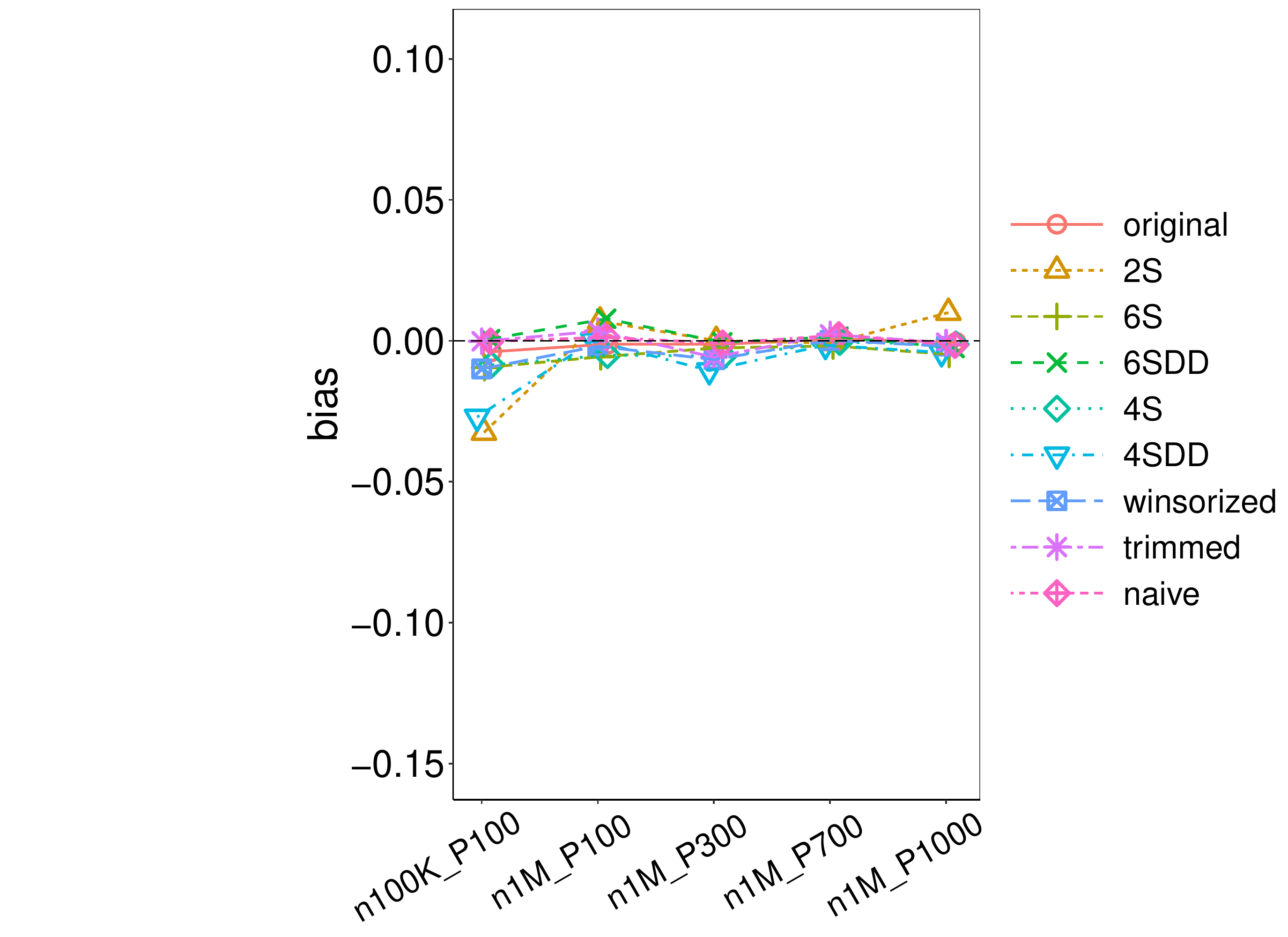}
\includegraphics[width=0.19\textwidth, trim={2.5in 0 2.6in 0},clip] {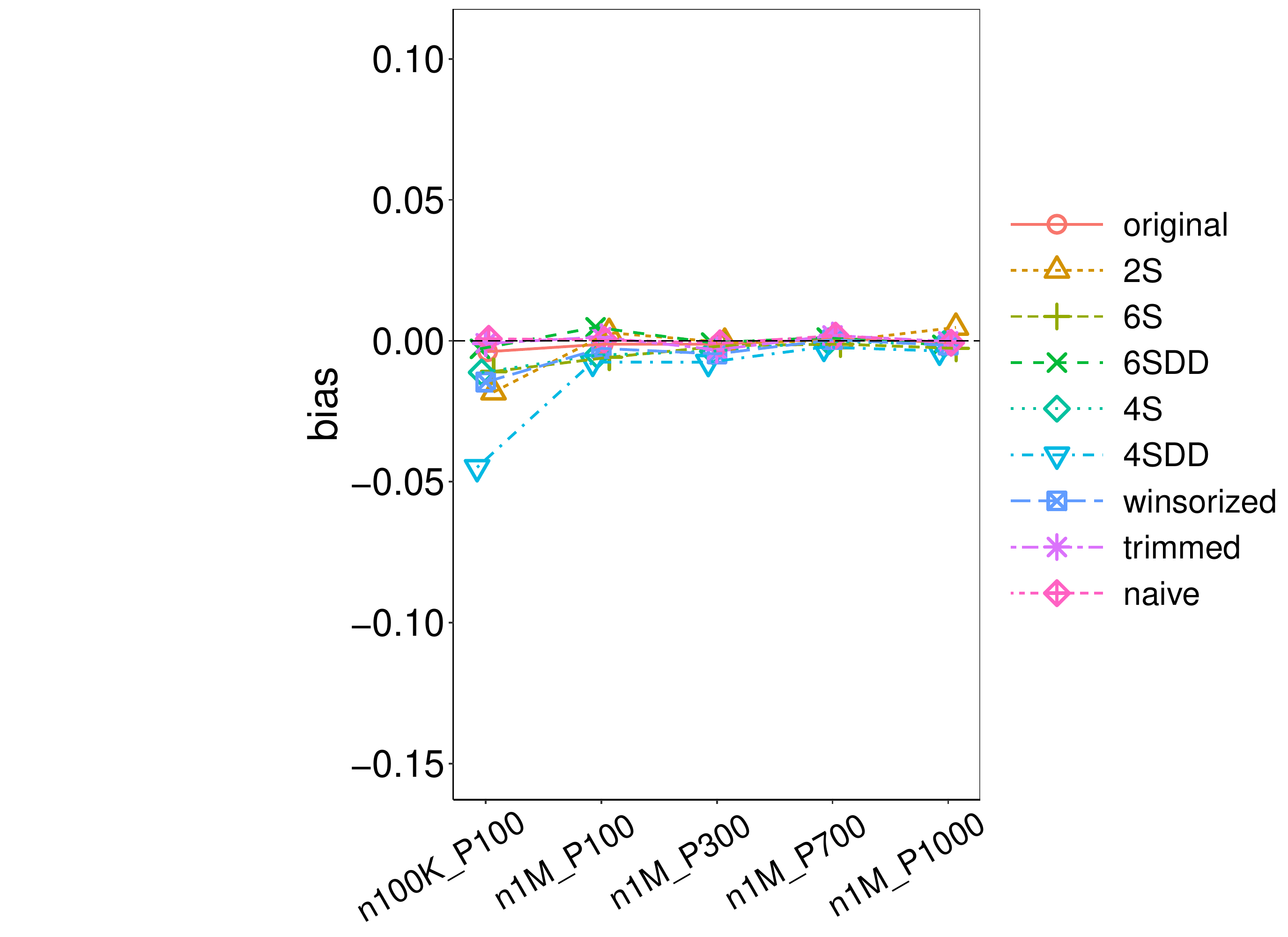}
\includegraphics[width=0.19\textwidth, trim={2.5in 0 2.6in 0},clip] {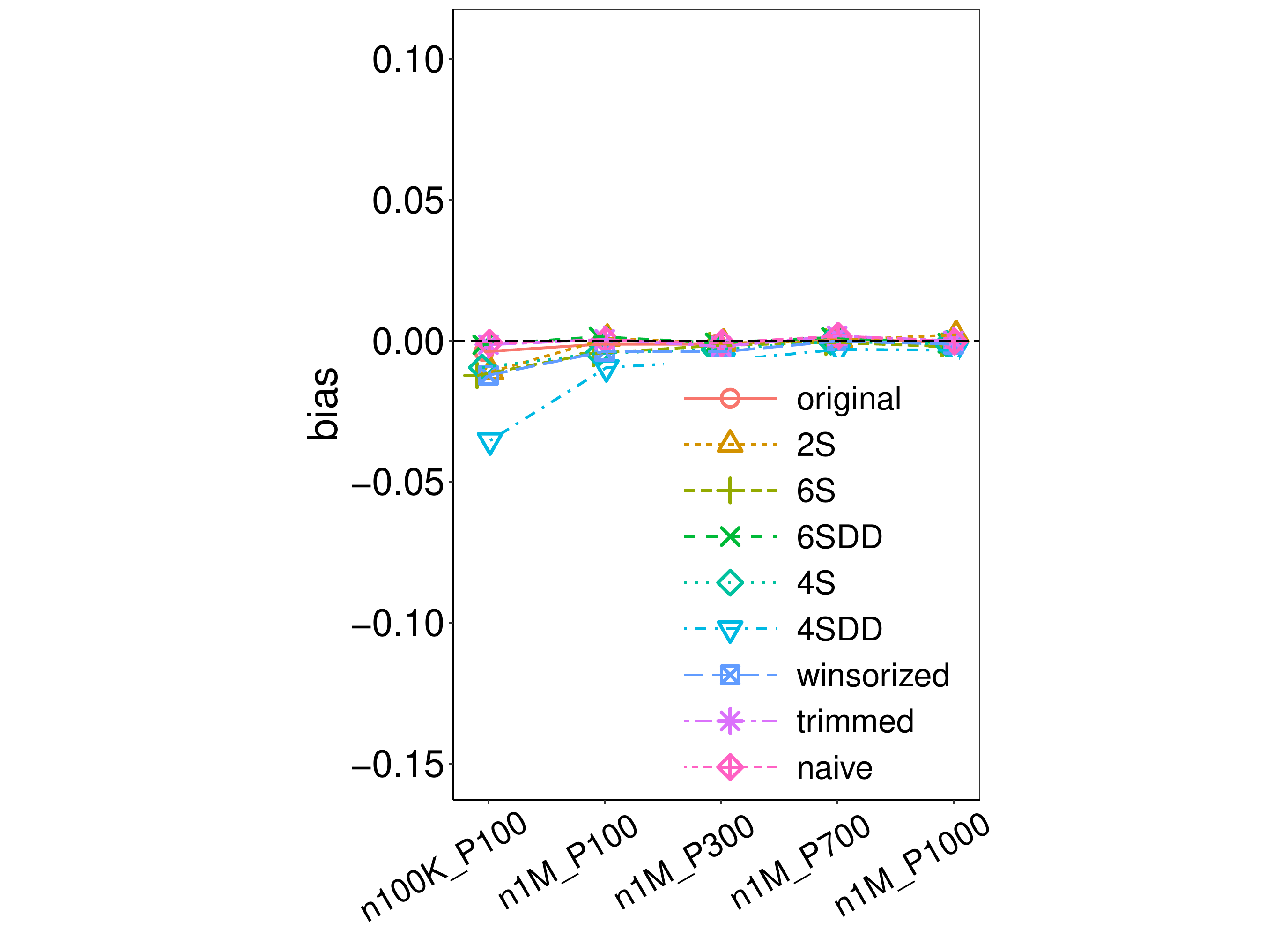}

\includegraphics[width=0.19\textwidth, trim={2.5in 0 2.6in 0},clip] {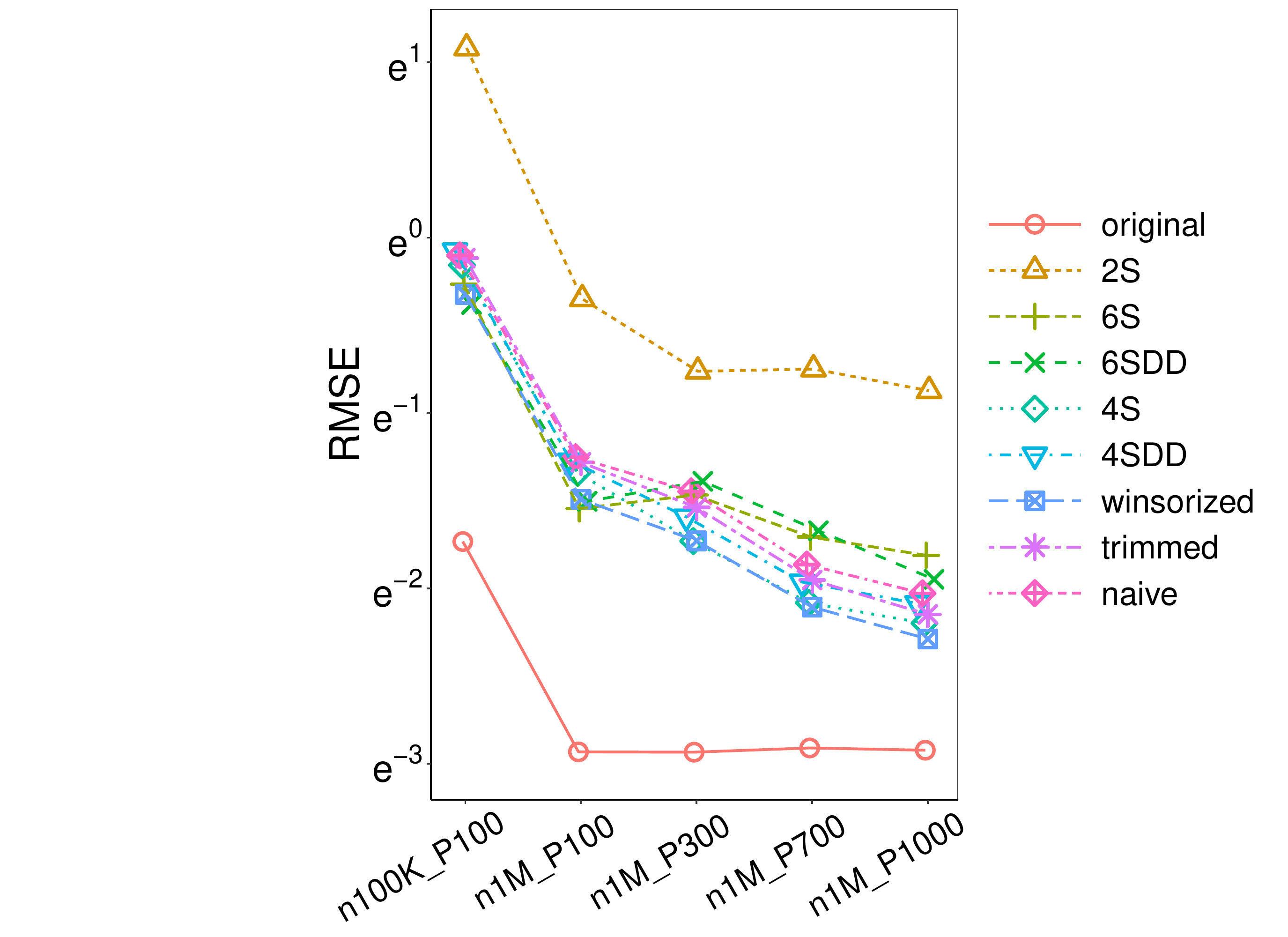}
\includegraphics[width=0.19\textwidth, trim={2.5in 0 2.6in 0},clip] {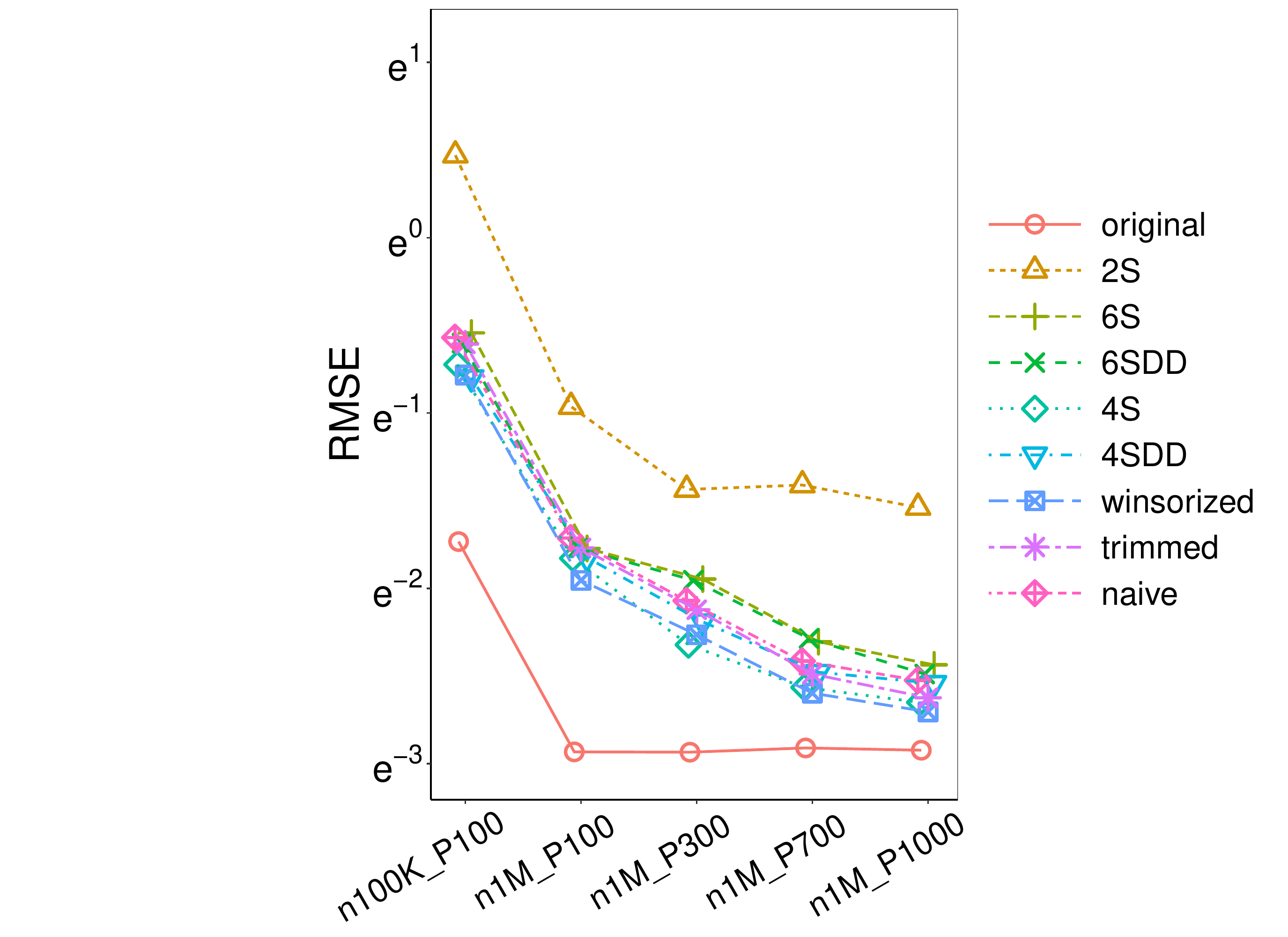}
\includegraphics[width=0.19\textwidth, trim={2.5in 0 2.6in 0},clip] {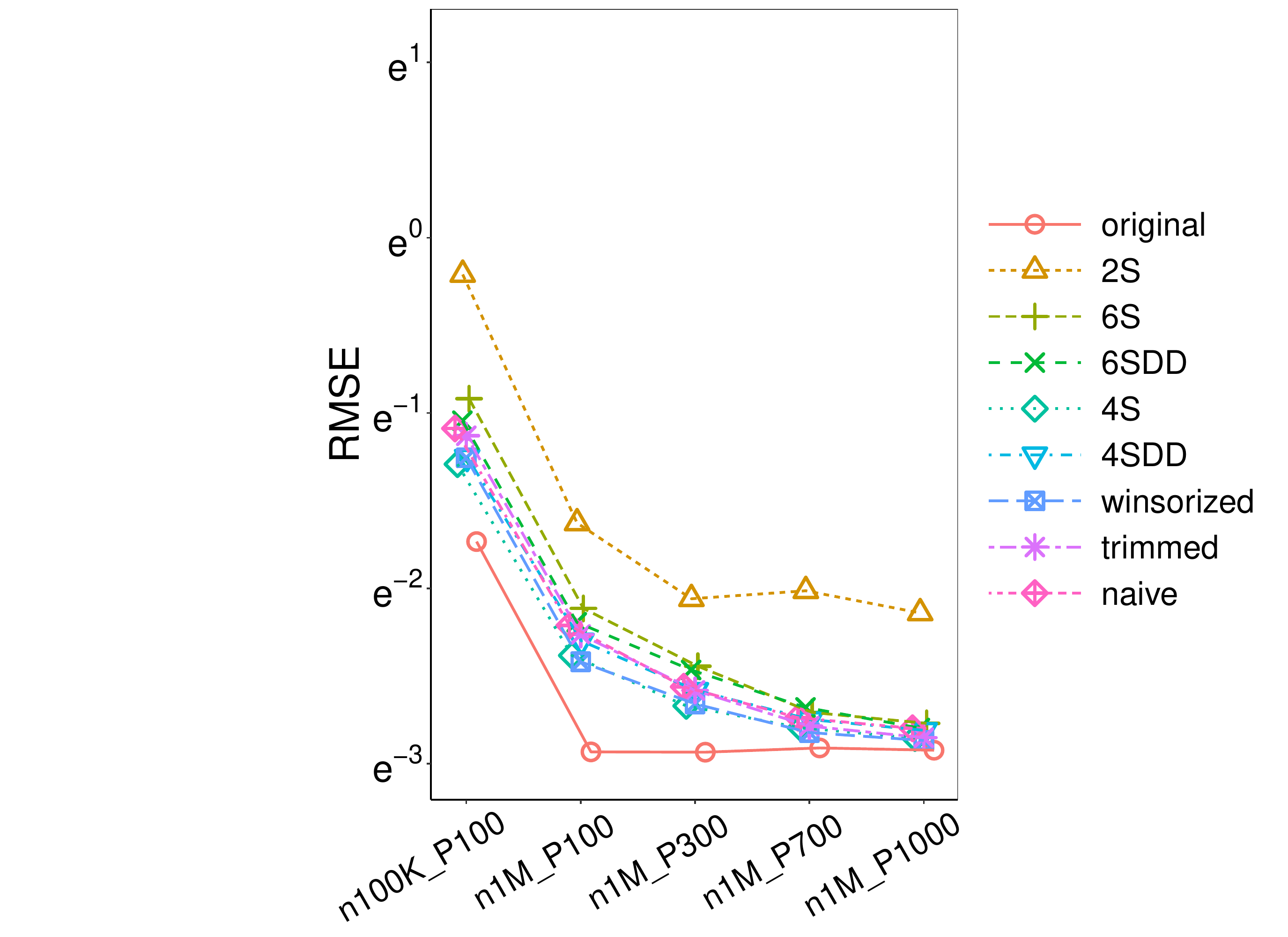}
\includegraphics[width=0.19\textwidth, trim={2.5in 0 2.6in 0},clip] {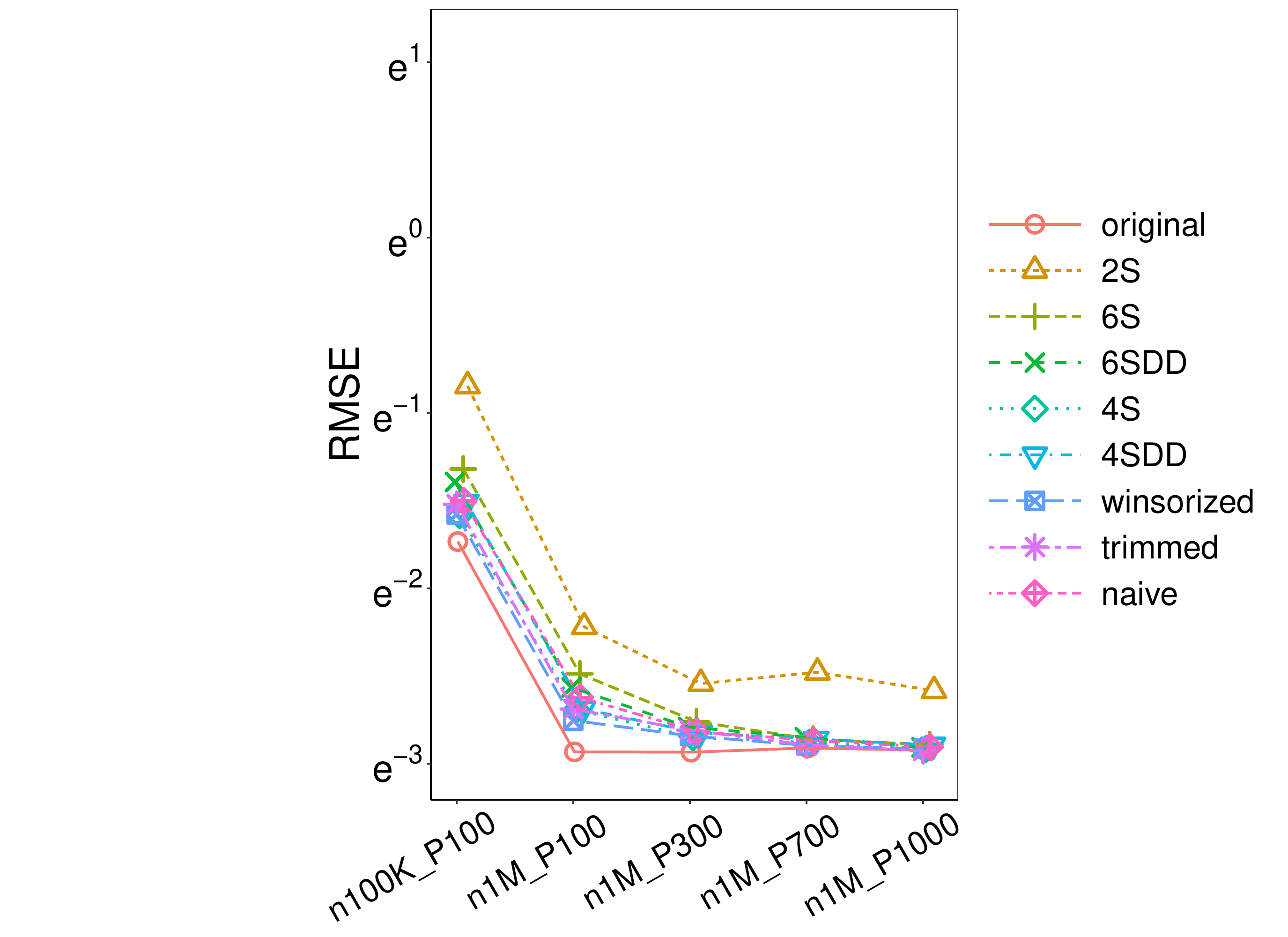}
\includegraphics[width=0.19\textwidth, trim={2.5in 0 2.6in 0},clip] {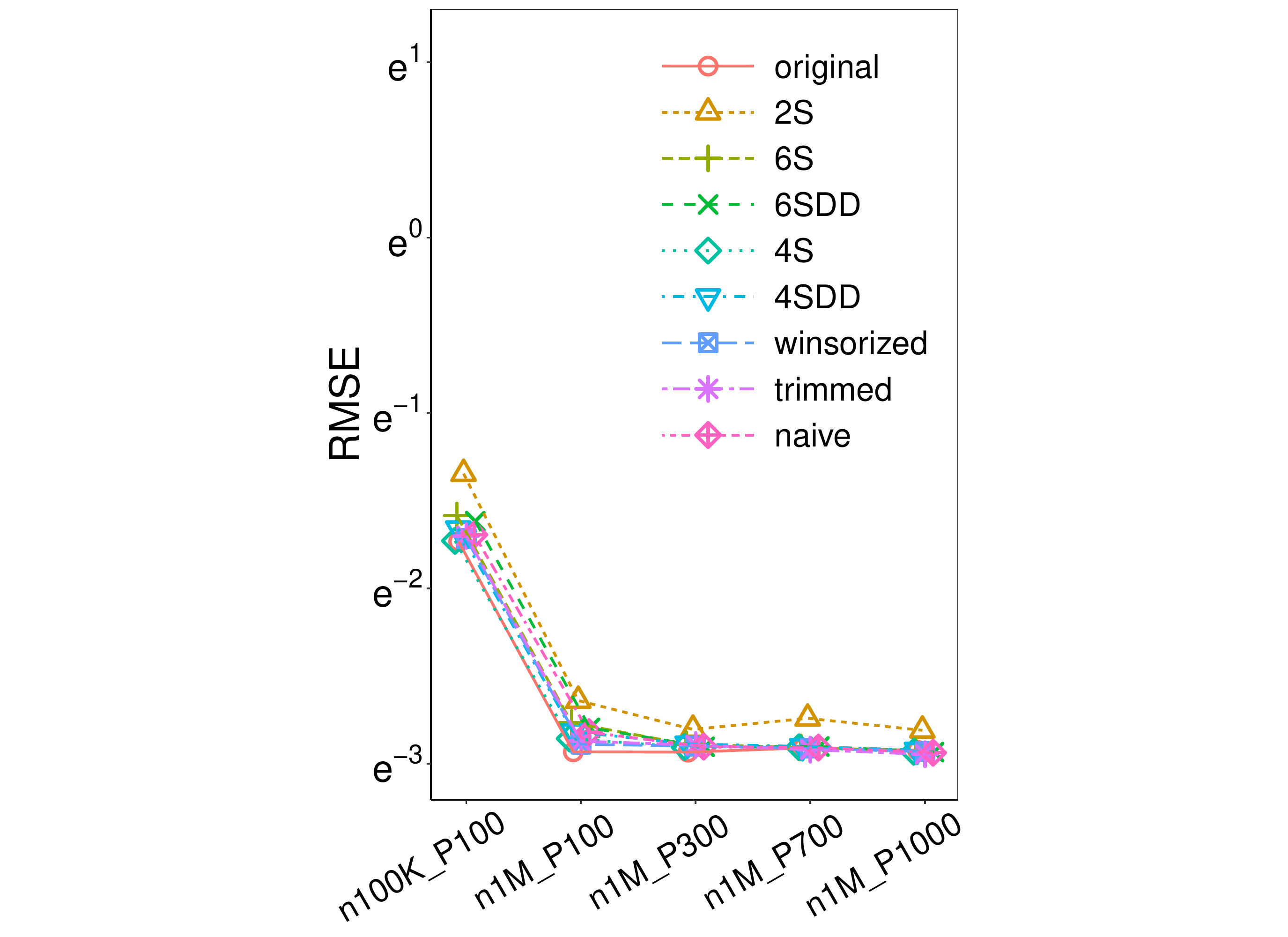}

\includegraphics[width=0.19\textwidth, trim={2.5in 0 2.6in 0},clip] {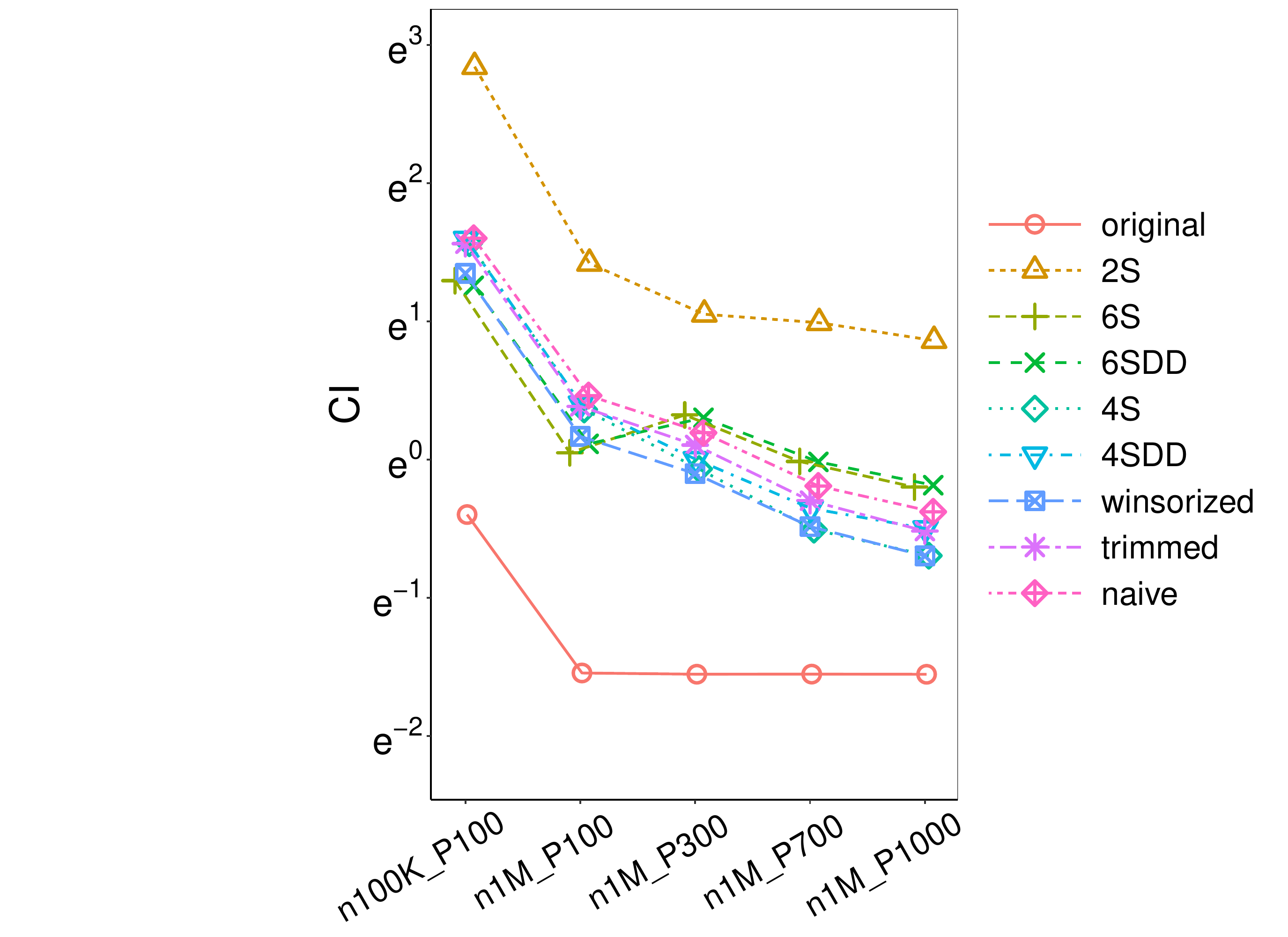}
\includegraphics[width=0.19\textwidth, trim={2.5in 0 2.6in 0},clip] {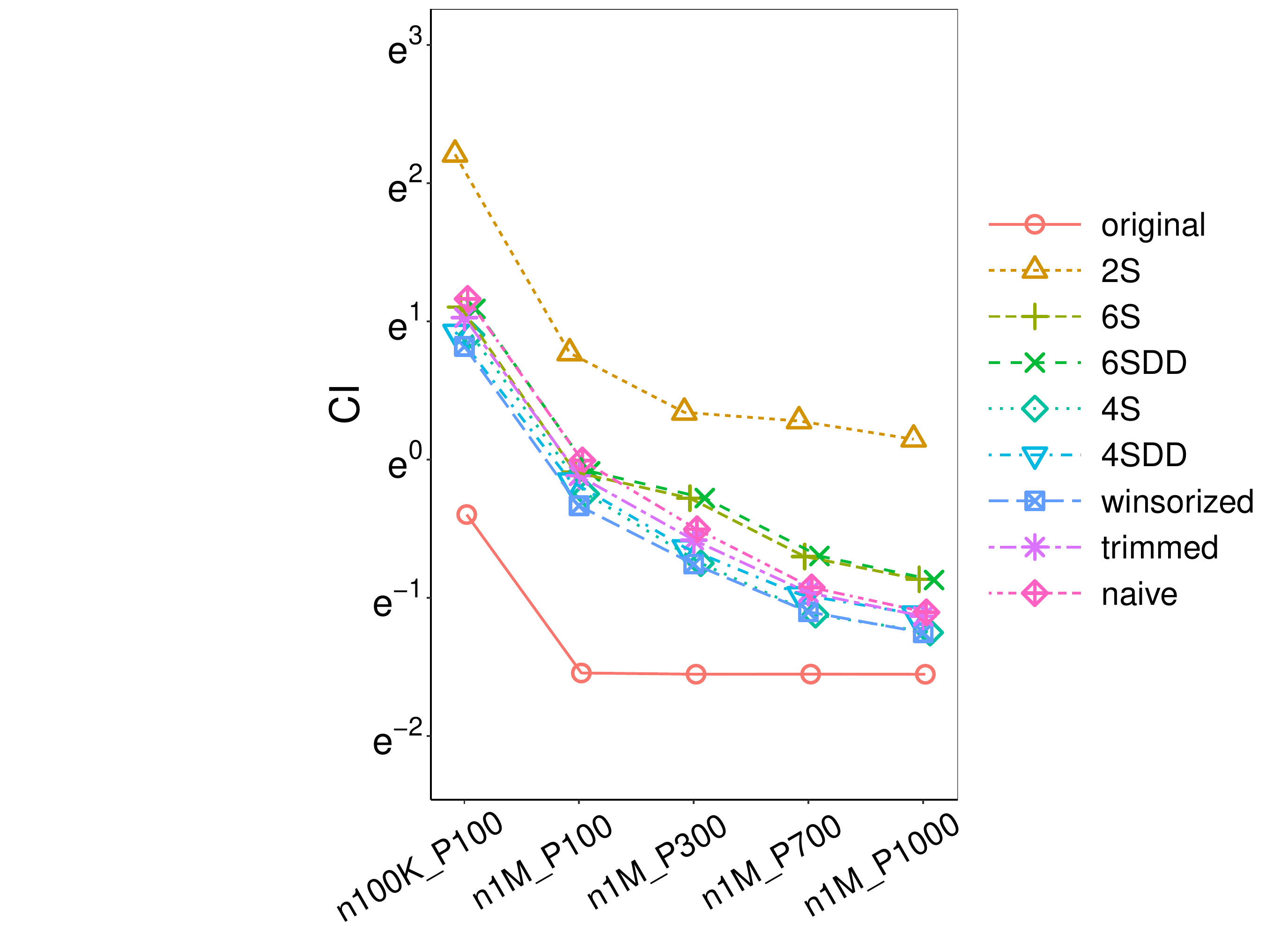}
\includegraphics[width=0.19\textwidth, trim={2.5in 0 2.6in 0},clip] {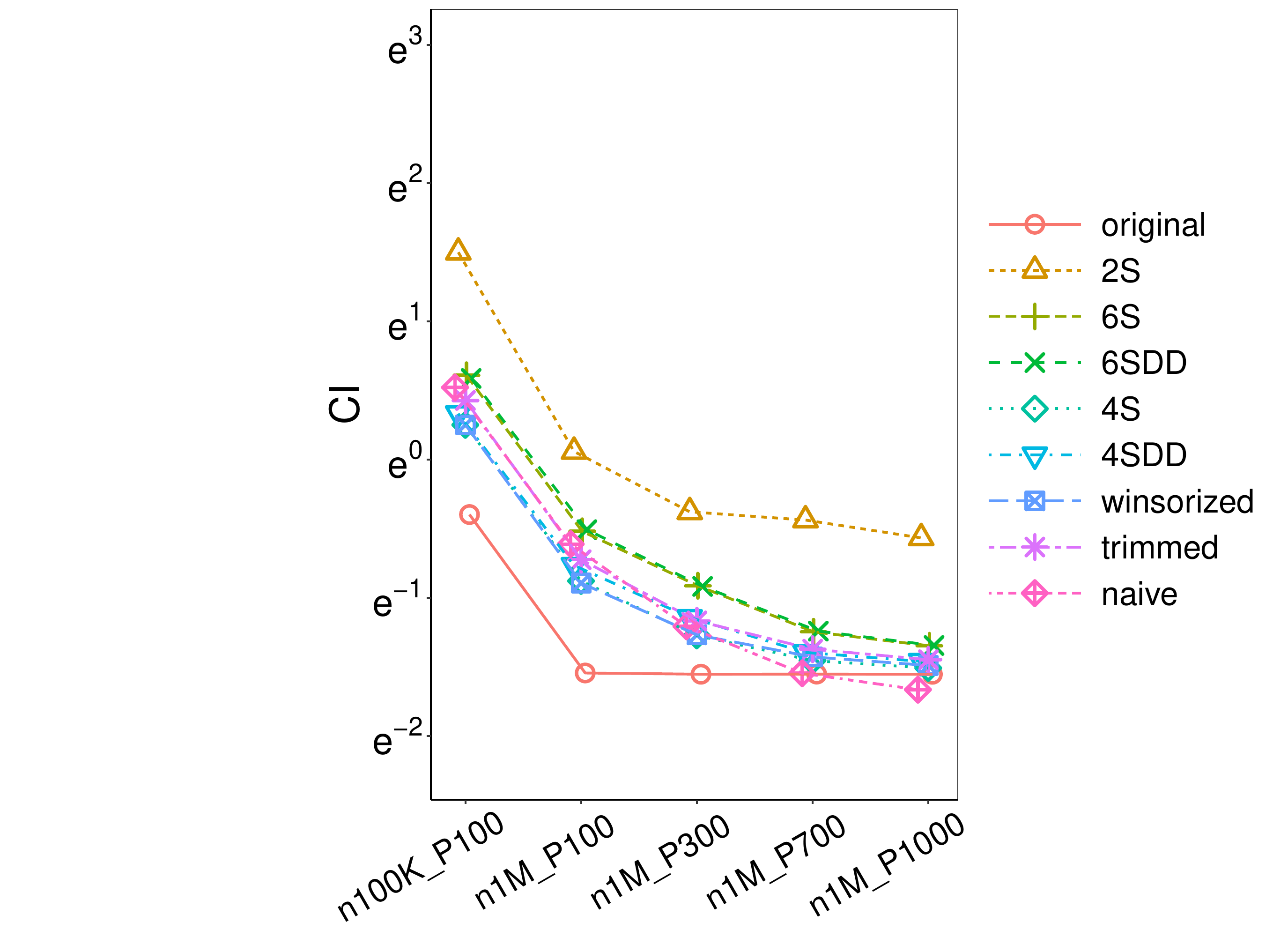}
\includegraphics[width=0.19\textwidth, trim={2.5in 0 2.6in 0},clip] {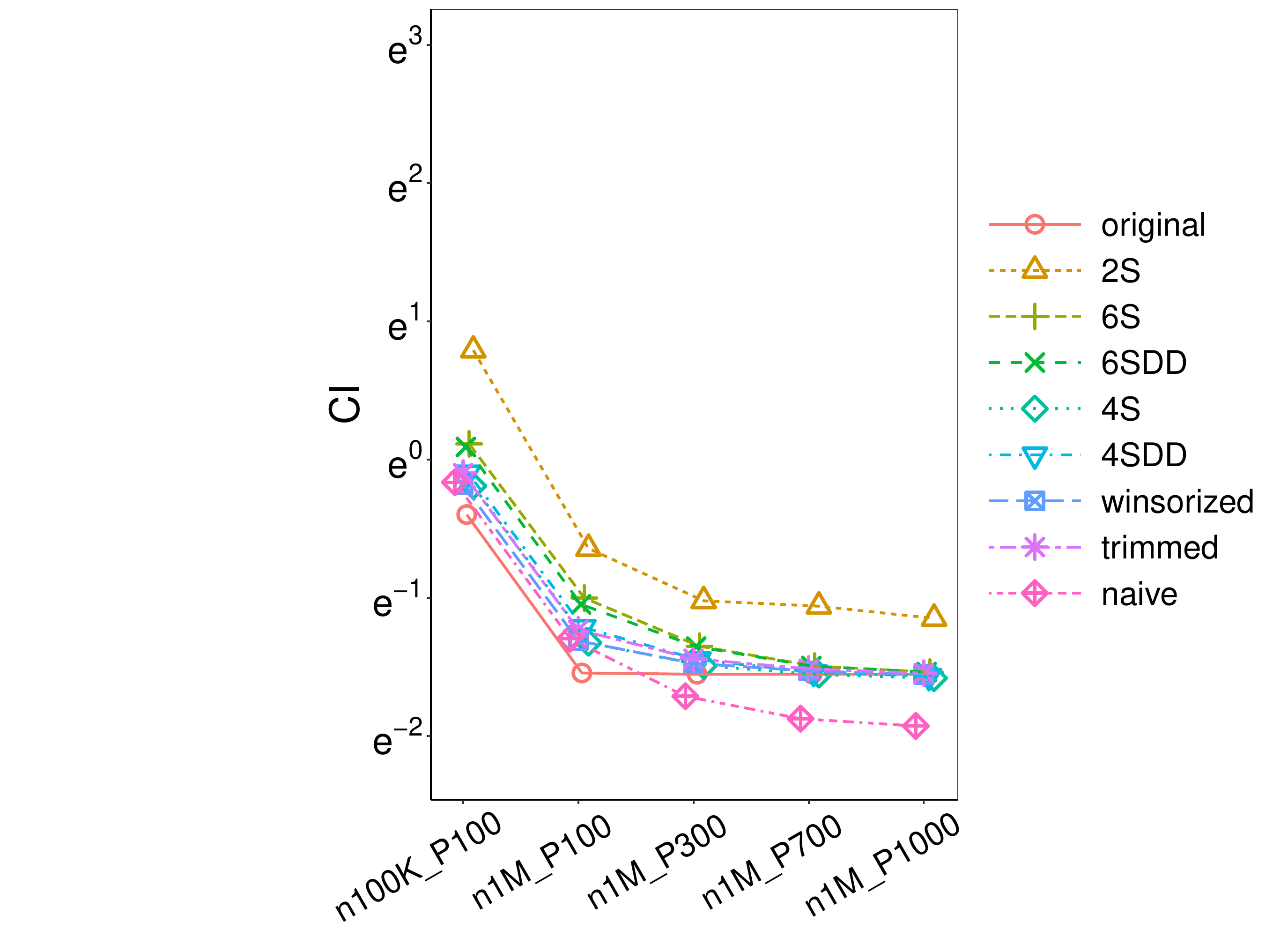}
\includegraphics[width=0.19\textwidth, trim={2.5in 0 2.6in 0},clip] {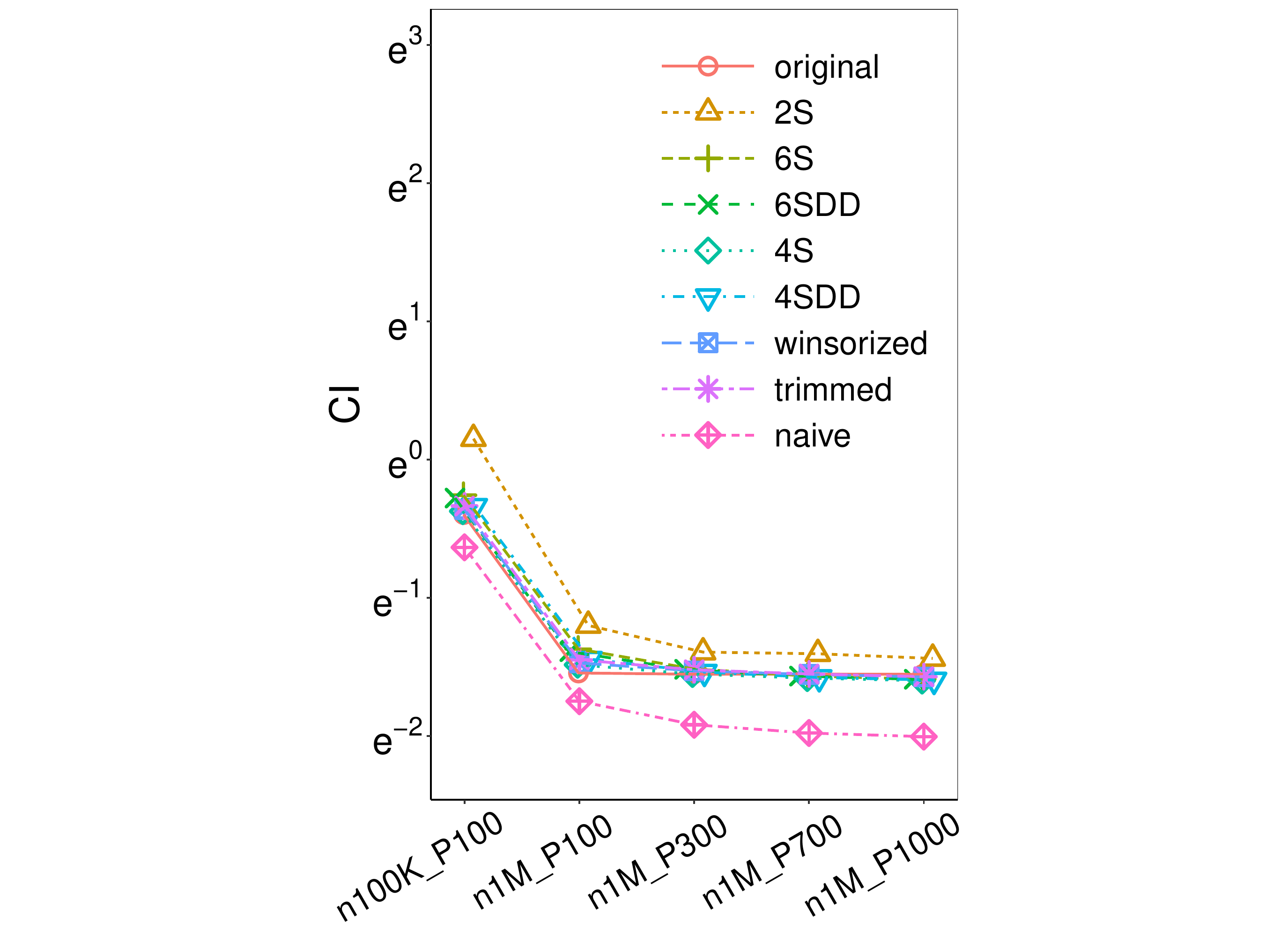}

\includegraphics[width=0.19\textwidth, trim={2.5in 0 2.6in 0},clip] {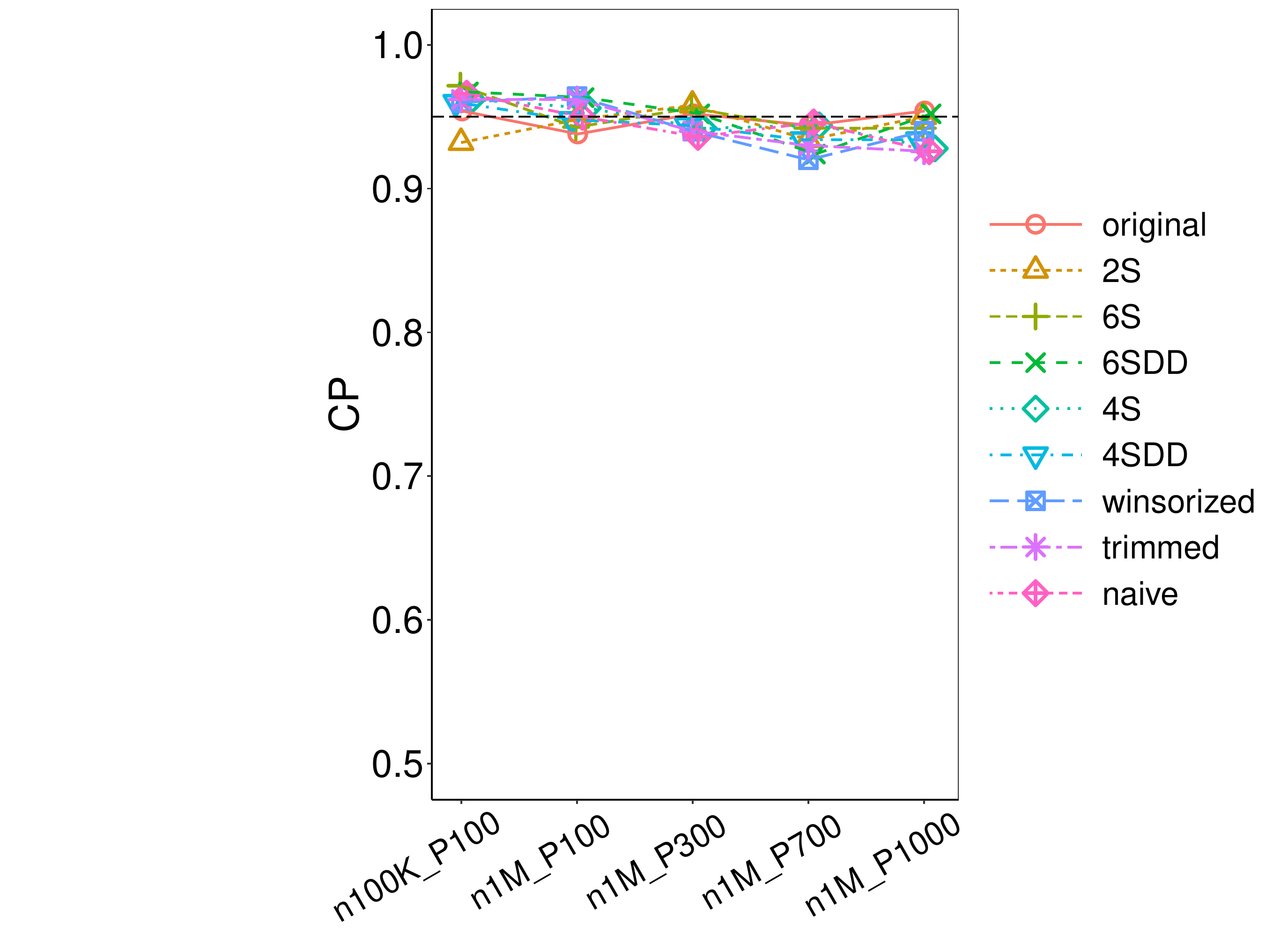}
\includegraphics[width=0.19\textwidth, trim={2.5in 0 2.6in 0},clip] {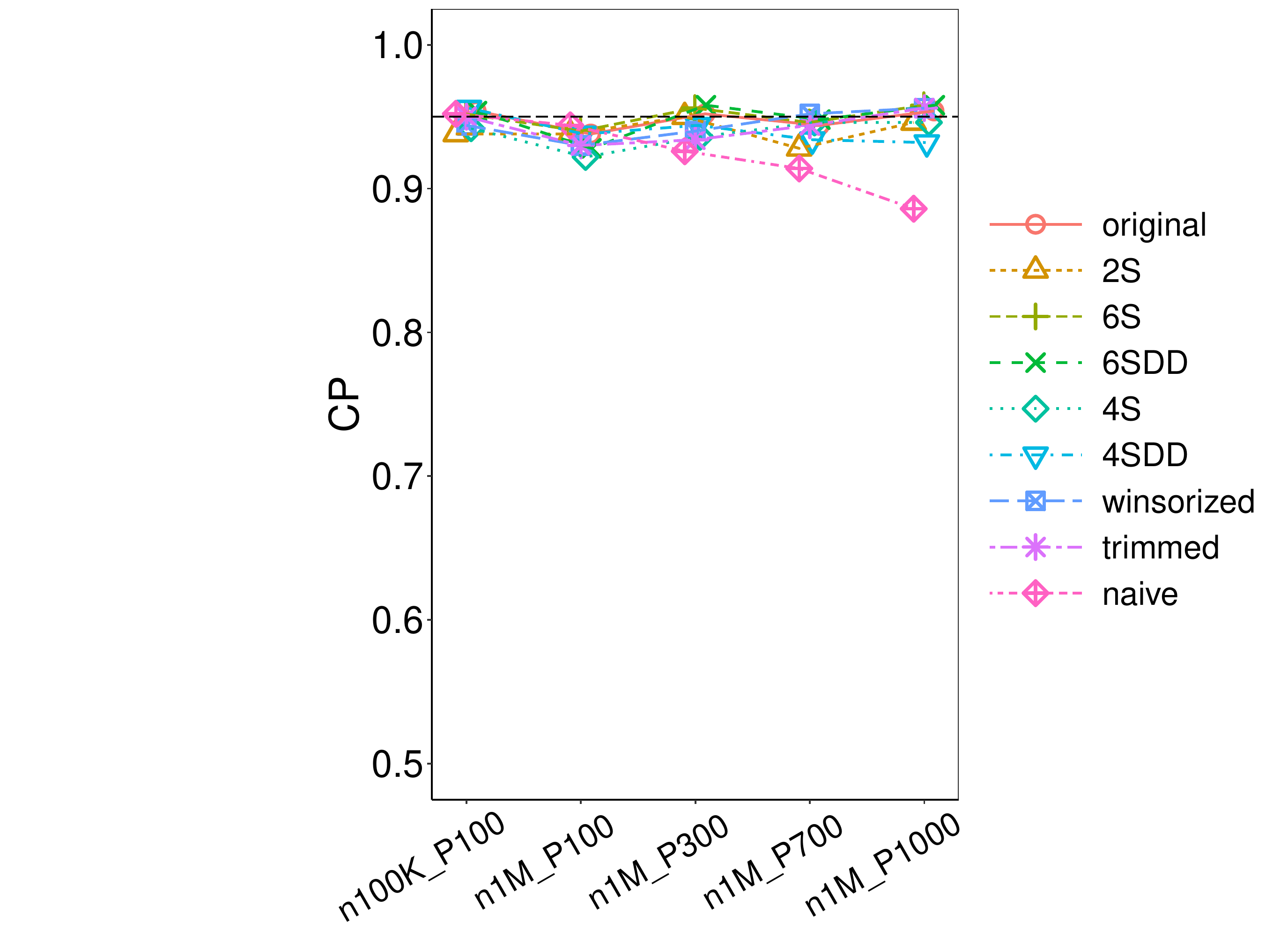}
\includegraphics[width=0.19\textwidth, trim={2.5in 0 2.6in 0},clip] {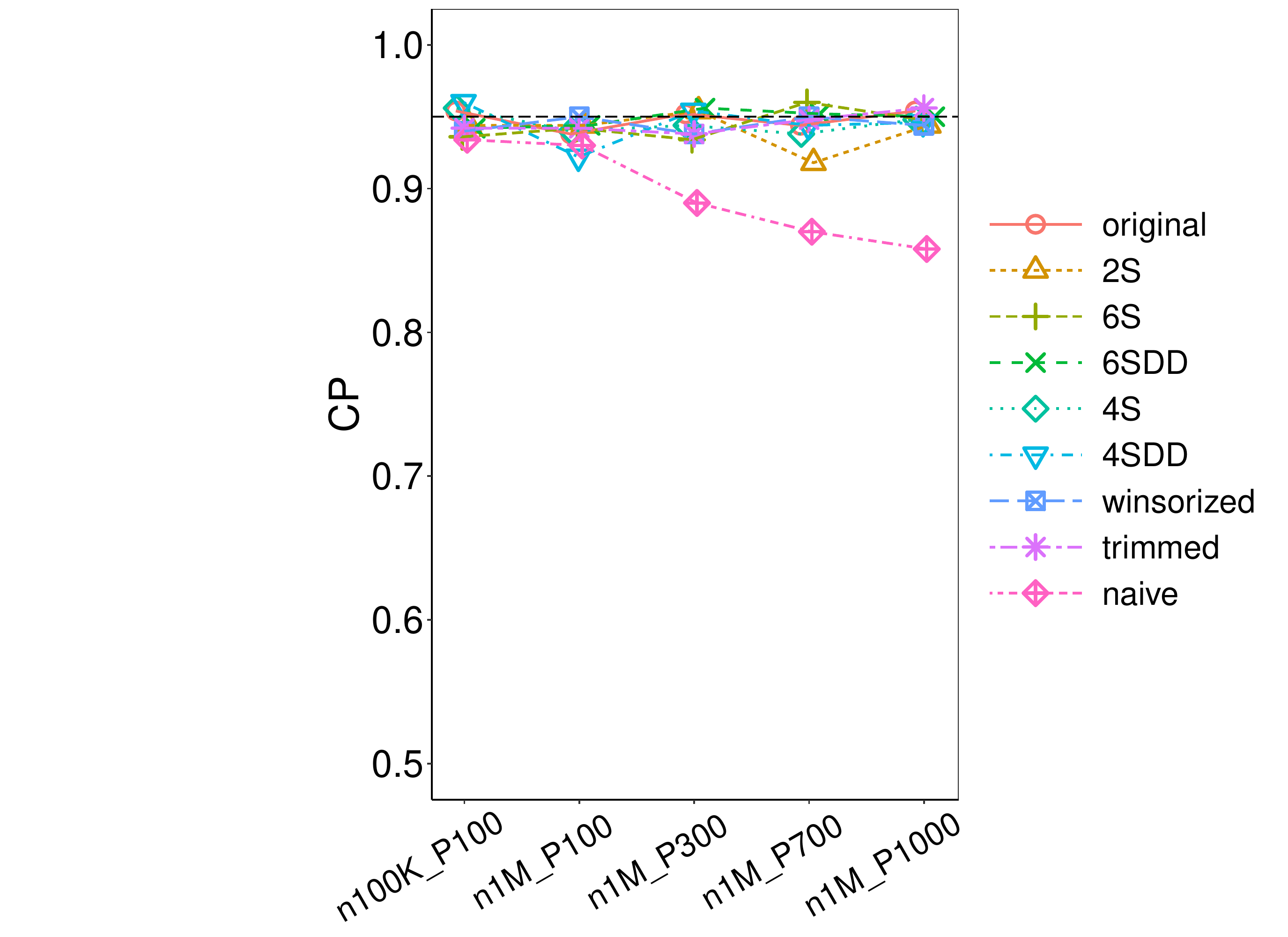}
\includegraphics[width=0.19\textwidth, trim={2.5in 0 2.6in 0},clip] {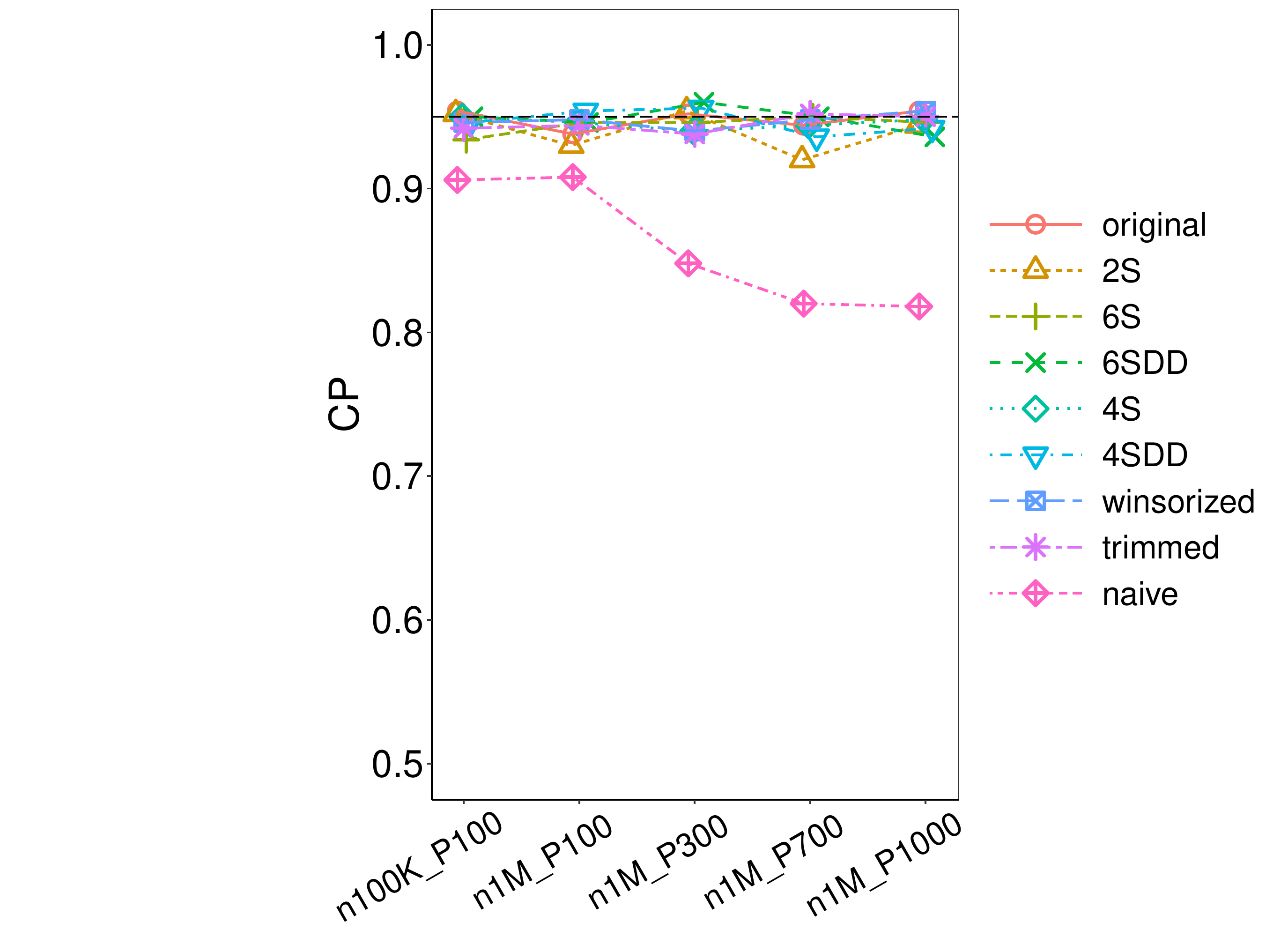}
\includegraphics[width=0.19\textwidth, trim={2.5in 0 2.6in 0},clip] {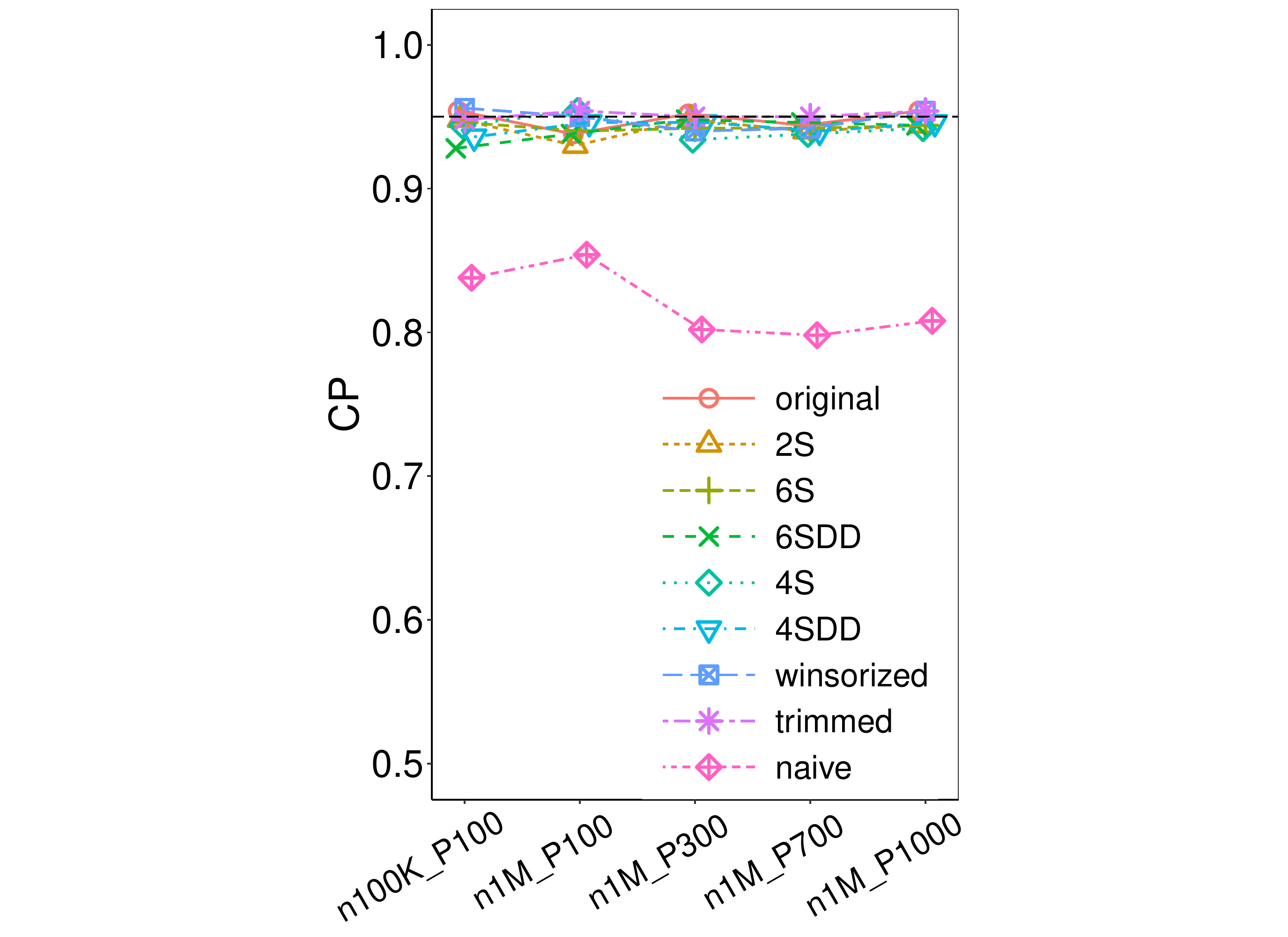}

\caption{Simulation results with $\rho$-zCDP for ZILN data with  $\alpha=\beta$ when $\theta=0$}
\label{fig:0szCDPZILN}
\end{figure}

The results are presented in Figures \ref{fig:0sDPZILN} and \ref{fig:0szCDPZILN} and are summarized  as follows. 4S is the best performer by all metrics and the na\"{i}ve method is the worst. In between the two, some approaches are better than others, depending on the evaluation metrics. In terms of bias, winsorized mean approaches 0 faster  than 6S, 6SDD, 4SDD, and 2S as $n$ or $P$ increases when $n=1m$ or as $\epsilon$ increases. As for RMSE and CI width,  winsorized mean and 4S are the best performers, followed by trimmed mean and 4SDD, and then 6SDD and 6SDD;  2S is the worst. For CP,  all methods, except for the  na\"{i}ve method, provide nominal-level coverage or slight over-coverage.  Trimmed mean is similar to winsorized mean, one might be slightly better than the other, depending on the $n, P,\epsilon$ values and the metrics. The Gaussian mechanism of $\rho$-zCDP does not seem to be advantageous in statistical utility compared to the Laplace mechanism of $\epsilon$-DP. 

\subsubsection{\texorpdfstring{$Y\sim$ ZILN}{}, \texorpdfstring{$\theta\ne0$ ($H_1$)}{H1 }, and \texorpdfstring{$\alpha=\beta$}{S}}\label{sec:1sZILN}
\begin{figure}[!htb]
\hspace{0.5in}$\epsilon=0.5$\hspace{0.8in}$\epsilon=1$\hspace{0.9in}$\epsilon=2$
\hspace{0.95in}$\epsilon=5$\hspace{0.9in}$\epsilon=50$

\includegraphics[width=0.19\textwidth, trim={2.5in 0 2.6in 0},clip] {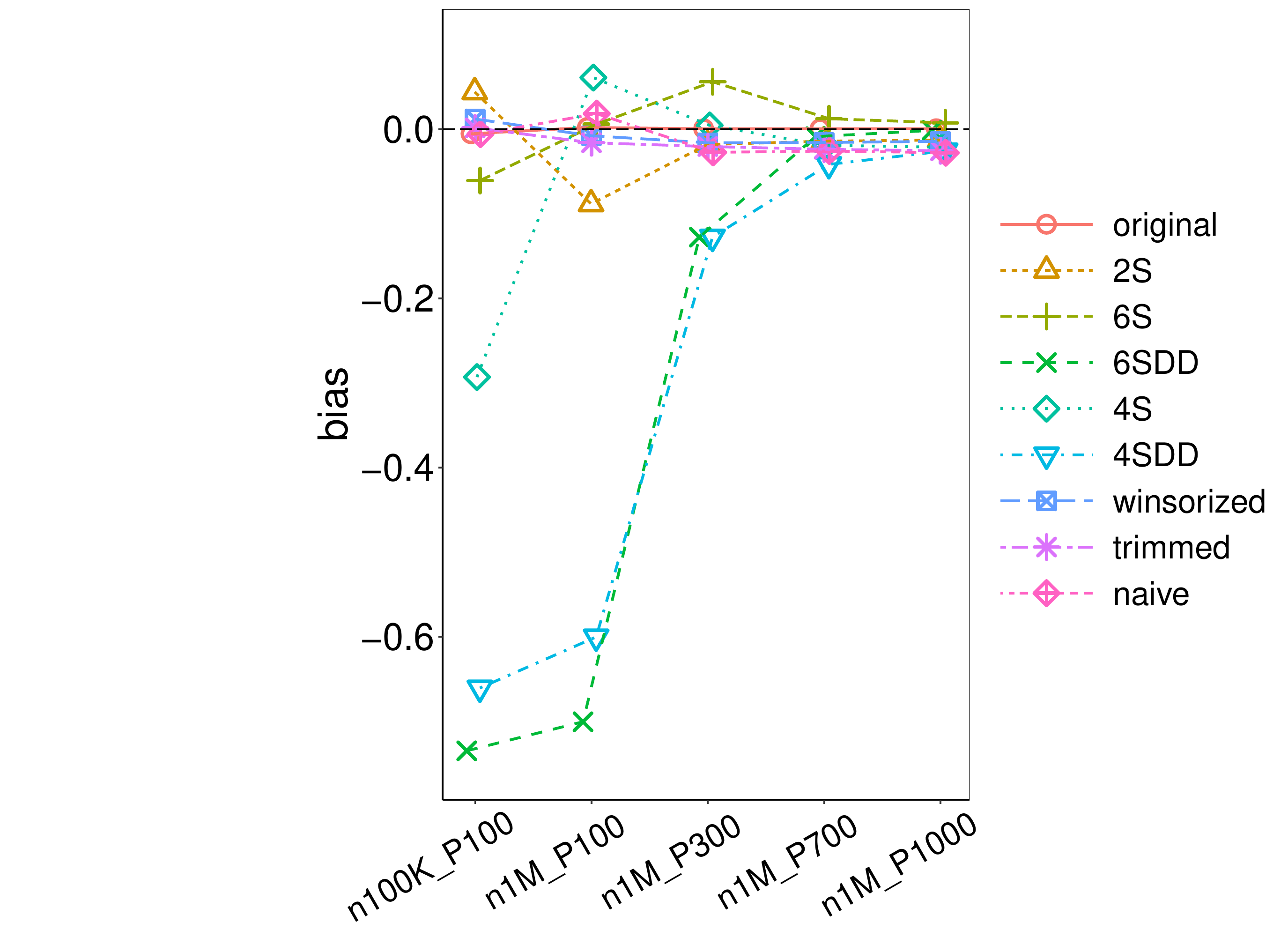}
\includegraphics[width=0.19\textwidth, trim={2.5in 0 2.6in 0},clip] {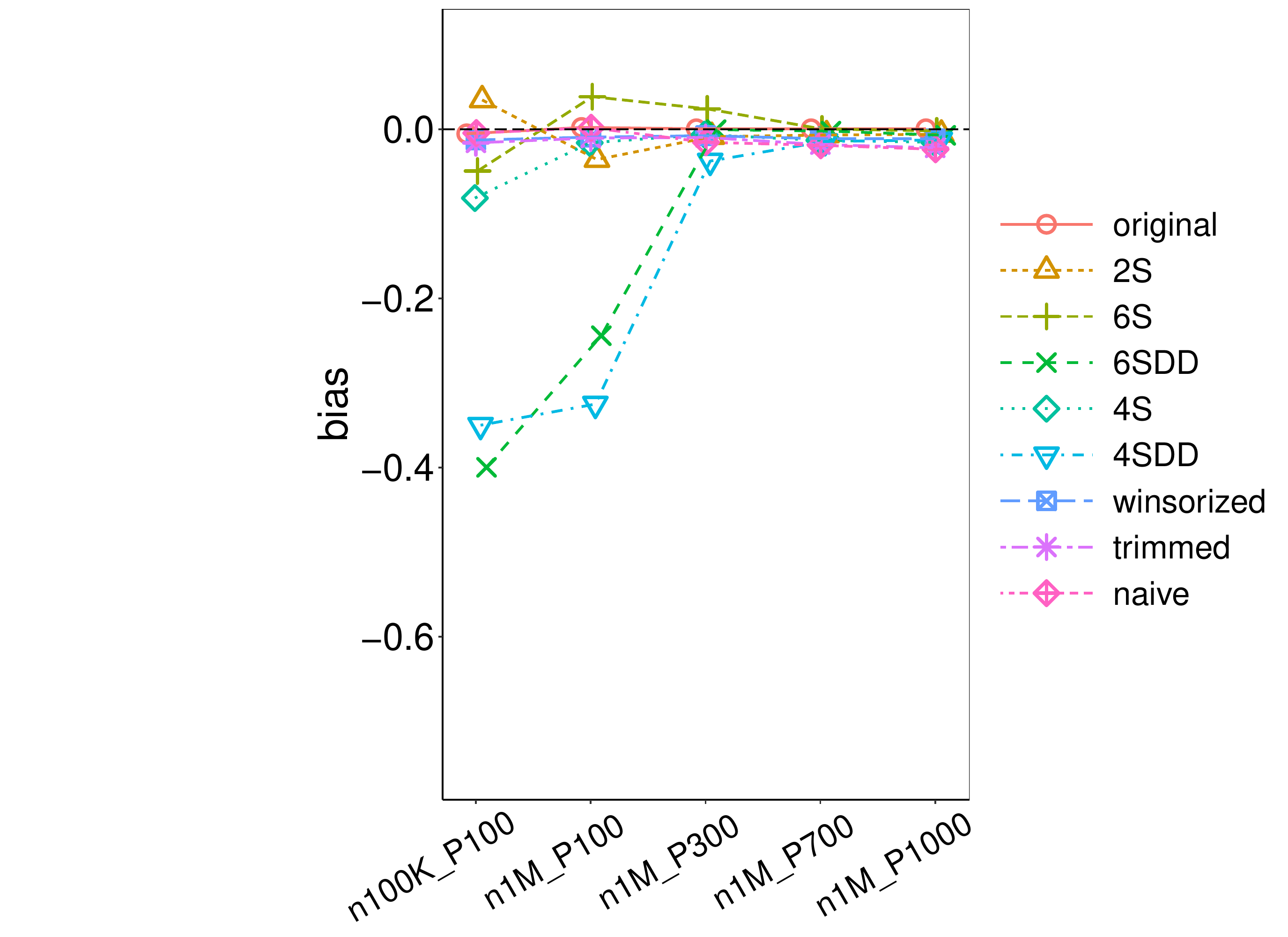}
\includegraphics[width=0.19\textwidth, trim={2.5in 0 2.6in 0},clip] {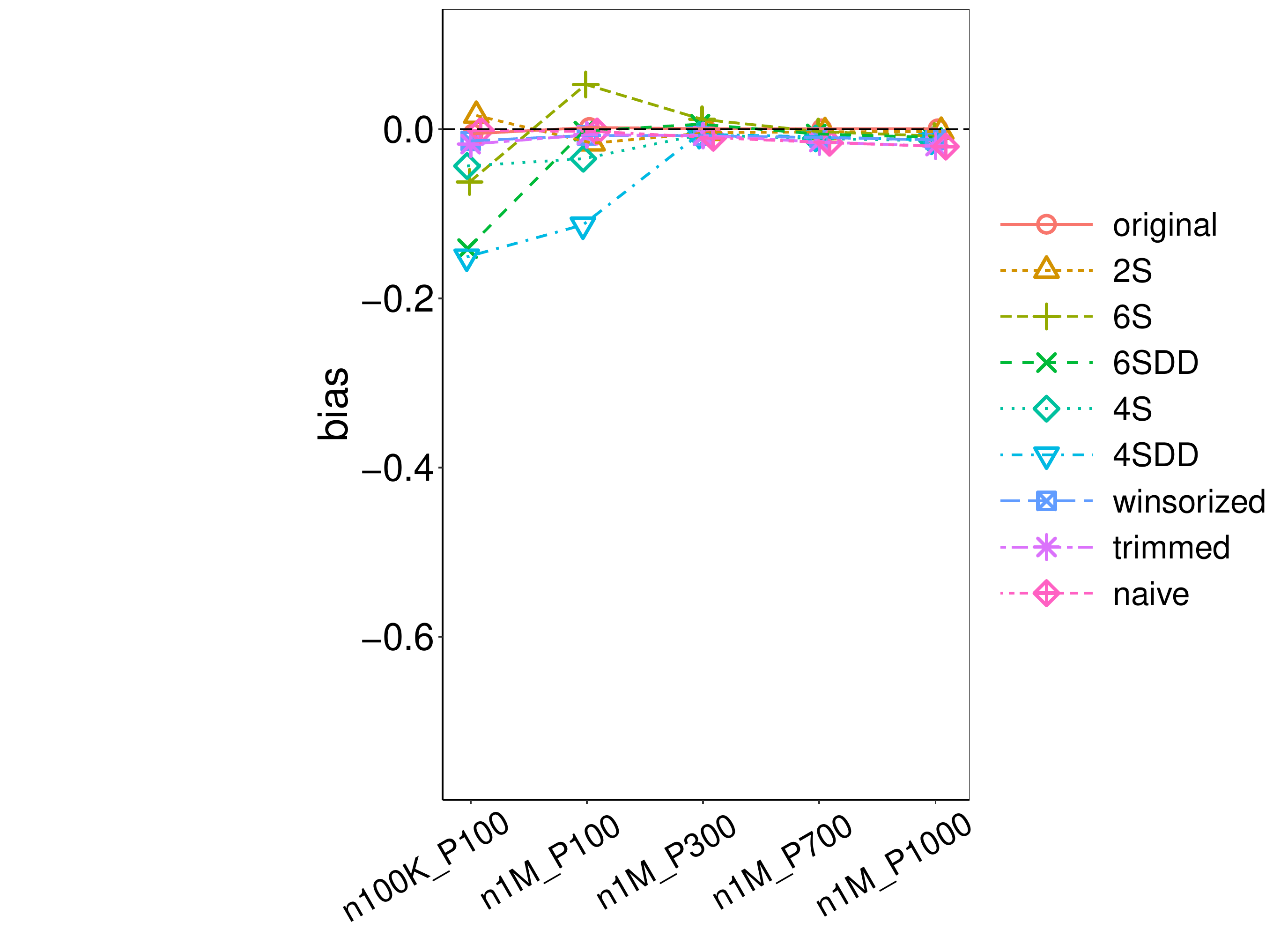}
\includegraphics[width=0.19\textwidth, trim={2.5in 0 2.6in 0},clip] {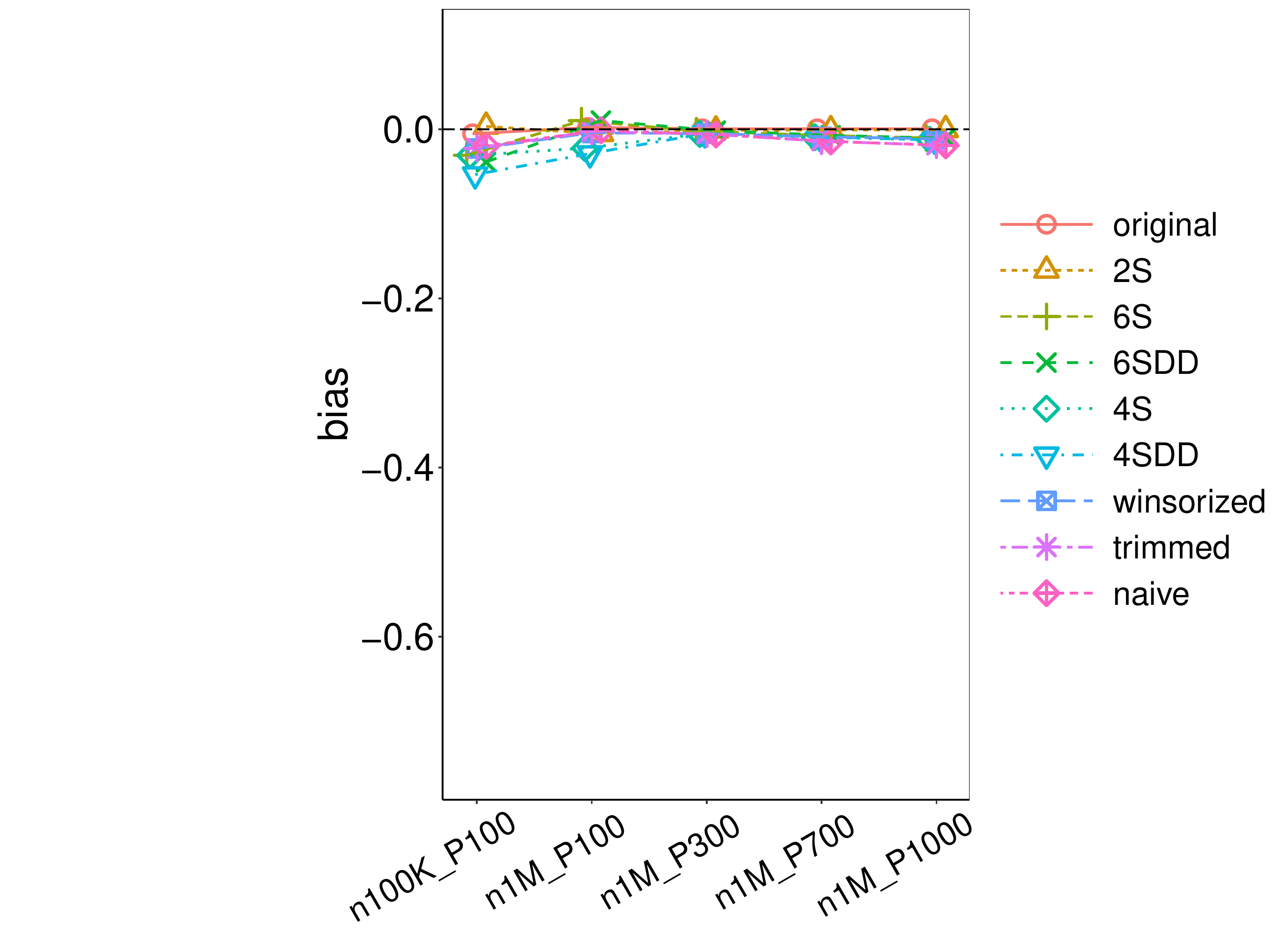}
\includegraphics[width=0.19\textwidth, trim={2.5in 0 2.6in 0},clip] {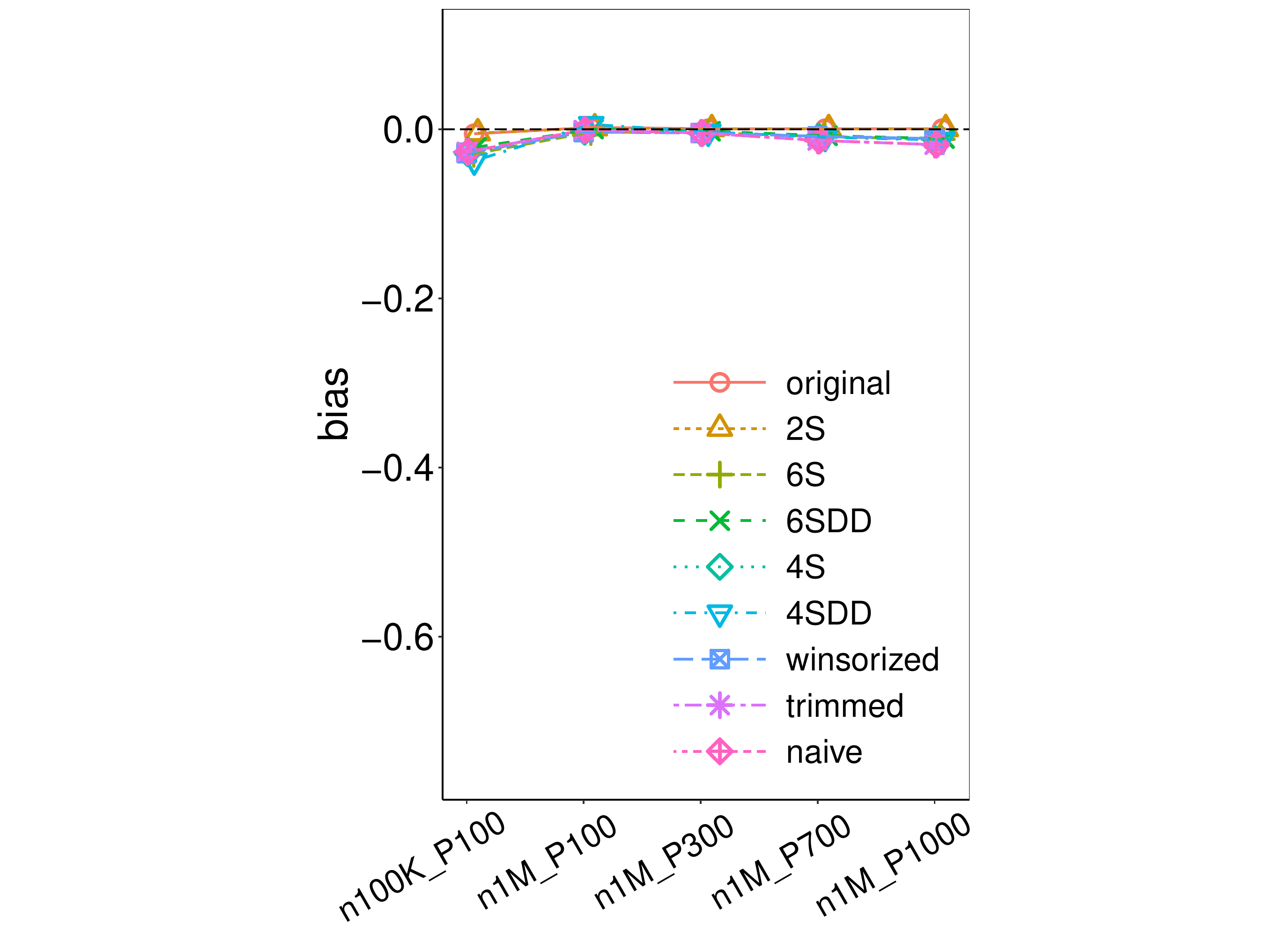}

\includegraphics[width=0.19\textwidth, trim={2.5in 0 2.6in 0},clip] {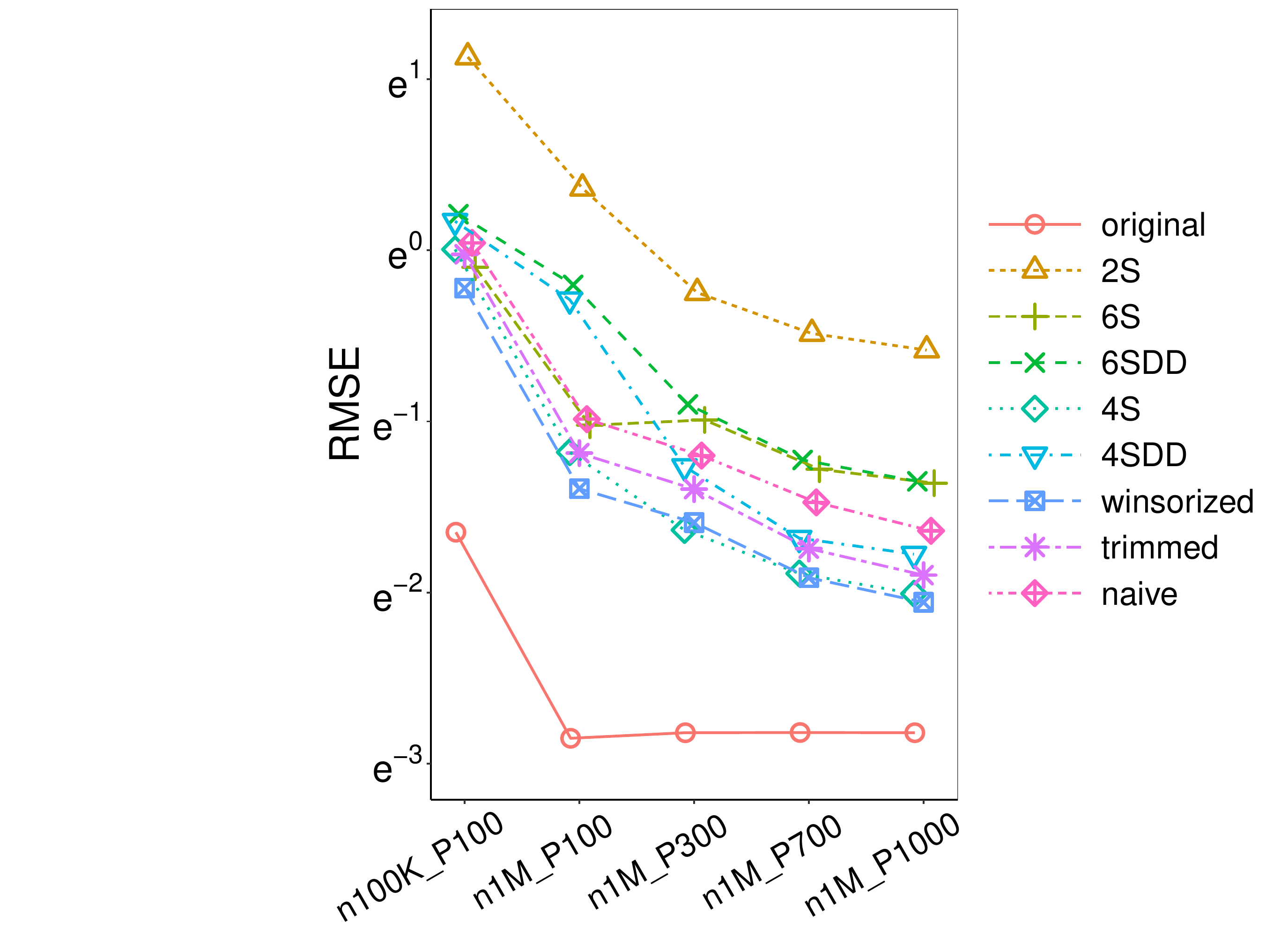}
\includegraphics[width=0.19\textwidth, trim={2.5in 0 2.6in 0},clip] {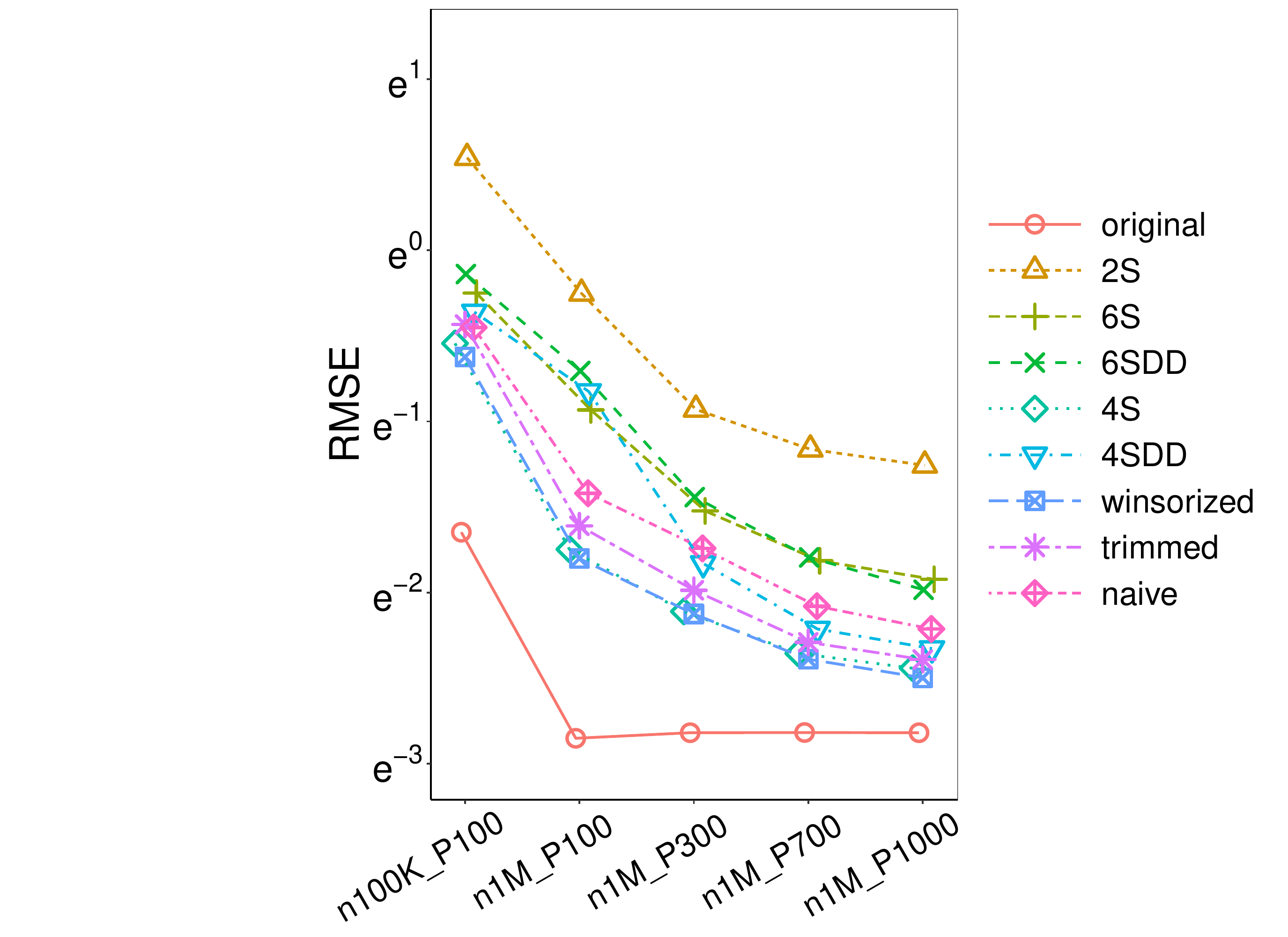}
\includegraphics[width=0.19\textwidth, trim={2.5in 0 2.6in 0},clip] {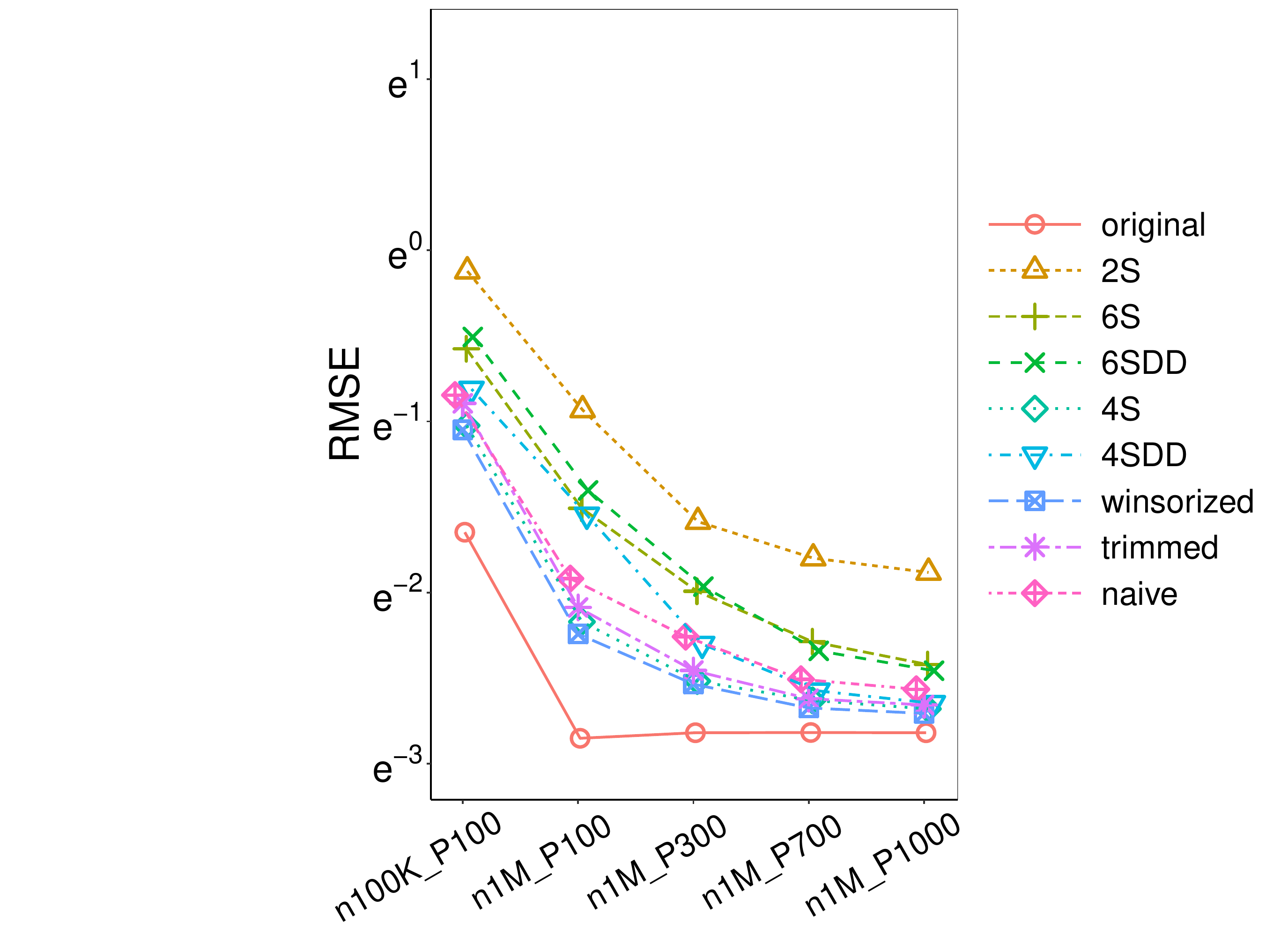}
\includegraphics[width=0.19\textwidth, trim={2.5in 0 2.6in 0},clip] {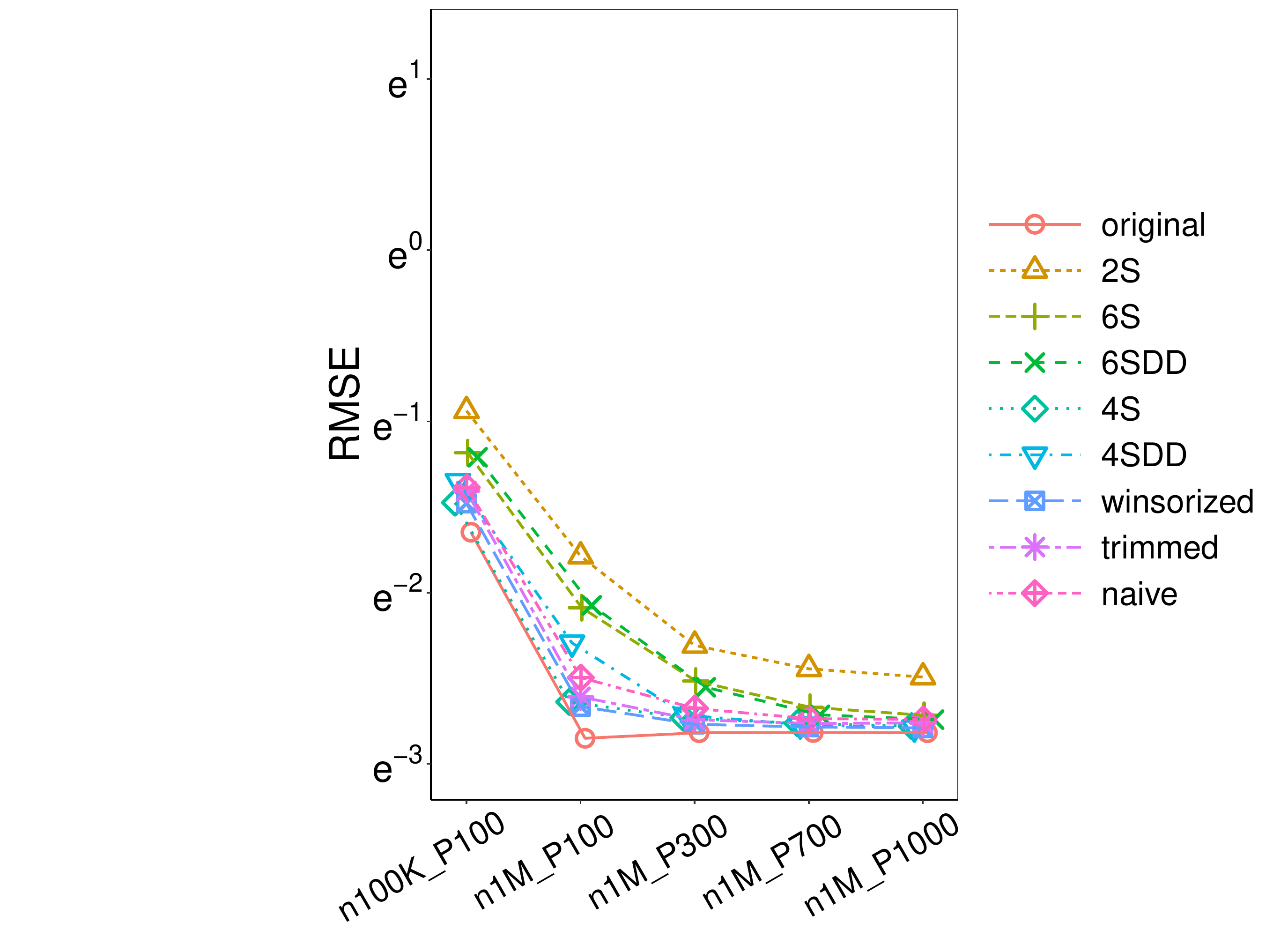}
\includegraphics[width=0.19\textwidth, trim={2.5in 0 2.6in 0},clip] {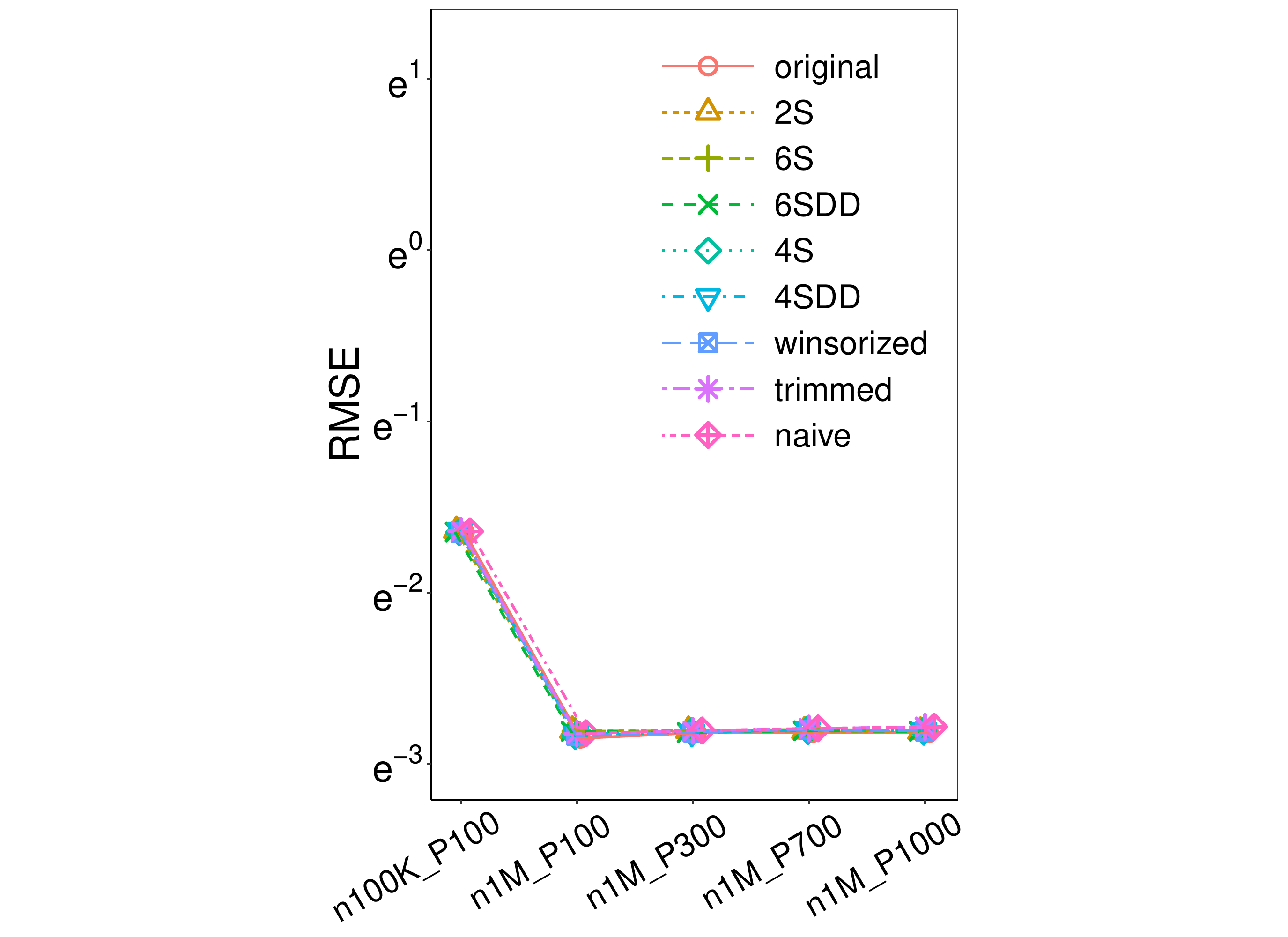}

\includegraphics[width=0.19\textwidth, trim={2.5in 0 2.6in 0},clip] {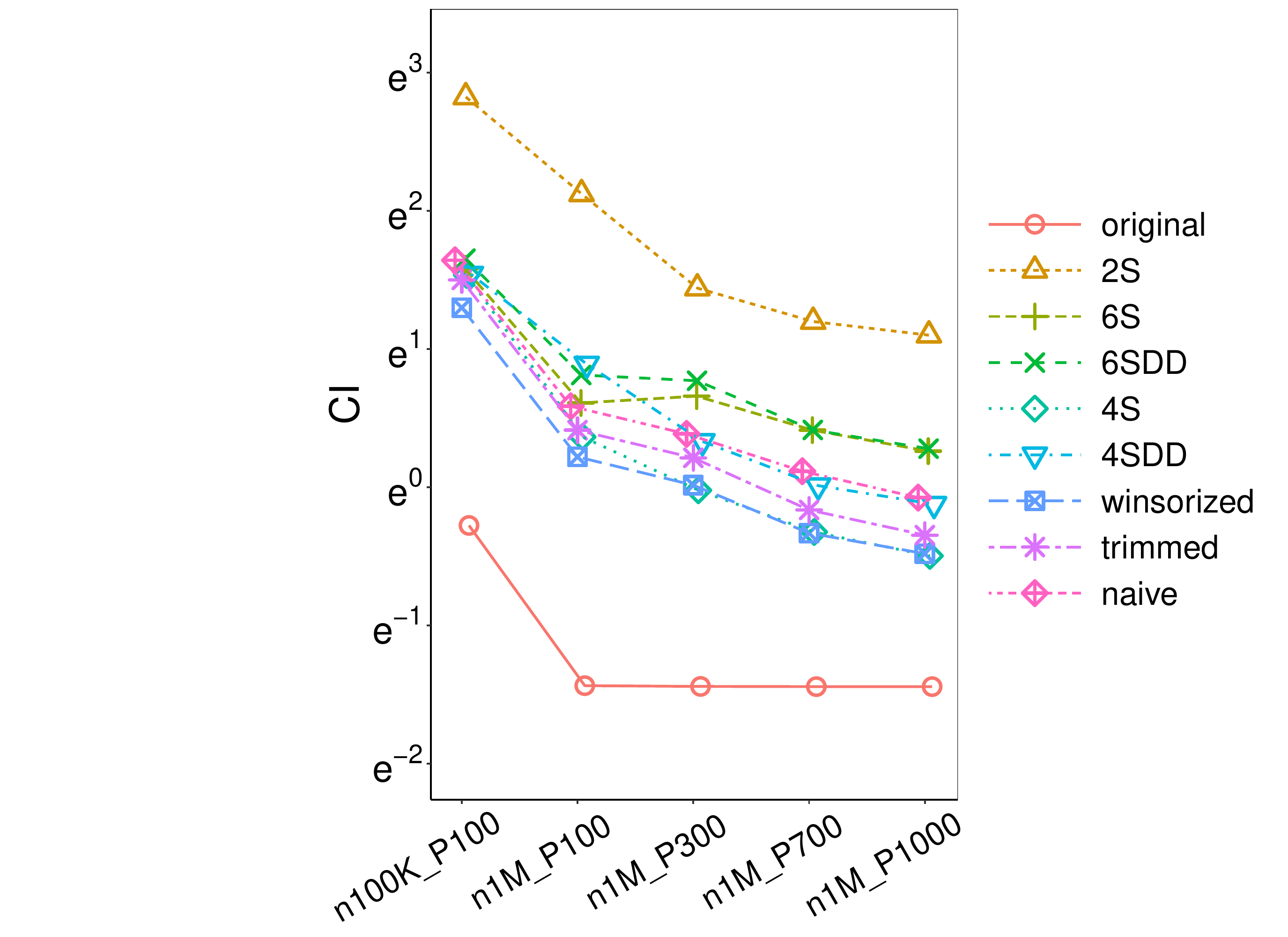}
\includegraphics[width=0.19\textwidth, trim={2.5in 0 2.6in 0},clip] {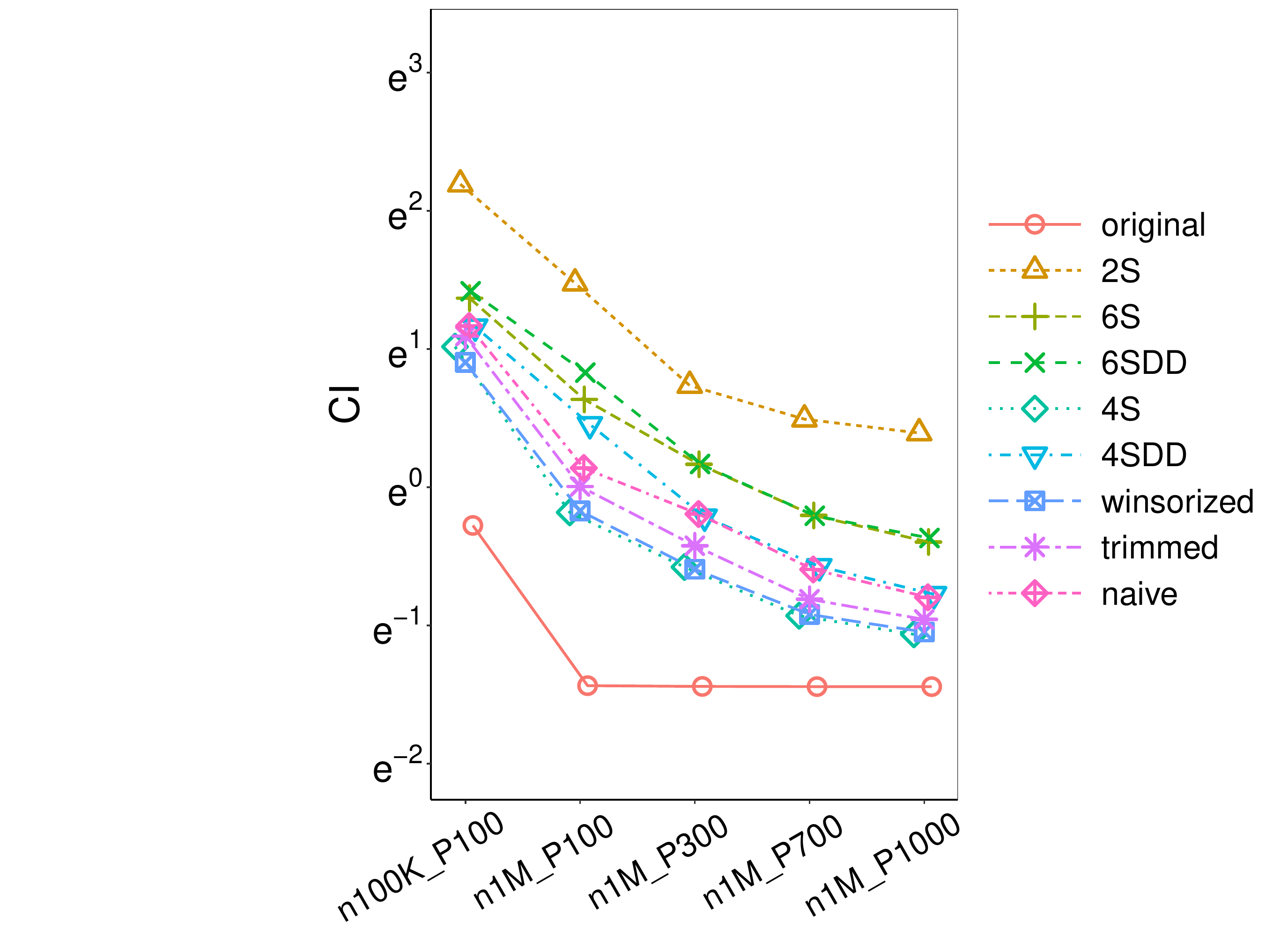}
\includegraphics[width=0.19\textwidth, trim={2.5in 0 2.6in 0},clip] {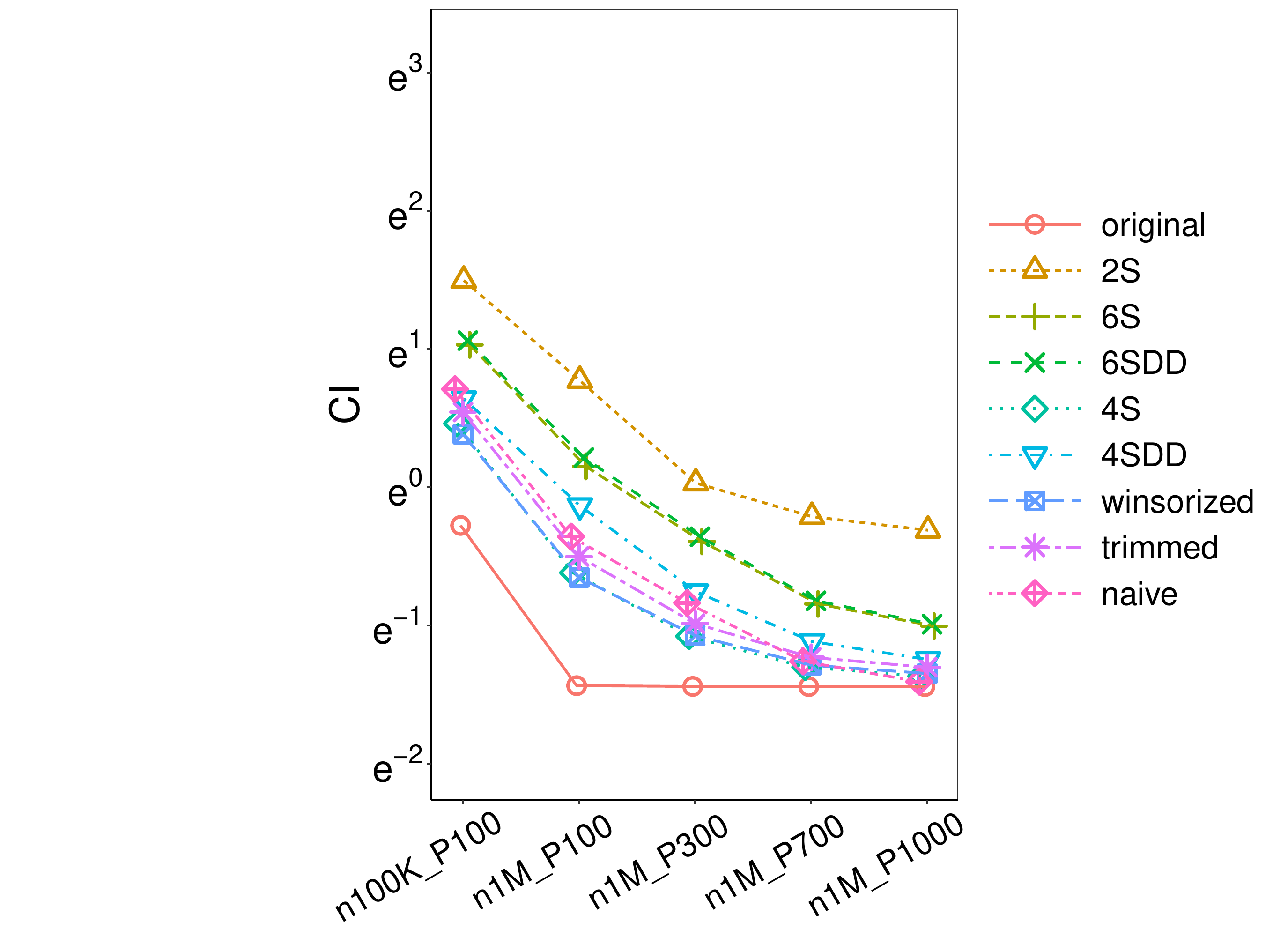}
\includegraphics[width=0.19\textwidth, trim={2.5in 0 2.6in 0},clip] {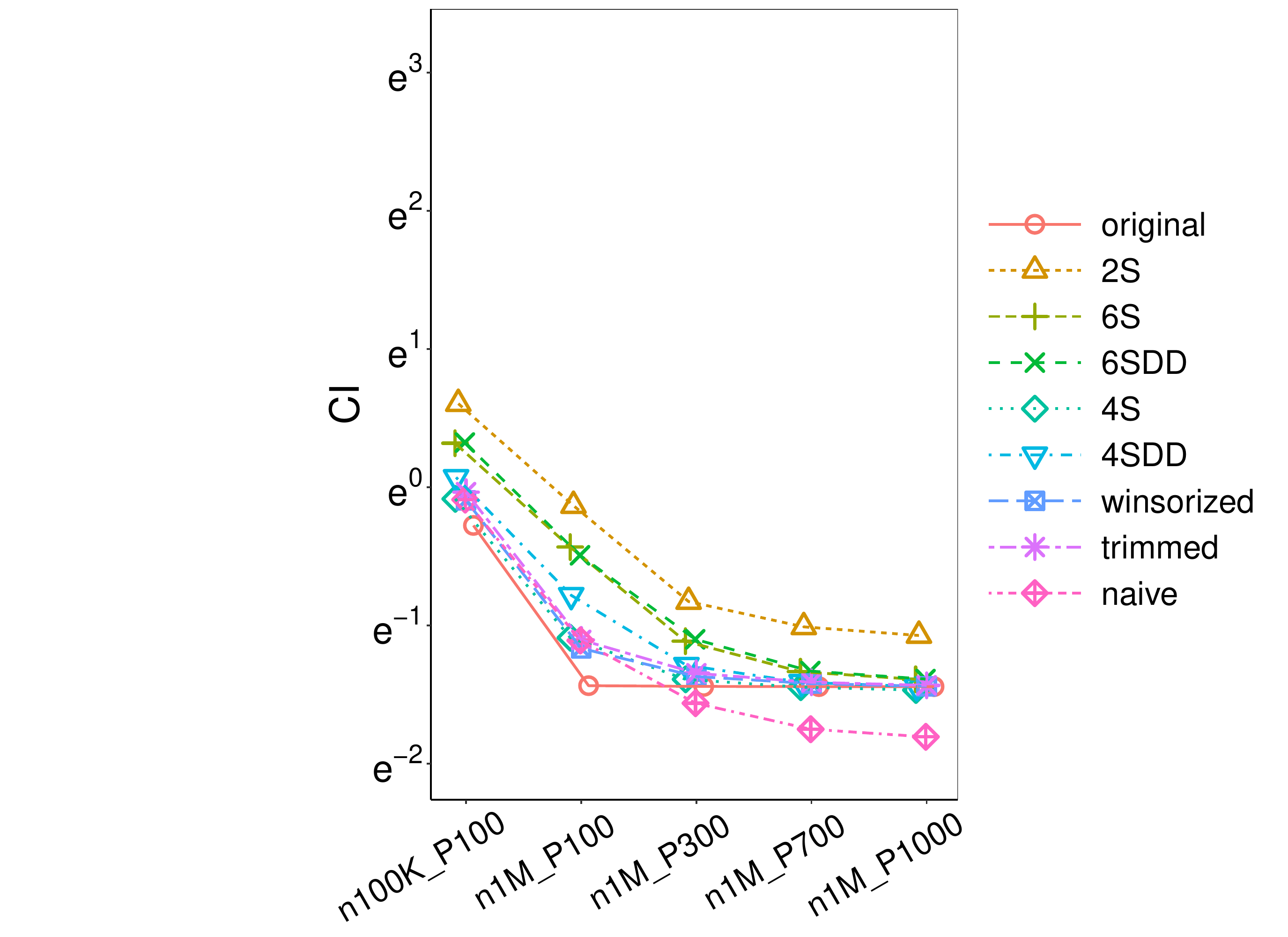}
\includegraphics[width=0.19\textwidth, trim={2.5in 0 2.6in 0},clip] {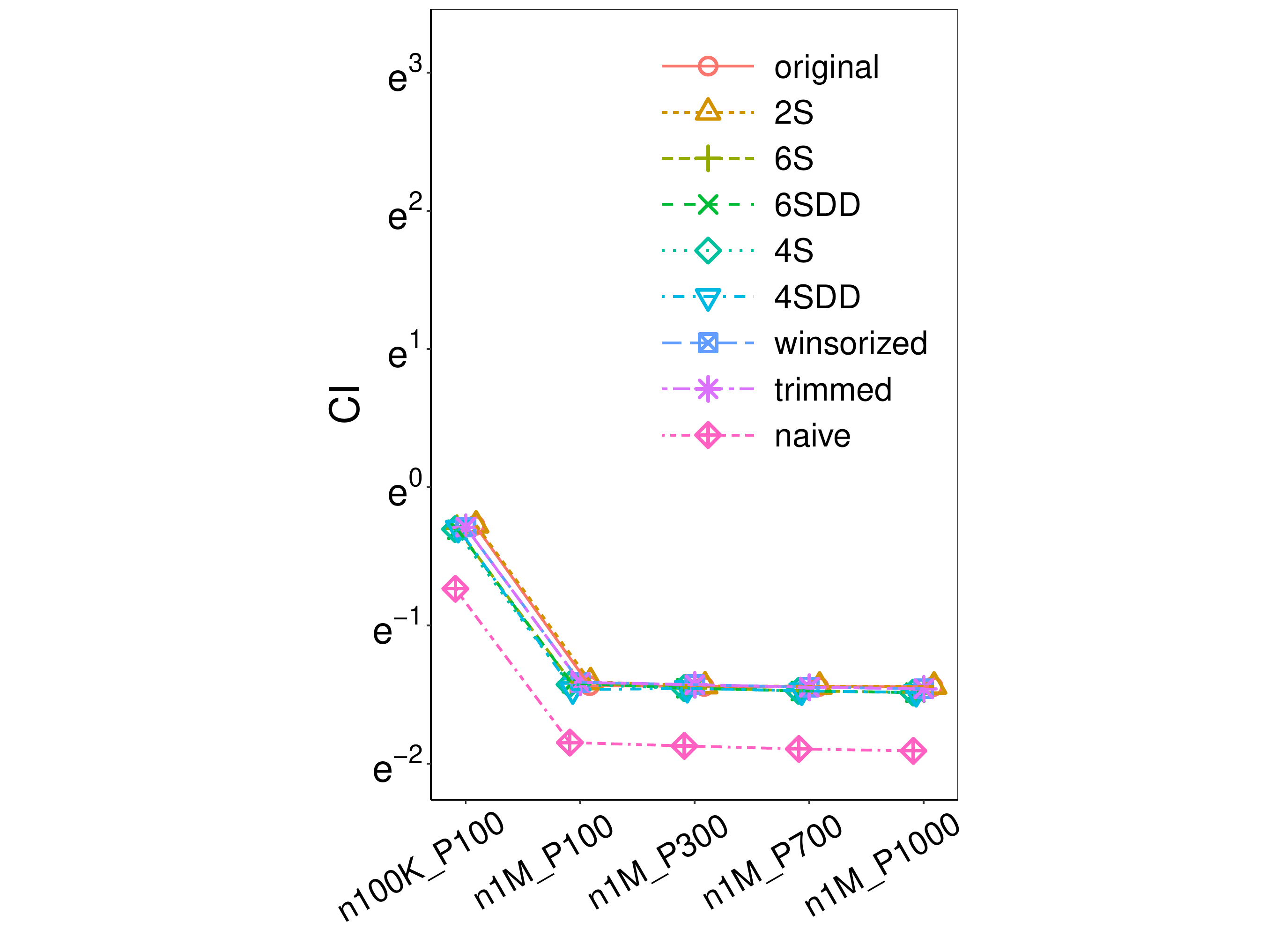}

\includegraphics[width=0.19\textwidth, trim={2.5in 0 2.6in 0},clip] {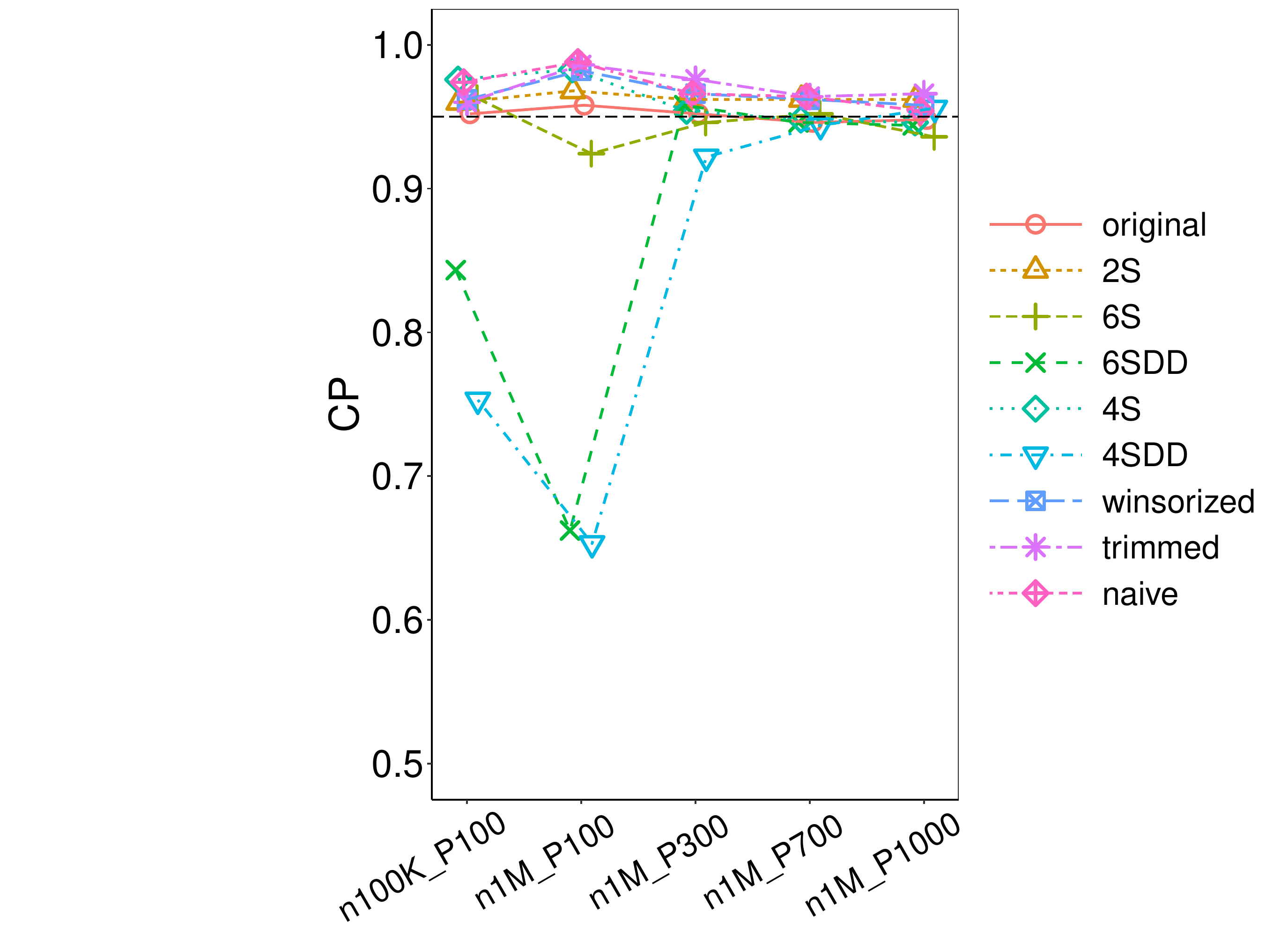}
\includegraphics[width=0.19\textwidth, trim={2.5in 0 2.6in 0},clip] {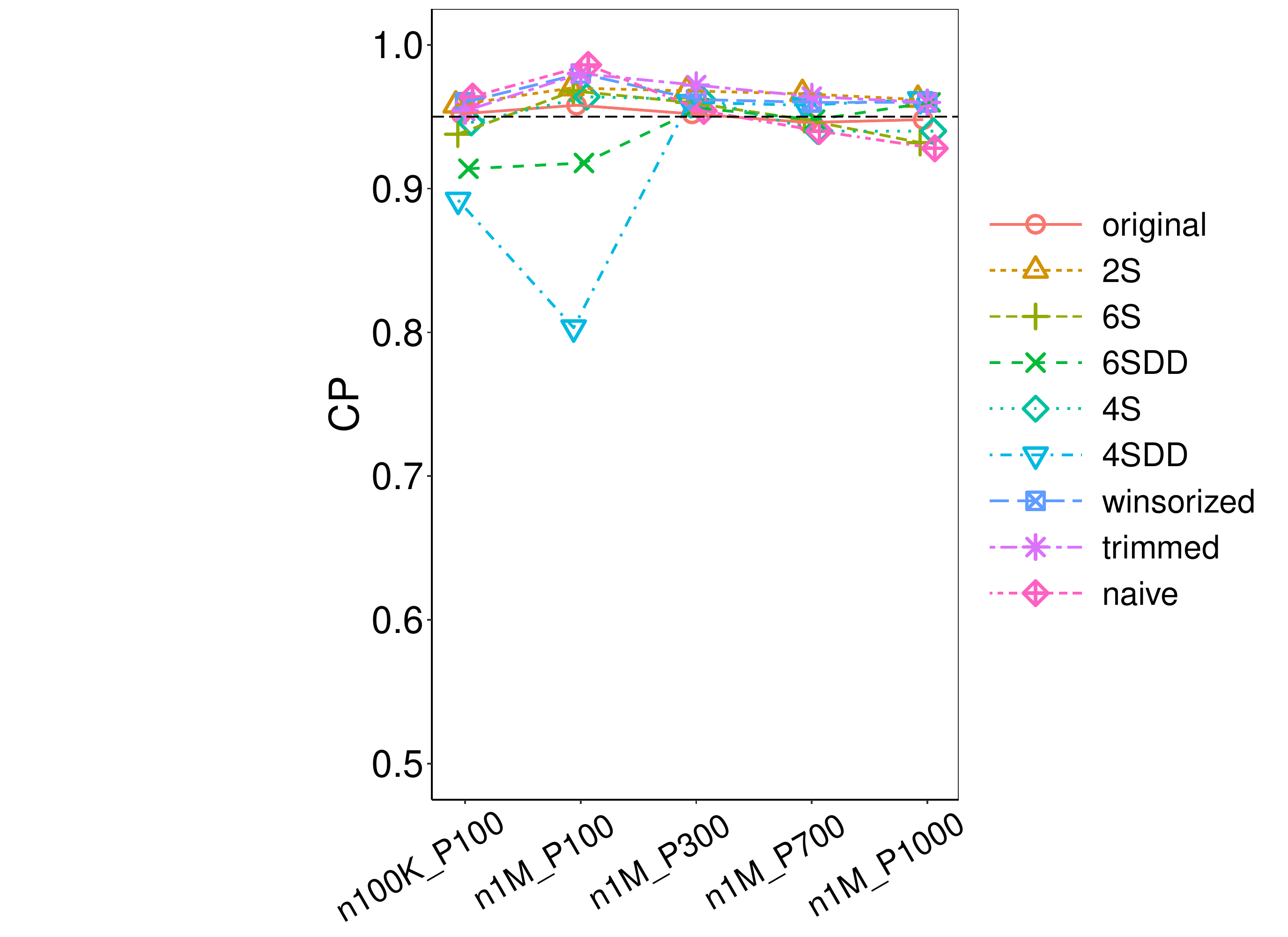}
\includegraphics[width=0.19\textwidth, trim={2.5in 0 2.6in 0},clip] {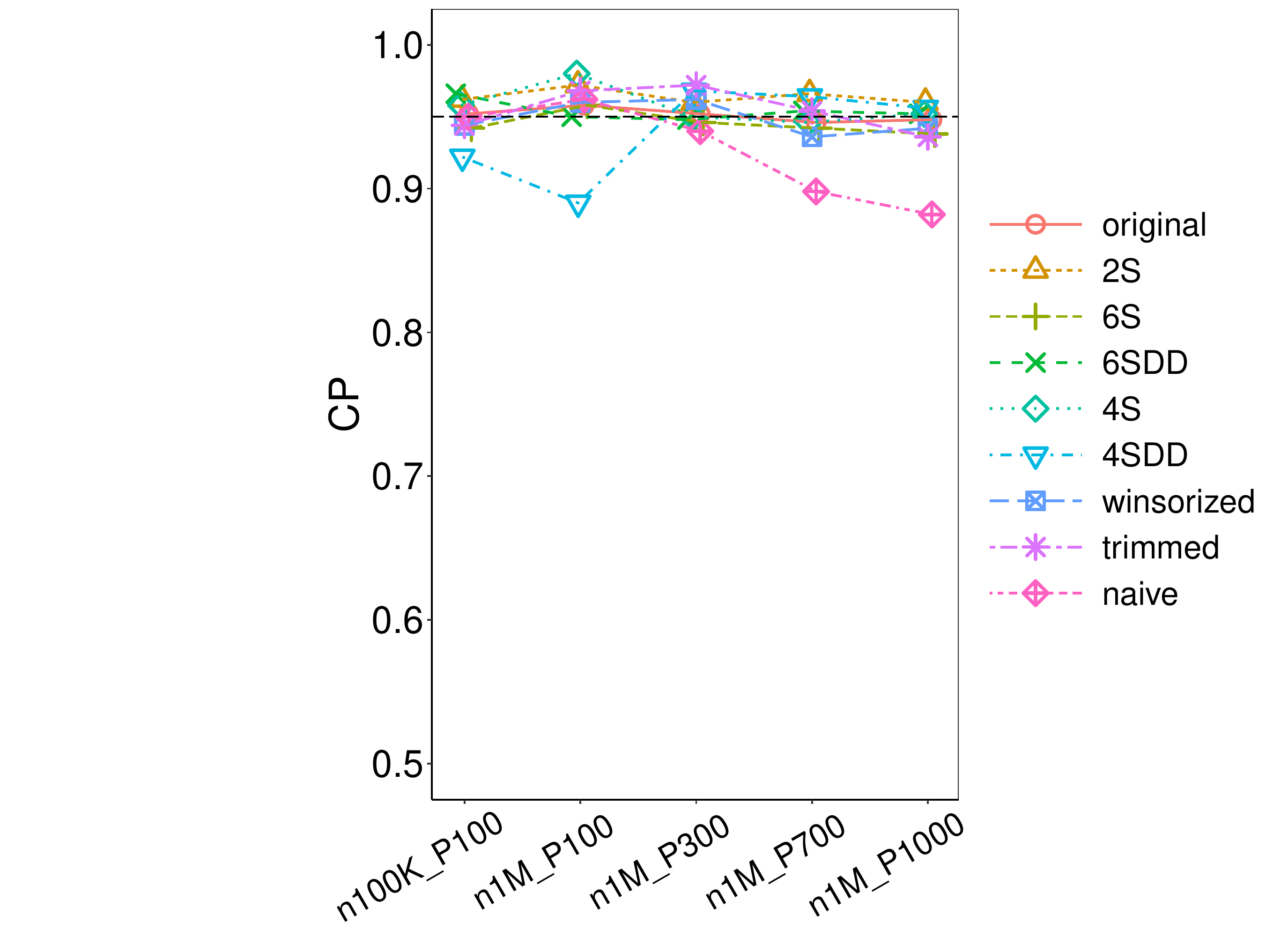}
\includegraphics[width=0.19\textwidth, trim={2.5in 0 2.6in 0},clip] {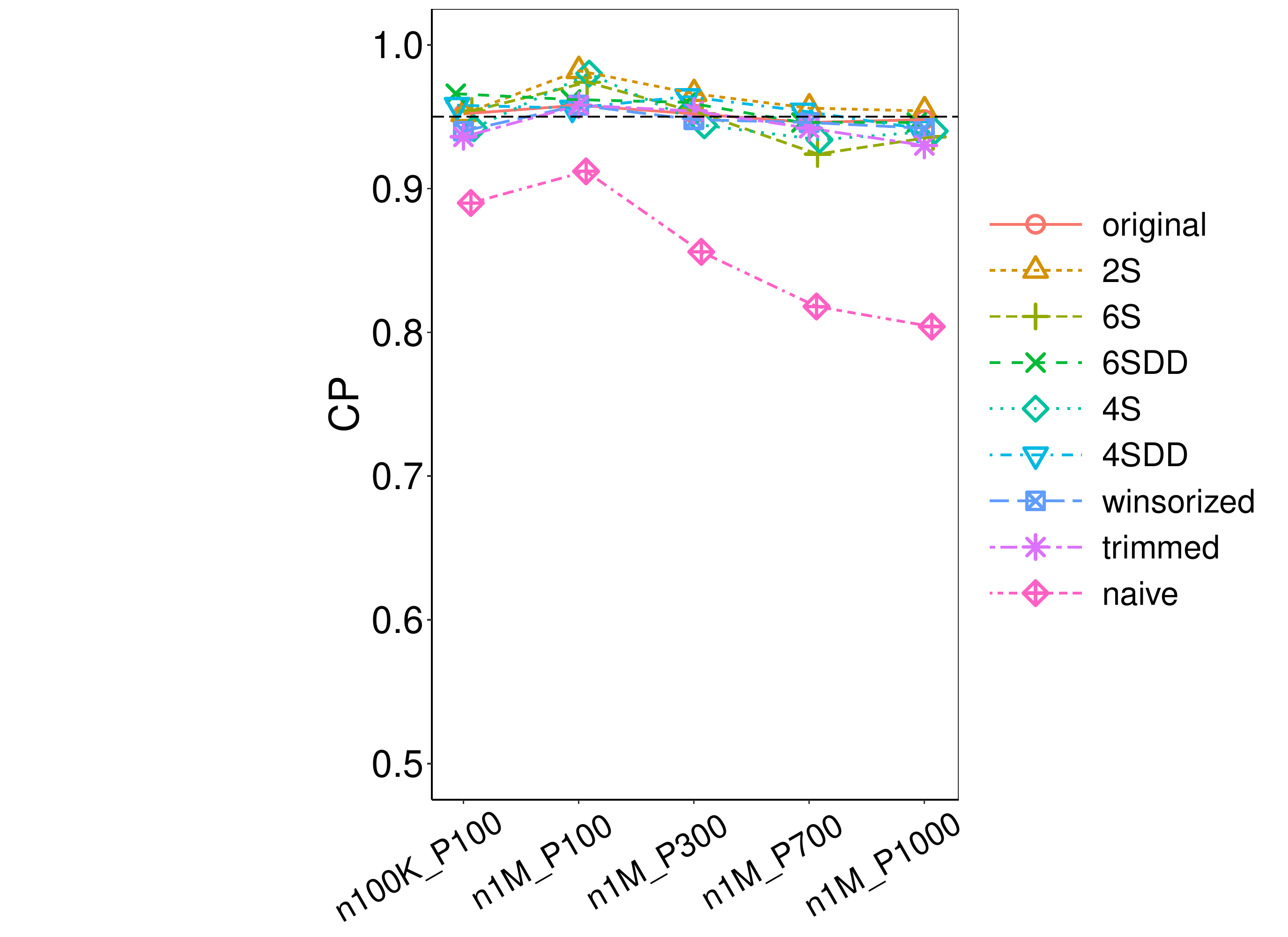}
\includegraphics[width=0.19\textwidth, trim={2.5in 0 2.6in 0},clip] {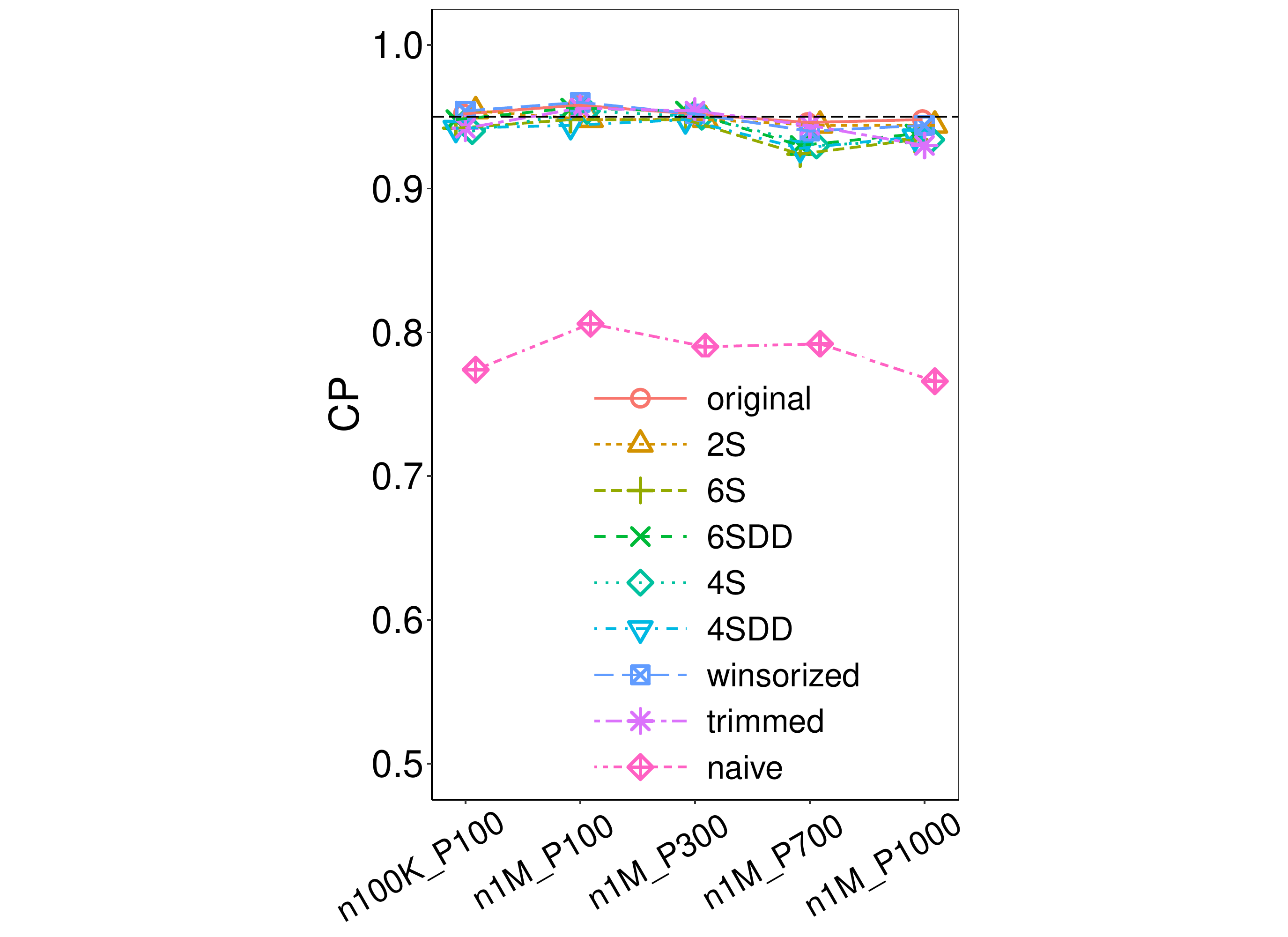}

\includegraphics[width=0.19\textwidth, trim={2.5in 0 2.6in 0},clip] {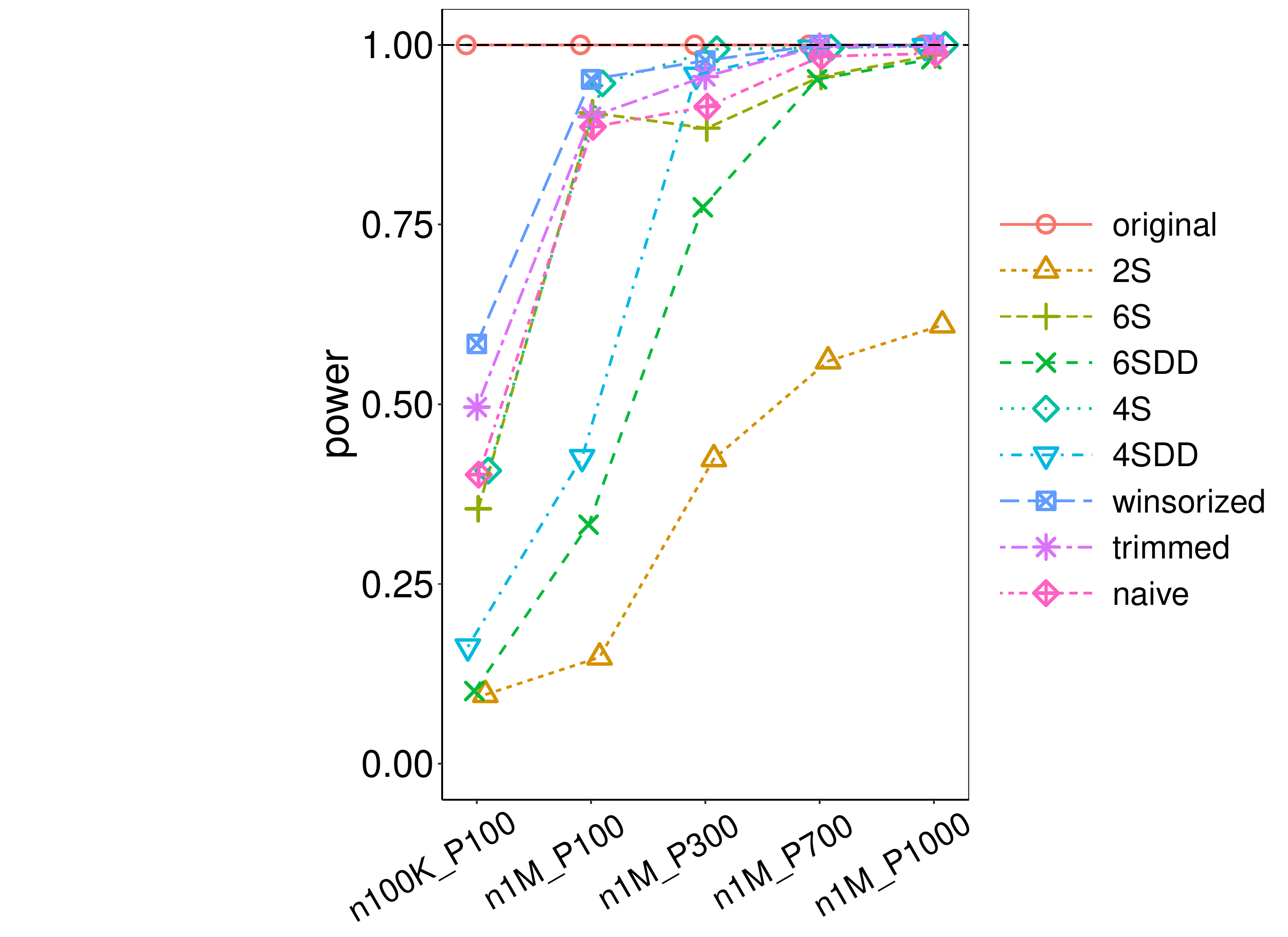}
\includegraphics[width=0.19\textwidth, trim={2.5in 0 2.6in 0},clip] {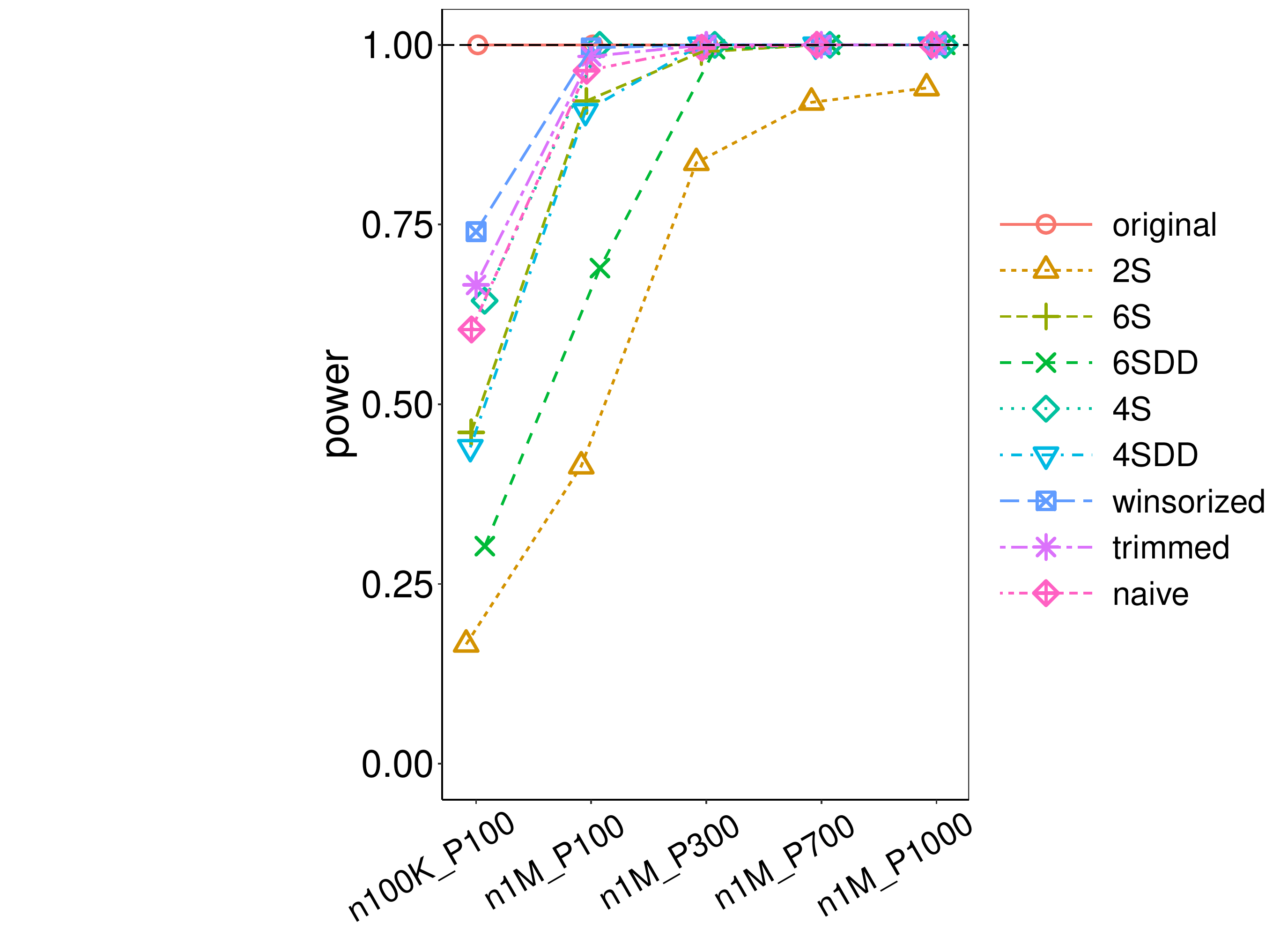}
\includegraphics[width=0.19\textwidth, trim={2.5in 0 2.6in 0},clip] {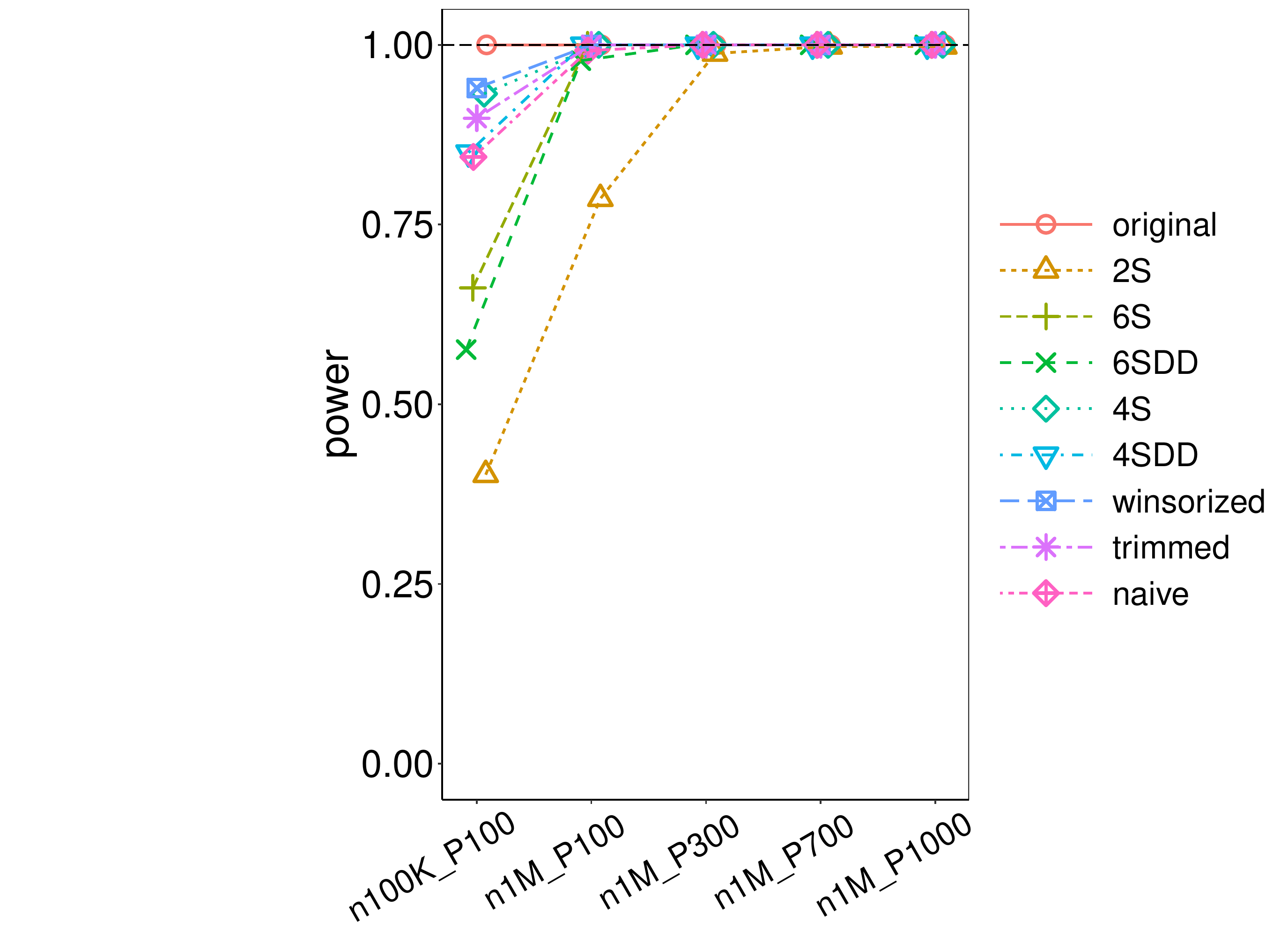}
\includegraphics[width=0.19\textwidth, trim={2.5in 0 2.6in 0},clip] {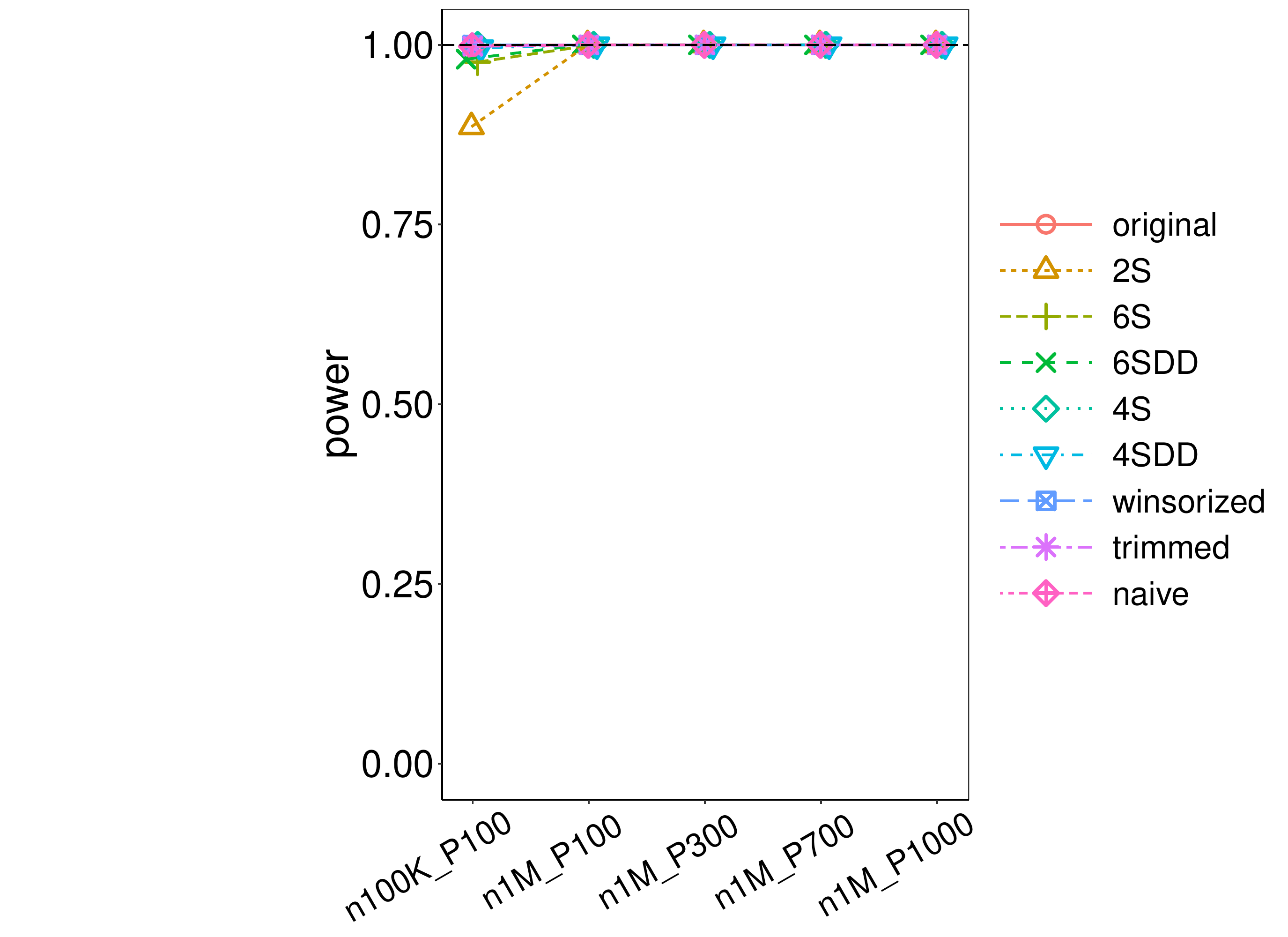}
\includegraphics[width=0.19\textwidth, trim={2.5in 0 2.6in 0},clip] {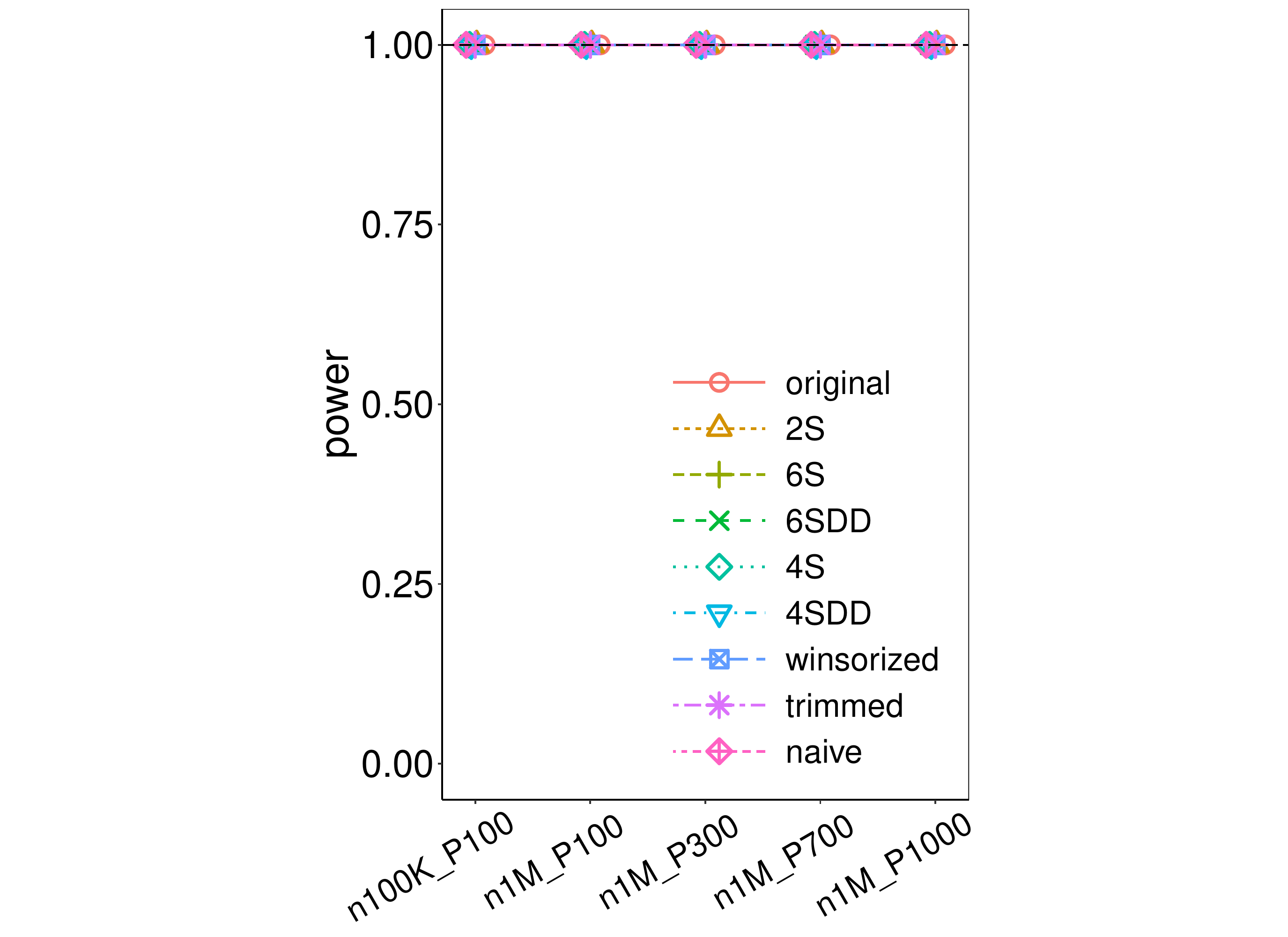}
\caption{Simulation results with $\epsilon$-DP for ZILN data with  $\alpha=\beta$ when $\theta\ne0$} \label{fig:1sDPZILN}
\end{figure}

\begin{figure}[!htb]
\hspace{0.45in}$\rho=0.005$\hspace{0.65in}$\rho=0.02$\hspace{0.65in}$\rho=0.08$
\hspace{0.65in}$\rho=0.32$\hspace{0.65in}$\rho=1.28$

\includegraphics[width=0.19\textwidth, trim={2.5in 0 2.6in 0},clip] {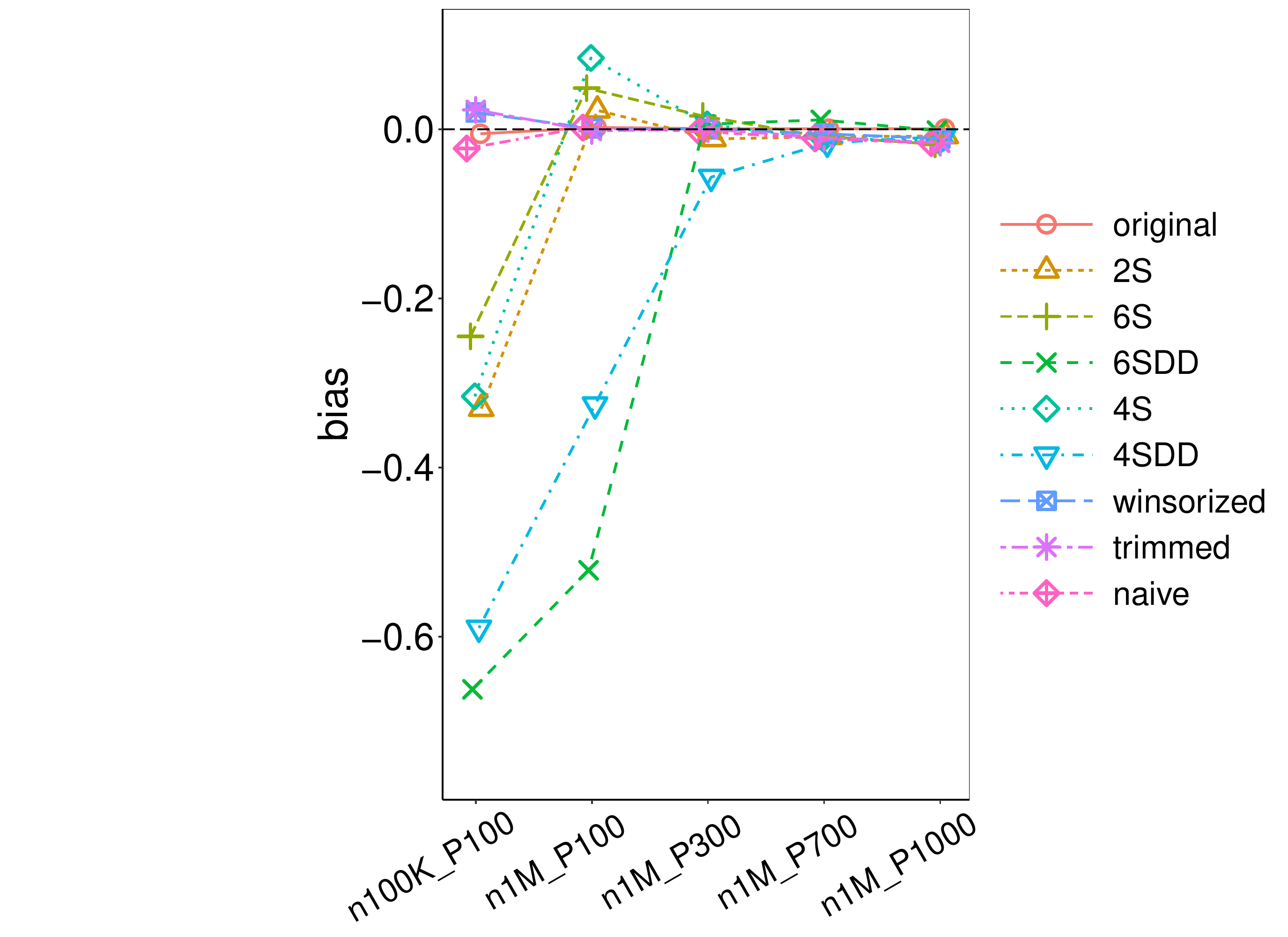}
\includegraphics[width=0.19\textwidth, trim={2.5in 0 2.6in 0},clip] {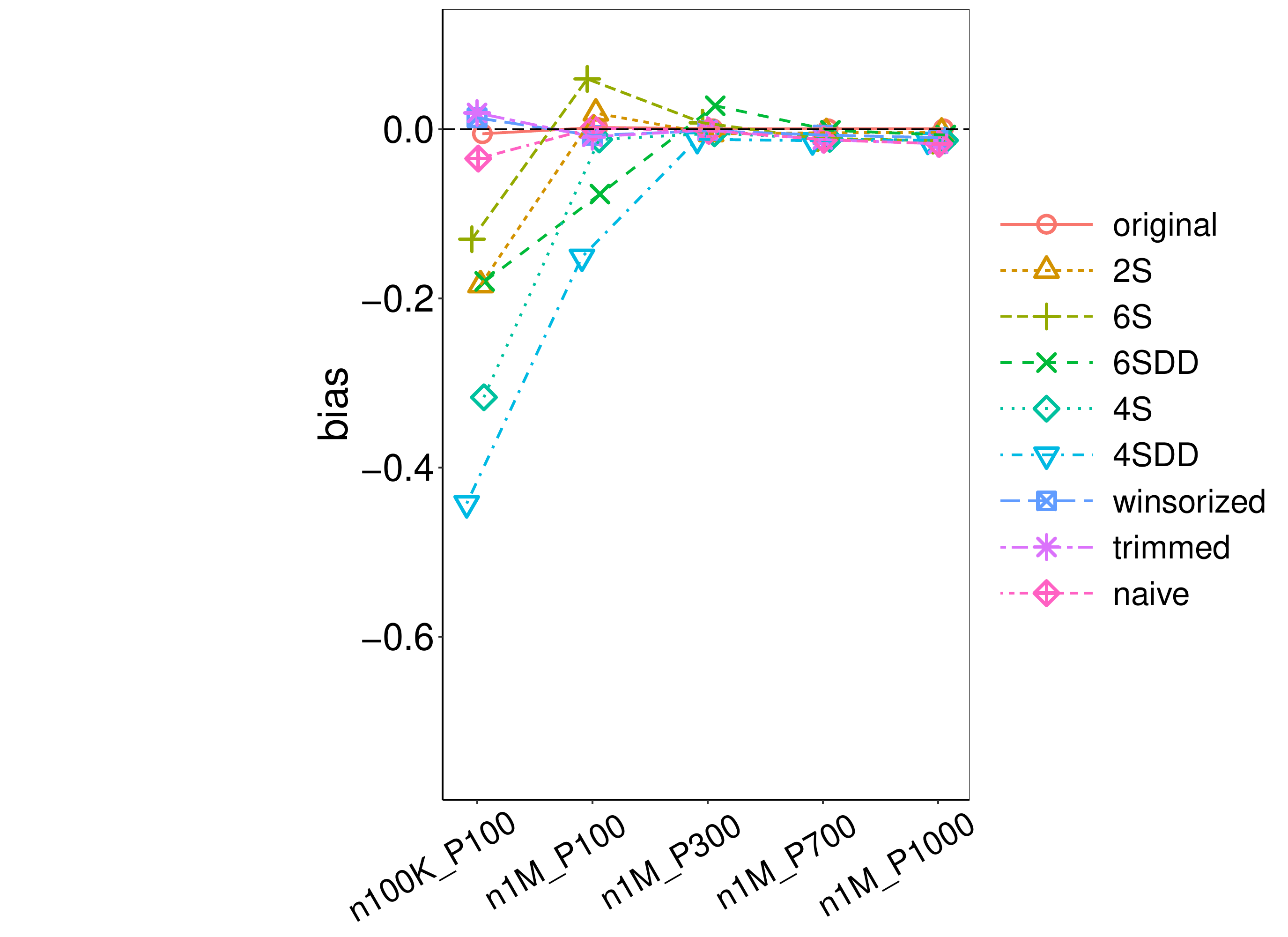}
\includegraphics[width=0.19\textwidth, trim={2.5in 0 2.6in 0},clip] {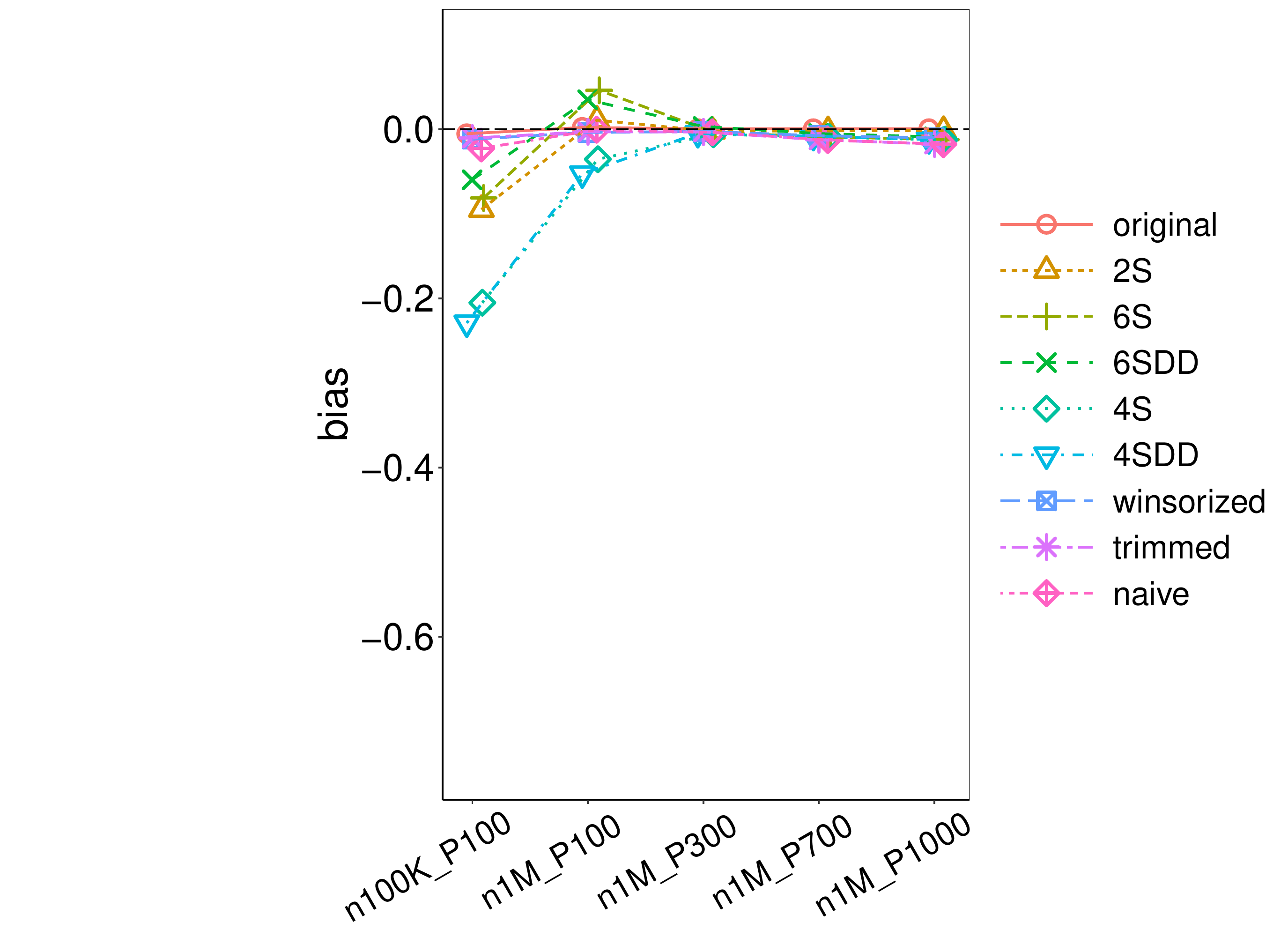}
\includegraphics[width=0.19\textwidth, trim={2.5in 0 2.6in 0},clip] {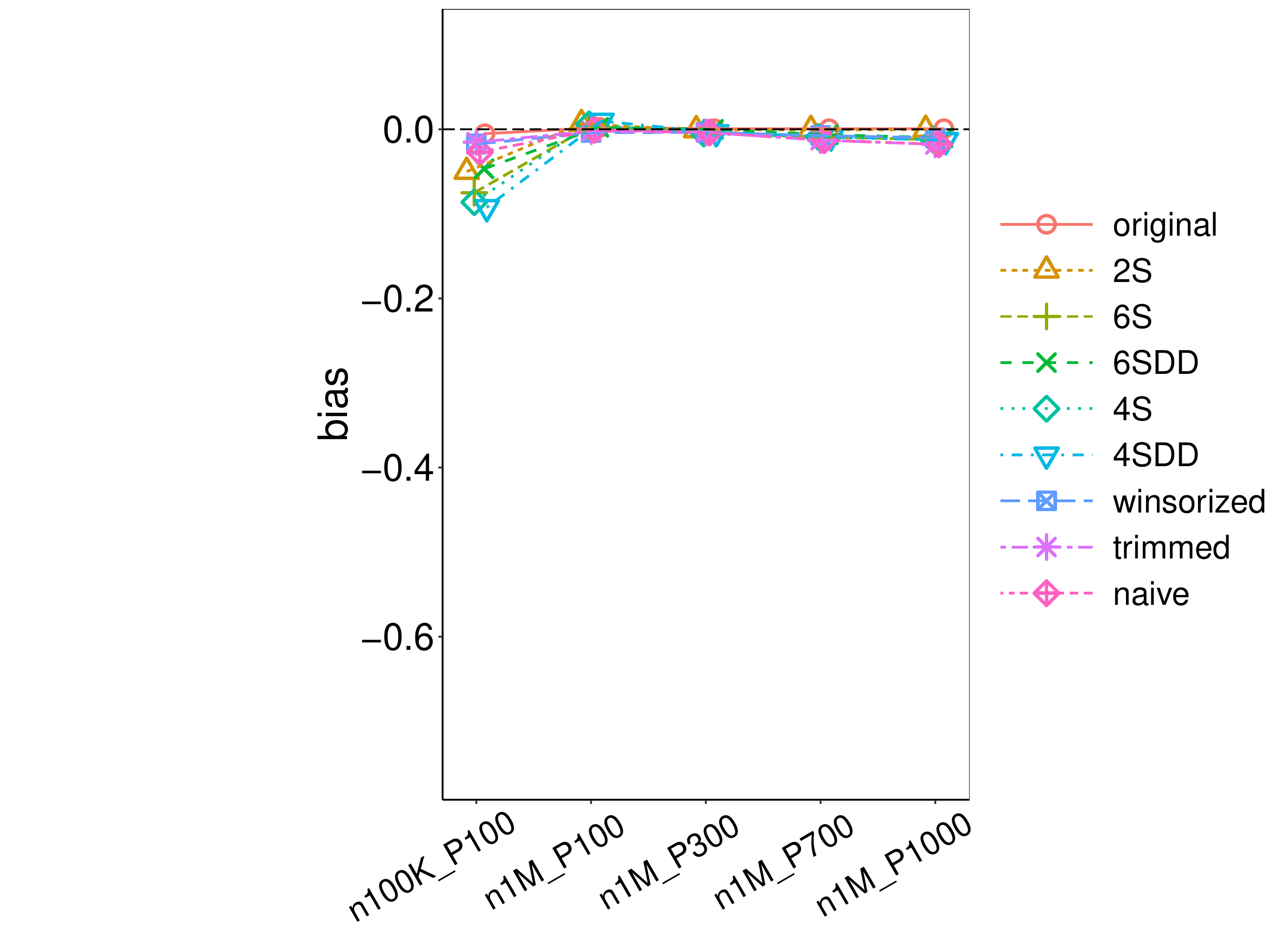}
\includegraphics[width=0.19\textwidth, trim={2.5in 0 2.6in 0},clip] {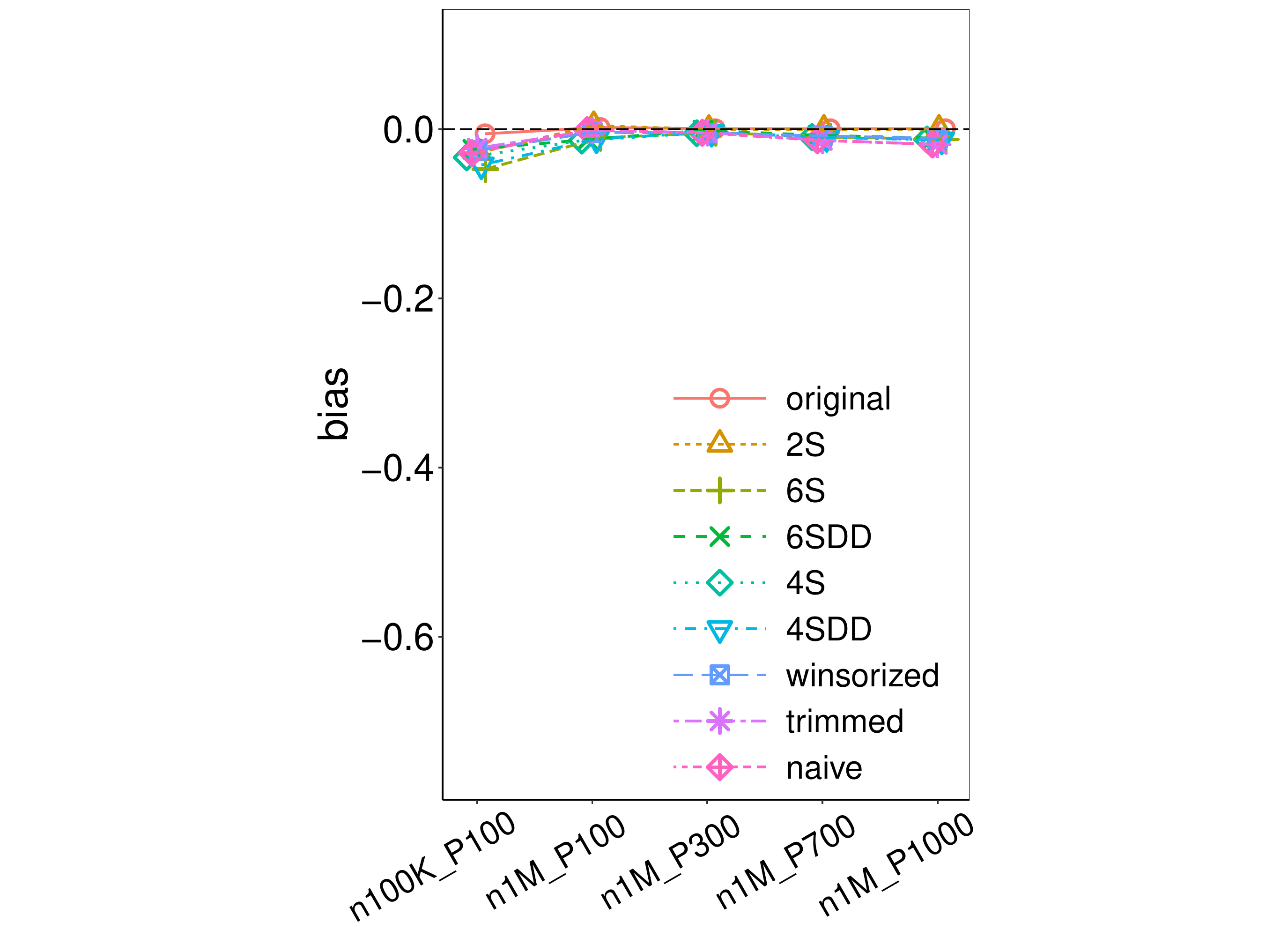}

\includegraphics[width=0.19\textwidth, trim={2.5in 0 2.6in 0},clip] {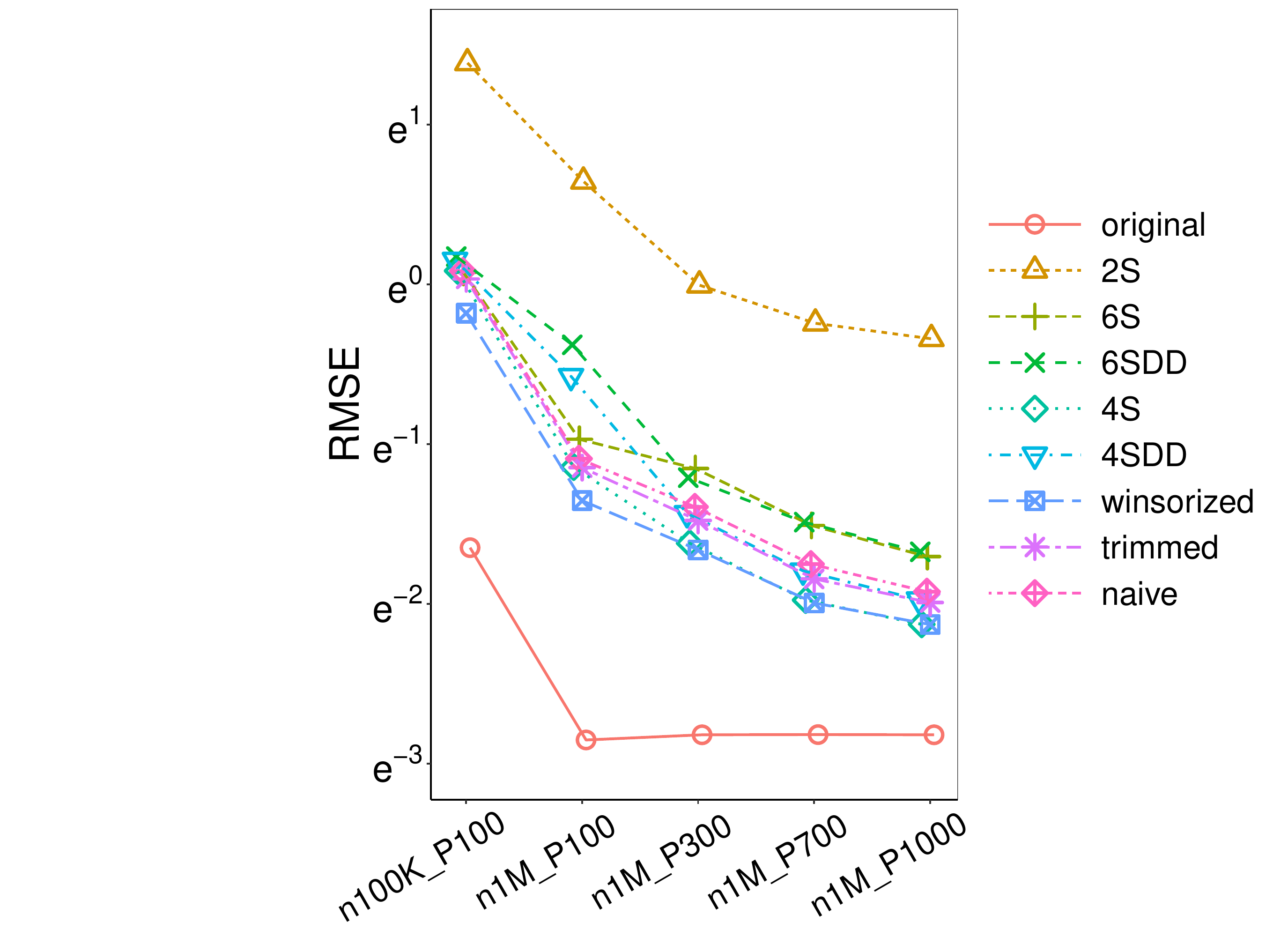}
\includegraphics[width=0.19\textwidth, trim={2.5in 0 2.6in 0},clip] {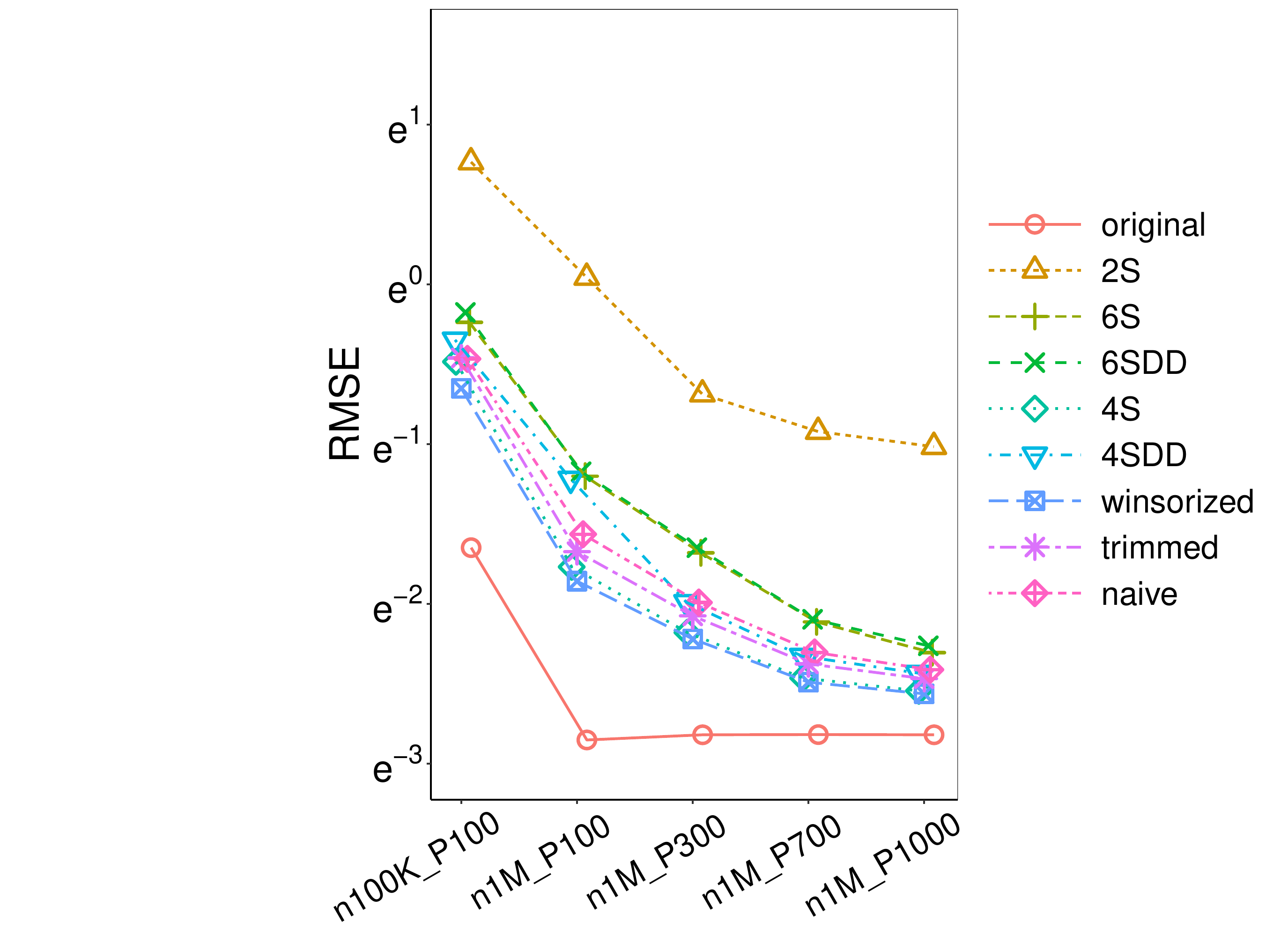}
\includegraphics[width=0.19\textwidth, trim={2.5in 0 2.6in 0},clip] {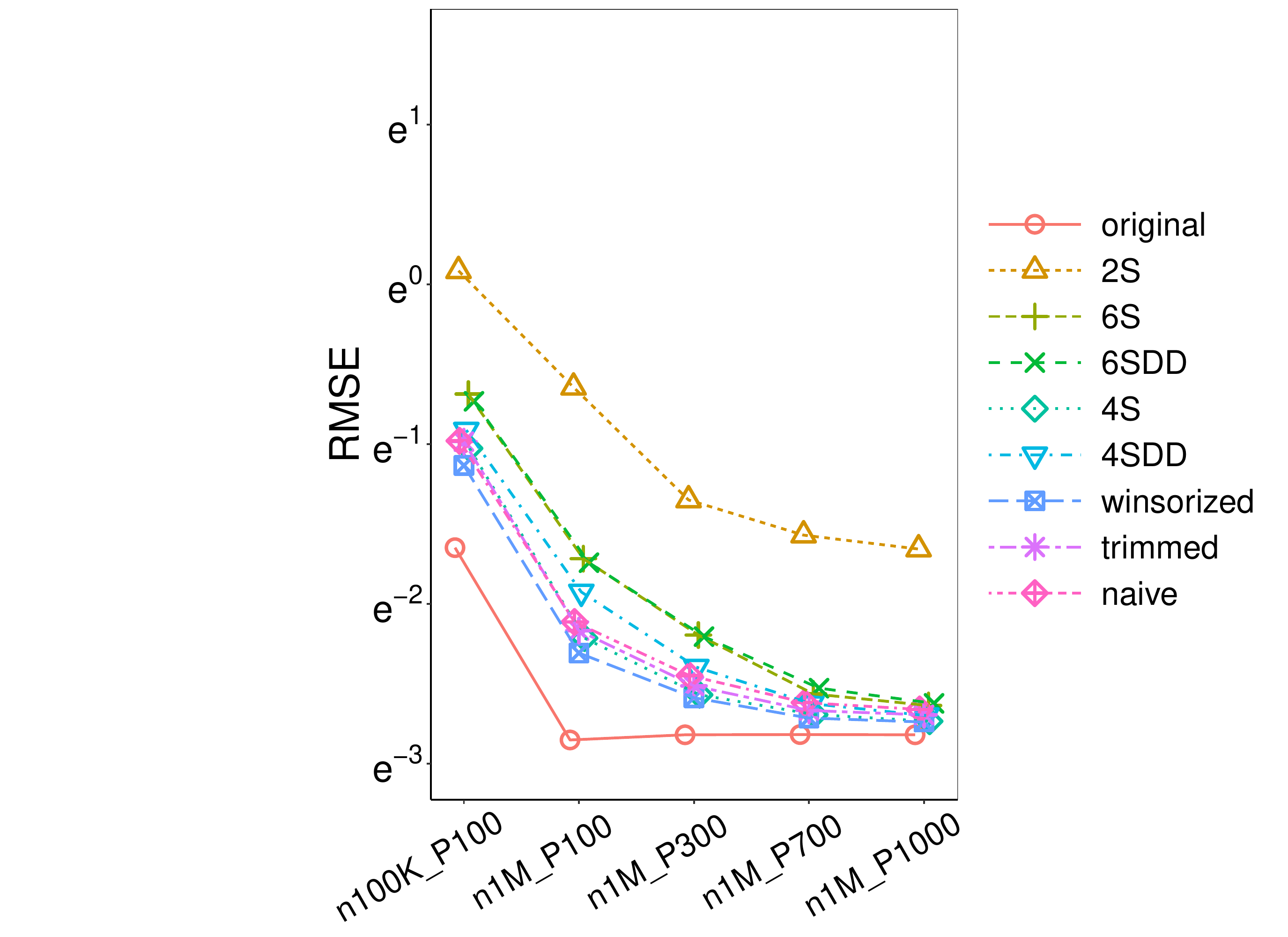}
\includegraphics[width=0.19\textwidth, trim={2.5in 0 2.6in 0},clip] {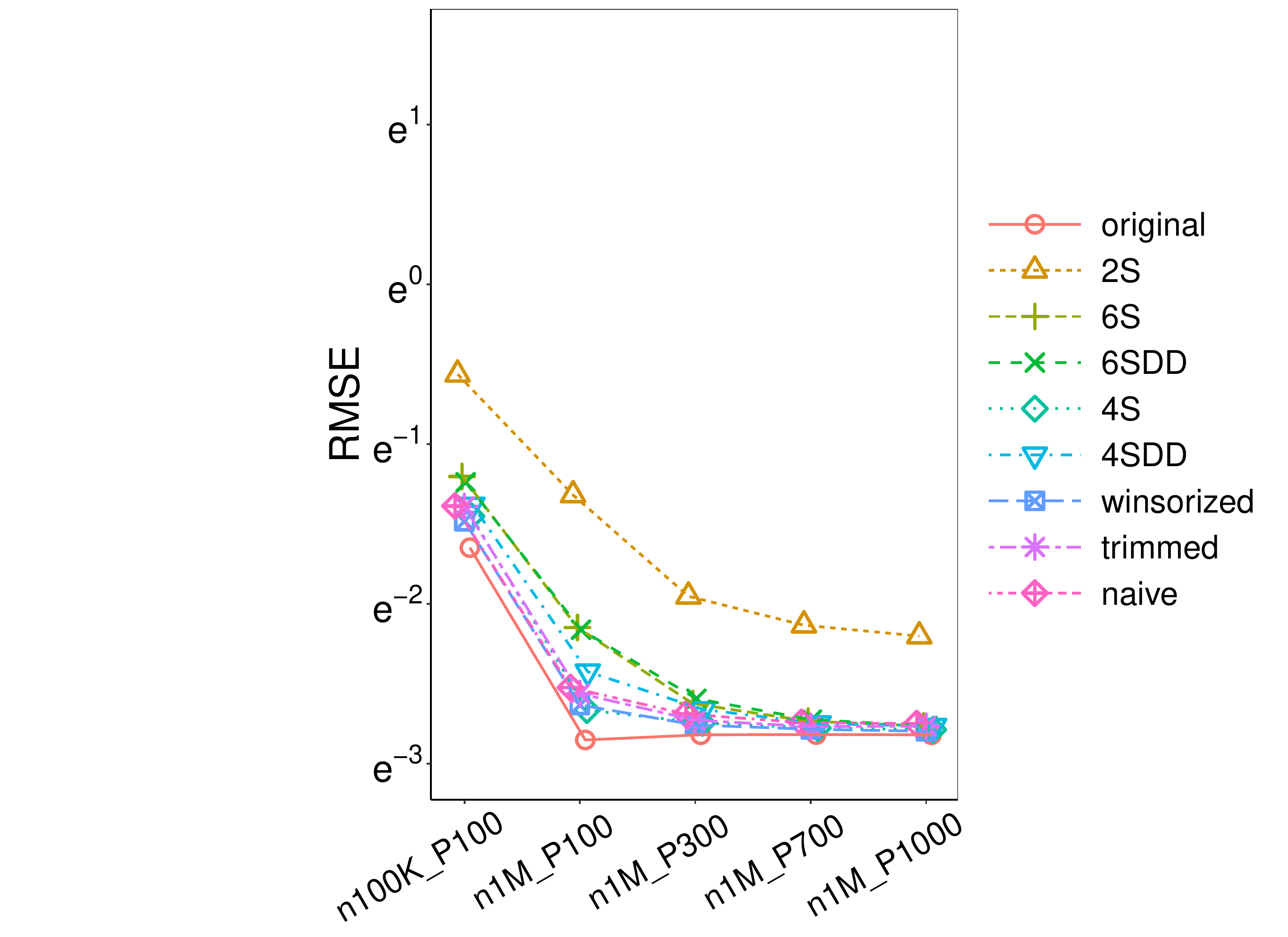}
\includegraphics[width=0.19\textwidth, trim={2.5in 0 2.6in 0},clip] {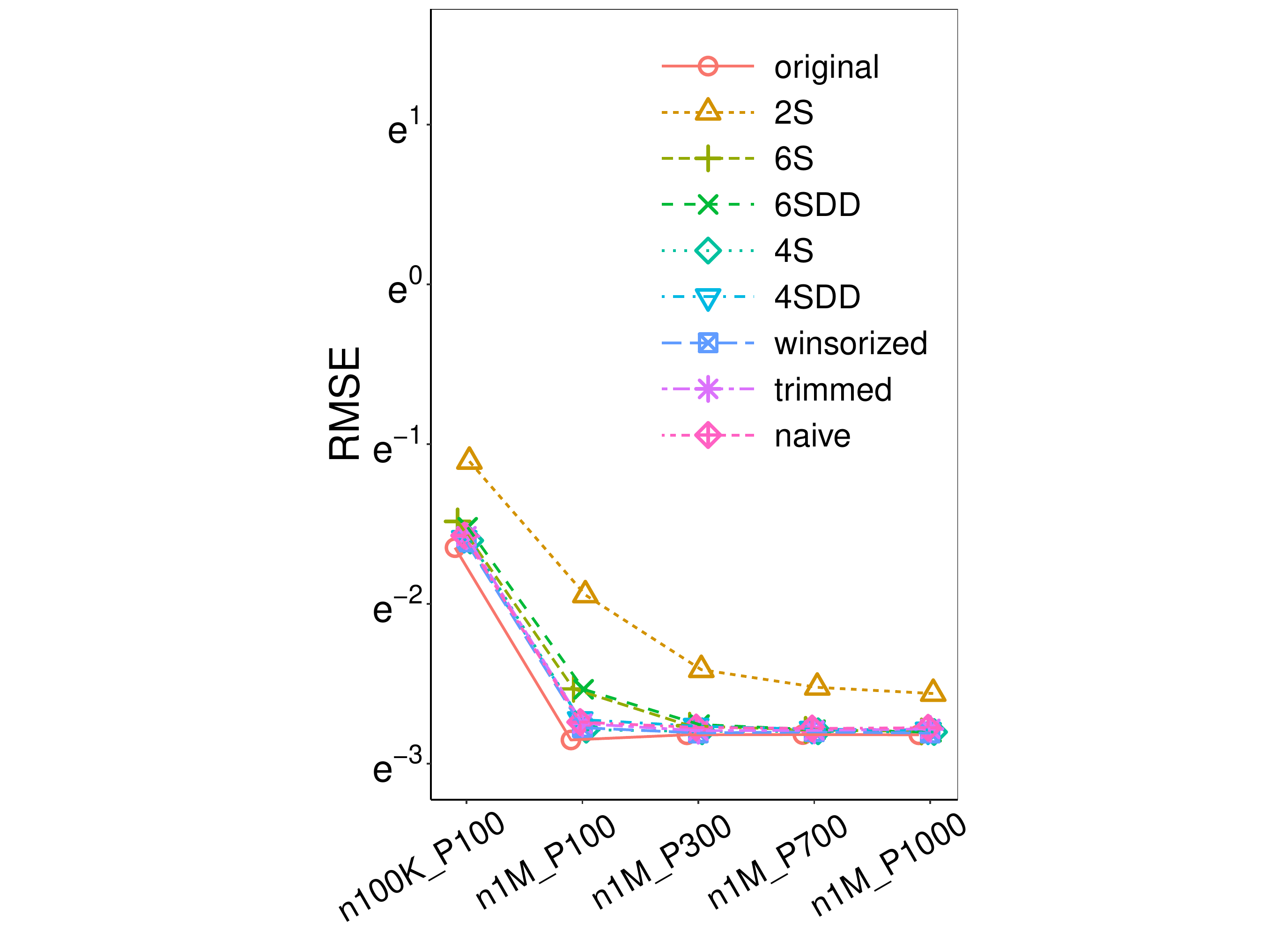}

\includegraphics[width=0.19\textwidth, trim={2.5in 0 2.6in 0},clip] {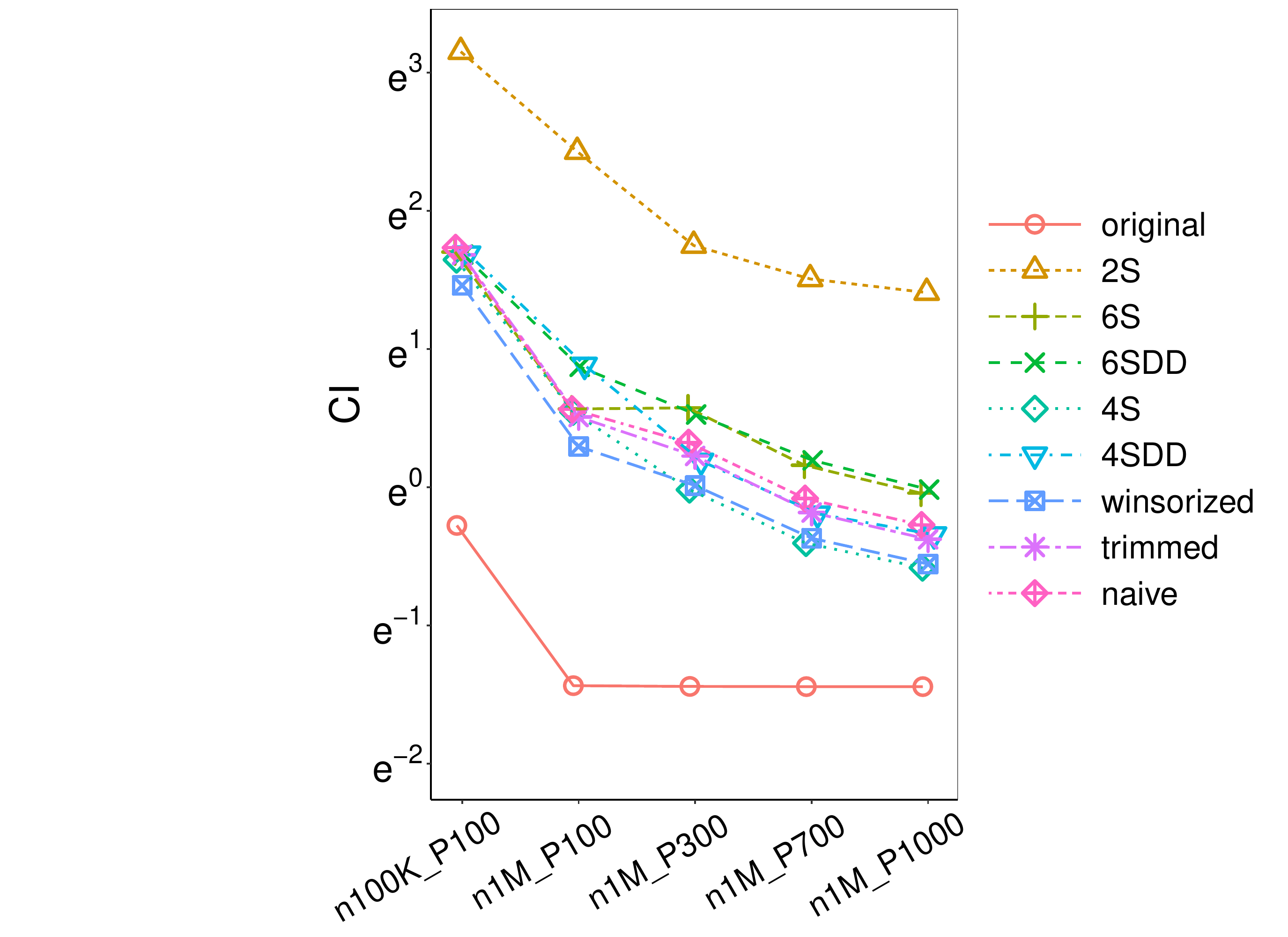}
\includegraphics[width=0.19\textwidth, trim={2.5in 0 2.6in 0},clip] {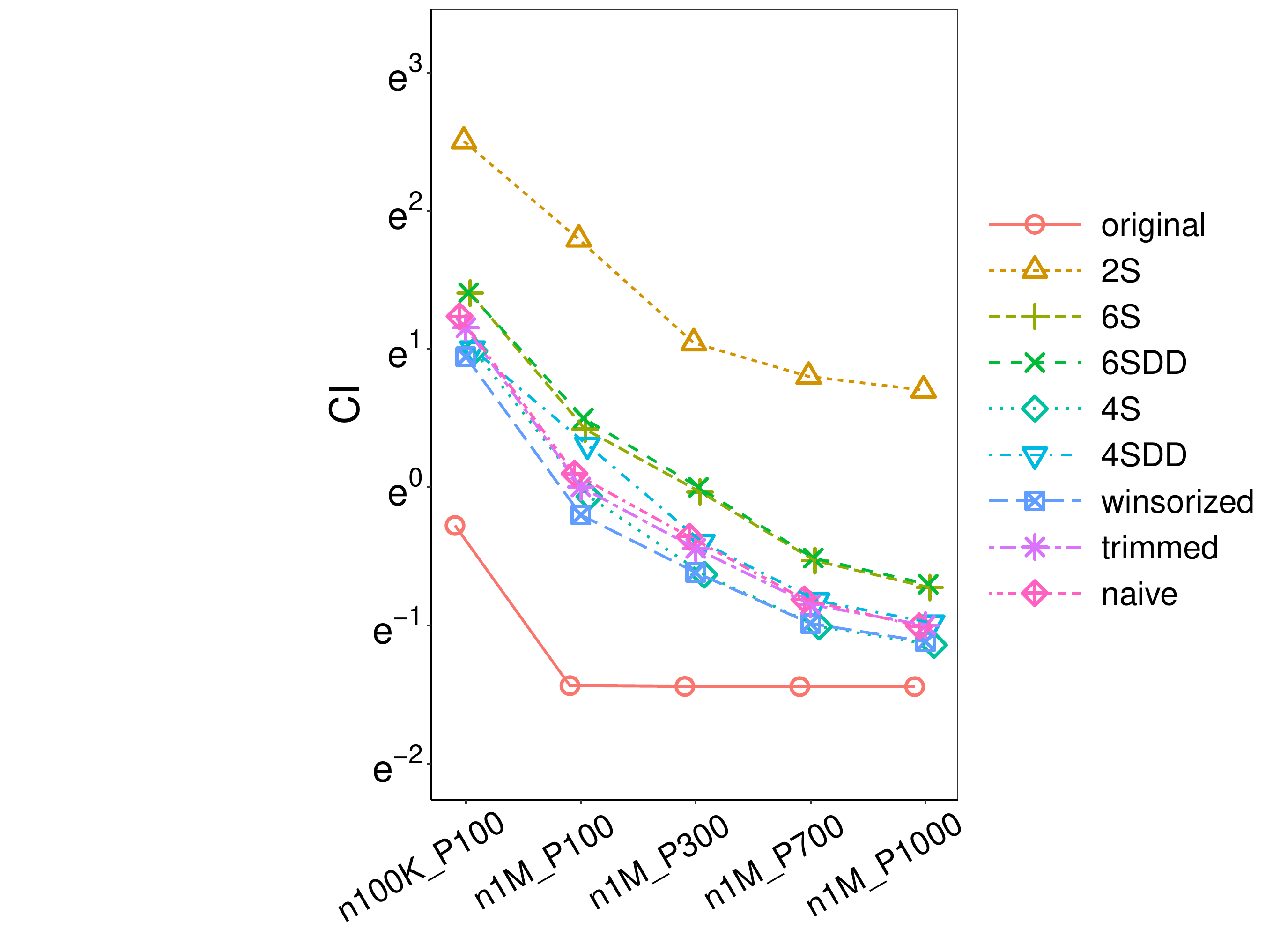}
\includegraphics[width=0.19\textwidth, trim={2.5in 0 2.6in 0},clip] {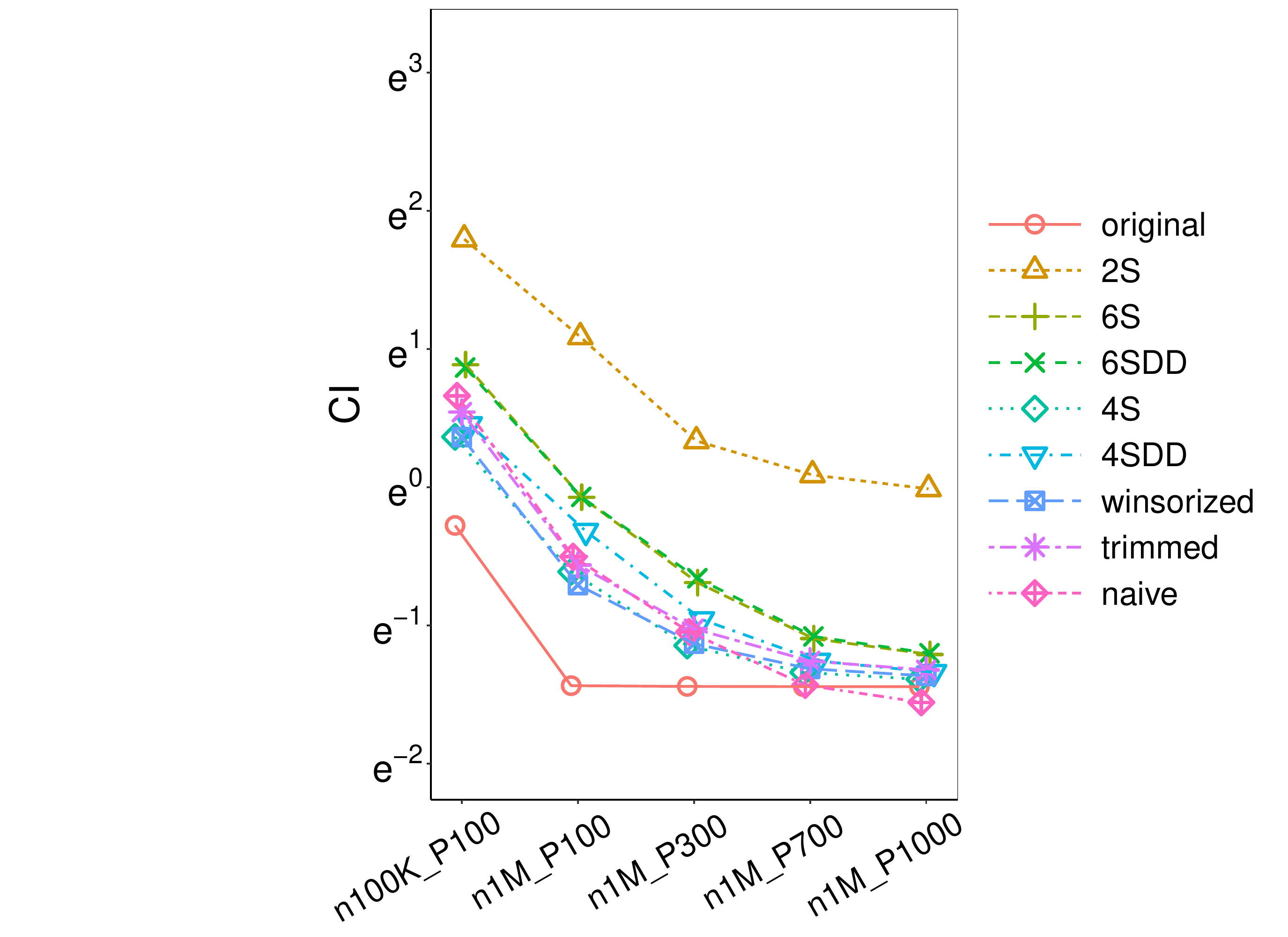}
\includegraphics[width=0.19\textwidth, trim={2.5in 0 2.6in 0},clip] {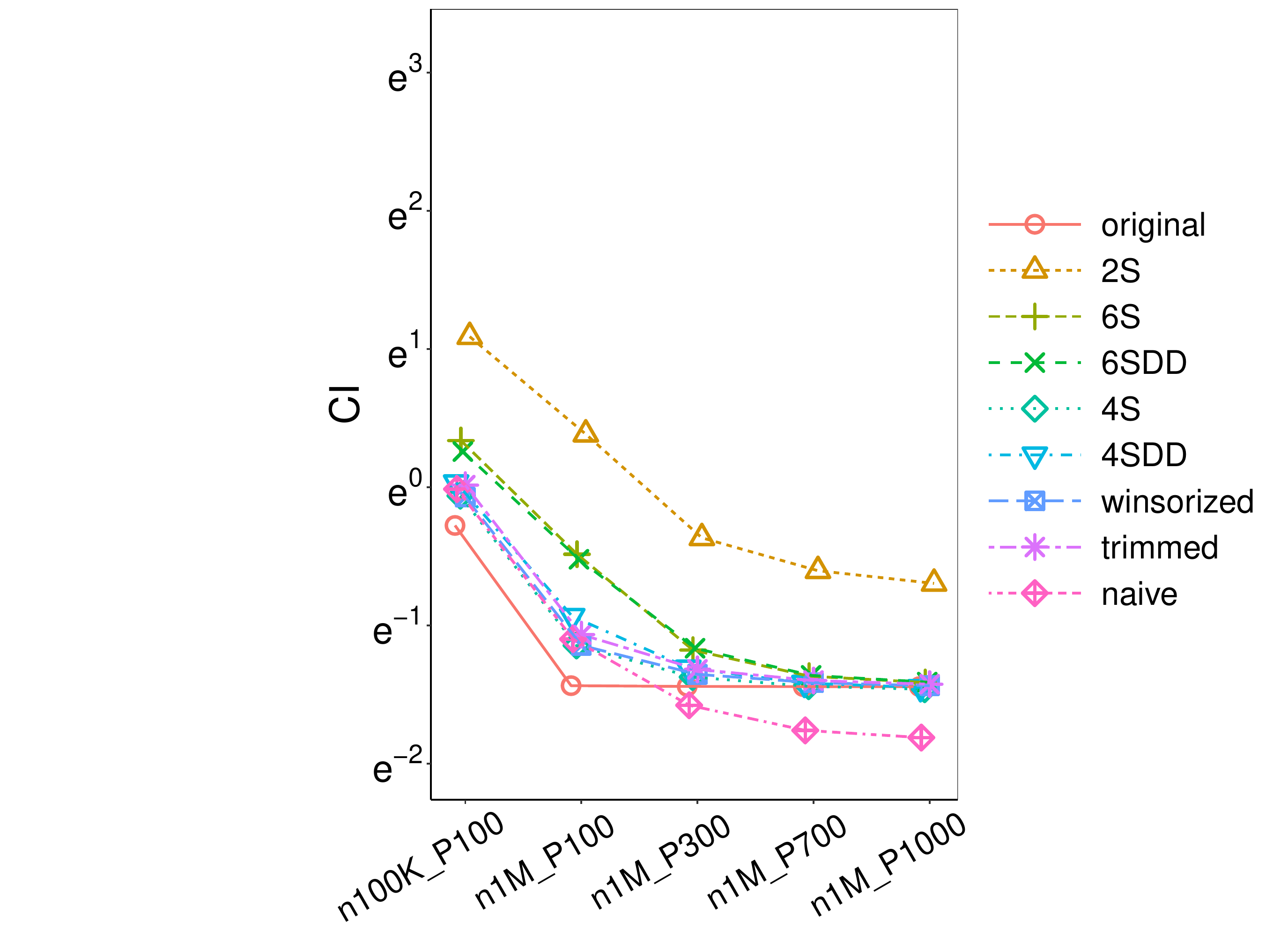}
\includegraphics[width=0.19\textwidth, trim={2.5in 0 2.6in 0},clip] {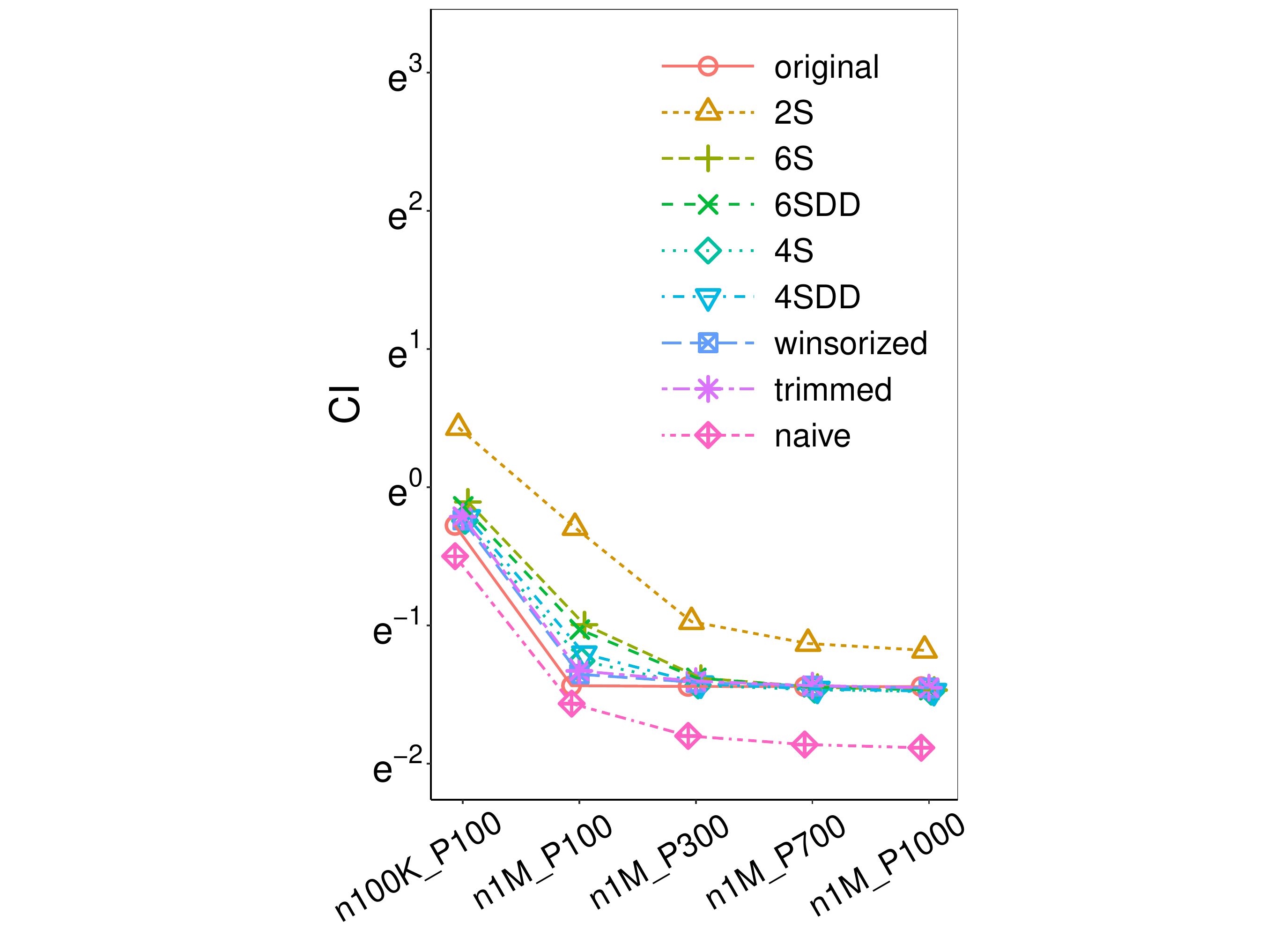}

\includegraphics[width=0.19\textwidth, trim={2.5in 0 2.6in 0},clip] {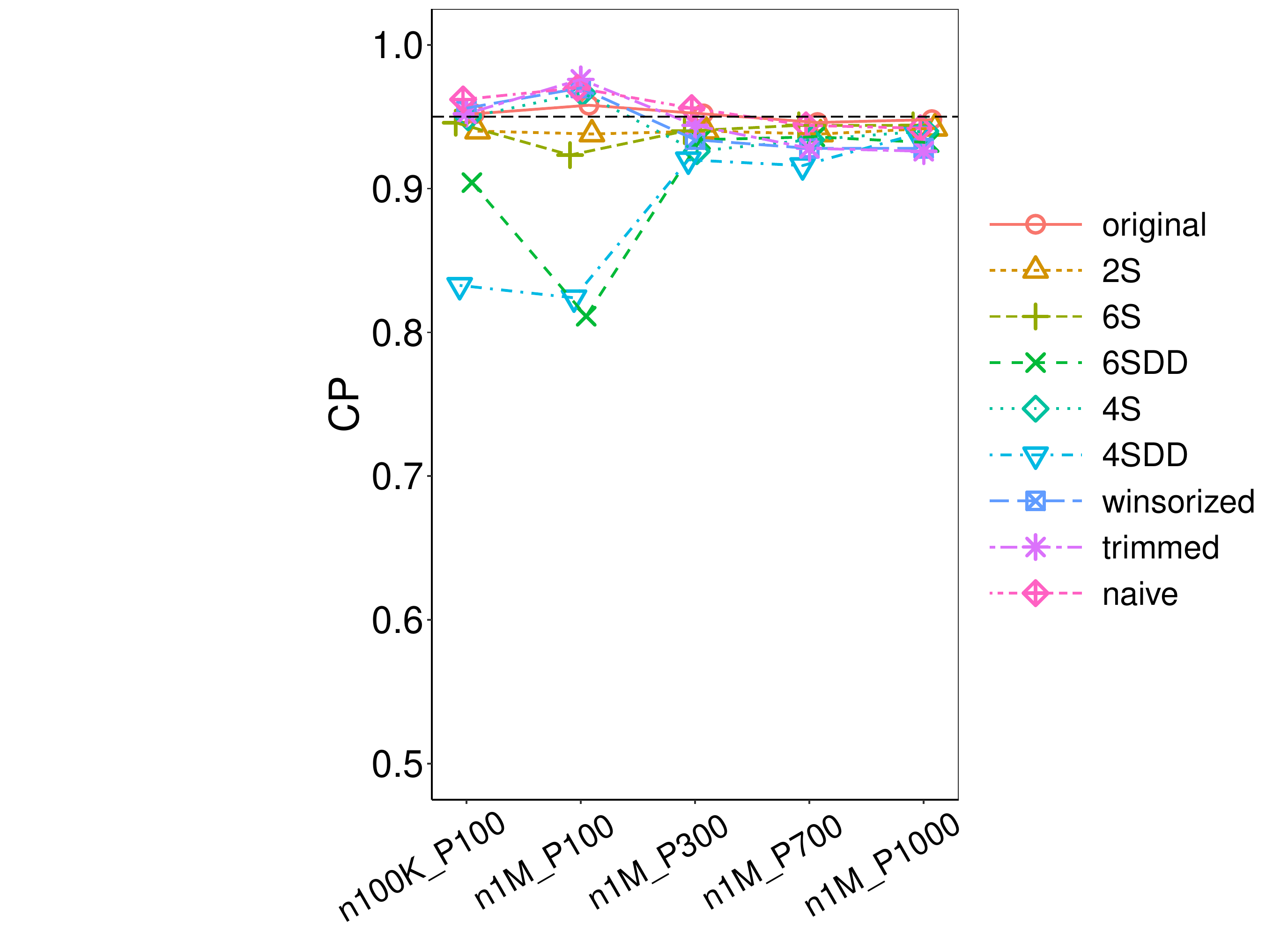}
\includegraphics[width=0.19\textwidth, trim={2.5in 0 2.6in 0},clip] {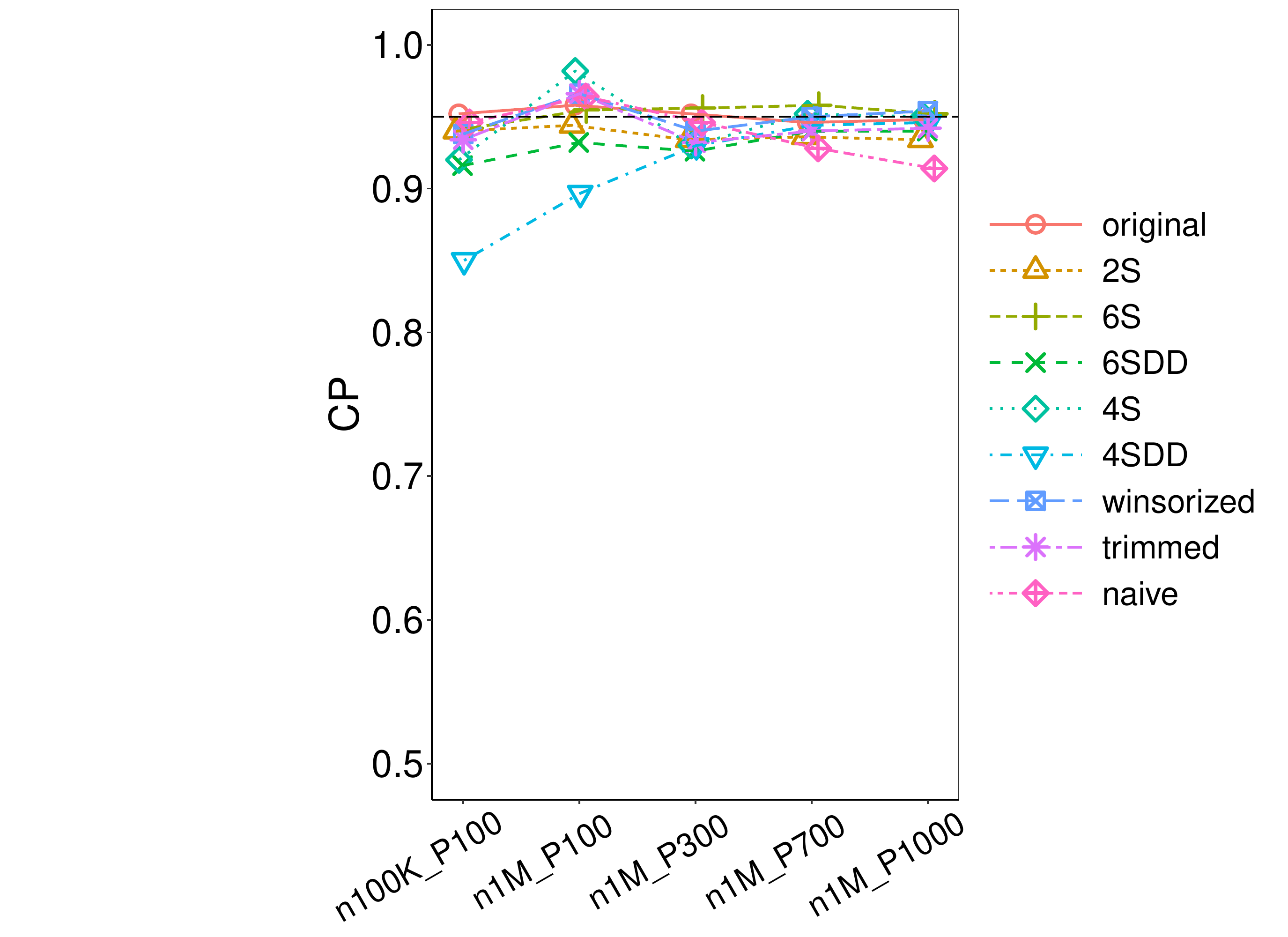}
\includegraphics[width=0.19\textwidth, trim={2.5in 0 2.6in 0},clip] {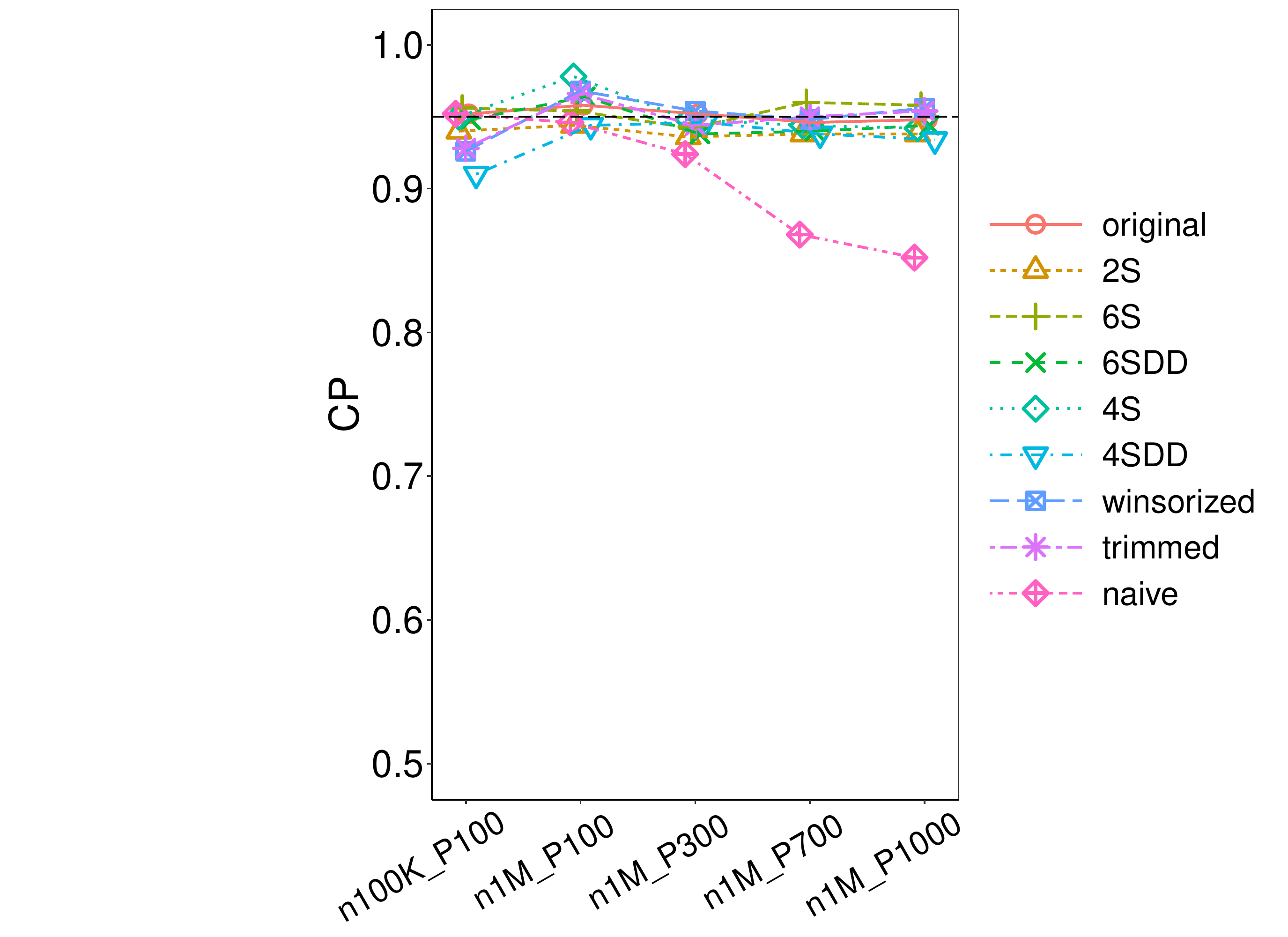}
\includegraphics[width=0.19\textwidth, trim={2.5in 0 2.6in 0},clip] {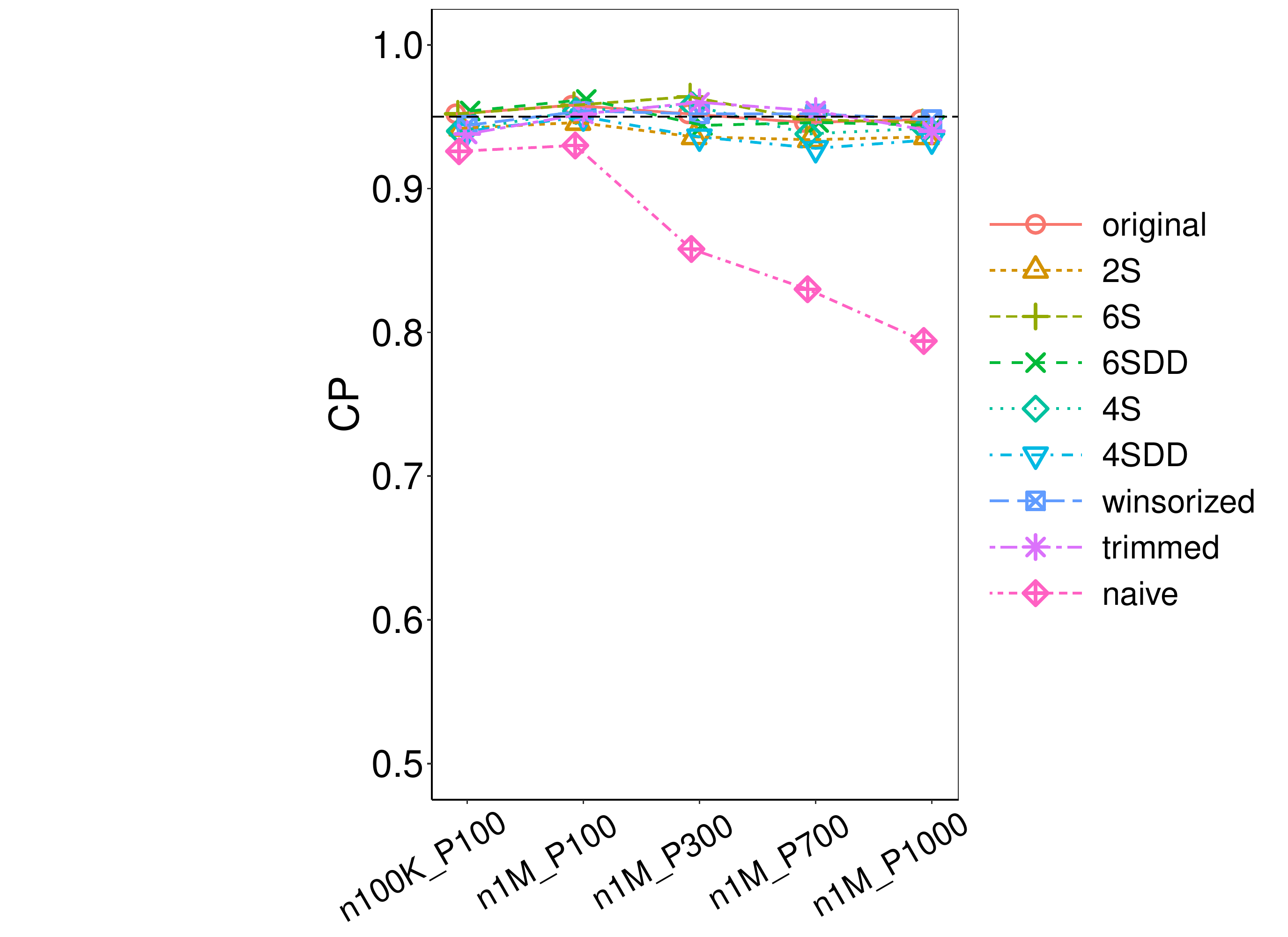}
\includegraphics[width=0.19\textwidth, trim={2.5in 0 2.6in 0},clip] {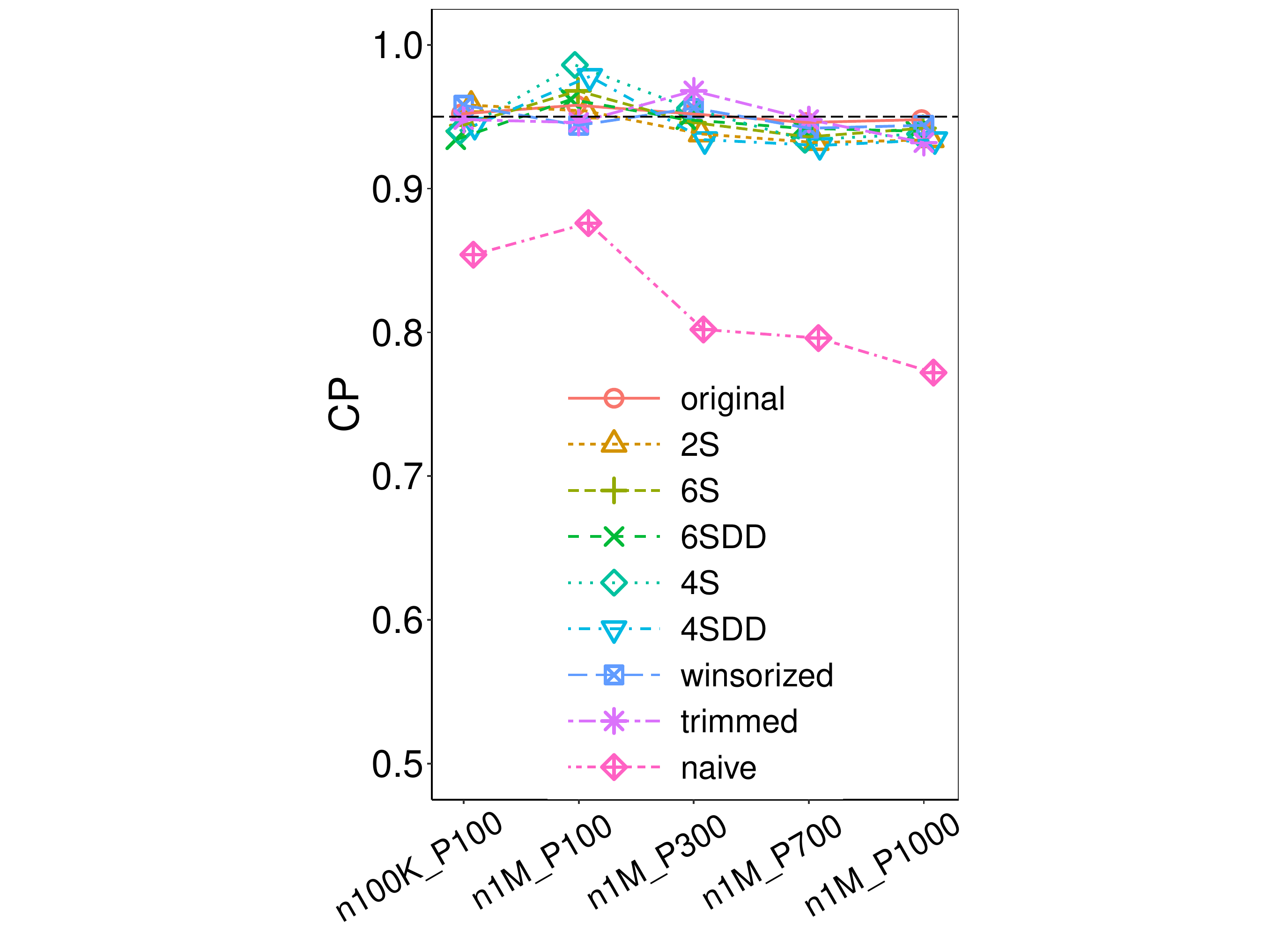}

\includegraphics[width=0.19\textwidth, trim={2.5in 0 2.6in 0},clip] {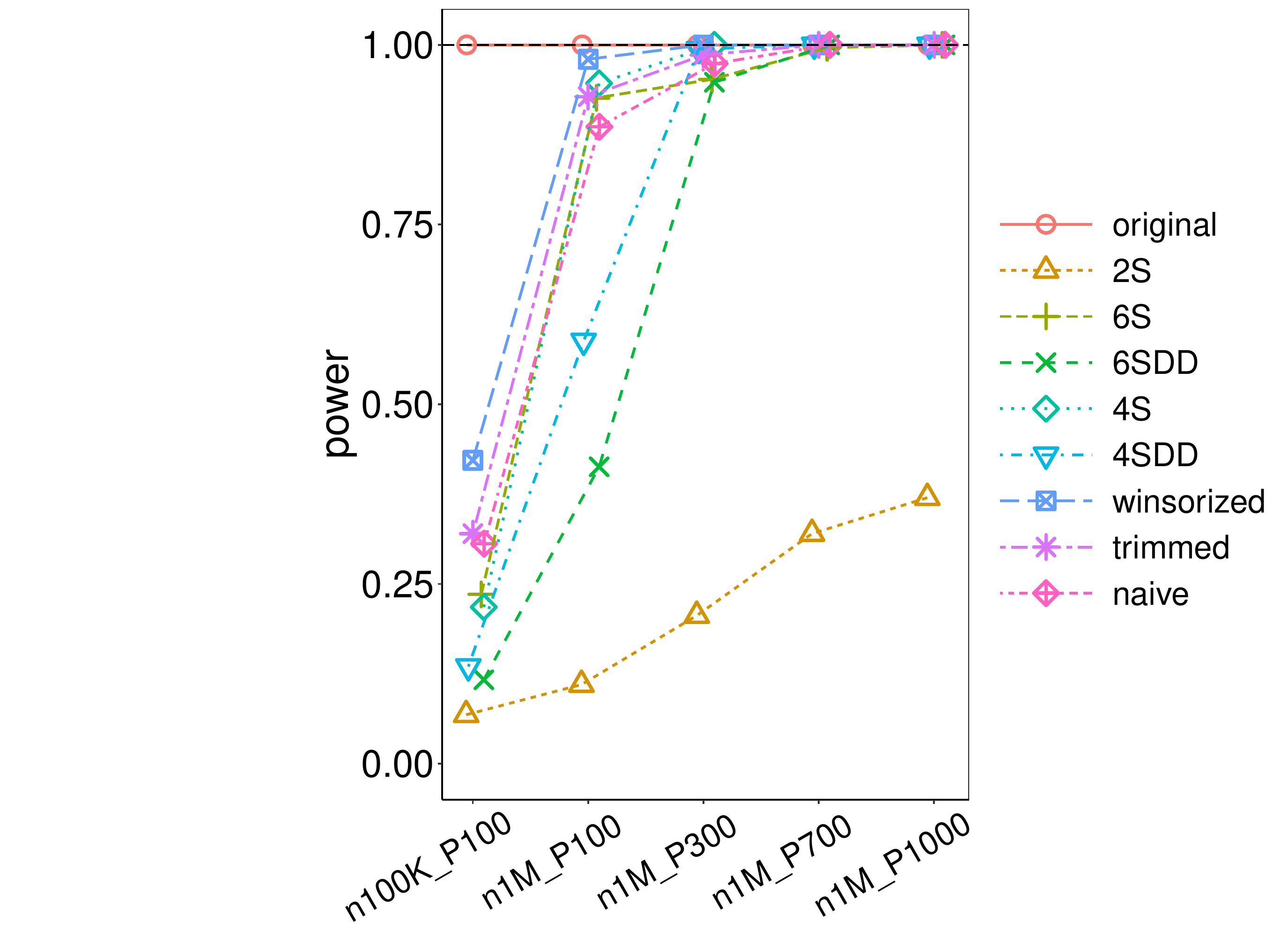}
\includegraphics[width=0.19\textwidth, trim={2.5in 0 2.6in 0},clip] {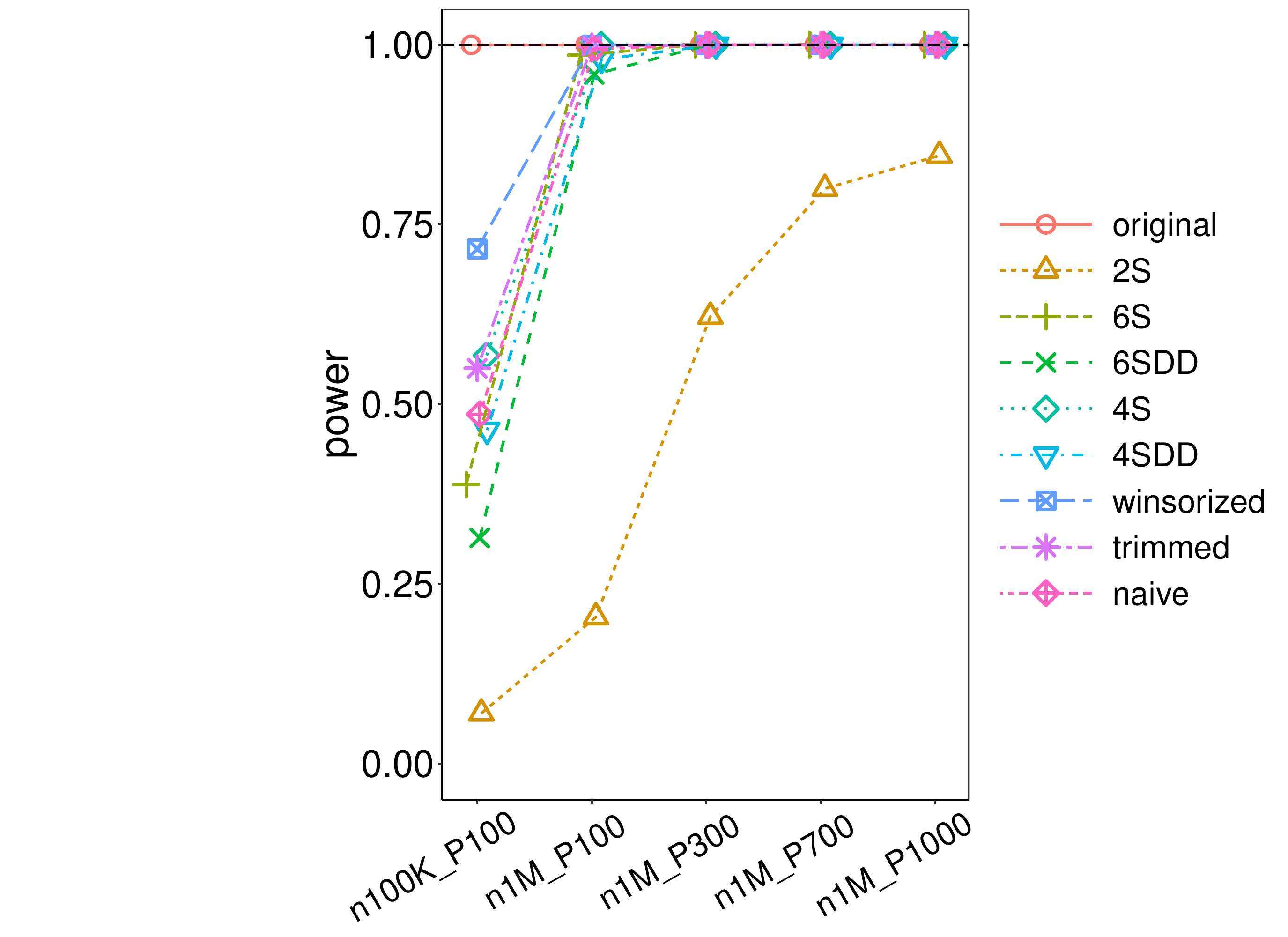}
\includegraphics[width=0.19\textwidth, trim={2.5in 0 2.6in 0},clip] {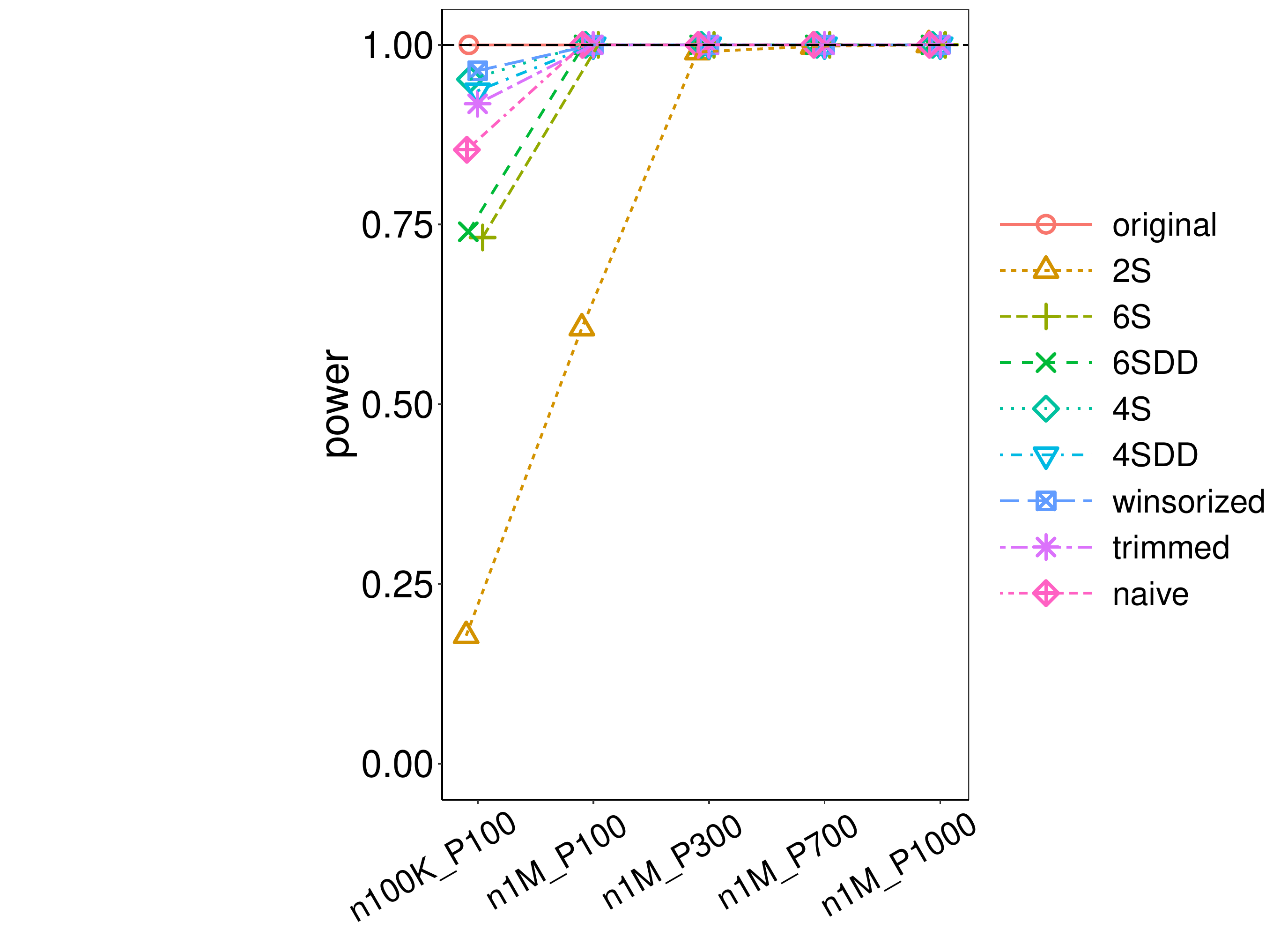}
\includegraphics[width=0.19\textwidth, trim={2.5in 0 2.6in 0},clip] {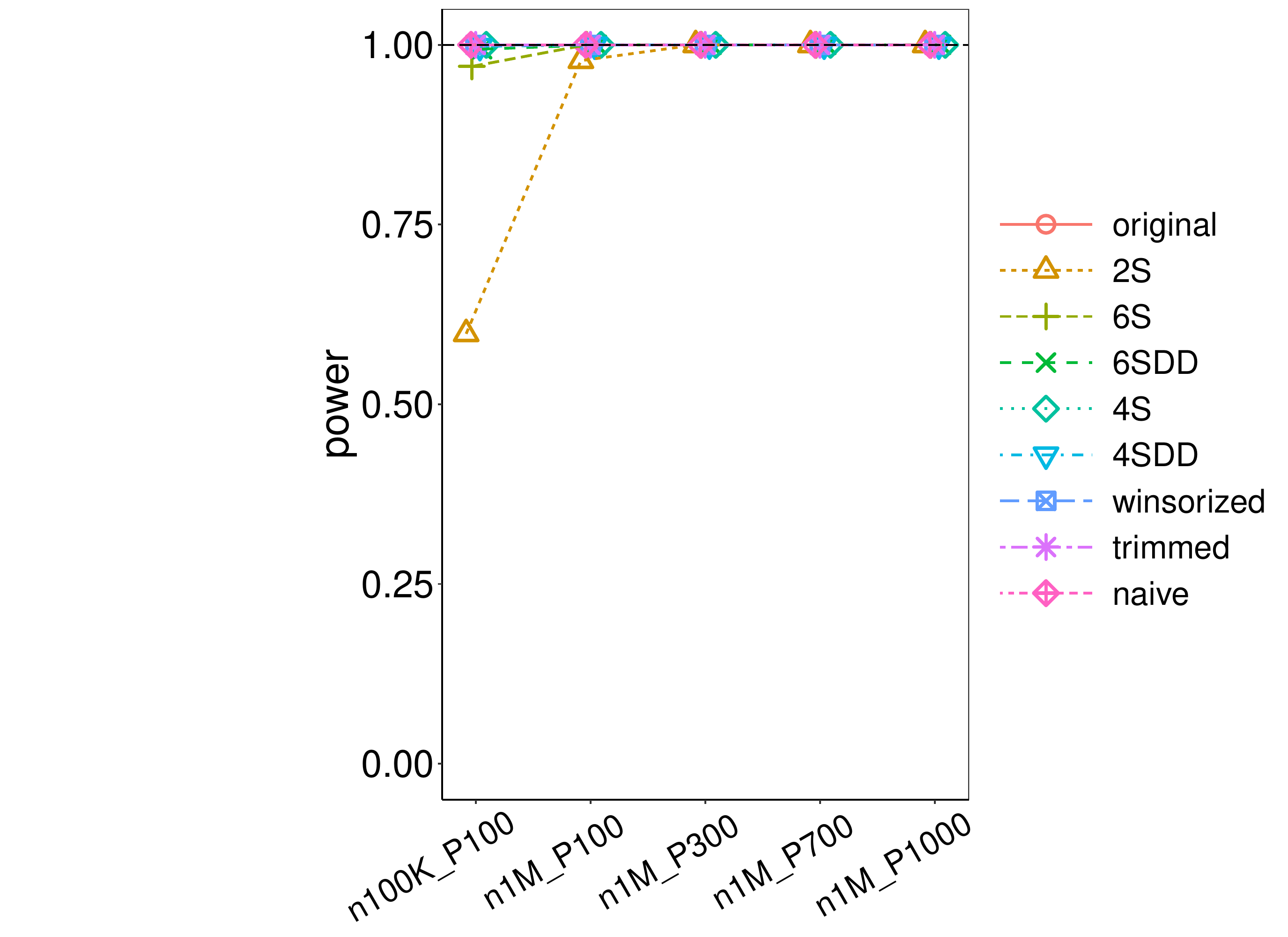}
\includegraphics[width=0.19\textwidth, trim={2.5in 0 2.6in 0},clip] {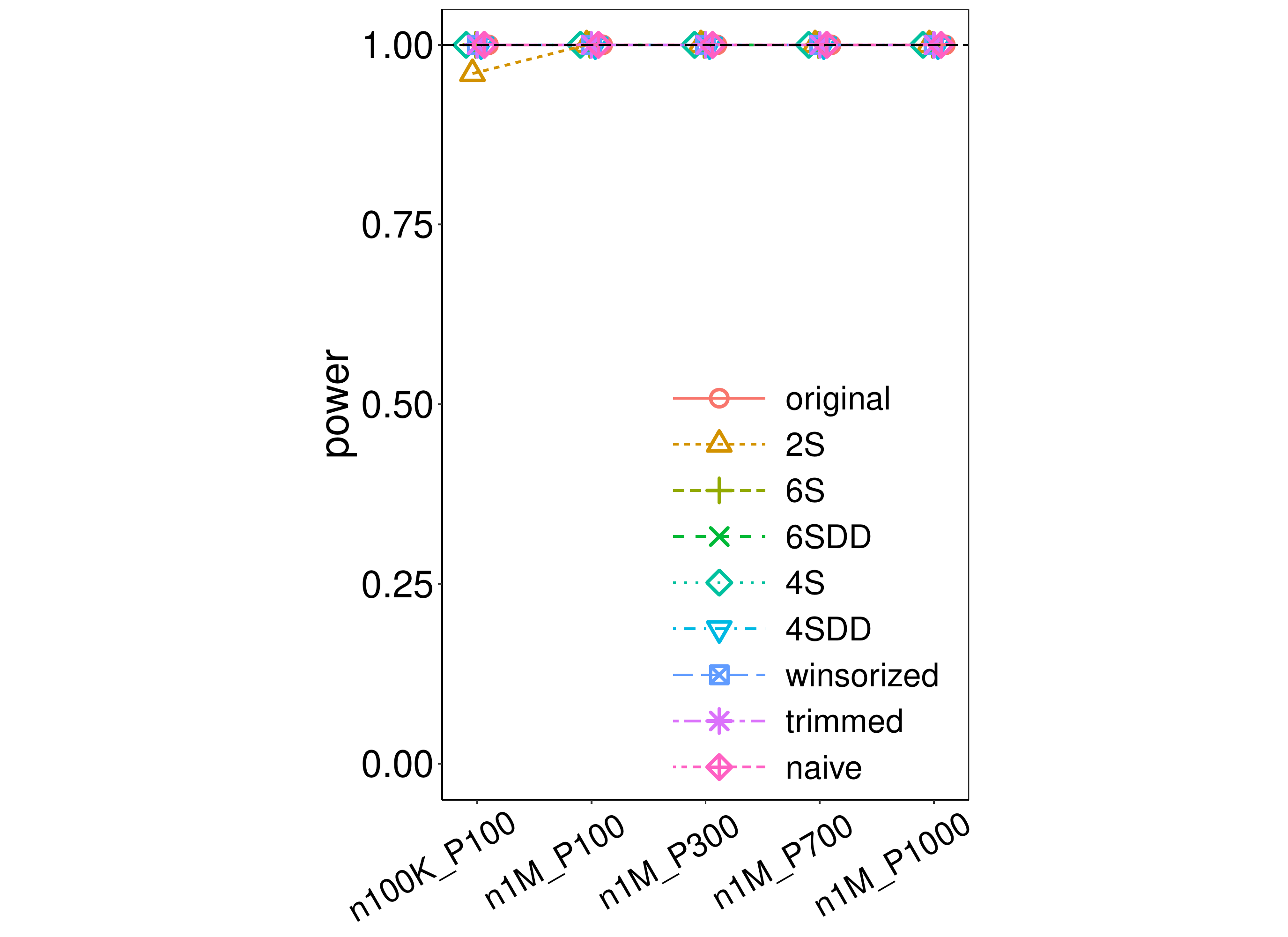}

\caption{Simulation results with $\rho$-zCDP for ZILN data with  $\alpha=\beta$ when $\theta\ne0$}
\label{fig:1szCDPZILN}
\end{figure}

The results are presented in Figures \ref{fig:1sDPZILN} and \ref{fig:1szCDPZILN} and are summarized  as follows.  Winsorized mean is the best performer by all metrics. In terms of bias, that in trimmed mean and the na\"{i}ve method is close to 0 as winsorized mean, so is that in 4S when $n=1m$. In terms of RMSE and CI width,  4S is similar to winsorized mean (the best group), followed by trimmed mean and the na\"{i}ve method (the second best group); 2S is the worst. Regarding CP, all methods except for 4SDD, 6SDD, and the na\"{i}ve method  provide nominal-level coverage in all cases (slight over-coverage at $n=100k/P=100$, $n=1m/P=100$ for $\epsilon\le1$). In terms of power, all methods provide power close to 1 except for 2S. The Gaussian mechanism of $\rho$-zCDP does not seem to be advantageous in statistical utility compared to the Laplace mechanism of $\epsilon$-DP.

\subsubsection{\texorpdfstring{$Y\sim$ ZILN}{}, \texorpdfstring{$\theta=0$ ($H_0$)}{H0 }, and \texorpdfstring{$\alpha\ne\beta$}{AS}}\label{sec:0asZILN}
\begin{figure}[!htb]
\hspace{0.5in}$\epsilon=0.5$\hspace{0.8in}$\epsilon=1$\hspace{0.9in}$\epsilon=2$
\hspace{0.95in}$\epsilon=5$\hspace{0.9in}$\epsilon=50$

\includegraphics[width=0.19\textwidth, trim={2.5in 0 2.6in 0},clip] {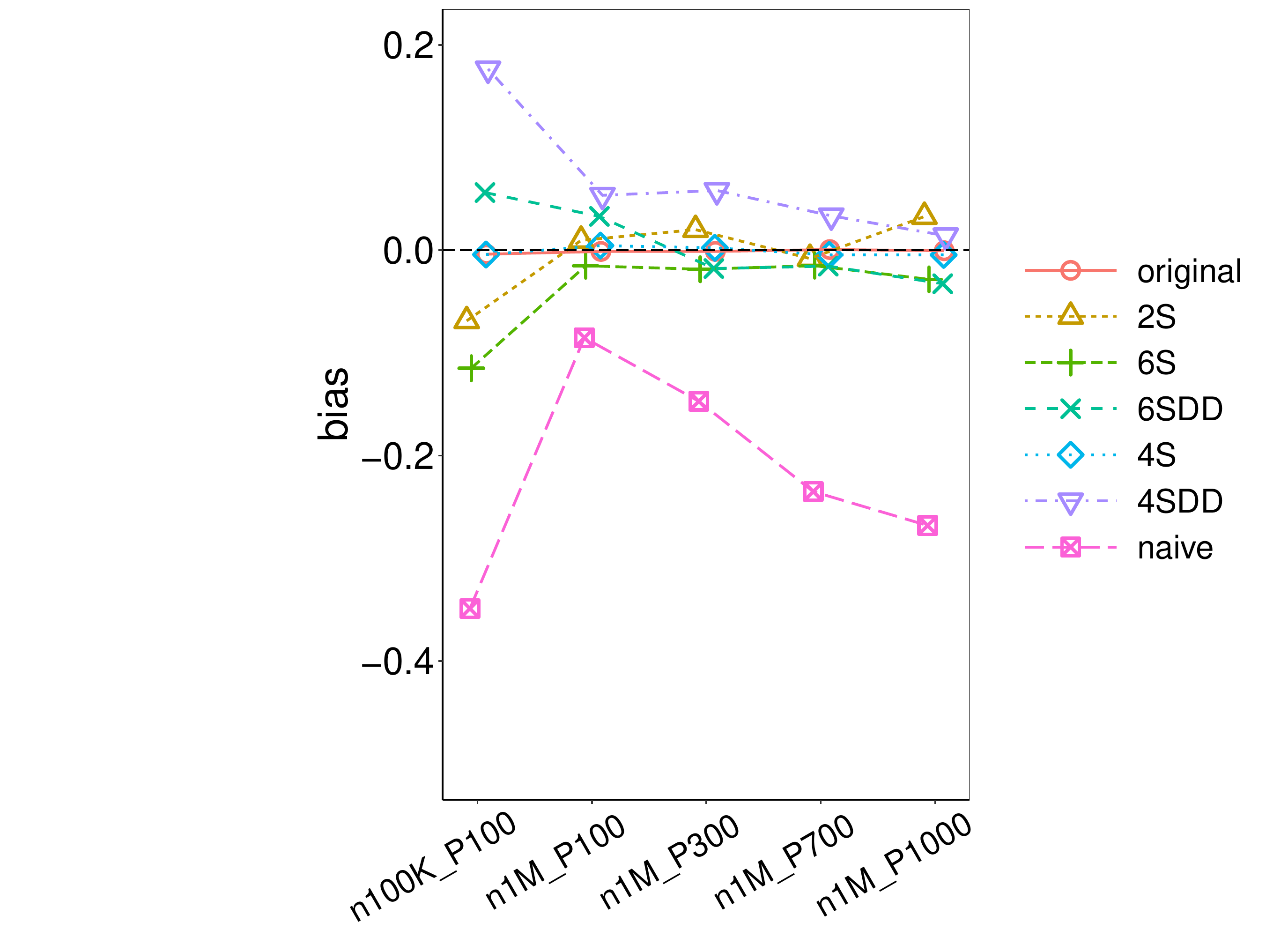}
\includegraphics[width=0.19\textwidth, trim={2.5in 0 2.6in 0},clip] {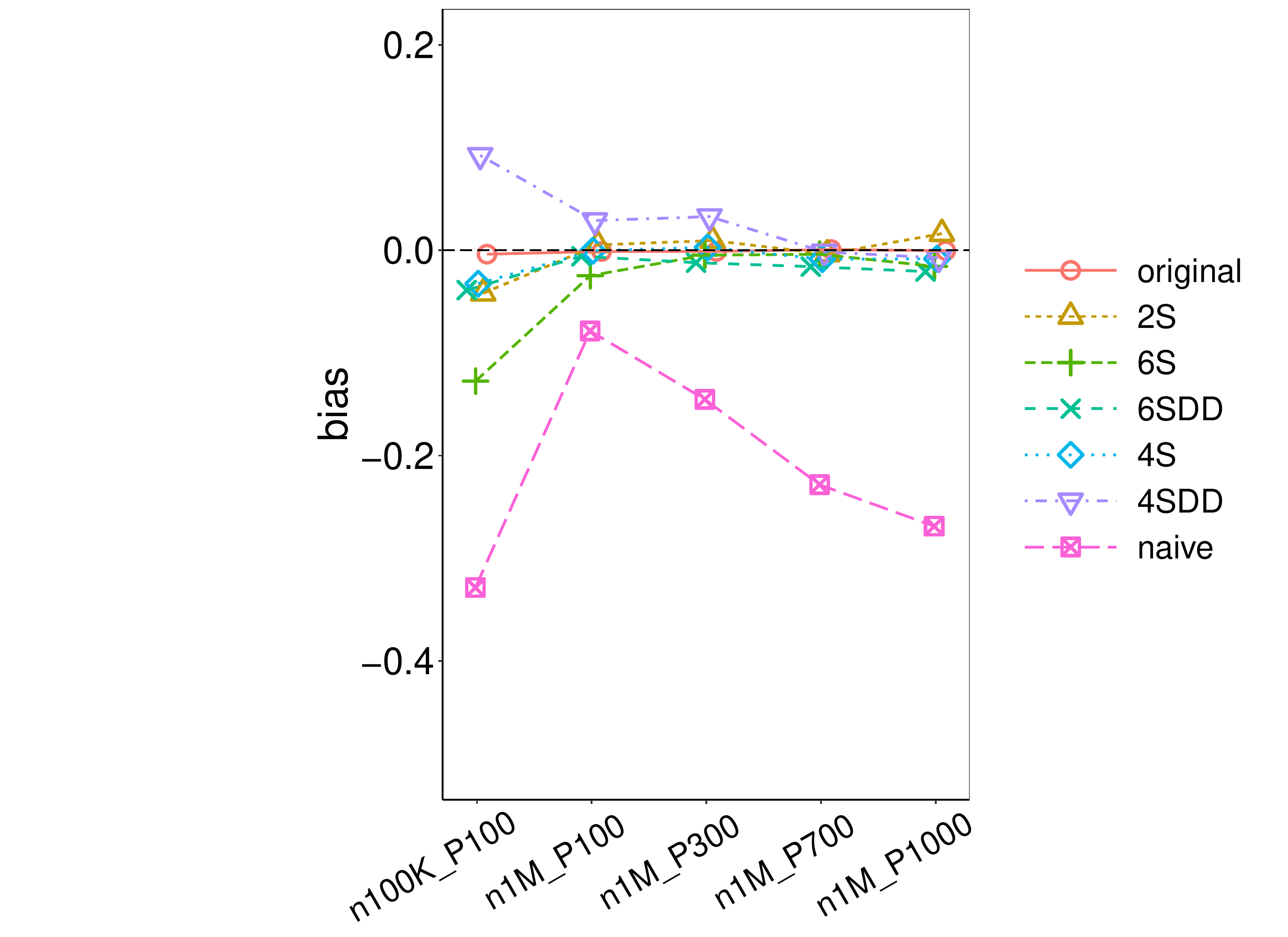}
\includegraphics[width=0.19\textwidth, trim={2.5in 0 2.6in 0},clip] {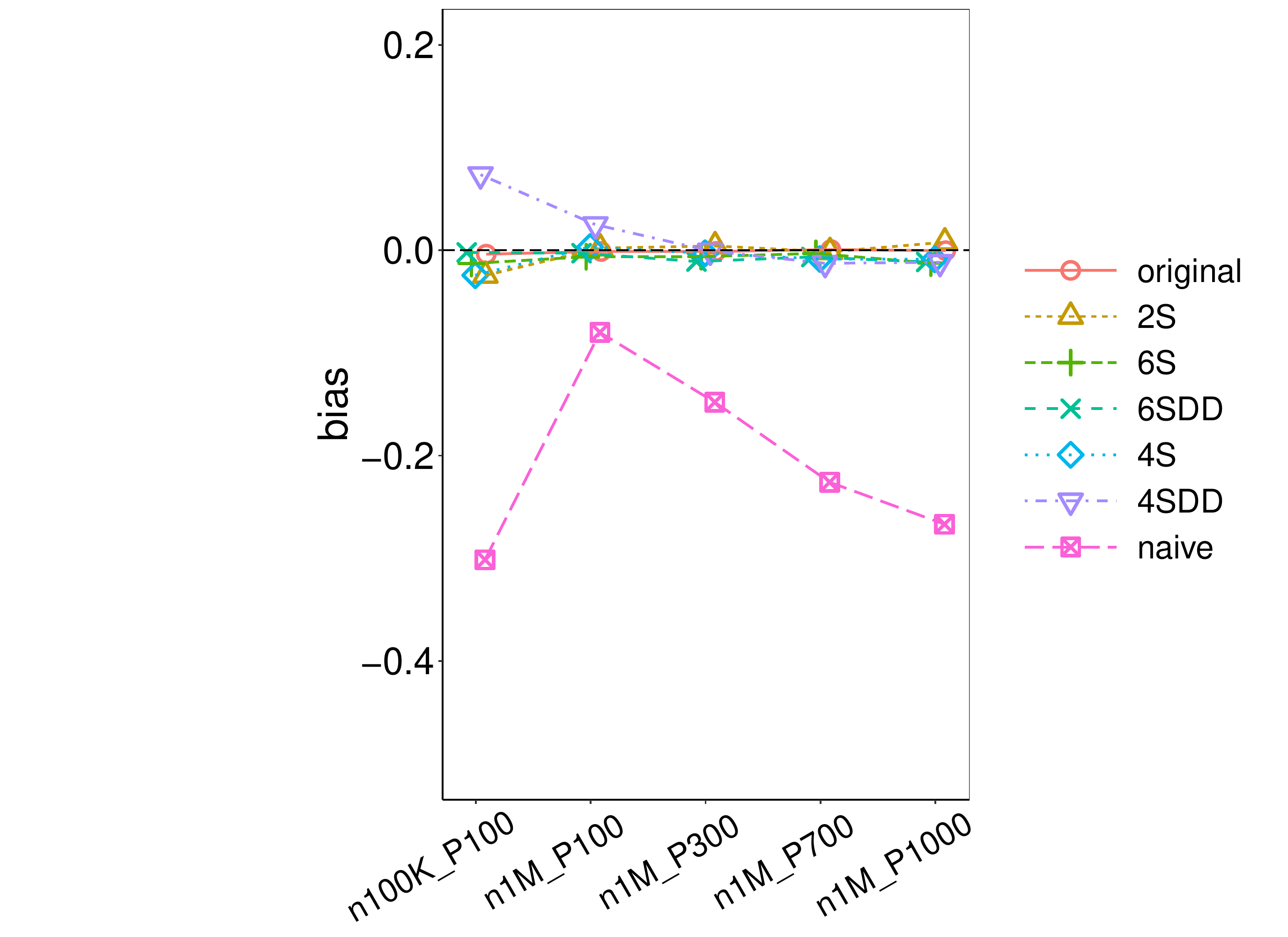}
\includegraphics[width=0.19\textwidth, trim={2.5in 0 2.6in 0},clip] {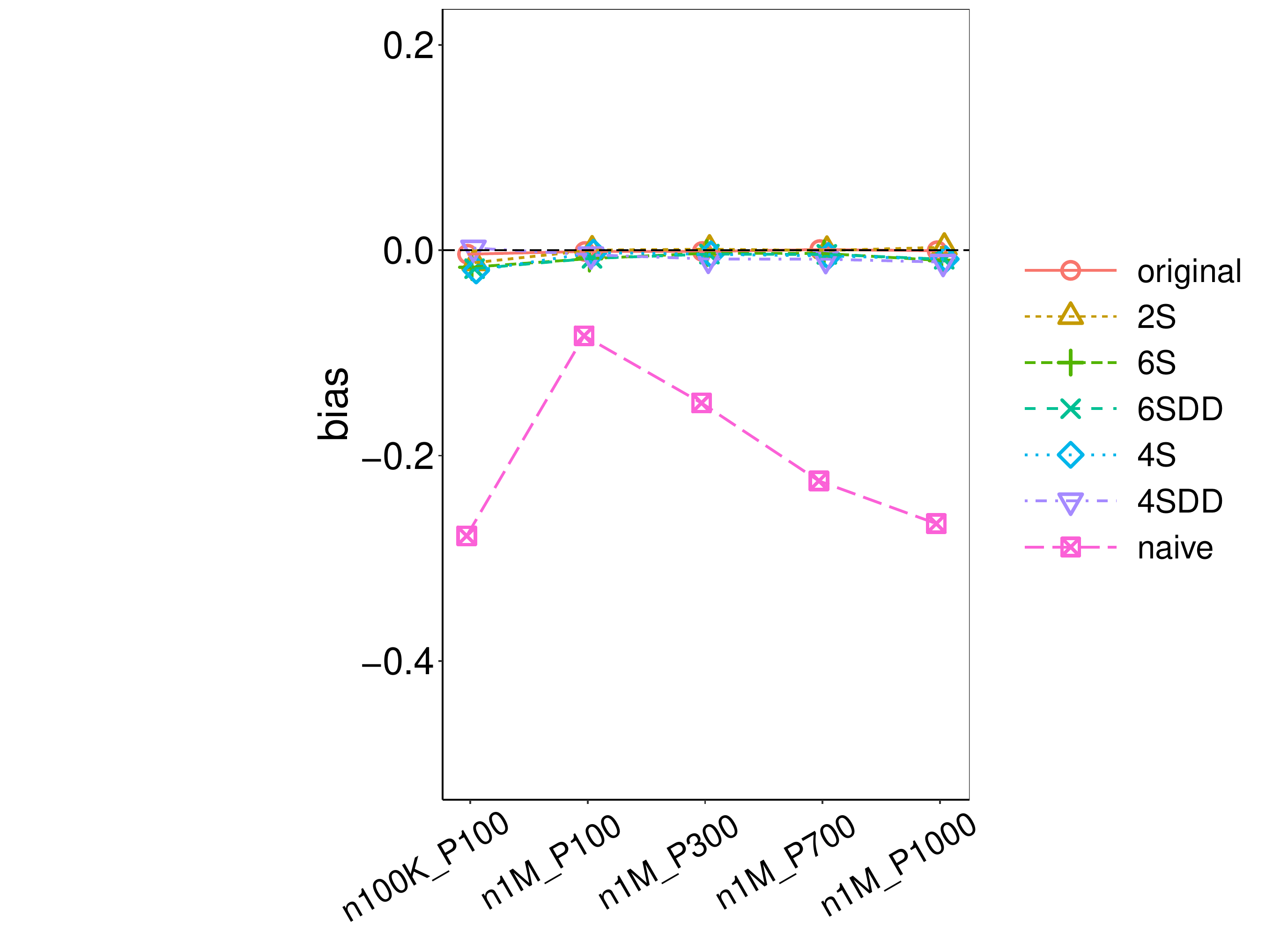}
\includegraphics[width=0.19\textwidth, trim={2.5in 0 2.6in 0},clip] {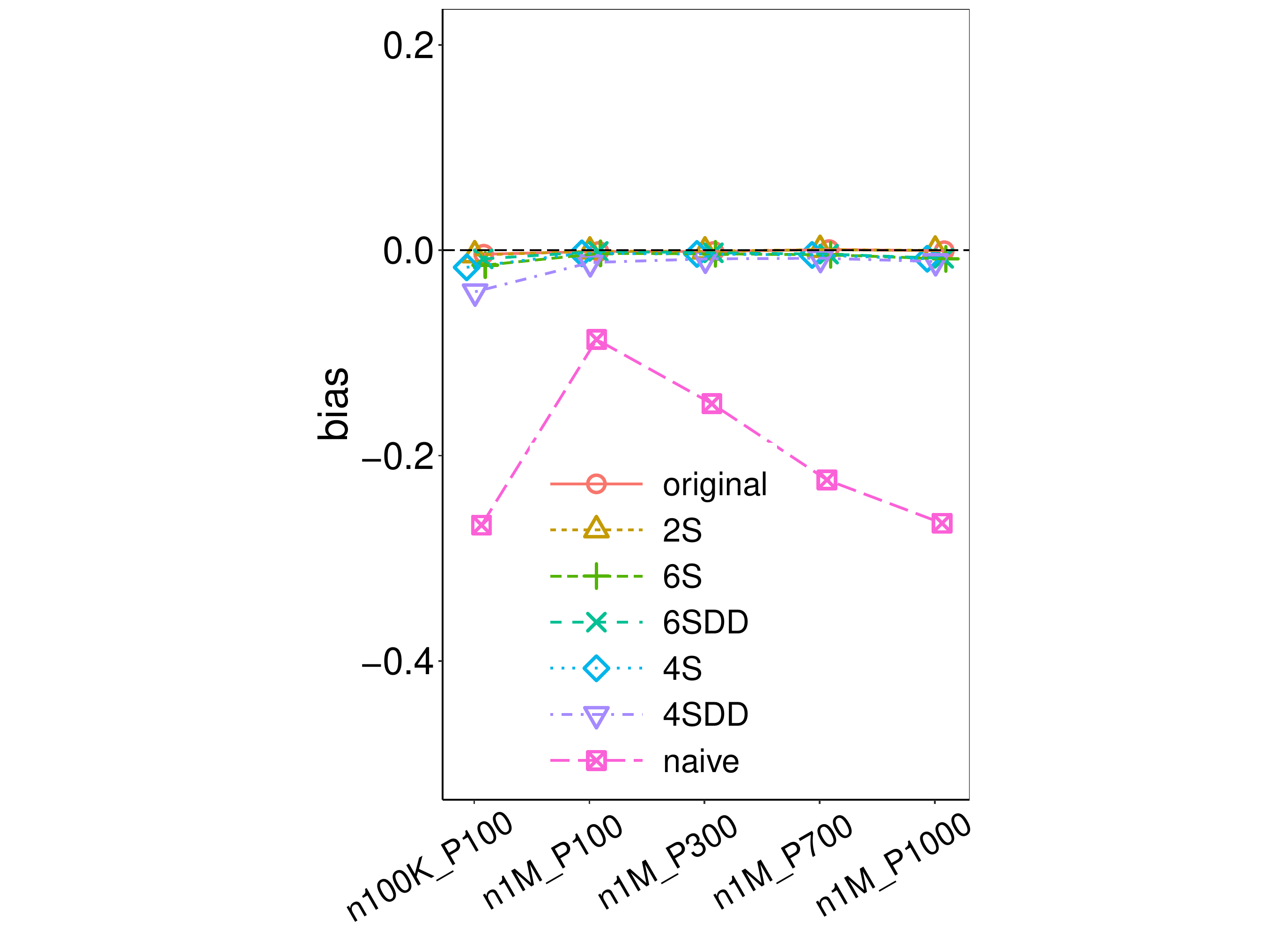}

\includegraphics[width=0.19\textwidth, trim={2.5in 0 2.6in 0},clip] {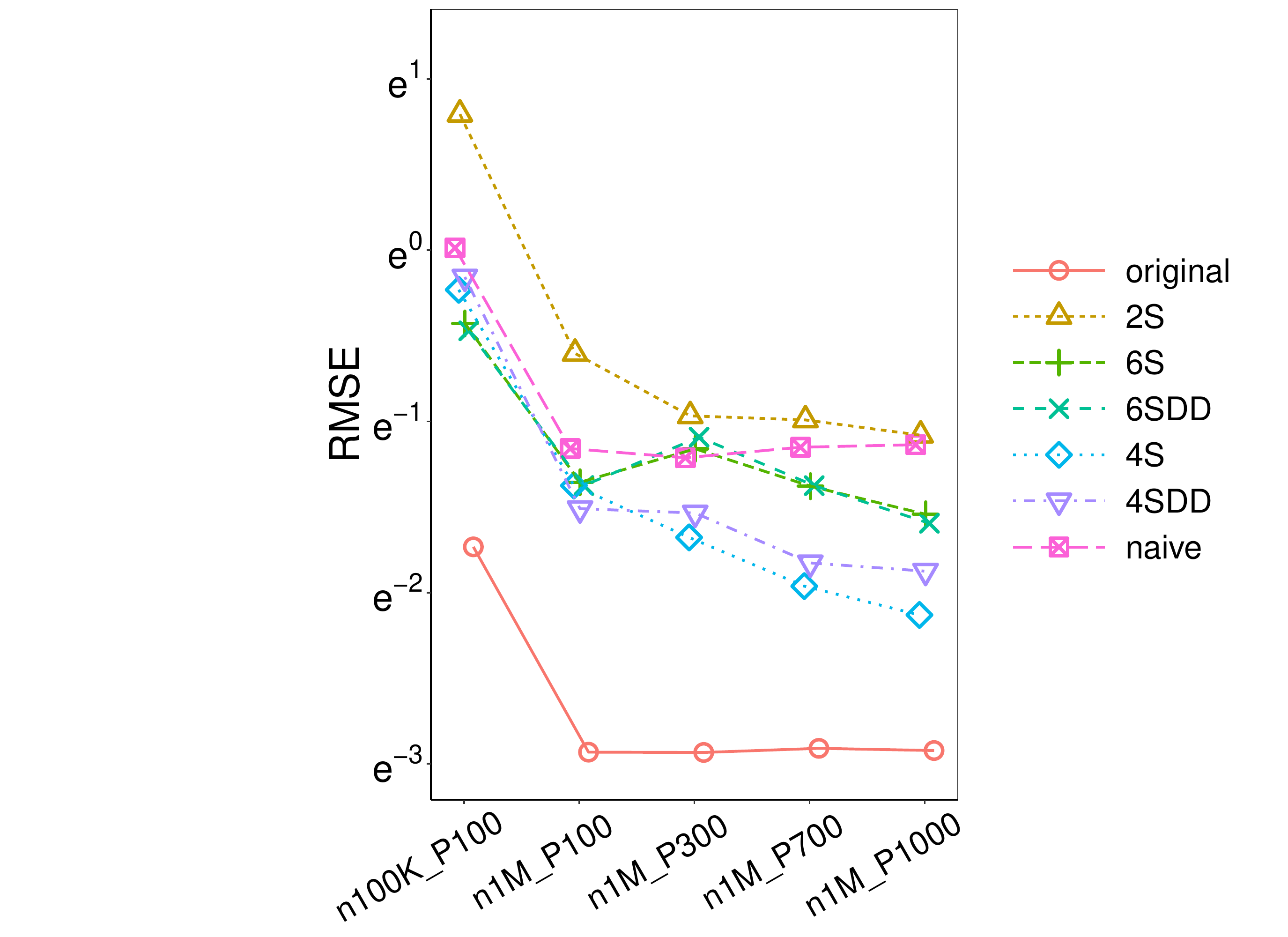}
\includegraphics[width=0.19\textwidth, trim={2.5in 0 2.6in 0},clip] {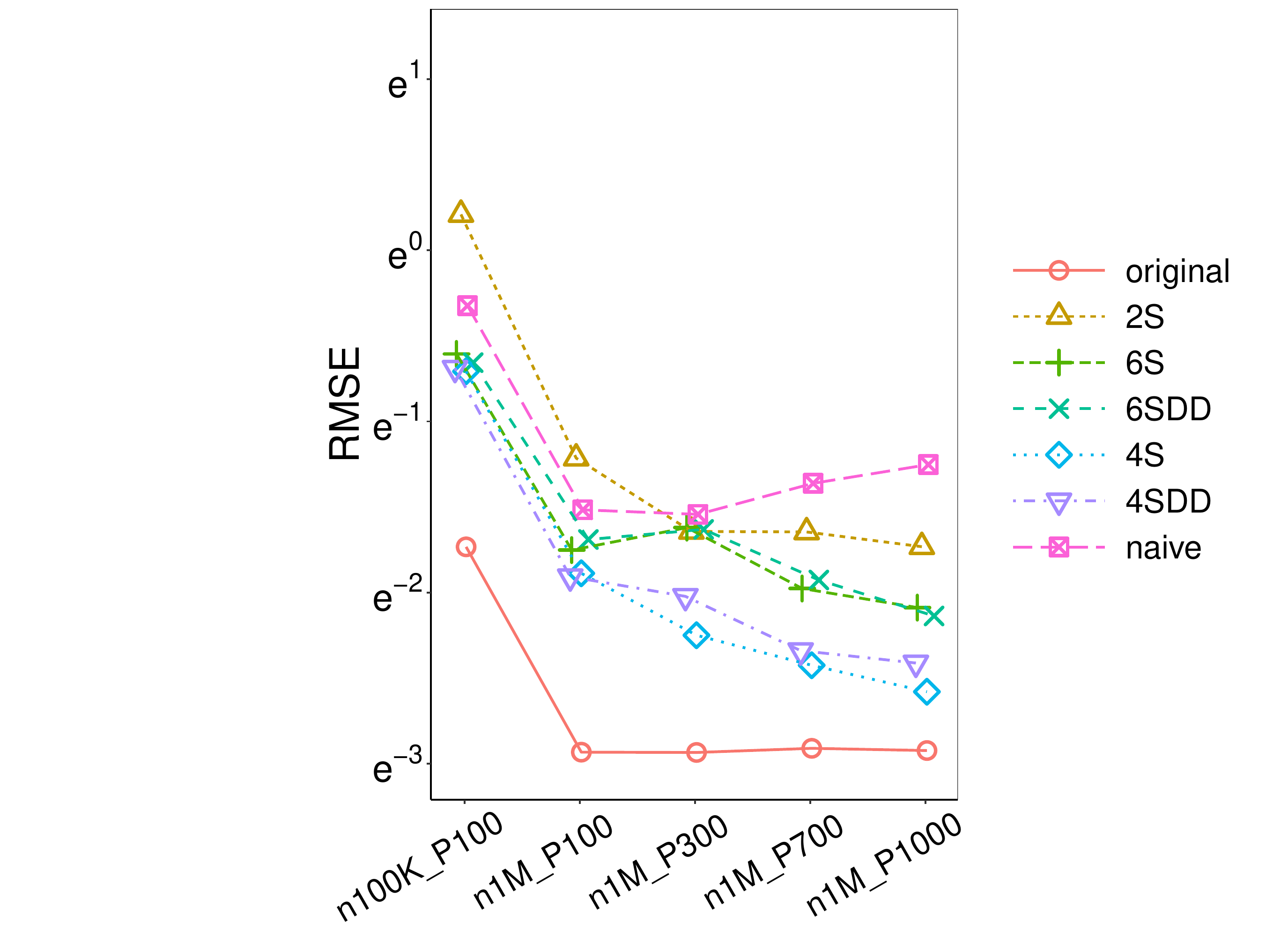}
\includegraphics[width=0.19\textwidth, trim={2.5in 0 2.6in 0},clip] {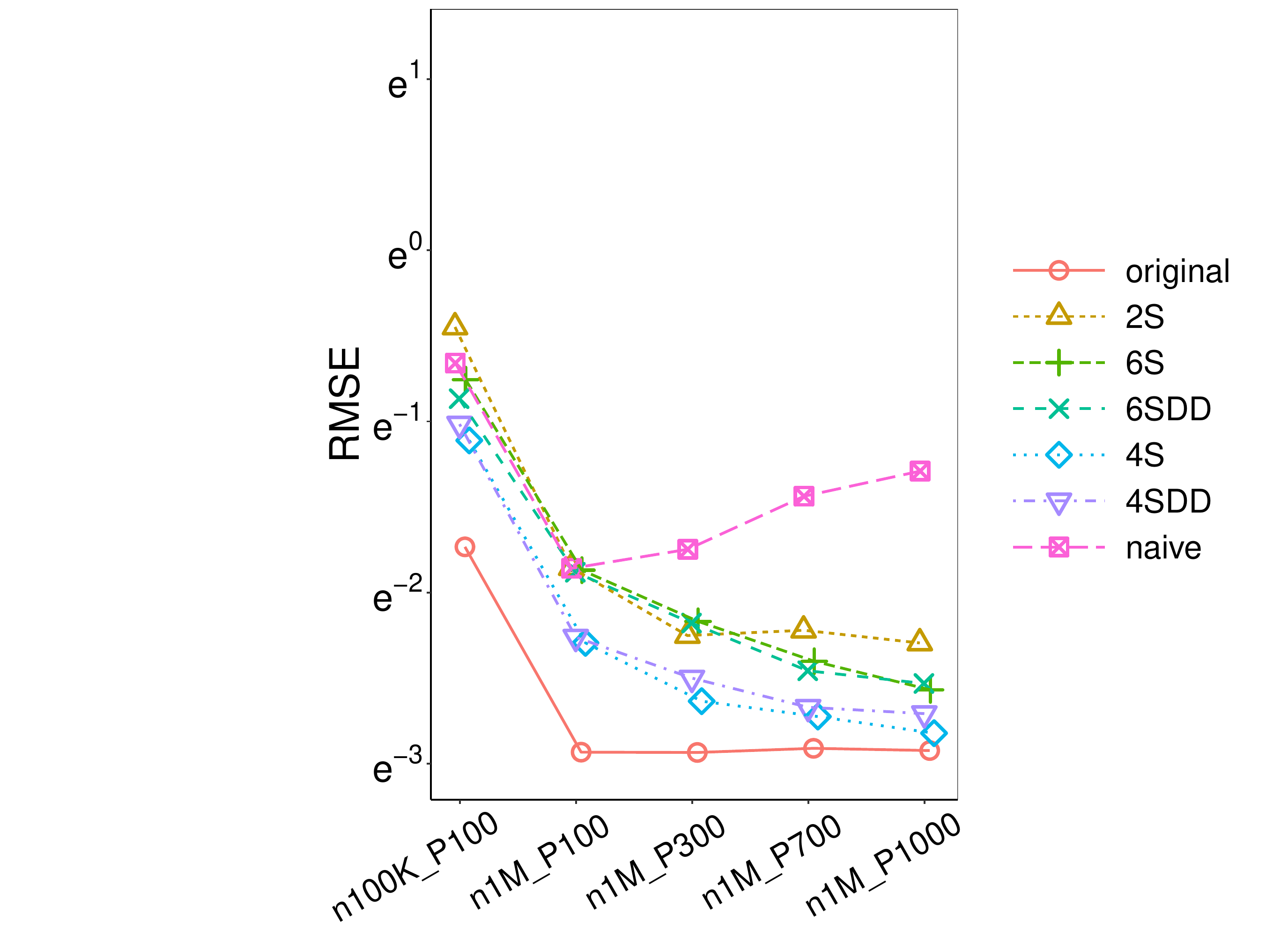}
\includegraphics[width=0.19\textwidth, trim={2.5in 0 2.6in 0},clip] {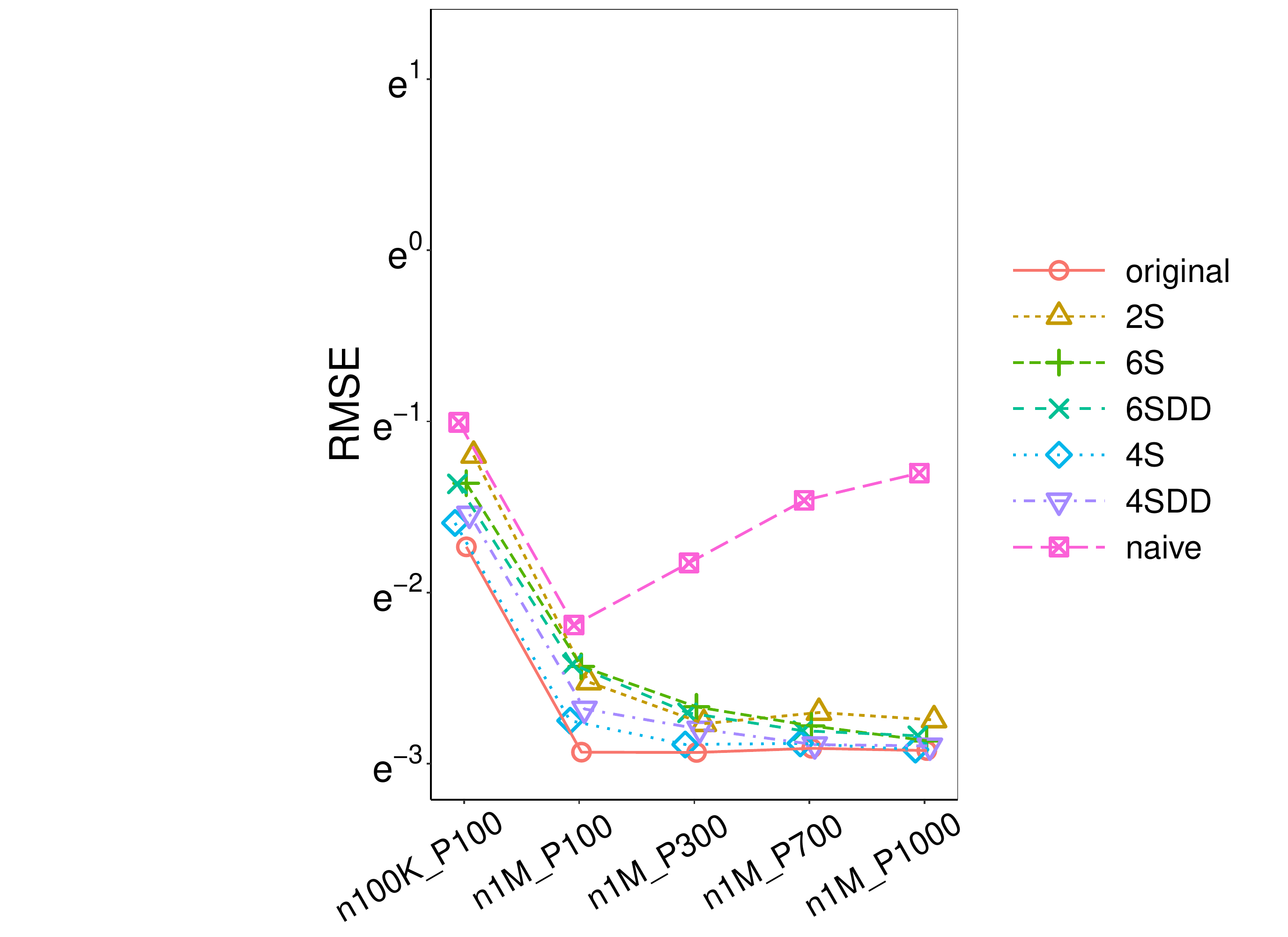}
\includegraphics[width=0.19\textwidth, trim={2.5in 0 2.6in 0},clip] {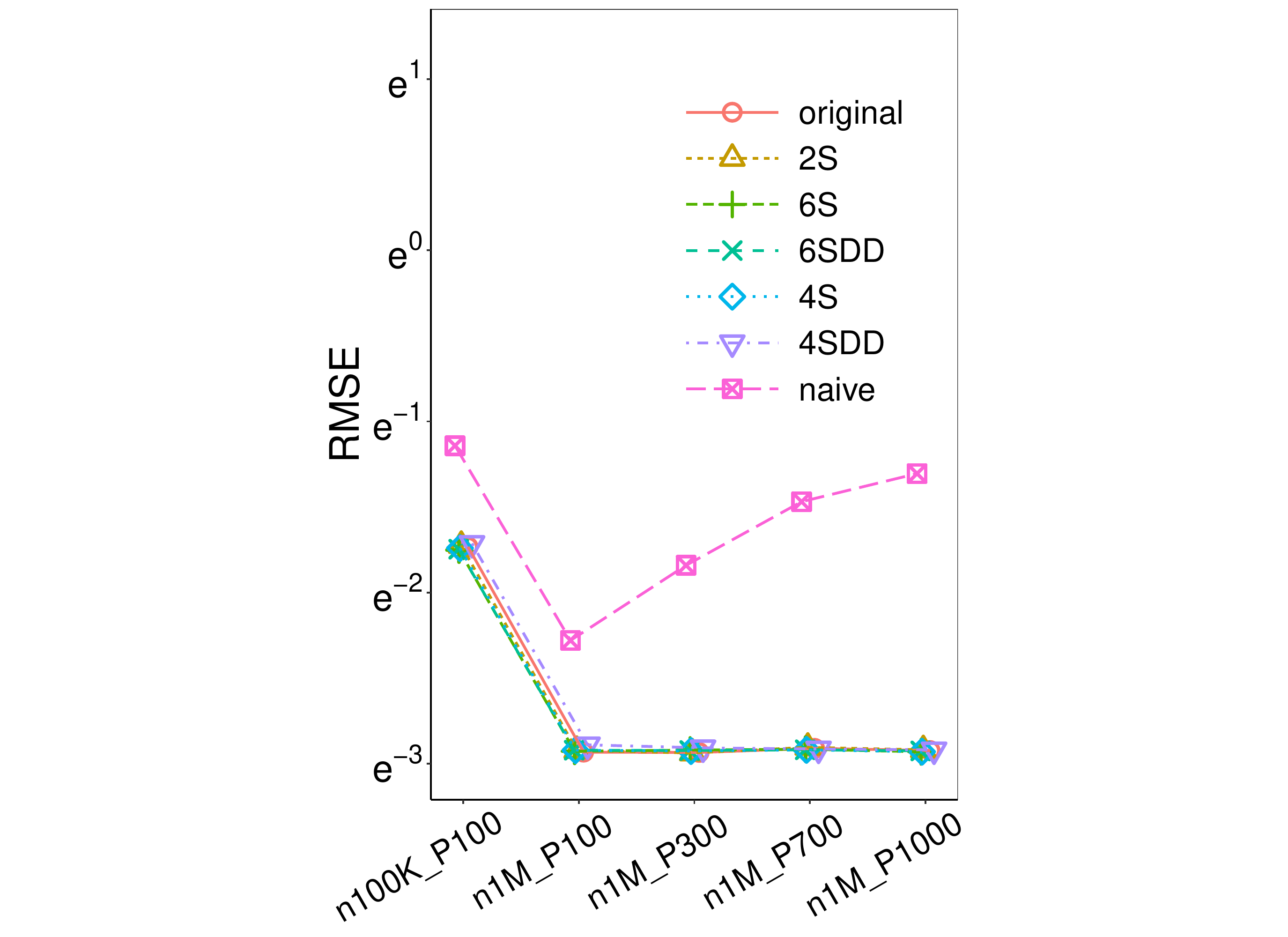}

\includegraphics[width=0.19\textwidth, trim={2.5in 0 2.6in 0},clip] {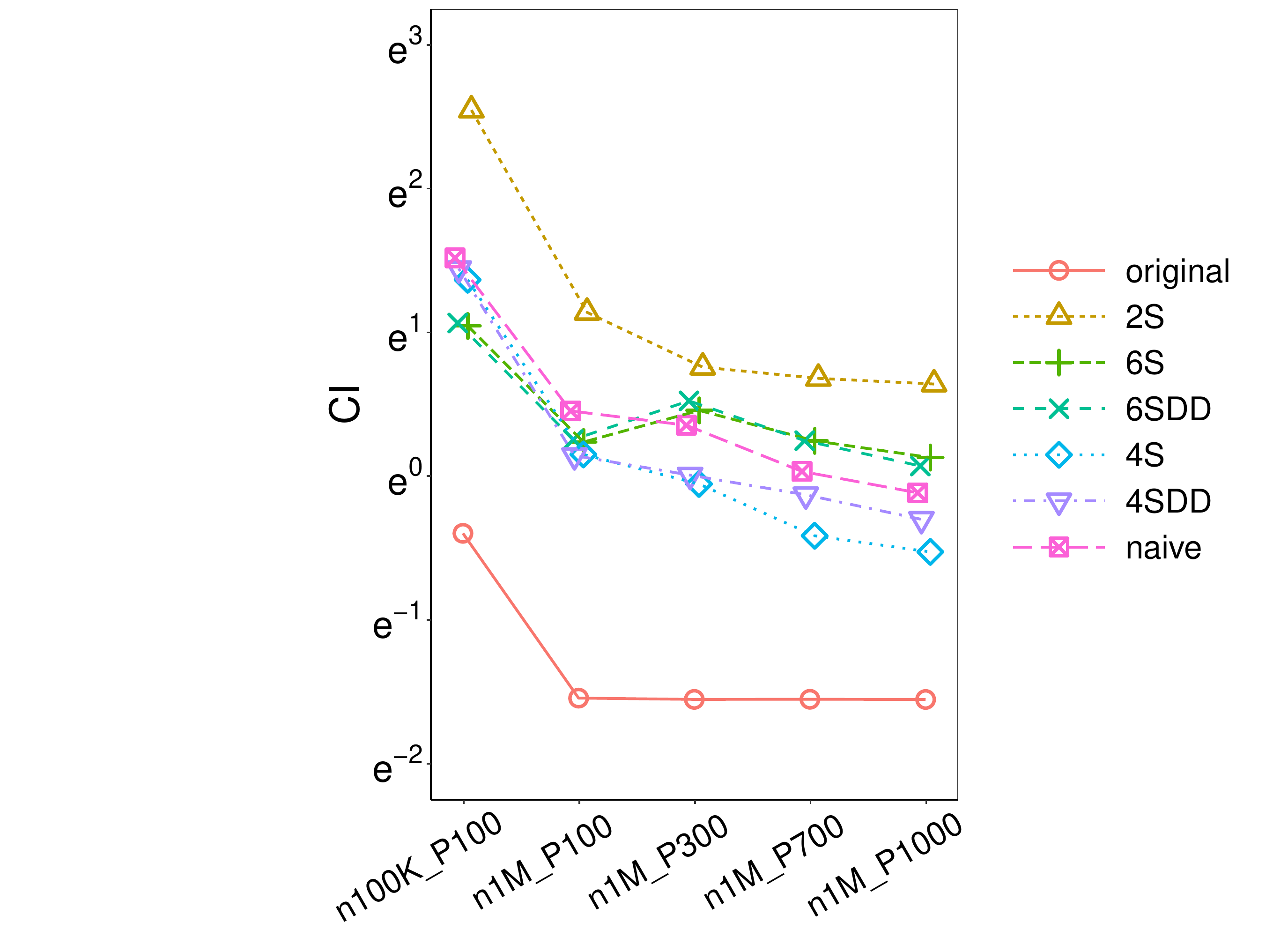}
\includegraphics[width=0.19\textwidth, trim={2.5in 0 2.6in 0},clip] {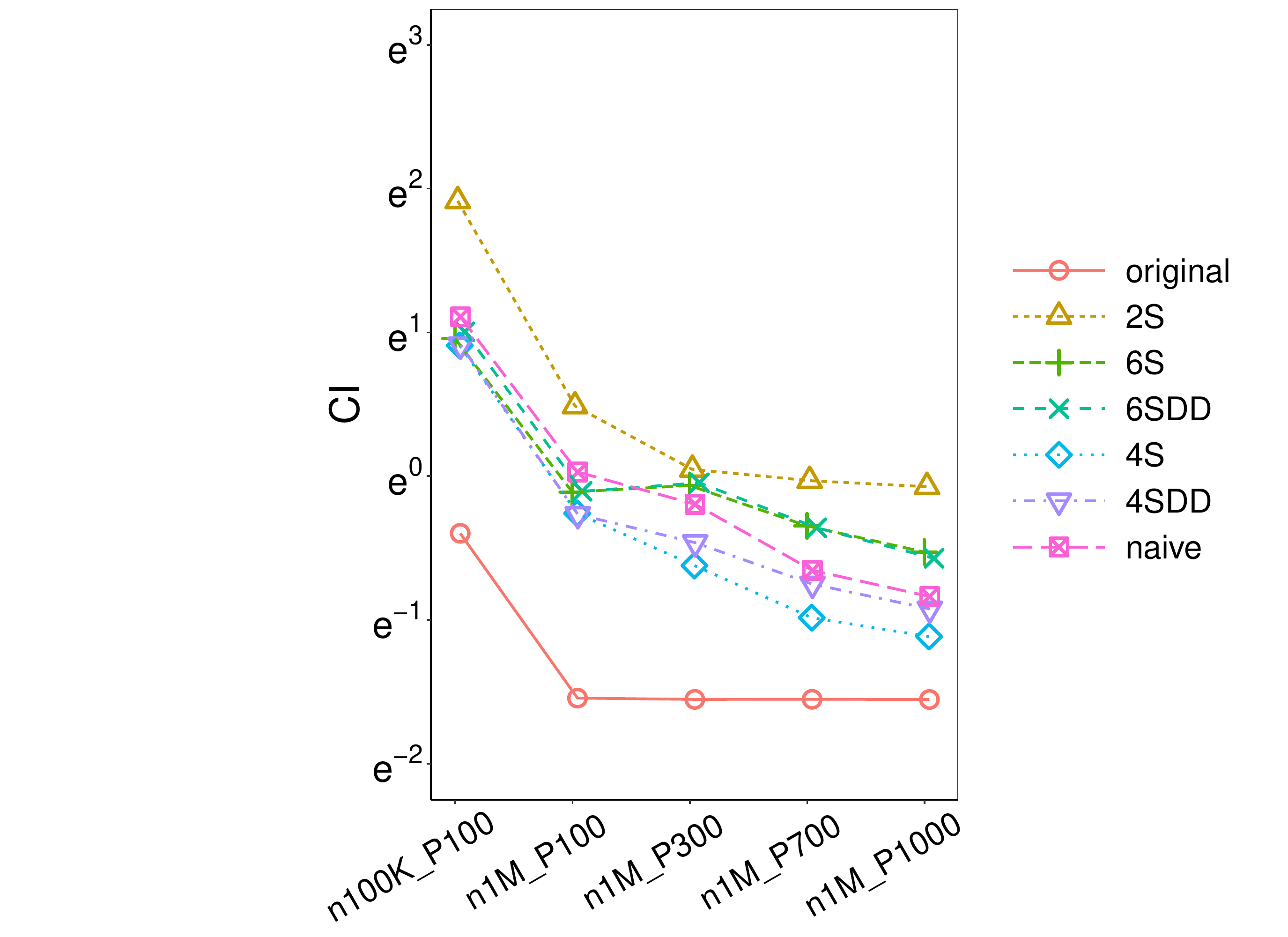}
\includegraphics[width=0.19\textwidth, trim={2.5in 0 2.6in 0},clip] {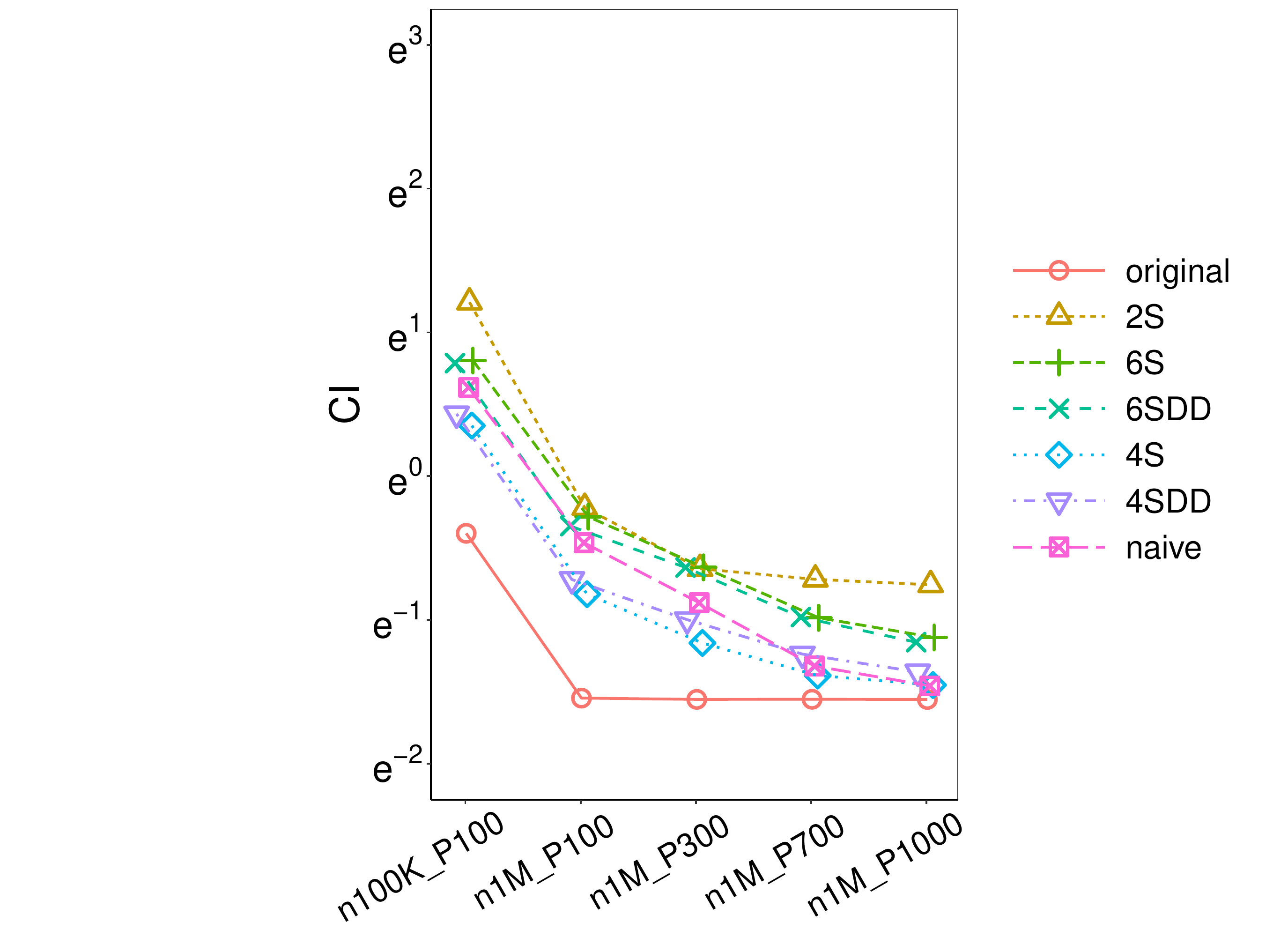}
\includegraphics[width=0.19\textwidth, trim={2.5in 0 2.6in 0},clip] {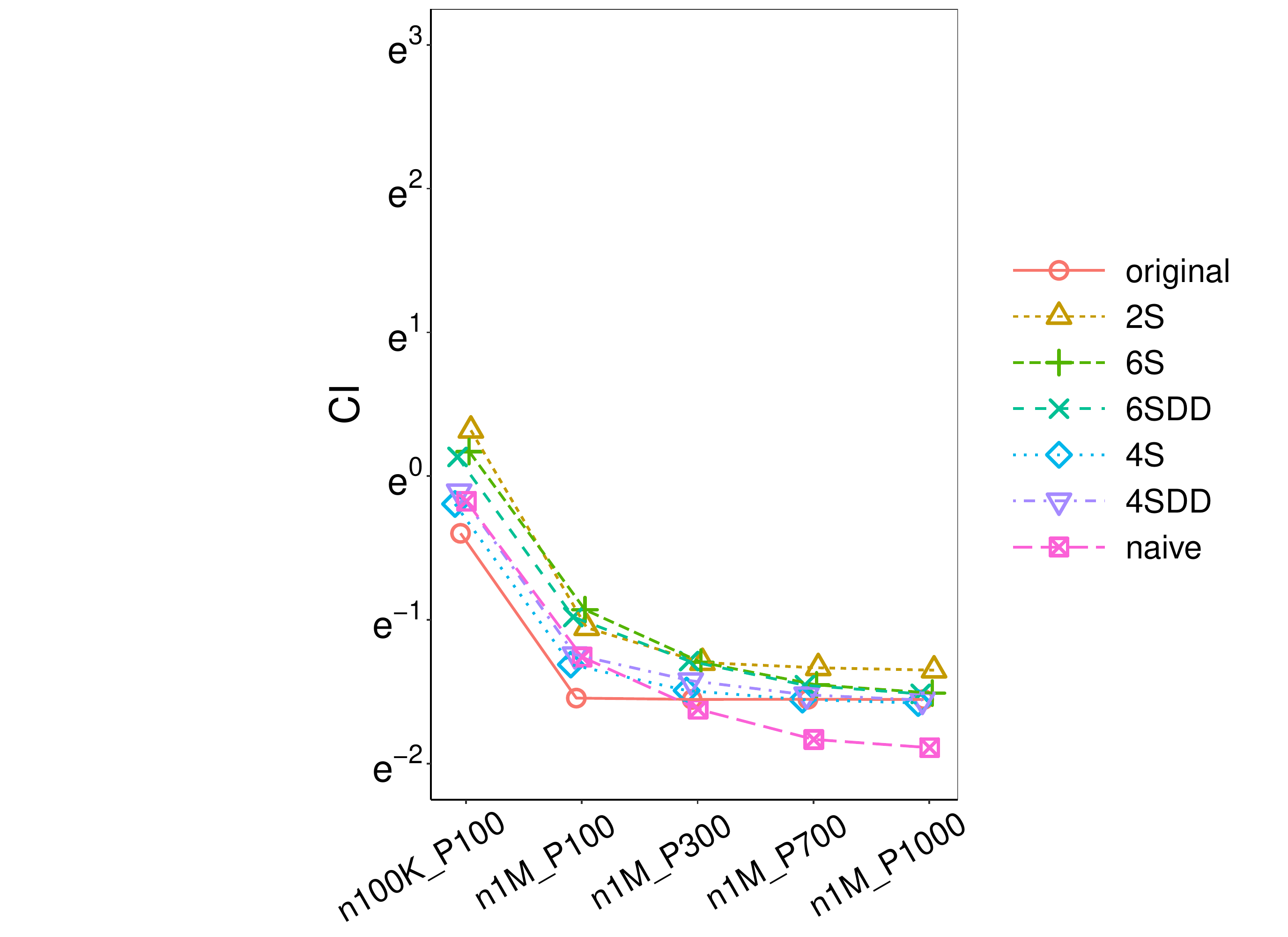}
\includegraphics[width=0.19\textwidth, trim={2.5in 0 2.6in 0},clip] {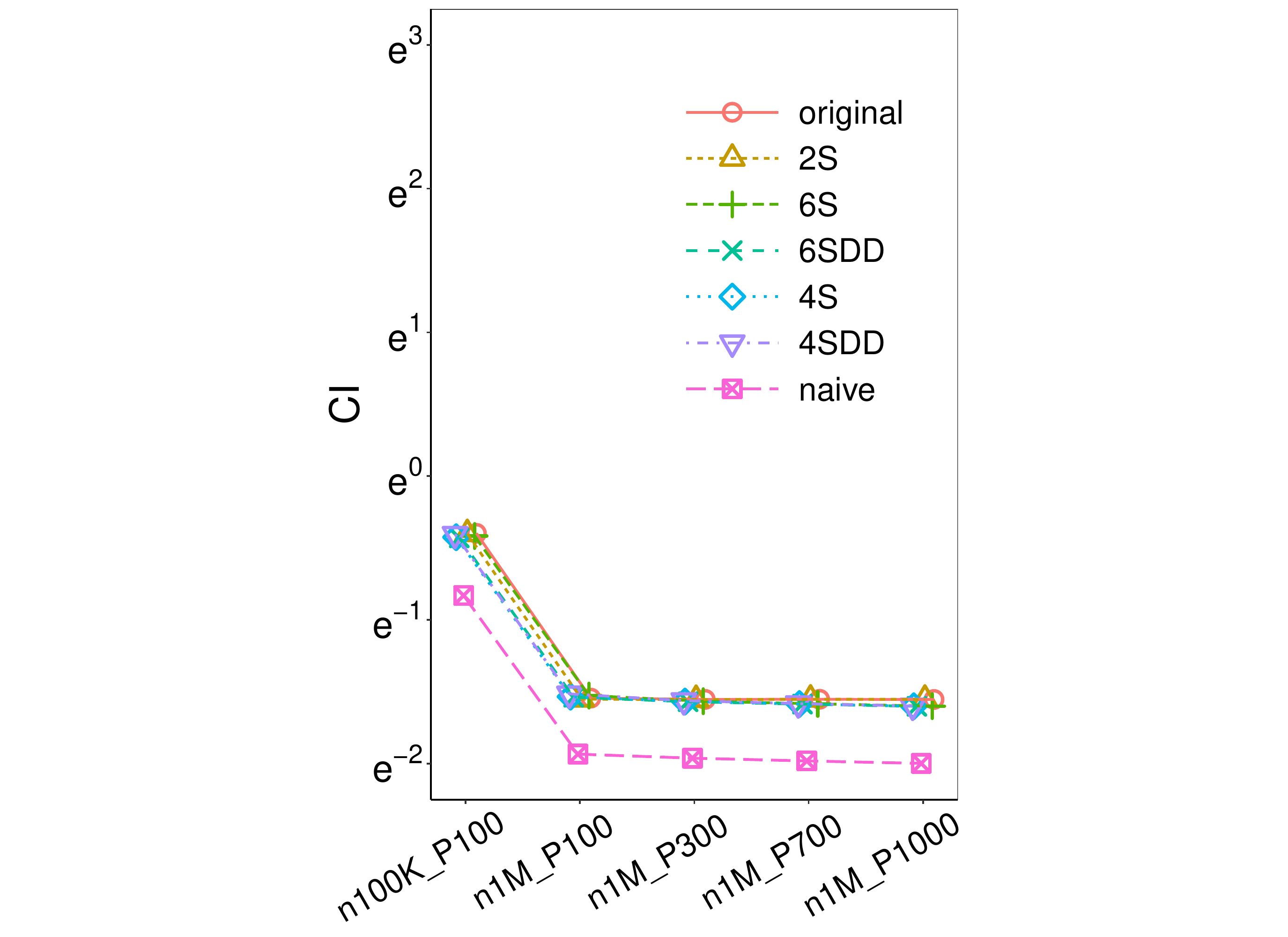}

\includegraphics[width=0.19\textwidth, trim={2.5in 0 2.6in 0},clip] {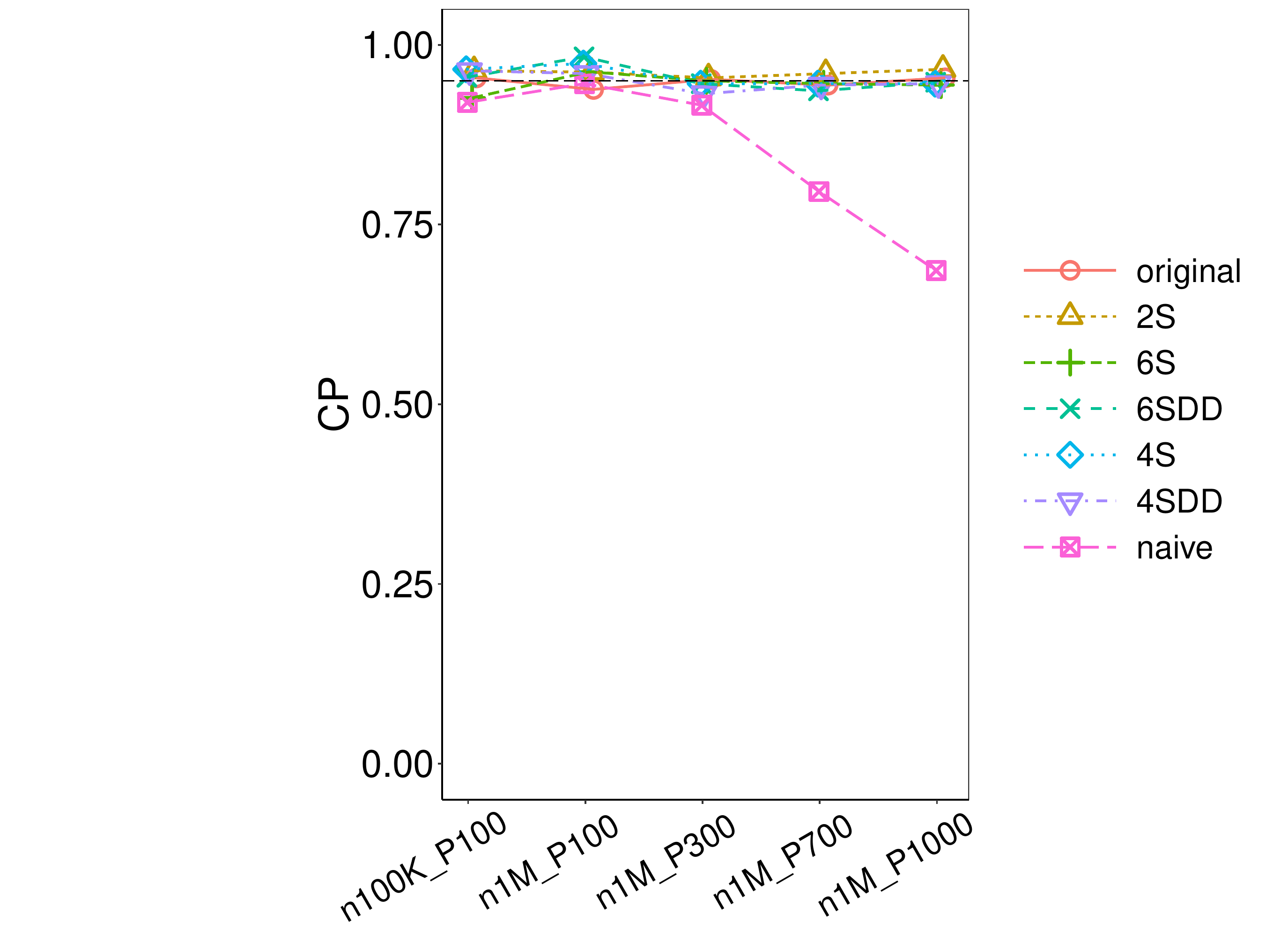}
\includegraphics[width=0.19\textwidth, trim={2.5in 0 2.6in 0},clip] {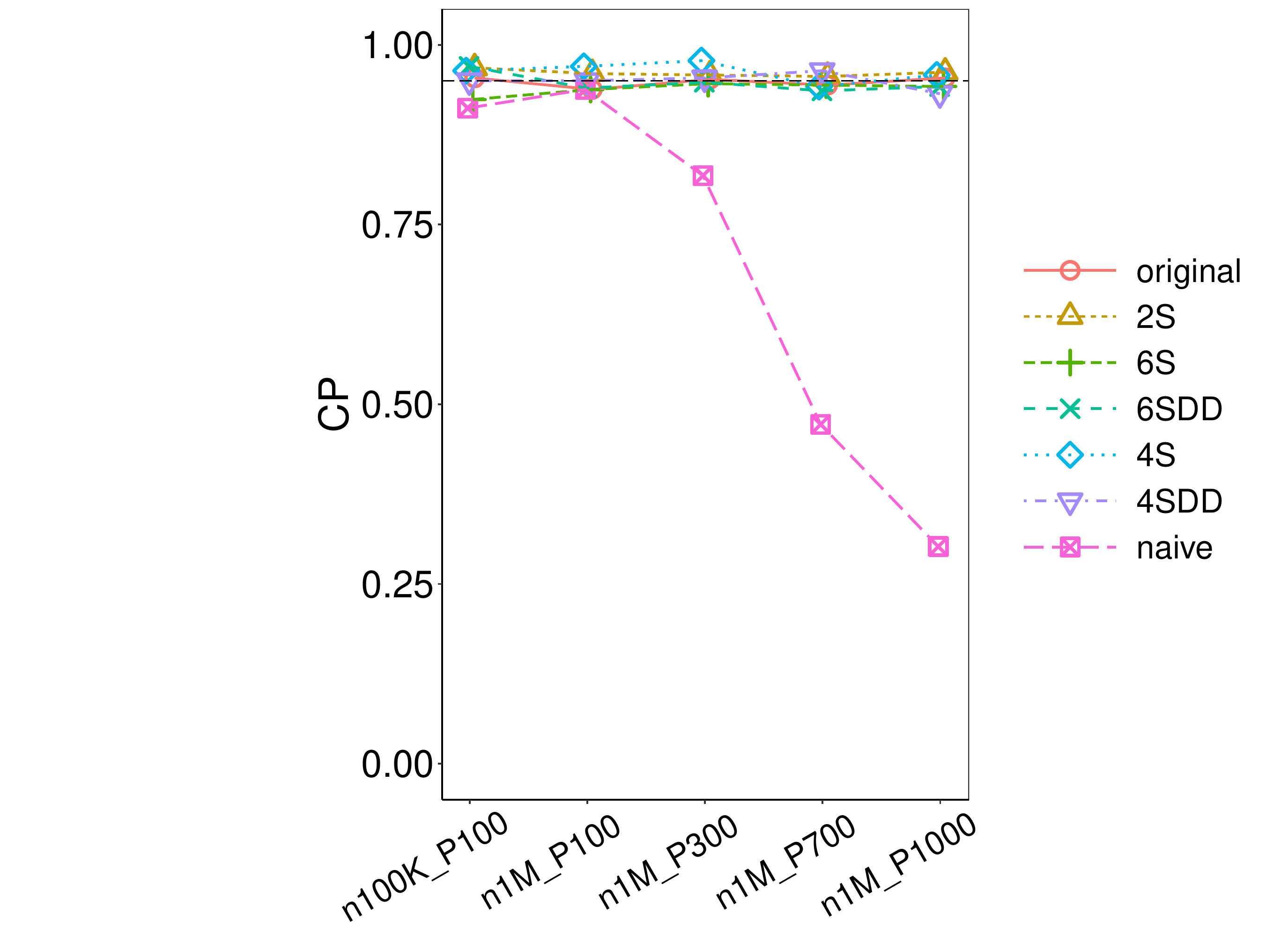}
\includegraphics[width=0.19\textwidth, trim={2.5in 0 2.6in 0},clip] {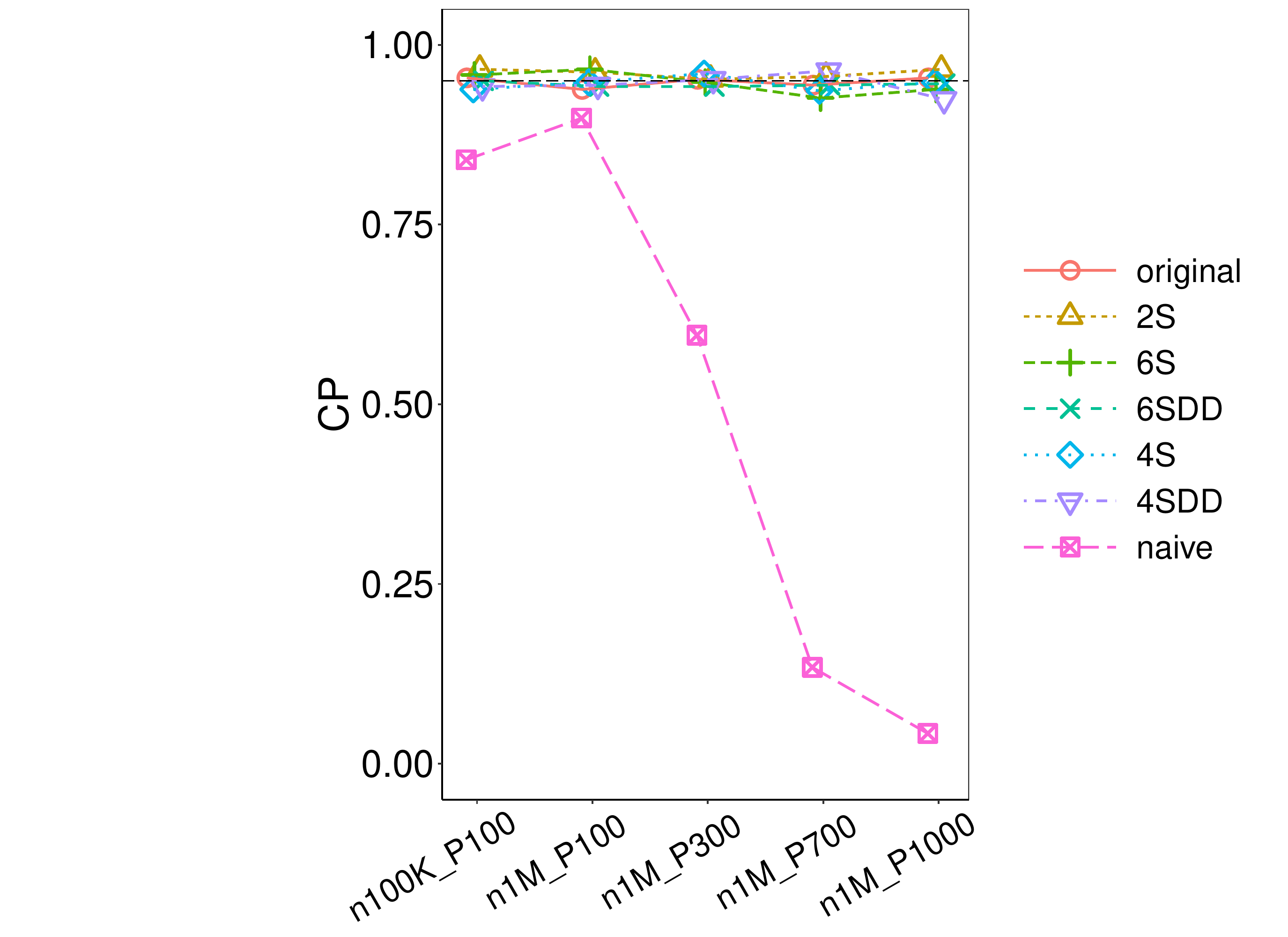}
\includegraphics[width=0.19\textwidth, trim={2.5in 0 2.6in 0},clip] {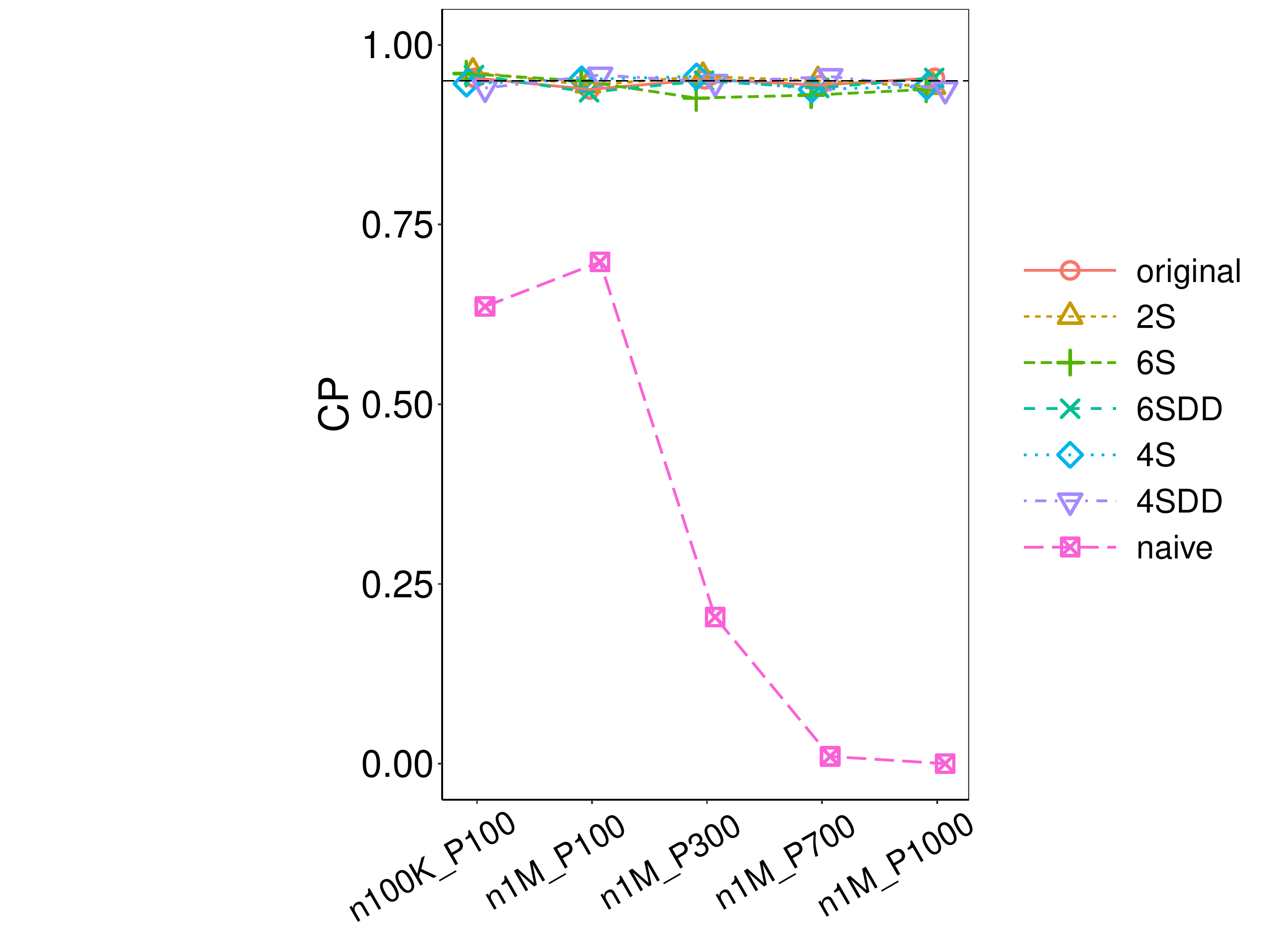}
\includegraphics[width=0.19\textwidth, trim={2.5in 0 2.6in 0},clip] {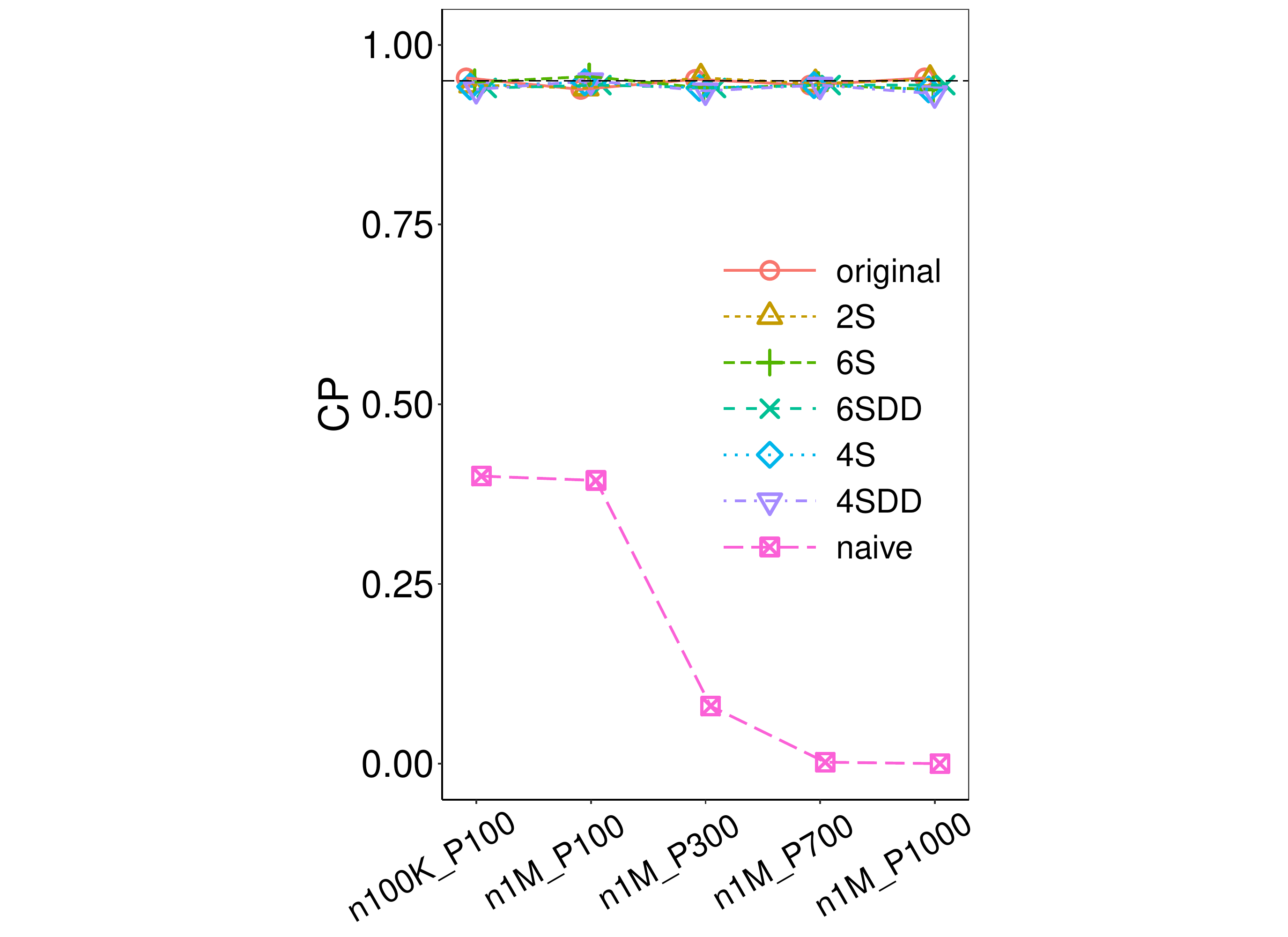}

\caption{Simulation results with $\epsilon$-DP for ZILN data with  $\alpha\ne\beta$ when $\theta=0$} \label{fig:0asDPZILN}
\end{figure}

\begin{figure}[!htb]
\hspace{0.45in}$\rho=0.005$\hspace{0.65in}$\rho=0.02$\hspace{0.65in}$\rho=0.08$
\hspace{0.65in}$\rho=0.32$\hspace{0.65in}$\rho=1.28$

\includegraphics[width=0.19\textwidth, trim={2.5in 0 2.6in 0},clip] {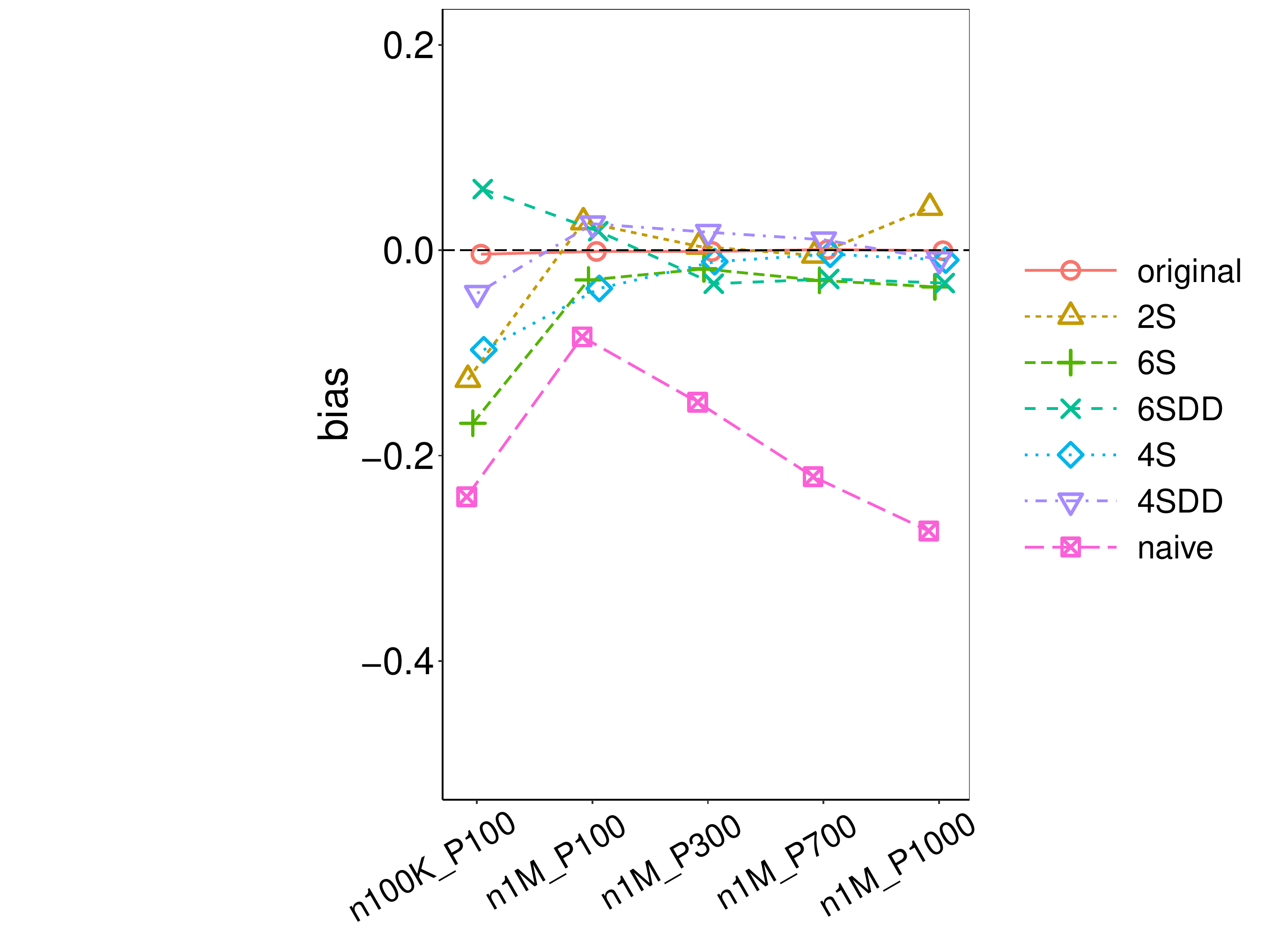}
\includegraphics[width=0.19\textwidth, trim={2.5in 0 2.6in 0},clip] {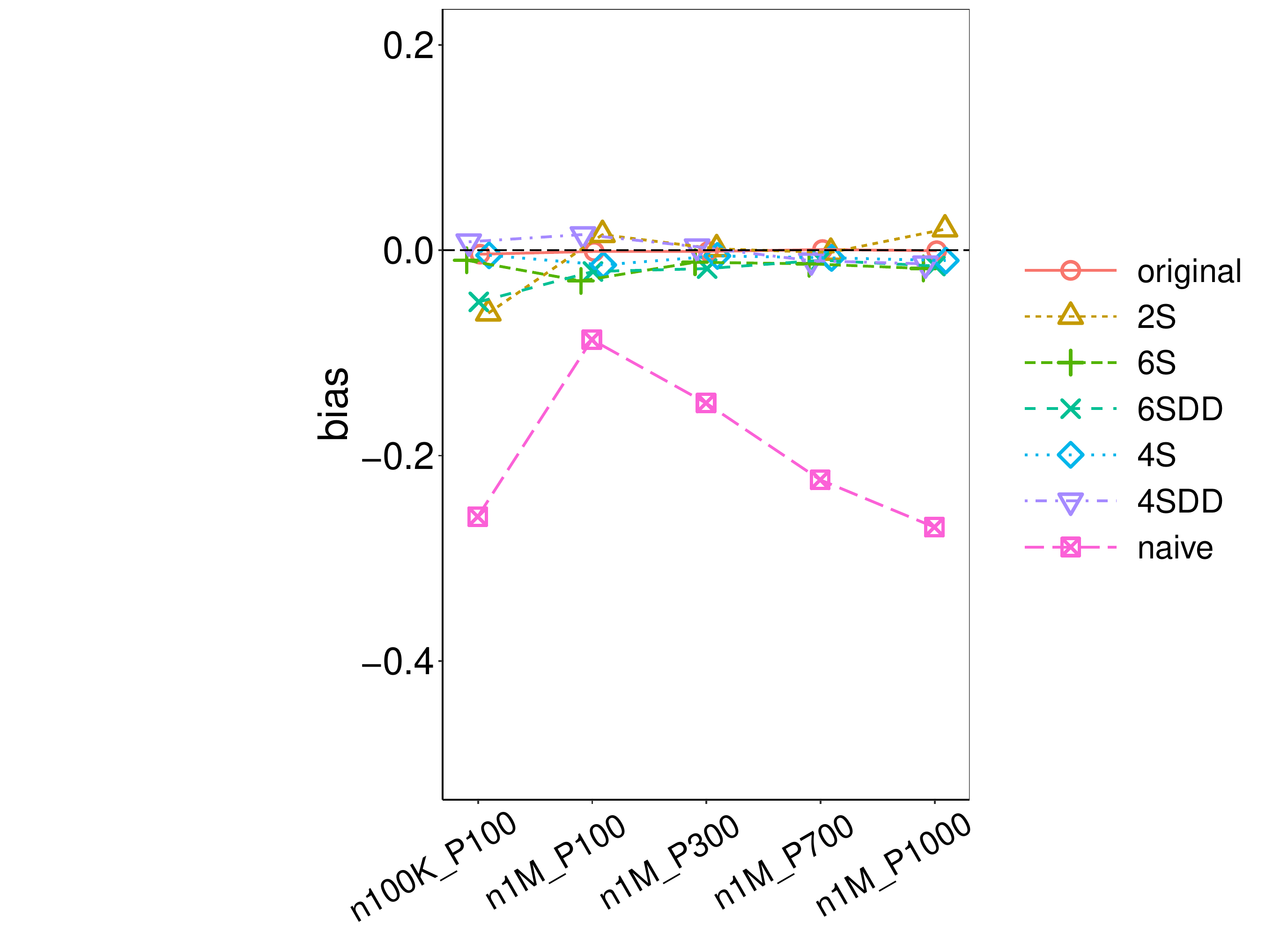}
\includegraphics[width=0.19\textwidth, trim={2.5in 0 2.6in 0},clip] {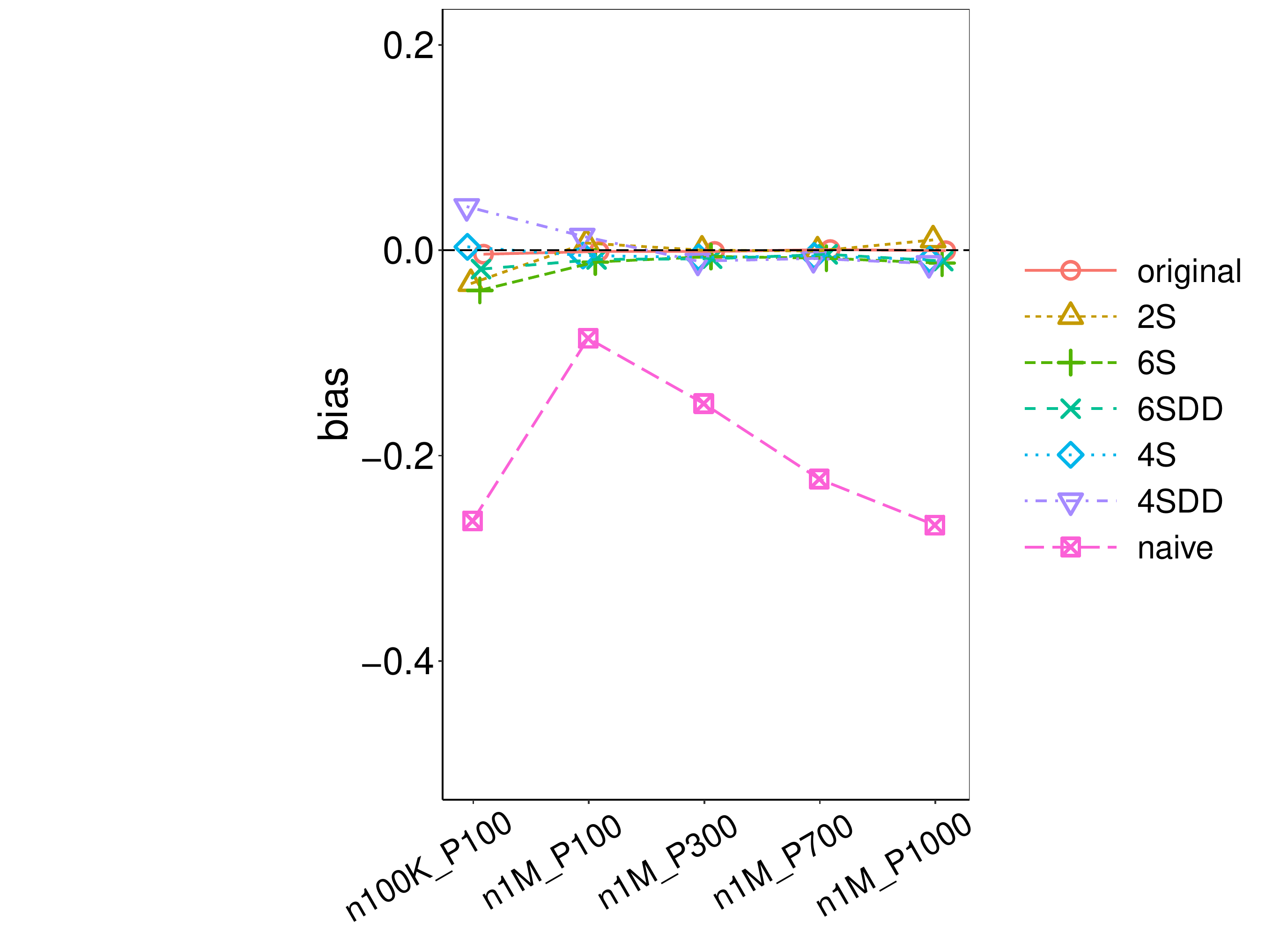}
\includegraphics[width=0.19\textwidth, trim={2.5in 0 2.6in 0},clip] {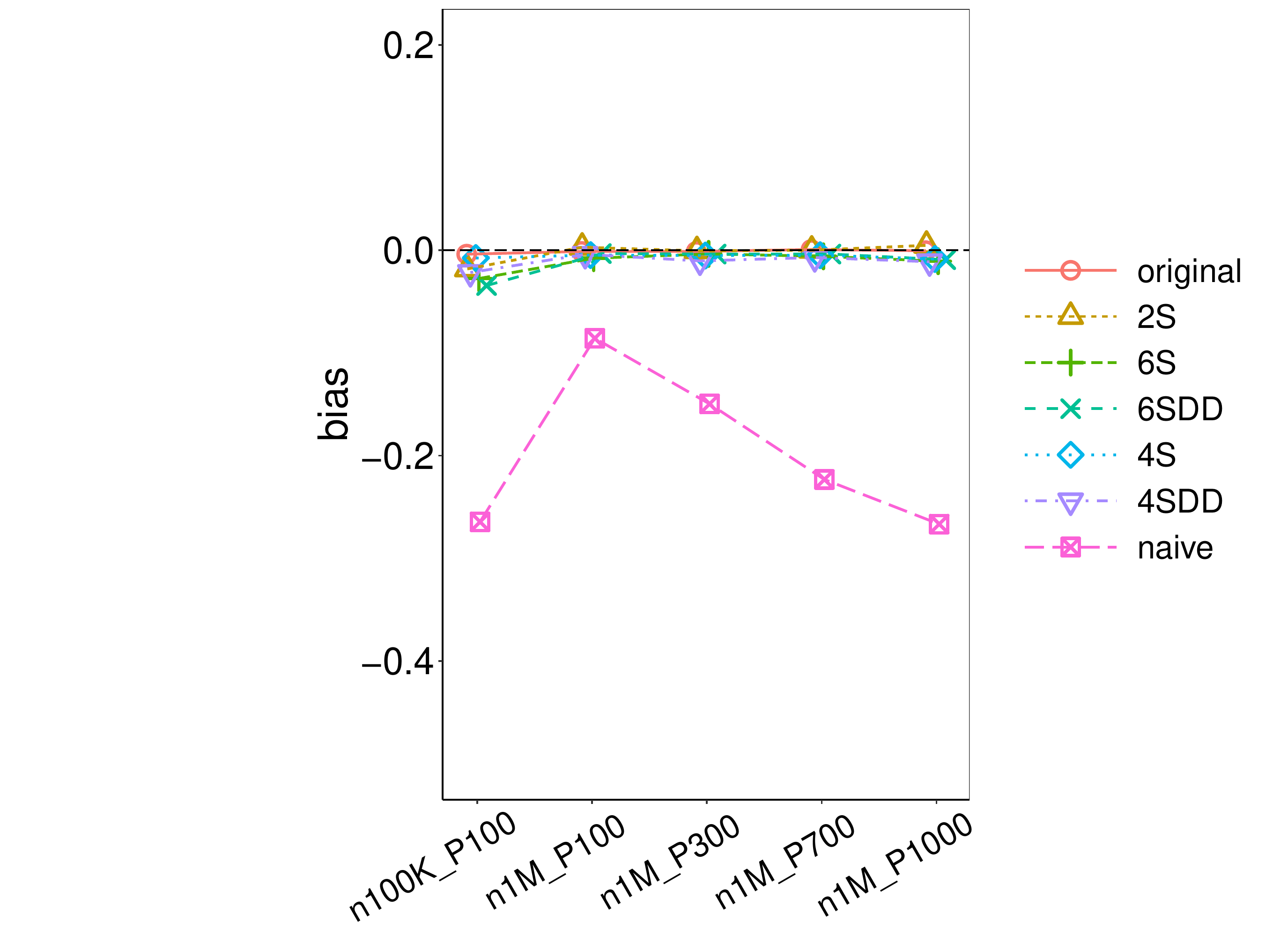}
\includegraphics[width=0.19\textwidth, trim={2.5in 0 2.6in 0},clip] {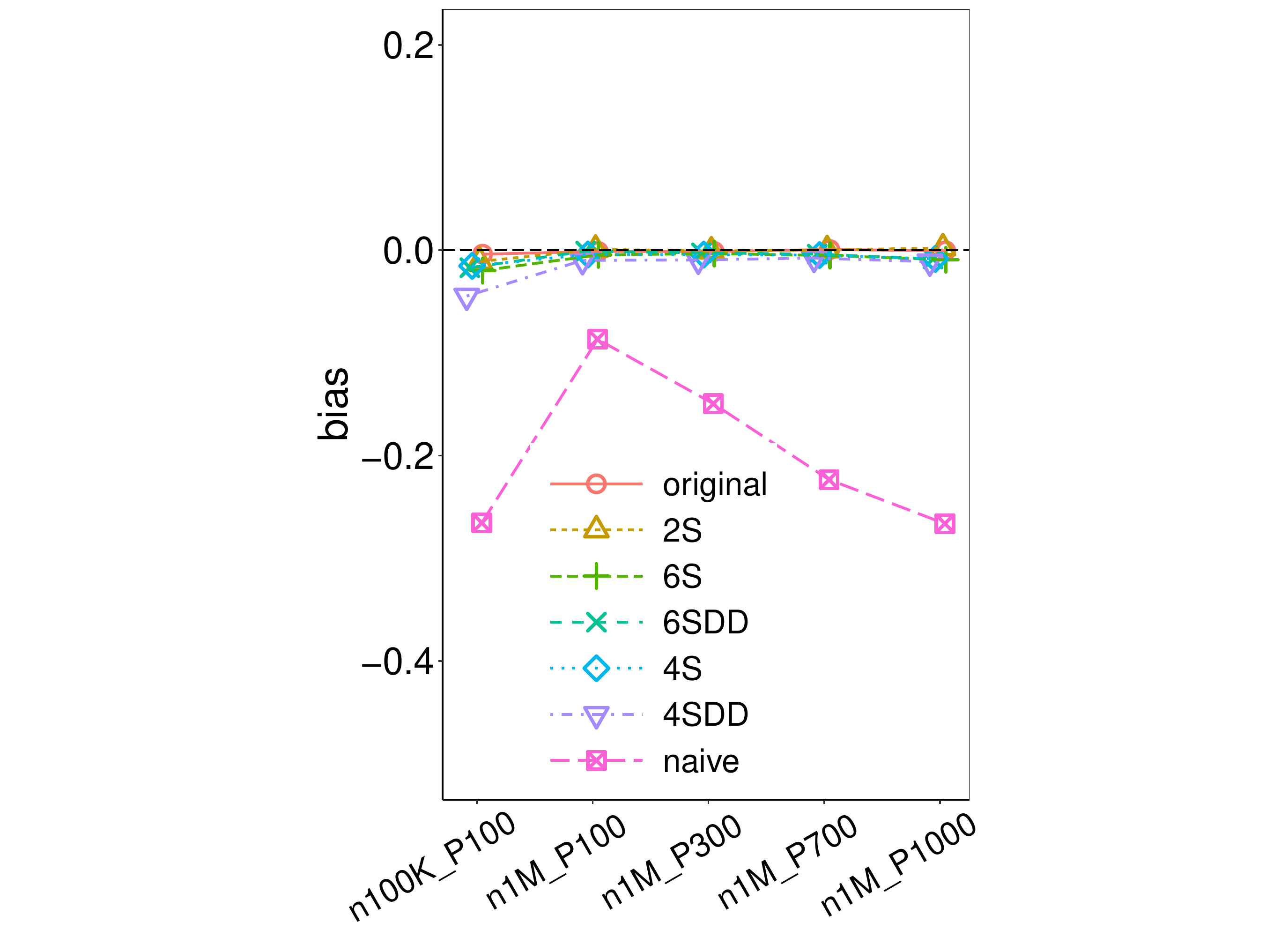}

\includegraphics[width=0.19\textwidth, trim={2.5in 0 2.6in 0},clip] {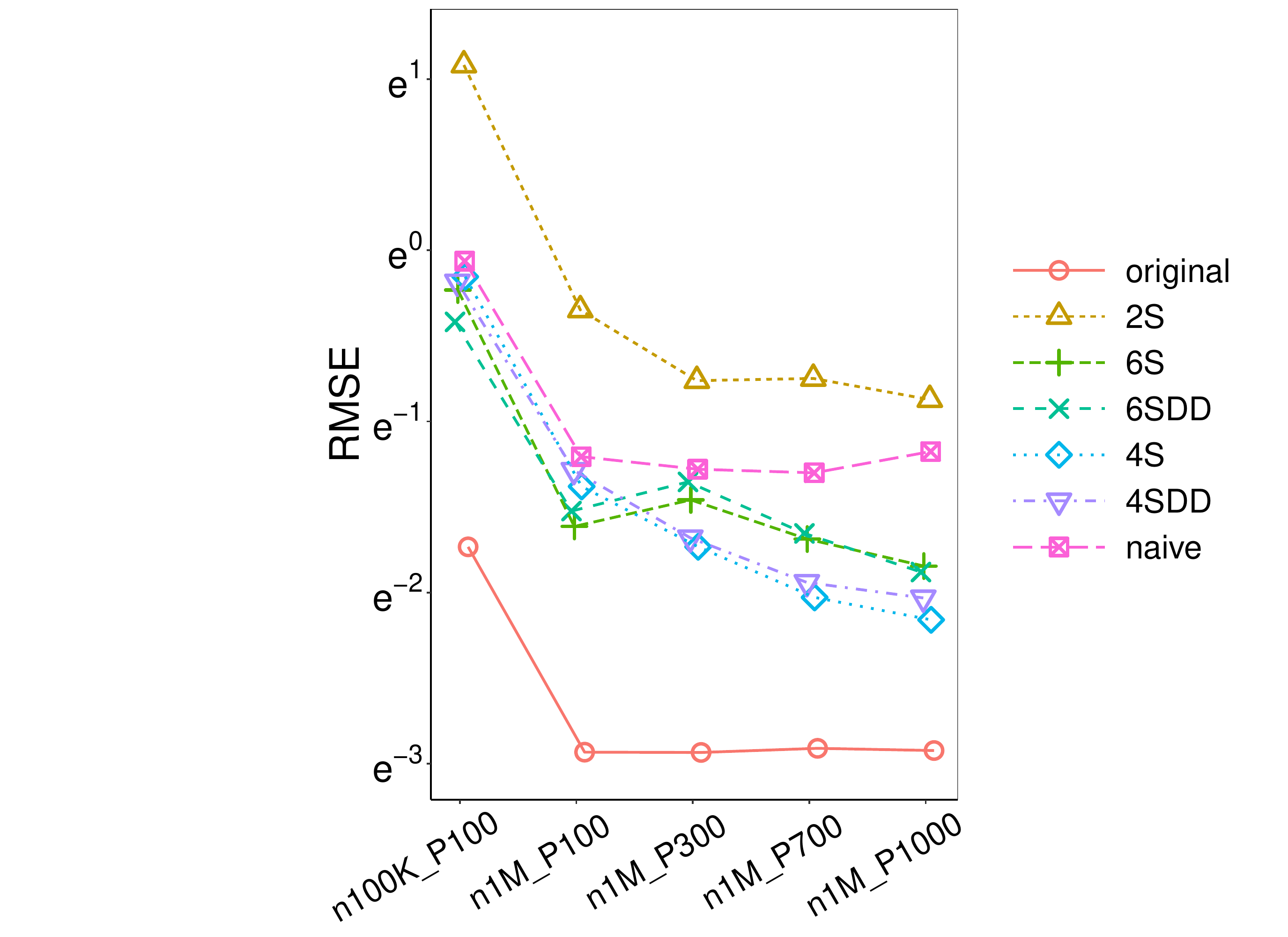}
\includegraphics[width=0.19\textwidth, trim={2.5in 0 2.6in 0},clip] {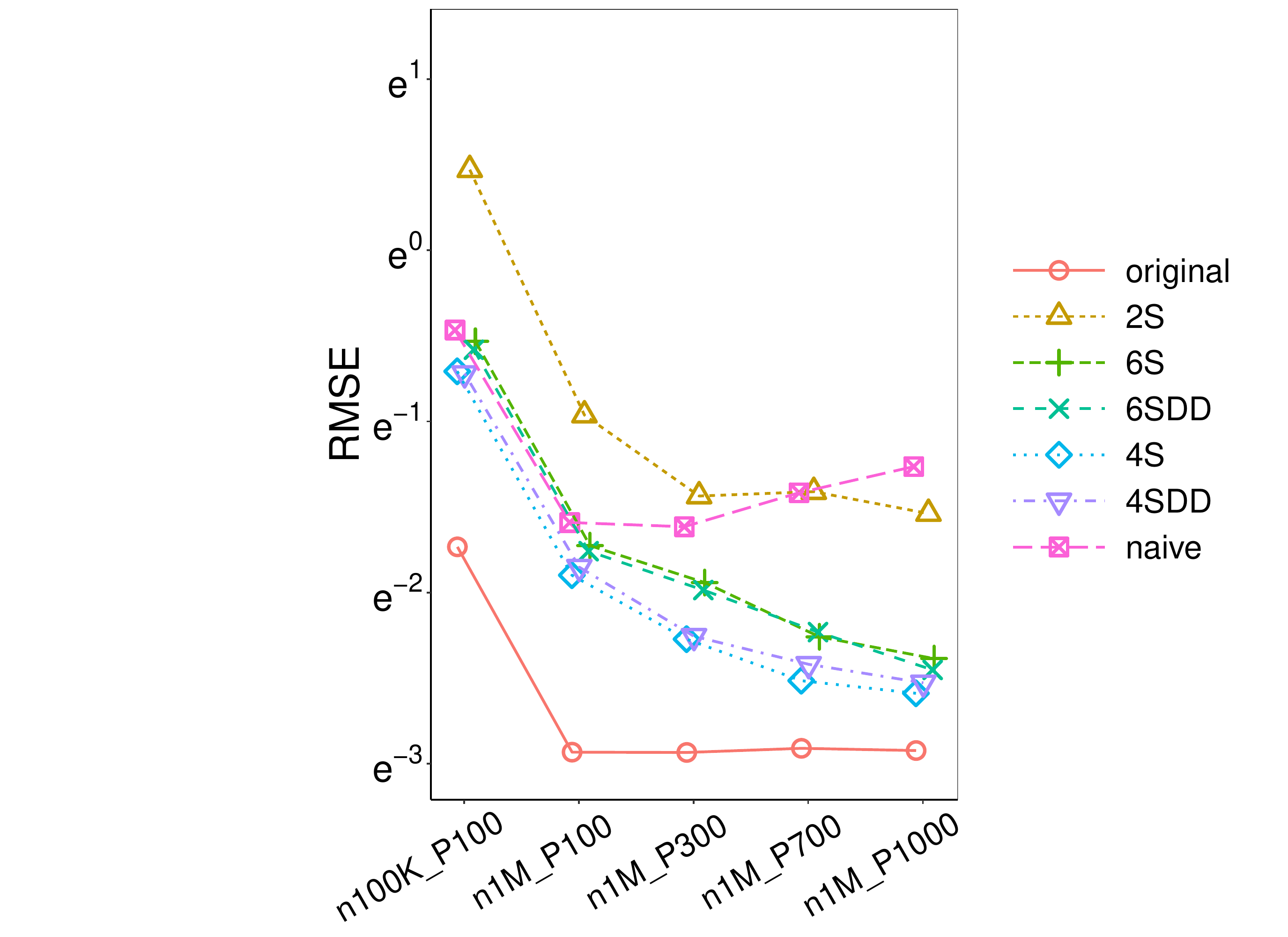}
\includegraphics[width=0.19\textwidth, trim={2.5in 0 2.6in 0},clip] {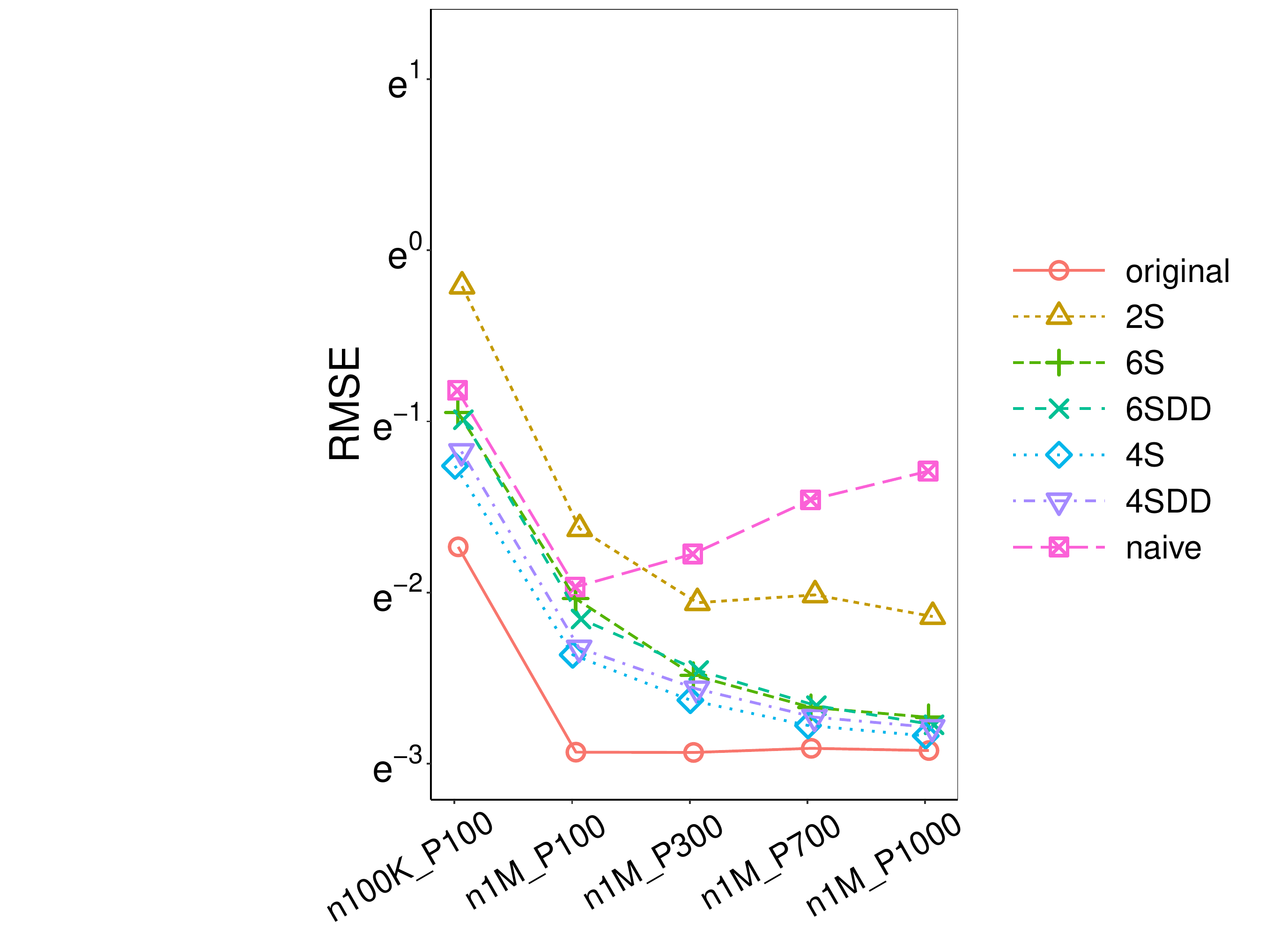}
\includegraphics[width=0.19\textwidth, trim={2.5in 0 2.6in 0},clip] {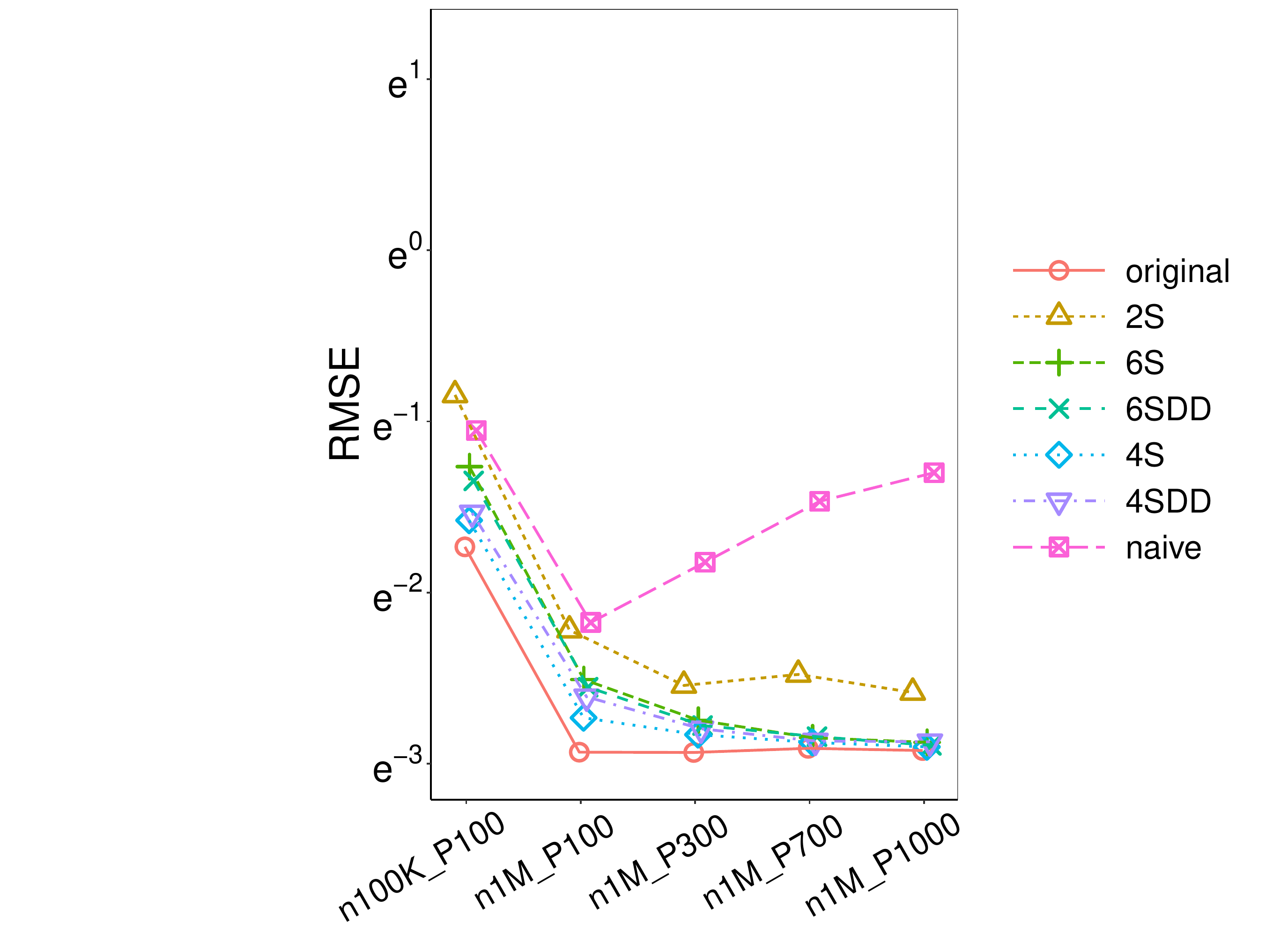}
\includegraphics[width=0.19\textwidth, trim={2.5in 0 2.6in 0},clip] {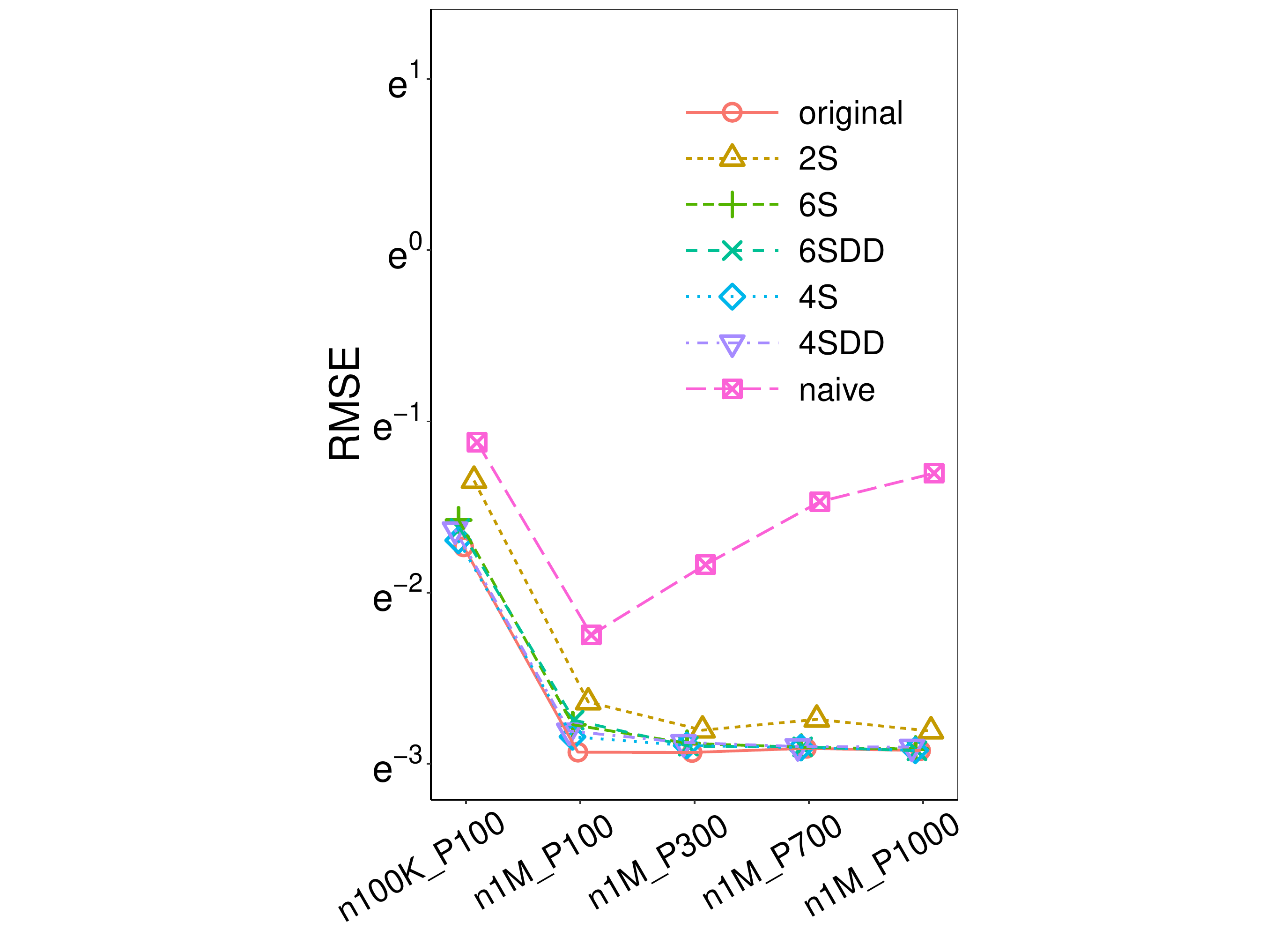}

\includegraphics[width=0.19\textwidth, trim={2.5in 0 2.6in 0},clip] {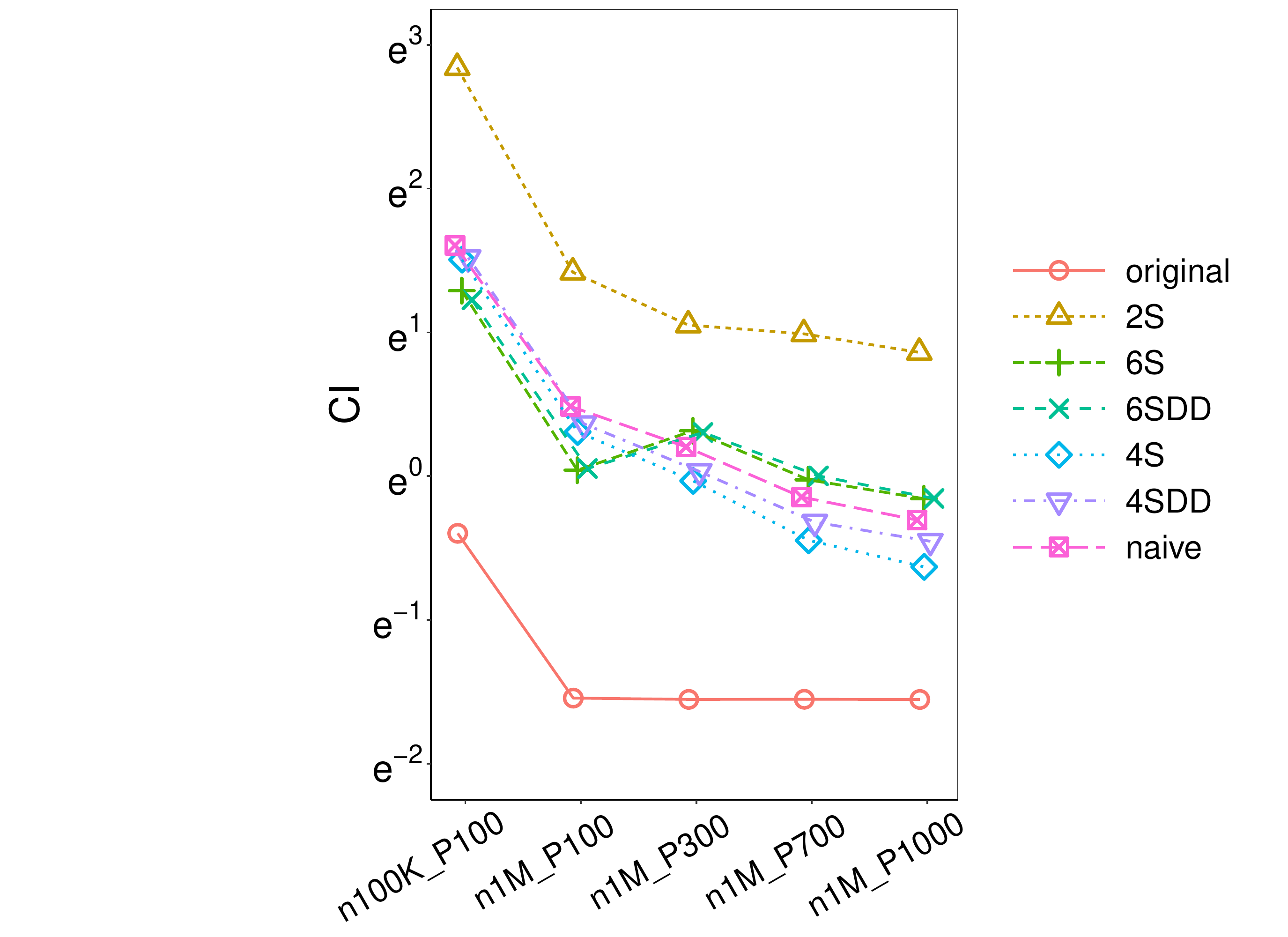}
\includegraphics[width=0.19\textwidth, trim={2.5in 0 2.6in 0},clip] {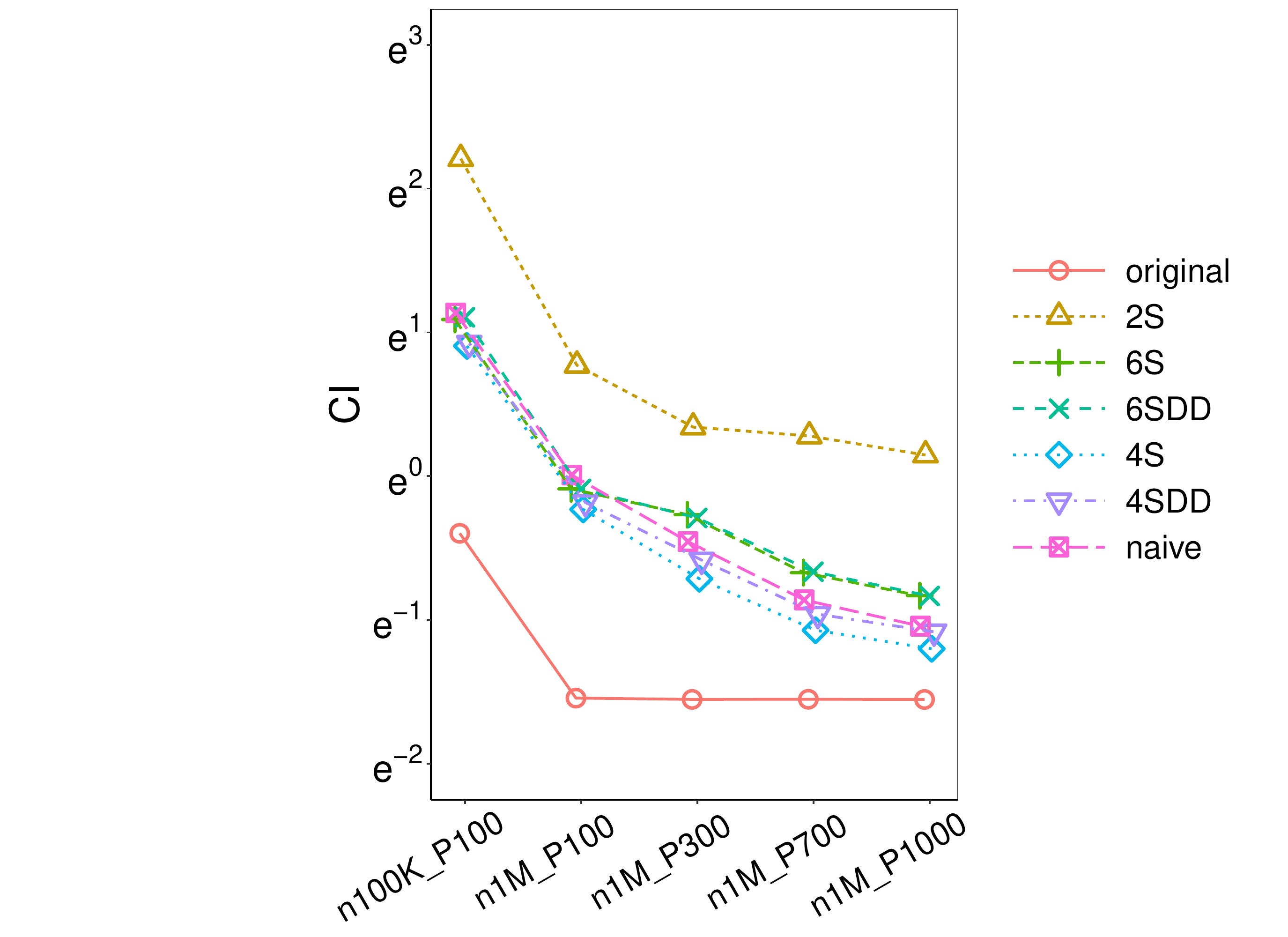}
\includegraphics[width=0.19\textwidth, trim={2.5in 0 2.6in 0},clip] {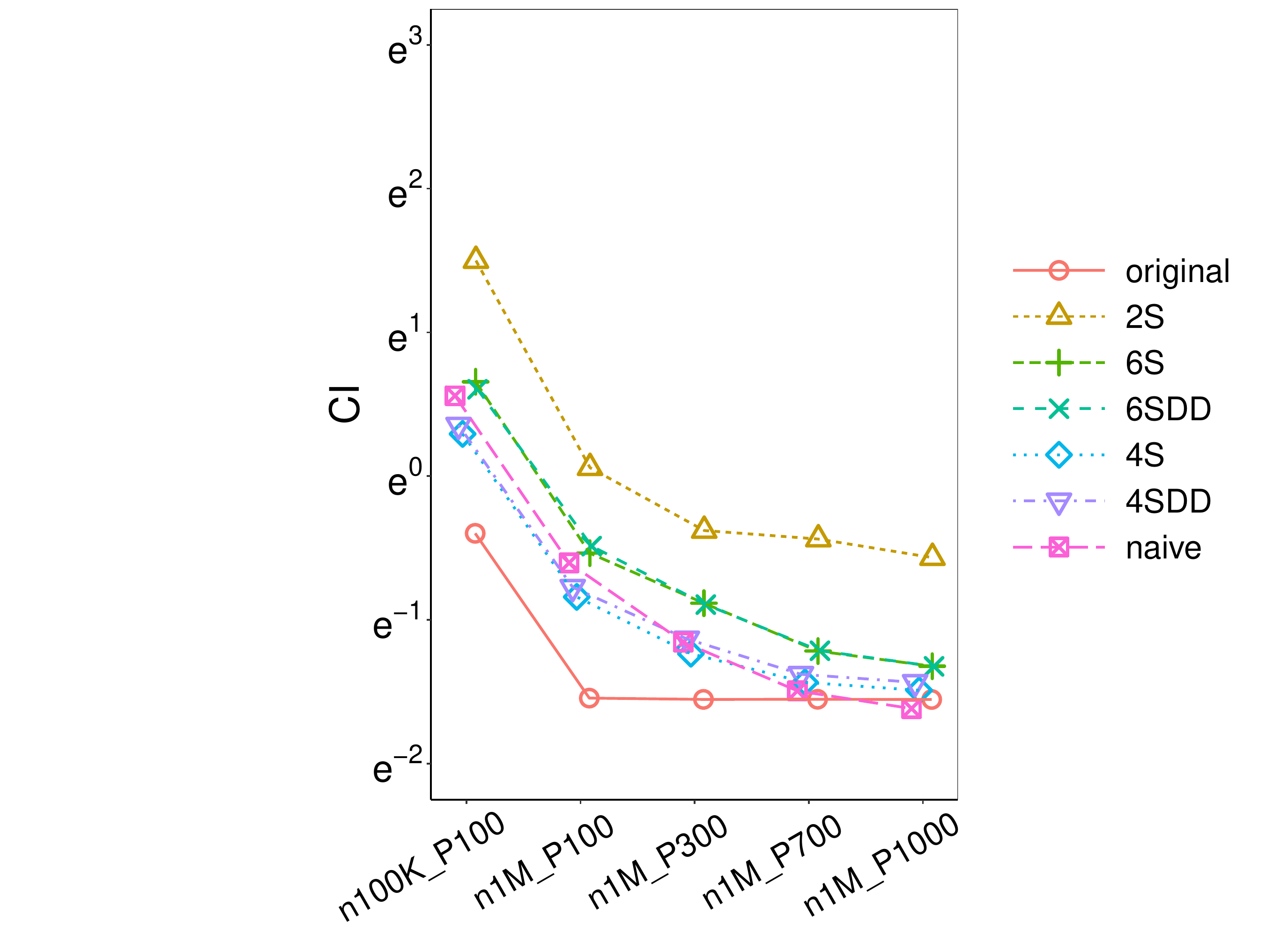}
\includegraphics[width=0.19\textwidth, trim={2.5in 0 2.6in 0},clip] {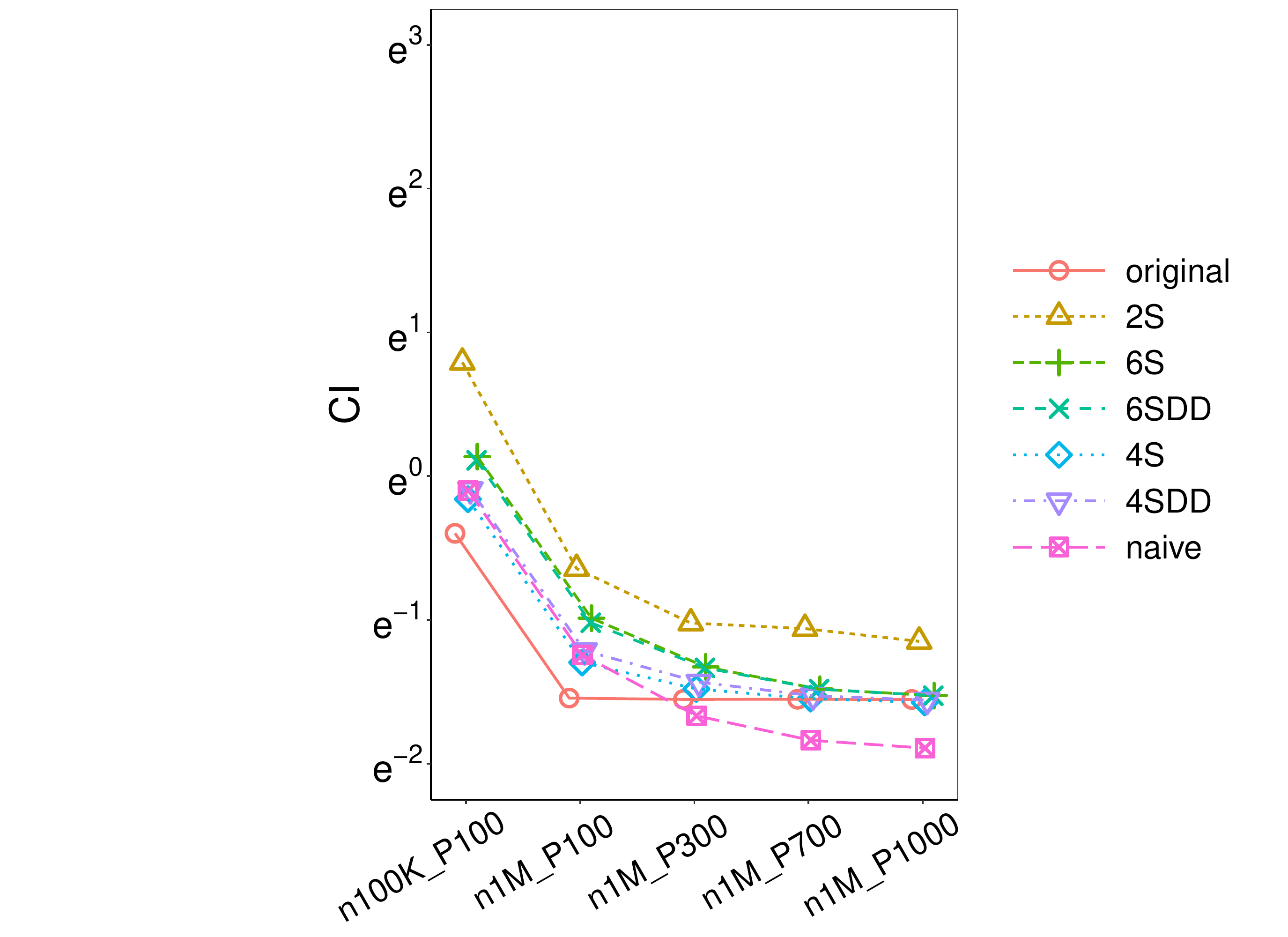}
\includegraphics[width=0.19\textwidth, trim={2.5in 0 2.6in 0},clip] {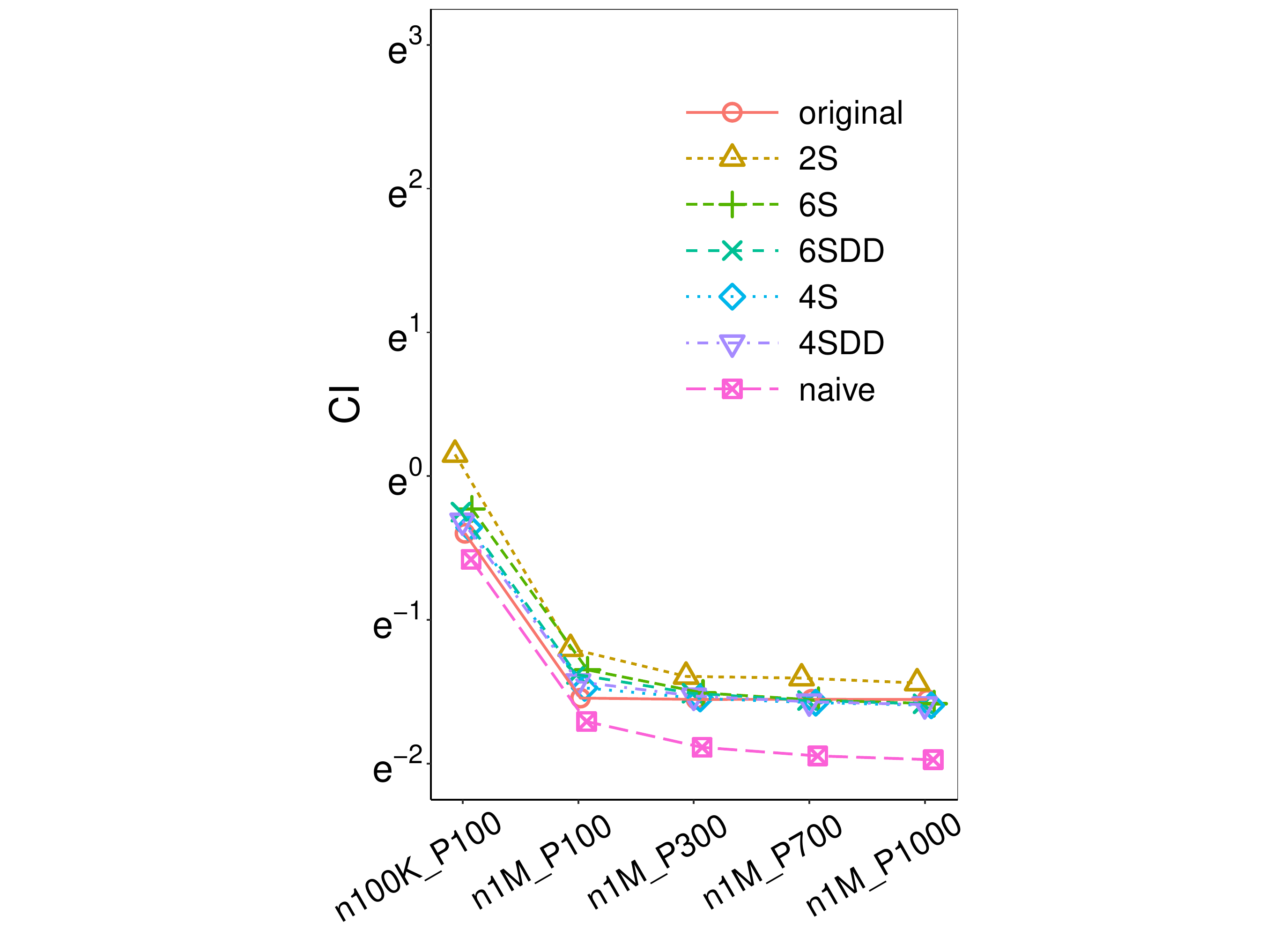}

\includegraphics[width=0.19\textwidth, trim={2.5in 0 2.6in 0},clip] {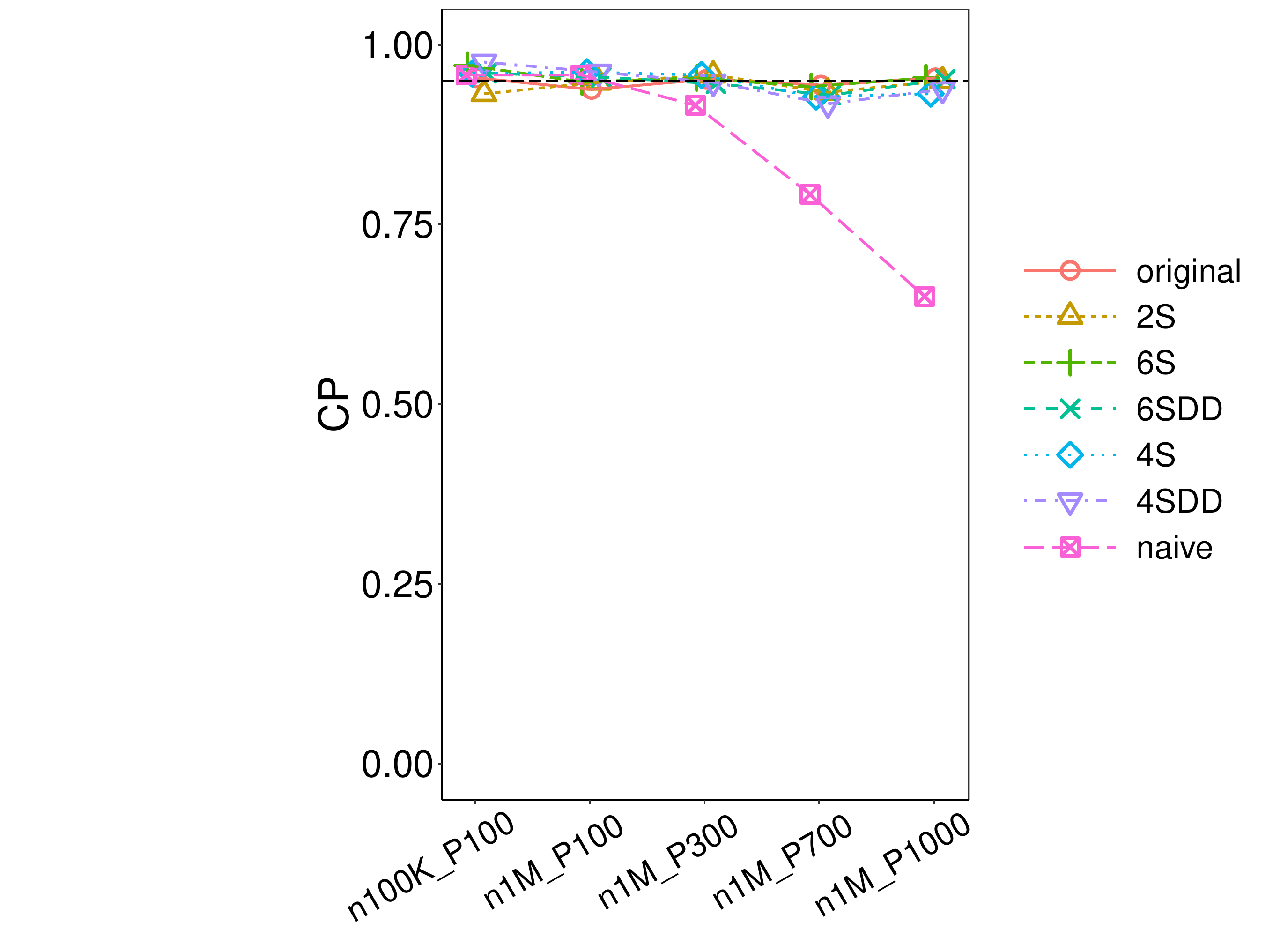}
\includegraphics[width=0.19\textwidth, trim={2.5in 0 2.6in 0},clip] {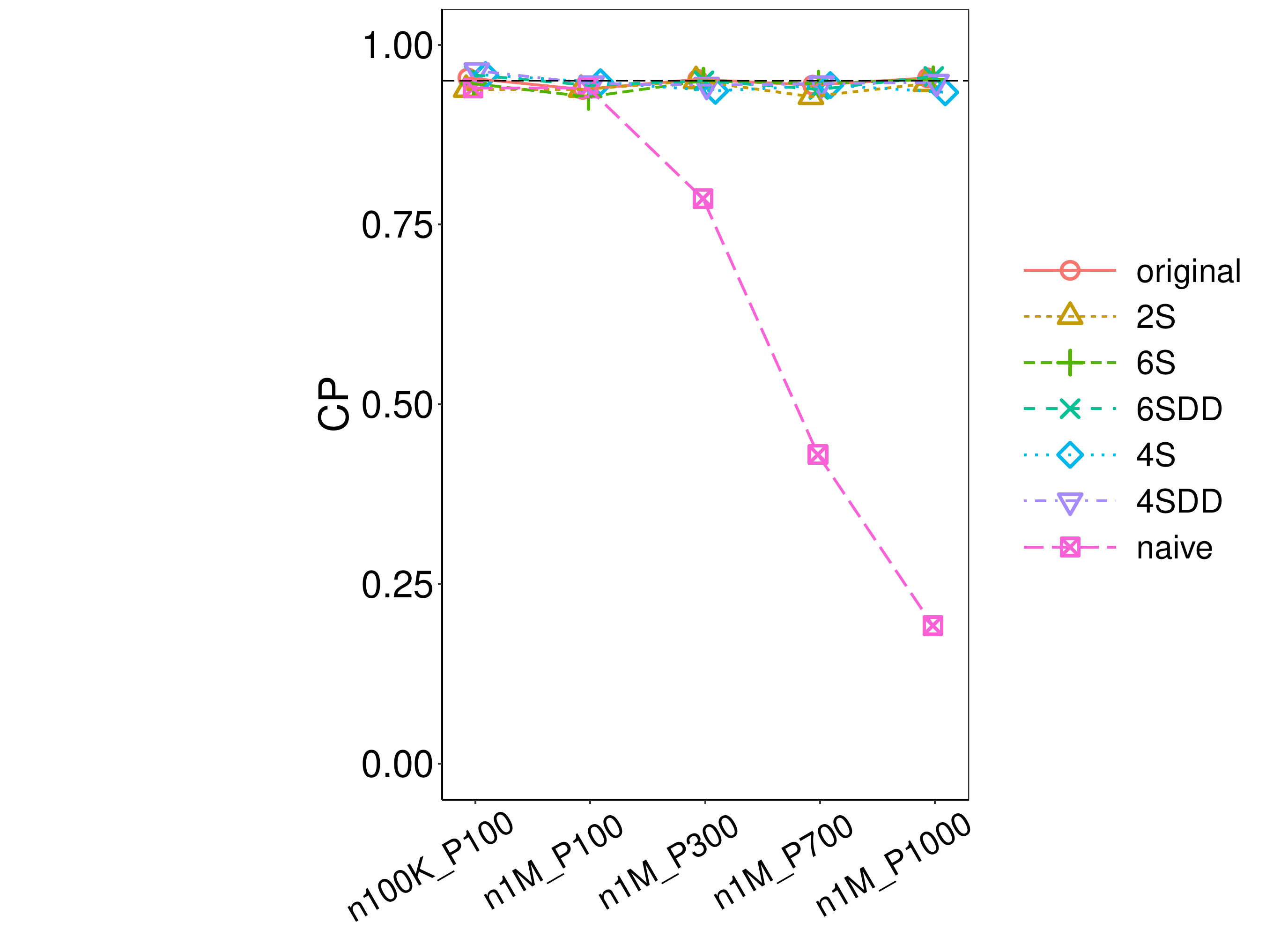}
\includegraphics[width=0.19\textwidth, trim={2.5in 0 2.6in 0},clip] {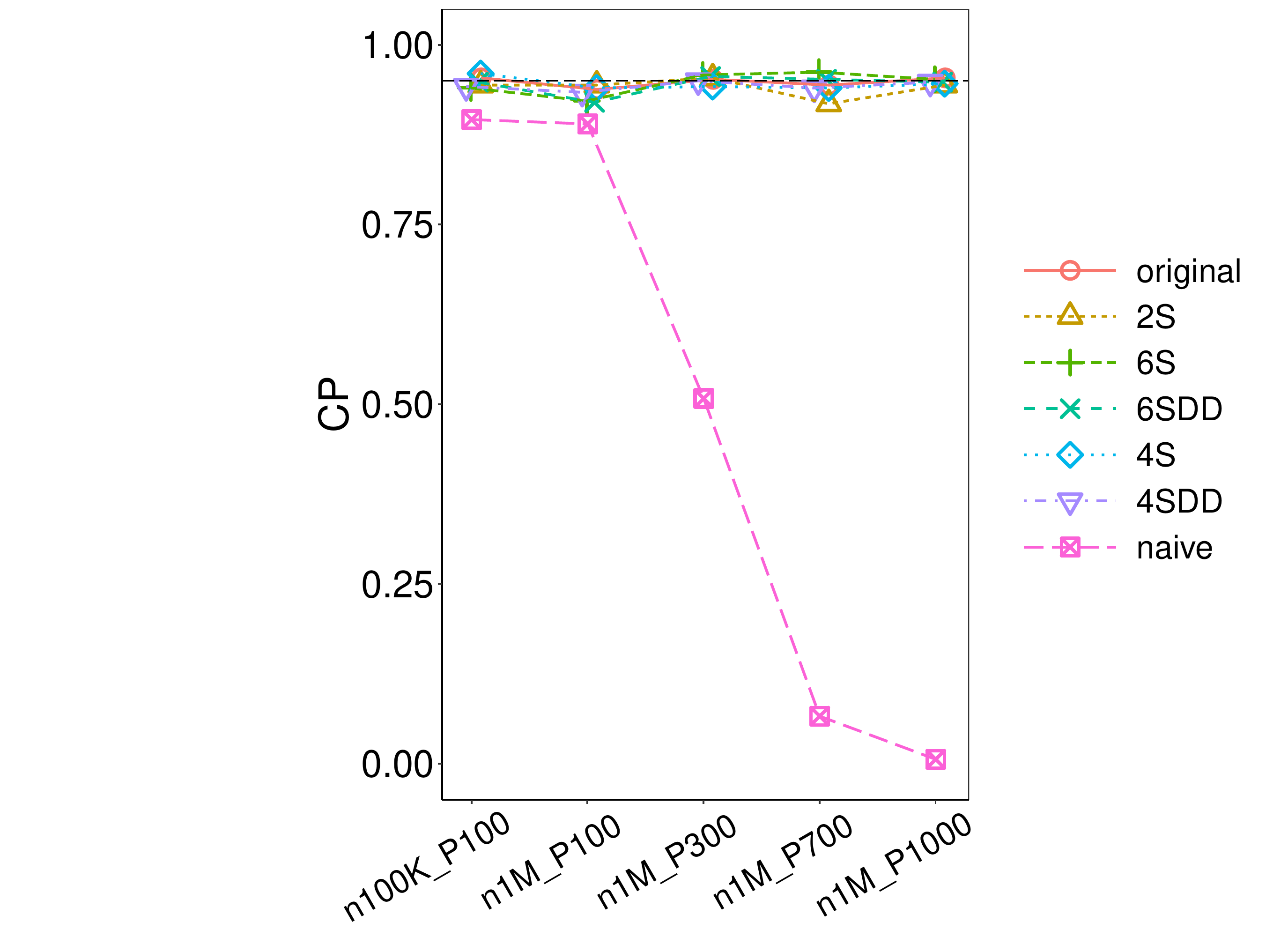}
\includegraphics[width=0.19\textwidth, trim={2.5in 0 2.6in 0},clip] {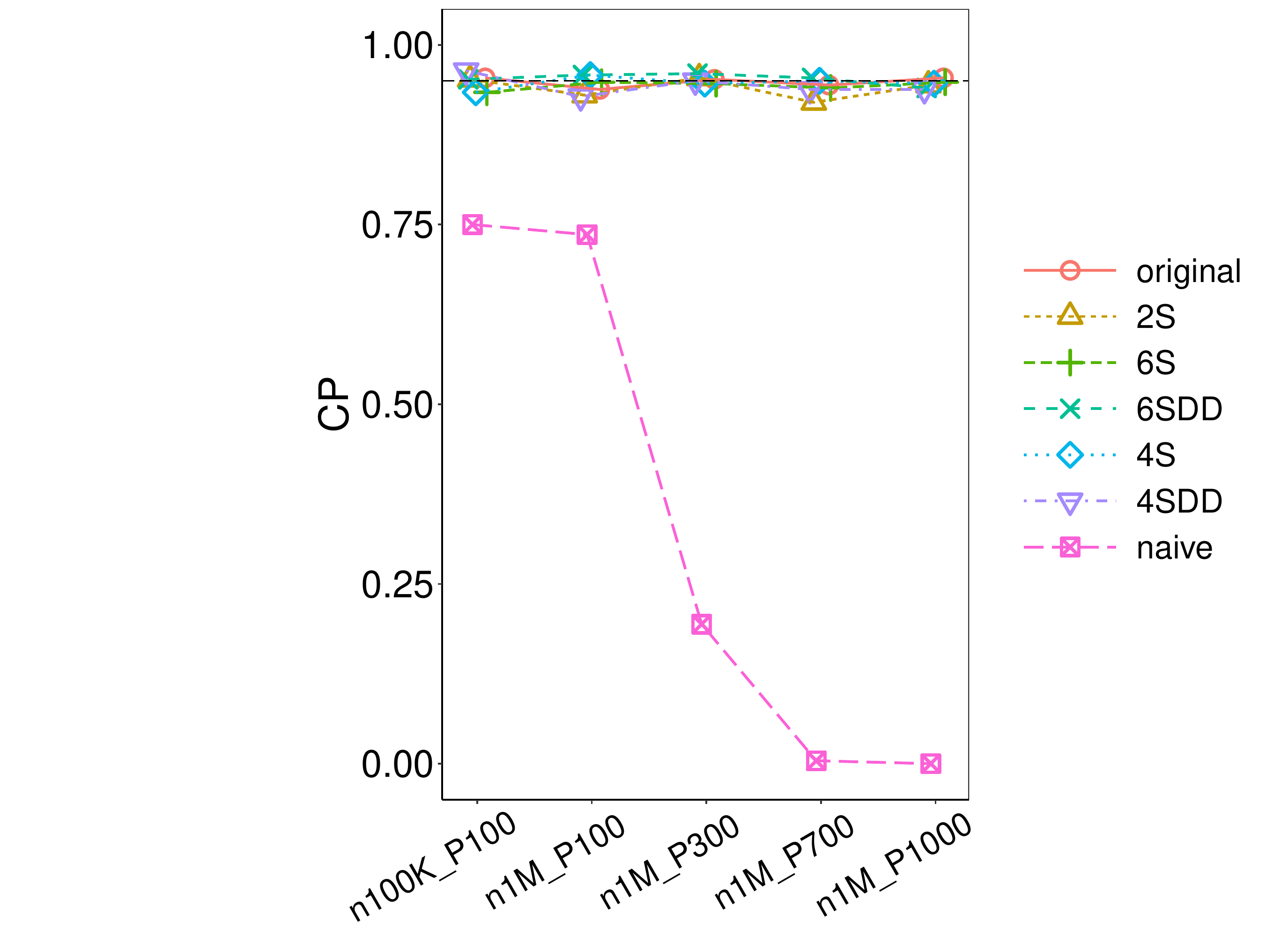}
\includegraphics[width=0.19\textwidth, trim={2.5in 0 2.6in 0},clip] {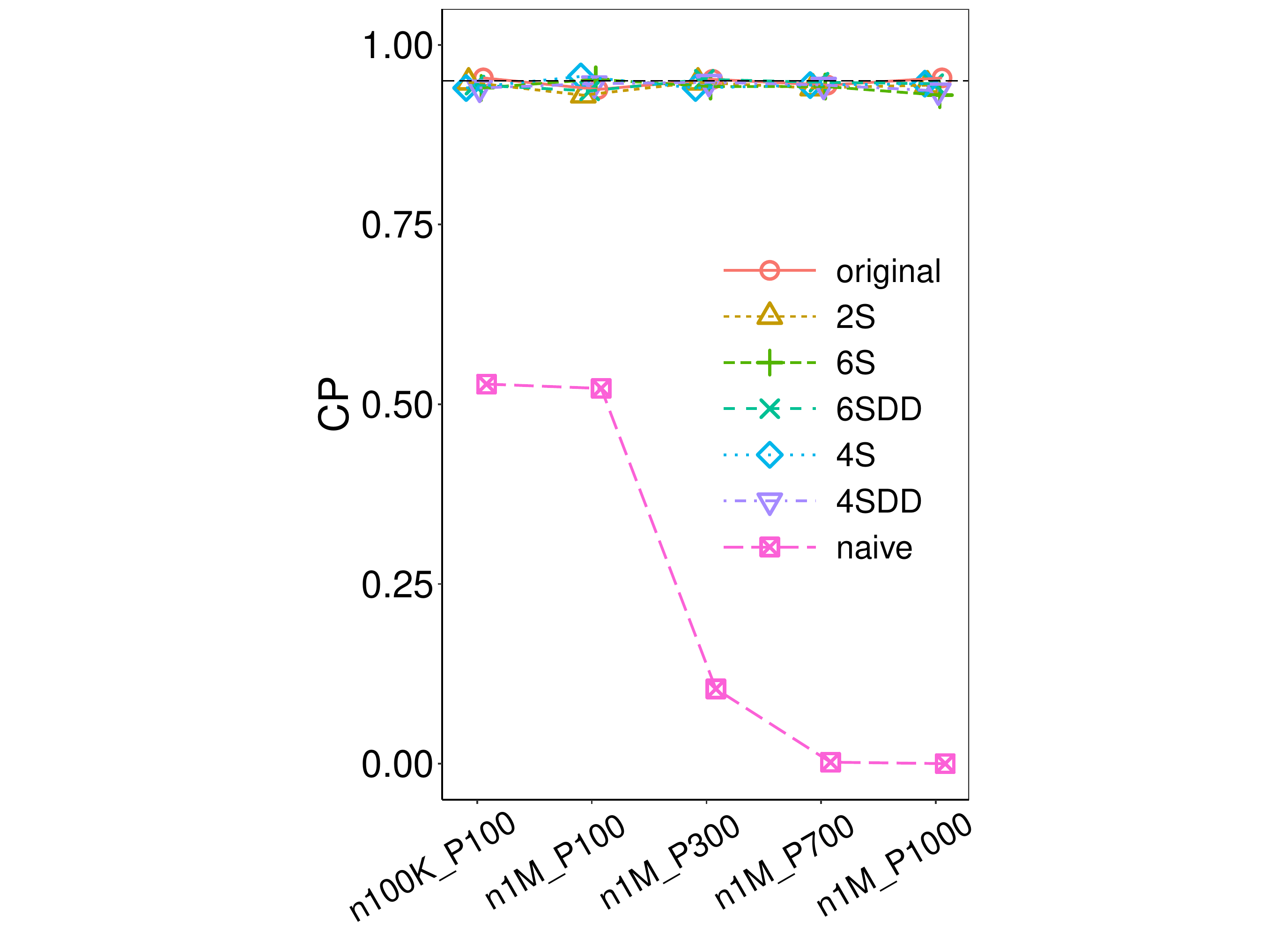}

\caption{Simulation results with $\rho$-zCDP for ZILN data with  $\alpha\ne\beta$ when $\theta=0$}
\label{fig:0aszCDPZILN}
\end{figure}

The results are presented in Figures \ref{fig:0asDPZILN} and \ref{fig:0aszCDPZILN} and are summarized  as follows. 4S is the best performer by all metrics and the na\"{i}ve method is the worst. In between the two,  the bias for 6SDD approaches 0 faster  than 6S, 2S, and 4SDD; as $n$ or $P$ increases when $n=1m$ or when $\epsilon$ increases. In terms of RMSE and CI width, 4S performs the best, followed closely by 4SDD, with the na\"{i}ve method the worst overall; all methods, except for the  na\"{i}ve method, provide nominal level CP. The Gaussian mechanism of $\rho$-zCDP does not seem to be advantageous in statistical utility compared to the Laplace mechanism of $\epsilon$-DP.

\subsubsection{\texorpdfstring{$Y\sim$ ZILN}{}, \texorpdfstring{$\theta\ne0$ ($H_1$)}{H1 }, and \texorpdfstring{$\alpha\ne\beta$}{AS}}\label{sec:1asZILN}
\begin{figure}[!htb]
\hspace{0.5in}$\epsilon=0.5$\hspace{0.8in}$\epsilon=1$\hspace{0.9in}$\epsilon=2$
\hspace{0.95in}$\epsilon=5$\hspace{0.9in}$\epsilon=50$

\includegraphics[width=0.19\textwidth, trim={2.5in 0 2.6in 0},clip] {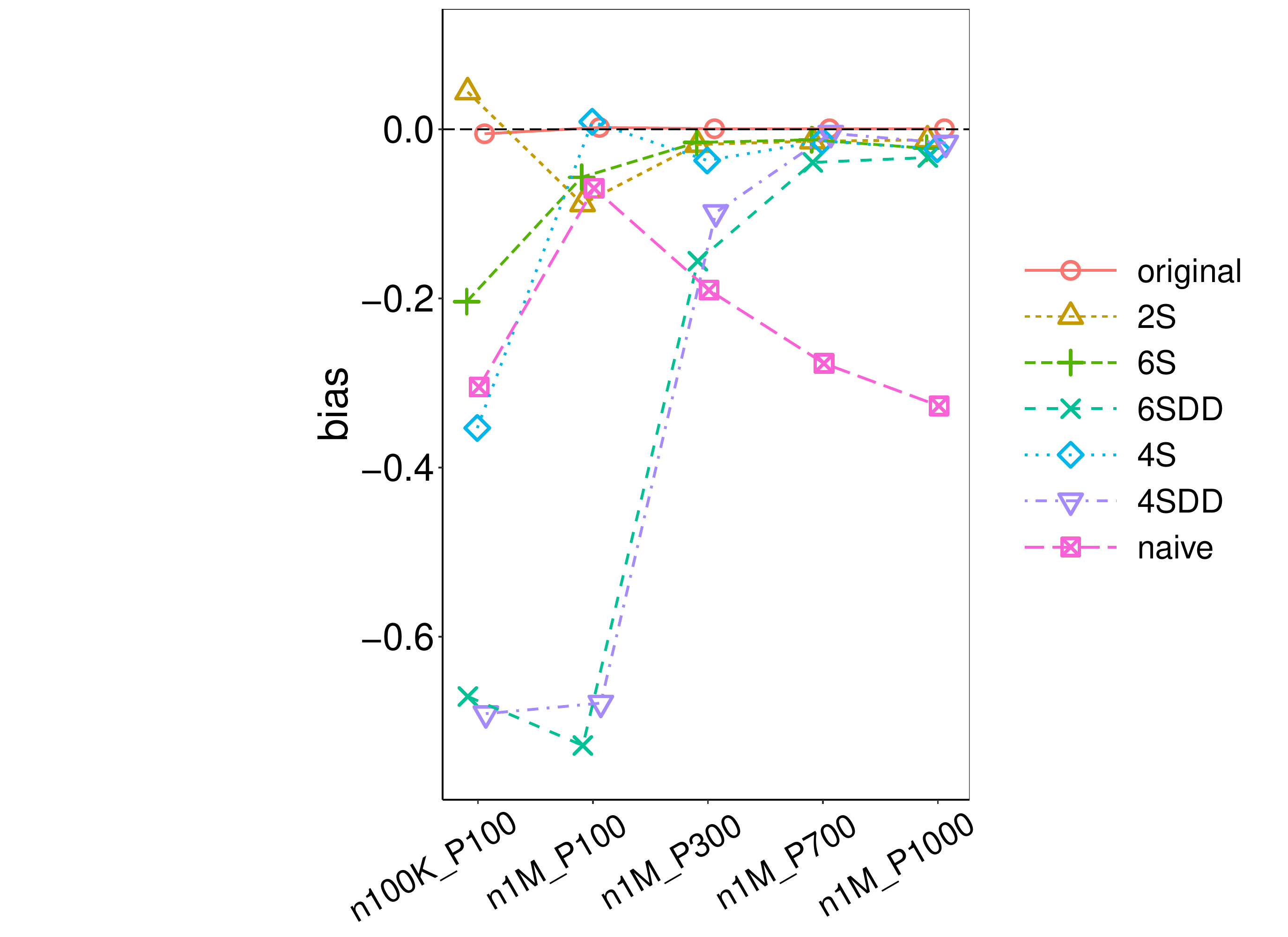}
\includegraphics[width=0.19\textwidth, trim={2.5in 0 2.6in 0},clip] {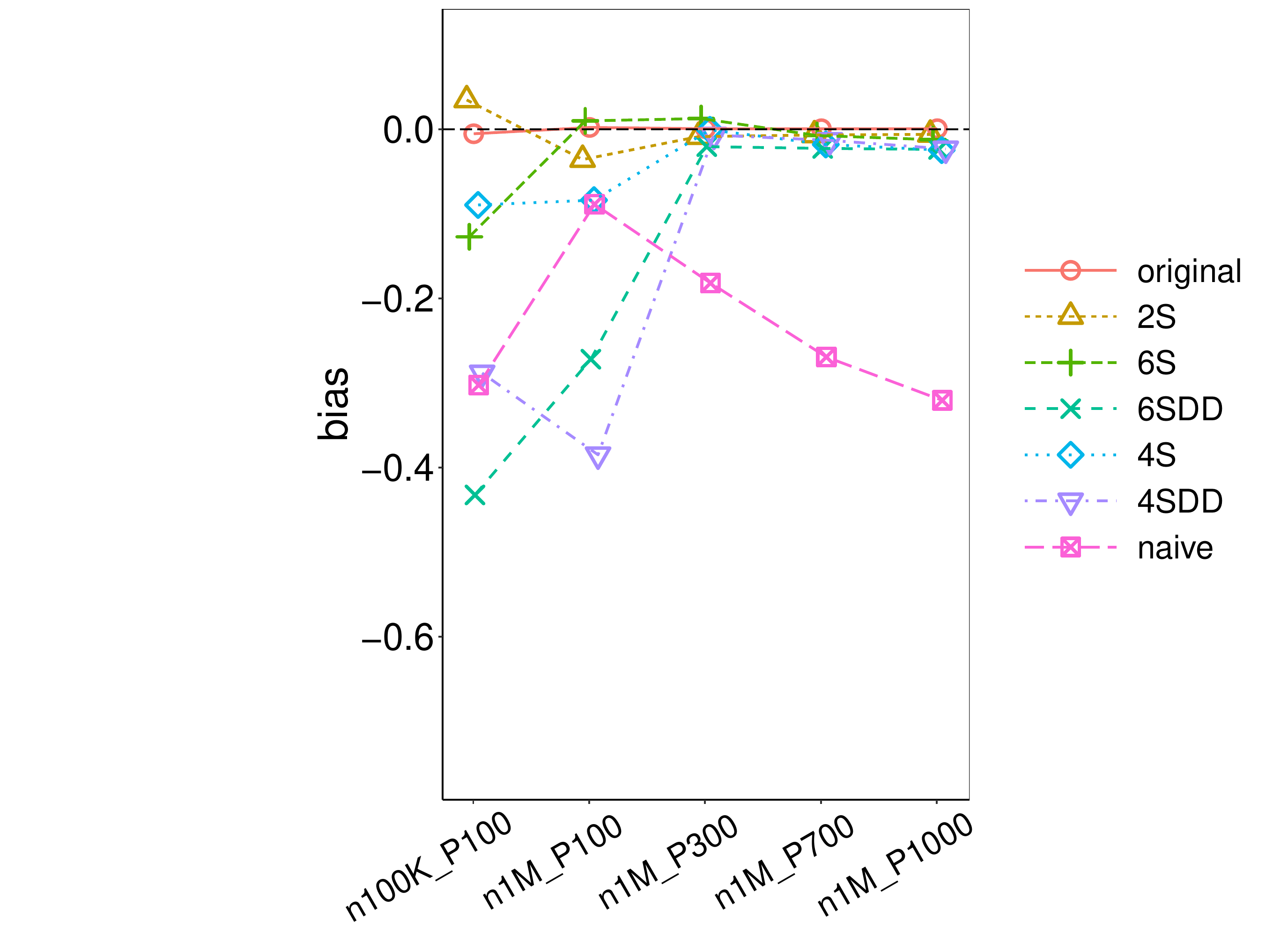}
\includegraphics[width=0.19\textwidth, trim={2.5in 0 2.6in 0},clip] {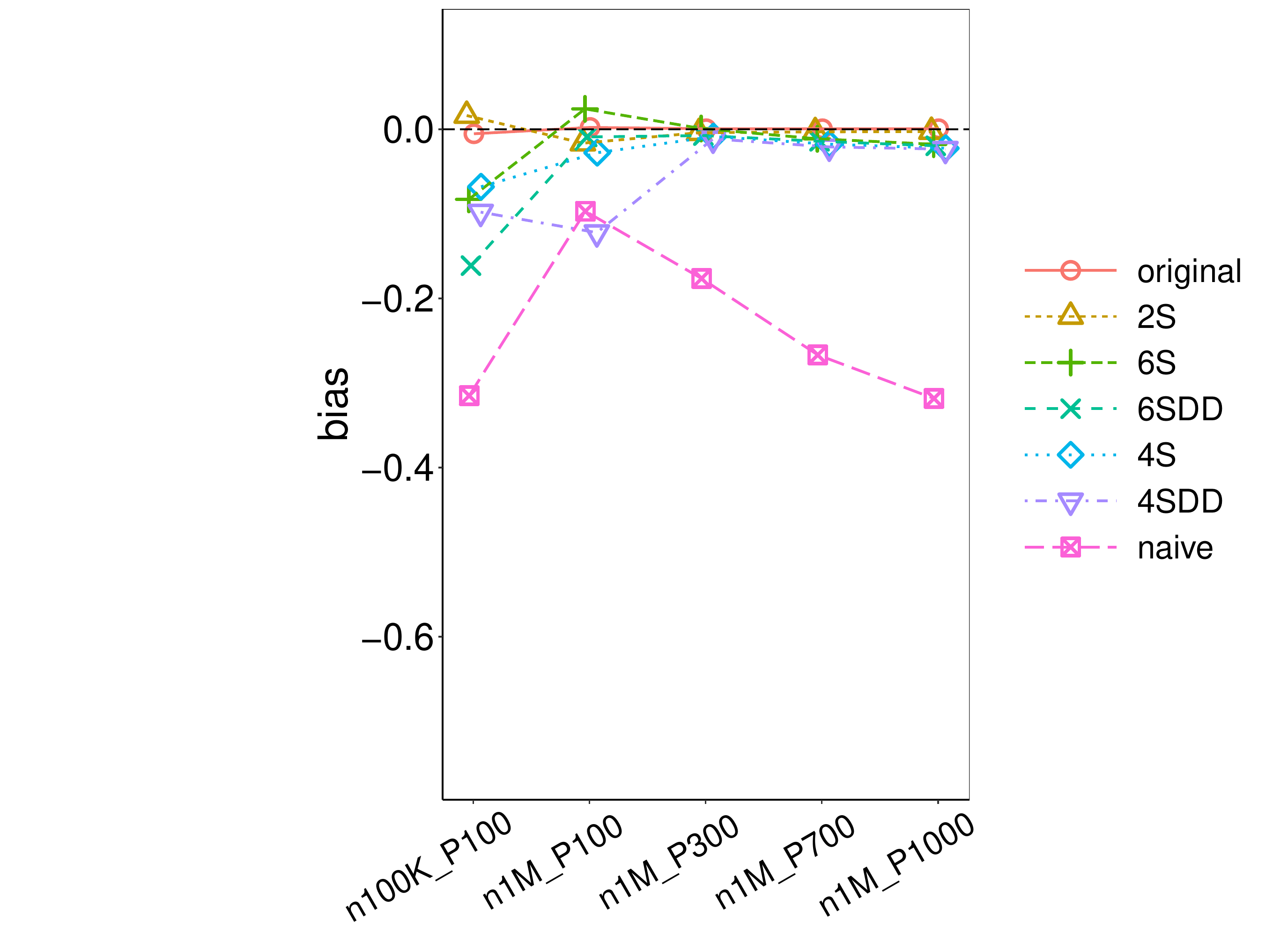}
\includegraphics[width=0.19\textwidth, trim={2.5in 0 2.6in 0},clip] {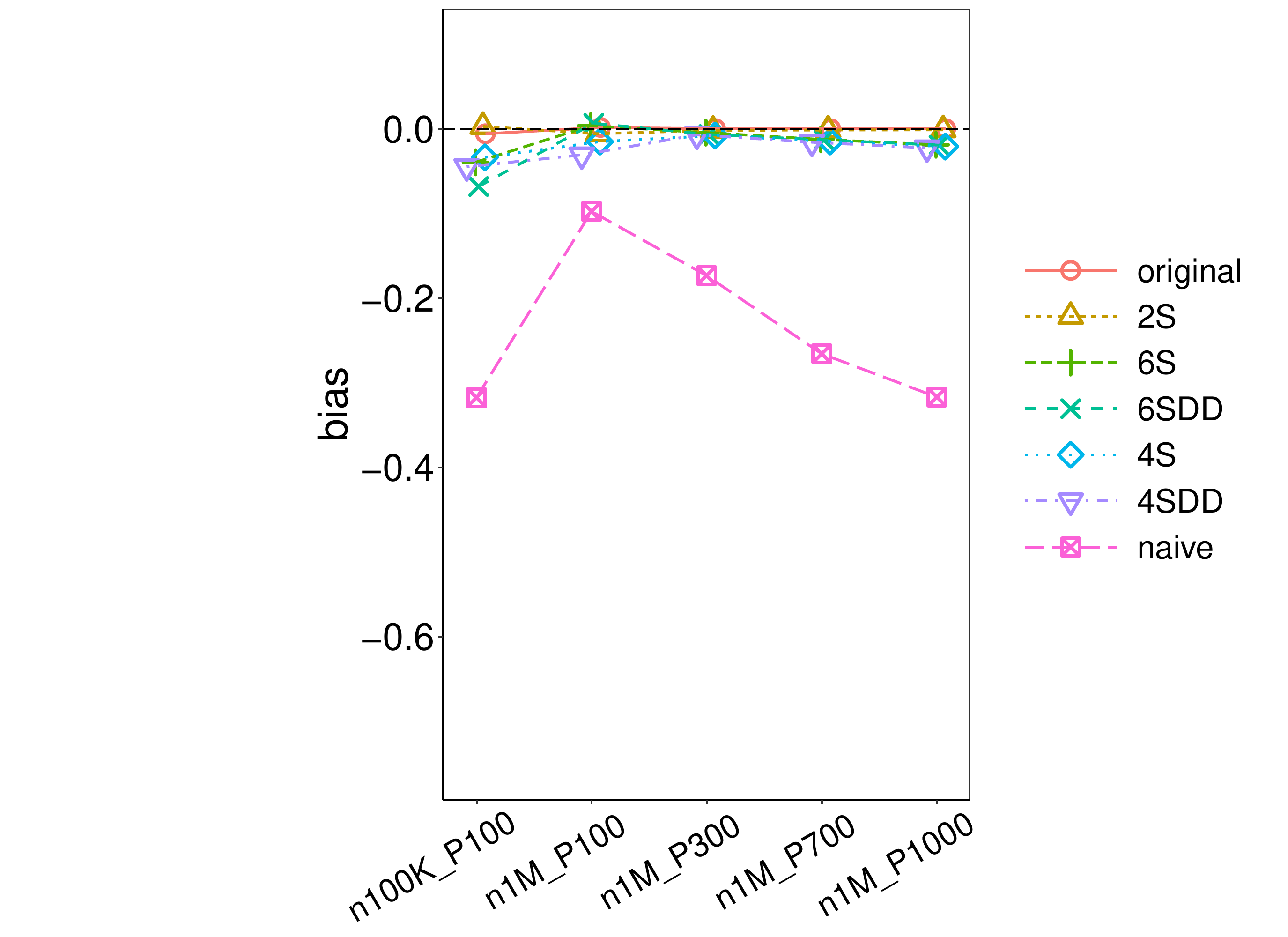}
\includegraphics[width=0.19\textwidth, trim={2.5in 0 2.6in 0},clip] {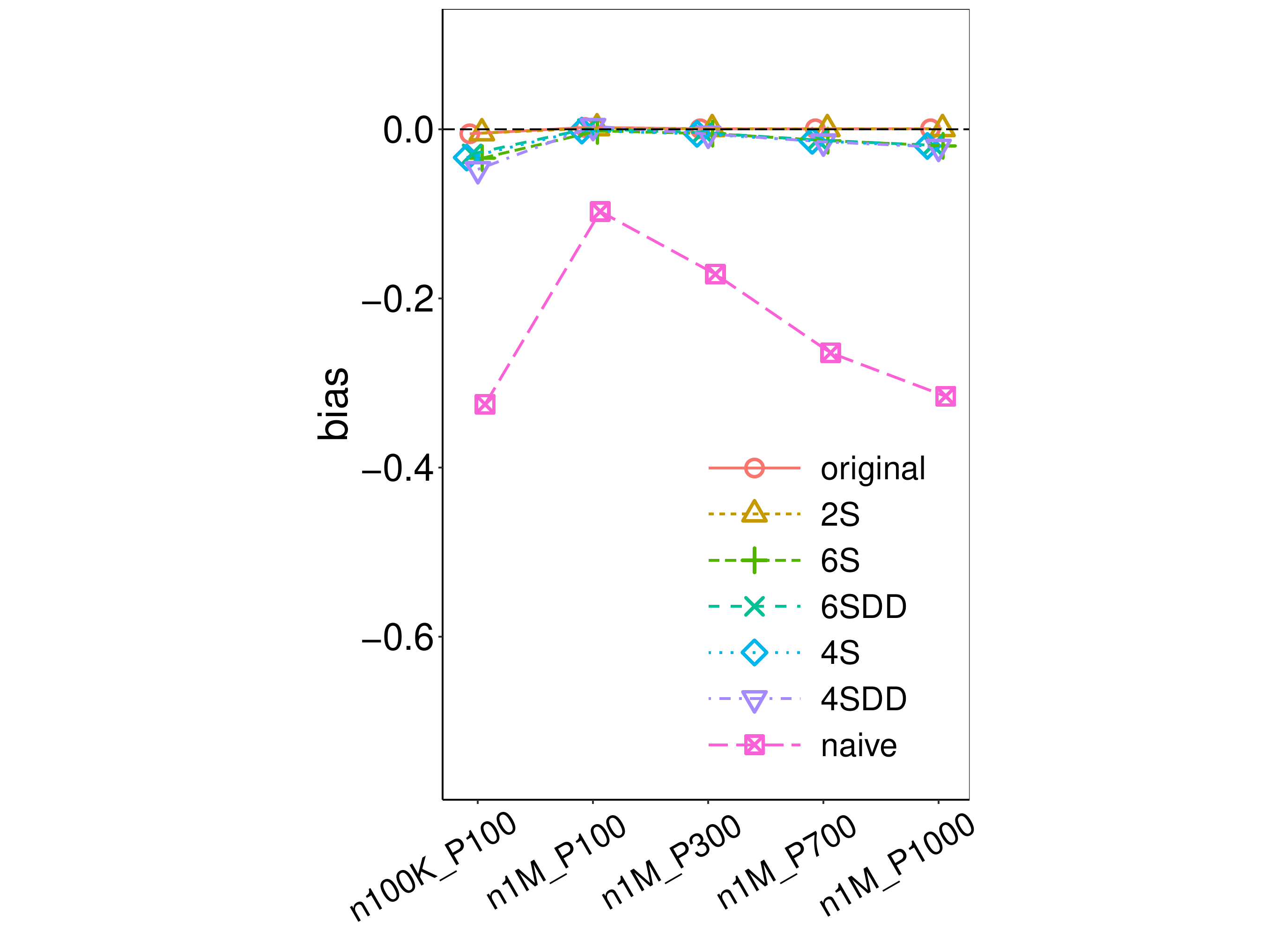}

\includegraphics[width=0.19\textwidth, trim={2.5in 0 2.6in 0},clip] {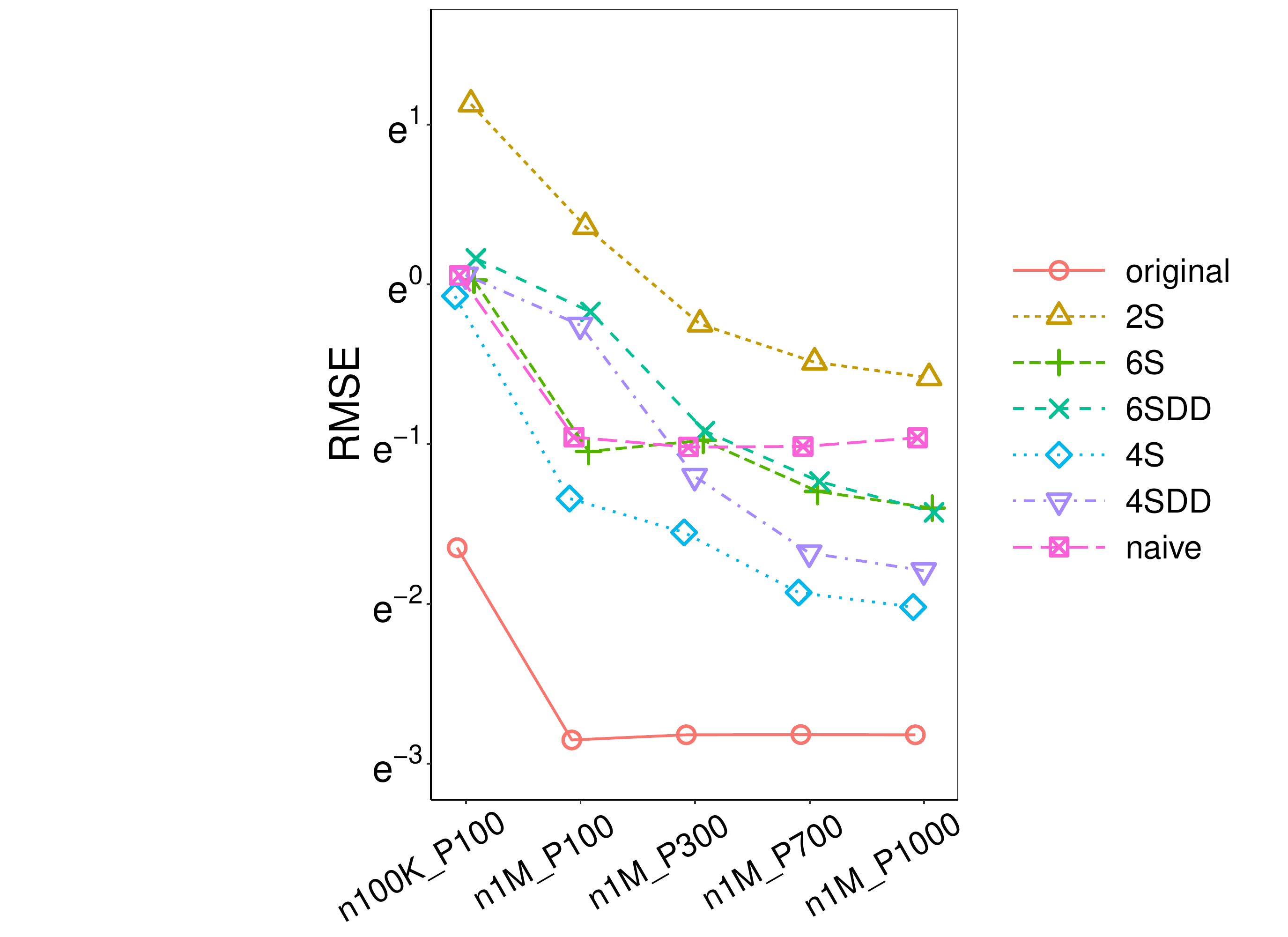}
\includegraphics[width=0.19\textwidth, trim={2.5in 0 2.6in 0},clip] {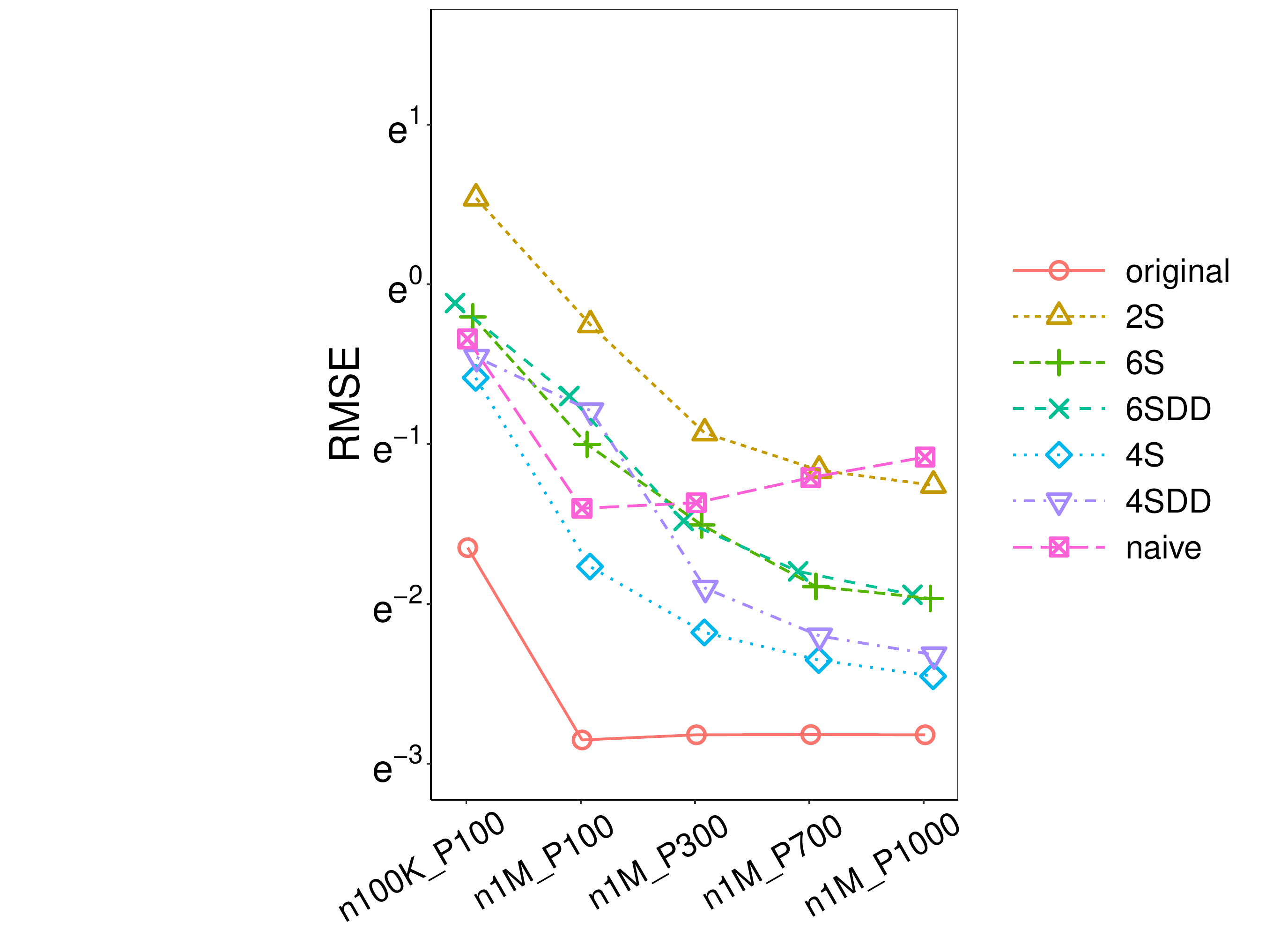}
\includegraphics[width=0.19\textwidth, trim={2.5in 0 2.6in 0},clip] {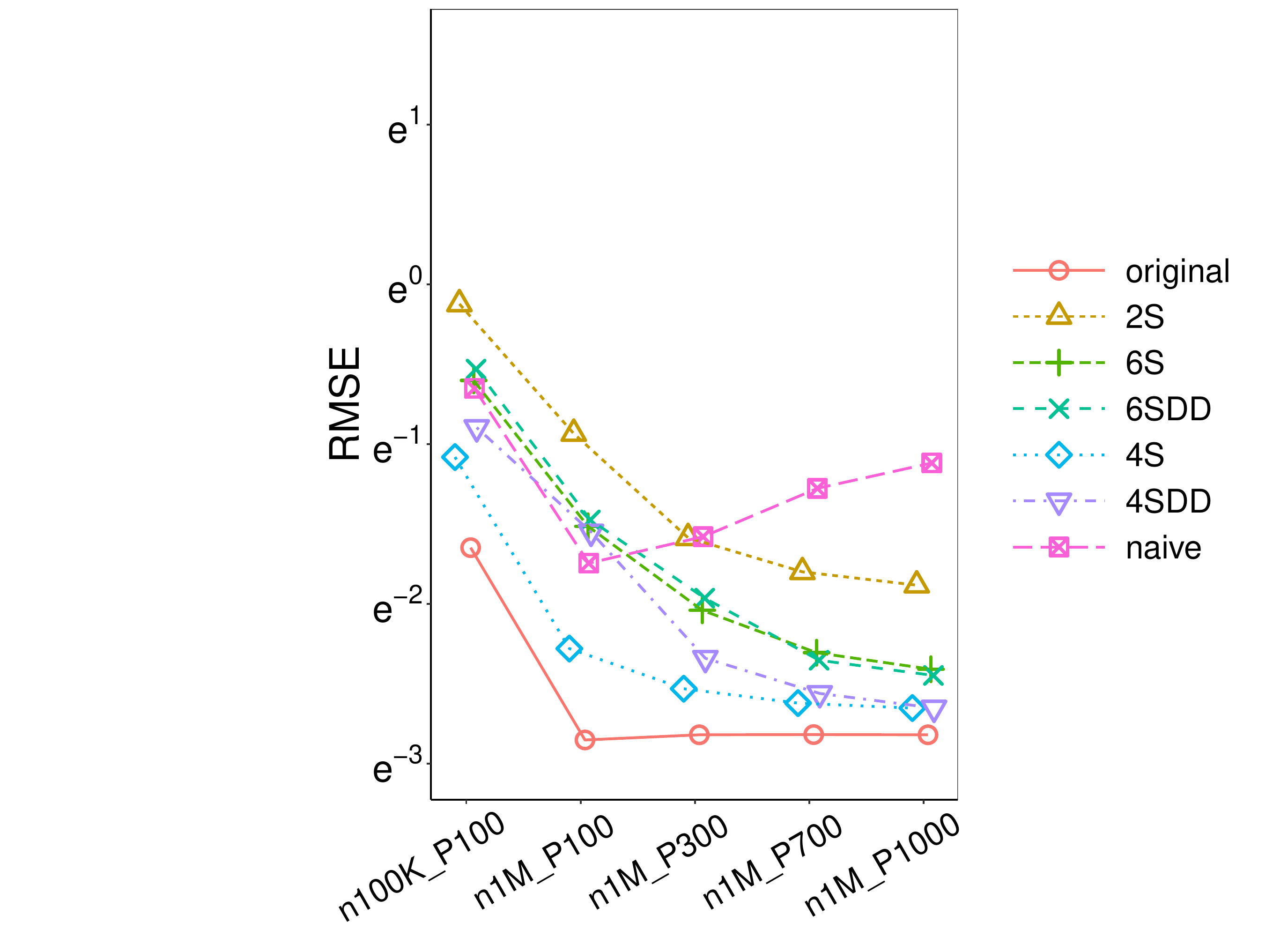}
\includegraphics[width=0.19\textwidth, trim={2.5in 0 2.6in 0},clip] {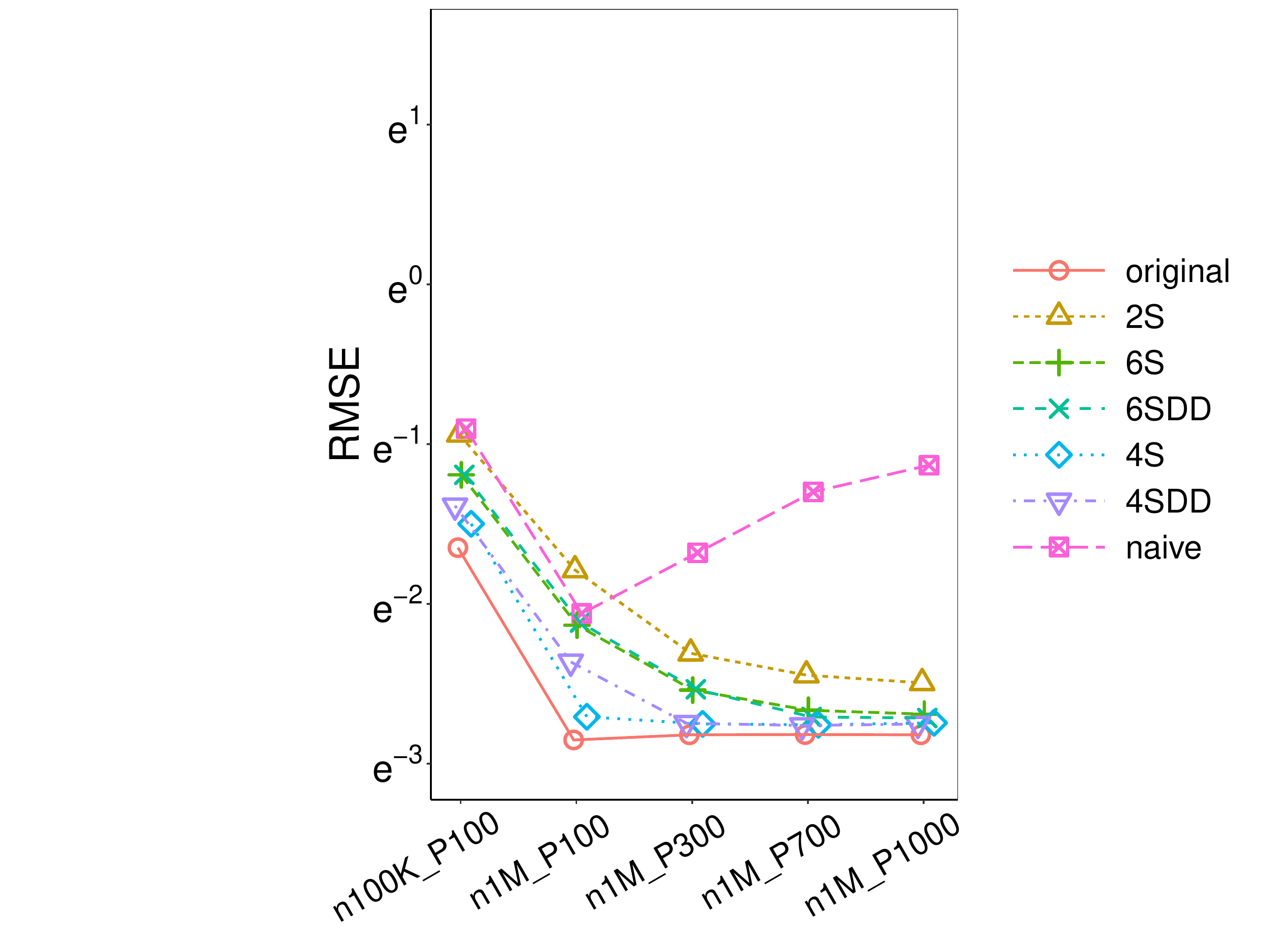}
\includegraphics[width=0.19\textwidth, trim={2.5in 0 2.6in 0},clip] {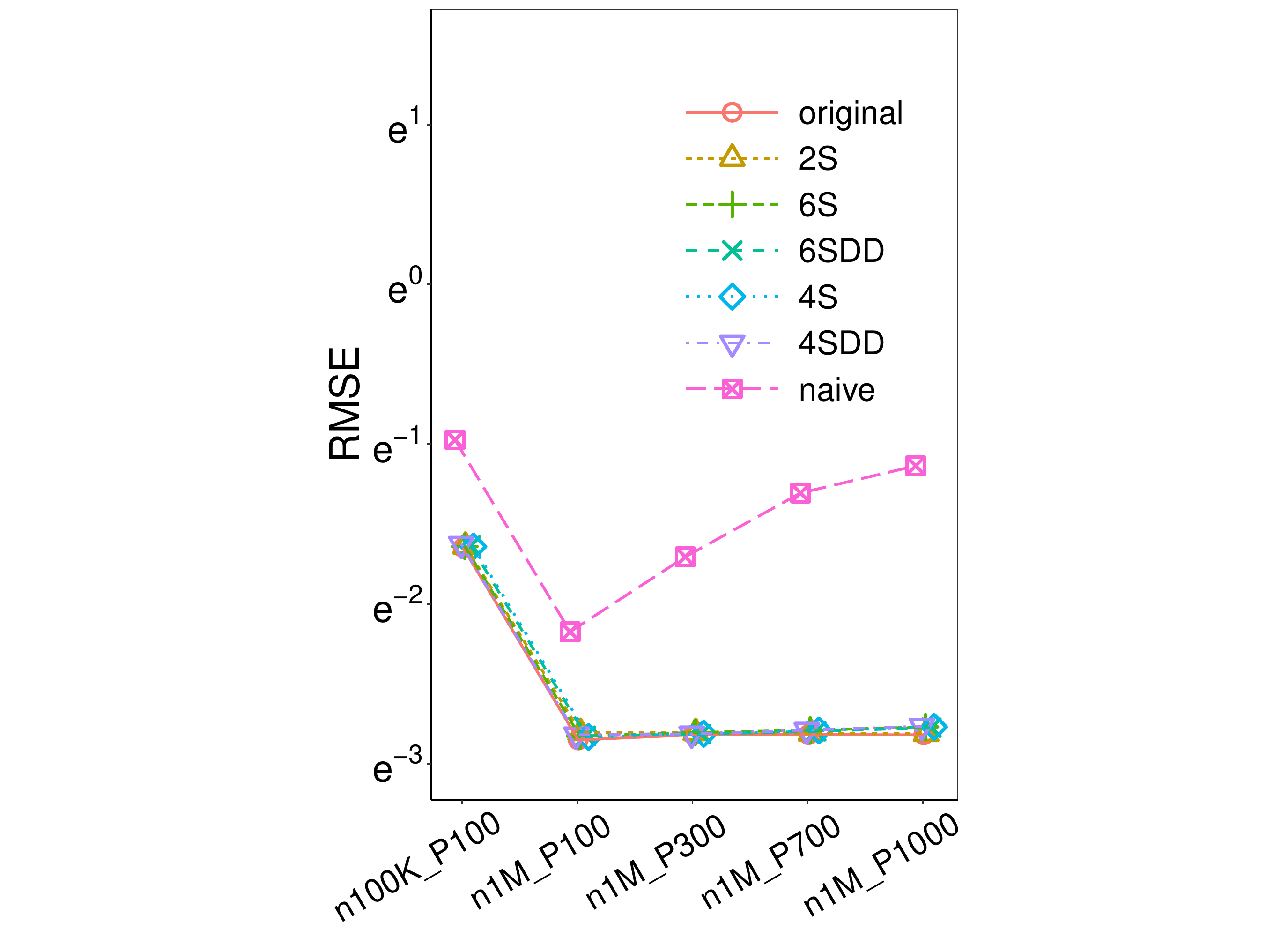}

\includegraphics[width=0.19\textwidth, trim={2.5in 0 2.6in 0},clip] {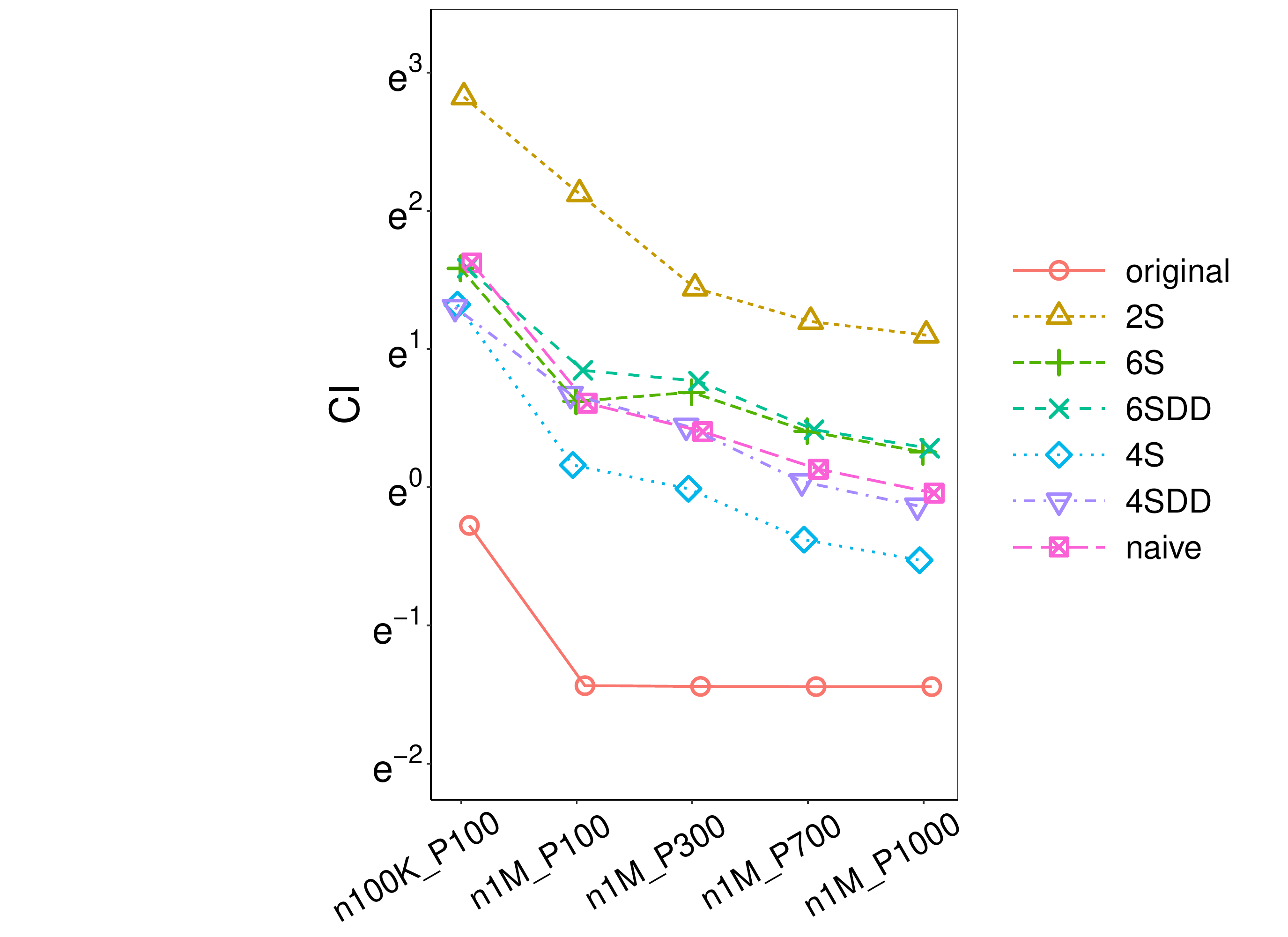}
\includegraphics[width=0.19\textwidth, trim={2.5in 0 2.6in 0},clip] {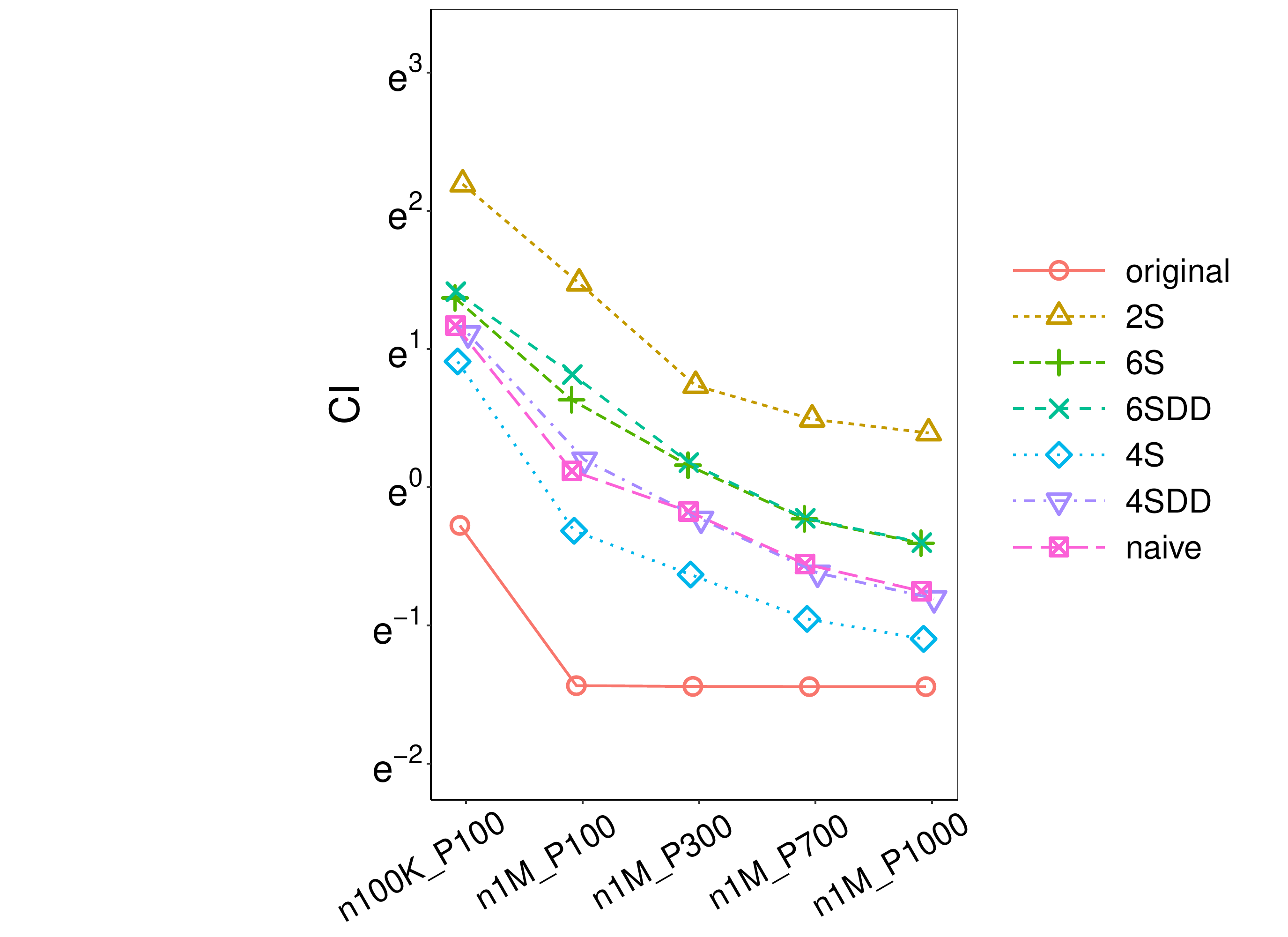}
\includegraphics[width=0.19\textwidth, trim={2.5in 0 2.6in 0},clip] {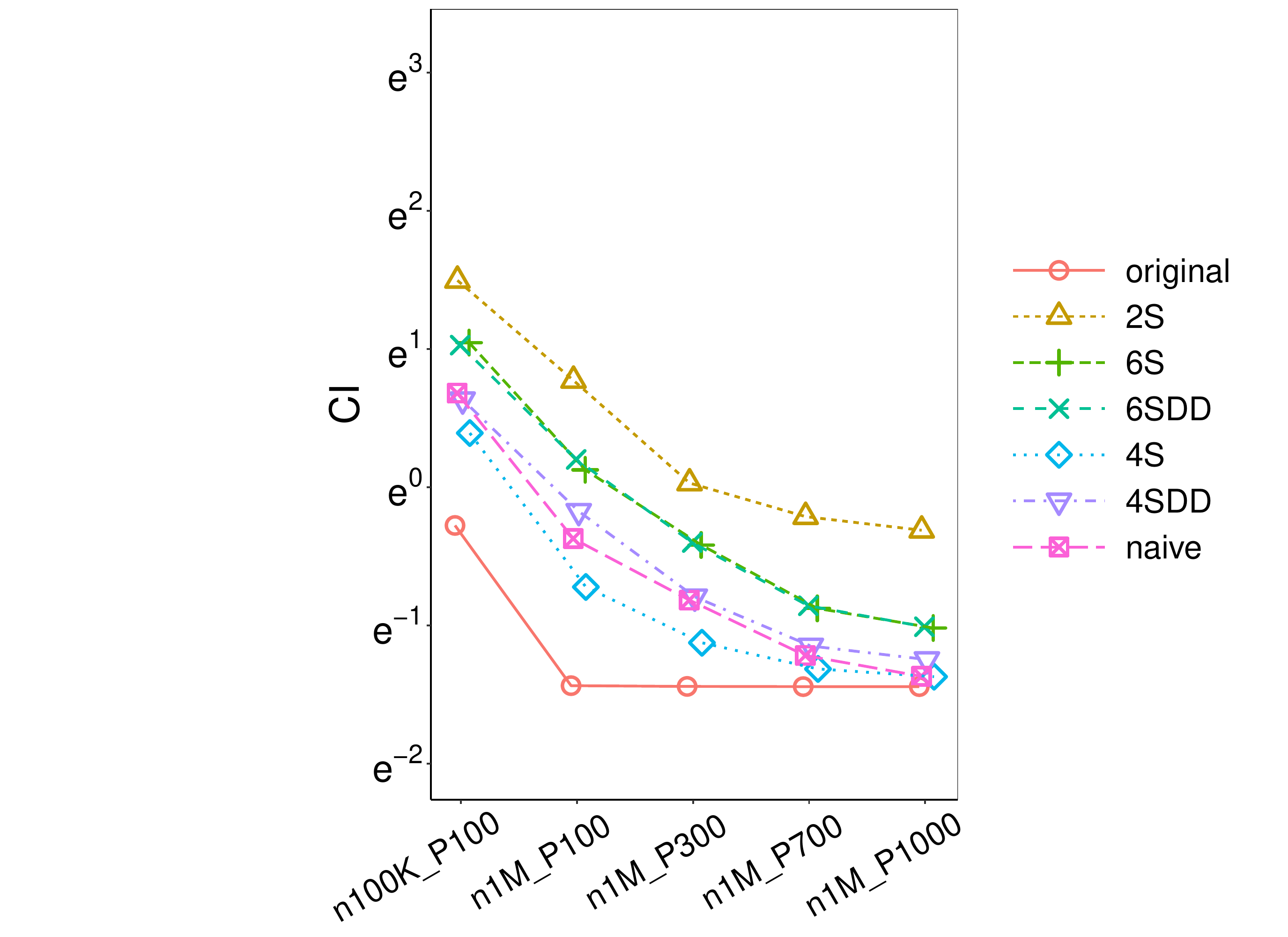}
\includegraphics[width=0.19\textwidth, trim={2.5in 0 2.6in 0},clip] {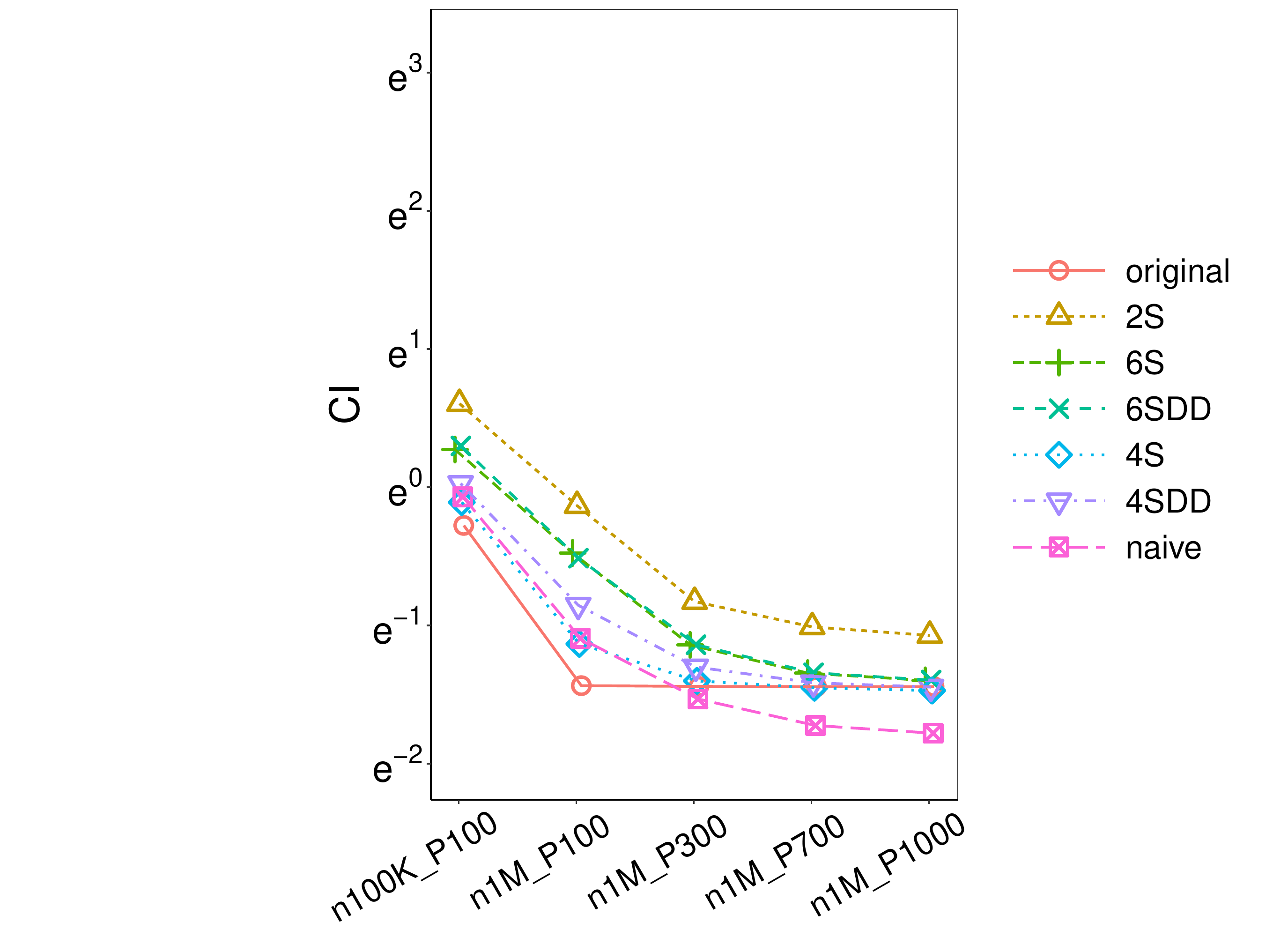}
\includegraphics[width=0.19\textwidth, trim={2.5in 0 2.6in 0},clip] {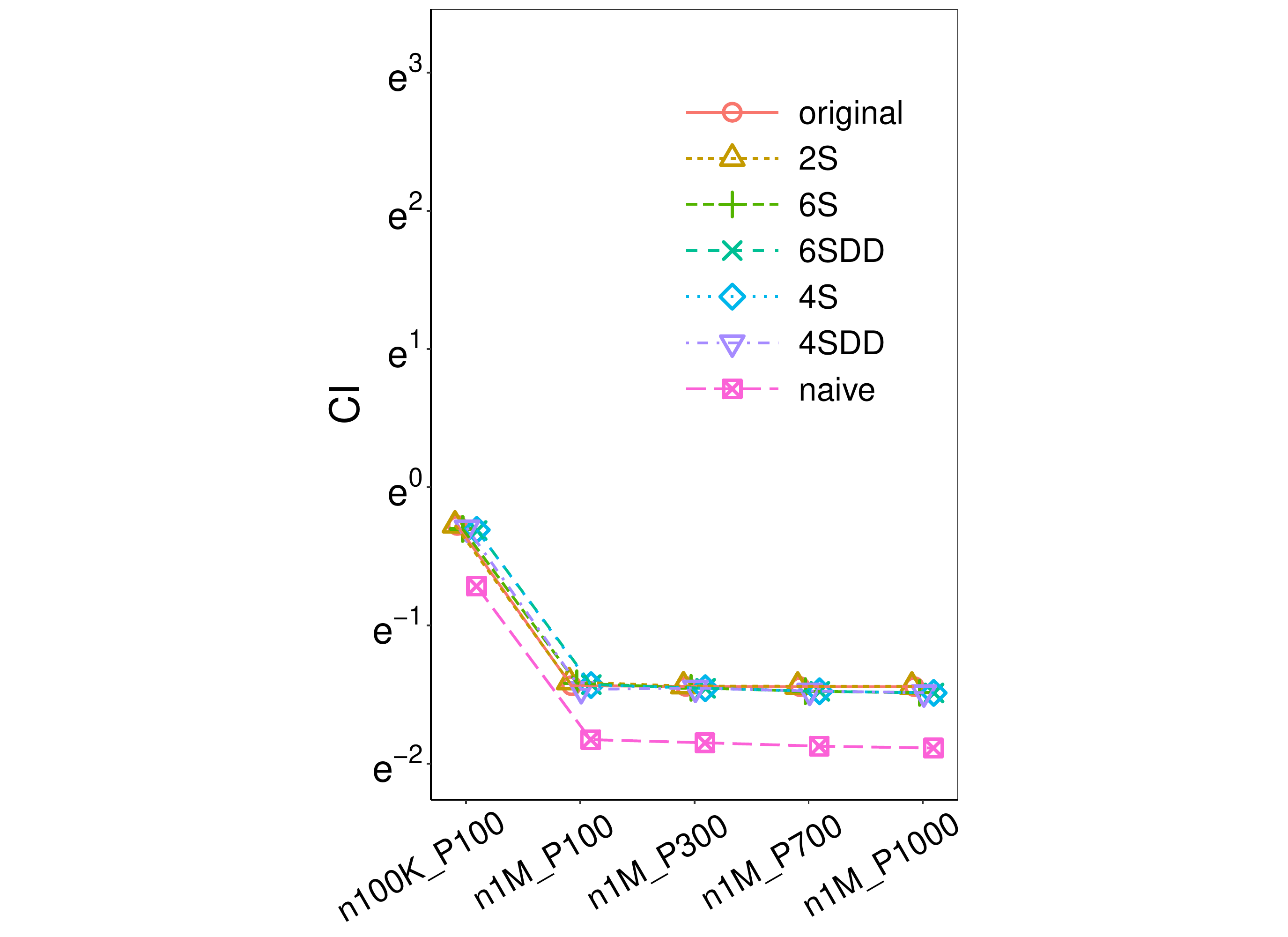}

\includegraphics[width=0.19\textwidth, trim={2.5in 0 2.6in 0},clip] {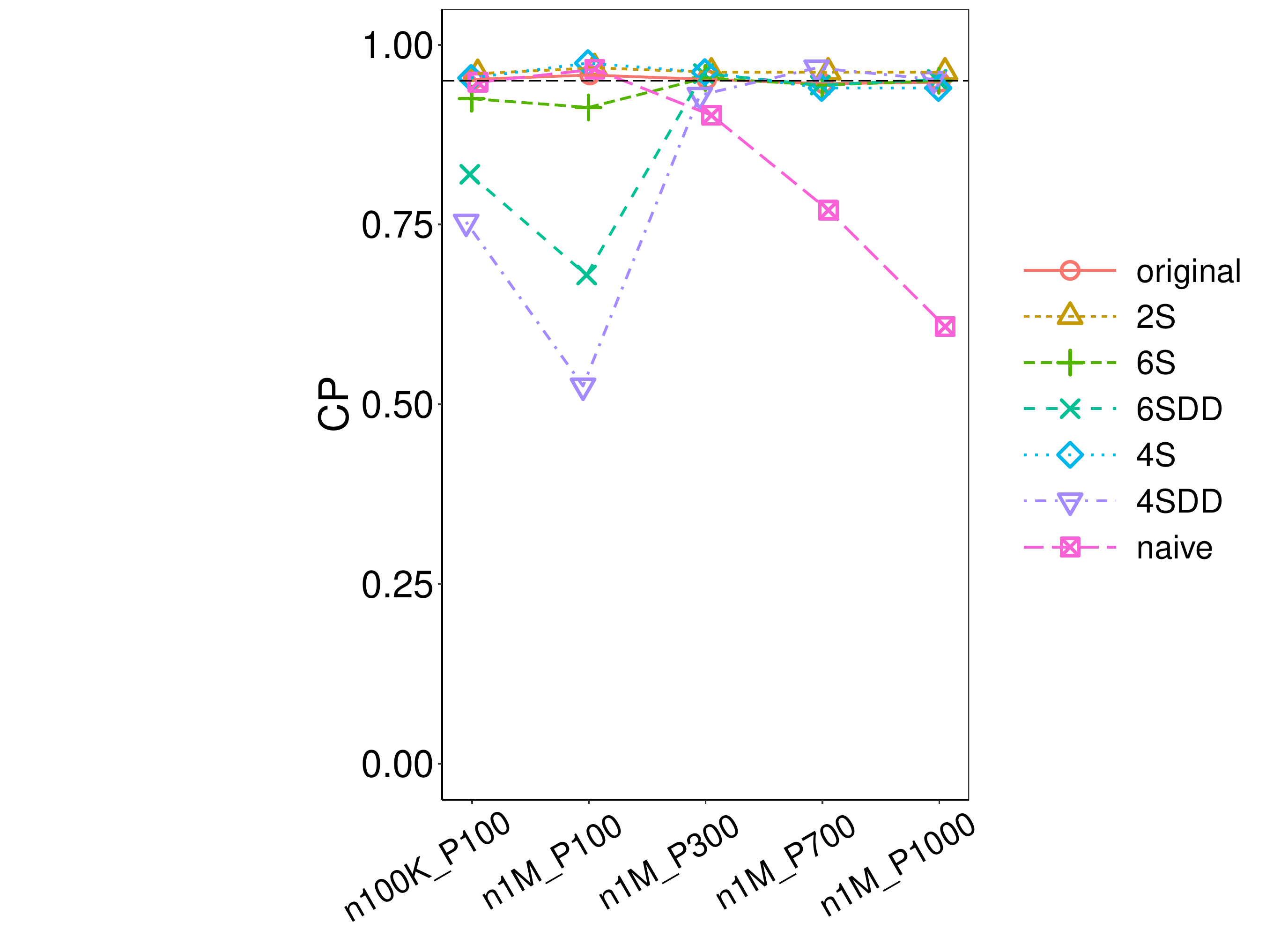}
\includegraphics[width=0.19\textwidth, trim={2.5in 0 2.6in 0},clip] {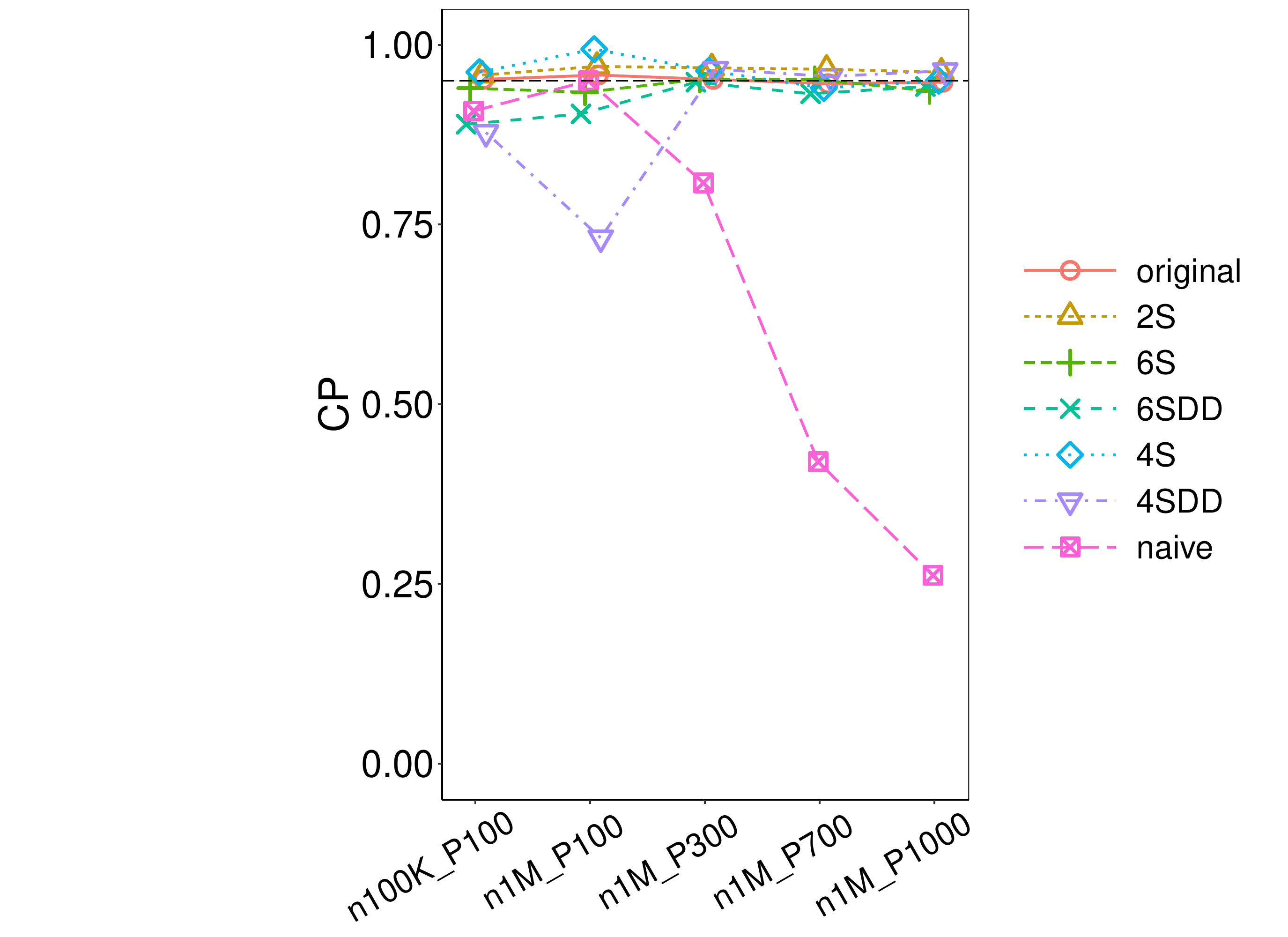}
\includegraphics[width=0.19\textwidth, trim={2.5in 0 2.6in 0},clip] {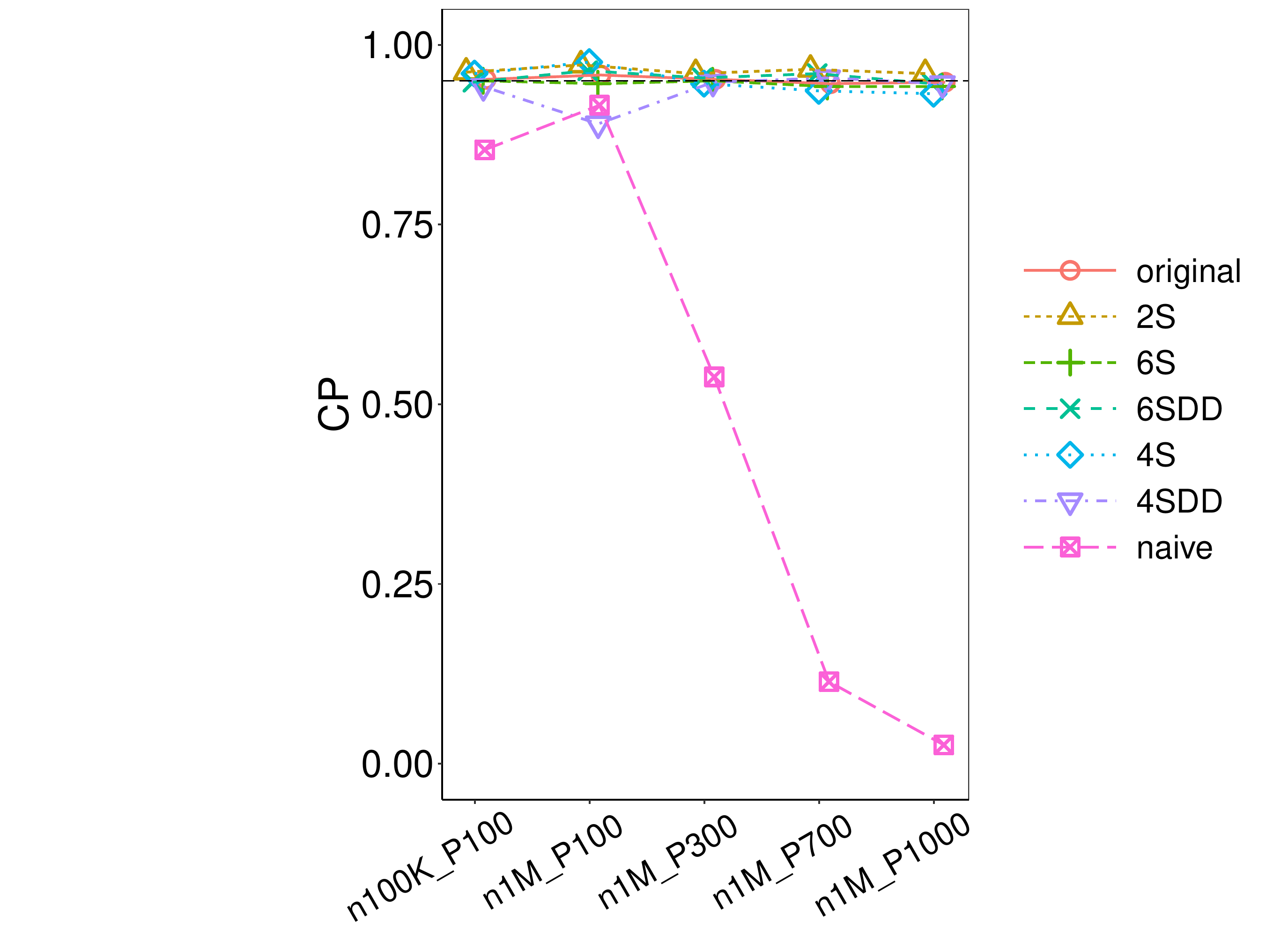}
\includegraphics[width=0.19\textwidth, trim={2.5in 0 2.6in 0},clip] {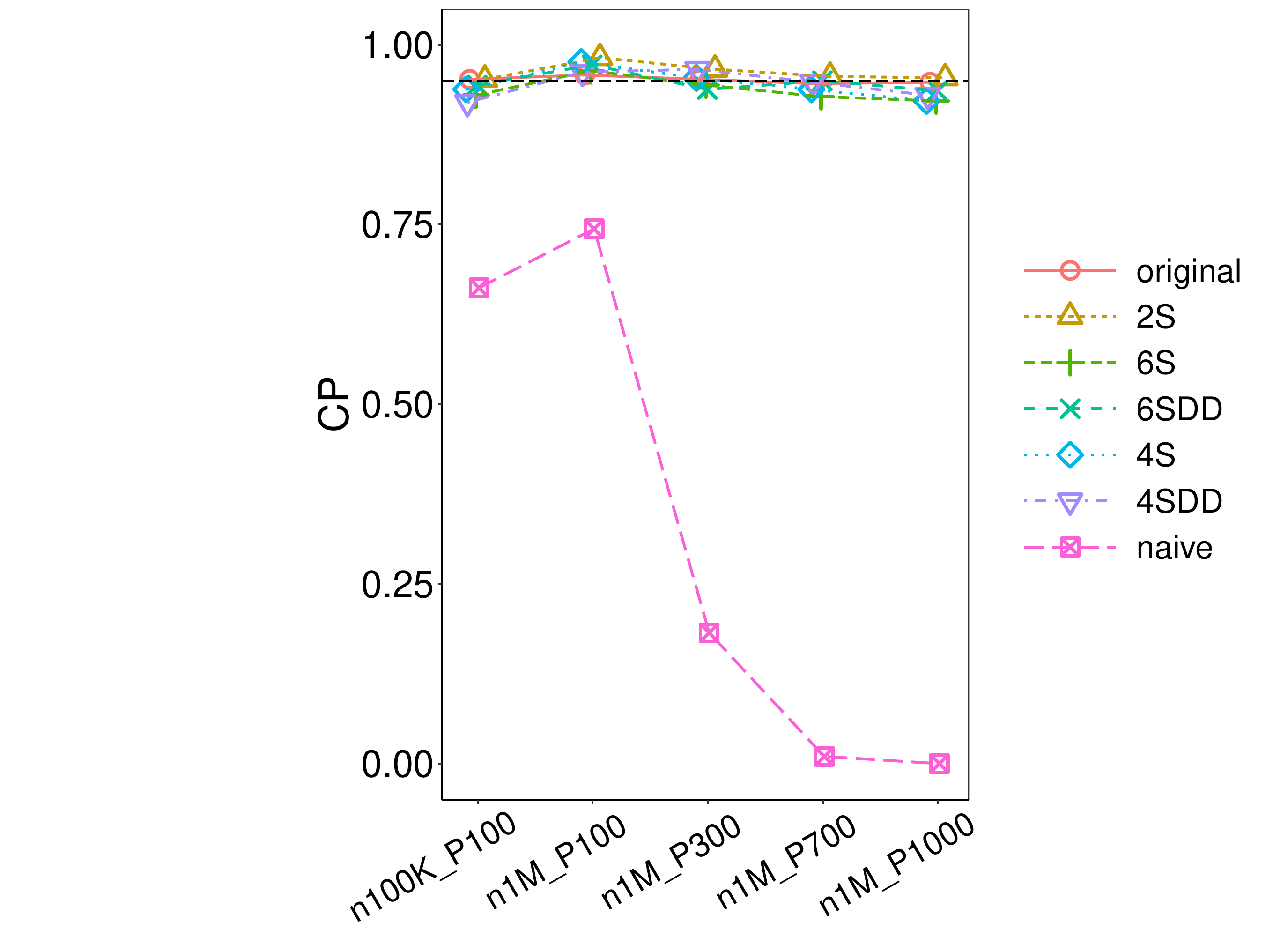}
\includegraphics[width=0.19\textwidth, trim={2.5in 0 2.6in 0},clip] {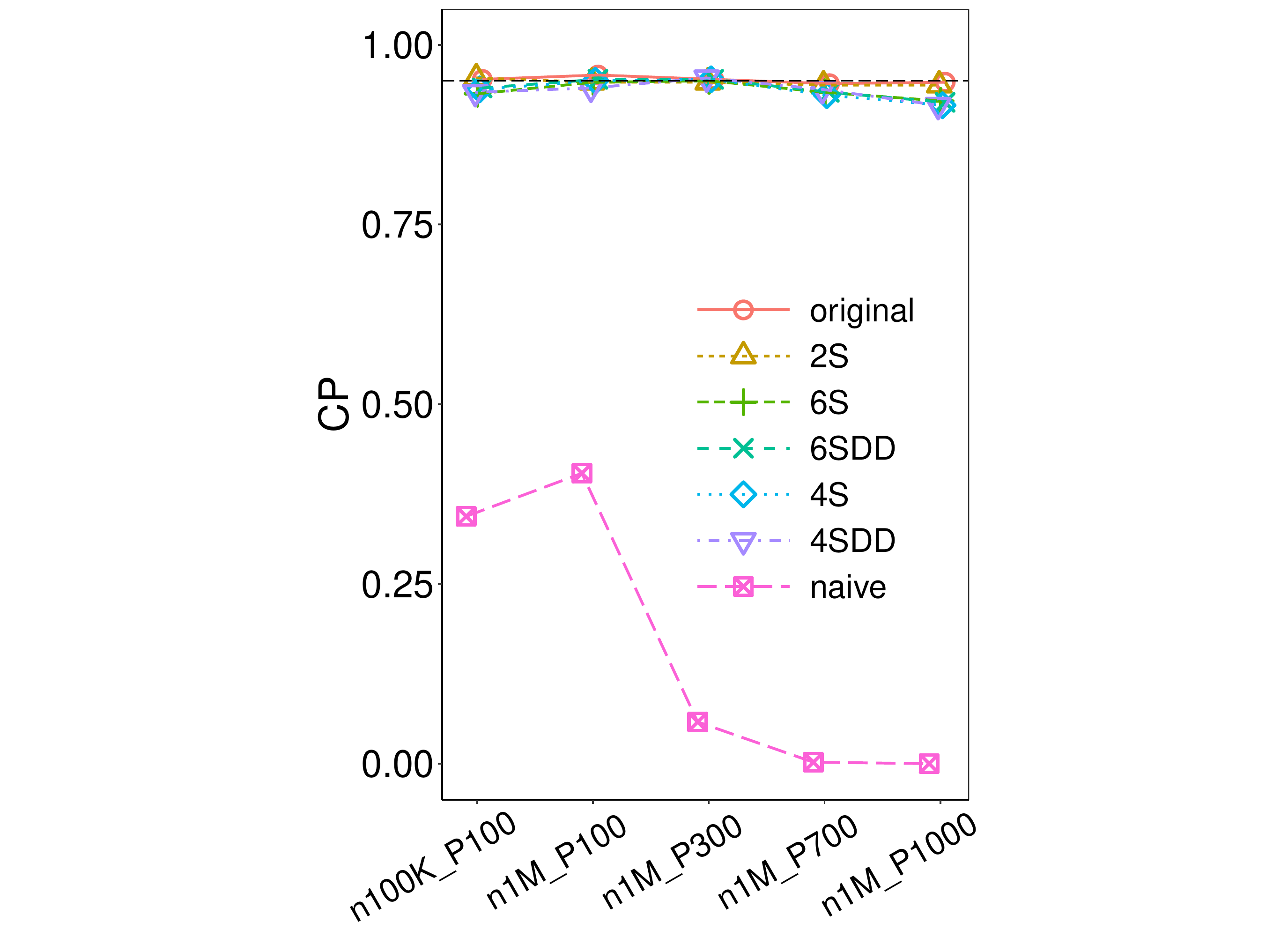}

\includegraphics[width=0.19\textwidth, trim={2.5in 0 2.6in 0},clip] {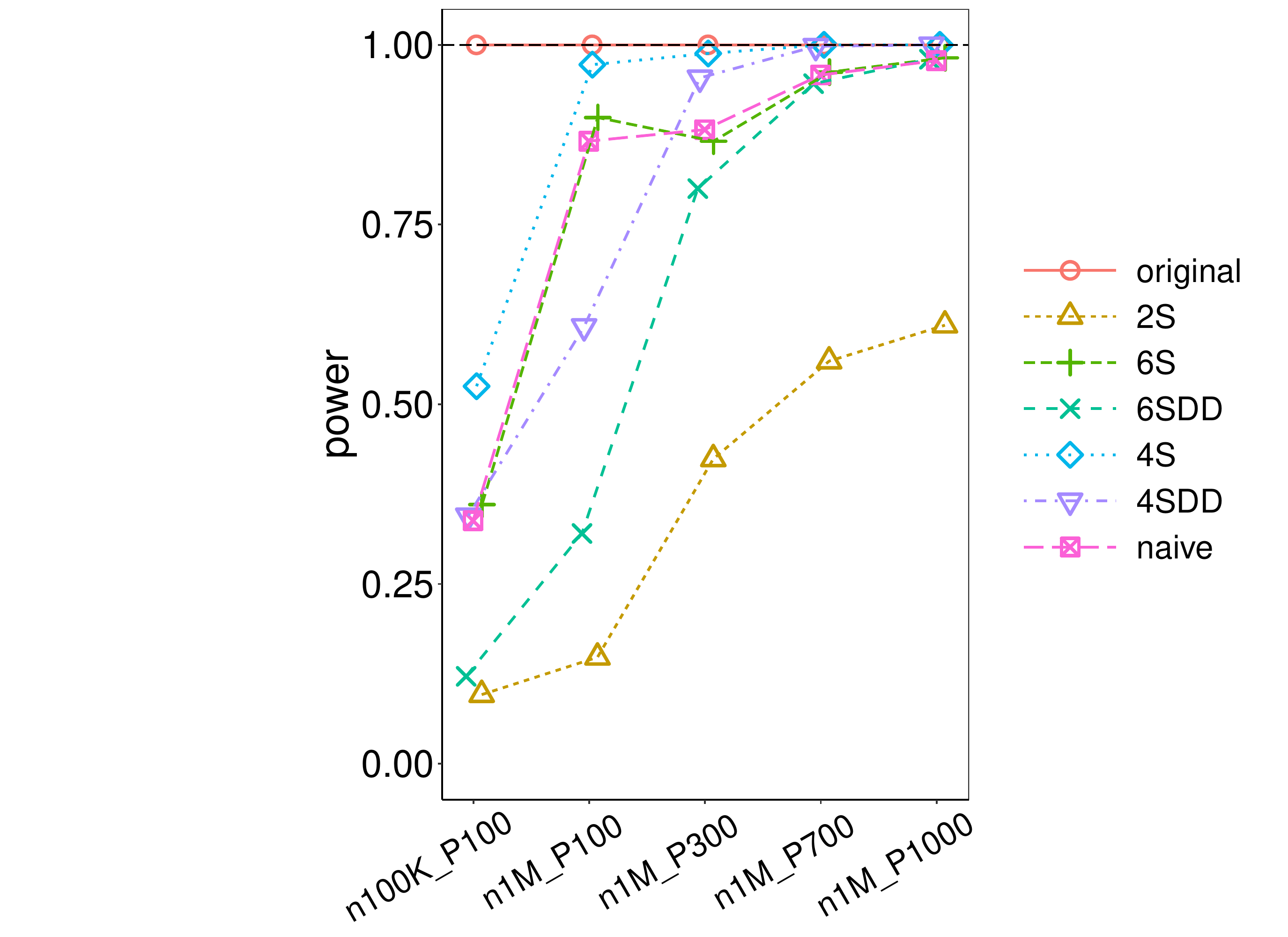}
\includegraphics[width=0.19\textwidth, trim={2.5in 0 2.6in 0},clip] {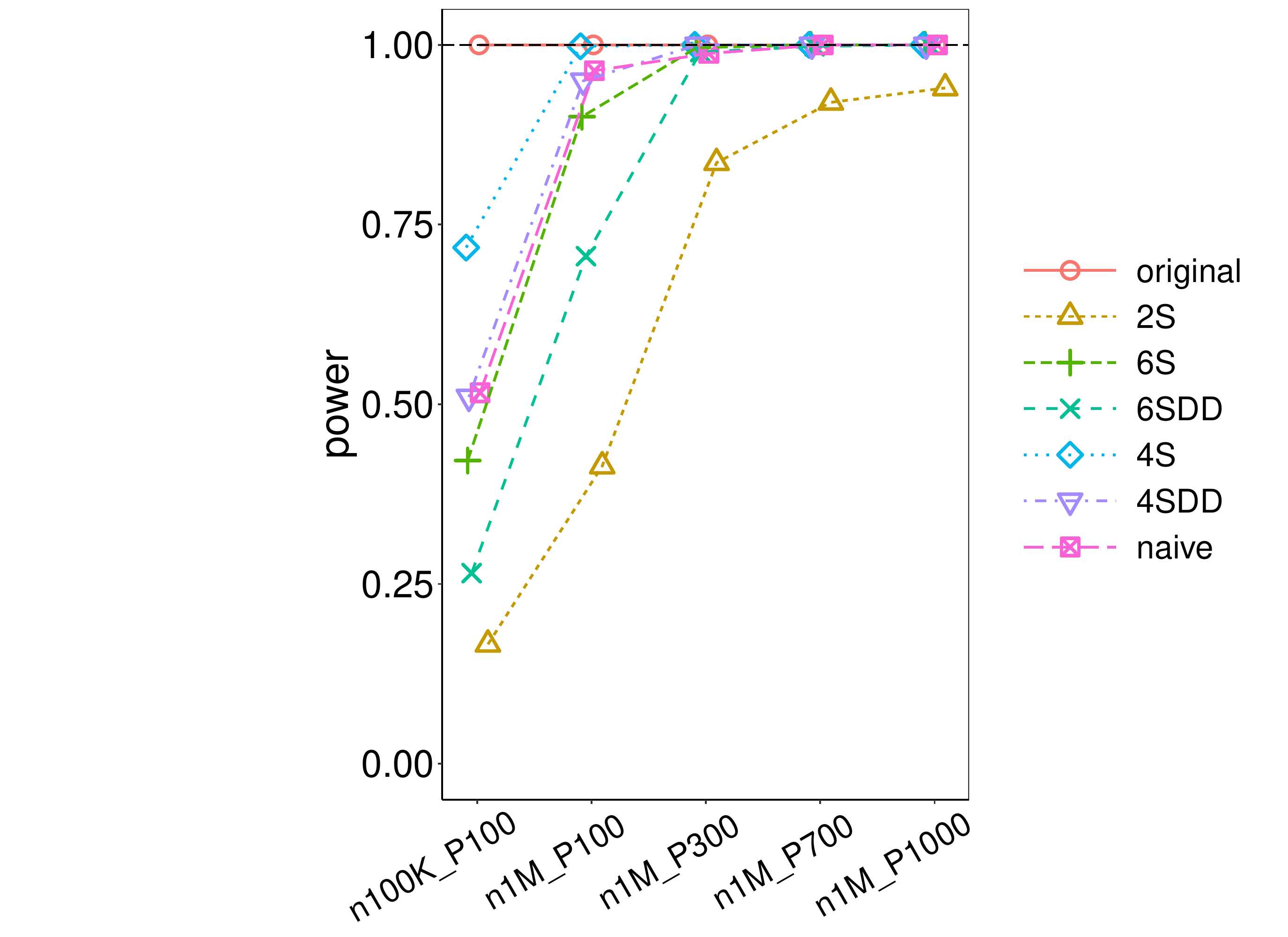}
\includegraphics[width=0.19\textwidth, trim={2.5in 0 2.6in 0},clip] {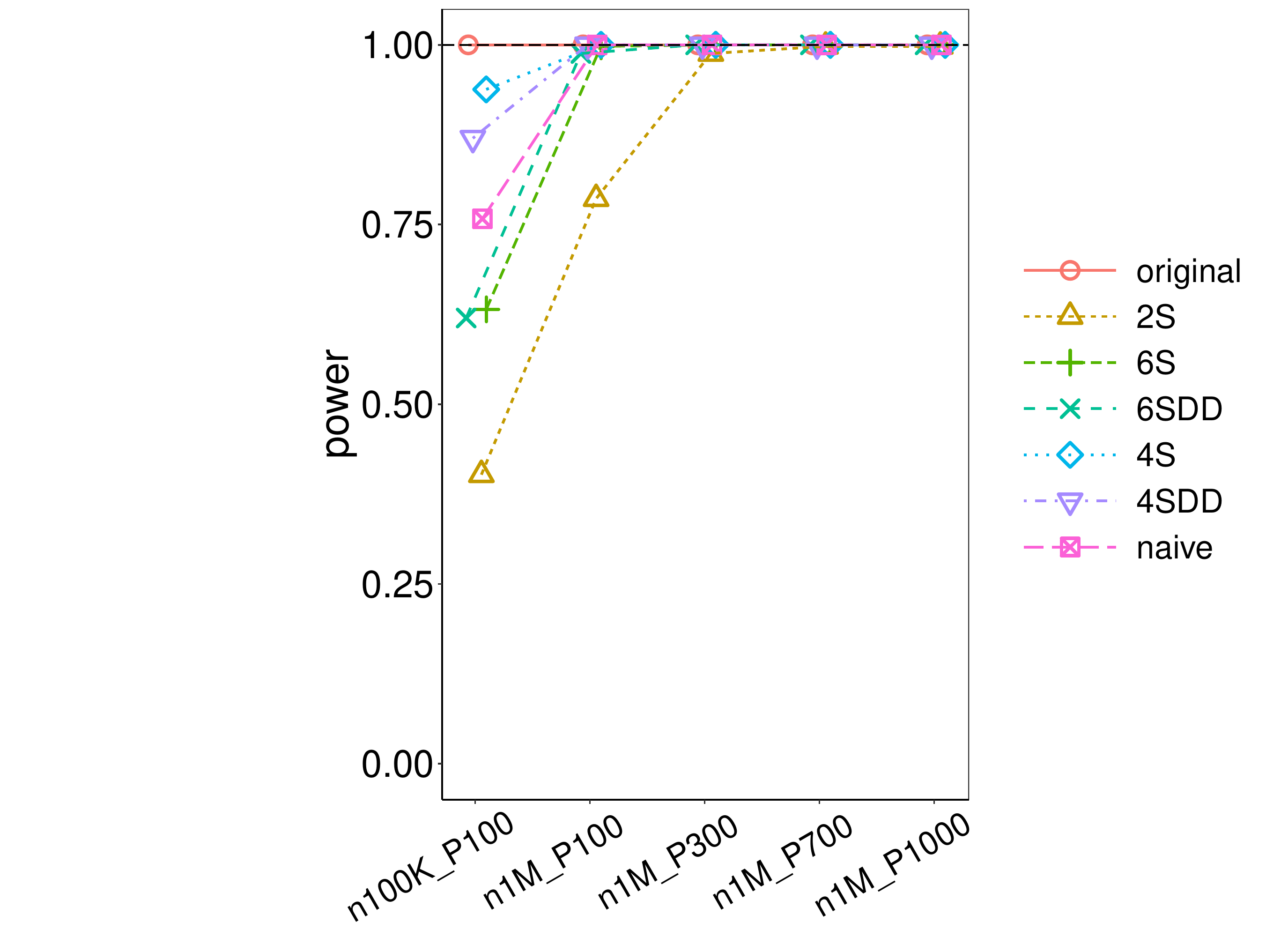}
\includegraphics[width=0.19\textwidth, trim={2.5in 0 2.6in 0},clip] {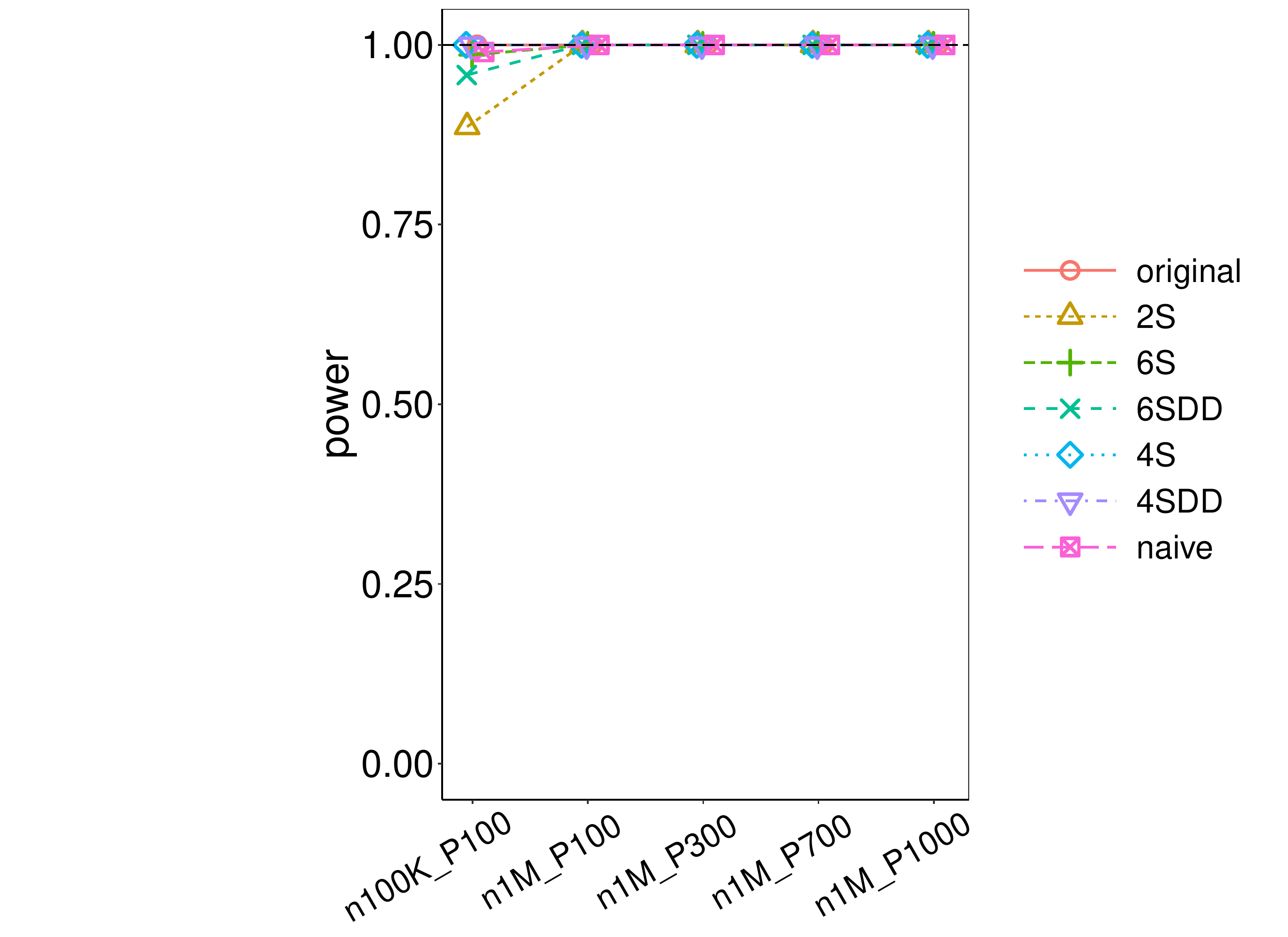}
\includegraphics[width=0.19\textwidth, trim={2.5in 0 2.6in 0},clip] {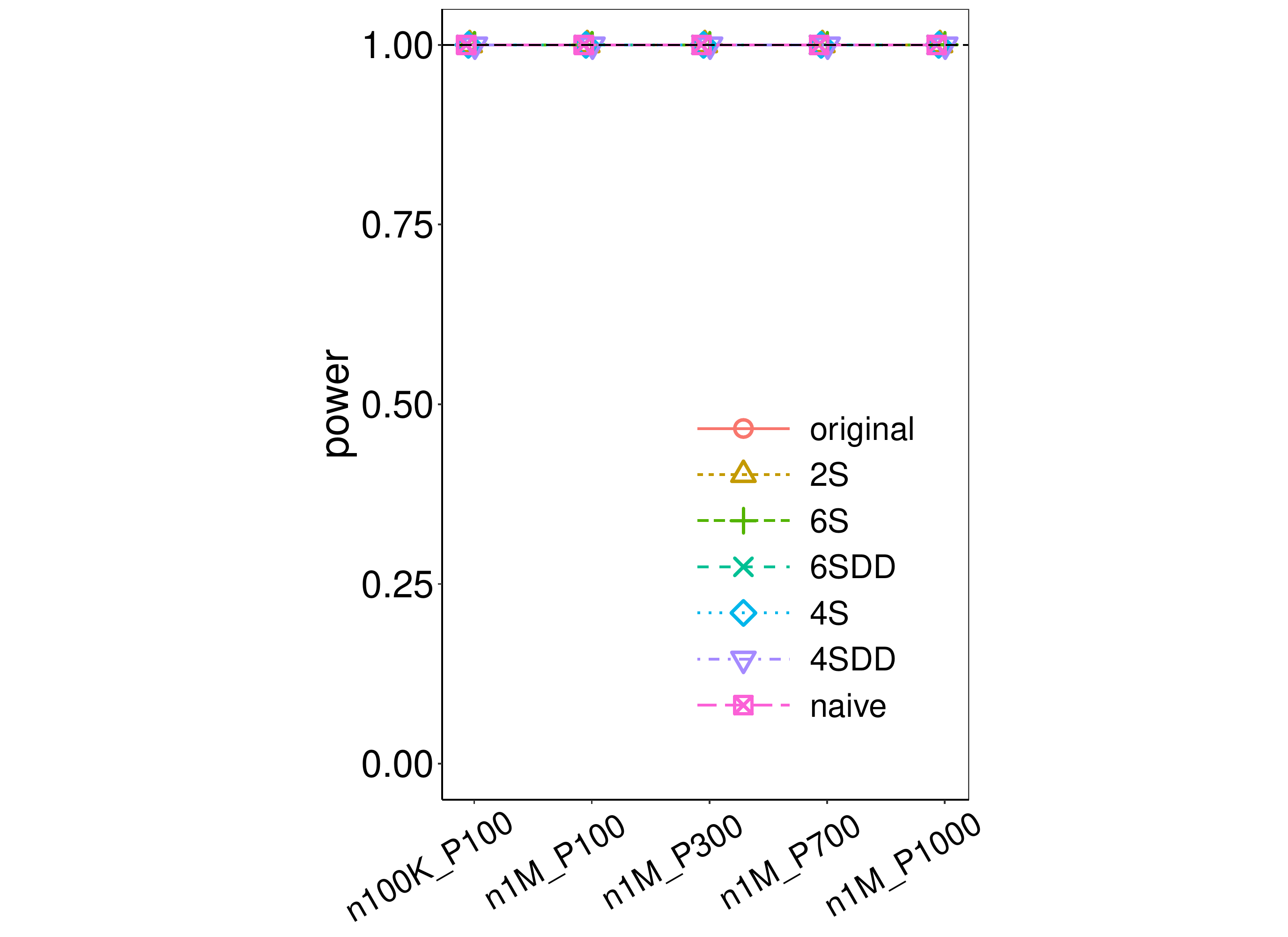}

\caption{Simulation results with $\epsilon$-DP for ZILN data with  $\alpha\ne\beta$ when $\theta\ne0$} \label{fig:1asDPZILN}
\end{figure}

\begin{figure}[!htb]
\hspace{0.45in}$\rho=0.005$\hspace{0.65in}$\rho=0.02$\hspace{0.65in}$\rho=0.08$
\hspace{0.65in}$\rho=0.32$\hspace{0.65in}$\rho=1.28$

\includegraphics[width=0.19\textwidth, trim={2.5in 0 2.6in 0},clip] {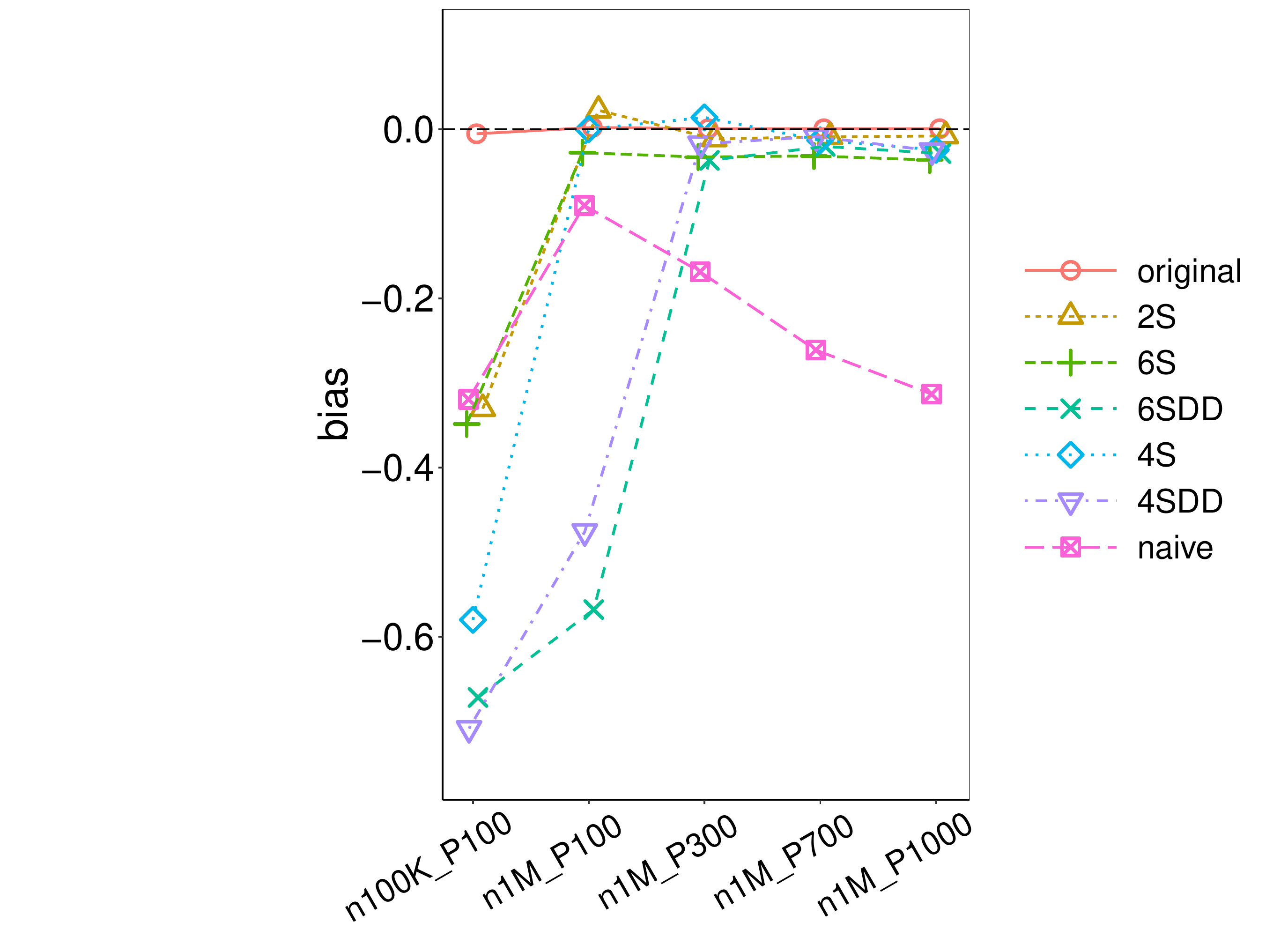}
\includegraphics[width=0.19\textwidth, trim={2.5in 0 2.6in 0},clip] {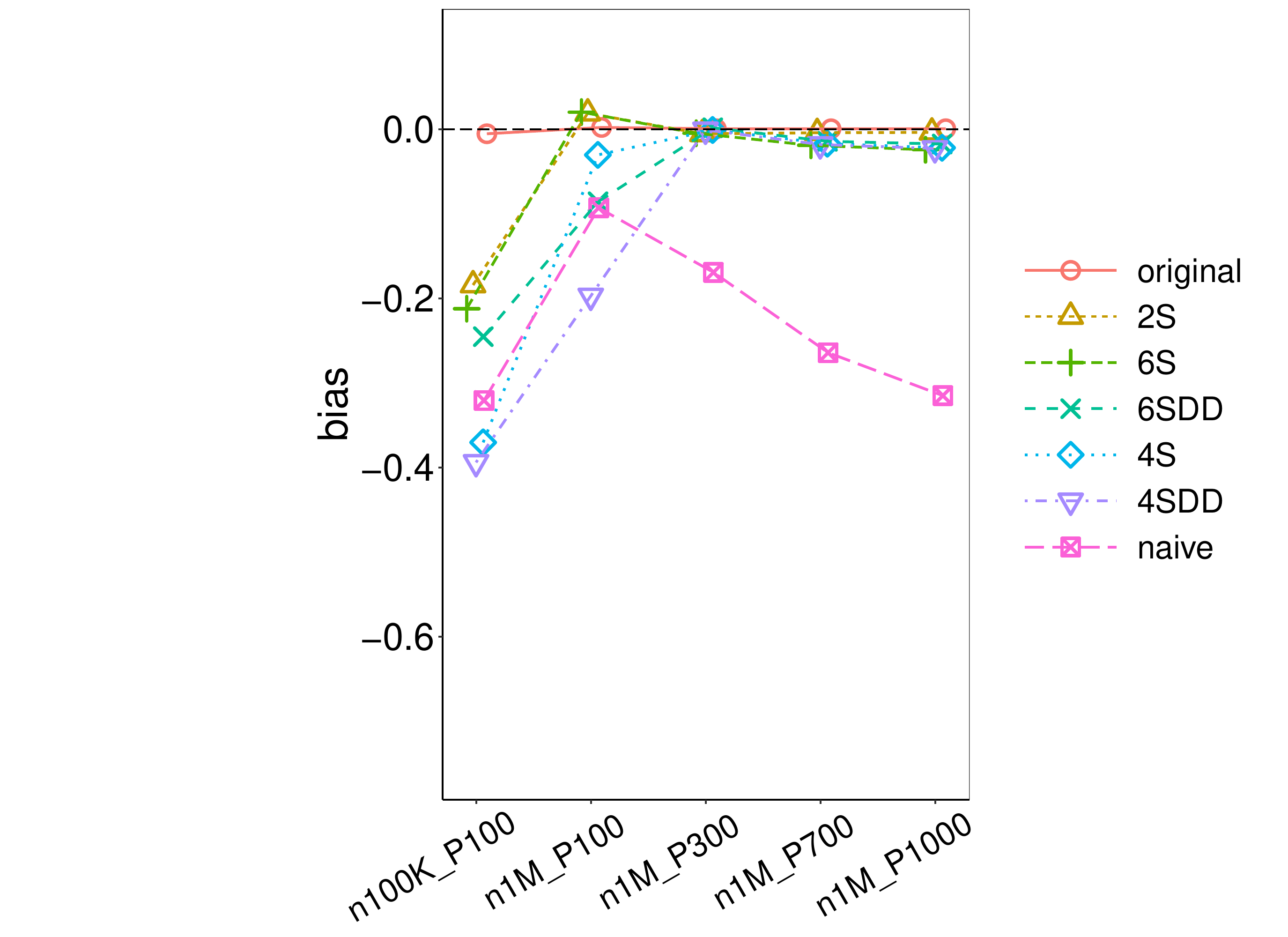}
\includegraphics[width=0.19\textwidth, trim={2.5in 0 2.6in 0},clip] {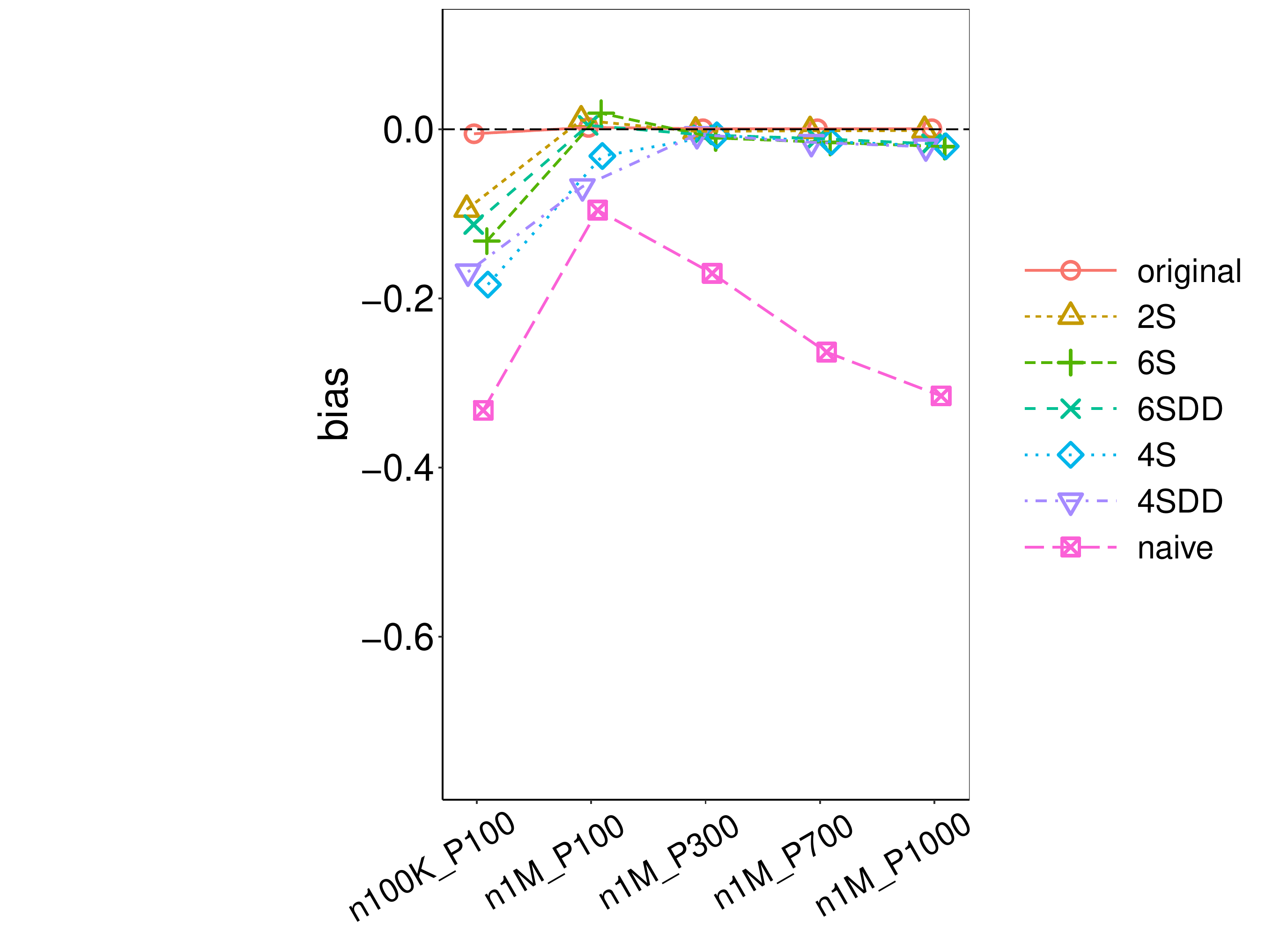}
\includegraphics[width=0.19\textwidth, trim={2.5in 0 2.6in 0},clip] {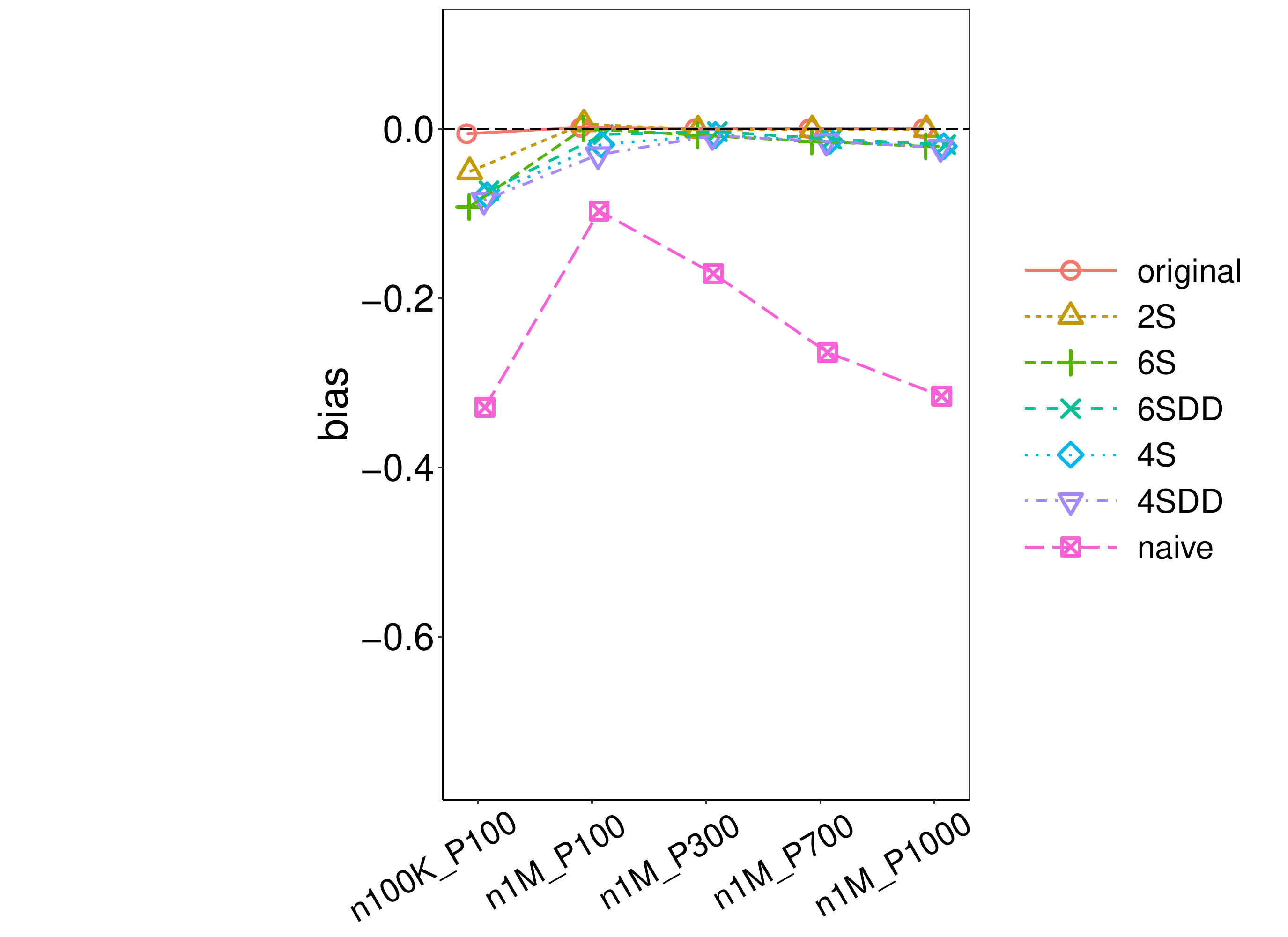}
\includegraphics[width=0.19\textwidth, trim={2.5in 0 2.6in 0},clip] {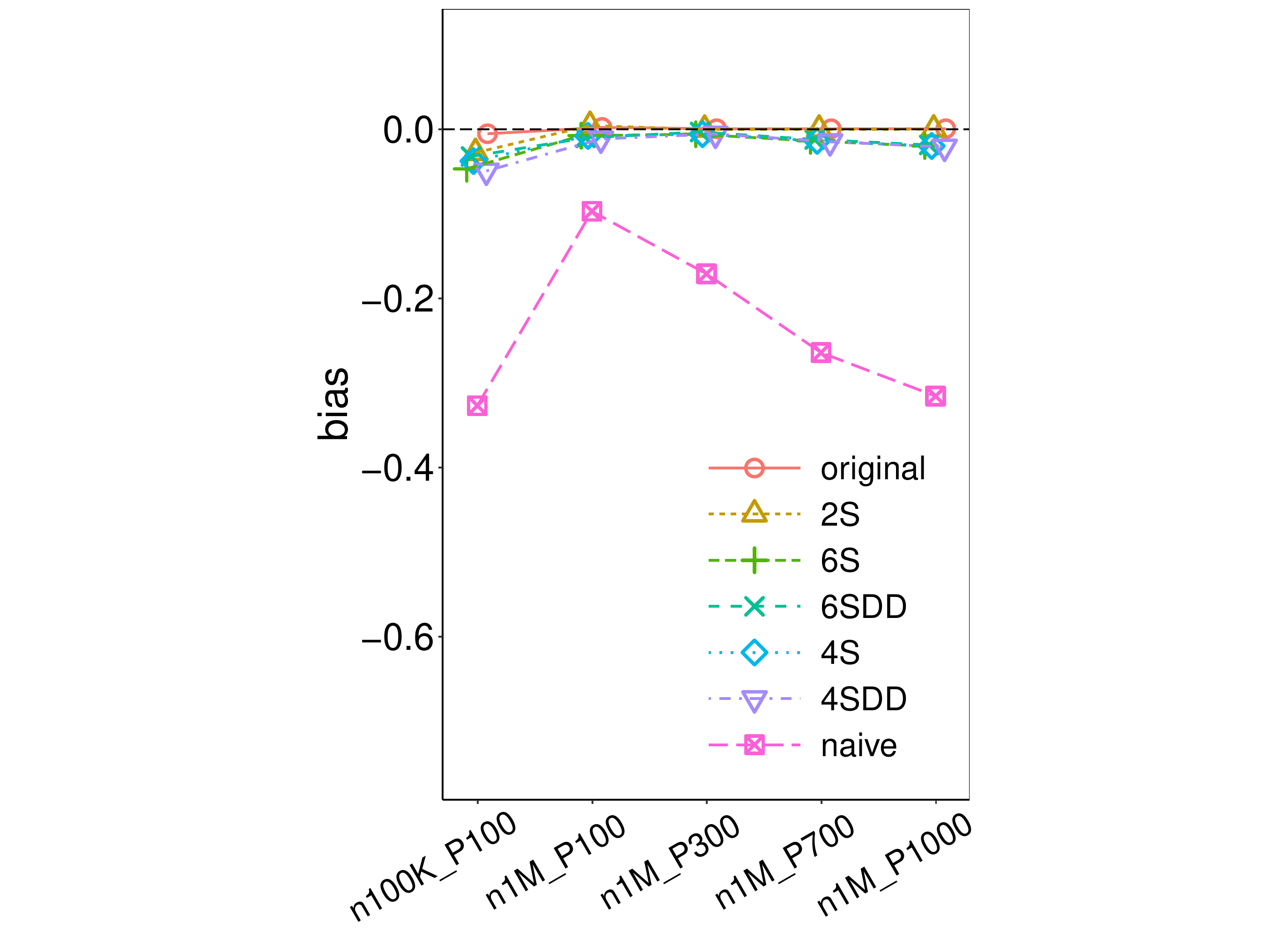}

\includegraphics[width=0.19\textwidth, trim={2.5in 0 2.6in 0},clip] {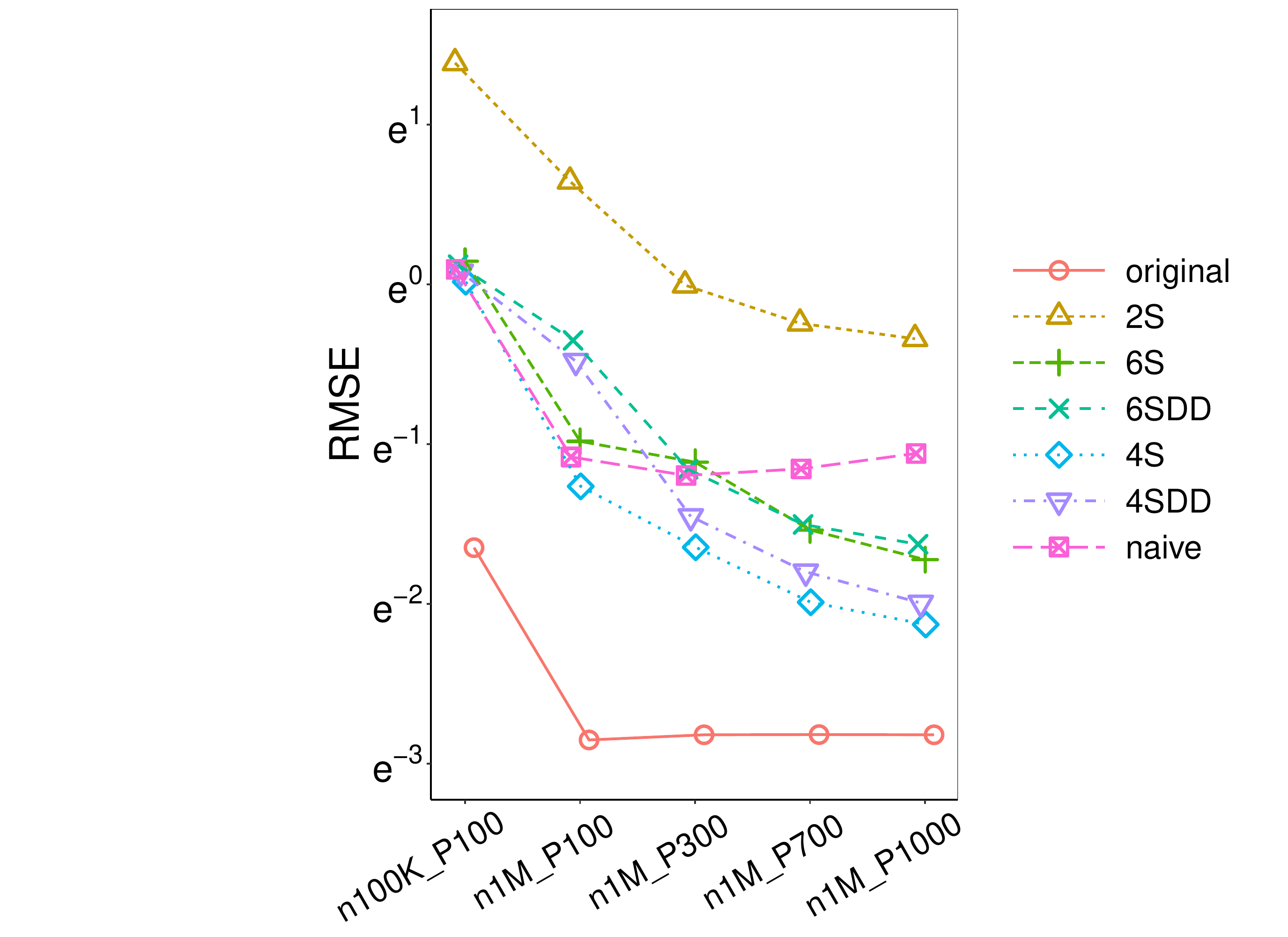}
\includegraphics[width=0.19\textwidth, trim={2.5in 0 2.6in 0},clip] {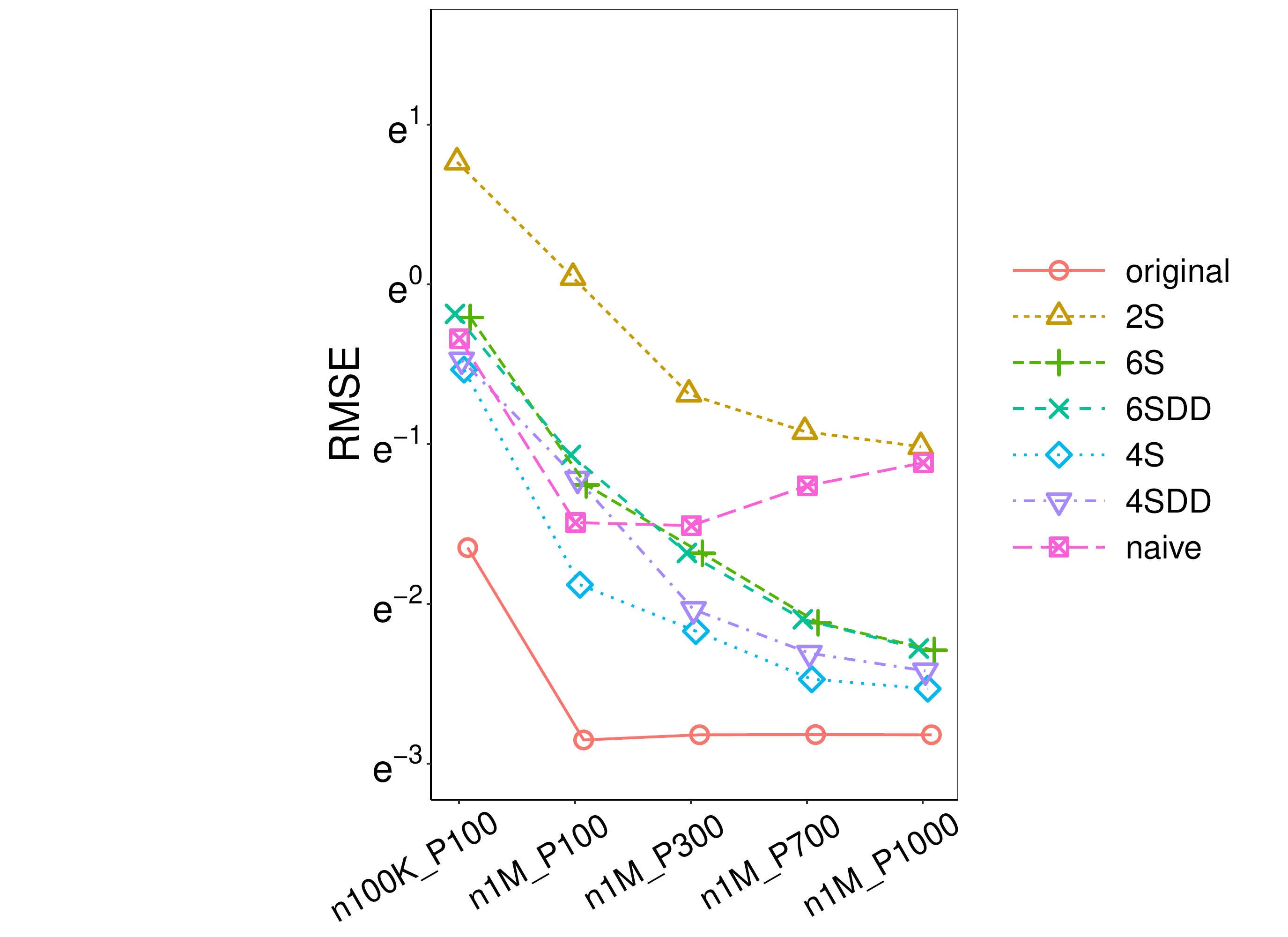}
\includegraphics[width=0.19\textwidth, trim={2.5in 0 2.6in 0},clip] {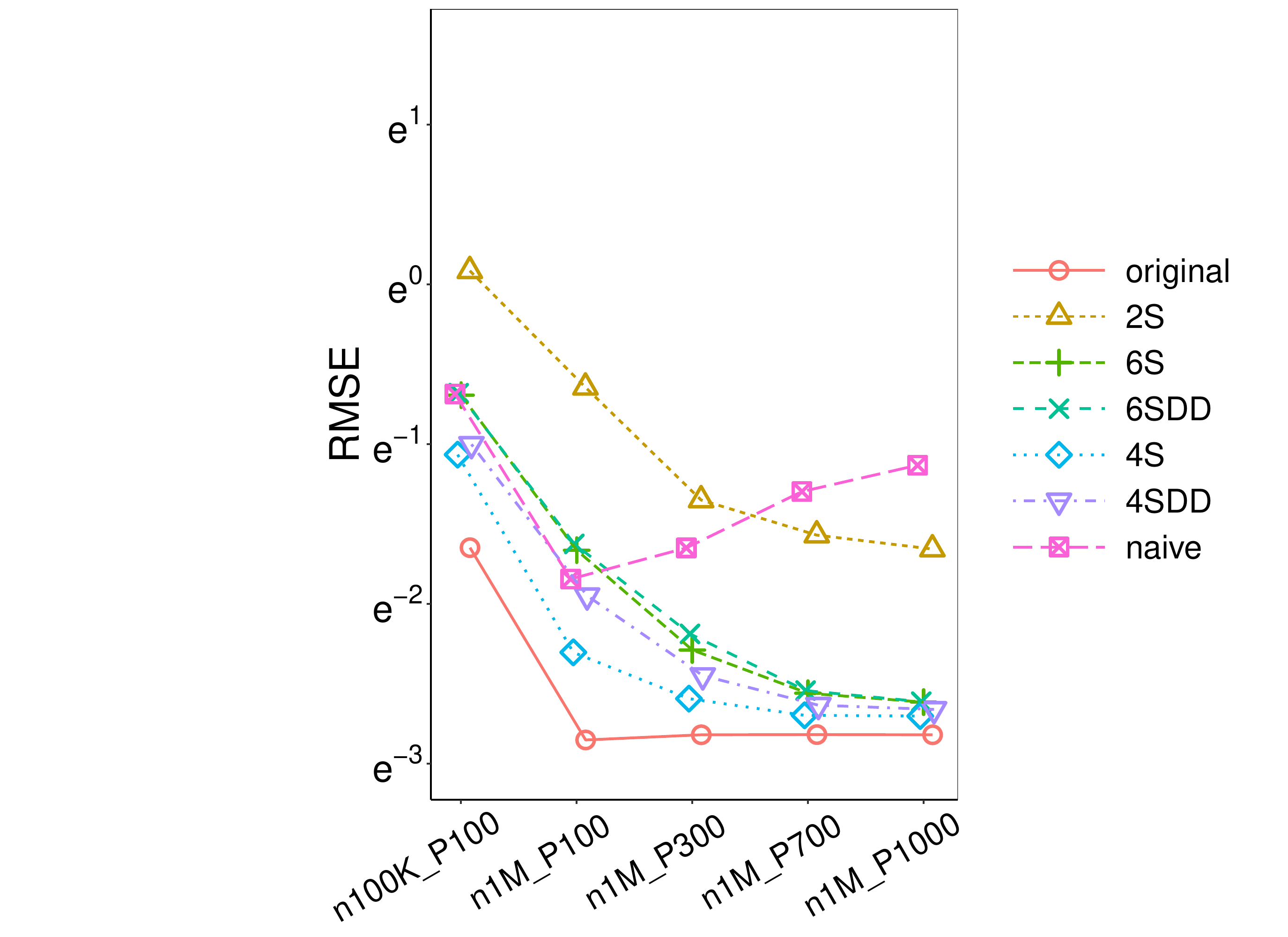}
\includegraphics[width=0.19\textwidth, trim={2.5in 0 2.6in 0},clip] {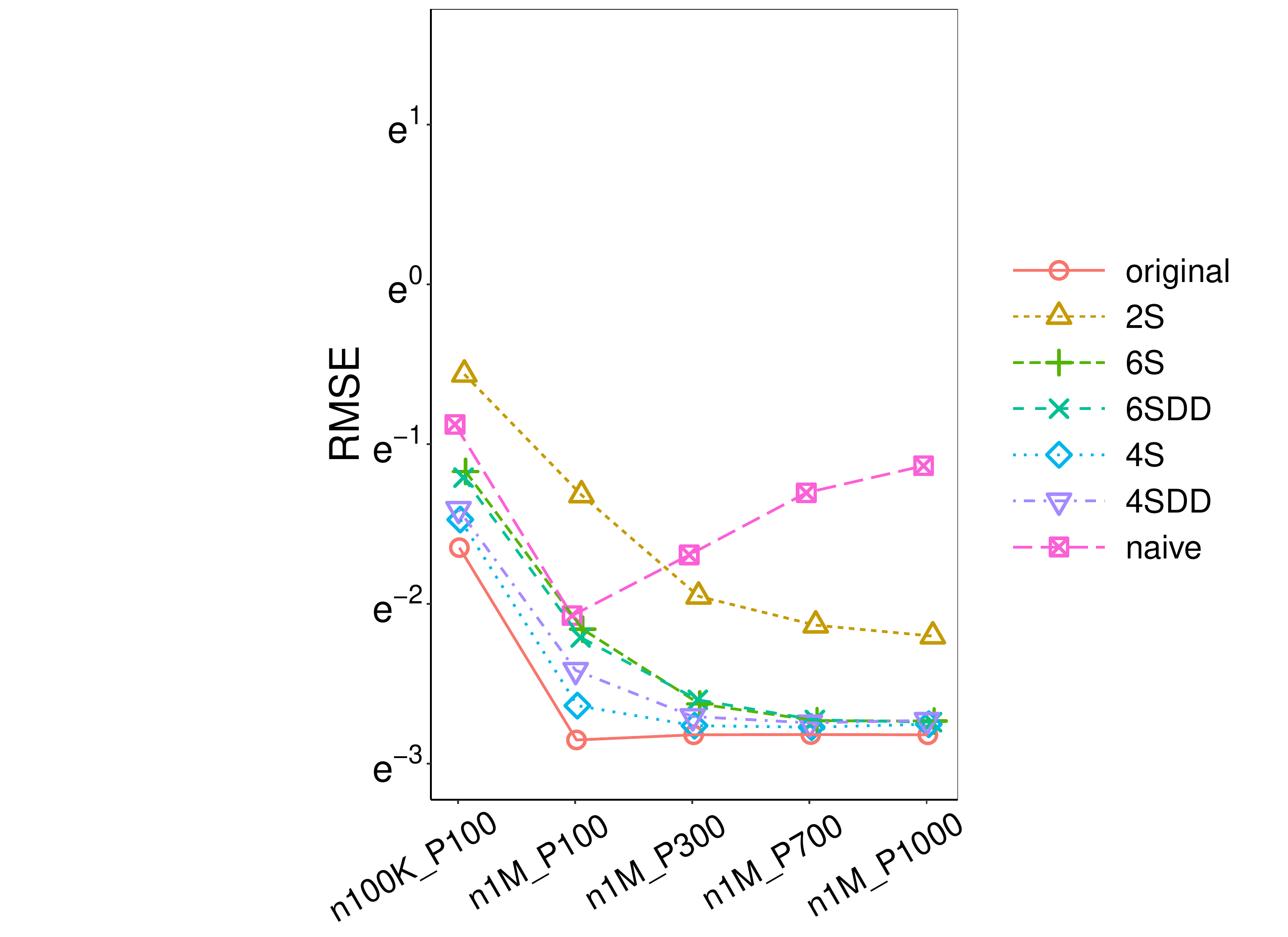}
\includegraphics[width=0.19\textwidth, trim={2.5in 0 2.6in 0},clip] {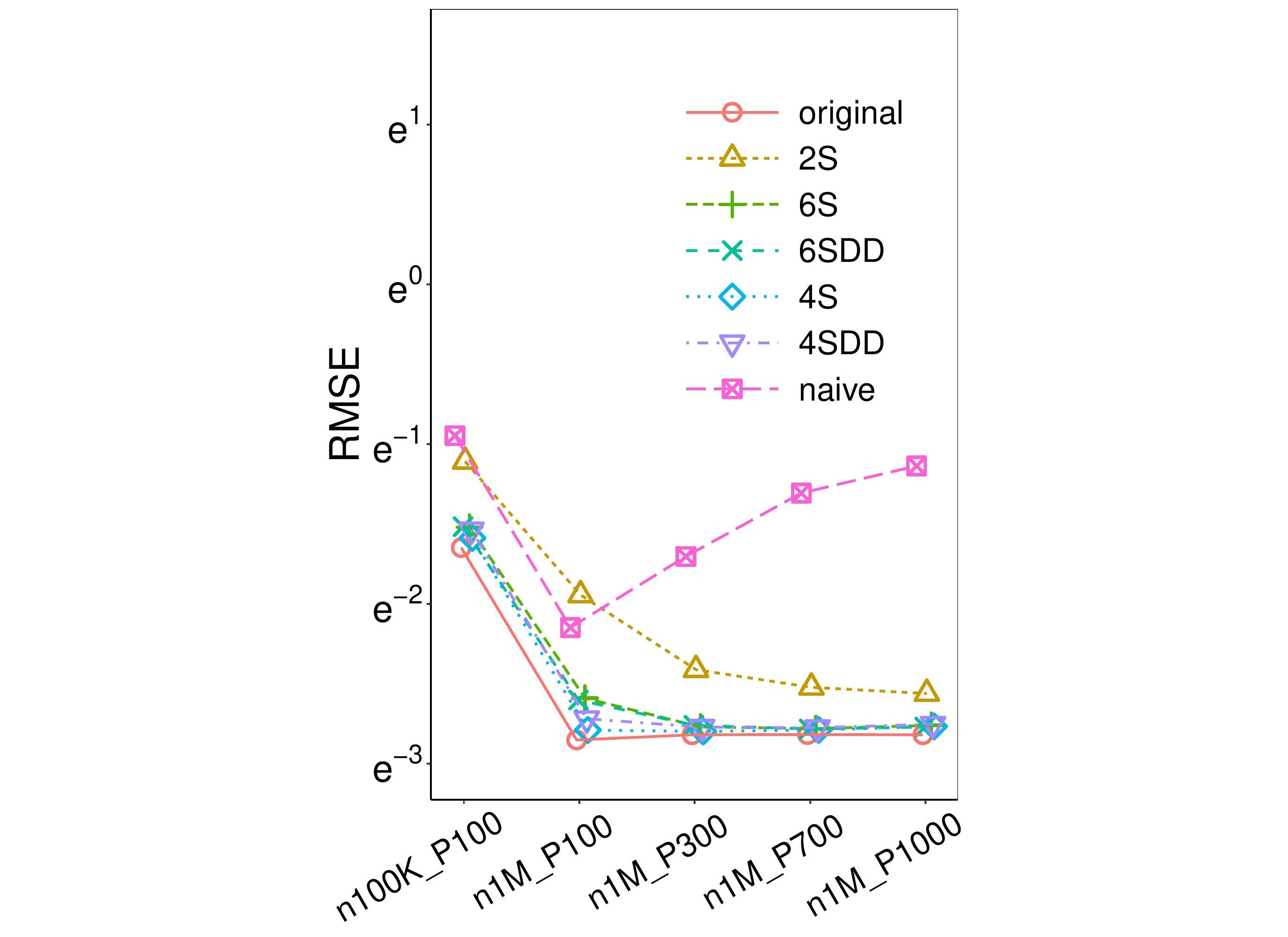}

\includegraphics[width=0.19\textwidth, trim={2.5in 0 2.6in 0},clip] {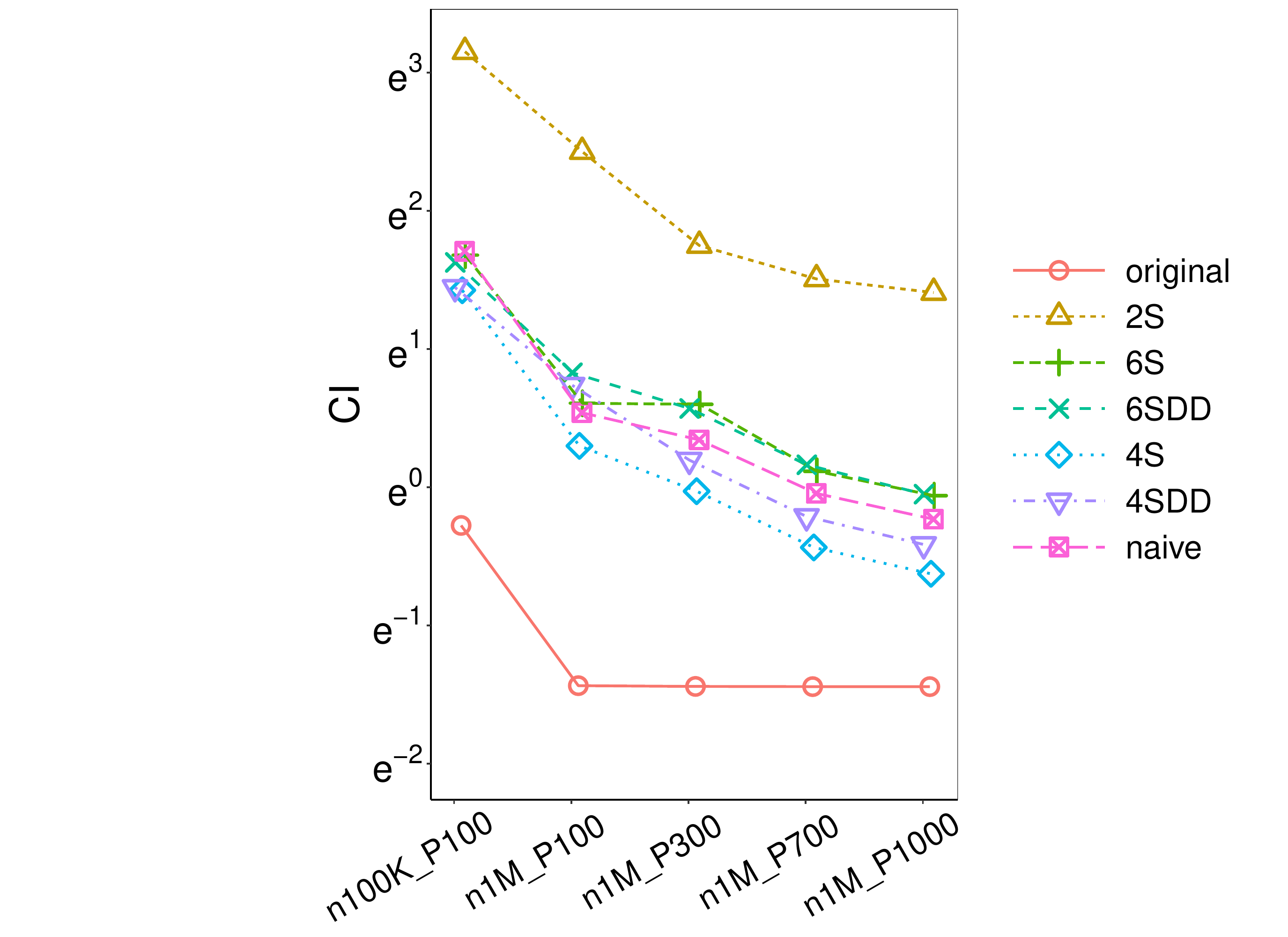}
\includegraphics[width=0.19\textwidth, trim={2.5in 0 2.6in 0},clip] {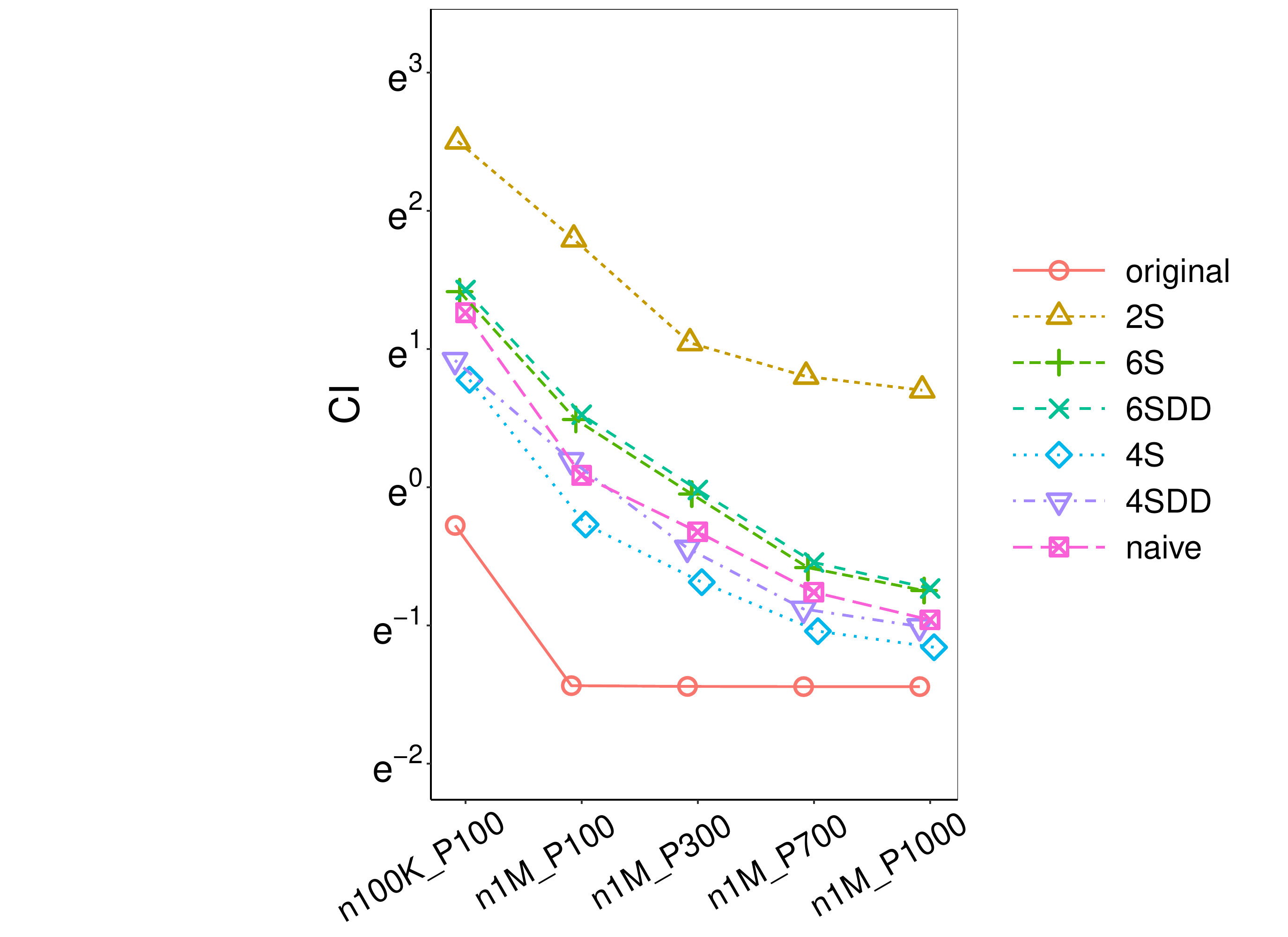}
\includegraphics[width=0.19\textwidth, trim={2.5in 0 2.6in 0},clip] {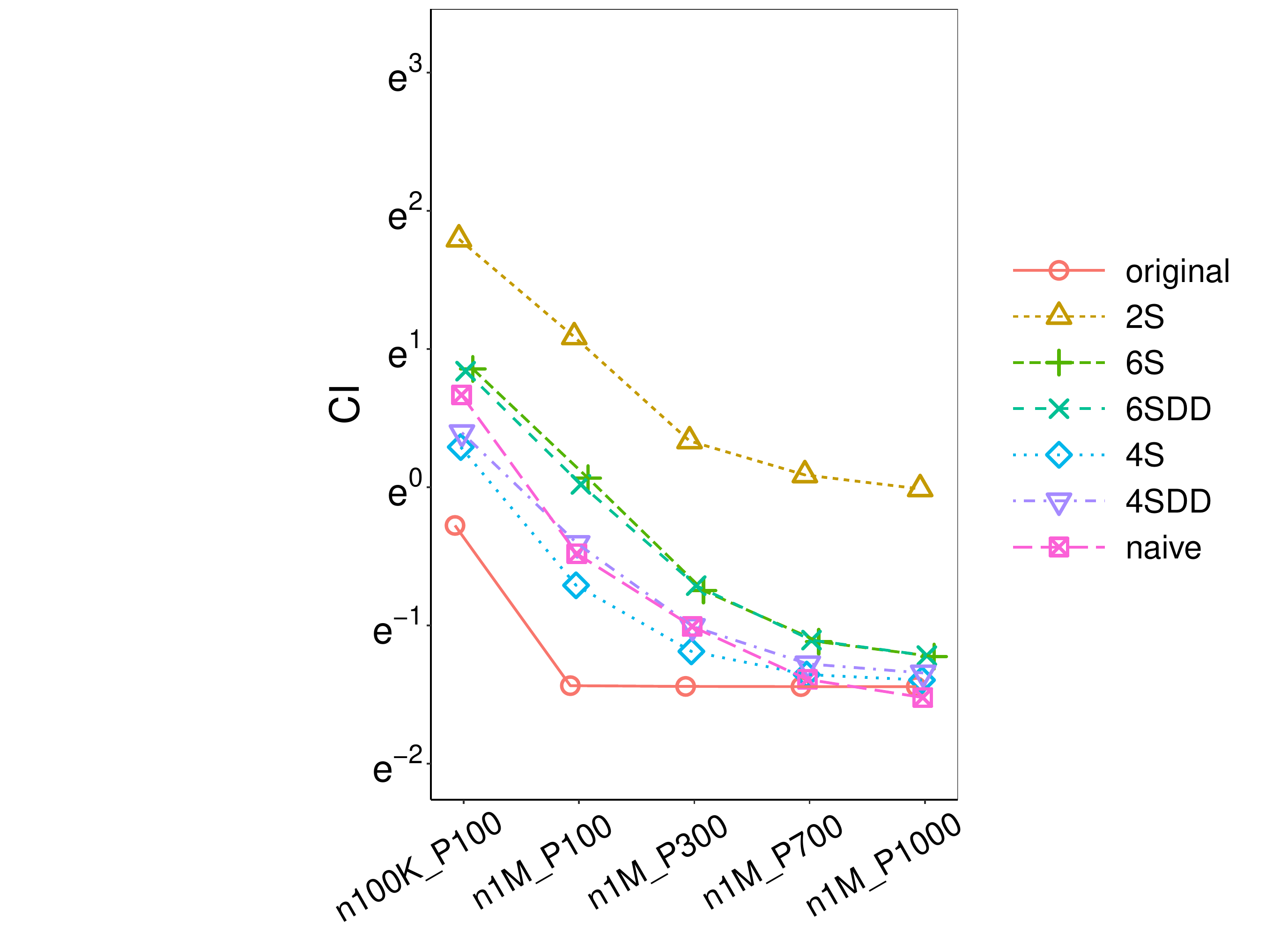}
\includegraphics[width=0.19\textwidth, trim={2.5in 0 2.6in 0},clip] {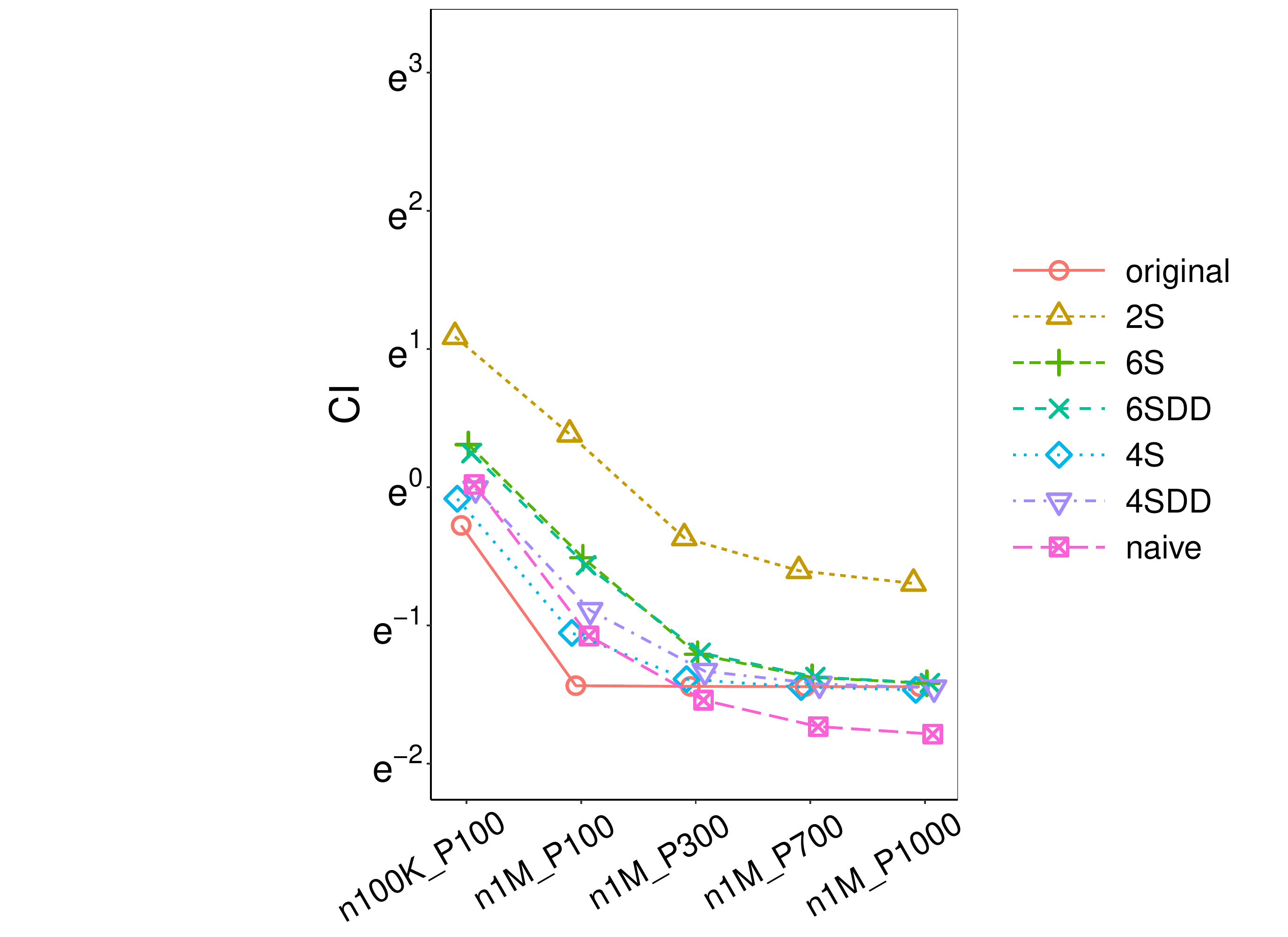}
\includegraphics[width=0.19\textwidth, trim={2.5in 0 2.6in 0},clip] {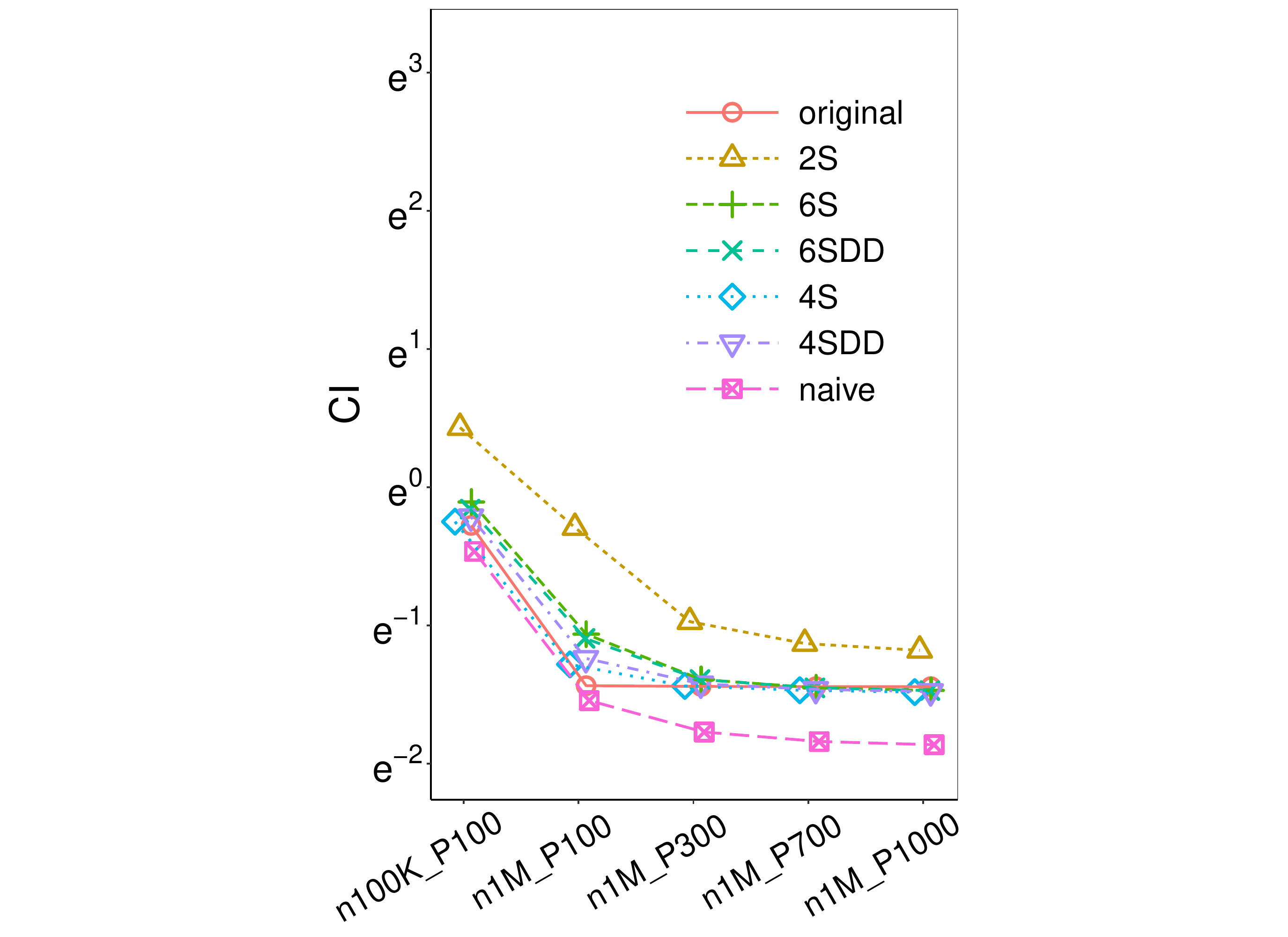}

\includegraphics[width=0.19\textwidth, trim={2.5in 0 2.6in 0},clip] {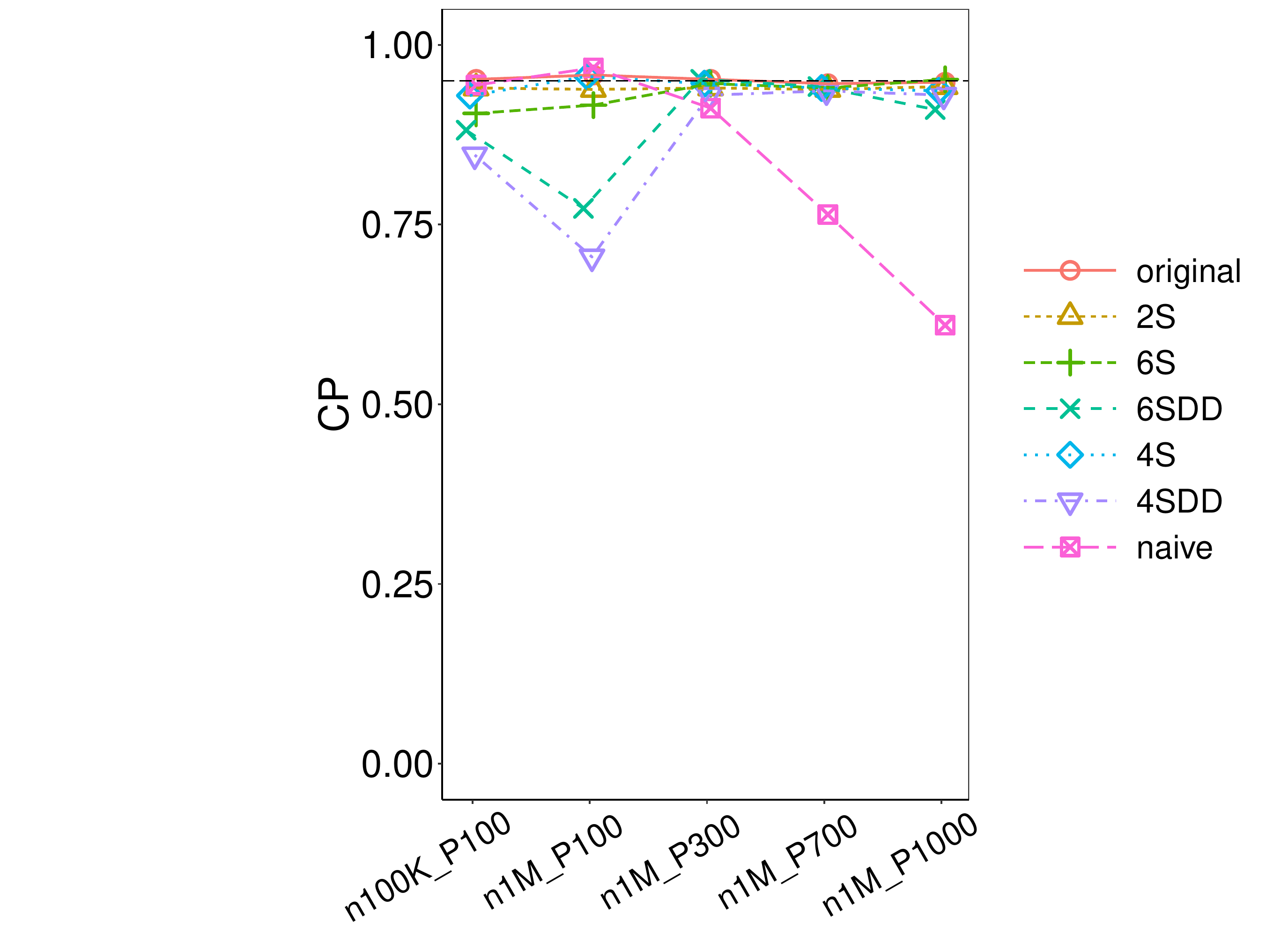}
\includegraphics[width=0.19\textwidth, trim={2.5in 0 2.6in 0},clip] {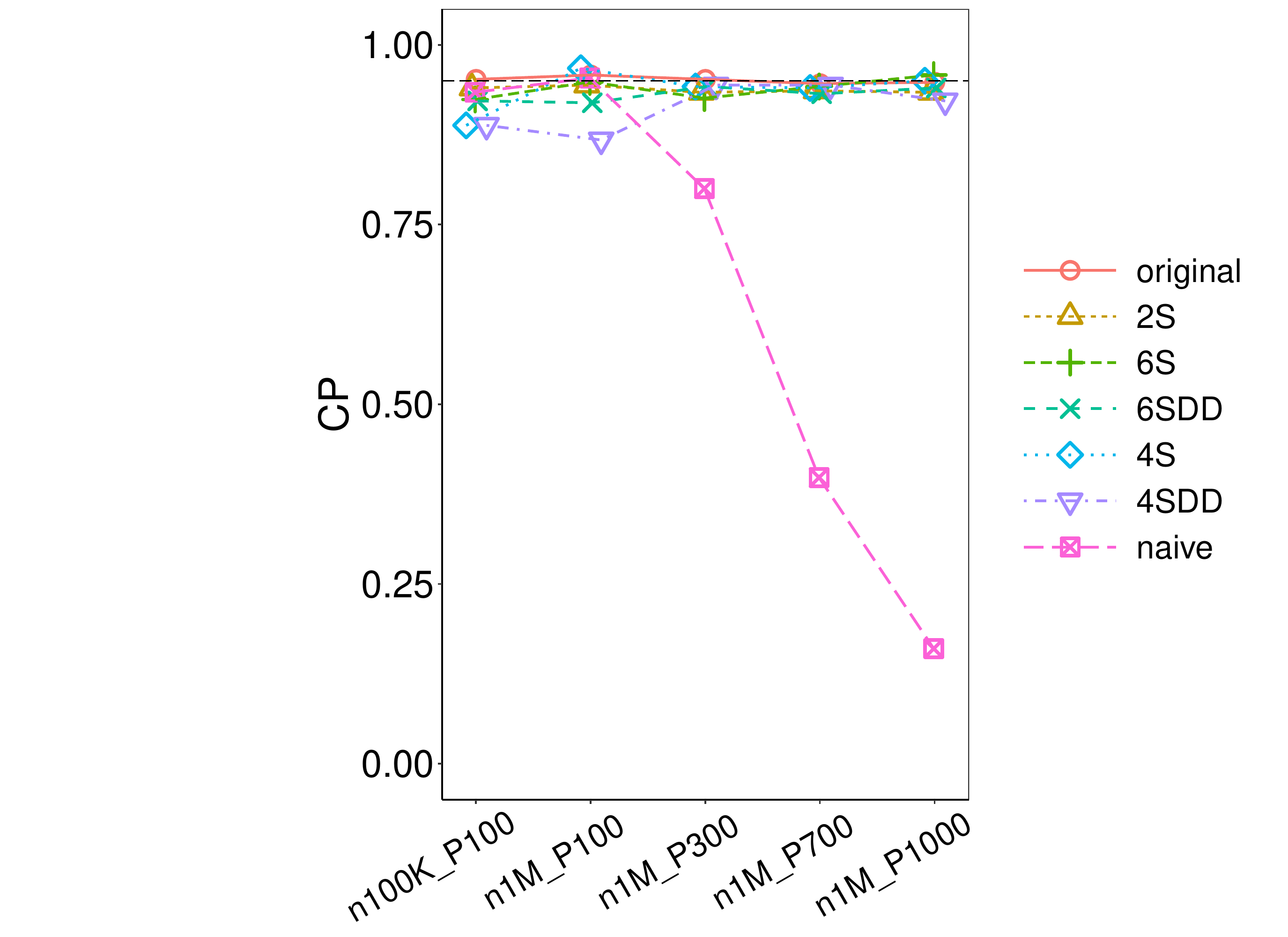}
\includegraphics[width=0.19\textwidth, trim={2.5in 0 2.6in 0},clip] {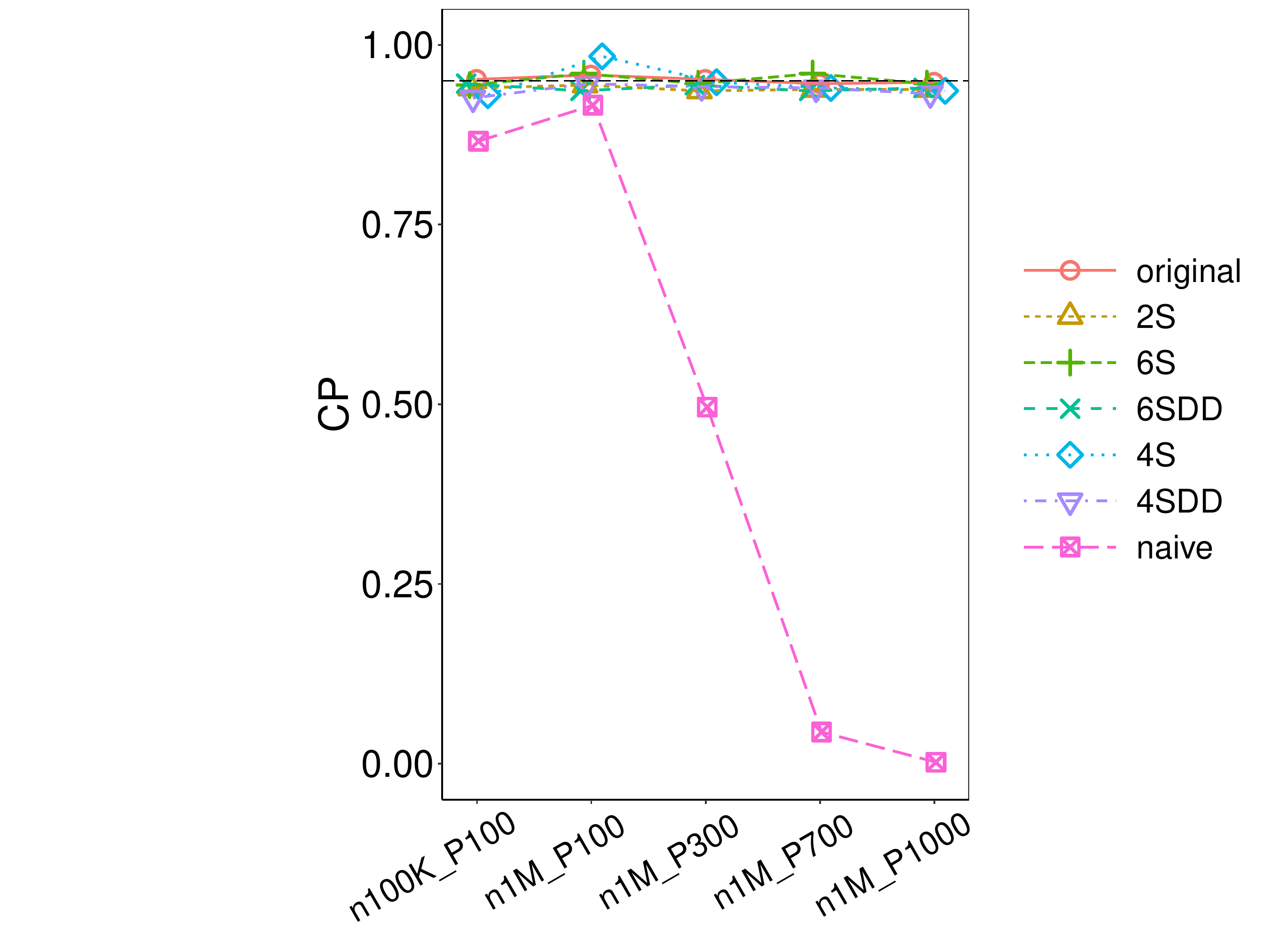}
\includegraphics[width=0.19\textwidth, trim={2.5in 0 2.6in 0},clip] {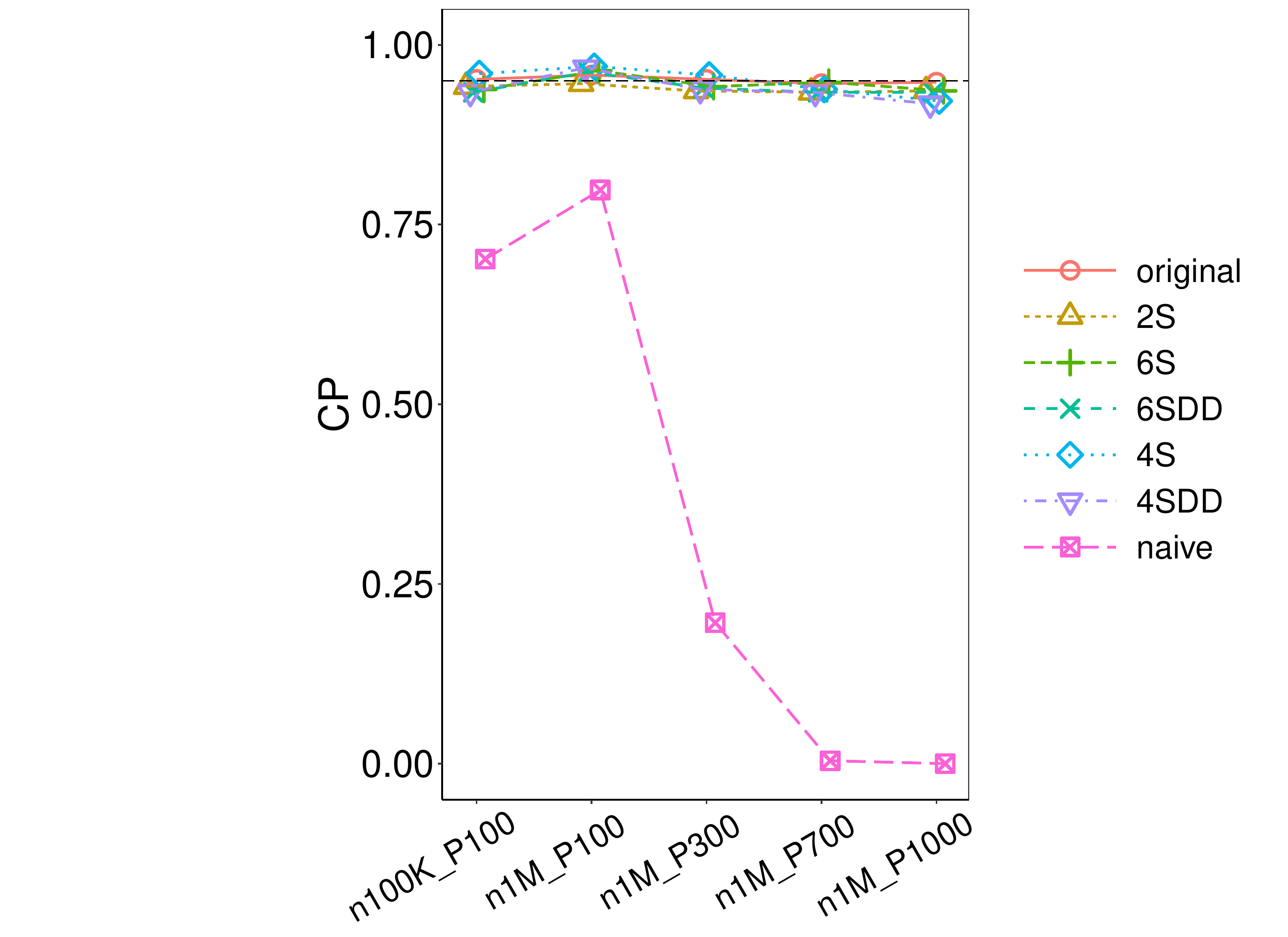}
\includegraphics[width=0.19\textwidth, trim={2.5in 0 2.6in 0},clip] {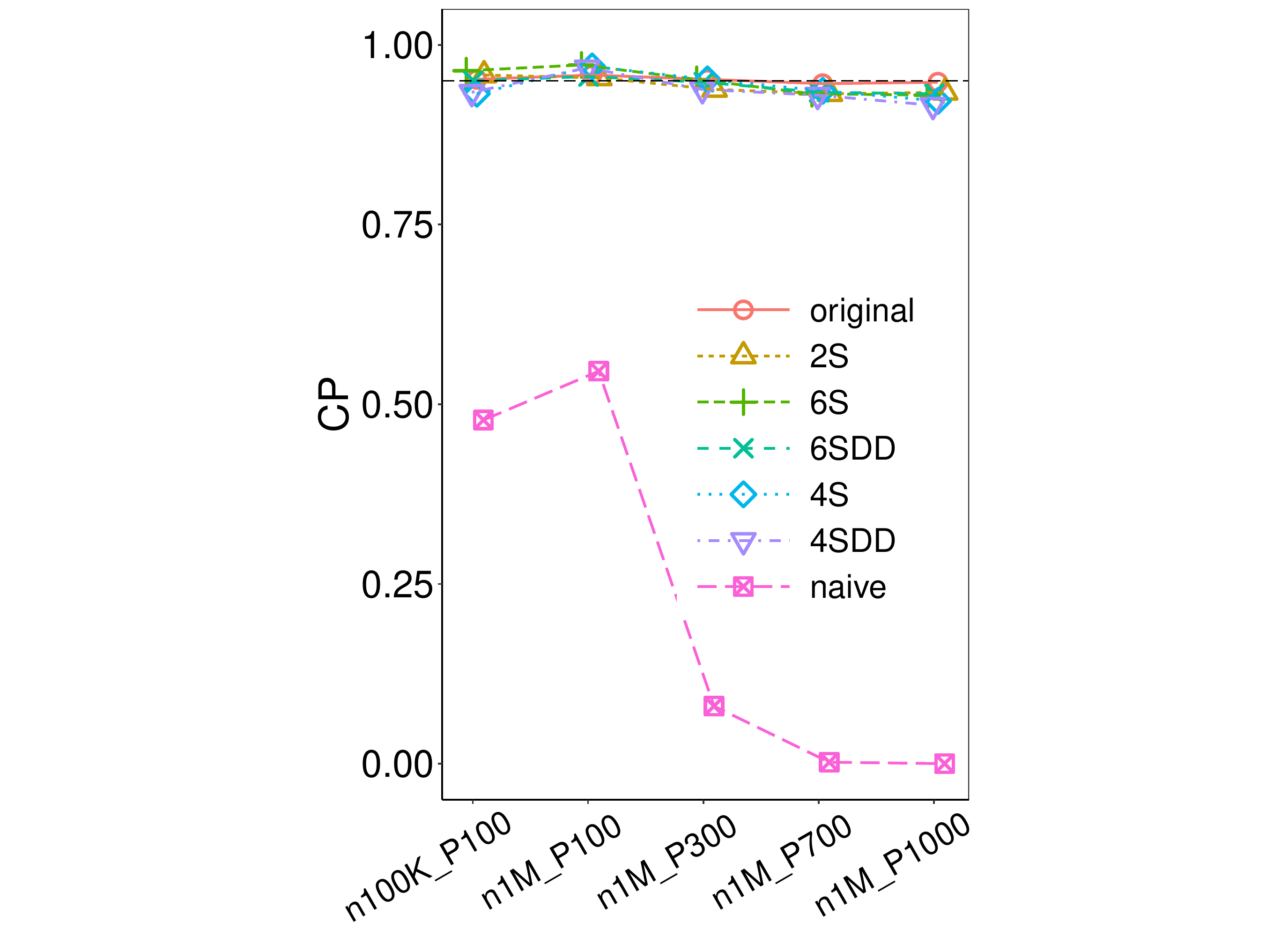}

\includegraphics[width=0.19\textwidth, trim={2.5in 0 2.6in 0},clip] {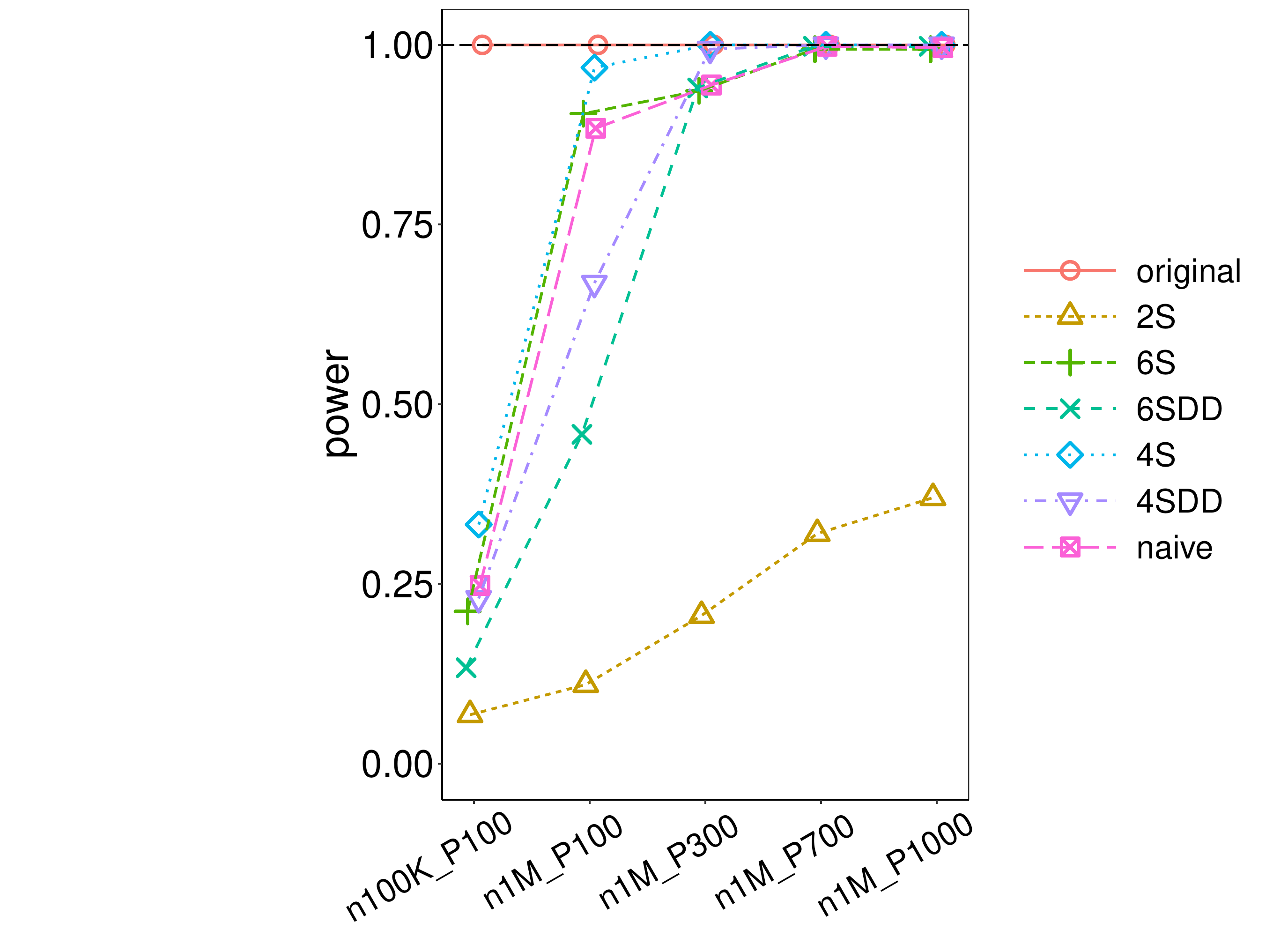}
\includegraphics[width=0.19\textwidth, trim={2.5in 0 2.6in 0},clip] {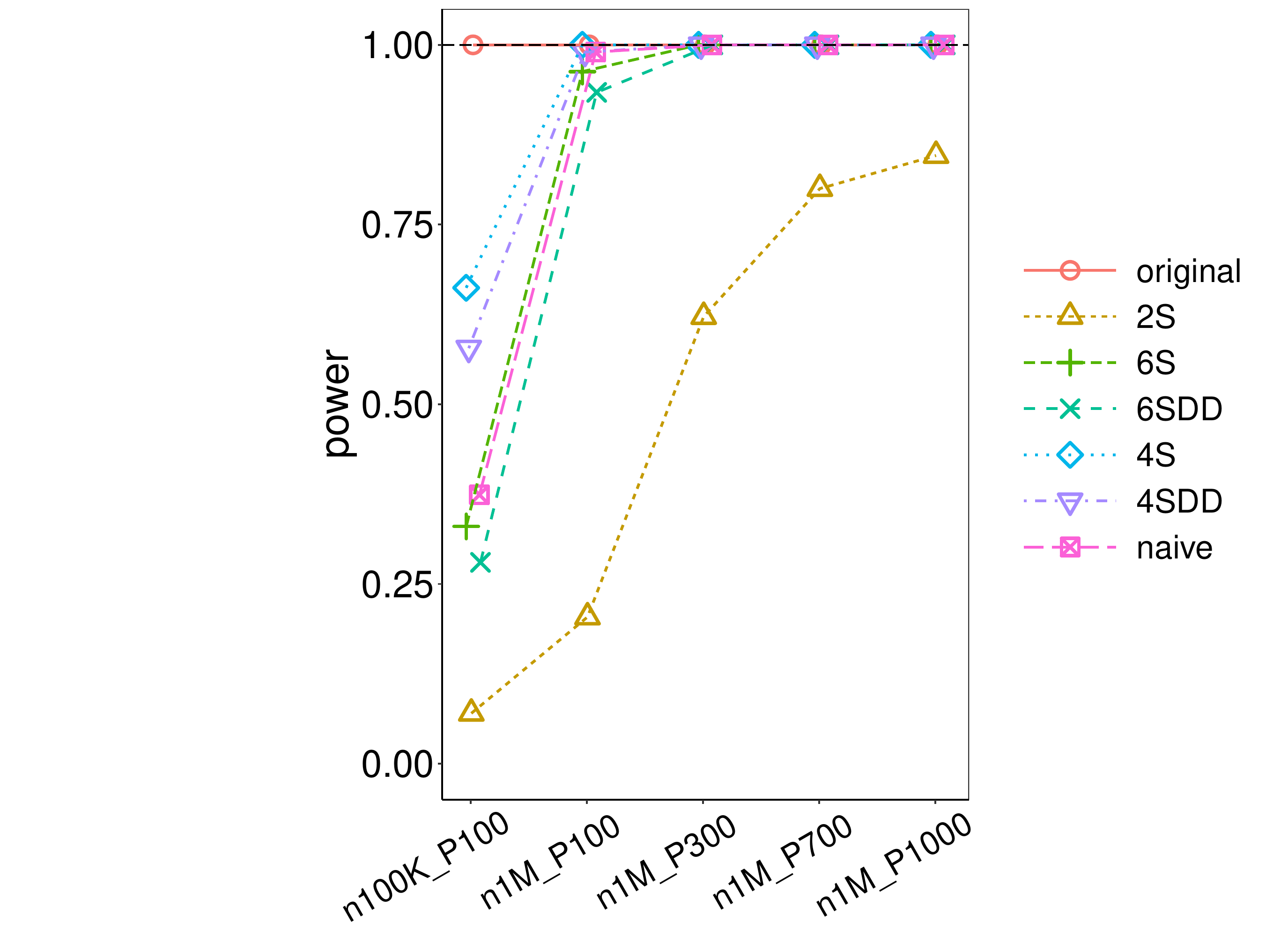}
\includegraphics[width=0.19\textwidth, trim={2.5in 0 2.6in 0},clip] {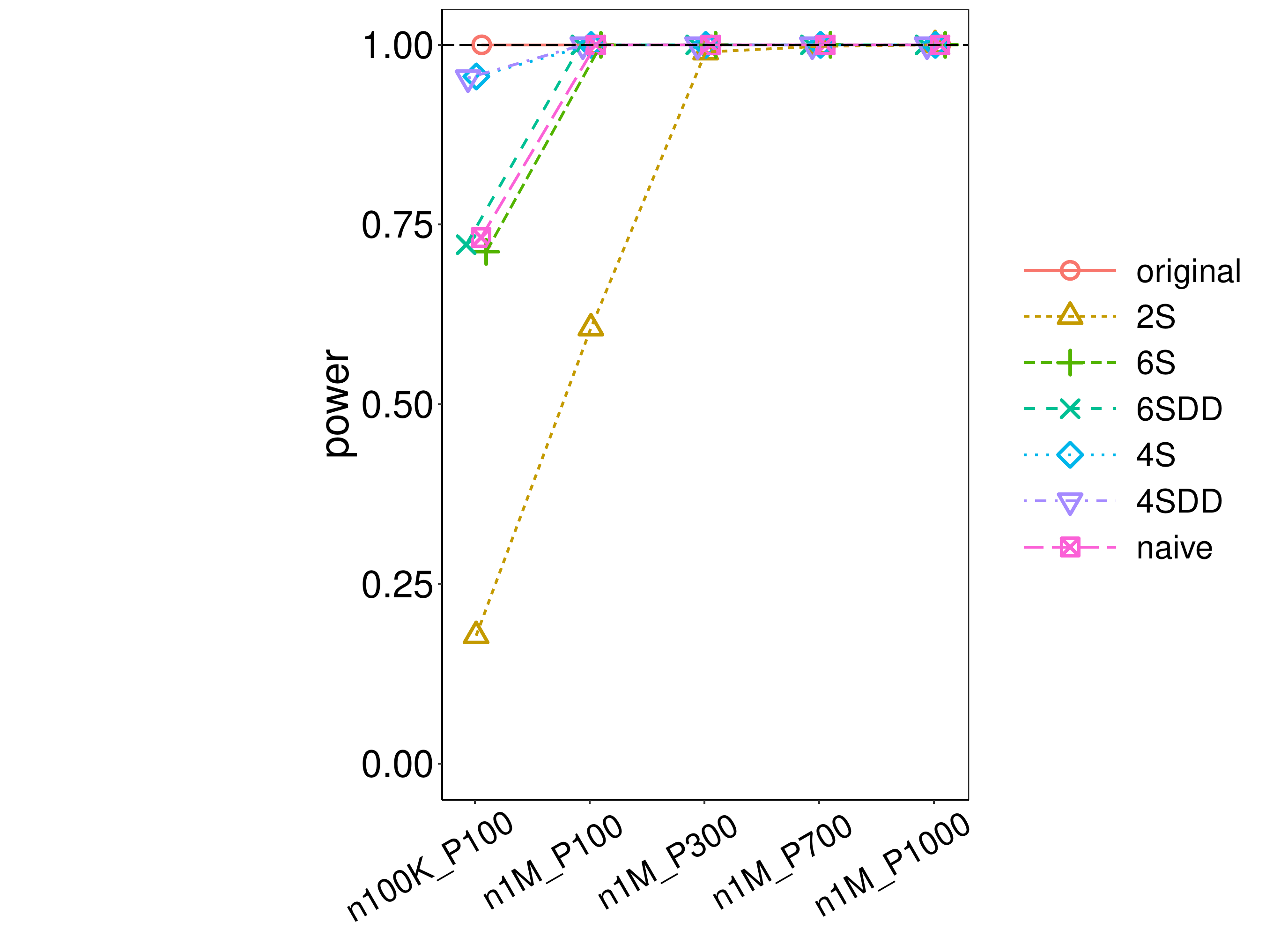}
\includegraphics[width=0.19\textwidth, trim={2.5in 0 2.6in 0},clip] {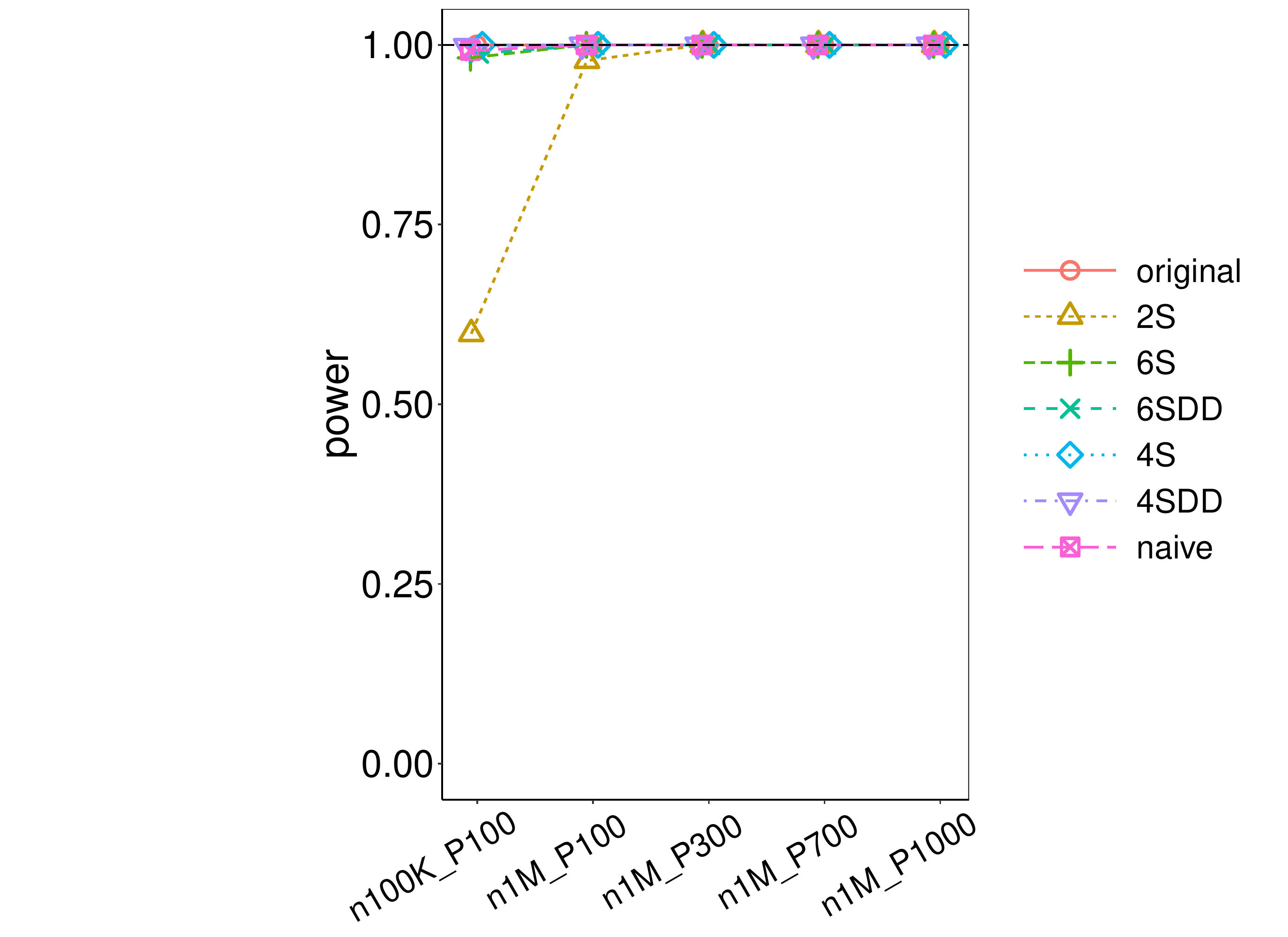}
\includegraphics[width=0.19\textwidth, trim={2.5in 0 2.6in 0},clip] {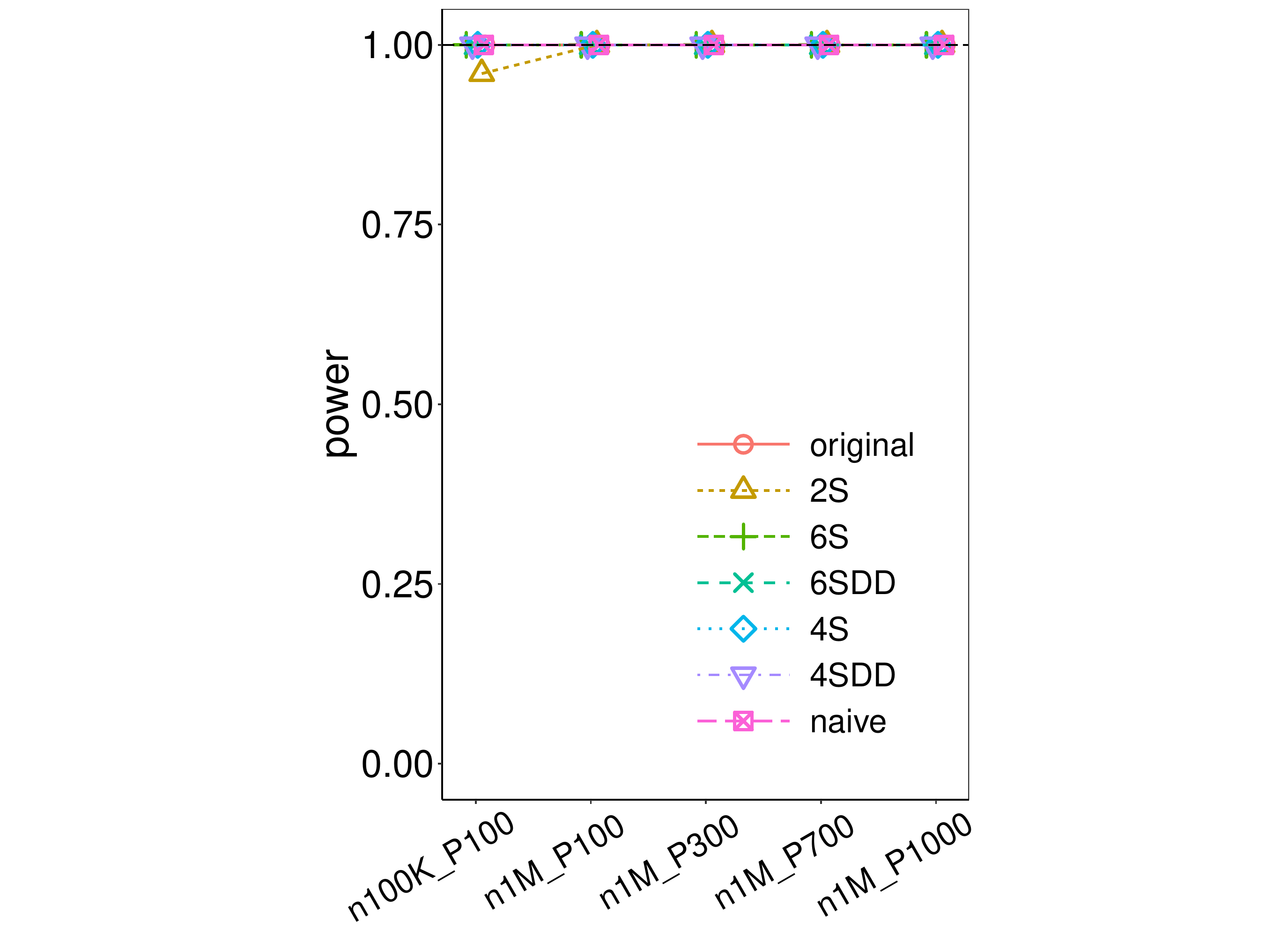}

\caption{Simulation results with $\rho$-zCDP for ZILN data with  $\alpha\ne\beta$ when $\theta\ne0$}\label{fig:1aszCDPZILN}
\end{figure}

The results are presented in Figures \ref{fig:1asDPZILN} and \ref{fig:1aszCDPZILN} and are summarized  as follows. 4S is the best performer by all metrics and the na\"{i}ve method is the worst. In between 4S and na\"{i}ve, in terms of bias, 6S approaches 0 faster as $n$ or $P$ increases when $n=1m$, or $\epsilon$ increases, than 2S, 4SDD, and 6SDD; in terms of RMSE and CI width, 4SDD, 6S and 6SDD perform better than others; regarding CP, all methods except for 4SDD, 6SDD and the na\"{i}ve method can provide nominal level CP for all cases; in terms of power, all methods can provide power much closer to 1 except for 2S, with 4S being the best. The Gaussian mechanism of $\rho$-zCDP does not seem to be advantageous in statistical utility compared to the Laplace mechanism of $\epsilon$-DP. 

\section{A Case Study}\label{sec:real}
We apply our proposed privacy-preserving inference approaches in Section \ref{sec:method} on a real dataset. The dataset  ``train.csv'' contains 10-day click-through data on some mobile ads 
and is available at \url{https://www.kaggle.com/datasets/wuyingwen06/avazu-ctr-train} 
The data comprises 41.4 million impressions in total (each presentation of ads is called an impression). Every impression was recorded for device IP, app ID, site domain, and 12 other individual-level factors. We used two variables from the data ``click'' and ``device\_ip''. The raw data on ``click'' are binary (0 or 1). We grouped clicks by $n=6,729,486$ ``device\_ip''. The count data after grouping ``device\_ip''are heavily right-skewed with 66\% zeros We then randomly split the data in half, each half representing a group (corresponding to the raw data $\mathbf{y}_0$ and $\mathbf{y}_1$). Hence the ground truth is the group mean difference, in this case, is 0.  We then applied our PAC-based methods to obtain privacy-preserving inference on the group mean difference. 

\begin{figure}[!htb]
\centering
\includegraphics[width=0.75\textwidth, trim={0.05in 0 0.4in 0.4in},clip]{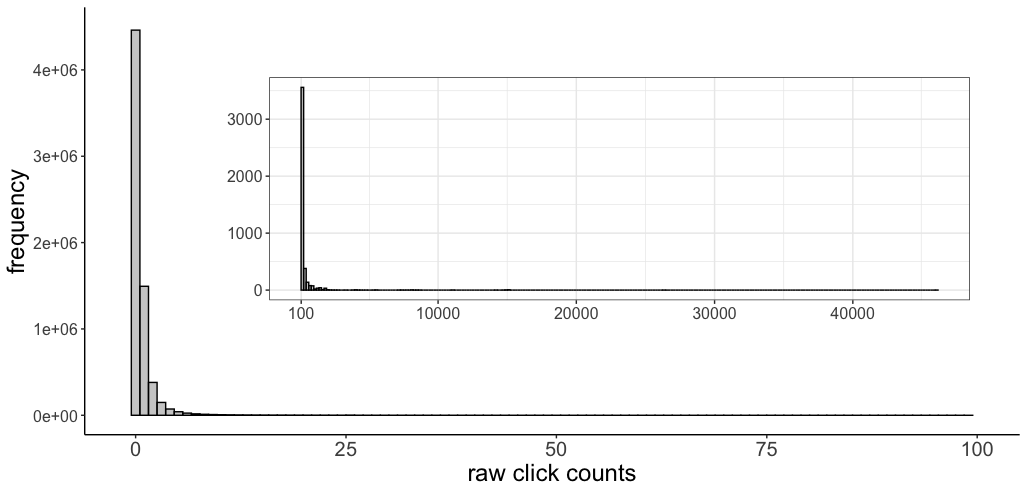}\\
(a) Histogram of raw click counts by IPs \\[6pt]
\includegraphics[width=0.75\textwidth, trim={0.1in 0 0in 0},clip]{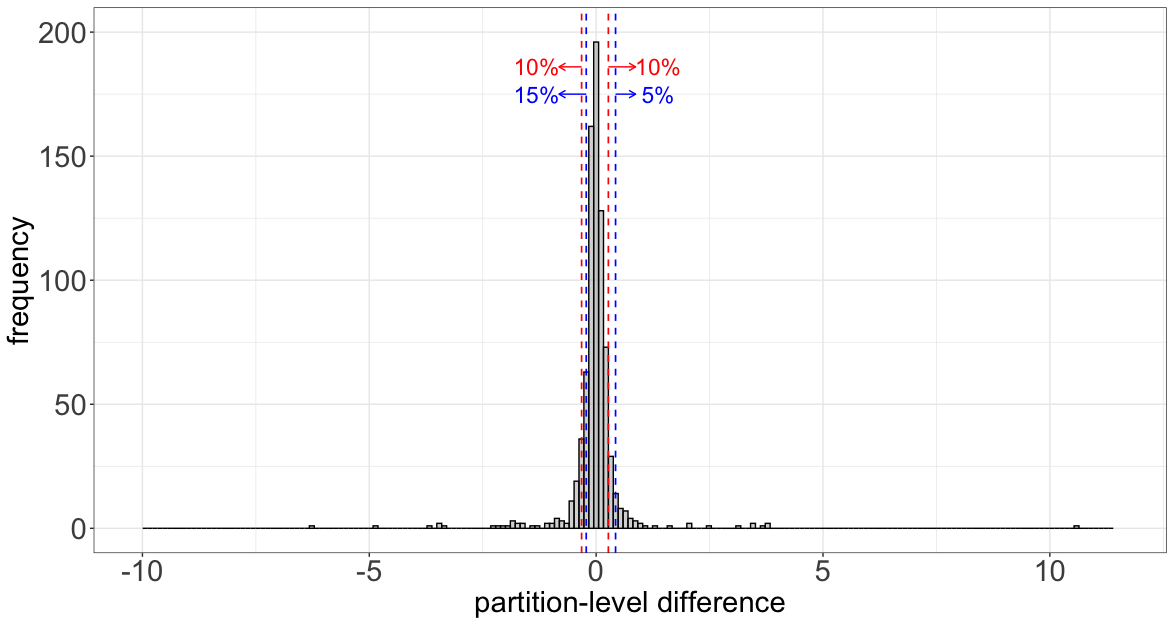}\\
(b) Histogram of partition-level differences between two groups
\vspace{-6pt}
\caption{Histograms in the case study (raw count data and partitioned-level difference)} \label{fig:Hist}\vspace{-12pt}
\end{figure}

We chose $P=800$ and employed the Laplace mechanism for achieving $\epsilon$-DP at $\epsilon=0.5, 1, 2, 5$, and 50, based on the simulation results in Section \ref{sec:simulation}. Figure \ref{fig:Hist} presents the histograms of the raw count data before PAC and of the partition-level differences between the two groups. The raw count is a typical zirs dataset -- a large of zeroes with extreme values on the right tail of the distribution.  After partitioning and differencing the group means  at the partition level, the distribution becomes roughly bell-shaped and symmetric, with a few outlying observations, implying the reasonableness of Gaussian assumptions of the partition-level data on which our privacy-preserving inferences are based. 

We examined  8 privacy-preserving methods (2S, 6S, 6SDD, 4S, 4SDD, winsorized, trimmed, and one na\"{i}ve) as listed in Section \ref{sec:setting} when $\alpha=\beta=0.1$ for the latter 7  PAC-based methods   and 6 privacy-preserving  methods (2S, 6S, 6SDD, 4S, 4SDD, and one na\'ive)  when $\alpha=0.15, \beta=0.05$  for the latter 5 PAC-based methods and used  $m=4$ in the MS approach outlined in Section \ref{sec:inference}. 

Table \ref{tab:casestudy} presents the privacy-preserving inference on the group mean difference from these different approaches.    In general, 4S is the best performer, and the 2S method, even in an overly optimistically unachievable setting, performs the worst. All methods based on the PAC data  -- 6S, 6SDD, 4S, 4SDD, winsorized, trimmed, even na\'{i}ve in this case -- yield comparable estimates on the group difference as the original when $\epsilon\ge 2$.  In addition,  when $\alpha=\beta=0.1$, the estimates in 4S, 4SDD, winsorized, and trimmed means are also comparable to the original inference  at $\epsilon = 1$. When $\epsilon<0.5$, the PAC-based methods yield similar point estimates, larger SEs, and wider CIs at $\epsilon = 0.5$; 6SDD is worse than 6S and yields similar performance to the na\'ive approach. In all cases, regardless of $\alpha,\beta,\epsilon$, the 95\% CIs in all methods contain 0, implying the privacy-preserving inferences in all cases are valid as the truth is $\theta=0$.

\begin{table}[!htb]
\caption{\centering Inference on group mean difference in the case study \newline ($\epsilon$-DP; $n=6,729,486; P=800; m=4$)} \label{tab:casestudy}\vspace{-8pt}
\centering
\resizebox{0.925\textwidth}{!}{
\begin{tabular}{c|c@{}|c|@{\hspace{2pt}}c@{\hspace{3pt}}c@{\hspace{2pt}}|c@{\hspace{3pt}}c@{\hspace{5pt}}c@{\hspace{5pt}}c@{\hspace{5pt}}c@{\hspace{5pt}}@{\hspace{5pt}}c@{\hspace{5pt}}c@{\hspace{5pt}}c@{\hspace{5pt}}c@{}}
\hline
$\alpha$&&& \multicolumn{2}{c|}{PnoC data } & \multicolumn{7}{c}{ PAC data}\\
\cline{3-14}
$\beta$ & Metric & $\epsilon$& original & 2S$^\dagger$ & \multicolumn{2}{c}{original} & 6S & 6SDD & 4S & 4SDD & wins$^\ddagger$.& trim.$^\ddagger$ & na\"ive\\
 &&&&& wins.$^\ddagger$ & trim.$^\ddagger$ &\\
\hline
&\multirow{5}{*}{ estimate } & 0.5 & & -3.534 && &-0.018 &-0.090 &-0.144 &-0.191  &0.088  &-0.102 &-0.066  \\ 
&&1 && 0.735 &&&-0.017  &0.053  &0.009  &0.006  &-0.039  &-0.047 &0.086\\ 
&&2& 0.007 &-0.517 & -0.005 & 0.001  &-0.003  &-0.001  &0.005 &0.006  &0.000  &-0.003 &-0.032\\ 
&&5&  & 0.403 &&&-0.003  &0.004  &0.002  &0.005  &0.005  &0.002 &0.000\\ 
&&50& & -0.007 &&&0.004  &0.004  &0.004  &0.003  &0.004   &0.001 &0.001\\
\cline{2-14}
&& 0.5&& 2400 &&& 98.06 & 60.12& 84.36& 204.3 &142.6 & 152.1 &150.4\\  
&& 1&& 2042 &&& 25.51 & 61.88 & 21.40 &  22.82 & 52.89 & 59.81 & 99.29\\
&SE & 2  & 25.46 & 921.2 &7.853 &7.852 & 27.14 & 12.26 &10.63  &12.15 &10.54 & 10.67 & 34.85\\ 
&($10^{-3}$) &5&& 489.8 &&& 10.09 & 8.083 & 8.905 & 8.422  & 9.523  &10.44  &4.766\\ 
0.1 &&50& & 66.31 &&& 7.770 & 7.811  &7.721  &7.717  &7.927   &7.927  &4.687\\
\cline{2-14}
0.1 &&0.5&& -11.17 && &-0.299 &-0.261 &-0.342 &-0.802 &-0.267 & -0.491 &-0.545\\
&&1& & -5.569 && &-0.069 &-0.129 &-0.033 &-0.040  &-0.154  &-0.185 &-0.230\\
&95\% CI &2& -0.043 & -3.449 & -0.020 & -0.015 &-0.081&-0.028&-0.018&-0.021&-0.022&-0.025&-0.140\\
&lower&5&&  -1.125 &&&-0.024 &-0.012 &-0.016 &-0.012  &-0.015  &-0.020 &-0.009\\ 
& bound&50& & -0.141 &&&-0.011 &-0.011 &-0.011 &-0.012 & -0.011 &-0.015 &-0.009 \\
\cline{2-14}
&&0.5&&  4.104 && &0.263 &0.081 &0.053 &0.420 &0.444 &0.286 &0.412 \\
&&1&  & 7.038 && &0.035 &0.234 &0.051 &0.052  &0.076 &0.090 &0.402\\
&95\% CI &2&0.057 & 2.414 & 0.011 & 0.016 &0.074 &0.026 &0.027 &0.033 &0.022 &0.019 &0.076 \\ 
&upper&5&&  1.931  &&&0.018 &0.020 &0.020 &0.022 &0.024 &0.024 &0.009\\
& bound&50& & 0.127 &&  &0.019&0.019 &0.019 &0.019  &0.020 &0.016 &0.010 \\
\hline 
&\multirow{5}{*}{ estimate } &0.5&& -2.876 &&&-0.115  &0.091 &-0.030 &-0.037 & & &  -0.048  \\ 
&&1& & -0.520 &&& 0.027 &-0.070 &0.018 &0.074 &  & & 0.014\\ 
&&2&0.007 & 1.095 &NA&NA&-0.002  &0.048  &0.028  &0.044  & NA& NA & 0.095\\ 
&&5& & -0.057 &&& 0.012  &0.011  &0.012  &0.013 &  & & 0.043\\ 
&&50& & 0.008 &&&0.014  &0.014 & 0.013  &0.014 &  & & 0.041\\
\cline{2-14}
&&0.5& & 3210  &&&53.93 & 82.99 & 23.57 & 31.02 &  & & 159.0\\ 
&&1& & 1482 &&&42.68 &57.26 &29.27  &54.98 &  & & 153.4\\
&SE &2& 25.46 & 1488&NA&NA &16.42 &80.28 &21.35 &29.86 & NA& NA& 120.8\\ 
&($10^{-3}$) &5&& 220.9 &&&11.69 &11.05 &8.825  &9.047 &  & & 5.314\\
0.15 & &50& & 59.25 & &&8.556 & 8.591 & 8.622 & 8.605 &  & & 5.350\\
\cline{2-14}
0.05 && 0.5 & & -12.95 &&& -0.263 &-0.163 &-0.086 &-0.115 &  & & -0.553\\ 
&&1&& -5.112&& &-0.081 &-0.238 &-0.042 &-0.066 &  & & -0.473\\ 
&95\% CI &2&-0.043 & -3.639 &NA&NA&-0.038 &-0.202 &-0.015 &-0.025 & NA& NA & -0.288\\ 
&lower &5& & -0.566 &&&-0.013 &-0.012 &-0.005 &-0.004 &  & & 0.030\\
&bound &50&& -0.114 &&&-0.002 &-0.003 &-0.004 &-0.003  & & &  0.031\\
\cline{2-14}
&& 0.5 & & 7.201 &&&0.034 &0.345 &0.025 &0.040 &  & & 0.457\\ 
&&1&& 4.072 &&&0.136 &0.099 &0.078 &0.214&  & & 0.501\\ 
&95\% CI &2&0.057 & 5.829 &NA&NA&0.034 &0.298 &0.071 &0.113 & NA& NA & 0.479\\ 
&upper &5& & 0.452 &&&0.036 &0.034 &0.029 &0.031 &  & & 0.053\\ 
&bound &50&& 0.130 &&&0.031 &0.031 &0.030 &0.030 & & &  0.051\\
\hline
\end{tabular}}
\resizebox{0.925\textwidth}{!}{
\begin{tabular}{l}
$^\dagger$ The implementation of 2S requires the specification of global bounds $U$ and $L$ for  the difference  between the two  \\
groups at the partition level. The local minimum and maximum of the difference  in  the dataset  are  -45987 and  \\
45987;  the global bounds can only be wider than the local bounds. We first run 2S  with global  bounds $L=-50K$ \\
and $U=-50K$, and the results were basically useless with extremely large  SE  values  and wide  CIs   for the group  \\
mean difference. For that reason, we used a much narrow global 
bound   $(L,U)= (-100,100)$, on which the results\\
in the  table are based. We understand the bounds are unrealistically  optimistic and unjustified,  and  represent\\
an absolutely best scenario that 2S could impossibly  achieve. The  inference in  this hypothetical scenario is even \\
worse than the PAC-based methods, implying data censoring is an effective way to  preserve  information in \\ 
this type of analysis with DP guarantees.\\
$^\ddagger$ wins. = winsorized; trim. = trimmed.\\
\hline
\end{tabular}}
\vspace{-6pt}
\end{table}

An interesting observation is that the width of the privacy-preserving CIs  and the SE values of the estimated mean difference at large $\epsilon$ values for the PAC-data-based methods are even smaller than those based on the original data without DP guarantees. This is because the data is rightly skewed with extreme-valued and outlying data points (Figure \ref{fig:Hist}(a)). The influence of these outlying observations, though mitigated after taking the averaging at the partition level, still exists at the partition-level data $Z$ (Figure \ref{fig:Hist}(b)). The trimmed mean and winsorized mean are well-known statistical methods that limit the effect of outliers or extreme values on estimates.in the case of $\alpha=\beta$, the SE values and the CI widths of trimmed mean and winsorized mean are smaller than those based on the uncensored data without DP (original). When $\epsilon>3$, 6S, 6SDD, 4S, and 4SDD are similar to the trimmed mean and winsorized mean and are robust to the influence of the outliers on the mean estimates. As $\epsilon$ decreases, the sanitization randomness starts to take over and inflates the SE and CI width in these methods.

\section{Discussion}\label{sec:discussion}
We present six privacy-preserving inferential approaches for group mean differences in zirs data. Four methods are likelihood-based and two are model-free. Our recommendation regarding the usage of these methods, based on theoretical analysis and empirical studies, is as follows. The 4S method would be the first choice in general. When censoring is symmetric ($\alpha=\beta$), $n\le100k$, and $\epsilon\le1$, winsorized mean performs slightly better than 4S and can be considered first. 

We establish MSE consistency for the privacy-preserving estimators as $P\rightarrow\infty$ with the regularity conditions of a large $n/P$ ratio for the CLT to hold in each partition and $P=o(n)$. How large is considered ``large enough'' for $n/P$ depends on the distribution of raw data. If the raw data is Gaussian, $n/P$ can be small as normality holds even without the CLT; if the raw data is as irregular as zirs data, our recommendation  is $n/P\ge10^3$, at least for the data similar to what's examined in our simulation studies and case study. As for $P$, we recommend $P\ge100$.  

Whether to use symmetric or asymmetric censoring in practical applications depends on several factors. If the distribution of partitioned-level differences is approximately normal,  the four likelihood-based privacy-preserving inferential approaches are valid regardless of symmetric or asymmetric censoring and the decision would be mostly driven by privacy consideration. Specifically, the default can be symmetric censoring; if there is more privacy concern regarding the observations on one tail of the distribution, a larger censoring percentage can be applied to the more sensitive tail. If the distribution has more outlying observations on one tail of the distribution, to mitigate the influence of the outliers and obtain a more robust estimate for the mean, we also recommend using asymmetric censoring and applying a larger censoring percentage to the fatter tail.

If partitioning is natural (e.g., each partition represents one local server), so are paired differences between the two groups in each partition. For manual partitioning, the raw data within each group can be randomly partitioned and the partitions between the two groups can be paired in a random manner to obtain the partition-level differences. The random partitioning and pairing would not bias the estimation of the group mean difference. In addition, the variability from random partitioning and pairing is rather small compared to the sampling error of the raw data and the sanitization uncertainty for privacy guarantees and is thus ignorable in the inferential process. On the other hand,  random partitioning and pairing are expected to yield some formal privacy guarantees though the associated privacy protection, if quantifiable, may not contribute significantly to the overall privacy guarantees. Nevertheless, this is an interesting topic that warrants further investigation. 

Our simulation studies suggest there is no significant difference in the utility of privacy-preserving inferences obtained via $\rho$-zCDP vs. $\epsilon$-DP at similar privacy guarantees. This may be due to that the number of sanitizations involved in each method is small (2 to 6), $P$ is relatively large, and we do not examine very small privacy loss ceases, where the advantage of  $\rho$-zCDP  over $\epsilon$-DP in privacy loss composition is not obvious.

A variation of the proposed PAC-based  methods is privacy-preserving inference based on the partition-level group mean data directly rather than the group differences in each partition, with or without censoring  (but censoring would help improve the utility of privacy-preserving inference with lowered global sensitivities for some statistics, as in the case of censoring the difference). The likelihood is formulated based on two groups of partition-level data, containing 4  unknown parameters (mean and variance in each of the two groups). The likelihood is then sanitized, from which MLE or Bayesian inferences n the mean difference can be obtained. The benefit of this approach is that it is easier to extend to the case comparing means of  multiple group ($>2$). Especially in the Bayesian framework, once the posterior samples are obtained for the group means, the mean difference estimates and the associated interval estimates can be obtained for any pair of groups. We conjecture the MSE consistency established in this work given the partition-level group mean difference data can apply directly  to the setting of partition-level group mean data, as the normality assumption is about partition-level means not on their differences. On the other hand, the number of sanitized statistics doubles with this approach, in the case of 2 groups, compared to the 6 approaches in this work. In addition, if the partition-level data in each group is not quite symmetric (e.g., when the raw data has a high zero-inflation percentage), differencing would help symmetrizing the data; in other words, the distribution of the partition-level group mean difference, compared to that of the partition-level mean in each group, would be more symmetric. We plan to explore this approach and compare it to the methods in this work  in the future.

The methods proposed in this work  assume independence among the raw data, among different partitions, and between the two groups for comparison. We plan to extend the methods to cases when the independence assumption does not hold. Especially for data collected from social media, where users are often related and form networks or graphs. For example, one user's  reaction to an ad may be viewable by his/her followers and may  thus influence their reactions to the same ad if they happen to participate in the same study. Besides the relations among the users in networks, users may also be clustered geographically, organizationally, etc. How to obtain privacy-preserving inference in data with complex relational information among individuals is of great interest and in need of valid and efficient approaches.

\small{ \section*{Acknowledgment and Disclaimer}
The work was motivated by the work done by Georgina Evans during her summer internship at Meta supervised by Jiming Paul Li and James Honaker.

Only non-Meta authors downloaded and accessed the case study dataset in Section \ref{sec:real}.  No data has resided on Meta’s servers or facilities.}

\newpage





\clearpage

\bibliographystyle{plainnat}
\bibliography{ref.bib}

\clearpage

\textbf{\LARGE{Supplementary Materials}}
\vspace{15pt}

\normalsize
The supplementary contains  two main parts. The first part contains the proofs of the lemmas, claims, and theorems in Section 4; the second part contains additional simulation results that supplement those in Section 5.

\vspace{15pt}
\section*{Part I: Proofs}
\subsection*{Proof of Theorem 1}
\begin{proof}
Since  $z_1, z_2, \ldots, z_P$ are a random sample from $\mathcal{N}(\theta, \frac{\sigma^2}{n/P})$,
\begin{equation}
\hat{\theta}
=\frac{\sum_{j=1}^{P}z_j}{P} = \frac{s_1}{P}\sim \mathcal{N}(\theta, \frac{\sigma^2}{n}).
\end{equation}
Therefore, $\mathbb{E}_{\mathbf{z}} \mathbb{E}_{\mathcal{M}|\mathbf{z}}(\hat{\theta}-\theta)^2
=\sigma^2/n$. If the Laplace mechanism is applied to sanitize $s_1$, then $s_1^*\sim \mbox{Lap}(s_1,\frac{U-L}{\epsilon/2})$ and
\begin{equation}
\mathbb{E}_{\mathcal{M}:Lap|\mathbf{z}}(\hat{\theta}^*-\hat{\theta})^2
= \frac{1}{P^2} \left(\frac{U-L}{\epsilon/2}\right)^2.
\end{equation}
Per the Cauchy-Schwarz inequality, 
\begin{equation*}
\begin{aligned}
\mathbb{E}_{\mathbf{z}} \mathbb{E}_{\mathcal{M}:Lap|\mathbf{z}}(\hat{\theta}^* - \theta)^2
&\leq \mathbb{E}_{\mathbf{z}} (\hat{\theta} - \theta)^2+ \mathbb{E}_{\mathbf{z}} \mathbb{E}_{\mathcal{M}:Lap|\mathbf{z}}(\hat{\theta}^* - \hat{\theta})^2  \\\
& \quad+ 2\sqrt{\mathbb{E}_{\mathbf{z}}\mathbb{E}_{\mathcal{M}:Lap|\mathbf{z}}(\hat{\theta} - \theta)^2 \cdot \mathbb{E}_{\mathbf{z}}(\hat{\theta}^*- \hat{\theta})^2}\\
& \leq \frac{\sigma^2}{n} + \frac{1}{P^2} \left(\frac{U-L}{\epsilon/2}\right)^2 + 2\sqrt{\frac{\sigma^2}{n} \cdot \frac{1}{P^2} \left(\frac{U-L}{\epsilon/2}\right)^2}\\
&=O\left(n^{-1}\right)+O\left(P^{-2}(U-L)^2\epsilon^{-2}\right)+O\left((U-L)n^{-1/2}P^{-1}\epsilon^{-1}\right).
\end{aligned}
\end{equation*}

If the Gaussian mechanism is applied instead to sanitize $s_1$, then $s^*_1 \sim \mathcal{N}(s_1,\frac{(U-L)^2}{2\rho})$ and 
\begin{align}
\mathbb{E}_{\mathcal{M}:Gaus|\mathbf{z}}(\hat{\theta}^*-\hat{\theta})^2
= \frac{1}{P^2}\frac{(U-L)^2}{2\rho}.
\end{align}
Similarly by the Cauchy-Schwarz inequality, 
\begin{align*}
\mathbb{E}_{\mathbf{z}} \mathbb{E}_{\mathcal{M}:Gaus|\mathbf{z}}(\hat{\theta}^* - \theta)^2
& \leq \frac{\sigma^2}{n} + \frac{1}{P^2}\frac{(U-L)^2}{2\rho} + 2\sqrt{\frac{\sigma^2}{n} \cdot \frac{1}{P^2}\frac{(U-L)^2}{2\rho}}\\
&=O\left(n^{-1}\right)+O\left(P^{-2}(U-L)^2\rho^{-1}\right)+O\left((U-L)n^{-1/2}P^{-1}\rho^{-1/2}\right).
\end{align*}
\end{proof}

\subsection*{Proof of Lemma 3}
\begin{proof}
Given a sample dataset $x_1,\ldots,x_P$, since $\lim _{P \rightarrow \infty}\nint{qP}/P=q \in (0,1)$, per Theorem 3 in \citet{smirnov1949limit}, 
\begin{equation}\label{eqn:smirnov}
    x_{(\nint{qP})} \stackrel{a . s.}{\longrightarrow} F_X^{-1}(q) \text{ as } P\rightarrow\infty
\end{equation}
at rate $P^{-1/2}$. Let $y_{(\nint{qP})}$ be the $\nint{qP}^{th}$ order statistic in a random sample of size $P$ from uniform(0, 1). Therefore, $y_{(\nint{qP}+1)}-y_{(\nint{qP})}\stackrel{a . s.}{\longrightarrow}0 \text{ as } P\rightarrow\infty$.

Per \citet{nagaraja2015spacings},
\begin{align}\label{eqn:uniform}
P \cdot(y_{([q P]+1)}-y_{(\nint{qP})})\stackrel{d}{\longrightarrow}\mbox{exp}(1),
\end{align} 
where $\mbox{exp}(1)$ represents an exponential random variable with rate parameter 1.

In addition, $\forall 1\leq i \leq P$, $x_{(i)} \stackrel{d}{=}F_X^{-1}\left(y_{(i)}\right)$. Then 
\begin{align}\label{eqn:gaptouniform}
(x_{(\nint{qP}+1)}-x_{(\nint{qP})}) &
\stackrel{d}{=}F_X^{-1}\left(y_{(\nint{qP}+1)}\right)-F_X^{-1},\notag\\
P\cdot(x_{(\nint{qP}+1)}-x_{(\nint{qP})})&
\stackrel{d}{=}\frac{F_X^{-1}\left(y_{(\nint{qP}+1)}\right)-F_X^{-1}\left(y_{(\nint{qP})}\right)}{\left(y_{(\nint{qP}+1)}-y_{(\nint{qP})}\right)} \cdot P \cdot\left(y_{(\nint{qP}+1)}-y_{(\nint{qP})}\right),
\end{align}
where $\stackrel{d}{=}$ stands for ``equal in distribution'', meaning two random variables have the same distribution. Per the definition of  pdf and the assumptions around $f_X$,
\begin{equation} \label{eqn:derivative}
\frac{F_X^{-1}\left(y_{([q P]+1)}\right)-F_X^{-1}\left(y_{([q P])}\right)}{\left(y_{(\nint{qP}+1)}-y_{(\nint{qP})}\right)} \stackrel{a . s.}{\longrightarrow} \frac{1}{f_X(F_X^{-1}(q))}
\end{equation}
Plugging Eqn \eqref{eqn:uniform} and Eqn \eqref{eqn:derivative} into Eqn \eqref{eqn:gaptouniform}, along with Slutsky's Theorem, we have
$$P\cdot(x_{(\nint{qP}+1)}-x_{(\nint{qP})}) \stackrel{d}{\longrightarrow}
\frac{1}{f_X(F_X^{-1}(q))}\mbox{exp}(1).$$
\end{proof}

\section*{Proof of Theorem 4}
\begin{proof}
Let $x_{(\nint{qP})}$ be the original quantile at $q$, 
\begin{align} 
& \quad \notag \mathbb{E}_{\mathcal{M},\mathbf{x}}\left(x_{(\nint{qP})}^*-F_X^{-1}(q)\right)^2= 
\mathbb{E}_{\mathbf{x}} \mathbb{E}_{\mathcal{M}|\mathbf{x}}\left(x_{(\nint{qP})}^*-F_X^{-1}(q)\right)^2 \\
& \notag =\mathbb{E}_{\mathbf{x}} \mathbb{E}_{\mathcal{M}|\mathbf{x}}\left(x_{(\nint{qP})}^* - x_{(\nint{qP})} + x_{(\nint{qP})}-F_X^{-1}(q)\right)^2\\
&= \mathbb{E}_{\mathbf{x}} \mathbb{E}_{\mathcal{M}|\mathbf{x}}\left(x_{(\nint{qP})}^*-x_{(\nint{qP})}\right)^2 + \mathbb{E}_{\mathbf{x}} \mathbb{E}_{\mathcal{M}|\mathbf{x}}\left(x_{(\nint{qP})}-F_X^{-1}(q) \right)^2\notag\\
& \quad +2\mathbb{E}_{\mathbf{x}} \mathbb{E}_{\mathcal{M}|\mathbf{x}}\left(x_{(\nint{qP})}^*-x_{(\nint{qP})}\right)\left(x_{(\nint{qP})}-F_X^{-1}(q) \right)\notag\\
&\leq\mathbb{E}_{\mathbf{x}} \mathbb{E}_{\mathcal{M}|\mathbf{x}}\left(x_{([q P])}^*-x_{(\nint{qP})}\right)^2 + \mathbb{E}_{\mathbf{x}}\left(x_{(\nint{qP})}-F_X^{-1}(q) \right)^2 \notag\\
& \quad +2\sqrt{\mathbb{E}_{\mathbf{x}} \mathbb{E}_{\mathcal{M}|\mathbf{x}}\left(x_{([q P])}^*-x_{(\nint{qP})}\right)^2\mathbb{E}_{\mathbf{x}} \left(x_{(\nint{qP})}-F_X^{-1}(q) \right)^2}.
\label{eqn:CS_inequality}
\end{align}
The inequality in Eqn \eqref{eqn:CS_inequality} holds per the Cauchy-Schwarz inequality. Based on Theorem 1 of \citet{walker1968note}, sample quantiles are Gaussian asymptotically; that is,
\begin{equation}\label{eqn:popuquantile0}
\sqrt{P}\left(x_{(\nint{qP})}-F_X^{-1}(q) \right) \stackrel{d}{\rightarrow} \mathcal{N}\left(0, \frac{q(1-q)}{\{f_X(F_X^{-1}(q))\}^2}\right),
\end{equation}
based on which, we obtain
\begin{equation}\label{eqn:popuquantile}
\mathbb{E}_{\mathbf{x}}\left(x_{(\nint{qP})}-F_X^{-1}(q) \right)^2 \rightarrow \frac{q(1-q)}{\{f_X(F_X^{-1}(q))\}^2} \cdot \frac{1}{P}.
\end{equation}
An intermediate step of the PrivateQuantile procedure is the sampling of index $j^*$ via the exponential mechanism with privacy loss $\epsilon$ (line 3 in Algorithm 1),
\begin{equation} \label{eqn:PrqP}
\Pr(j^*)\propto (x_{(j^*+1)} - x_{(j^*)})\exp(-\epsilon|j^*-\nint{qP}|).
\end{equation}
As $P\rightarrow \infty$ or $\epsilon\rightarrow\infty$, $\epsilon|j^*-\nint{qP}|\rightarrow \infty$ when $j^*\ne \nint{qP}$, whereas $\epsilon|j^*-\nint{qP}|=0$ when  $j^*= \nint{qP}$. Therefore,  as $P\rightarrow \infty$ or $\epsilon\rightarrow\infty$, $ \Pr(j^*=\nint{qP})\rightarrow 1$ and 
\begin{align}
\Pr(x_{(\nint{qP})}^* \sim \mbox{Unif}(x_{(\nint{qP})},x_{(\nint{qP}+1)}))\rightarrow 1\label{Eqn:limitdist}.
\end{align}
Eqns \eqref{eqn:PrqP} and \eqref{Eqn:limitdist} imply the limiting distribution of $x_{(\nint{qP})}^*$ is a uniform distribution from $x_{(\nint{qP})}$ to $x_{(\nint{qP}+1)}$, achieved at the rate of $e^{O(P\epsilon)}$. Define $h\triangleq x_{(\nint{qP})}^*-x_{(\nint{qP})}$, then 
\begin{equation}
e^{P\epsilon}h \stackrel{d}{\rightarrow} \mbox{Unif}(0, x_{(\nint{qP}+1)}-x_{(\nint{qP})}).
\end{equation}
Therefore, as $P\rightarrow \infty$ 
\begin{equation}
\begin{aligned} \label{eqn:sanitizetopublic}
&\mathbb{E}_{\mathbf{x}}\mathbb{E}_{\mathcal{M}|\mathbf{x}} (x_{(\nint{qP})}^*-x_{(\nint{qP})})^2= \mathbb{E}_{\mathbf{x}}\mathbb{E}_{\mathcal{M}|\mathbf{x}}(h^ 2)=\mathbb{E}_{\mathbf{x}} \{\mathbb{V}_{\mathcal{M}|\mathbf{x}}(h)+ (\mathbb{E}_{\mathcal{M}|\mathbf{x}}(h))^2\} \\
\rightarrow &\; e^{-2P\epsilon}\mathbb{E}_{\mathbf{x}}\left[\frac{(x_{(\nint{qP}+1)}-x_{(\nint{qP})})^2}{12} + \frac{(x_{(\nint{qP}+1)}-x_{(\nint{qP})})^2}{4}\right]\\ 
=&\; \mathbb{E}_{\mathbf{x}}\left[\frac{(x_{(\nint{qP}+1)}-x_{(\nint{qP})})^2}{3}\right]
\rightarrow\frac{2/3}{e^{2P\epsilon}(Pf_X(F_X^{-1}(q)))^2}.
\end{aligned}
\end{equation}

Plugging in Eqns  \eqref{eqn:popuquantile} and \eqref{eqn:sanitizetopublic} on the right-hand side of Eqn \eqref{eqn:CS_inequality}, we have 
\begin{equation}
\begin{aligned}
& \qquad \mathbb{E}_{\mathbf{x}} \mathbb{E}_{\mathcal{M}|\mathbf{x}}\left(x_{([q P])}^*-F_X^{-1}(q)\right)^2 \\
& \leq \frac{2/3}{e^{2P\epsilon}(Pf_X(F_X^{-1}(q)))^2} + \frac{q(1-q)}{\{f_X(F_X^{-1}(q))\}^2} \frac{1}{P}+ 2\sqrt{\frac{1}{3} \frac{2}{e^{2P\epsilon}(Pf_X(F_X^{-1}(q)))^2}  \frac{q(1-q)}{\{f_X(F_X^{-1}(q))\}^2}  \frac{1}{P}}\\
& = O(P^{-1})+O(e^{-P\epsilon}P^{-3/2}).
\end{aligned}
\end{equation}
\end{proof}

\vspace{-12pt}\subsection*{Proof of Corollary 5}
\begin{proof}
Per Eqn \eqref{Eqn:limitdist}, the sanitized quantile at $q$ via PrivateQuantile  is \\ 
$x^*_{(\nint{qP})}\xrightarrow{d}$ Unif$(x_{(\nint{qP})},x_{(\nint{qP}+1)})$, and
\begin{align}
l^*-l=z^*_{(\nint{P\alpha})} - z_{(\nint{P\alpha})}&\xrightarrow{d}\mbox{Unif}(0,z_{(\nint{P\alpha}+1)}-z_{(\nint{P\alpha})}),\label{eqn:l}\\
u^*-u=z^*_{(\nint{P(1-\beta)})}-z_{(\nint{P(1-\beta)})}&\xrightarrow{d}\mbox{Unif}(0,z_{(\nint{P(1-\beta)}+1)}-z_{(\nint{P(1-\beta)})}),\label{eqn:u}
\end{align}
as $P\rightarrow\infty$ at rate $e^{O(P\epsilon)}$. Therefore,
\begin{equation}
\begin{aligned}
\label{eqn:(1)(2)}
\mathbb{E}_{\mathbf{z}} \mathbb{E}_{\mathcal{M}:PQ|\mathbf{z}}(l^*-l)
\rightarrow &\; e^{-P\epsilon} \mathbb{E}_{\mathbf{x}}\left[\frac{(z_{(\nint{P\alpha}+1)}-z_{(\nint{P\alpha})})}{2}\right] 
\rightarrow\frac{1}{2e^{P\epsilon}Pf_X(F_X^{-1}(\alpha))},\\
\mathbb{E}_{\mathbf{z}} \mathbb{E}_{\mathcal{M}:PQ|\mathbf{z}}(l^*-l)^2
=&\; \mathbb{E}_{\mathbf{z}}\left[\left(\mathbb{E}_{\mathcal{M}:PQ|\mathbf{z}}(l^*-l)\right)^2+ \mathbb{V}_{\mathcal{M}:PQ|\mathbf{z}}(l^*-l)\right]\\
\rightarrow &\; e^{-2P\epsilon}e^{-2P\epsilon}\mathbb{E}_{\mathbf{x}}\left[\frac{(z_{(\nint{P\alpha}+1)}-z_{(\nint{P\alpha})})^2}{12} + \frac{(z_{(\nint{P\alpha}+1)}-z_{(\nint{P\alpha})})^2}{4}\right]\\
=&\;e^{-2P\epsilon}\mathbb{E}_{\mathbf{x}}\left[\frac{(z_{(\nint{P\alpha}+1)}-z_{(\nint{P\alpha})})^2}{3}\right]
\rightarrow\frac{2/3}{e^{2P\epsilon}(Pf_X(F_X^{-1}(\alpha)))^2}.
\end{aligned}
\end{equation}
The convergence results for $u^*-u$ can be obtained in a similar manner by replacing $\alpha$ with $1-\beta$.
Let $PQ$ standards for PrivateQuanitile procedure. Based on Eqns \eqref{eqn:l} and \eqref{eqn:u}, 
\begin{equation}
\begin{aligned}
\label{eqn:EMzu*l*}
\mathbb{E}_{\mathcal{M}:PQ|\mathbf{z}} (u^*-l^*)
&= \mathbb{E}_{\mathcal{M}:PQ|\mathbf{z}} (u^*-u+u-l^*+l-l)\\
&= \mathbb{E}_{\mathcal{M}:PQ|\mathbf{z}} (u^*-u) - \mathbb{E}_{\mathcal{M}:PQ|\mathbf{z}} (l^*-l )+u-l\\
&\xrightarrow{} \frac{z_{(\nint{P(1-\beta)}+1)}-z_{(\nint{P(1-\beta)})}}{2} - \frac{z_{(\nint{P\alpha}+1)}-z_{(\nint{P\alpha})}}{2} + u-l.
\end{aligned}
\end{equation}
Per Eqn \eqref{eqn:popuquantile0}, as $P \rightarrow \infty$,
\begin{equation}\label{eqn:Ezul}
\mathbb{E}_{\mathbf{z}}(u-l)= F_Z^{-1}(1-\beta)-F_Z^{-1}(\alpha) + O(P^{-1/2}).
\end{equation}
Lemma 3, Eqn \eqref{eqn:EMzu*l*}, and Eqn \eqref{eqn:Ezul} taken together, 
\begin{align}\label{eqn:EzMzu*l*}
&\mathbb{E}_{\mathbf{z}} \mathbb{E}_{\mathcal{M}:PQ|\mathbf{z}}(u^*-l^*)
\xrightarrow{}\frac{1}{2e^{P\epsilon}Pf_Z(F^{-1}_{Z}(1-\beta))} - \frac{1}{2e^{P\epsilon}Pf_Z(F^{-1}_{Z}(\alpha))} + \mathbb{E}_{\mathbf{z}}(u-l)\notag\\
\rightarrow &\;\frac{1}{2e^{P\epsilon}Pf_Z(F^{-1}_{Z}(1-\beta))} - \frac{1}{2e^{P\epsilon}Pf_Z(F^{-1}_{Z}(\alpha))} +F_Z^{-1}(1-\beta)-F_Z^{-1}(\alpha)+ O(P^{-1/2})\\
= &\;F_Z^{-1}(1-\beta)-F_Z^{-1}(\alpha) + O(e^{-P\epsilon}P^{-1}) + O(P^{-1/2}).\notag
\end{align}
\begin{align*}
&\mathbb{E}_{\mathcal{M}:PQ|\mathbf{z}} (u^*-l^*)^2 = \mathbb{E}_{\mathcal{M}:PQ|\mathbf{z}} (u^*-u+u-l^*+l-l)^2\\
=&\; \mathbb{E}_{\mathcal{M}:PQ|\mathbf{z}} (u^*-u)^2 + \mathbb{E}_{\mathcal{M}:PQ|\mathbf{z}} (l^*-l )^2+(u-l)^2- 2\mathbb{E}_{\mathcal{M}:PQ|\mathbf{z}}(u^*-u)\mathbb{E}_{\mathcal{M}:PQ|\mathbf{z}}(l^*-l )\\
&+ 2(u-l)\mathbb{E}_{\mathcal{M}:PQ|\mathbf{z}} (u^*-u) - 2(u-l)\mathbb{E}_{\mathcal{M}:PQ|\mathbf{z}} (l^*-l )\\
\xrightarrow{}&\; \frac{(z_{(\nint{P(1-\beta)}+1)}-z_{(\nint{P(1-\beta)})})^2}{3e^{2P\epsilon}} + \frac{(z_{(\nint{P\alpha}+1)}-z_{(\nint{P\alpha})})^2}{3e^{2P\epsilon}} + (u-l)^2\\
& - \frac{2(z_{(\nint{P(1-\beta)}+1)}-z_{(\nint{P(1-\beta)})})(z_{(\nint{P\alpha}+1)}-z_{(\nint{P\alpha})})}{2e^{2P\epsilon}}\\
& + (u-l)\frac{z_{(\nint{P(1-\beta)}+1)}-z_{(\nint{P(1-\beta)})}-(z_{(\nint{P\alpha}+1)}-z_{(\nint{P\alpha})})}{e^{P\epsilon}}.
\end{align*}
Based on Eqns \eqref{eqn:popuquantile0} and \eqref{eqn:popuquantile},
\begin{equation}
\begin{aligned}\label{eqn:Ezul2}
&\mathbb{E}_{\mathbf{z}}(u-l)^2 = \left(\mathbb{E}_{\mathbf{z}}(u-l)\right)^2+\mathbb{V}_{\mathbf{z}}(u-l)\\
= &\left(\mathbb{E}_{\mathbf{z}}(u-l)\right)^2+\mathbb{V}_{\mathbf{z}}\left(u-F_Z^{-1}(1-\beta) + F_Z^{-1}(1-\beta) -l +F_Z^{-1}(\alpha)-F_Z^{-1}(\alpha)\right)\\
= &\left(\mathbb{E}_{\mathbf{z}}(u-l)\right)^2+\mathbb{V}_{\mathbf{z}}\left(u-F_Z^{-1}(1-\beta)\right) +\mathbb{V}_{\mathbf{z}}\left(l -F_Z^{-1}(\alpha)\right))\\
\rightarrow & (F_Z^{-1}(1-\beta)-F_Z^{-1}(\alpha) + O(P^{-1/2}))^2+ \frac{\beta(1-\beta)}{\{f_Z(F_Z^{-1}(1-\beta))\}^2}\frac{1}{P} + \frac{\alpha(1-\alpha)}{\{f_Z(F_Z^{-1}(\alpha))\}^2}\frac{1}{P}\\
\end{aligned}
\end{equation}

Lemma 3, Eqns \eqref{eqn:sanitizetopublic}, \eqref{eqn:Ezul}, \eqref{eqn:EzMzu*l*} and \eqref{eqn:Ezul2} taken together, as $P\rightarrow \infty$,
\begin{align}
&\mathbb{E}_{\mathbf{z}} \mathbb{E}_{\mathcal{M}:PQ|\mathbf{z}}(u^*-l^*)^2\notag\\
\rightarrow&\frac{2/3}{(Pe^{P\epsilon}f_Z(F_Z^{-1}(1-\beta)))^2} + \frac{2/3}{(Pe^{P\epsilon}f_Z(F_Z^{-1}(\alpha)))^2}-\frac{1}{2e^{2P\epsilon}P^2f_Z(F^{-1}_{Z}(\alpha))f_Z(F^{-1}_{Z}(1-\beta))}\notag\\
&+ (F_Z^{-1}(1-\beta)-F_Z^{-1}(\alpha) + O(P^{-1/2}))^2+ \frac{\beta(1-\beta)}{f_Z(F_Z^{-1}(1-\beta))\}^2}\frac{1}{e^{P\epsilon}P} + \frac{\alpha(1-\alpha)}{\{f_Z(F_Z^{-1}(\alpha))\}^2}\frac{1}{e^{P\epsilon}P}\notag\\
& + \left(F_Z^{-1}(1-\beta)-F_Z^{-1}(\alpha) + O(P^{-1/2})\right) e^{-P\epsilon}\bigg(\frac{1}{Pf_Z(F^{-1}_{Z}(1-\beta))}-\frac{1}{Pf_Z(F^{-1}_{Z}(\alpha))}\bigg)\\
= &\; O(P^{-2}e^{-2P\epsilon})+(F_Z^{-1}(1-\beta)-F_Z^{-1}(\alpha))^2 + O(P^{-1/2})+O(P^{-1})+O(P^{-1}e^{-P\epsilon})+O(P^{-3/2}e^{-P\epsilon})\notag\\
= &\; (F_Z^{-1}(1-\beta)-F_Z^{-1}(\alpha))^2 + O(P^{-1/2})++O(P^{-1}e^{-P\epsilon})\notag
\end{align}  
\end{proof}

\subsection*{Proof of Theorem 6}
\begin{proof}
Let $\hat{\theta}= g(\mathbf{s}'_6)$ denote the original MLE or the posterior mean of $\theta$ given the censored Gaussian likelihood and the Jeffreys' prior. Per the Cauchy-Schwarz inequality, we have
\begin{equation}
\begin{aligned}\label{eqn:CS_inequality2}
&\quad \mathbb{E}_{\mathbf{z}} \mathbb{E}_{\mathcal{M}|\mathbf{z}} (\hat{\theta}^*-\theta)^2 
= \mathbb{E}_{\mathbf{z}} \mathbb{E}_{\mathcal{M}|\mathbf{z}} (\hat{\theta}^* - \hat{\theta} + \hat{\theta}-\theta)^2\\
&=\mathbb{E}_{\mathbf{z}} \mathbb{E}_{\mathcal{M}|\mathbf{z}} (\hat{\theta}^*-\hat{\theta})^2 + \mathbb{E}_{\mathbf{z}} (\hat{\theta}-\theta)^2 + 2\mathbb{E}_{\mathbf{z}}\left[(\hat{\theta}-\theta) \mathbb{E}_{\mathcal{M}|\mathbf{z}} (\hat{\theta}^*-\hat{\theta})\right]\\
& \leq \mathbb{E}_{\mathbf{z}} \mathbb{E}_{\mathcal{M}|\mathbf{z}} (\hat{\theta}^*-\hat{\theta})^2 + \mathbb{E}_{\mathbf{z}} (\hat{\theta}-\theta)^2 + 
2\sqrt{\mathbb{E}_{\mathbf{z}} \mathbb{E}_{\mathcal{M}|\mathbf{z}} (\hat{\theta}^*-\hat{\theta})^2 \mathbb{E}_{\mathbf{z}} (\hat{\theta}-\theta)^2},
\end{aligned}
\end{equation}
Applying the asymptotic normality of MLE or the posterior mean as $P\rightarrow\infty$, i.e.,$\sqrt{P}(\hat{\theta}-\theta) \stackrel{d}{\longrightarrow} N\left(0, \sigma^2_\theta\right)$, we have
\begin{equation}\label{eqn:MLE}
 \mathbb{E}_{\mathbf{z}}(\hat{\theta}-\theta)^2
=(\mathbb{E}_{\mathbf{z}}(\hat{\theta}-\theta))^2+
 \mathbb{V}_{\mathbf{z}}(\hat{\theta}-\theta) \rightarrow \sigma^2_\theta/P.
\end{equation}

Applying the first-order Taylor expansion of $g(\mathbf{s}'^*_6)$ around $\mathbf{s}'_6$,
\begin{align}\label{eqn:1storder}
g(\mathbf{s}'^{*}_6)&=g\left(\mathbf{s}'_6\right)+\left(\mathbf{s}'^{*}_6-\mathbf{s}'_6\right)^{\top} \nabla g\left(\mathbf{s}'_6\right)+O\left(\left(\mathbf{s}'^{*}_6-\mathbf{s}'_6\right)^2\right),\mbox{ that is,}\notag\\
\hat{\theta}^* - \hat{\theta}&=\left(\mathbf{s}'^{*}_6-\mathbf{s}'_6\right)^{\top} \nabla g\left(\mathbf{s}'_6\right)+O\left(\left(\mathbf{s}'^{*}_6-\mathbf{s}'_6\right)^2\right)\notag\\
\mathbb{E}_{\mathbf{z}}\mathbb{E}_{\mathcal{M}|\mathbf{z}} (\hat{\theta}^*-\hat{\theta})^2 
&\approx \mathbb{E}_{\mathbf{z}}\mathbb{E}_{\mathcal{M}|\mathbf{z}} ((\mathbf{s}'^{*}_6-\mathbf{s}'_6)^{\top} \nabla g\left(\mathbf{s}'_6\right))^2.
\end{align}
WLOG, suppose the Laplace mechanism is used to sanitize $P_l,P_u,s'_1,s'_2$ in $\mathbf{s}_6'$ (PrivateQuantile sanitizes $l$ and $u$) and the total privacy budget $\epsilon$ is split equally to sanitize each of 6 elements in  $\mathbf{s}_6'$. Then
$P^*_l \sim \mbox{Lap}(P_l,1/(\epsilon/6))$, 
$P^*_u \sim\mbox{Lap}(P_u, 1/(\epsilon/6))$, 
$s'^*_1 \sim \mbox{Lap}(s'_1,(u^*-l^*)/(\epsilon/6))$, and 
$s'^*_2 \sim \mbox{Lap}(s'_2,\max\{u^{2*},l^{2*}\}/(\epsilon/6))$. Since each of the elements in $\mathbf{s}'_6=(l,u,P_l,P_u,s'_1,s'_2)$ is  independently sanitized, the 14 cross-product terms that involve at least one  statistic out of  $P_l,P_u,s'_1,s'_2$ in the expectation over $\mathcal{M}|\mathbf{z}$ in Eqn \eqref{eqn:1storder} are 0 and we have
\begin{align}
\label{eqn:expansion}
&\mathbb{E}_{\mathbf{z}}\mathbb{E}_{\mathcal{M}|\mathbf{z}} ((\mathbf{s}'^{*}_6-\mathbf{s}'_6)^{\top} \nabla g\left(\mathbf{s}'_6\right))^2 \notag \\
= & \mathbb{E}_{\mathbf{z}}\mathbb{E}_{\mathcal{M}|\mathbf{z}} \left( \frac{\partial g(\mathbf{s}'_6)}{\partial l}(l^*-l)\right)^2 + \mathbb{E}_{\mathbf{z}}\mathbb{E}_{\mathcal{M}|\mathbf{z}} \left(\frac{\partial g(\mathbf{s}'_6)}{\partial u}(u^*-u)\right)^2 + \mathbb{E}_{\mathbf{z}}\mathbb{E}_{\mathcal{M}|\mathbf{z}}\left((u^*-u)(l^*-l)\frac{\partial g(\mathbf{s}'_6)}{\partial l}\frac{\partial g(\mathbf{s}'_6)}{\partial u}\right)\notag \\ 
& + \mathbb{E}_{\mathbf{z}}\mathbb{E}_{\mathcal{M}|\mathbf{z}} \left(\frac{\partial g(\mathbf{s}'_6)}{\partial P_l}(P_l^*-P_l)\right)^2 + \mathbb{E}_{\mathbf{z}}\mathbb{E}_{\mathcal{M}|\mathbf{z}} \left(\frac{\partial g(\mathbf{s}'_6)}{\partial P_u}(P_u^*-P_u)\right)^2+ \mathbb{E}_{\mathbf{z}}\mathbb{E}_{\mathcal{M}|\mathbf{z}} \left(\frac{\partial g(\mathbf{s}'_6)}{\partial s'_1}(s'^*_1-s'_1)\right)^2 \notag \\
& + \mathbb{E}_{\mathbf{z}}\mathbb{E}_{\mathcal{M}|\mathbf{z}} \left(\frac{\partial g(\mathbf{s}'_6)}{\partial s'_2}(s'^*_2-s'_2)\right)^2\notag \\
=& \mathbb{E}_{\mathbf{z}}\bigg(\frac{\partial g(\mathbf{s}'_6)}{\partial l}\bigg)^2\mathbb{E}_{\mathbf{z}}\mathbb{E}_{\mathcal{M}|\mathbf{z}} (l^*-l)^2 + \mathbb{E}_{\mathbf{z}}\bigg(\frac{\partial g(\mathbf{s}'_6)}{\partial u}\bigg)^2\mathbb{E}_{\mathbf{z}}\mathbb{E}_{\mathcal{M}|\mathbf{z}} (u^*-u)^2 \notag \\
&+\mathbb{E}_{\mathbf{z}}\mathbb{E}_{\mathcal{M}|\mathbf{z}}\left(u^*-u)\mathbb{E}_{\mathbf{z}}\mathbb{E}_{\mathcal{M}|\mathbf{z}}(l^*-l\right)\mathbb{E}_{\mathbf{z}}\bigg(\frac{\partial g(\mathbf{s}'_6)}{\partial l}\bigg)\mathbb{E}_{\mathbf{z}}\bigg(\frac{\partial g(\mathbf{s}'_6)}{\partial u}\bigg)\notag \\
&+\mathbb{E}_{\mathbf{z}}\bigg(\frac{\partial g(\mathbf{s}'_6)}{\partial P_l}\bigg)^2\frac{2}{(\epsilon/6)^{2} } +
\mathbb{E}_{\mathbf{z}}\bigg(\frac{\partial g(\mathbf{s}'_6)}{\partial P_u}\bigg)^2\cdot \frac{2}{(\epsilon/6)^{2} }  + \mathbb{E}_{\mathbf{z}}\bigg(\frac{\partial g(\mathbf{s}'_6)}{\partial s'_1}\bigg)^2\frac{2\mathbb{E}_{\mathbf{z}}\mathbb{E}_{\mathcal{M}|\mathbf{z}} (u^*-l^*)^2}{(\epsilon/6)^2}\notag \\
& + \mathbb{E}_{\mathbf{z}}\bigg(\frac{\partial g(\mathbf{s}'_6)}{\partial s'_2}\bigg)^2\frac{2\mathbb{E}_{\mathbf{z}}\mathbb{E}_{\mathcal{M}|\mathbf{z}}(\max\{u^{2*},l^{2*}\})}{(\epsilon/6)^2}.
\end{align}

To evaluate $\nabla_{\mathbf{s}'_6}g(\mathbf{s}'_6)$, we follow a similar approach as the proof of Lemma 3.1 in \citet{gould2016differentiating}. WLOG, let $\hat{\theta}=g(\mathbf{s}'_6)= \operatorname{argmax}_{\theta \in \mathbb{R}} ll(\theta|\mathbf{s}'_6)$  be the MLE (posterior mean and MLE are asymptotically equivalent as $P\rightarrow \infty$), where 
\begin{equation}
\begin{aligned}
ll(\theta|\mathbf{s}'_6)  \propto 
& P_l \log \Phi\left(\frac{l-\theta}{\sigma}\right)+P_u \log \left(1-\Phi\left(\frac{u-\theta}{\sigma}\right)\right)-\frac{P-P_l-P_u}{2 \sigma^2} \theta^2 \notag\\
&-\frac{1}{2 \sigma^2} s_2^{\prime}+\frac{\theta}{\sigma^2} s_1^{\prime}+(P_l+P_u-P) \log \sigma.
\end{aligned}
\end{equation}
By the definition of MLE, $\frac{\partial  ll(\theta|\mathbf{s}'_6)}{\partial\theta}|_{\theta=g(\mathbf{s}'_6)} =0,$
differentiating both sides of which, we have
\begin{align}\label{eqn:mle02}
\nabla_{\mathbf{s}'_6}\frac{\partial  ll(\theta|\mathbf{s}'_6)}{\partial \theta}|_{\theta=g(\mathbf{s}'_6)} =0, 
\end{align}
and applying the chain rule to Eqn \eqref{eqn:mle02}, focusing on  one element of $\mathbf{s}$ at a time, say, on $u$, we have
\begin{equation}\label{eqn:mle03}
\begin{aligned}
\frac{d}{du}(\frac{\partial  ll(\theta|\mathbf{s}'_6)}{\partial \theta}|_{\theta=g(\mathbf{s}'_6)})
&=\frac{\partial^2  ll(\theta|\mathbf{s}'_6)}{\partial \theta^2}|_{\theta=g(\mathbf{s}'_6)}\frac{\partial g(\mathbf{s}'_6)}{\partial u} + \frac{\partial^2 ll(\theta|\mathbf{s}'_6)}{\partial u \partial \theta}|_{\theta=g(\mathbf{s}'_6)}\\
& = \frac{\partial^2  ll(g(\mathbf{s}'_6)|\mathbf{s}'_6)}{\partial \theta^2}\frac{\partial g(\mathbf{s}'_6)}{\partial u} + \frac{\partial^2 ll(g(\mathbf{s}'_6)|\mathbf{s}'_6)}{\partial u \partial \theta}=0.
\end{aligned}
\end{equation}
Rearranging the terms, we have
\begin{align}
\label{eqn:gradientu}
\frac{\partial g(\mathbf{s}'_6)}{\partial u}
=-\bigg(\frac{\partial^2  ll(g(\mathbf{s}'_6)|\mathbf{s}'_6)}{\partial \theta^2}\bigg)^{-1}\frac{\partial^2 ll(g(\mathbf{s}'_6)|\mathbf{s}'_6)}{\partial u \partial \theta}.
\end{align}
Applying  Eqns \eqref{eqn:gradientu} and  \eqref{eqn:mle03} for $u$ in a similar manner for all the elements in $\mathbf{s}$, we have
\begin{equation}\label{eqn:gradient}
\begin{aligned}
\!\!\!
\nabla_{\mathbf{s}'_6}g(\mathbf{s}'_6)
\!=\!-(\frac{\partial^2  ll(g(\mathbf{s}'_6)|\mathbf{s}'_6)}{\partial \theta^2})^{-1}\nabla_{\mathbf{s}'_6}\frac{\partial ll(g(\mathbf{s}'_6)|\mathbf{s}'_6)}{\partial \theta}\!=\!-(\frac{\partial^2  ll(g(\mathbf{s}'_6)|\mathbf{s}'_6)}{\partial \theta^2})^{-1}\frac{\partial}{\partial \theta}\nabla_{\mathbf{s}'_6}ll(g(\mathbf{s}'_6)|\mathbf{s}'_6).
\end{aligned}\!
\end{equation}

The first-order and the second-order derivatives of the log-likelihood function w.r.t $\theta$ are,
\begin{equation}
\begin{aligned}
\label{eqn:1stterm}
\frac{\partial  ll(\theta|\mathbf{s}'_6)}{\partial \theta}
=&-P_l \frac{\phi(\frac{l-\theta}{\sigma})}{\sigma \Phi(\frac{l-\theta}{\sigma})} + P_u\frac{\phi(\frac{u-\theta}{\sigma})}{\sigma (1-\Phi(\frac{u-\theta}{\sigma}))} - \frac{P-P_l-P_u}{\sigma^2}\theta + \frac{s'_1}{\sigma^2},\\
\frac{\partial^2  ll(\theta|\mathbf{s}'_6)}{\partial \theta^2}
=&-\frac{P_l}{\sigma^2} \left((\frac{\phi(\frac{l-\theta}{\sigma})}{\Phi(\frac{l-\theta}{\sigma})})^2-\frac{\phi(\frac{l-\theta}{\sigma})}{\Phi(\frac{l-\theta}{\sigma})}\right)-\frac{P_u}{\sigma^2}\left((\frac{\phi(\frac{u-\theta}{\sigma})}{1-\Phi(\frac{u-\theta}{\sigma})})^2+\frac{\phi(\frac{u-\theta}{\sigma})}{1-\Phi(\frac{u-\theta}{\sigma})}\right) \\
& - \frac{P-P_l-P_u}{\sigma^2}.
\end{aligned}
\end{equation}

Take the first derivative of $ll(\theta|\mathbf{s}'_6)$ with regard to $\mathbf{s}'_6=(l,u,P_l,P_u,s'_1,s'_2)$, we have
\begin{align}
\nabla_{\mathbf{s}'_6} ll(\theta|\mathbf{s}'_6) 
=&\begin{pmatrix}
 P_l\left(\Phi\left(\frac{l-\theta}{\sigma}\right)\right)^{-1}
\phi\left(\frac{l-\theta}{\sigma}\right)\frac{1}{\sigma}\\
-P_u\left(1-\Phi\left(\frac{u-\theta}{\sigma}\right)\right)^{-1}
\phi\left(\frac{u-\theta}{\sigma}\right)\frac{1}{\sigma}\\
\log\Phi\left(\frac{l-\theta}{\sigma}\right)+\frac{\theta^2}{2 \sigma^2}+ \log\sigma\\ 
\log\left(1-\Phi\left(\frac{u-\theta}{\sigma}\right)\right)+\frac{\theta^2}{2 \sigma^2} + \log\sigma\\ 
\frac{\theta}{\sigma^2}\\
-\frac{1}{2 \sigma^2} \\  
\end{pmatrix}
\end{align}
and its partial derivative w.r.t $\theta$  is
\begin{equation}
\begin{aligned}
\label{eqn:2ndterm}
\frac{\partial}{\partial \theta}\nabla_{\mathbf{s}'_6}ll(\theta|\mathbf{s}'_6)
=&\begin{pmatrix}
P_l \left(\phi^2(\frac{l-\theta}{\sigma})\Phi^{-2}(\frac{l-\theta}{\sigma})-\phi(\frac{l-\theta}{\sigma})\Phi^{-1}(\frac{l-\theta}{\sigma})\right)\sigma^{-2}\\
P_u\left(\phi^2(\frac{u-\theta}{\sigma})(1-\Phi(\frac{u-\theta}{\sigma}))^{-2}+\phi(\frac{u-\theta}{\sigma})(1-\Phi(\frac{u-\theta}{\sigma}))^{-1}\right)\sigma^{-2}\\
-\phi(\frac{l-\theta}{\sigma})(\sigma \Phi(\frac{l-\theta}{\sigma}))^{-1} + \theta \sigma^{-2}\\ 
\phi(\frac{u-\theta}{\sigma})(\sigma (1-\Phi(\frac{u-\theta}{\sigma}))^{-1} + \theta \sigma^{-2}\\ 
\sigma^{-2}\\
0 
\end{pmatrix}.
\end{aligned}
\end{equation}
Let $A=\phi(\frac{l-\hat{\theta}}{\sigma})(\Phi(\frac{l-\hat{\theta}}{\sigma}))^{-1}\text{ and } B=\phi(\frac{u-\hat{\theta}}{\sigma})(1-\Phi(\frac{u-\hat{\theta}}{\sigma}))^{-1}$. Plugging $\hat{\theta}=g(\mathbf{s}'_6)$ in Eqns \eqref{eqn:1stterm} and \eqref{eqn:2ndterm},  $\nabla_{\mathbf{s}'_6}g(\mathbf{s}'_6)$ in Eqn \eqref{eqn:gradient} can be written as
\begin{equation}\label{eqn:gs6gradient}
\!\!\begin{aligned}
\nabla_{\mathbf{s}'_6}g(\mathbf{s}'_6)
=\begin{pmatrix}
\frac{\partial g(\mathbf{s}'_6)}{\partial l}\\
\frac{\partial g(\mathbf{s}'_6)}{\partial u}\\
\frac{\partial g(\mathbf{s}'_6)}{\partial P_l}\\
\frac{\partial g(\mathbf{s}'_6)}{\partial P_u}\\
\frac{\partial g(\mathbf{s}'_6)}{\partial s'_1}\\
\frac{\partial g(\mathbf{s}'_6)}{\partial s'_2}
\end{pmatrix}\!
=\begin{pmatrix}
P_l(A^2-A)\left(P_l (A^2-A)+P_u(B^2+B) + P-P_l-P_u\right)^{-1}\\
P_u(B^2+B)\left(P_l (A^2-A)+P_u(B^2+B) + P-P_l-P_u\right)^{-1}\\
(-A\sigma + \hat{\theta})\left(P_l (A^2-A)+P_u(B^2+B) + P-P_l-P_u\right)^{-1}\\
(B\sigma + \hat{\theta})\left(P_l (A^2-A)+P_u(B^2+B) + P-P_l-P_u\right)^{-1}\\
\left(P_l (A^2-A)+P_u(B^2+B) + P-P_l-P_u\right)^{-1}\\
0
\end{pmatrix}.
\end{aligned}
\end{equation}
Given that data $\mathbf{z}'$ is bounded by $[l,u]$, so is the mean estimate $\hat{\theta} \in [l,u]$. Therefore,
\begin{align}
    \phi(\frac{l-\hat{\theta}}{\sigma}) &\in [\phi(\frac{l-u}{\sigma}),\phi(0)], 
    \qquad \Phi(\frac{l-\hat{\theta}}{\sigma})\in [\Phi(\frac{l-u}{\sigma}),0.5].\\
    \phi(\frac{u-\hat{\theta}}{\sigma}) &\in [\phi(\frac{l-u}{\sigma}),\phi(0)], 
    \qquad \Phi(\frac{u-\hat{\theta}}{\sigma})\in [0.5,1-\Phi(\frac{l-u}{\sigma})],
\end{align}
leading to $ A, B \in [2\phi(\frac{l-u}{\sigma}),(\sqrt{2\pi}\Phi(\frac{l-u}{\sigma}))^{-1}].$
Note that  $A^2-A\geq-1/4$ and $B^2+B>0$ (since $B>0$). In addition, since $\Phi((l-u)/\sigma)<0.5$, then $(\sqrt{2\pi}\Phi((l-u)/\sigma)^{-1}>\sqrt{2/\pi} \approx 0.798 >0.5$. Let $P_l\approx \alpha P$ and $P_u\approx \beta P$; assume $5\alpha/4 +\beta<1$,  then the denominator that appears in the first 5 elements in $\nabla_{\mathbf{s}'_6}g(\mathbf{s}'_6)$ in Eqn \eqref{eqn:gs6gradient}
$$\left(P_l (A^2-A)+P_u(B^2+B) + P-P_l-P_u\right)^{-1}< P^{-1}(-\frac{1}{4}\alpha + 1-\alpha-\beta)^{-1}=P^{-1}(1-\frac{5}{4}\alpha -\beta)^{-1},$$
and the 5th element is
\begin{equation}
    0<\frac{\partial g(\mathbf{s}'_6)}{\partial s'_1}
    < P^{-1}(1-\frac{5}{4}\alpha -\beta)^{-1}.
\end{equation}

For the 1st and 2nd elements in Eqn \eqref{eqn:gs6gradient}, there are two scenarios;

(1) when $A^2-A\geq0$, 
$$0\leq P_l(A^2-A)\left(P_l (A^2-A)+P_u(B^2+B) + P-P_l-P_u\right)^{-1}<1,$$
$$0<P_u(B^2+B)\left(P_l (A^2-A)+P_u(B^2+B) + P-P_l-P_u\right)^{-1}<1.$$

(2) when $-1/4\leq A^2-A<0$,
$$-\frac{1}{4}P\alpha \leq P_l(A^2-A)<0,$$
$$P_l (A^2-A)+P_u(B^2+B) + P-P_l-P_u > P(1-\frac{5}{4}\alpha-\beta)>0.$$

Therefore,
\begin{equation}
\begin{aligned}
    \frac{\alpha}{5\alpha+4\beta-4}&<  P_l(A^2-A)\left(P_l (A^2-A)+P_u(B^2+B) + P-P_l-P_u\right)^{-1}<0,\\
    0&<P_u(B^2+B)\left(P_l (A^2-A)+P_u(B^2+B) + P-P_l-P_u\right)^{-1}<1;
\end{aligned}
\end{equation}
and
\begin{equation}
\begin{aligned}
\frac{\alpha}{5\alpha+4\beta-4}
<  &\frac{\partial g(\mathbf{s}'_6)}{\partial l}<1,\\ 
0<&\frac{\partial g(\mathbf{s}'_6)}{\partial u}<1. 
\end{aligned}
\end{equation}
The numerators in the 3rd and 4th elements in Eqn \eqref{eqn:gs6gradient} satisfy
$$-A\sigma + \hat{\theta} \in \left[l-\sigma(\sqrt{2\pi}\Phi(\frac{l-u}{\sigma}))^{-1}, u-2\sigma \phi(\frac{l-u}{\sigma})\right],$$
$$
\quad B\sigma + \hat{\theta} \in \left[l+2\sigma\phi(\frac{l-u}{\sigma}),u+\sigma(\sqrt{2\pi}\Phi(\frac{l-u}{\sigma}))^{-1}\right],$$
respectively.  Therefore,
\begin{equation}
\begin{aligned}
    \left(\frac{\partial g(\mathbf{s}'_6)}{\partial P_l}\right)^2 
    &< P^{-2}(1-\frac{5}{4}\alpha -\beta)^{-2}\max\{(l-\sigma(\sqrt{2\pi}\Phi(\frac{l-u}{\sigma}))^{-1})^2,(u-2\sigma \phi(\frac{l-u}{\sigma}))^{2}\},\\
    \left(\frac{\partial g(\mathbf{s}'_6)}{\partial P_u}\right)^2 
    &<  P^{-2}(1-\frac{5}{4}\alpha -\beta)^{-2}\max\{(l+2\sigma\phi(\frac{l-u}{\sigma}))^2,(u+\sigma(\sqrt{2\pi}\Phi(\frac{l-u}{\sigma}))^{-1})^{2}\}.\\
\end{aligned}
\end{equation}

Combing all the results above with Corollary 5, Eqn \eqref{eqn:expansion} can be written as
\begin{align}
&\mathbb{E}_{\mathbf{z}}\mathbb{E}_{\mathcal{M}:Lap|\mathbf{z}} ((\mathbf{s}'^{*}_6-\mathbf{s}'_6)^{\top} \nabla g\left(\mathbf{s}'_6\right))^2 \notag\\
=& \mathbb{E}_{\mathbf{z}}\bigg(\frac{\partial g(\mathbf{s}'_6)}{\partial l}\bigg)^2\mathbb{E}_{\mathbf{z}}\mathbb{E}_{\mathcal{M}:PQ|\mathbf{z}} (l^*-l)^2 + \mathbb{E}_{\mathbf{z}}\bigg(\frac{\partial g(\mathbf{s}'_6)}{\partial u}\bigg)^2\mathbb{E}_{\mathbf{z}}\mathbb{E}_{\mathcal{M}:PQ|\mathbf{z}} (u^*-u)^2 \notag\\
&+\mathbb{E}_{\mathbf{z}}\mathbb{E}_{\mathcal{M}:PQ|\mathbf{z}}\left(u^*-u)\mathbb{E}_{\mathbf{z}}\mathbb{E}_{\mathcal{M}:PQ|\mathbf{z}}(l^*-l\right)\mathbb{E}_{\mathbf{z}}\bigg(\frac{\partial g(\mathbf{s}'_6)}{\partial l}\bigg)\mathbb{E}_{\mathbf{z}}\bigg(\frac{\partial g(\mathbf{s}'_6)}{\partial u}\bigg) \notag\\
&+\mathbb{E}_{\mathbf{z}}\bigg(\frac{\partial g(\mathbf{s}'_6)}{\partial P_l}\bigg)^2 \frac{2}{(\epsilon/6)^{2} } +
\mathbb{E}_{\mathbf{z}}\bigg(\frac{\partial g(\mathbf{s}'_6)}{\partial P_u}\bigg)^2 \frac{2}{(\epsilon/6)^{2} }  + \mathbb{E}_{\mathbf{z}}\bigg(\frac{\partial g(\mathbf{s}'_6)}{\partial s'_1}\bigg)^2\frac{2\mathbb{E}_{\mathbf{z}}\mathbb{E}_{\mathcal{M}:PQ|\mathbf{z}} (u^*-l^*)^2}{(\epsilon/6)^2} \notag\\
& + \mathbb{E}_{\mathbf{z}}\bigg(\frac{\partial g(\mathbf{s}'_6)}{\partial s'_2}\bigg)^2  \frac{2\mathbb{E}_{\mathbf{z}}\mathbb{E}_{\mathcal{M}:PQ|\mathbf{z}}(\max\{u^{2*},l^{2*}\})}{(\epsilon/6)^2} \notag\\
\leq & \max\bigg\{\bigg(\frac{\alpha}{5\alpha+4\beta-4}\bigg)^2,1\bigg\} \frac{2/3}{(Pf_Z(F_Z^{-1}(\alpha)))^2} + \frac{2/3}{(Pf_Z(F_Z^{-1}(1-\beta)))^2}\notag\\
& + \frac{1}{2Pf_Z(F^{-1}_{Z}(1-\beta))}\cdot \frac{1}{2Pf_Z(F^{-1}_{Z}(\alpha))} \max\bigg\{\frac{\alpha}{5\alpha+4\beta-4},1\bigg\}\notag\\
&+P^{-2}\bigg(1-\frac{5}{4}\alpha -\beta\bigg)^{-2}\max\bigg\{\bigg(l-\sigma\bigg(\sqrt{2\pi}\Phi\bigg(\frac{l-u}{\sigma}\bigg)\bigg)^{-1}\bigg)^2,\bigg(u-2\sigma \phi\bigg(\frac{l-u}{\sigma}\bigg)\bigg)^{2}\bigg\} \frac{2}{(\epsilon/6)^{2} }\notag \\
&+P^{-2}\bigg(1-\frac{5}{4}\alpha -\beta\bigg)^{-2}\max\bigg\{\bigg(l+2\sigma\phi\bigg(\frac{l-u}{\sigma}\bigg)\bigg)^2,\bigg(u+\sigma\bigg(\sqrt{2\pi}\Phi\bigg(\frac{l-u}{\sigma}\bigg)\bigg)^{-1}\bigg)^{2}\bigg\} \frac{2}{(\epsilon/6)^{2} }\notag\\
&+ P^{-2}\bigg(1-\frac{5}{4}\alpha -\beta\bigg)^{-2}\frac{2\left((F_Z^{-1}(1-\beta)-F_Z^{-1}(\alpha))^2 + O(P^{-1/2})\right)}{(\epsilon/6)^2} +0\notag\\
&=O(P^{-2} + P^{-2} \epsilon^{-2})\label{eqn:likelhood2}.
\end{align}
Plugging Eqns \eqref{eqn:likelhood2} and \eqref{eqn:MLE} in Eqn \eqref{eqn:CS_inequality2}, we have 
\begin{align}
\mathbb{E}_{\mathbf{z}}\mathbb{E}_{\mathcal{M}:Lap|\mathbf{z}} (\hat{\theta}^*-\theta)^2 &\leq O(P^{-1} + P^{-2}+P^{-2} \epsilon^{-2} + P^{-3/2} + P^{-3/2}\epsilon^{-1})\notag\\
&=O(P^{-1}+P^{-3/2}\epsilon^{-1})
\end{align}

If the Gaussian mechanism is applied to sanitize $P_l, P_u, s'_1, s'_2$ in $\mathbf{s}'_6$. Then,
$P^*_l \sim \mathcal{N}(P_l,1/2(\rho/6))$, 
$P^*_u \sim \mathcal{N}(P_l,1/2(\rho/6))$, 
$s'^*_1 \sim \mathcal{N}(s'_1,(u^*-l^*)^2/2(\rho/6))$, and 
$s'^*_2 \sim \mathcal{N}(s'_2,(\max\{u^{2*},l^{2*}\})^2/2(\rho/6))$.
Therefore,$$\mathbb{E}_{\mathbf{z}}\mathbb{E}_{\mathcal{M}:Gaus|\mathbf{z}}((\mathbf{s}'^{*}_6-\mathbf{s}'_6)^{\top} \nabla g\left(\mathbf{s}'_6\right))^2 = O(P^{-2} + P^{-2}\rho^{-1}).$$
Per the Cauchy-Schwarz inequality in Eqn \eqref{eqn:CS_inequality2}
\begin{align}
\mathbb{E}_{\mathbf{z}}\mathbb{E}_{\mathcal{M}:Gaus|\mathbf{z}} (\hat{\theta}^*-\theta)^2 &\leq O(P^{-1} + P^{-2}+P^{-2} \rho^{-1} + P^{-3/2} + P^{-3/2}\rho^{-1/2})\notag \\
&=O(P^{-1}+P^{-3/2}\rho^{-1/2})
\end{align}
\end{proof}

\subsection*{Proof of Theorem 7}
\begin{proof}
Part 1: The trimmed mean is $\hat{\theta}_t = (P-P_l-P_u)^{-1}\sum_{j=1}^{P-P_l-P_u}z'_j$ and its sanitized version is $\hat{\theta}^*_{\text{t}}=s'^*_1/(P-P\alpha-P\beta)$. Per the Cauchy-Schwarz inequality, 
\begin{equation}\label{eqn:CS_inequalityt}
\begin{aligned}
 \mathbb{E}_{\mathbf{z}} \mathbb{E}_{\mathcal{M}|\mathbf{z}}(\hat{\theta}^*_t - \theta)^2
 &\leq \mathbb{E}_{\mathbf{z}} \mathbb{E}_{\mathcal{M}|\mathbf{z}}(\hat{\theta}^*_t - \hat{\theta}_t)^2 \!+\!\mathbb{E}_{\mathbf{z}} (\hat{\theta}_t - \theta)^2\!+\!2\sqrt{\mathbb{E}_{\mathbf{z}} \mathbb{E}_{\mathcal{M}|\mathbf{z}}(\hat{\theta}_t - \theta)^2 \mathbb{E}_{\mathbf{z}} (\hat{\theta}^*_t - \hat{\theta}_t)^2}.\!\!\\
\end{aligned}
\end{equation}
Theorem 3.1 in \citet{bickel1965some} suggests that, when $\alpha=\beta$,
\begin{equation}\label{asym_trim2}
\begin{aligned}
\sqrt{P} \cdot (\hat{\theta}_t -\theta)
&\xrightarrow{d} N\left(0, \sigma_t^2(\alpha)\right), \\
\text{where }\sigma_t^2(\alpha)
=(1-2\alpha)^{-2}&\left(\int_{F^{-1}_Z(\alpha)}^{F^{-1}_Z(1-\alpha)} z^2 \mathrm{~d} F_{Z}(z)+2 \alpha(F^{-1}_Z(\alpha))^2\right),
\end{aligned}
\end{equation}
implying that
\begin{equation}\label{eqn:t1}
\mathbb{E}_{\mathbf{z}} (\hat{\theta}_t - \theta)^2\rightarrow P^{-1}\sigma_t^2(\alpha) \mbox{ as } P\rightarrow\infty.
\end{equation}
WLOG, suppose the Laplace mechanism is used, then $s'^*_1-s'_1 \sim \mbox{Lap}(\frac{u^*-l^*}{\epsilon/4})$ and 
\begin{equation}\label{eqn:t2}
\mathbb{E}_{\mathbf{z}} \mathbb{E}_{\mathcal{M}|\mathbf{z}}(\hat{\theta}^*_t - \hat{\theta}_t)^2 =\frac{2}{P^2(1-\alpha-\beta)^2}\frac{\mathbb{E}_{\mathbf{z}} \mathbb{E}_{\mathcal{M}:PQ|\mathbf{z}}(u^*-l^*)^2}{(\epsilon/4)^2}
\end{equation}

Applying (4) in Corollary 5 to Eqn \eqref{eqn:t2}, we have
\begin{equation}\label{eqn:t4}
\mathbb{E}_{\mathbf{z}}\mathbb{E}_{\mathcal{M}|\mathbf{z}} (\hat{\theta}^*_t - \hat{\theta}_t)^2 \xrightarrow{P\rightarrow\infty} \frac{2}{P^2(1-\alpha-\beta)^2}\frac{(F_Z^{-1}(1-\beta)-F_Z^{-1}(\alpha))^2 + O(P^{-1/2})}{(\epsilon/4)^2},
\end{equation}

Plugging Eqns \eqref{eqn:t1} and \eqref{eqn:t4} on the right side of Eqn \eqref{eqn:CS_inequalityt}, 
\begin{equation}
\begin{aligned}
\mathbb{E}_{\mathbf{z}} \mathbb{E}_{\mathcal{M}|\mathbf{z}}(\hat{\theta}^*_t - \theta)^2
& \leq \frac{2}{P^2(1-\alpha-\beta)^2}\frac{(F_Z^{-1}(1-\beta)-F_Z^{-1}(\alpha))^2 + O(P^{-1/2})}{(\epsilon/4)^2} +\frac{\sigma_t^2(\alpha)}{P}\\
&+2\sqrt{\frac{\sigma_t^2(\alpha)}{P}\frac{2}{P^2(1-\alpha-\beta)^2}\frac{(F_Z^{-1}(1-\beta)-F_Z^{-1}(\alpha))^2 + O(P^{-1/2})}{(\epsilon/4)^2}}\\
&=O(P^{-2}\epsilon^{-2}+P^{-1}+P^{-3/2}\epsilon^{-1})
=O(P^{-1}+P^{-3/2}\epsilon^{-1}).
\end{aligned}
\end{equation}

If the Gaussian mechanism is used, then $s'^*_1-s'_1 \sim \mathcal{N}(0,\frac{(u^*-l^*)^2}{2(\rho/4)})$, similar to Eqn \eqref{eqn:t4},
\begin{equation}\label{eqn:t5}
\mathbb{E}_{\mathbf{z}}\mathbb{E}_{\mathcal{M}|\mathbf{z}} (\hat{\theta}^*_t - \hat{\theta}_t)^2 \xrightarrow{P\rightarrow\infty} \frac{2}{P^2(1-\alpha-\beta)^2}\frac{(F_Z^{-1}(1-\beta)-F_Z^{-1}(\alpha))^2 + O(P^{-1})}{2(\rho/4)},
\end{equation}
Therefore,
\begin{equation}
\mathbb{E}_{\mathbf{z}} \mathbb{E}_{\mathcal{M}|\mathbf{z}}(\hat{\theta}^*_t - \theta)^2 \leq O(P^{-2}\rho^{-1}+P^{-1}+P^{-3/2}\rho^{-1/2})
=O(P^{-1}+P^{-3/2}\rho^{-1/2}).
\end{equation}
Part 2: The winsorized mean is $\hat{\theta}_w = \alpha l + \beta u + \sum_{j=1}^{P-P_l-P_u}z'_j/P$ and its sanitized counterpart is 
$\hat{\theta}^*_{\text{w}}=\alpha l^*+\beta u^*+s'^*_1/P$. Applying the Cauchy-Schwarz inequality, 
\begin{equation}\label{eqn:CS_inequalityw}
\begin{aligned}
    \mathbb{E}_{\mathbf{z}} \mathbb{E}_{\mathcal{M}|\mathbf{z}}(\hat{\theta}^*_w - \theta)^2
    &\leq \mathbb{E}_{\mathbf{z}} \mathbb{E}_{\mathcal{M}|\mathbf{z}}(\hat{\theta}^*_w - \hat{\theta}_w)^2 \!+\! 
    \mathbb{E}_{\mathbf{z}} (\hat{\theta}_w\!-\! \theta)^2 \!+ \!2\sqrt{\mathbb{E}_{\mathbf{z}} (\hat{\theta}_w \!-\! \theta)^2 \mathbb{E}_{\mathbf{z}} \mathbb{E}_{\mathcal{M}|\mathbf{z}}(\hat{\theta}^*_w \!-\! \hat{\theta}_w)^2}.\\
\end{aligned}
\end{equation}
Theorem 3.2 in \citet{bickel1965some} derives the asymptotics of the winsorized mean when $\alpha=\beta$,
\begin{equation}\label{eqn:asym_winsor}
\begin{aligned}
    \sqrt{P} \cdot (\hat{\theta}_w -\theta)
    &\xrightarrow{d}N\left(0, \sigma_w^2(\alpha)\right)\\
    \text{where }\sigma_w^2(\alpha)
=\int_{F^{-1}_Z(\alpha)}^{F^{-1}_Z(1-\alpha)} z^2 \mathrm{~d} F_{Z}(z)&+2 \alpha[F^{-1}_Z(1-\alpha)+\alpha / f_{Z}(F^{-1}_Z(\alpha))]^2, 
\end{aligned}
\end{equation}
implying that
\begin{equation}\label{MLE_winsor}
\begin{aligned}
\mathbb{E}_{\mathbf{z}}(\hat{\theta}_w - \theta)^2\rightarrow \sigma_w^2(\alpha)/P \mbox{ as } P\rightarrow\infty.\\
\end{aligned}
\end{equation}

WLOG, suppose that the Laplace mechanism is used, then 
\begin{equation}\label{eqn:EM_Lap}
\mathbb{E}_{\mathcal{M}:Lap|\mathbf{z}}(s_1'^{*}-s_1')=0;
\qquad
\mathbb{E}_{\mathcal{M}:Lap|\mathbf{z}}(s_1'^{*}-s_1')^2=2(\frac{u^*-l^*}{\epsilon/4})^2.
\end{equation}
Applying Corollary 5 and  Lemma 3 to expand the first square term in Eqn \eqref{eqn:CS_inequalityw}, 
\begin{align}
\label{eqn:EM_winsor}
&\mathbb{E}_{\mathbf{z}}\mathbb{E}_{\mathcal{M}|\mathbf{z}}(\hat{\theta}^*_w - \hat{\theta}_w)^2 \notag\\
=& \alpha^2 \mathbb{E}_{\mathbf{z}}\mathbb{E}_{\mathcal{M}:PQ|\mathbf{z}}(l^*-l)^2 + \beta^2 \mathbb{E}_{\mathbf{z}}\mathbb{E}_{\mathcal{M}:PQ|\mathbf{z}}(u^*-u)^2 + P^{-2} \mathbb{E}_{\mathbf{z}}\mathbb{E}_{\mathcal{M}|\mathbf{z}}(s_1'^{*}-s_1')^2 \notag\\
&+ 2\alpha\beta  \mathbb{E}_{\mathbf{z}}\mathbb{E}_{\mathcal{M}:PQ|\mathbf{z}}(l^*-l) \cdot   \mathbb{E}_{\mathbf{z}}\mathbb{E}_{\mathcal{M}:PQ|\mathbf{z}}(u^*-u) + 2\alpha P^{-1} \mathbb{E}_{\mathbf{z}}\mathbb{E}_{\mathcal{M}:PQ|\mathbf{z}}(l^*-l)\cdot  \mathbb{E}_{\mathbf{z}} \mathbb{E}_{\mathcal{M}|\mathbf{z}}(s_1'^{*}-s_1')\notag\\
& +2\beta P^{-1}\mathbb{E}_{\mathbf{z}}\mathbb{E}_{\mathcal{M}:PQ|\mathbf{z}}(u^*-u) \cdot \mathbb{E}_{\mathbf{z}}\mathbb{E}_{\mathcal{M}|\mathbf{z}}(s_1'^{*}-s_1') \notag\\
\rightarrow& \frac{2\alpha^2}{3(Pf_Z(F_Z^{-1}(\alpha)))^2} +  \frac{2\beta^2}{3(Pf_Z(F_Z^{-1}(1-\beta)))^2} + 2P^{-2}\mathbb{E}_{\mathbf{z}}\mathbb{E}_{\mathcal{M}:PQ|\mathbf{z}}(\frac{u^*-l^*}{\epsilon/4})^2\notag\\
&+2\alpha\beta \frac{1}{2Pf_Z(F_Z^{-1}(\alpha))\cdot 2Pf_Z(F_Z^{-1}(1-\beta))}\notag\\
\rightarrow& \frac{2\alpha^2}{3(Pf_Z(F_Z^{-1}(\alpha)))^2} +  \frac{2\beta^2}{3(Pf_Z(F_Z^{-1}(1-\beta)))^2} + \frac{32}{P^2\epsilon^2}\left((F_Z^{-1}(1-\beta)-F_Z^{-1}(\alpha))^2 + O(P^{-1/2})\right)\notag\\
&+ \frac{\alpha\beta}{2P^2f_Z(F_Z^{-1}(\alpha))\cdot f_Z(F_Z^{-1}(1-\beta))}\notag\\
=&O(P^{-2} + P^{-2}\epsilon^{-2}).
\end{align}

Plugging Eqns \eqref{MLE_winsor} and \eqref{eqn:EM_winsor} into Eqn \eqref{eqn:CS_inequalityw}, we have
\begin{equation}
\begin{aligned}
\mathbb{E}_{\mathbf{z}} \mathbb{E}_{\mathcal{M}|\mathbf{z}}(\hat{\theta}^*_w - \theta)^2
\leq & O(P^{-1}) + O(P^{-2} + P^{-2}\epsilon^{-2})+ O(P^{-3/2} + P^{-3/2}\epsilon^{-1})\\
= & O(P^{-1} + P^{-3/2}\epsilon^{-1}).
\end{aligned}
\end{equation}

If the Gaussian mechanism is used instead, then
\begin{equation}\label{eqn:EM_Gaus}
\mathbb{E}_{\mathcal{M}:Gaus|\mathbf{z}}(s_1'^{*}-s_1')=0;
\qquad
\mathbb{E}_{\mathcal{M}:Gaus|\mathbf{z}}(s_1'^{*}-s_1')^2=\frac{(u^*-l^*)^2}{2(\rho/4)},
\end{equation}
therefore, still follow the derivation in Eqn \eqref{eqn:EM_winsor}
\begin{align}
\mathbb{E}_{\mathbf{z}}\mathbb{E}_{\mathcal{M}:Gaus|\mathbf{z}}(\hat{\theta}^*_w - \hat{\theta}_w)^2 &= O(P^{-2} + P^{-2}\rho^{-1})\notag\\
\mathbb{E}_{\mathbf{z}} \mathbb{E}_{\mathcal{M}:Gaus|\mathbf{z}}(\hat{\theta}^*_w - \theta)^2 
&\leq O(P^{-1}) + O(P^{-2} + P^{-2}\rho^{-1})+ O(P^{-3/2} + P^{-3/2}\rho^{-1/2})\notag\\
&= O(P^{-1} + P^{-3/2}\rho^{-1/2}).
\end{align}

\end{proof}

\section*{II. Additional Simulation Results}
\begin{landscape}
\subsection*{Gaussian, $\theta=0$ and $\alpha=\beta$}

\begin{figure}[!htb]
\hspace{0.6in}$\epsilon=0.5$\hspace{1.3in}$\epsilon=1$\hspace{1.4in}$\epsilon=2$
\hspace{1.4in}$\epsilon=5$\hspace{1.4in}$\epsilon=50$\\
\includegraphics[width=0.26\textwidth, trim={2.2in 0 2.2in 0},clip] {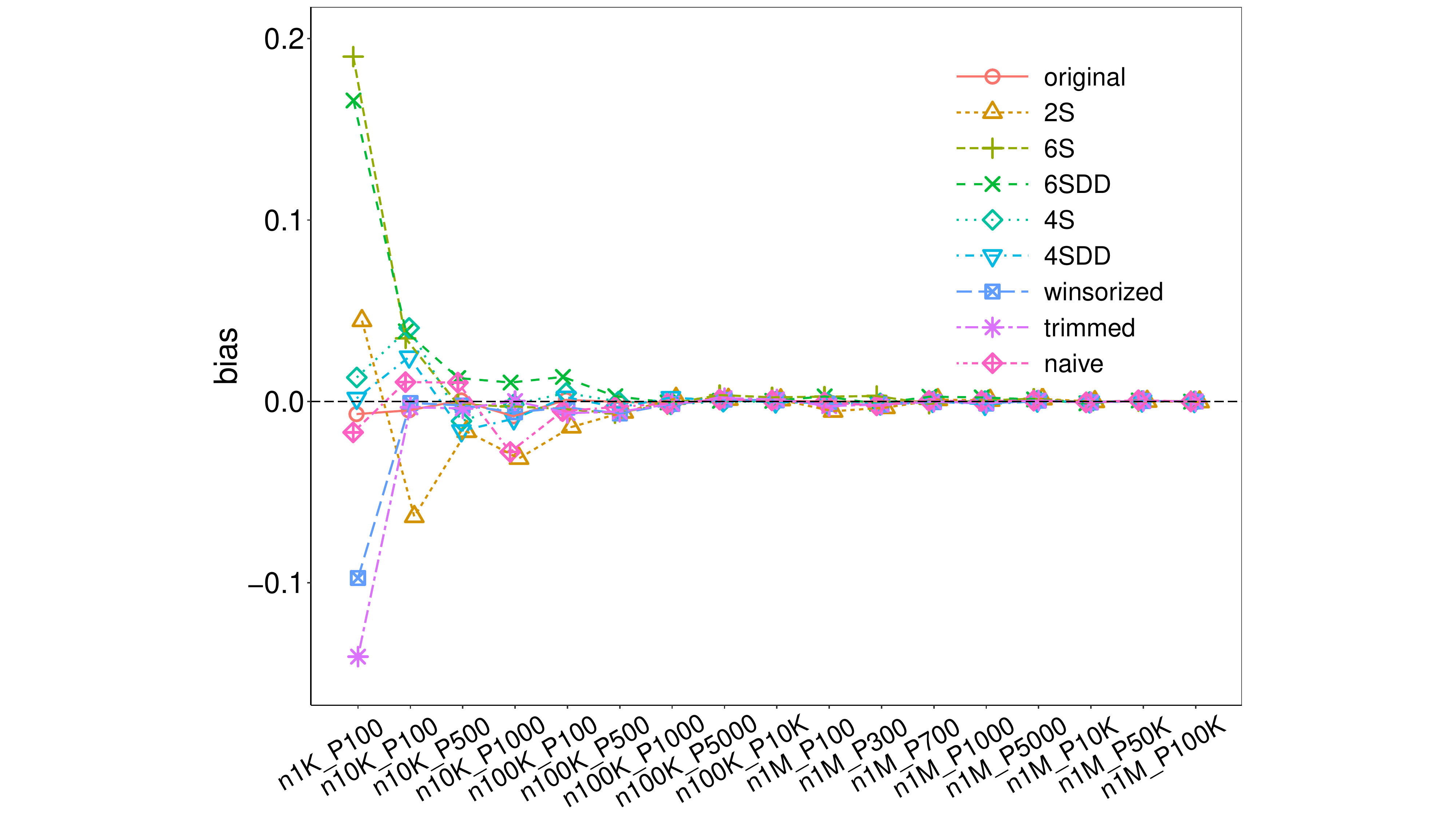}
\includegraphics[width=0.26\textwidth, trim={2.2in 0 2.2in 0},clip] {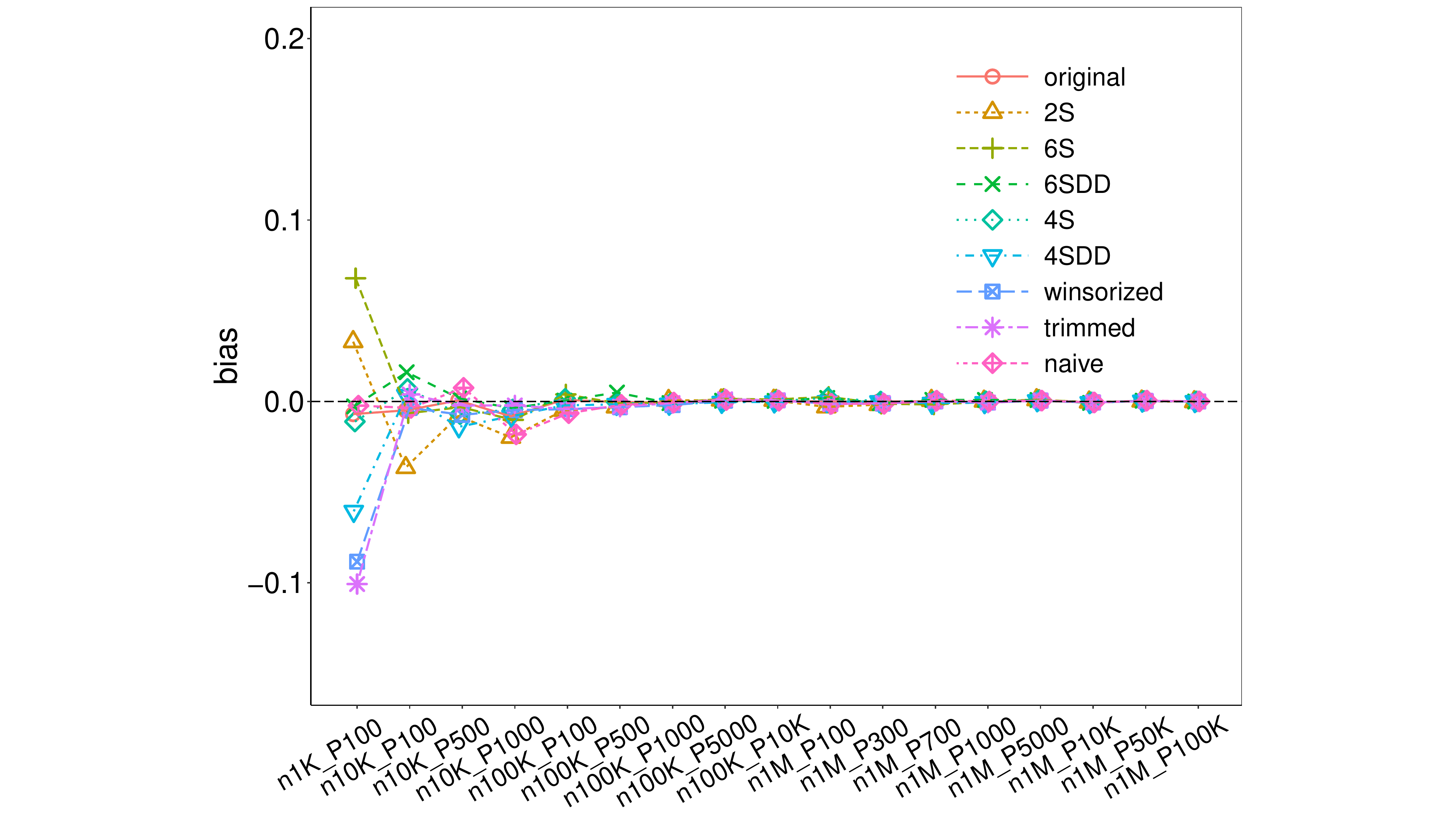}
\includegraphics[width=0.26\textwidth, trim={2.2in 0 2.2in 0},clip] {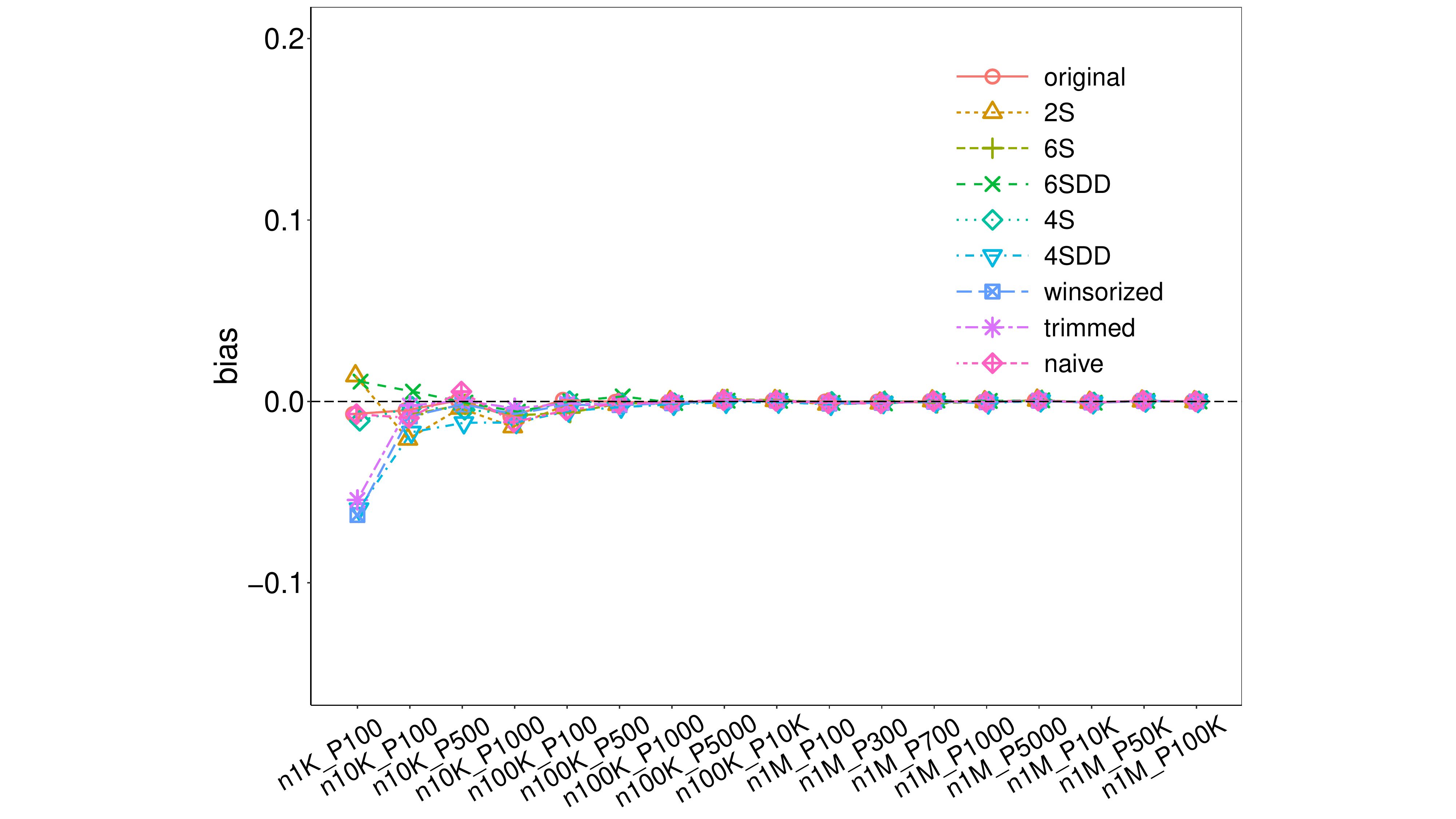}
\includegraphics[width=0.26\textwidth, trim={2.2in 0 2.2in 0},clip] {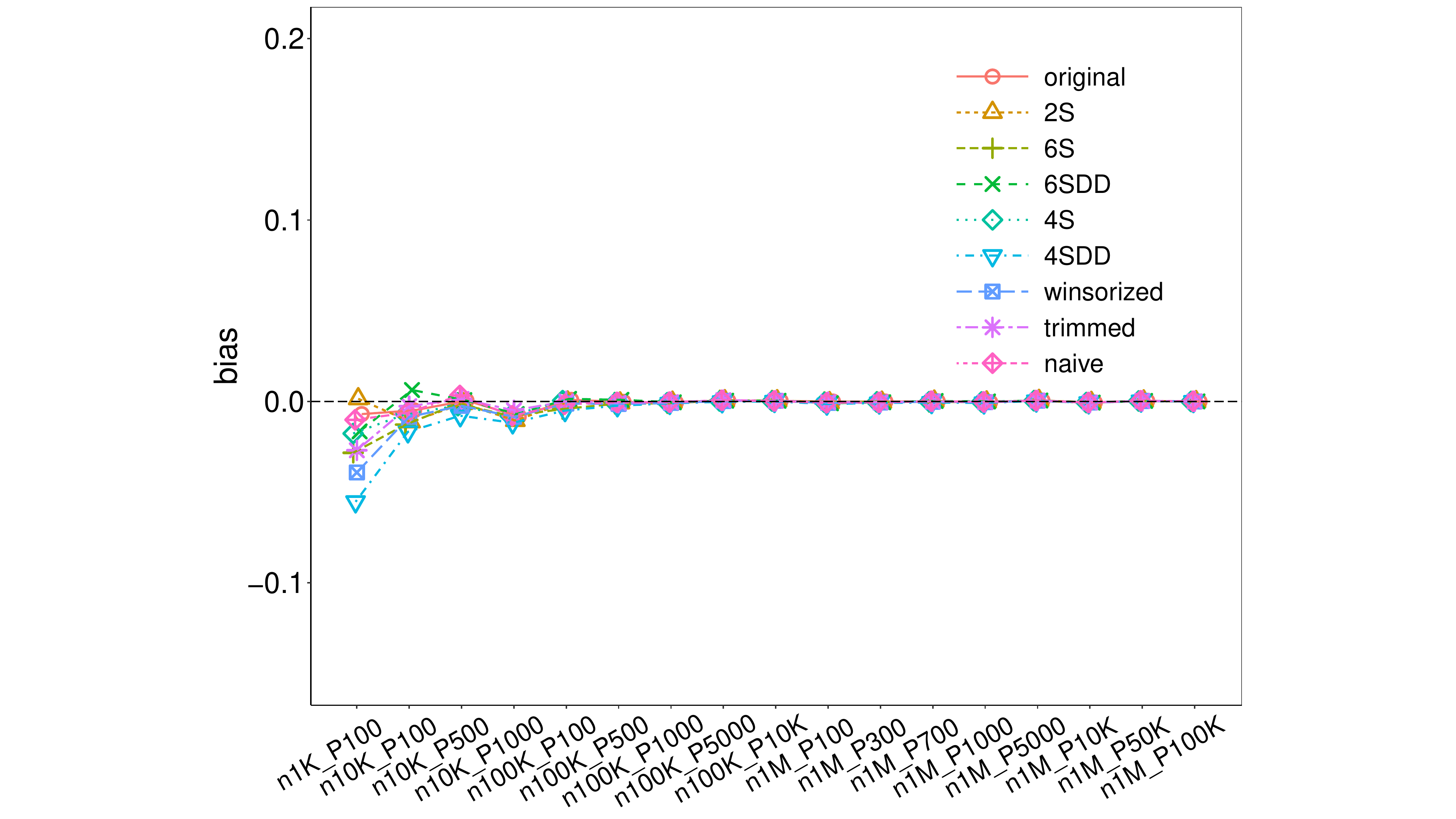}
\includegraphics[width=0.26\textwidth, trim={2.2in 0 2.2in 0},clip] {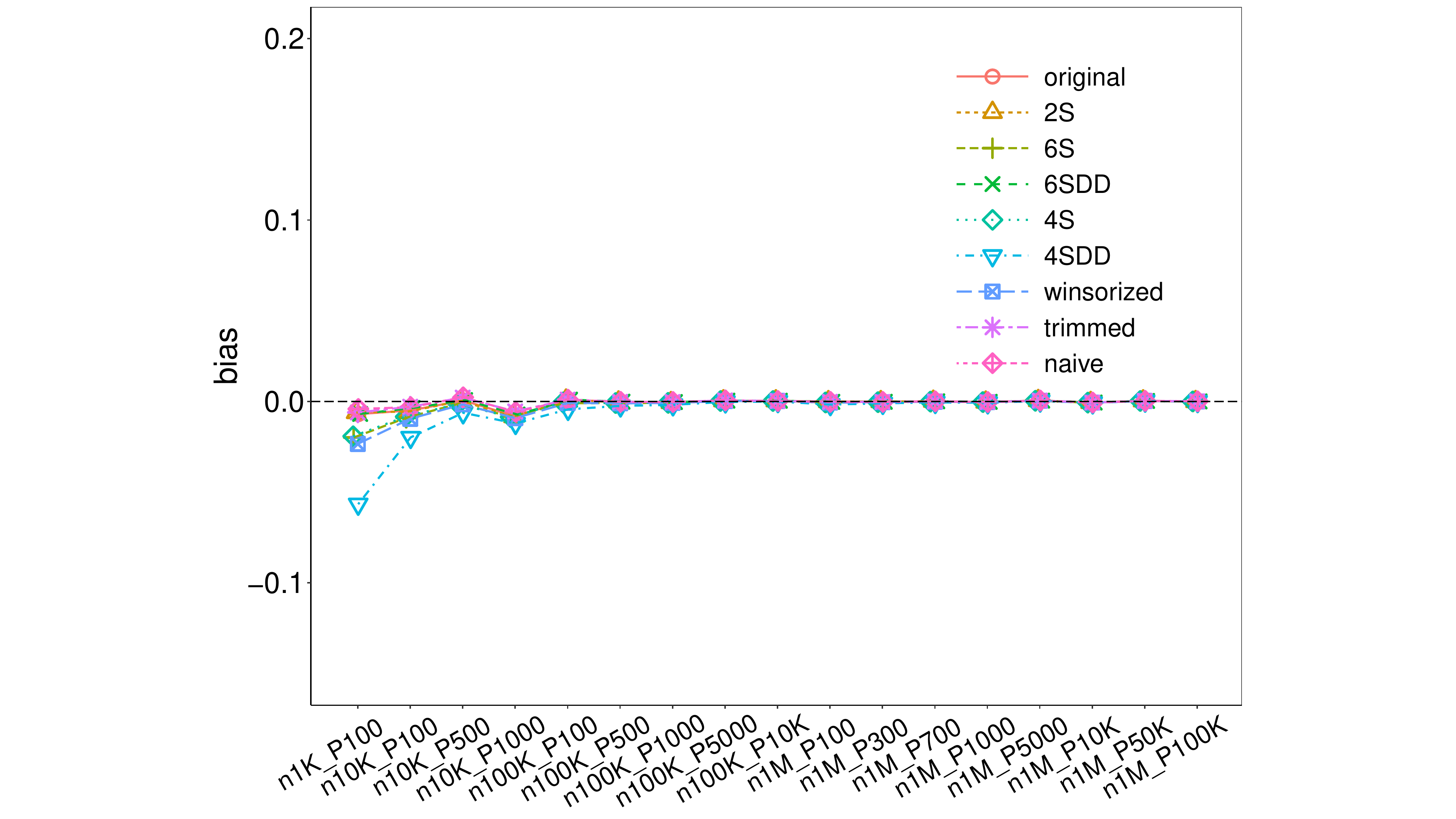}\\
\includegraphics[width=0.26\textwidth, trim={2.2in 0 2.2in 0},clip] {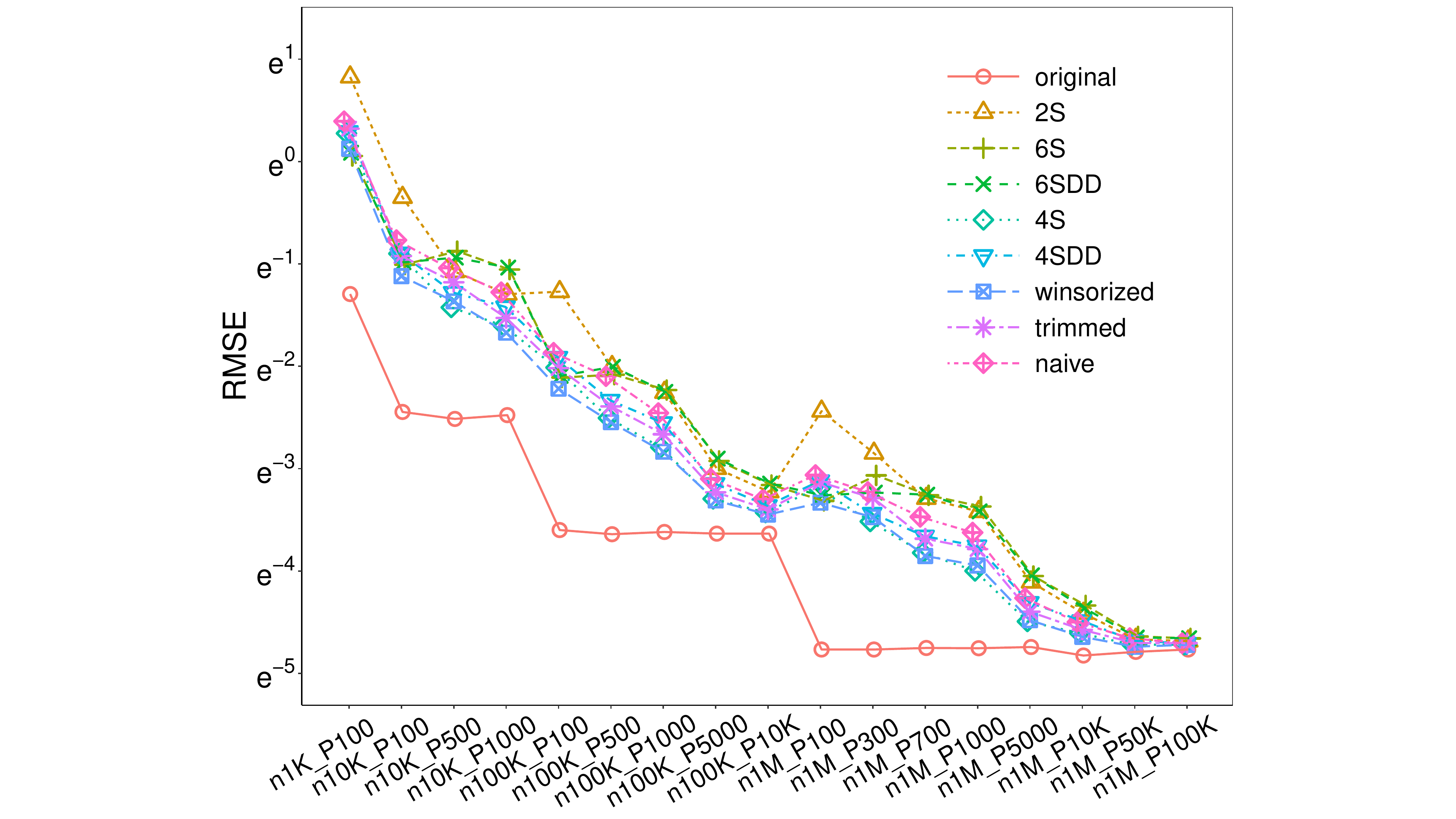}
\includegraphics[width=0.26\textwidth, trim={2.2in 0 2.2in 0},clip] {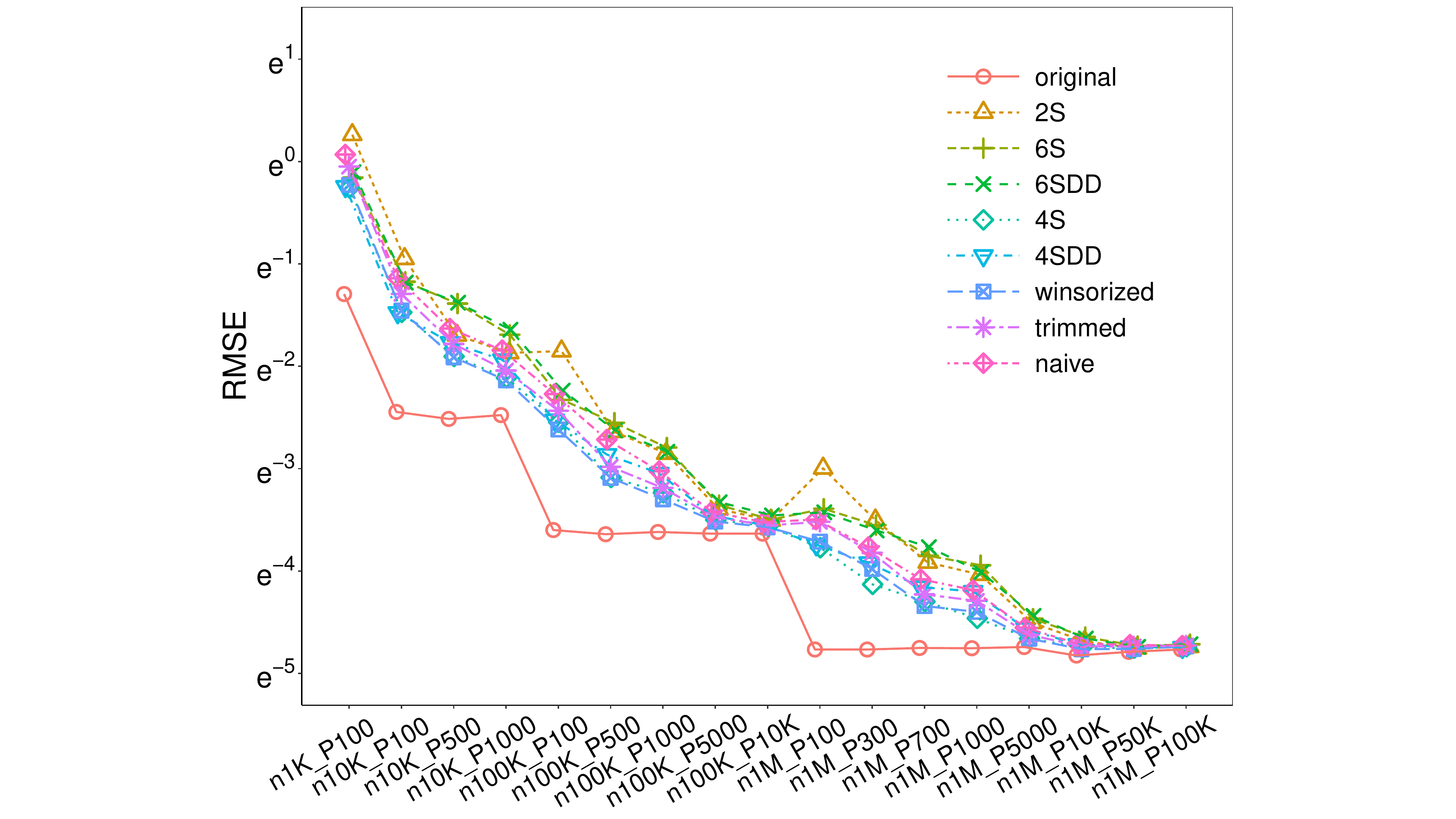}
\includegraphics[width=0.26\textwidth, trim={2.2in 0 2.2in 0},clip] {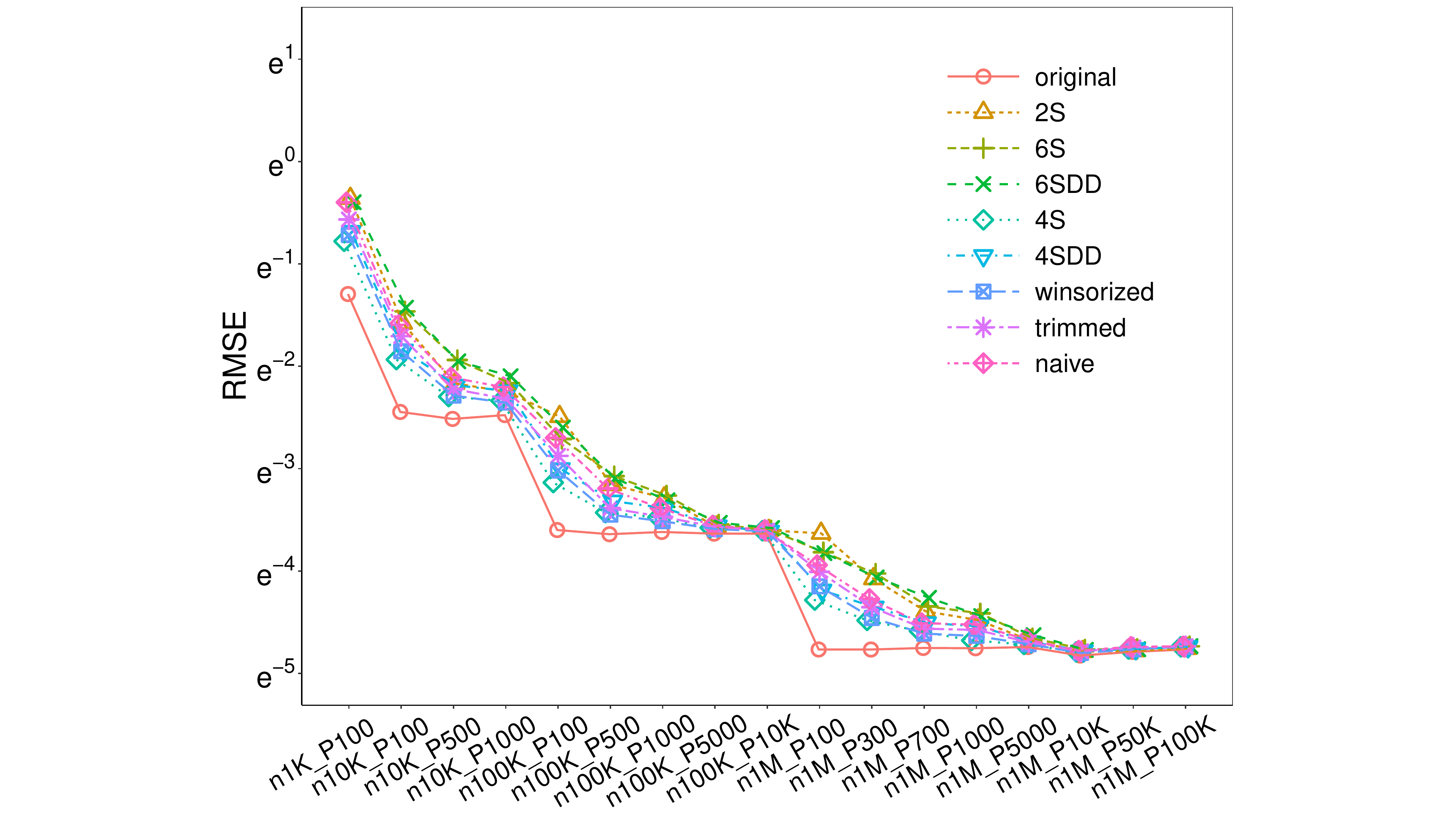}
\includegraphics[width=0.26\textwidth, trim={2.2in 0 2.2in 0},clip] {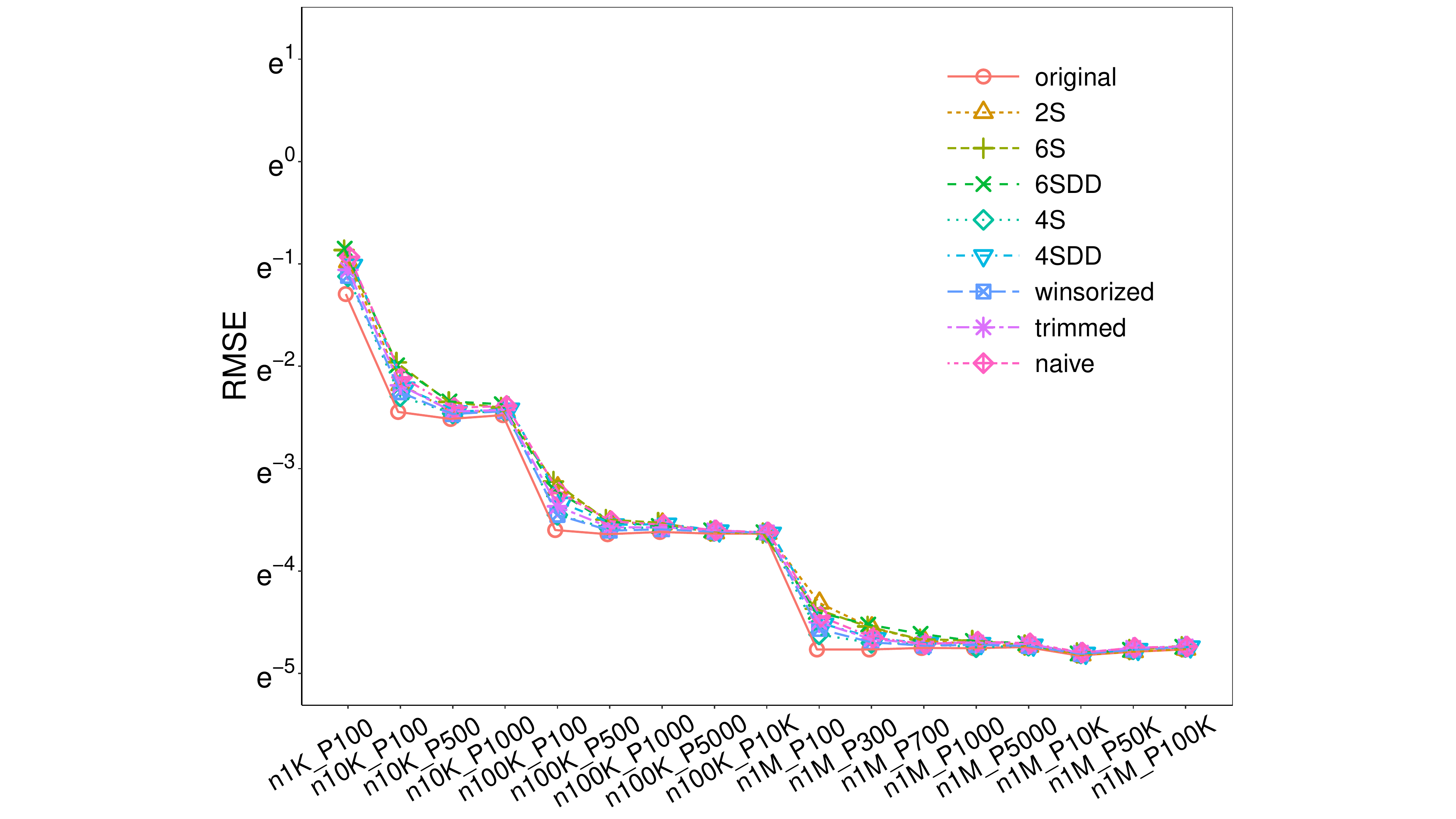}
\includegraphics[width=0.26\textwidth, trim={2.2in 0 2.2in 0},clip] {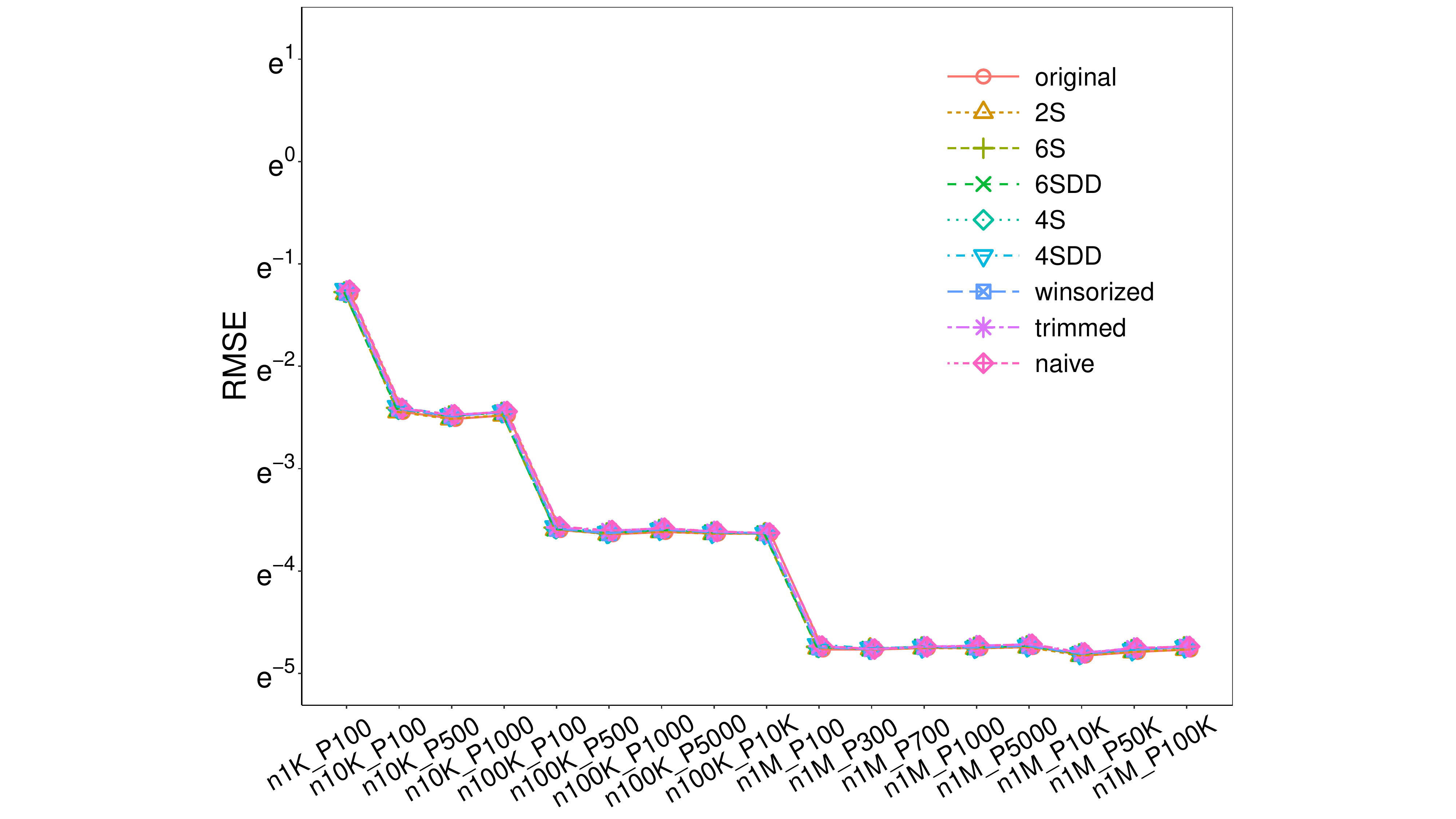}\\
\includegraphics[width=0.26\textwidth, trim={2.2in 0 2.2in 0},clip] {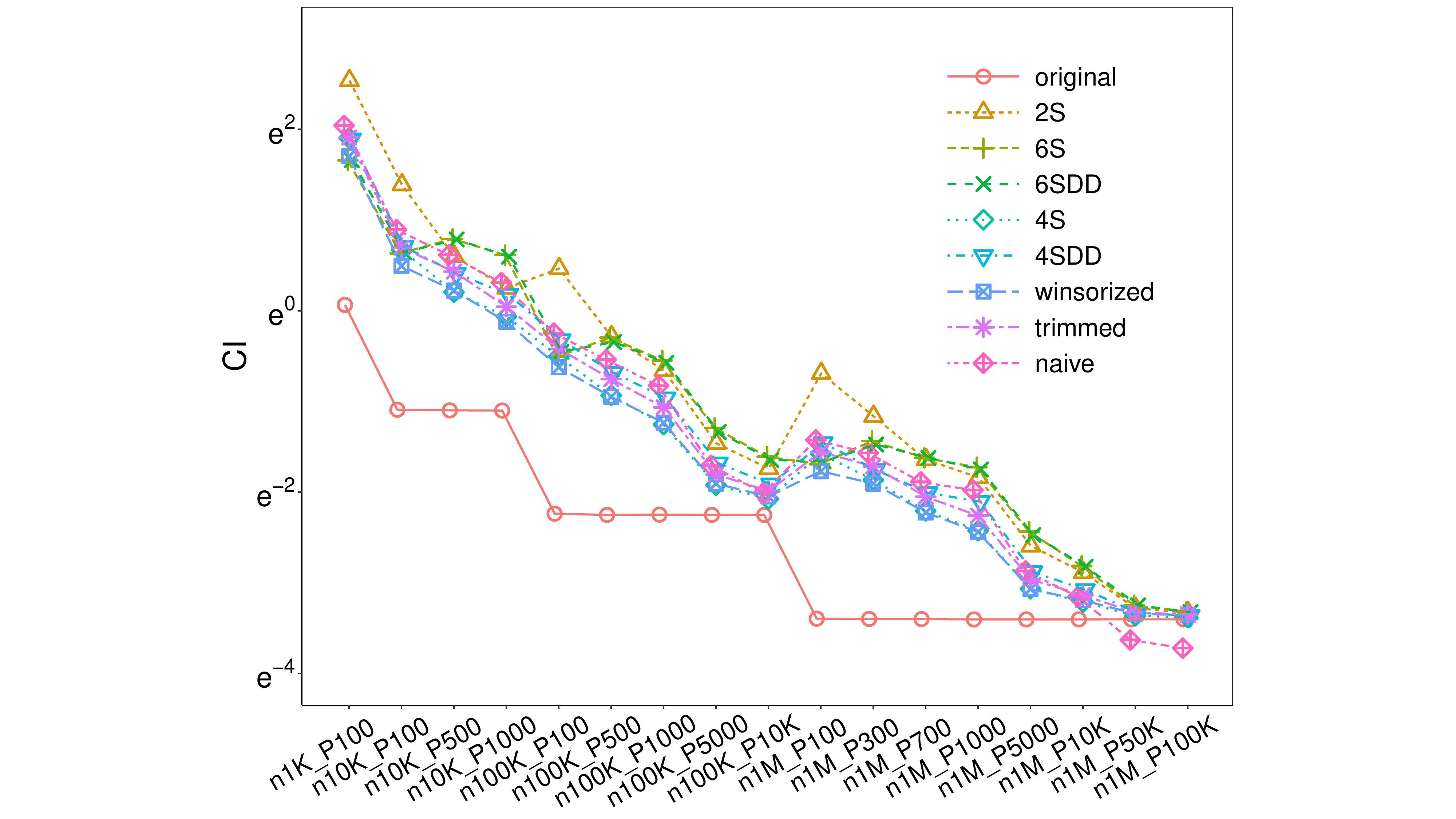}
\includegraphics[width=0.26\textwidth, trim={2.2in 0 2.2in 0},clip] {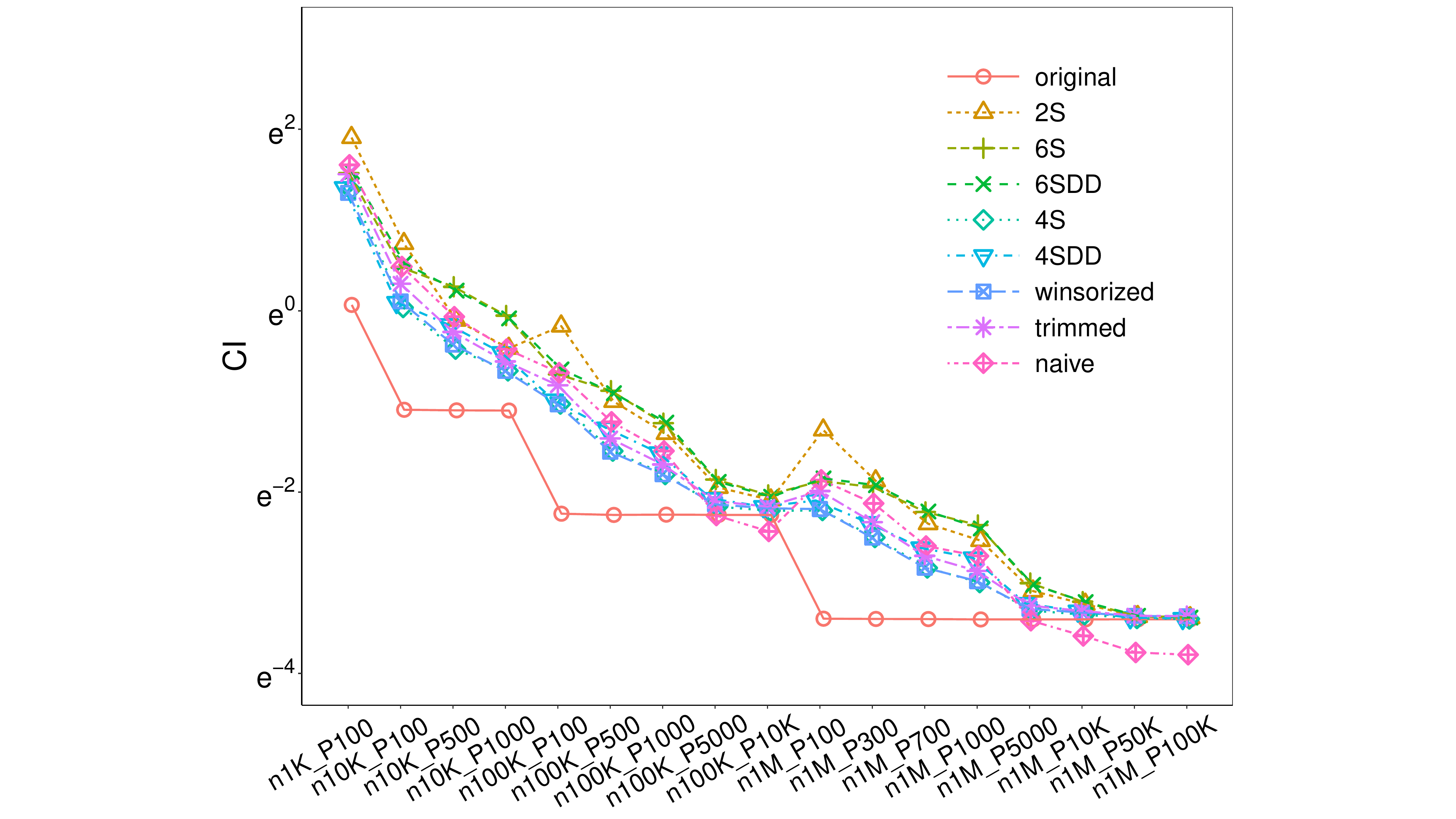}
\includegraphics[width=0.26\textwidth, trim={2.2in 0 2.2in 0},clip] {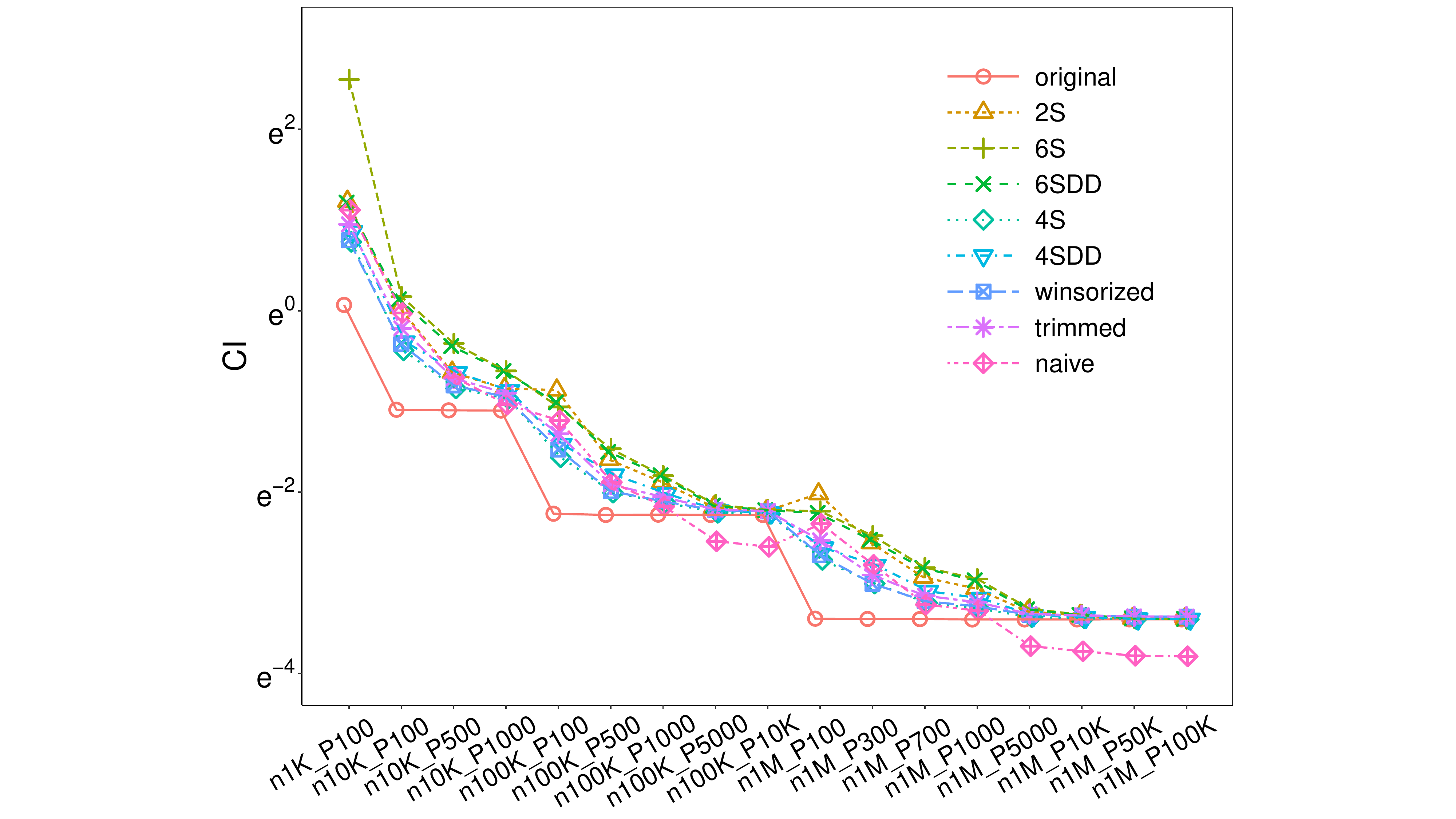}
\includegraphics[width=0.26\textwidth, trim={2.2in 0 2.2in 0},clip] {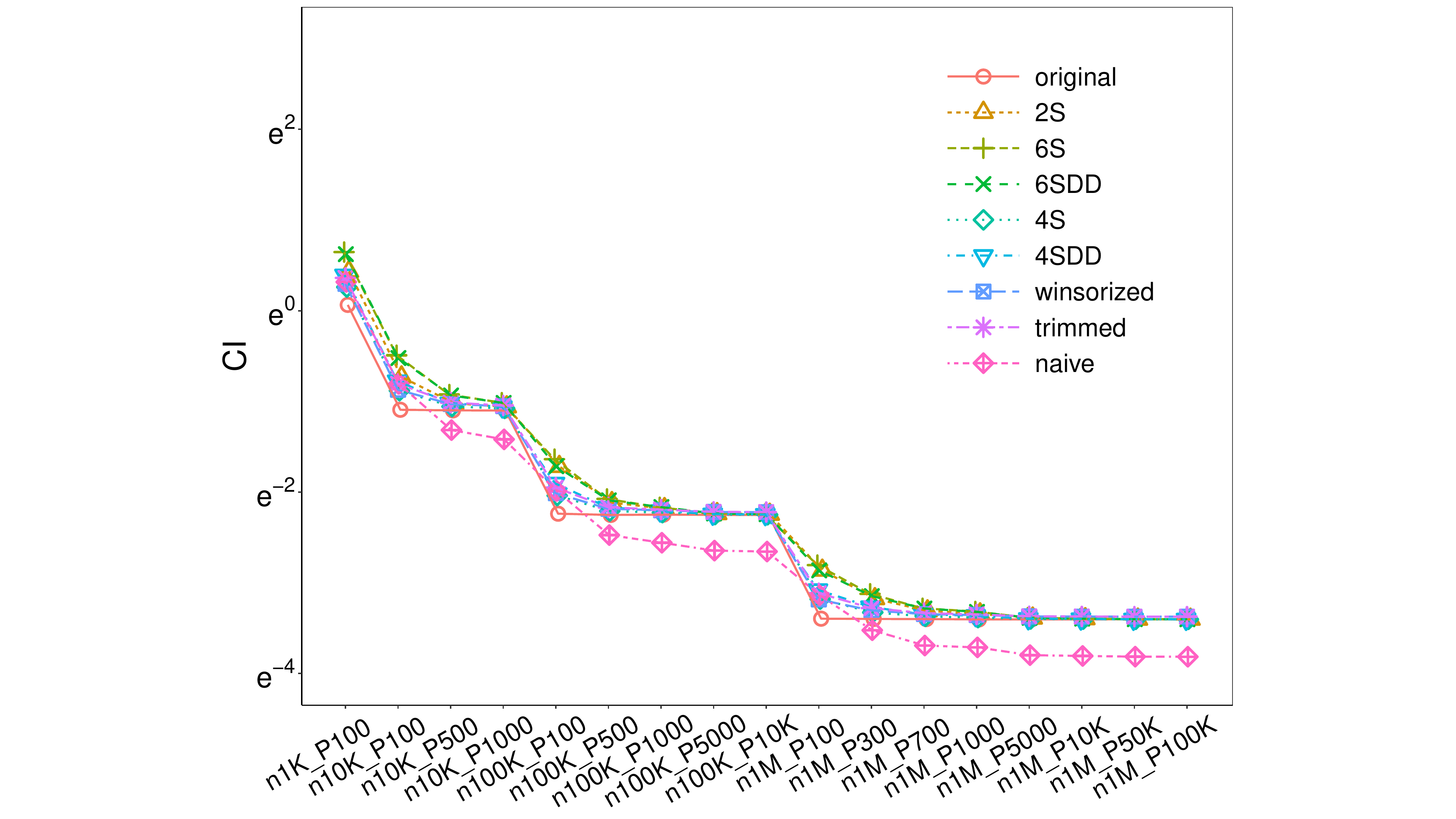}
\includegraphics[width=0.26\textwidth, trim={2.2in 0 2.2in 0},clip] {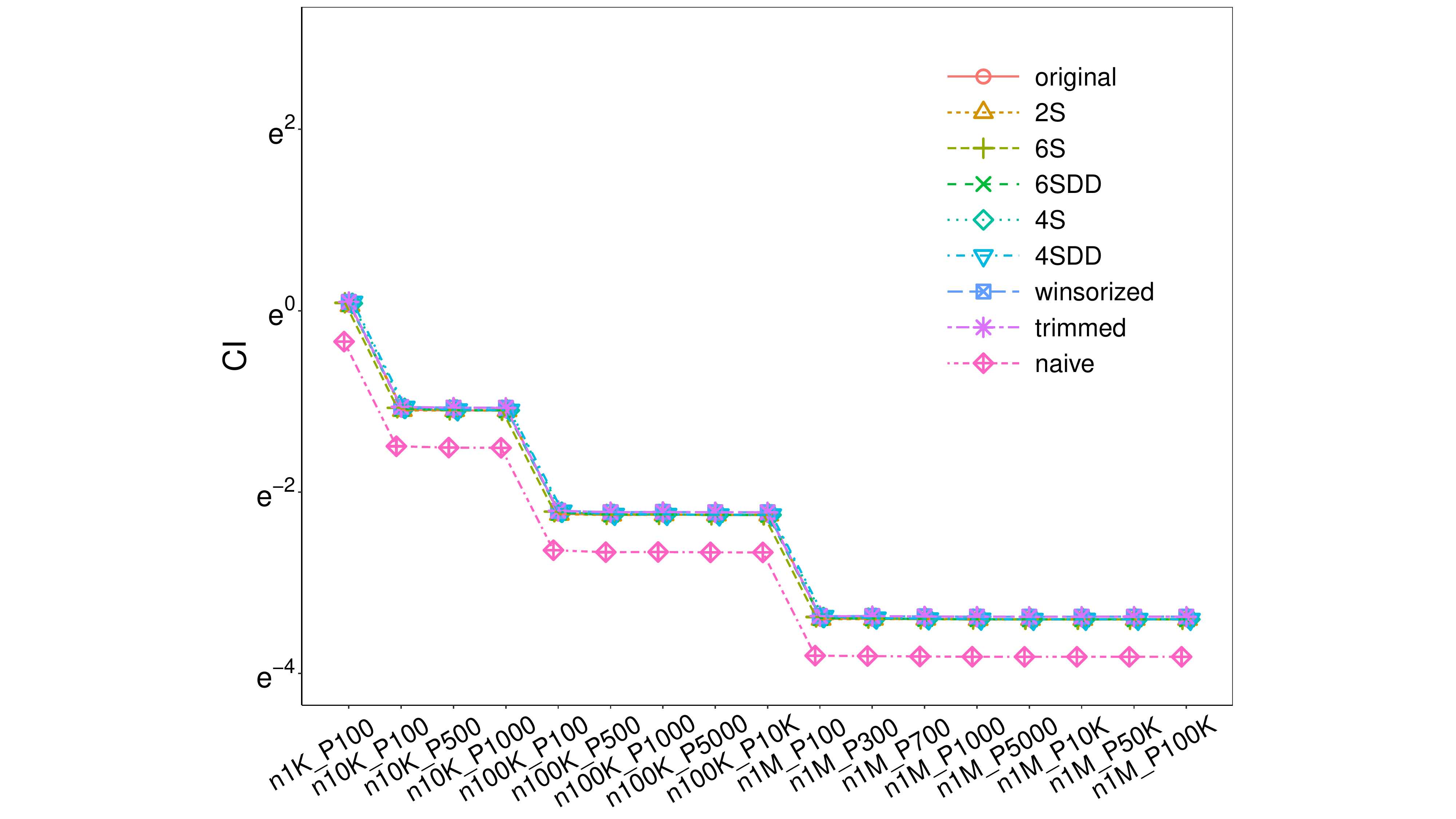}\\
\includegraphics[width=0.26\textwidth, trim={2.2in 0 2.2in 0},clip] {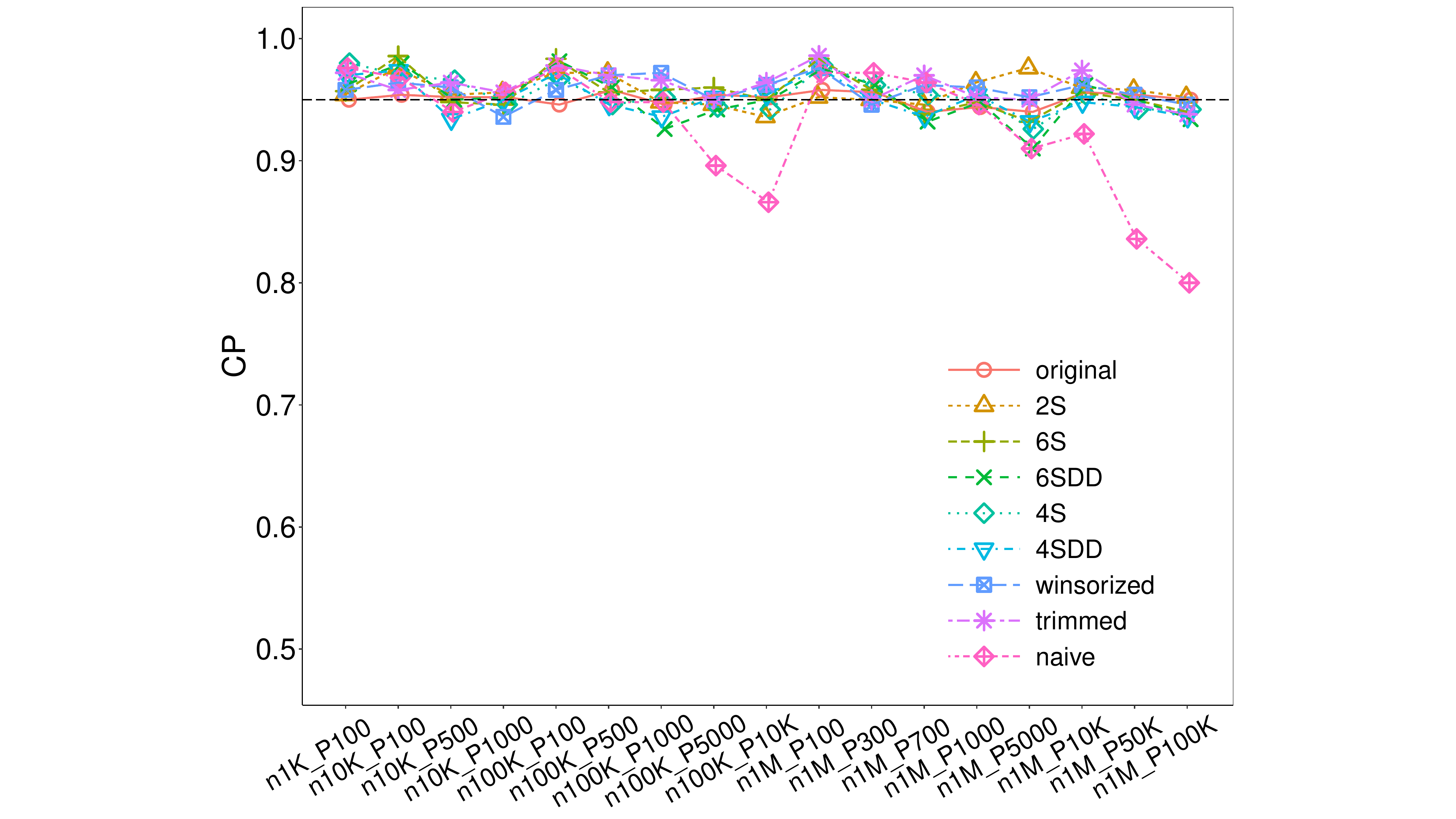}
\includegraphics[width=0.26\textwidth, trim={2.2in 0 2.2in 0},clip] {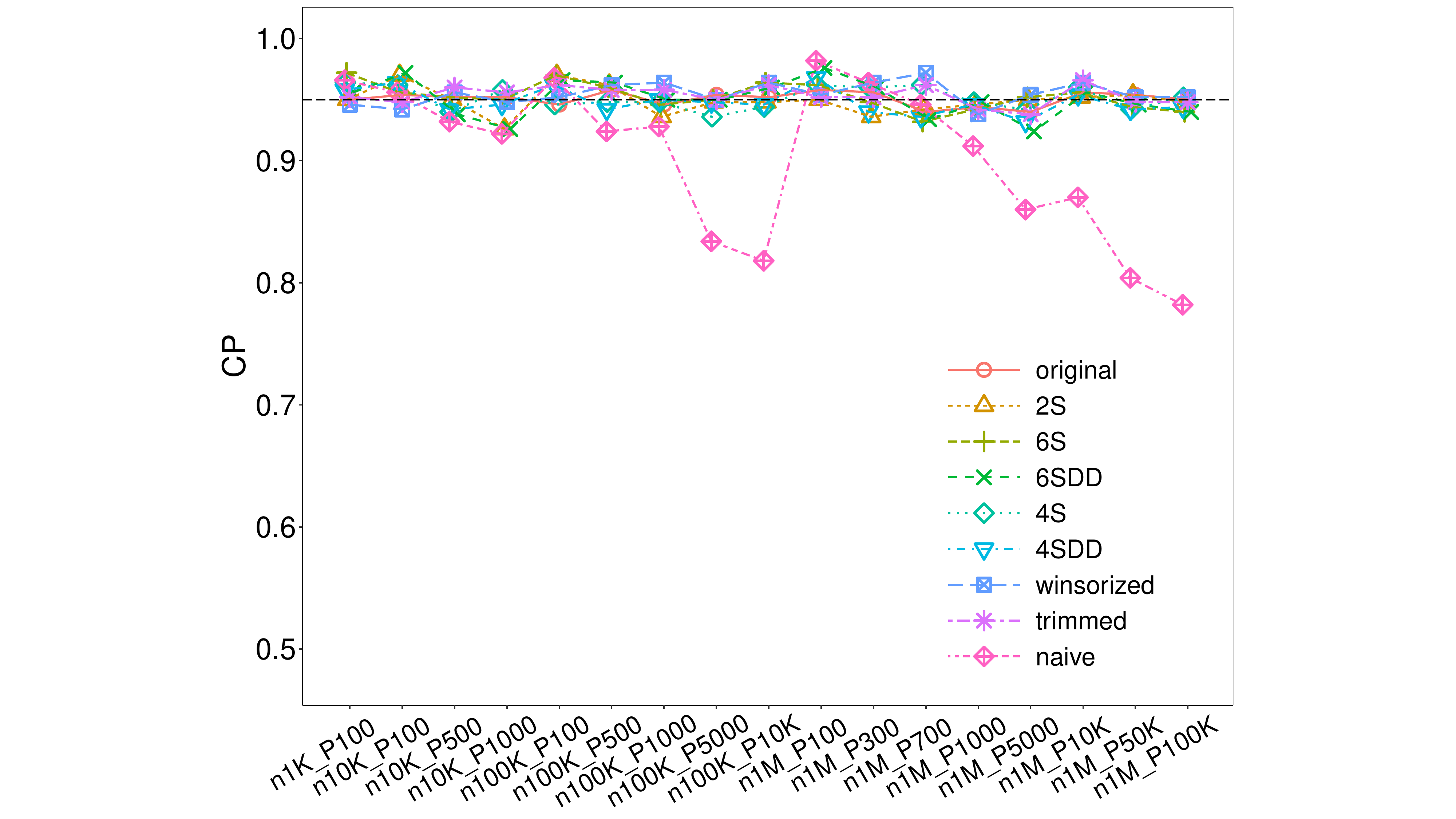}
\includegraphics[width=0.26\textwidth, trim={2.2in 0 2.2in 0},clip] {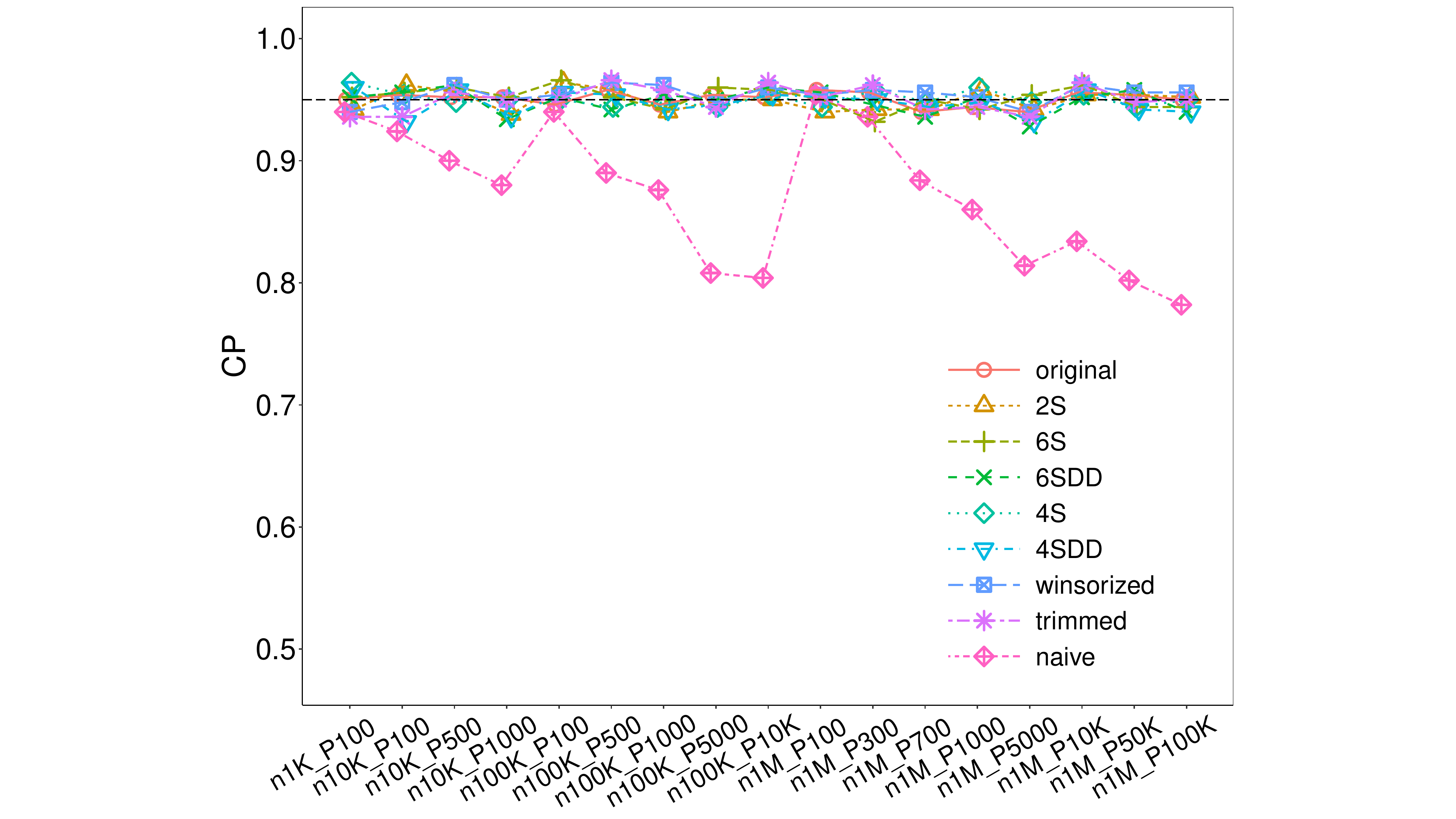}
\includegraphics[width=0.26\textwidth, trim={2.2in 0 2.2in 0},clip] {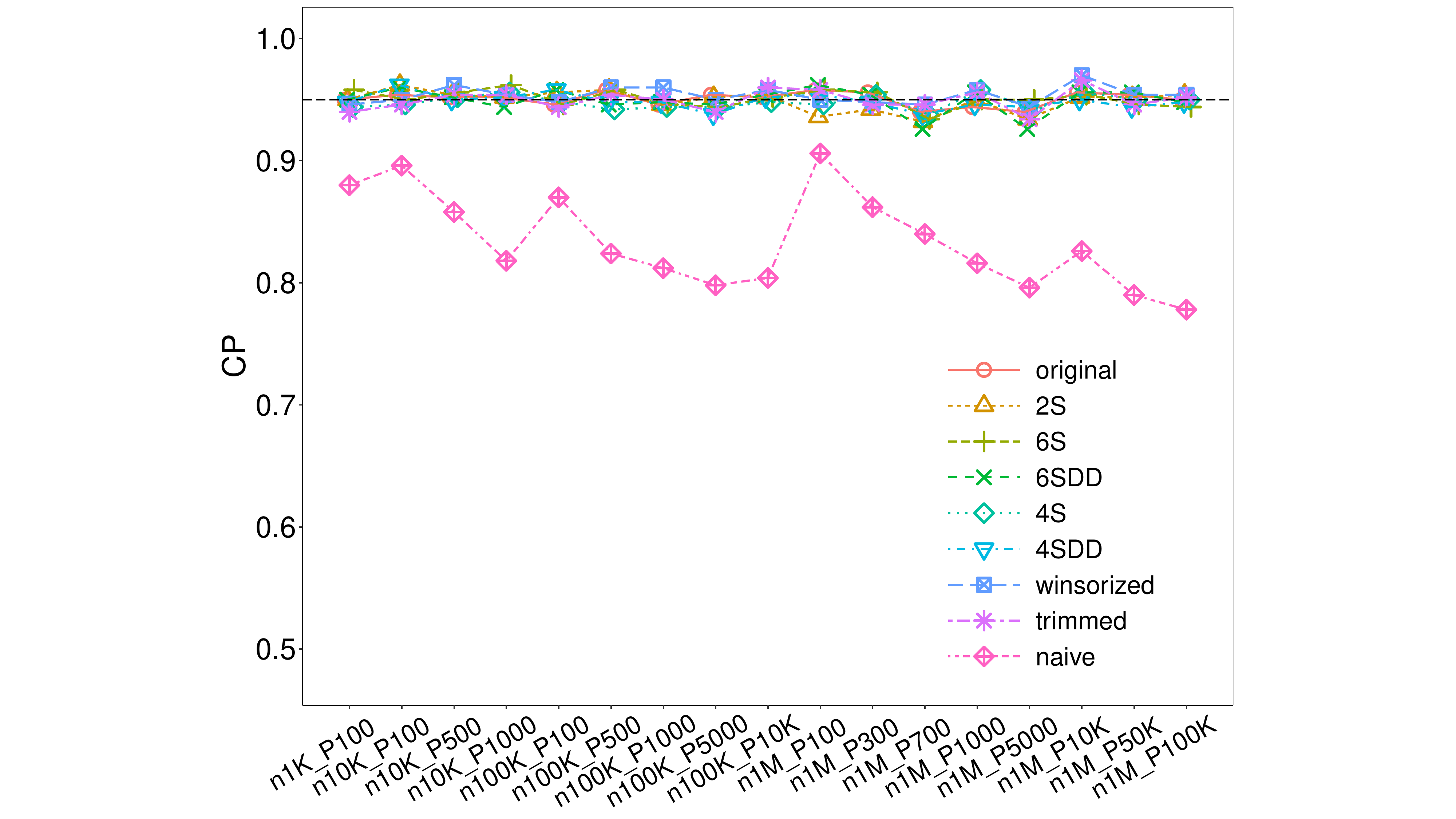}
\includegraphics[width=0.26\textwidth, trim={2.2in 0 2.2in 0},clip] {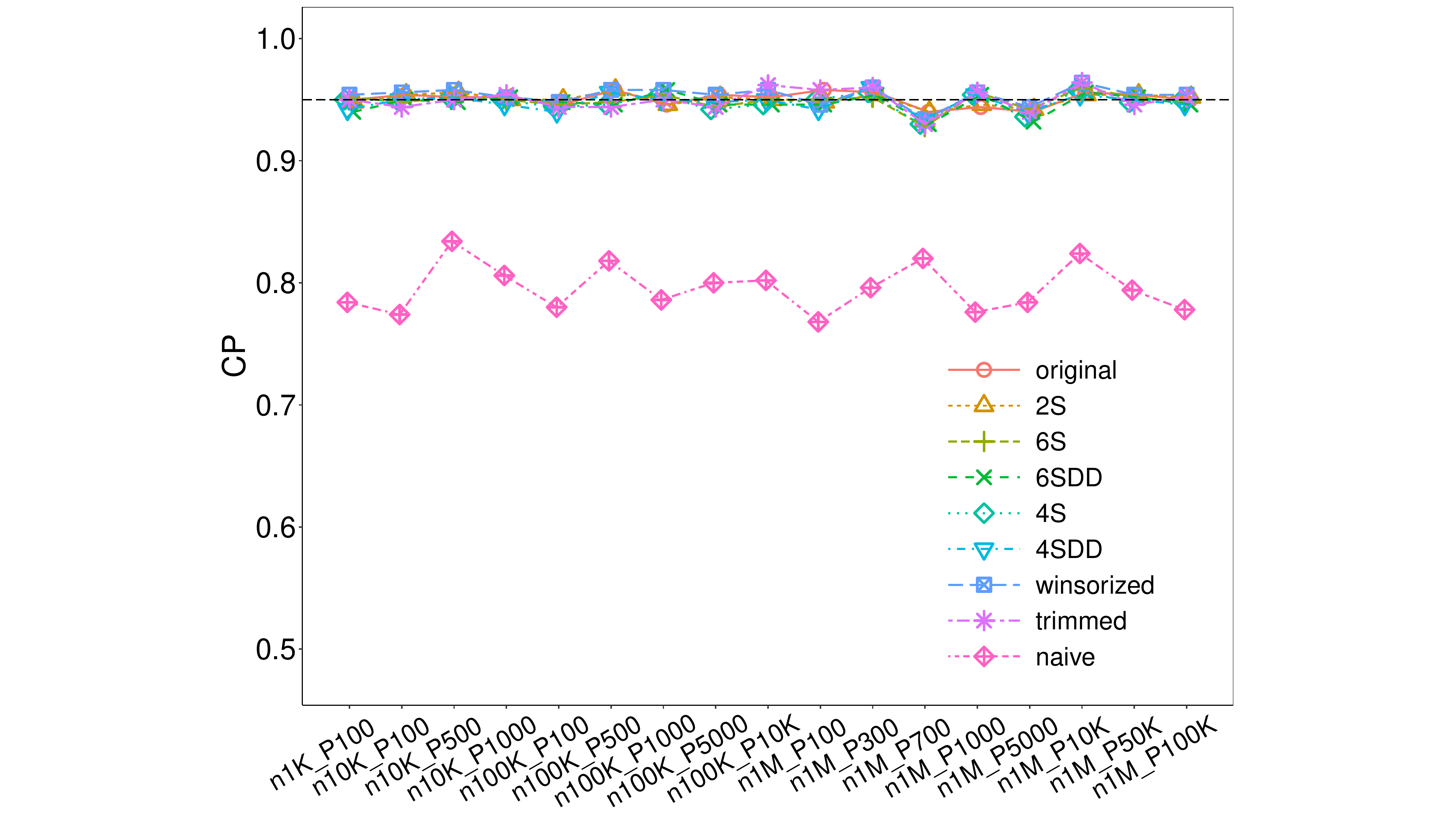}\\

\caption{Gaussian data; $\epsilon$-DP; $\theta=0$ and $\alpha=\beta$} \label{fig:0sDP}
\end{figure}
\end{landscape}

\begin{landscape}
\begin{figure}[!htb]
\hspace{0.6in}$\rho=0.005$\hspace{1in}$\rho=0.02$\hspace{1.2in}$\rho=0.08$
\hspace{1.1in}$\rho=0.32$\hspace{1.2in}$\rho=1.28$\\
\includegraphics[width=0.26\textwidth, trim={2.2in 0 2.2in 0},clip] {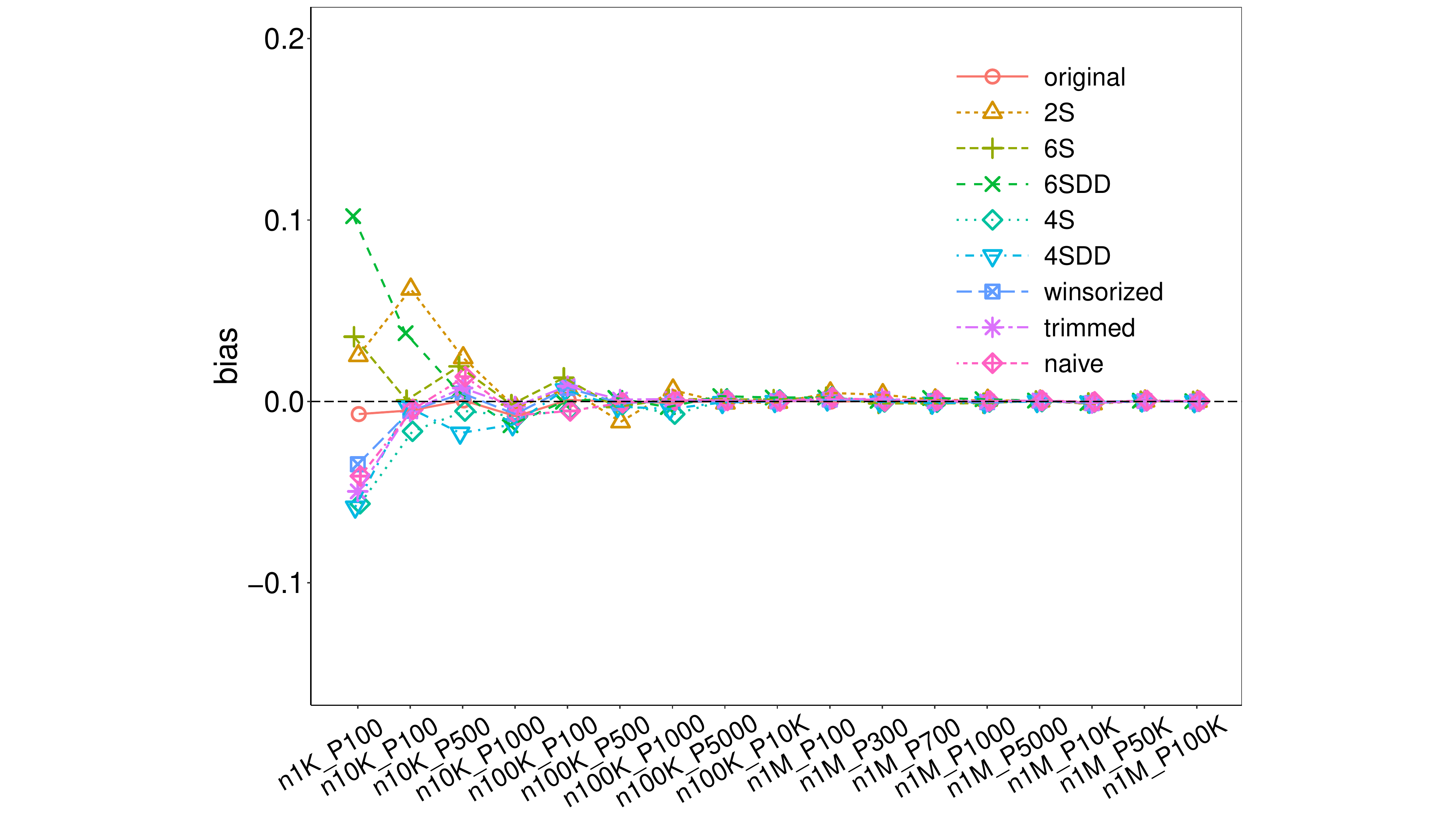}
\includegraphics[width=0.26\textwidth, trim={2.2in 0 2.2in 0},clip] {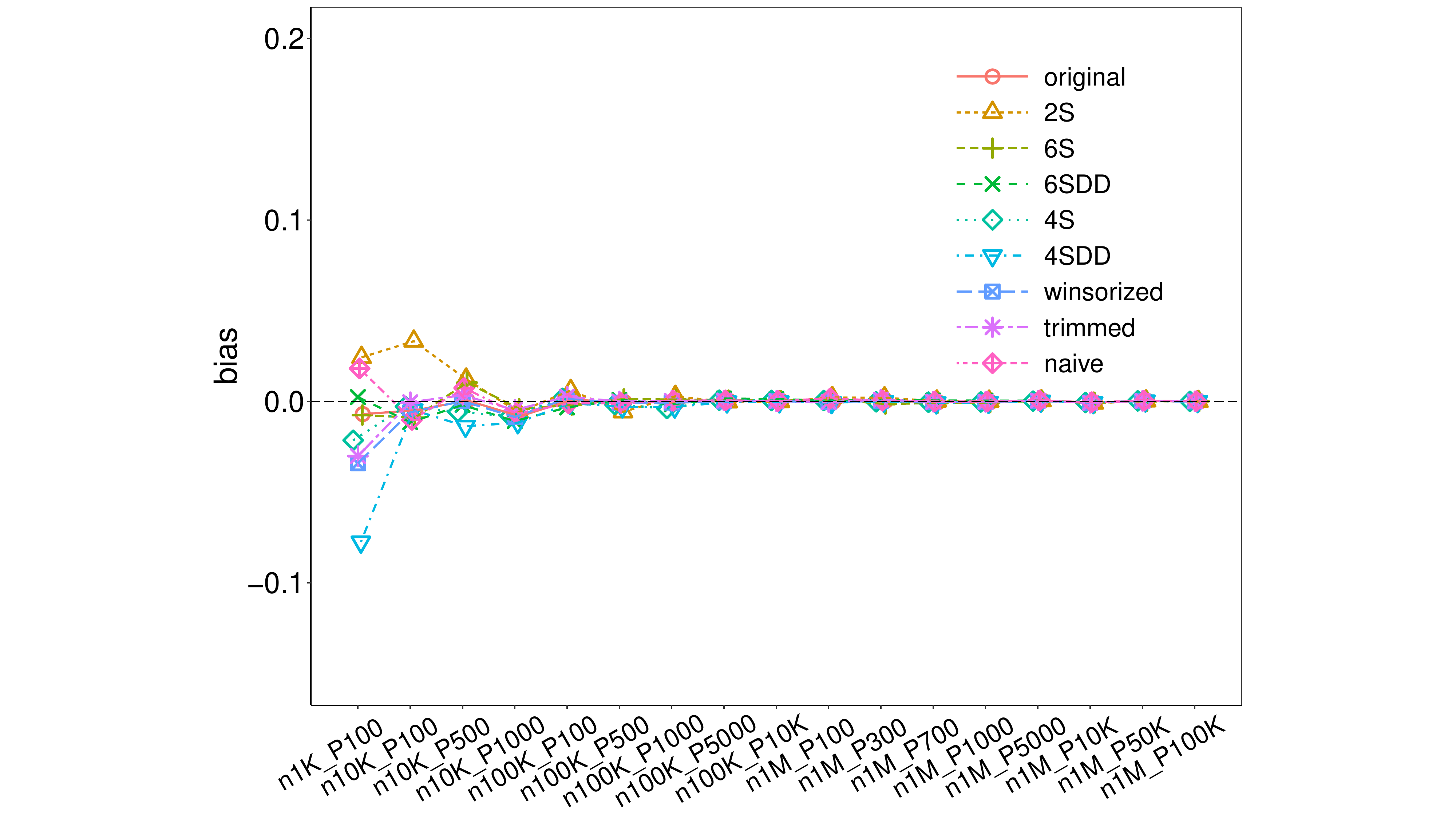}
\includegraphics[width=0.26\textwidth, trim={2.2in 0 2.2in 0},clip] {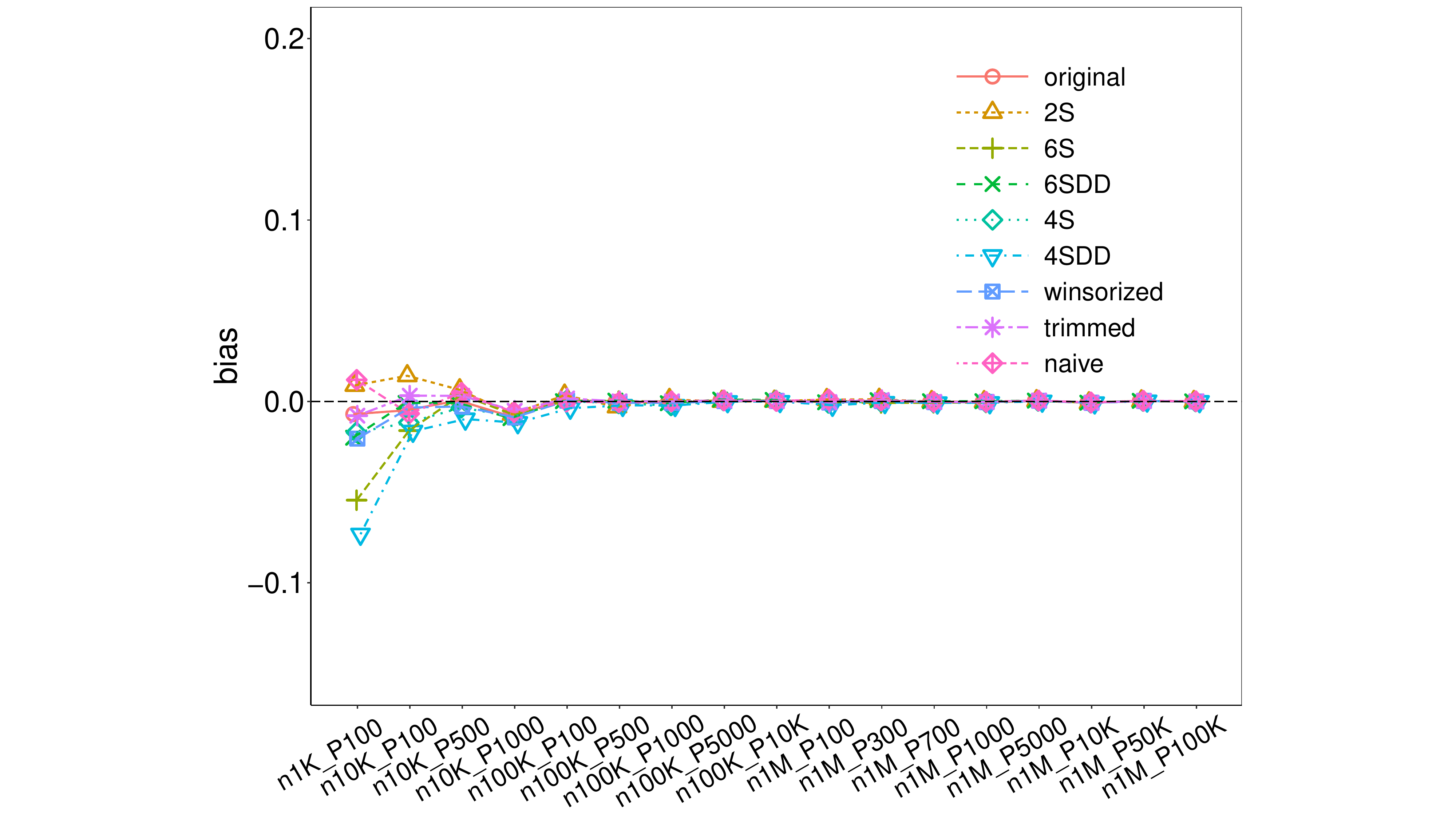}
\includegraphics[width=0.26\textwidth, trim={2.2in 0 2.2in 0},clip] {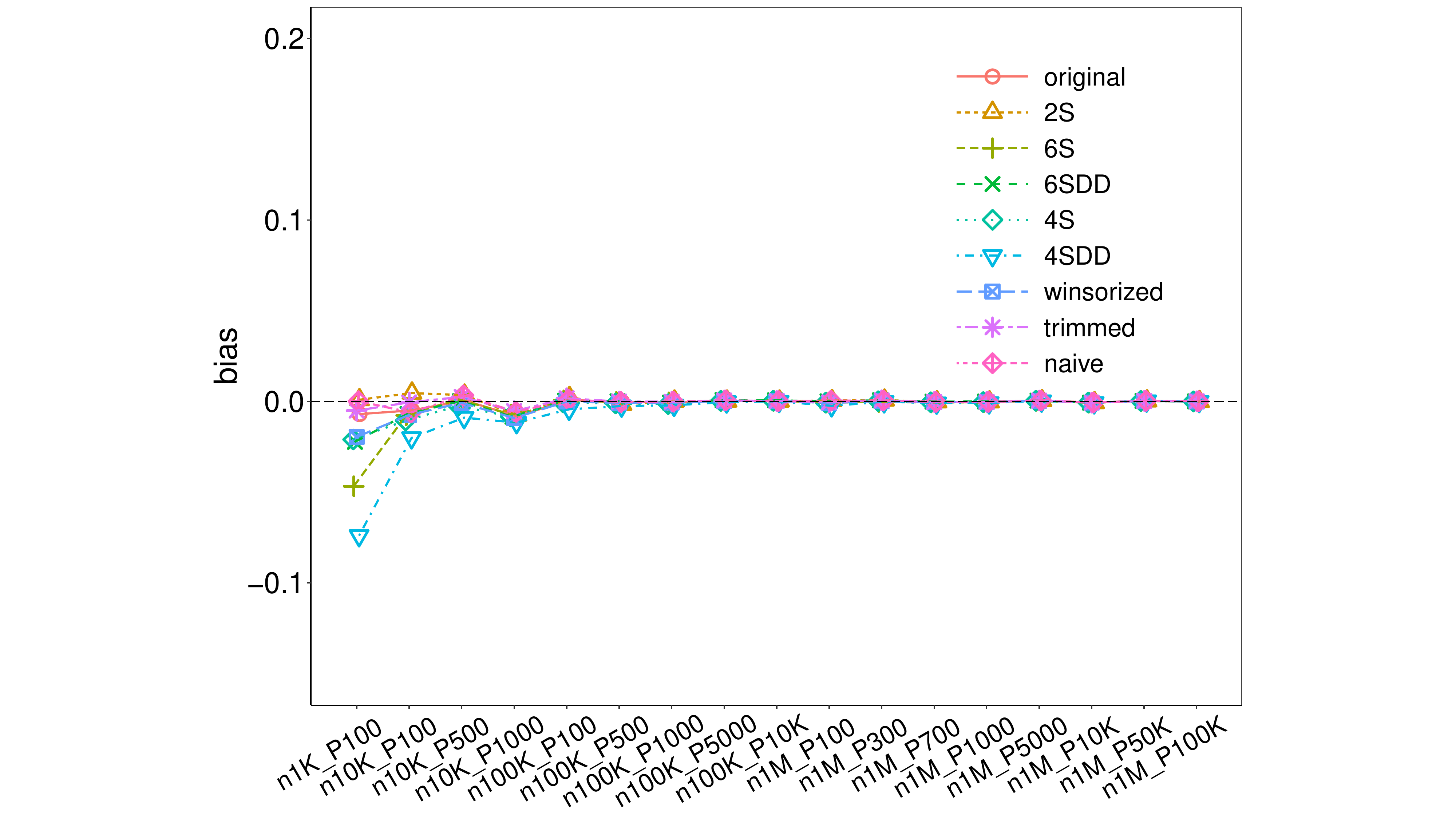}
\includegraphics[width=0.26\textwidth, trim={2.2in 0 2.2in 0},clip] {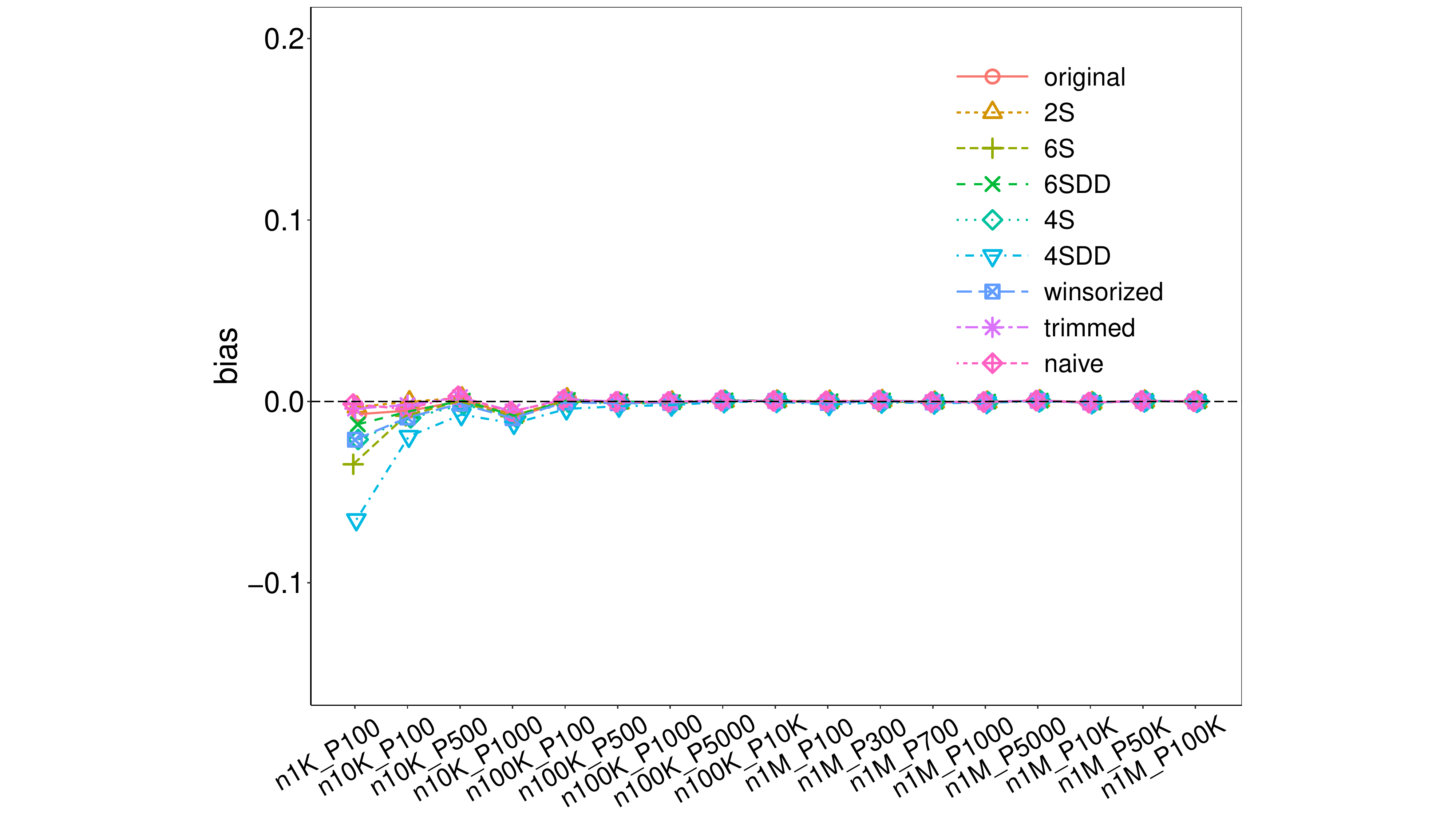}\\
\includegraphics[width=0.26\textwidth, trim={2.2in 0 2.2in 0},clip] {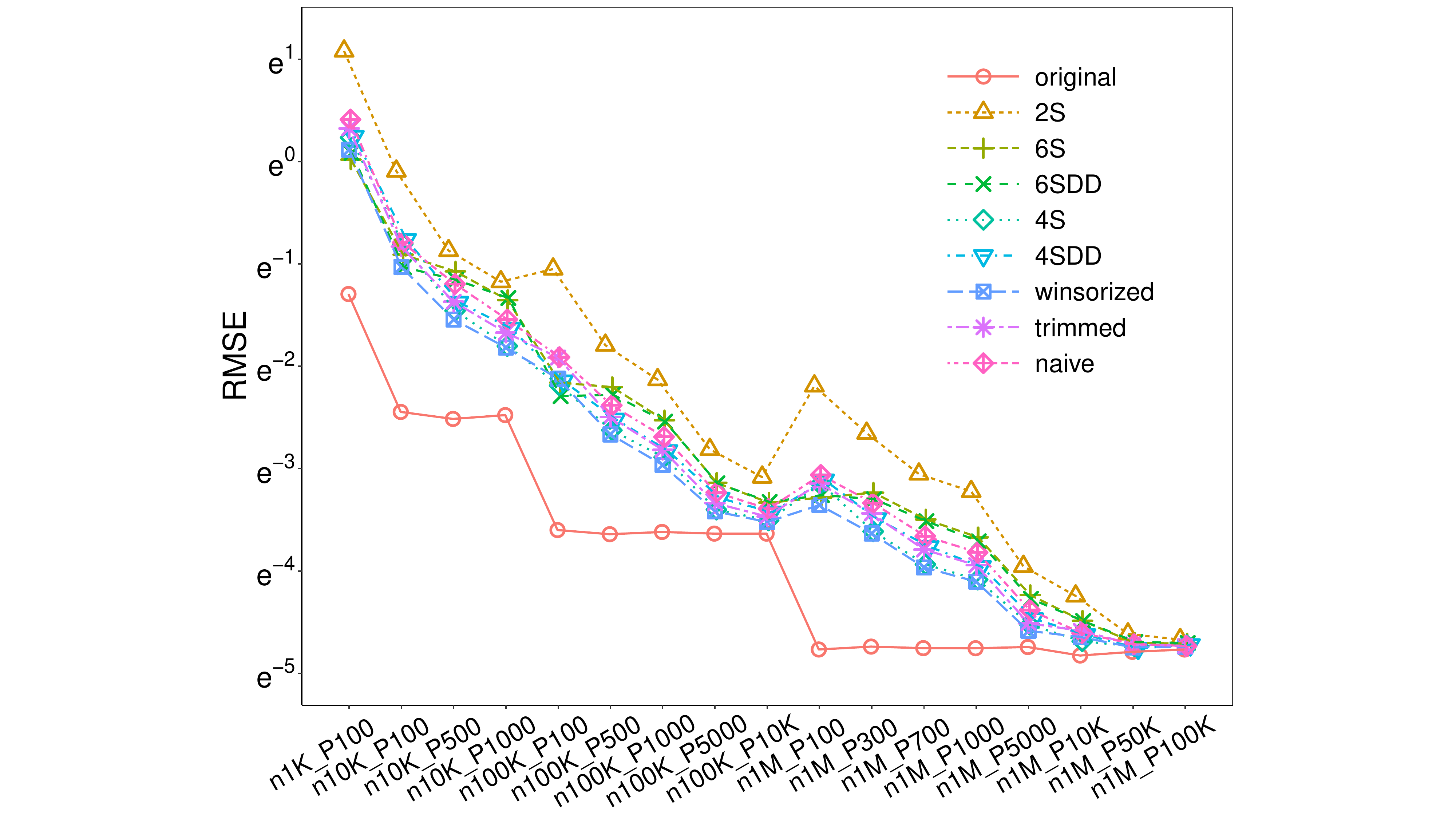}
\includegraphics[width=0.26\textwidth, trim={2.2in 0 2.2in 0},clip] {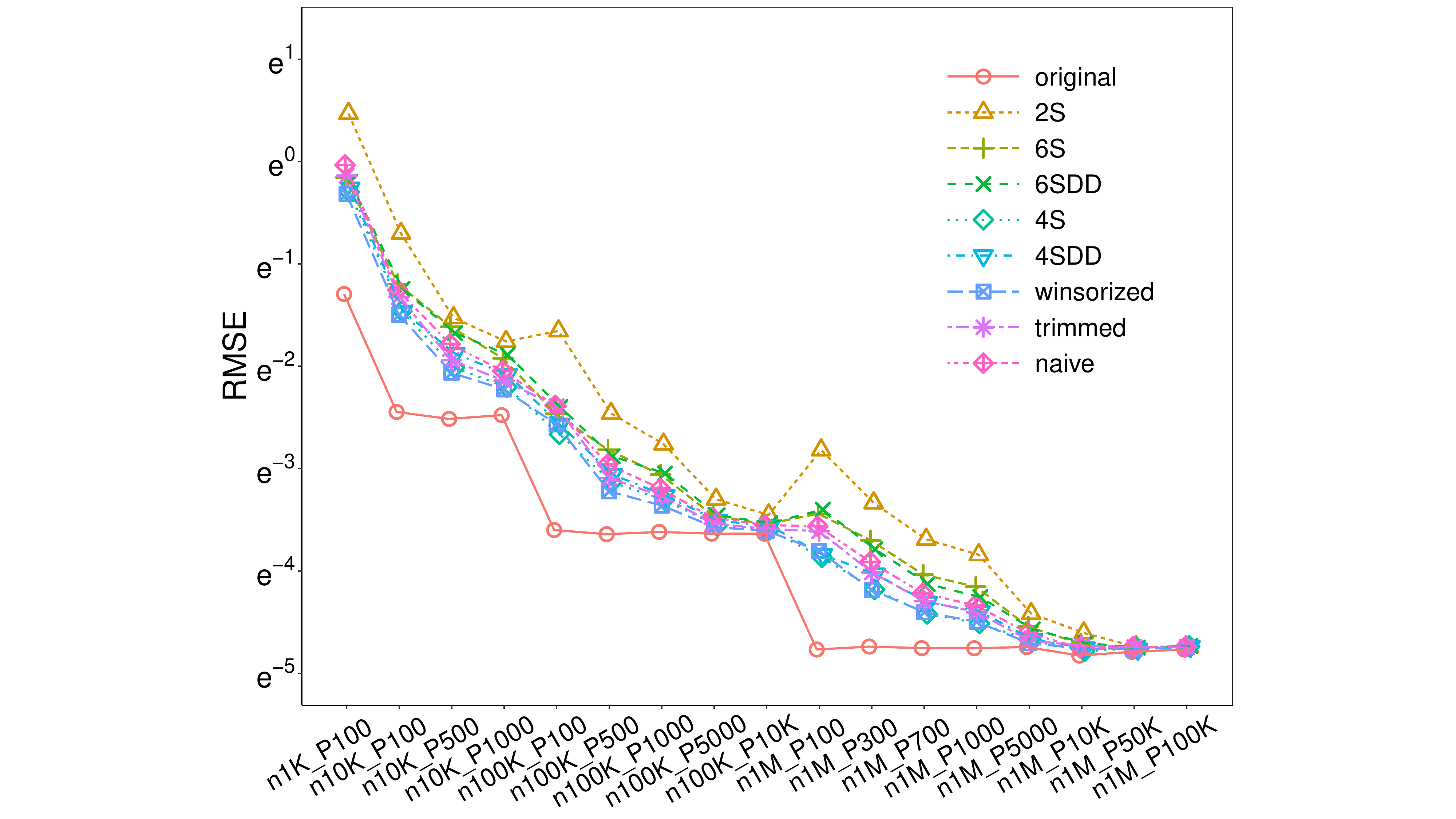}
\includegraphics[width=0.26\textwidth, trim={2.2in 0 2.2in 0},clip] {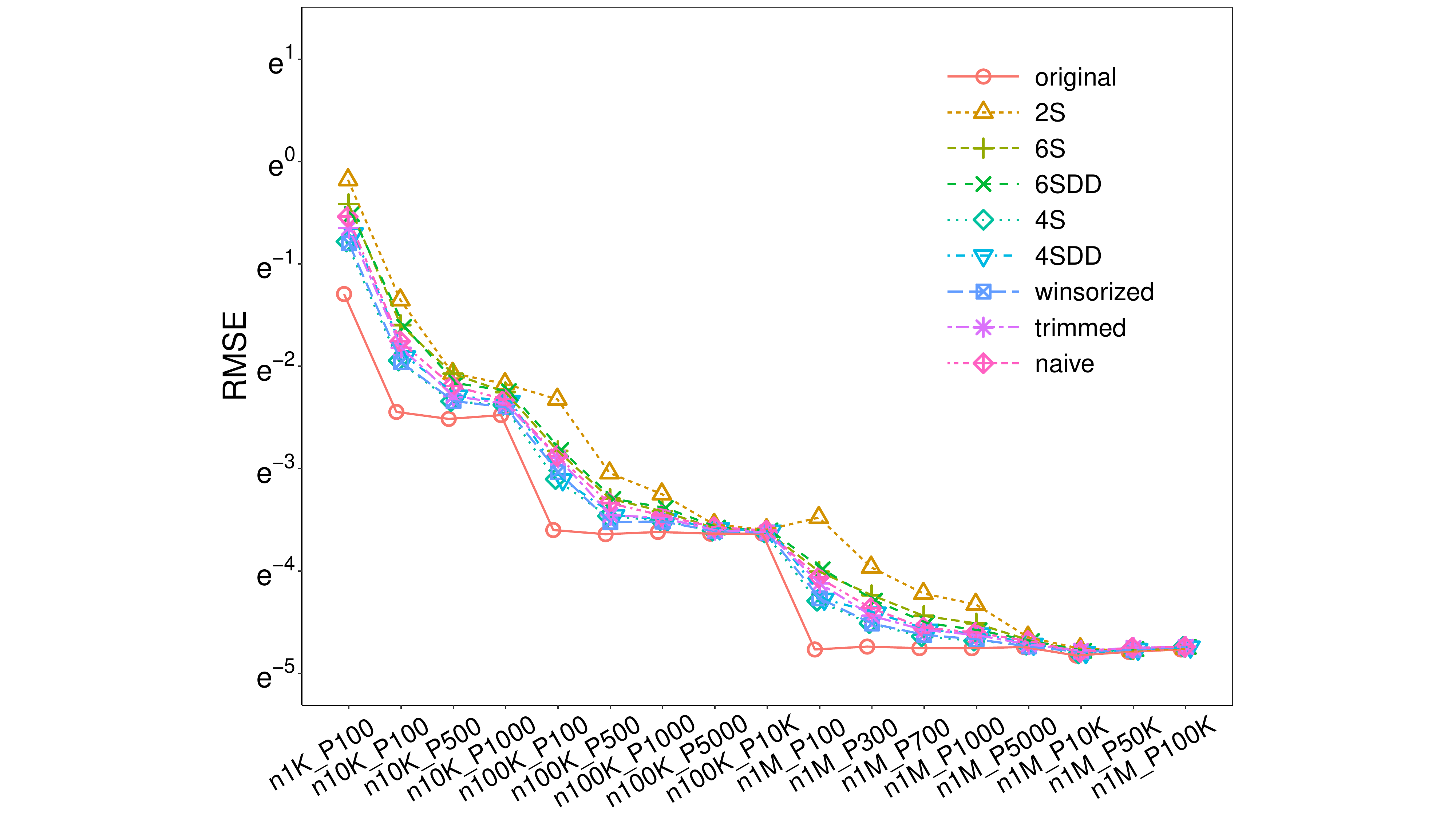}
\includegraphics[width=0.26\textwidth, trim={2.2in 0 2.2in 0},clip] {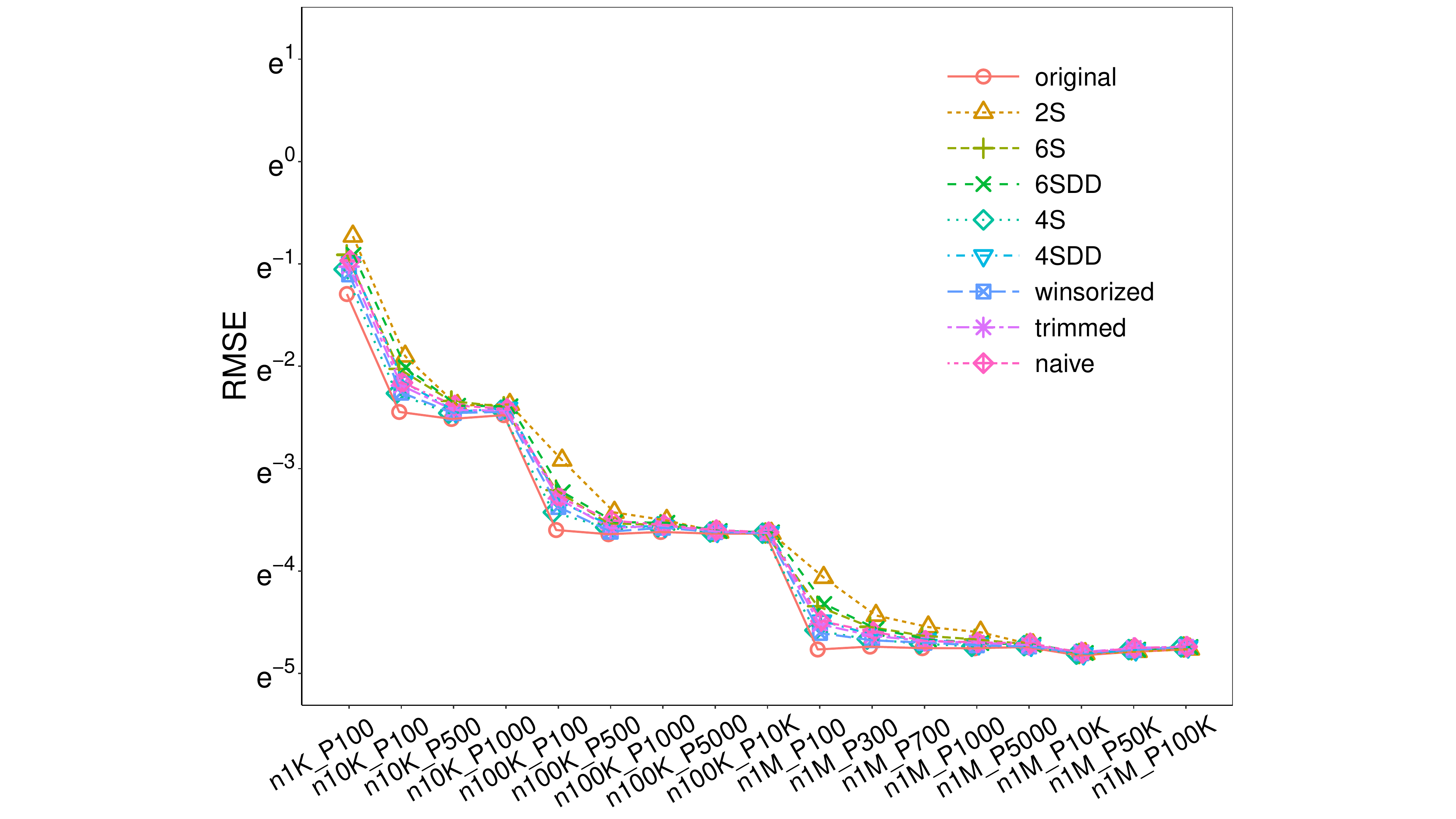}
\includegraphics[width=0.26\textwidth, trim={2.2in 0 2.2in 0},clip] {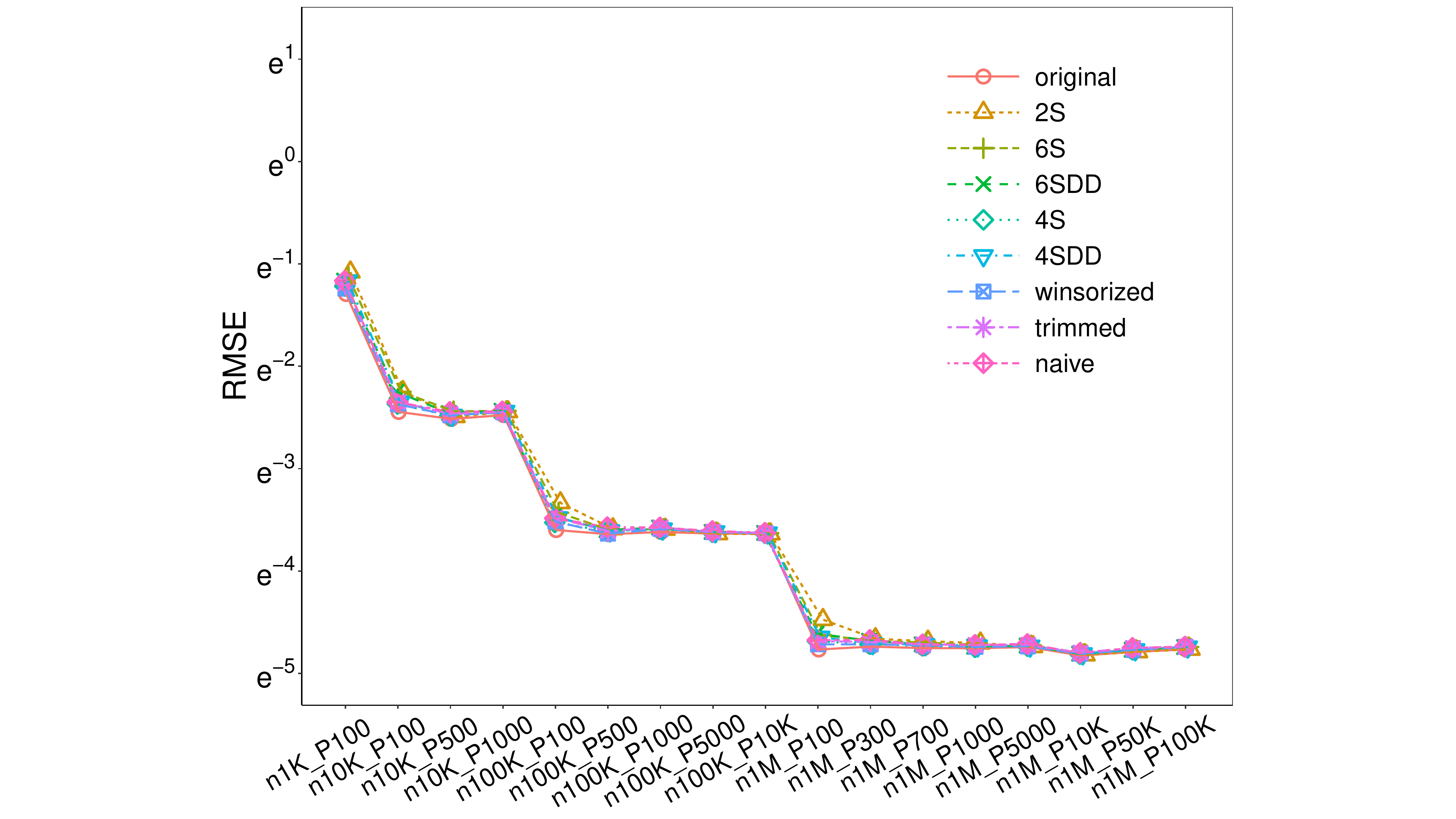}\\
\includegraphics[width=0.26\textwidth, trim={2.2in 0 2.2in 0},clip] {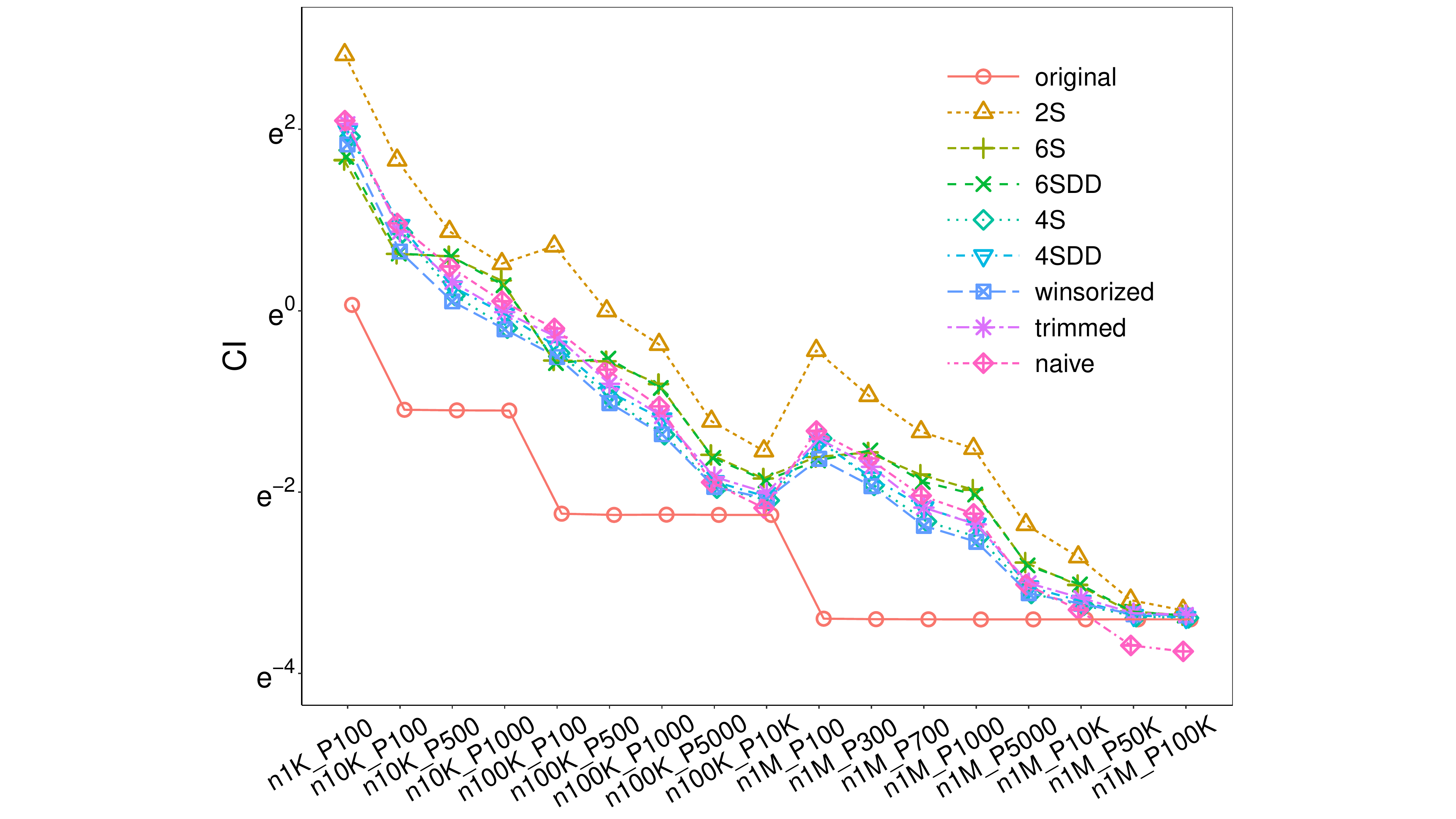}
\includegraphics[width=0.26\textwidth, trim={2.2in 0 2.2in 0},clip] {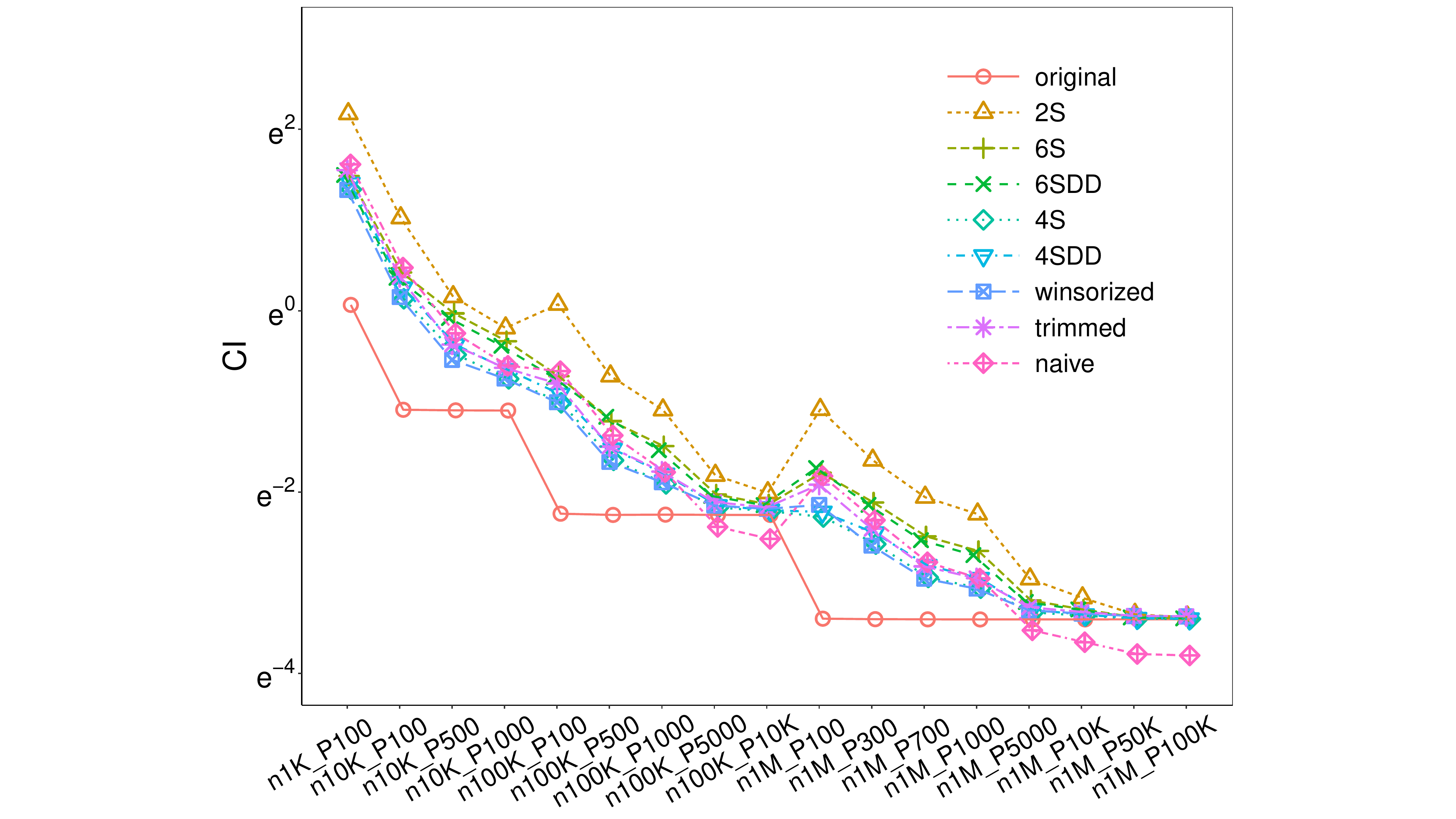}
\includegraphics[width=0.26\textwidth, trim={2.2in 0 2.2in 0},clip] {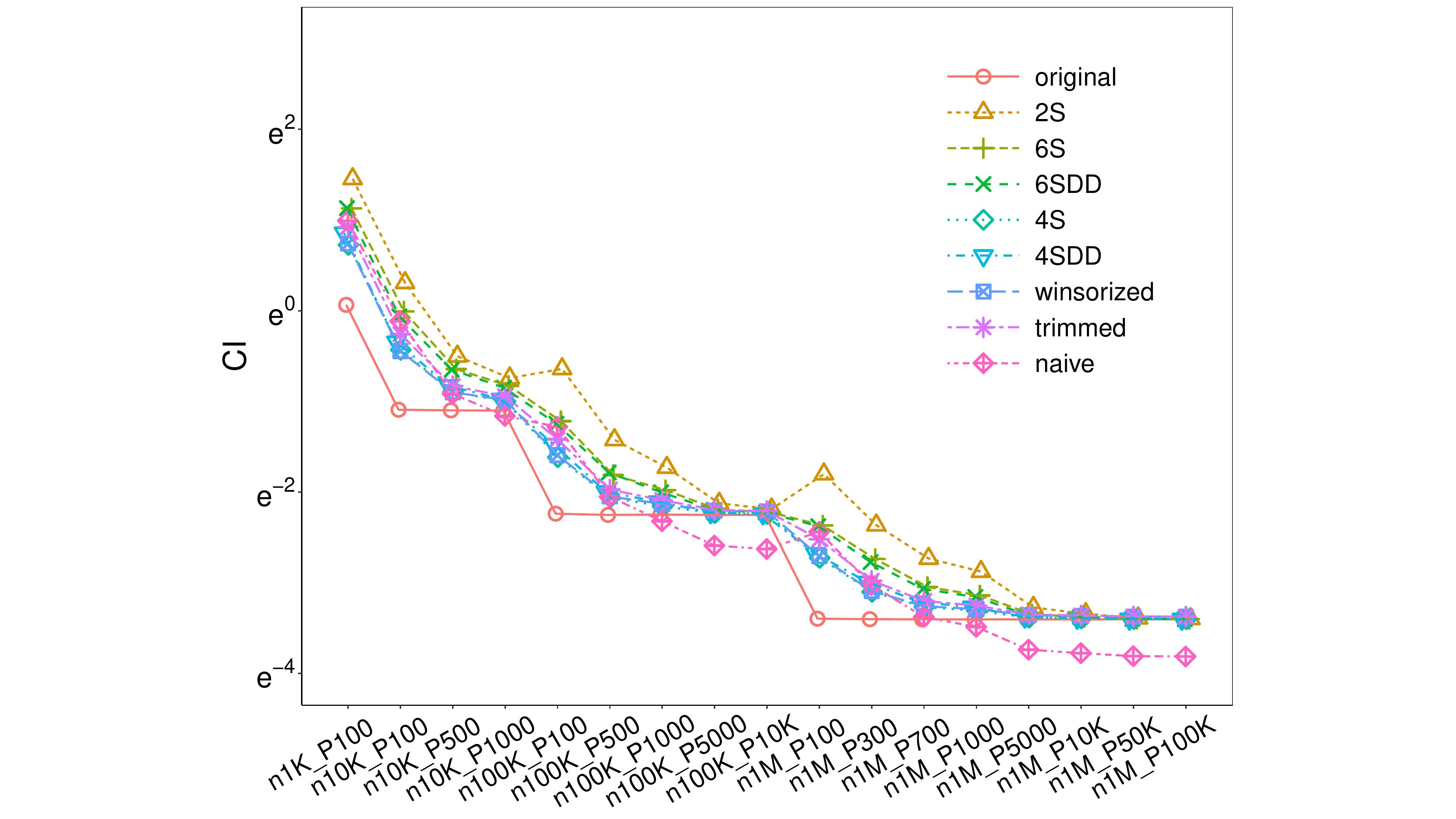}
\includegraphics[width=0.26\textwidth, trim={2.2in 0 2.2in 0},clip] {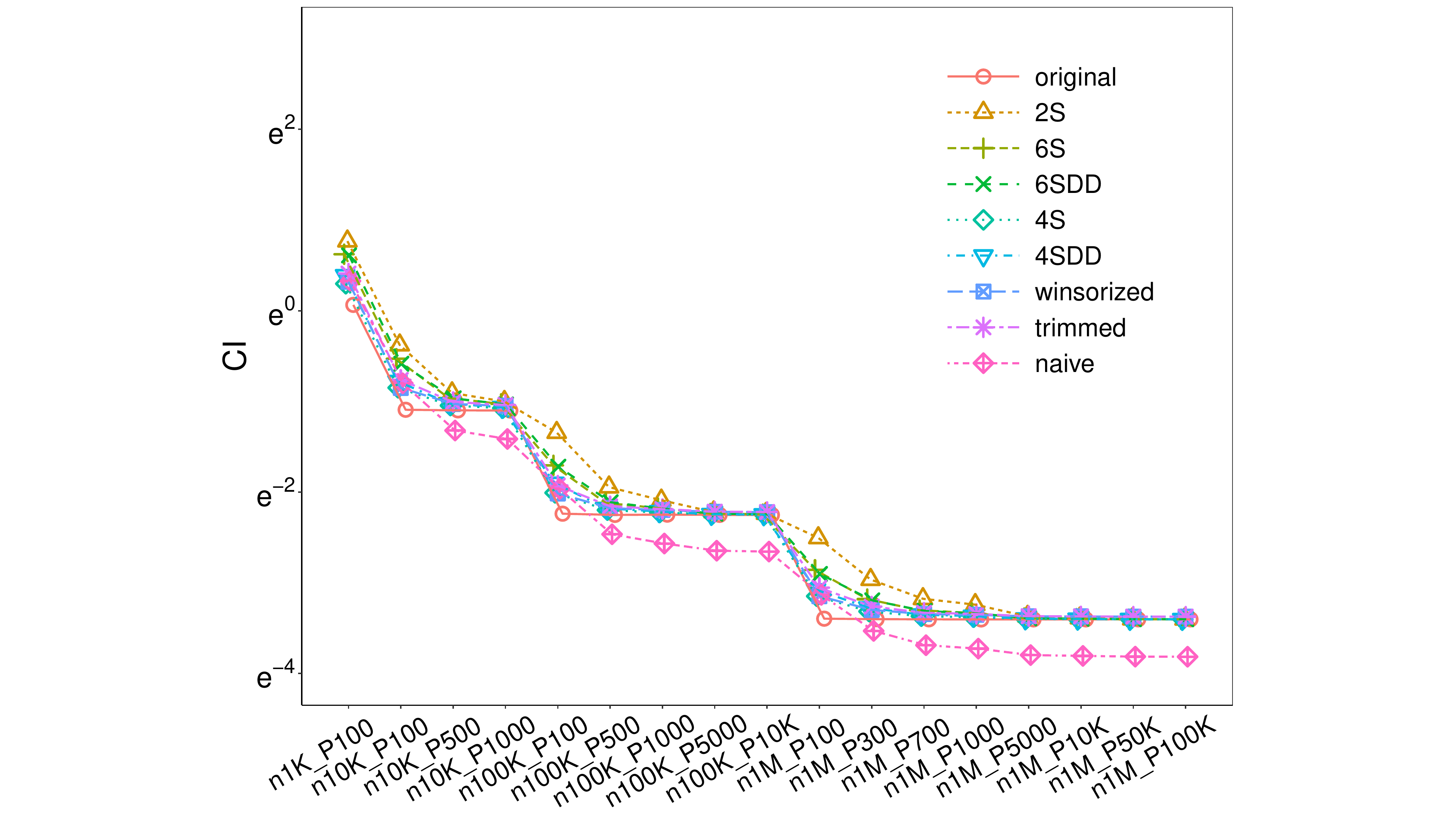}
\includegraphics[width=0.26\textwidth, trim={2.2in 0 2.2in 0},clip] {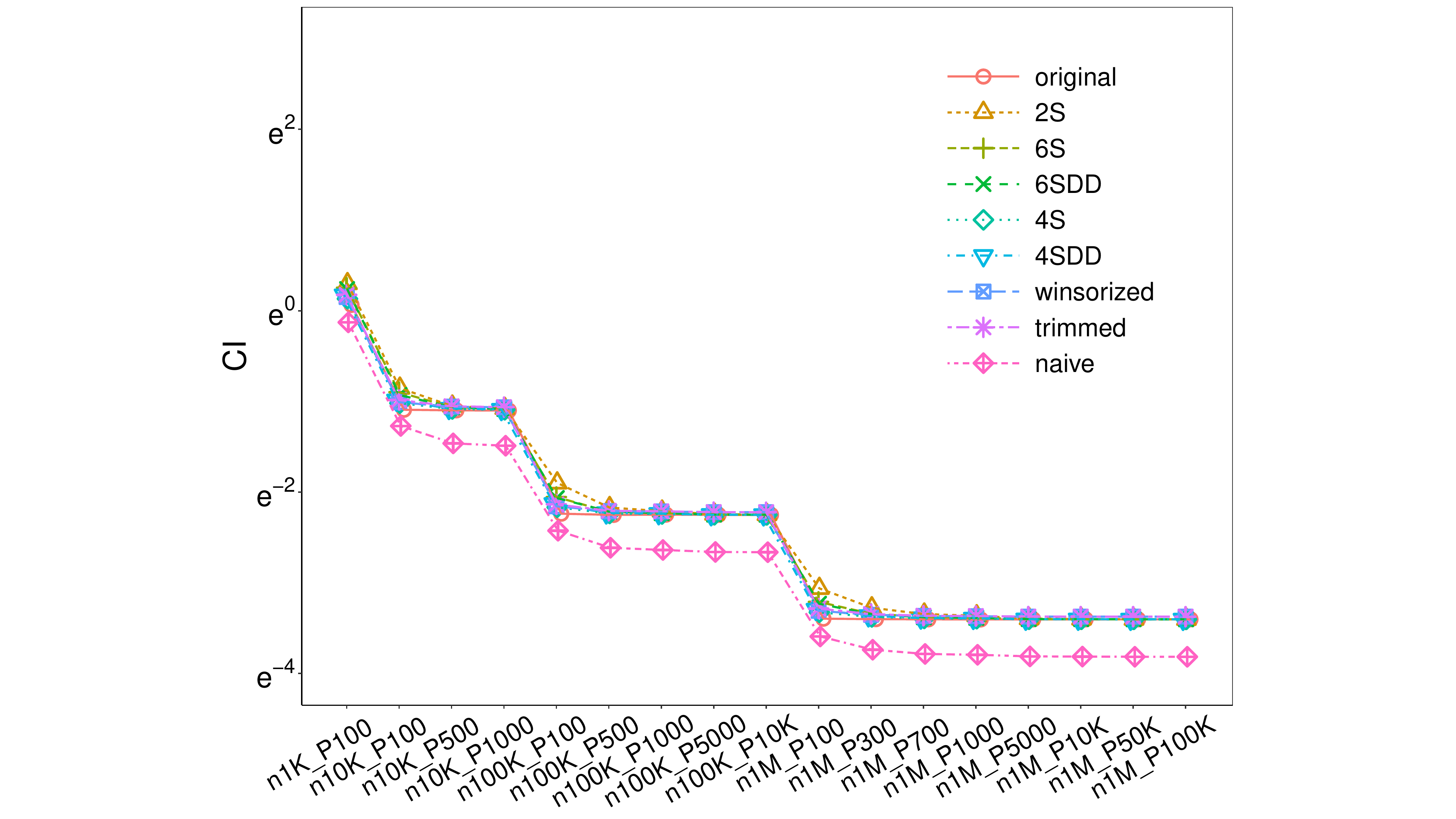}\\
\includegraphics[width=0.26\textwidth, trim={2.2in 0 2.2in 0},clip] {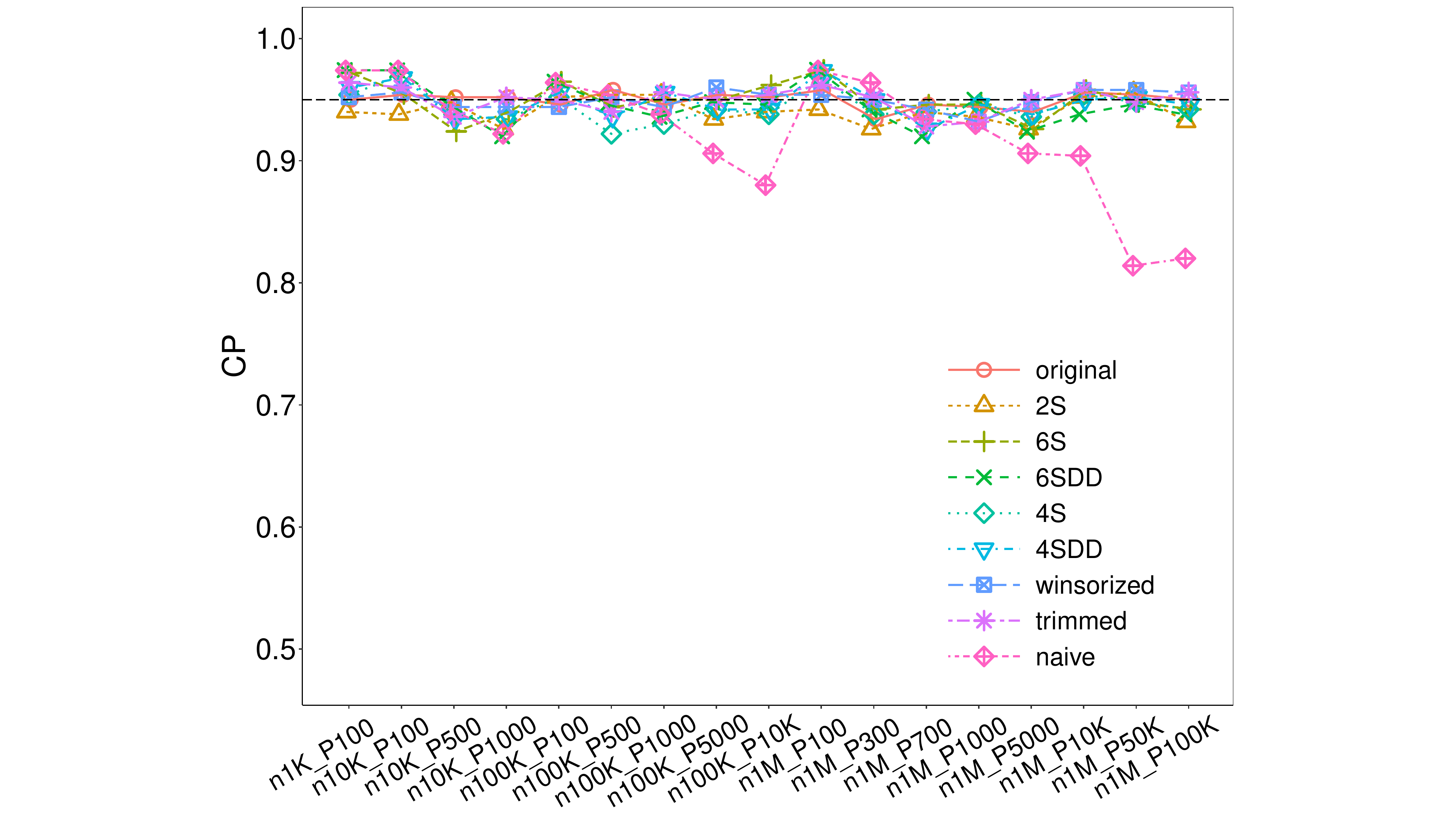}
\includegraphics[width=0.26\textwidth, trim={2.2in 0 2.2in 0},clip] {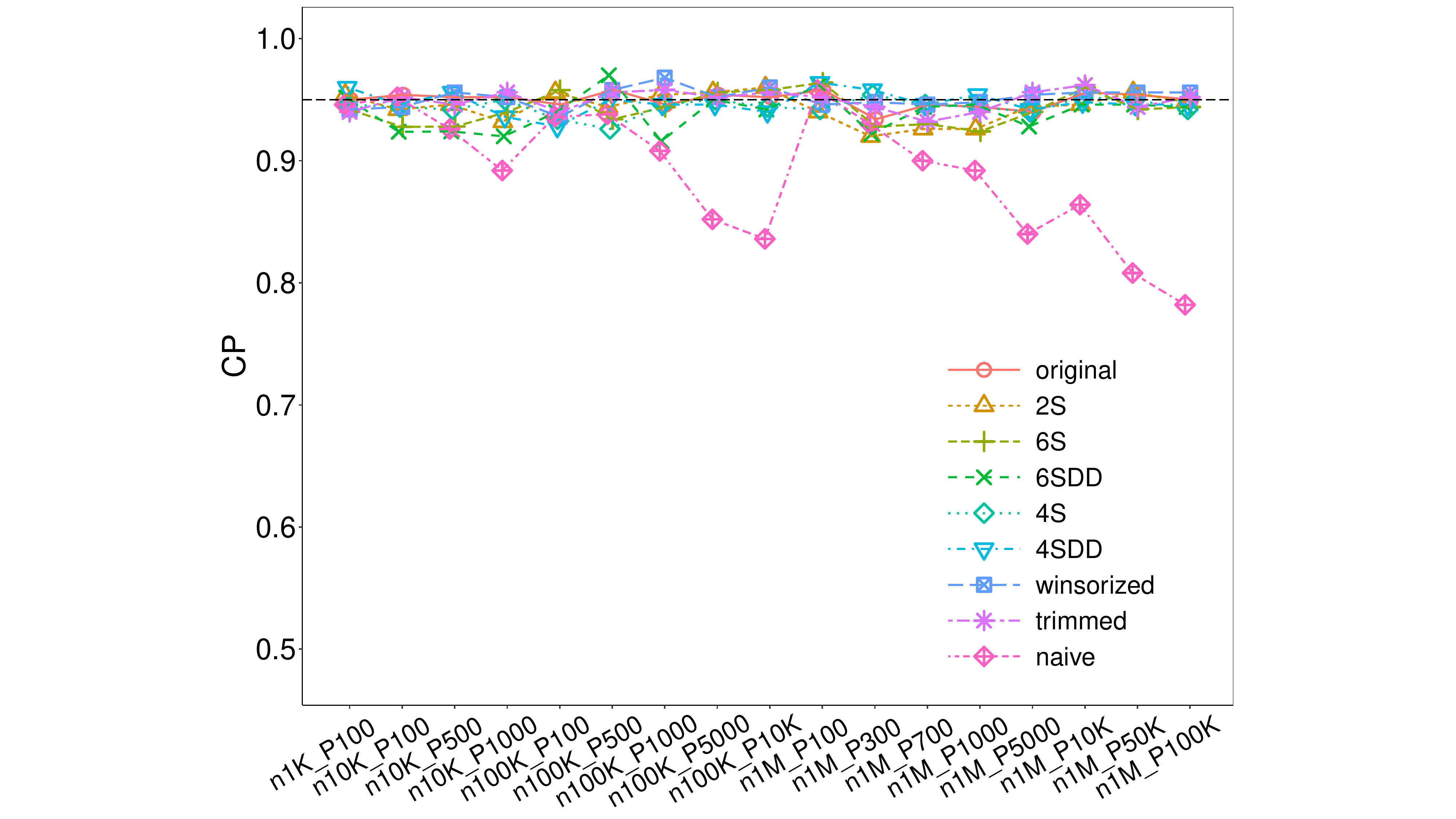}
\includegraphics[width=0.26\textwidth, trim={2.2in 0 2.2in 0},clip] {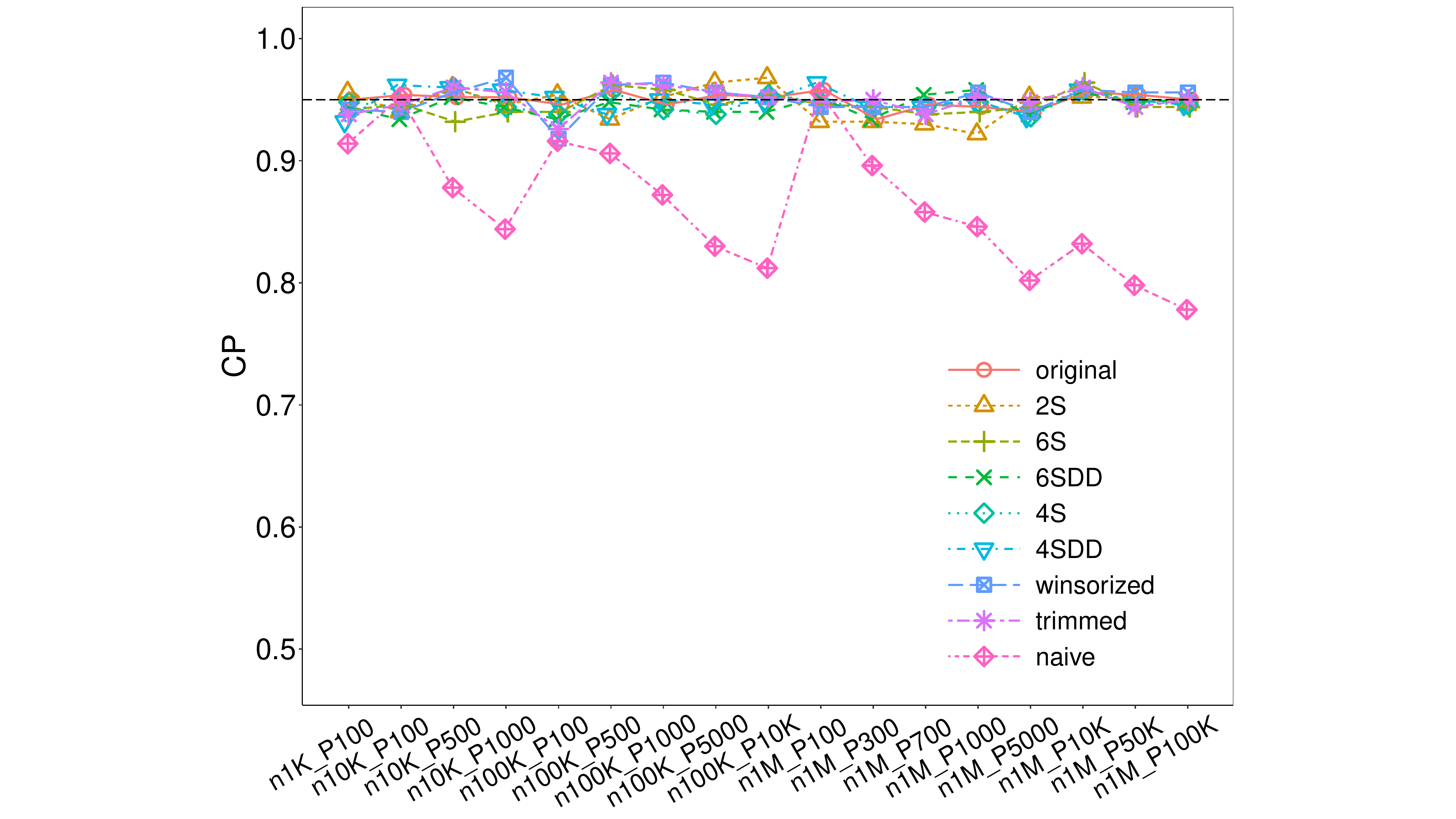}
\includegraphics[width=0.26\textwidth, trim={2.2in 0 2.2in 0},clip] {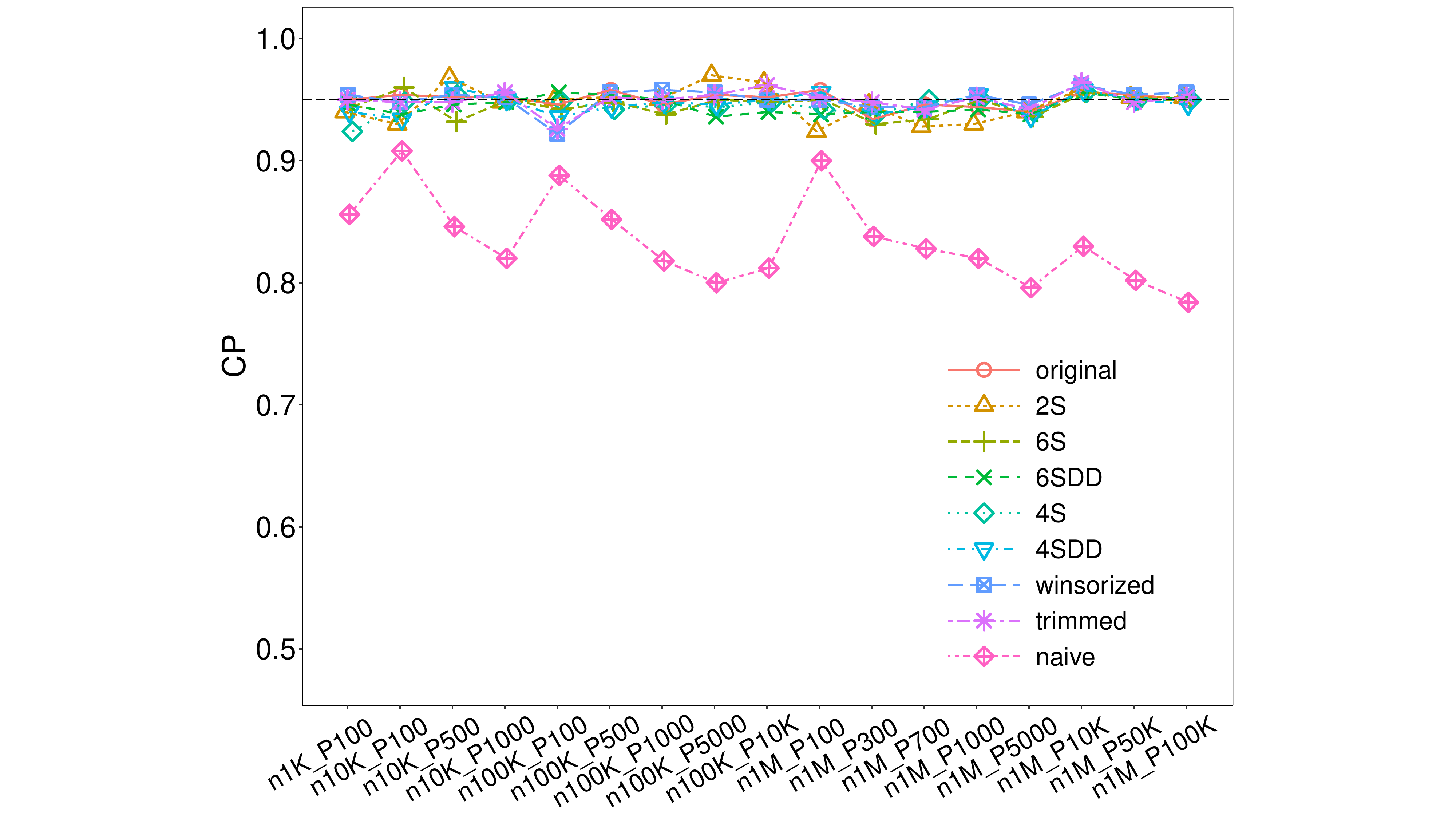}
\includegraphics[width=0.26\textwidth, trim={2.2in 0 2.2in 0},clip] {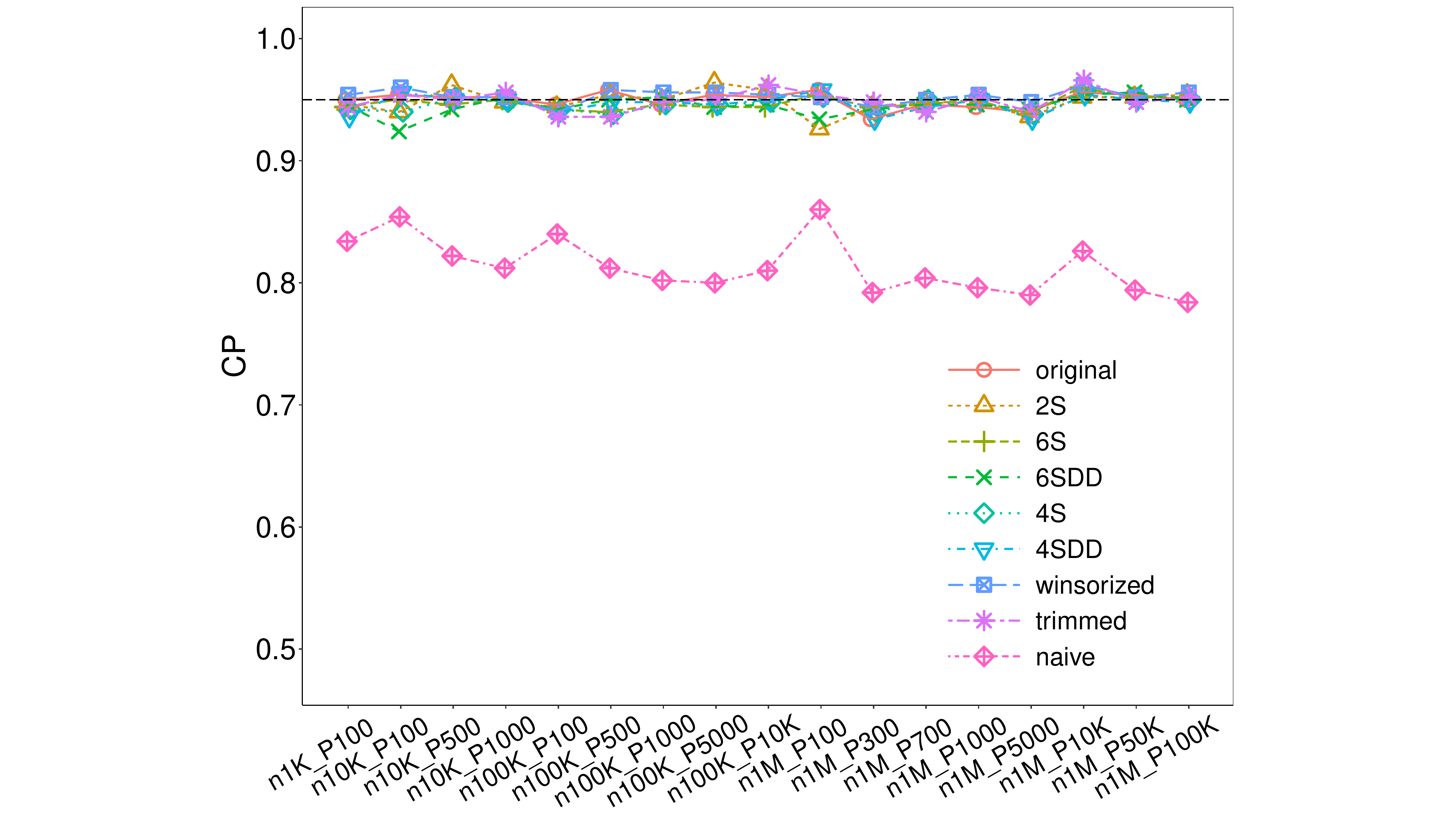}\\

\caption{Gaussian data; $\rho$-zCDP; $\theta=0$ and $\alpha=\beta$}
\label{fig:0szCDP}
\end{figure}

\end{landscape}

\begin{landscape}

\subsection*{Gaussian, $\theta\ne0$ and $\alpha=\beta$}
\begin{figure}[!htb]
\centering
$\epsilon=0.5$\hspace{0.9in}$\epsilon=1$\hspace{1in}$\epsilon=2$
\hspace{1in}$\epsilon=5$\hspace{0.9in}$\epsilon=50$\\
\includegraphics[width=0.215\textwidth, trim={2.2in 0 2.2in 0},clip] {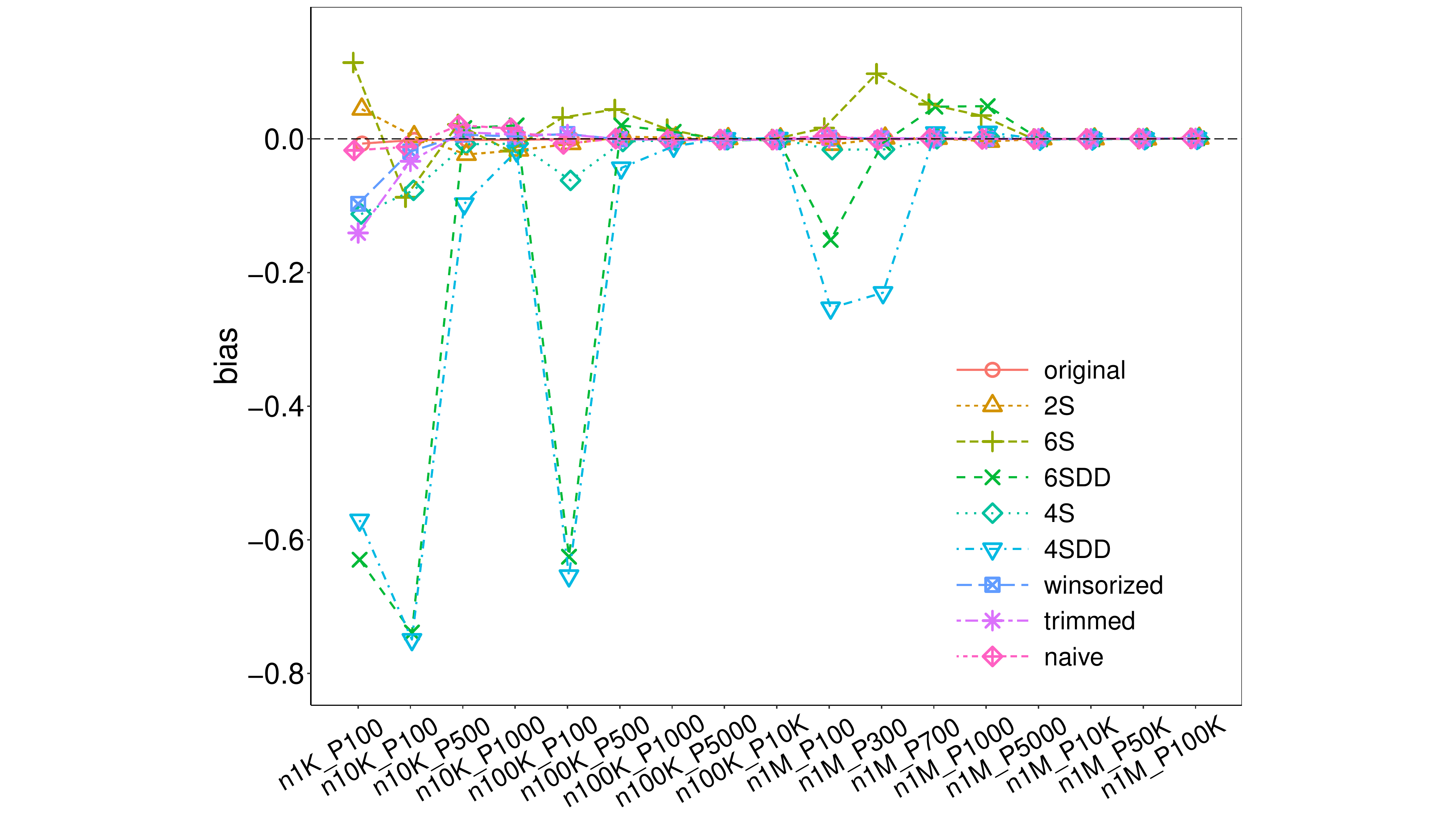}
\includegraphics[width=0.215\textwidth, trim={2.2in 0 2.2in 0},clip] {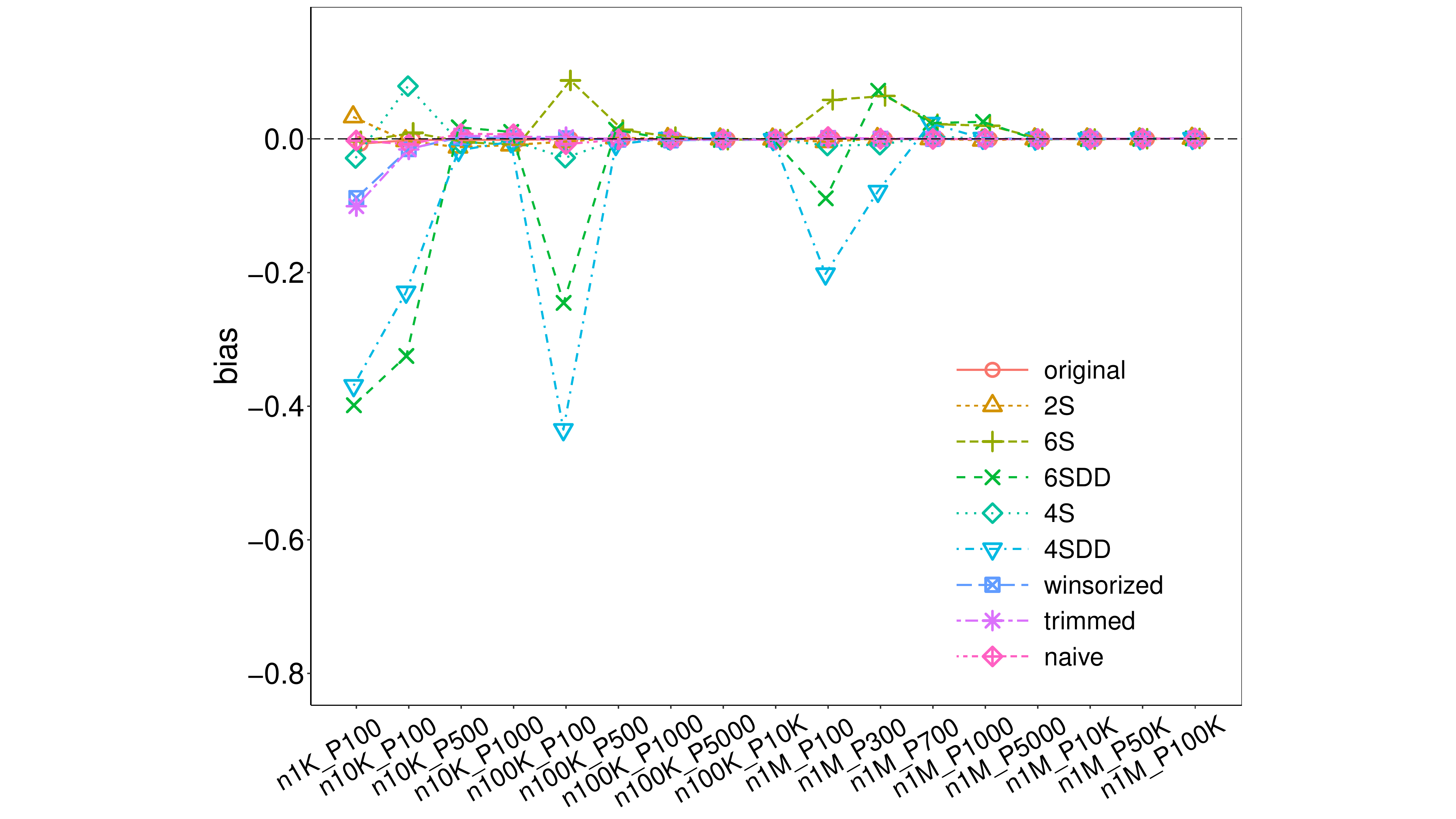}
\includegraphics[width=0.215\textwidth, trim={2.2in 0 2.2in 0},clip] {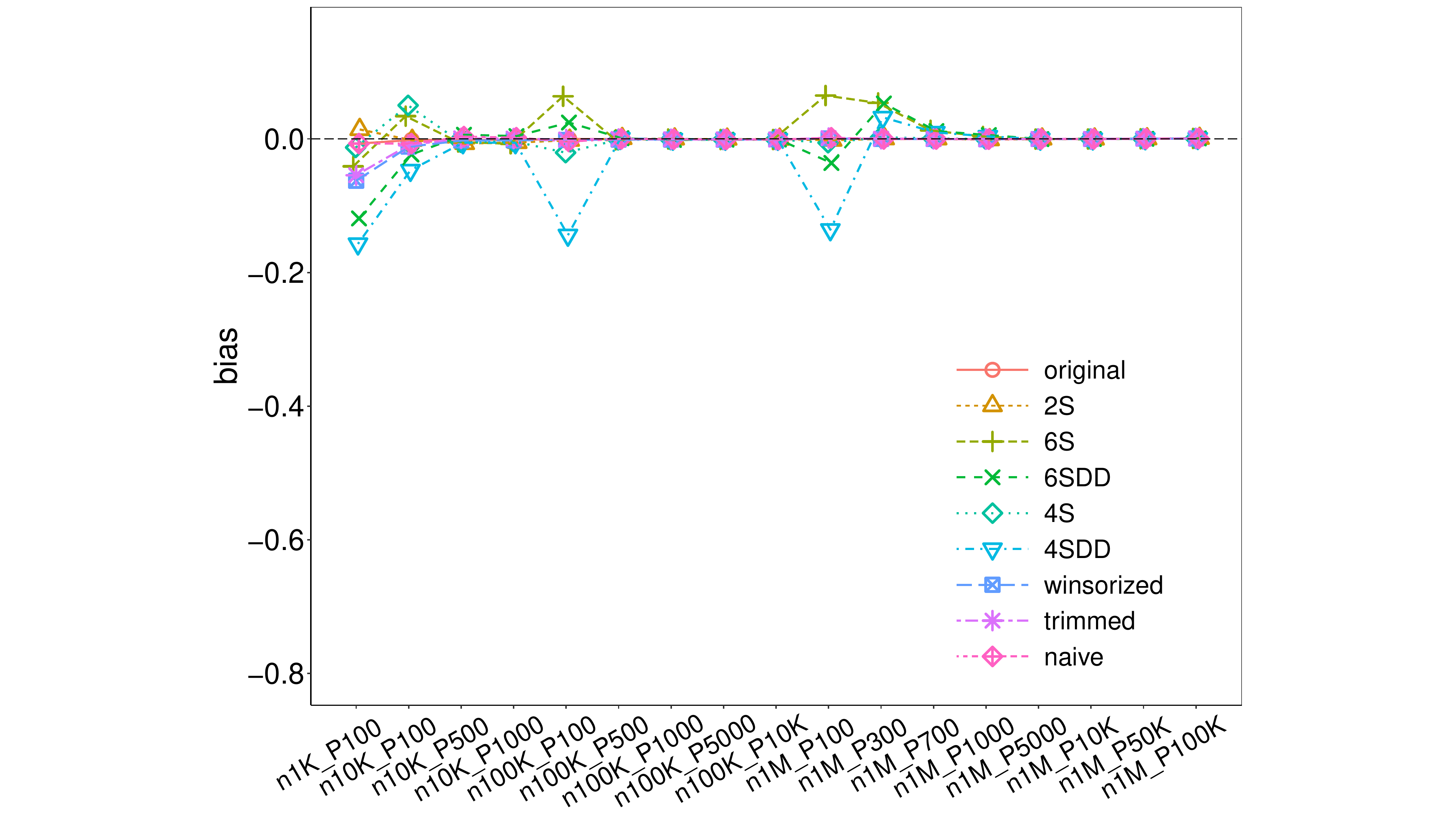}
\includegraphics[width=0.215\textwidth, trim={2.2in 0 2.2in 0},clip] {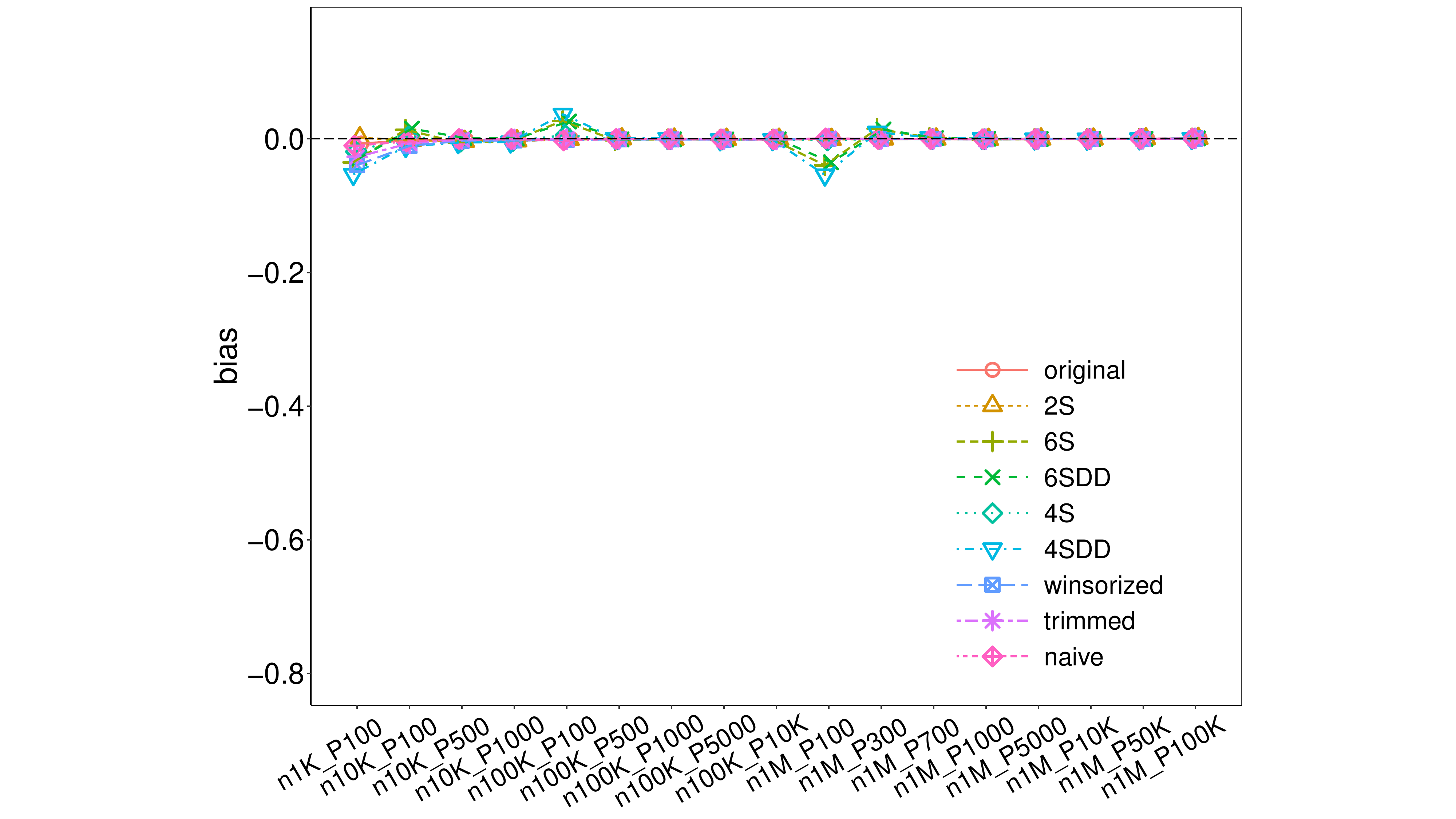}
\includegraphics[width=0.215\textwidth, trim={2.2in 0 2.2in 0},clip] {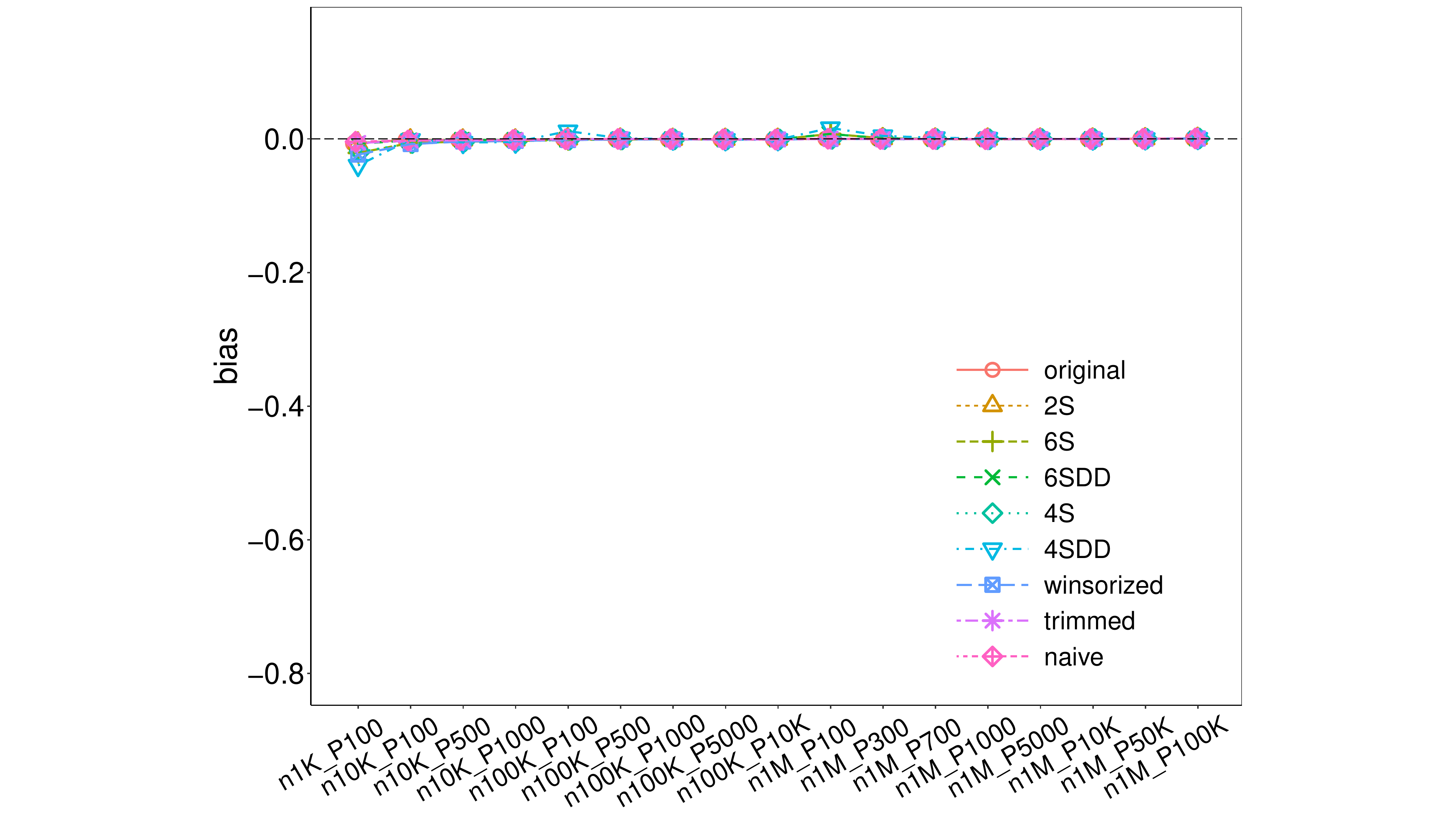}\\
\includegraphics[width=0.215\textwidth, trim={2.2in 0 2.2in 0},clip] {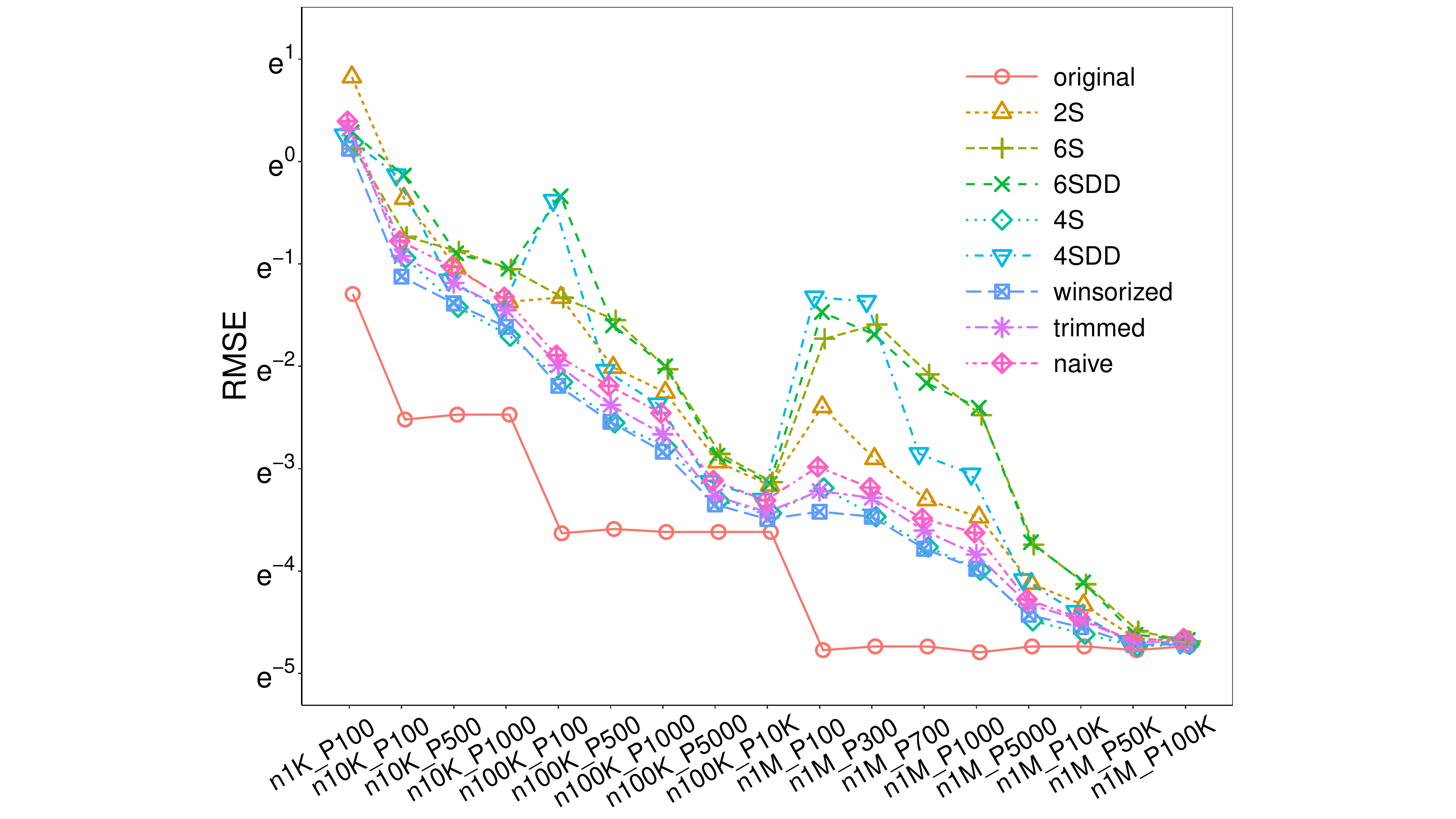}
\includegraphics[width=0.215\textwidth, trim={2.2in 0 2.2in 0},clip] {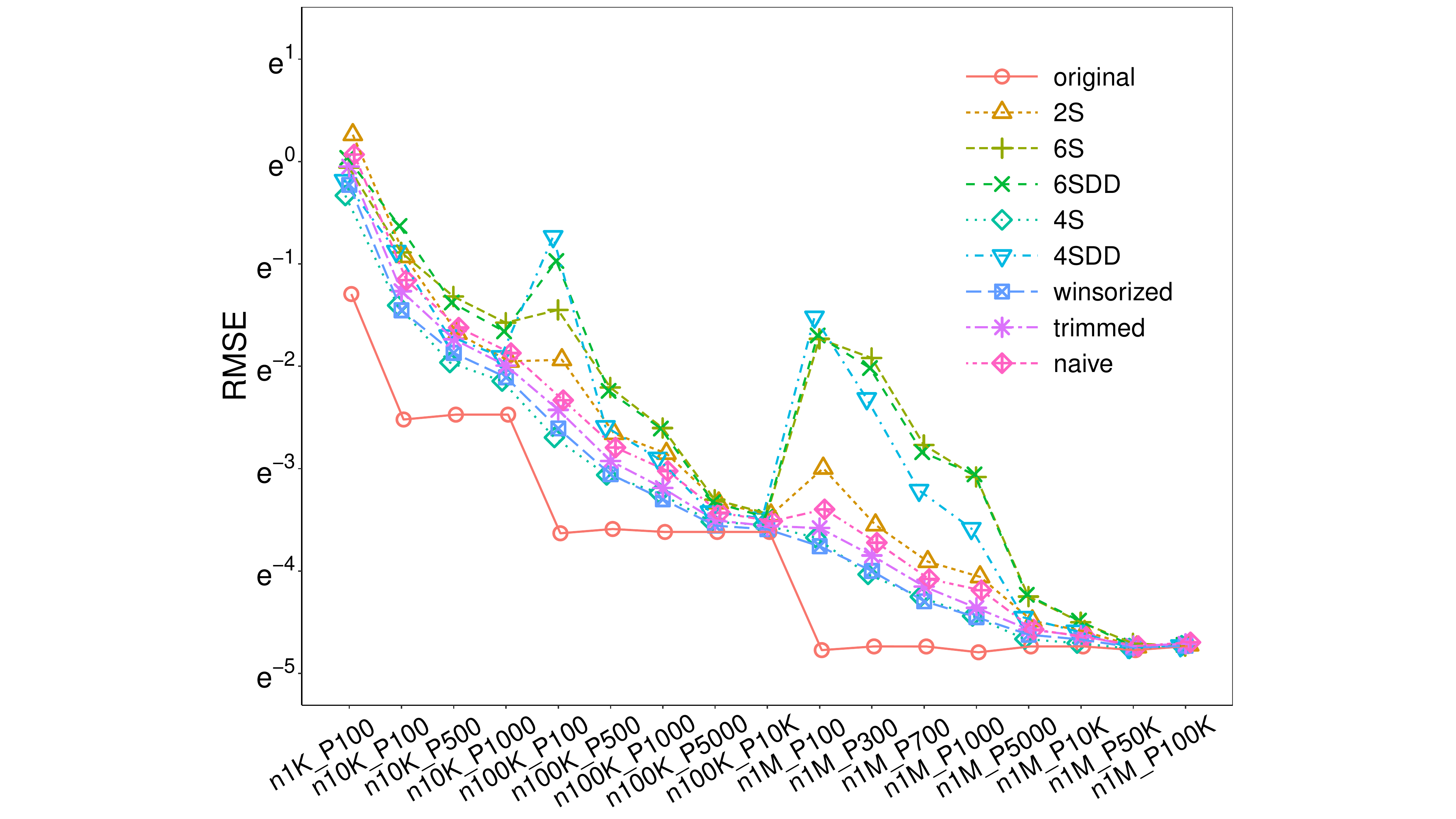}
\includegraphics[width=0.215\textwidth, trim={2.2in 0 2.2in 0},clip] {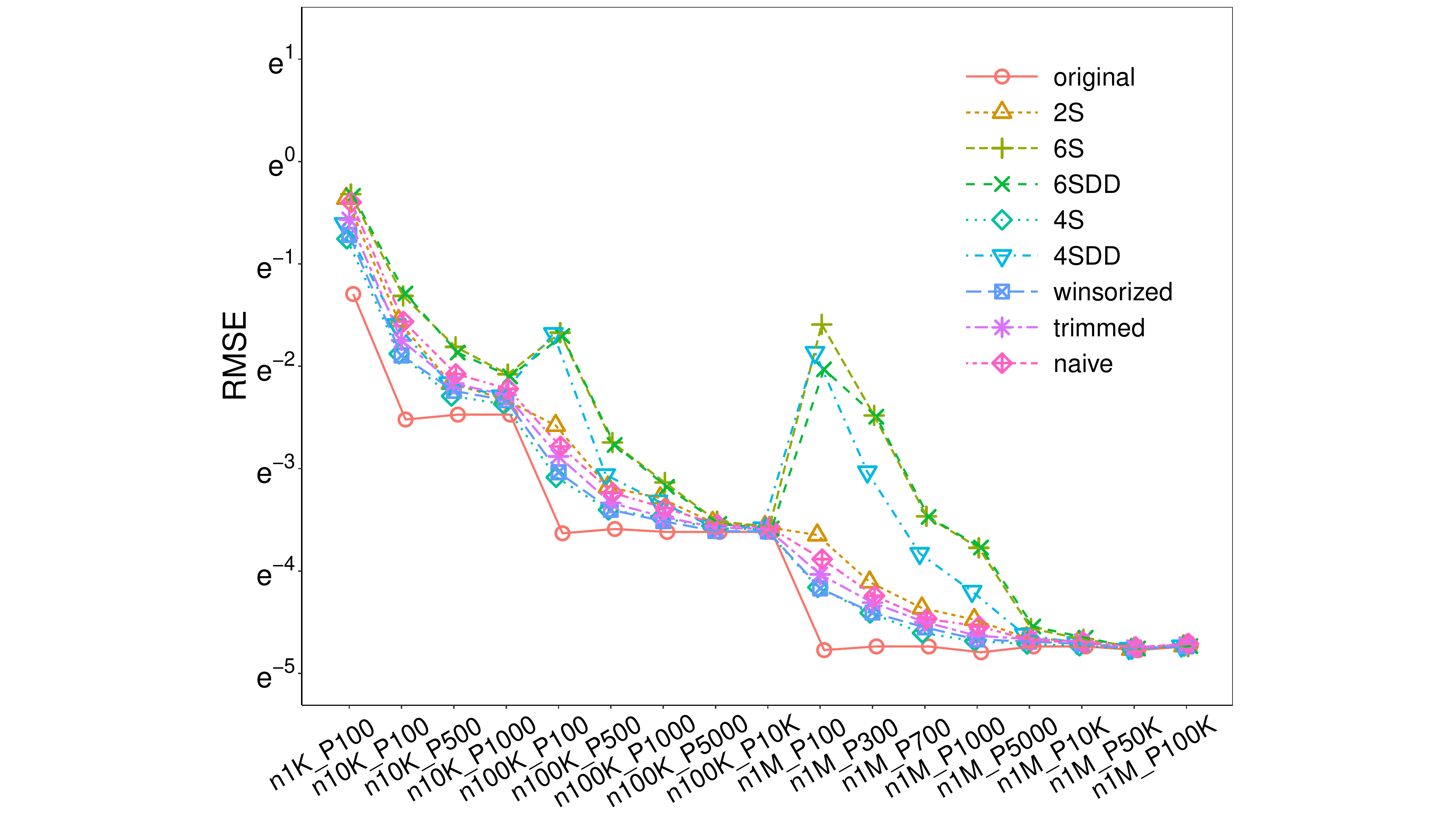}
\includegraphics[width=0.215\textwidth, trim={2.2in 0 2.2in 0},clip] {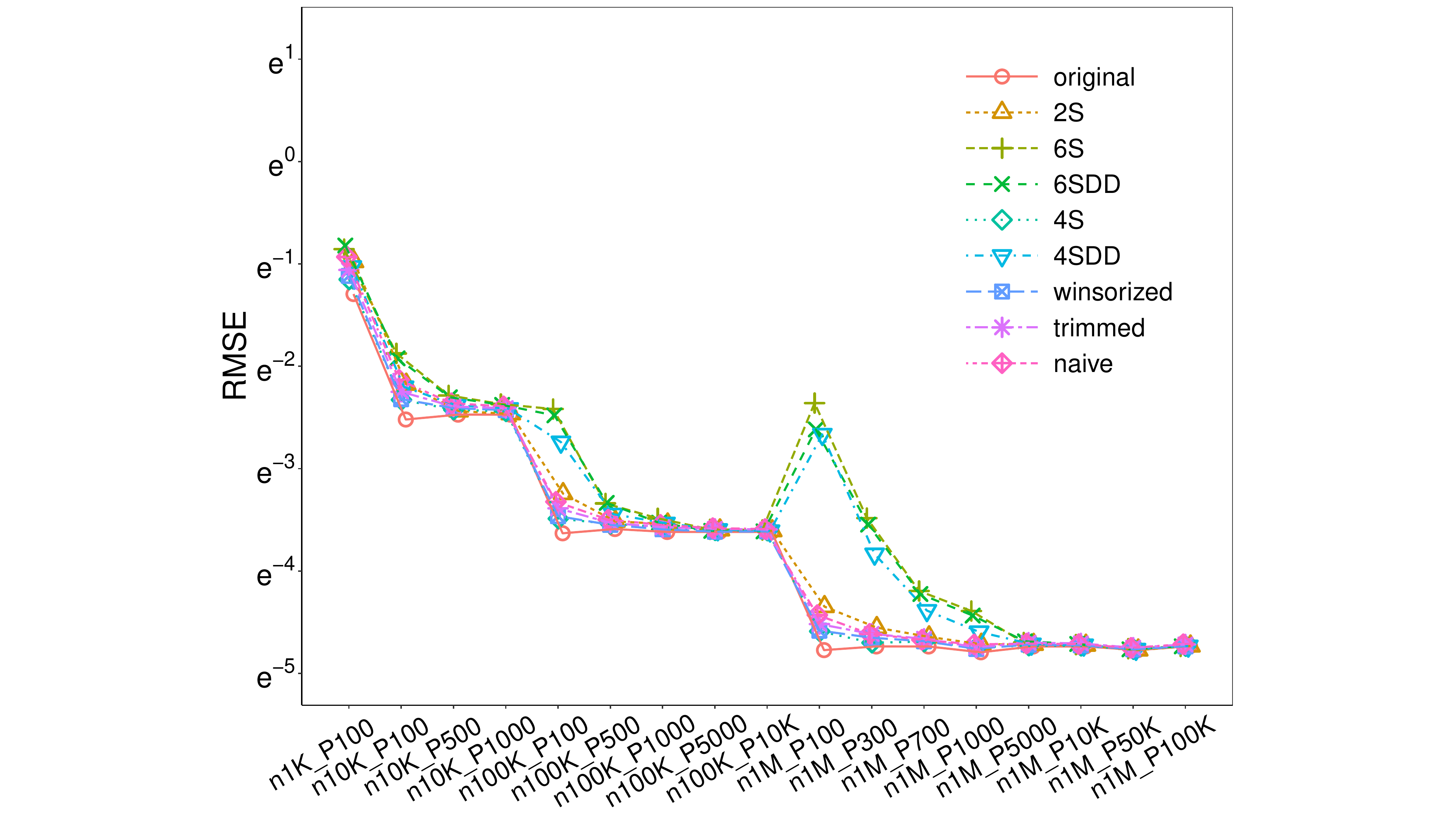}
\includegraphics[width=0.215\textwidth, trim={2.2in 0 2.2in 0},clip] {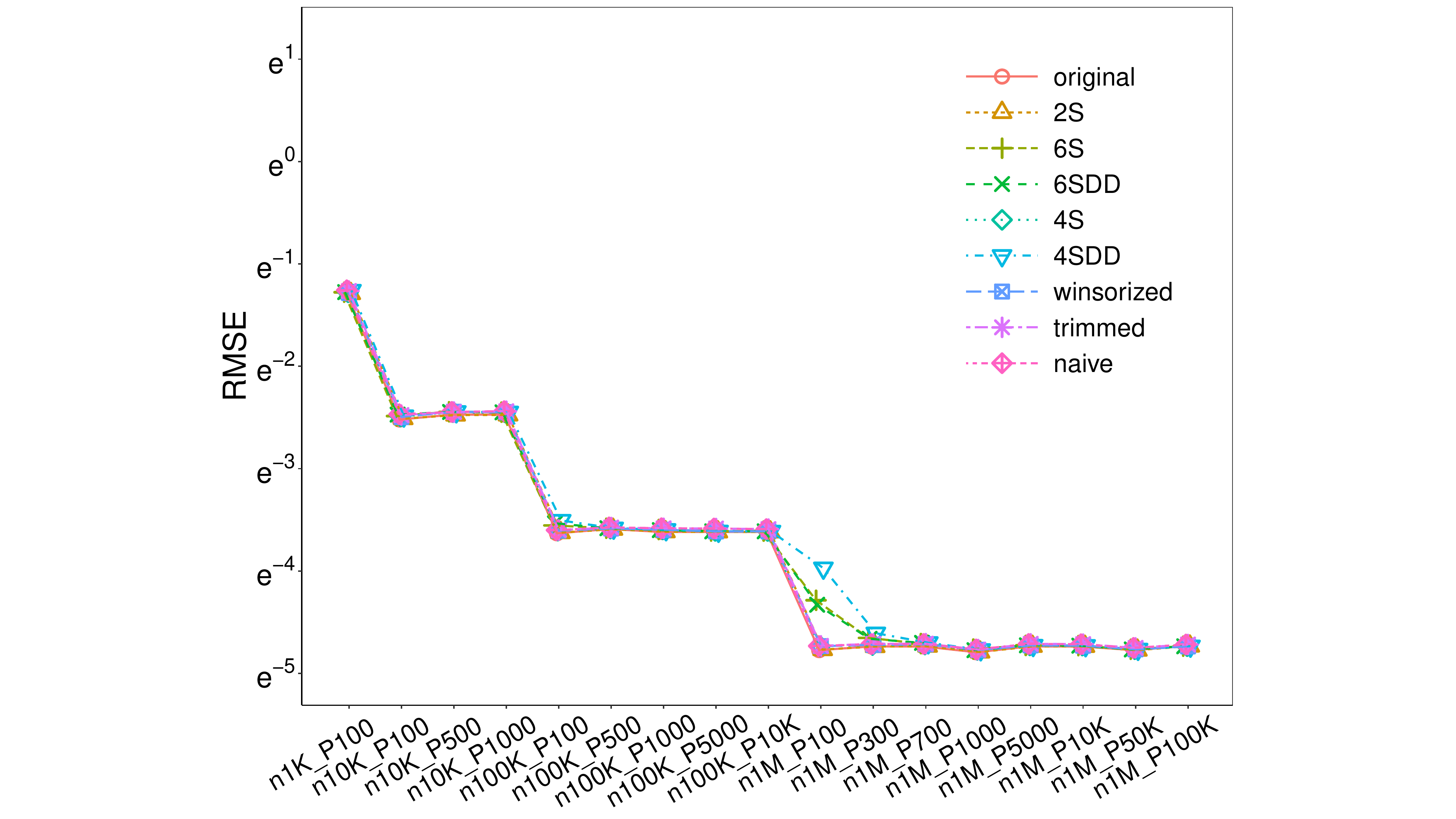}\\
\includegraphics[width=0.215\textwidth, trim={2.2in 0 2.2in 0},clip] {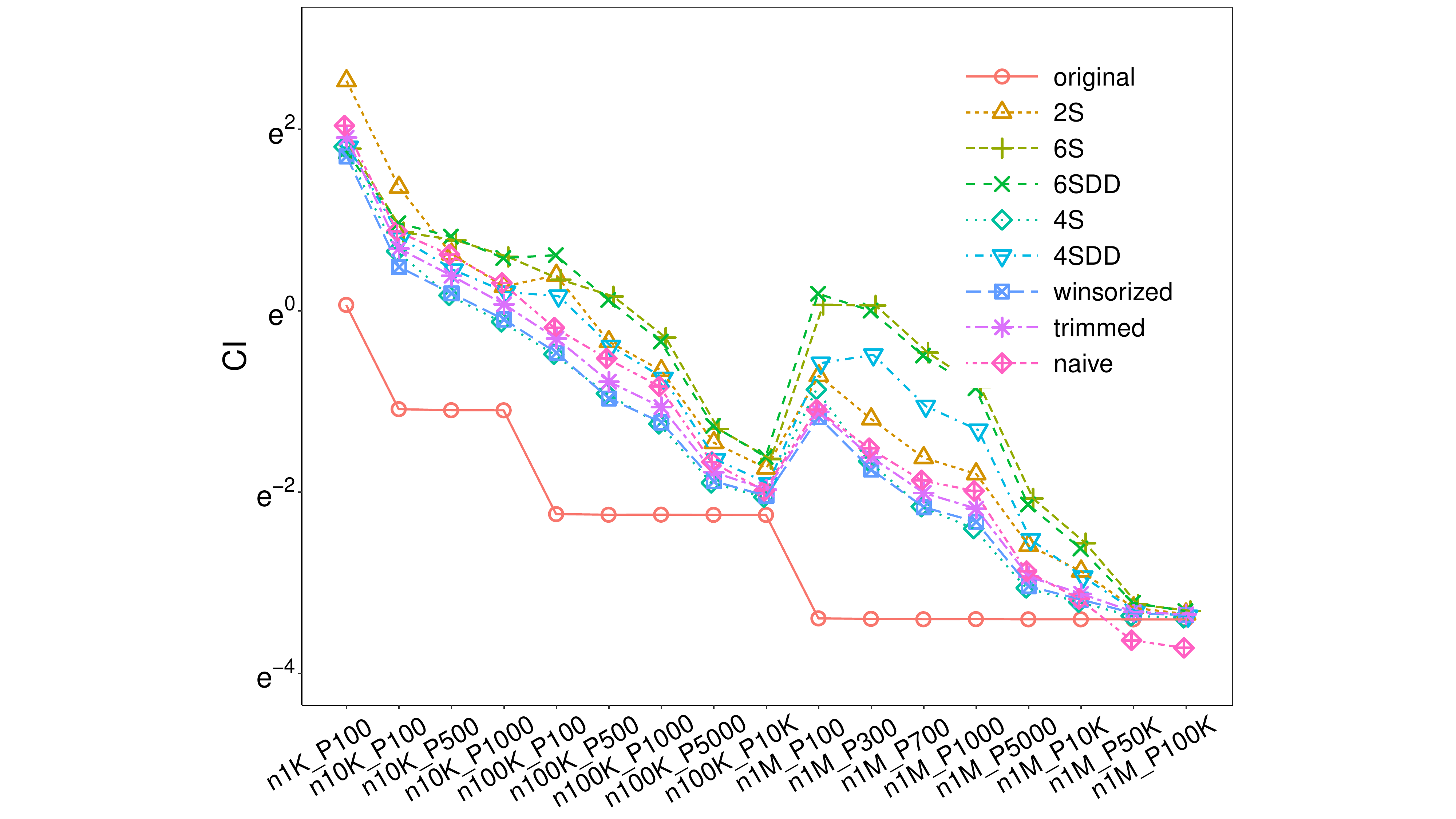}
\includegraphics[width=0.215\textwidth, trim={2.2in 0 2.2in 0},clip] {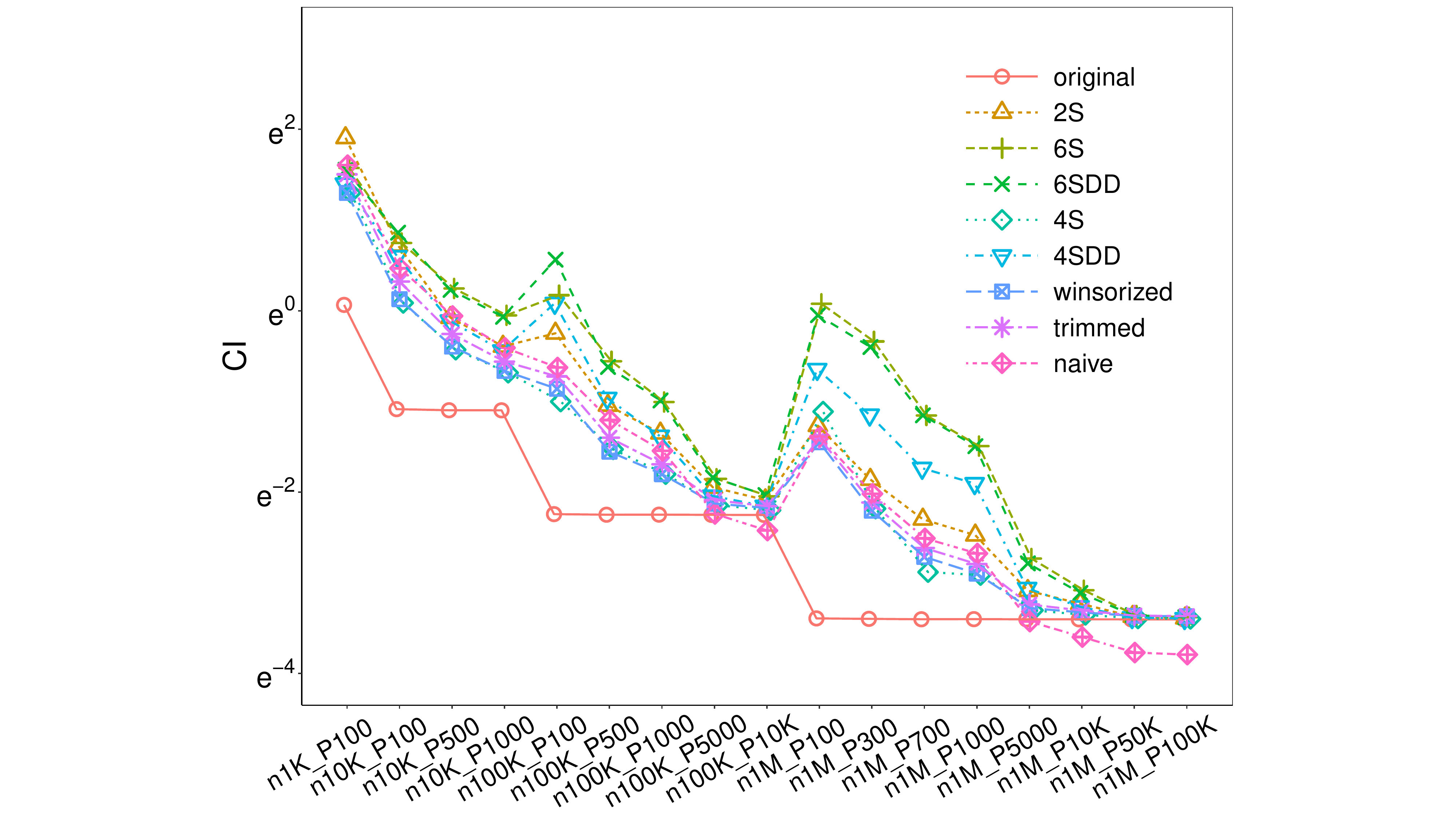}
\includegraphics[width=0.215\textwidth, trim={2.2in 0 2.2in 0},clip] {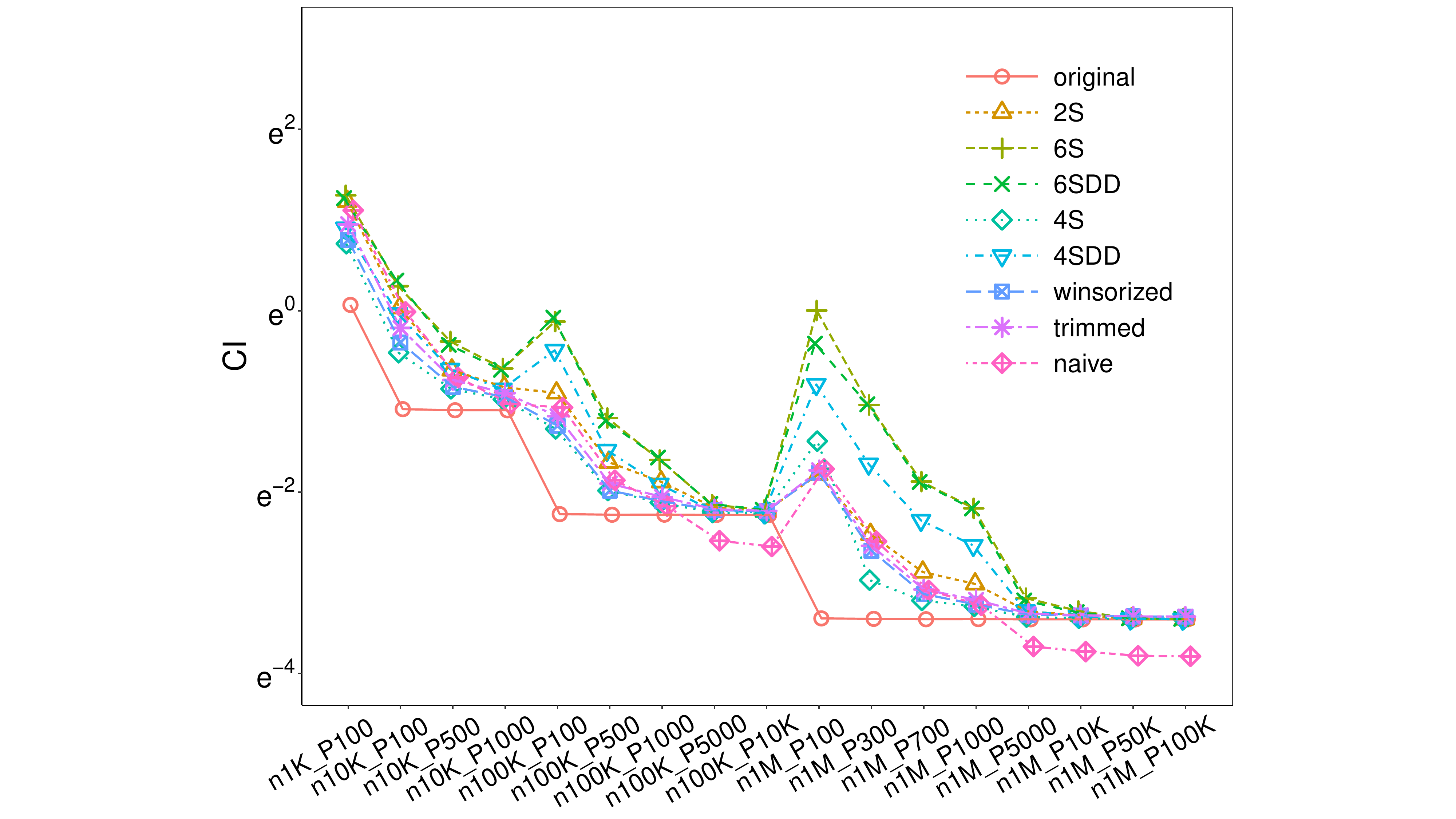}
\includegraphics[width=0.215\textwidth, trim={2.2in 0 2.2in 0},clip] {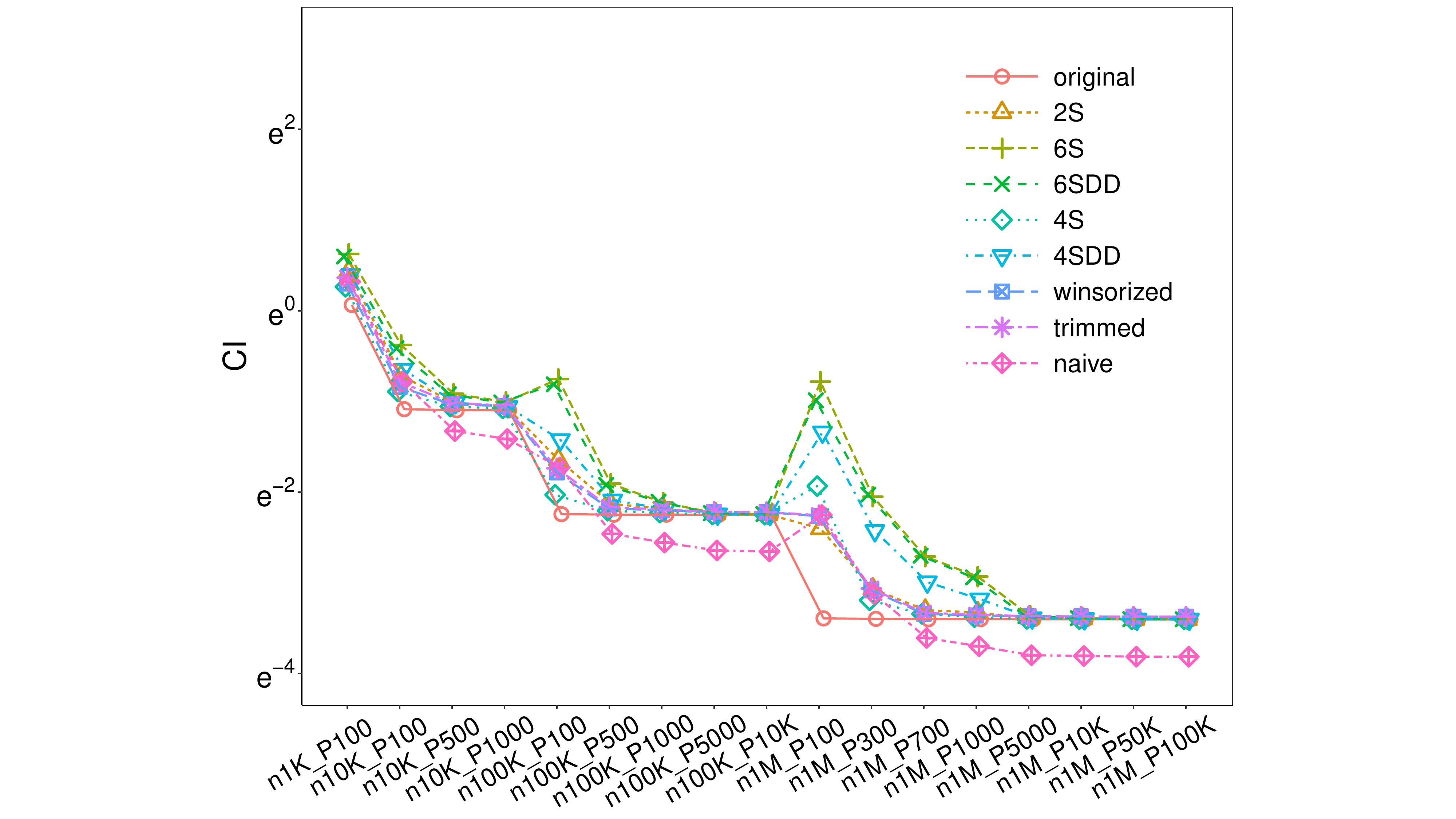}
\includegraphics[width=0.215\textwidth, trim={2.2in 0 2.2in 0},clip] {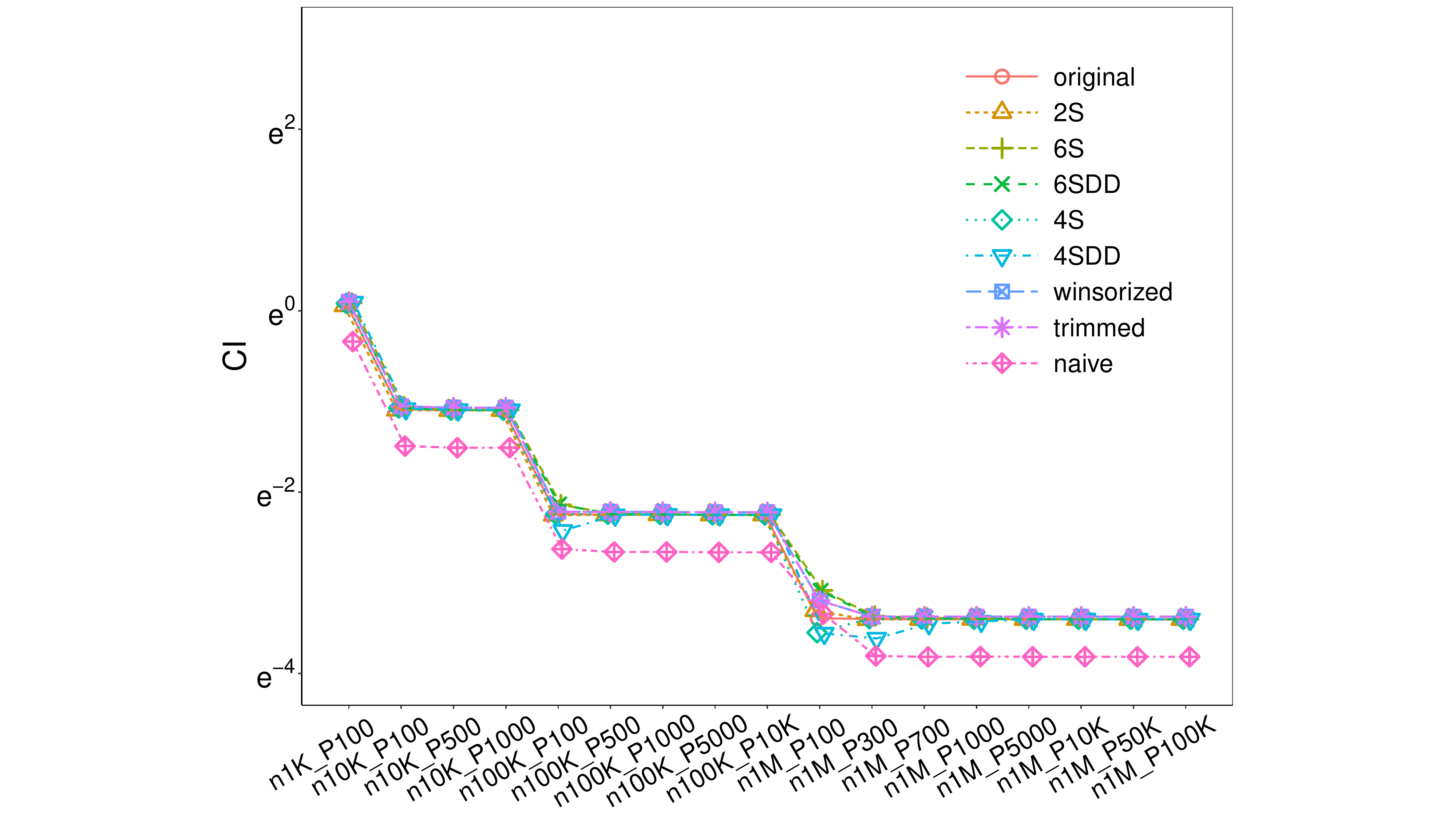}\\
\includegraphics[width=0.215\textwidth, trim={2.2in 0 2.2in 0},clip] {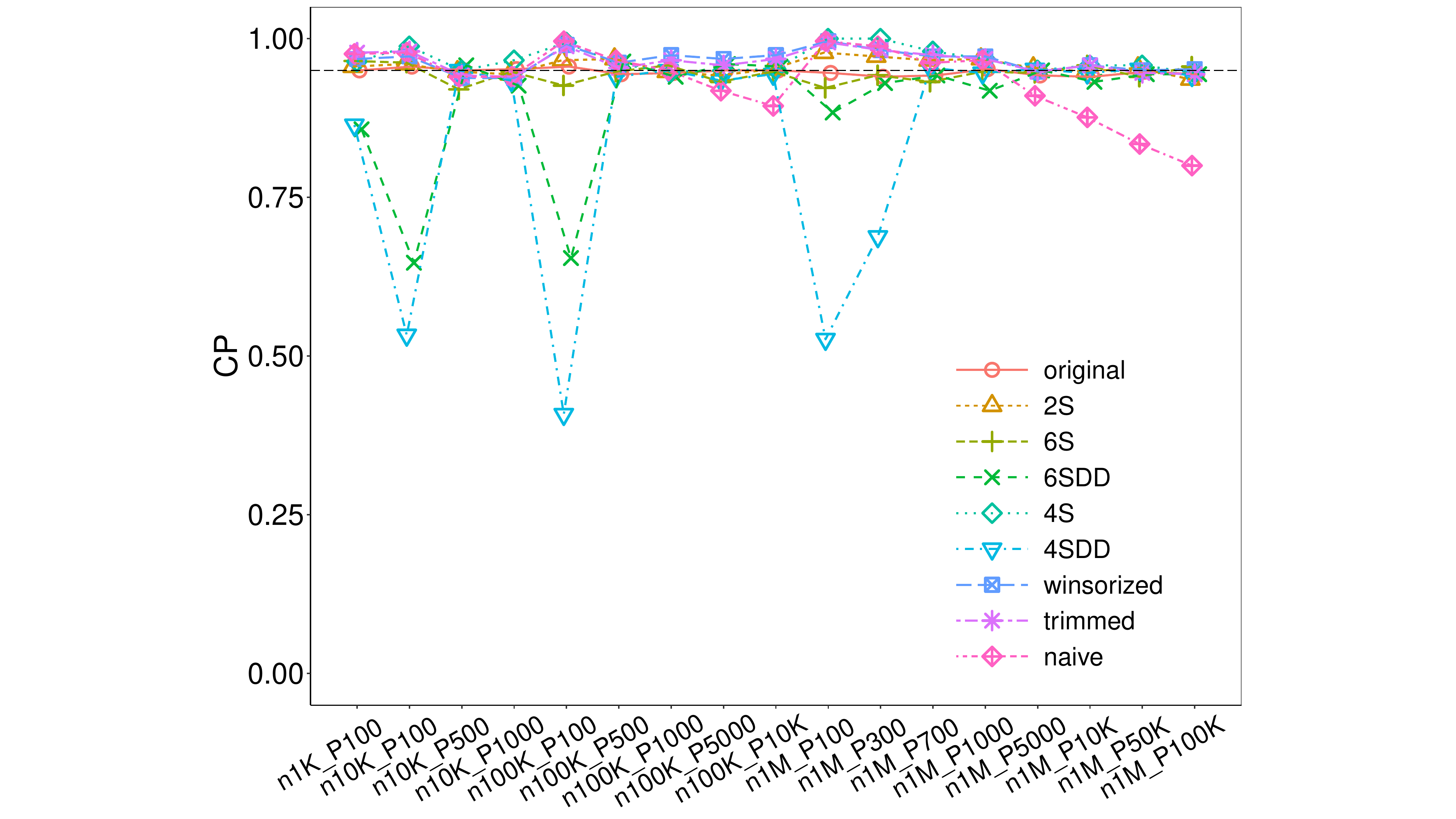}
\includegraphics[width=0.215\textwidth, trim={2.2in 0 2.2in 0},clip] {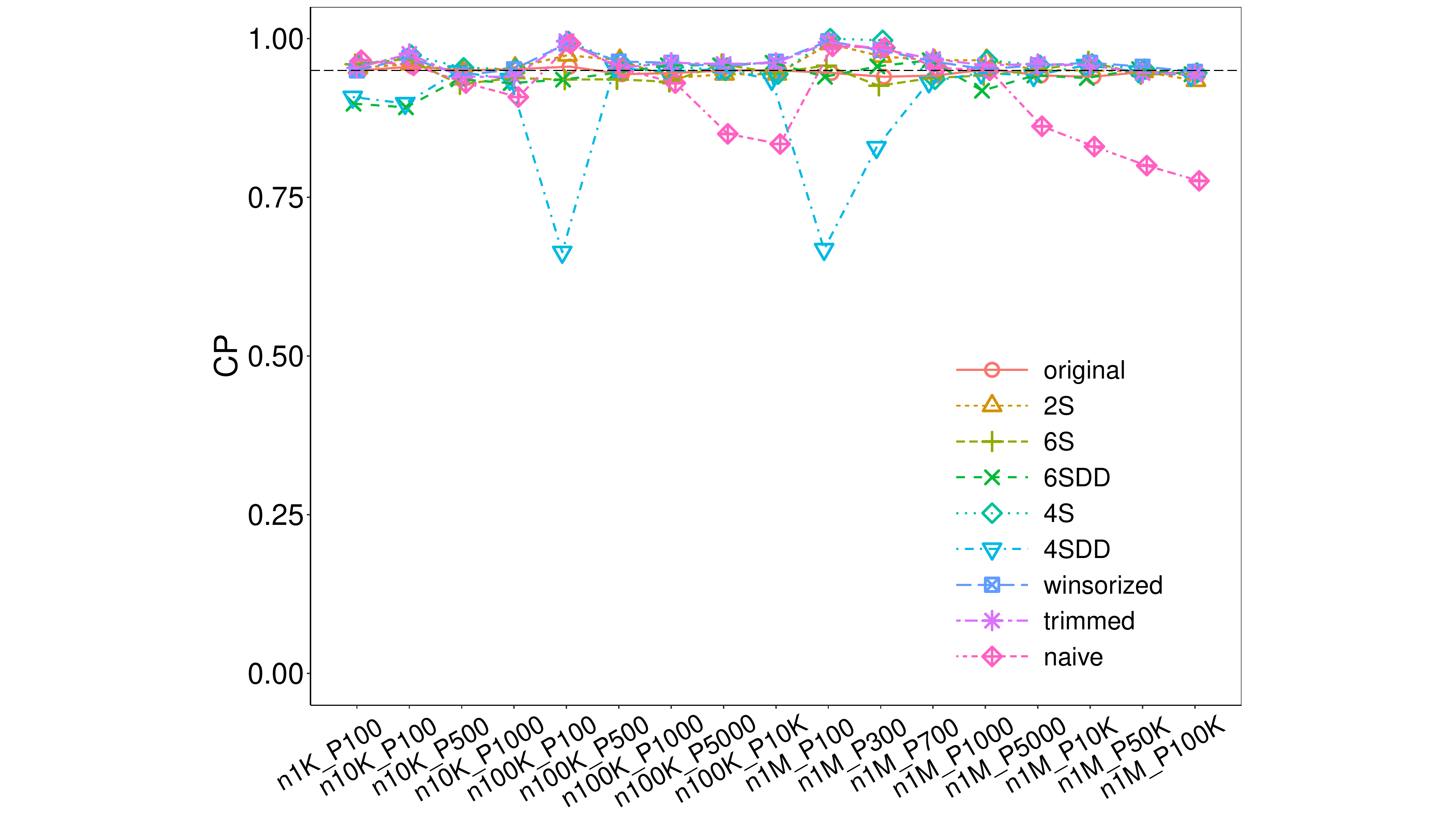}
\includegraphics[width=0.215\textwidth, trim={2.2in 0 2.2in 0},clip] {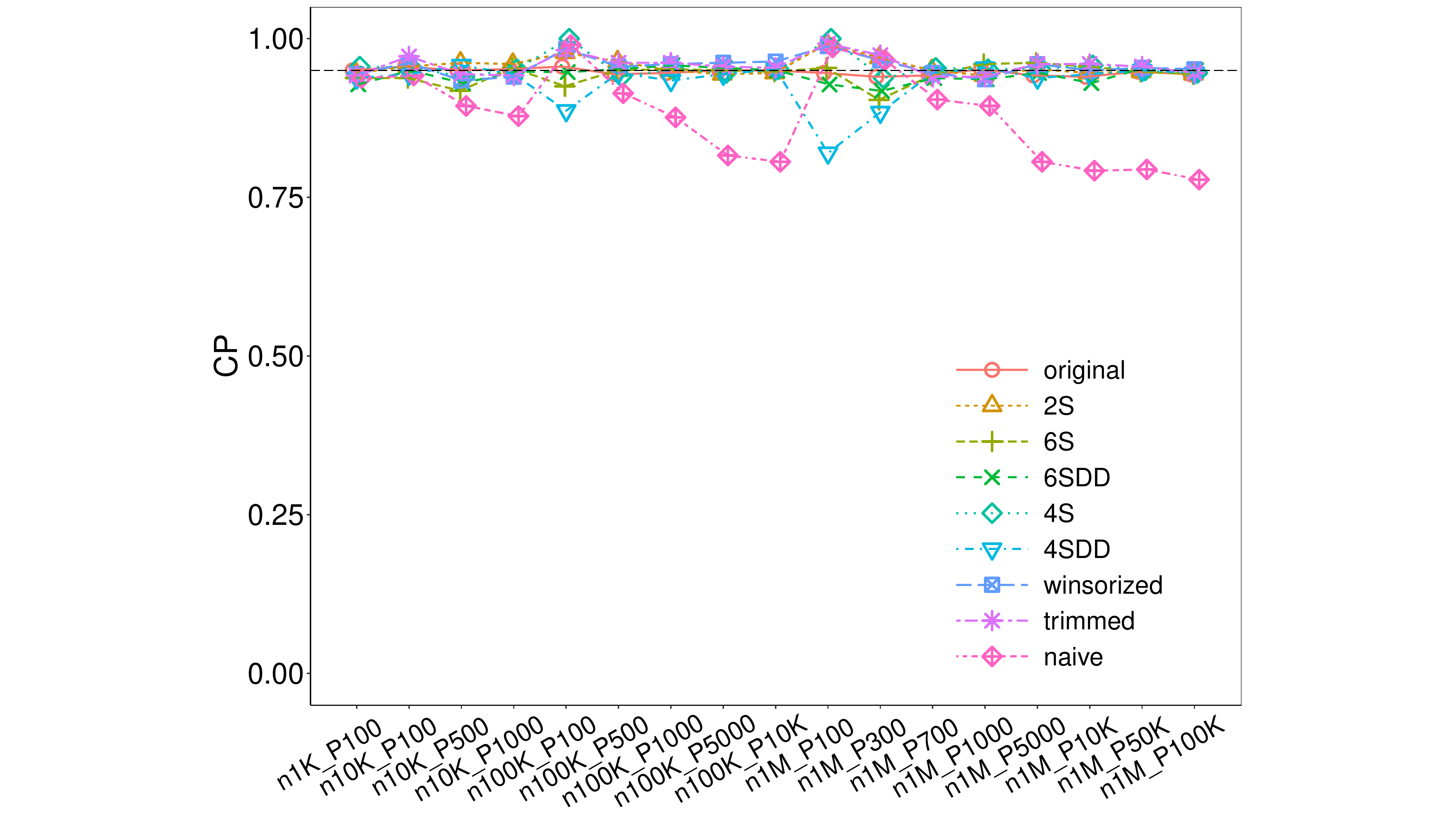}
\includegraphics[width=0.215\textwidth, trim={2.2in 0 2.2in 0},clip] {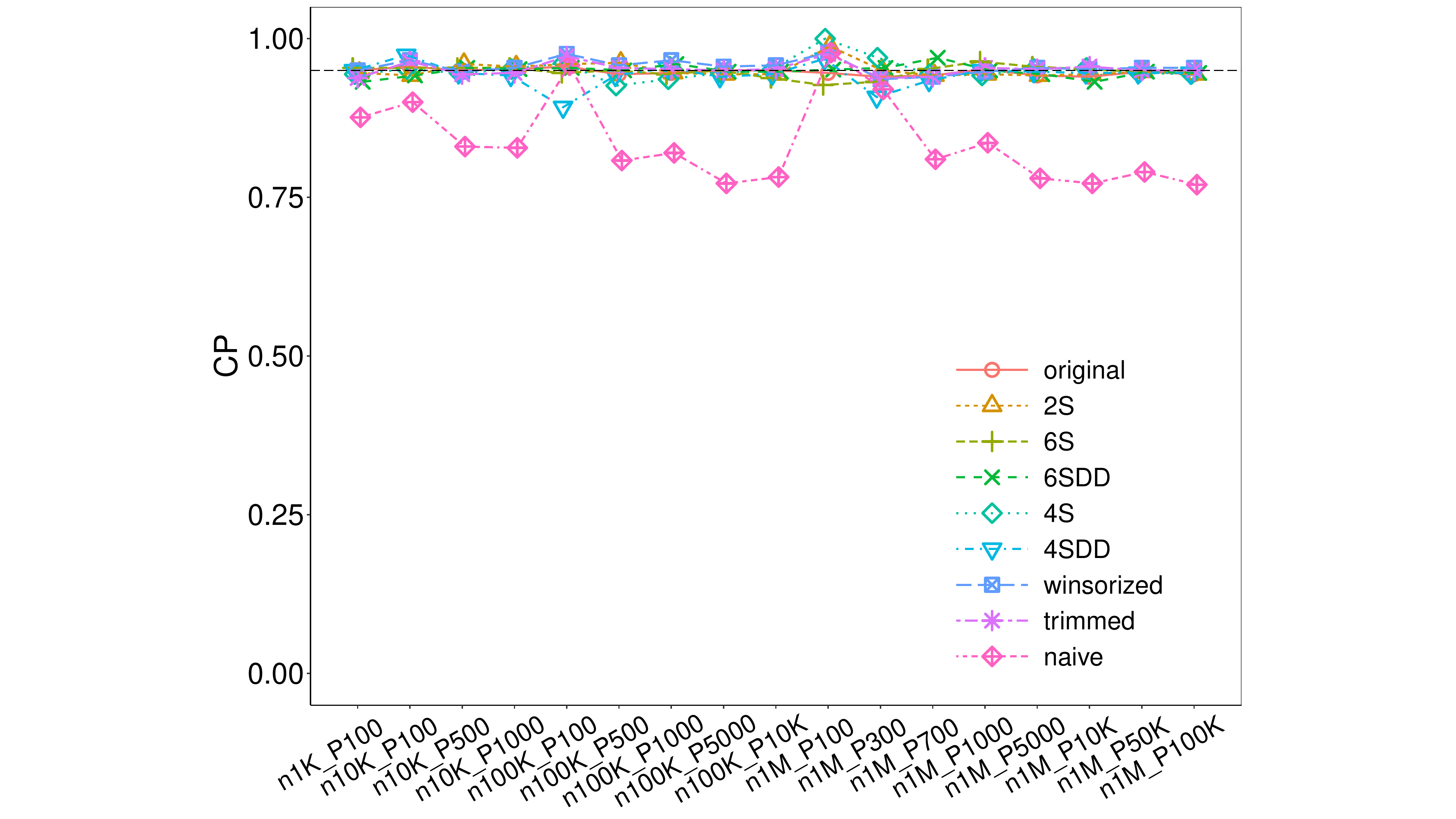}
\includegraphics[width=0.215\textwidth, trim={2.2in 0 2.2in 0},clip] {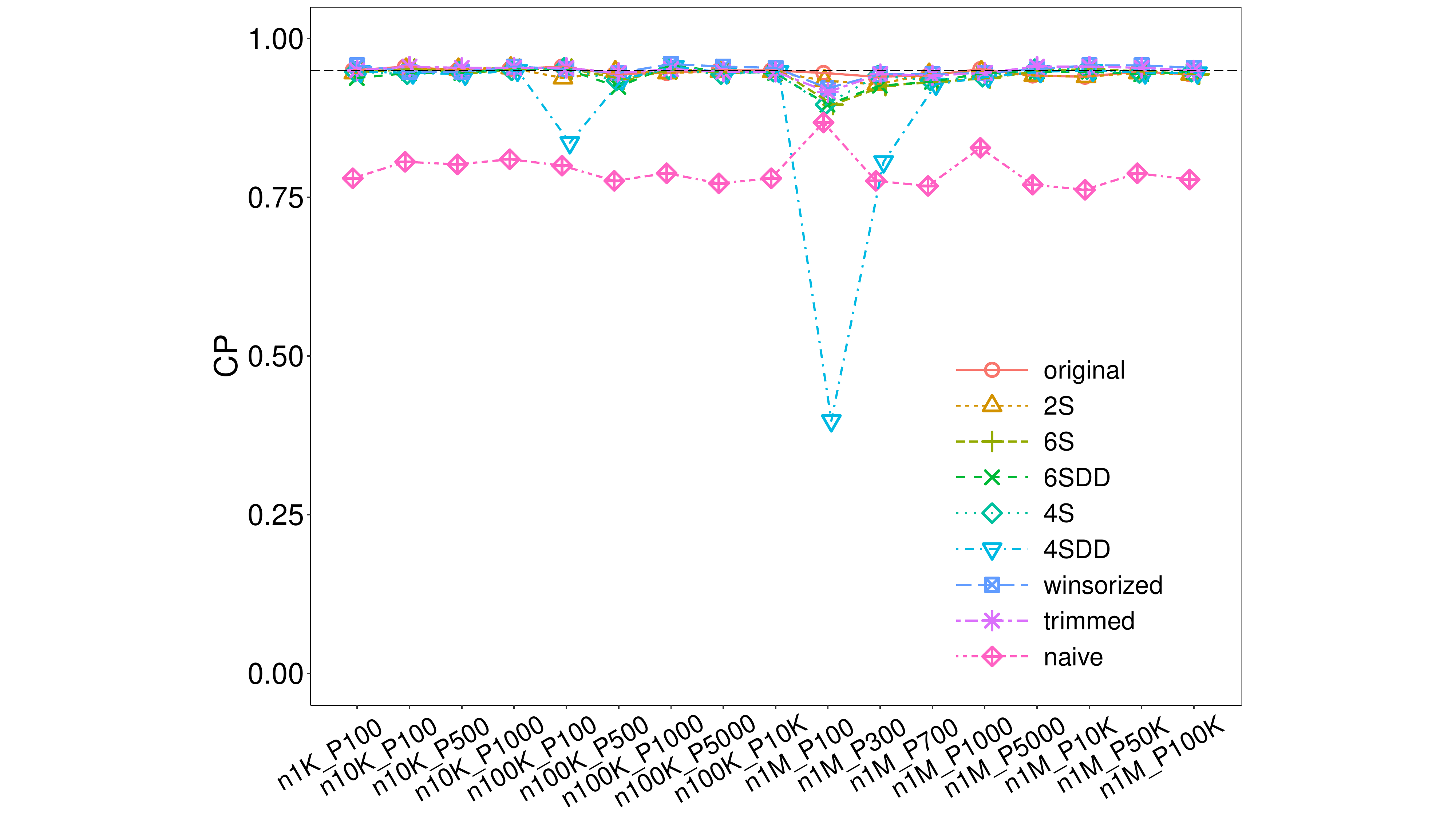}\\
\includegraphics[width=0.215\textwidth, trim={2.2in 0 2.2in 0},clip] {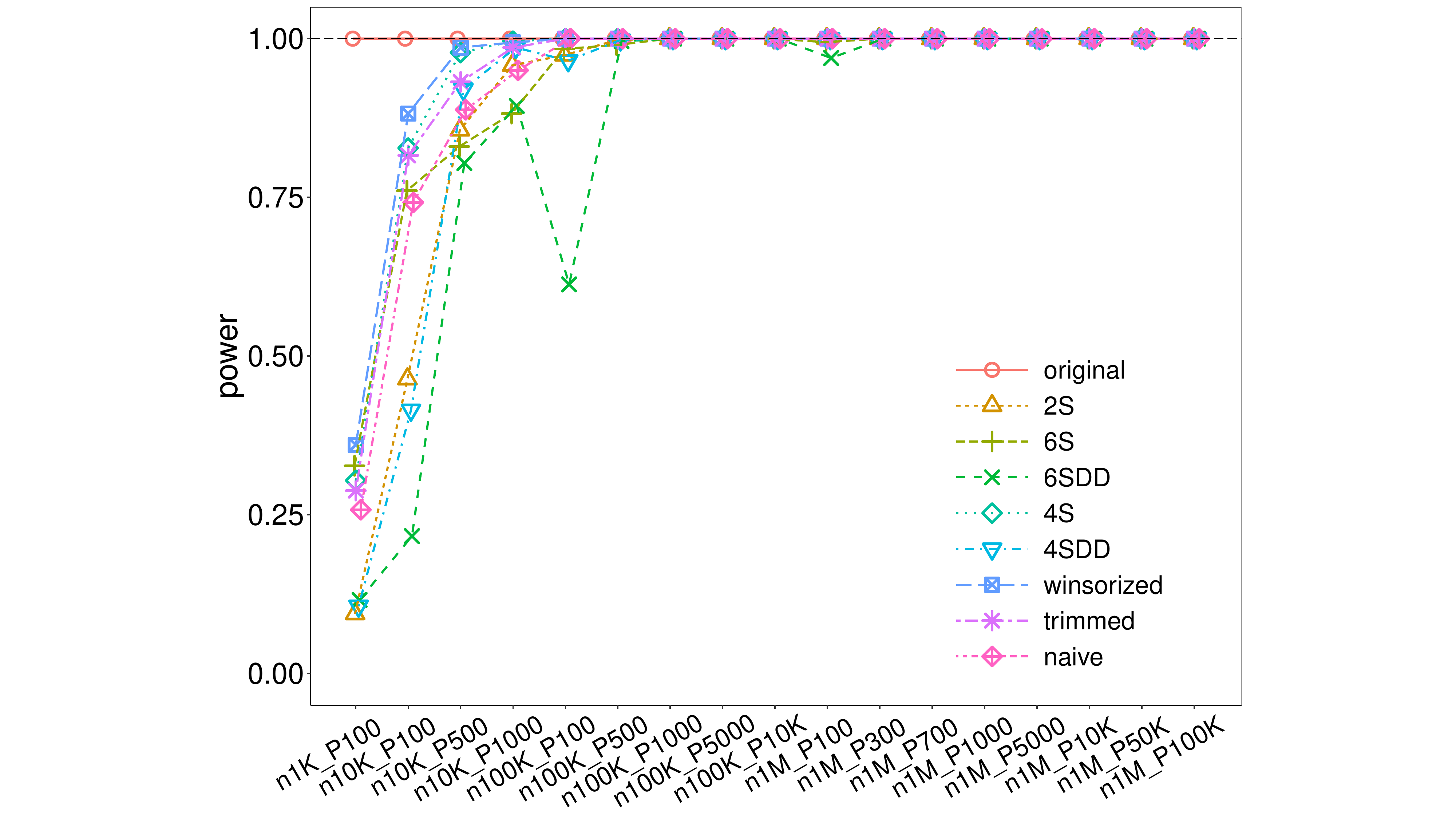}
\includegraphics[width=0.215\textwidth, trim={2.2in 0 2.2in 0},clip] {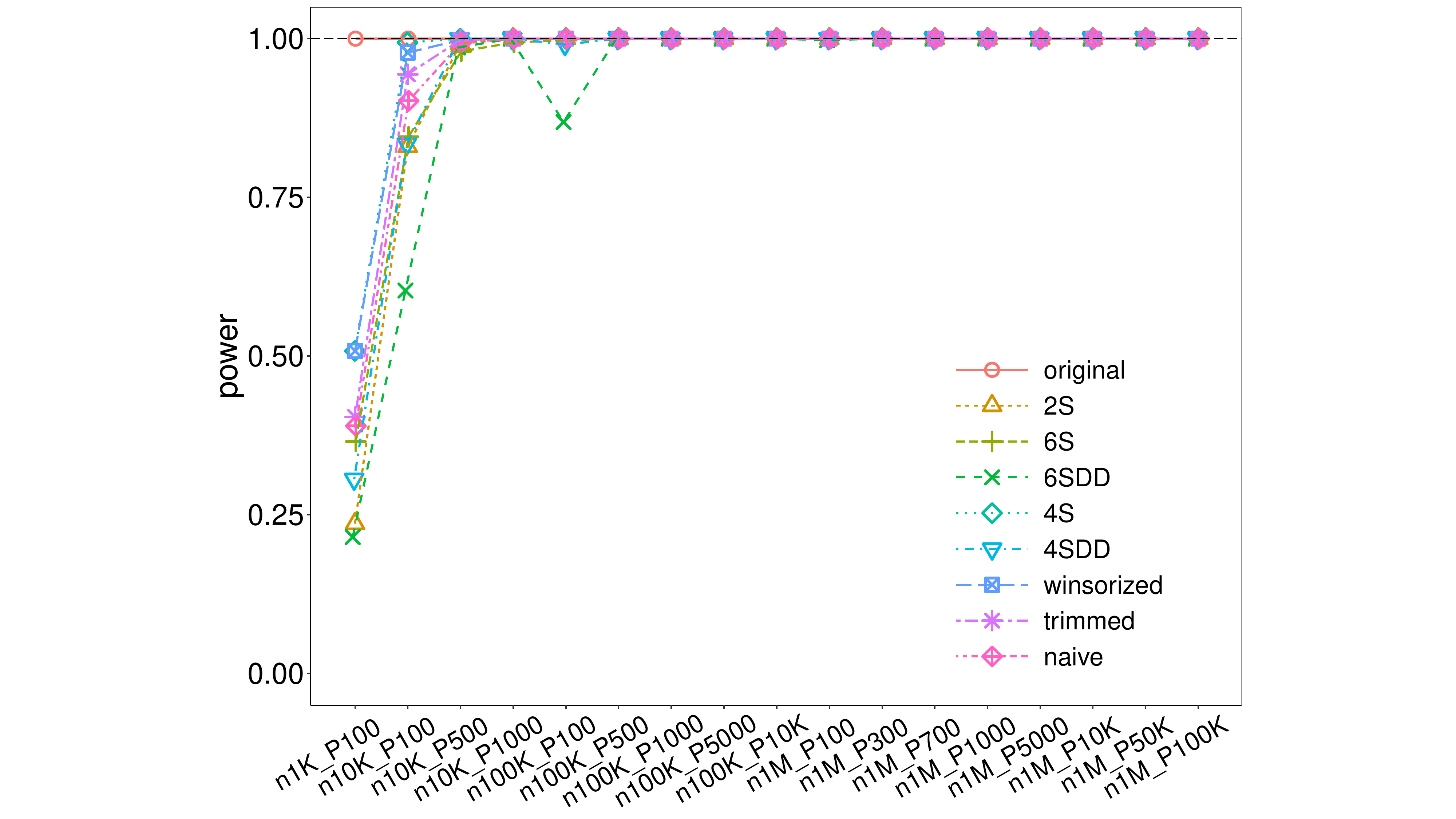}
\includegraphics[width=0.215\textwidth, trim={2.2in 0 2.2in 0},clip] {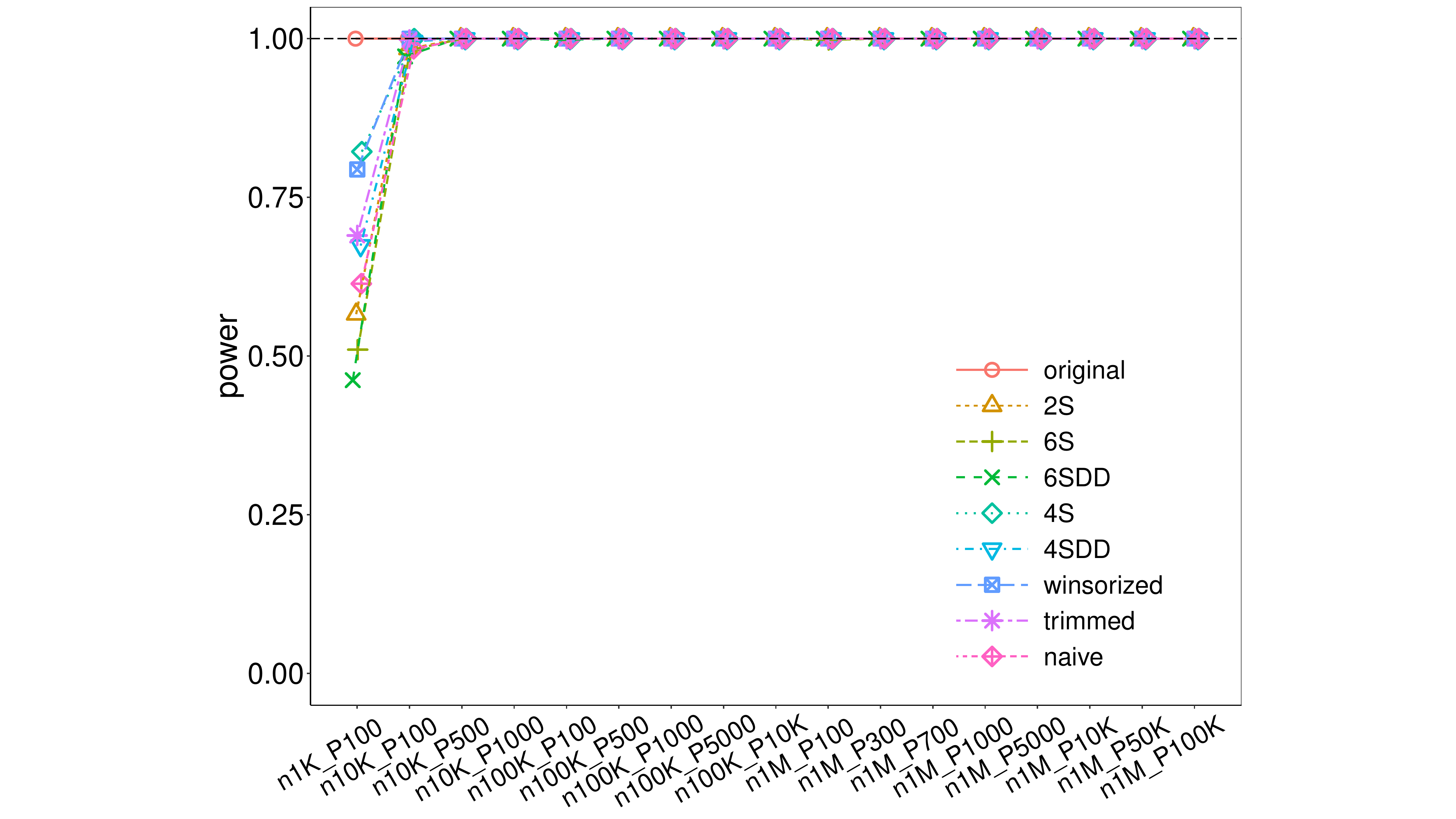}
\includegraphics[width=0.215\textwidth, trim={2.2in 0 2.2in 0},clip] {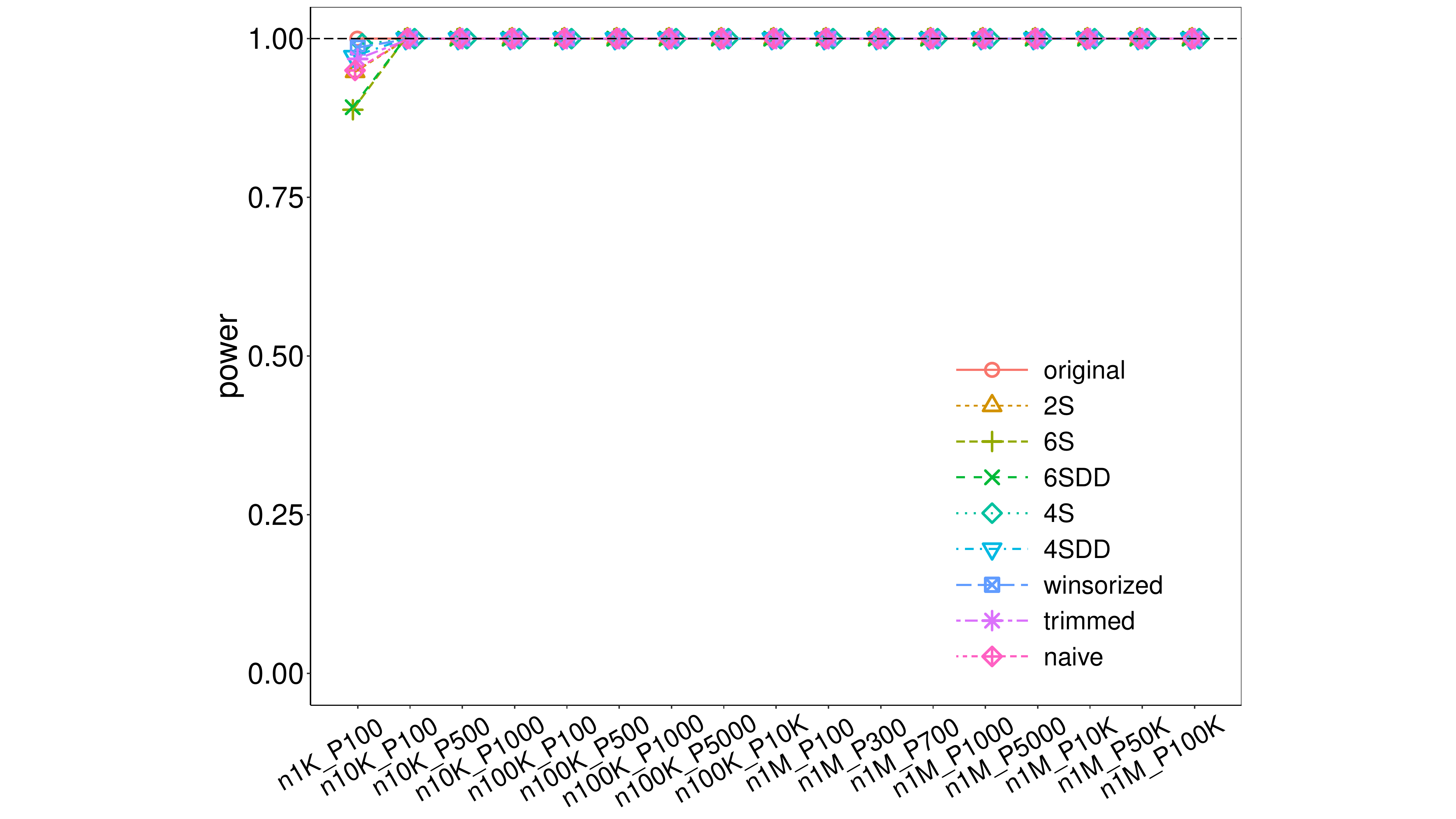}
\includegraphics[width=0.215\textwidth, trim={2.2in 0 2.2in 0},clip] {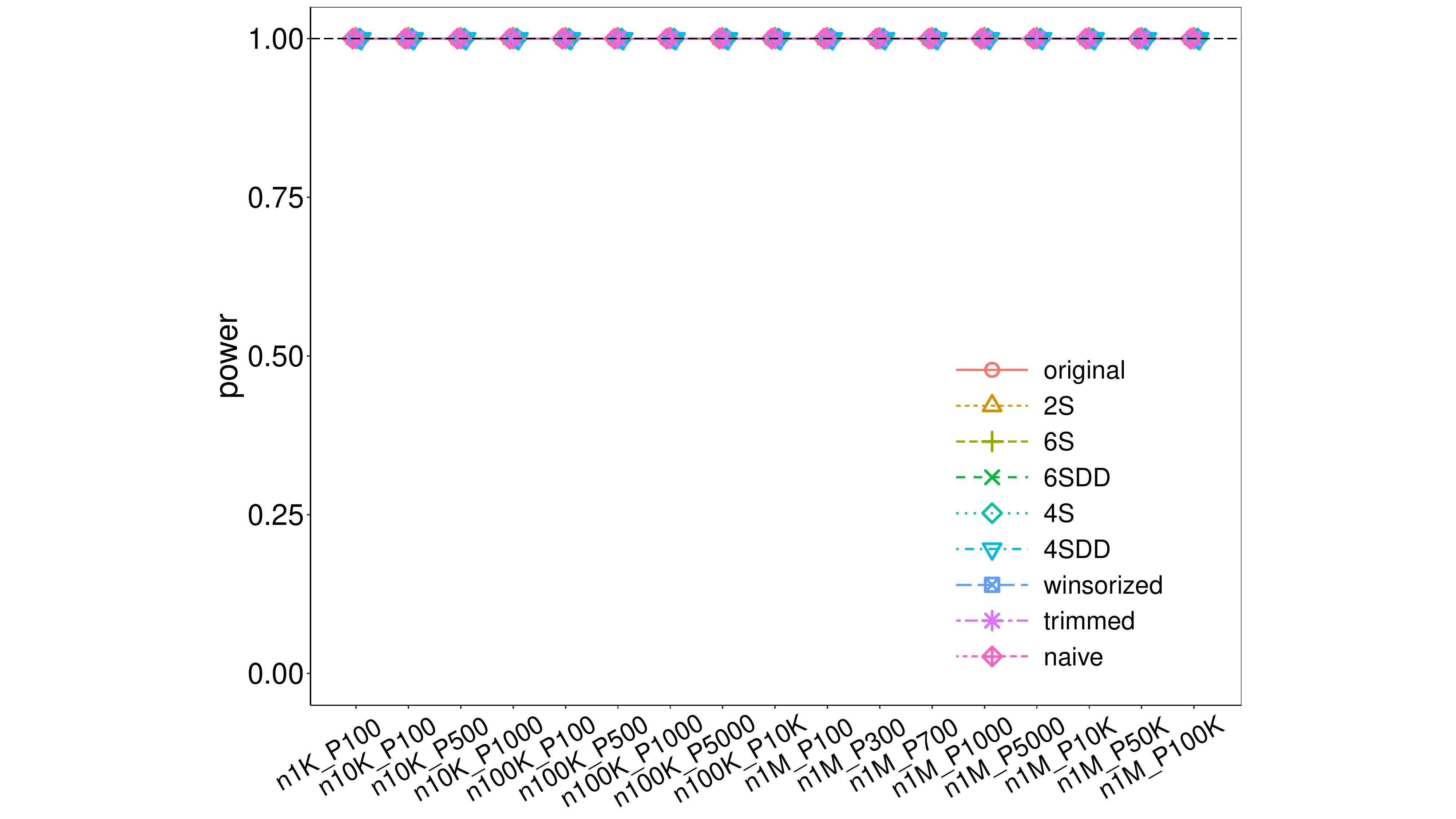}\\
\caption{Gaussian data; $\epsilon$-DP; $\theta\ne0$ and $\alpha=\beta$} \label{fig:1sDP}
\end{figure}

\end{landscape}

\begin{landscape}

\begin{figure}[!htb]
\centering
$\rho=0.005$\hspace{1in}$\rho=0.02$\hspace{1in}$\rho=0.08$
\hspace{1in}$\rho=0.32$\hspace{0.8in}$\rho=1.28$\\
\includegraphics[width=0.24\textwidth, trim={2.2in 0 2.2in 0},clip] {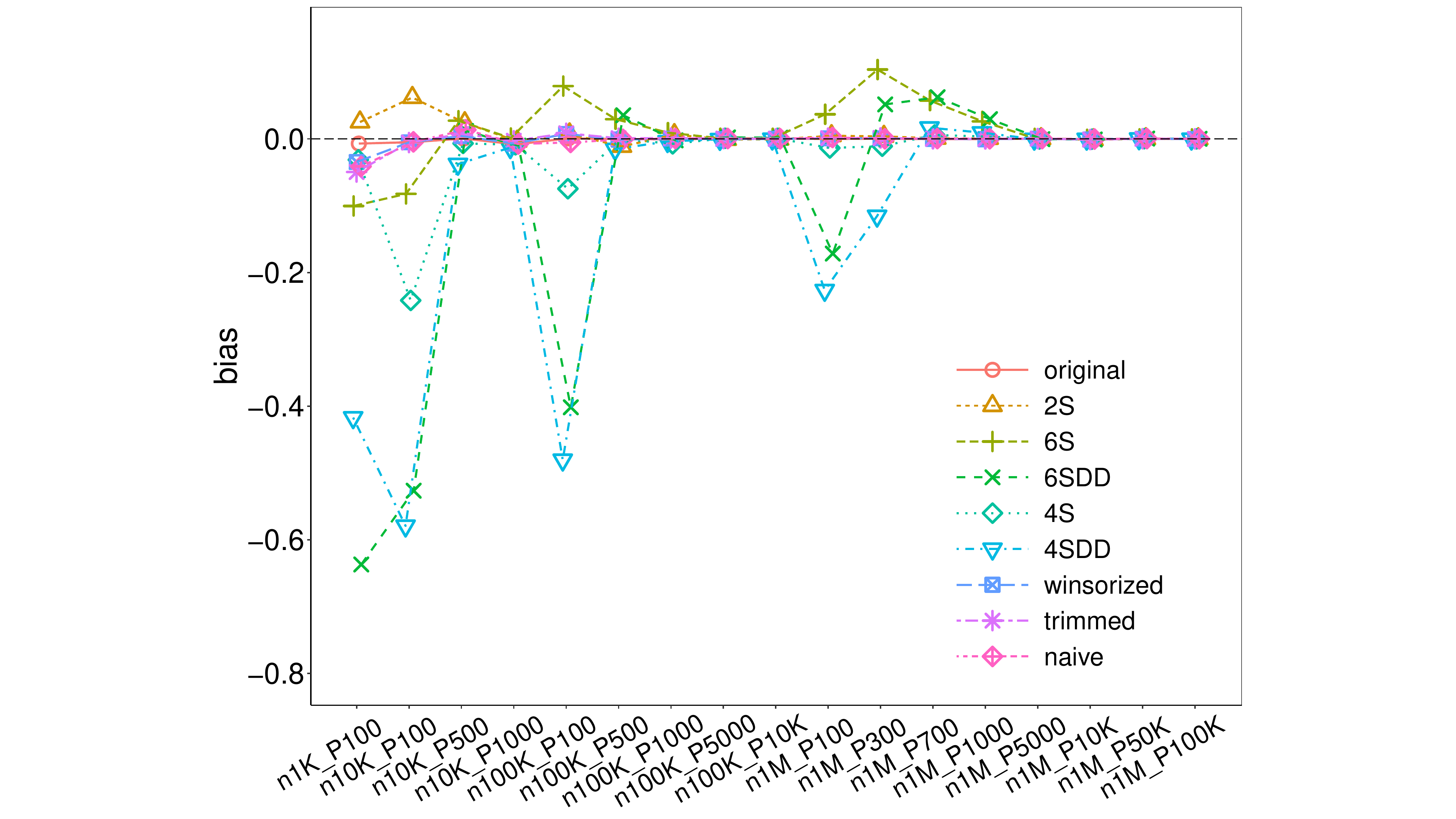}
\includegraphics[width=0.24\textwidth, trim={2.2in 0 2.2in 0},clip] {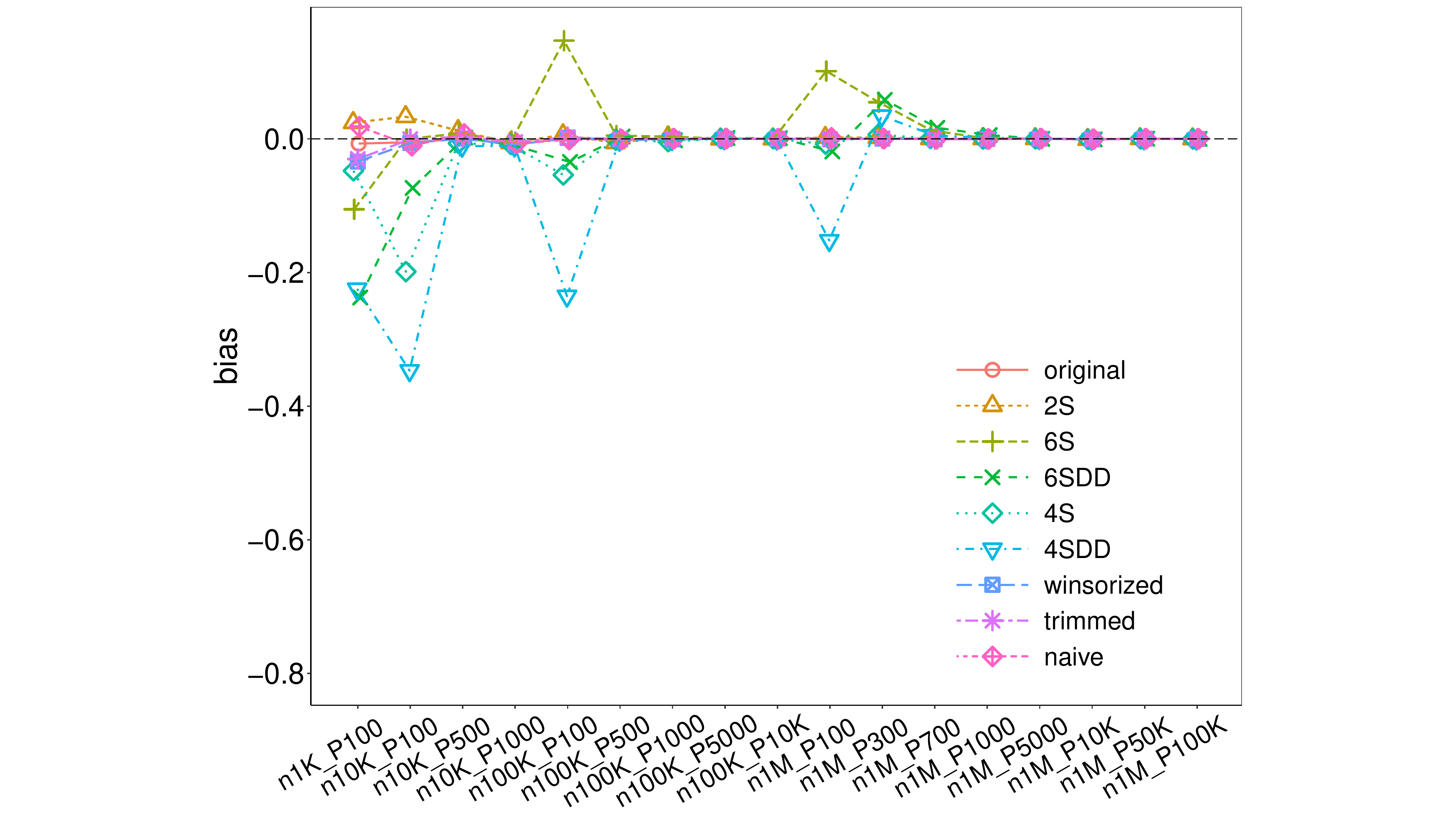}
\includegraphics[width=0.24\textwidth, trim={2.2in 0 2.2in 0},clip] {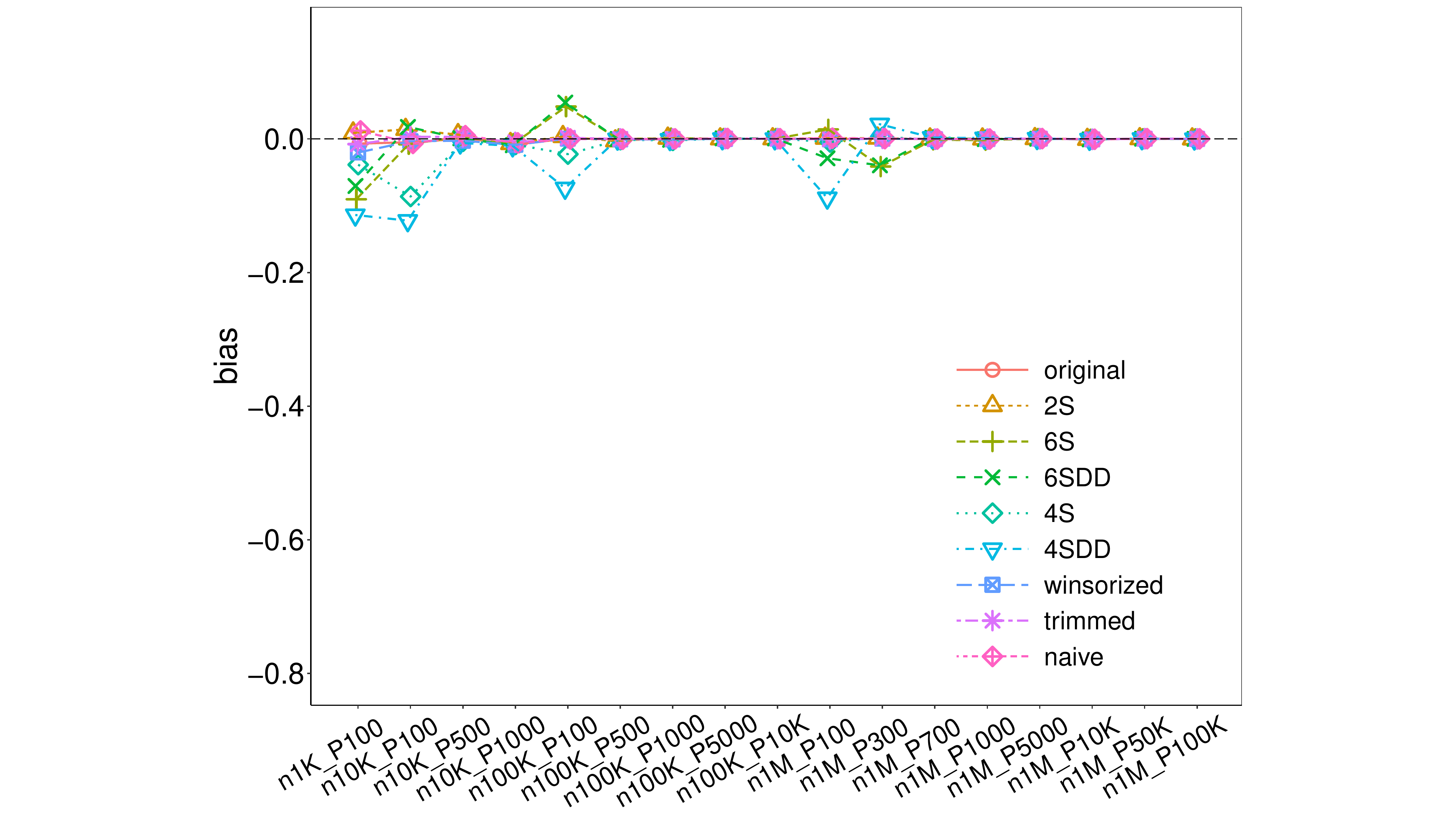}
\includegraphics[width=0.24\textwidth, trim={2.2in 0 2.2in 0},clip] {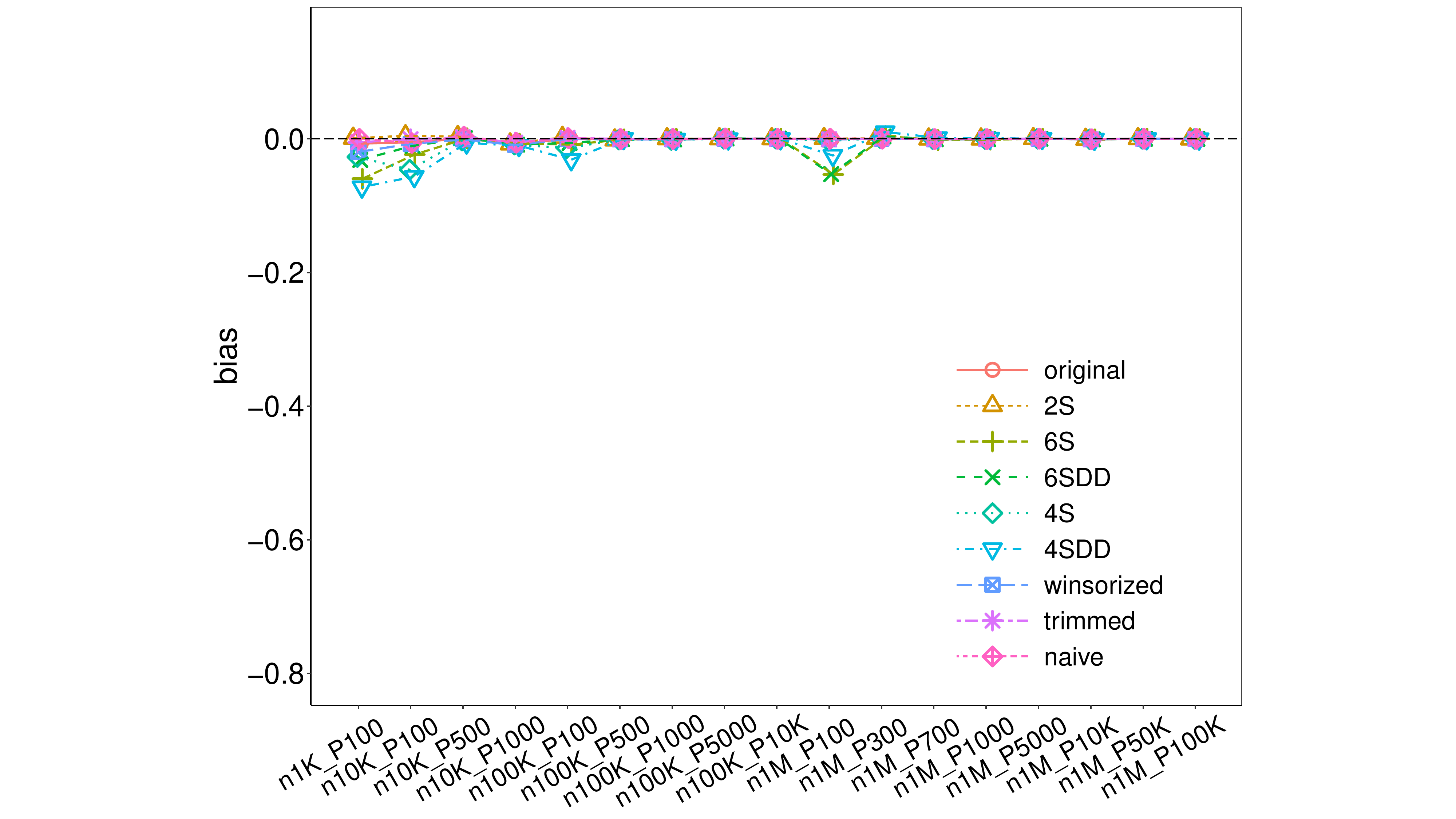}
\includegraphics[width=0.24\textwidth, trim={2.2in 0 2.2in 0},clip] {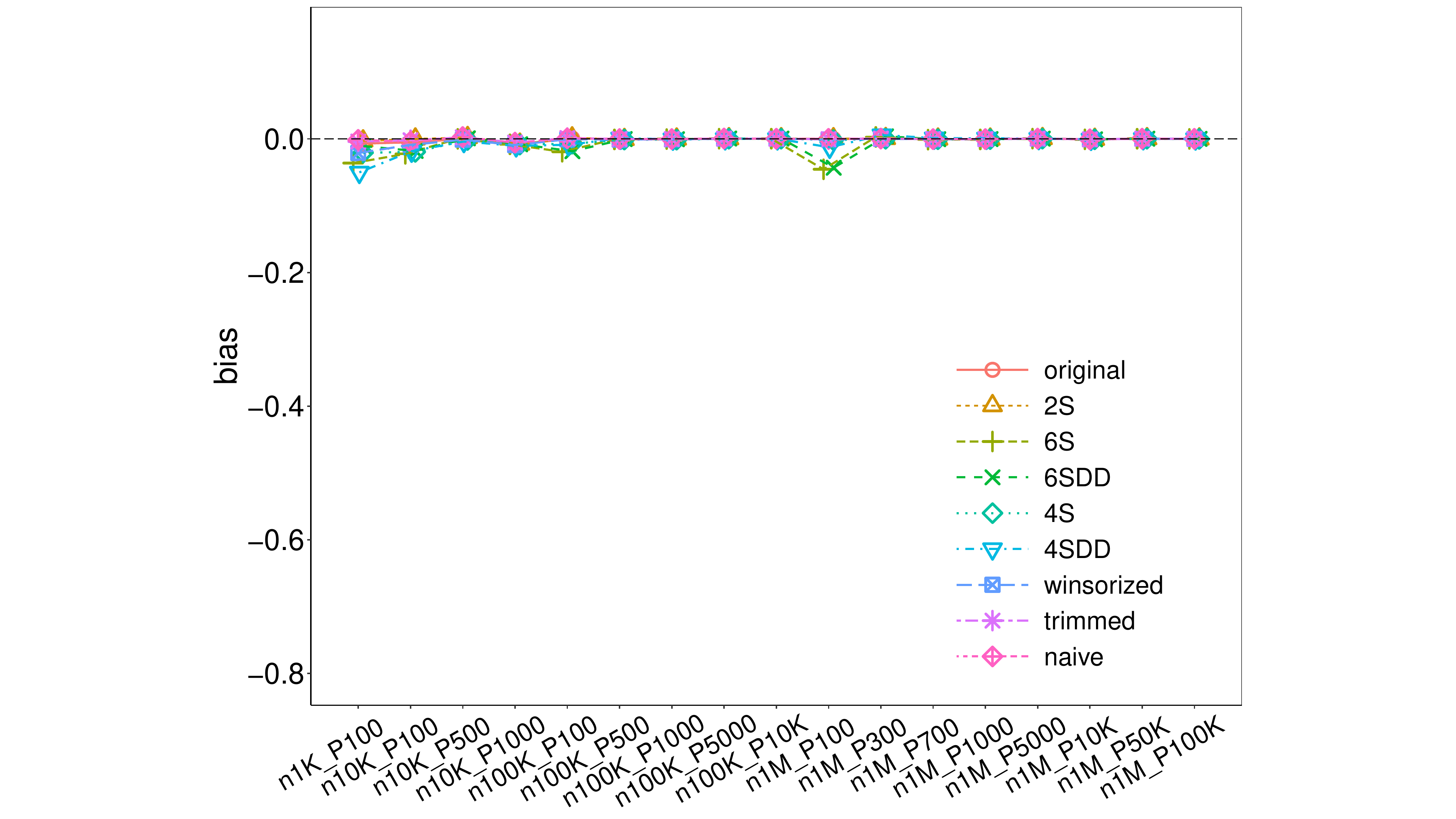}\\
\includegraphics[width=0.24\textwidth, trim={2.2in 0 2.2in 0},clip] {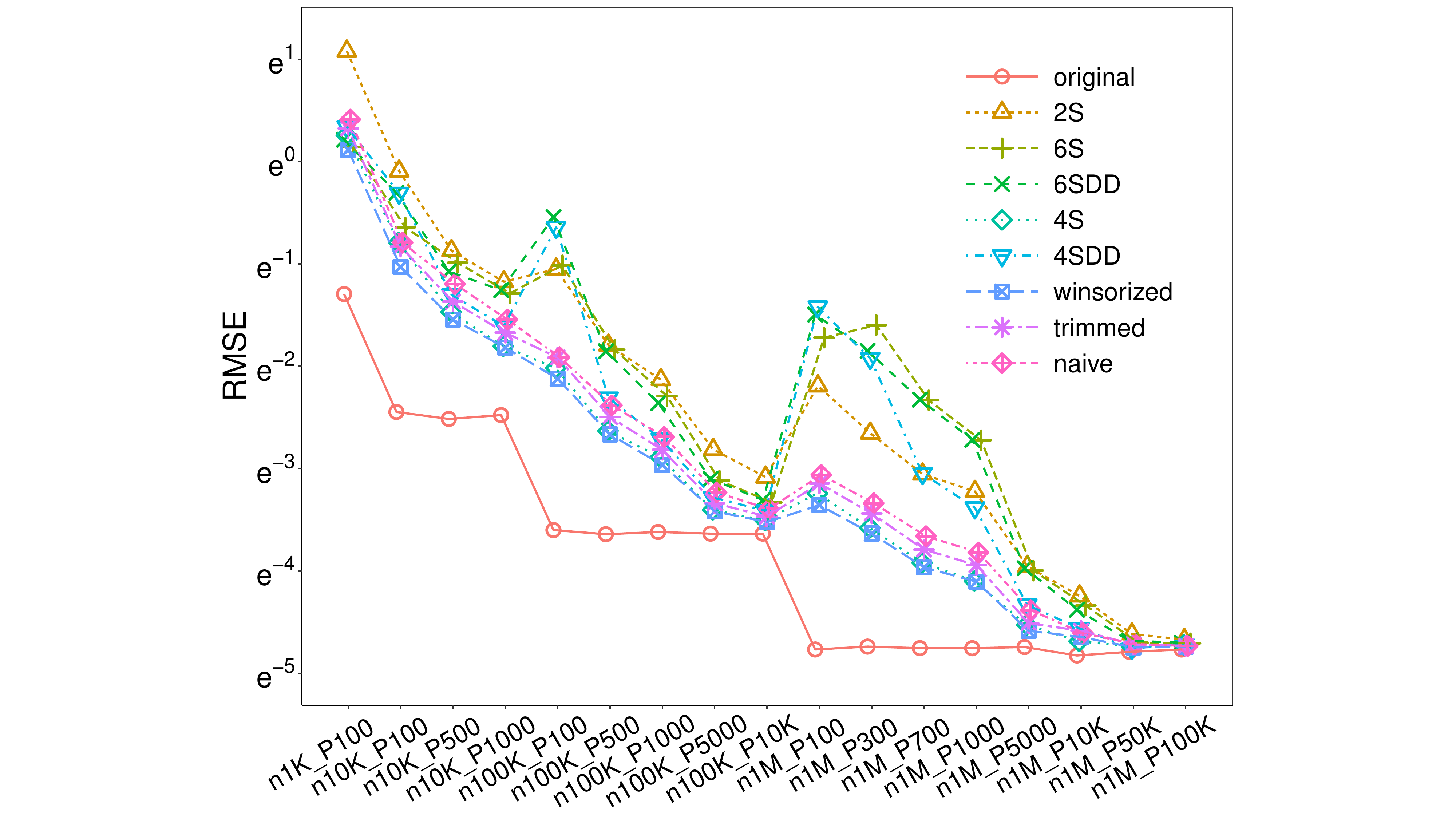}
\includegraphics[width=0.24\textwidth, trim={2.2in 0 2.2in 0},clip] {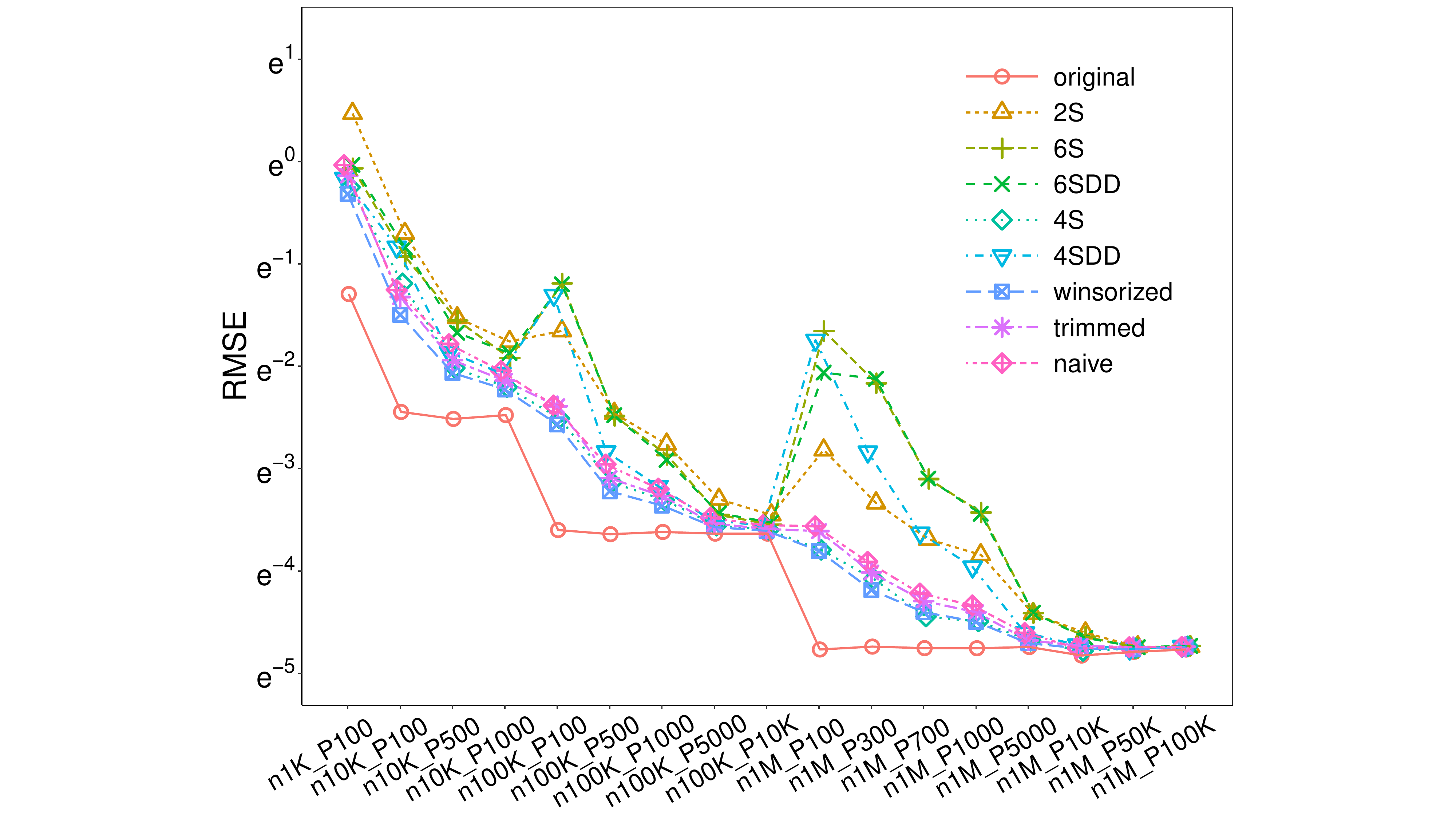}
\includegraphics[width=0.24\textwidth, trim={2.2in 0 2.2in 0},clip] {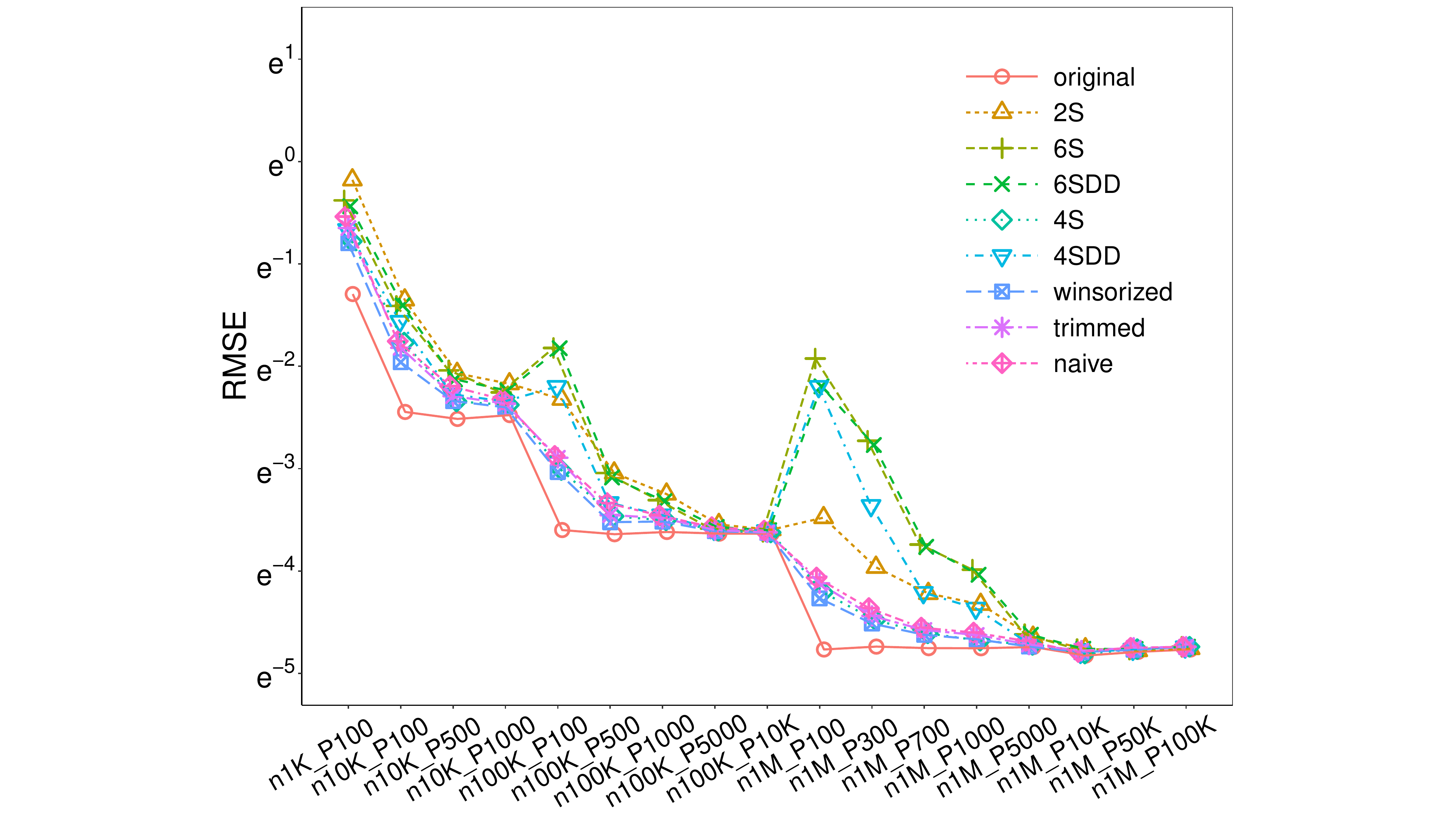}
\includegraphics[width=0.24\textwidth, trim={2.2in 0 2.2in 0},clip] {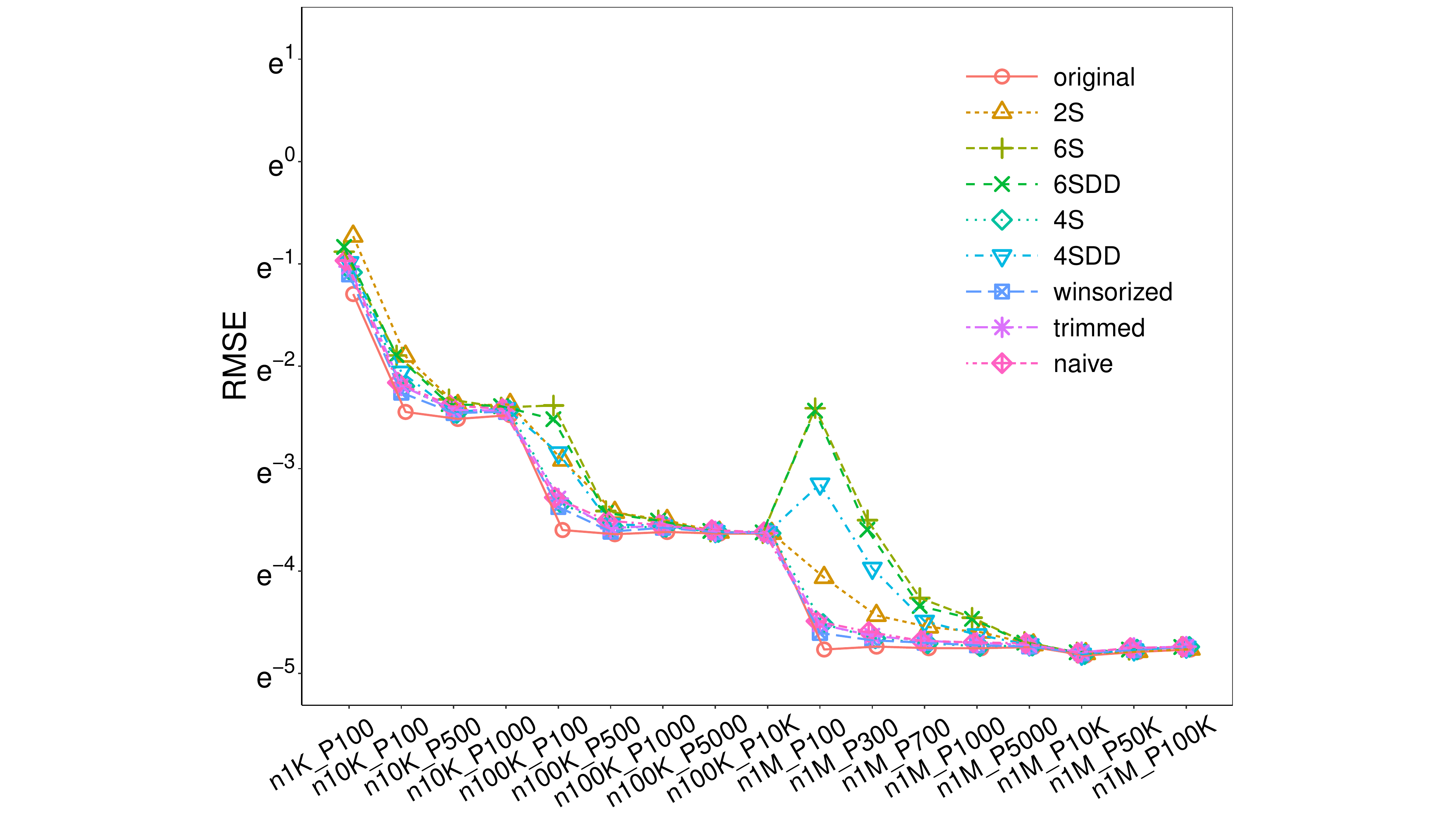}
\includegraphics[width=0.24\textwidth, trim={2.2in 0 2.2in 0},clip] {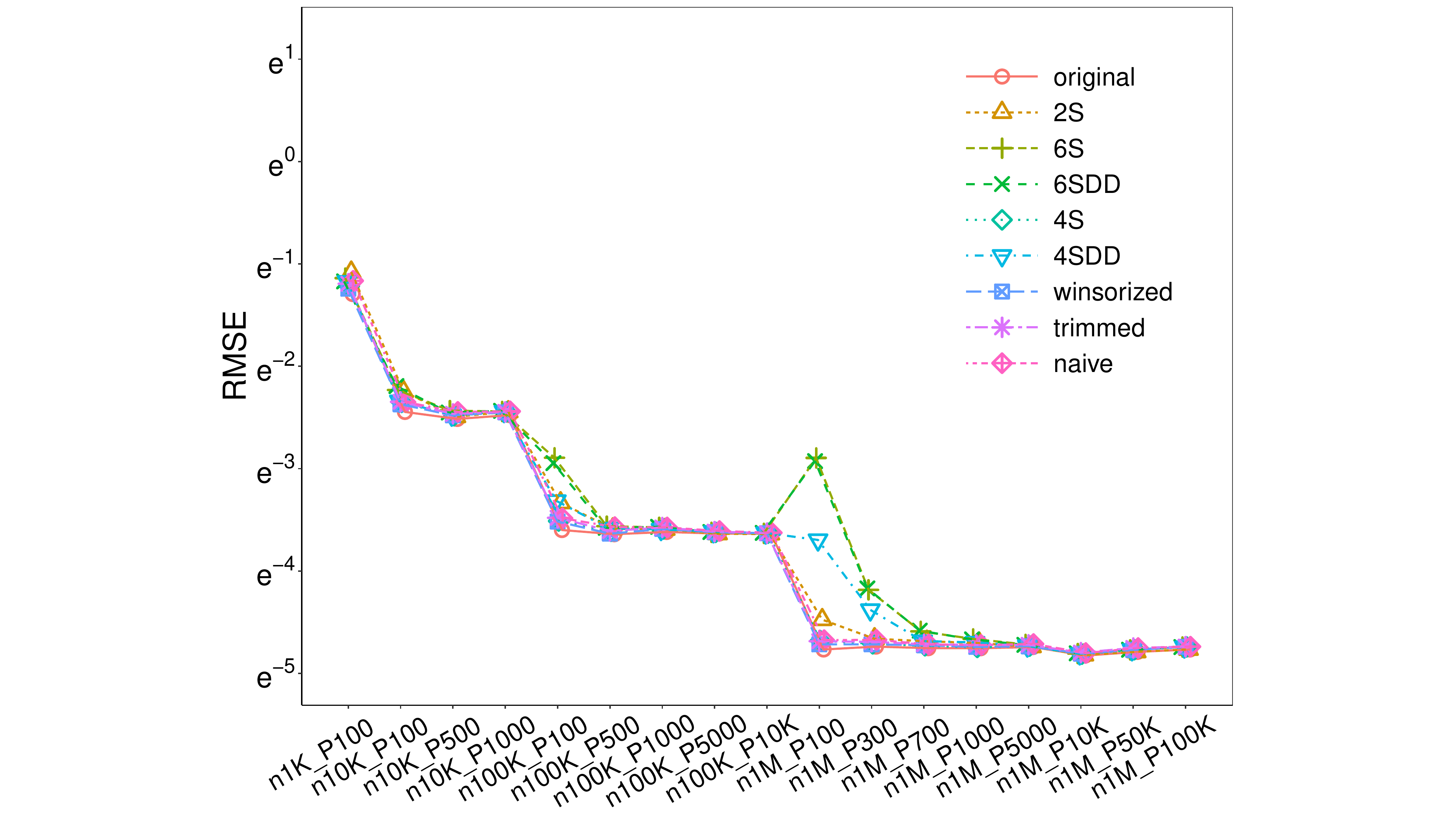}\\
\includegraphics[width=0.24\textwidth, trim={2.2in 0 2.2in 0},clip] {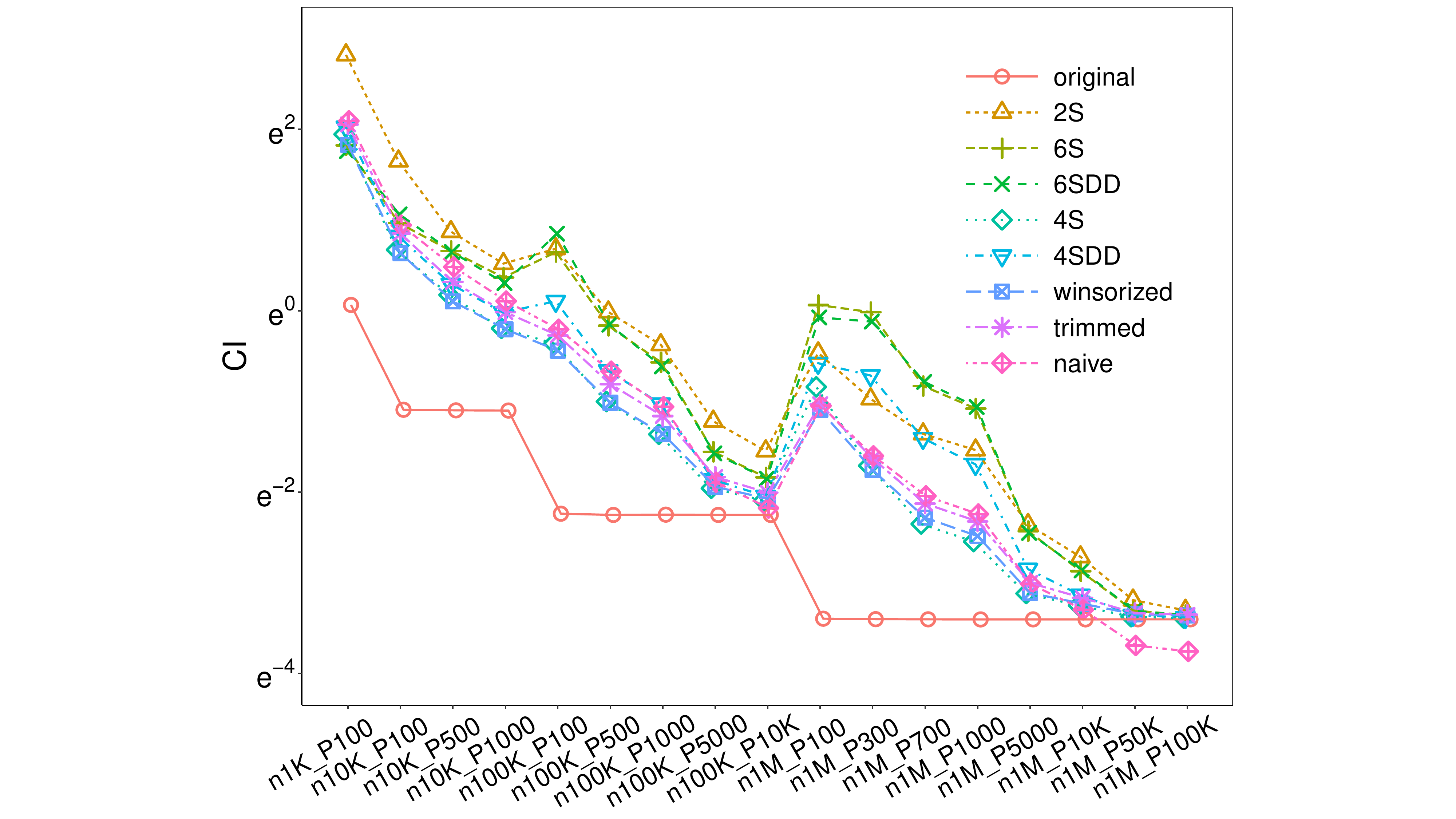}
\includegraphics[width=0.24\textwidth, trim={2.2in 0 2.2in 0},clip] {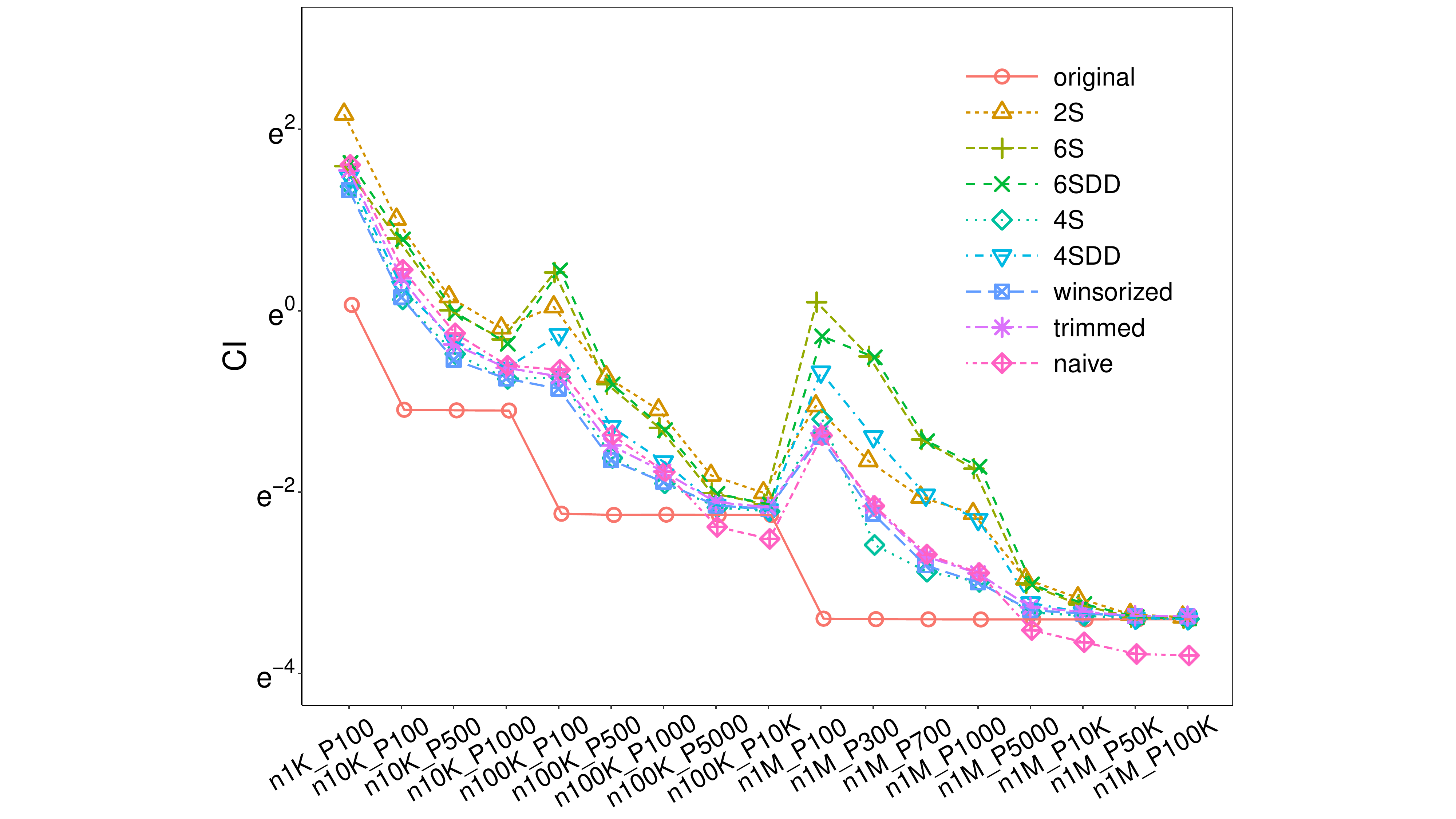}
\includegraphics[width=0.24\textwidth, trim={2.2in 0 2.2in 0},clip] {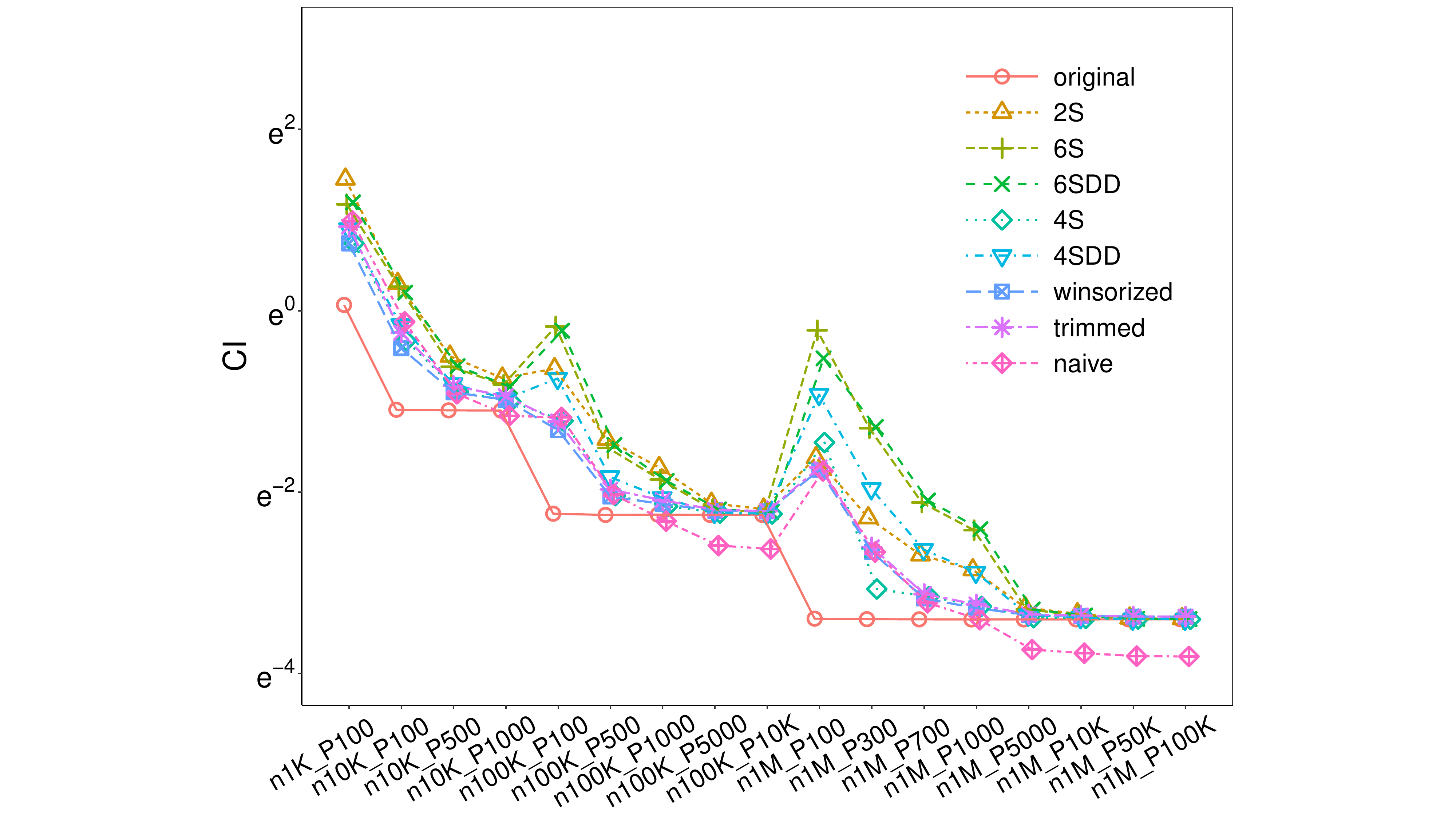}
\includegraphics[width=0.24\textwidth, trim={2.2in 0 2.2in 0},clip] {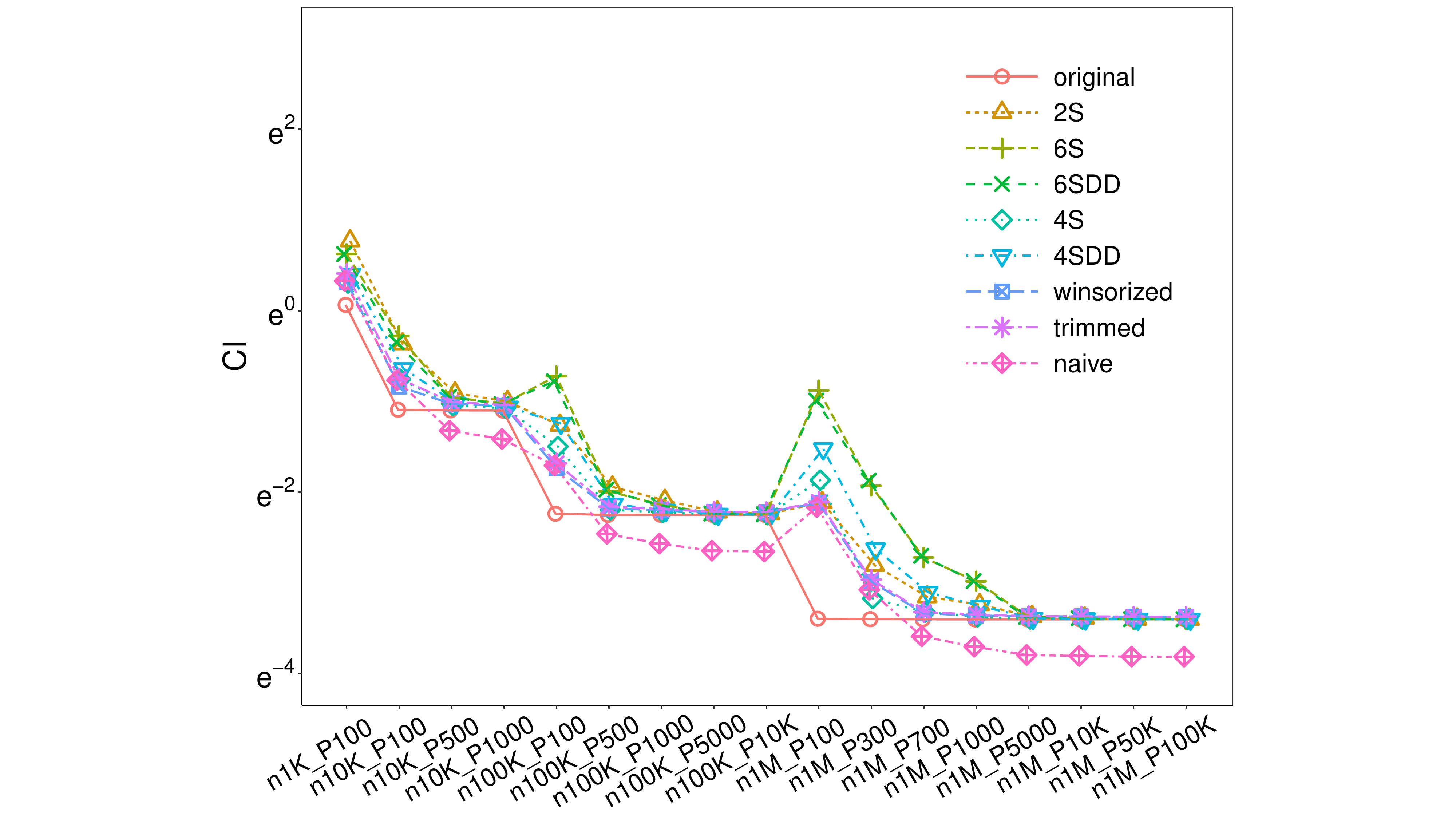}
\includegraphics[width=0.24\textwidth, trim={2.2in 0 2.2in 0},clip] {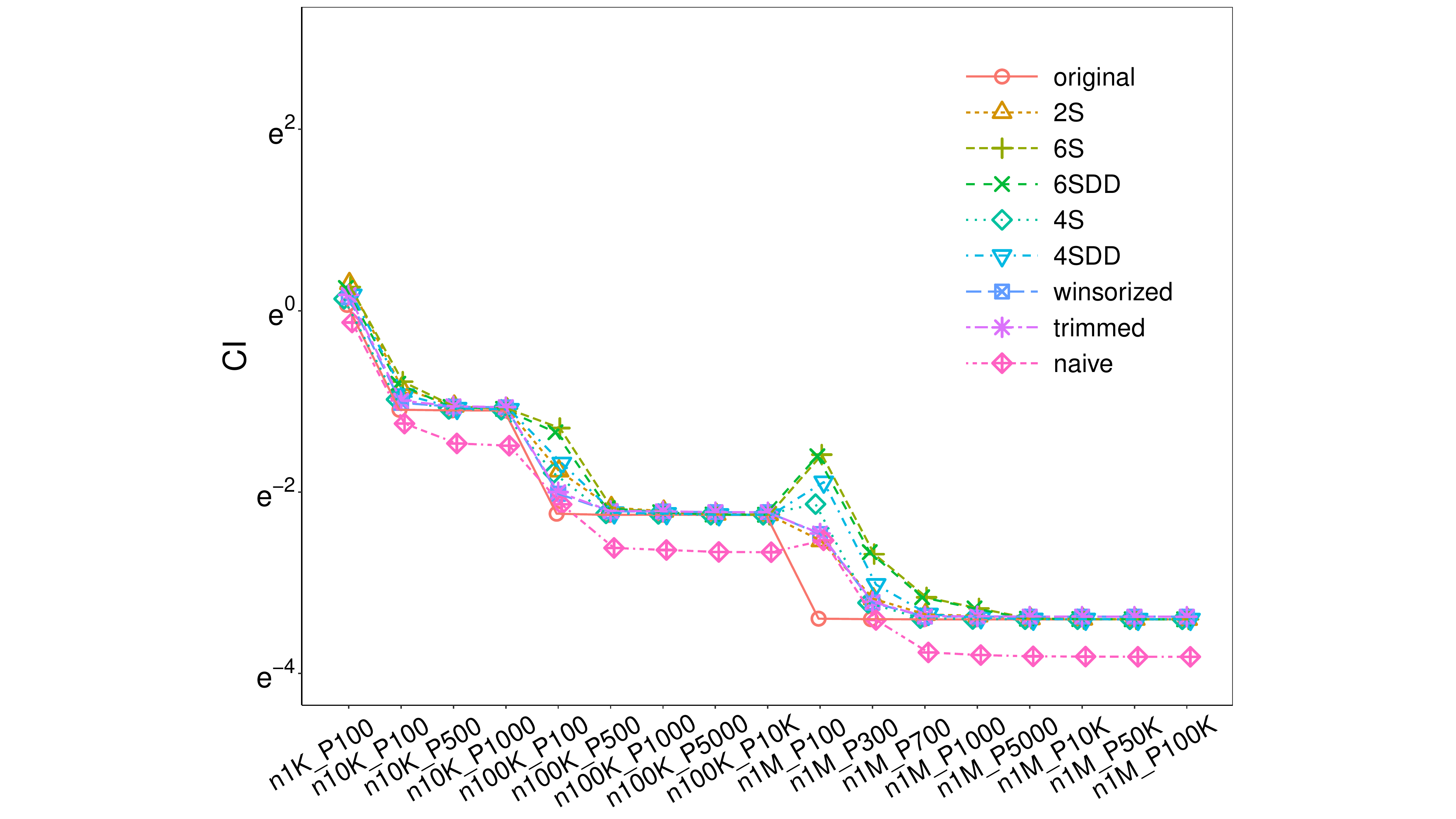}\\
\includegraphics[width=0.24\textwidth, trim={2.2in 0 2.2in 0},clip] {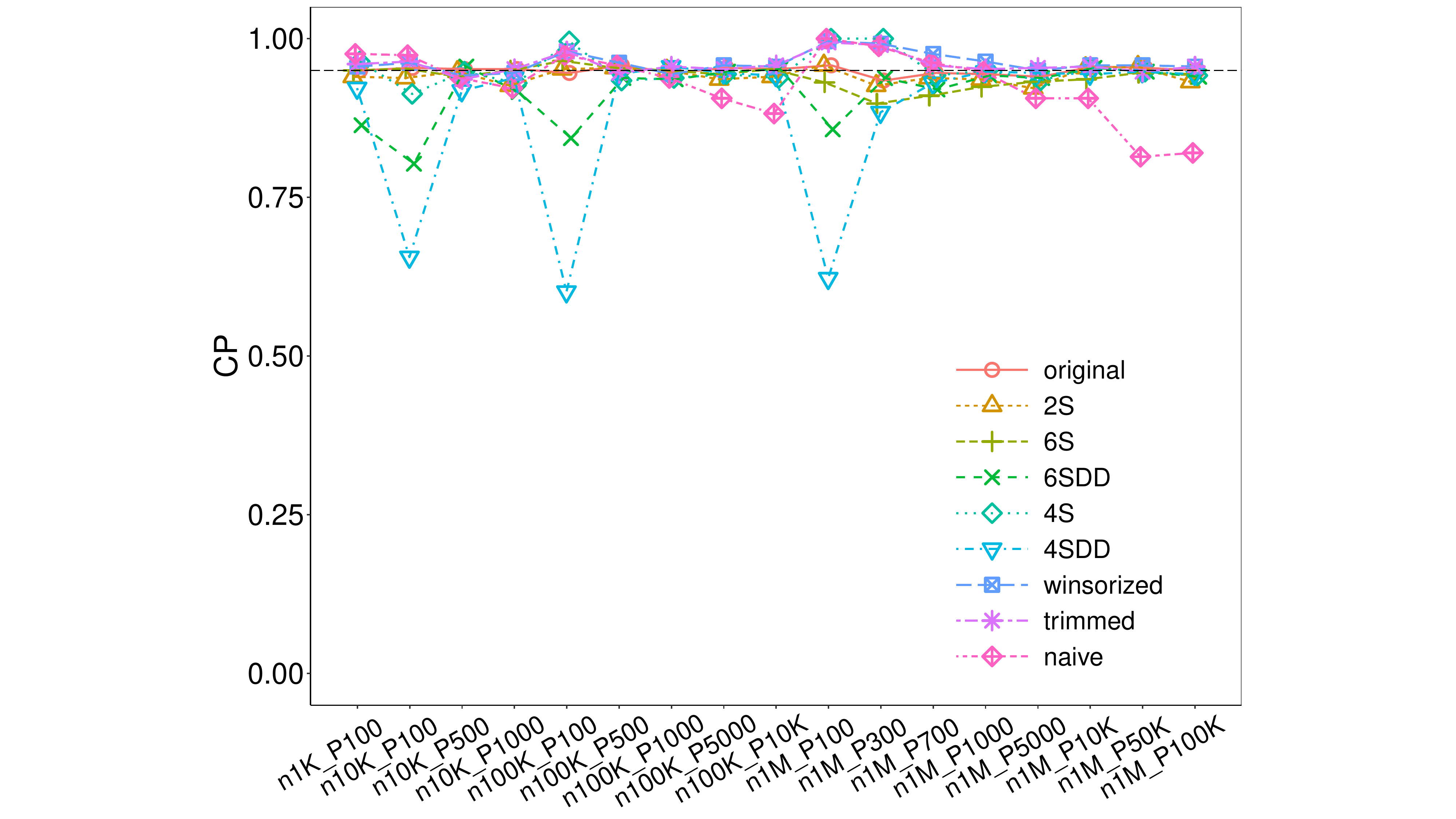}
\includegraphics[width=0.24\textwidth, trim={2.2in 0 2.2in 0},clip] {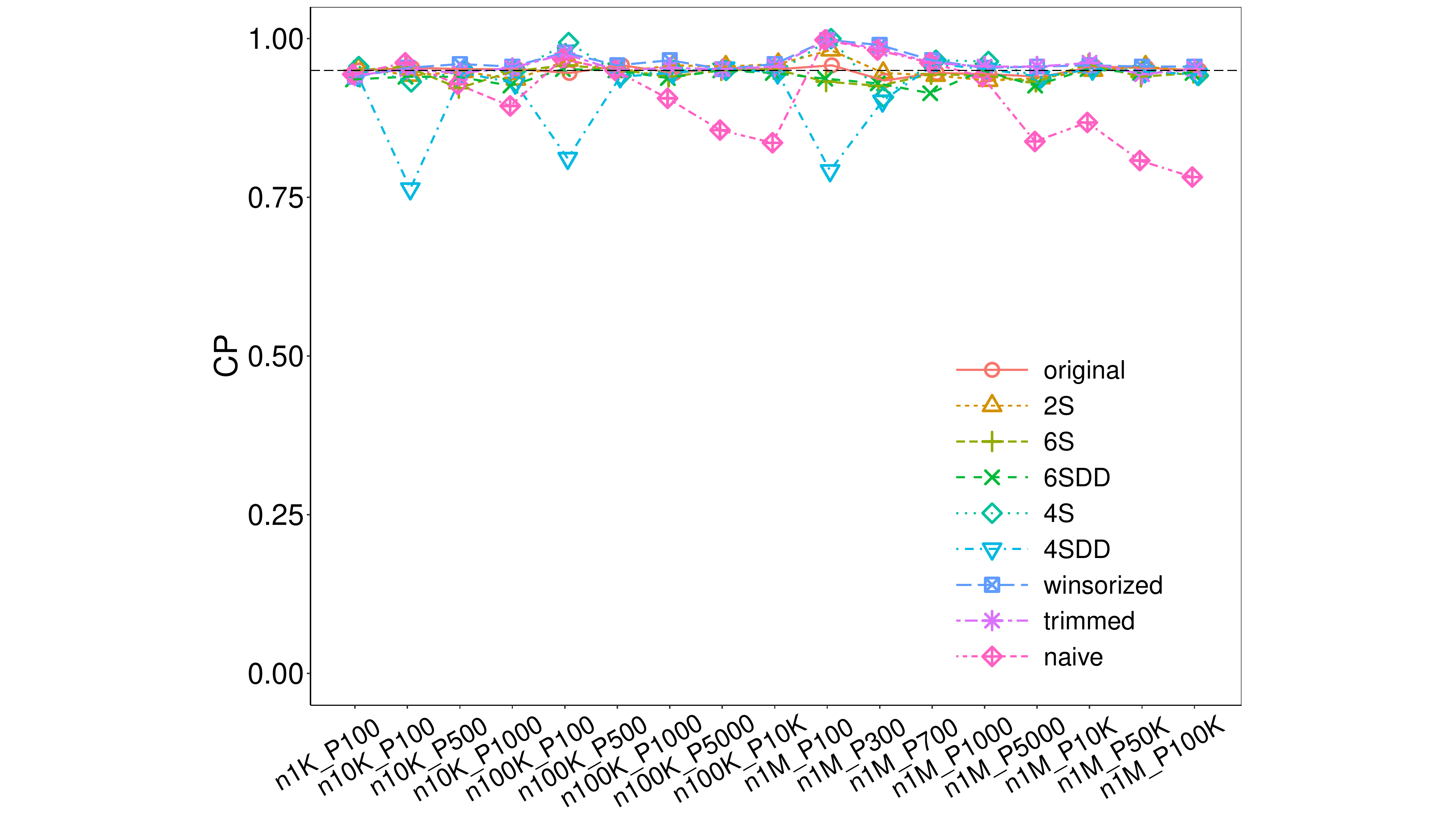}
\includegraphics[width=0.24\textwidth, trim={2.2in 0 2.2in 0},clip] {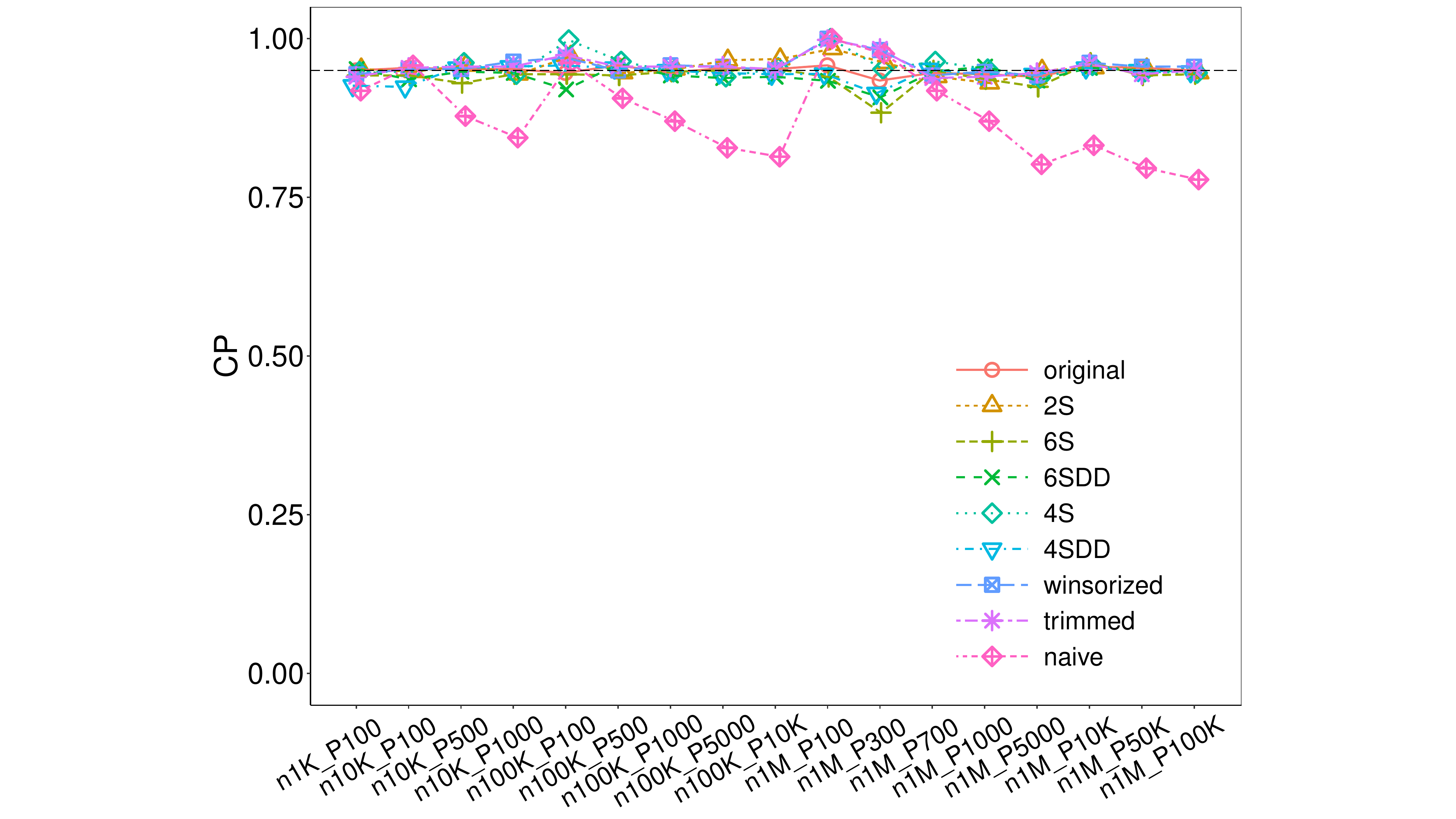}
\includegraphics[width=0.24\textwidth, trim={2.2in 0 2.2in 0},clip] {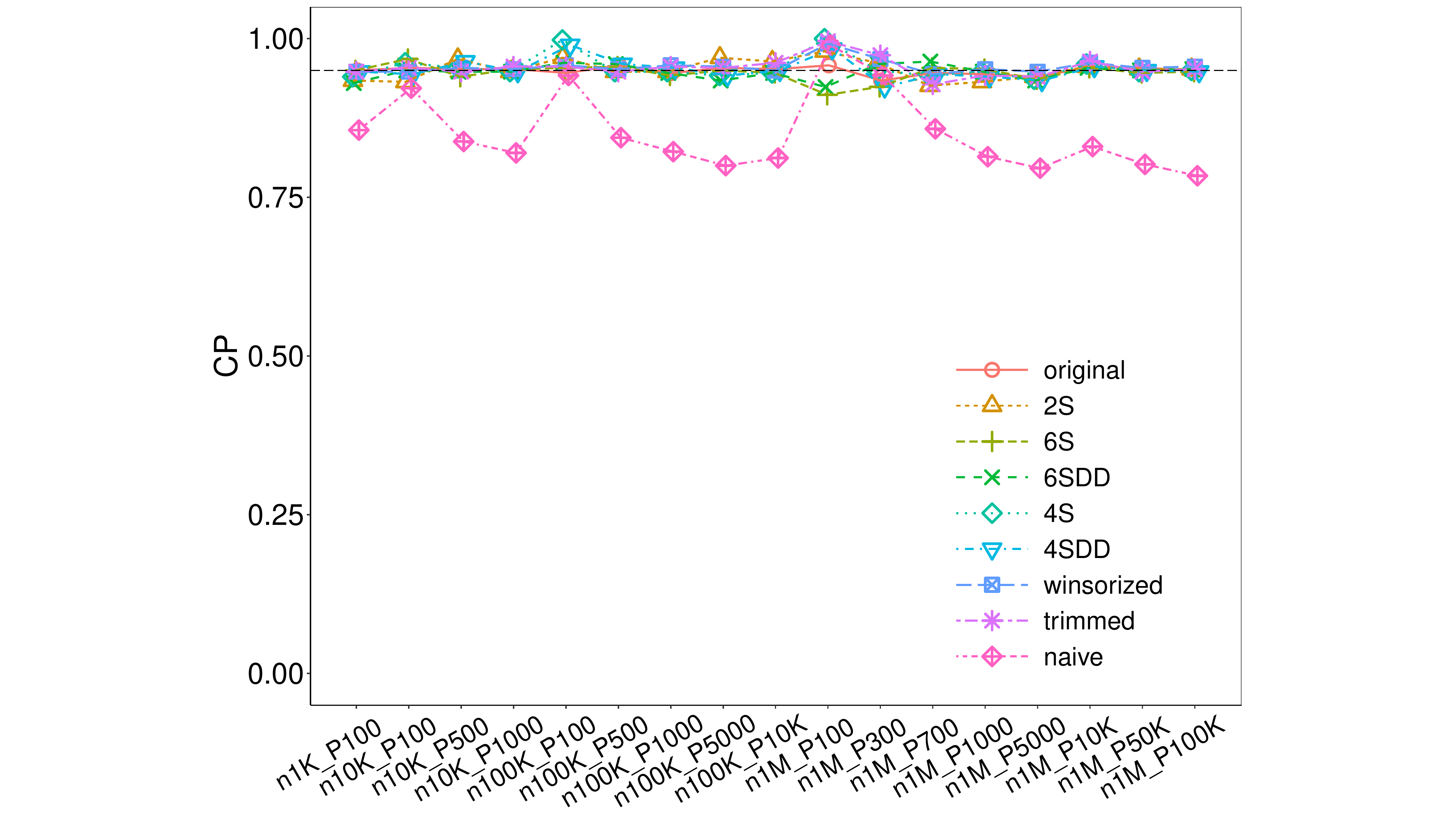}
\includegraphics[width=0.24\textwidth, trim={2.2in 0 2.2in 0},clip] {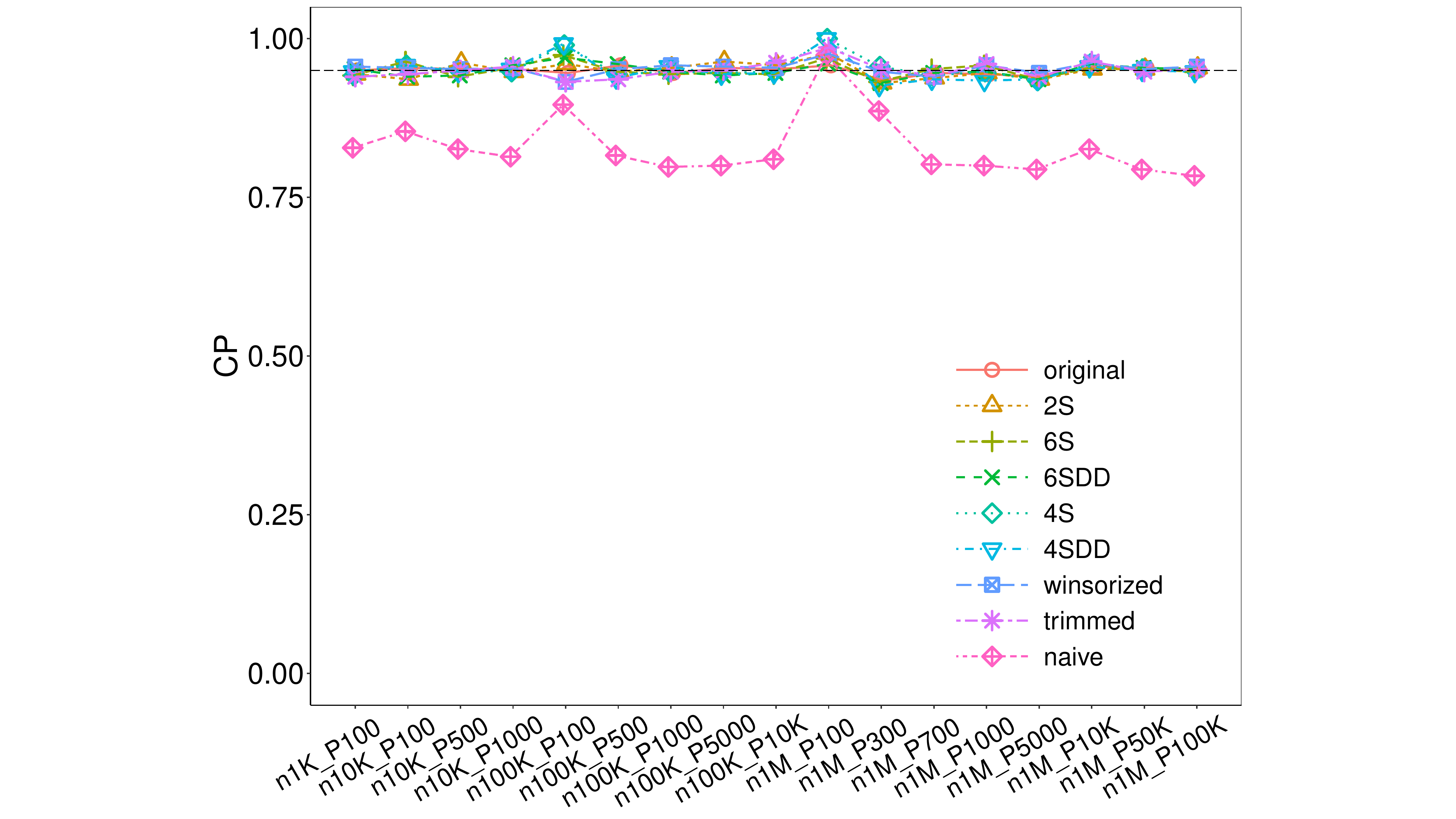}\\
\includegraphics[width=0.24\textwidth, trim={2.2in 0 2.2in 0},clip] {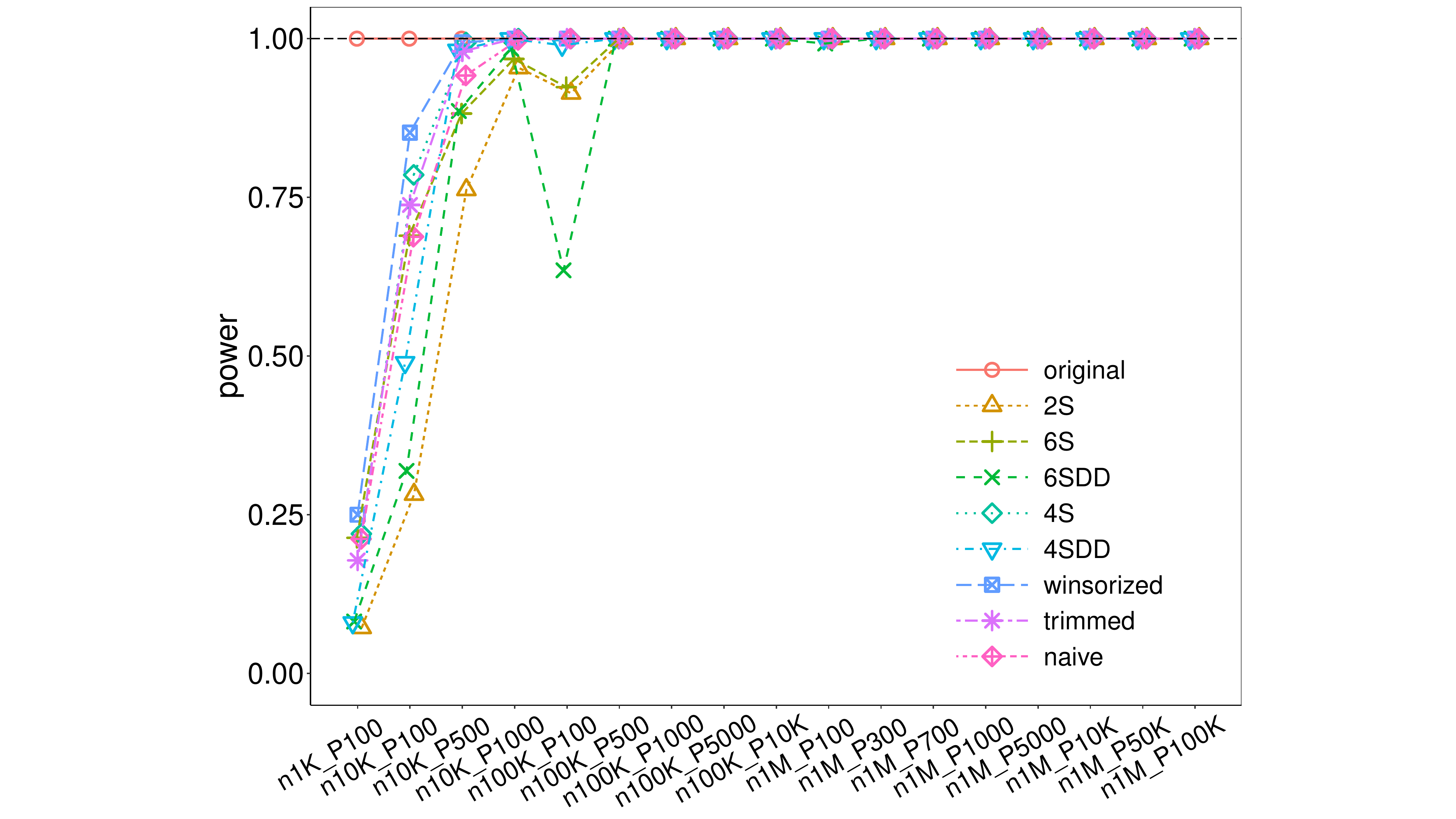}
\includegraphics[width=0.24\textwidth, trim={2.2in 0 2.2in 0},clip] {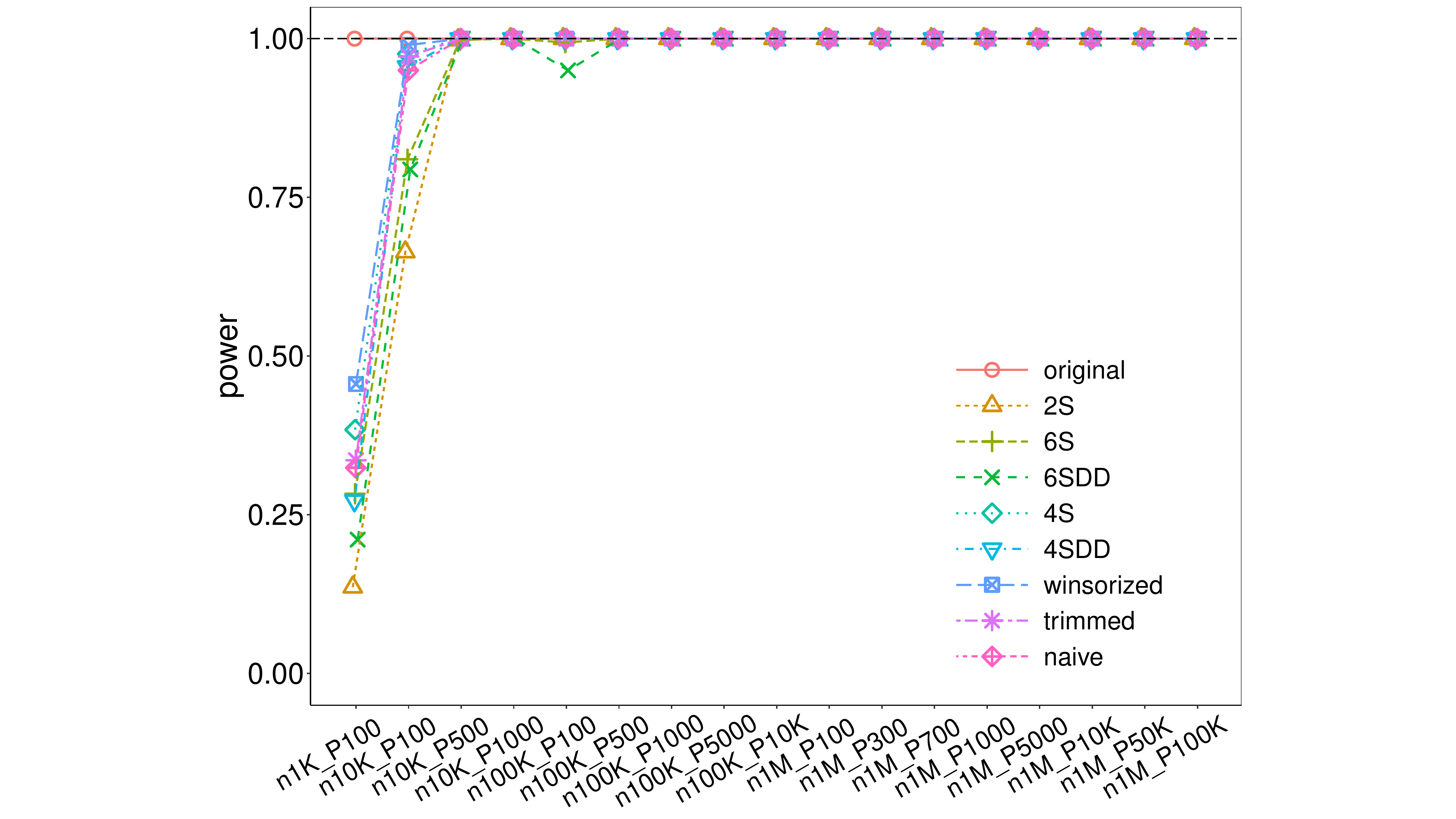}
\includegraphics[width=0.24\textwidth, trim={2.2in 0 2.2in 0},clip] {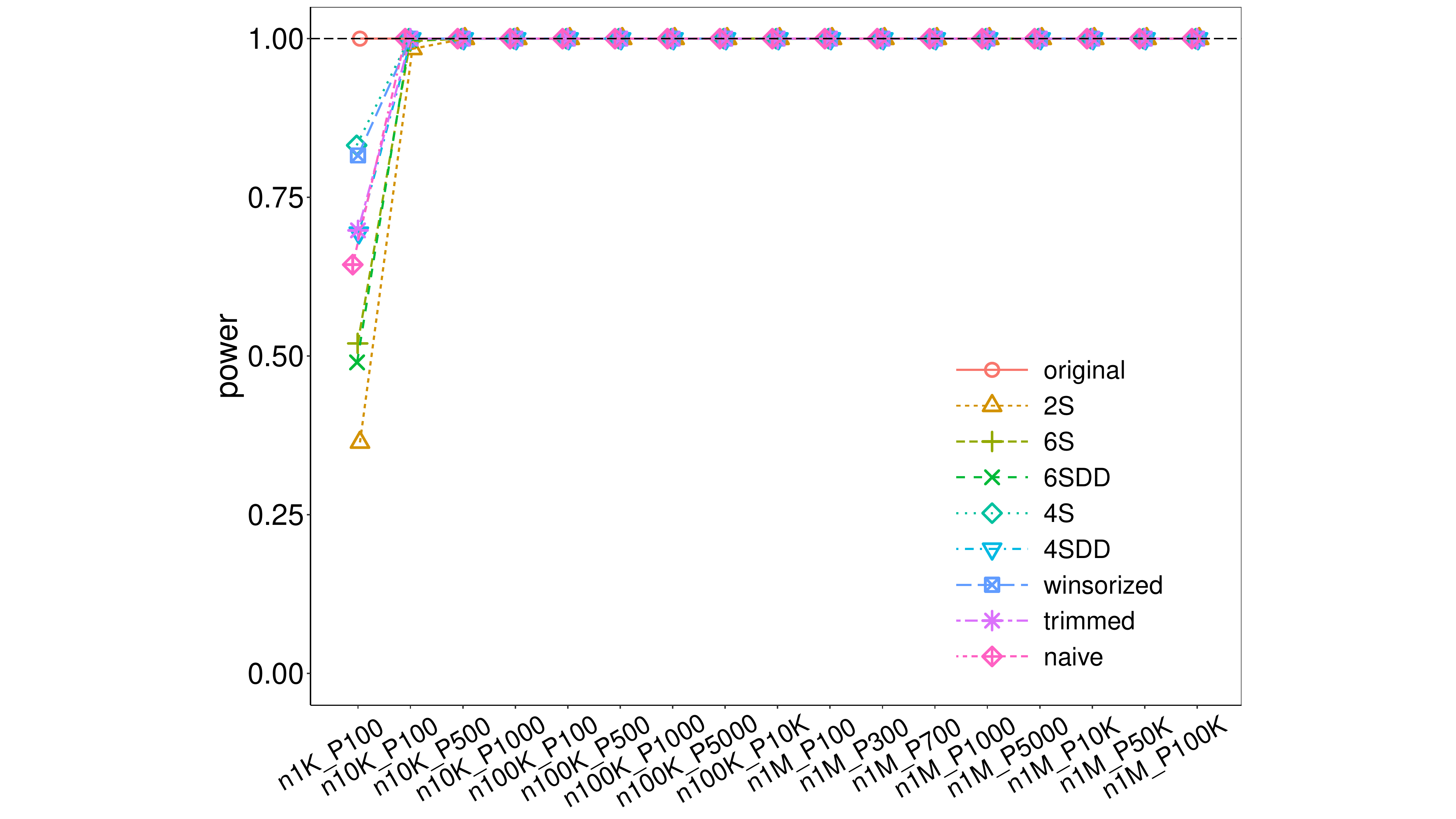} 
\includegraphics[width=0.24\textwidth, trim={2.2in 0 2.2in 0},clip] {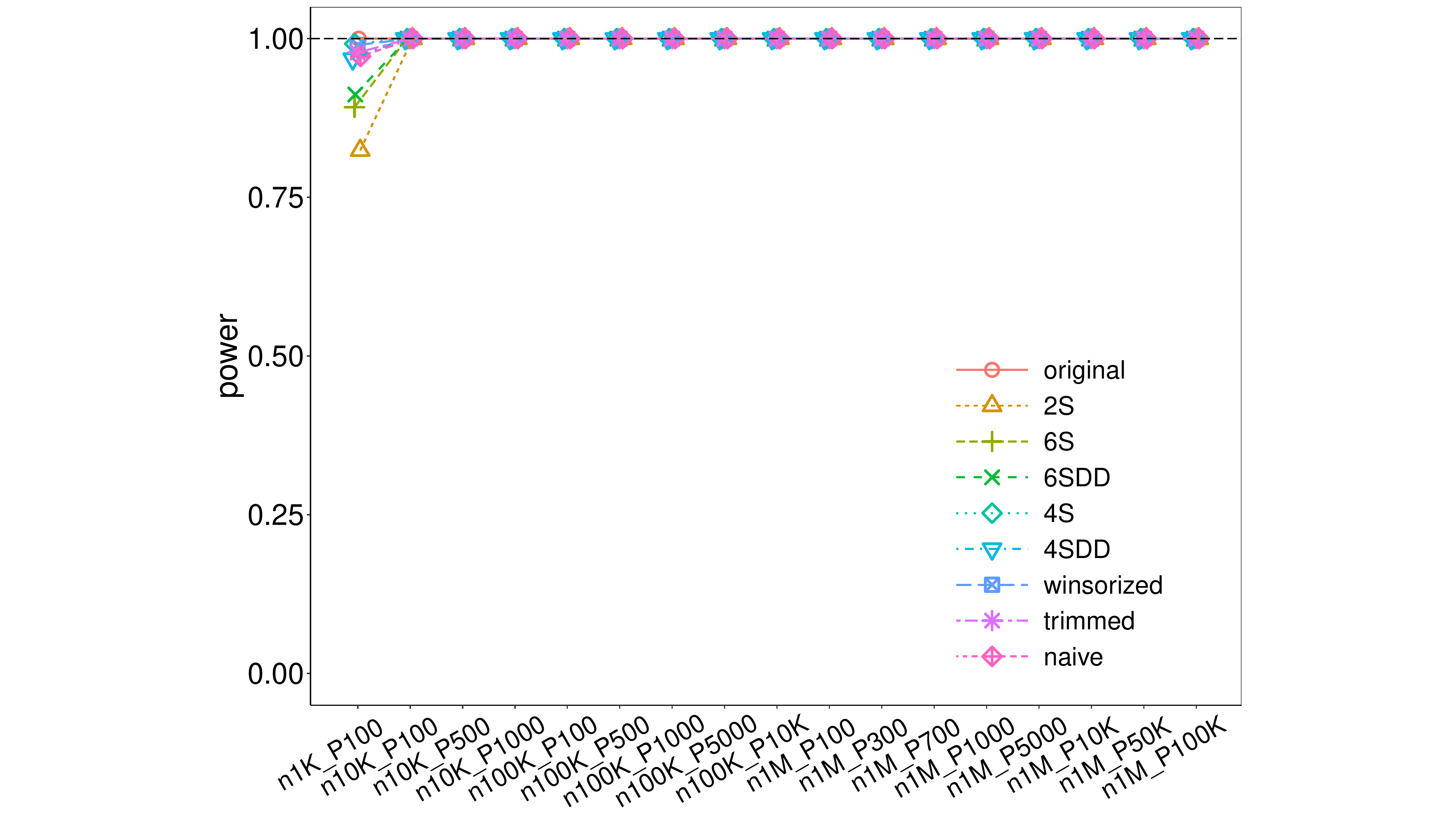}
\includegraphics[width=0.24\textwidth, trim={2.2in 0 2.2in 0},clip] {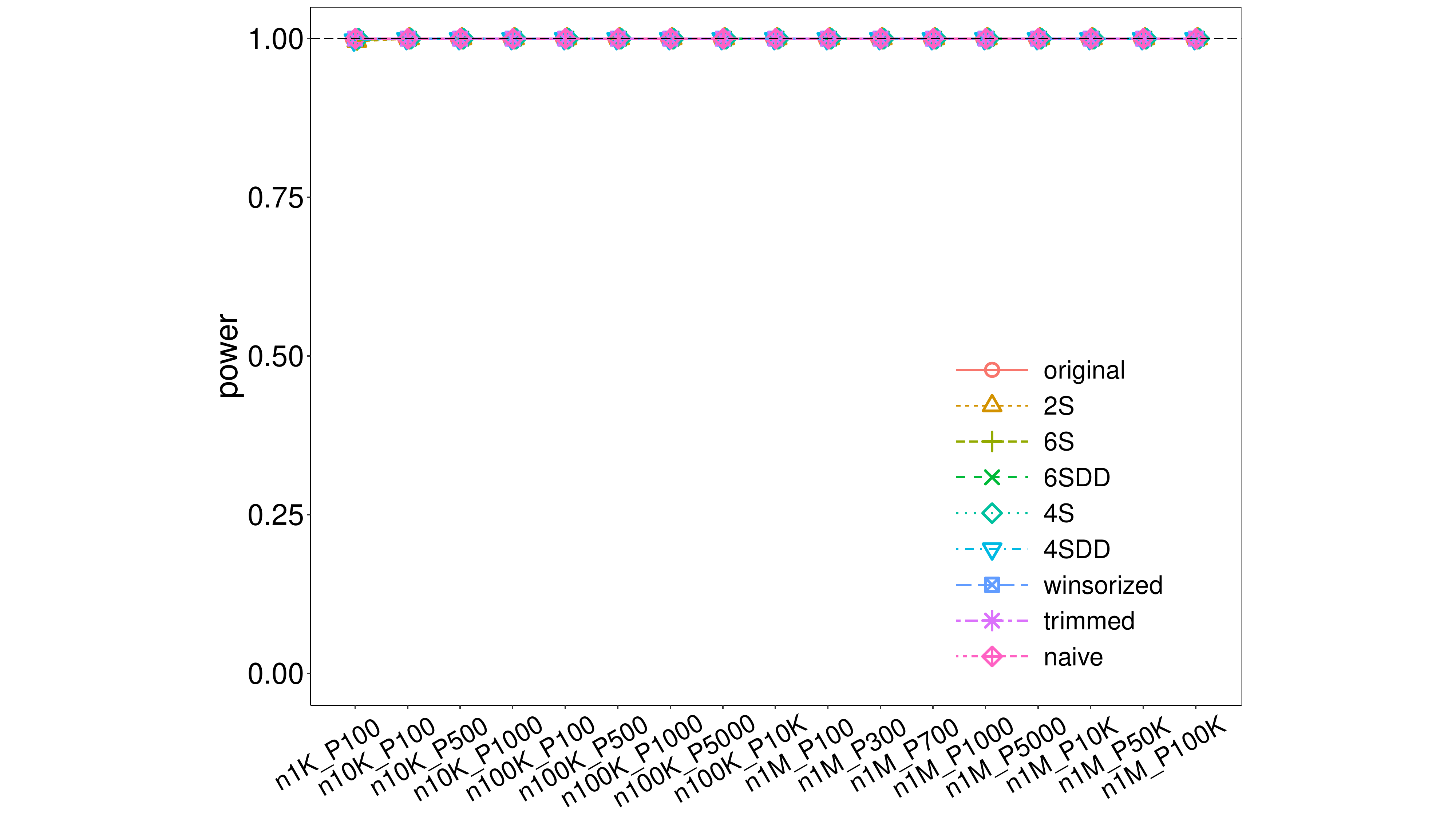}\\
\caption{Gaussian data; $\rho$-zCDP; $\theta\ne0$ and $\alpha=\beta$}
\label{fig:1szCDP}
\end{figure}

\end{landscape}

\begin{landscape}
\subsection*{Gaussian, $\theta=0$ and $\alpha\ne\beta$}
\begin{figure}[!htb]
\hspace{0.6in}$\epsilon=0.5$\hspace{1.3in}$\epsilon=1$\hspace{1.4in}$\epsilon=2$
\hspace{1.4in}$\epsilon=5$\hspace{1.4in}$\epsilon=50$\\
\includegraphics[width=0.26\textwidth, trim={2.2in 0 2.2in 0},clip] {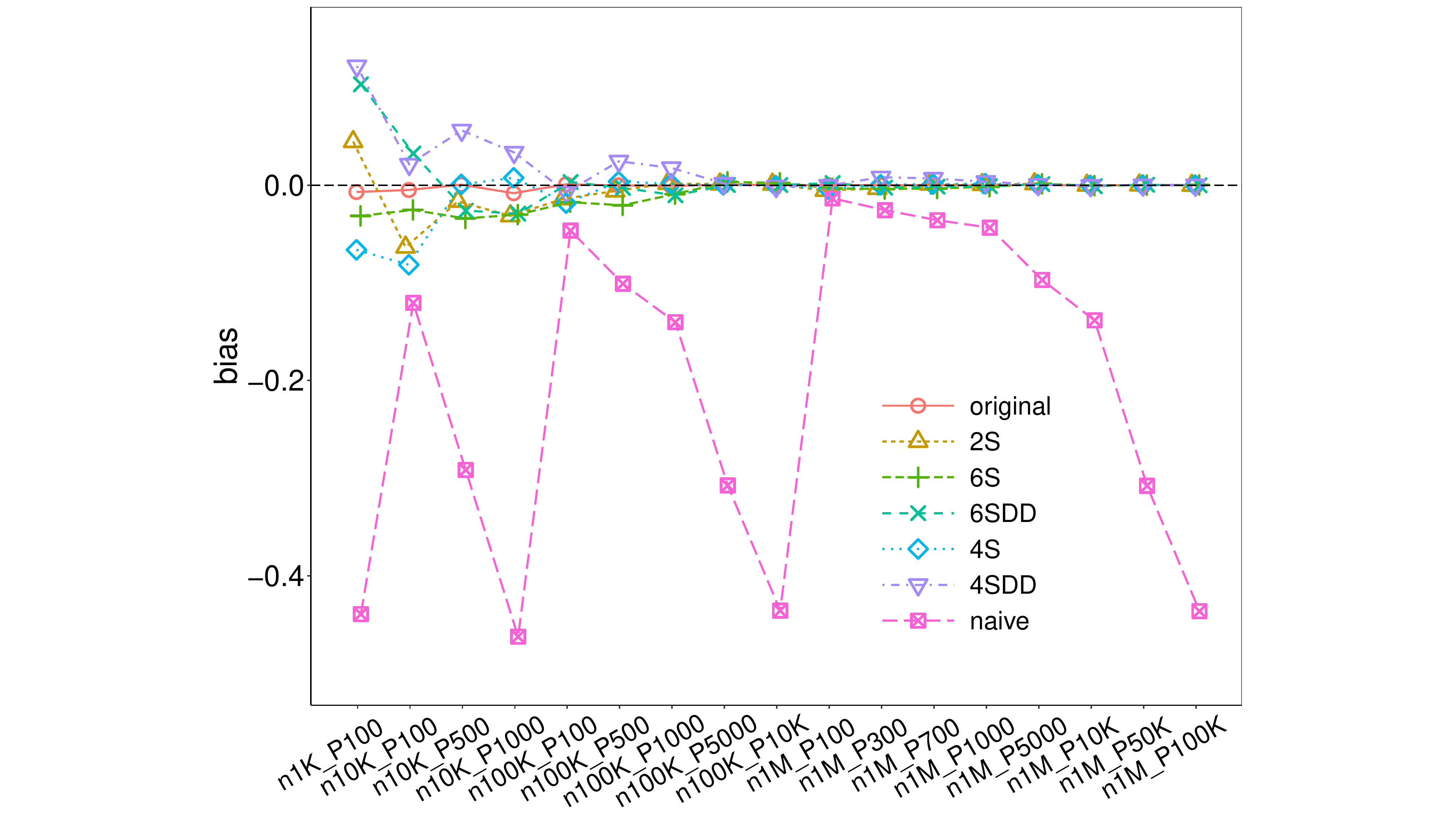}
\includegraphics[width=0.26\textwidth, trim={2.2in 0 2.2in 0},clip] {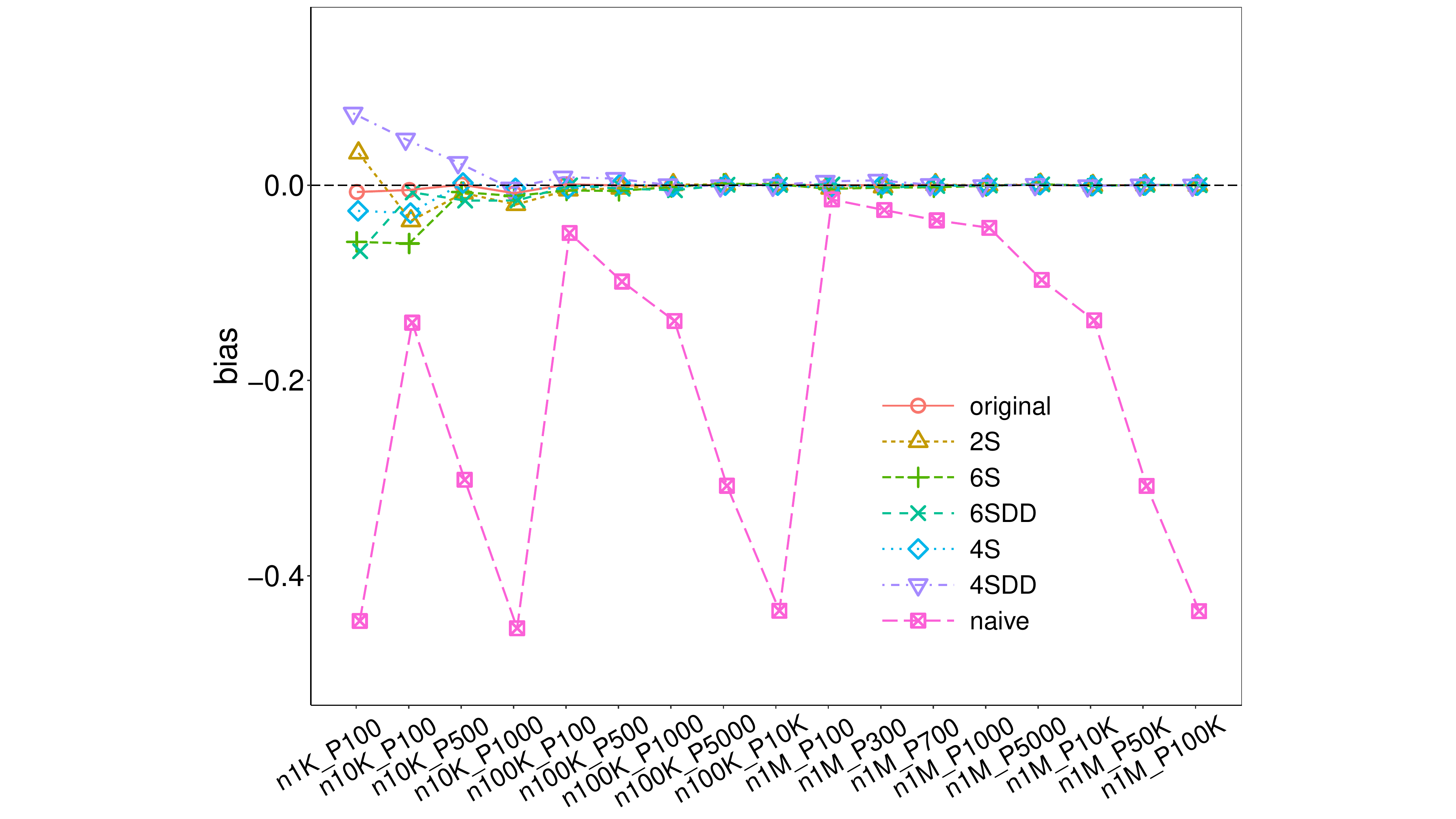}
\includegraphics[width=0.26\textwidth, trim={2.2in 0 2.2in 0},clip] {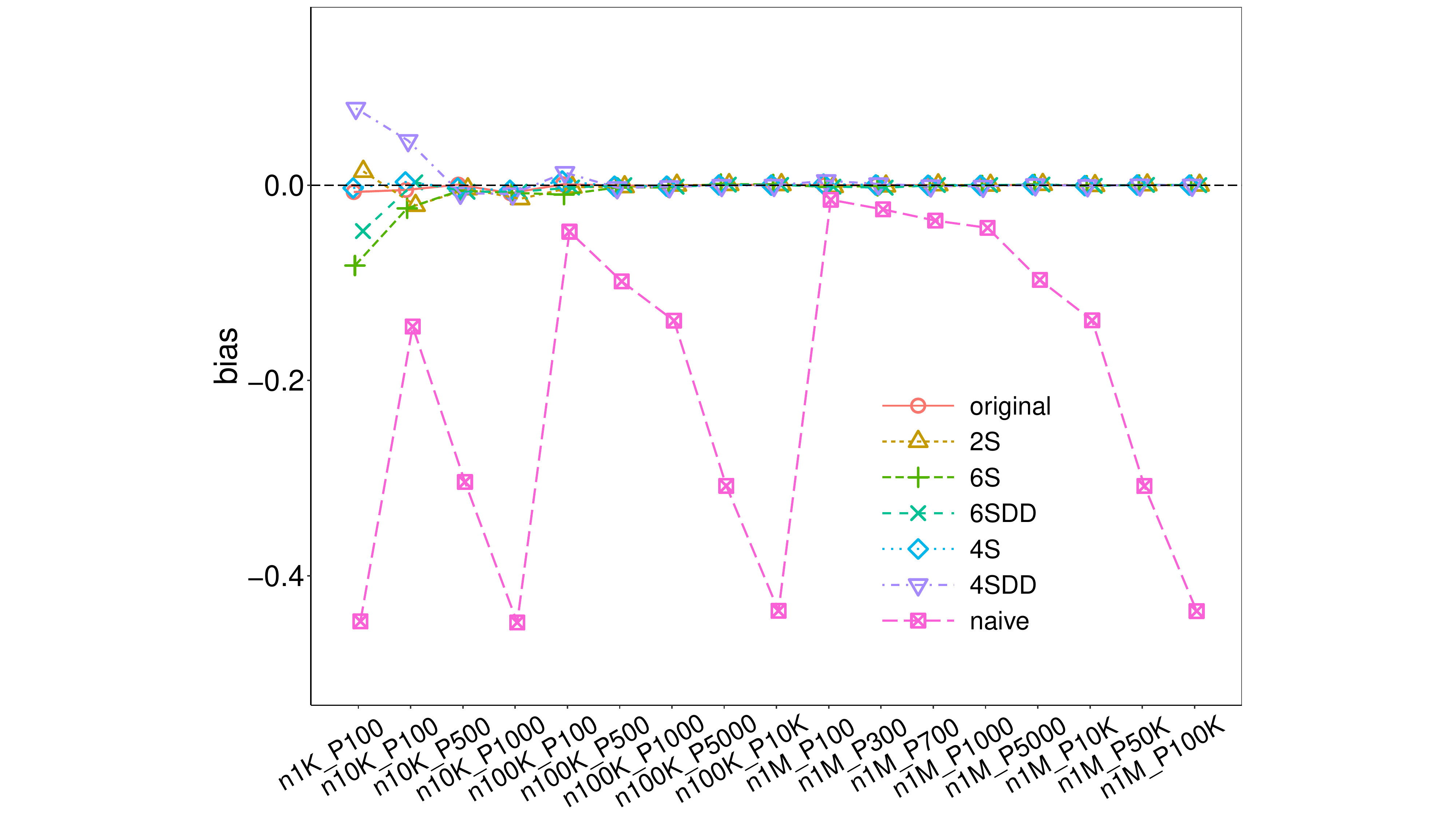}
\includegraphics[width=0.26\textwidth, trim={2.2in 0 2.2in 0},clip] {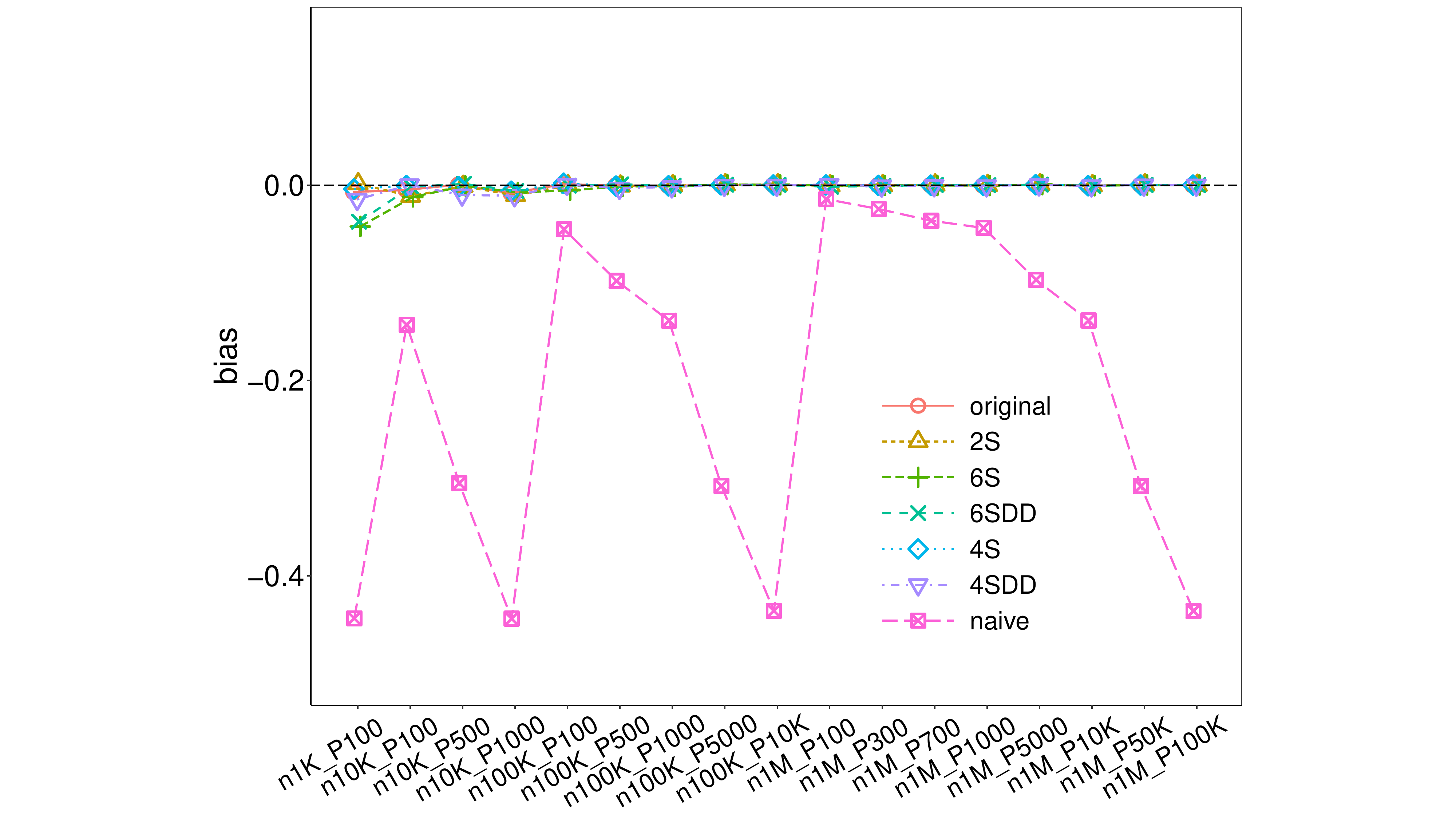}
\includegraphics[width=0.26\textwidth, trim={2.2in 0 2.2in 0},clip] {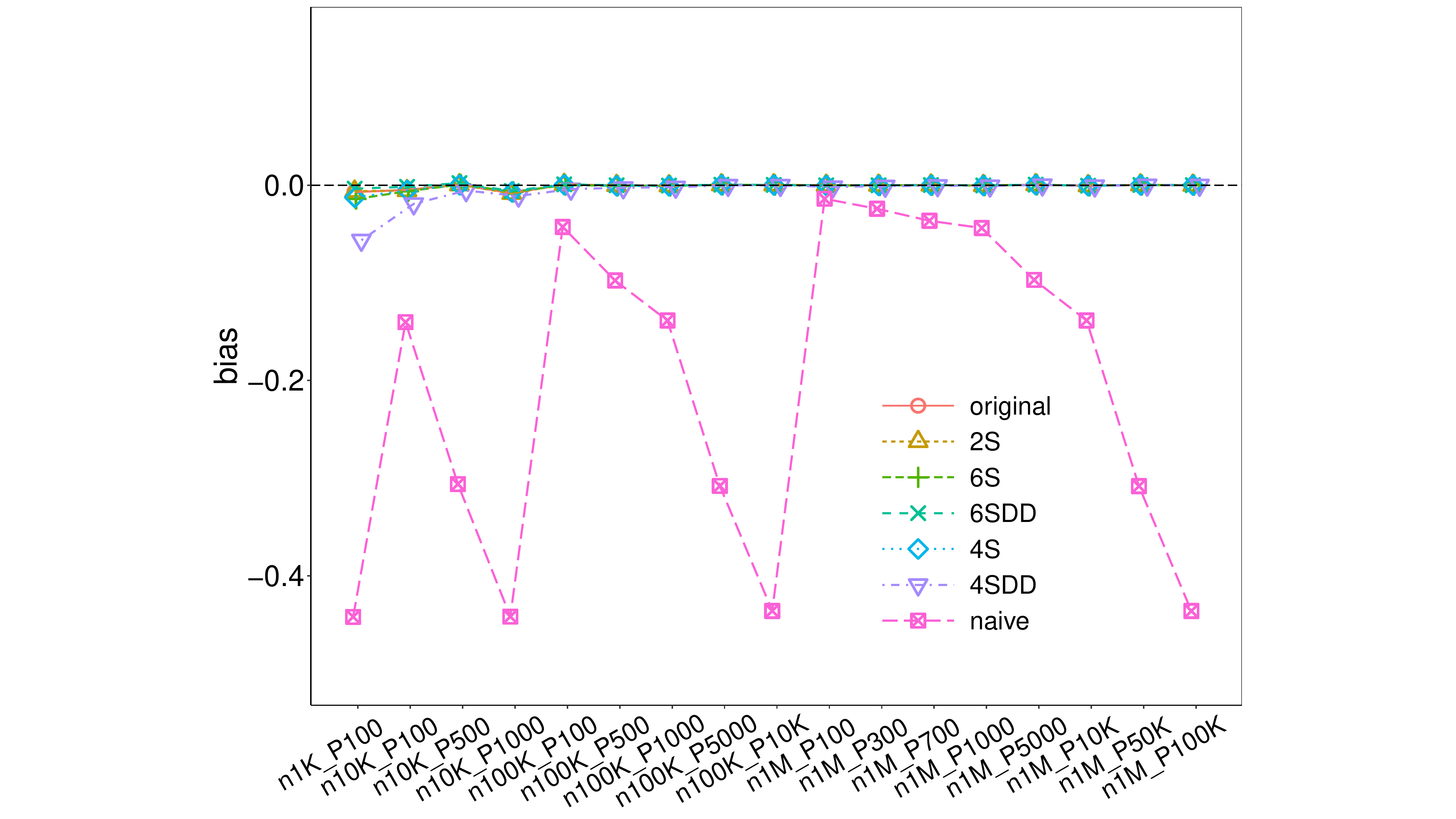}\\
\includegraphics[width=0.26\textwidth, trim={2.2in 0 2.2in 0},clip] {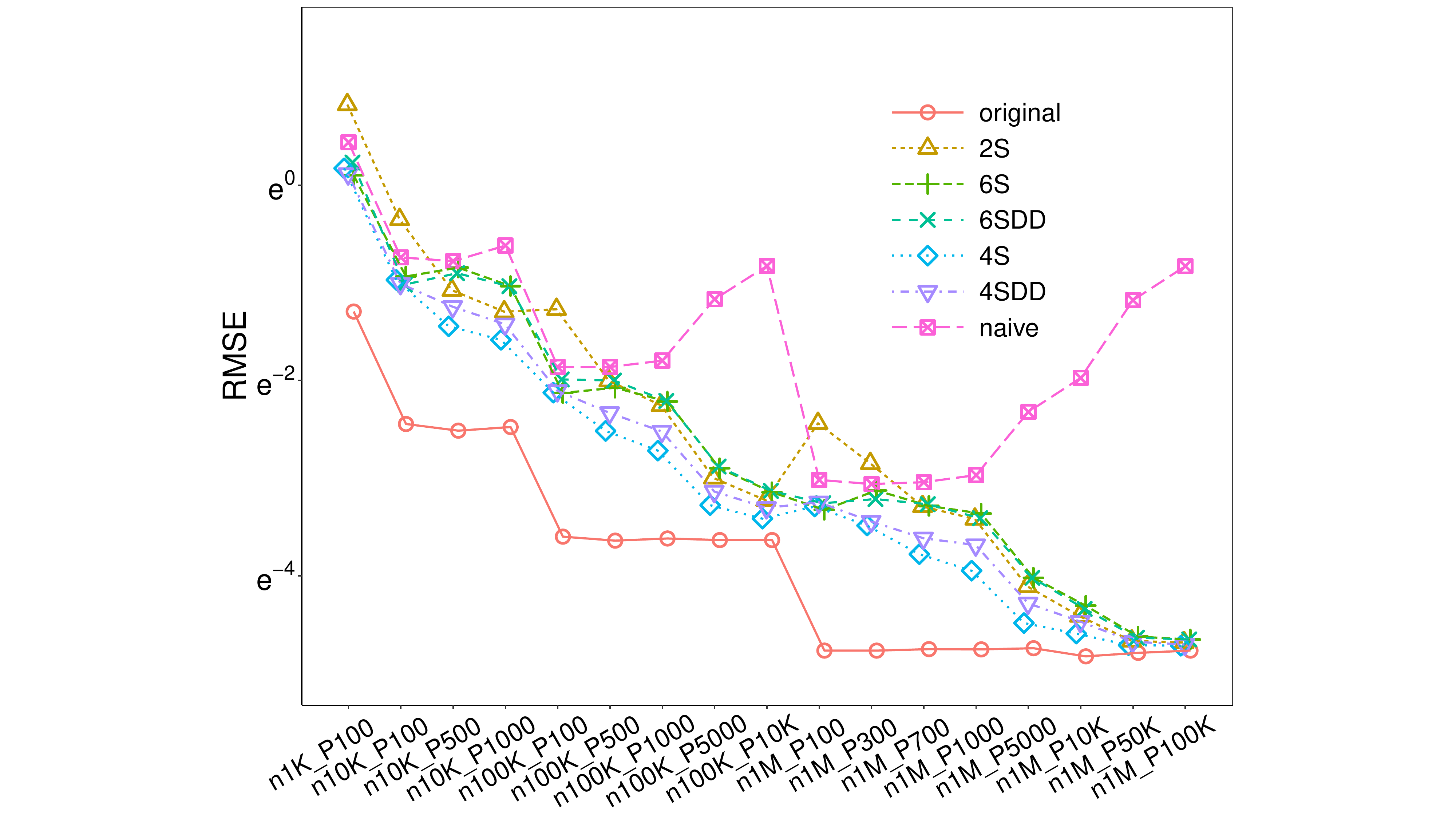}
\includegraphics[width=0.26\textwidth, trim={2.2in 0 2.2in 0},clip] {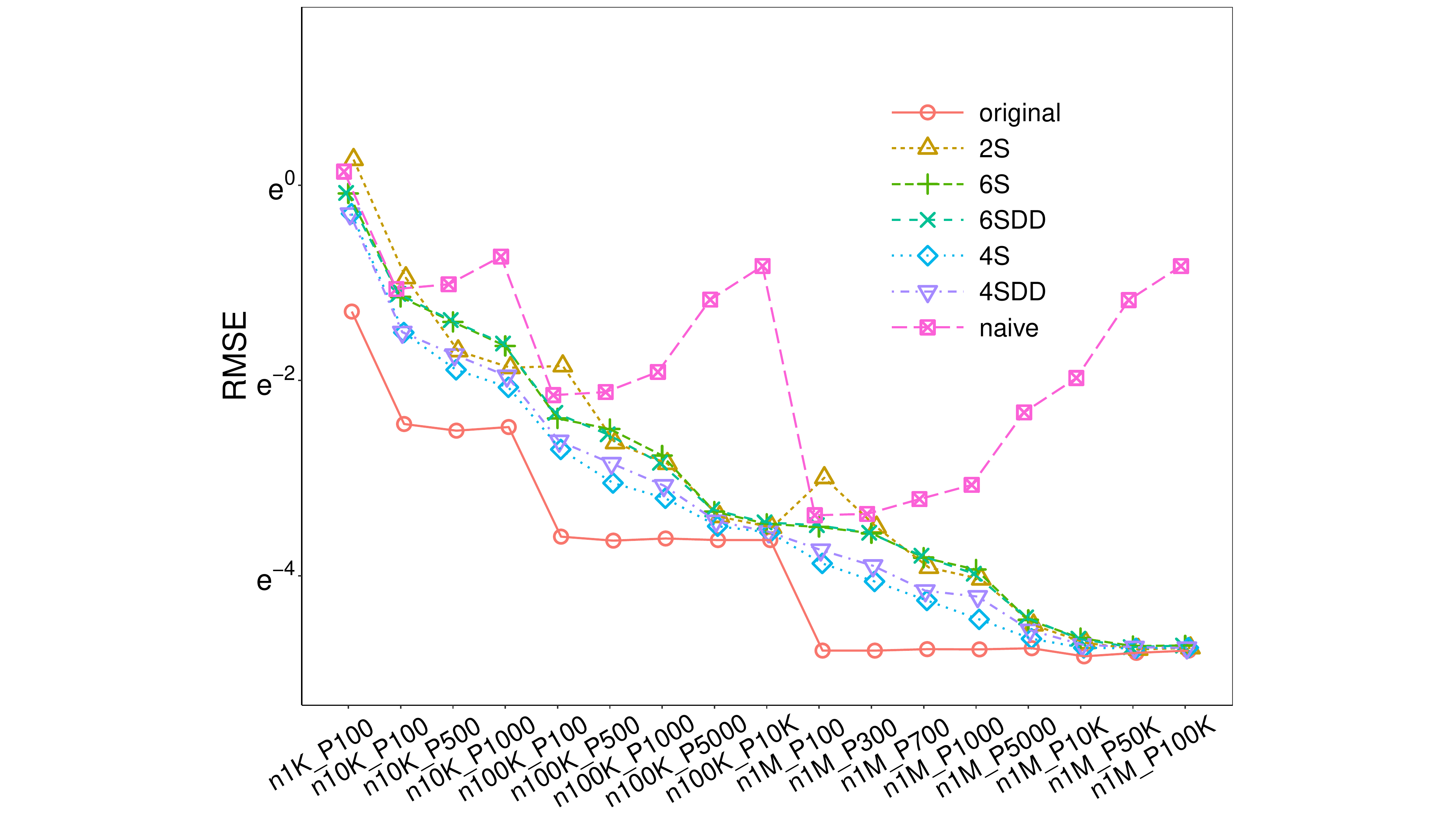}
\includegraphics[width=0.26\textwidth, trim={2.2in 0 2.2in 0},clip] {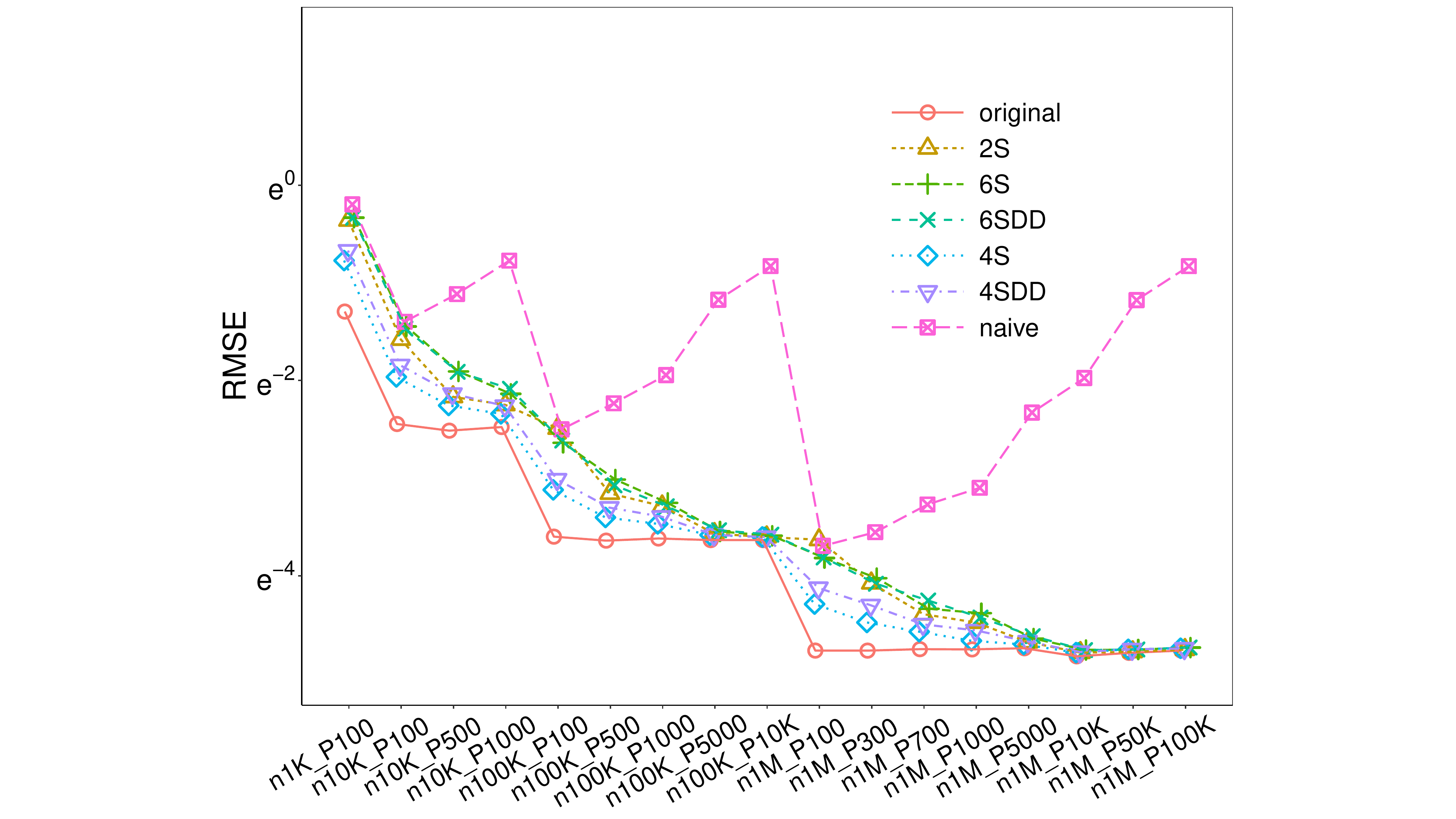}
\includegraphics[width=0.26\textwidth, trim={2.2in 0 2.2in 0},clip] {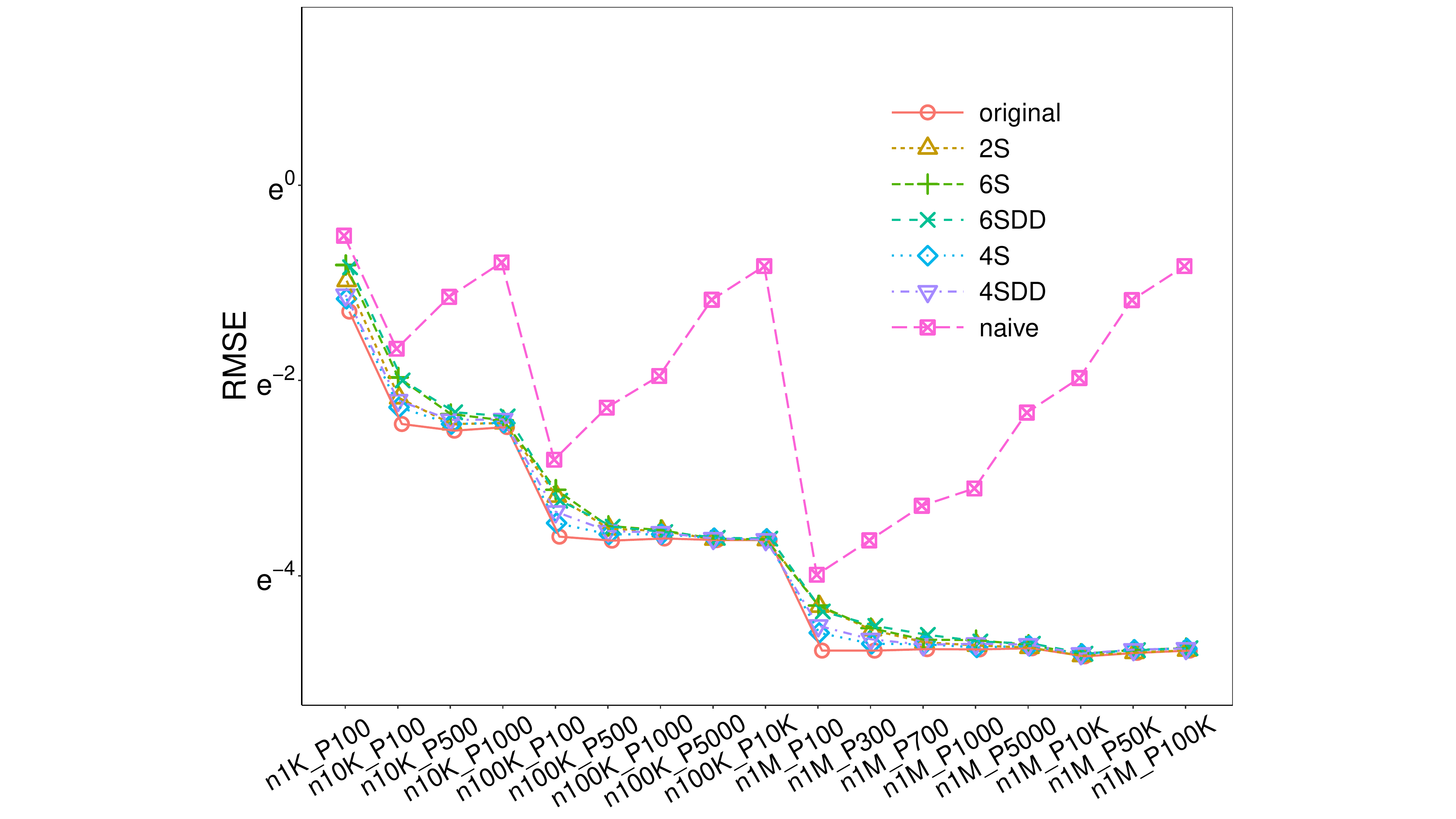}
\includegraphics[width=0.26\textwidth, trim={2.2in 0 2.2in 0},clip] {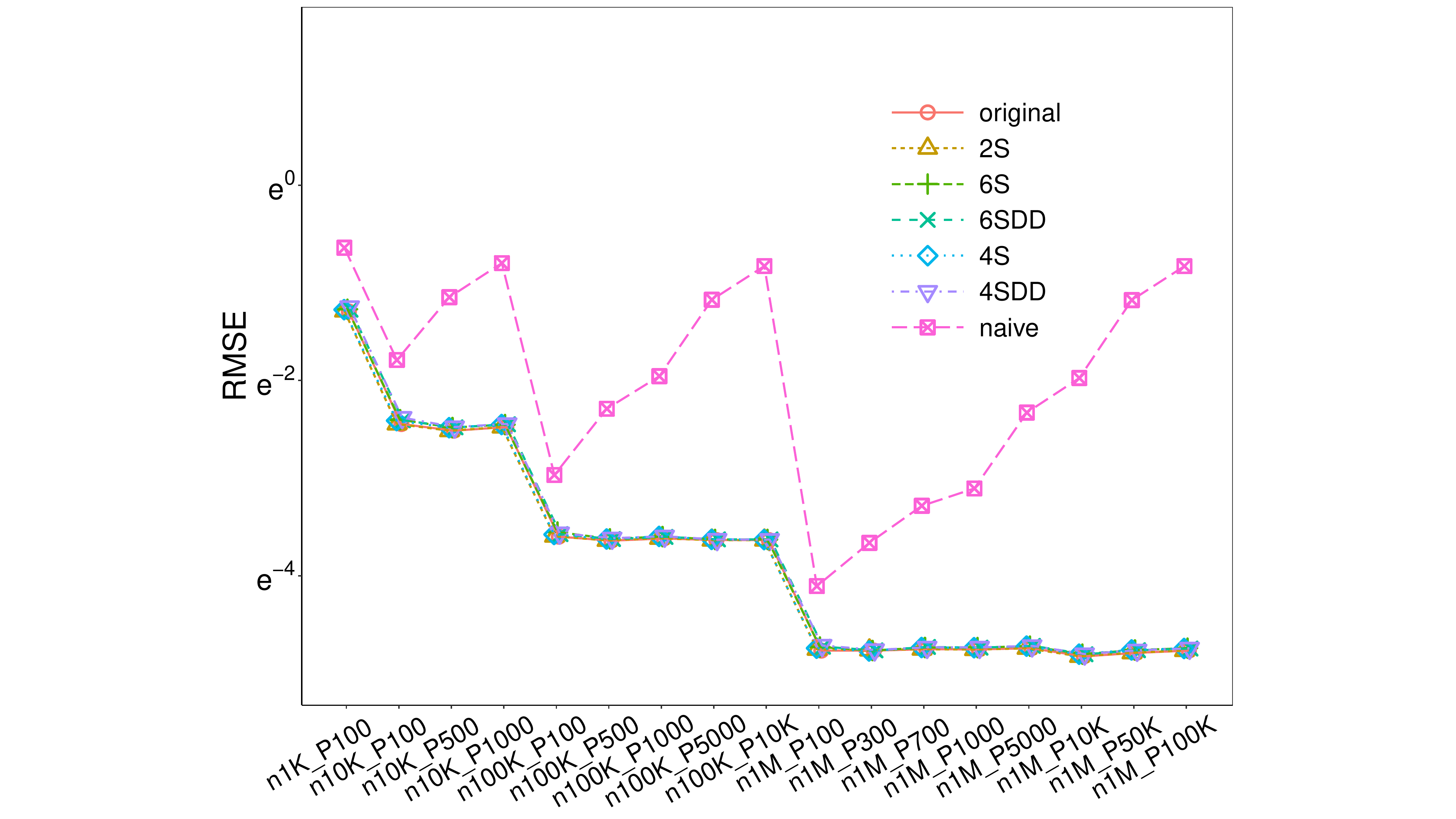}\\
\includegraphics[width=0.26\textwidth, trim={2.2in 0 2.2in 0},clip] {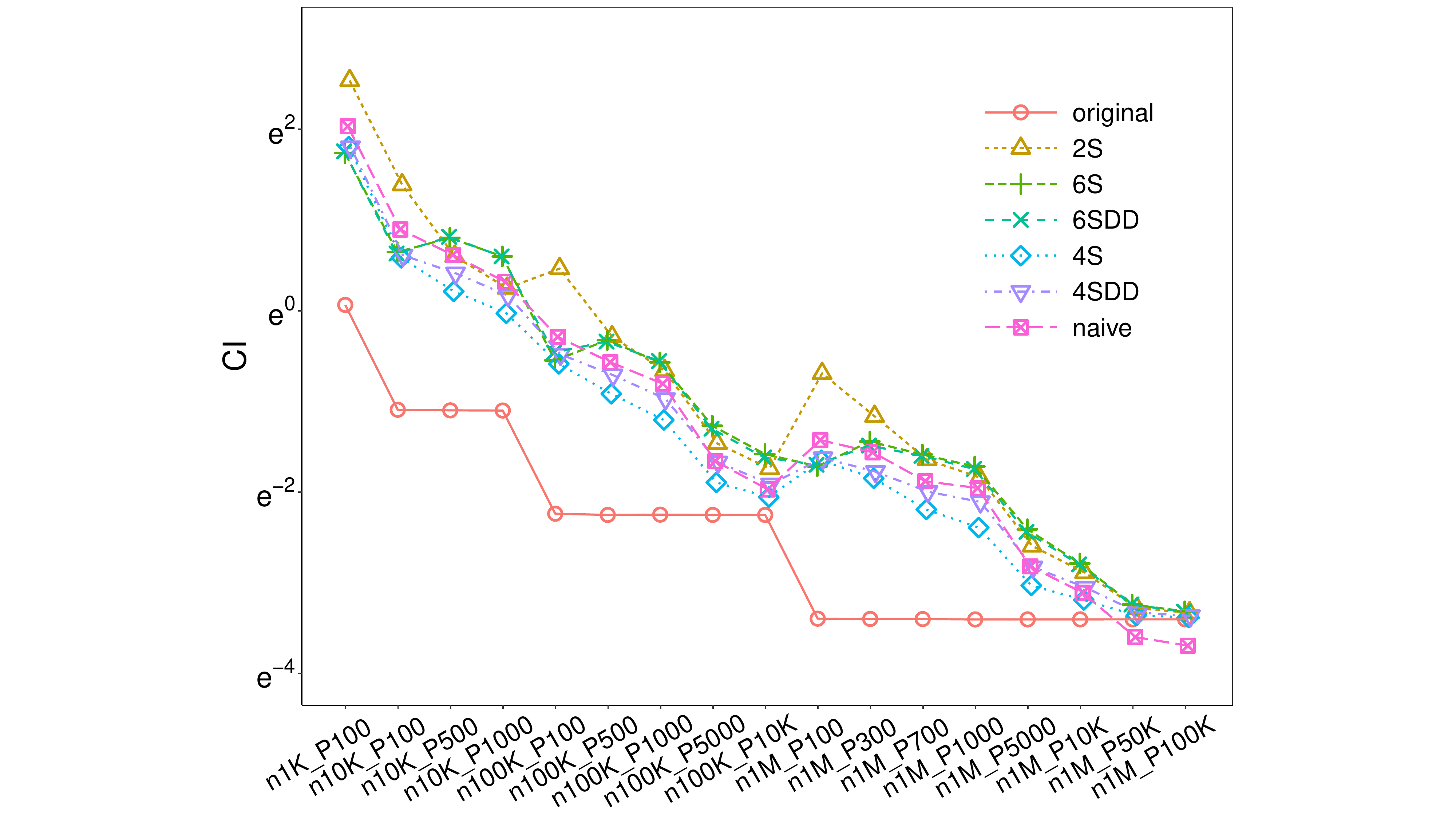}
\includegraphics[width=0.26\textwidth, trim={2.2in 0 2.2in 0},clip] {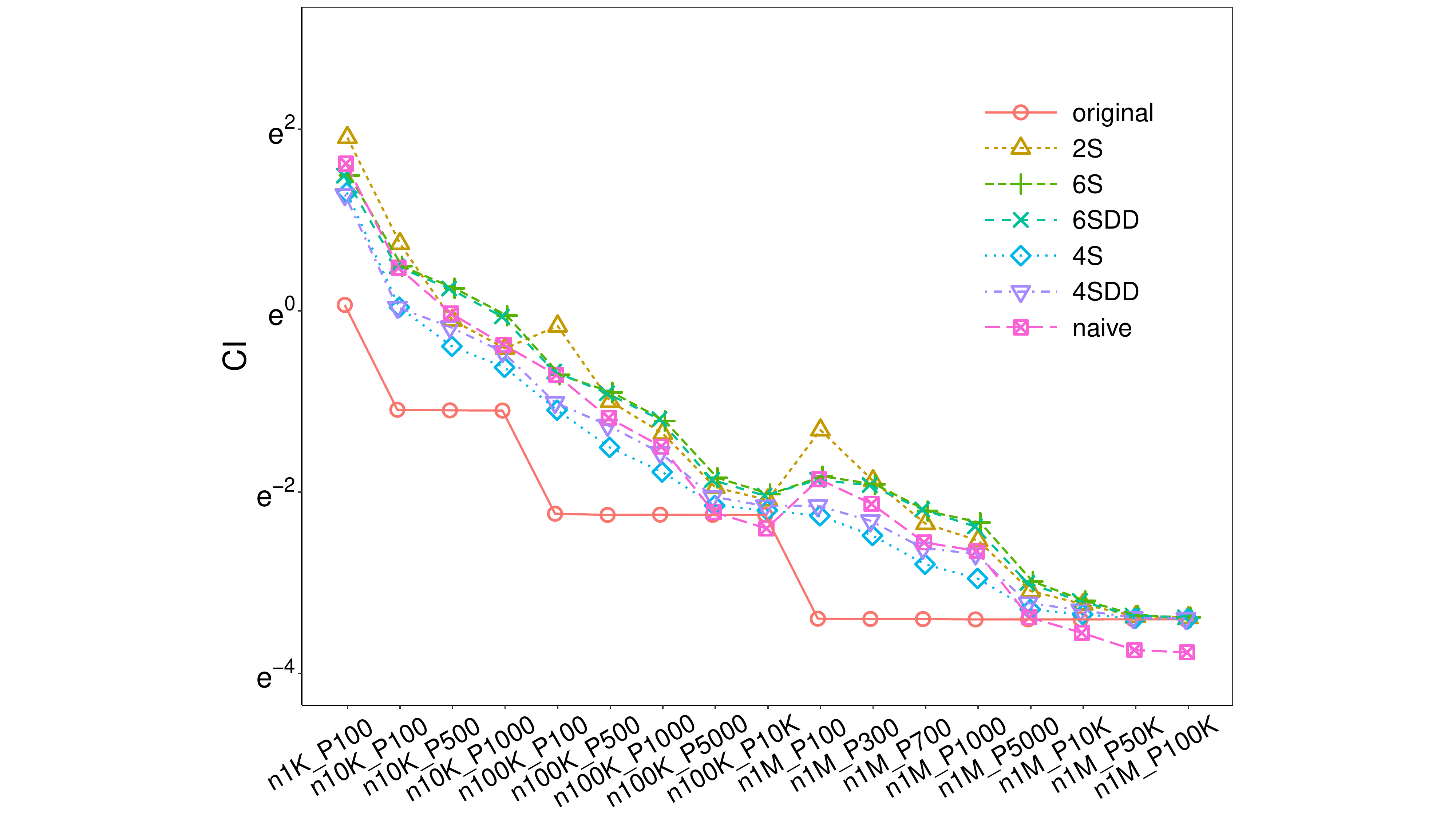}
\includegraphics[width=0.26\textwidth, trim={2.2in 0 2.2in 0},clip] {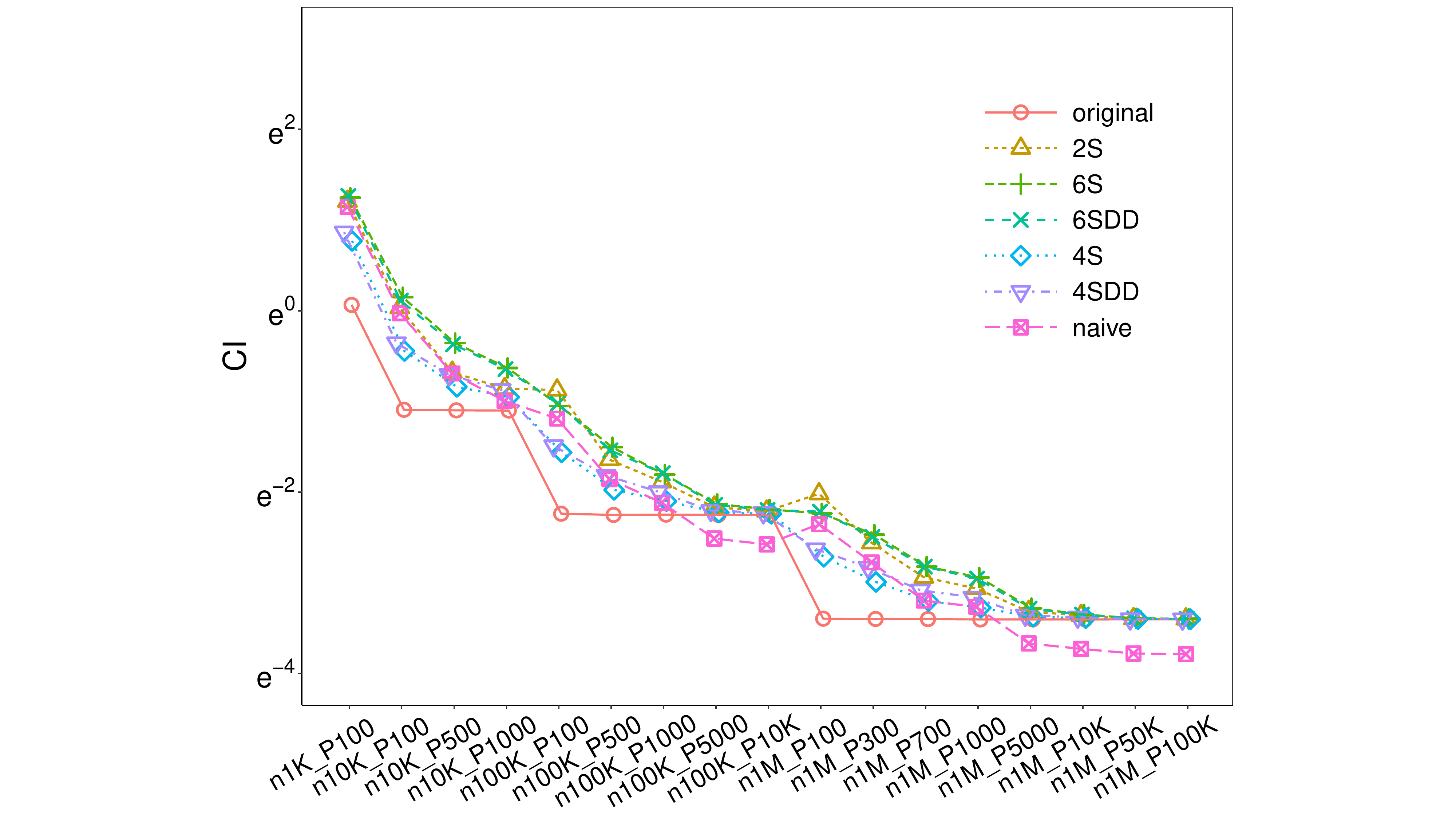}
\includegraphics[width=0.26\textwidth, trim={2.2in 0 2.2in 0},clip] {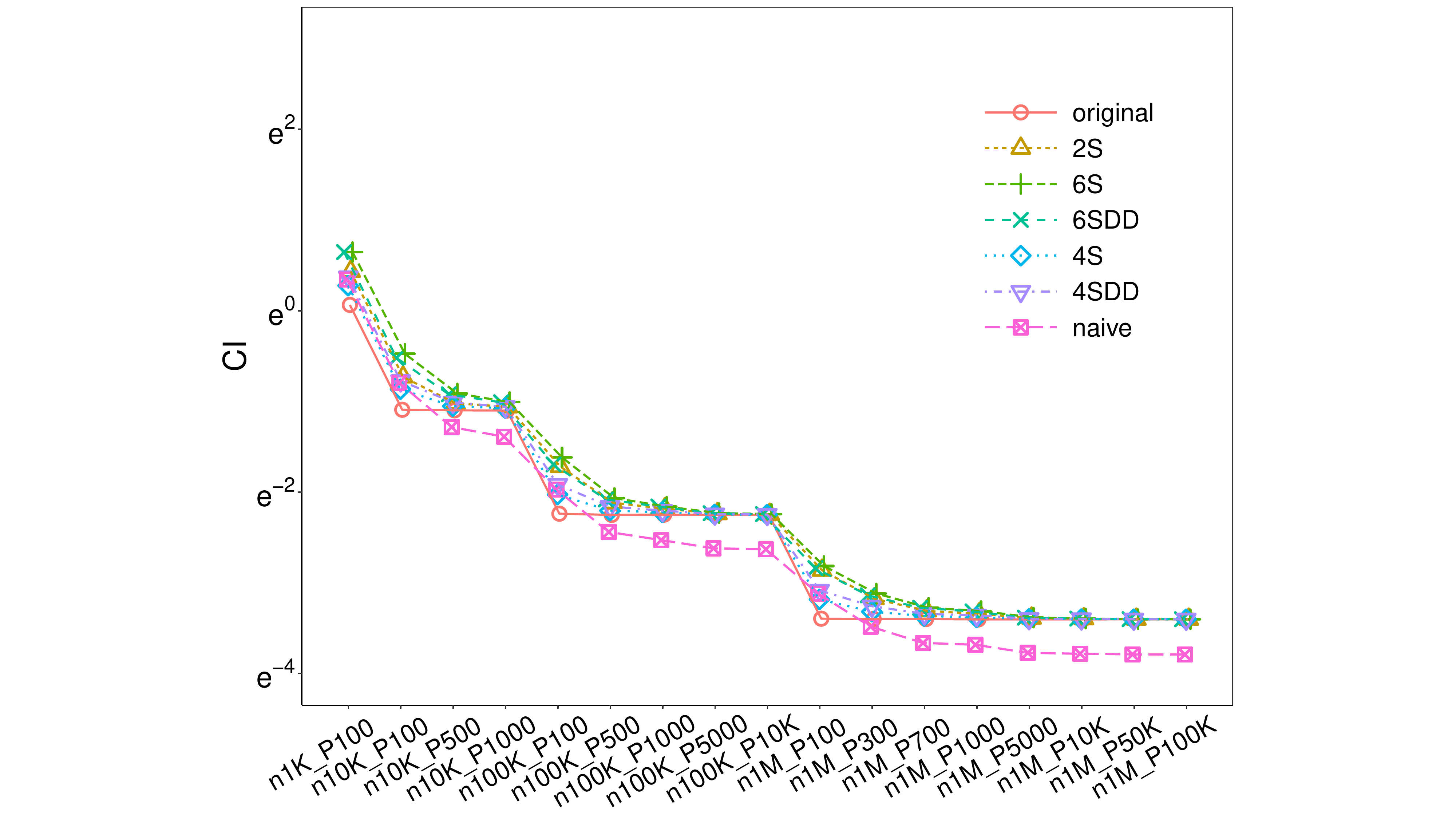}
\includegraphics[width=0.26\textwidth, trim={2.2in 0 2.2in 0},clip] {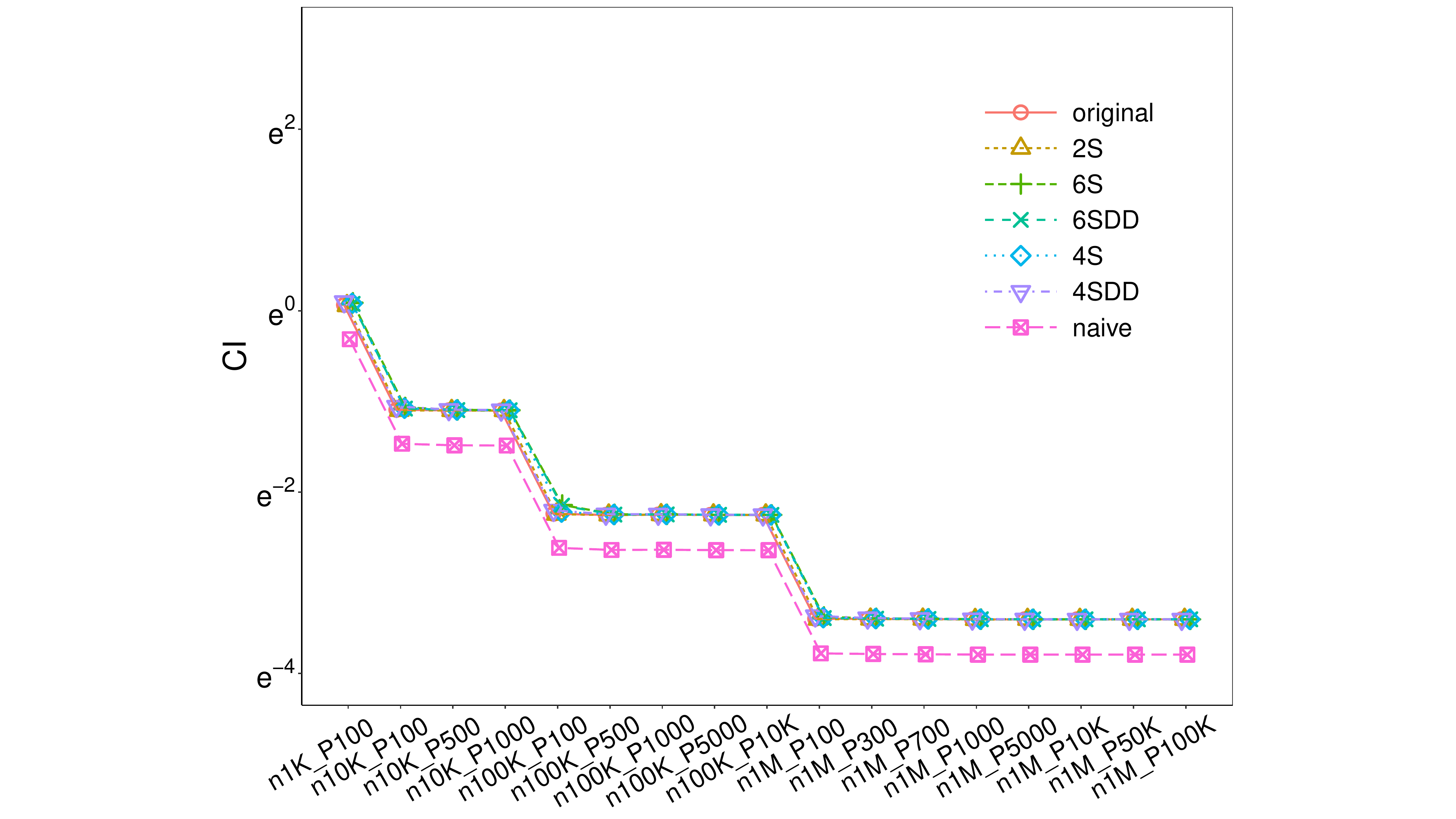}\\
\includegraphics[width=0.26\textwidth, trim={2.2in 0 2.2in 0},clip] {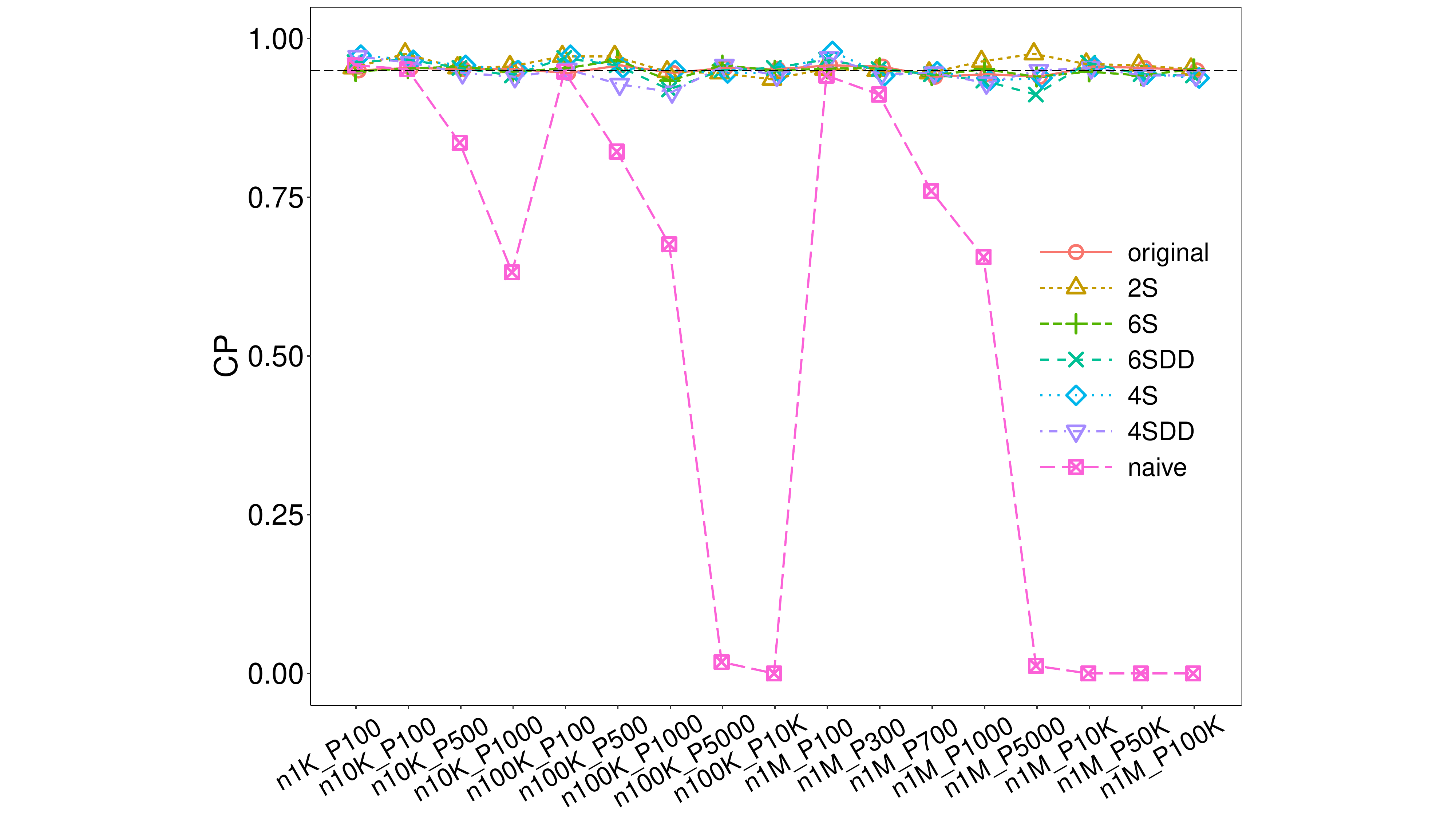}
\includegraphics[width=0.26\textwidth, trim={2.2in 0 2.2in 0},clip] {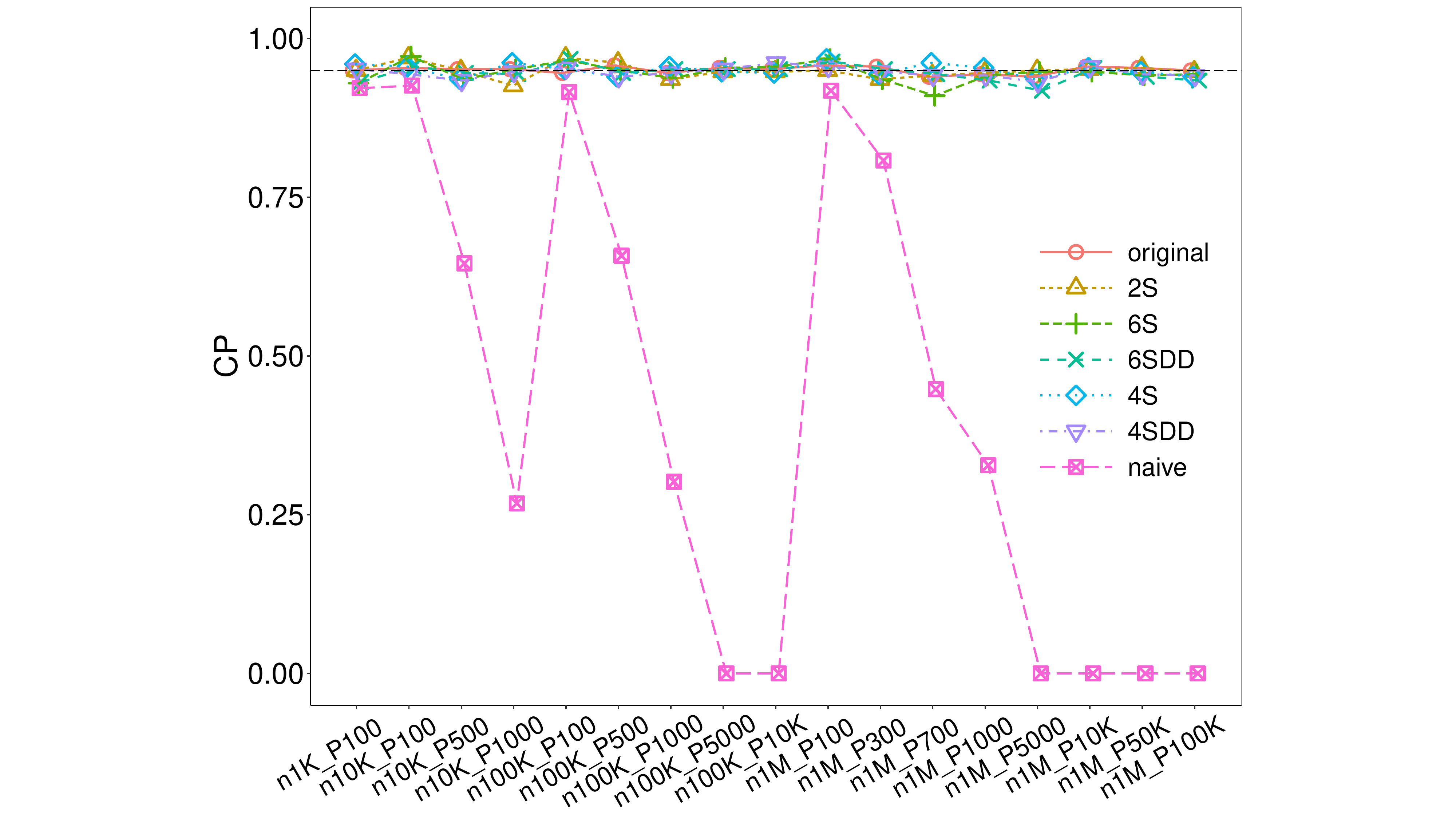}
\includegraphics[width=0.26\textwidth, trim={2.2in 0 2.2in 0},clip] {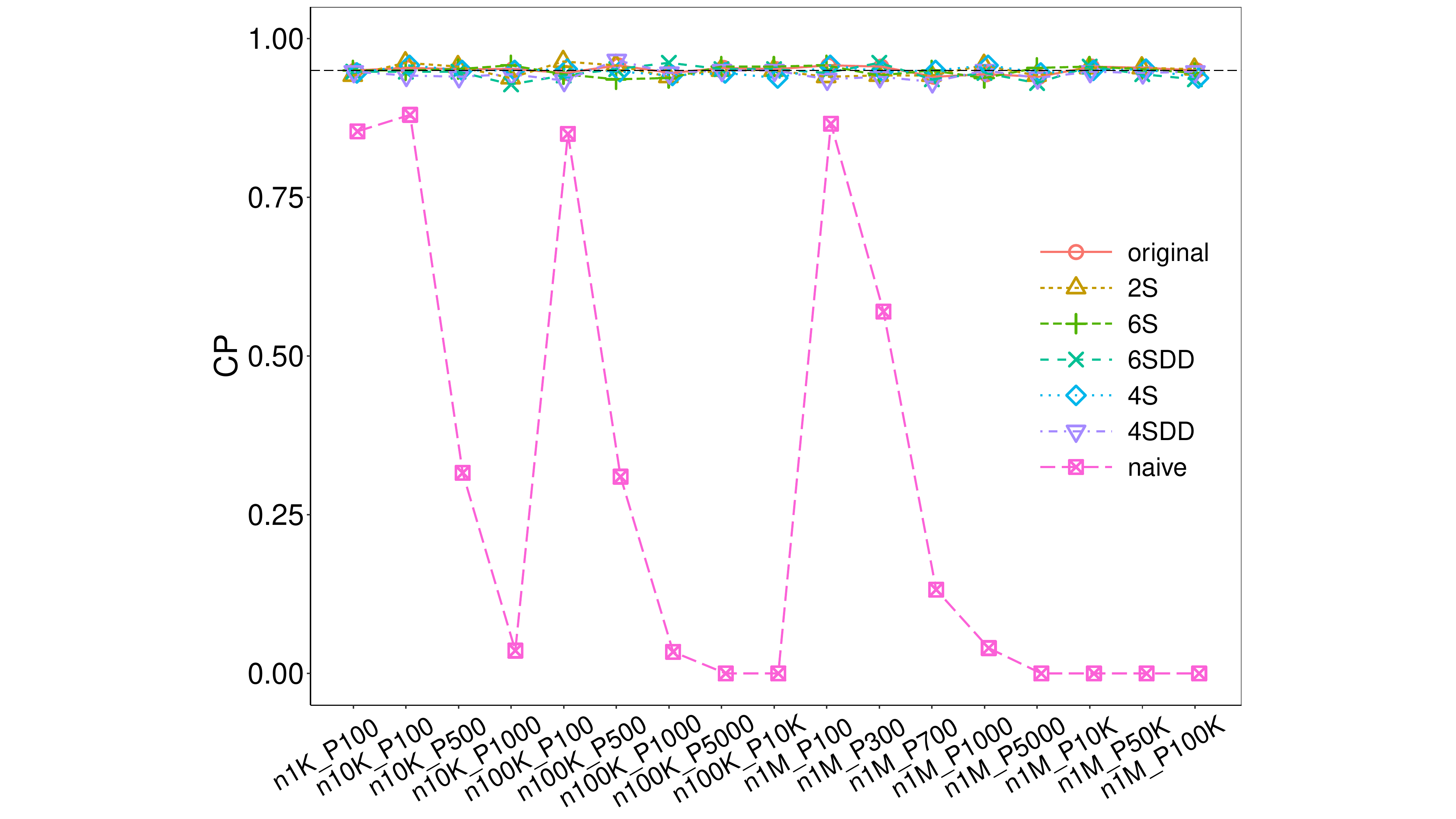}
\includegraphics[width=0.26\textwidth, trim={2.2in 0 2.2in 0},clip] {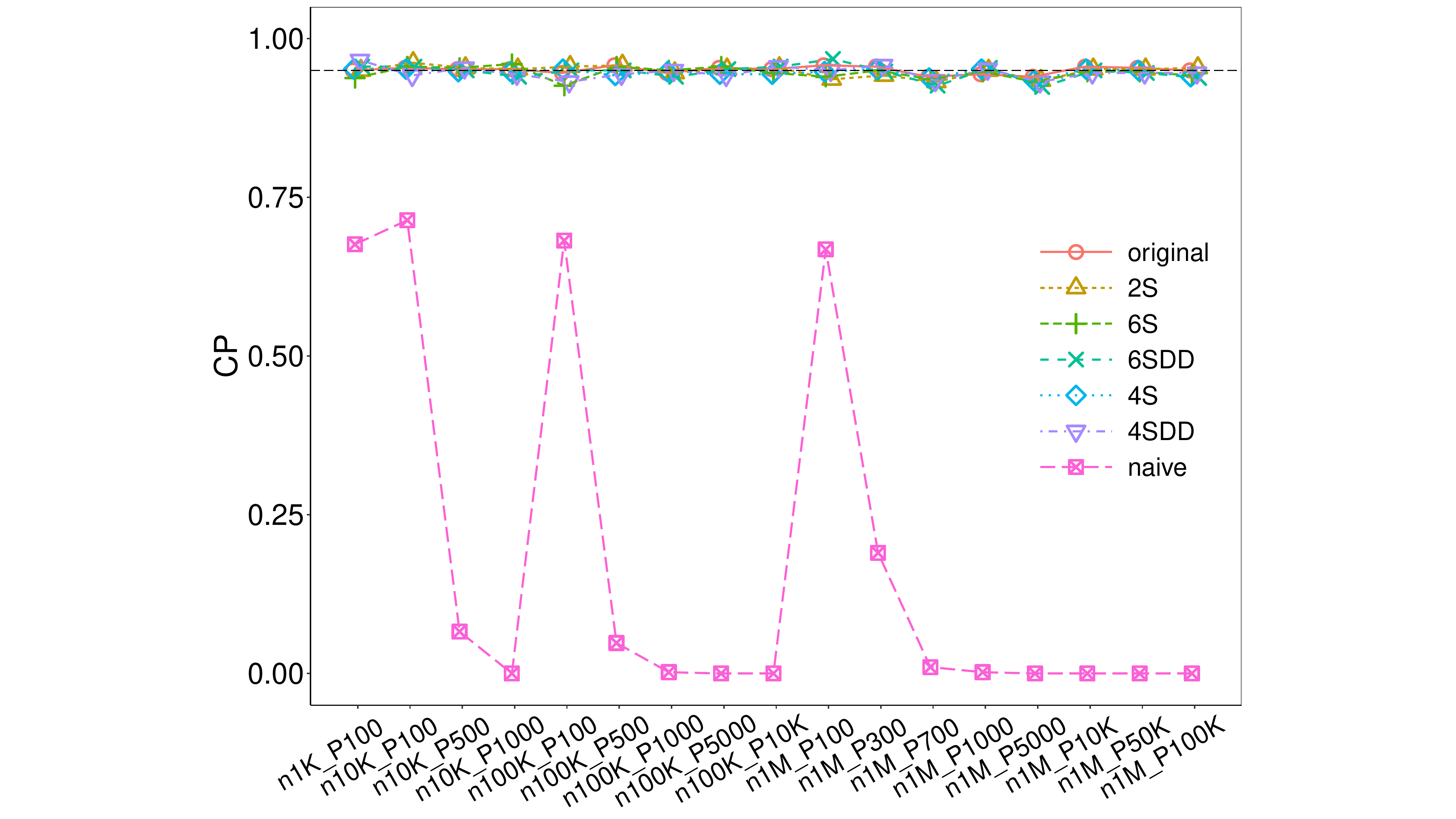}
\includegraphics[width=0.26\textwidth, trim={2.2in 0 2.2in 0},clip] {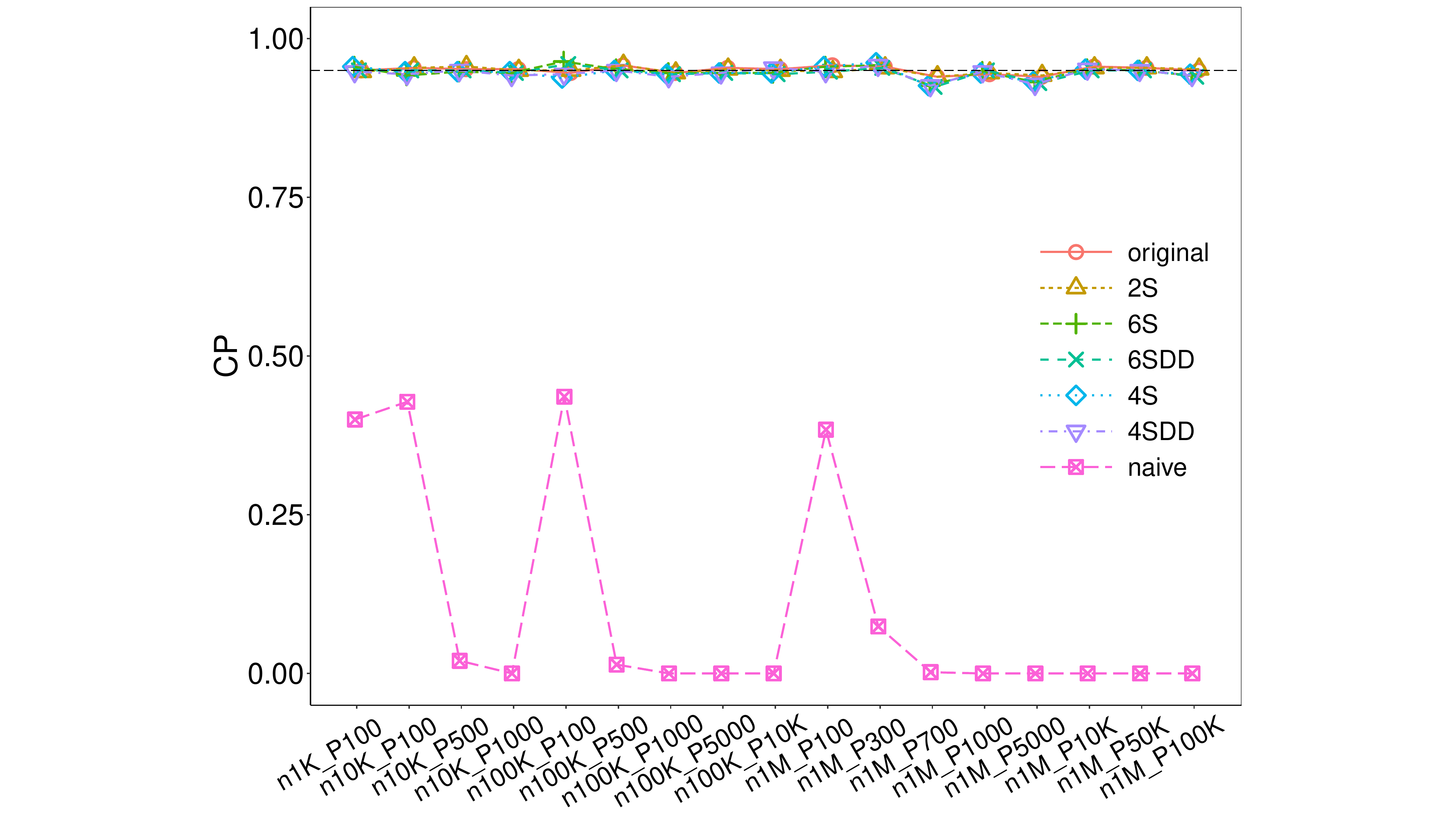}\\
\caption{Gaussian data; $\epsilon$-DP; $\theta=0$ and $\alpha\ne\beta$} \label{fig:0asDP}
\end{figure}
\end{landscape}

\begin{landscape}
\begin{figure}[!htb]
\hspace{0.6in}$\rho=0.005$\hspace{1in}$\rho=0.02$\hspace{1.2in}$\rho=0.08$
\hspace{1.1in}$\rho=0.32$\hspace{1.2in}$\rho=1.28$\\
\includegraphics[width=0.26\textwidth, trim={2.2in 0 2.2in 0},clip] {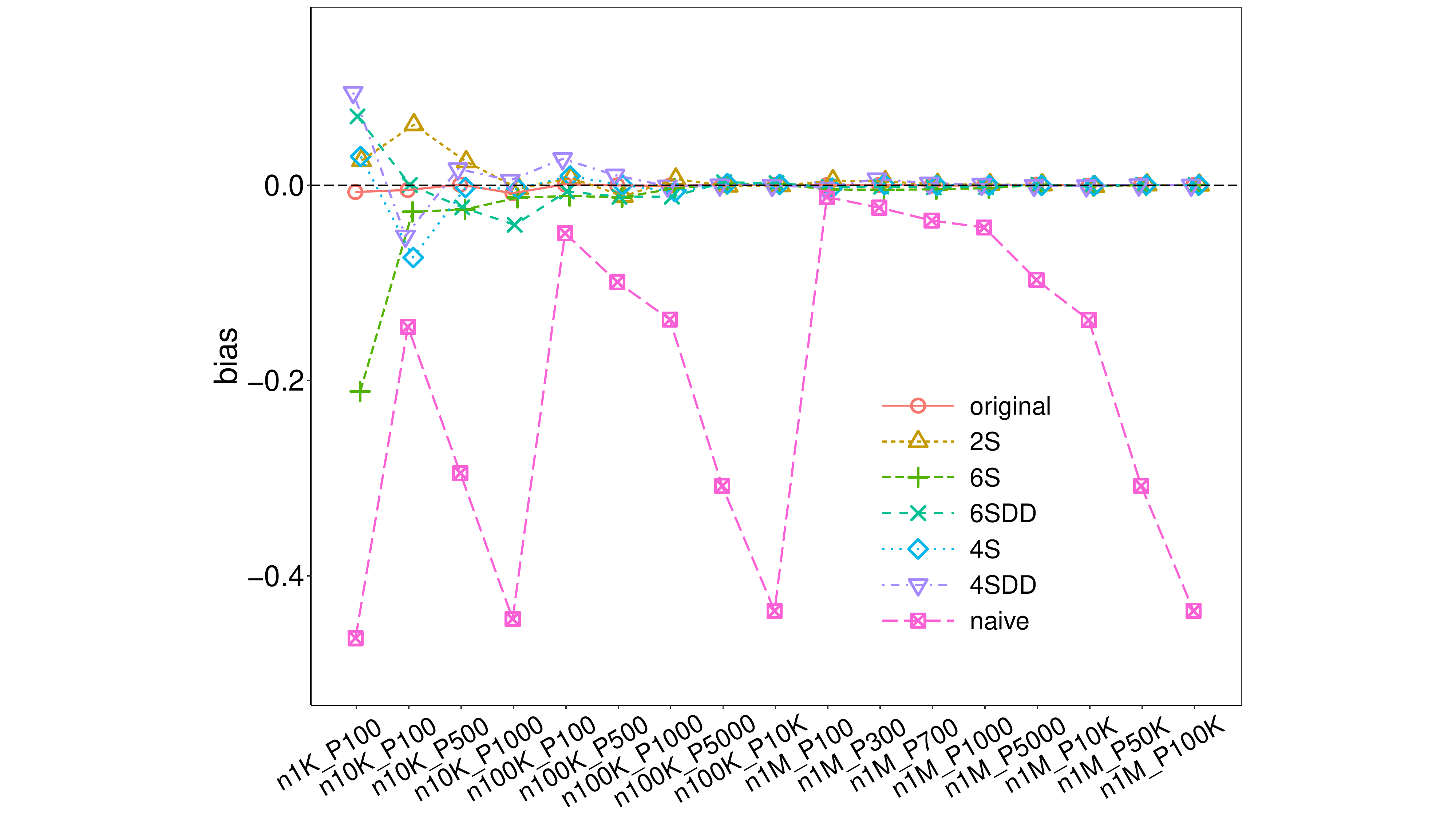}
\includegraphics[width=0.26\textwidth, trim={2.2in 0 2.2in 0},clip] {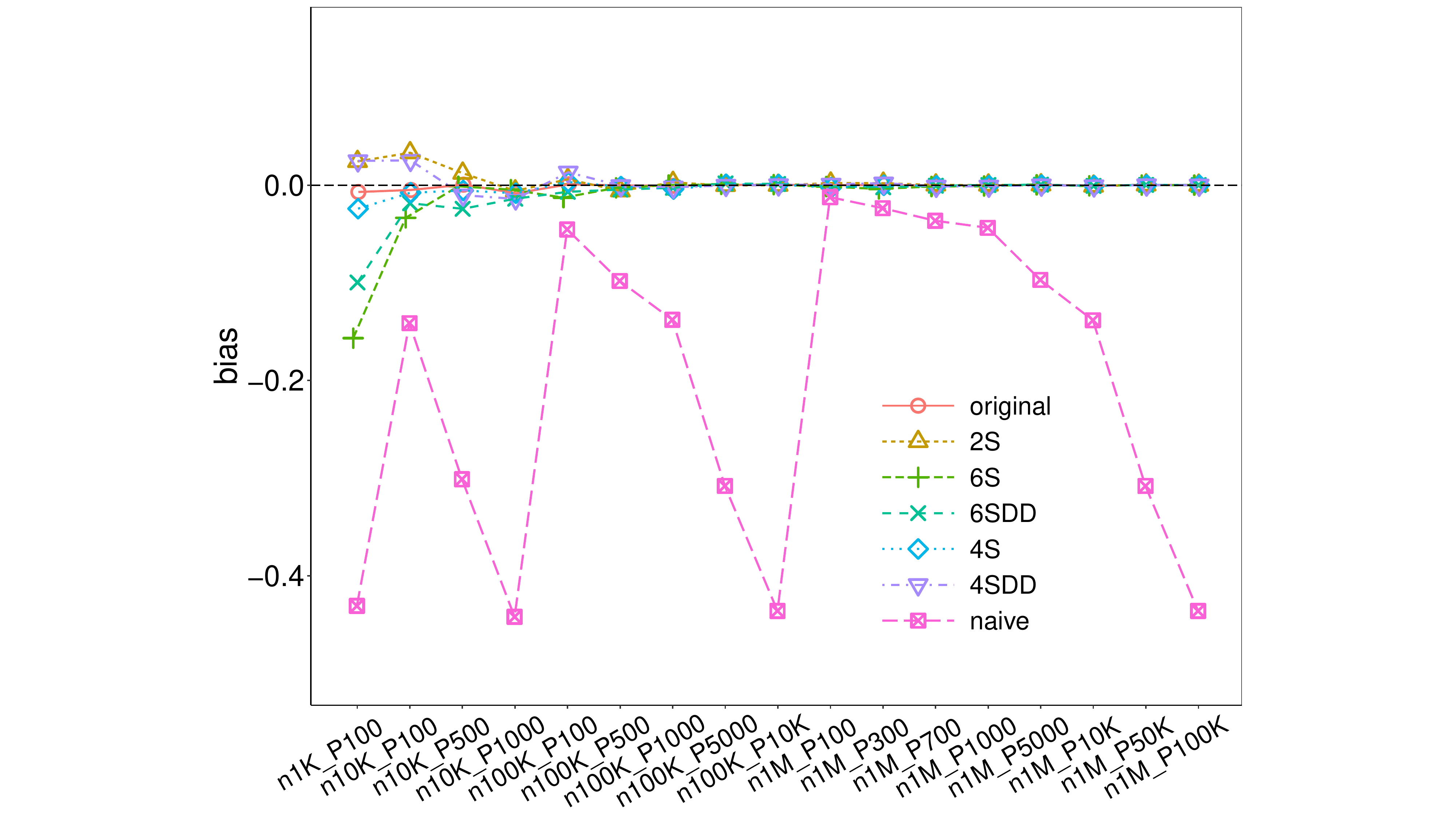}
\includegraphics[width=0.26\textwidth, trim={2.2in 0 2.2in 0},clip] {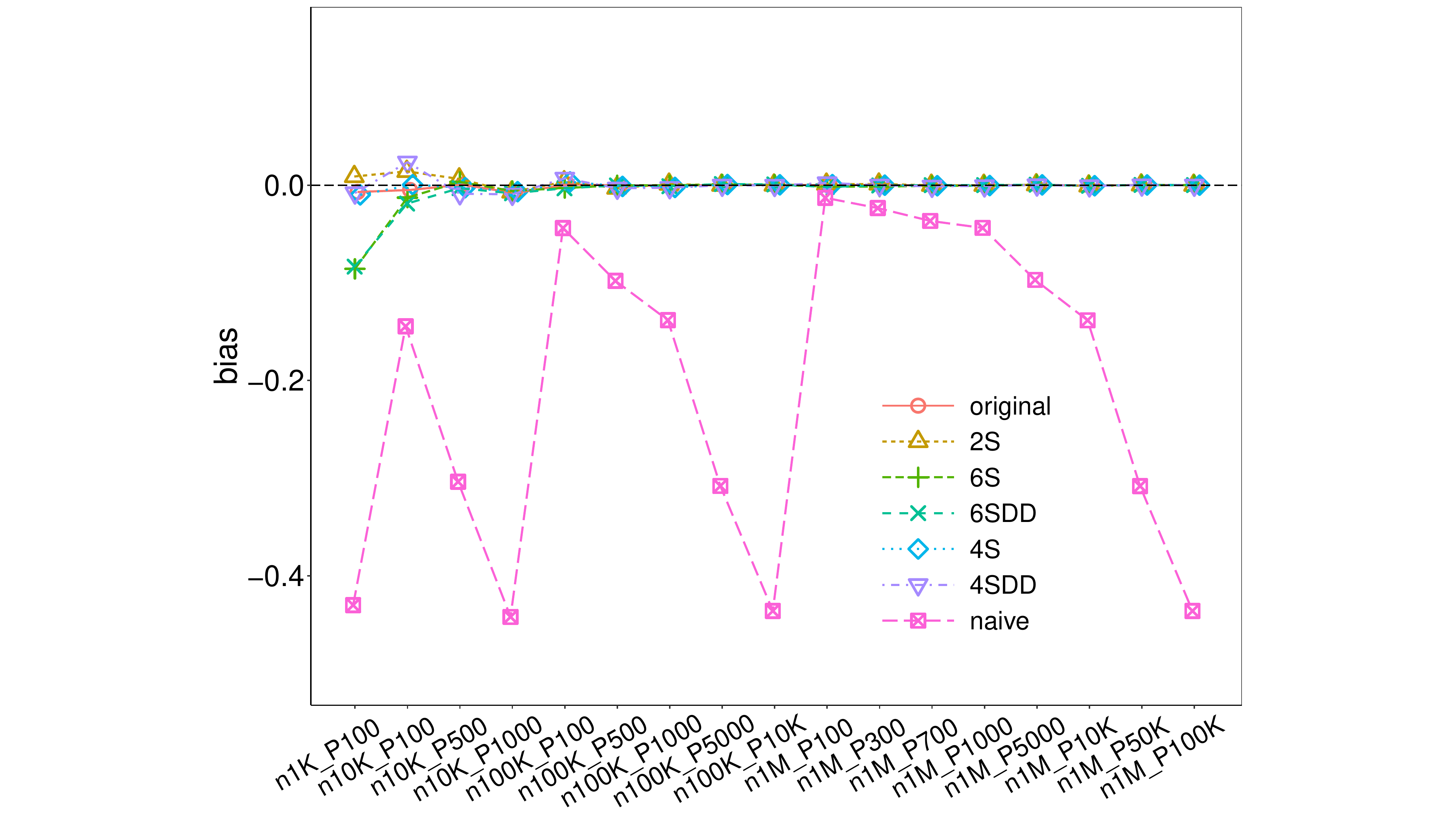}
\includegraphics[width=0.26\textwidth, trim={2.2in 0 2.2in 0},clip] {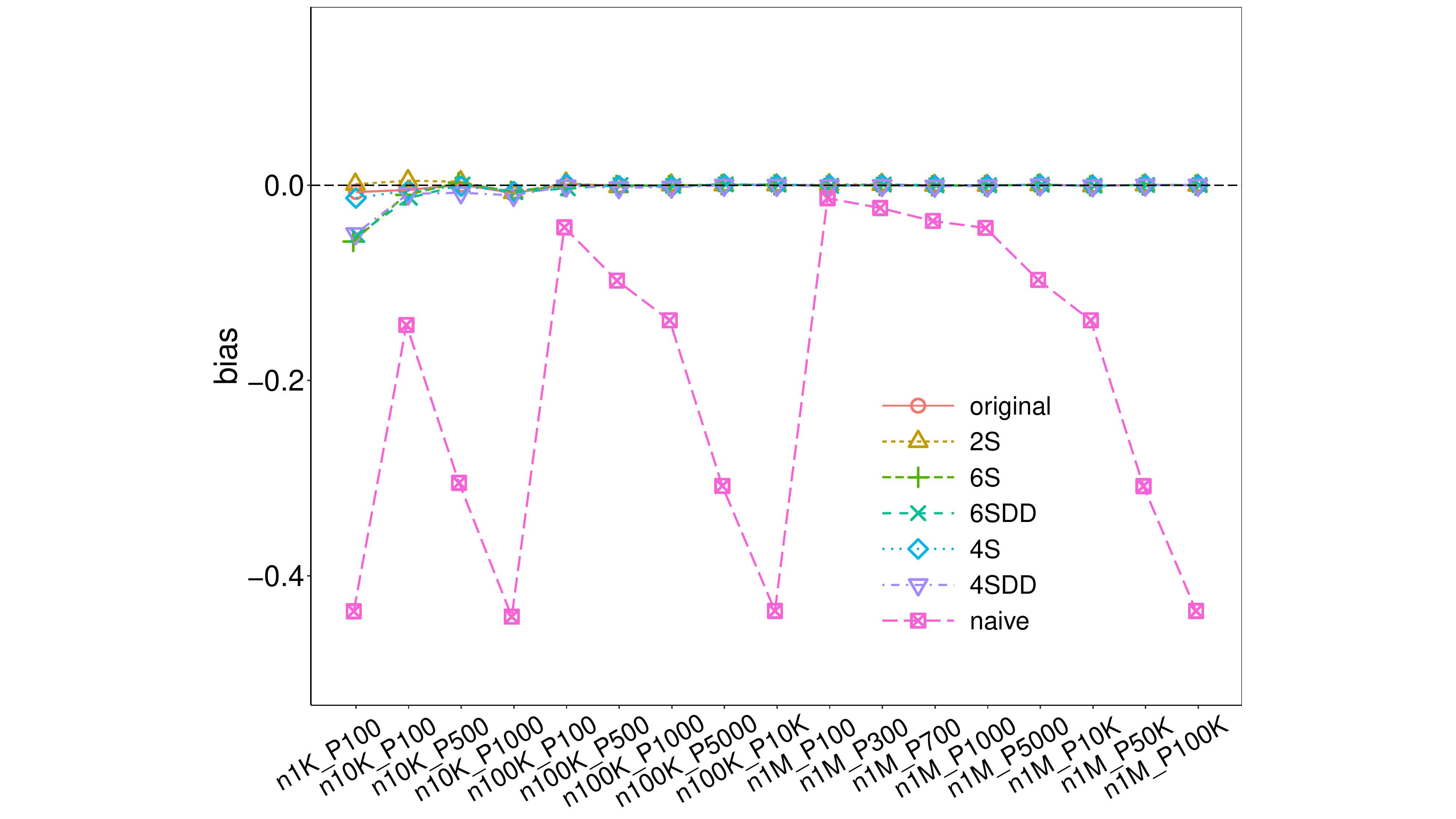}
\includegraphics[width=0.26\textwidth, trim={2.2in 0 2.2in 0},clip] {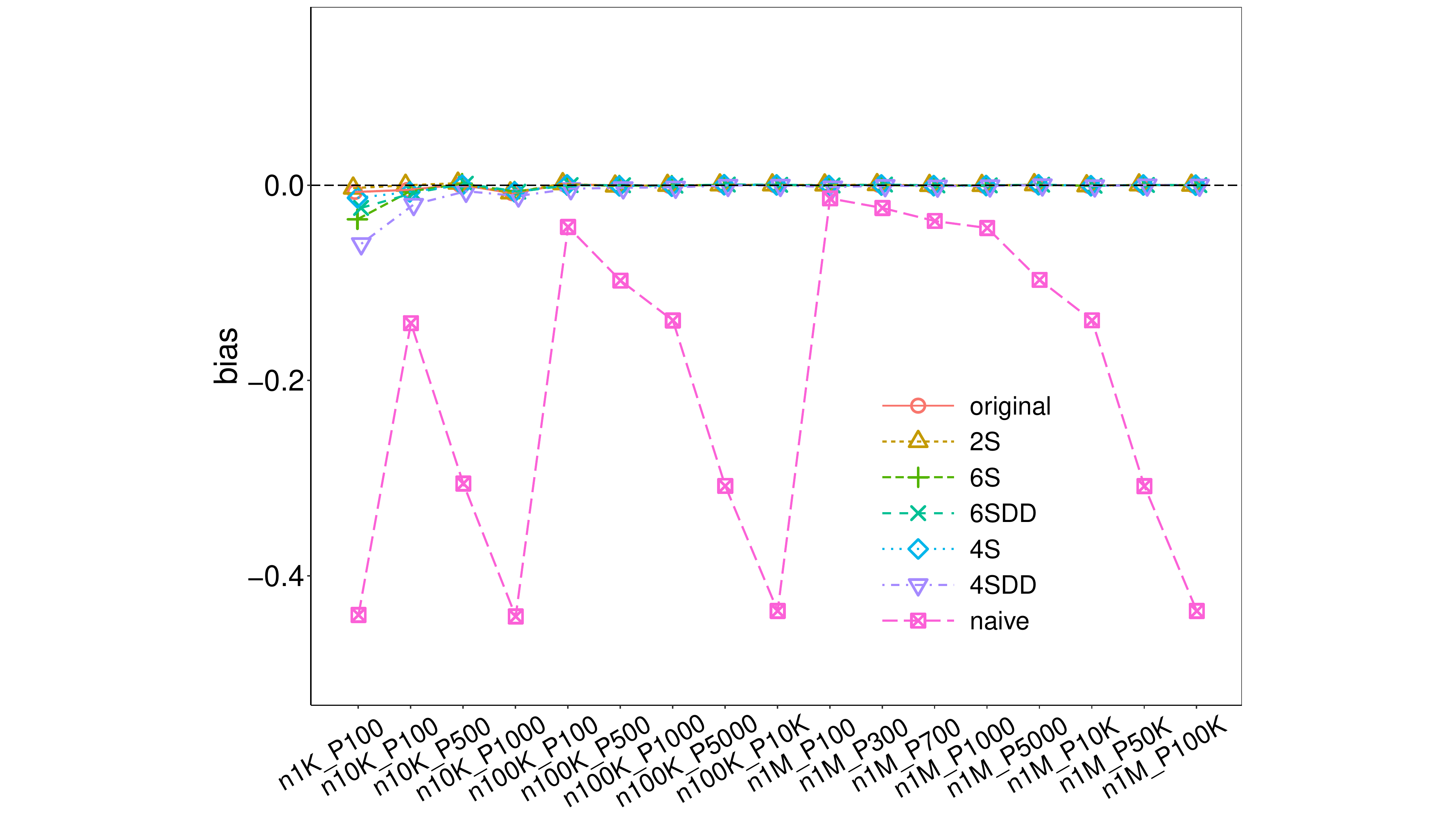}\\
\includegraphics[width=0.26\textwidth, trim={2.2in 0 2.2in 0},clip] {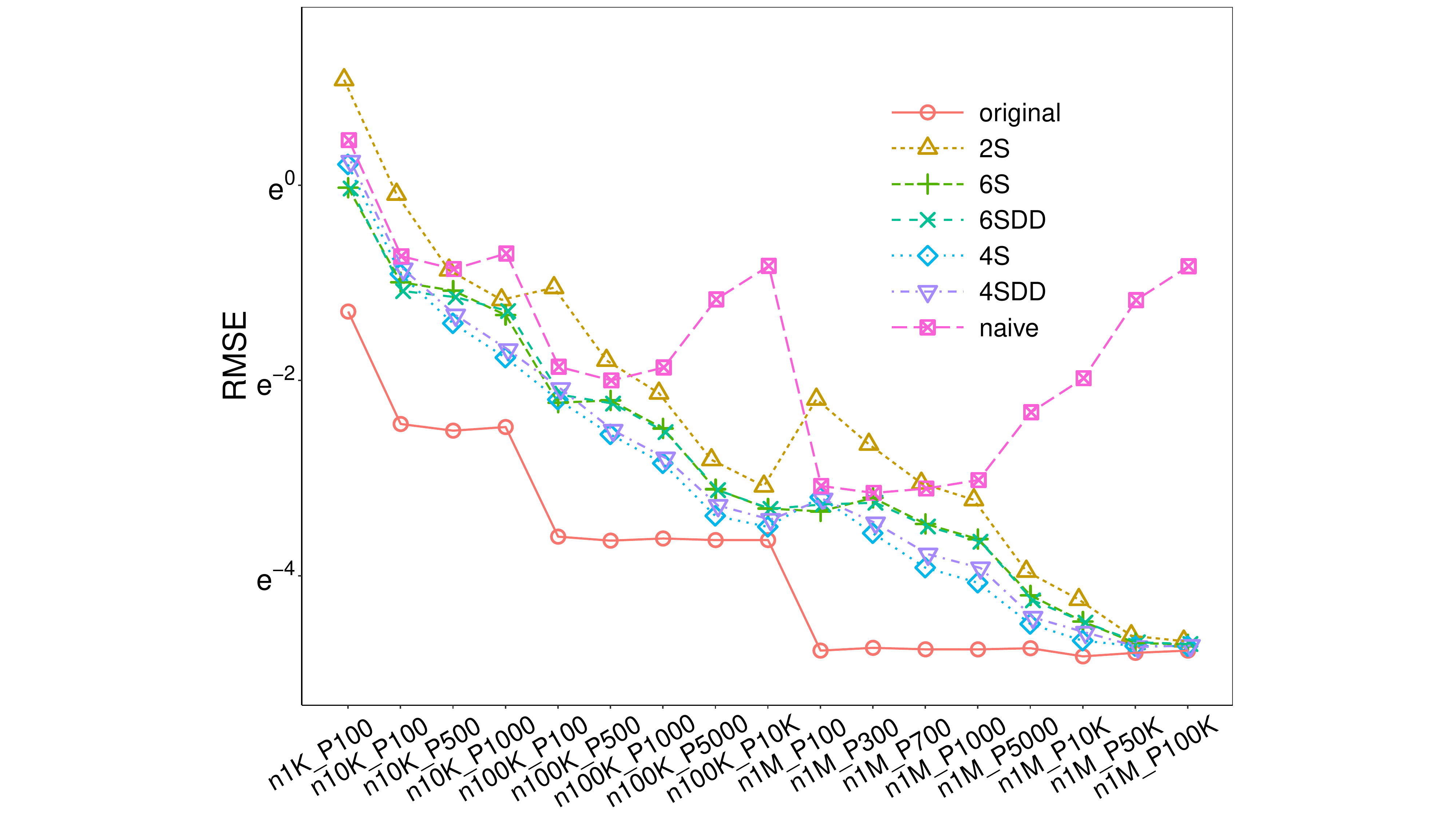}
\includegraphics[width=0.26\textwidth, trim={2.2in 0 2.2in 0},clip] {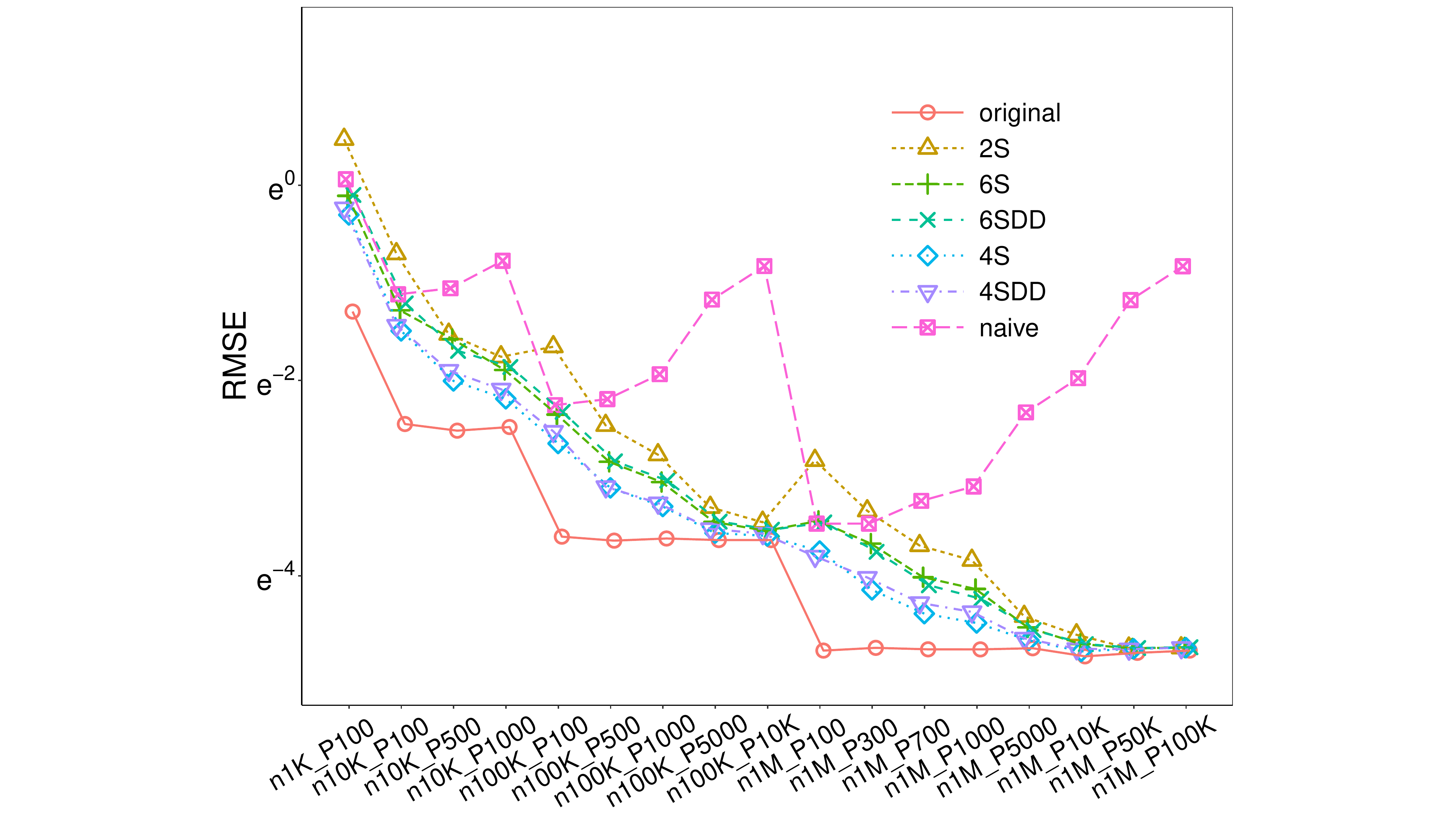}
\includegraphics[width=0.26\textwidth, trim={2.2in 0 2.2in 0},clip] {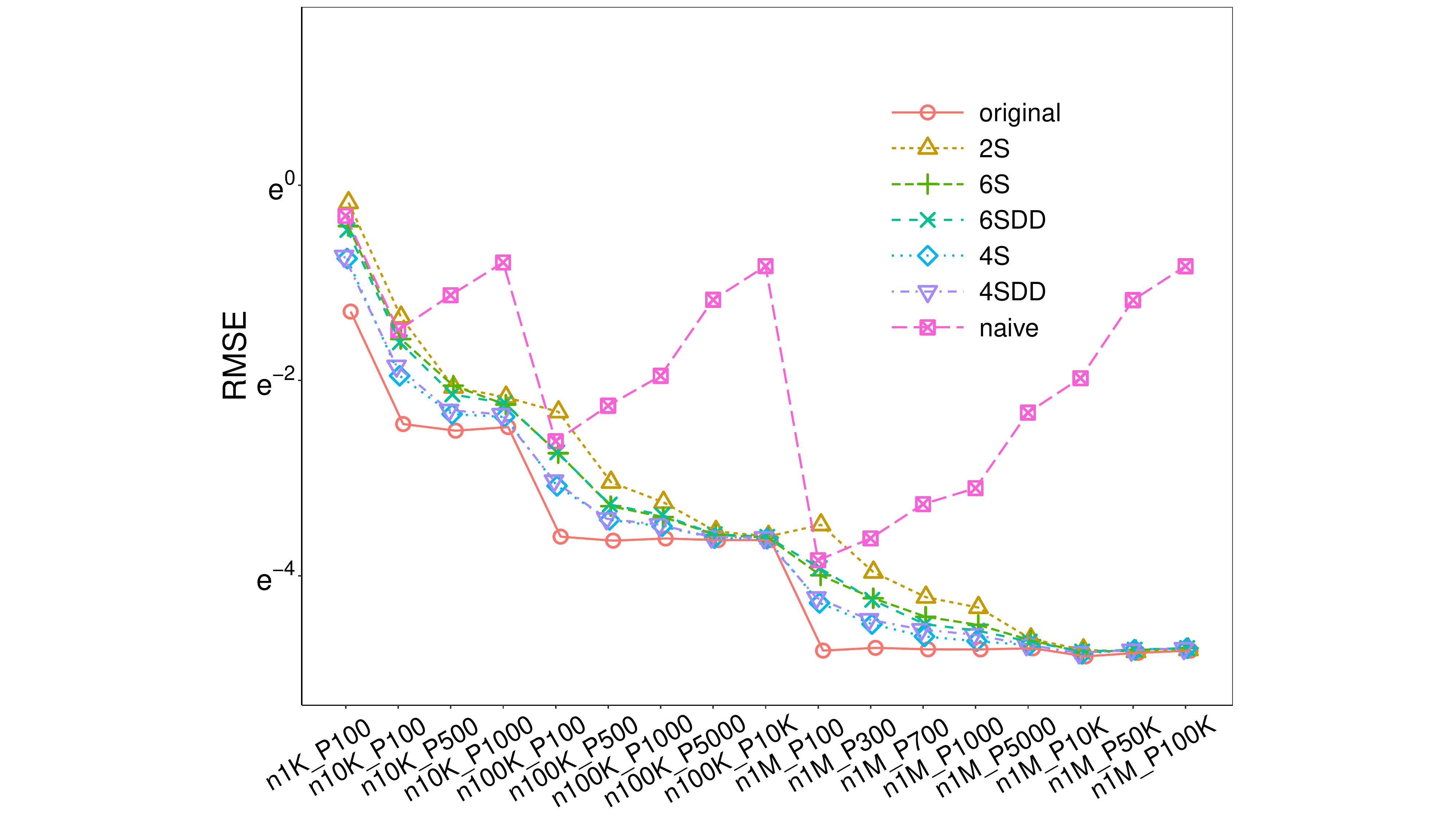}
\includegraphics[width=0.26\textwidth, trim={2.2in 0 2.2in 0},clip] {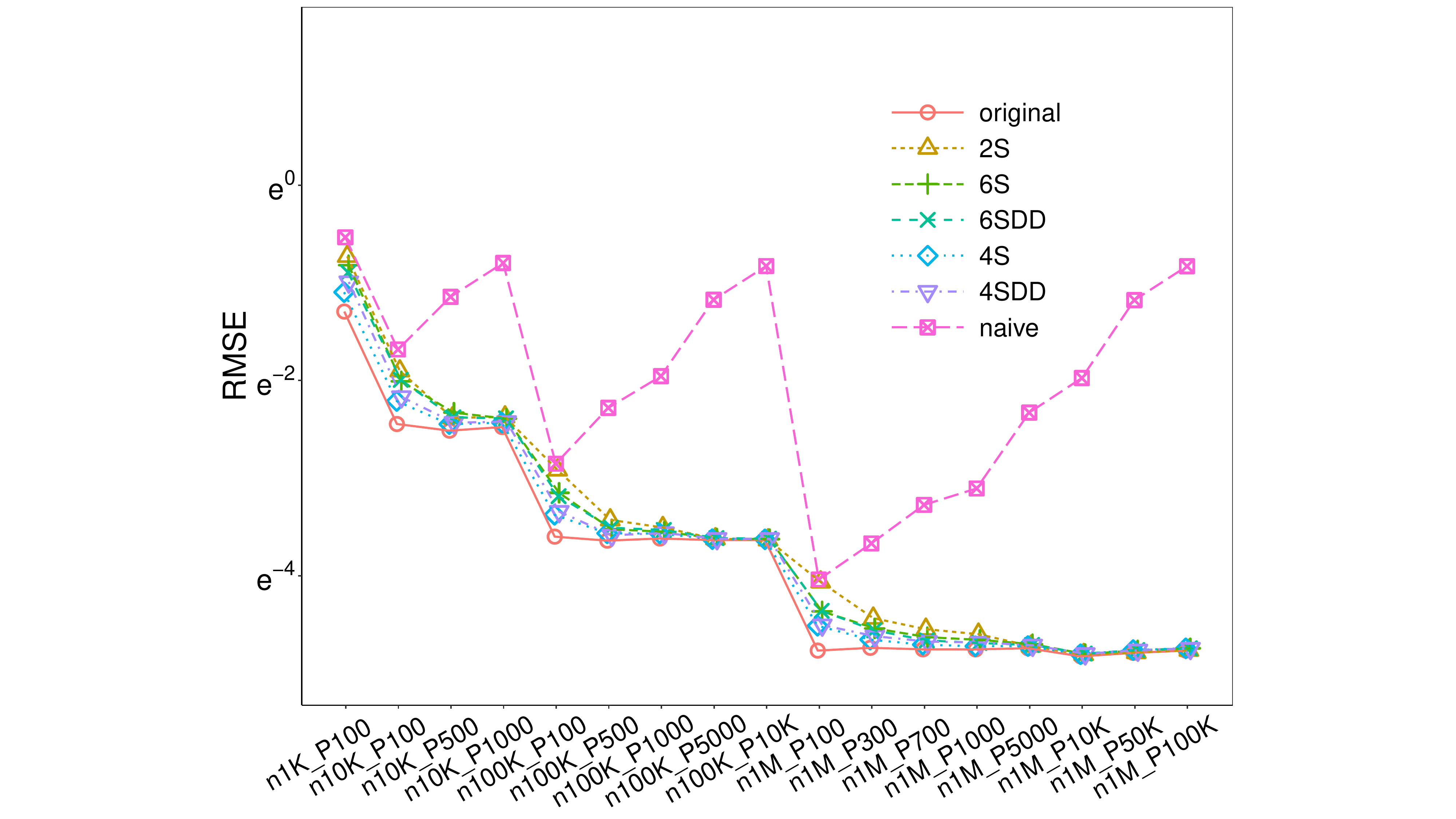}
\includegraphics[width=0.26\textwidth, trim={2.2in 0 2.2in 0},clip] {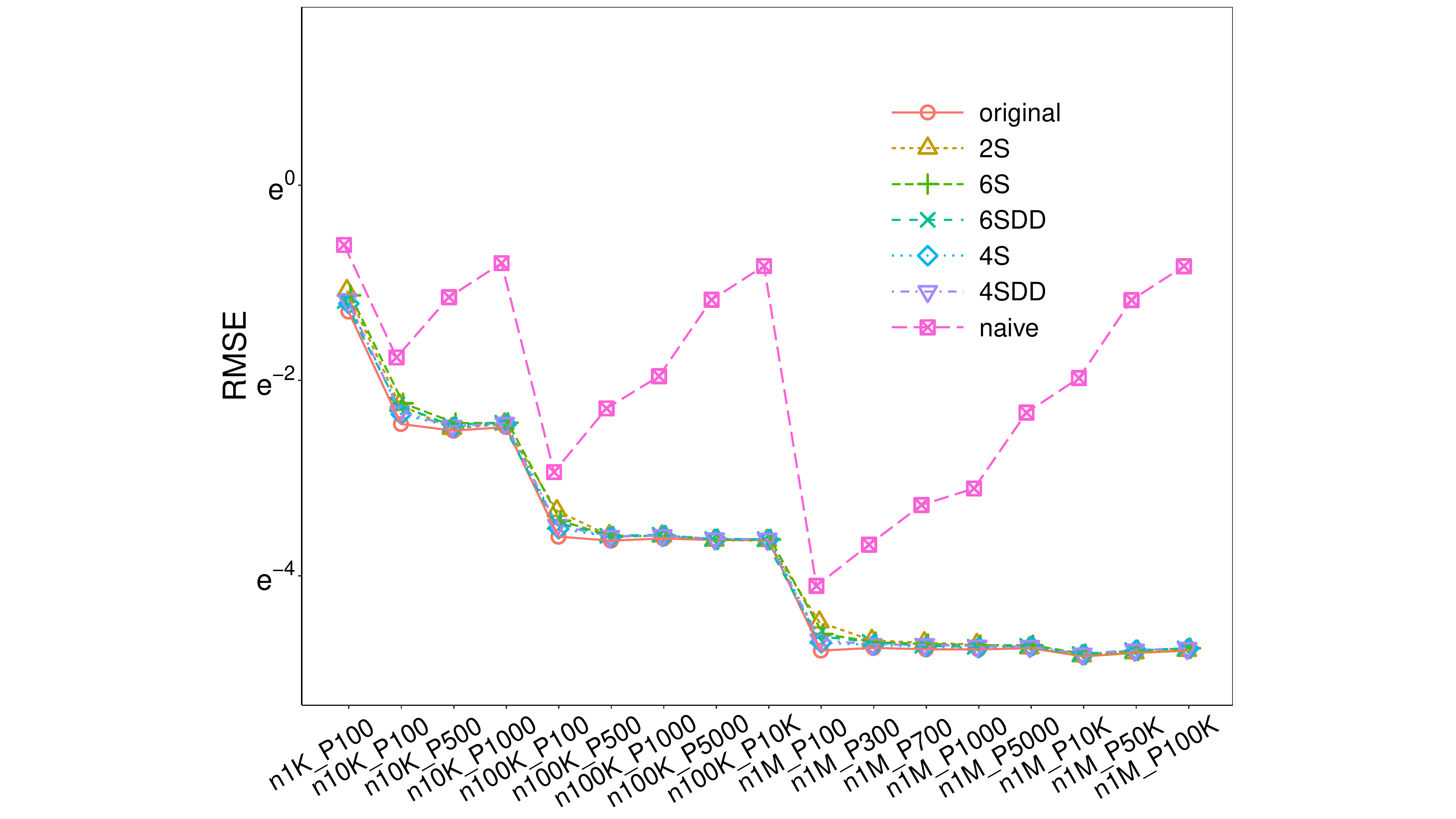}\\
\includegraphics[width=0.26\textwidth, trim={2.2in 0 2.2in 0},clip] {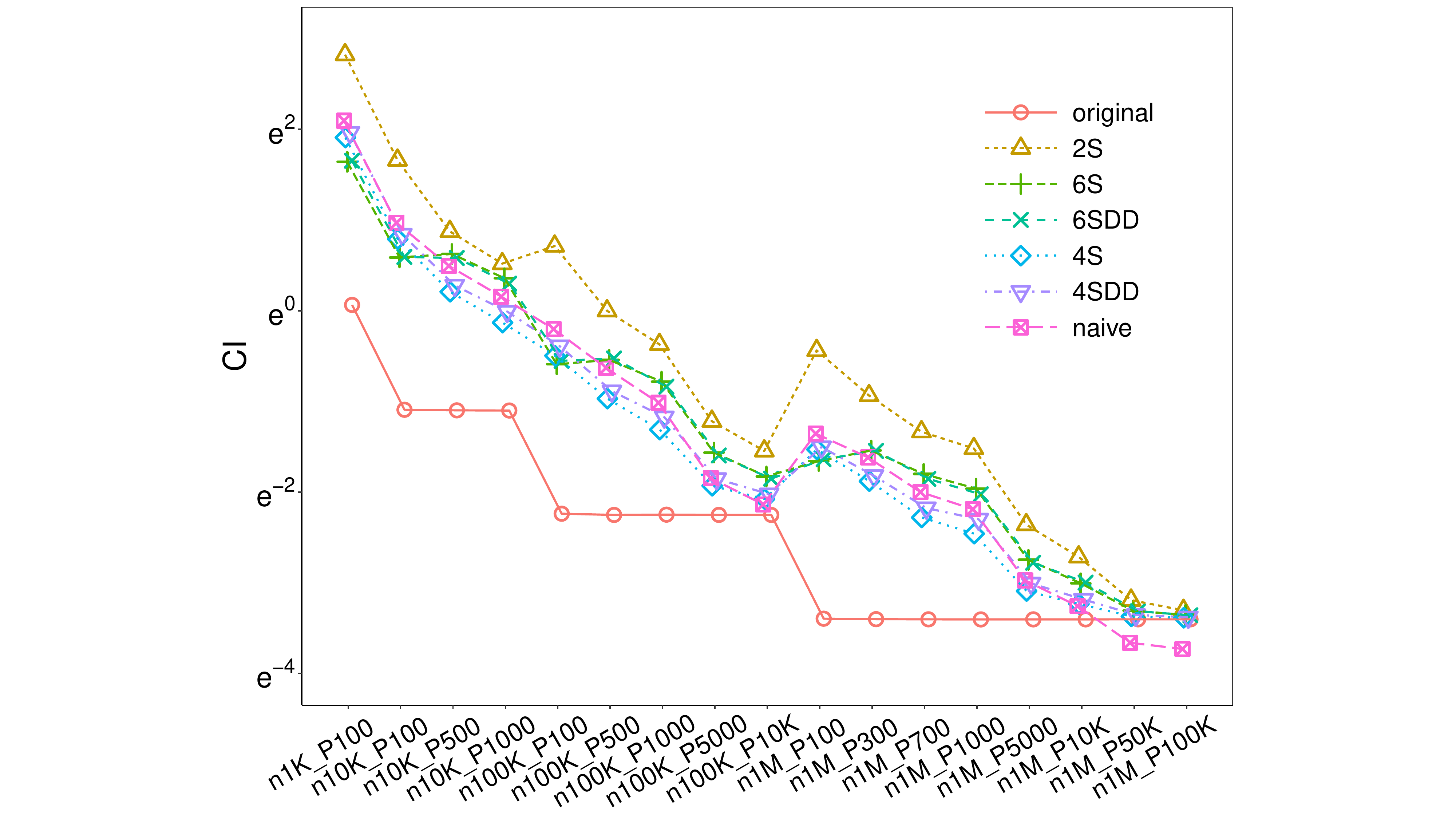}
\includegraphics[width=0.26\textwidth, trim={2.2in 0 2.2in 0},clip] {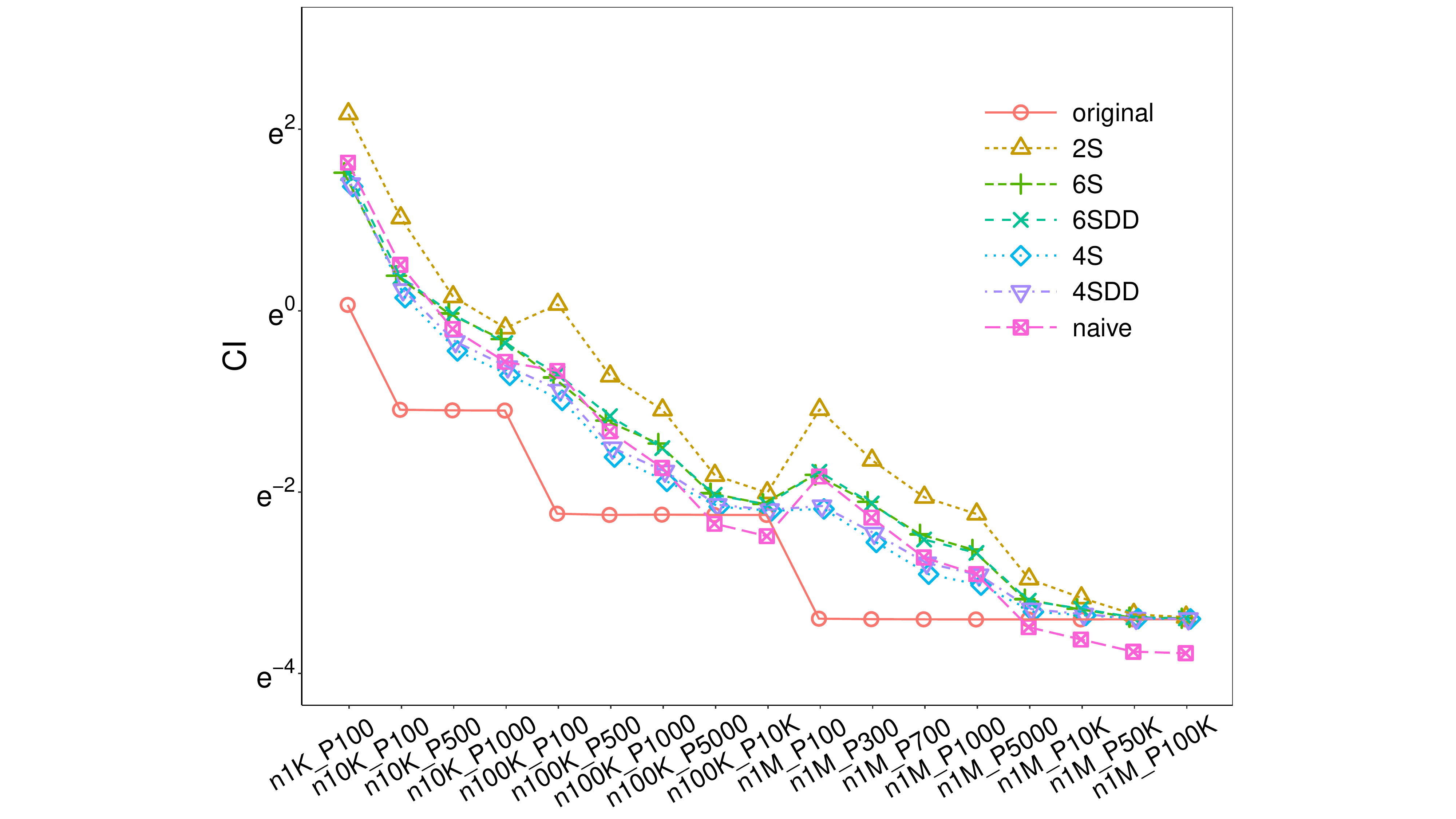}
\includegraphics[width=0.26\textwidth, trim={2.2in 0 2.2in 0},clip] {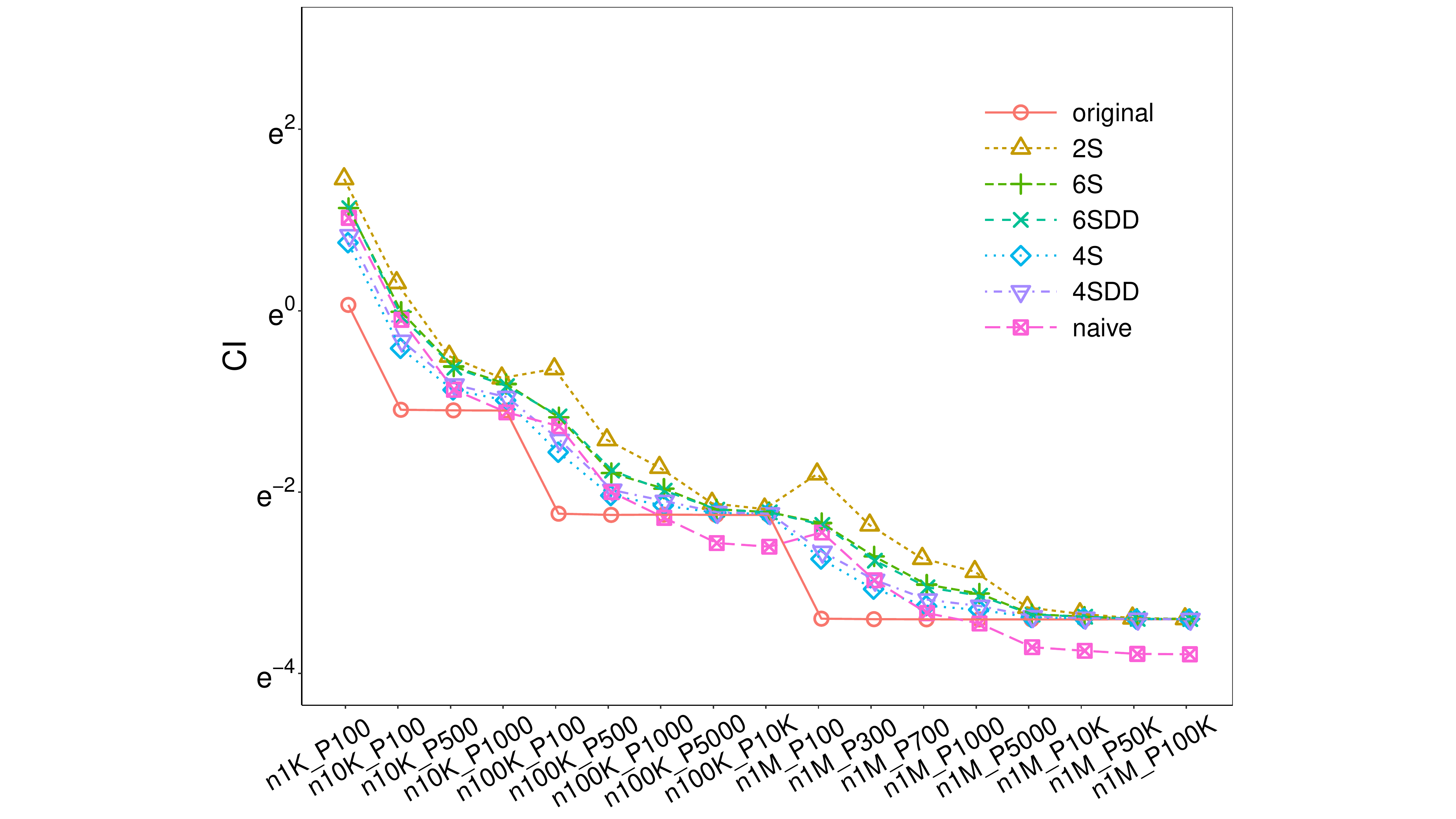}
\includegraphics[width=0.26\textwidth, trim={2.2in 0 2.2in 0},clip] {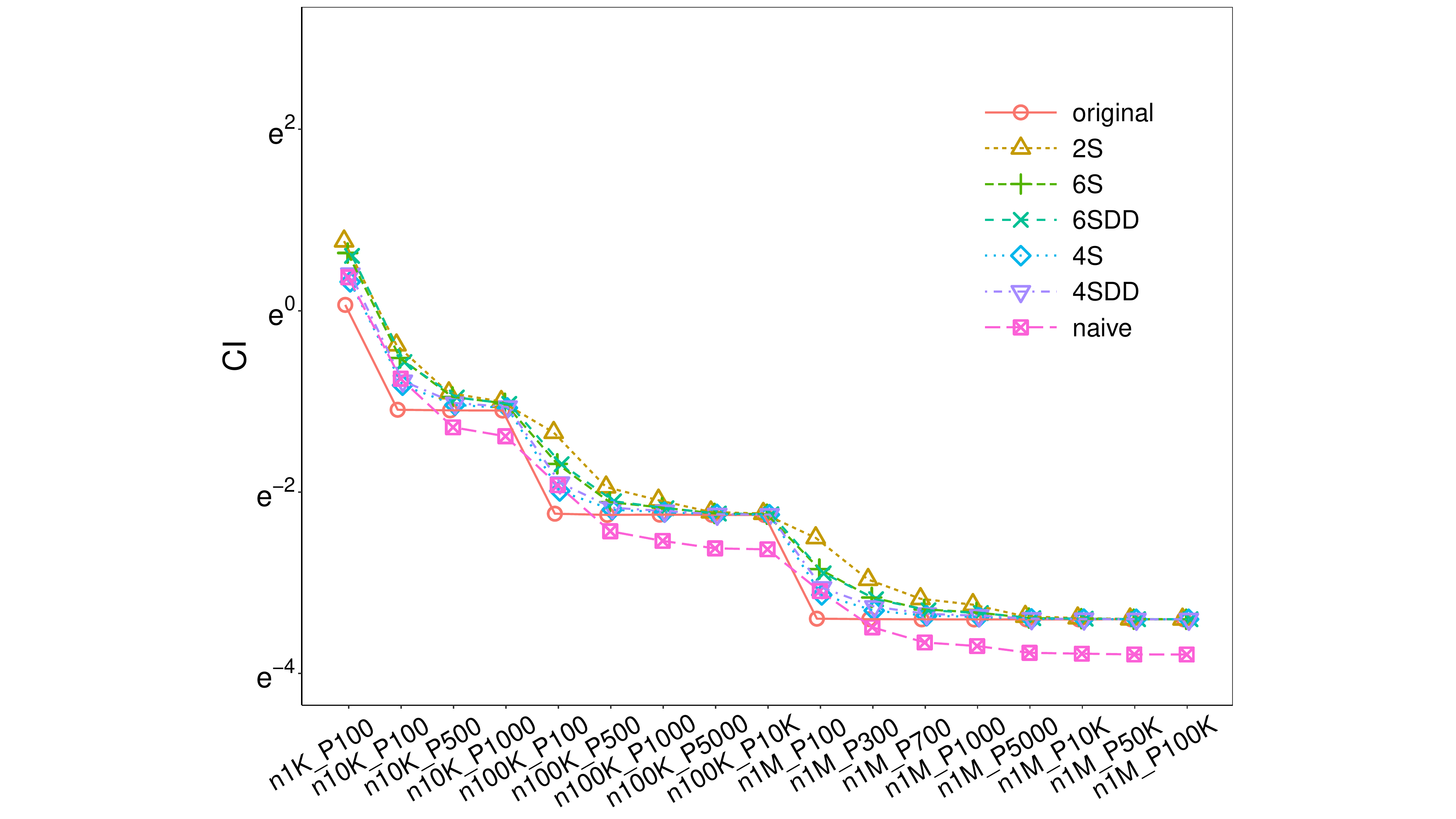}
\includegraphics[width=0.26\textwidth, trim={2.2in 0 2.2in 0},clip] {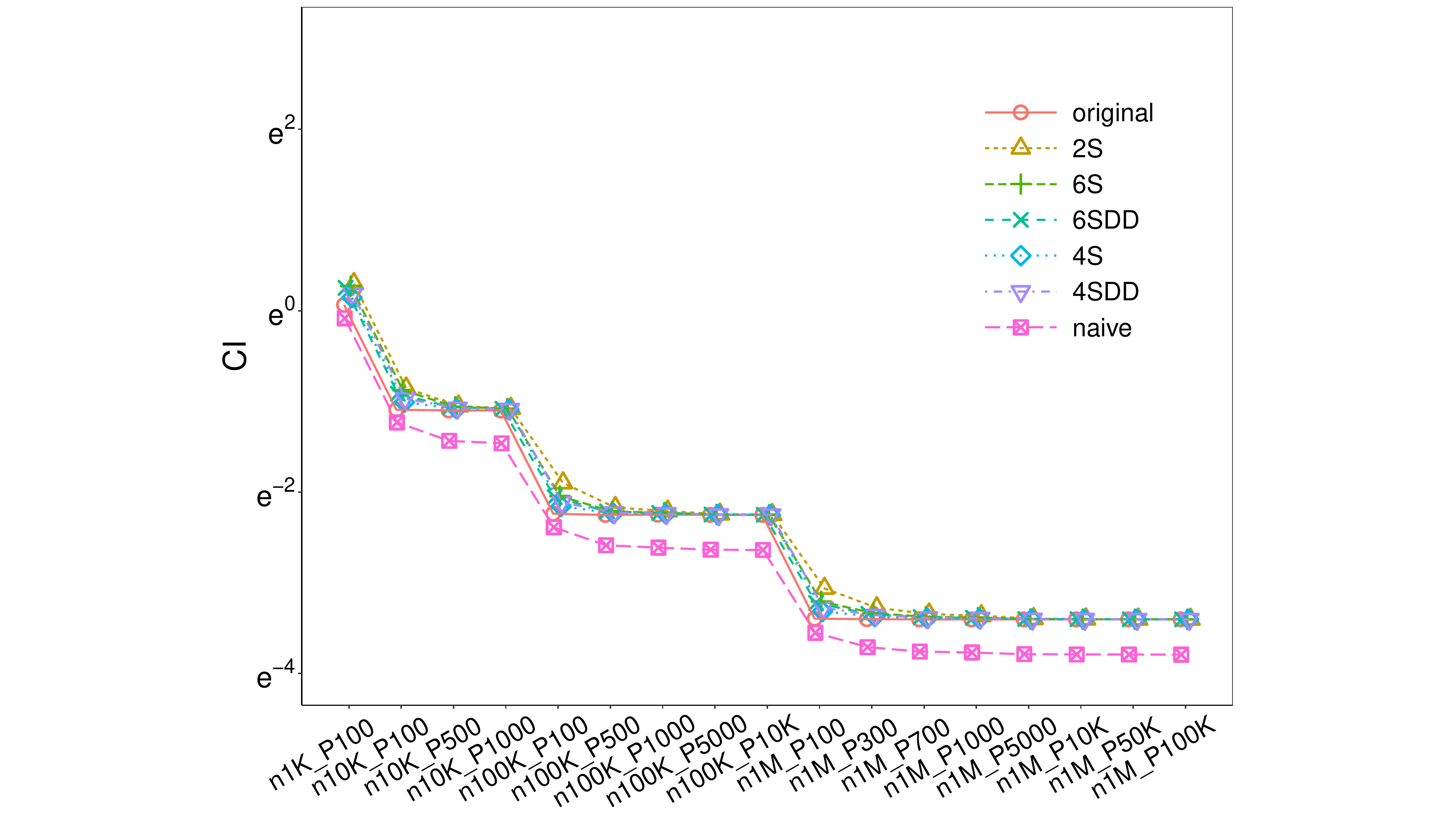}\\
\includegraphics[width=0.26\textwidth, trim={2.2in 0 2.2in 0},clip] {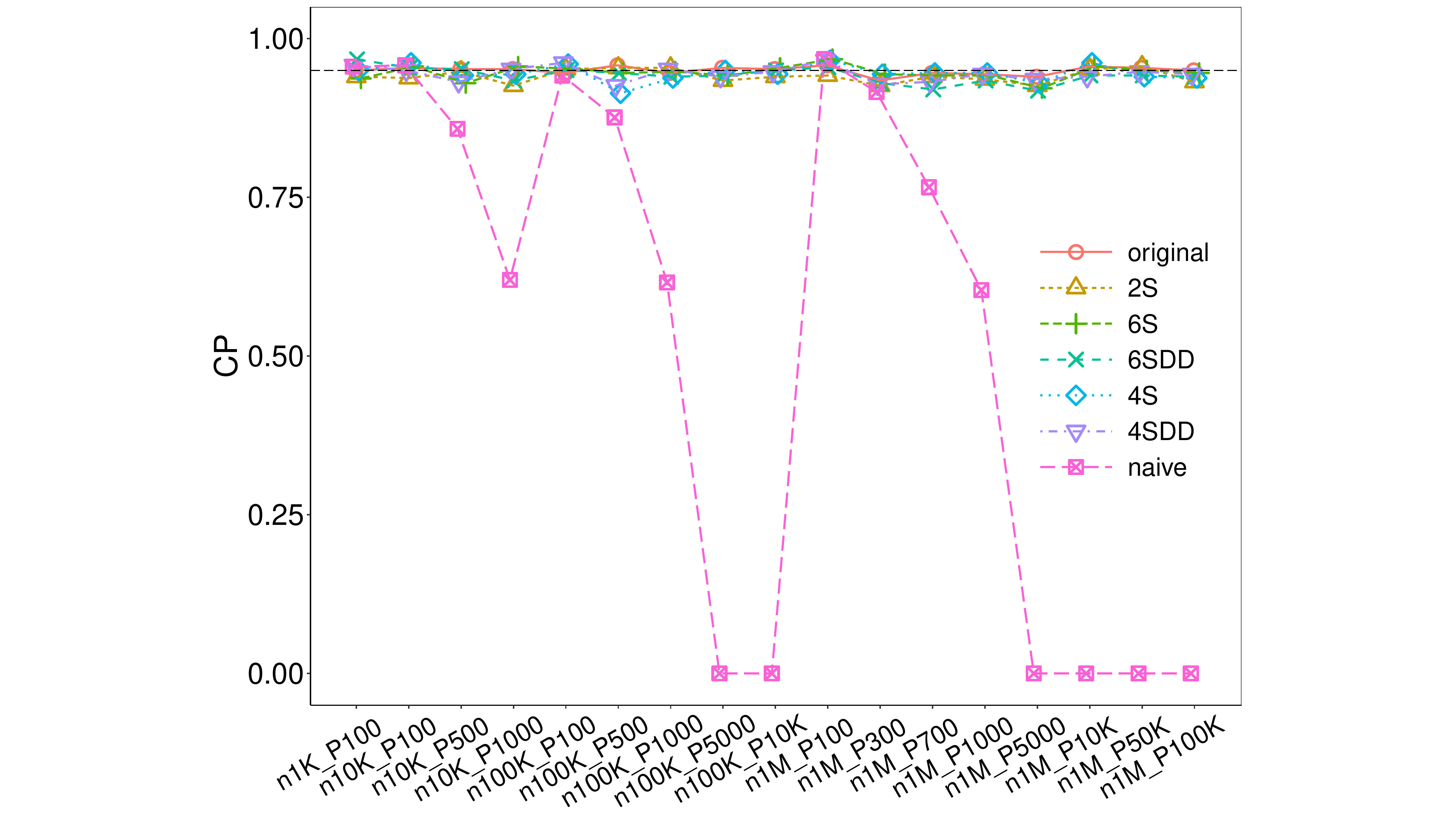}
\includegraphics[width=0.26\textwidth, trim={2.2in 0 2.2in 0},clip] {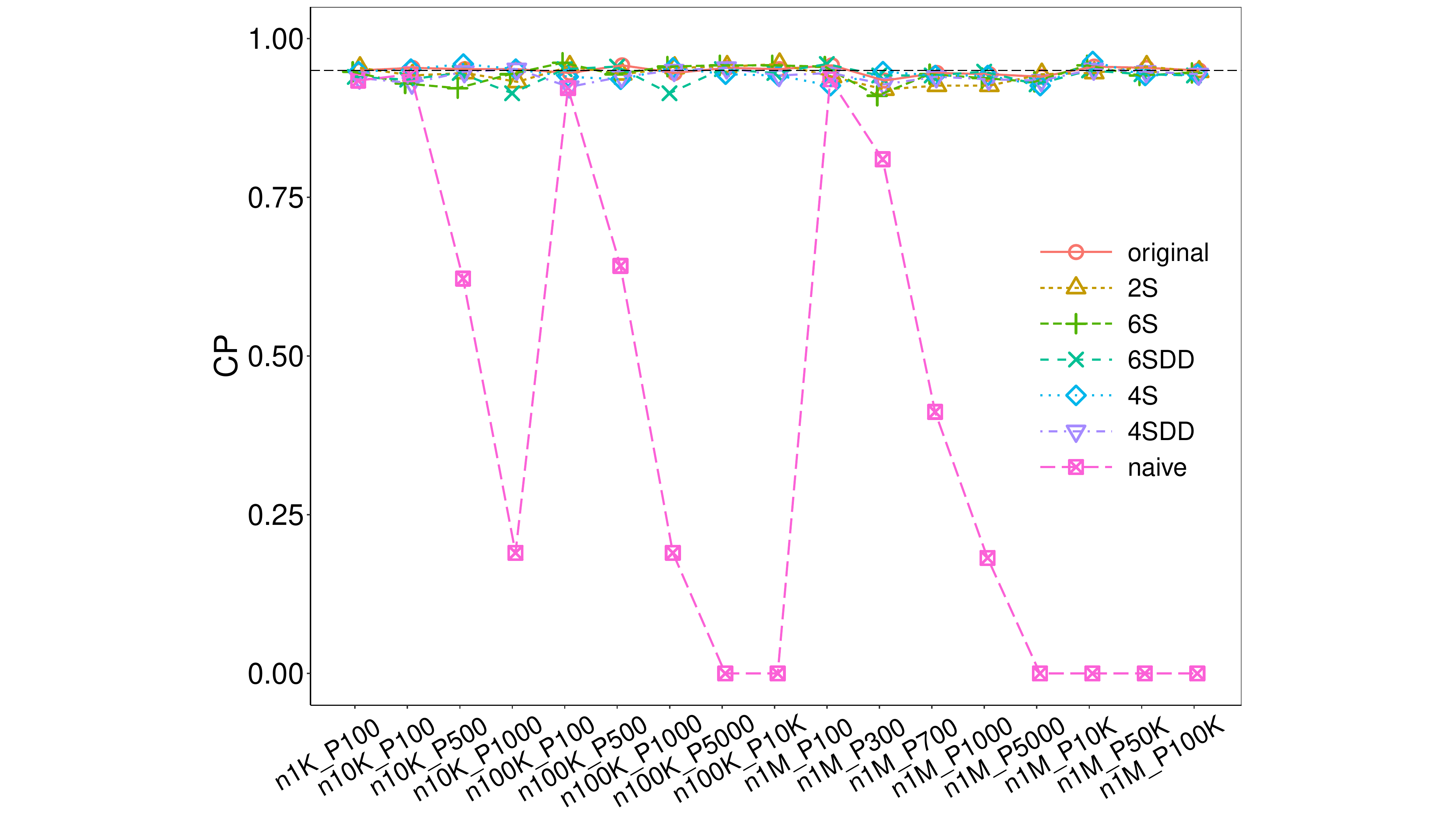}
\includegraphics[width=0.26\textwidth, trim={2.2in 0 2.2in 0},clip] {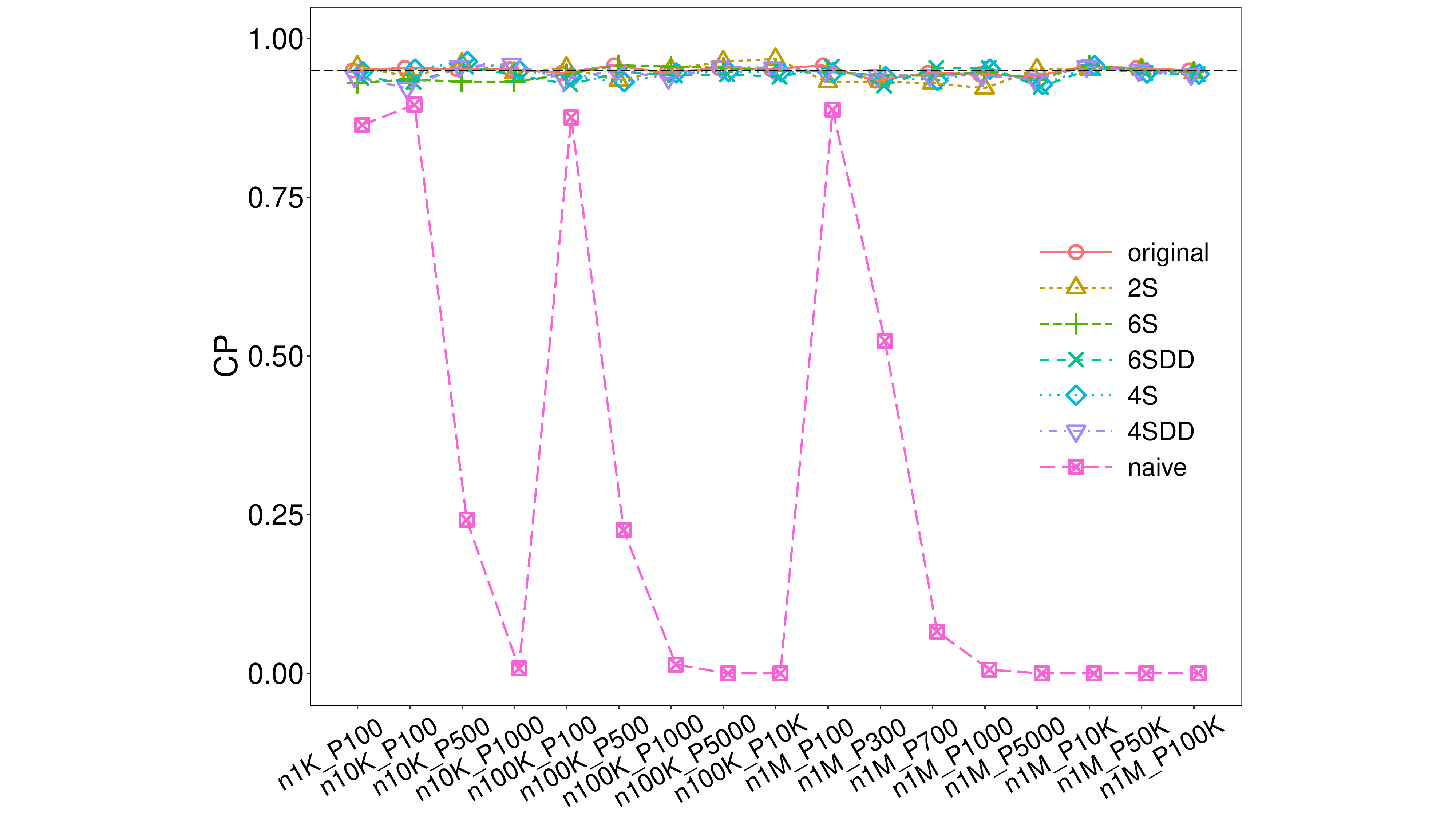}
\includegraphics[width=0.26\textwidth, trim={2.2in 0 2.2in 0},clip] {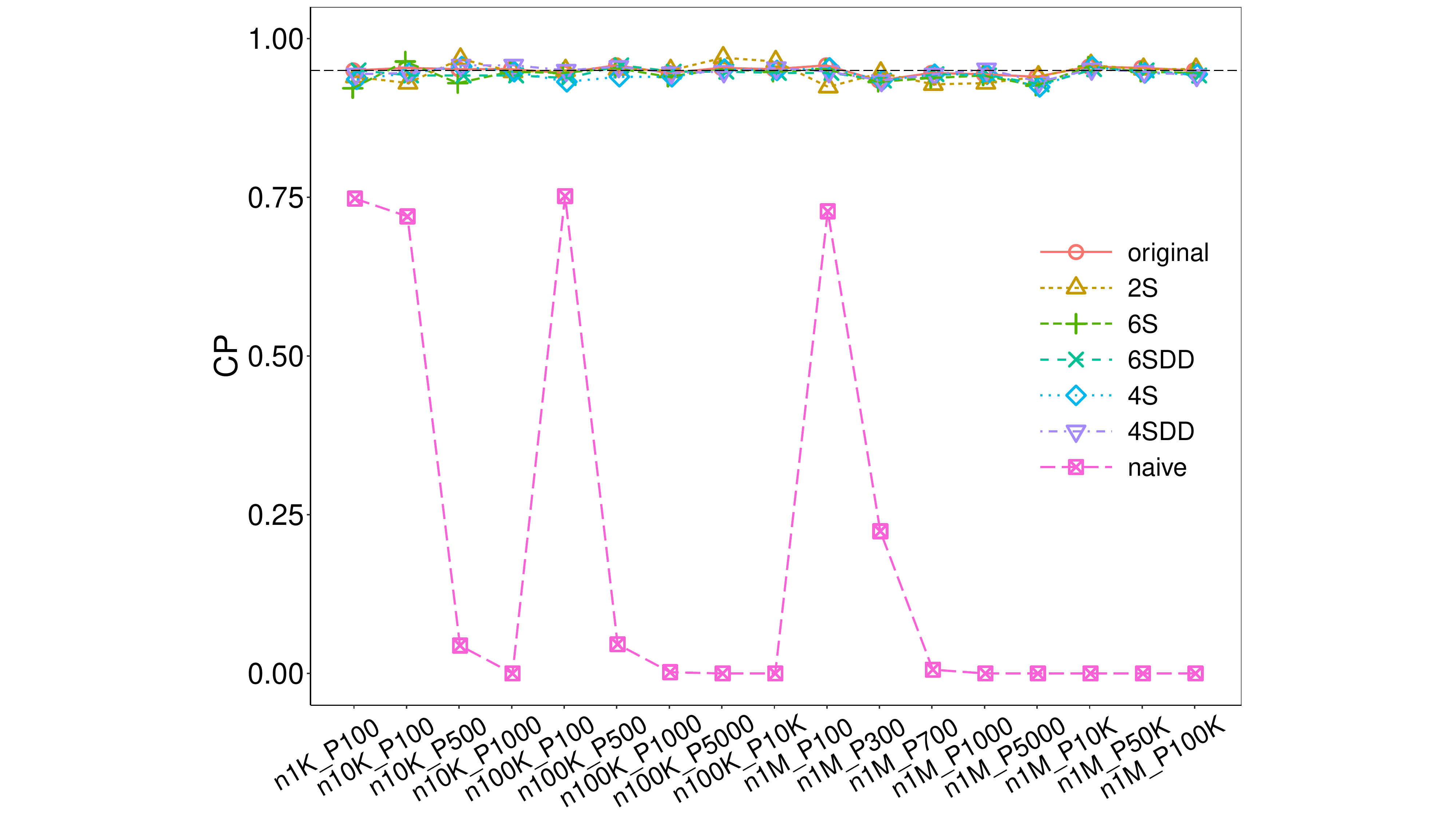}
\includegraphics[width=0.26\textwidth, trim={2.2in 0 2.2in 0},clip] {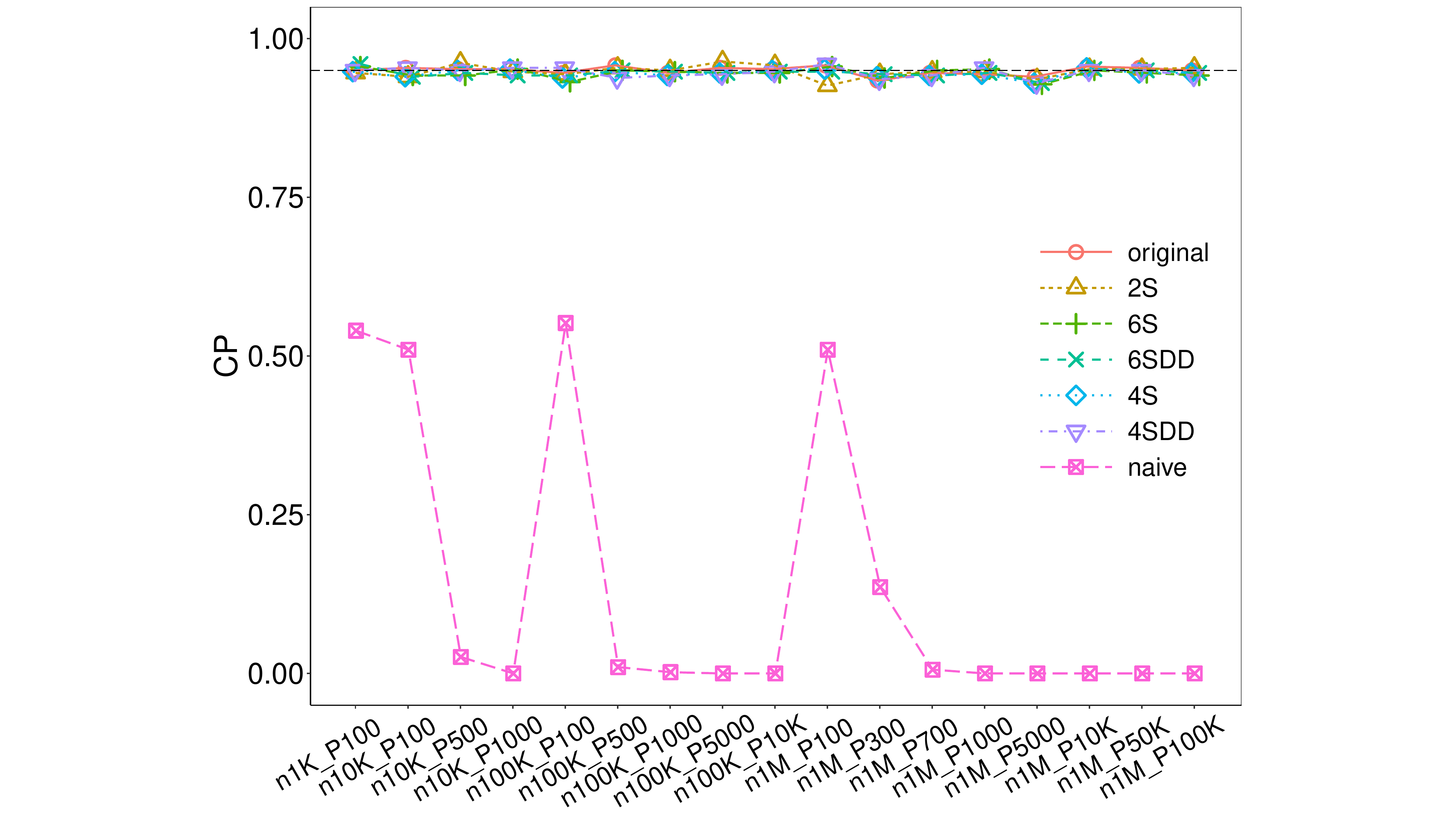}\\
\caption{Gaussian data; $\rho$-zCDP; $\theta=0$ and $\alpha\ne\beta$}
\label{fig:0aszCDP}
\end{figure}
\end{landscape}

\begin{landscape}
\subsection*{Gaussian, $\theta\ne0$ and $\alpha\ne\beta$}
\begin{figure}[!htb]
\centering
\centering
$\epsilon=0.5$\hspace{0.9in}$\epsilon=1$\hspace{1in}$\epsilon=2$
\hspace{1in}$\epsilon=5$\hspace{0.9in}$\epsilon=50$\\
\includegraphics[width=0.215\textwidth, trim={2.2in 0 2.2in 0},clip] {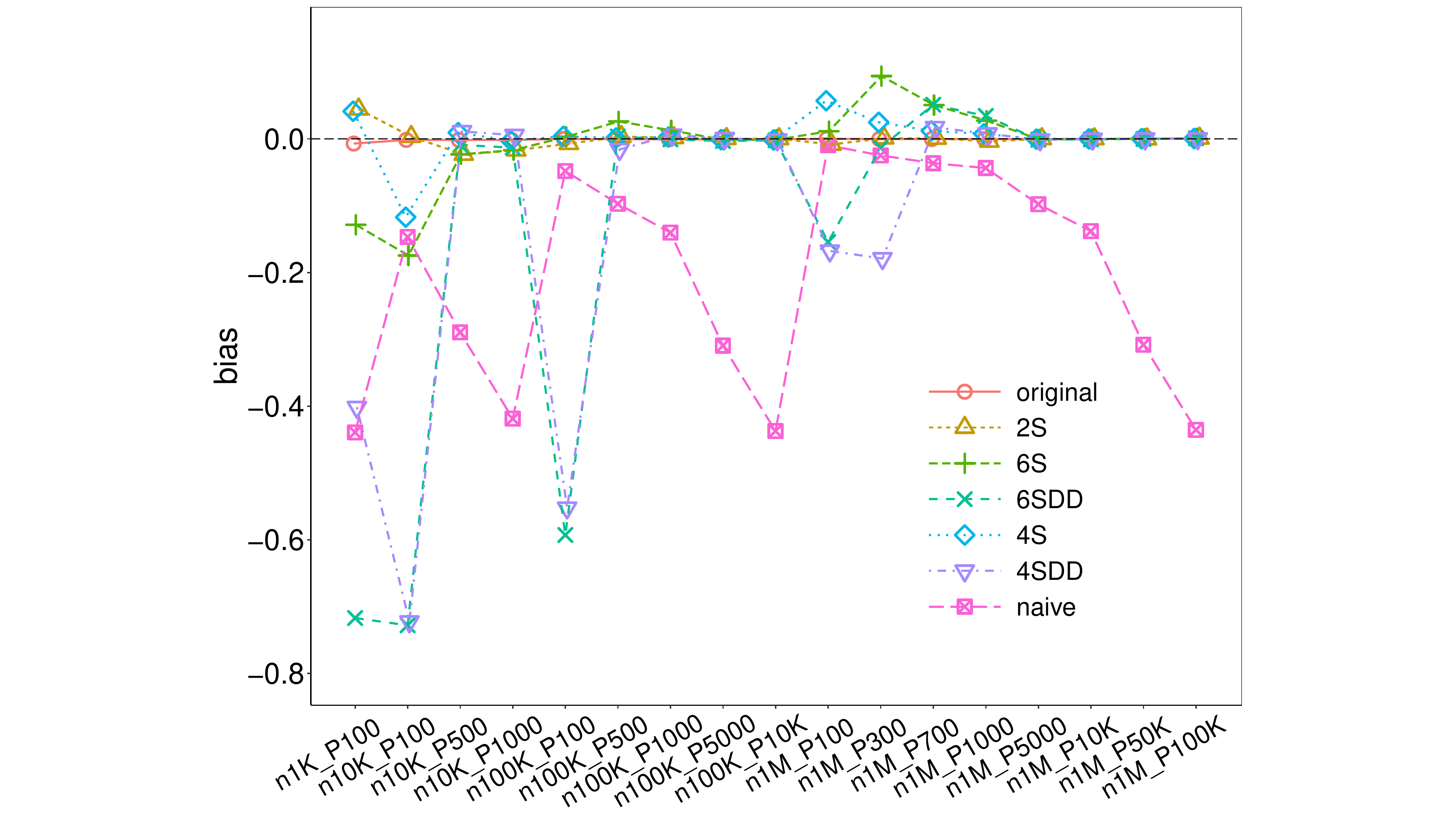}
\includegraphics[width=0.215\textwidth, trim={2.2in 0 2.2in 0},clip] {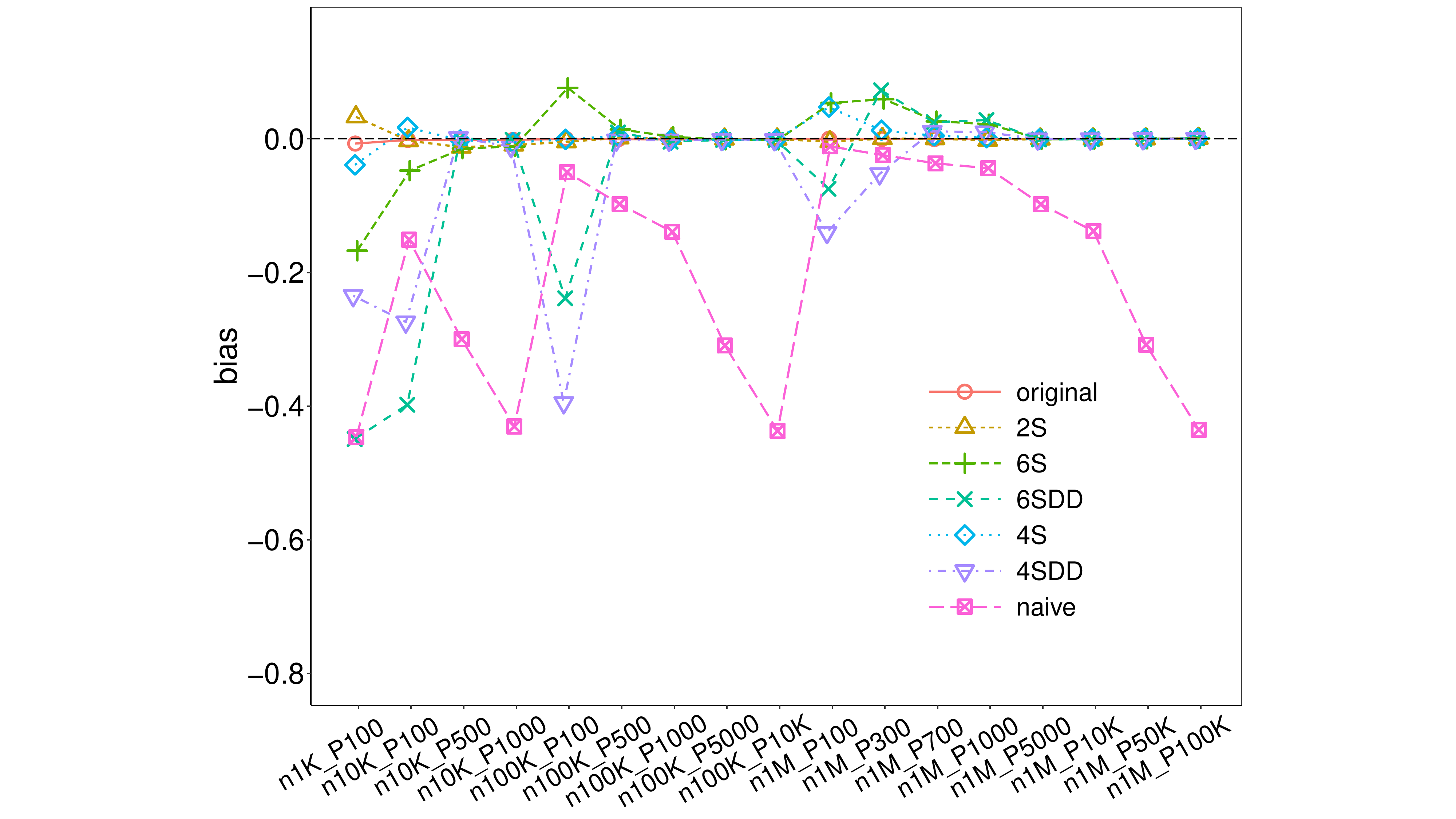}
\includegraphics[width=0.215\textwidth, trim={2.2in 0 2.2in 0},clip] {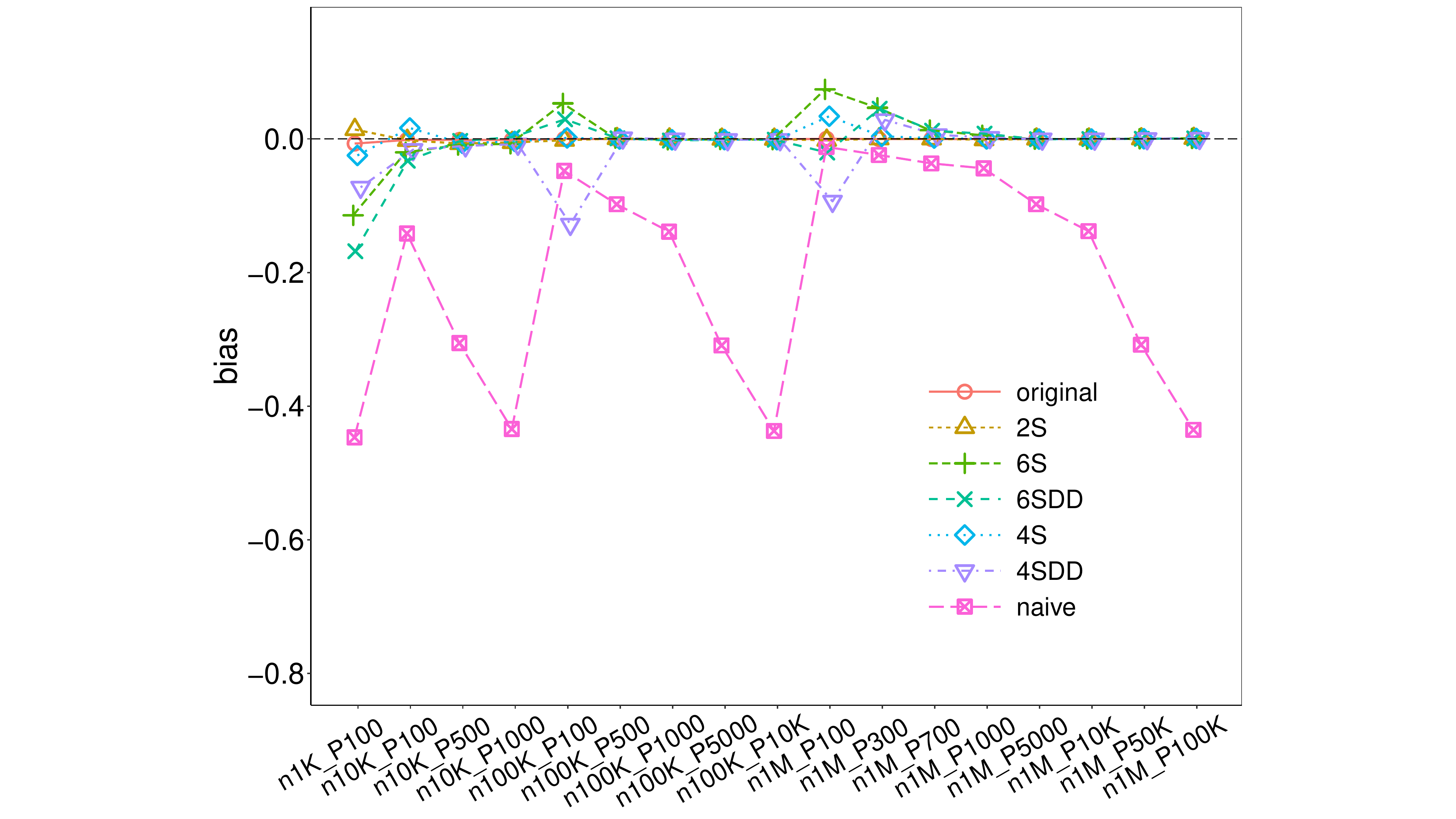}
\includegraphics[width=0.215\textwidth, trim={2.2in 0 2.2in 0},clip] {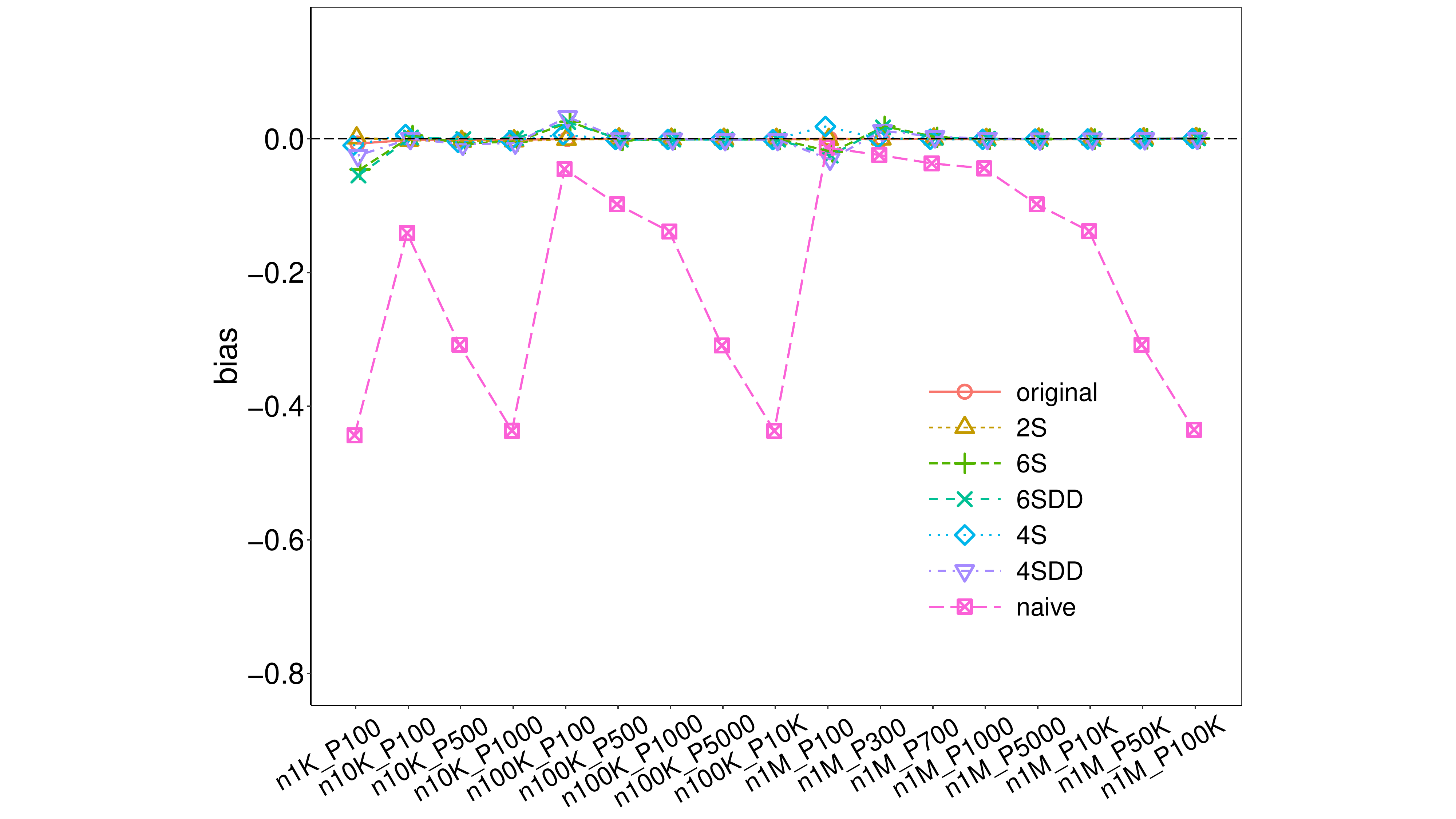}
\includegraphics[width=0.215\textwidth, trim={2.2in 0 2.2in 0},clip] {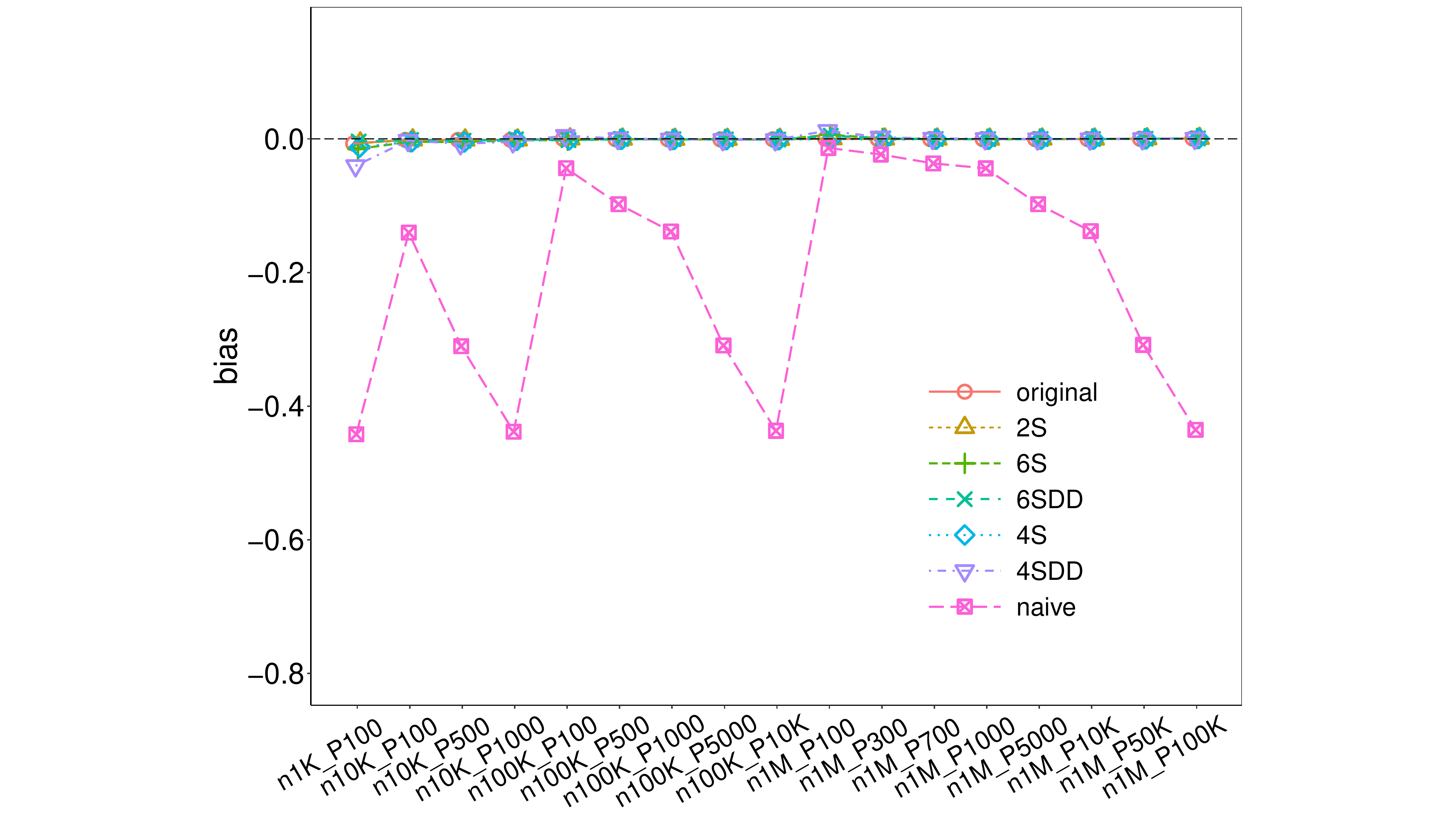}\\
\includegraphics[width=0.215\textwidth, trim={2.2in 0 2.2in 0},clip] {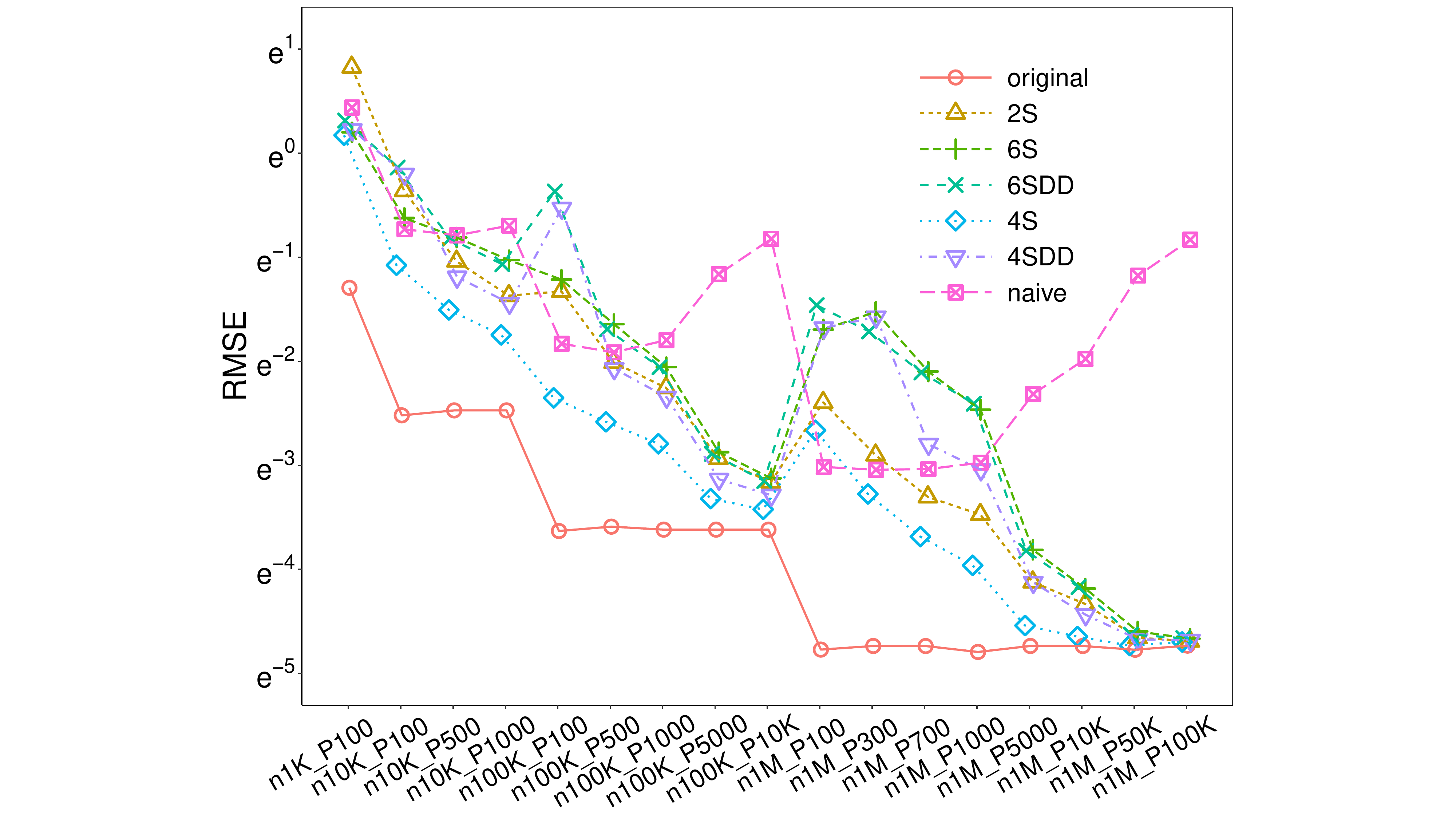}
\includegraphics[width=0.215\textwidth, trim={2.2in 0 2.2in 0},clip] {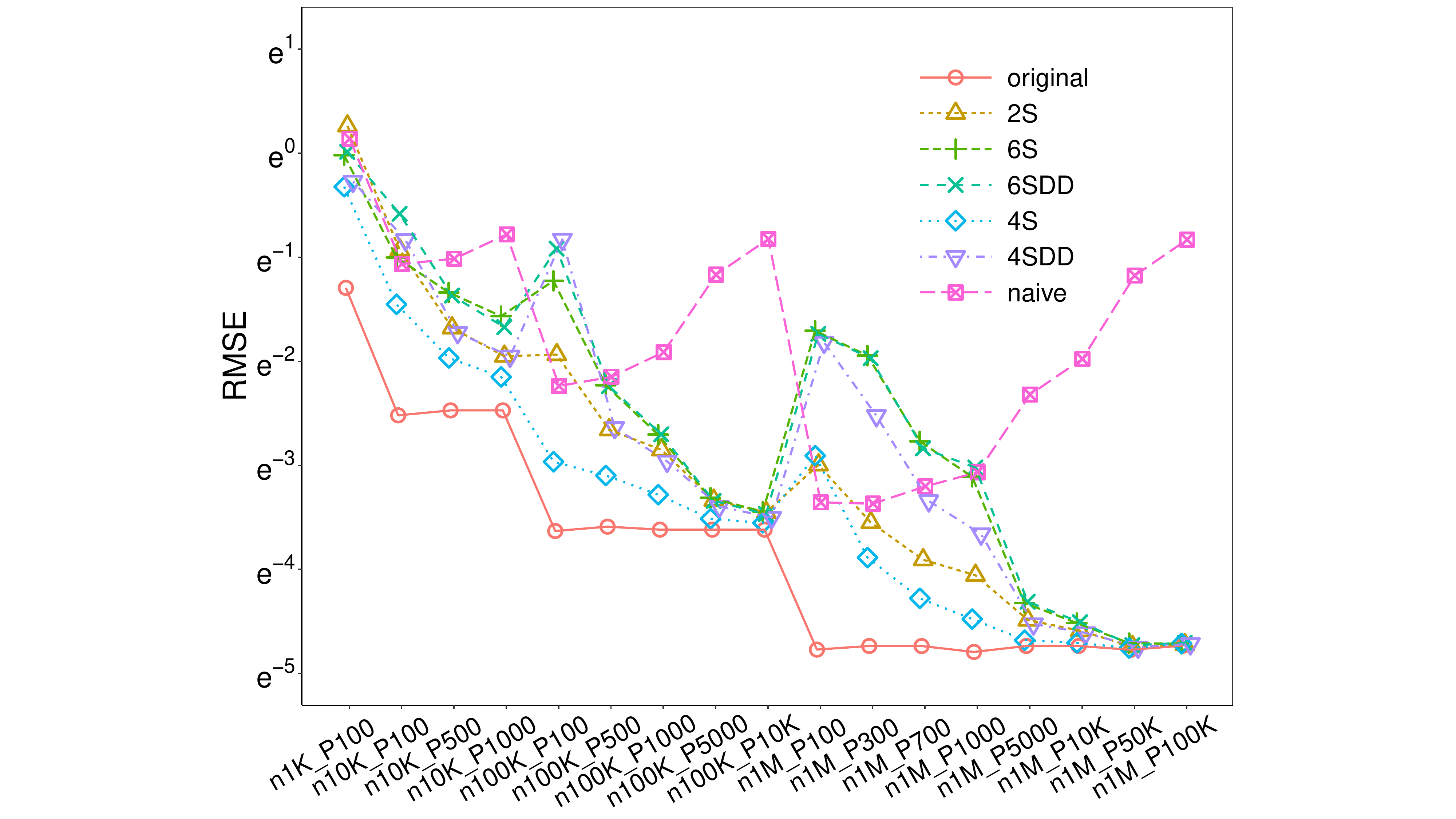}
\includegraphics[width=0.215\textwidth, trim={2.2in 0 2.2in 0},clip] {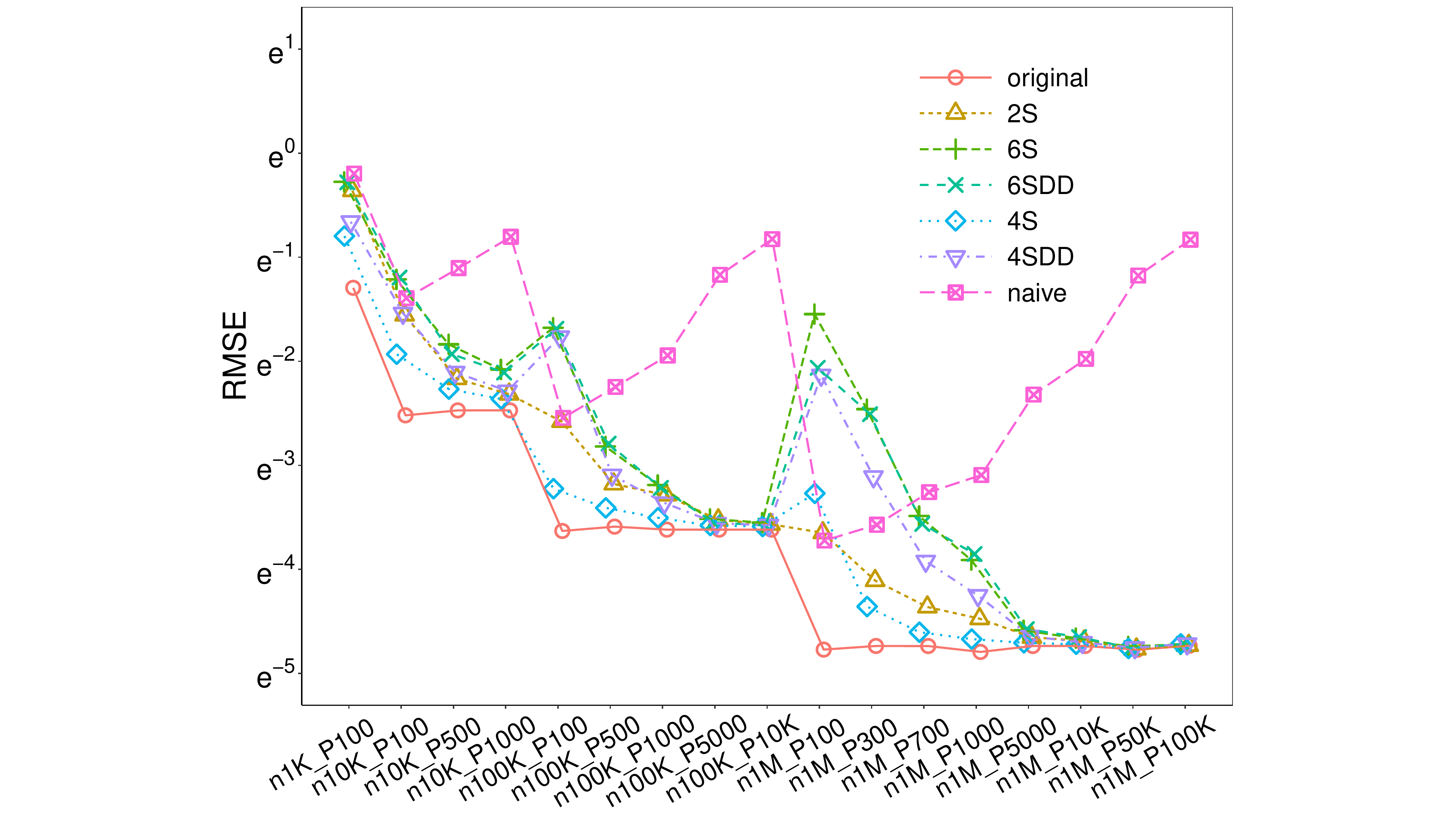}
\includegraphics[width=0.215\textwidth, trim={2.2in 0 2.2in 0},clip] {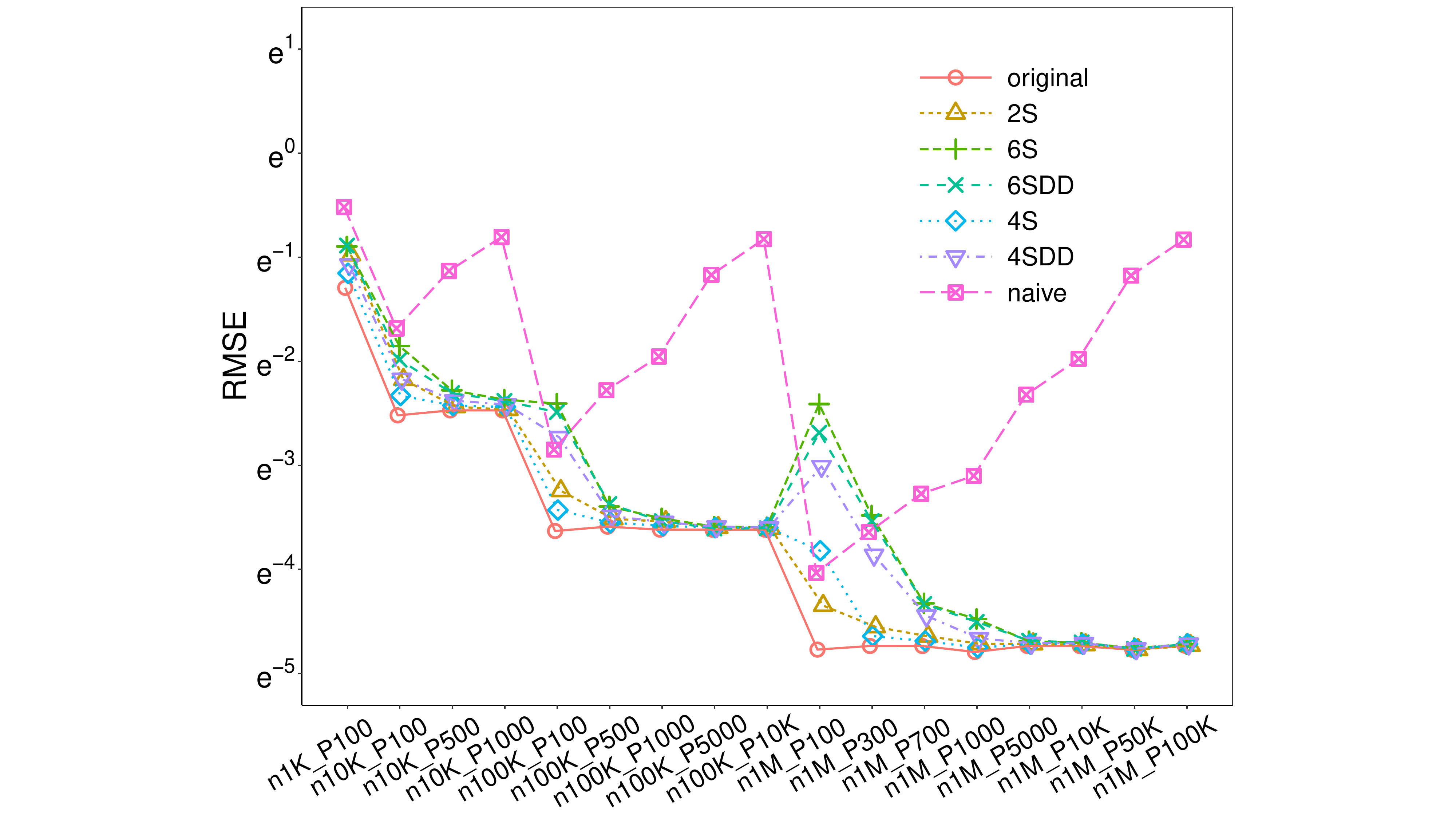}
\includegraphics[width=0.215\textwidth, trim={2.2in 0 2.2in 0},clip] {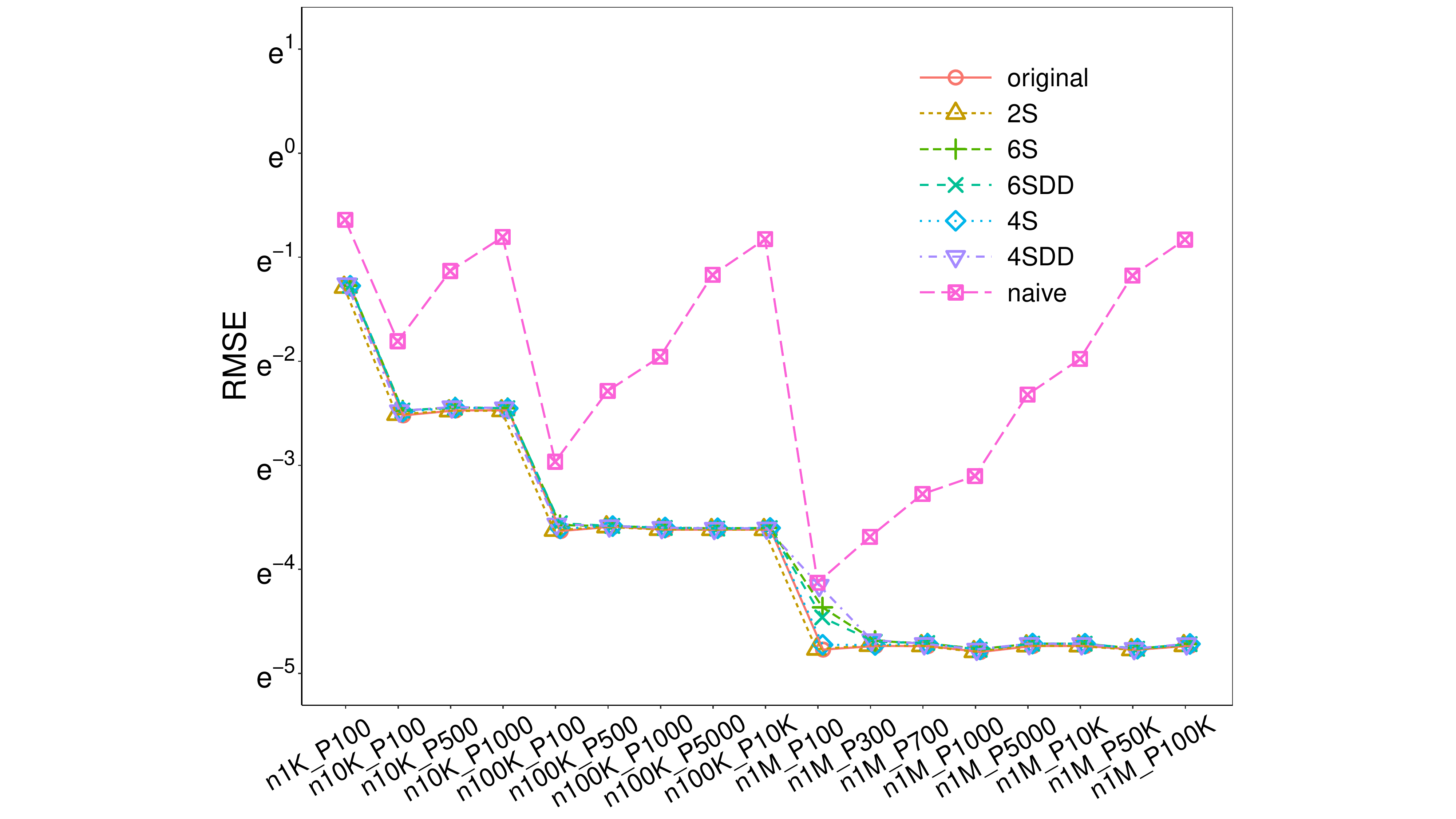}\\
\includegraphics[width=0.215\textwidth, trim={2.2in 0 2.2in 0},clip] {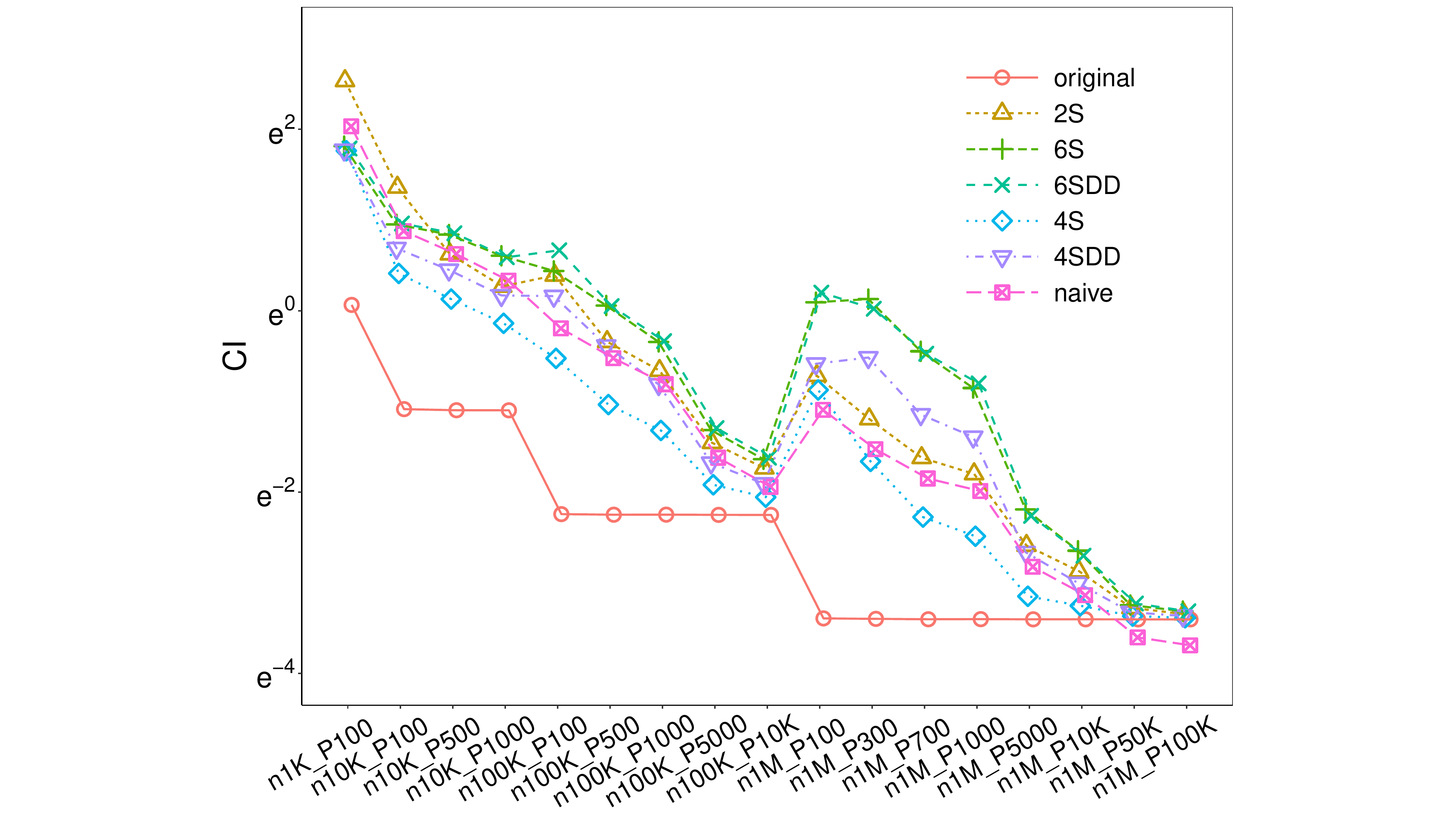}
\includegraphics[width=0.215\textwidth, trim={2.2in 0 2.2in 0},clip] {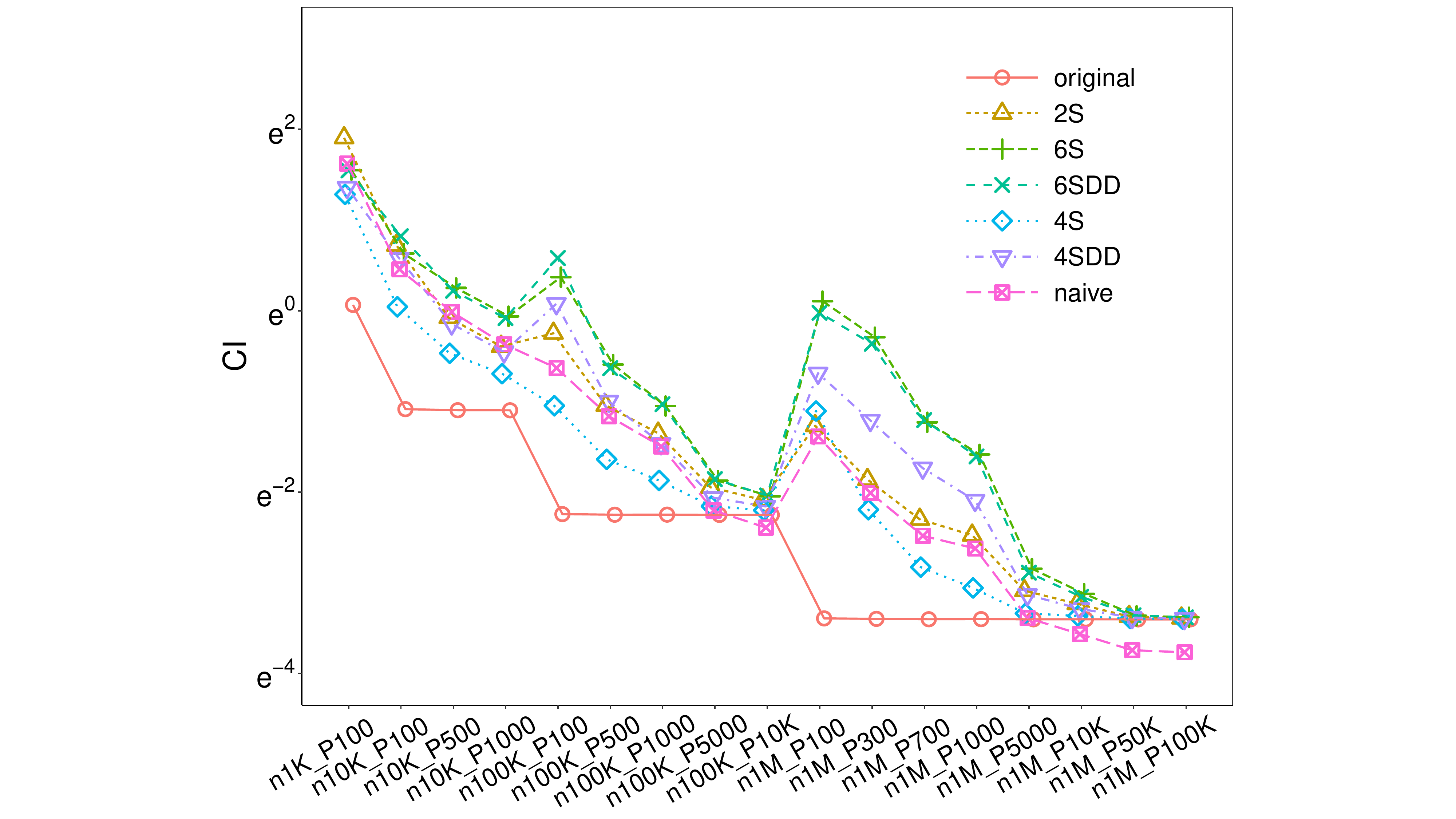}
\includegraphics[width=0.215\textwidth, trim={2.2in 0 2.2in 0},clip] {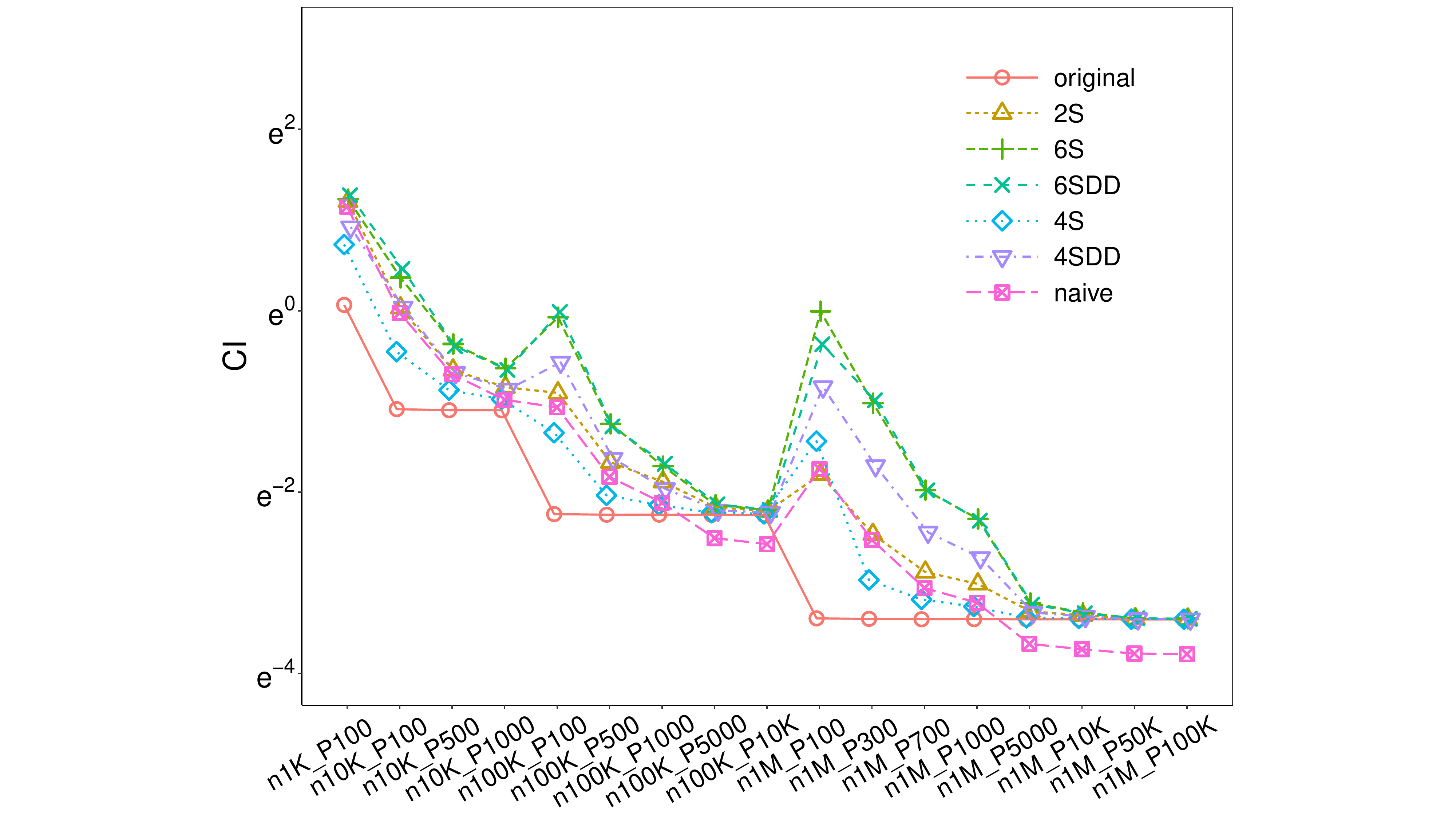}
\includegraphics[width=0.215\textwidth, trim={2.2in 0 2.2in 0},clip] {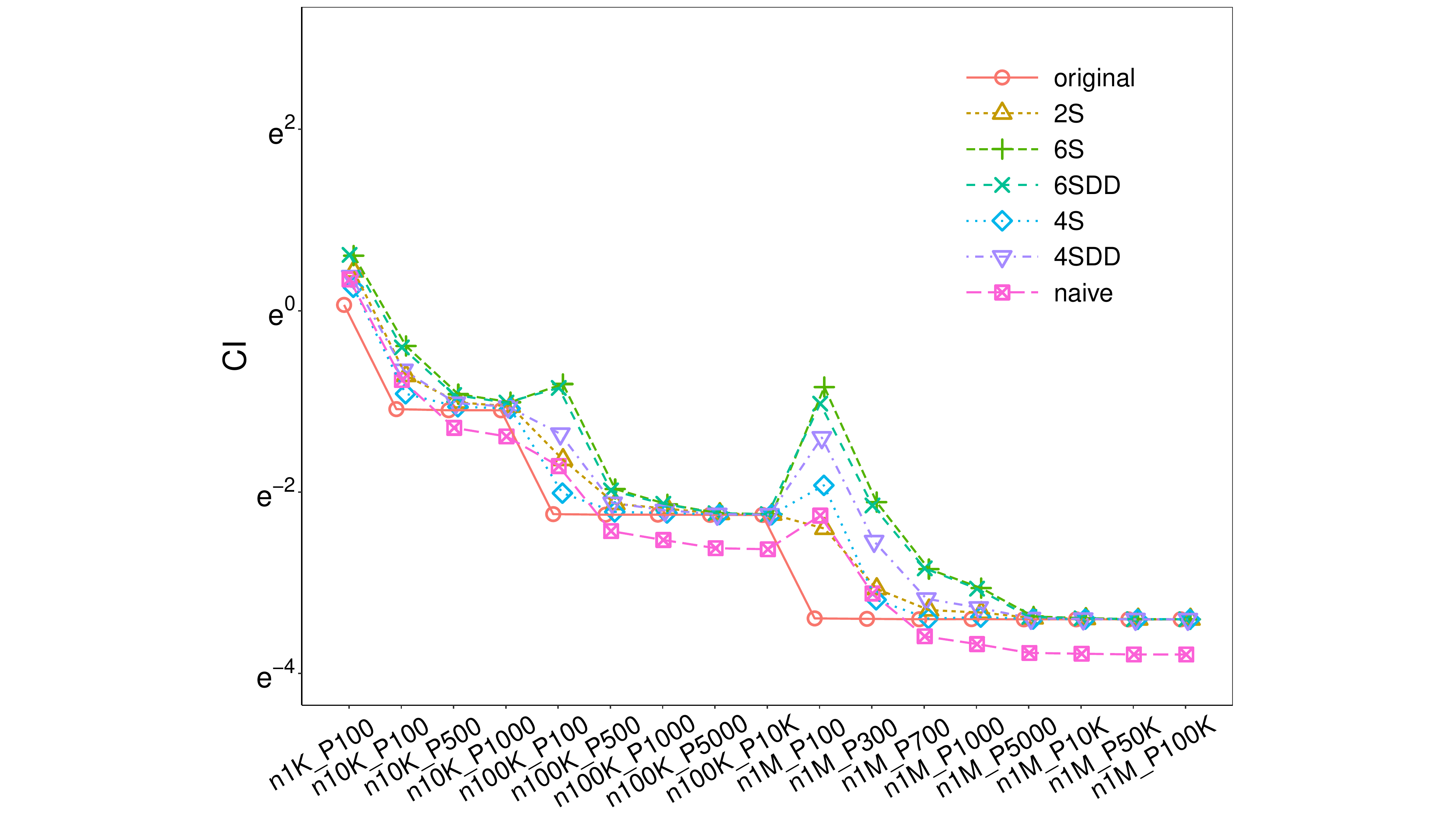}
\includegraphics[width=0.215\textwidth, trim={2.2in 0 2.2in 0},clip] {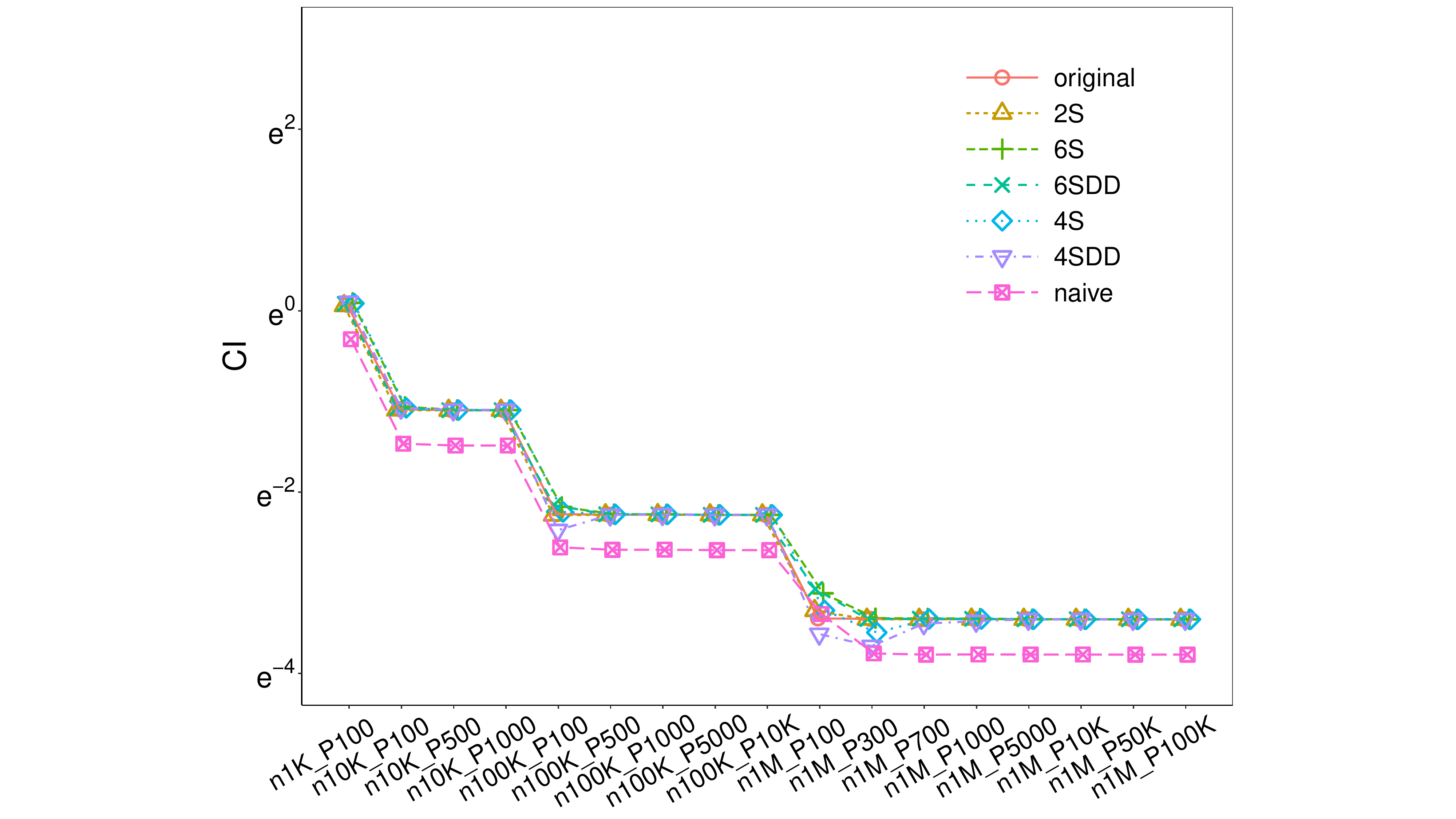}\\
\includegraphics[width=0.215\textwidth, trim={2.2in 0 2.2in 0},clip] {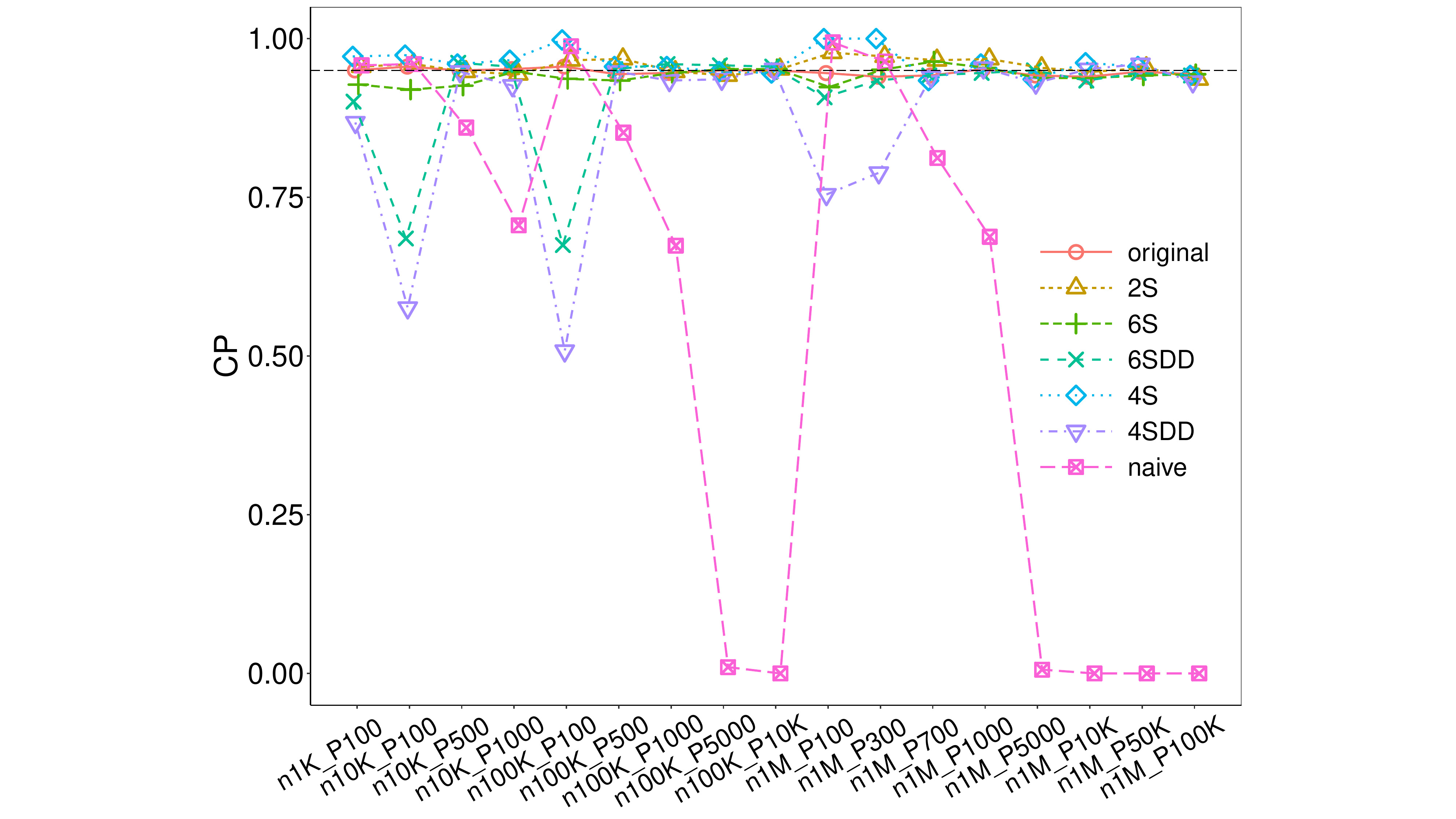}
\includegraphics[width=0.215\textwidth, trim={2.2in 0 2.2in 0},clip] {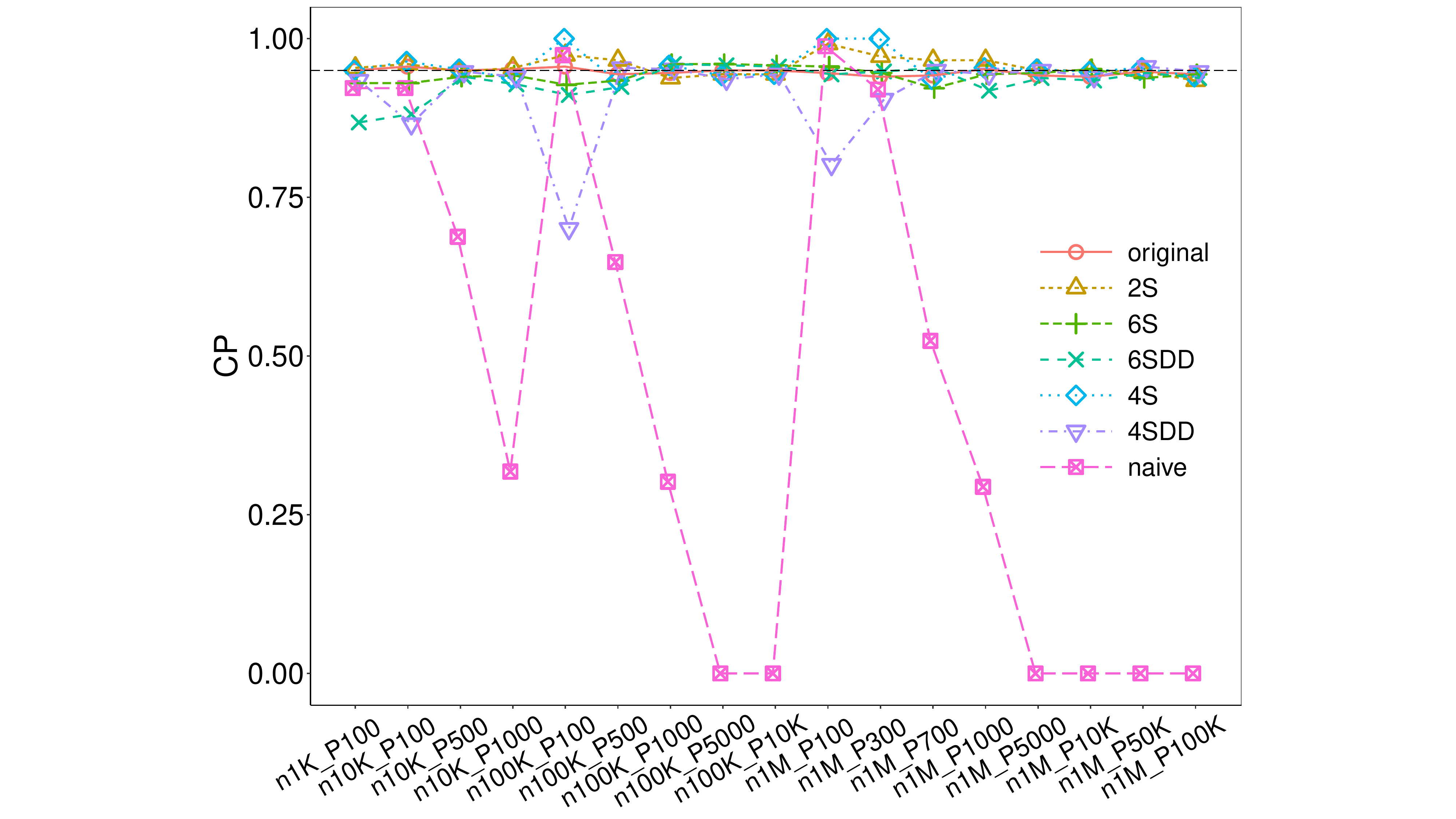}
\includegraphics[width=0.215\textwidth, trim={2.2in 0 2.2in 0},clip] {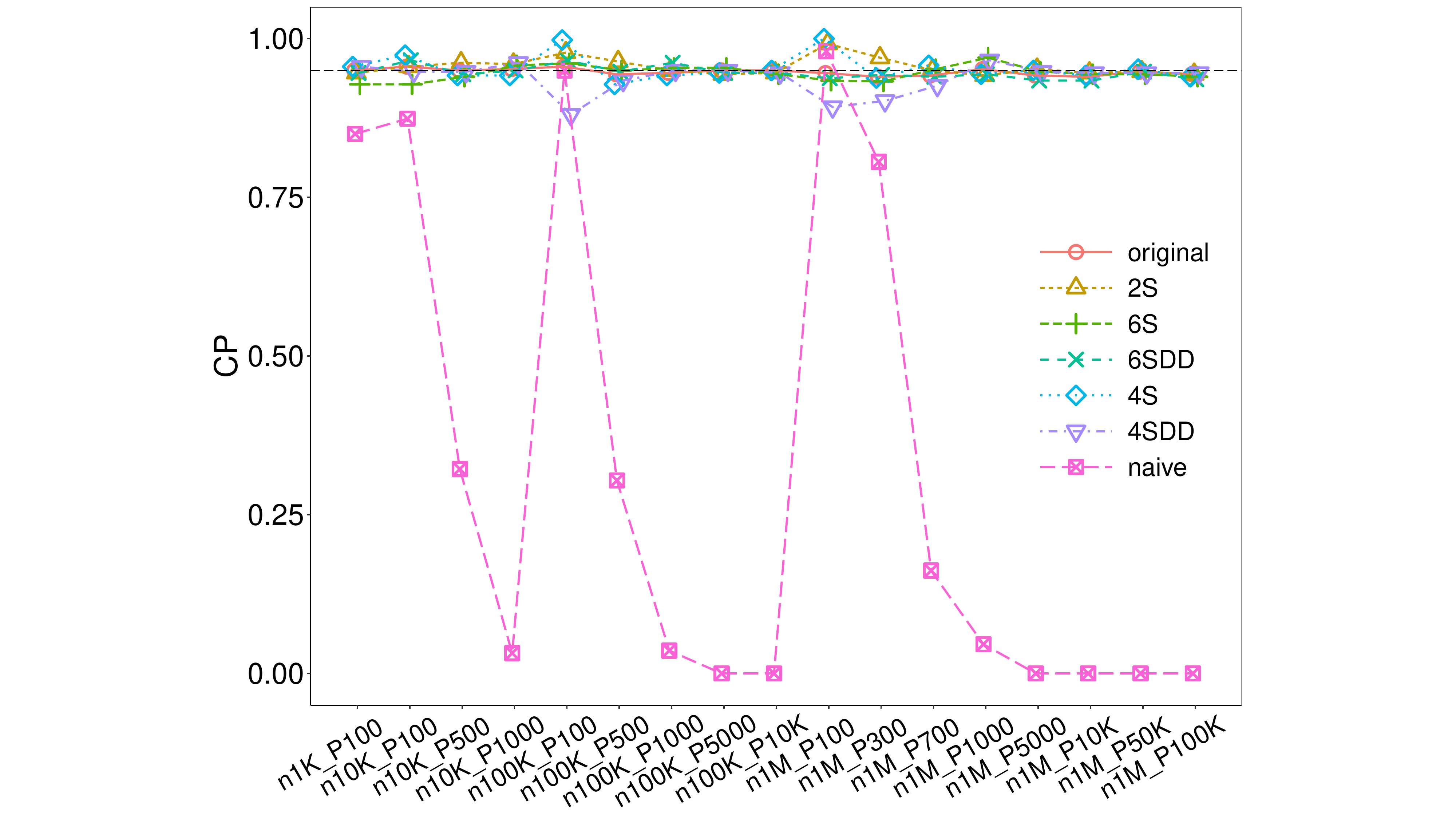}
\includegraphics[width=0.215\textwidth, trim={2.2in 0 2.2in 0},clip] {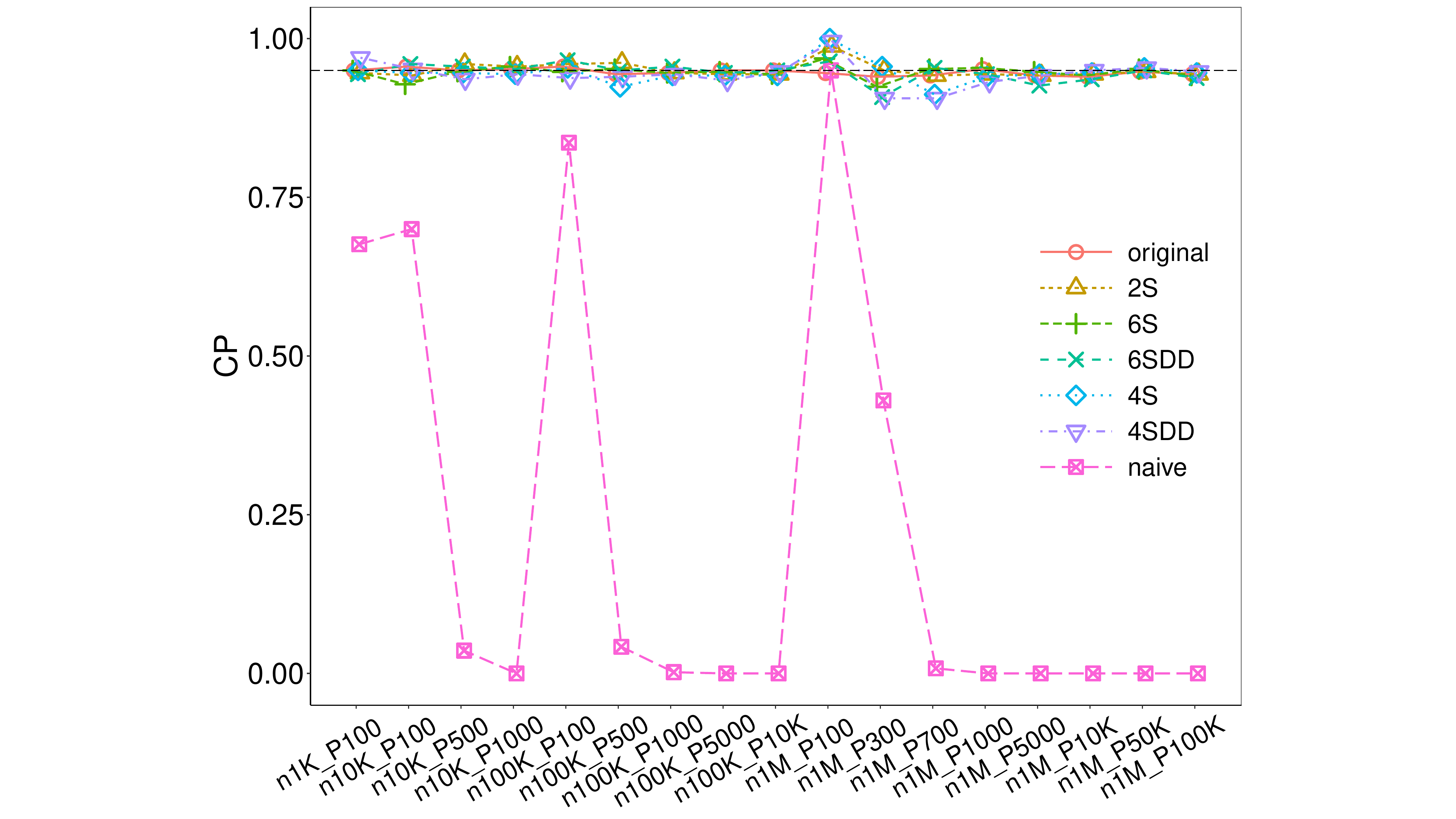}
\includegraphics[width=0.215\textwidth, trim={2.2in 0 2.2in 0},clip] {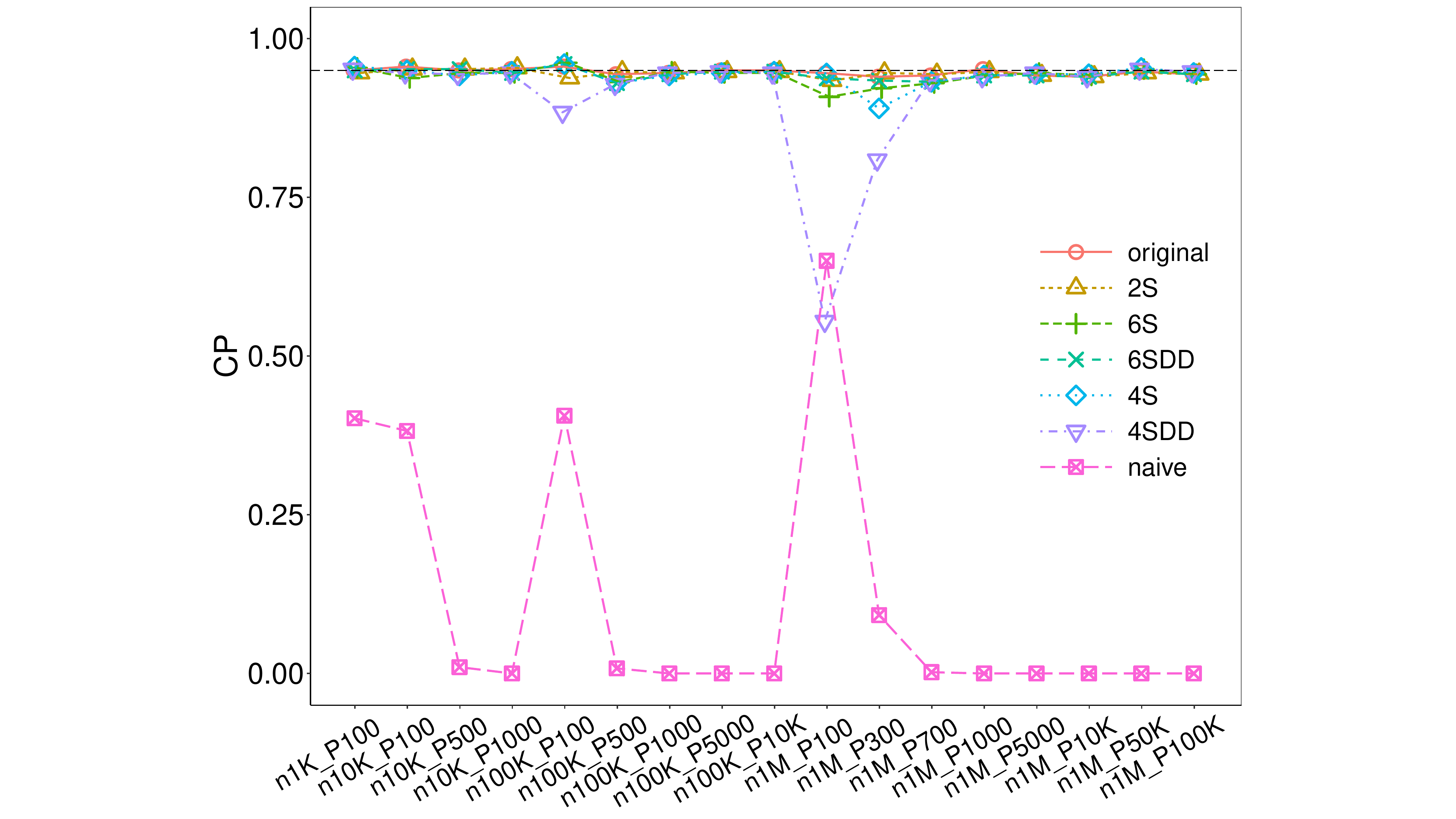}\\
\includegraphics[width=0.215\textwidth, trim={2.2in 0 2.2in 0},clip] {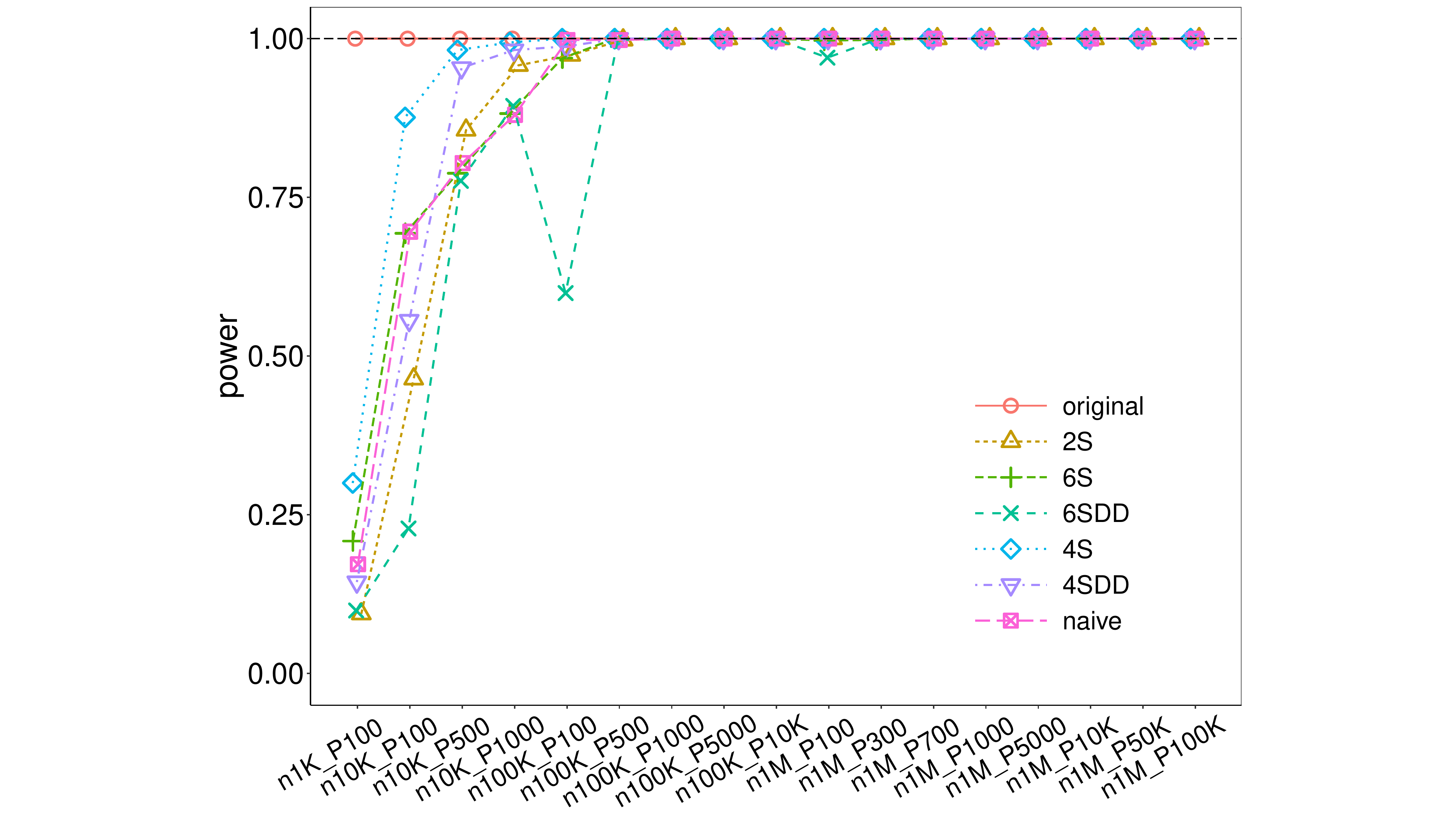}
\includegraphics[width=0.215\textwidth, trim={2.2in 0 2.2in 0},clip] {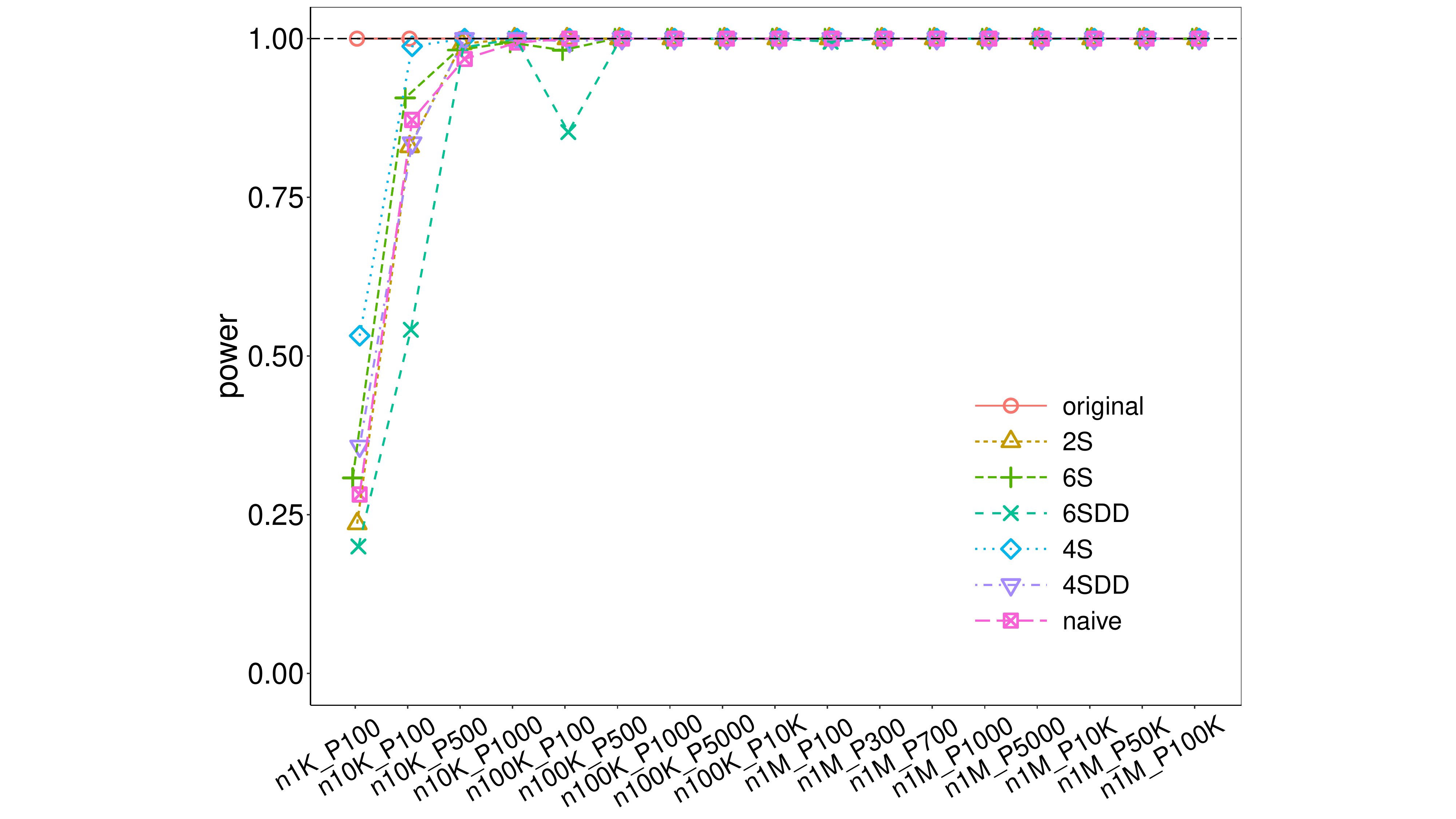}
\includegraphics[width=0.215\textwidth, trim={2.2in 0 2.2in 0},clip] {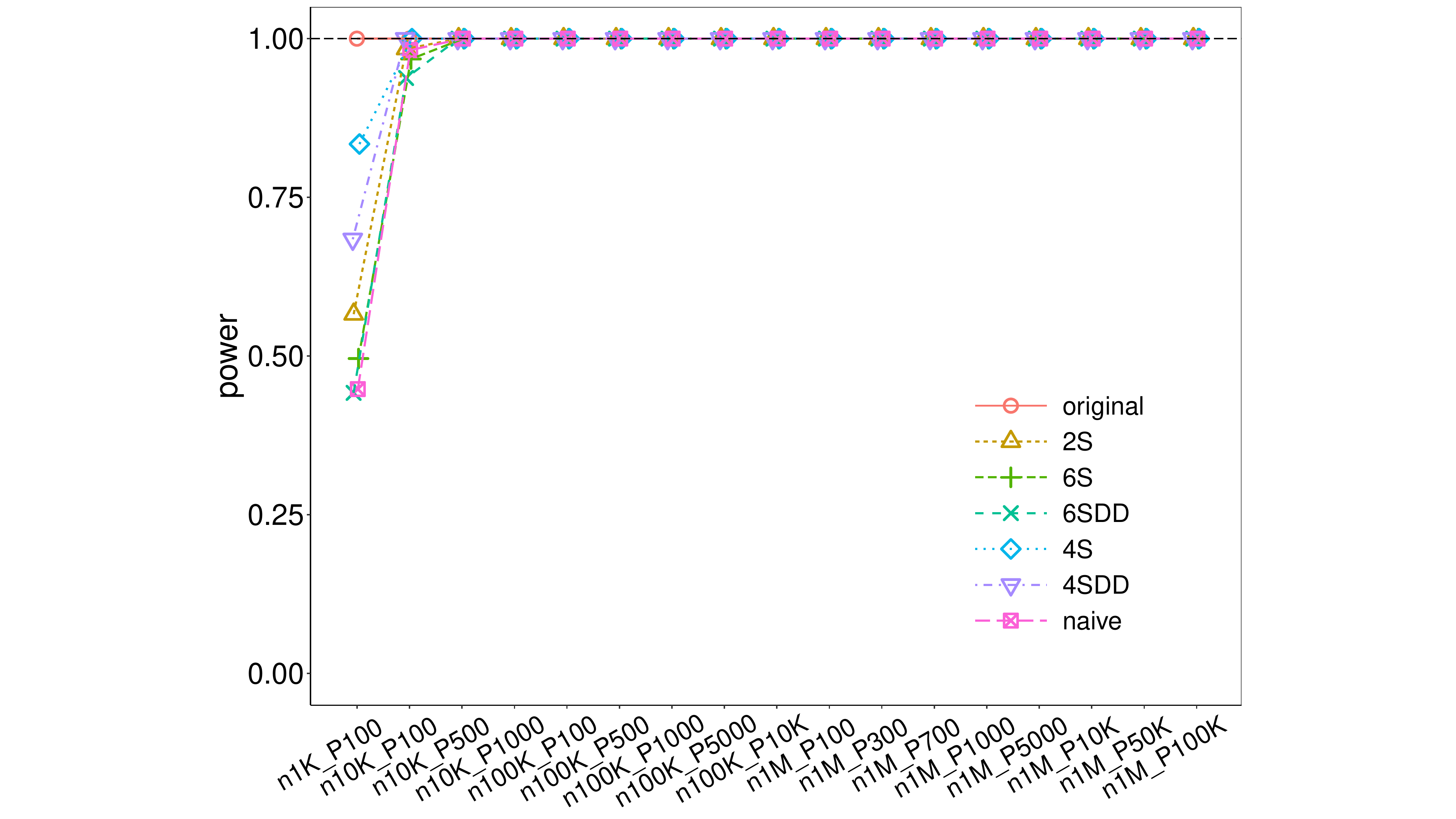}
\includegraphics[width=0.215\textwidth, trim={2.2in 0 2.2in 0},clip] {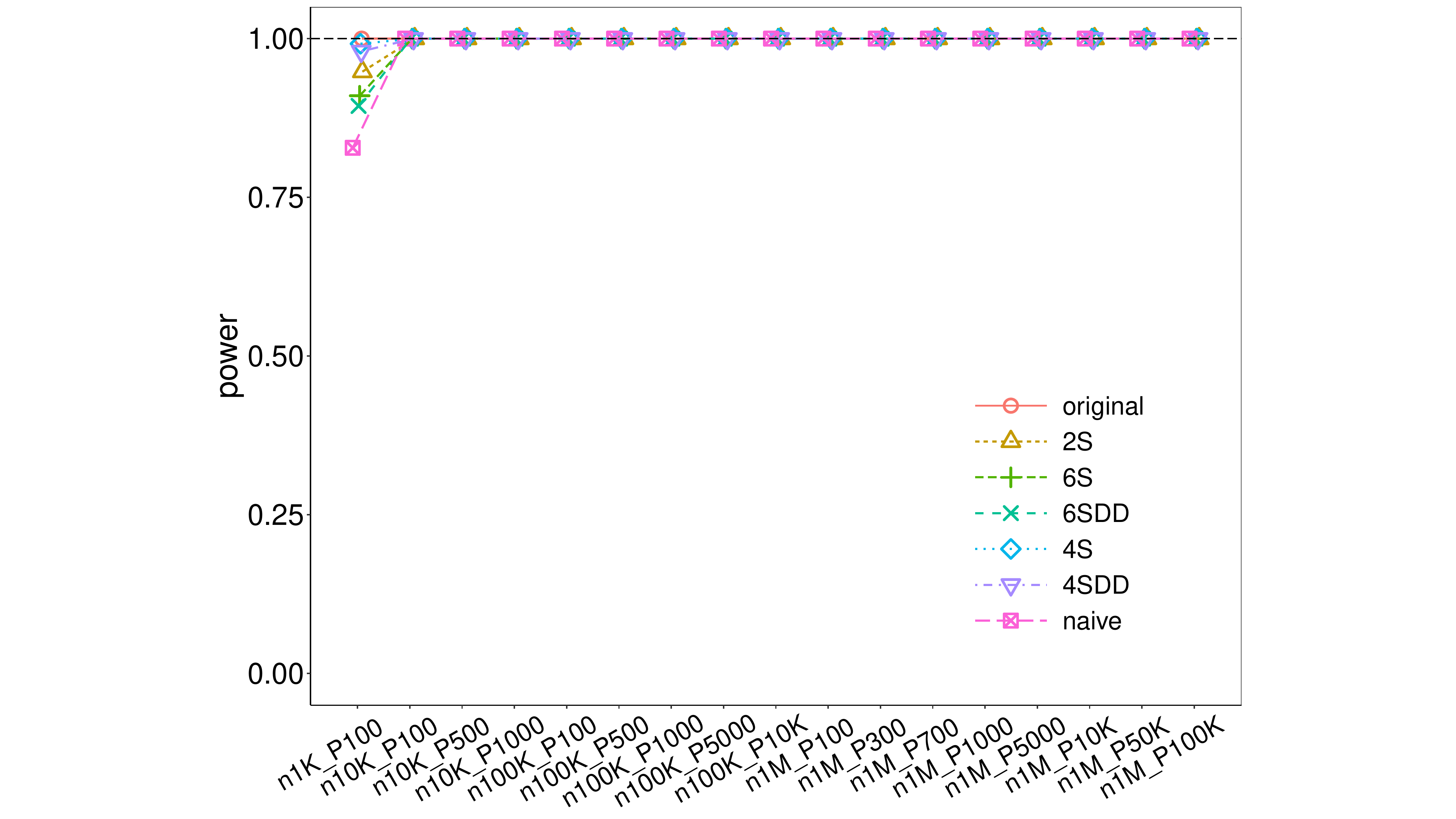}
\includegraphics[width=0.215\textwidth, trim={2.2in 0 2.2in 0},clip] {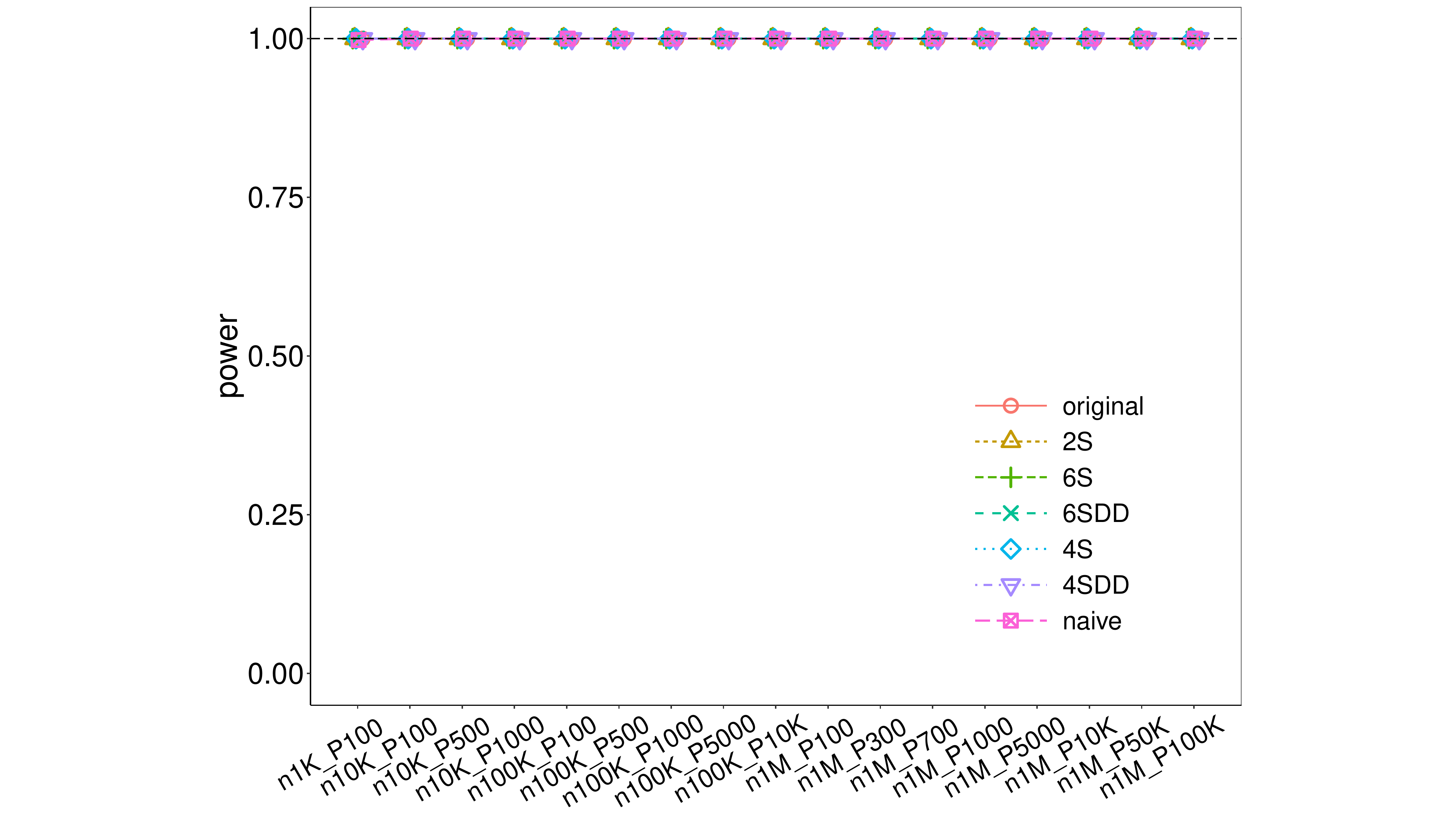}\\
\caption{Gaussian data; $\epsilon$-DP; $\theta\ne0$ and $\alpha\ne\beta$} \label{fig:1asDP}
\end{figure}
\end{landscape}

\begin{landscape}
\begin{figure}[!htb]
\centering
$\rho=0.005$\hspace{1in}$\rho=0.02$\hspace{1in}$\rho=0.08$
\hspace{1in}$\rho=0.32$\hspace{0.8in}$\rho=1.28$\\
\includegraphics[width=0.24\textwidth, trim={2.2in 0 2.2in 0},clip] {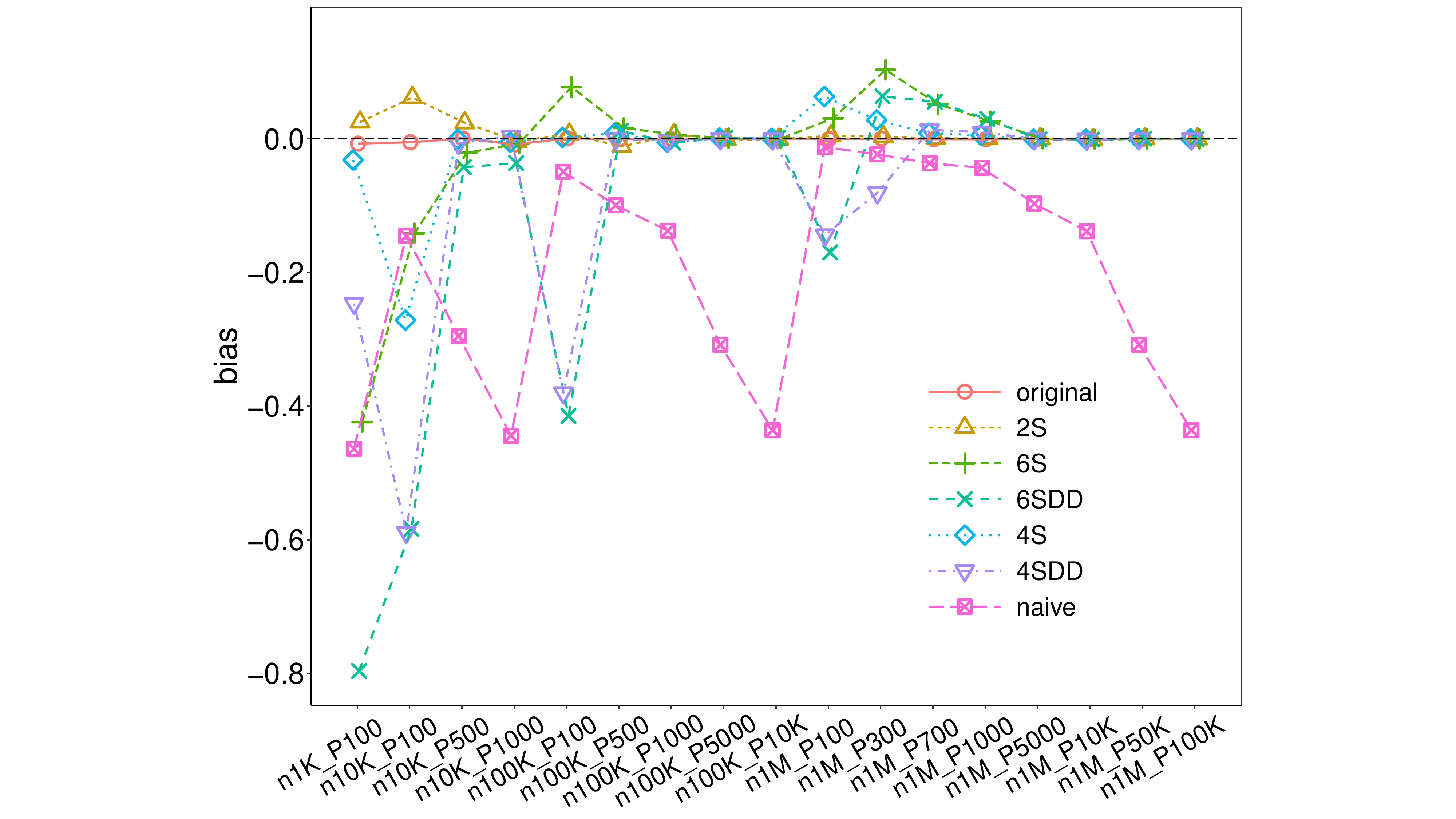}
\includegraphics[width=0.24\textwidth, trim={2.2in 0 2.2in 0},clip] {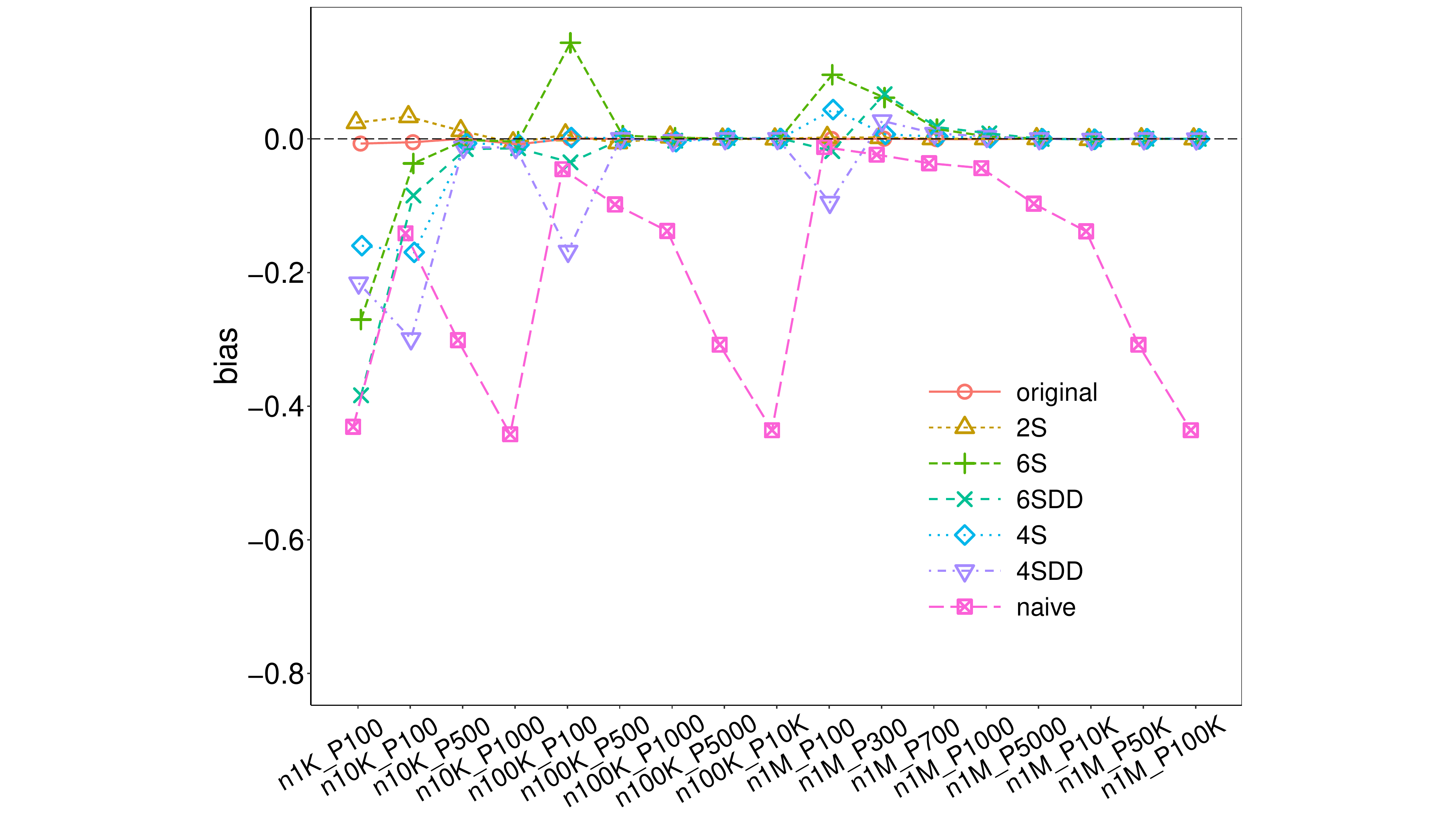}
\includegraphics[width=0.24\textwidth, trim={2.2in 0 2.2in 0},clip] {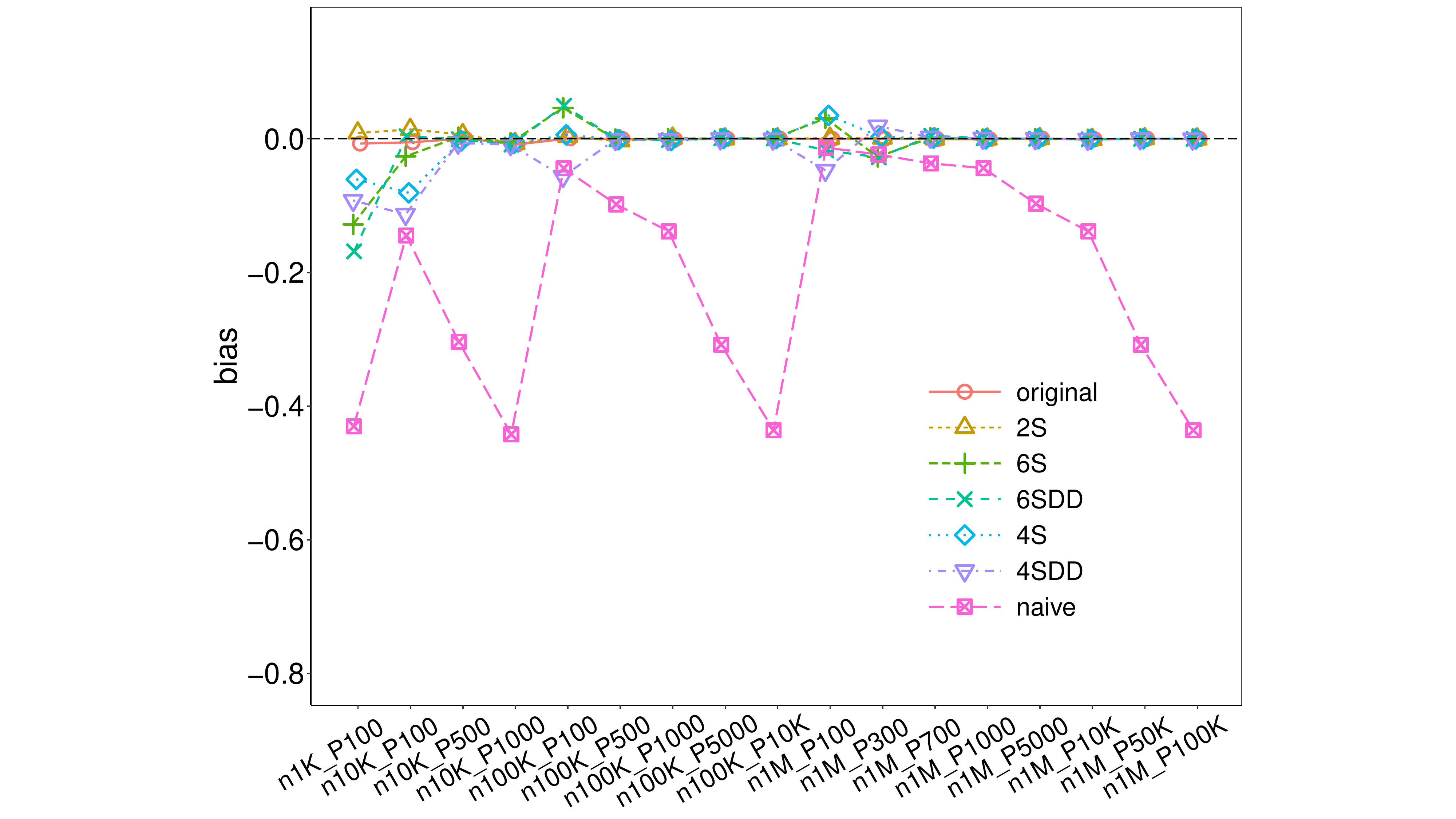}
\includegraphics[width=0.24\textwidth, trim={2.2in 0 2.2in 0},clip] {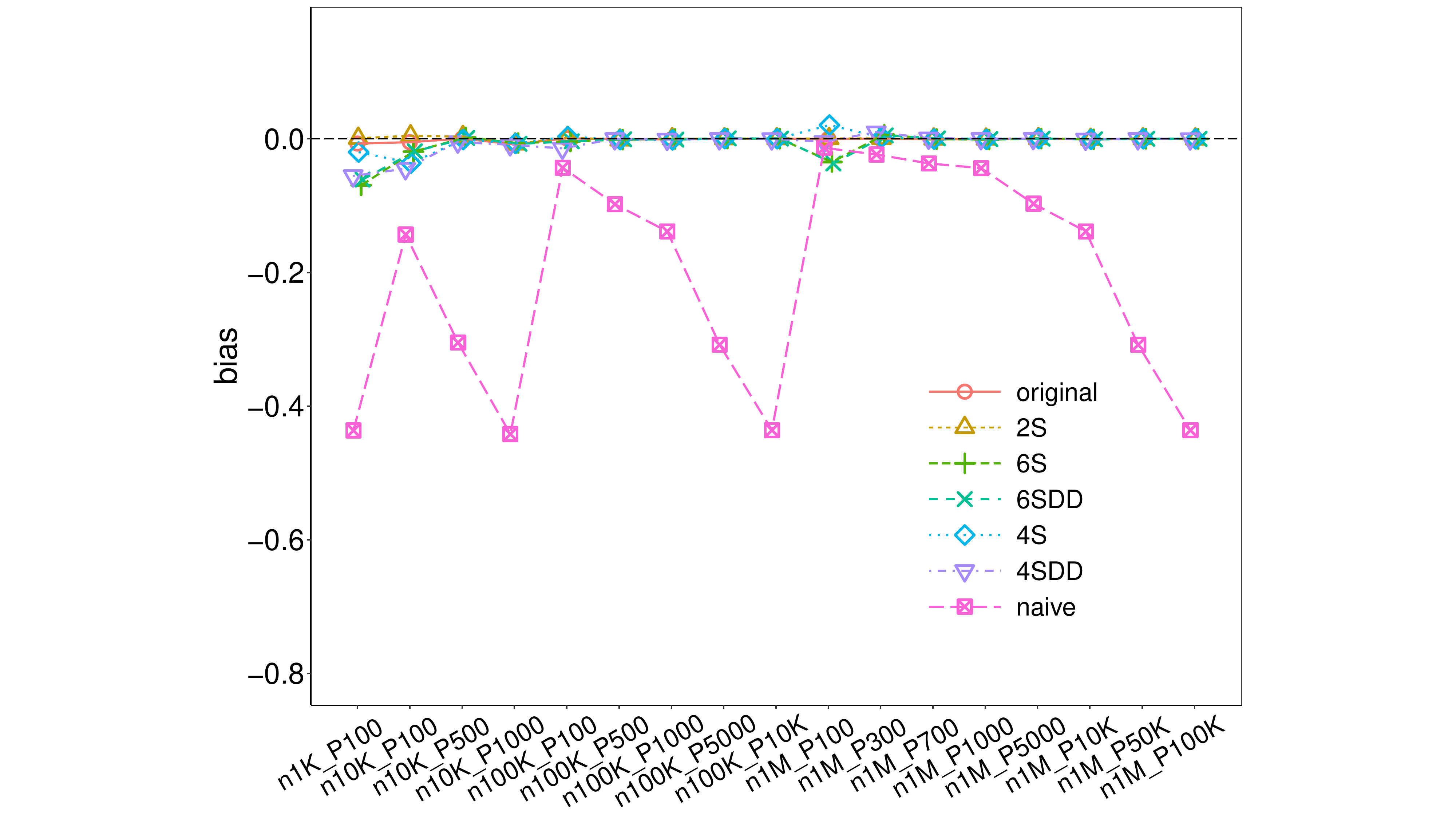}
\includegraphics[width=0.24\textwidth, trim={2.2in 0 2.2in 0},clip] {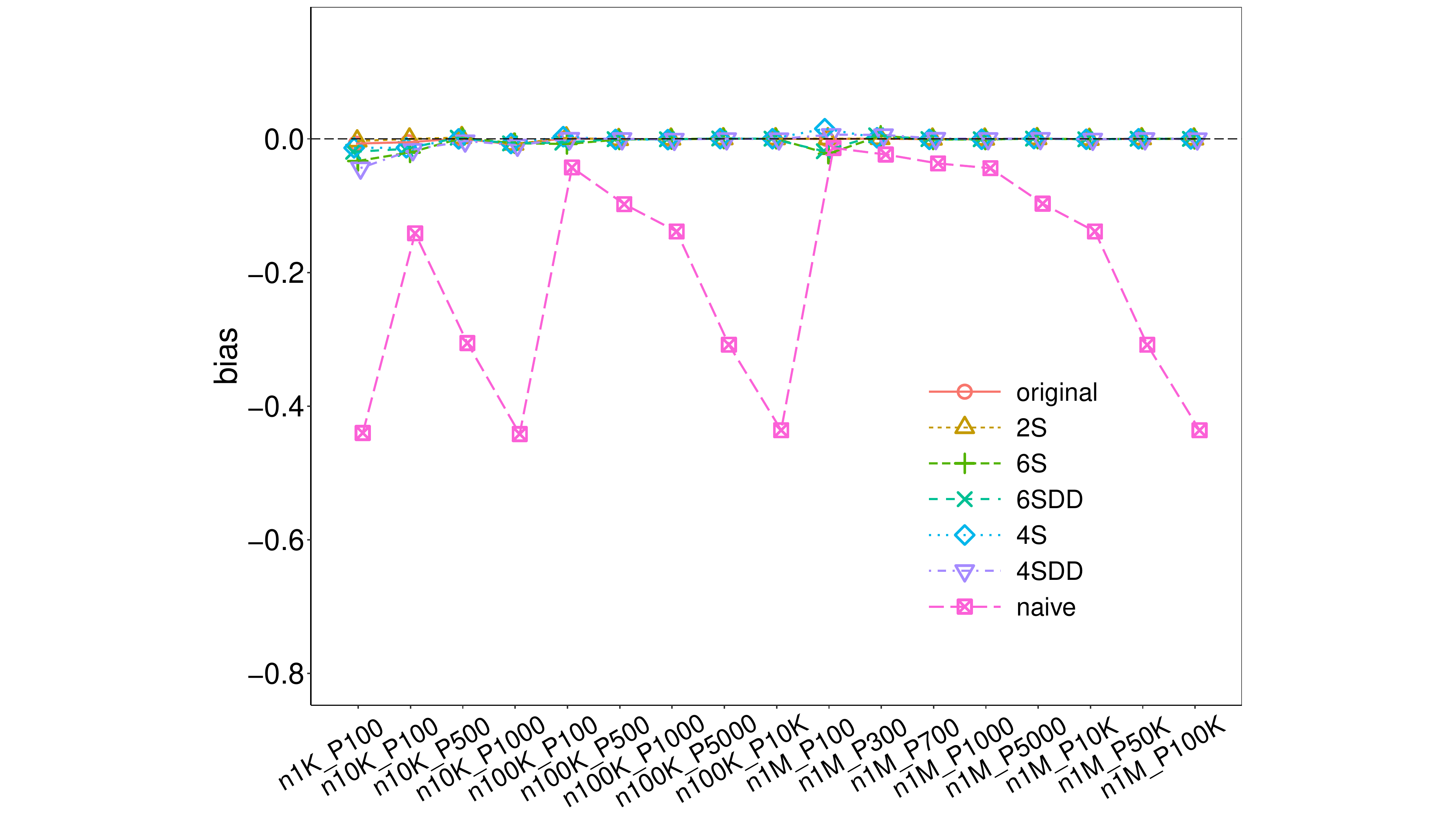}\\
\includegraphics[width=0.24\textwidth, trim={2.2in 0 2.2in 0},clip] {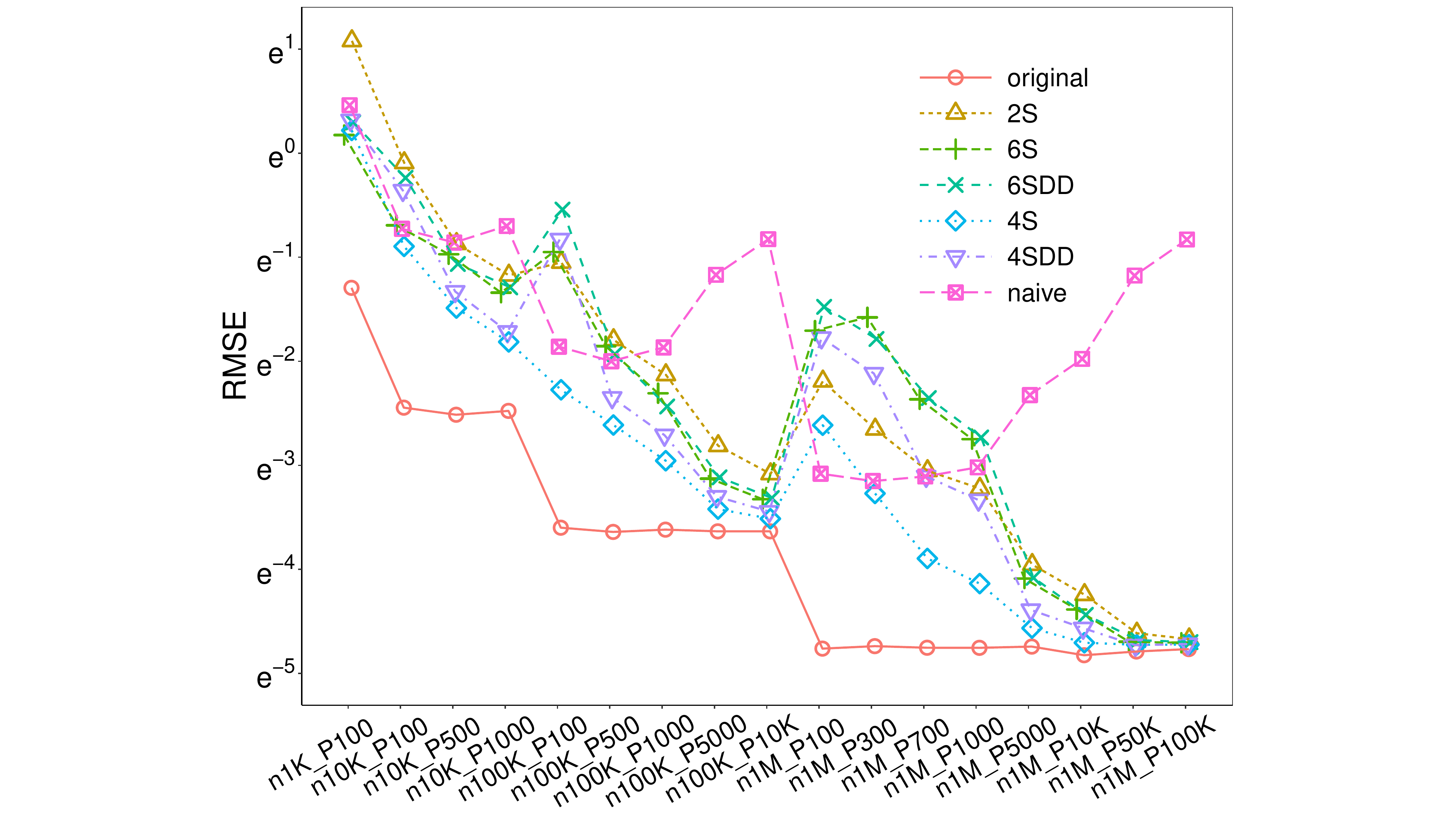}
\includegraphics[width=0.24\textwidth, trim={2.2in 0 2.2in 0},clip] {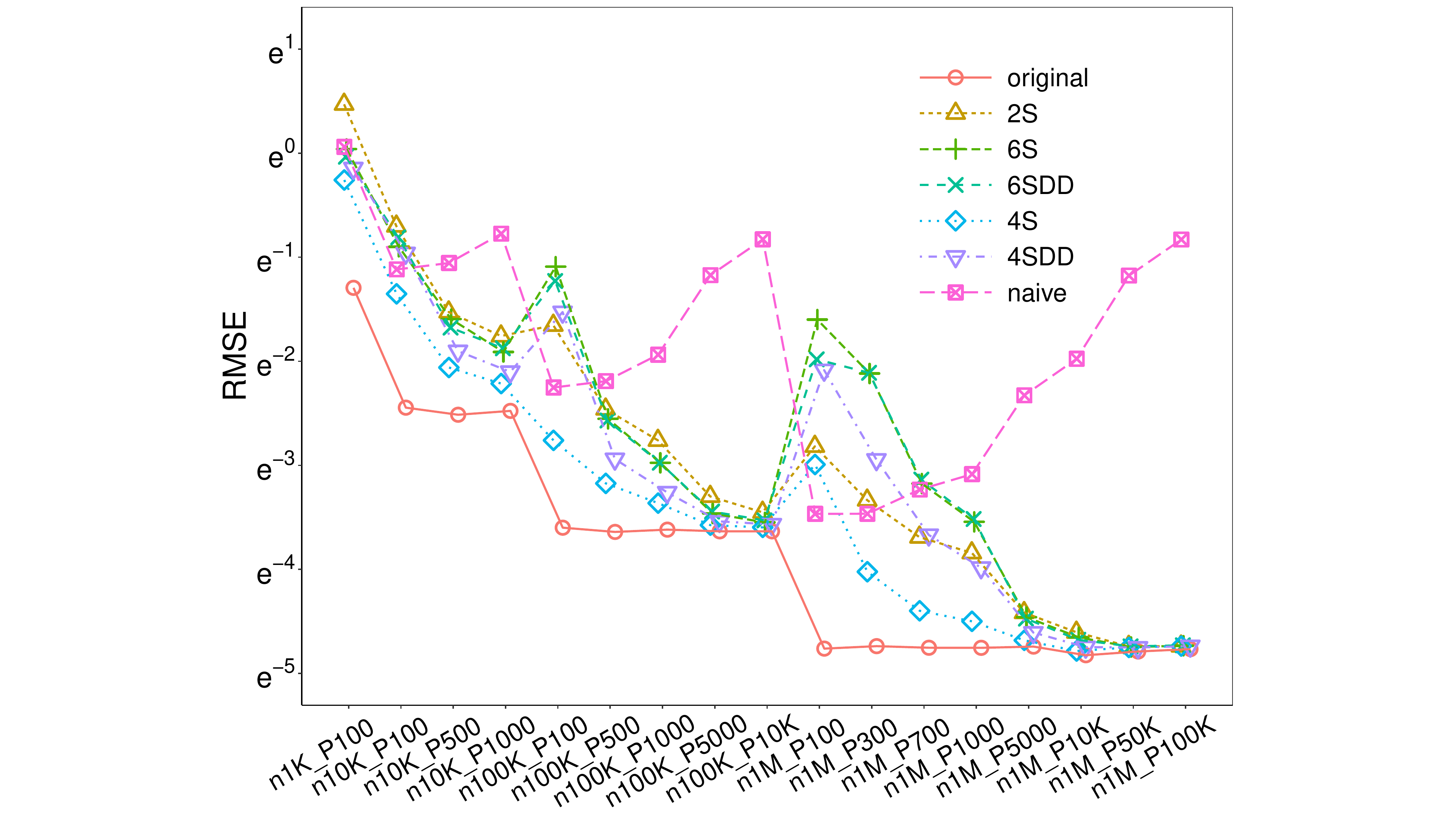}
\includegraphics[width=0.24\textwidth, trim={2.2in 0 2.2in 0},clip] {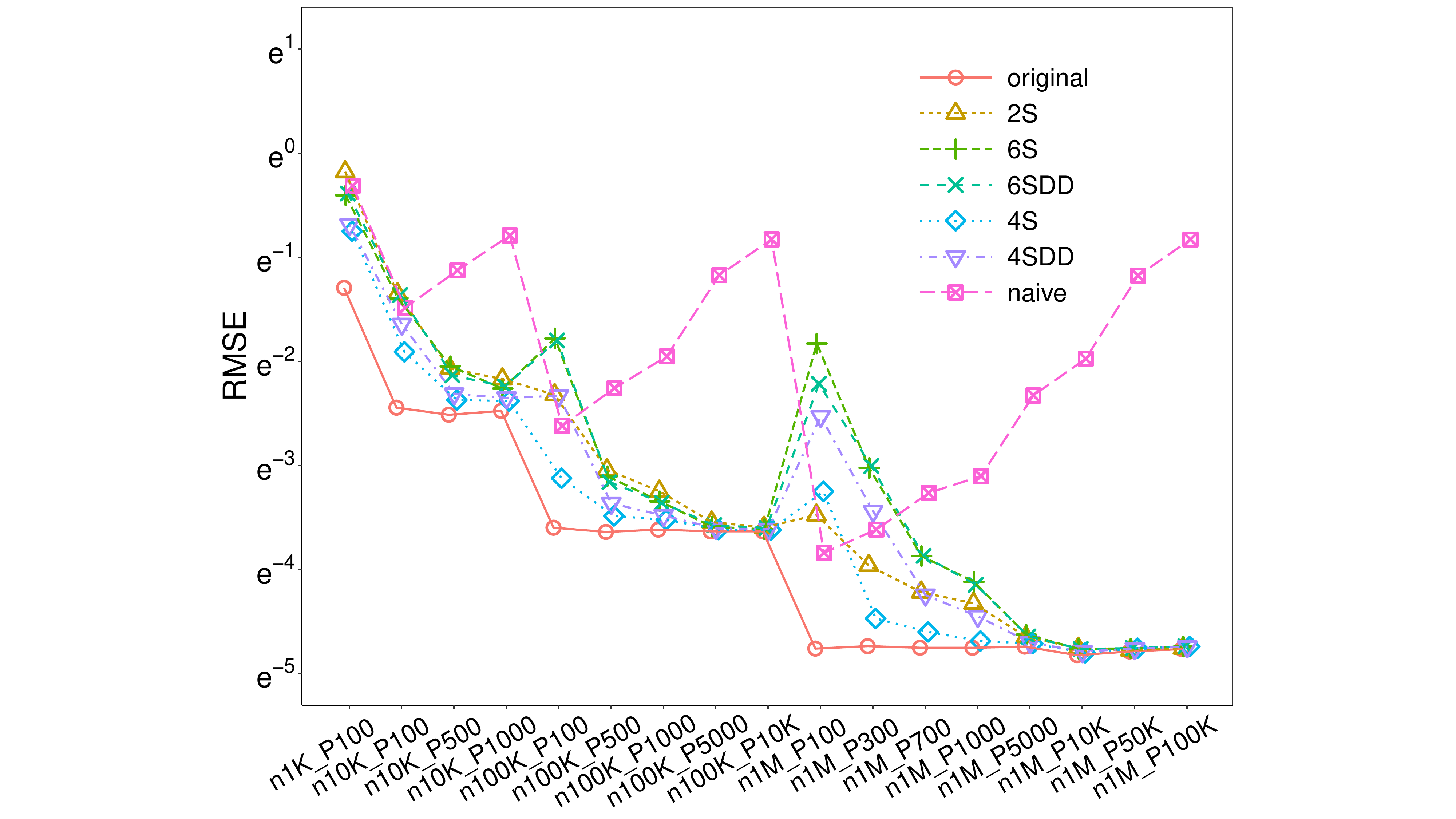}
\includegraphics[width=0.24\textwidth, trim={2.2in 0 2.2in 0},clip] {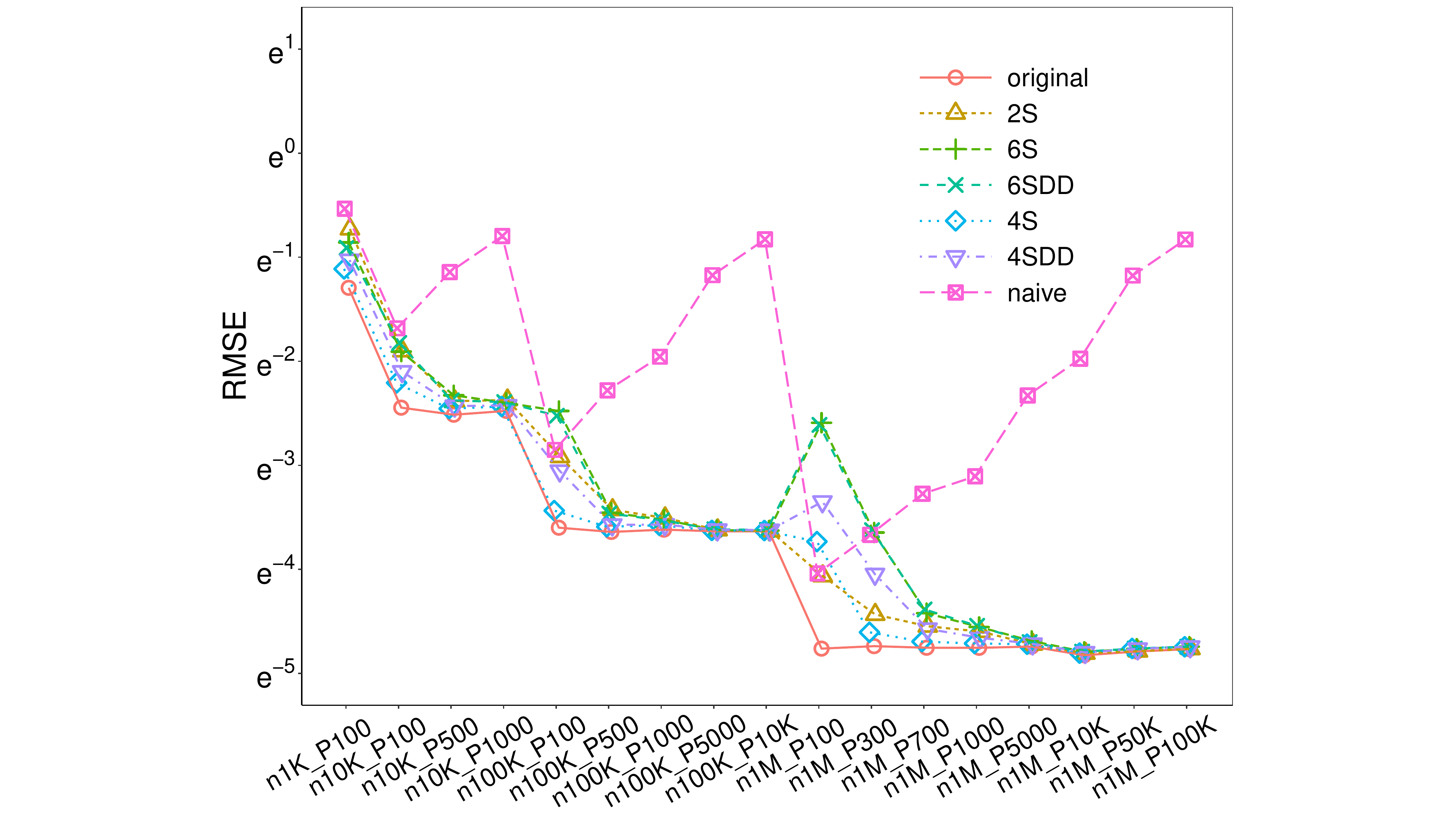}
\includegraphics[width=0.24\textwidth, trim={2.2in 0 2.2in 0},clip] {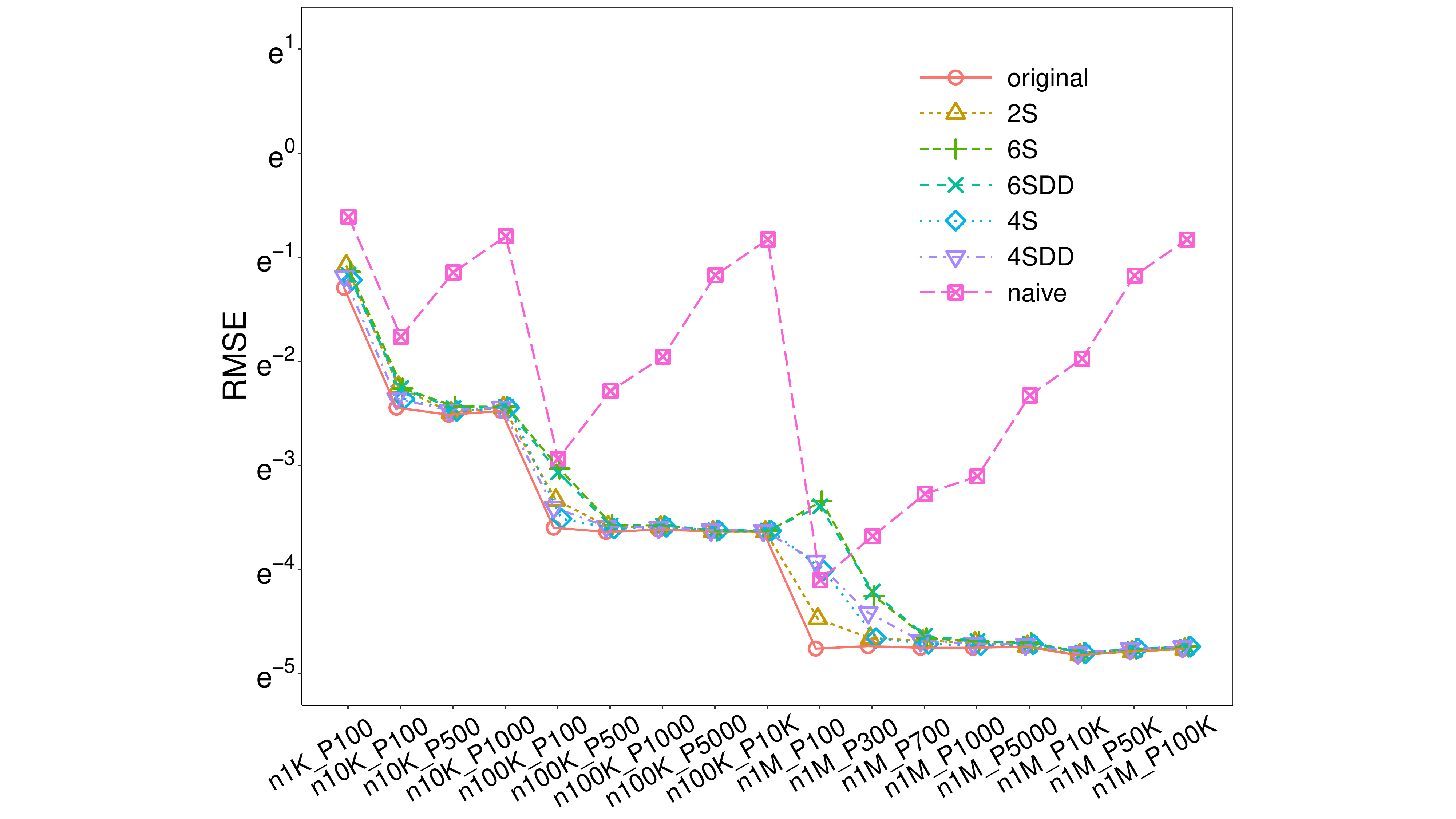}\\
\includegraphics[width=0.24\textwidth, trim={2.2in 0 2.2in 0},clip] {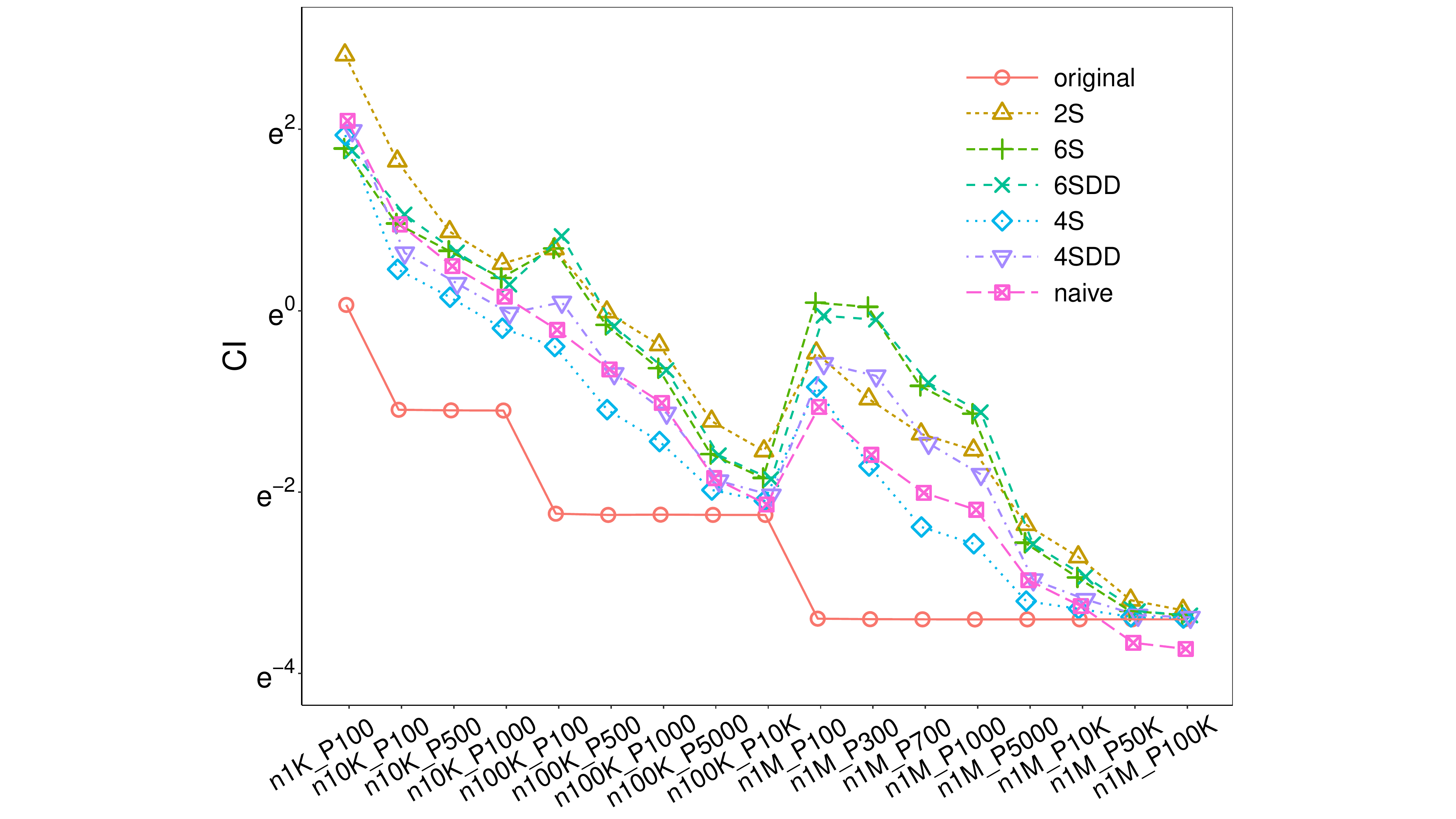}
\includegraphics[width=0.24\textwidth, trim={2.2in 0 2.2in 0},clip] {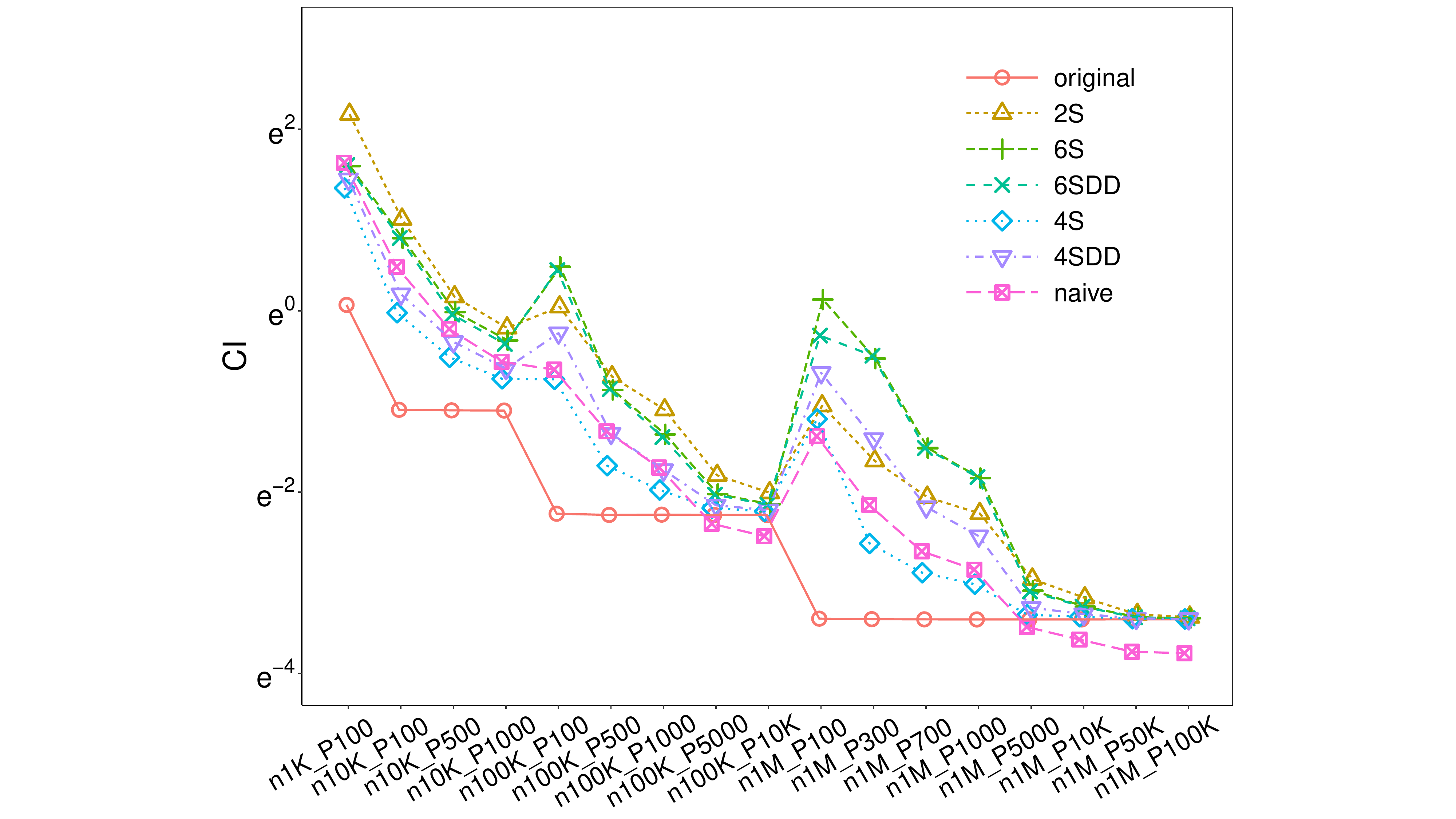}
\includegraphics[width=0.24\textwidth, trim={2.2in 0 2.2in 0},clip] {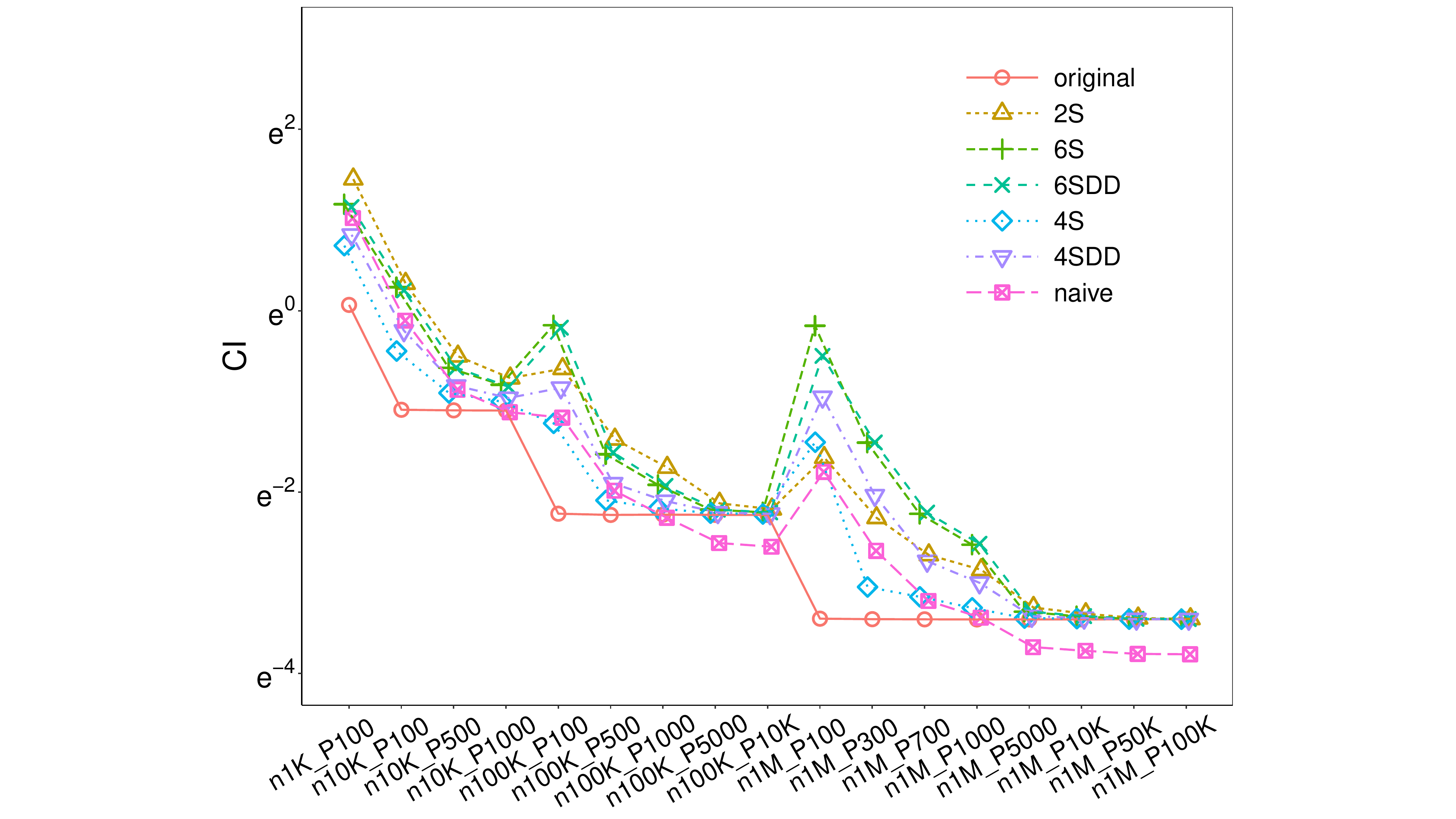}
\includegraphics[width=0.24\textwidth, trim={2.2in 0 2.2in 0},clip] {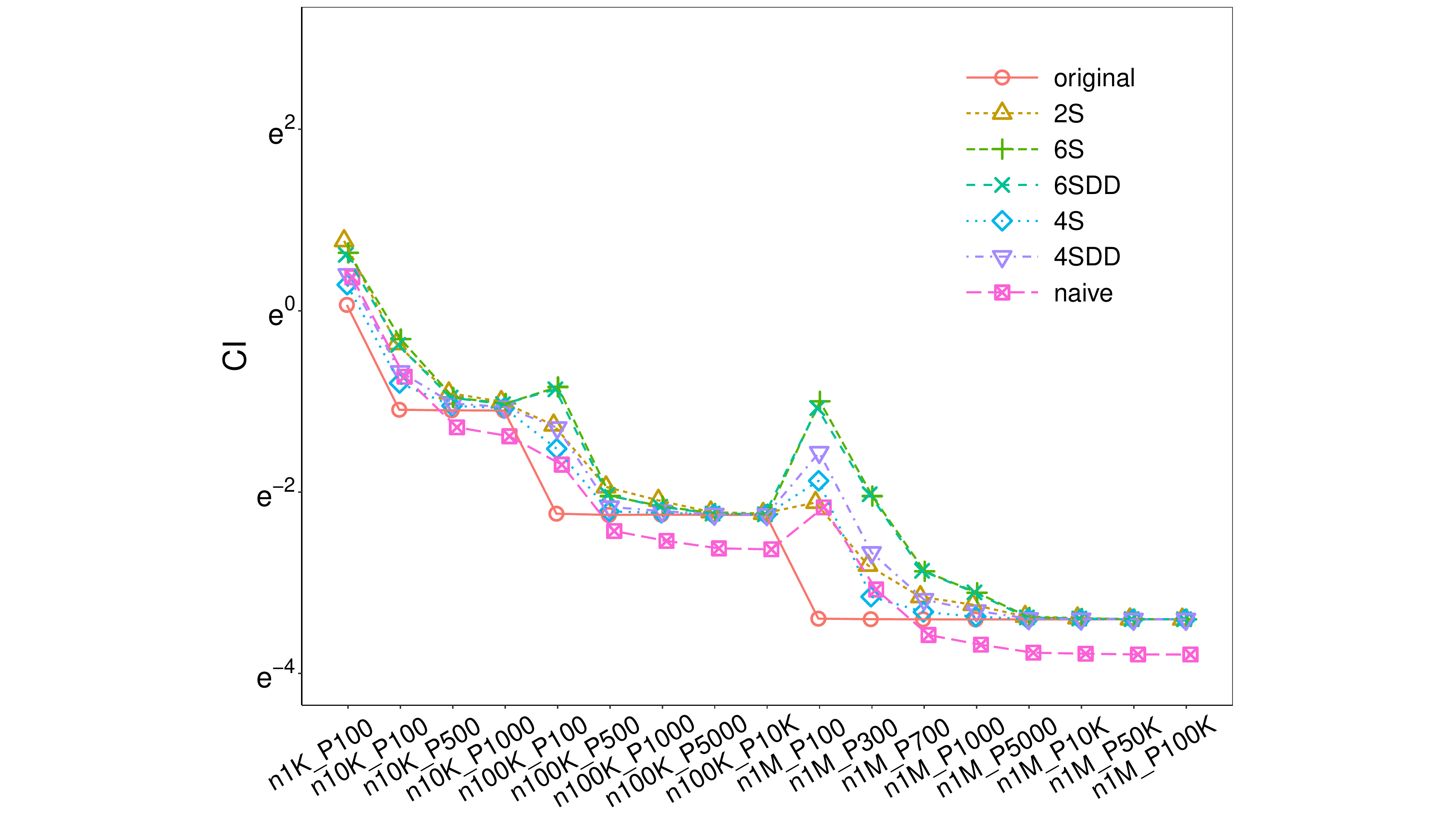}
\includegraphics[width=0.24\textwidth, trim={2.2in 0 2.2in 0},clip] {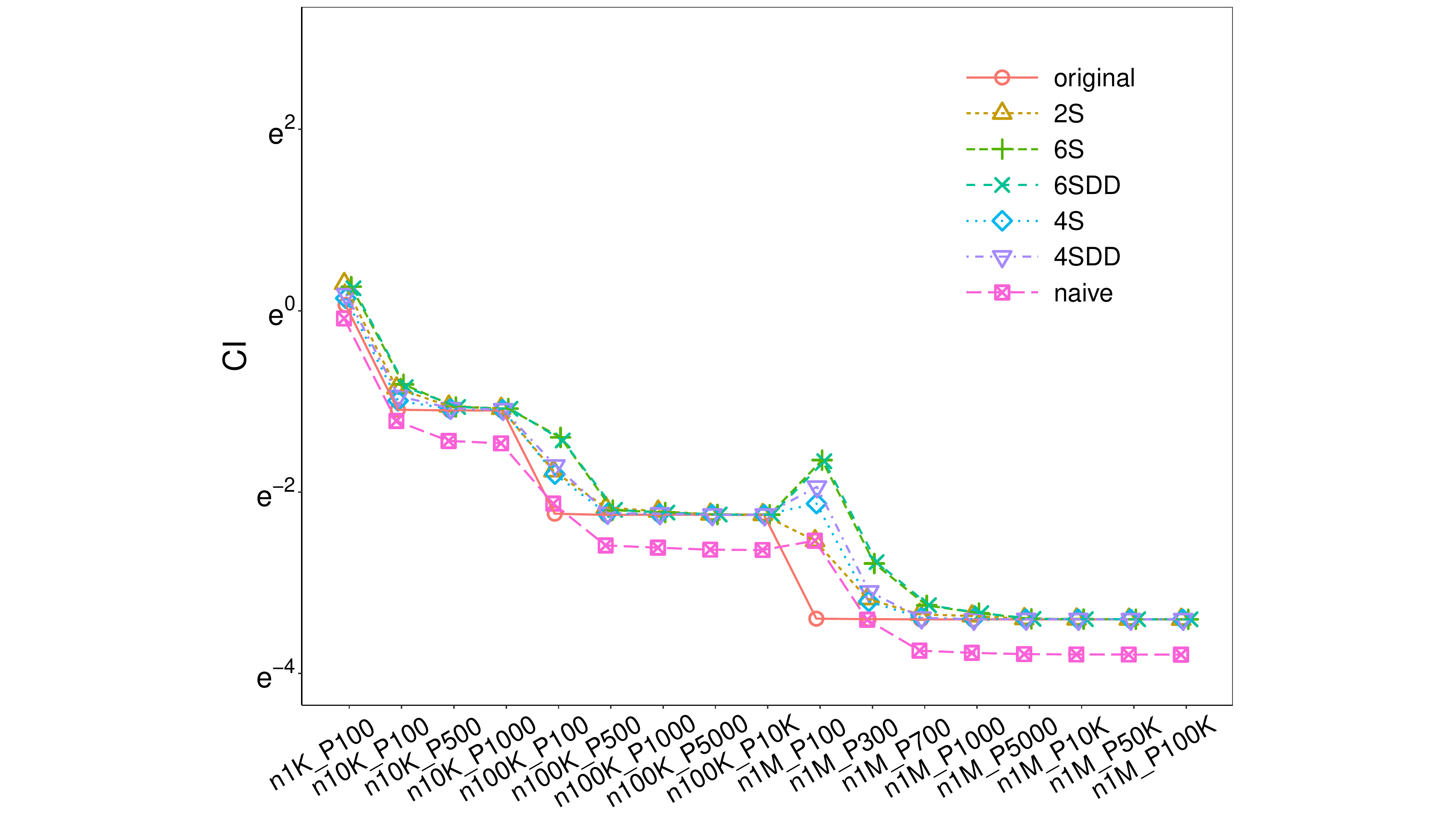}\\
\includegraphics[width=0.24\textwidth, trim={2.2in 0 2.2in 0},clip] {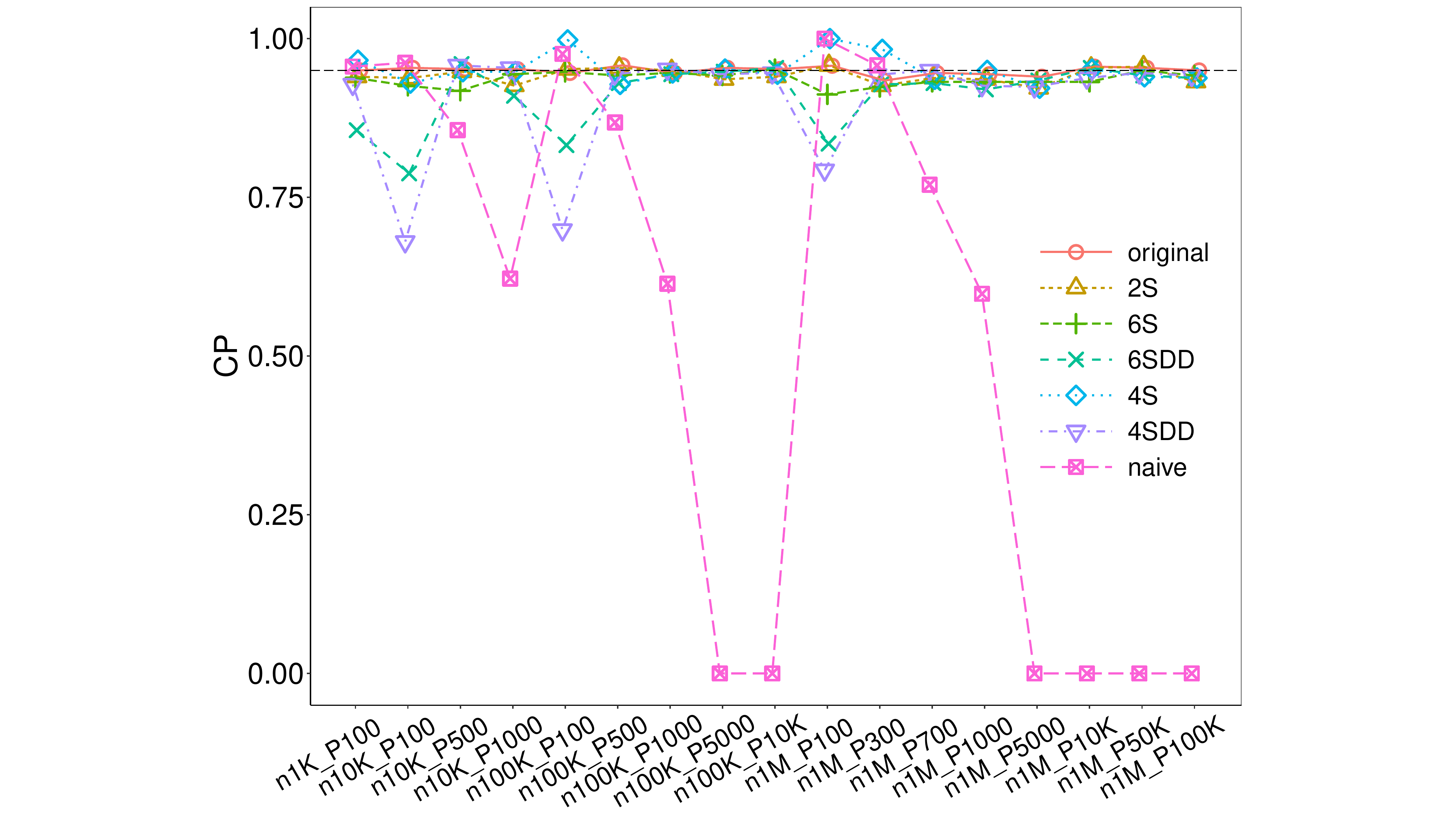}
\includegraphics[width=0.24\textwidth, trim={2.2in 0 2.2in 0},clip] {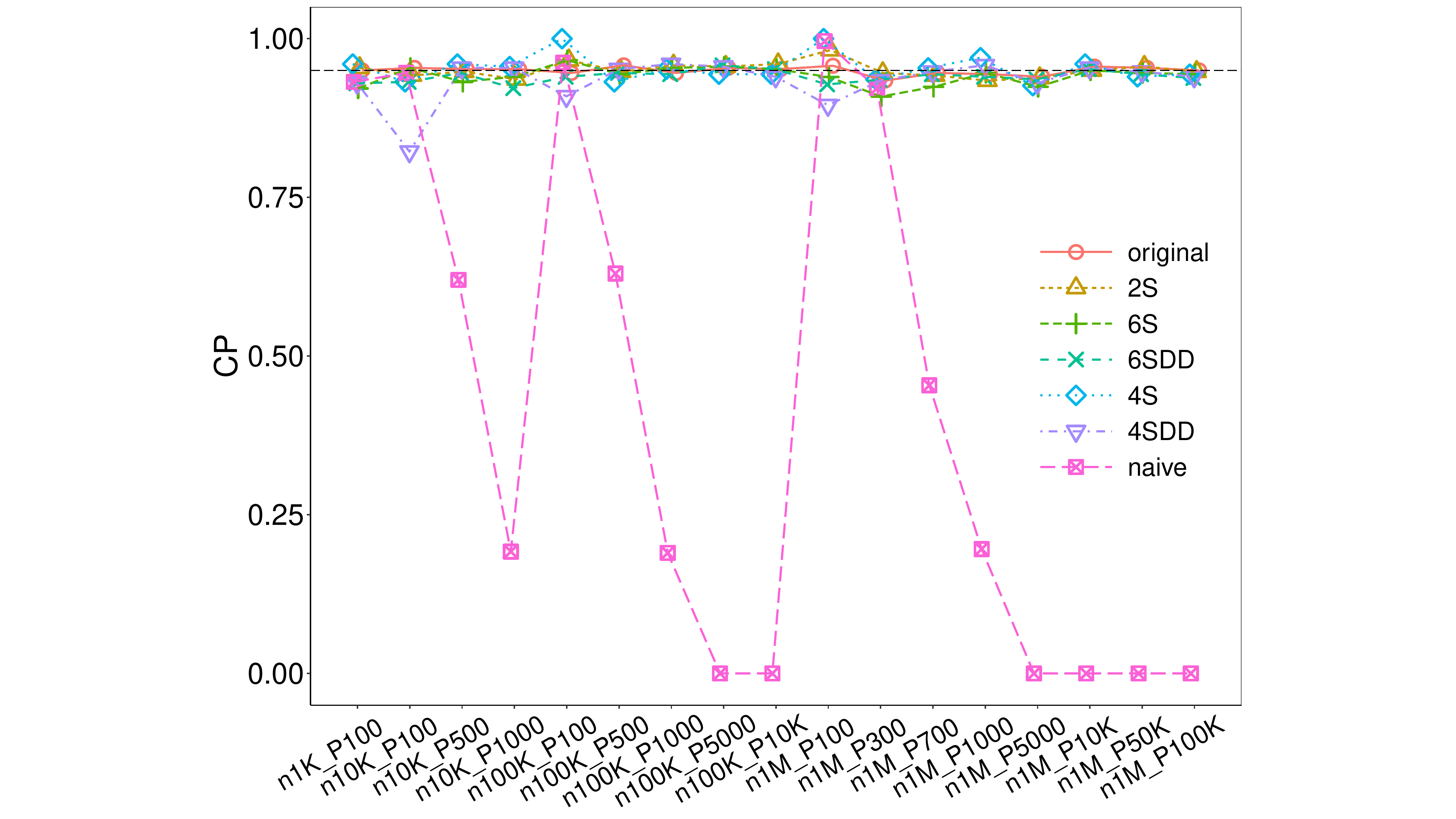}
\includegraphics[width=0.24\textwidth, trim={2.2in 0 2.2in 0},clip] {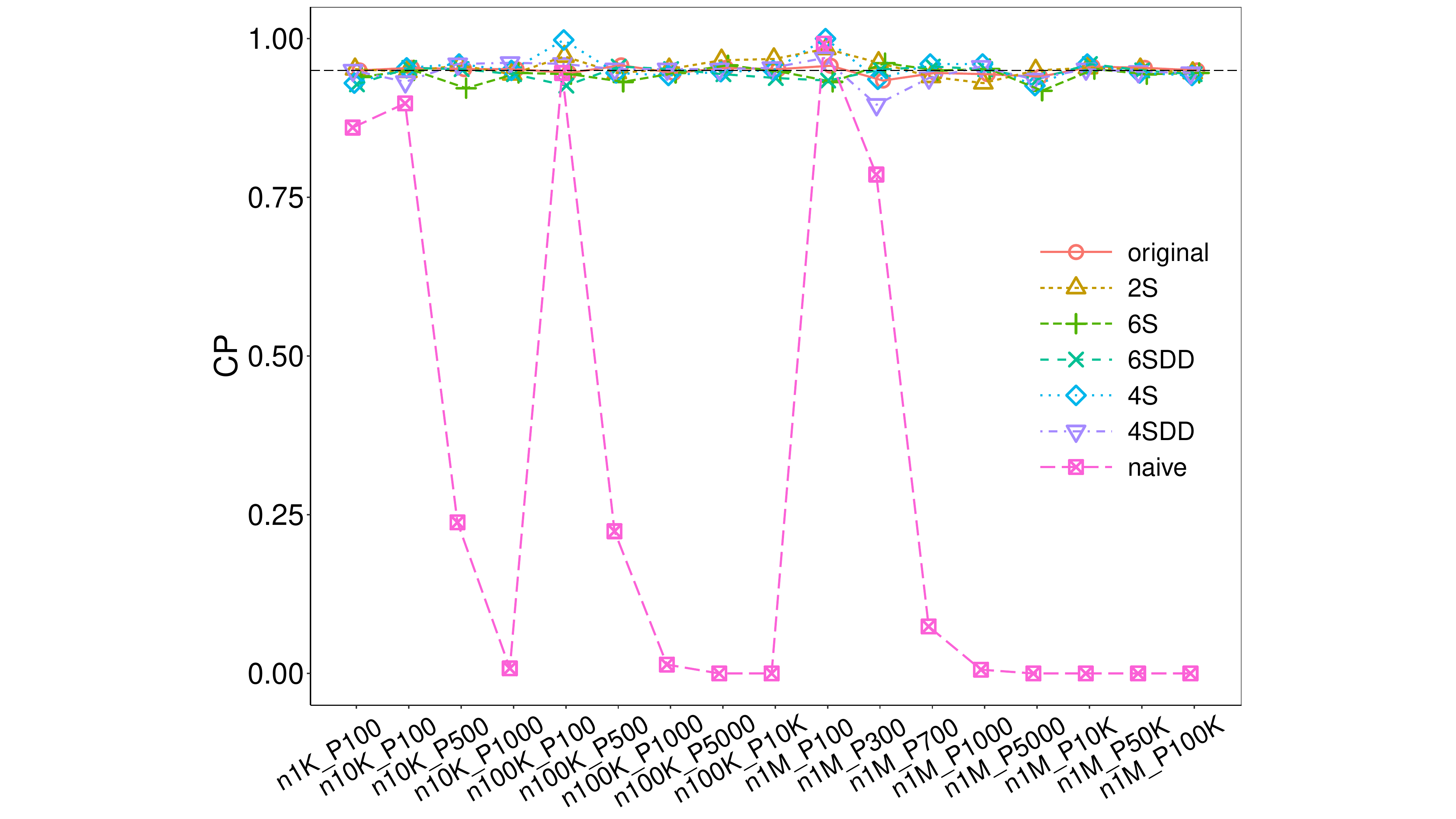}
\includegraphics[width=0.24\textwidth, trim={2.2in 0 2.2in 0},clip] {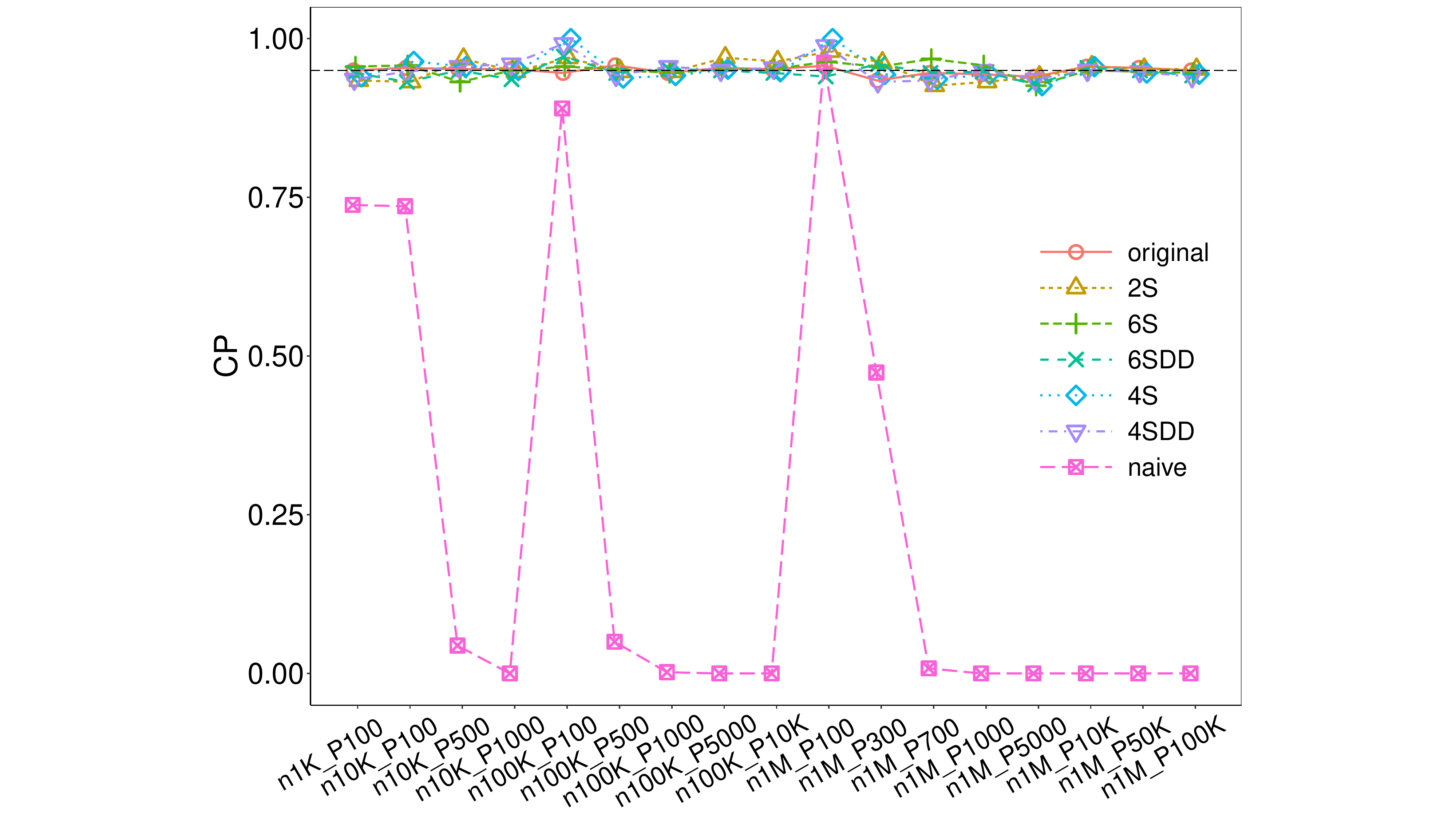}
\includegraphics[width=0.24\textwidth, trim={2.2in 0 2.2in 0},clip] {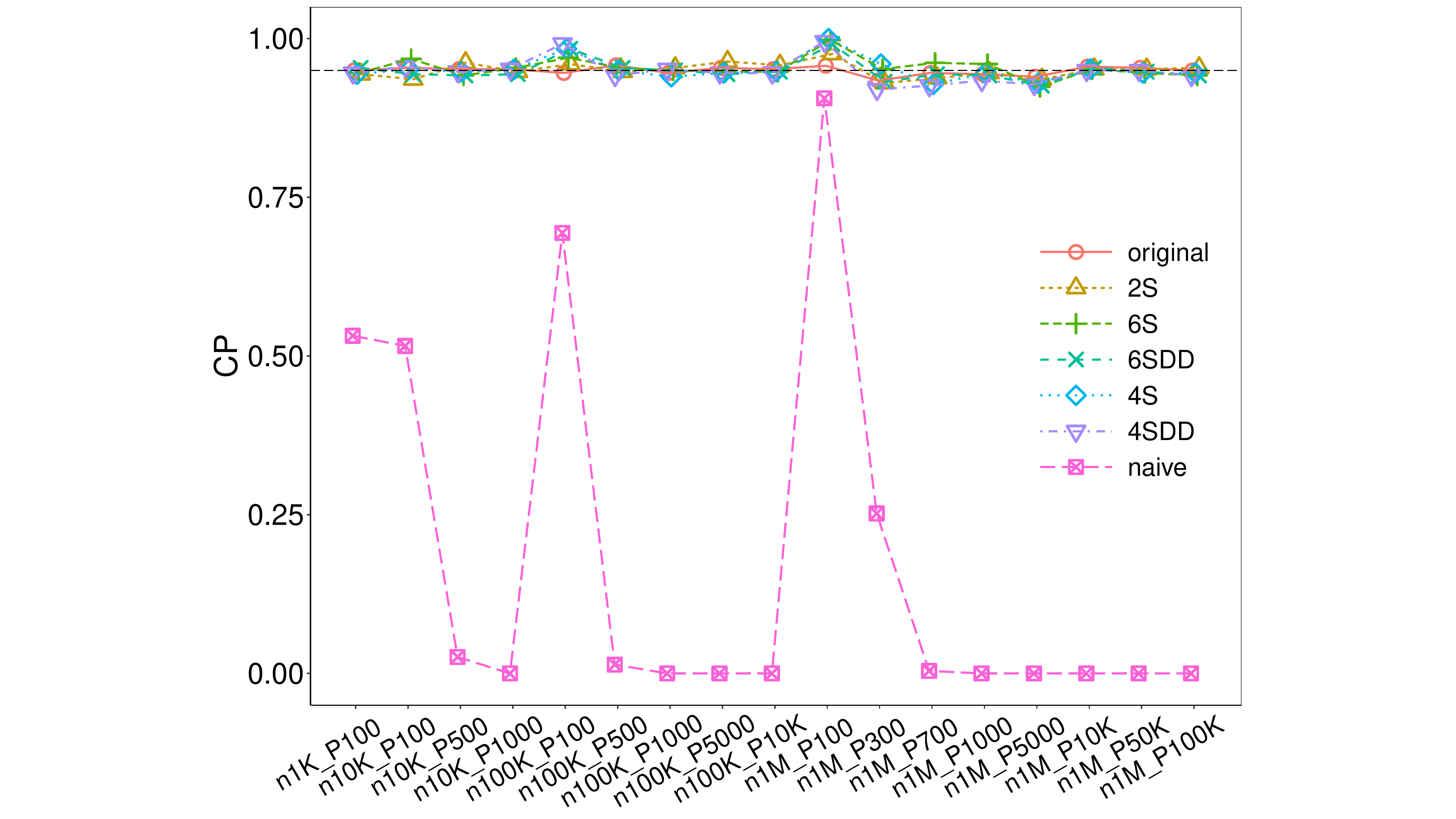}\\
\includegraphics[width=0.24\textwidth, trim={2.2in 0 2.2in 0},clip] {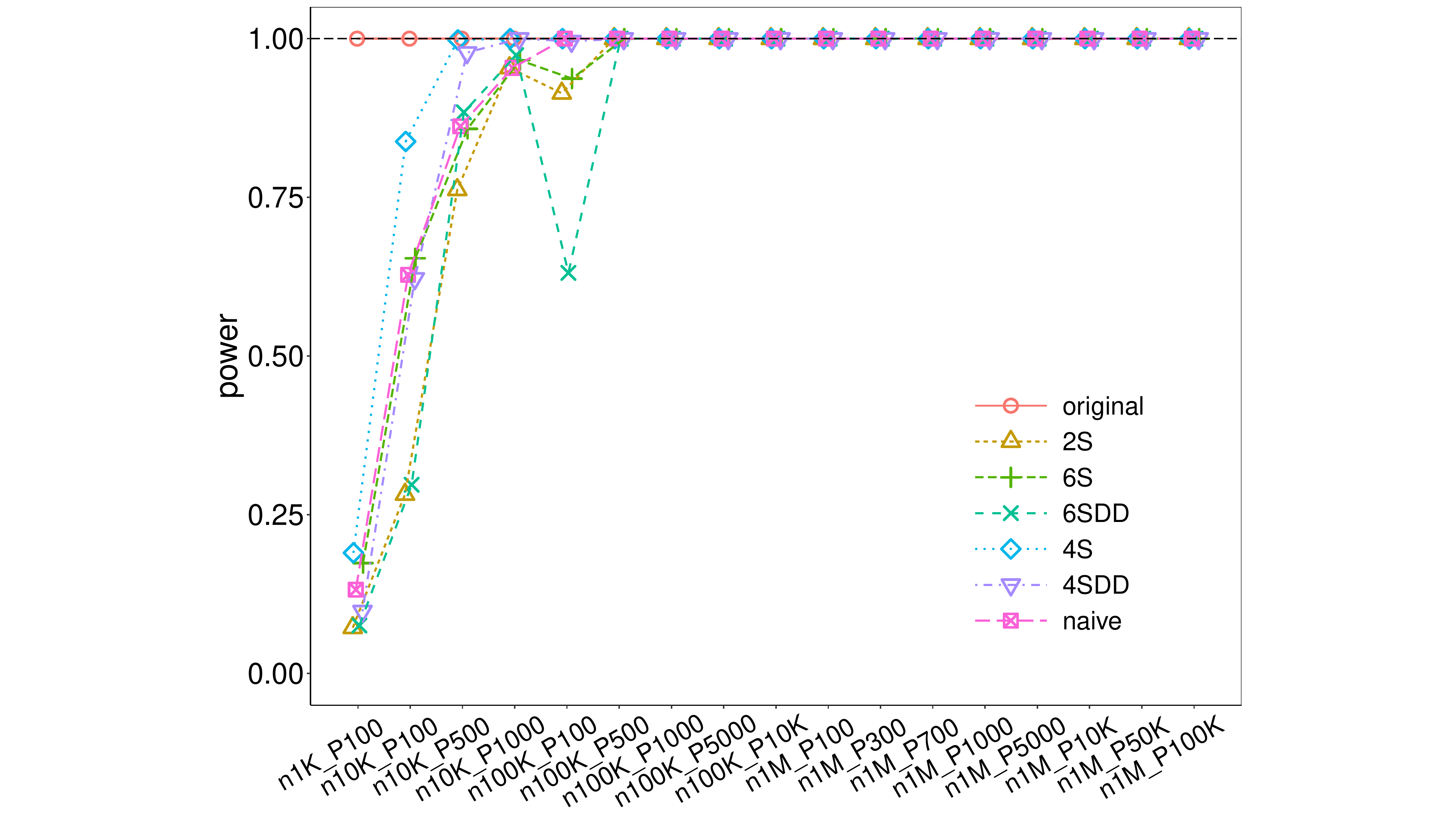}
\includegraphics[width=0.24\textwidth, trim={2.2in 0 2.2in 0},clip] {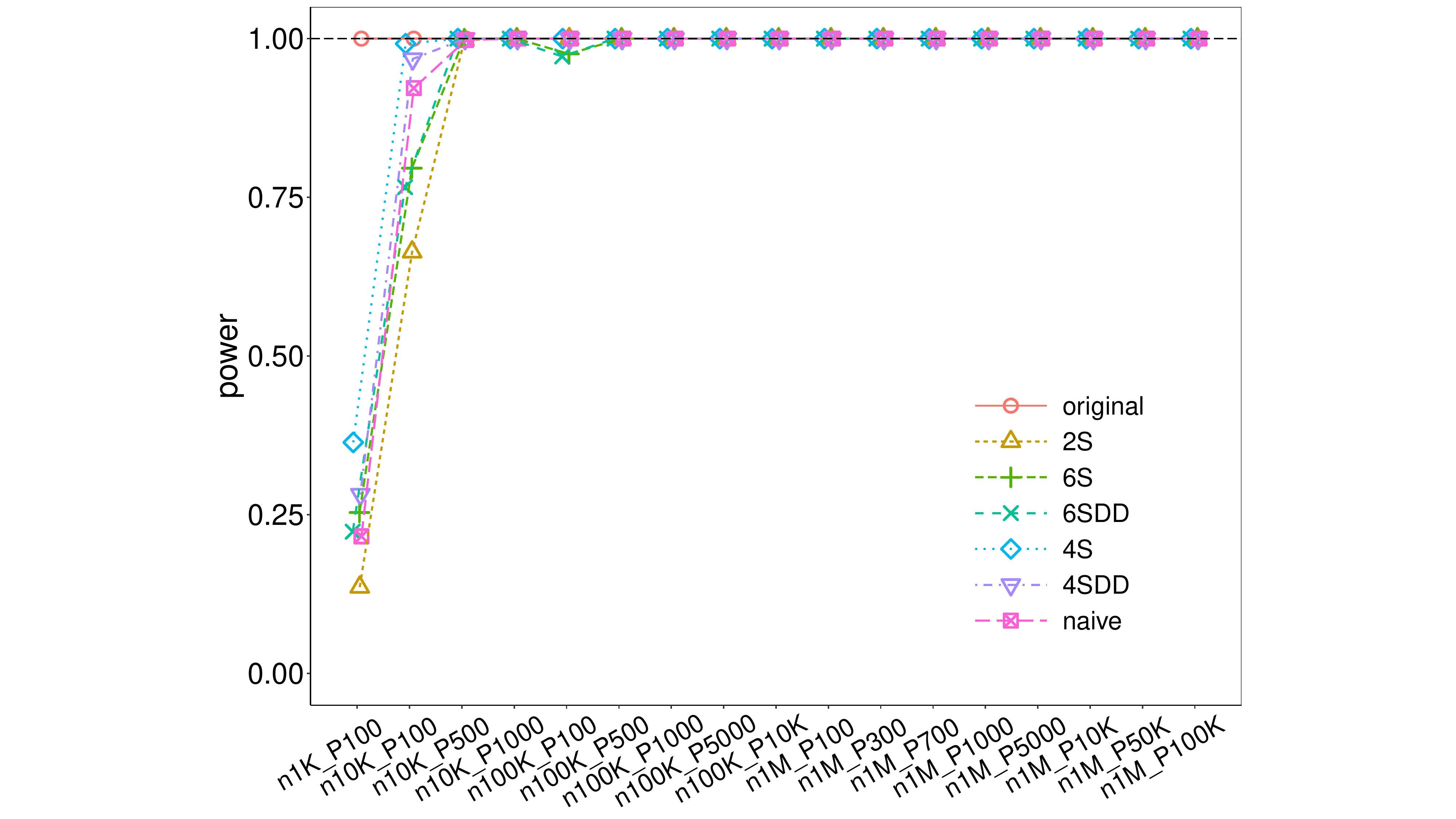}
\includegraphics[width=0.24\textwidth, trim={2.2in 0 2.2in 0},clip] {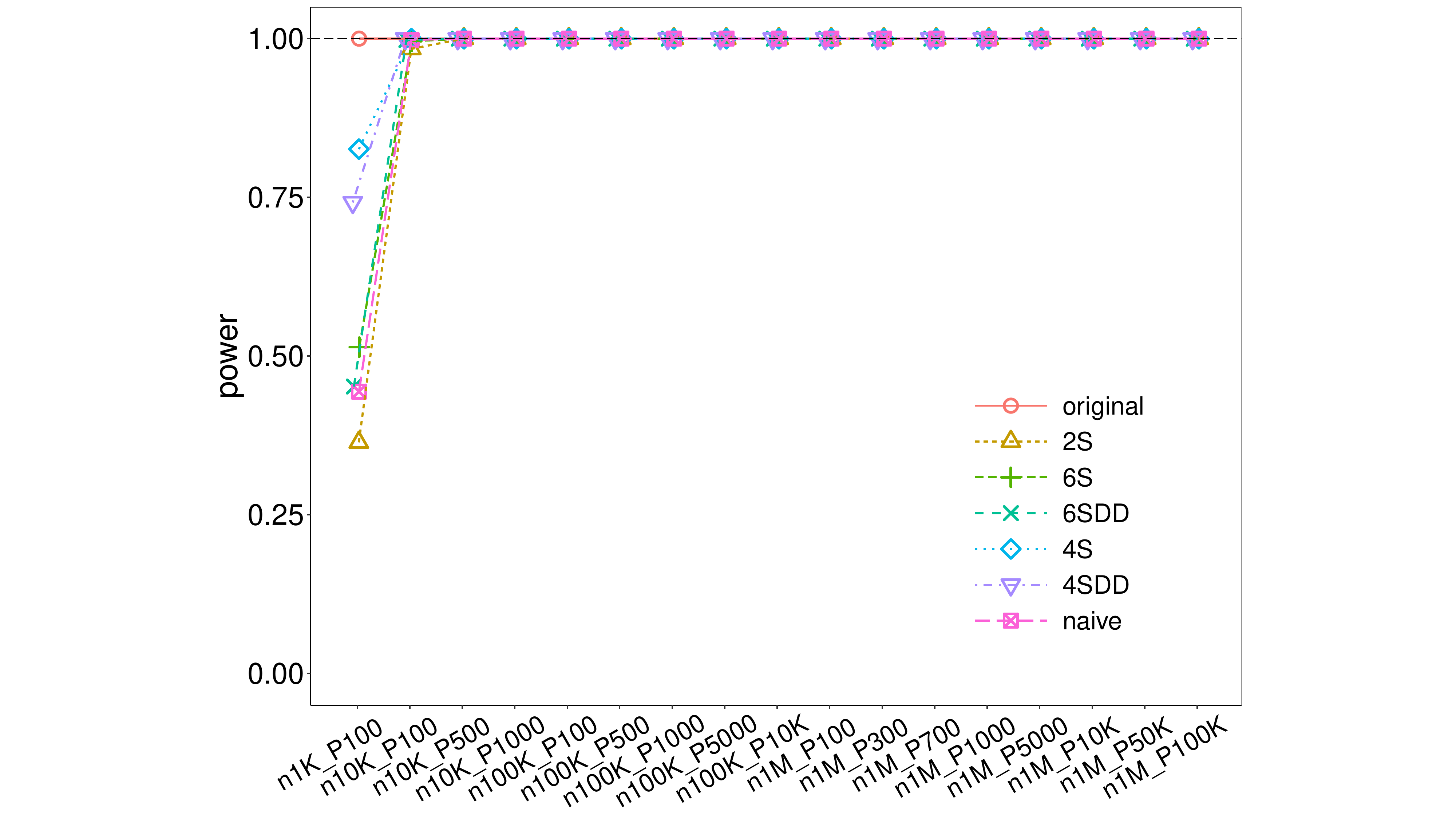}
\includegraphics[width=0.24\textwidth, trim={2.2in 0 2.2in 0},clip] {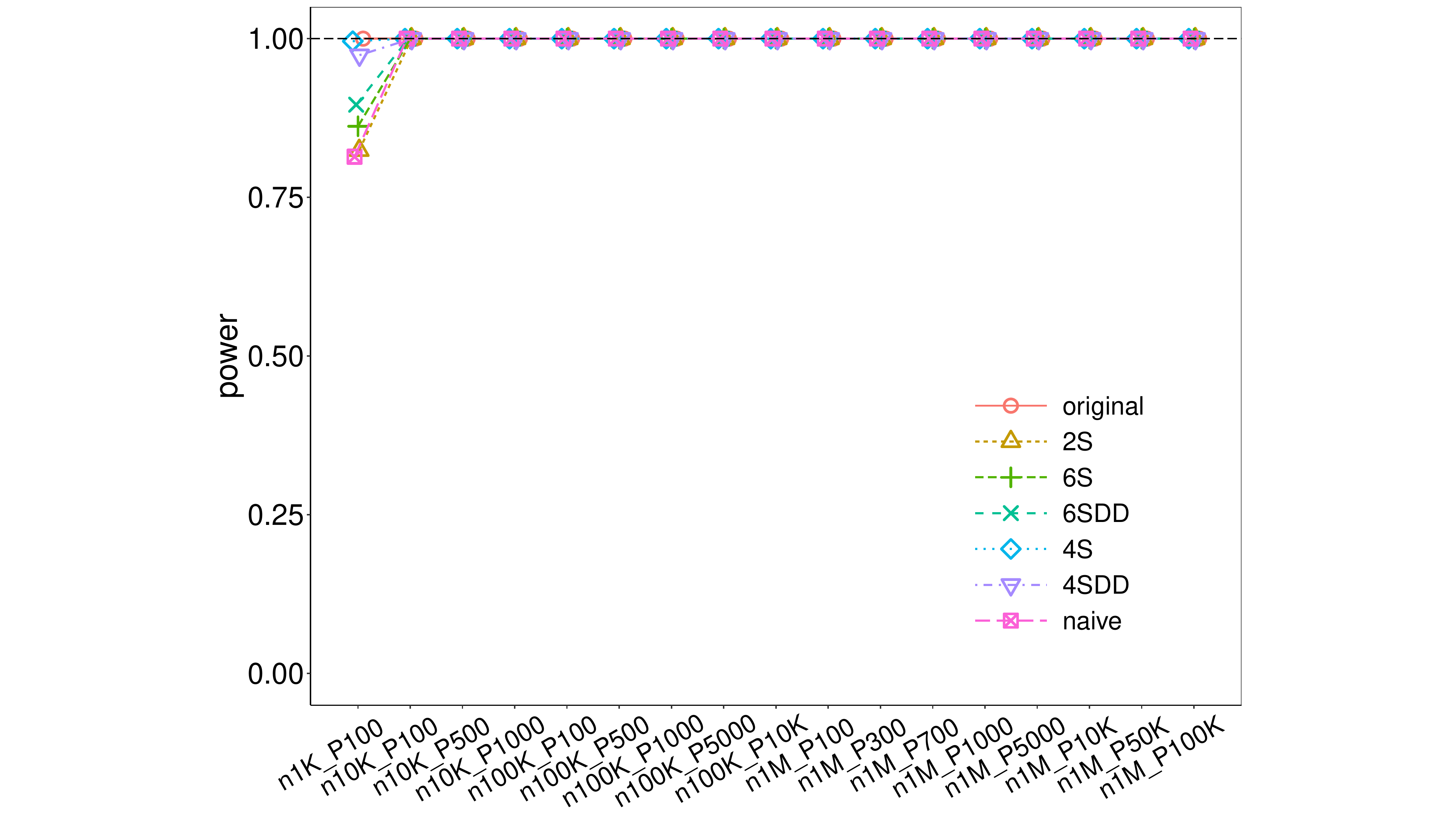}
\includegraphics[width=0.24\textwidth, trim={2.2in 0 2.2in 0},clip] {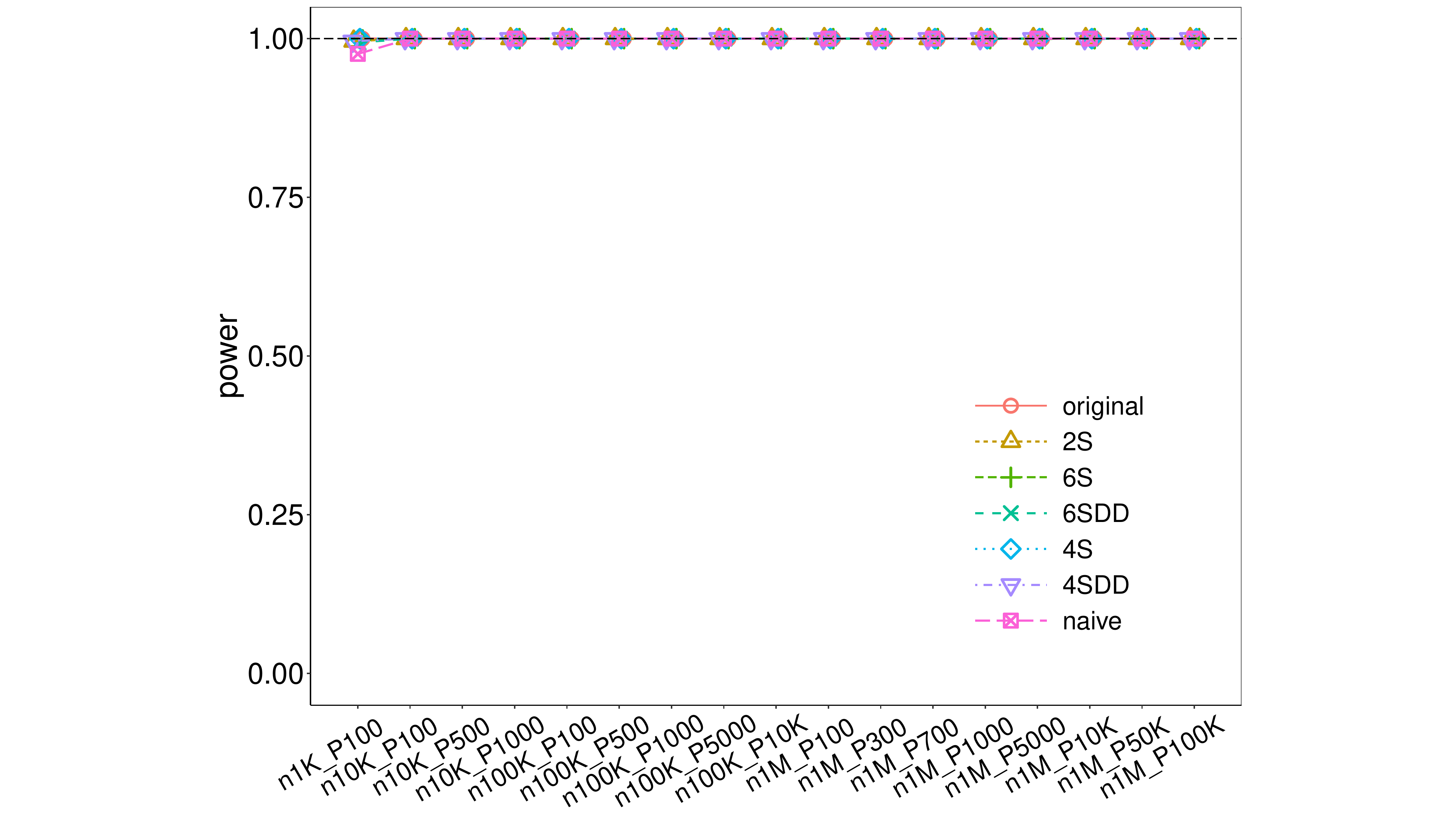}\\
\caption{Gaussian data; $\rho$-zCDP; $\theta\ne0$ and $\alpha\ne\beta$}
\label{fig:1aszCDP}
\end{figure}
\end{landscape}

\begin{landscape}
\subsection*{ZINB, $\theta=0$ and $\alpha=\beta$}

\begin{figure}[!htb]
\hspace{0.6in}$\epsilon=0.5$\hspace{1.3in}$\epsilon=1$\hspace{1.4in}$\epsilon=2$
\hspace{1.4in}$\epsilon=5$\hspace{1.4in}$\epsilon=50$\\
\includegraphics[width=0.26\textwidth, trim={2.2in 0 2.2in 0},clip] {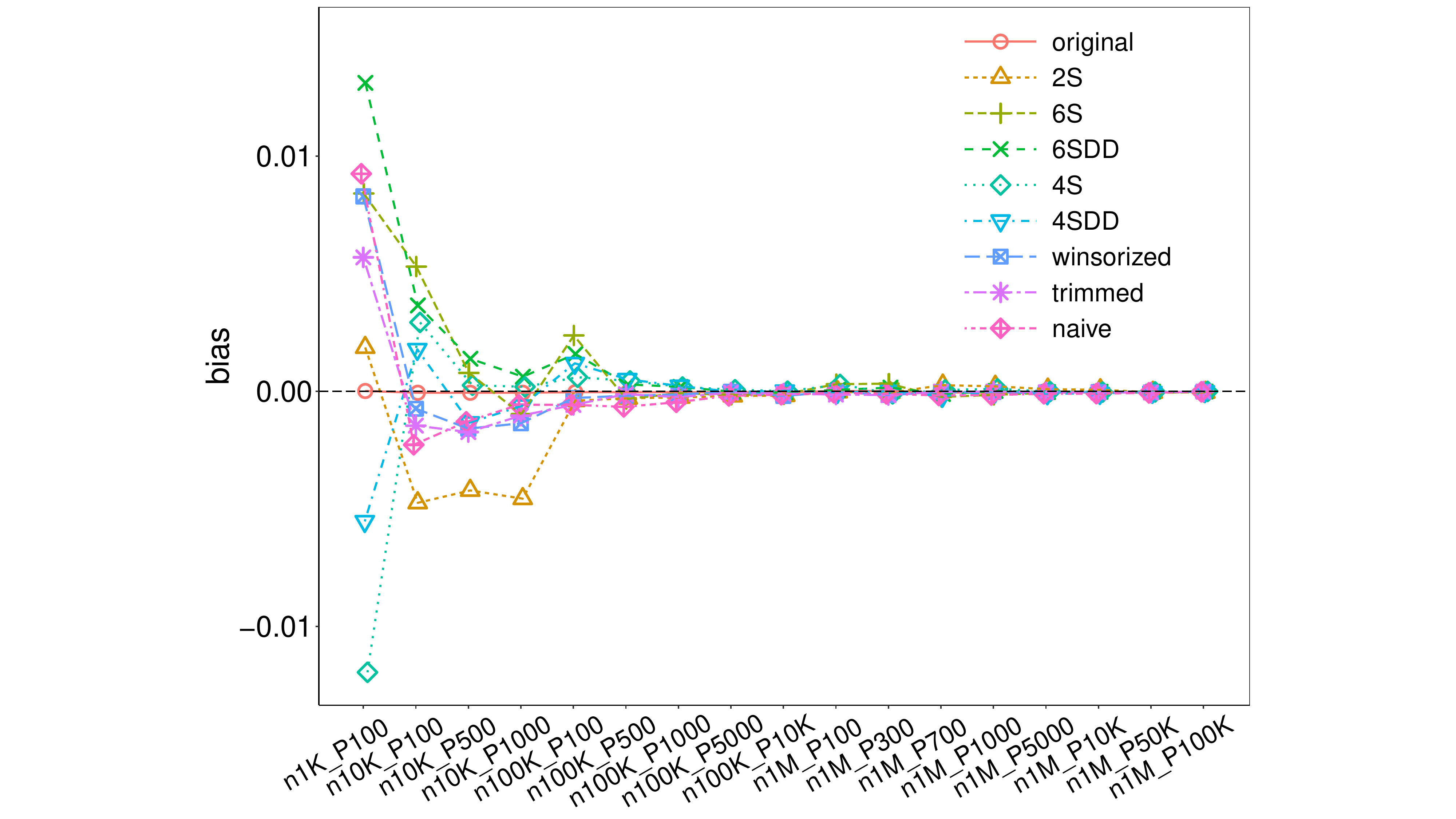}
\includegraphics[width=0.26\textwidth, trim={2.2in 0 2.2in 0},clip] {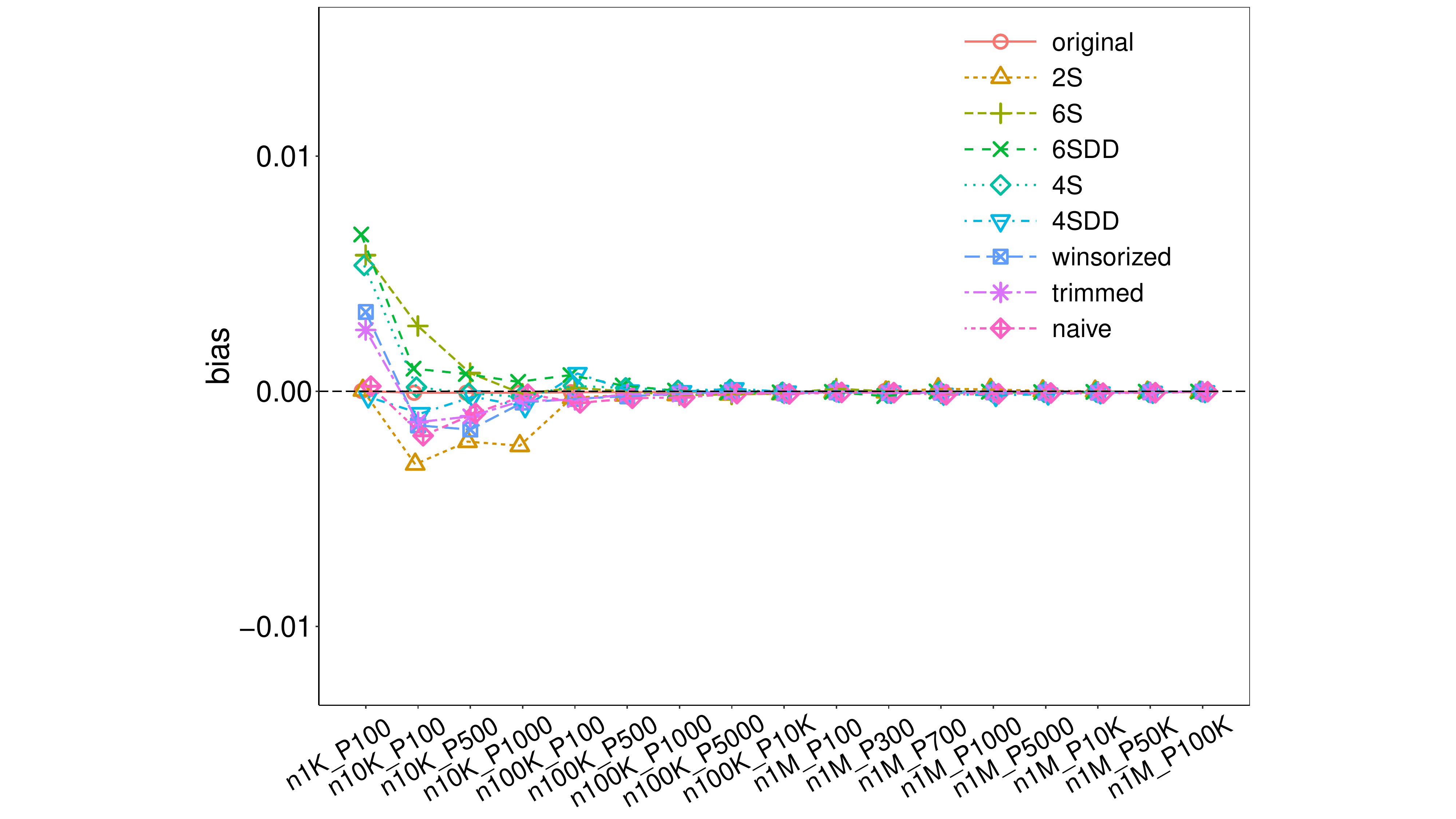}
\includegraphics[width=0.26\textwidth, trim={2.2in 0 2.2in 0},clip] {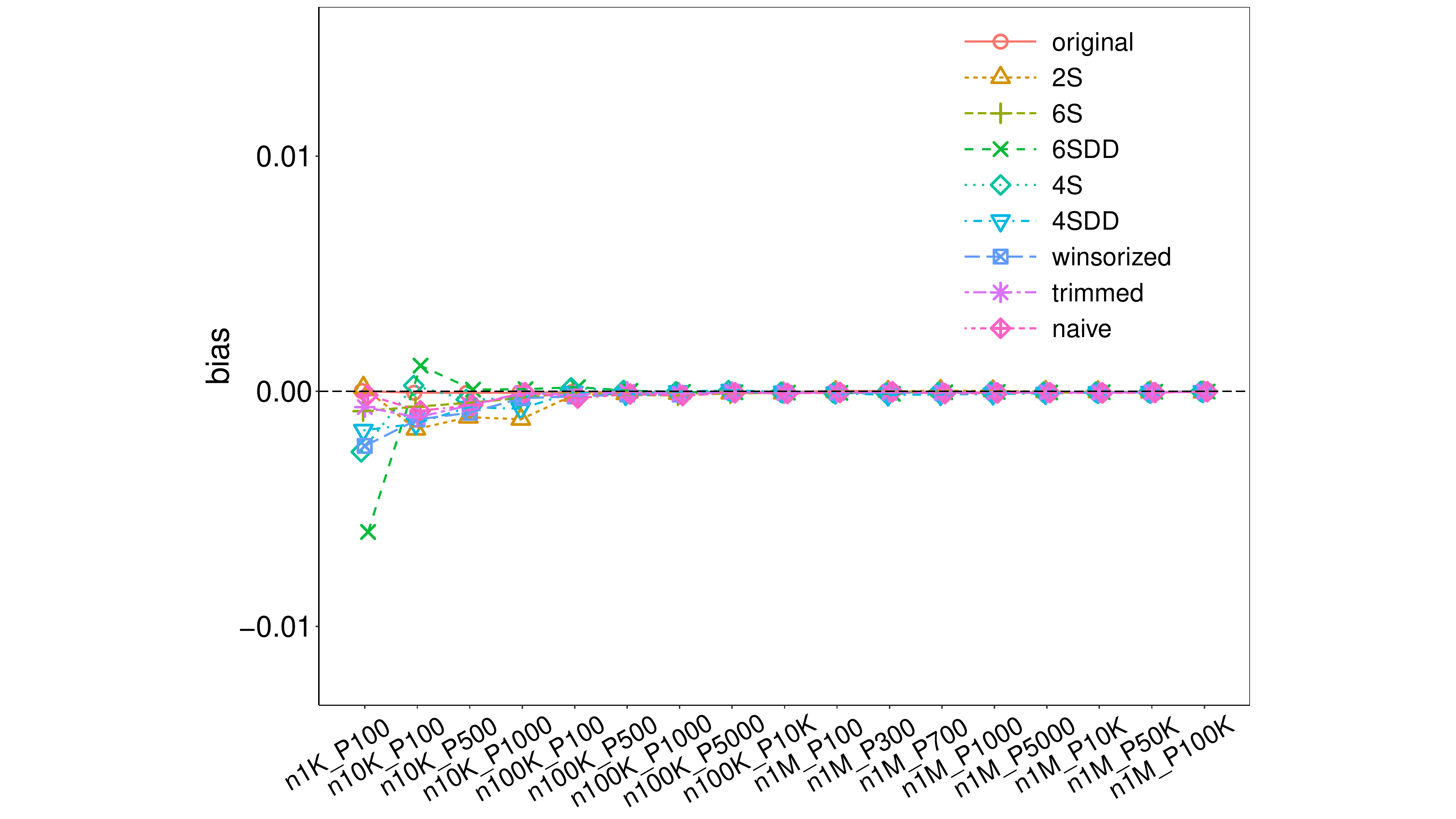}
\includegraphics[width=0.26\textwidth, trim={2.2in 0 2.2in 0},clip] {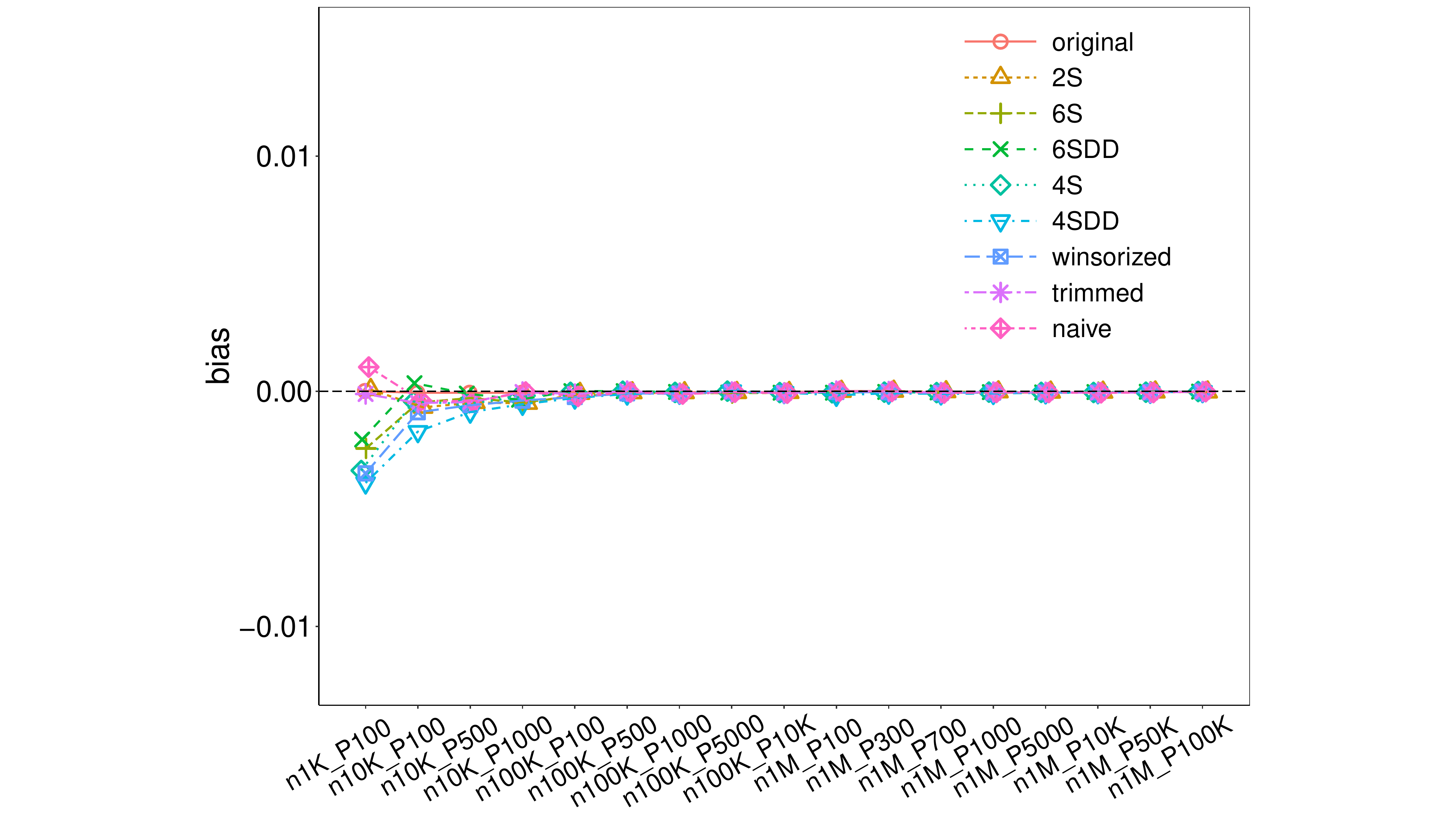}
\includegraphics[width=0.26\textwidth, trim={2.2in 0 2.2in 0},clip] {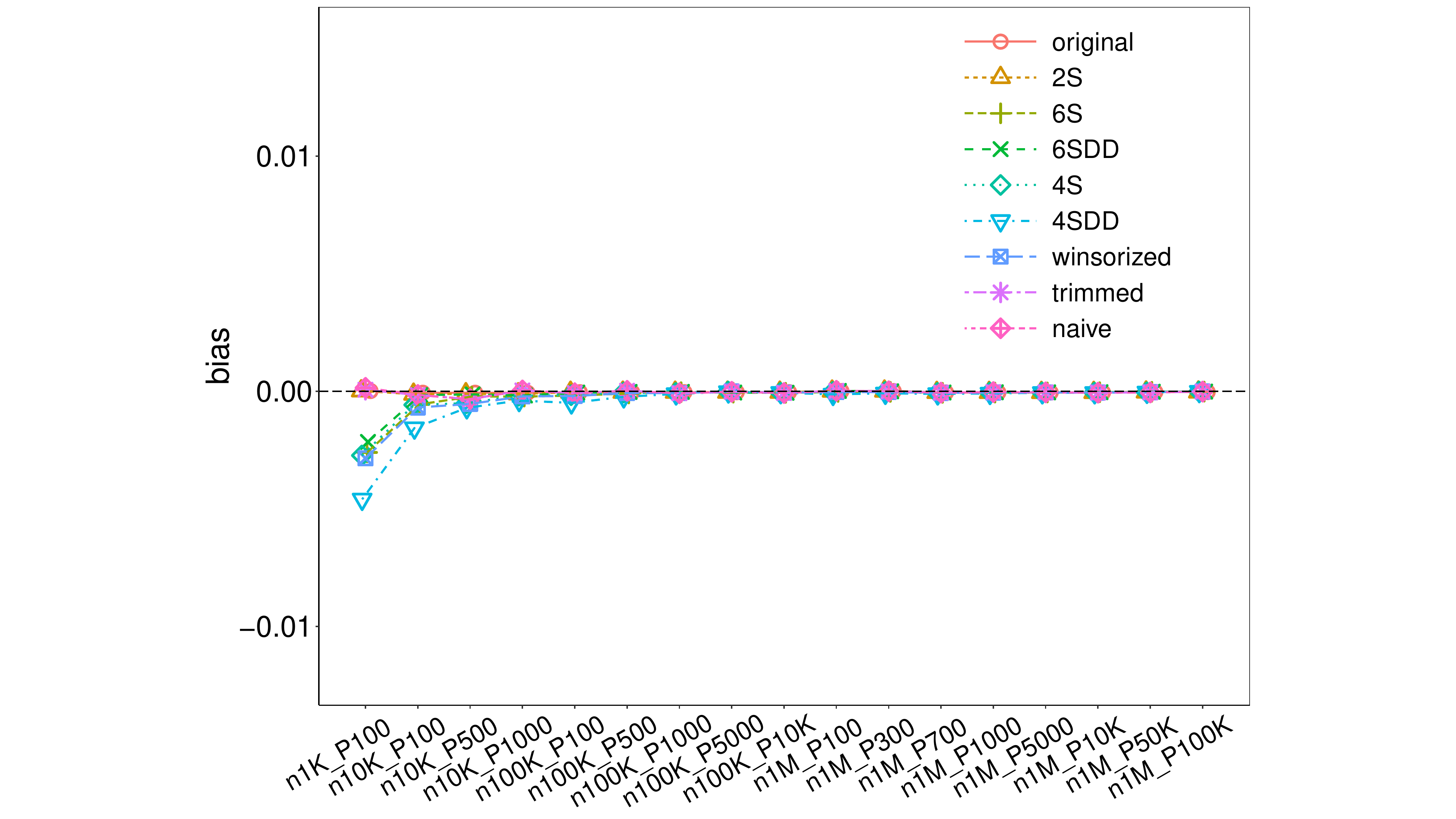}\\
\includegraphics[width=0.26\textwidth, trim={2.2in 0 2.2in 0},clip] {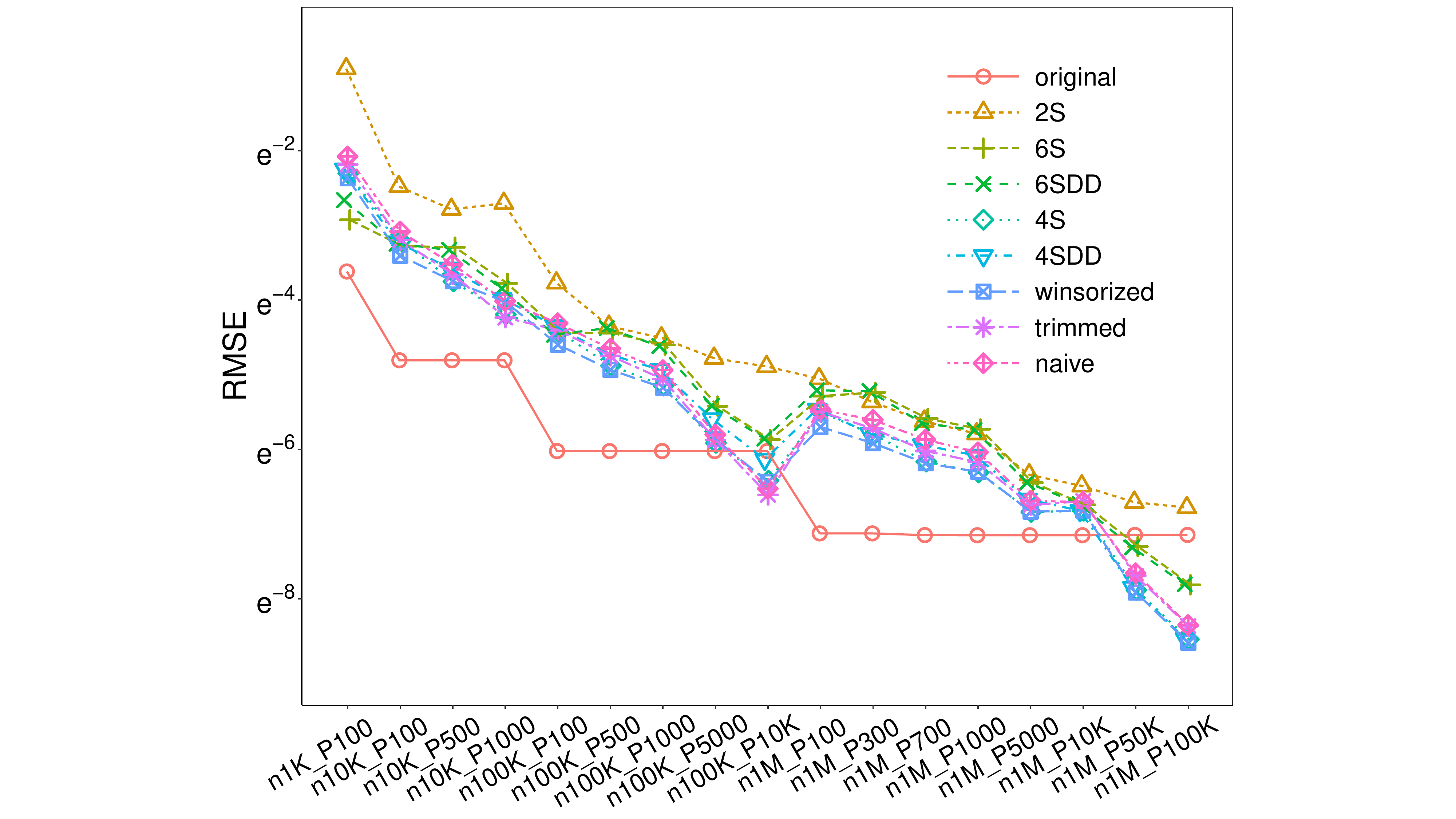}
\includegraphics[width=0.26\textwidth, trim={2.2in 0 2.2in 0},clip] {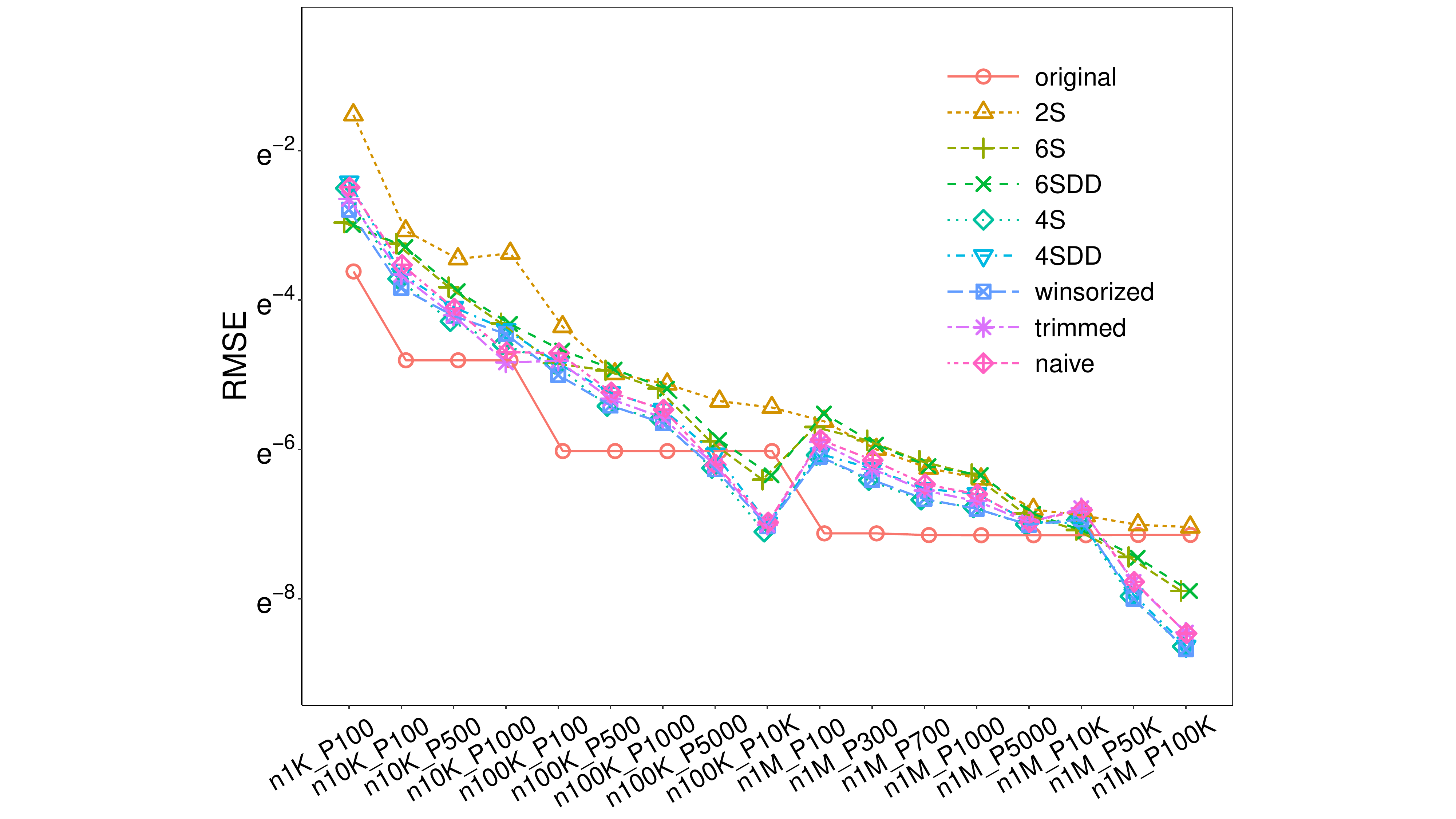}
\includegraphics[width=0.26\textwidth, trim={2.2in 0 2.2in 0},clip] {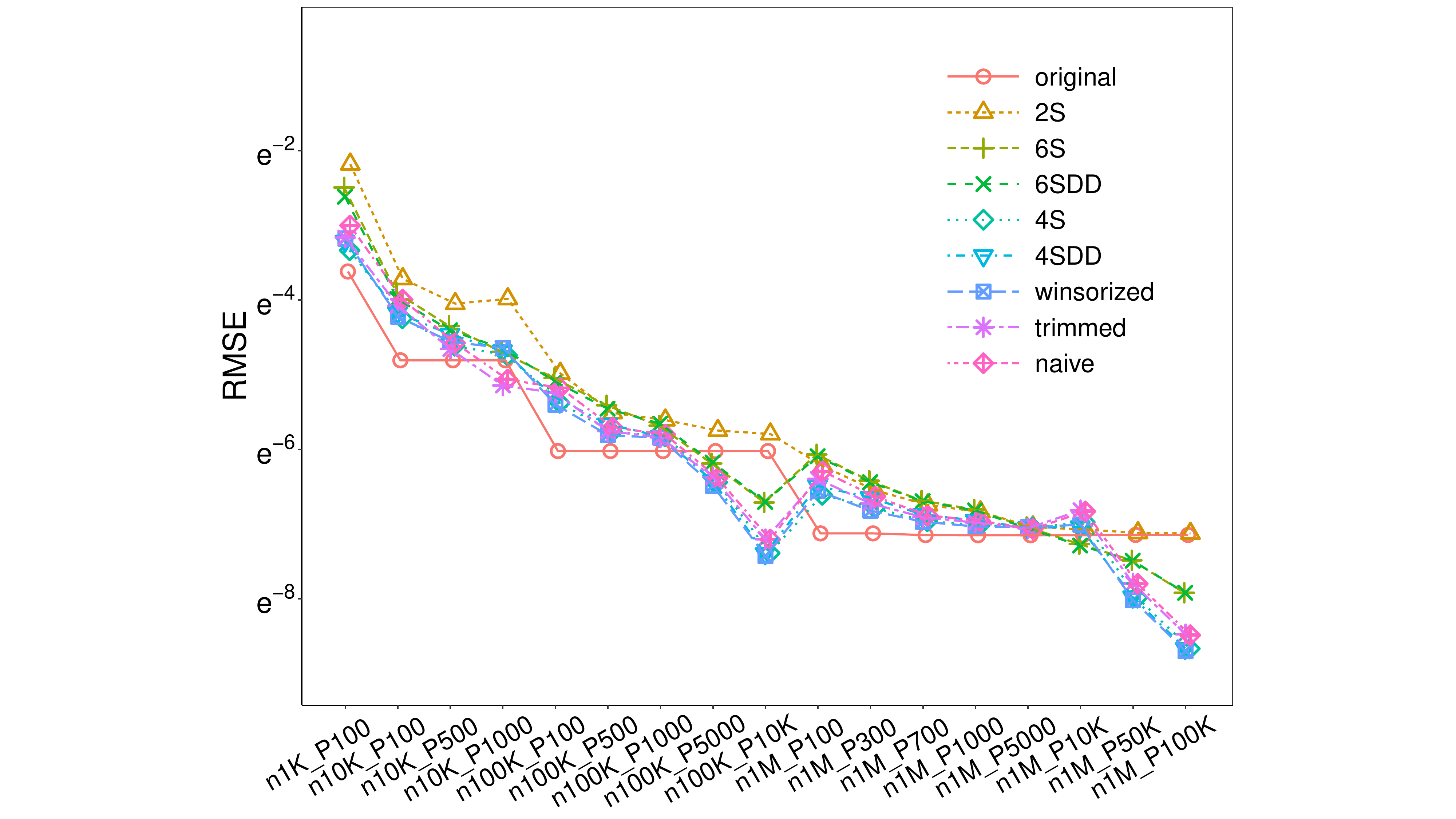}
\includegraphics[width=0.26\textwidth, trim={2.2in 0 2.2in 0},clip] {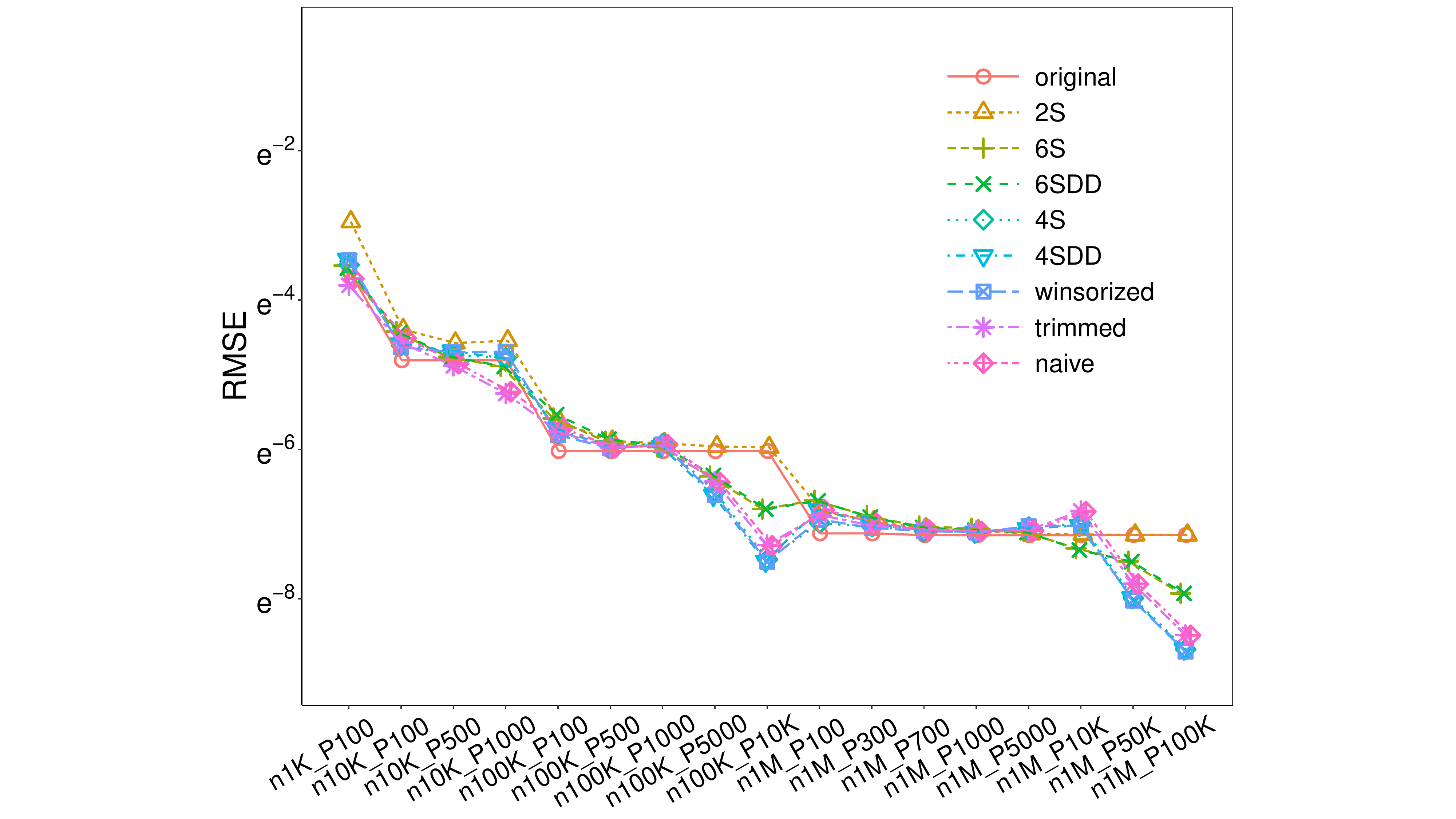}
\includegraphics[width=0.26\textwidth, trim={2.2in 0 2.2in 0},clip] {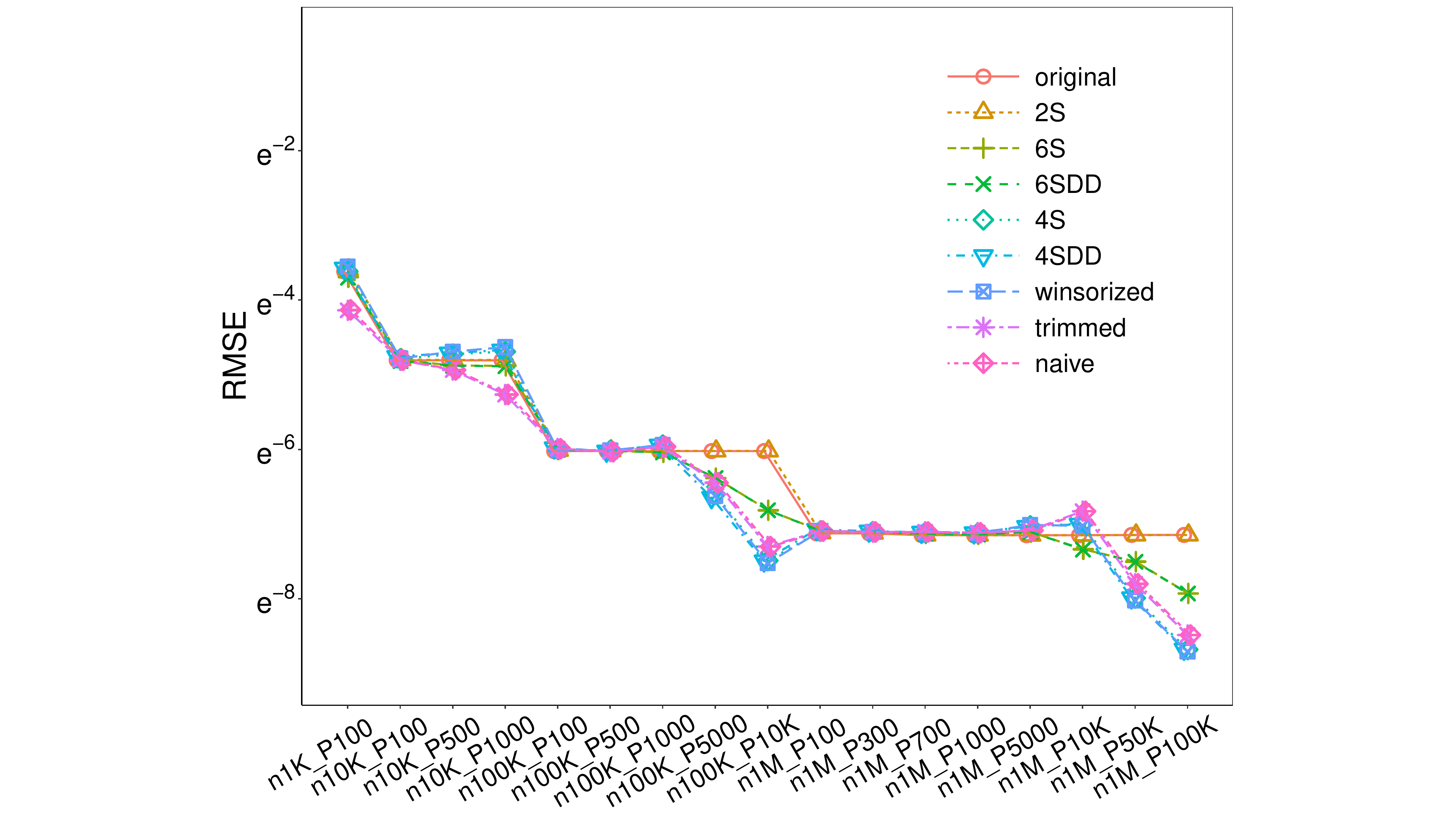}\\
\includegraphics[width=0.26\textwidth, trim={2.2in 0 2.2in 0},clip] {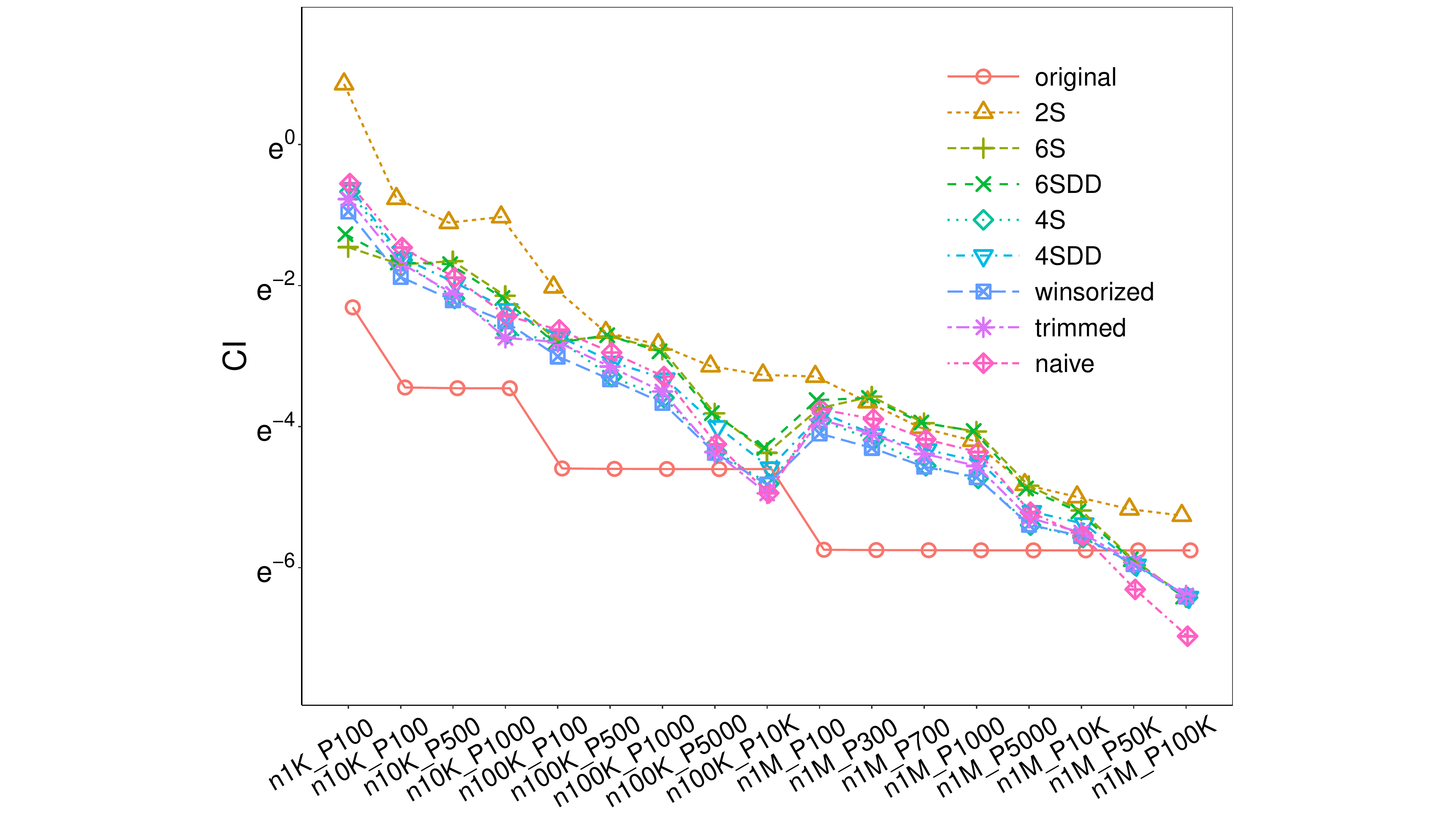}
\includegraphics[width=0.26\textwidth, trim={2.2in 0 2.2in 0},clip] {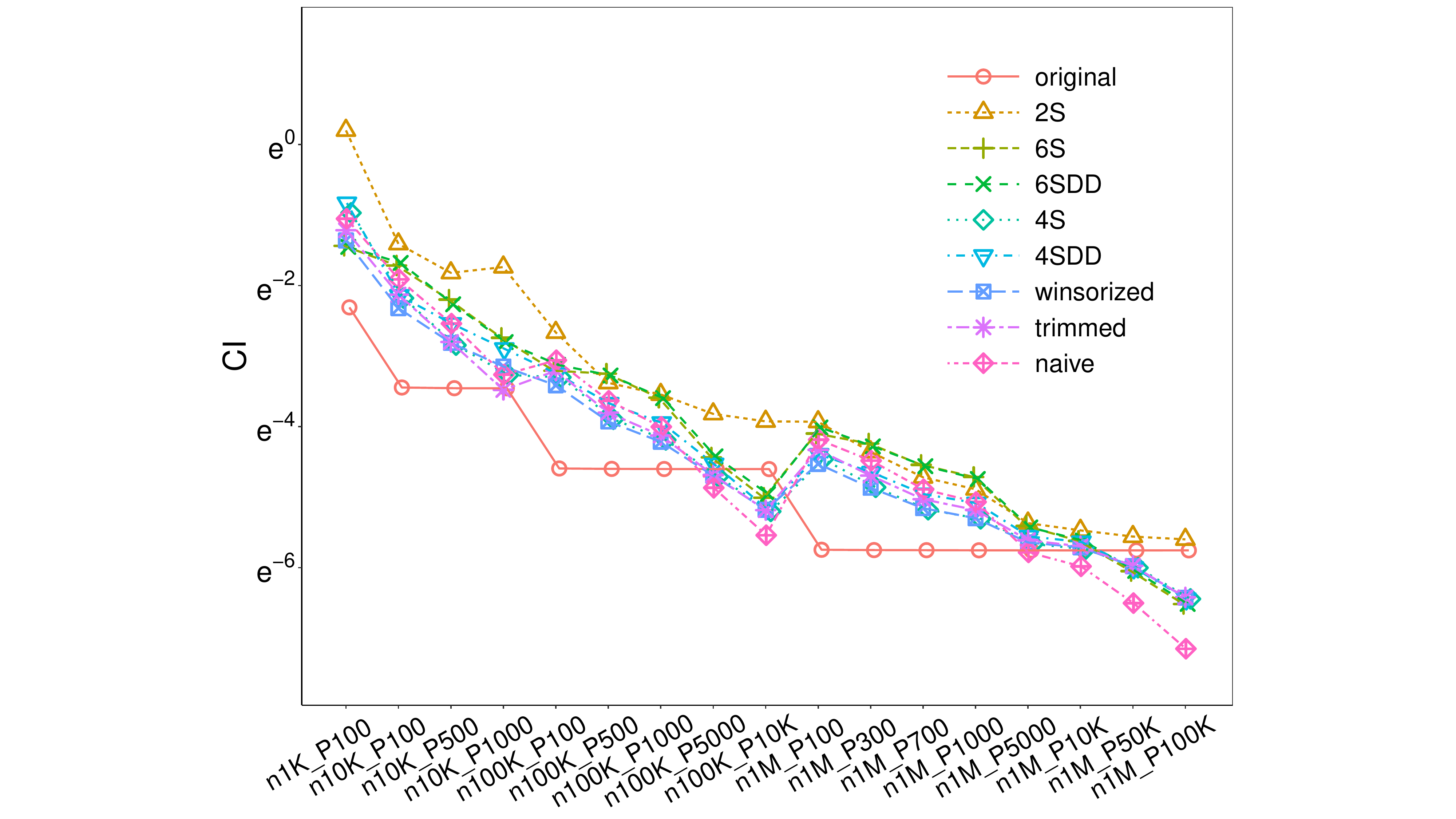}
\includegraphics[width=0.26\textwidth, trim={2.2in 0 2.2in 0},clip] {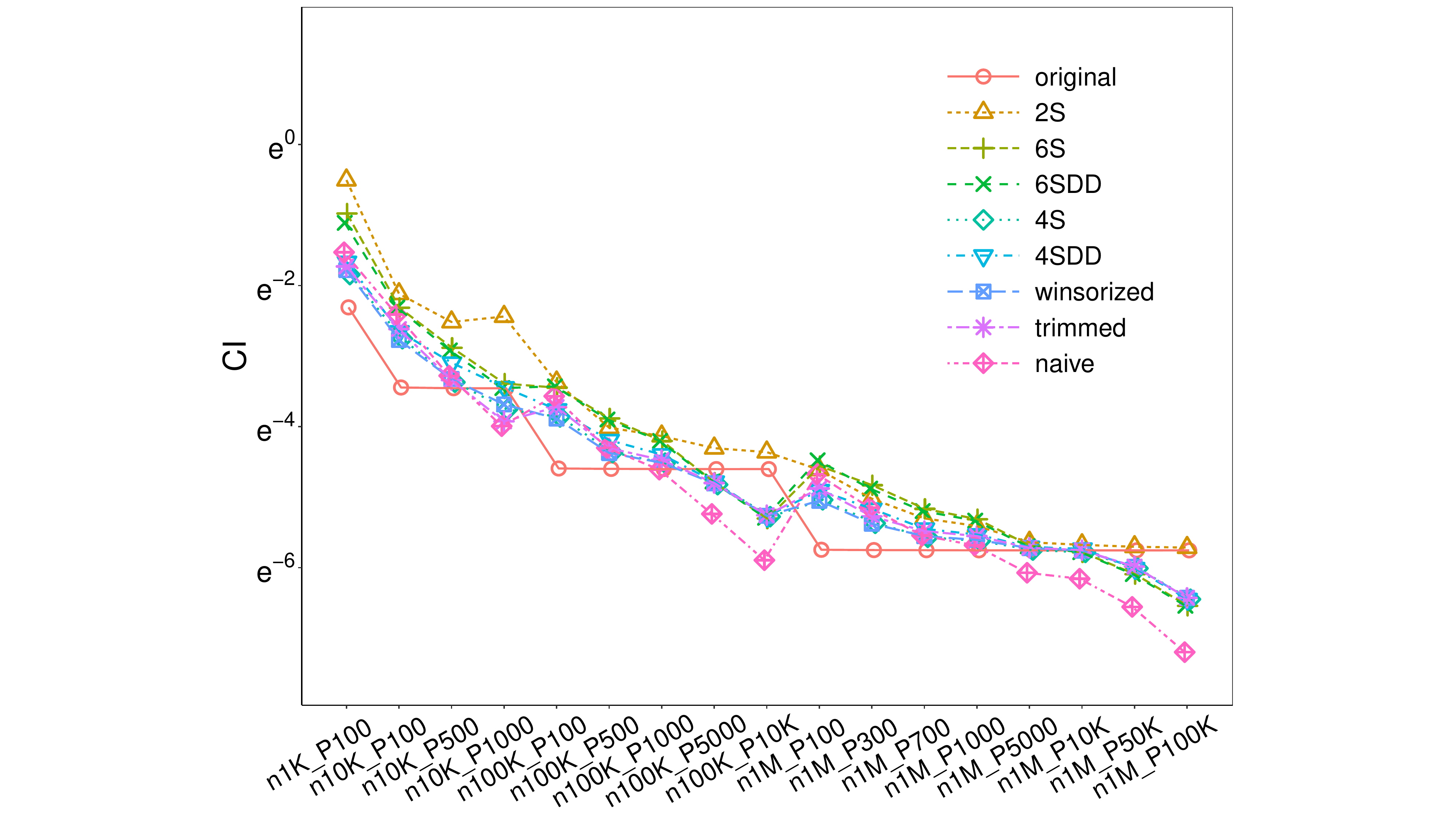}
\includegraphics[width=0.26\textwidth, trim={2.2in 0 2.2in 0},clip] {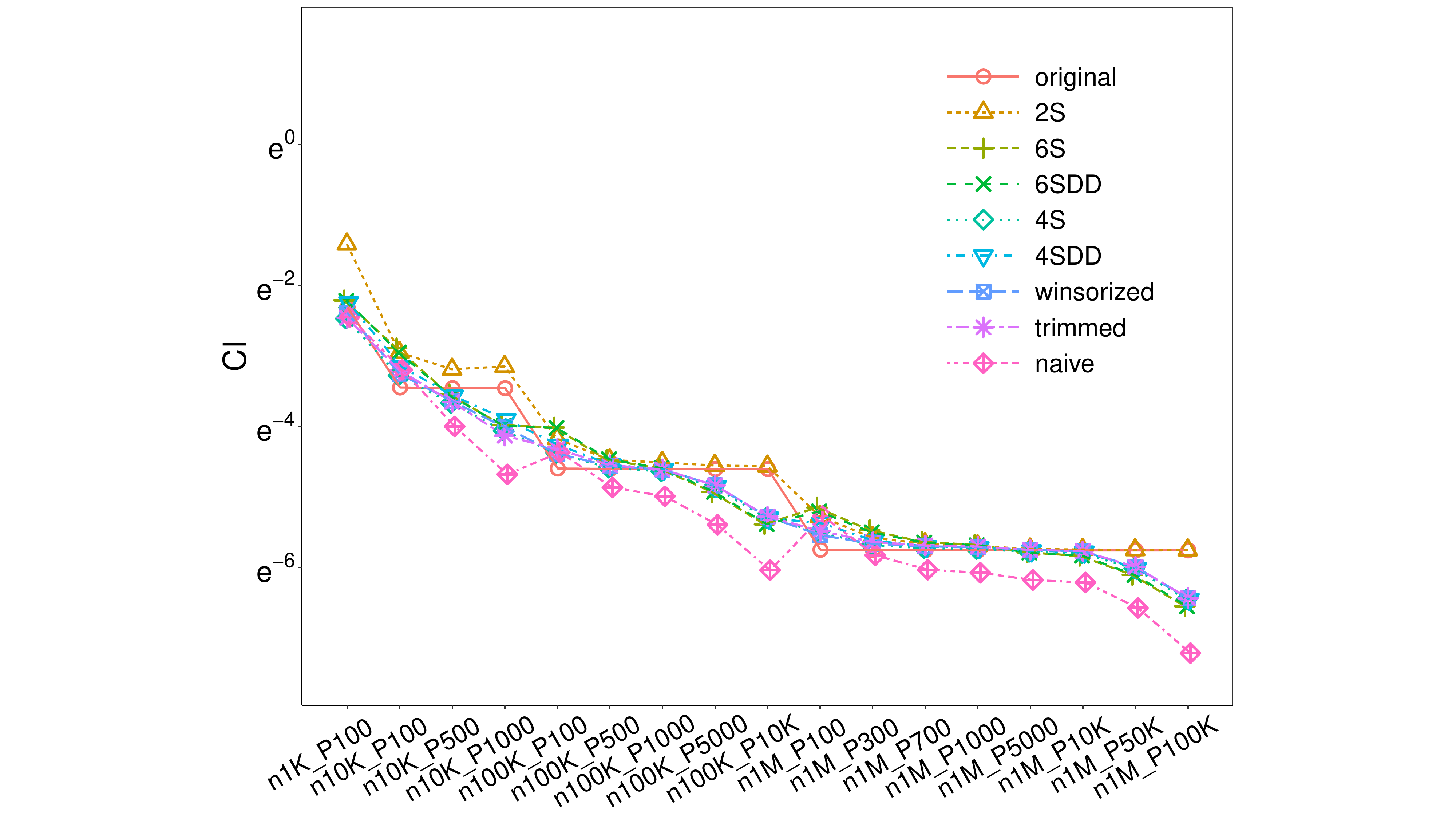}
\includegraphics[width=0.26\textwidth, trim={2.2in 0 2.2in 0},clip] {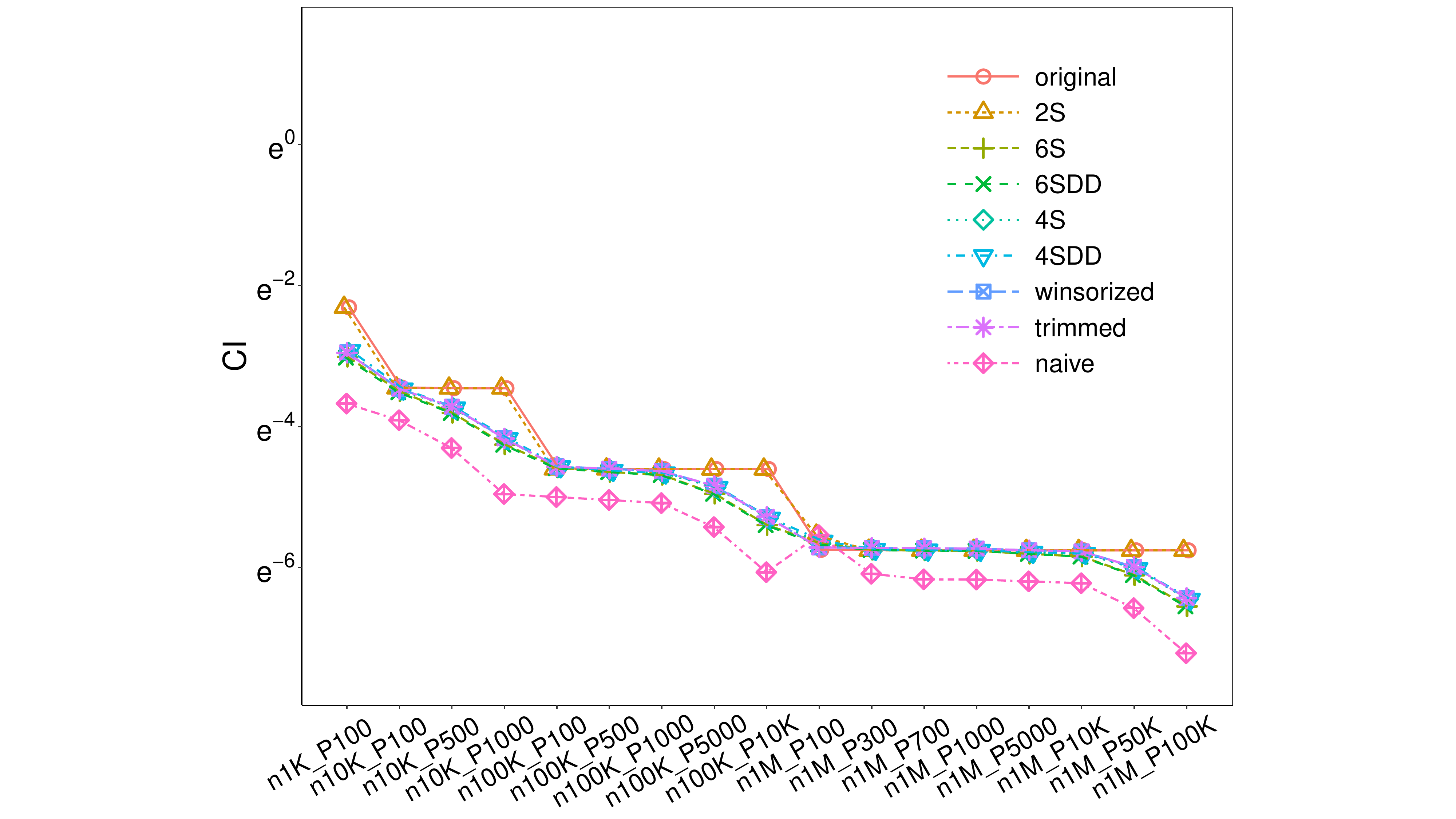}\\
\includegraphics[width=0.26\textwidth, trim={2.2in 0 2.2in 0},clip] {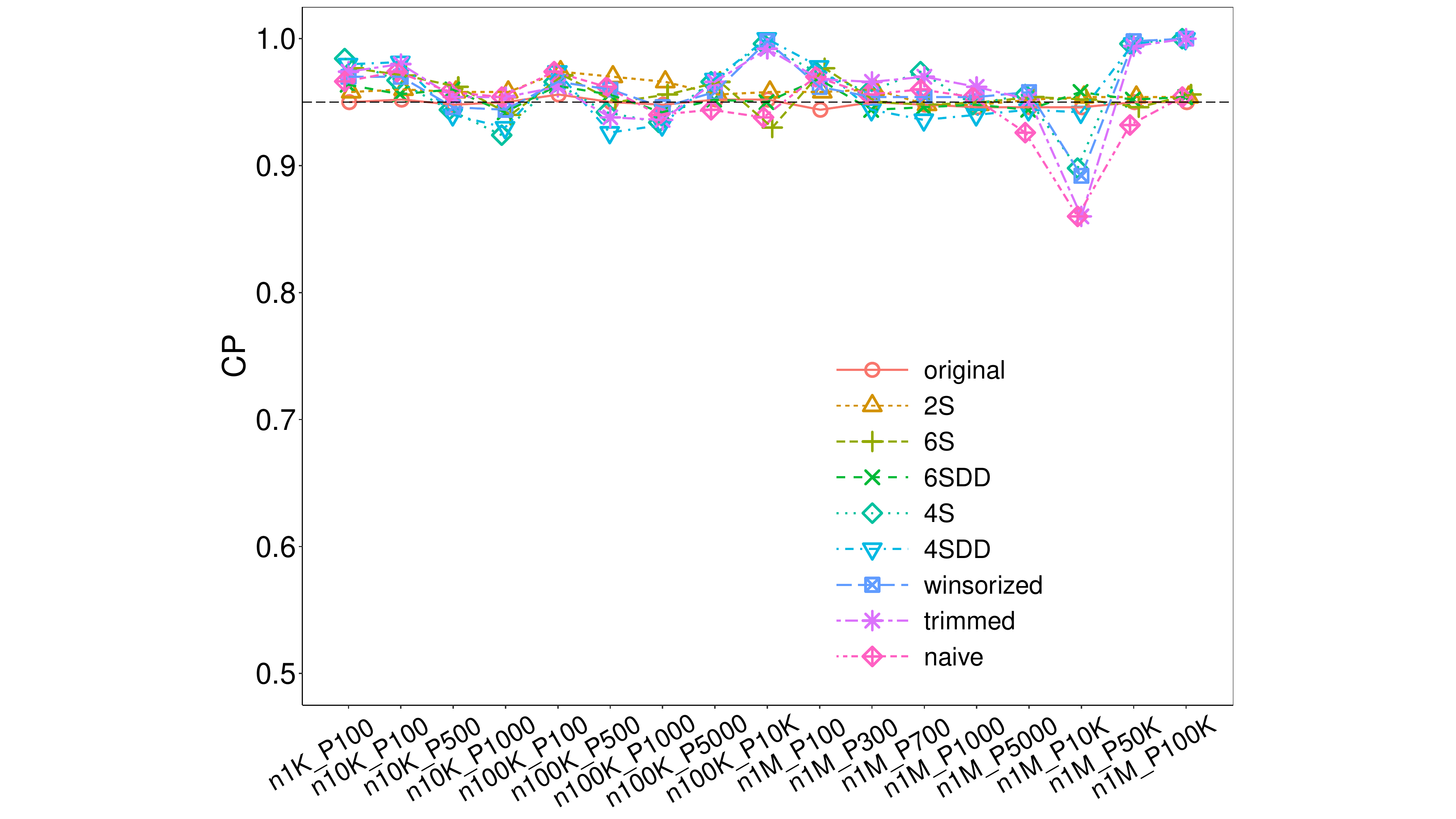}
\includegraphics[width=0.26\textwidth, trim={2.2in 0 2.2in 0},clip] {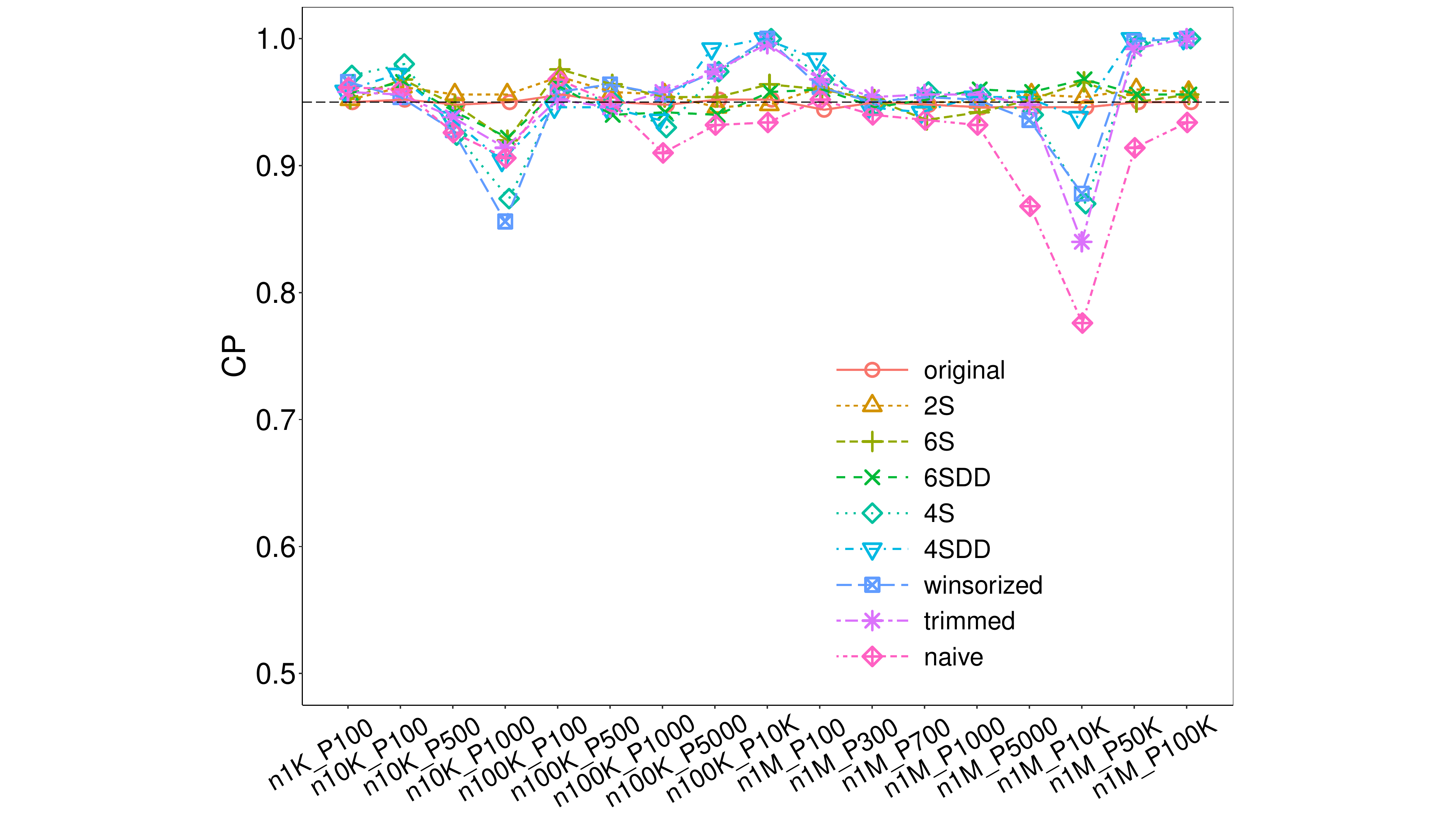}
\includegraphics[width=0.26\textwidth, trim={2.2in 0 2.2in 0},clip] {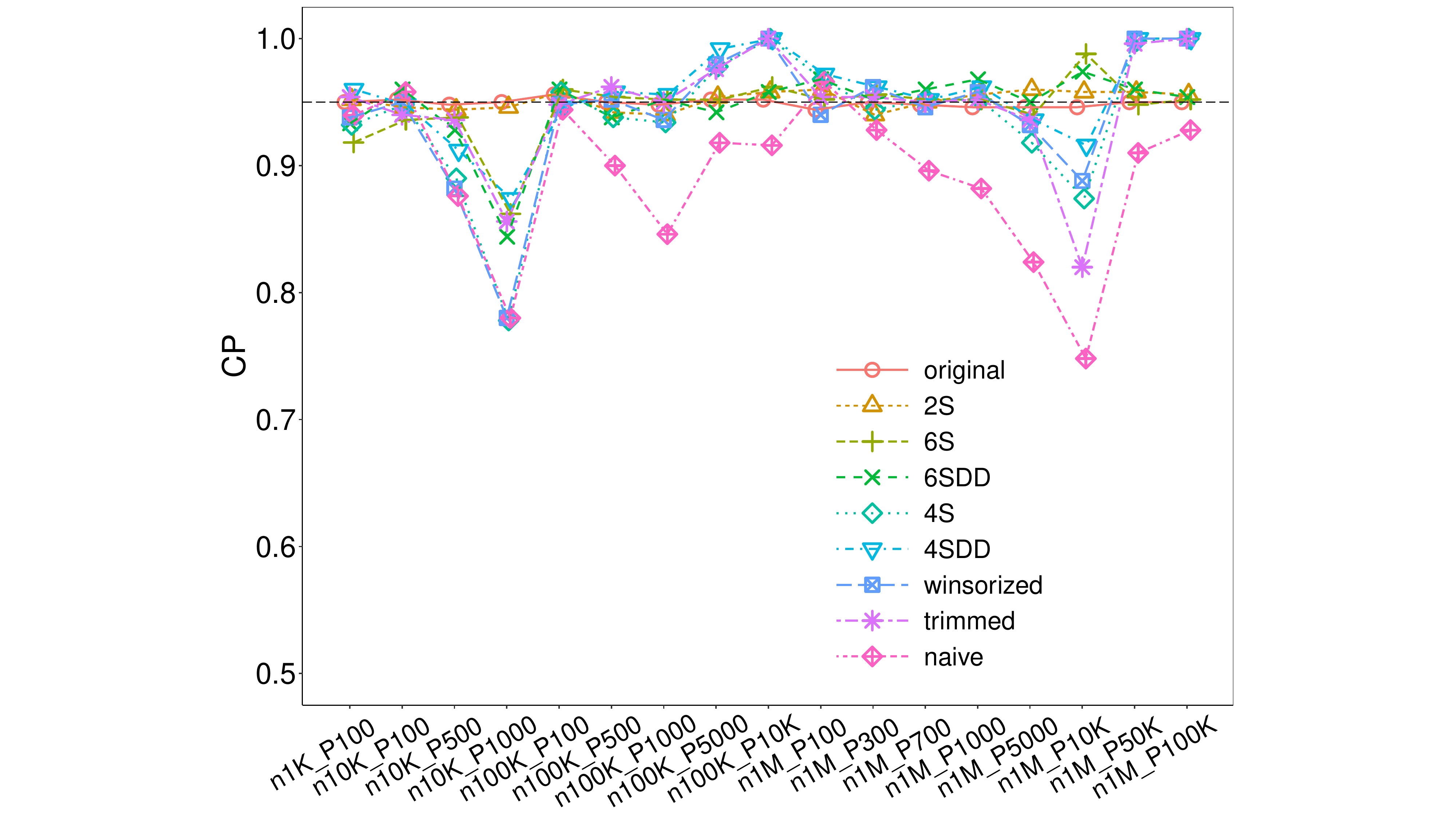}
\includegraphics[width=0.26\textwidth, trim={2.2in 0 2.2in 0},clip] {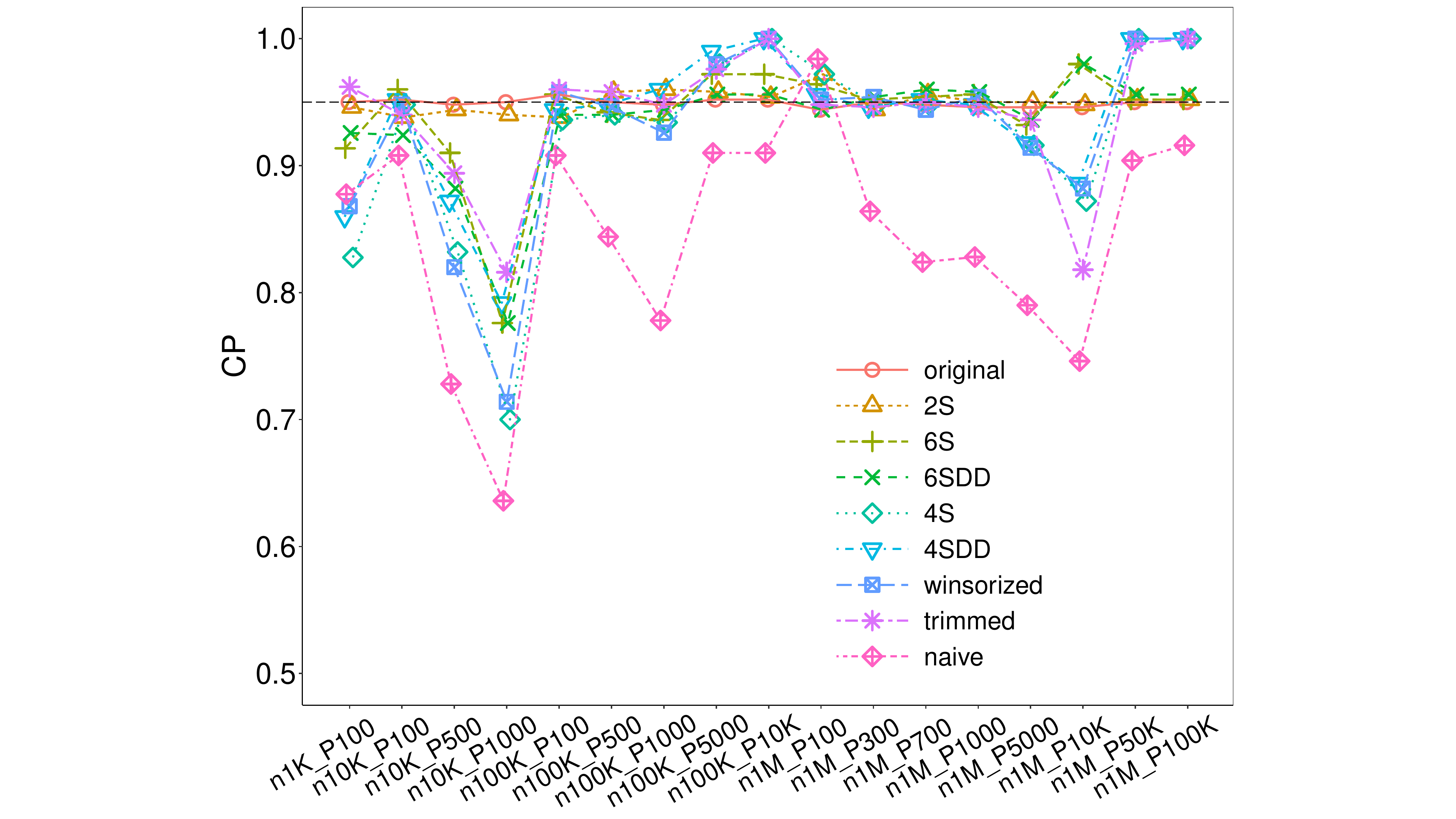}
\includegraphics[width=0.26\textwidth, trim={2.2in 0 2.2in 0},clip] {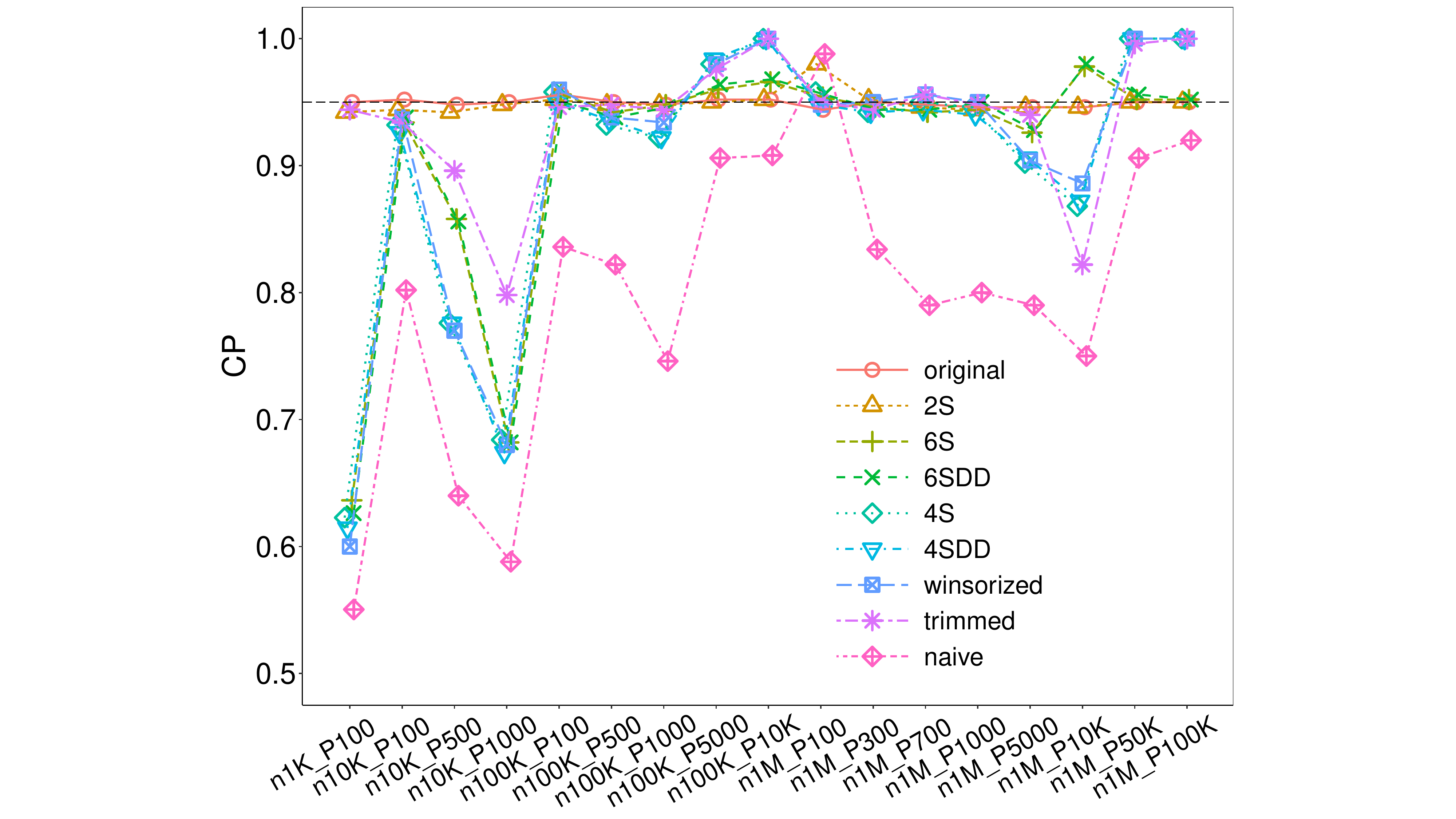}\\

\caption{ZINB data; $\epsilon$-DP; $\theta=0$ and $\alpha=\beta$} \label{fig:0sDPzinb}
\end{figure}
\end{landscape}

\begin{landscape}
\begin{figure}[!htb]
\hspace{0.6in}$\rho=0.005$\hspace{1in}$\rho=0.02$\hspace{1.2in}$\rho=0.08$
\hspace{1.1in}$\rho=0.32$\hspace{1.2in}$\rho=1.28$\\
\includegraphics[width=0.26\textwidth, trim={2.2in 0 2.2in 0},clip] {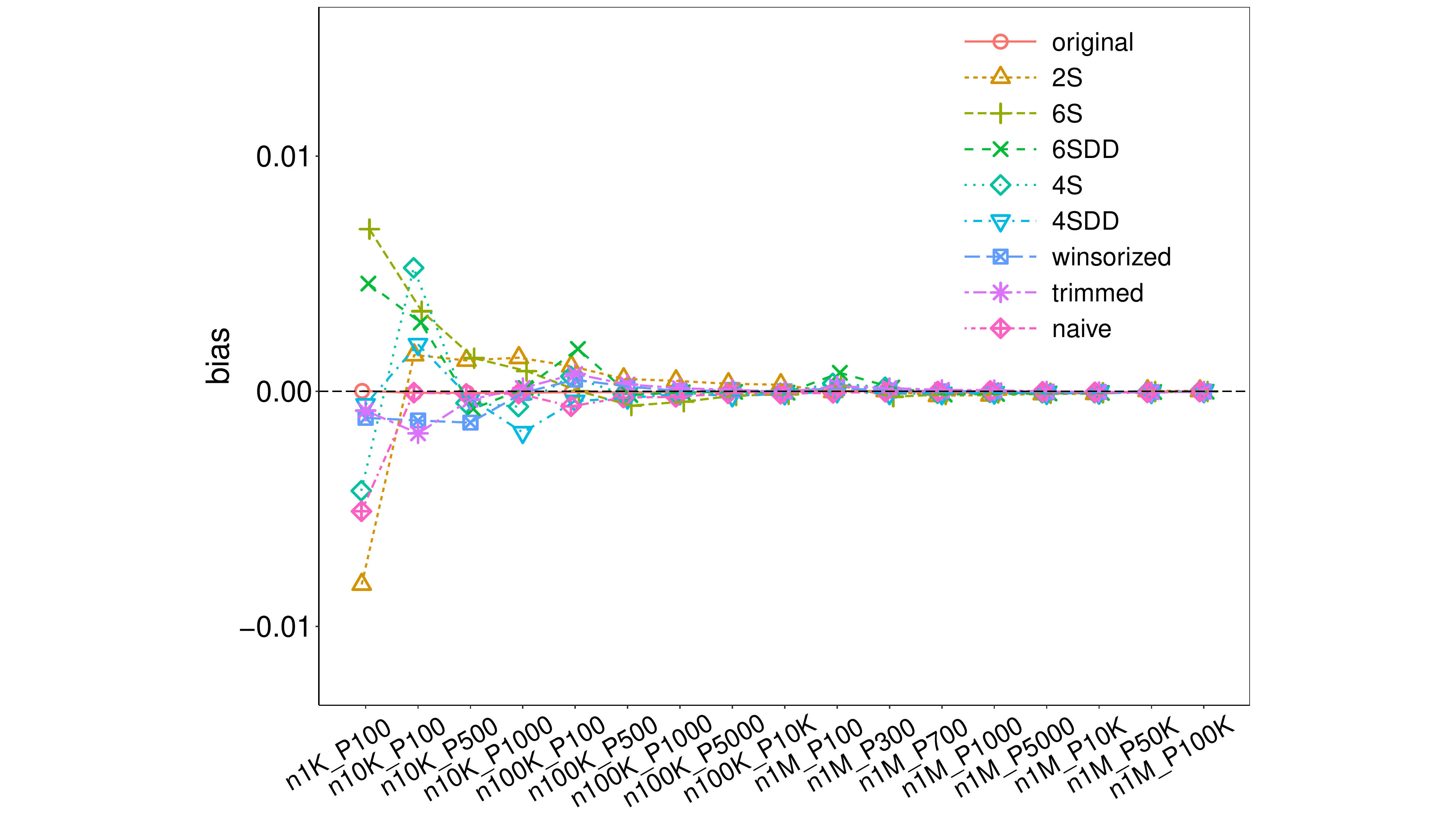}
\includegraphics[width=0.26\textwidth, trim={2.2in 0 2.2in 0},clip] {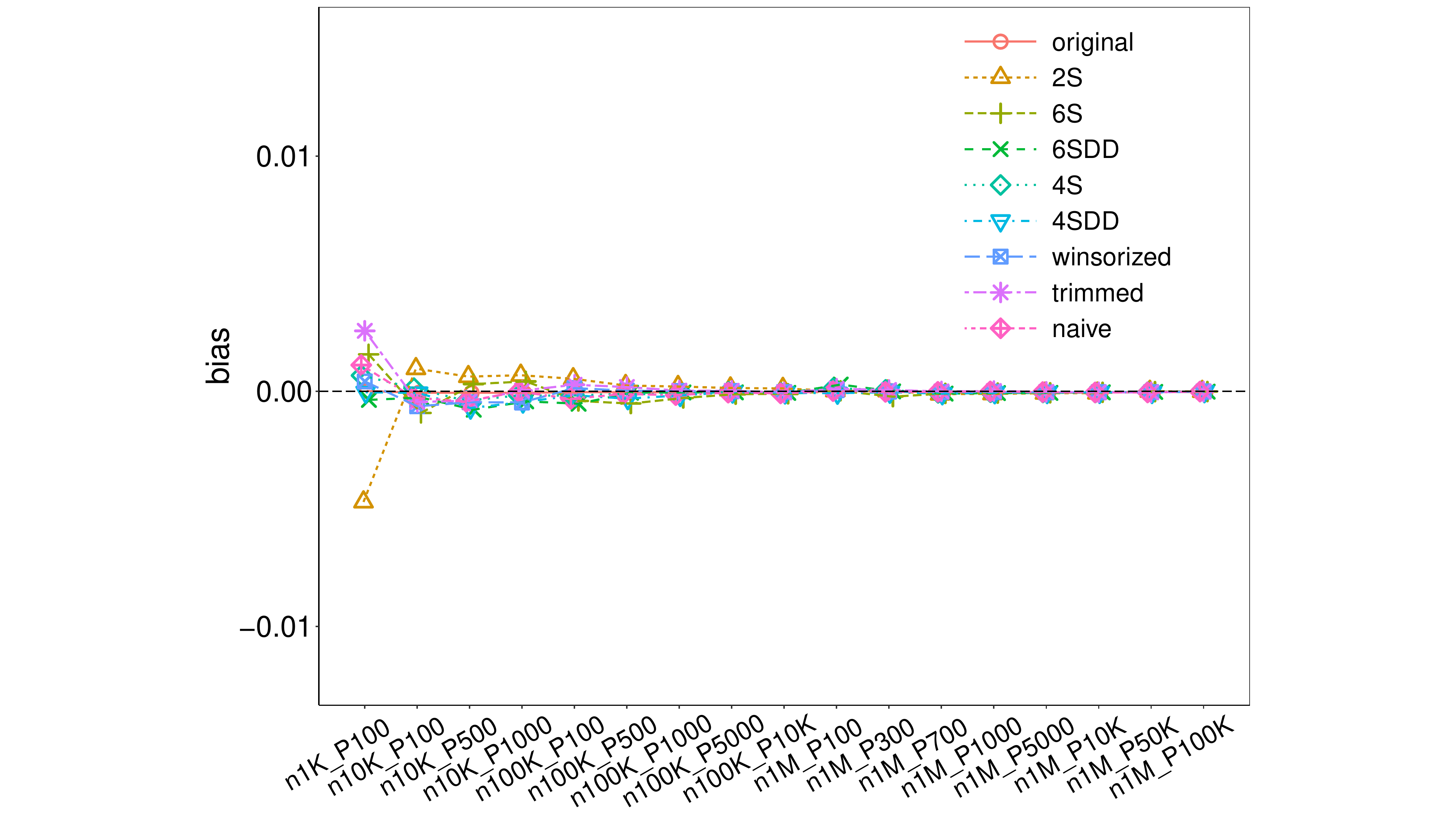}
\includegraphics[width=0.26\textwidth, trim={2.2in 0 2.2in 0},clip] {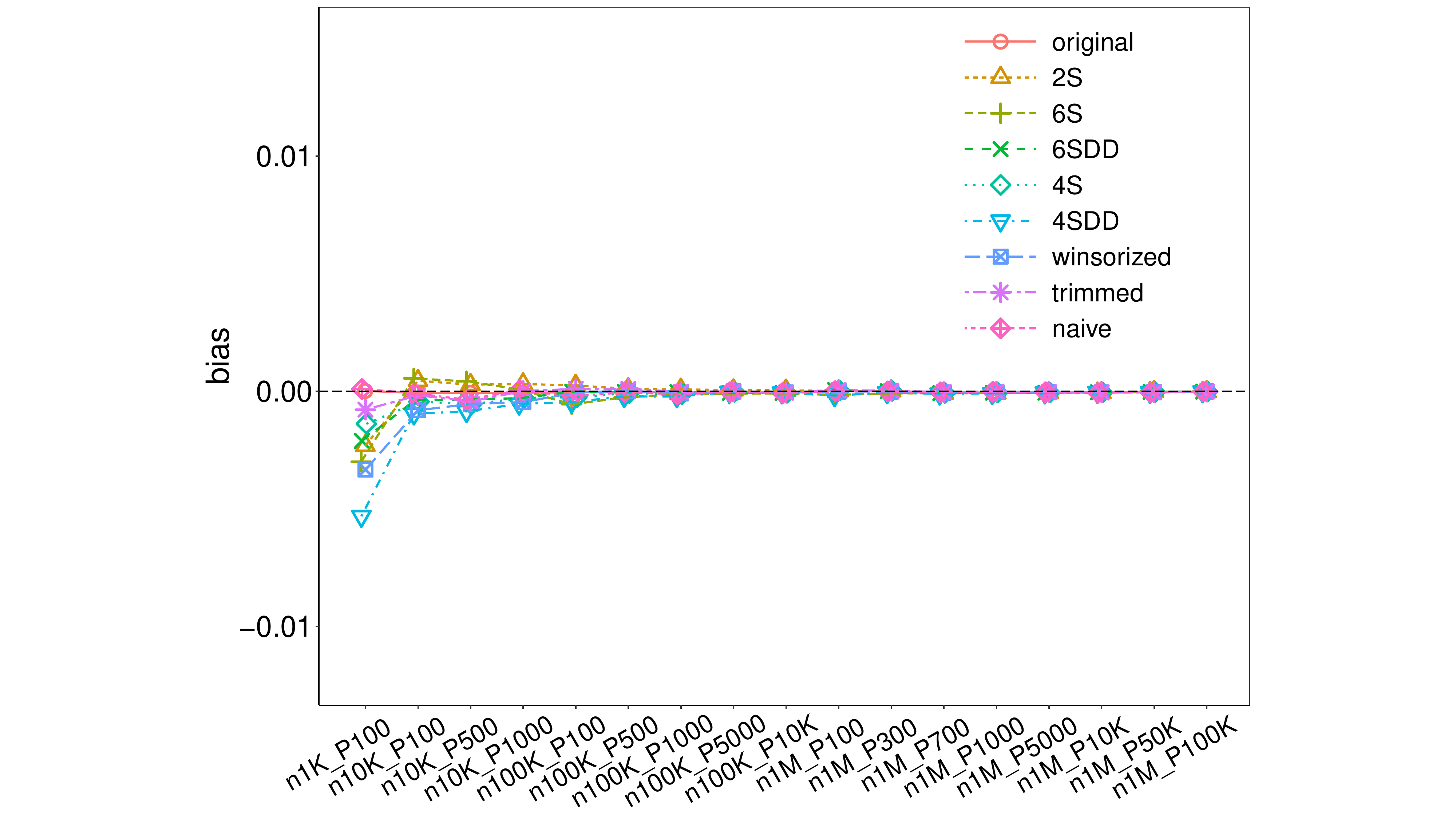}
\includegraphics[width=0.26\textwidth, trim={2.2in 0 2.2in 0},clip] {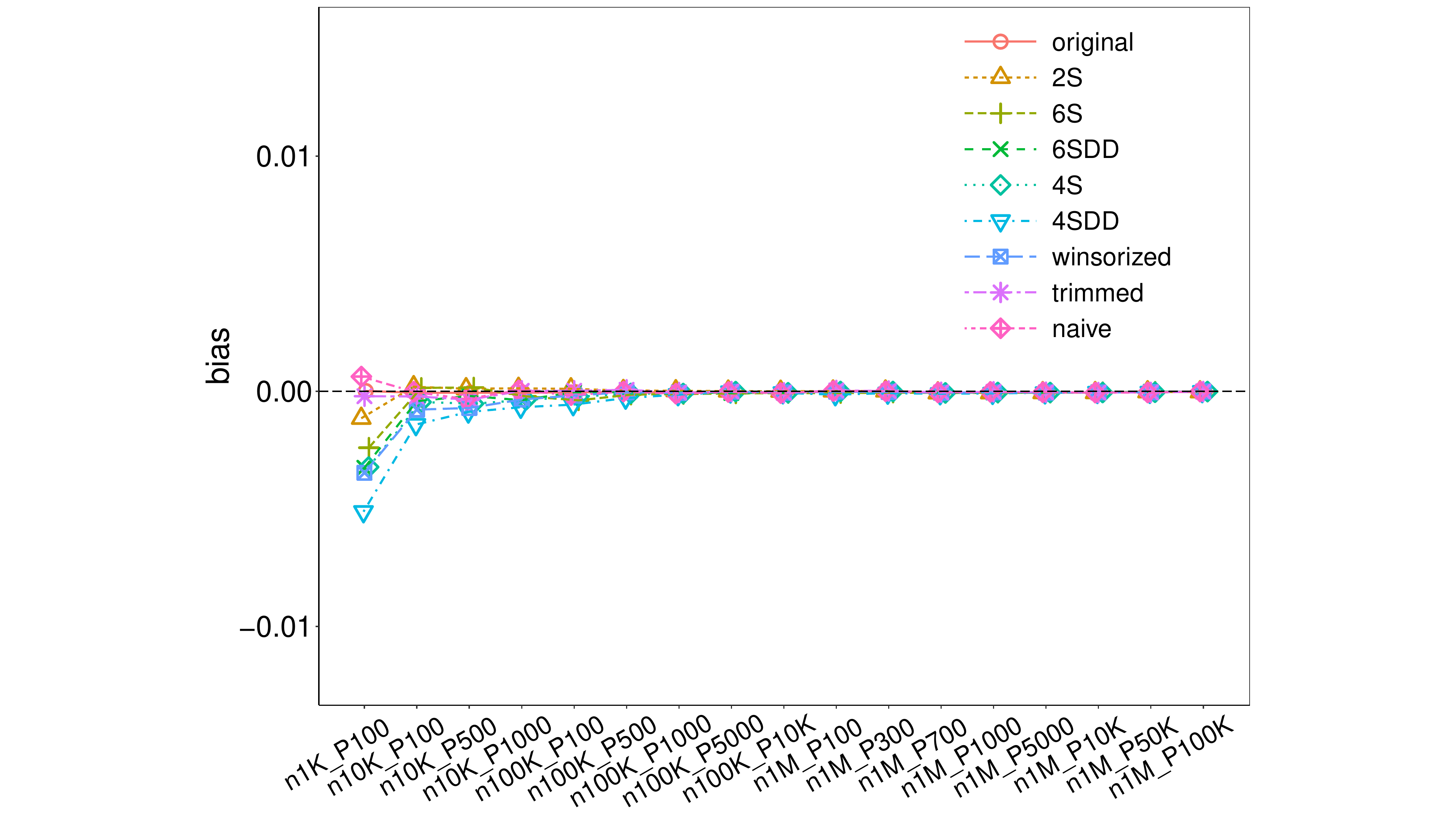}
\includegraphics[width=0.26\textwidth, trim={2.2in 0 2.2in 0},clip] {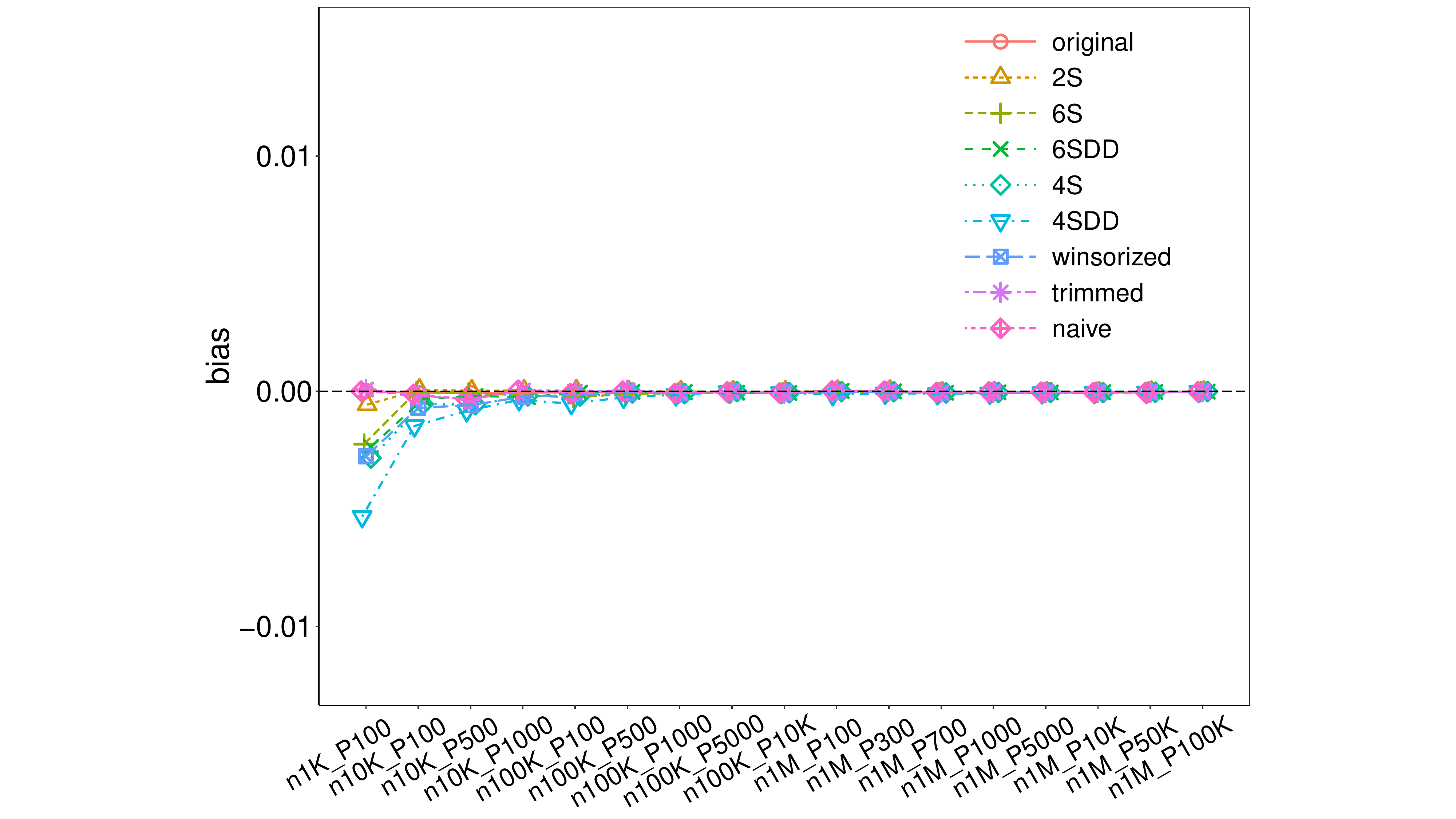}\\
\includegraphics[width=0.26\textwidth, trim={2.2in 0 2.2in 0},clip] {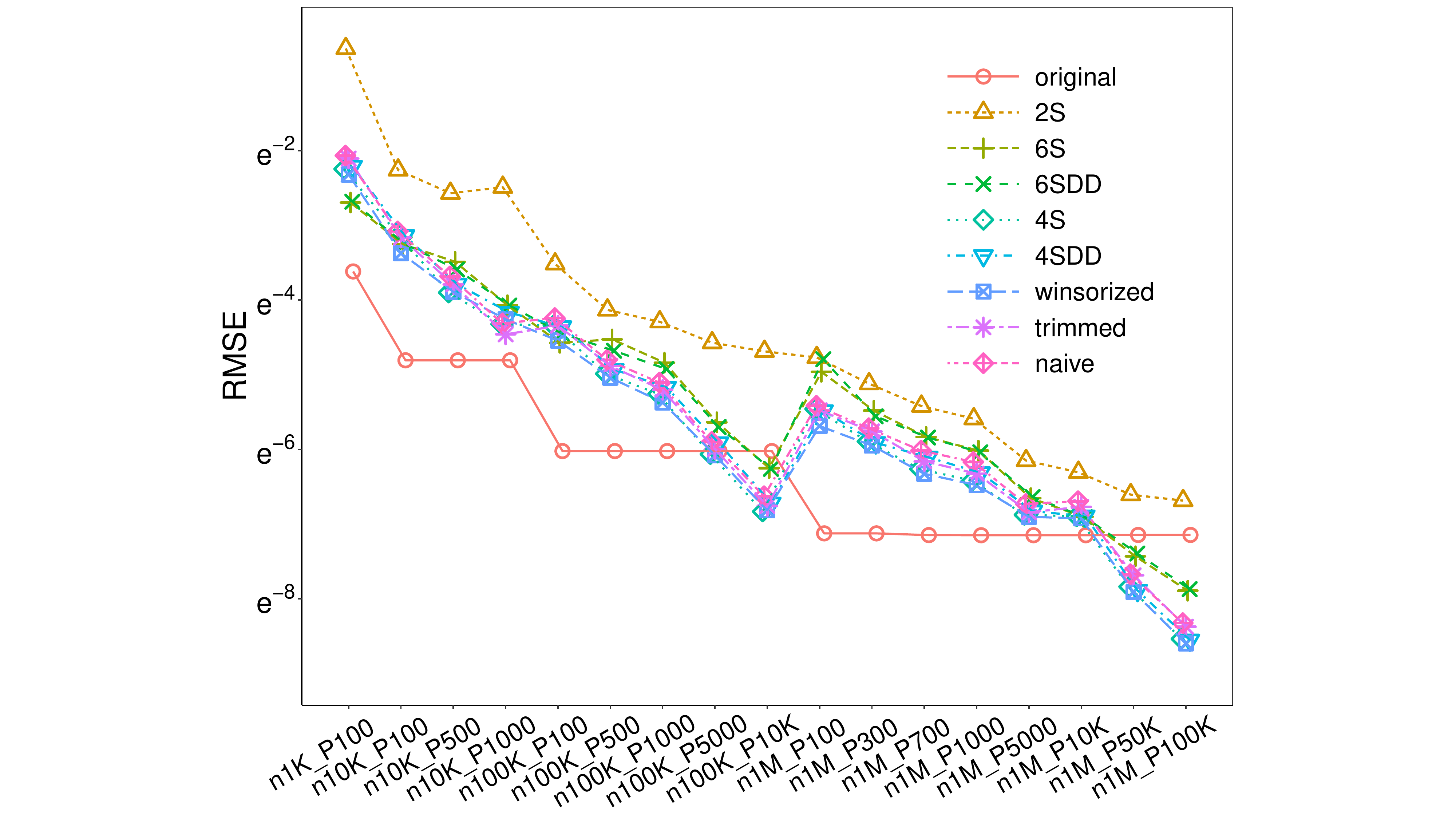}
\includegraphics[width=0.26\textwidth, trim={2.2in 0 2.2in 0},clip] {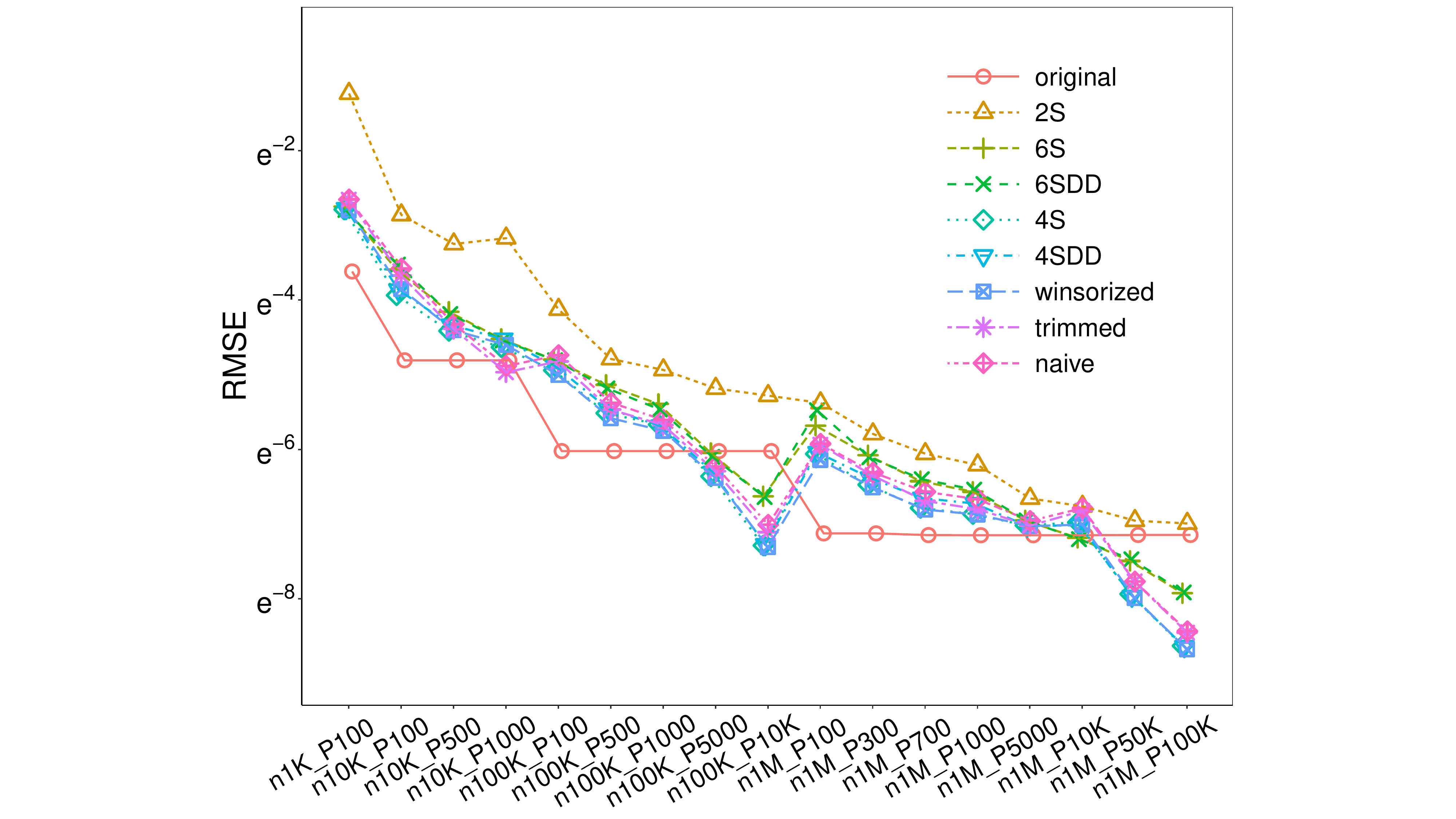}
\includegraphics[width=0.26\textwidth, trim={2.2in 0 2.2in 0},clip] {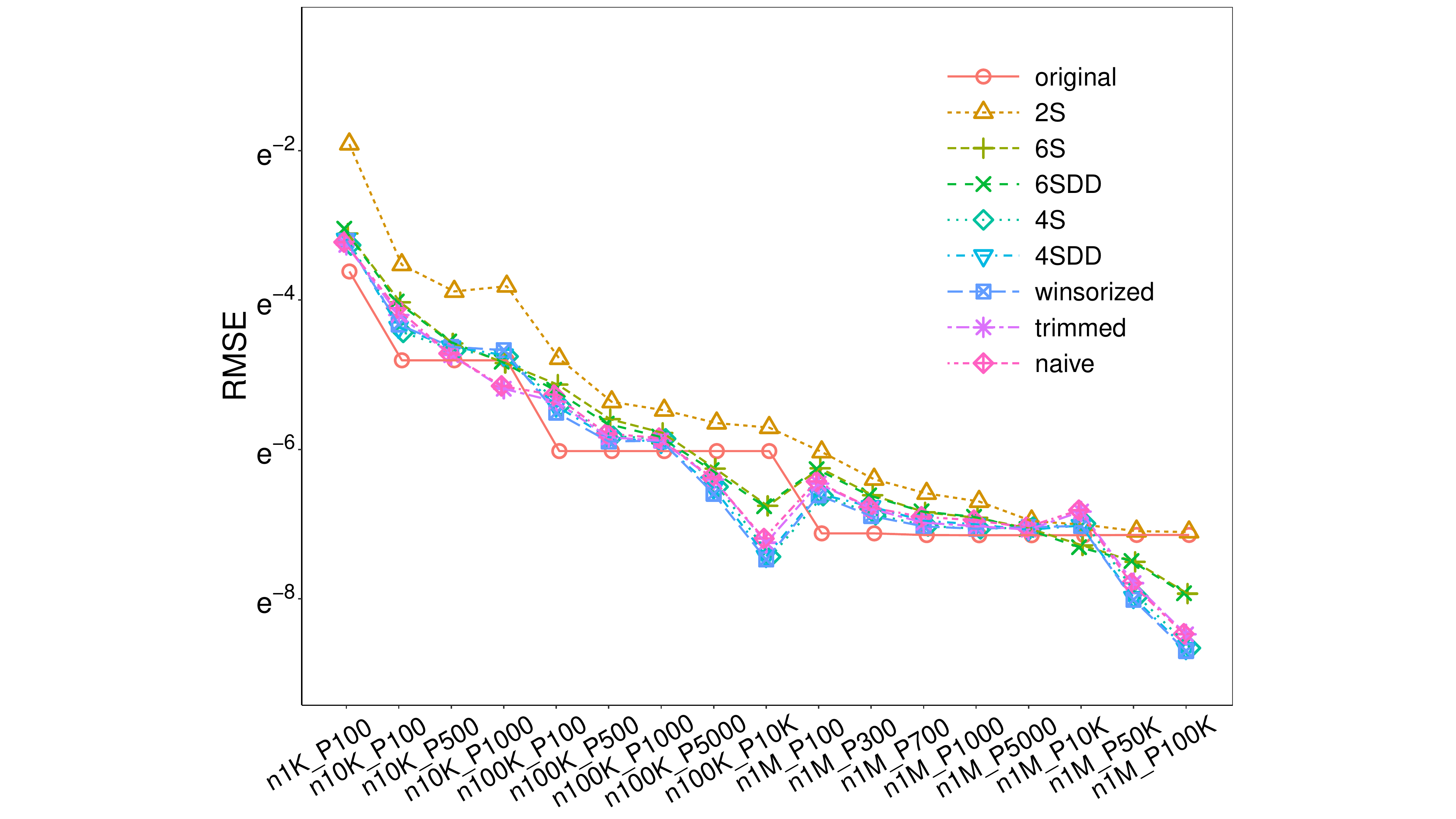}
\includegraphics[width=0.26\textwidth, trim={2.2in 0 2.2in 0},clip] {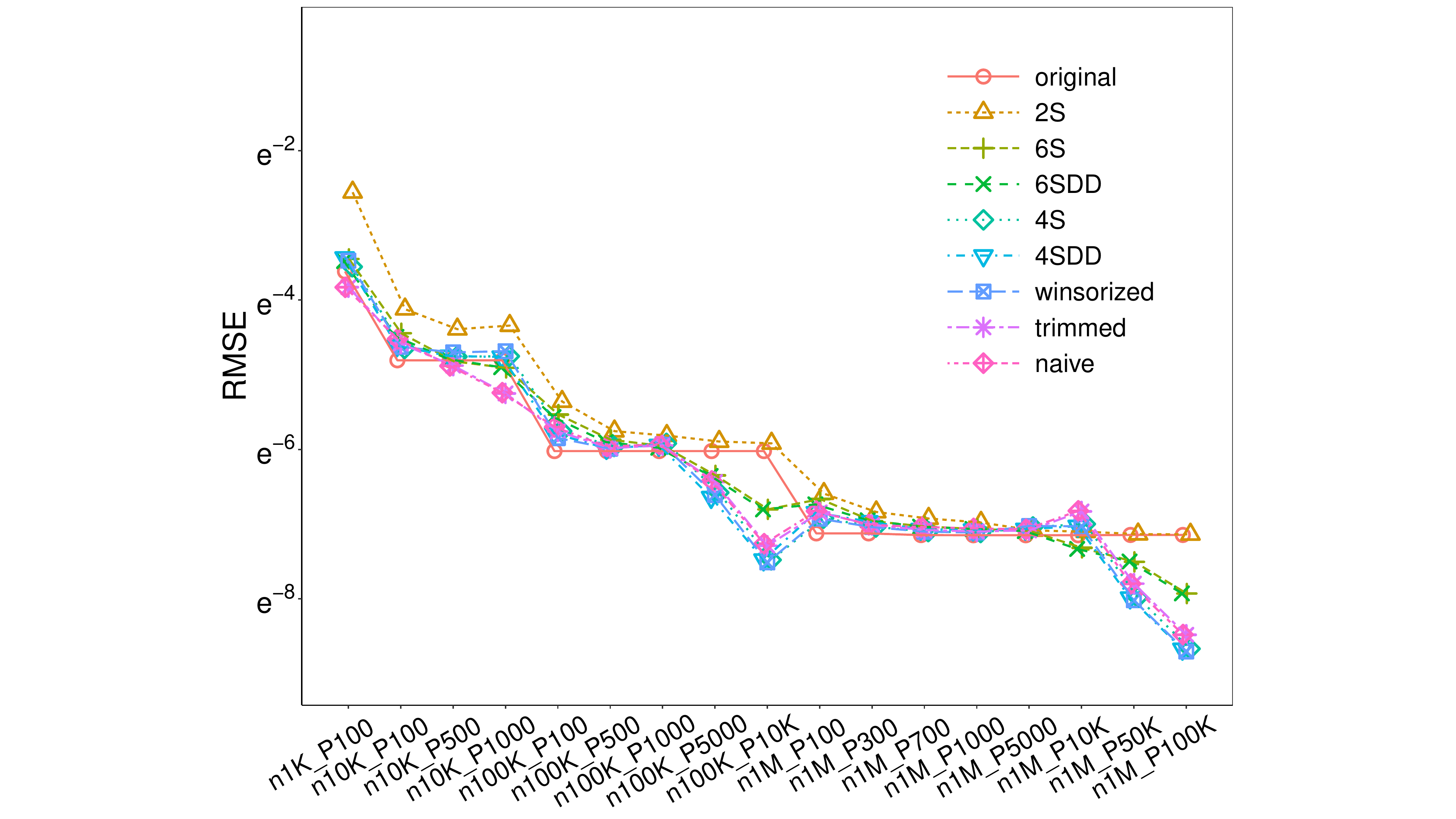}
\includegraphics[width=0.26\textwidth, trim={2.2in 0 2.2in 0},clip] {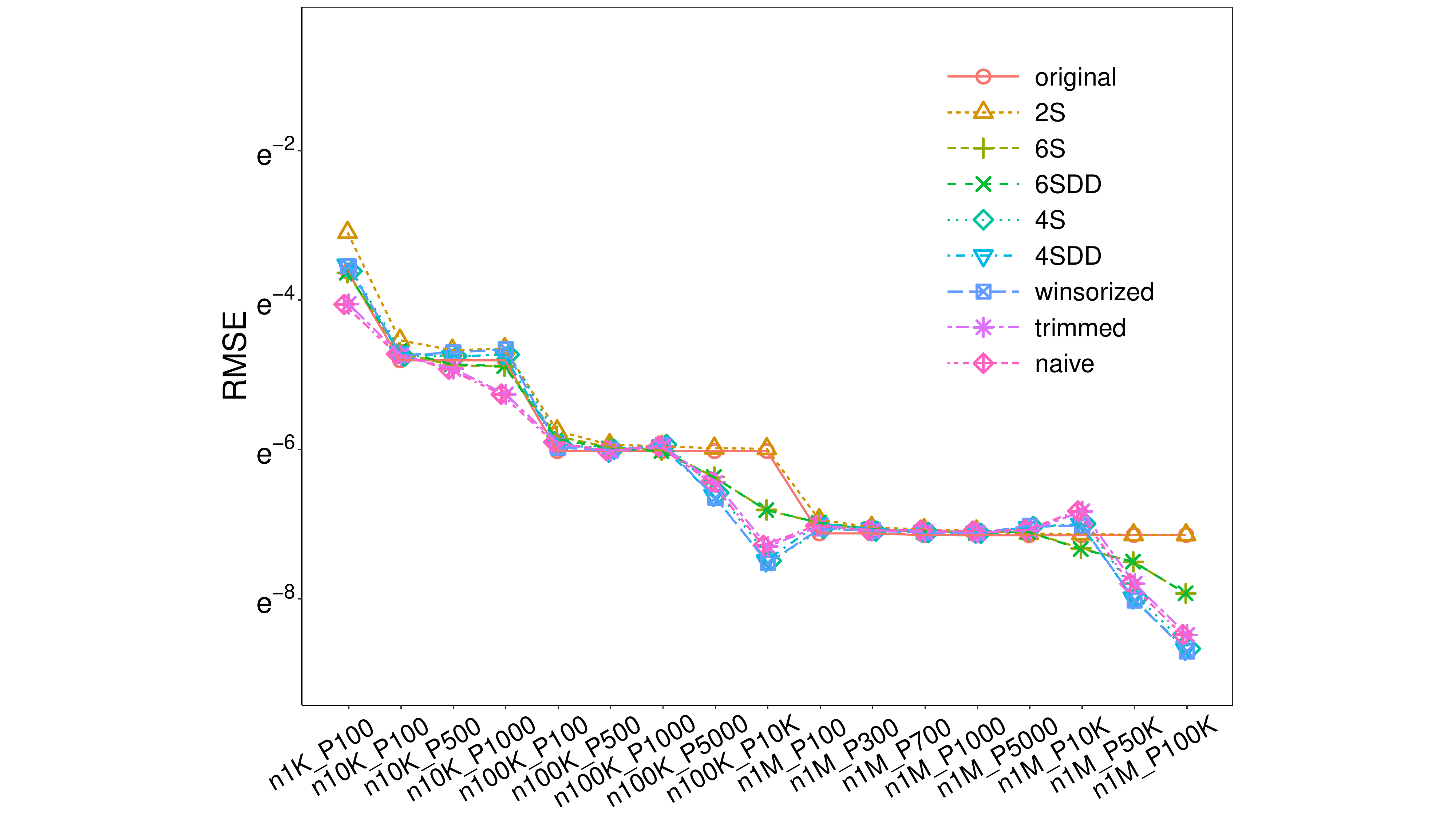}\\
\includegraphics[width=0.26\textwidth, trim={2.2in 0 2.2in 0},clip] {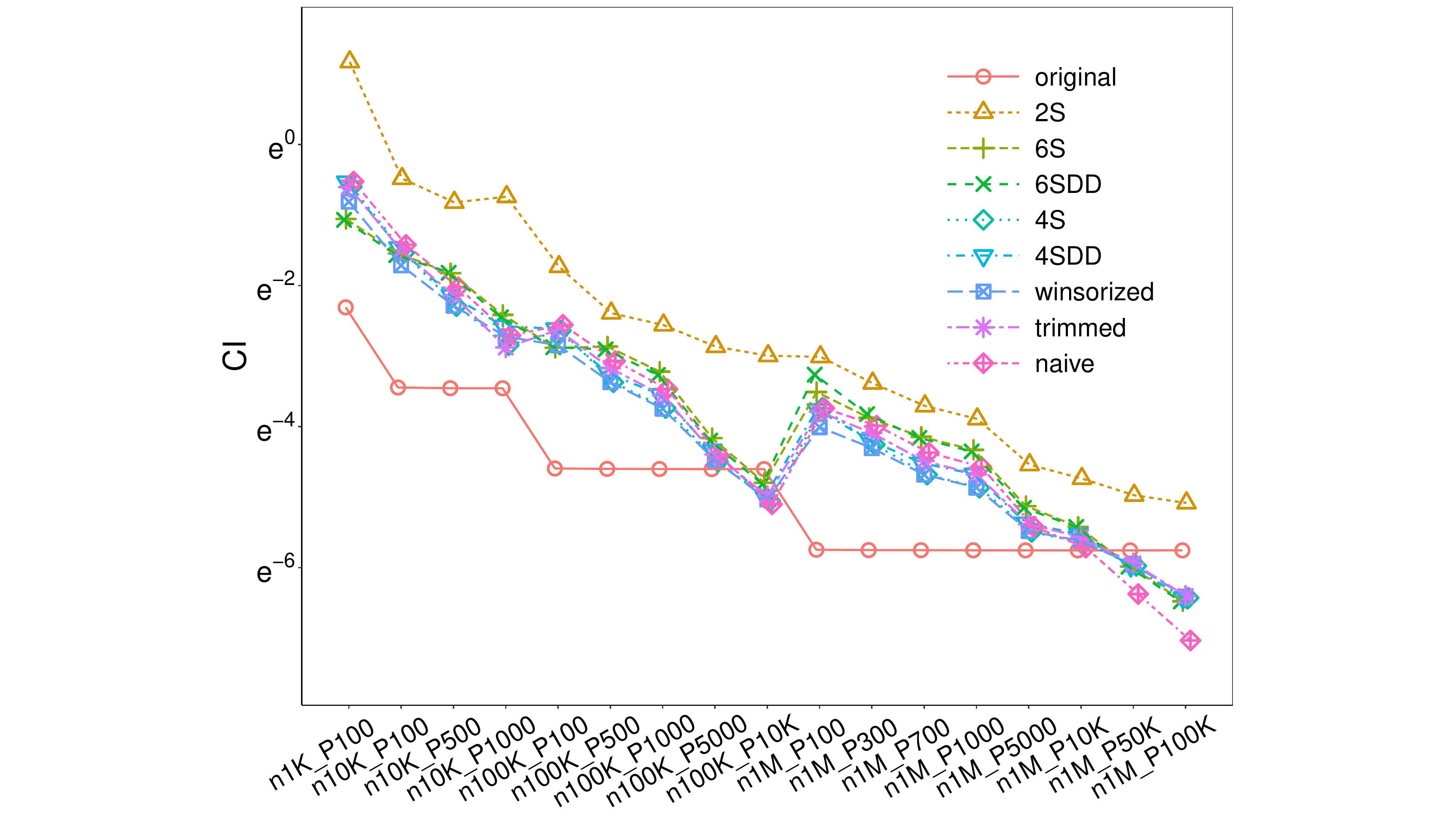}
\includegraphics[width=0.26\textwidth, trim={2.2in 0 2.2in 0},clip] {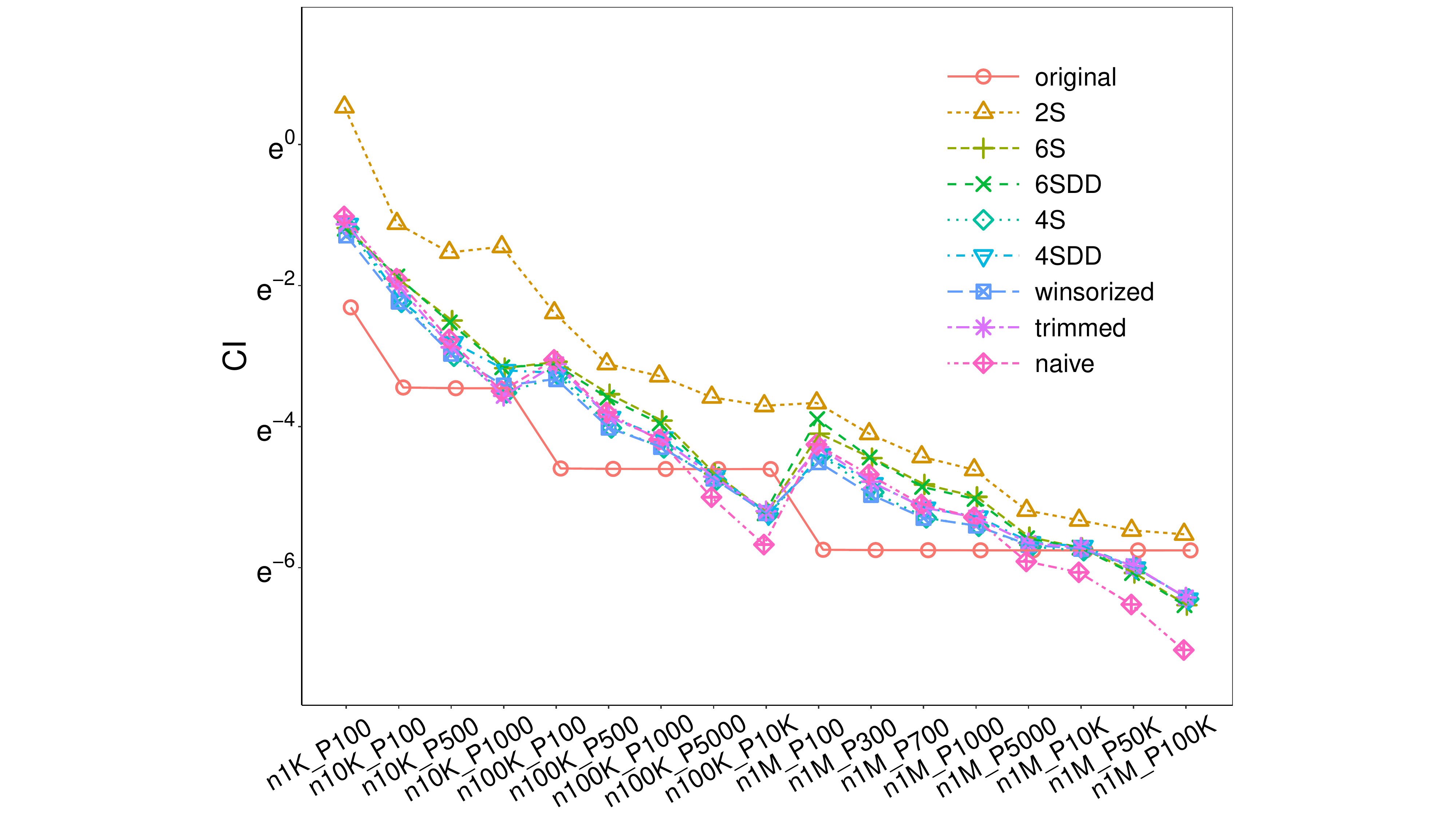}
\includegraphics[width=0.26\textwidth, trim={2.2in 0 2.2in 0},clip] {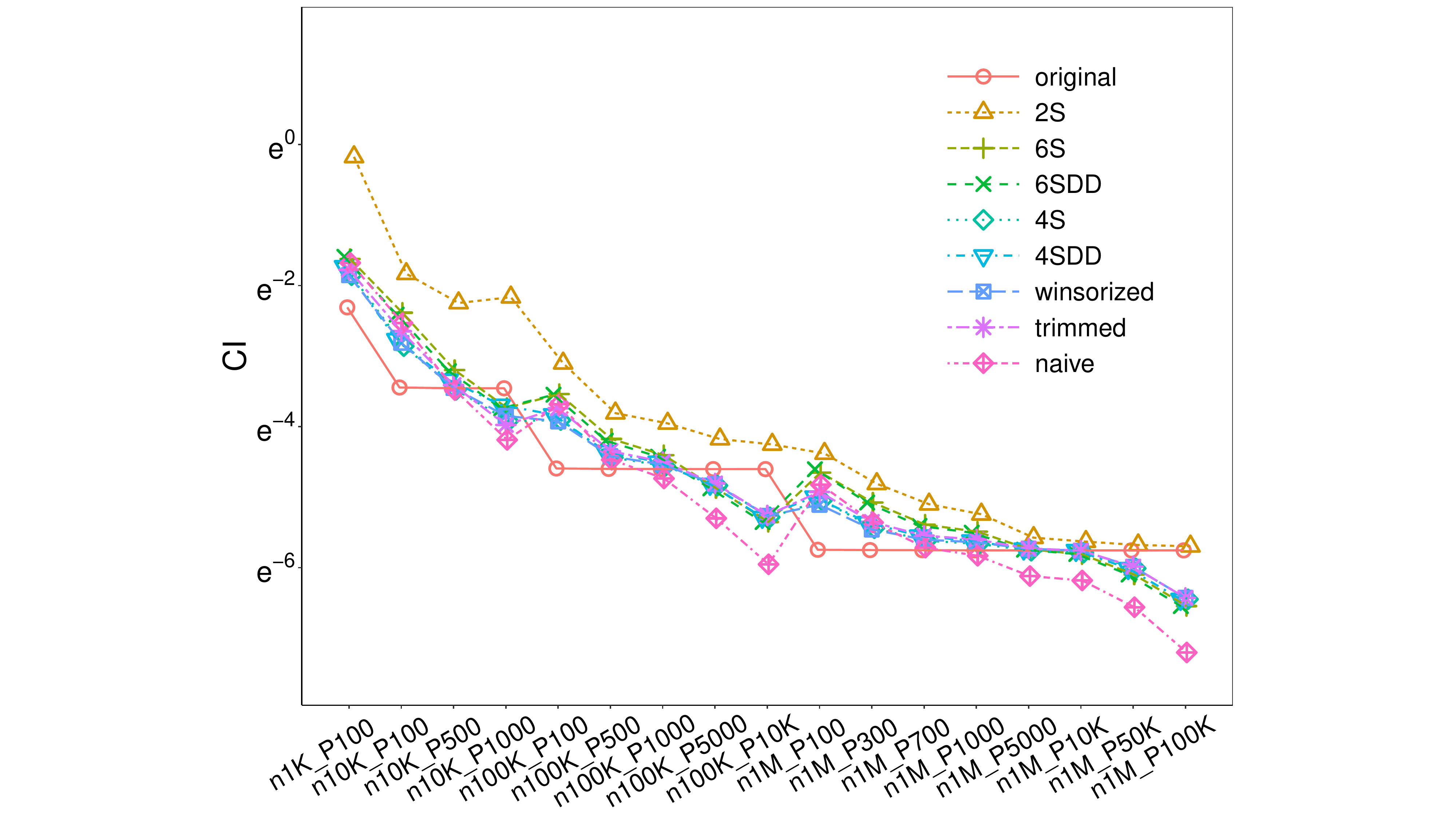}
\includegraphics[width=0.26\textwidth, trim={2.2in 0 2.2in 0},clip] {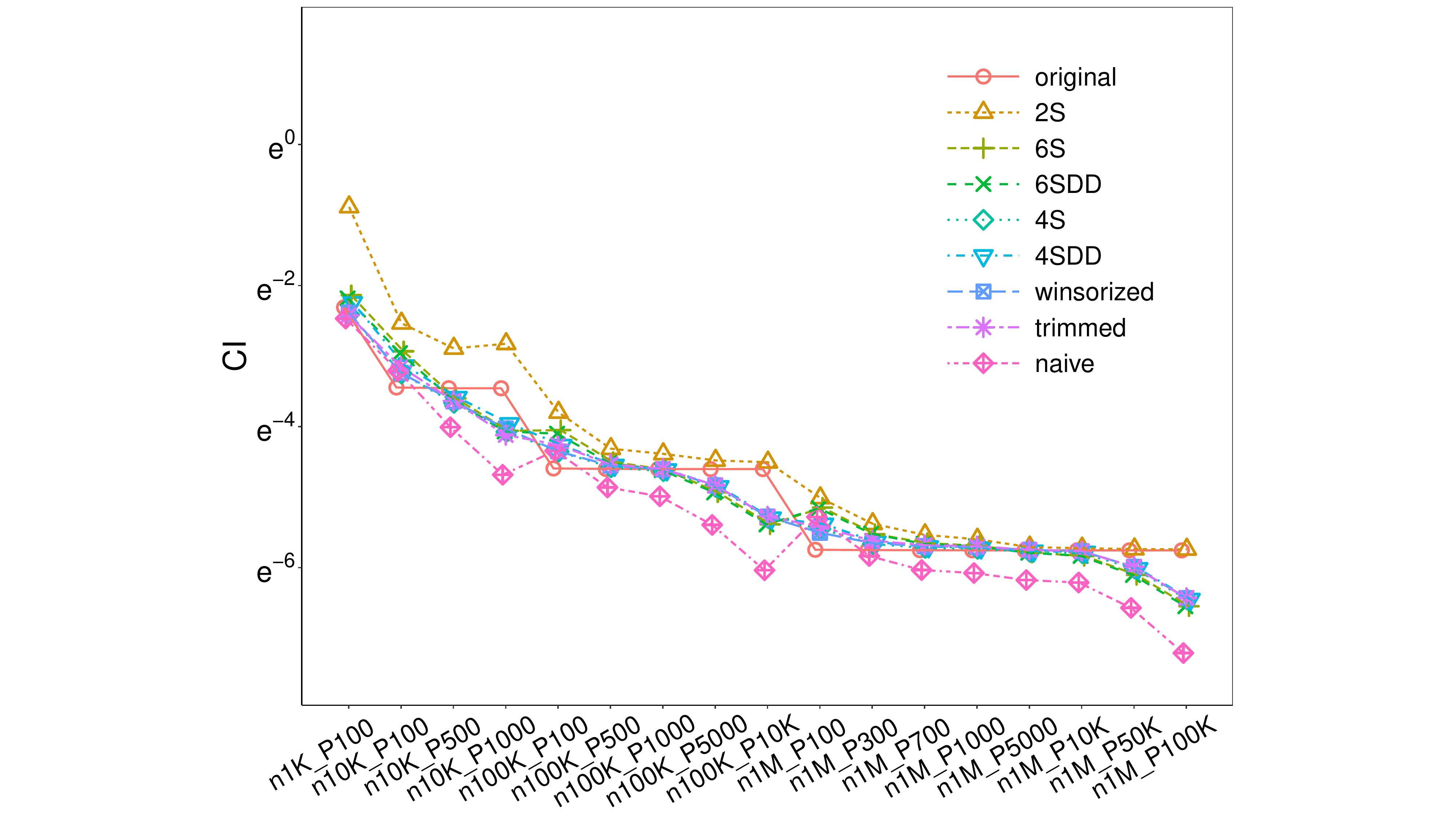}
\includegraphics[width=0.26\textwidth, trim={2.2in 0 2.2in 0},clip] {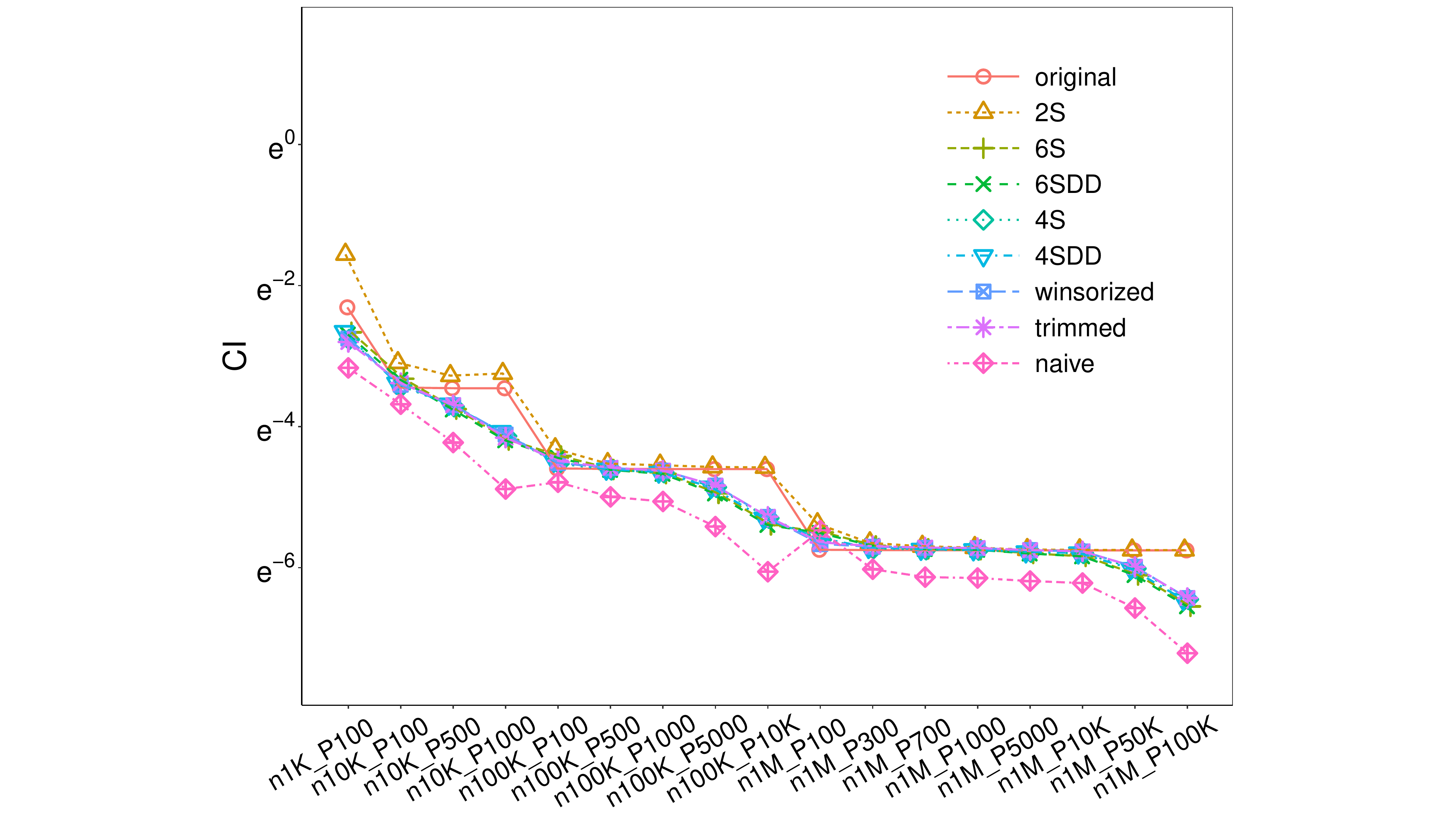}\\
\includegraphics[width=0.26\textwidth, trim={2.2in 0 2.2in 0},clip] {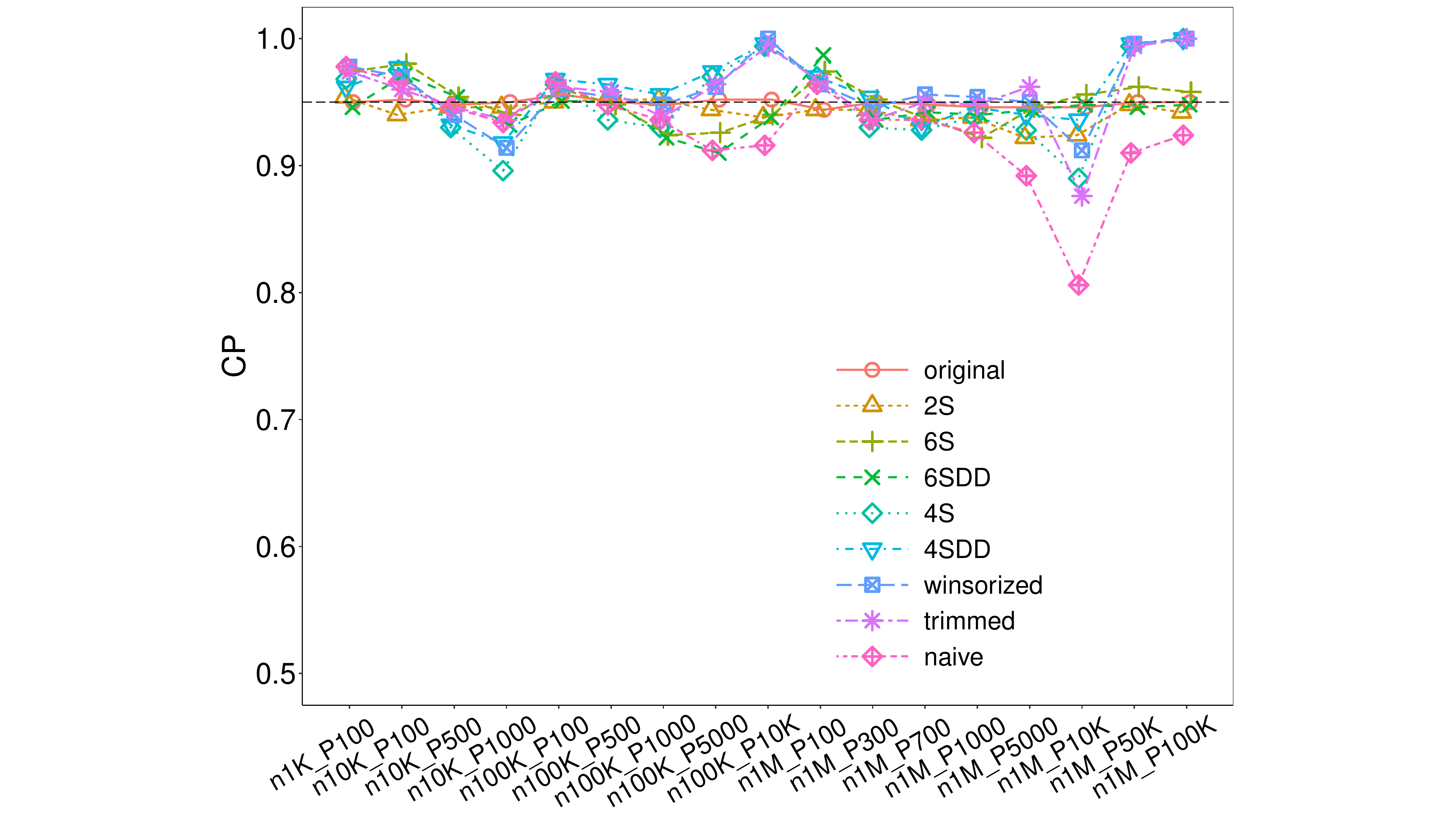}
\includegraphics[width=0.26\textwidth, trim={2.2in 0 2.2in 0},clip] {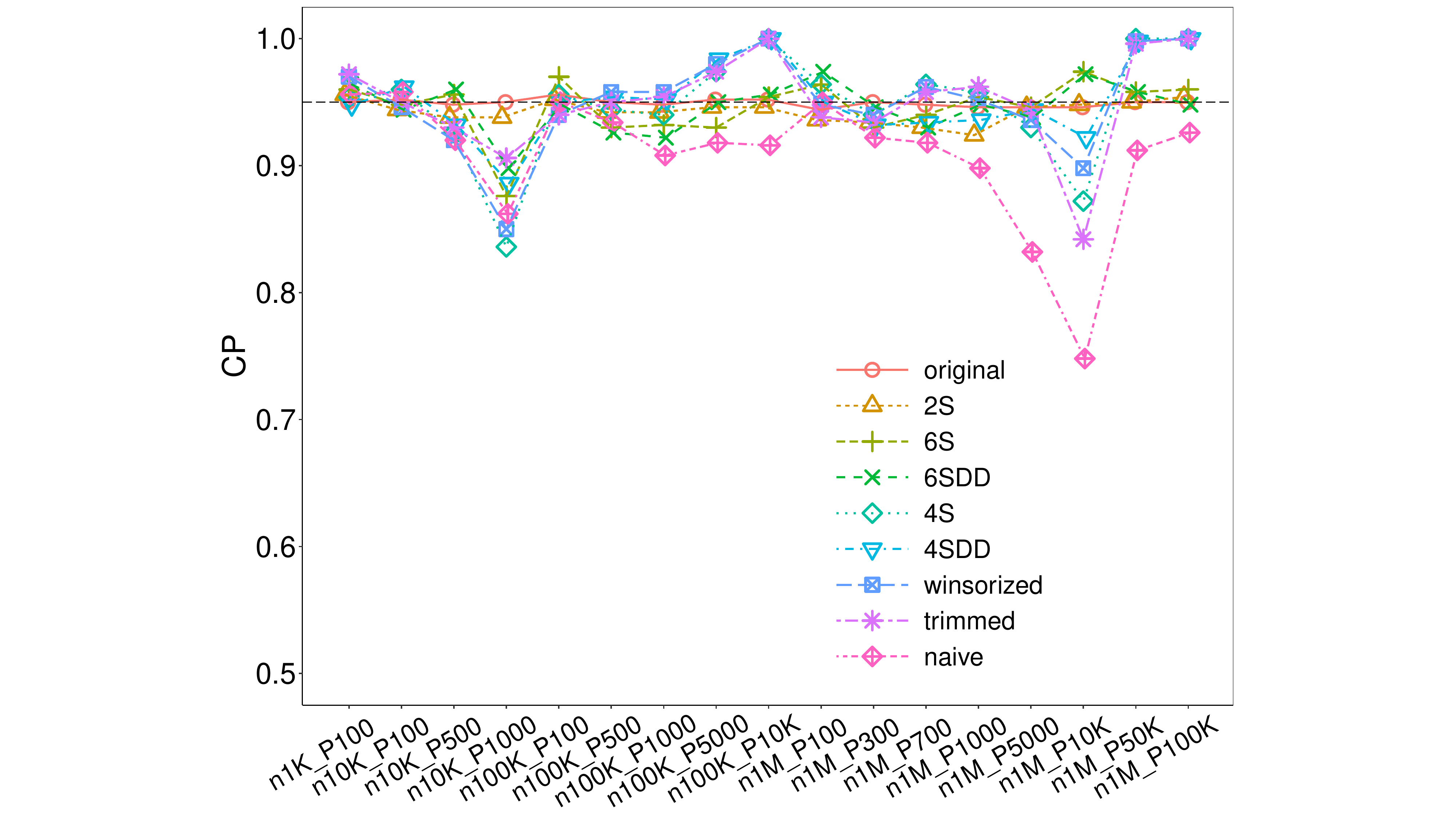}
\includegraphics[width=0.26\textwidth, trim={2.2in 0 2.2in 0},clip] {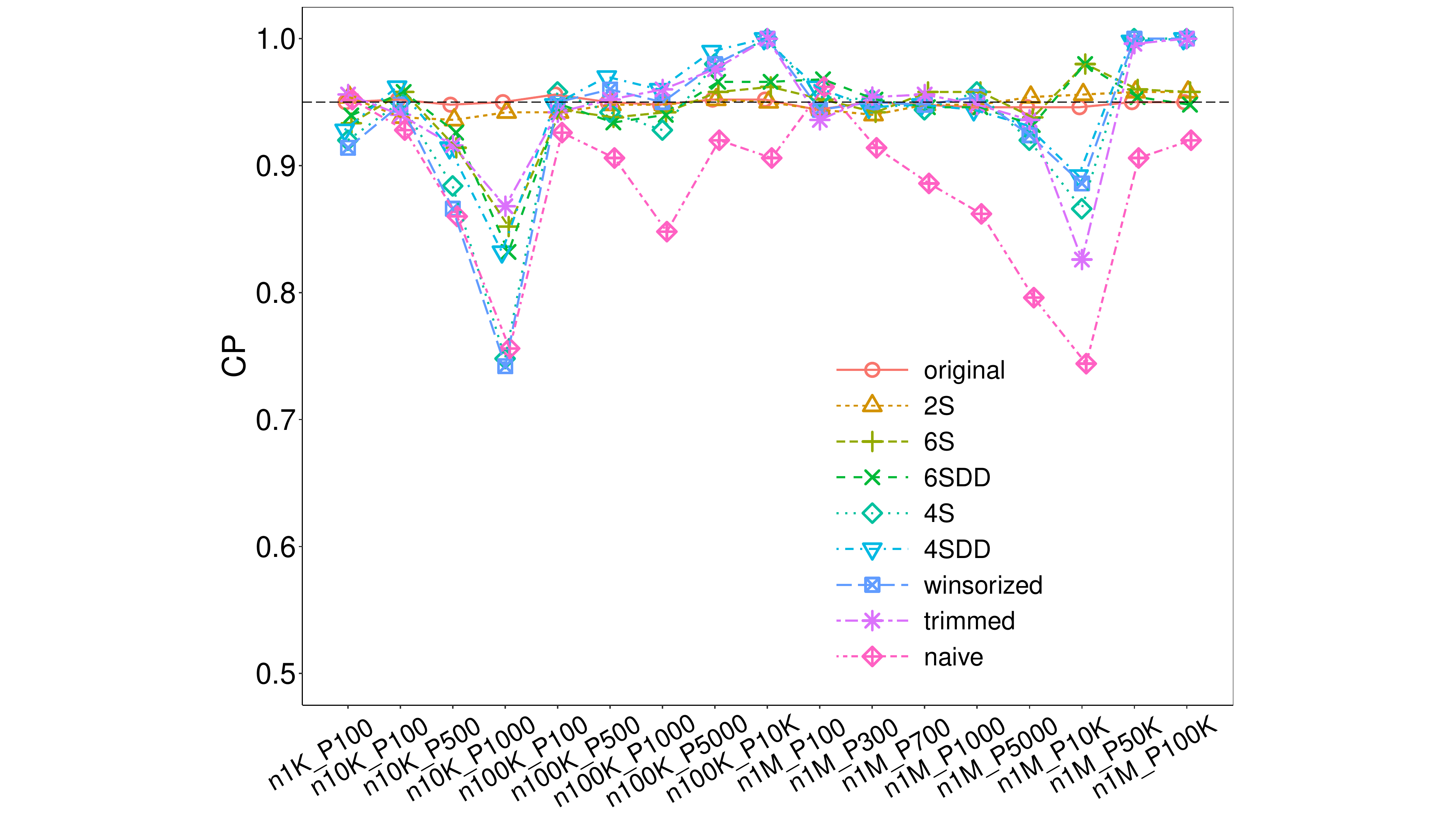}
\includegraphics[width=0.26\textwidth, trim={2.2in 0 2.2in 0},clip] {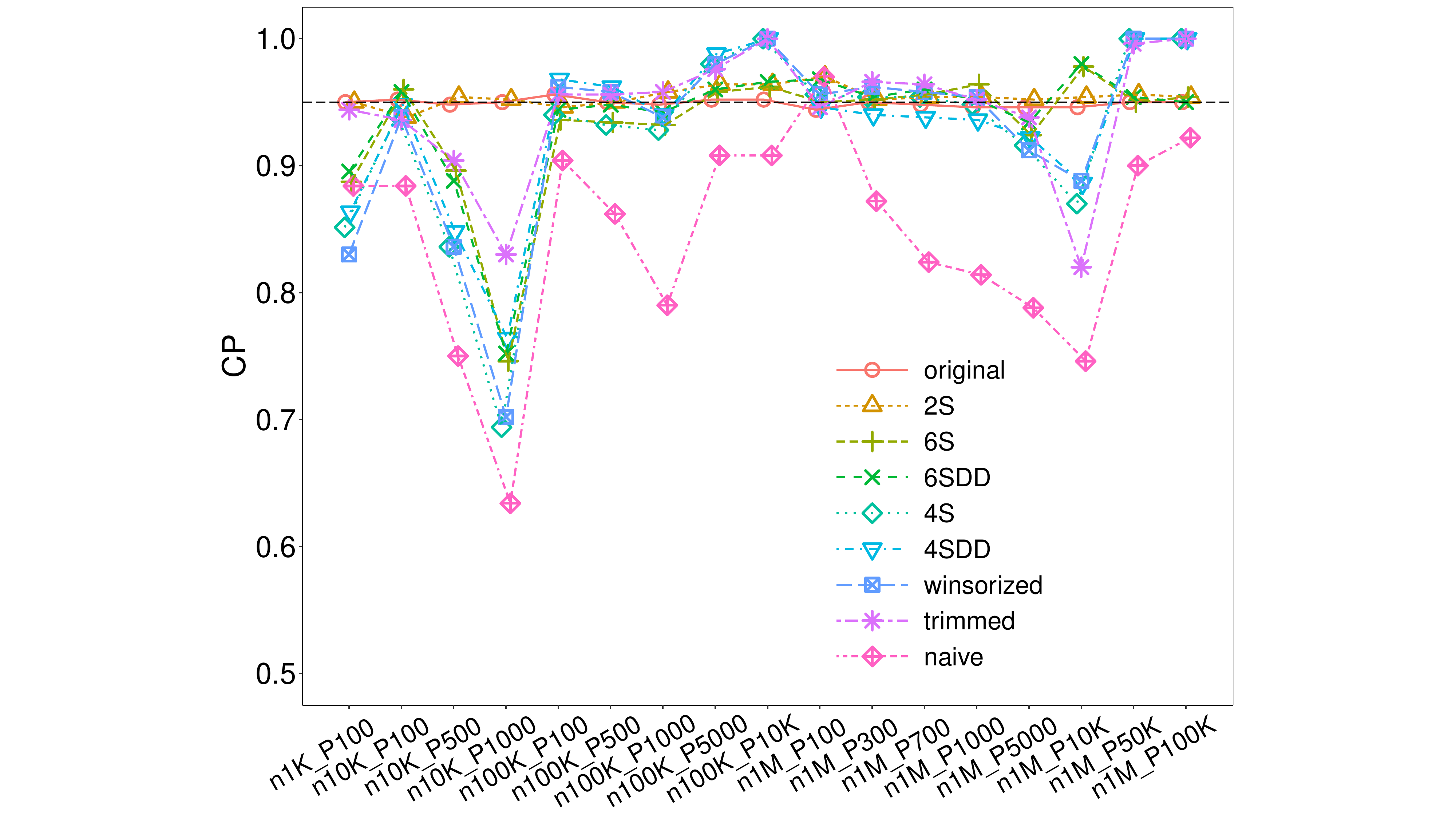}
\includegraphics[width=0.26\textwidth, trim={2.2in 0 2.2in 0},clip] {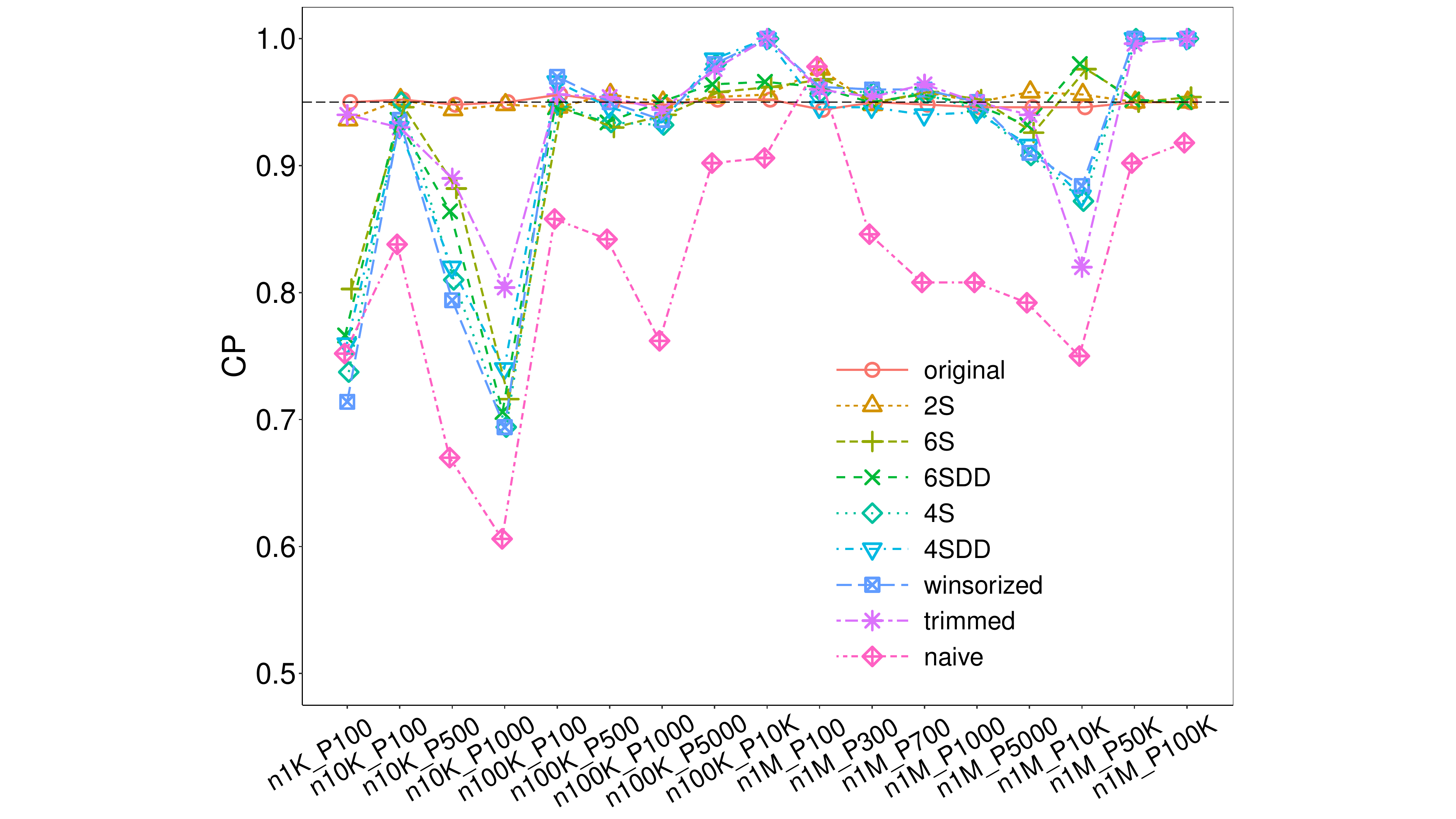}\\

\caption{ZINB data; $\rho$-zCDP; $\theta=0$ and $\alpha=\beta$}
\label{fig:0szCDPzinb}
\end{figure}

\end{landscape}

\begin{landscape}

\subsection*{ZINB, $\theta\ne0$ and $\alpha=\beta$}
\begin{figure}[!htb]
\centering
$\epsilon=0.5$\hspace{0.9in}$\epsilon=1$\hspace{1in}$\epsilon=2$
\hspace{1in}$\epsilon=5$\hspace{0.9in}$\epsilon=50$\\
\includegraphics[width=0.215\textwidth, trim={2.2in 0 2.2in 0},clip] {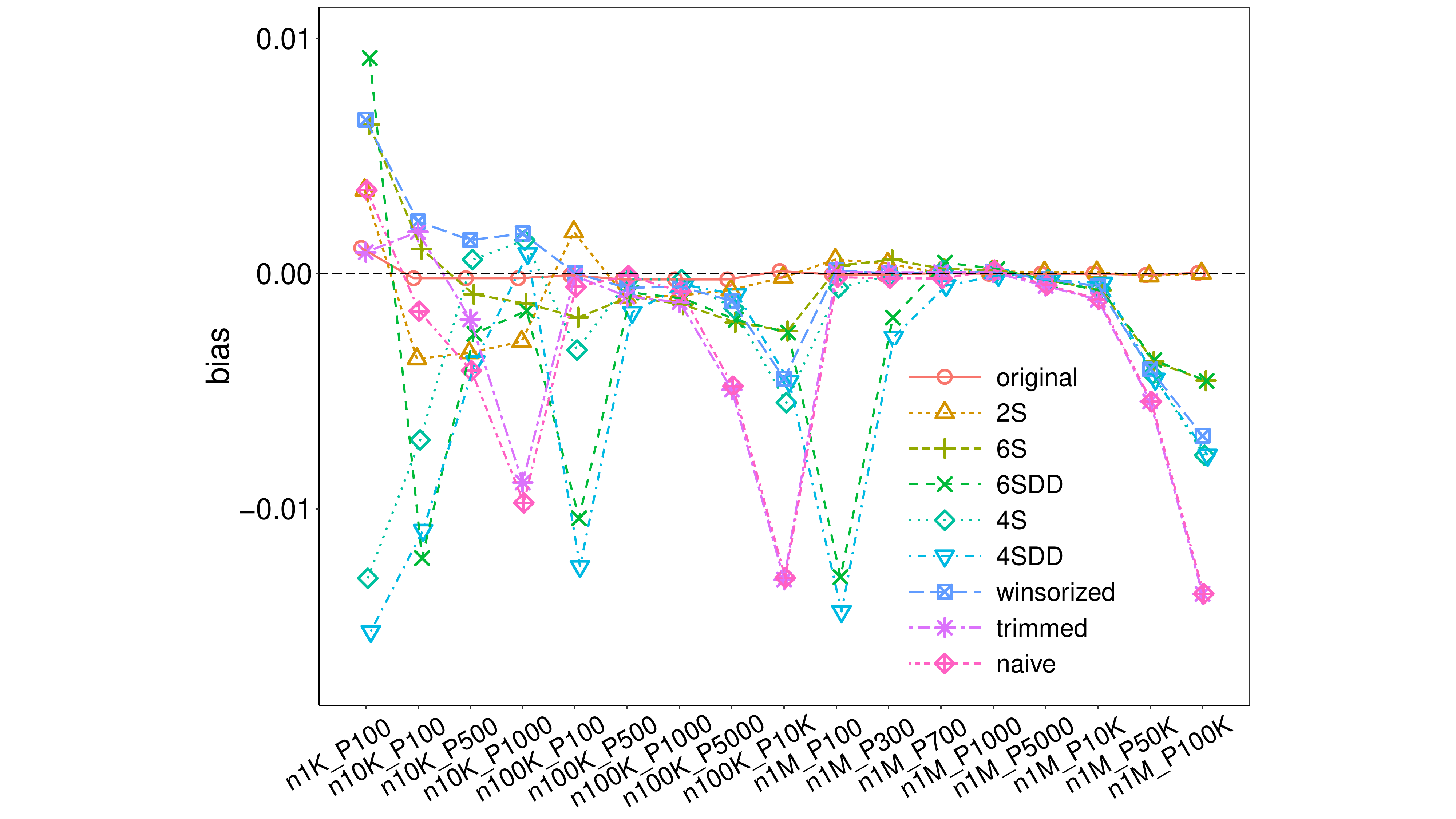}
\includegraphics[width=0.215\textwidth, trim={2.2in 0 2.2in 0},clip] {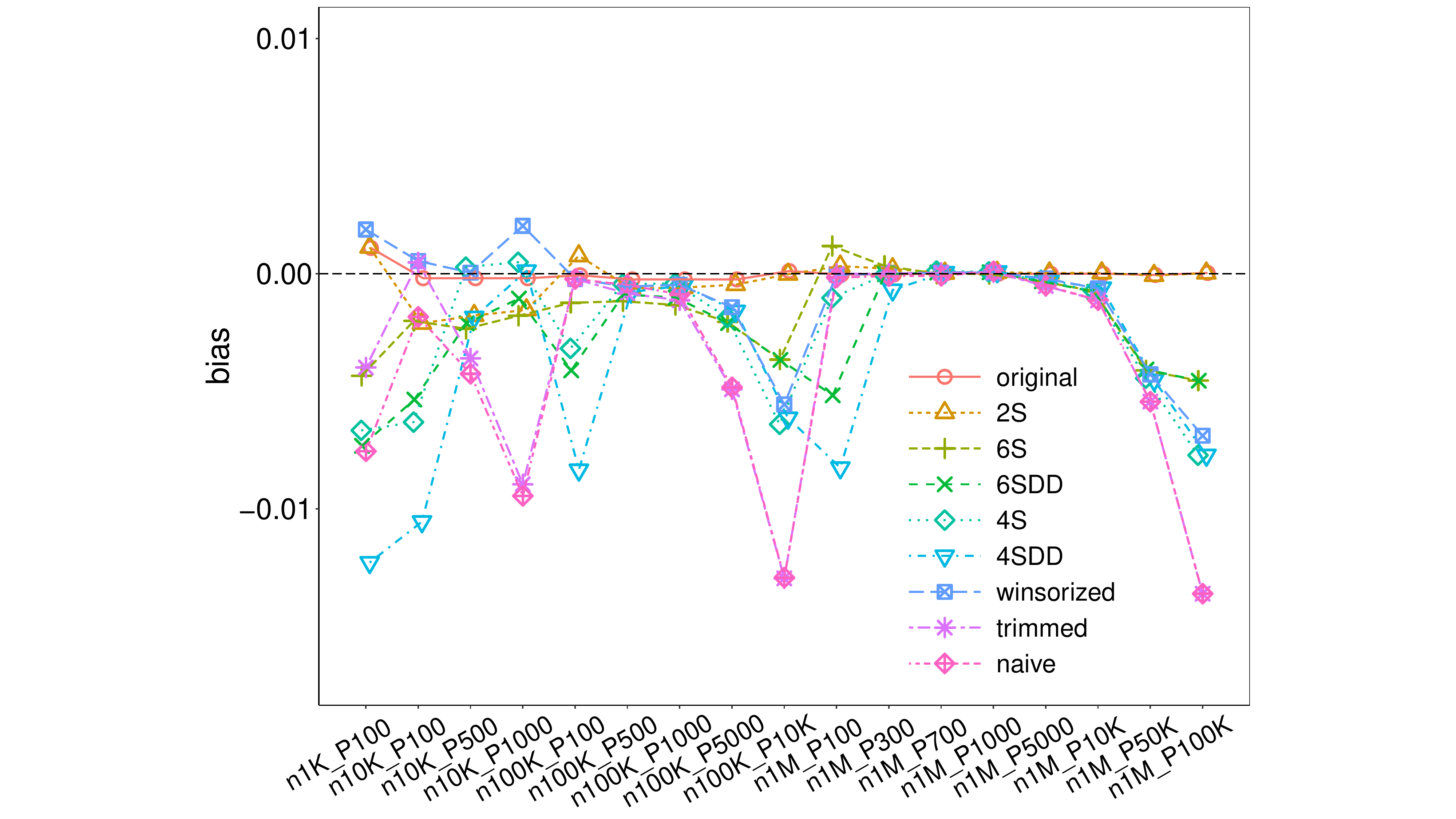}
\includegraphics[width=0.215\textwidth, trim={2.2in 0 2.2in 0},clip] {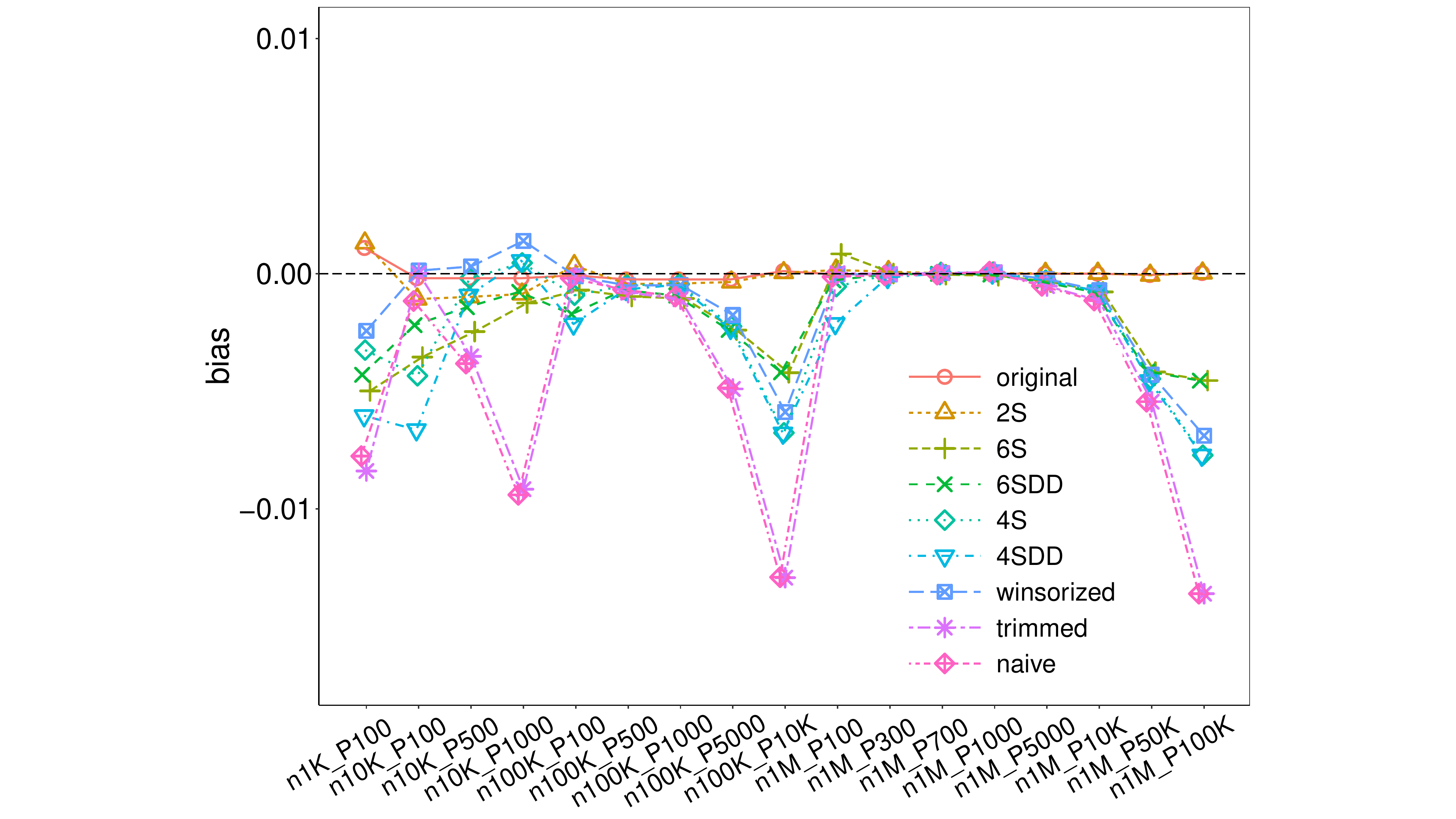}
\includegraphics[width=0.215\textwidth, trim={2.2in 0 2.2in 0},clip] {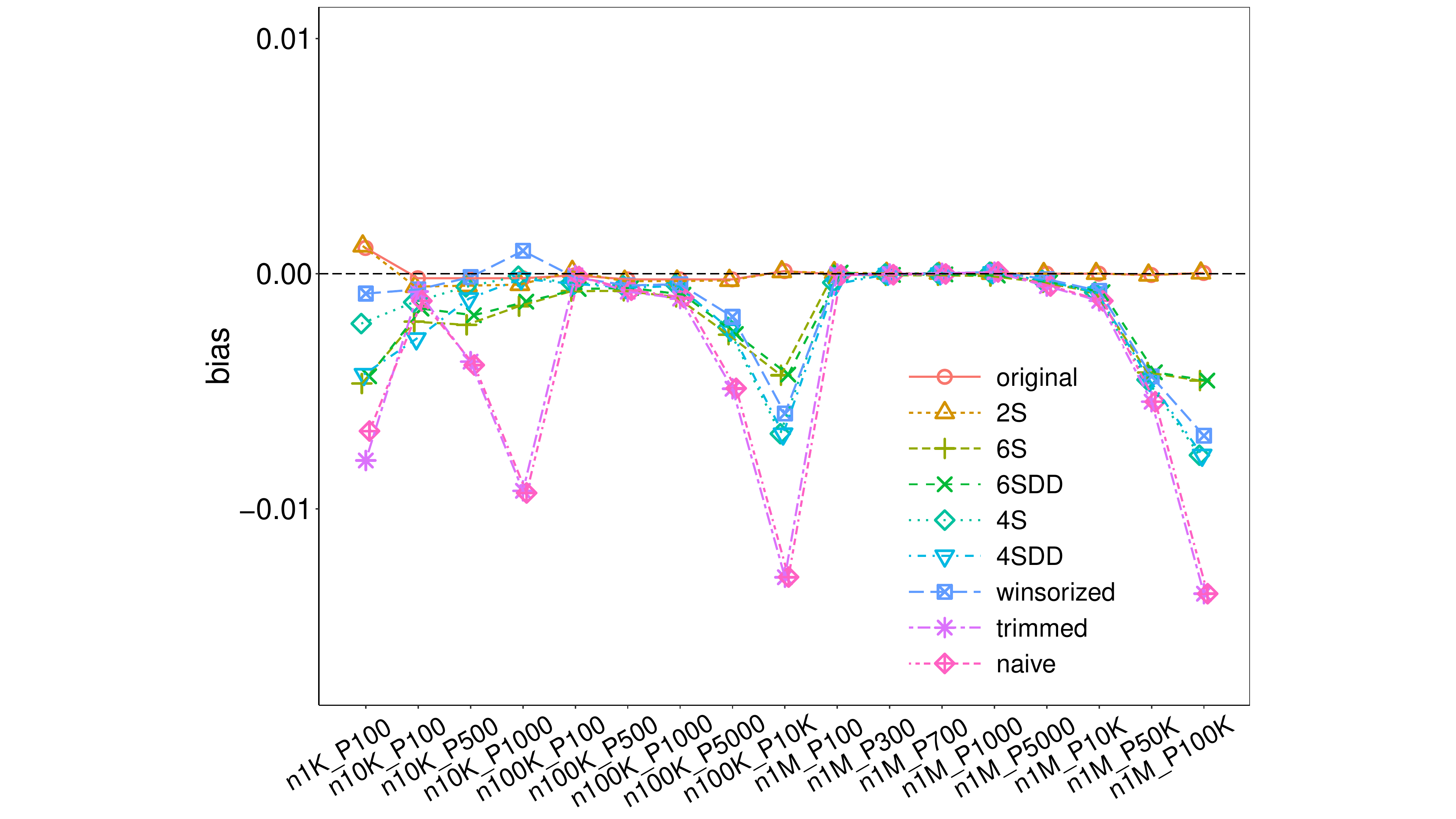}
\includegraphics[width=0.215\textwidth, trim={2.2in 0 2.2in 0},clip] {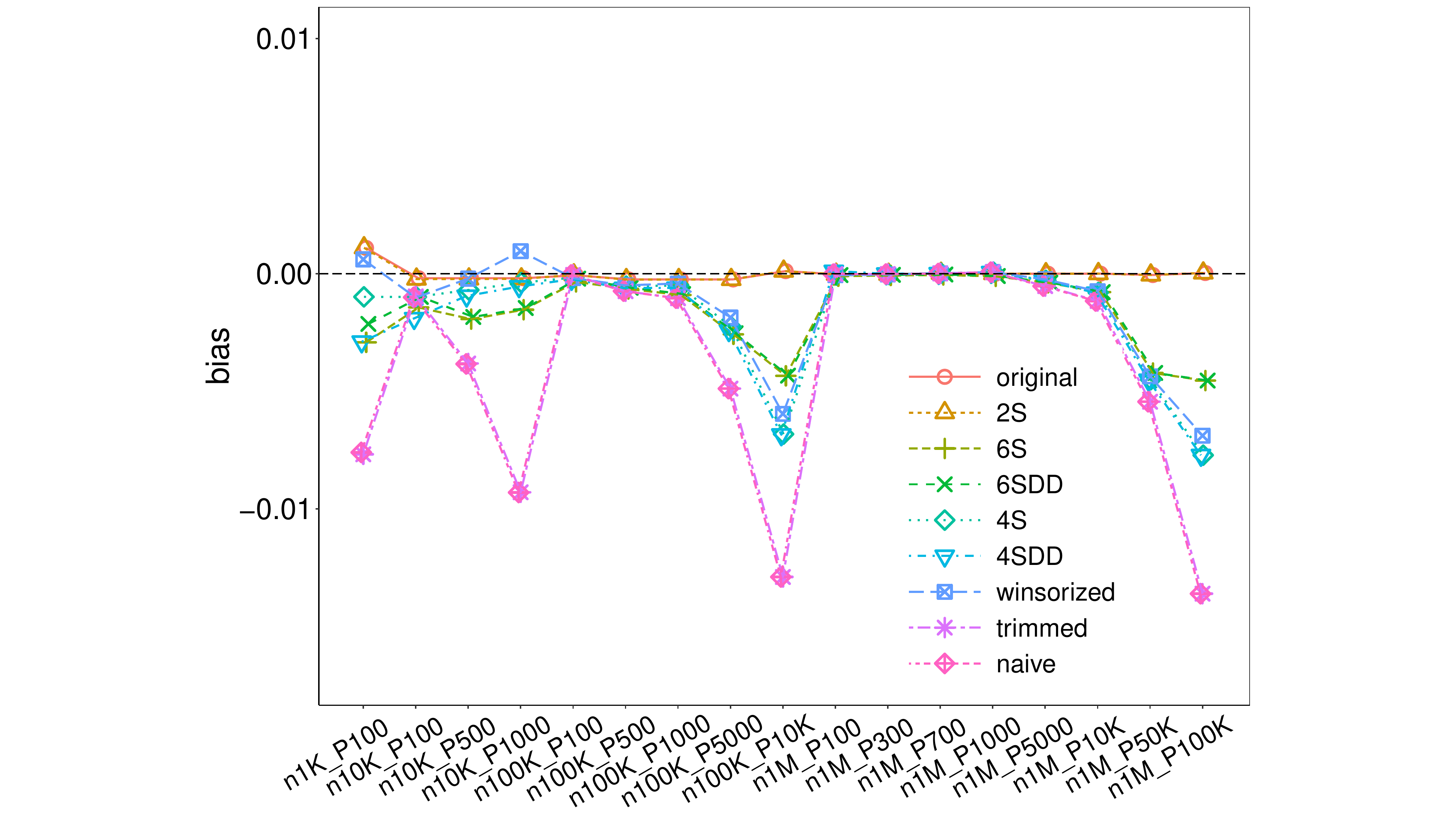}\\
\includegraphics[width=0.215\textwidth, trim={2.2in 0 2.2in 0},clip] {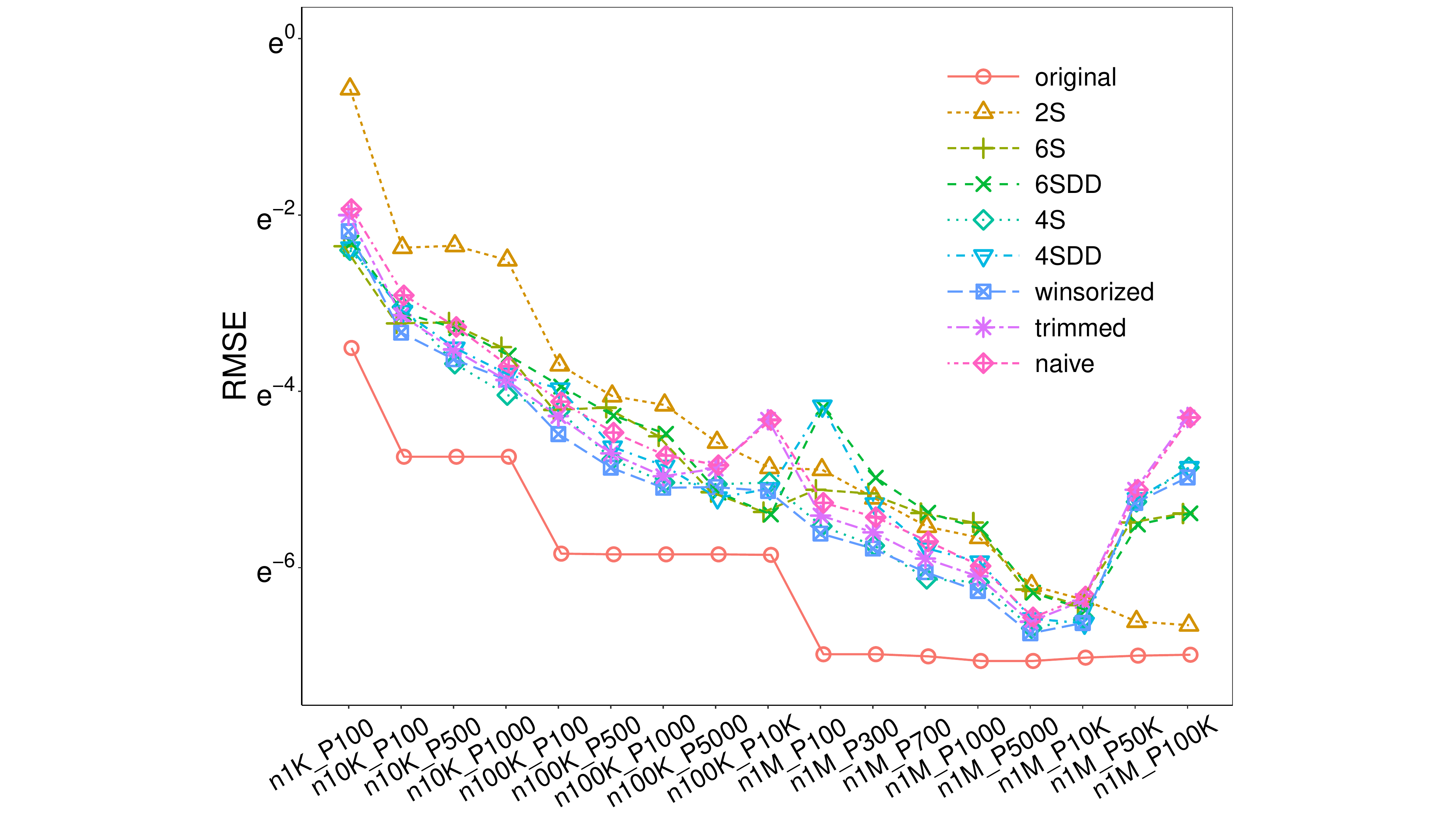}
\includegraphics[width=0.215\textwidth, trim={2.2in 0 2.2in 0},clip] {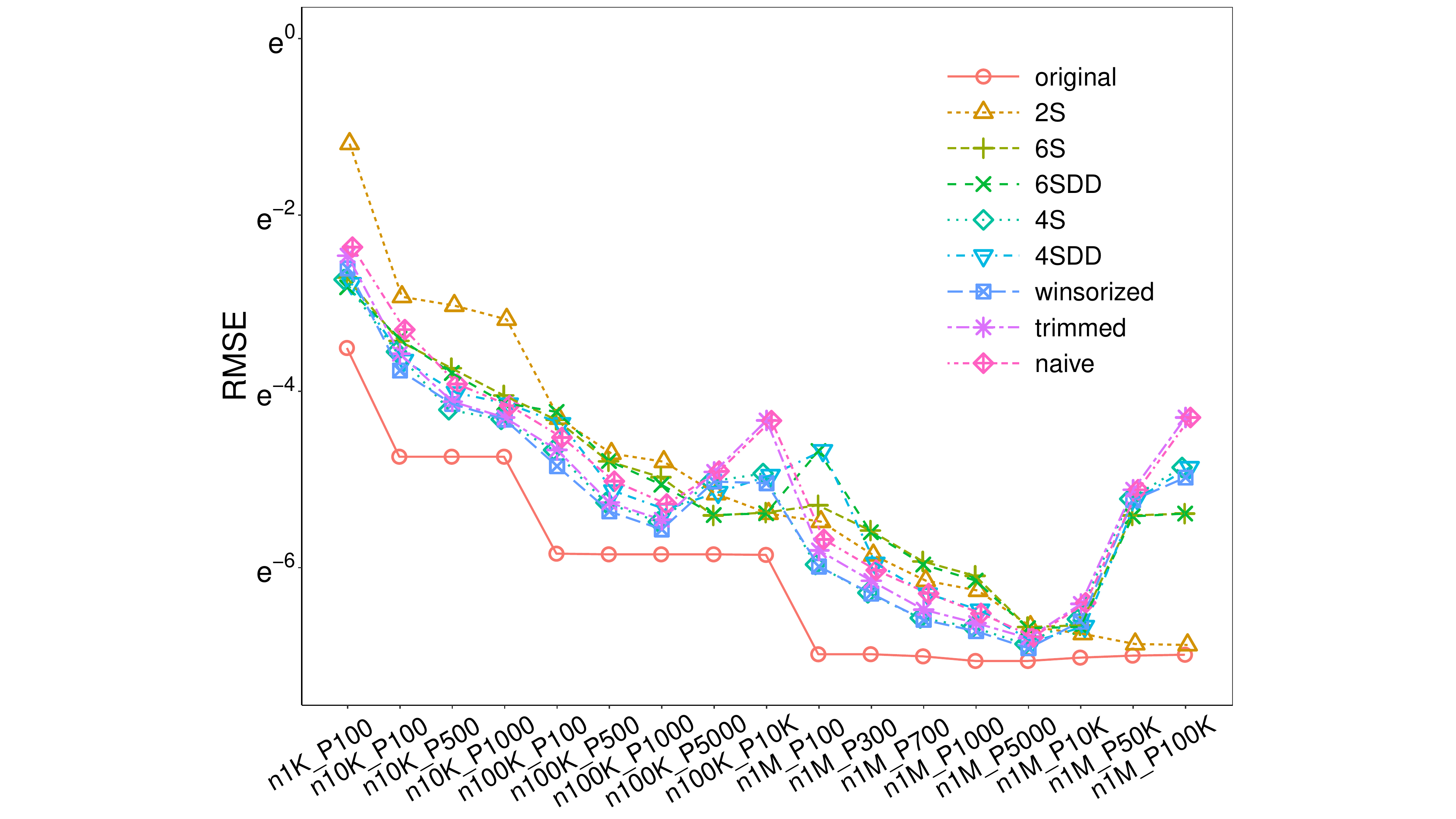}
\includegraphics[width=0.215\textwidth, trim={2.2in 0 2.2in 0},clip] {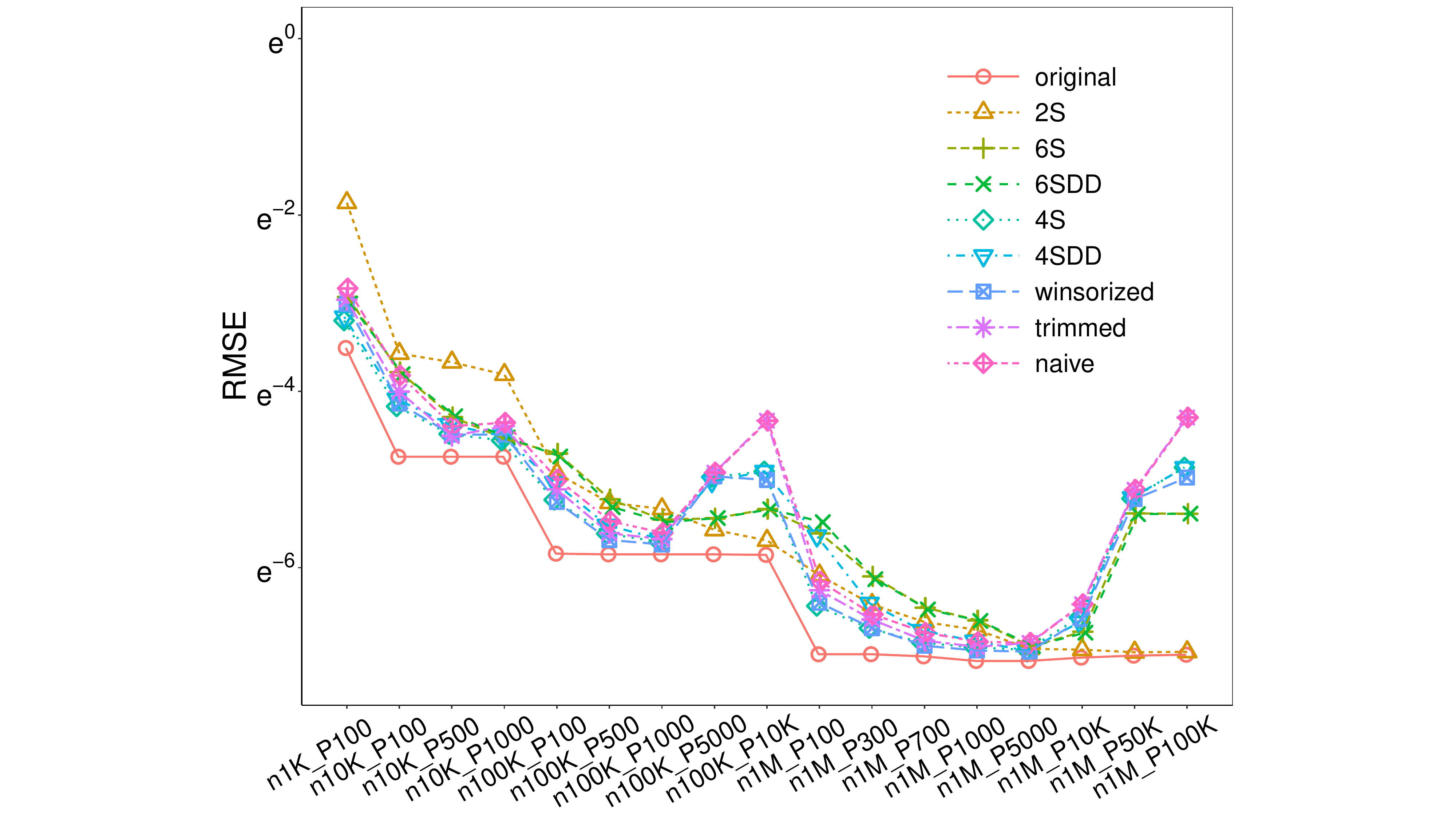}
\includegraphics[width=0.215\textwidth, trim={2.2in 0 2.2in 0},clip] {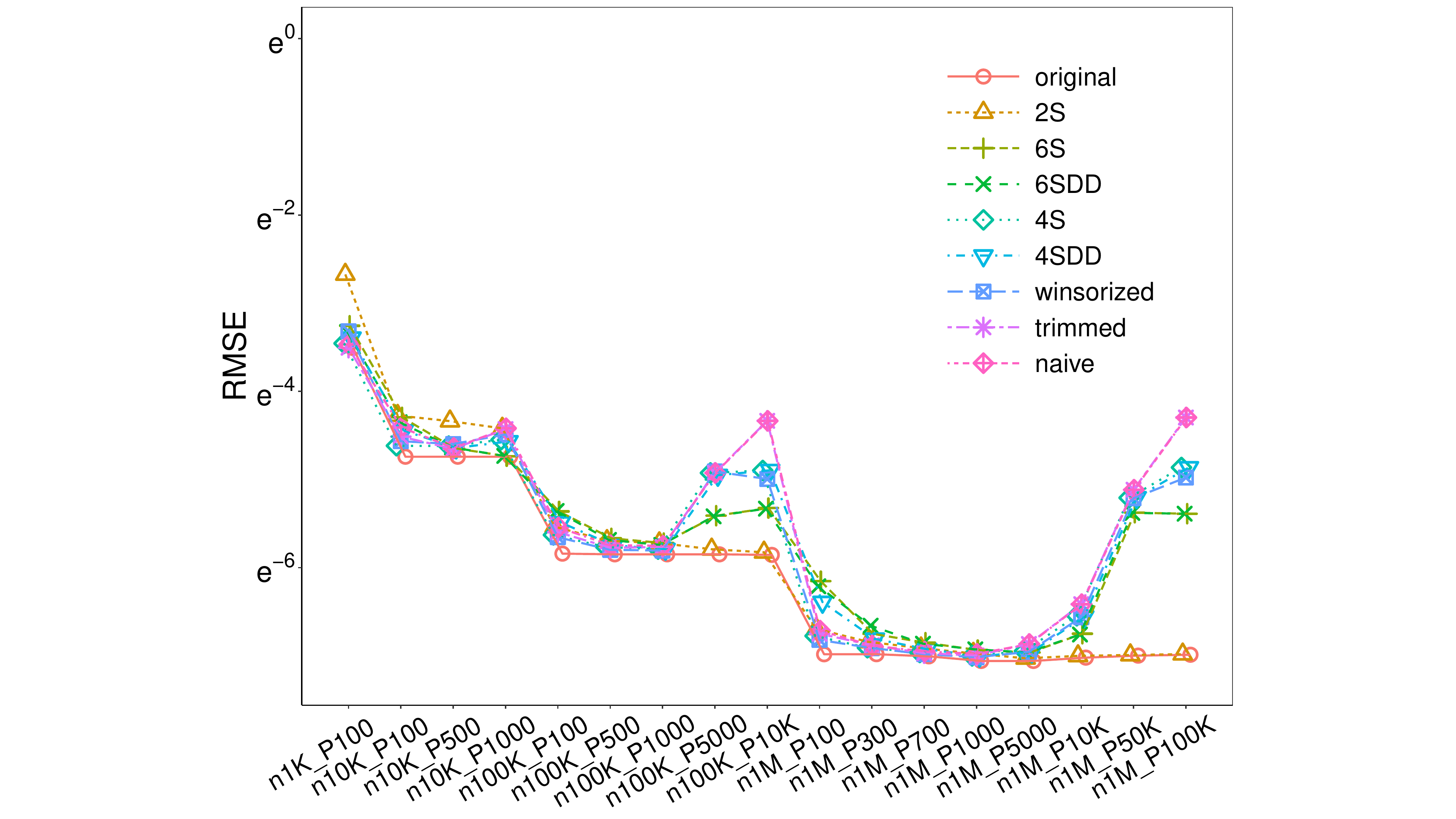}
\includegraphics[width=0.215\textwidth, trim={2.2in 0 2.2in 0},clip] {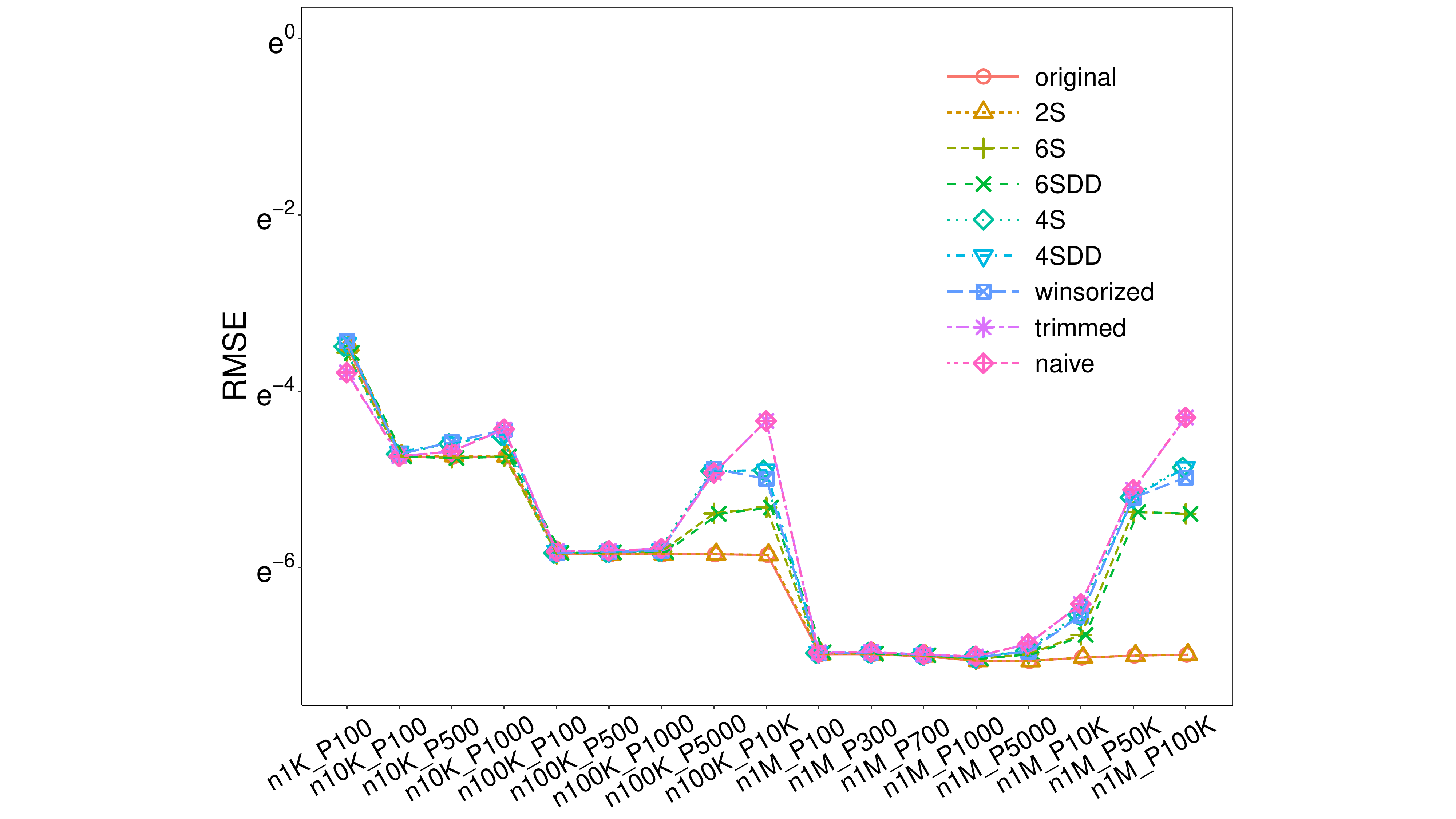}\\
\includegraphics[width=0.215\textwidth, trim={2.2in 0 2.2in 0},clip] {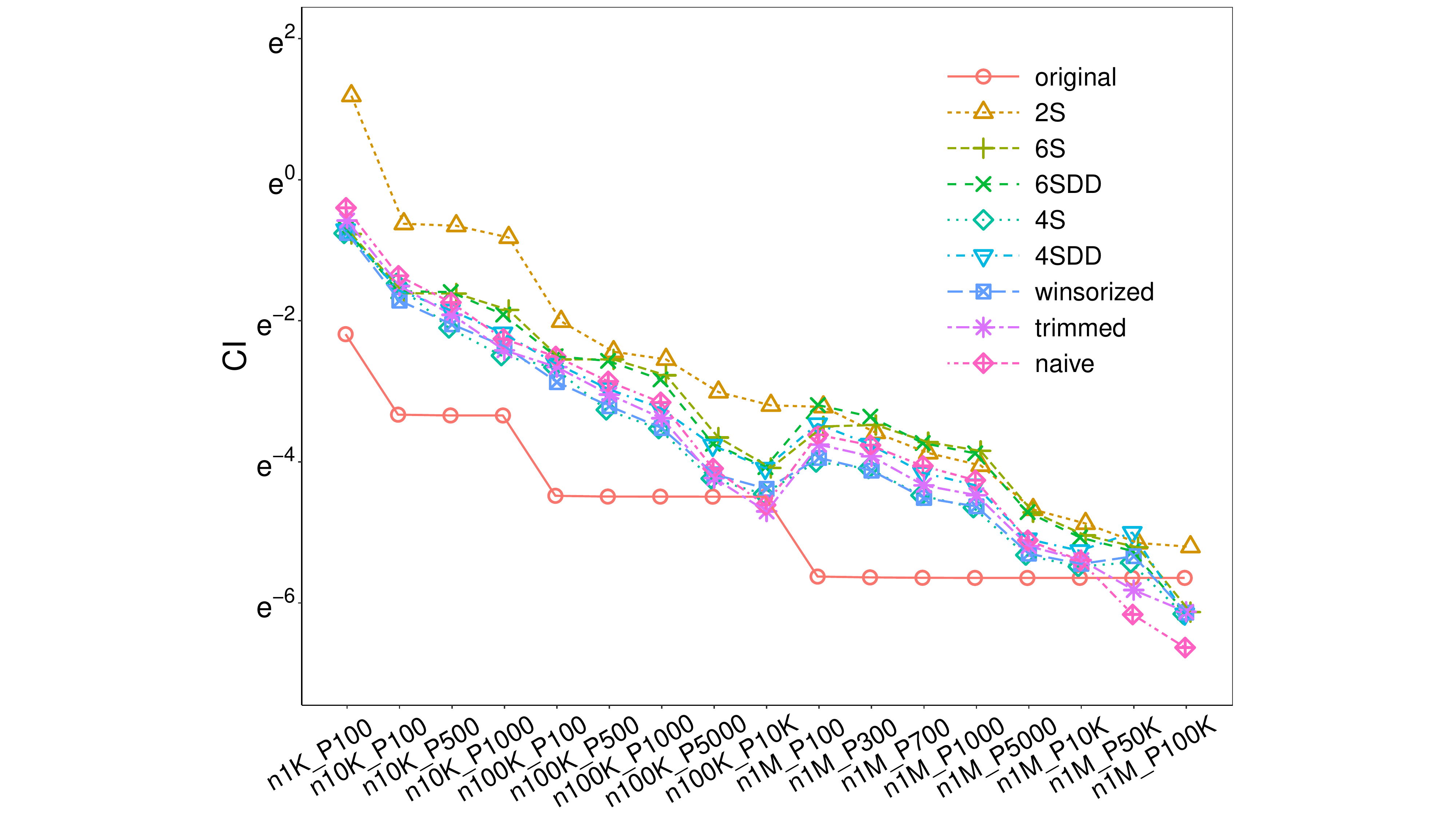}
\includegraphics[width=0.215\textwidth, trim={2.2in 0 2.2in 0},clip] {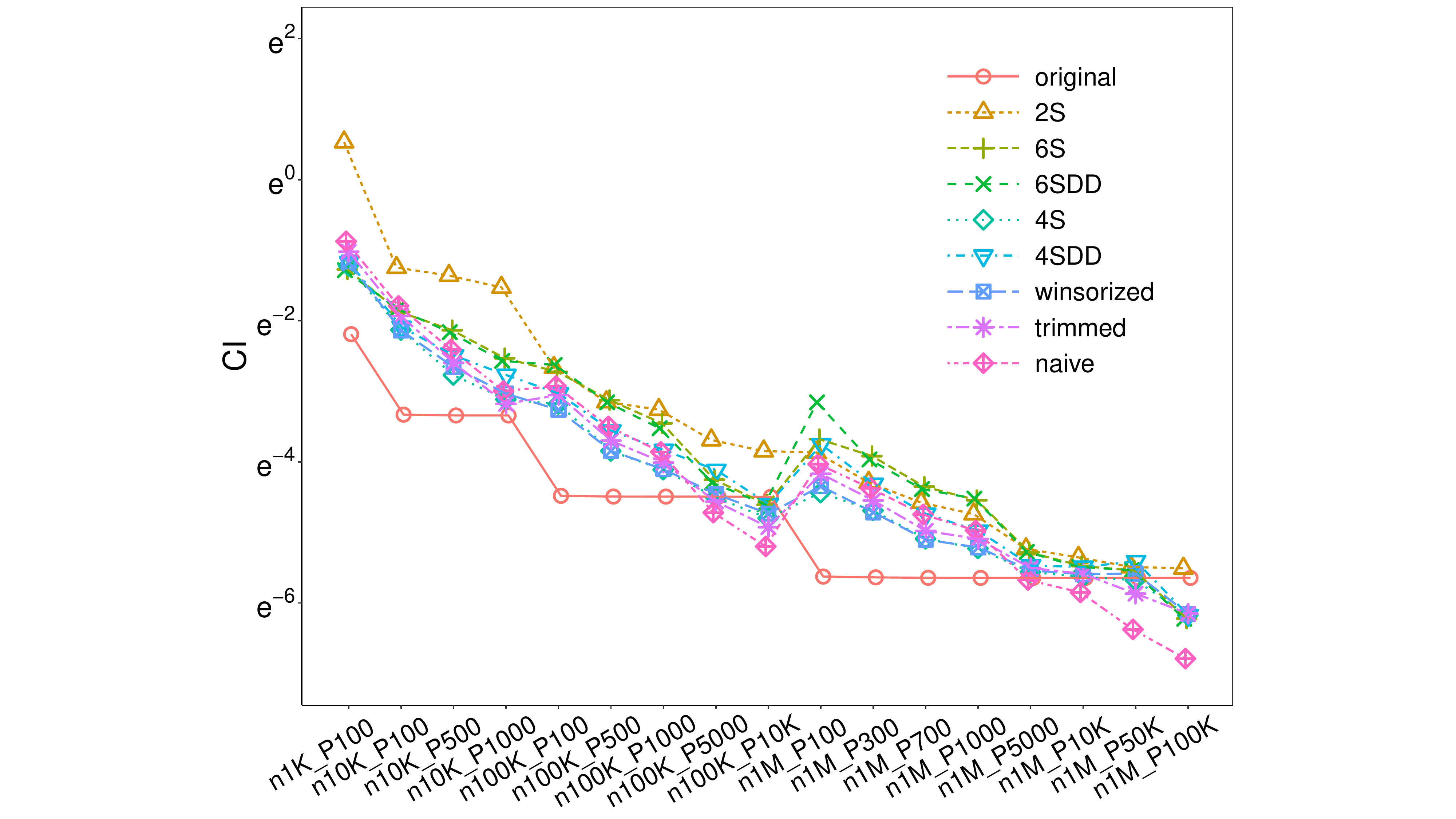}
\includegraphics[width=0.215\textwidth, trim={2.2in 0 2.2in 0},clip] {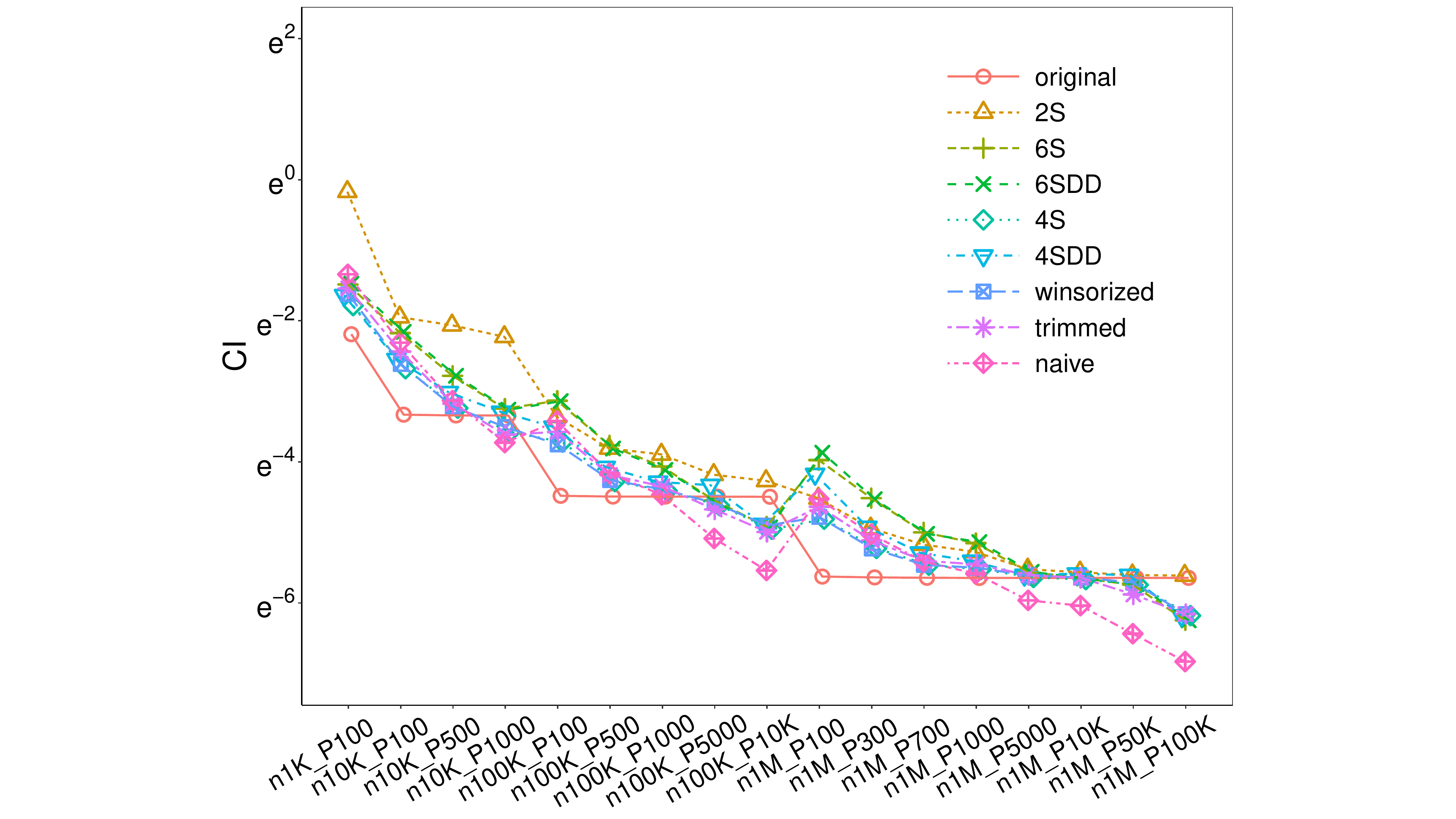}
\includegraphics[width=0.215\textwidth, trim={2.2in 0 2.2in 0},clip] {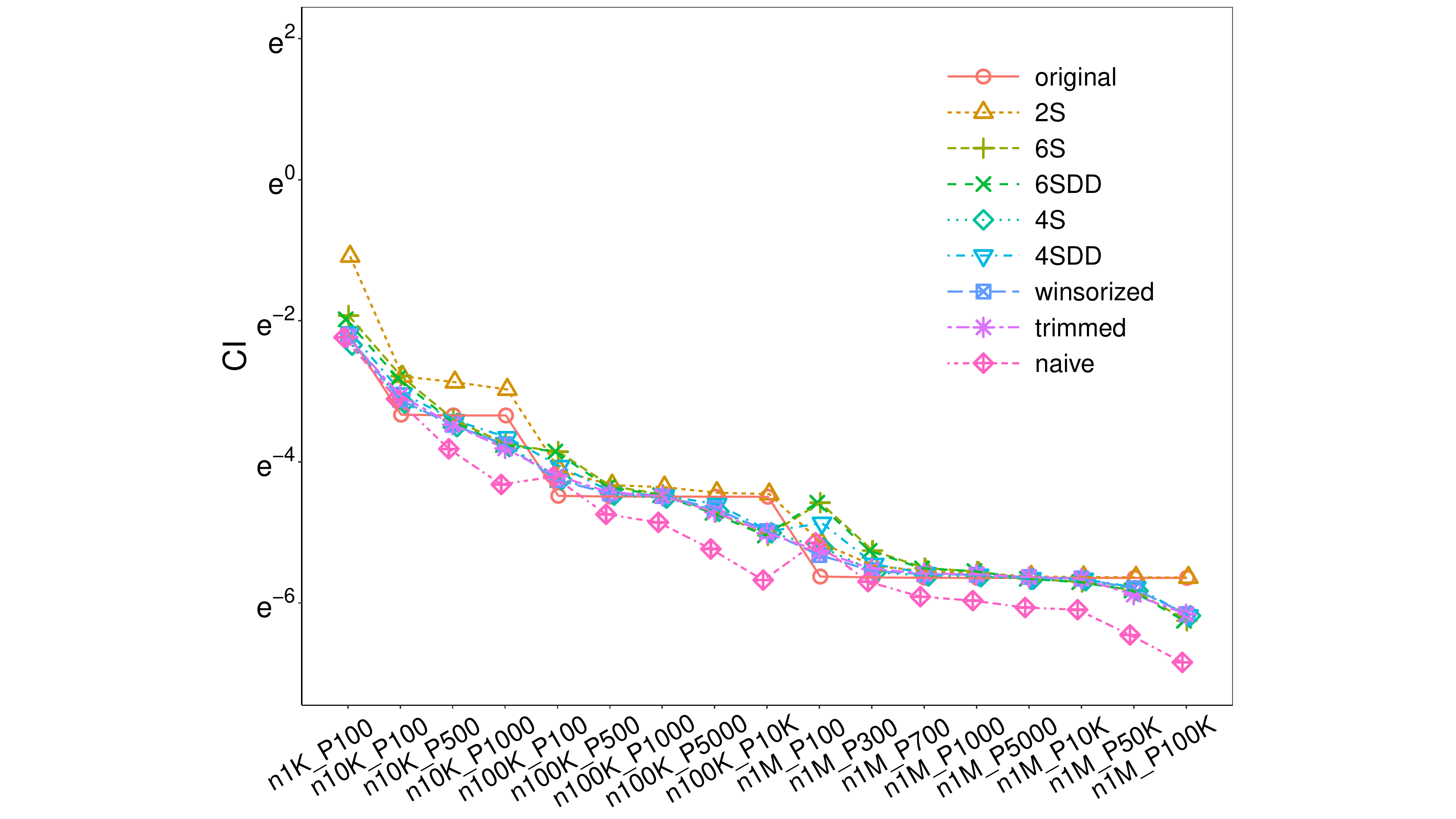}
\includegraphics[width=0.215\textwidth, trim={2.2in 0 2.2in 0},clip] {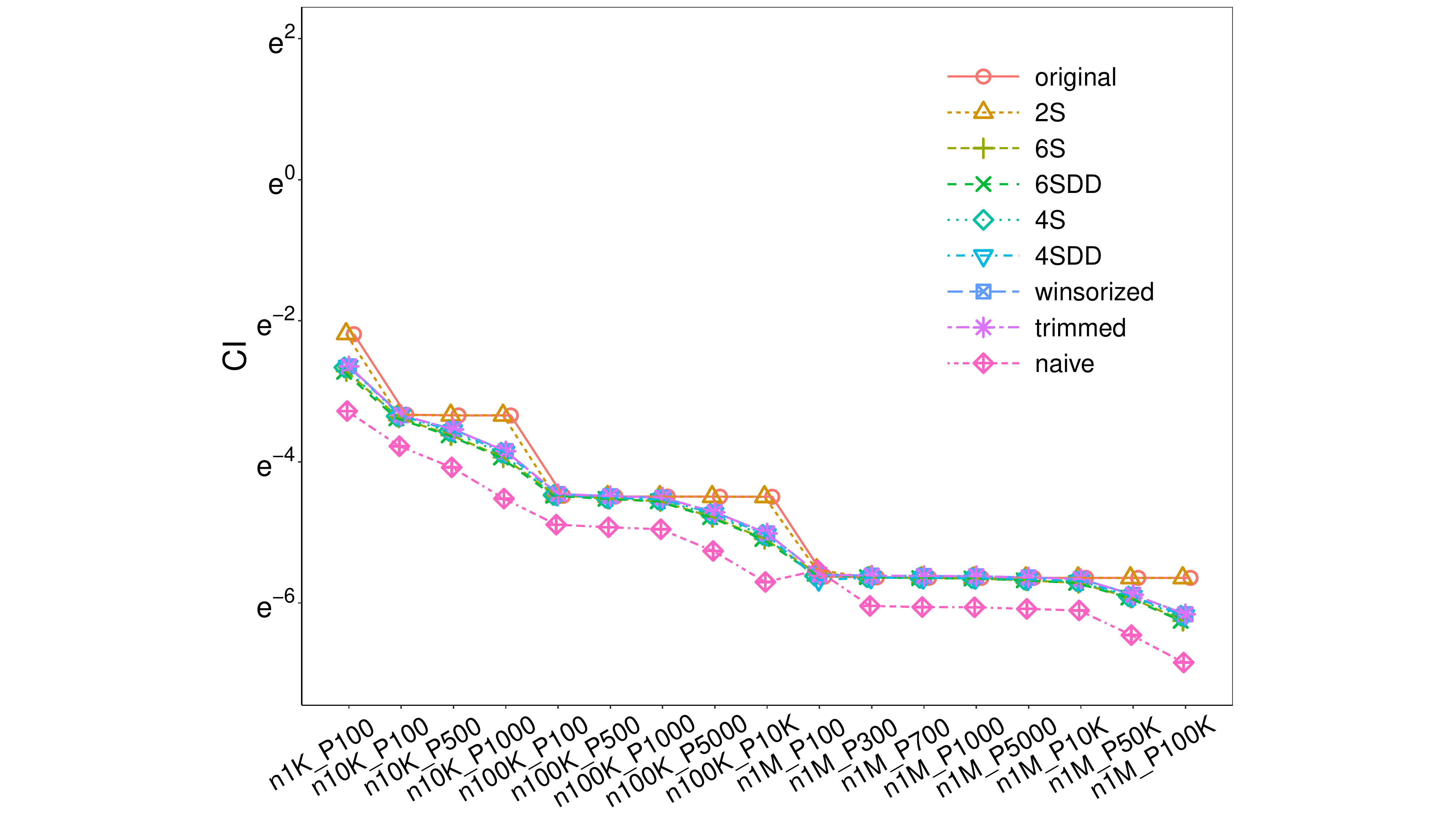}\\
\includegraphics[width=0.215\textwidth, trim={2.2in 0 2.2in 0},clip] {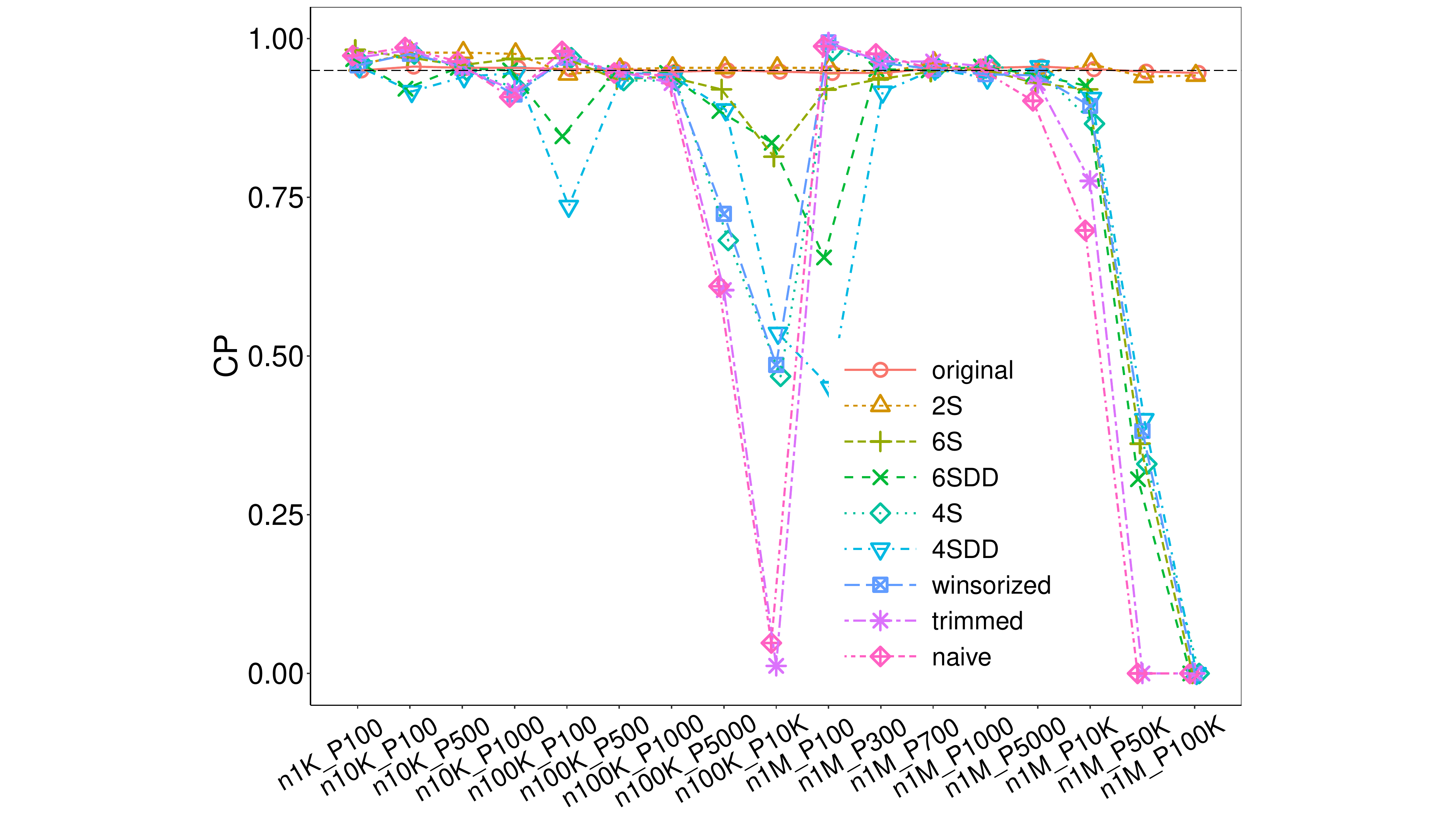}
\includegraphics[width=0.215\textwidth, trim={2.2in 0 2.2in 0},clip] {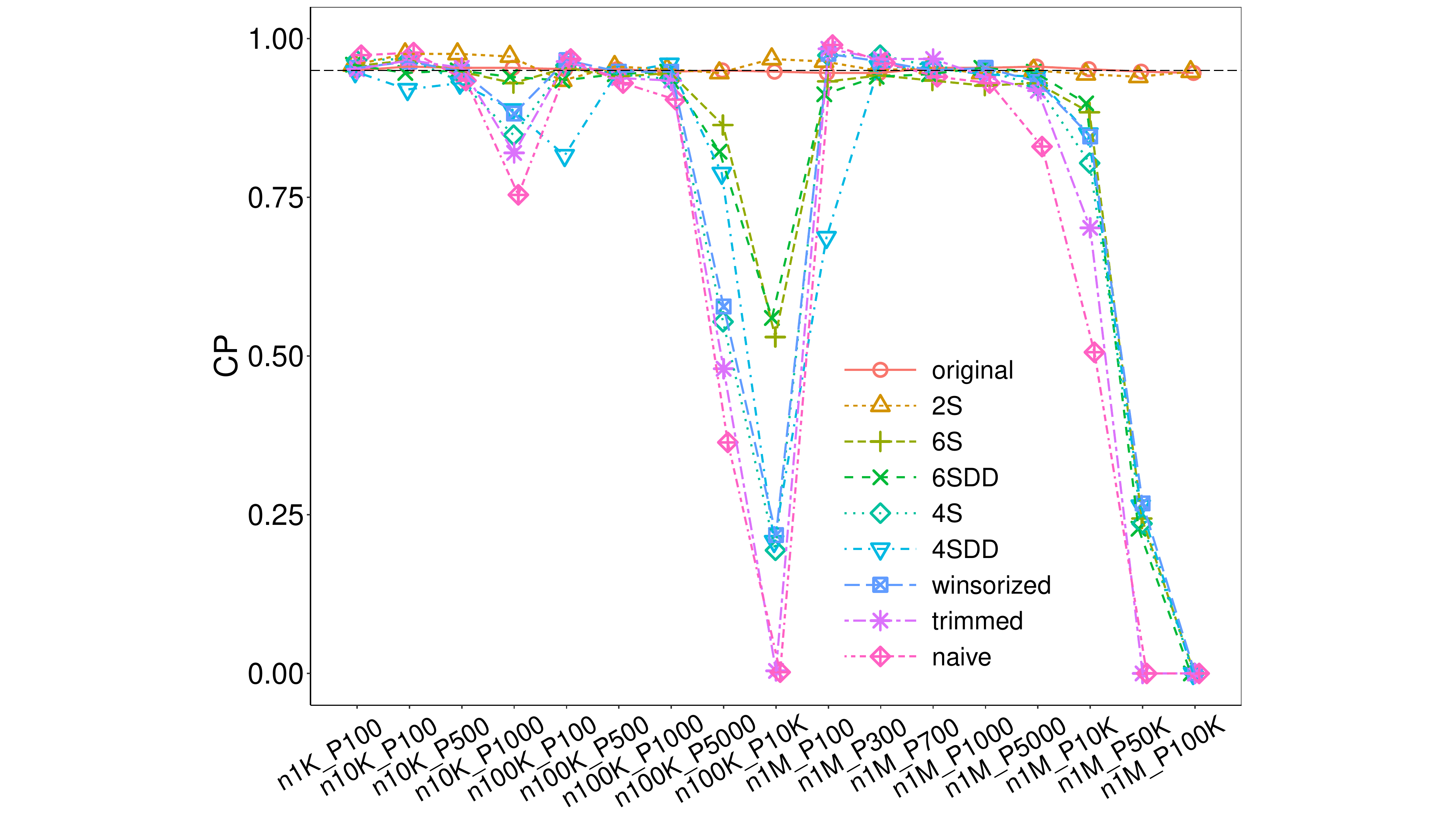}
\includegraphics[width=0.215\textwidth, trim={2.2in 0 2.2in 0},clip] {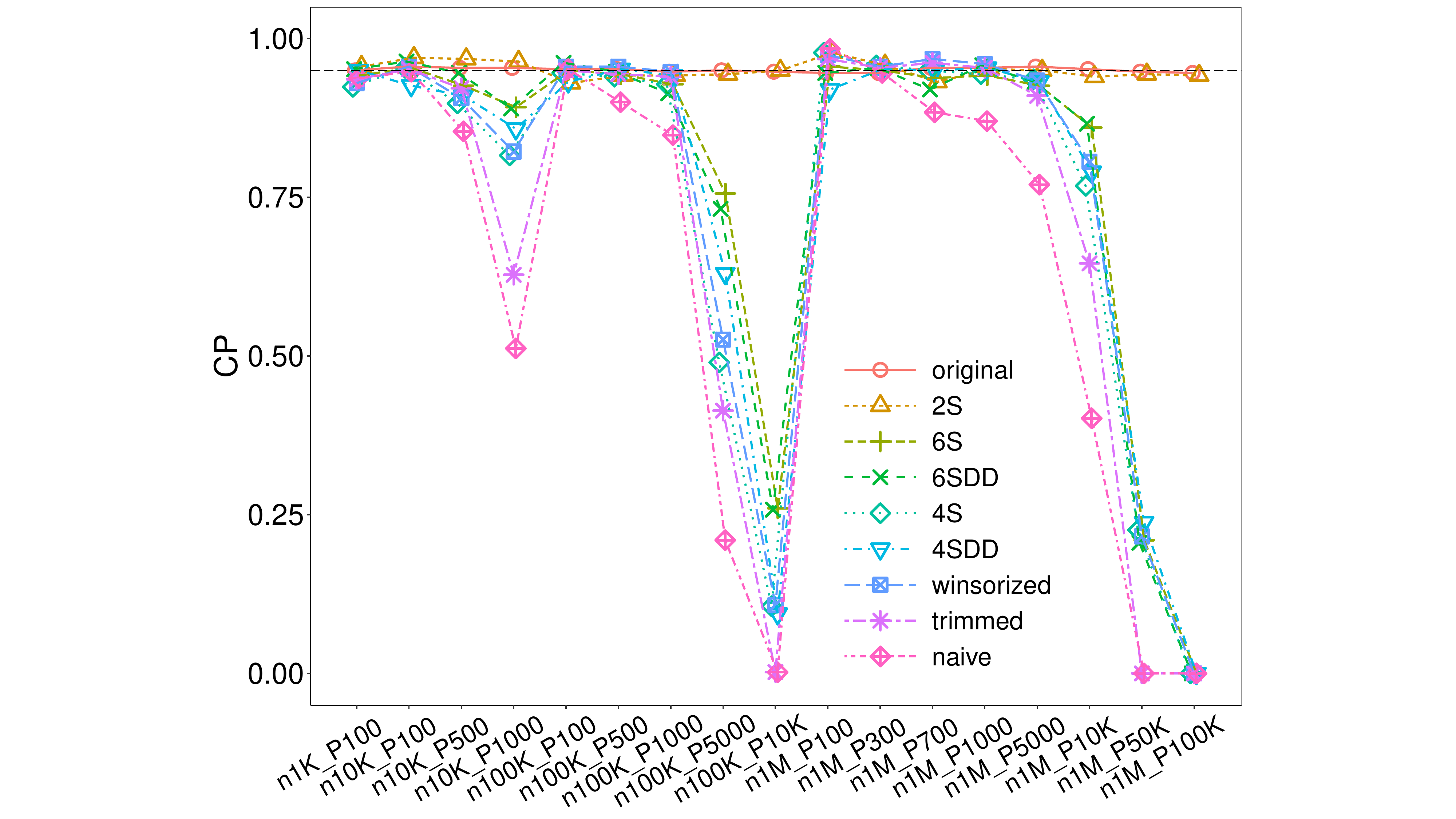}
\includegraphics[width=0.215\textwidth, trim={2.2in 0 2.2in 0},clip] {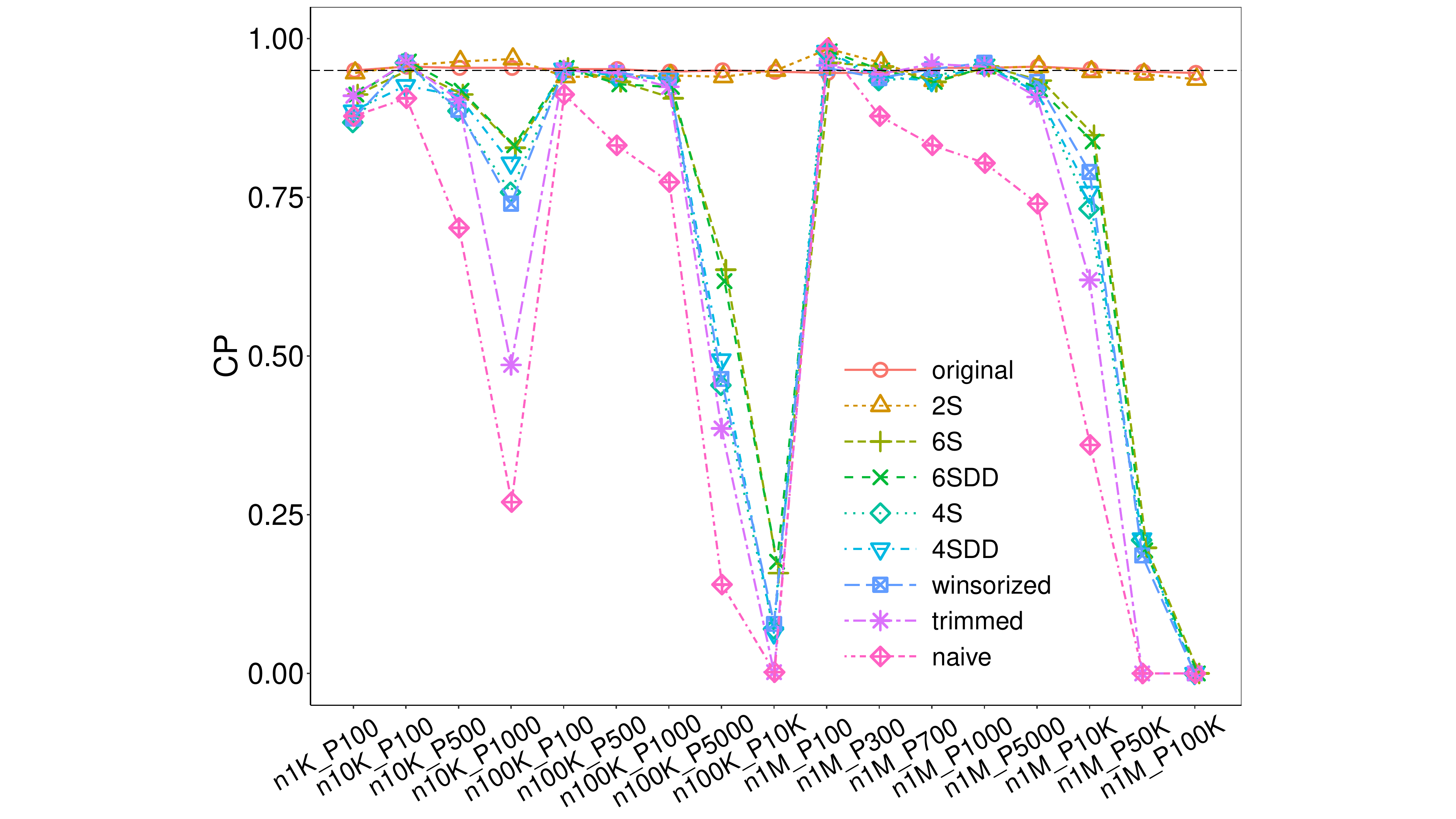}
\includegraphics[width=0.215\textwidth, trim={2.2in 0 2.2in 0},clip] {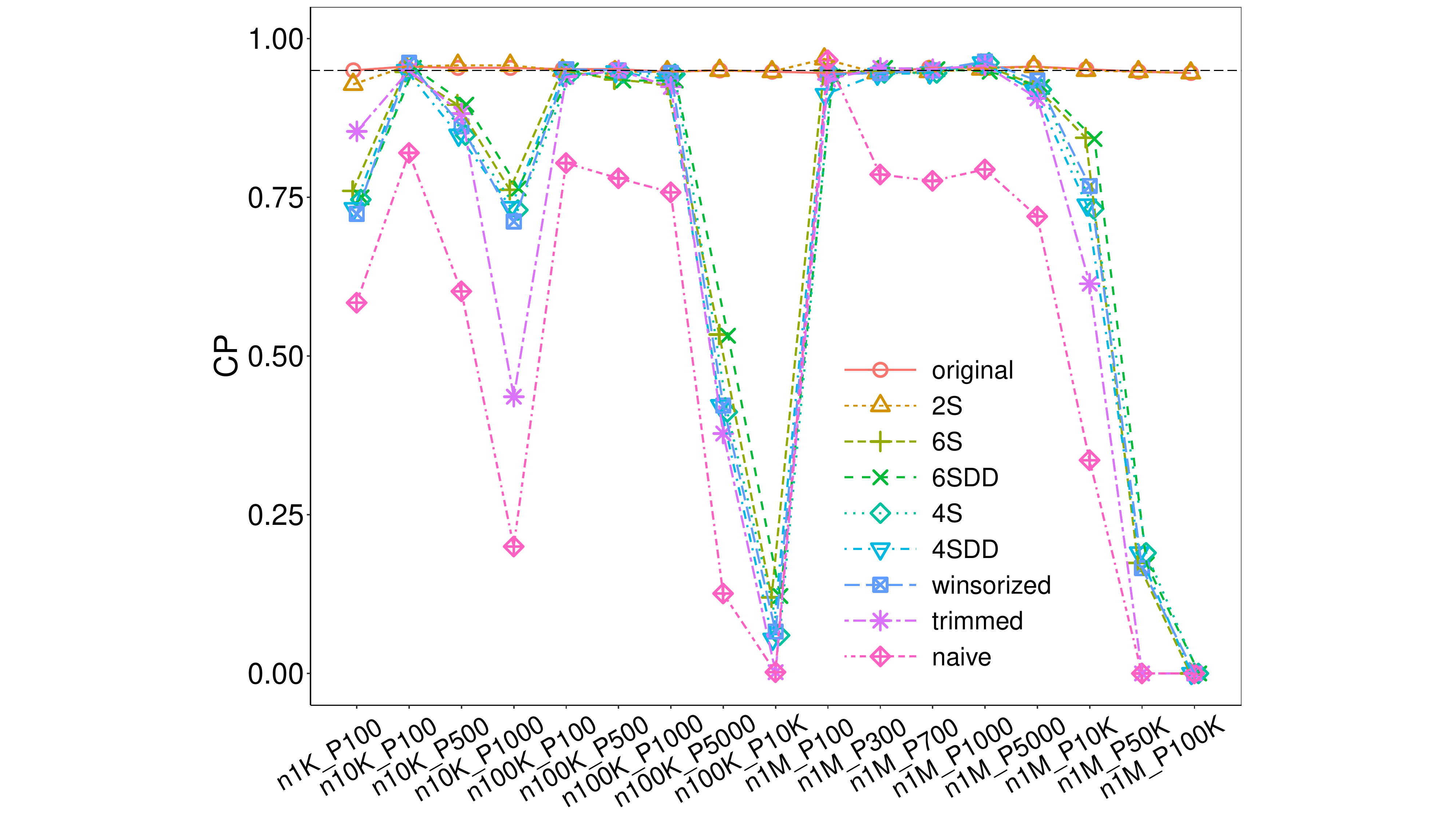}\\
\includegraphics[width=0.215\textwidth, trim={2.2in 0 2.2in 0},clip] {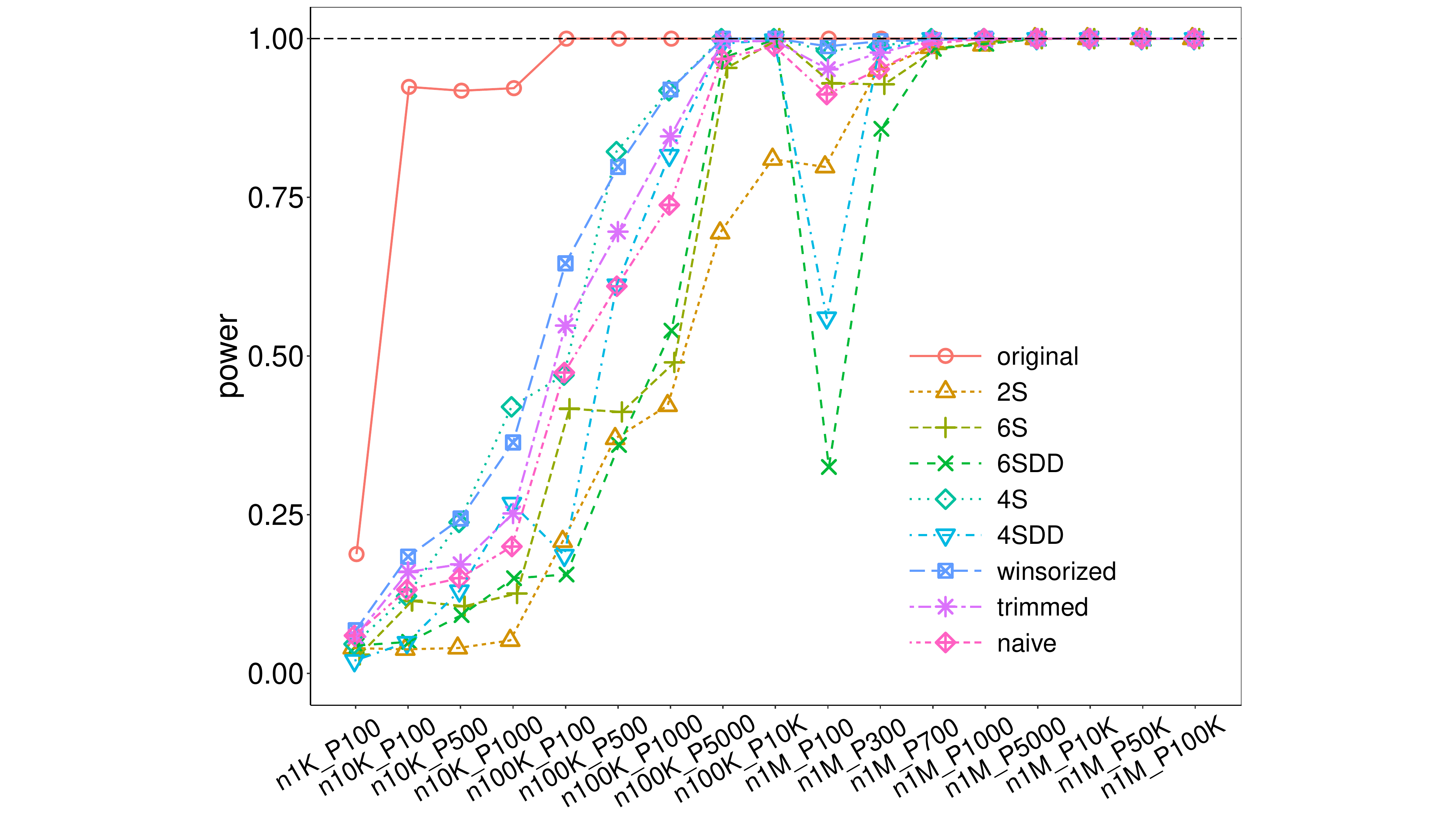}
\includegraphics[width=0.215\textwidth, trim={2.2in 0 2.2in 0},clip] {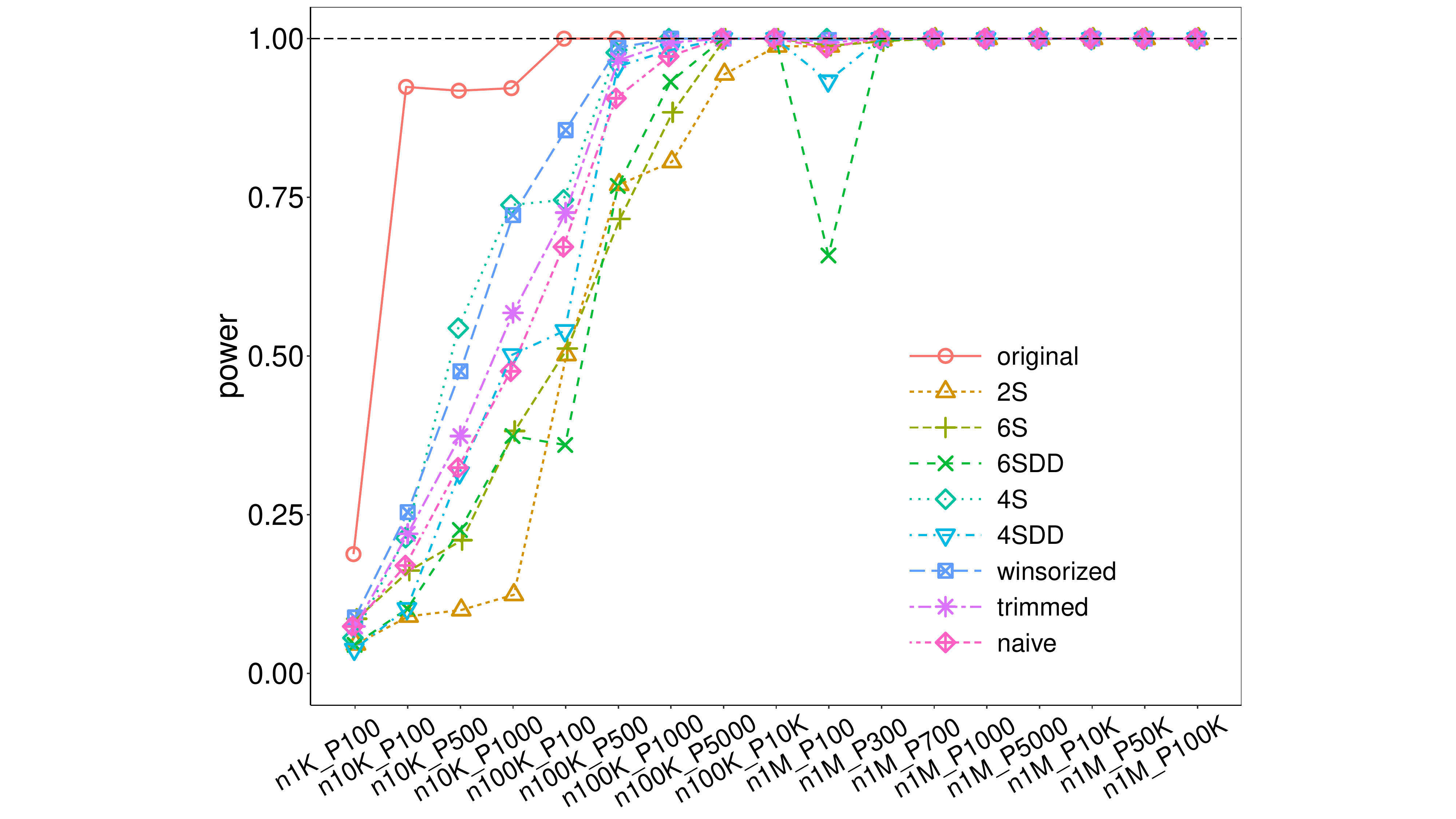}
\includegraphics[width=0.215\textwidth, trim={2.2in 0 2.2in 0},clip] {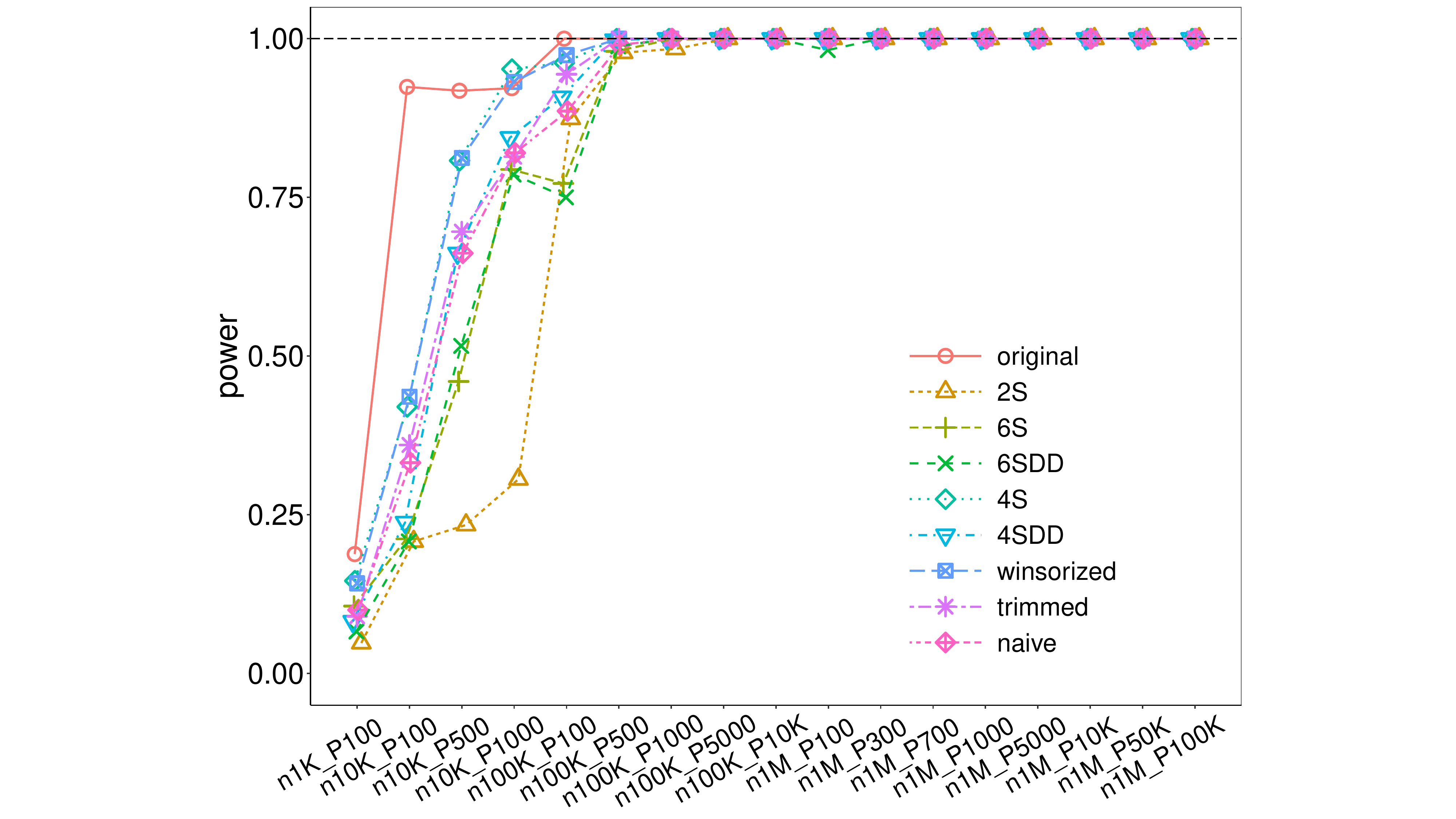}
\includegraphics[width=0.215\textwidth, trim={2.2in 0 2.2in 0},clip] {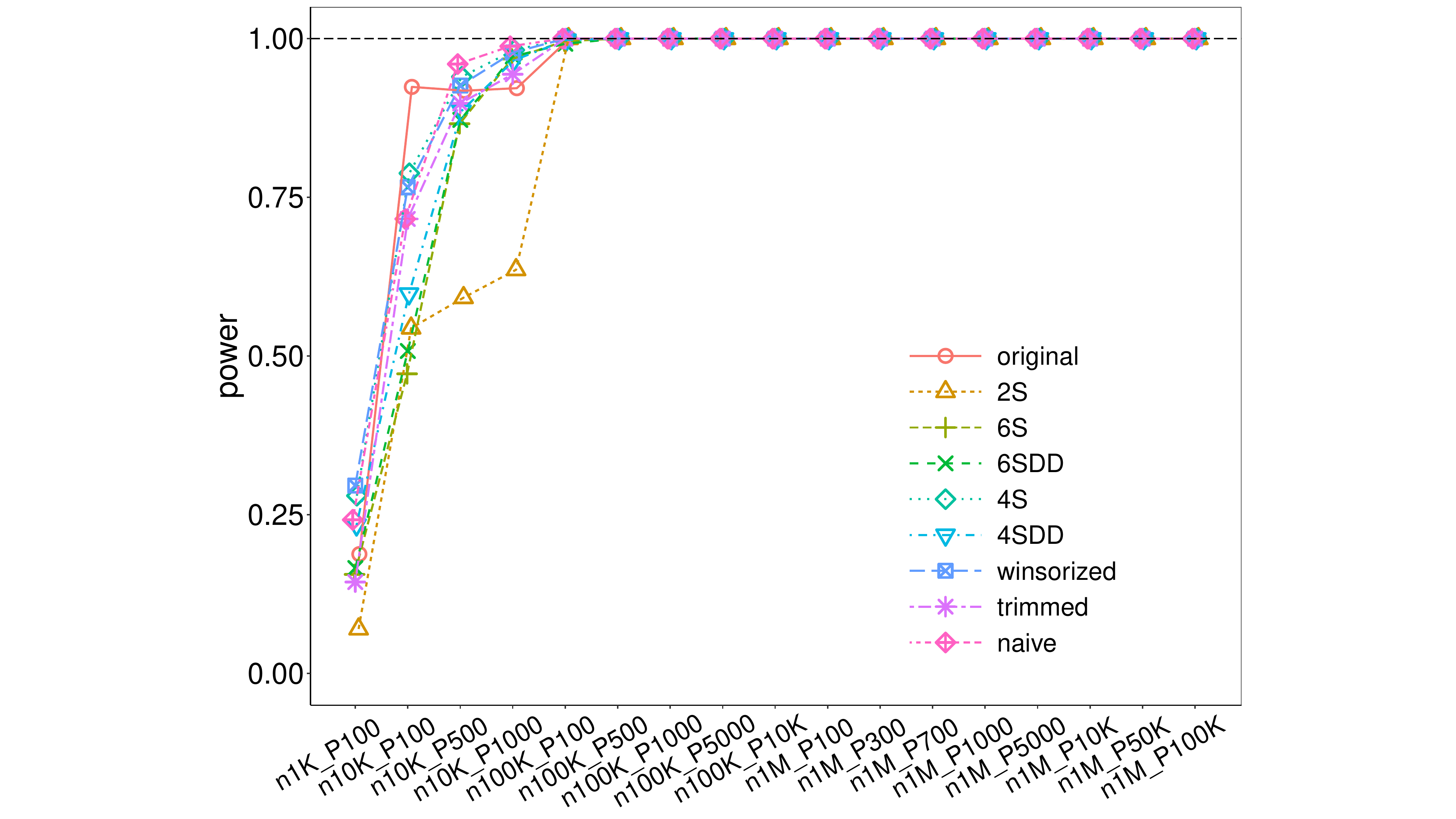}
\includegraphics[width=0.215\textwidth, trim={2.2in 0 2.2in 0},clip] {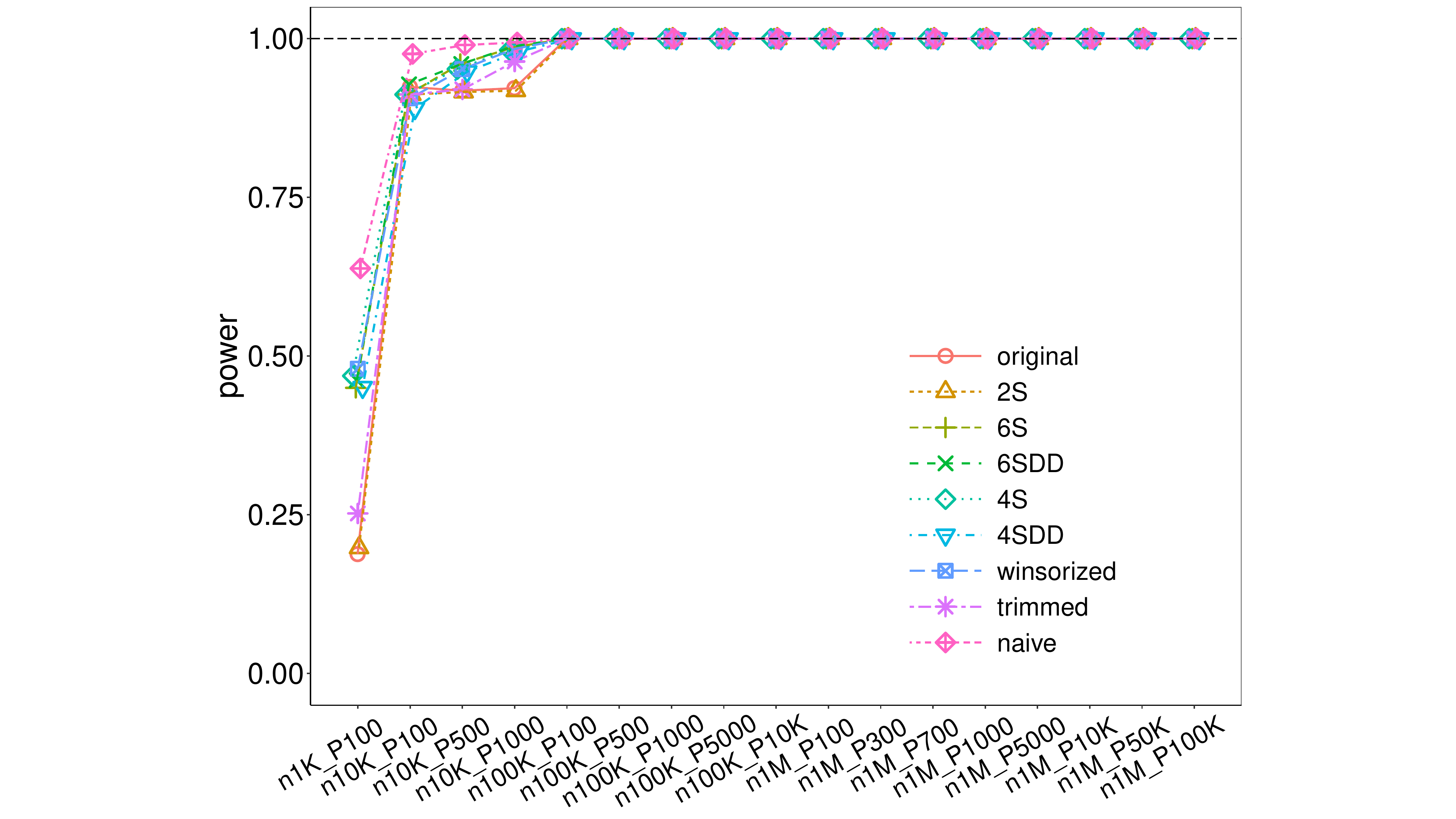}\\
\caption{ZINB data; $\epsilon$-DP; $\theta\ne0$ and $\alpha=\beta$} \label{fig:1sDPzinb}
\end{figure}

\end{landscape}

\begin{landscape}

\begin{figure}[!htb]
\centering
$\rho=0.005$\hspace{1in}$\rho=0.02$\hspace{1in}$\rho=0.08$
\hspace{1in}$\rho=0.32$\hspace{0.8in}$\rho=1.28$\\
\includegraphics[width=0.24\textwidth, trim={2.2in 0 2.2in 0},clip] {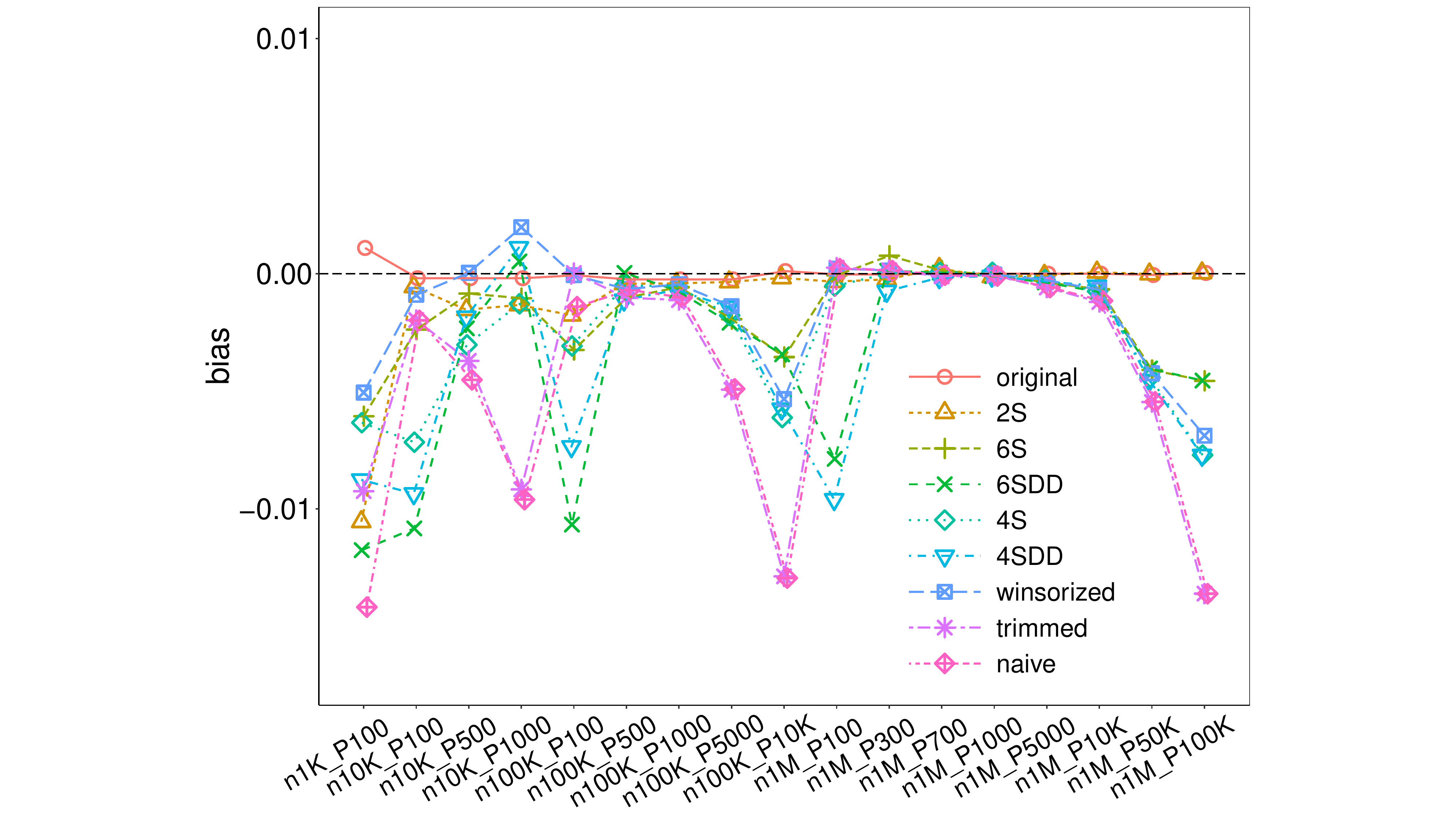}
\includegraphics[width=0.24\textwidth, trim={2.2in 0 2.2in 0},clip] {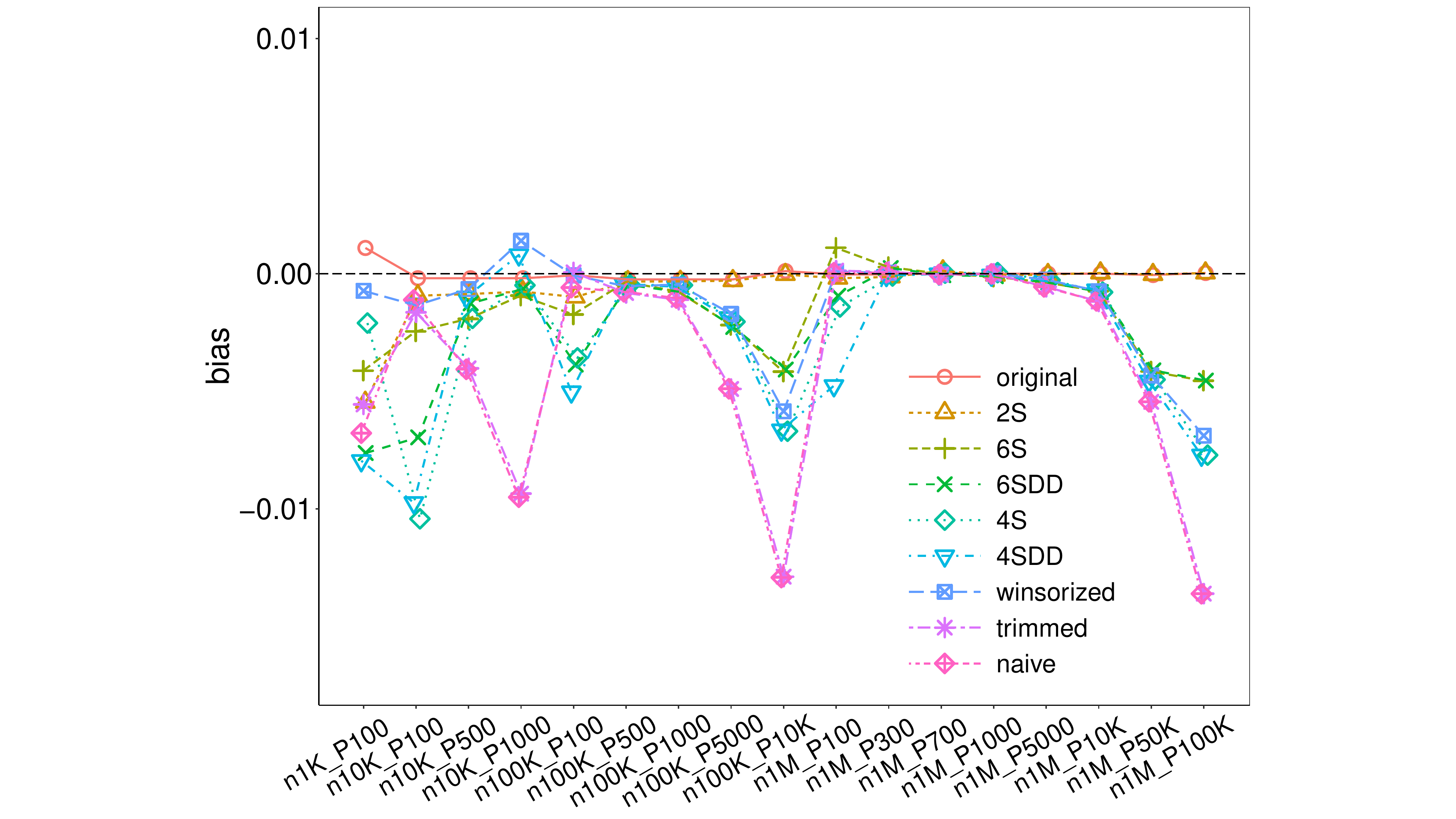}
\includegraphics[width=0.24\textwidth, trim={2.2in 0 2.2in 0},clip] {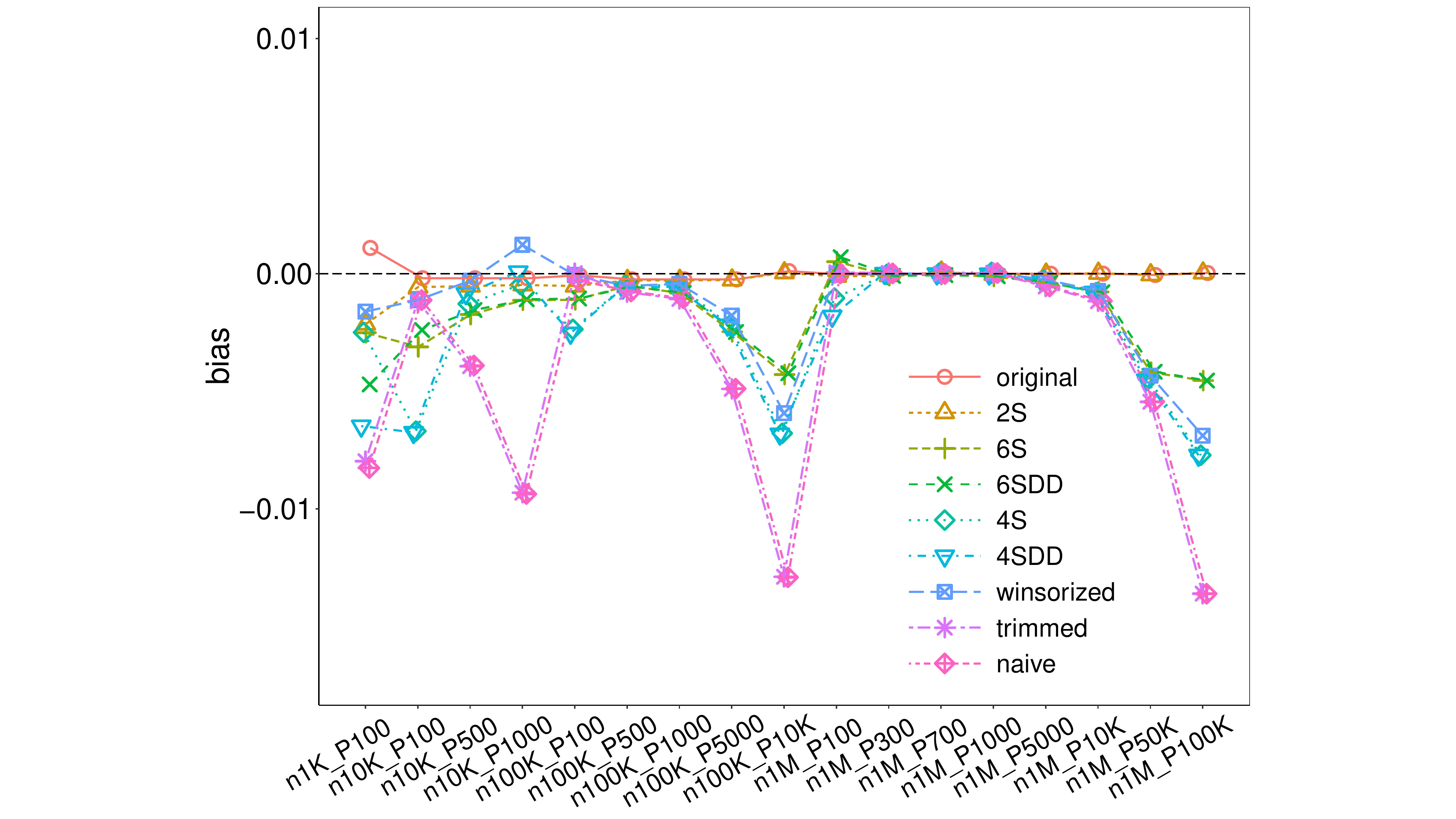}
\includegraphics[width=0.24\textwidth, trim={2.2in 0 2.2in 0},clip] {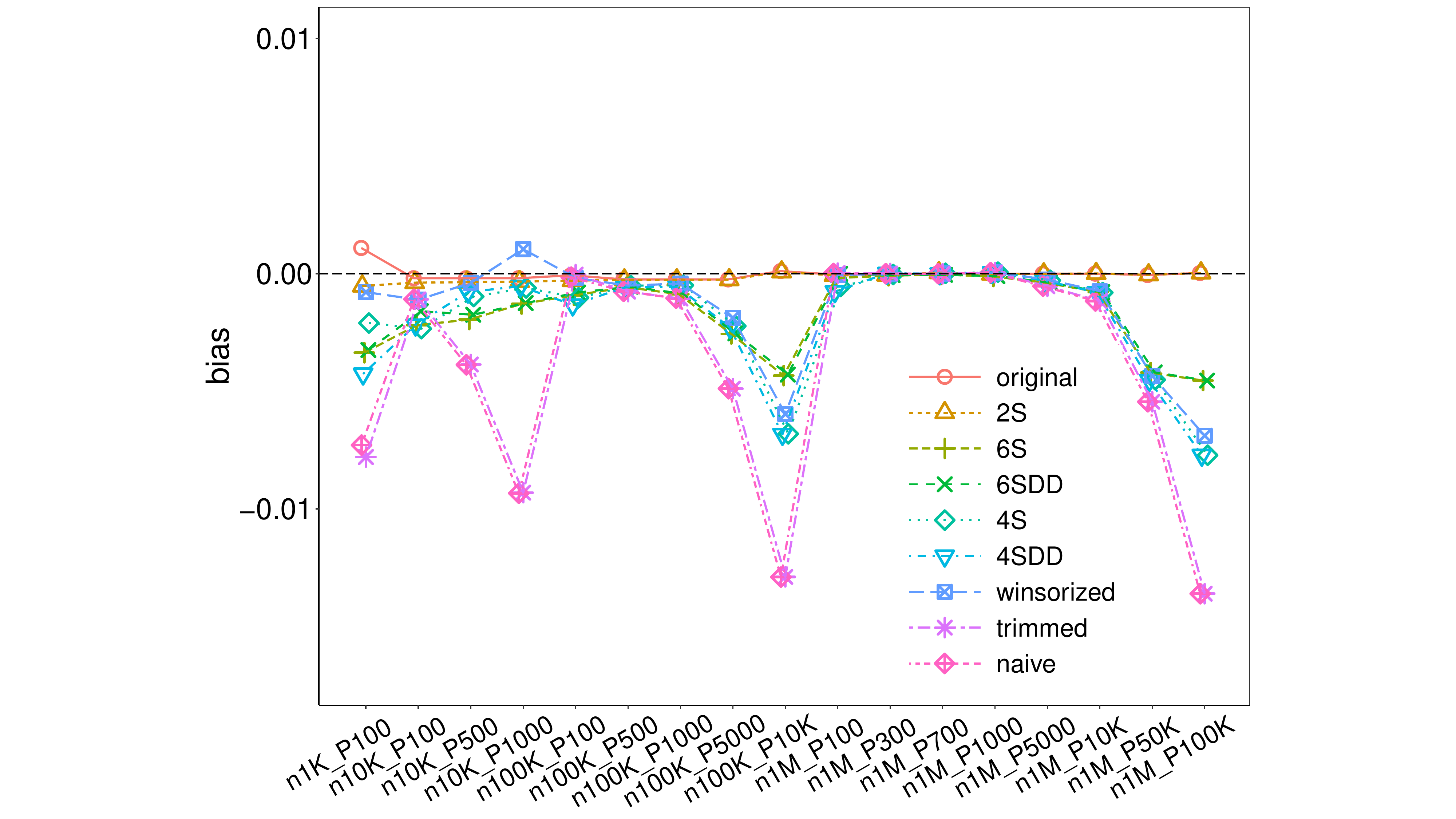}
\includegraphics[width=0.24\textwidth, trim={2.2in 0 2.2in 0},clip] {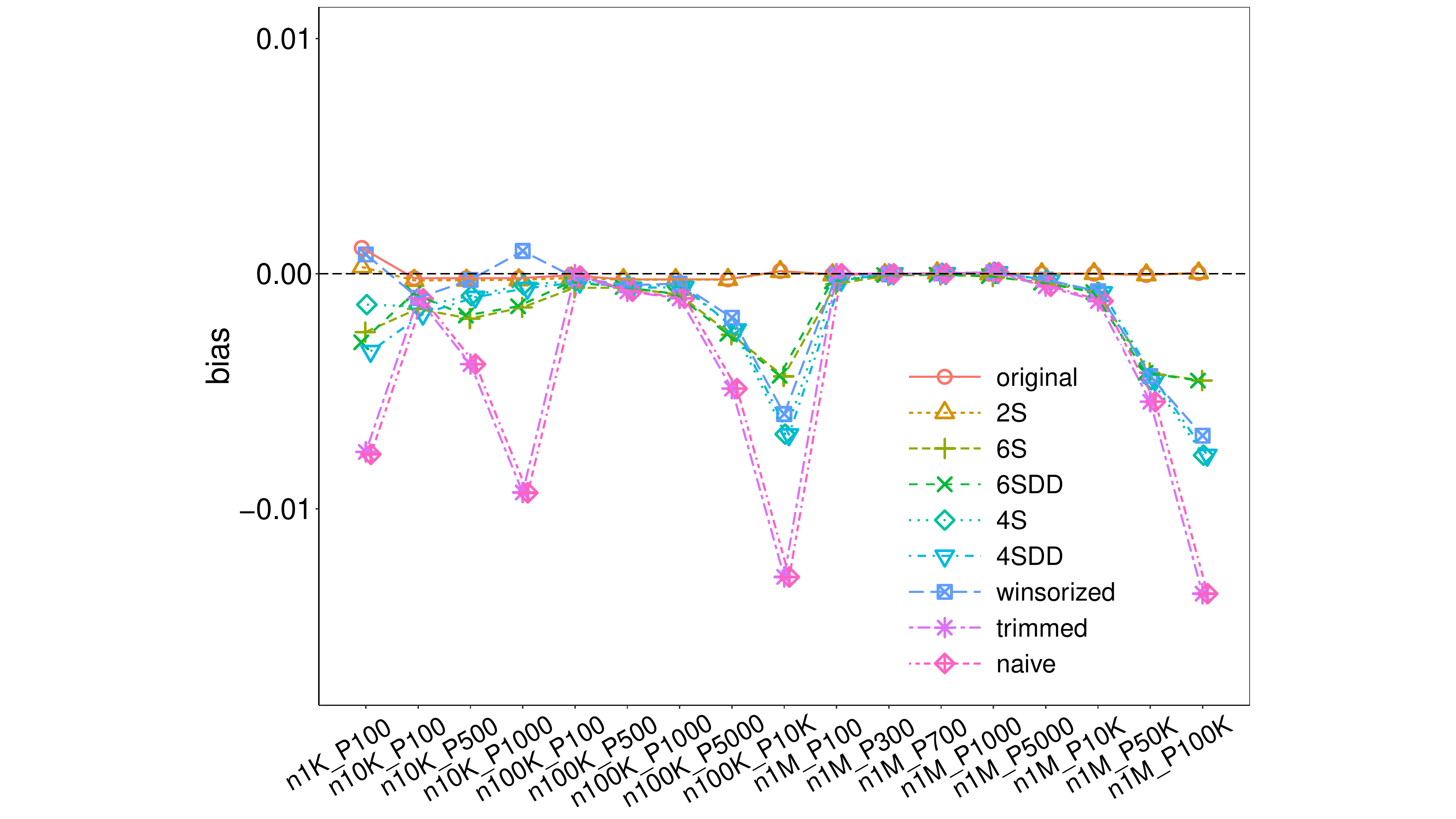}\\
\includegraphics[width=0.24\textwidth, trim={2.2in 0 2.2in 0},clip] {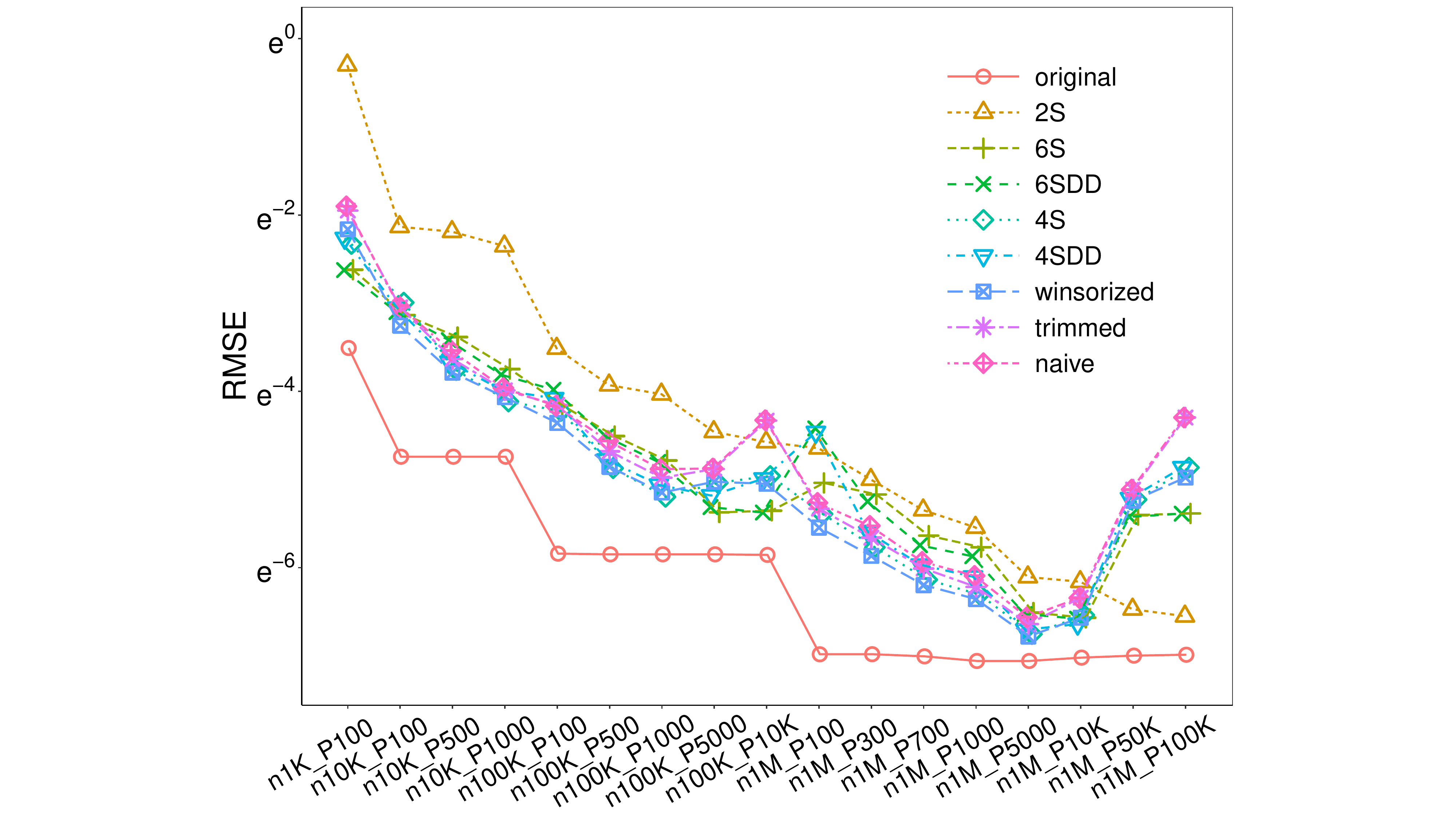}
\includegraphics[width=0.24\textwidth, trim={2.2in 0 2.2in 0},clip] {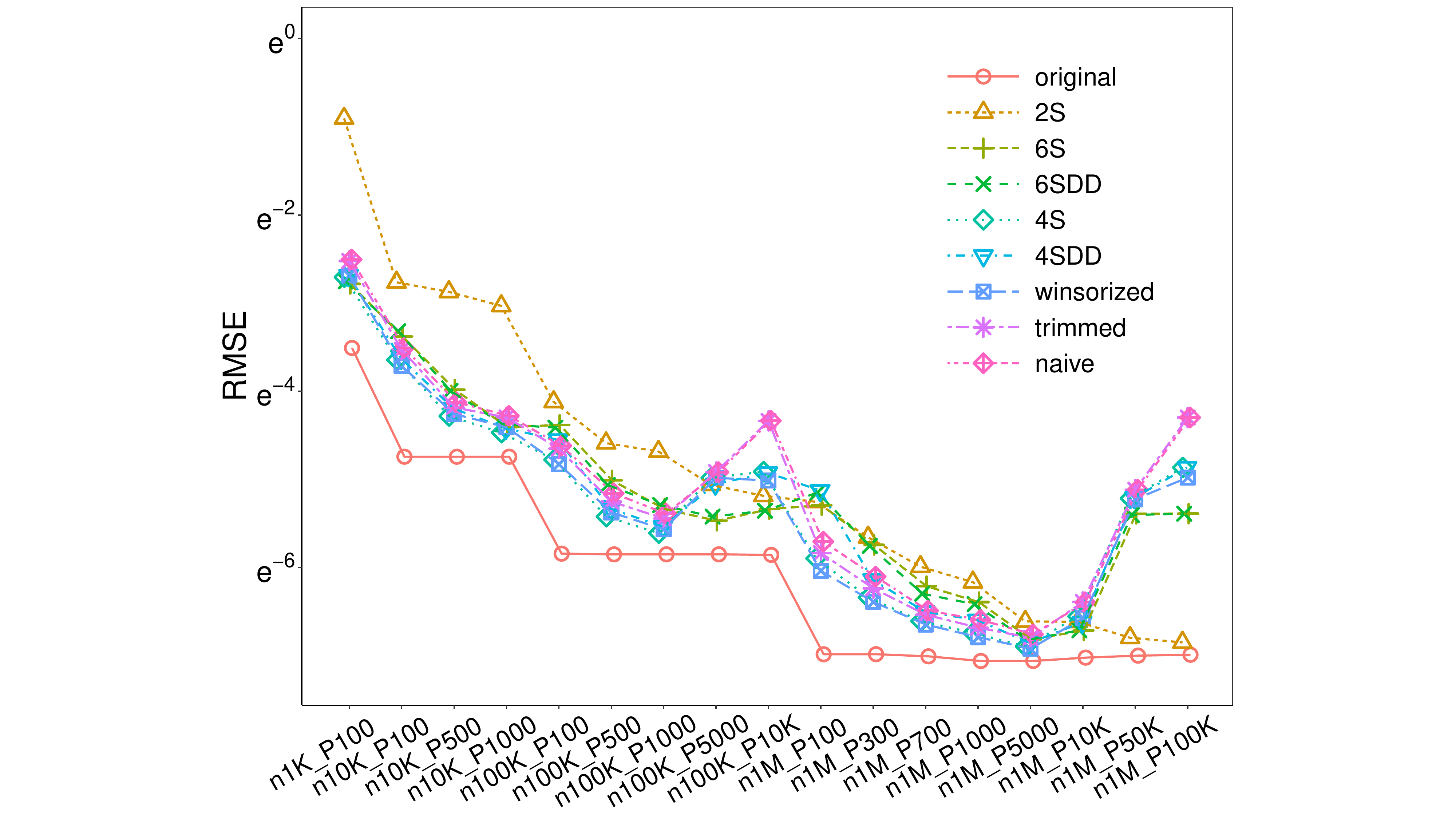}
\includegraphics[width=0.24\textwidth, trim={2.2in 0 2.2in 0},clip] {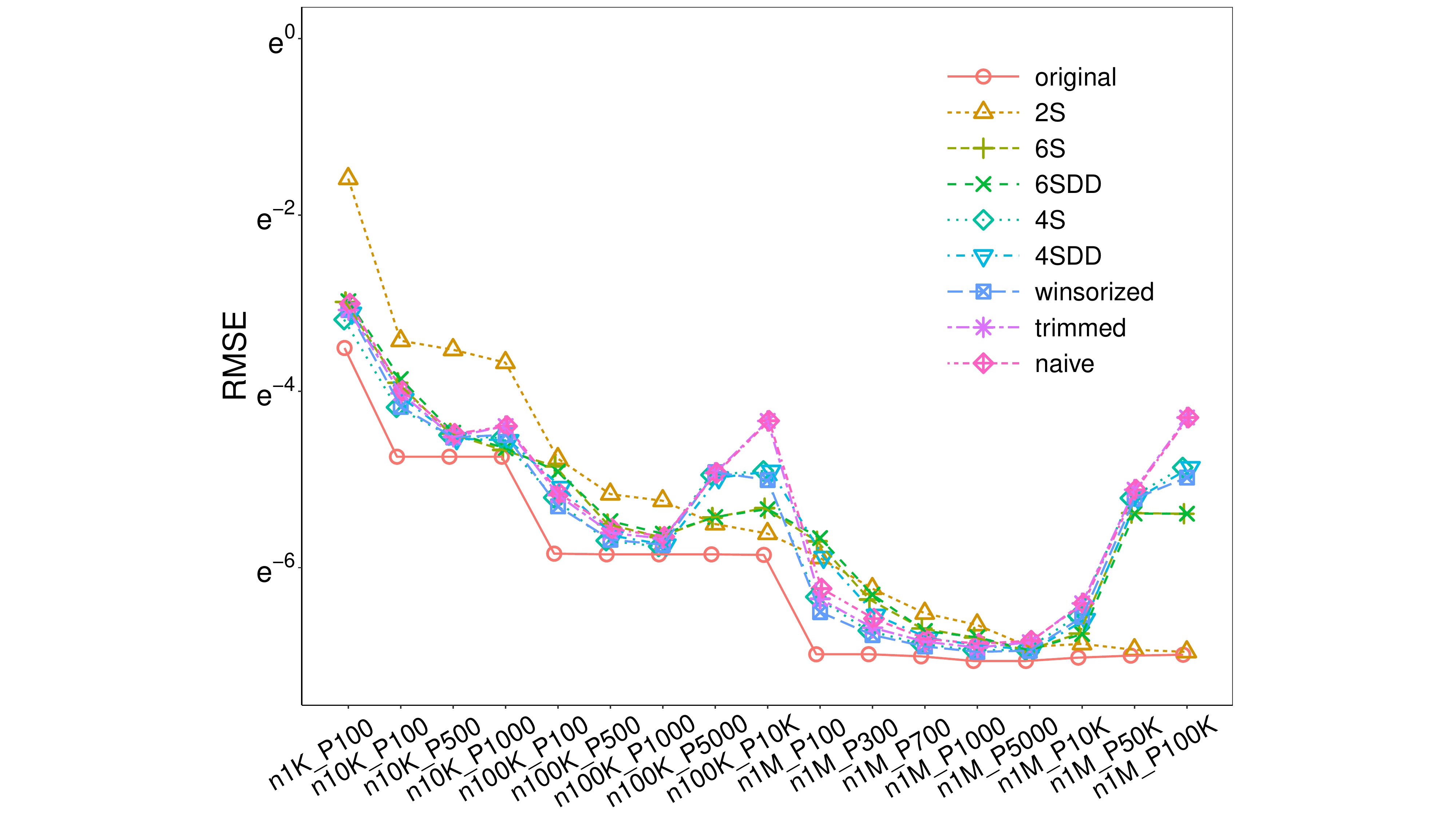}
\includegraphics[width=0.24\textwidth, trim={2.2in 0 2.2in 0},clip] {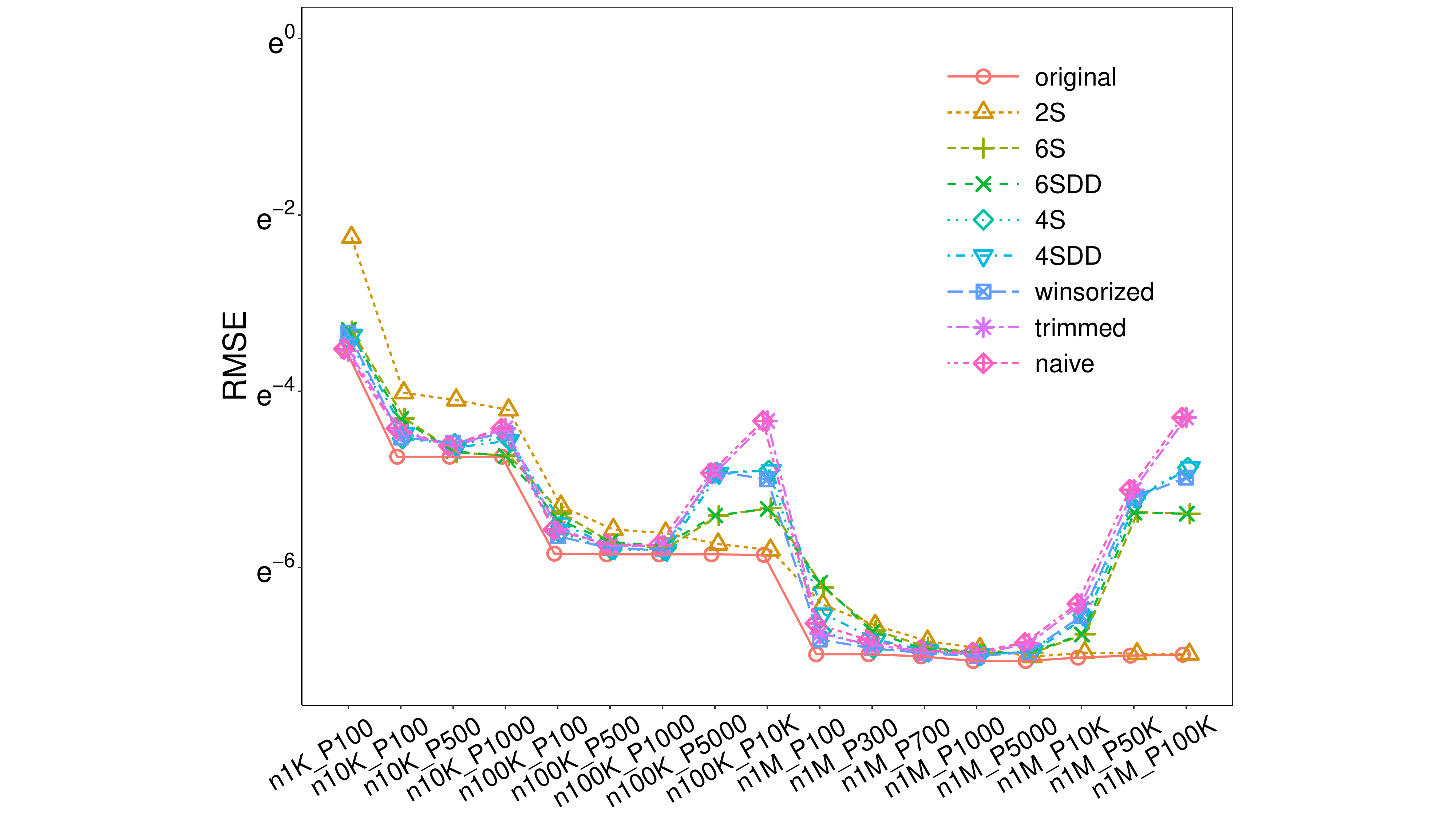}
\includegraphics[width=0.24\textwidth, trim={2.2in 0 2.2in 0},clip] {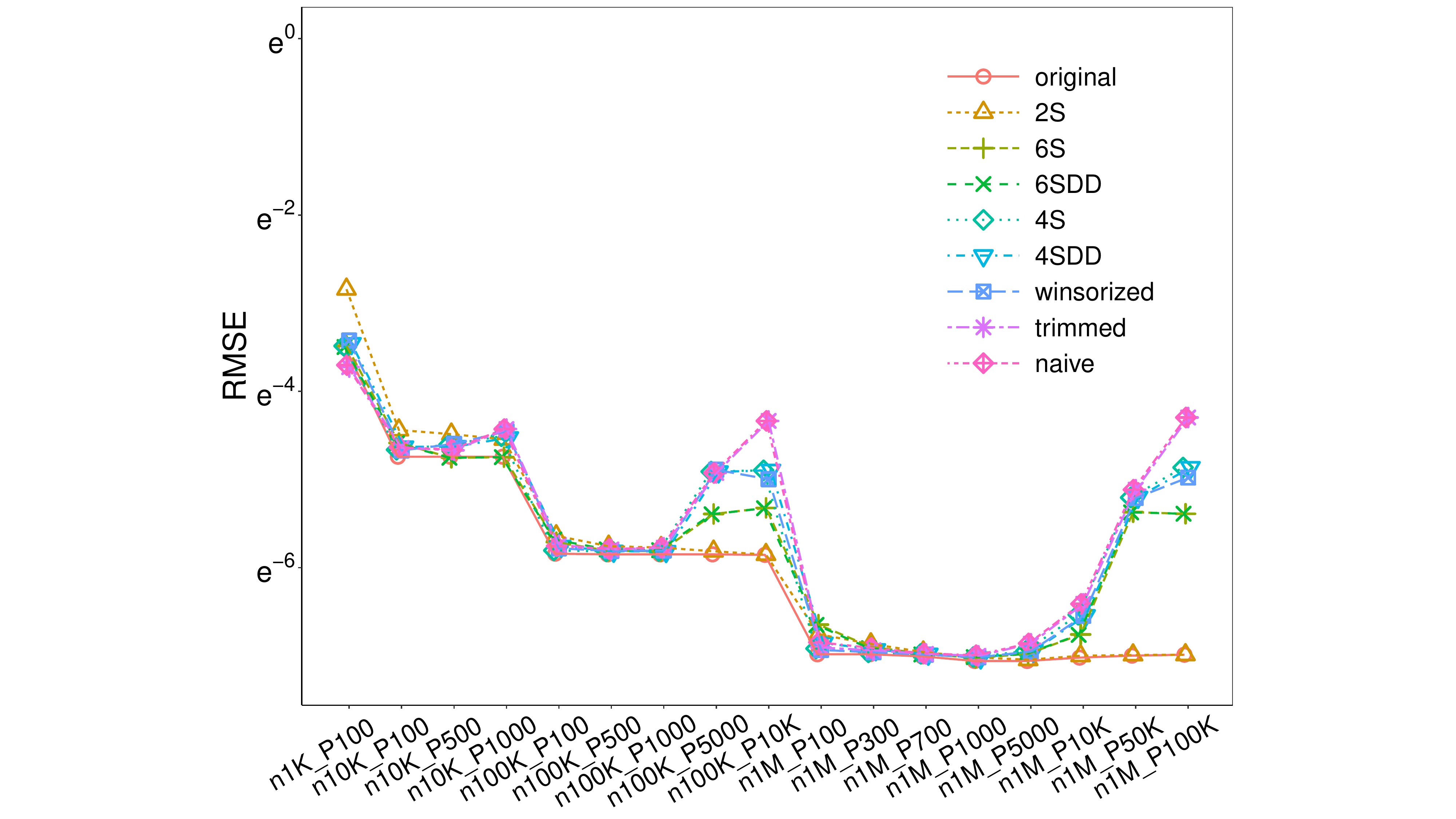}\\
\includegraphics[width=0.24\textwidth, trim={2.2in 0 2.2in 0},clip] {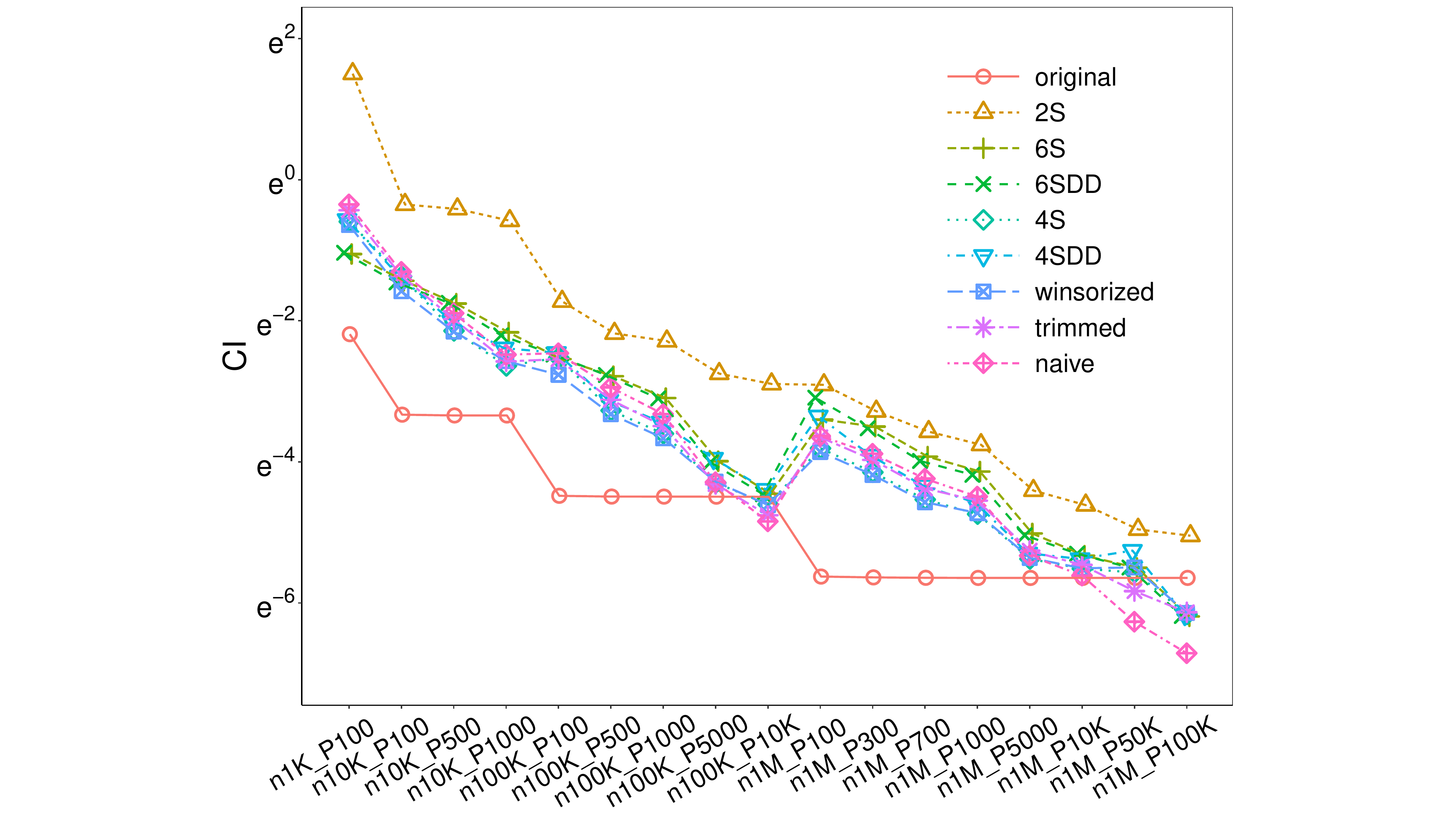}
\includegraphics[width=0.24\textwidth, trim={2.2in 0 2.2in 0},clip] {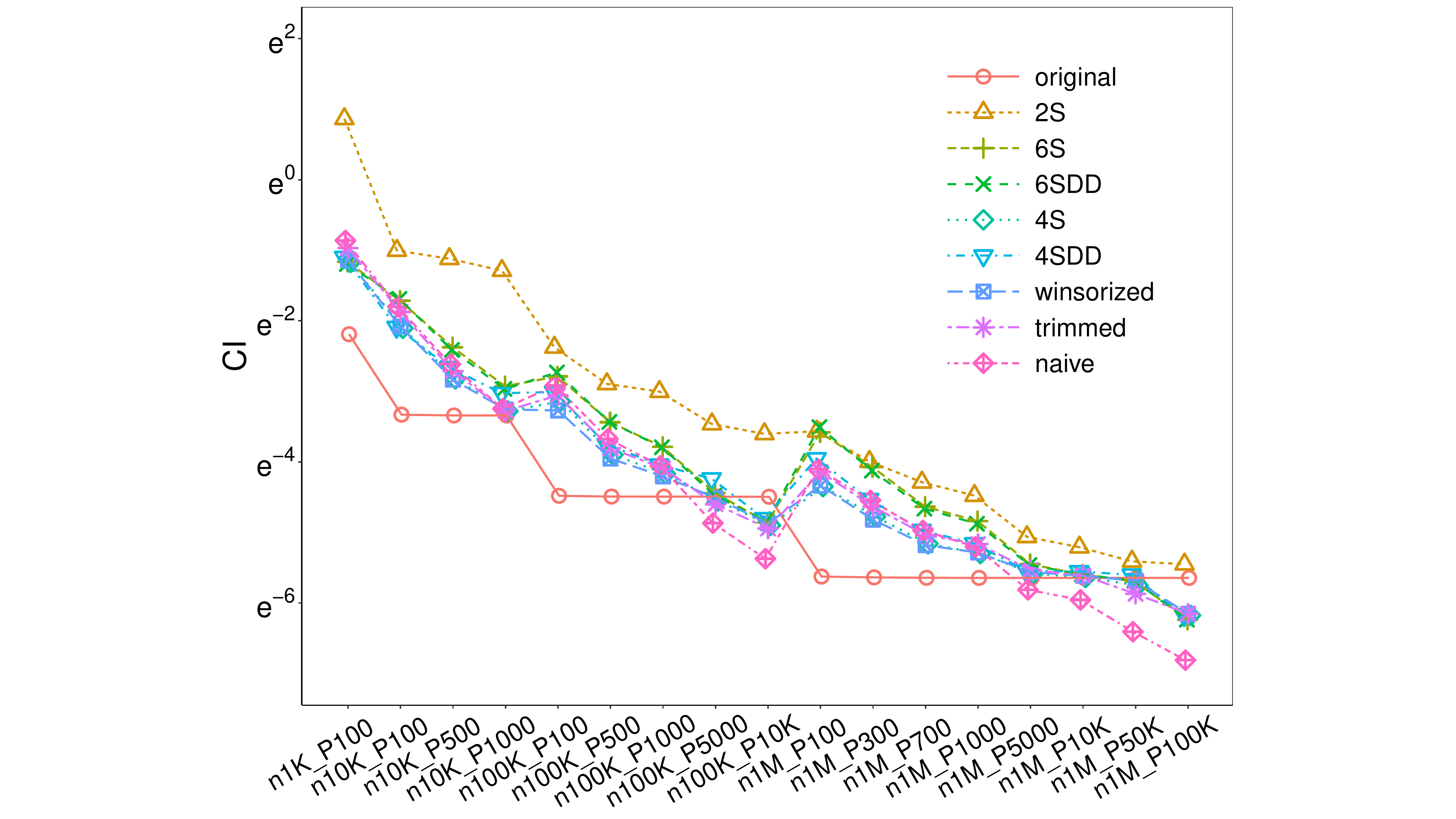}
\includegraphics[width=0.24\textwidth, trim={2.2in 0 2.2in 0},clip] {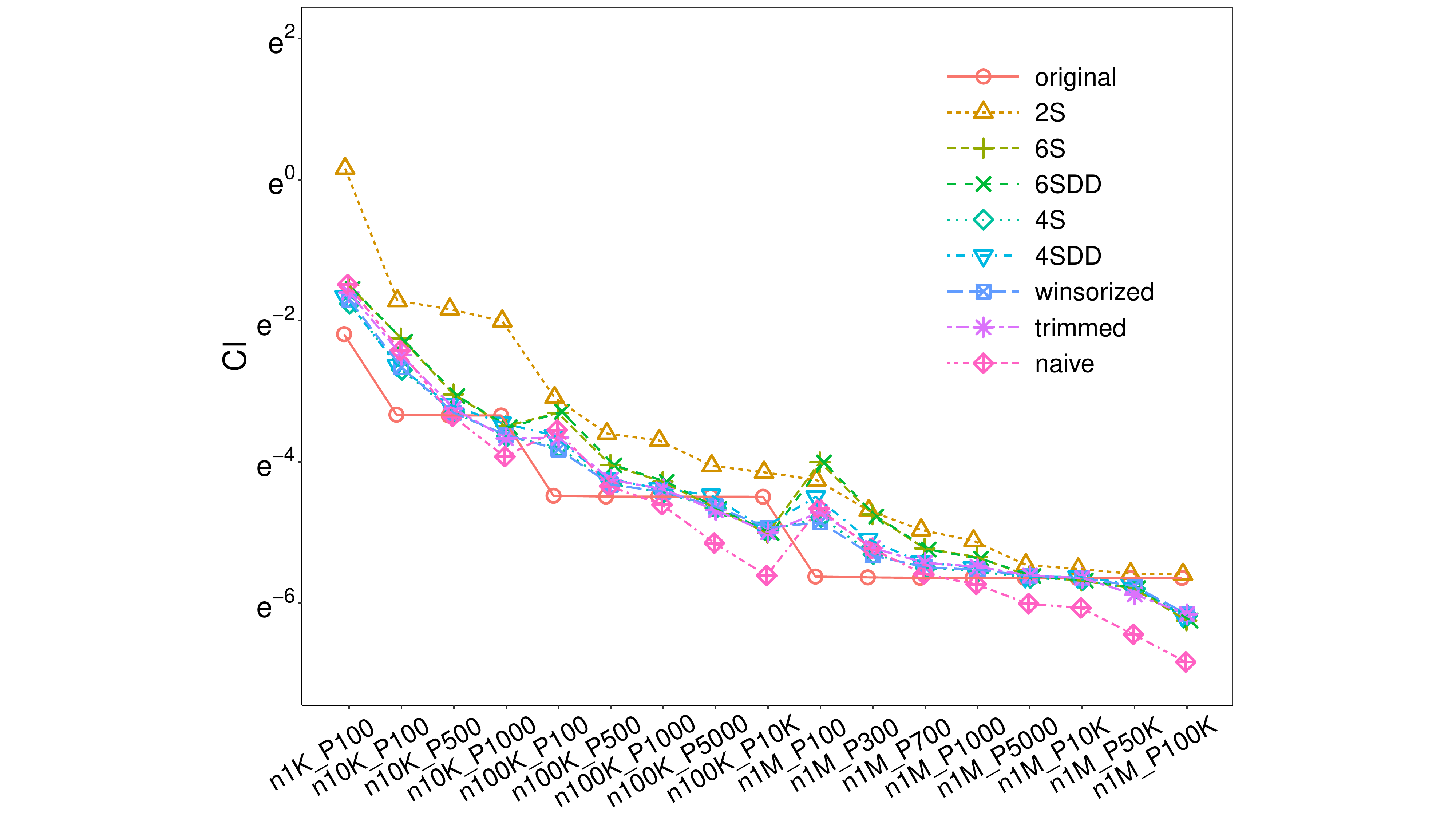}
\includegraphics[width=0.24\textwidth, trim={2.2in 0 2.2in 0},clip] {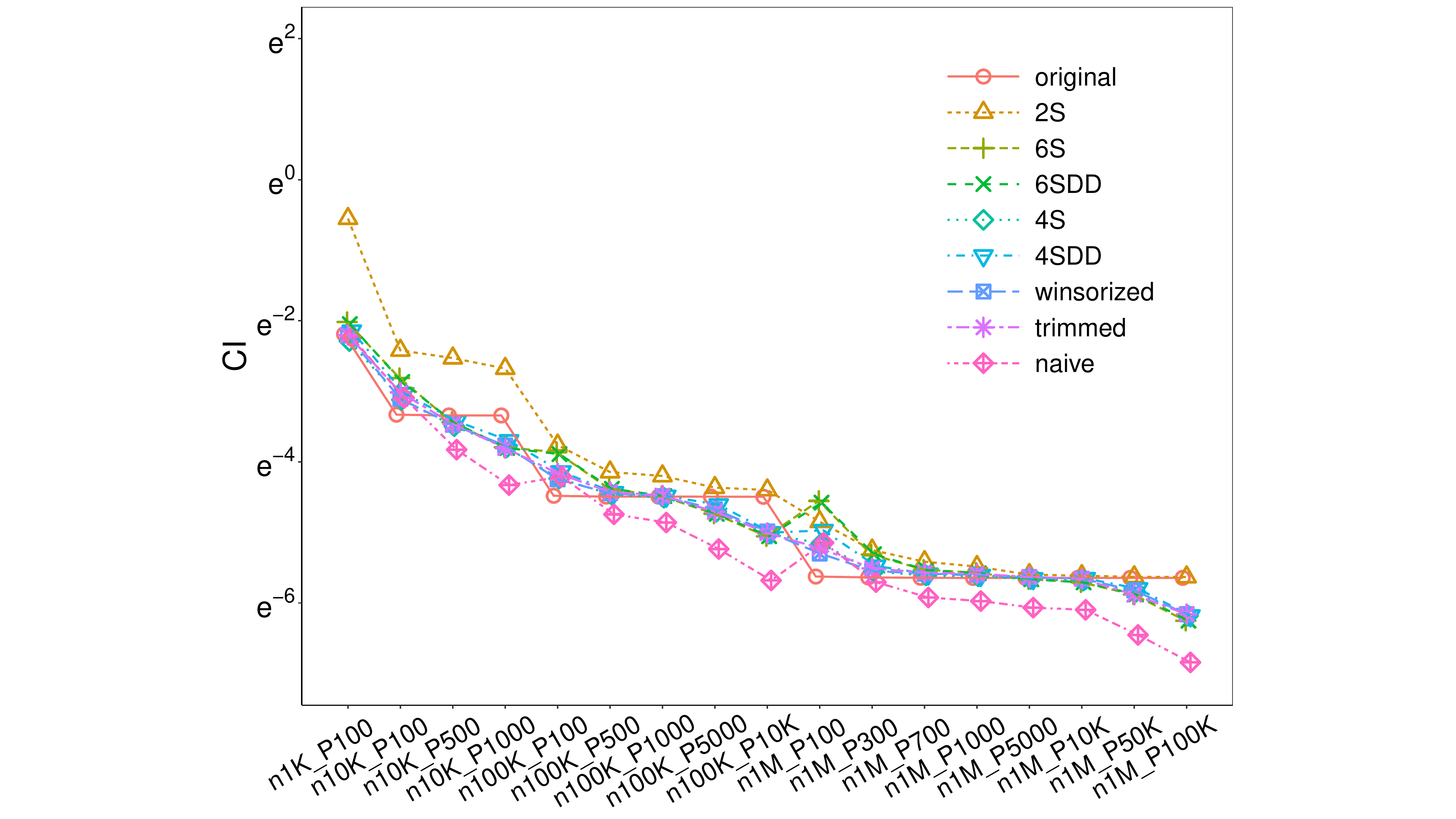}
\includegraphics[width=0.24\textwidth, trim={2.2in 0 2.2in 0},clip] {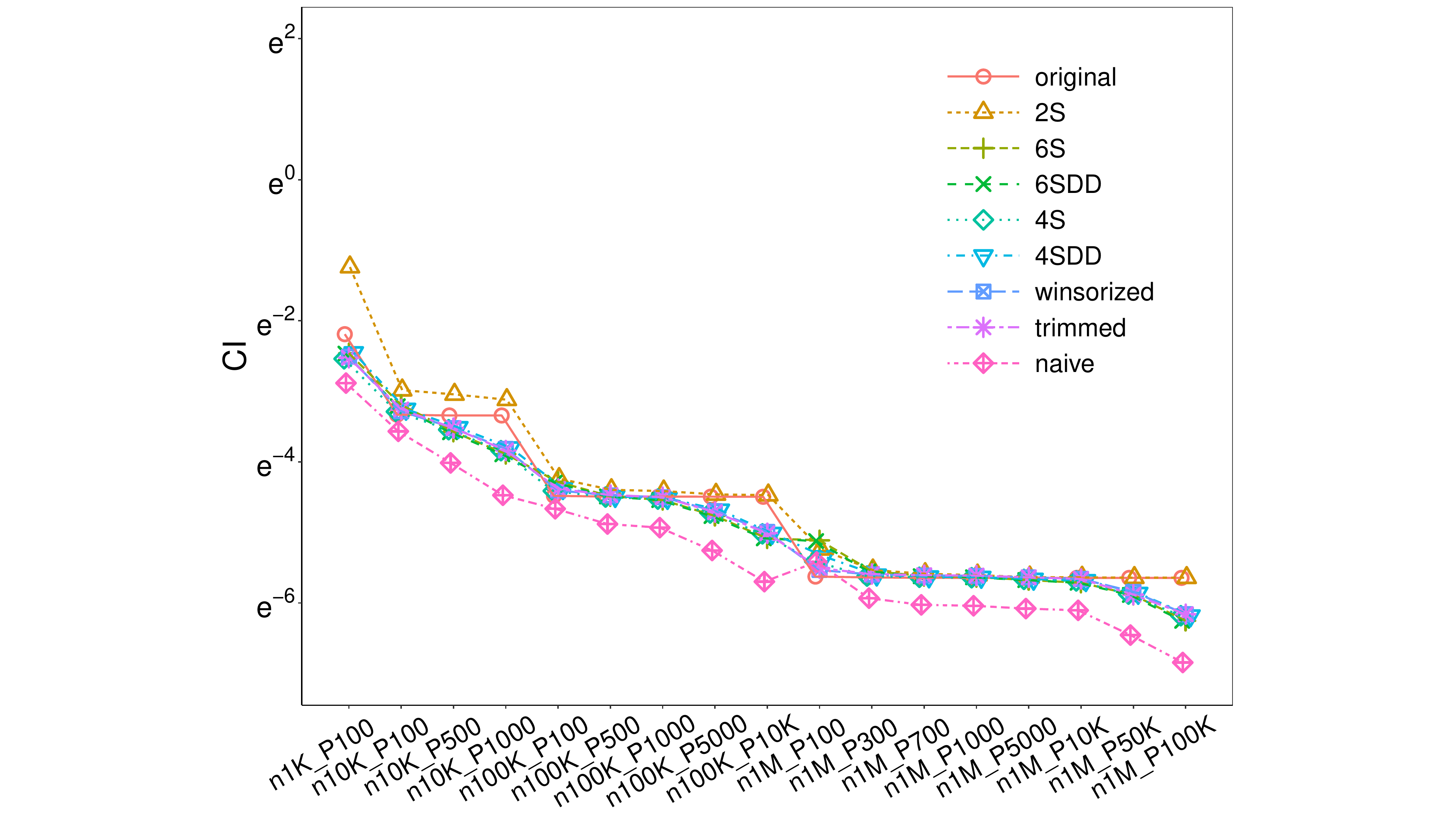}\\
\includegraphics[width=0.24\textwidth, trim={2.2in 0 2.2in 0},clip] {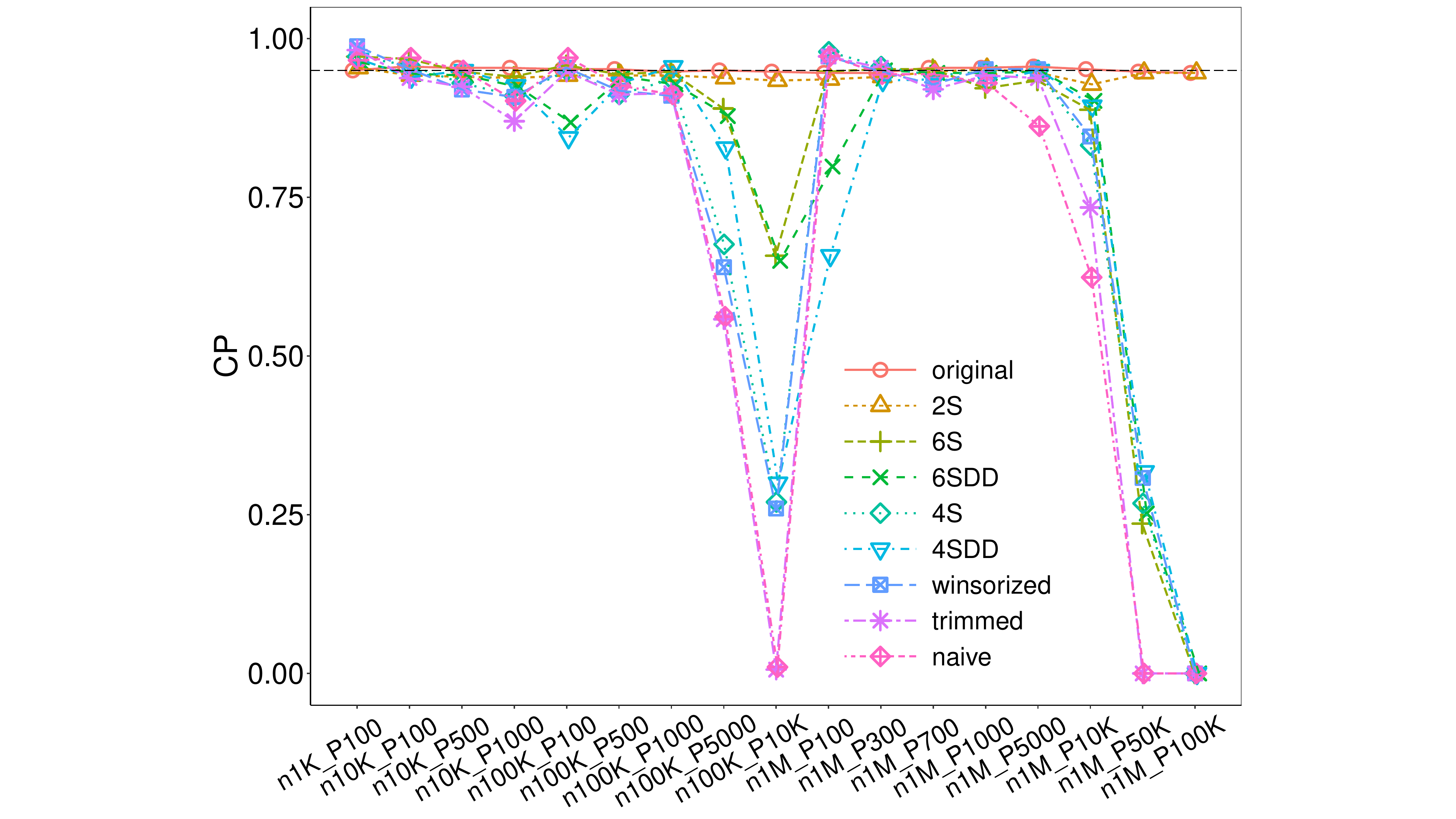}
\includegraphics[width=0.24\textwidth, trim={2.2in 0 2.2in 0},clip] {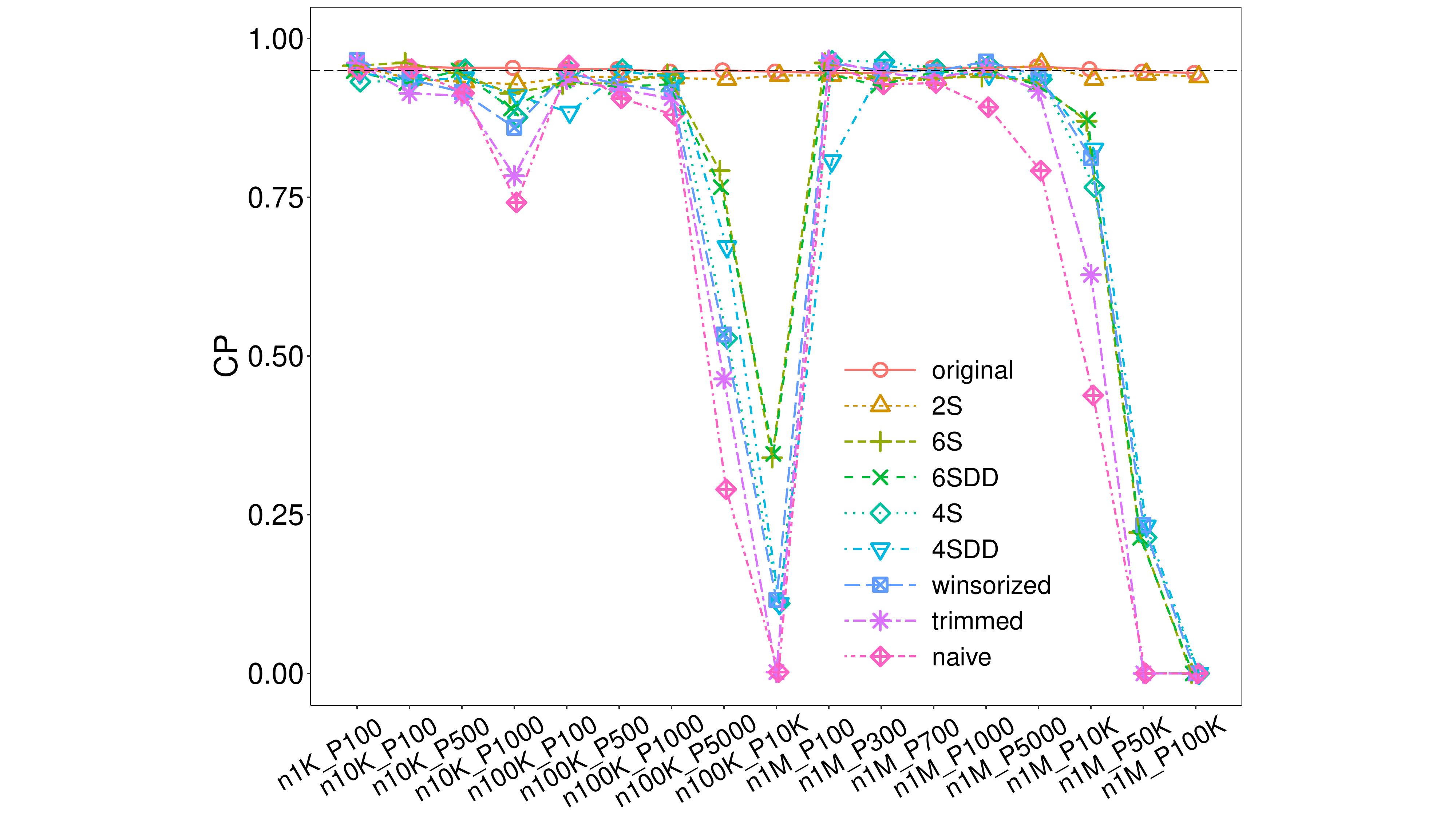}
\includegraphics[width=0.24\textwidth, trim={2.2in 0 2.2in 0},clip] {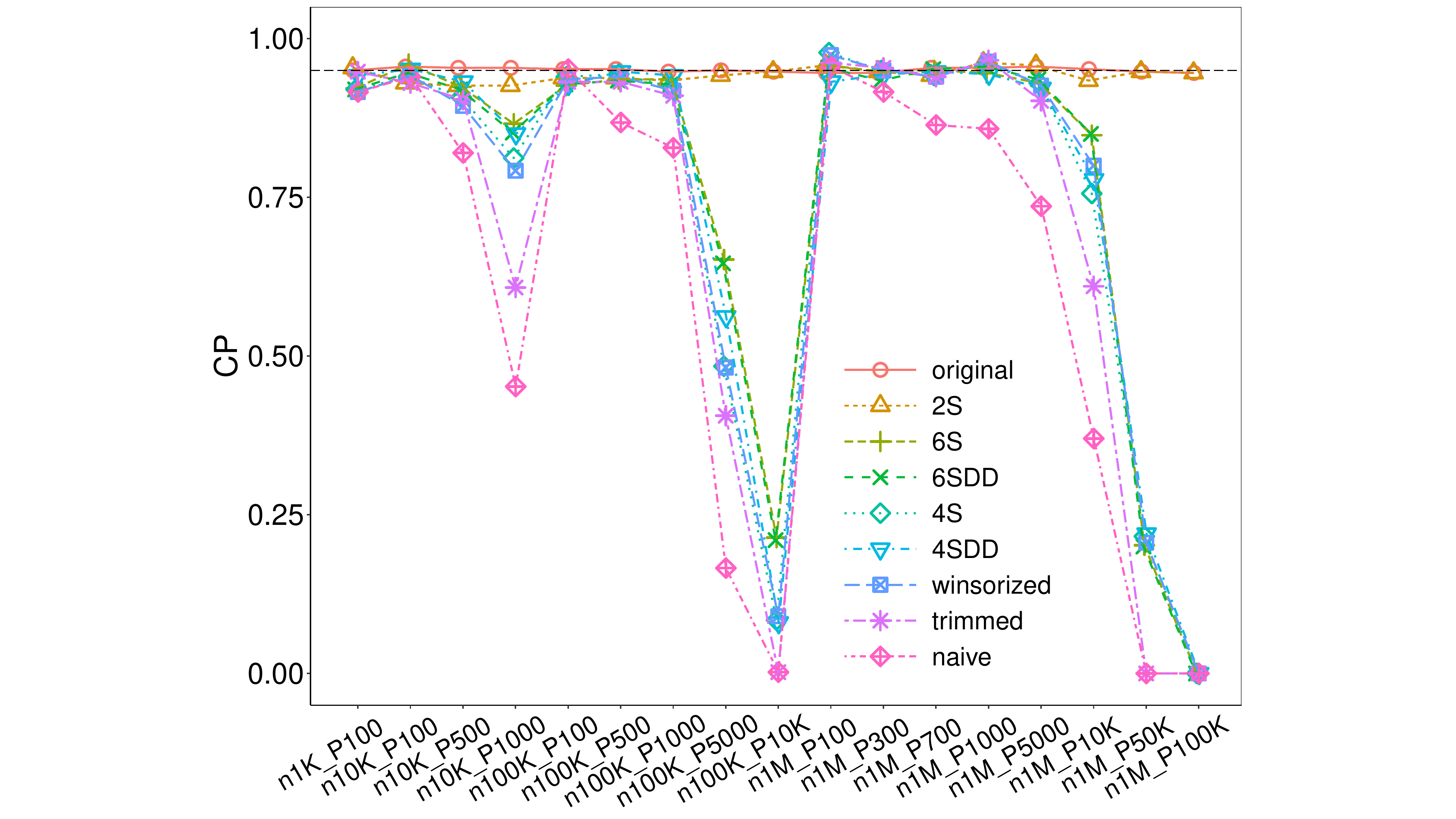}
\includegraphics[width=0.24\textwidth, trim={2.2in 0 2.2in 0},clip] {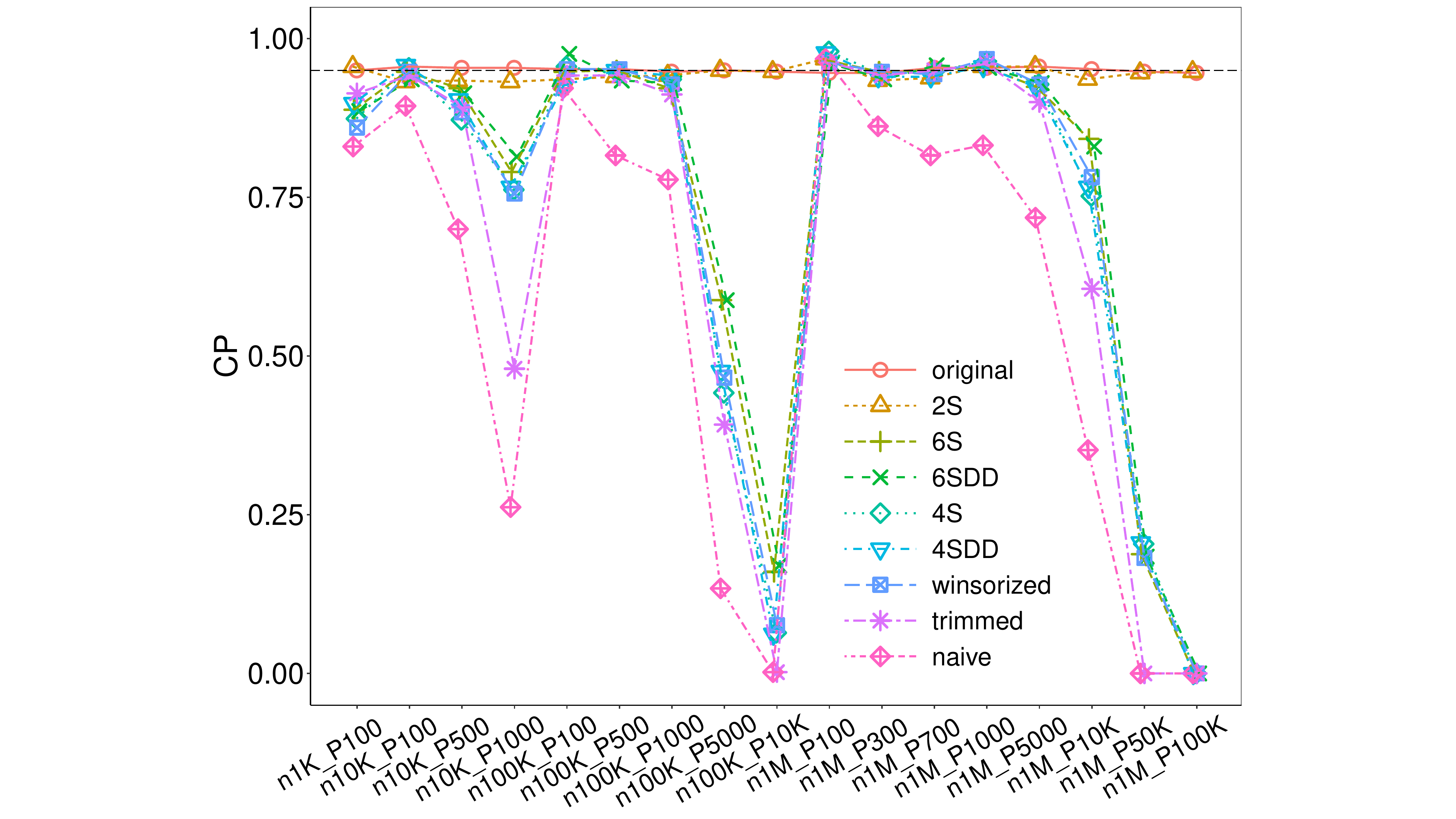}
\includegraphics[width=0.24\textwidth, trim={2.2in 0 2.2in 0},clip] {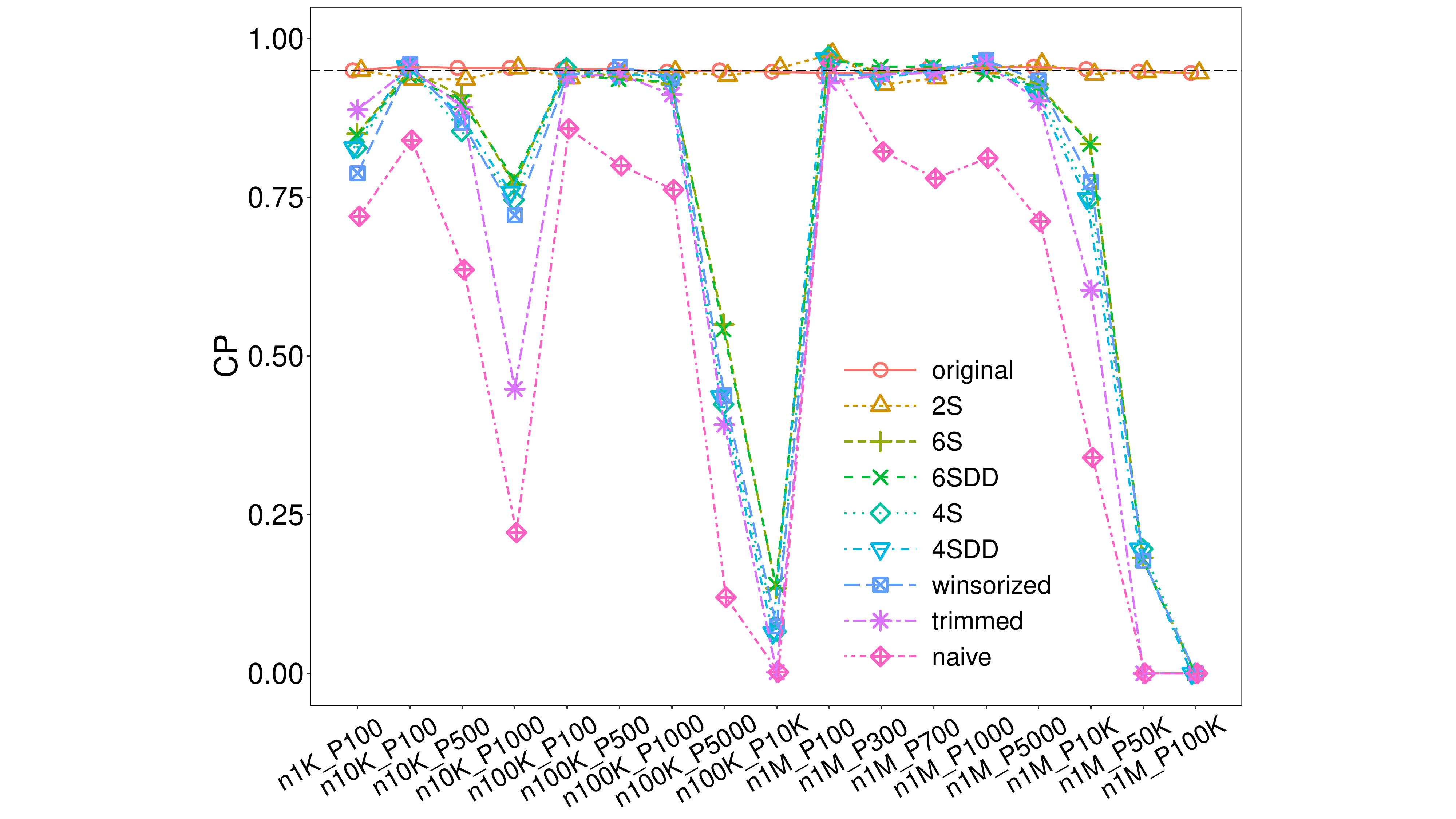}\\
\includegraphics[width=0.24\textwidth, trim={2.2in 0 2.2in 0},clip] {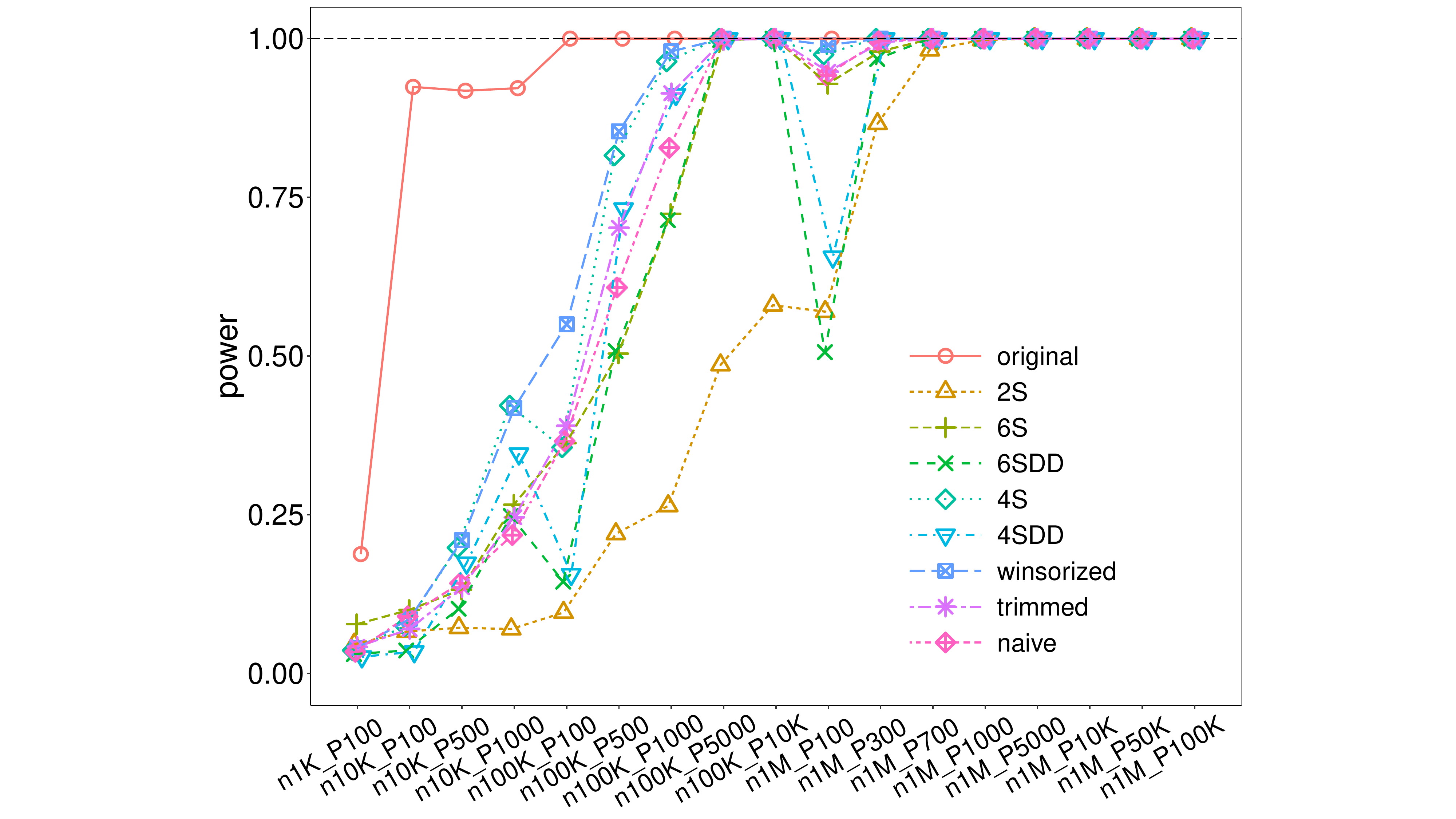}
\includegraphics[width=0.24\textwidth, trim={2.2in 0 2.2in 0},clip] {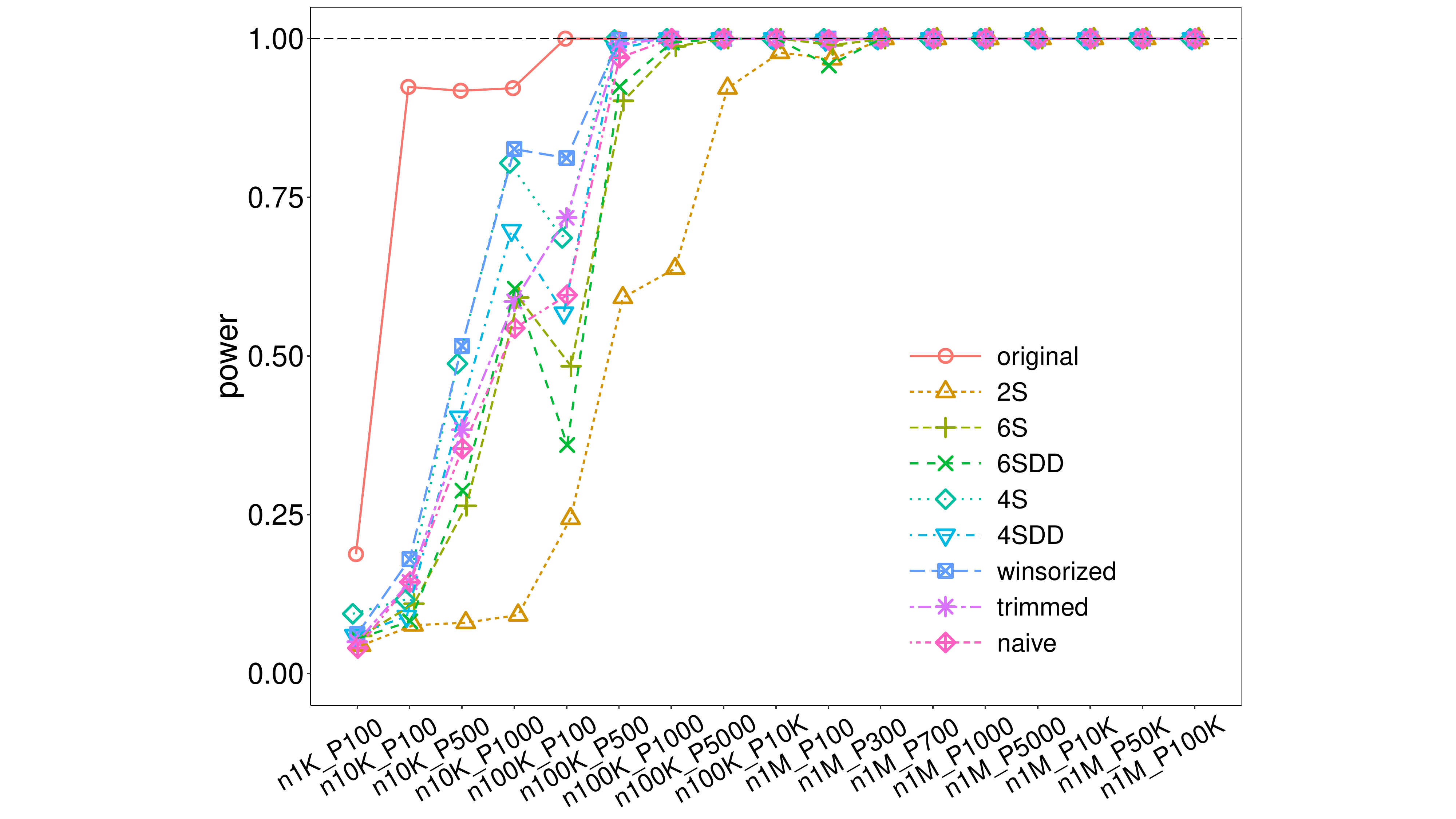}
\includegraphics[width=0.24\textwidth, trim={2.2in 0 2.2in 0},clip] {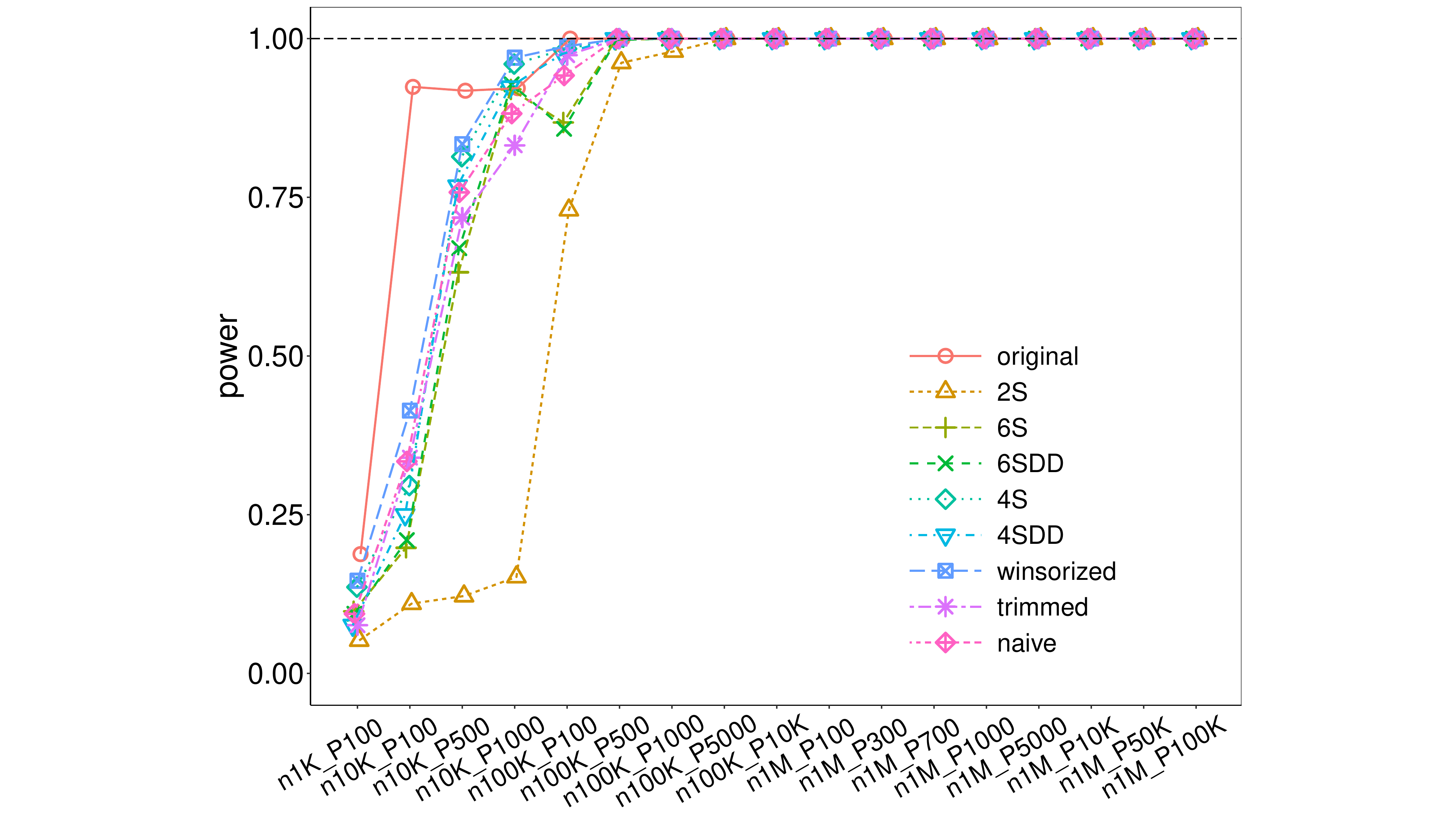}
\includegraphics[width=0.24\textwidth, trim={2.2in 0 2.2in 0},clip] {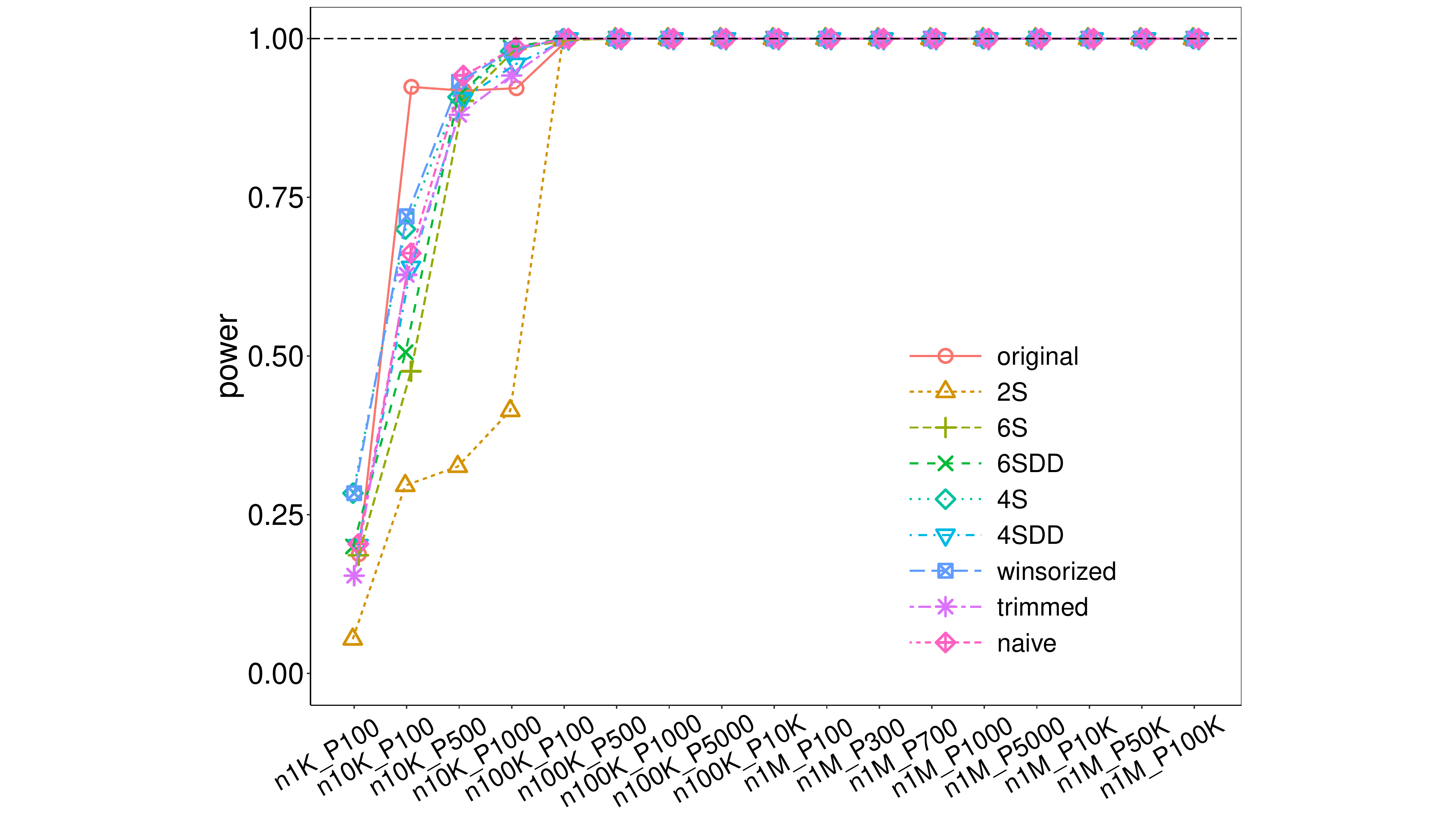}
\includegraphics[width=0.24\textwidth, trim={2.2in 0 2.2in 0},clip] {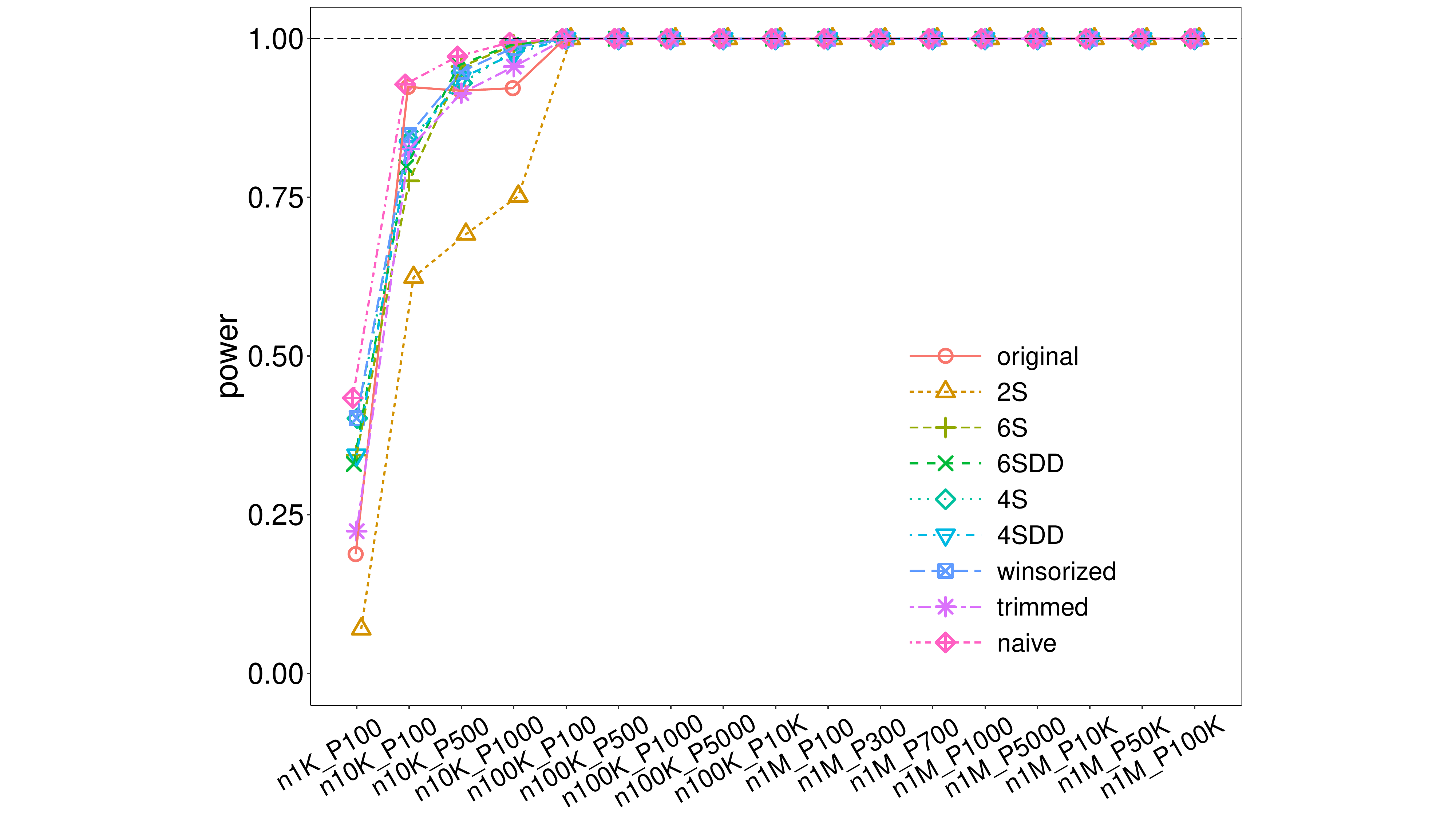}\\
\caption{ZINB data; $\rho$-zCDP; $\theta\ne0$ and $\alpha=\beta$}
\label{fig:1szCDPzinb}
\end{figure}

\end{landscape}

\begin{landscape}
\subsection*{ZINB, $\theta=0$ and $\alpha\ne\beta$}
\begin{figure}[!htb]
\hspace{0.6in}$\epsilon=0.5$\hspace{1.3in}$\epsilon=1$\hspace{1.4in}$\epsilon=2$
\hspace{1.4in}$\epsilon=5$\hspace{1.4in}$\epsilon=50$\\
\includegraphics[width=0.26\textwidth, trim={2.2in 0 2.2in 0},clip] {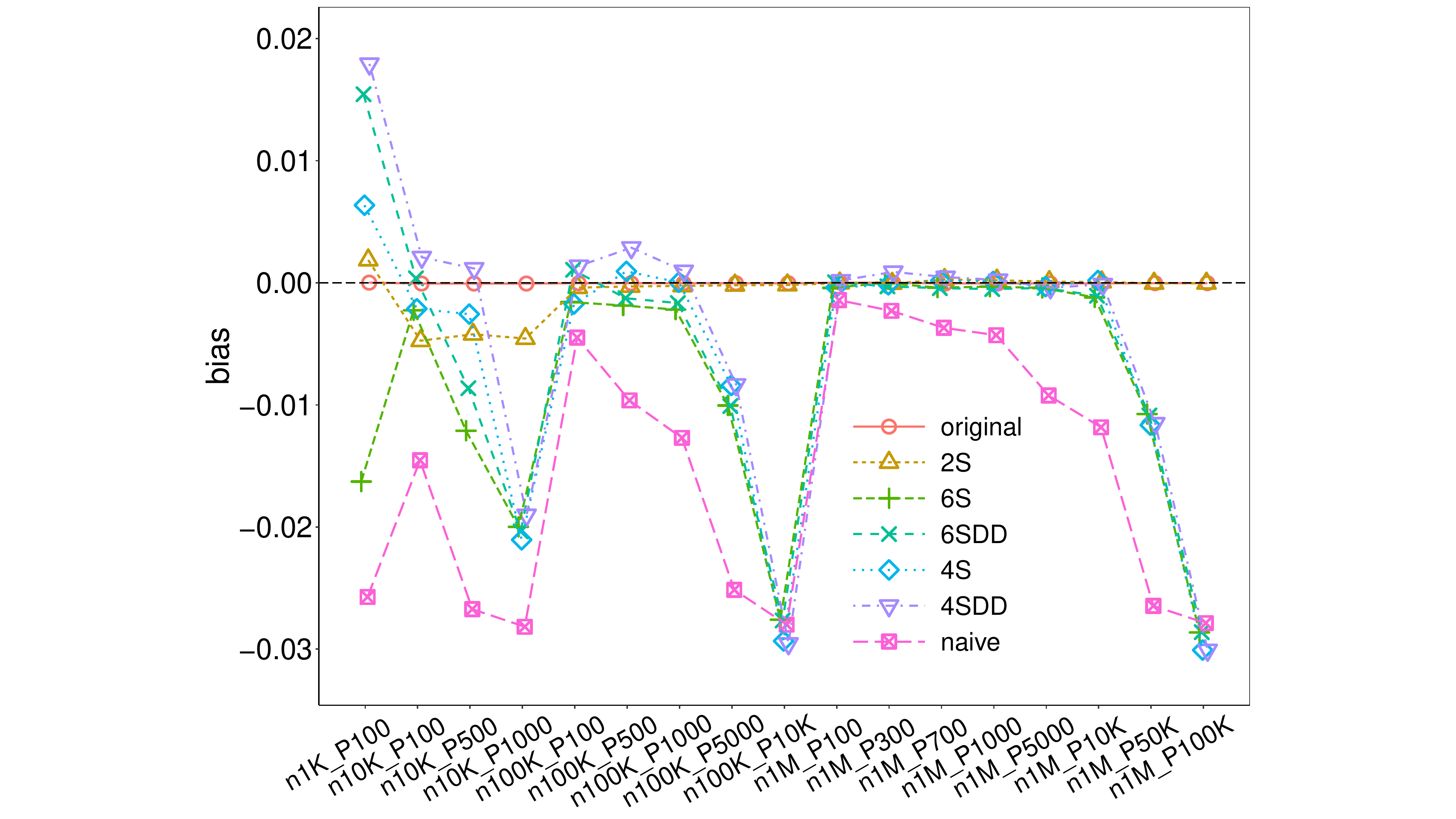}
\includegraphics[width=0.26\textwidth, trim={2.2in 0 2.2in 0},clip] {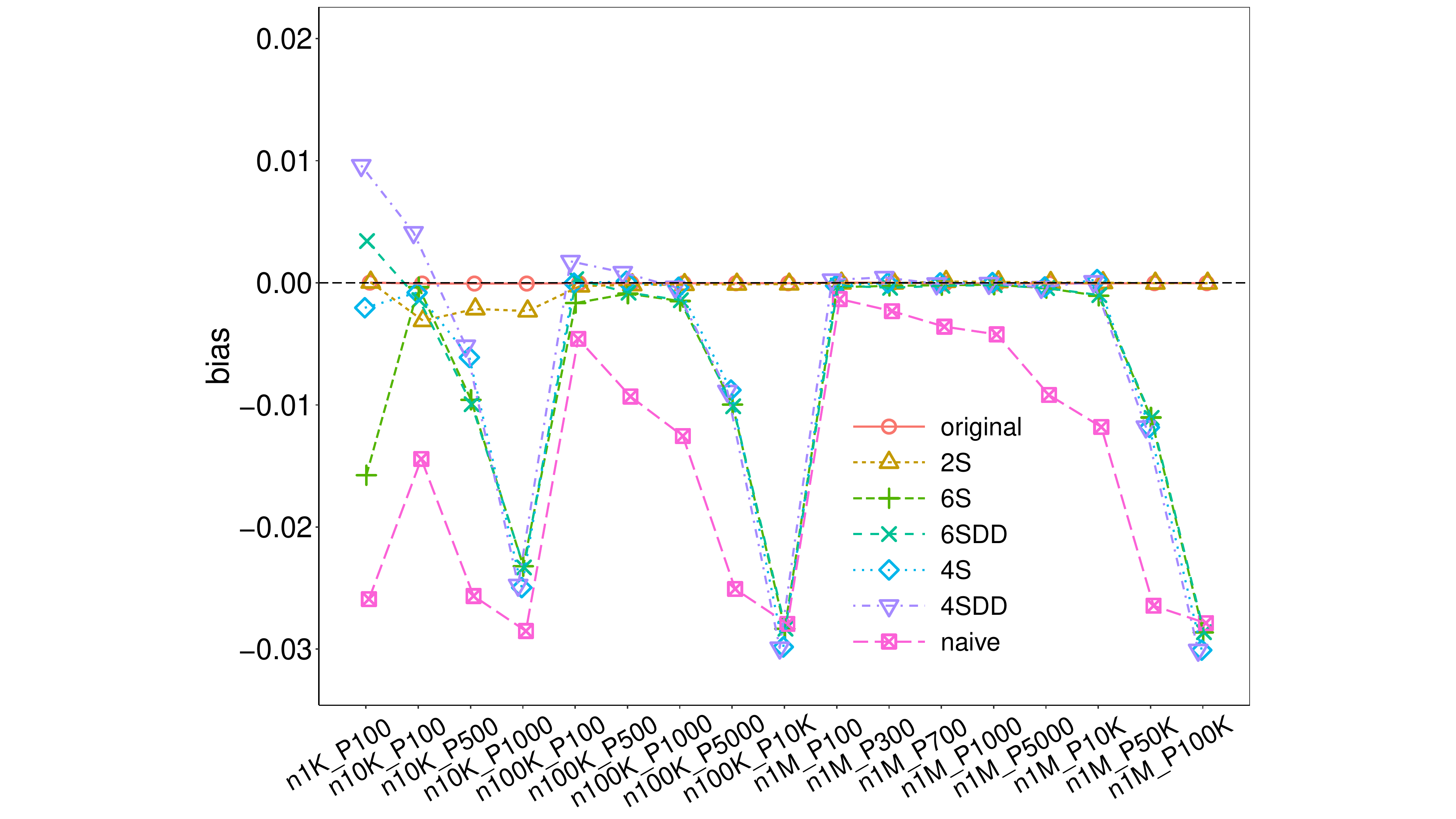}
\includegraphics[width=0.26\textwidth, trim={2.2in 0 2.2in 0},clip] {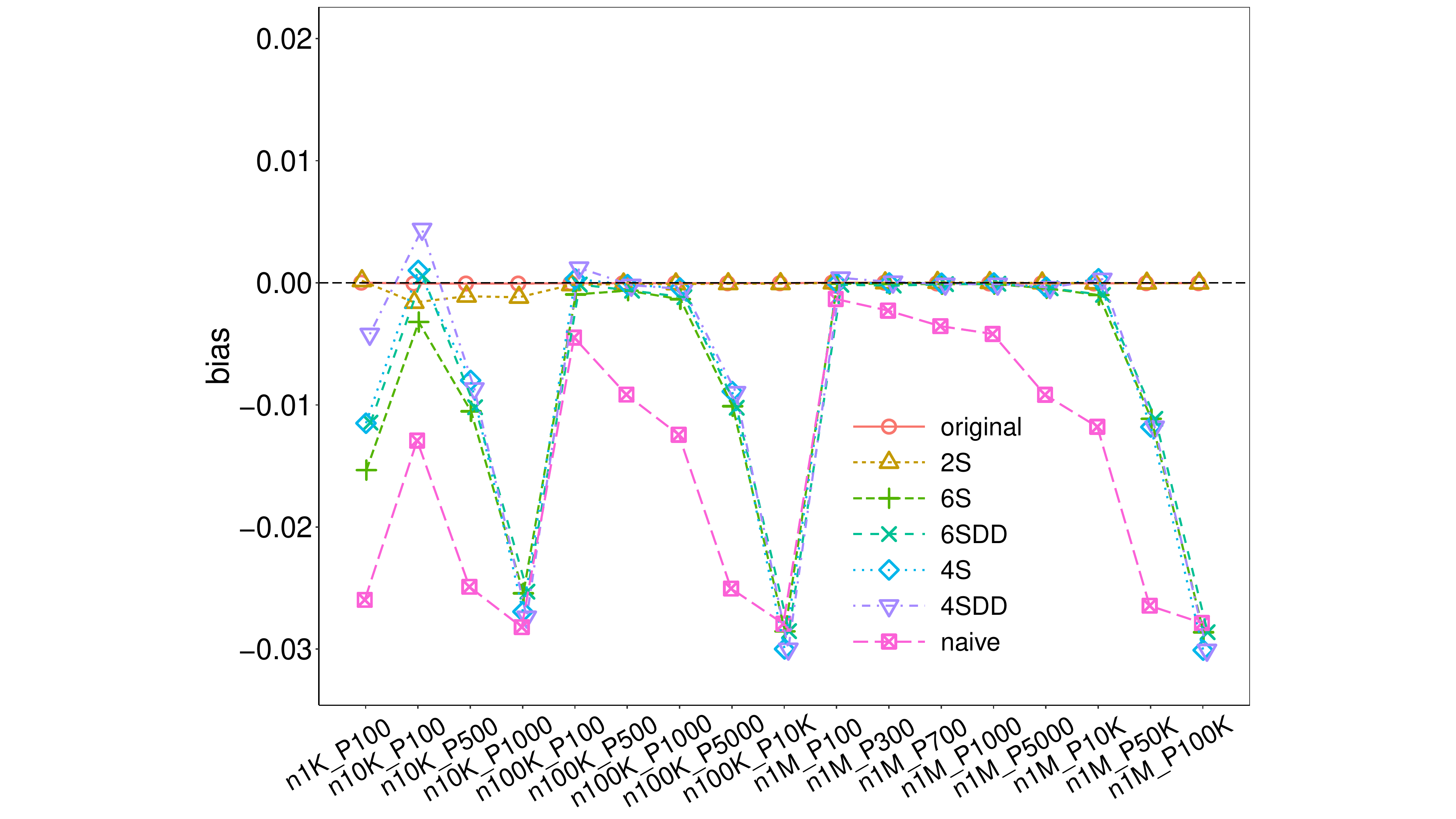}
\includegraphics[width=0.26\textwidth, trim={2.2in 0 2.2in 0},clip] {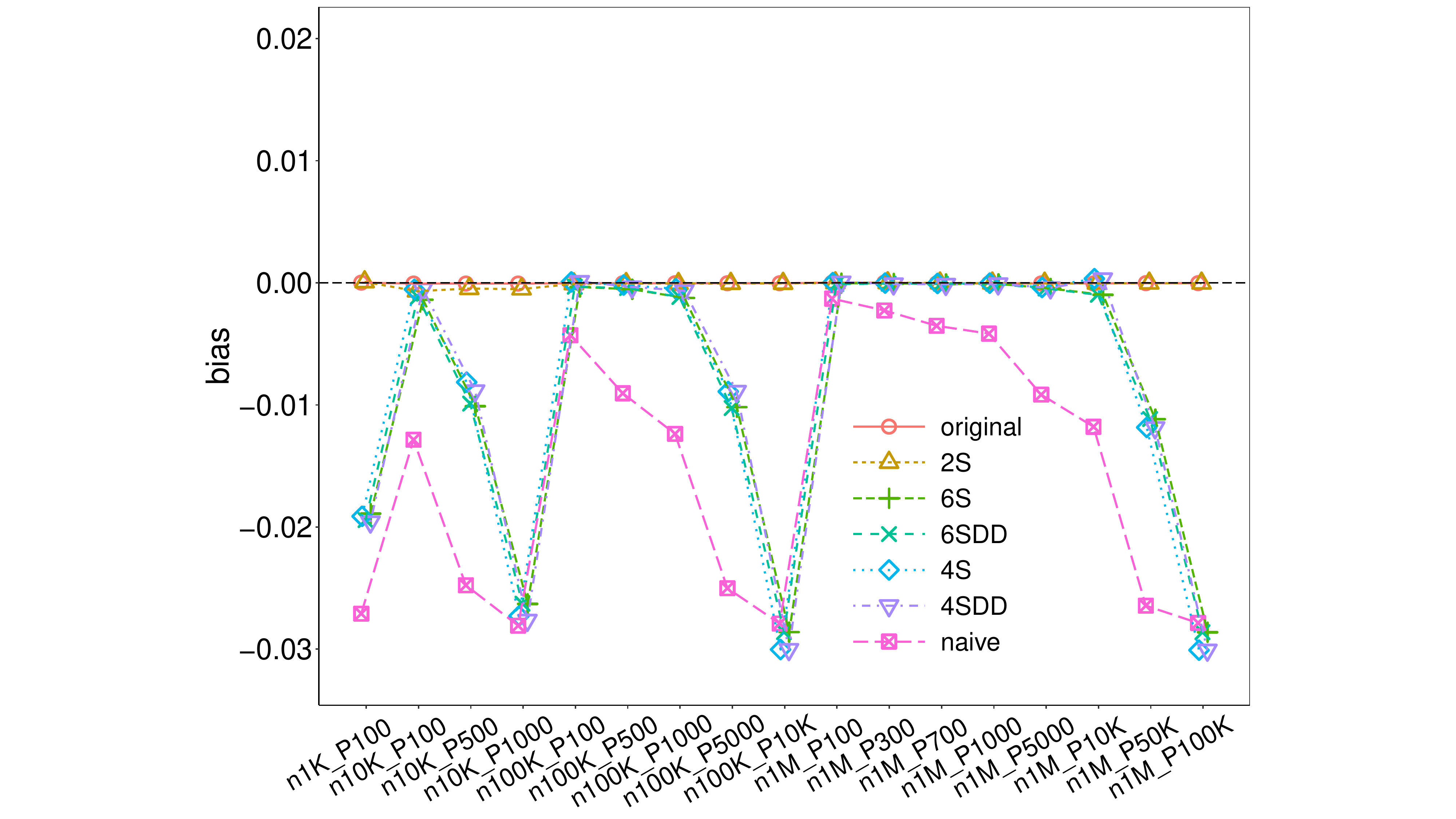}
\includegraphics[width=0.26\textwidth, trim={2.2in 0 2.2in 0},clip] {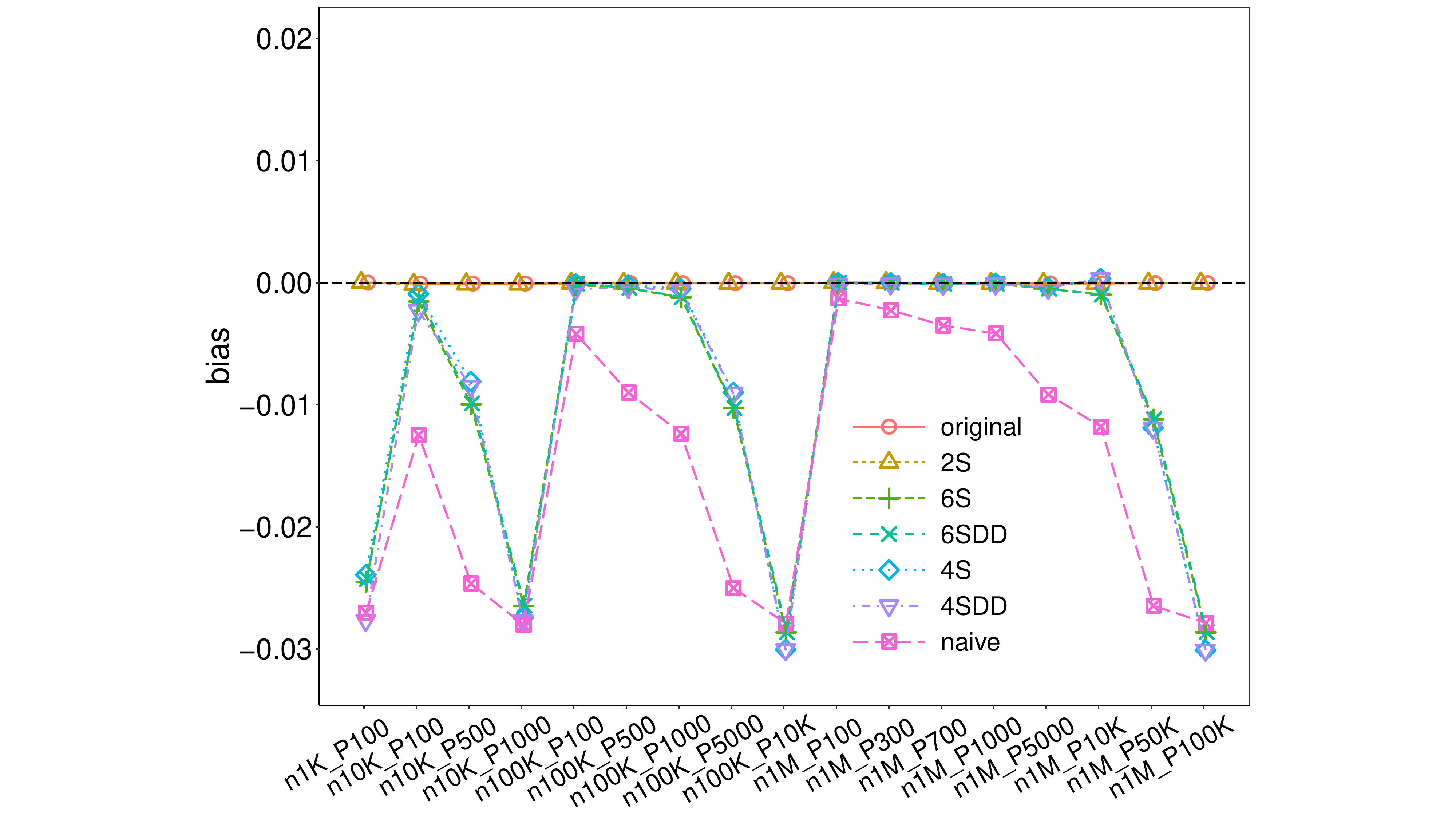}\\
\includegraphics[width=0.26\textwidth, trim={2.2in 0 2.2in 0},clip] {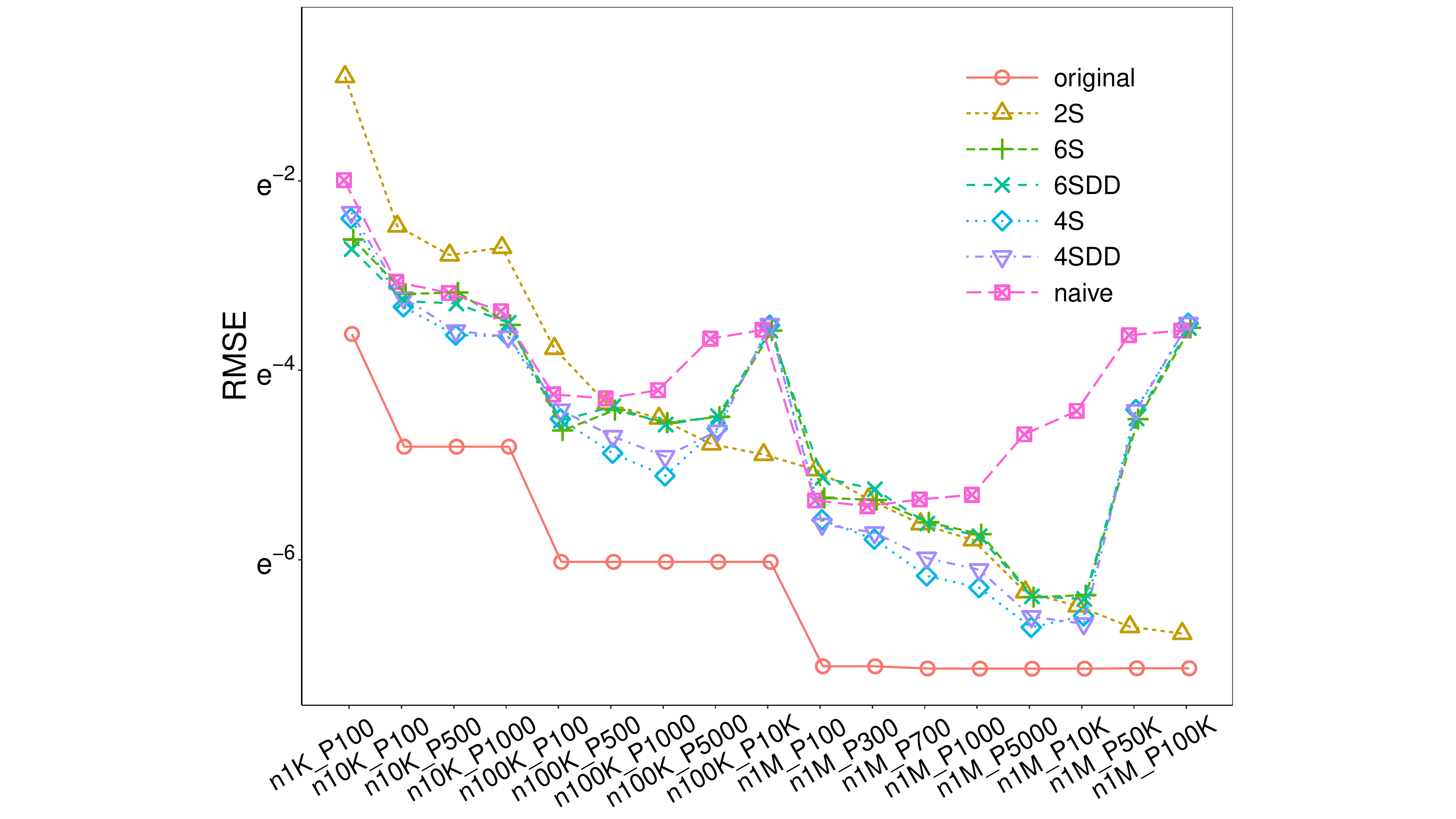}
\includegraphics[width=0.26\textwidth, trim={2.2in 0 2.2in 0},clip] {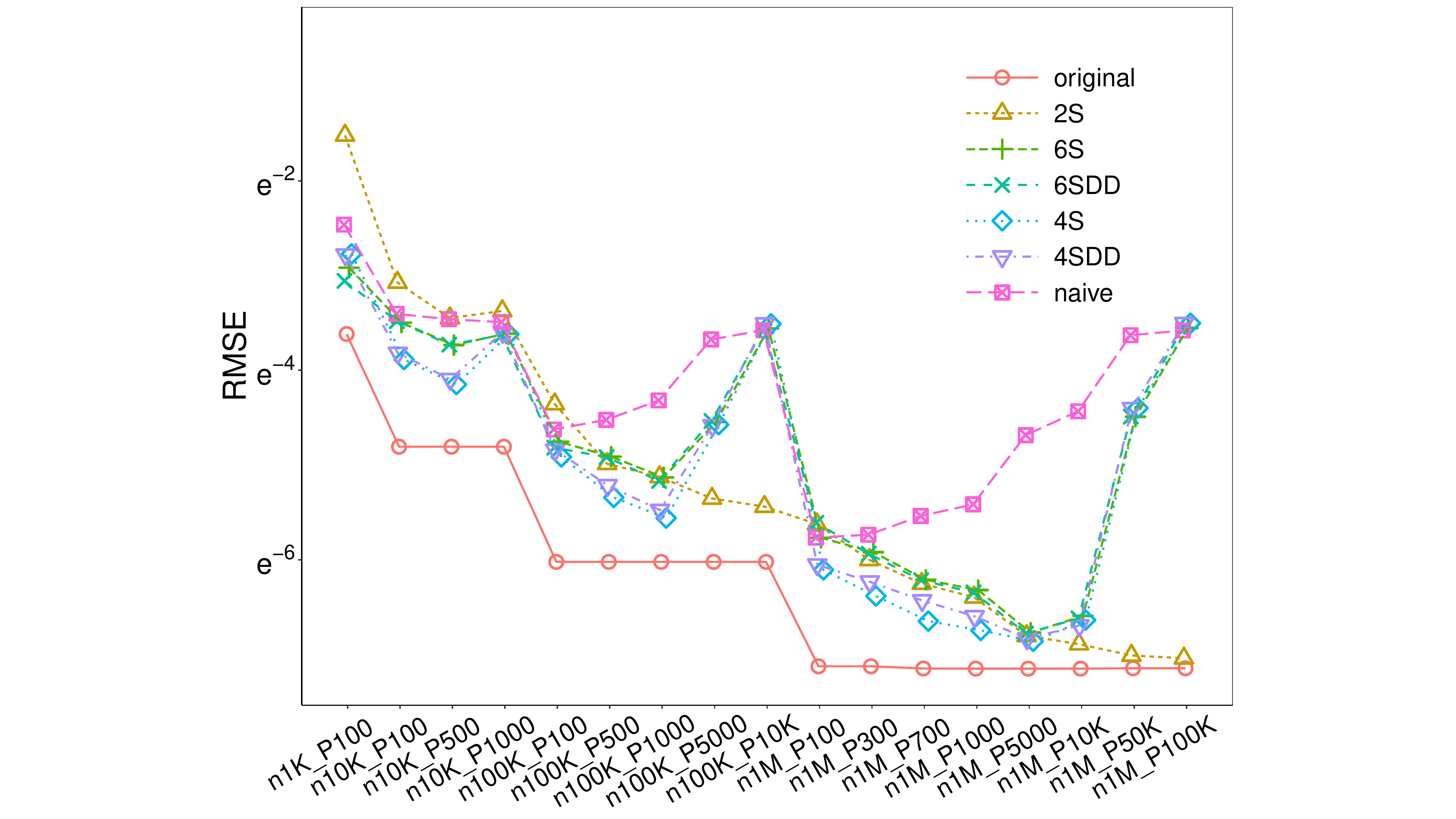}
\includegraphics[width=0.26\textwidth, trim={2.2in 0 2.2in 0},clip] {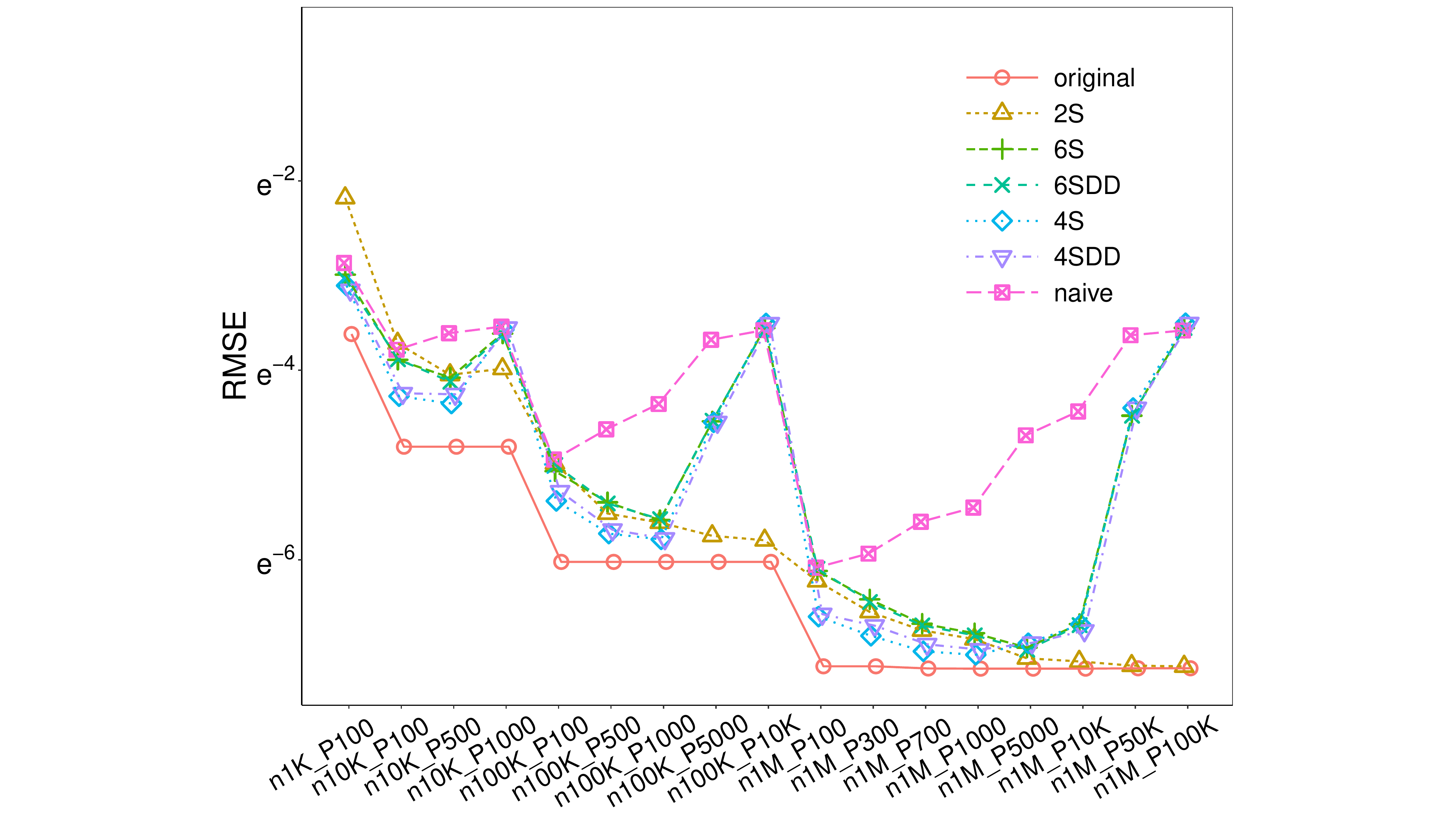}
\includegraphics[width=0.26\textwidth, trim={2.2in 0 2.2in 0},clip] {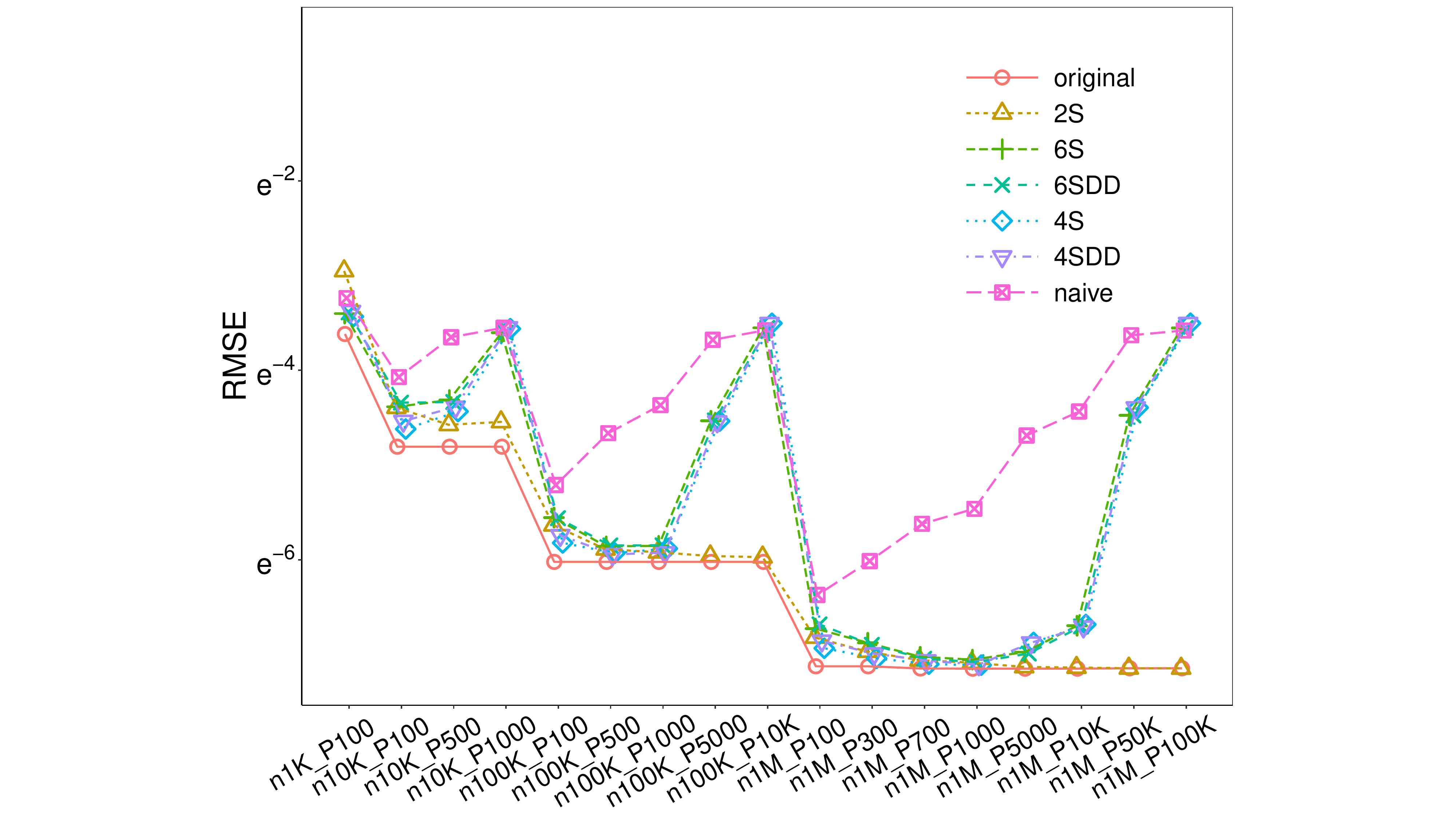}
\includegraphics[width=0.26\textwidth, trim={2.2in 0 2.2in 0},clip] {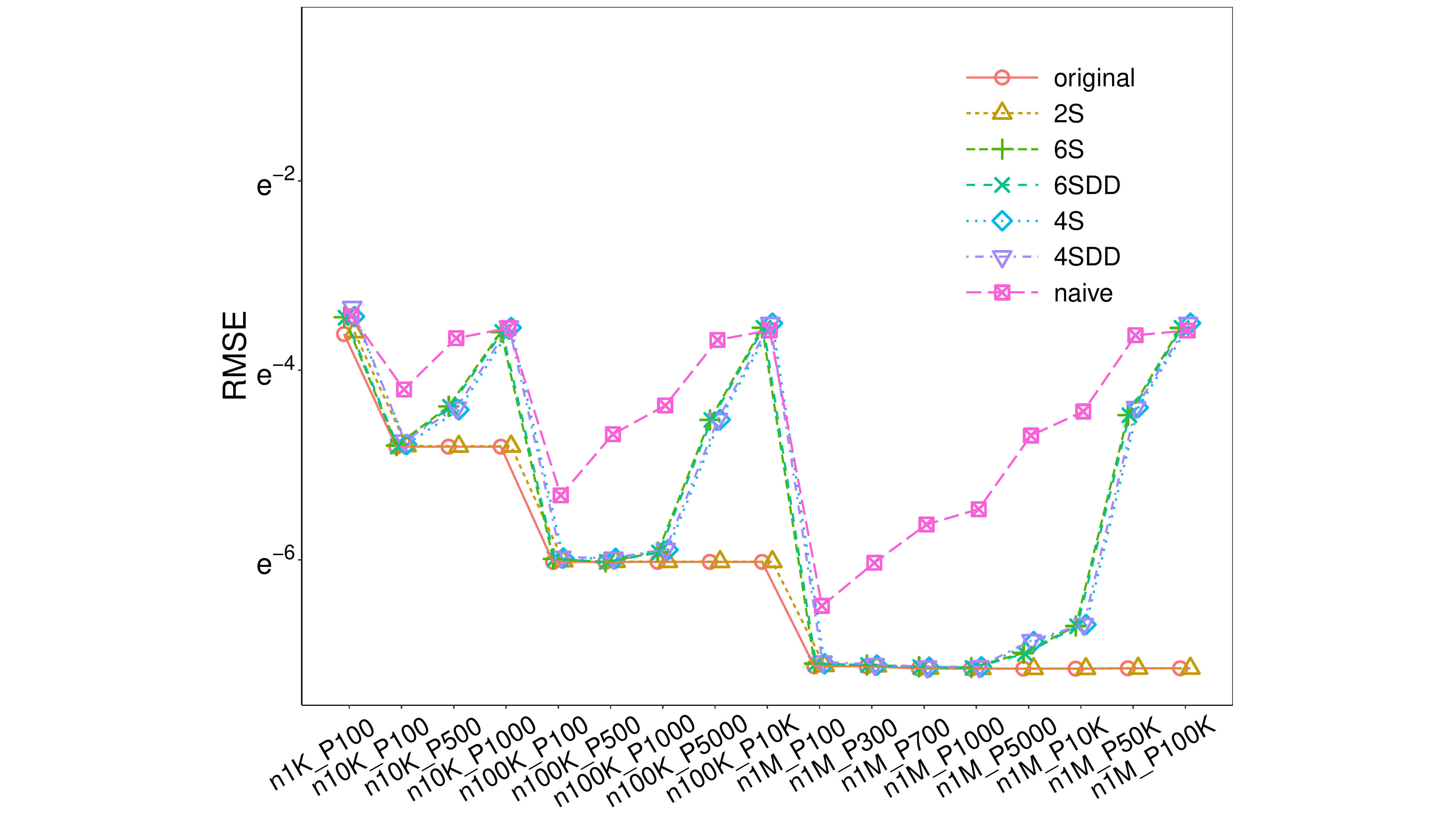}\\
\includegraphics[width=0.26\textwidth, trim={2.2in 0 2.2in 0},clip] {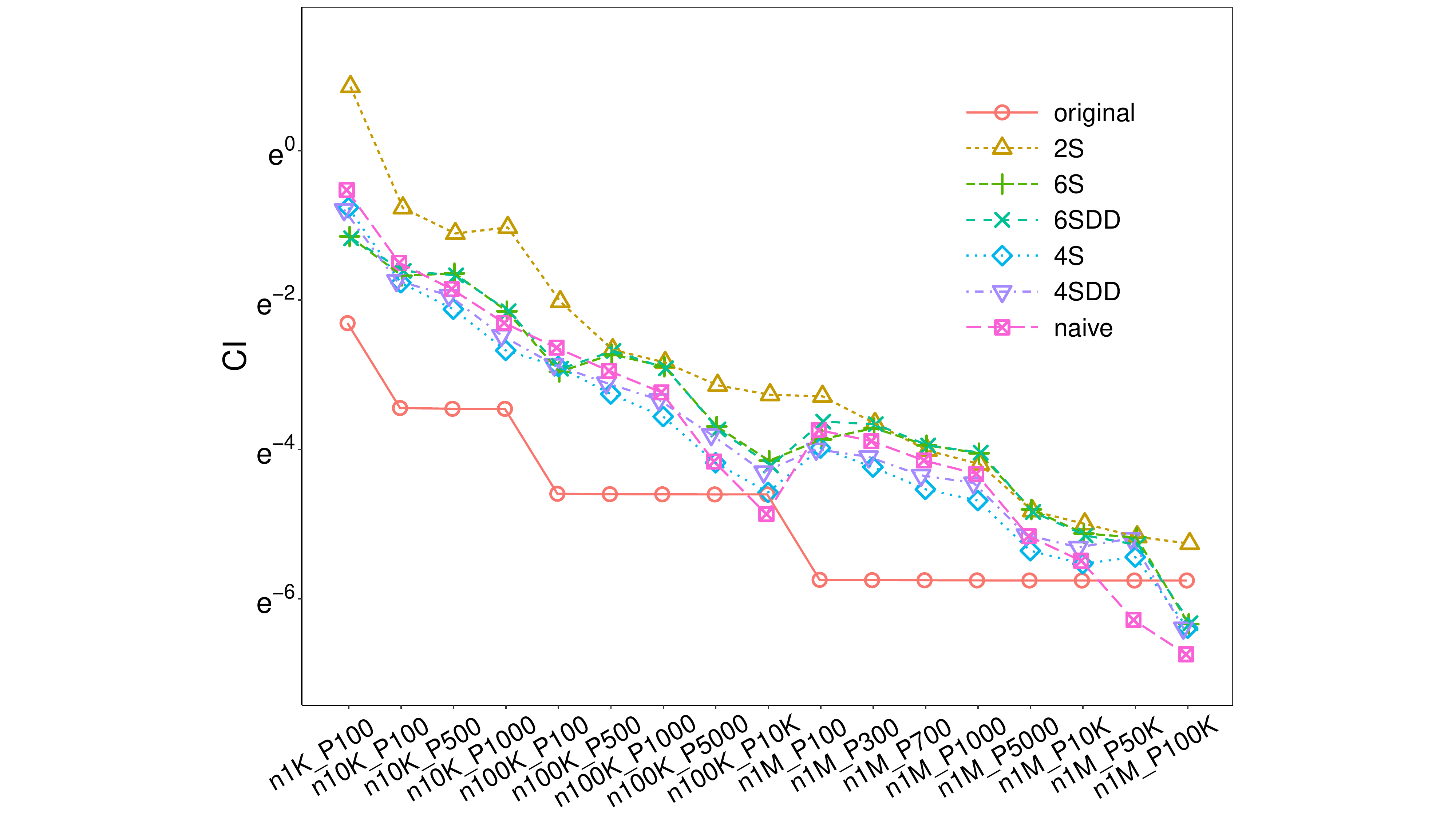}
\includegraphics[width=0.26\textwidth, trim={2.2in 0 2.2in 0},clip] {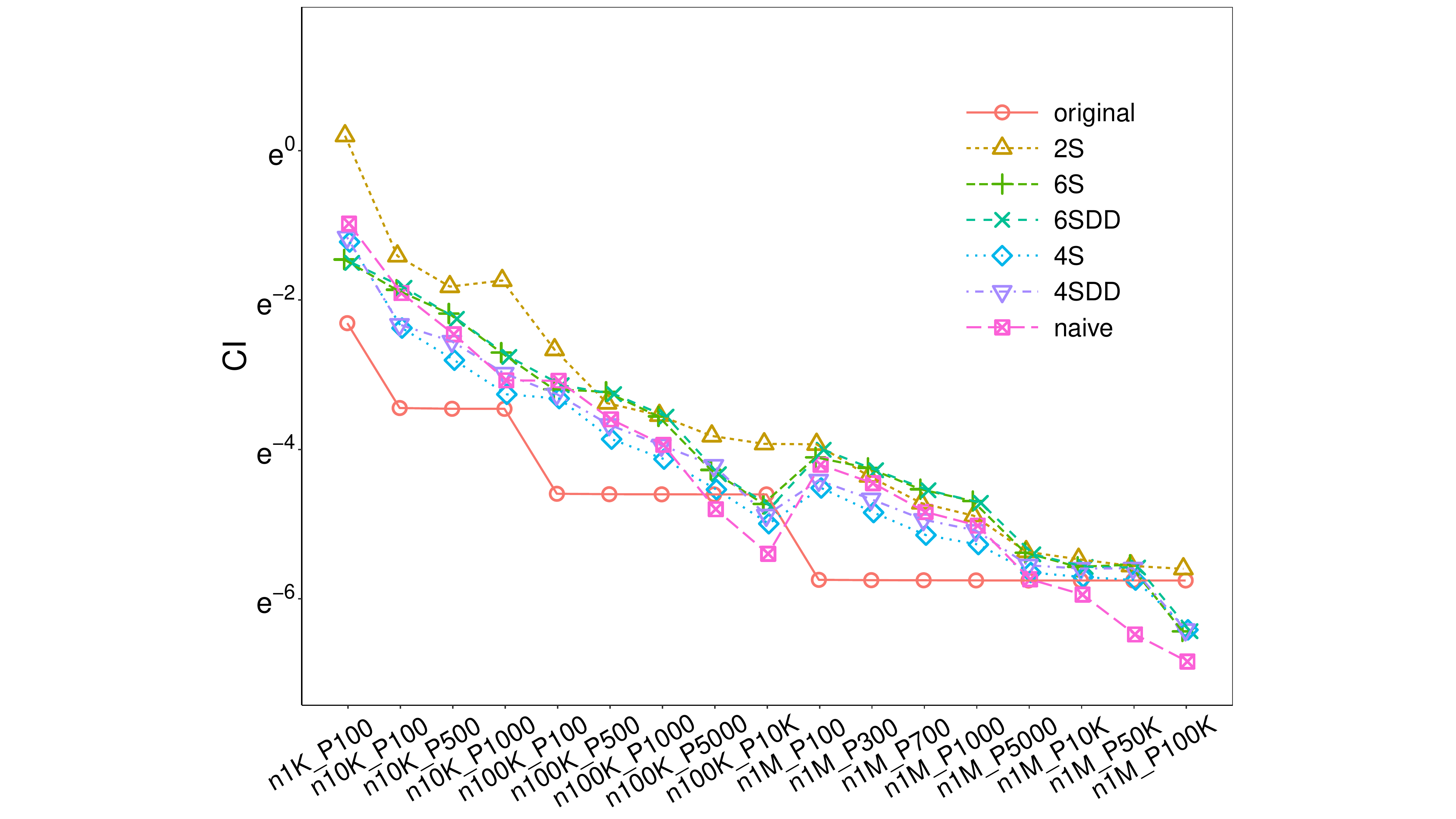}
\includegraphics[width=0.26\textwidth, trim={2.2in 0 2.2in 0},clip] {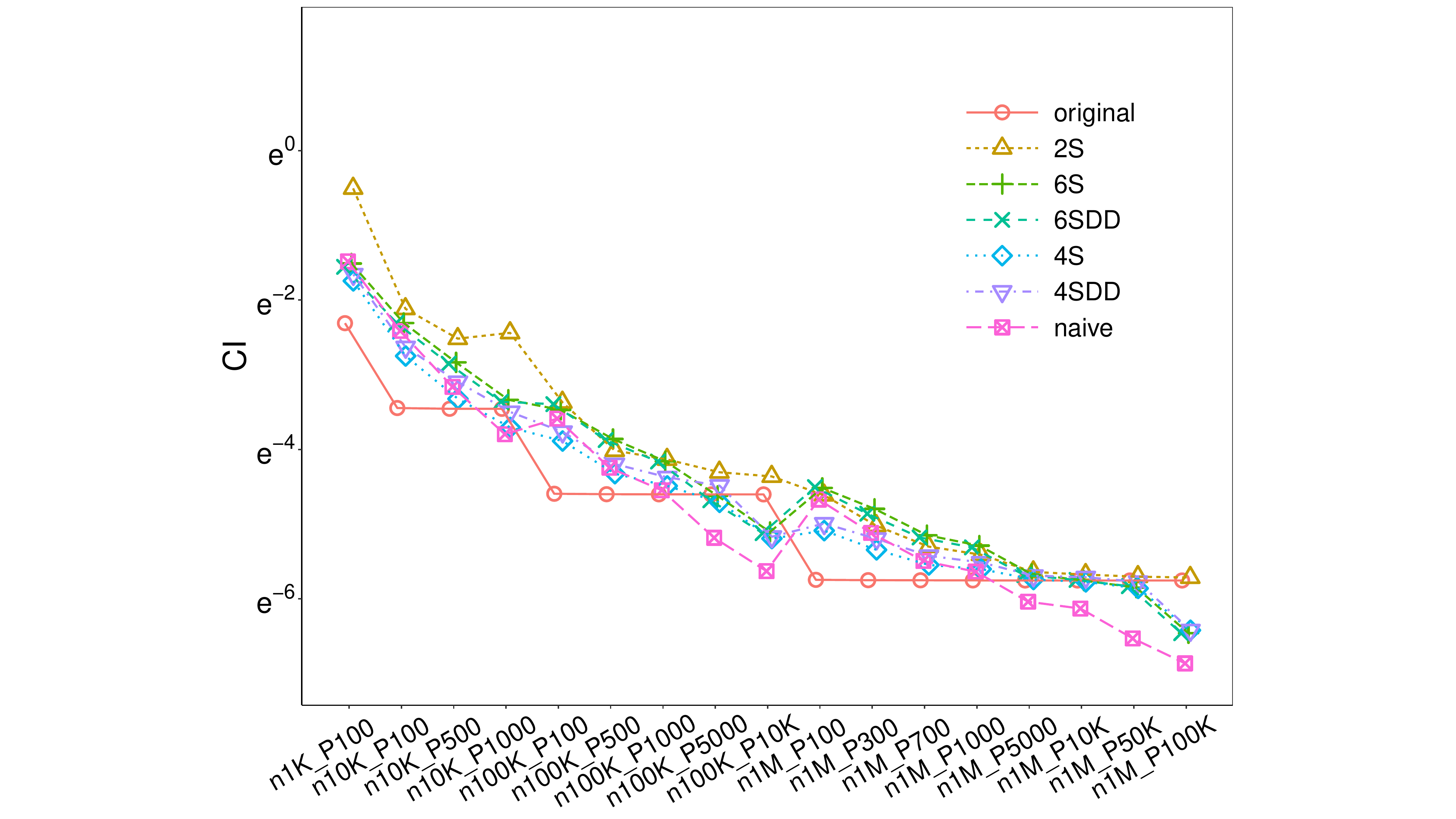}
\includegraphics[width=0.26\textwidth, trim={2.2in 0 2.2in 0},clip] {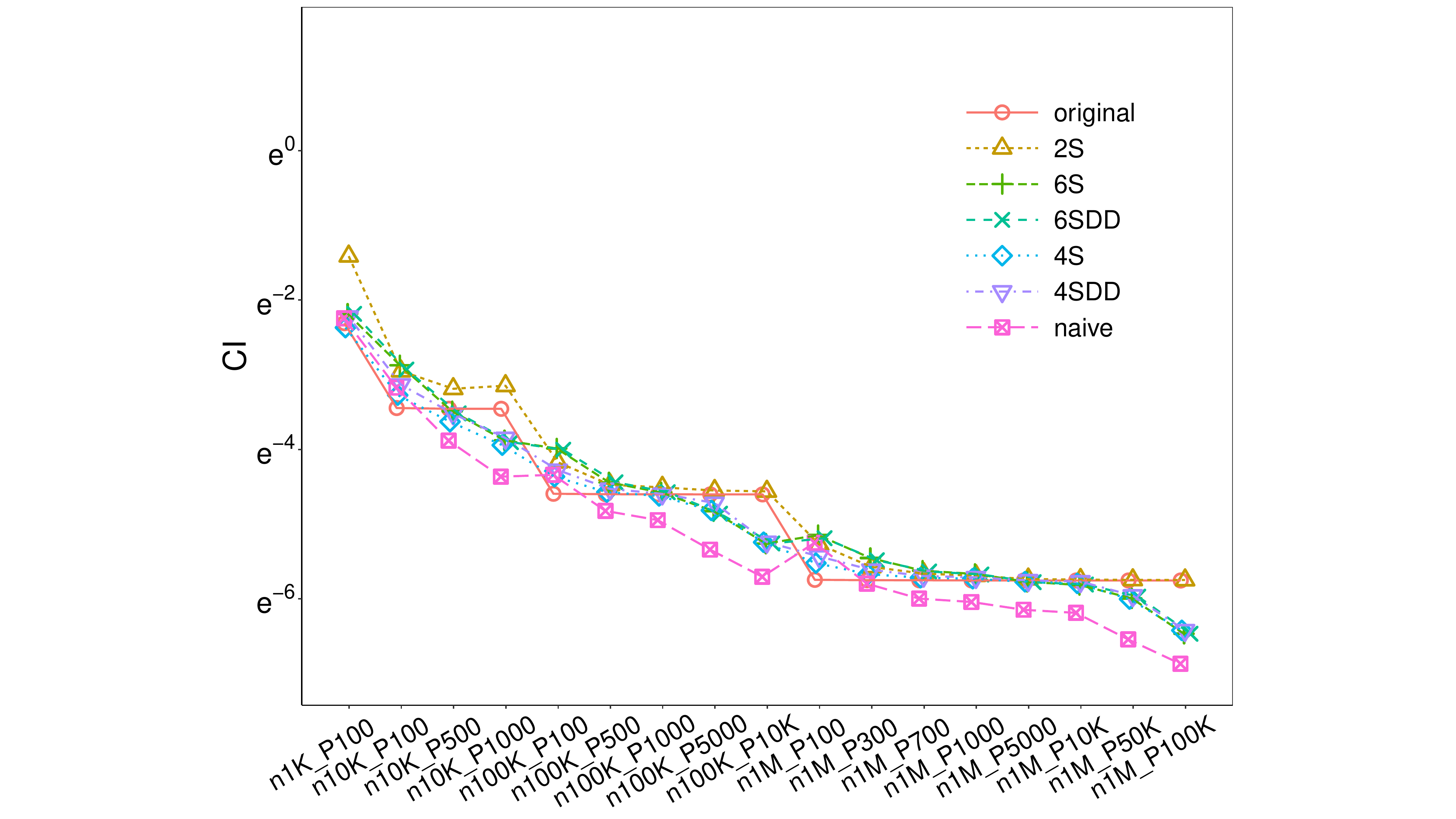}
\includegraphics[width=0.26\textwidth, trim={2.2in 0 2.2in 0},clip] {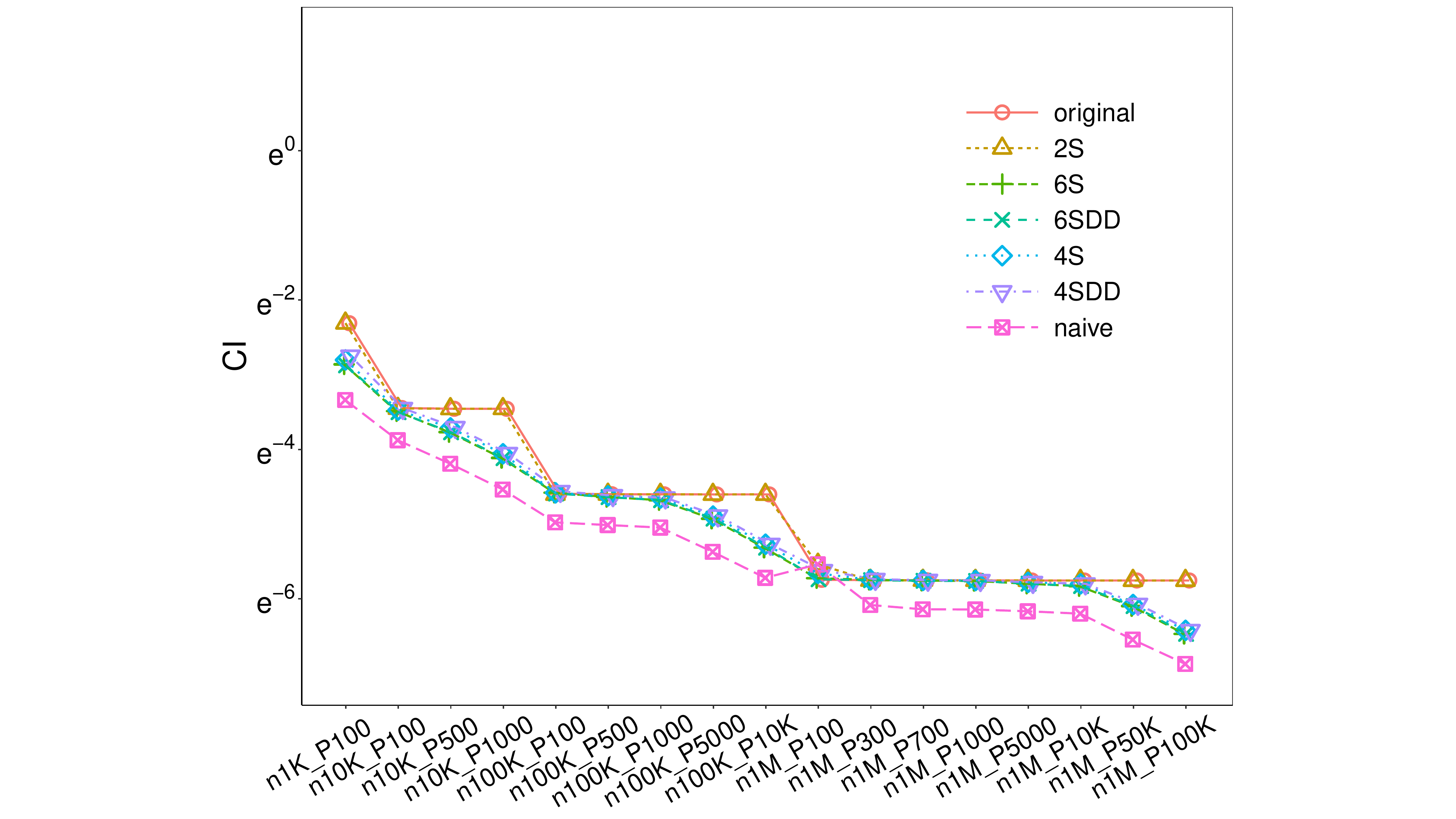}\\
\includegraphics[width=0.26\textwidth, trim={2.2in 0 2.2in 0},clip] {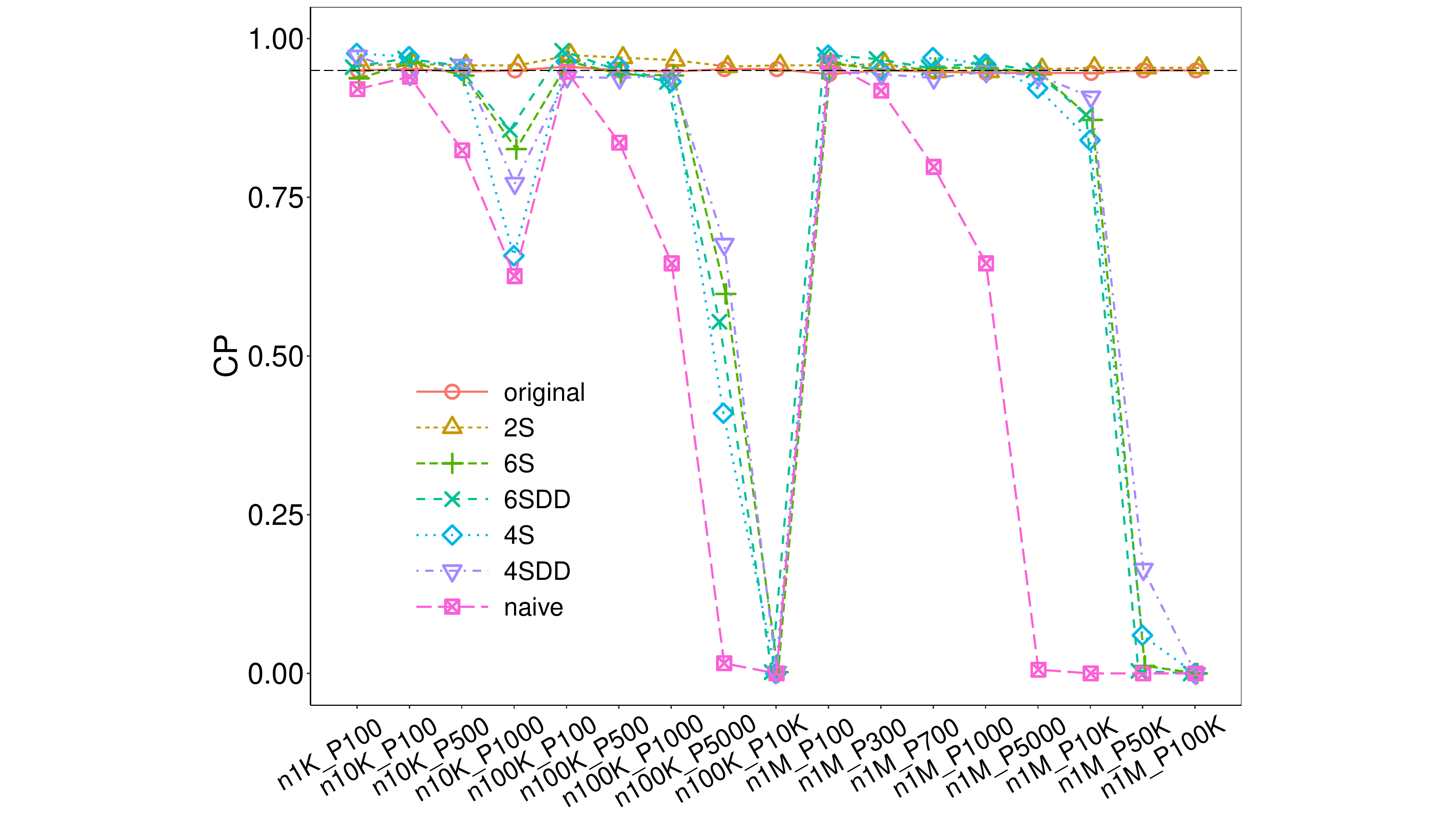}
\includegraphics[width=0.26\textwidth, trim={2.2in 0 2.2in 0},clip] {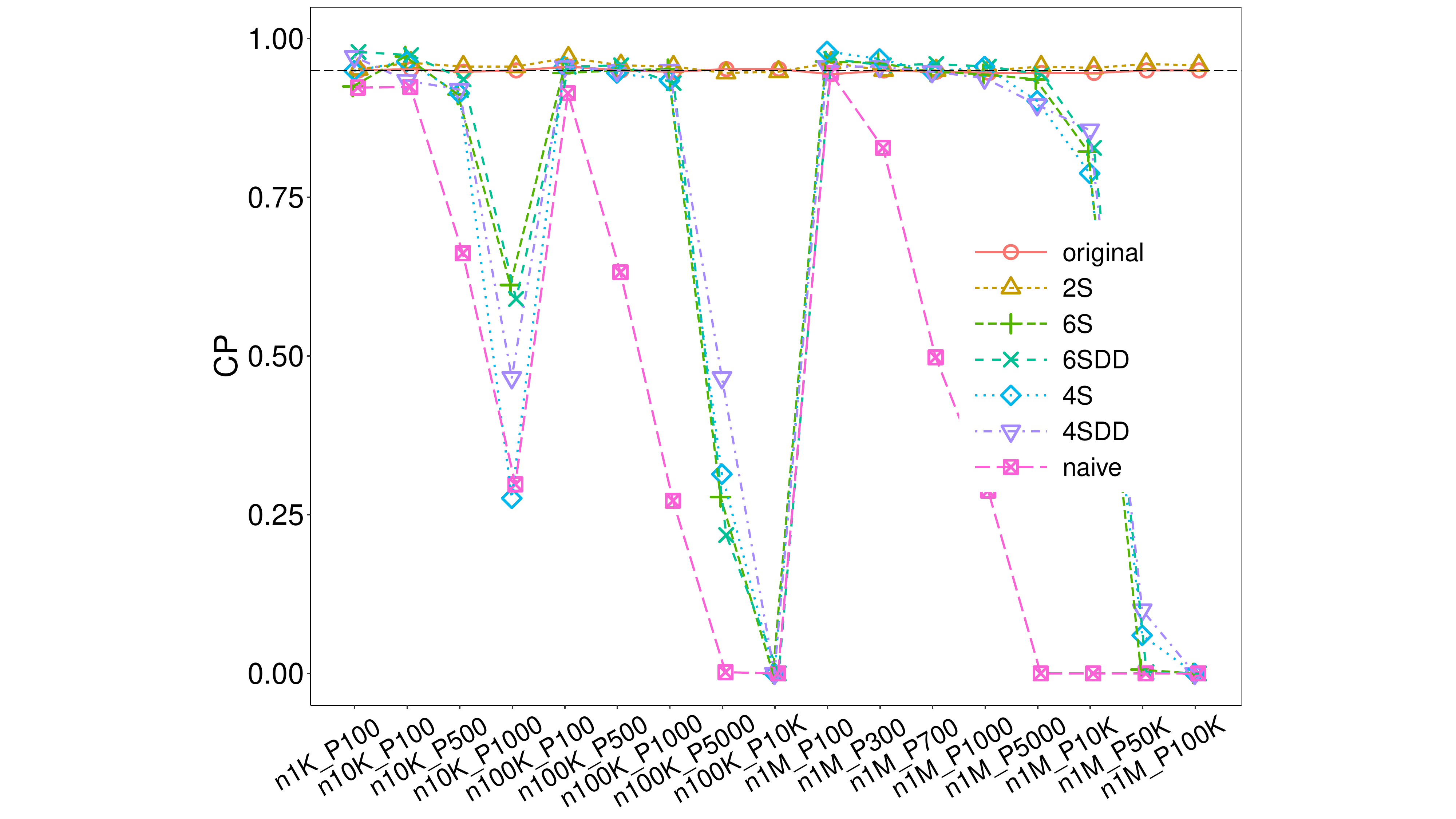}
\includegraphics[width=0.26\textwidth, trim={2.2in 0 2.2in 0},clip] {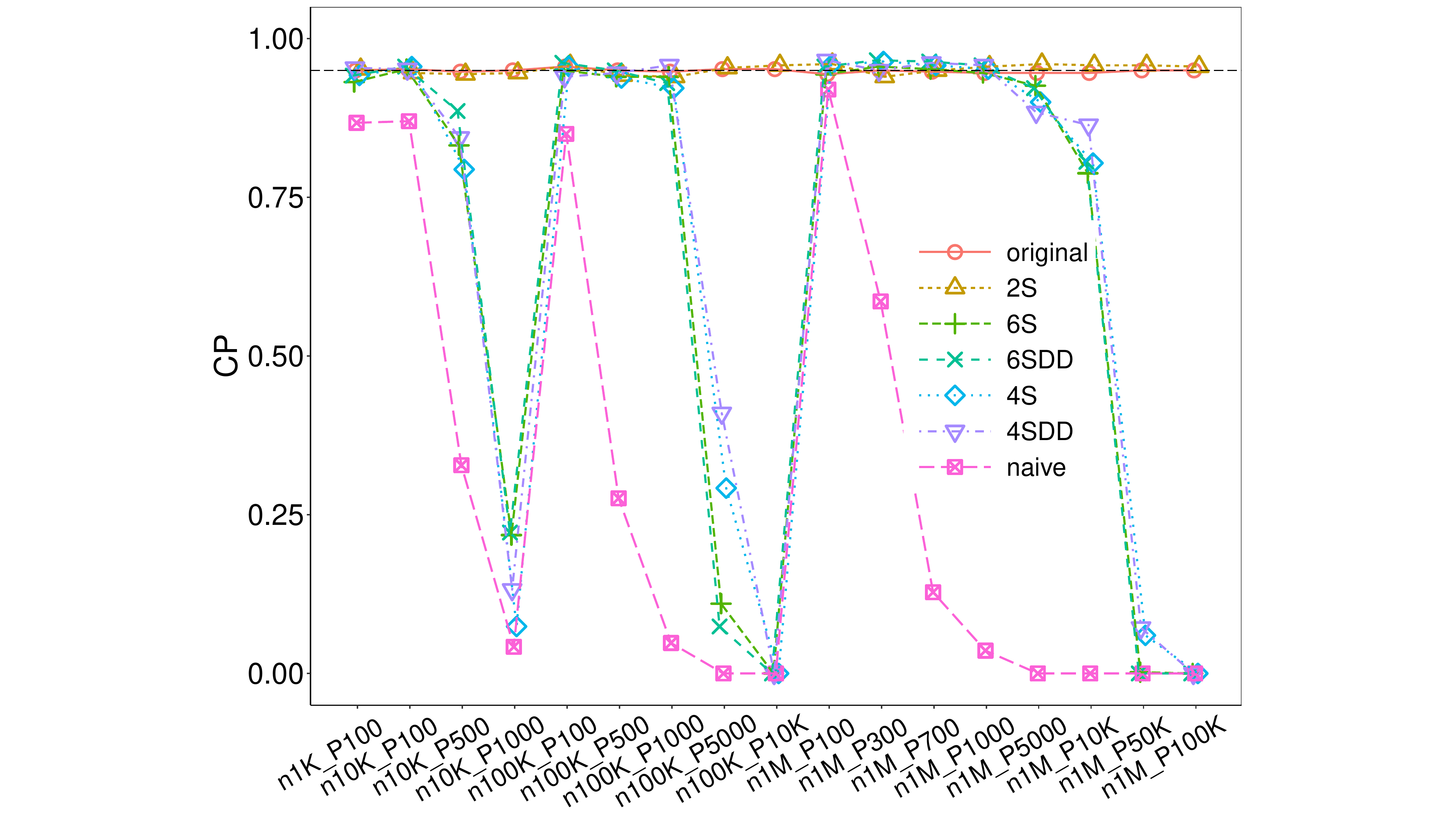}
\includegraphics[width=0.26\textwidth, trim={2.2in 0 2.2in 0},clip] {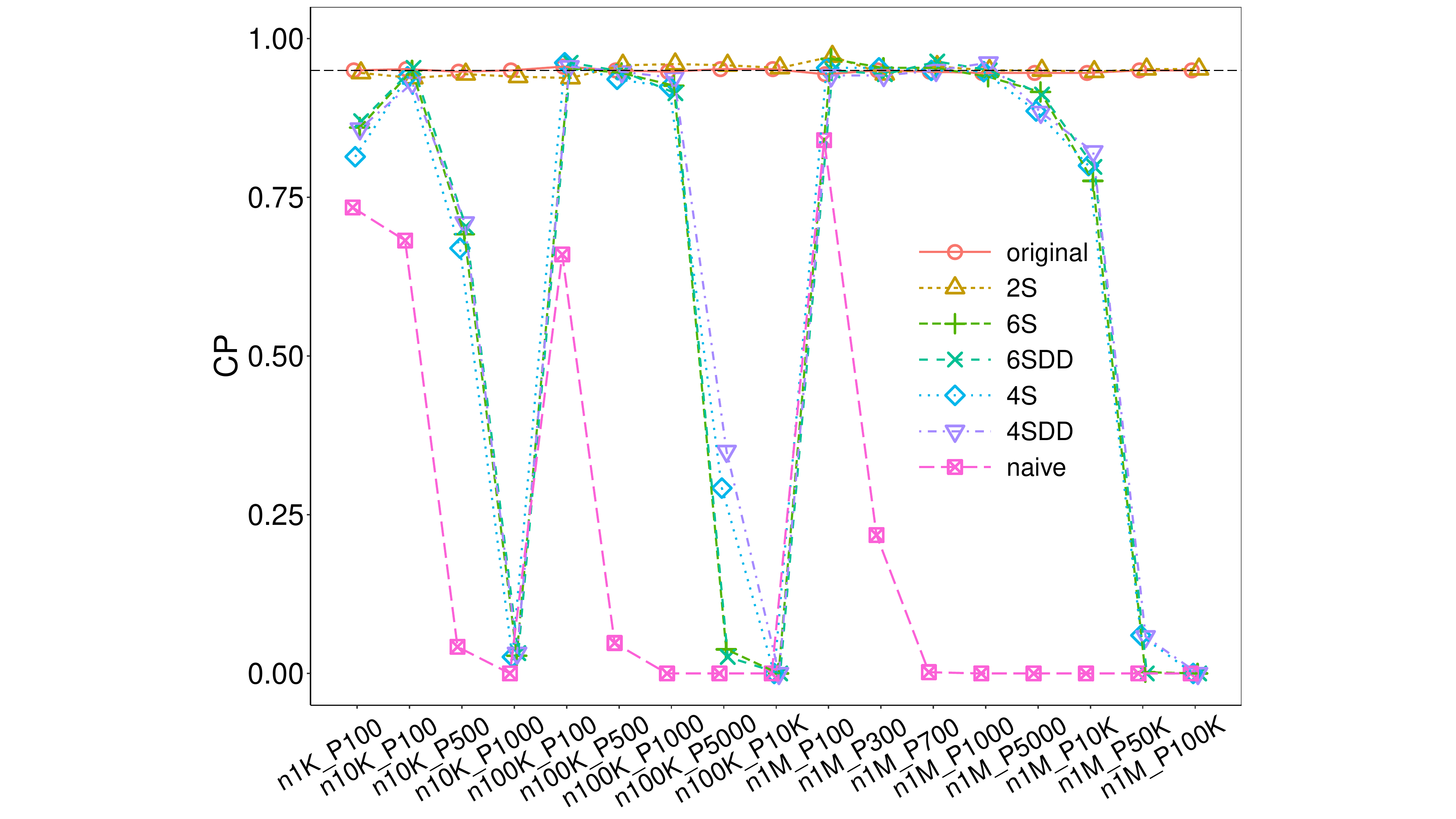}
\includegraphics[width=0.26\textwidth, trim={2.2in 0 2.2in 0},clip] {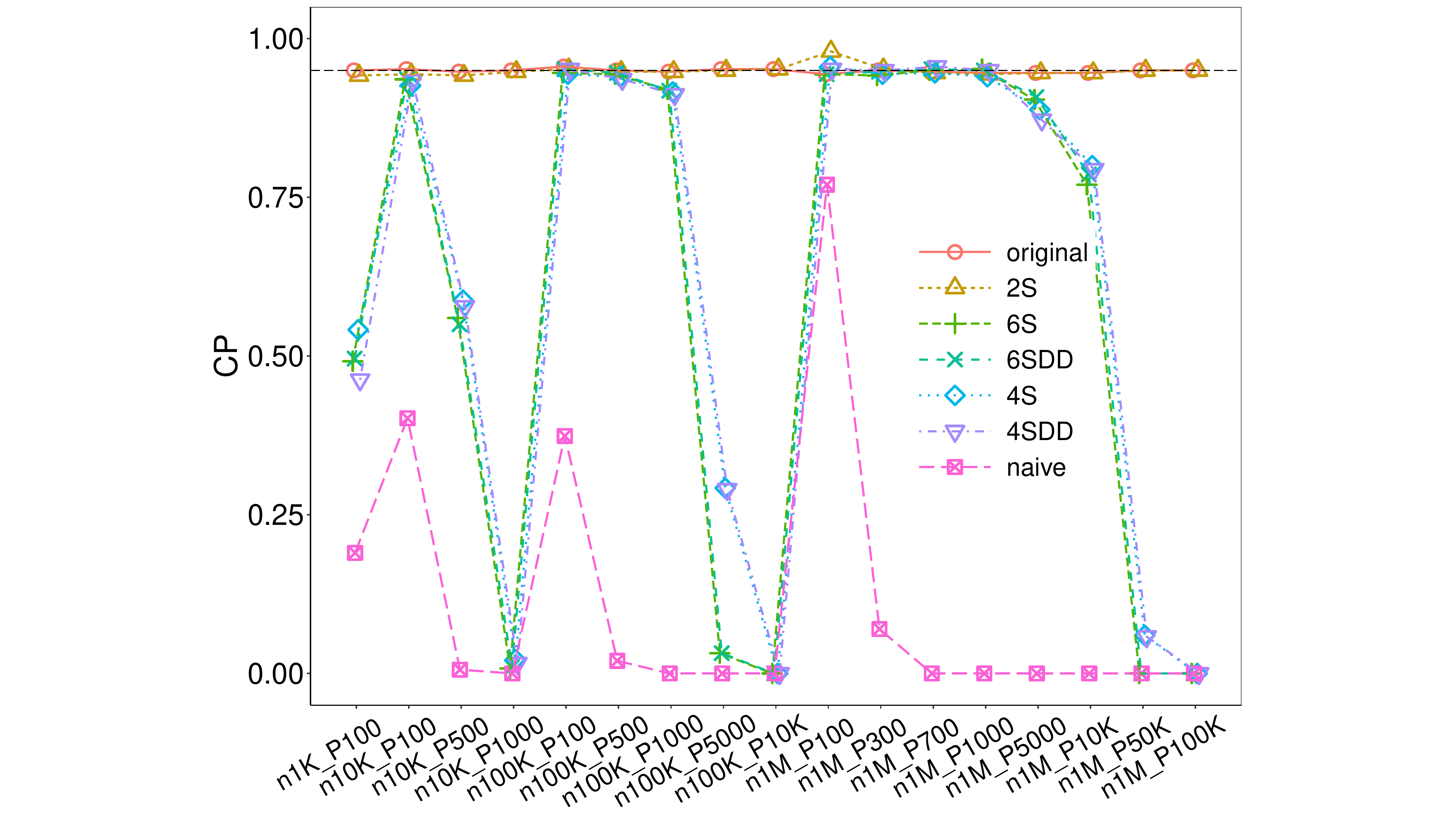}\\
\caption{ZINB data; $\epsilon$-DP; $\theta=0$ and $\alpha\ne\beta$} \label{fig:0asDPzinb}
\end{figure}
\end{landscape}

\begin{landscape}
\begin{figure}[!htb]
\hspace{0.6in}$\rho=0.005$\hspace{1in}$\rho=0.02$\hspace{1.2in}$\rho=0.08$
\hspace{1.1in}$\rho=0.32$\hspace{1.2in}$\rho=1.28$\\
\includegraphics[width=0.26\textwidth, trim={2.2in 0 2.2in 0},clip] {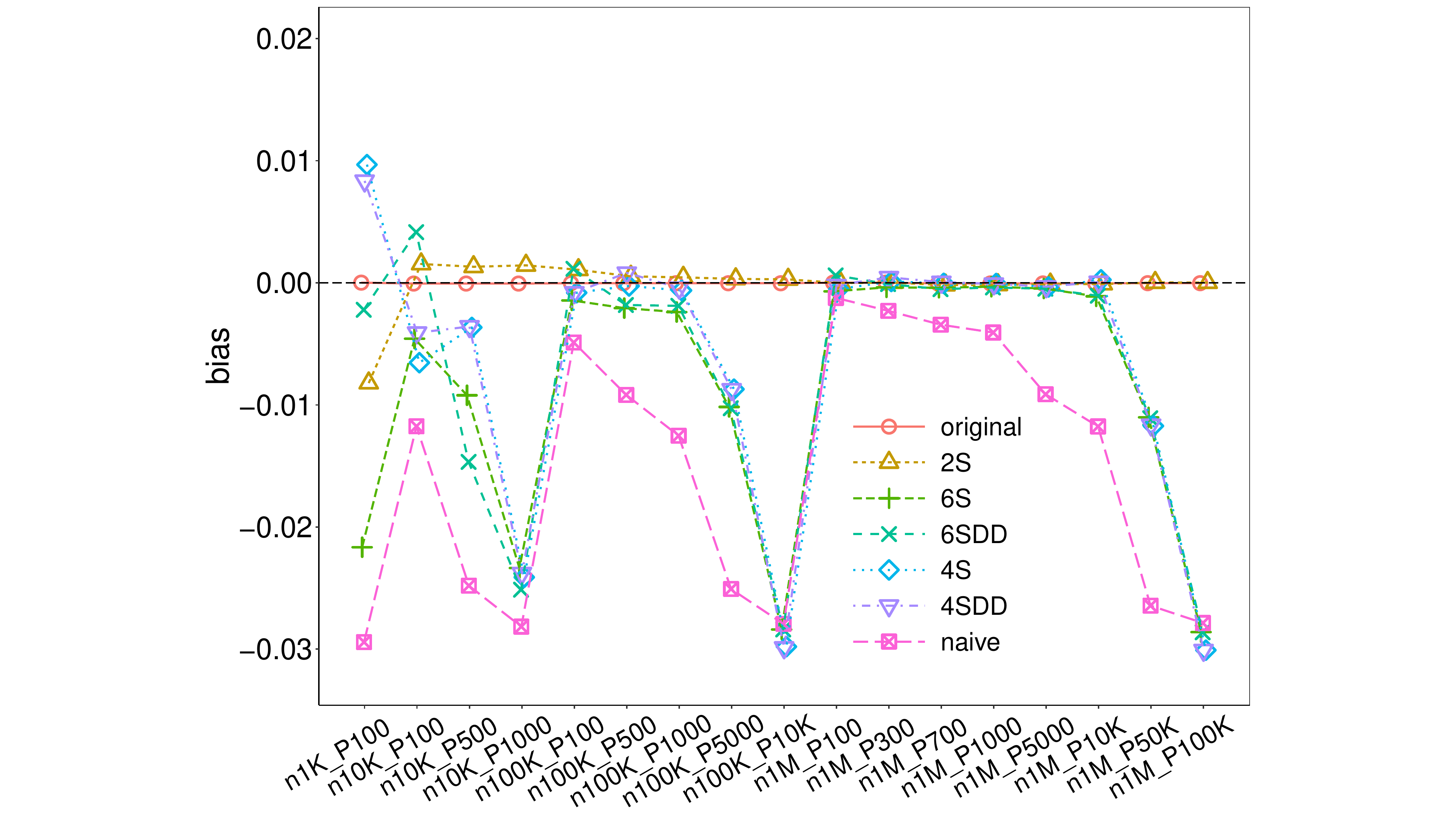}
\includegraphics[width=0.26\textwidth, trim={2.2in 0 2.2in 0},clip] {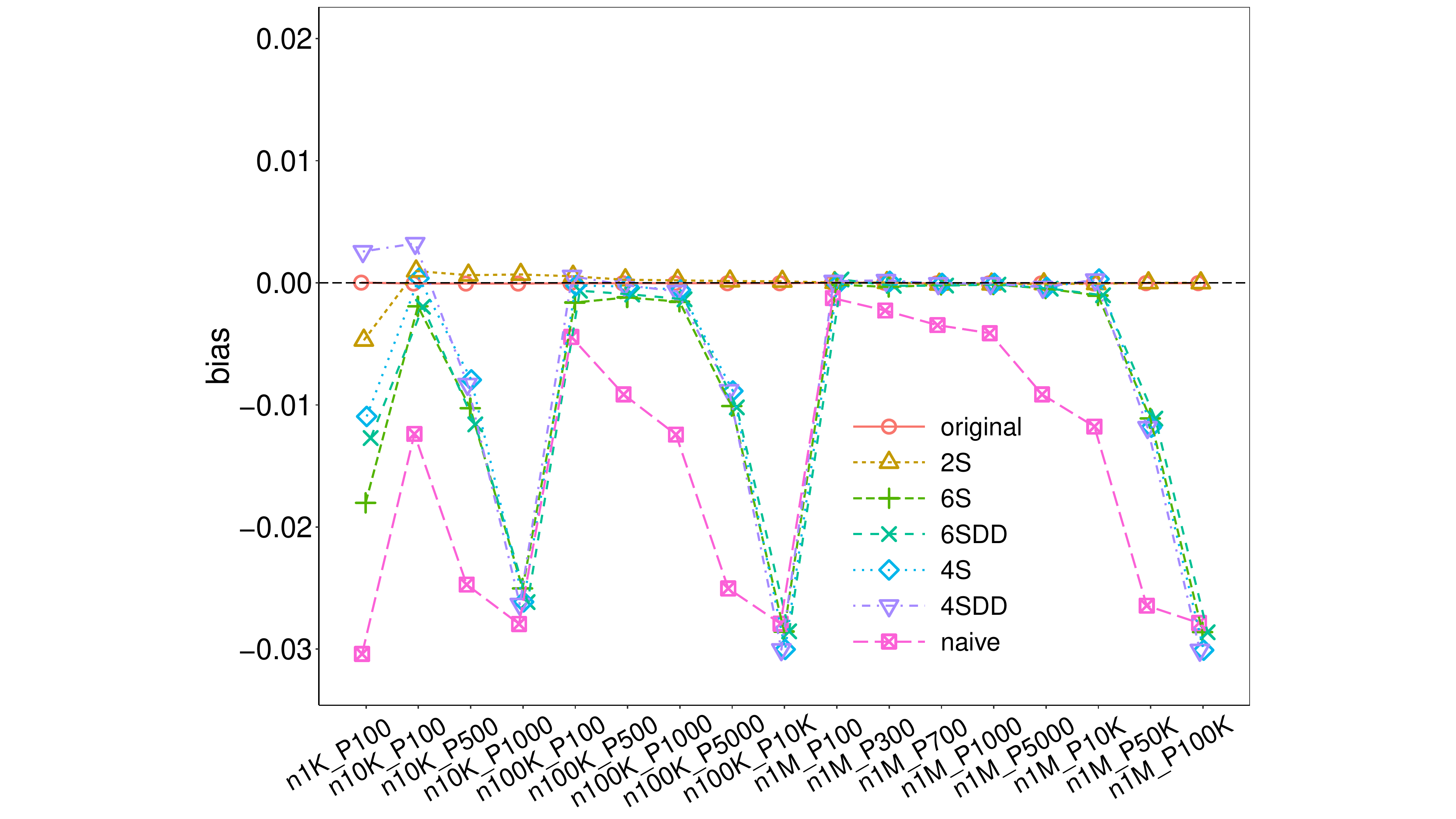}
\includegraphics[width=0.26\textwidth, trim={2.2in 0 2.2in 0},clip] {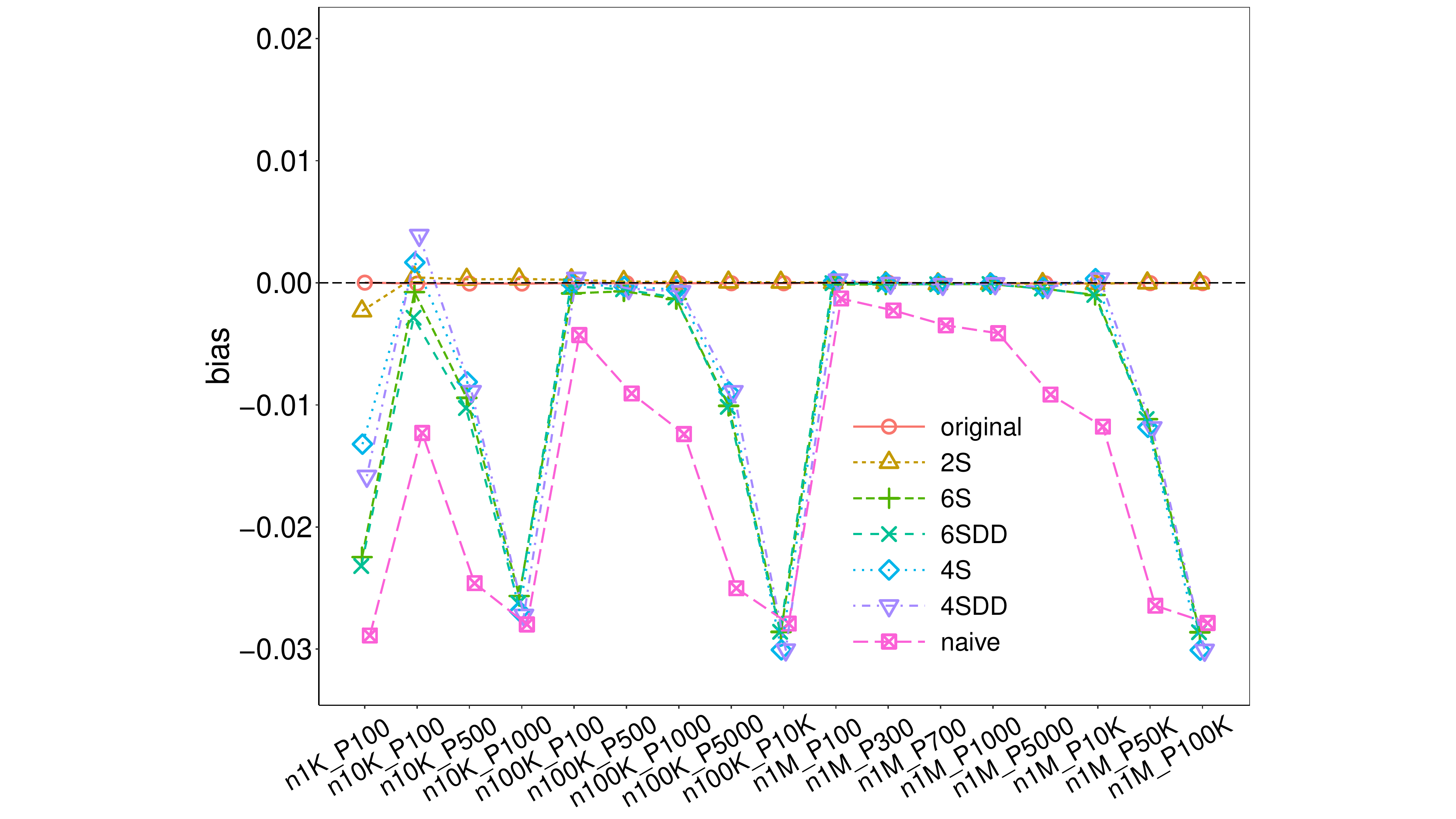}
\includegraphics[width=0.26\textwidth, trim={2.2in 0 2.2in 0},clip] {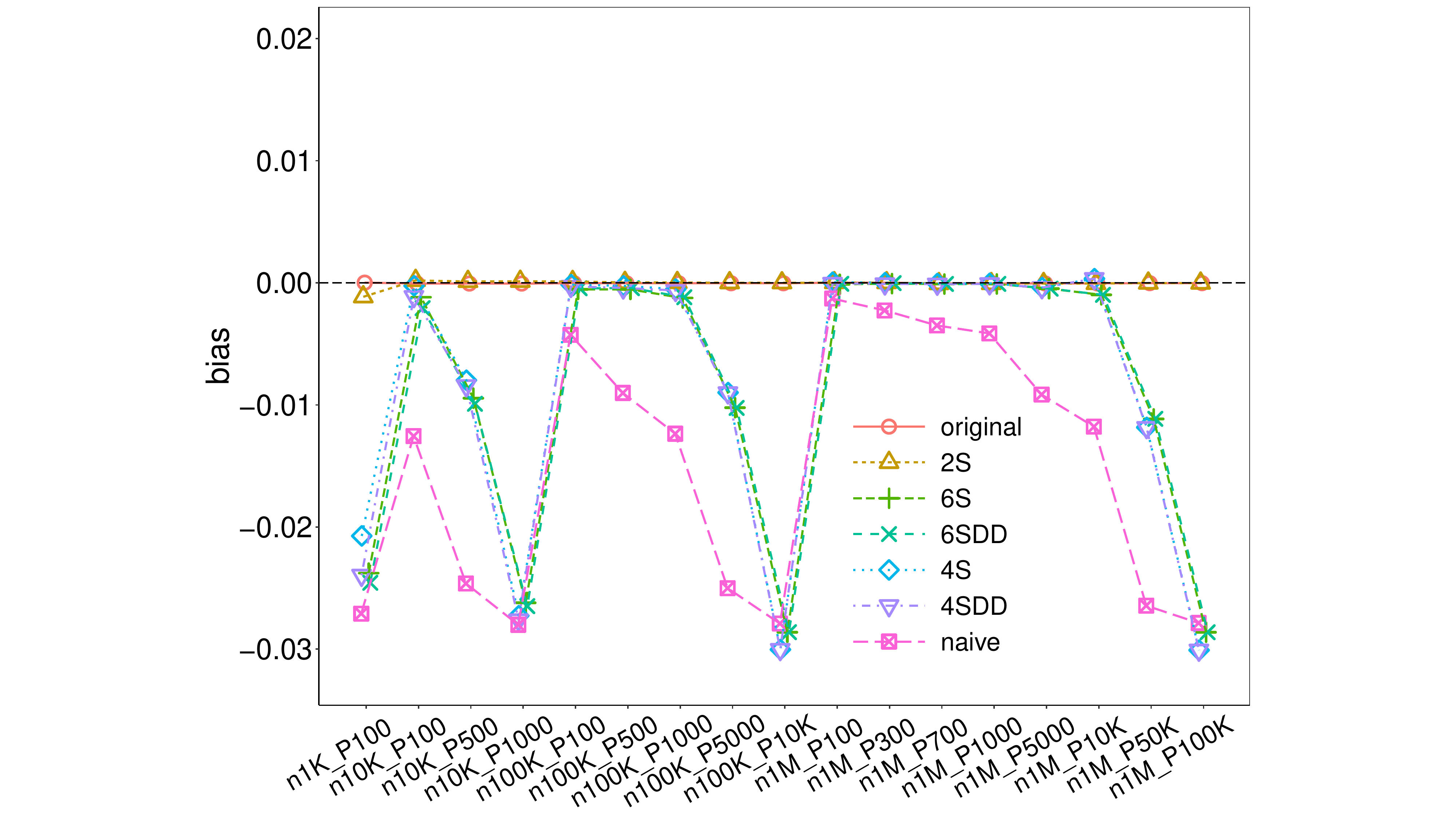}
\includegraphics[width=0.26\textwidth, trim={2.2in 0 2.2in 0},clip] {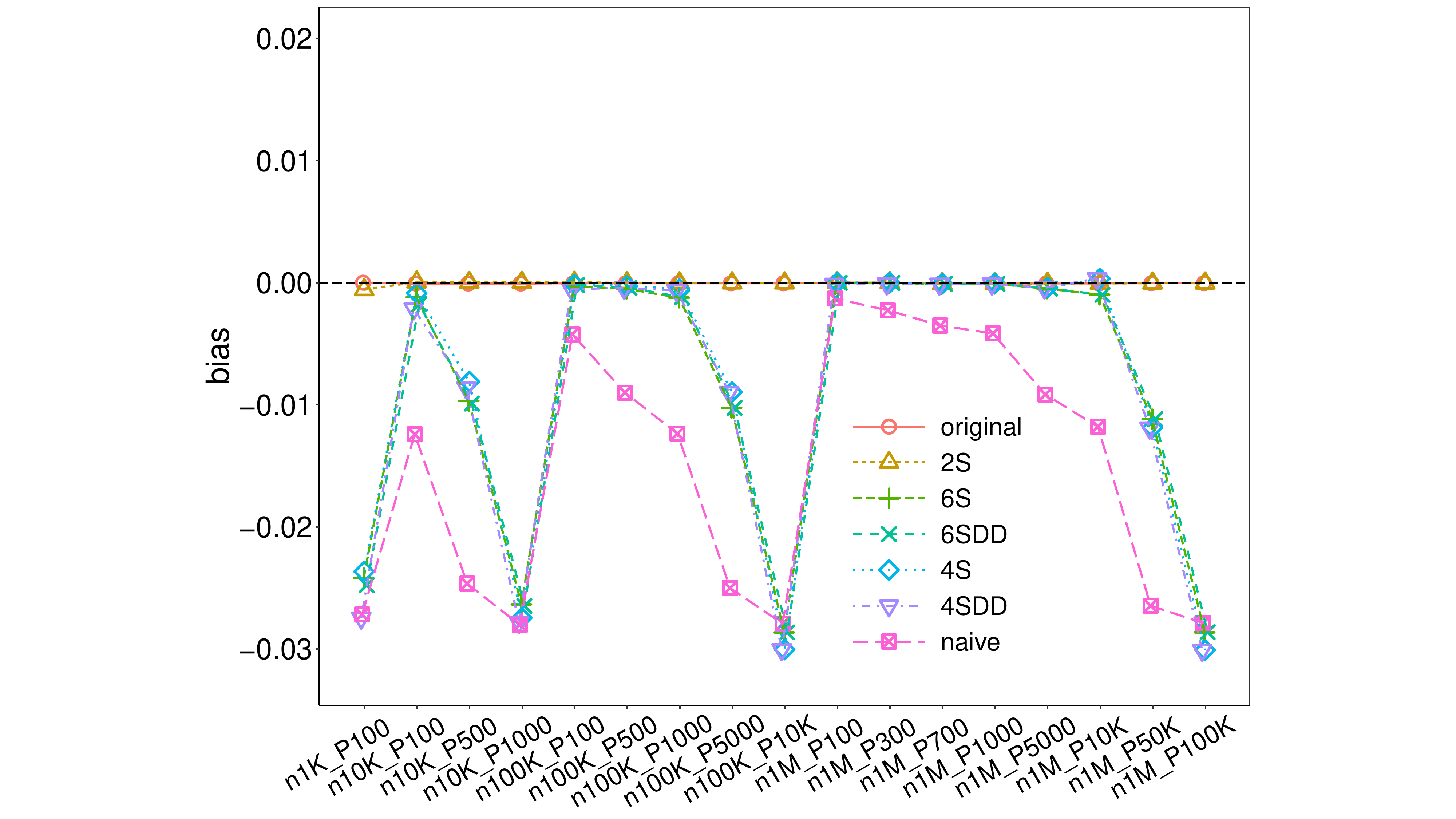}\\
\includegraphics[width=0.26\textwidth, trim={2.2in 0 2.2in 0},clip] {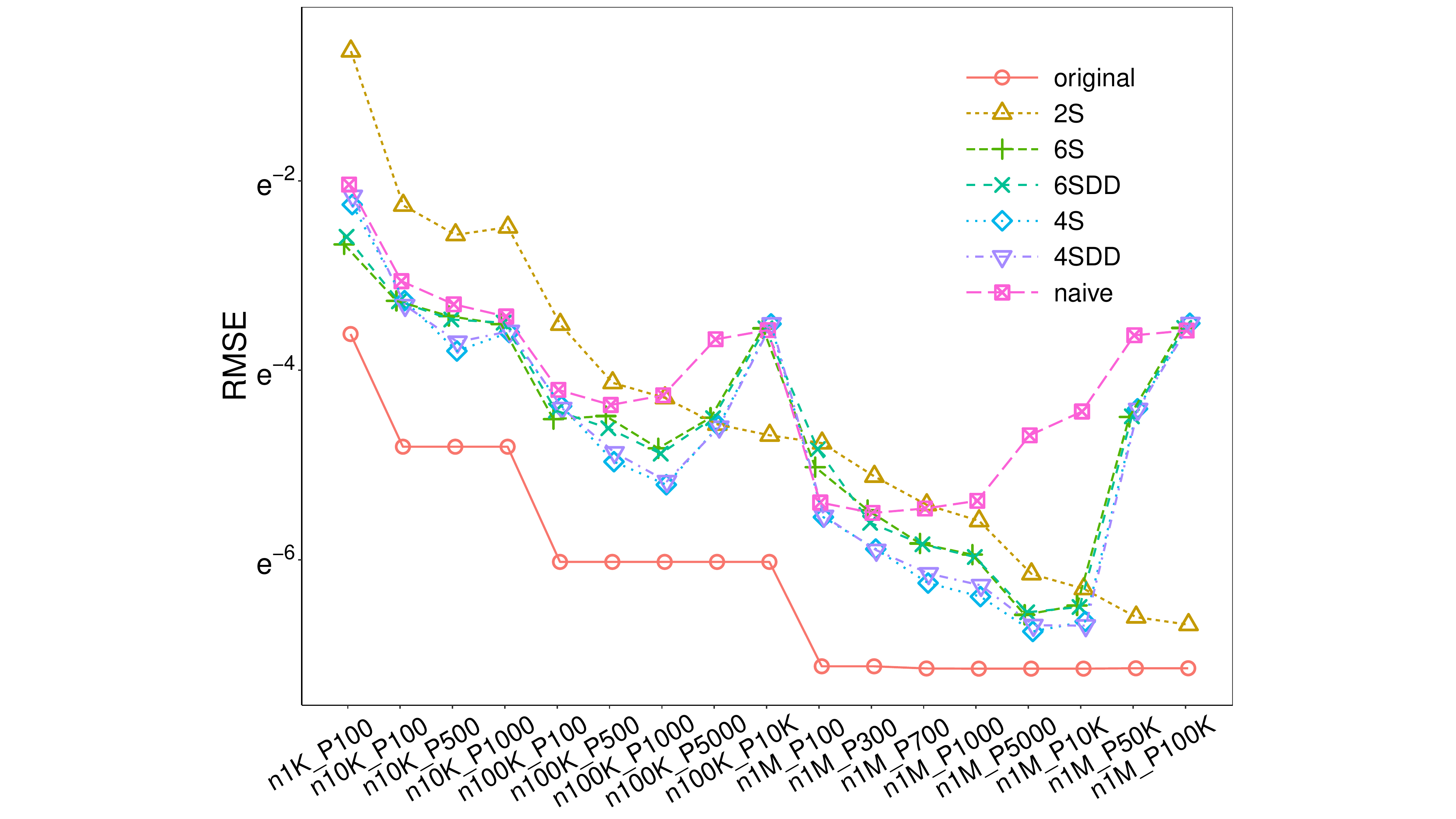}
\includegraphics[width=0.26\textwidth, trim={2.2in 0 2.2in 0},clip] {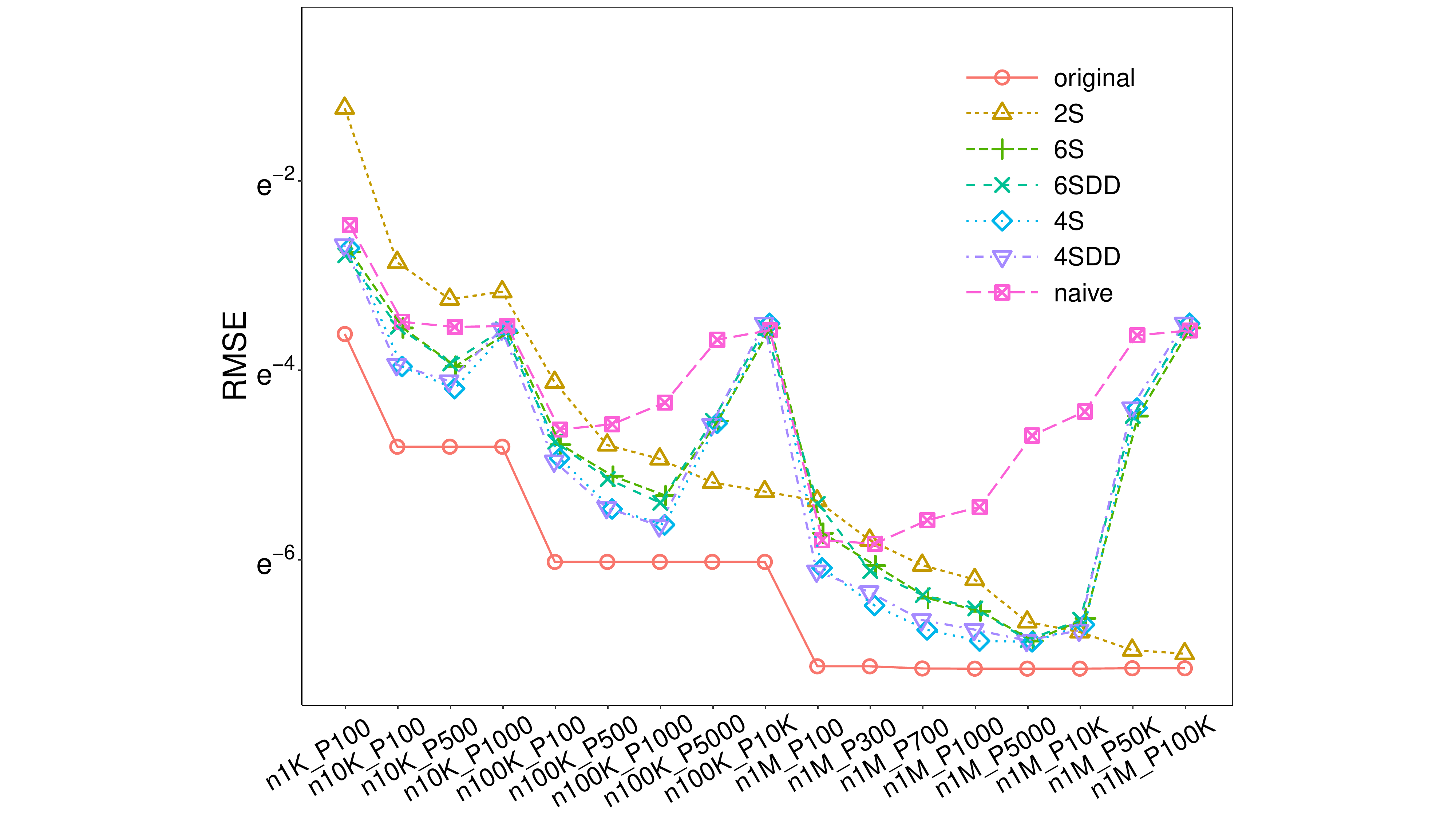}
\includegraphics[width=0.26\textwidth, trim={2.2in 0 2.2in 0},clip] {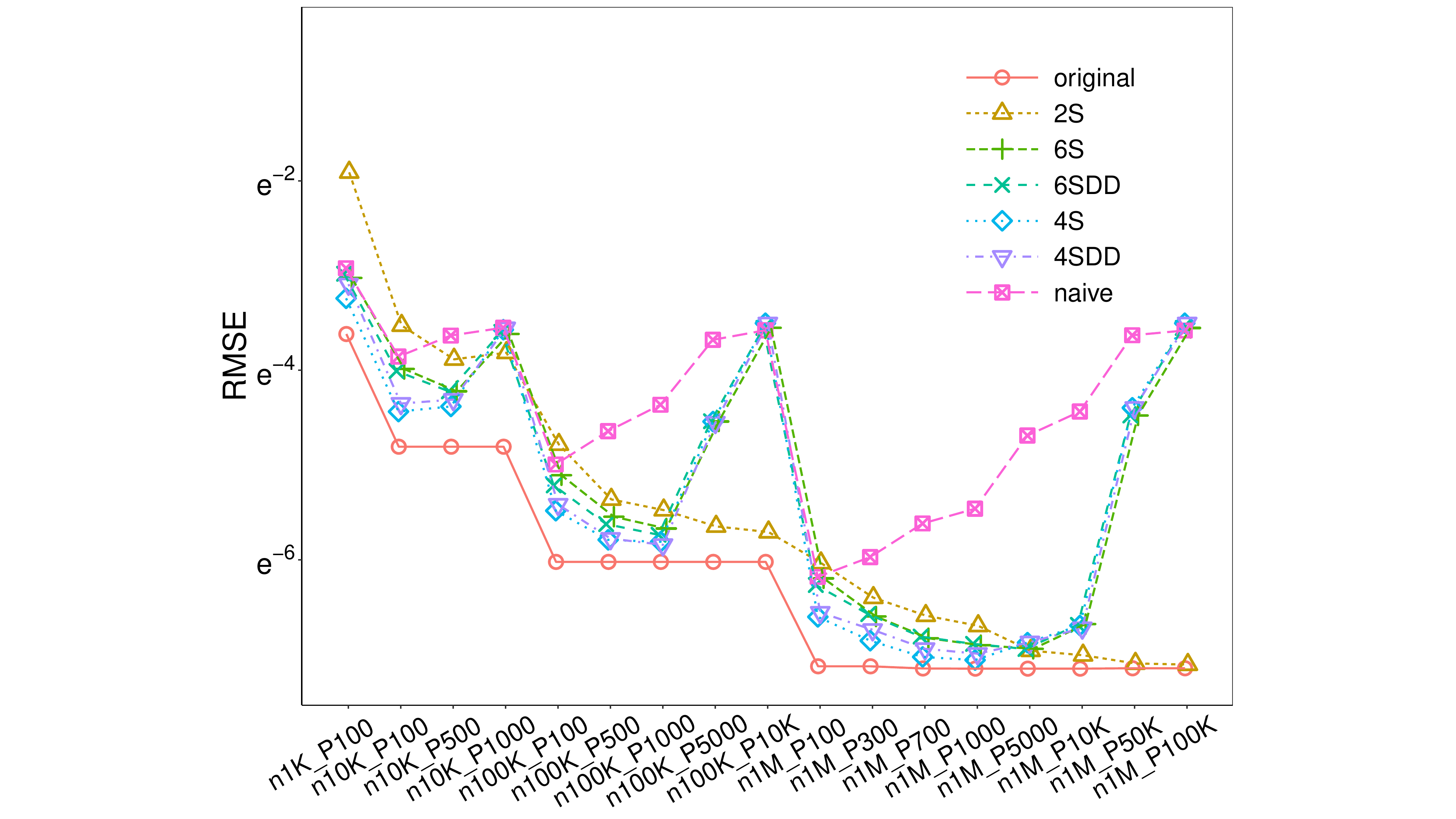}
\includegraphics[width=0.26\textwidth, trim={2.2in 0 2.2in 0},clip] {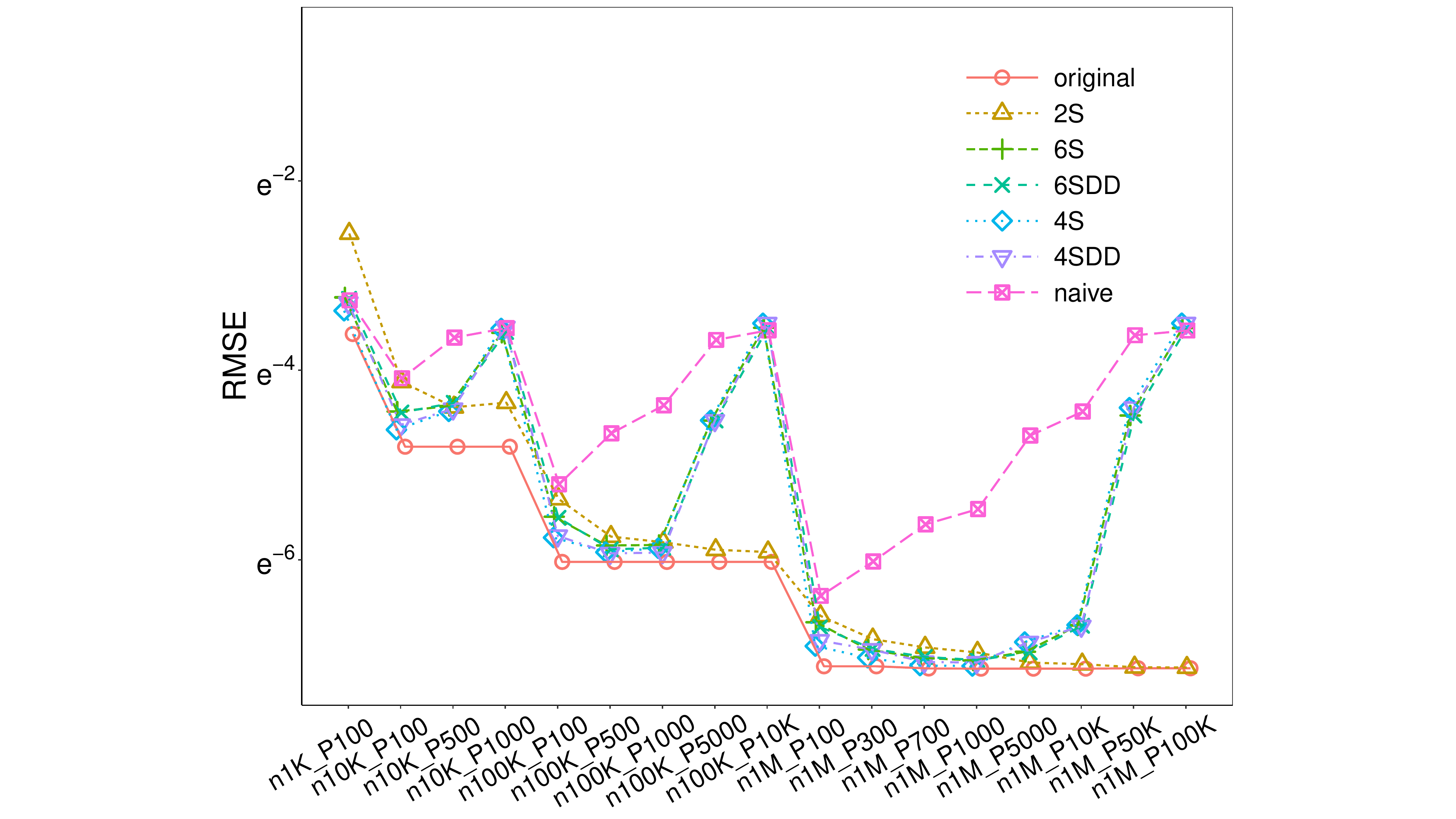}
\includegraphics[width=0.26\textwidth, trim={2.2in 0 2.2in 0},clip] {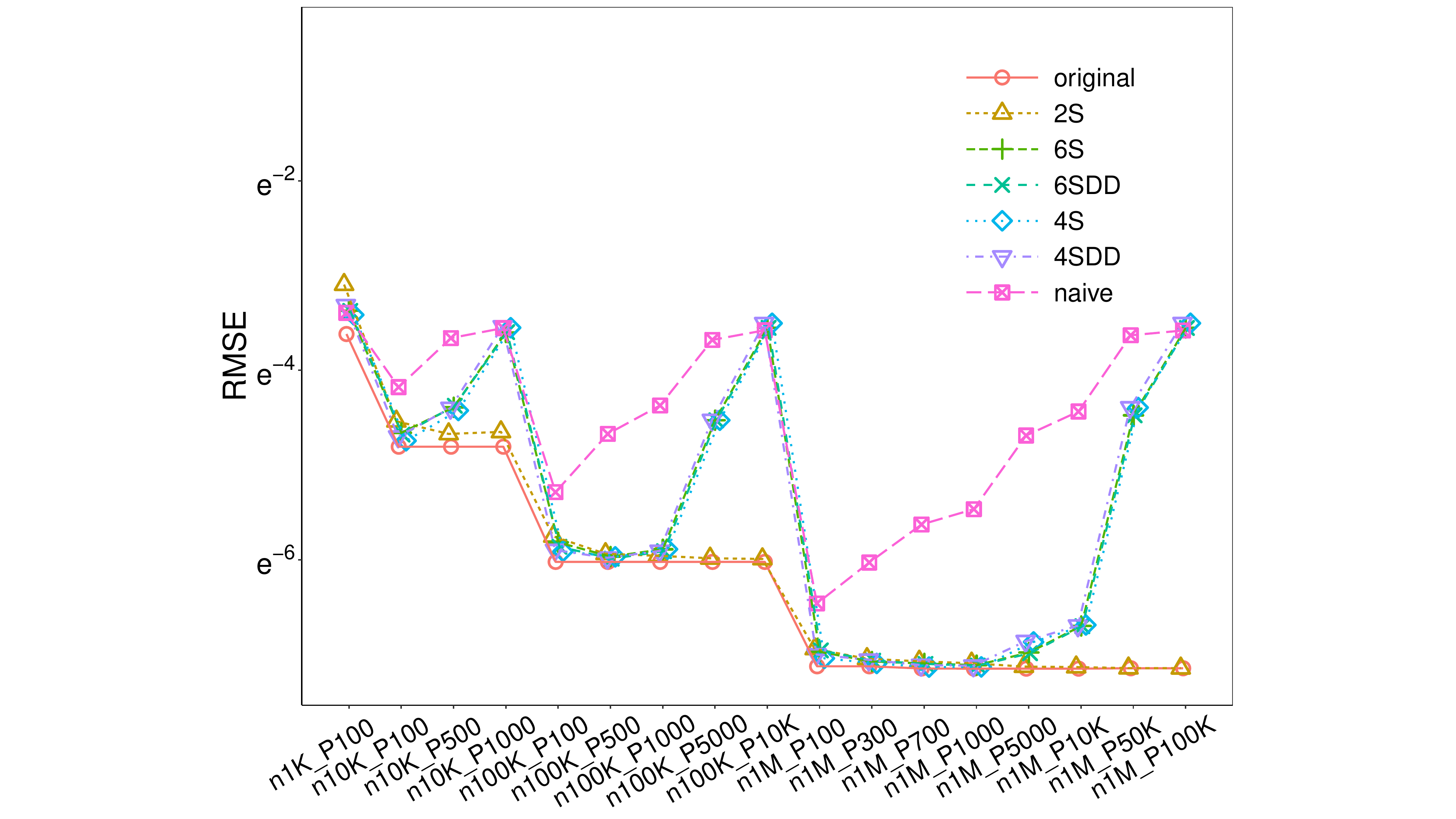}\\
\includegraphics[width=0.26\textwidth, trim={2.2in 0 2.2in 0},clip] {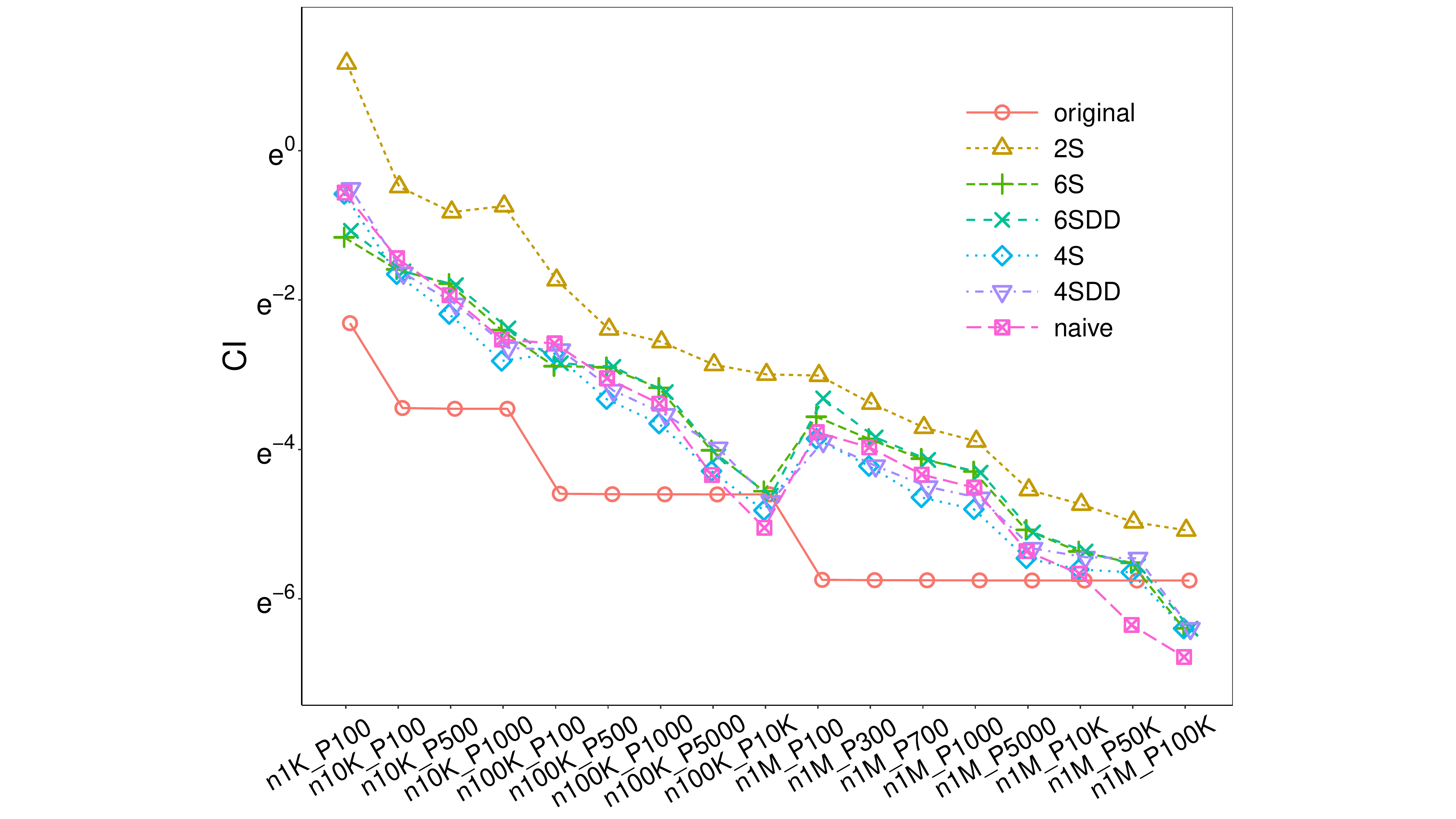}
\includegraphics[width=0.26\textwidth, trim={2.2in 0 2.2in 0},clip] {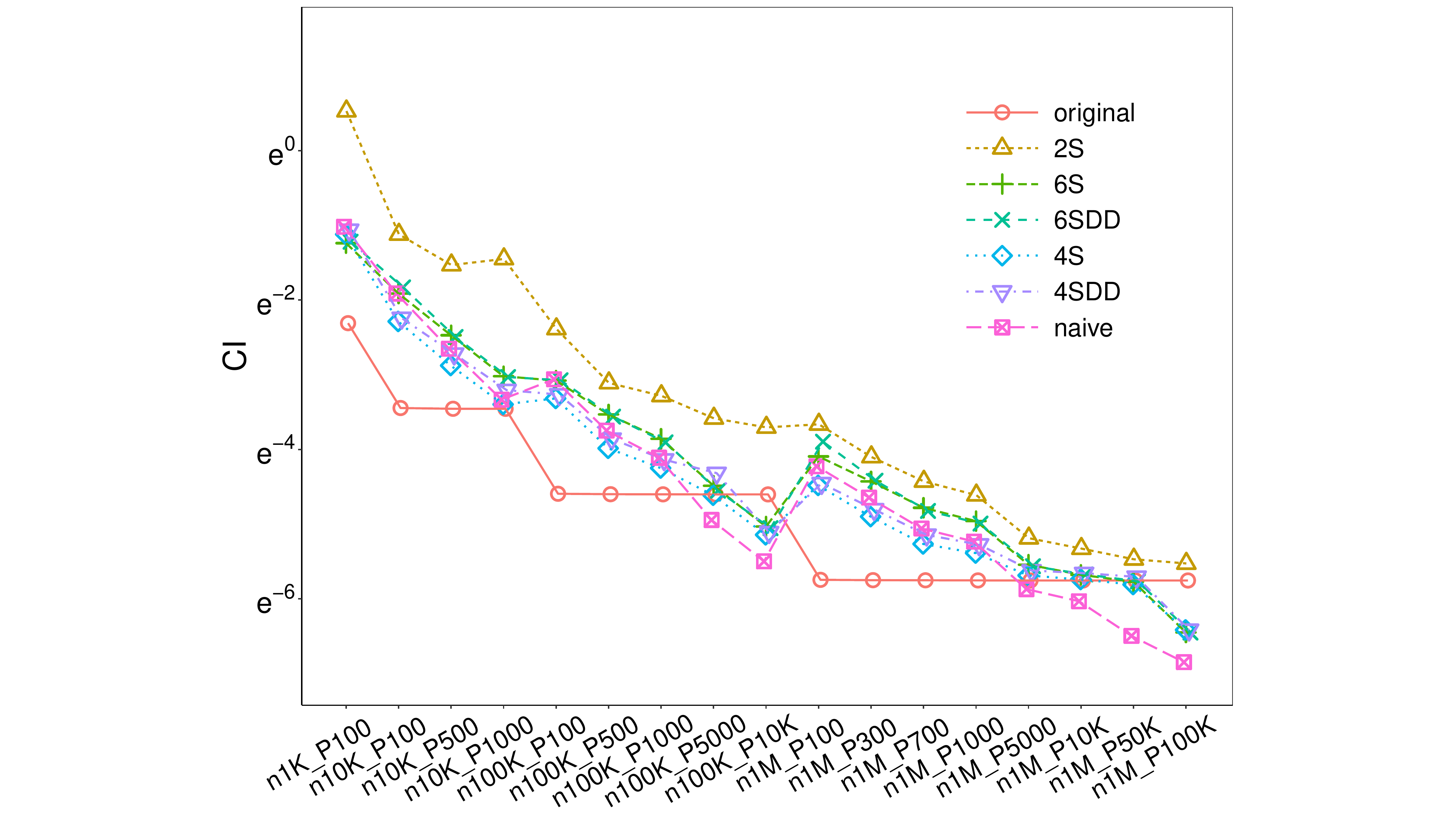}
\includegraphics[width=0.26\textwidth, trim={2.2in 0 2.2in 0},clip] {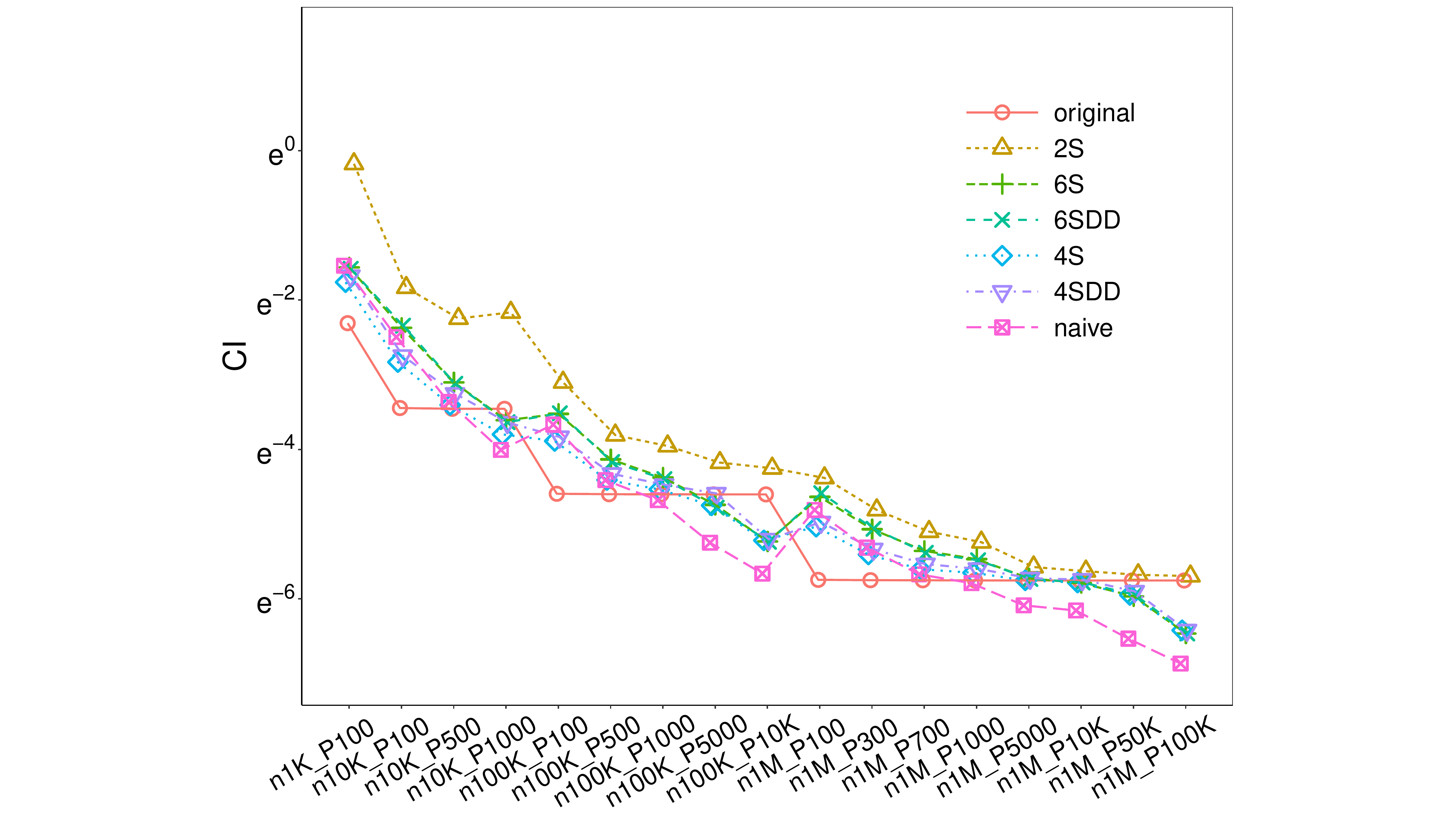}
\includegraphics[width=0.26\textwidth, trim={2.2in 0 2.2in 0},clip] {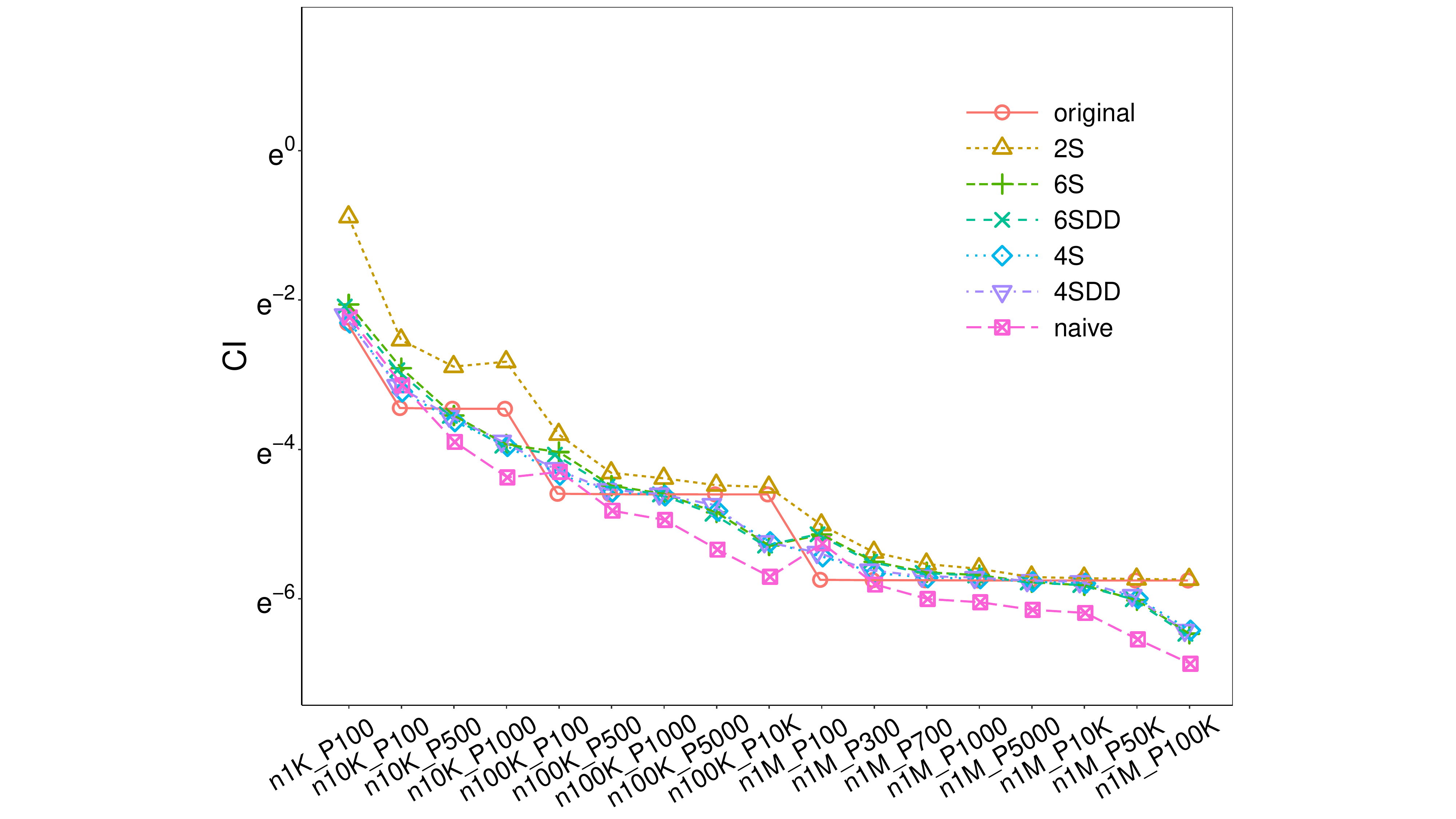}
\includegraphics[width=0.26\textwidth, trim={2.2in 0 2.2in 0},clip] {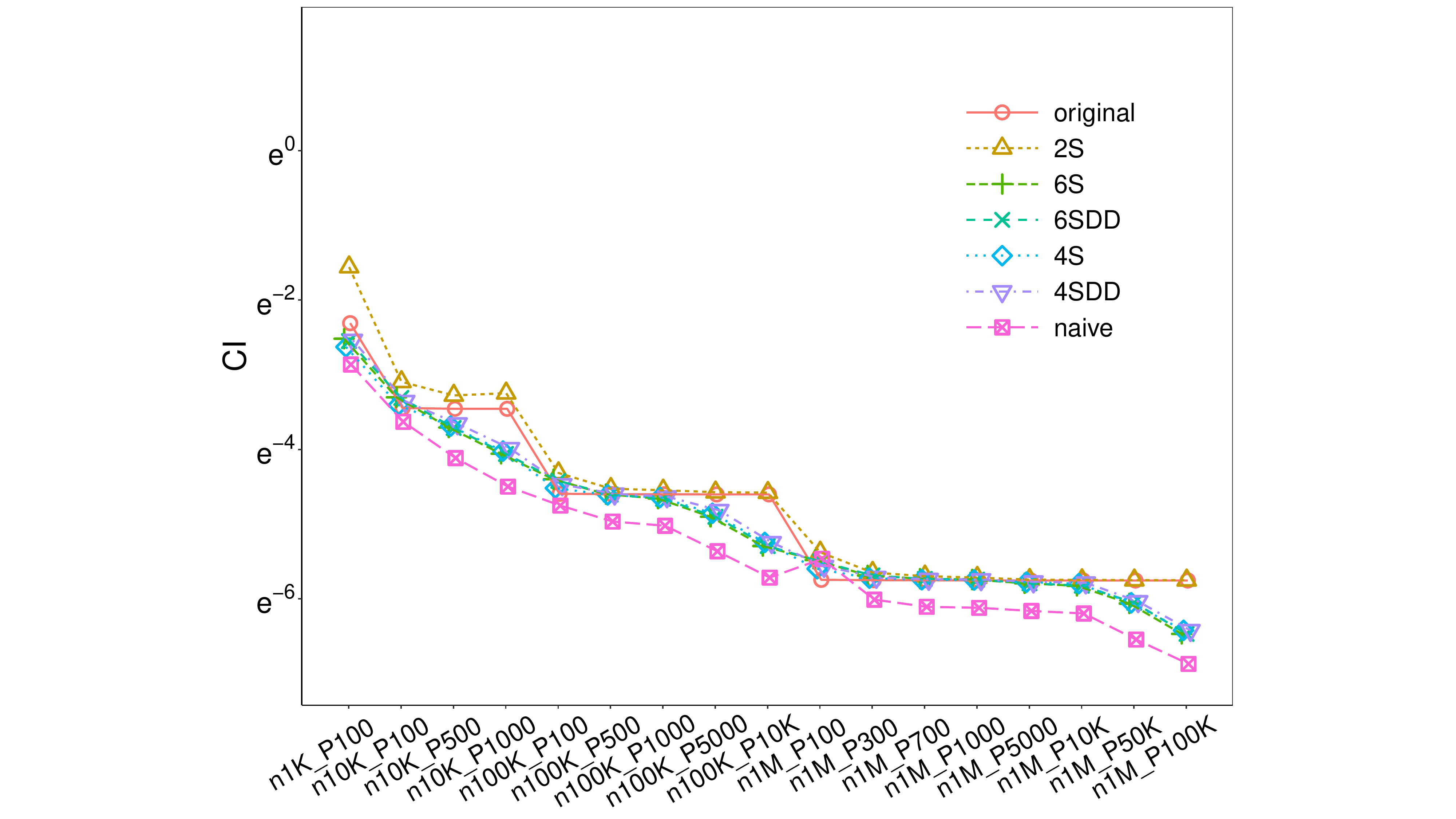}\\
\includegraphics[width=0.26\textwidth, trim={2.2in 0 2.2in 0},clip] {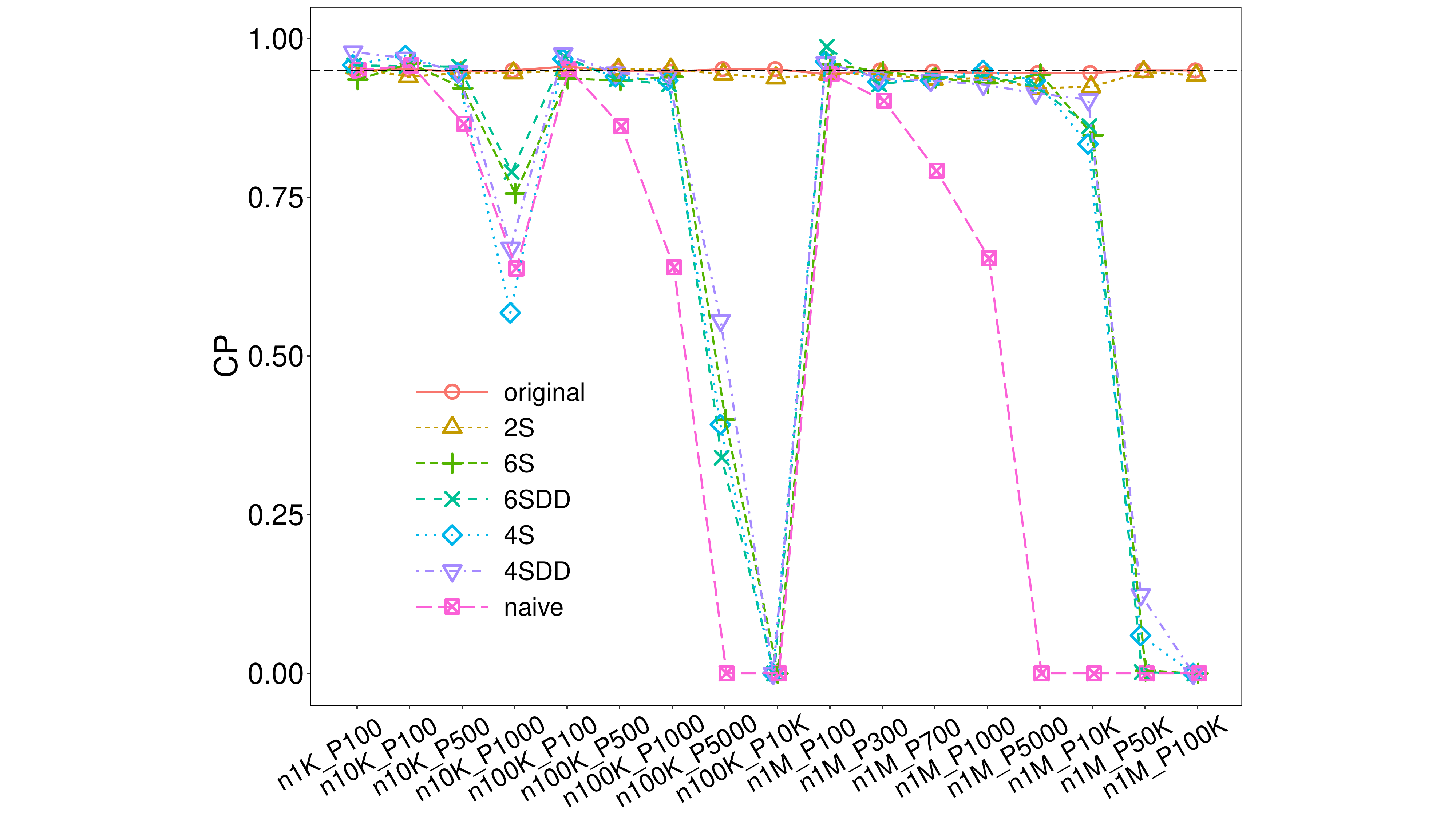}
\includegraphics[width=0.26\textwidth, trim={2.2in 0 2.2in 0},clip] {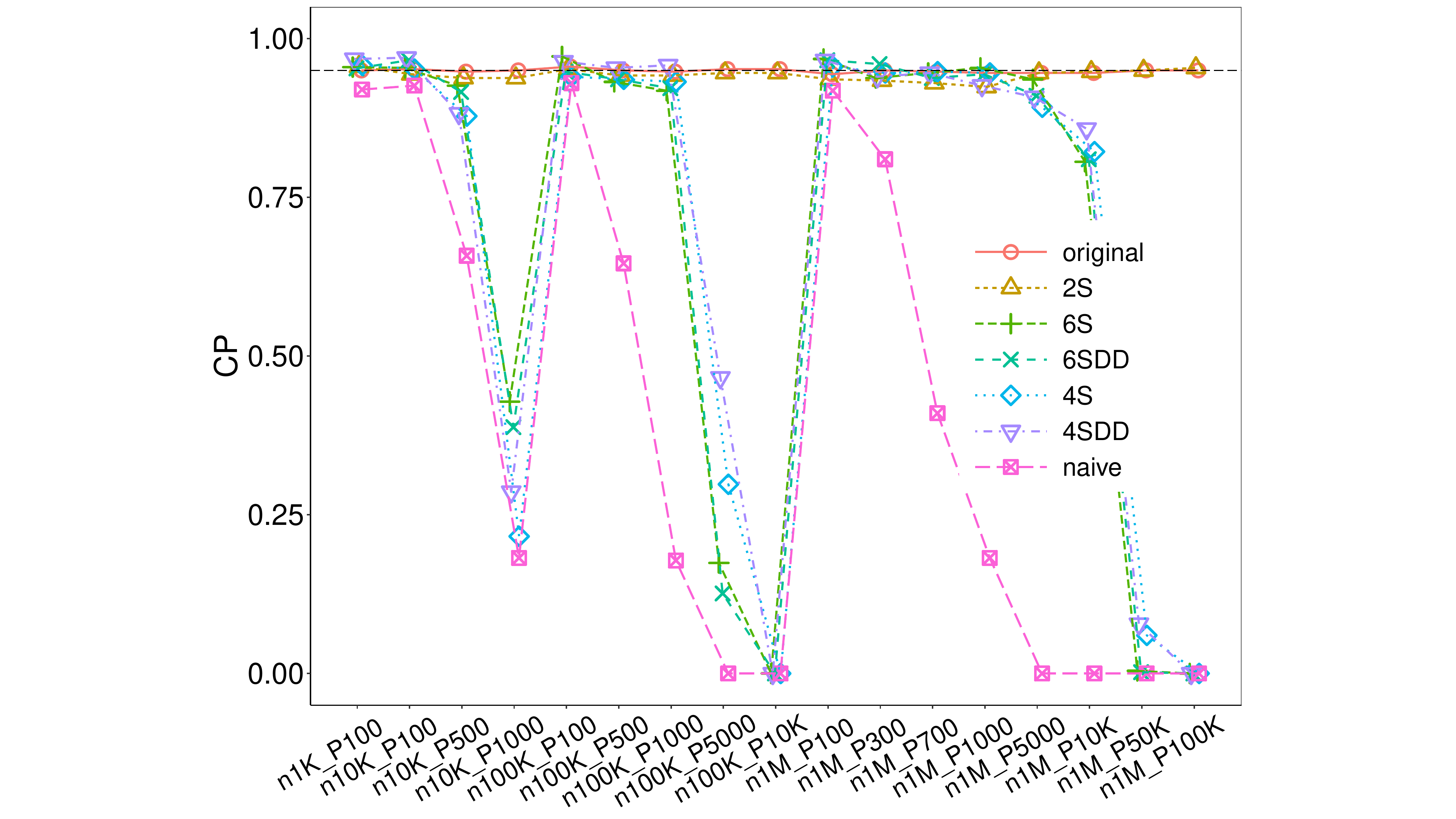}
\includegraphics[width=0.26\textwidth, trim={2.2in 0 2.2in 0},clip] {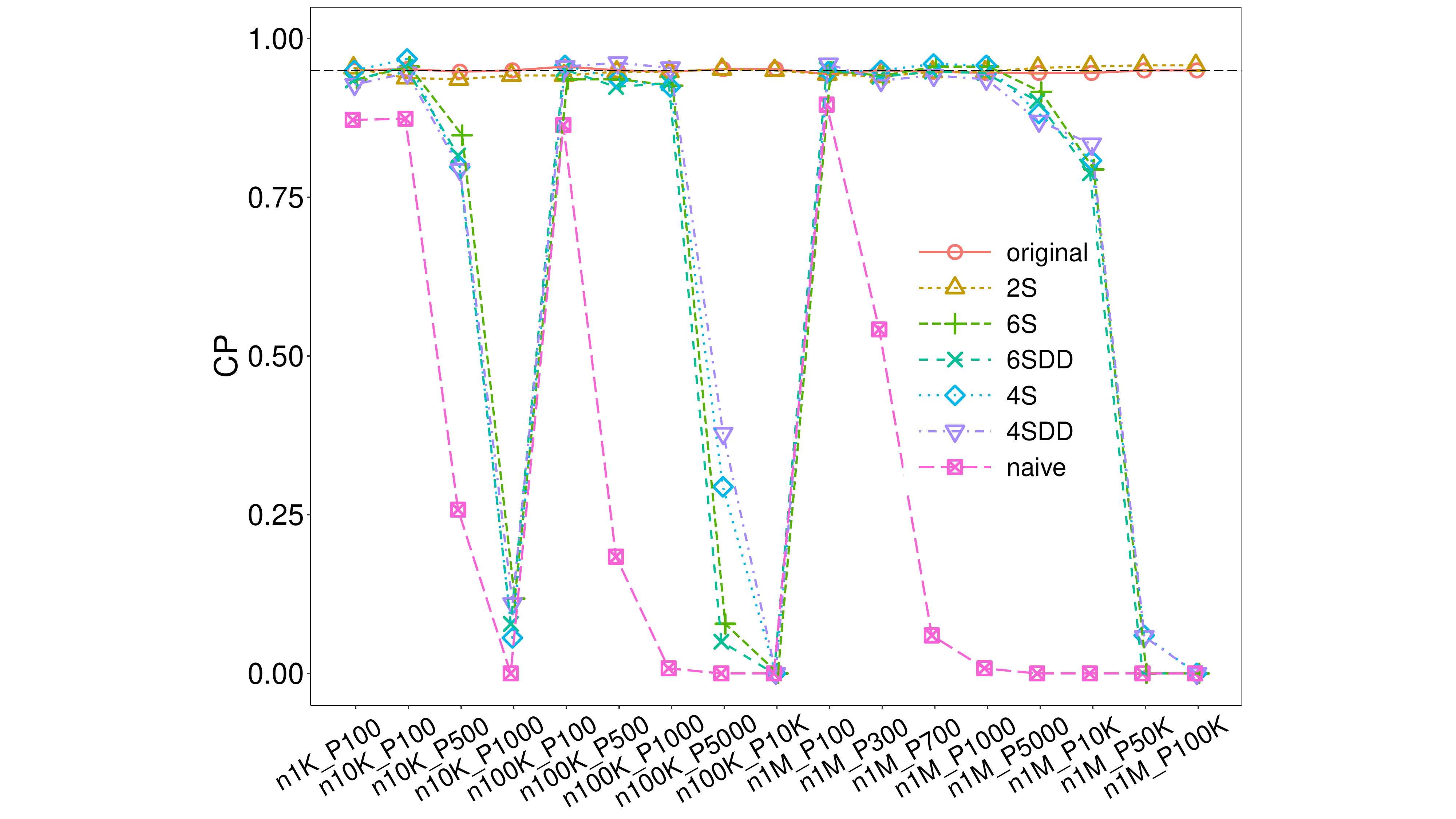}
\includegraphics[width=0.26\textwidth, trim={2.2in 0 2.2in 0},clip] {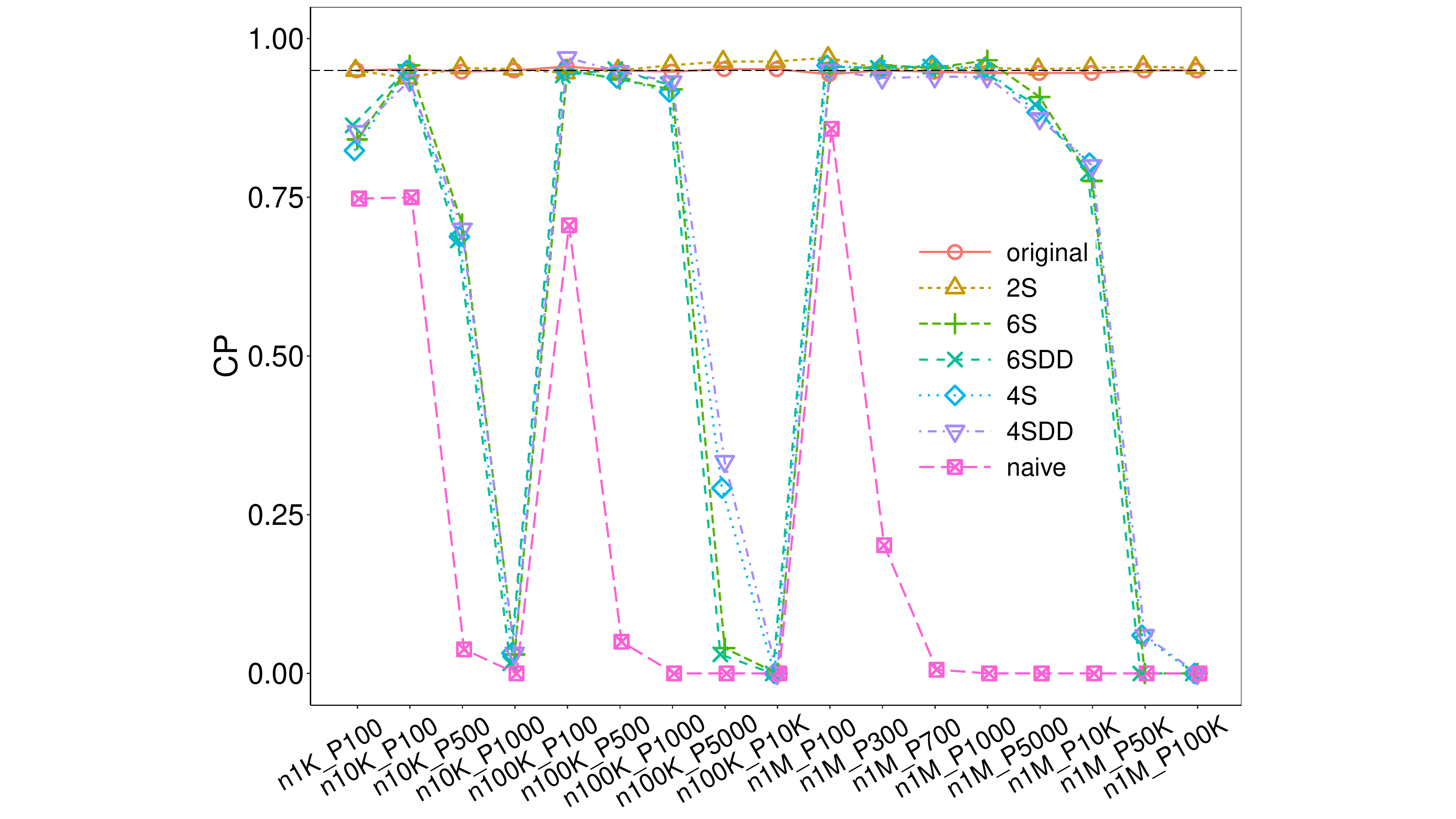}
\includegraphics[width=0.26\textwidth, trim={2.2in 0 2.2in 0},clip] {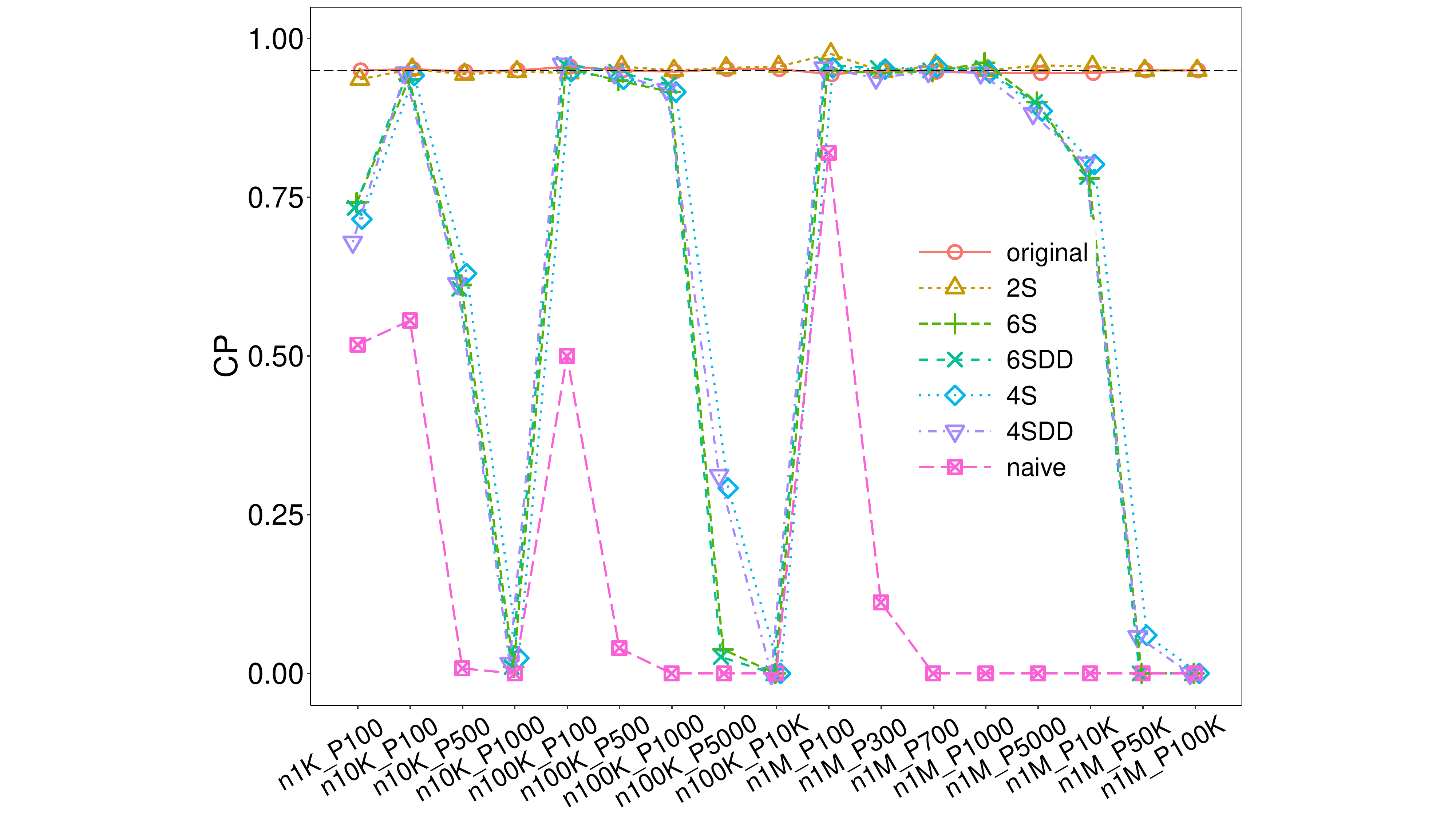}\\
\caption{ZINB data; $\rho$-zCDP; $\theta=0$ and $\alpha\ne\beta$}
\label{fig:0aszCDPzinb}
\end{figure}
\end{landscape}

\begin{landscape}
\subsection*{ZINB, $\theta\ne0$ and $\alpha\ne\beta$}
\begin{figure}[!htb]
\centering
\centering
$\epsilon=0.5$\hspace{0.9in}$\epsilon=1$\hspace{1in}$\epsilon=2$
\hspace{1in}$\epsilon=5$\hspace{0.9in}$\epsilon=50$\\
\includegraphics[width=0.215\textwidth, trim={2.2in 0 2.2in 0},clip] {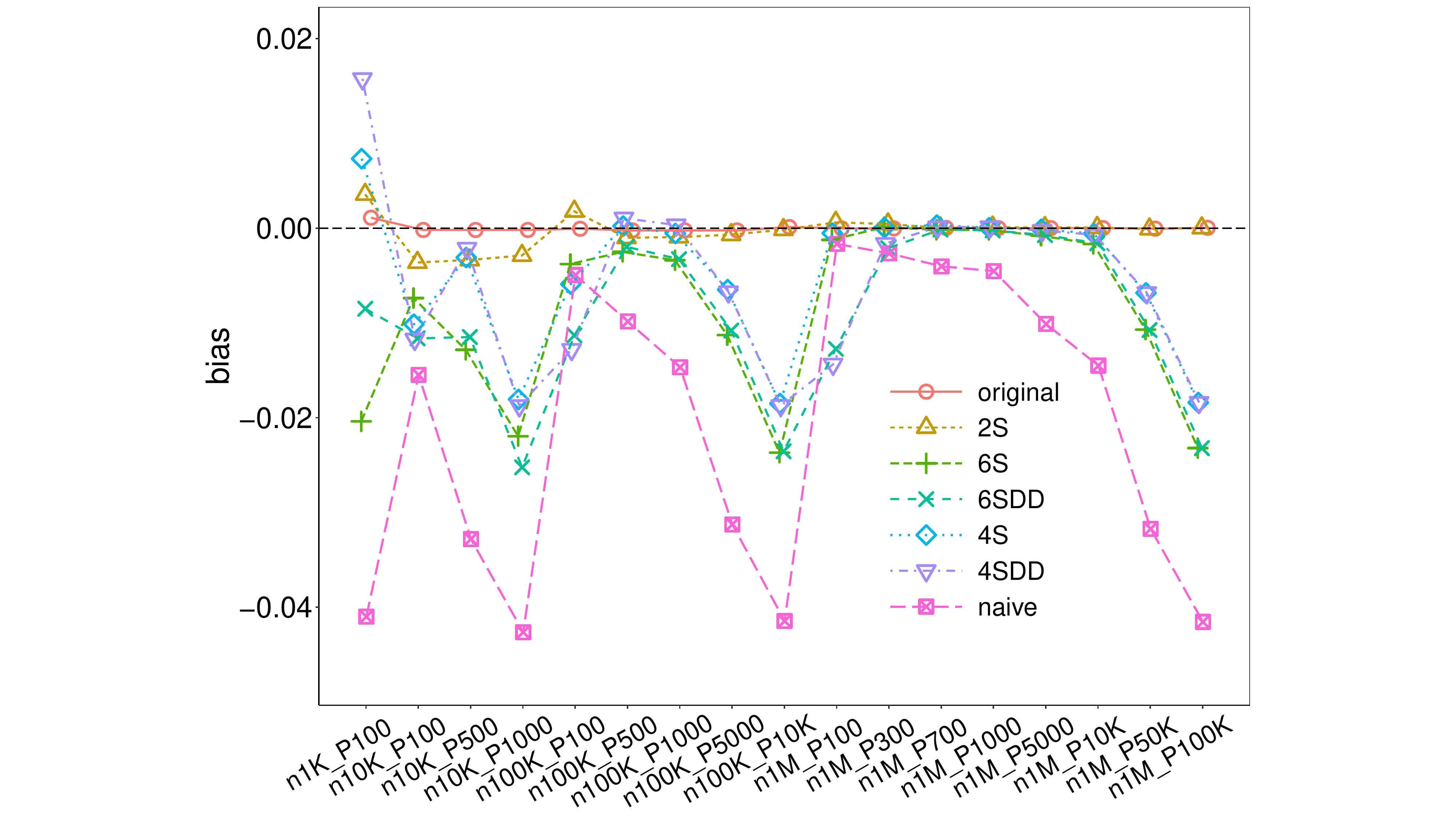}
\includegraphics[width=0.215\textwidth, trim={2.2in 0 2.2in 0},clip] {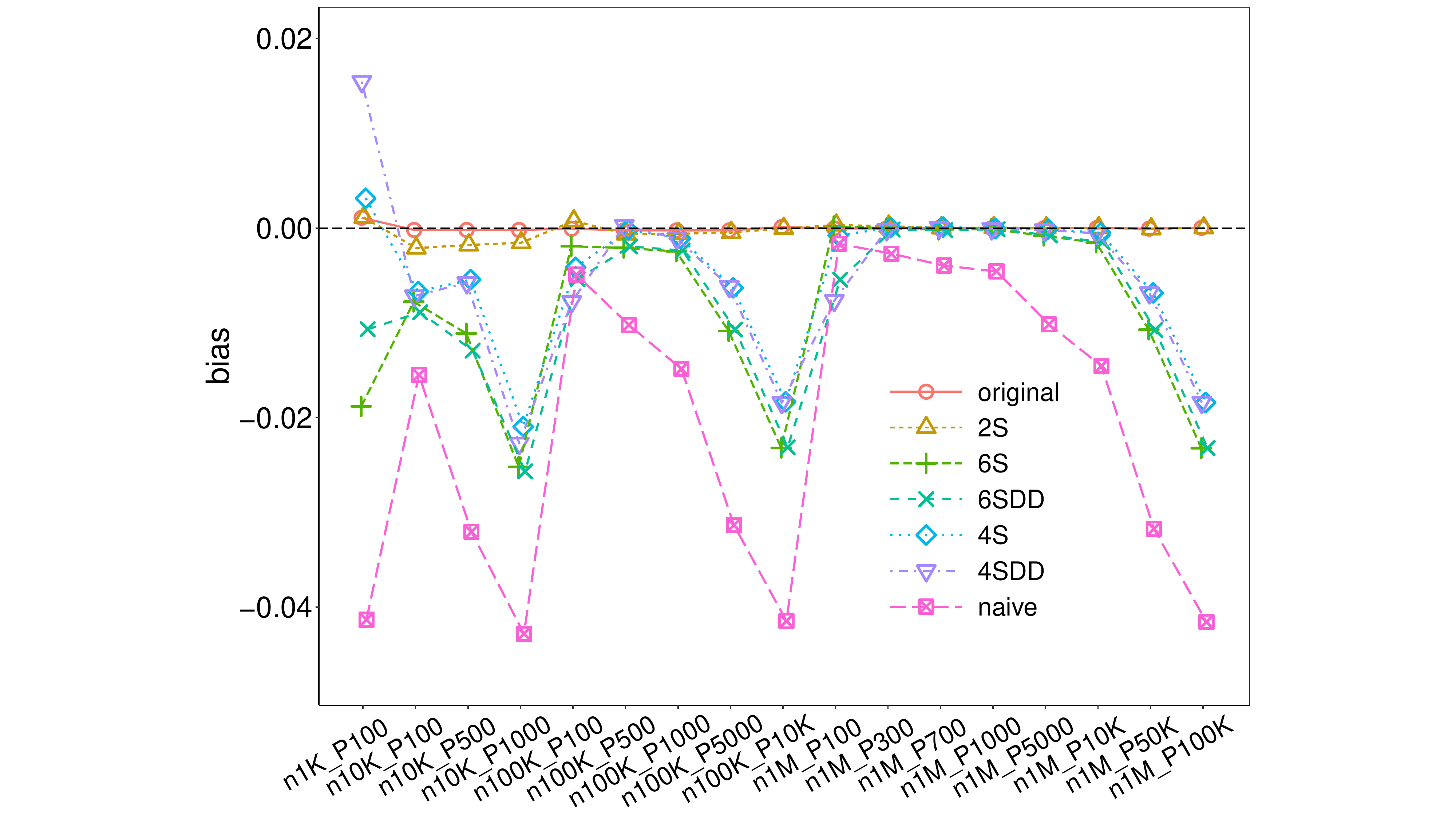}
\includegraphics[width=0.215\textwidth, trim={2.2in 0 2.2in 0},clip] {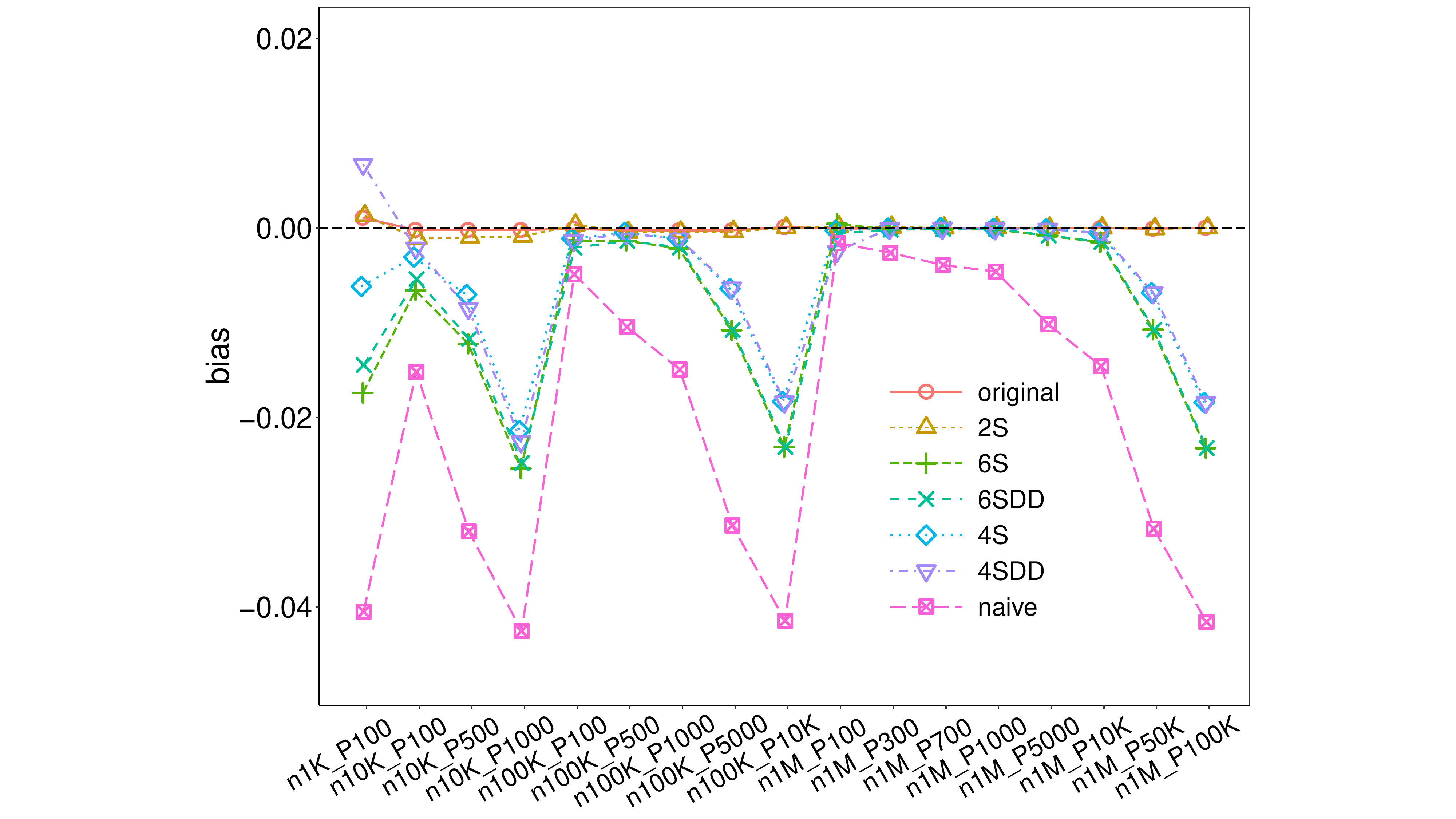}
\includegraphics[width=0.215\textwidth, trim={2.2in 0 2.2in 0},clip] {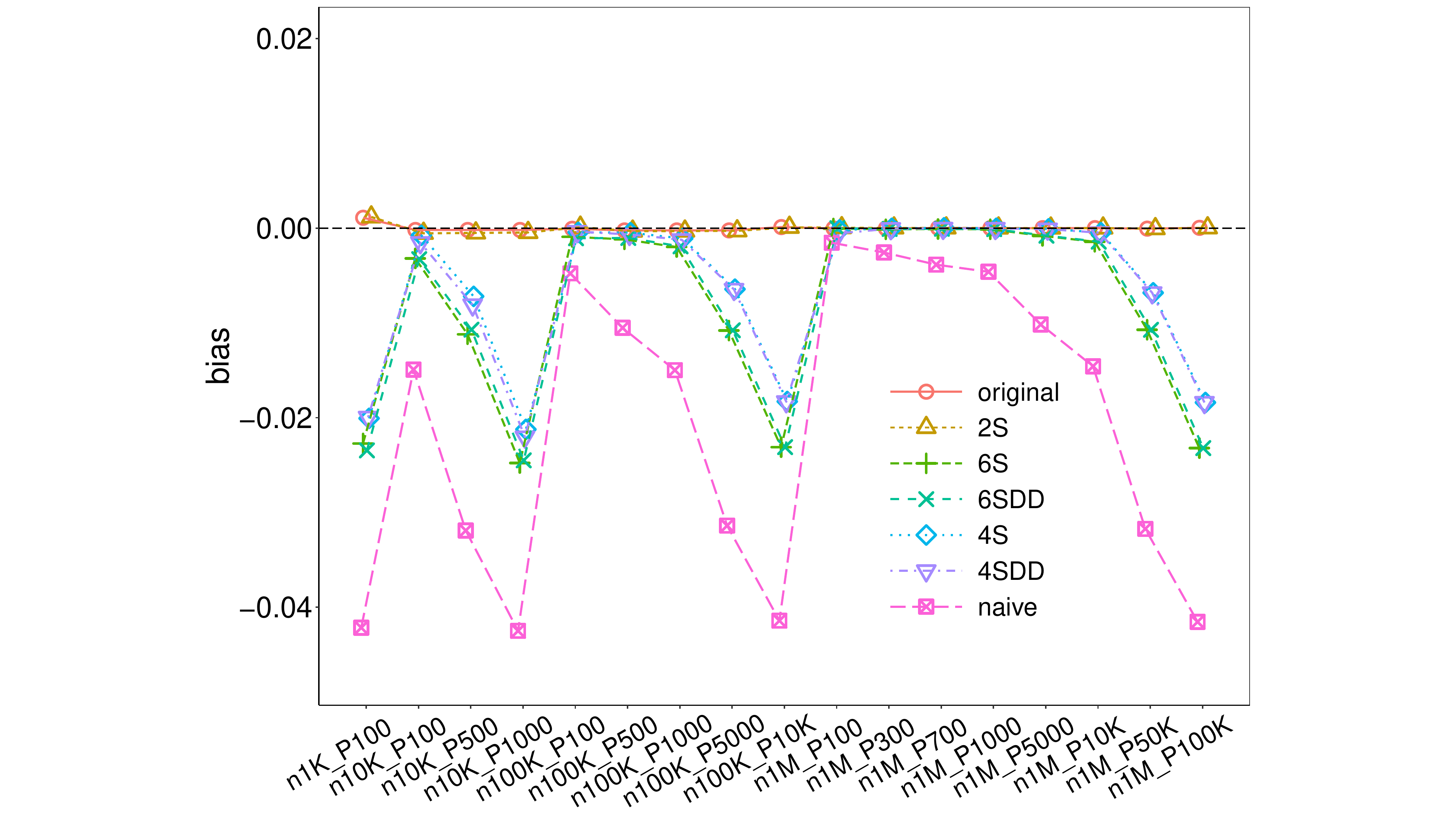}
\includegraphics[width=0.215\textwidth, trim={2.2in 0 2.2in 0},clip] {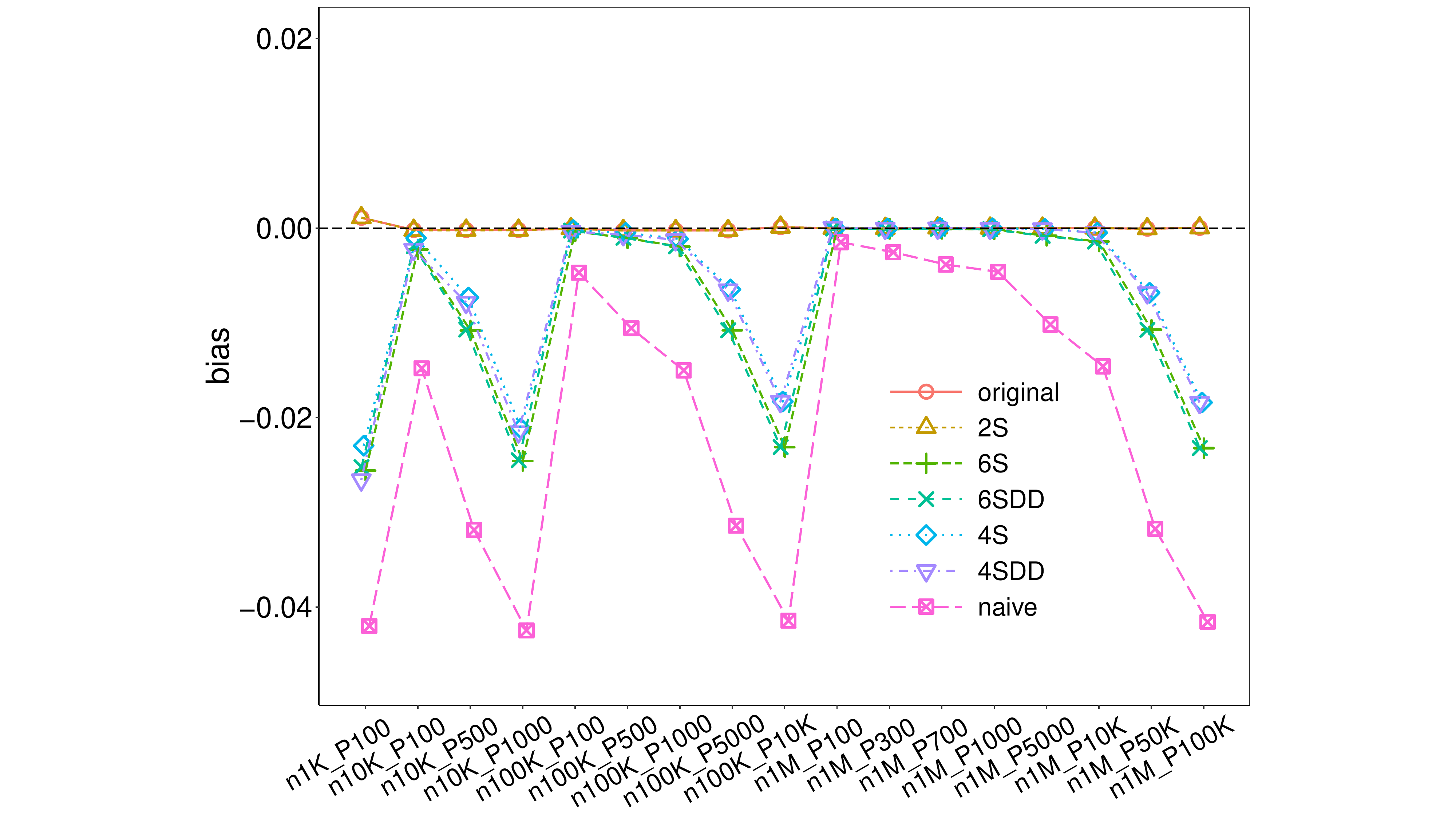}\\
\includegraphics[width=0.215\textwidth, trim={2.2in 0 2.2in 0},clip] {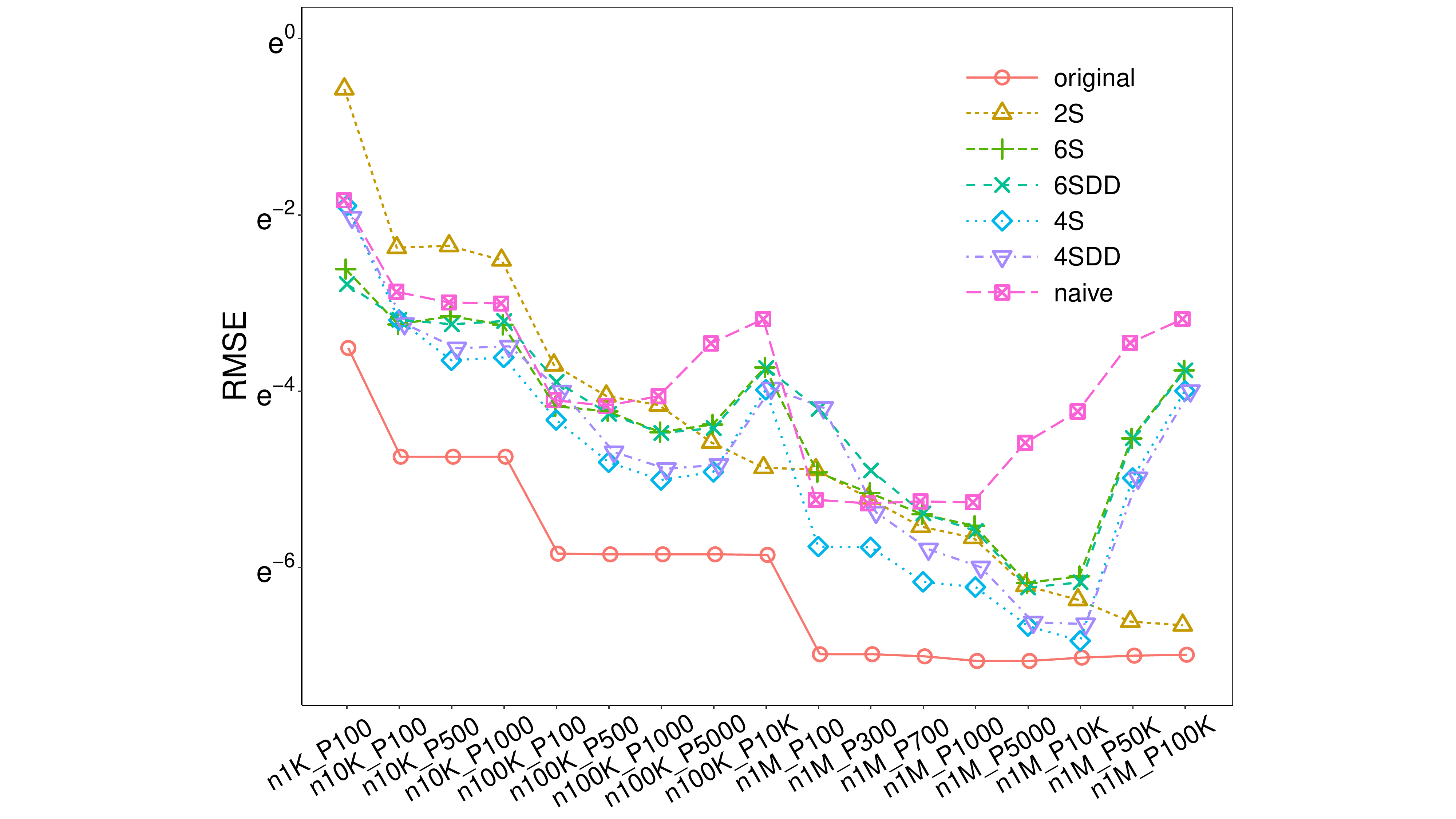}
\includegraphics[width=0.215\textwidth, trim={2.2in 0 2.2in 0},clip] {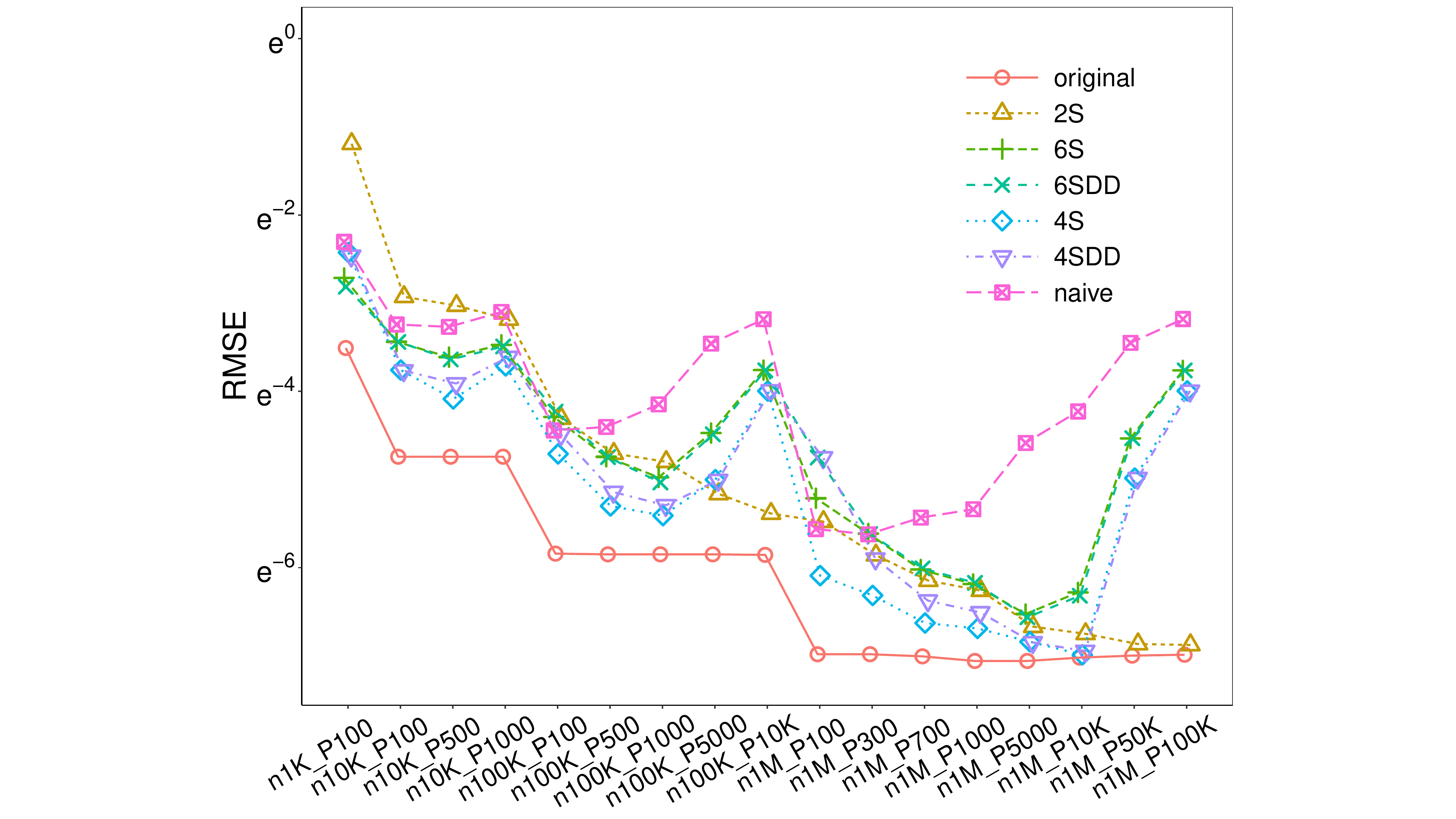}
\includegraphics[width=0.215\textwidth, trim={2.2in 0 2.2in 0},clip] {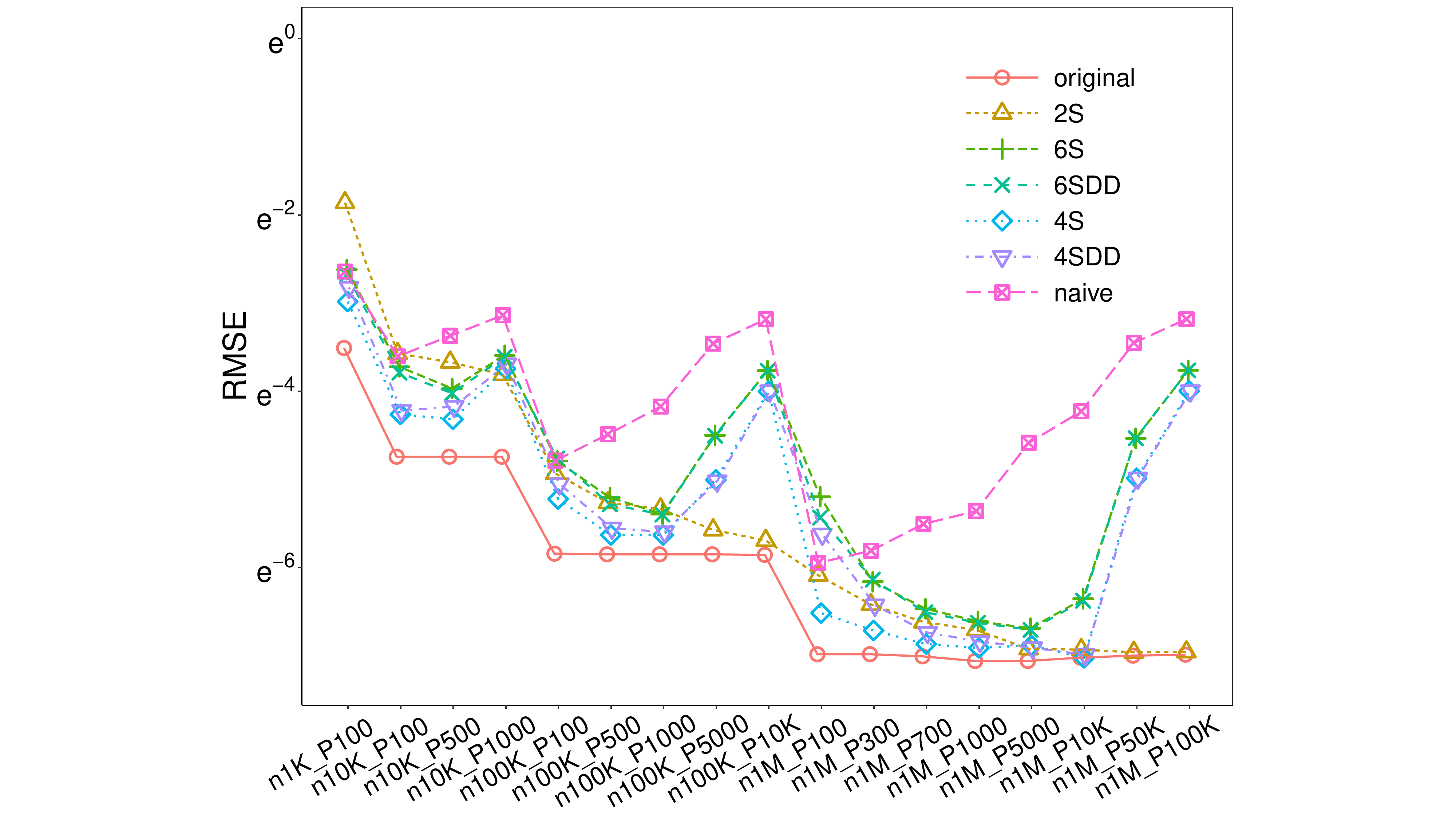}
\includegraphics[width=0.215\textwidth, trim={2.2in 0 2.2in 0},clip] {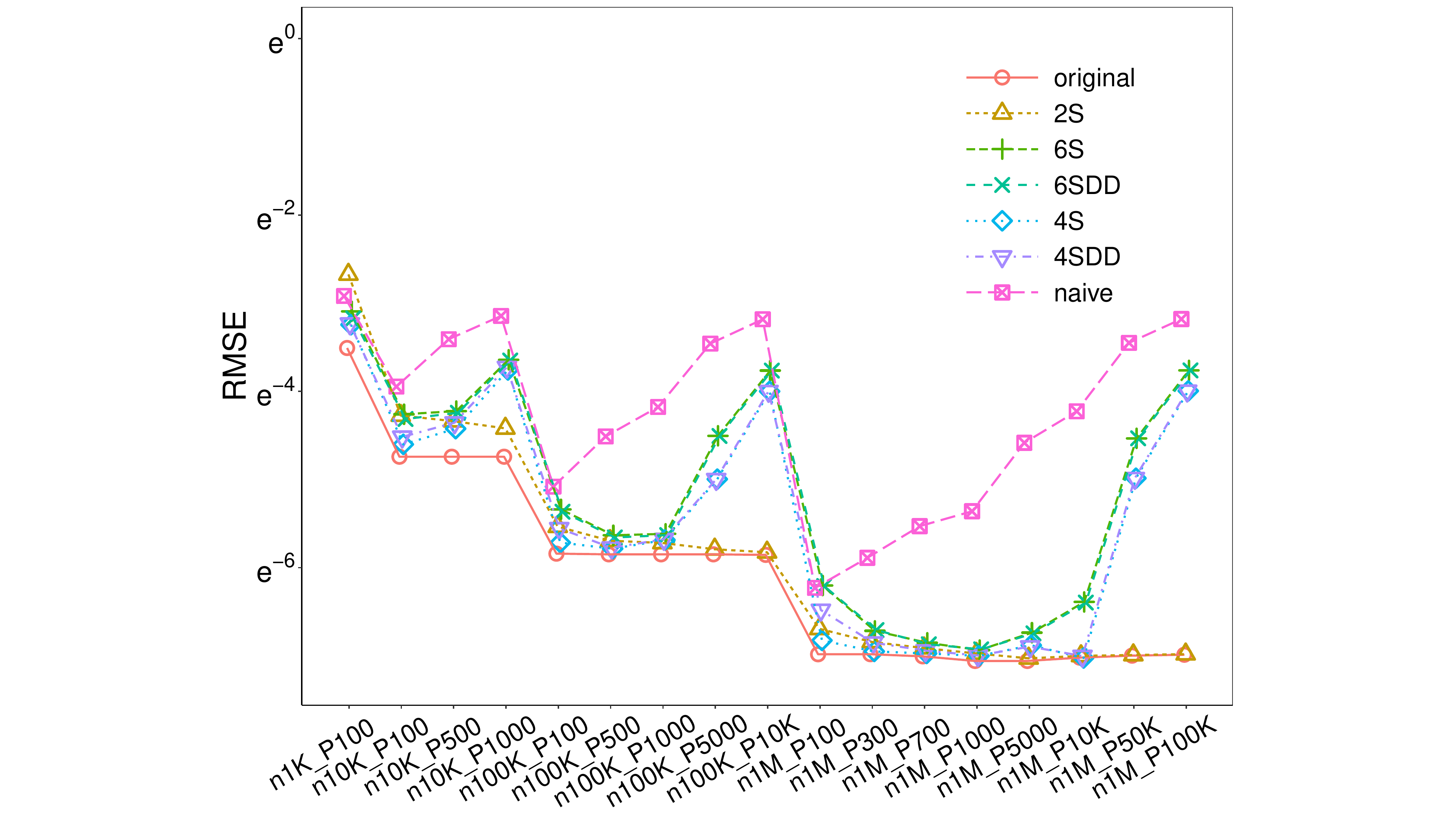}
\includegraphics[width=0.215\textwidth, trim={2.2in 0 2.2in 0},clip] {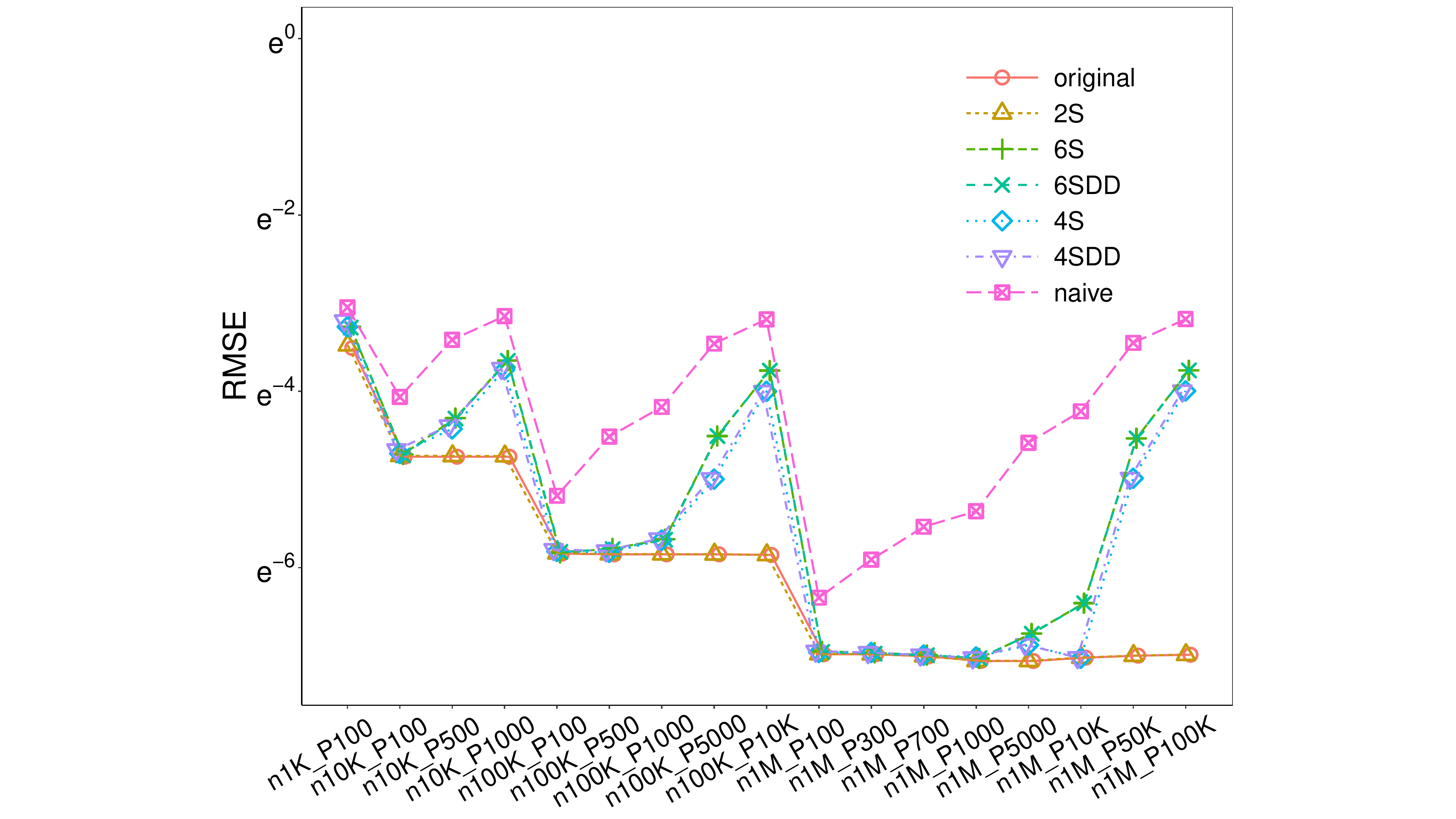}\\
\includegraphics[width=0.215\textwidth, trim={2.2in 0 2.2in 0},clip] {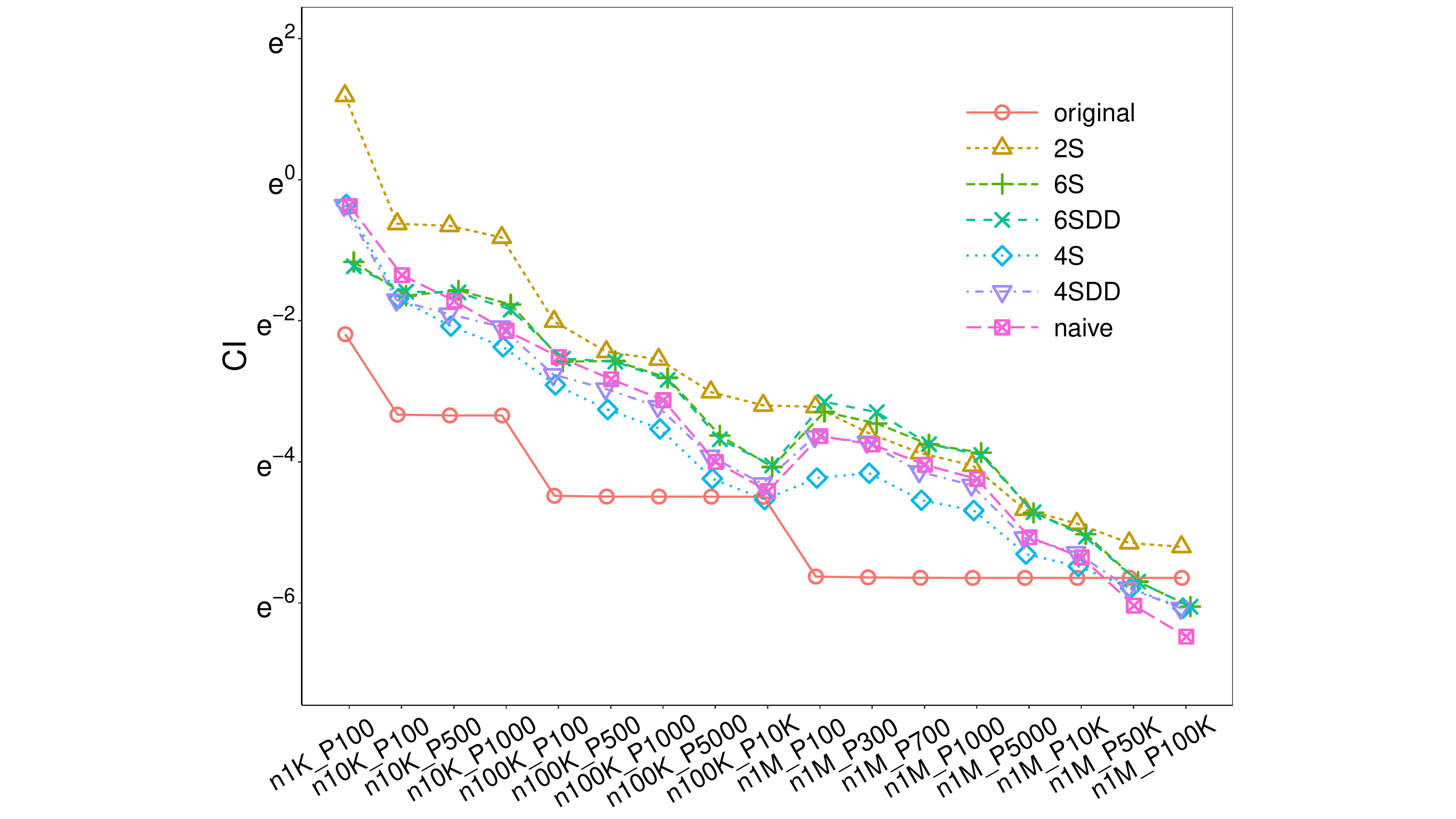}
\includegraphics[width=0.215\textwidth, trim={2.2in 0 2.2in 0},clip] {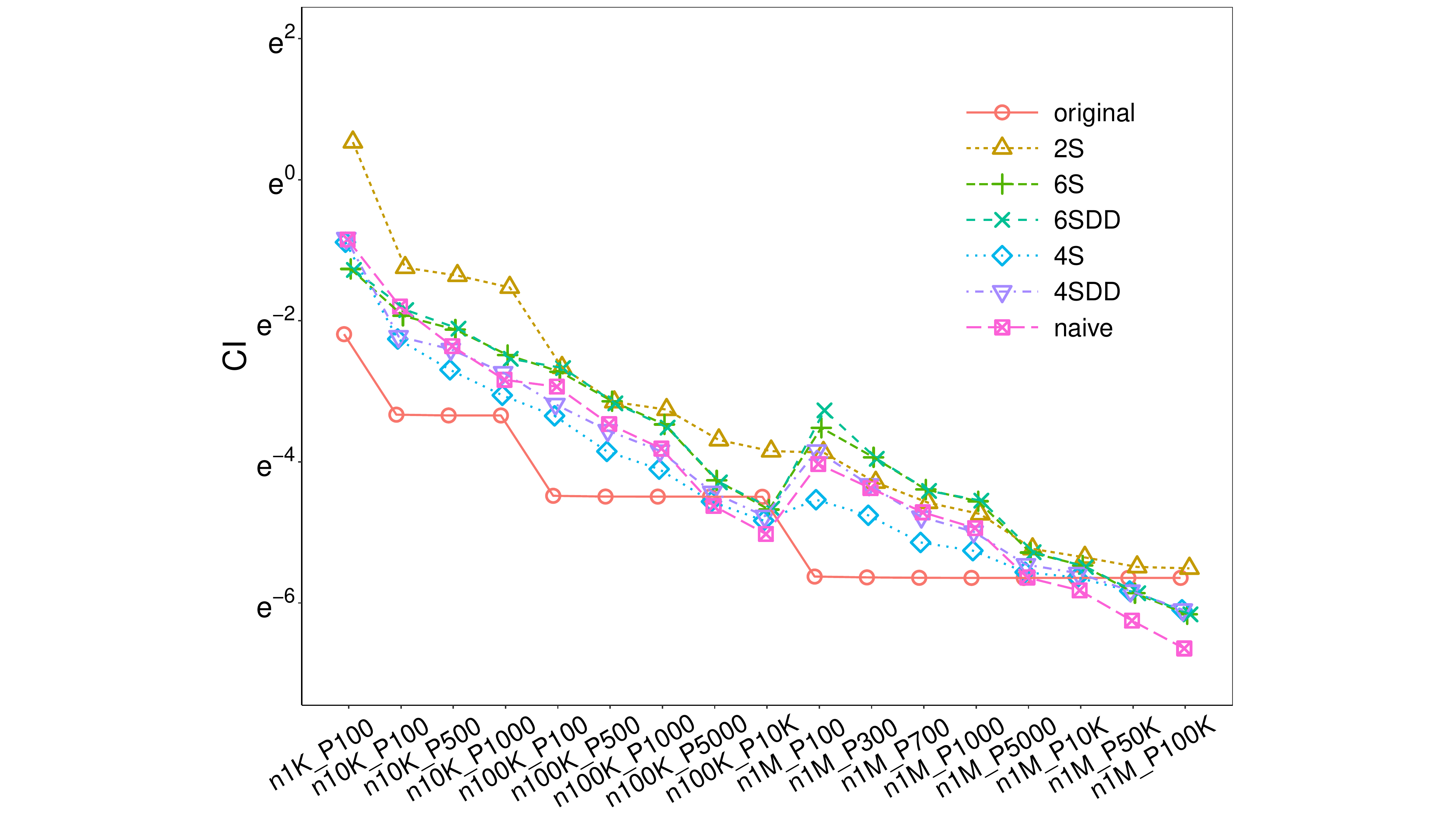}
\includegraphics[width=0.215\textwidth, trim={2.2in 0 2.2in 0},clip] {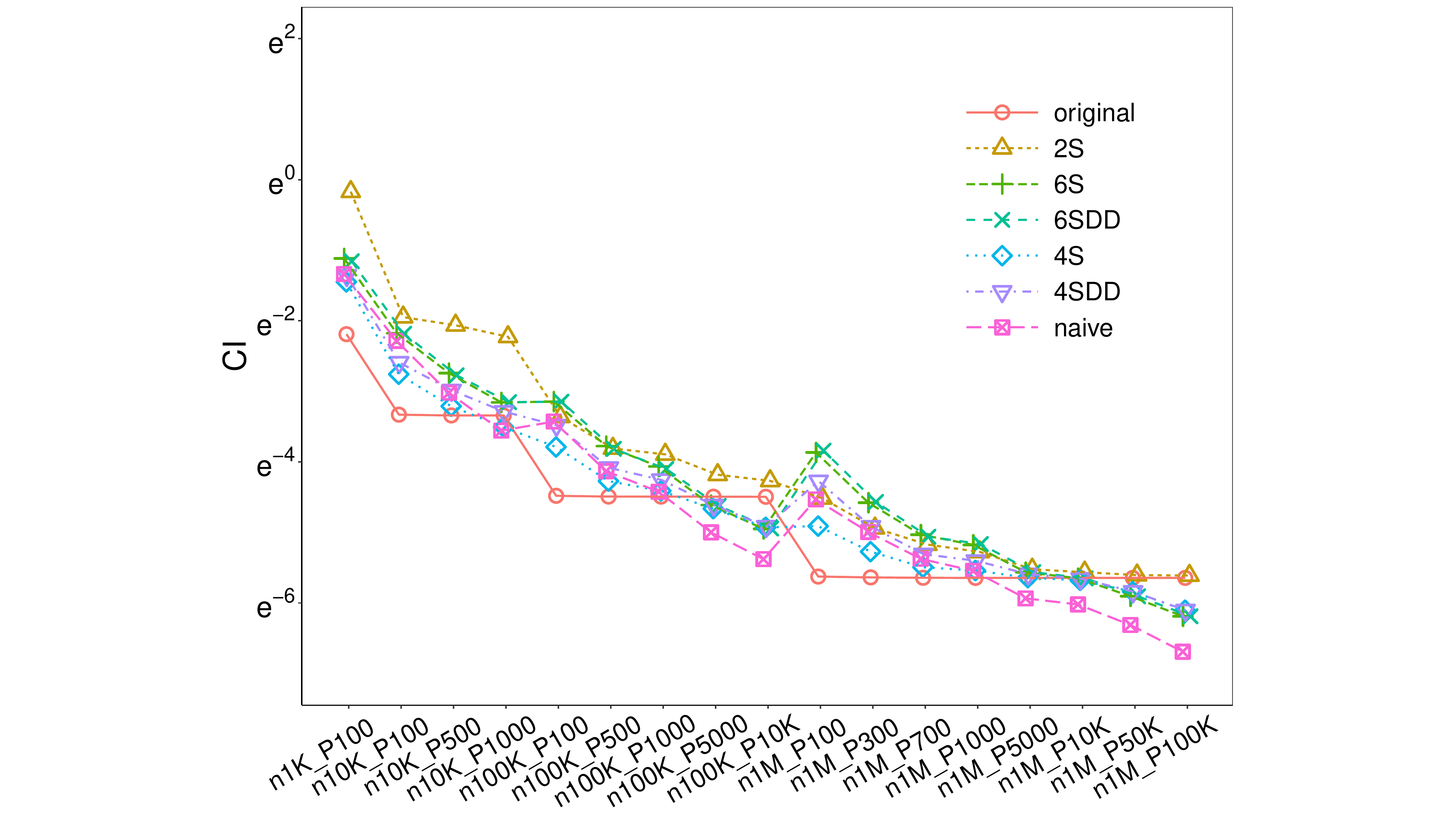}
\includegraphics[width=0.215\textwidth, trim={2.2in 0 2.2in 0},clip] {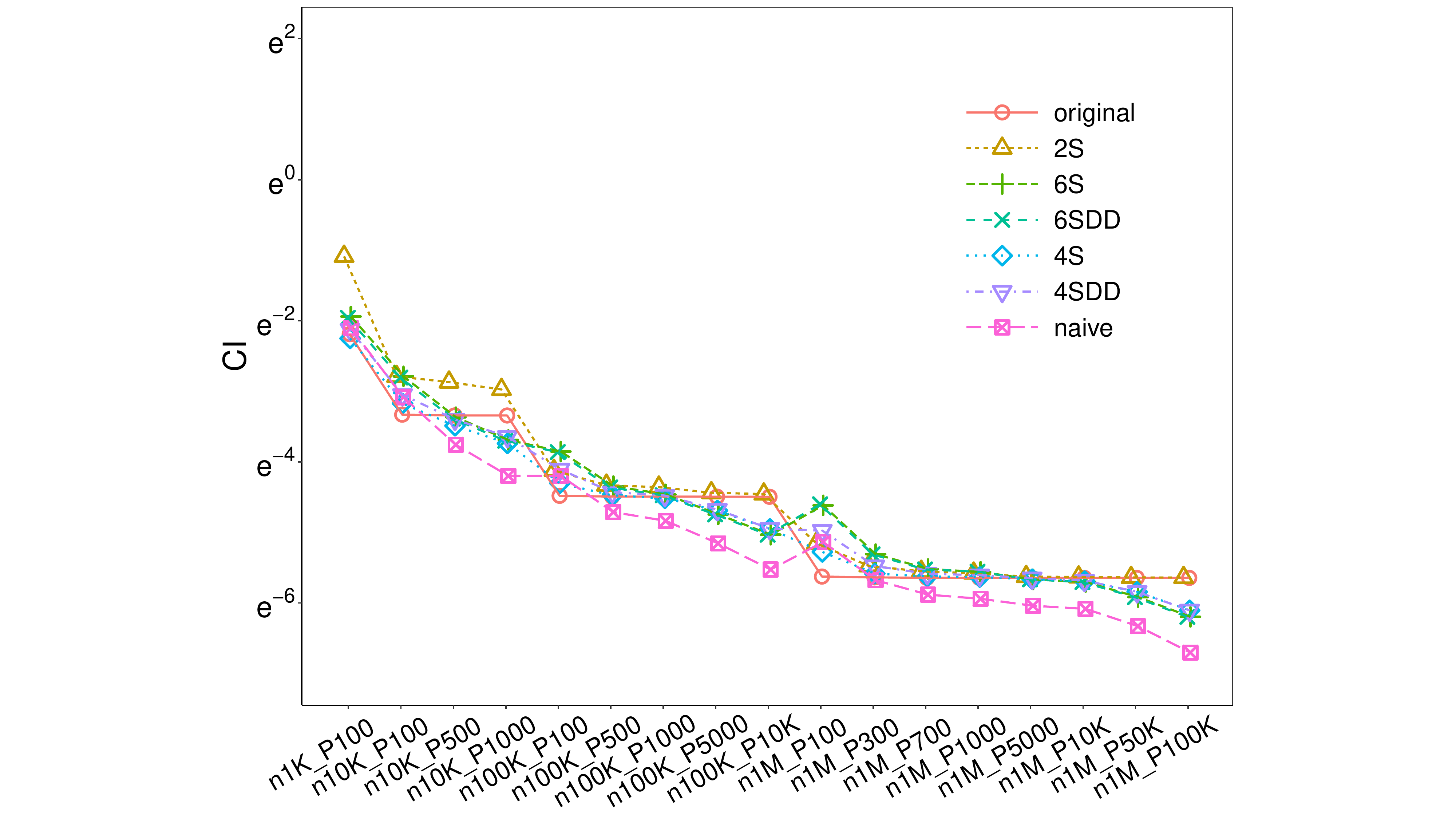}
\includegraphics[width=0.215\textwidth, trim={2.2in 0 2.2in 0},clip] {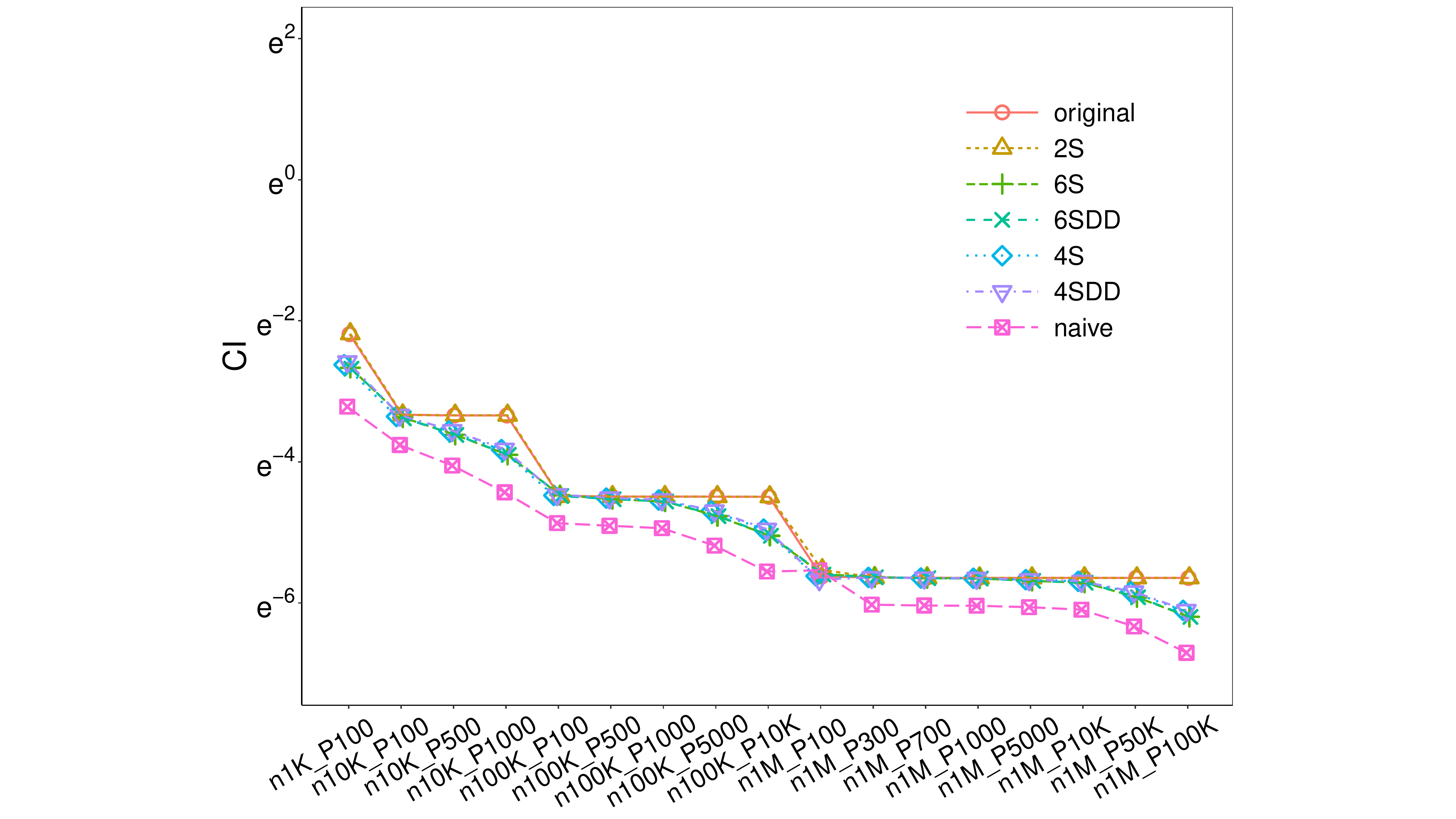}\\
\includegraphics[width=0.215\textwidth, trim={2.2in 0 2.2in 0},clip] {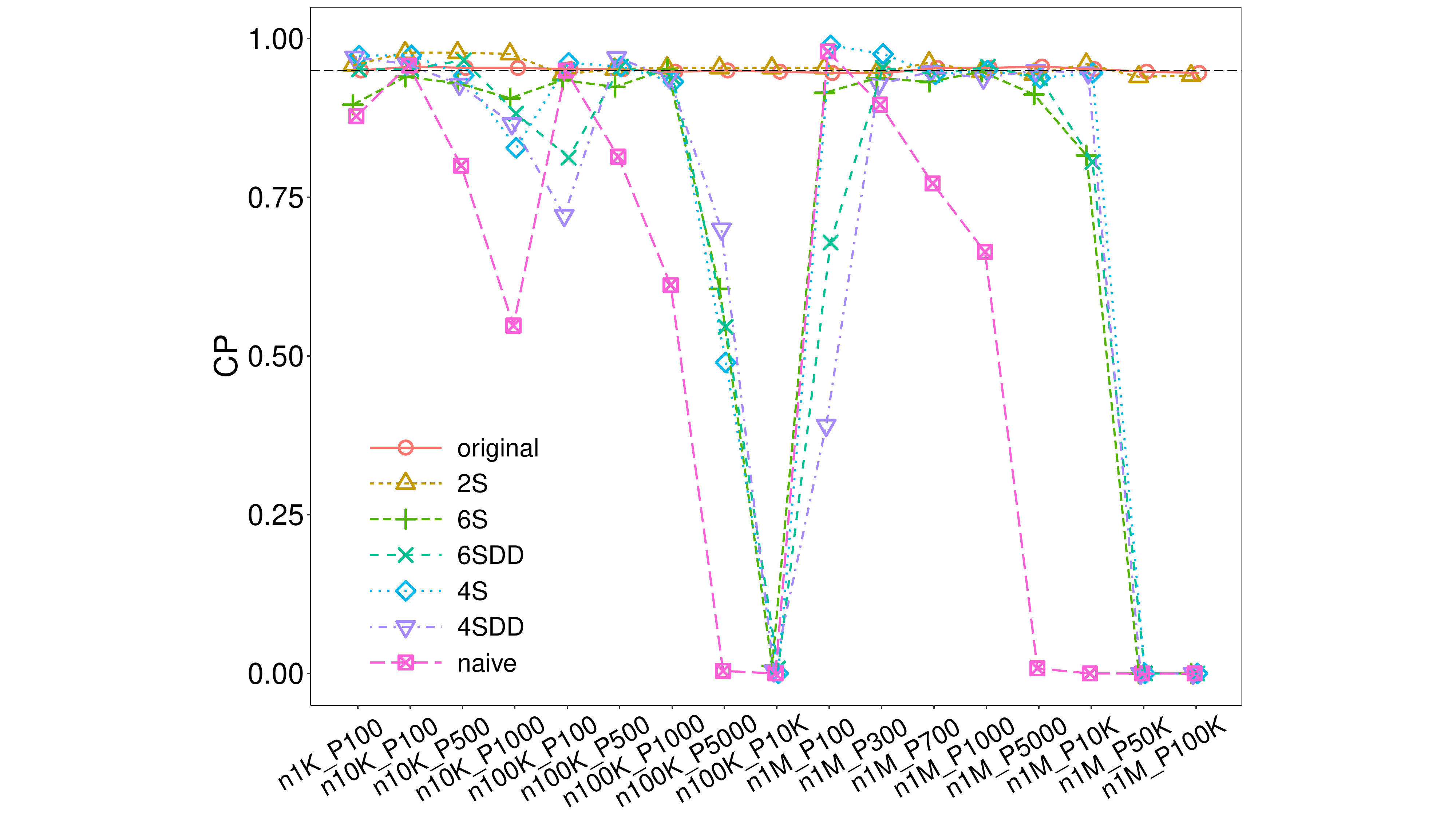}
\includegraphics[width=0.215\textwidth, trim={2.2in 0 2.2in 0},clip] {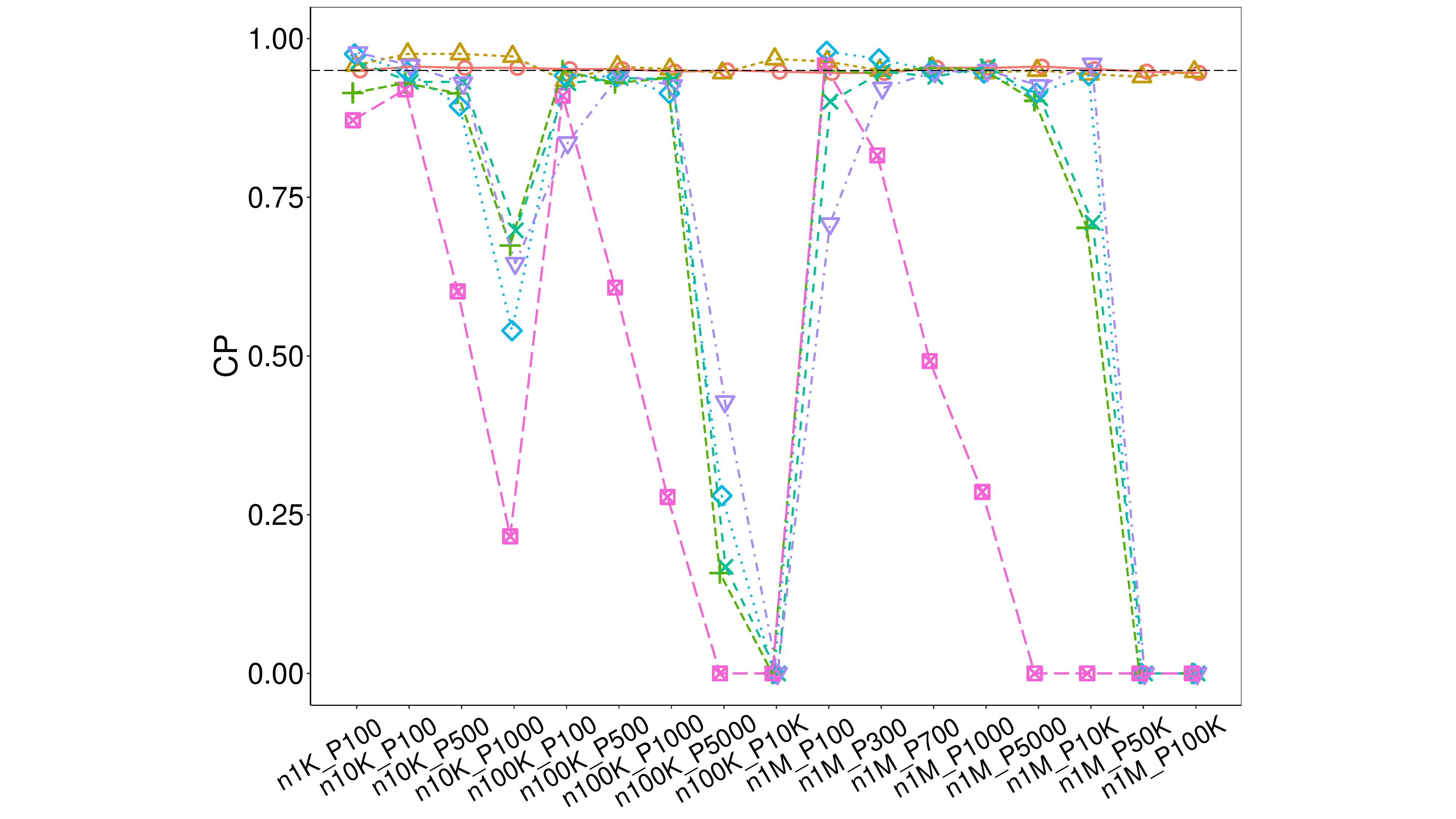}
\includegraphics[width=0.215\textwidth, trim={2.2in 0 2.2in 0},clip] {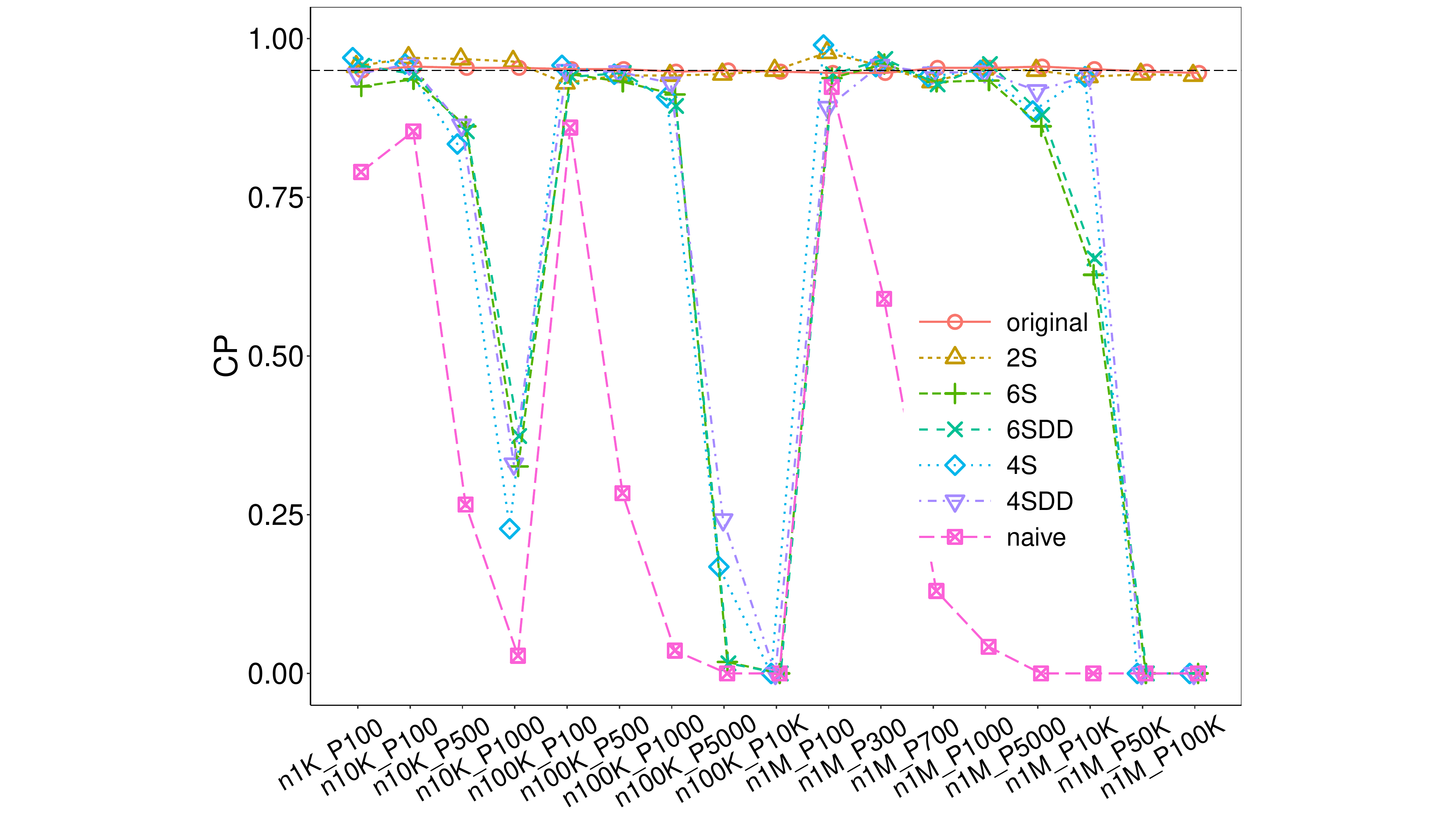}
\includegraphics[width=0.215\textwidth, trim={2.2in 0 2.2in 0},clip] {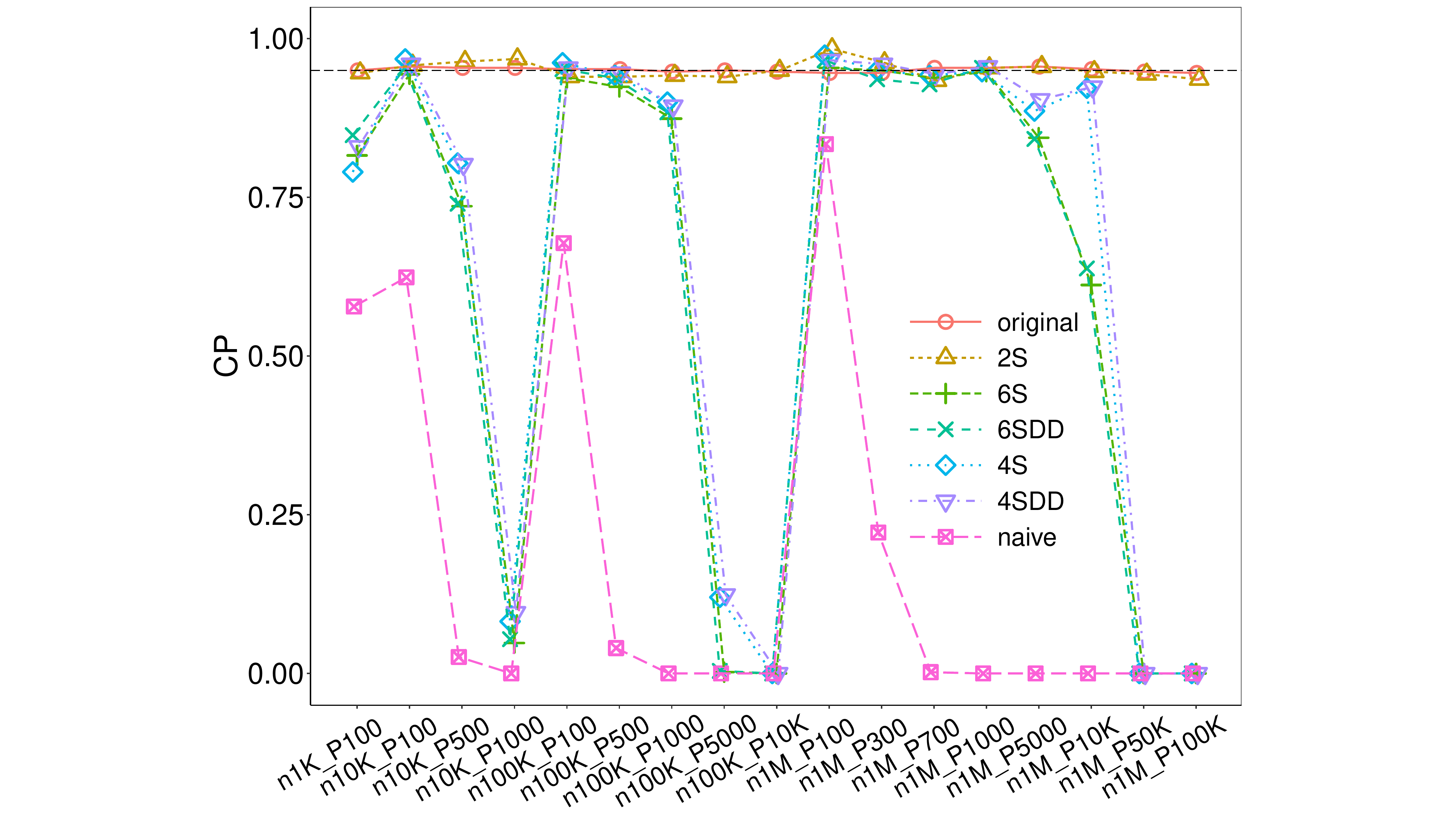}
\includegraphics[width=0.215\textwidth, trim={2.2in 0 2.2in 0},clip] {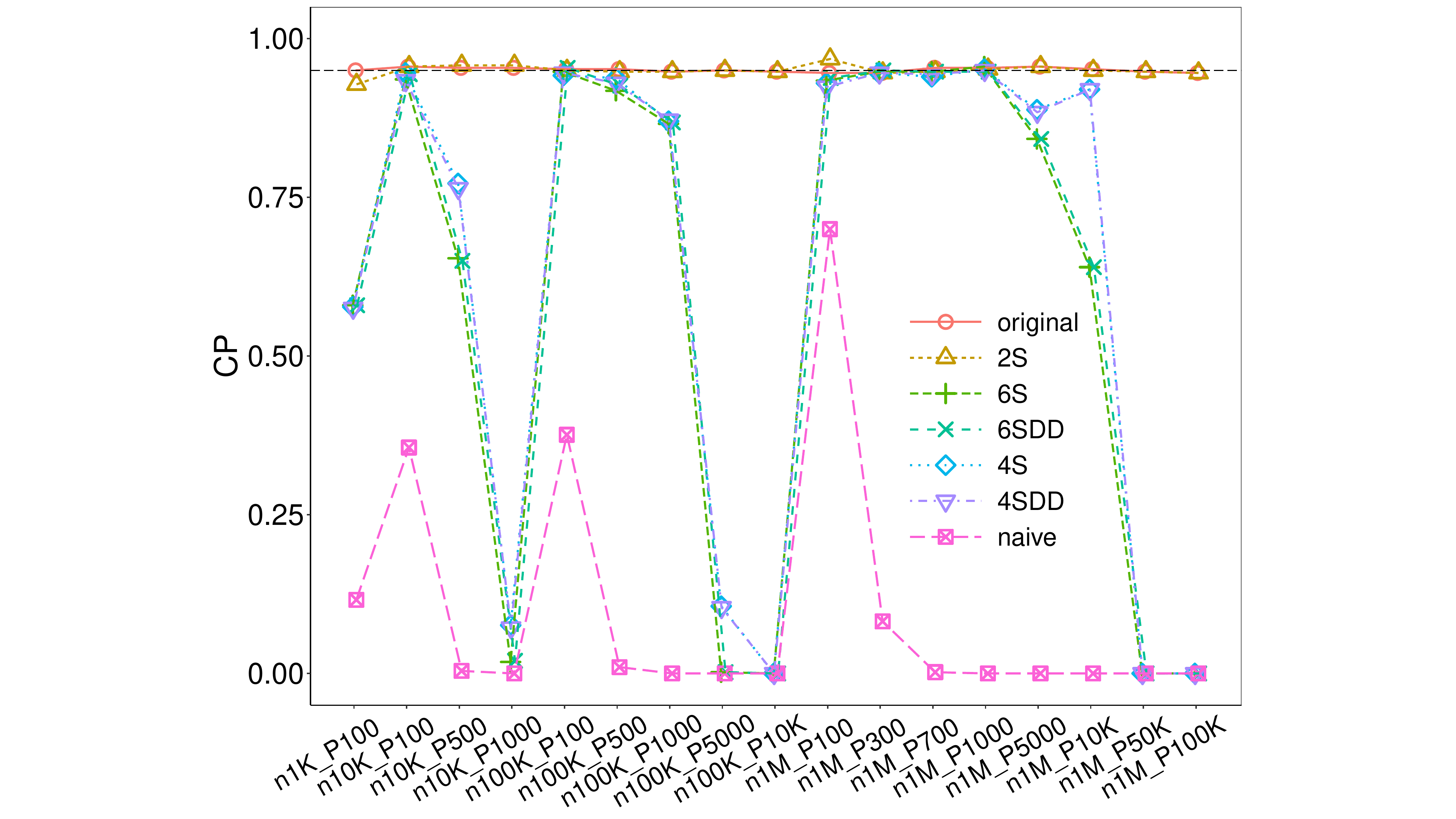}\\
\includegraphics[width=0.215\textwidth, trim={2.2in 0 2.2in 0},clip] {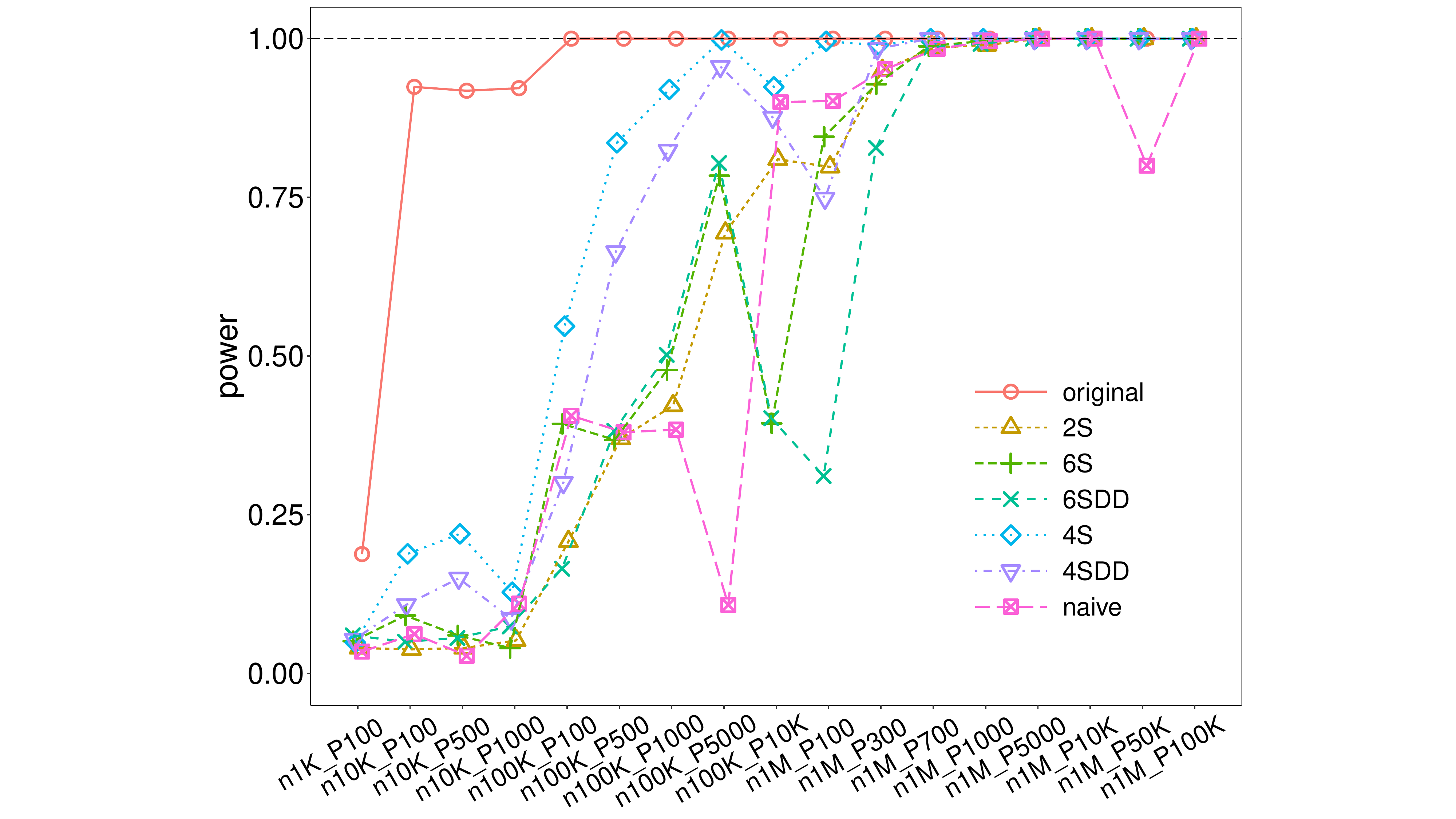}
\includegraphics[width=0.215\textwidth, trim={2.2in 0 2.2in 0},clip] {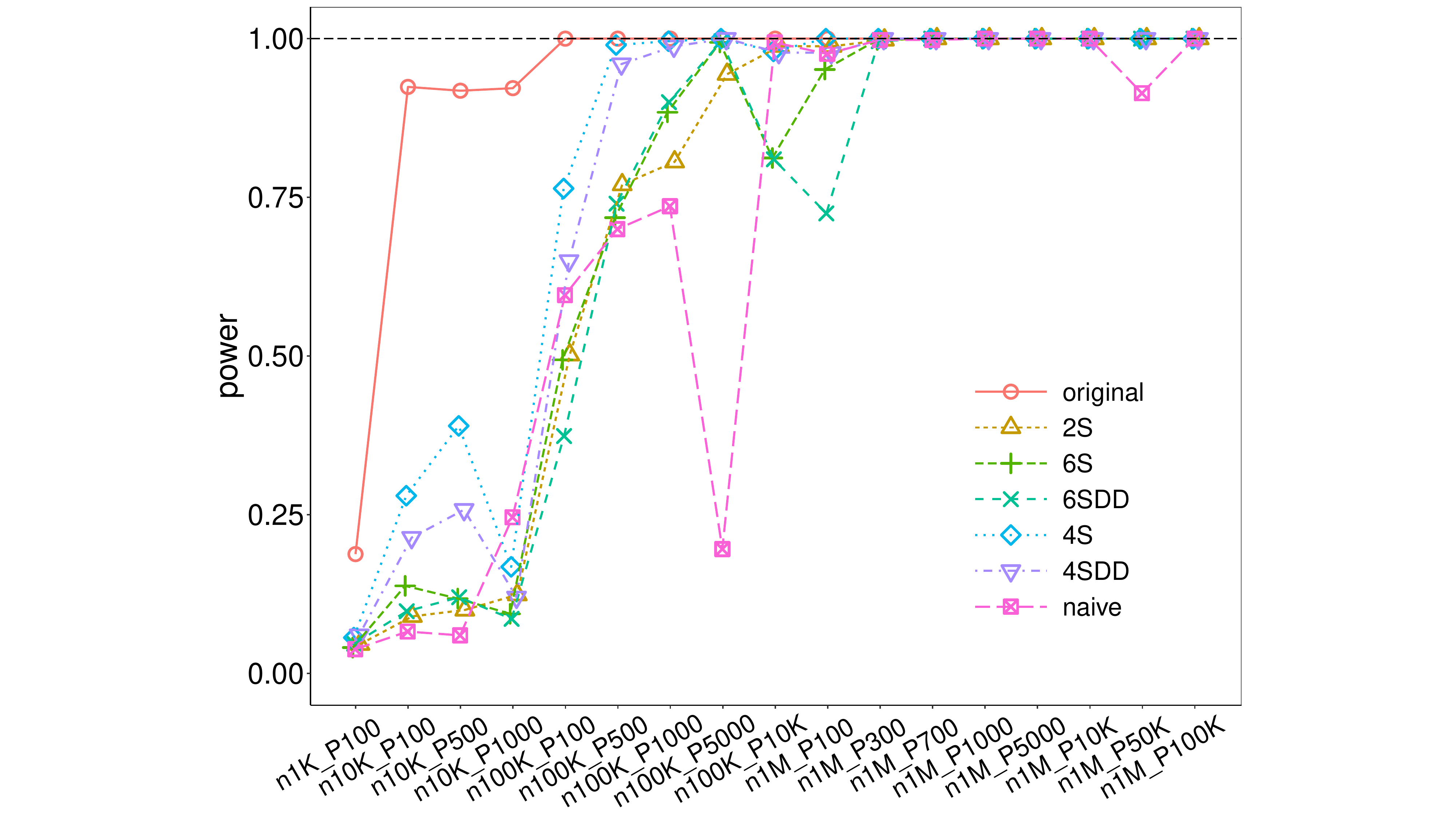}
\includegraphics[width=0.215\textwidth, trim={2.2in 0 2.2in 0},clip] {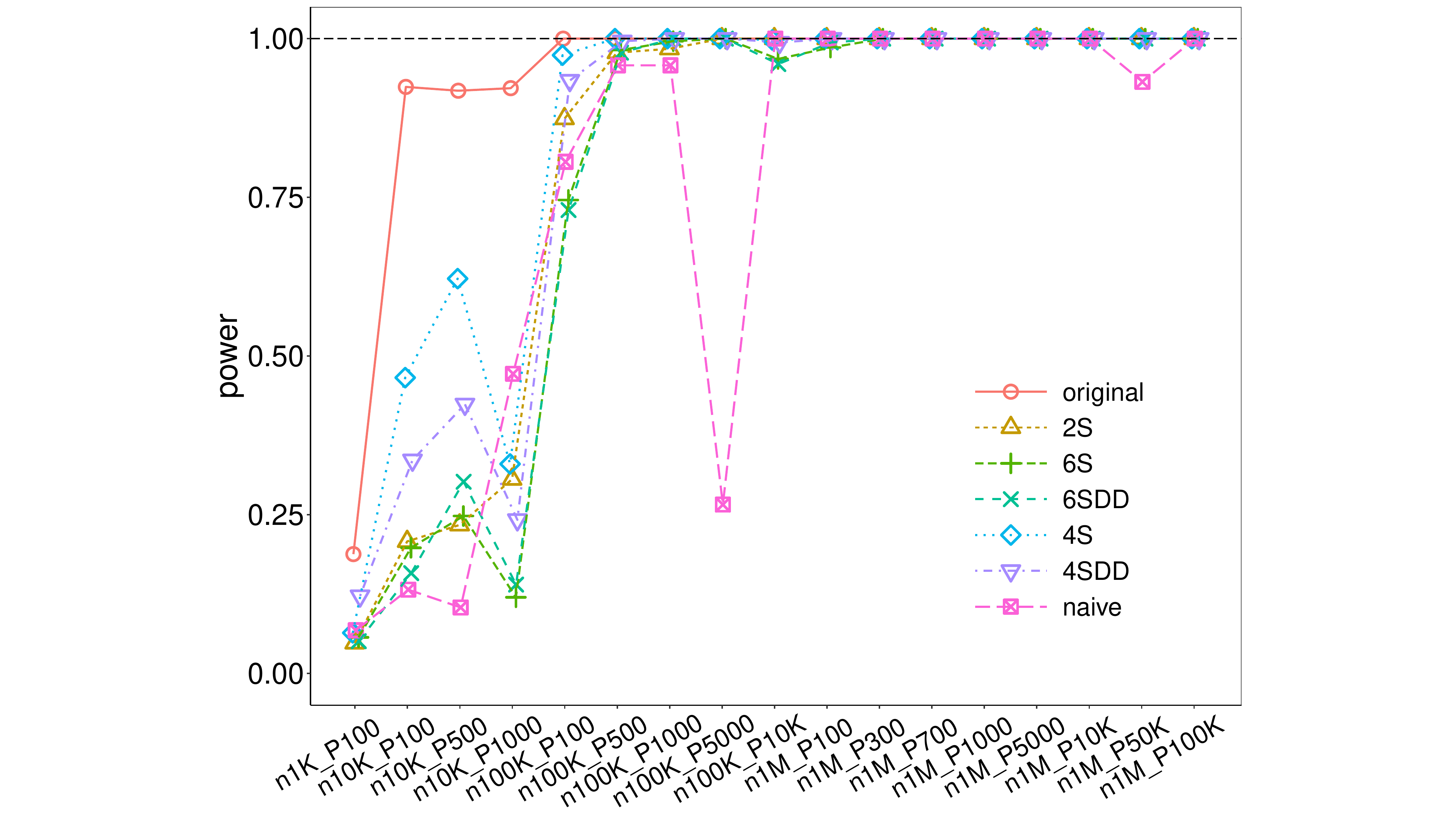}
\includegraphics[width=0.215\textwidth, trim={2.2in 0 2.2in 0},clip] {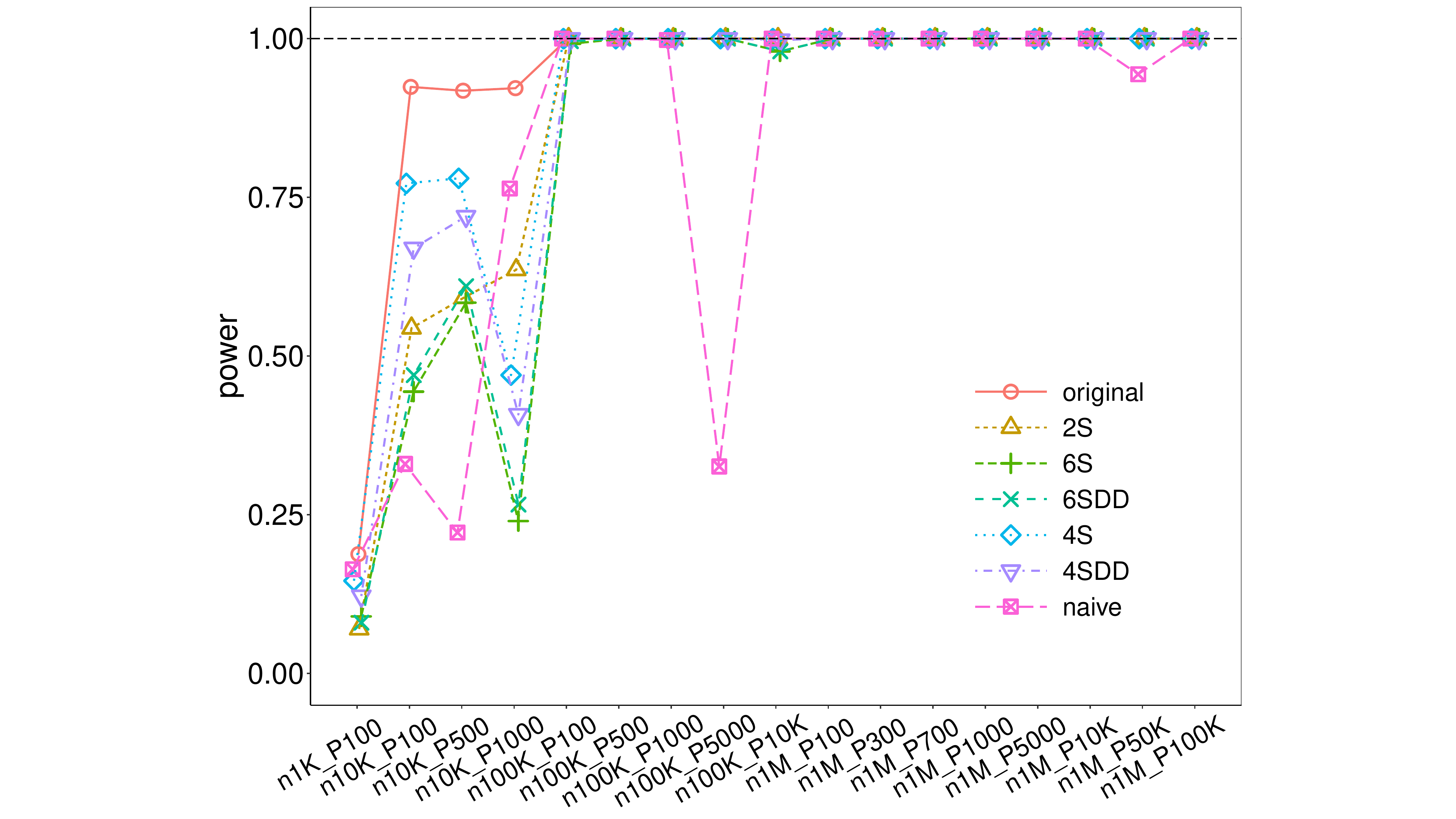}
\includegraphics[width=0.215\textwidth, trim={2.2in 0 2.2in 0},clip] {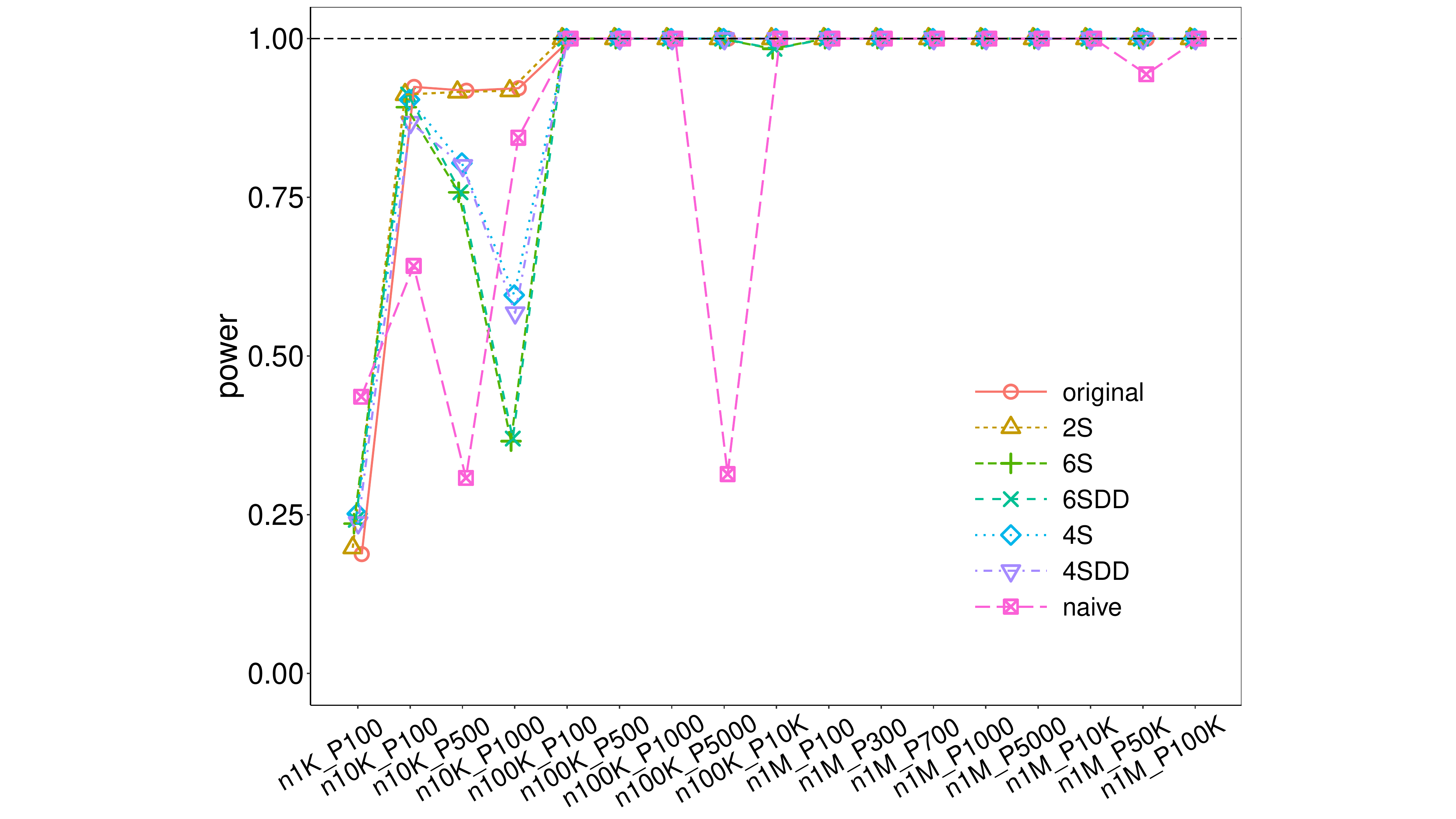}\\
\caption{ZINB data; $\epsilon$-DP; $\theta\ne0$ and $\alpha\ne\beta$} \label{fig:1asDPzinb}
\end{figure}
\end{landscape}

\begin{landscape}
\begin{figure}[!htb]
\centering
$\rho=0.005$\hspace{1in}$\rho=0.02$\hspace{1in}$\rho=0.08$
\hspace{1in}$\rho=0.32$\hspace{0.8in}$\rho=1.28$\\
\includegraphics[width=0.24\textwidth, trim={2.2in 0 2.2in 0},clip] {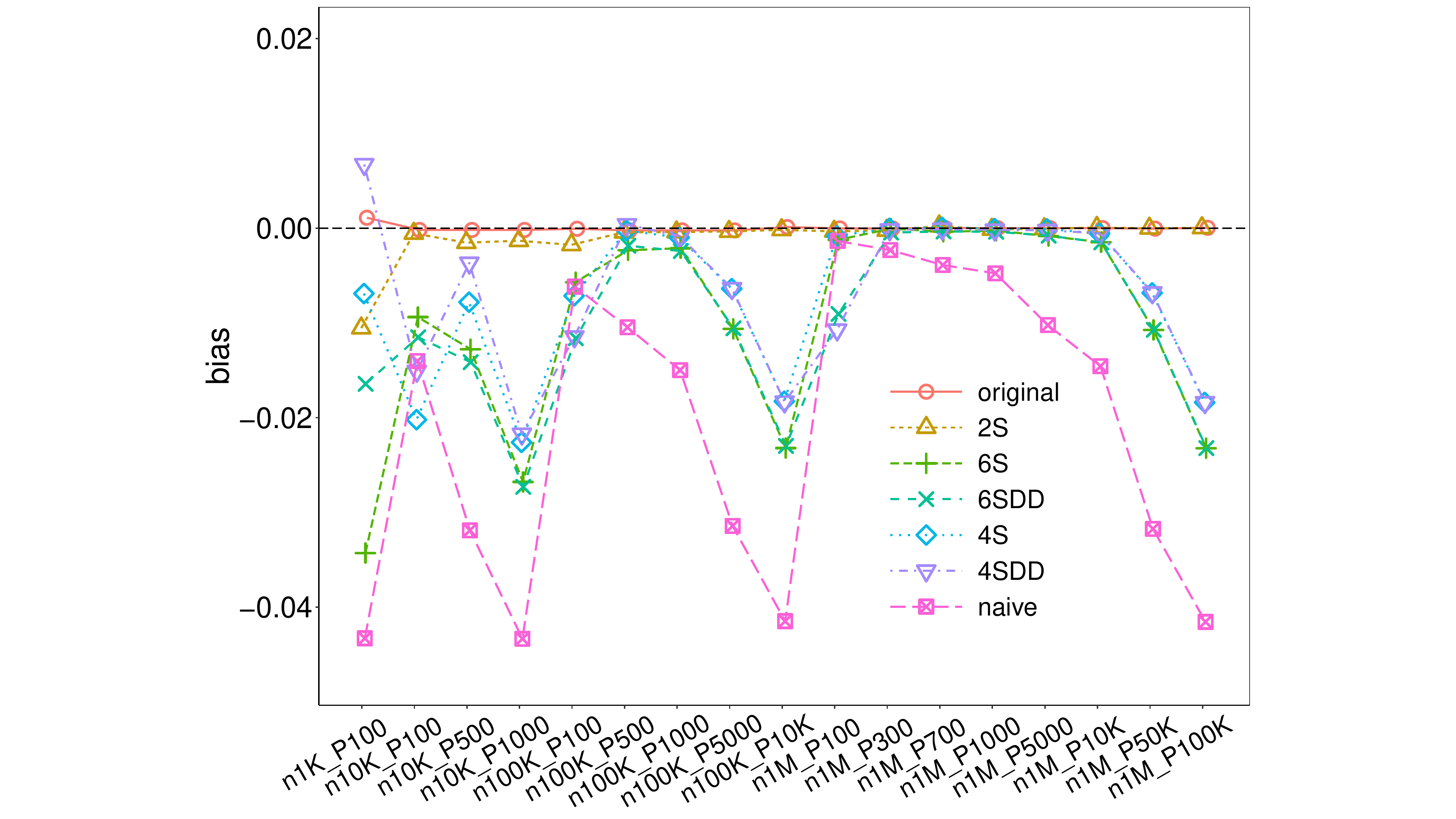}
\includegraphics[width=0.24\textwidth, trim={2.2in 0 2.2in 0},clip] {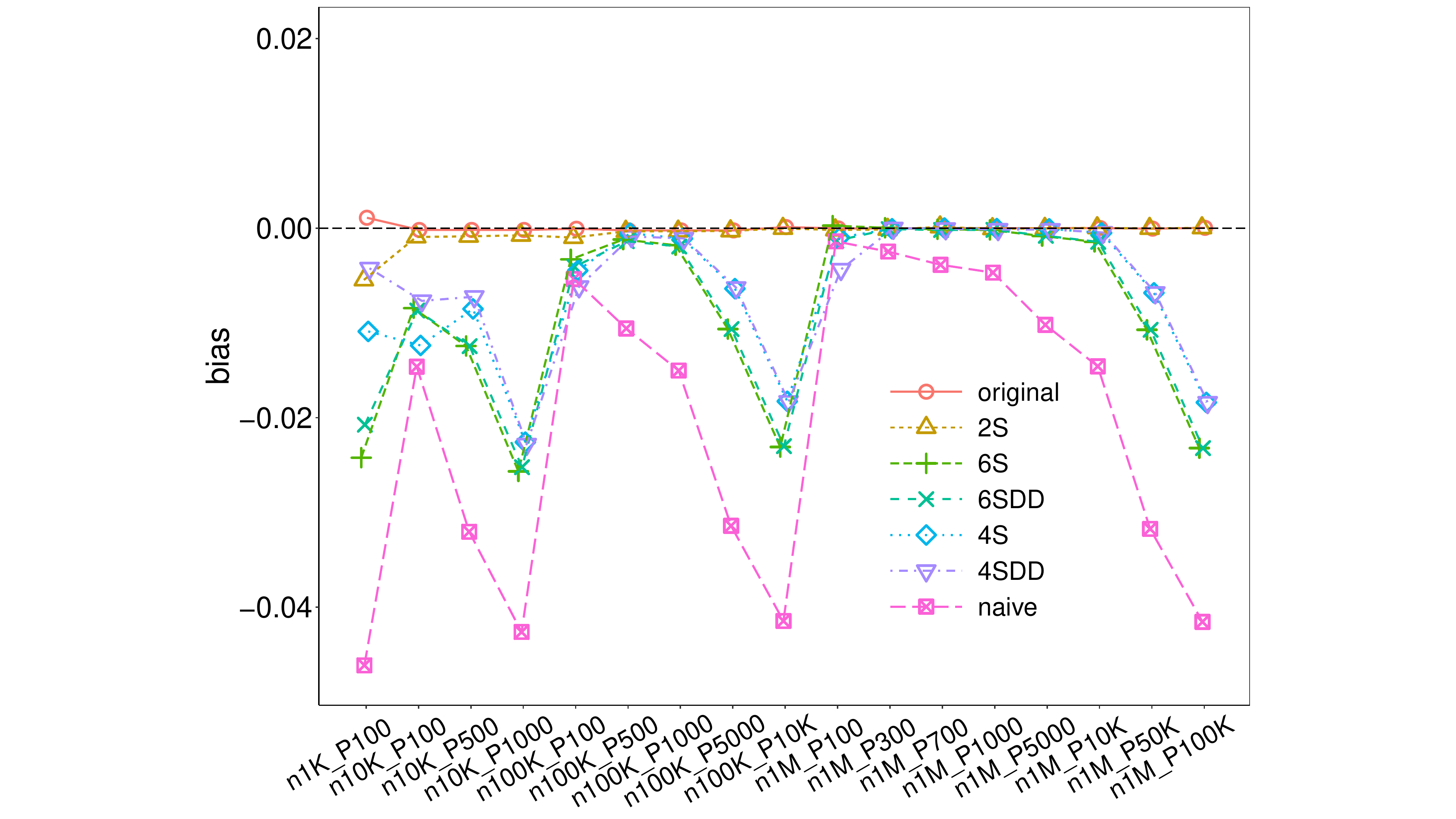}
\includegraphics[width=0.24\textwidth, trim={2.2in 0 2.2in 0},clip] {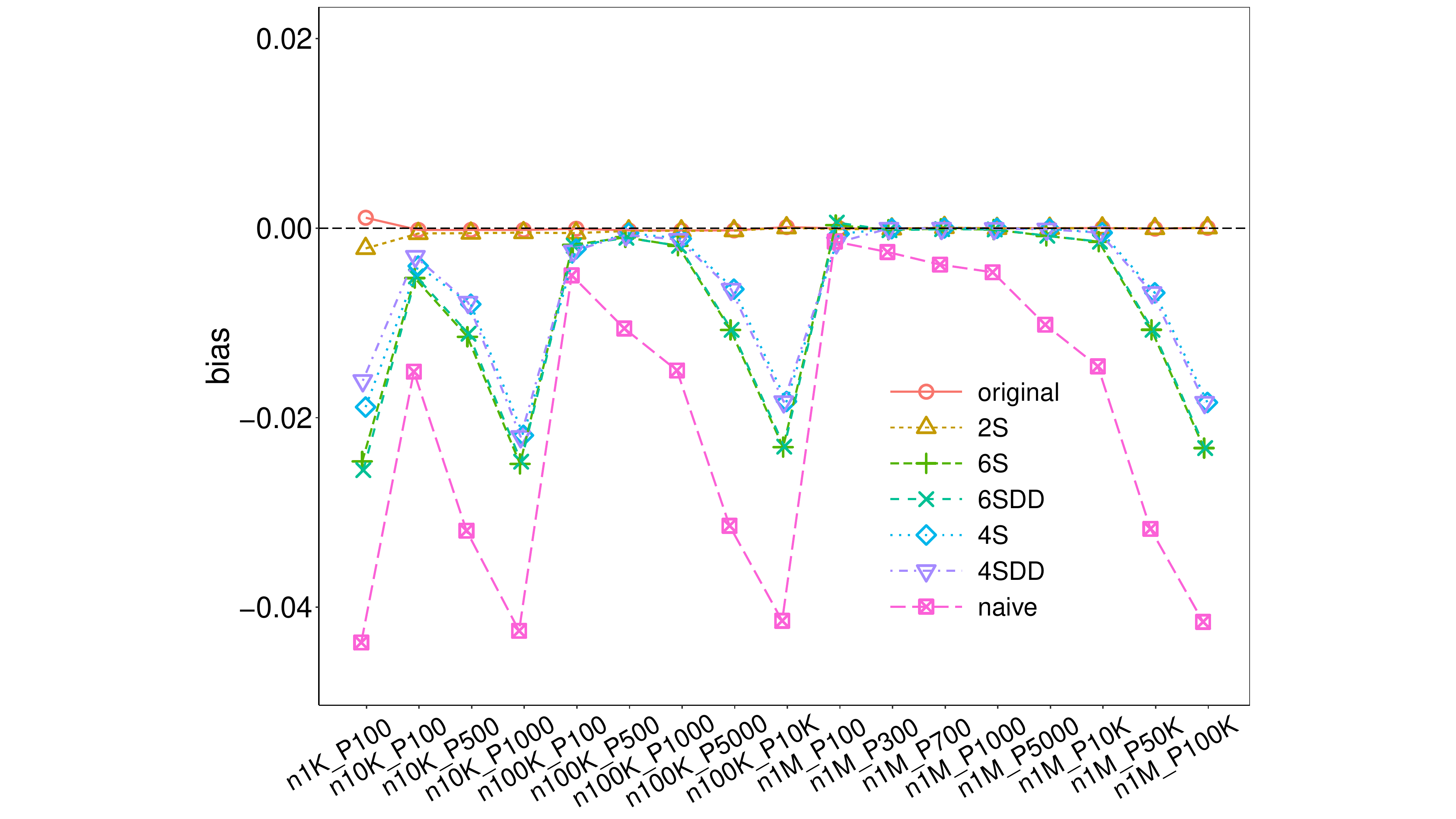}
\includegraphics[width=0.24\textwidth, trim={2.2in 0 2.2in 0},clip] {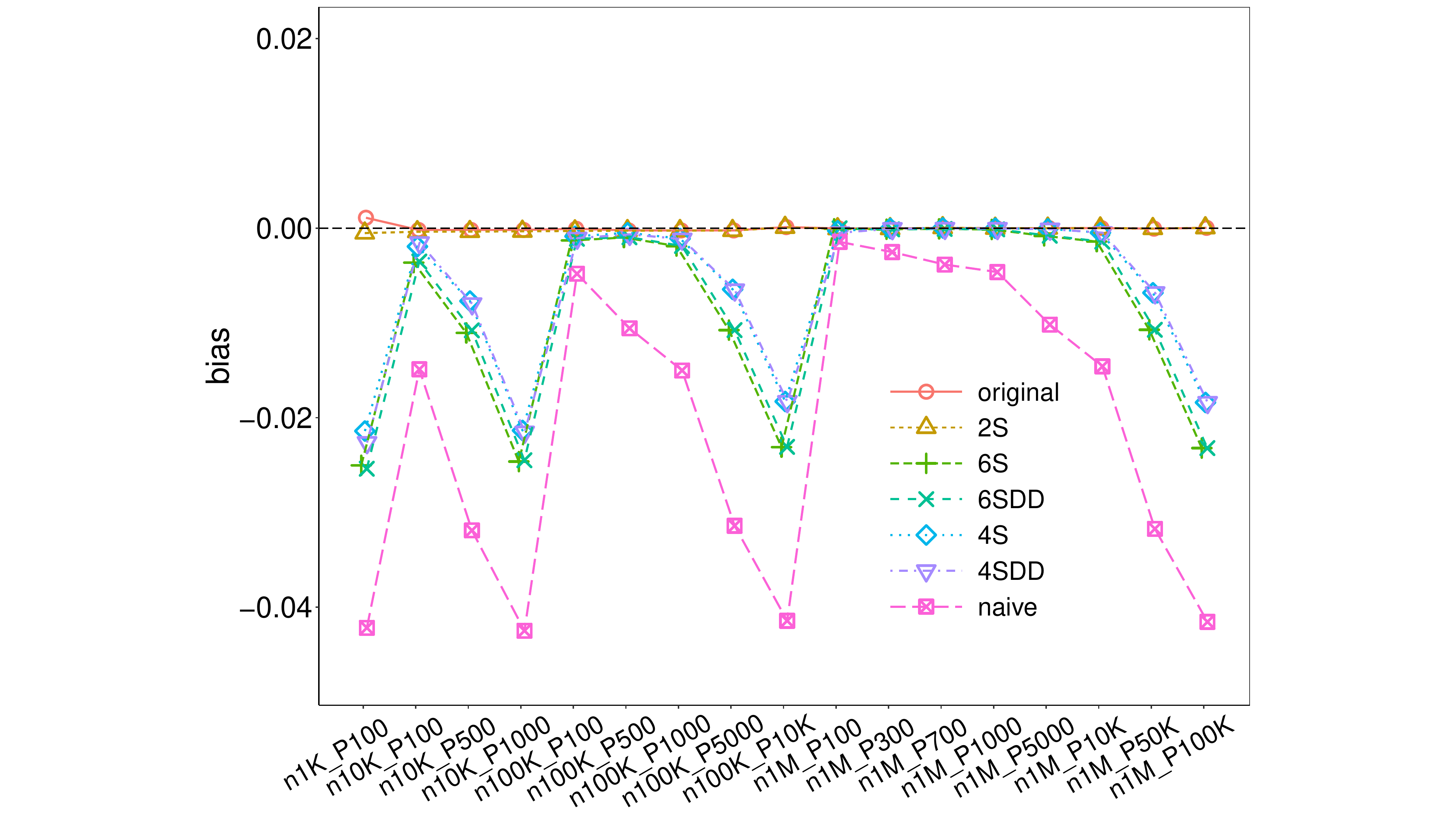}
\includegraphics[width=0.24\textwidth, trim={2.2in 0 2.2in 0},clip] {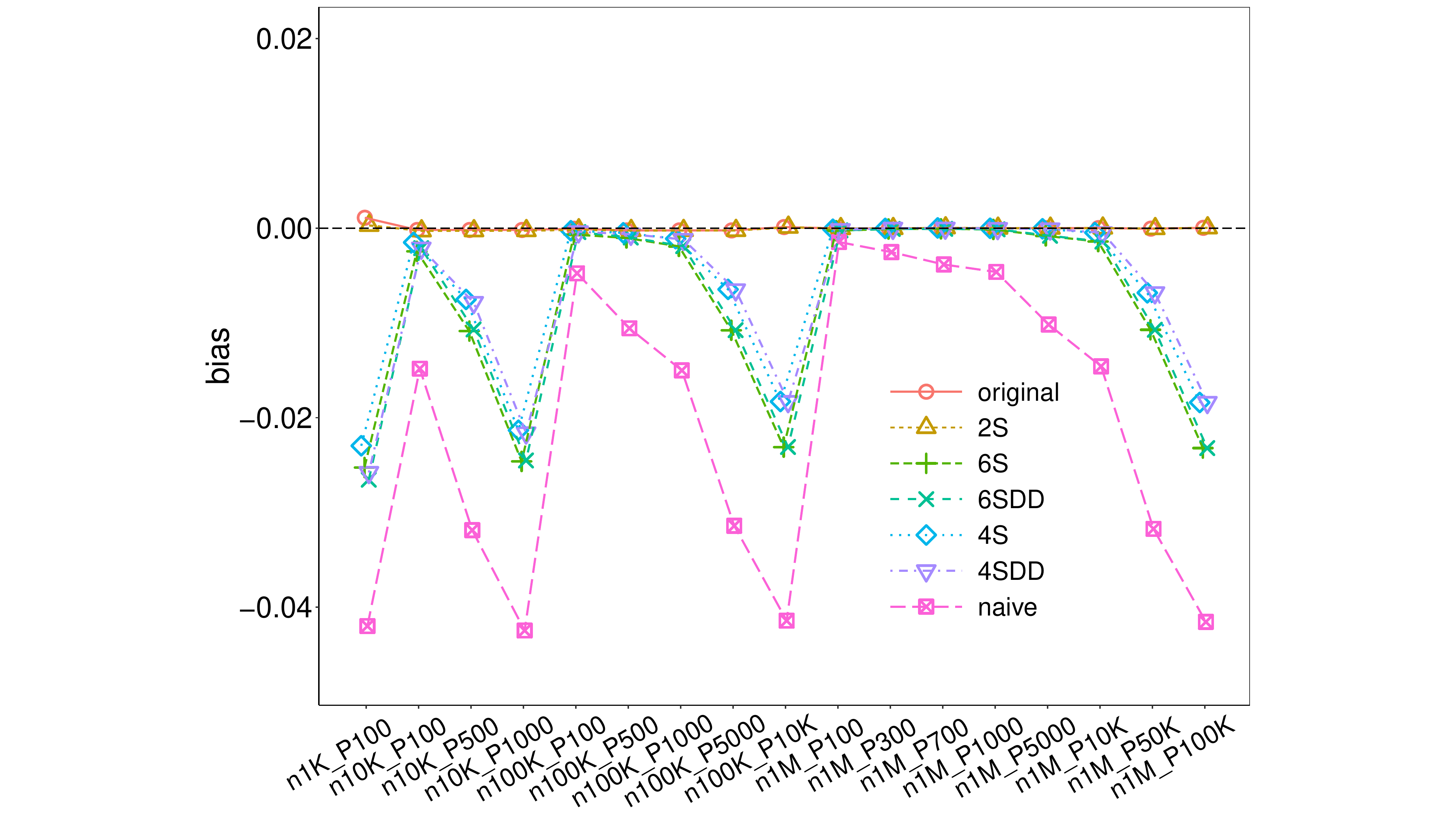}\\
\includegraphics[width=0.24\textwidth, trim={2.2in 0 2.2in 0},clip] {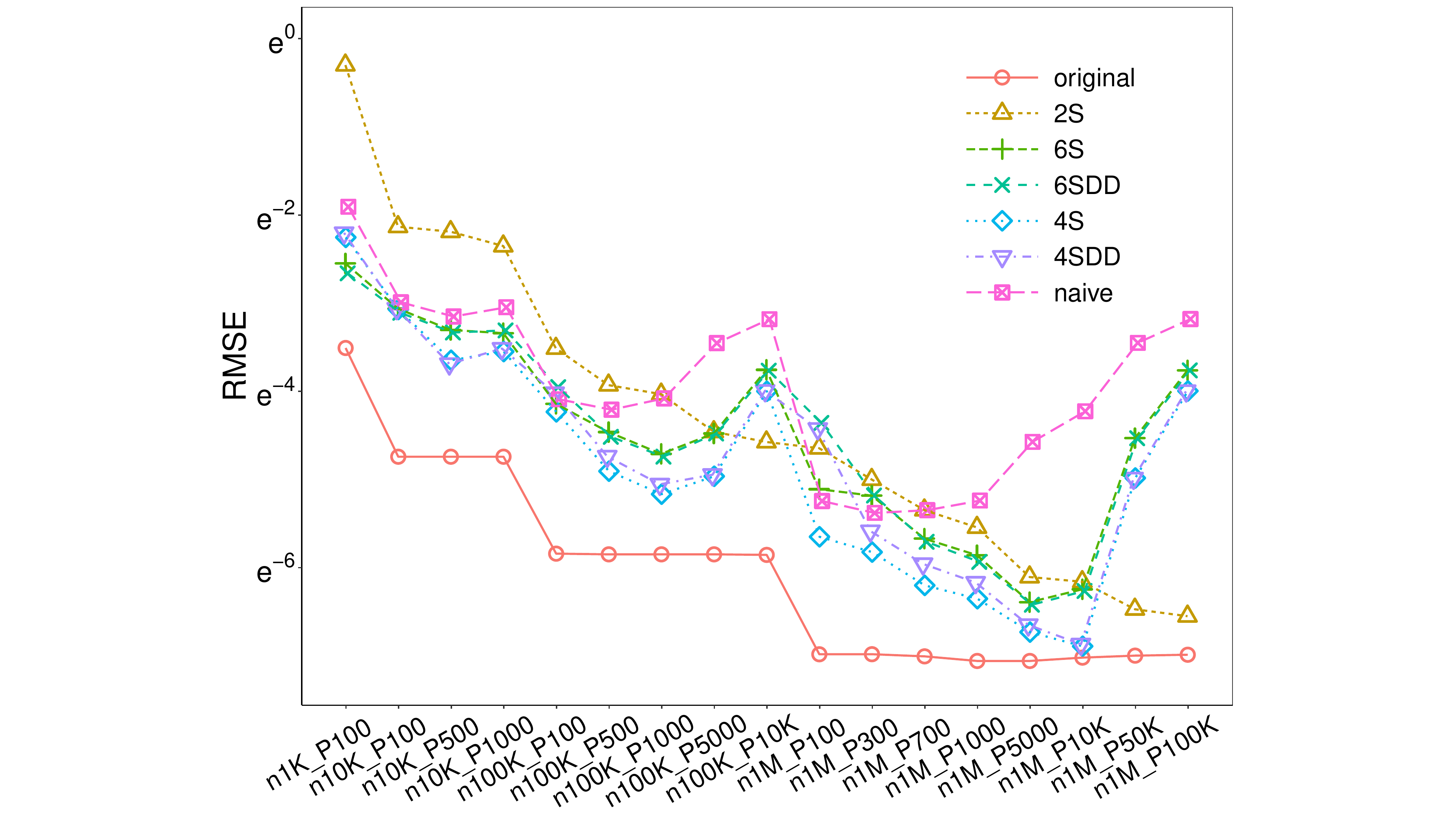}
\includegraphics[width=0.24\textwidth, trim={2.2in 0 2.2in 0},clip] {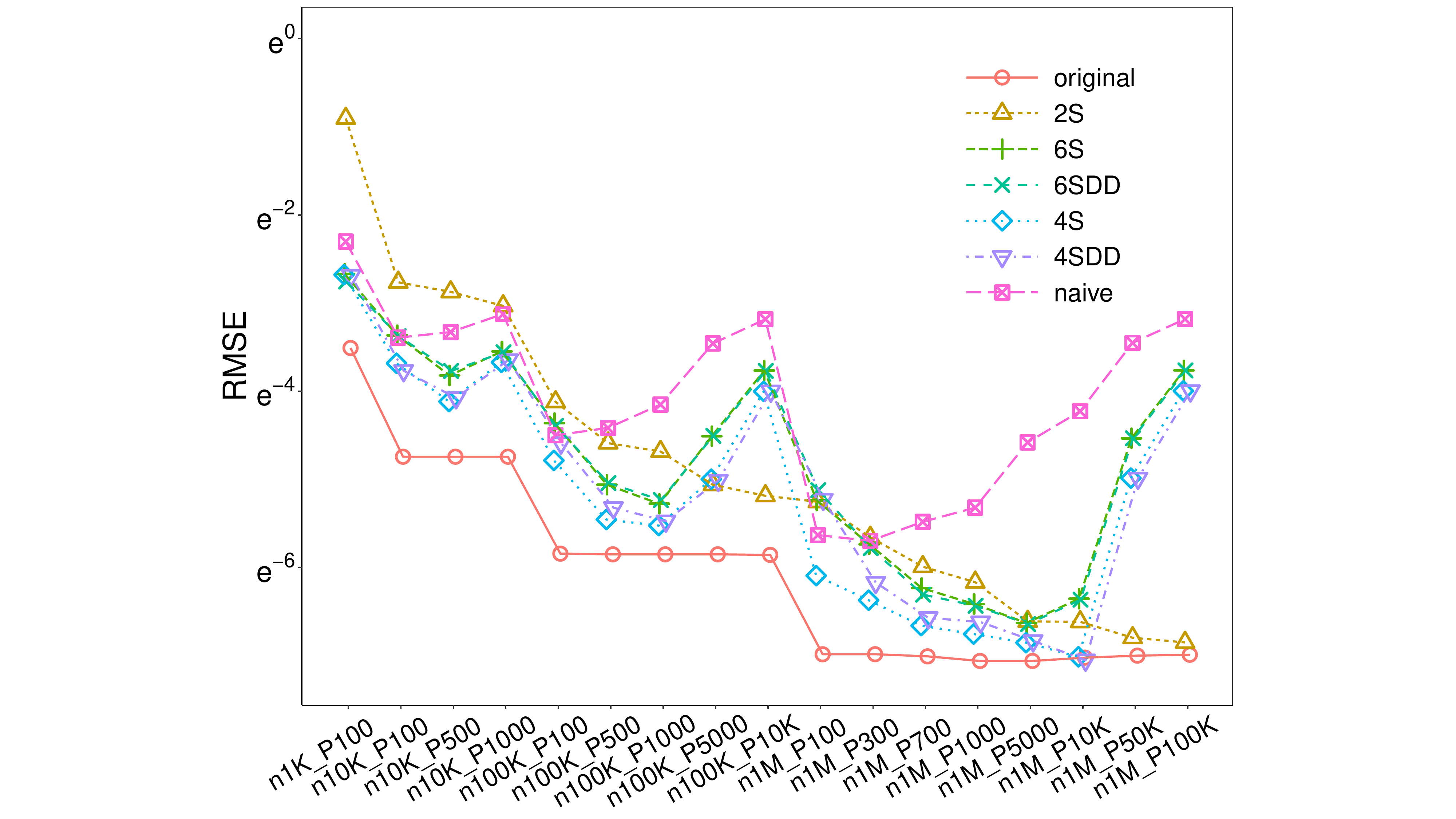}
\includegraphics[width=0.24\textwidth, trim={2.2in 0 2.2in 0},clip] {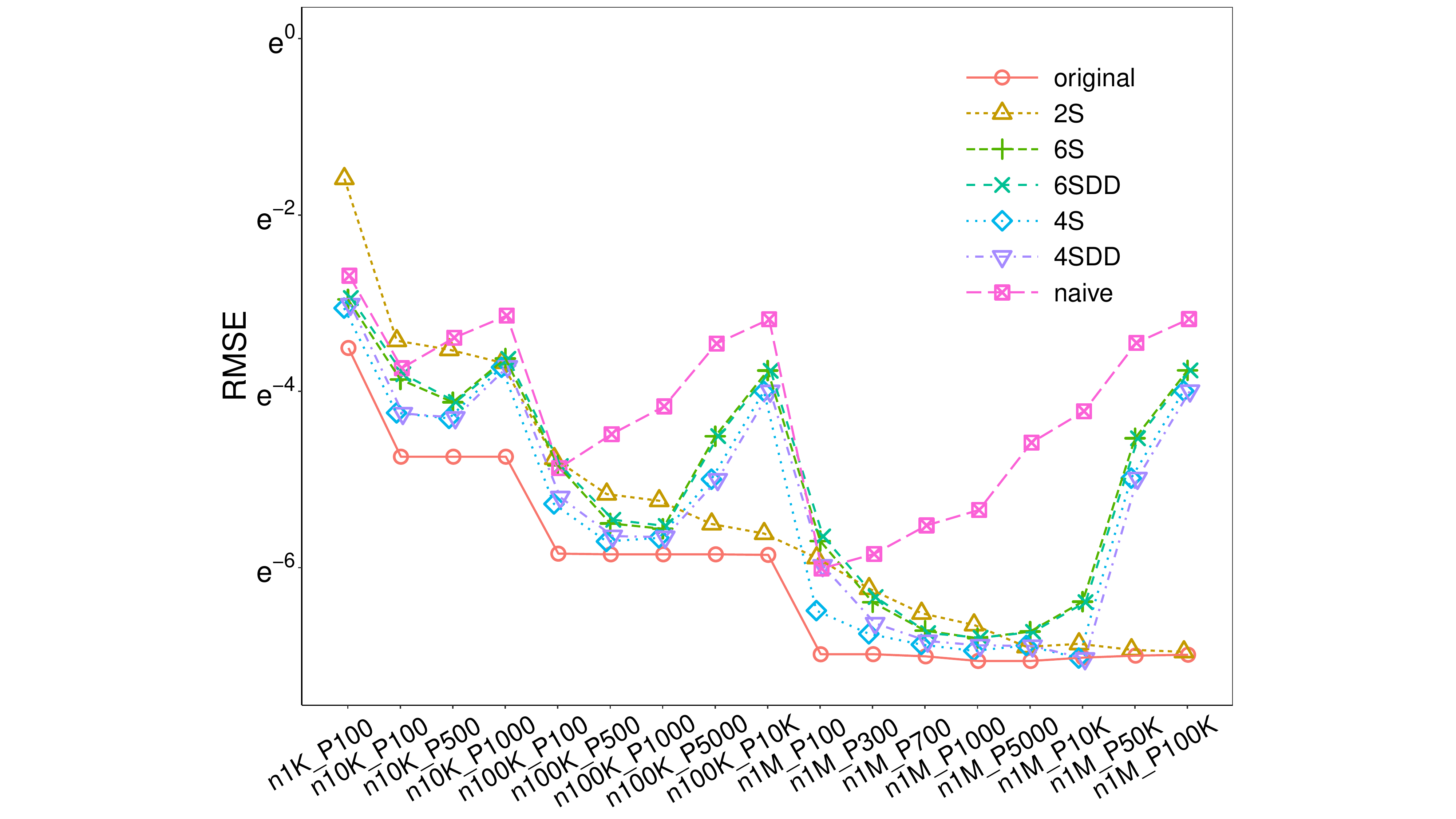}
\includegraphics[width=0.24\textwidth, trim={2.2in 0 2.2in 0},clip] {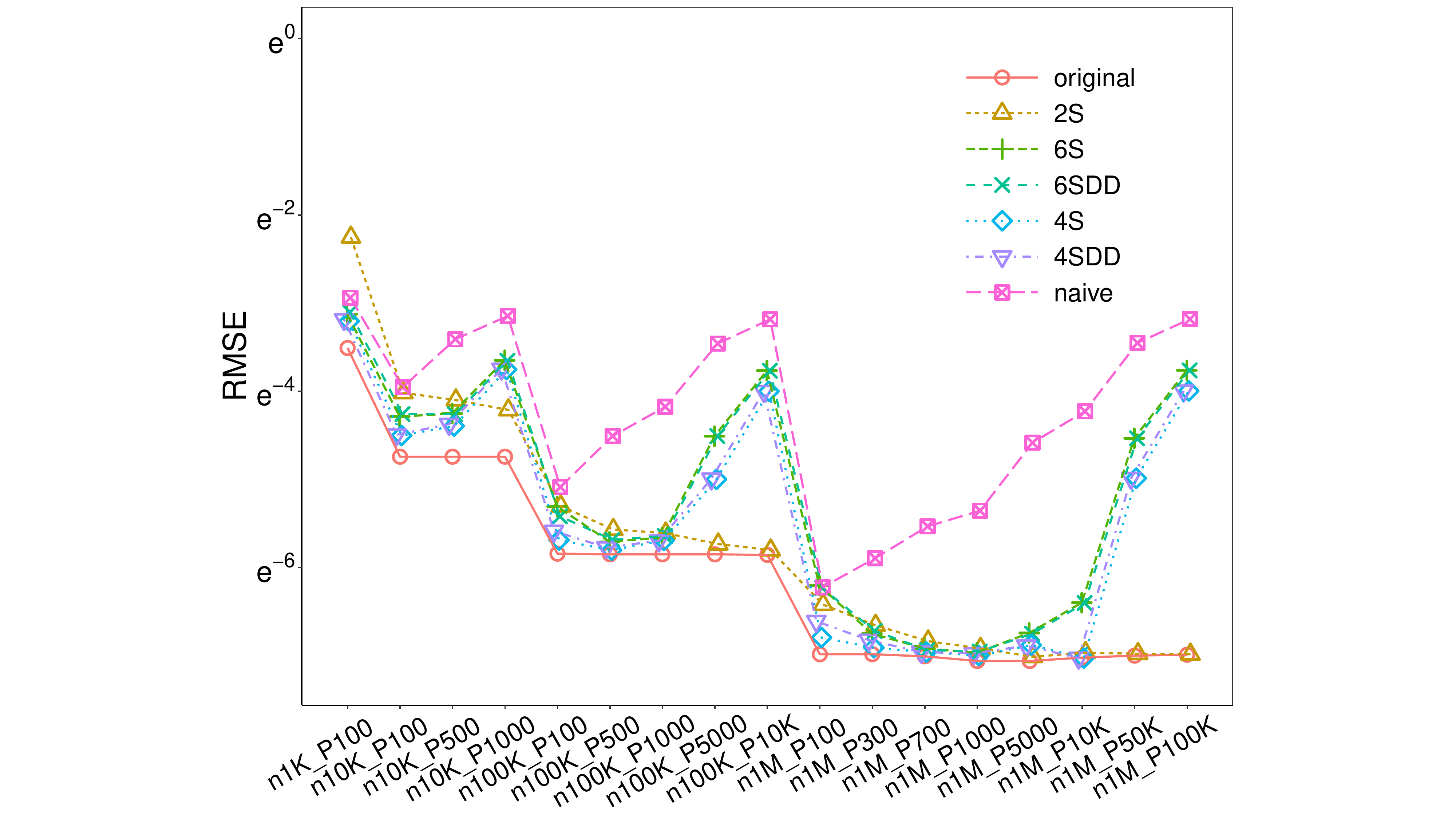}
\includegraphics[width=0.24\textwidth, trim={2.2in 0 2.2in 0},clip] {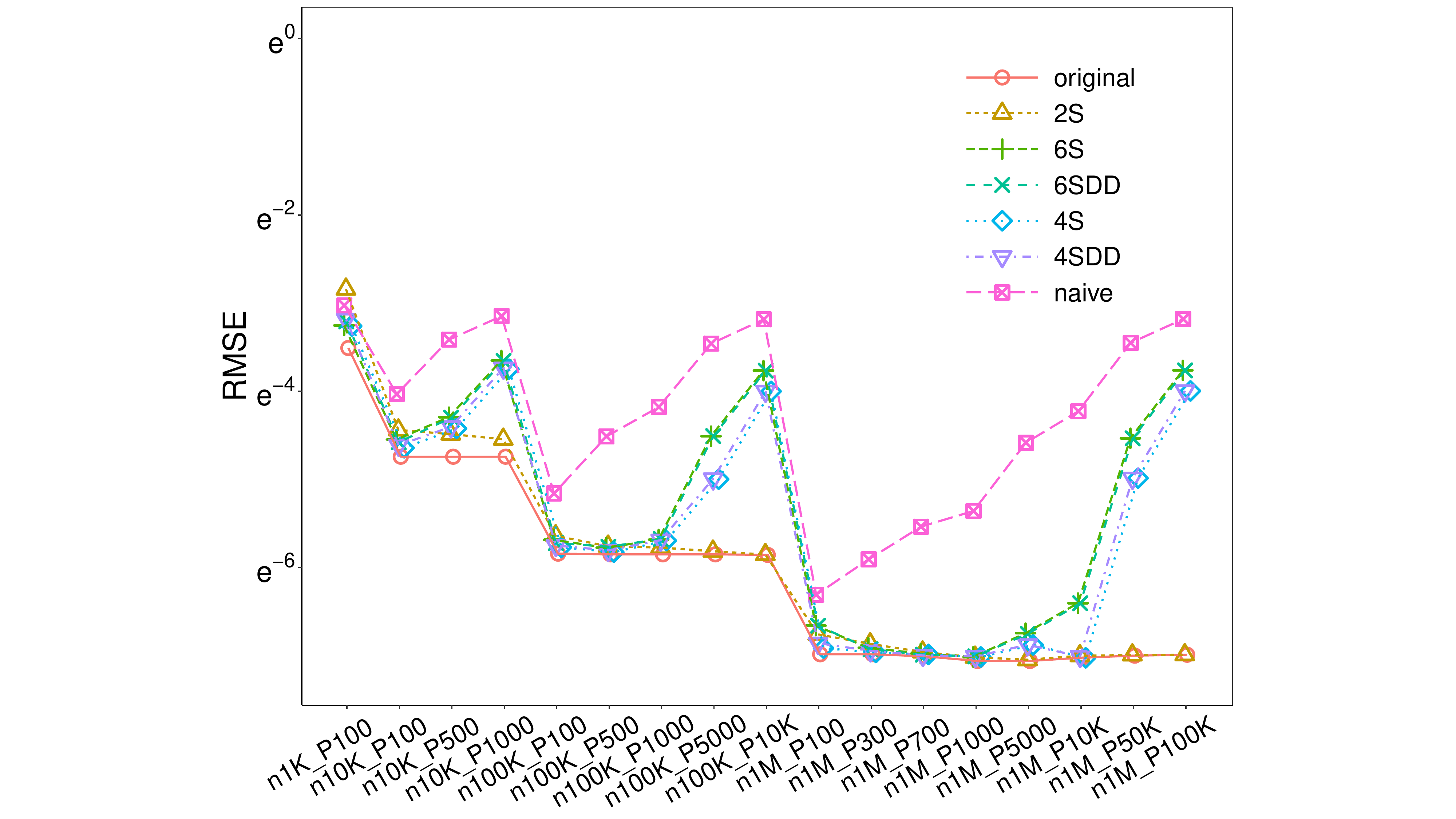}\\
\includegraphics[width=0.24\textwidth, trim={2.2in 0 2.2in 0},clip] {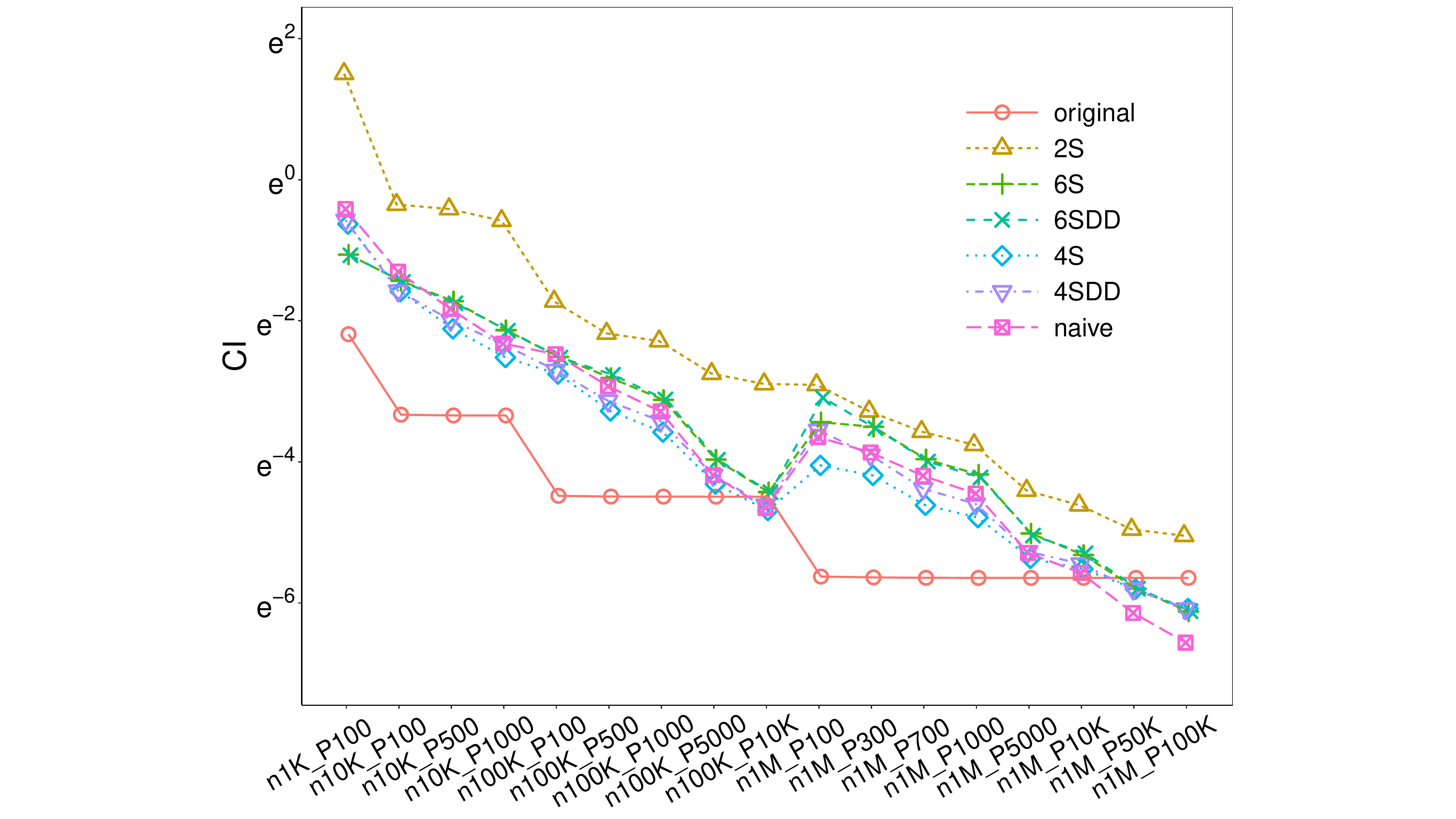}
\includegraphics[width=0.24\textwidth, trim={2.2in 0 2.2in 0},clip] {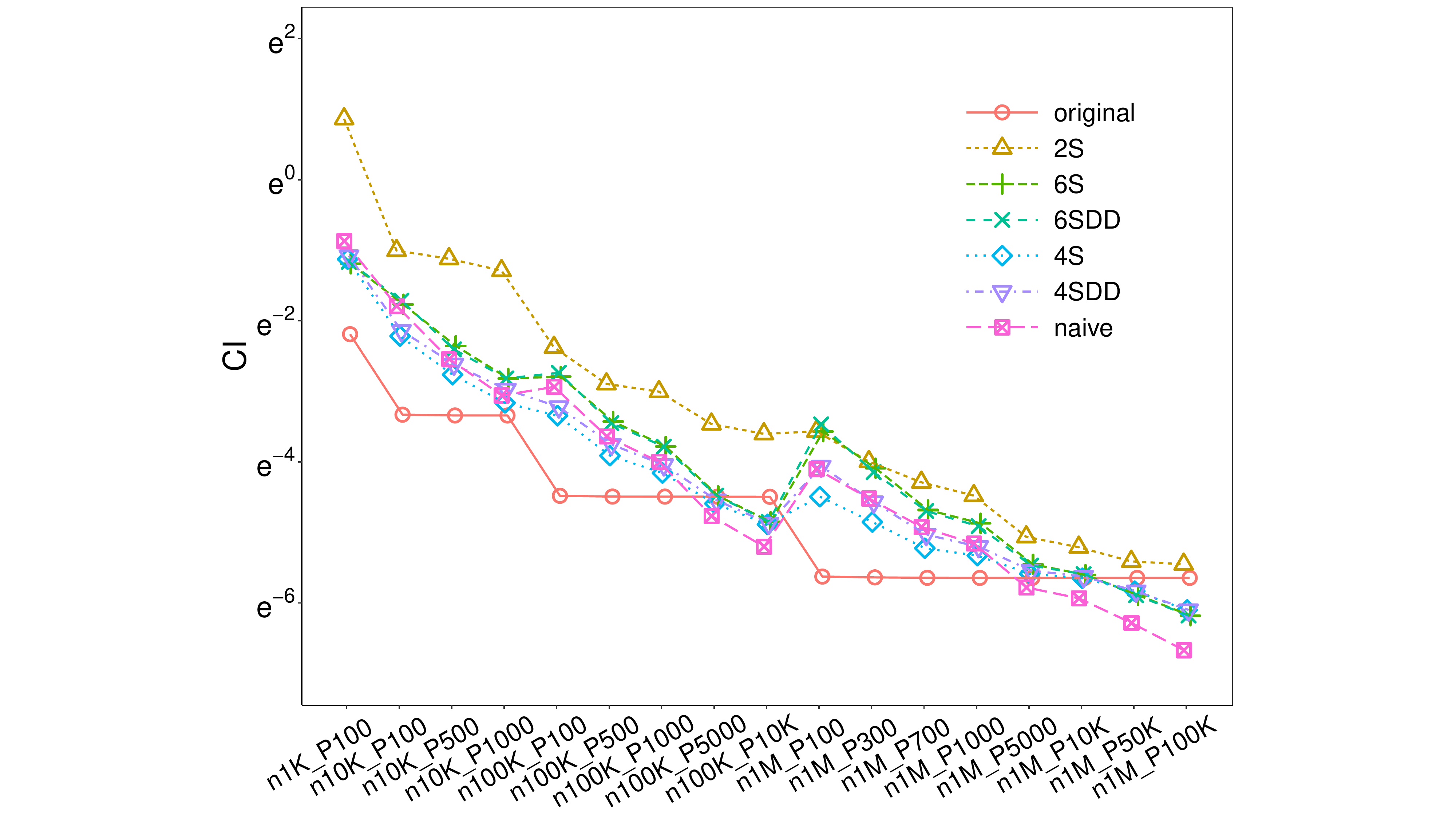}
\includegraphics[width=0.24\textwidth, trim={2.2in 0 2.2in 0},clip] {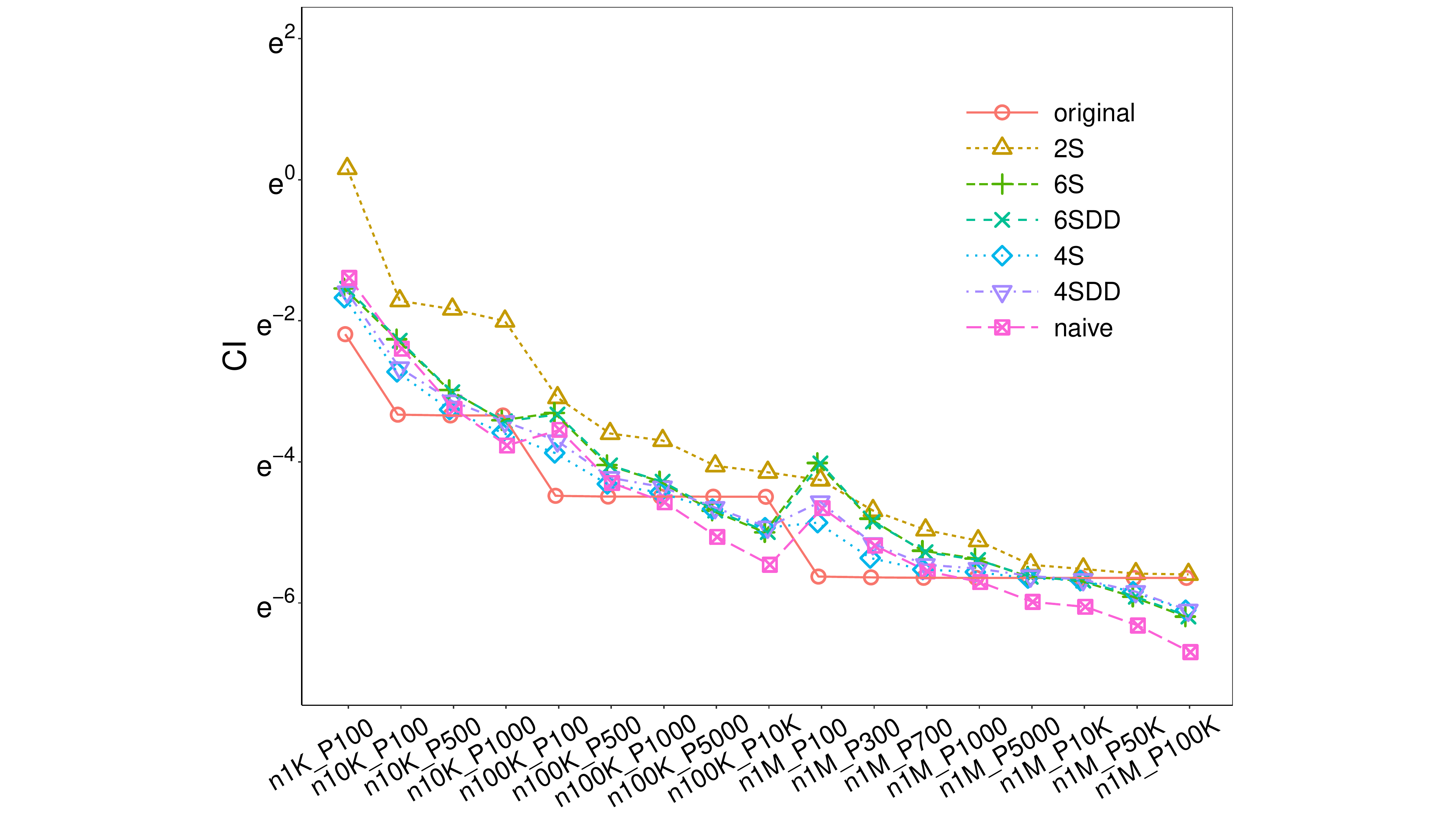}
\includegraphics[width=0.24\textwidth, trim={2.2in 0 2.2in 0},clip] {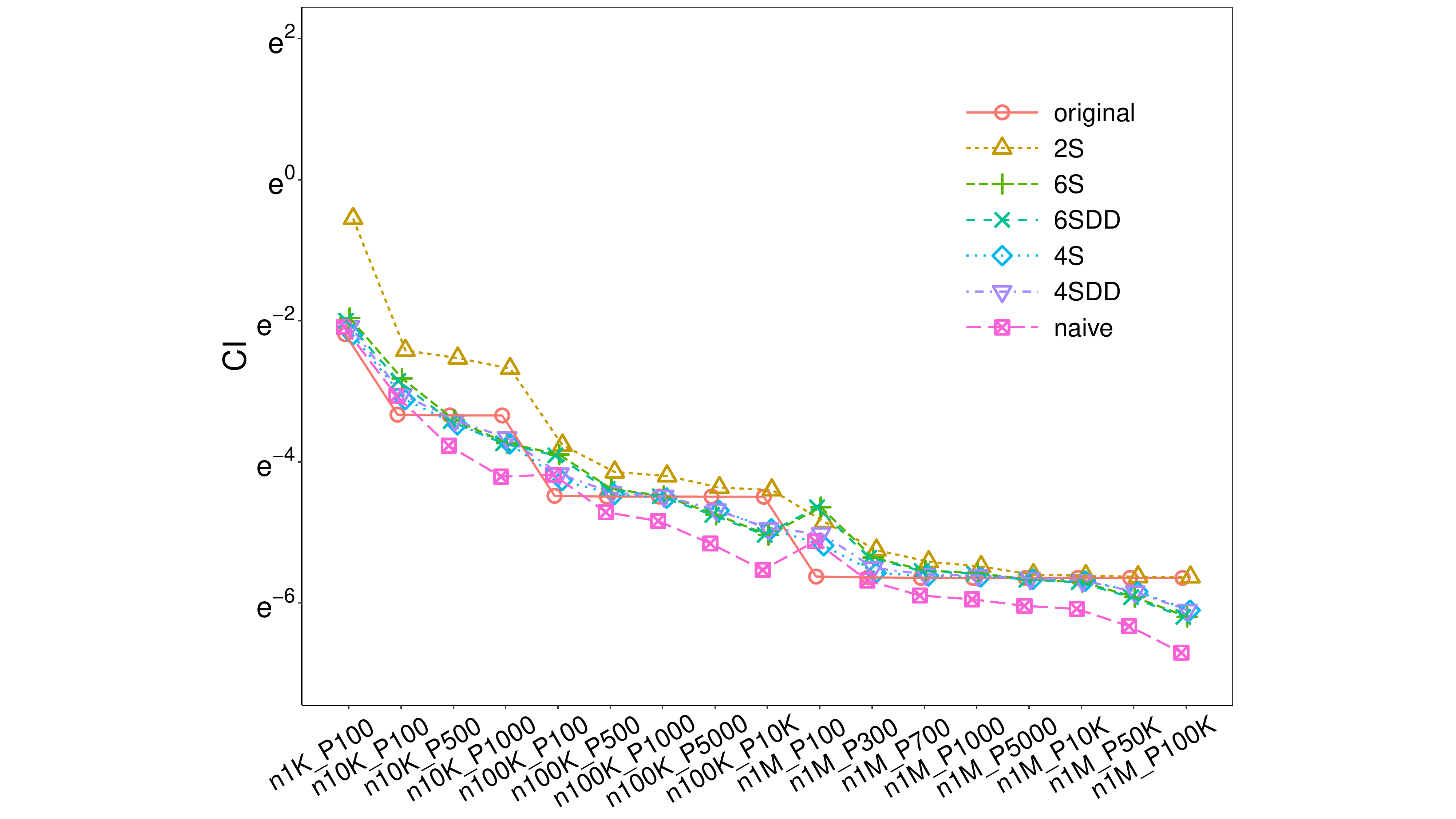}
\includegraphics[width=0.24\textwidth, trim={2.2in 0 2.2in 0},clip] {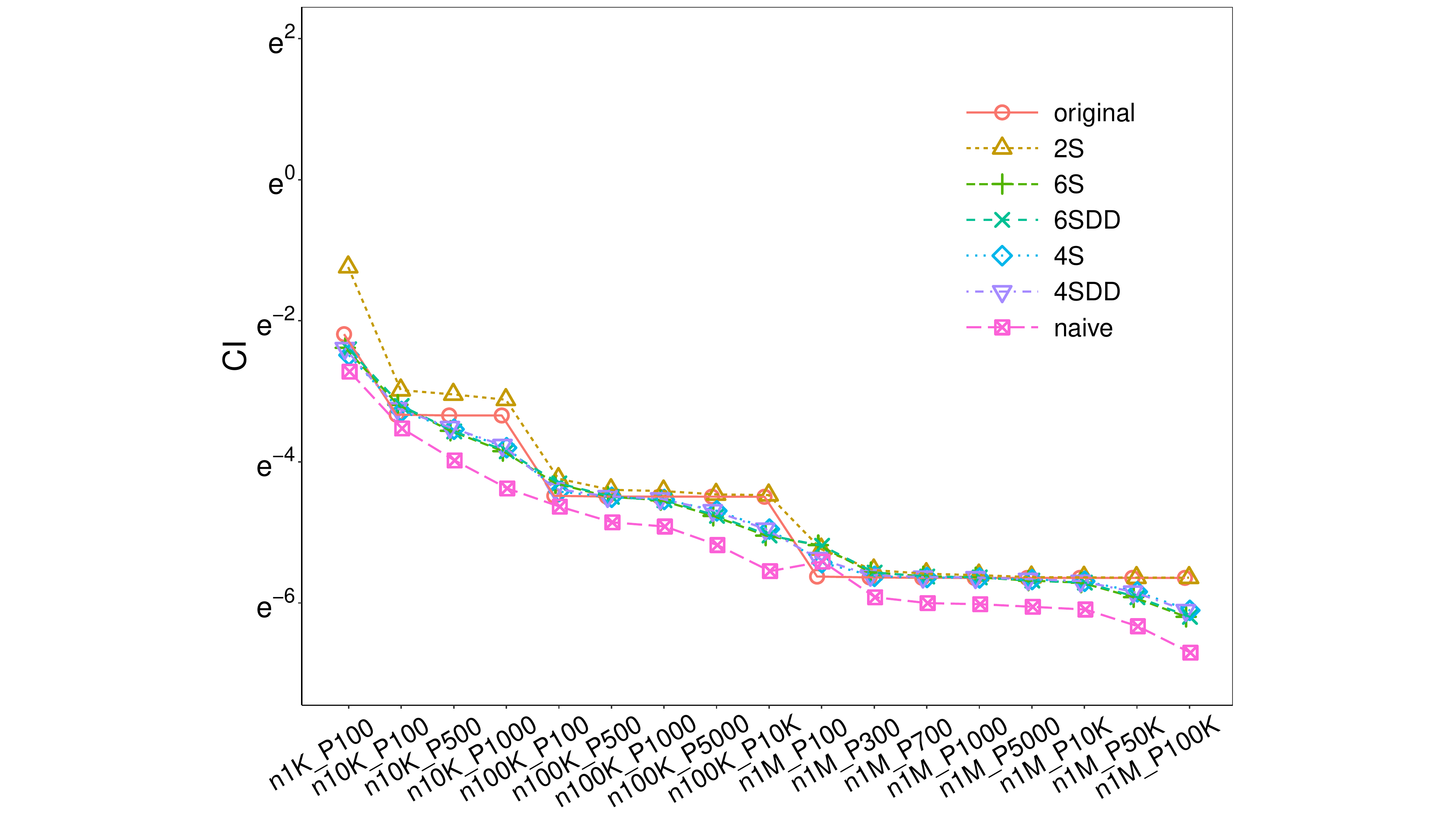}\\
\includegraphics[width=0.24\textwidth, trim={2.2in 0 2.2in 0},clip] {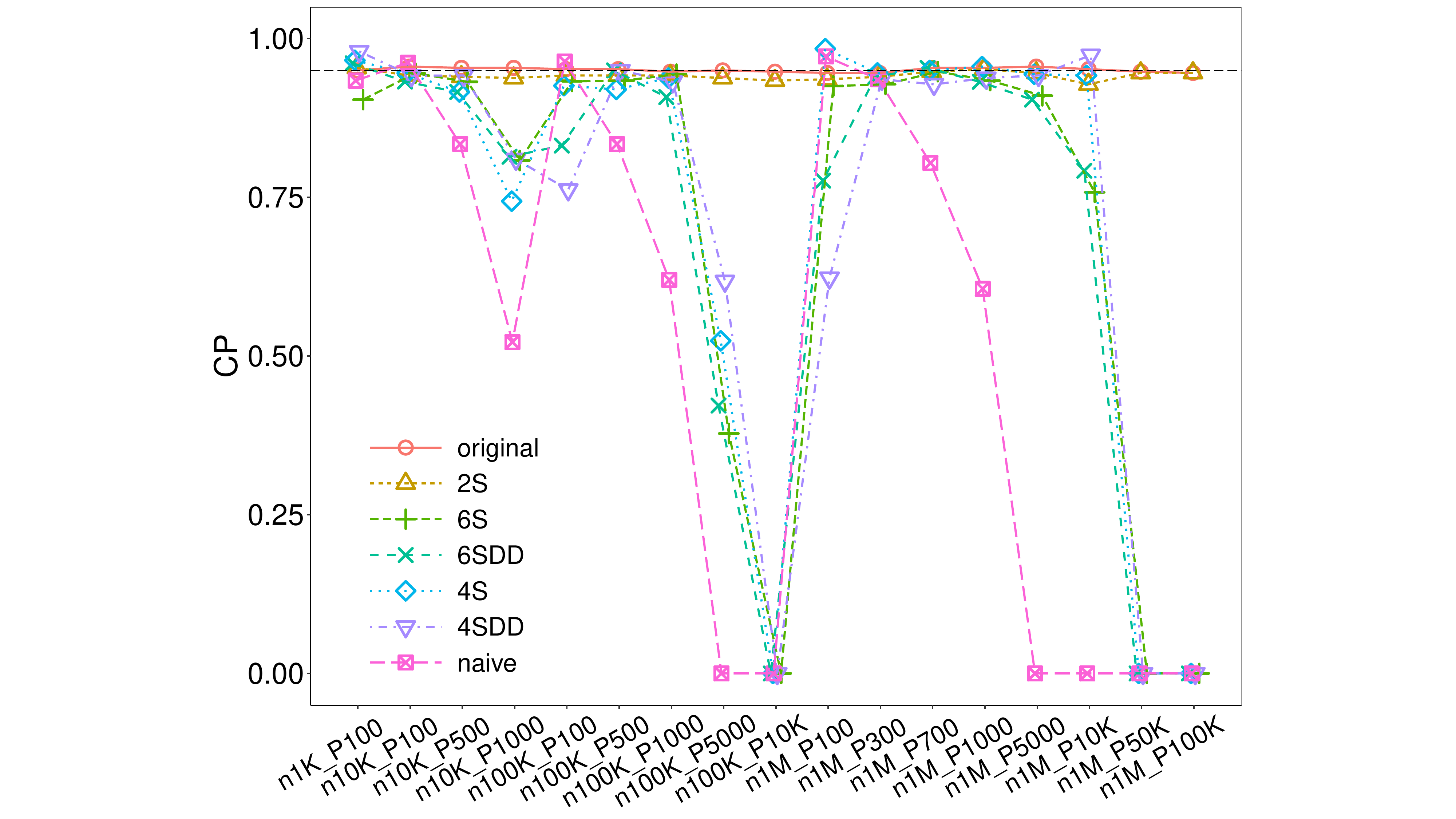}
\includegraphics[width=0.24\textwidth, trim={2.2in 0 2.2in 0},clip] {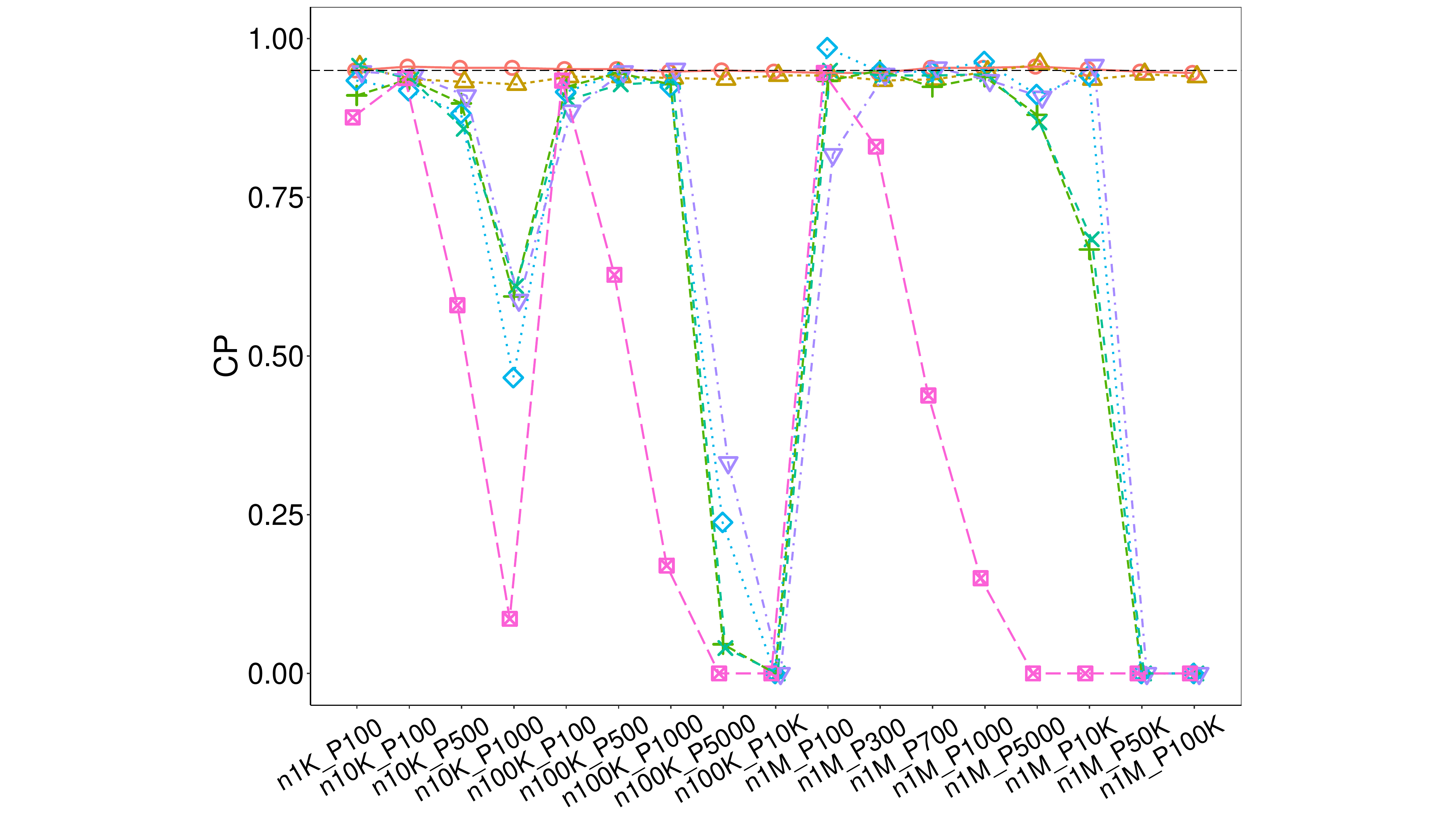}
\includegraphics[width=0.24\textwidth, trim={2.2in 0 2.2in 0},clip] {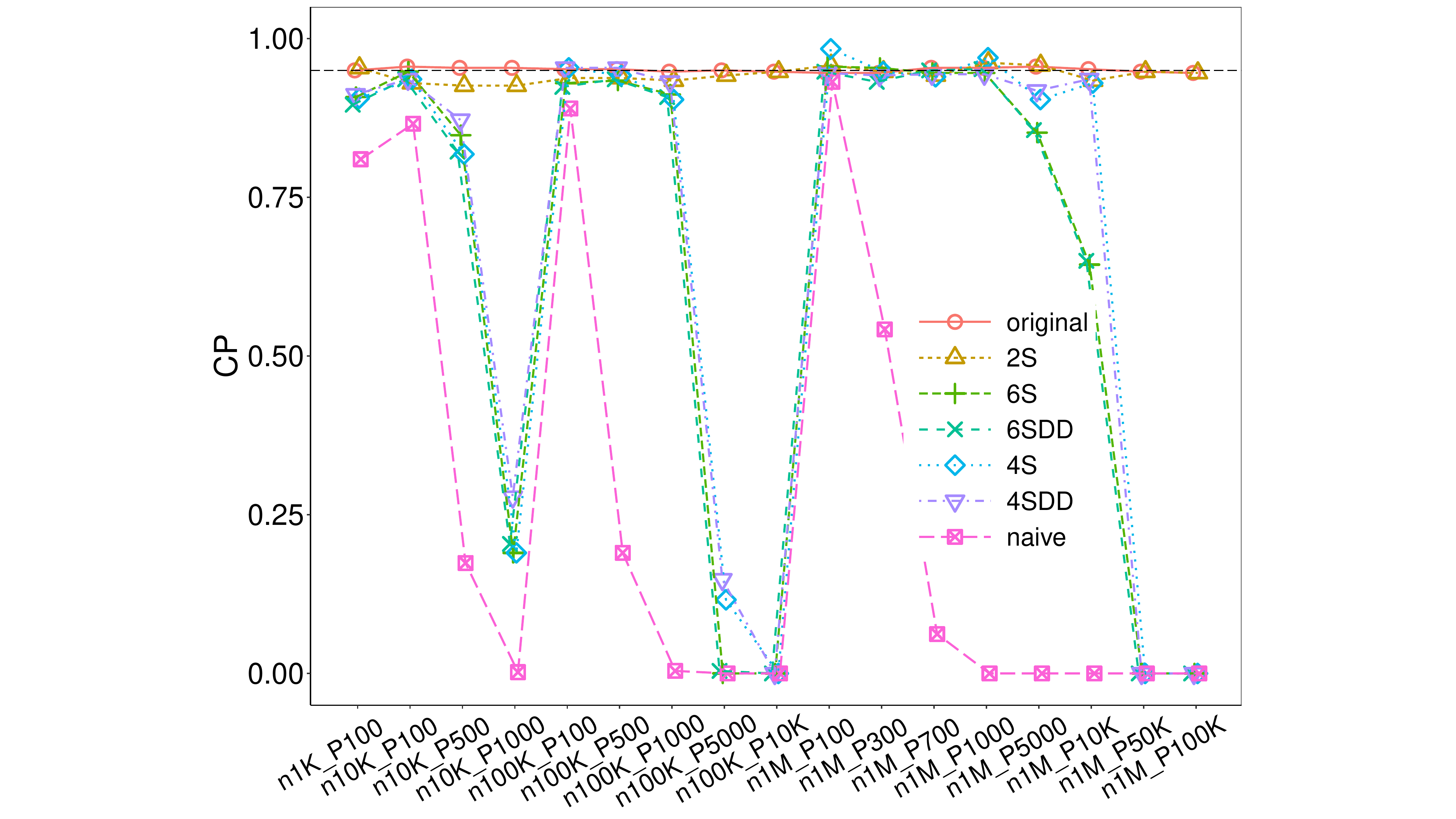}
\includegraphics[width=0.24\textwidth, trim={2.2in 0 2.2in 0},clip] {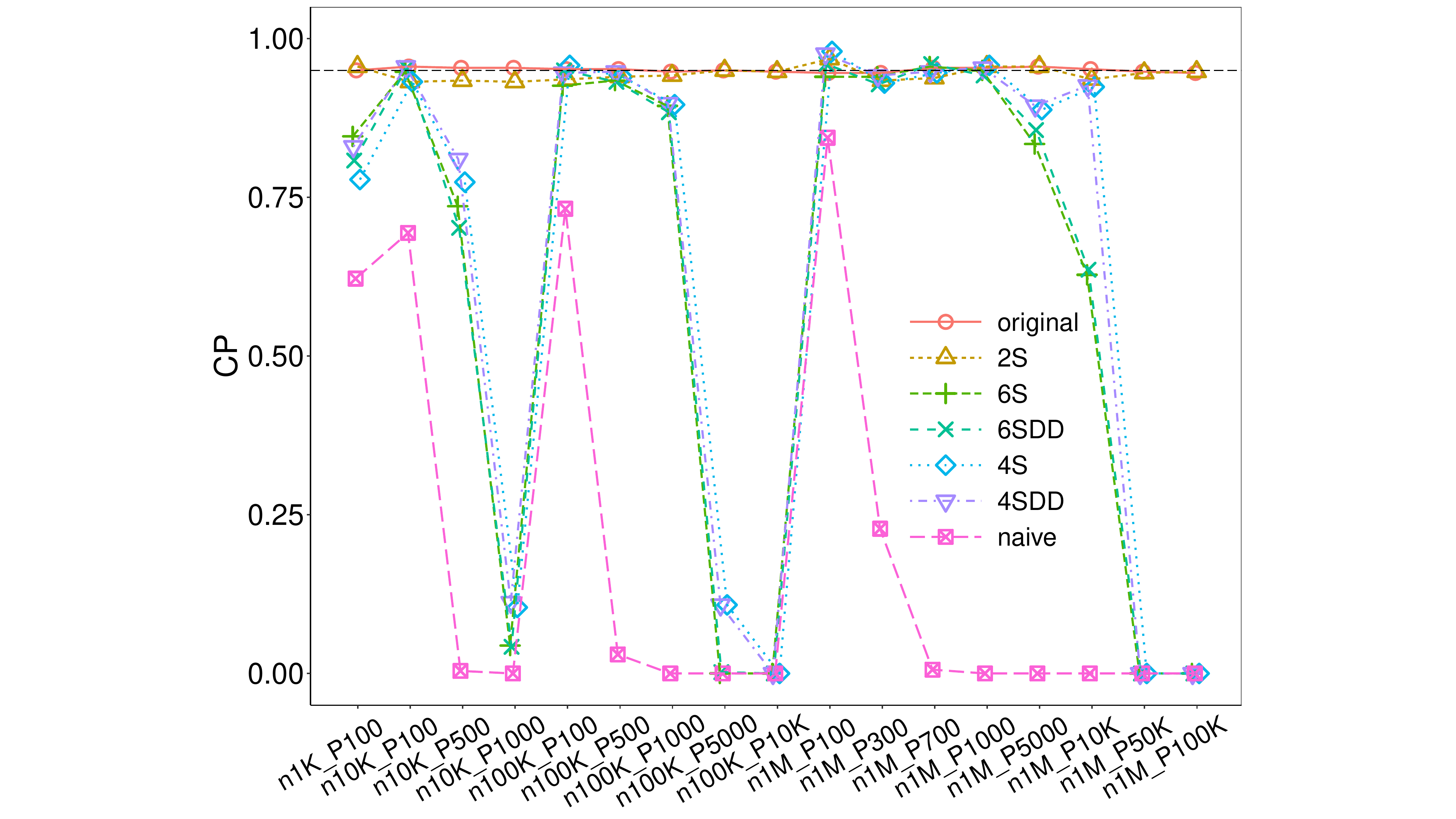}
\includegraphics[width=0.24\textwidth, trim={2.2in 0 2.2in 0},clip] {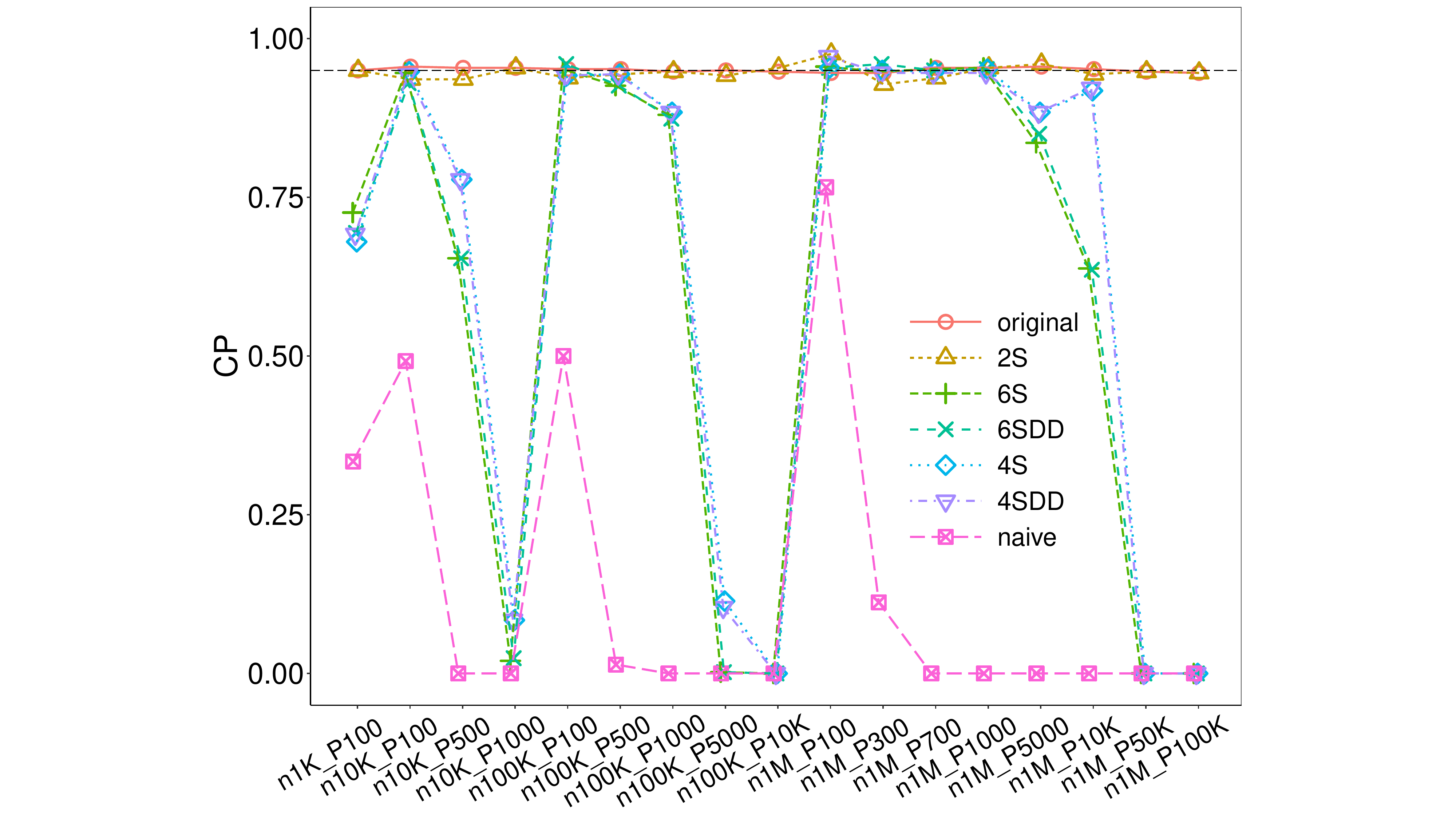}\\
\includegraphics[width=0.24\textwidth, trim={2.2in 0 2.2in 0},clip] {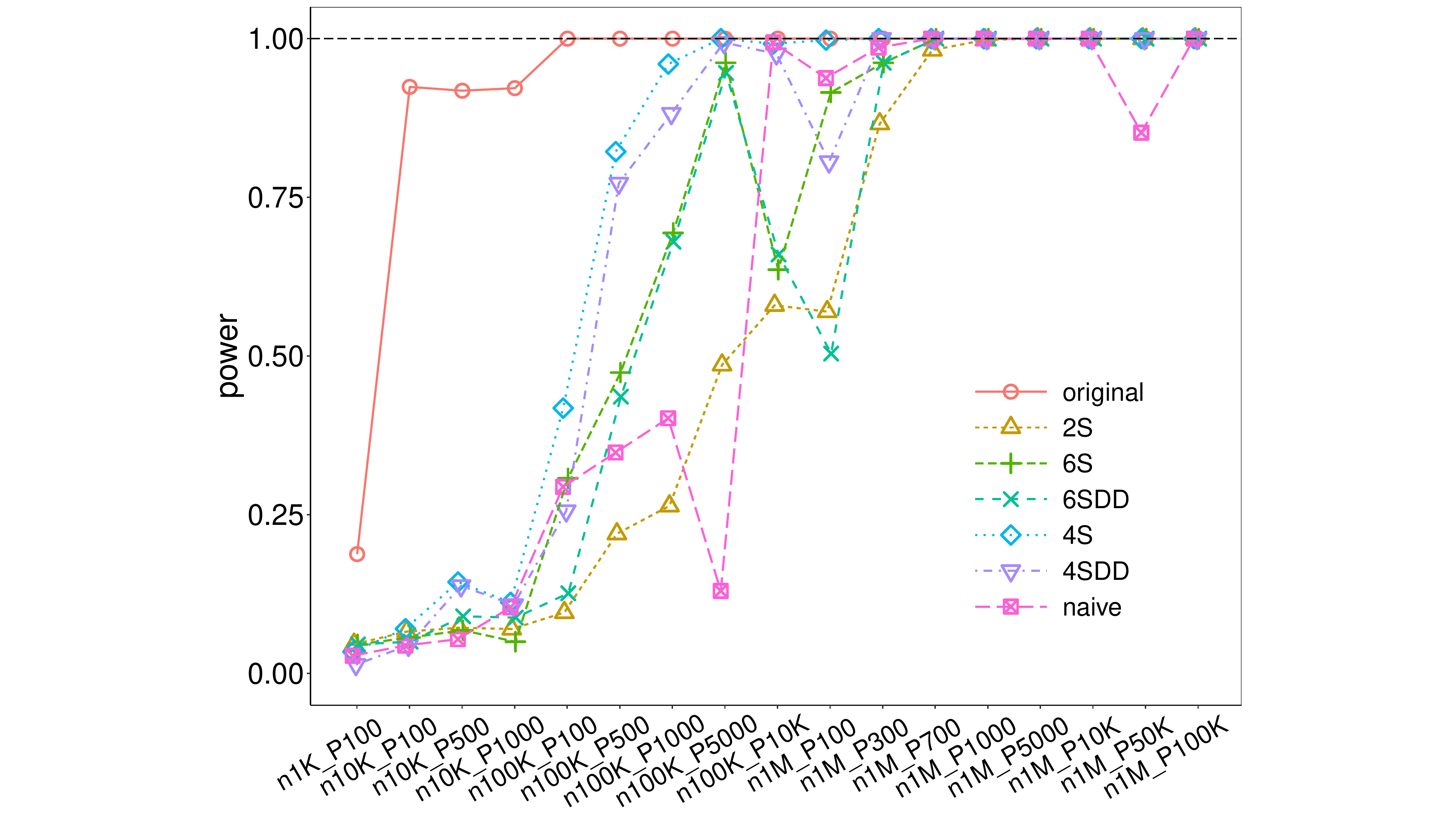}
\includegraphics[width=0.24\textwidth, trim={2.2in 0 2.2in 0},clip] {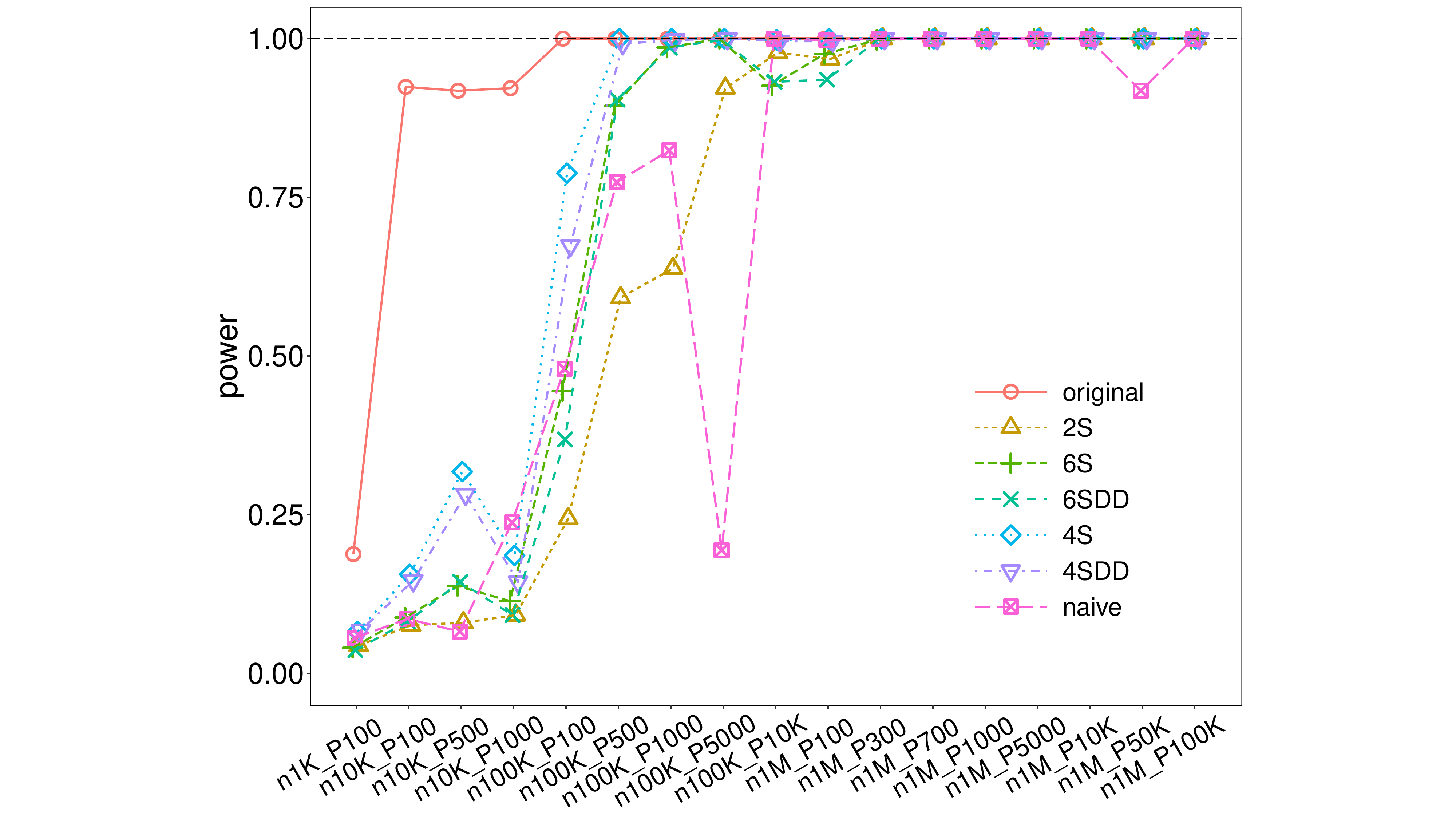}
\includegraphics[width=0.24\textwidth, trim={2.2in 0 2.2in 0},clip] {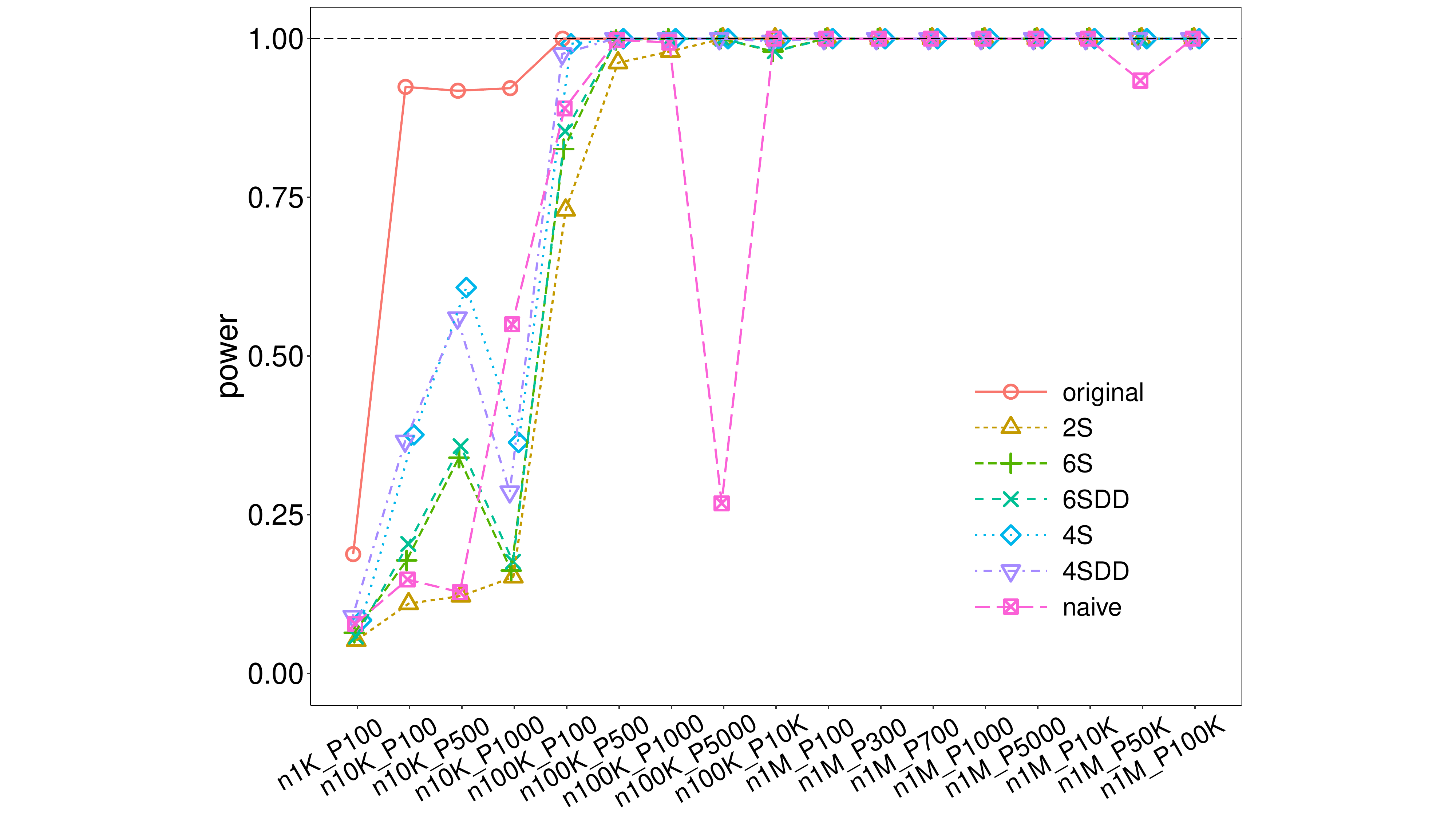}
\includegraphics[width=0.24\textwidth, trim={2.2in 0 2.2in 0},clip] {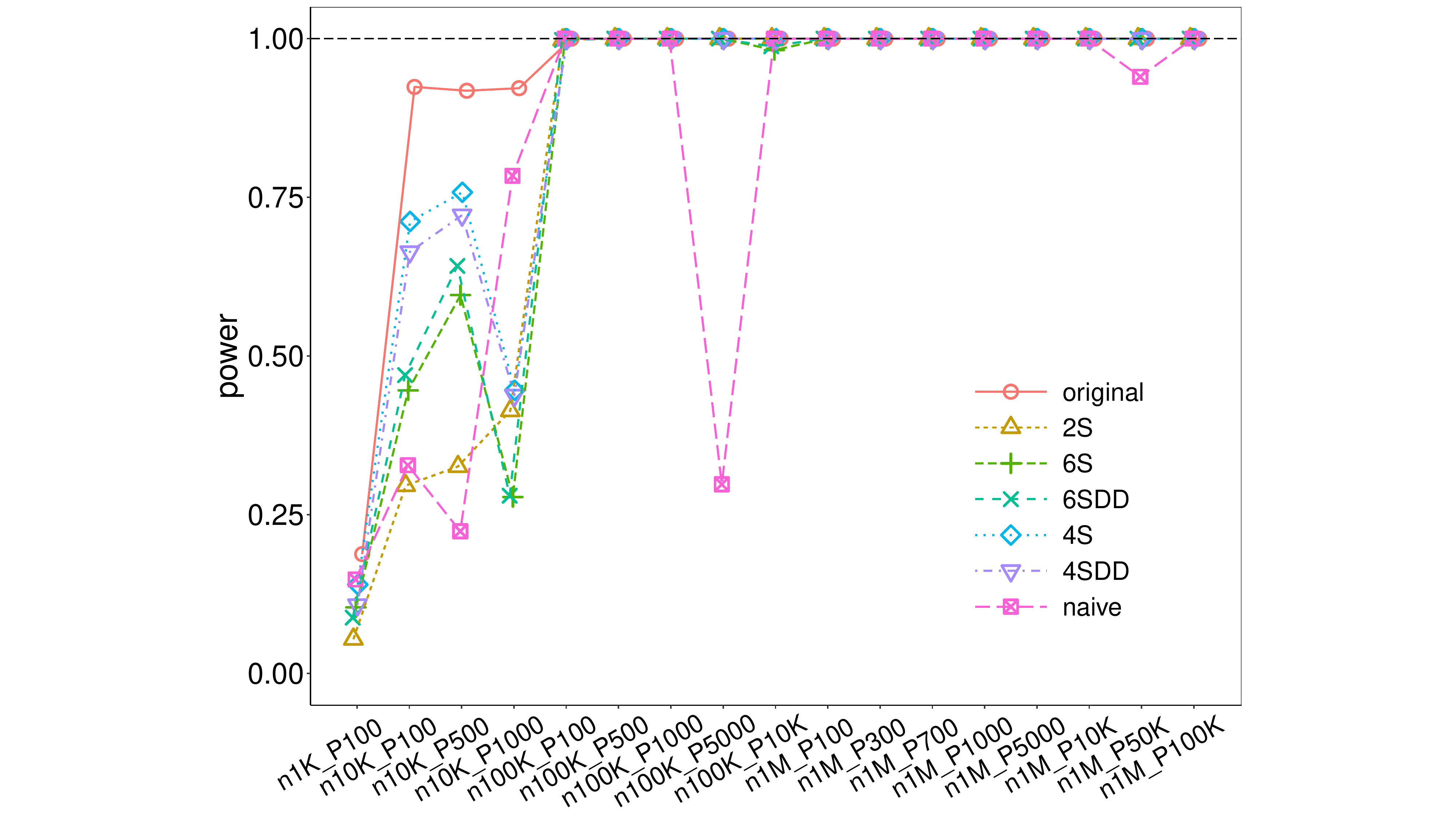}
\includegraphics[width=0.24\textwidth, trim={2.2in 0 2.2in 0},clip] {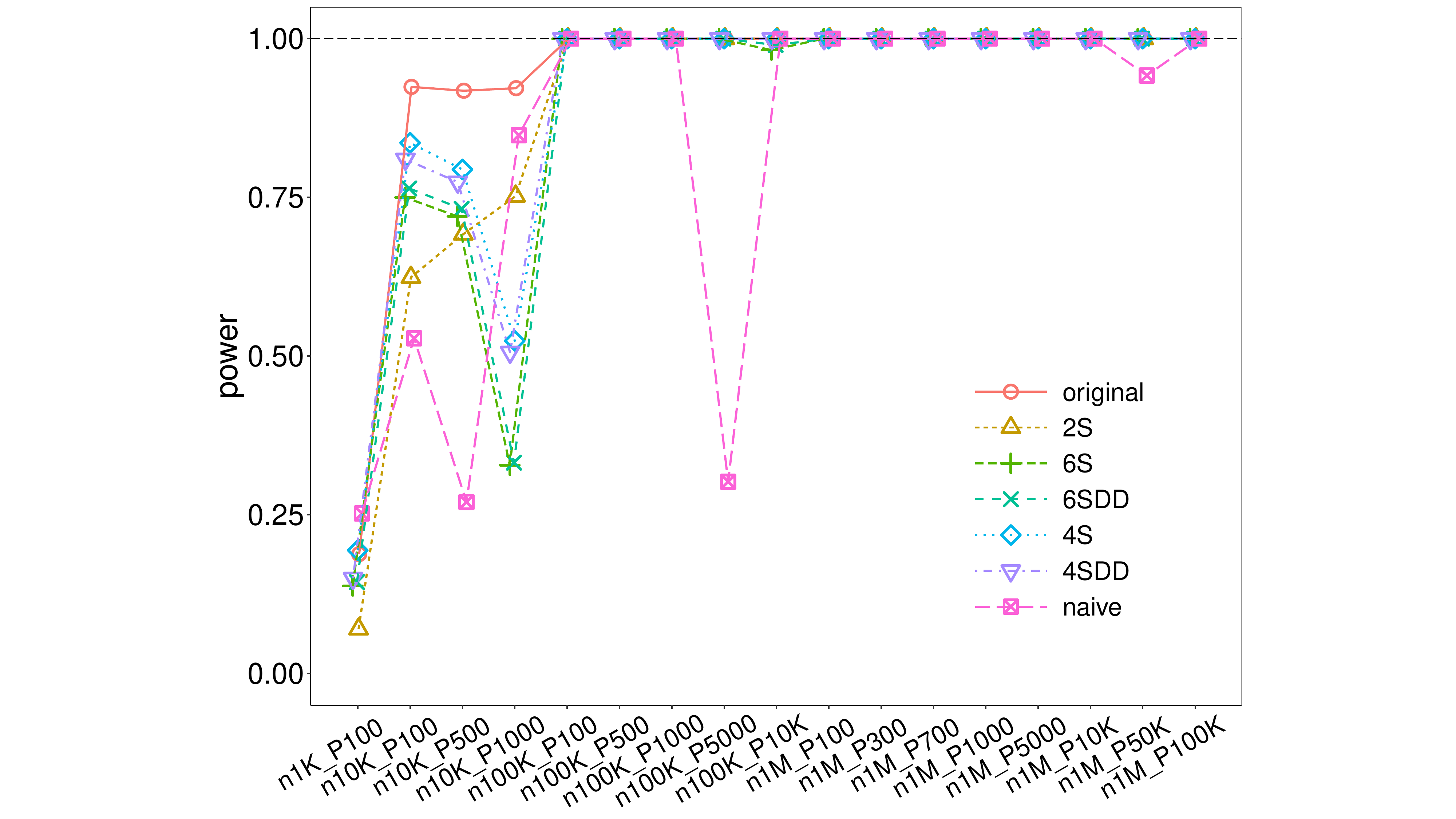}\\
\caption{ZINB data; $\rho$-zCDP; $\theta\ne0$ and $\alpha\ne\beta$}
\label{fig:1aszCDPzinb}
\end{figure}
\end{landscape}

\begin{landscape}
\subsection*{ZILN, $\theta=0$ and $\alpha=\beta$}

\begin{figure}[!htb]
\hspace{0.6in}$\epsilon=0.5$\hspace{1.3in}$\epsilon=1$\hspace{1.4in}$\epsilon=2$
\hspace{1.4in}$\epsilon=5$\hspace{1.4in}$\epsilon=50$\\
\includegraphics[width=0.26\textwidth, trim={2.2in 0 2.2in 0},clip] {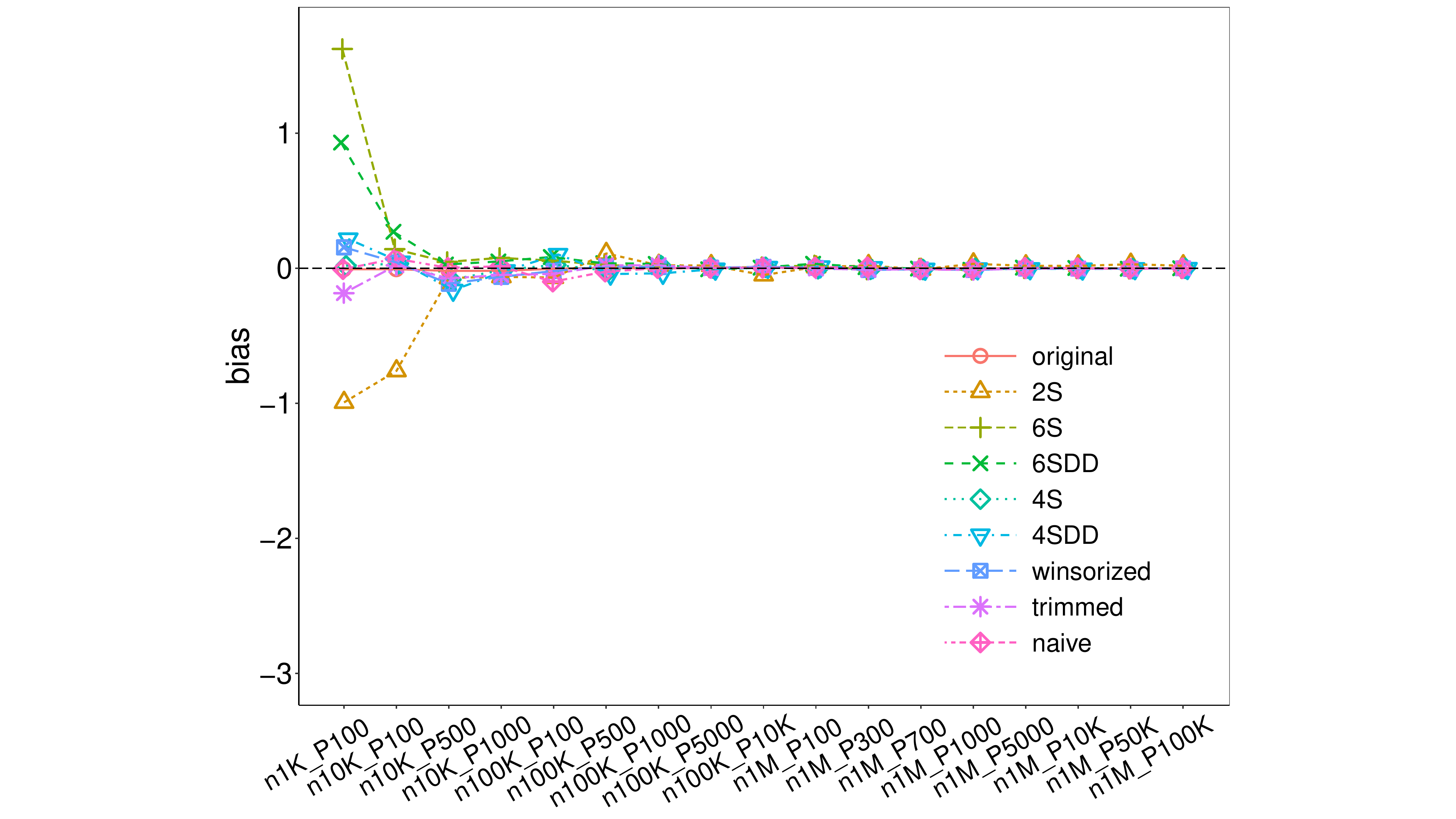}
\includegraphics[width=0.26\textwidth, trim={2.2in 0 2.2in 0},clip] {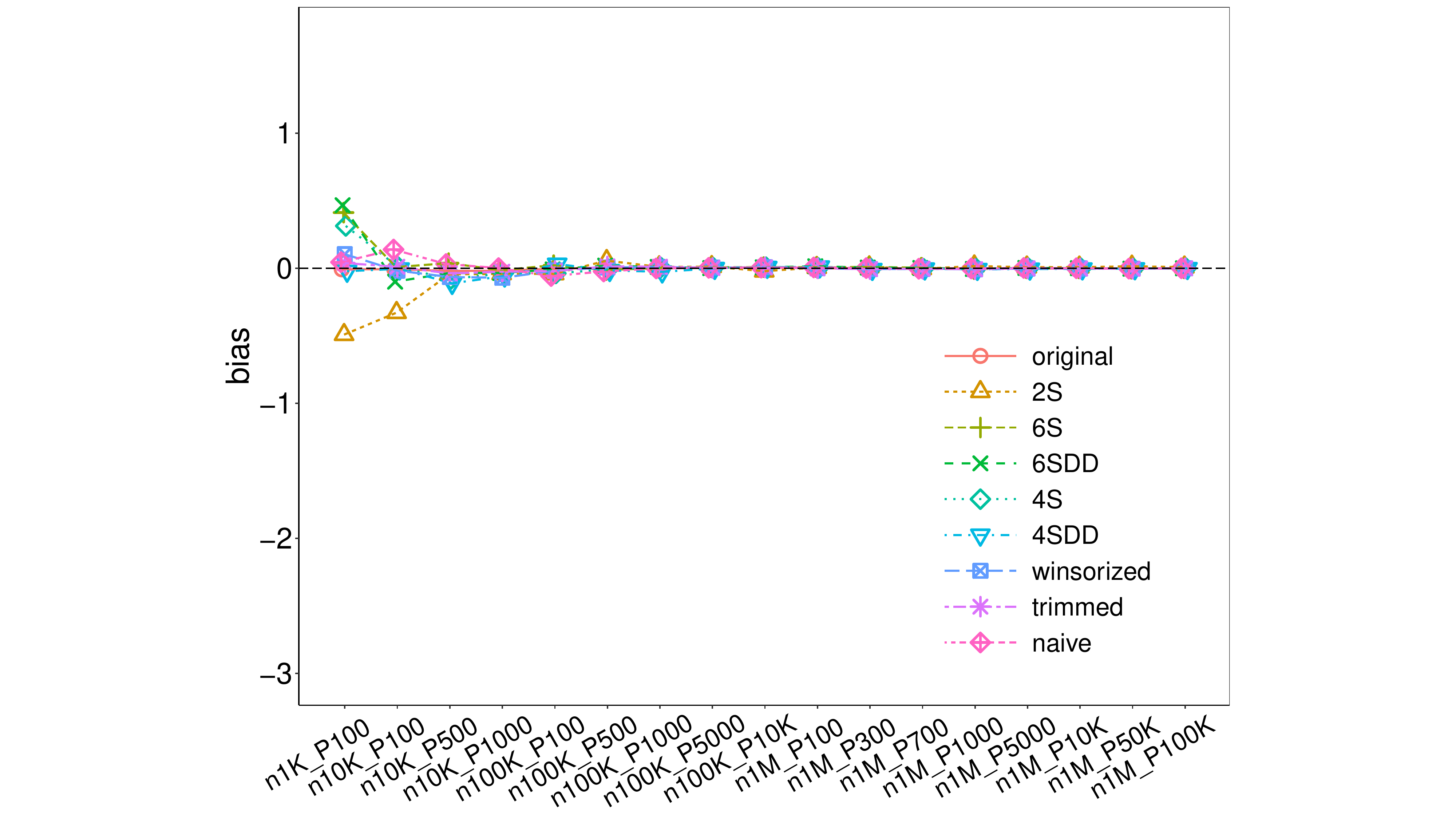}
\includegraphics[width=0.26\textwidth, trim={2.2in 0 2.2in 0},clip] {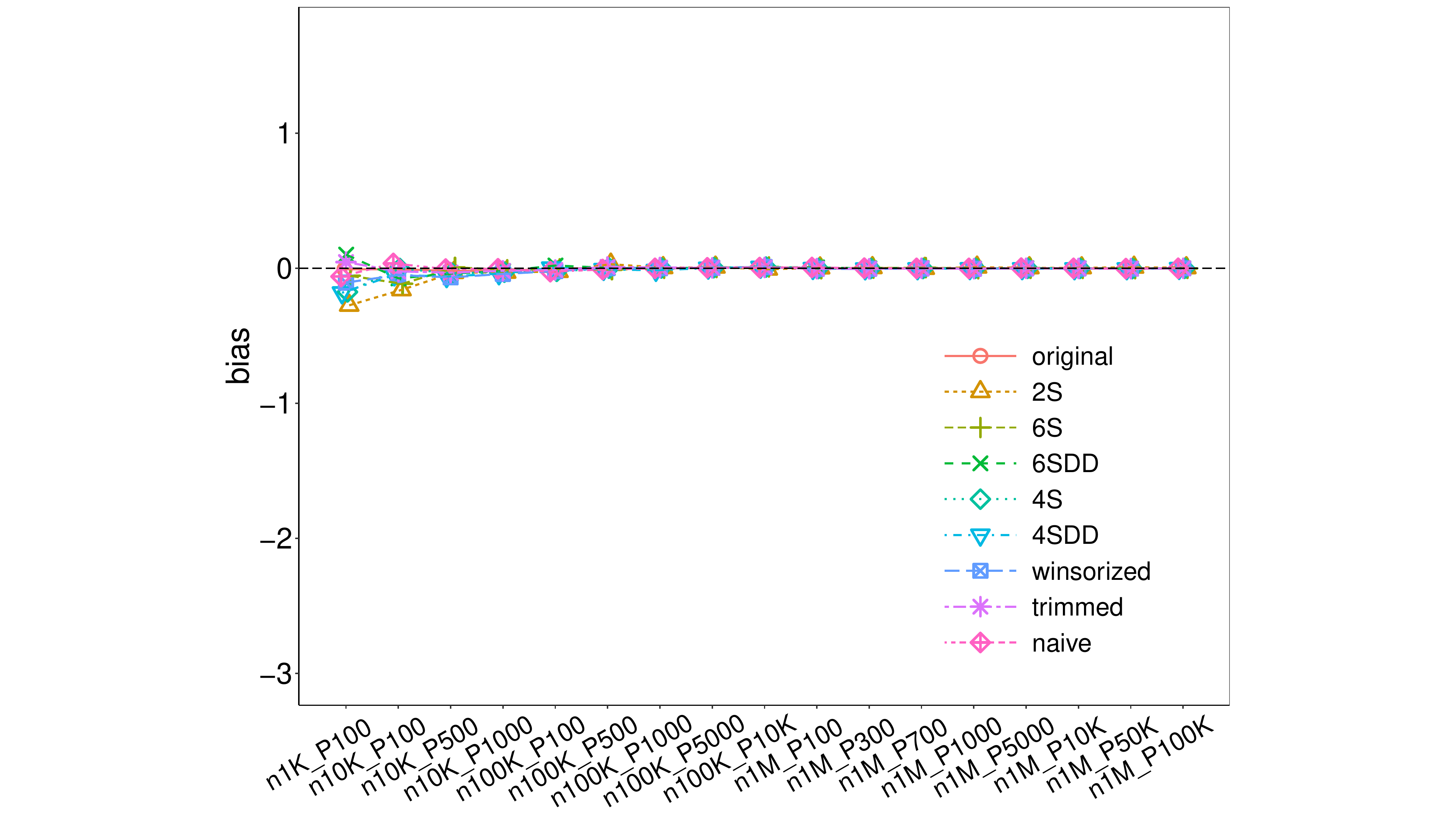}
\includegraphics[width=0.26\textwidth, trim={2.2in 0 2.2in 0},clip] {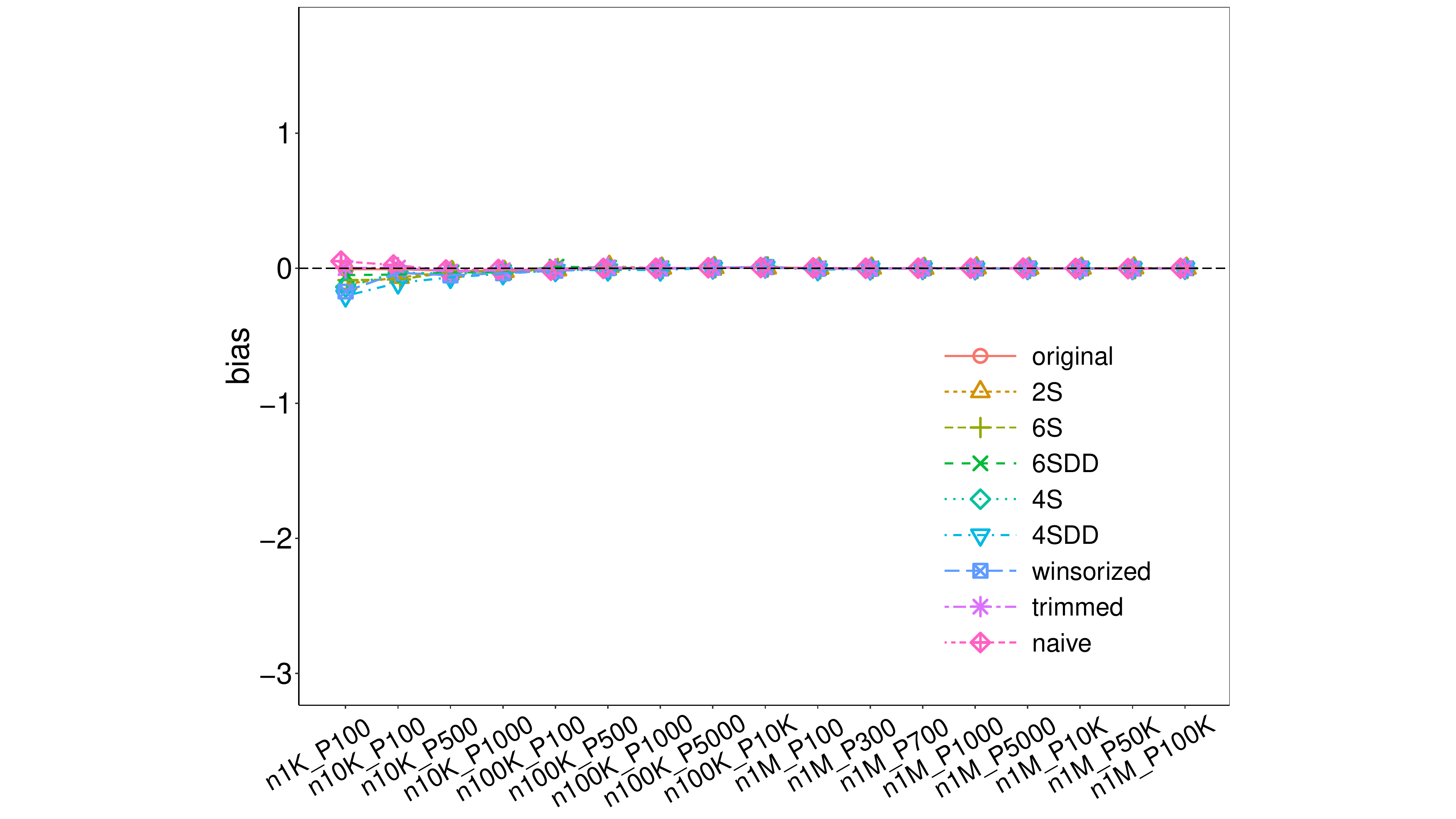}
\includegraphics[width=0.26\textwidth, trim={2.2in 0 2.2in 0},clip] {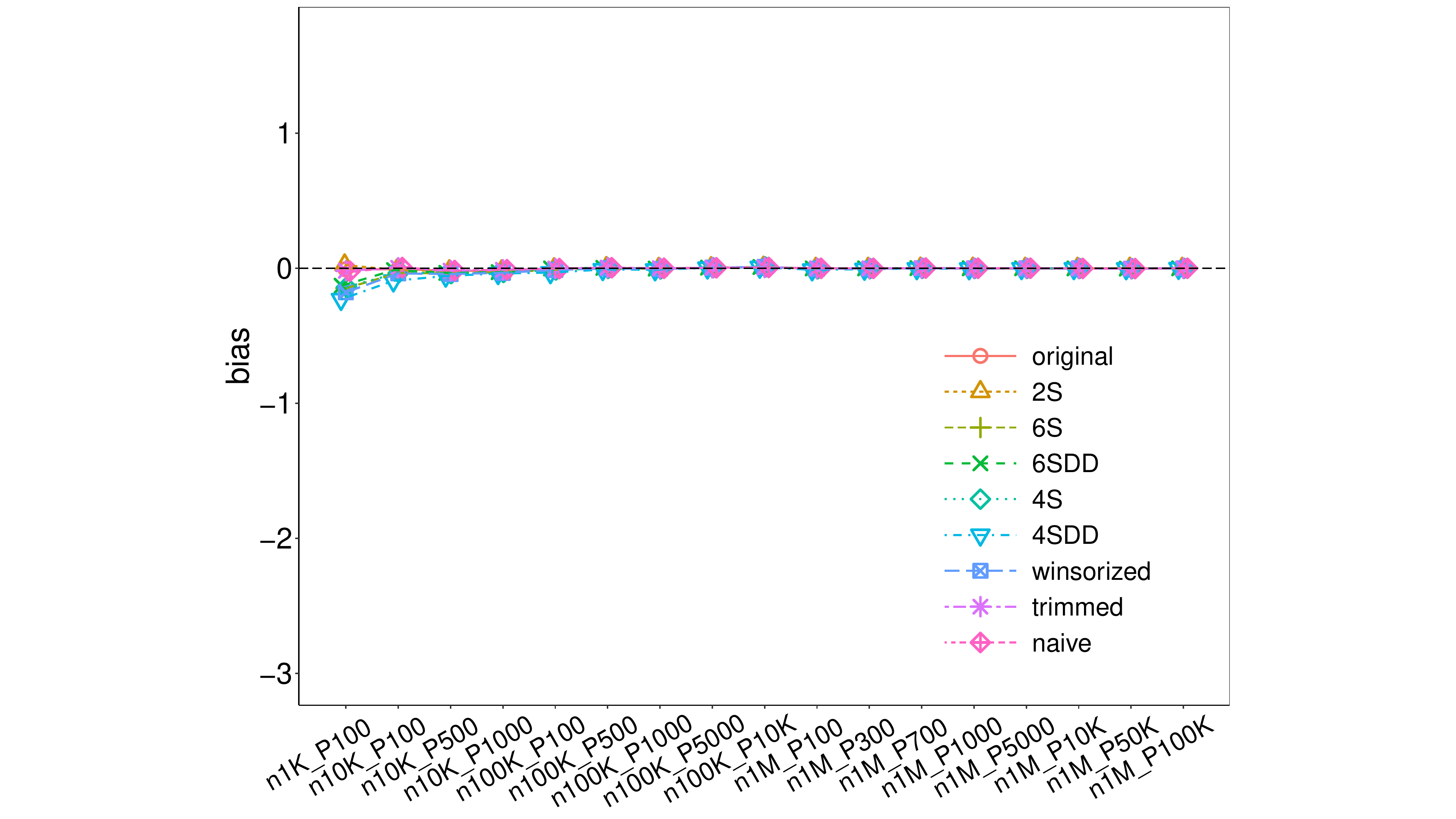}\\
\includegraphics[width=0.26\textwidth, trim={2.2in 0 2.2in 0},clip] {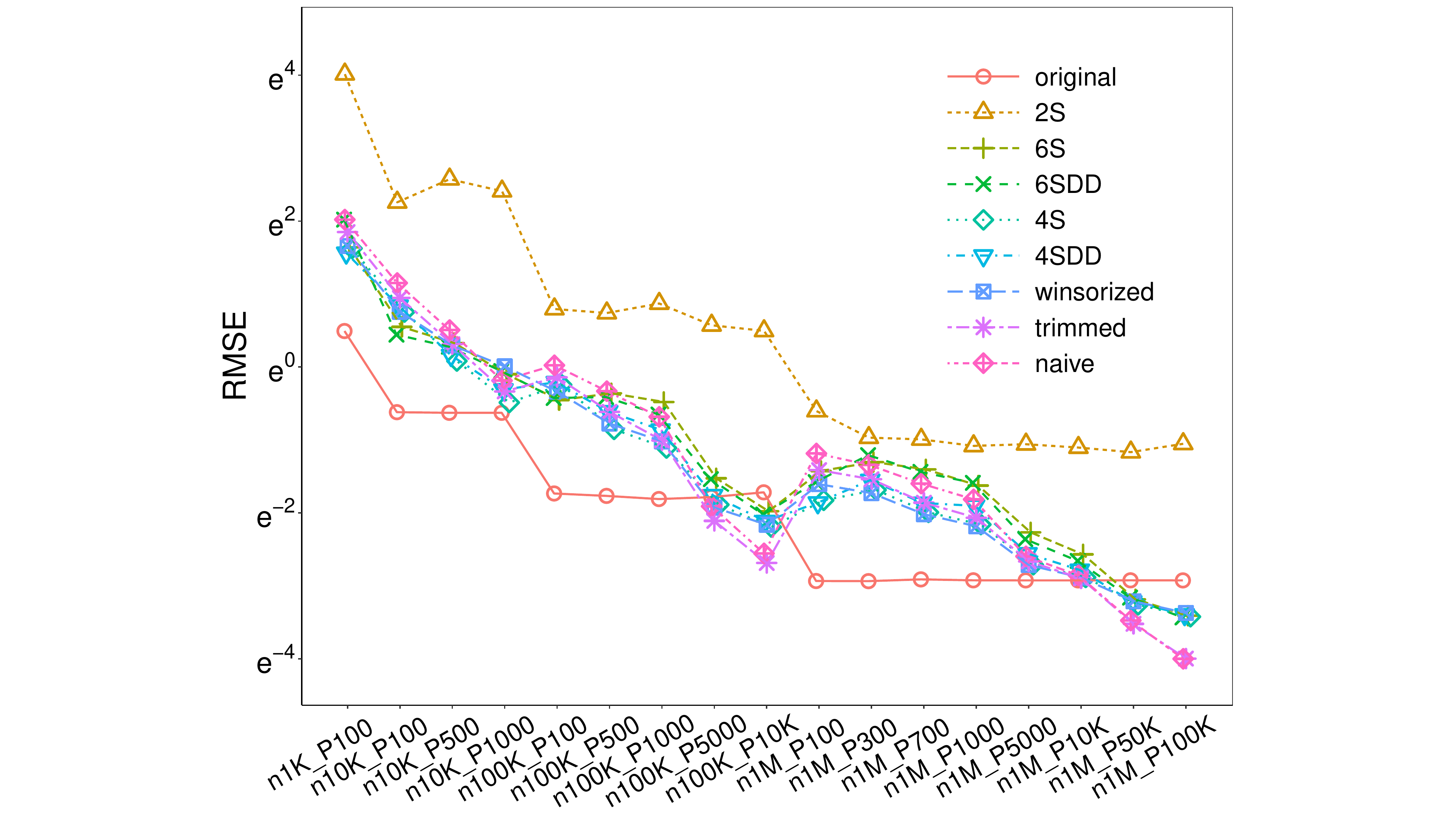}
\includegraphics[width=0.26\textwidth, trim={2.2in 0 2.2in 0},clip] {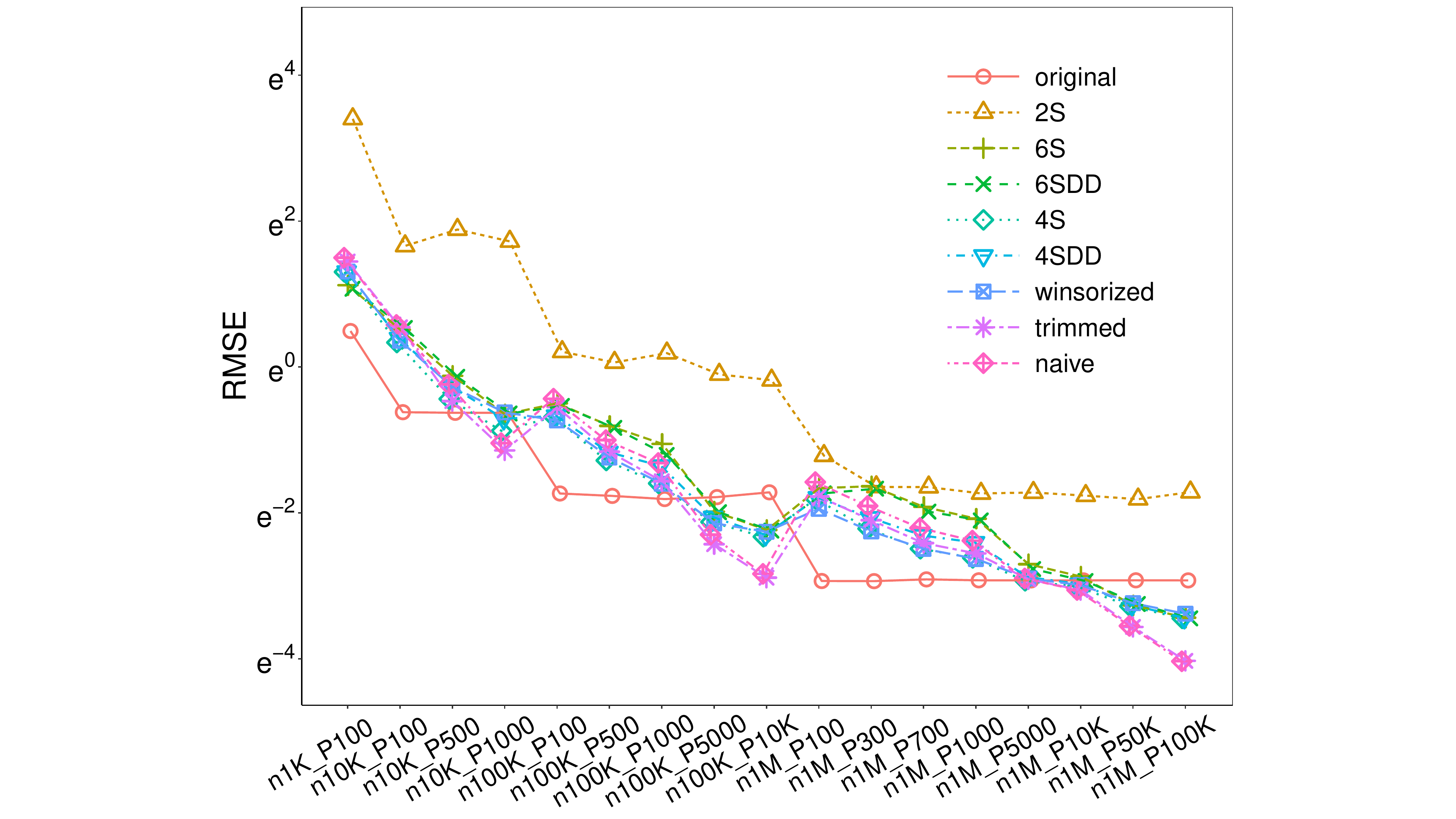}
\includegraphics[width=0.26\textwidth, trim={2.2in 0 2.2in 0},clip] {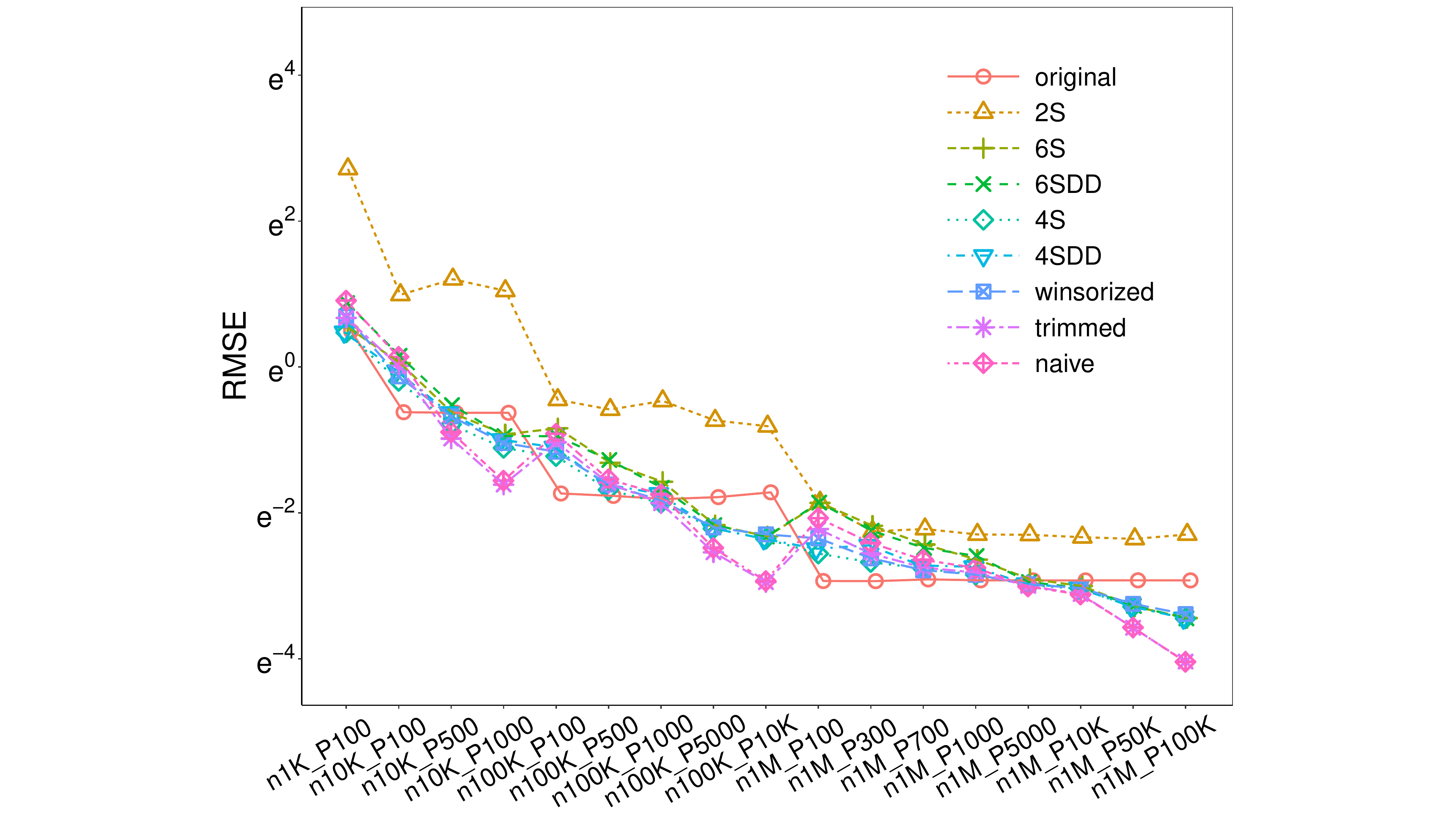}
\includegraphics[width=0.26\textwidth, trim={2.2in 0 2.2in 0},clip] {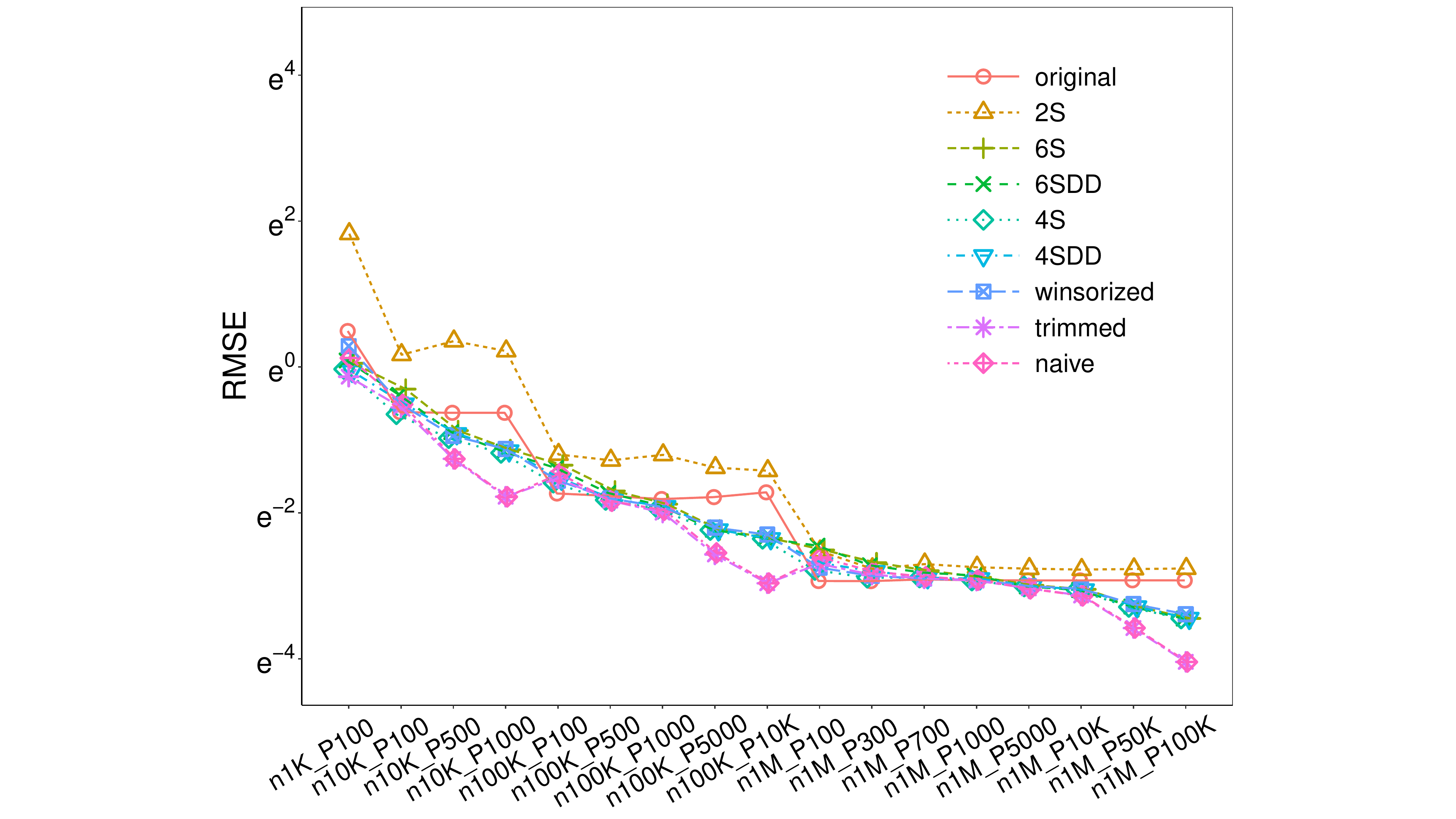}
\includegraphics[width=0.26\textwidth, trim={2.2in 0 2.2in 0},clip] {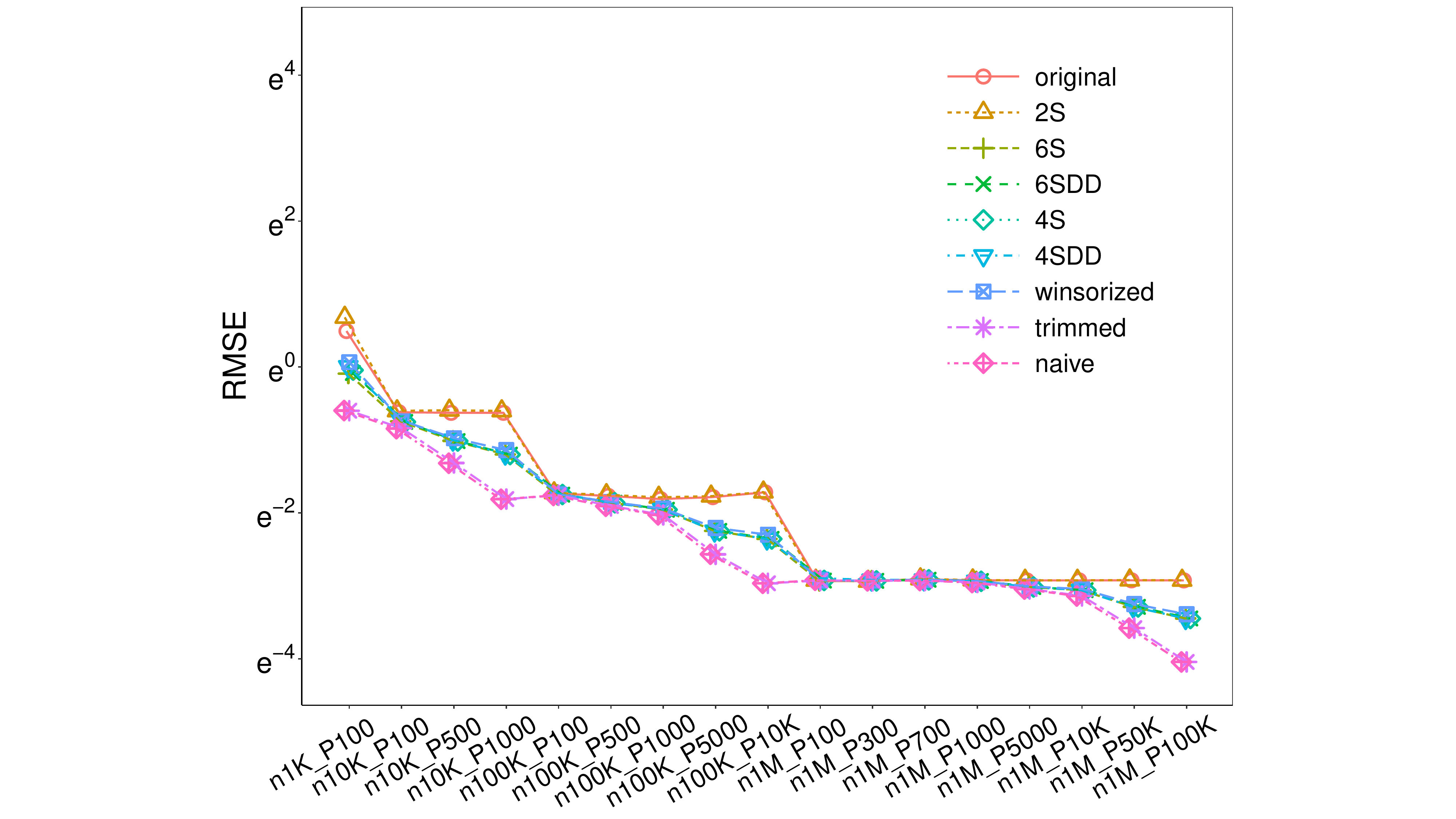}\\
\includegraphics[width=0.26\textwidth, trim={2.2in 0 2.2in 0},clip] {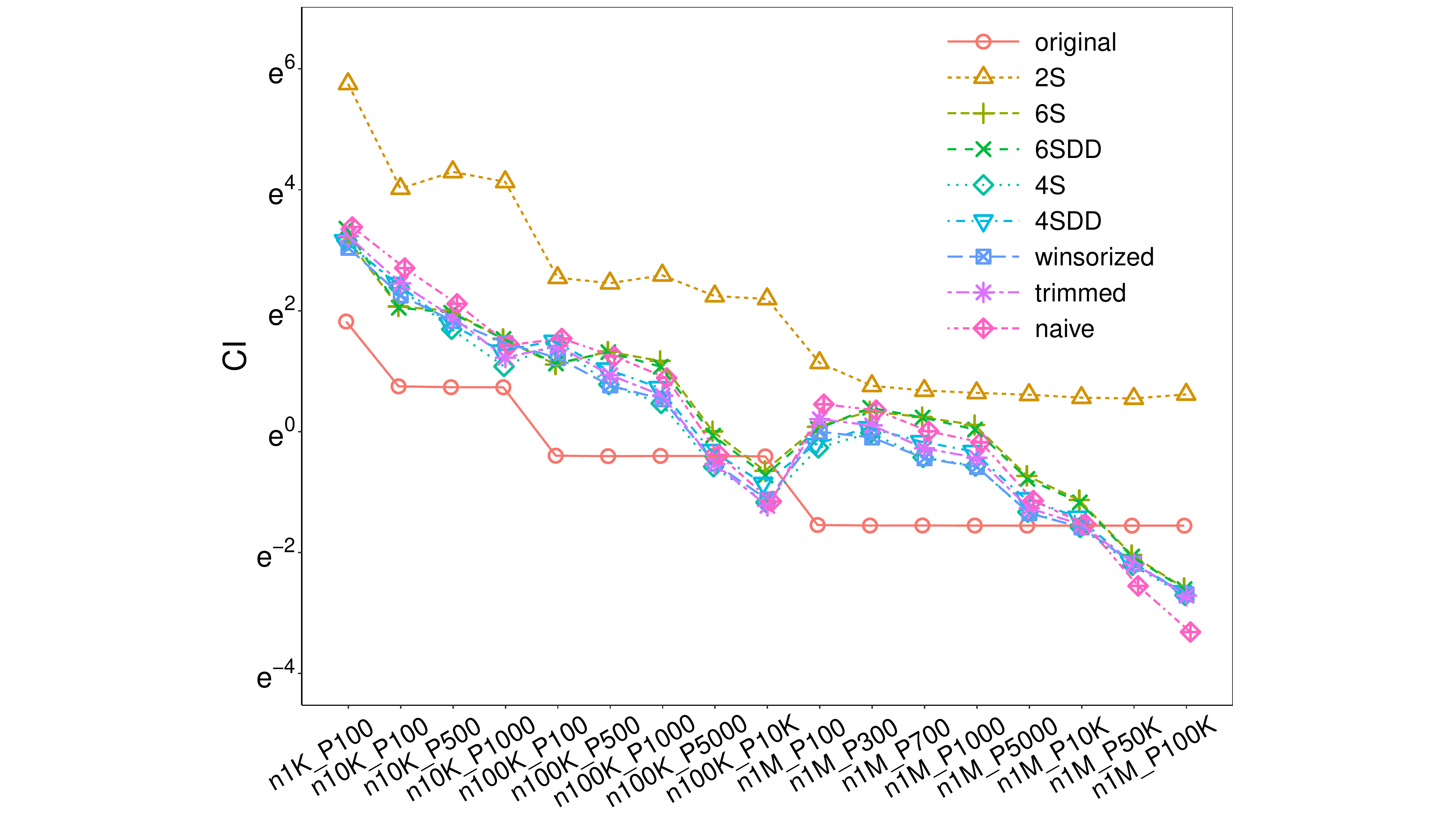}
\includegraphics[width=0.26\textwidth, trim={2.2in 0 2.2in 0},clip] {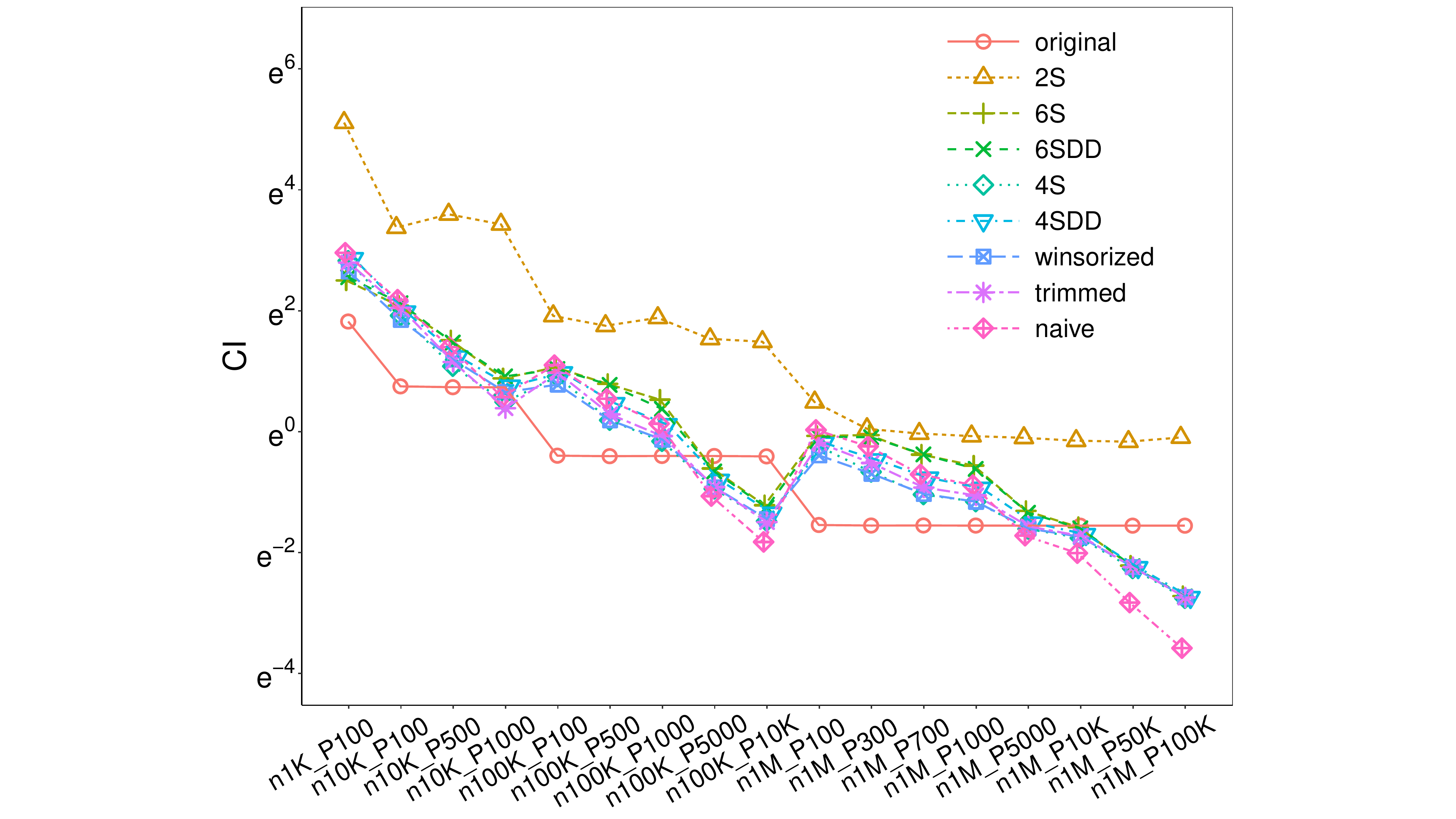}
\includegraphics[width=0.26\textwidth, trim={2.2in 0 2.2in 0},clip] {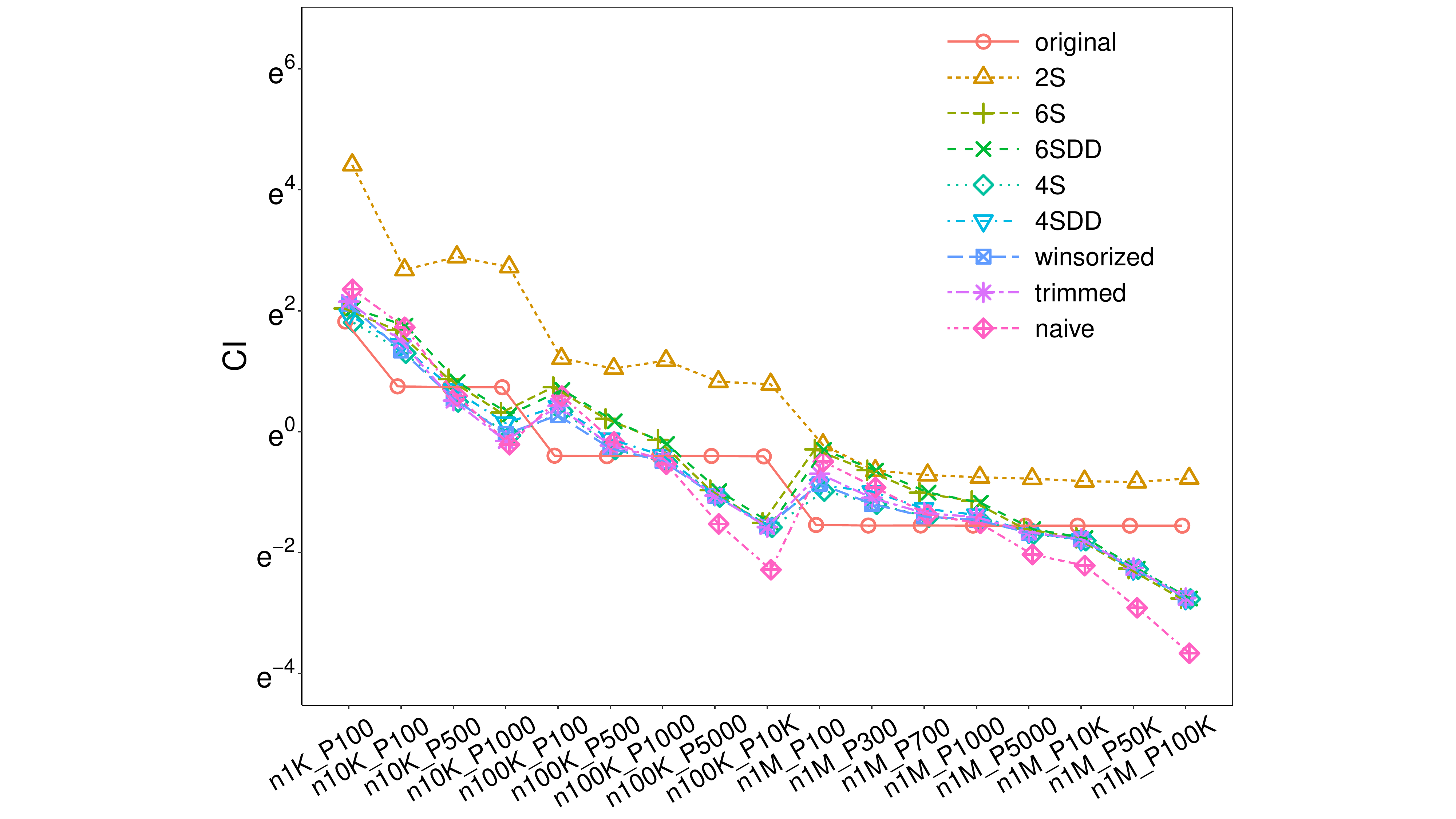}
\includegraphics[width=0.26\textwidth, trim={2.2in 0 2.2in 0},clip] {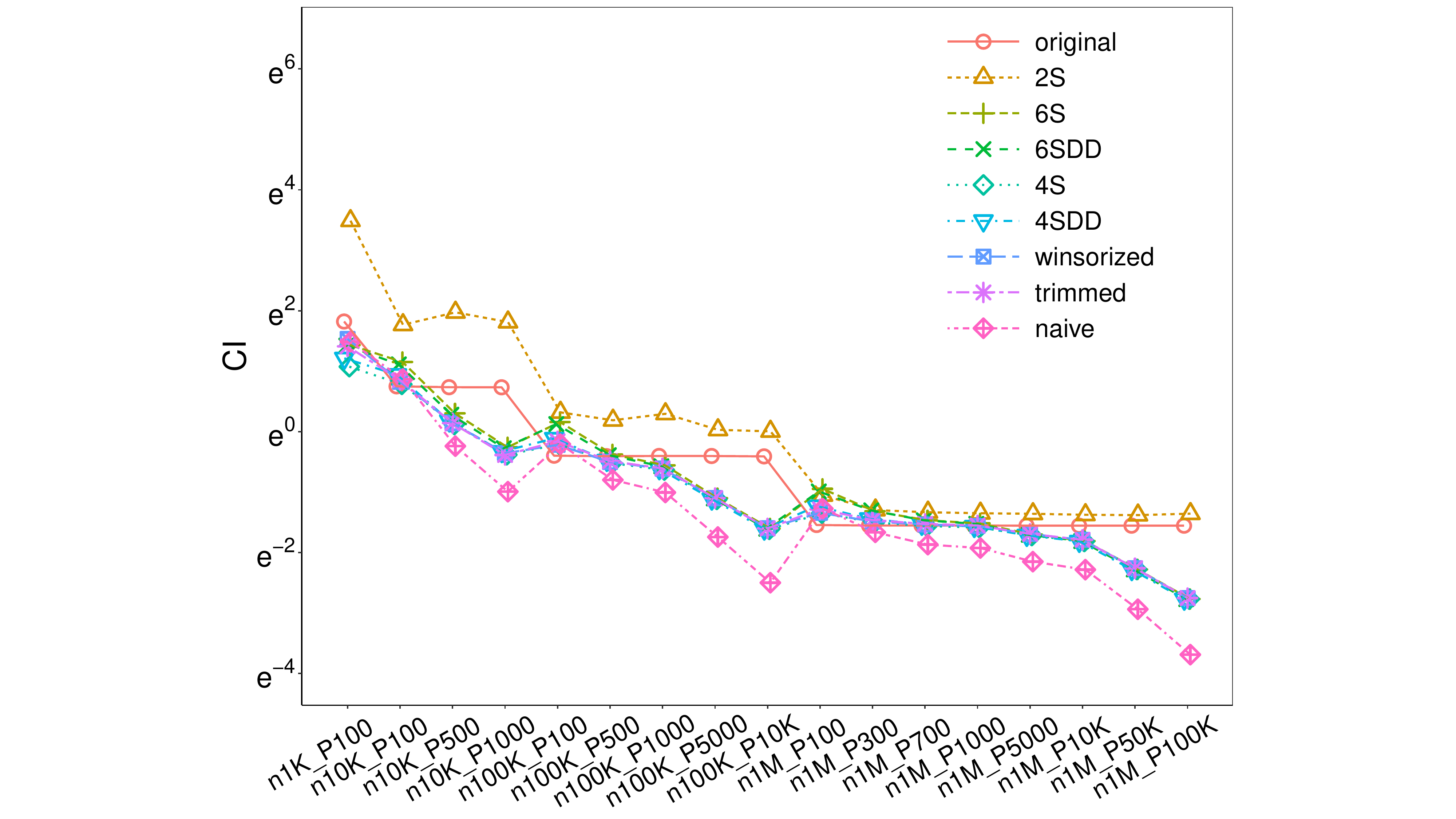}
\includegraphics[width=0.26\textwidth, trim={2.2in 0 2.2in 0},clip] {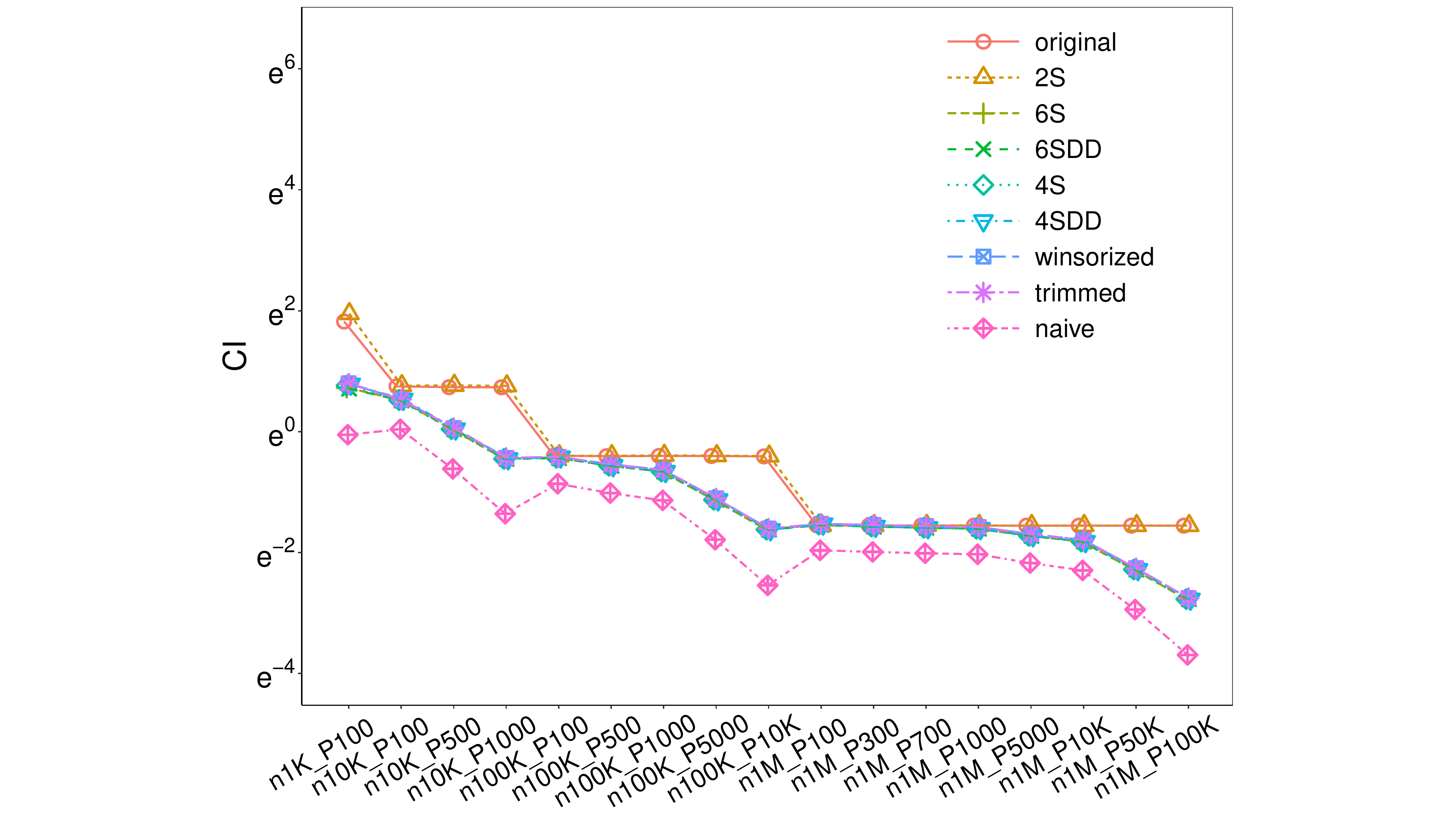}\\
\includegraphics[width=0.26\textwidth, trim={2.2in 0 2.2in 0},clip] {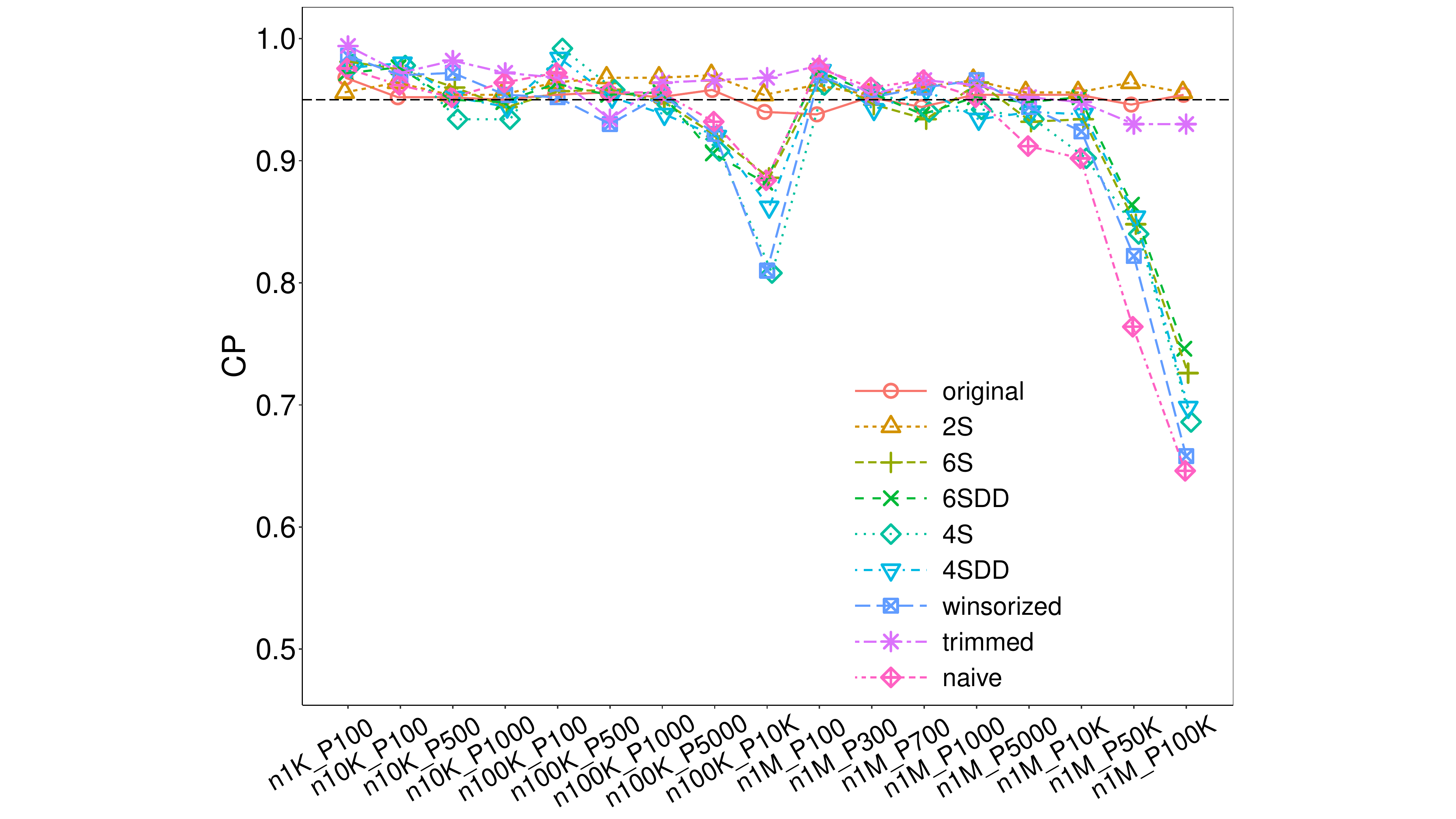}
\includegraphics[width=0.26\textwidth, trim={2.2in 0 2.2in 0},clip] {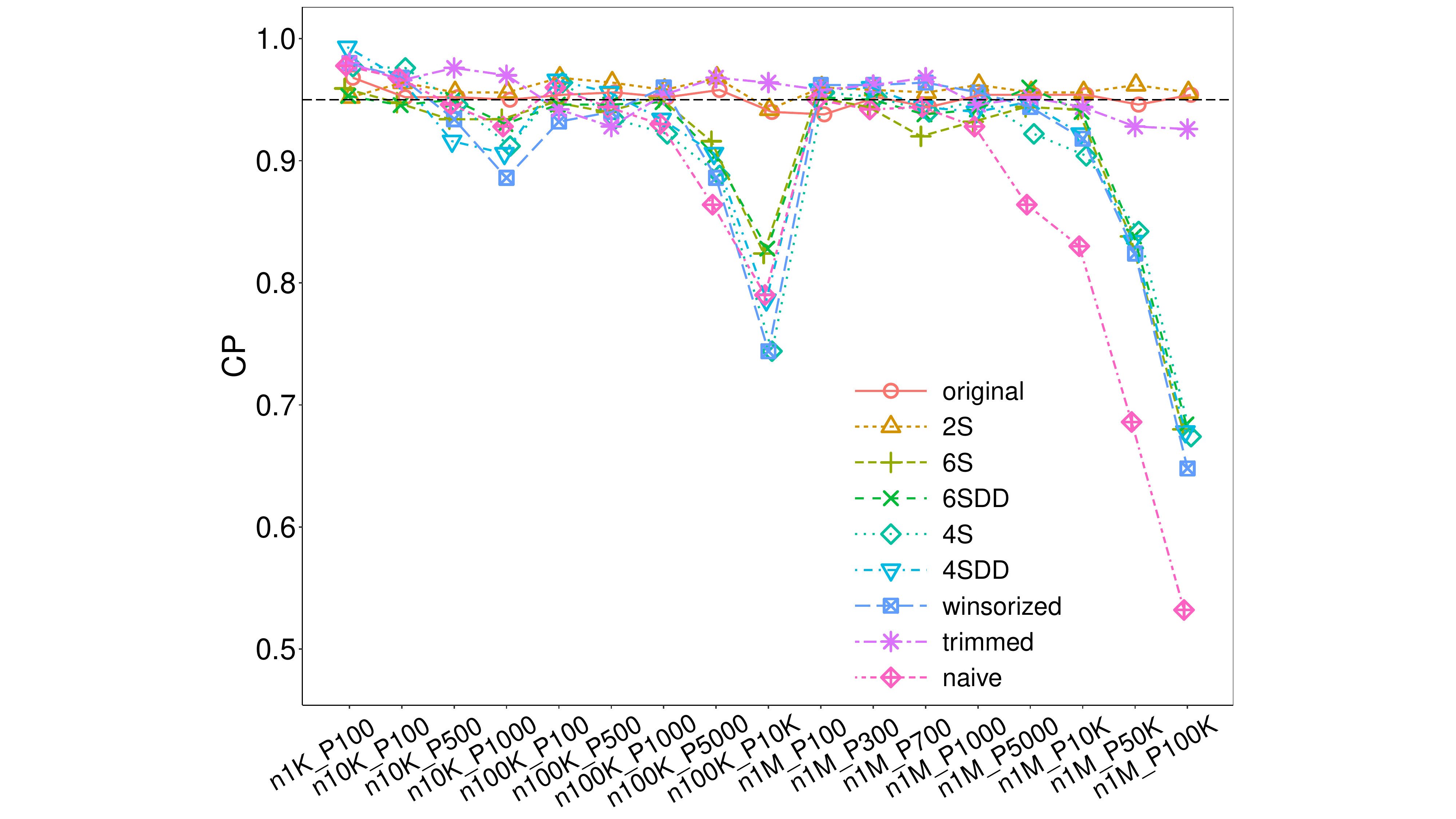}
\includegraphics[width=0.26\textwidth, trim={2.2in 0 2.2in 0},clip] {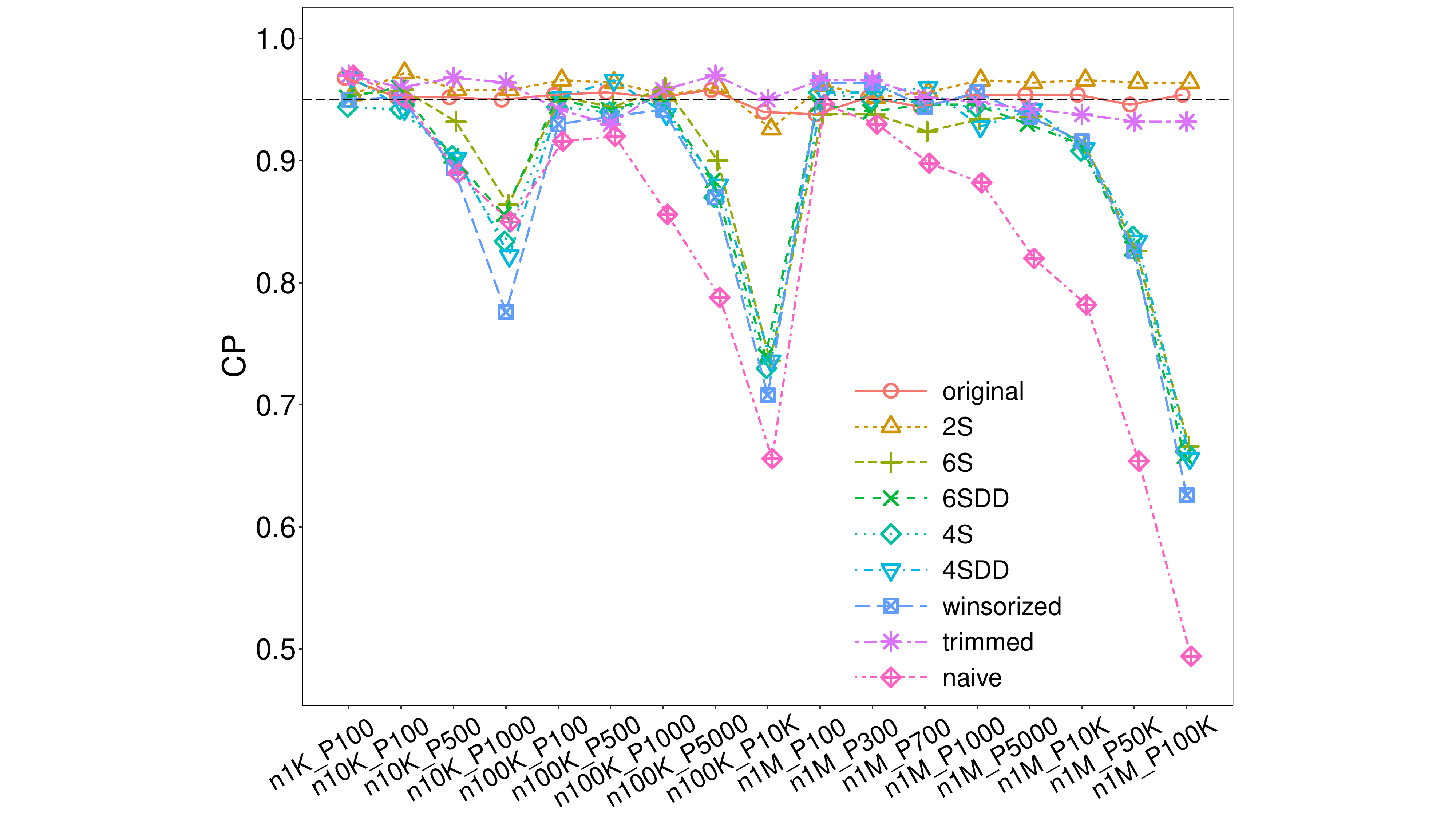}
\includegraphics[width=0.26\textwidth, trim={2.2in 0 2.2in 0},clip] {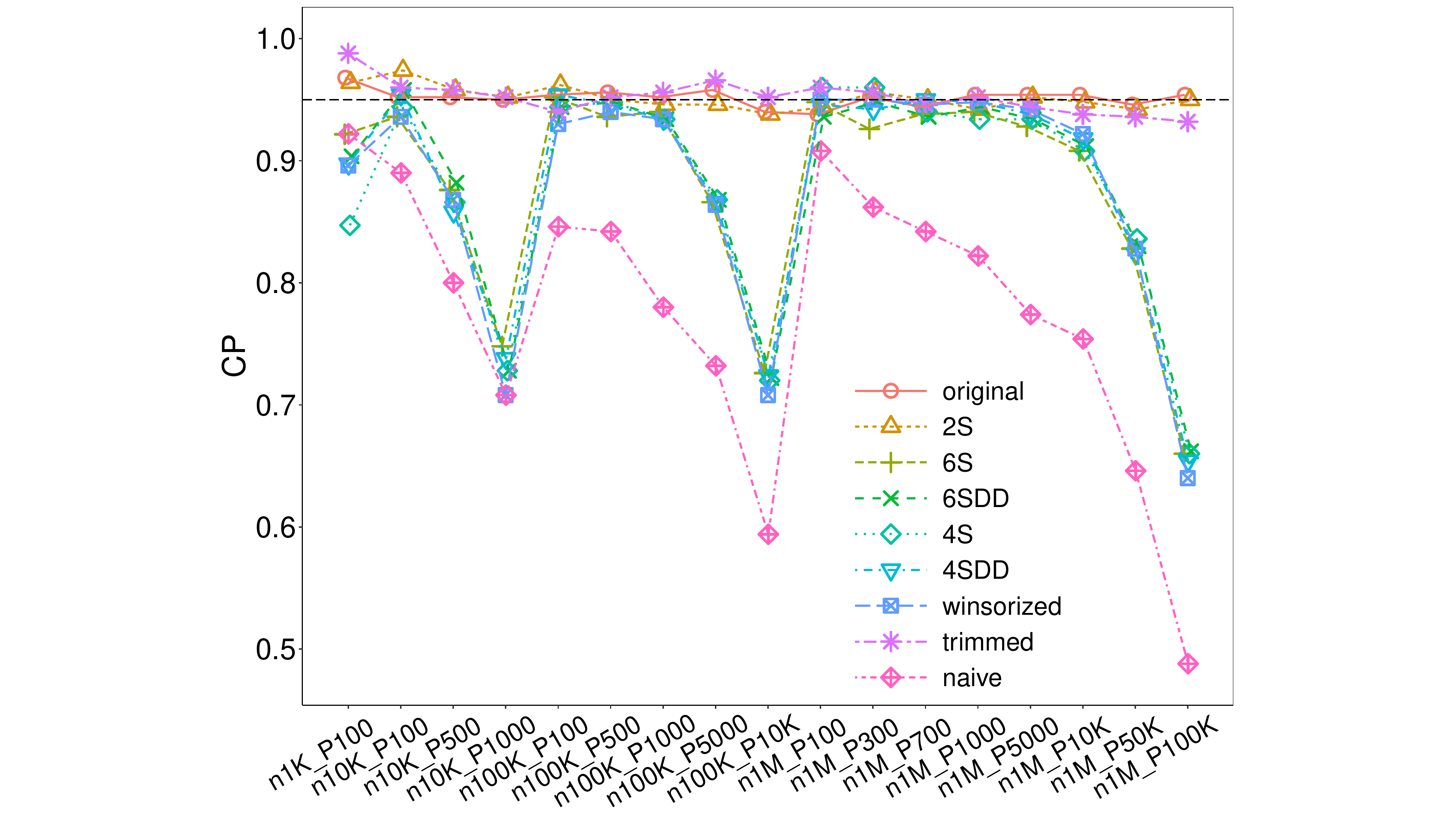}
\includegraphics[width=0.26\textwidth, trim={2.2in 0 2.2in 0},clip] {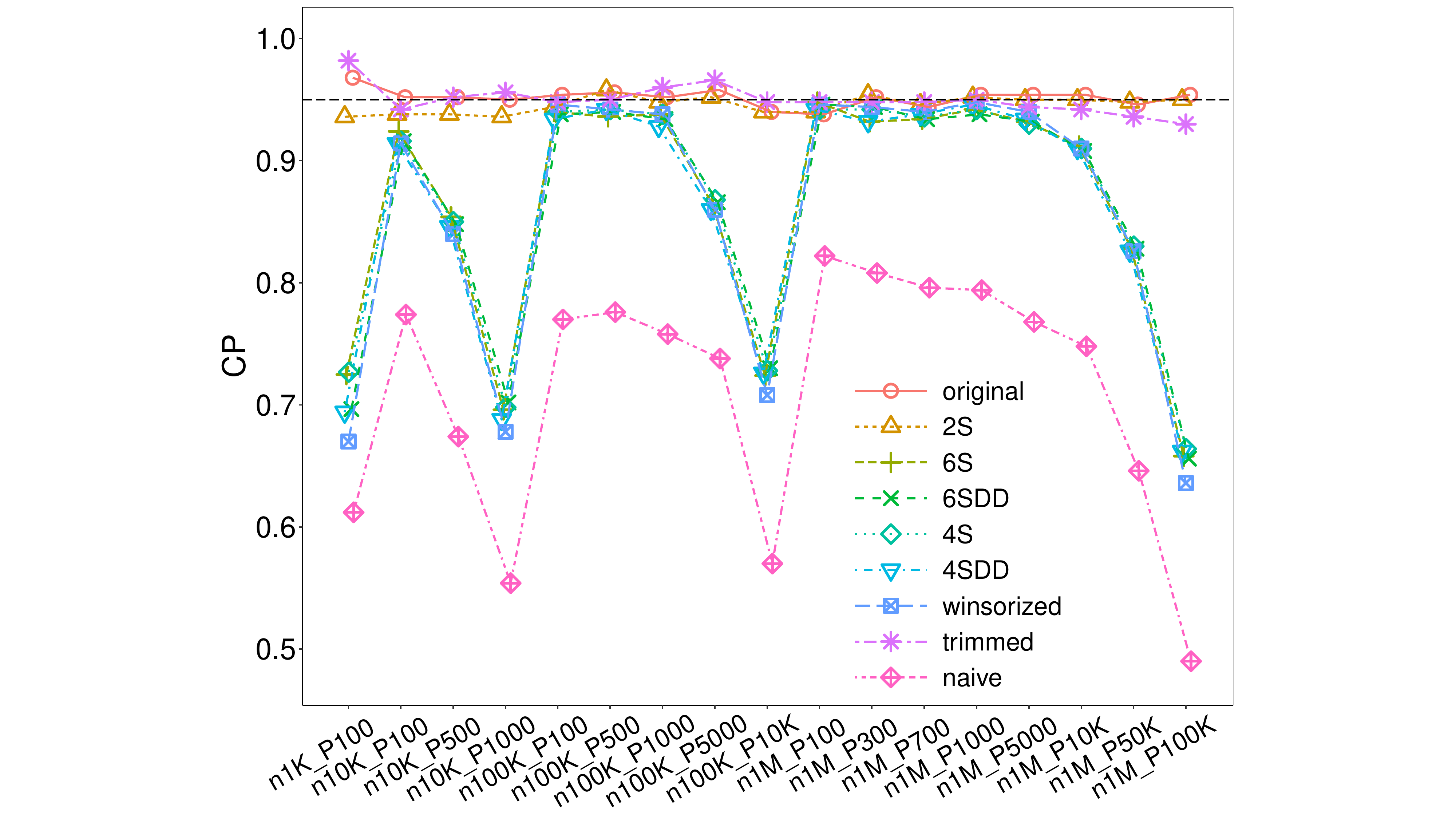}\\

\caption{ZILN data; $\epsilon$-DP; $\theta=0$ and $\alpha=\beta$} \label{fig:0sDPziln}
\end{figure}
\end{landscape}

\begin{landscape}
\begin{figure}[!htb]
\hspace{0.6in}$\rho=0.005$\hspace{1in}$\rho=0.02$\hspace{1.2in}$\rho=0.08$
\hspace{1.1in}$\rho=0.32$\hspace{1.2in}$\rho=1.28$\\
\includegraphics[width=0.26\textwidth, trim={2.2in 0 2.2in 0},clip] {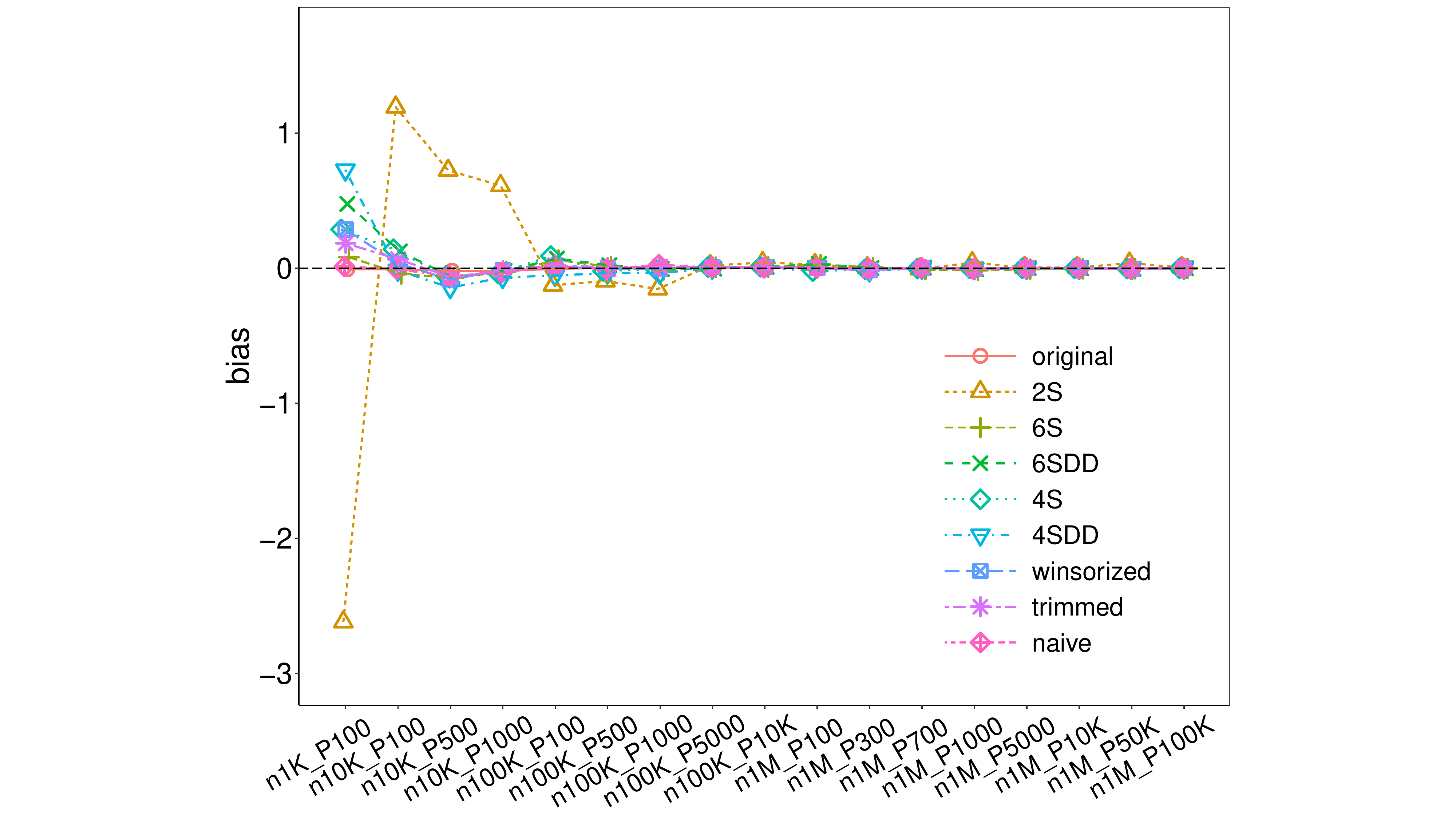}
\includegraphics[width=0.26\textwidth, trim={2.2in 0 2.2in 0},clip] {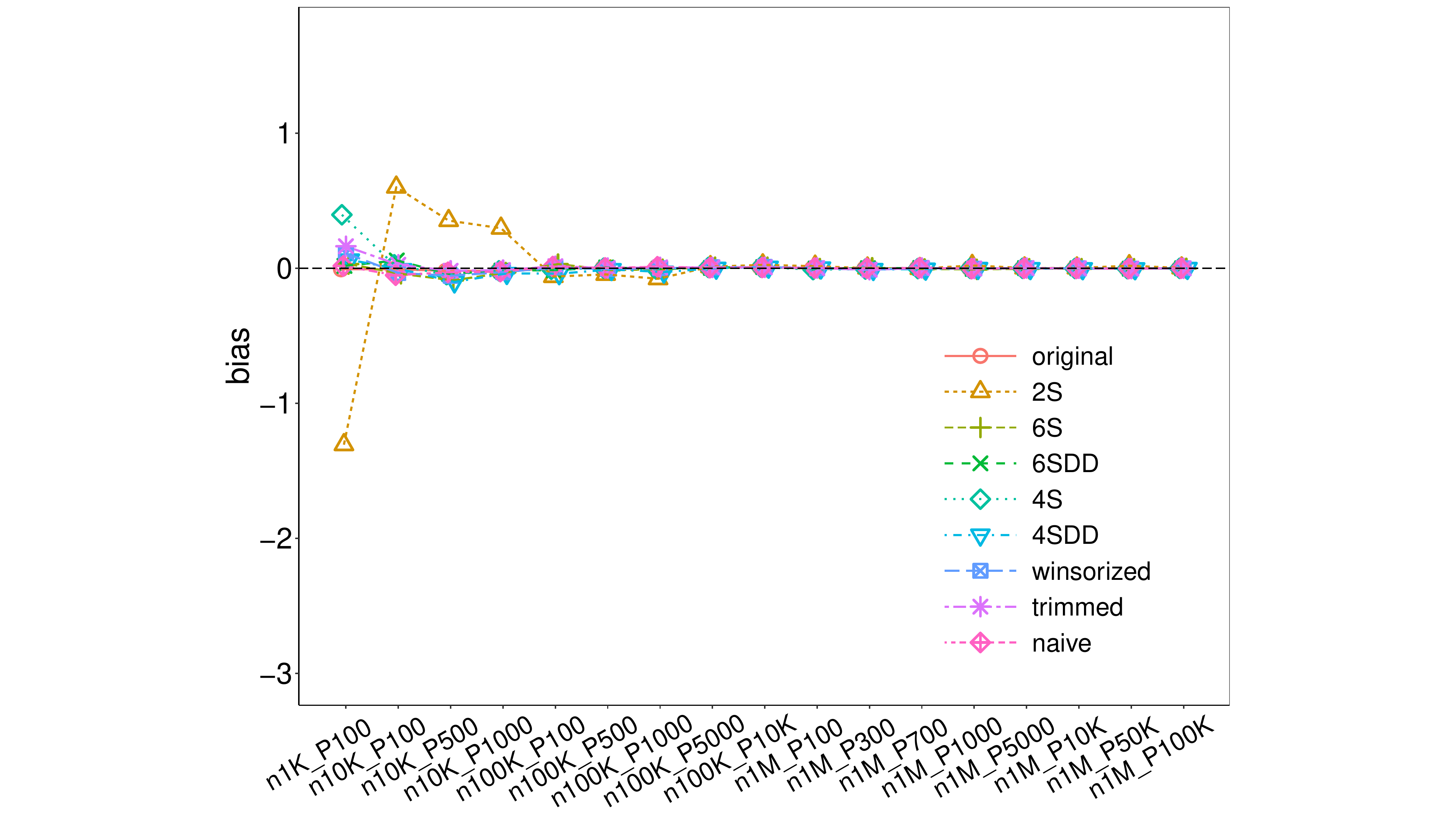}
\includegraphics[width=0.26\textwidth, trim={2.2in 0 2.2in 0},clip] {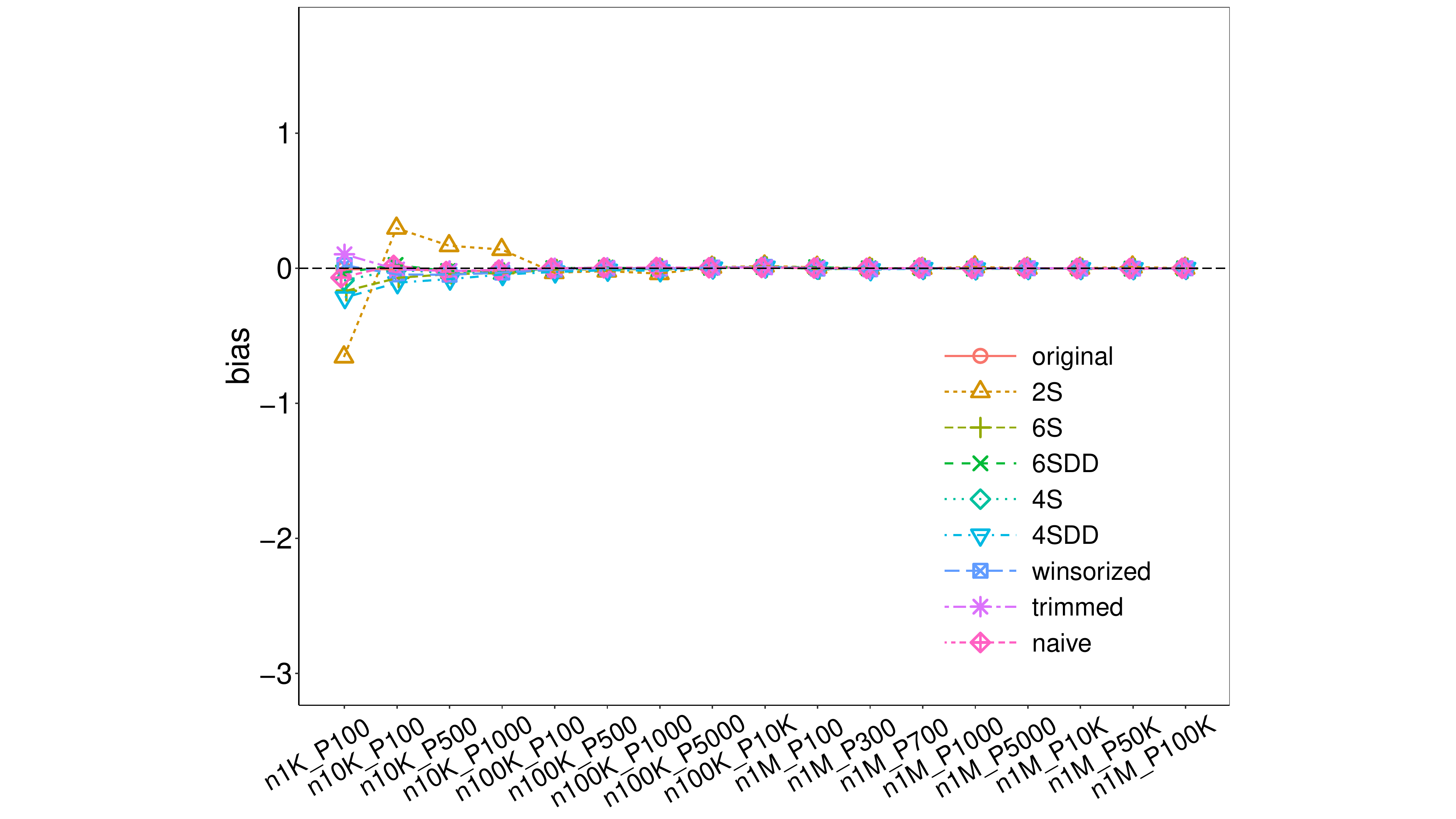}
\includegraphics[width=0.26\textwidth, trim={2.2in 0 2.2in 0},clip] {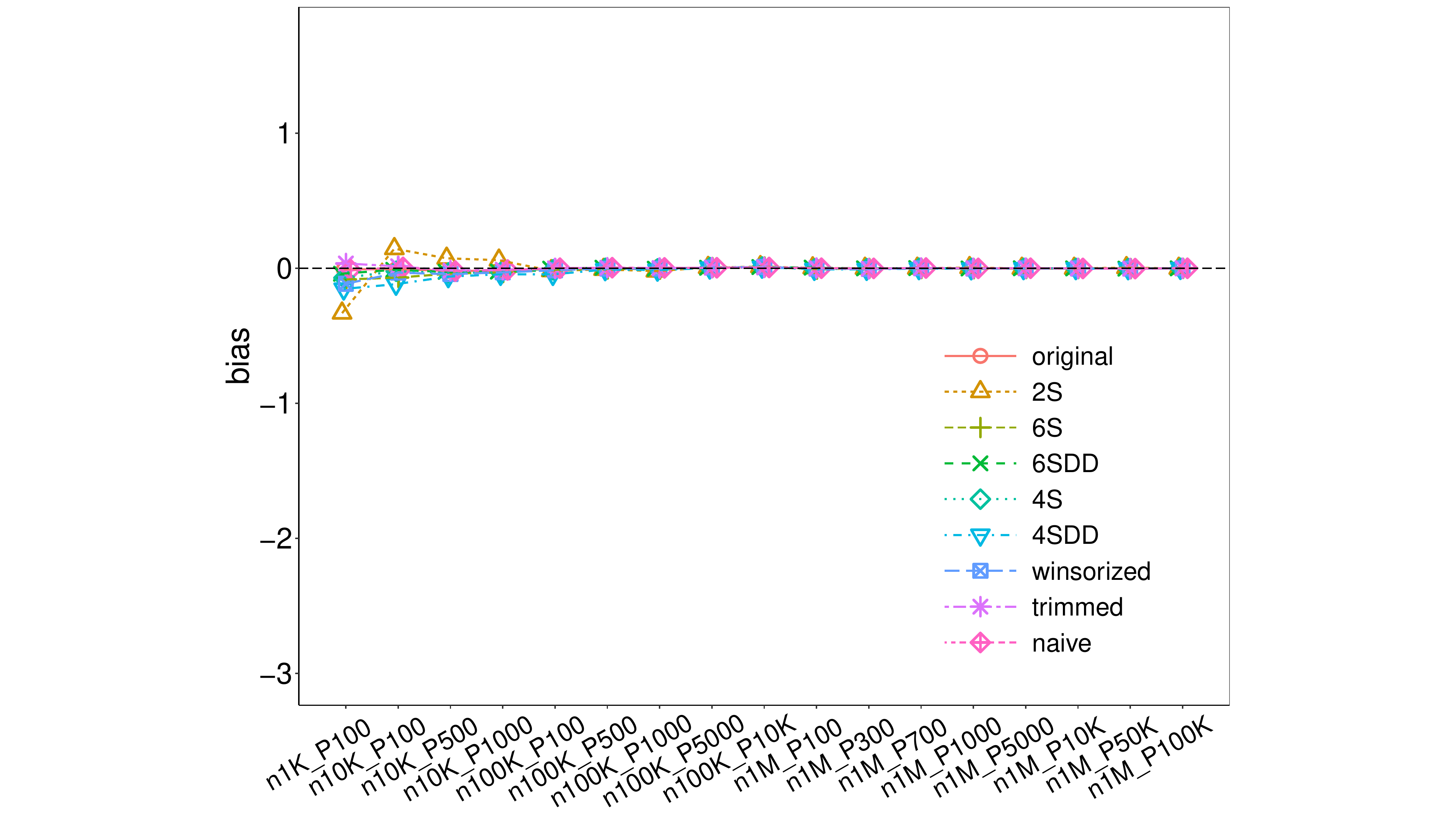}
\includegraphics[width=0.26\textwidth, trim={2.2in 0 2.2in 0},clip] {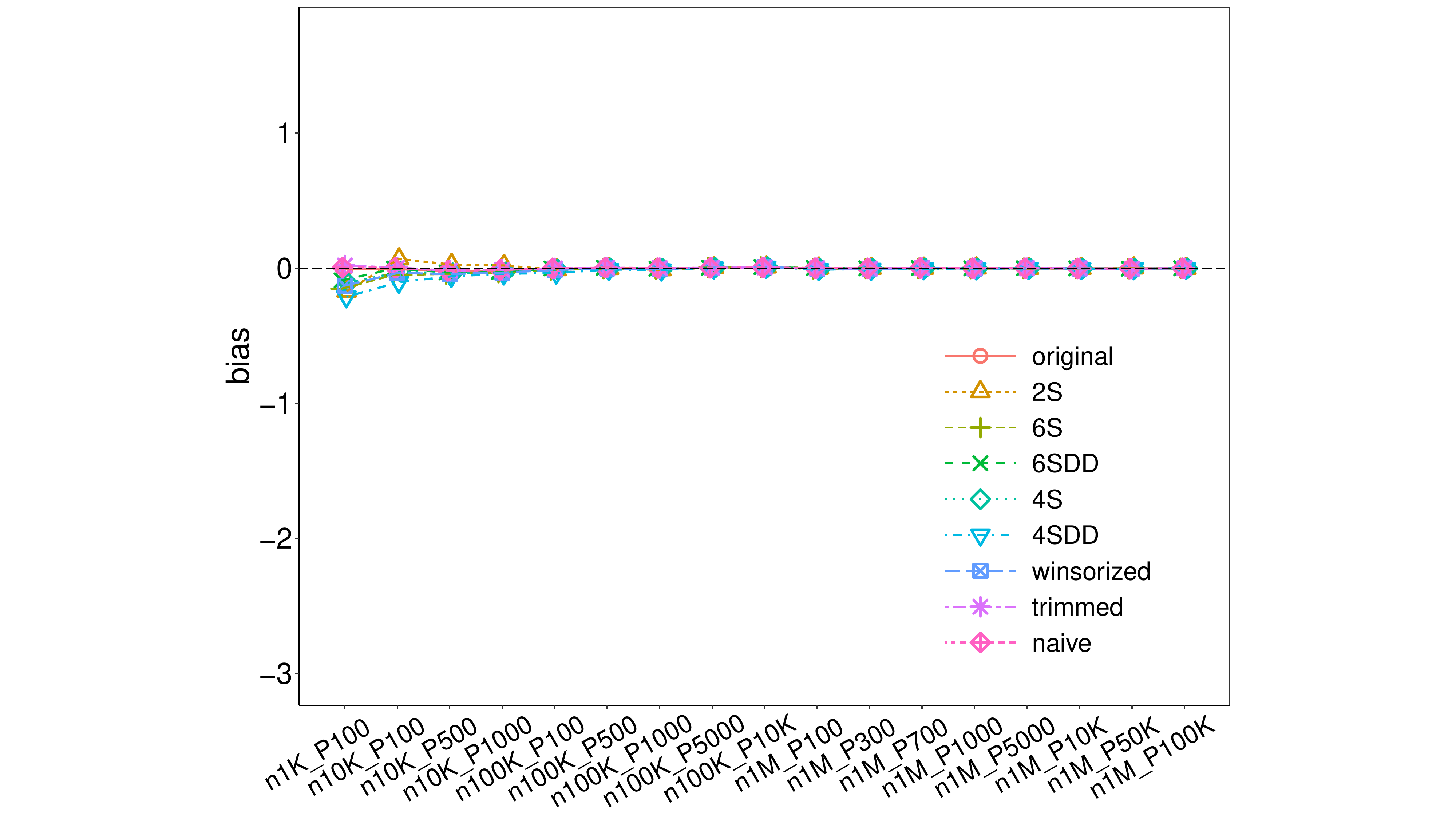}\\
\includegraphics[width=0.26\textwidth, trim={2.2in 0 2.2in 0},clip] {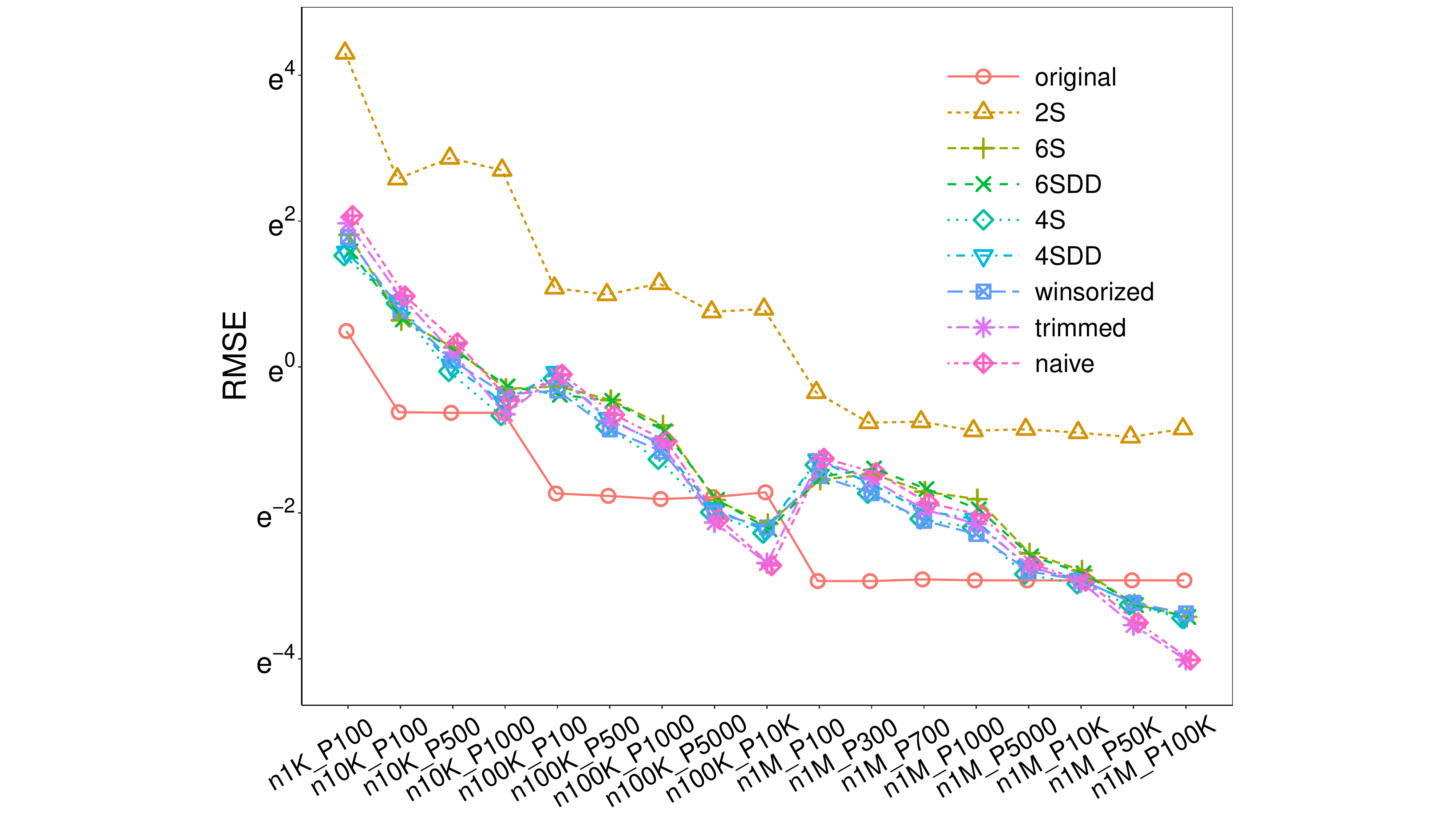}
\includegraphics[width=0.26\textwidth, trim={2.2in 0 2.2in 0},clip] {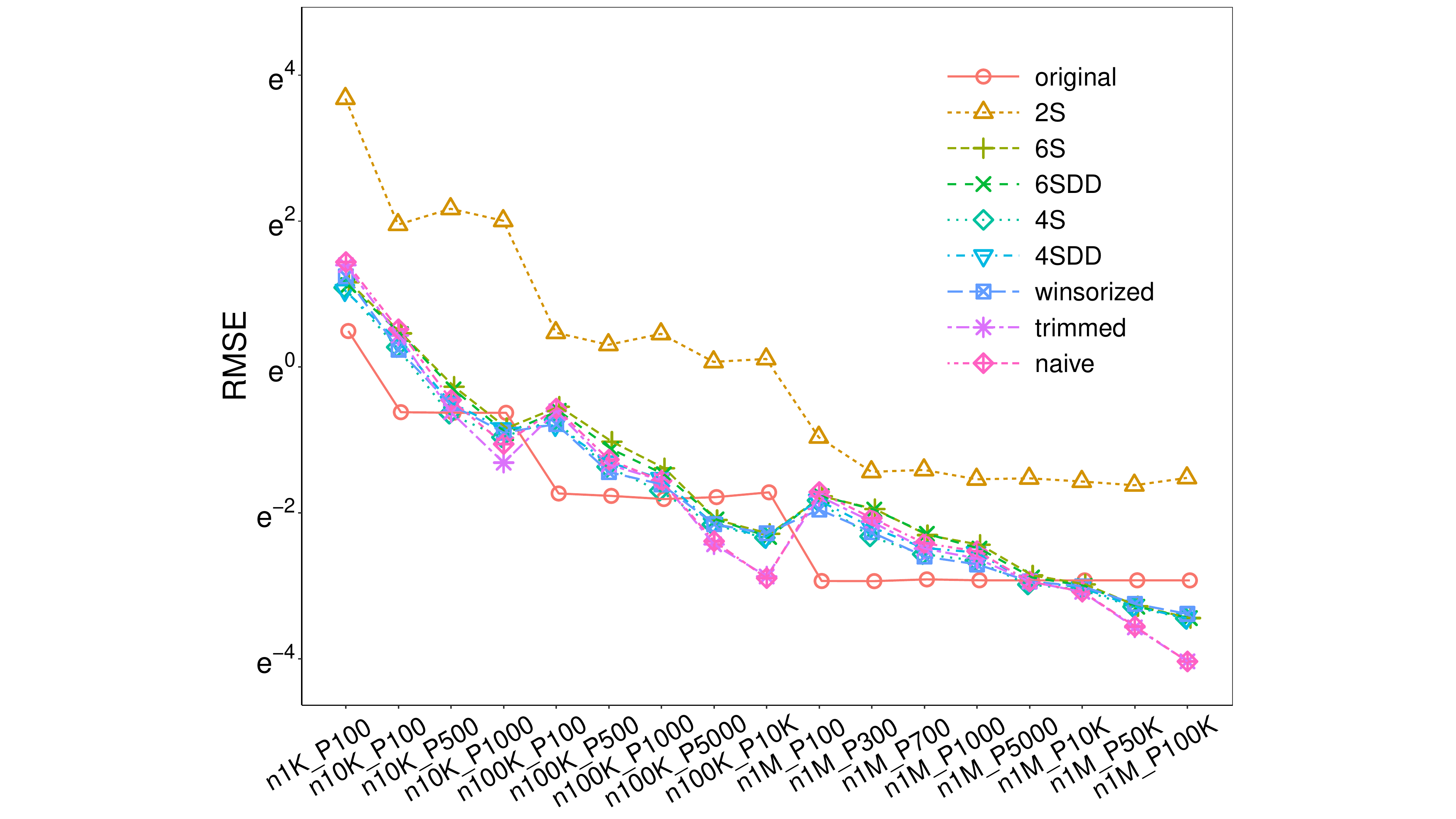}
\includegraphics[width=0.26\textwidth, trim={2.2in 0 2.2in 0},clip] {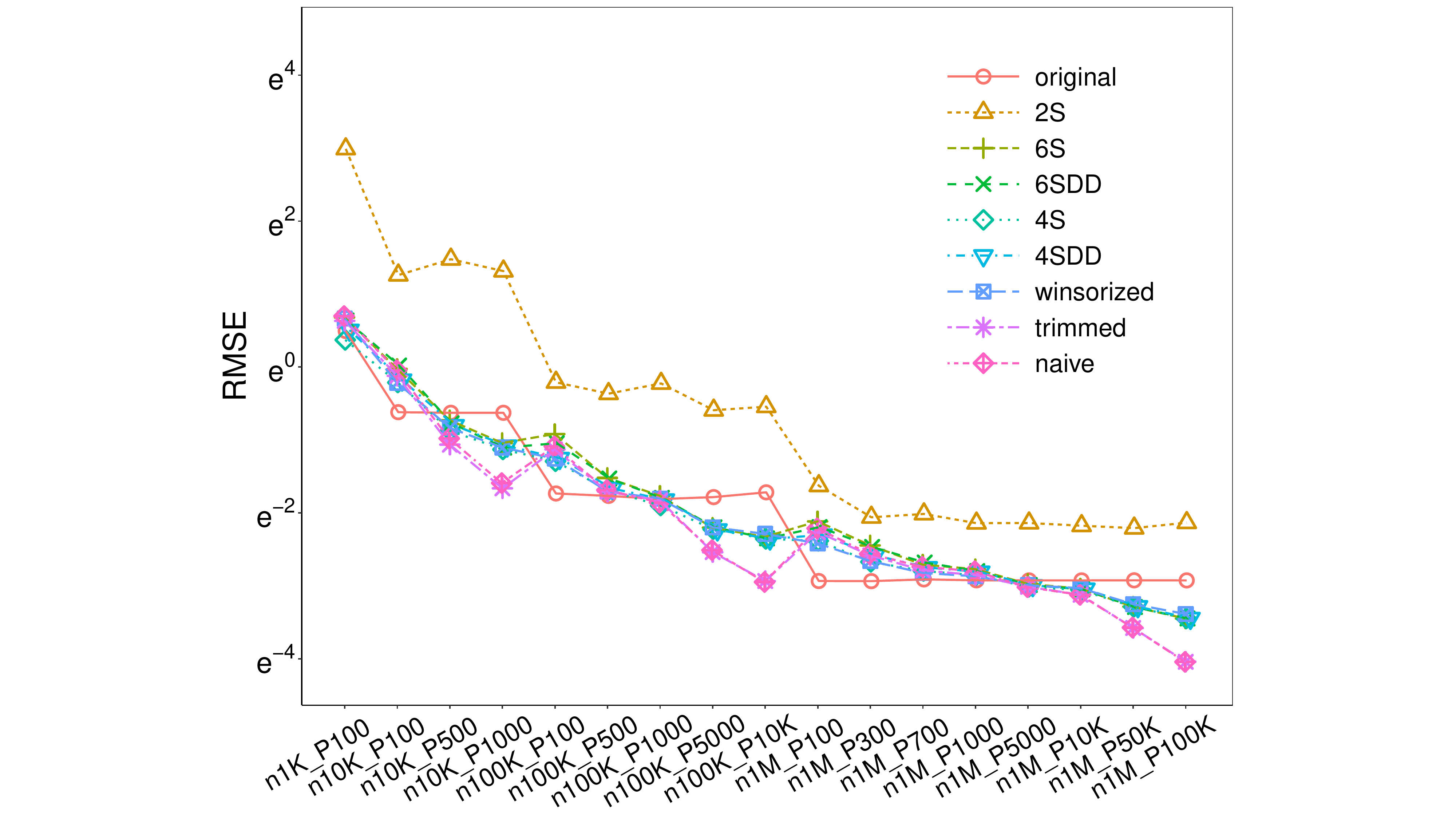}
\includegraphics[width=0.26\textwidth, trim={2.2in 0 2.2in 0},clip] {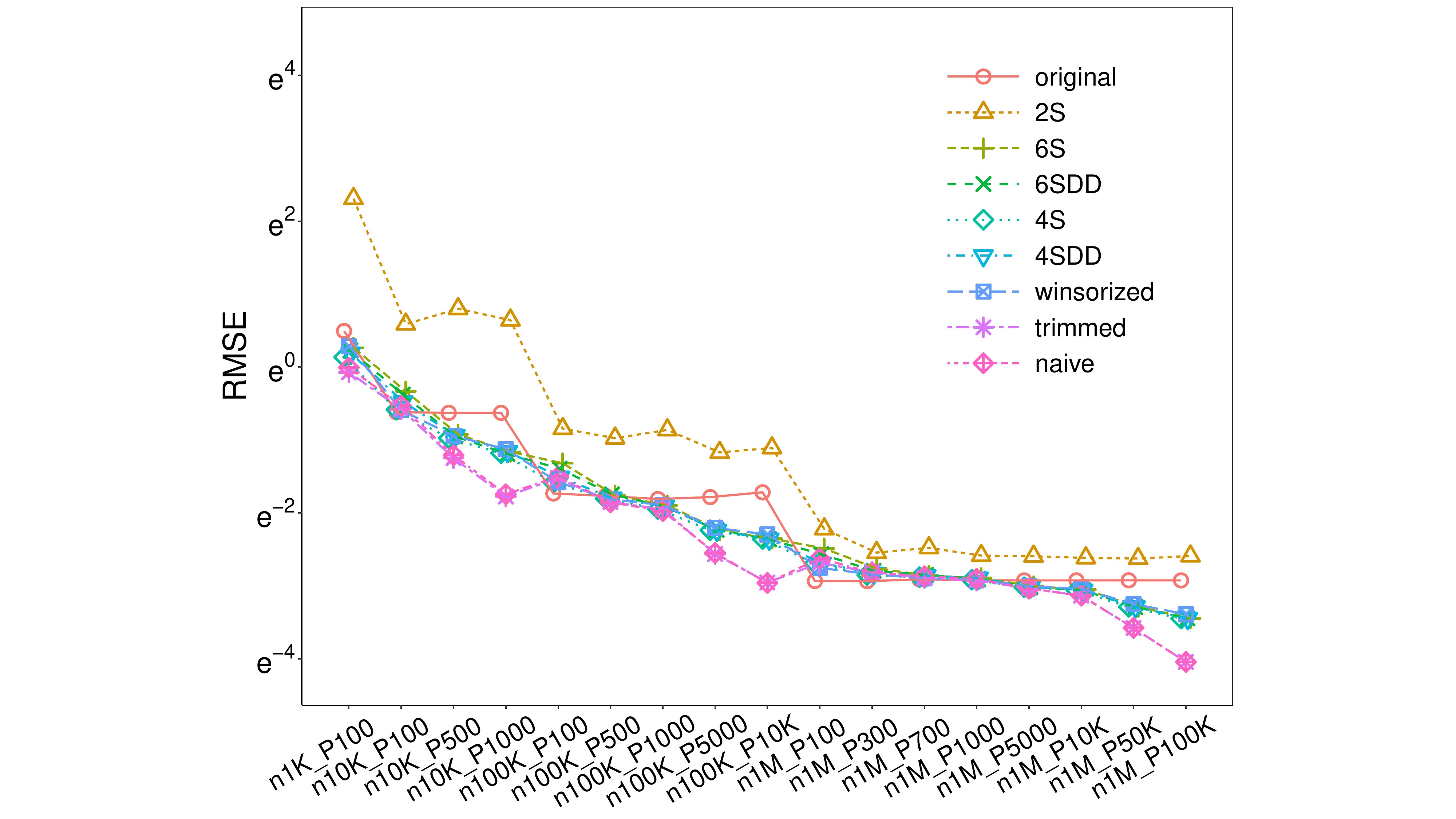}
\includegraphics[width=0.26\textwidth, trim={2.2in 0 2.2in 0},clip] {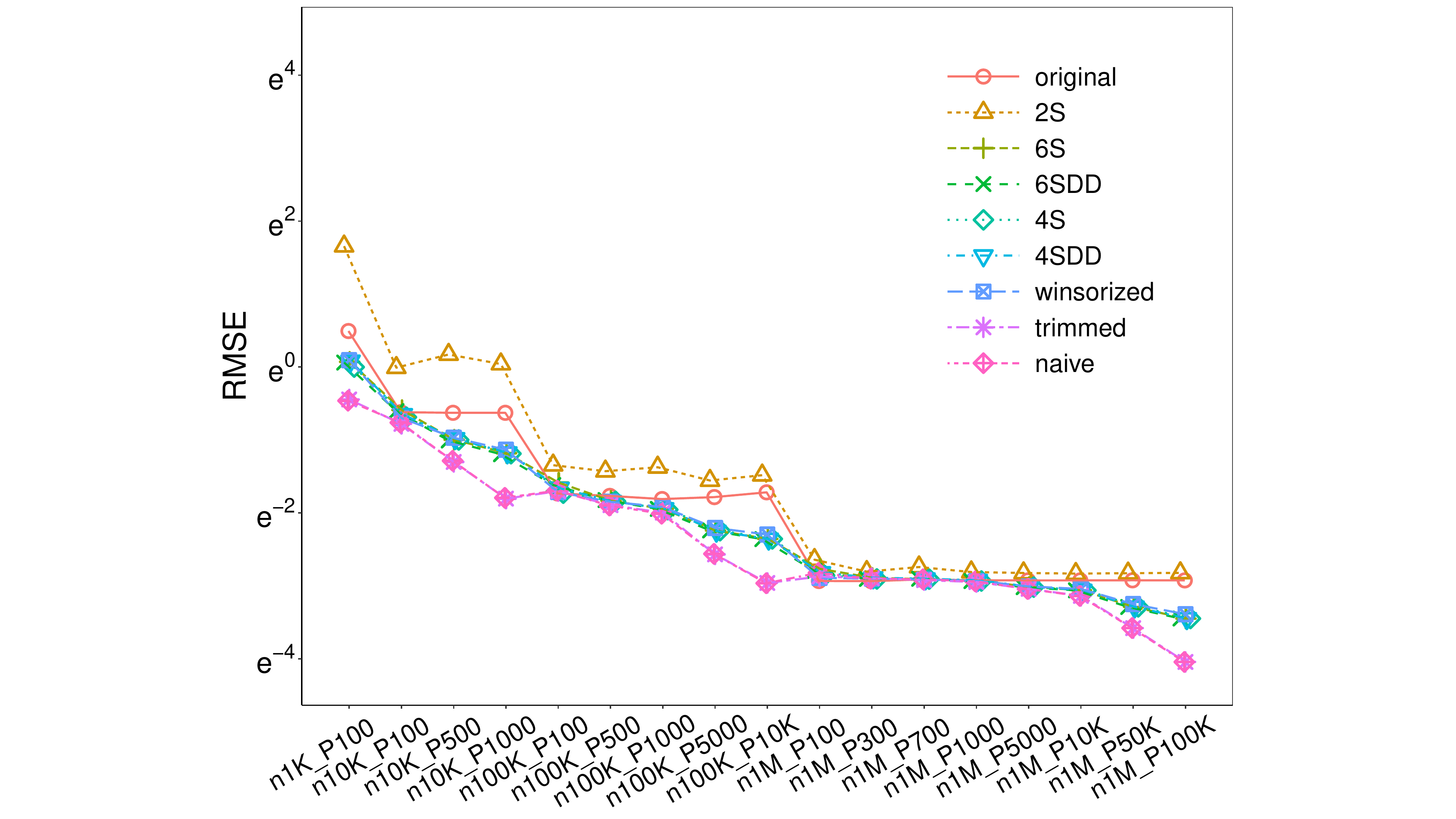}\\
\includegraphics[width=0.26\textwidth, trim={2.2in 0 2.2in 0},clip] {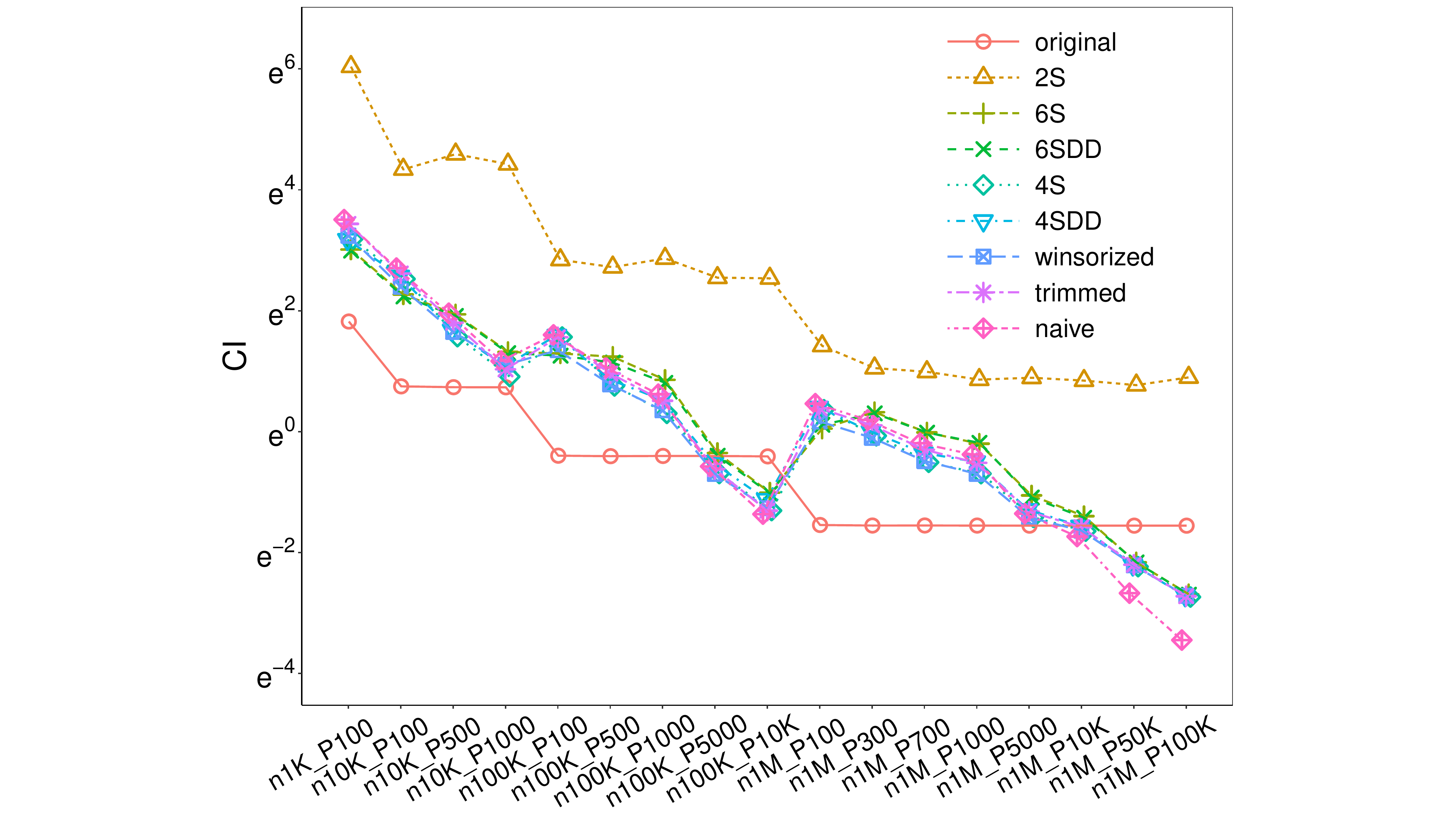}
\includegraphics[width=0.26\textwidth, trim={2.2in 0 2.2in 0},clip] {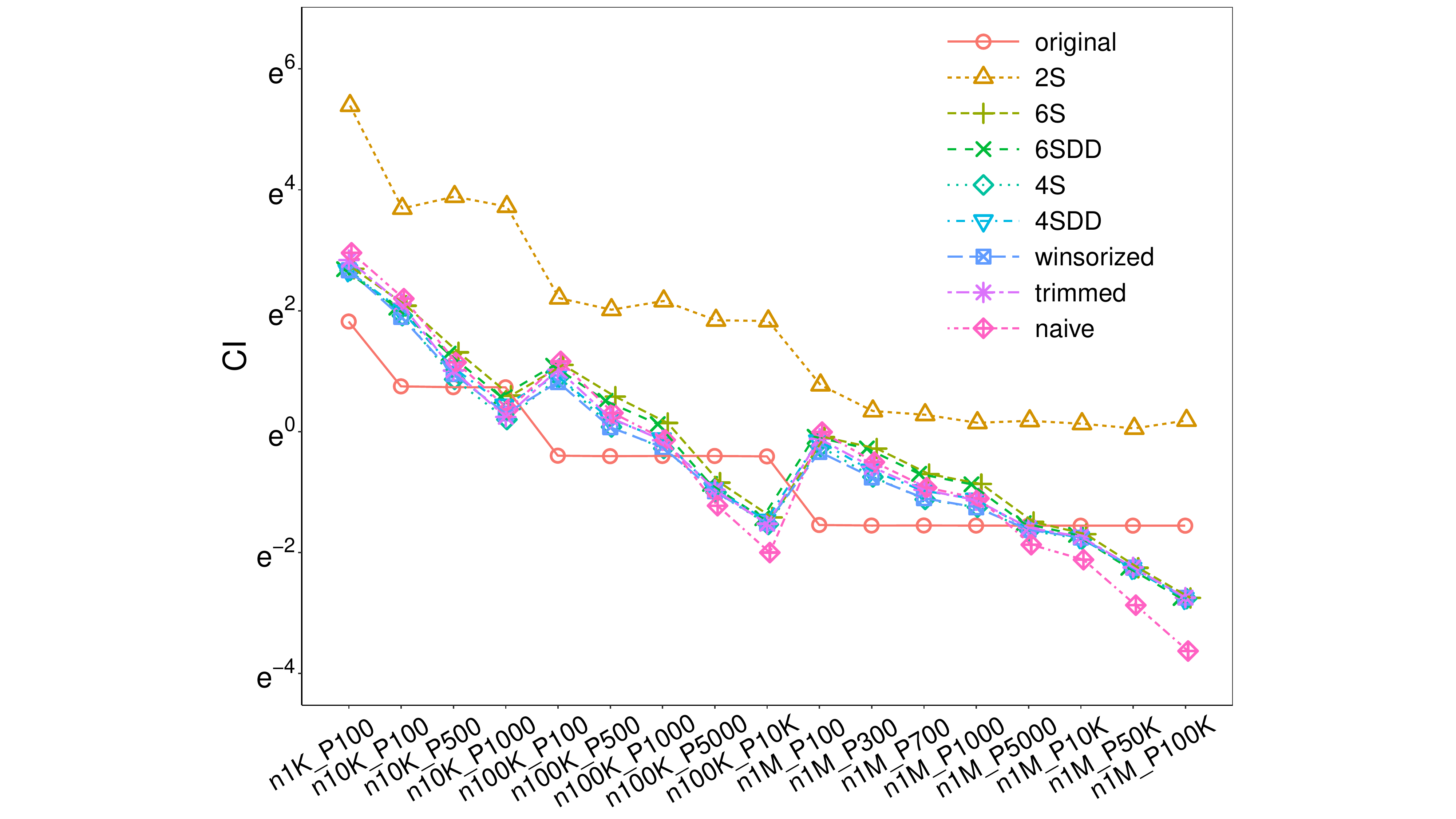}
\includegraphics[width=0.26\textwidth, trim={2.2in 0 2.2in 0},clip] {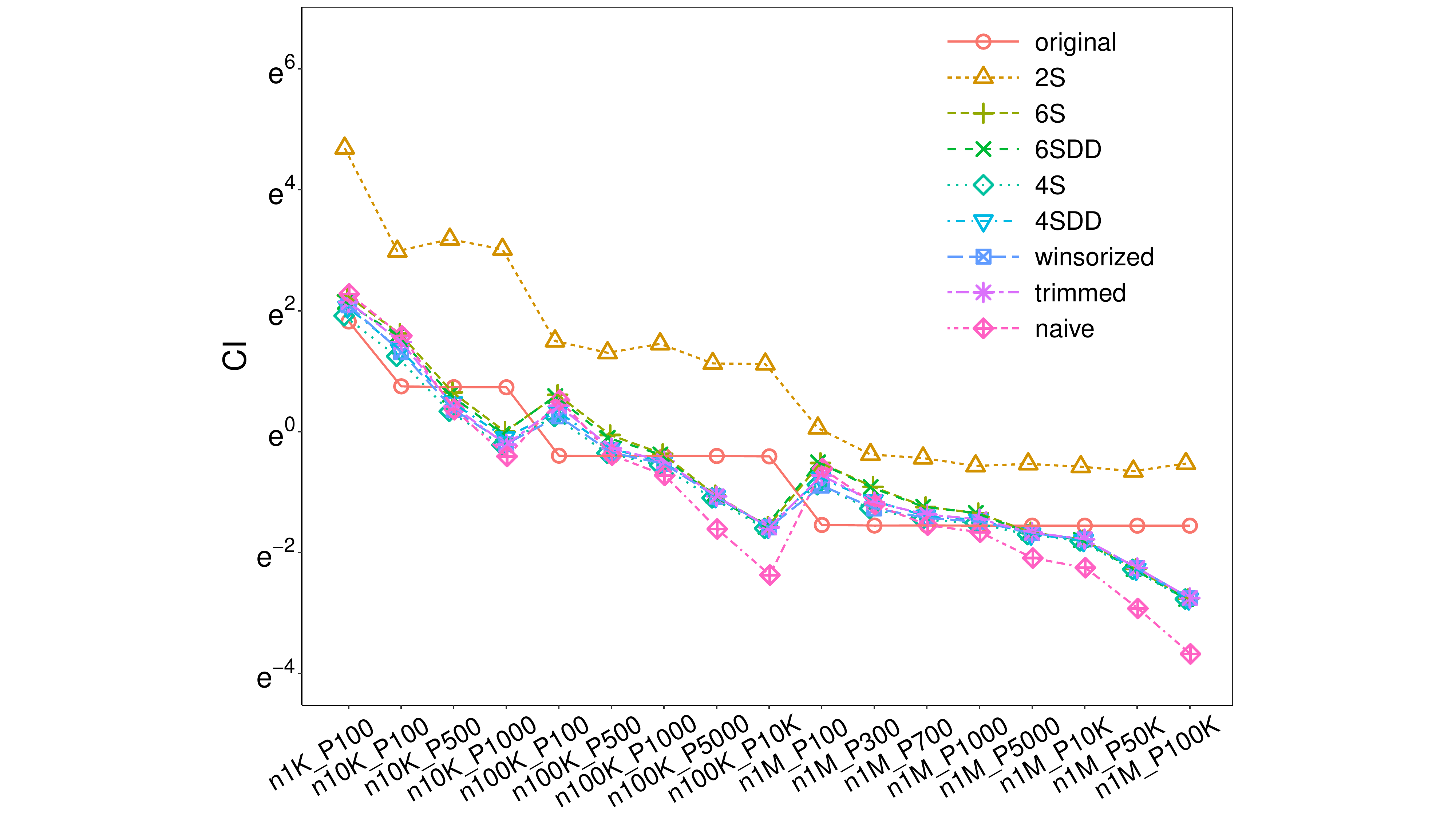}
\includegraphics[width=0.26\textwidth, trim={2.2in 0 2.2in 0},clip] {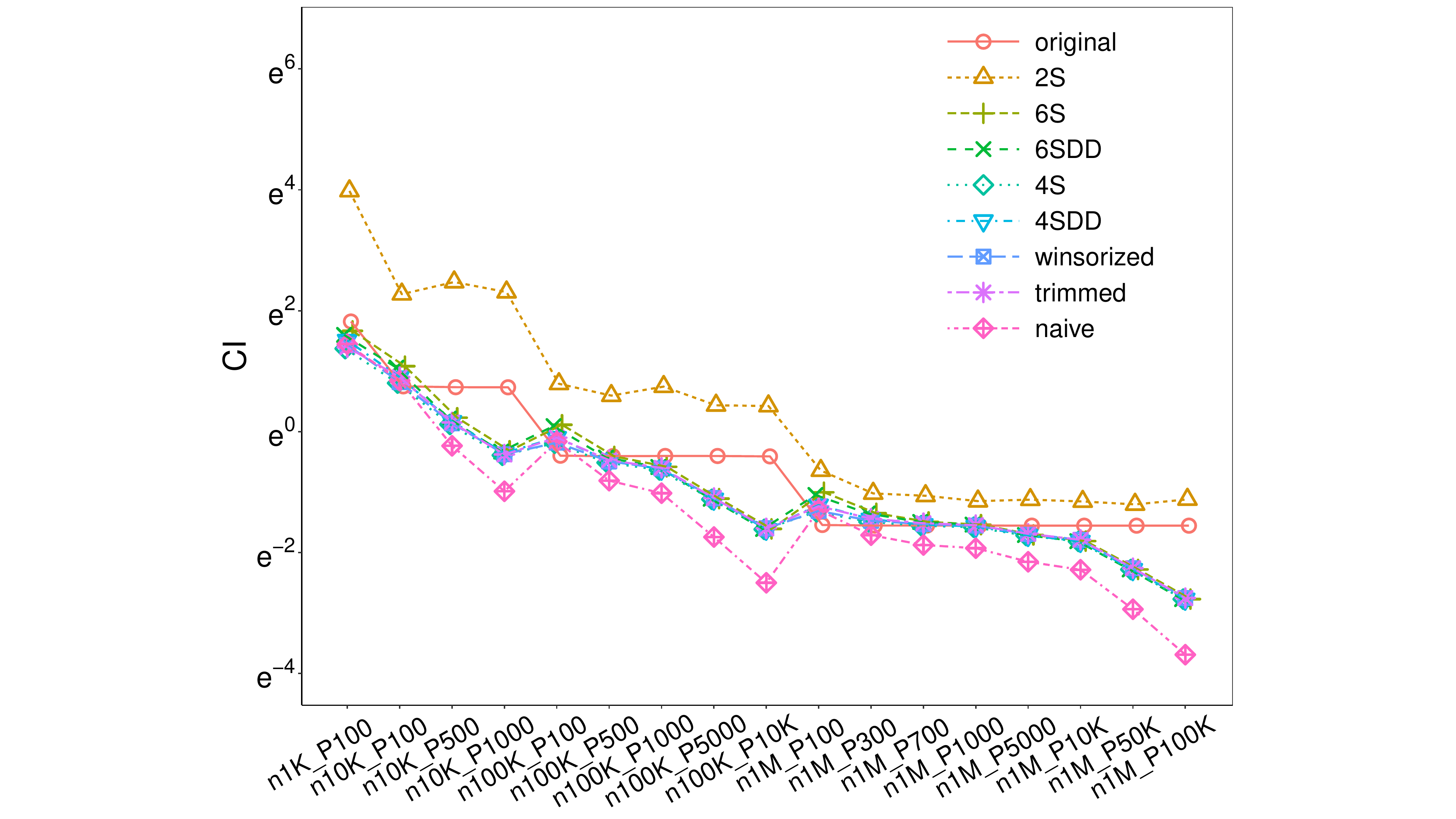}
\includegraphics[width=0.26\textwidth, trim={2.2in 0 2.2in 0},clip] {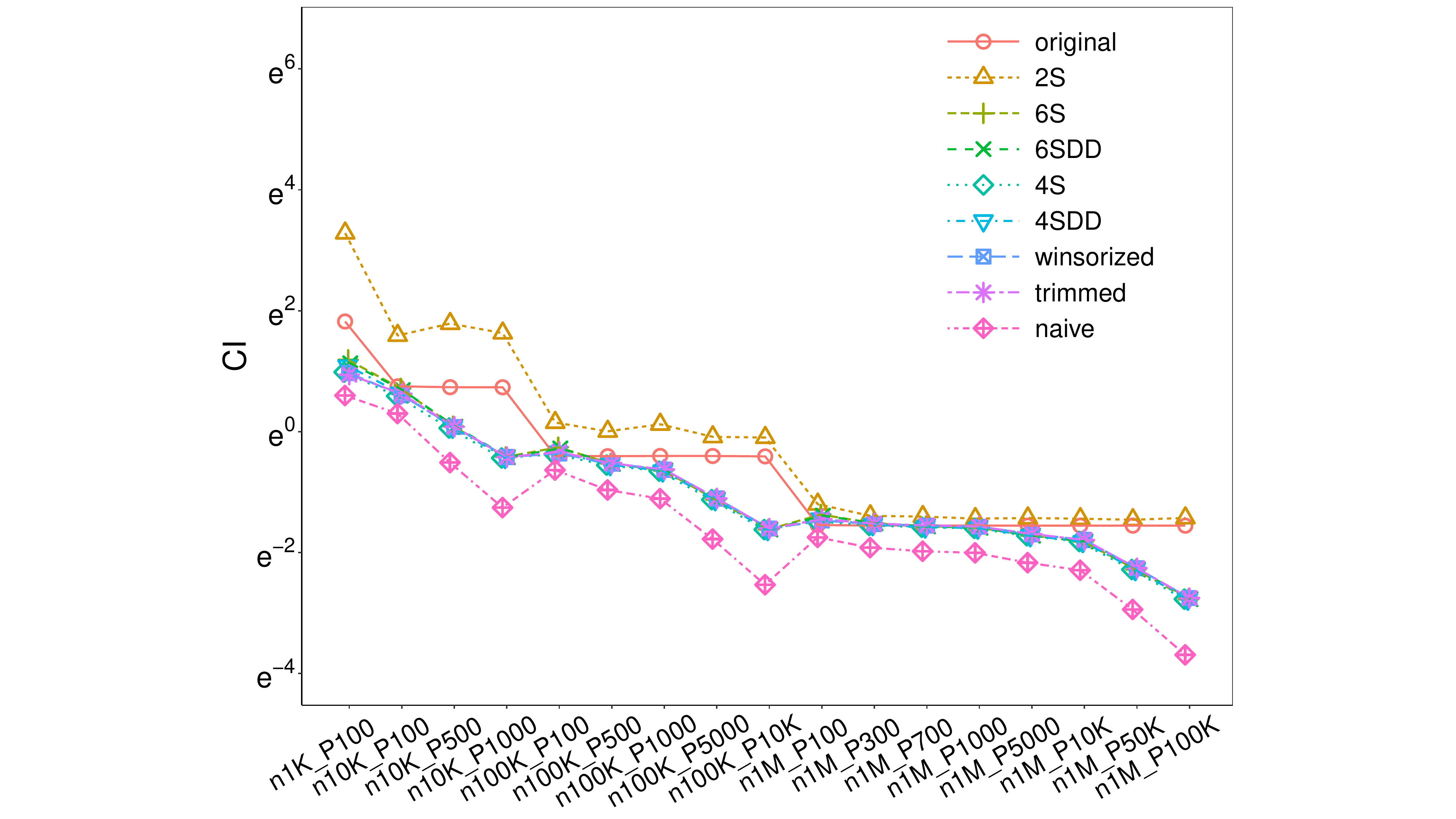}\\
\includegraphics[width=0.26\textwidth, trim={2.2in 0 2.2in 0},clip] {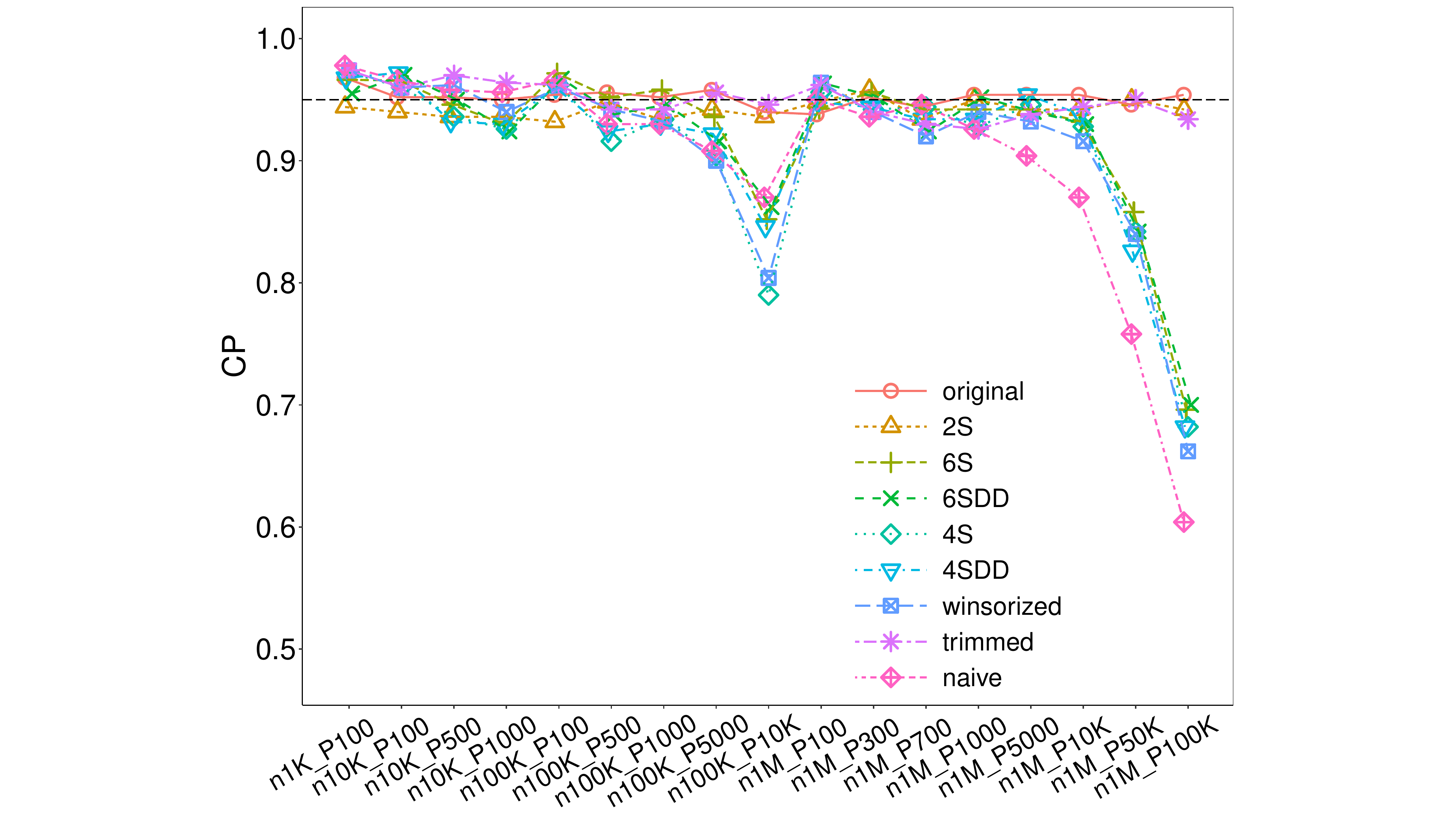}
\includegraphics[width=0.26\textwidth, trim={2.2in 0 2.2in 0},clip] {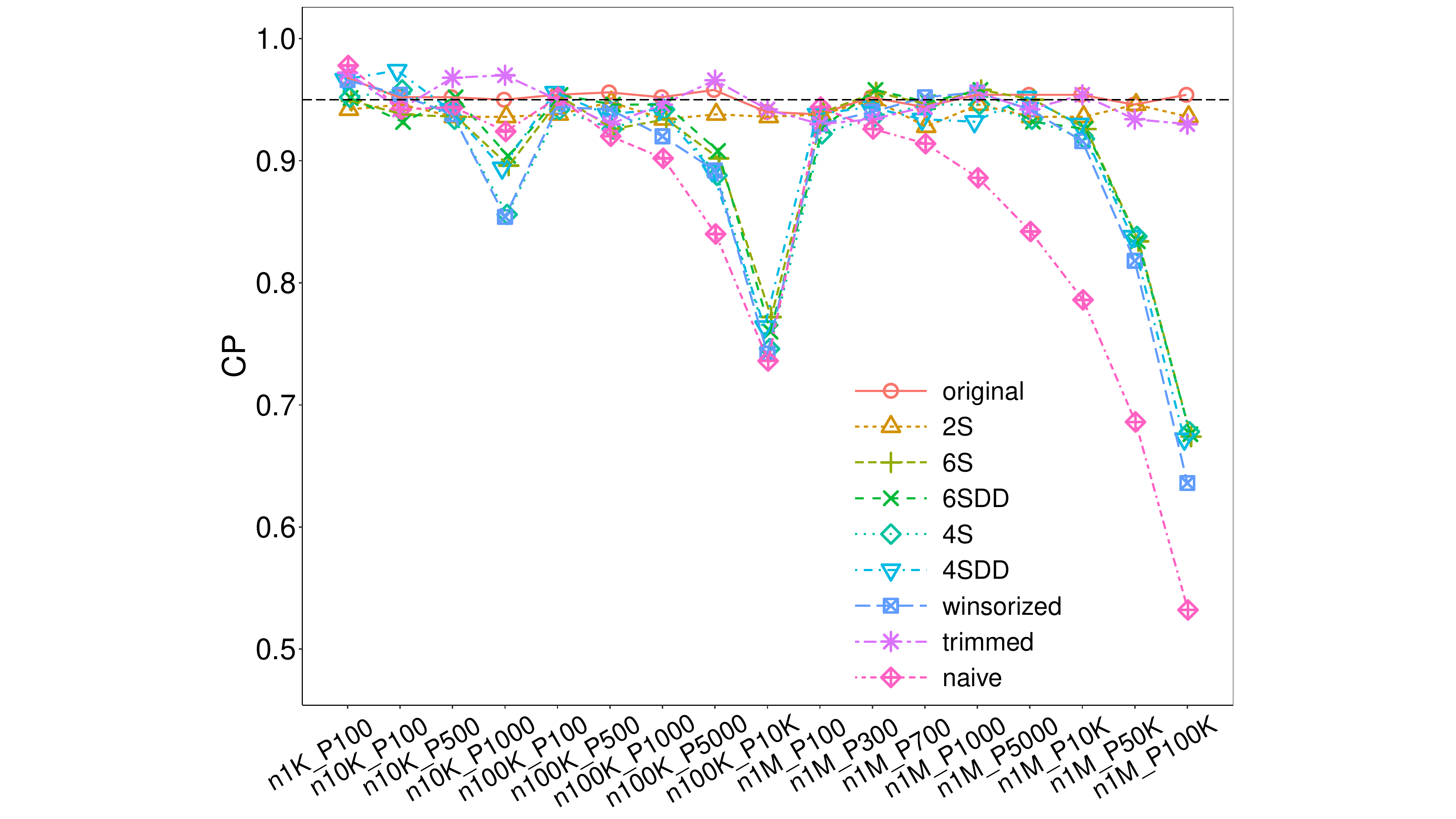}
\includegraphics[width=0.26\textwidth, trim={2.2in 0 2.2in 0},clip] {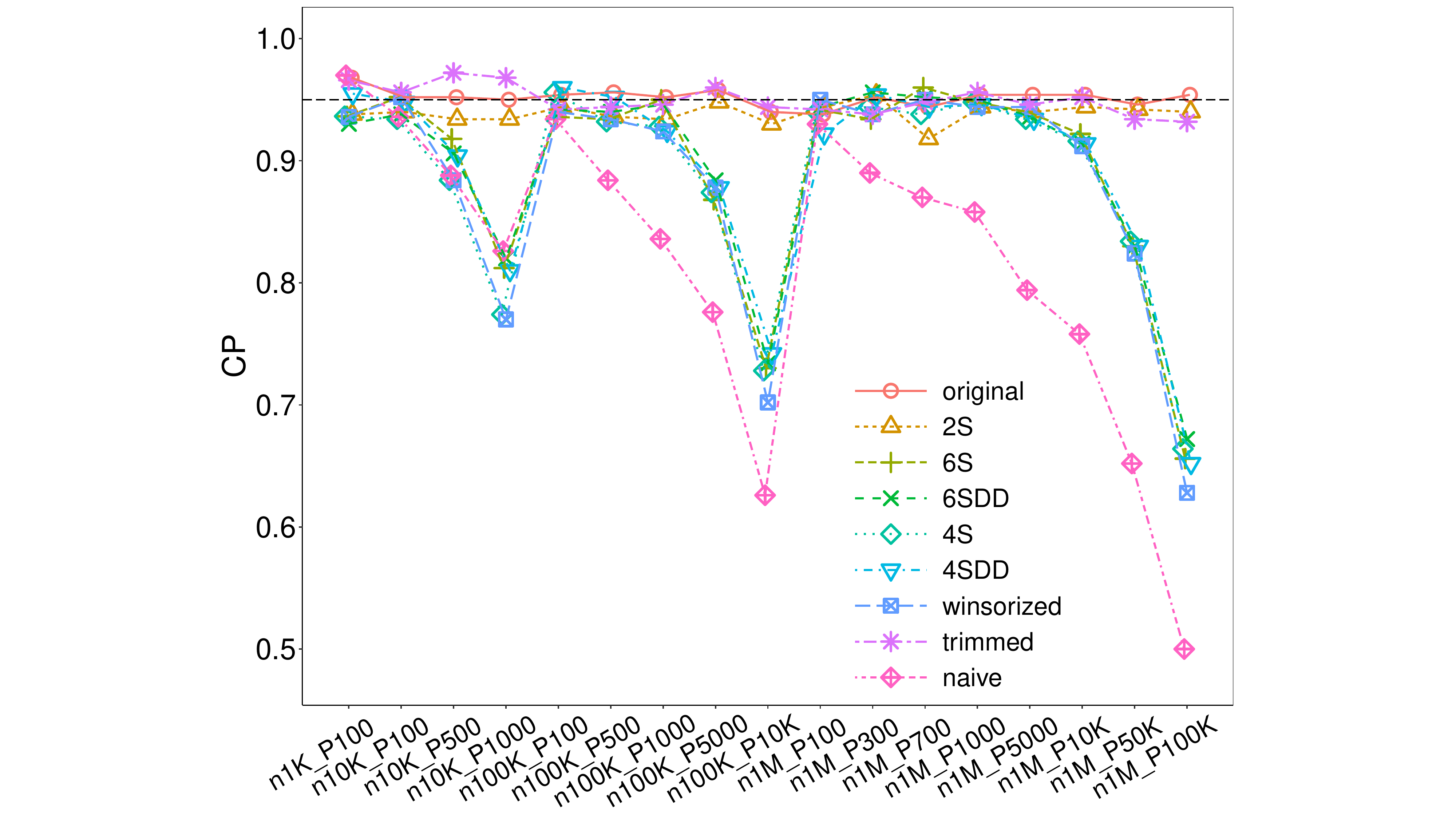}
\includegraphics[width=0.26\textwidth, trim={2.2in 0 2.2in 0},clip] {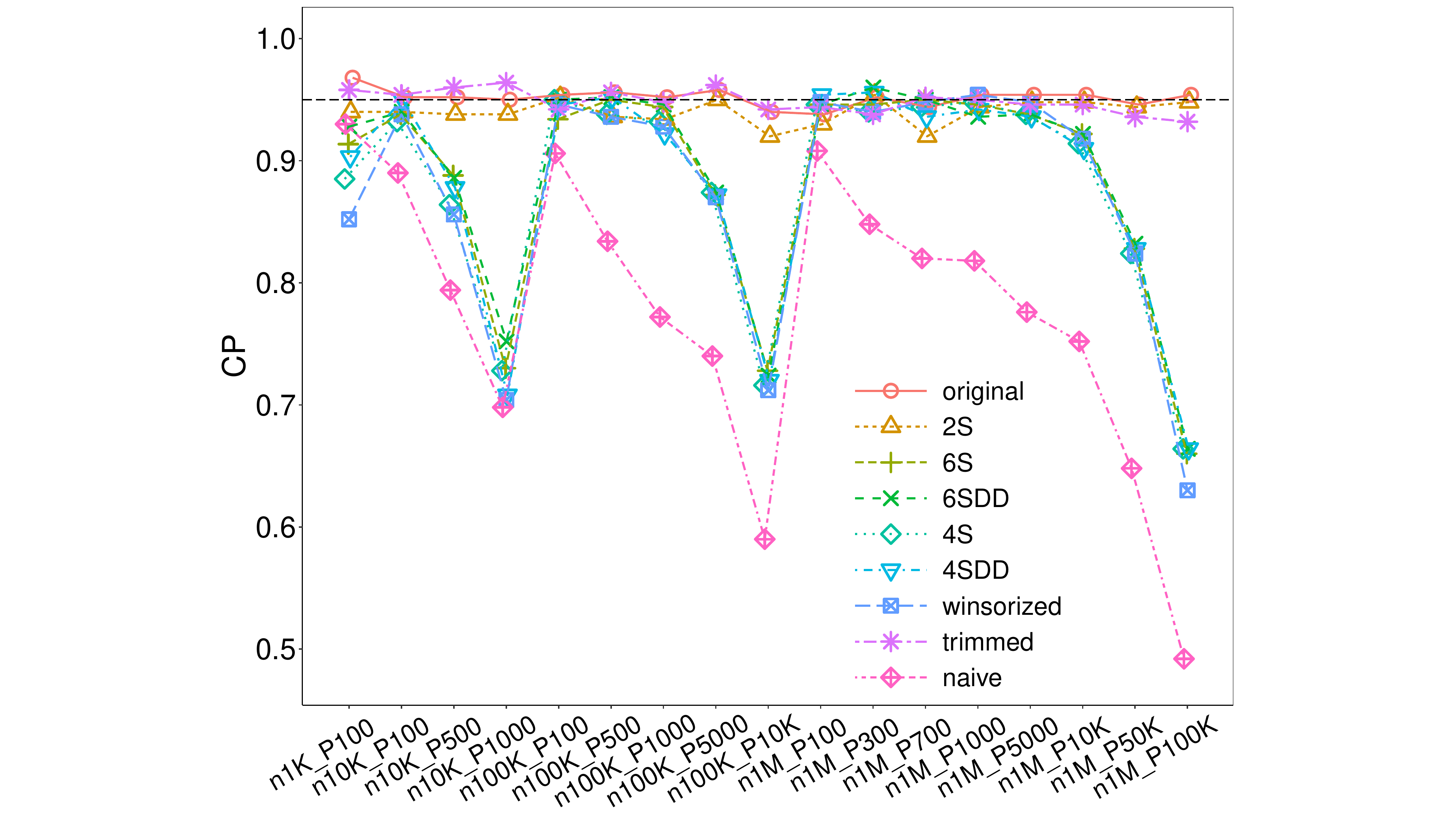}
\includegraphics[width=0.26\textwidth, trim={2.2in 0 2.2in 0},clip] {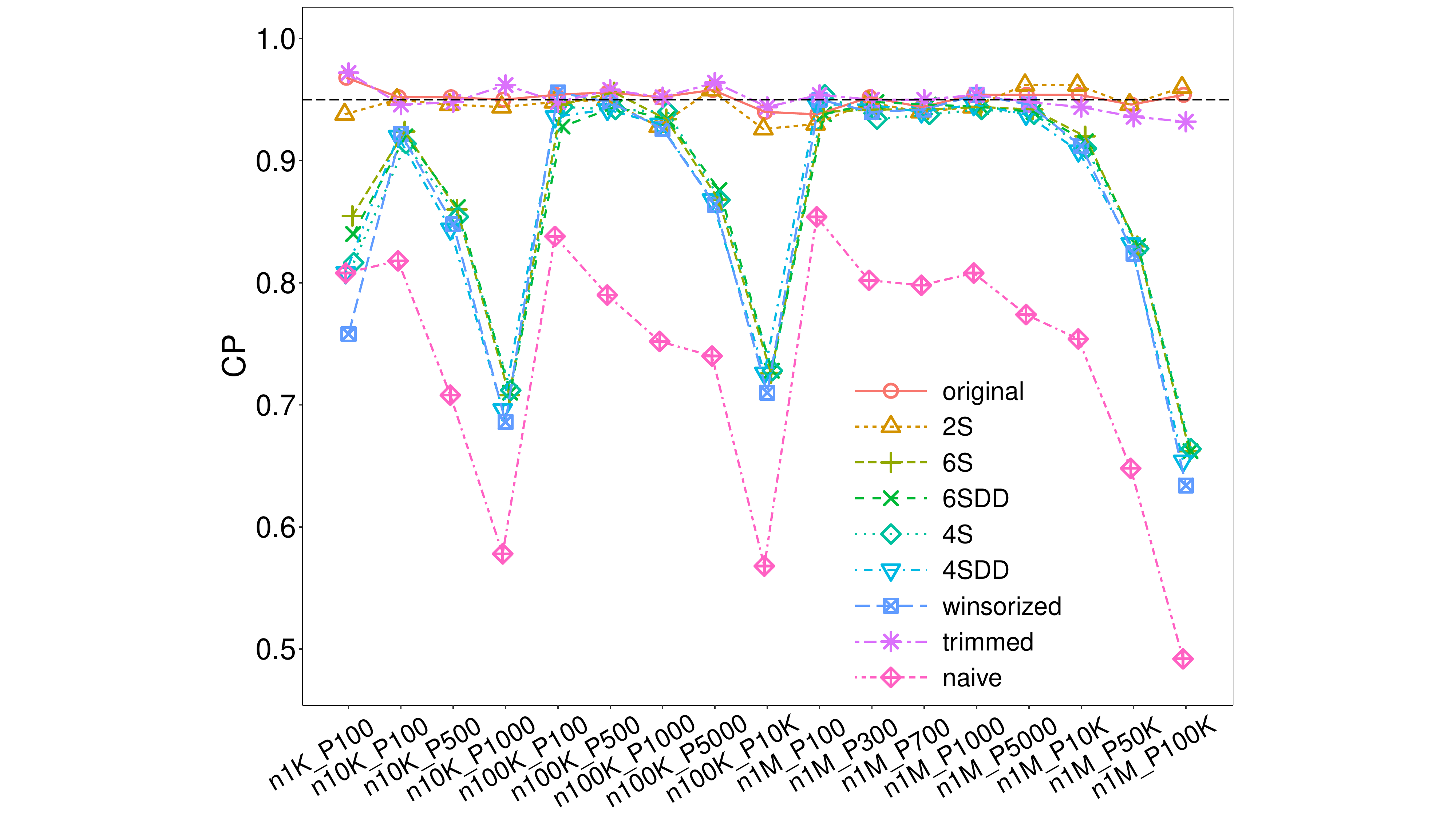}\\

\caption{ZILN data; $\rho$-zCDP; $\theta=0$ and $\alpha=\beta$}
\label{fig:0szCDPziln}
\end{figure}

\end{landscape}

\begin{landscape}

\subsection*{ZILN, $\theta\ne0$ and $\alpha=\beta$}
\begin{figure}[!htb]
\centering
$\epsilon=0.5$\hspace{0.9in}$\epsilon=1$\hspace{1in}$\epsilon=2$
\hspace{1in}$\epsilon=5$\hspace{0.9in}$\epsilon=50$\\
\includegraphics[width=0.215\textwidth, trim={2.2in 0 2.2in 0},clip] {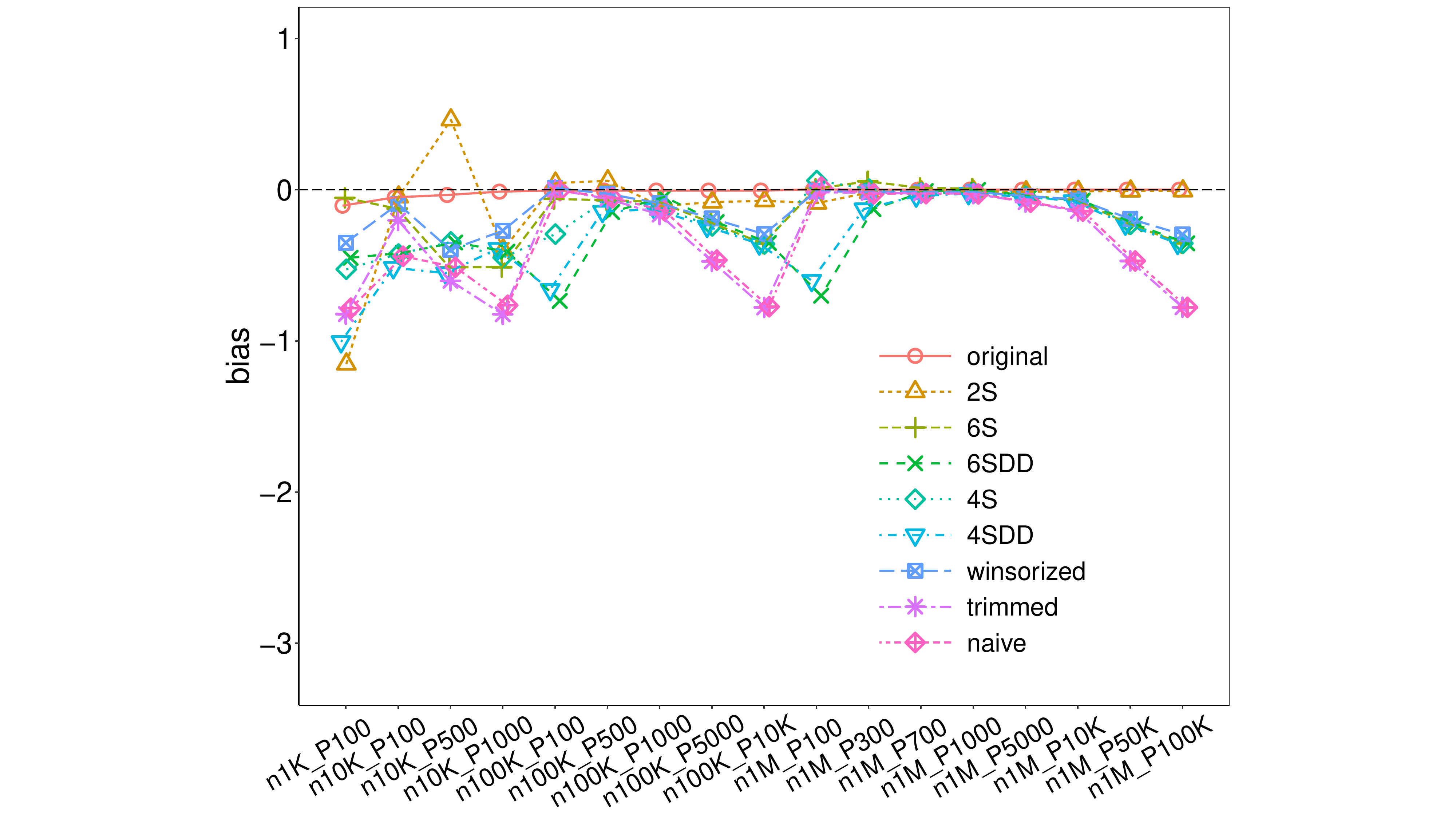}
\includegraphics[width=0.215\textwidth, trim={2.2in 0 2.2in 0},clip] {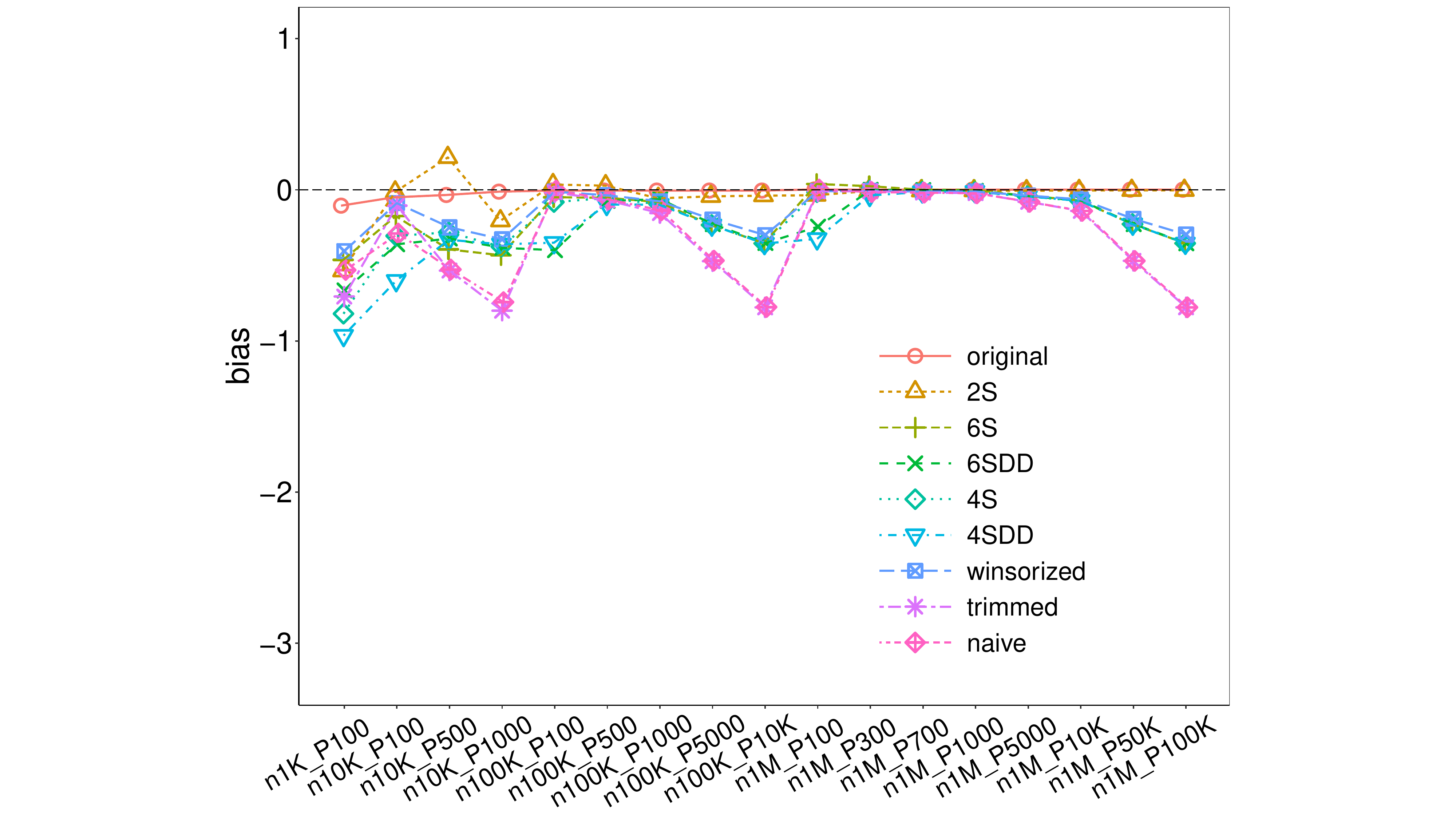}
\includegraphics[width=0.215\textwidth, trim={2.2in 0 2.2in 0},clip] {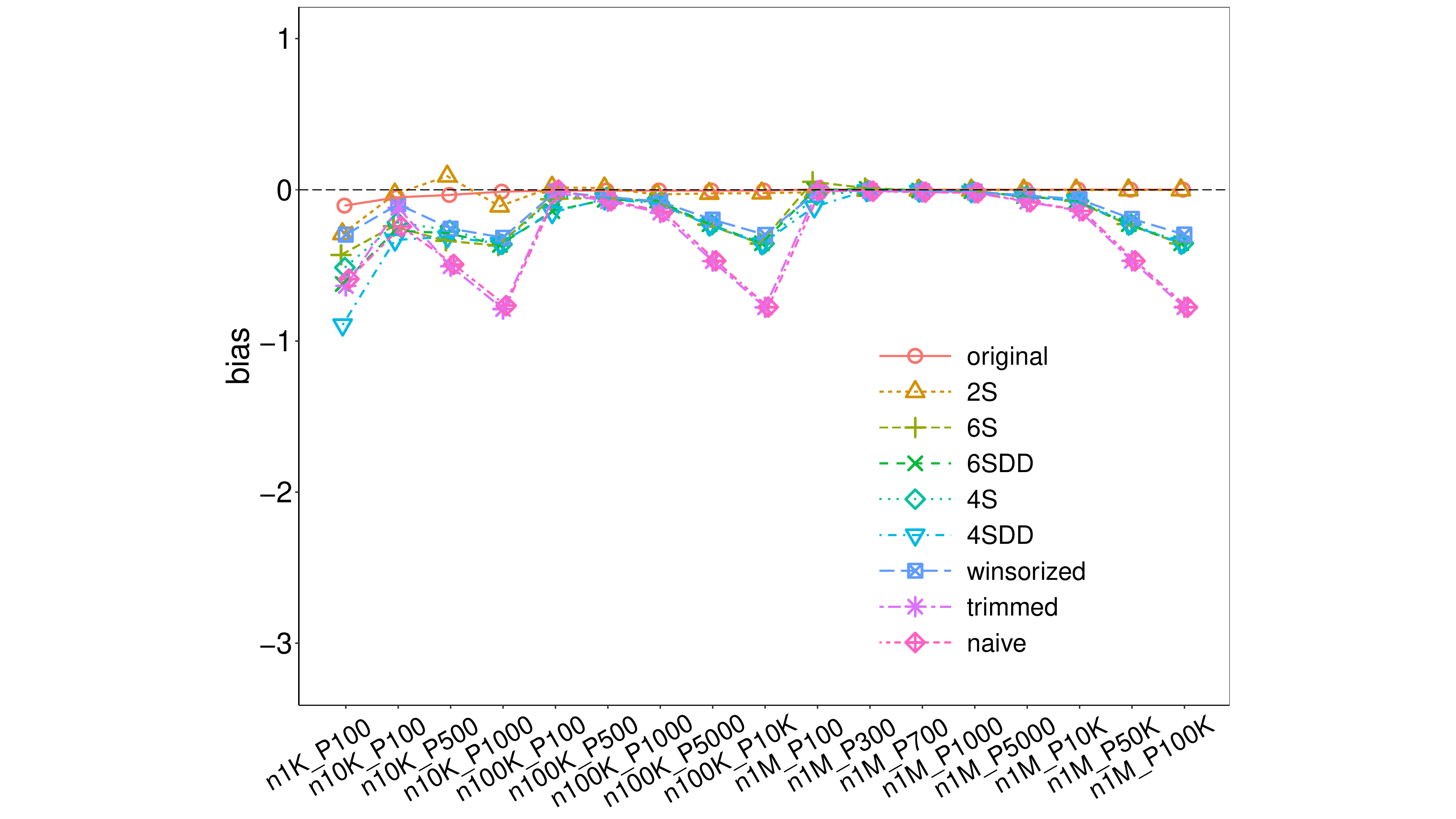}
\includegraphics[width=0.215\textwidth, trim={2.2in 0 2.2in 0},clip] {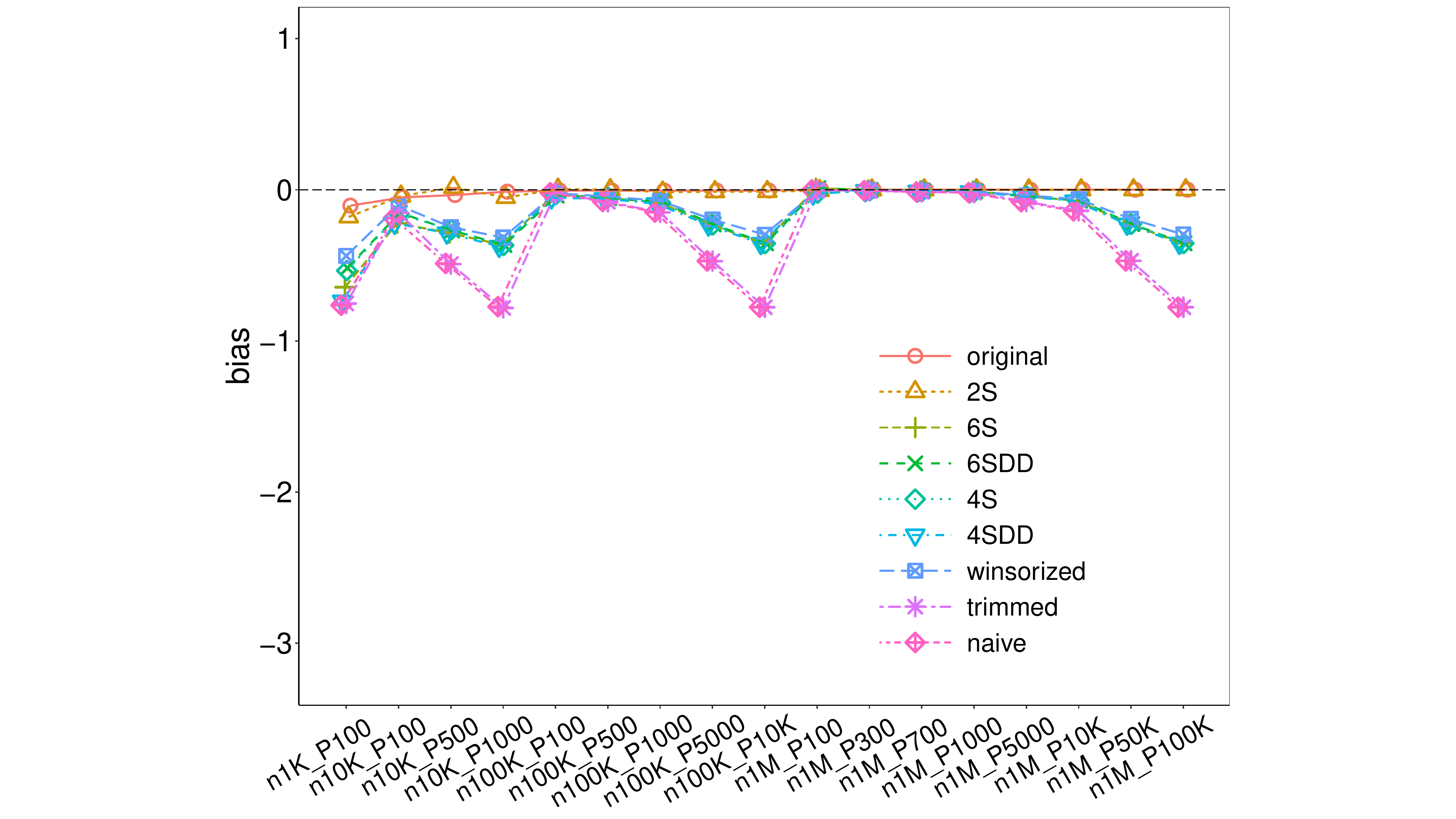}
\includegraphics[width=0.215\textwidth, trim={2.2in 0 2.2in 0},clip] {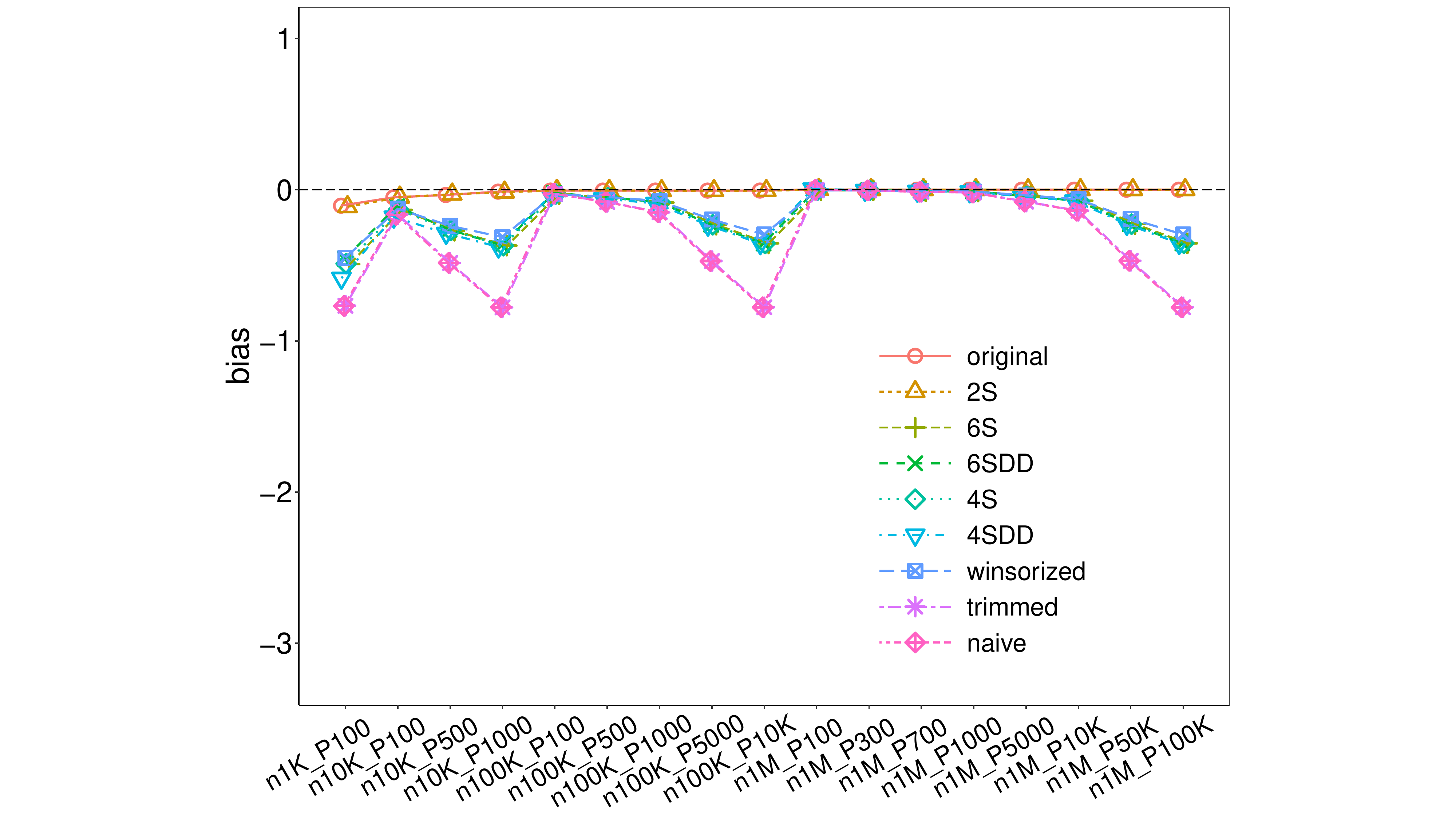}\\
\includegraphics[width=0.215\textwidth, trim={2.2in 0 2.2in 0},clip] {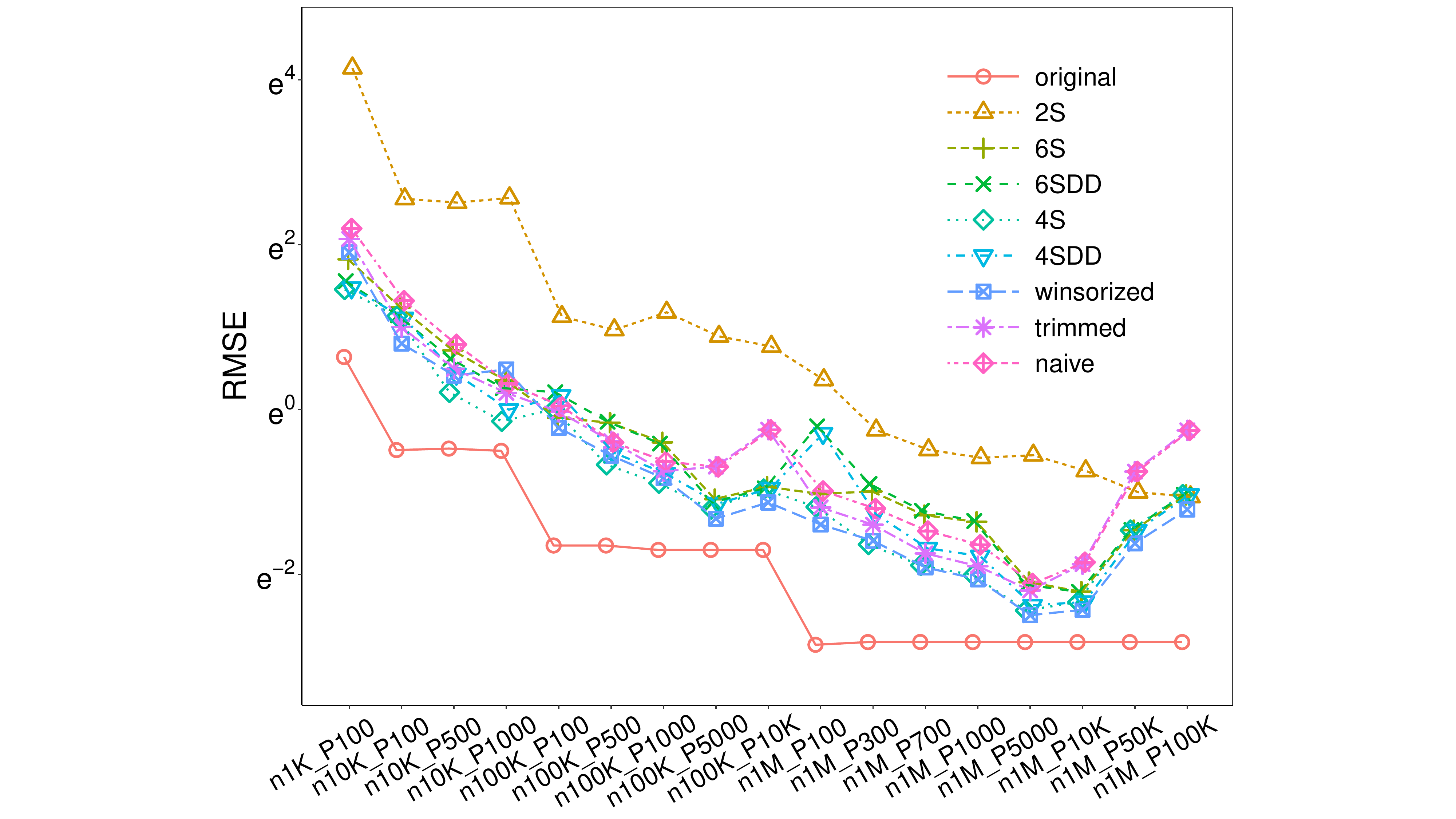}
\includegraphics[width=0.215\textwidth, trim={2.2in 0 2.2in 0},clip] {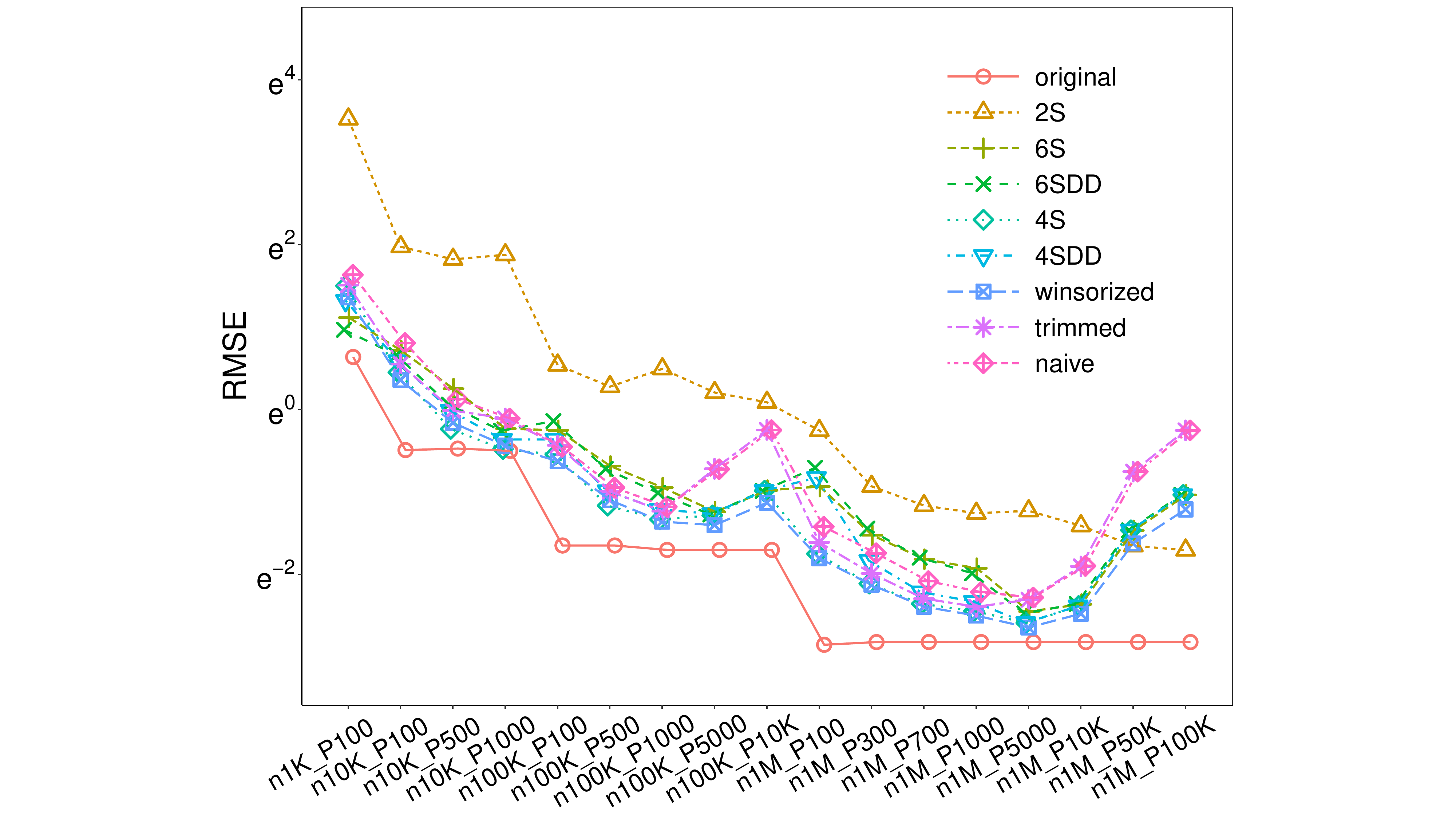}
\includegraphics[width=0.215\textwidth, trim={2.2in 0 2.2in 0},clip] {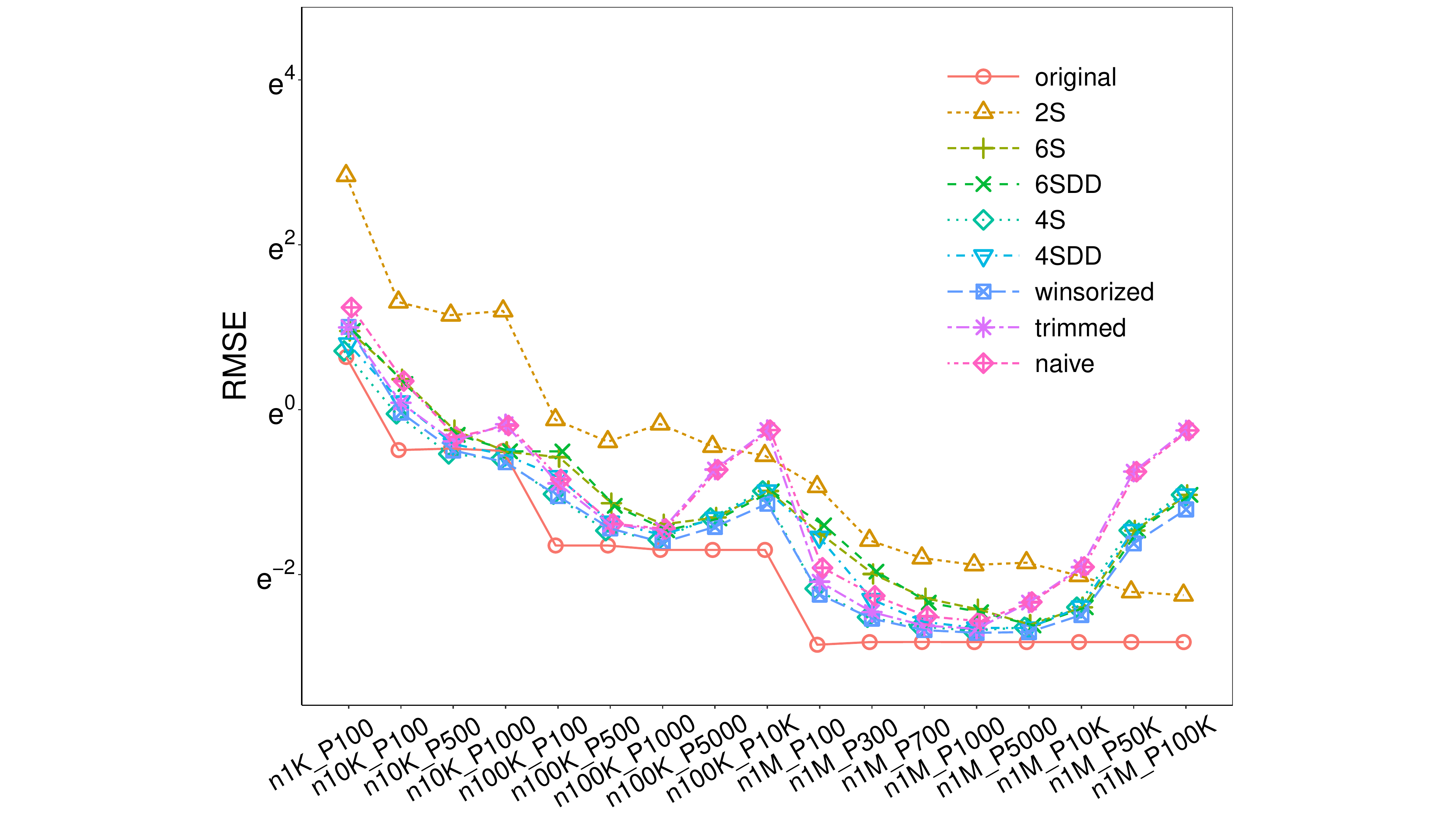}
\includegraphics[width=0.215\textwidth, trim={2.2in 0 2.2in 0},clip] {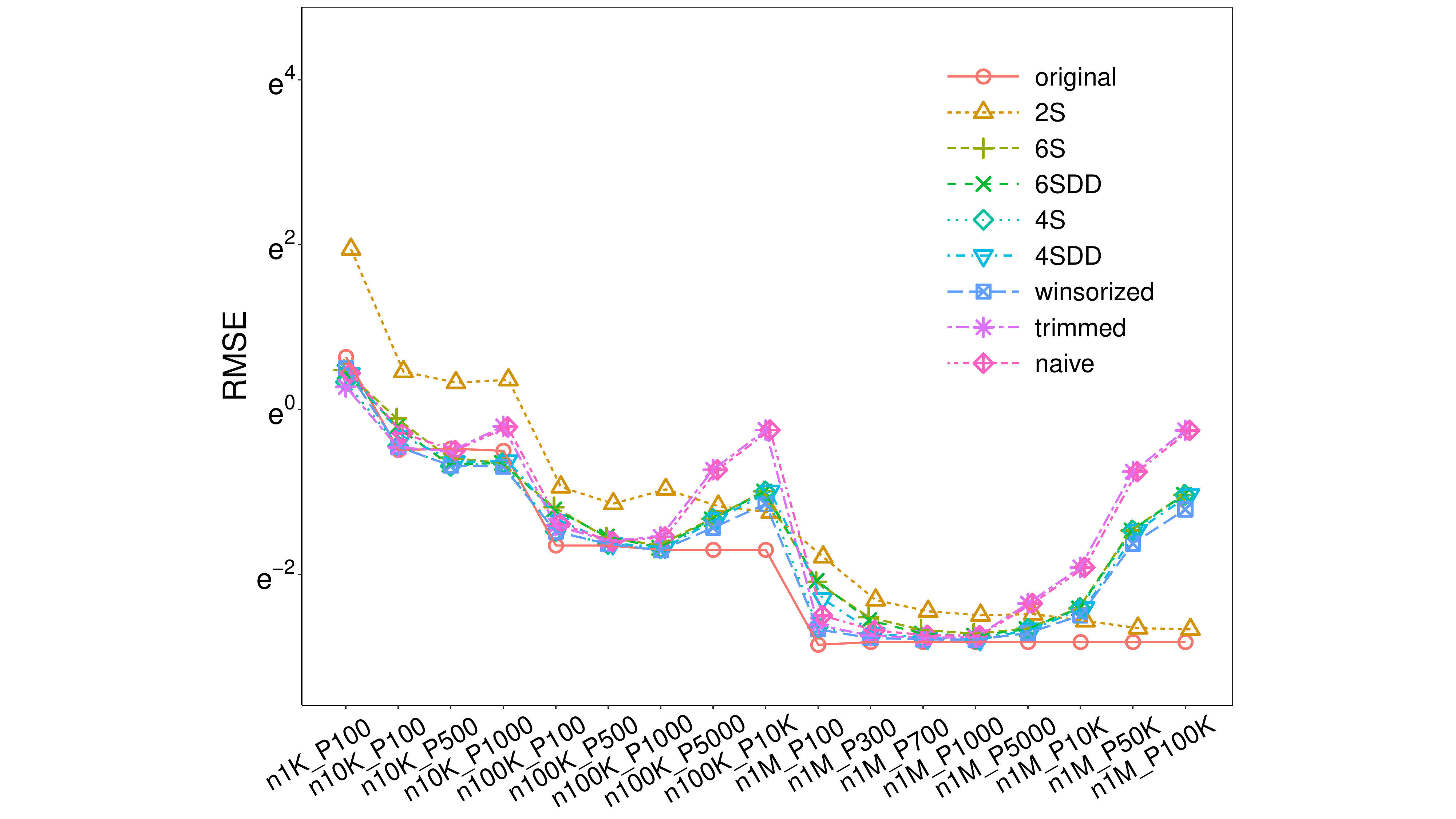}
\includegraphics[width=0.215\textwidth, trim={2.2in 0 2.2in 0},clip] {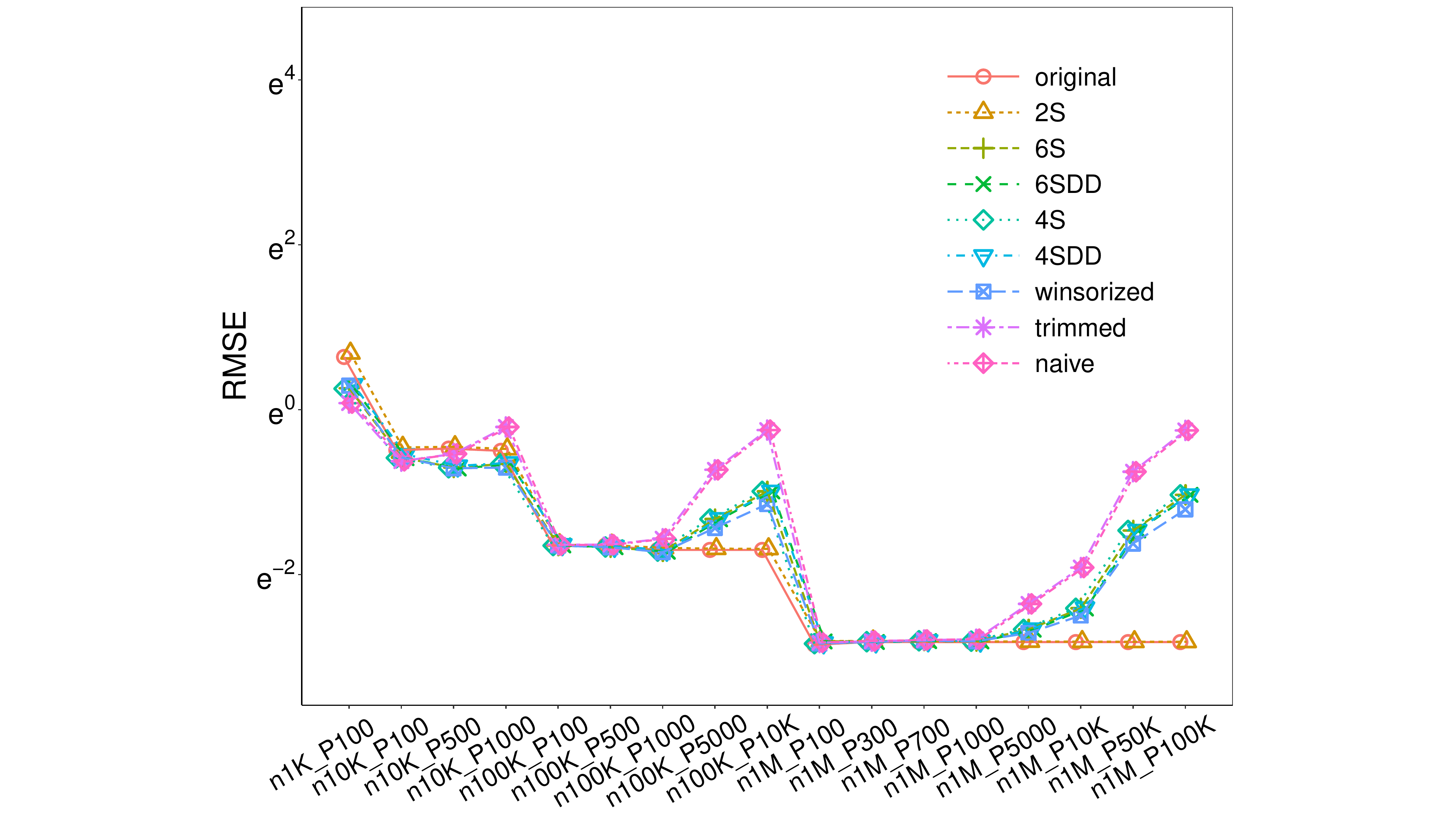}\\
\includegraphics[width=0.215\textwidth, trim={2.2in 0 2.2in 0},clip] {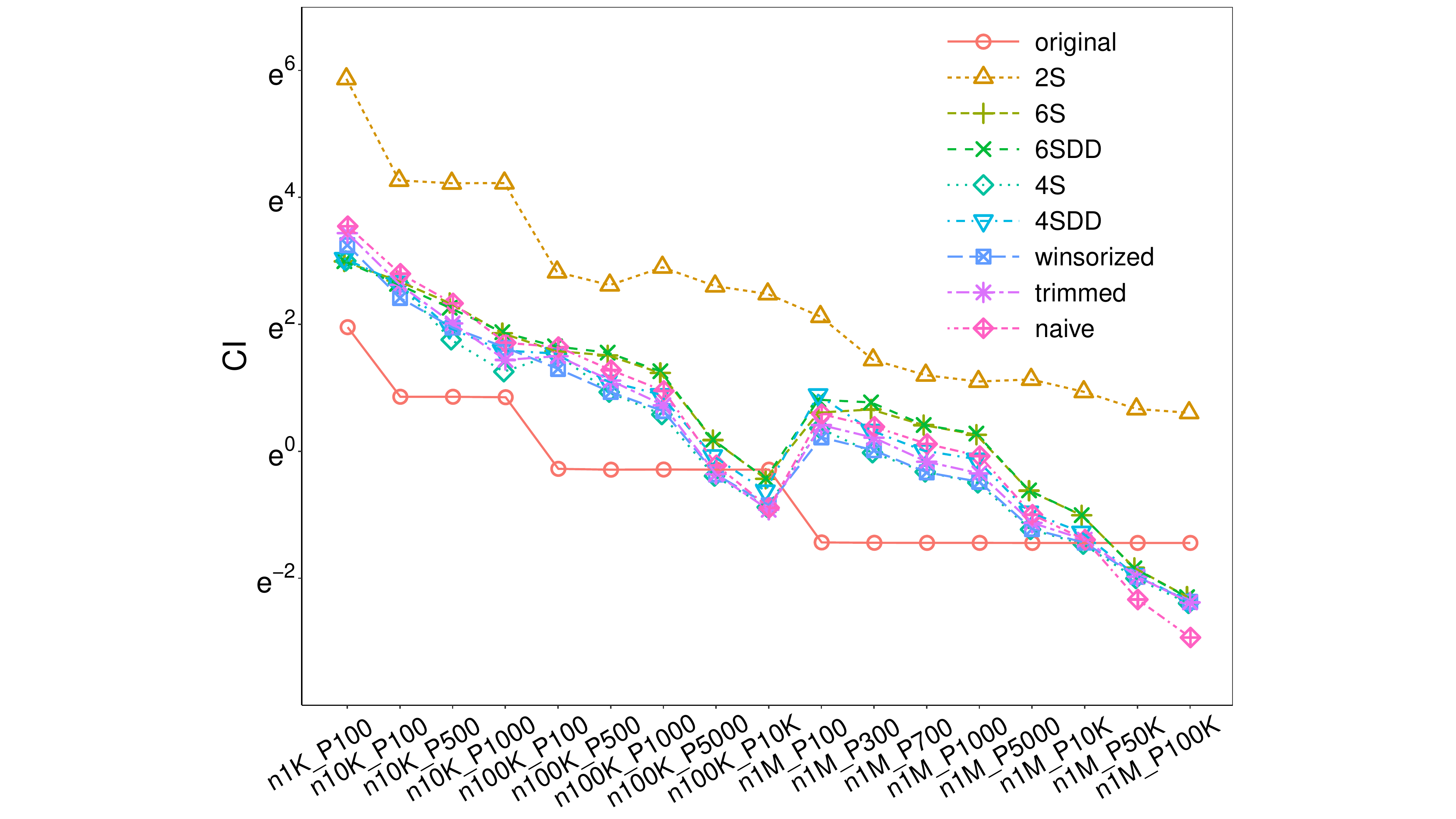}
\includegraphics[width=0.215\textwidth, trim={2.2in 0 2.2in 0},clip] {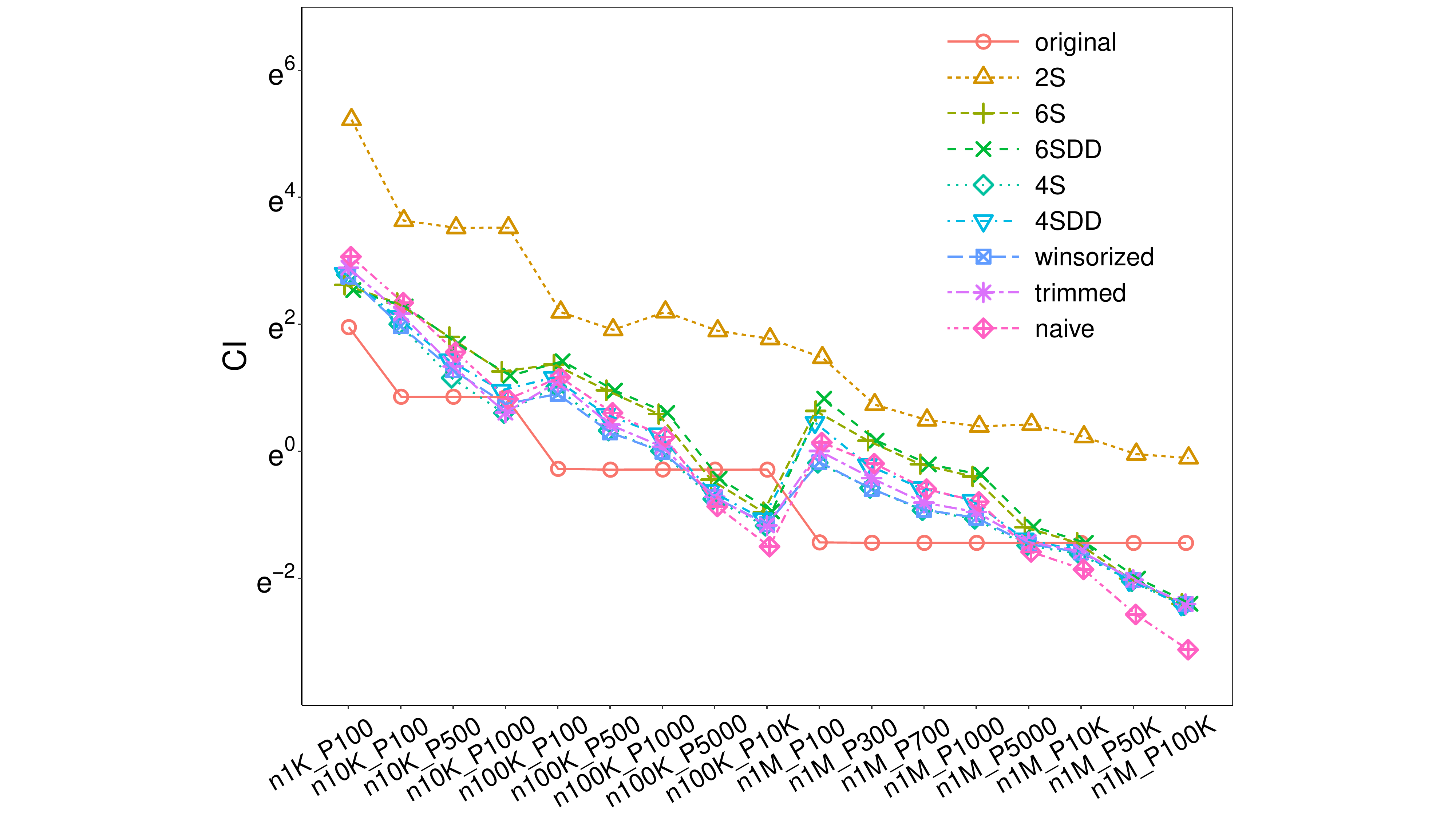}
\includegraphics[width=0.215\textwidth, trim={2.2in 0 2.2in 0},clip] {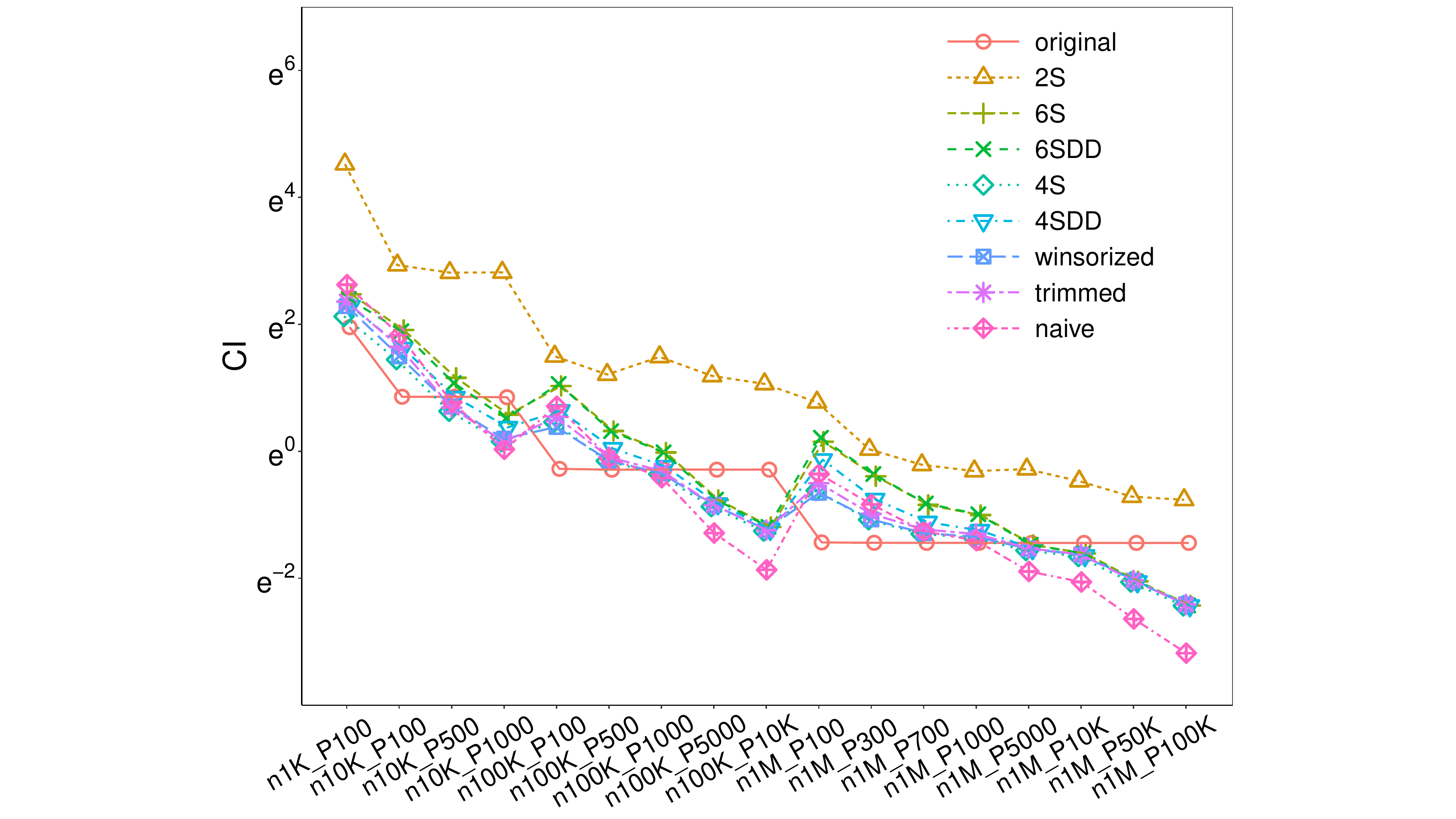}
\includegraphics[width=0.215\textwidth, trim={2.2in 0 2.2in 0},clip] {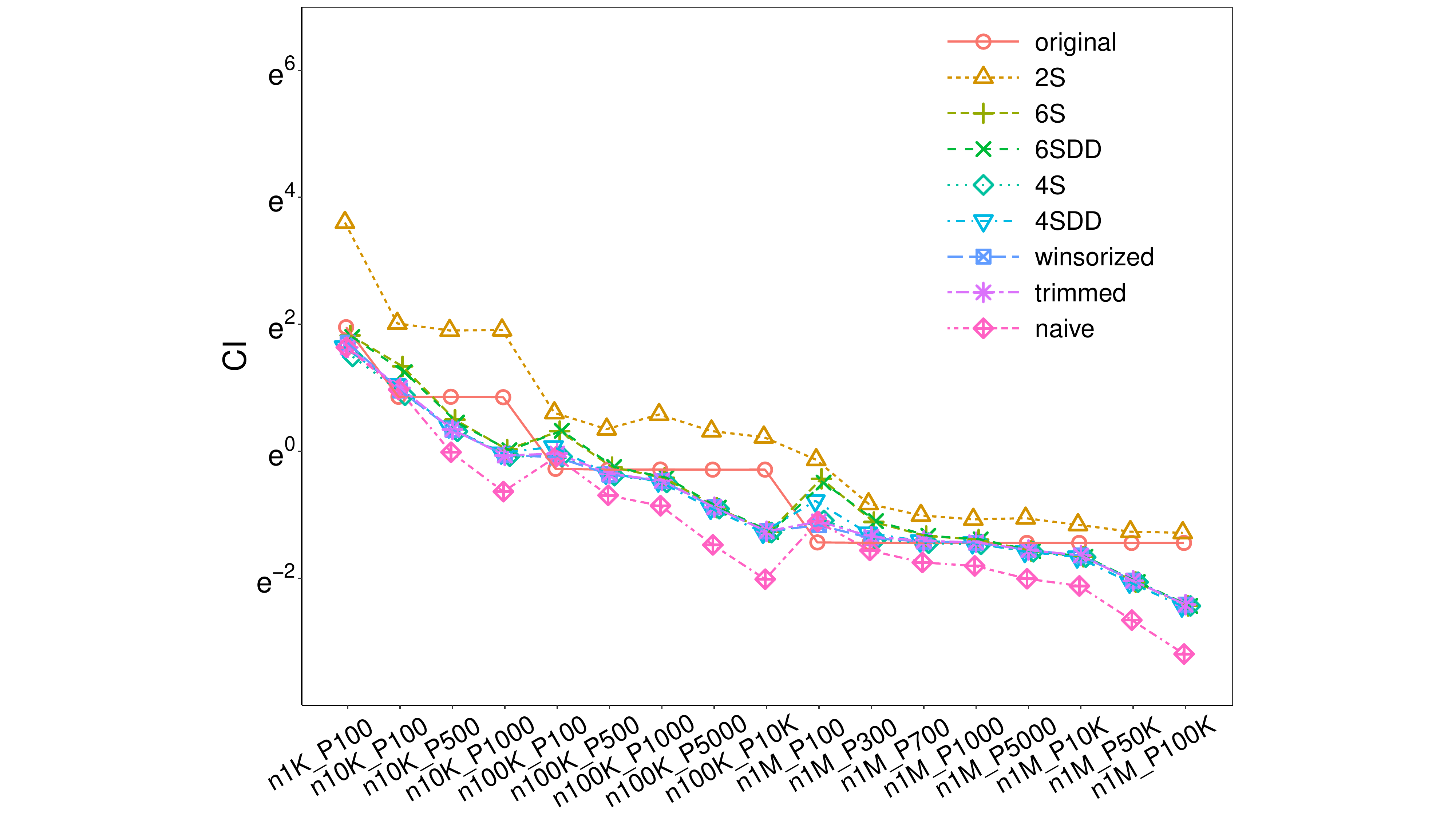}
\includegraphics[width=0.215\textwidth, trim={2.2in 0 2.2in 0},clip] {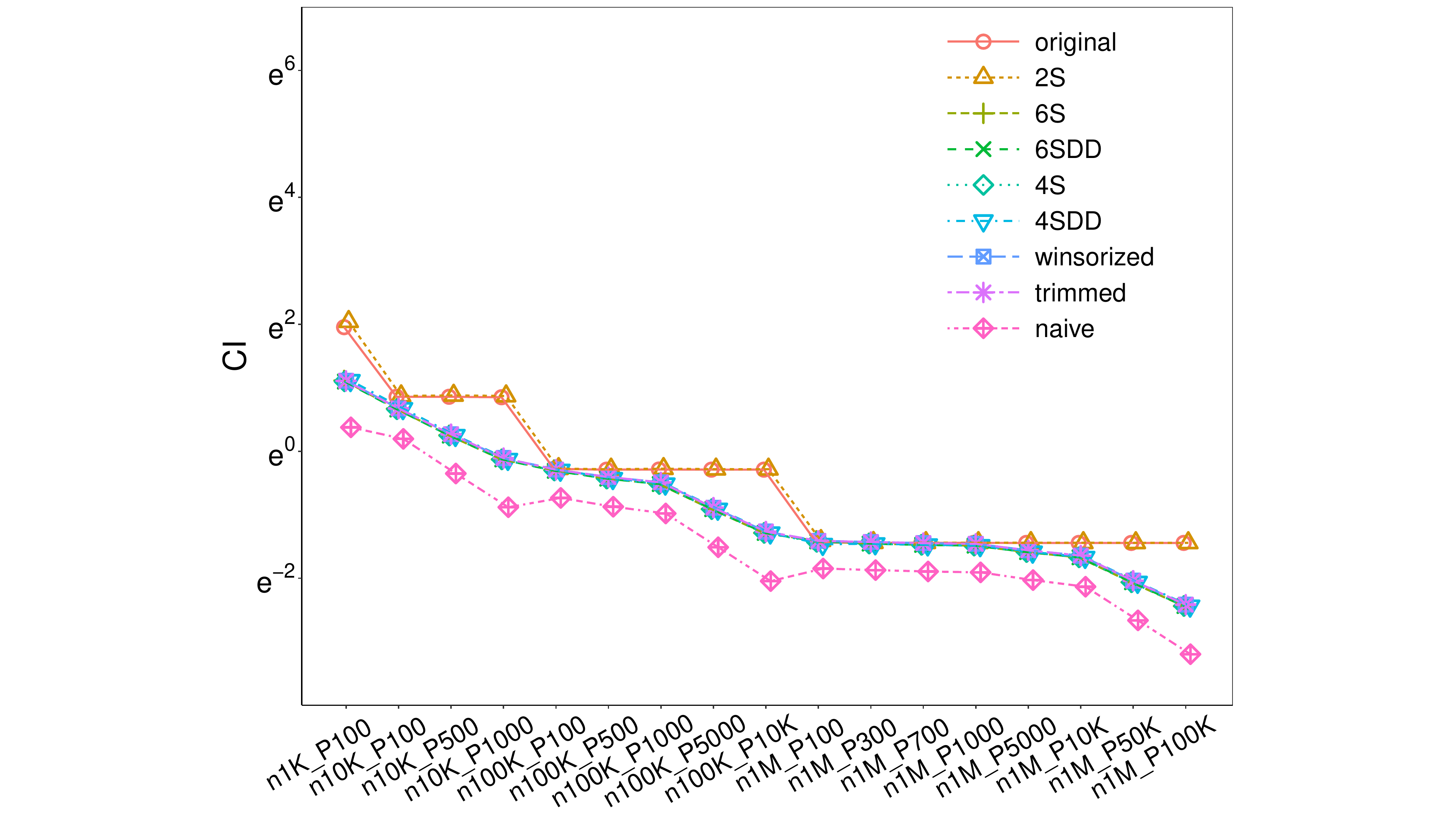}\\
\includegraphics[width=0.215\textwidth, trim={2.2in 0 2.2in 0},clip] {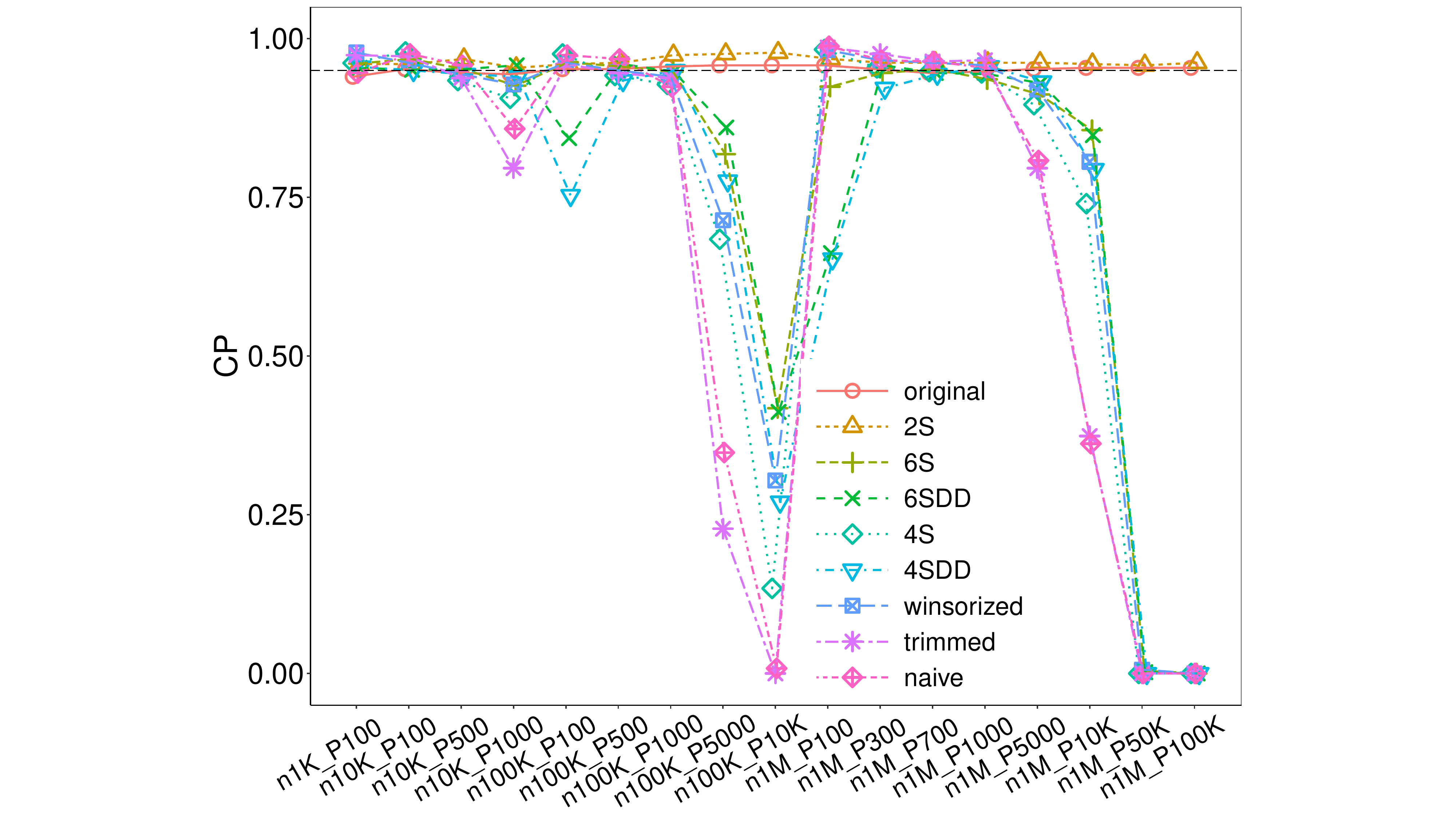}
\includegraphics[width=0.215\textwidth, trim={2.2in 0 2.2in 0},clip] {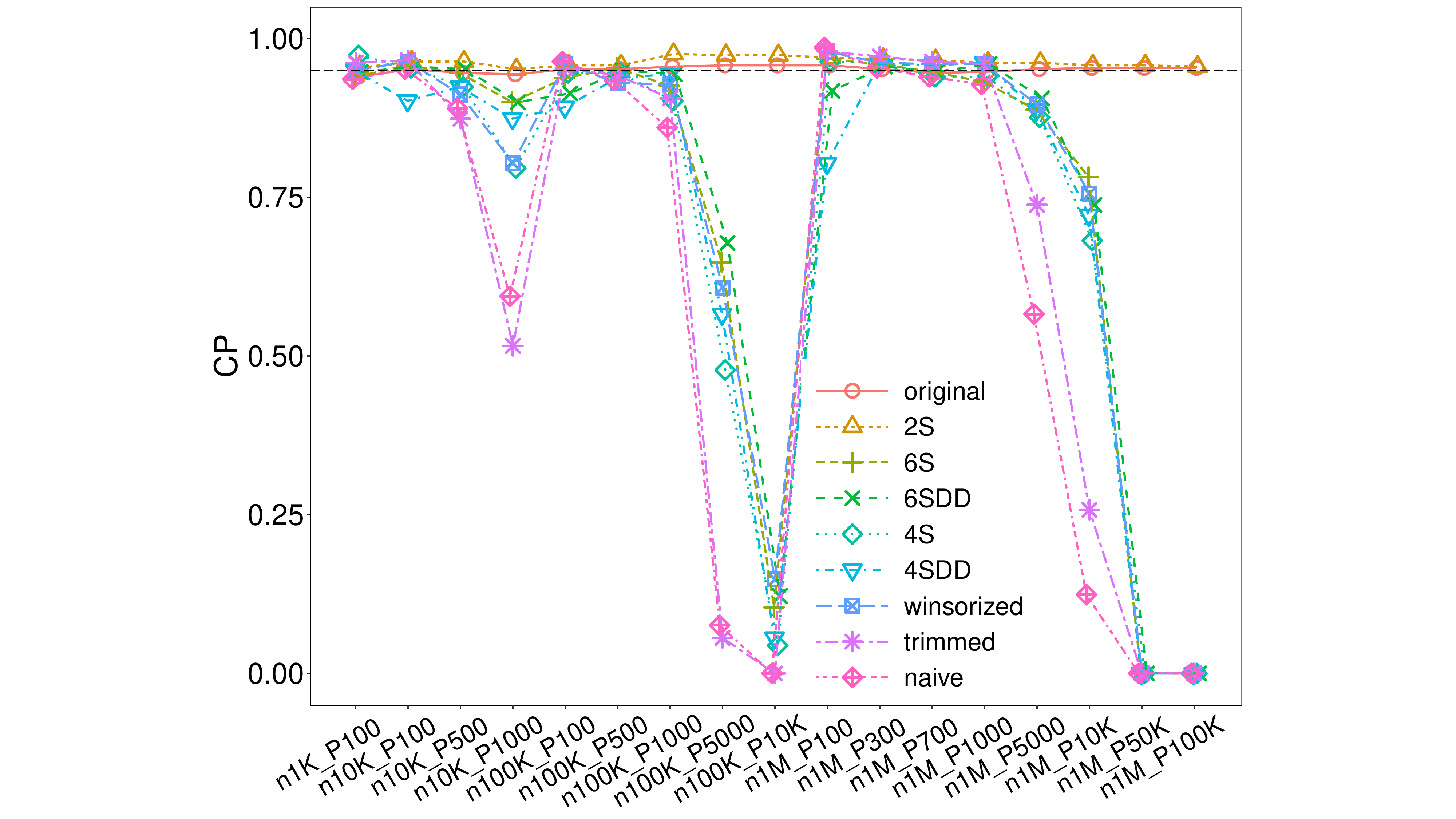}
\includegraphics[width=0.215\textwidth, trim={2.2in 0 2.2in 0},clip] {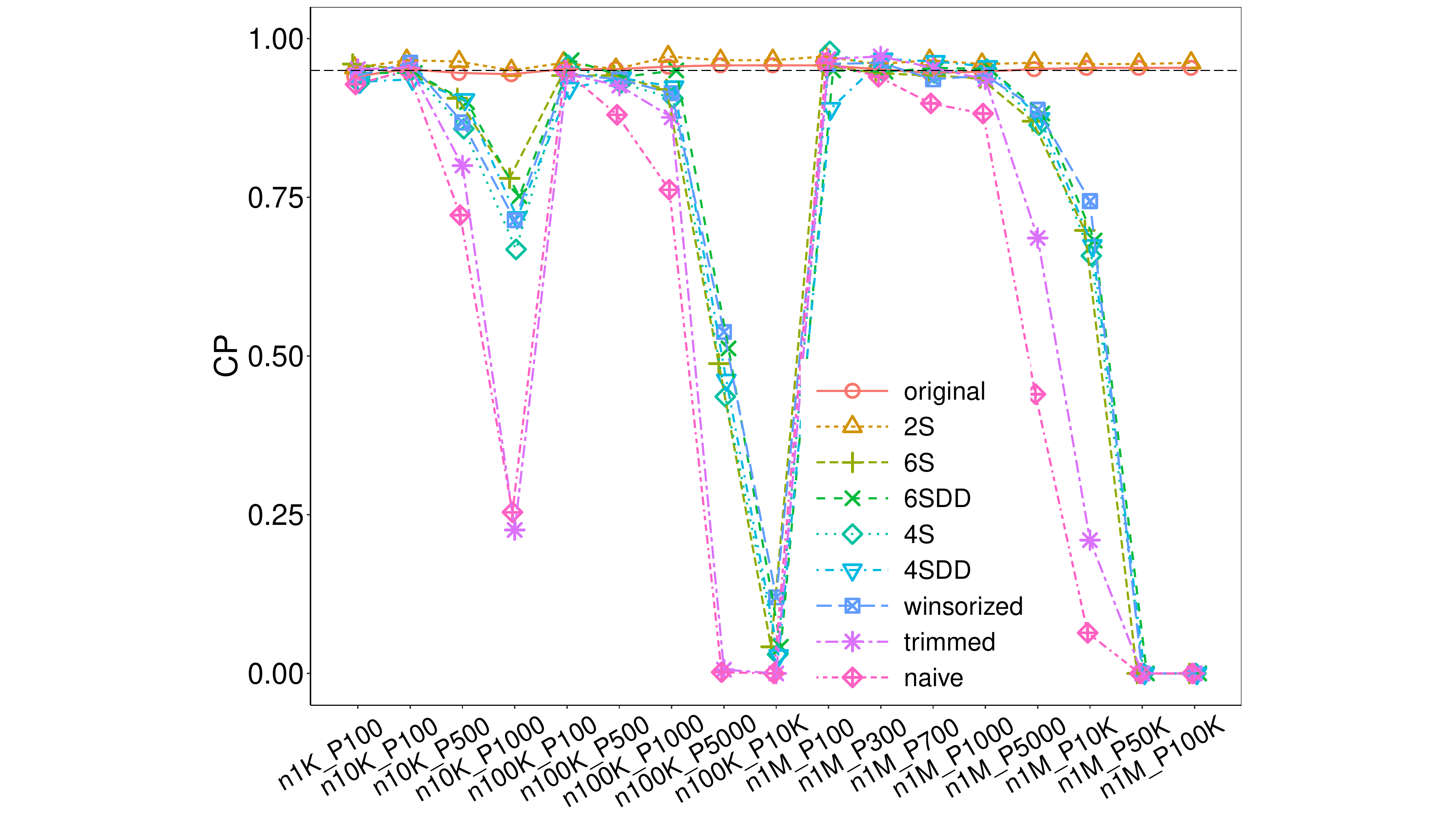}
\includegraphics[width=0.215\textwidth, trim={2.2in 0 2.2in 0},clip] {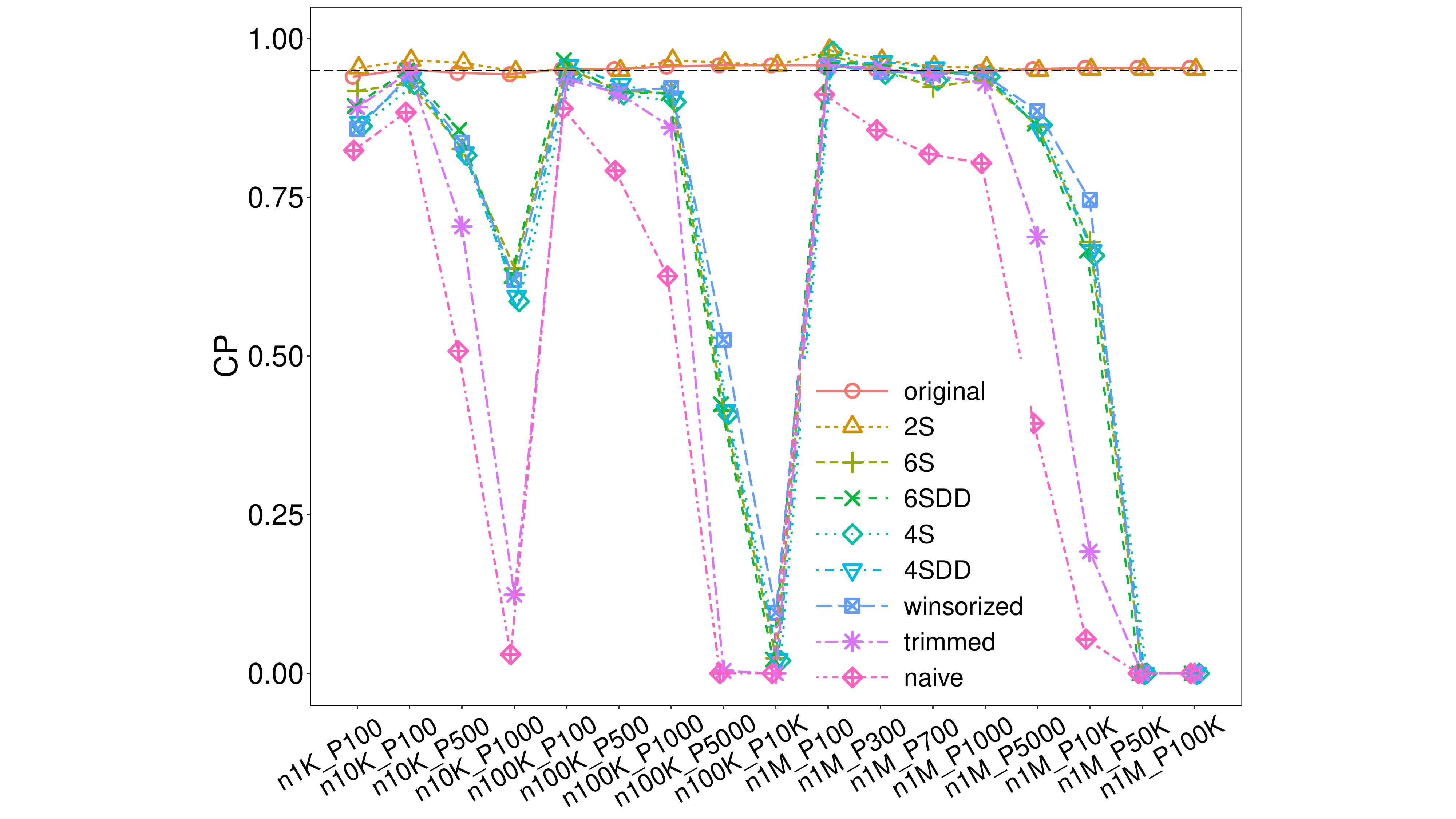}
\includegraphics[width=0.215\textwidth, trim={2.2in 0 2.2in 0},clip] {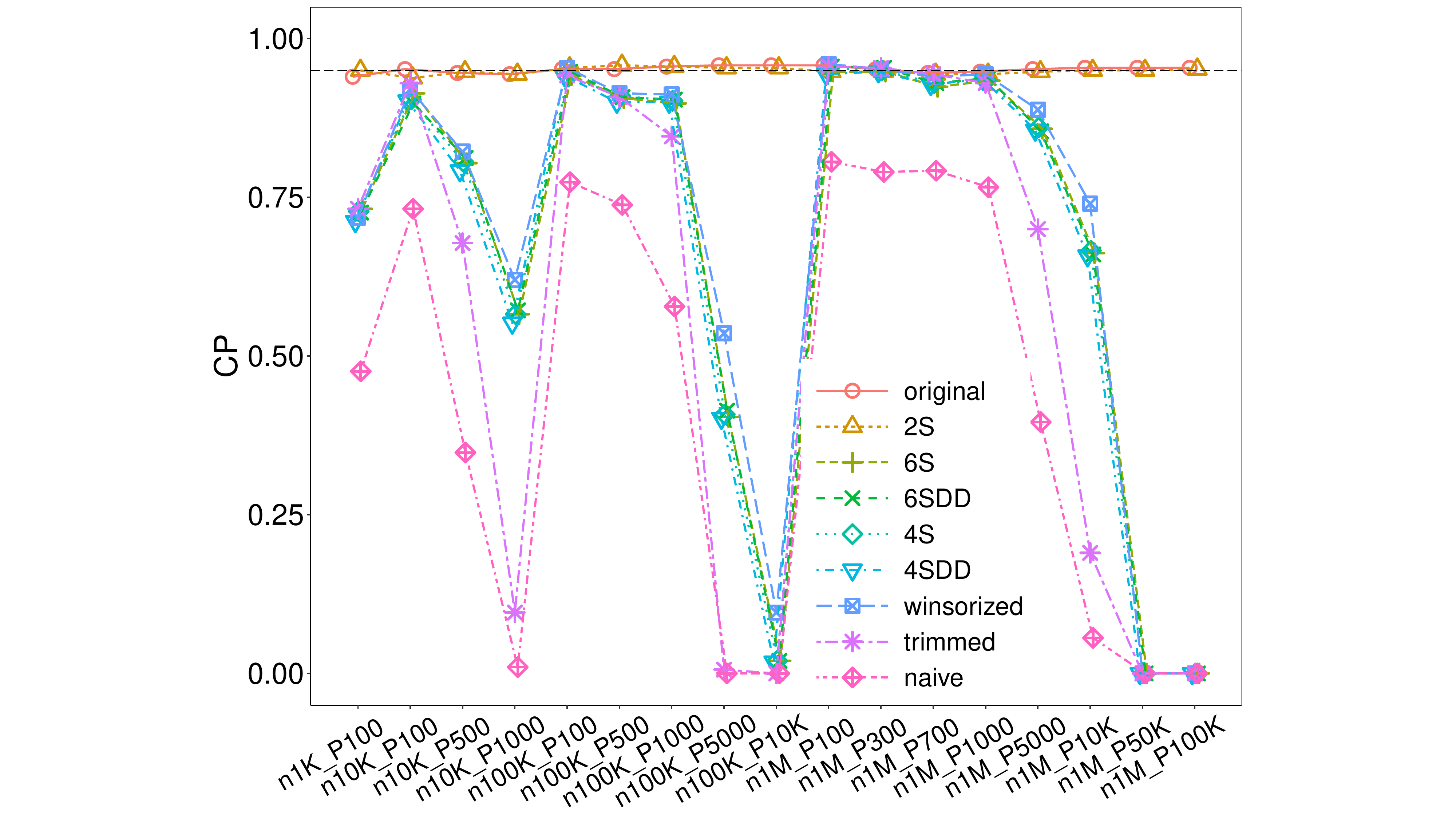}\\
\includegraphics[width=0.215\textwidth, trim={2.2in 0 2.2in 0},clip] {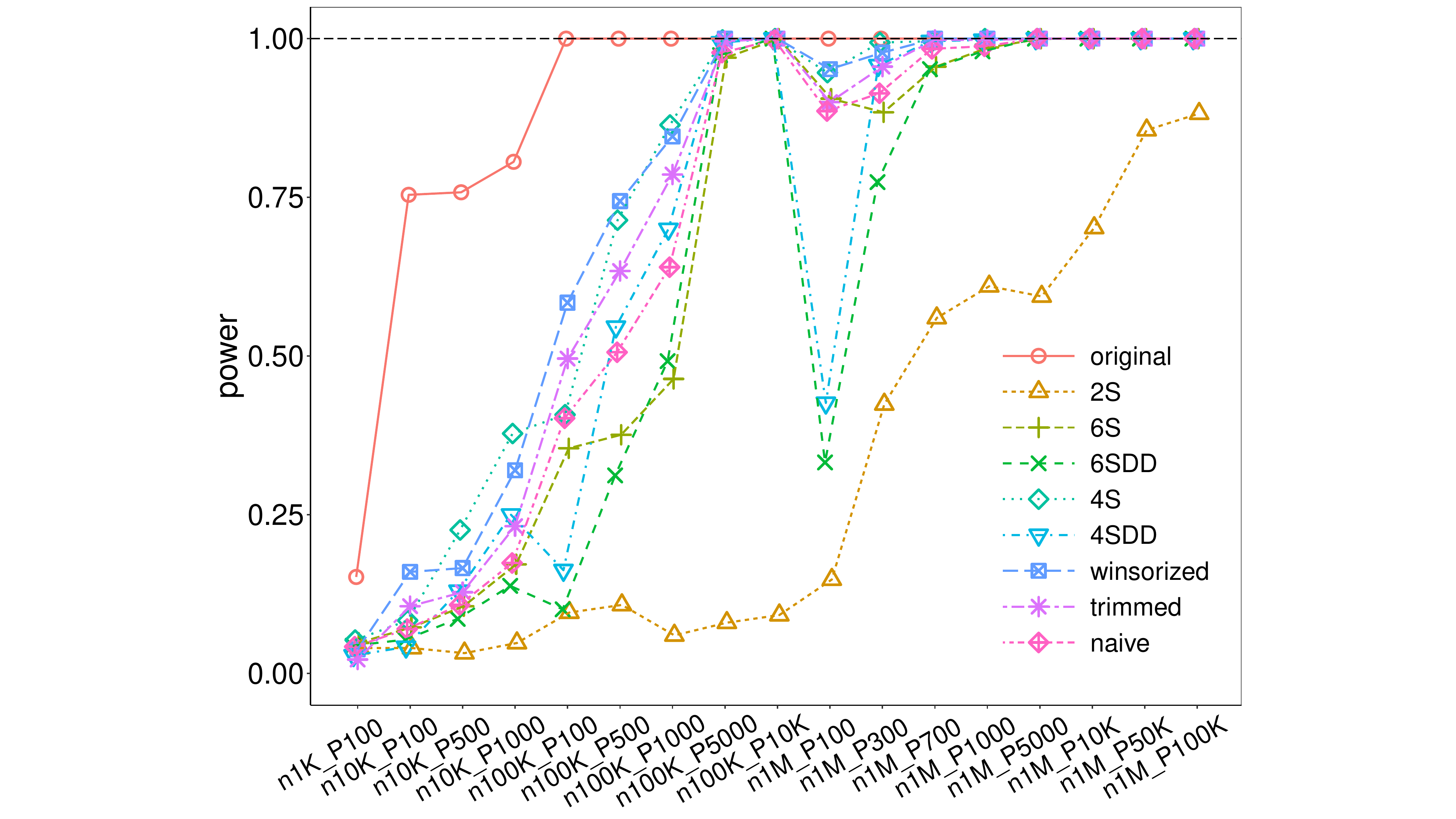}
\includegraphics[width=0.215\textwidth, trim={2.2in 0 2.2in 0},clip] {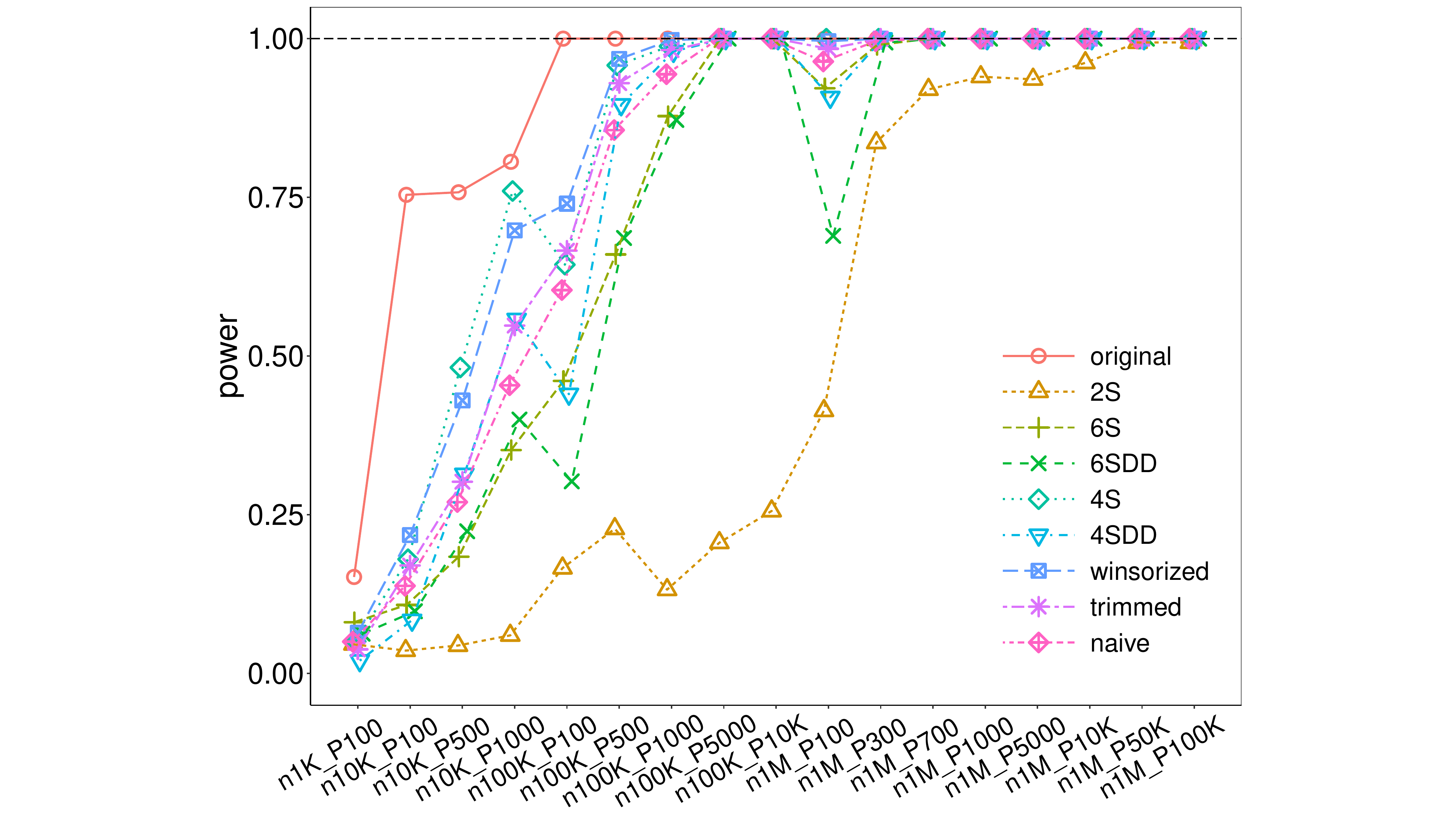}
\includegraphics[width=0.215\textwidth, trim={2.2in 0 2.2in 0},clip] {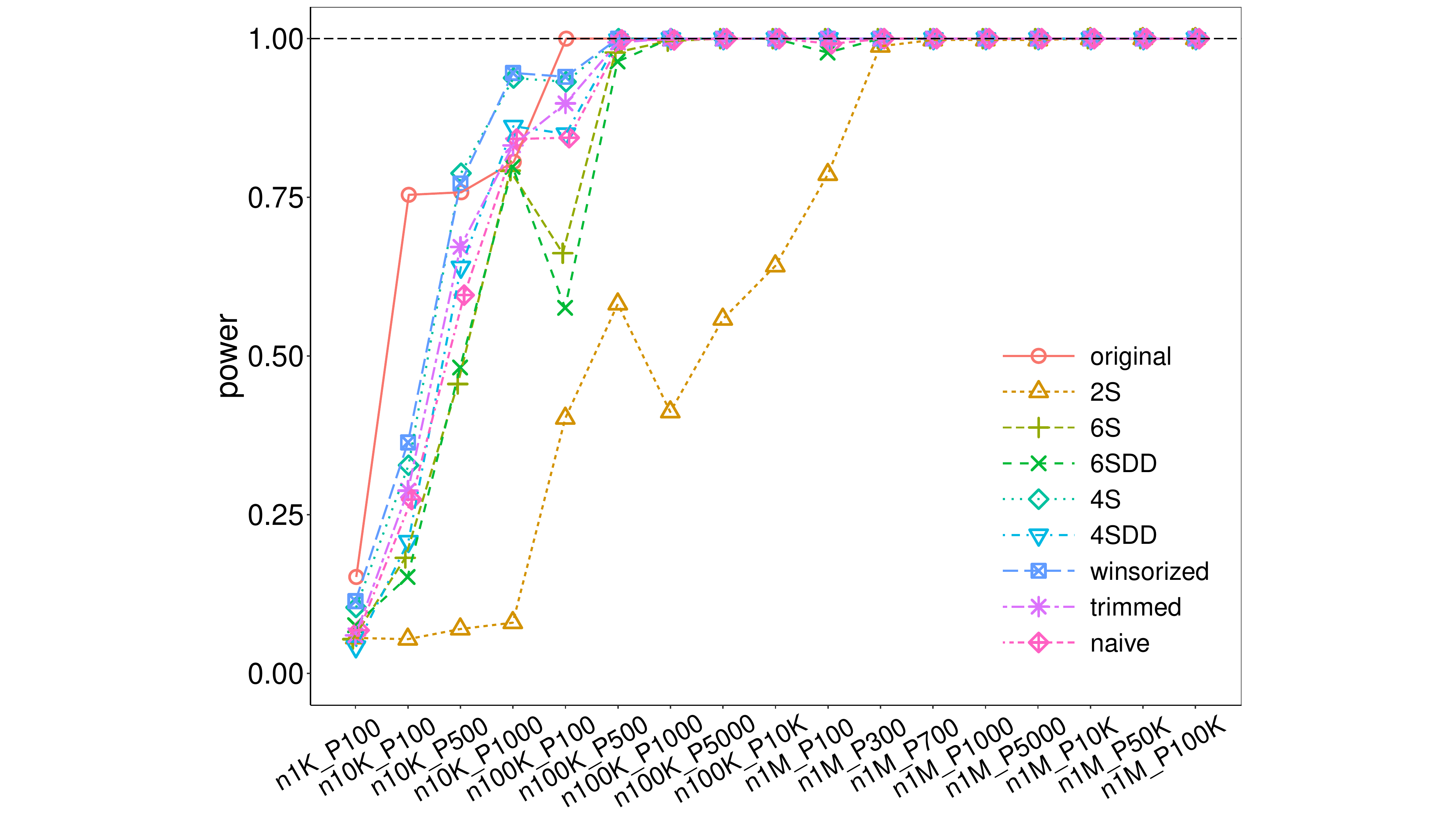}
\includegraphics[width=0.215\textwidth, trim={2.2in 0 2.2in 0},clip] {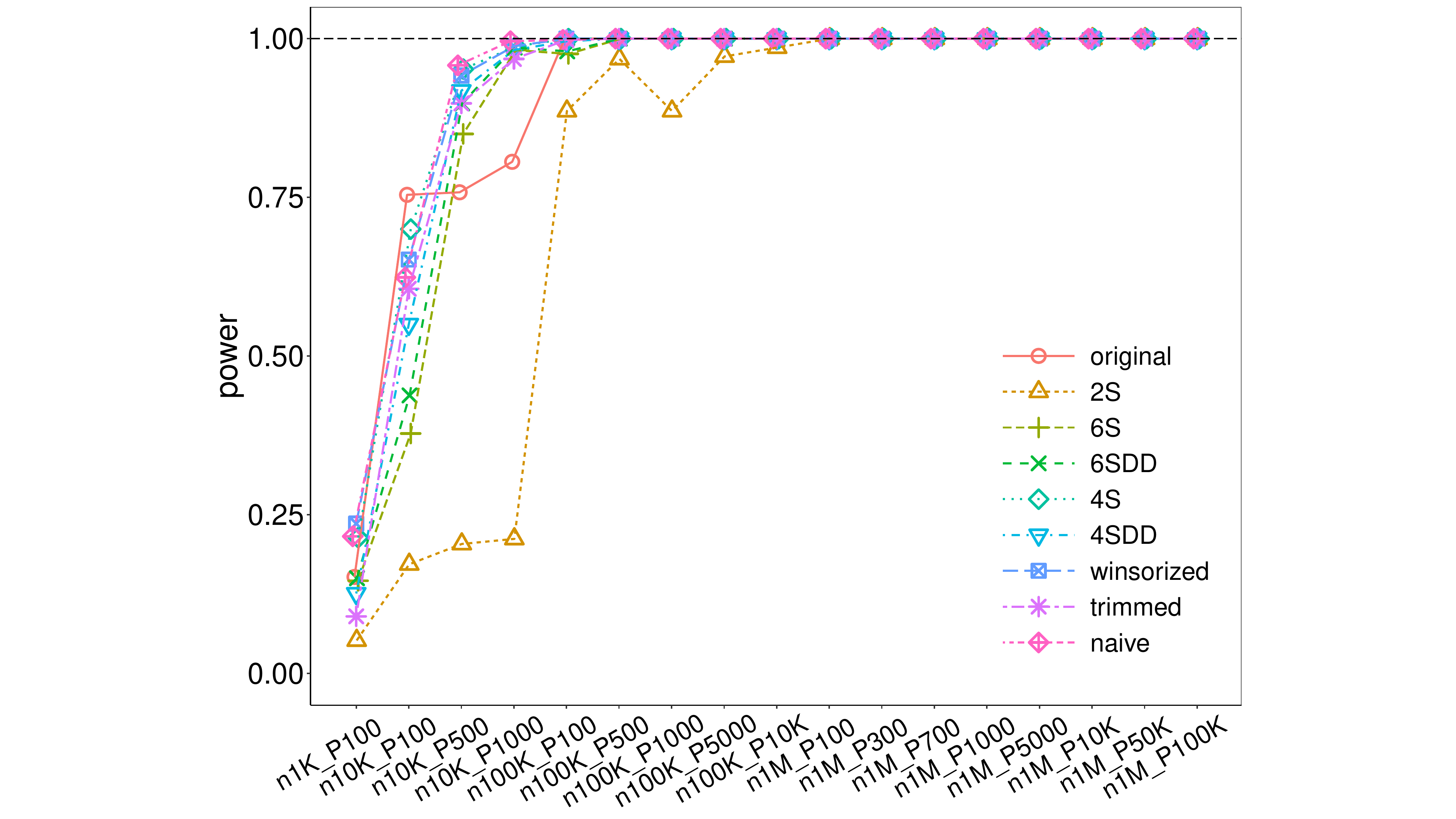}
\includegraphics[width=0.215\textwidth, trim={2.2in 0 2.2in 0},clip] {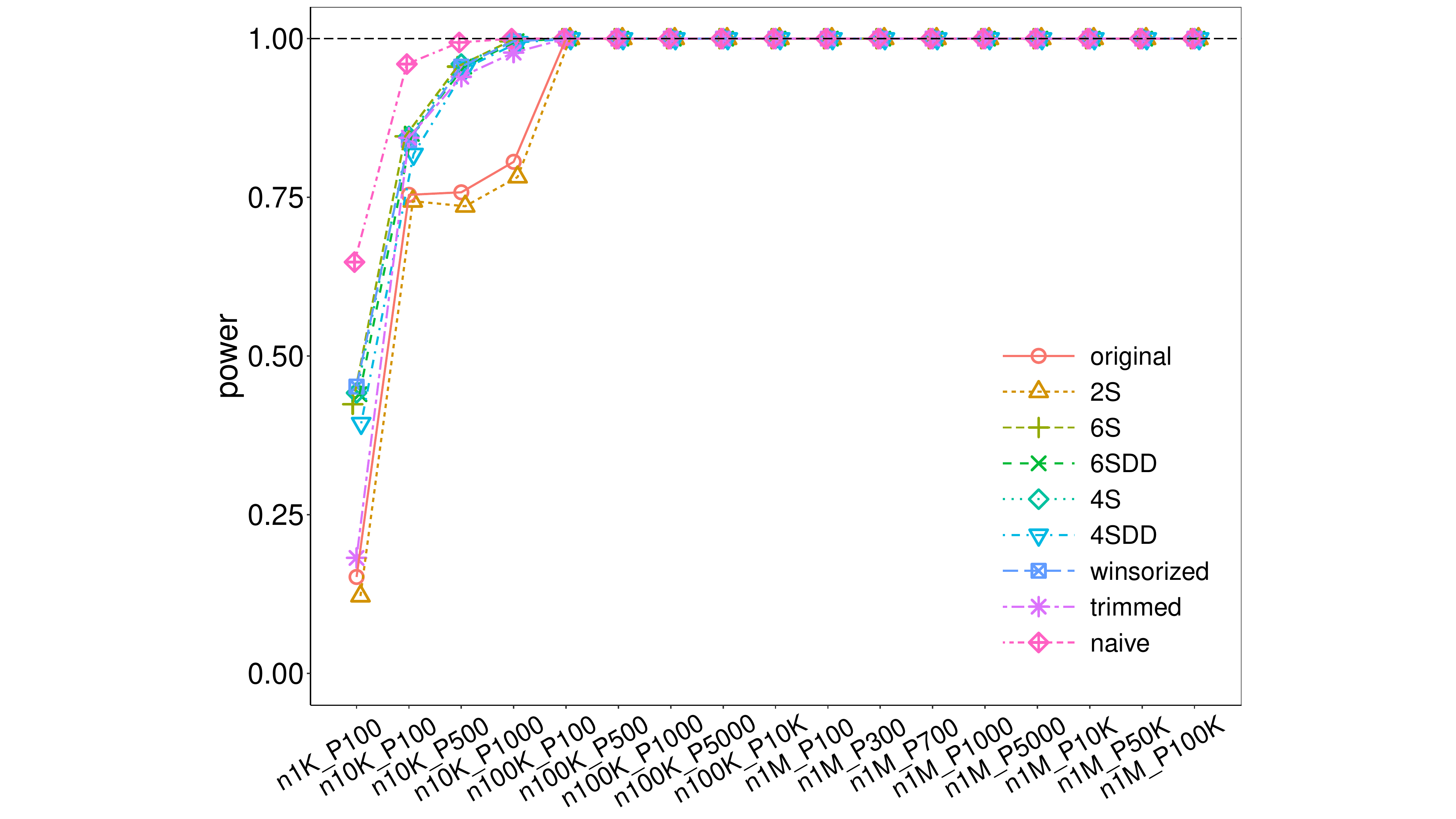}\\
\caption{ZILN data; $\epsilon$-DP; $\theta\ne0$ and $\alpha=\beta$} \label{fig:1sDPziln}
\end{figure}

\end{landscape}

\begin{landscape}

\begin{figure}[!htb]
\centering
$\rho=0.005$\hspace{1in}$\rho=0.02$\hspace{1in}$\rho=0.08$
\hspace{1in}$\rho=0.32$\hspace{0.8in}$\rho=1.28$\\
\includegraphics[width=0.24\textwidth, trim={2.2in 0 2.2in 0},clip] {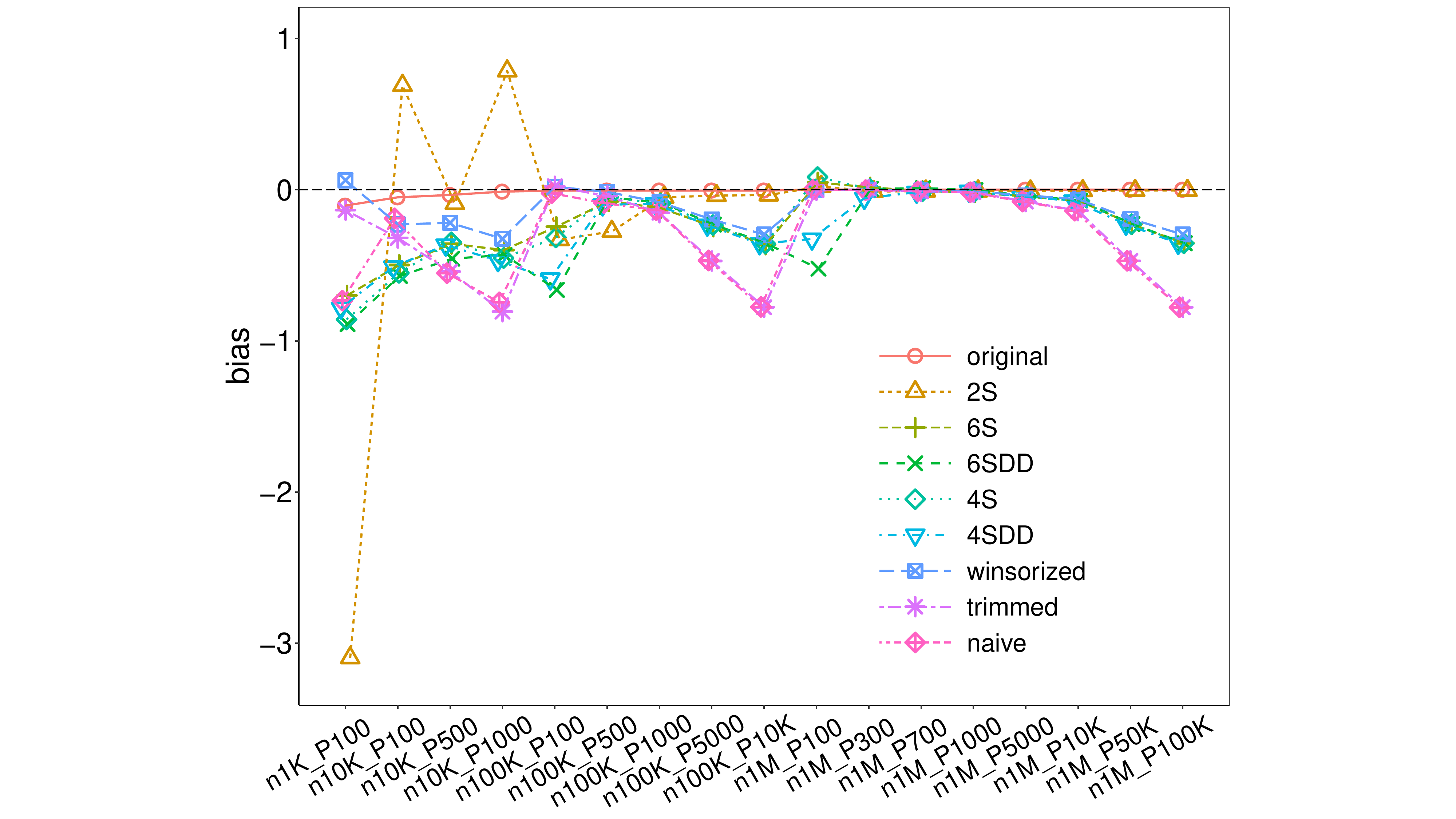}
\includegraphics[width=0.24\textwidth, trim={2.2in 0 2.2in 0},clip] {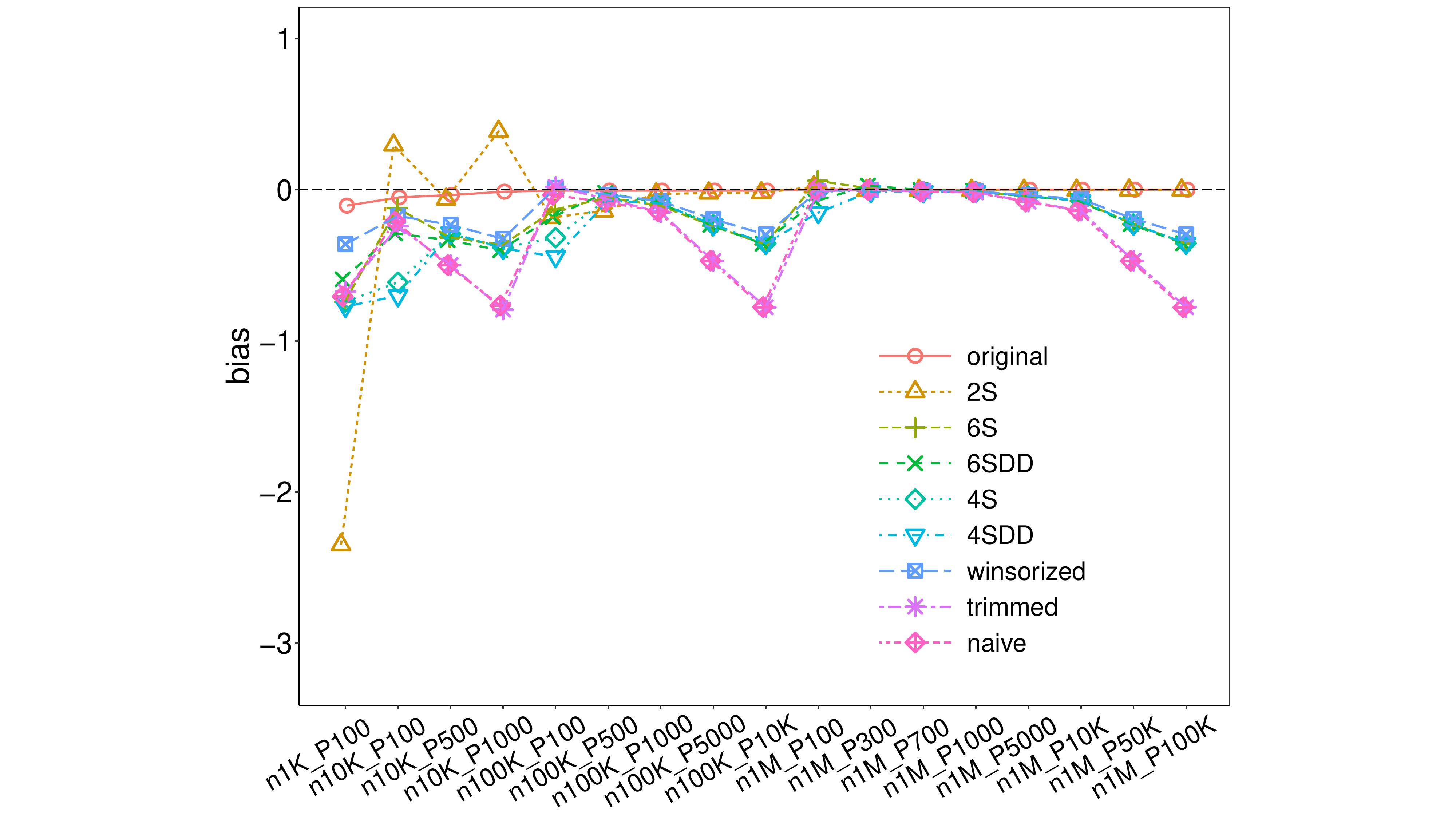}
\includegraphics[width=0.24\textwidth, trim={2.2in 0 2.2in 0},clip] {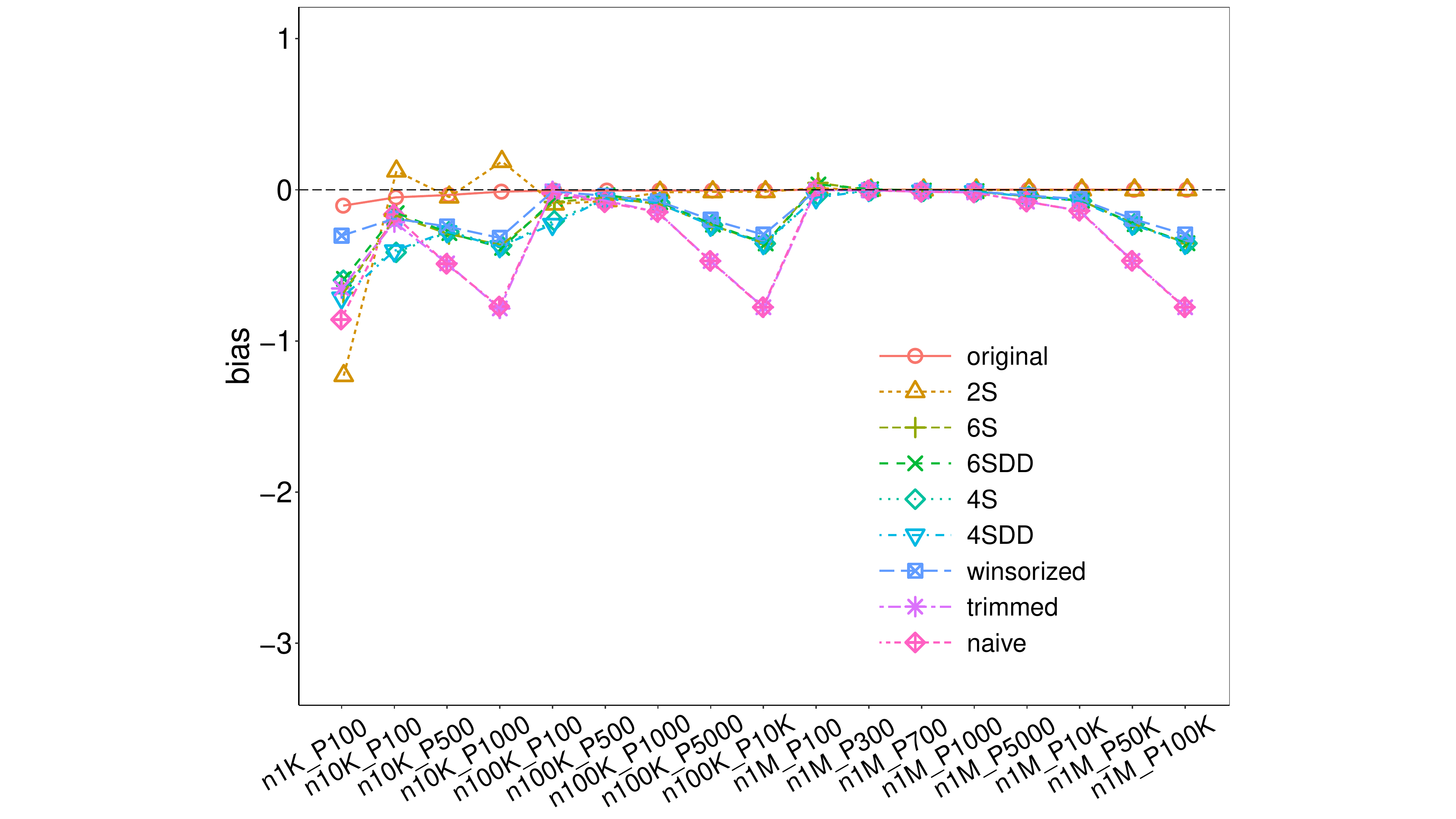}
\includegraphics[width=0.24\textwidth, trim={2.2in 0 2.2in 0},clip] {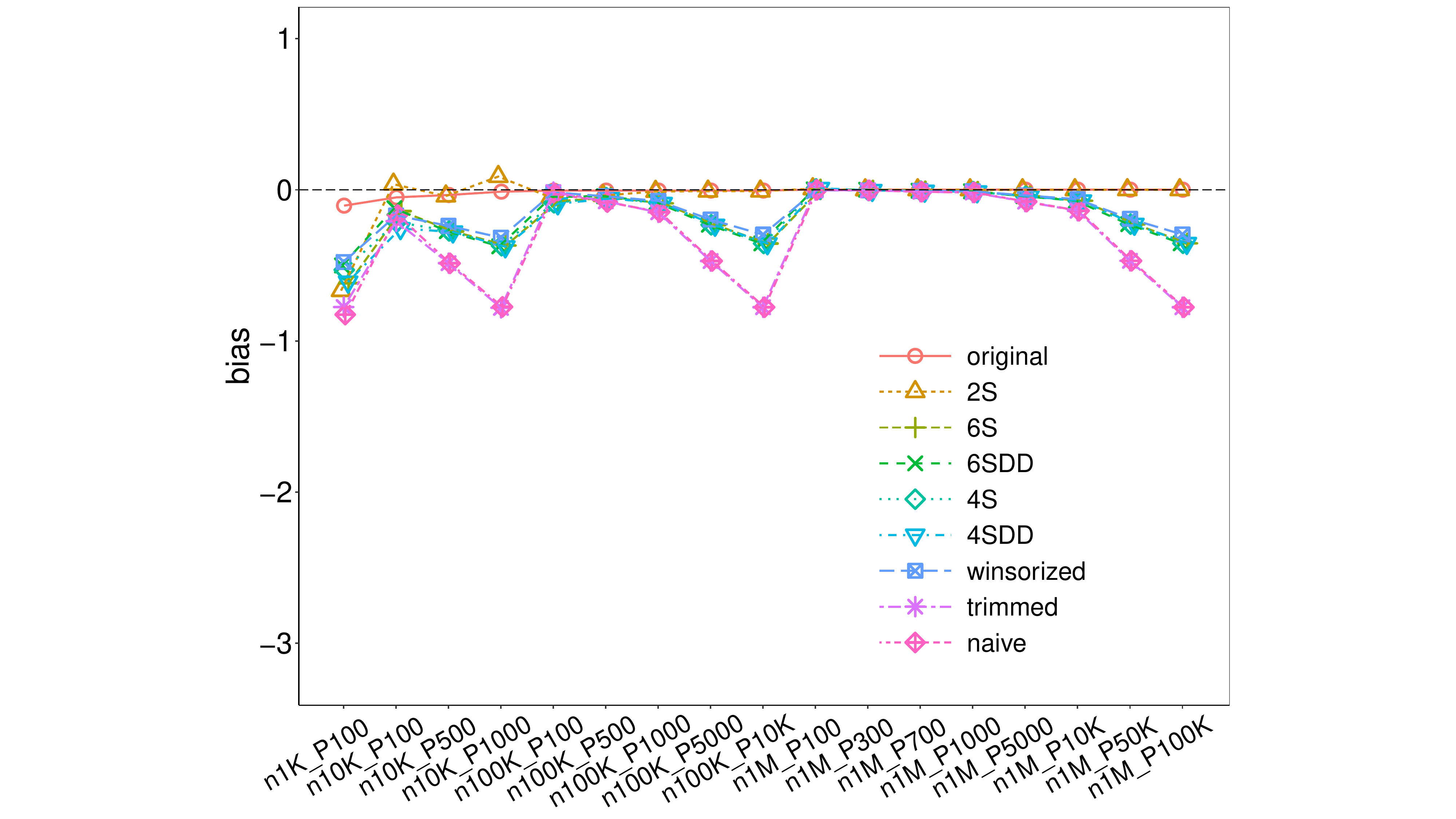}
\includegraphics[width=0.24\textwidth, trim={2.2in 0 2.2in 0},clip] {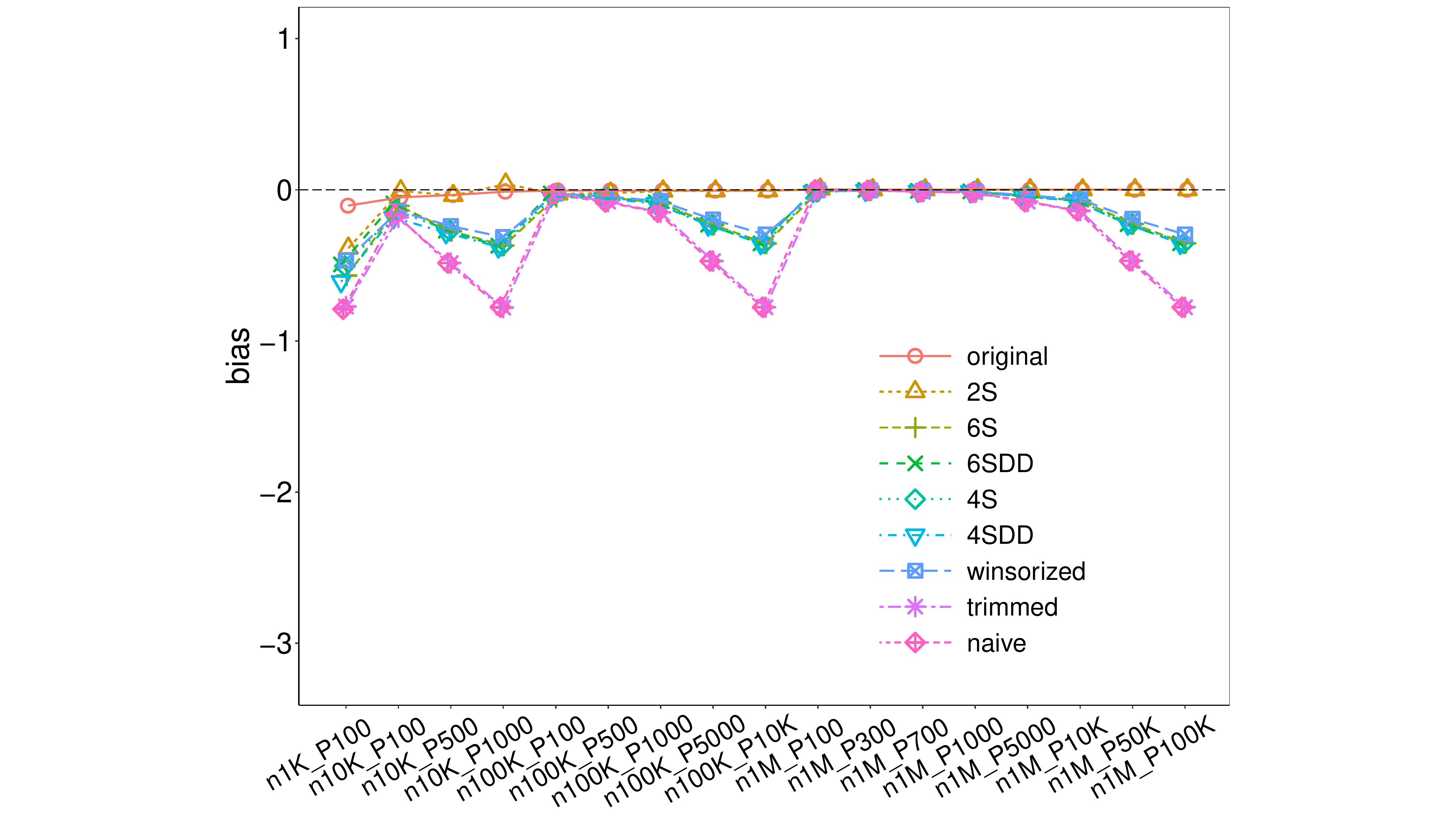}\\
\includegraphics[width=0.24\textwidth, trim={2.2in 0 2.2in 0},clip] {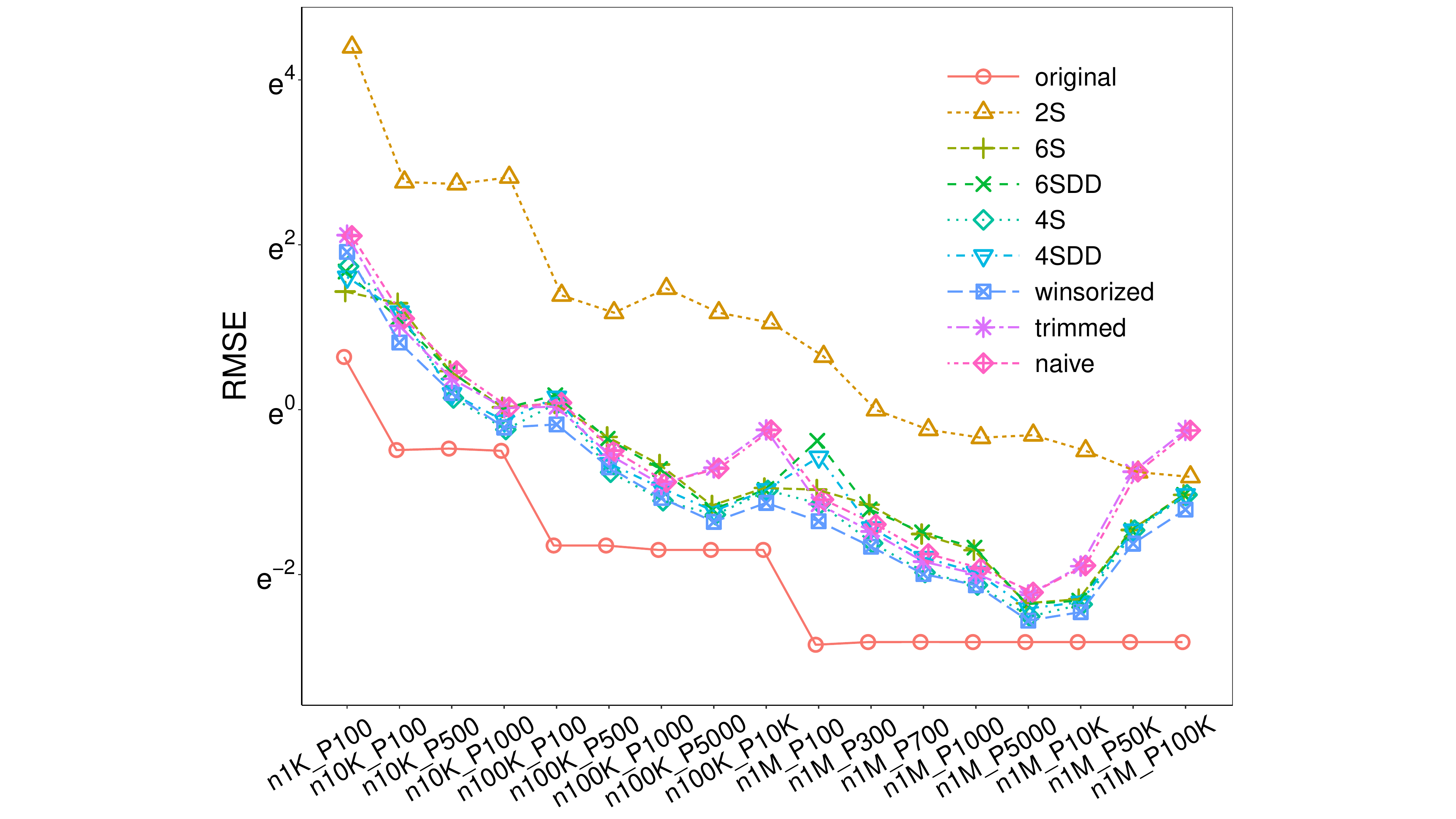}
\includegraphics[width=0.24\textwidth, trim={2.2in 0 2.2in 0},clip] {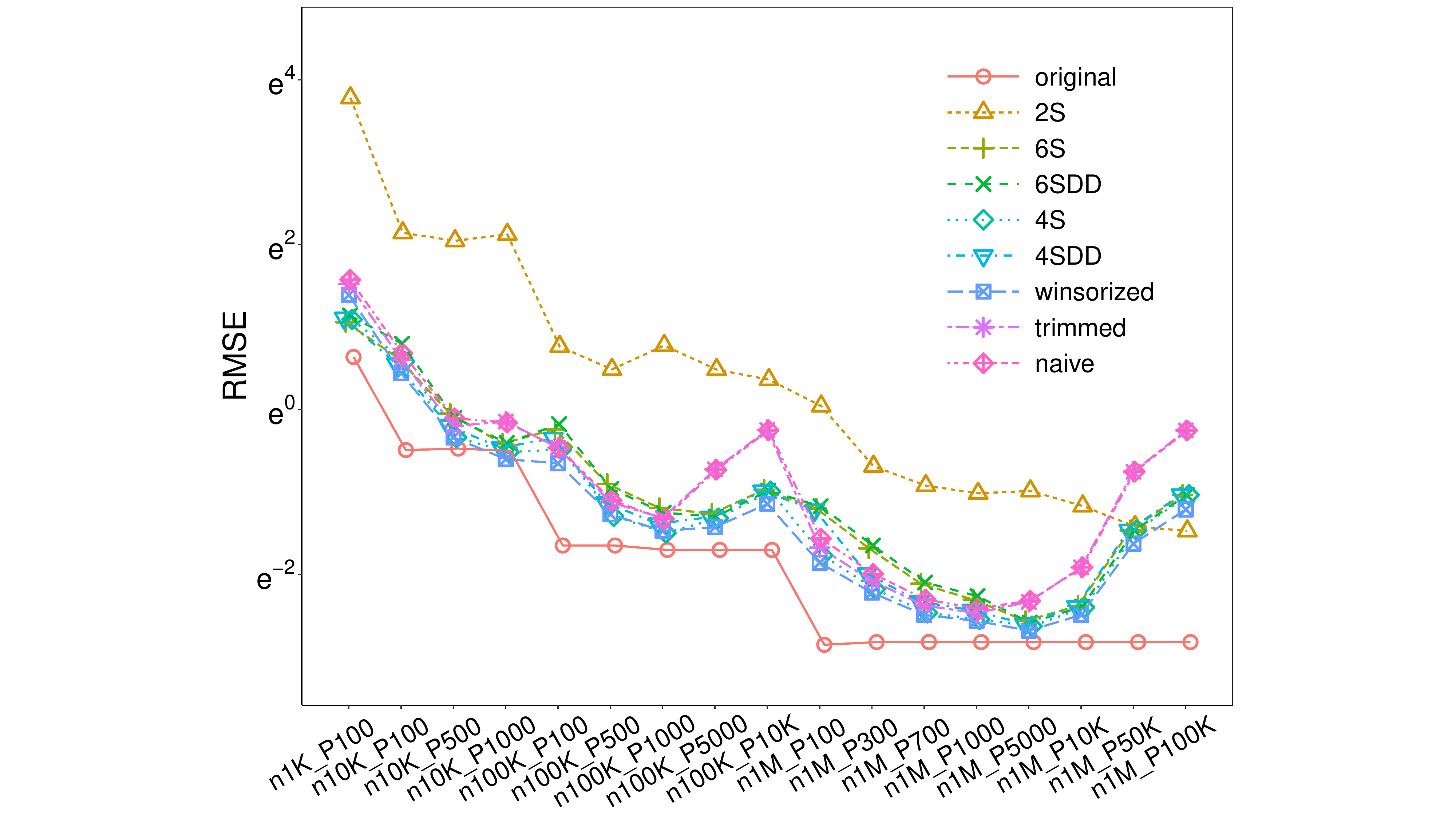}
\includegraphics[width=0.24\textwidth, trim={2.2in 0 2.2in 0},clip] {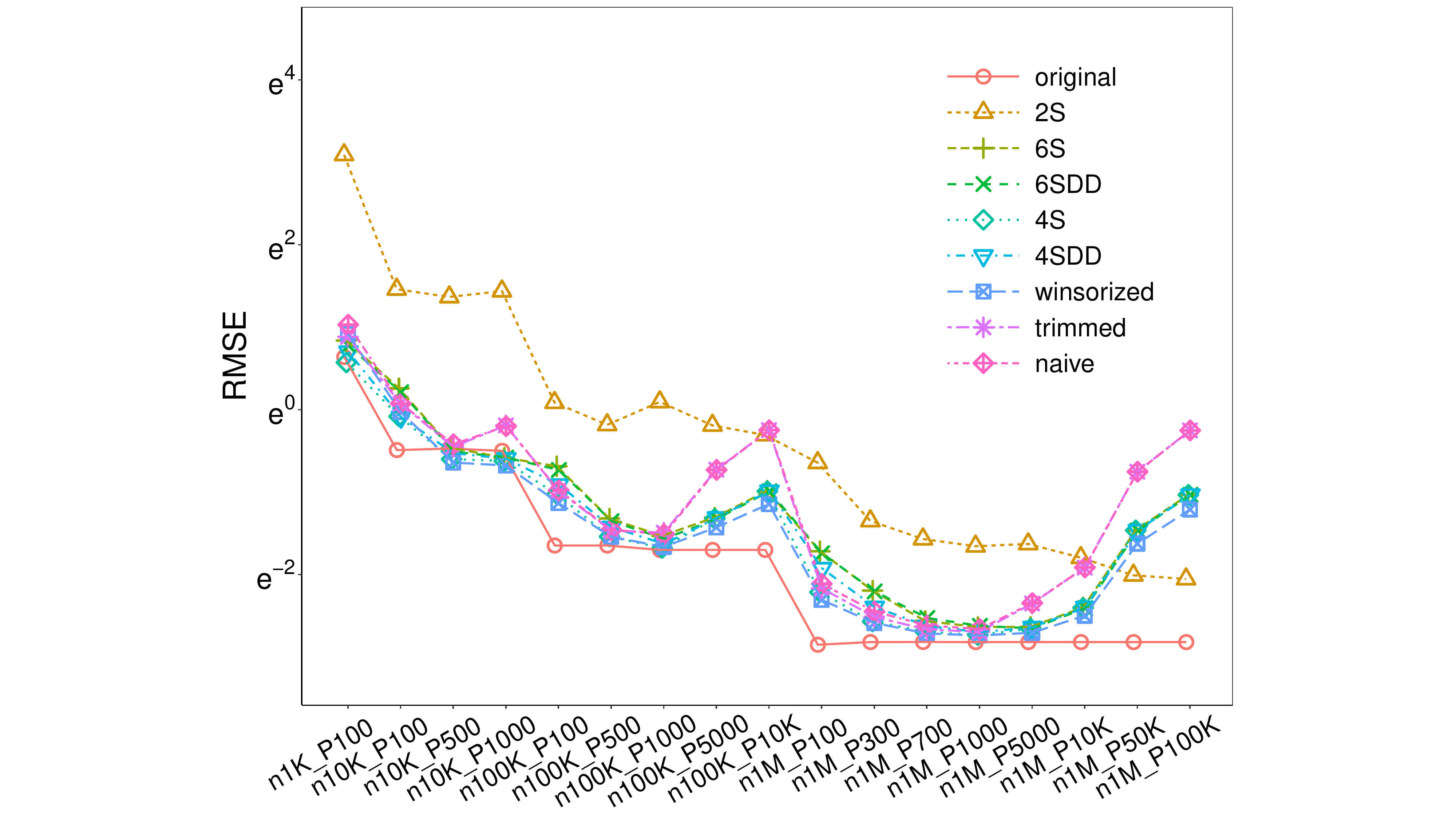}
\includegraphics[width=0.24\textwidth, trim={2.2in 0 2.2in 0},clip] {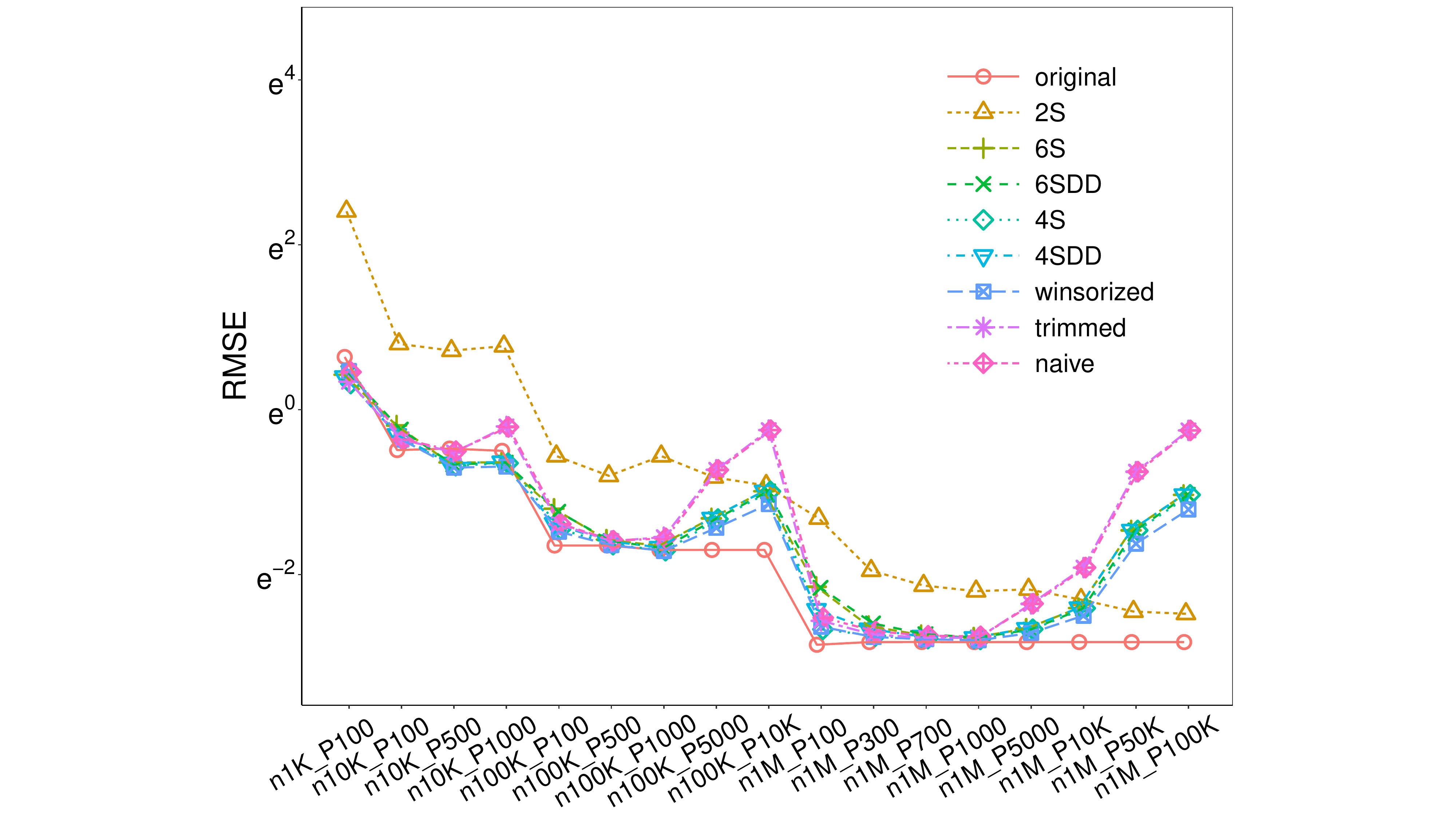}
\includegraphics[width=0.24\textwidth, trim={2.2in 0 2.2in 0},clip] {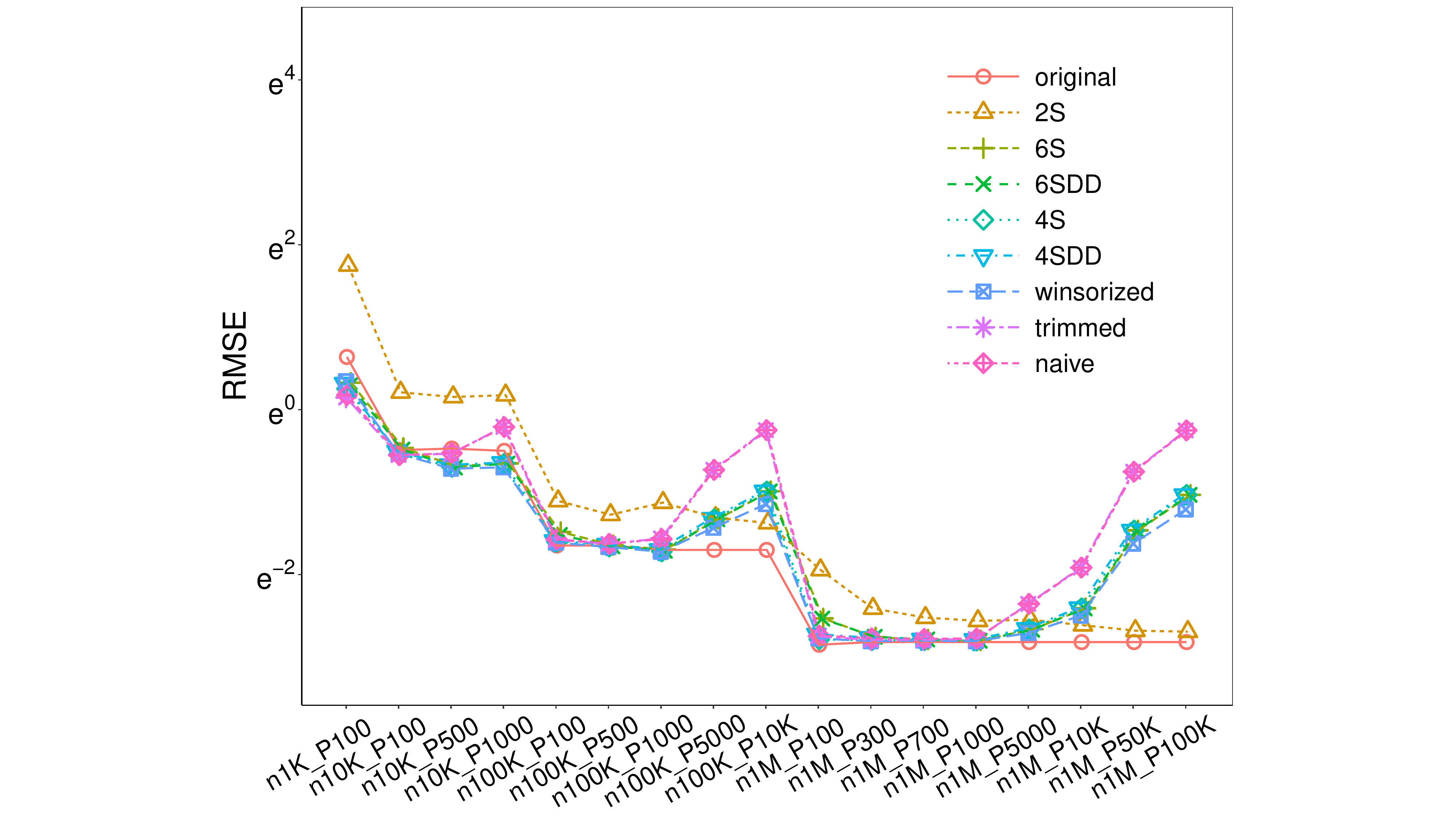}\\
\includegraphics[width=0.24\textwidth, trim={2.2in 0 2.2in 0},clip] {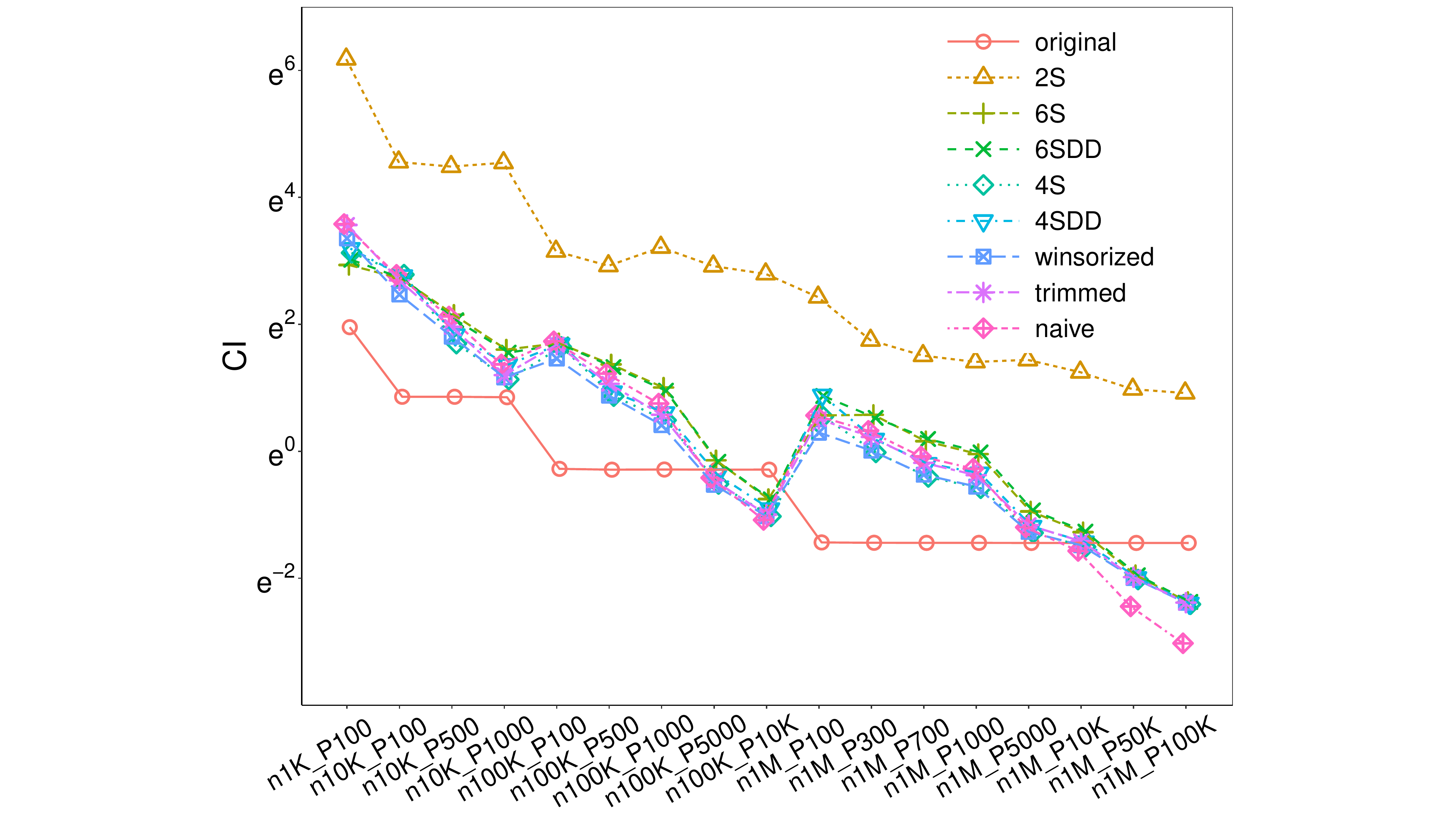}
\includegraphics[width=0.24\textwidth, trim={2.2in 0 2.2in 0},clip] {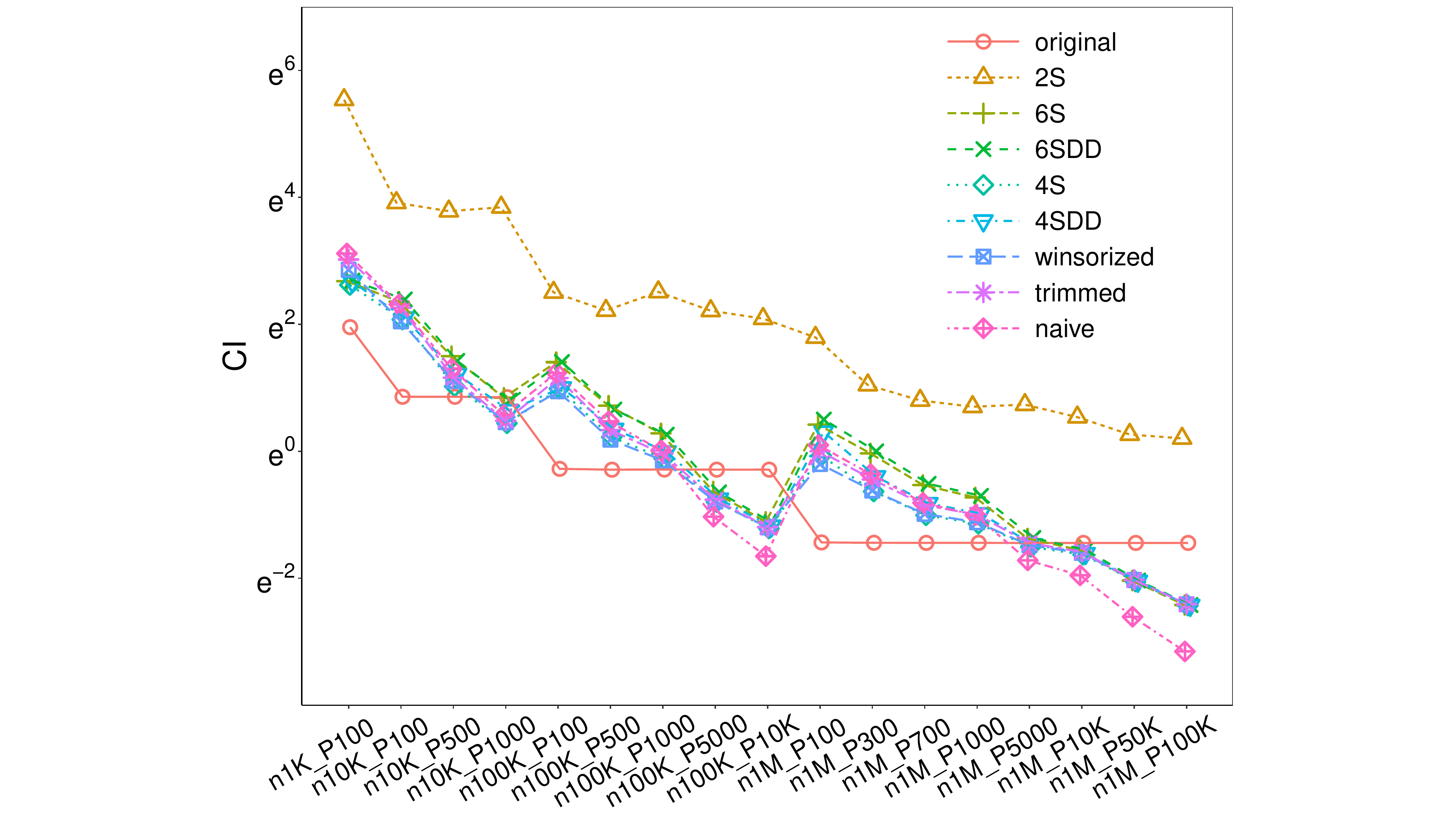}
\includegraphics[width=0.24\textwidth, trim={2.2in 0 2.2in 0},clip] {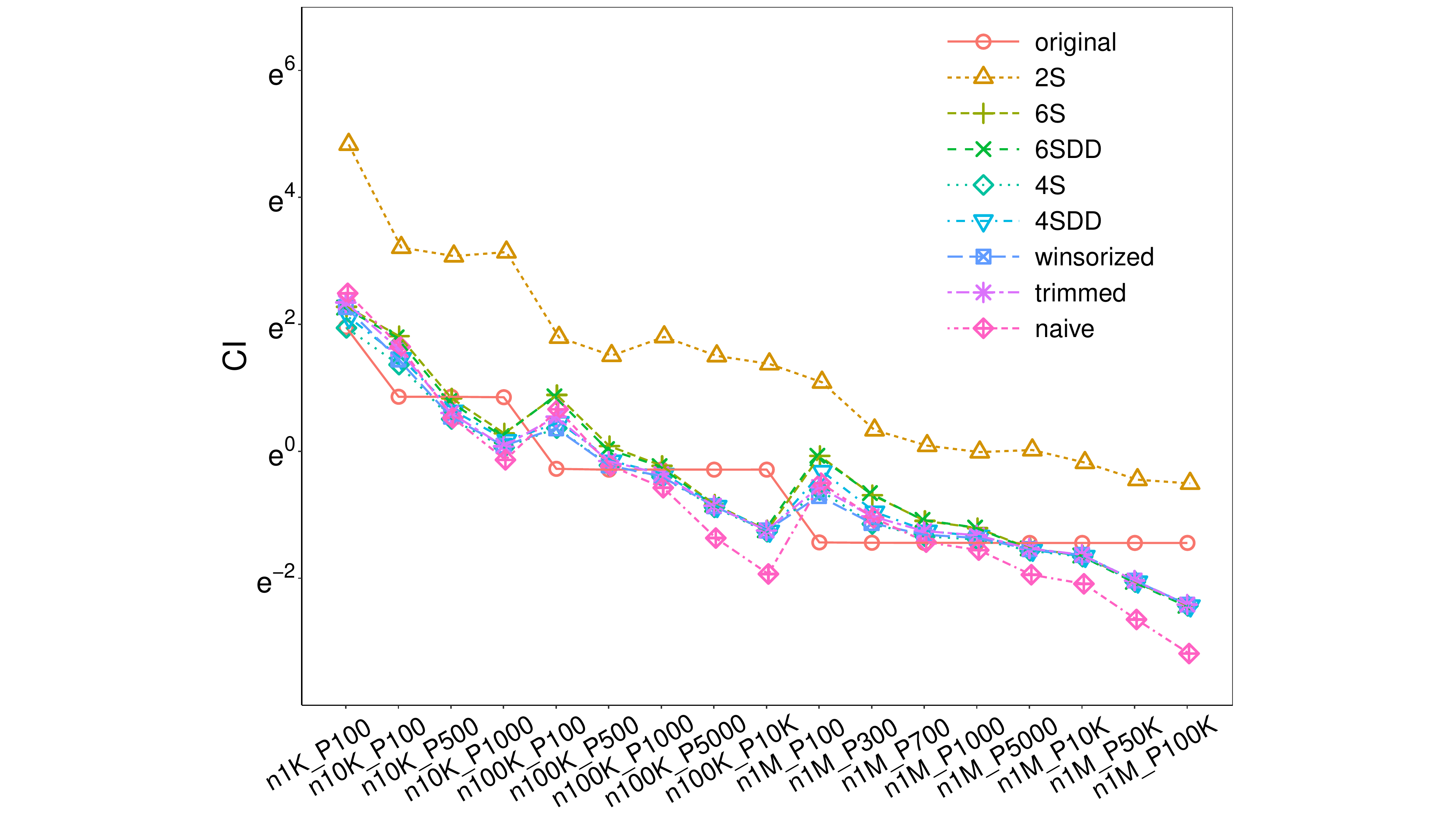}
\includegraphics[width=0.24\textwidth, trim={2.2in 0 2.2in 0},clip] {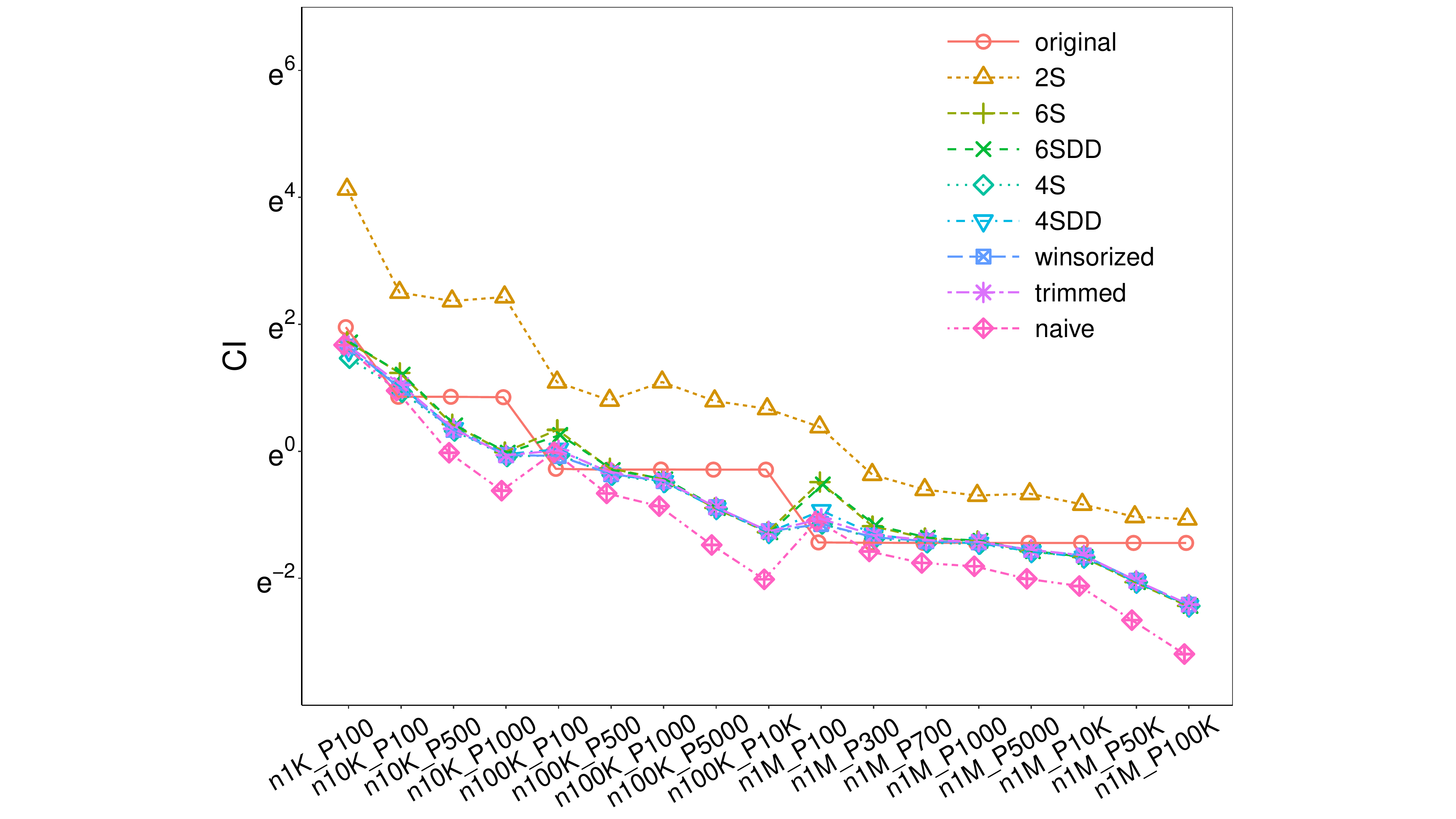}
\includegraphics[width=0.24\textwidth, trim={2.2in 0 2.2in 0},clip] {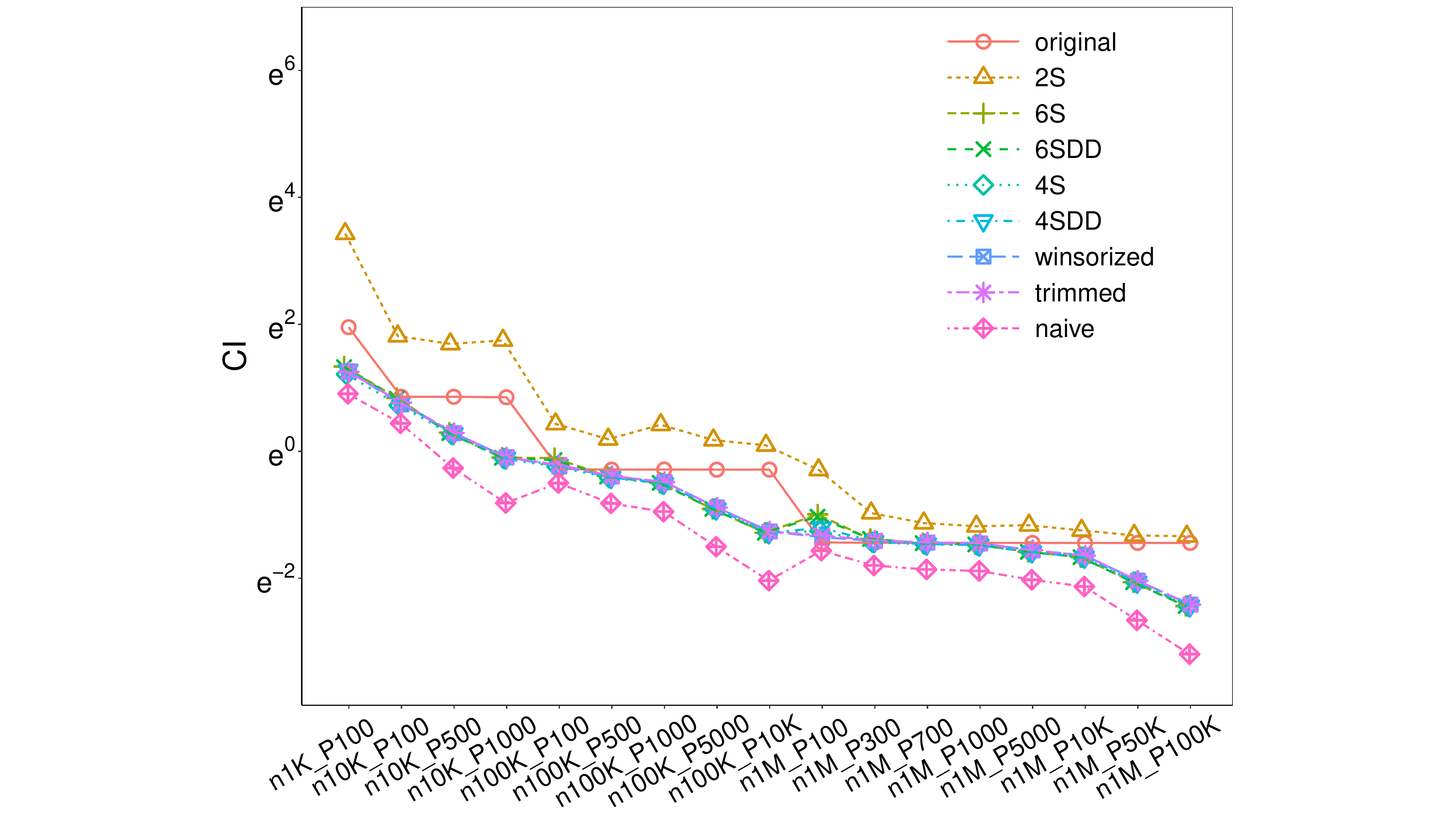}\\
\includegraphics[width=0.24\textwidth, trim={2.2in 0 2.2in 0},clip] {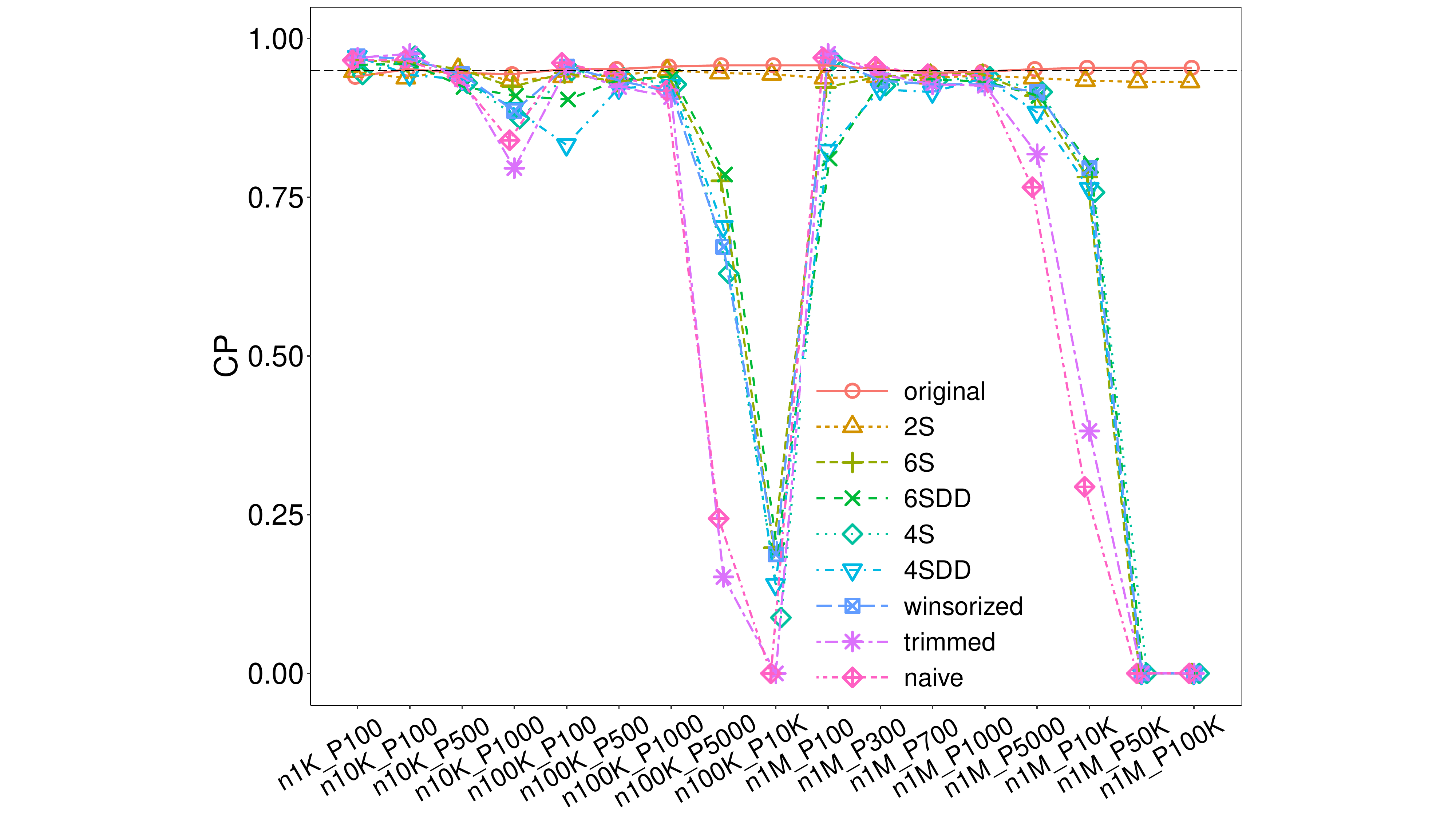}
\includegraphics[width=0.24\textwidth, trim={2.2in 0 2.2in 0},clip] {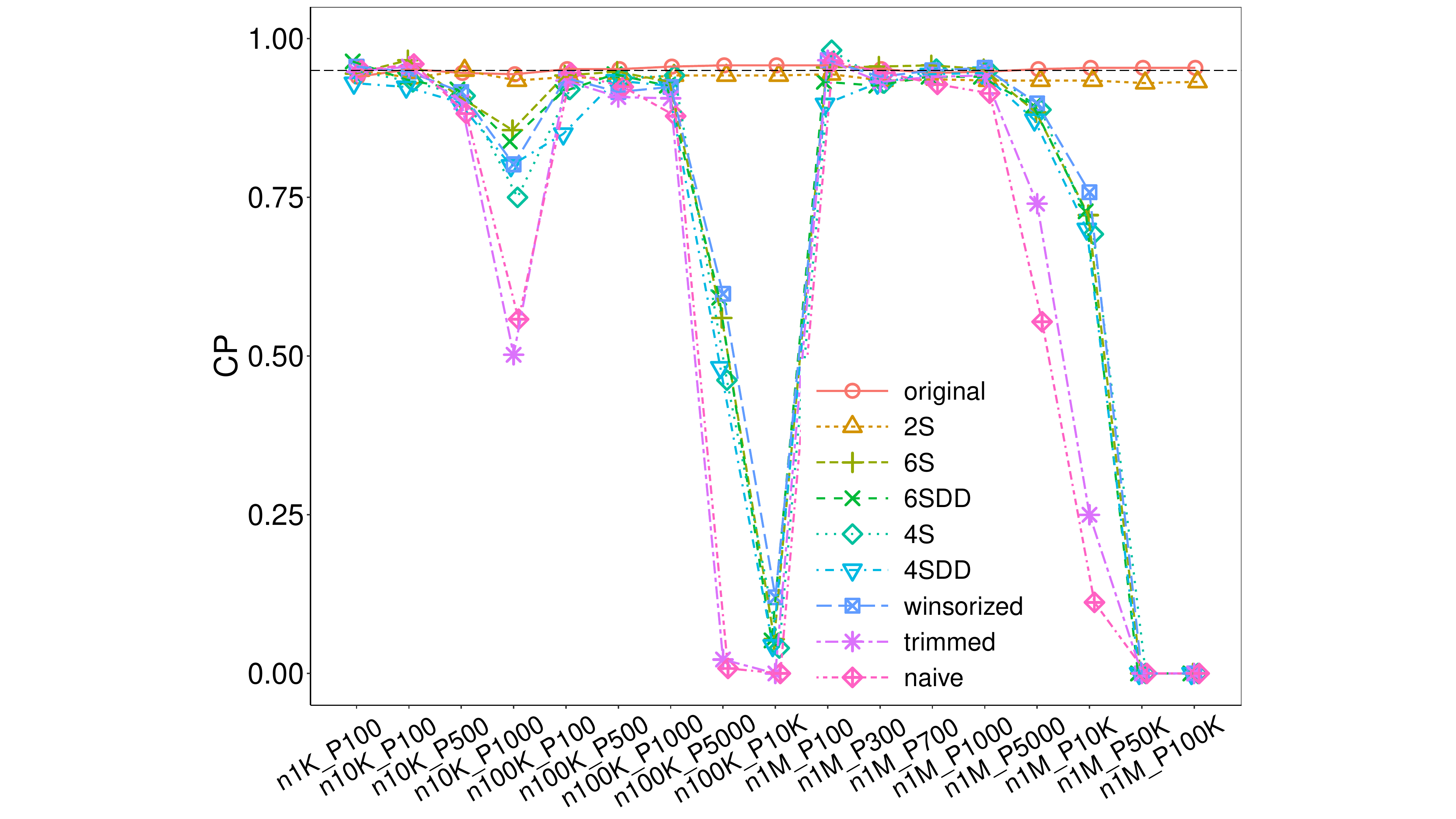}
\includegraphics[width=0.24\textwidth, trim={2.2in 0 2.2in 0},clip] {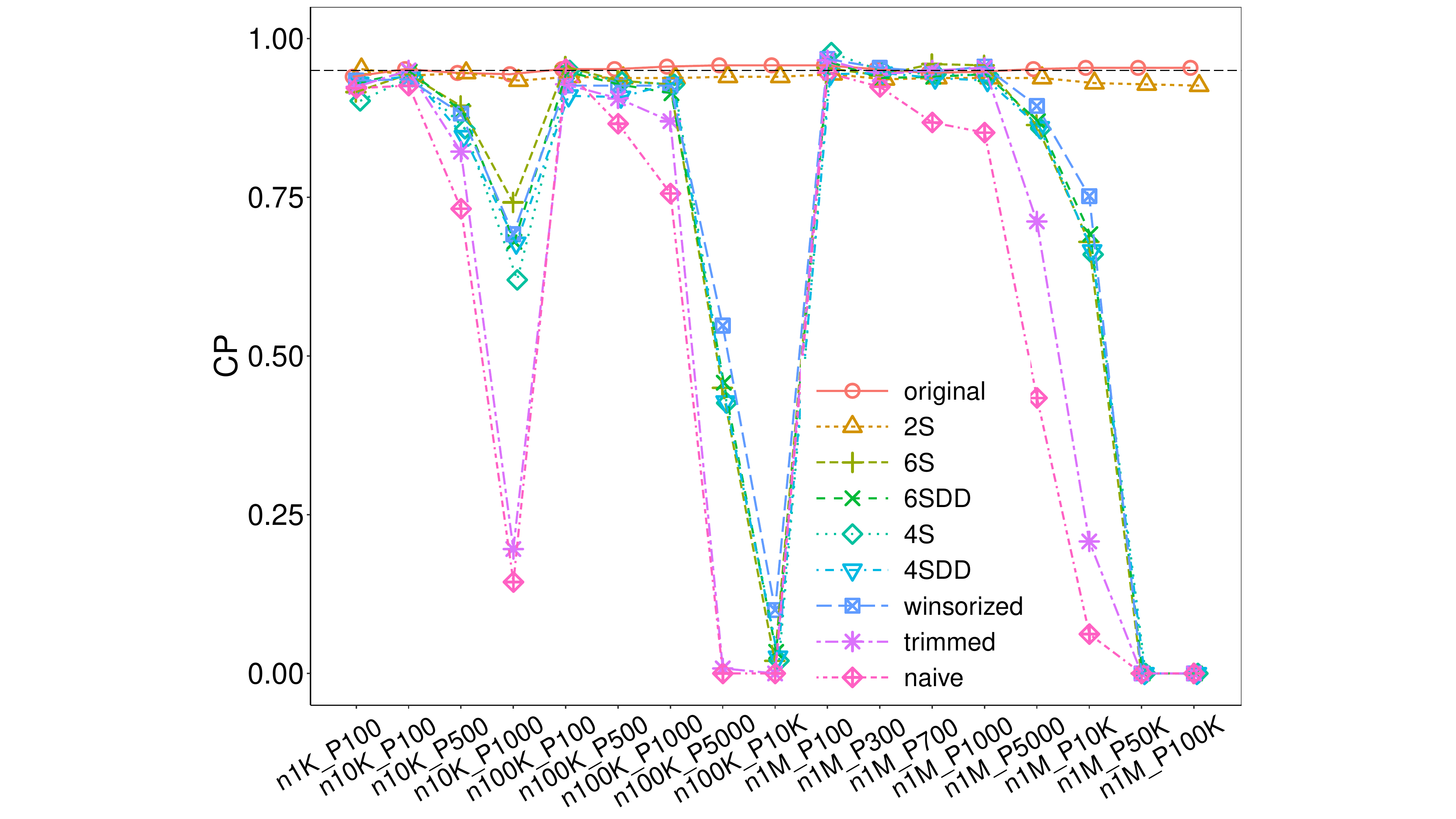}
\includegraphics[width=0.24\textwidth, trim={2.2in 0 2.2in 0},clip] {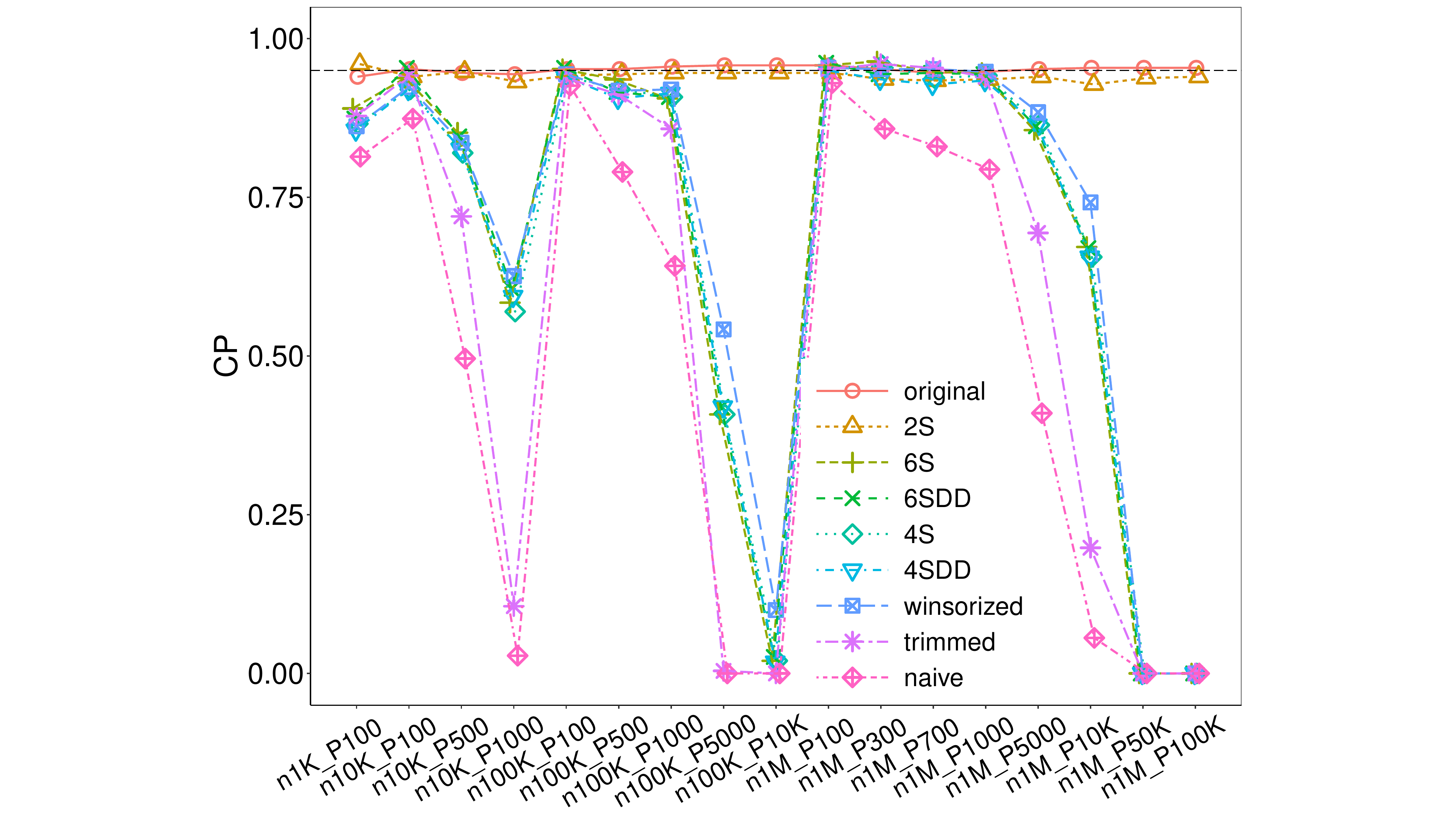}
\includegraphics[width=0.24\textwidth, trim={2.2in 0 2.2in 0},clip] {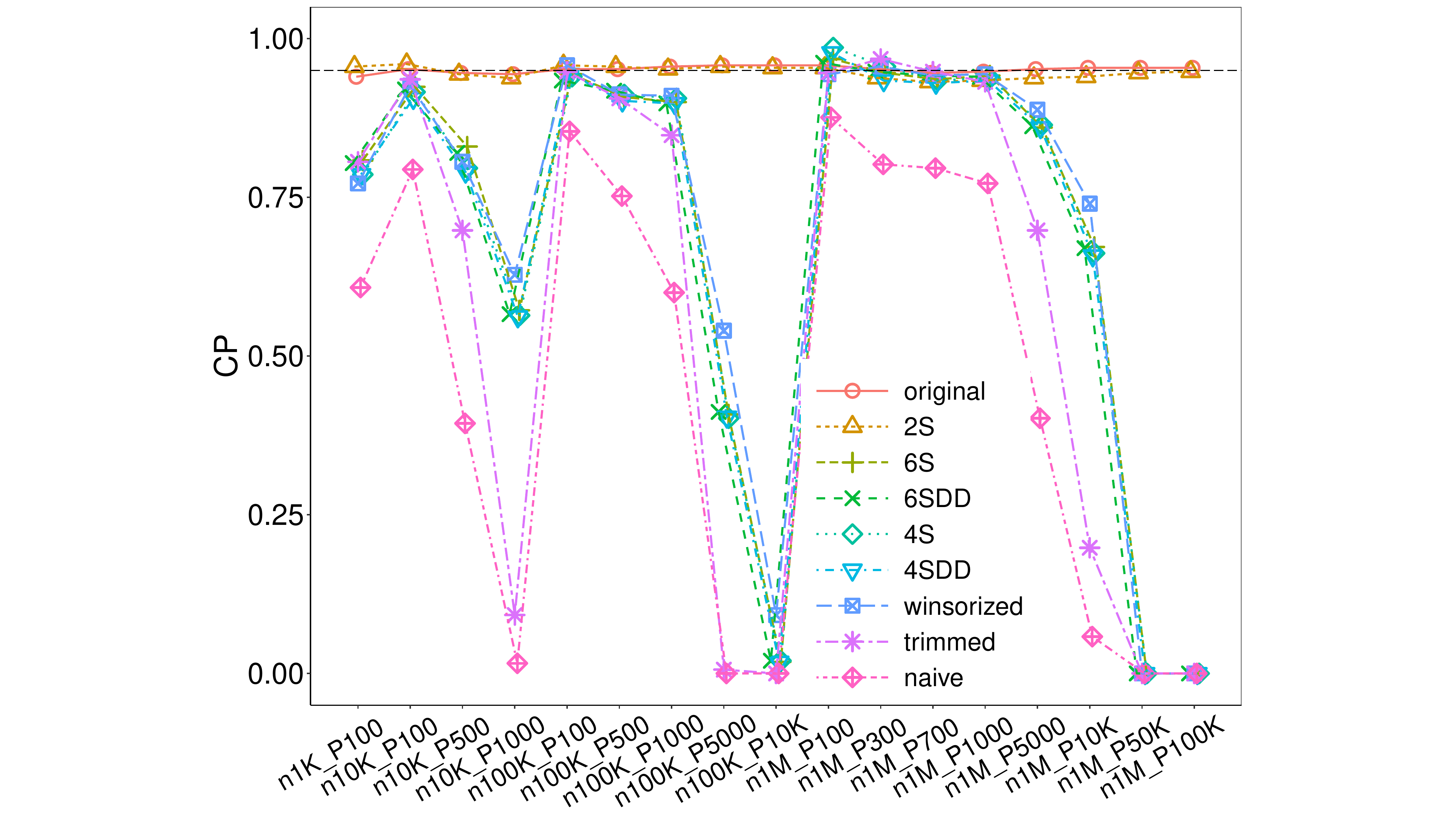}\\
\includegraphics[width=0.24\textwidth, trim={2.2in 0 2.2in 0},clip] {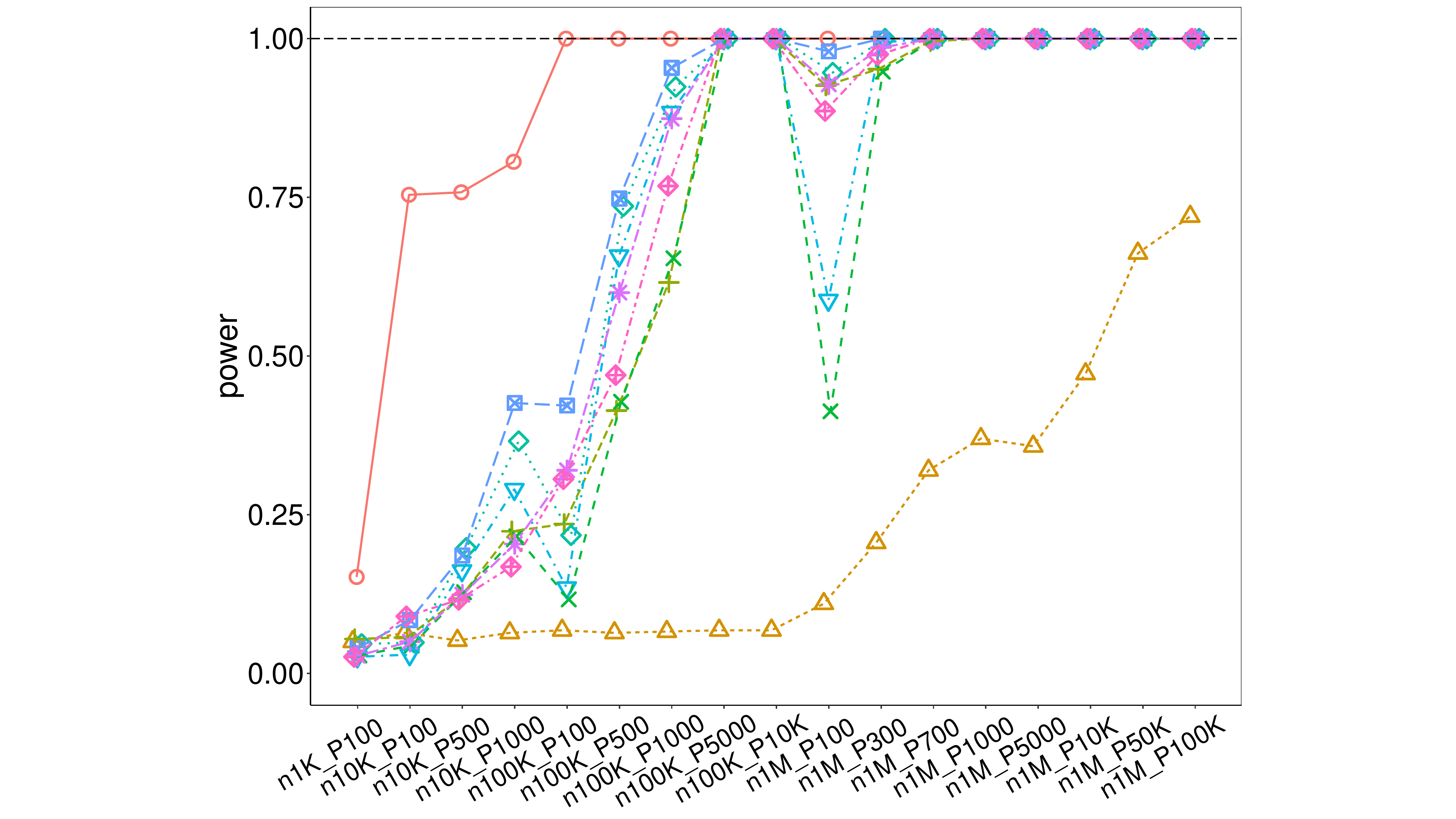}
\includegraphics[width=0.24\textwidth, trim={2.2in 0 2.2in 0},clip] {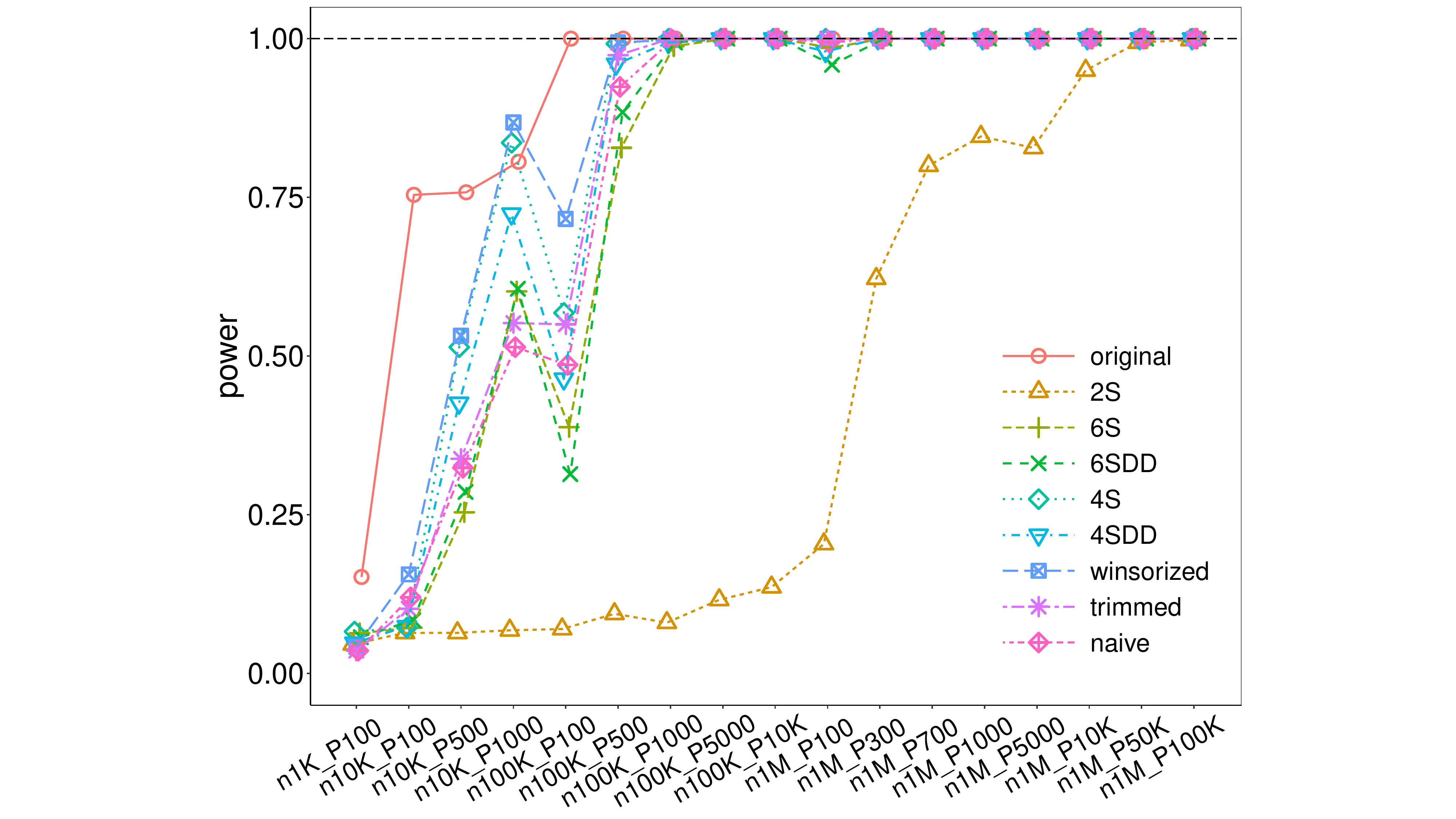}
\includegraphics[width=0.24\textwidth, trim={2.2in 0 2.2in 0},clip] {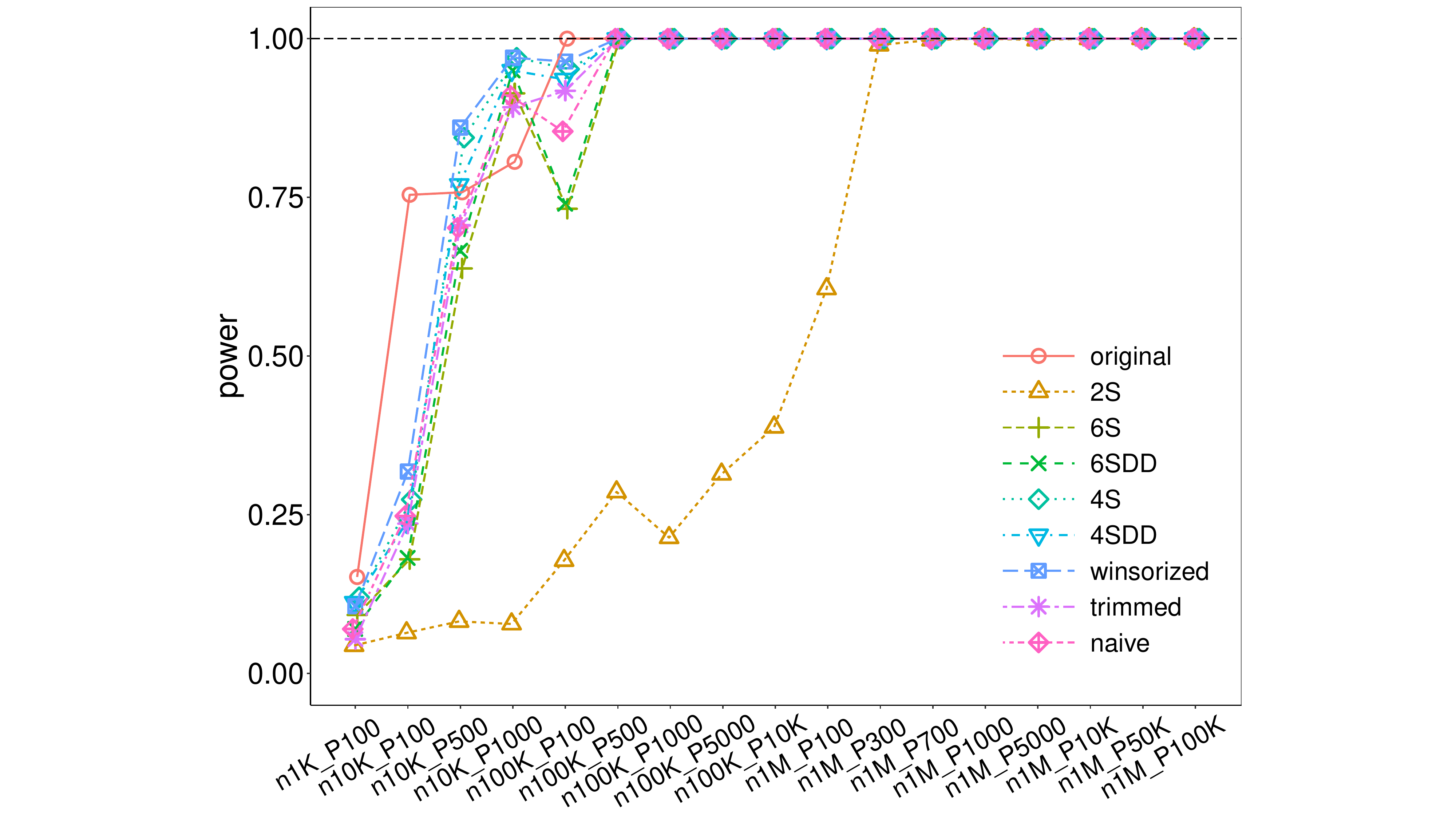}
\includegraphics[width=0.24\textwidth, trim={2.2in 0 2.2in 0},clip] {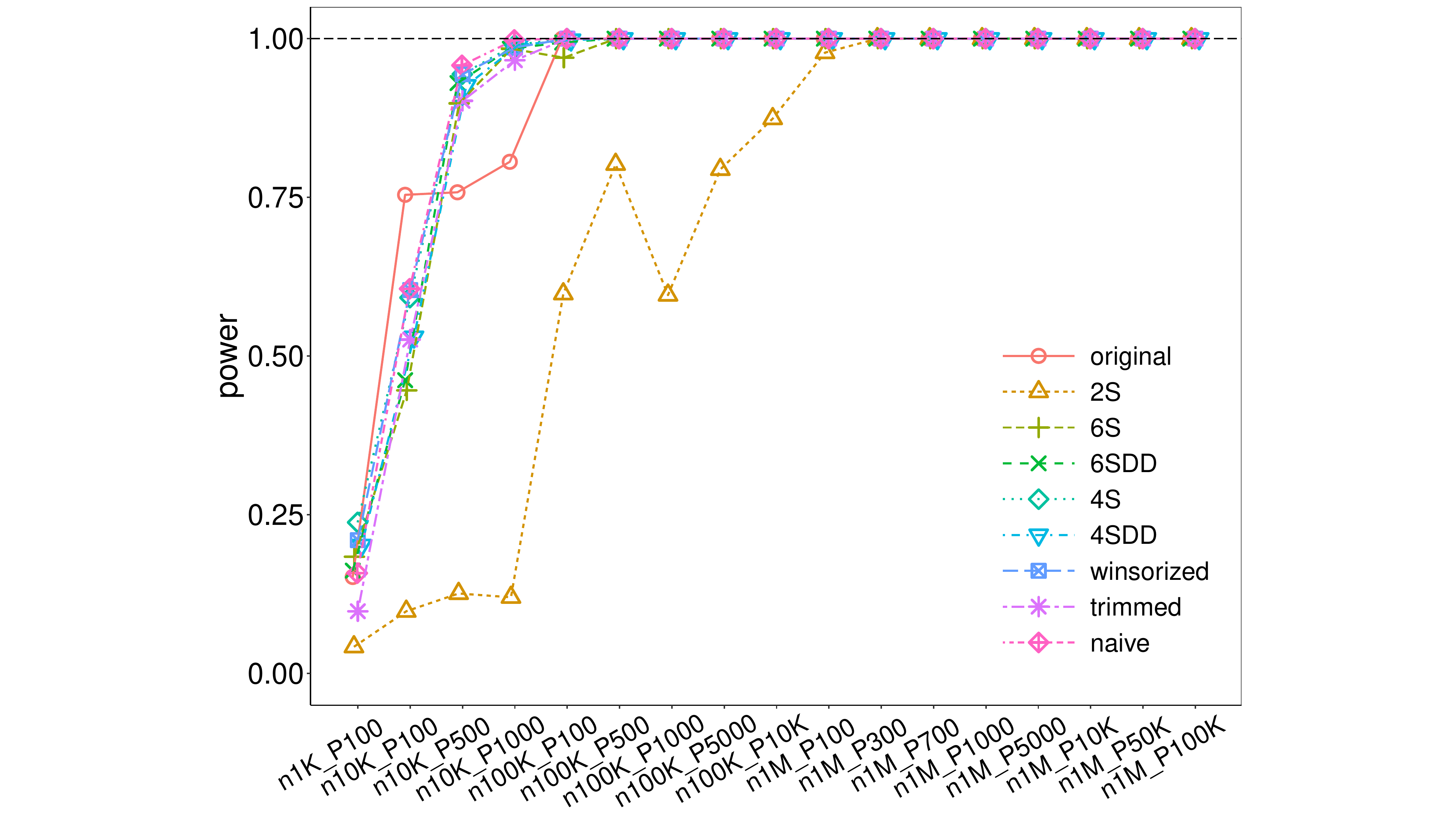}
\includegraphics[width=0.24\textwidth, trim={2.2in 0 2.2in 0},clip] {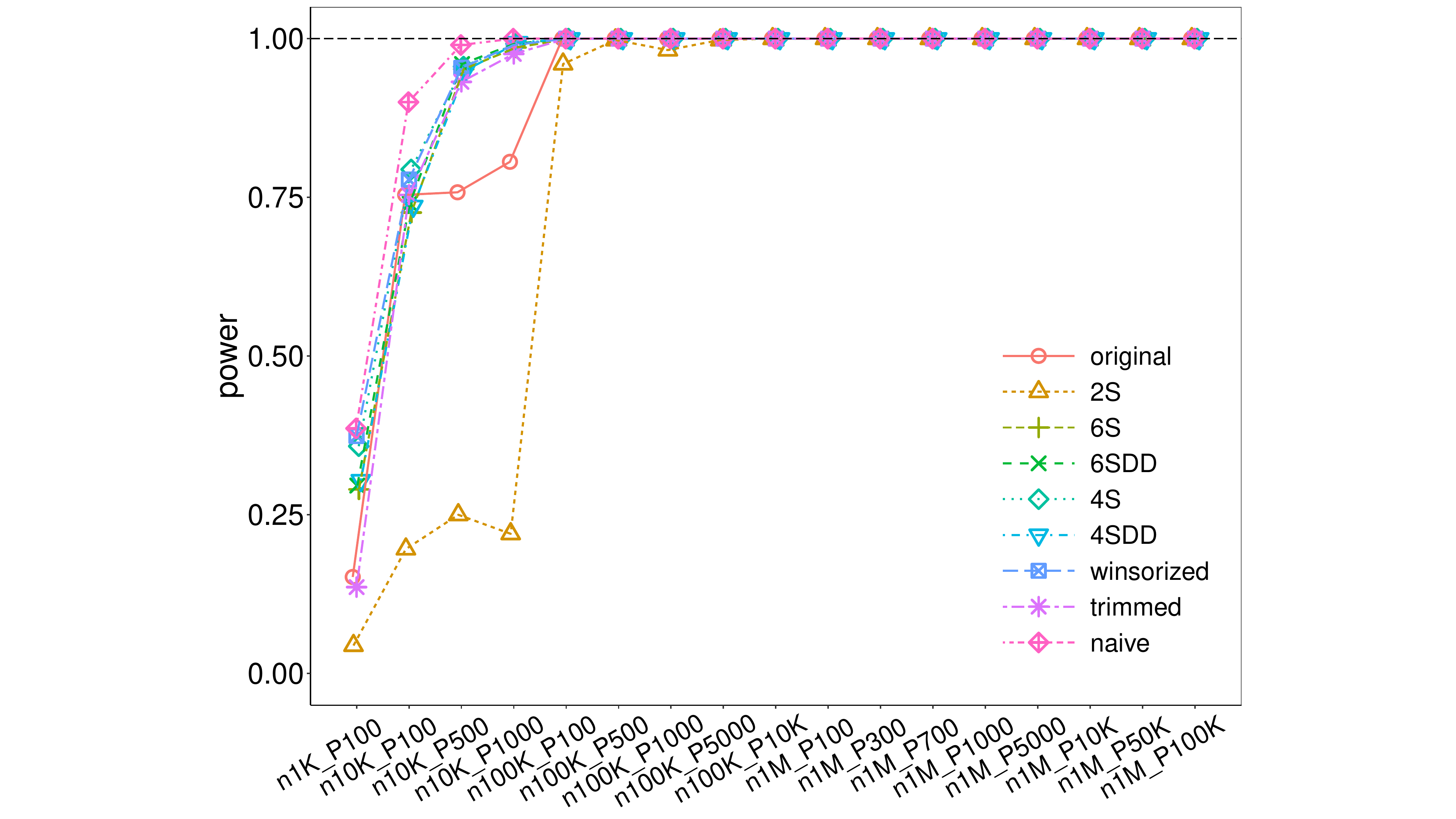}\\
\caption{ZILN data; $\rho$-zCDP; $\theta\ne0$ and $\alpha=\beta$}
\label{fig:1szCDPziln}
\end{figure}

\end{landscape}

\begin{landscape}
\subsection*{ZILN, $\theta=0$ and $\alpha\ne\beta$}
\begin{figure}[!htb]
\hspace{0.6in}$\epsilon=0.5$\hspace{1.3in}$\epsilon=1$\hspace{1.4in}$\epsilon=2$
\hspace{1.4in}$\epsilon=5$\hspace{1.4in}$\epsilon=50$\\
\includegraphics[width=0.26\textwidth, trim={2.2in 0 2.2in 0},clip] {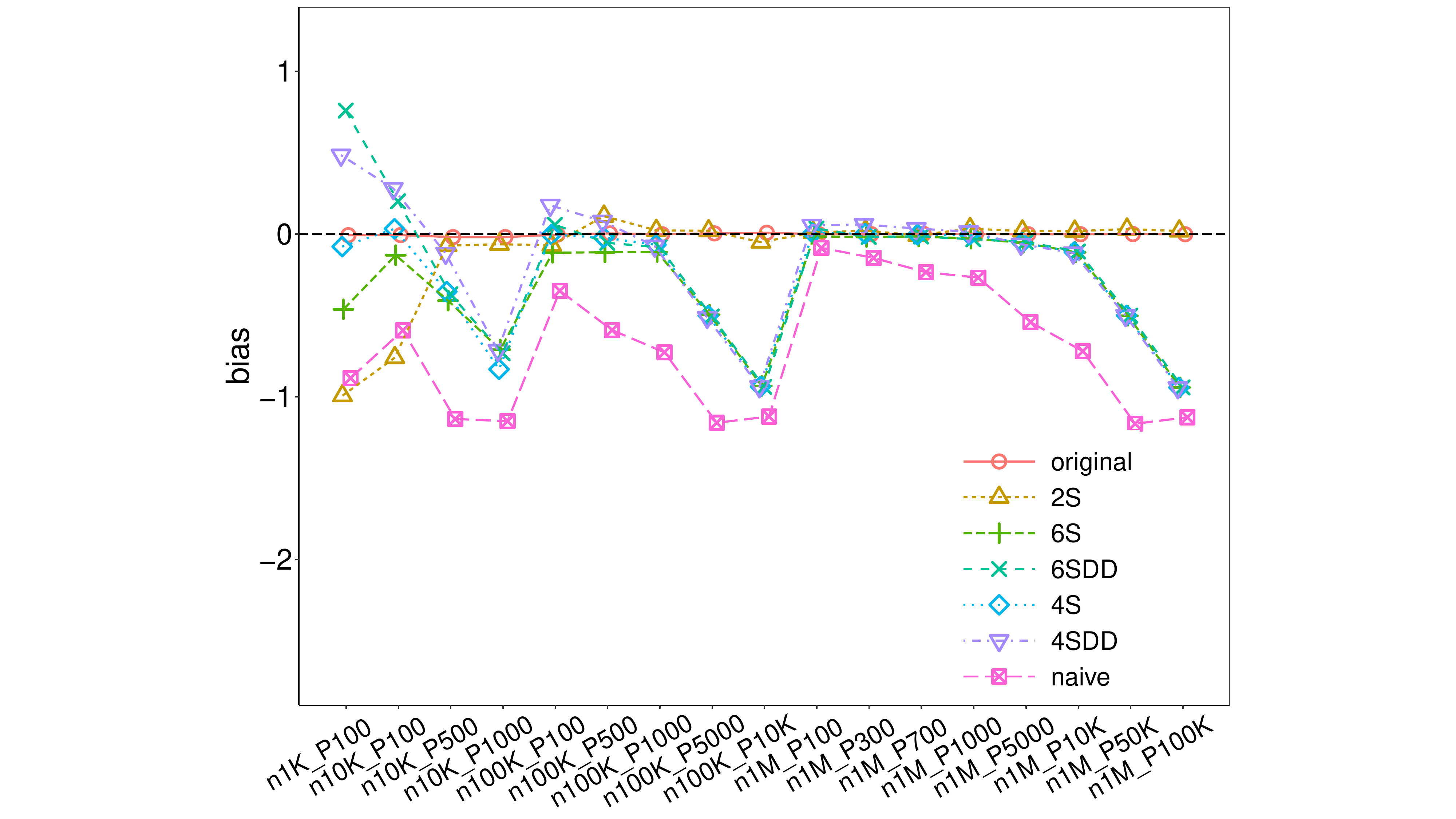}
\includegraphics[width=0.26\textwidth, trim={2.2in 0 2.2in 0},clip] {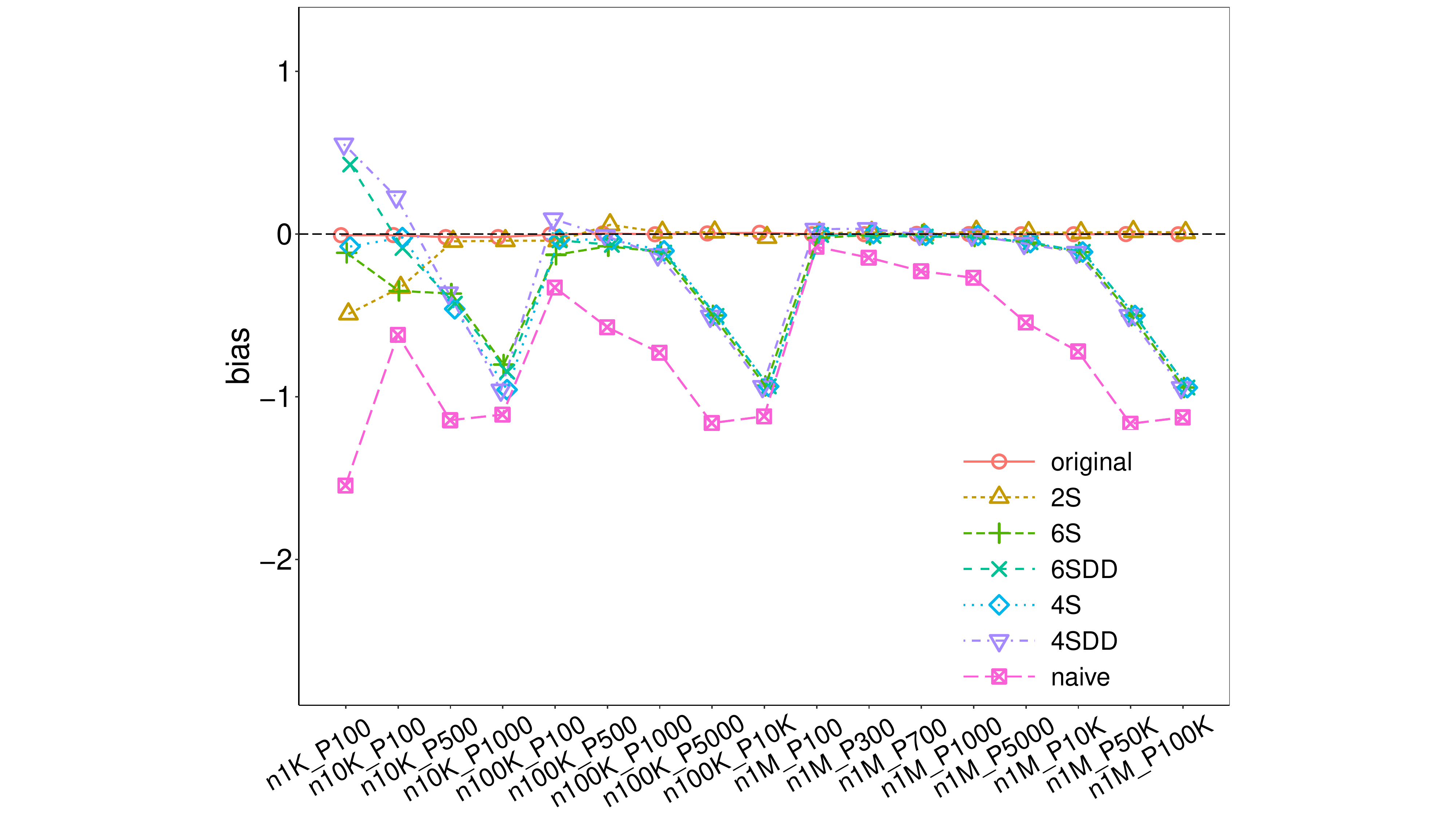}
\includegraphics[width=0.26\textwidth, trim={2.2in 0 2.2in 0},clip] {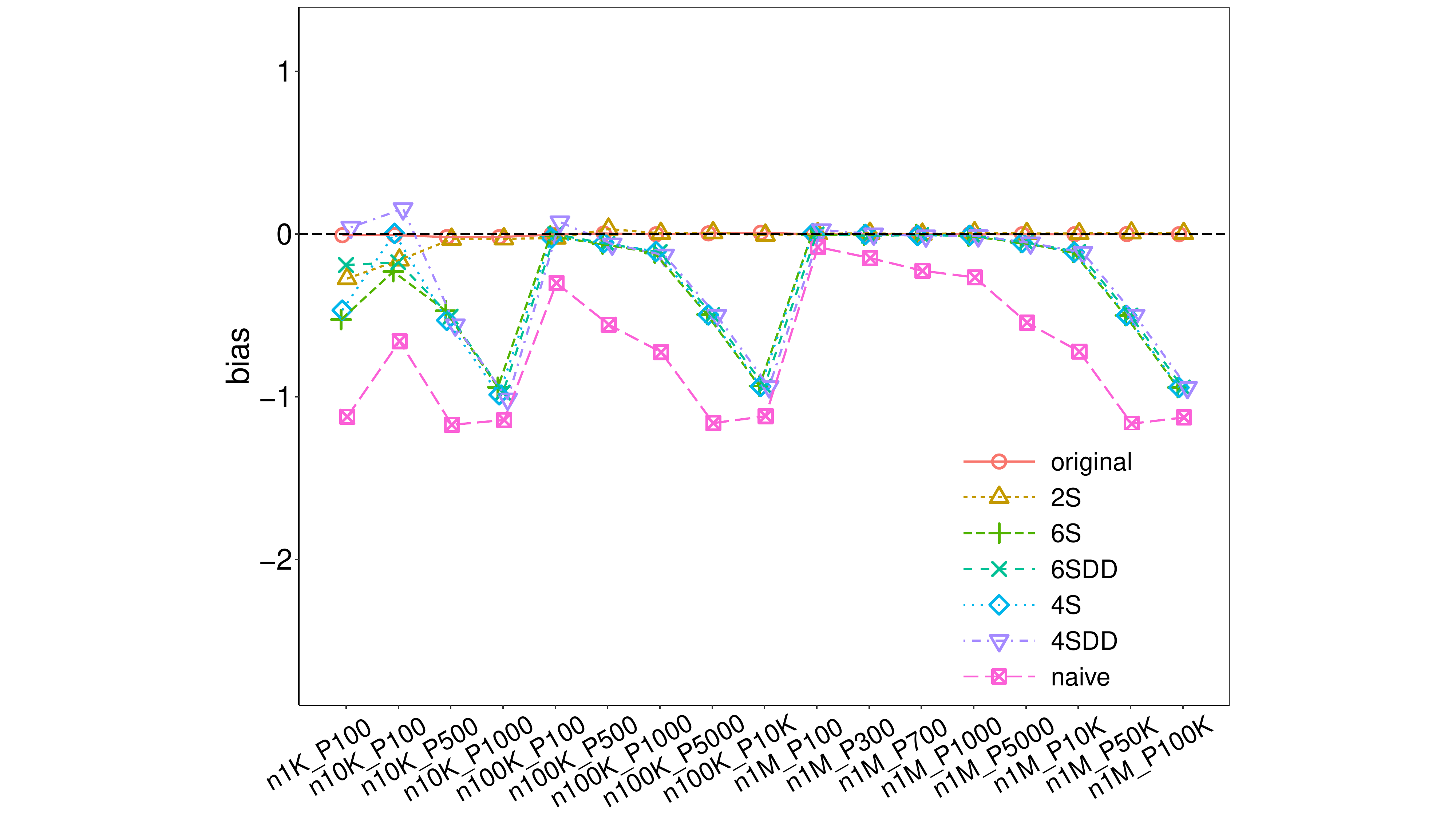}
\includegraphics[width=0.26\textwidth, trim={2.2in 0 2.2in 0},clip] {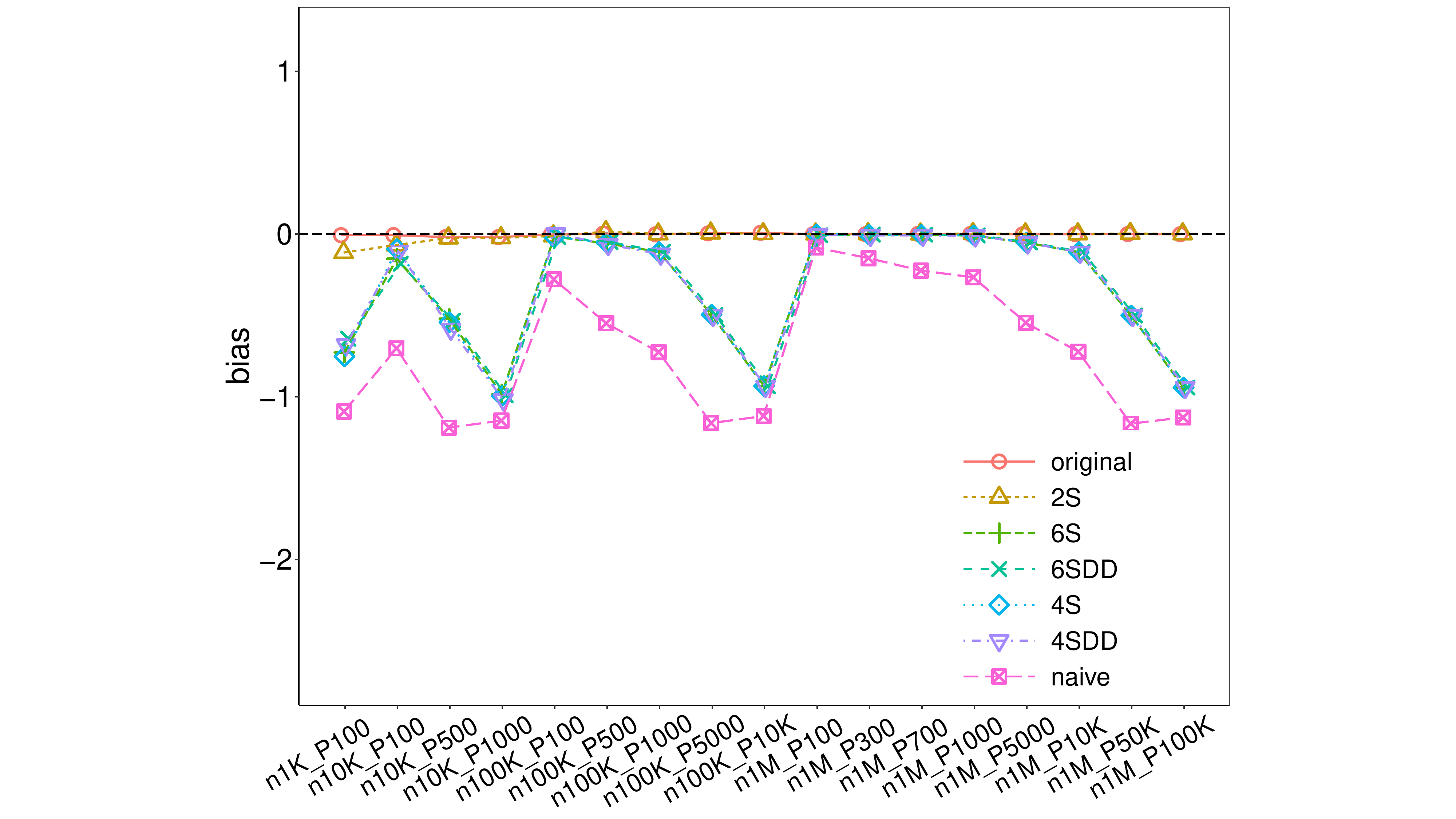}
\includegraphics[width=0.26\textwidth, trim={2.2in 0 2.2in 0},clip] {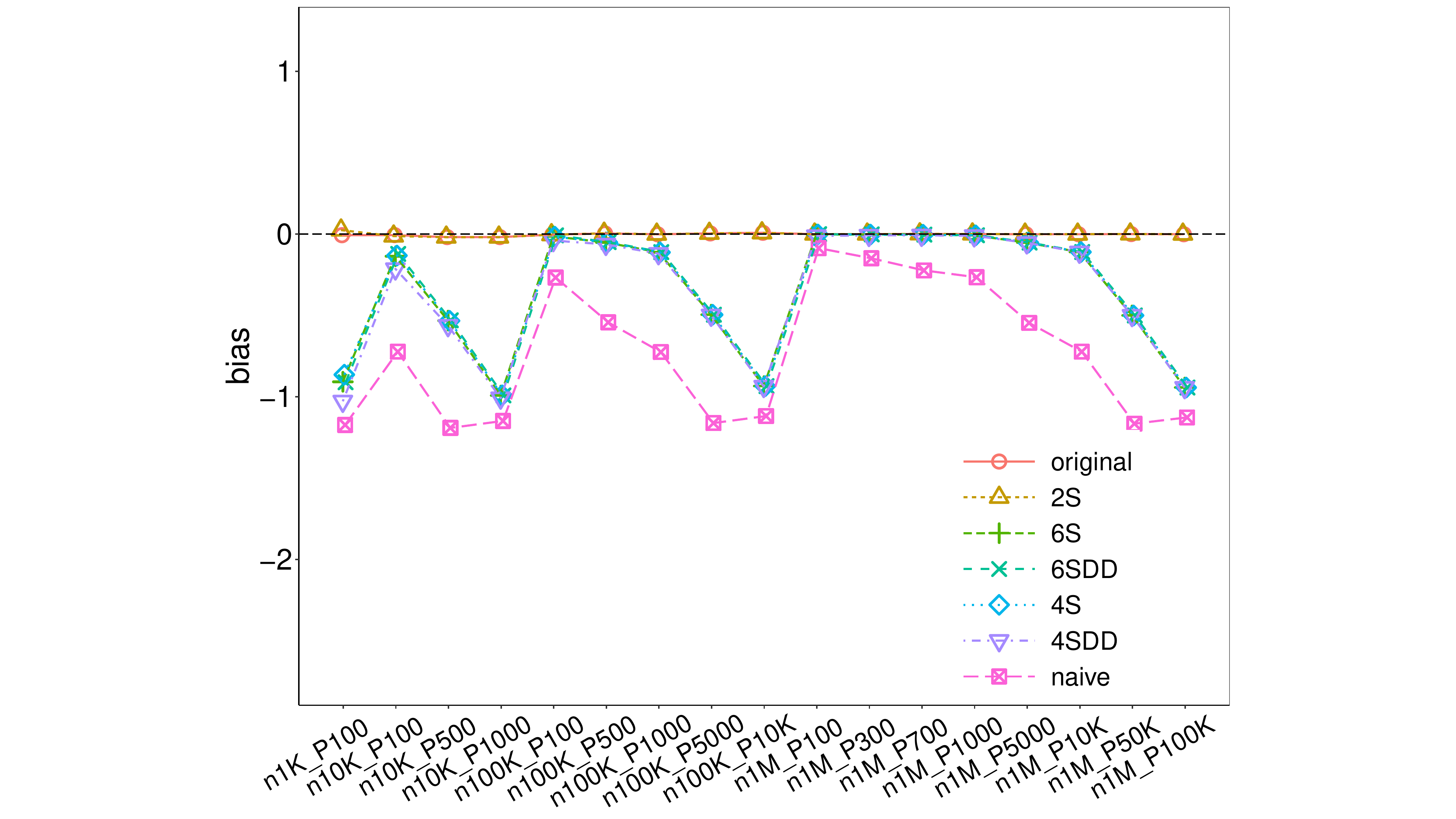}\\
\includegraphics[width=0.26\textwidth, trim={2.2in 0 2.2in 0},clip] {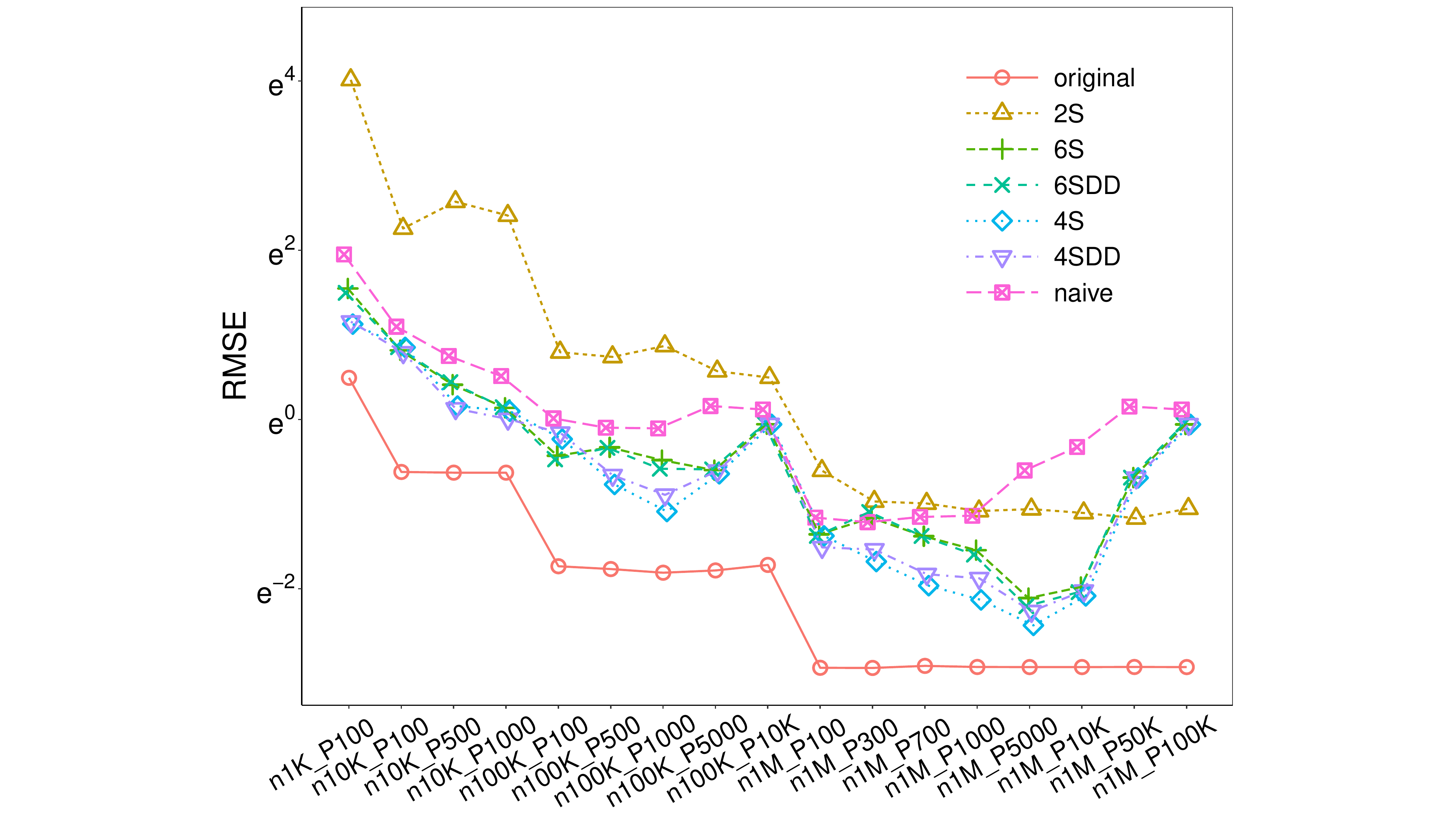}
\includegraphics[width=0.26\textwidth, trim={2.2in 0 2.2in 0},clip] {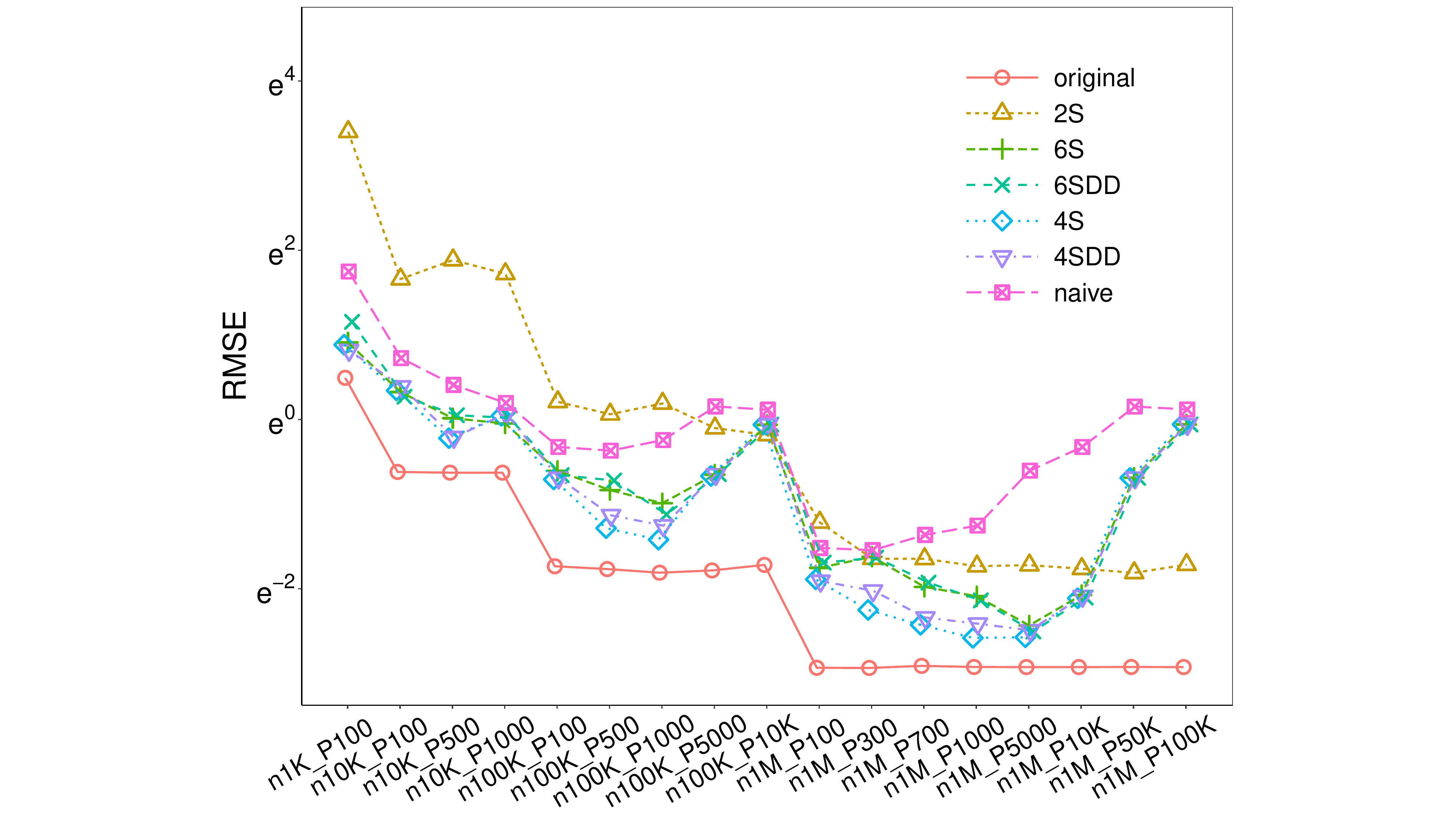}
\includegraphics[width=0.26\textwidth, trim={2.2in 0 2.2in 0},clip] {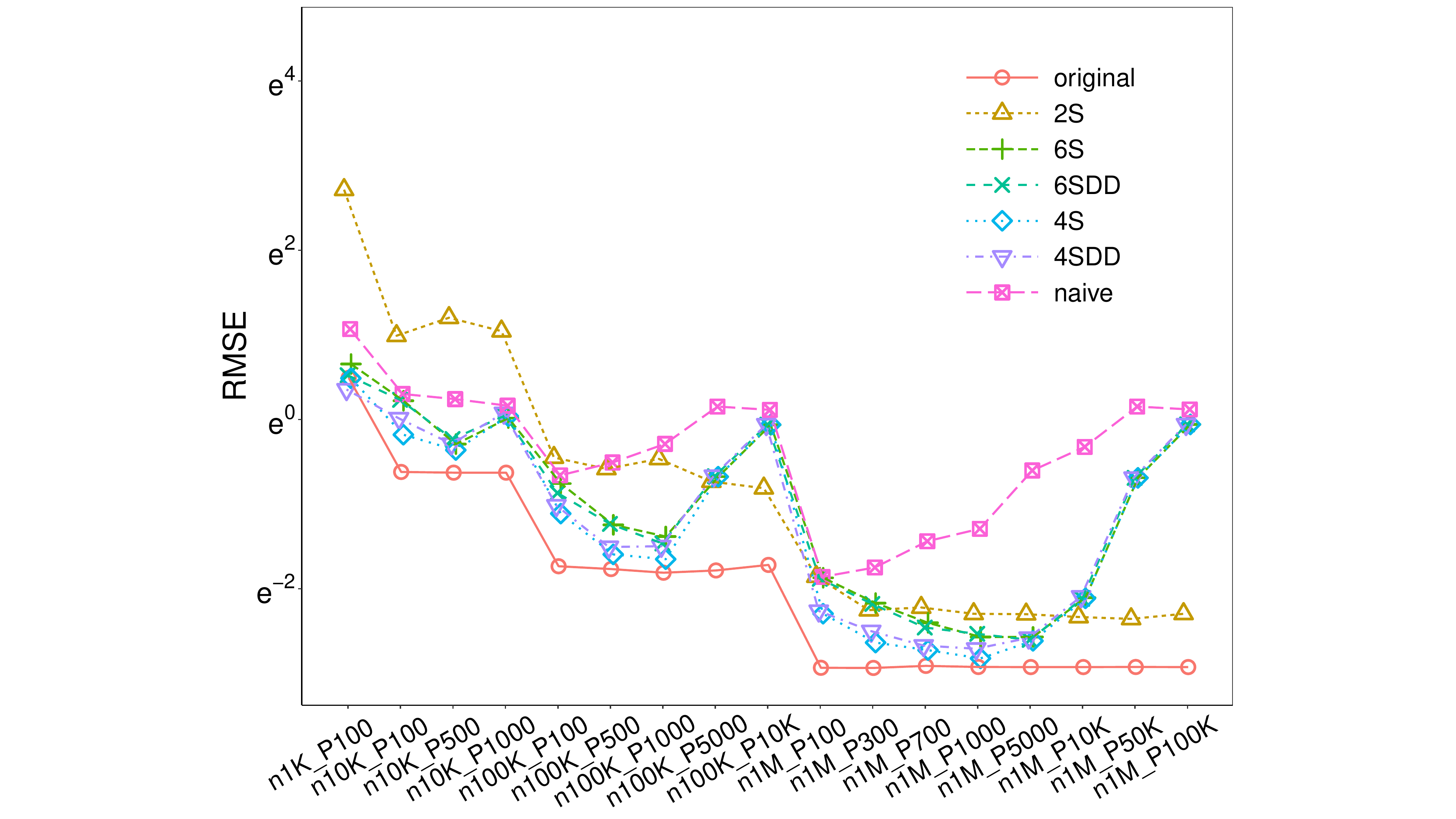}
\includegraphics[width=0.26\textwidth, trim={2.2in 0 2.2in 0},clip] {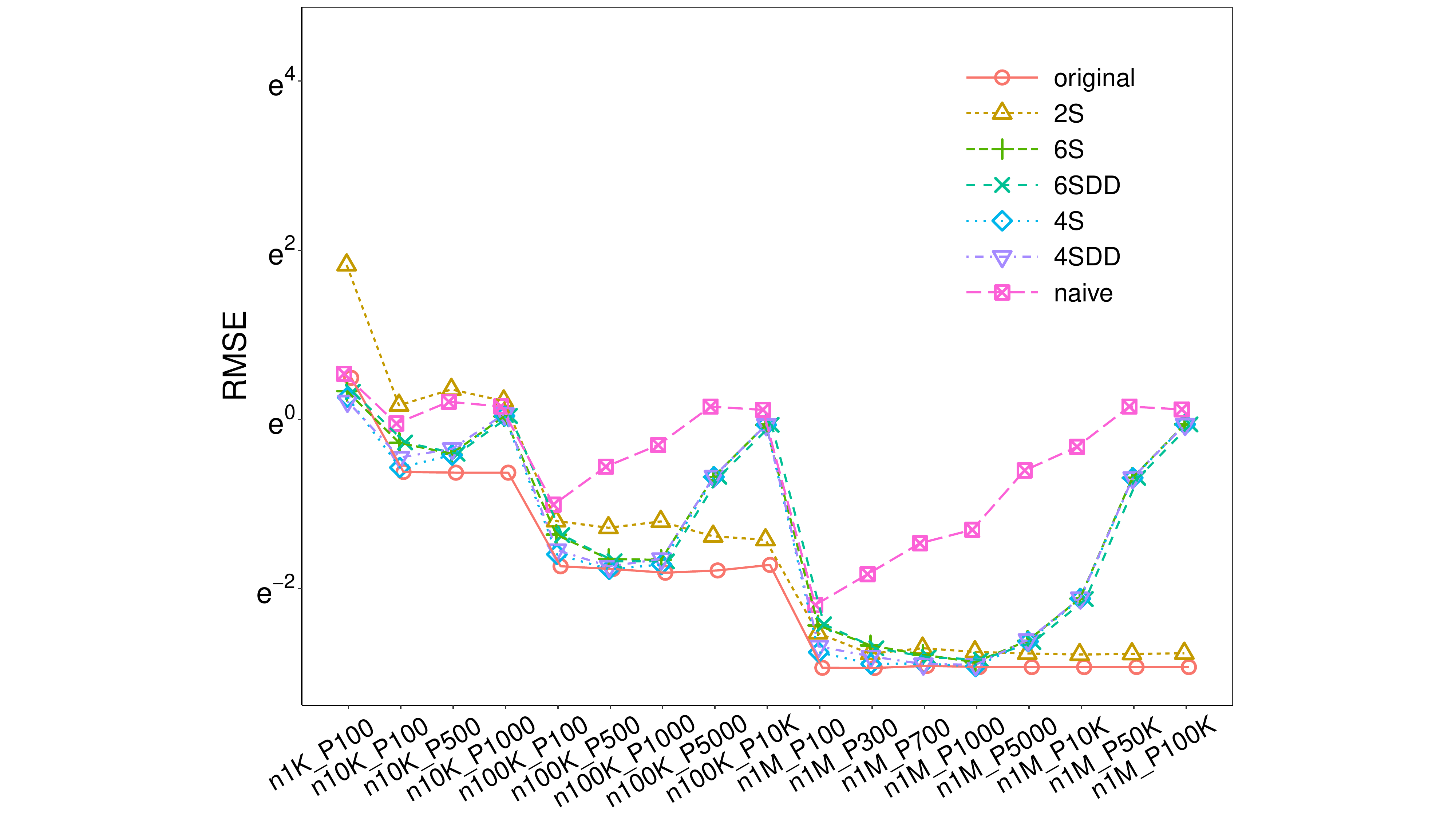}
\includegraphics[width=0.26\textwidth, trim={2.2in 0 2.2in 0},clip] {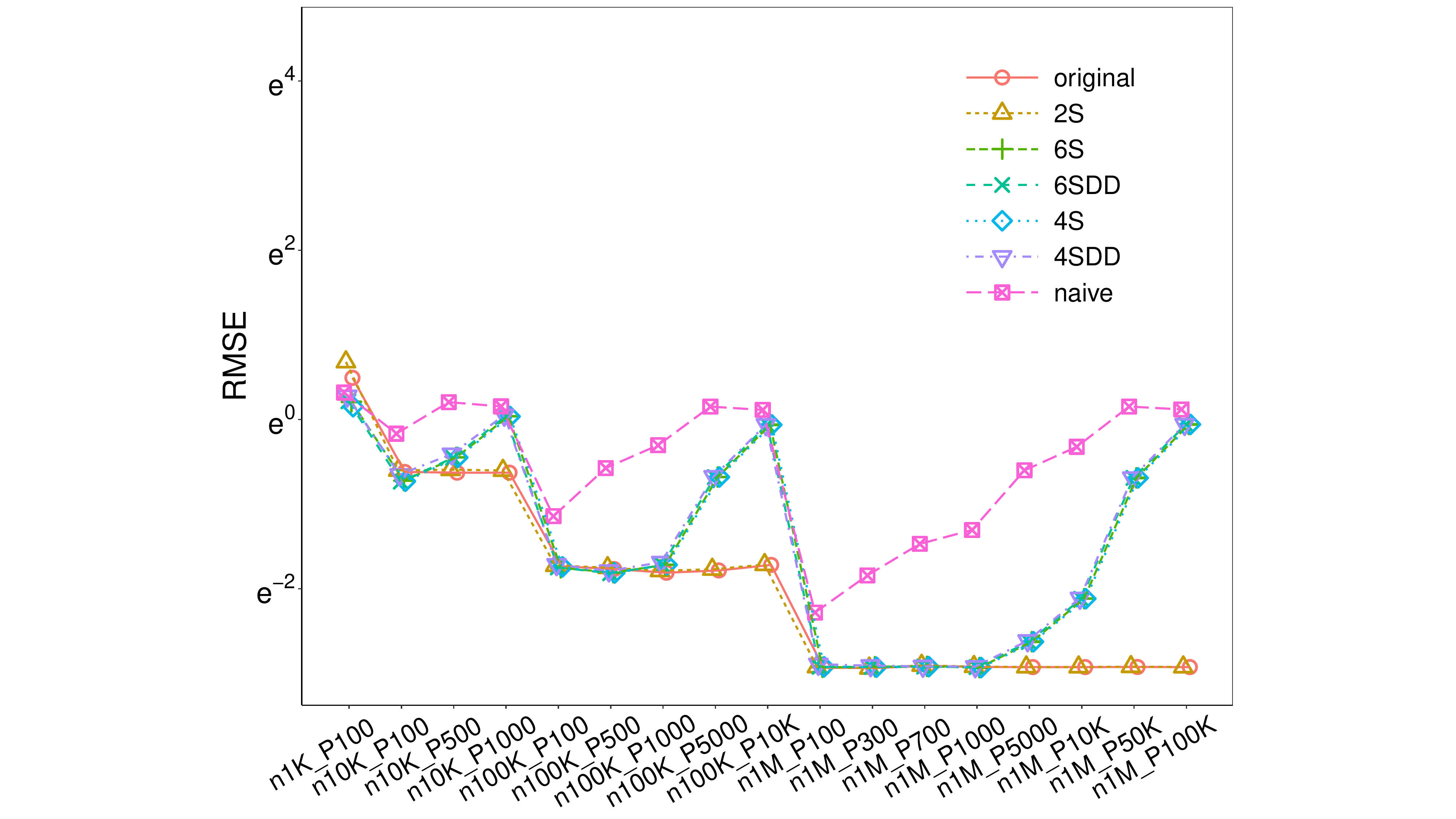}\\
\includegraphics[width=0.26\textwidth, trim={2.2in 0 2.2in 0},clip] {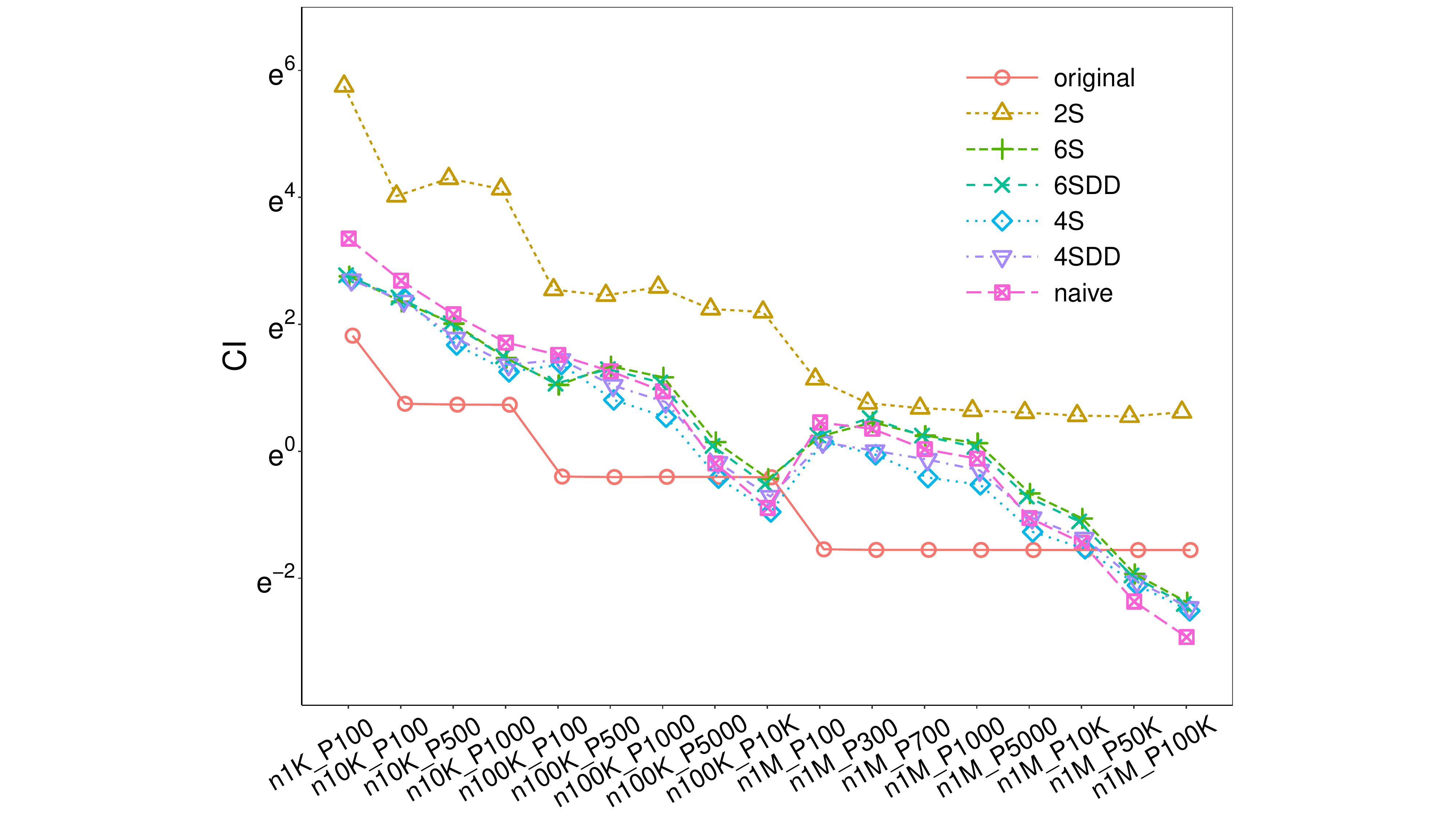}
\includegraphics[width=0.26\textwidth, trim={2.2in 0 2.2in 0},clip] {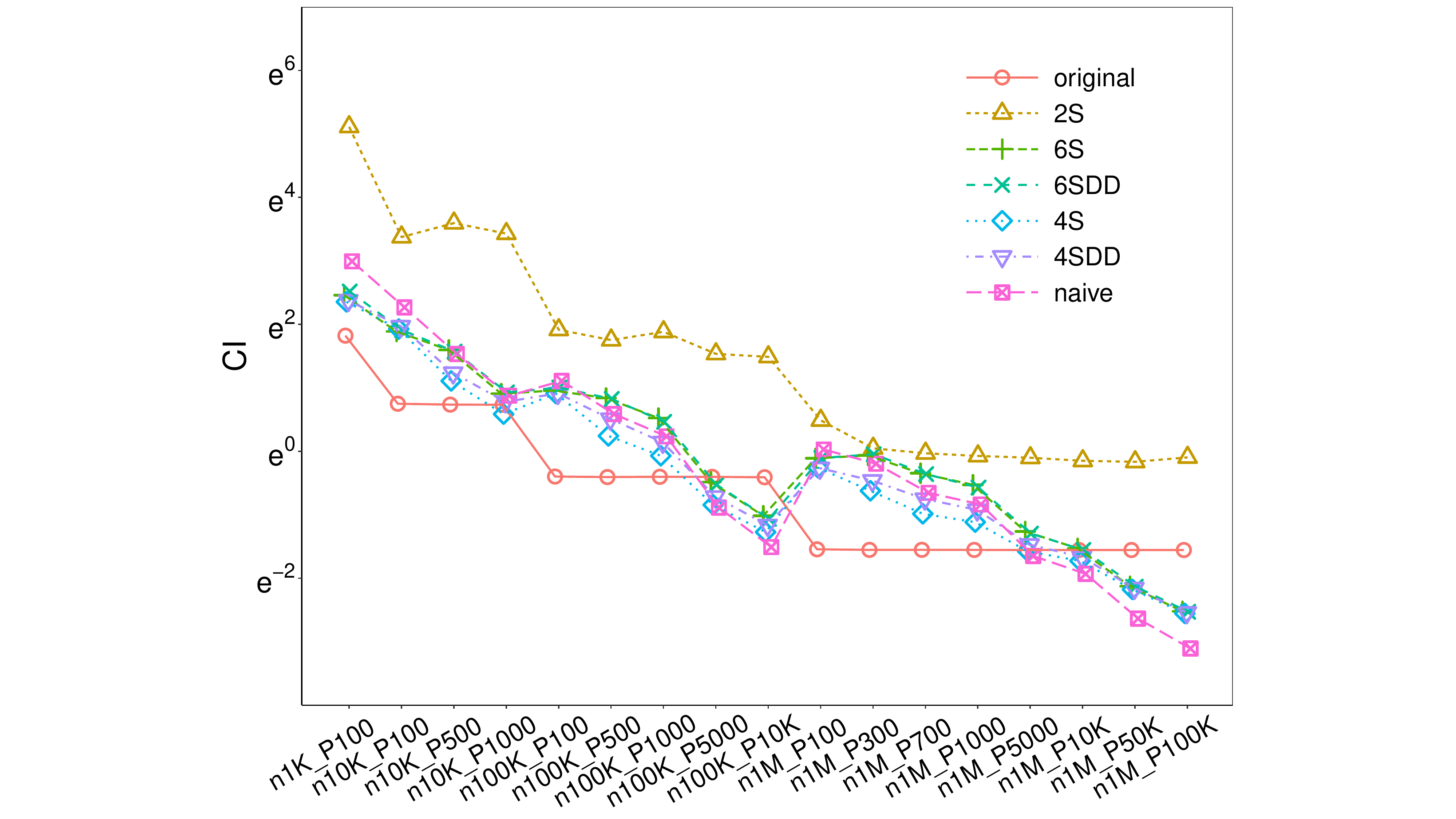}
\includegraphics[width=0.26\textwidth, trim={2.2in 0 2.2in 0},clip] {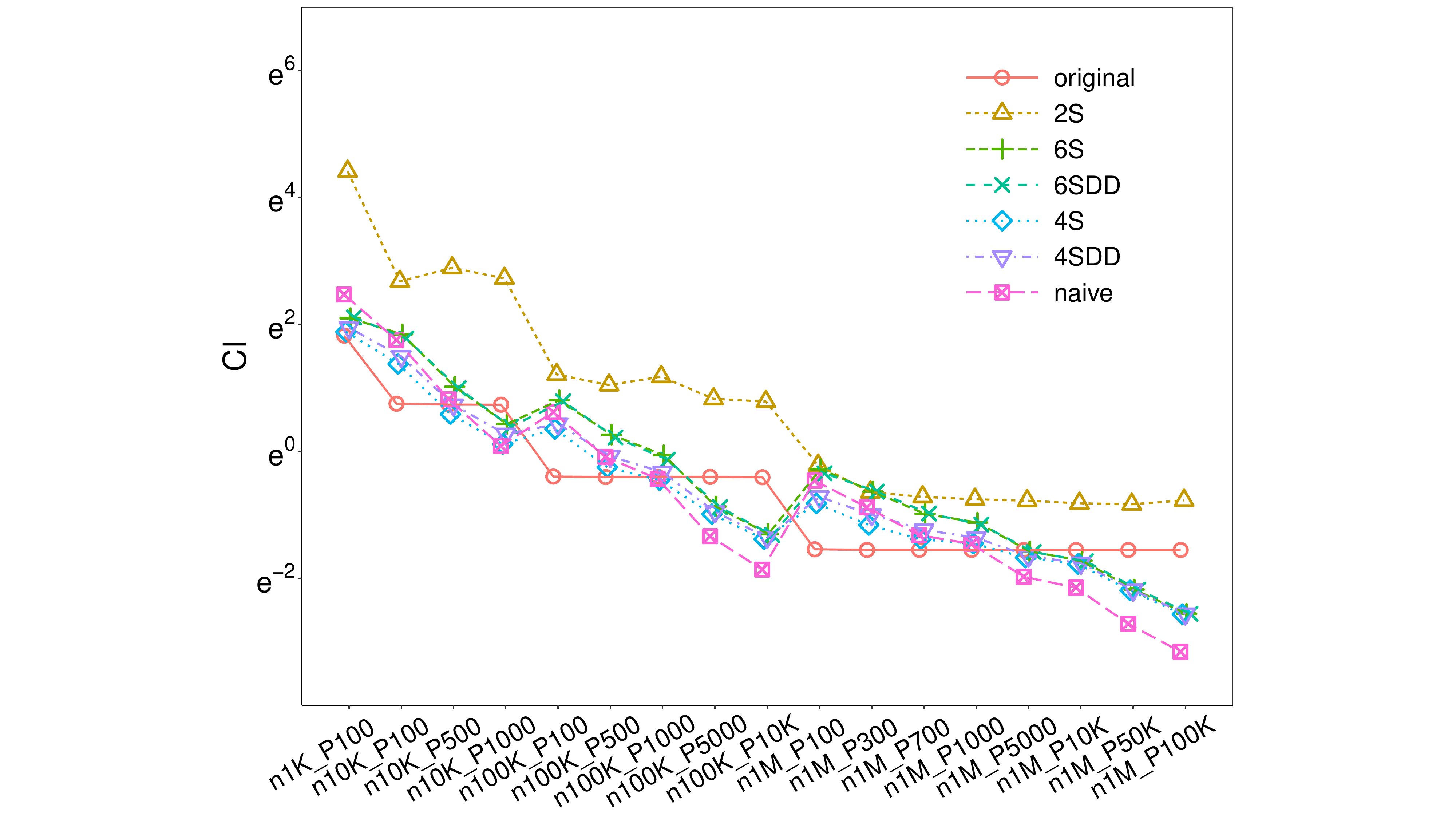}
\includegraphics[width=0.26\textwidth, trim={2.2in 0 2.2in 0},clip] {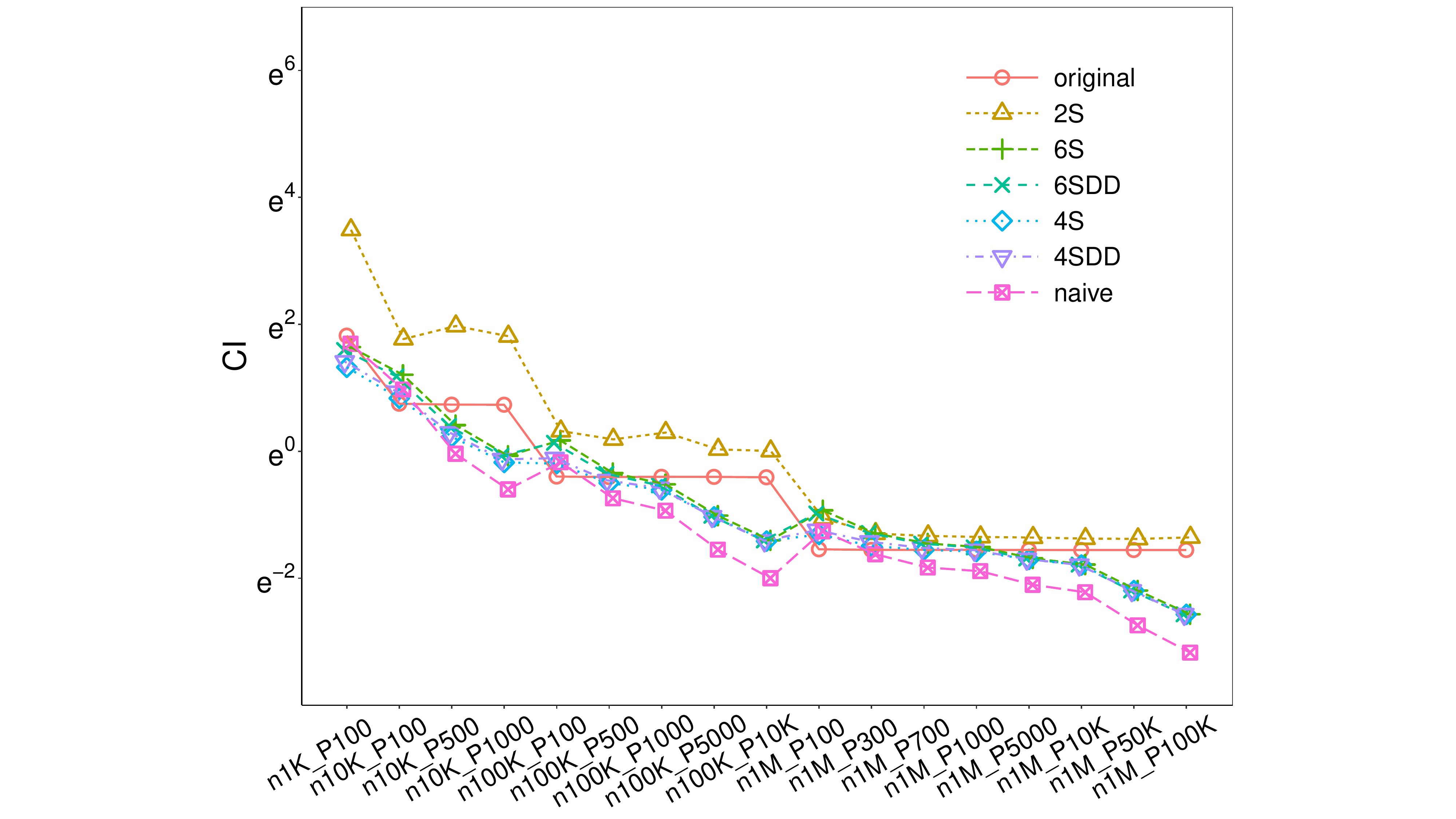}
\includegraphics[width=0.26\textwidth, trim={2.2in 0 2.2in 0},clip] {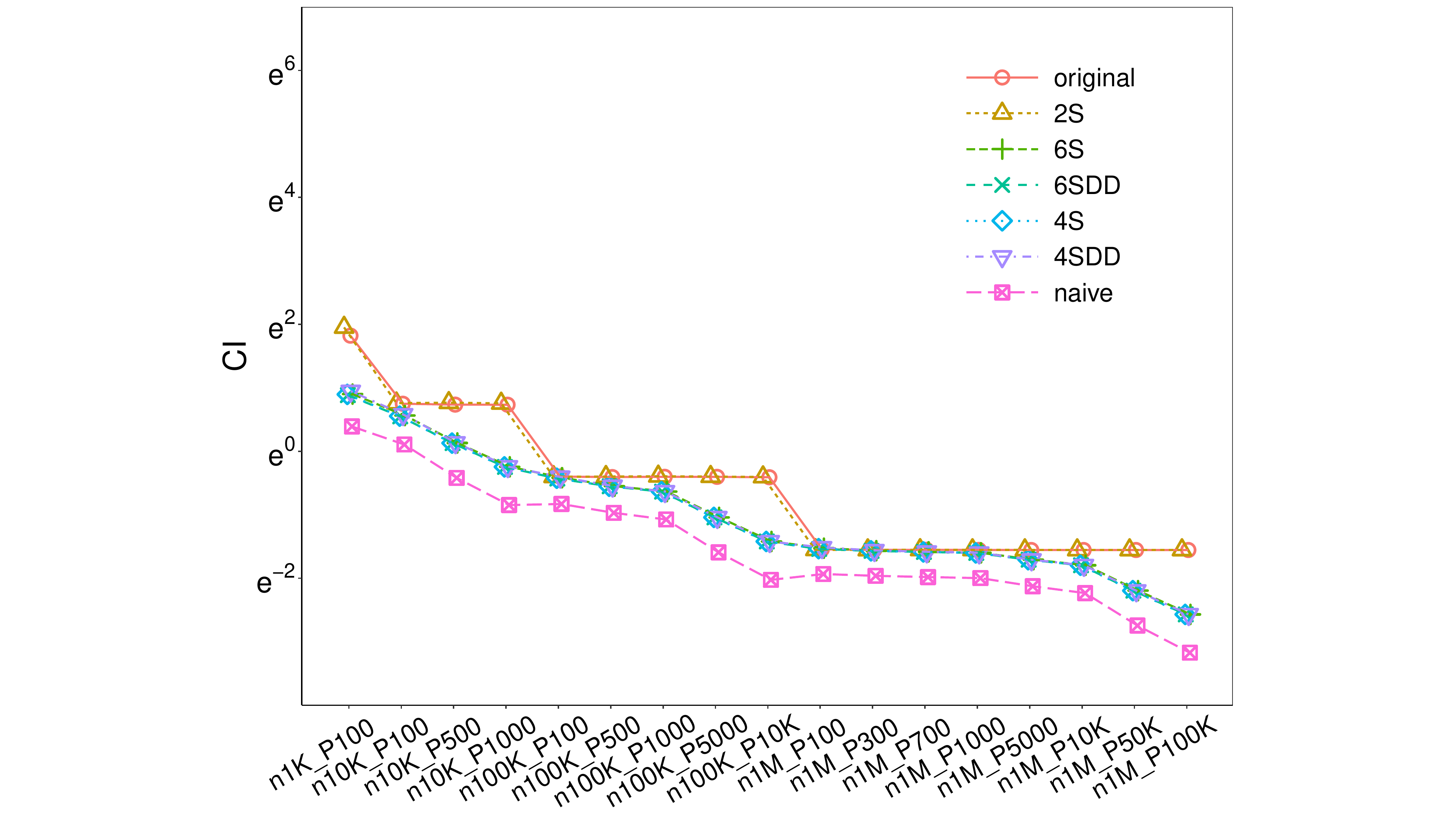}\\
\includegraphics[width=0.26\textwidth, trim={2.2in 0 2.2in 0},clip] {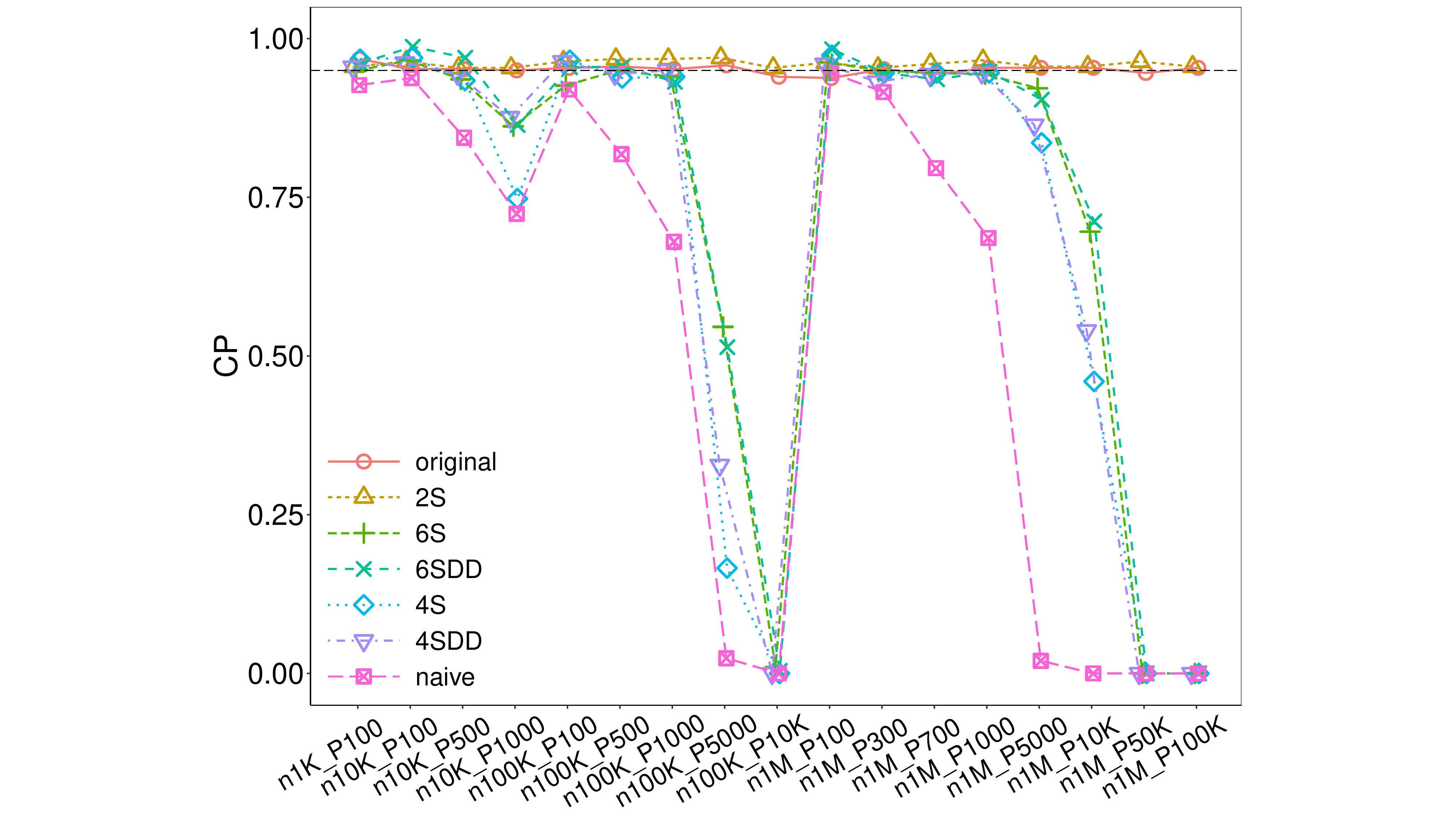}
\includegraphics[width=0.26\textwidth, trim={2.2in 0 2.2in 0},clip] {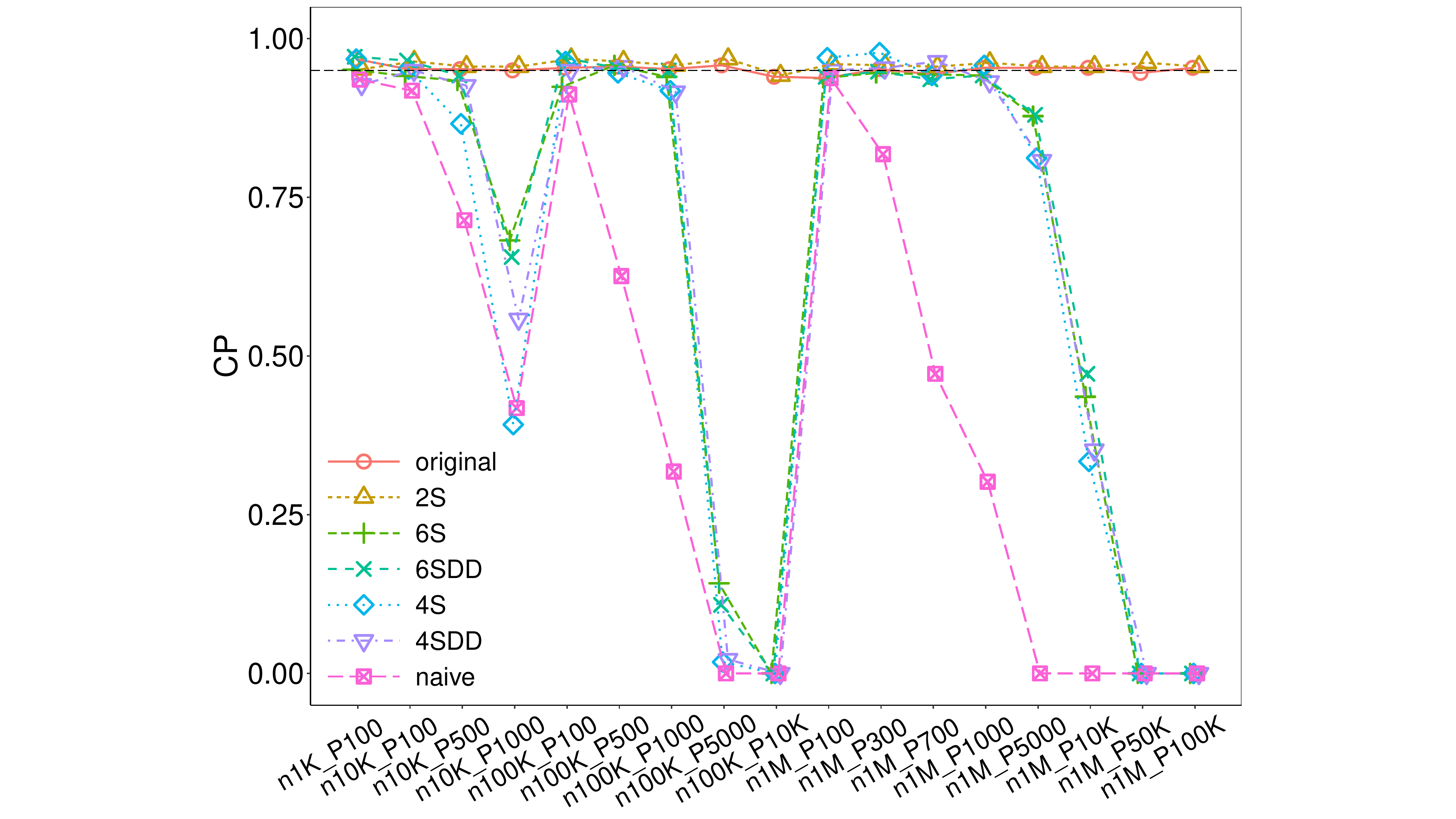}
\includegraphics[width=0.26\textwidth, trim={2.2in 0 2.2in 0},clip] {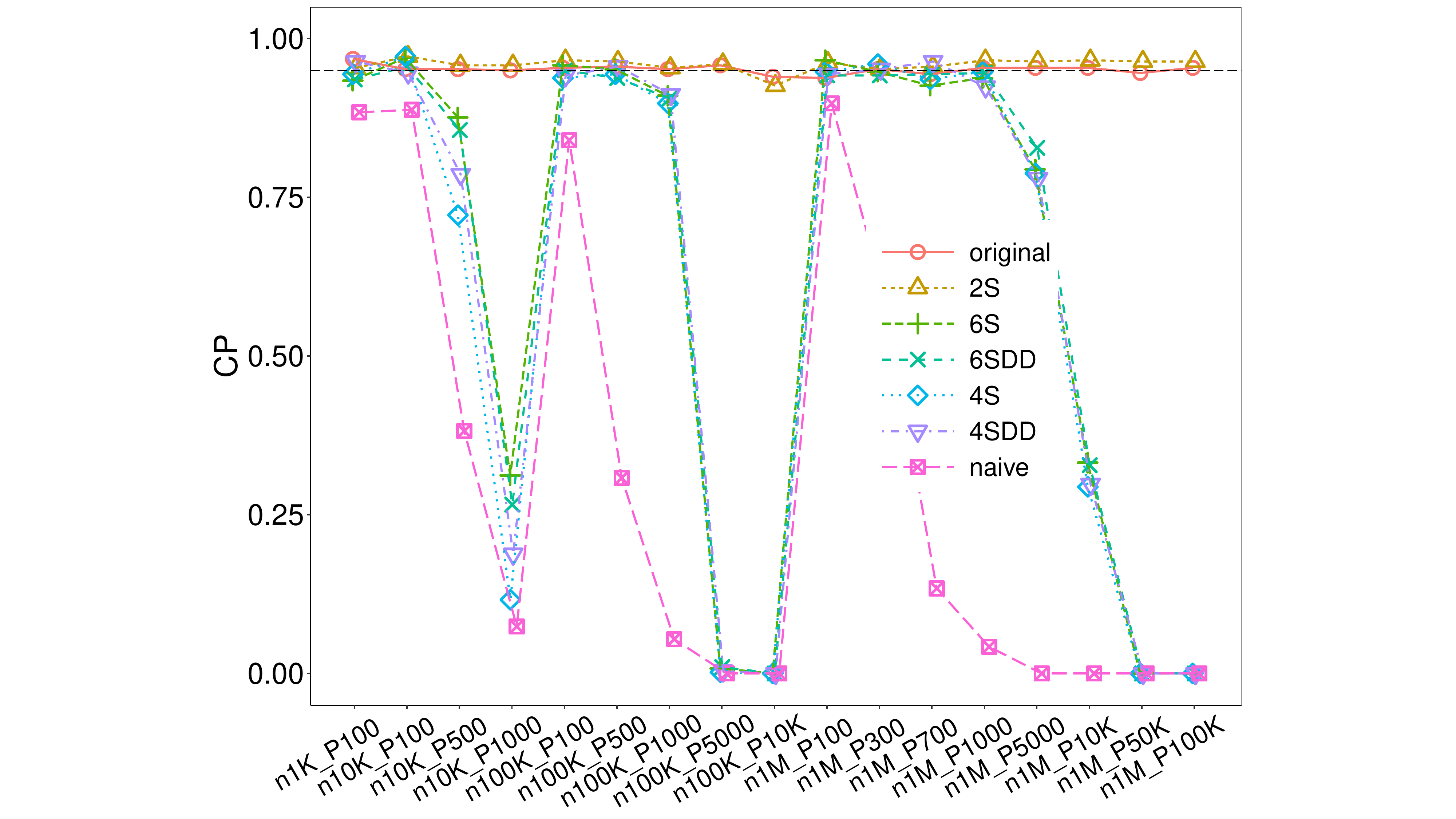}
\includegraphics[width=0.26\textwidth, trim={2.2in 0 2.2in 0},clip] {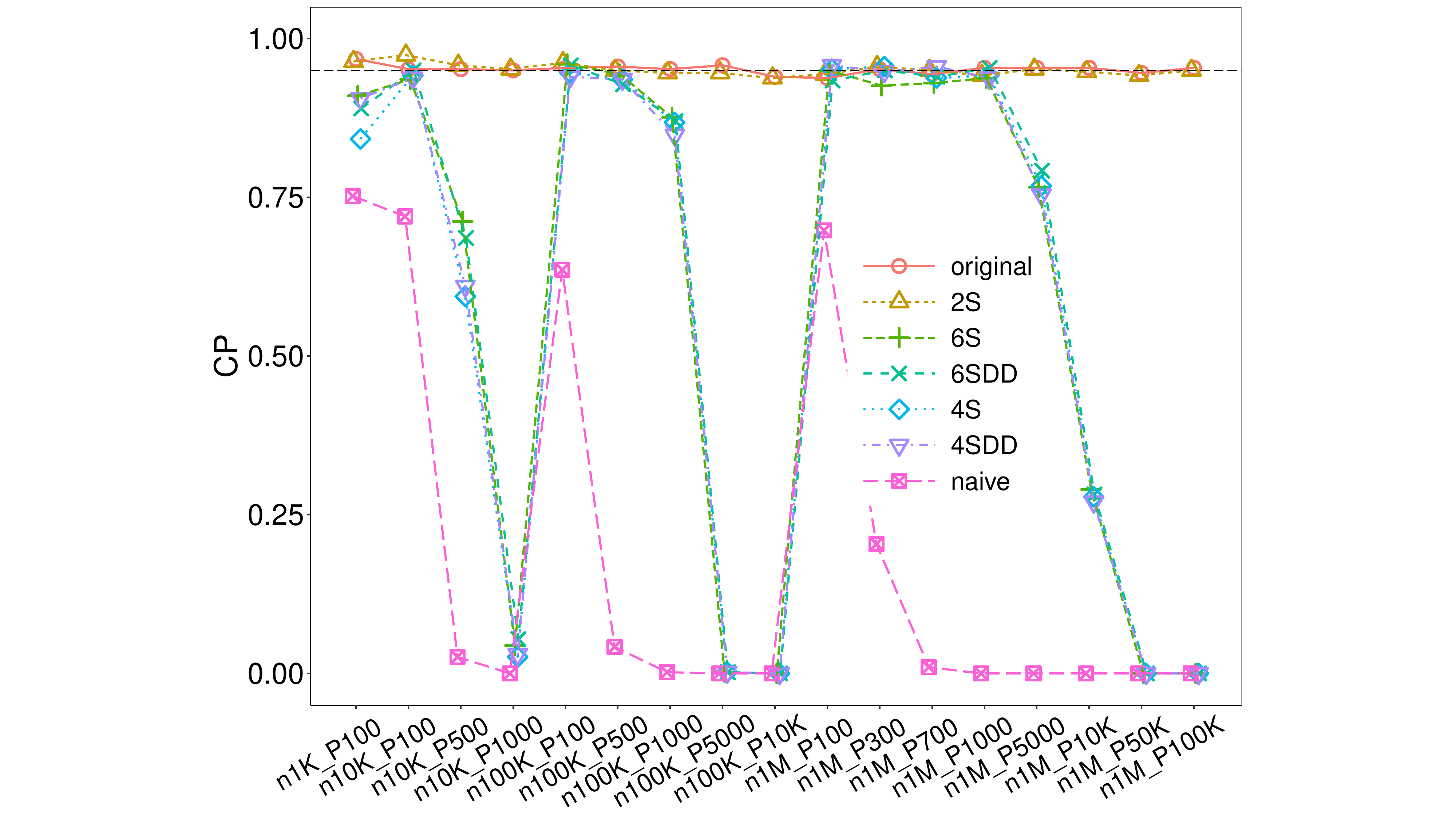}
\includegraphics[width=0.26\textwidth, trim={2.2in 0 2.2in 0},clip] {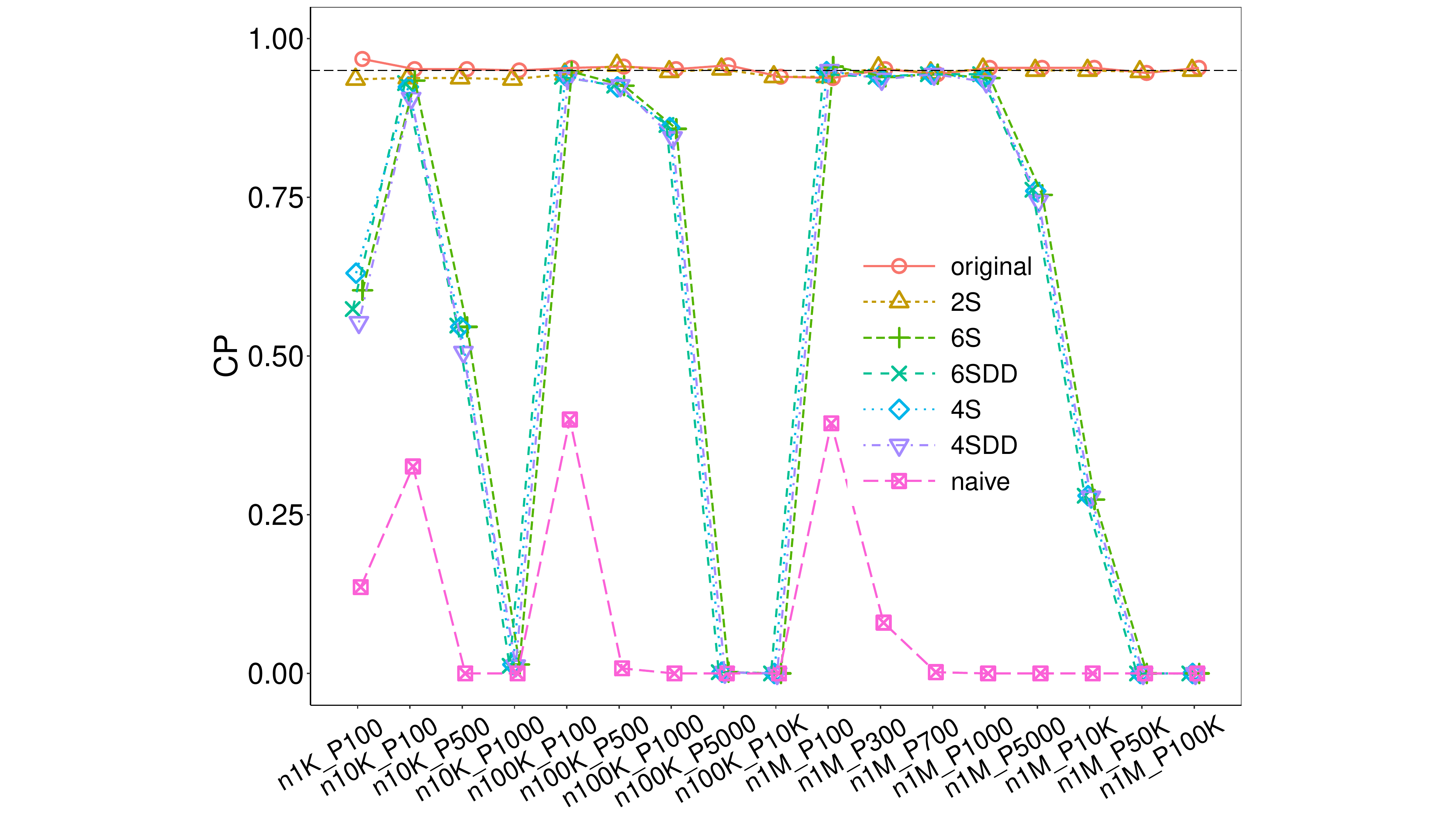}\\
\caption{ZILN data; $\epsilon$-DP; $\theta=0$ and $\alpha\ne\beta$} \label{fig:0asDPziln}
\end{figure}
\end{landscape}

\begin{landscape}
\begin{figure}[!htb]
\hspace{0.6in}$\rho=0.005$\hspace{1in}$\rho=0.02$\hspace{1.2in}$\rho=0.08$
\hspace{1.1in}$\rho=0.32$\hspace{1.2in}$\rho=1.28$\\
\includegraphics[width=0.26\textwidth, trim={2.2in 0 2.2in 0},clip] {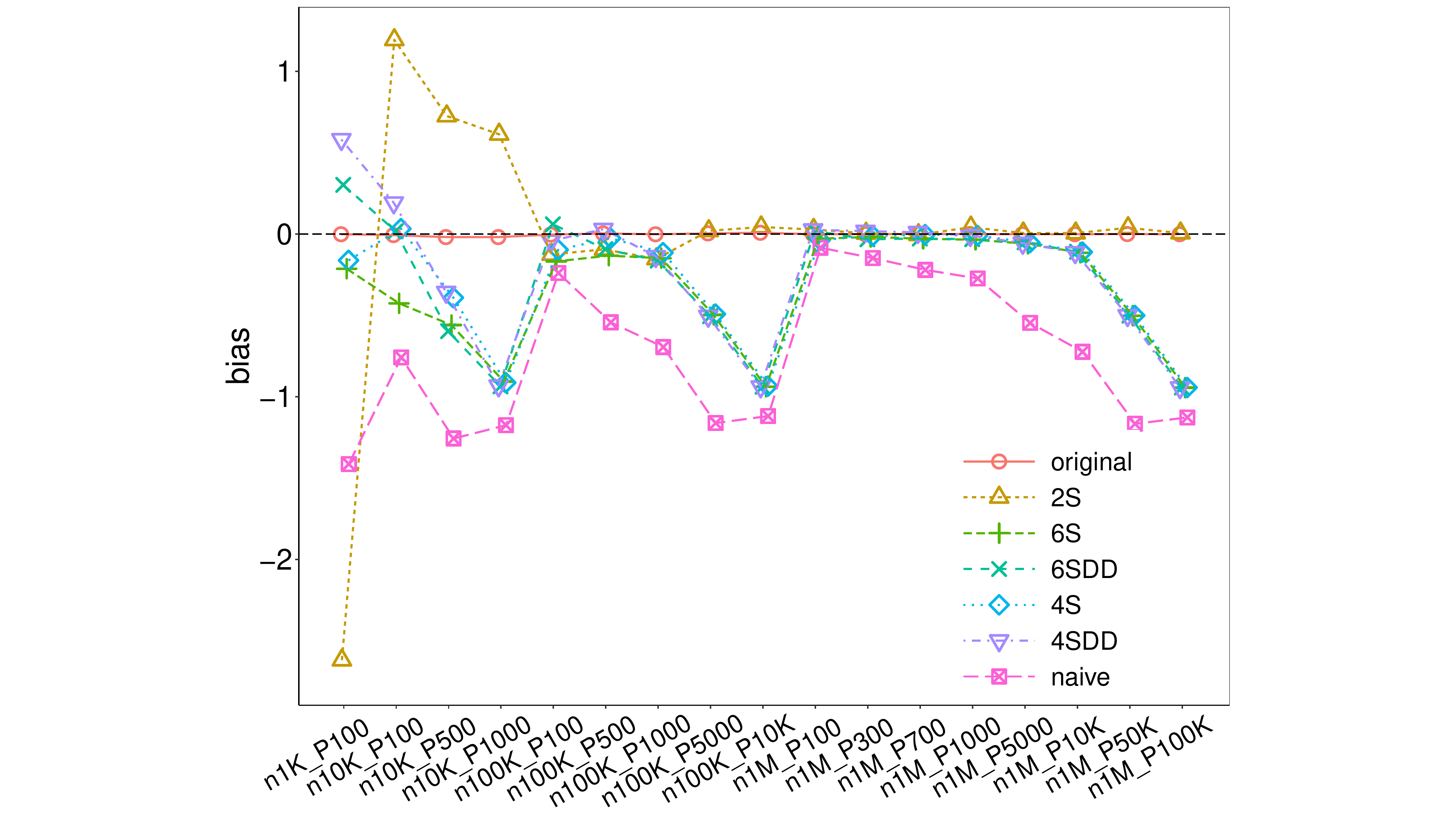}
\includegraphics[width=0.26\textwidth, trim={2.2in 0 2.2in 0},clip] {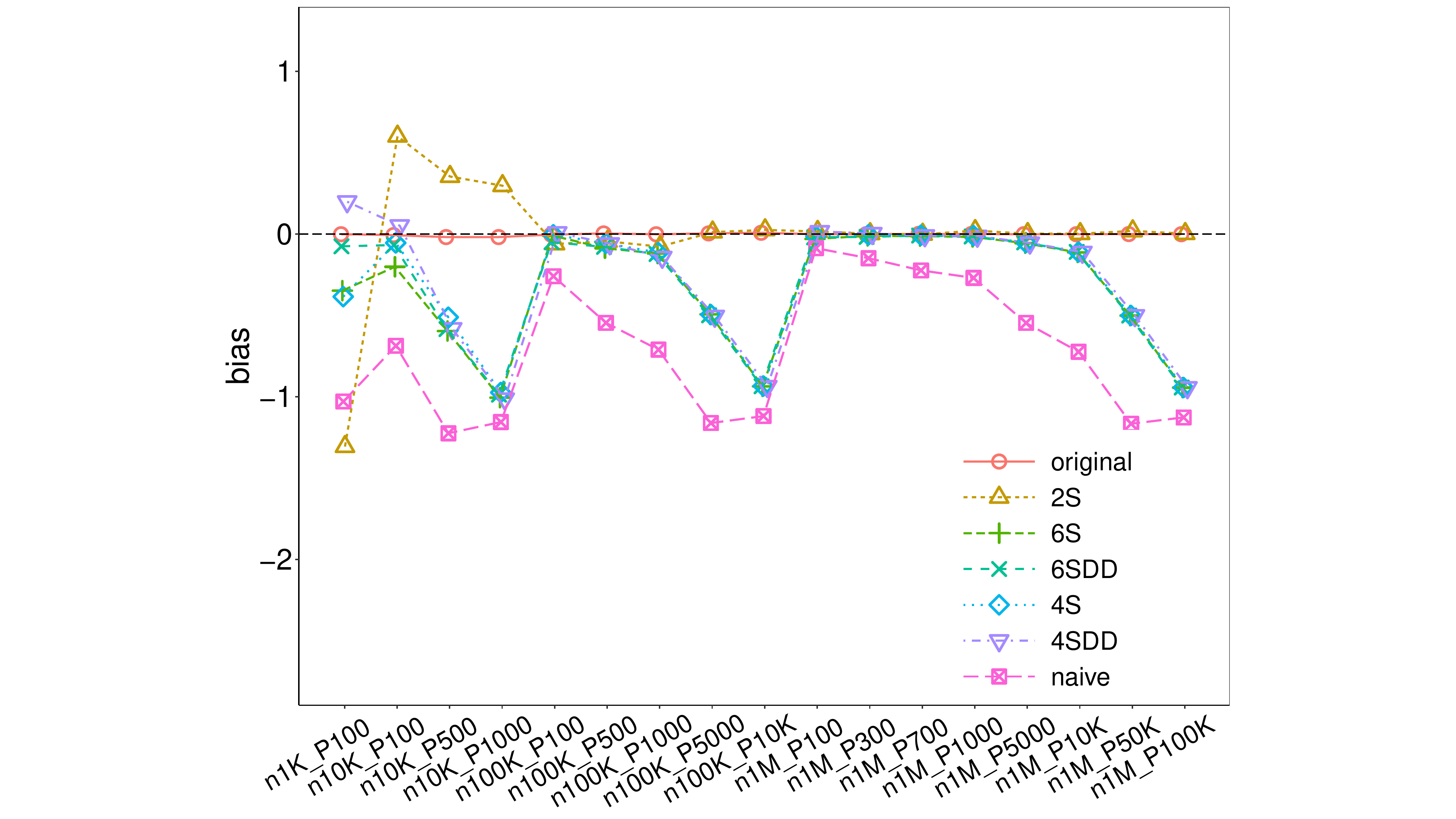}
\includegraphics[width=0.26\textwidth, trim={2.2in 0 2.2in 0},clip] {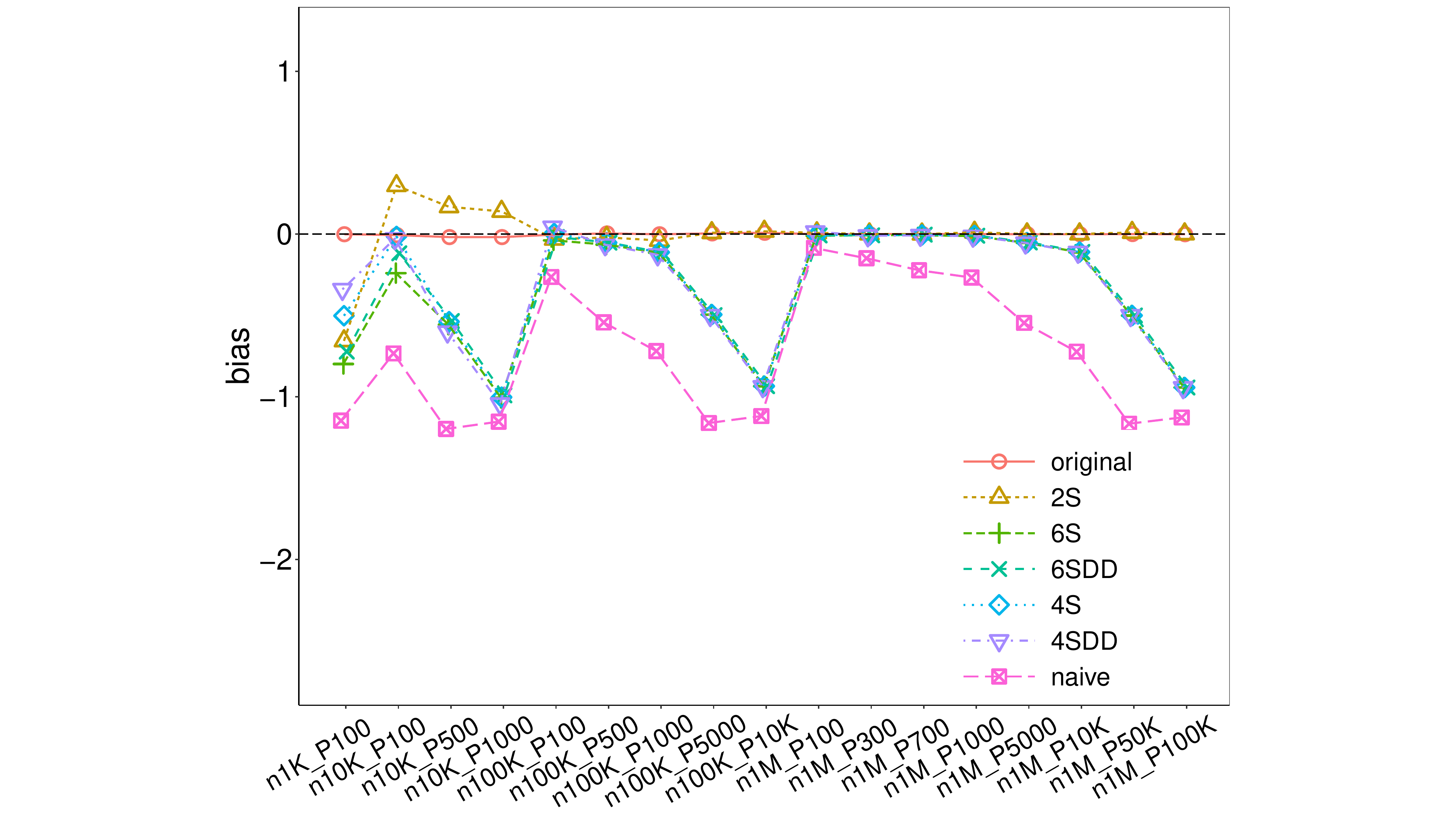}
\includegraphics[width=0.26\textwidth, trim={2.2in 0 2.2in 0},clip] {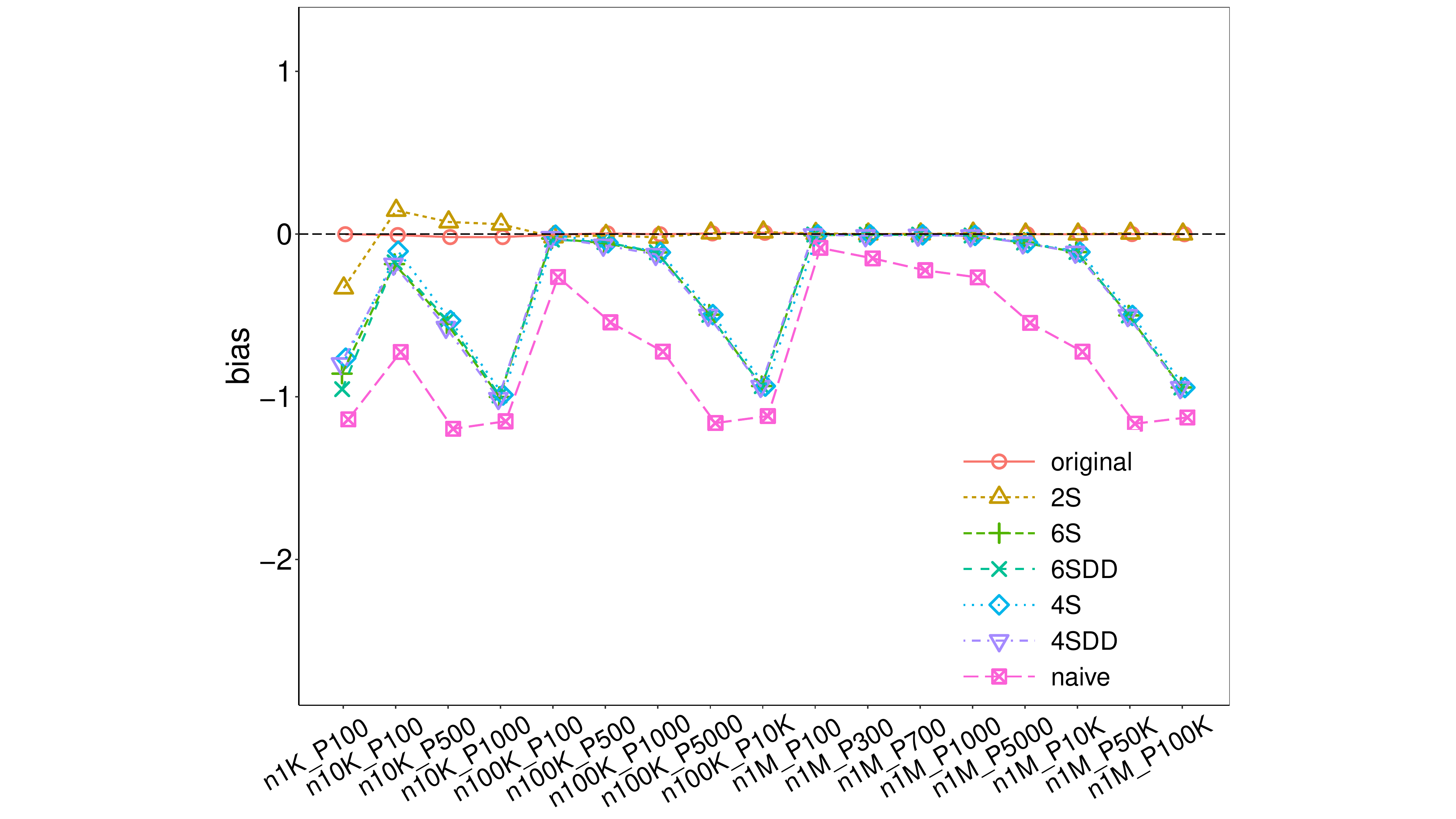}
\includegraphics[width=0.26\textwidth, trim={2.2in 0 2.2in 0},clip] {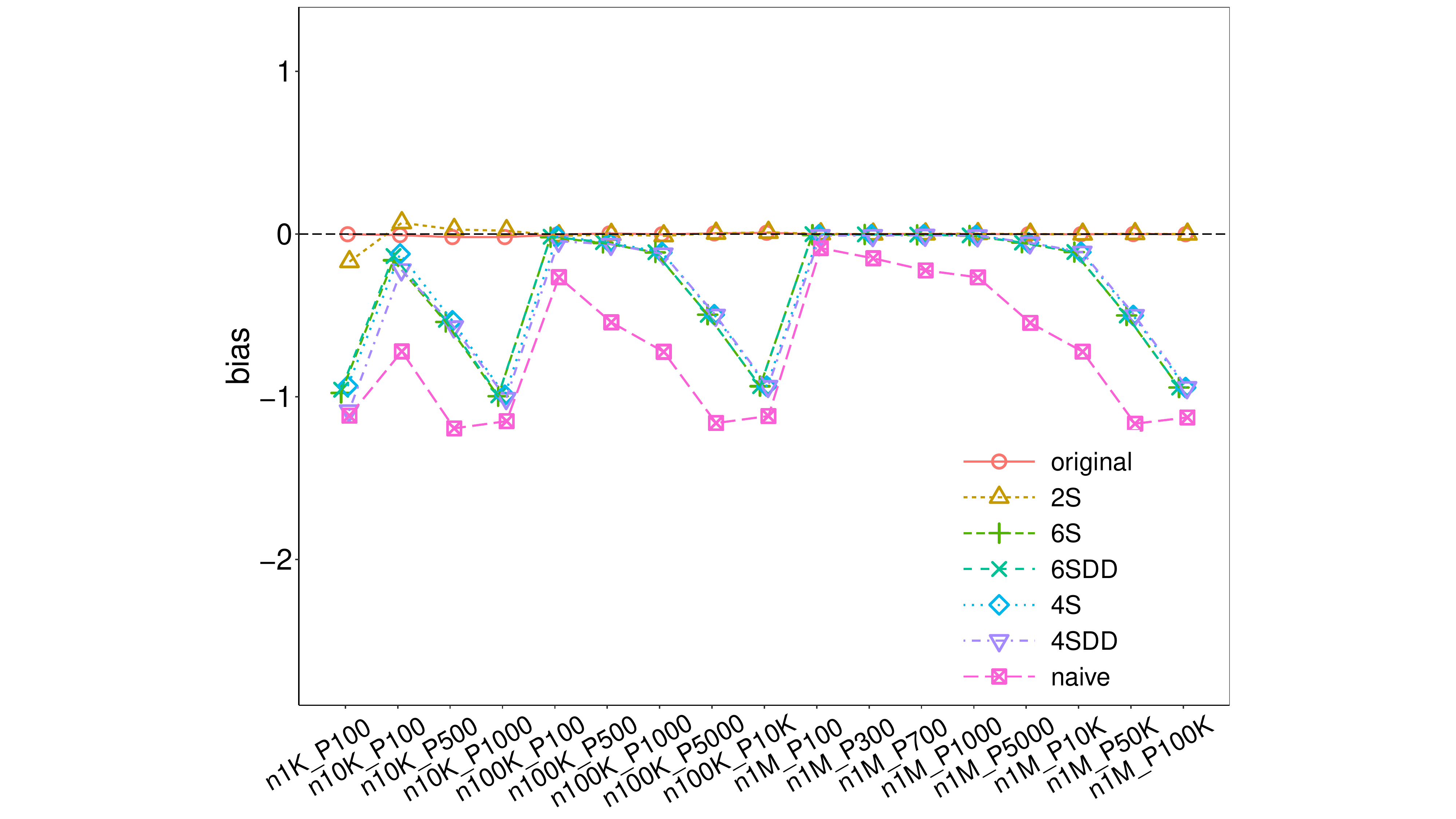}\\
\includegraphics[width=0.26\textwidth, trim={2.2in 0 2.2in 0},clip] {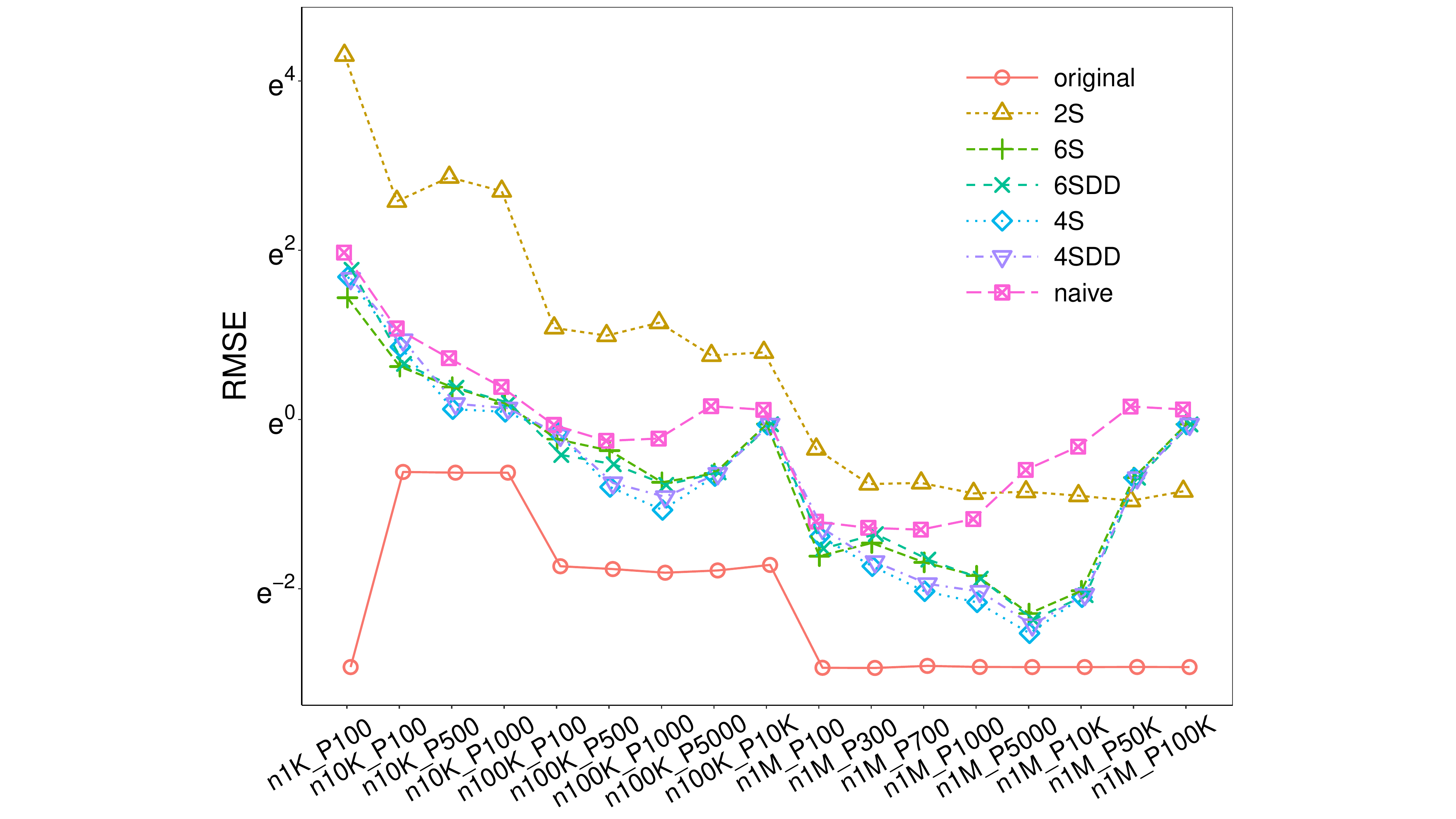}
\includegraphics[width=0.26\textwidth, trim={2.2in 0 2.2in 0},clip] {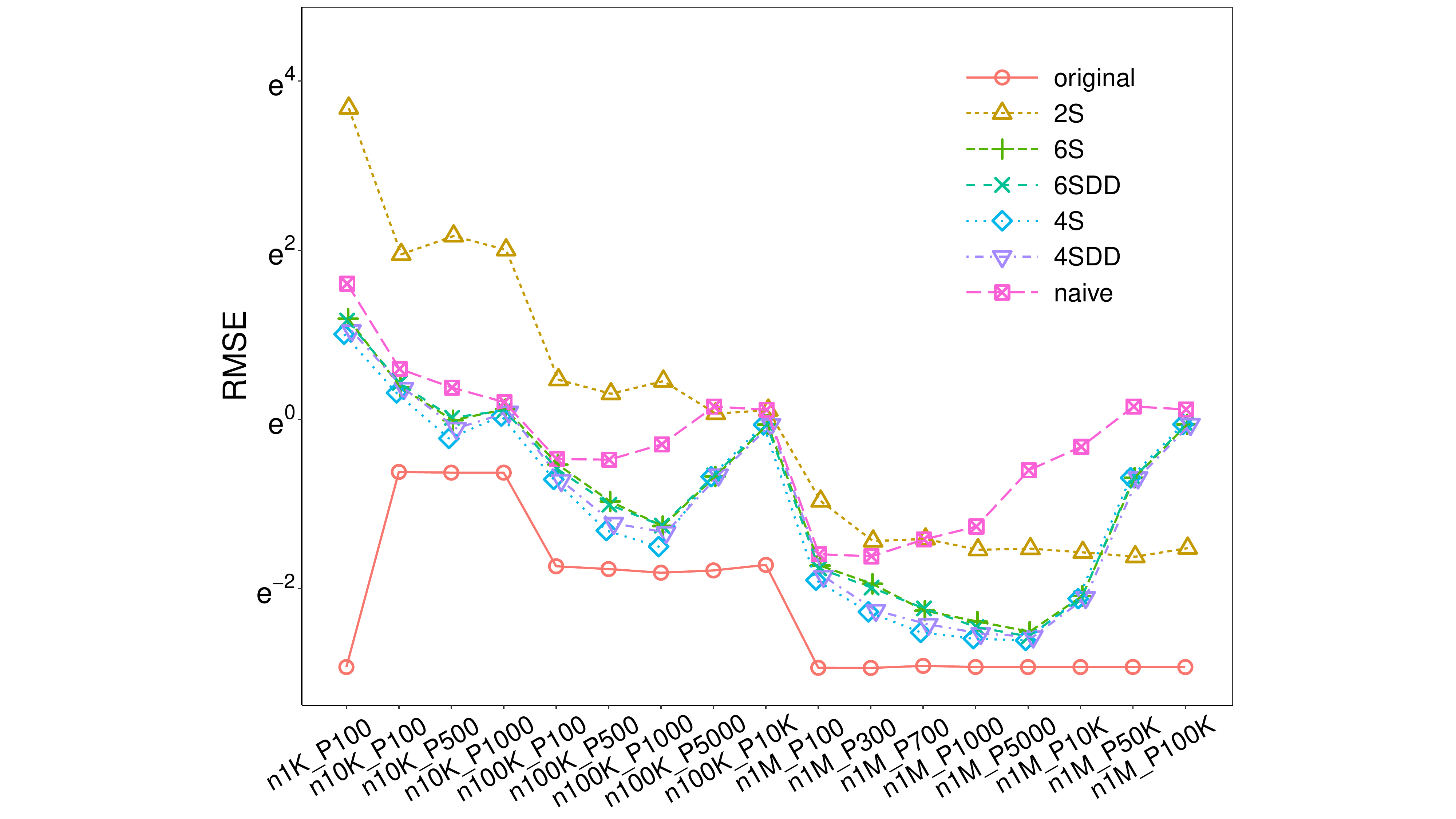}
\includegraphics[width=0.26\textwidth, trim={2.2in 0 2.2in 0},clip] {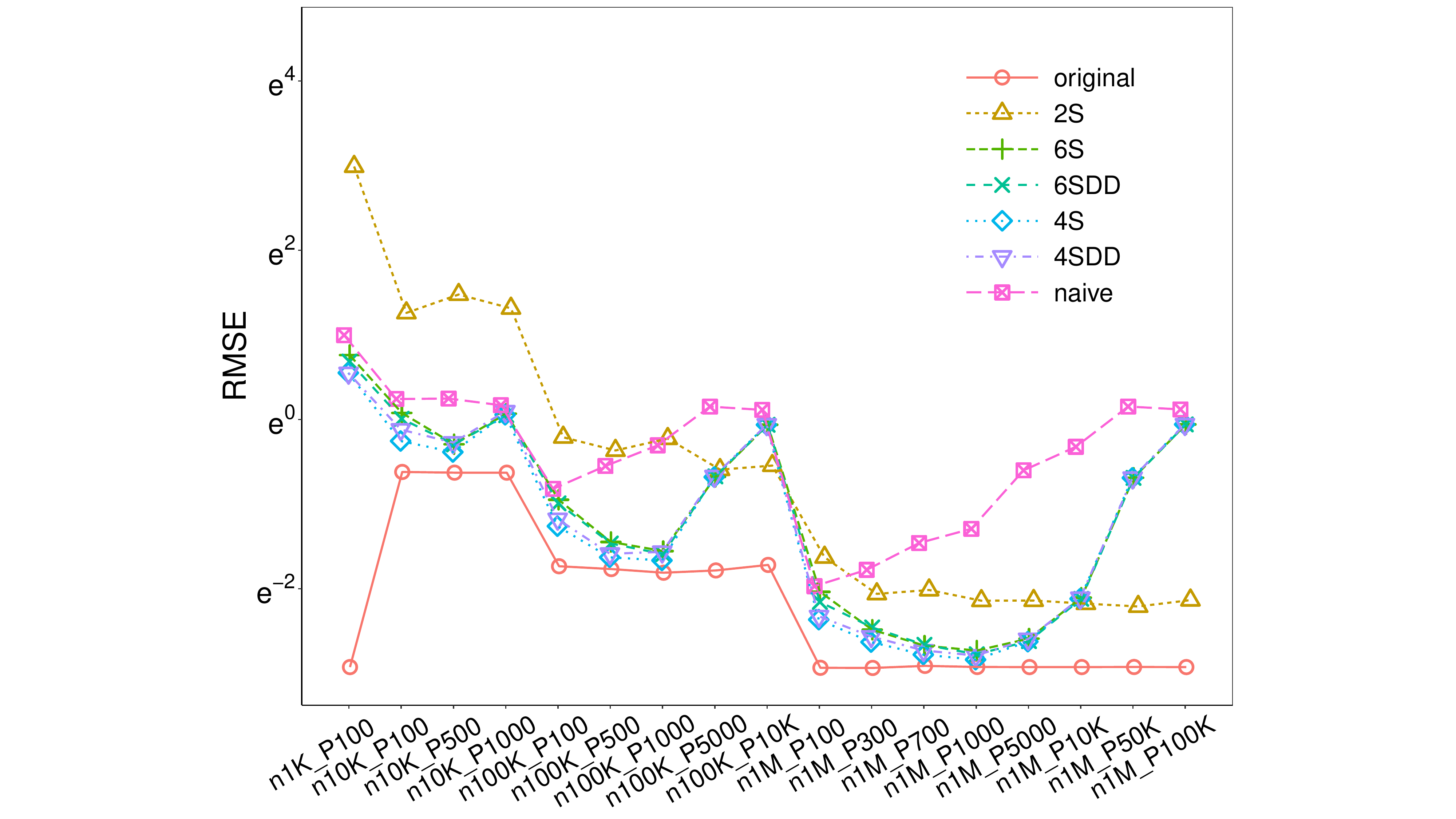}
\includegraphics[width=0.26\textwidth, trim={2.2in 0 2.2in 0},clip] {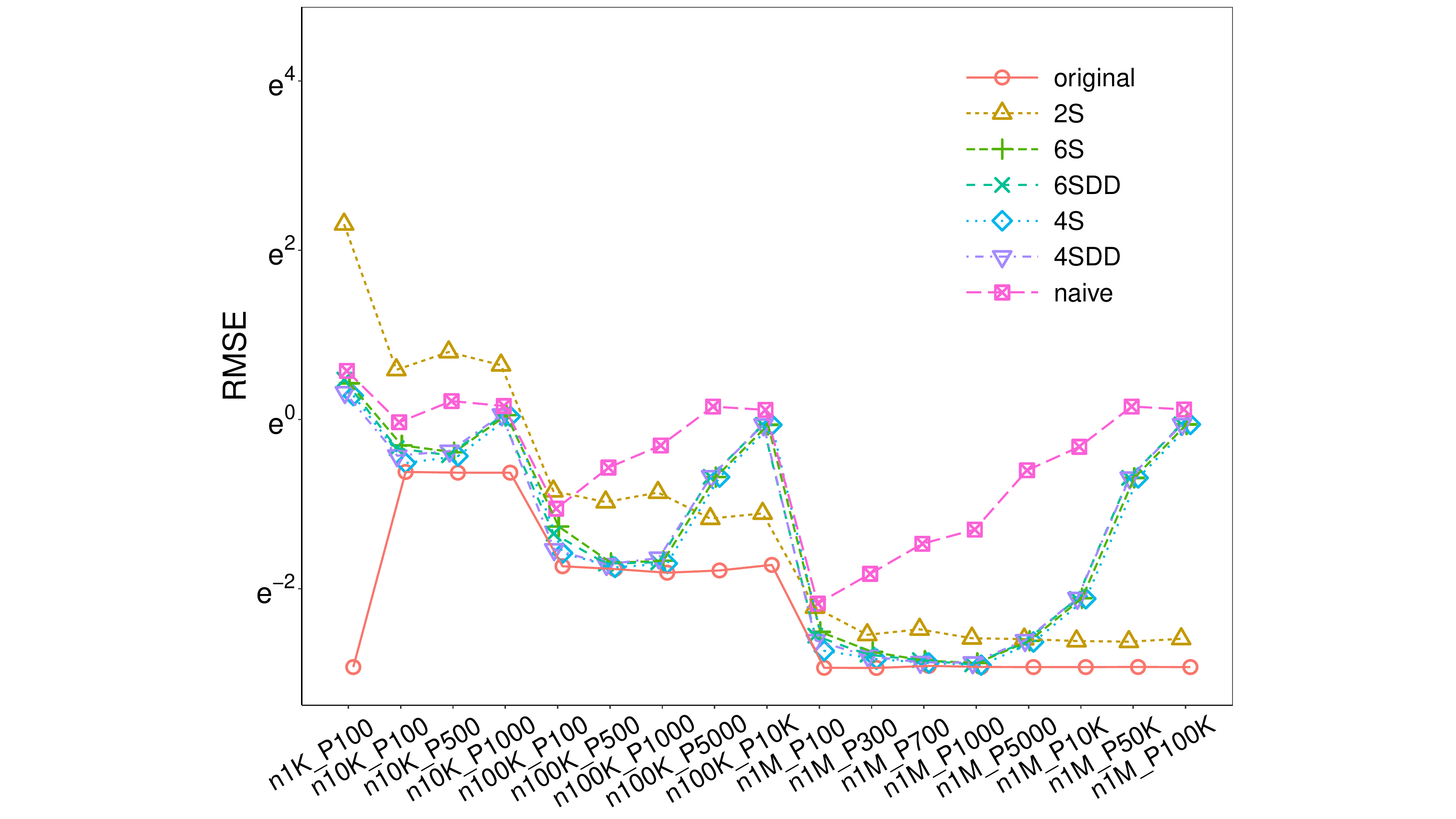}
\includegraphics[width=0.26\textwidth, trim={2.2in 0 2.2in 0},clip] {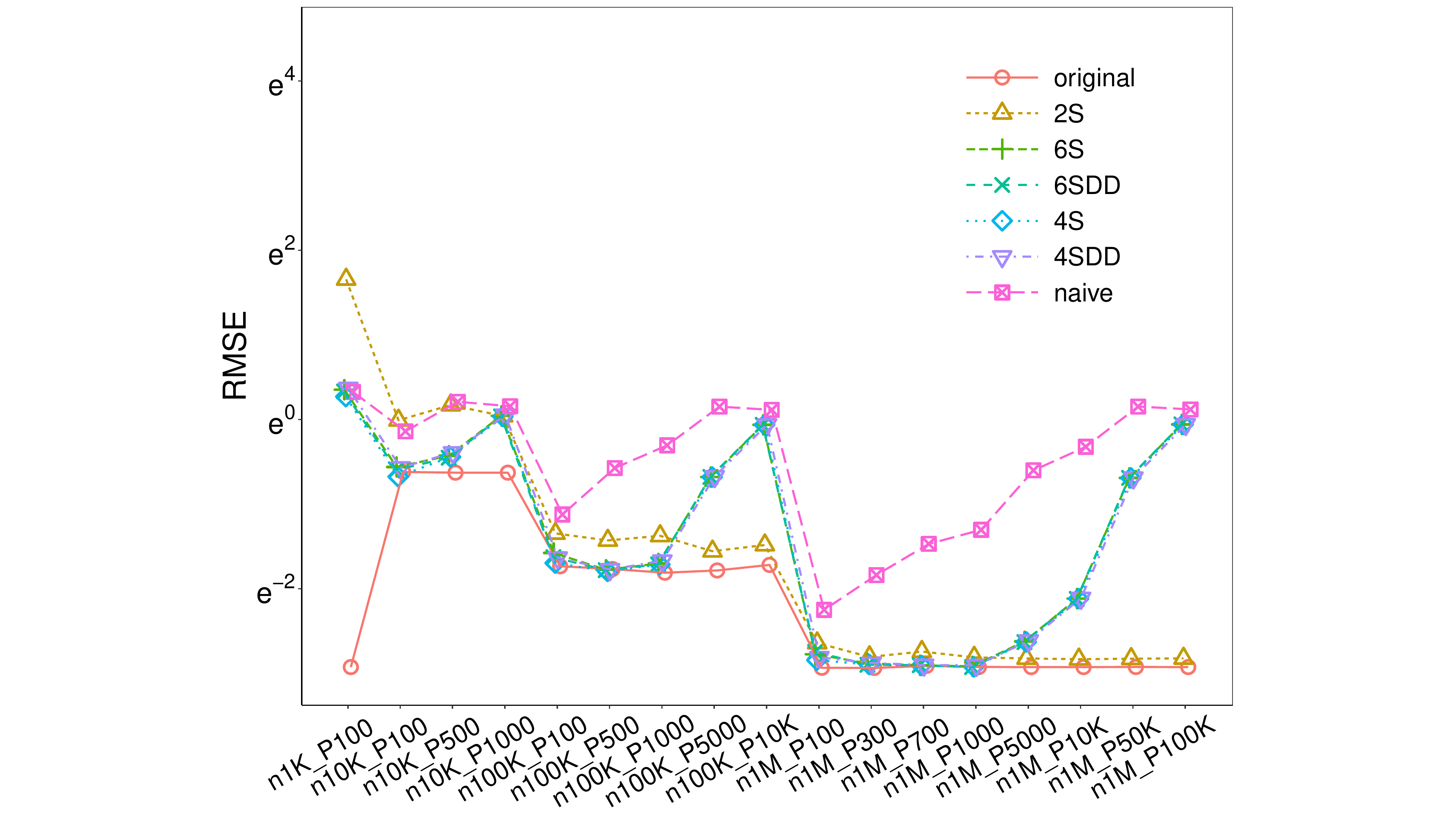}\\
\includegraphics[width=0.26\textwidth, trim={2.2in 0 2.2in 0},clip] {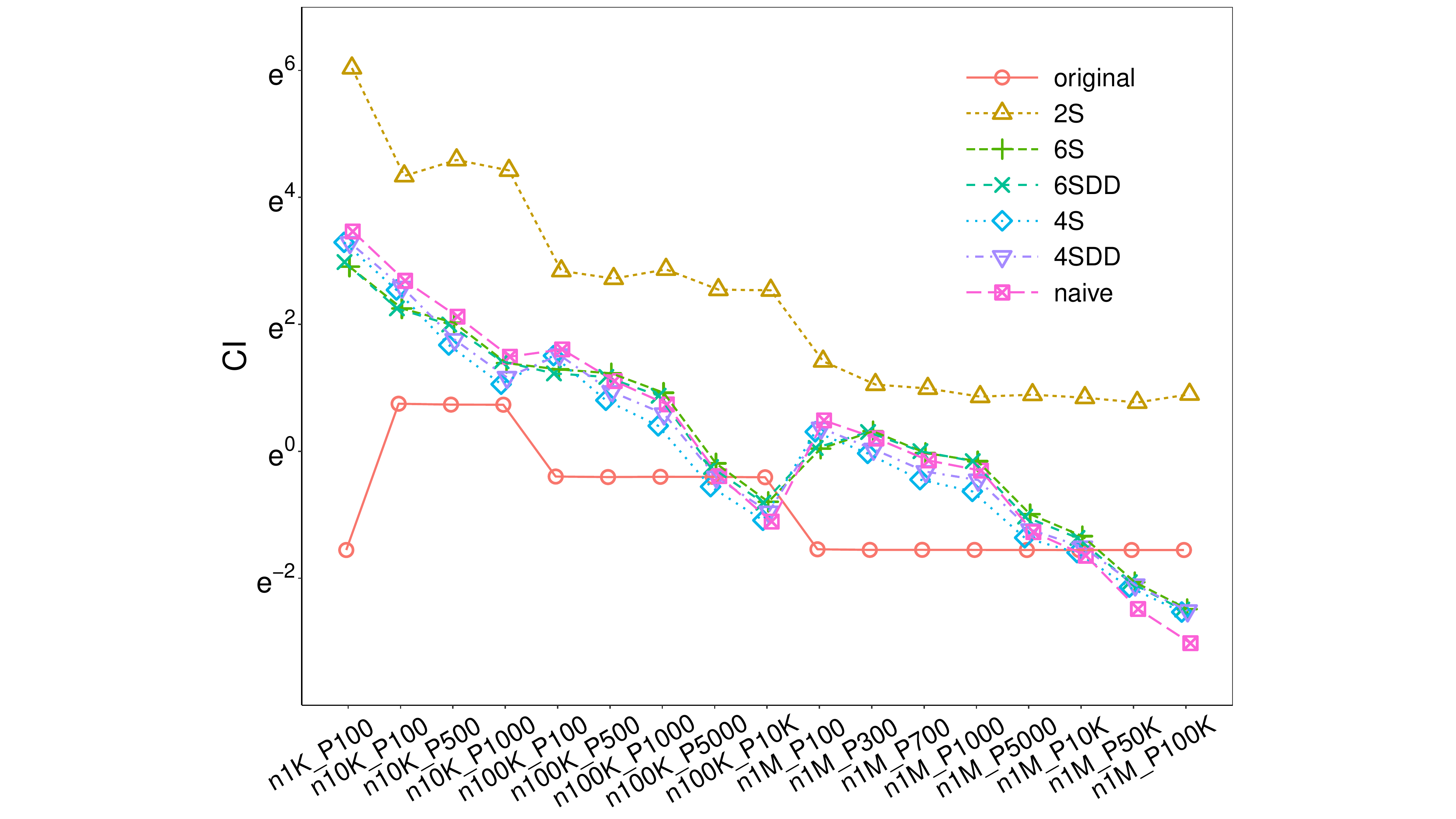}
\includegraphics[width=0.26\textwidth, trim={2.2in 0 2.2in 0},clip] {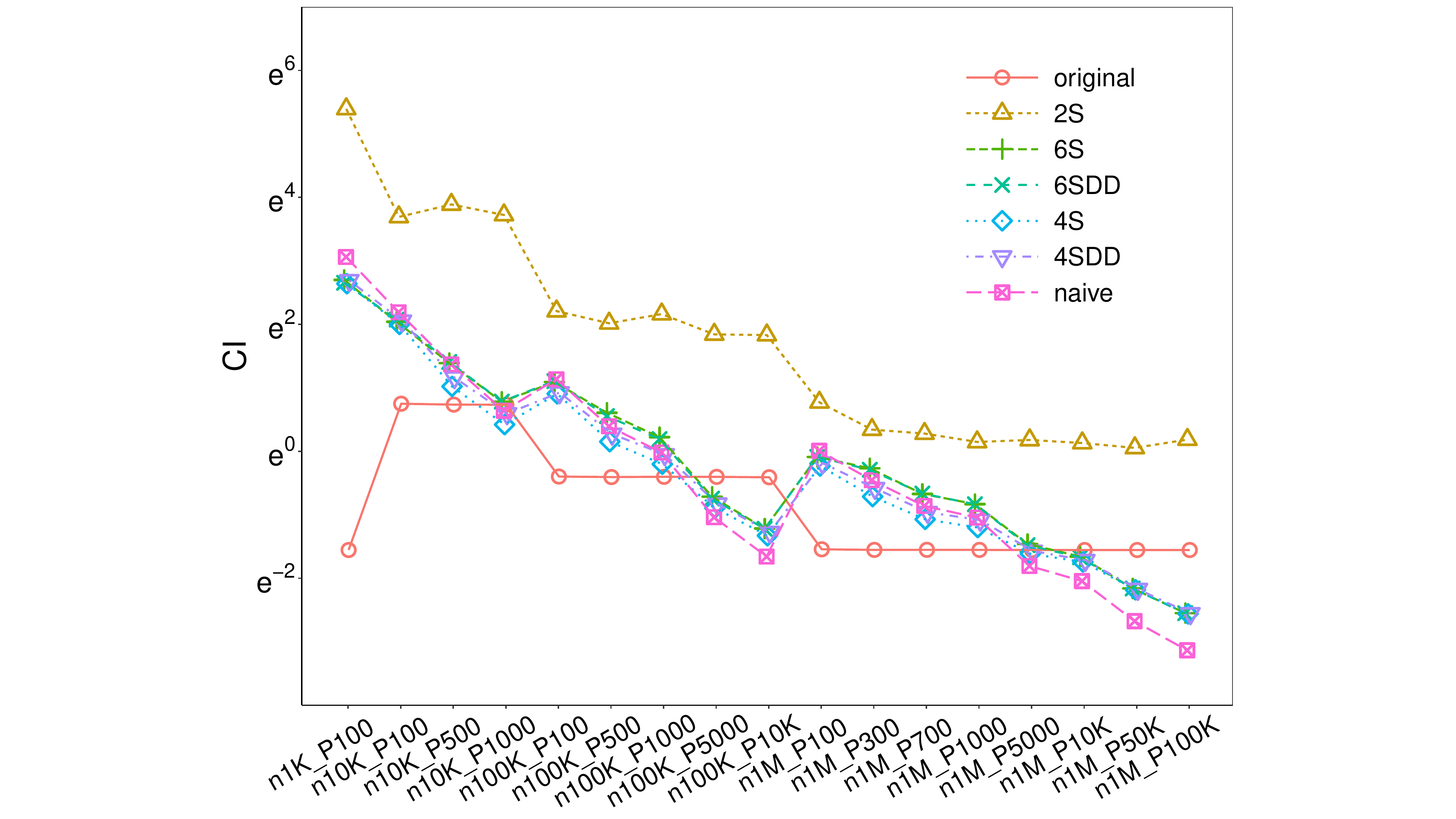}
\includegraphics[width=0.26\textwidth, trim={2.2in 0 2.2in 0},clip] {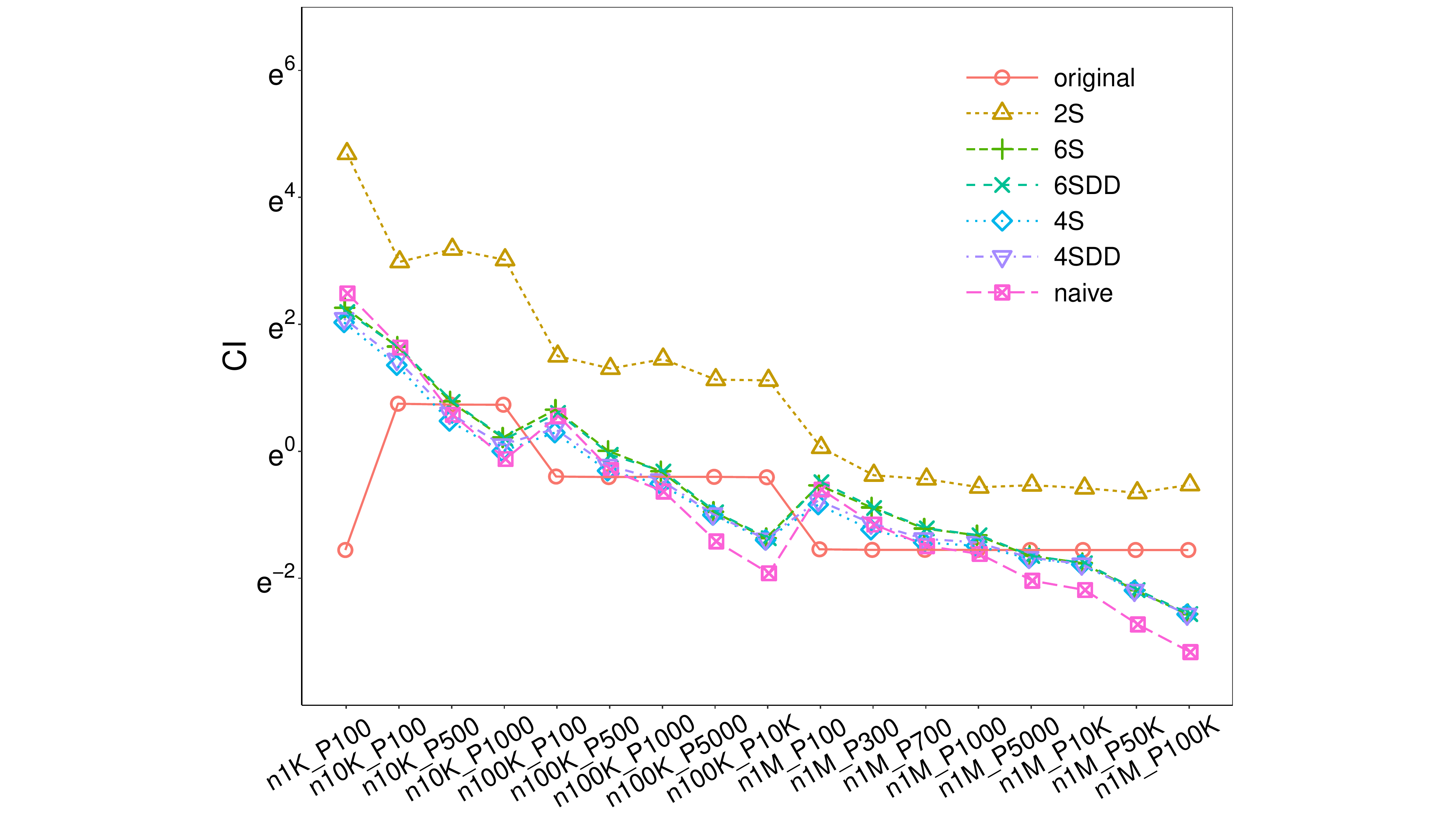}
\includegraphics[width=0.26\textwidth, trim={2.2in 0 2.2in 0},clip] {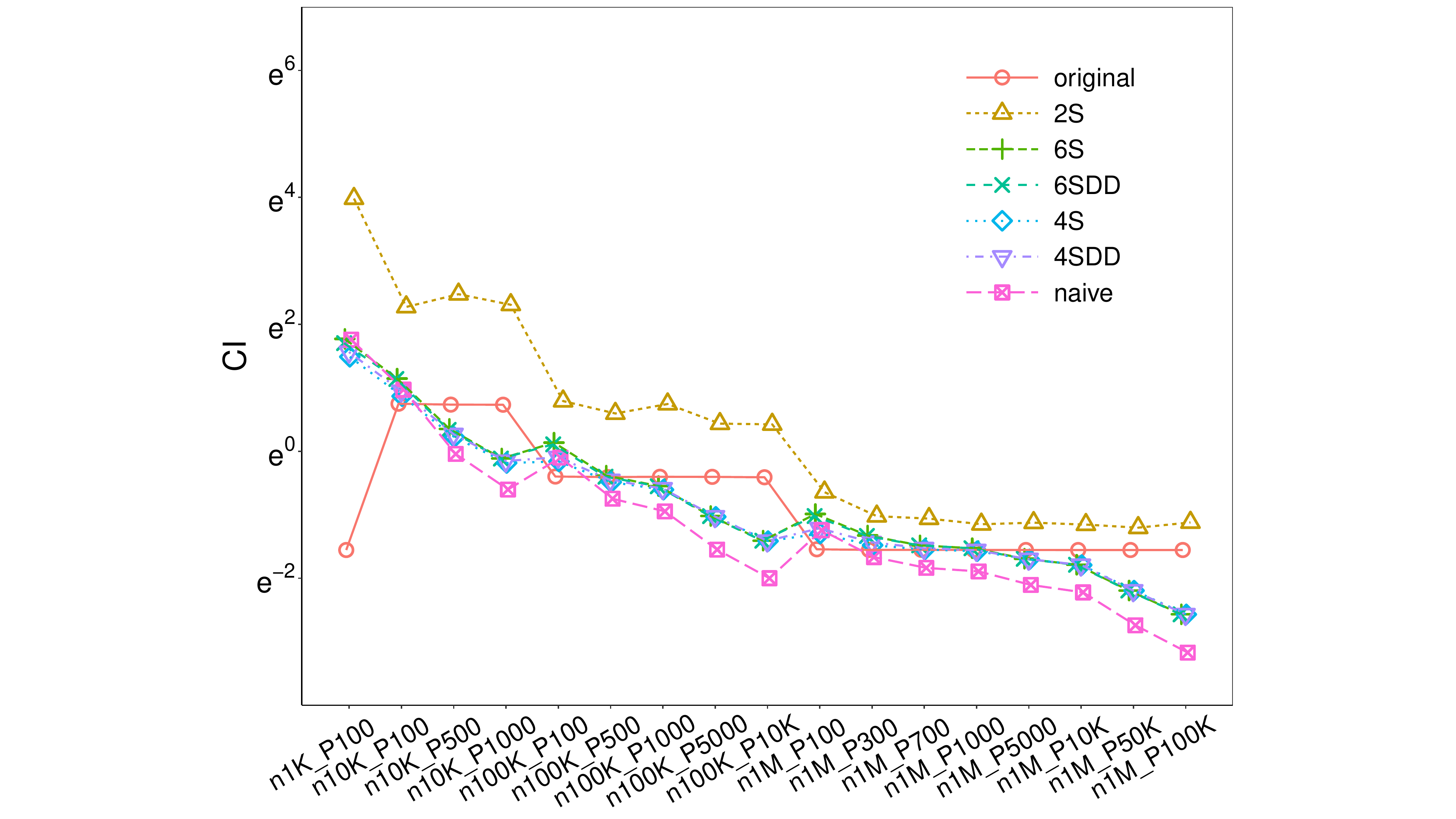}
\includegraphics[width=0.26\textwidth, trim={2.2in 0 2.2in 0},clip] {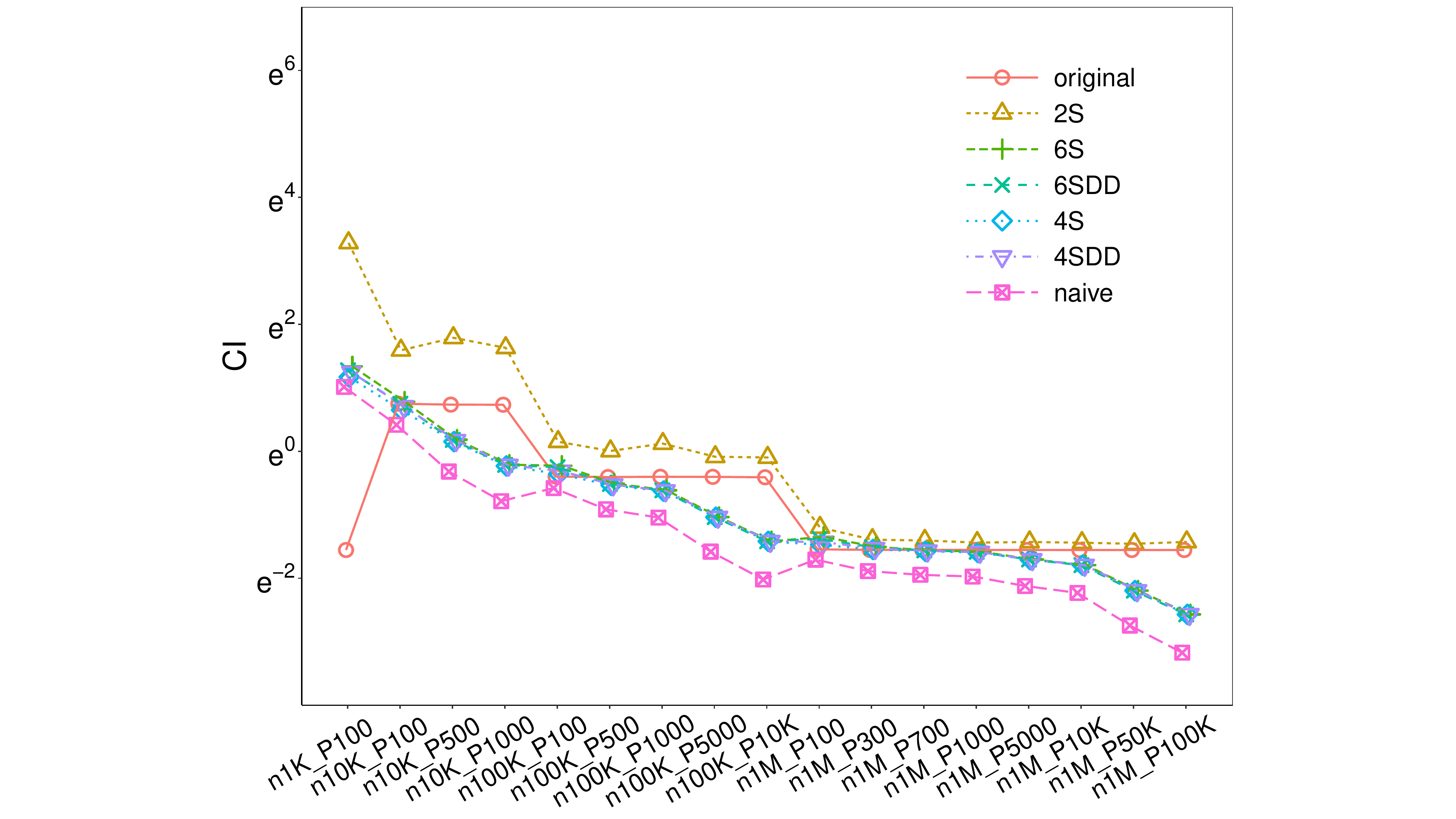}\\
\includegraphics[width=0.26\textwidth, trim={2.2in 0 2.2in 0},clip] {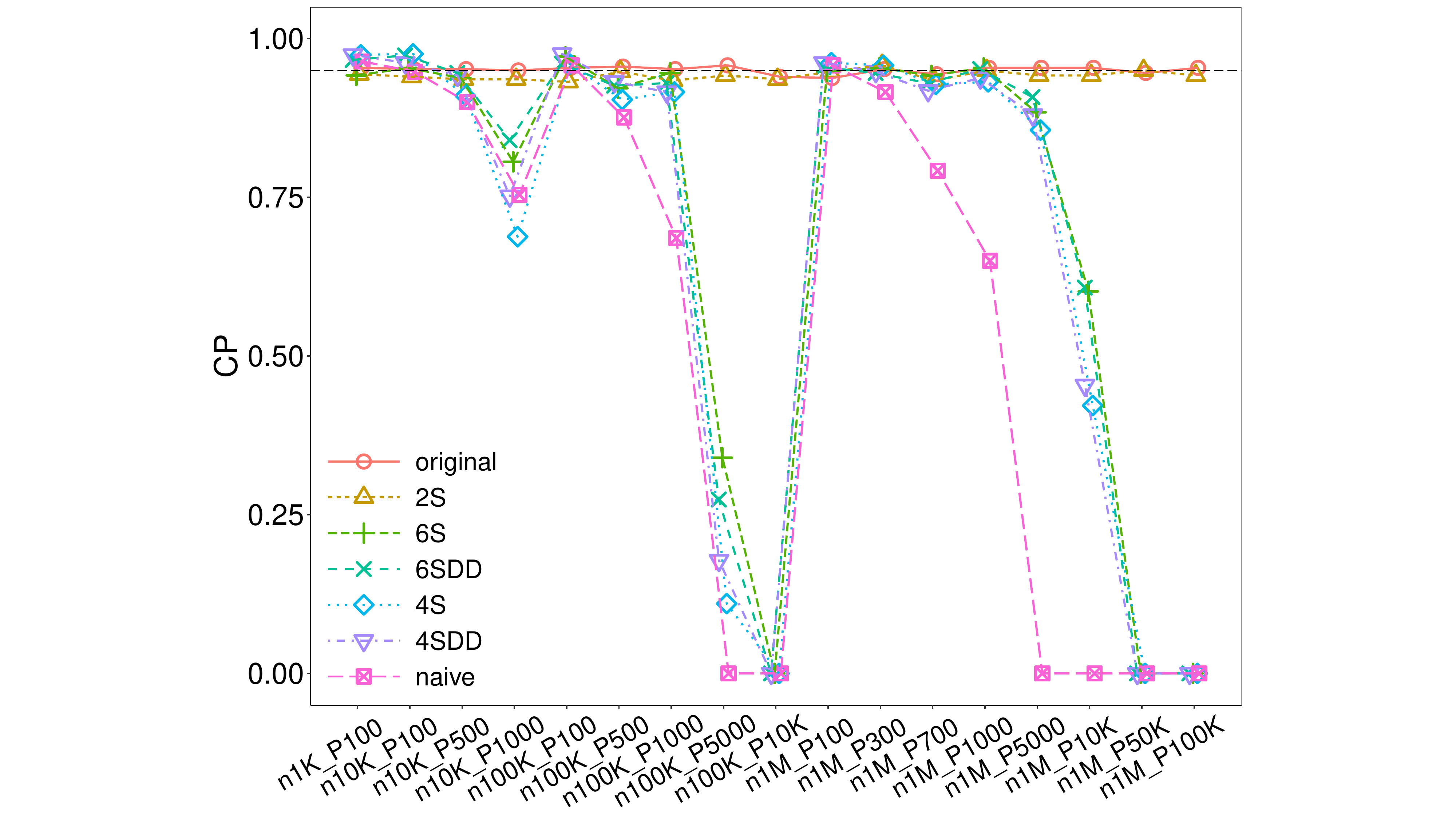}
\includegraphics[width=0.26\textwidth, trim={2.2in 0 2.2in 0},clip] {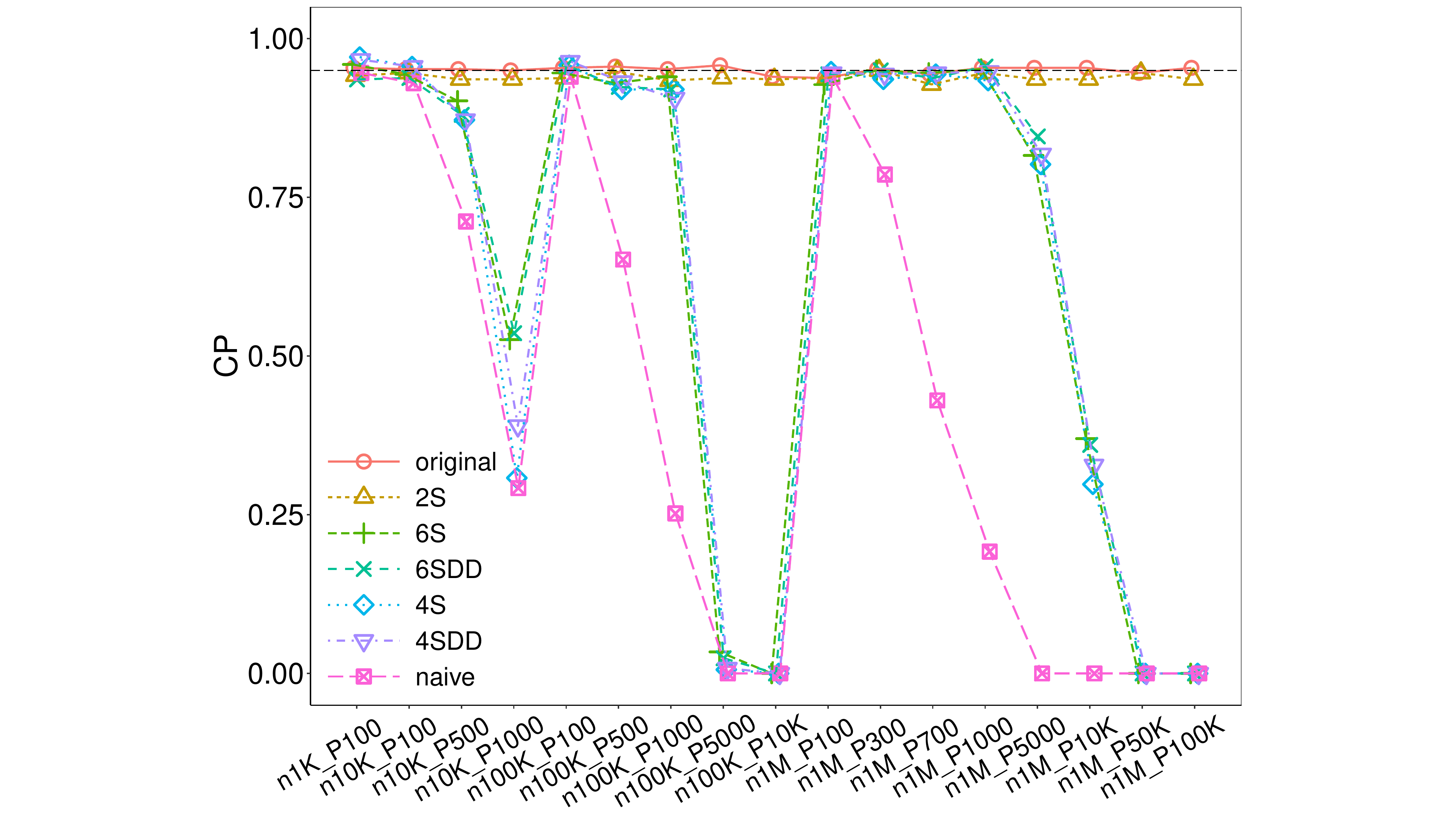}
\includegraphics[width=0.26\textwidth, trim={2.2in 0 2.2in 0},clip] {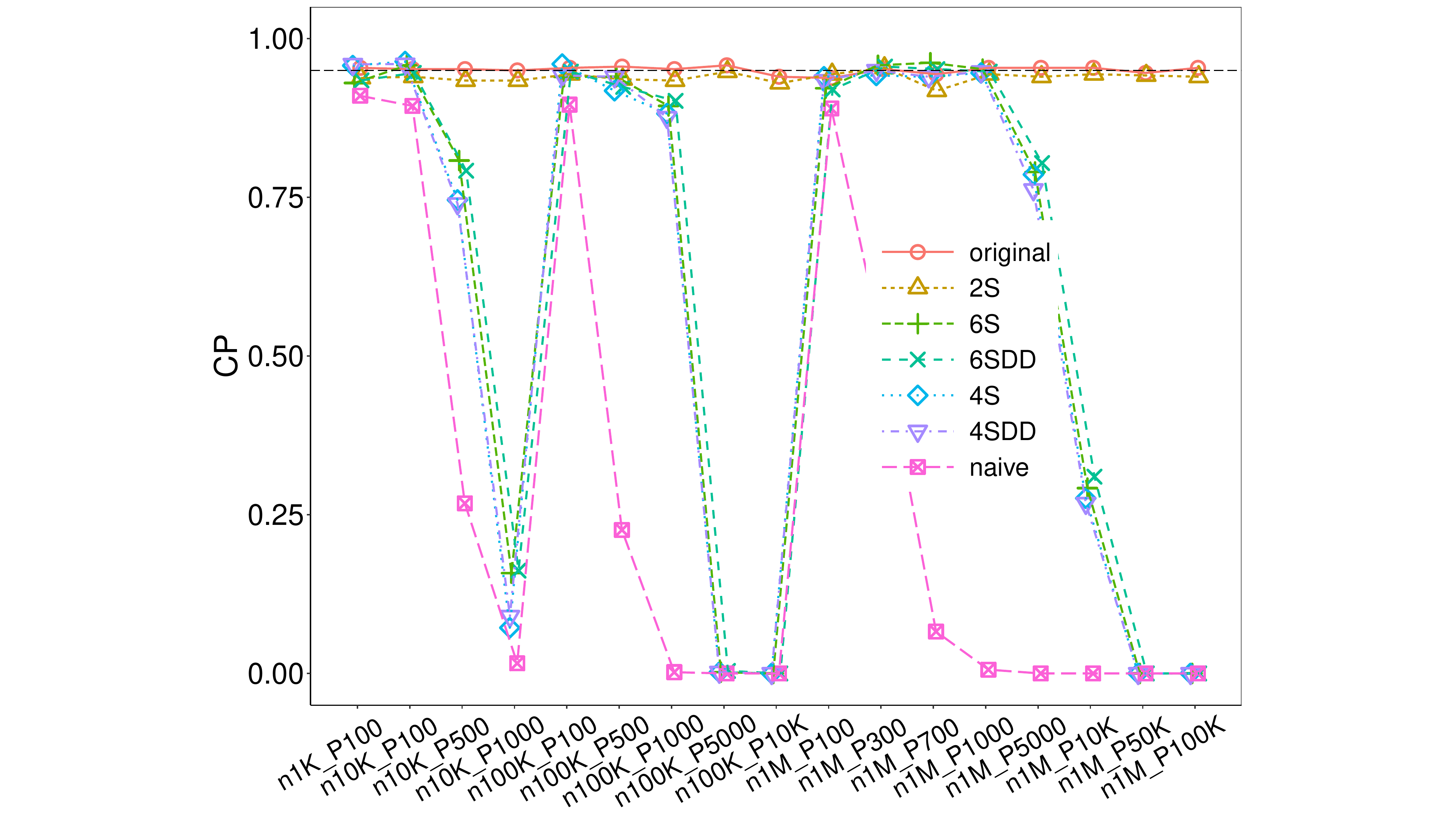}
\includegraphics[width=0.26\textwidth, trim={2.2in 0 2.2in 0},clip] {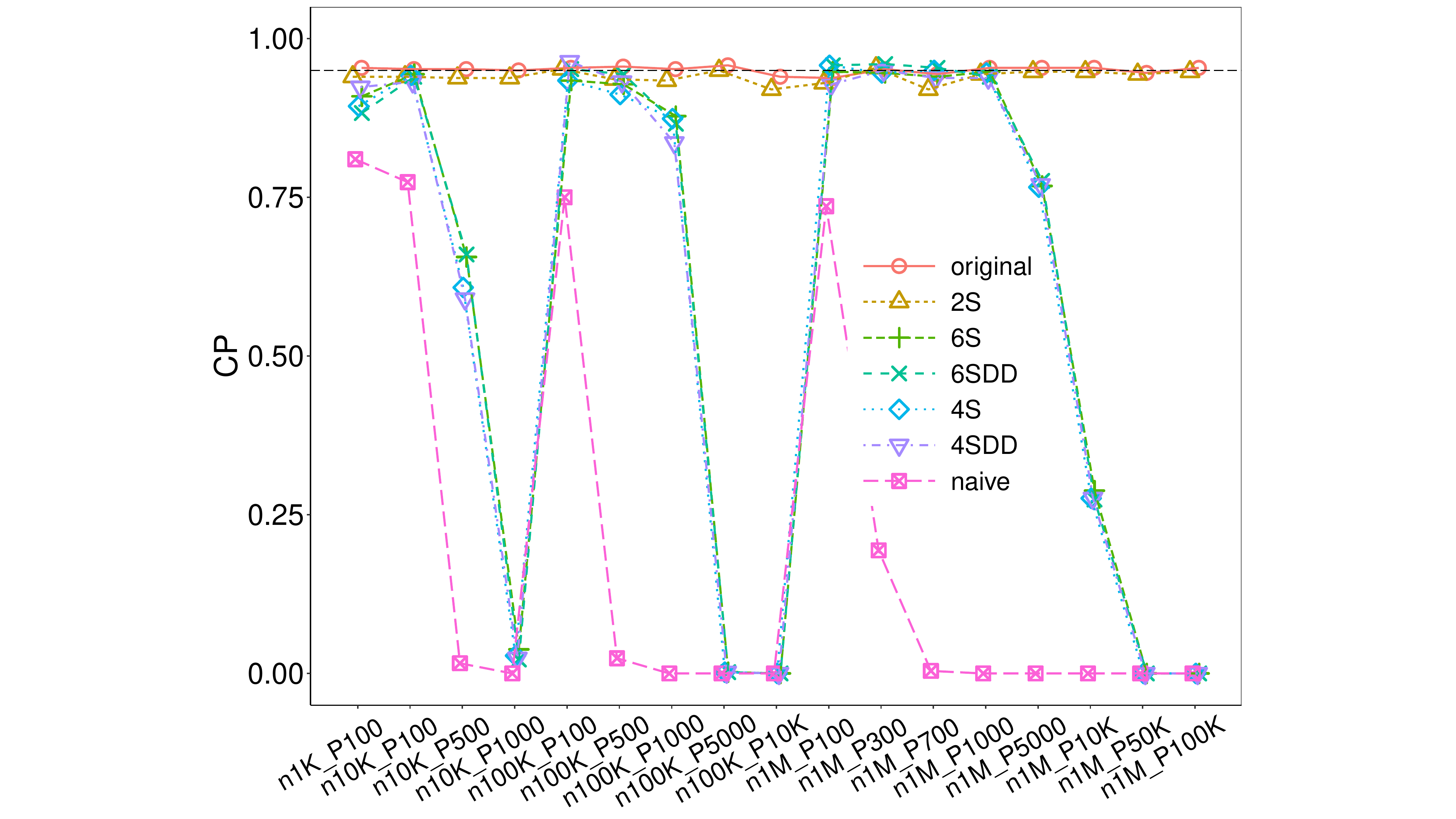}
\includegraphics[width=0.26\textwidth, trim={2.2in 0 2.2in 0},clip] {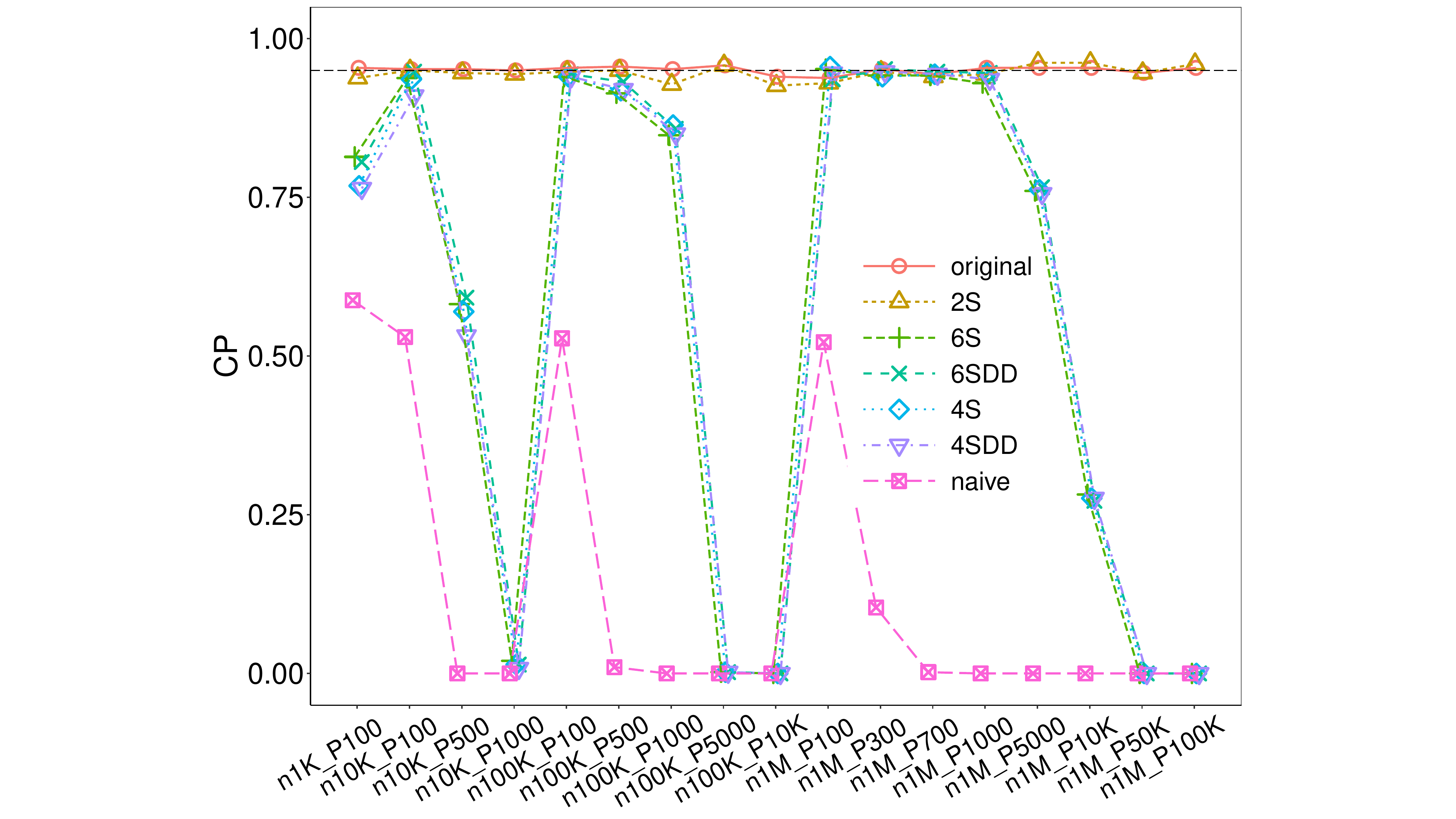}\\
\caption{ZILN data; $\rho$-zCDP; $\theta=0$ and $\alpha\ne\beta$}
\label{fig:0aszCDPziln}
\end{figure}
\end{landscape}

\begin{landscape}
\subsection*{ZILN, $\theta\ne0$ and $\alpha\ne\beta$}
\begin{figure}[!htb]
\centering
\centering
$\epsilon=0.5$\hspace{0.9in}$\epsilon=1$\hspace{1in}$\epsilon=2$
\hspace{1in}$\epsilon=5$\hspace{0.9in}$\epsilon=50$\\
\includegraphics[width=0.215\textwidth, trim={2.2in 0 2.2in 0},clip] {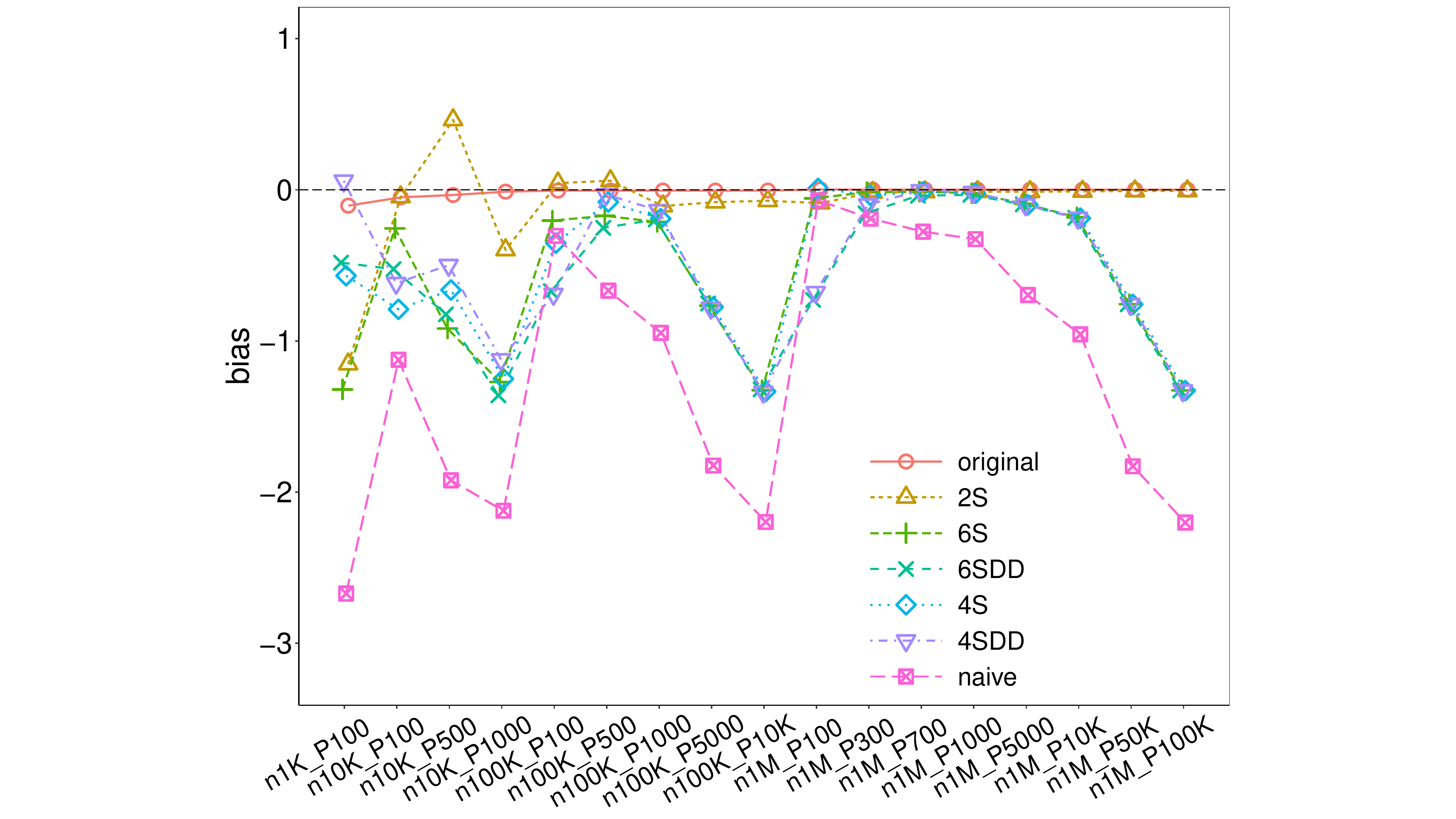}
\includegraphics[width=0.215\textwidth, trim={2.2in 0 2.2in 0},clip] {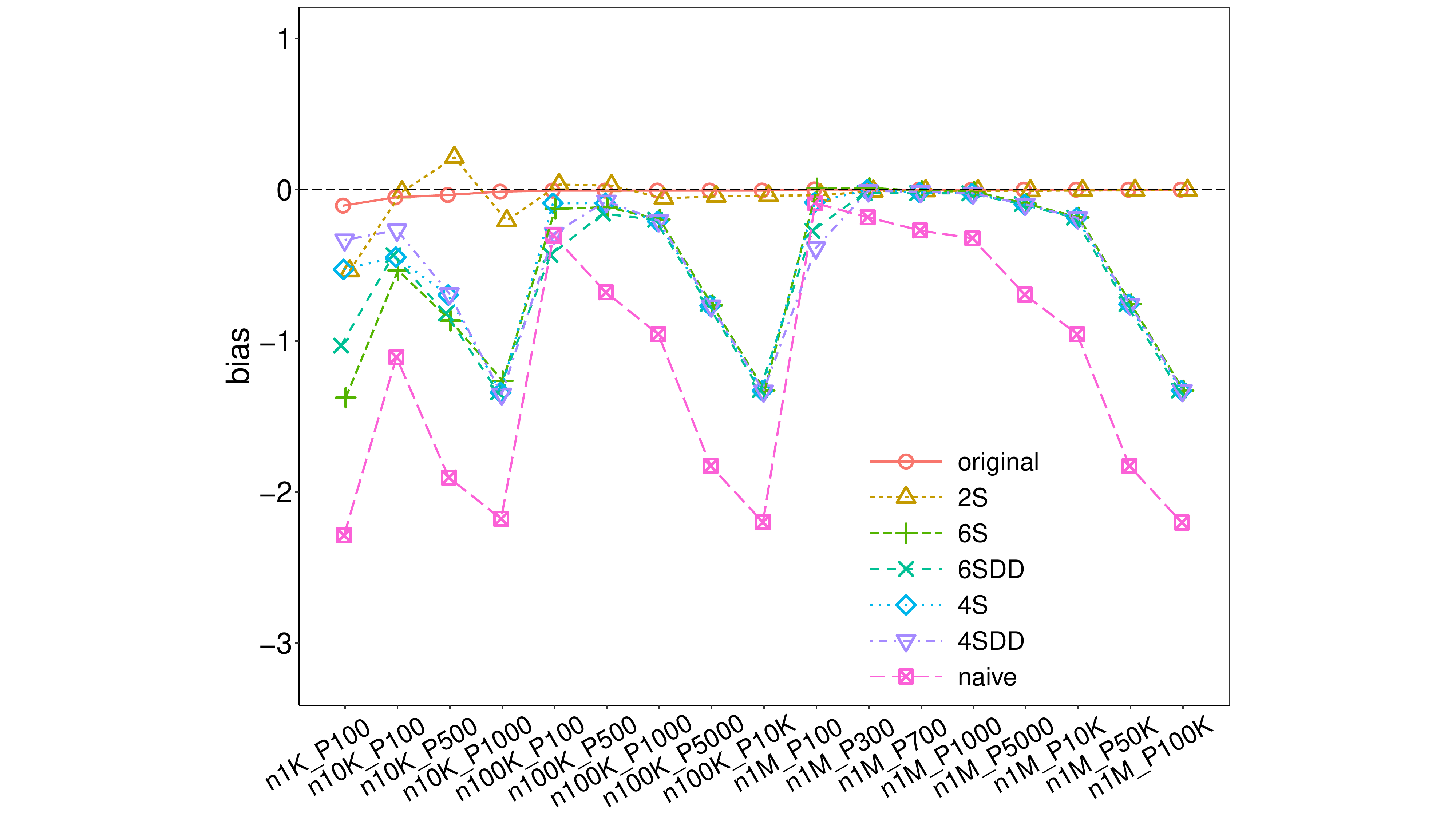}
\includegraphics[width=0.215\textwidth, trim={2.2in 0 2.2in 0},clip] {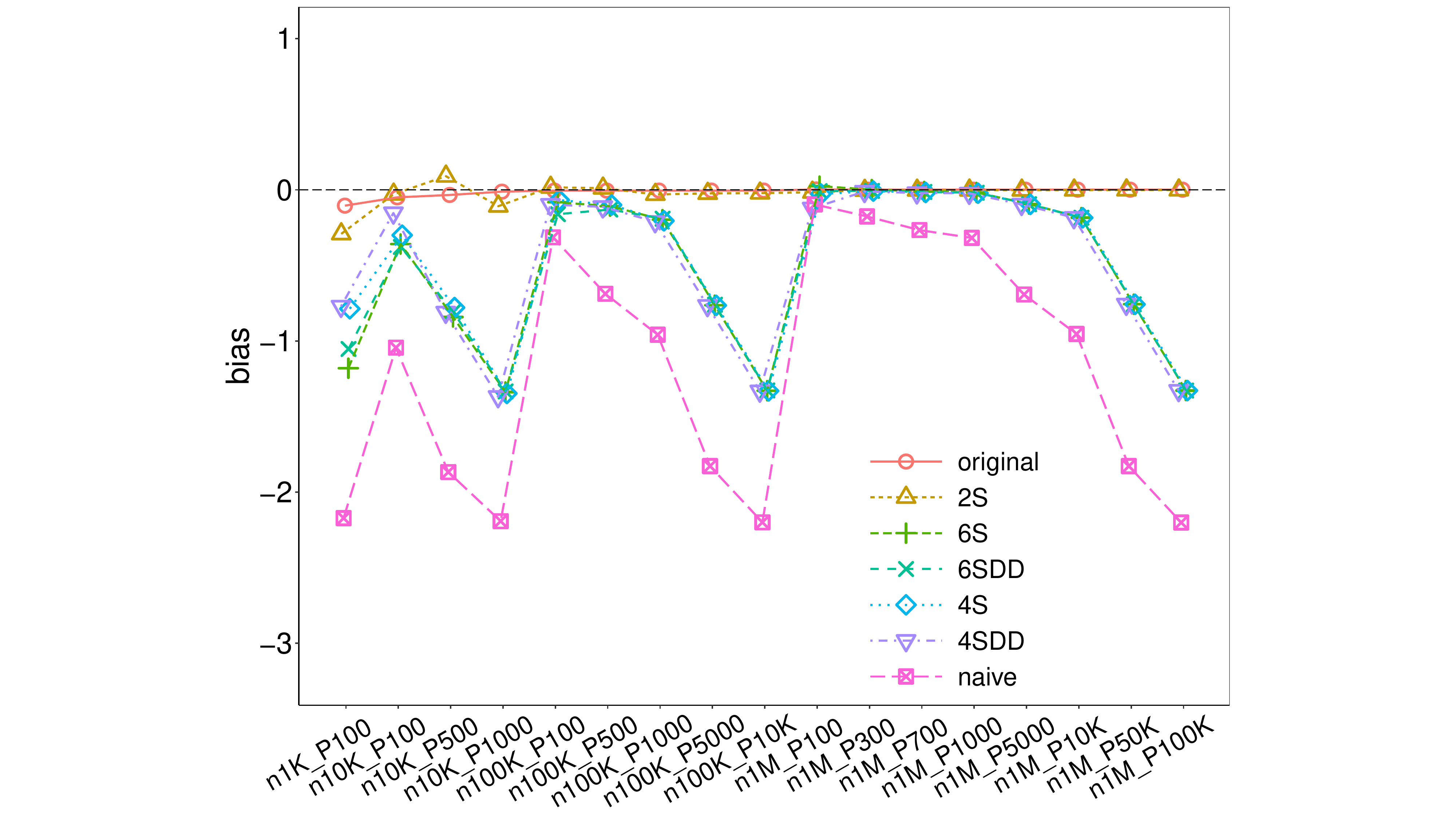}
\includegraphics[width=0.215\textwidth, trim={2.2in 0 2.2in 0},clip] {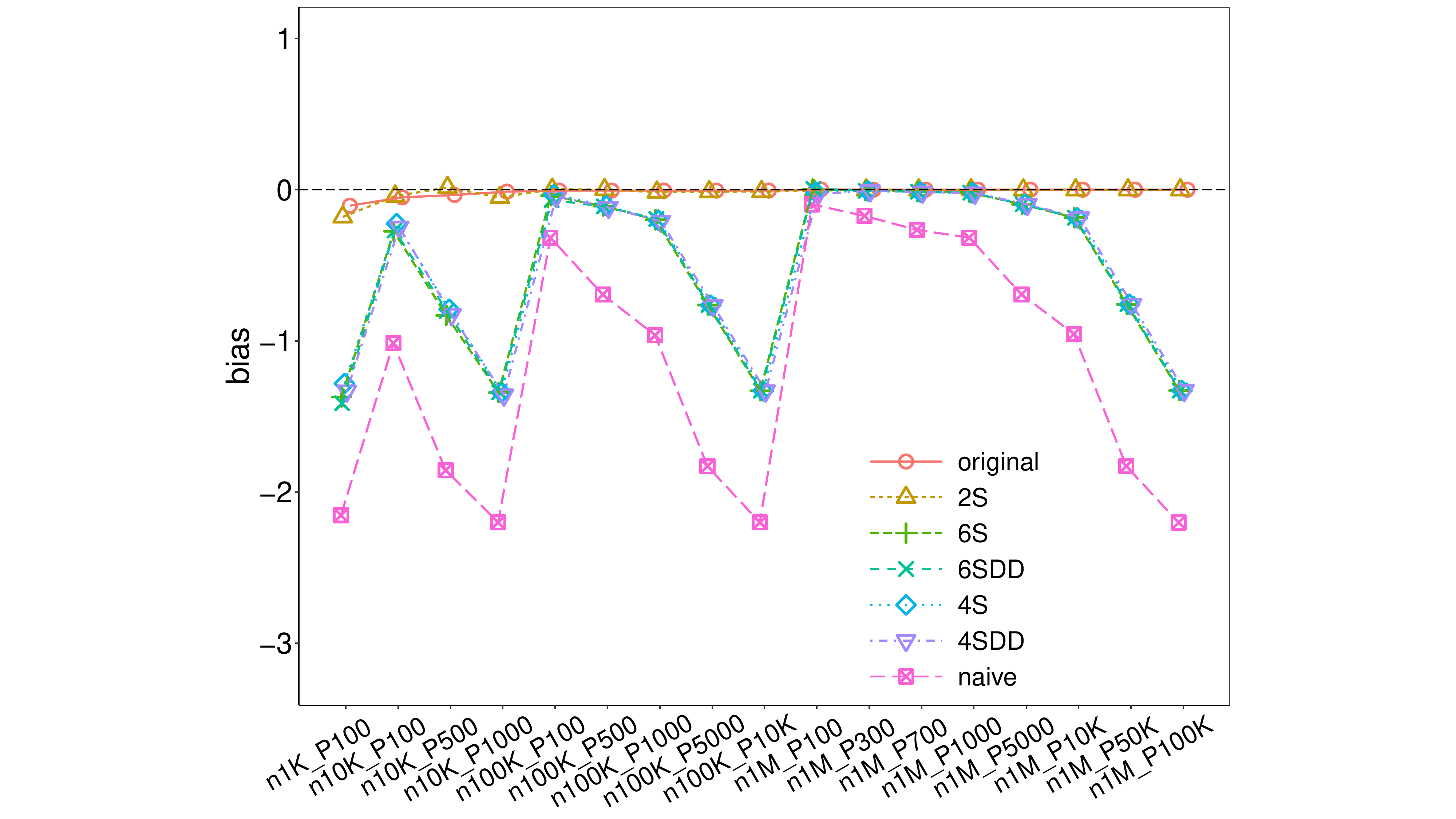}
\includegraphics[width=0.215\textwidth, trim={2.2in 0 2.2in 0},clip] {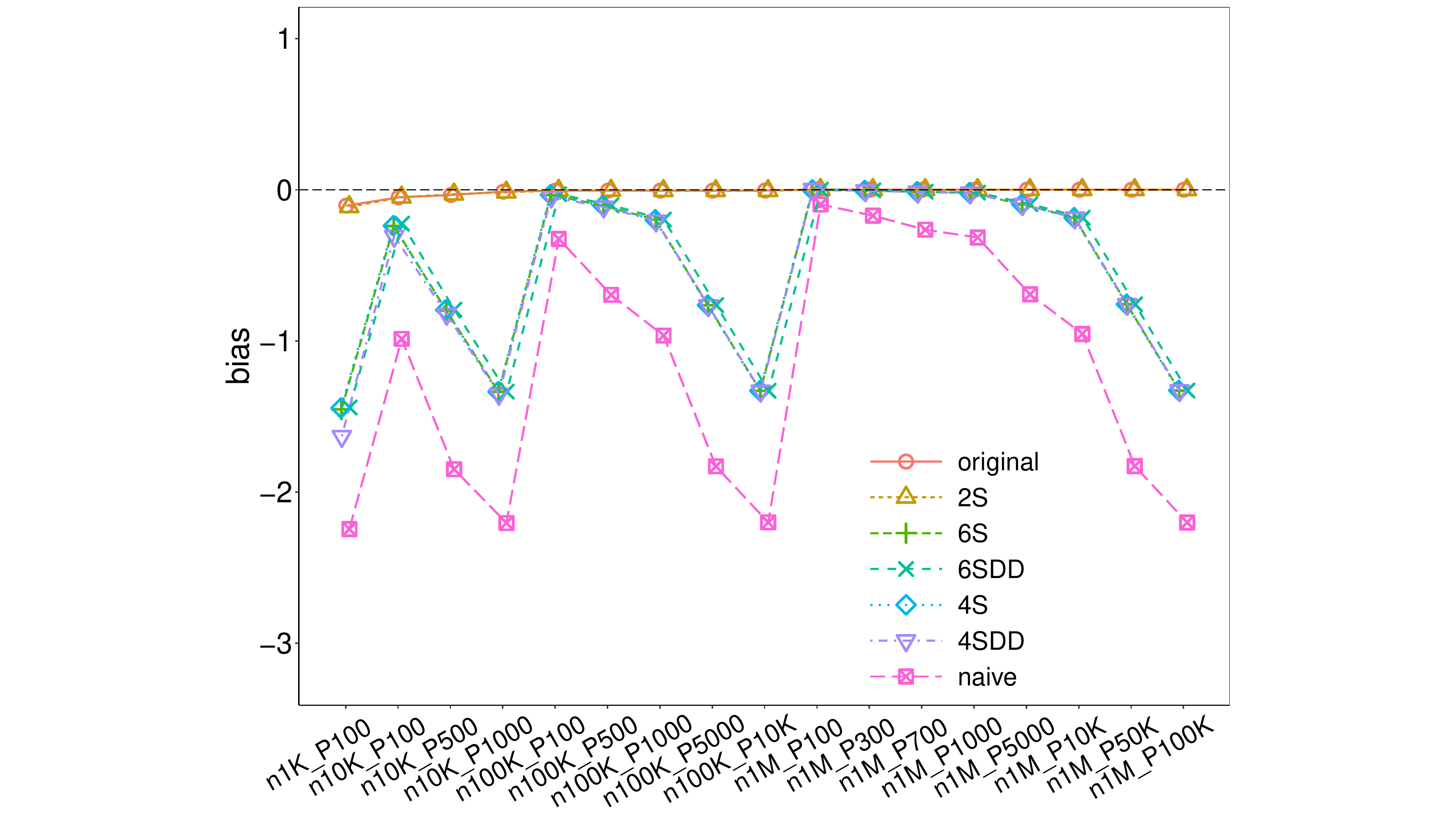}\\
\includegraphics[width=0.215\textwidth, trim={2.2in 0 2.2in 0},clip] {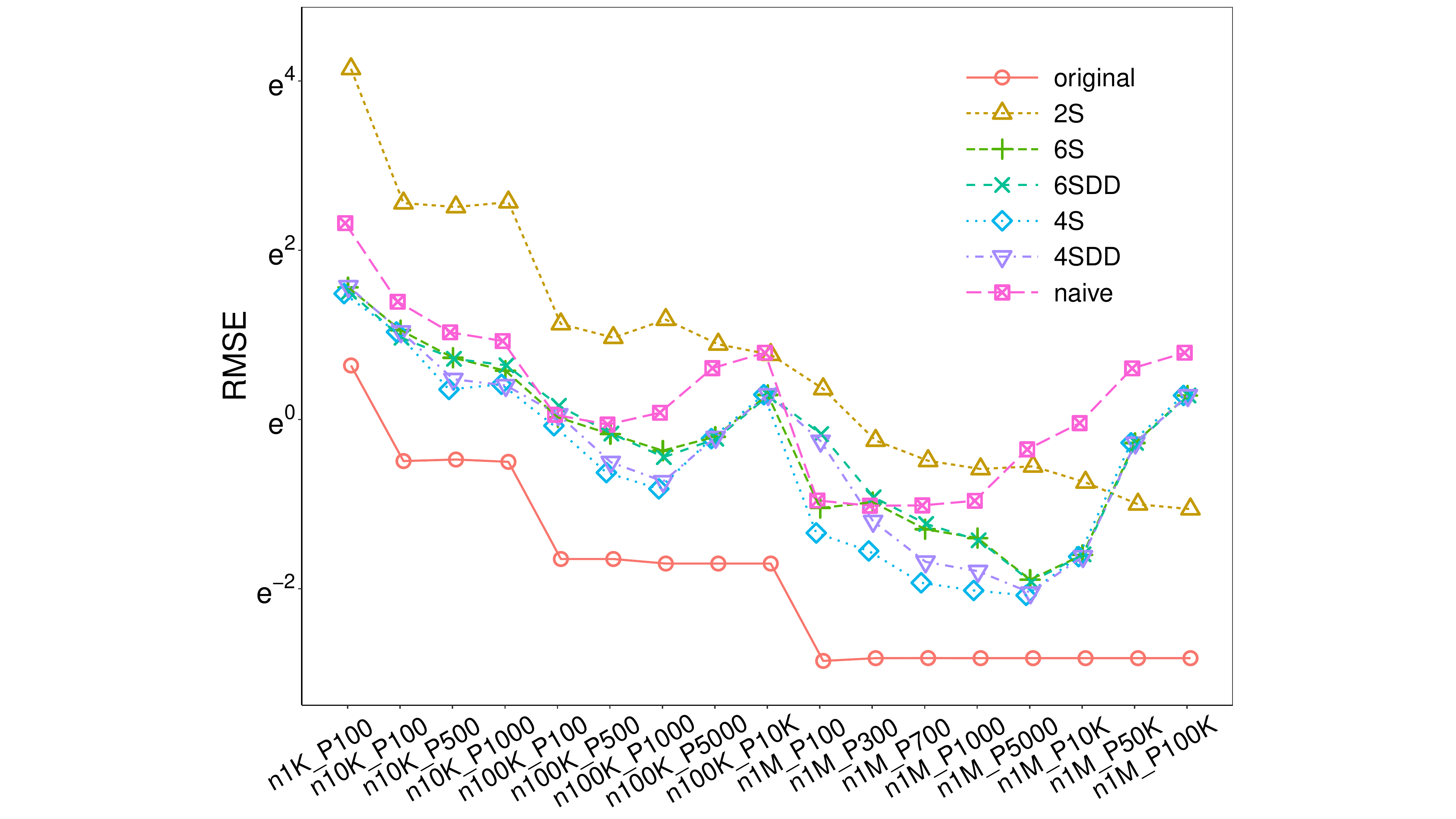}
\includegraphics[width=0.215\textwidth, trim={2.2in 0 2.2in 0},clip] {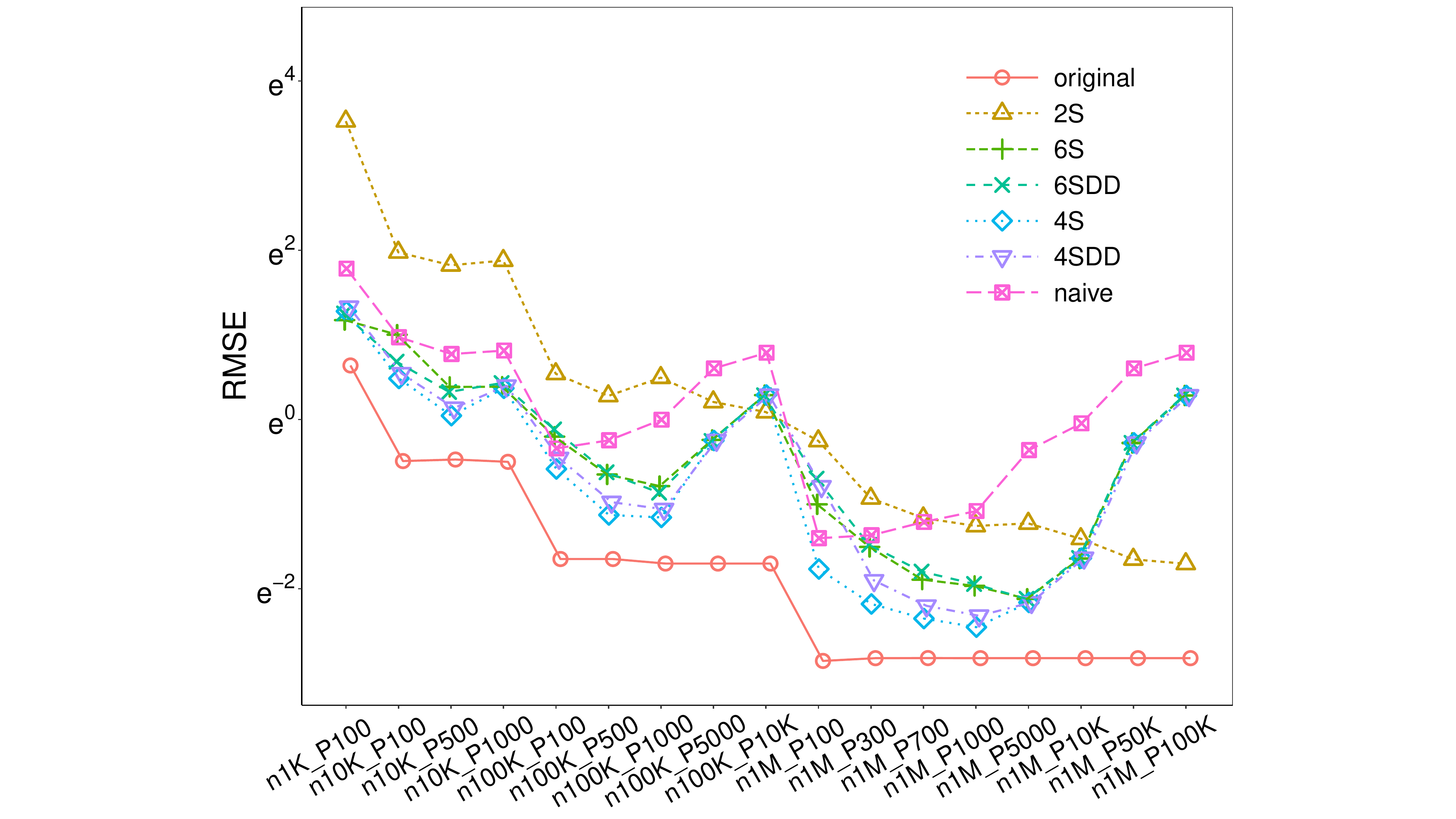}
\includegraphics[width=0.215\textwidth, trim={2.2in 0 2.2in 0},clip] {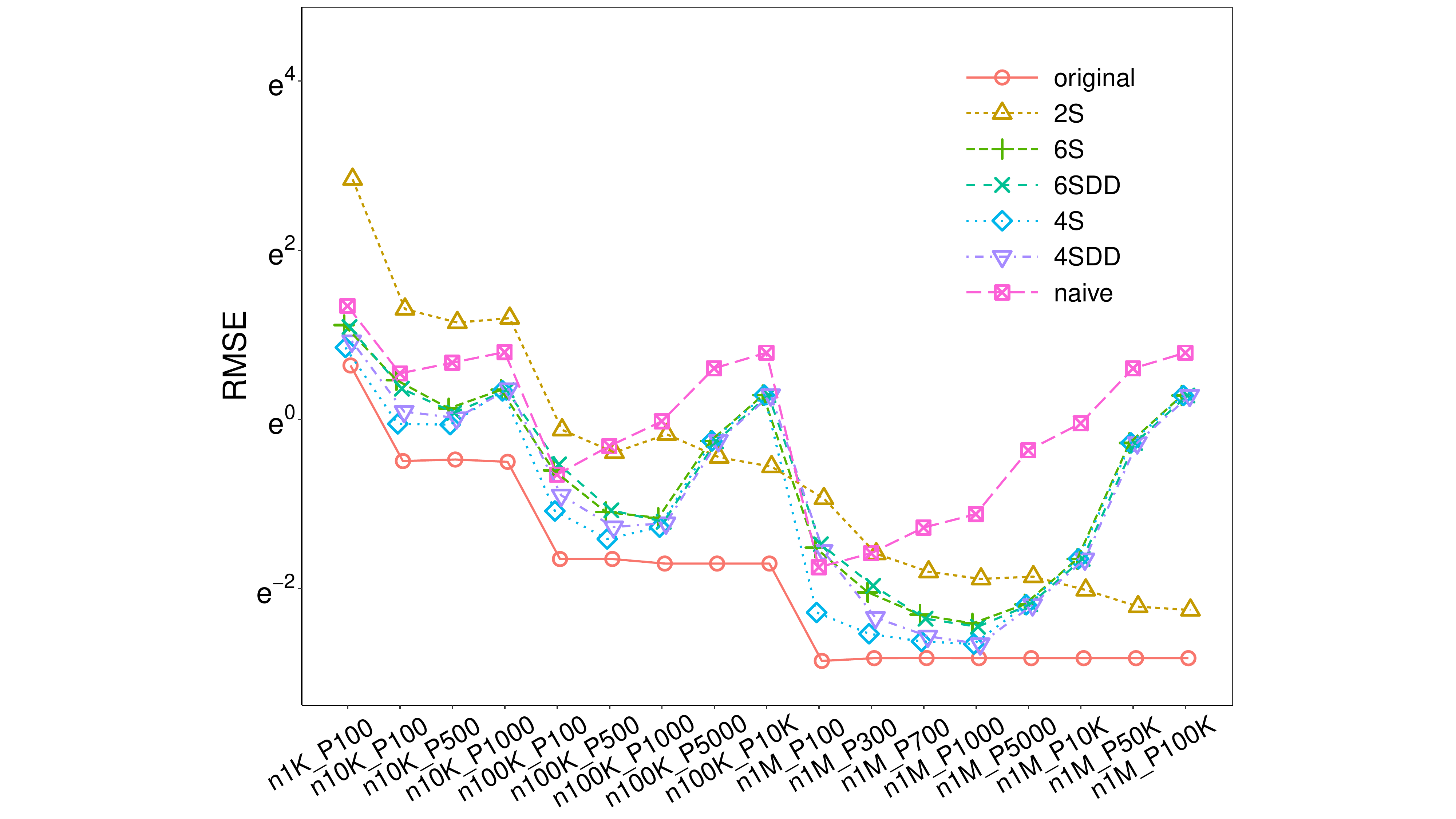}
\includegraphics[width=0.215\textwidth, trim={2.2in 0 2.2in 0},clip] {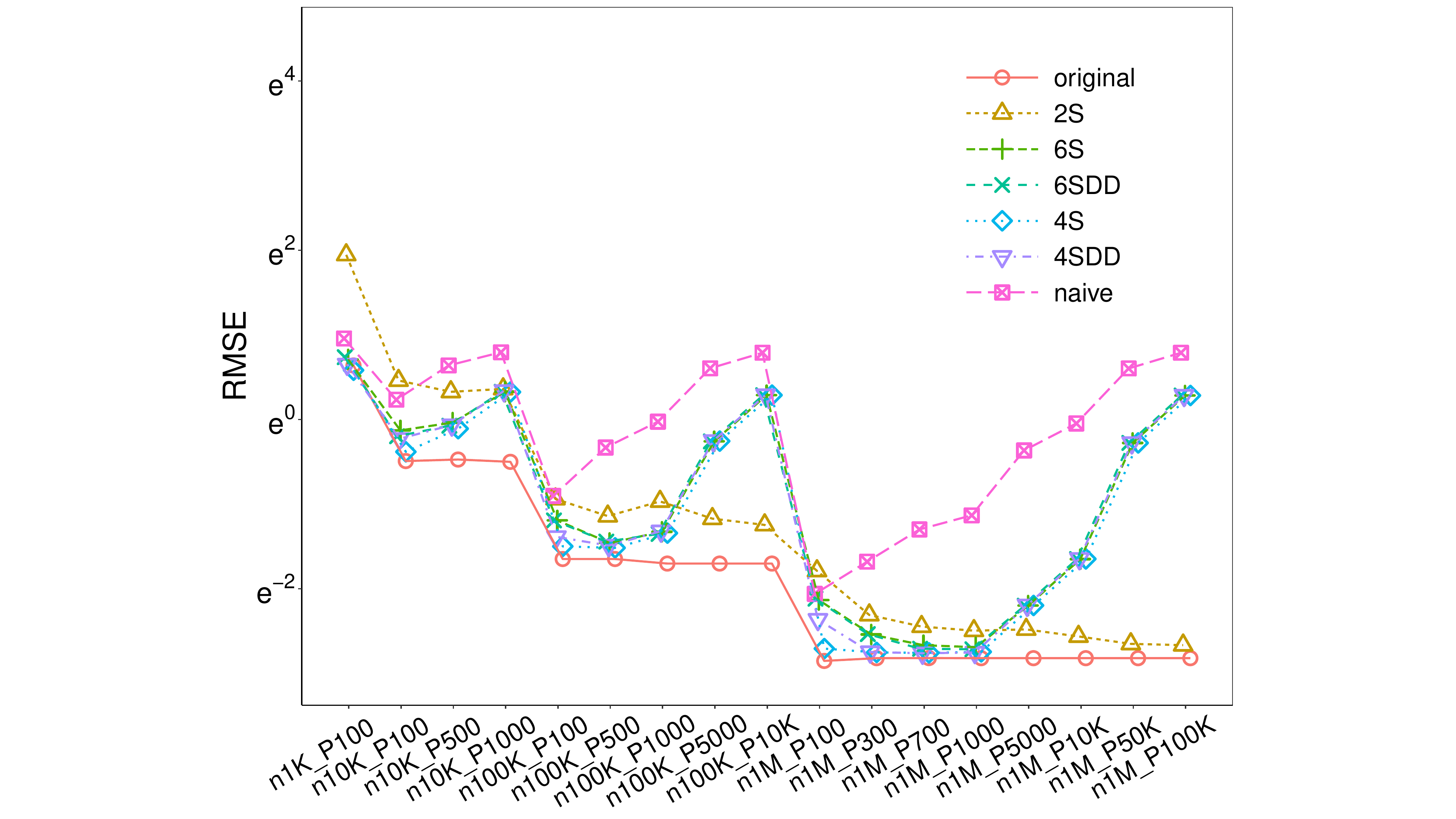}
\includegraphics[width=0.215\textwidth, trim={2.2in 0 2.2in 0},clip] {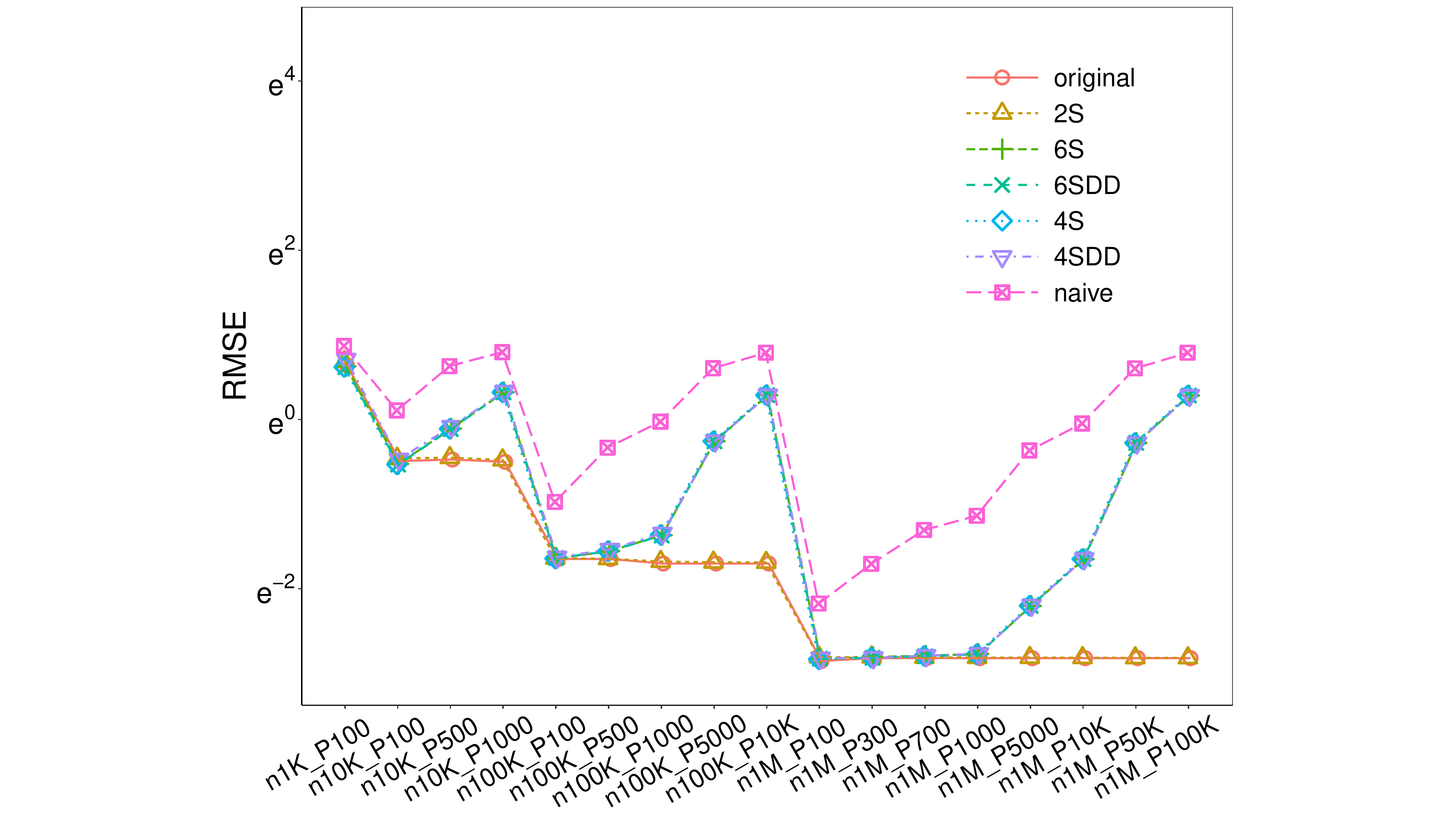}\\
\includegraphics[width=0.215\textwidth, trim={2.2in 0 2.2in 0},clip] {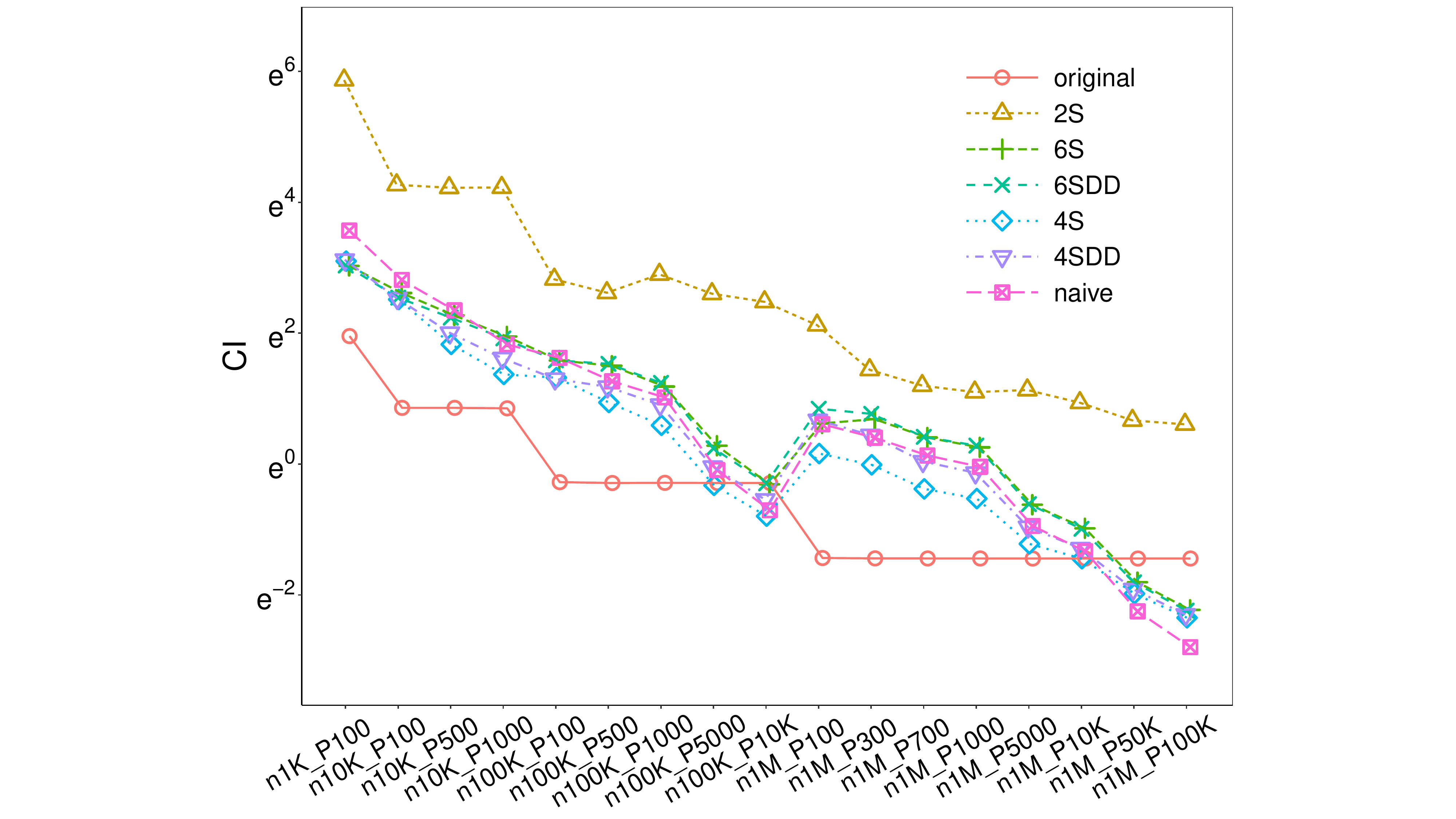}
\includegraphics[width=0.215\textwidth, trim={2.2in 0 2.2in 0},clip] {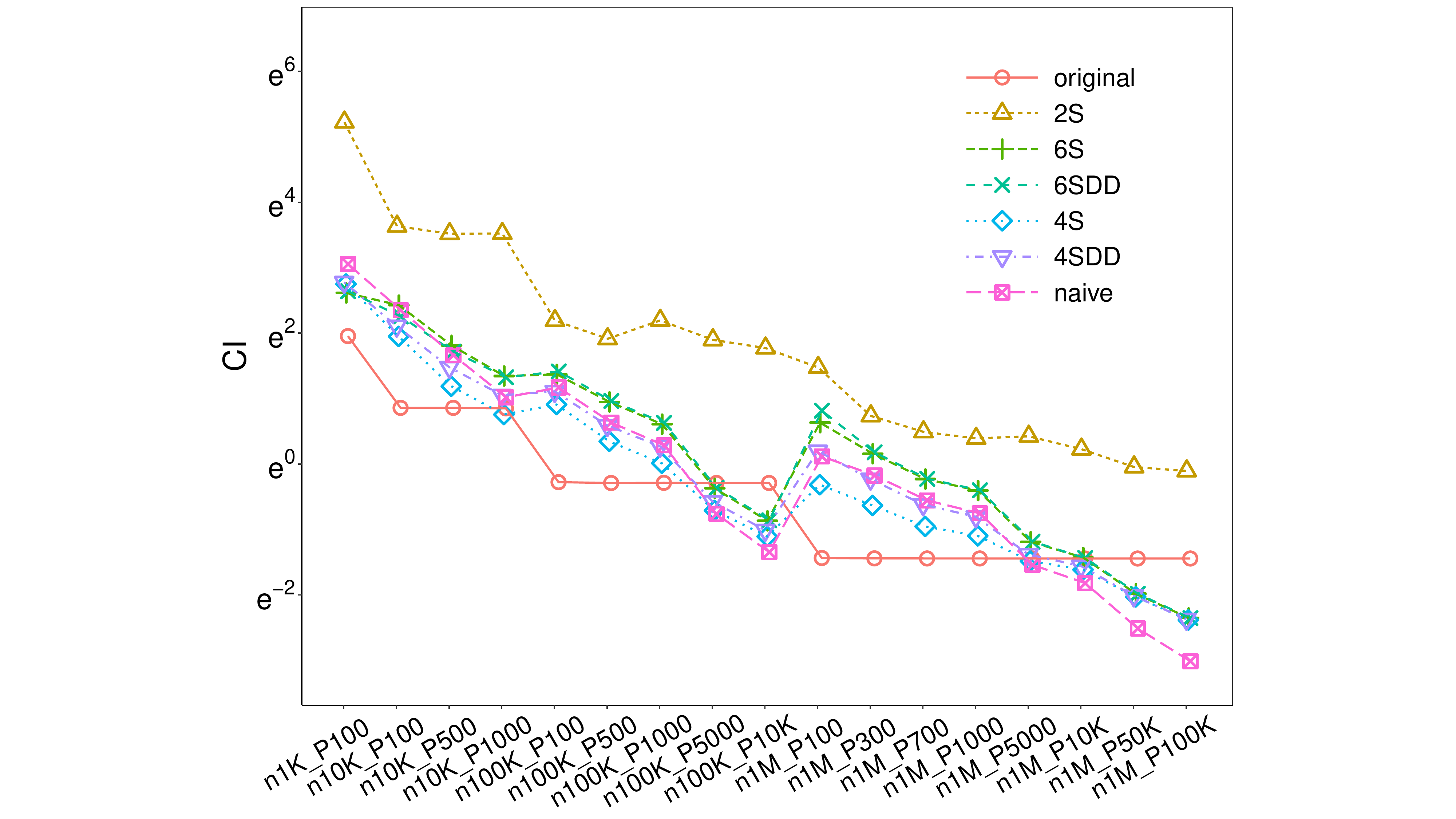}
\includegraphics[width=0.215\textwidth, trim={2.2in 0 2.2in 0},clip] {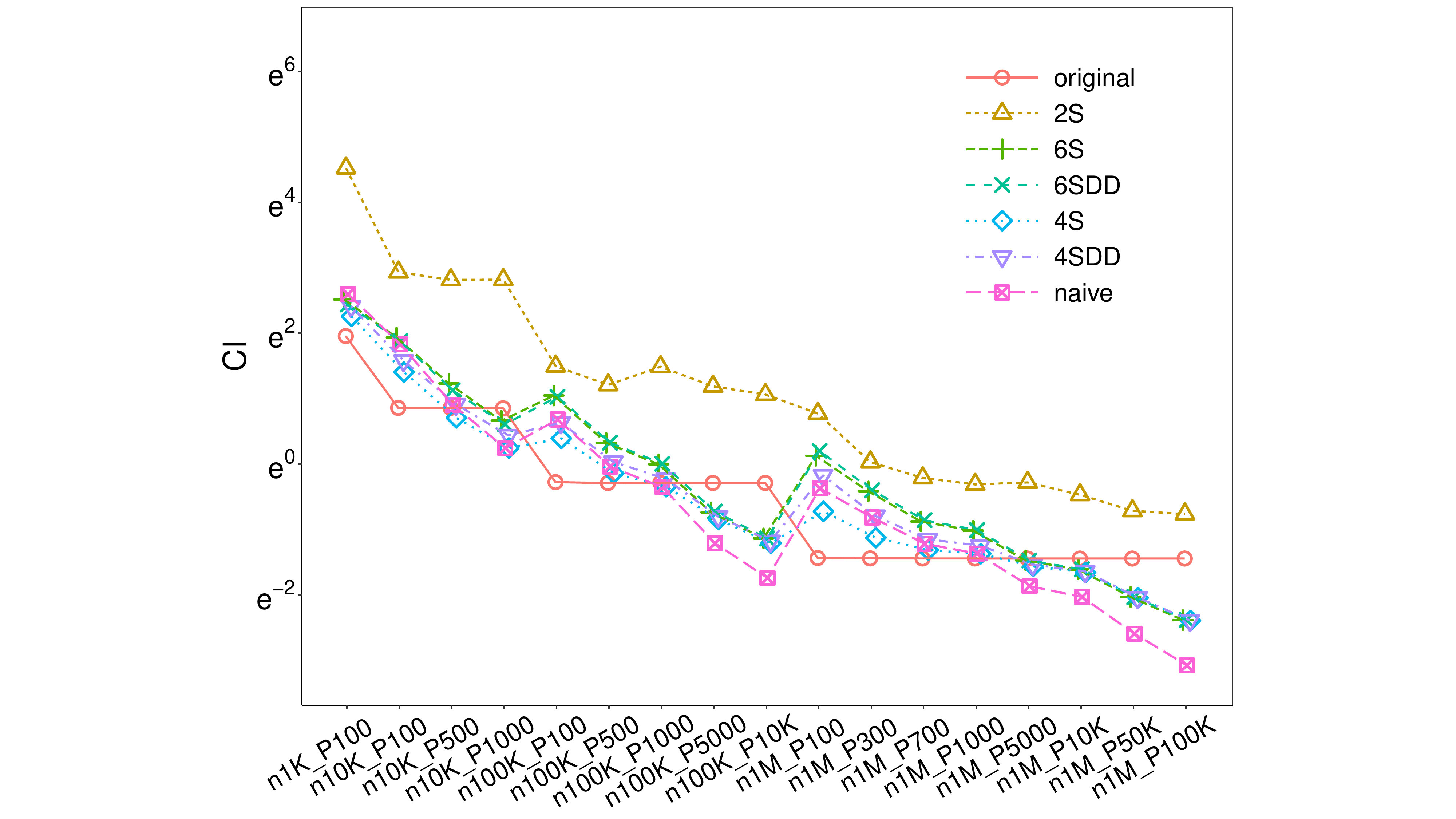}
\includegraphics[width=0.215\textwidth, trim={2.2in 0 2.2in 0},clip] {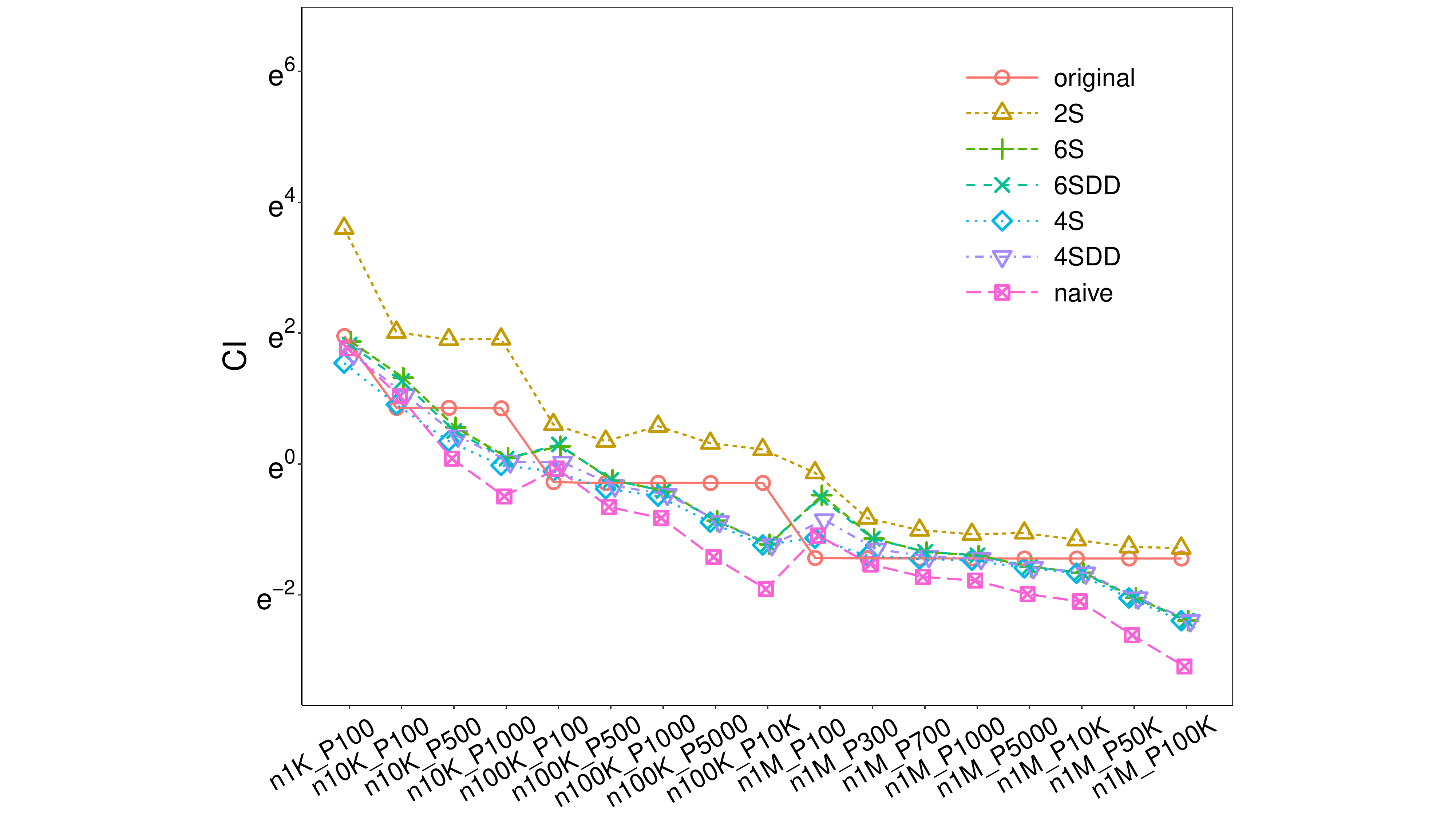}
\includegraphics[width=0.215\textwidth, trim={2.2in 0 2.2in 0},clip] {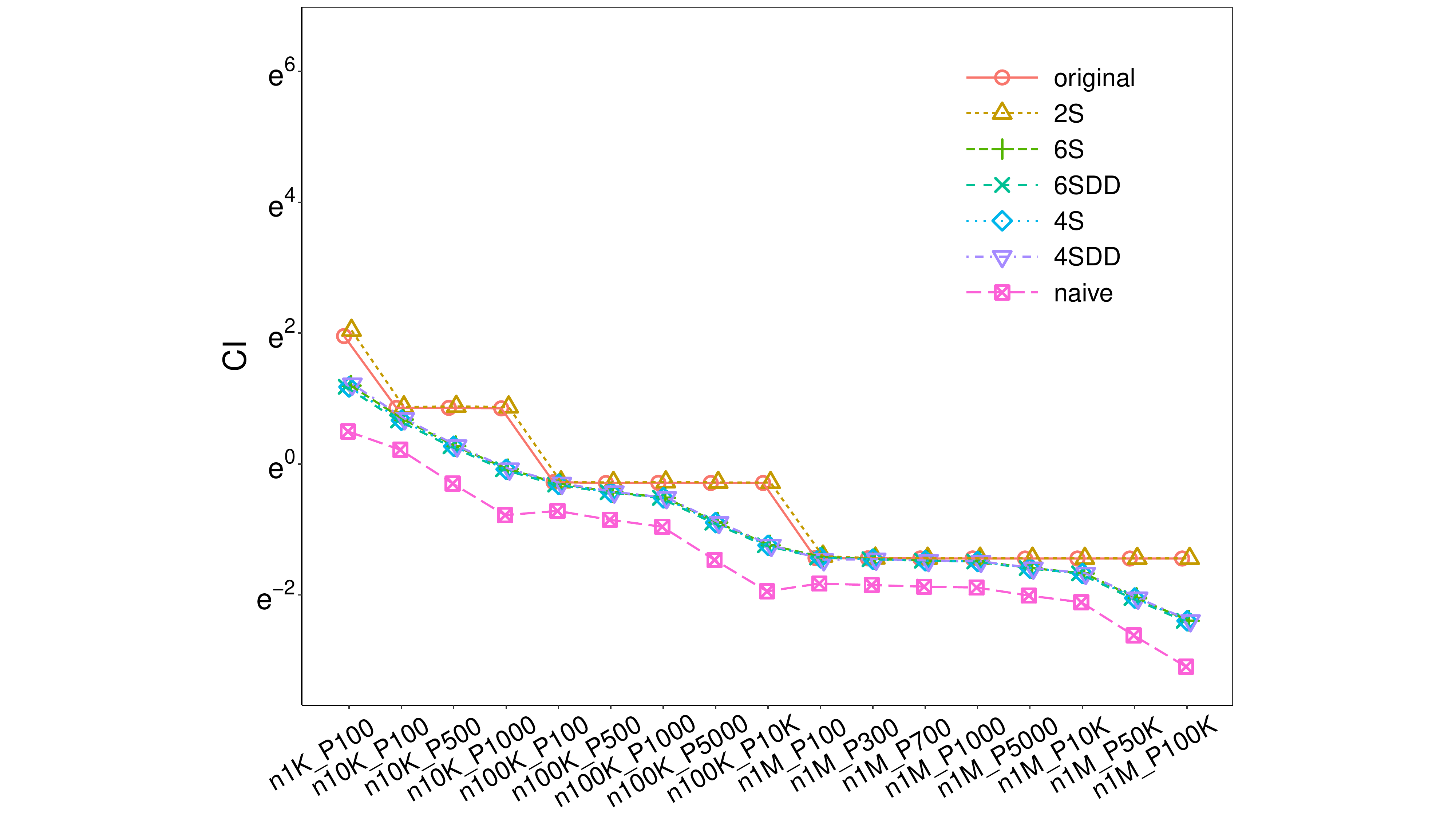}\\
\includegraphics[width=0.215\textwidth, trim={2.2in 0 2.2in 0},clip] {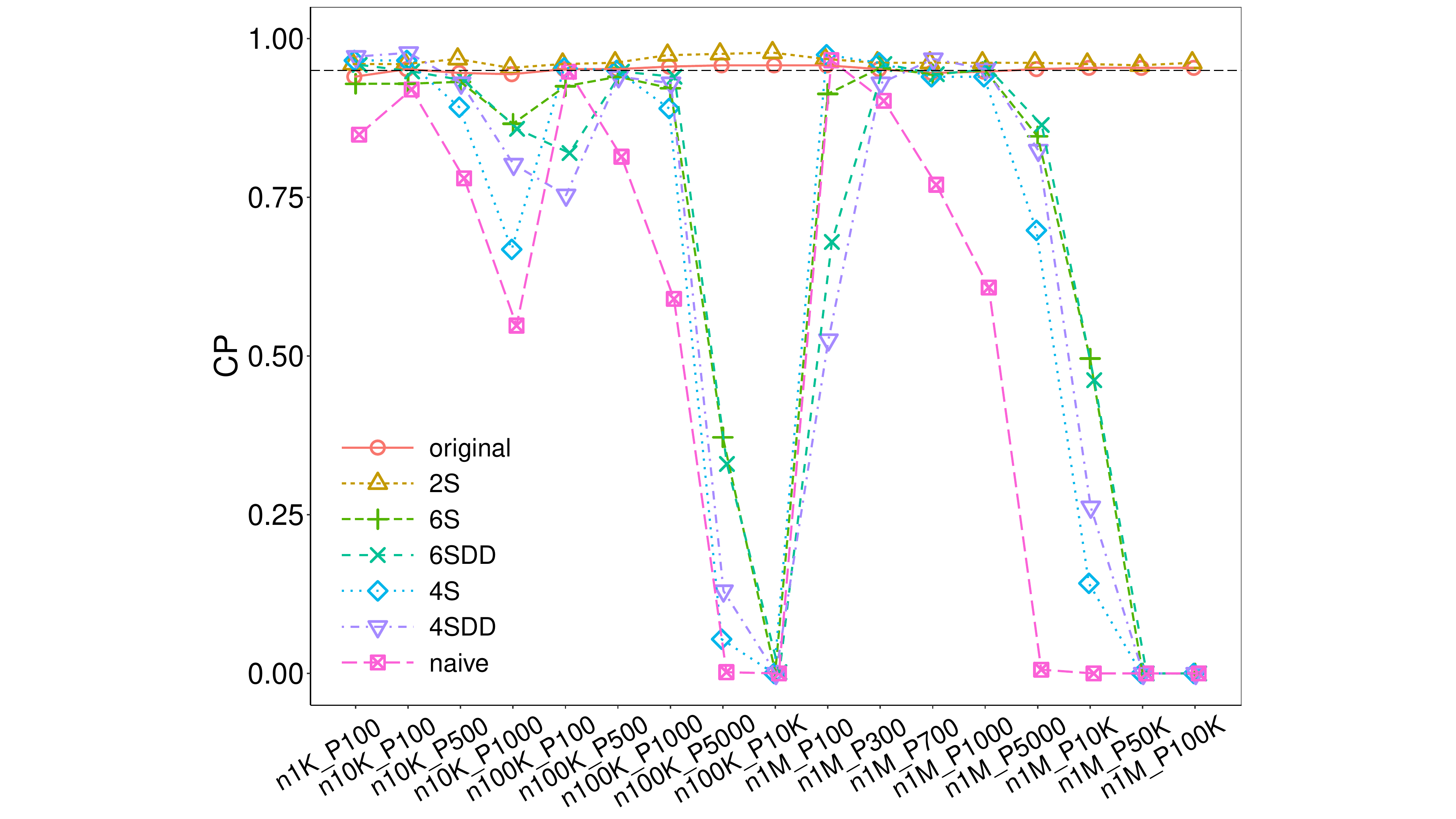}
\includegraphics[width=0.215\textwidth, trim={2.2in 0 2.2in 0},clip] {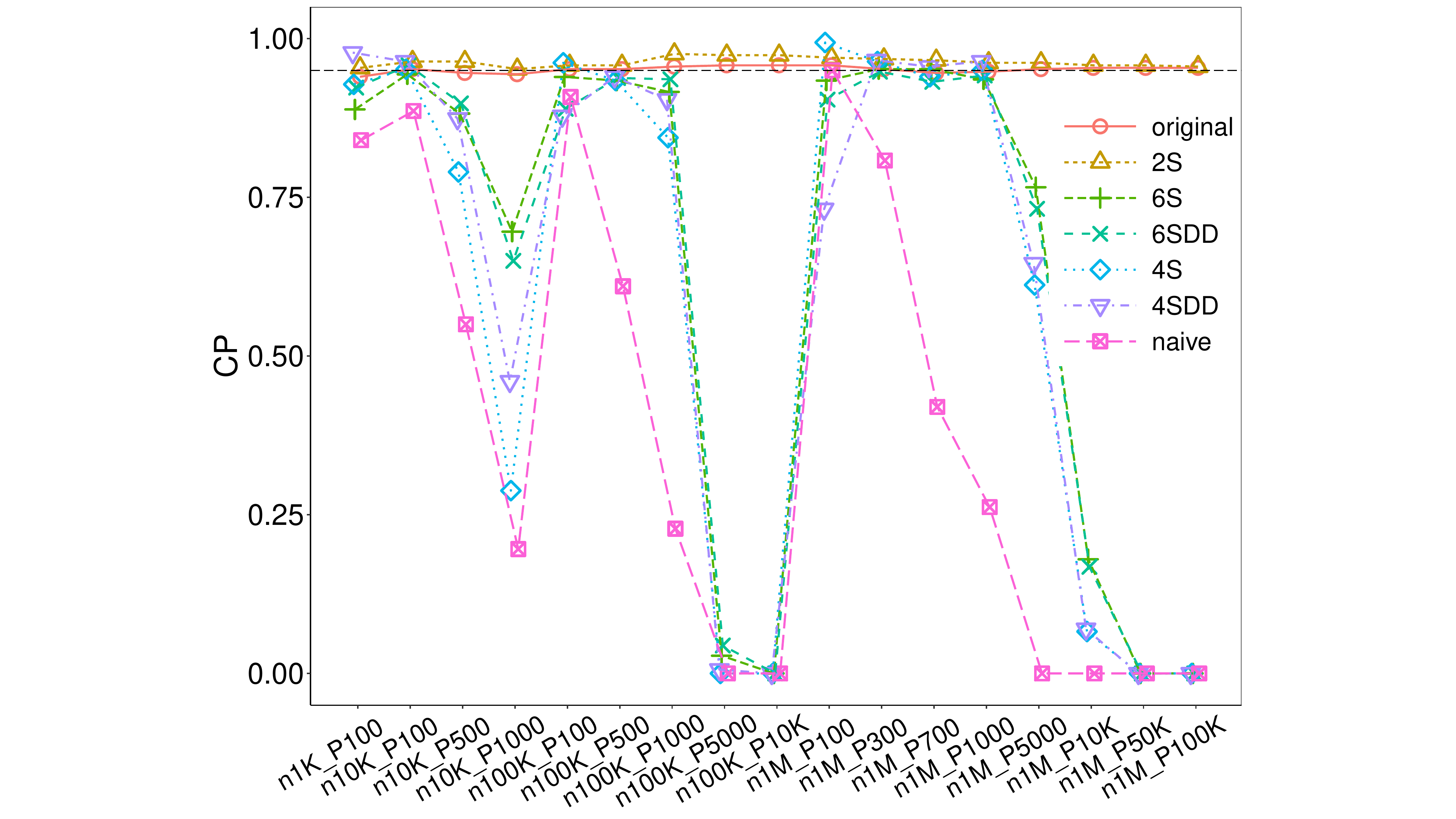}
\includegraphics[width=0.215\textwidth, trim={2.2in 0 2.2in 0},clip] {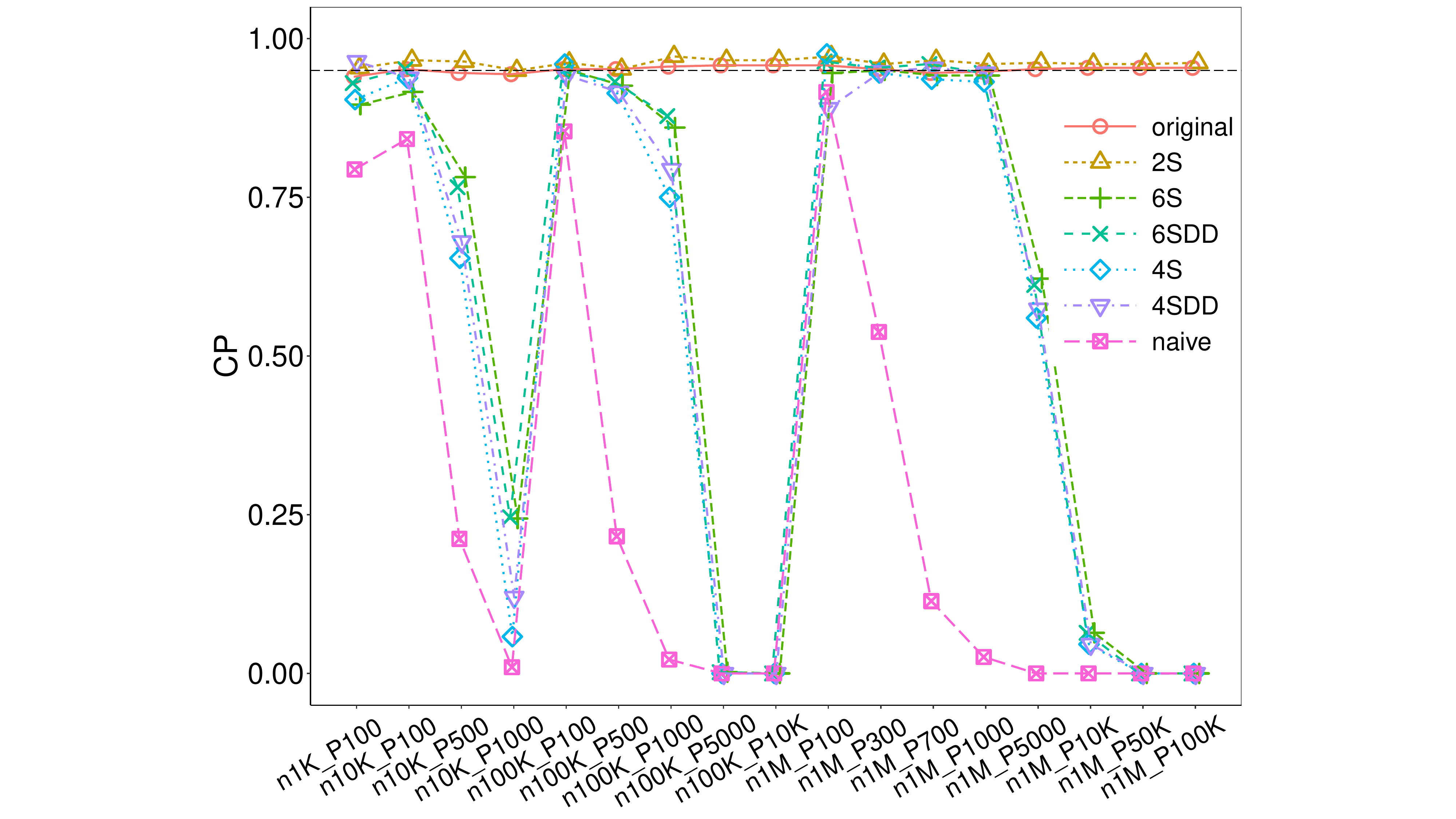}
\includegraphics[width=0.215\textwidth, trim={2.2in 0 2.2in 0},clip] {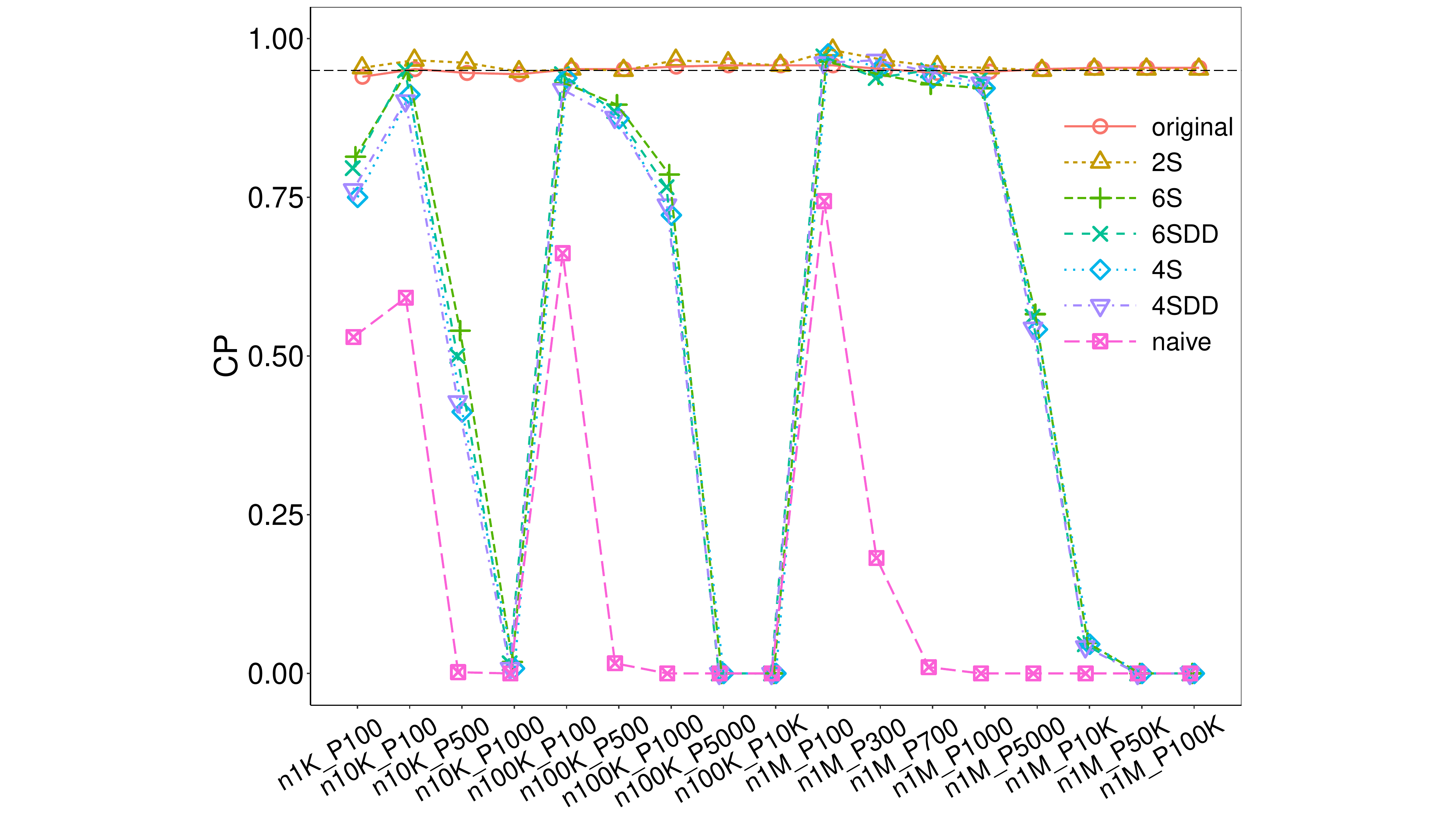}
\includegraphics[width=0.215\textwidth, trim={2.2in 0 2.2in 0},clip] {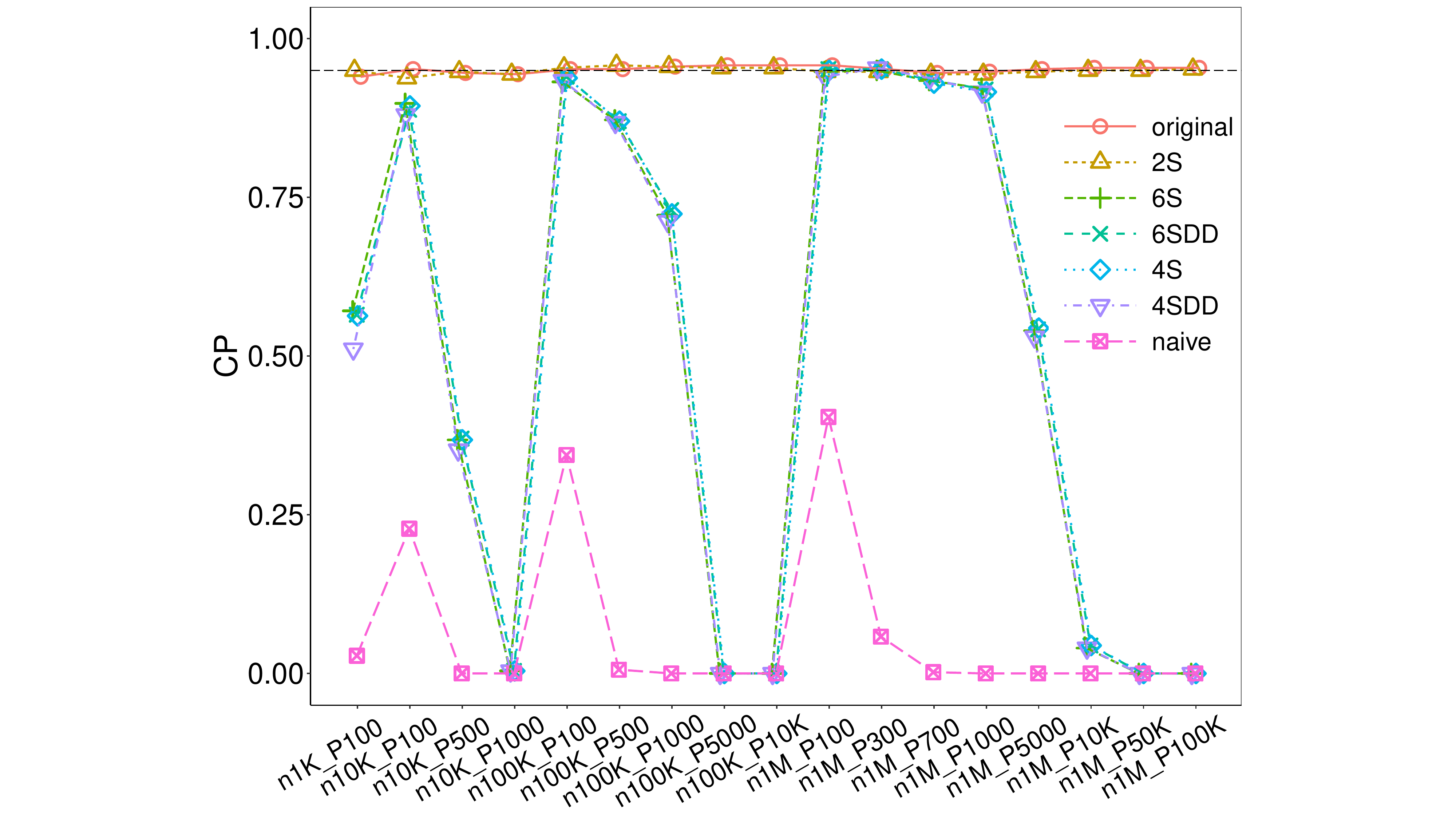}\\
\includegraphics[width=0.215\textwidth, trim={2.2in 0 2.2in 0},clip] {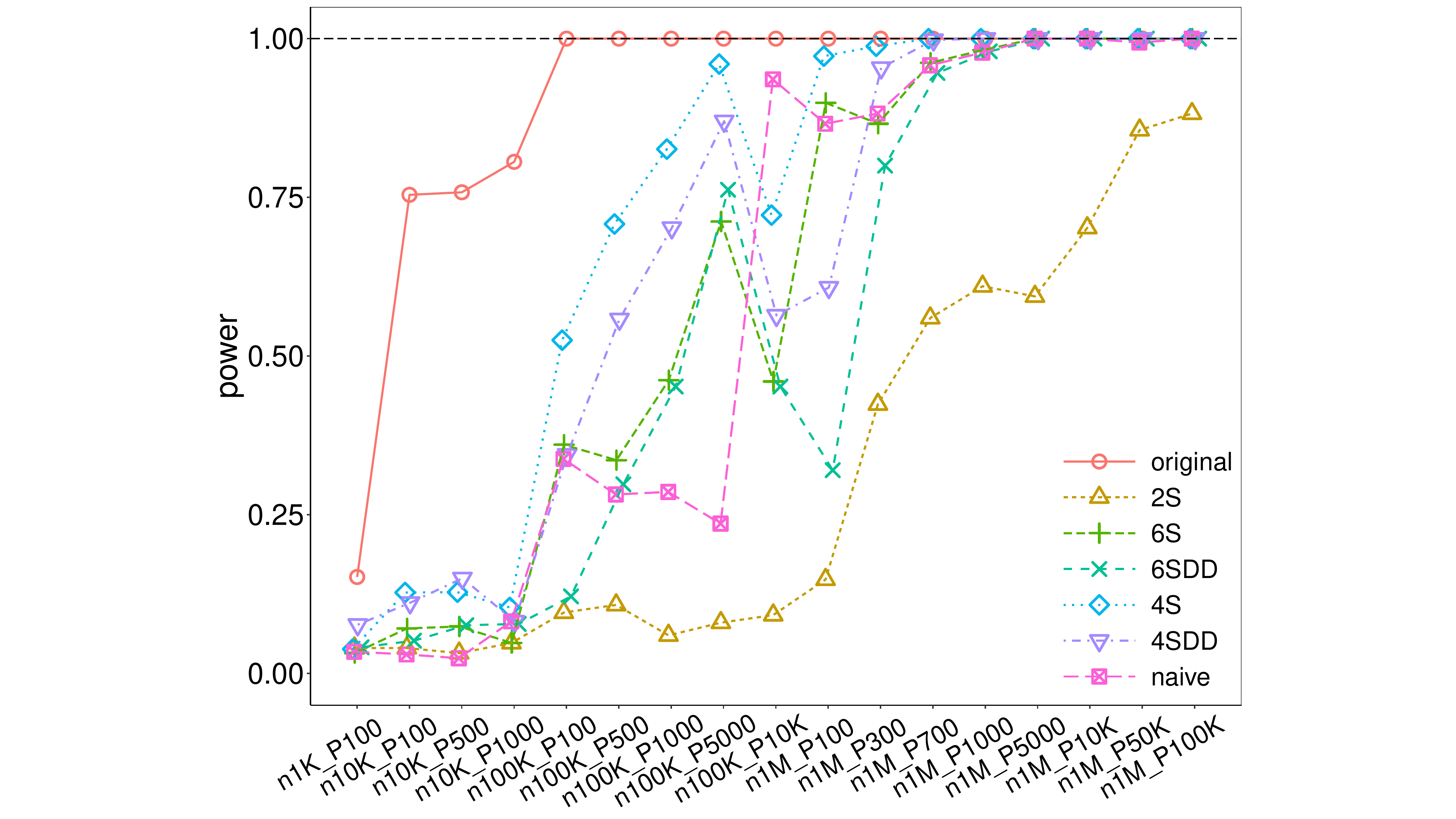}
\includegraphics[width=0.215\textwidth, trim={2.2in 0 2.2in 0},clip] {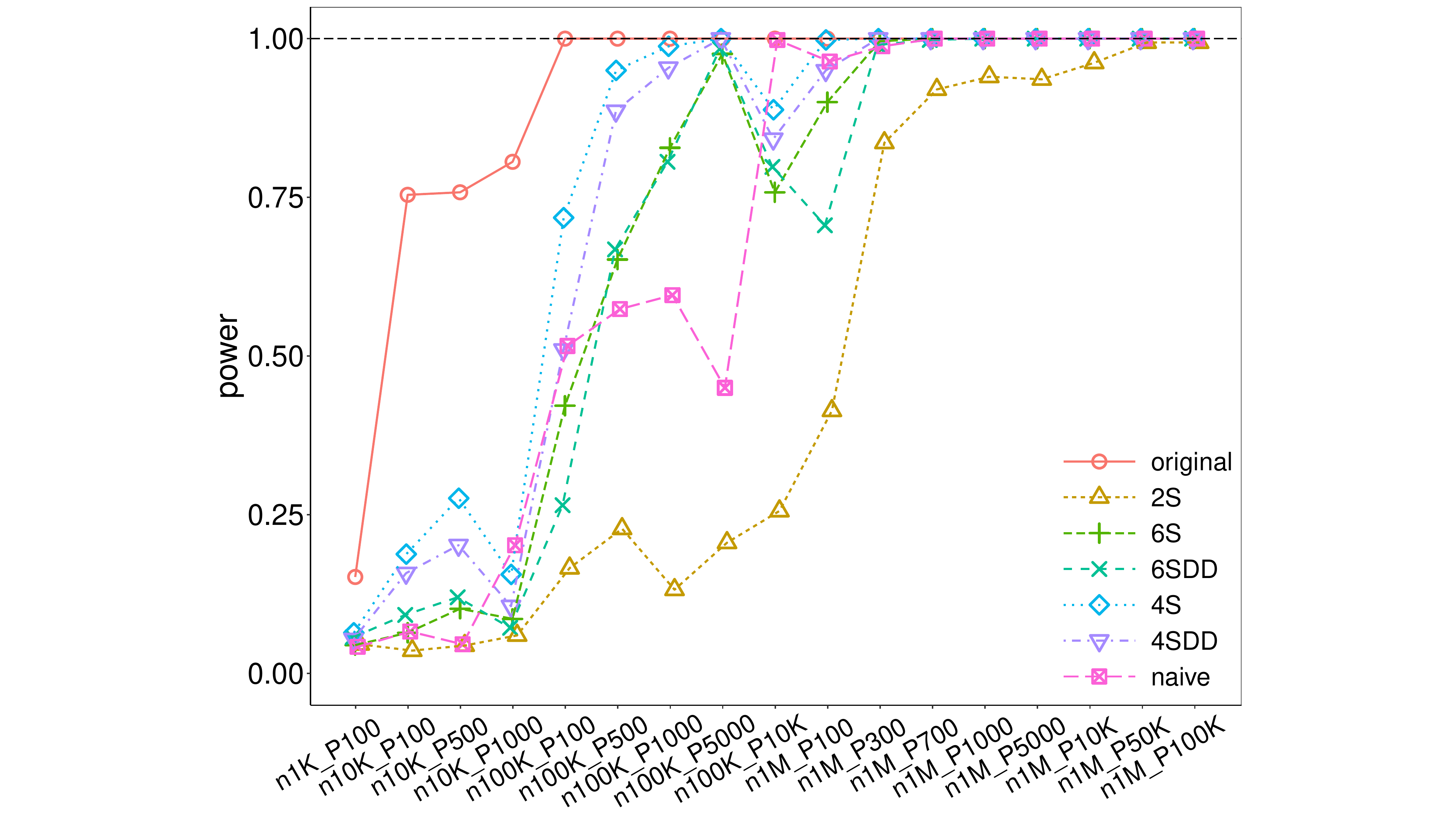}
\includegraphics[width=0.215\textwidth, trim={2.2in 0 2.2in 0},clip] {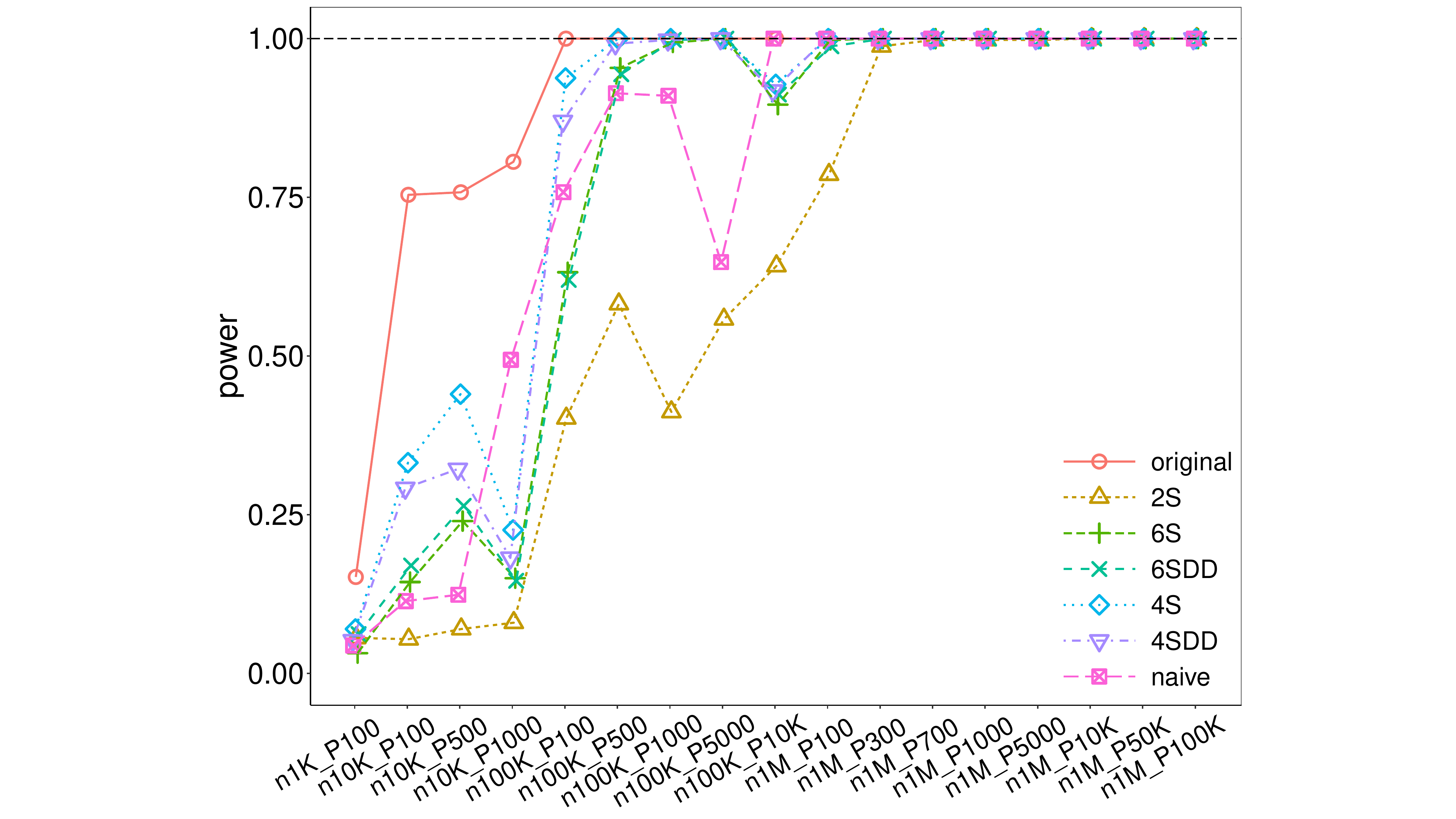}
\includegraphics[width=0.215\textwidth, trim={2.2in 0 2.2in 0},clip] {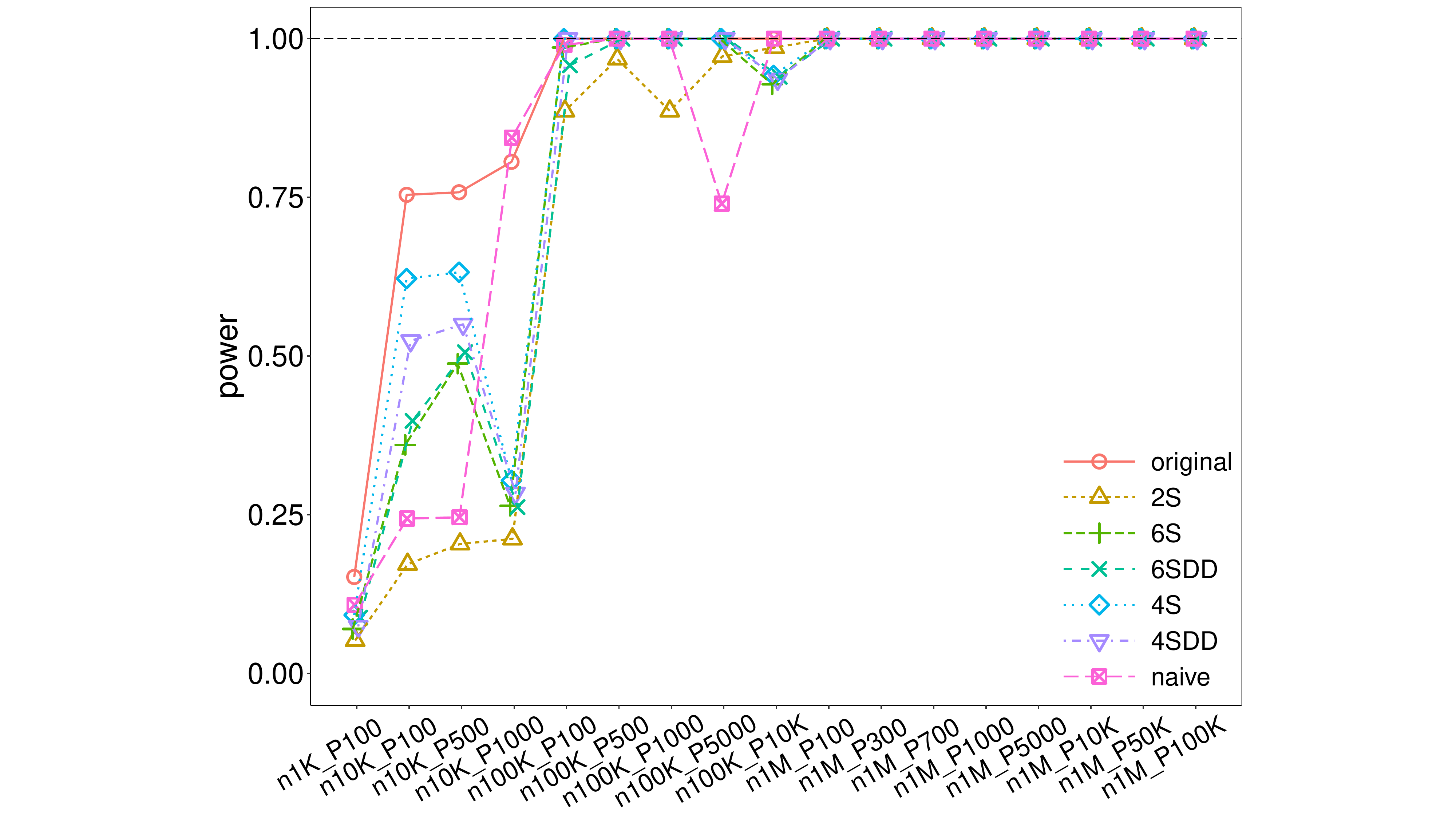}
\includegraphics[width=0.215\textwidth, trim={2.2in 0 2.2in 0},clip] {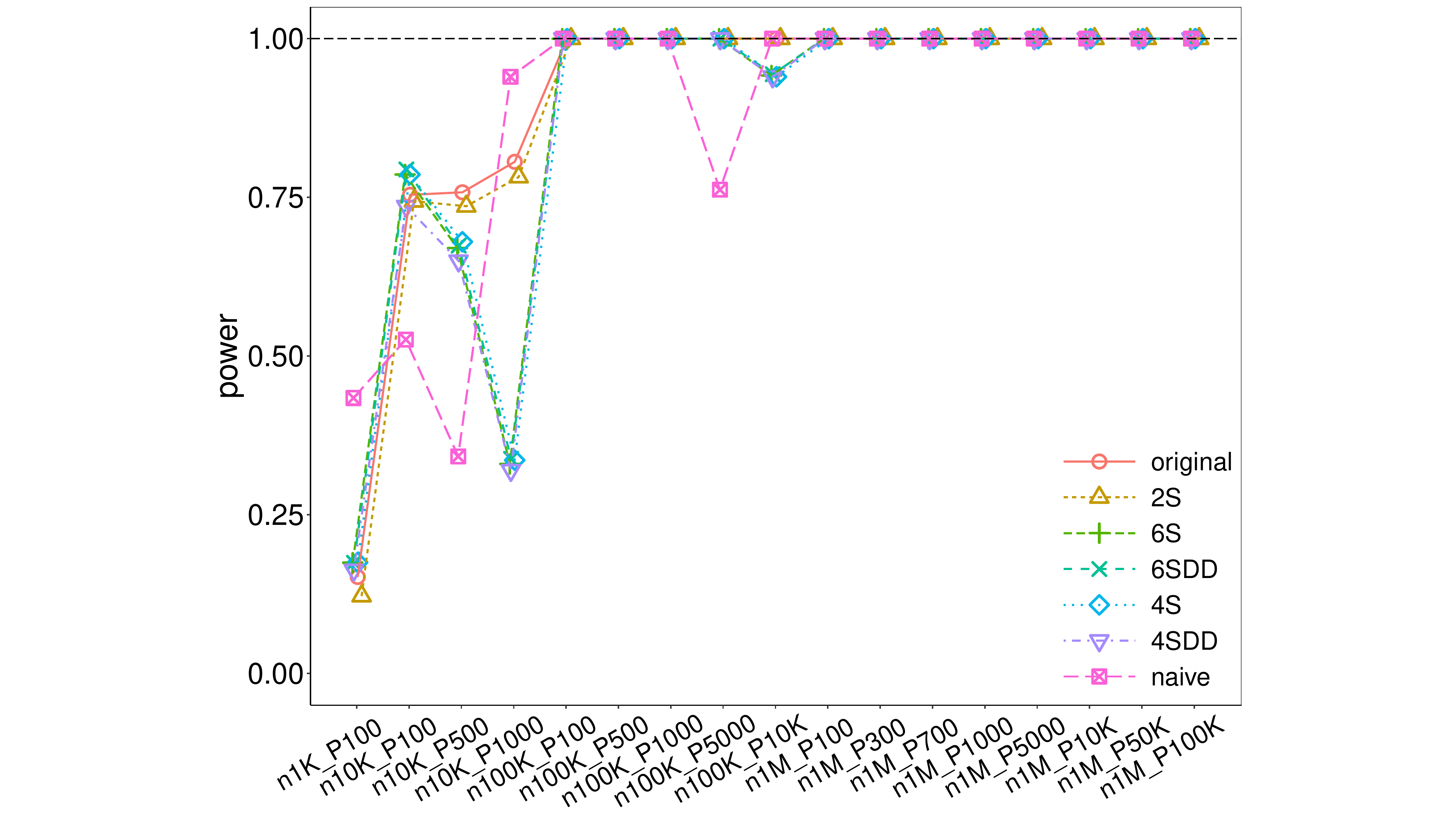}\\
\caption{ZILN data; $\epsilon$-DP; $\theta\ne0$ and $\alpha\ne\beta$} \label{fig:1asDPziln}
\end{figure}
\end{landscape}

\begin{landscape}
\begin{figure}[!htb]
\centering
$\rho=0.005$\hspace{1in}$\rho=0.02$\hspace{1in}$\rho=0.08$
\hspace{1in}$\rho=0.32$\hspace{0.8in}$\rho=1.28$\\
\includegraphics[width=0.24\textwidth, trim={2.2in 0 2.2in 0},clip] {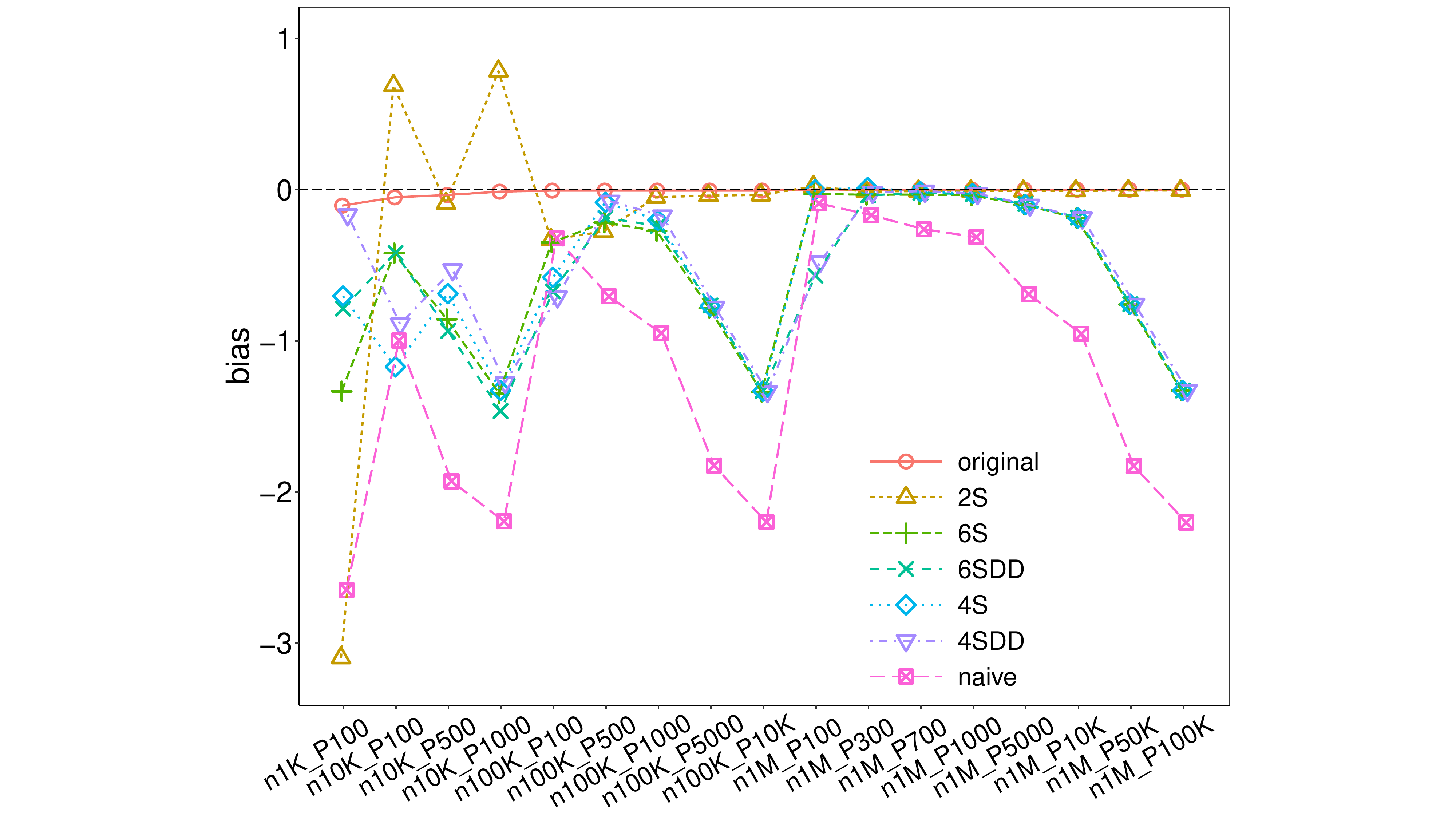}
\includegraphics[width=0.24\textwidth, trim={2.2in 0 2.2in 0},clip] {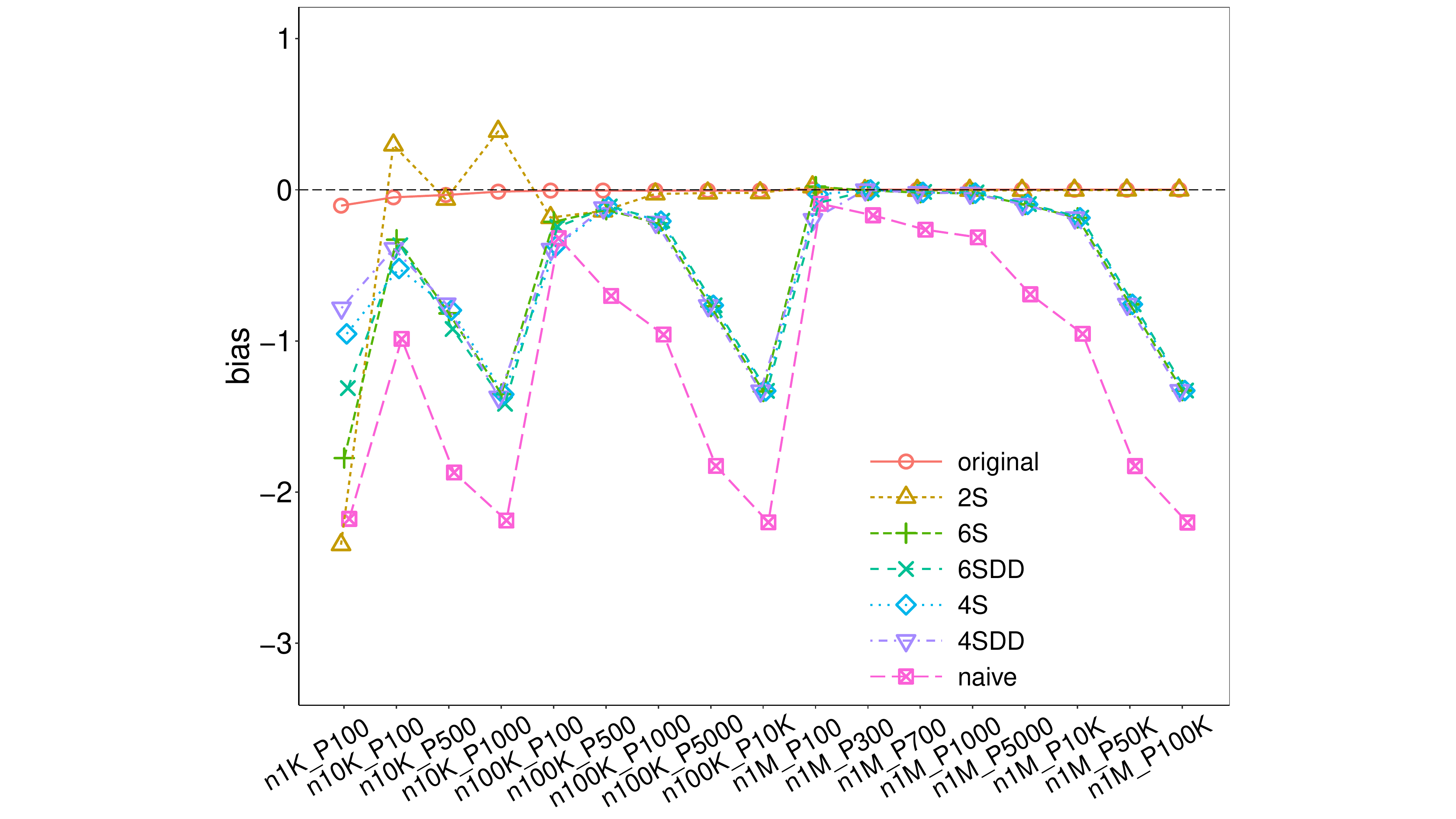}
\includegraphics[width=0.24\textwidth, trim={2.2in 0 2.2in 0},clip] {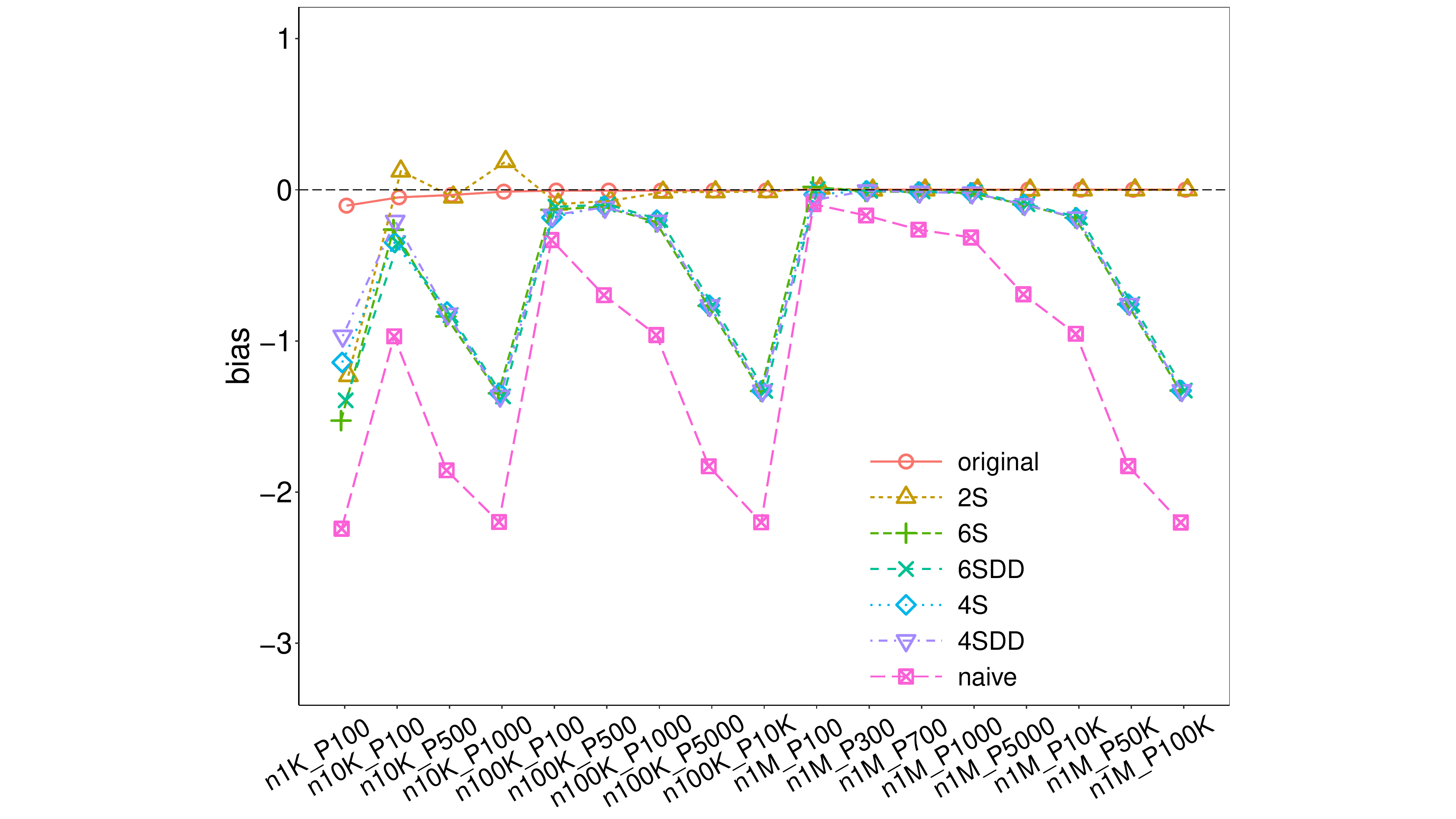}
\includegraphics[width=0.24\textwidth, trim={2.2in 0 2.2in 0},clip] {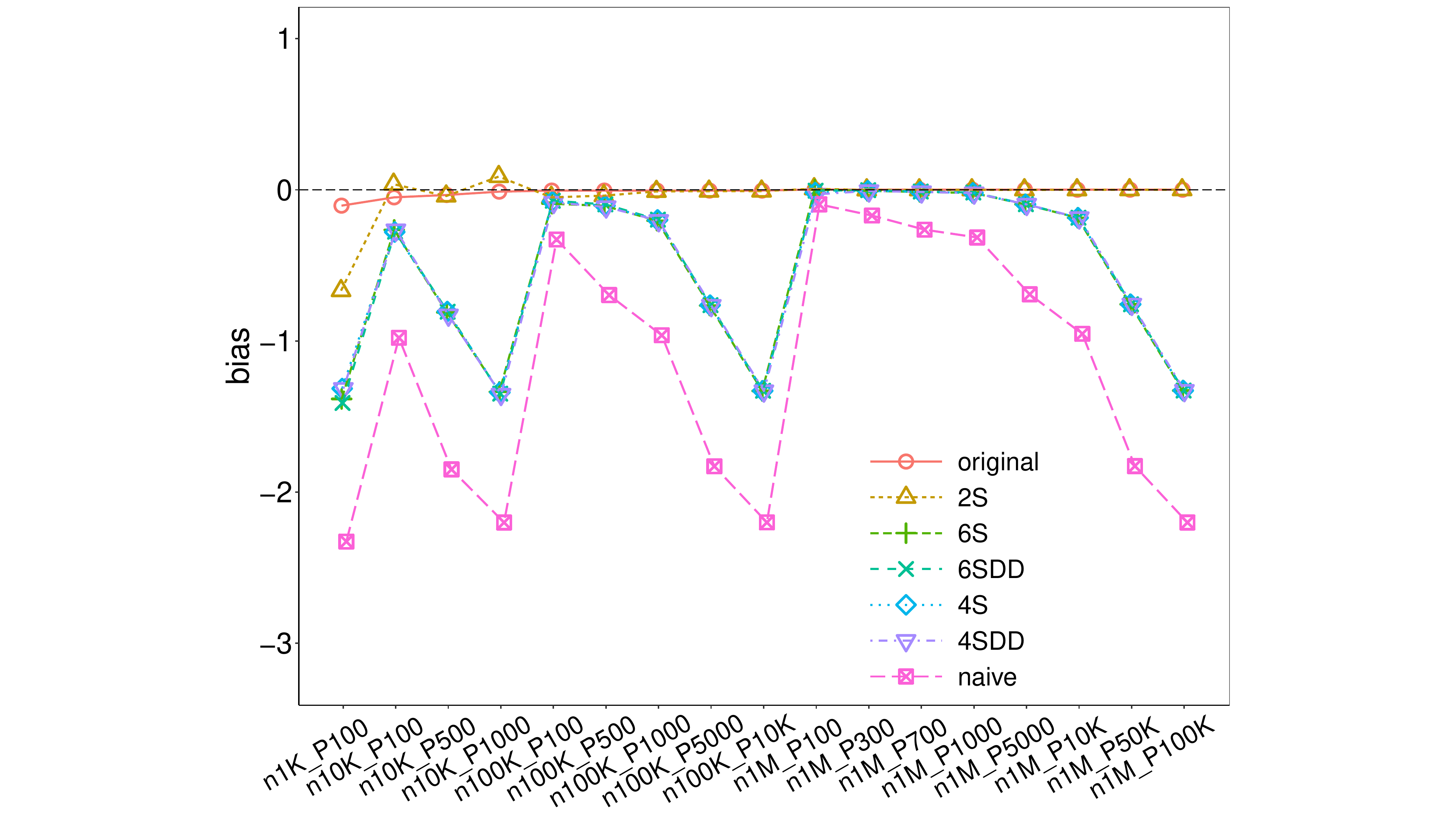}
\includegraphics[width=0.24\textwidth, trim={2.2in 0 2.2in 0},clip] {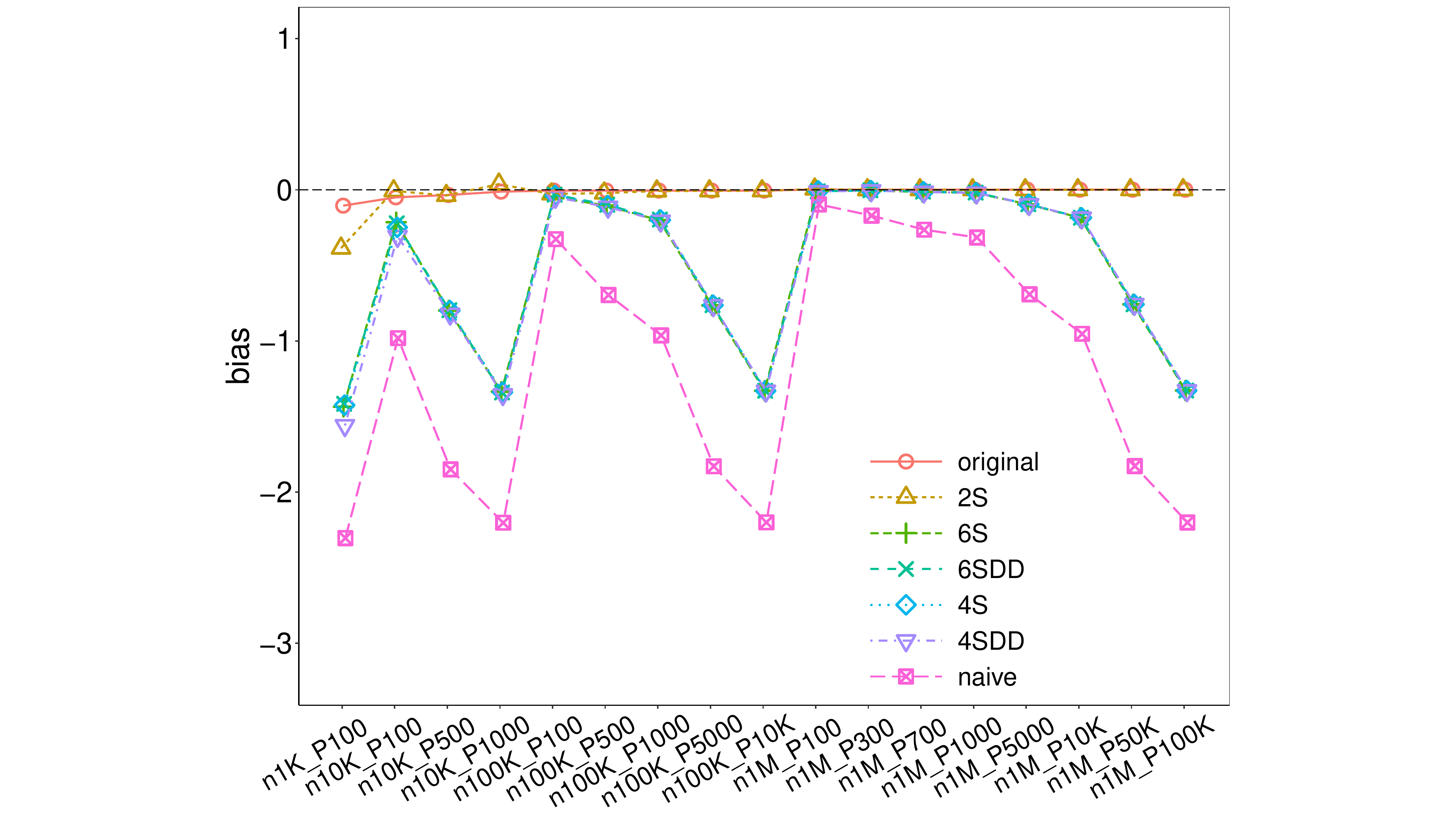}\\
\includegraphics[width=0.24\textwidth, trim={2.2in 0 2.2in 0},clip] {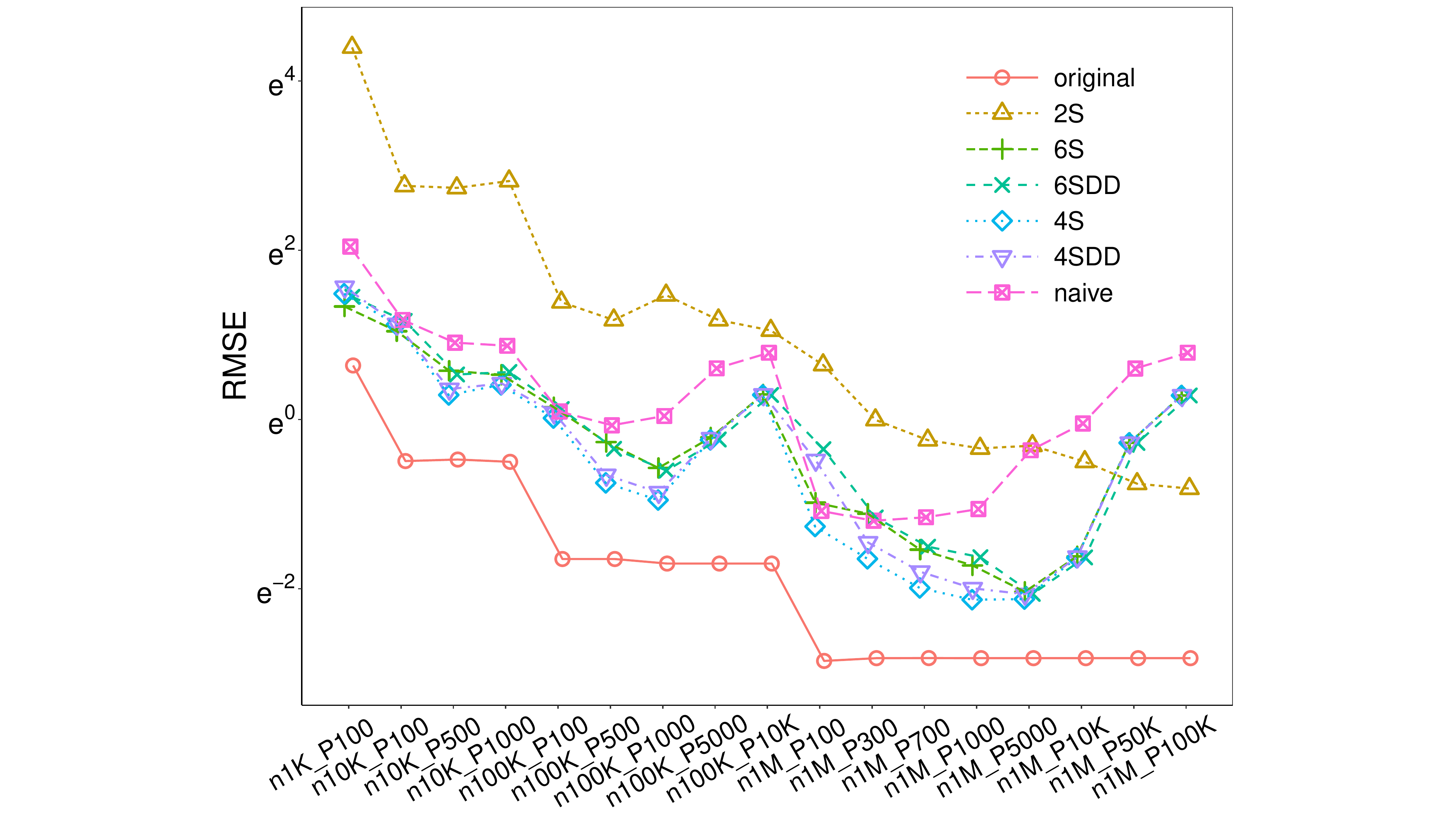}
\includegraphics[width=0.24\textwidth, trim={2.2in 0 2.2in 0},clip] {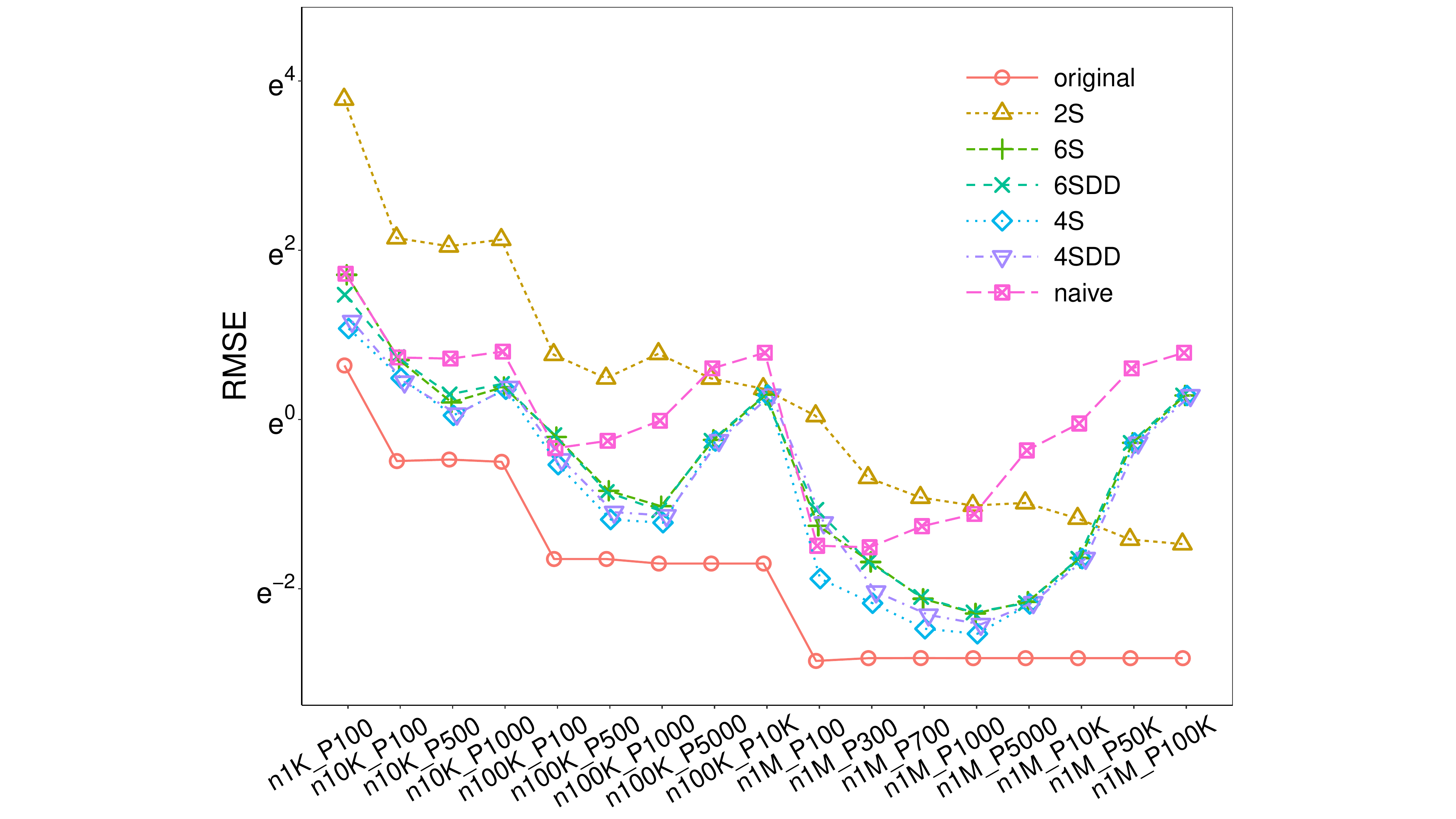}
\includegraphics[width=0.24\textwidth, trim={2.2in 0 2.2in 0},clip] {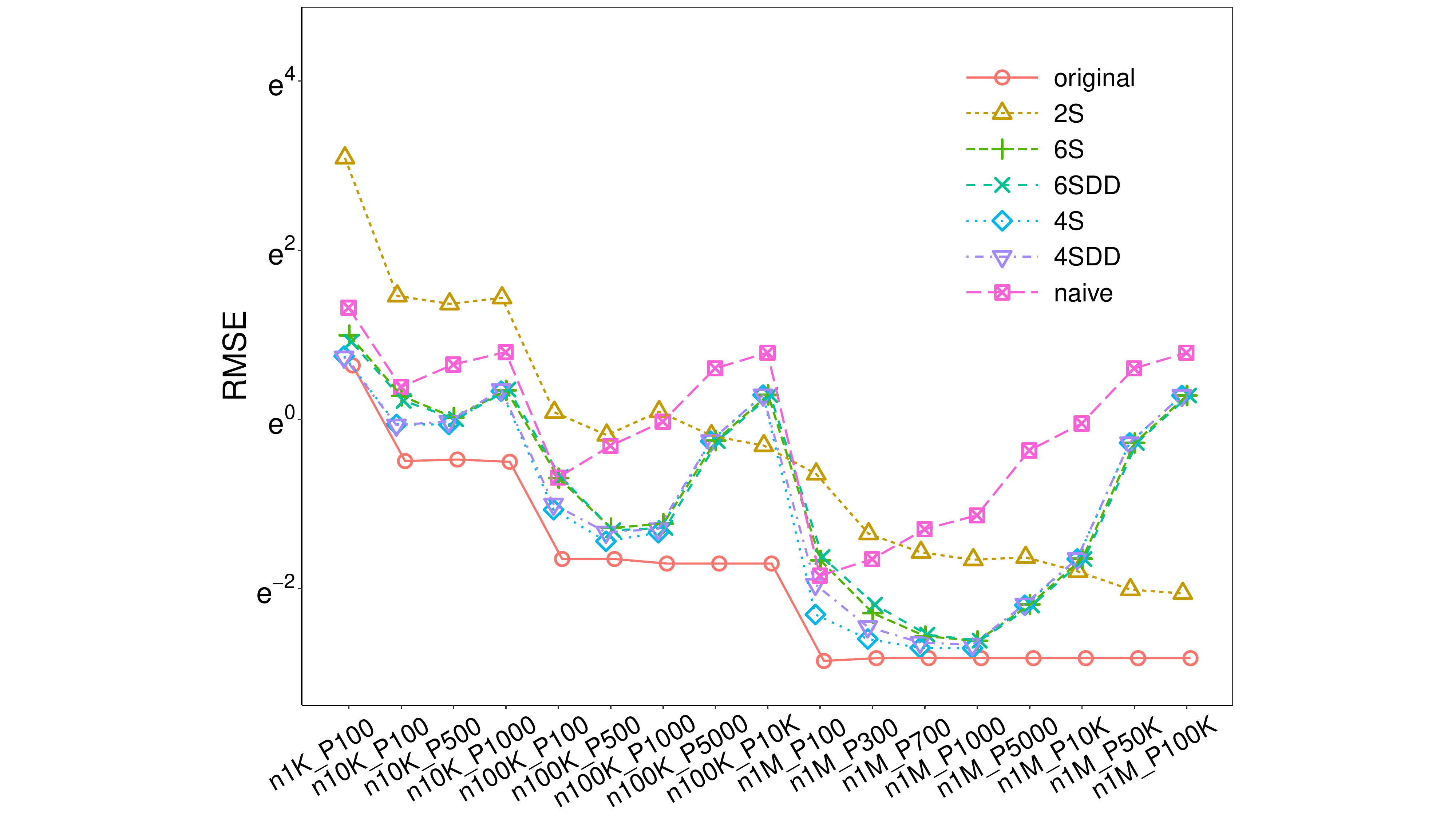}
\includegraphics[width=0.24\textwidth, trim={2.2in 0 2.2in 0},clip] {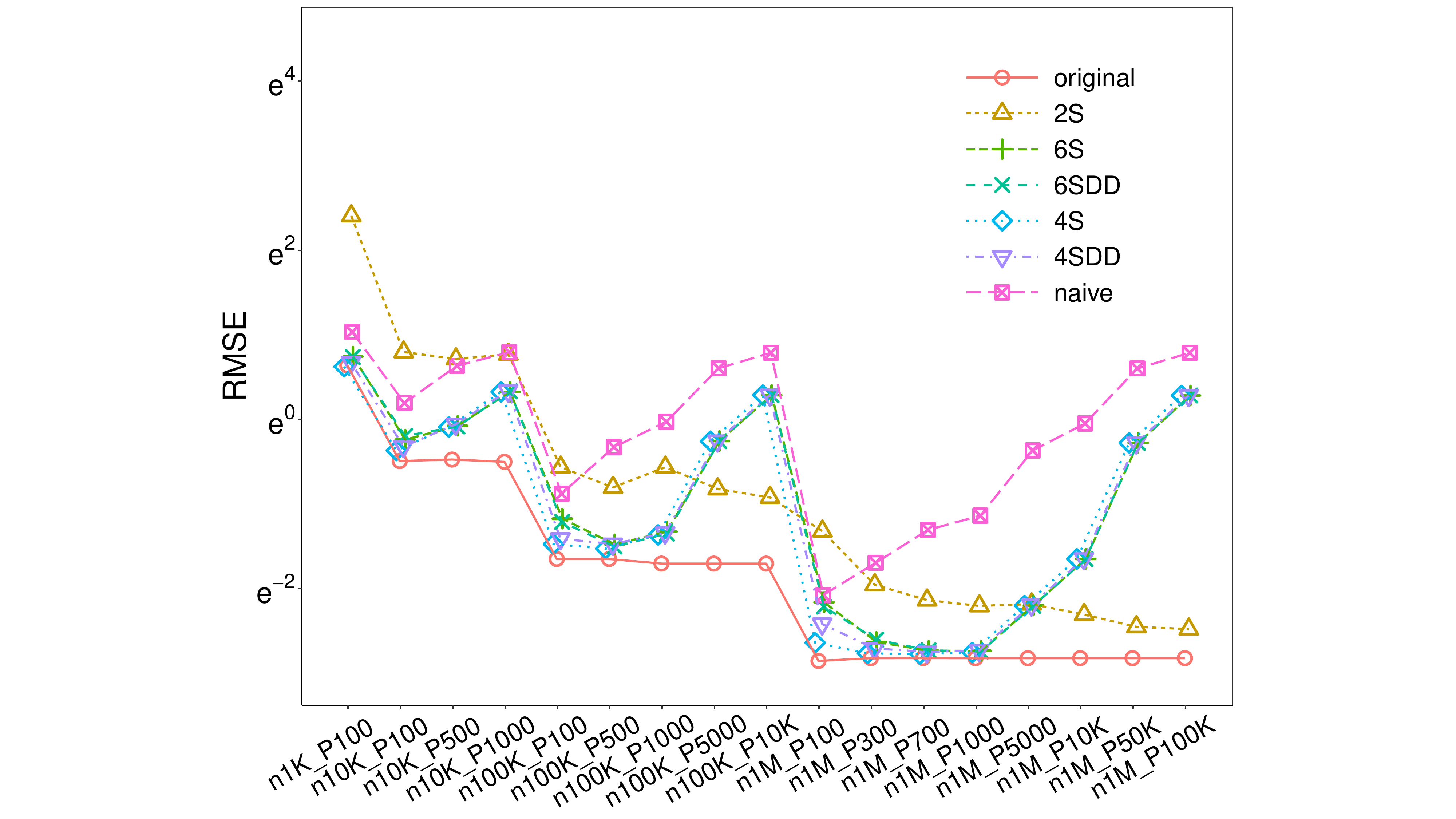}
\includegraphics[width=0.24\textwidth, trim={2.2in 0 2.2in 0},clip] {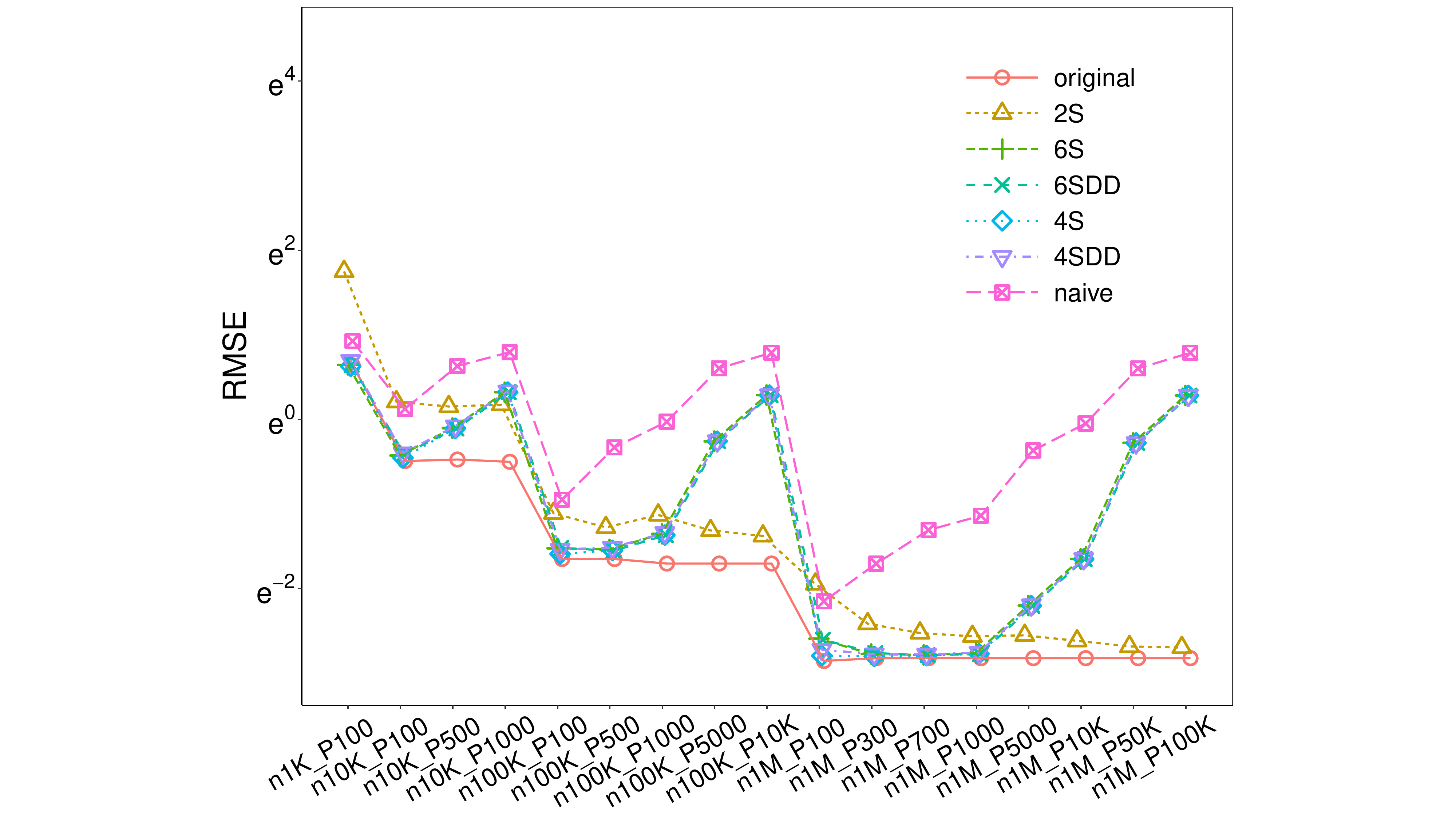}\\
\includegraphics[width=0.24\textwidth, trim={2.2in 0 2.2in 0},clip] {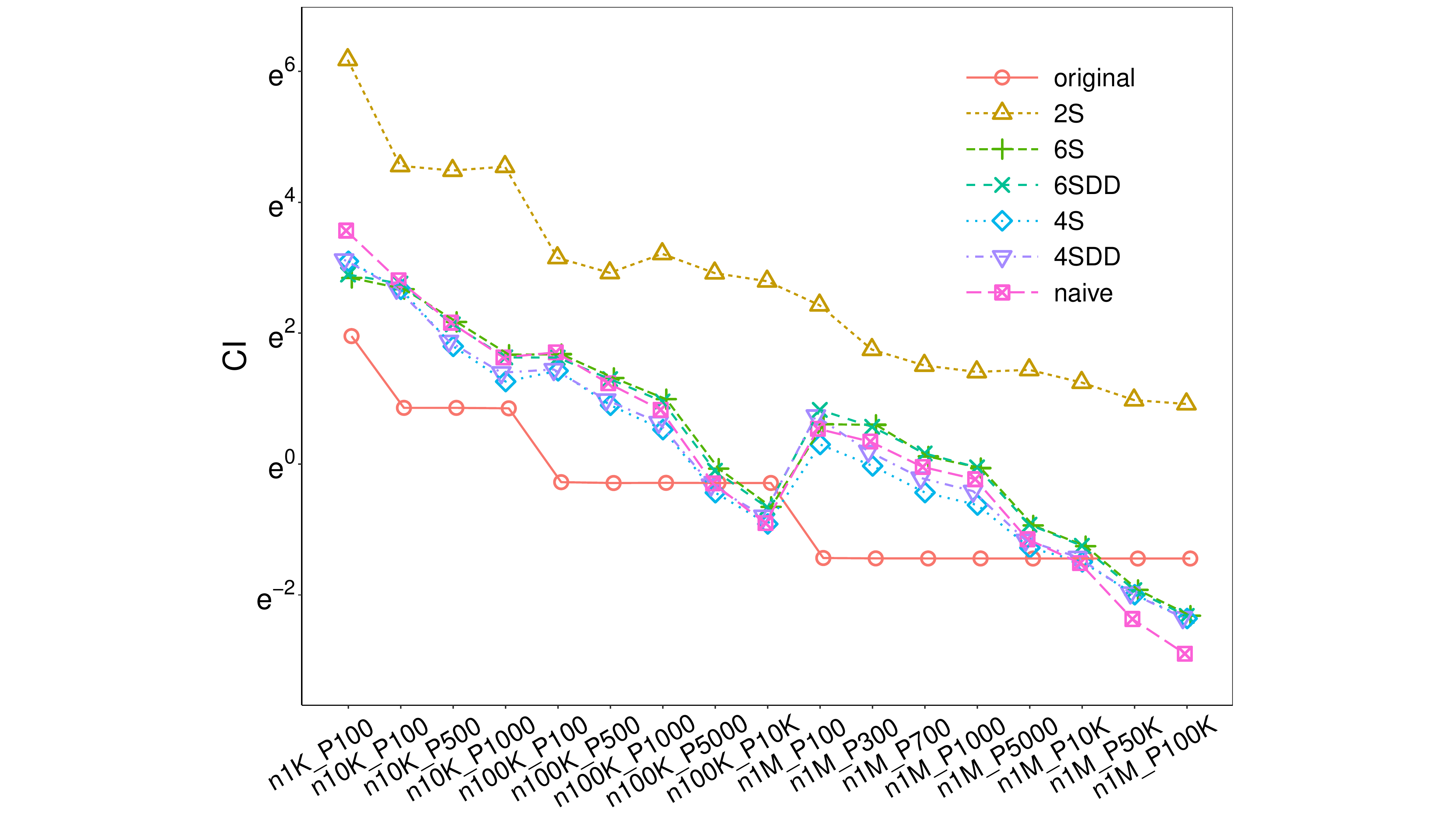}
\includegraphics[width=0.24\textwidth, trim={2.2in 0 2.2in 0},clip] {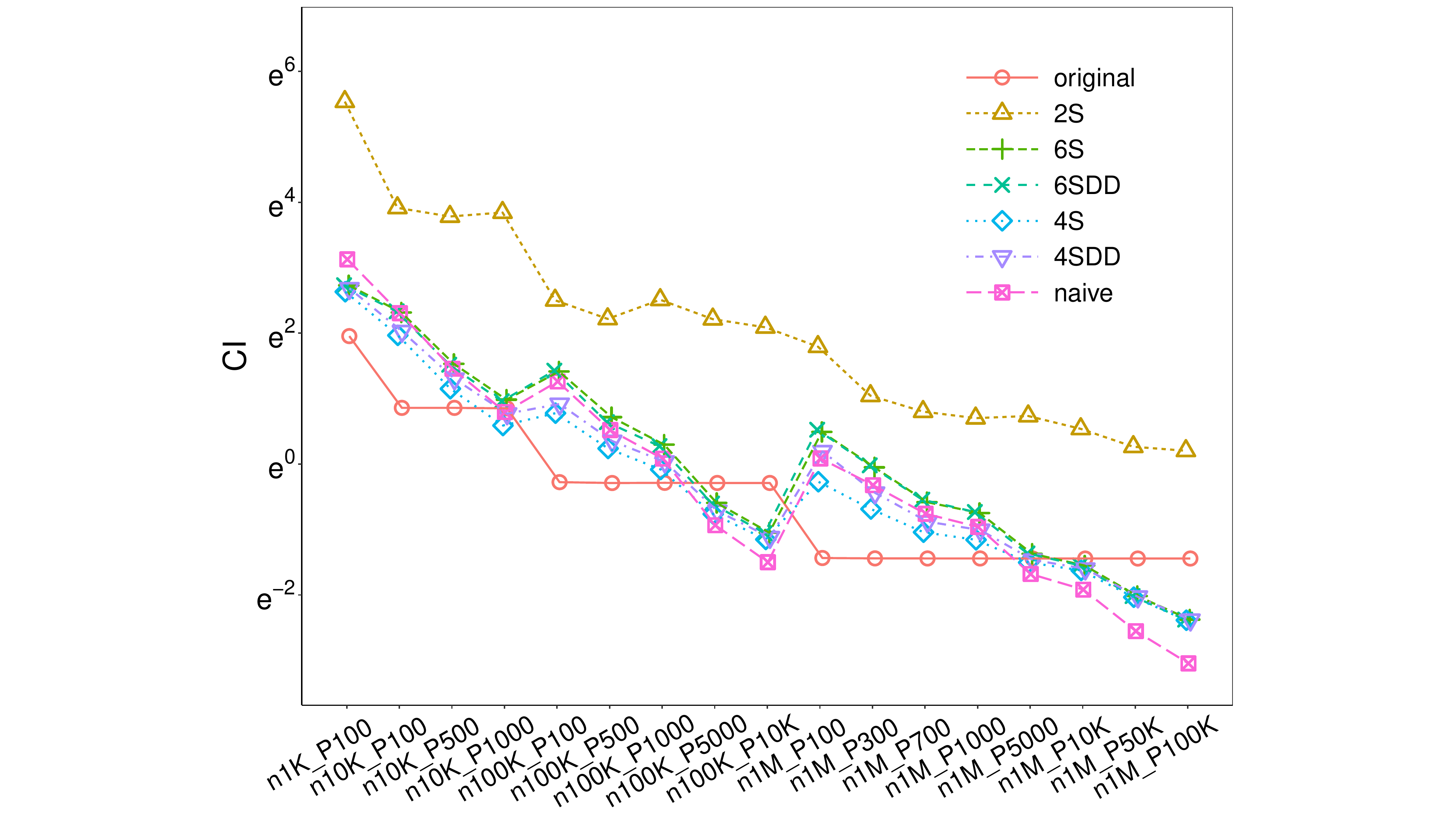}
\includegraphics[width=0.24\textwidth, trim={2.2in 0 2.2in 0},clip] {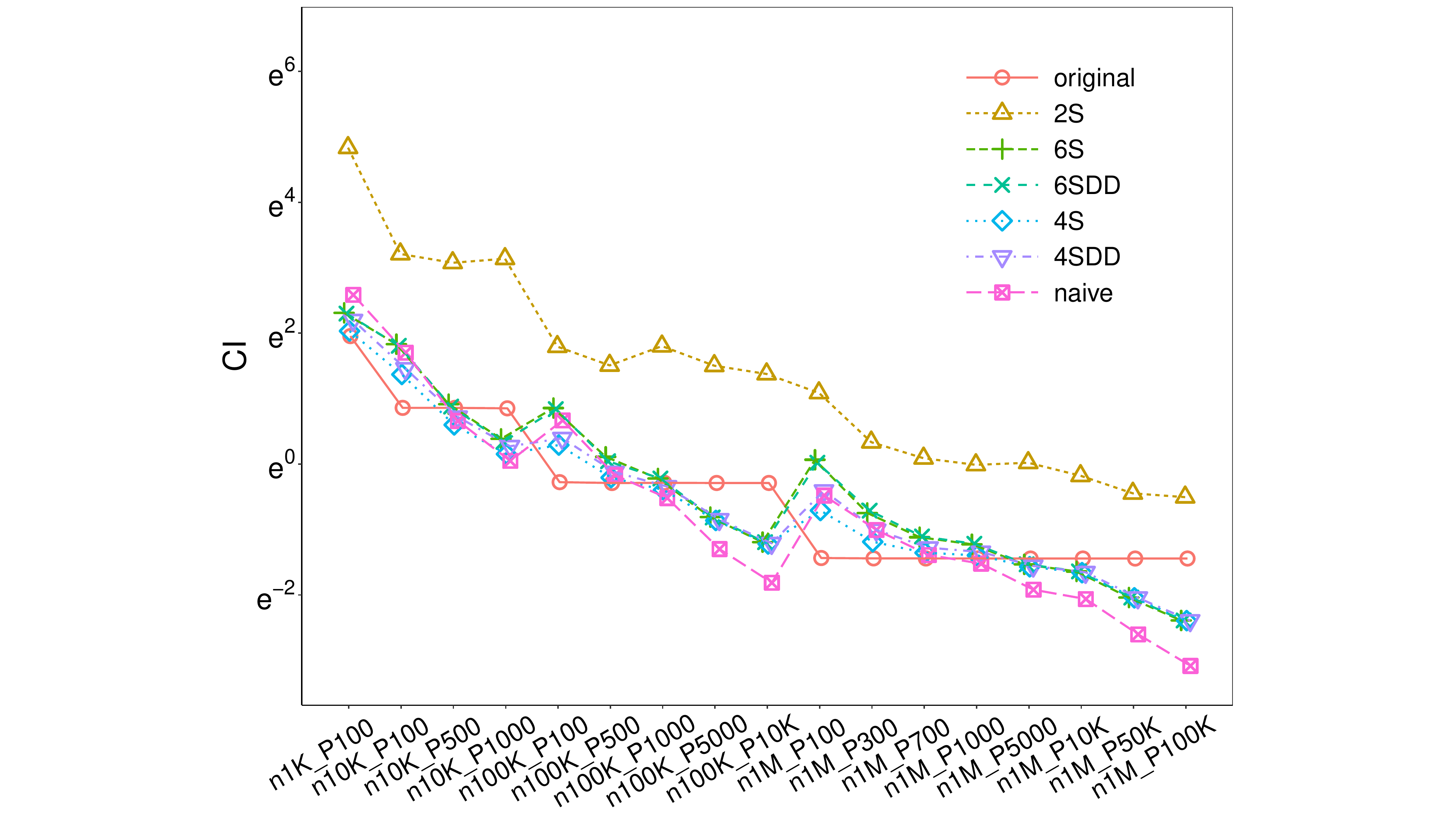}
\includegraphics[width=0.24\textwidth, trim={2.2in 0 2.2in 0},clip] {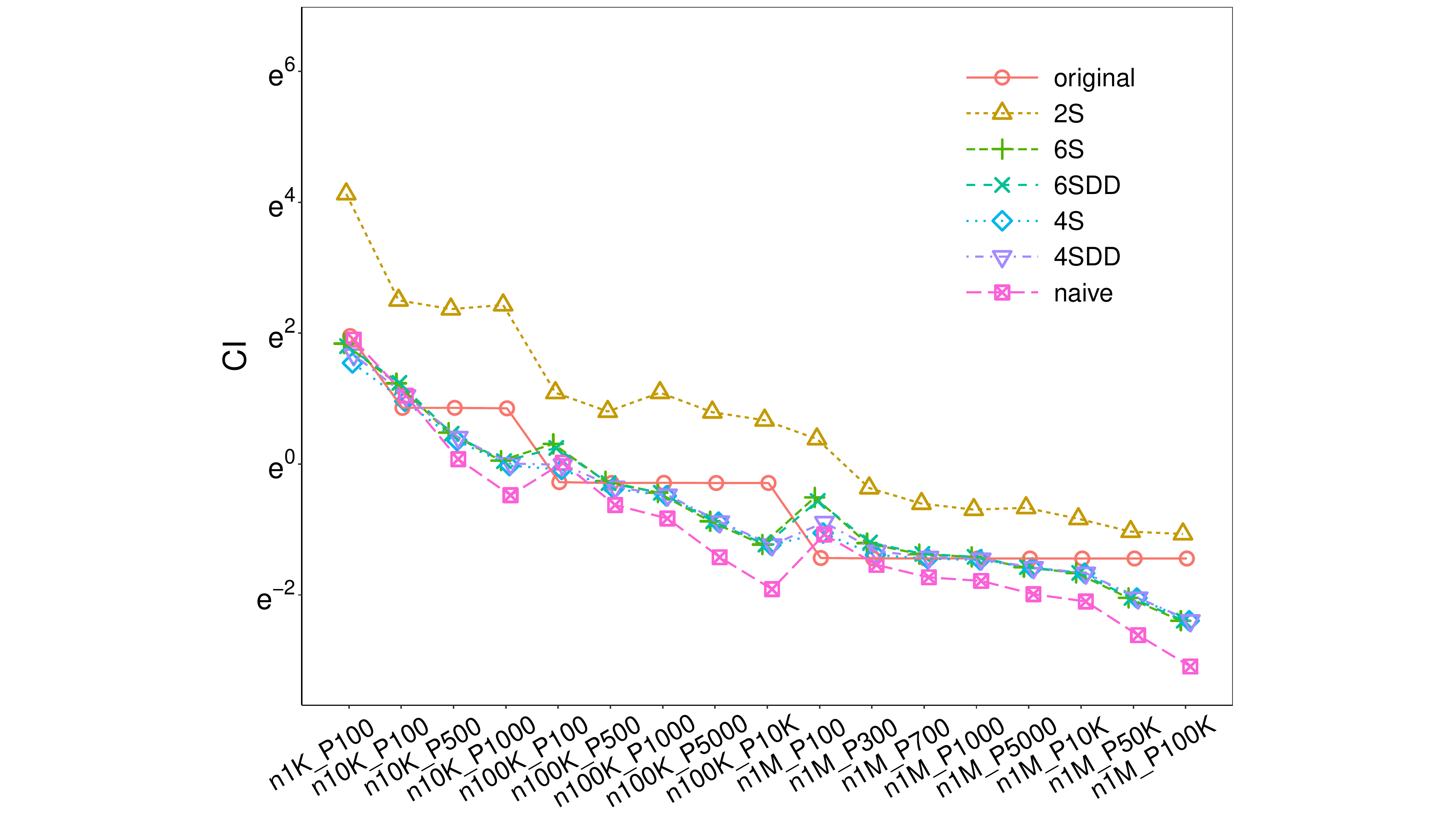}
\includegraphics[width=0.24\textwidth, trim={2.2in 0 2.2in 0},clip] {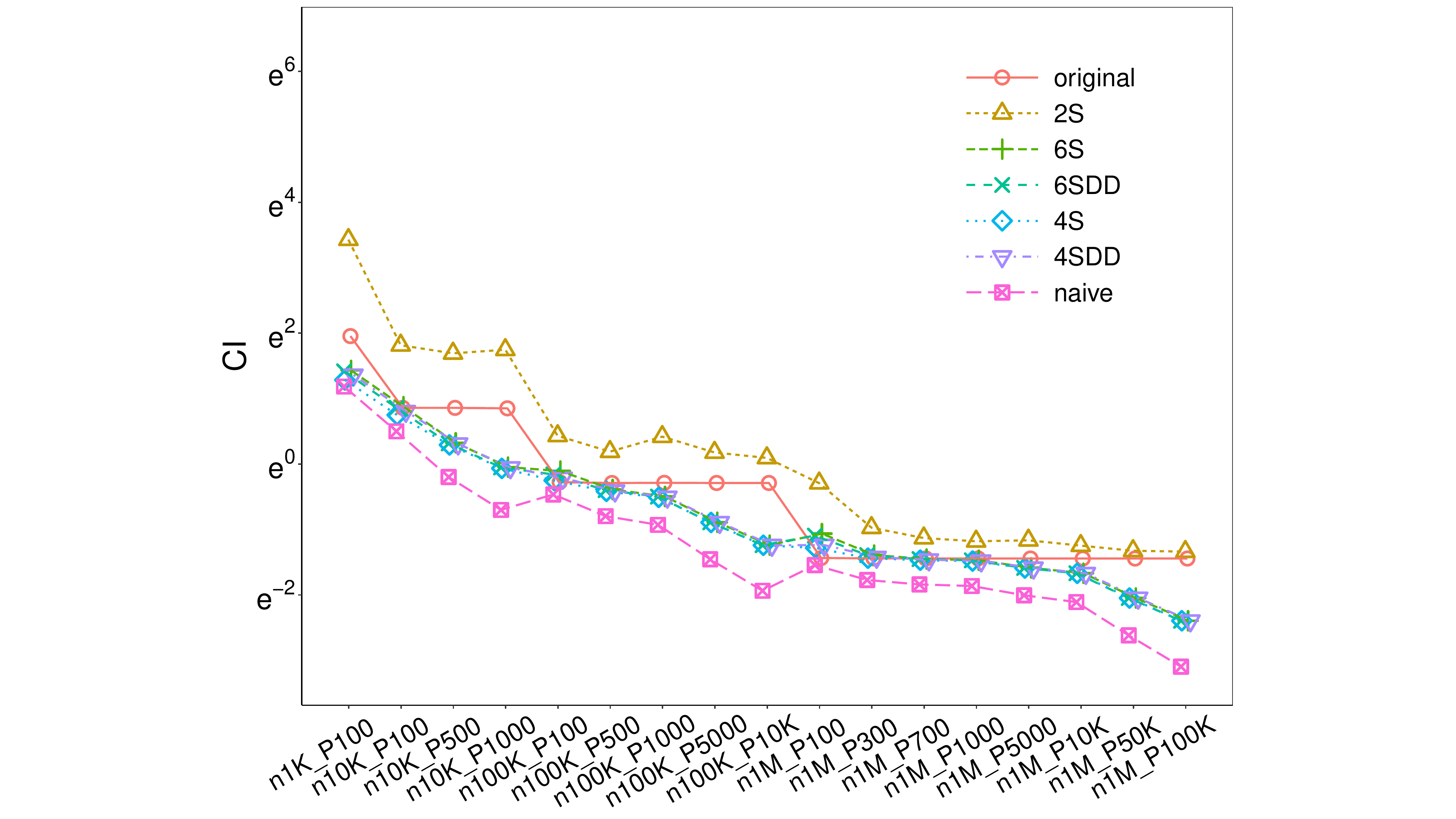}\\
\includegraphics[width=0.24\textwidth, trim={2.2in 0 2.2in 0},clip] {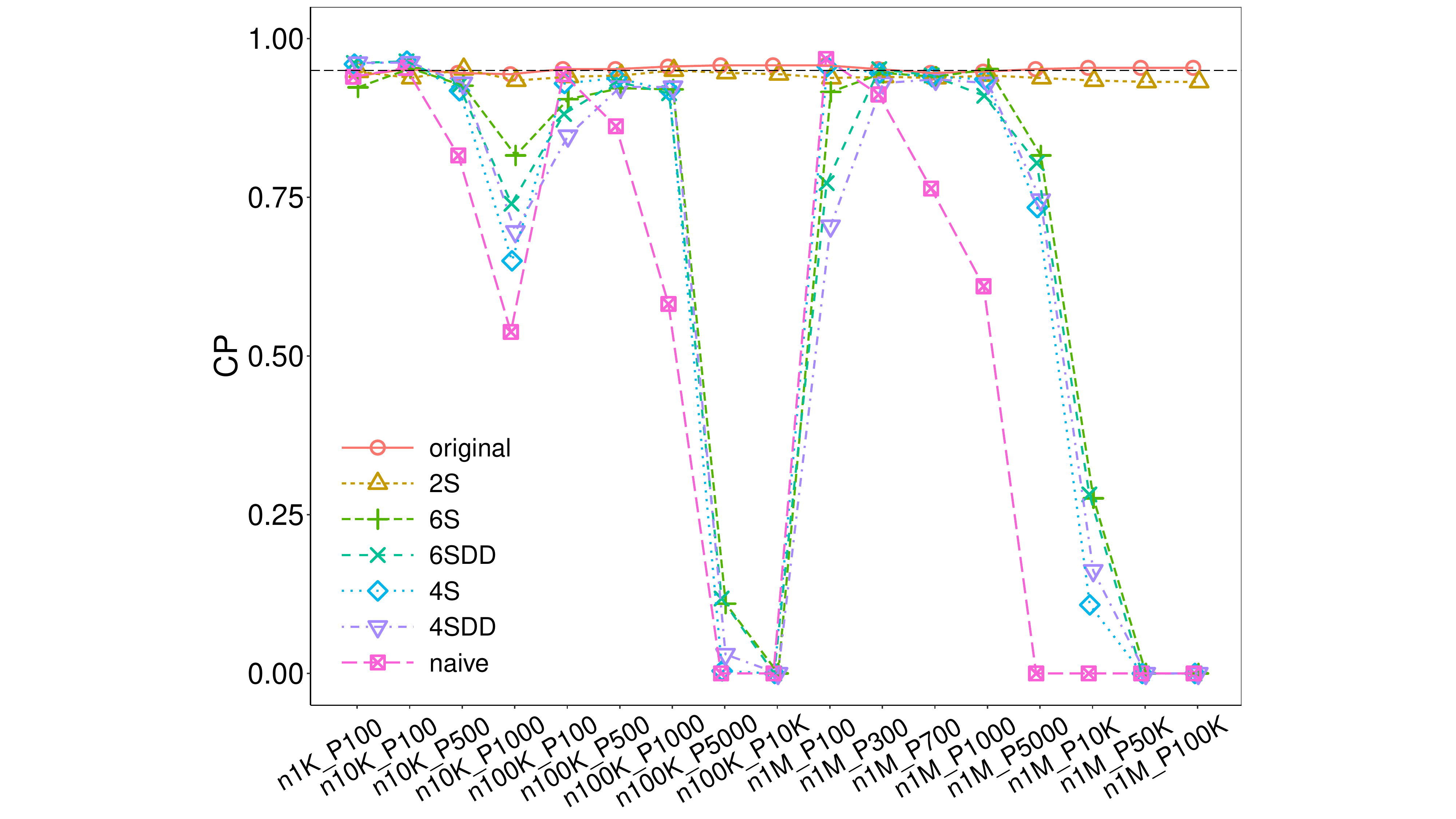}
\includegraphics[width=0.24\textwidth, trim={2.2in 0 2.2in 0},clip] {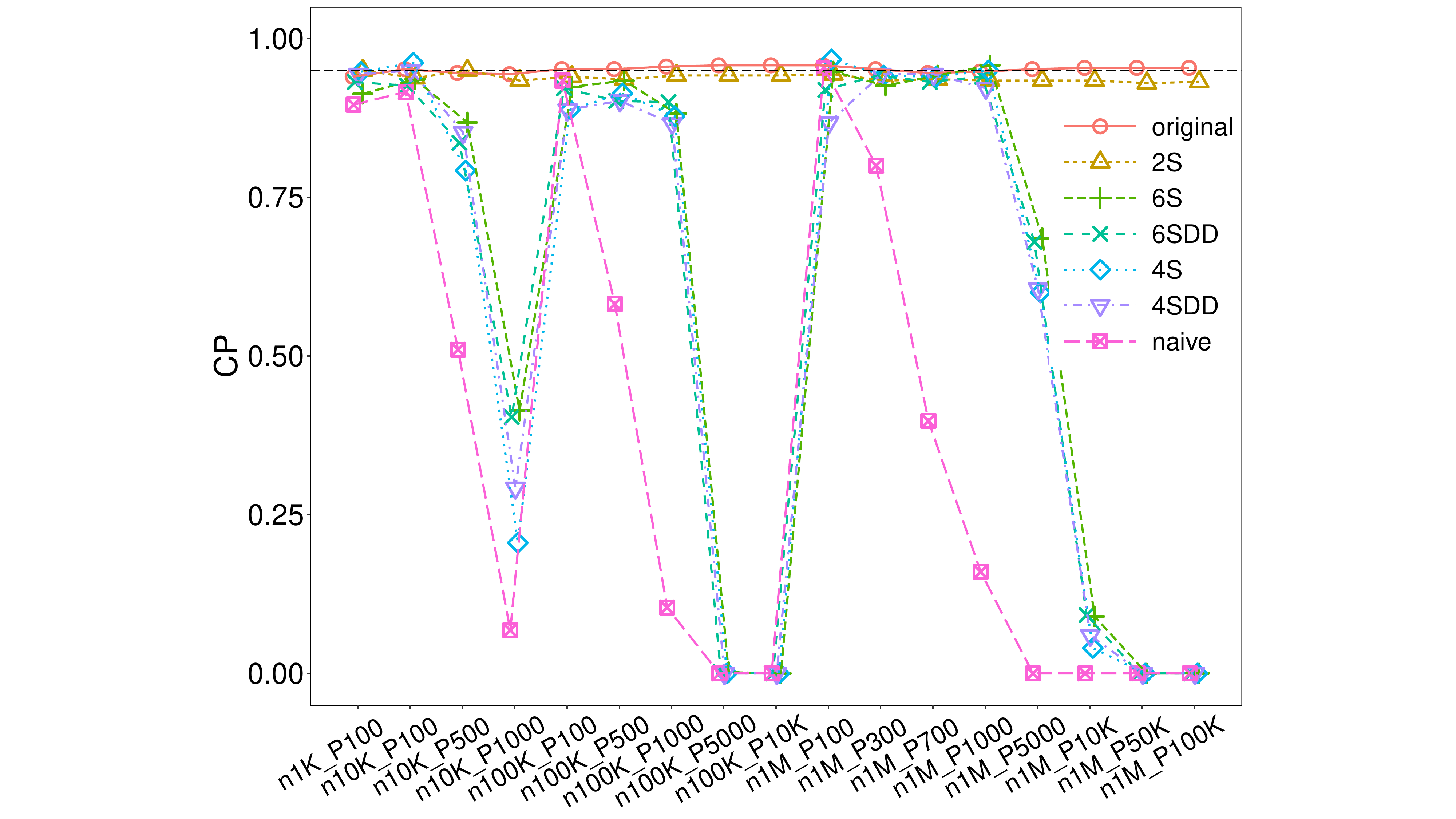}
\includegraphics[width=0.24\textwidth, trim={2.2in 0 2.2in 0},clip] {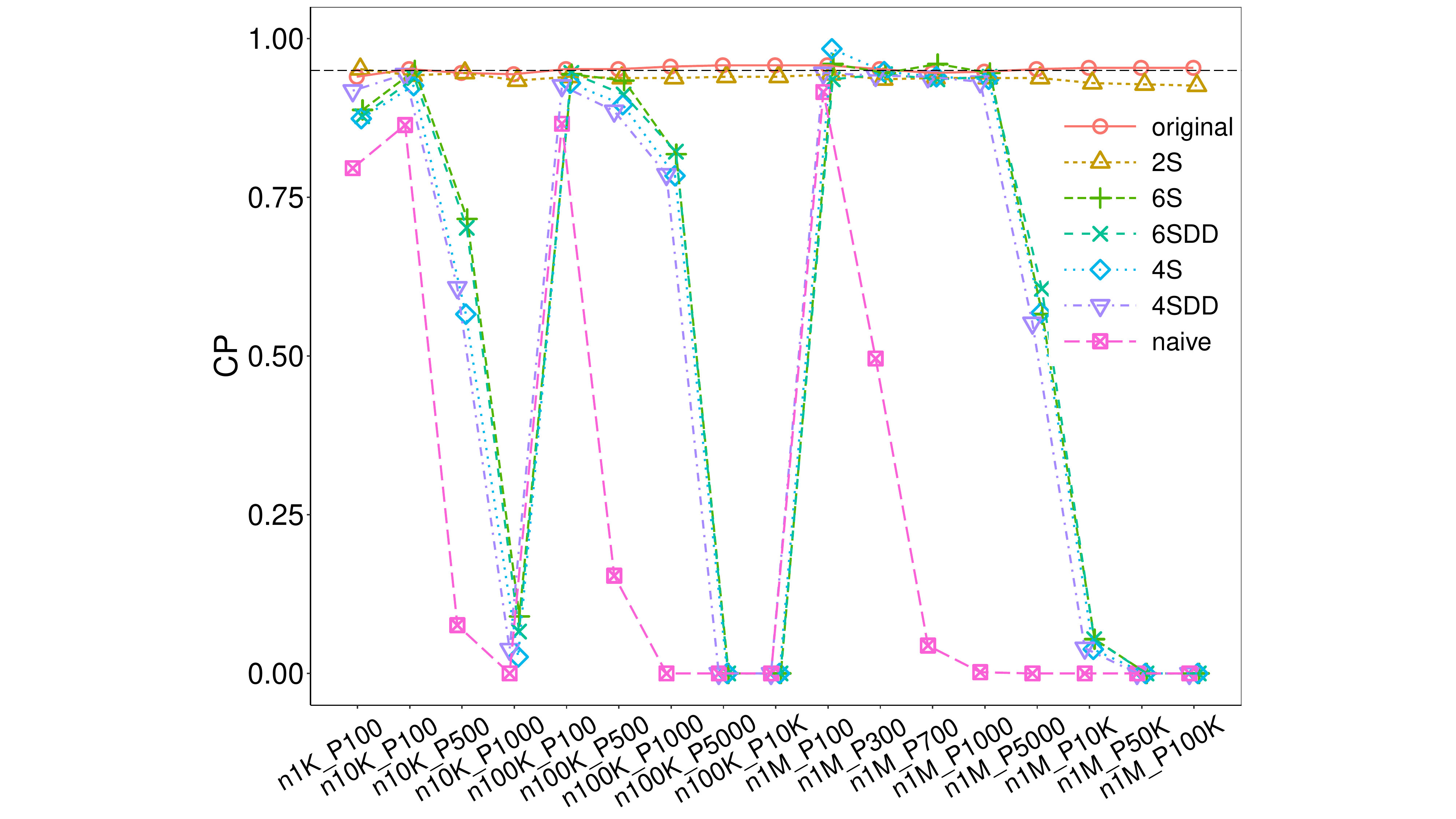}
\includegraphics[width=0.24\textwidth, trim={2.2in 0 2.2in 0},clip] {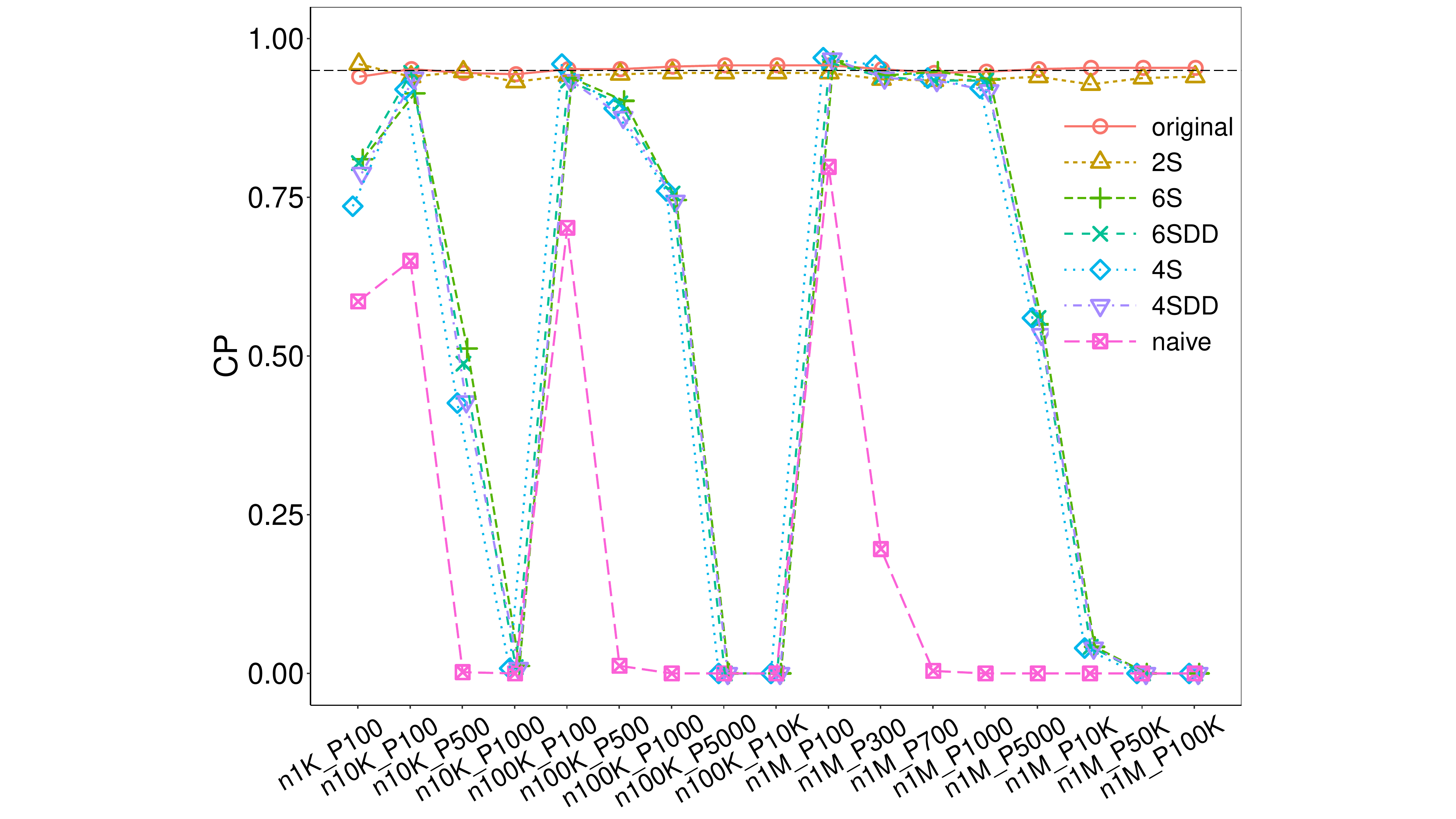}
\includegraphics[width=0.24\textwidth, trim={2.2in 0 2.2in 0},clip] {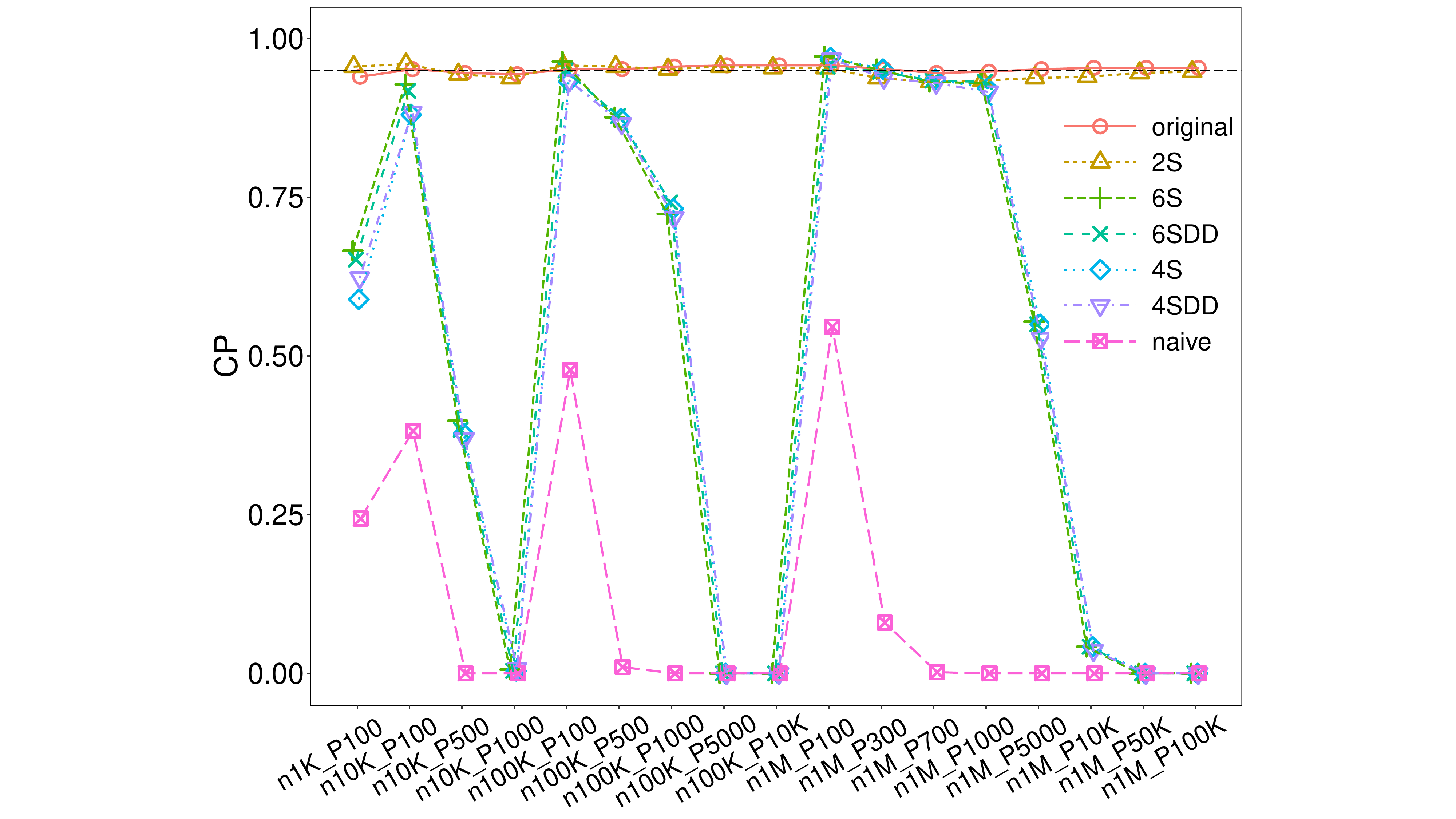}\\
\includegraphics[width=0.24\textwidth, trim={2.2in 0 2.2in 0},clip] {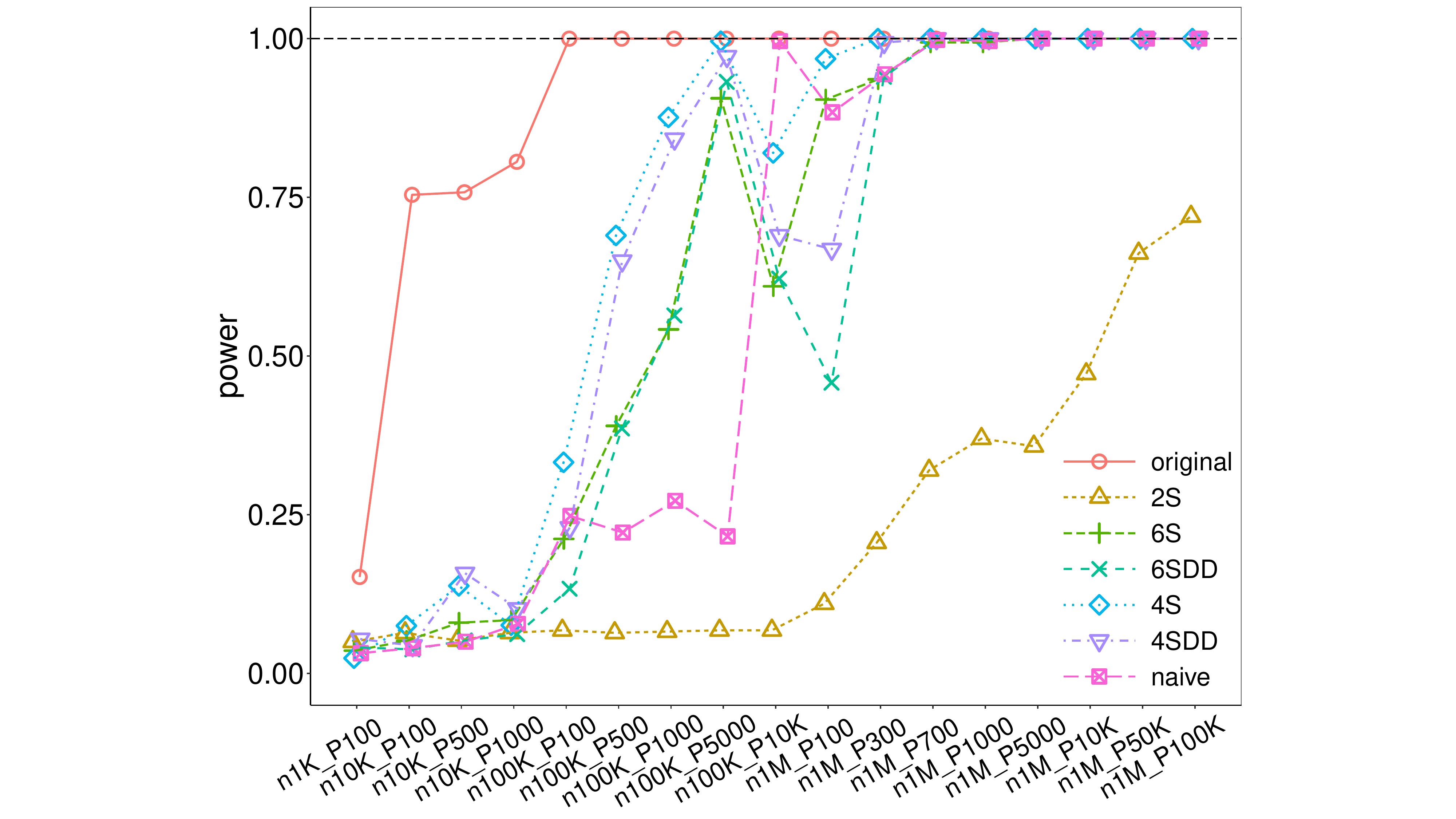}
\includegraphics[width=0.24\textwidth, trim={2.2in 0 2.2in 0},clip] {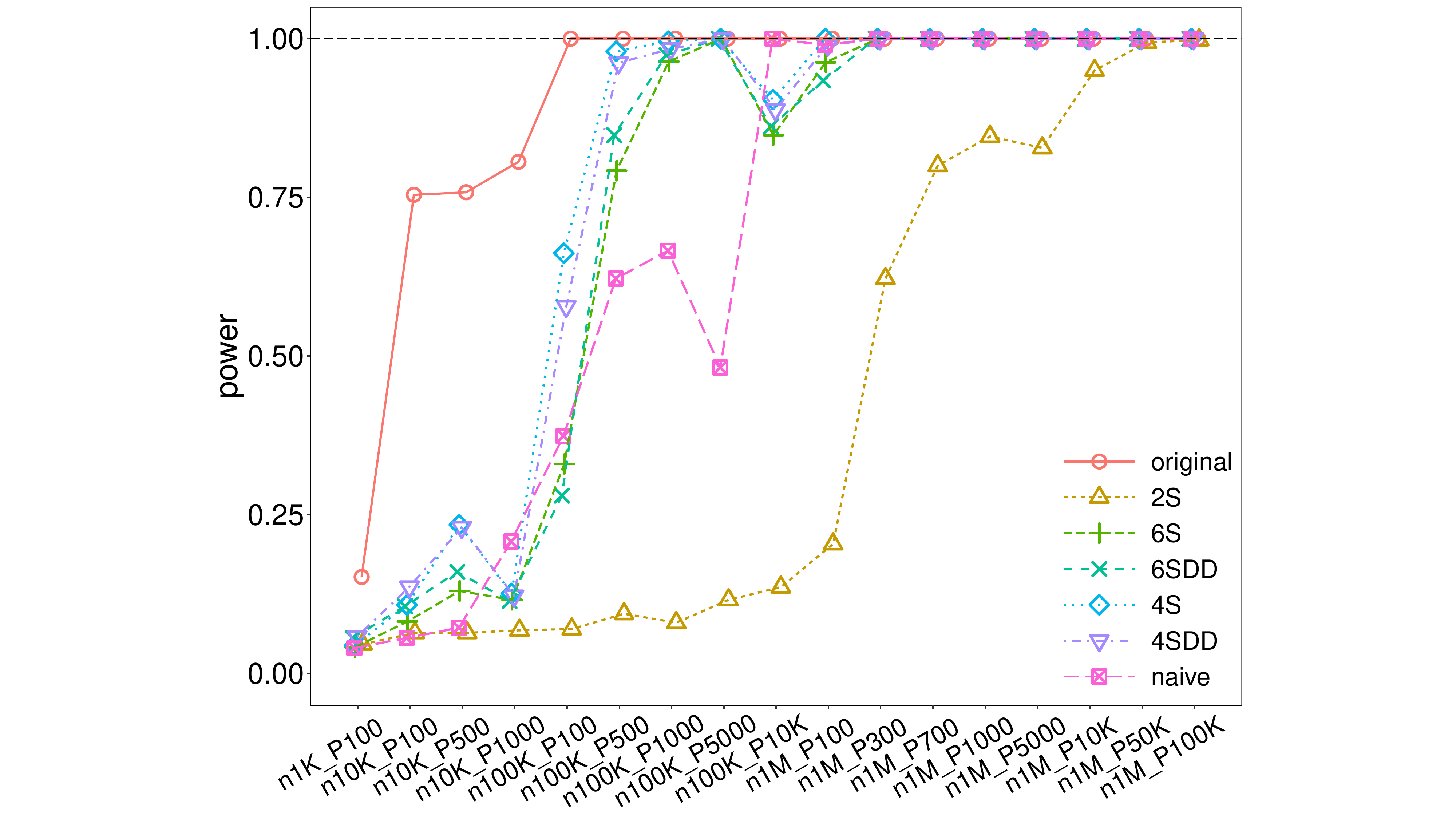}
\includegraphics[width=0.24\textwidth, trim={2.2in 0 2.2in 0},clip] {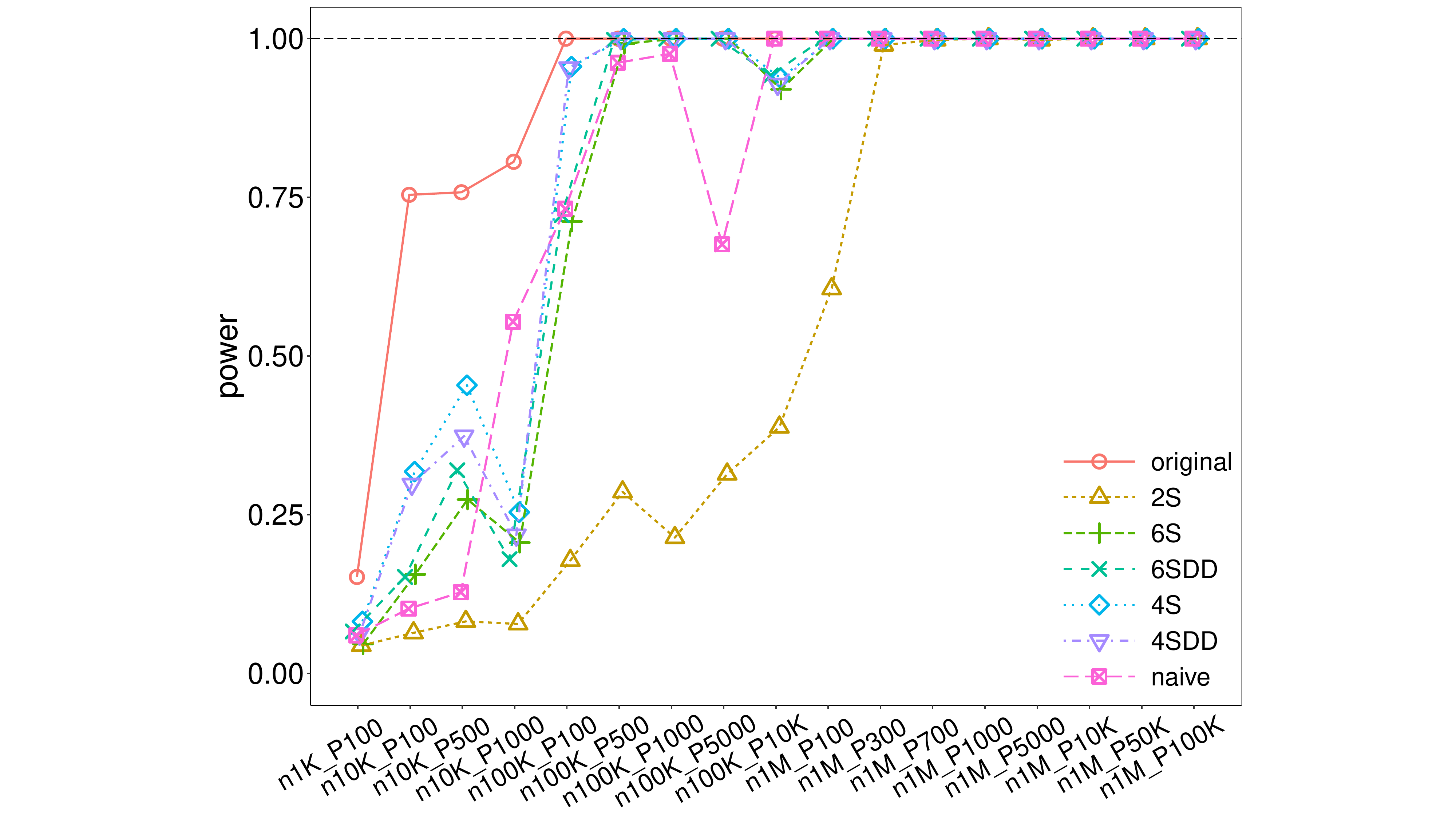}
\includegraphics[width=0.24\textwidth, trim={2.2in 0 2.2in 0},clip] {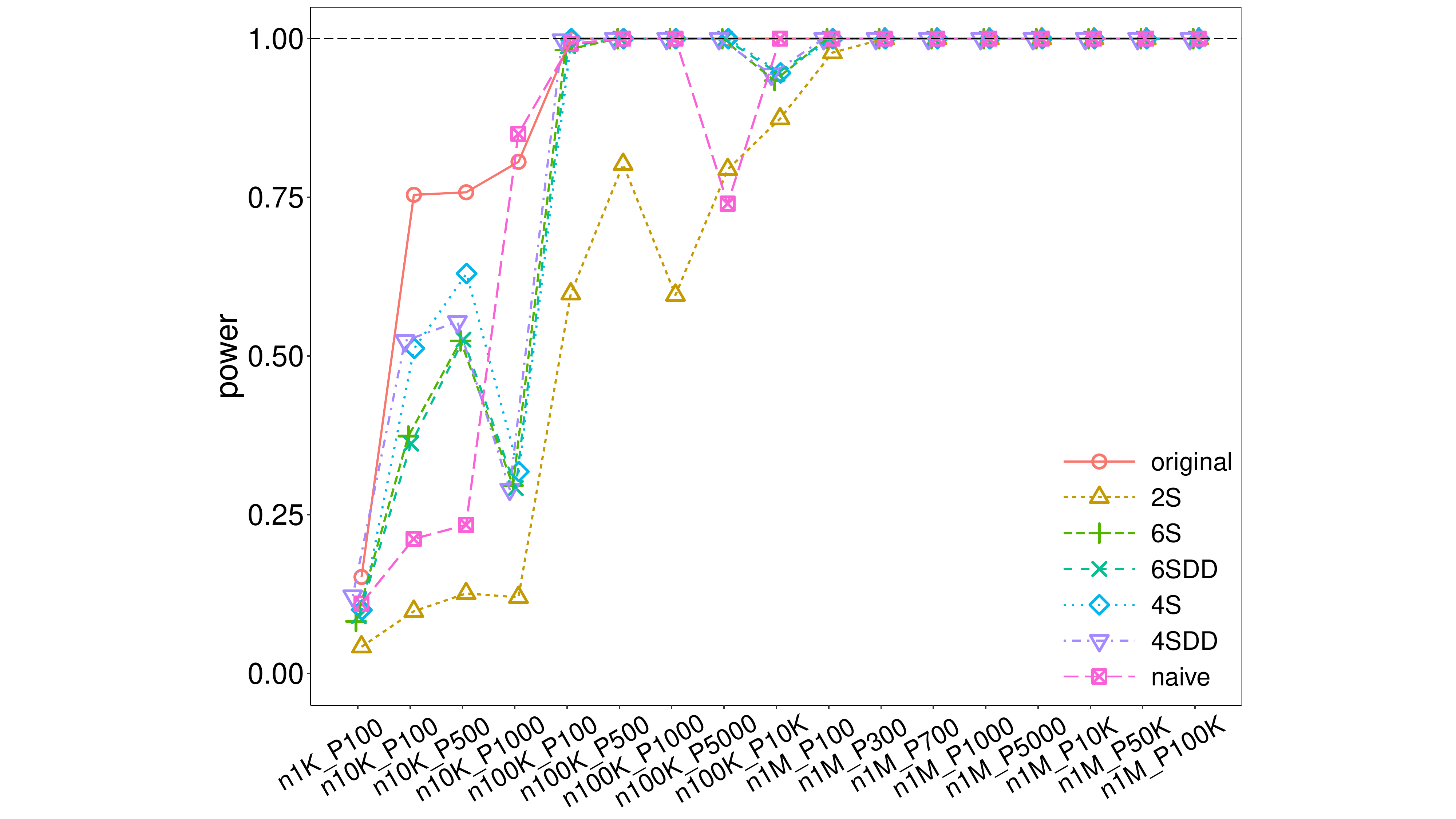}
\includegraphics[width=0.24\textwidth, trim={2.2in 0 2.2in 0},clip] {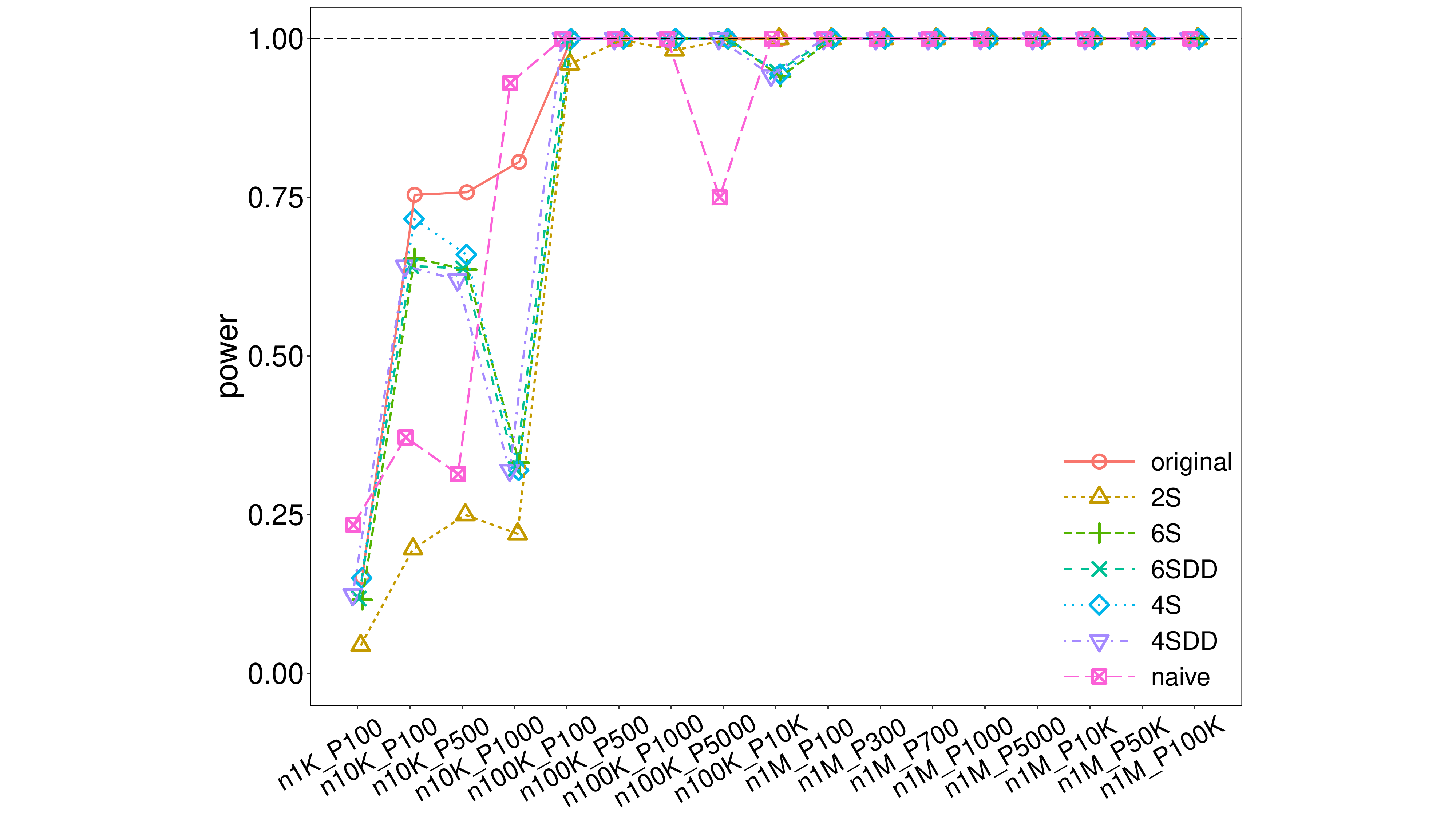}\\
\caption{ZILN data; $\rho$-zCDP; $\theta\ne0$ and $\alpha\ne\beta$}
\label{fig:1aszCDPziln}
\end{figure}
\end{landscape}

\end{document}